# Report on the Physics at the HL-LHC and Perspectives for the HE-LHC

Collection of notes from ATLAS and CMS

The ATLAS and CMS Collaborations

# Preface

The “Workshop on the Physics of HL-LHC and Perspectives at HE-LHC” [1], which took place between October 2017 and December 2018 at CERN, represented an LHC-wide effort of experimentalists and theorists with the aim to review and further refine the understanding of the physics potential of the High-Luminosity LHC (HL-LHC), and to prepare the exploitation of the HL-LHC data to the fullest possible extent. The workshop also provided an opportunity to begin a more systematic study of the physics at the High-Energy LHC, a possible new pp collider project in the LHC ring with a centre-of-mass energy of about 27 TeV.

The HL/HE-LHC workshop studies benefitted from the experience gained with the data analysis and physics simulation of the LHC Runs 1 and 2. The results extend and further refine previous studies produced for the Update of the European Strategy of Particle Physics in 2012–2013 [2], the ECFA HL-LHC workshops in 2013, 2014 and 2016 [3], as well as the Snowmass Workshop on the planning for the Future of U.S. Particle Physics in 2013 [4].

The workshop was organized in five working groups, on QCD, electroweak and top quark physics (WG1), Higgs boson and electroweak symmetry breaking (WG2), Beyond the Standard-Model physics (WG3), flavour physics (WG4), and high-density QCD physics (WG5). The reports from the five working groups are available on the arXiv [5]. The most important results were summarized in two ten-page documents, submitted to the European Strategy Group in December 2018 [6].

This book collects the original notes from the ATLAS and CMS Collaborations, used as input to the workshop and to the reports of the working groups [5].

The ATLAS and CMS Collaborations

# Contents

# Detector Performance

# Expected performance of the ATLAS detector at the High-Luminosity LHC

The ATLAS Collaboration

The High-Luminosity LHC (HL-LHC) will deliver proton-proton collisions at a center-of-mass energy of $\sqrt{s} = 14$ TeV with a baseline instantaneous luminosity of $5 \cdot 10^{34}$ cm$^{-2}$s$^{-1}$ and an ultimate achievable instantaneous luminosity of $7.5 \cdot 10^{34}$ cm$^{-2}$s$^{-1}$. The ATLAS detector is being upgraded for the HL-LHC running conditions to support a broad physics program in the presence of significantly increased pileup and more than a decade of data-taking. A comprehensive campaign to understand the physics reach of the experiment at the HL-LHC and a possible higher energy LHC (HE-LHC) is underway. This note provides a reference for the ATLAS detector performance for the physics projections that are included in the HL/HE-LHC Yellow Report, including documenting the assumptions made regarding the reconstruction and identification of physics objects and systematic uncertainties for the full anticipated dataset.

# 1 Introduction

The LHC physics program has just completed the Run 2 data-taking period and is heading, after a Phase-I upgrade, towards its Run 3 data-taking, as shown in Figure 1. A Phase-II upgrade is scheduled following Run 3 to further develop the LHC into the High-Luminosity LHC (HL-LHC). [1]

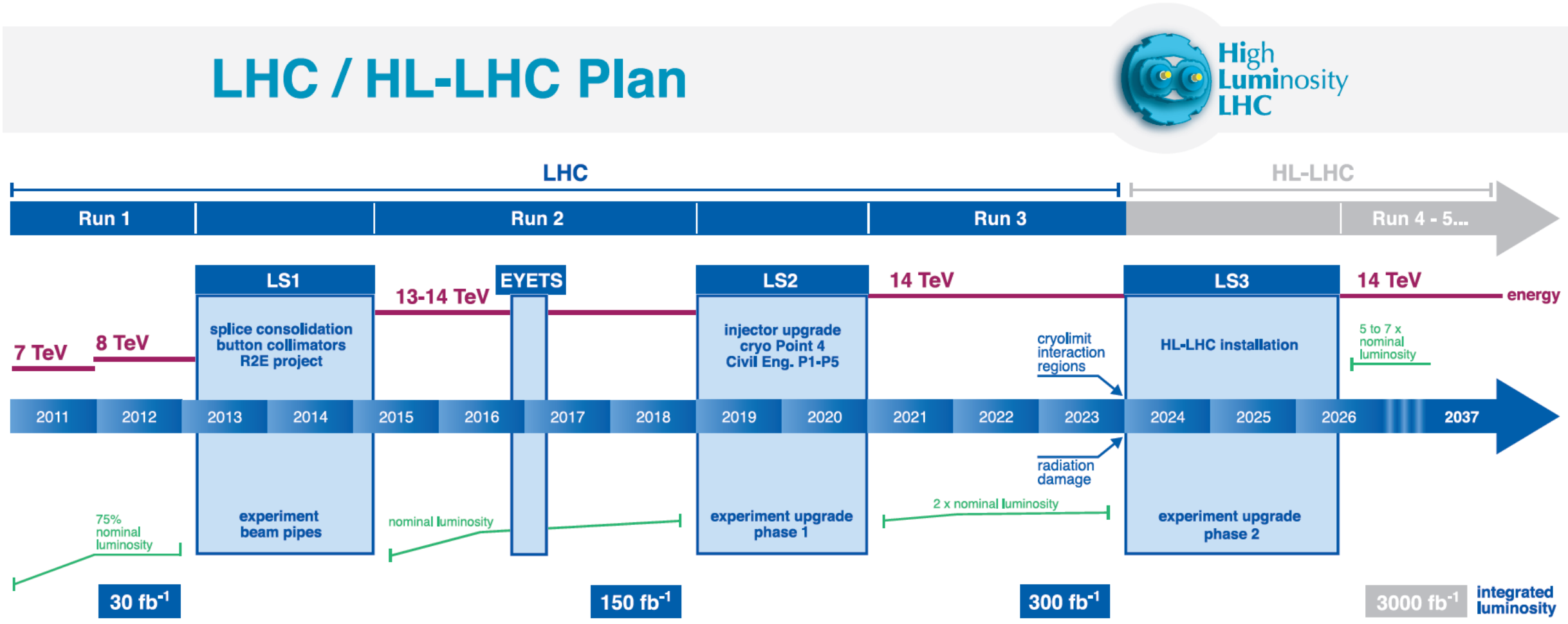

Figure 1: Timeline for the LHC accelerator operation and planned upgrades.

The upgraded HL-LHC will deliver proton-proton collisions at a center-of-mass energy of $\sqrt{s} = 14$ TeV with a baseline instantaneous luminosity of $5 \cdot 10^{34}$ cm$^{-2}$s$^{-1}$ and an ultimate achievable instantaneous luminosity of $7.5 \cdot 10^{34}$ cm$^{-2}$s$^{-1}$. This will potentially increase the average pileup $\langle\mu\rangle$, or the number of collisions per bunch crossing, to approximately 200. The HL-LHC will enable the ATLAS experiment to increase the collected integrated luminosity by approximately an order of magnitude throughout its operation, reaching an integrated luminosity of about 3000 fb$^{-1}$ (4000 fb$^{-1}$ in the "ultimate" scenario). This dataset holds tremendous potential for advances to precision measurements of Standard Model (SM) processes, with particular emphasis on probing the Higgs and electroweak sectors, and to searches for physics Beyond the Standard Model (BSM). Realizing this potential requires upgrades to the ATLAS experiment in the form of the Phase-II upgrade [2, 3] to enable sufficient performance in the face of the more challenging experimental conditions expected at the HL-LHC while also increasing radiation hardness and replacing aging detector components.

The planned ATLAS upgrades have been driven by the physics goals of the collaboration to optimize physics output. A comprehensive campaign to understand the physics reach of the experiment in the face of HL-LHC conditions is underway. The work began with the design of the detector upgrades, and a significant amount of performance projections can therefore be found in the various Technical Design Reports that the ATLAS Collaboration has produced to document the design and performance of upgraded components to the detector. Expected performance estimates of both the HL-LHC and, further, the hypothetical High-Energy LHC (HE-LHC) upgrade with an assumed center-of-mass energy of $\sqrt{s} = 27$ TeV and a total integrated luminosity of 15 ab$^{-1}$ [4] comprise the CERN HL/HE-LHC Yellow Report. While the HL-LHC was the focus of the ATLAS studies, a limited set of projections include an estimate for the HE-LHC.

The purpose of this note is to provide a reference for ATLAS physics projections that are included in

2

this Yellow Report, documenting the performance assumptions made regarding the reconstruction and identification of physics objects and systematic uncertainties.

The organization of this document is as follows:

- Section 2 includes brief descriptions of the upgrades outlined in the ATLAS Technical Design Reports that correspond to the ATLAS detector at the HL-LHC.
- Section 3 includes a description of the various strategies that are used for physics projections and the strategy employed for estimating systematic uncertainties.
- Section 4 describes the expected performance for ATLAS at the HL-LHC for various physics objects and event-level quantities that are used in analysis projections, including the assumed systematic uncertainties.

# 2 ATLAS detector upgrades

## 2.1 Inner Tracker

The ATLAS Inner Tracker will be completely replaced for Phase-II operations to provide excellent tracking in the face of the high-pile environment expected at the HL-LHC. The new silicon-only design (ITk) will achieve improved momentum resolution for reconstructed tracks and extend the $|\eta|$ coverage from $|\eta| < 2.5$ to $|\eta| < 4.0$ with a lower material budget than in Run 2. A silicon pixel detector composed of 5 barrel layers will be placed closest to the beamline. A silicon strip detector with 4 barrel layers will extend tracking out to higher radii. A series of rings will extend coverage to the forward region. These upgrades are described in detail in the Pixel Detector Technical Design Report [5] and the Silicon Strip Detector Technical Design Report [6]. The inner tracker layout named "Inclined Duals" was the baseline for the Pixel Technical Design Report and is widely used for performance studies presented in this note. The pixel pitch size was set to $50 \times 50\ \mu\text{m}^2$.

## 2.2 Calorimeters

The ATLAS Liquid Argon (LAr) Calorimeter will have entirely new frontend and readout electronics optimized to withstand radiation conditions for the duration of Phase-II running. The electronics architecture is designed to output full-granularity digitized signals at 40 MHz. These upgrades will combat Phase-II conditions with active pileup correction techniques using nearby bunch crossings to maintain an excellent energy resolution over a wide dynamic range. These upgrades are described in detail in the Liquid Argon Calorimeter Technical Design Report [7].

The ATLAS Tile Calorimeter will use new frontend and readout electronics, power supplies, and optical link interface boards to withstand increased radiation conditions for the duration of Phase-II running. These upgrades are described in detail in the Tile Calorimeter Technical Design Report [8].

### 2.3 Muon Spectrometer

A large fraction of the ATLAS Muon Spectrometer frontend and on- and off-detector readout and trigger electronics will be replaced to enable higher trigger rates and longer latencies. Additional muon chambers will be installed to maintain muon identification and reconstruction performance, increase trigger acceptance, and suppress the rate of random coincidences. The possibility to extend the muon acceptance to $|\eta| < 4$ is still under study (high-$\eta$ tagger), although most performance results presented to date for HL-LHC studies do not yet take possible improvements from this extension into account in their projections. These upgrades are described in detail in the Muon Spectrometer Technical Design Report [9].

### 2.4 Trigger & Data Acquisition

The detector upgrades present new requirements and new opportunities for the Trigger and Data Acquisition (TDAQ) systems. ATLAS will use a two-level TDAQ design as a baseline; a 'Level-0' hardware trigger leads to detector readout of 1 MHz for luminosities up to $7.5 \cdot 10^{34}$ cm$^{-2}$s$^{-1}$ and a processing farm, the 'Event filter' (EF), reduces the output data rate to 10 kHz. The design supports an evolved architecture with track-based triggers running at 4 MHz. The HL-LHC TDAQ system is described in detail in the TDAQ Technical Design Report [10].

The hardware trigger system is largely redesigned and allows for higher data granularity and enhanced flexibility beyond what will be afforded during the Run 3 data taking. Increased tracking functionality allows single object trigger thresholds to be kept low and assists pileup mitigation for the very challenging hadronic signatures at the HL-LHC. The baseline TDAQ architecture includes a Hardware Tracker (HTT) sitting in parallel to the processing farm in the EF; the HTT provides the EF with tracks that would not have been reconstructible otherwise due to the required computing resources. The HTT works in two modes: one reconstructs tracks in regions of interest and one performs full-event tracking. Both implementations use the same hardware that is customised according to the needs of each.

### 2.5 High-Granularity Timing Detector

The ATLAS High-Granularity Timing Detector (HGTD), which will precisely measure the timings of charged particles, will be installed covering $2.4 < |\eta| < 4.0$ in front of the LAr calorimeter to reduce background from pileup jets, as the increased pileup expected in high-luminosity running will require additional mitigation strategies. A timing resolution of 30 ps for minimum-ionizing particles is expected. These upgrades are described in detail in the HGTD Technical Proposal [11]. Most performance results presented to date for HL-LHC studies do not yet take possible improvements from the HGTD into account in their projections.

## 3 Projection strategies

Different approaches have been used to assess the sensitivity of the ATLAS experiment at the HL-LHC and HE-LHC. For some of the projections, a mix of the approaches described below is used, in order to deliver the most realistic result. The total integrated luminosity for the HL-LHC dataset is assumed to be 3000 fb$^{-1}$ at a center-of-mass energy of $\sqrt{s}$ = 14 TeV. For HE-LHC studies the same expected

4

detector performance is assumed as the Phase-II ATLAS detector, but in a hypothetical accelerator with an assumed center-of-mass energy of $\sqrt{s}$ = 27 TeV and total integrated luminosity of 15 ab$^{-1}$.

The effect of systematic uncertainties is taken into account based on the studies performed for the Run 2 analyses and using common guidelines for projecting the expected improvements that are foreseen owing to the large dataset and upgraded detectors, as described in Section 3.1.

**Detailed simulations** are used to assess the performance of reconstructed objects in the upgraded detectors and HL-LHC conditions, as described in Section 2. For some of the projections, such simulations are directly interfaced to different event generators, parton showering (PS) and hadronisation generators. Monte Carlo (MC) generated events are used for SM and BSM processes, and are employed in the various projections to estimate the expected contributions of each process.

**Extrapolations** rely on existing results with event statistics scaled to the HL-LHC luminosity to estimate the expected sensitivity. The increased center-of-mass energy and the performance of the upgraded detectors are taken into account for most of the extrapolations using scale factors on the individual processes contributing to the signal regions. Such scale factors are derived from the expected cross sections and from detailed simulation studies. This technique benefits from the full complexity of the existing analysis, which often includes data-driven background methods and has been optimized for performance. However, relying on current signal and control region selections, efficiencies, acceptances, object reconstruction and identification, etc. does not fully account for possible improvements and challenges expected with an upgraded detector and HL-LHC conditions.

**Parametric simulations** are used for some of the projections to allow a full re-optimization of the analysis selections that profit from the larger available datasets without requiring all samples to be simulated in HL-LHC conditions, which is computationally expensive. Particle-level definitions are used for electrons, photons, muons, taus, jets and missing transverse momentum. These are constructed from stable particles of the MC event record with a lifetime larger than $0.3 \cdot 10^{-10}$ s within the observable pseudorapidity range. Jets are reconstructed using the anti-$k_t$ algorithm [12] implemented in the FastJet [13] package, with a radius parameter of $R = 0.4$. All stable final-state particles are used to reconstruct jets except for the neutrinos, leptons and photons associated to $W$ or $Z$ boson or $\tau$ lepton decays. The effects of an upgraded ATLAS detector are taken into account by applying energy smearing, efficiencies and fake rates to generator-level quantities, following parameterisations based on detector performance studies with the detailed simulations. The effect of high pileup at the HL-LHC is incorporated by overlaying minimum-bias events with $\langle\mu\rangle = 200$ onto the hard-scatter events. Jets from pileup are then randomly selected as jets to be considered for analysis.

### 3.1 Systematic uncertainties

It is a significant challenge to predict the expected systematic uncertainties of physics results at the end of HL-LHC running. In almost all cases it would be pessimistic to assume a similar performance as seen in Run 2 given the very large increase in integrated luminosity, resulting in vastly larger data samples of the control processes used to measure the energy scales, resolutions and efficiencies of the different physics objects. In addition, it is reasonable to anticipate improvements to techniques of determining systematic uncertainties over an additional decade of data-taking. To estimate the expected performance, experts in the various physics objects and detector systems have studied current limitations to systematic uncertainties in detail to determine which contributions are limited by statistics and where there are more fundamental limitations. Predictions were made taking into account the increased integrated luminosity and expected potential gains in technique. These recommendations, often referred to in projections as the "baseline", were then harmonized with CMS to take advantage of a wider array of expert opinions and to allow the experiments to make sensitivity predictions on equal footing. In some cases there were additional sets of assumptions explored, referred to as "optimistic" scenarios, to reflect particular potential improvements that could be foreseen. The expected systematic uncertainties are reported along with object performance in the following sections.

Several general principles were defined for assessing the expected statistical, theoretical, and experimental uncertainties:

- Uncertainties due to statistics of available Monte Carlo simulation are set to zero for projections. As with other sources of uncertainty, the level of available Monte Carlo statistics in 2035 is difficult to predict. A clearer understanding of the fundamental potential of ATLAS in the HL-LHC can be found by de-coupling this potential source of uncertainty. In some cases, where experience from current running has shown the level of Monte Carlo simulation statistics to be a significant concern, a comparison is done between the “baseline” scenario with zero uncertainty, and a scenario assuming an effective Monte Carlo luminosity (number of events) equal to 1.5 times what will be available for the data.
- The intrinsic statistical uncertainty in measurements for extrapolated analyses scales with $1/\sqrt{\mathrm{L}}$, where L is the projection’s integrated luminosity divided by that of the reference Run 2 analysis.
- If predictions from theory do not change from current precision, systematic uncertainties from modeling would dominate for many of the HL-LHC projections. In some cases theorists have provided a detailed description of expected performance, such as for parton distribution functions. In other cases, analyses are performed making simple assumptions, with a default decision to divide the theory uncertainties, both inclusive cross-sections as well as modeling uncertainties, by a factor of two. Results are shown with theory and experimental systematic uncertainties defined so that the impact of the decisions can be clearly seen.
- Systematics driven by intrinsic detector limitations are left unchanged, or revised according to detailed simulation studies of the upgraded detector.
- Uncertainties on methods, as for instance non-statistical uncertainties on data-driven techniques, are kept at the same value as in the latest public results available, assuming that the harsher HL-LHC conditions will be compensated for by improved techniques for evaluating systematic uncertainties.
- In the case where a parametric simulation is done, only the leading sources of systematic uncertainties are often considered. For the extrapolations based on Run 2 analyses, a more complete set of nuisance parameters is available, though, again, a focus is placed on the largest sources of uncertainty.

#### 3.1.1 Parton distribution functions

For analyses where an accurate knowledge of the proton Parton Distribution Functions (PDFs) makes a significant difference in sensitivity, scale factors are used to estimate the expected HL-LHC PDF uncertainties achievable by the end of the HL-LHC physics program. The projected PDFs have been estimated from assumptions on the measurement uncertainties achievable after HL-LHC on key SM processes and re-evaluating the resulting PDFs. A set of PDFs with reduced uncertainties as well as a set of scale factors to apply as a ratio of the current uncertainties (PDF4LHC15) versus expected uncertainties are provided in Ref. [14]. There are two scenarios given: the conservative scenario assumes that there will be no reduction in the experimental systematic errors and the optimistic scenario assumes a reduction by a factor of 2.5 in the experimental systematic errors. The obtained scale factors are reported in Table 1.

Table 1: Expected scale factors for PDF uncertainties are given for two scenarios. The conservative scenario assumes that no improvements will be achieved in experimental systematic errors, and the optimistic scenario (in parentheses) assumes a reduction in experimental systematic errors by a factor of 2.5. [14]

| PDFs: HL-LHC/Current | 10 GeV < $M_X$ < 40 GeV | 40 GeV < $M_X$ < 1 TeV | 1 TeV < $M_X$ < 6 TeV |
|---|---|---|---|
| gluon-gluon luminosity | 0.58 (0.49) | 0.41 (0.29) | 0.38 (0.24) |
| quark-gluon luminosity | 0.71 (0.65) | 0.49 (0.42) | 0.39 (0.29) |
| quark-quark luminosity | 0.78 (0.73) | 0.46 (0.37) | 0.60 (0.45) |
| quark-antiquark luminosity | 0.73 (0.70) | 0.40 (0.30) | 0.61 (0.50) |
| up-strange luminosity | 0.73 (0.67) | 0.38 (0.27) | 0.42 (0.38) |

# 4 Expected performance

## 4.1 Luminosity

The peak instantaneous luminosity for the HL-LHC dataset is expected to be $\approx 5 \times 10^{34}$ cm$^{-2}$s$^{-1}$, with a corresponding average of approximately 140 interactions per bunch crossing [1]. The HL-LHC is expected to produce a total integrated luminosity of 250 fb$^{-1}$ per year and 3000 fb$^{-1}$ in its 12-year lifetime [1]. An ultimate instantaneous luminosity of $\approx 7.5 \times 10^{34}$ cm$^{-2}$s$^{-1}$, corresponding to approximately 200 interactions per bunch-crossing, is foreseen as ultimately achievable, which makes this higher level of pileup the appropriate target for the ATLAS upgrades.

Physics analyses would benefit from an uncertainty on the full dataset integrated luminosity as low as 1–1.5%. An ambitious goal of 1% has been assumed in the physics studies for the Yellow Report, compared with about 2% typically achieved at Run 1 or Run 2. This target uncertainty is extremely challenging, taking into account the more difficult experimental conditions (particularly the average pileup $\langle\mu\rangle$ of 200) expected at HL-LHC. It will be pursued profiting from the experience from previous runs, hardware upgrades to the Beam Conditions Monitor (BCM) and Luminosity Cherenkov Integrating Detector (LUCID), and the new HGTD. The new BCM will be mounted on a ring within the pixel detector of the ITk and will have smaller sensor pads to accommodate higher occupancy levels at $\langle\mu\rangle \approx 200$. The LUCID-3 detector is foreseen to use quartz fibre bundles in place of quartz counters. The HGTD will have a bunch-by-bunch luminosity capability, and should have excellent linearity owing to the relatively low occupancy. In addition to these detectors, the LAr and Tile calorimeters, and measurements based on track-counting and reconstructed Z-boson counting, will be used to monitor the long-term stability of the various luminosity measurements, and the linearity between the low-luminosity VdM scans used to establish the absolute calibration and the physics data-taking regime.

Table 2: Representative trigger menu for ATLAS operations at the HL-LHC. The offline $p_T$ thresholds indicate the momentum above which a typical analysis would use the data. Where multiple object triggers are described only one threshold is given if both objects are required to be at the same $p_T$; otherwise, each threshold is given with the two values separated by a comma. In the case of the $e - \mu$ trigger in Run 2, two sets of thresholds were used depending on running period, and both are listed. This table is a subset of Table 6.4 from the TDAQ TDR [10].

| Trigger Selection | Run 1 Offline $p_T$ Threshold [GeV] | Run 2 (2017) Offline $p_T$ Threshold [GeV] | Planned HL-LHC Offline $p_T$ Threshold [GeV] |
|---|---|---|---|
| isolated single $e$ | 25 | 27 | 22 |
| isolated single $\mu$ | 25 | 27 | 20 |
| single $\gamma$ | 120 | 145 | 120 |
| forward $e$ | | | 35 |
| di-$\gamma$ | 25 | 25 | 25 |
| di-$e$ | 15 | 18 | 10 |
| di-$\mu$ | 15 | 15 | 10 |
| $e - \mu$ | 17,6 | 8,25 / 18,15 | 10 |
| single $\tau$ | 100 | 170 | 150 |
| di-$\tau$ | 40,30 | 40,30 | 40,30 |
| single $b$-jet | 200 | 235 | 180 |
| single jet | 370 | 460 | 400 |
| large-$R$ jet | 470 | 500 | 300 |
| four-jet (w/ $b$-tags) | | 45(1-tag) | 65(2-tags) |
| four-jet | 85 | 125 | 100 |
| $H_T$ | 700 | 700 | 375 |
| $E_T^{miss}$ | 150 | 200 | 210 |
| VBF inclusive (di-jets) | | | 2x75 w/ ($\Delta\eta > 2.5$ & $\Delta\phi < 2.5$) |

## 4.2 Trigger

An initial baseline trigger menu (see Table 2) has been developed to enable a diverse physics program at the HL-LHC that supports precision measurements at the electroweak scale and a wide array of BSM searches. The menu includes reasonably low-momentum electrons and muons, coupled with a comprehensive set of hadronically-decaying tau lepton triggers, missing-transverse-momentum ($E_T^{miss}$) triggers, and jet triggers, including massive large radius (large-$R$) jet triggers, all built into a flexible menu with contingencies to allow for new ideas. Generally, trigger selections are planned to have thresholds similar to, or below, what we have in the current data taking with a notable exception being the multi-jet and $E_T^{miss}$ triggers, which become particularly challenging in high-pileup environments. In addition to the items listed in the menu, several dedicated selections have been explored. For example, most B-physics trigger signatures are based on 6 GeV dimuon triggers, as in Run 2, with additional mass and vertex selections.

There are a number of examples where significant improvements can be found in trigger performance due to the HL-LHC TDAQ upgrades. There are significant improvements in muon trigger efficiencies due to increased resistive plate chamber coverage, with single muon trigger efficiencies going from $\approx 70\%$ (Run 2) to $\approx 90\%$ (HL-LHC) for $|\eta| < 1.05$. The coverage maps are shown in Figure 2(a) and Figure 2(b) for Phase-I and the HL-LHC respectively. This change brings a particularly large benefit to analyses that rely on multi-muon triggers. Additional examples of improvements include multijet triggers and single electrons, which benefit from the architecture changes that include the introduction of a new global trigger in the first trigger level and/or the presence of the HTT.

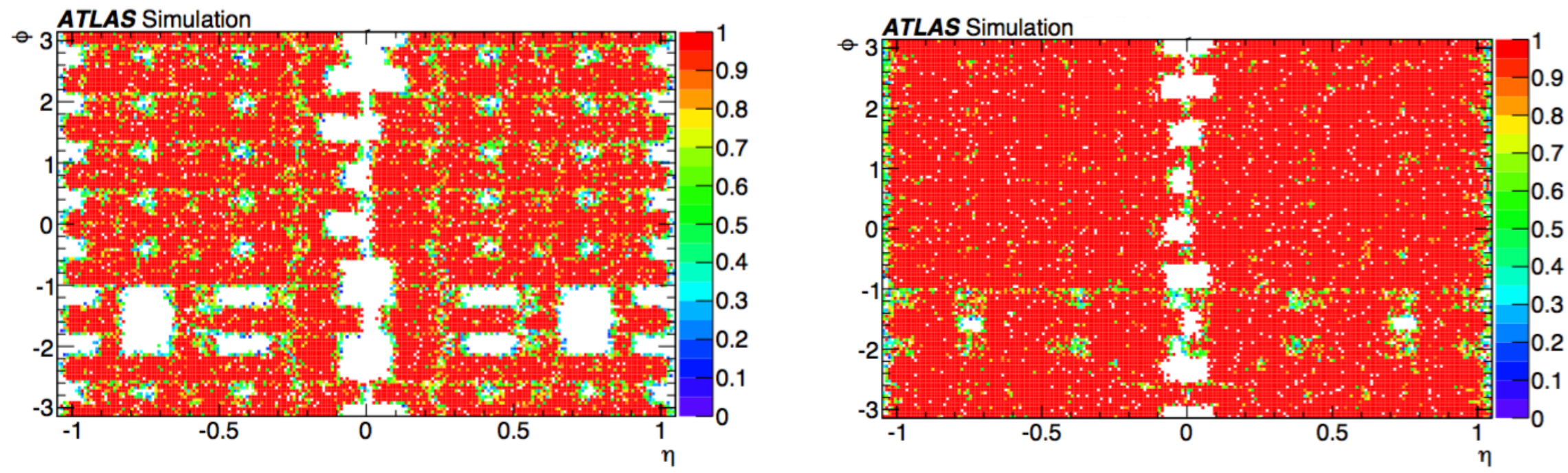

(a) (b)

Figure 2: Muon trigger coverage in the barrel region ($|\eta| < 1$) using (a) the Phase-I system in HL-LHC conditions and (b) the Phase-II system with resistive plate chambers operated with a two-station coincidence. Figures 6.5 (a,c) from the TDAQ TDR [10].

### 4.3 Track reconstruction

The tracking performance benefits from the entirely new all-silicon detector that will be installed for HL-LHC running. This detector extends the tracking range in $\eta$ from $|\eta| < 2.5$ in Run 2 to $|\eta| < 4.0$. The new tracker has a relatively low material budget and provides excellent tracking efficiency and resolution. The tracking efficiency for 10 GeV muons, pions and electrons is shown in Figure 3. Transverse momentum ($q/p_{\mathrm{T}}$) resolution and impact parameter ($d_0$) resolution for muons of representative transverse momentum ($p_{\mathrm{T}}$) values are shown in Figures 4(a) and 4(b) respectively.

Recent studies have shown that the material budget of the ITk detector was underestimated at the time of the TDR writing, so these results may be optimistic. A careful re-tuning and re-optimization is underway. Some analyses, such as lifetime measurements, analyses with a strong reliance on b-tagging, and other B-physics projections, are particularly sensitive to tracking and vertexing performance. In many of these cases the results have been evaluated with both the TDR-predicted performance and Run 2 performance in order to quantify the sensitivity to tracking and vertexing.

### 4.4 Electrons

Electron reconstruction and identification benefit from the expected excellent track reconstruction performance of the new inner tracker and its lower material budget as well as its extension to higher $|\eta|$. The identification requirements have been re-tuned for the new inner tracker and expected HL-LHC conditions and were studied in Ref. [7] and [5]. The $p_{\mathrm{T}}$-dependent electron reconstruction and identification efficiencies measured with the ITk for the three identification working points of loose, medium and tight are shown in Figure 5(a) for the central ($0 < |\eta| < 2.5$) region. The charge mis-identification probability for central electrons as a function of $\eta$ is shown in Figure 5(b), where the effect of a tight identification requirement and the Run 2 performance are also shown for comparison. Furthermore, the performance of an artificial neural network for forward electron identification is shown in Figures 6(a) ($Z \to ee$ efficiency) and 6(b) (truth jet fake rates for loose, medium, and tight working points).

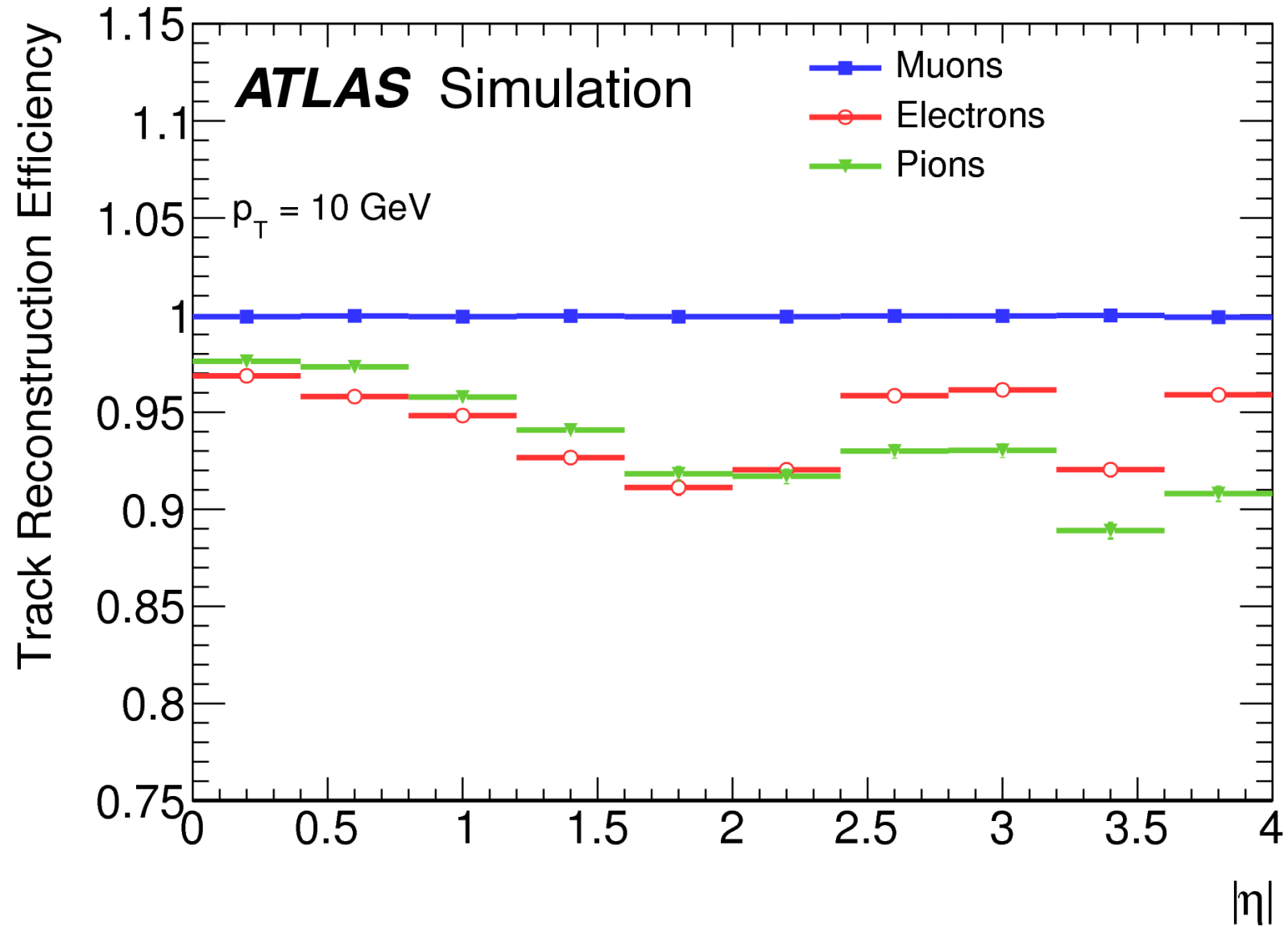

Figure 3: Track reconstruction efficiency for single muons, pions and electrons with a constant transverse momentum of $p_T$ = 10 GeV. Figure 3.3(a) from the Pixel Detector TDR [5].

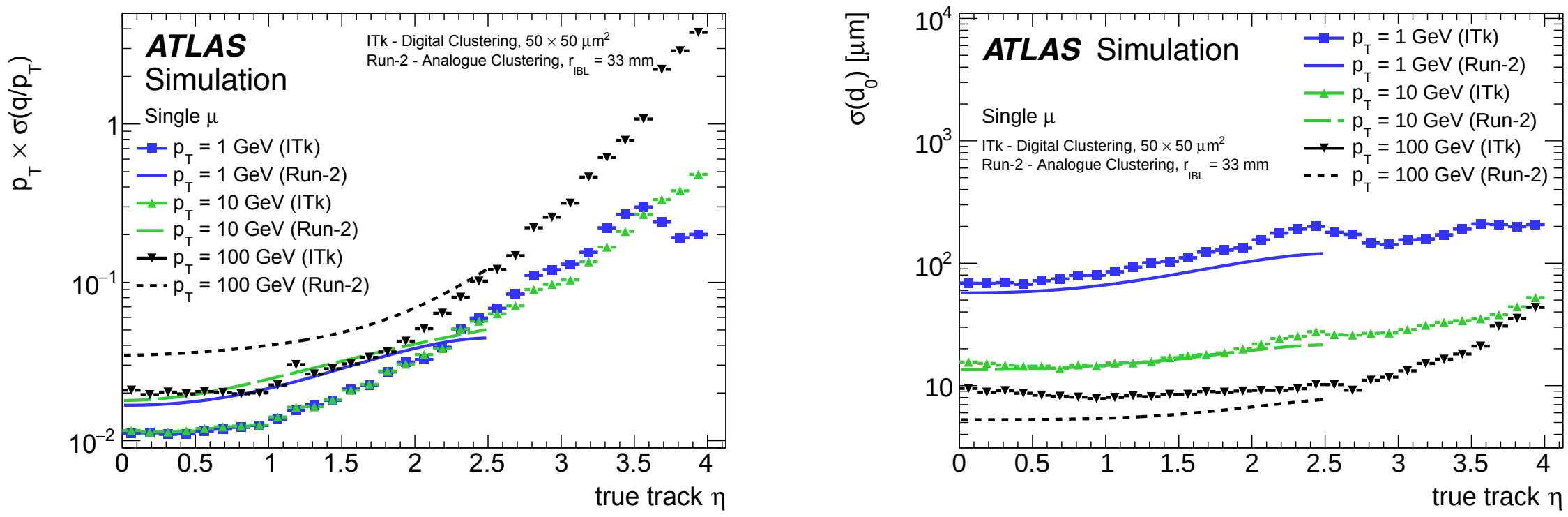

(a) (b)

Figure 4: (a) Track parameter resolution in $q/p_T$ as a function of $\eta$ for a single muon sample. Overlaid are the results for the current Run 2 detector. Figure 3.6(e) from the Pixel Detector TDR [5]. (b) $d_0$ resolution as a function of $\eta$ for a single muon sample. Overlaid are the results for the current Run 2 detector. Figure 3.6(a) from the Pixel Detector TDR [5].

The baseline systematic uncertainty assumption for electrons is that they will remain stable despite the harsher conditions of the HL-LHC, yielding to similar uncertainties as in Run 2. Uncertainties on isolation are expected to slightly decrease due to better understanding of the methods and detectors and yielding a

10

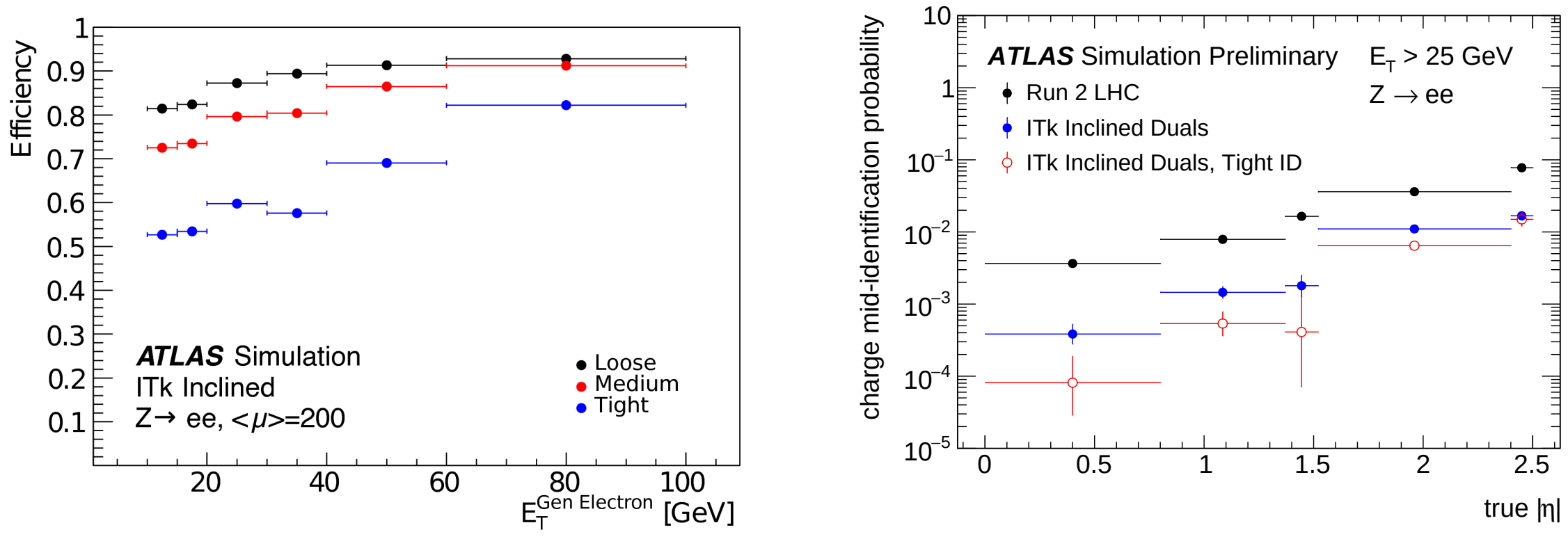

(a) (b)

Figure 5: (a) Electron efficiency for the various working points for a $Z \to ee$ simulated sample with $\langle \mu \rangle = 200$ in the region $|\eta| < 2.5$. Figure 3.26(b) from the Pixel Detector TDR [5]. (b) Electron charge mis-identification probability as a function of $|\eta|$.

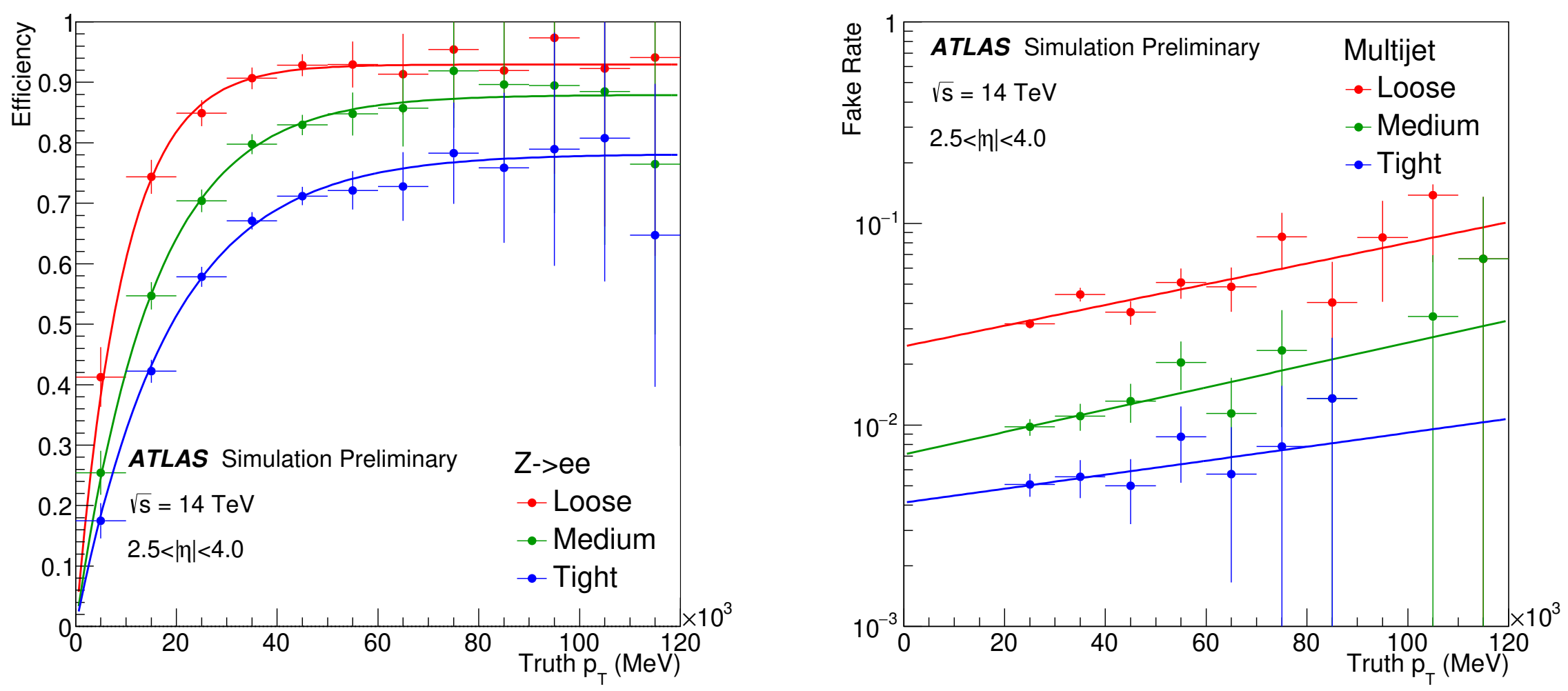

(a) (b)

Figure 6: Forward ($2.5 < |\eta| < 4.0$) electron identification neural network performance as a function of truth $p_T$: (a) efficiency of electrons from simulated $Z \to ee$ events and (b) fake rate of simulated jets.

lower uncertainty on the combination of the reconstruction, identification and isolation efficiency, from present values based on studies performed by CMS. Representative values of uncertainties are shown in Table 3.

Table 3: Representative values for systematic uncertainties for electrons at the HL-LHC. These uncertainties are consistent with the Run 2 uncertainties with the exception of the combination of the reconstruction, identification and isolation efficiency at high $p_T$ (above 200 GeV), where dedicated studies at CMS were used as an ATLAS approximation. [15]

| Electron Parameter | Range | Uncertainty |
|---|---|---|
| Energy scale | $p_T \approx 45$ GeV | 0.1% |
| | high $p_T$, up to 200 GeV | 0.3% |
| Reconstruction + Identification Efficiency (ID) | $p_T \approx 45$ GeV | 0.5% |
| Reconstruction + ID + Isolation Efficiency | $p_T > 200$ GeV | 2% |

### 4.5 Muons

Improvements to the inner tracker and increased coverage of the muon detectors result in a higher acceptance for combined muons and improved resolution for low-to-medium $p_T$ muons. The muon momentum resolution is shown in Figure 7(a) and the improvement to the invariant mass resolution for Higgs decays to two and four muons is shown in Figure 7(b). In parametric simulations, the impact of isolation was established by imposing isolation on the particles in the Monte Carlo "truth" record. Expected track-based isolation efficiencies for prompt and secondary muons are shown in 8(a) and 8(b) as a function of $|\eta|$ and $p_T$ of the muon, though the isolation used was not fully tuned for high-pileup so further improvements can be expected. The reconstruction efficiency is taken from single muon Monte Carlo simulated with Run 2 reconstruction algorithms [16] running on a geometry that includes the Phase-II ITk with the Run 2 muon spectrometer. The ITk's extended $\eta$ range allows the “combined muon” category, which matches a muon track or stub in the muon spectrometer to a track in the inner detector, to extend from the Run 2 value of $|\eta| < 2.5$ to $|\eta| < 2.7$.

Uncertainties in muon reconstruction, identification, isolation efficiency, momentum scale, and momentum resolution are very well under control already. It is expected that the same accuracy can be maintained for the large HL-LHC dataset. Systematic uncertainties on muon-related performance from Run 2, which are used for HL-LHC projections, are summarized in Table 4 within the $|\eta| < 2.5$ range.

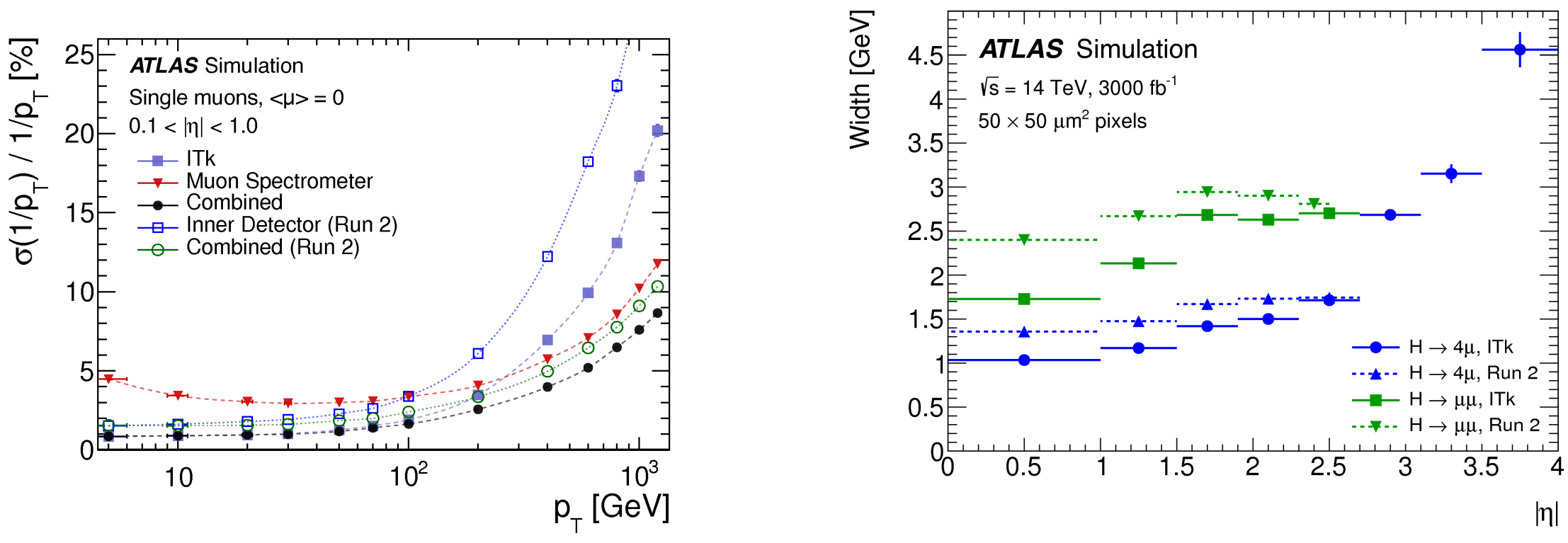

(a) (b)

Figure 7: (a) Combined muon momentum resolution and the individual contributions from the ITk and the upgraded Muon Spectrometer, with the Run 2 comparison included. Figure 3.26(a) from the Pixel Detector TDR [5]. (b) Di-muon (green) and four-muon (blue) mass resolution for Higgs decays to muons. Figure 3.31 from the Pixel Detector TDR [5].

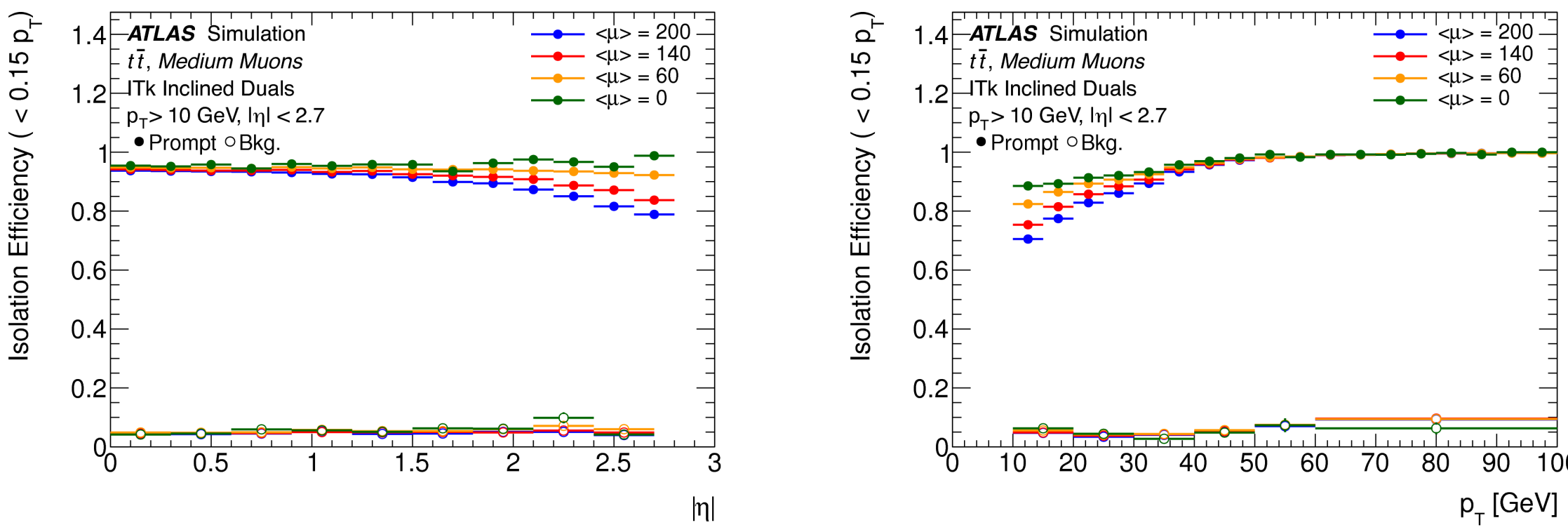

(a) (b)

Figure 8: Efficiency for a track-based isolation requirement that is $p_{\mathrm{T}}$-dependent for prompt muons and secondary backgrounds within $|\eta| < 2.7$ versus (a) $|\eta|$ and (b) $p_{\mathrm{T}}$. Figure 3.28 in the Pixel Detector TDR [5].

Table 4: Run 2 systematic uncertainties for muons, which are also assumed for ATLAS running at the HL-LHC. [15]

| Muon Parameter | Range | Run 2 Uncertainty |
|---|---|---|
| Reconstruction + Identification Efficiency | $p_{\mathrm{T}}$ < 200 GeV | 0.1% |
| | 200 GeV < $p_{\mathrm{T}}$ < 1 TeV | 2-20% |
| Resolution | $p_{\mathrm{T}}$ < 200 GeV | 5% |
| | 200 GeV < $p_{\mathrm{T}}$ < 1 TeV | 10-20% |
| Energy Scale | $p_{\mathrm{T}}$ < 200 GeV | 0.05% |
| Isolation Efficiency | All working points | 0.5% |

## 4.6 Taus

The reconstruction and identification of tau leptons that decay semi-hadronically ($\tau_{\text{had-vis}}$) benefits from the ITk detector, with its excellent tracking performance and extension to higher $\eta$ ranges. The identification algorithms have been re-optimized for the upgrade detector and studied in Ref. [5]. Since then, a more accurate assessment of the expected performance has been carried out, which has been used for the studies in the Yellow Report. The identification efficiency for 1-prong (one charged track) and 3-prong (multiple charged tracks) $\tau_{\text{had-vis}}$ candidates using simulated $Z \to \tau\tau$ events are shown in Figures 9(a) and 9(b) respectively for the loose, medium, and tight working points. The jet rejection is shown for both 1-prong and 3-prong $\tau_{\text{had-vis}}$ candidates at the various working points in Figures 10(a) and 10(b) respectively, as a function of efficiency for candidates above a $p_{\text{T}}$ of 20 GeV. Current optimizations show the rejection in the HL-LHC optimization out-performing Run 2 in all eta regions except the far-forward region where there is no Run 2 comparison point available.

The systematic uncertainties for $\tau_{\text{had-vis}}$ candidates have been estimated from Run 2 systematics by scaling down the sources of uncertainty that are driven by statistics, which will improve at the HL-LHC, and making educated assumptions about how the theory and modeling uncertainties are likely to change.

For analyses that are using truth-based projections, the uncertainty on the $\tau_{\text{had-vis}}$ identification efficiency is taken as 5%, where an optimistic scenario of 2.5% has also been defined. The energy scale uncertainty is conservatively assumed to be at the level of 2-3%.

For projections coming from current analyses, the following scale factors for adjusting the systematic uncertainties have been provided:

- The scale factor to apply to Run 2 systematic uncertainties on tau identification efficiency for 1-prong taus is 0.9 (0.45) in the baseline (optimistic) scenario.

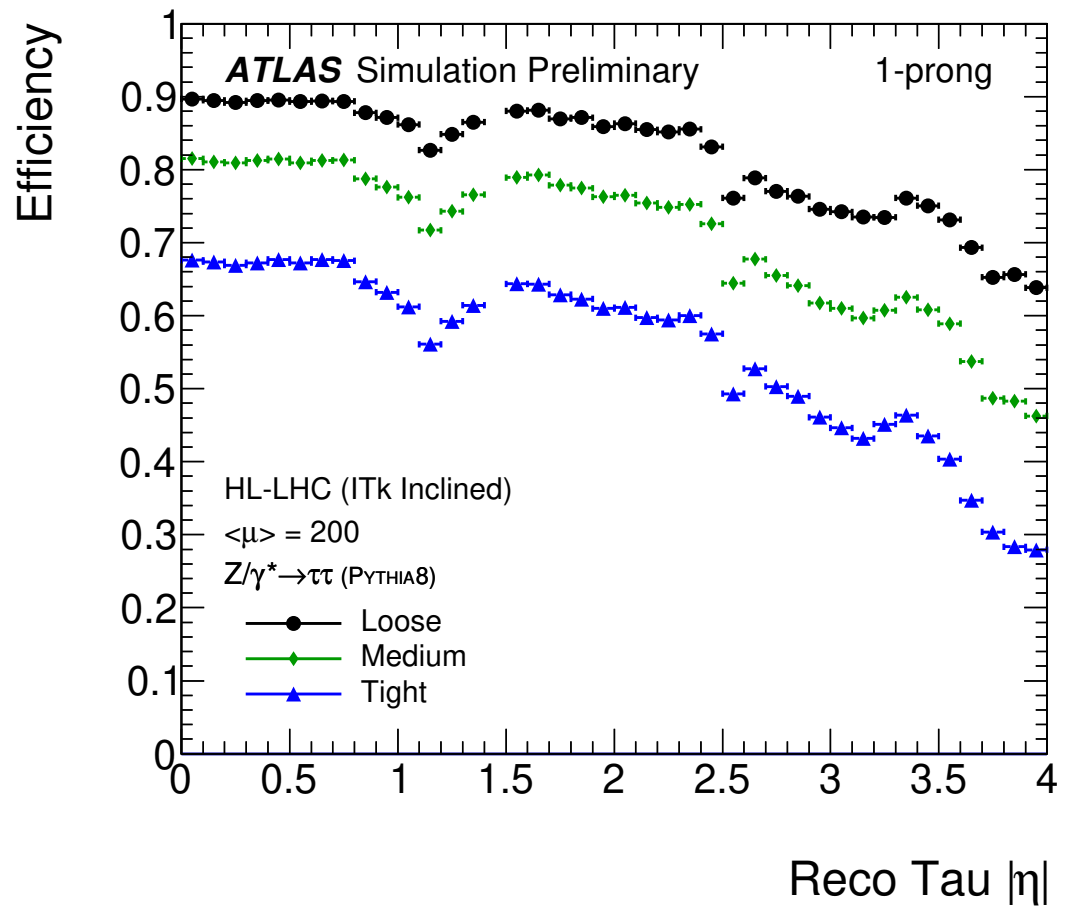

(a)

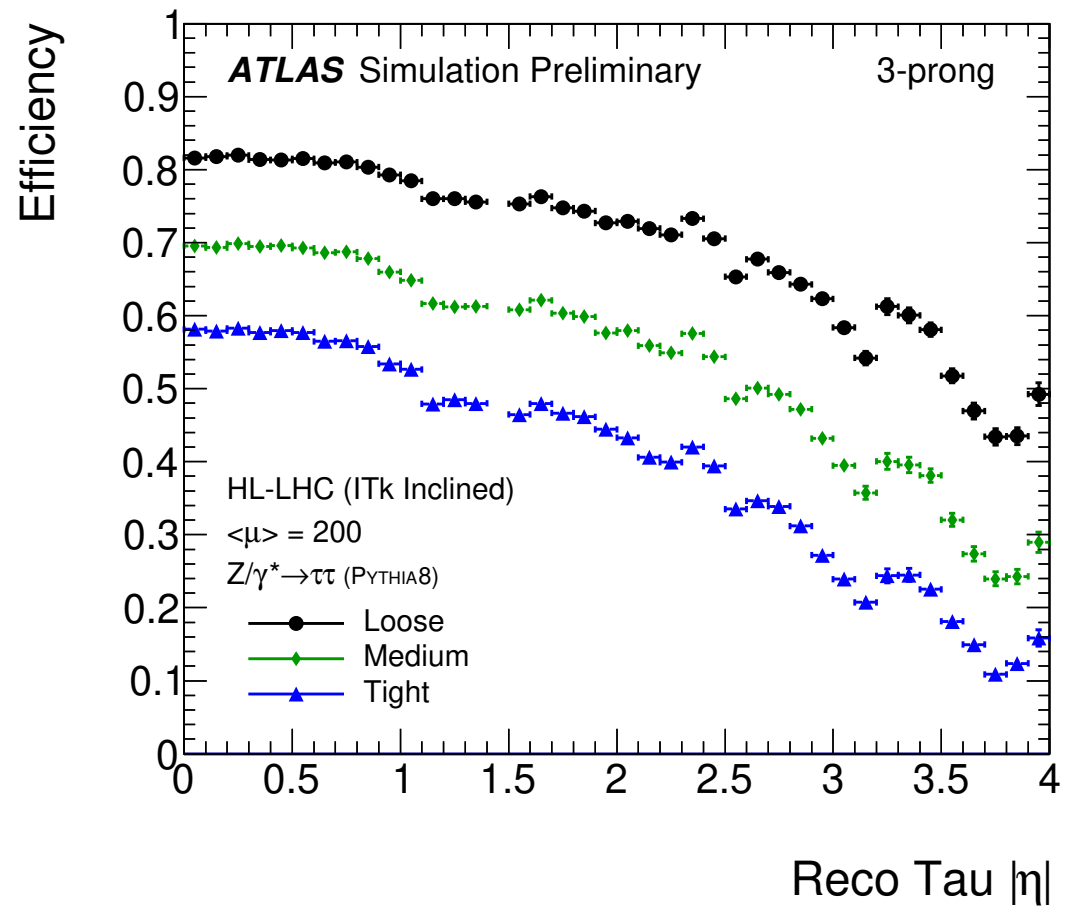

(b)

Figure 9: Tau identification efficiency for the three working points (Loose, Medium, and Tight) as a function of $\eta$ for reconstructed $\tau_{\text{had-vis}}$ candidates, shown for (a) one-prong and (b) three-prong tau leptons.

14

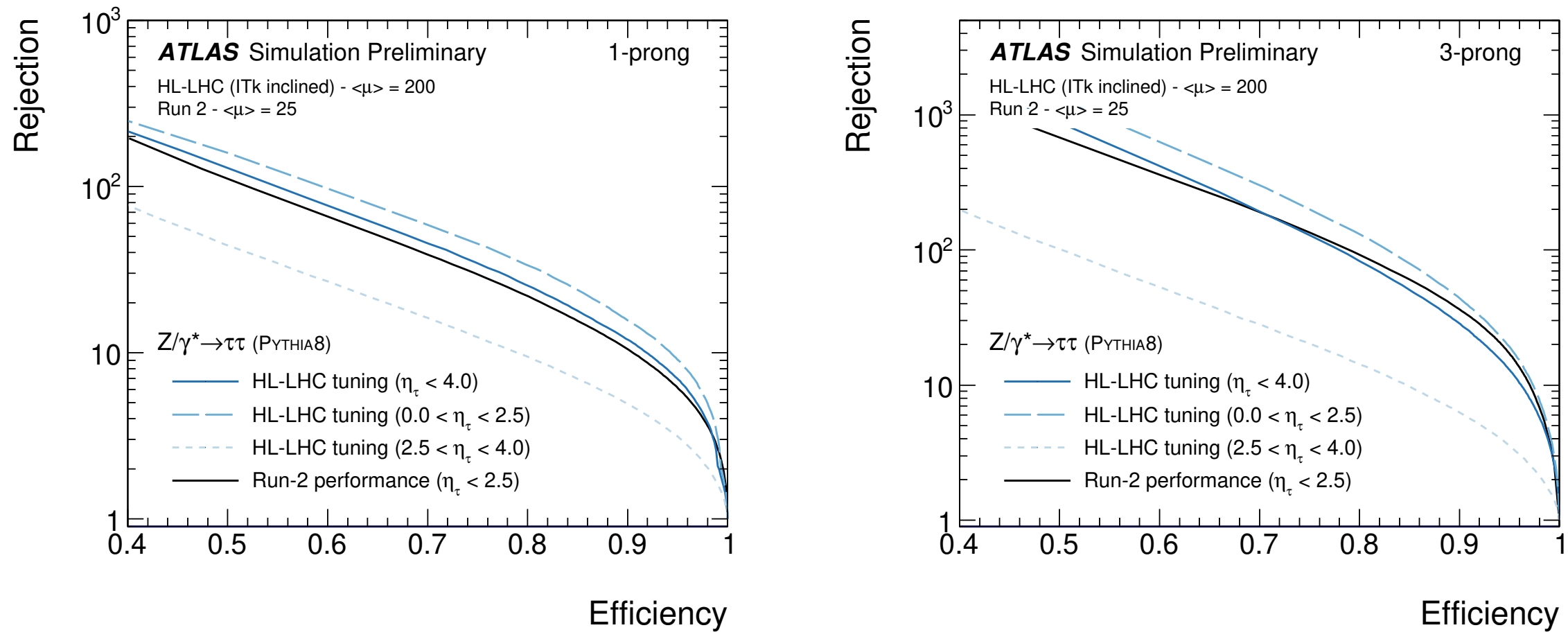

(a) (b)

Figure 10: Jet rejection as a function of $\tau_{\text{had-vis}}$ efficiency for the algorithm optimized for HL-LHC detector and conditions ("HL-LHC tuning") for $\tau_{\text{had-vis}}$ candidates with a $p_{\text{T}}$ above 20 GeV and within $|\eta| < 4.0$ (solid blue), $|\eta| < 2.5$ (dark dashed blue), and $2.5 < |\eta| < 4.0$ (dashed light blue), compared to the Run 2 performance optimized for the Run 2 detector and conditions ("Run 2 performance") for $\tau_{\text{had-vis}}$ candidates within $|\eta| < 2.5$ (solid black), shown for (a) one-prong and (b) three-prong tau leptons.

- The scale factor on the *in situ* uncertainty on the tau energy scale is 0.6, which is found by taking the current measurements and setting the sources of statistical uncertainty equal to zero.

Other tau lepton-related systematic uncertainties are expected to remain similar to what they are in Run 2.

### 4.7 Photons

Figure 11 illustrates the expected energy resolution of photons under $\langle\mu\rangle = 0$ and $\langle\mu\rangle = 200$ pileup conditions, assuming the same reconstruction techniques as those currently employed in Run 2. The resolution is shown only for unconverted photons in the barrel region of the detector ($|\eta| < 0.8$). The level of electronics noise simulated is that of the existing LAr readout. The photon resolution curves obtained at $\langle\mu\rangle = 0$ and $\langle\mu\rangle = 200$ are subtracted in quadrature in order to illustrate the size of the pileup-only contribution to the photon resolution.

The expected energy resolution is further quantified for the benchmark physics process $H \to \gamma\gamma$ in Figure 12, showing the expected di-photon mass resolution. Figure 12(a) shows the effect of pileup on the expected resolution, as well as the comparison with Run 2. Two scenarios for energy resolution, an optimistic one and a pessimistic one, are considered. The optimistic scenario assumes that the statistics available with the HL-LHC will allow for the global constant term to be at 0.7%, which is its design value, while the pessimistic scenario uses the constant term found with 2015 data at 1% in the barrel and 1.4% in the endcap. The scenarios also differ in their treatment of pileup noise with the pessimistic approach using a value consistent with untuned current reconstruction algorithms with full simulation of $\langle\mu\rangle = 200$ and the optimistic approach assuming that future offline corrections can reduce this to the equivalent of the performance of full simulation with $\langle\mu\rangle = 75$. It is worth noting that the lower material budget of the upgraded inner detector results in more unconverted photons, which have a better energy resolution than the converted ones. Figure 12(b) compares different hard-scatter vertex selection strategies to show the robustness against the performance of such identification algorithms; the pointing capability of the LAr calorimeter allows for a good mass resolution to be preserved in spite of the high level of pileup.

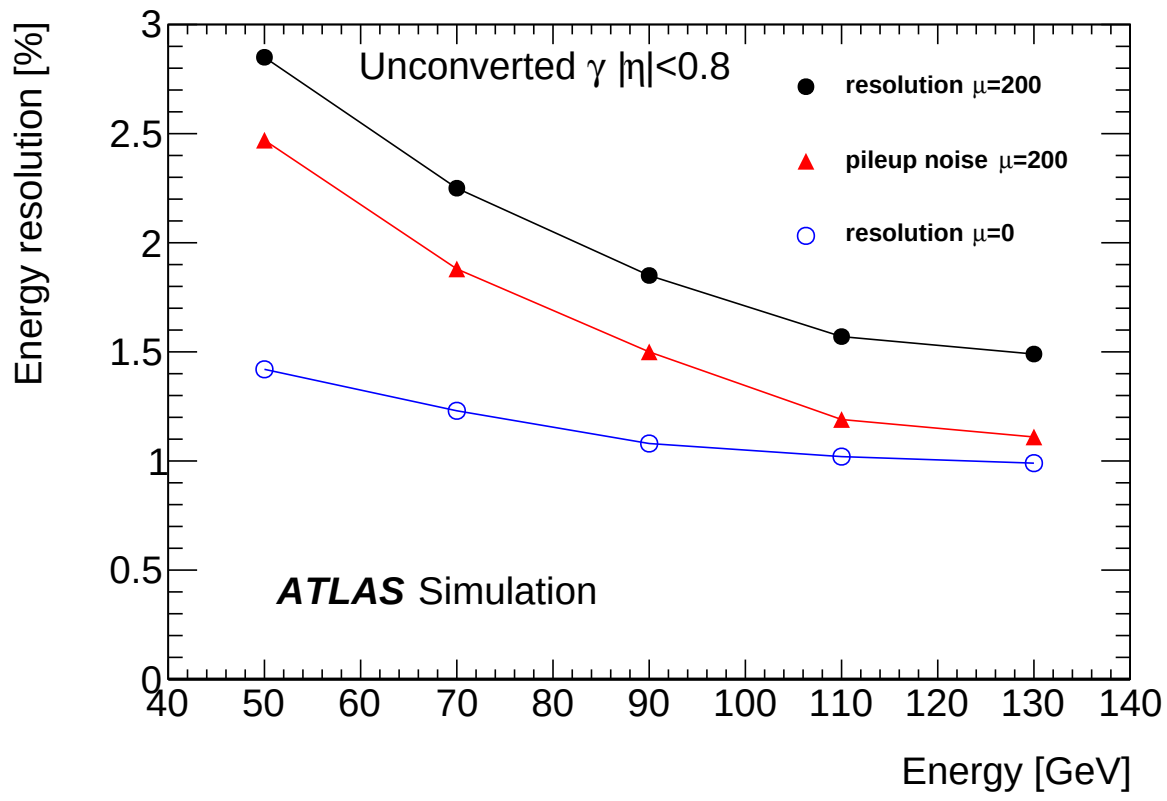

Figure 11: Photon energy resolution expected under different pileup conditions, and contributions of pileup-only noise to the energy resolution. (Chapter 4, Figure 9 from the LAr TDR) [7].

The baseline systematic uncertainty assumption for photons is that they will remain unchanged from Run 2 values at the HL-LHC, with the exception of the combination of the reconstruction, identification and isolation efficiency, which is reduced from the Run 2 value with a scale factor of 0.8. The reduction comes from expected improvements in the understanding of the current methodology and, to a smaller degree, the increased dataset available. Representative values of uncertainties are shown in Table 5. In analyses where uncertainties due to photons dominate, the impact of halving the uncertainties on the photon resolution and scales was explored.

Table 5: Some representative values for systematic uncertainties for photons at the HL-LHC. These uncertainties are consistent with the Run 2 uncertainties with the exception of the combination of the reconstruction, identification and isolation, where a scale factor of 0.8 has been applied. [15]

| Photon Parameter | Range | Uncertainty |
|---|---|---|
| Energy scale | $p_T \approx 60$ GeV | 0.3% |
| | high $p_T$, up to 200 GeV | 0.5% |
| Resolution | $p_T \approx 60$ GeV | 10% |
| Reconstruction + ID + Isolation | $p_T < 200$ GeV | 2% |

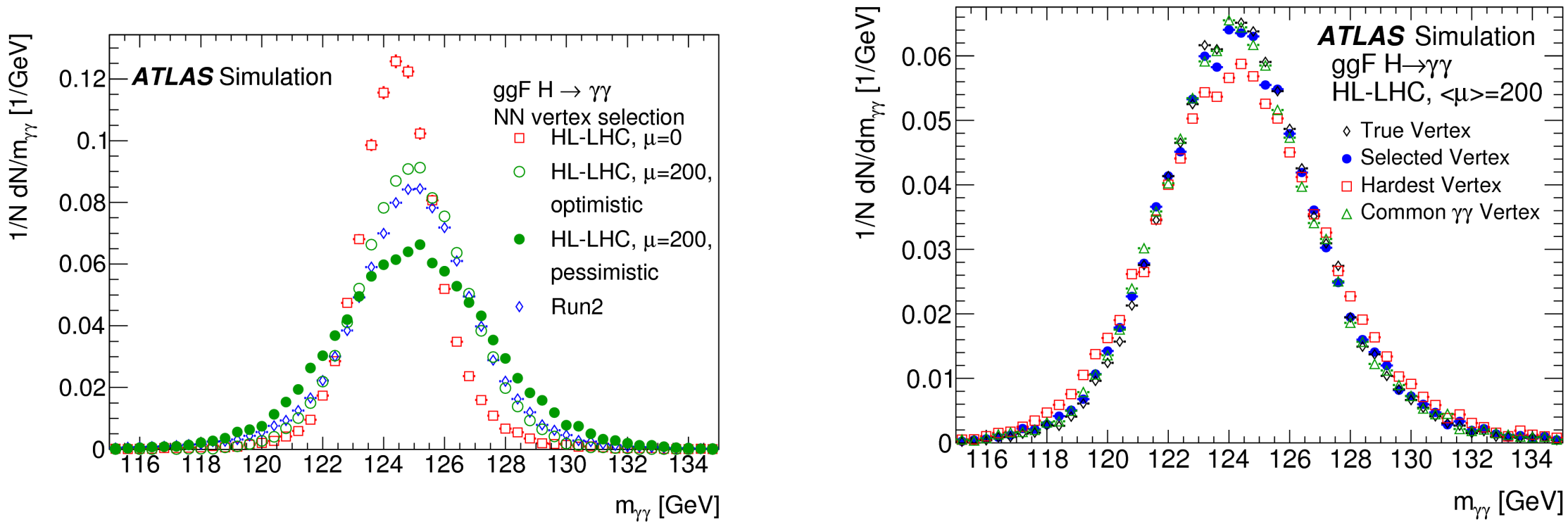

(a) (b)

Figure 12: Diphoton invariant mass for $H \to \gamma\gamma$ events (a) obtained using data in Run 2, $\langle\mu\rangle = 0$ simulation and $\langle\mu\rangle = 200$ simulation at HL-LHC using the optimistic and pessimistic photon resolution scenarios, (b) for different algorithms used to choose the hard-scatter interaction primary vertex (Chapter 4, Figures 16 and 15b from the LAr TDR) [7].

## 4.8 Jets

Jet reconstruction, as well as the separation between pileup and hard-scatter jets, benefits from the excellent tracking capabilities and extended $\eta$ range of the ITk detector. At $\langle\mu\rangle = 200$, each event is expected to have on the order of 5 jets with $p_{\rm T} > 30$ GeV produced in pileup interactions. Several techniques to suppress such pileup jets based on tracking information have been developed in Run 1 and 2. The results presented in this section are based on the $R_{p_{\rm T}}$ discriminant, which is defined as the sum of the transverse momentum of tracks associated to the jet that originates from the hard-scatter vertex over the jet $p_{\rm T}$ (as measured by the calorimeter). Pileup jets will tend to have $R_{p_{\rm T}}$ close to zero, while jets originating from the hard-scatter tend to have higher $R_{p_{\rm T}}$ values. A pileup jet mitigation $R_{p_{\rm T}}$ selection is applied for jets with $p_{\rm T}$ <100 GeV and $|\eta| < 3.8$ that has a 2% selection efficiency for pileup jets. The expected number of pileup jets before and after this selection has been applied is shown as a function of $\eta$ in Figure 13(a). The efficiency for jets originating from the hard-scatter interaction is shown versus pileup jet efficiency in Figure 13(b). More advanced taggers are expected to be developed in the HL-LHC timescale, likely enhancing significantly the pileup jets rejection capabilities.

The estimated relative jet $p_{\rm T}$ resolution, which is to a very good approximation the same as the relative jet energy resolution, is presented in Figure 14(a) and the fractional jet mass resolution for trimmed, large radius jets (anti-$k_T$, $R = 1.0$) is presented in Figure 14(b). Projections use the expected performance for calorimeter jets; however Figure 14(a) shows that particle-flow jets, currently under study, have the potential to have better resolution in the low $p_{\rm T}$ regime.

Each of the main components of the overall jet energy scale (JES) uncertainty are expected to remain constant or decrease in the transition from Run 2 to the HL-LHC. Two estimates are presented, a default labelled “baseline” and an “optimistic” estimate that assumes an improved understanding of the MC modelling of jet fragmentation and improved understanding of the effects of pileup on the JES. Figures 15(a) and 15(b) and Table 6 summarize the “baseline” and “optimistic” scenarios for the fractional uncertainties of the various components of the JES uncertainty.

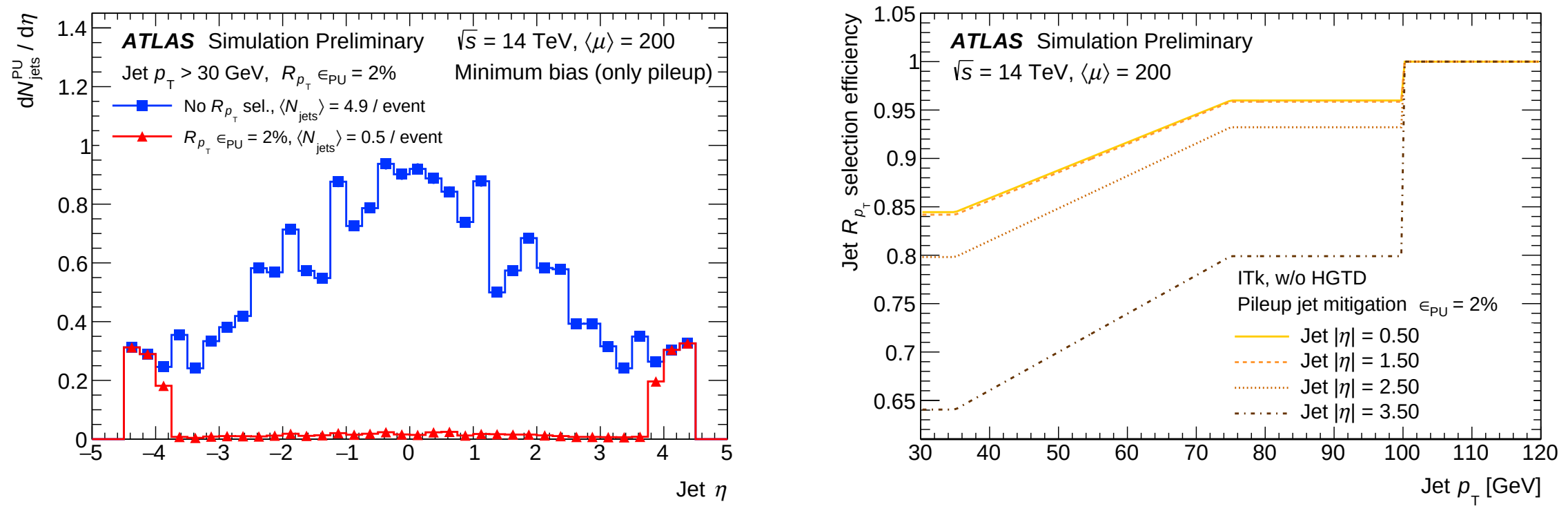

(a) (b)

Figure 13: (a) Expected number of pileup jets per unit of pseudorapidity before and after pileup jet suppression. (b) Efficiency for jets originating from the hard scatter using the $R_{p_T}$ tagger. In both Figures, a selection based on the $R_{p_T}$ tagger is applied that achieves a 98% rejection of pileup jets ($\epsilon_{PU} = 2\%$) in the region of tracking coverage: $|\eta| < 3.8$.

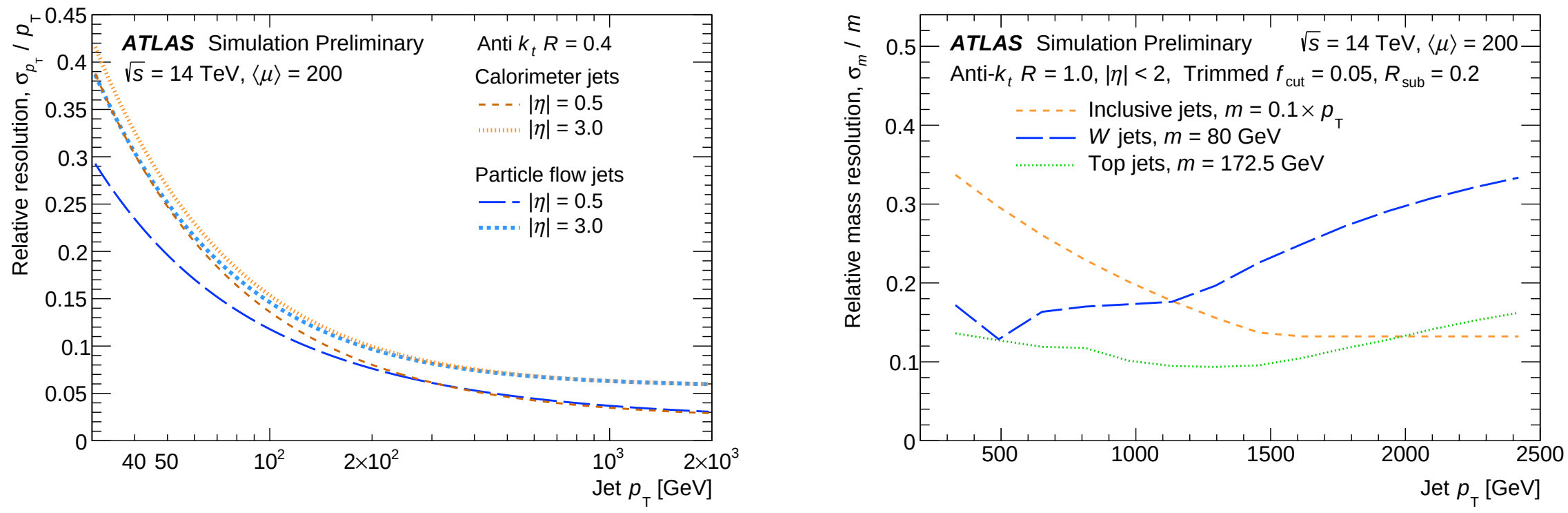

(a) (b)

Figure 14: (a) Relative jet $p_T$ resolution. (b) Fractional jet mass resolution for trimmed, large radius jets.

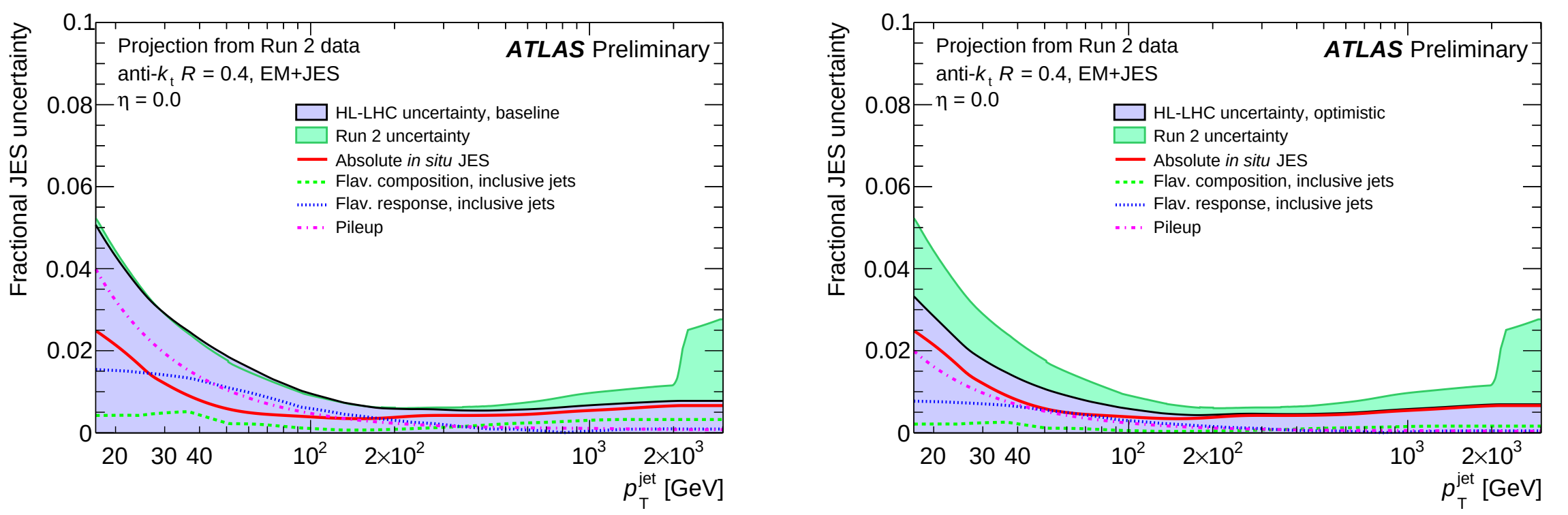

(a) (b)

Figure 15: (a) Baseline and (b) optimistic scenarios for HL-LHC jet energy scale uncertainties with a dijet-like flavour composition.

Table 6: Expected jet energy scale (JES) uncertainties at the HL-LHC in the "baseline" and "optimistic" scenarios.

| Uncertainity component | Percentage Uncertainty (Baseline Estimate) | Percentage Uncertainty (Optimistic Estimate) |
|---|---|---|
| Absolute JES scale | 1% - 2% | 1% - 2% |
| Pileup | 0 - 4% | 0 - 2% |
| JET flavour composition | 0 - 1% | 0 - 0.5% |
| JET flavour response | 0 - 1.5% | 0 - 0.8% |

## 4.9 Flavour tagging

Flavour tagging benefits from the excellent tracking and $\eta$ coverage of the ITk detector. A multivariate algorithm [17] has been re-tuned for the expected ATLAS Phase-II detector and its performance assessed. The light-jet rejection versus $b$−jet efficiency is shown in various slices of $\eta$ in Figures 16(a) and 16(b) along with a comparison with Run 2 performance.

The performance in $t\bar{t}$ with $\langle\mu\rangle = 200$ is shown for light-jet rejection and c-jet, b-jet and pileup-jet efficiency in Figure 17 for the working point with an average b-jet efficiency of 70%. In the benchmark channel with $HH \to \gamma\gamma bb$, the purity of b-jets when both jets are tagged is at the level of 97%.

The expected flavour tagging uncertainties have been derived extrapolating current performance and taking into account new methods that may be used in the future, especially at high-$p_T$ and large $\eta$. The expected uncertainties on identification efficiency for $b$-jets, $c$-jets and light-jets are summarized in Table 7.

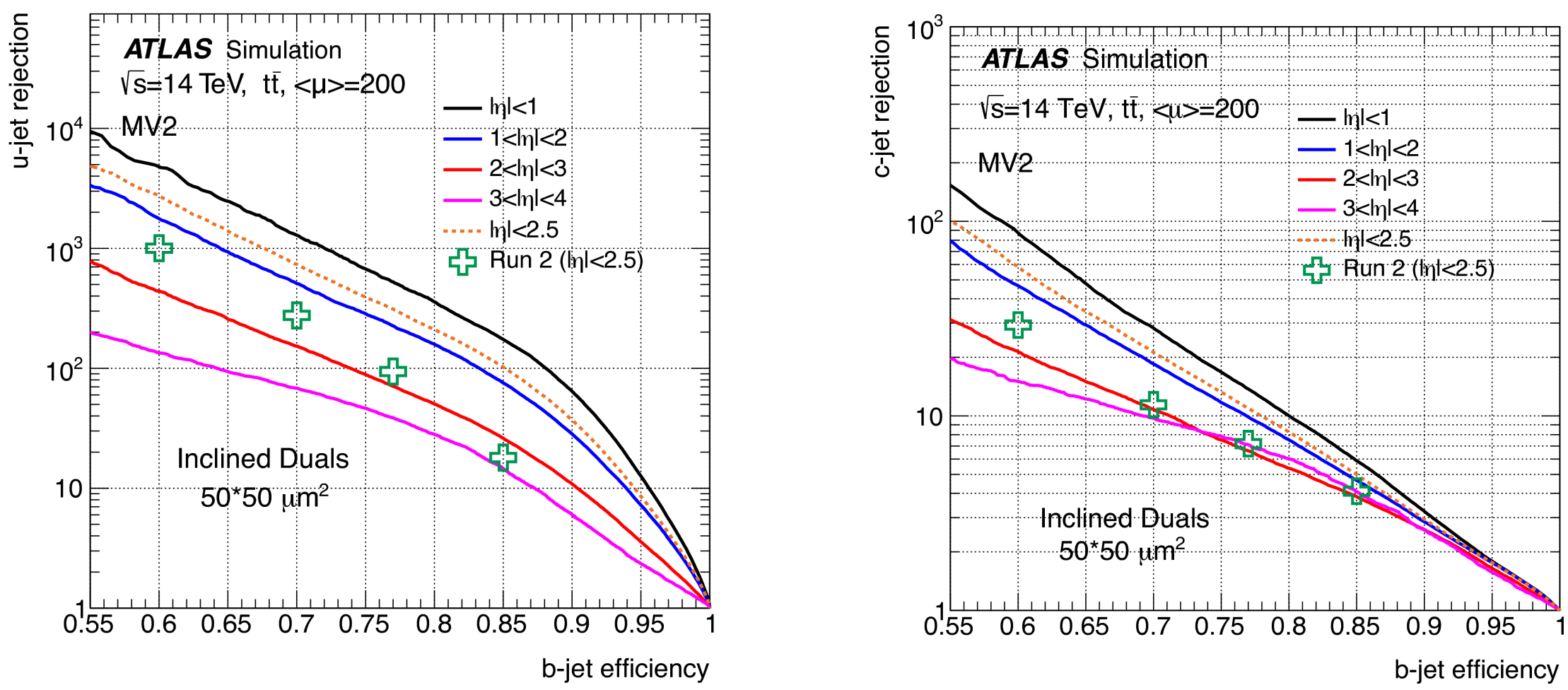

(a) (b)

Figure 16: Performance of the MV2 b-tagging algorithms in $t\bar{t}$ events with 200 pileup for the ITk layout. Results are shown for 50×50 $\mu$m$^2$ pixels, using digital clustering in the reconstruction. For comparison purposes, the performance for ATLAS during Run 2 with an average of 30 pileup events is shown as crosses. The rejection of (a) light-flavour jets and (b) c-jets for different $\eta$ regions is shown as a function of b-jet efficiency. Figures 3.21(b) and 3.23(b) from the Pixel Detector TDR. [5]

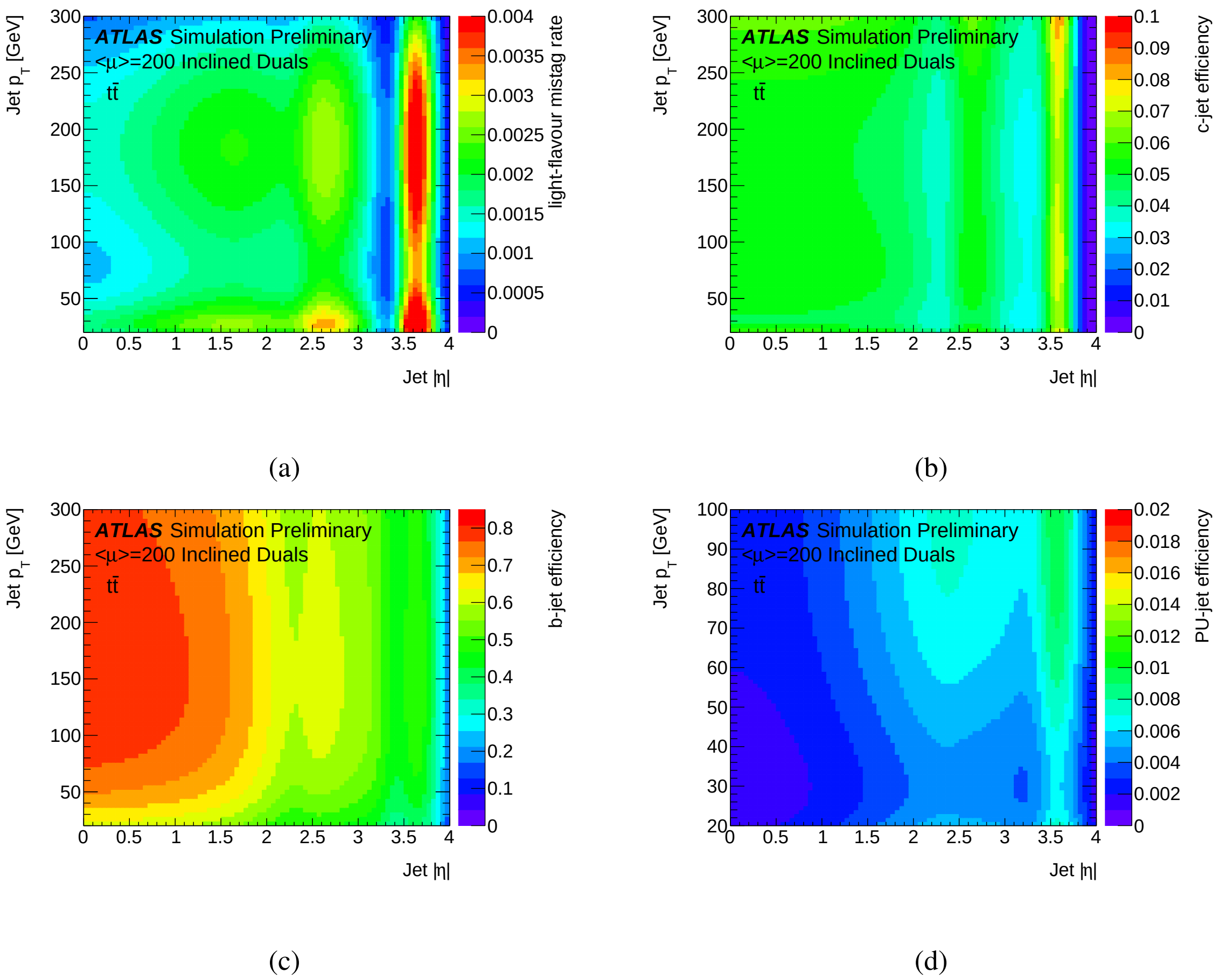

Figure 17: MV2 algorithm performance in $t\bar{t}$ events with a $b$-jet efficiency of 70% and $\langle\mu\rangle = 200$ of (a) the light flavour mistag rate, (b) c-jet efficiency, (c) b-jet efficiency, and (d) pileup jet efficiency.

Table 7: Representative values for systematic uncertainties for flavour tagging at the HL-LHC. [15]

| Uncertainty | Expected value at HL-LHC | Comments |
|---|---|---|
| $b$-jet efficiency | 1% | $30 < p_{\mathrm{T}} < 300$ GeV |
| $b$-jet efficiency | 2-6% | $p_{\mathrm{T}} > 300$ GeV |
| $c$-jet efficiency | 2% | all working points |
| light-jet mistag | 5 - 15% | working-point dependent |

## 4.10 Missing Transverse Energy ($E_T^{\mathrm{miss}}$)

The event $E_T^{\mathrm{miss}}$ is computed as the negative value of the vectorial sum of calibrated high-$p_T$ particles and jets, together with a soft-term. The soft-term is computed from reconstructed charged particles that are not associated to high-$p_T$ objects and are compatible with originating from the hard-scatter interaction.

Pileup jets are suppressed using the same tagger described in Section 4.8. The event $E_T^{\text{miss}}$ resolution depends strongly on the final state of the event in question. Detailed studies in $t\bar{t}$ events with HL-LHC conditions were performed and the expected resolution of $E_T^{\text{miss}}$ in such events is shown in Figure 18, illustrating that forward tracking capabilities used in forward pileup jets rejection are crucial for $E_T^{\text{miss}}$ resolution.

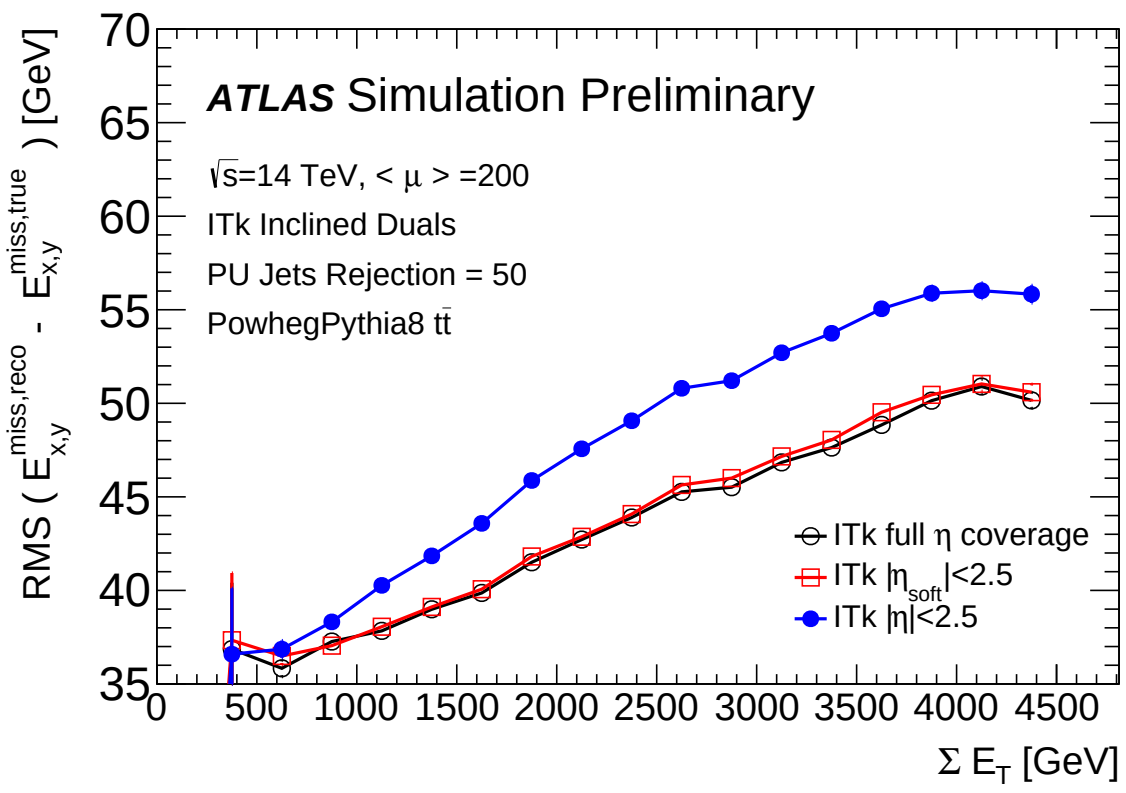

Figure 18: The resolutions of $E_T^{\text{miss}}$ in Monte Carlo $t\bar{t}$ events with an average of 200 pileup events. The resolutions are shown as a function of the scalar sum of the event transverse energy. Three variations of the $E_T^{\text{miss}}$ calculation are shown: first, only tracks within $|\eta| < 2.5$ are used for both the pile-up jet rejection and the track soft term (blue line); second, tracks are used for the full $\eta$ coverage to reject pileup jets (red line); and third, forward tracks are used for both the pileup jet rejection and the track soft term (black line).

The systematic uncertainties on all the hard objects used to form the $E_{\text{T}}^{\text{miss}}$ are propagated through the $E_{\text{T}}^{\text{miss}}$ calculation. These form the dominant systematic uncertainties on this quantity and are highly process- and analysis-dependent.

### 4.11 Heavy ions

The Heavy Ion physics program is expected to continue at least throughout Run 4, and possibly beyond. The upgraded ATLAS detector is well equipped to take full advantage of such a dataset using dedicated tuning and reconstruction algorithms. The replacement of the ATLAS tracking detector, which extends the $\eta$ coverage significantly ($|\eta| < 2.5$ becomes $|\eta| < 4.0$ for charged tracks), results in significant improvements for these measurements. Figures 19 and 20 show the expected charged particle reconstruction efficiency and track parameter resolution in minimum-bias (0–100% centrality) Pb+Pb collisions.

Additional improvements will be provided by the HGTD and Zero Degree Calorimeter (ZDC), but these improvements have not yet been taken into account in the HL-LHC studies.

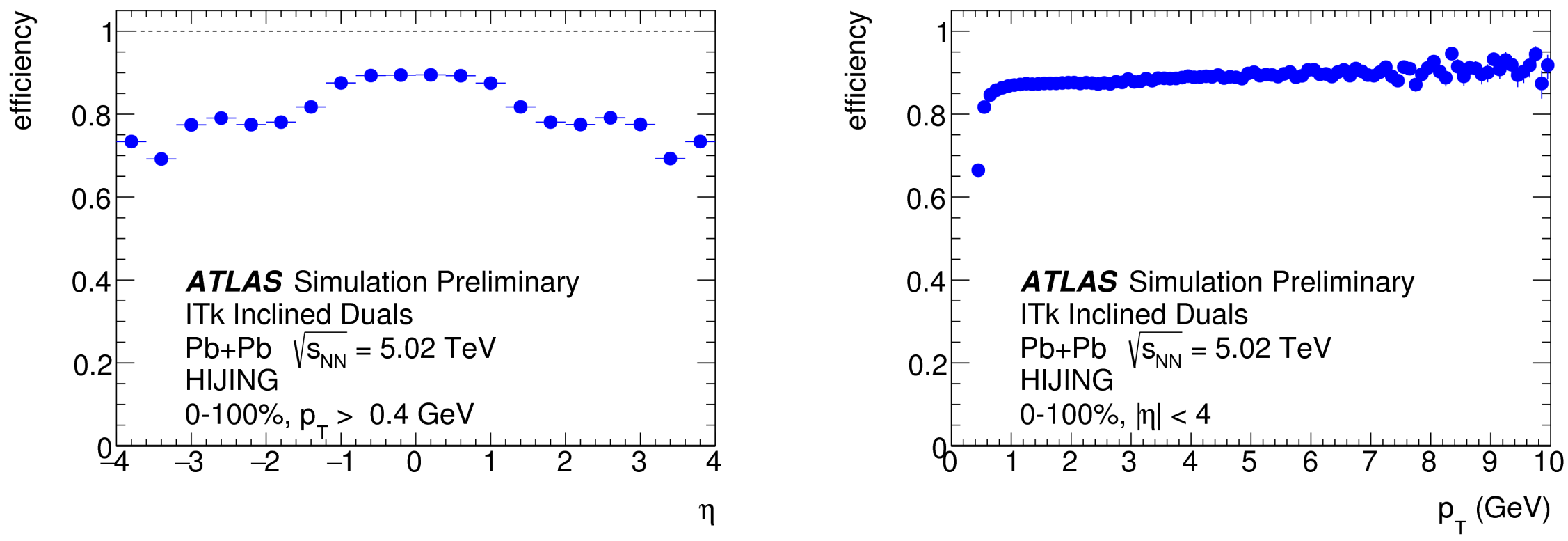

(a) (b)

Figure 19: Track reconstruction efficiency as a function of (a) $\eta$ and (b) $p_T$ in minimum bias (0–100% centrality) Pb+Pb collisions with the ITk upgrade. Figure 2 from Ref. [18].

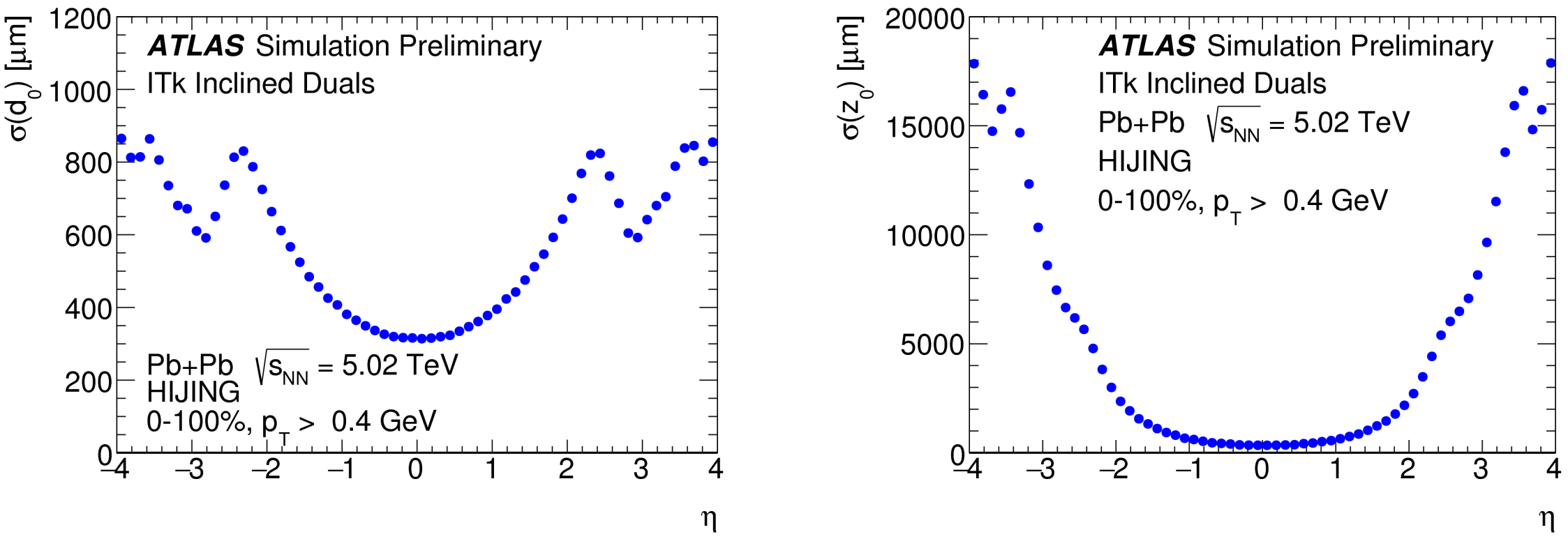

(a) (b)

Figure 20: Resolution of (a) track parameters $d_0$ and(b) $z_0$ as a function of $\eta$ for a minimum track $p_T$ threshold of 0.4 GeV in minimum bias (0–100% centrality) Pb+Pb collisions with the ITk upgrade. Figure 3 from Ref. [18].

# 5 Conclusion

The HL-LHC will provide an unprecedented amount of integrated luminosity to the ATLAS experiment, which enables a wide range of physics to be explored. The ATLAS detector is well positioned to take full advantage of this dataset thanks to a series of upgrades to its sub-detectors.

In this note we summarize and reference the baseline expected performance of the upgraded ATLAS detector. Such performance assumptions are used in recent physics projection studies and will be a baseline reference for future ones. These studies heavily rely on the recent Phase-II Technical Design Reports (TDRs) but include some more recent developments that were not available at the time of the TDRs.

Additionally, many physics projections will be significantly limited by systematic uncertainties. Advancements in detector and theoretical understanding, together with the usage of such a large dataset in *in situ* techniques, are expected to improve our knowledge and consequently reduce some of these uncertainties.

General guidelines, harmonized with the CMS Collaboration, as well as specific recommendations in terms of expected systematic uncertainties, have been presented.

# Expected performance of the physics objects with the upgraded CMS detector at the HL-LHC

The CMS Collaboration

**Abstract**

In this note, the performance of the physics objects (electrons, photons, taus, jets, and missing energy), as expected after the CMS Phase-2 detector upgrade, is presented. The performance studies use the full simulation of the CMS Phase-2 detector with a mean number of proton-proton interactions per bunch crossing of 200. In addition, an evaluation of the systematic uncertainties for HL-LHC studies are presented.

# 1 Introduction

The upgraded CERN High-Luminosity LHC (HL-LHC) will deliver peak instantaneous luminosities of $5 \times 10^{34}\,\mathrm{cm}^{-2}\,\mathrm{s}^{-1}$, or even $7.5 \times 10^{34}\,\mathrm{cm}^{-2}\,\mathrm{s}^{-1}$ in the ultimate performance scenario [1]. This performance can be contrasted with the current LHC, which provided instantaneous luminosities up to $1.5 \times 10^{34}\,\mathrm{cm}^{-2}\,\mathrm{s}^{-1}$ in 2016. With this increase in instantaneous luminosity, the total pileup (PU), or number of proton-proton interactions per bunch crossing, is expected to increase from a mean PU of 27 with the LHC in 2016 to 140 or even 200 PU at the HL-LHC. Similarly, the levels of radiation are expected to significantly increase in all regions of the detector, in particular in its forward regions.

The CMS detector [2] will be substantially upgraded in order to exploit the physics potential provided by the increase in luminosity at the HL-LHC, and to cope with the demanding operational conditions at the HL-LHC [3]. This upgrade is referred to as the CMS Phase-2 Upgrade. The increase in radiation levels requires improved radiation hardness, while the larger PU and associated increase in particle density require higher detector granularity to reduce the average channel occupancy, increased bandwidth to accommodate higher data rates, and improved trigger capability to keep the trigger rate at an acceptable level without compromising physics potential.

The upgrade of the first level hardware trigger (L1) will allow for an increase of L1 rate and latency to about 750 kHz and 12.5 $\mu$s, respectively. The upgraded L1 will also feature inputs from the silicon tracker, allowing for real-time track fitting and particle-flow (PF) reconstruction [4] of objects at the trigger level. The upgrade of the high-level software trigger (HLT) will allow the HLT rate to be increased to 7.5 kHz.

The entire silicon tracking system, which consists of pixel and strip detectors, will be replaced. The new tracker will feature extended geometrical coverage and provide efficient tracking up to pseudorapidities of about $|\eta| = 4$, improved radiation hardness, higher granularity, and compatibility with higher data rates and a longer trigger latency. In addition, the tracker will provide information on tracks above a configurable transverse momentum threshold to the L1 trigger, information presently only available at the HLT. It will also allow for including tracks with low momentum ($\approx$3 GeV). This will allow the trigger rates to be kept at a sustainable level without sacrificing physics potential. The Phase-2 tracker will include an Inner Tracker based on silicon pixel modules and an Outer Tracker made from silicon modules with strip and macro-pixel sensors.

In the barrel, the electromagnetic calorimeter (ECAL) uses lead-tungstate crystals read out with avalanche photodiodes (APDs). The crystals will be cooled to lower temperatures than currently used to mitigate noise in the APDs due to radiation damage, and the front-end electronics will be improved in order to cope with the trigger latency and bandwidth requirements. The upgraded readout will also provide precision timing information. New front-end electronics will allow the exploitation of the information from single crystals in the L1 trigger, while the present system integrates the same information only in groups of $5 \times 5$ crystals. The hadronic calorimeter (HCAL) consists in the barrel region of brass absorber plates and plastic scintillator layers, read out by hybrid photodiodes (HPDs), which will be replaced with silicon photomultipliers (SiPMs). The scintillator tiles close to the beam line will be replaced. The object performance in the central region assumes a barrel calorimeter aging corresponding to an integrated luminosity of 1000 $\mathrm{fb}^{-1}$.

The electromagnetic and hadronic endcap calorimeters will be replaced with a new combined electromagnetic and hadronic sampling calorimeter (HGCal) based primarily on silicon pad sensors. Plastic scintillator tiles, read out by SiPMs, will be used at large distances from the beam line in the hadronic section. With silicon pad cell sizes of 0.5–1 $cm^2$ and 28 (12) sampling layers in the electromagnetic (hadronic) sections, this detector will provide high transverse and longitudinal granularity, as well as high-precision timing information of the high energy showers, leading to improved PU rejection and identification of electrons, photons, tau leptons, and jets.

While the muon chambers are expected to cope with the increased particle rates, the front-end electronics for the drift tube chambers (DTs) and cathode strip chambers (CSCs) will be replaced with improved versions to increase radiation tolerance, readout speed, and performance. In the forward region, the muon system will be enhanced, both with improved resistive plate chambers (RPCs) and new chambers based on the gas electron multiplier (GEM) technique. The new chambers add redundancy, improve the triggering and reconstruction performance, and increase the acceptance in the forward detector region up to about $|\eta| = 2.8$.

In addition, a minimum ionizing particle (MIP) timing detector (MTD) [5] will be added between the tracker and the ECAL, providing timing measurements up to $|\eta| = 3.0$ for the charged particle tracks that cross it. Timing at this nominal resolution allows for 4-dimensional reconstruction of vertices and significantly offsets the performance degradation due to high PU. Unless otherwise specified, the studies shown in this note assume a MTD timing resolution of 30 ps.

As a result of these upgrades, the lepton acceptance will extend to pseudorapidities of $|\eta| < 3.0$ and the jet acceptance, including b jets, will extend to $|\eta| < 4.0$. A detailed overview of the CMS detector upgrade program is presented in Refs. [3, 6–9].

PU mitigation in CMS relies upon PF event reconstruction [4], which removes charged particle tracks that are inconsistent with the vertex of interest, and upon statistical inference techniques like pileup-per-particle-identification (PUPPI) [10]. The PF algorithm aims to reconstruct and identify each individual particle in an event, with an optimized combination of information from the various elements of the CMS detector. PUPPI mitigation is an algorithm designed to remove PU using both global event information and local information to identify PU at the particle level.

This note describes the physics object performance expected, given the Phase-2 upgrades described above, in Section 2. The expected performance in Phase-2 is often shown below as compared to the performance in Run 2, which refers to studies done with the 2016 dataset unless otherwise specified. In addition to showing the CMS object performance as a function of familiar quantities such as the object transverse momentum $p_T$, we also show the performance as a function of PU density. PU density is the number of PU interactions per millimeter. We consider the longitudinal profile of the beam spot as a Gaussian shape with a width of 4.4 cm. We study the dependence of physics objects on the PU density, instead of only the total PU, in order to gain insight into the best way of delivering luminosity from the HL-LHC. In Section 3, we describe the projected systematic uncertainties for HL-LHC studies.

The studies presented in this note are mostly documented in Refs. [3, 6–9] and collated here to give a coherent overview of the performance evaluation for all objects. A series of workshops on the physics of the HL-LHC and perspectives at the High Energy LHC (HE-LHC) have been held, and the results of these workshops are being documented in a Yellow Report, which will be submitted for the next review of the European strategy for particle physics. The analyses in the Yellow Report are based on the object performance and systematic uncertainties presented below.

# 2 Object performance

## 2.1 Tracking and vertexing performance

We will first describe the expected tracking and vertexing performance with the upgraded CMS detector [6]. Figure 1 shows the tracking efficiency for single muons with different PU scenarios, where the efficiency is stable and close to 100% for the entire $\eta$ range, in both PU scenarios. Figure 2 shows the tracking efficiency and fake rate for charged tracks in $t\bar{t}$ events with different PU scenarios. The efficiency is around 90% in the central region, dropping off at $|\eta| > 3.8$, while the fake rate is lower than 2% in the entire range of $\eta$ for PU 140.

Figure 3 shows the tracking efficiency in the cores of jets as a function of the distance between tracks and their nearest neighbors, $\Delta R = \sqrt{(\Delta\phi)^2 + (\Delta\eta)^2}$, for the current tracker and the Phase-2 tracker. In the current Phase-1 reconstruction, a special algorithm to split clusters has been implemented, as well as a special iteration to perform robust tracking in jet cores. Although this has not yet been ported to the Phase-2 reconstruction, a significant improvement can already be seen for small values of $\Delta R$ thanks to the higher granularity of the new detector. Further improvement is expected for large values of $\Delta R$ as well, after applying a similar tuning.

In addition, Fig. 4 shows the $p_T$ and the transverse impact parameter ($d_0$) resolutions for the current tracker and the Phase-2 tracker. The $p_T$ resolution deteriorates for large $\eta$ because of the shorter lever arm in the projection to the bending plane. Still, the better hit resolution of the Phase-2 tracker and the reduction of the material budget results in a significantly improved $p_T$ resolution, as shown in the figure. The transverse impact parameter resolution is also improved with respect to the Phase-1 detector, ranging from below 10 $\mu$m in the central region to about 20 $\mu$m at the edge of the acceptance.

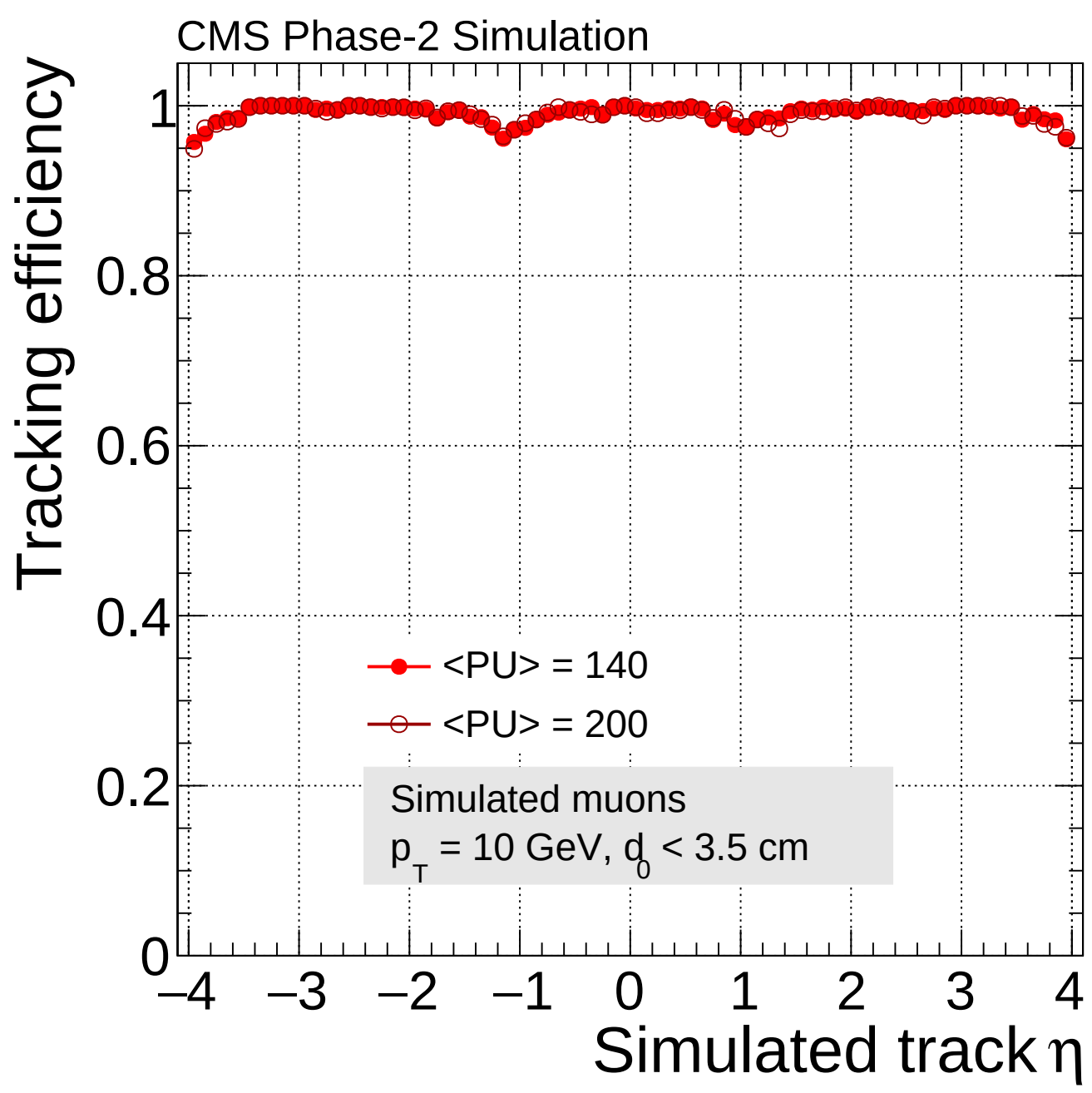

Figure 1: The tracking efficiency as a function of $\eta$ for single muons with $p_T$ equal to 10 GeV, with 140 (full circles) and 200 (open circles) PU. The efficiency is shown for tracks produced within a radius of 3.5 cm from the center of the luminous region. Taken from Ref. [6].

**4**

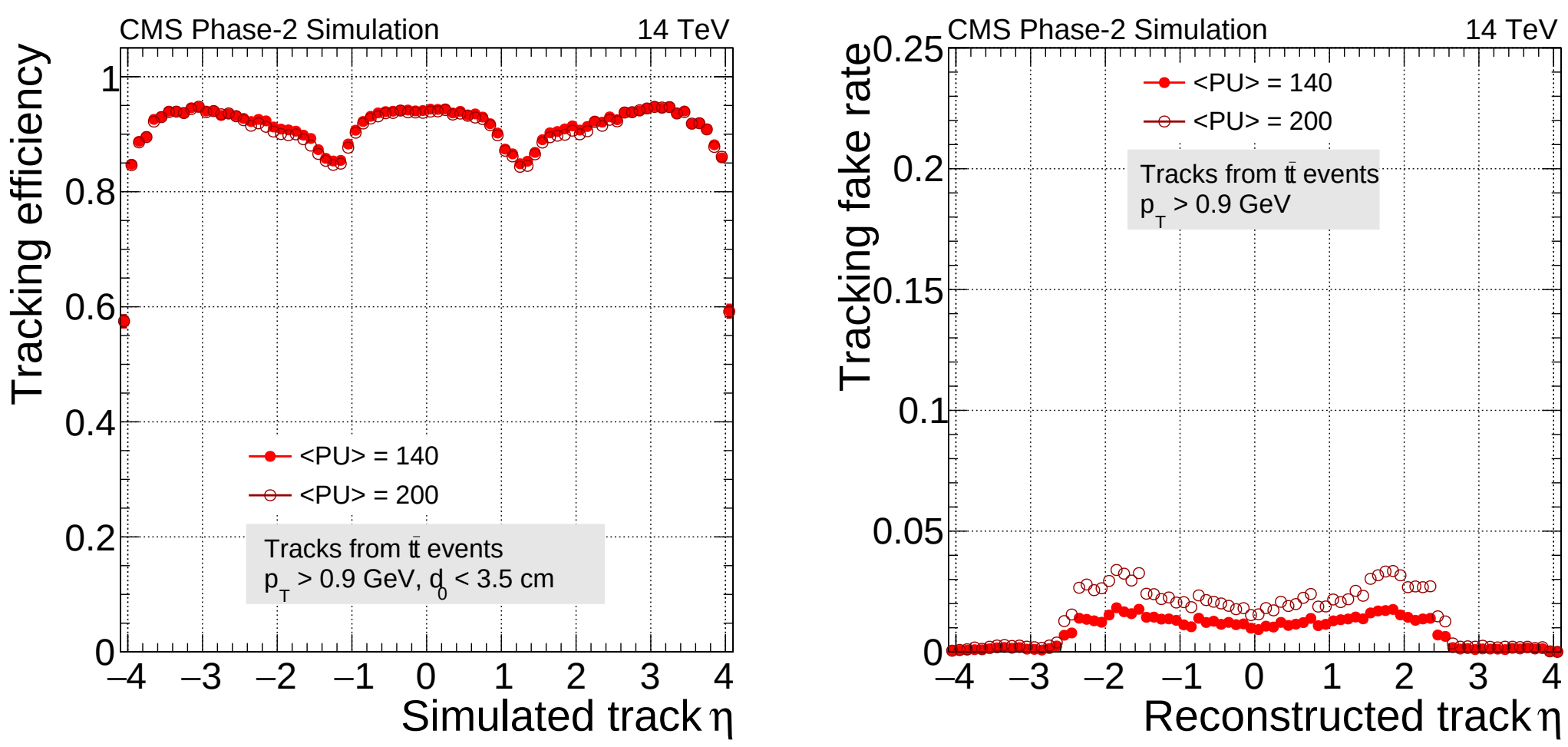

Figure 2: The tracking efficiency (left) and fake rate (right) as a function of $\eta$ for $t\bar{t}$ events with 140 (full circles) and 200 (open circles) PU. The tracks are required to have $p_{\mathrm{T}} > 0.9\,\mathrm{GeV}$. The efficiency is shown for tracks produced within a radius of 3.5 cm from the center of the luminous region. Taken from Ref. [6].

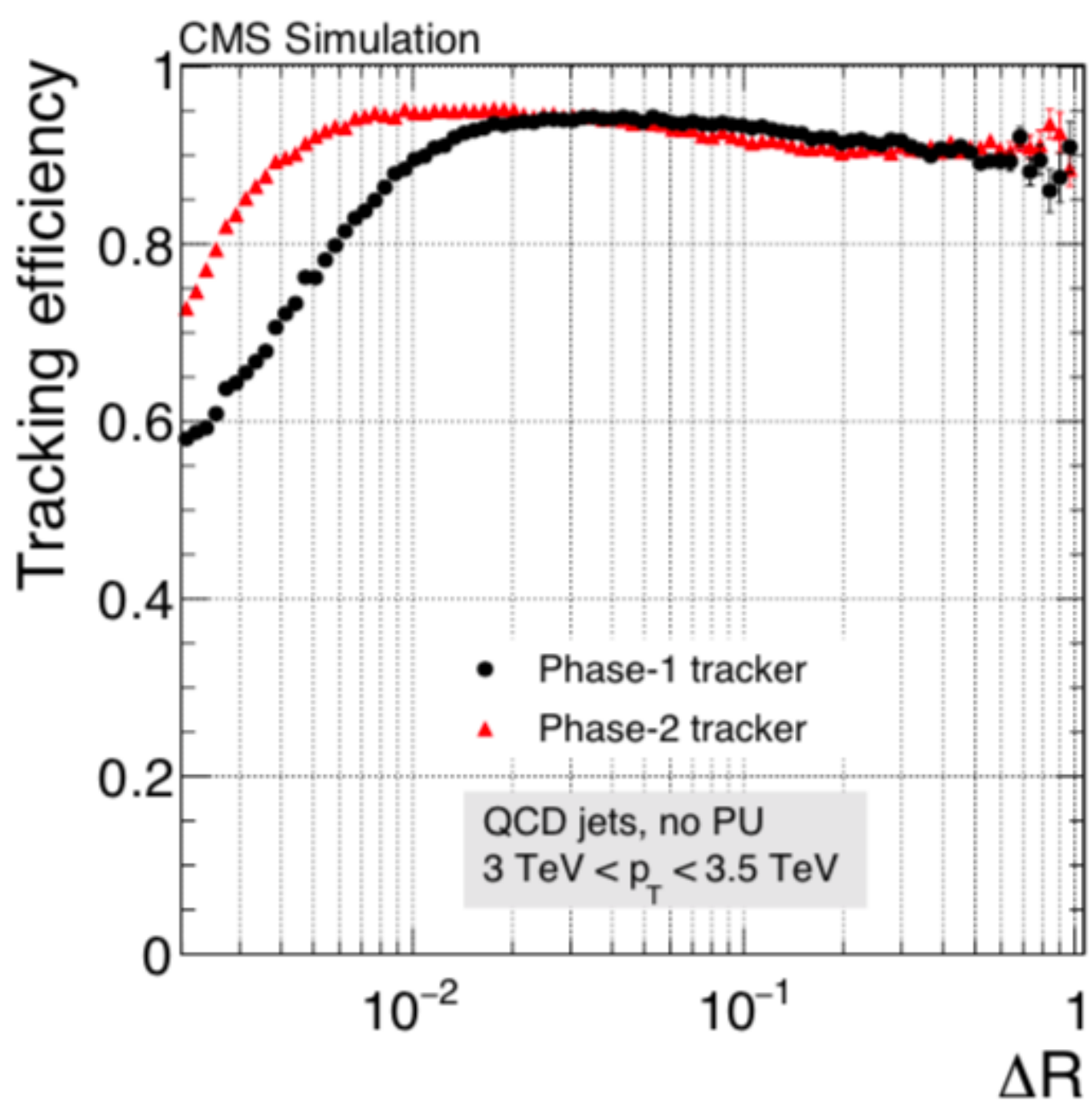

Figure 3: The tracking efficiency in the cores of jets with $3 < p_{\mathrm{T}} < 3.5\,\mathrm{TeV}$ as a function of the distance between a simulated track and its nearest neighbor, $\Delta R$, for the current tracker (black) and the Phase-2 tracker (red), without PU. Taken from Ref. [6].

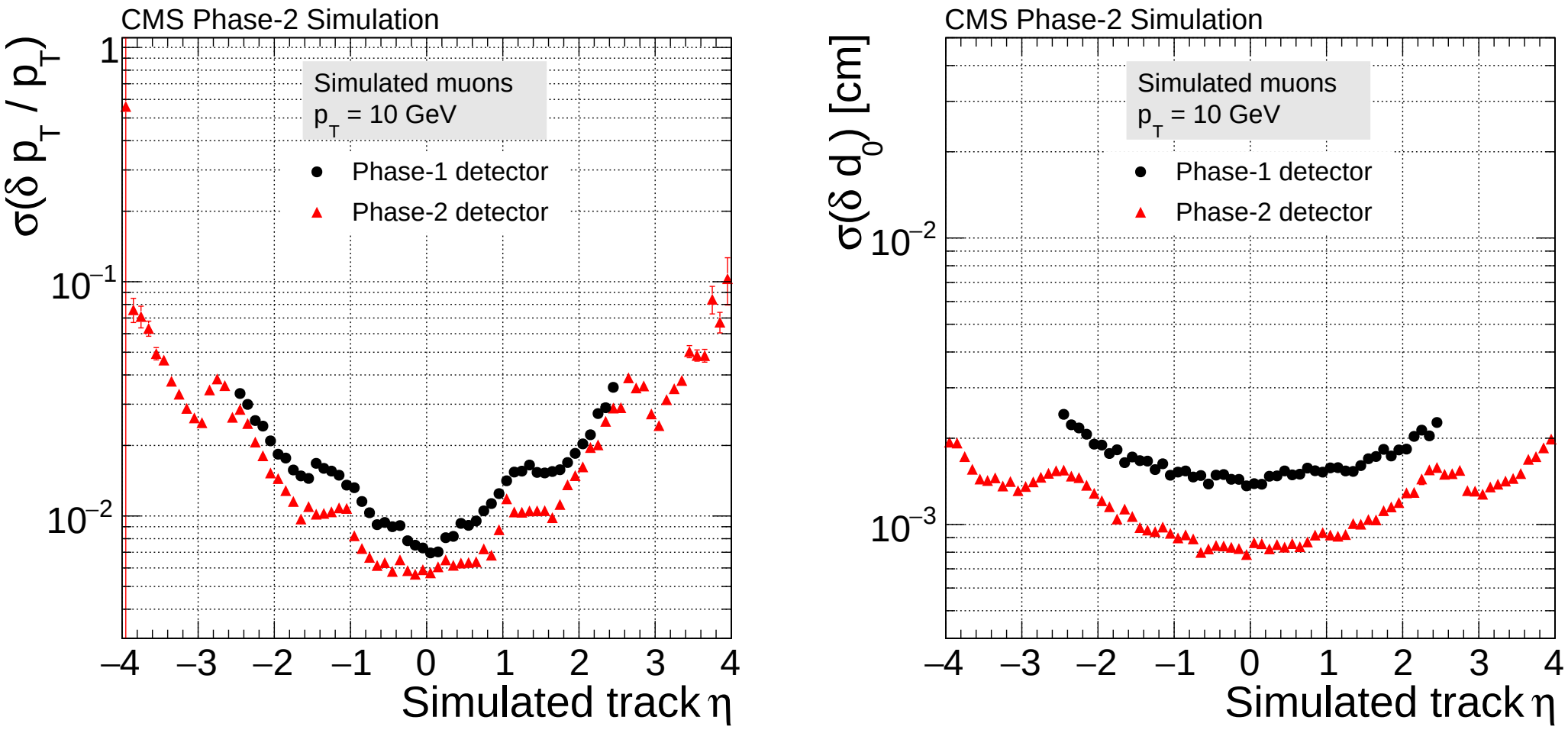

Figure 4: The relative resolution of the $p_T$ (left) and the $d_0$ resolution as a function of $\eta$ for the current tracker (black dots) and the upgraded tracker (red triangles), using single isolated muons with a $p_T$ of 10 GeV. Taken from Ref. [6].

The vertexing performance of the Phase-2 CMS detector is shown in Figs. 5 and 6. Figure 5 shows the vertex position resolution as a function of the number of tracks associated to the vertex, for different PU scenarios. The vertex position resolution is almost independent of the amount of PU in the event and the longitudinal resolution is only 50% worse than the transverse one, as expected given the pixel dimensions of the Inner Tracker modules. Furthermore, Fig. 6 shows the efficiency to reconstruct and identify the primary vertex (PV) as a function of the highest $p_T$ jet in simulated multijet events. As expected, the efficiency increases with the jet momentum due to the presence of higher momentum tracks, and it is smaller at high PU, especially in the forward region due to tracks from overlapping PU jets.

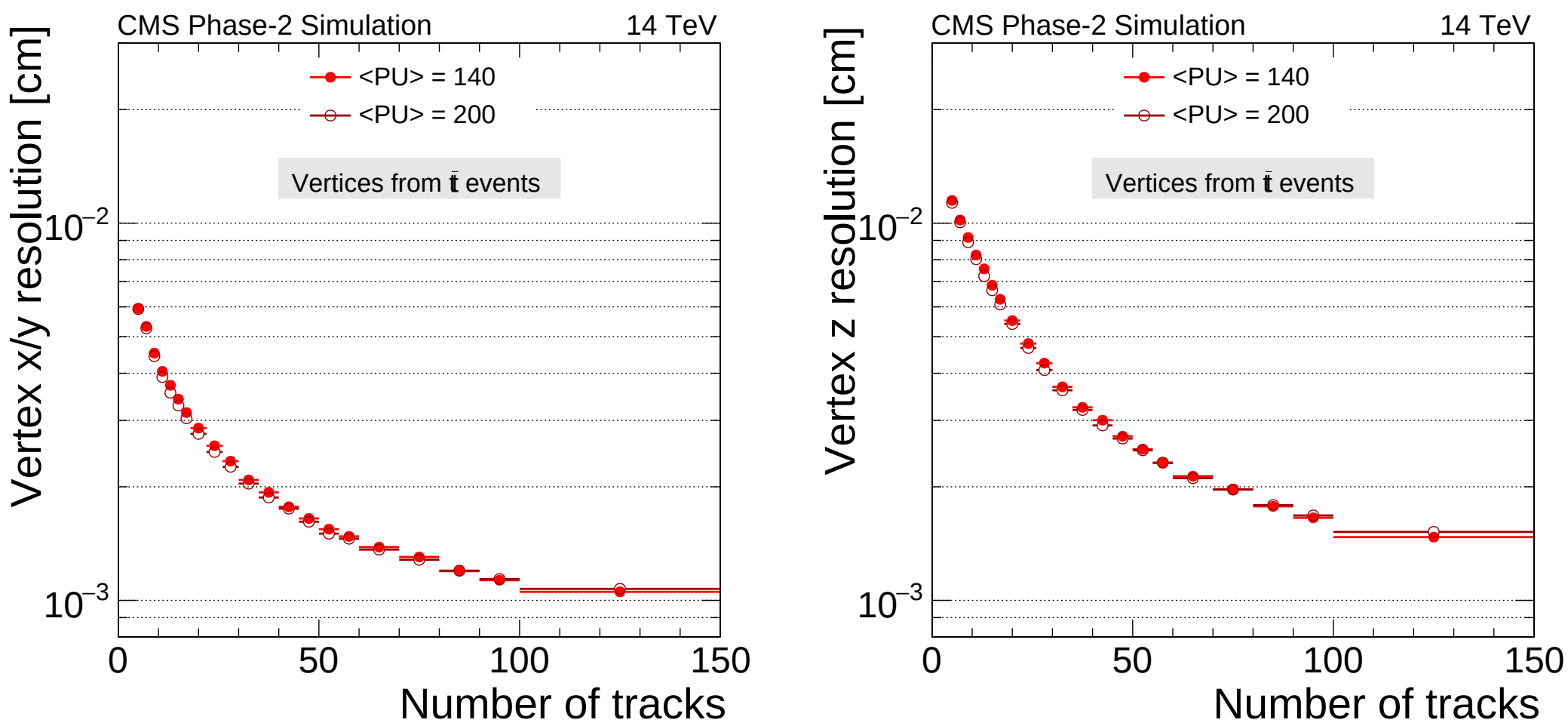

Figure 5: The vertex position resolution in $x$ and $y$ (left) and $z$ (right) as a function of the number of associated tracks to the vertex, for $t\bar{t}$ events with 140 (full circles) and 200 (open circles) PU. Taken from Ref. [6].

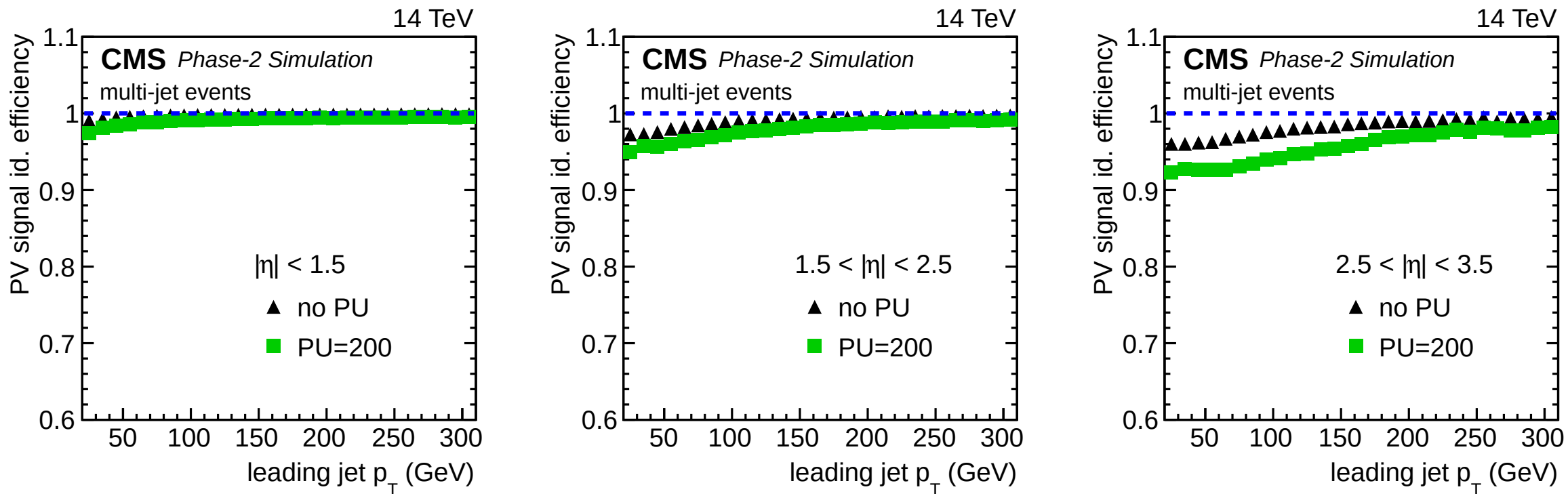

Figure 6: The efficiency to reconstruct the hard interaction vertex and to identify it correctly as the PV, as a function of the leading jet $p_{\mathrm{T}}$ in simulated multijet events with $\geq 2$ jets. The leading jet, i.e. the jet with the highest $p_{\mathrm{T}}$, is contained in the $|\eta|$ range 0–1.5 (left), 1.5–2.5 (middle), or 2.5–3.5 (right). The identification efficiency for PV signal jets increases with the leading jet $p_{\mathrm{T}}$. Compared to events without PU (black triangles), it is slightly lower at 200 PU (green squares). Taken from Ref. [8].

## 2.2 Electron performance

We next describe the electron reconstruction performance [7, 8]. Figures 7 and 8 show distributions of a few key electron shower variables, for signal and background, with and without PU. These figures show the signal to background discrimination power of these variables and their stability with respect to PU.

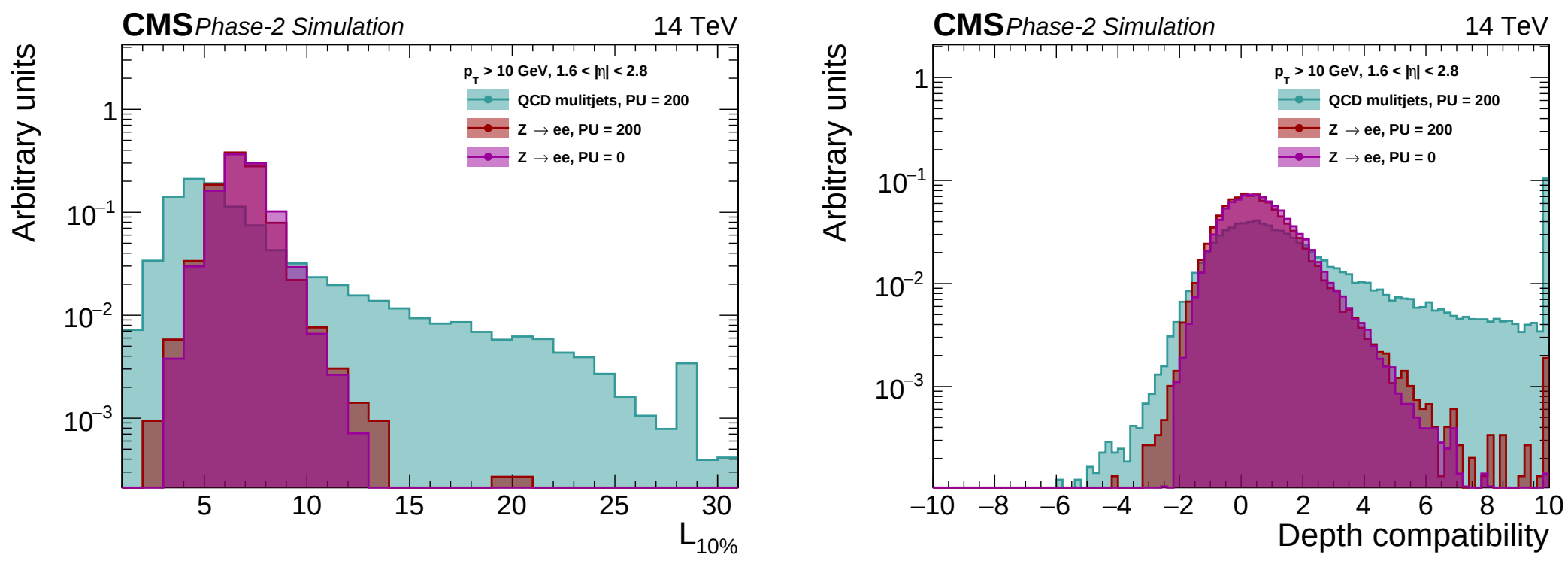

Figure 7: For signal and background electron candidates in the presence of PU as well as electrons without PU, two example variables sensitive to the shower longitudinal development are shown: the layer number for which the accumulated energy reach 10% of the ECAL endcap energy ($L_{10\%}$) (left), and the shower depth compatibility (right). Taken from Ref. [8].

In Fig. 9, we show the background rejection as a function of the electron reconstruction efficiency for different sets of input variables in the Boosted Decision Tree (BDT) multivariate estimator, trained on $Z \rightarrow ee$ events with PU. For a 95% signal efficiency, the background efficiency is 1% for $p_{\mathrm{T}} > 20\,\mathrm{GeV}$, and 10% for $10 < p_{\mathrm{T}} < 20\,\mathrm{GeV}$. In the same plots, equivalent BDT trainings are presented with reduced sets of input variables: beginning with track-based variables, a sizeable gain in performance is achieved through the energy momentum comparison; and the addition of principal component analysis (PCA)-based variables leads to further improvement in performance. Finally, the addition of extra information related to the longitudinal develop-ment, such as the compatibility in shower depth, or in the layer of the 10% cumulative ECAL endcap fraction (L10%), improves the performance only for low $p_T$ electrons, and is therefore only important in the $10 < p_T < 20$ GeV range.

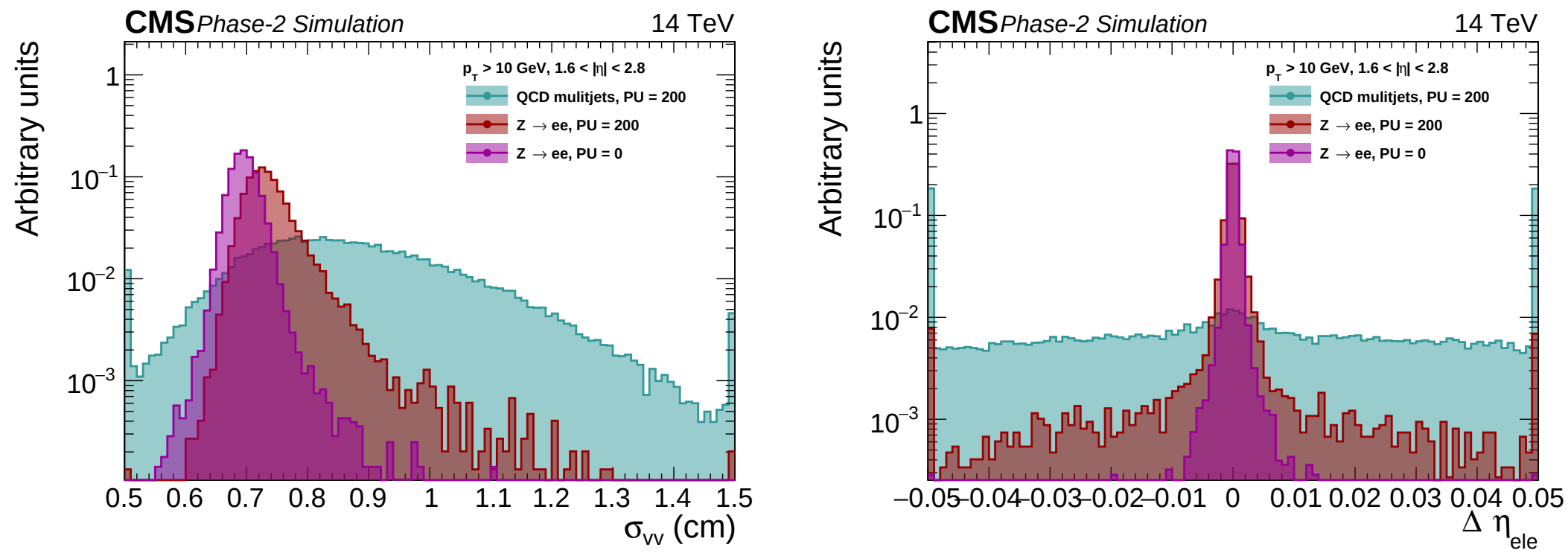

Figure 8: For signal and background electron candidates, respectively from $Z \rightarrow ee$ and multijet events, with and without PU, the shower spread along the radial direction ($\sigma_{VV}$) (left) and the distance in $\eta$ between the electron cluster and the track extrapolation ($\Delta\eta_{ele}$) (right) are shown. Taken from Ref. [8].

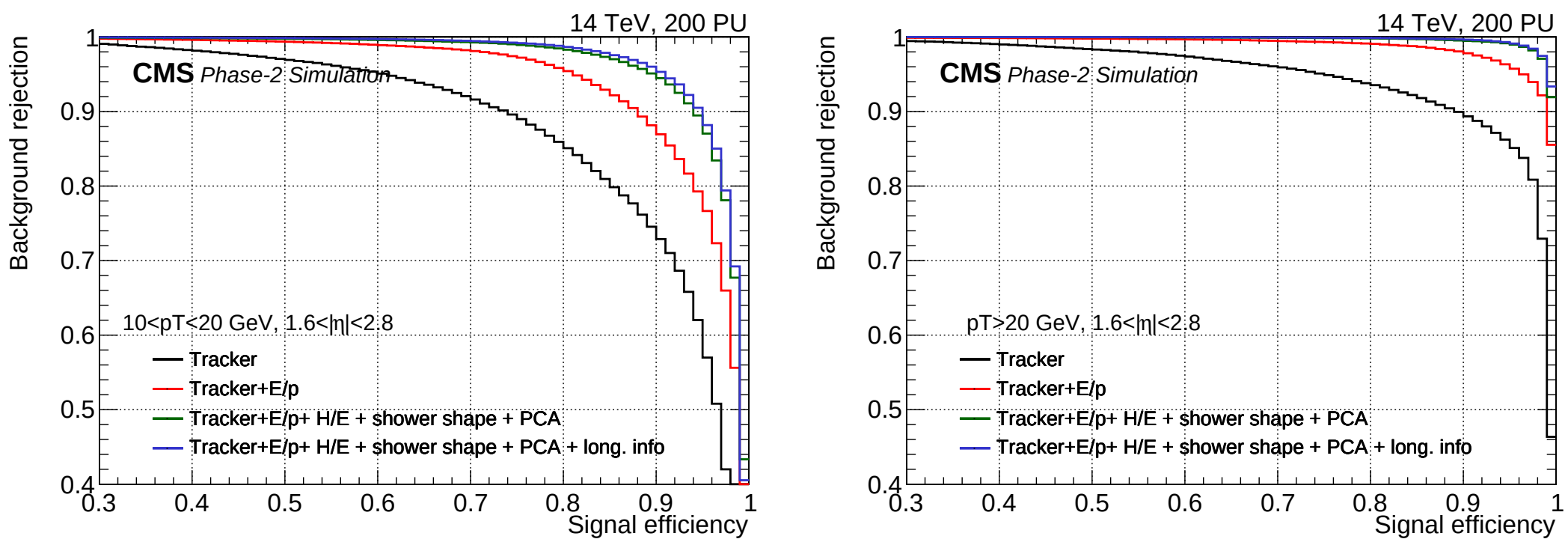

Figure 9: The purity as a function of the efficiency for electrons with $10 < p_T < 20$ GeV (left) and with $p_T > 20$ GeV (right), for different sets of input variables in the BDT multivariate estimator. Taken from Ref. [8].

The electron reconstruction efficiency as a function of $p_T$ and $\eta$ is shown in Fig. 10. An increase in efficiency can be observed from 92% at $|\eta| = 1.5$ to 98% at $|\eta| = 3$. The background efficiency also tends to increase at high $|\eta|$.

The electron reconstruction efficiency as a function of PU is shown in Fig. 11 for the full acceptance of the Phase-2 tracker and in Fig. 12 for the HGCal acceptance, using a BDT. Almost no dependence is observed on the number of vertices, which shows the robustness against PU conditions. No dependence against PU density is confirmed.

Finally, Fig. 13 shows the electron reconstruction efficiency for several ECAL barrel aging conditions. The performance is maintained with age, despite the preliminary tuning of the clustering parameters, to which the electron efficiency at low $p_T$ is quite sensitive.

**8**

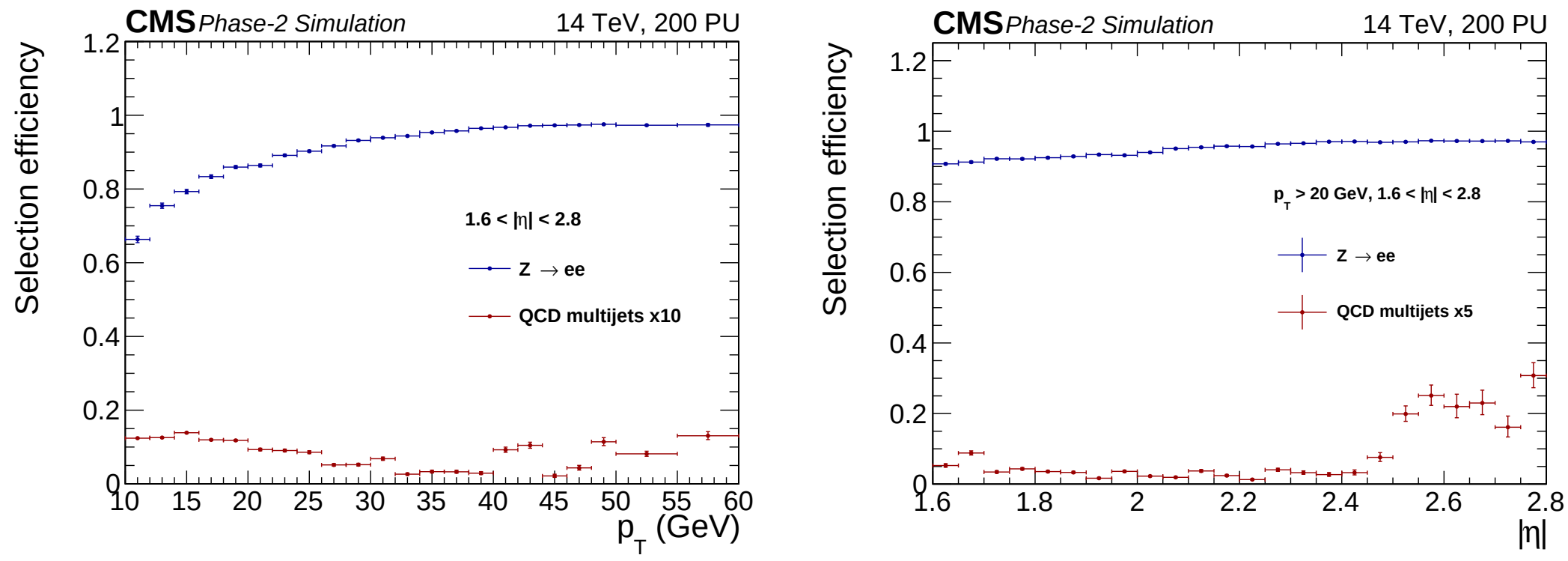

Figure 10: The evolution of the signal (blue) and background (red) efficiency, as a function of $p_T$ (left) and as a function of the cluster $|\eta|$ (right), for a high-efficiency selection of electrons with $p_T > 20$ GeV. Taken from Ref. [8].

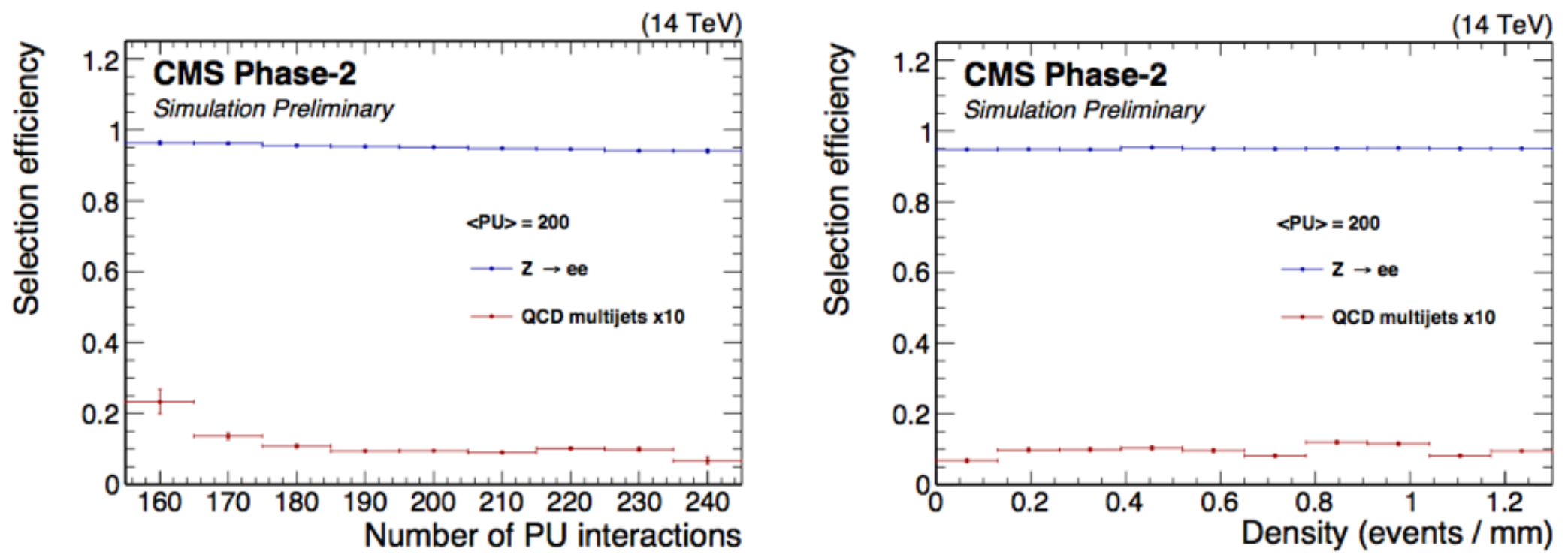

Figure 11: The electron (blue) and QCD multijet misidentification efficiency (red) reconstruction efficiency for $p_T > 20$ GeV is shown as a function of the number of PU interactions (left) and as a function of the PU density (right). Taken from Ref. [8].

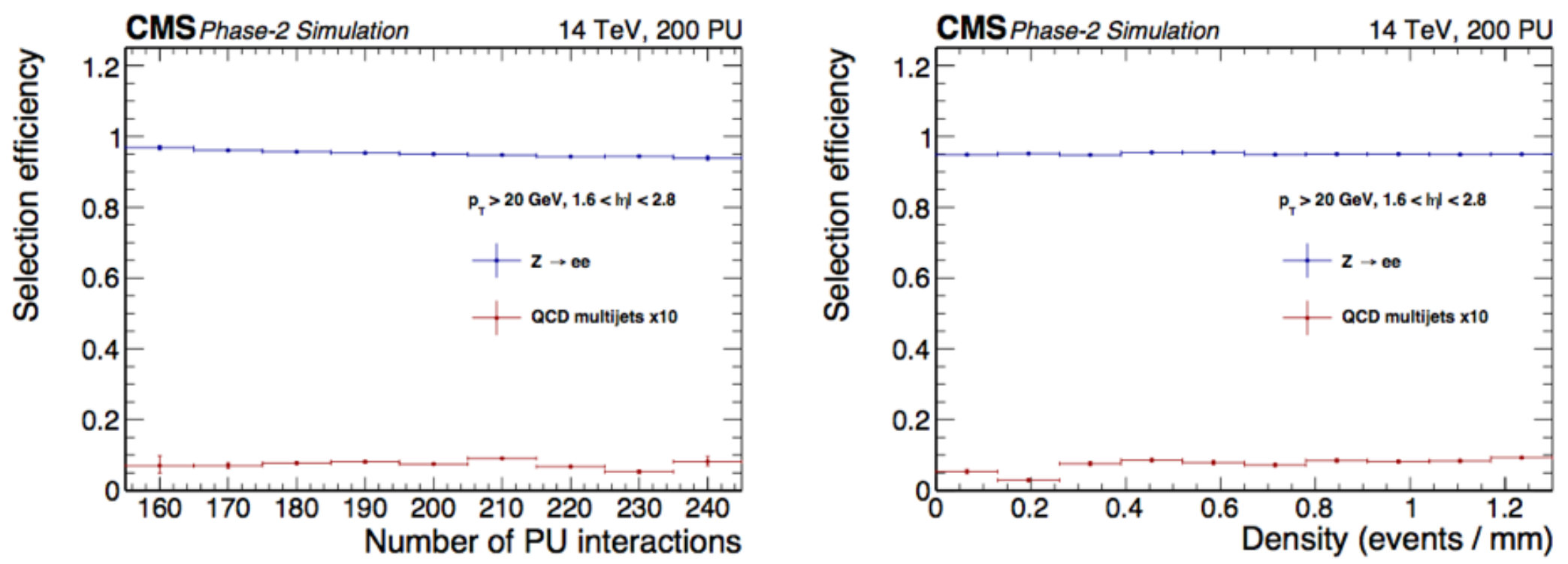

Figure 12: The electron (blue) and QCD multijet misidentification efficiency (red) reconstruction efficiency for $p_T > 20$ GeV in the HGCal region is shown as a function of the number of PU interactions (left) and as a function of the PU density (right). Taken from Ref. [8].

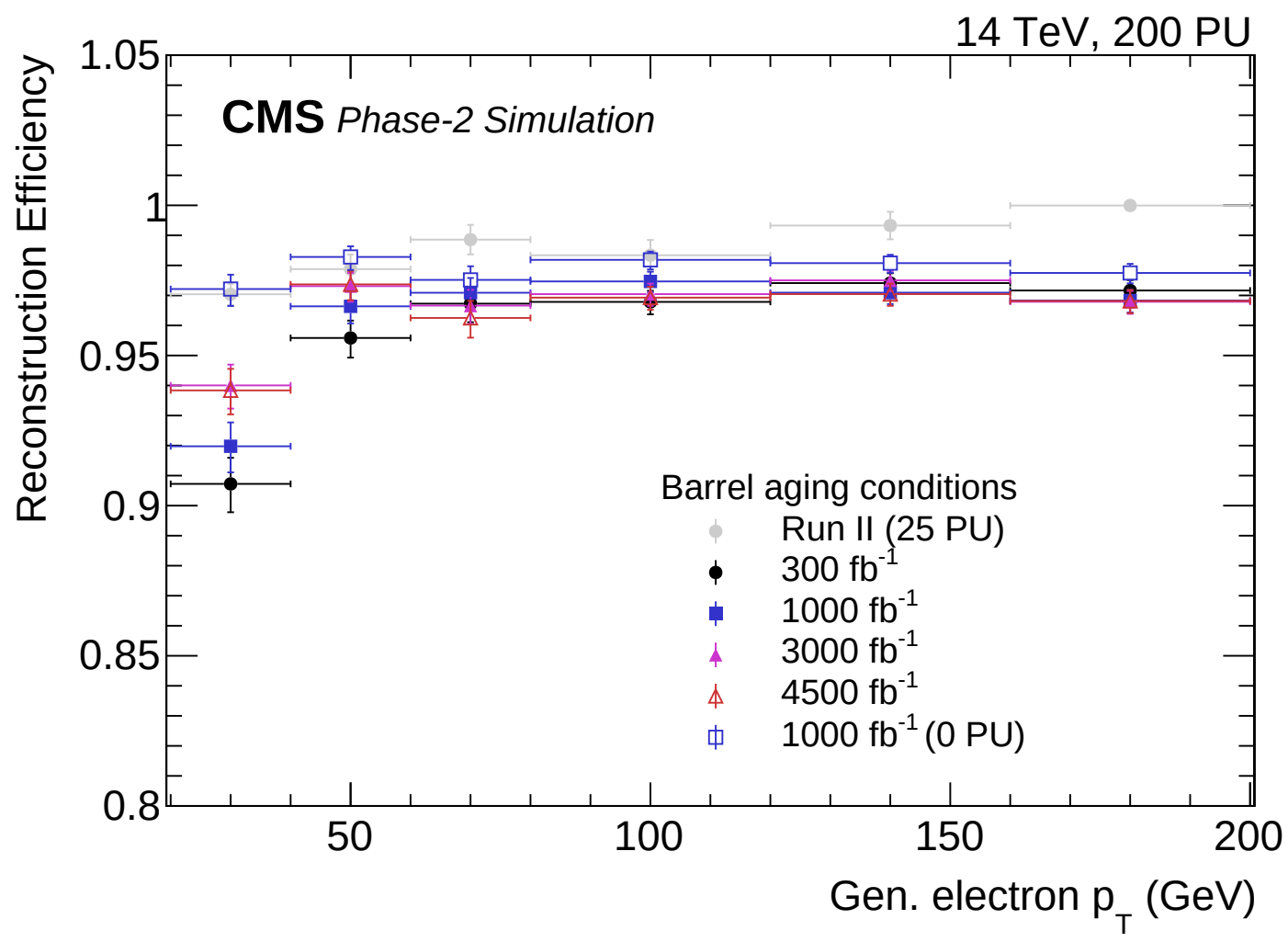

Figure 13: The electron reconstruction efficiency for several ECAL barrel aging conditions. The efficiency is defined as the number of reconstructed electrons matched within $\Delta R(\eta, \phi) < 0.1$ of a generated electron, divided by the number of generated electrons within the acceptance region $|\eta| < 1.4$. The electrons were generated with a uniform distribution in $p_{\mathrm{T}}$. Taken from Ref. [7].

## 2.3 Photon performance

In this section, the photon performance is discussed [7, 8]. Figure 14 shows the photon efficiency as a function of the photon misidentification probability. Based on this figure, several working points are defined. Figure 15 shows the photon reconstruction efficiency, identification efficiency, and photon misidentification probability as a function of the generated photon $|\eta|$ and $p_{\mathrm{T}}$, for the working points defined by Fig. 14. Figure 16 shows the photon reconstruction efficiency for several ECAL barrel aging conditions. The impact of PU and aging can be further mitigated with the optimization of the clustering algorithm.

## 2.4 Muon performance

Here we describe the muon performance in the CMS Phase-2 detector [5, 9]. Figures 17 and 18 show the muon reconstruction and identification efficiency and the muon background multiplicity, respectively, as a function of $|\eta|$. One can see the efficiency of the upgraded muon system is expected to be remarkably resilient to the HL-LHC adverse conditions. Muon reconstruction in the extended pseudorapidity range, $2.4 < |\eta| < 2.8$, is also highly efficient and robust. The rate of background muons in the full pseudorapidity range, including $2.4 < |\eta| < 2.8$, remains fairly independent of the PU conditions.

Figures 19 and 20 show the efficiency to identify prompt muons from Drell-Yan events and nonprompt muons from $\mathrm{t\bar{t}}$ events using charged isolation. The effect with and without precision timing is shown. Tracks entering the isolation sum are associated to the signal vertex within a window of $|D_z| < 1$ mm, and $|D_t| < 3\sigma_t$ in the case of precision timing, where the nominal timing resolution is 30 ps. A clear benefit can be seen in terms of a reduced nonprompt efficiency in the with-timing case when the prompt efficiency is greater than 80%. Furthermore, the impact of precision timing is evident at high event densities, with an acceptance gain of about 6% at the average event density of 1.4 mm with PU 200. The isolation efficiency at 200 PU with timing is equivalent to the isolation efficiency of current-era LHC PU densities without timing,

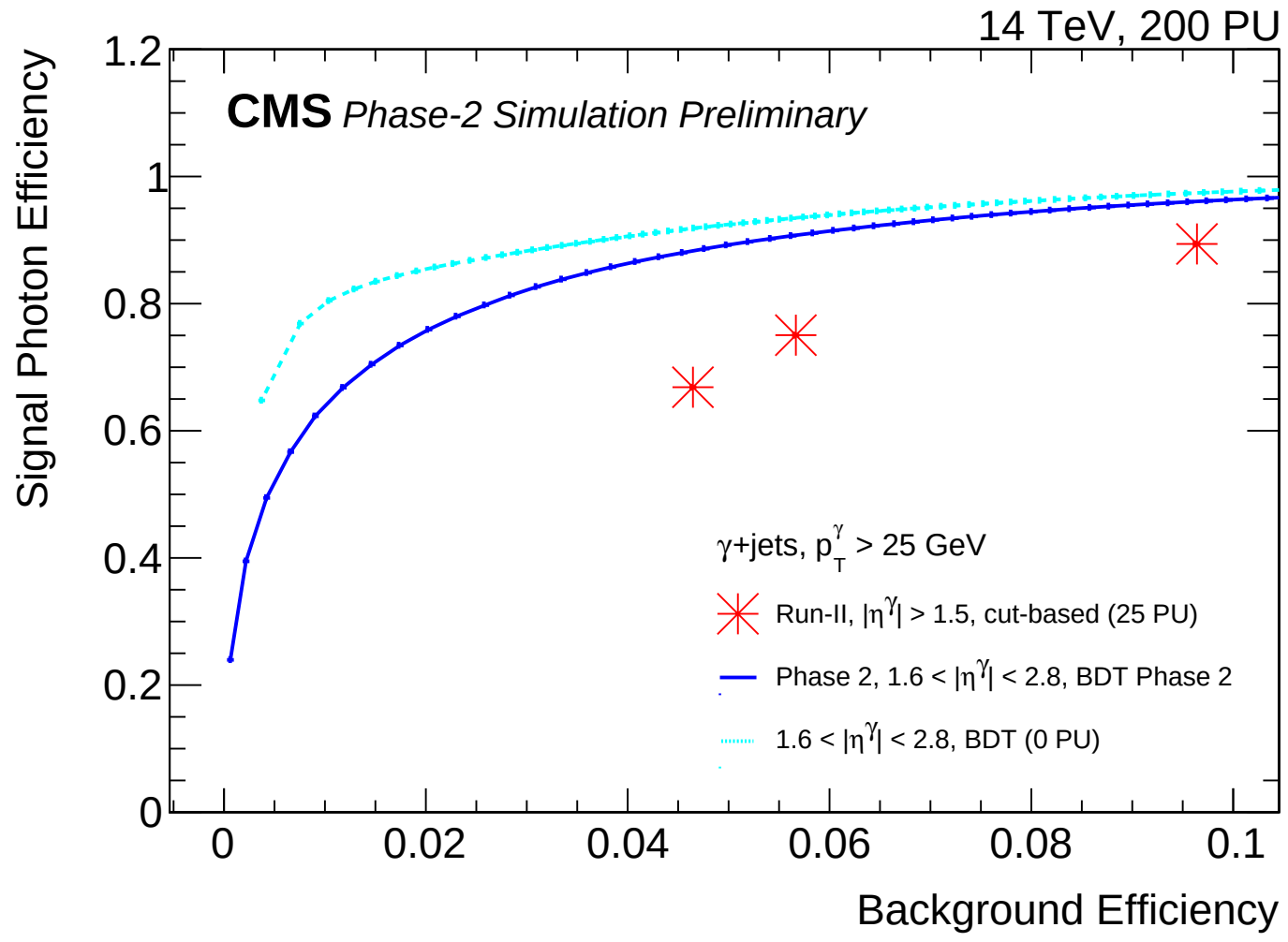

Figure 14: The photon efficiency as a function of the photon-misidentification probability in simulated $\gamma+$ jets events for the BDT training described in the text. Signal photons are matched within $\Delta R < 0.1$ to isolated photons generated within the kinematic phase space $p_{\mathrm{T}} > 25\,\mathrm{GeV}$ and $1.6 < |\eta| < 2.8$. Misidentified photons are defined as reconstructed photons found in the same kinematic phase space but not matched to an isolated generated photon. The performance of a Run 2 selection criteria-based identification is also presented, evaluated on a similar sample of $\gamma+$ jets events produced using the Run 2 conditions (25 PU at $\sqrt{s} = 13\,\mathrm{TeV}$). Taken from Ref. [8].

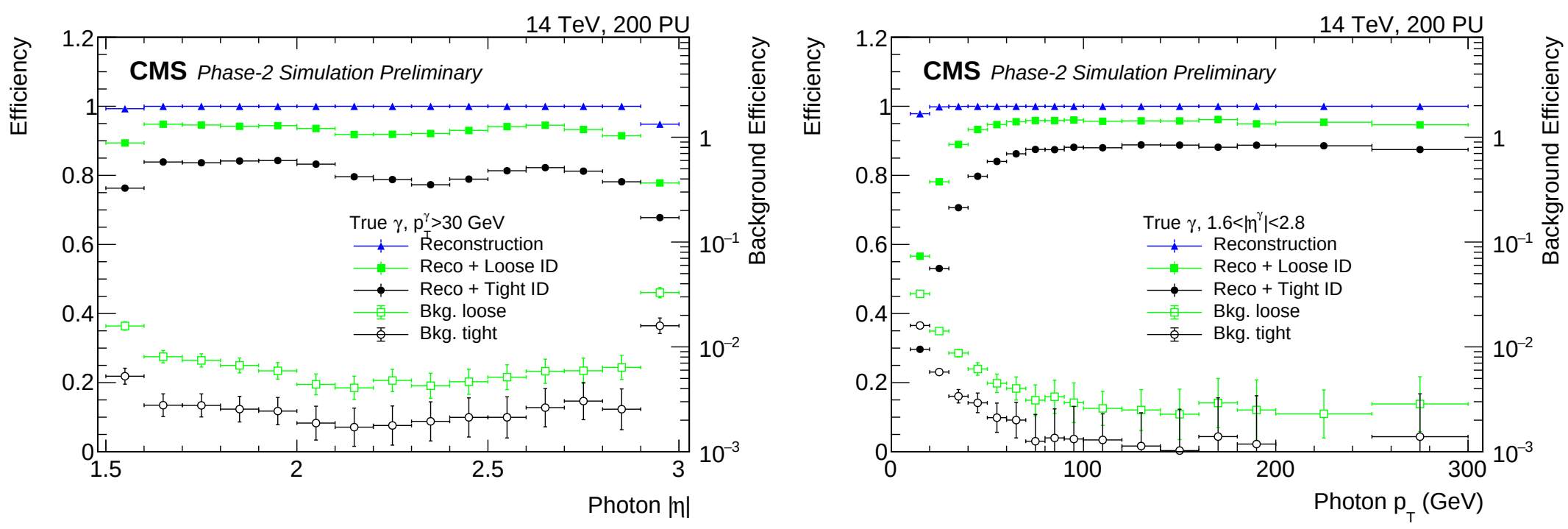

Figure 15: The photon reconstruction efficiency, identification efficiency, and misidentification probability, for two identification working points, as a function of the generated photon $|\eta|$ (left) and $p_{\mathrm{T}}$ (right). The photon reconstruction efficiency is defined as the efficiency for which a reconstructed photon is found within $\Delta R < 0.1$ of a generated prompt photon. Identification efficiencies for signal photons are relative to the generated prompt photon. Misidentified photons are defined as reconstructed photons not matched to an isolated generated photon. Taken from Ref. [8].

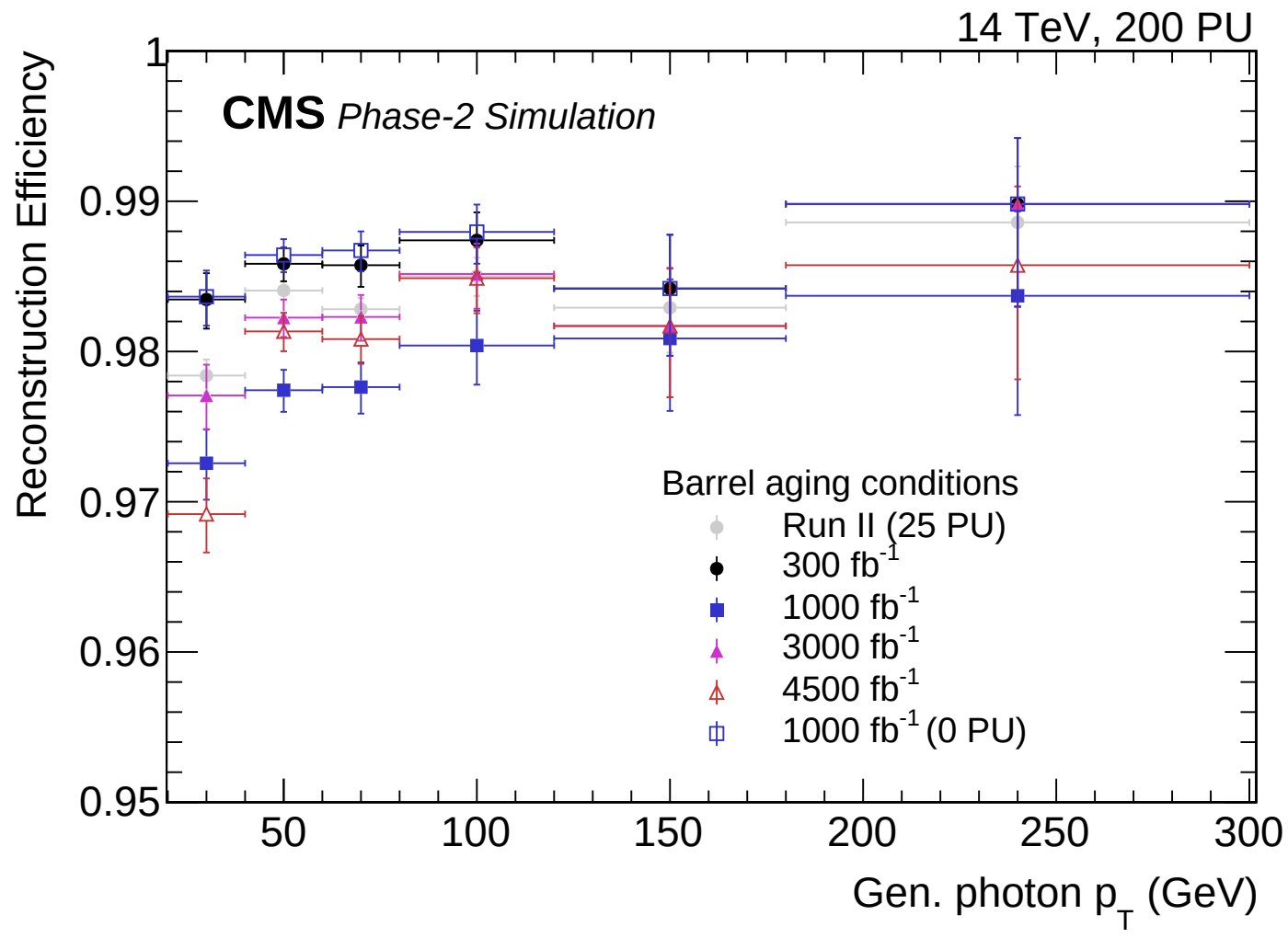

Figure 16: The photon reconstruction efficiency for several ECAL barrel aging conditions. The efficiency is defined as the number of reconstructed photons matched within $\Delta R < 0.1$ of a generated prompt photon from decays of a Higgs boson to two photons, divided by the number of generated photons within the acceptance region $|\eta| < 1.4$. Taken from Ref. [7].

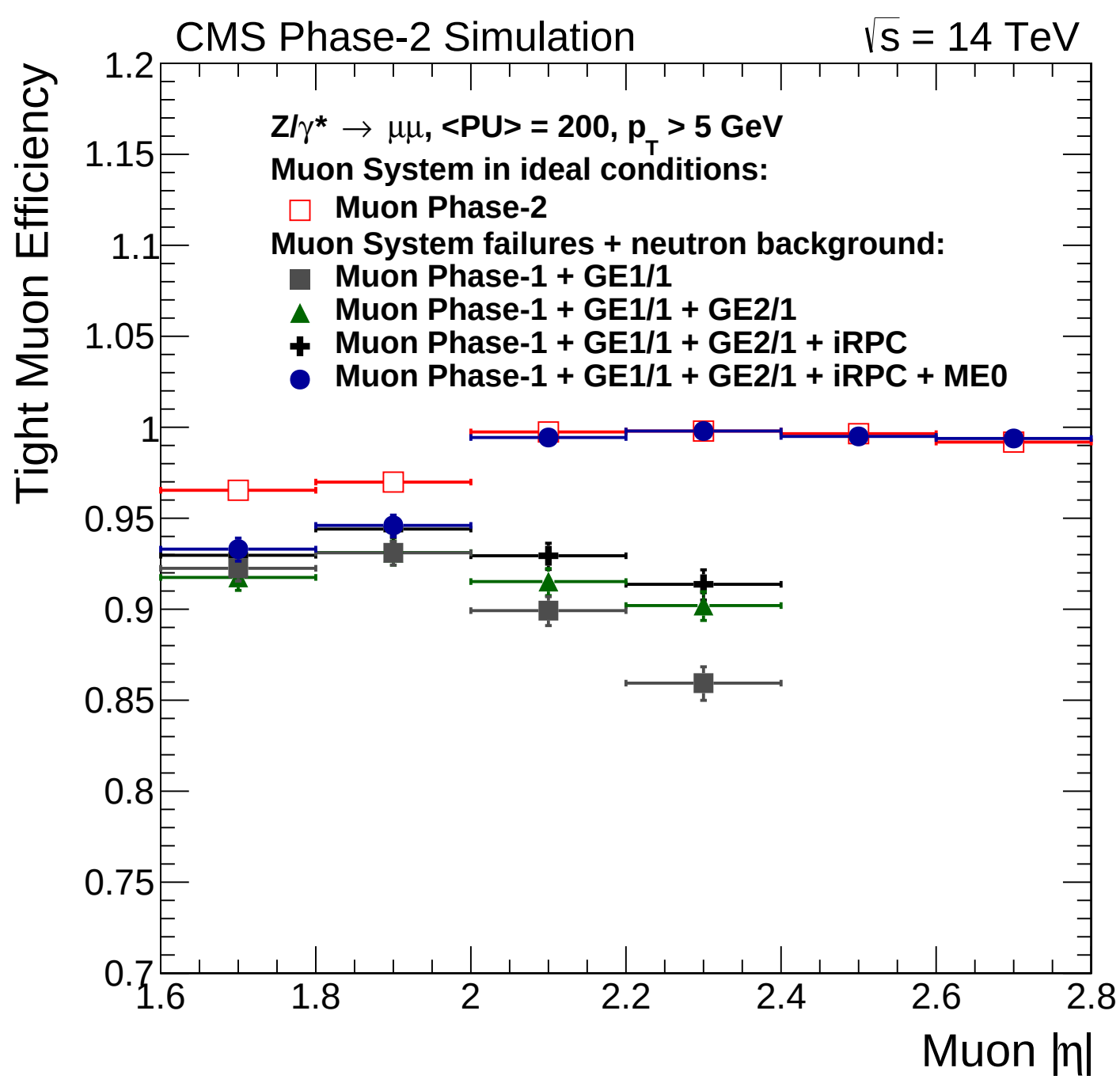

Figure 17: The muon reconstruction and identification efficiency with statistical uncertainty in Drell-Yan events, as a function of a simulated muon's $|\eta|$, for tight muon selection criteria. Results for different detector configurations are shown. The solid points assume 200 PU, HL-LHC neutron background, and a model of the muon system aging. The open squares show the results for the un-aged muon system and without the neutron-induced background. Taken from Ref. [9].

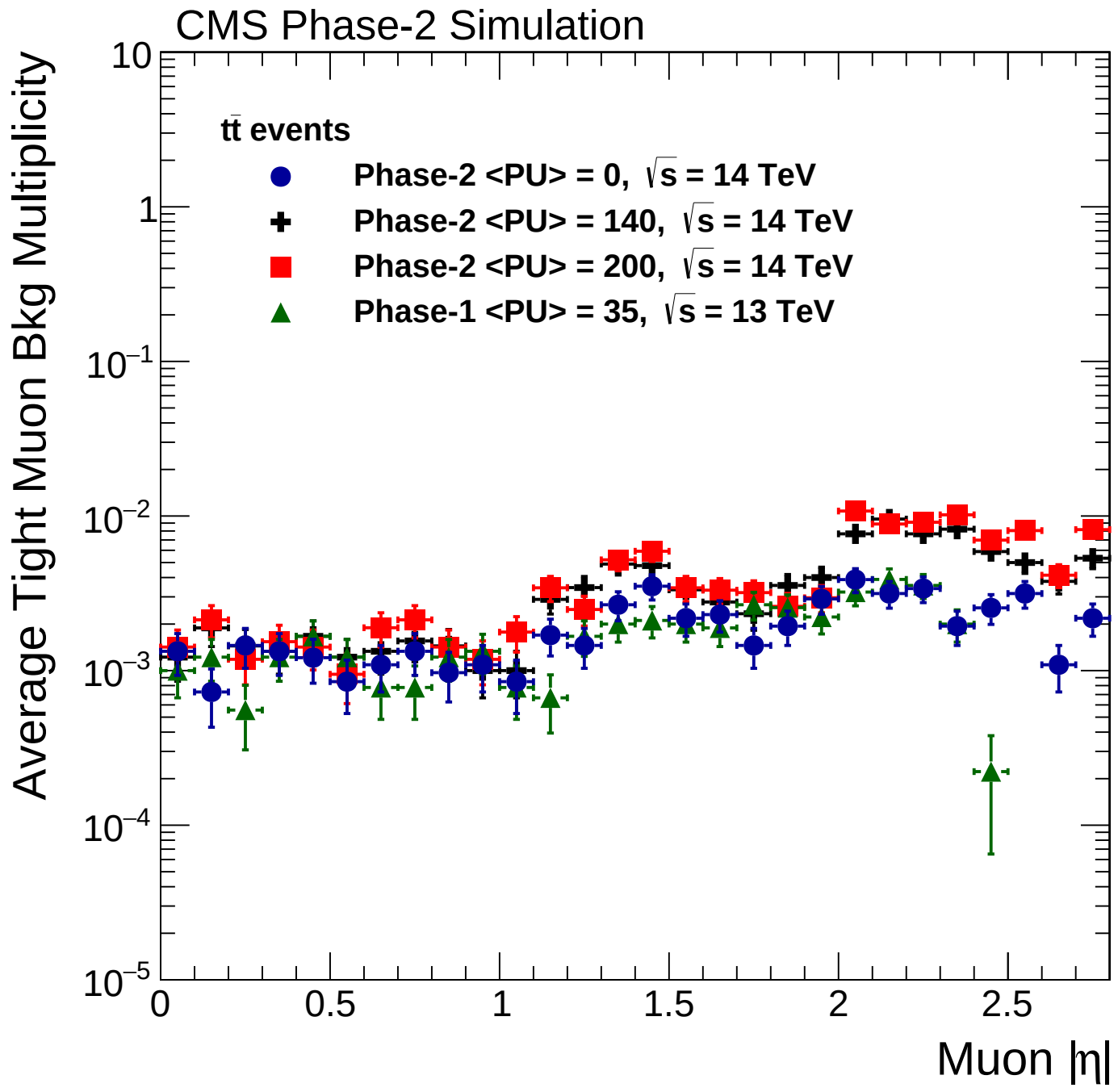

Figure 18: The average background-muon multiplicity in $t\bar{t}$ events as a function of muon $|\eta|$ for the Phase-2 detector in three PU scenarios, compared to the performance of the Phase-1 detector. Taken from Ref. [9].

at constant background.

## 2.5 Tau lepton performance

Here we describe the tau lepton performance in the CMS Phase-2 detector [5, 8]. Figures 21 and 22 show the efficiency and misidentification probability for $\tau$ leptons that decay into hadrons ($\tau_{had}$) as a function of $\eta$ and $p_T$, respectively. The reconstruction efficiency is stable, and does not depend on running conditions or the physical process that produces for $\tau$ leptons. The misidentification probability of QCD multijets as $\tau_{had}$ leptons increases with $p_T$ because high $p_T$ jets are more collimated. The performance of the $\tau_{had}$ reconstruction is similar to that achieved in the recent Run 2 CMS data taking.

Figures 23 and 24 show the performance of isolated $\tau_{had}$ leptons. In terms of the charged isolation efficiency for real tau leptons, there is an improvement of performance at 200 PU with timing that exceeds that of the current-era LHC PU densities without timing. In addition, one can see the overall benefits in terms of recovered prompt candidate efficiency tracks with time resolution. The efficiency gain is still sizeable at 50 ps resolution.

## 2.6 Jet performance

The jet performance [5, 8] is shown in this section. Jets are reconstructed offline from the energy deposits in the calorimeter towers and clustered using the anti-$k_T$ algorithm [11, 12] with a distance parameter of 0.4.

Figures 25 and 26 show the corrected jet response resolution as a function of the generated jet $p_T$ ($p_T^{Gen}$) and as a function of the PU density, respectively. Only modest degradation of the jet resolution are observed relative to the central part of the detector for jets with $1.7 < |\eta| < 2.8$,

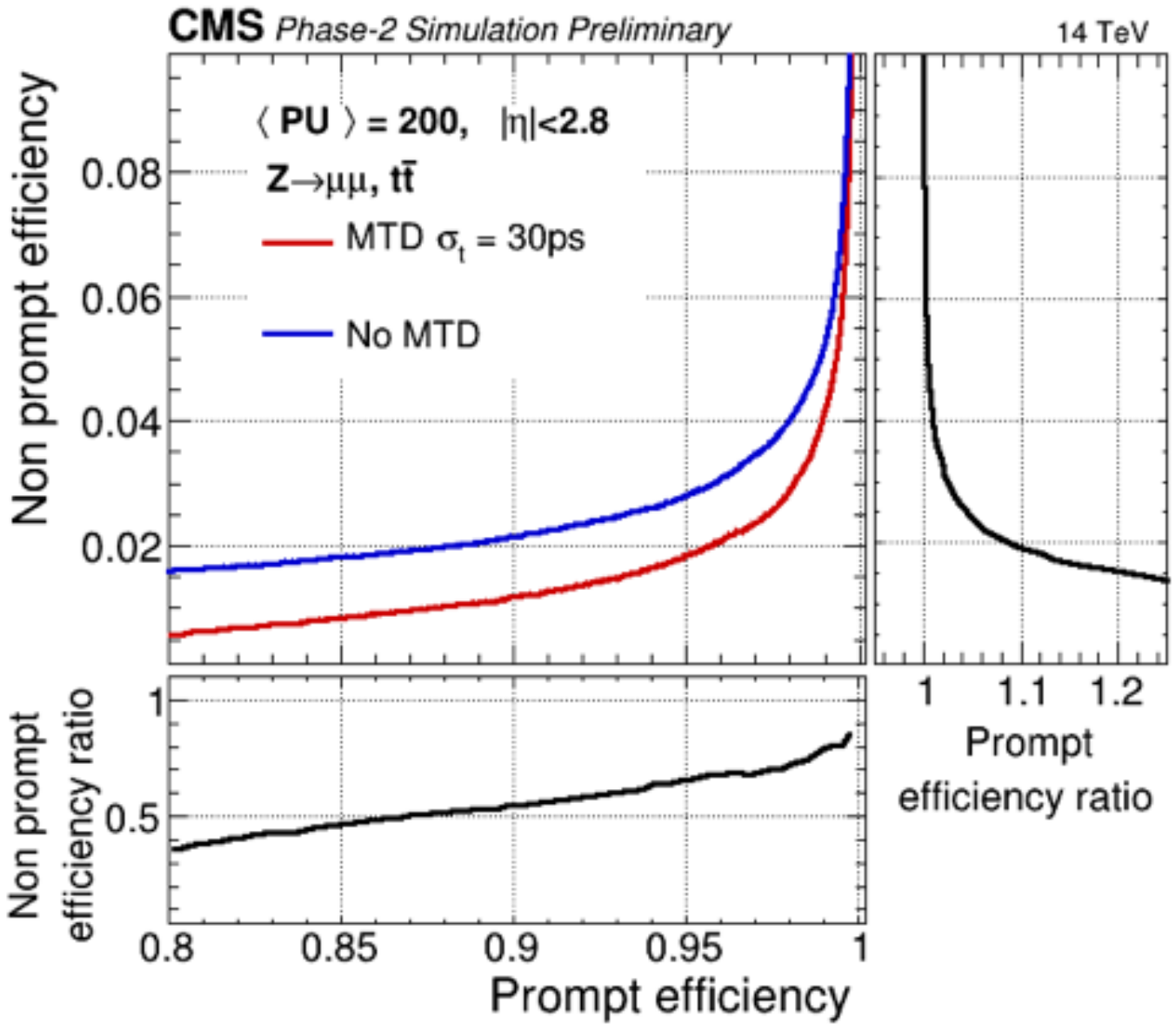

Figure 19: The efficiency for identifying prompt muons from Drell-Yan events and nonprompt muons from $t\bar{t}$ events using charged isolation is shown, with and without precision timing from the MTD for charged particles. Tracks entering the isolation sum are associated to the signal vertex within a window of $|D_z| < 1$ mm, and $|D_t| < 3\sigma_t$ in the case of precision timing, where the nominal timing resolution is 30 ps. The bottom panel shows the ratio of the nonprompt efficiency with the MTD precision timing divided by without, at constant prompt muon efficiency. The right panel shows the ratio of the prompt muon efficiency with the MTD precision timing divided by without, at constant nonprompt muon efficiency. Taken from Ref. [5].

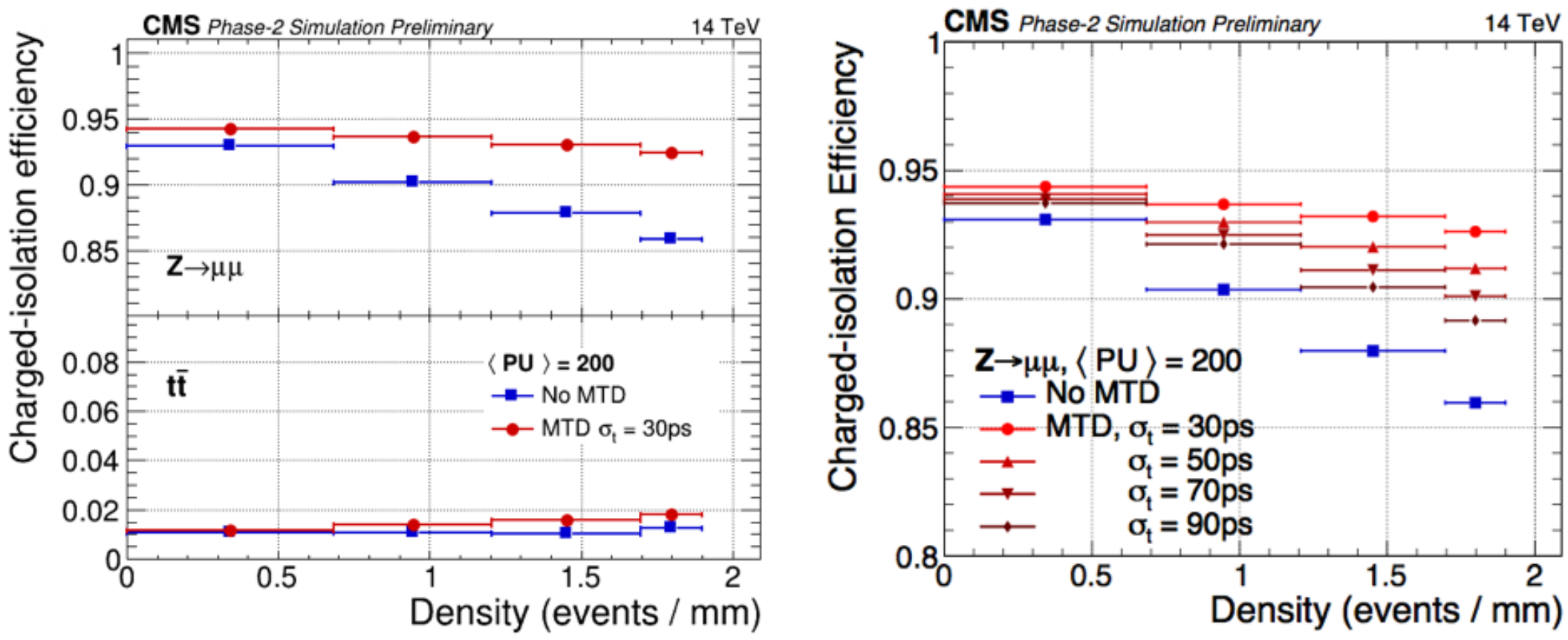

Figure 20: Left: The efficiency for identifying prompt muons from Drell-Yan events and nonprompt muons from $t\bar{t}$ events using charged isolation is shown as a function of PU density, with and without precision timing for charged particles. Tracks entering the isolation sum are associated to the signal vertex within a window of $|D_z| < 1$ mm, and $|D_t| < 3\sigma_t$ in the case of MTD precision timing, where the nominal timing resolution is 30 ps. Right: The efficiency for identifying prompt muons with different assumptions for the MTD precision timing resolution is shown, where the track-vertex association criteria with timing is always $|D_t| < 3\sigma_t$. Taken from Ref. [5].

**14**

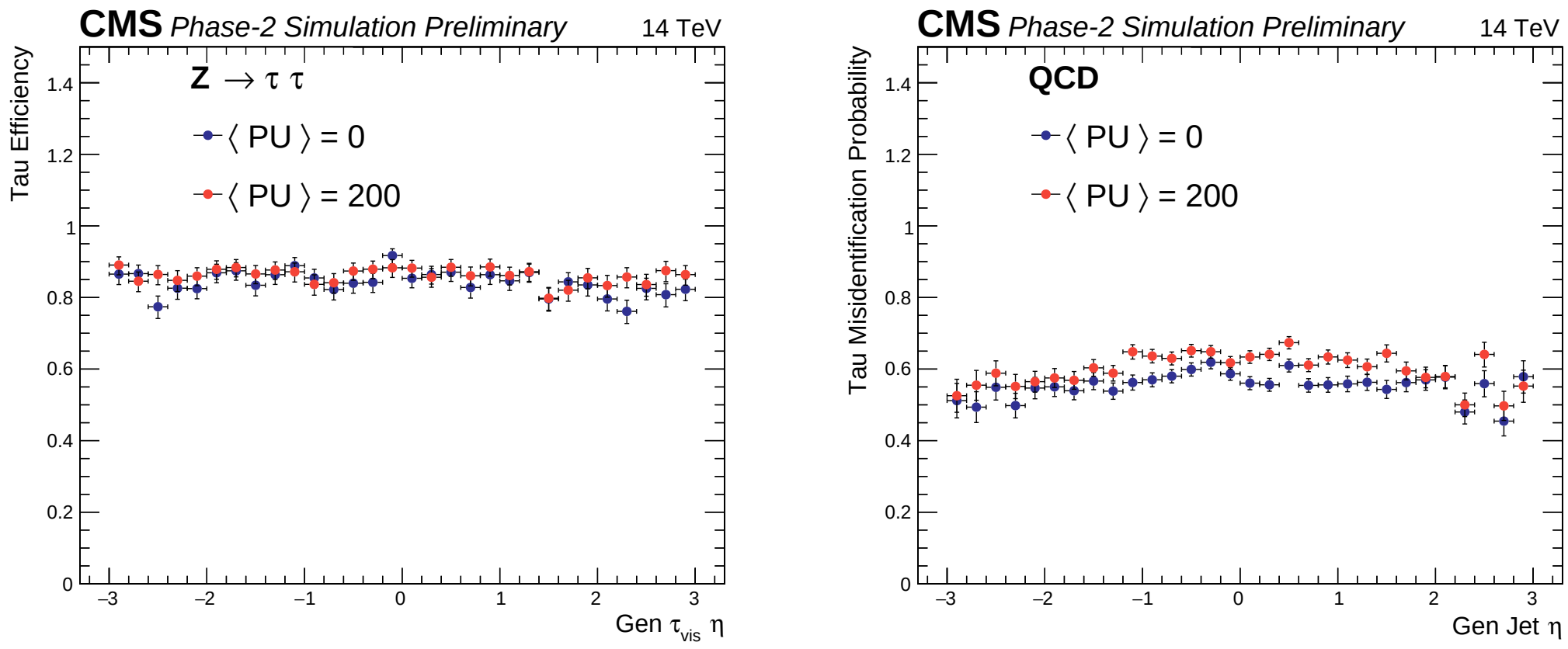

Figure 21: The $\tau_{had}$ efficiency (left) and QCD multijet misidentification probability (right) as a function of $\eta$. Taken from Ref. [8].

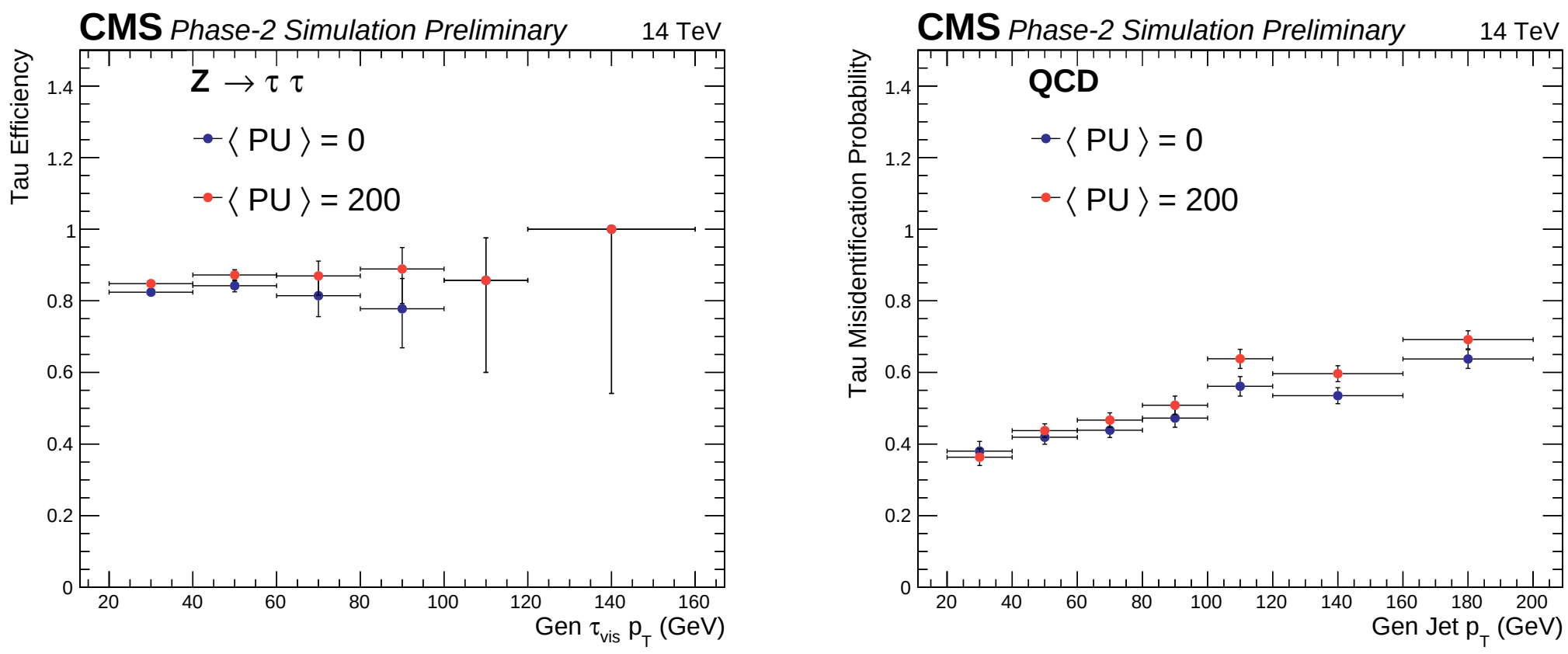

Figure 22: The $\tau_{had}$ efficiency (left) and QCD multijet misidentification probability (right) as a function of $p_{\mathrm{T}}$. Taken from Ref. [8].

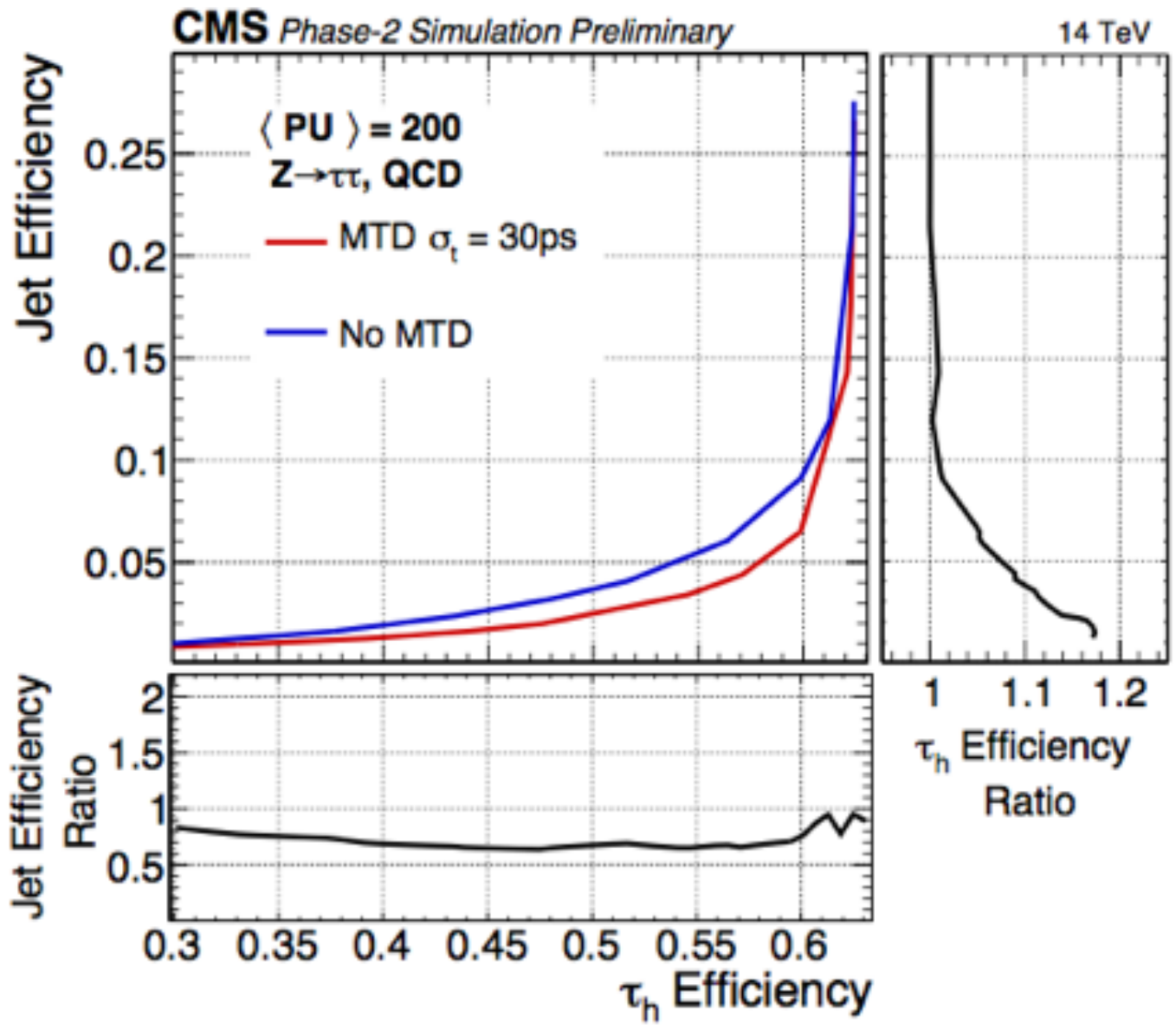

Figure 23: The performance of $\tau_{had}$ charged isolation in simulated $Z/\gamma* \to \tau\tau$ and QCD multi-jet events, expressed as misidentification probability for generator-matched jets as a function of the $\tau_{had}$ identification efficiency for events with 200 PU with (red) and without (blue) the MIP timing window requirement. The timing detector is assumed to have a resolution of 30 ps. The bottom panel shows the ratio of the QCD multijet efficiency with divided by without the MTD precision timing, at constant $\tau_{had}$ efficiency. The right panel shows the ratio of the $\tau_{had}$ efficiency with divided by without the MTD precision timing, at constant QCD multijet efficiency. Taken from Ref. [5].

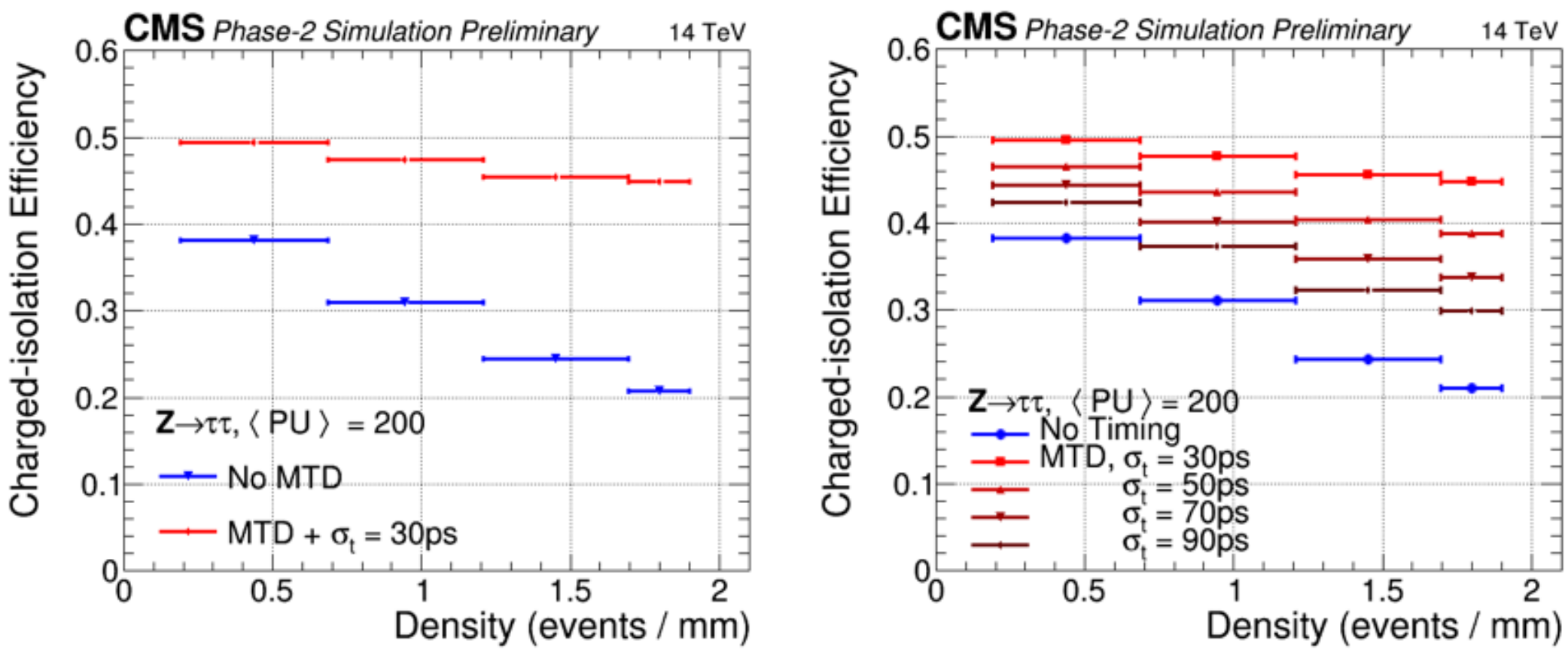

Figure 24: The efficiency of identifying isolated $\tau_{had}$ lepton decays as a function of the PU density for $p_T > 20$ GeV and $|\eta| < 2.4$ is shown. The efficiency (left) is shown with and without the MTD precision timing, where the nominal timing resolution is 30 ps. The efficiency (right) is shown without precision timing and with several different values for the nominal timing resolution. The efficiency is computed using $Z/\gamma* \to \tau\tau$ events for the Phase-2 detector configuration with a fixed cut at charged-isolation $<$2.5 GeV. Taken from Ref. [5].

and only a small degradation in the jet energy resolution is observed as a function of the PU density.

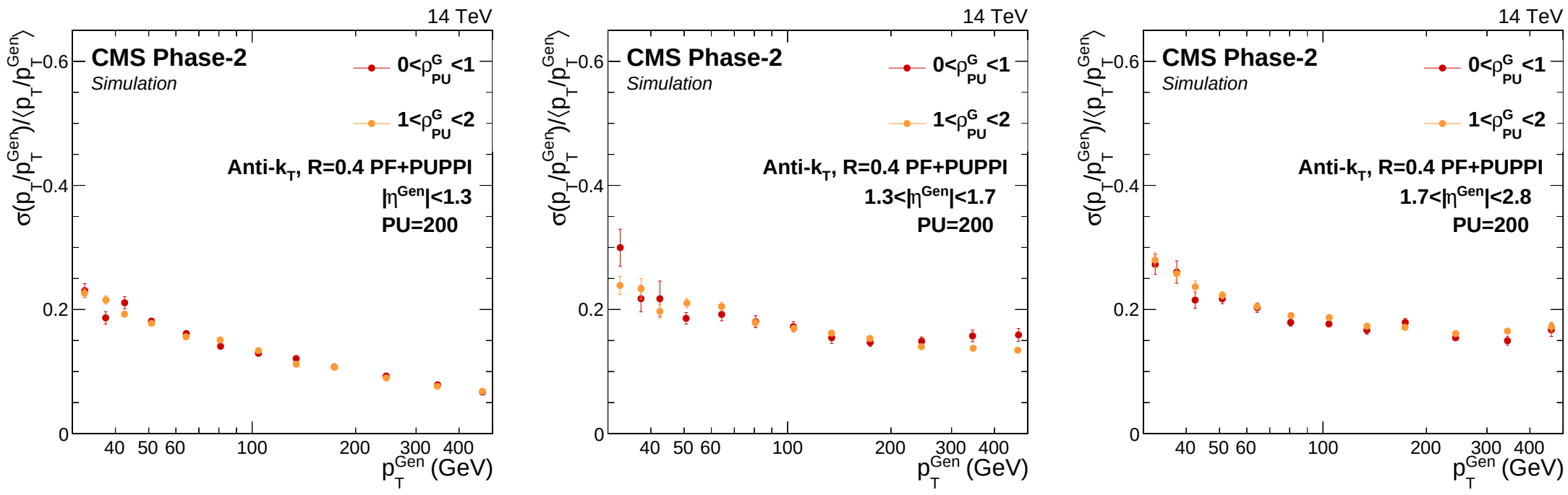

Figure 25: The corrected jet response resolution for $|\eta| < 1.3$ (left), $1.3 < |\eta| < 1.7$ (middle), and $1.7 < |\eta| < 2.8$ (right) as a function of $p_T^{Gen}$ for PF+PUPPI jets with 200 PU. Taken from Ref. [8].

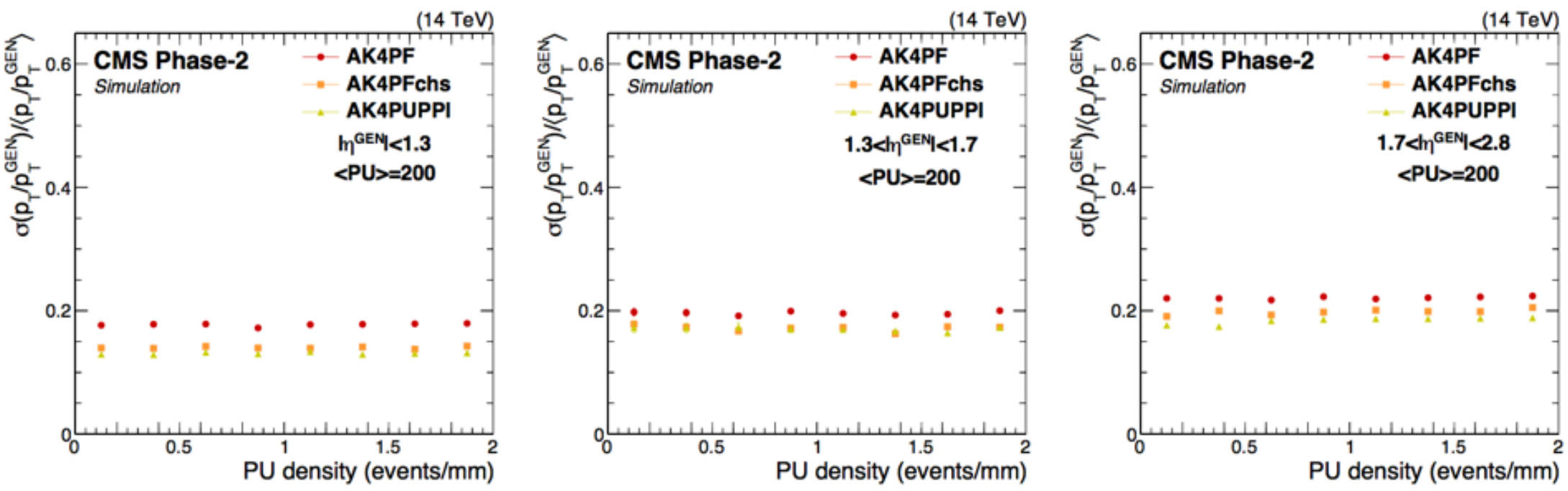

Figure 26: The corrected jet response resolution for $|\eta| < 1.3$ (left), $1.3 < |\eta| < 1.7$ (middle), and $1.7 < |\eta| < 2.8$ (right) as a function of the PU density for different jet algorithms with 200 PU. The jet algorithms shown are PF jets, PF jets with charged hadron subtraction, and PUPPI jets. All jets have been matched to a particle-level jet with $90\,\mathrm{GeV} < p_T^{Gen} < 120\,\mathrm{GeV}$. Taken from Ref. [8].

Particles originating from PU interactions may accidentally be clustered into overlapping low-$p_T$ jets that combine to form a single high-$p_T$ jet, referred to as a PU jet. The rate of these PU jets is quantified by the ratio of the average number of jets in a given $p_T$ bin to the average number of reconstructed jets matched to a particle-level jet. Figure 27 shows the fraction of the number of jets out of the number of jets matched to a generator level jet with $p_T > 10\,\mathrm{GeV}$ as a function of PU. Figure 28 shows the rate of signal and PU jets, both of which are reconstructed with the PUPPI algorithm, with and without the MTD precision timing. The PU jet rate for jets with $1.3 < |\eta| < 3.0$ is only moderately degraded relative to the central barrel part of the detector.

## 2.7 b tagging performance

The b tagging efficiency as a function of the jet $p_T$ is shown in Fig. 29 [8]. Compared to events without PU, the b jet tagging efficiency remains large at high PU in all $p_T$ and $|\eta|$ regions of interest.

The secondary vertex tagging misidentification probability as a function of the b tagging efficiency is shown in Fig. 30 [5]. In very high PU conditions, secondary vertex b tagging is degraded by the formation of spurious secondary vertices caused by PU tracks, reducing the ability to distinguish signal from background. In order to mitigate this problem, the secondary vertexing algorithms were updated to be aware of timing information from the MTD. By requiring tracks to be within $3.5\sigma_t$ of the primary vertex, the number of spurious reconstructed secondary vertices was reduced by 30%. This causes the ROC curves in Fig. 30 to improve significantly, especially for tighter working points where near-zero PU performance is achieved and the dependence of b tagging efficiency on the PU density is removed.

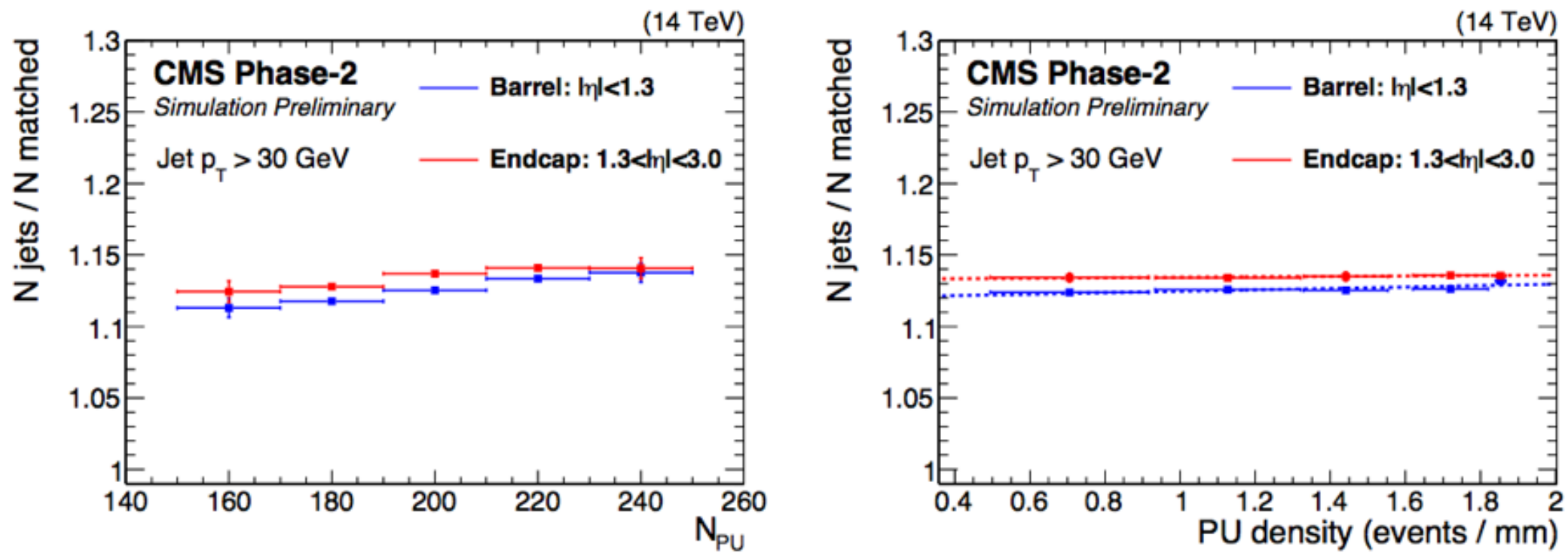

Figure 27: (N jets)/(N jets matched to a generator level jet with $p_T > 10$ GeV) as a function of the number of PU interactions (left) and as a function of the PU density (right) for PUPPI jets. The PU rate in the endcap is only slightly degraded, compared to that in the barrel. Taken from Ref. [8].

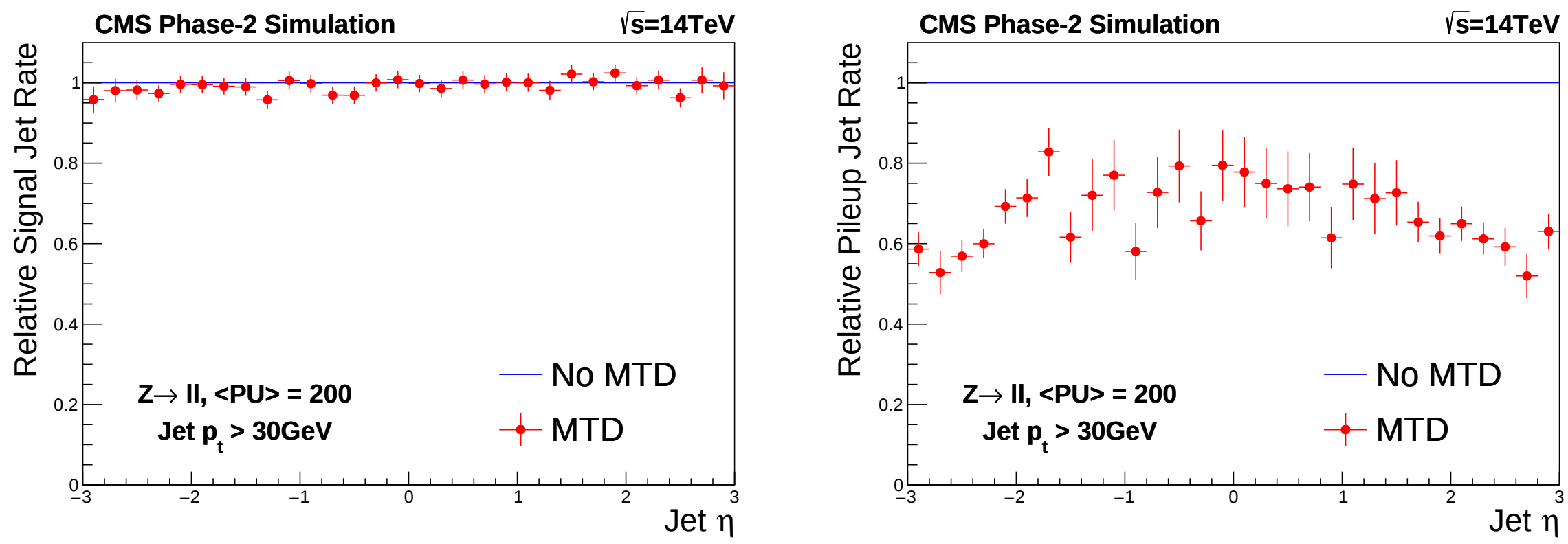

Figure 28: The rate of signal jets (left) and PU jets (right) reconstructed with the PUPPI algorithm and with $p_T > 30$ GeV, with and without the MTD precision timing. Taken from Ref. [5].

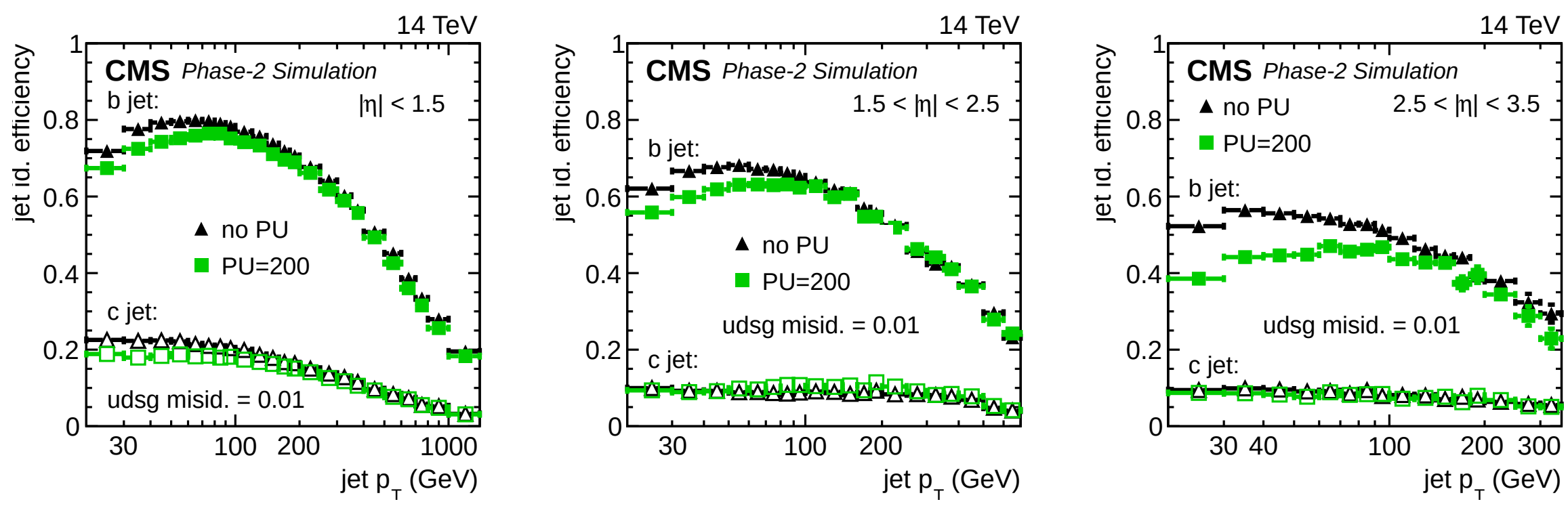

Figure 29: The tagging efficiencies for prompt b jets (filled symbols) and prompt b jets (open symbols) as a function of the jet $p_T$ in simulated multijet events. The tagging efficiencies are evaluated for an average misidentification probability of 0.01 for light parton jets (udsg), and shown for 0 PU (black triangles) and 200 PU (green squares). The tagging efficiencies are shown for three $|\eta|$ ranges: 0–1.5 (left), 1.5–2.5 (middle), and 2.5–3.5 (right). Taken from Ref. [8].

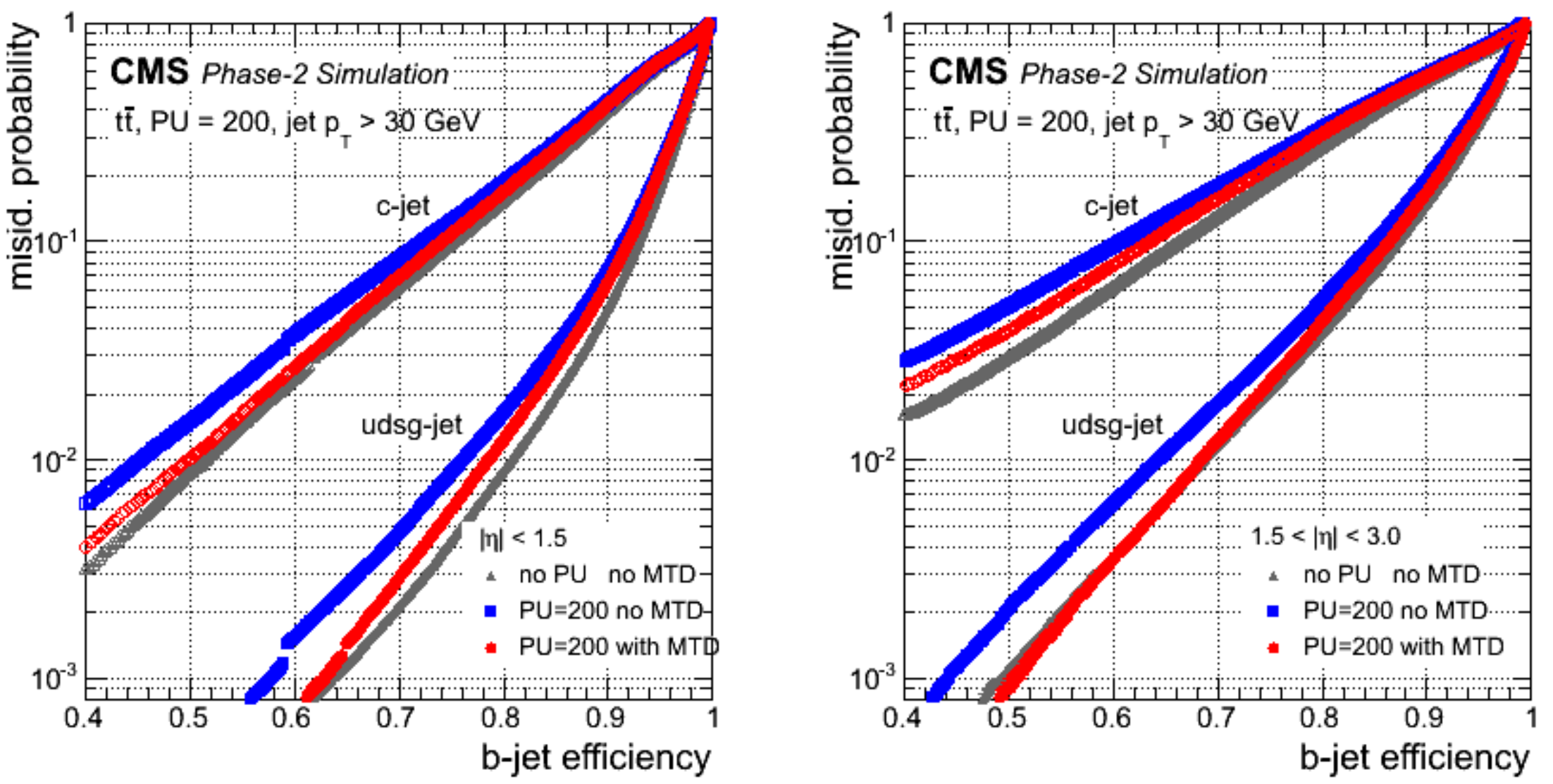

Figure 30: The secondary vertex tagging misidentification probability as a function of the b tagging efficiency, for light and charm jets for $|\eta| < 1.5$ (left) and for $1.5 < |\eta| < 3.0$ (right). Results with and without the MTD precision timing are compared to the 0 PU case. Taken from Ref. [5].

The b tagging performance as a function of the PU density is shown in Fig. 31. A moderate decrease of the b jet tagging efficiency without the MTD precision timing can be observed. With the MTD precision timing, the b jet tagging efficiency is improved and the dependence on PU density is removed.

## 2.8 Jet substructure performance

In this section, the jet substructure performance is shown [8]. Figure 32 shows some jet substructure observables with the current detector and with the Phase-2 detector. The number of jet constituents from quark and gluon jets in simulated QCD multijet events in Fig. 32 left demonstrates the ability to reconstruct identification observables for quark and gluon jet. In the barrel region, the number of constituents decreases slightly in Phase-2 compared to Phase-0. However, an increase is observed in constituents in the HGCal region, which may be attributed to the higher granularity of the endcap calorimeter and the higher number of PU interactions. The $\tau_3/\tau_2$ distributions for top quark jets in high mass resonant $t\bar{t}$ production and quark or gluon jets in multijet simulation in Fig. 32 right demonstrate excellent performance of the HGCal in identifying subjets for highly boosted W, Z, and Higgs bosons, and top quarks, at a level of quality similar to that of the barrel calorimeter.

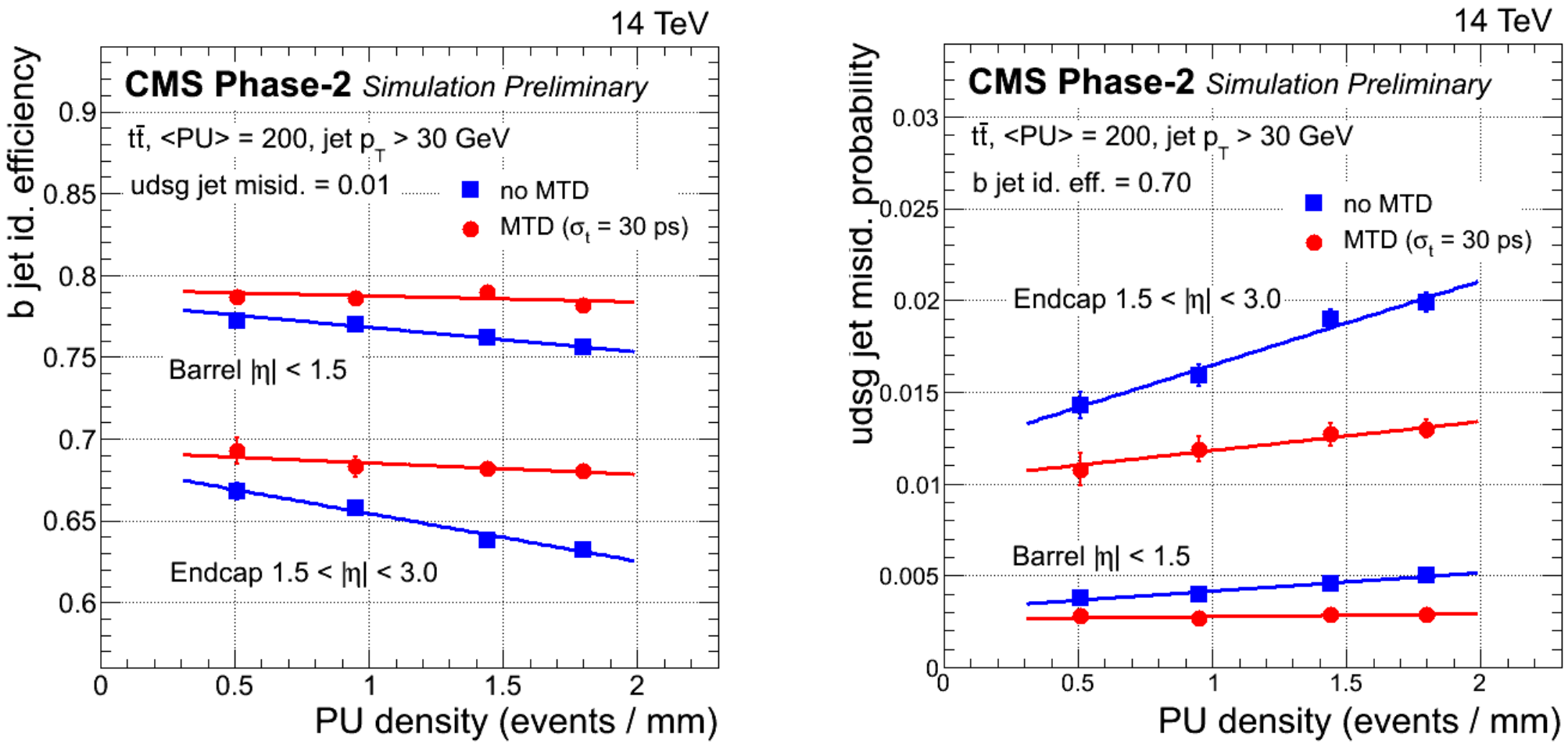

Figure 31: The efficiency of b jet tagging (left) and the light jet misidentification probability (right) are shown as a function of PU density, with and without the MTD precision timing, assuming a timing resolution of 30 ps. The efficiency is computed on $t\bar{t}$ events for a fixed misidentification probability on QCD multijet events of light parton jets (udsg) of 0.01. The misidentification probability is shown for a fixed b jet identification efficiency of 0.70. Linear fits are superimposed for the barrel and endcap pseudorapidity regions. Taken from Ref. [5].

In Fig. 33, the background efficiency as a function of the signal efficiency is shown for jet substructure observables. The discrimination power achievable by the endcap of the Phase-2 detector for boosted W, Z, and Higgs bosons, and top jets against quark/gluon jets is found to be similar or better than in the barrel region.

### 2.9 Missing transverse momentum performance

The missing transverse momentum vector is defined as the projection onto the plane perpendicular to the beam axis of the negative vector sum of the momenta of all reconstructed PF objects in an event. Its magnitude is referred to as missing transverse momentum ($p_{\mathrm{T}}^{\mathrm{miss}}$). The $p_{\mathrm{T}}^{\mathrm{miss}}$ performance is shown in Figs. 34 and 35 [8]. A resolution of about 25 GeV is achieved in the perpendicular component using PUPPI, with the upgraded detector in events containing a mean PU of 200. For comparison, the corresponding resolution in Run 2 is indicated by a dotted magenta line. There is a modest degradation of the resolution with increasing PU density.

### 2.10 MIP timing performance

Here we describe the additional performance of the MTD [5]. Figures 36 and 37 show the number of PU tracks incorrectly associated with the primary vertex as a function of PU density, with and without the MTD precision timing. These results suggest a generic reduction in the effective amount of PU by a factor of approximately four to five for physics quantities constructed from charged particles.

## 3 Systematic uncertainties

The large HL-LHC dataset will enable accurate measurements and unprecedented sensitivity to very rare phenomena. As a result, the current understanding of systematic uncertainties will become a limiting factor for more and more analyses. We attempt to define a set of common

**20**

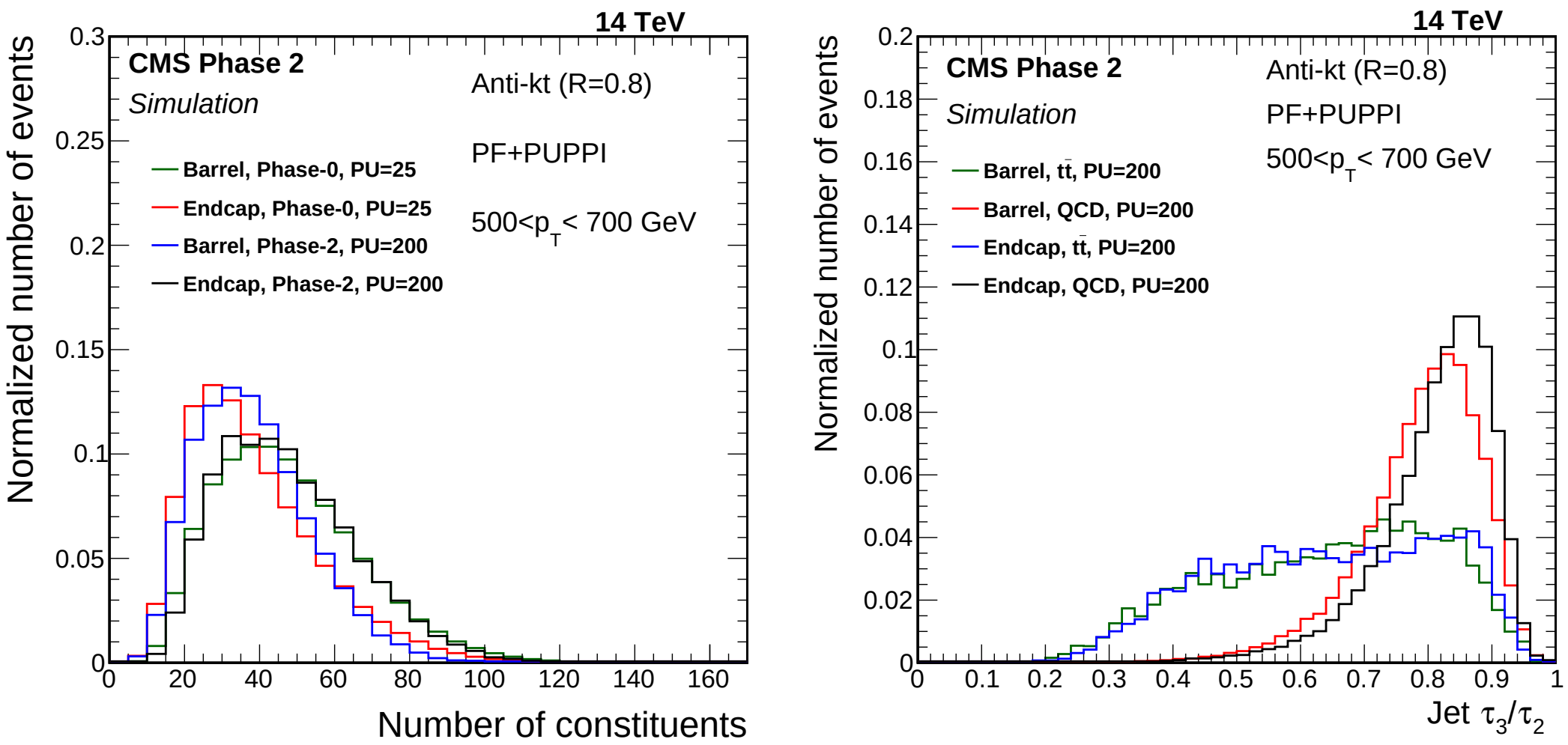

Figure 32: A comparison of jet substructure observables for barrel ($|\eta| < 0.7$) and endcap ($1.9 < |\eta| < 2.4$) calorimeters in Phase-0 (25 PU) and Phase-2 (200 PU). The number of jet constituents from quark or gluon jets in QCD multijet simulation is shown (left), along with the $\tau_3/\tau_2$ for top quark jets in high mass resonant $t\bar{t}$ production and quark or gluon jets in QCD multijet simulation (right). Taken from Ref. [8].

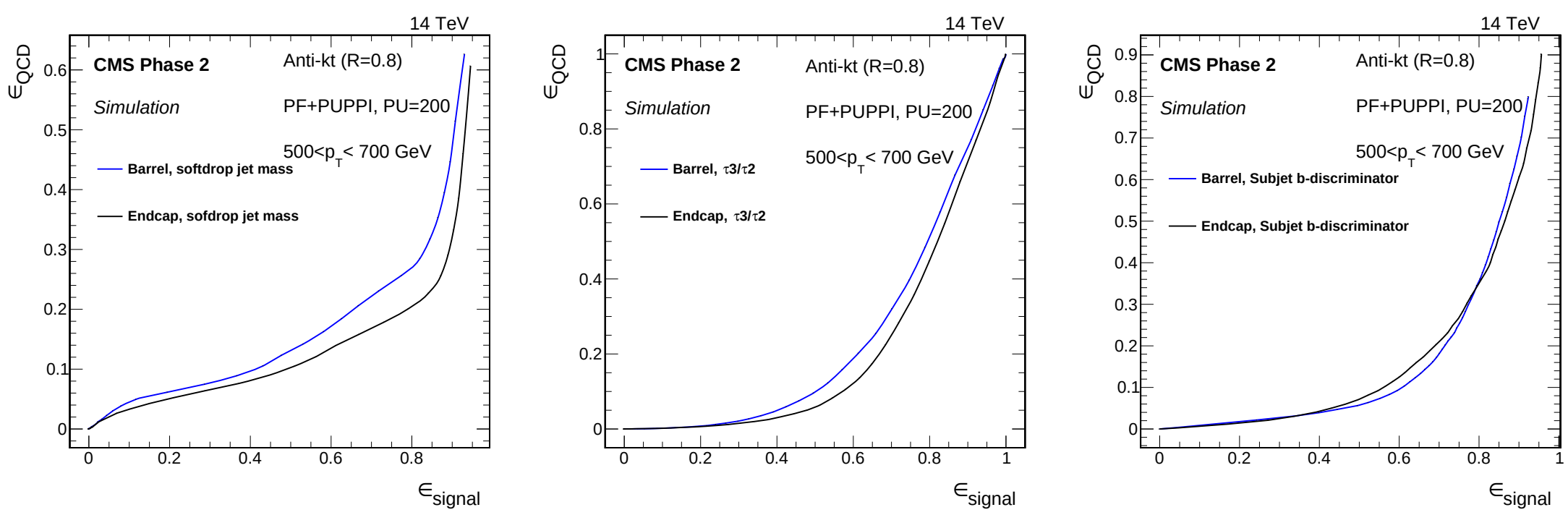

Figure 33: The background efficiency ($\epsilon_{\mathrm{QCD}}$) as a function of the signal identification efficiency ($\epsilon_{\mathrm{signal}}$) for common jet substructure observables in Phase-2 (200 PU) for barrel ($|\eta| < 0.7$) and endcap ($1.9 < |\eta| < 2.4$) calorimeters. The softdrop jet mass (left), $\tau_3/\tau_2$ (middle), and subjet b tagging (right) are shown. Taken from Ref. [8].

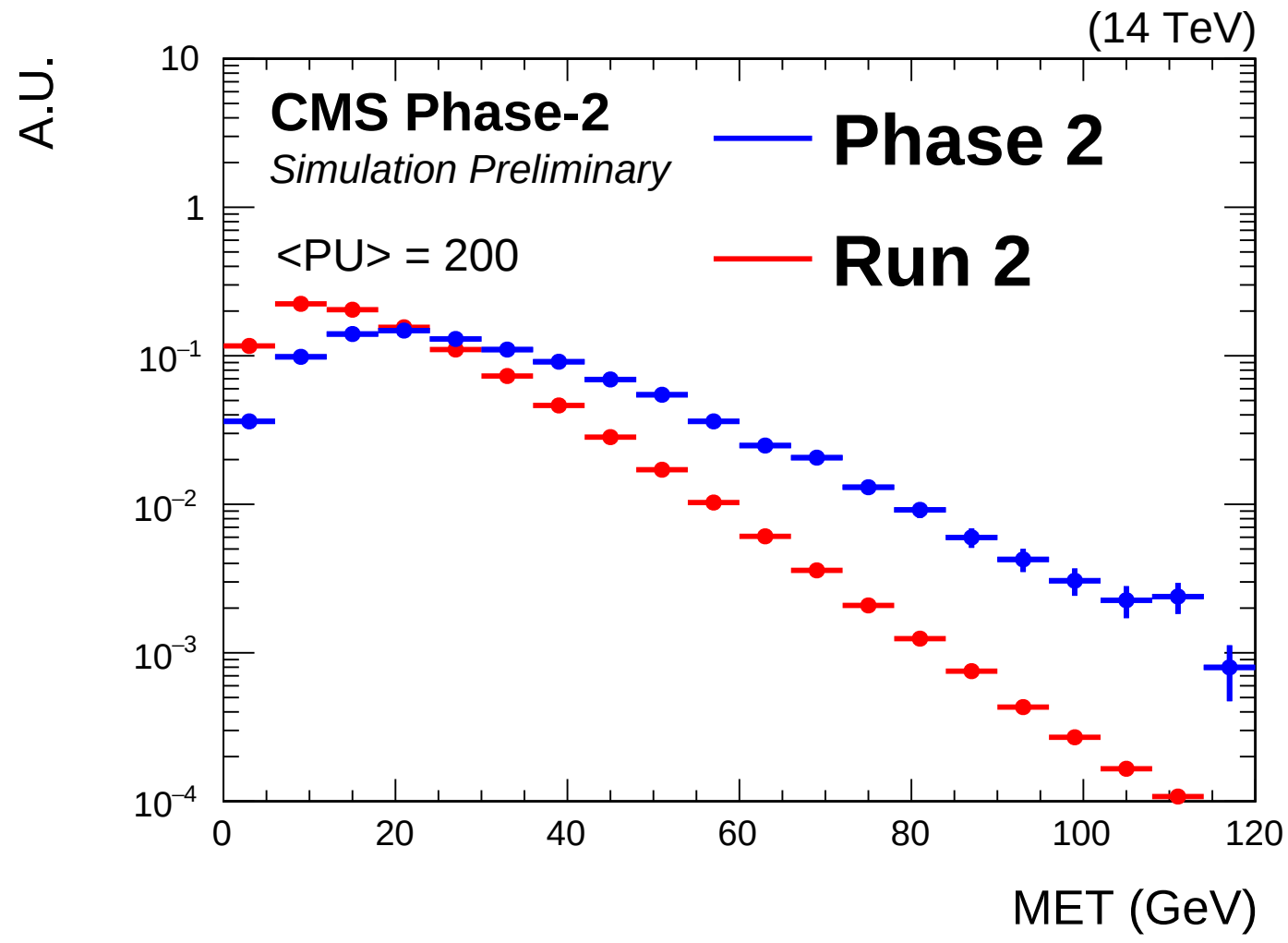

Figure 34: The PUPPI $p_T^{miss}$ distribution for the Phase-2 detector (with PU 200) in $Z \to \mu\mu$ events. The PUPPI $p_T^{miss}$ distribution in Run 2 is shown in red. Taken from Ref. [8].

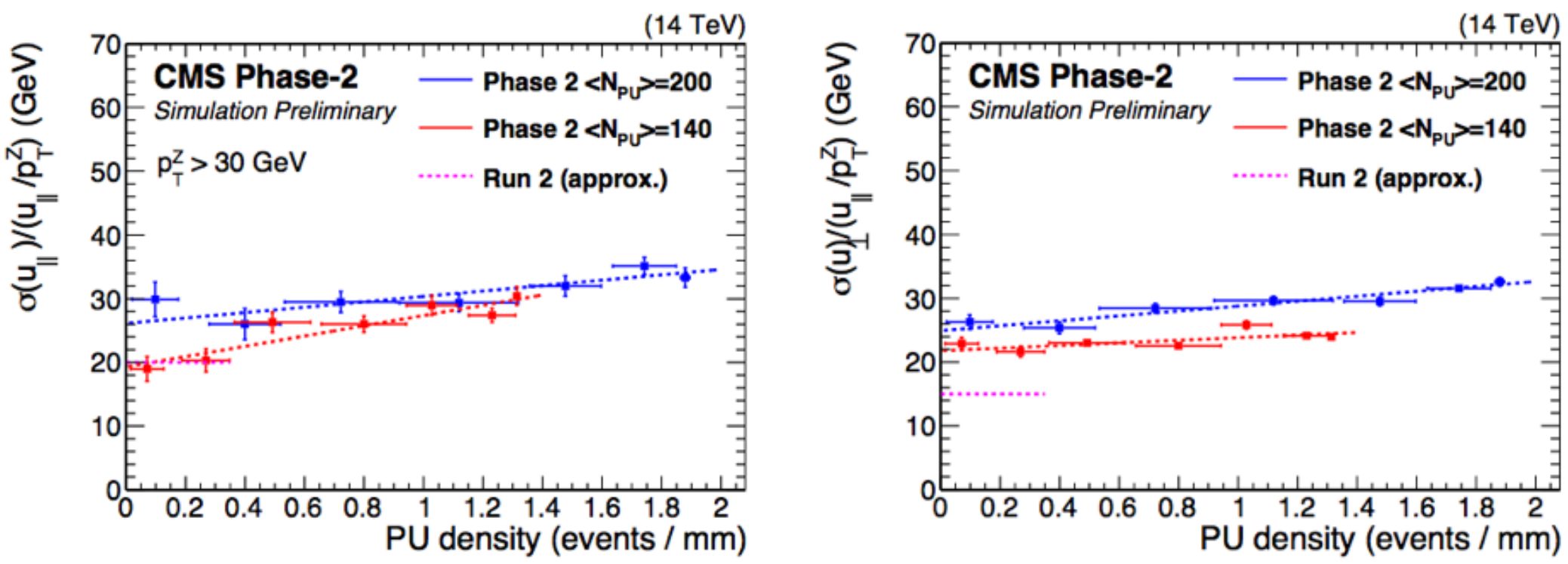

Figure 35: Parallel (left) and perpendicular (right) $p_T^{miss}$ resolution is shown as a function of PU density in $Z \to \mu\mu$ events using the PUPPI mitigation. The blue points indicate the Phase-2 performance with a mean 200 PU, the red points indicate the Phase-2 performance with a mean 140 PU, and the pink dashed line indicates the Run 2 performance with a mean 27 PU. A mild degradation in performance is observed for Phase-2. Taken from Ref. [8].

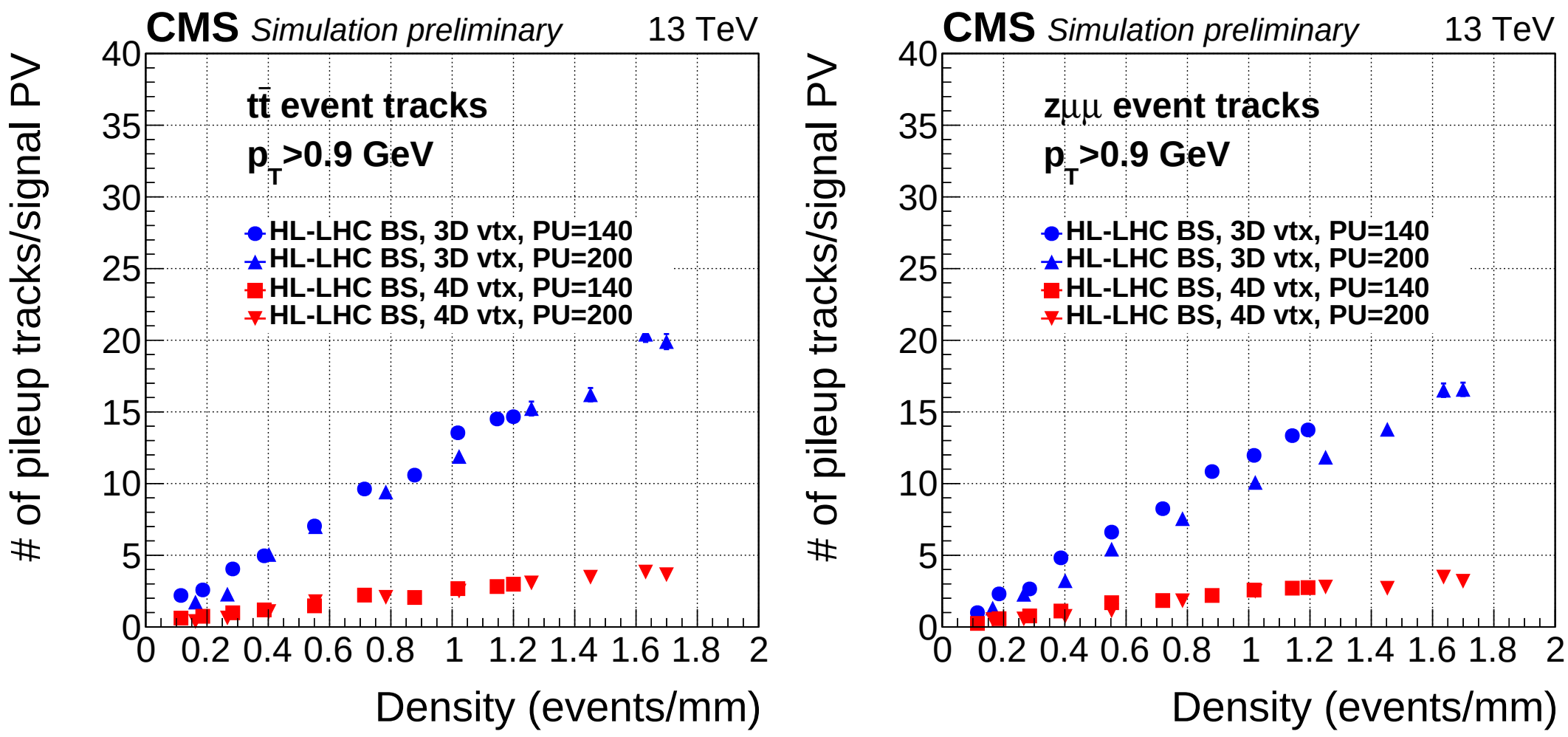

Figure 36: The number of PU tracks incorrectly associated with the primary vertex in $t\bar{t}$ (left) and $Z \to \mu\mu$ (right) events as a function of the PU density, shown with (4D vertex) and without (3D vertex) the MTD precision timing. Taken from Ref. [5].

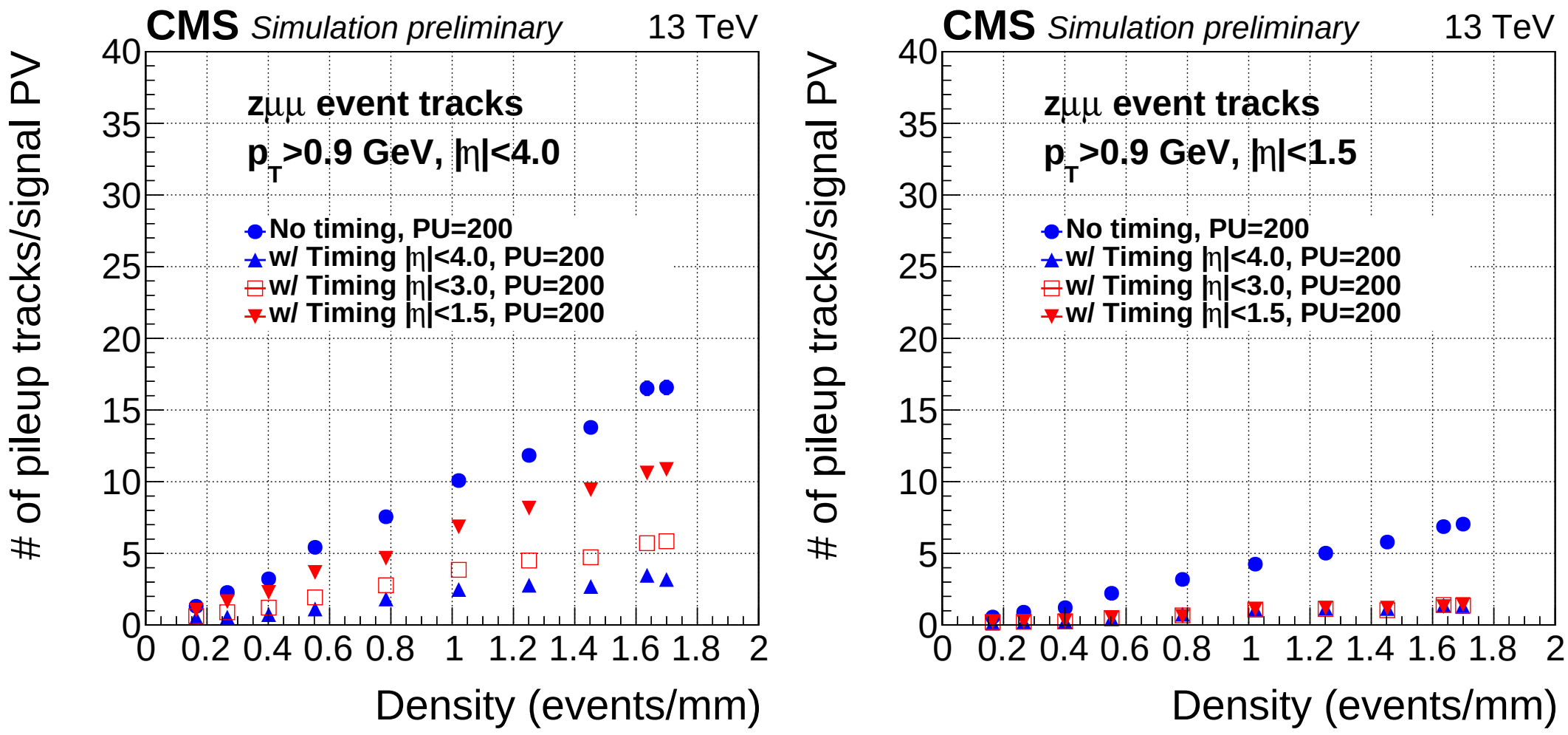

Figure 37: The number of PU tracks in $Z \to \mu\mu$ events incorrectly associated with the primary vertex as a function of PU density, shown without and with the MTD precision timing for several different acceptance scenarios, considering tracks within the full Tracker acceptance (left) and just in the central part (right) of the detector. Taken from Ref. [5].

systematic uncertainties for all analyses, aiming for a realistic projection while starting from the experience in Run 2. The aim is to achieve coherence among different analyses. However, there are practical limits to the goal of coherence, and there are many nonuniversal and analysis-specific aspects that are hard or impossible to generalize, so these uncertainties should be considered as guidelines only.

Extrapolating to the HL-LHC era from Run 2 conditions is not straightforward. We rely here on the same methods. At the HL-LHC, we shall benefit from the increase in integrated luminosity ($> 3000\,\mathrm{fb}^{-1}$ to be compared to $\sim 40\,\mathrm{fb}^{-1}$ in Run 2 measurements from 2016). Some of the components of the systematic uncertainties, which are currently limited by the relatively small available data sample, will benefit from the increase of the number of collected events and may be reduced to much smaller levels. Furthermore, the estimates of the systematic uncertainties will benefit from at least 10 years of further measurements, with some expected improvements in the tuning of the Monte Carlo generators and in the description of the detectors in the simulation (not speaking of refined analysis techniques). The main sources of the considered systematic uncertainties and their projected values are described below.

## 3.1 Electron and photon uncertainties

For the Run 2 analyses (2016–2017 dataset), an uncertainty of 0.2–2% (depending on $\eta$) is assigned to electron and photon reconstruction, identification, and isolation efficiency [13, 14]. The sources of uncertainty are signal and background modeling in simulation, the use of different generators, the event selection, and the tag electron selection in the tag-and-probe technique used for the efficiency measurement. For the HL-LHC, with the increased dataset and upgraded detectors, the effects due to background modeling, initial-state radiation, and signal resolution may decrease. However, the effects due to PU, especially for isolation, may lead to increased systematic uncertainties. As a result, the current studies indicate a projected systematic uncertainty of 0.5% for electrons, including isolation. Thus, from current studies, a 0.5% systematic uncertainty is projected for photon reconstruction and identification efficiency. For photon isolation efficiencies, due to the challenging PU environment, a 2% systematic uncertainty is assumed; however, this does not take into account the PU mitigation due to the timing detectors and hence could be reduced considerably. Thus, the overall projected uncertainty is kept at the level of Run 2.

The electron energy scale systematic uncertainty ranges between 0.1% to 0.3%, depending on the pseudorapidity difference between the nominal and measured value of the Z boson mass peak in the data, as shown in Fig. 38 left. It is difficult to reduce this uncertainty much further. We keep the same systematic uncertainty for the HL-LHC because the larger dataset will help in monitoring detector stability, and we expect to be able to mitigate the effects from PU.

The energy resolution for photons has been studied in the upgraded CMS upgraded detector. The performance of the barrel calorimeter is studied as a function of PU and aging effects with different integrated luminosities, as shown in Fig 39. The variation in performance will lead to systematic uncertainties in analyses where the diphoton mass resolutions plays a major role, e.g., Higgs boson decays to diphotons.

## 3.2 Muon uncertainties

The uncertainty in muon reconstruction and identification efficiency in Run 2 analyses is estimated to be 0.1–0.5% [13], depending on the pseudorapidity of the muon. The uncertainty in the muon isolation variable is around $\sim$0.5%. For the HL-LHC, we have examined the muon reconstruction and identification efficiency as a function of PU collisions and find that it is fairly stable and robust against PU. With a larger data sample and the upgraded detector, the uncertainty due to the background modeling may decrease, while the dependence of isolation on PU may lead to increased systematic uncertainties. However, as for the other objects, the expected improvements from the timing detectors are not included. Thus, the projected systematic uncertainties in the muon reconstruction, identification, and isolation will remain the same as the Run 2 uncertainties.

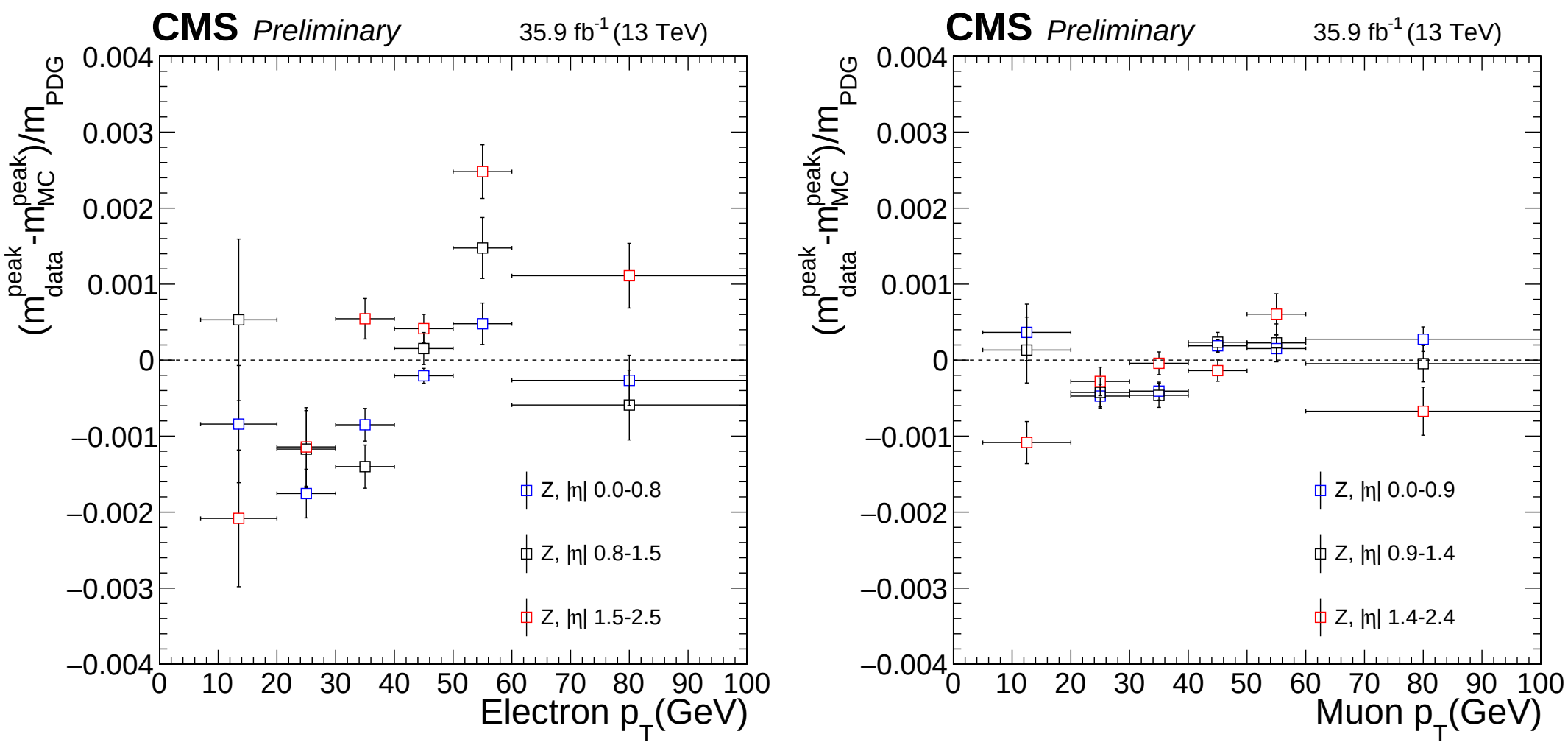

Figure 38: The difference between the $Z \to \ell\ell$ mass peak positions in data and simulation, normalized by the nominal Z boson mass, obtained as a function of the $p_T$ and $|\eta|$ of one of the leptons, regardless of the second, for electrons (left) and muons (right). Taken from Ref [15].

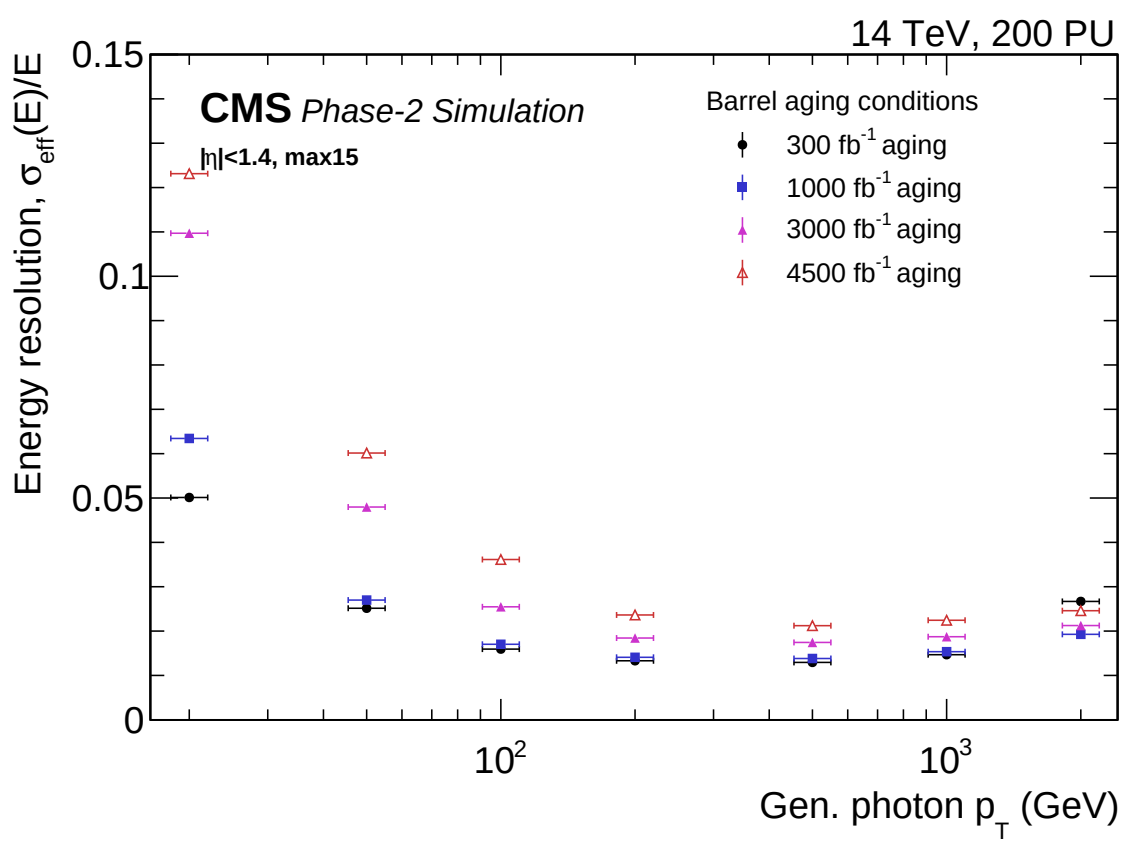

Figure 39: The single photon energy resolution as a function of $p_T$ and aging scenario, for simulated photon gun samples with 200 PU. The photon energy is estimated using the sum of the energy of the 15 most energetic crystals in the photon supercluster. Taken from Ref [7].

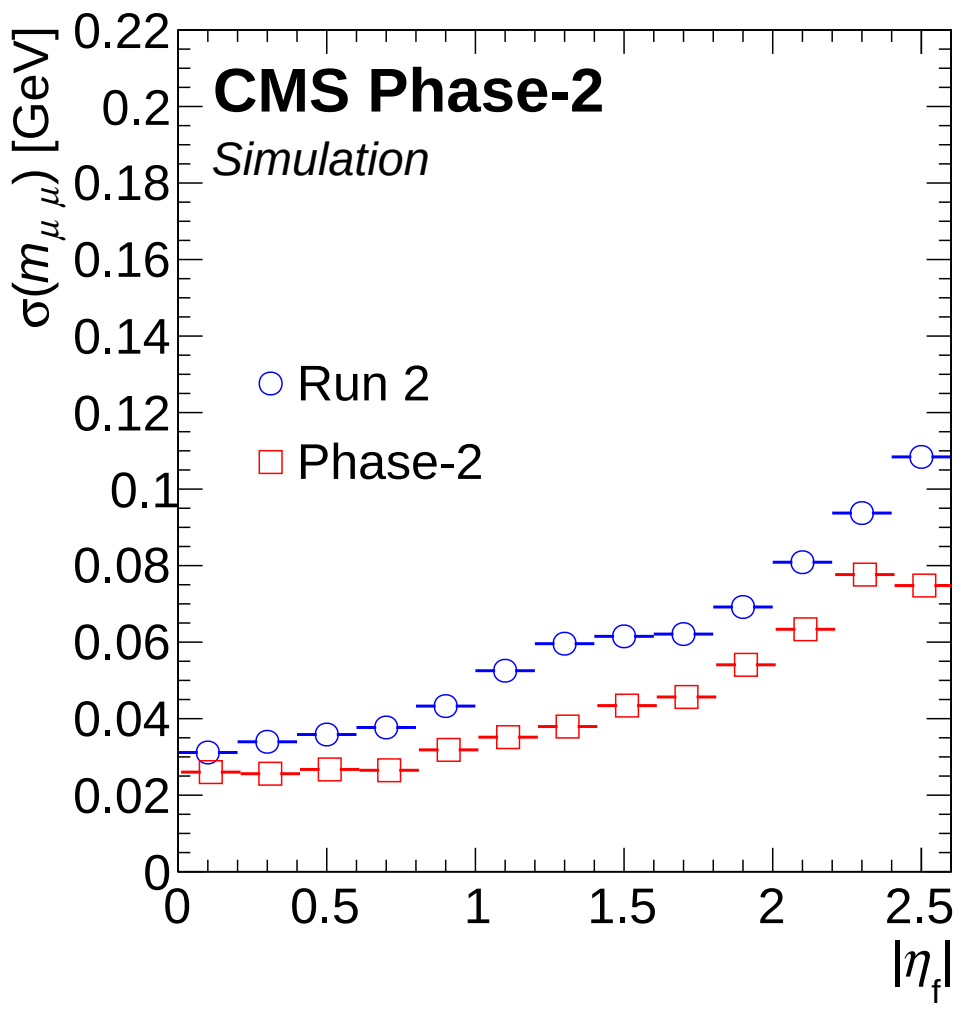

Figure 40: The dimuon mass resolution as a function of $|\eta_\mathrm{f}|$ for $\mathrm{B^0_s} \to \mu\mu$ in Run 2 (blue) and Phase-2 (red). Taken from Ref [6].

Figure 40 shows how the dimuon mass resolution improves with the upgraded tracker detector. The uncertainty in the resolution is expected to be around 5% for muons with $p_\mathrm{T}$ below 200 GeV, and around 10–20% for TeV muons.

While the energy scale will continue to be determined with high precision, we currently assume a value of 0.05%, similar to that in Run 2 (see Fig. 38 right).

### 3.3 Hadronic tau uncertainties

The uncertainty in hadronic tau $\tau_h$ reconstruction, identification and isolation efficiency for Run 2 analyses is determined to be 4–7% [13]. The main sources for this uncertainty are $\tau_h$ simulation modeling, multiplicity of charged hadrons in hadronization of quark/gluon jets, and tracking efficiencies, especially for low $p_\mathrm{T}$ tracks. For the HL-LHC, the dominant uncertainty due to low $p_\mathrm{T}$ tracks is expected to improve significantly with the upgraded tracker. The effect of PU on the isolation of the $\tau_h$ will be challenging, and may possibly dominate the uncertainty. Thus, we keep the same uncertainty as in Run 2 of $\sim$5% per $\tau_h$. As improvements may be expected from further developments such as advanced machine learning for identification and PU mitigation, for the analyses which have a high impact from this uncertainty, the result with half the uncertainty, i.e., 2.5%, was also quoted.

The $\tau_h$ energy scale systematic uncertainty for Run 2 is around 1.5–3%, depending on $\eta$. This uncertainty is dominated by theory modeling and detector effects. It is expected that advances in methods may further reduce the uncertainty from in situ calibration of the $\tau_h$ energy scale.

### 3.4 Flavor tagging uncertainties

The expected flavor tagging uncertainties have been derived by extrapolating the Run 2 performance [16], taking into account new methods that may be used in the future, especially at high $p_{\mathrm{T}}$ and large $\eta$. The projected uncertainty is shown in Fig. 41 and the details are discussed below.

b **jet tagging efficiency:**
Measurements from the data in Run 2 rely on $\mathrm{t\bar{t}}$ events and on multijet events with a muon from semileptonic b hadron decay (muon-jet). Several methods are used for each event sample, and their combination provides a reduction in the overall uncertainty. The $\mathrm{t\bar{t}}$ methods are used for measurements with a typical jet $p_{\mathrm{T}}$ range from 30 to 300 GeV. The muon-jet methods provide coverage of a broad $p_{\mathrm{T}}$ range from 20 to about 1000 GeV in Run 2.

- Some systematic uncertainties are common, or partially common, in both sets of methods: b quark fragmentation, branching fractions of b and c hadrons, jet energy scale and resolution, and PU modeling.
- Some systematic uncertainties are specific to the $\mathrm{t\bar{t}}$ methods: factorization and renormalization scales, modeling of the $\mathrm{t\bar{t}}$ generator and simulation, physics background yield, tagging of non-b jets, $p_{\mathrm{T}}^{\mathrm{miss}}$ modeling, and identification and isolation of lepton from W boson decay.
- Some other systematic uncertainties are more specific to the muon-jet methods: fraction of gluon splitting into b quark pairs, muon selection, calibration and contribution from non-b jets, b jet template, and request of another tagged b jet in the events for some methods.

In the jet $p_{\mathrm{T}}$ range of $\mathrm{t\bar{t}}$ events, $\mathrm{t\bar{t}}$ and muon-jet methods provide compatible b jet tagging efficiencies, within a precision of 1%. The systematic uncertainty in Run 2 is 4-6% for a jet $p_{\mathrm{T}}$ of 1000 GeV. At the HL-LHC, although challenging, we assume that all systematic uncertainties in the b jet tagging efficiency will be reduced by a factor of two, with the increased data sample. A parametrization of the overall uncertainty is derived as a function of the b jet $p_{\mathrm{T}}$, with a minimum set at 1% around 100 GeV.

c **jet tagging efficiency:** Measurements from the data in Run 2 [16] rely on single lepton $\mathrm{t\bar{t}}$ events and on W + c events.

- As for b jet tagging, some systematic uncertainties are common or partially common in both methods: parton distribution function, factorization and renormalization scales, b quark fragmentation, identification and isolation of leptons from W boson decay, jet energy scale and resolution, and PU modeling.
- Some systematic uncertainties are specific to the $\mathrm{t\bar{t}}$ method: cross section of the simulated processes, integrated luminosity, and tagging of light flavor and b jets.
- Some other systematic uncertainties are specific to the W + c method: $\mathrm{D} \to \mu$ branching fraction, soft muon requirement, number of tracks in the jet, background estimate, and $p_{\mathrm{T}}^{\mathrm{miss}}$ modeling.

The overall systematic uncertainty in the c tagging efficiency is typically a factor two to three larger for c jets than for b jets. As for b jets, we assume that the systematic uncertainties in the b jet tagging efficiency will be reduced by a factor of two at the HL-LHC.

**Misidentification probability of light flavor jets (mistag rate):** The main systematic uncertainties in the negative tag method are: the sign flip probability, which is significant for the loose operation point and with jet $p_T < 100\,\text{GeV}$ for the medium and tight operating points, and the fraction of b and c jets in multijet events, which is significant for the medium and tight operating points with $p_T > 100\,\text{GeV}$. Other systematic uncertainties are due to the fraction of gluon jets in the multijet sample, the contribution from $K^0_S$ and $\lambda$ decays, the secondary interactions in the detector material, the fraction of mismeasured tracks, the event sample dependence, and the PU modeling. The most significant systematic uncertainties can be directly estimated from data measurements. We therefore assume that they will be reduced by a factor two at the HL-LHC, leading to 5, 10, and 15% uncertainty for the operating points with 10, 1, and 0.1% mistag rates, respectively.

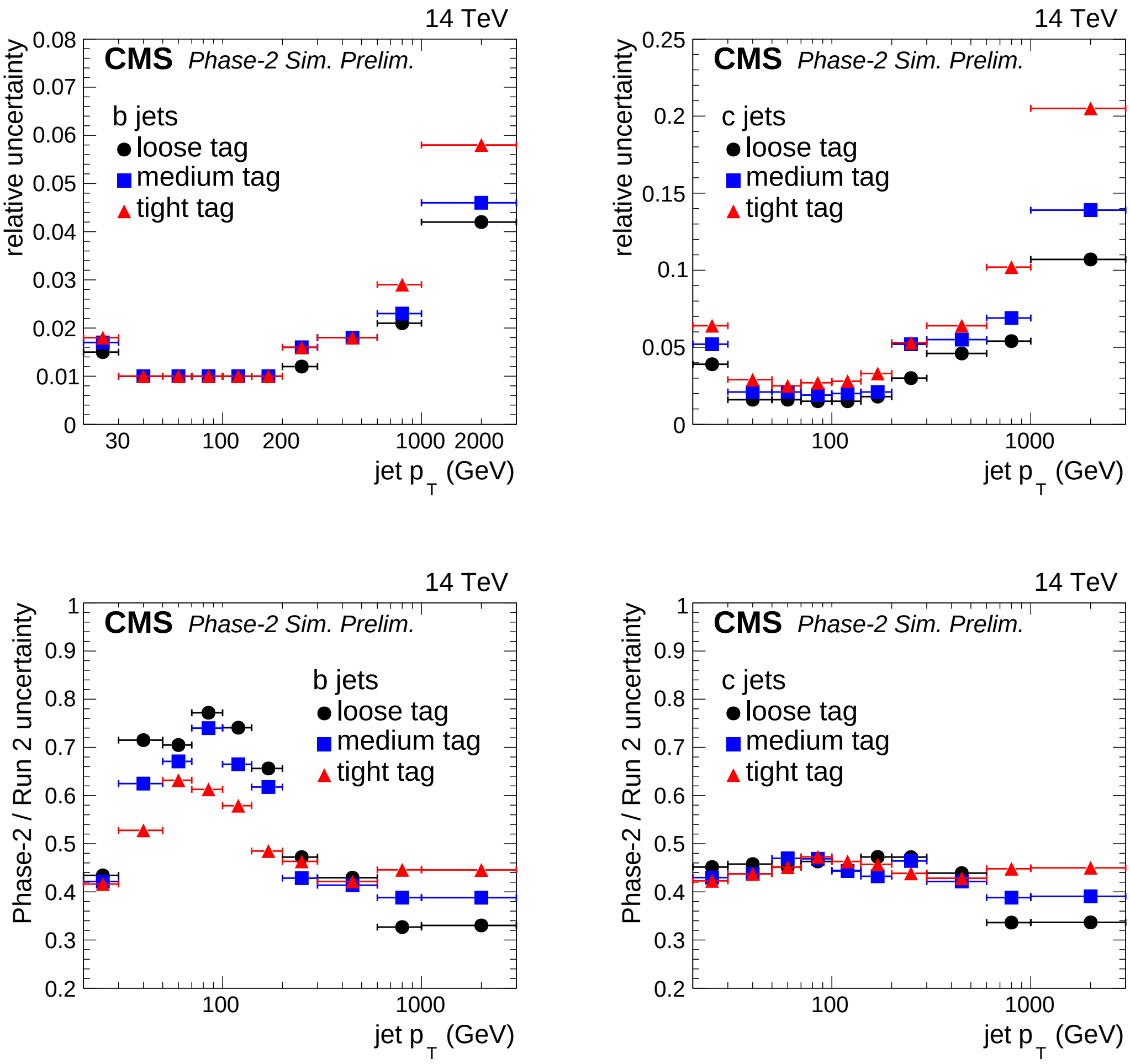

Figure 41: The projected systematic uncertainty in flavor tagging efficiencies for the HL-LHC. The uncertainty in b (c) tagging is shown on the left (right). The expected reduction in the uncertainties compared to Run 2 is shown in the bottom row.

## 3.5 Jet and missing transverse energy uncertainties

To extrapolate the uncertainties for the jet energy corrections (JEC), we examine the current uncertainties from each of the individual sources of JEC for Run 2 [17], as shown in Fig. 42. The absolute jet energy scale uncertainty scales with the statistical precision of the $Z \to \mu\mu$+jet samples and will benefit from updated methods to mitigate inefficiencies at low $p_T$ and at high PU. Thus, we expect the absolute scale uncertainty to be reduced from its current value of 0.5% to 0.1–0.2%. The relative jet energy scale uncertainty and its $p_T$ dependence is expected to improve due to better modeling of the ECAL response. In addition, the larger samples of Z+jet and $\gamma$+jet will also help to constrain the low $p_T$ jet response, leading to a reduction of the uncertainties from 3 to 0.5% at high $\eta$. The jet flavor dependence of the uncertainty is expected to be reduced by a factor of two by modifying the current method to use a mixture of PYTHIA and HERWIG [18] as the reference Monte Carlo generator. Further improvements are possible by developing methods based on using data control samples. The component of the uncertainty from PU is kept the same, as with updated techniques, we expect the effect of additional PU could be mitigated. The two other components, "method and sample" and "time stability," are likely to be addressed through time-dependent simulation, and are not currently considered for the HL-LHC projections. Furthermore, the total JEC uncertainty expected for the HL-LHC is half of its value in Run 2 and approximately 1% or less for jets with $p_T > 30$ GeV. For boosted jets with a distance parameter of 0.8, the JEC systematic uncertainties scale similarly.

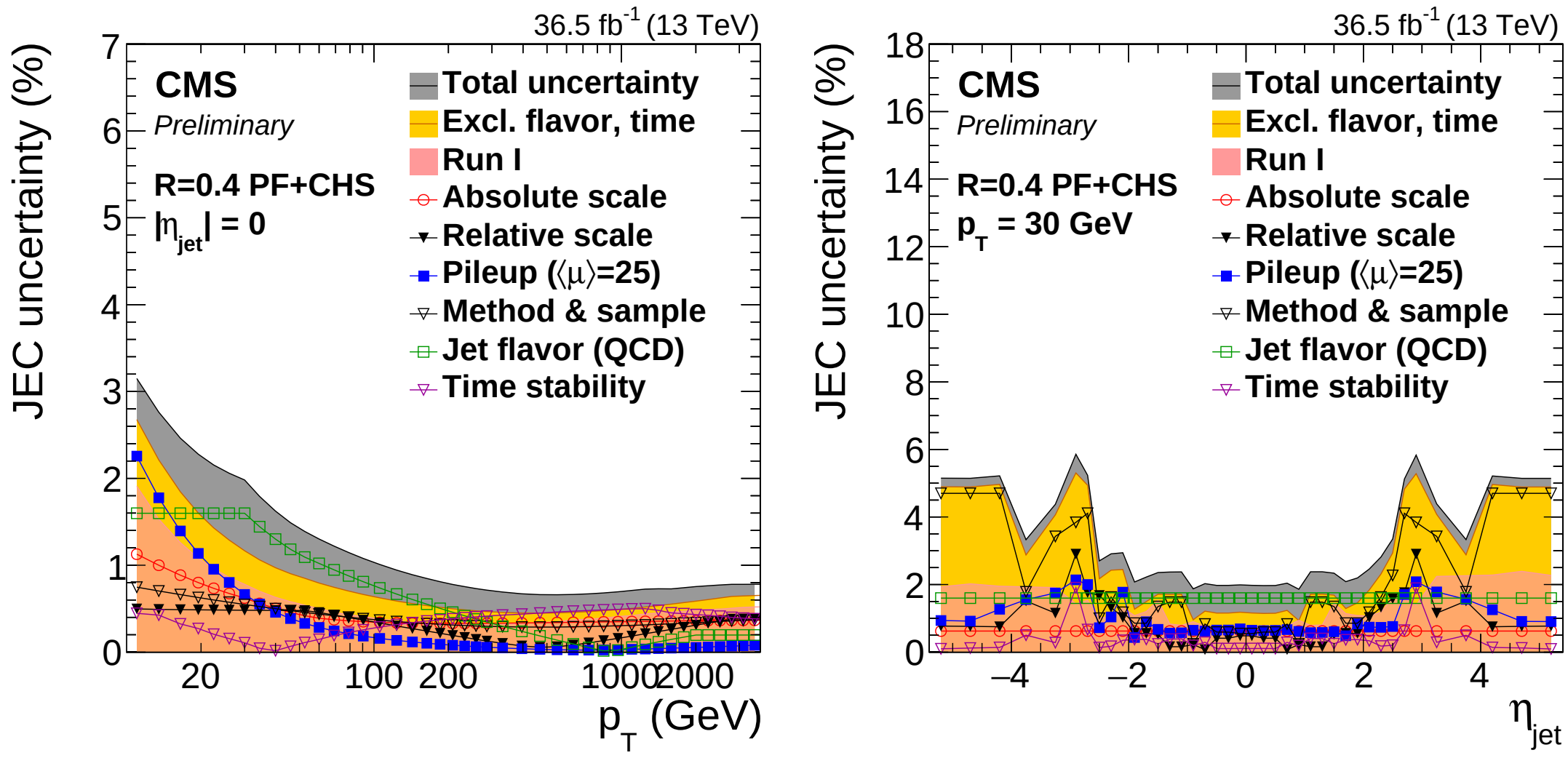

Figure 42: The systematic uncertainty in the jet energy scale, as measured for Run 2. The individual sources of the systematic uncertainty are shown together with the total uncertainty. The total uncertainty is obtained by adding all uncertainties in quadrature. Taken from Ref. [17].

The systematic uncertainties in the W, Z, and Higgs boson jet tagging variables stay the same as in Run 2.

The uncertainties in the jet energy scale resolution (JER) are currently dominated by the methods used to derive them and have the potential for large improvements. We expect to achieve Run 1 performance at the HL-LHC, despite the harsher conditions, and hence we expect the uncertainty to be half of the Run 2 values for HL-LHC analyses.

The $p_T^{miss}$ systematic uncertainties are driven by the object scale and resolution uncertainties. These systematic uncertainties are correlated with the high $p_T$ objects in the event and are expected to scale accordingly. A large fraction of this uncertainty comes from the jets. The JEC and JER uncertainties for each jet is propagated to the $p_T^{miss}$ uncertainty. The component of the uncertainty from unclustered energy in the event is expected to be subdominant, and we propose it to be 10% of the unclustered energy, using the same method as in Run 1.

### 3.6 Optimally achievable "floor" experimental uncertainties

Table 1 shows the achievable "floor" experimental systematic uncertainties. For some objects, different values are given for different working points (WPs). The object identification (ID)

Table 1: The “floor” systematic uncertainties for the HL-LHC.

| Uncertainty | Working point/ component | Value |
|---|---|---|
| Electron ID | All WPs, $p_{\mathrm{T}} > 20\,\mathrm{GeV}$ | 0.5% |
| | All WPs, $10 < p_{\mathrm{T}} < 20\,\mathrm{GeV}$ | 2.5% |
| Photon ID | | 2% |
| Muon ID | All WPs | 0.5% |
| Tau ID | All WPs | 2.5% |
| Jet energy scale | Total | 1–2.5% |
| | Absolute scale | 0.1–0.2% |
| | Relative scale | 0.1–0.5% |
| | PU | 0–2% |
| | Jet flavor | 0.75% |
| Jet energy resolution | | 3–5% as a function of $\eta$ |
| b-tagging | b jets (all WPs) | 1% |
| | c jets (all WPs) | 2% |
| | Light jets, loose WP | 5% |
| | Light jets, medium WP | 10% |
| | Light jets, tight WP | 15% |
| | Subjet b tagging | 1% |
| | Double c tagging | |
| $p_{\mathrm{T}}^{\mathrm{miss}}$ | Propagate jet energy corrections uncertainties (must) | |
| | Propagate jet energy resolution uncertainties (recommended) | |
| | Vary unclustered energy by 10% (recommended) | |
| Integrated luminosity | | 1% |

includes isolation. The details of the choices in this table are explained below.

### 3.7 Extrapolation scenarios

In analyses using the full CMS Phase-2 detector simulation or the fast-simulation package DELPHES [19] and an integrated luminosity at the HL-LHC of $3000\,\mathrm{fb}^{-1}$, the experimental systematic uncertainties are the “floor” values as described above and summarized in Table 1. No uncertainty is included for possible statistical limitations of Monte Carlo simulations.

For extrapolations from Run 2, the strategy of applying experimental uncertainties to HL-LHC analyses closely follows the strategy used and documented in Ref. [20]. Three scenarios are considered: “Run 2 uncertainty”, “YR18 uncertainty” and “Stat-Only”.

The “Run 2 uncertainty” scenario, which is referred to as “S1” in Ref. [20], is useful for direct comparison with Run 2 analyses. As such, it is a sanity check. In this scenario, we assume that detector performance stays approximately constant because the detector simulation advances and operational experience may compensate for limitations such as increased PU and detector aging. The experimental, theoretical, and integrated luminosity systematic uncertainties are kept constant with integrated luminosity, while the statistical uncertainty of the data is scaled with $1/\sqrt{L}$, where $L$ is the integrated luminosity.

Another scenario is the “YR18 uncertainty” scenario, referred to as “S2” in Ref. [20]. This sce-nario reflects uncertainties that we consider achievable at the HL-LHC, from the current perspective. In this scenario, the statistical uncertainty and intrinsic detector limitations are treated as in the Run 2 uncertainty scenario. Theory uncertainties, from both cross section normalization and modeling, are scaled by a factor of 1/2. In extrapolations from Run 2, experimental systematic uncertainties are scaled down from the Run 2 values by the square root of the integrated luminosity until the "floor" values as described above and summarized in Table 1 are reached.

The final scenario considered is the "Stat-Only" scenario. This statistical uncertainty-only scenario indicates the ultimate precision limit, assuming no systematic uncertainties.

In all of the above scenarios, no uncertainty is included for possible statistical limitations of Monte Carlo simulations.

## 4 Summary

The performance of the physics objects with Phase-2 upgrade of the CMS detector at the High-Luminosity LHC with an average of 200 PU interactions has been collated in this note. Furthermore, the expected systematic uncertainties for the physics objects to be used by the physics sensitivity studies and projections at the High-Luminosity LHC have been described.

# Standard Model Physics

# CMS Physics Analysis Summary

Contact: cms-phys-conveners-ftr@cern.ch 

## High-$p_{\mathrm{T}}$ jet measurements at the HL-LHC

The CMS Collaboration

### Abstract

Processes containing jets with high transverse momenta are studied for the upgraded CMS Phase-2 detector design at the High-Luminosity LHC assuming a center-of-mass energy of 14 TeV and an integrated luminosity of 3 $\mathrm{ab}^{-1}$. The high luminosity allows to fully exploit high transverse momentum jets (boosted jets) and to differentiate between various jet types. Inclusive jet production, the production of jets originating from b or t quarks, as well as from W bosons are studied, with emphasis on the transverse momentum spectrum of the jets and angular correlations between the two jets with highest transverse momenta.

# 1 Introduction

The theory of quantum chromodynamics (QCD) is the underlying theory to describe interactions among quarks and gluons, i.e., partons. Inclusive jet production is a QCD process that allows to probe perturbative QCD calculations and the proton structure at the highest accessible scales. With the expected integrated luminosity of 3 $ab^{-1}$ at the High Luminosity LHC (HL-LHC) [1] the accessible range in transverse momentum $p_T$ can reach a few TeV, the highest $p_T$ ever reached in a collider. A wide collection of inclusive jet measurements was carried out at the LHC by the ATLAS and CMS collaborations at center-of-mass energies $\sqrt{s} = 2.76$ [2, 3], 7 TeV [4–8], 8 TeV [9, 10] and 13 TeV [11, 12], and at lower $\sqrt{s}$ by experiments at other hadron colliders [13–17]. Measurements of inclusive jet cross sections are generally in agreement with theoretical calculations at next-to-leading order (NLO), or at next-to-next-to-leading order (NNLO) or NLO including resummation of next-to-leading logarithmic soft gluon terms. The jet cross sections play a crucial role in the determination of parton density functions and the strong coupling $\alpha_S$, especially at the highest scales.

The improved tracking and b tagging performance at the HL-LHC [18, 19] and jet substructure analysis techniques will allow to discriminate jets of different origin. In this document, we study kinematic distributions of jets in inclusive jet production, as well as in final states containing bottom quark (b), top quark (t) jets, and W boson jets. In addition to the cross section as a function of the transverse momentum $p_T$, angular correlations between the jets with highest $p_T$ are investigated. Higher order QCD radiation affects the distribution of the angular correlation, and especially the region where the jets are back-to-back in the transverse plane is sensitive to multiple "soft" gluon contributions, treated by all-order resummation and parton showers. This region is of particular interest since soft-gluon interference effects between the initial and final state can be significant [20, 21]. The azimuthal correlations in $t\bar{t}$ production is of particular interest because of color interference effects [22, 23].

In inclusive jet production at 13 TeV [11] jet transverse momenta of up to about 2 TeV were reached. The main uncertainties in the high-$p_T$ ($p_T > 800$ GeV) region come from the jet energy calibration and statistical accuracy. Measurements of jets originating from b quarks are important to investigate the heavy-flavor contribution to the total jet cross section and to study the agreement of the measurement with available theoretical predictions. In particular, inclusive b production is very sensitive to higher-order corrections and to parton showers. By exploiting the long lifetime of the B hadrons produced by b quarks, one can identify b jets. Since the b tagging algorithm strongly relies on the tracking information, only jets within the tracker acceptance can be considered. Measurements of inclusive b jet cross sections were already performed at the Tevatron [24, 25] and at HERA [26, 27]. They exhibited a large disagreement between data and theory and helped to improve our understanding of the b quark production and fragmentation. Measurements performed at $\sqrt{s} = 7$ TeV by the ATLAS [28, 29] and CMS [30, 31] collaborations show a reasonable agreement with theoretical calculations.

In top quark production processes, t jets can be defined when the top quark decays hadronically and all decay products can be clustered into a single jet. The production of W bosons is studied in the high-$p_T$ region, where the W boson decays hadronically and are reconstructed as jets. We apply jet substructure techniques [32] to discriminate the jets originating from top quarks and W bosons from the QCD background. Measurements of t-jet cross sections were performed at $\sqrt{s} = 8$ TeV in Ref. [33] and at $\sqrt{s} = 13$ TeV in Refs. [34, 35] where jets with $p_T$ up to 1 TeV were observed.

Angular correlations between the two leading $p_T$ jets and their dependency on the production process are also investigated. The analysis technique is inspired by previous analyses on azimuthal correlations in high-$p_T$ dijet production [36, 37] .

With the luminosity expected at HL-LHC, measurements of cross sections of jet production can reach transverse momenta of a few TeV with reasonable precision. The program of jet physics will substantially profit from the HL-LHC data since higher scales can be reached and the region of very low partonic momentum fractions $x$ can be accessed, where the parton density becomes large.

# 2 Analysis strategy

All results discussed in this note are based on PYTHIA 8 [38] with tune CUETP8M1 [39] supplemented with the Delphes simulation [40] of the CMS Phase-2 detector, except the study of boosted W bosons, where particle level distributions are presented. In inclusive jet and b jet production, the size of the higher order corrections are estimated using the POWHEG generator [41] and were found to be of the order of 20%. For $t\bar{t}$ jet production the size of the higher order corrections can be even larger. For example, a 20% difference in the cross section will lead to a difference of up to 10% in the predicted statistical uncertainty.

The higher luminosity at the HL-LHC will allow to extract jet energy corrections and b tagging scale factors at high $p_T$ with much higher precision, leading to smaller systematic uncertainties. The extended tracker coverage up to a pseudorapidity of $|\eta| = 4$, better tracking performance and expected progress in machine learning (ML) techniques will especially improve the jet flavor tagging based on jet substructure. With the extended tracker coverage, b jets can be measured in the forward region, which is currently inaccessible.

However, even the jet energy calibration can benefit from these methods, for example in the extraction of the flavor-dependent jet energy corrections. The analysis of inclusive jet production can also benefit from the extended tracker coverage since jets reconstructed from particle-flow [42–44] objects incorporating tracks are typically much more precise than jets reconstructed from only calorimeter objects. In Run 2 this was visible in both the size of jet energy resolution in the central and forward direction and in the uncertainties of the jet energy scale and jet energy resolution corrections. Since the jet energy corrections are extracted from in-situ measurements, such as dijet or $\gamma$-jet final states, their precision is expected to improve with increasing luminosity.

# 3 Systematic uncertainties

## 3.1 The b tagging at Phase-2 and related systematic uncertainties

Most of the presented studies rely on b tagging. The cross section of b jet production is about 3–4% of inclusive jet production cross section. In order to achieve sufficiently high purity of the measured b-tagged jets, the light-flavor (udsg) tagging efficiency (referred to as mistagging efficiency) must be as low as possible. For analyses presented in this note, the DeepCSV b tagging algorithm [45] trained for the HL-LHC is used.

The b tagging efficiencies predicted by the simulation are slightly different compared to that measured in data. To correct for this difference, so-called scale factors (SF) are introduced, which are defined as the ratio between the b tagging efficiency in data and simulation. These scale factors are obtained from measurements of b jet enhanced processes [45]. The efficiencies of b tagging, c tagging and light-flavor tagging are corrected by the corresponding scale factors. In this note, we assume that the b tagging scale factors are equal to one, but with uncertainties according to the ones obtained in Ref. [46].

In our studies we use a tight working point, defined by a light-flavor (udsg) mistag rate of 0.1% (for a medium working point, the mistag rate is 1 %, leading to a much higher background contribution). The expected uncertainty of the b tagging scale factor is 15% [46] as shown in Fig. 1. The b tagging uncertainty grows towards higher $p_T$, since it is more difficult to reconstruct a secondary vertex as the tracks become nearly collinear. An overview of the systematic uncertainties in b tagging is given in Table 1, more details are given in Ref. [46].

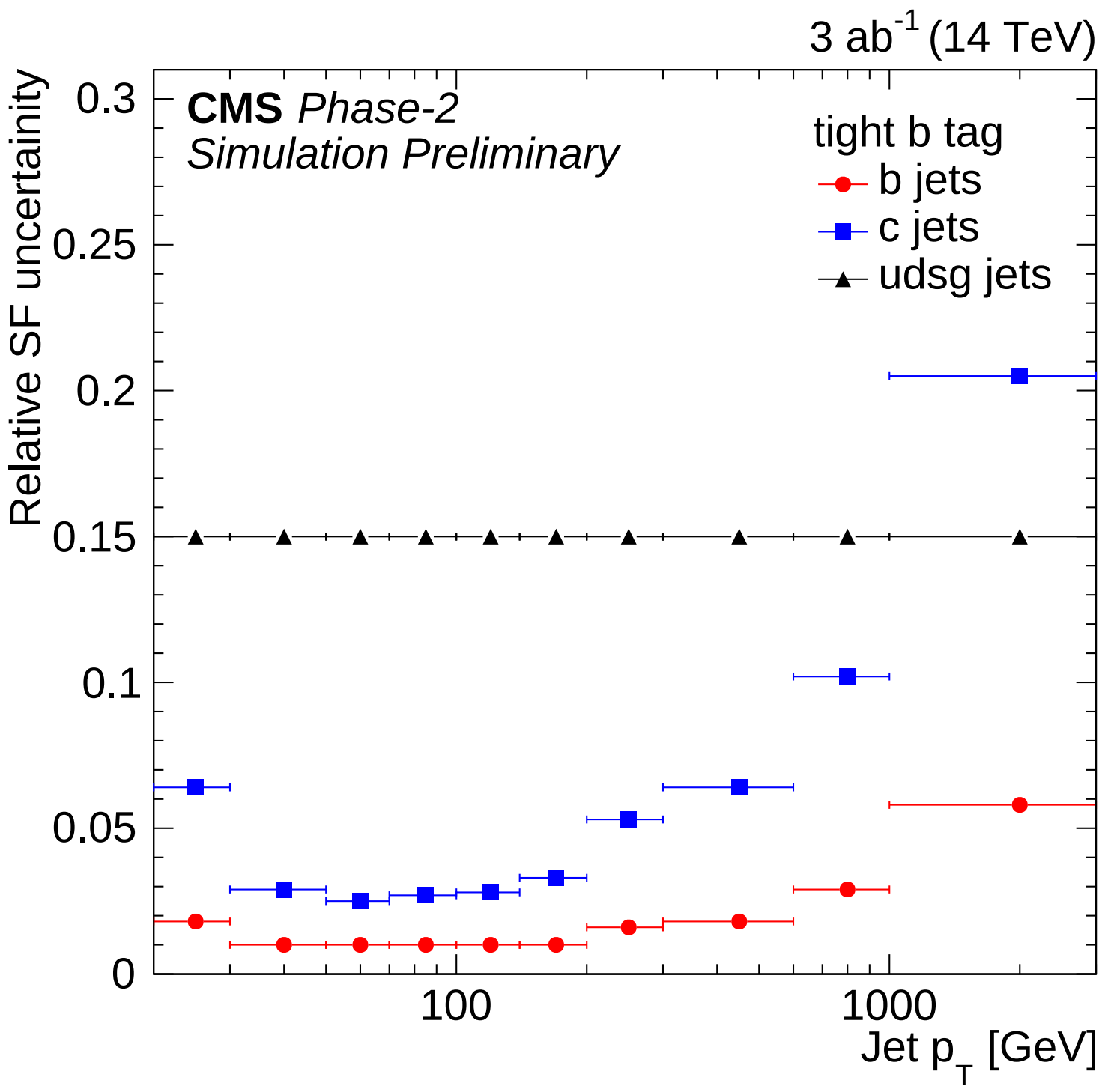

Figure 1: Expected b- tagging scale factor uncertainties as a function of jet $p_T$ for the tight working point [46].

Table 1: The b tagging scale factor (SF) uncertainties for several $p_T$ values [46]. The scale factor uncertainties for jets with $R = 0.4$ and $R = 0.8$ are assumed to be identical.

| $p_T$ [GeV] | 100 | 500 | 2000 |
|---|---|---|---|
| b tagging SF unc. | 1% | 2% | 6% |
| c tagging SF unc. | 3% | 7% | 20% |
| light-flavor tagging SF unc. | 15% | 15% | 15% |

The tagging efficiencies, as obtained from the Delphes simulation, and the related flavor composition of the b-tagged sample of inclusive jets are shown in Fig. 2.

Figure 2 (left) shows the tagging efficiencies as a function of the jet $p_T$. The b tagging efficiency decreases from $\sim$ 70% at $p_T = 100$ GeV to about 20% at $p_T = 1$ TeV, which leads to a larger light-flavor contamination of the b-tagged sample as shown in Fig. 2 (right). Jets containing charm hadrons have similar properties as jets with a B hadron, e.g., the presence of a displaced vertex, and there is a non-negligible probability to misidentify a c jet as a b jet. This probability is rather constant as a function of $p_T$, as shown in Fig. 2 (right).

To evaluate the expected systematic uncertainties from b tagging, we assume, for simplicity,

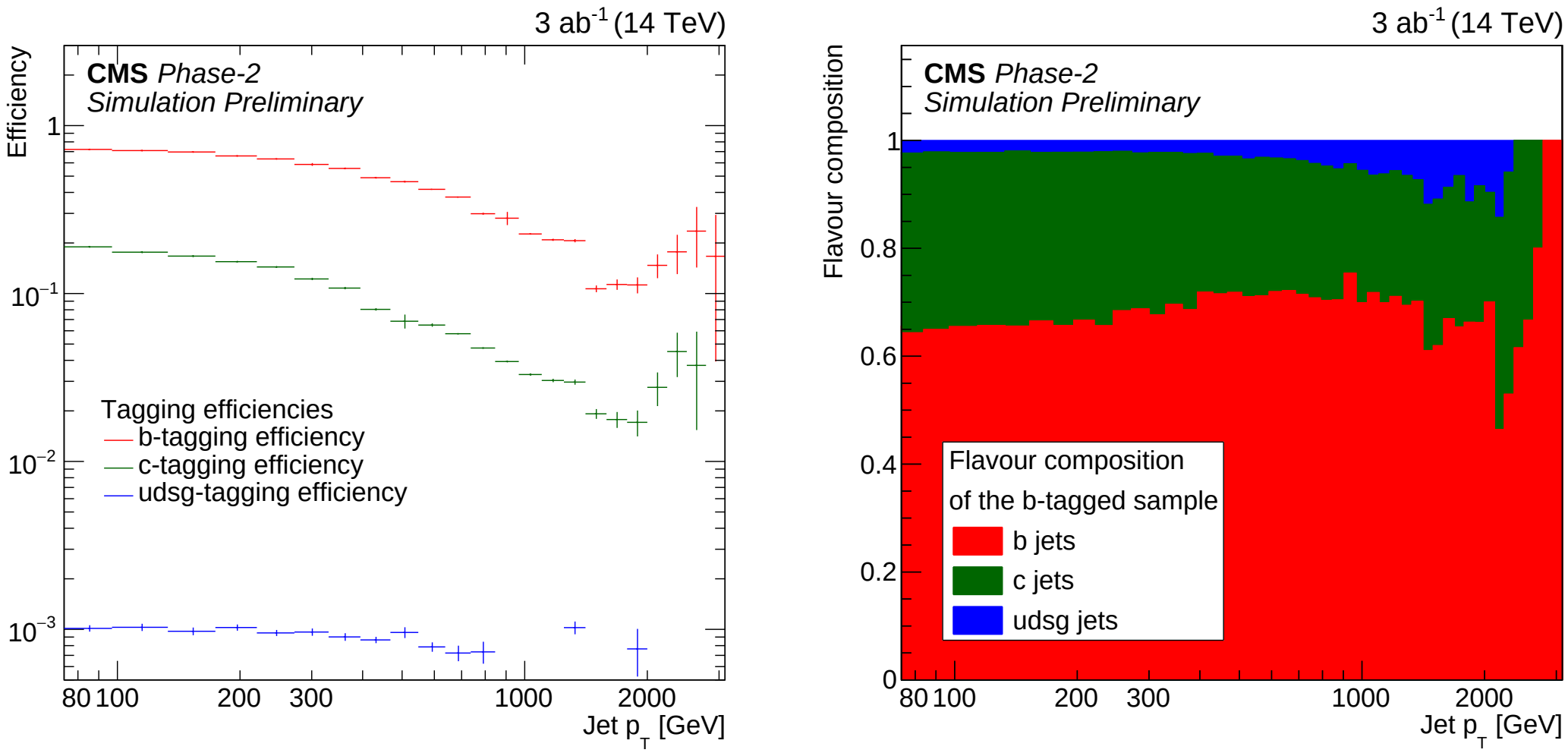

Figure 2: Predicted b-tagging efficiencies with the tight working point for jets with $R = 0.4$ (left). Predicted flavor composition of the b-tagged sample (right).

that only the b-tagged events are used to obtain the cross section:

$$\sigma_b^{\text{data}} = \frac{\sigma_{\text{b tag}}^{\text{data}}}{\sigma_{\text{b tag}}^{\text{MC}}} \sigma_b^{\text{MC}}, \tag{1}$$

where $\sigma_b^{\text{data,MC}}$ is the b jet cross section in the data and Monte Carlo simulation (MC), respectively. The cross section of b-tagged jet production in the MC simulation $\sigma_{\text{b tag}}^{\text{MC}}$ can be calculated as:

$$\sigma_{\text{b tag}}^{\text{MC}} = \sigma_b^{\text{MC}} \epsilon_b + \sigma_c^{\text{MC}} \epsilon_c + \sigma_l^{\text{MC}} \epsilon_l \, , \tag{2}$$

where $\epsilon_{b,c,l}$ are the probabilities that b jet, c jet or light-flavor jet is b tagged and $\sigma_{b,c,l}^{\text{MC}}$ are the b jets, c jets and light-flavor jet cross sections in the Monte Carlo simulation.

In Eq. (1), the background from wrongly tagged b jets is implicitly subtracted. This background fraction increases the resulting statistical uncertainty of the true level cross section:

$$\frac{\Delta\sigma}{\sigma} = \frac{\sqrt{N_b + N_{bg}}}{N_b} \tag{3}$$

where $N_b$ is the number of events with tagged b jets, i.e., the signal, and $N_{bg}$ is the number of events in which other flavors were mistagged, i.e., the background.

In the calculation of the resulting systematic uncertainty of the predicted cross section, the b tagging and c tagging SF uncertainties are assumed to be correlated (as treated in Run 2), whereas light flavor tagging is taken as uncorrelated with the other two.

The expected uncertainty of the inclusive b jet cross section as a function of $p_{\text{T}}$ shown in Fig. 3. The uncertainty coming from the uncertainty of the light-flavor and heavy-flavor SF varies between 2% at low $p_{\text{T}}$ and 10% at large $p_{\text{T}}$. The b tagging systematic uncertainty is dominated by the b+c SF uncertainty in the high-$p_{\text{T}}$ region.

The b tagging performance is also crucial for top quark tagging, since a b-tagged subjet is required (Section 4.4). The b tagging performance of jets with larger cone size is comparable to one of the jets with $R = 0.4$. It is important to mention that (in case of dijet production) requiring one jet to be b tagged increases the probability that the other jet is also tagged and the background contamination is lower compared to inclusive jet production.

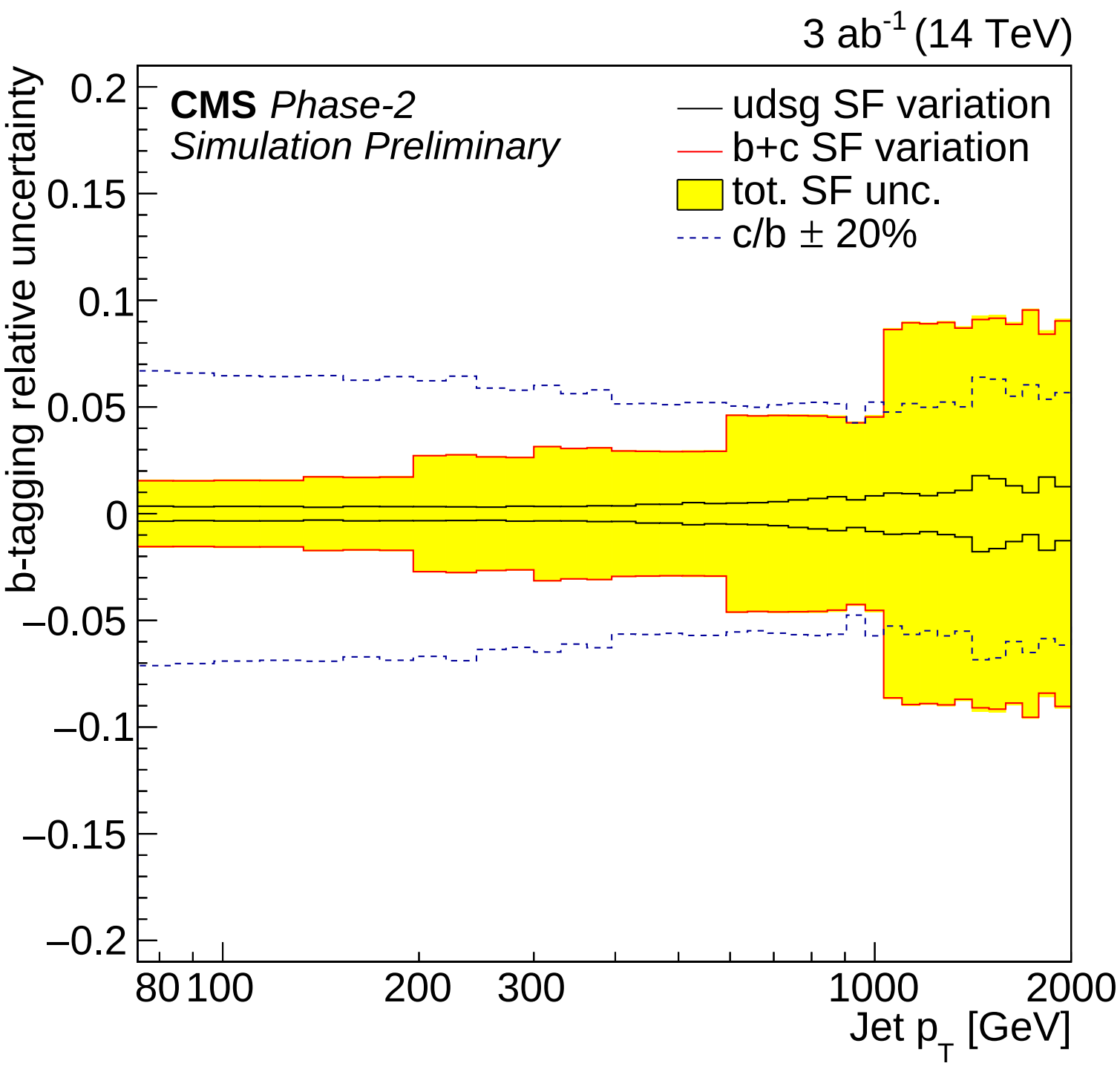

Figure 3: Expected b-tagging systematic uncertainty of the inclusive b-jet cross section.

The model uncertainty of the b tagging is related to differences in jet flavor composition in MC and data. This can affect the predicted amount of background from c and light flavors and, consequently, the measured particle-level cross section. To evaluate this model dependence, the flavor composition in PYTHIA 8 and in HERWIG ++ was compared in Ref. [45]. The flavor fractions b/c were found to differ maximally by 20% and this value is considered as a model uncertainty (as indicated in Fig. 3). The amount of light-flavor jets is well constrained by the inclusive jet cross section and, therefore, no model dependence is considered.

### 3.2 Other sources of systematic uncertainties

In addition to the uncertainties from b tagging, the uncertainties related to the jet energy calibration can significantly contribute. Based on previous experience [47], they can be about 1–2% within the tracker acceptance, where the 2% value is expected at lower $p_T$ mainly due to the uncertainty introduced by the subtraction of effects from additional proton-proton collisions (pileup). In the high-$p_T$ region the dominant component in the jet energy scale uncertainty (JES) is due to the jet flavor dependence of the detector response, which is slightly different for quark- and gluon-induced jets. A 1% shift in the energy calibration leads to about 5% change of the cross section $d\sigma/dp_T$ if the cross section falls as $\propto p_T^{-5}$.

The uncertainty in the measured integrated luminosity is assumed to be 1%.

# 4 Results

## 4.1 Inclusive jet production

The inclusive jet cross section at particle level, without any flavor requirement, is shown as a function of $p_T$ for a rapidity range of $|y| < 0.5$ in Fig. 4 (left). The statistical uncertainty, visible in the ratio, corresponds to an integrated luminosity of 3 $ab^{-1}$. The systematic uncertainty (shown as the grey band) is dominated by the jet energy scale uncertainty (JEC). Also shown is the expected inclusive jet cross section at $\sqrt{s} = 13$ TeV with uncertainties corresponding to an integrated luminosity of 150 $fb^{-1}$.

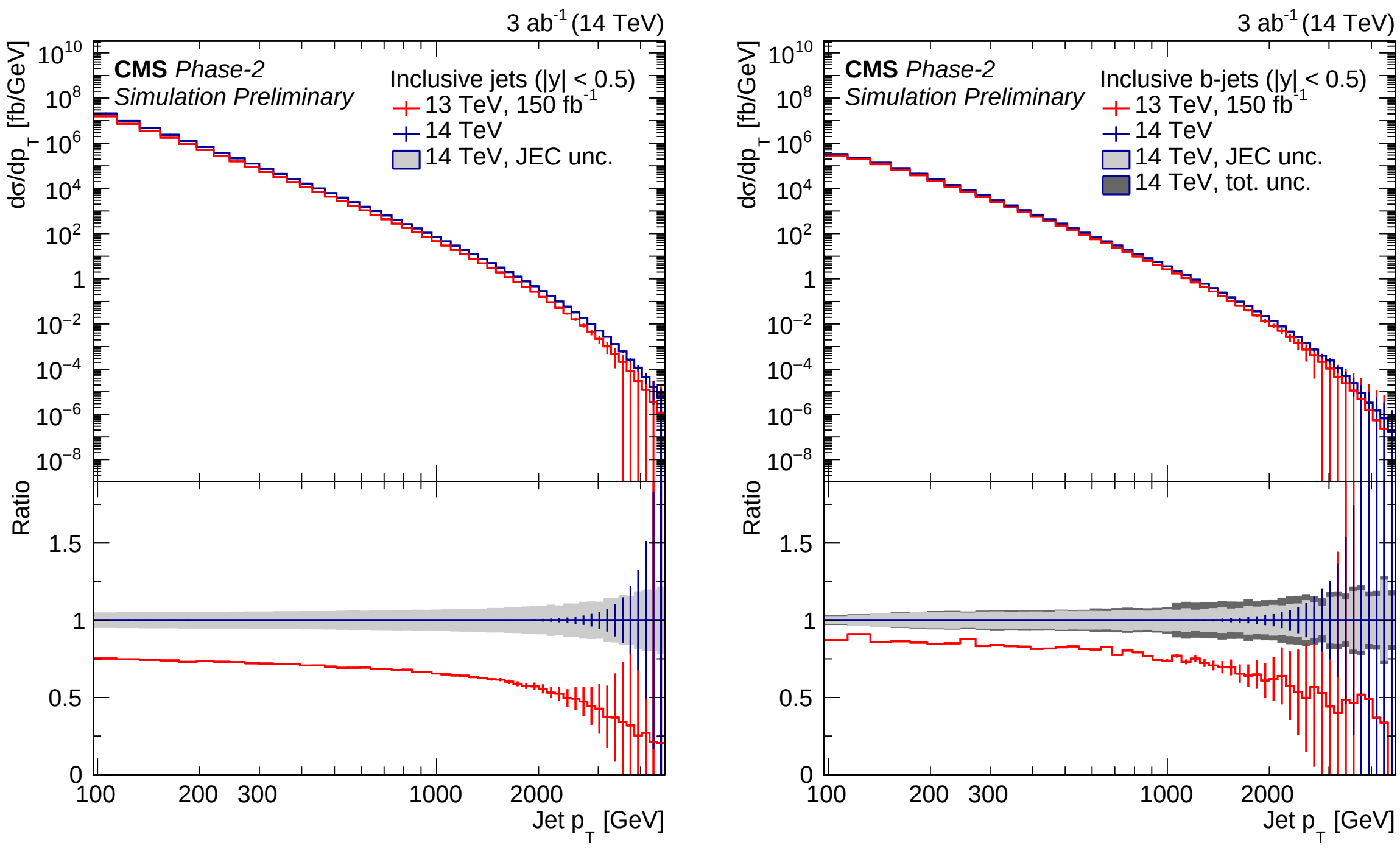

Figure 4: Comparison of the 13 TeV and 14 TeV cross sections for inclusive jet (left) and inclusive b jet (right) production at particle level as a function of $p_T$ in $|y| < 0.5$. The lower panel shows the ratio to the jet cross section at 14 TeV. The uncertainties in the ratio represent the expected statistical uncertainty assuming 150 $fb^{-1}$ and 3 $ab^{-1}$, respectively. The systematic uncertainty is shown for 14 TeV and is dominated by the jet energy scale uncertainty for inclusive jet production, and by the jet energy scale uncertainty and by the uncertainties from b tagging for the inclusive b jets.

Compared to Run 2 measurements at $\sqrt{s} = 13$ TeV the increase of the center-of-mass energy leads to about twice larger cross section at highest $p_T$. Taking into account the much higher luminosity and the higher cross section, the statistical uncertainty is expected to be around six times smaller, compared to the analysis of the Run 2 data. A measurement of the inclusive jet cross section up to $p_T \sim 4$ TeV can be performed with about 10 events above this threshold.

## 4.2 Inclusive b jet production

In Fig. 4 (right), the inclusive b jet cross section at particle level as a function of $p_T$ for $|y| < 0.5$ is shown. The statistical uncertainty corresponds to an integrated luminosity of 3 $ab^{-1}$, where the b tagging efficiency, as described in Section 3.1, is included. The systematic uncertainty of around 5% in the low-$p_T$ region rising to around 10% at high-$p_T$ includes uncertainties from jet energy scale calibration as well as uncertainties from b tagging. For comparison, also the expected cross section at 13 TeV with uncertainties corresponding to an integrated luminosity of 150 $fb^{-1}$ is shown. Compared to Run 2 measurements at $\sqrt{s} = 13\,\text{TeV}$, the increase of the center-of-mass energy leads to about twice larger cross section at largest $p_T$. A measurement of the inclusive b jet cross section can reach transverse momenta of $p_T \sim 3\,\text{TeV}$ with about 30 events above this threshold, where the details depend crucially on the b tagging performance at highest $p_T$ (the b tagging SF uncertainties were derived only up to 2 TeV [46] and the uncertainty is expected to increase with $p_T$).

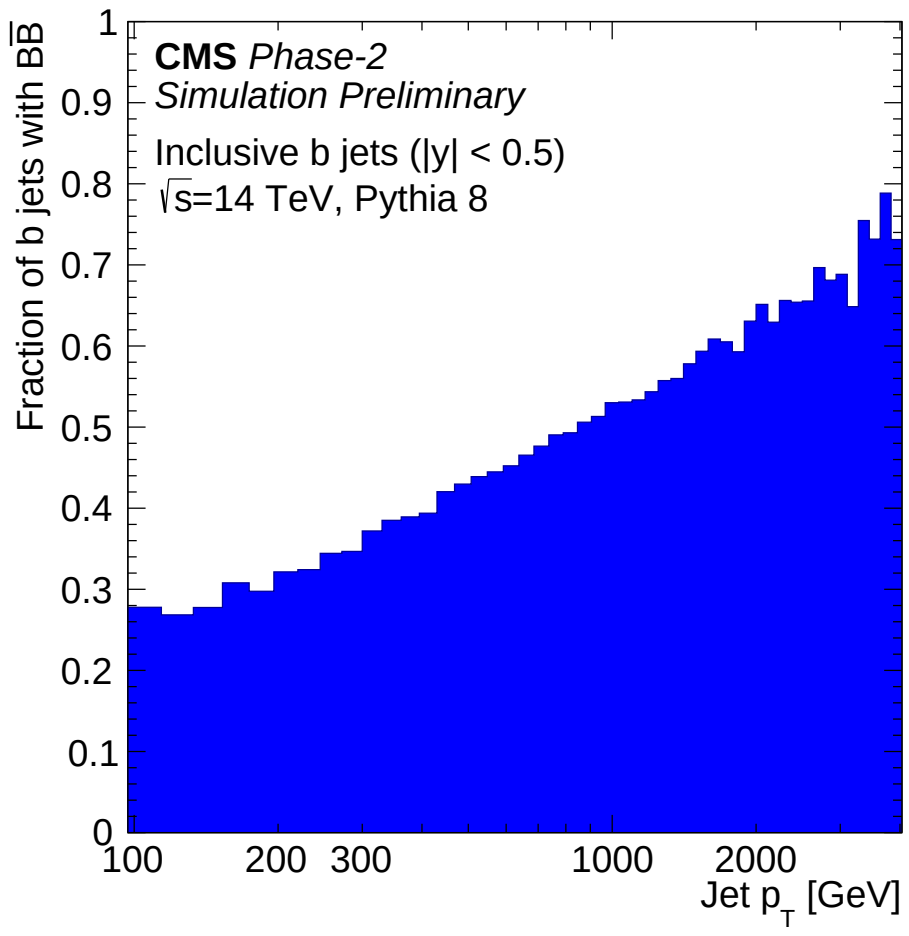

Figure 5: Fraction of b jets containing both a B and a $\overline{\text{B}}$ hadron as a function of the jet $p_T$.

In the high-$p_T$ region, the mass of the b quark becomes negligible with respect to the jet momentum. This leads to a high probability that the b quark is not only produced in the hard subprocess, but also during further QCD radiation, simulated with a parton shower. In such cases, a pair of B hadrons inside the b jet can be observed, where one consists of a b quark, and the second of a $\overline{\text{b}}$ quark. The fraction of such jets as a function of $p_T$, as predicted by PYTHIA 8, is shown in Fig. 5.

### 4.3 High-$p_T$ b$\overline{\text{b}}$ jets

The angular correlations $\Delta\phi = |\phi_2 - \phi_1|$ and $|\Delta y| = |y_2 - y_1|$ between the two leading $p_T$ jets are studied. The flavor dependence of the angular correlations are investigated by selecting dijet events with at least one or two b-jets. The leading jet $p_T$ must satisfy $400 < p_T < 800\,\text{GeV}$ or $p_T > 1600\,\text{GeV}$ while the subleading jet is required to be above 200 GeV. The event selection follows closely the Run 1 and Run 2 measurements [36, 48, 49].

The angular resolution is found to be 0.07 rad for $|\Delta\phi|$, obtained from the Delphes simulation (and consistent with the resolution found in Run 2 [36]). The resolution in $|y|$ has a similar size. The systematic uncertainties are treated as in the previous section and are dominated by the jet energy scale and b tagging scale factors uncertainties.

In Fig. 6, the particle-level cross section as a function of $\Delta\phi$ is shown. The statistical uncertainty corresponds to an integrated luminosity of 3 $ab^{-1}$ including b tagging as described in Section 3.1. The systematic uncertainty includes uncertainties from jet energy scale calibration as well as uncertainties from b tagging. It is around 5% in the low-$p_T$ region and rises to 10% at high $p_T$.

**8**

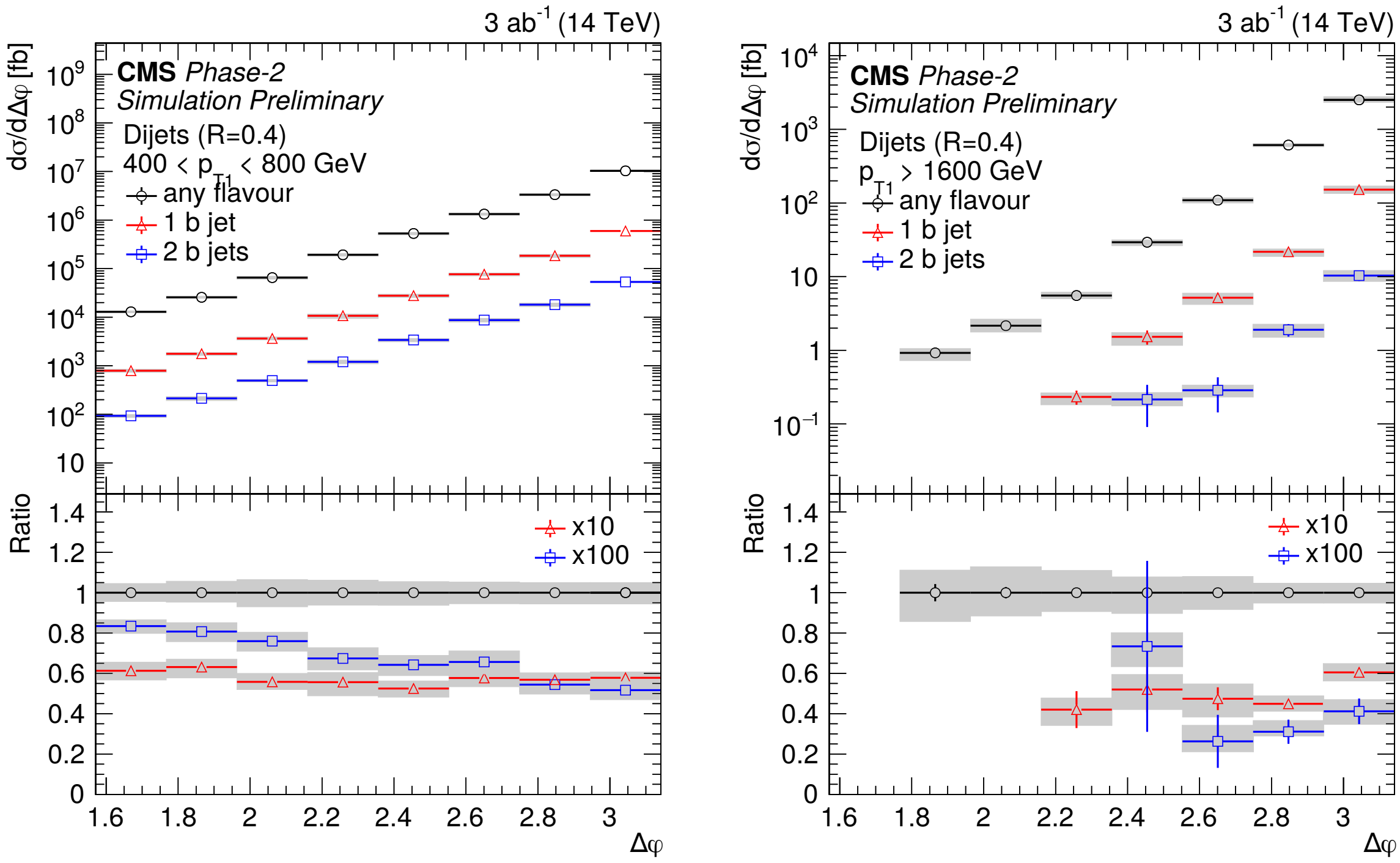

Figure 6: Distribution of the azimuthal correlation $\Delta\phi$ between two leading jets at the particle level for leading jet $p_T$ between 400 GeV and 800 GeV (left) and above 1600 GeV (right). The uncertainties represent the expected statistical uncertainty assuming 3 ab$^{-1}$. The systematic uncertainty includes the jet energy scale uncertainty (JEC) and uncertainties from b tagging.

The shape of the $\Delta\phi$ distribution of inclusive dijet production differs from the one of $\mathrm{b\bar{b}}$ jet production. When both leading jets are required to be b jets, the dominant production channel is $gg \to \mathrm{b\bar{b}}$. Since the gluons in the initial state radiate more than quarks, the $p_{\mathrm{T}}$ of the $\mathrm{b\bar{b}}$ system is expected to be higher and, consequently, the jets are more decorrelated in $\Delta\phi$. At larger $p_{\mathrm{T}}$ ($p_{\mathrm{T}} > 1600\,\mathrm{GeV}$) this effect becomes less visible, also because of the restricted range in $\Delta\phi$ due to statistics. There is no apparent difference between single b jet production and the inclusive cross section. The figures in this section include the ratio with respect to the jet+jet differential cross section (the relative uncertainties shown in the lower panel correspond the statistical and systematic uncertainties of the production cross sections) to visualize the size of the uncertainties and the difference in shape.

In Fig. 7 the particle-level cross section as a function of $|\Delta y|$ is shown, with statistical uncertainties corresponding to an integrated luminosity of 3 ab$^{-1}$ including b tagging as described in Section 3.1. Larger differences between the cross sections of different flavors can be seen, where the b jets are preferably produced in the central region. The main reason for this observation is the suppression of the b quark density in the proton with respect to the light flavors at high $x$. In Run 2 similar distributions were studied for inclusive dijet production [37].

In conclusion, different regions in rapidity and $\Delta\phi$ are sensitive to the different parton-level processes and thus can provide constraints on the parton densities, especially when the jet flavor is measured.

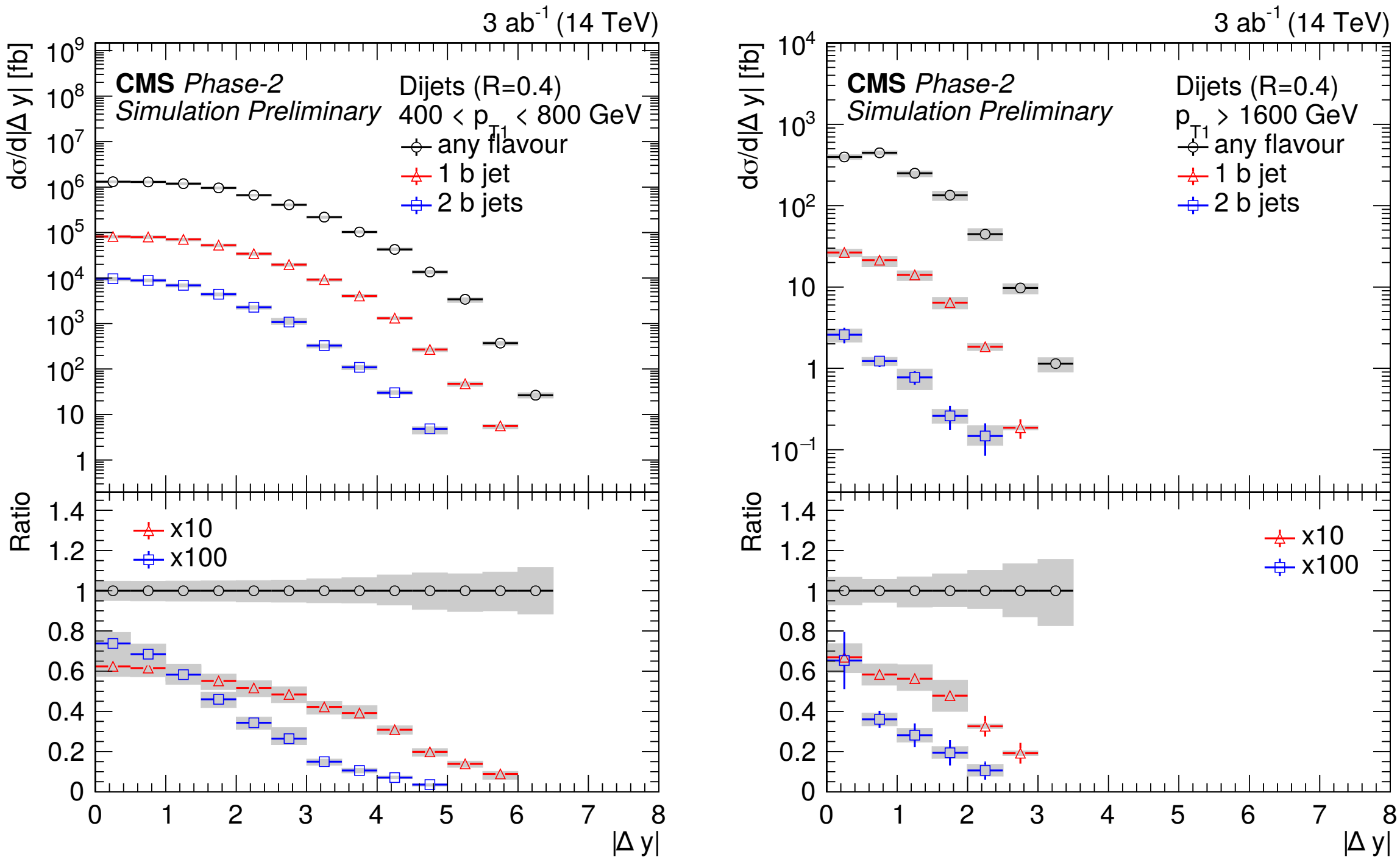

Figure 7: Distribution of the rapidity difference $|\Delta y|$ between two leading jets at the particle level for leading jet $p_T$ between 400 GeV and 800 GeV (left) and above 1600 GeV (right). The uncertainties represent the expected statistical uncertainty assuming 3 $ab^{-1}$. The systematic uncertainty includes the jet energy scale uncertainty (JEC) and uncertainties from b tagging.

### 4.4 High-$p_T$ $t\bar{t}$-jets

Jets originating from t quarks provide further information on the flavor dependence of QCD cross sections. The t jets are defined in the fully hadronic decay mode, where the t quark decays into a W boson and a b quark with the W boson decaying hadronically. The measurement can be efficiently performed in the boosted region, with jet $p_T > 400$ GeV. In contrast to the inclusive and b jet measurements, a jet radius of $R = 0.8$ is used to ensure all decay products are clustered into one jet. We use a particle level definition for the t jet, i.e., the jet must contain a B hadron as well as two subjets, where the subjet with largest $p_T$ must have a mass of $50 < m_{subjet} < 150$ GeV and can be identified as a W boson candidate. The subjets are found by applying the soft-drop algorithm [50] which also suppresses the contribution from soft partons, as well as from underlying event and (at detector level) pileup.

Of particular interest are the azimuthal correlations between $t\bar{t}$ jets in the back-to-back region in the transverse plane, as they might be subject to significant corrections due to color correlations between initial- and final-state soft gluons [22, 23].

Top quark jets can be distinguished from the dominant background of QCD multijets through substructure techniques at the detector level: the soft-drop algorithm (with $z_{cut} = 0.1$ and $\beta = 0$) is applied to remove the contribution from soft partons [50]. The soft-drop mass is required to be around the top quark mass and the N-subjettiness variables $\tau_N$ are used to suppress the QCD background [51]. Since the b quark should be present in the jet, the b tagging technique can be used to further suppress QCD background. Only leading and subleading jets with $p_T > 400$ GeV and $|\eta| < 2.5$, $m_{SD} > 105$ GeV, and $\tau_3/\tau_2 < 0.68$ together with a b tag (with tight working point) are kept as $t\bar{t}$ jets candidates at the detector level. These selection criteria are based on the experience from Run 2 analyses [33], giving confidence on good signal selection and significant background rejection.

In Fig. 8 (left), the particle level cross section for $t\bar{t}$ jets is shown as a function of the leading jet transverse momentum. The statistical uncertainties correspond to an integrated luminosity of 3 $ab^{-1}$ including efficiencies for selecting t jets at the detector level. The efficiency for selecting $t\bar{t}$ jets ranges from 25 % at $p_T \sim 500$ GeV to about 5 % at $p_T > 1.5$ TeV, as obtained from the Delphes simulation. Systematic uncertainties originate from b tagging, jet energy scale, and the uncertainty related to the jet substructure, i.e., to the jet mass scale and the jet mass resolution. Both of them affect the shape of the $m_{SD}$ distribution. Based on the analyses from Run 2, the jet mass scale uncertainty in the barrel region is around 1% and the jet mass resolution uncertainty is around 10%.

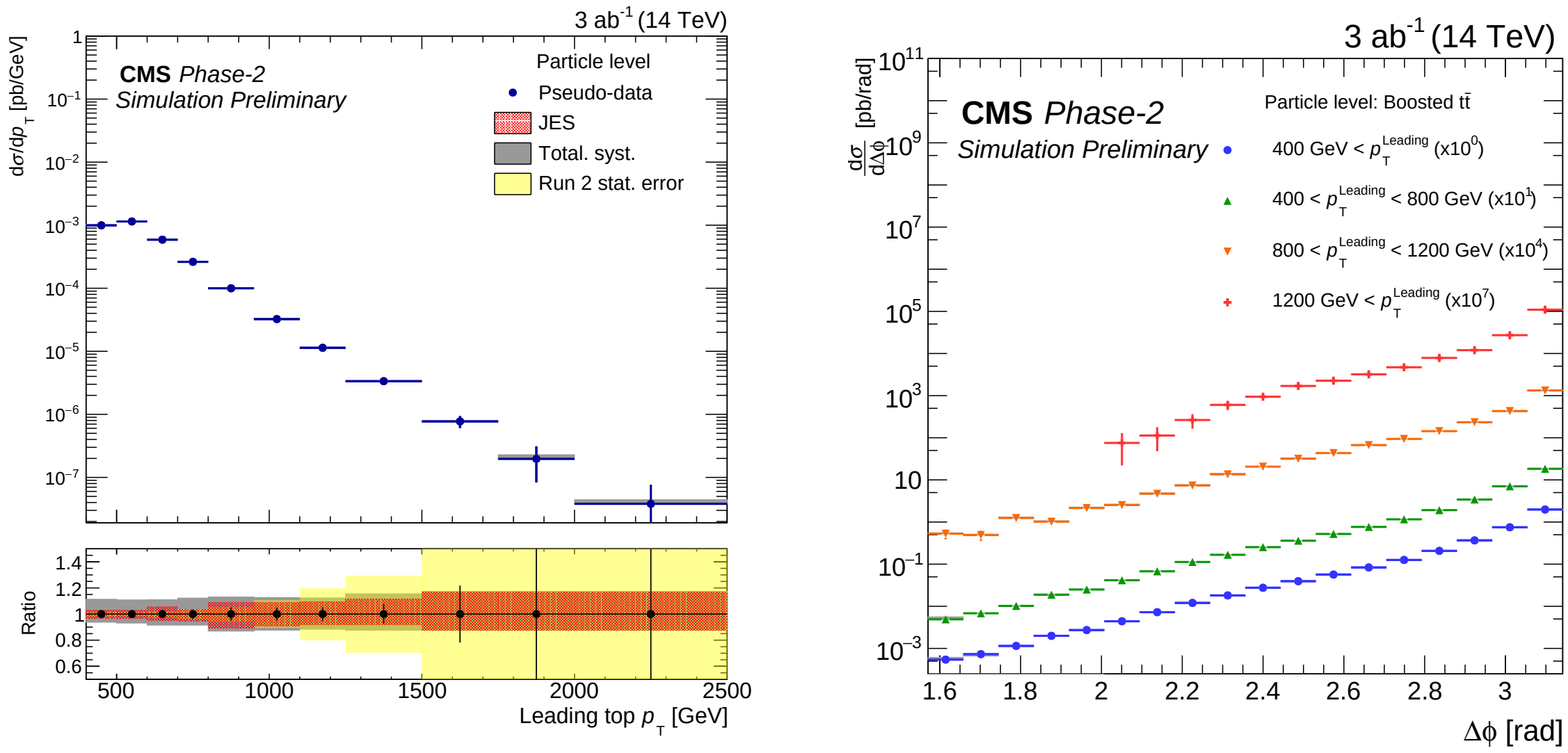

Figure 8: The cross section at particle level as a function of the leading-jet $p_T$ in $t\bar{t}$ events (left), and as a function of $\Delta\phi$ between the two leading $t\bar{t}$ jets (right). The statistical uncertainties correspond to an integrated luminosity of 3 $ab^{-1}$, including efficiencies from the selection of t jets at detector level. The systematic uncertainties are described in the main text.

In Fig. 8 (right), the azimuthal correlation for $t\bar{t}$ jets is shown for various ranges of the leading jet $p_T$. The uncertainties are obtained in the same way as for Fig. 8 (left). The efficiency for selecting $t\bar{t}$ jets ranges from 10% at small $\Delta\phi$ to about 20% at $\Delta\phi \sim \pi$, as obtained from the Delphes simulation.

### 4.5 W boson production at large $p_T$

Jets originating from hadronic decays of W and Z bosons form also a contribution to inclusive jet cross sections. For simplicity, we consider here only W boson production which has a hadronic branching fraction of $\sim$70%. As in the case of the t jet, jets with a radius of $R = 0.8$ have to be considered to ensure that all decay products of the W boson are included in the jet. Of particular interest are again the azimuthal correlations between a highly boosted, high-$p_T$ W boson decaying hadronically and the recoiling jet. The kinematic situation is very similar as in the case of $t\bar{t}$ jets, with the difference that the jet from the hadronically decaying vector boson has no color connection to the initial-state partons, and thus the azimuthal correlation does not suffer from color correlations between initial and final-state partons.

The W boson jets are identified by anti-$k_T$ jets with $R = 0.8$, where the hadronic decay products of the W boson are fully contained inside the jet. The major background is coming from the QCD multijets. To suppress this background, the soft-drop mass of the jet is required to be close to the W mass, namely $65 < m_{SD} < 105$ GeV. The particle-level cross section as a function of the $p_T$ of the W boson candidates of W+jet events where the W boson jet is required to have a $p_T > 400$ GeV and $|\eta| < 2.5$ is shown in Fig. 9 (left). In Fig. 9 (right) the azimuthal correlation between the jet originating from the W boson and the recoil jet is shown for several intervals of the W boson transverse momentum. The statistical uncertainties do not include any correction from efficiencies, since the background from QCD processes is large and would need further studies.

One of the interesting features of this process is the absence of color connection between the W boson jet and the initial and/or final state, in contrast to dijets or $t\bar{t}$ jets.

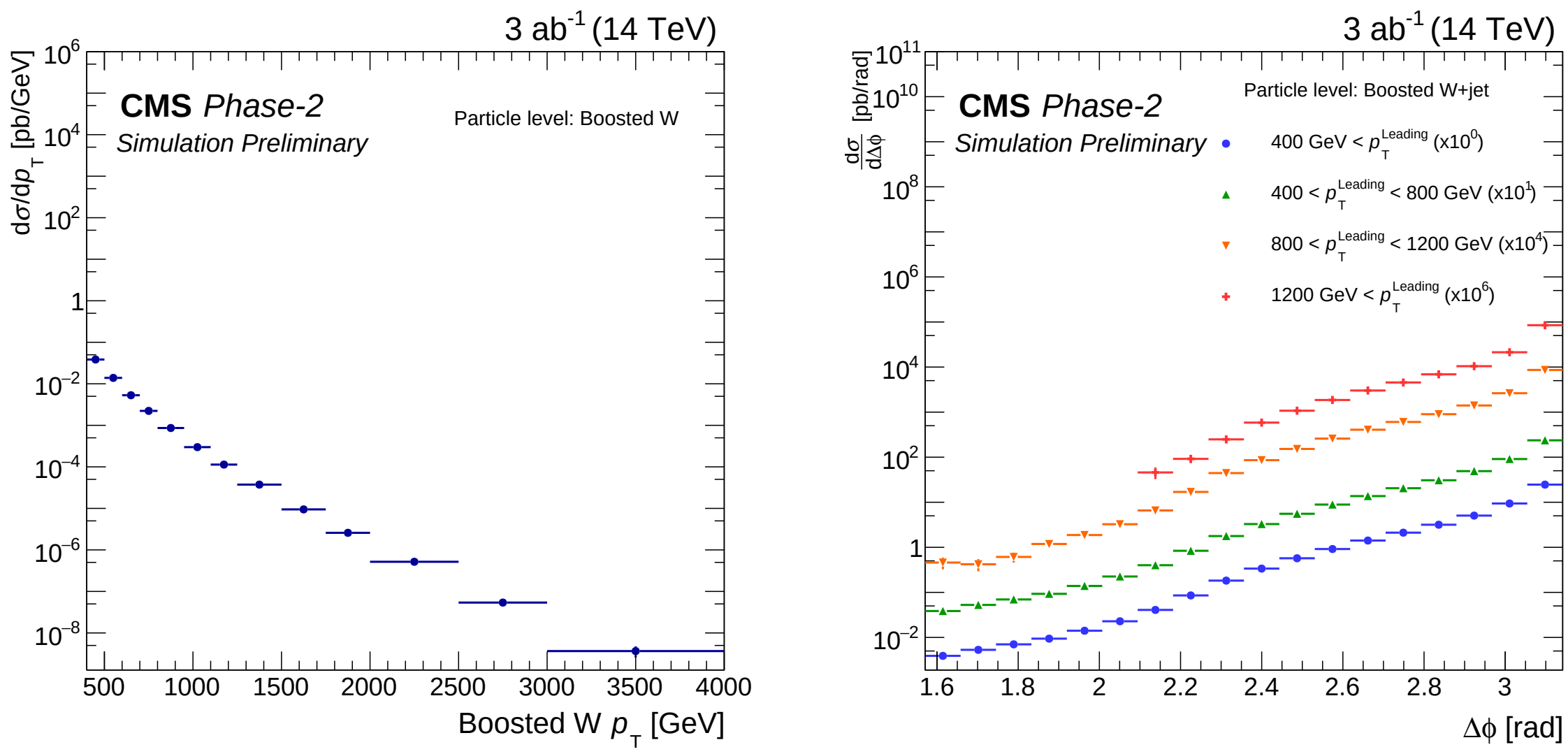

Figure 9: The cross section as a function of $p_T$ for hadronically decaying W bosons (left), and as a function of $\Delta\phi$ between the jet originating from the W boson and the recoil jet (right). The statistical uncertainties do not include selection efficiencies.

## 4.6 Overview of the jet measurements

In Fig. 10 we show a comparison of the jet cross sections as a function of $p_T$ and as a function of $\Delta\phi$ for the different processes discussed above. For comparison, here all use $R = 0.8$. In Fig. 10 (left) the inclusive b jet cross section is shown (for comparison with the inclusive jet cross section), while in Fig. 10 (right) the two-b-jet cross section is shown. Except for the cross section for W production, the statistical uncertainties shown correspond to an integrated luminosity of 3 $ab^{-1}$ including efficiencies due to b tagging and selection at the detector level, estimated from the Delphes simulation.

It can be seen that the shapes of the $p_T$ spectra are comparable but in the normalization the $t\bar{t}$ cross section is about ten thousand times smaller than the inclusive jet cross section. The ratio to the inclusive dijet cross section as a function of $\Delta\phi$ illustrates the differences in shape of the $\Delta\phi$ distribution of the different processes (all processes are normalized at $\Delta\phi = \pi$), which depend on the partonic configuration of the initial state.

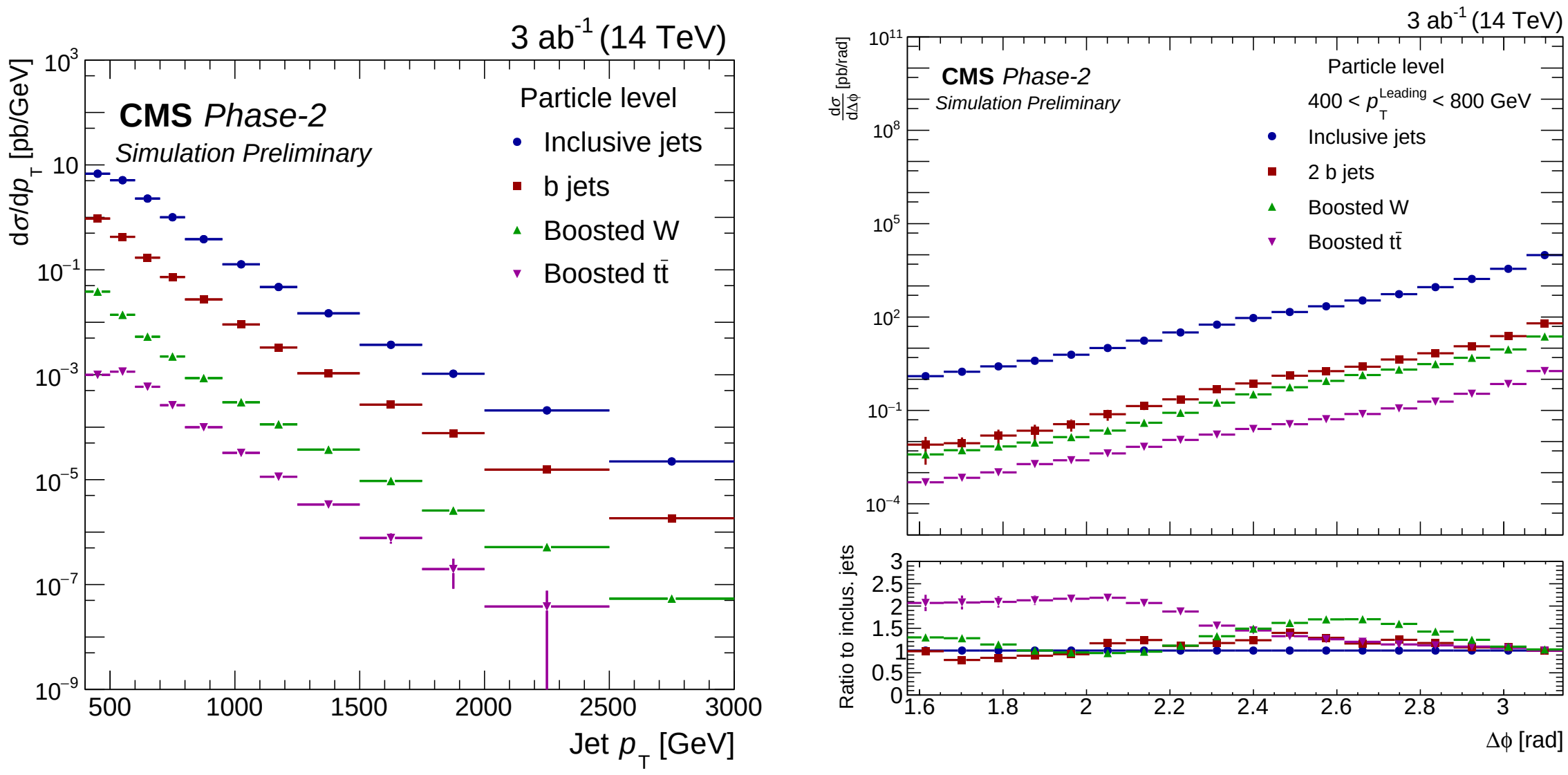

Figure 10: The overview of the particle-level differential jet cross sections (with $R = 0.8$) as a function of $p_{\mathrm{T}}$ (left) and $\Delta\phi$ (right) for various processes. In the left plot the inclusive b jet cross section is shown (for comparison with the inclusive jet cross section), while for $\Delta\phi$ the two-b-jet cross section is shown. For the ratio the normalization is fixed arbitrarily at $\Delta\phi = \pi$. The cross section of W production does not include statistical uncertainties corrected for efficiencies and background subtraction.

# 5 Conclusion

We have determined the expected reach in $p_{\mathrm{T}}$ for inclusive jets and b jets at the HL-LHC. The HL-LHC data will allow to probe the proton structure and perturbative QCD in general at the highest ever achieved scales. The inclusive b jet production is a process, which can be identified with high accuracy. We show that at high $p_{\mathrm{T}}$, the b jets are more and more affected by gluon splitting.

The angular correlation between the two leading $p_{\mathrm{T}}$ jets is evaluated as a function of the $\Delta\phi$ and $|\Delta y|$ variables. It is demonstrated that these variables together with the possible b-jet requirement enhance the sensitivity to the different partonic content of the proton. The studies are complemented with a particle-level study of boosted W+jet events. The angular correlation variables are sensitive to perturbative soft-gluon radiation and are important for calculations involving soft gluon resummation.

The boosted $\mathrm{t\bar{t}}$ cross section in the high $p_{\mathrm{T}}$ region is studied, where even the top quark mass becomes negligible. Consequently, the top quark pair is produced at a rate comparable to that of light quarks. However, the prominent process at high $p_{\mathrm{T}}$ is the quark-quark scattering which makes the top quark pair production still suppressed, as the probability to produce top quarks within the QCD evolution (in the shower) is low. This is in contrast to the case of b quarks, which at high $p_{\mathrm{T}}$ typically are produced within the QCD evolution, i.e., in the initial-state shower.

With an integrated luminosity of 3 $\mathrm{ab}^{-1}$, inclusive jet cross section measurements can reach a $p_{\mathrm{T}} \sim 4$ TeV, inclusive b jet measurements can reach a $p_{\mathrm{T}} \sim 3$ TeV, jets originating from hadronic top quarks can reach a $p_{\mathrm{T}} \sim 2$ TeV, and boosted hadronically decaying W bosons can access the region of $p_{\mathrm{T}} \sim 2.5$ TeV.

# ATLAS PUB Note

ATL-PHYS-PUB-2018-051

17th December 2018

# Prospects for jet and photon physics at the HL-LHC and HE-LHC

The ATLAS Collaboration

In this note prospects for the measurement of the inclusive jet, dijet, inclusive prompt photon and photon+jet production cross sections in proton-proton collisions at 14 and 27 TeV are presented. Double differential predictions for the inclusive jet cross sections as a function of the absolute jet rapidity and jet transverse momentum and the dijet spectrum as a function of half the absolute rapidity separation between the two highest transverse momentum jets and the invariant mass of these two jets are evaluated. Relevant uncertainties, including the individual contributions to the jet energy scale uncertainty, are calculated for jets with $p_{\mathrm{T}} > 100$ GeV within jet rapidity $|y| < 3$. Expectations for inclusive isolated photons are presented in terms of cross sections differentially in photon transverse energy in different ranges of photon pseudorapidity. Estimations for photon+jet events are described in terms of distributions in photon transverse energy, jet transverse momentum, invariant mass of the photon+jet system and $|\cos\theta^{*}|$. The study covers the region of photon transverse energies above 400 GeV and jet transverse momenta in excess of 300 GeV. A good understanding of these processes is of relevance for searches for new phenomena beyond the Standard Model. The sensitivity of these processes to the parton distribution functions in the proton is also shown.

# 1 Introduction

Precise measurements of jet and photon production cross sections are crucial in understanding physics at hadron colliders. Inclusive jet production ($p + p \to \text{jet} + X$) cross sections, dijet production ($p + p \to \text{jet} + \text{jet} + X$) cross sections as well as inclusive photon production ($p + p \to \gamma + X$) cross sections and cross sections for associated photon and jet production ($p + p \to \gamma + \text{jet} + X$) provide valuable information about the strong coupling constant ($\alpha_s$) and the parton density functions (PDFs) of the proton. Furthermore, events with jets and photons in the final-state represent a background to many other processes at the Large Hadron Collider (LHC) [1]. A good understanding of the photon and jet production processes is therefore relevant in many searches for new physics.

The LHC provided $pp$ collisions at centre-of-mass energies $\sqrt{s}$ = 7, 8 and 13 TeV and delivered more than 385 fb$^{-1}$ to the ALICE, ATLAS, CMS and LHCb experiments during the Run-1 and Run-2 data-taking periods. The high-luminosity phase of the Large Hadron Collider (HL-LHC) is expected to start in 2026 with $pp$ collisions at $\sqrt{s}$ = 14 TeV and will deliver a total integrated luminosity of about 6000 fb$^{-1}$ to all experiments. The High-Energy LHC (HE-LHC) is expected to use the existing LHC tunnel and provide $pp$ collisions at $\sqrt{s}$ = 27 TeV to collect more than 15000 fb$^{-1}$ of data over 20 years of operation.

Production of jets and photons in $pp$ collisions are among the processes directly testing the smallest experimentally accessible space-time distances. Next-to-leading-order (NLO) perturbative QCD calculations [2, 3] give quantitative predictions of the jet production cross sections. Progress in next-to-next-to-leading-order (NNLO) QCD calculations has been made recently [4–6]. As fixed-order QCD calculations only make predictions for the quarks and gluons associated with the short-distance scattering process, corrections for the fragmentation of these partons to particles need to be evaluated.

The production of prompt photons inclusively and in association with at least one jet in $pp$ collisions provides a testing ground for perturbative QCD with a hard colourless probe. All photons produced in $pp$ collisions that are not from hadron decays are considered as “prompt”. Two processes contribute to prompt-photon production: the direct process, in which the photon originates directly from the hard interaction, and the fragmentation process, in which the photon is emitted in the fragmentation of a high transverse momentum parton [7, 8].

Measurements of the cross sections for inclusive isolated-photon production and associated photon and jet production at the highest photon transverse energies ($E_{\rm T}^{\gamma}$) and jet transverse momenta ($p_{\rm T}^{\rm jet}$) as well as jet production at highest jet transverse momentum and dijet invariant mass allow for tests of the Standard Model predictions in a regime beyond what has been explored so far. They represent a wealth of data to test the fixed order calculations as well as investigate novel approaches to the description of parton radiation and evaluate the importance of electroweak corrections in pure QCD production processes calculations. In addition, since the dominant photon production mechanism in $pp$ collisions at the LHC proceeds via the $qg \to q\gamma$ channel and the jet production goes via $gg$ and $qg$ scatterings (with $qq$ channel providing a large contribution in the high-$p_{\rm T}$ range), those measurements are sensitive to the gluon density in the proton [9–12]. Furthermore, those measurements validate the modelling used for background studies in searches for physics beyond the Standard Model that involve photons and jets, such as the search for new phenomena in final states with a photon and a jet [13, 14].

The dynamics of the underlying photon+jet production processes in $2 \to 2$ hard collinear scattering can be investigated using the variable $\theta^*$, where $\cos\theta^* \equiv \tanh(\Delta y/2)$ and $\Delta y$ is the difference between the rapidities of the two final-state particles. The variable $\theta^*$ coincides with the scattering polar angle in the centre-of-mass frame for collinear scattering of massless particles, and its distribution is sensitive to the spin of the exchanged particle. The distribution of the invariant mass of the leading photon and the leading jet ($m^{\gamma-\mathrm{jet}}$) is also used to study the event dynamics since it is predicted in QCD to be monotonically decreasing for increasing values of $m^{\gamma-\mathrm{jet}}$ in the absence of resonances that decay into a photon and a jet.

Prospects are presented for prompt photon and jet production in $pp$ collisions at $\sqrt{s}$ = 14 TeV and $\sqrt{s}$ = 27 TeV in terms of cross sections for inclusive isolated photons and for photon+jet events as well as inclusive jet and dijet production cross sections. For inclusive isolated photons, expectations for the cross section differentially in $E_{\mathrm{T}}^{\gamma}$ in different ranges in photon pseudorapidity ($\eta^{\gamma}$)[1] are presented. For photon+jet events, estimations for the cross section differentially in $E_{\mathrm{T}}^{\gamma}$, $p_{\mathrm{T}}^{\mathrm{jet}}$, $\cos\theta^{*}$ and $m^{\gamma-\mathrm{jet}}$ are presented. The jet production study is presented in terms of double differential cross sections for inclusive jet transverse momentum and the dijet system mass binned in jet rapidity and half absolute rapidity difference between the two leading jets, respectively. The upper-end reach of the energetic observables, such as $E_{\mathrm{T}}^{\gamma}$, $p_{\mathrm{T}}^{\mathrm{jet}}$, $m^{\gamma-\mathrm{jet}}$ and $m_{\mathrm{jj}}$, is determined and the extension with respect to the latest measurements by the ATLAS Collaboration is emphasized [15–17].

In addition, this note presents a study of the uncertainties in the inclusive jet cross sections related to the uncertainties in the measurement of the jet energies in proton-proton collisions at $\sqrt{s}$ = 14 TeV for jets with $p_{\mathrm{T}} > 100$ GeV and within $|y| < 3$.

## 2 The ATLAS detector and the High-Luminosity and High-Energy LHC

The ATLAS experiment [18] at the LHC is a multi-purpose particle detector with a forward-backward symmetric cylindrical geometry and a near $4\pi$ coverage in solid angle. It consists of an inner tracking detector surrounded by a thin superconducting solenoid providing a 2 T axial magnetic field, electromagnetic and hadron calorimeters, and a muon spectrometer. The inner tracking detector covers the pseudorapidity range $|\eta| < 2.5$. It consists of silicon pixel, silicon micro-strip, and transition radiation tracking detectors. Lead/liquid-argon (LAr) sampling calorimeters provide electromagnetic (EM) energy measurements with high granularity. A hadron (steel/scintillator-tile) calorimeter covers the central pseudorapidity range ($|\eta| < 1.7$). The end-cap and forward regions are instrumented with LAr calorimeters for both EM and hadronic energy measurements up to $|\eta| = 4.9$. The muon spectrometer surrounds the calorimeters and is based on three large air-core toroidal superconducting magnets with eight coils each. The field integral of the toroids ranges between 2.0 and 6.0 T m across most of the detector. The muon spectrometer includes a system of precision tracking chambers and fast detectors for triggering. A two-level trigger system, custom hardware followed by a software-based level, is used for online event selection and to reduce the event rate to about 1 kHz for offline reconstruction and storage.

The HL-LHC will operate at an instantaneous luminosity up to $7.5 \times 10^{34}$ cm$^{-2}$s$^{-1}$ that corresponds to an average number of inelastic proton-proton collisions per bunch-crossing $\langle\mu\rangle$ of 200. The HL-LHC conditions will demand a performance from the ATLAS detector that is well beyond the original design. An upgrade of all major ATLAS sub-detectors is needed before the start of this new phase to cope with the high-radiation environment and the large increase in pileup. The new Inner Tracker (ITk) [19, 20] will extend the ATLAS tracking capabilities to pseudorapidity $|\eta| < 4.0$. The upgraded Muon Spectrometer [21] with a forward muon tagger included will also provide lepton identification capabilities to $|\eta| < 4.0$. The new high granularity timing detector (HGTD) [22] designed to mitigate the pileup effects is foreseen in the forward region of $2.4 < |\eta| < 4.0$. The electronics of both LAr [23] and Tile [24] calorimeters will be upgraded to cope with longer latencies needed by the trigger system at such harsh pileup conditions. An upgraded TDAQ system [25] based on a hardware trigger with a maximum rate of 1 MHz and 10 ms latency and software-based reconstruction will send event data into storage at up to 10 kHz rate. A study of the expected performance of the upgraded ATLAS detector at the HL-LHC is reported in Ref. [26].

[1] ATLAS uses a right-handed coordinate system with its origin at the nominal interaction point (IP) in the centre of the detector and the $z$-axis along the beam pipe. The $x$-axis points from the IP to the centre of the LHC ring, and the $y$-axis points upwards. Cylindrical coordinates $(r, \phi)$ are used in the transverse plane, $\phi$ being the azimuthal angle around the $z$-axis. The pseudorapidity is defined in terms of the polar angle $\theta$ as $\eta = -\ln\tan(\theta/2)$. Angular distance is measured in units of $\Delta R \equiv \sqrt{(\Delta\eta)^2 + (\Delta\phi)^2}$.

The HE-LHC target luminosity is $2.5\times10^{35}$ cm$^{-1}$s$^{-1}$. The HE-LHC will employ the dipole magnets with a field of 16 T developed in the framework of the Future Circular Collider project [27]. The HE-LHC could accommodate two high-luminosity interaction-points at the locations of the ATLAS and CMS experiments [28]. It will allow to study new physics scenarios beyond the reach of the 14 TeV collider.

# 3 Analysis

## 3.1 Photon Analysis

The study of photon production is done via the analysis of inclusive isolated photons and that of photon production in association with at least one jet. In both analyses the photon is required to have a transverse energy in excess of 400 GeV and the pseudorapidity to lie in the range $|\eta^\gamma| < 2.37$ excluding the region $1.37 < |\eta^\gamma| < 1.56$. The photon is required to be isolated by imposing an upper limit on the amount of transverse energy inside a cone of size $\Delta R = 0.4$ in the $\eta$–$\phi$ plane around the photon, excluding the photon itself: $E_{\mathrm{T}}^{\mathrm{iso}} < E_{\mathrm{T,max}}^{\mathrm{iso}}$.

In the inclusive photon analysis, the goal is the measurement of the differential cross section as a function of $E_{\mathrm{T}}^{\gamma}$ in four regions of the photon pseudorapidity: $|\eta^\gamma| < 0.6$, $0.6 < |\eta^\gamma| < 1.37$, $1.56 < |\eta^\gamma| < 1.81$ and $1.81 < |\eta^\gamma| < 2.37$. Photon isolation is enforced by requiring $E_{\mathrm{T}}^{\mathrm{iso}} < 4.2 \cdot 10^{-3} \cdot E_{\mathrm{T}}^{\gamma} + 4.8$ GeV.

In the photon+jet analysis, jets are reconstructed using the anti-$k_t$ algorithm [29] with a radius parameter $R = 0.4$. Jets overlapping with the photon are not considered if the jet axis lies within a cone of size $\Delta R = 0.8$ from the photon. The leading jet is required to have transverse momentum above 300 GeV and rapidity in the range $|y^{\mathrm{jet}}| < 2.37$. No additional condition is used for the differential cross sections as functions of $E_{\mathrm{T}}^{\gamma}$ and $p_{\mathrm{T}}^{\mathrm{jet}}$. For the differential cross sections as functions of the invariant mass of the photon+jet system and $|\cos\theta^*|$, additional constraints are imposed: $m^{\gamma-\mathrm{jet}} > 1.45$ TeV, $|\cos\theta^*| < 0.83$ and $|\eta^\gamma \pm y^{\mathrm{jet}}| < 2.37$. These additional constraints are imposed to remove the bias due to the rapidity and transverse-momentum requirements on the photon and the leading jet [30, 31]. Photon isolation is enforced by requiring $E_{\mathrm{T}}^{\mathrm{iso}} < 4.2 \cdot 10^{-3} \cdot E_{\mathrm{T}}^{\gamma} + 10$ GeV.

The yields of inclusive isolated photons and of photon+jet events are estimated using the program JETPHOX 1.3.1_2 [32, 33]. This program includes a full next-to-leading-order QCD calculation of both the direct-photon and fragmentation contributions to the cross sections for the $pp \to \gamma$+X and $pp \to \gamma$+jet+X reactions. The number of massless quark flavours is set to five. The renormalisation ($\mu_{\mathrm{R}}$), factorisation ($\mu_{\mathrm{F}}$) and fragmentation ($\mu_{\mathrm{f}}$) scales are chosen to be $\mu_{\mathrm{R}} = \mu_{\mathrm{F}} = \mu_{\mathrm{f}} = E_{\mathrm{T}}^{\gamma}$. The calculations are performed using the MMHT2014 [34] parameterisations of the proton parton distribution functions (PDFs) and the BFG set II of parton-to-photon fragmentation functions at NLO [35]. Predictions are also obtained with other PDF sets, namely CT14 [36], HERAPDF2.0 [37], NNPDF3.0 [38] and PDF4LHC HL-LHC [39]. The strong coupling constant $\alpha_s(m_Z)$ is set to the value assumed in the fit to determine the PDFs.

The reliability of the estimated yields using the program JETPHOX is supported by the high purity of the signal photons and the fact that the NLO QCD predictions describe adequately the measurements of these processes using $pp$ collisions at $\sqrt{s} = 13$ TeV [15, 16].

### 3.2 Jet Analysis

#### 3.2.1 Experimental analysis

Jets are reconstructed using the anti-$k_t$ algorithm with distance parameter R = 0.4 as implemented in the FastJet software package [40]. Jets are calibrated following the procedure described in [41]. The four momenta of the jets are recalculated to originate from the hard-scatter vertex rather than from the centre of the detector. The jet energy is corrected for the effect of pile-up using jet area-based correction together with residual number of primary vertices ($N_{\text{PV}}$)- and $\langle\mu\rangle$-dependent correction as described in [42]. In addition, a jet energy- and $\eta$-dependent correction is applied. It is derived from Monte Carlo (MC) simulation and is designed to lead to agreement in energy and direction between reconstructed jets and particle jets on average. Further jet calibration steps applied in Run-2 measurements, namely the Global Sequential Calibration (GSC) [43] and the in situ calibration [41] are not derived and used in the current study. The GSC reduces effects from fluctuations in the composition of particles forming a jet and fluctuations in the hadronic shower caused by interactions of the hadrons with dead material in the calorimeter. An in situ correction is applied on data to remove residual differences in energy response between data and MC simulation evaluated using techniques where the $p_{\text{T}}$ of the jet is balanced against well-measured objects, for example in photon+jet and $Z$-boson+jet events.

The total jet energy scale (JES) uncertainty in Run-2 measurements is compiled from 88 sources that all need to be propagated through the analysis in order to correctly account for uncertainty correlations in the jet calibration.

A reduced set of uncertainty components (nuisance parameters) is derived from eigenvectors and eigenvalues of the diagonalised total JES covariance matrix on the jet-level. The globally reduced configuration with 19 nuisance parameters (NPs) is used in this study. Eight NPs coming from the in situ techniques are related to the detector description, physics modelling and measurements of the $Z/\gamma$ energies in ATLAS calorimeters. Three describe the physics modelling and the statistics of the dijet MC sample and the non-closure of the method, used to derive the $\eta$-intercalibration [41]. The single-hadron response studies [44] are used to describe the JES uncertainty in the high-$p_{\text{T}}$ jet regions, where the in situ studies have limited statistics. Four NPs are due to the pile-up corrections of the jet kinematics, that take into account mis-modelling of $N_{\text{PV}}$ and $\langle\mu\rangle$ distributions, dependence of the average energy density, $\rho$, on the pileup activity in a given event, $\rho$-topology, and the residual $p_{\text{T}}$ dependence. Finally, two uncertainty components take into account the difference in the calorimeter response to the quark- and gluon-initiated jets (flavour response) and the jet flavour composition, and one uncertainty in the correction for the energy leakage beyond the calorimeter, the ”punch-through“ effect.

In order to estimate the precision in the jet cross section measurements at the HL-LHC three scenarios of possible uncertainties in the jet energy scale calibration are defined.

In all three scenarios, the high-$p_T$ uncertainty, the punch-through uncertainty and the flavour composition uncertainty are considered to be negligible. The JES uncertainty in the high-$p_T$ range will be accessed using the multi-jet balance (MJB) method, rather than single hadron response measurements, since the high statistics at the HL-LHC will allow precision JES measurements in the high-$p_T$ region. Flavour composition and flavour response uncertainties are driven by the MC generator differences. With advances in the MC models and tuning of their parameters these uncertainties could be significantly reduced. The flavour composition uncertainty is therefore ignored to study the maximal impact of the expected improvements on the modelling of parton showers and hadronisation on precision jet energy measurements. The flavour response uncertainties are kept at the same value as in Run-2 or reduced by a factor of two in conservative and optimistic scenarios, respectively.

The pile-up uncertainties, except the $\rho$-topology uncertainty, are considered to be negligible. Current small uncertainties in the JES due to mis-modelling of $N_{PV}$ and $\langle\mu\rangle$ distributions and the residual $p_T$ dependence lead to very small uncertainties at the HL-LHC conditions. With the advances of new pile-up rejection techniques the $\rho$-topology uncertainty could be maintained at a level comparable to the one in Run-2 or reduced by a factor of two. This is addressed in conservative and optimistic scenarios.

Since the Run-2 jet energy resolution (JER) uncertainty estimation is conservative, the final Run-2 JER uncertainty is expected (based on Run-1 experience) to be about twice as small as the current one. Therefore, the JER uncertainty is estimated to be half of that in Run-2.

The rest of uncertainty sources are fixed in different scenarios as follows:

- Conservative scenario:
  - All in situ components are kept at the same value as in Run-2, except the uncertainties related to the photon energy measurement in the high-$E_T$ range and the MJB method uncertainties. These uncertainties are reduced by a factor of two, since they are expected to be improved at the HL-LHC.
  - MC modelling uncertainty in the $\eta$-intercalibration is reduced by a factor of two while the other two are neglected. Currently, MC modelling uncertainty is derived by comparison of leading-order (LO) pQCD generators. In future, with advances in MC generators development, this uncertainty is expected to be reduced.
  - Flavour response uncertainty is set to the Run-2 value;
  - $\rho$-topology uncertainty is unchanged compared to Run-2 results;
- Optimistic scenario:
  - All in situ components are treated identically to the conservative scenario;
  - All three uncertainty sources in the $\eta$-intercalibration method are ignored;
  - Flavour response uncertainty is reduced by a factor of two compared to Run-2 results;
  - $\rho$-topology uncertainty is two times smaller as in Run-2;
- Pessimistic scenario:
  - same as optimistic scenario, but all uncertainty sources of in situ methods are retained from Run-2.

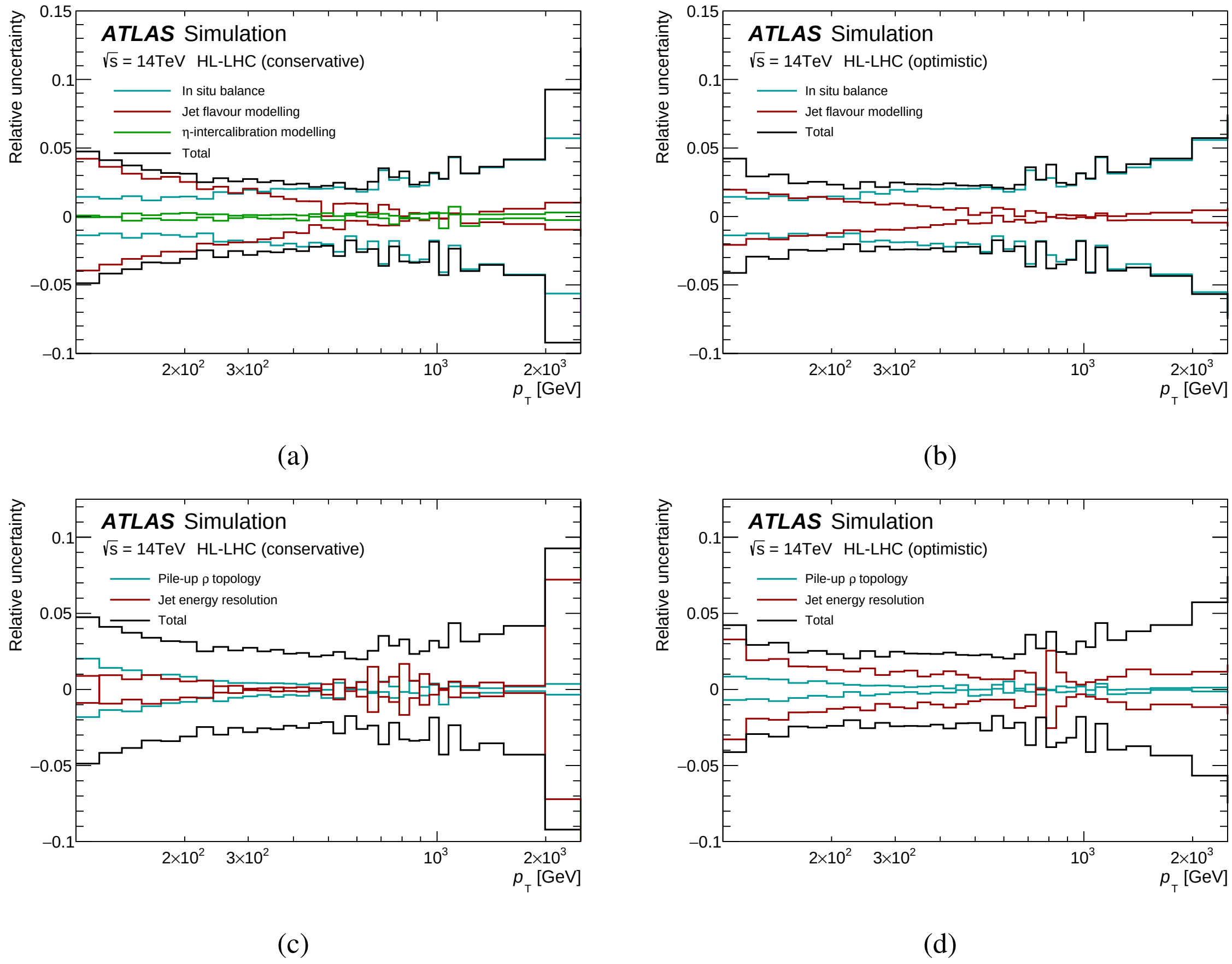

Figure 1: Relative systematic uncertainty for (a,c) conservative and (b,d) optimistic scenarios in the inclusive jet cross sections as a function of jet $p_T$ in the $|y| < 3$ rapidity region. The individual uncertainty components are shown in different colours. The total systematic uncertainty, calculated by adding the individual uncertainties in quadrature, is shown as a black line.

All components of the JES uncertainty are propagated from the jet-level to the cross section level as follows. The jet $p_T$ is scaled up and down by one standard deviation of each source of uncertainty. The difference between the nominal reco-level spectrum and the systematically shifted one is taken as a systematic uncertainty. All JES uncertainties are treated as bin-to-bin correlated and independent from each other in this procedure. The unfolding of the detector-level distributions to the particle-level spectrum is not performed is this study. A possible modification of the shapes of uncertainty components during the unfolding procedure is expected to be small and neglected in this study.

The inclusive jet cross sections in this section are studied as a function of the jet transverse momentum for jets with $p_T > 100$ GeV and within $|y| < 3$.

The estimation of the JES uncertainty in the measurements of inclusive jet cross section at the HL-LHC for three JES uncertainty scenarios are presented in Figure 1.

### 3.2.2 Fixed-order predictions and PDF sensitivity

Theoretical predictions at NLO QCD are calculated using MCFM [45] interfaced to APPLgrid [46] for fast convolution with PDFs. The renormalisation and factorisation scales are set to $\mu_{\mathrm{R}} = \mu_{\mathrm{F}} = p_{\mathrm{T}}^{\mathrm{jet}}$ for the inclusive jet cross section and $\mu_{\mathrm{R}} = \mu_{\mathrm{F}} = m_{\mathrm{jj}}$ for the dijet mass distribution. The predictions are calculated using CT14nnlo [36] PDF set provided by the LHAPDF6 [47].

The main uncertainties in the NLO predictions come from uncertainties associated with the PDFs, the choice of renormalisation and factorisation scales, and the uncertainty in the value of the strong coupling. PDF uncertainties are defined at the 68% CL and propagated through the calculations following the prescription given by the PDF set authors. The nominal scales are independently varied up or down by a factor of two in both directions excluding opposite variations of $\mu_R$ and $\mu_F$. The envelope of resulting variations of the predicted cross section is taken as the total scale uncertainty. The uncertainty from $\alpha_s$ is evaluated by calculating the cross sections using two PDF sets that differ only on the value of the strong coupling at $M_Z$ and then scaling the cross section difference corresponding to an uncertainty $\Delta\alpha_s = 0.0015$ [48].

The inclusive jet cross sections are studied double-differentially as a function of the jet transverse momentum and absolute jet rapidity while the dijet production cross sections are presented as a function of the invariant mass of the dijet system and as a function of half the absolute rapidity separation between the two highest-$p_{\mathrm{T}}$ jets satisfying $|y| < 3$, denoted $y^*$. In both analyses the leading jet is required to be within $|y| < 3$ and to have $p_{\mathrm{T}} > 100$ GeV. The other jets are required to be in the same rapidity range with $p_{\mathrm{T}} > 75$ GeV.

Figures 2 and 3 shows the theoretical uncertainties, calculated using CT14 [36] PDF set, in the inclusive jet and dijet cross sections for representative phase-space regions at $\sqrt{s} = 14$ and 27 TeV, respectively. The total uncertainty is about 5 % in the low- and intermediate $p_{\mathrm{T}}$ and $m_{\mathrm{jj}}$ regions, growing to 20–40% in the high-$p_{\mathrm{T}}$ and dijet mass ranges.

Measurements of weak boson [49], top quark [50], photon, jet productions [51] (and many others) at the LHC have been already used as inputs to global PDF fits [34, 36, 52, 53] in determination of the proton structure. High precision LHC data have allowed to further constrain the knowledge of the proton content by extending the coverage of PDF-related phase space in measurements and to significantly reduce PDF uncertainties.

A study to estimate the impact of future PDF-sensitive measurements at the HL-LHC on the precision of PDFs determination was performed in Ref. [39]. Three possible scenarios for the experimental systematic uncertainties were considered. This study concluded that HL-LHC measurements will further reduce the PDF uncertainties and published the dedicated PDF sets, PDF4LHC HL-LHC, where the HL-LHC pseudo-data were included in the fits.

Figure 4 and 5 present the comparison of inclusive jet and dijet production cross sections calculated using different PDF sets at $\sqrt{s} = 14$ and 27 TeV, respectively. It shows 5–10% difference between central values in the low- and intermediate-$p_{\mathrm{T}}$ and $m_{\mathrm{jj}}$ regions, however these predictions are still compatible with the quoted PDF uncertainty. The differences between various PDF sets predictions in the high-$p_{\mathrm{T}}$ and $m_{\mathrm{jj}}$ range highlights the expected constraining power of future measurements at the HL-LHC and HE-LHC.

Figure 6 and 7 depict the comparison of PDF uncertainties in the inclusive jet and dijet production cross sections for CT14 and PDF4LHC HL-LHC (optimistic scenario) in the $pp$ collisions at $\sqrt{s} = 14$ and

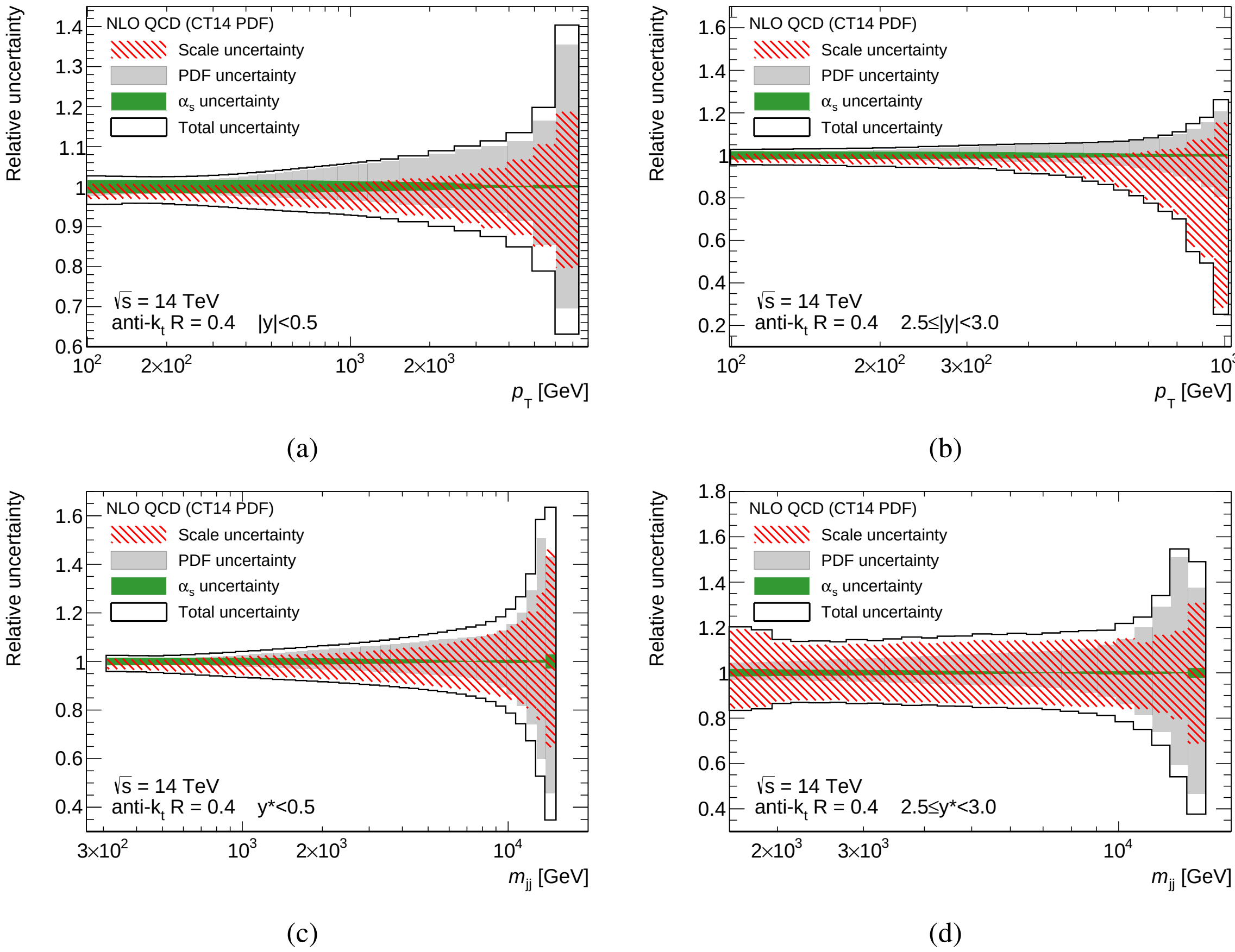

Figure 2: Relative NLO QCD uncertainties in the (a,b) inclusive jet and (c,d) dijet cross sections calculated using the CT14 PDF set at $\sqrt{s}$ = 14 TeV. Panels (a,c) and (b,d) correspond to the first and last $|y|$ and $y^*$ bins in measurements, respectively. The uncertainties due to the renormalisation and factorisation scale, $\alpha_s$, PDF and the total uncertainty are shown. The total uncertainty, calculated by adding the individual uncertainties in quadrature, is shown as a black line.

27 TeV. A significant improvement in PDF extraction is expected with the inclusion of PDF-sensitive measurements at the HL-LHC in PDF fits.

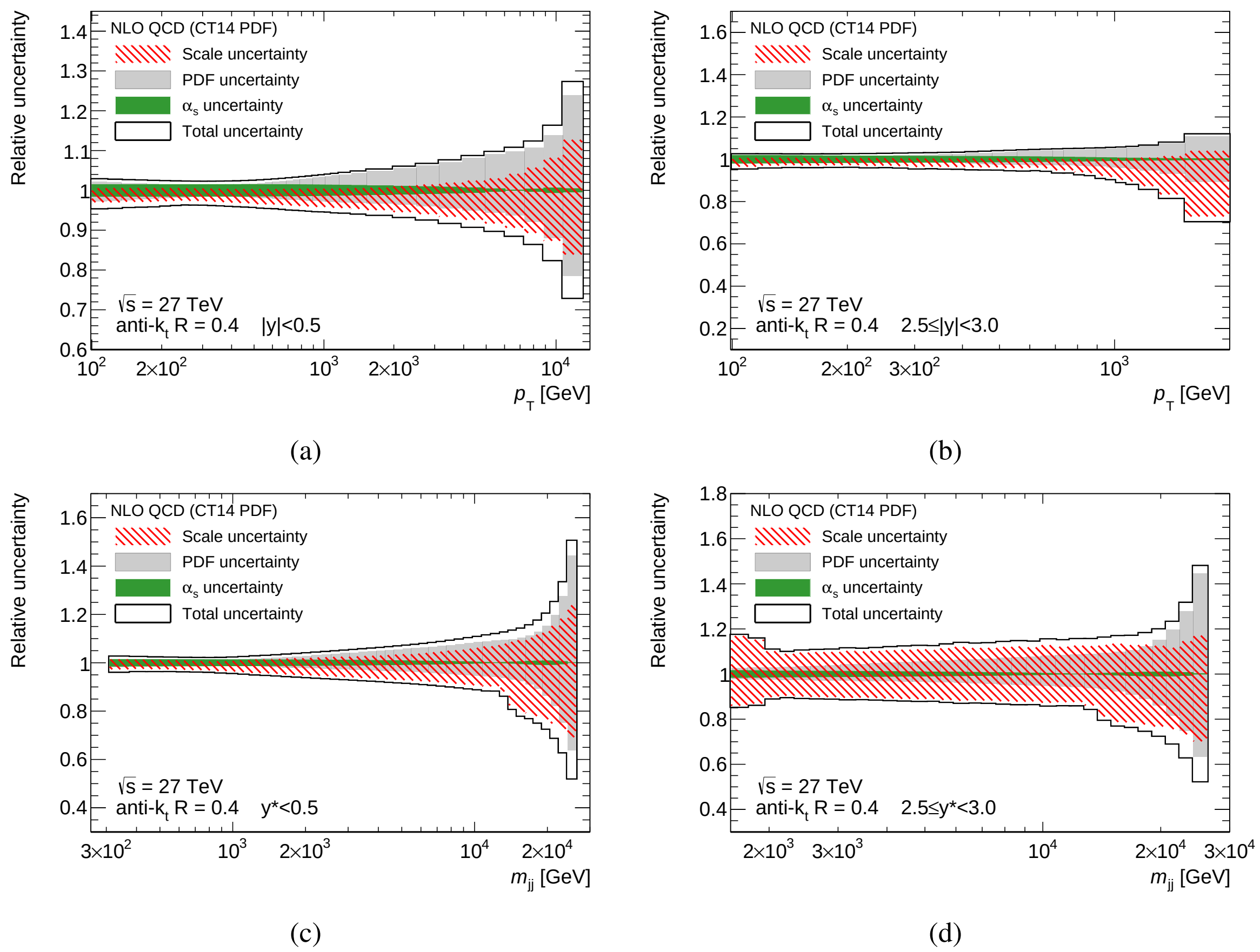

Figure 3: Relative NLO QCD uncertainties in the (a,b) inclusive jet and (c,d) dijet cross sections calculated using the CT14 PDF set at $\sqrt{s}$ = 27 TeV. Panels (a,c) and (b,d) correspond to the first and last $|y|$ and $y^*$ bins in measurements, respectively. The uncertainties due to the renormalisation and factorisation scale, $\alpha_s$, PDF and the total uncertainty are shown. The total uncertainty, calculated by adding the individual uncertainties in quadrature, is shown as a black line.

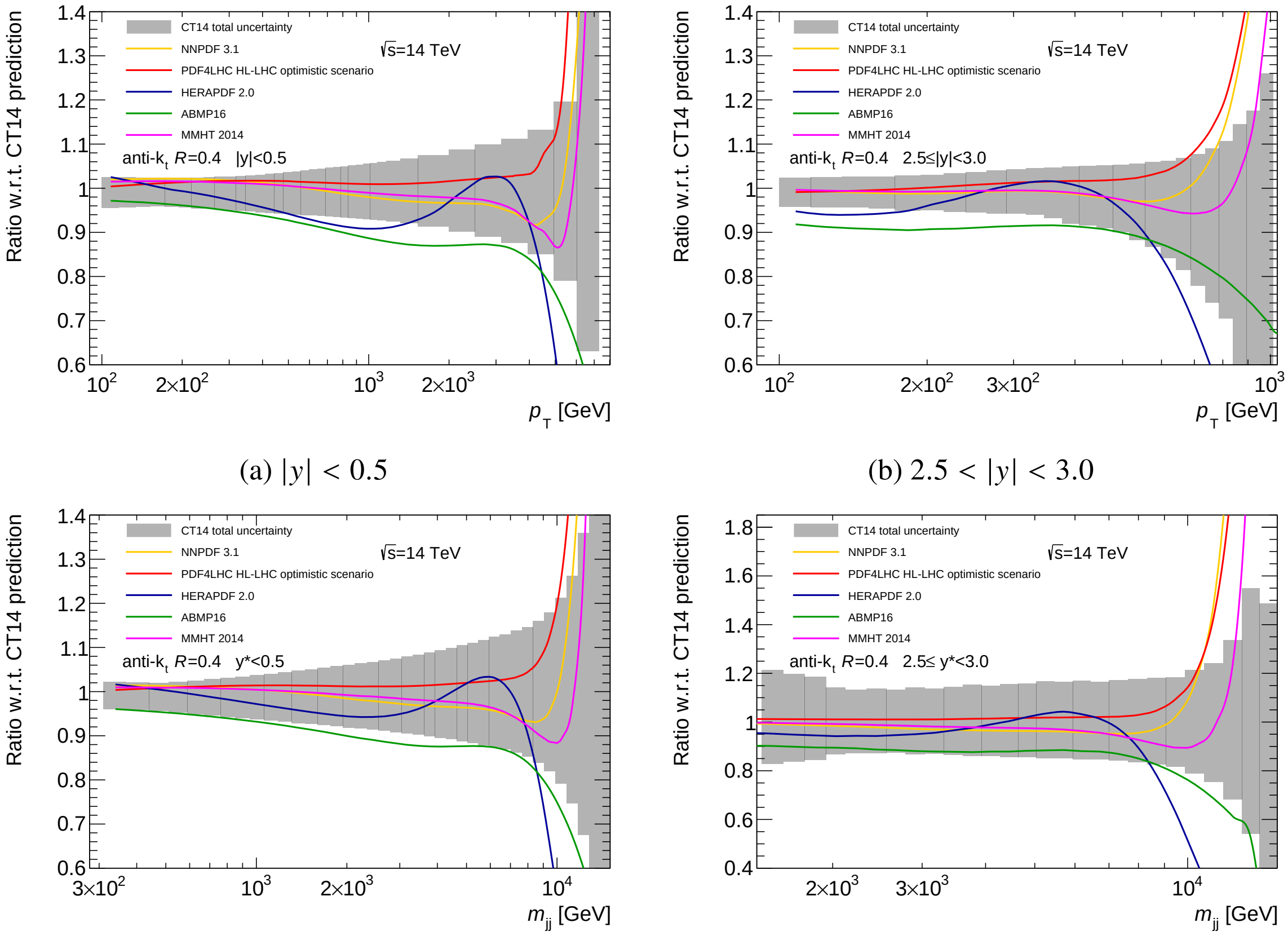

(a) $|y| < 0.5$ (b) $2.5 < |y| < 3.0$

(c) $y^* < 0.5$ (d) $2.5 < y^* < 3.0$

Figure 4: Ratio of cross sections calculated using NNPDF 3.1 [52], MMHT 2014 [34], ABMP 16 [53], PDF4LHC HL-LHC [39], to one using CT14 [36] PDFs in the (a,b) inclusive jet and (c,d) dijet cross sections at $\sqrt{s}$ = 14 TeV. Panels (a,c) and (b,d) correspond to the first and last $|y|$ and $y^*$ bins in measurements, respectively. The gray band depicts the total NLO pQCD uncertainty in cross section calculated using CT14 [36] PDF set.

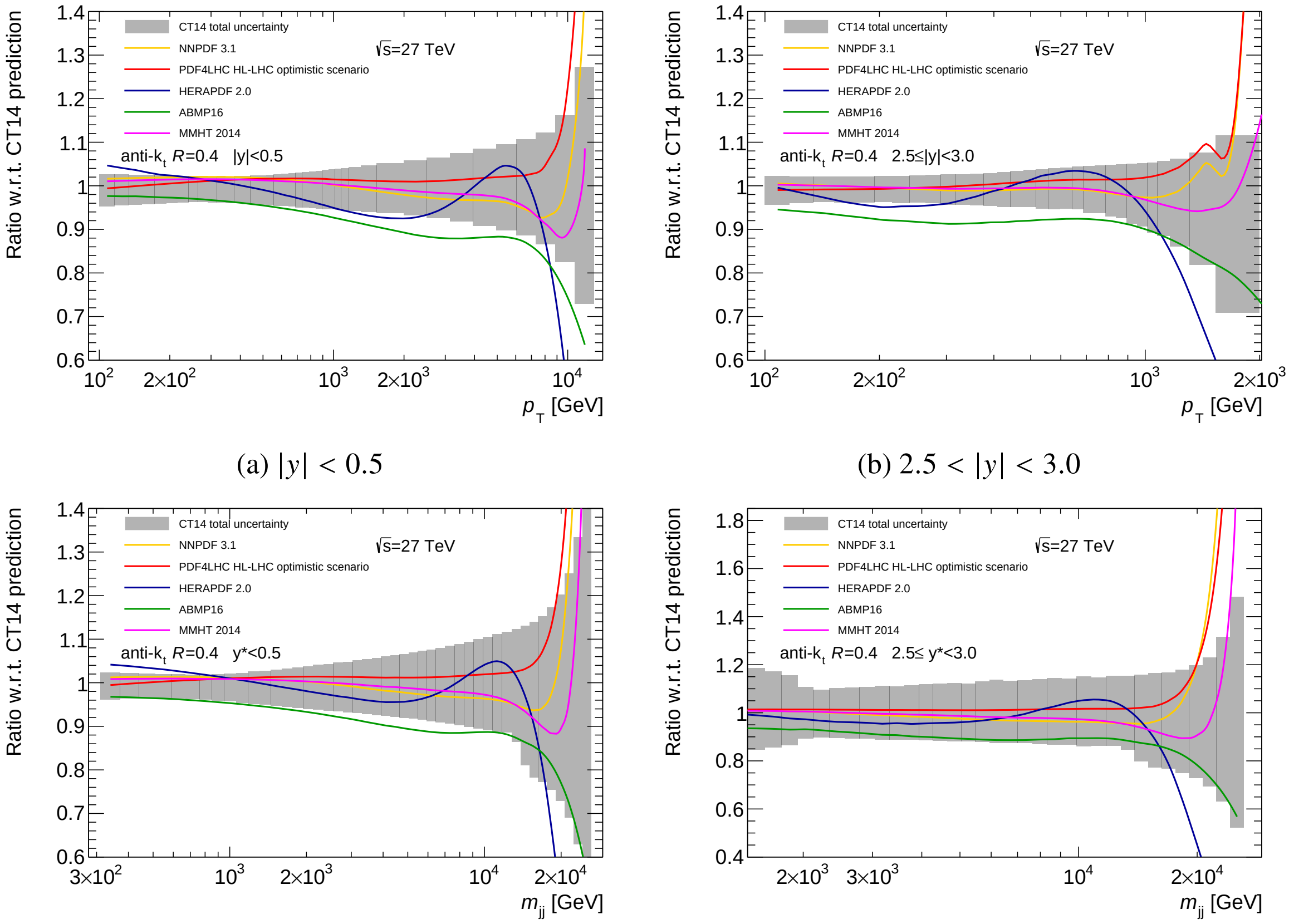

(a) $|y| < 0.5$

(b) $2.5 < |y| < 3.0$

(c) $y^* < 0.5$

(d) $2.5 < y^* < 3.0$

Figure 5: Ratio of cross sections calculated using NNPDF 3.1 [52], MMHT 2014 [34], ABMP 16 [53], PDF4LHC HL-LHC [39], to one using CT14 [36] PDFs in the (a,b) inclusive jet and (c,d) dijet cross sections at $\sqrt{s}$ = 27 TeV. Panels (a,c) and (b,d) correspond to the first and last $|y|$ and $y^*$ bins in measurements, respectively. The gray band depicts the total NLO pQCD uncertainty in cross section calculated using CT14 PDF set.

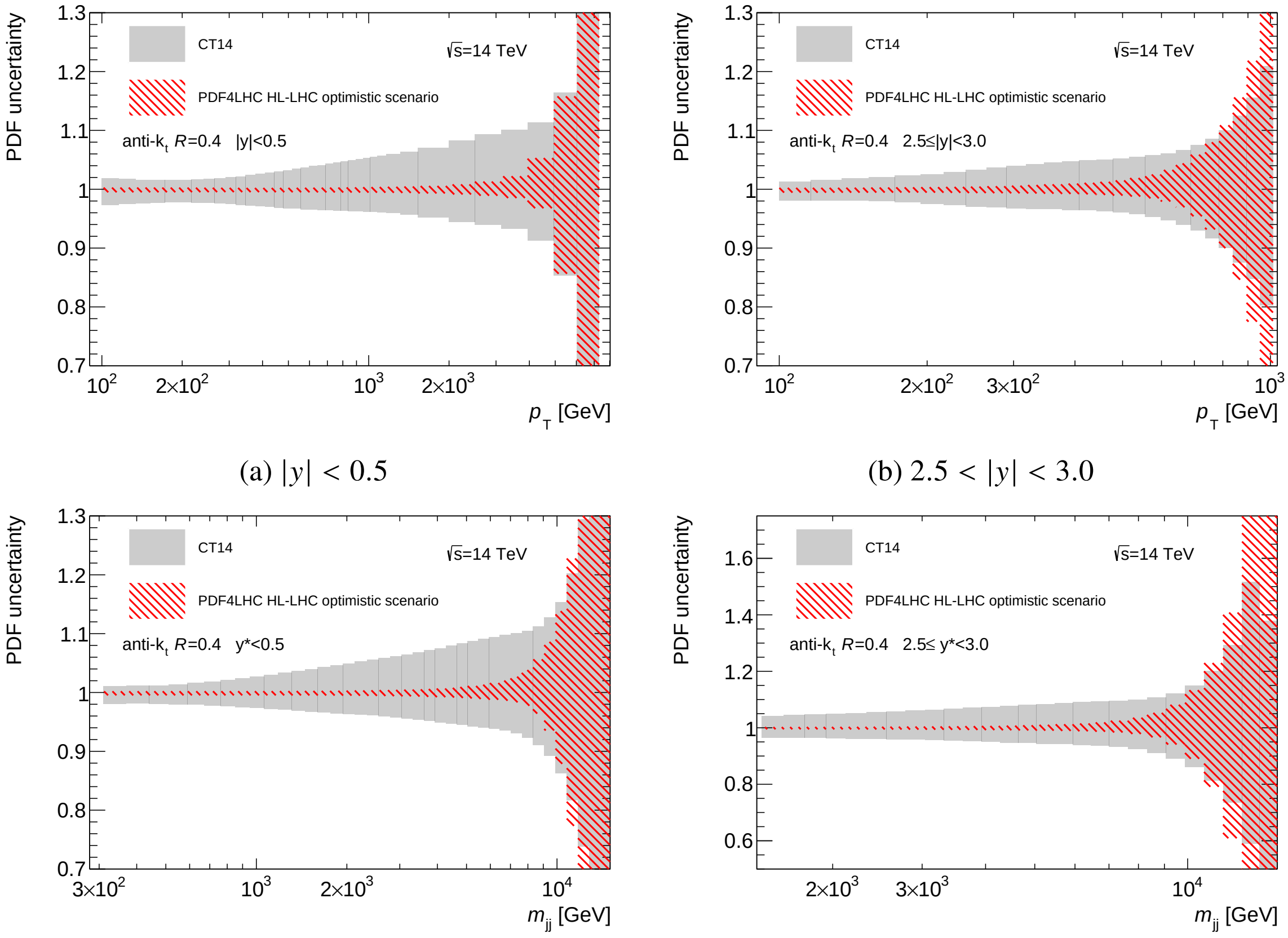

(a) $|y| < 0.5$

(b) $2.5 < |y| < 3.0$

(c) $y^* < 0.5$

(d) $2.5 < y^* < 3.0$

Figure 6: Comparison of the PDF uncertainty in the (a,b) inclusive jet and (c,d) dijet cross sections calculated using the CT14 PDF and PDF4LHC HL-LHC [39] sets at $\sqrt{s} = 14$ TeV. Panels (a,c) and (b,d) correspond to the first and last $|y|$ and $y^*$ bins in measurements, respectively.

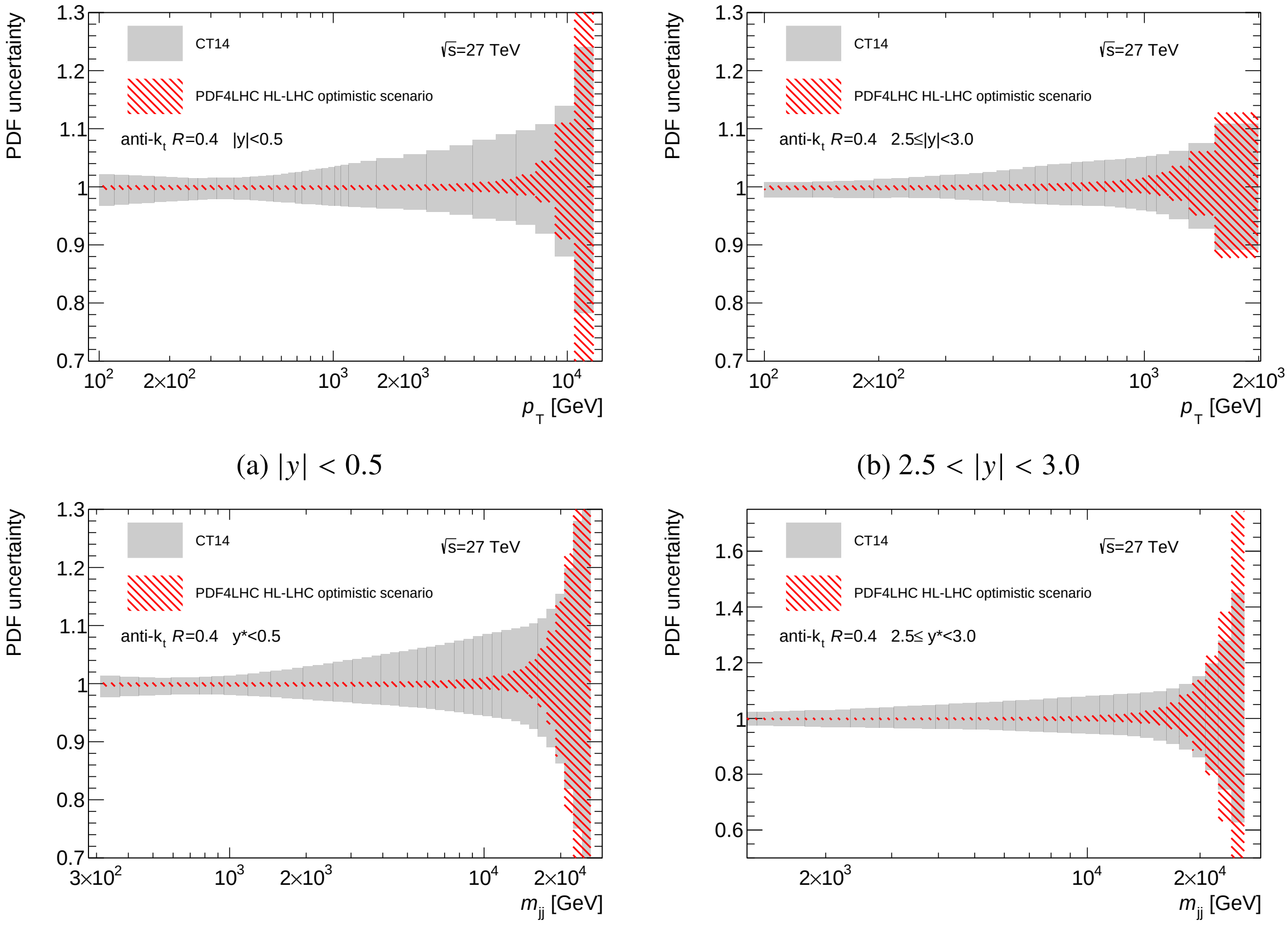

(a) $|y| < 0.5$ (b) $2.5 < |y| < 3.0$

(c) $y^* < 0.5$ (d) $2.5 < y^* < 3.0$

Figure 7: Comparison of the PDF uncertainty in the (a,b) inclusive jet and (c,d) dijet cross sections calculated using the CT14 PDF and PDF4LHC HL-LHC [39] sets at $\sqrt{s}$ = 27 TeV. Panels (a,c) and (b,d) correspond to the first and last $|y|$ and $y^*$ bins in measurements, respectively.

### 3.2.3 Non-perturbative effects

The fixed-order predictions are obtained at the parton-level. The non-perturbative corrections (NPCs) are applied to bring the theoretical predictions from parton-level to particle-level in order to allow a comparison with the measured cross sections in data. The NPC are evaluated using Pythia v8.210 MC [54] generator with A14 [55] underlying event tune and the tune variations are used to evaluate the uncertainty in the NPC due to the differences in hadronisation and underlying event modelling.

Figures 8 to 11 show separate corrections for the hadronisation, underlying event as well as the total non-perturbative correction to the inclusive jet and dijet production cross section in $pp$ collisions at $\sqrt{s} = 14$ and 27 TeV.

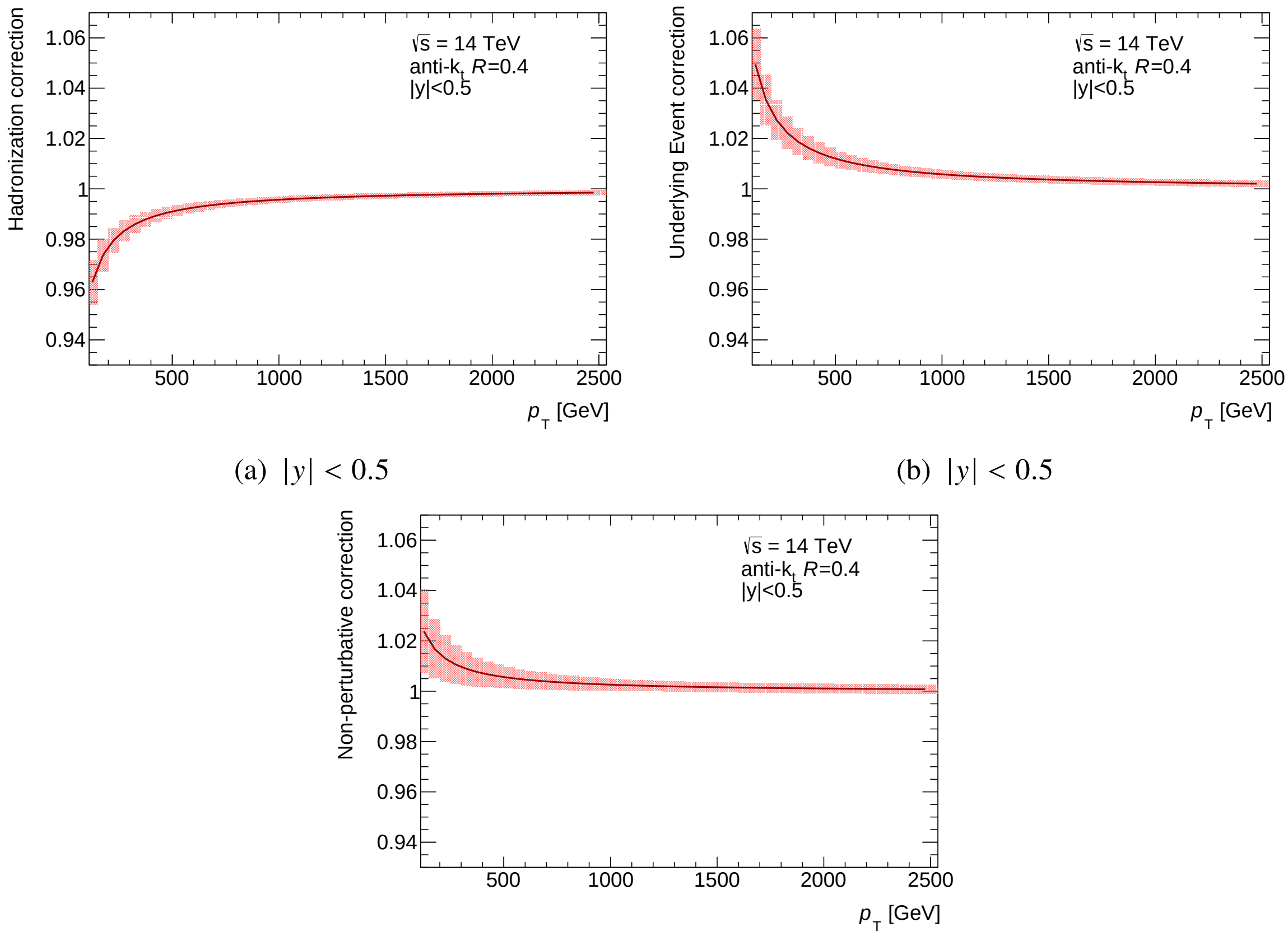

(a) $|y| < 0.5$

(b) $|y| < 0.5$

(c) $|y| < 0.5$

Figure 8: Non-perturbative corrections for the inclusive jet production cross section at $\sqrt{s} = 14$ TeV in the $|y| < 0.5$ rapidity range. Separate (a) corrections for the hadronisation, (b) underlying event and (c) the total non-perturbative correction are shown.

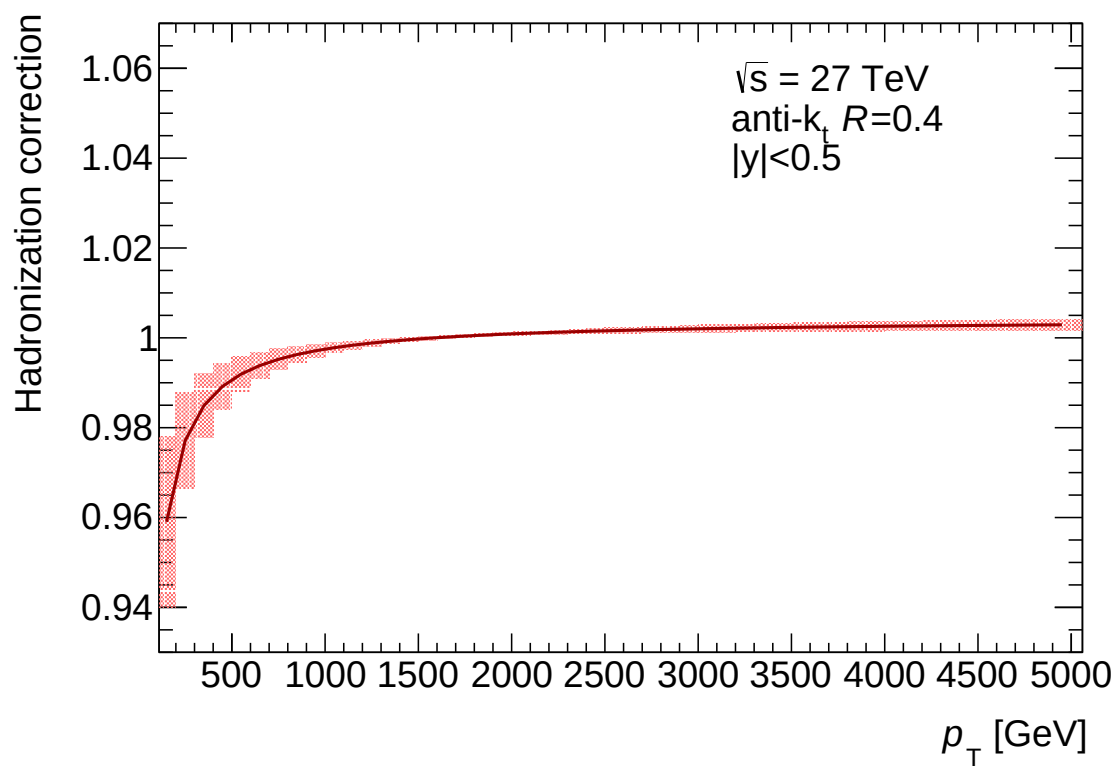

(a) $|y| < 0.5$

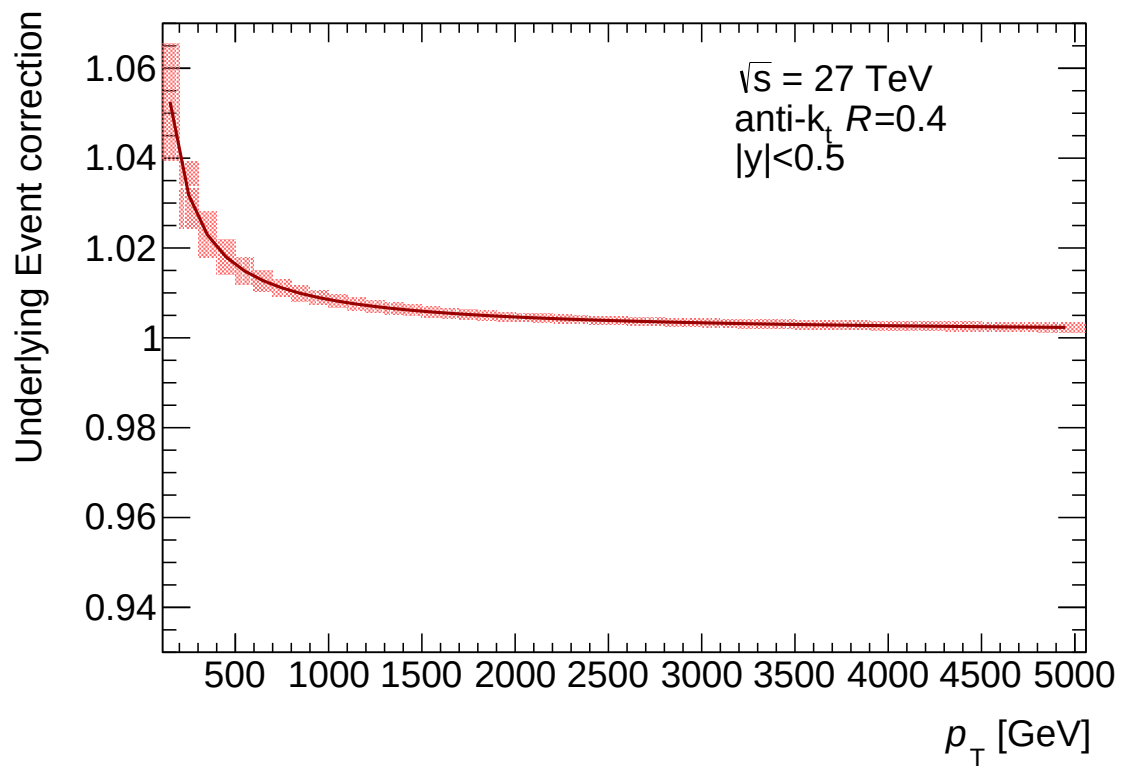

(b) $|y| < 0.5$

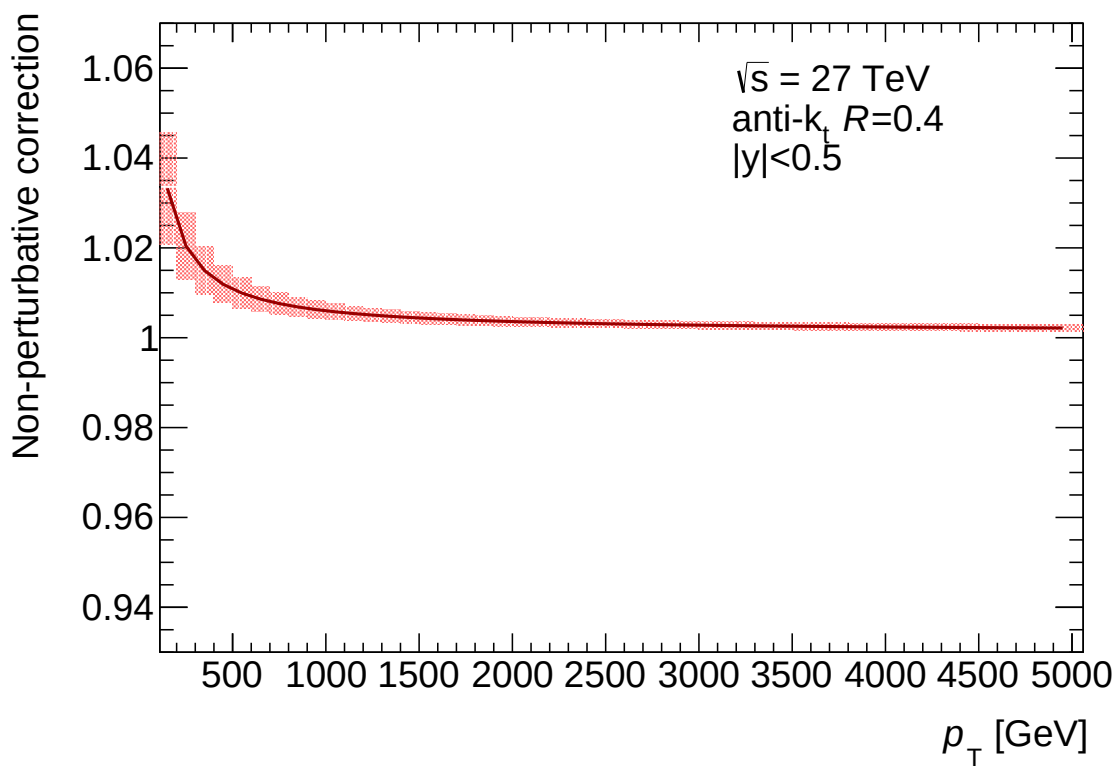

(c) $|y| < 0.5$

Figure 9: Non-perturbative corrections for the inclusive jet production cross section at $\sqrt{s} = 27$ TeV in the $|y| < 0.5$ rapidity range. Separate (a) corrections for the hadronisation, (b) underlying event and (c) the total non-perturbative correction are shown.

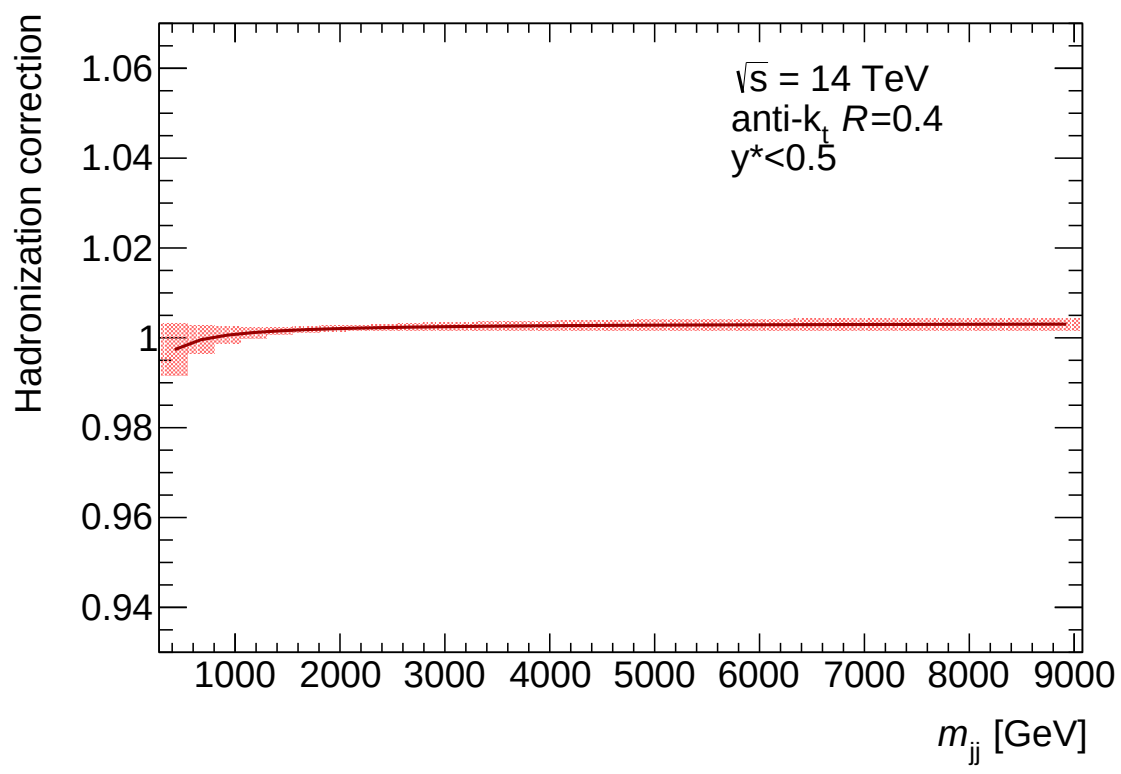

(a) $y^* < 0.5$

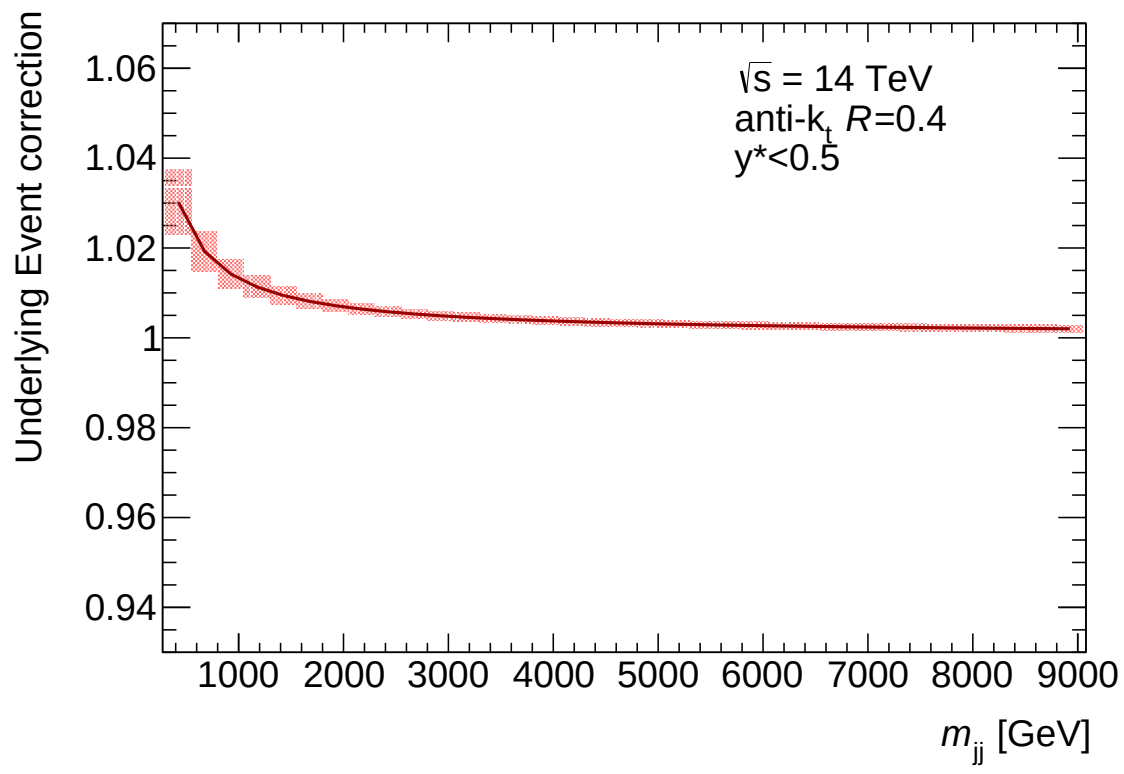

(b) $y^* < 0.5$

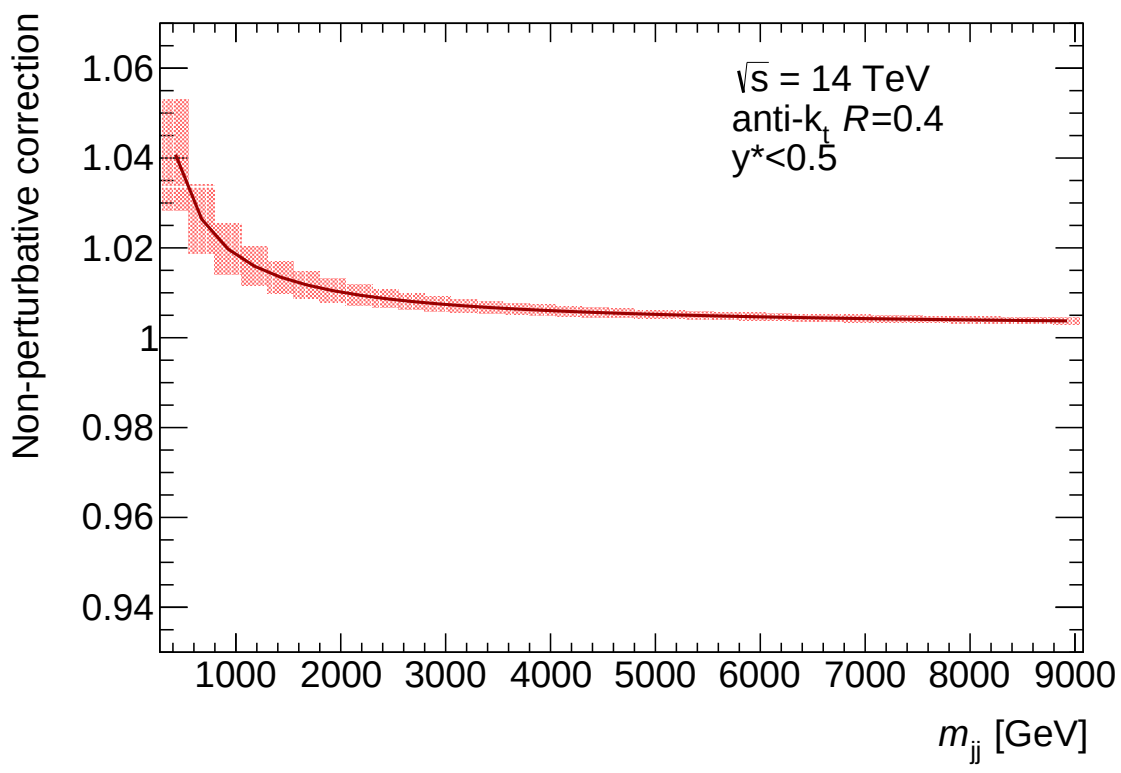

(c) $y^* < 0.5$

Figure 10: Non-perturbative corrections for the dijet production cross section at $\sqrt{s}$ = 14 TeV in the $y^* < 0.5$ rapidity range. Separate (a) corrections for the hadronisation, (b) underlying event and (c) the total non-perturbative correction are shown.

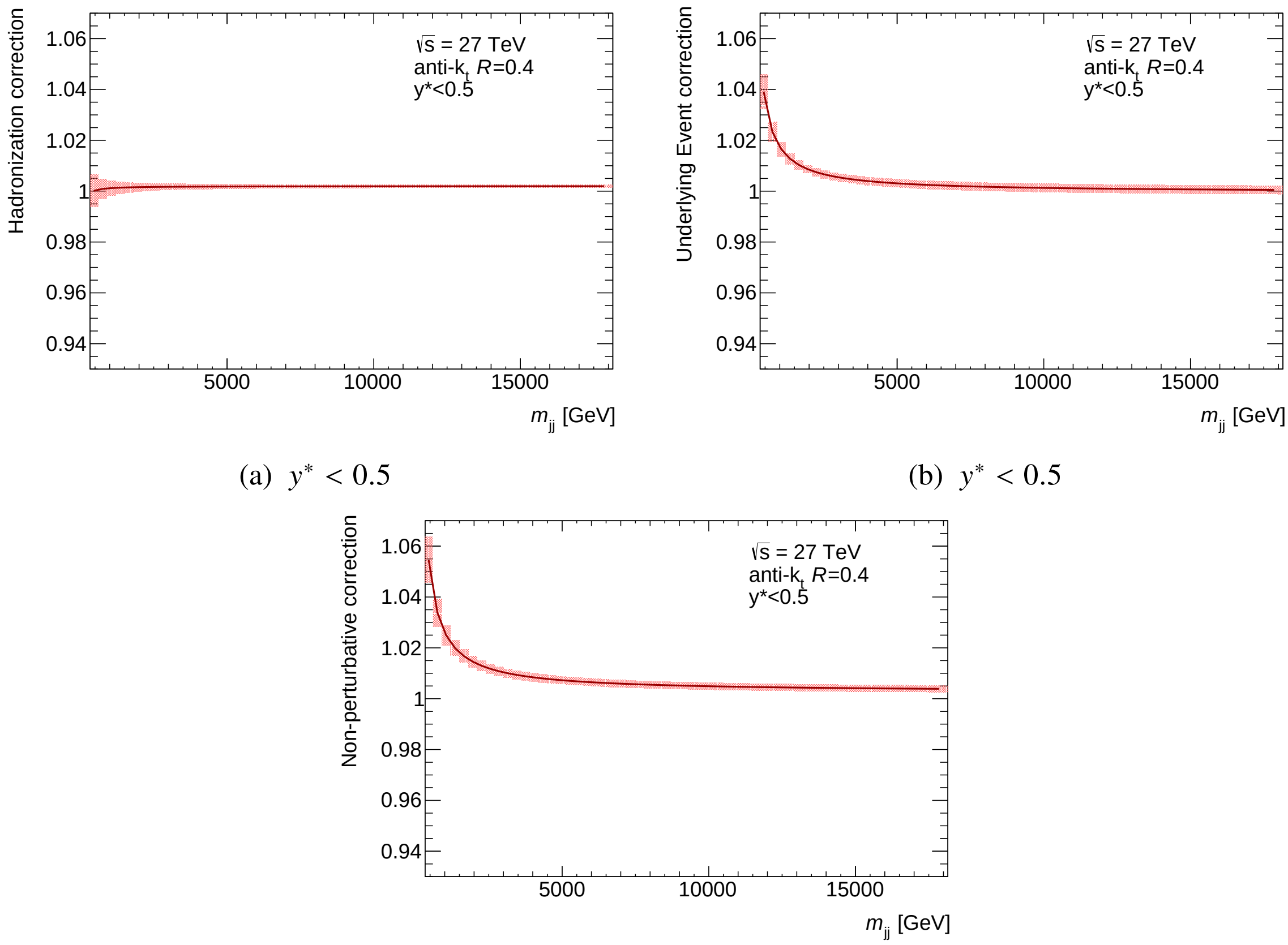

(a) $y^* < 0.5$

(b) $y^* < 0.5$

(c) $y^* < 0.5$

Figure 11: Non-perturbative corrections for the dijet production cross section at $\sqrt{s}$ = 27 TeV in the $y^* < 0.5$ rapidity range. Separate (a) correction for the hadronisation, (b) underlying event and (c) the total non-perturbative correction are shown.

The weak radiative corrections in the dijet production at $\sqrt{s}$ =14 TeV are calculated in Ref. [56]. This corrections can be of the order of several per-cents in the tails of kinematic distributions due to the Sudakov-type logarithms. The impact of these effects on the inclusive jet and dijet cross section predictions is not considered in this note.

# 4 Results

## 4.1 Photon Results

The predicted number of inclusive isolated photon events as a function of $E_T^\gamma$ in the different ranges of $|\eta^\gamma|$ assuming an integrated luminosity of 3 ab$^{-1}$ of $pp$ collision data at $\sqrt{s}$ = 14 TeV is shown in Fig. 12. The predicted number of events above an $E_T^\gamma$ threshold is shown in Fig. 13. The reach in $E_T^\gamma$ is (a) 3–3.5 TeV for $|\eta^\gamma| < 0.6$, (b) 2.5–3 TeV for $0.6 < |\eta^\gamma| < 1.37$, (c) 1.5–2 TeV for $1.56 < |\eta^\gamma| < 1.81$ and (d) 1–1.5 TeV for $1.81 < |\eta^\gamma| < 2.37$. This represents a significant extension of the region measured so far with $pp$ collisions at $\sqrt{s}$ = 13 TeV [15]; as an example, the $E_T^\gamma$ reach is extended from 1.5 TeV to 3–3.5 TeV for $|\eta^\gamma| < 0.6$. The projected cross sections as a function of $E_T^\gamma$ together with Run-2 results at $\sqrt{s}$ = 13 TeV [15] are shown in Fig. 14.

The sensitivity to the proton PDFs is studied in the ratio between the predicted cross sections with CT14, NNPDF3.0 and HERAPDF2.0 and those using MMHT2014. The ratios are shown in Fig. 15 and differences of up to 30% are seen. The predicted relative statistical uncertainty on the number of inclusive isolated photon events as a function of $E_T^\gamma$ assuming an integrated luminosity of 3 ab$^{-1}$ of collision data at $\sqrt{s}$ = 14 TeV in different ranges of photon pseudorapidity is shown in Fig. 16. A relative statistical uncertainty below 10% is achieved for photon transverse energies up to 2.5 TeV (1.5 TeV) for $|\eta^\gamma| < 0.6$ and $0.6 < |\eta^\gamma| < 1.37$ ($1.56 < |\eta^\gamma| < 1.81$ and $1.81 < |\eta^\gamma| < 2.37$).

The photon energy scale and resolution represent the dominant source of systematic uncertainty for the measurement of the inclusive isolated-photon cross section $d\sigma/dE_T^\gamma$ in $pp$ collisions at $\sqrt{s}$ = 13 TeV [15]. The size of this systematic uncertainty as estimated in Run-2 using 3.2 fb$^{-1}$ of $pp$ collision data is shown in Table 1 for selected regions. The aforementioned estimations of the systematic uncertainties due to the photon energy scale and resolution will possibly be improved by using Run-2 and Run-3 data. Furthermore, improvements in the systematic uncertainties are also expected from the HL-LHC data thanks to the increased statistics for the photon energy calibration and in situ determination of the photon identification and isolation efficiencies.

Table 1: Systematic uncertainty due to the photon energy scale ($\gamma$-ES) and resolution ($\gamma$-ER) for the measurement of the inclusive isolated-photon cross section $d\sigma/E_T^\gamma$ in $pp$ collisions at $\sqrt{s}$ = 13 TeV in different regions of $|\eta^\gamma|$ [15].

| $E_T^\gamma$ [GeV] | $\gamma$-ES and $\gamma$-ER systematic uncertainty (in %) | | | |
|---|---|---|---|---|
| | $\lvert\eta^\gamma\rvert < 0.6$ | $0.6 < \lvert\eta^\gamma\rvert < 1.37$ | $1.56 < \lvert\eta^\gamma\rvert < 1.81$ | $1.81 < \lvert\eta^\gamma\rvert < 2.37$ |
| 400–470 | +2.2, −2.2 | +3.0, −2.9 | +11, −9.3 | +4.5, −4.4 |
| 750–900 | +3.0, −2.8 | +3.8, −3.8 | +16, −15 | +6.9, −6.5 |
| 900–1100 | +3.3, −2.9 | +4.1, −4.1 | +18, −18 | |
| 1100–1500 | +4.0, −3.1 | +4.6, −4.6 | | |

The predicted number of photon+jet events as a function of $E_{\mathrm{T}}^{\gamma}$, $p_{\mathrm{T}}^{\mathrm{jet}}$, $m^{\gamma-\mathrm{jet}}$ and $|\cos\theta^{*}|$, assuming an integrated luminosity of 3 ab$^{-1}$ of $pp$ collision data at $\sqrt{s}$ = 14 TeV, is shown in Fig. 17. The predictions show that the reach in $E_{\mathrm{T}}^{\gamma}$ and $p_{\mathrm{T}}^{\mathrm{jet}}$ is 3.5 TeV and that the reach in $m^{\gamma-\mathrm{jet}}$ is 7 TeV. The predicted relative statistical uncertainty on the number of photon+jet events as a function of the different observables, assuming an integrated luminosity of 3 ab$^{-1}$ of $pp$ collision data at $\sqrt{s}$ = 14 TeV, is shown in Fig. 18. The relative statistical uncertainty is below 10% for (a) $E_{\mathrm{T}}^{\gamma}$ up to 2.5 TeV, (b) $p_{\mathrm{T}}^{\mathrm{jet}}$ up to 3 TeV and (c) $m^{\gamma-\mathrm{jet}}$ up to 6 TeV; for $|\cos\theta^{*}|$ the relative statistical uncertainty is below 1% for the entire range considered. In comparison to the latest ATLAS measurements at $\sqrt{s}$ = 13 TeV with 3.2 fb$^{-1}$ of integrated luminosity [16], the projections presented here extend significantly the reach in several observables: for $E_{\mathrm{T}}^{\gamma}$ and $p_{\mathrm{T}}^{\mathrm{jet}}$ from 1.5 TeV to 3.5 TeV and for $m^{\gamma-\mathrm{jet}}$ from 3.3 TeV to 7 TeV.

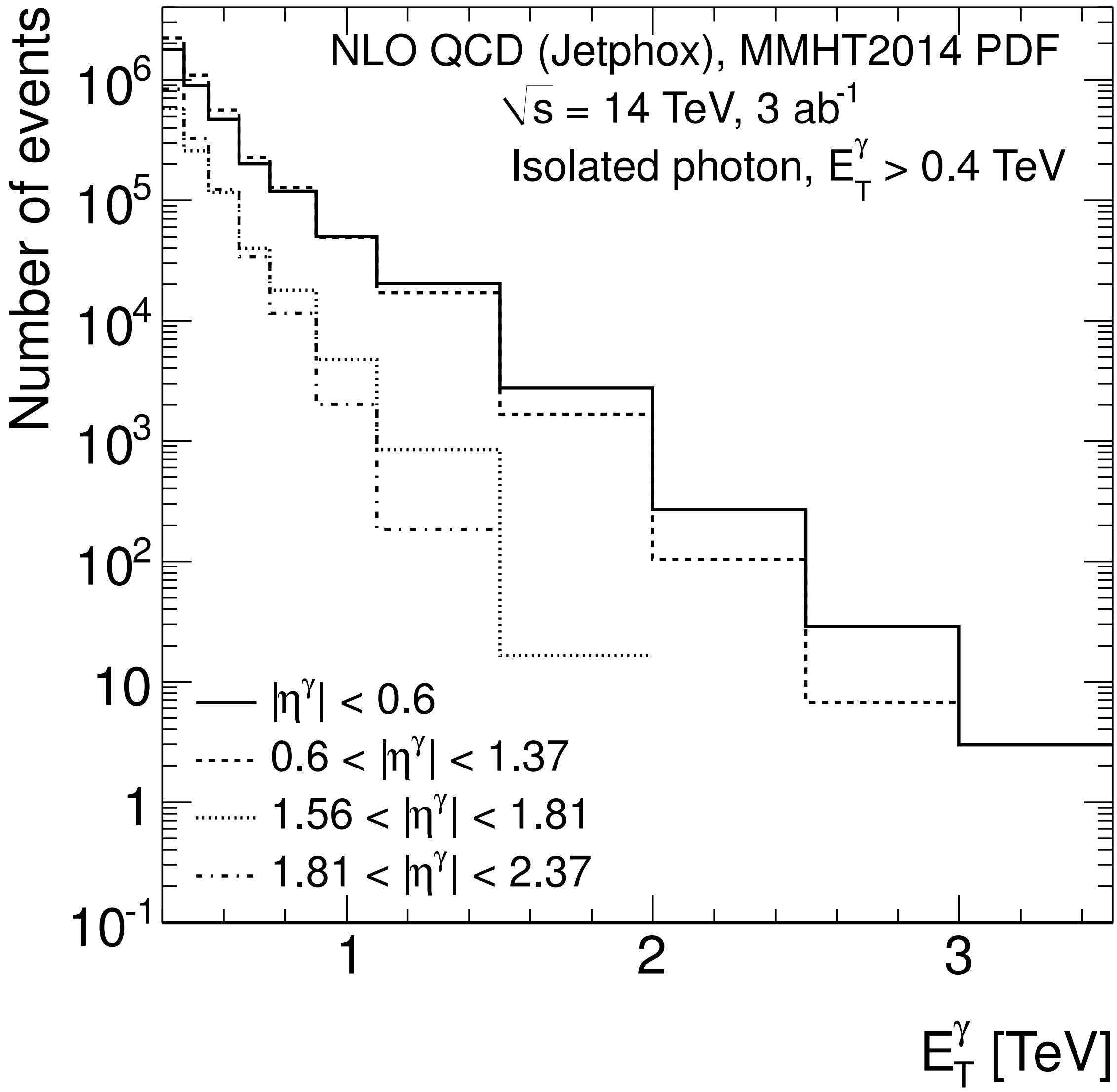

Figure 12: Predicted number of inclusive isolated photon events as a function of $E_T^\gamma$ assuming an integrated luminosity of 3 ab$^{-1}$ of $pp$ collision data at $\sqrt{s}$ = 14 TeV in different ranges of photon pseudorapidity: $|\eta^\gamma| < 0.6$ (solid histogram), $0.6 < |\eta^\gamma| < 1.37$ (dashed histogram), $1.56 < |\eta^\gamma| < 1.81$ (dotted histogram) and $1.81 < |\eta^\gamma| < 2.37$ (dot-dashed histogram).

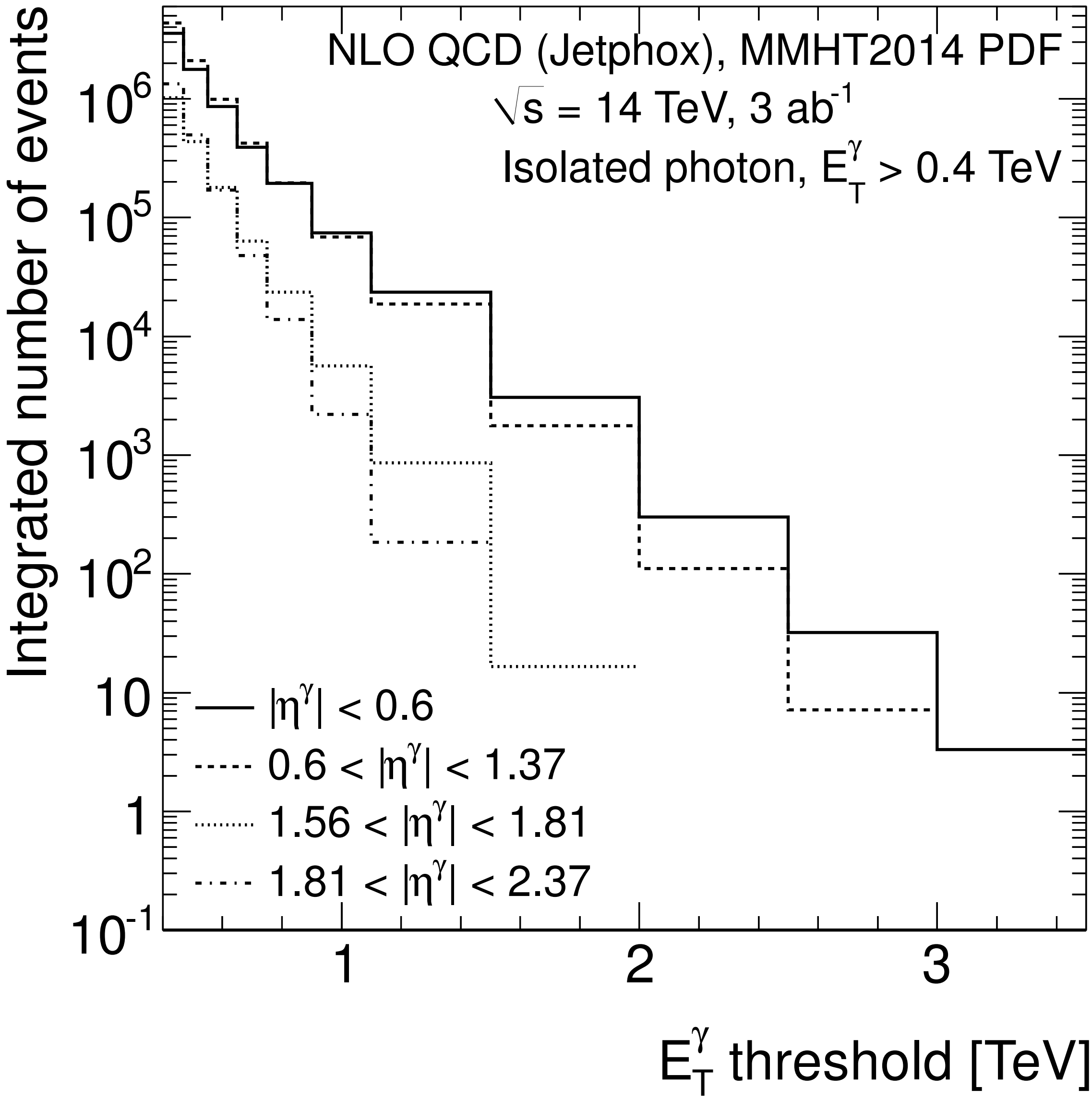

Figure 13: Predicted number of inclusive isolated photon events above an $E_T^\gamma$ threshold assuming an integrated luminosity of 3 ab$^{-1}$ of $pp$ collision data at $\sqrt{s}$ = 14 TeV in different ranges of photon pseudorapidity: $|\eta^\gamma| < 0.6$ (solid histogram), $0.6 < |\eta^\gamma| < 1.37$ (dashed histogram), $1.56 < |\eta^\gamma| < 1.81$ (dotted histogram) and $1.81 < |\eta^\gamma| < 2.37$ (dot-dashed histogram).

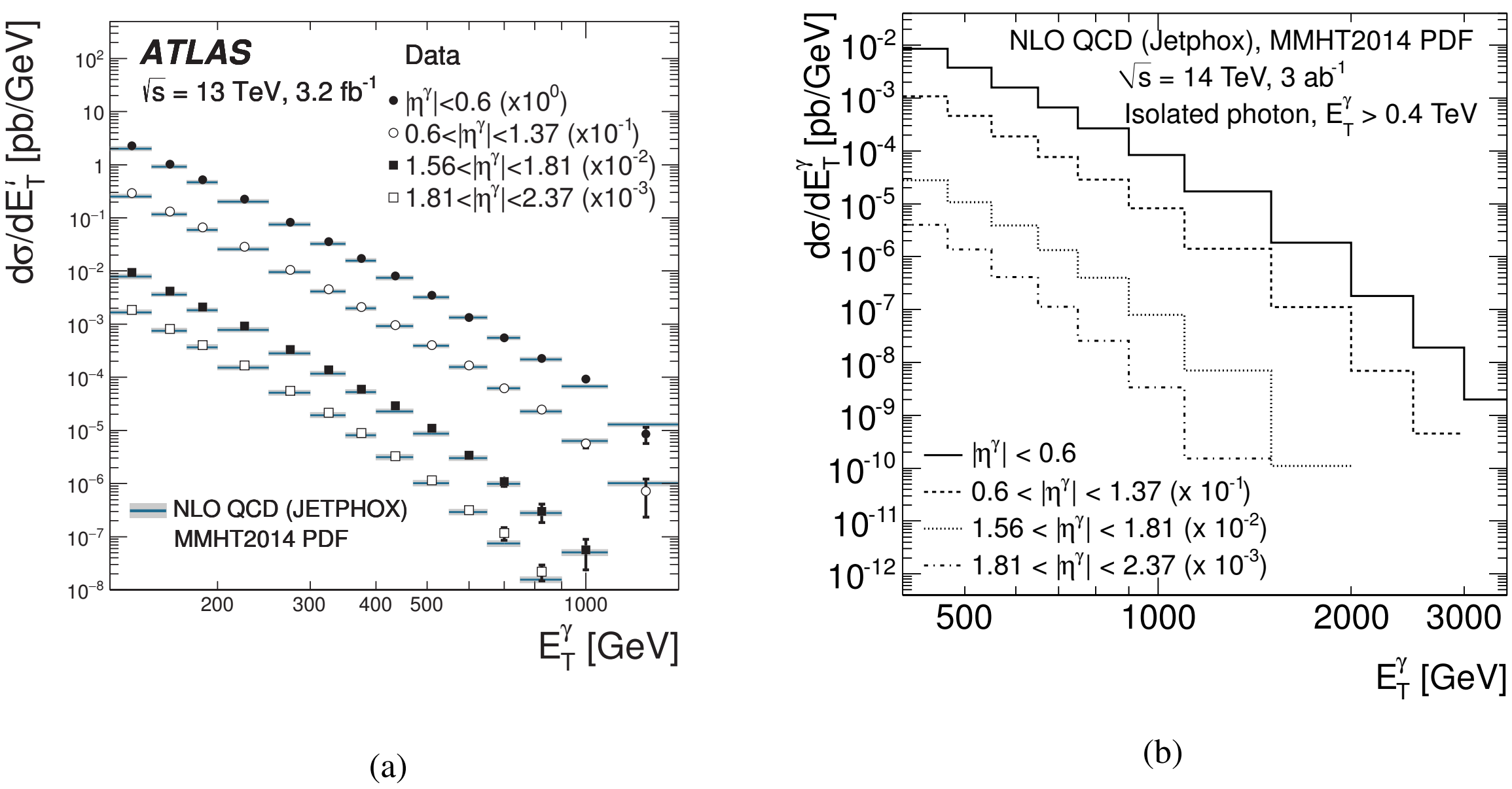

Figure 14: (a) Measured cross sections in $pp$ collisions at $\sqrt{s} = 13$ TeV for isolated-photon production as functions of $E_T^\gamma$ in $|\eta^\gamma| < 0.6$ (black dots), $0.6 < |\eta^\gamma| < 1.37$ (open circles), $1.56 < |\eta^\gamma| < 1.81$ (black squares) and $1.81 < |\eta^\gamma| < 2.37$ (open squares). The NLO QCD predictions from Jetphox based on the MMHT2014 PDFs (solid lines) are also shown. The measurements and the predictions are normalised by the factors shown in parentheses to aid visibility. The error bars represent the statistical and systematic uncertainties added in quadrature. The shaded bands display the theoretical uncertainty. (b) Predicted cross section in $pp$ collisions at $\sqrt{s} = 14$ TeV in $|\eta^\gamma| < 0.6$ (solid histogram), $0.6 < |\eta^\gamma| < 1.37$ (dashed histogram), $1.56 < |\eta^\gamma| < 1.81$ (dotted histogram) and $1.81 < |\eta^\gamma| < 2.37$ (dot-dashed histogram).

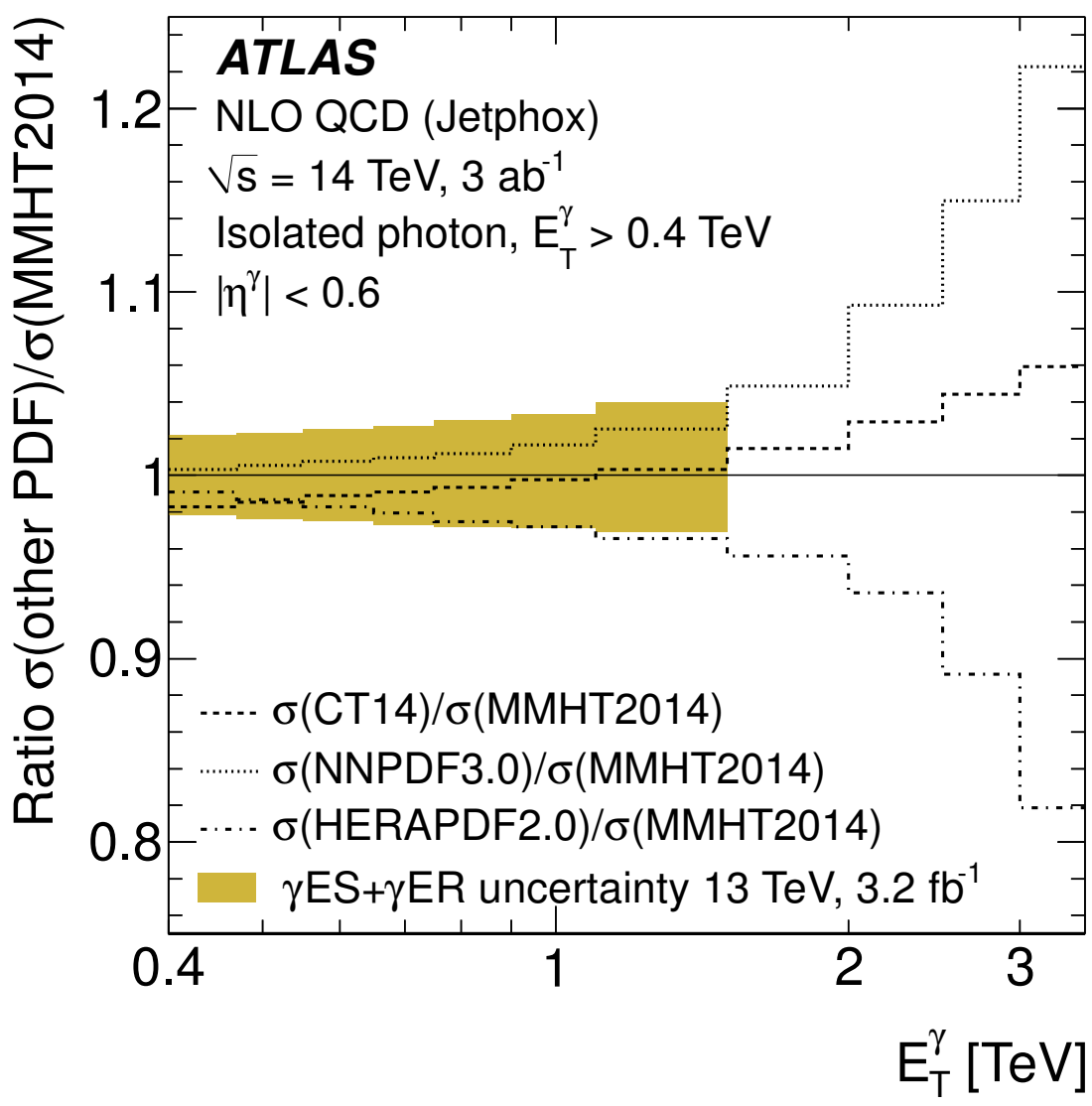

(a)

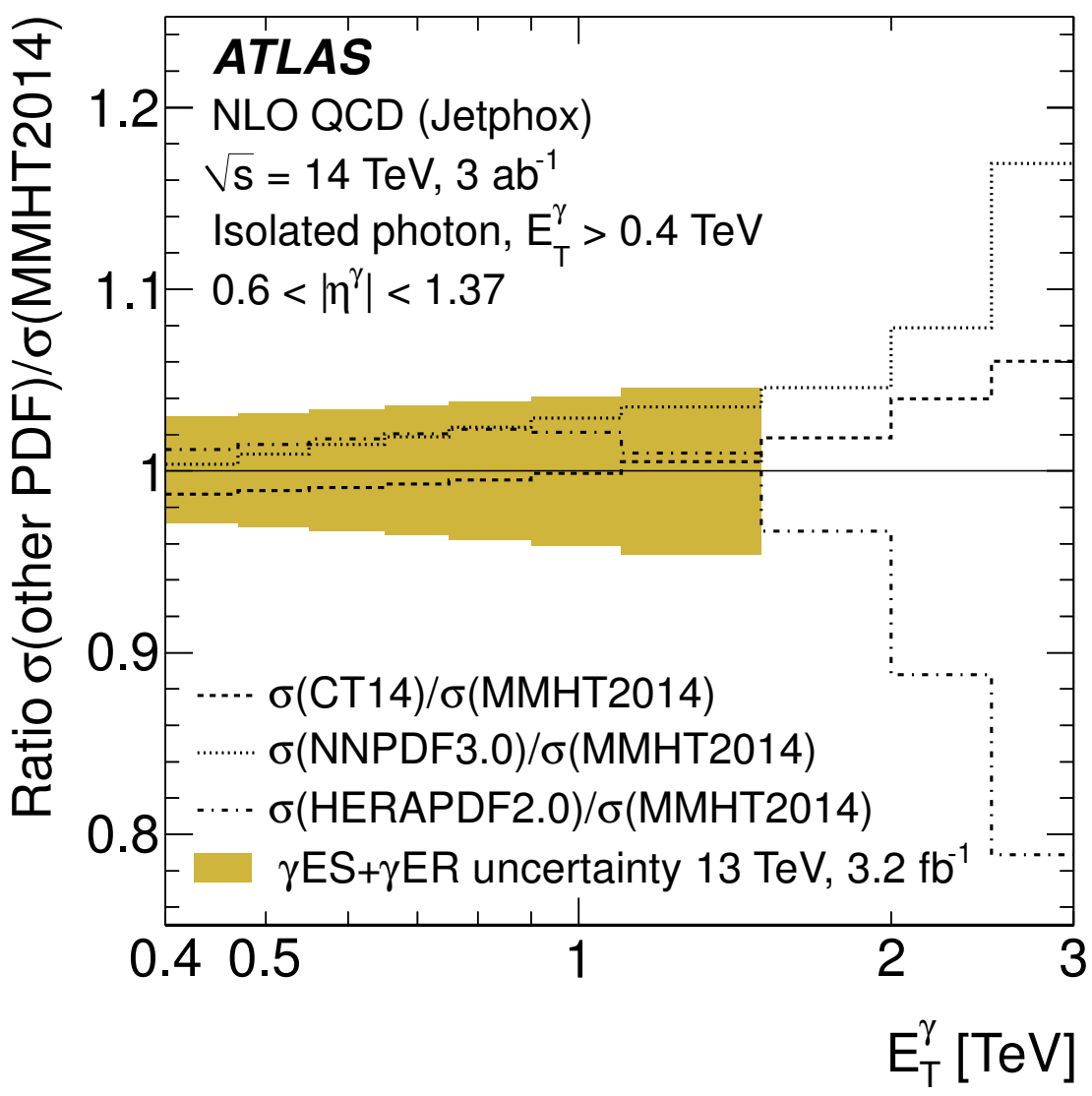

(b)

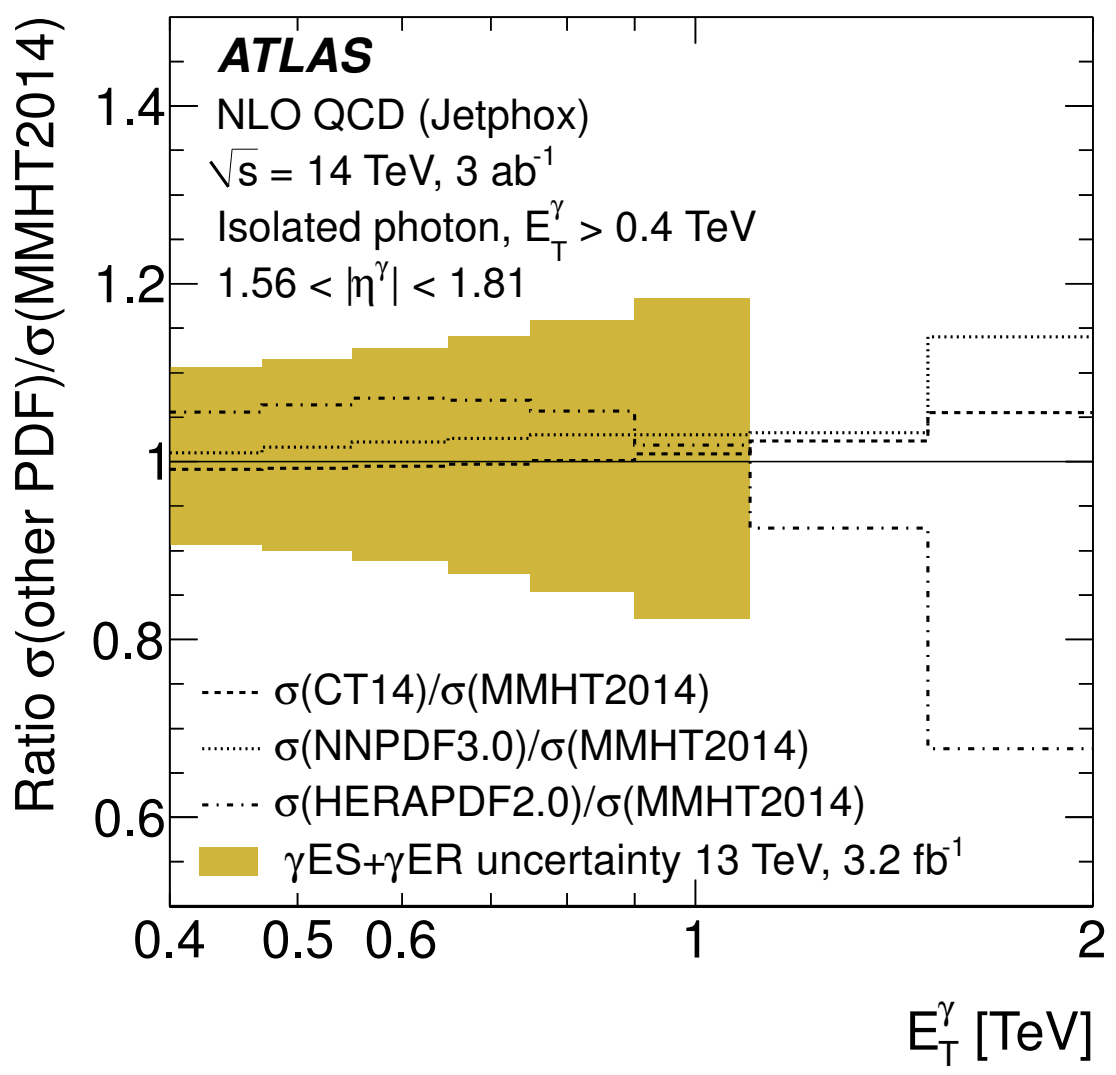

(c)

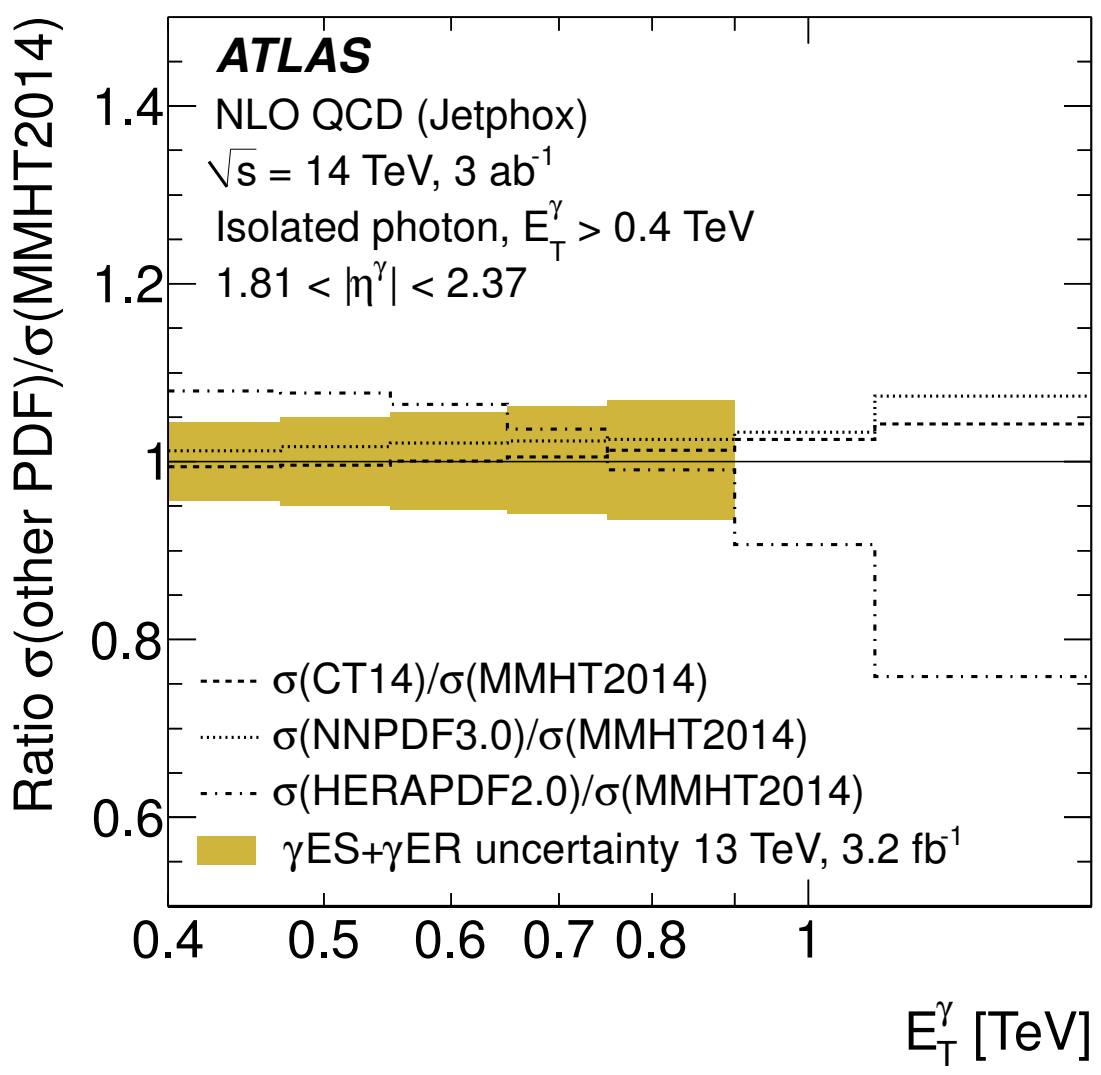

(d)

Figure 15: Ratios of the predicted cross sections of inclusive isolated photon events using different PDFs as functions of $E_{\mathrm{T}}^{\gamma}$ in $pp$ collisions at $\sqrt{s}$ = 14 TeV in different ranges of photon pseudorapidity: (a) $|\eta^{\gamma}| < 0.6$, (b) $0.6 < |\eta^{\gamma}| < 1.37$, (c) $1.56 < |\eta^{\gamma}| < 1.81$ and (d) $1.81 < |\eta^{\gamma}| < 2.37$. The ratios of the predictions using CT14 (dashed lines), NNPDF3.0 (dotted lines) and HERAPDF2.0 (dot-dashed lines) over those using MMHT2014 are shown. The shaded band represents the relative systematic uncertainty due to the photon energy scale ($\gamma$ES) and resolution ($\gamma$ER) estimated with 3.2 fb$^{-1}$ of $pp$ collisions at $\sqrt{s}$ = 13 TeV [15].

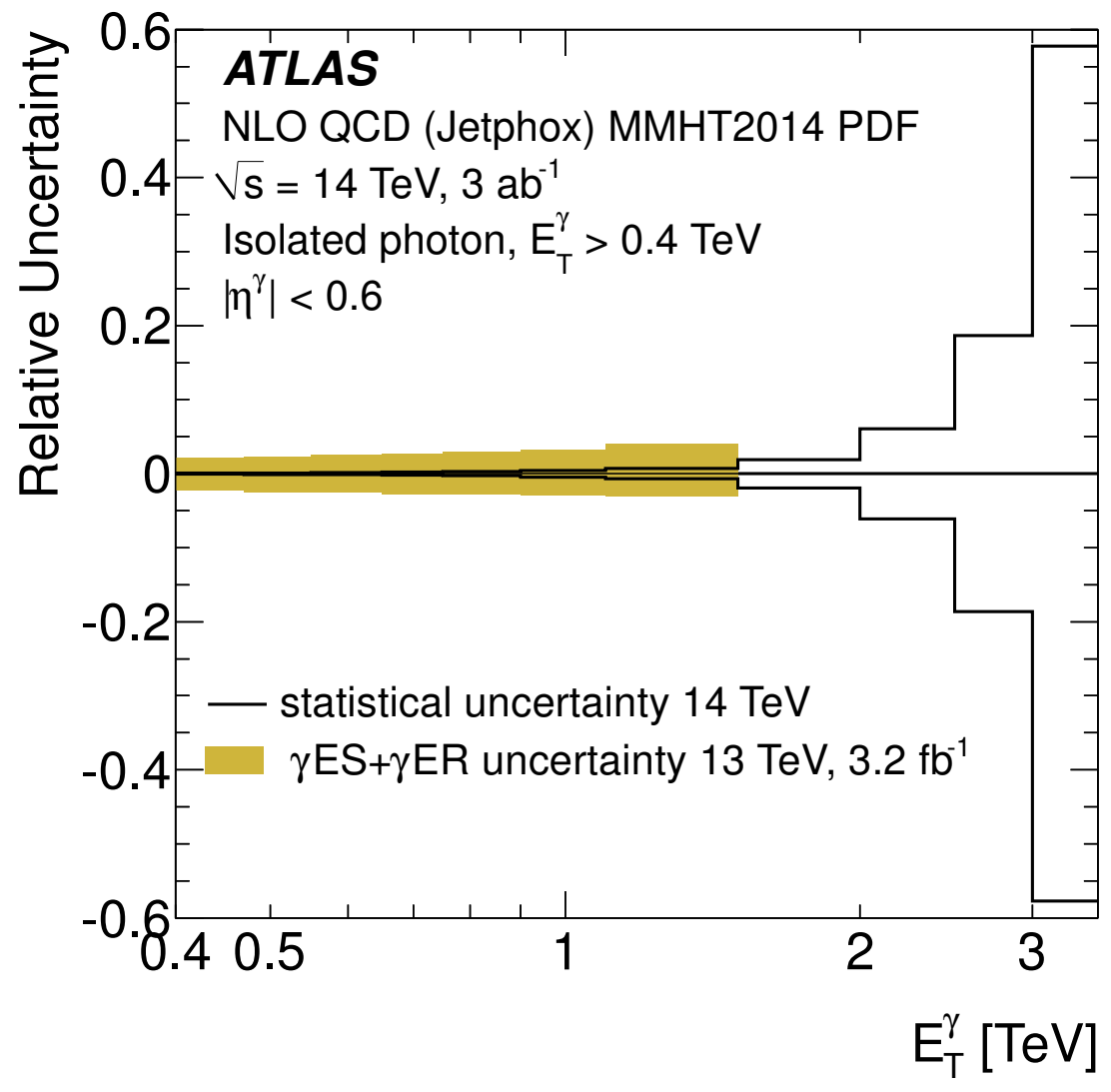

(a)

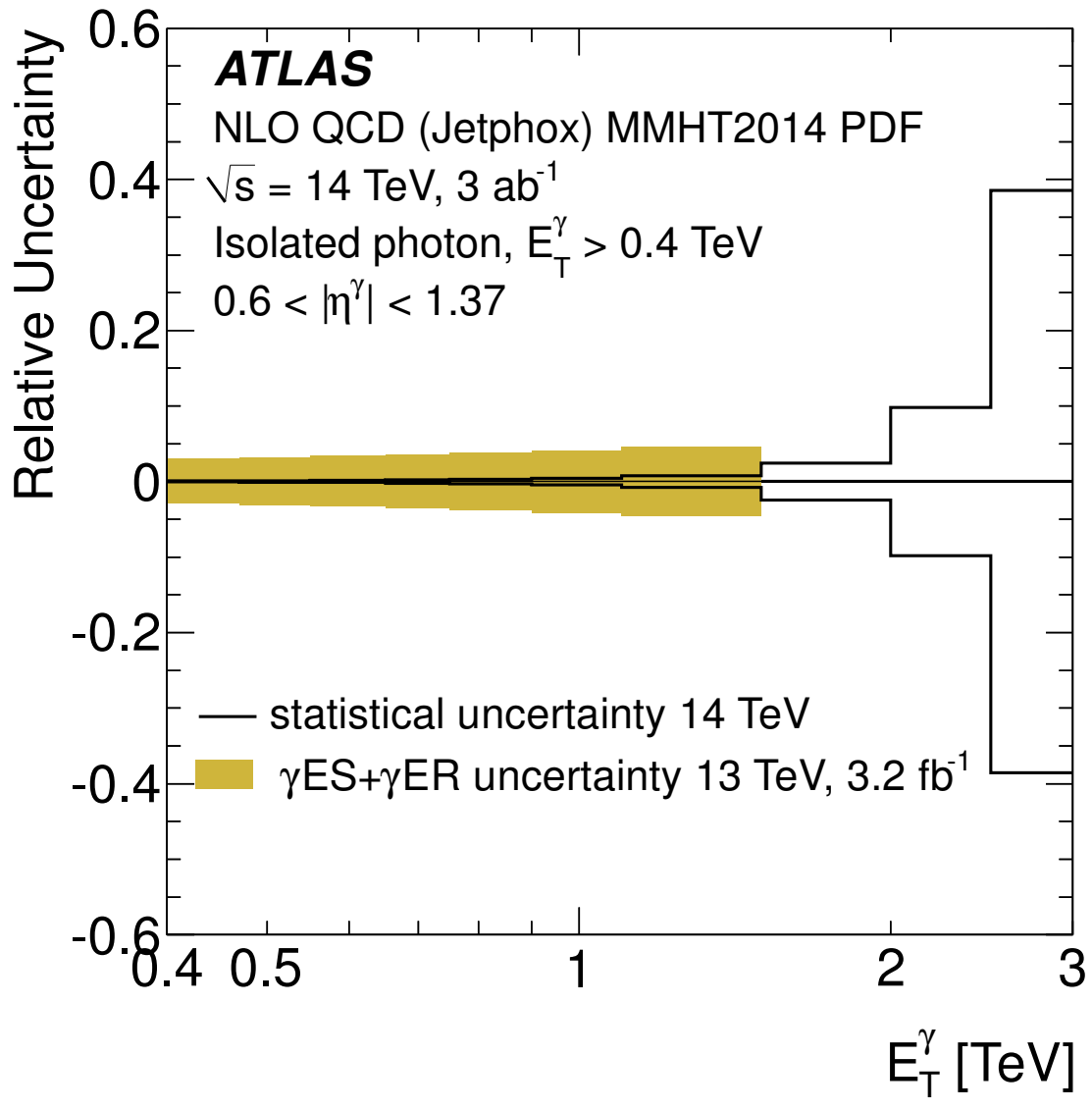

(b)

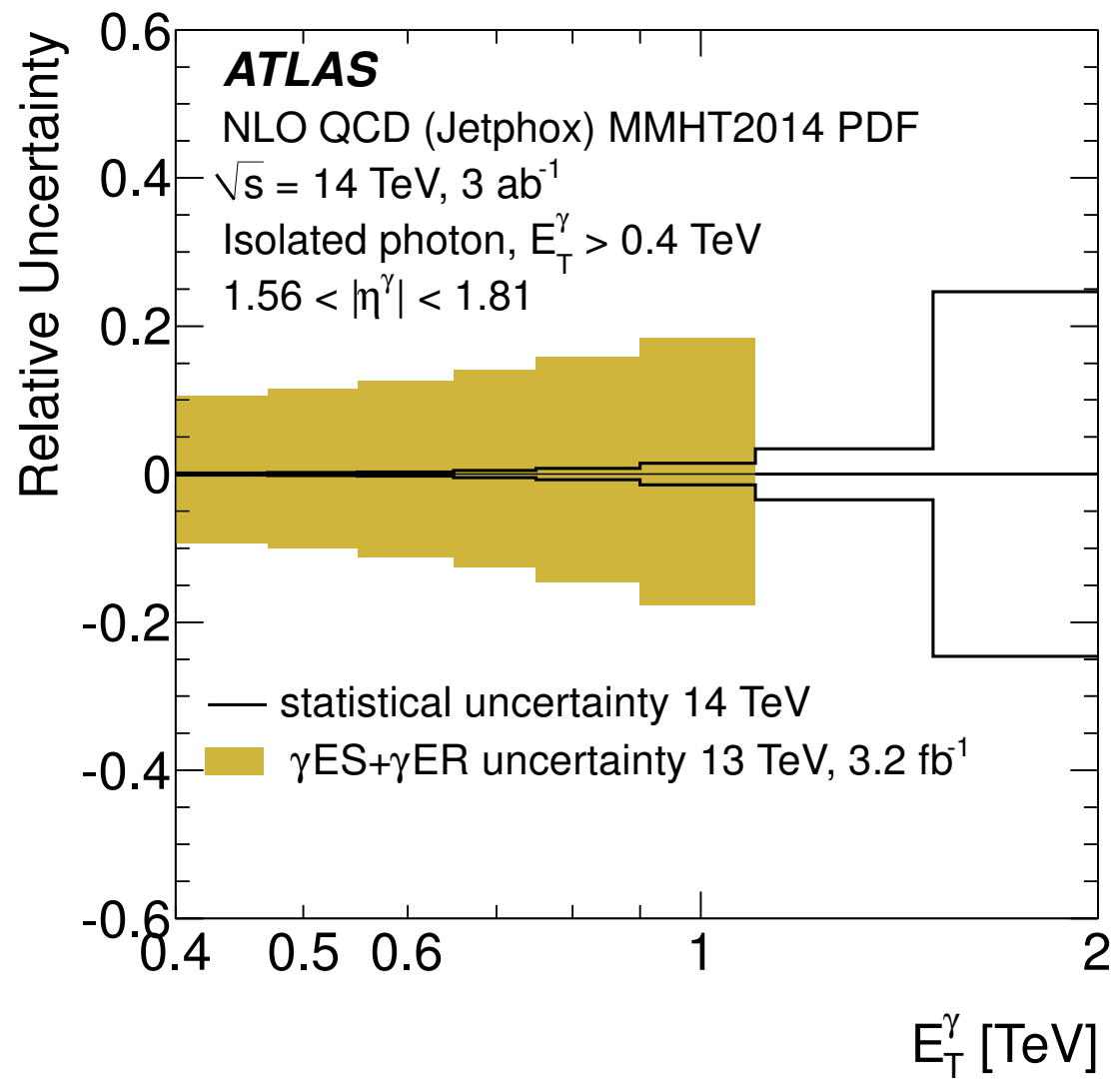

(c)

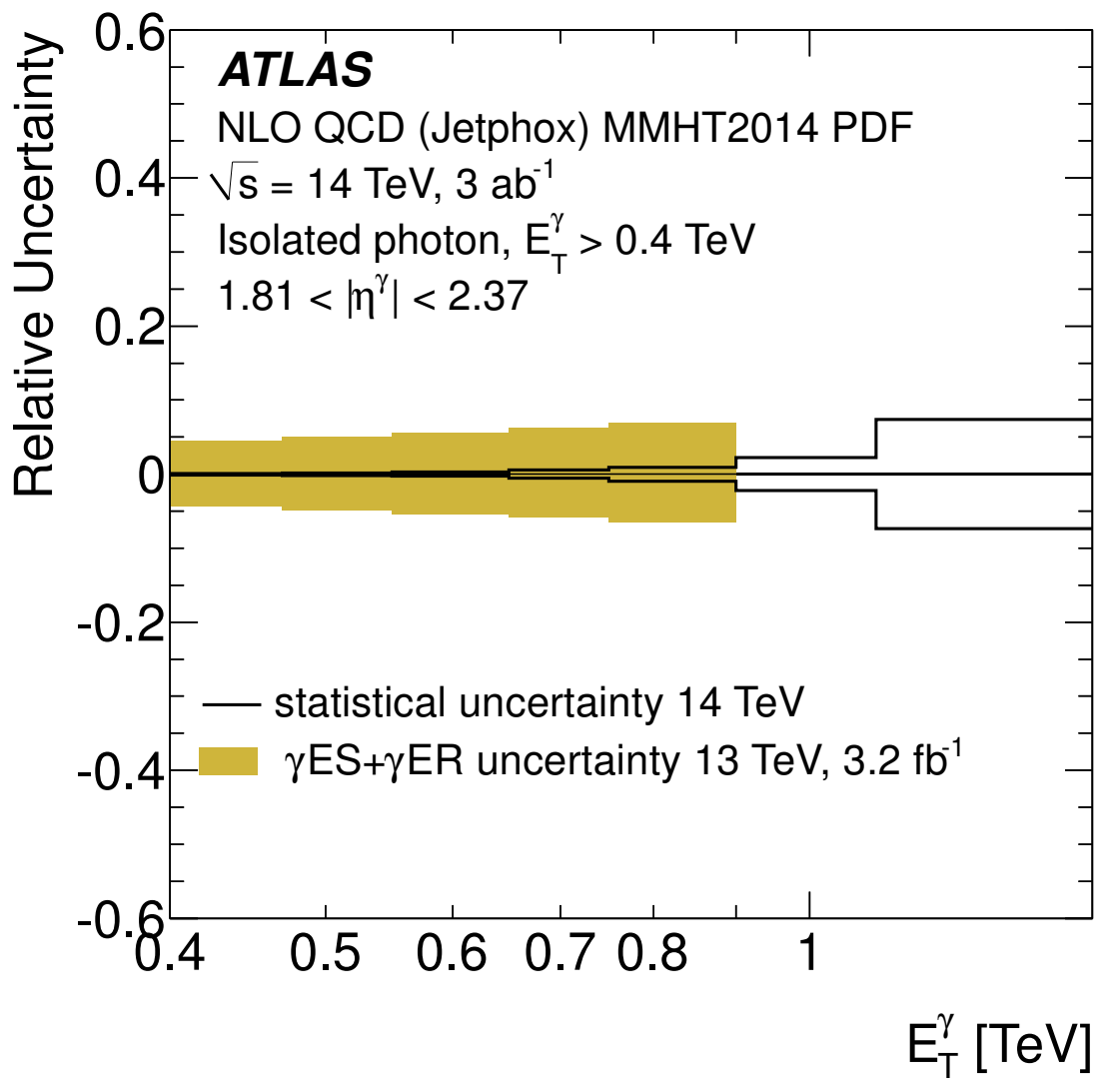

(d)

Figure 16: Predicted relative statistical uncertainty on the number of inclusive isolated photon events as a function of $E_{\mathrm{T}}^{\gamma}$ assuming an integrated luminosity of 3 ab$^{-1}$ of $pp$ collision data at $\sqrt{s}$ = 14 TeV in different ranges of photon pseudorapidity: (a) $|\eta^{\gamma}| < 0.6$, (b) $0.6 < |\eta^{\gamma}| < 1.37$, (c) $1.56 < |\eta^{\gamma}| < 1.81$ and (d) $1.81 < |\eta^{\gamma}| < 2.37$. The shaded band represents the relative systematic uncertainty due to the photon energy scale ($\gamma$ES) and resolution ($\gamma$ER) estimated with 3.2 fb$^{-1}$ of $pp$ collisions at $\sqrt{s}$ = 13 TeV [15].

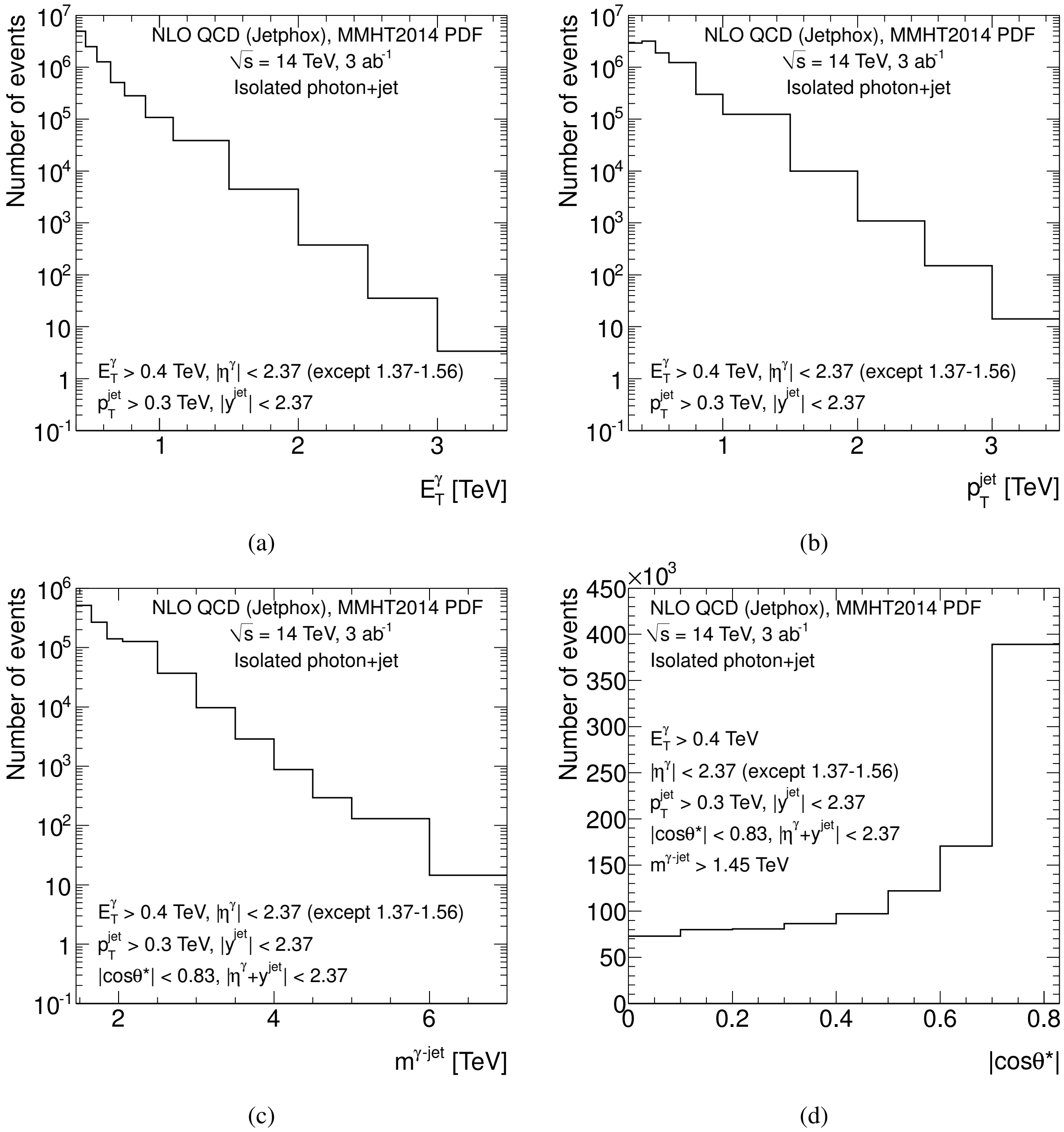

Figure 17: Predicted number of photon+jet events assuming an integrated luminosity of 3 ab$^{-1}$ of collision data at $\sqrt{s}$ = 14 TeV as a function of different observables: (a) $E_T^\gamma$, (b) $p_T^{\text{jet}}$, (c) $m^{\gamma-\text{jet}}$ and (d) $|\cos\theta^*|$.

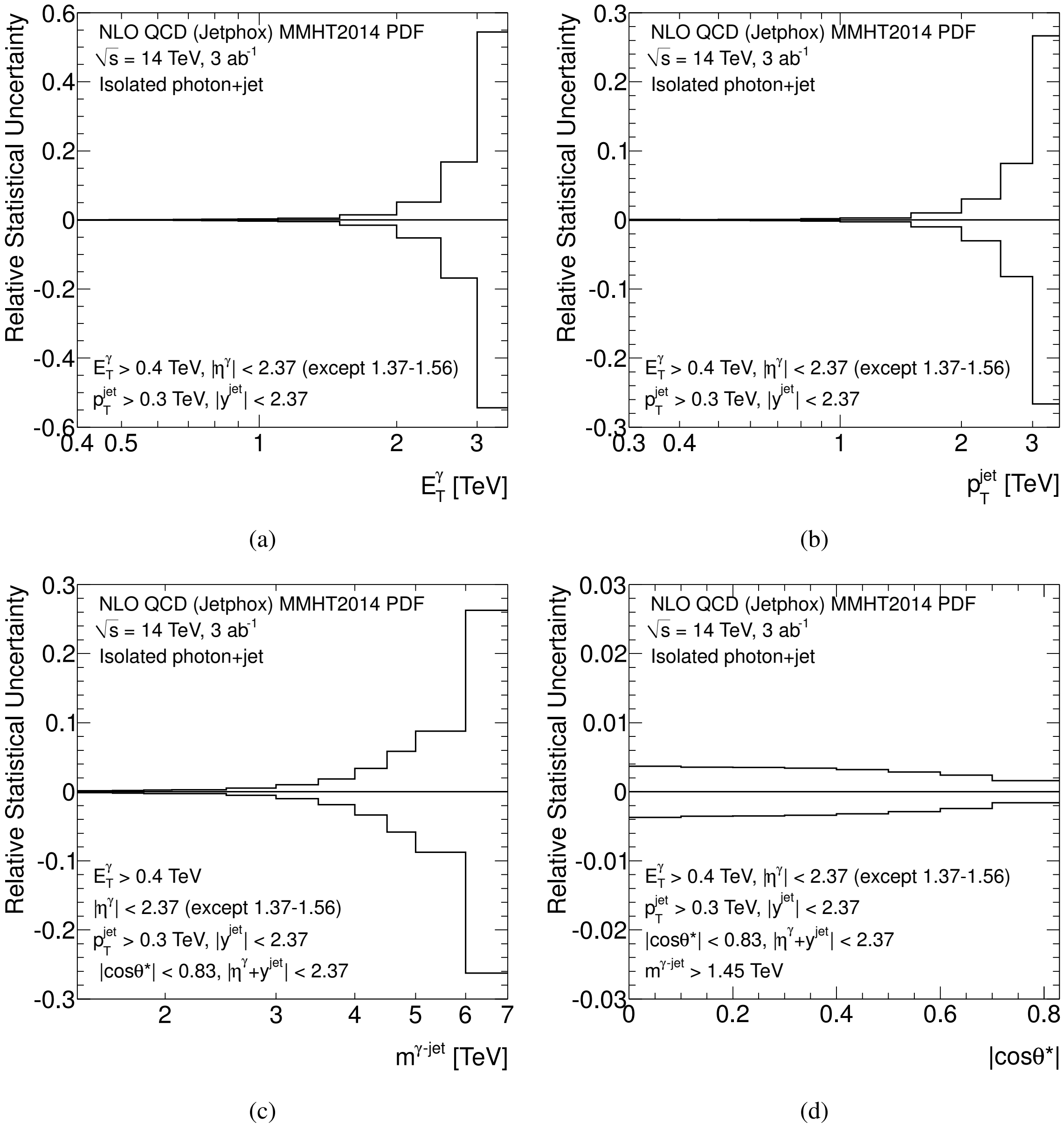

Figure 18: Predicted relative statistical uncertainty on the number of photon+jet events assuming an integrated luminosity of 3 ab$^{-1}$ of $pp$ collision data at $\sqrt{s}$ = 14 TeV as a function of different observables: (a) $E_T^\gamma$, (b) $p_T^{\text{jet}}$, (c) $m^{\gamma-\text{jet}}$ and (d) $|\cos\theta^*|$.

### 4.1.1 Impact of inclusive photon measurements at HL-LHC on the proton PDFs

The impact of the proposed measurements of inclusive isolated photon production in *pp* collisions at $\sqrt{s}$ = 14 TeV in different ranges of $|\eta^{\gamma}|$ on the proton PDFs is illustrated as follows. The uncertainty in the theoretical predictions due to the uncertainties in the proton PDFs has been evaluated using the studies listed below:

- The MMHT2014 analysis. The uncertainty in the predictions due to the current knowledge of the proton PDFs is estimated by repeating the calculations using the 50 sets from the MMHT2014 error analysis and applying the Hessian method for the evaluation of the PDF uncertainties.
- The Ultimate PDF analysis [39]. This analysis includes the expectations of several measurements at the HL-LHC to quantify their impact on the proton PDFs. It considers measurements of inclusive isolated photon measurements as well as measurements of the production of jets, electroweak gauge bosons and top quark pair production at the HL-LHC. Three scenarios are analysed depending on the assumptions on possible improvements on the experimental systematic uncertainties at HL-LHC. Scenario 1 is conservative, scenario 3 is optimistic and scenario 2 represents an intermediate stage. The resulting profiled PDF sets can be used for phenomenology studies by employing the uncertainty prescription of symmetric Hessian sets, as it is done here.

The relative uncertainty in the predictions due to the uncertainties in the PDFs is shown in Fig. 19 for the MMHT2014 analysis as well as for the three scenarios of the Ultimate PDF analysis. In comparison to the current estimate of the uncertainty using MMHT2014, the measurements at the HL-LHC lead to a significant reduction, which in certain regions such as $E_{\mathrm{T}}^{\gamma} \sim$ 1–2 TeV and $|\eta^{\gamma}| < 0.6$ is as large as a factor 4.

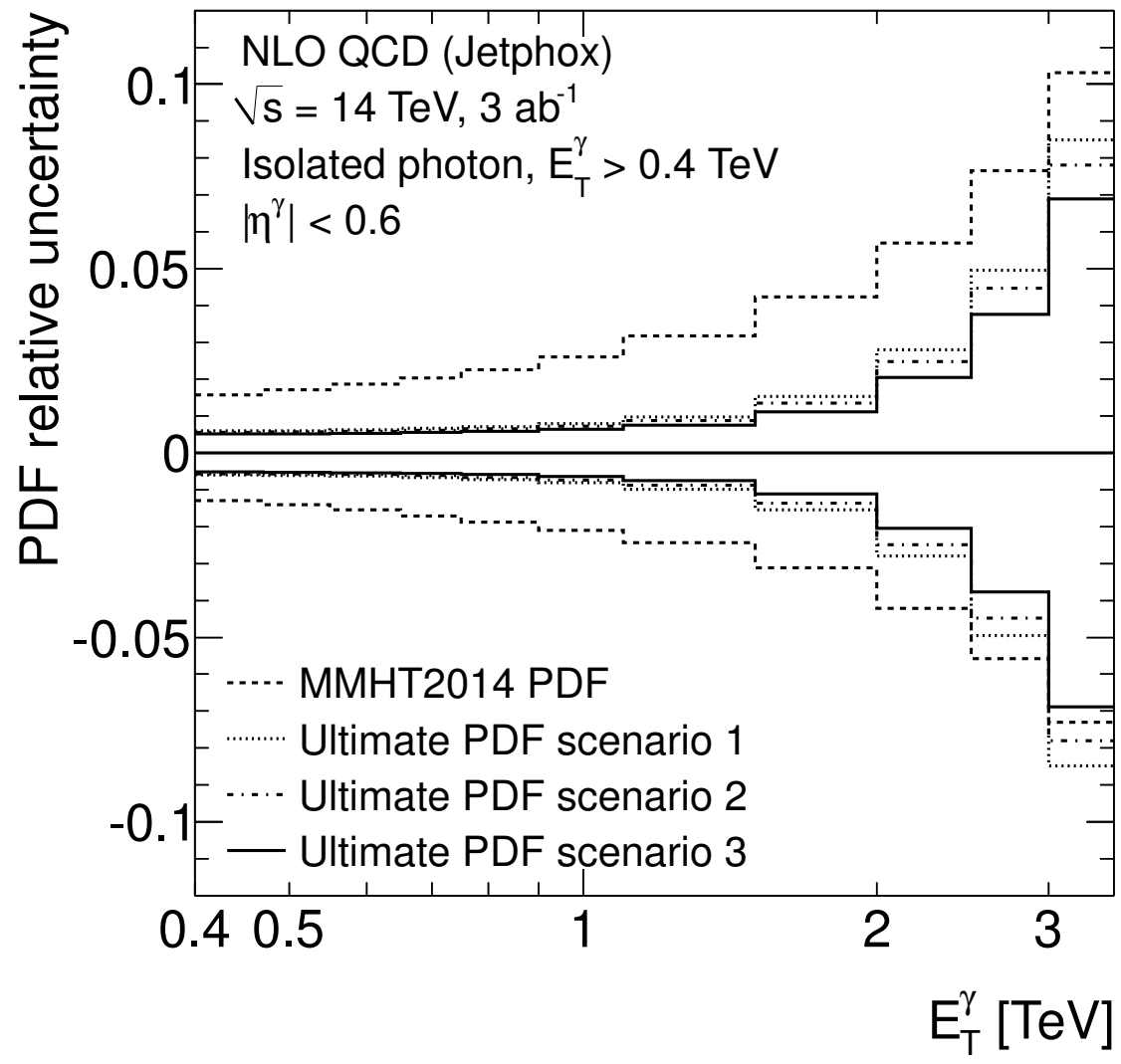

(a)

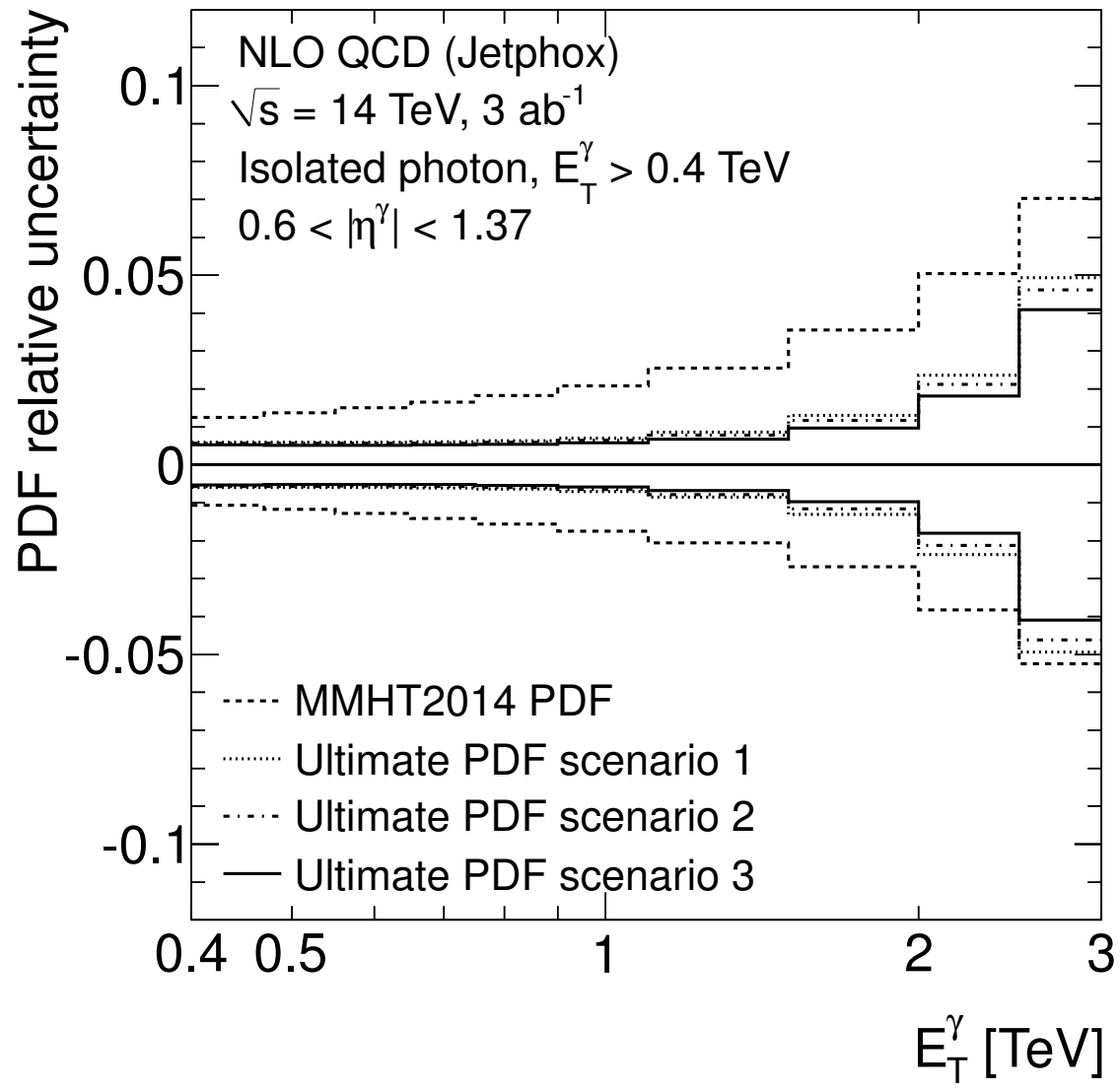

(b)

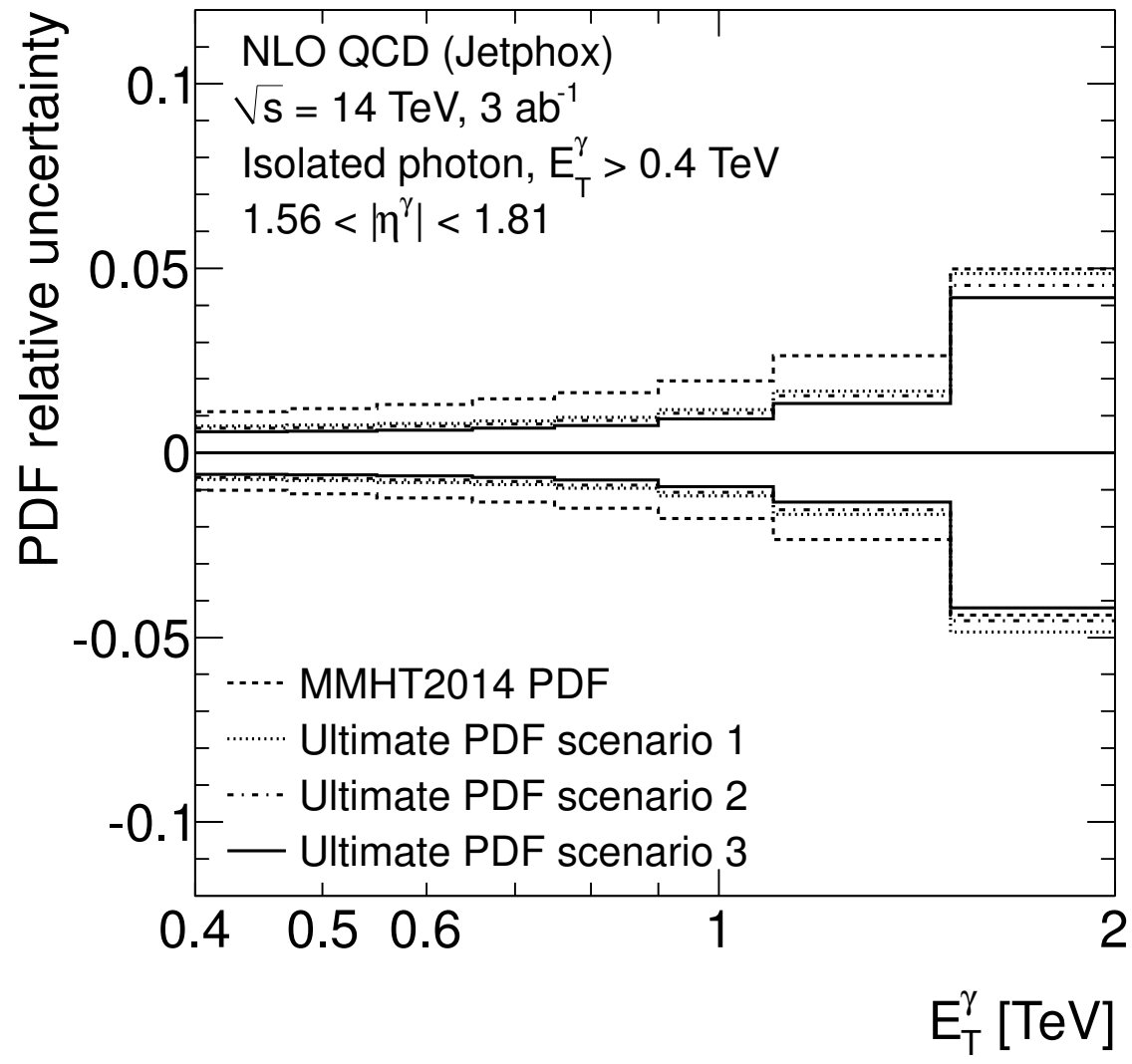

(c)

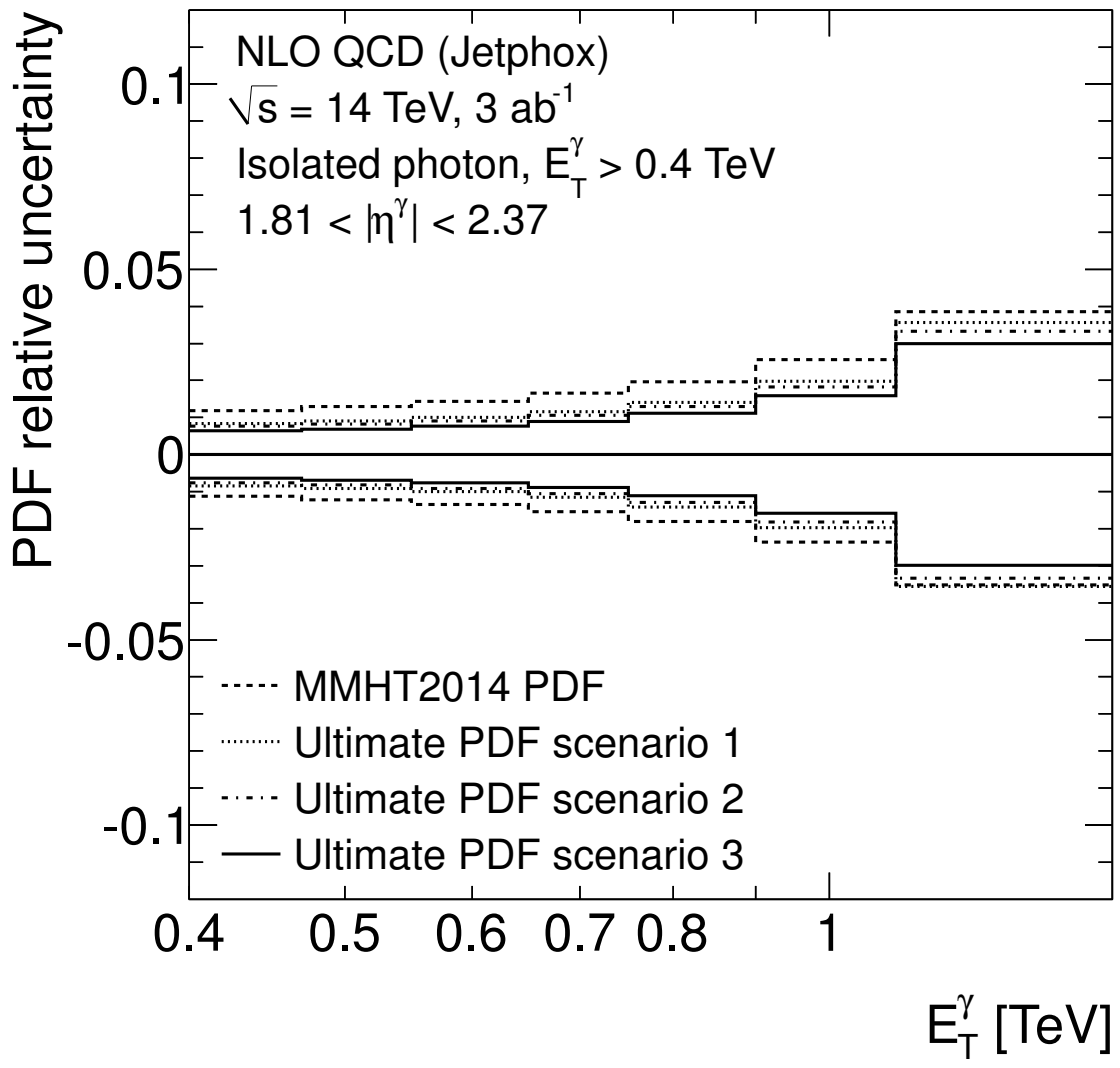

(d)

Figure 19: Relative uncertainty in the predicted number of inclusive isolated photon events due to the uncertainties in the PDFs as a function of $E_T^\gamma$ in $pp$ collisions at $\sqrt{s} = 14$ TeV in different ranges of photon pseudorapidity: (a) $|\eta^\gamma| < 0.6$, (b) $0.6 < |\eta^\gamma| < 1.37$, (c) $1.56 < |\eta^\gamma| < 1.81$ and (d) $1.81 < |\eta^\gamma| < 2.37$. The relative uncertainty due to the PDFs is shown for different PDF sets: the MMHT2014 PDF set (dashed lines) as well as the Ultimate PDF set in scenario 1 (dotted lines), 2 (dot-dashed lines) and 3 (solid lines).

### 4.1.2 Photon Results at HE-LHC

Prospects are also obtained for inclusive isolated photon and photon+jet production in $pp$ collisions at $\sqrt{s} = 27$ TeV assuming an integrated luminosity of 15 ab$^{-1}$. The predicted number of inclusive isolated photon events as a function of $E_T^{\gamma}$ in the different ranges of $|\eta^{\gamma}|$ is shown in Fig. 20. The reach in $E_T^{\gamma}$ is (a) 5 TeV for $|\eta^{\gamma}| < 0.6$ and $0.6 < |\eta^{\gamma}| < 1.37$, (b) 3–3.5 TeV for $1.56 < |\eta^{\gamma}| < 1.81$ and (c) 2.5–3 TeV for $1.81 < |\eta^{\gamma}| < 2.37$. The predicted cross sections are shown in Fig. 21. The ratios of the predictions based on CT14, NNPDF3.0 and HERAPDF2.0 over those using MMHT2014 are shown in Fig. 22 and differences of up to 40% are seen. The predicted relative statistical uncertainty on the number of inclusive isolated photon events as a function of $E_T^{\gamma}$ in different ranges of photon pseudorapidity is shown in Fig. 23. A relative statistical uncertainty below 10% is achieved for photon transverse energies up to (a) 4.5 TeV for $|\eta^{\gamma}| < 0.6$, (b) 4 TeV for $0.6 < |\eta^{\gamma}| < 1.37$, (c) 3 TeV for $1.56 < |\eta^{\gamma}| < 1.81$ and (d) 2.5 TeV for $1.81 < |\eta^{\gamma}| < 2.37$.

The predicted number of photon+jet events as a function of $E_T^{\gamma}$, $p_T^{\text{jet}}$, $m^{\gamma-\text{jet}}$ and $|\cos\theta^*|$ is shown in Fig. 24. The predictions show that the reach in $E_T^{\gamma}$ and $p_T^{\text{jet}}$ is 5 TeV and the reach in $m^{\gamma-\text{jet}}$ is 12 TeV. The predicted relative statistical uncertainty on the number of photon+jet events as a function of the different observables is shown in Fig. 25. The relative statistical uncertainty is below 10% for (a) $E_T^{\gamma}$ up to 4.5 TeV, (b) $p_T^{\text{jet}}$ up to 5 TeV and (c) $m^{\gamma-\text{jet}}$ up to 10 TeV; for $|\cos\theta^*|$ the relative statistical uncertainty is below 0.1% for the entire range considered.

## 4.2 Jet Results

The predicted inclusive jet and dijet cross sections are shown in Figures 26 and 27 in the proton-proton collisions at $\sqrt{s} = 14$ and 27 TeV, respectively. The cross section values are calculated at NLO pQCD accuracy. The inclusive jet cross sections are calculated as a function of the jet $p_T$ in six equal-size bins of absolute jet rapidity for jets in the $|y| < 3$ range with $p_T > 100$ GeV. The dijet cross sections are calculated as a function of the invariant mass of the dijet system ($m_{jj}$) in six equal-size bins of half absolute rapidity separation between two leading in $p_T$ jets.

The predicted number of inclusive jet events as a function of jet $p_T$ in the different ranges of the jet rapidity and dijet events as a function of $m_{jj}$ in the $pp$ collisions at $\sqrt{s} = 14$ and 27 TeV are shown in Fig. 28 and 29. The reach in $p_T$ for the inclusive jet cross section measurements is 5.5 TeV in the $|y| < 0.5$ region and 1 TeV in the $2.5 < |y| < 3.0$ region at HL-LHC. For the dijet production the $m_{jj}$ reach is 9 TeV in the $y^* < 0.5$ region and 11.5 TeV in the $2.5 < y^* < 3.0$ region at HL-LHC. In the case of HE-LHC the inclusive jet cross sections can be measured up to 10 (2.2) TeV in the $|y| < 0.5$ $(2.5 < |y| < 3.0)$ region and the dijet production can reach dijet invariant masses of 17 (22) TeV in the $y < 0.5$ $(2.5 < y^* < 3.0)$ region.

The predicted relative statistical uncertainty in the number of inclusive jet and dijet events as a function of the jet $p_T$ in the $|y| < 0.5$ range and $m_{jj}$ in the $y^* < 0.5$ bin assuming an integrated luminosity of 3 (15) ab$^{-1}$ of $pp$ collision data at $\sqrt{s} = 14$ (27) TeV for HL(HE)-LHC is shown in Fig. 30. The relative statistical uncertainty is well below 1% everywhere, except for the highest $p_T$ and $m_{jj}$ bins of the measurements.

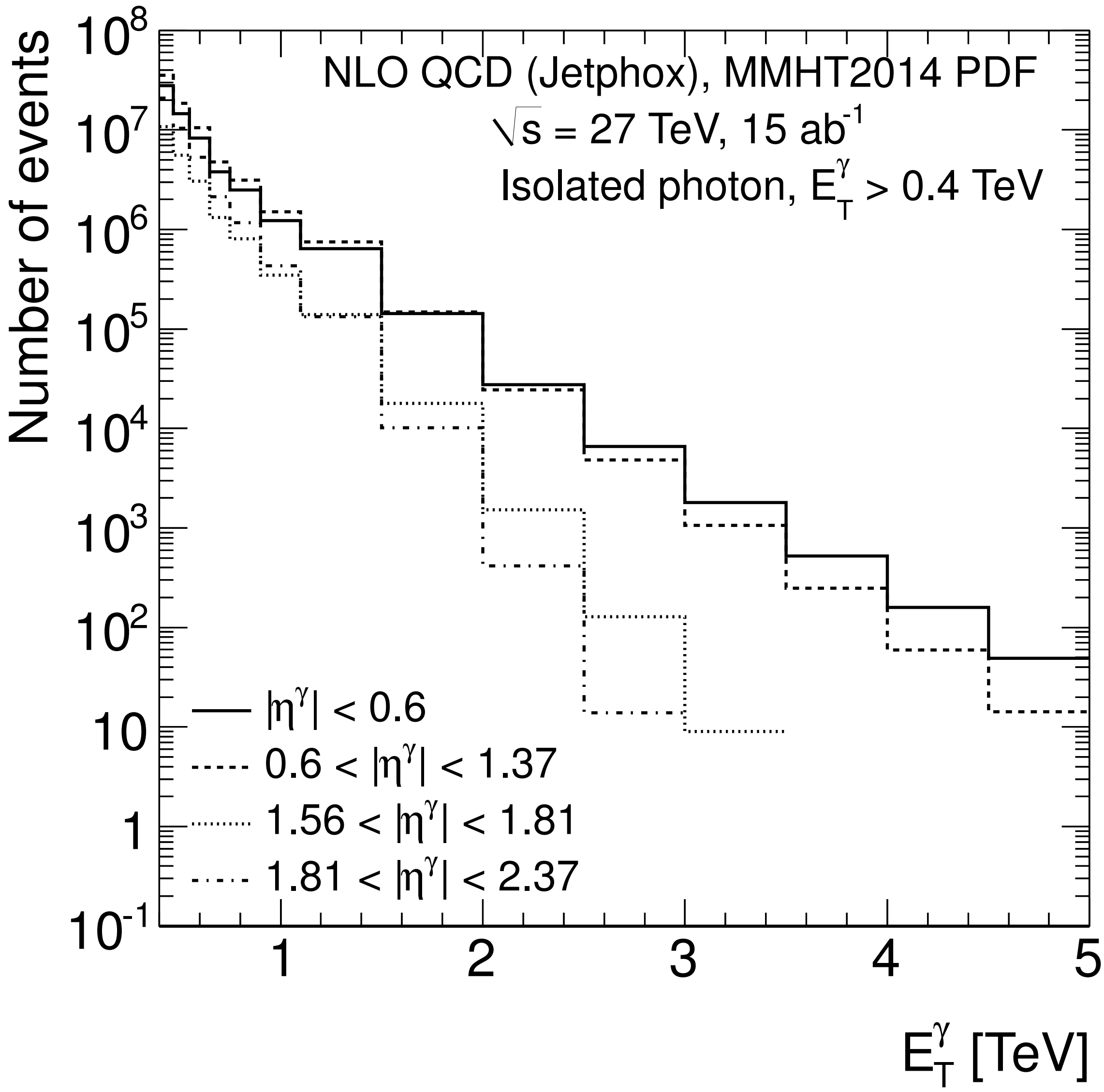

Figure 20: Predicted number of inclusive isolated photon events as a function of $E_T^\gamma$ assuming an integrated luminosity of 15 ab$^{-1}$ of $pp$ collision data at $\sqrt{s}$ = 27 TeV in different ranges of photon pseudorapidity: $|\eta^\gamma| < 0.6$ (solid histogram), $0.6 < |\eta^\gamma| < 1.37$ (dashed histogram), $1.56 < |\eta^\gamma| < 1.81$ (dotted histogram) and $1.81 < |\eta^\gamma| < 2.37$ (dot-dashed histogram).

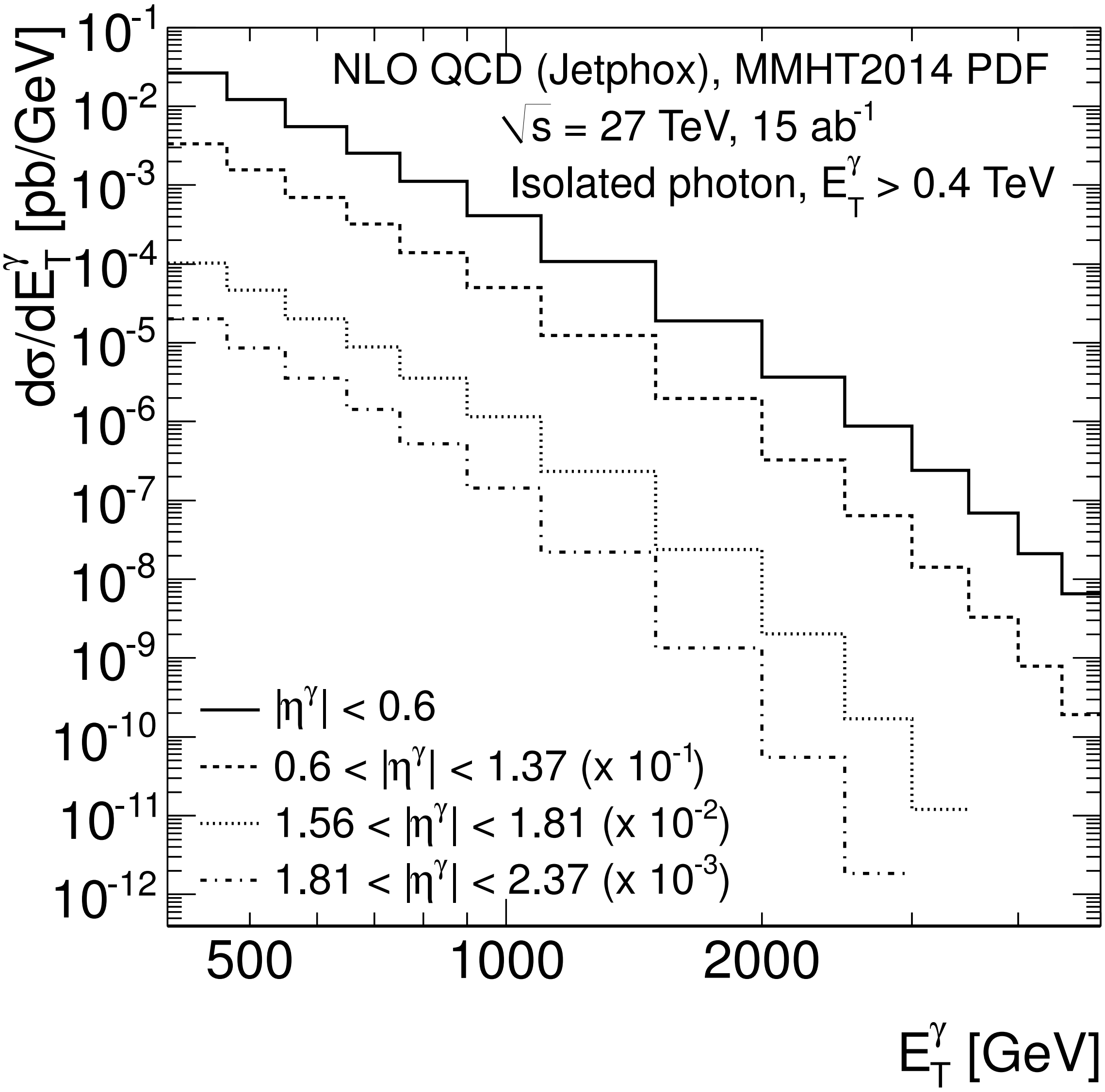

Figure 21: Predicted cross sections in $pp$ collisions at $\sqrt{s} = 27$ TeV in $|\eta^\gamma| < 0.6$ (solid histogram), $0.6 < |\eta^\gamma| < 1.37$ (dashed histogram), $1.56 < |\eta^\gamma| < 1.81$ (dotted histogram) and $1.81 < |\eta^\gamma| < 2.37$ (dot-dashed histogram).

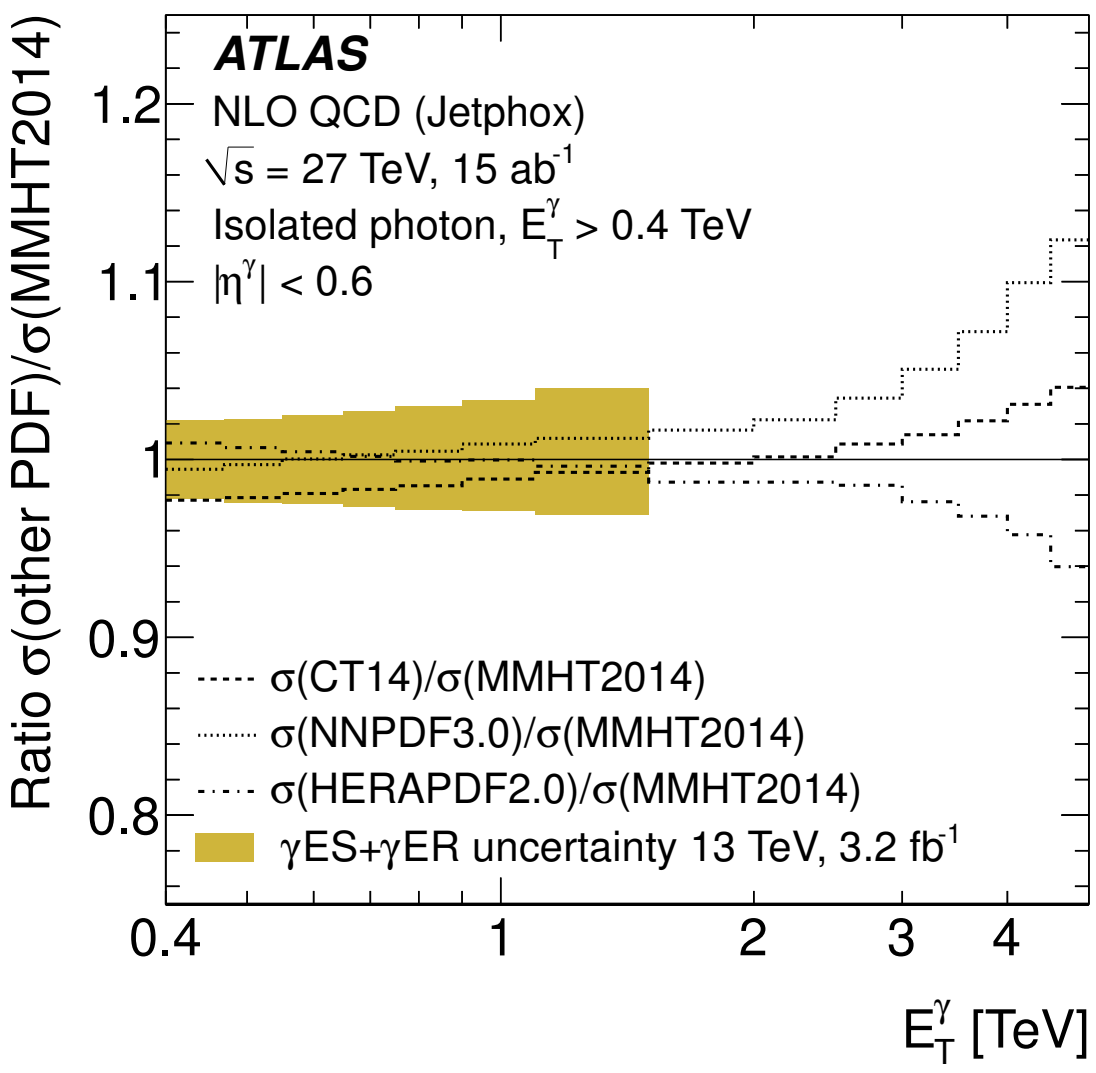

(a)

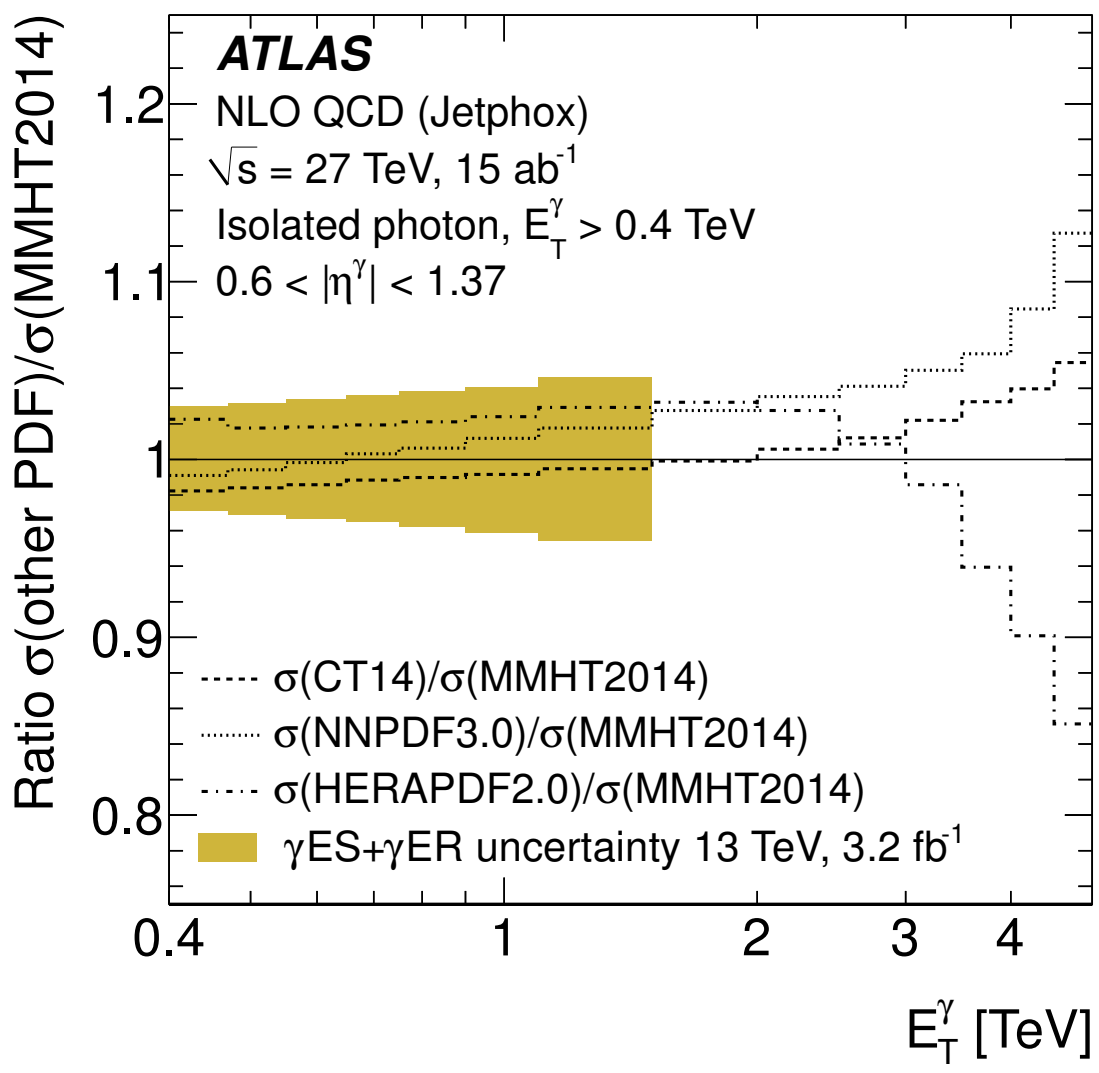

(b)

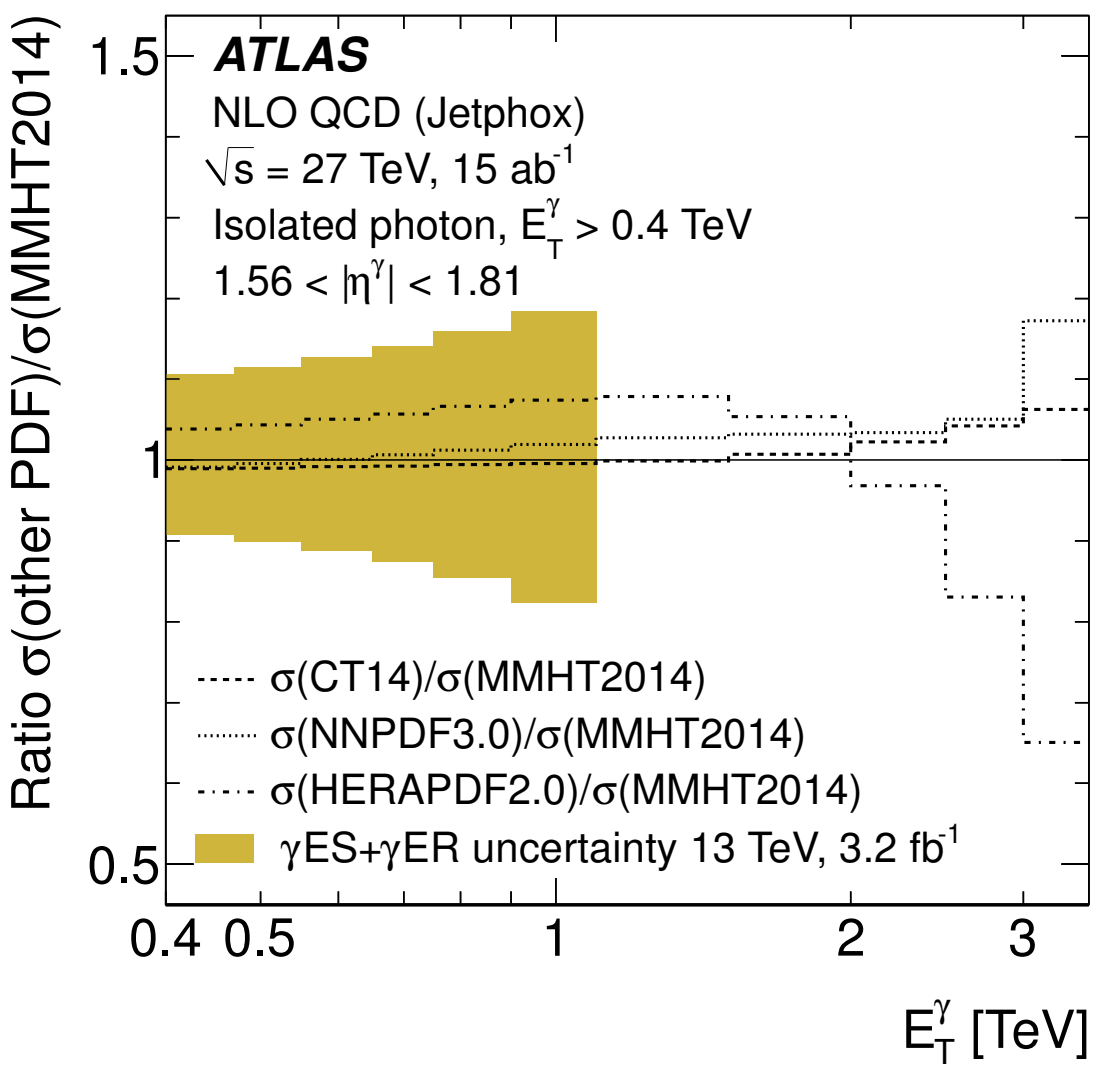

(c)

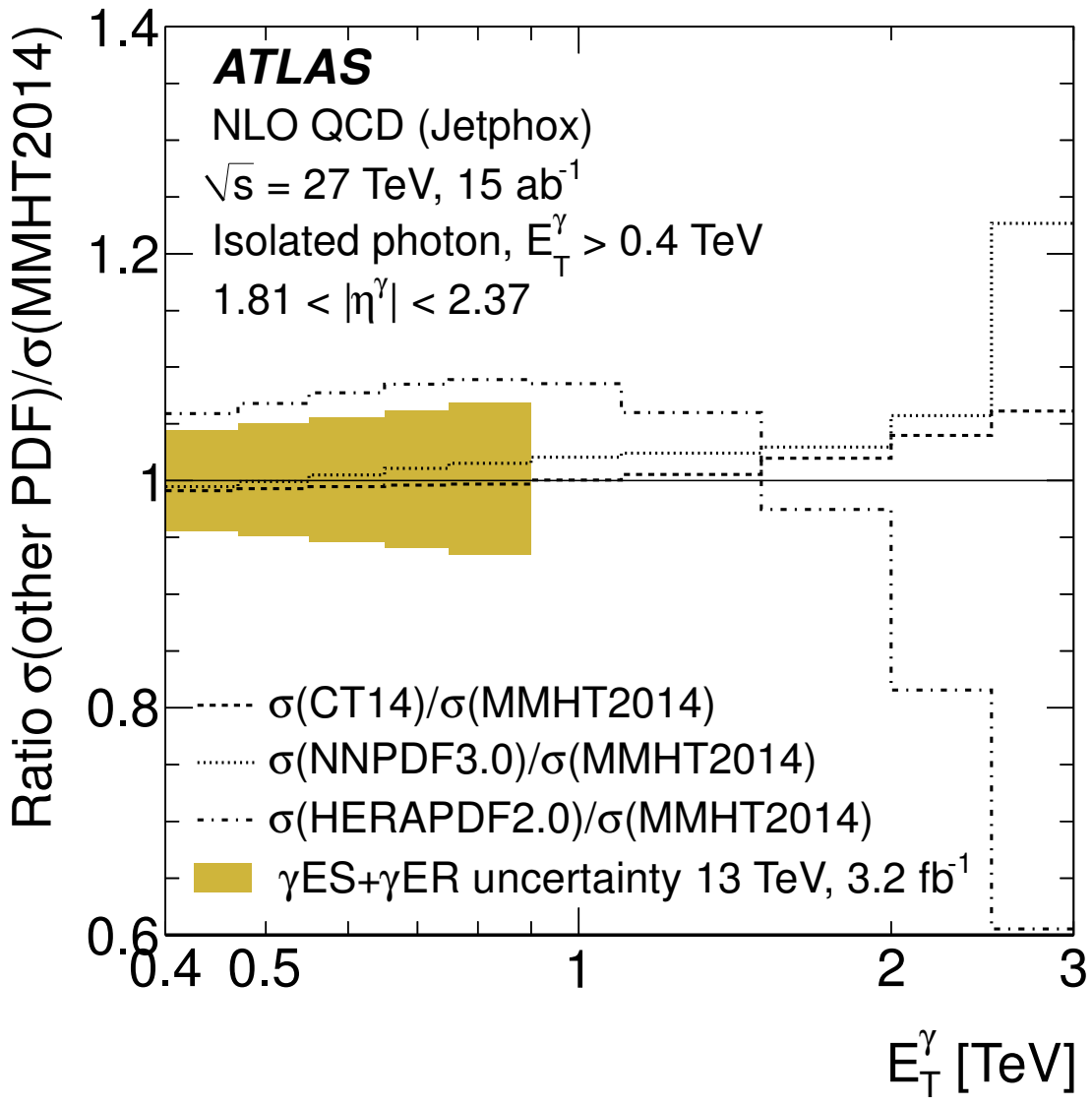

(d)

Figure 22: Ratios of the predicted number of inclusive isolated photon events using different PDFs as functions of $E_T^\gamma$ in $pp$ collisions at $\sqrt{s} = 27$ TeV in different ranges of photon pseudorapidity: (a) $|\eta^\gamma| < 0.6$, (b) $0.6 < |\eta^\gamma| < 1.37$, (c) $1.56 < |\eta^\gamma| < 1.81$ and (d) $1.81 < |\eta^\gamma| < 2.37$. The ratios of the predictions using CT14 (dashed lines), NNPDF3.0 (dotted lines) and HERAPDF2.0 (dot-dashed lines) over those using MMHT2014 are shown. The shaded band represents the relative systematic uncertainty due to the photon energy scale ($\gamma$ES) and resolution ($\gamma$ER) estimated with 3.2 fb$^{-1}$ of $pp$ collisions at $\sqrt{s} = 13$ TeV [15].

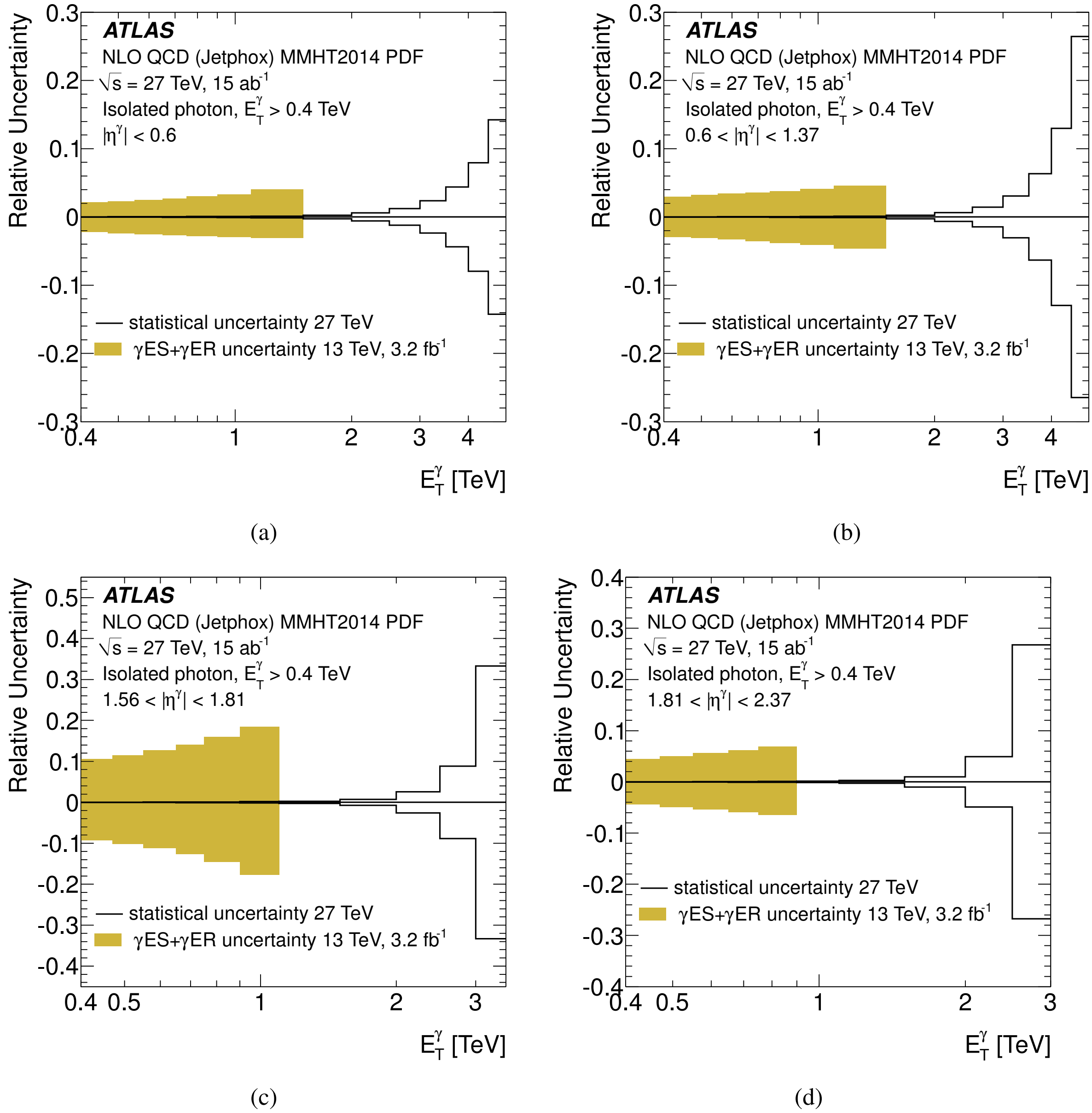

Figure 23: Predicted relative statistical uncertainty on the number of inclusive isolated photon events as a function of $E_{\mathrm{T}}^{\gamma}$ assuming an integrated luminosity of 15 ab$^{-1}$ of $pp$ collision data at $\sqrt{s}$ = 27 TeV in different ranges of photon pseudorapidity: (a) $|\eta^{\gamma}| < 0.6$, (b) $0.6 < |\eta^{\gamma}| < 1.37$, (c) $1.56 < |\eta^{\gamma}| < 1.81$ and (d) $1.81 < |\eta^{\gamma}| < 2.37$. The shaded band represents the relative systematic uncertainty due to the photon energy scale ($\gamma$ES) and resolution ($\gamma$ER) estimated with 3.2 fb$^{-1}$ of $pp$ collisions at $\sqrt{s}$ = 13 TeV [15].

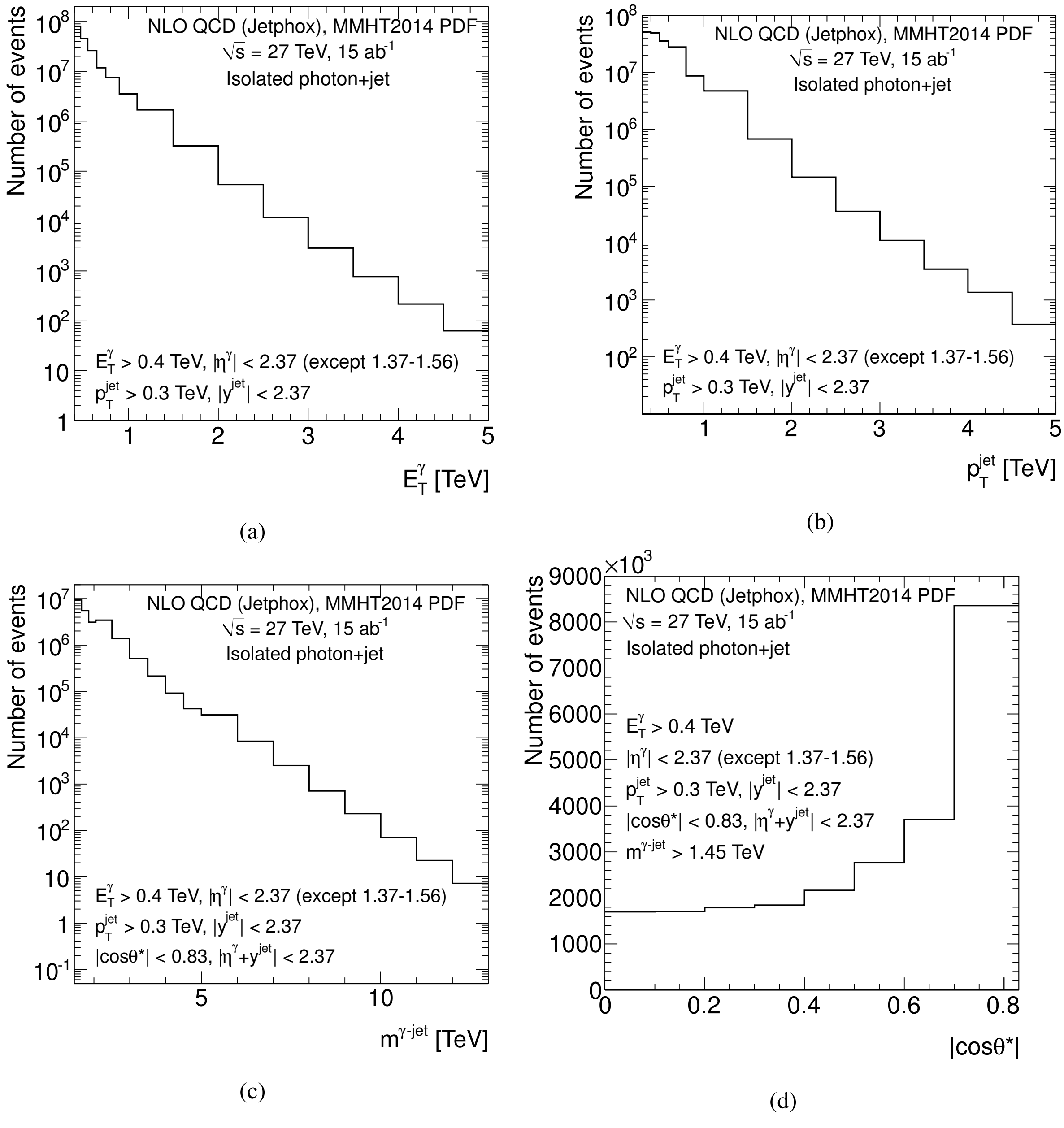

Figure 24: Predicted number of photon+jet events assuming an integrated luminosity of 15 ab$^{-1}$ of $pp$ collision data at $\sqrt{s}$ = 27 TeV as a function of different observables: (a) $E_T^{\gamma}$, (b) $p_T^{\text{jet}}$, (c) $m^{\gamma-\text{jet}}$ and (d) $|\cos\theta^*|$.

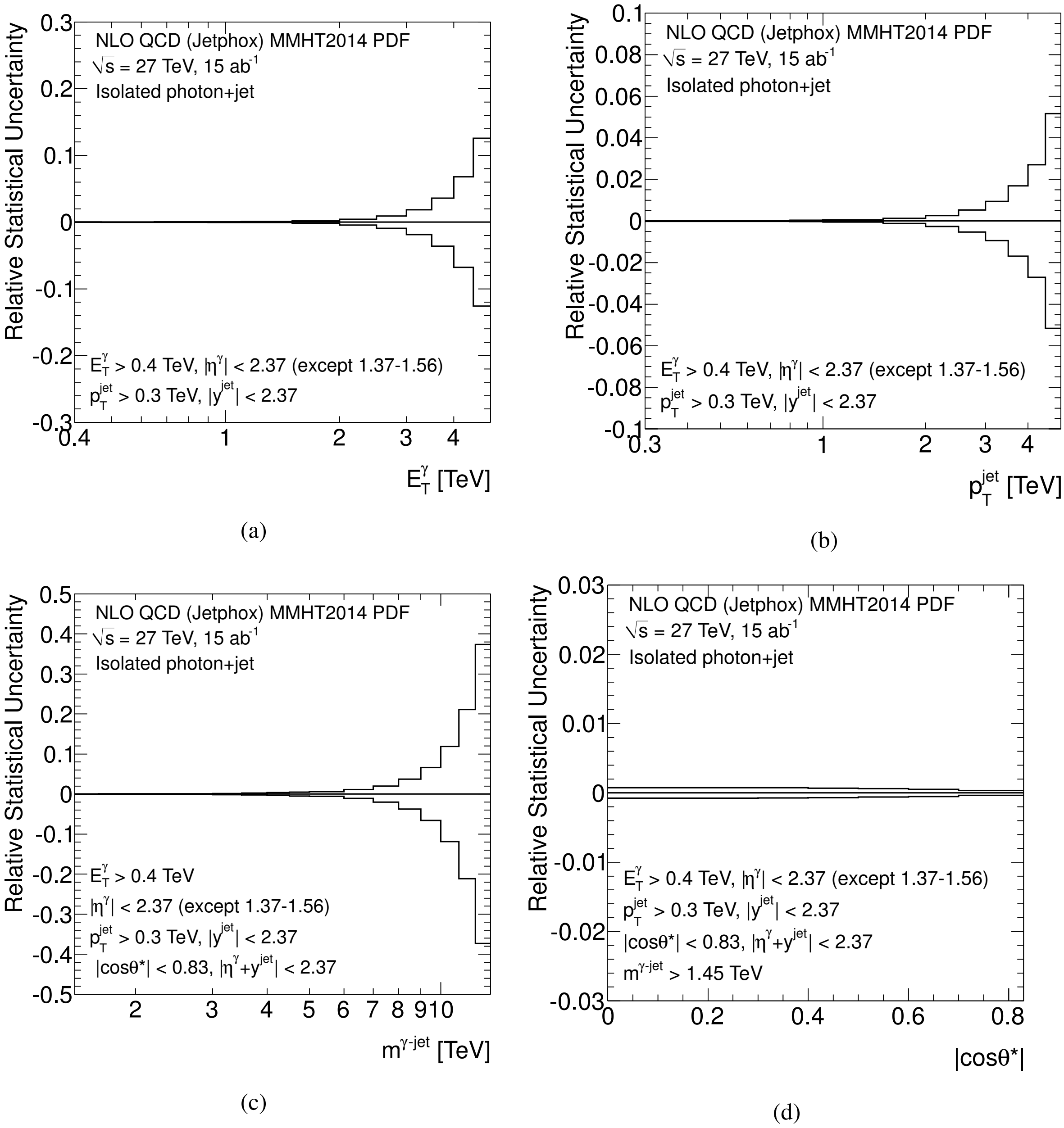

Figure 25: Predicted relative statistical uncertainty on the number of photon+jet events assuming an integrated luminosity of 15 ab$^{-1}$ of $pp$ collision data at $\sqrt{s}$ = 27 TeV as a function of different observables: (a) $E_{\mathrm{T}}^{\gamma}$, (b) $p_{\mathrm{T}}^{\mathrm{jet}}$, (c) $m^{\gamma-\mathrm{jet}}$ and (d) $|\cos\theta^{*}|$.

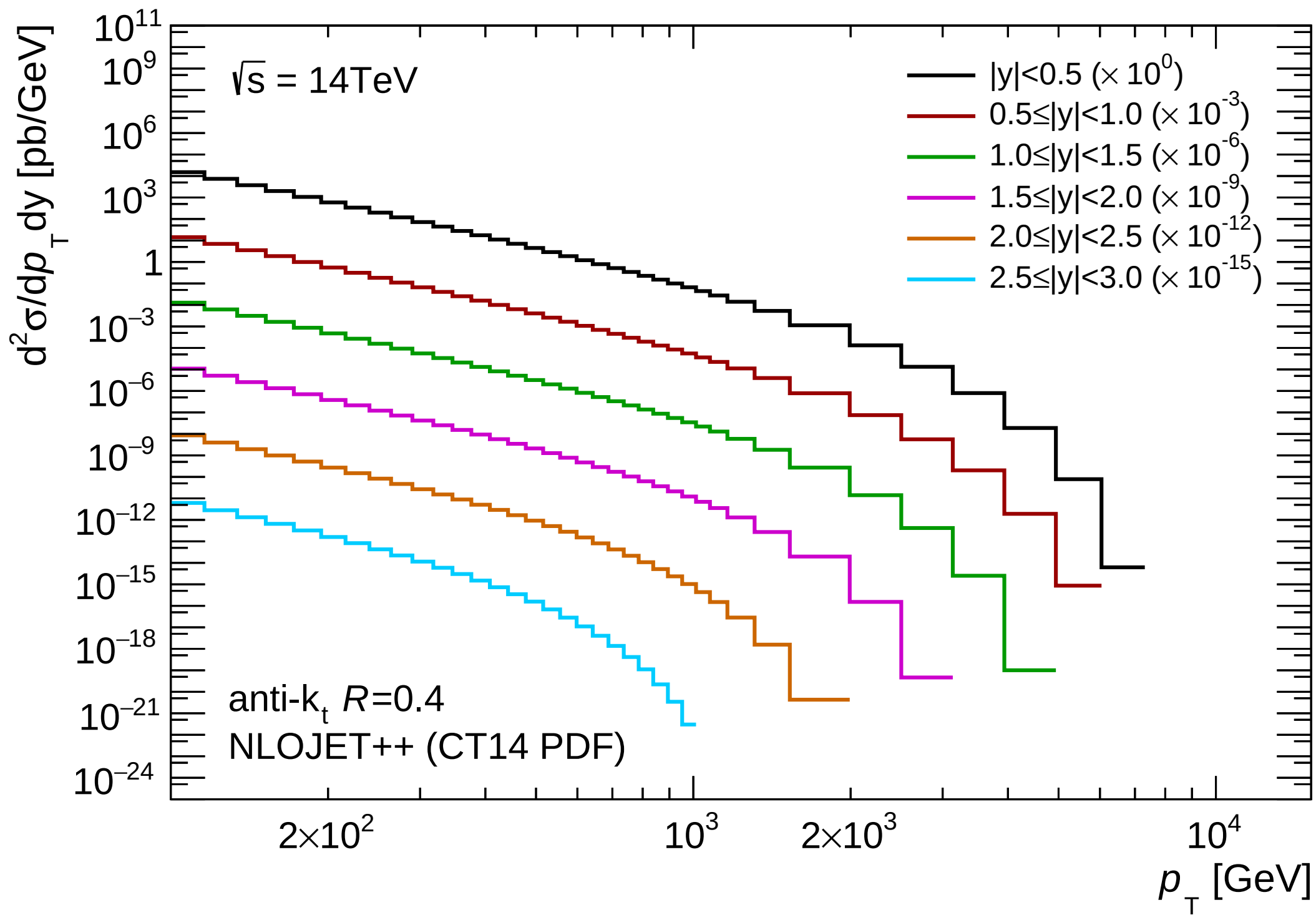

(a) Inclusive jet cross sections

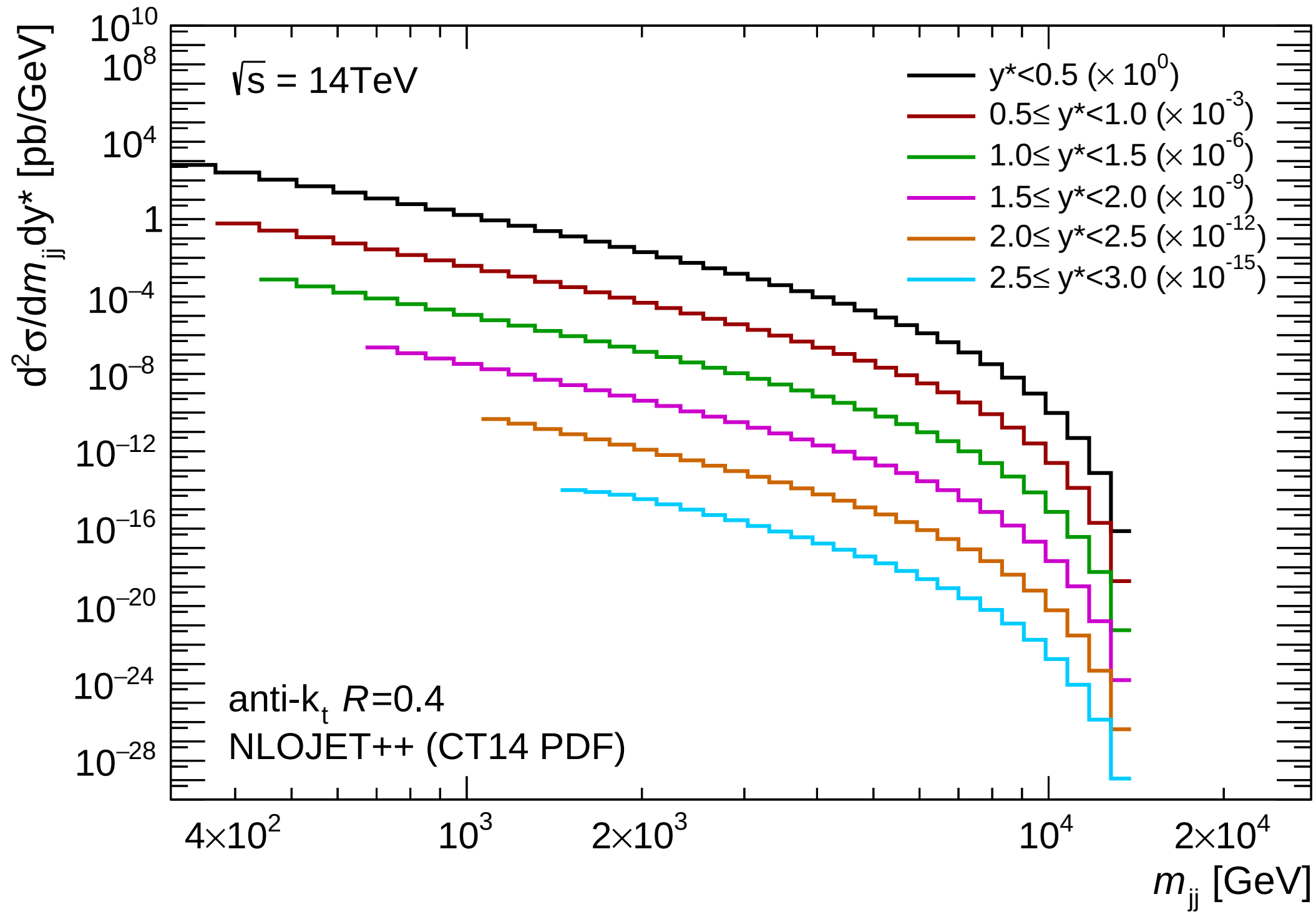

(b) Dijet cross sections

Figure 26: NLO pQCD theory predictions for (a) inclusive jet and (b) dijet cross sections at $\sqrt{s}$ = 14 TeV

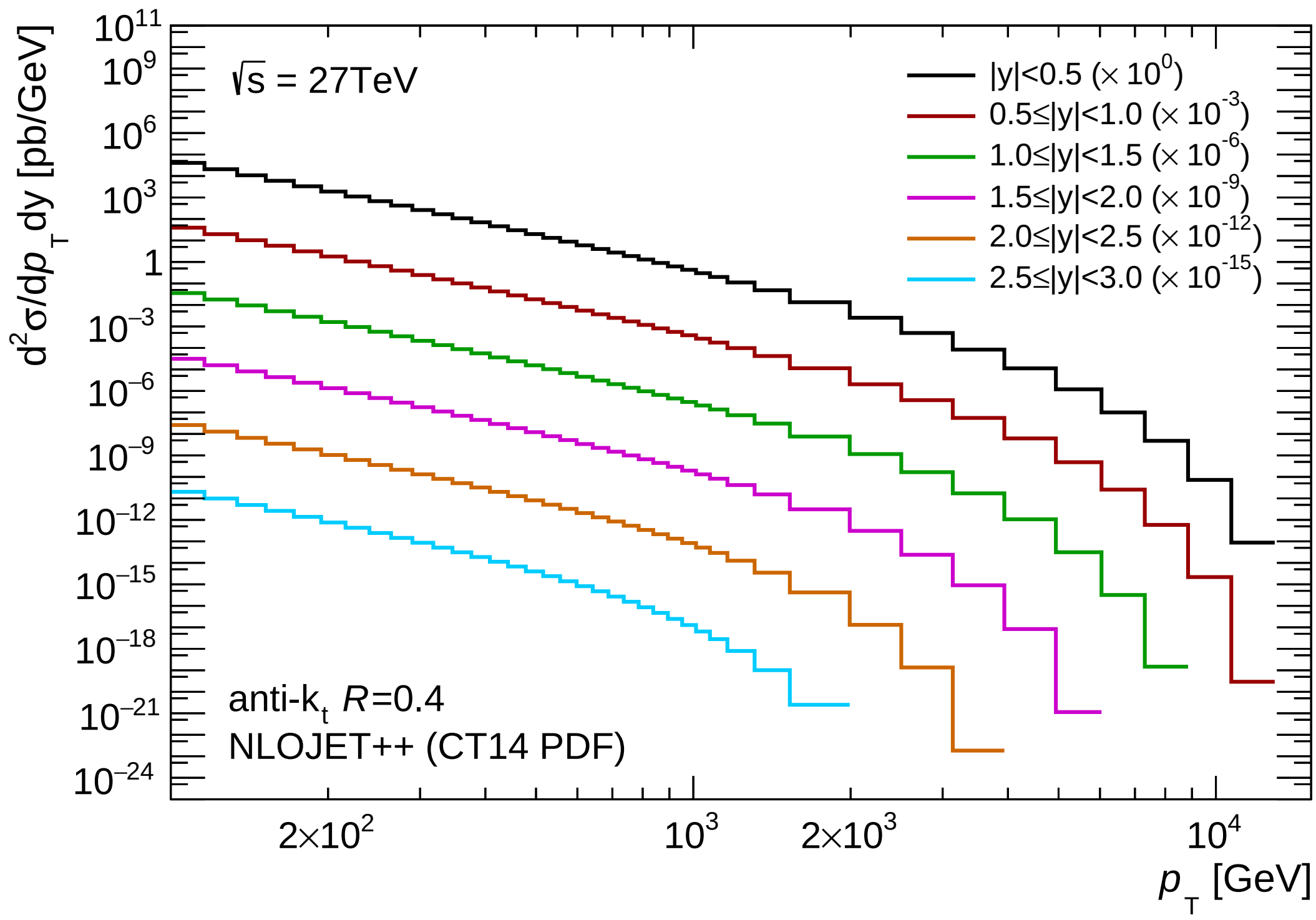

(a) Inclusive jet cross sections

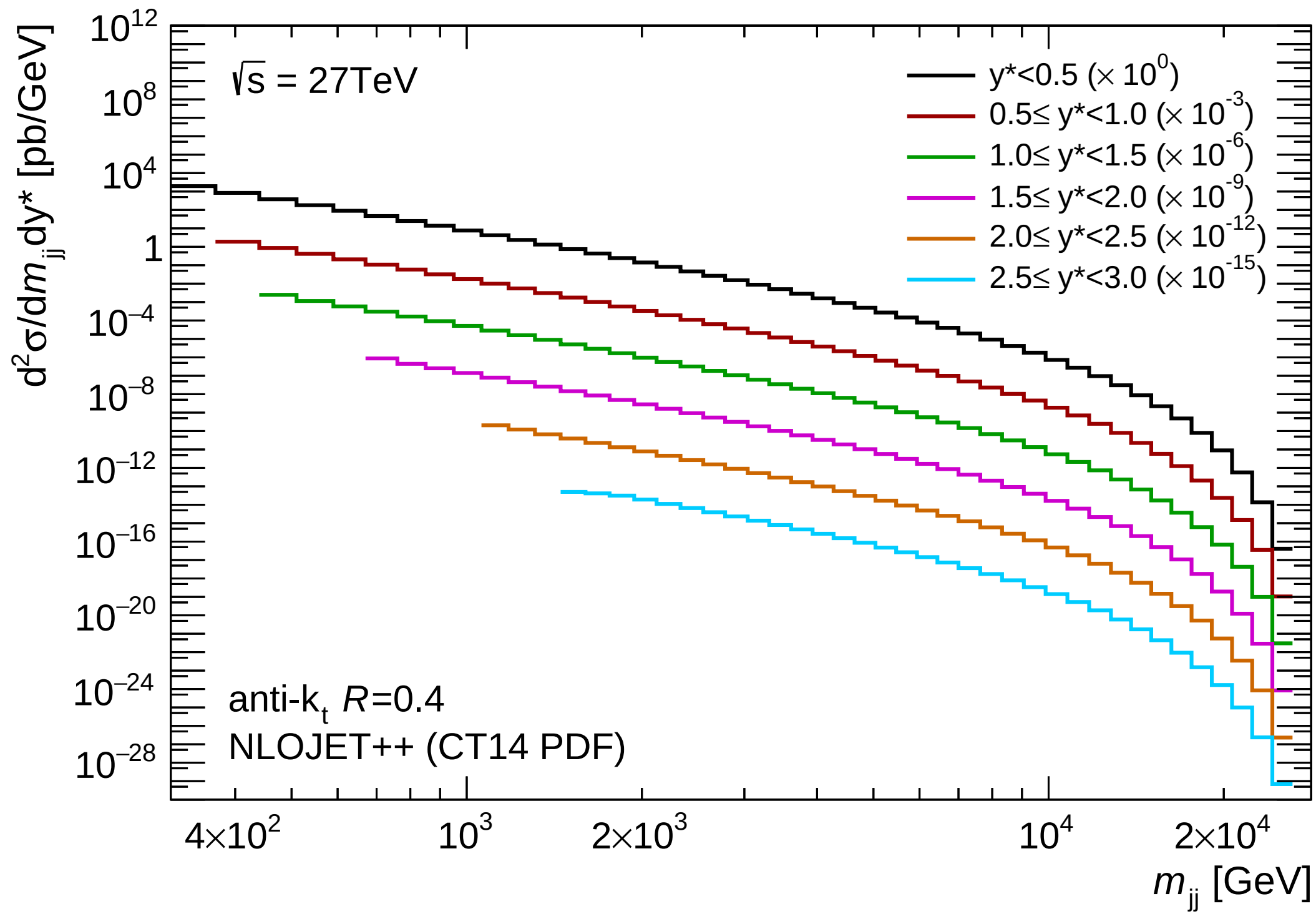

(b) Dijet cross sections

Figure 27: NLO pQCD theory predictions for (a) inclusive jet and (b) dijet cross sections at $\sqrt{s} = 27$ TeV

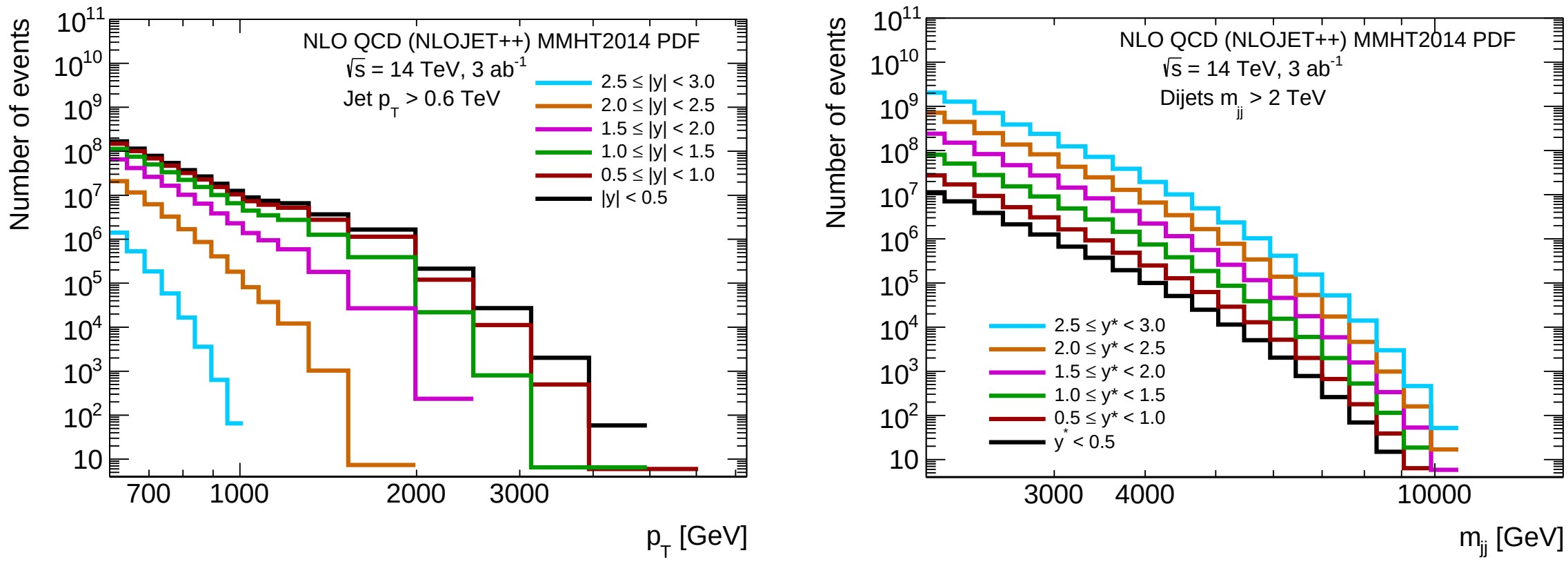

(a) Inclusive jet event yield

(b) Dijet event yield

Figure 28: Predicted number of inclusive jet and dijet events as a function of jet $p_{\mathrm{T}}$ and $m_{\mathrm{jj}}$ assuming an integrated luminosity of 3 ab$^{-1}$ of $pp$ collision data at $\sqrt{s} = 14$ TeV in different ranges of $|y|$ and $y^{*}$.

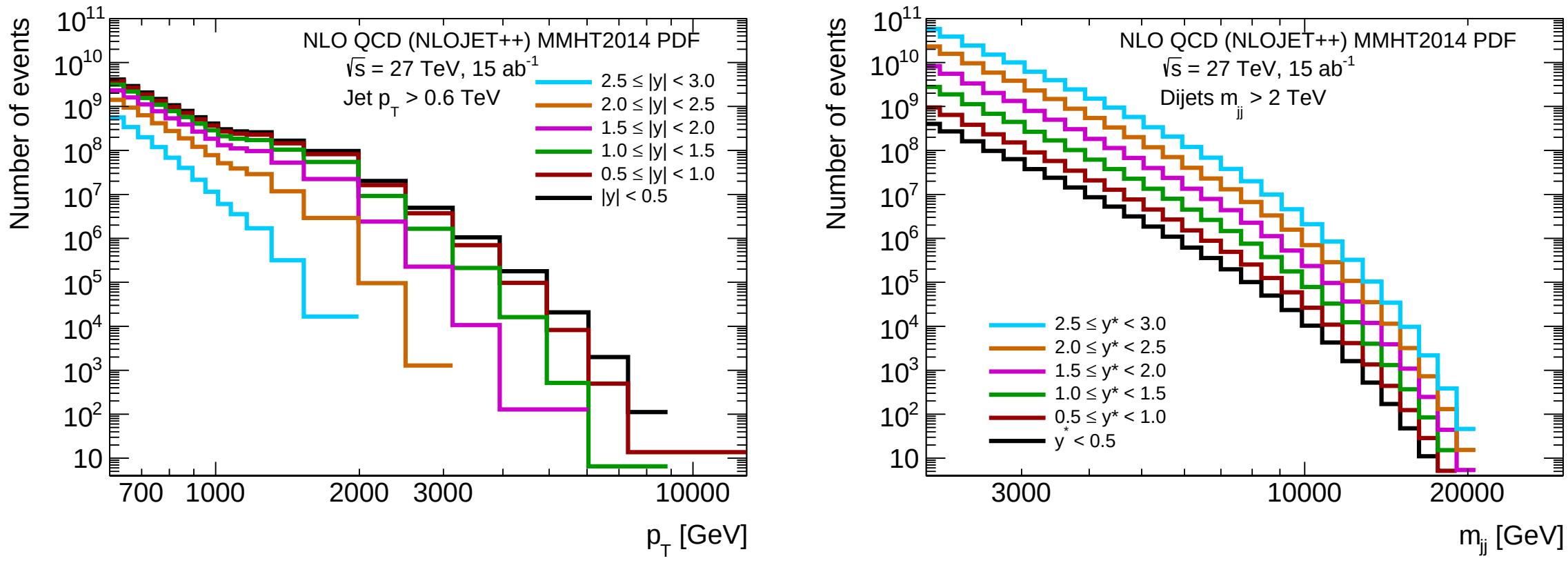

(a) Inclusive jet event yield

(b) Dijet event yield

Figure 29: Predicted number of inclusive jet and dijet events as a function of jet $p_{\mathrm{T}}$ and $m_{\mathrm{jj}}$ assuming an integrated luminosity of 15 ab$^{-1}$ of $pp$ collision data at $\sqrt{s} = 27$ TeV in different ranges of $|y|$ and $y^{*}$.

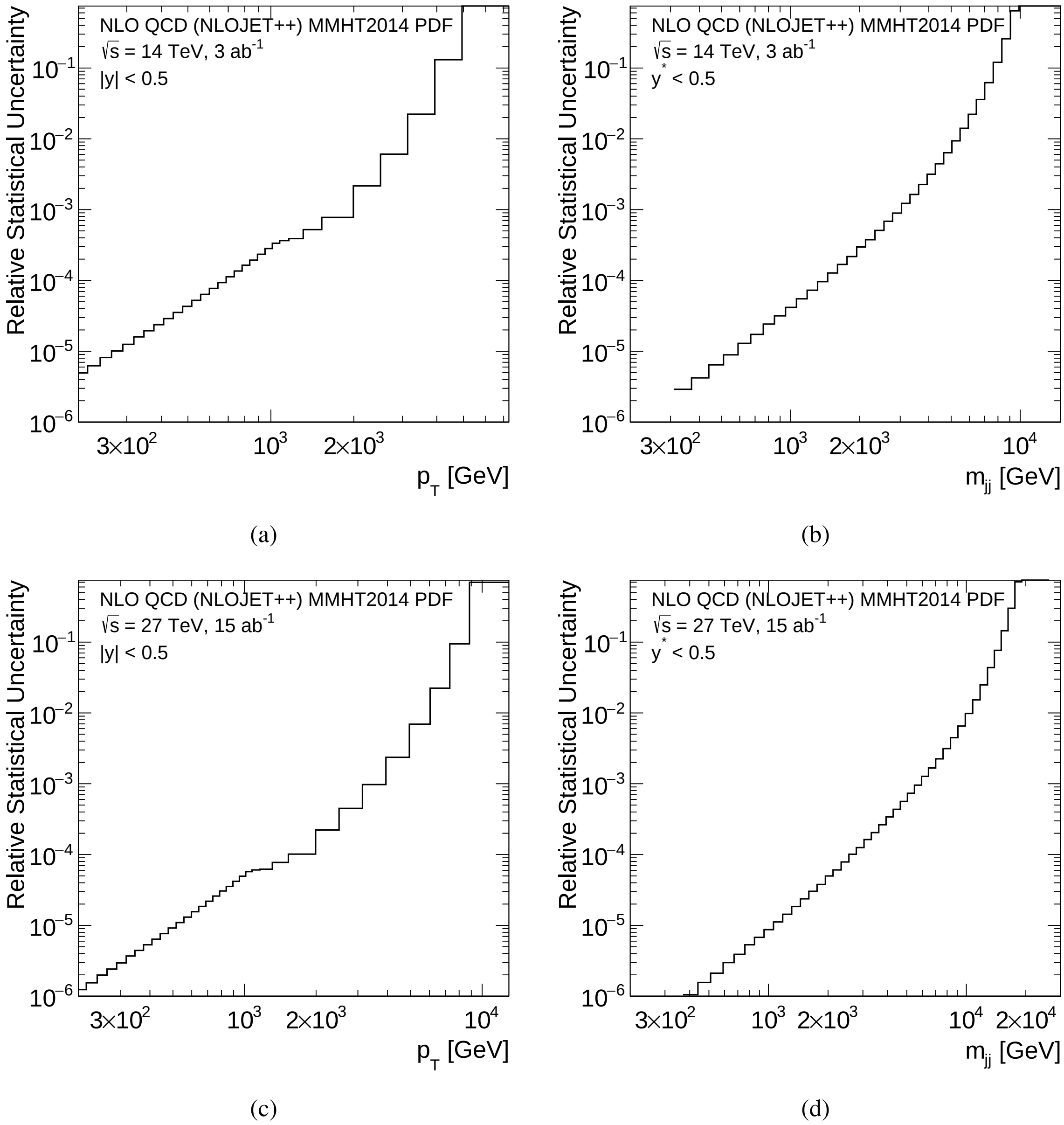

Figure 30: Predicted relative statistical uncertainty in the number of inclusive jet and dijet events as a function of jet $p_T$ and $m_{jj}$ assuming an integrated luminosity of 3 (15) $ab^{-1}$ of $pp$ collision data at $\sqrt{s}$ = 14 (27) TeV in $|y| < 0.5$ and $y^* < 0.5$ ranges.

The total predicted JES uncertainty in the inclusive jet cross section measurement for the three HL-LHC scenarios is illustrated in Figure 31 and compared to the total JES uncertainty estimate for the Run-2 jet cross section measurements. Total JES uncertainty in the low $p_T$ range is same as in Run-2 and is about 2 % lower in the high-$p_T$ region. In conservative and pessimistic scenarios JES uncertainties in the cross section are very similar in the intermediate and high-$p_T$ range, while JES uncertainty is about 1 % lower in the low-$p_T$ range for the optimistic scenario.

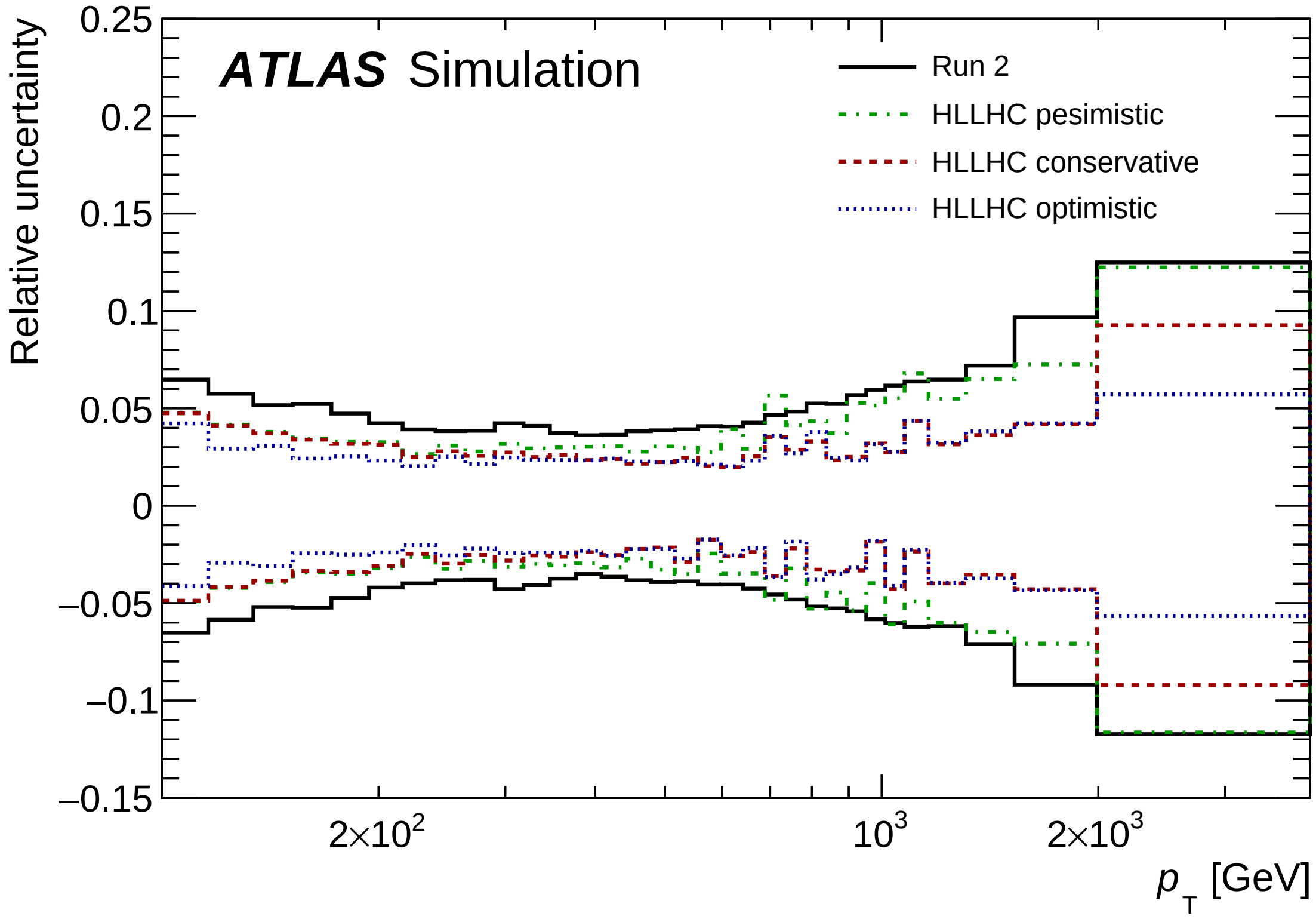

Figure 31: Relative uncertainties in the inclusive jet cross section measurements at the HL-LHC due the JES uncertainties. Three HL-LHC scenarios are compared to the Run-2 performance. Black line corresponds to the Run-2 performance. Green, red and blue lines represent pessimistic, conservative and optimistic scenarios, respectively.

# 5 Conclusion

Prospects for isolated-photon production inclusively and in association with at least one jet as well as for the inclusive jet and dijet production measurements at the HL-LHC and HE-LHC are presented.

The production of inclusive isolated photons is studied for photon transverse energies above 400 GeV in four ranges of photon pseudorapidity, namely $|\eta^{\gamma}| < 0.6$, $0.6 < |\eta^{\gamma}| < 1.37$, $1.56 < |\eta^{\gamma}| < 1.81$ and $1.81 < |\eta^{\gamma}| < 2.37$. The reach in $E_{\mathrm{T}}^{\gamma}$ is extended significantly with respect to recent measurements by the ATLAS Collaboration: for the most central region, $|\eta^{\gamma}| < 0.6$, the $E_{\mathrm{T}}^{\gamma}$ reach is extended from 1.5 TeV to 3–3.5 TeV (5 TeV) assuming an integrated luminosity of 3 ab$^{-1}$ (15 ab$^{-1}$) of collision data at $\sqrt{s} = 14$ TeV (27 TeV). For photon+jet events, expectations are shown for the distributions in $E_{\mathrm{T}}^{\gamma}$, $p_{\mathrm{T}}^{\mathrm{jet}}$, $m^{\gamma-\mathrm{jet}}$ and $|\cos\theta^{*}|$. Jets are required to have $p_{\mathrm{T}}^{\mathrm{jet}} > 300$ GeV and $|y^{\mathrm{jet}}| < 2.37$. An integrated luminosity of 3 ab$^{-1}$ (15 ab$^{-1}$) of collision data at $\sqrt{s} = 14$ TeV (27 TeV) leads to significant extensions of phase space in comparison with recent measurements at $\sqrt{s} = 13$ TeV: for $E_{\mathrm{T}}^{\gamma}$ and $p_{\mathrm{T}}^{\mathrm{jet}}$ from 1.5 TeV to 3.5 TeV (5 TeV) and for $m^{\gamma-\mathrm{jet}}$ from 3.3 TeV to 7 TeV (12 TeV).

The inclusive jet production cross sections at NLO pQCD accuracy for jets with $p_{\mathrm{T}} > 100$ GeV within $|y| < 3$ in six bins of the absolute jet rapidity are calculated. The non-perturbative effects are taken into account as multiplicative factors. The reach in jet transverse momentum in the central rapidity range in comparison to the recent ATLAS measurements [17] is extended from 3.5 TeV to 5.5 (10) TeV for the inclusive jet $p_{\mathrm{T}}$ and from 9 TeV to 11.5 (22) TeV for the dijet invariant mass at the HL-LHC (HE-LHC).

The expected experimental uncertainties in the inclusive jet measurements are studied using three possible scenarios for the precision in the jet energy measurements. In all considered scenarios the inclusive jet cross section measurements will improve compared to Run-2 results precision. In the optimistic scenario, the expected precision will be almost two times better than one in the corresponding Run-2 measurements.

The impact of non-perturbative effects in the high transverse momentum range is small, around 1–2%, allowing to directly test the perturbative QCD predictions at the energy frontiers set by HL/HE-LHC.

A study of PDF sensitivity of the photon and jet production cross sections based on current PDF sets such as MMHT2014, CT14, NNPDF3.0 and HERAPDF2.0 show differences between predictions of up to 30%. That will allow to further constrain the PDFs by performing the photon and jet production measurements at the HL-LHC and HE-LHC. The expected impact on the determination of the proton PDFs of these measurements together with those of other processes such as the production of electroweak gauge bosons and top quark pairs at HL-LHC is illustrated with the estimations of the PDF induced uncertainties based on the PDF4LHC HL-LHC PDF set.

# CMS Physics Analysis Summary

Contact: cms-phys-conveners-ftr@cern.ch

## Projection of differential $t\bar{t}$ production cross section measurements in the e/$\mu$+jets channels in pp collisions at the HL-LHC

The CMS Collaboration

### Abstract

A study of the resolved reconstruction of top quark pairs in the e/$\mu$+jets channels and a projection of differential $t\bar{t}$ cross section measurements at the HL-LHC with an integrated luminosity of 3 $ab^{-1}$ at 14 TeV are presented. The analysis techniques are based on previous measurements of differential $t\bar{t}$ cross sections at 13 TeV. It is shown that such a measurement is feasible at the HL-LHC despite the expected large number of pileup interactions. The precision of the differential cross section will profit from the enormous amount of data and the extended $\eta$-range of the Phase-2 CMS detector. The results are used to estimate the improvement of constraints on parton distribution functions.

# 1 Introduction

Precision measurements of top quark properties present an important test of the standard model (SM). As the heaviest particle in the SM, the top quark plays an important role for the electroweak symmetry breaking and becomes a sensitive probe for physics beyond the SM. Therefore, $\mathrm{t\bar{t}}$ or top quark reconstruction is also important for many searches.

Based on analysis techniques used in previous measurements of differential $\mathrm{t\bar{t}}$ cross sections [1, 2] at 13 TeV, we present a study of the feasibility and performance of such a measurement at the CERN HL-LHC, which is planned to be operated from 2026. It is expected to collect an integrated luminosity of about $3\,\mathrm{ab}^{-1}$ at a center-of-mass energy of 14 TeV. The high instantaneous luminosity will result in up to 200 pp interactions per bunch crossing (pileup). Therefore, effective pileup mitigation techniques like PUPPI [3] are essential for a good performance of the $\mathrm{t\bar{t}}$ reconstruction.

A detailed description of the CMS detector, including a definition of the coordinate system and kinematic variables, can be found in Ref. [4]. Most subsystems of the CMS detector will be improved or replaced in order to cope with the high pileup condition (Phase-2 upgrade) [5–8].

We provide projections of the measurements of differential $\mathrm{t\bar{t}}$ cross sections at parton level. This includes an estimate of the expected signal yield and its uncertainty based on simulations of the Phase-2 CMS detector. We study the distributions of the transverse momentum $p_{\mathrm{T}}$ and rapidity $y$ of the hadronically (labeled $\mathrm{t_h}$) and leptonically (labeled $\mathrm{t}_\ell$) decaying top quarks and the mass $M$, $p_{\mathrm{T}}$, and $y$ of the $\mathrm{t\bar{t}}$ system. In addition, the normalized double-differential cross section as function of $M(\mathrm{t\bar{t}})$ *vs.* $|y(\mathrm{t\bar{t}})|$ is used to study its constraints on the parton distribution functions (PDF).

# 2 Simulation

The Monte Carlo generator POWHEG [9–12] (v2,hvq) is used to simulate the production of $\mathrm{t\bar{t}}$ events with next-to-leading-order (NLO) accuracy in QCD. The renormalization $\mu_{\mathrm{r}}$ and factorization $\mu_{\mathrm{f}}$ scales are set to the transverse mass $m_{\mathrm{T}} = \sqrt{m_{\mathrm{t}}^2 + p_{\mathrm{T}}^2}$ of the top quark, where $p_{\mathrm{T}}$ is the transverse momentum of the top quark and a top quark mass $m_{\mathrm{t}} = 172.5\,\mathrm{GeV}$ is used. The result is combined with the parton shower simulations of PYTHIA8 [13, 14] (v8.219) using the underlying event tune CUETP8M2T4 [15, 16]. The simulation is normalized to an inclusive $\mathrm{t\bar{t}}$ production cross section of 985 pb [17]. This value is calculated with next-to-NLO accuracy, including the resummation of next-to-next-to-leading-logarithmic soft-gluon terms.

A sample of 200 million generated events are interfaced to a fast simulation of the Phase-2 CMS detector based on the DELPHES [18] detector simulation framework. The sample includes a simulation of an average pileup of 200 pp interactions per bunch crossing.

We do not use simulations of non-$\mathrm{t\bar{t}}$ backgrounds. According to the 13 TeV analysis of 2016 data [2], the total background contribution is about 4.5%. A contribution of 2.7% from single top quark production is subtracted according to the SM expectation. The remaining background, 1.8%, consists of multijet, Drell–Yan, and W boson events. A common shape of the distribution of these backgrounds is extracted from a b jet reduced control region. Its normalization is based on the simulated ratio of events in the signal and in the control regions. Since this method, involving a control region in data, is not applicable in studies based on simulation only, we use the related systematic uncertainties obtained in the 2016 analysis.

# 3 Physics objects and event selection

The signal signature in the e/$\mu$+jets channels consists of a single isolated electron or muon, a neutrino, and a b-jet from the decay of $t_\ell$. In addition, 3 jets, one of which is a b-jet, are expected from the decay of $t_h$. Hence, events with exactly one isolated electron or muon with $p_T > 30\,\text{GeV}$ and $|\eta| < 2.8$ are selected. Events with additional isolated electrons or muons with $p_T > 15\,\text{GeV}$ and $|\eta| < 2.8$ are rejected. At least 4 jets with $p_T > 30\,\text{GeV}$ and $|\eta| < 4.0$ are required. Due to the extended $\eta$ coverage of the Phase-2 tracker, the requirement of $|\eta| < 2.4$ for all objects in [1, 2] can be relaxed. At least 2 of the jets have to be identified as b jets, i.e., fulfilling a requirement of the DeepCSV b-tagger [19] with a b-jet selection efficiency of about 70% and rejection of about 95% for other jets in $t\bar{t}$ events. For an effective b tagging $|\eta| < 3.5$ is required for the two b jets. Charged hadron subtracted jets, being the standard in most analyses, cannot be used under HL-LHC conditions, since a large number of jets from pileup is expected. In contrast to previous analyses the PUPPI algorithm [3] is used for pileup mitigation. It assigns a weight to each reconstructed particle flow (PF) object [20] according to the probability that it originates from the leading primary vertex, which, among the reconstructed primary vertices, is the one with the largest value of summed physics-object $p_T^2$. When the jets are clustered and the $\vec{p}_T^{\text{miss}}$ is calculated, the momenta of the particles are scaled by these weights. Jets are clustered from the weighted PF objects using the anti-$k_T$ jet algorithm with a distance parameter of 0.4 implemented in the FASTJET package [21]. Jets within $\Delta R = 0.4$ of the isolated electrons or muons are rejected, where $\Delta R = \sqrt{(\Delta\phi)^2 + (\Delta\eta)^2}$ with $\Delta\phi$ and $\Delta\eta$ are the differences in azimuthal angle (in radians) and pseudorapidity between the directions of two objects. The missing transverse momentum $\vec{p}_T^{\text{miss}}$ is calculated as the negative of the vectorial sum of transverse momenta of all weighted PF candidates in the event. Jet energy corrections are propagated to improve the measurement of $\vec{p}_T^{\text{miss}}$.

# 4 Reconstruction of the $t\bar{t}$ system

A detailed description of the $t\bar{t}$ reconstruction is presented in Refs. [1, 2]. For the reconstruction all possible permutations of assigning detector-level jets to the corresponding $t\bar{t}$ decay products are tested and a likelihood $\lambda$ that a certain permutation is correct is evaluated. Permutations are considered only if the two jets with the highest b identification probabilities are the two b jet candidates. In each event, the permutation with the highest value of $\lambda$ is selected. The likelihood $\lambda$ is constructed from the two dimensional probability density $m_t$–$m_W$ of correctly assigned jets for the hadronically decaying top quark and the probability density of $D_{\nu,\text{min}}$ obtained for a correct b jet from a leptonically decaying top quark. The variable $D_{\nu,\text{min}}$ is obtained in the calculation of the neutrino momentum [22]. The probability densities used in the reconstruction are extracted from the Phase-2 $t\bar{t}$ simulation.

In Fig. 1 the distributions of $\lambda$ and the reconstructed $m_t$ of the hadronically decaying top quarks are shown for the Phase-2 simulation. The $t\bar{t}$ simulation is divided into the categories of events with correctly reconstructed top quarks (right reco), at least one wrong jet assignment (wrong reco), at least one missing decay product (nonreconstructable), and events that are not signal events in the $\ell$+jets channels (nonsignal). Comparisons of these distributions with those in the 2016 analysis [2] show that a similar fraction of correct reconstructed top quark pairs is expected and the resolution of the $m_t$ peak is comparable despite the harsh pileup conditions.

In Figs. 2 and 3 the expected signal yields are shown for the distributions of $p_T(t_h)$ and $|y(t_h)|$. Corresponding distributions for $p_T(t_\ell)$ and $|y(t_\ell)|$ and of the $t\bar{t}$ system $M(t\bar{t})$, $p_T(t\bar{t})$, and $|y(t\bar{t})|$ are shown in Appendix A. In addition, properties of the migration matrices, representing the relations between parton level and detector level are shown. These properties are the purity defined as the fraction of parton-level top quarks in the same bin at the detector level, the stability defined as the fraction of detector-level top quarks in the same bin at the parton level, and the bin efficiency defined as the ratio of the number of events found in a certain bin at detector level and the number of events found at parton-level in the same bin. In an ideal case, for diagonal migration matrices, purity and stability are equal to one and the bin efficiency corresponds to the acceptance for each bin. While purity and stability remain almost unchanged with respect to the 2016 analysis, the bin efficiency is increased especially in the high rapidity regions due to the extended $\eta$-range of the Phase-2 CMS detector. In Fig. 4 the migration matrix and its properties for the double-differential measurements as a function of $M(\mathrm{t\bar{t}})$ *vs.* $|y(\mathrm{t\bar{t}})|$ are shown. The D'Agostini method [23] is used for the unfolding of the detector-level distributions. A detailed discussion about the selected number of iterations, that control the level of regularization, is presented in the 2016 analysis [2].

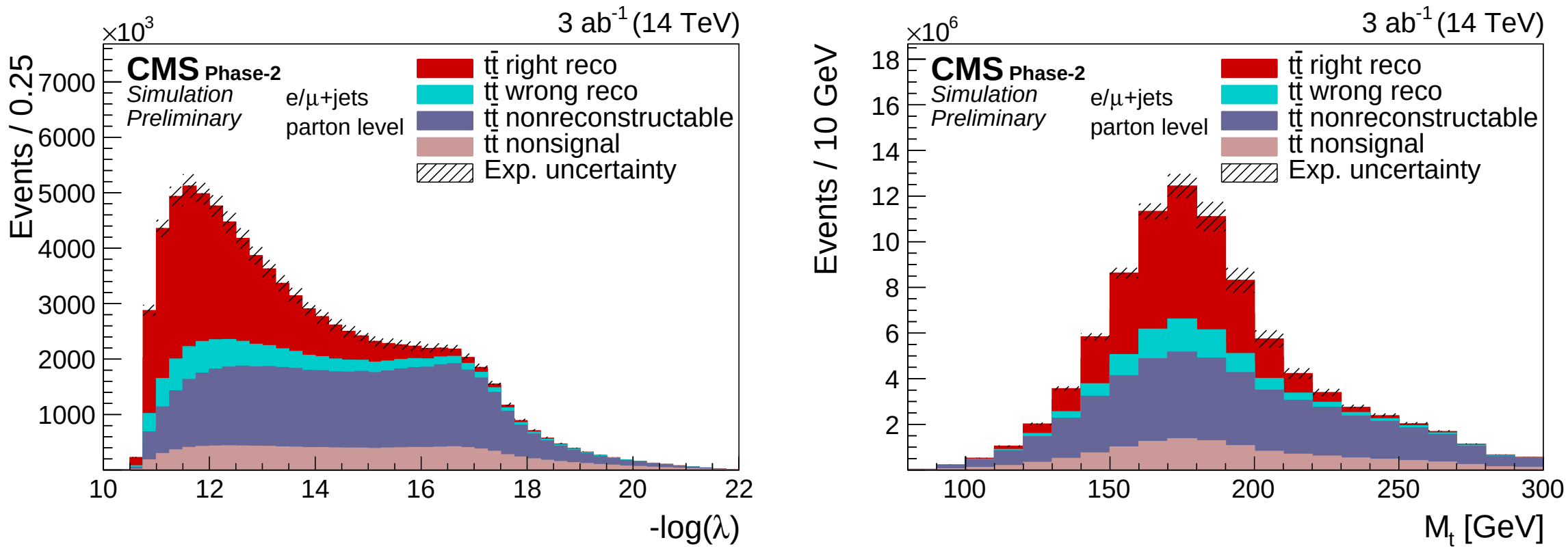

Figure 1: Distributions of $\lambda$ and the reconstructed $m_{\mathrm{t}}$ of the hadronically decaying top quarks are shown for the Phase-2 simulation.

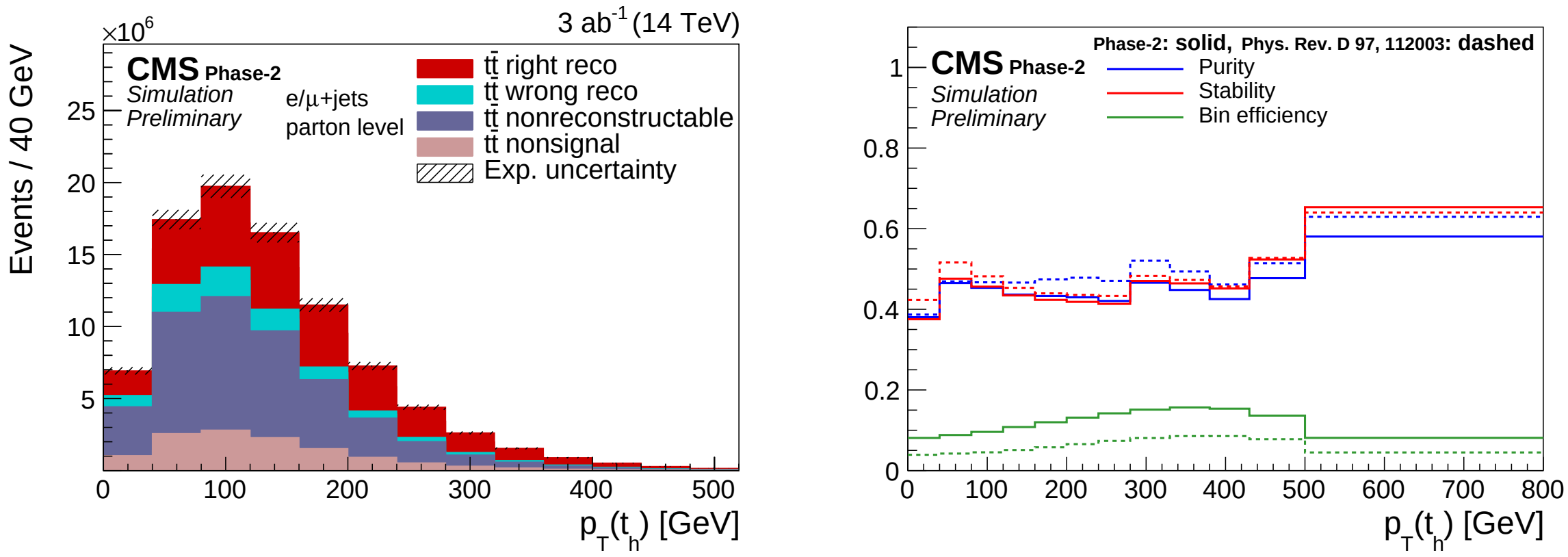

Figure 2: Expected signal yields (left) and properties of the migration matrix (right) for the measurement of $p_{\mathrm{T}}(\mathrm{t_h})$. For comparison the properties are also shown for the 2016 analysis [2].

**4**

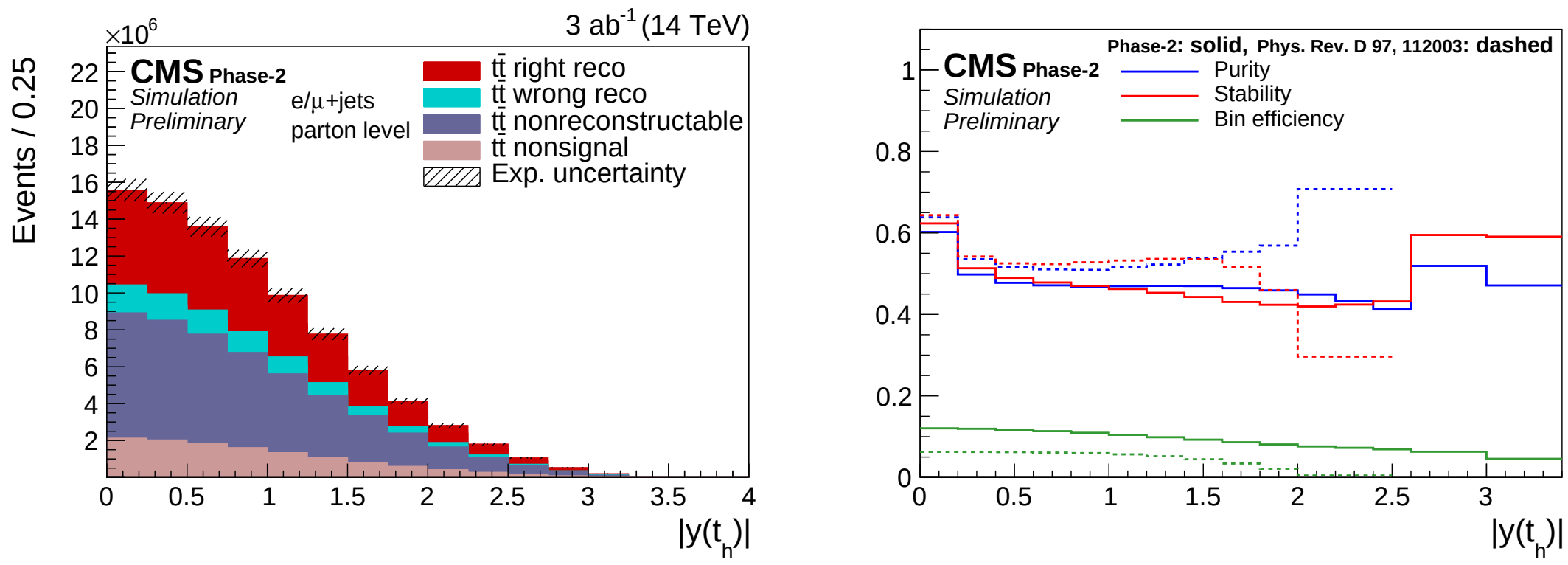

Figure 3: Expected signal yields (left) and properties of the migration matrix (right) for the measurement of $|y(\mathrm{t_h})|$. For comparison the properties are also shown for the 2016 analysis [2].

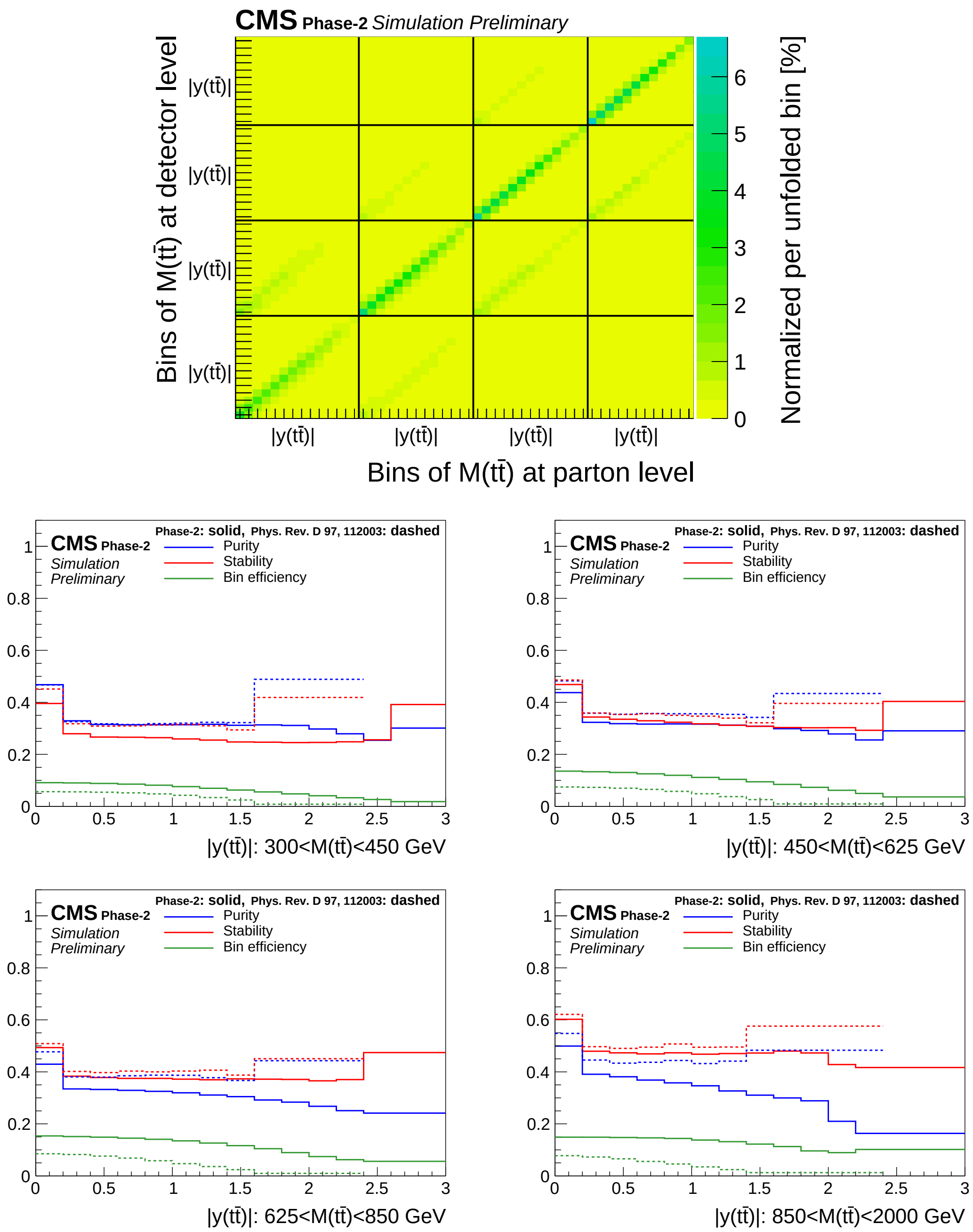

Figure 4: Migration matrix (upper) and its properties (middle, lower) for the double-differential measurements as a function of $M(t\bar{t})$ *vs.* $|y(t\bar{t})|$. There are four $|y(t\bar{t})|$ distributions for different regions of $M(t\bar{t})$. The large off-diagonal structures in the migration matrix correspond to migrations among $M(t\bar{t})$ regions. For comparison the properties are also shown for the 2016 analysis [2].

# 5 Uncertainties

The following experimental uncertainties are estimated based on the expected performance of the Phase-2 CMS detector. For comparison the typical uncertainties in the 2016 analysis are given in parentheses:

- luminosity measurement: 1% (2.5% [24])
- muon reconstruction and identification: 0.5% (1–2% [25])
- electron reconstruction and identification: 1% (1–2% [26])
- b-tagging efficiency: $p_{\mathrm{T}}$ dependent 1–5% (1–3% [19])
- b-tagging mistagging efficiency: $p_{\mathrm{T}}$ dependent about 10% (8%) for u, d, s, and gluon jets and 2–14% (2–6%) for c jets.
- jet energy calibration: for jets in the typical $p_{\mathrm{T}}$-range of $30 < p_{\mathrm{T}} < 300\,\mathrm{GeV}$, the uncertainty of the jet energy decreases from 1.7% (2.7% [27]) to 0.45% (0.5%). The reduced uncertainties are cause by improvements of the Phase-2 detector and a reduction of uncertainties in the calibration methods due to the high amount of data.
- jet energy resolution: $\eta$ dependent 3% (5%) in the central region and 5% (12%) in the forward region.
- the $\vec{p}_{\mathrm{T}}^{\mathrm{miss}}$: the variations in the jet energy scale are propagated to the $\vec{p}_{\mathrm{T}}^{\mathrm{miss}}$.

To propagate these uncertainties to the cross section results the Phase-2 simulation is either reweighted or the momenta of certain objects are rescaled in order to follow the desired variation. Afterwards, the nominal distributions are unfolded with the migration matrices and correction factors obtained from the varied simulations. The resulting differences in the unfolded yields are taken as uncertainties.

Previous analyses showed that theoretical and modeling uncertainties make a significant contribution to the overall uncertainty. The uncertainty sources considered are the initial-/final-state parton shower scales, parton shower matching scale $h_{\mathrm{damp}}$, PDF variations, parton shower tuning, $m_{\mathrm{t}}$, leptonic b-decay branching ratio, b-fragmentation, renormalization/factorization scales, color reconnection model, and the background subtraction, which is mostly based on SM predictions. These uncertainty sources have been studied in the 2016 analysis [2] and are taken from there. For the extended rapidity range, the uncertainties in the highest available rapidity bins of the 2016 analysis are used. These theoretical uncertainties are reduced by a factor of two, since several improvements of the theoretical predictions are expected and further measurements can reduce modeling uncertainties. In addition, the measurable portion of the phase space is increased, resulting in a reduction of the theory based uncertainties in the extrapolation to the full phase space.

# 6 Cross section results

In Figs. 5–7 the projection of the differential cross sections with the expected uncertainties are shown. For comparison we also show the uncertainties in the 2016 analysis. The normalized differential cross sections can be found in Appendix B.

The normalized double-differential cross section as a function of $M(\mathrm{t\bar{t}})$ *vs.* $|y(\mathrm{t\bar{t}})|$ is presented in Figs 8 and 9. This is further used for PDF constraints in Section 7.

In the bulk of the distributions, where the uncertainties of the 2016 analysis have a negligible statistical component, a gain of precision can only be reached by a reduction of systematic

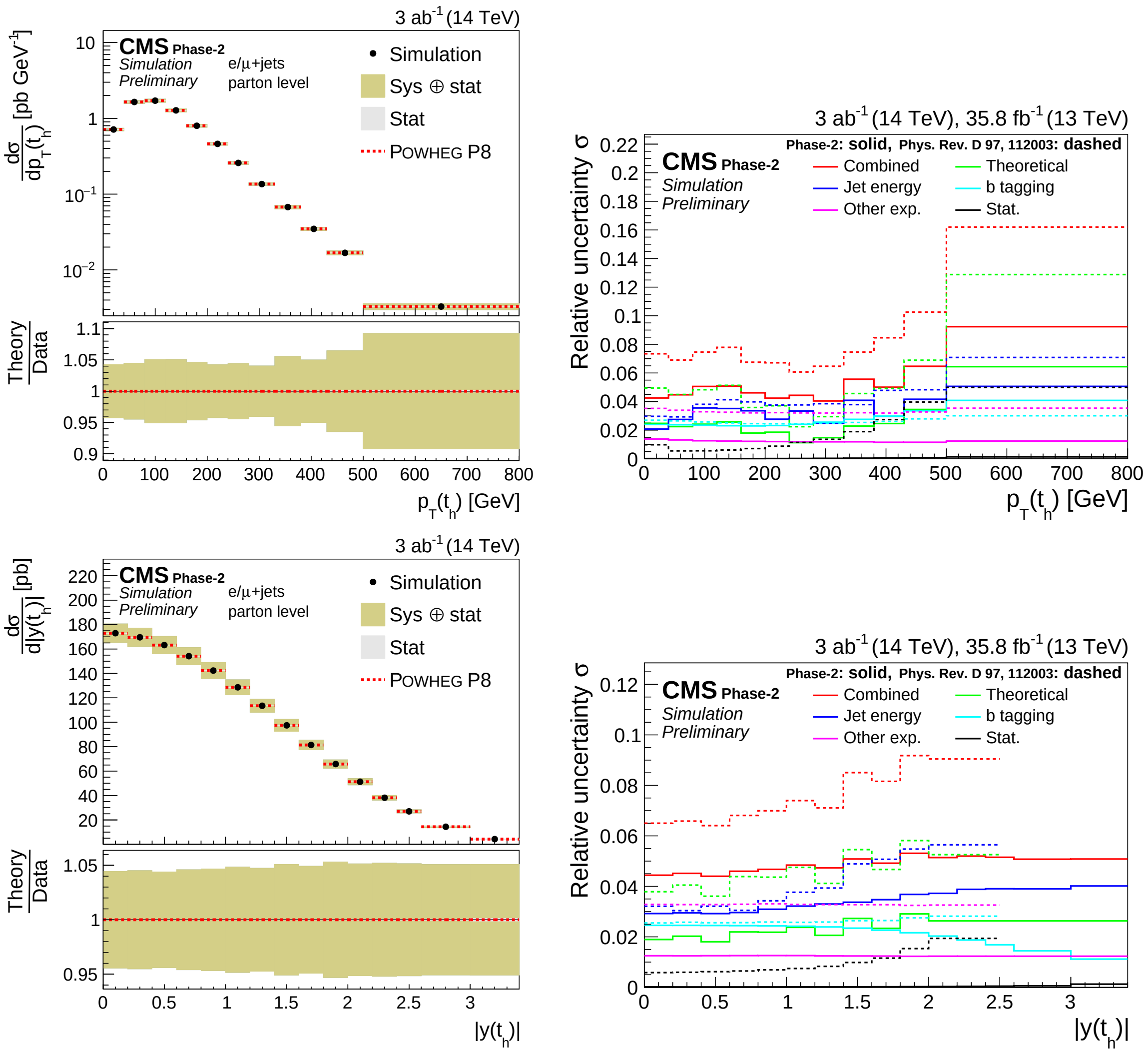

Figure 5: Differential cross sections (left) as a function of $p_{\mathrm{T}}(\mathrm{t_h})$ (upper) and $|y(\mathrm{t_h})|$ (lower). The corresponding relative uncertainties (right) in the Phase-2 projections are compared to the uncertainties in the 2016 measurements [2].

**8**

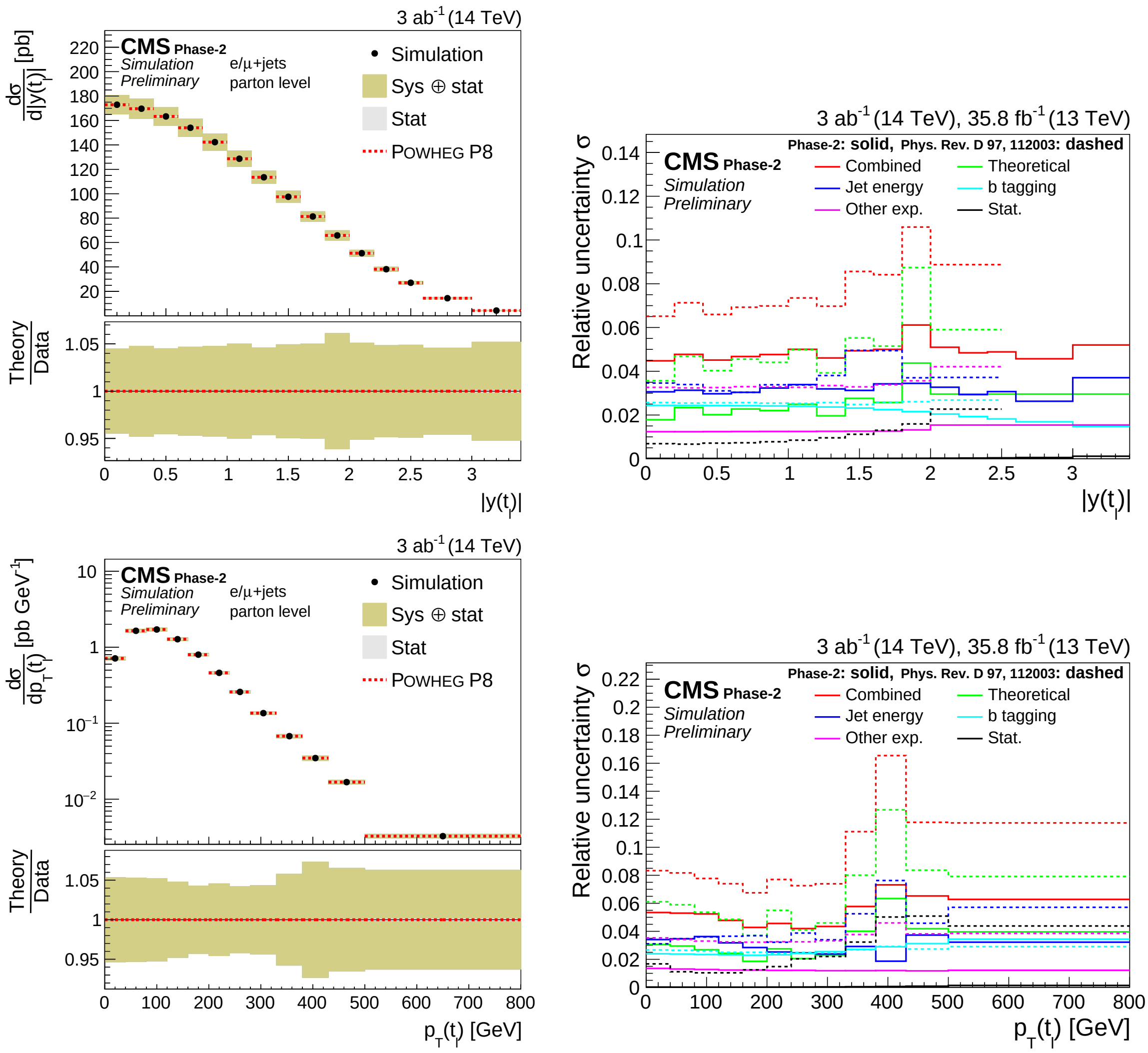

Figure 6: Differential cross sections (left) as a function of $p_T(t_\ell)$ (upper) and $|y(t_\ell)|$ (lower). The corresponding relative uncertainties (right) in the Phase-2 projections are compared to the uncertainties in the 2016 measurements [2].

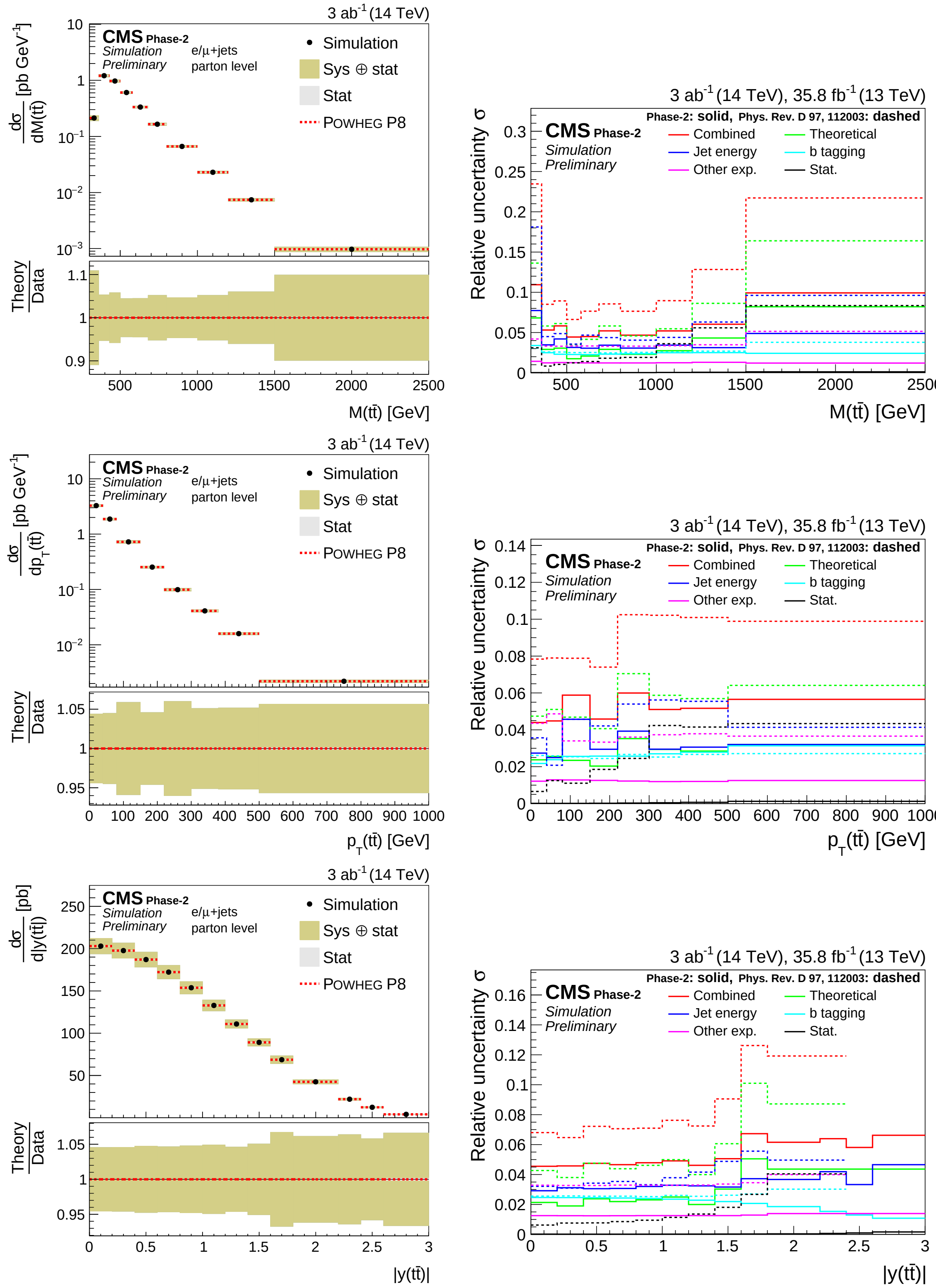

Figure 7: Differential cross sections (left) as a function of $M(\mathrm{t\bar{t}})$ (upper), $p_{\mathrm{T}}(\mathrm{t\bar{t}})$ (middle), and $|y(\mathrm{t\bar{t}})|$ (lower). The corresponding relative uncertainties (right) in the Phase-2 projections are compared to the uncertainties in the 2016 measurements [2].

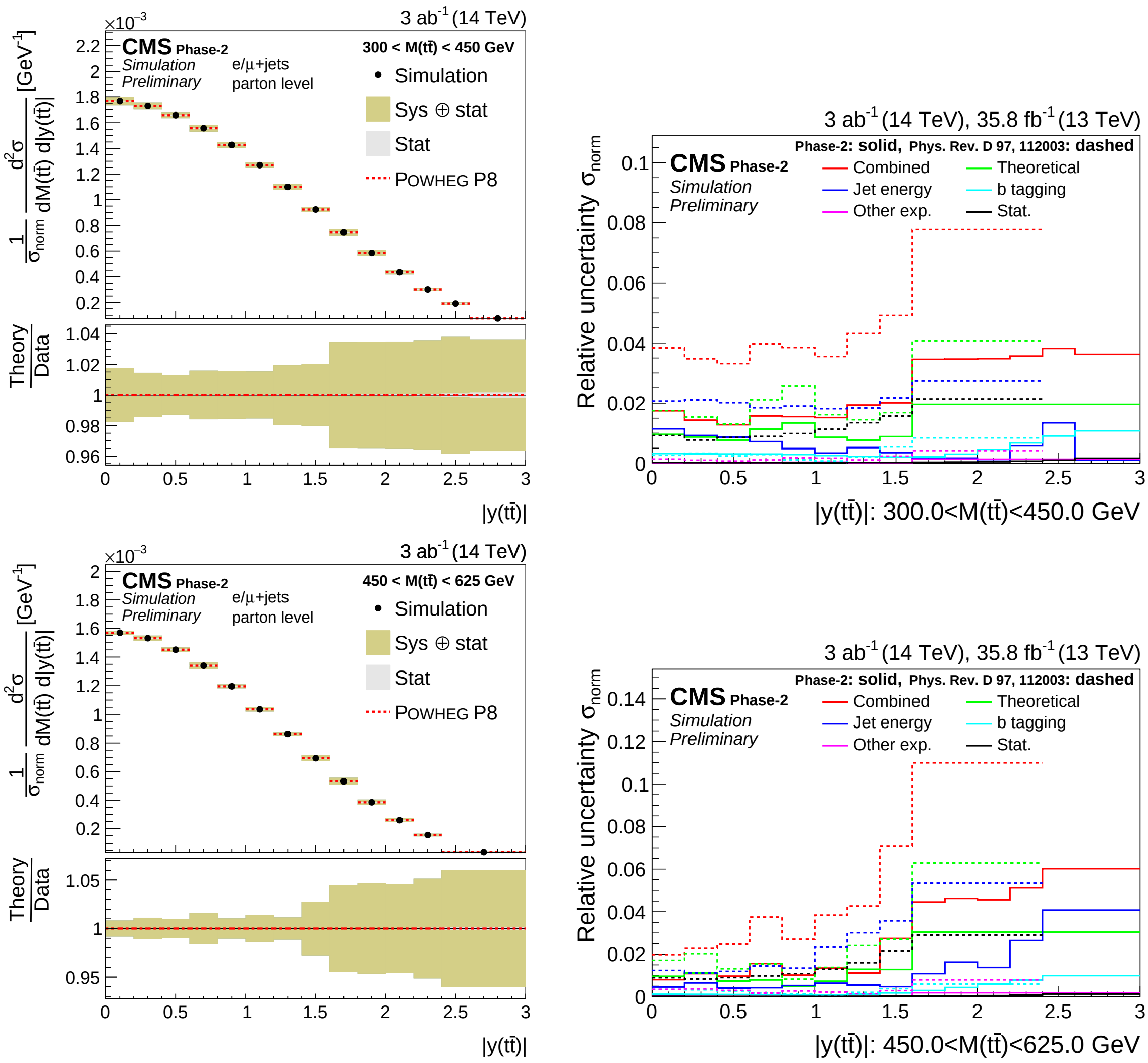

Figure 8: Projections of the normalized double-differential cross section as a function of $M(\mathrm{t\bar{t}})$ *vs.* $|y(\mathrm{t\bar{t}})|$ (left). The corresponding relative uncertainties (right) in the Phase-2 projections are compared to the uncertainties in the 2016 measurements [2].

uncertainties. On the experimental side, the current uncertainties of about 5–8% are expected to be further reduced by a few percent mainly because of the improved jet energy calibration and b jet identification. In low populated regions of the phase space, e.g., at high rapidity and $M(\mathrm{t\bar{t}})$, a significant reduction of the overall uncertainty can be achieved due to the decreasing statistical uncertainties with the large amount of data. This reduction of the statistical uncertainty also allows for more precise measurements in an increased number of bins as demonstrated in the projection of the double-differential cross section. The extended $\eta$-coverage of the phase-2 detector enables measurements at high rapidity that are not possible with the current detector.

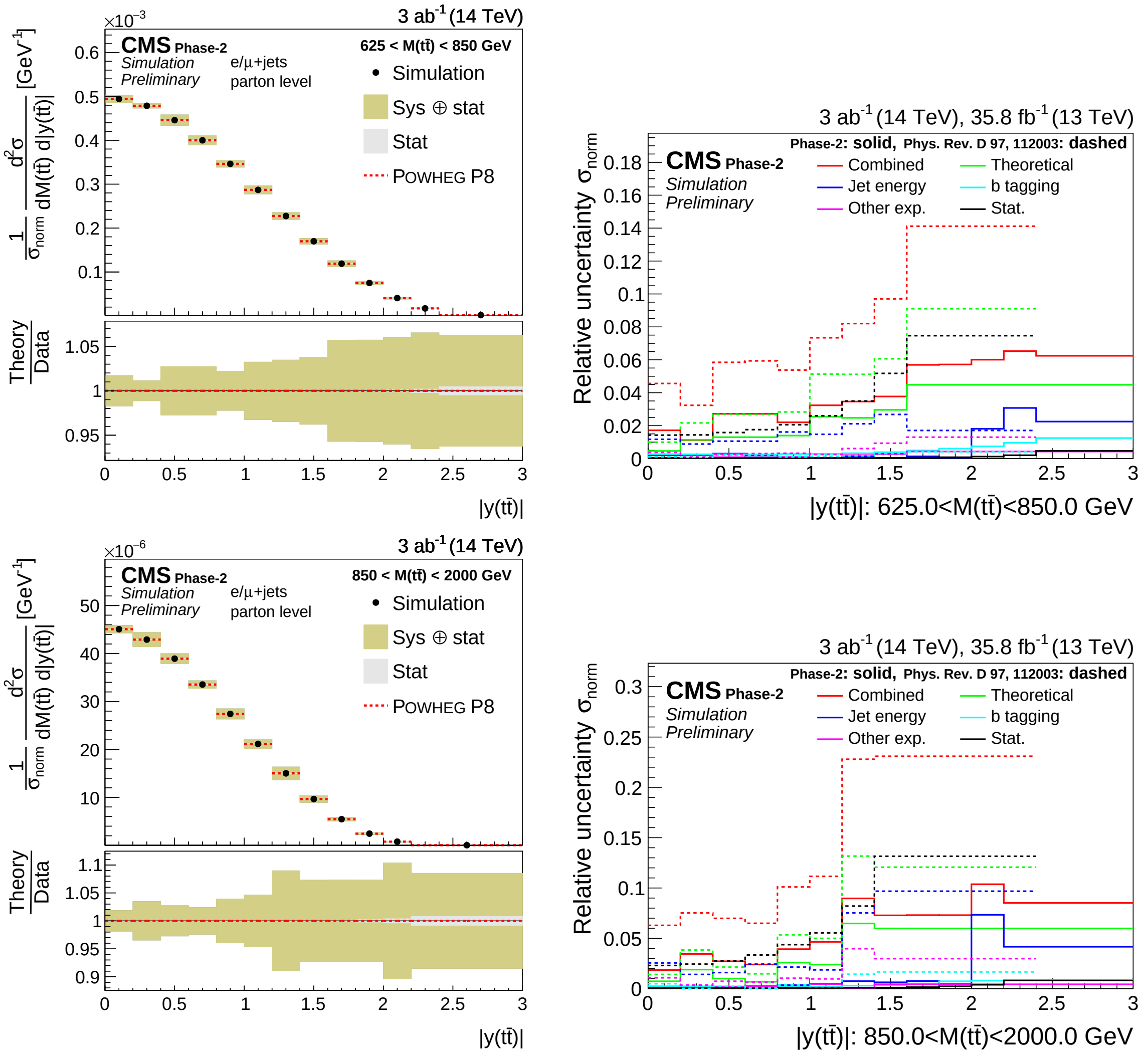

Figure 9: Projections of the normalized double-differential cross section as a function of $M(t\bar{t})$ *vs.* $|y(t\bar{t})|$ (left). The corresponding relative uncertainties (right) in the Phase-2 projections are compared to the uncertainties in the 2016 measurements [2].

## 7 PDF constraints from double-differential cross sections

The impact of differential $\mathrm{t\bar{t}}$ cross section measurements at the HL-LHC on the proton PDFs is quantitatively estimated using a profiling technique [28]. This technique is based on minimizing $\chi^2$ between data and theoretical predictions taking into account both experimental and theoretical uncertainties arising from PDF variations. Three NLO PDF sets were chosen for this study: ABMP16 [29], CT14 [30] and NNPDF3.1 [31] available via the LHAPDF interface (version 6.1.5) [32]. All PDF sets are provided with uncertainties in the form of eigenvectors. No $\mathrm{t\bar{t}}$ data were used to determine the CT14 PDF set, only total $\mathrm{t\bar{t}}$ production cross section measurements were used to determine the ABMP16 PDFs, and both total and differential (from Run-I LHC) $\mathrm{t\bar{t}}$ cross sections were used in the NNPDF3.1 extraction. The PDF uncertainties of CT14, evaluated at 90% CL, are rescaled to 68% CL.

For this study, the normalized double-differential $\mathrm{t\bar{t}}$ production cross sections as a function of $M(\mathrm{t\bar{t}})$ *vs.* $|y(\mathrm{t\bar{t}})|$ are used, which are expected to impose stringent constraints on the gluon distribution [33]. The typical $x$ values probed can be estimated using the LO kinematic relation $x = (M(\mathrm{t\bar{t}})/\sqrt{s}) \exp\left[\pm y(\mathrm{t\bar{t}})\right]$. Hence the $\mathrm{t\bar{t}}$ measurements are expected to be sensitive to $x$ values in the region $0.002 \lesssim x \lesssim 0.5$, as estimated using the highest or lowest $|y(\mathrm{t\bar{t}})|$ or $M(\mathrm{t\bar{t}})$ bins and taking the low or high bin edge where the cross section is largest.

The study is performed using the xFITTER program (version 2.0.0) [34], an open-source QCD fit framework for PDF determination. The theoretical predictions for the $\mathrm{t\bar{t}}$ cross sections are calculated at NLO QCD using the MG5_aMC@NLO (version 2.6.0) [35] framework, interfaced with the aMCFAST (version 1.3.0) [36] and APPLGRID (version 1.4.70) [37] programs. The number of active flavors is set to $n_f = 5$, the top quark pole mass $m_\mathrm{t} = 172.5\,\mathrm{GeV}$ is used and the strong coupling strength is set to $\alpha_s(M_Z) = 0.118$. The renormalization and factorization scales are chosen to be $\mu_\mathrm{r} = \mu_\mathrm{f} = H'/2, H' = \sum_i m_{\mathrm{T},i}$, where the sum runs over all final-state partons (t, $\mathrm{\bar{t}}$ and up to one light parton) and $m_\mathrm{T}$ is transverse mass $m_\mathrm{T} = \sqrt{m^2 + p_\mathrm{T}^2}$.

The $\chi^2$ value is calculated using the full covariance matrix representing the statistical and systematic uncertainties of the data. The PDF uncertainties are treated through nuisance parameters. For each nuisance parameter a penalty term is added to the $\chi^2$, representing the prior knowledge of the parameter. The values of these nuisance parameters at the minimum are interpreted as optimized, or profiled, PDFs, while their uncertainties determined using the tolerance criterion of $\Delta\chi^2 = 1$ correspond to the new PDF uncertainties. The profiling approach assumes that the new data are compatible with theoretical predictions using the existing PDFs, such that no modification of the PDF fitting procedure is needed. Under this assumption, the central values of the measured cross sections are set to the central values of the theoretical predictions.

The original and profiled ABMP16, CT14, and NNPDF3.1 PDF uncertainties are shown in Fig. 10–12, respectively. The uncertainties of the gluon, valence quark, and sea quark distributions are presented at the scale $\mu_\mathrm{f}^2 = 30\,000\,\mathrm{GeV}^2 \simeq m_\mathrm{t}^2$ relevant for $\mathrm{t\bar{t}}$ production. A consistent impact of the $\mathrm{t\bar{t}}$ data on the PDFs is observed for all PDF sets. In particular, the uncertainties of the gluon distribution are drastically reduced once the $\mathrm{t\bar{t}}$ data are included in the fit. The improvement is about one order of magnitude at $x \approx 0.5$ which is the edge of kinematic reach of the $\mathrm{t\bar{t}}$ data. In this region the gluon distribution lacks direct constraints in the present PDF fits. A small improvement is also observed for the sea and valence quark distributions.

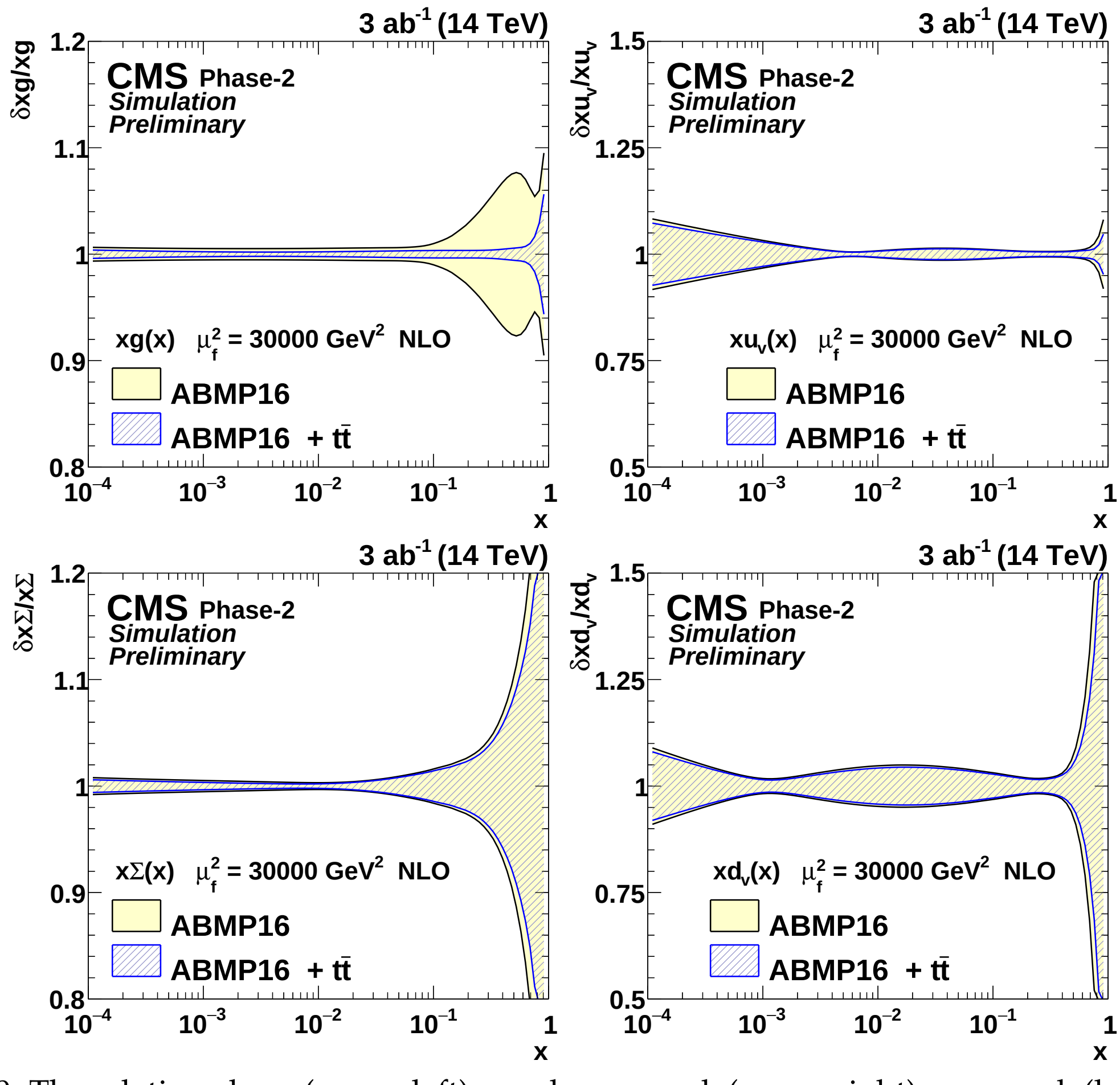

Figure 10: The relative gluon (upper left), u valence quark (upper right), sea quark (lower left), and d valence quark (lower right) uncertainties of the original and profiled ABMP16 PDF set.

**14**

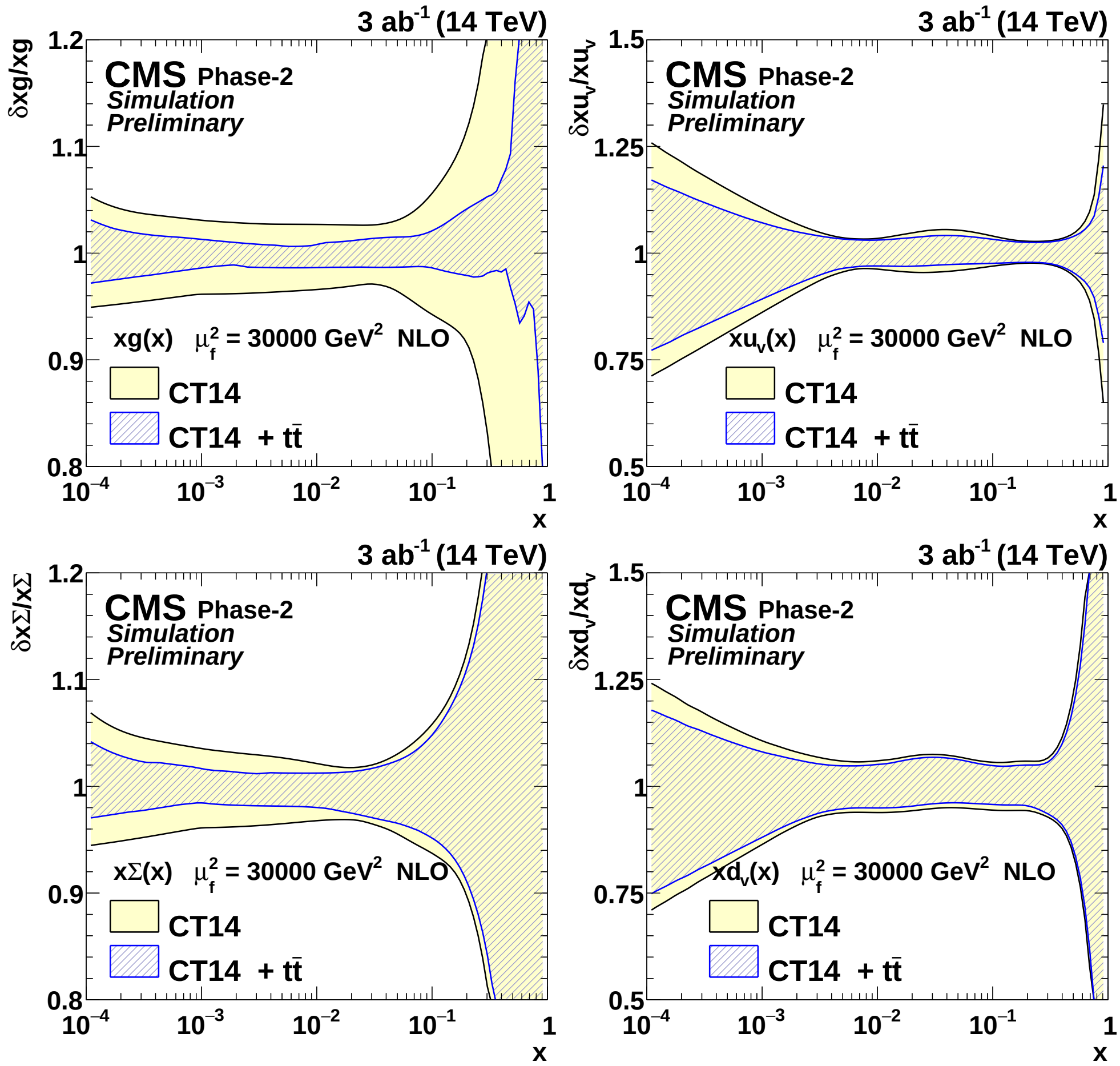

Figure 11: The relative gluon (upper left), u valence quark (upper right), sea quark (lower left), and d valence quark (lower right) uncertainties of the original and profiled CT14 PDF set.

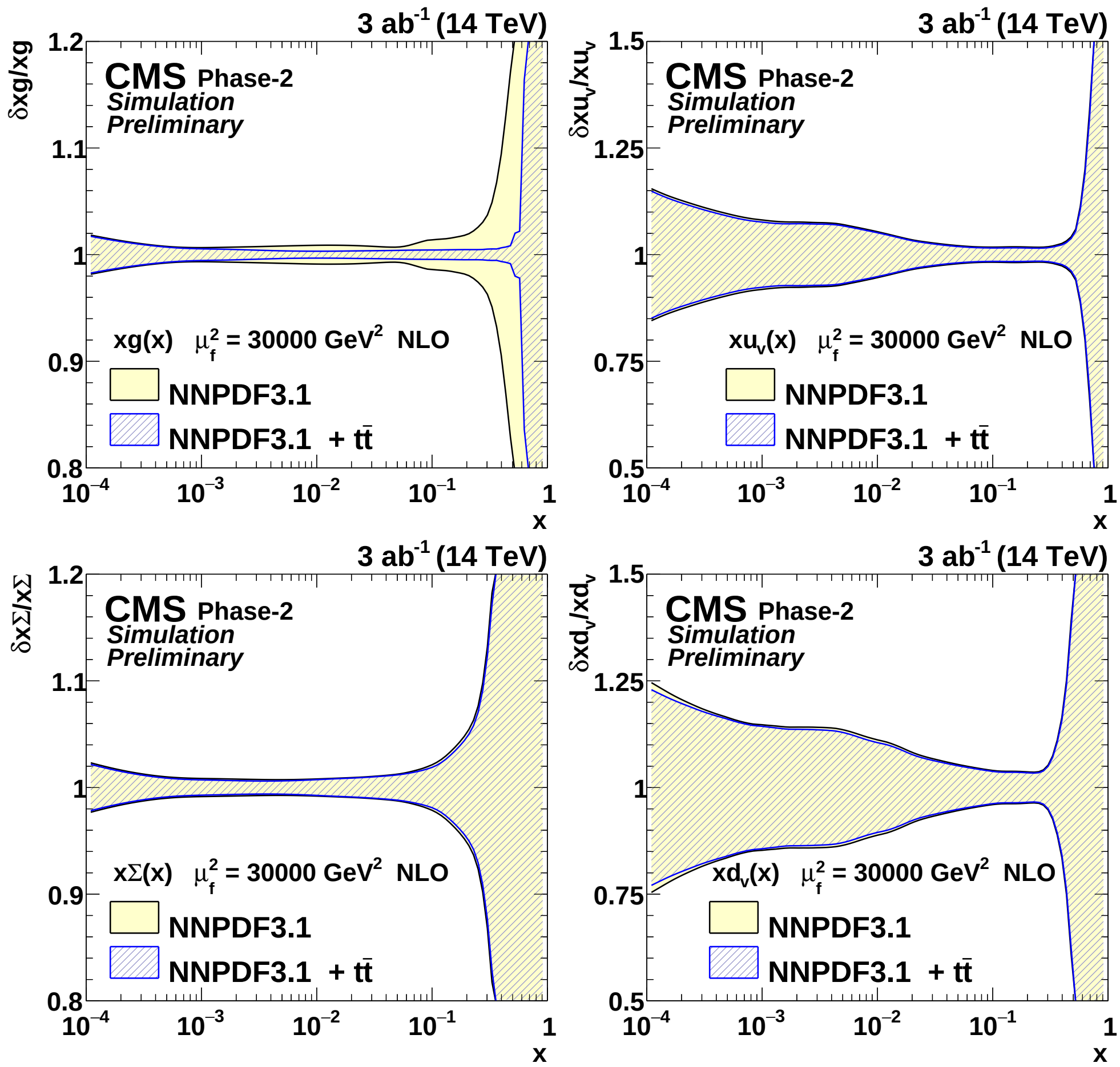

Figure 12: The relative gluon (upper left), u valence quark (upper right), sea quark (lower left), and d valence quark (lower right) uncertainties of the original and profiled NNPDF3.1 PDF set.

# 8 Summary

A projection of differential $t\bar{t}$ cross section measurements at the HL-LHC has been shown. Using pileup mitigation techniques like PUPPI these measurements become feasible in an environment of 200 pileup events. The high amount of data and the extended $\eta$-range of the Phase-2 detector allow for fine-binned measurements in phase-space regions — especially at high rapidity — that are not accessible in current measurements. The most significant reduction of uncertainty is expected because of an improved jet energy calibration and a reduced uncertainty in the b jet identification. It is demonstrated that the projected differential $t\bar{t}$ cross sections have a strong impact on the gluon distribution in the proton. Overall, this measurement will profit from both, the improved Phase-2 CMS detector and the high amount of data expected at the HL-LHC.

# A Reconstruction properties of various kinematic quantities

In Figs 13–17 the expected signal yields in the different categories, the migration matrices, and their properties are shown for various the kinematic quantities of the $t_\ell$ and the $t\bar{t}$ system.

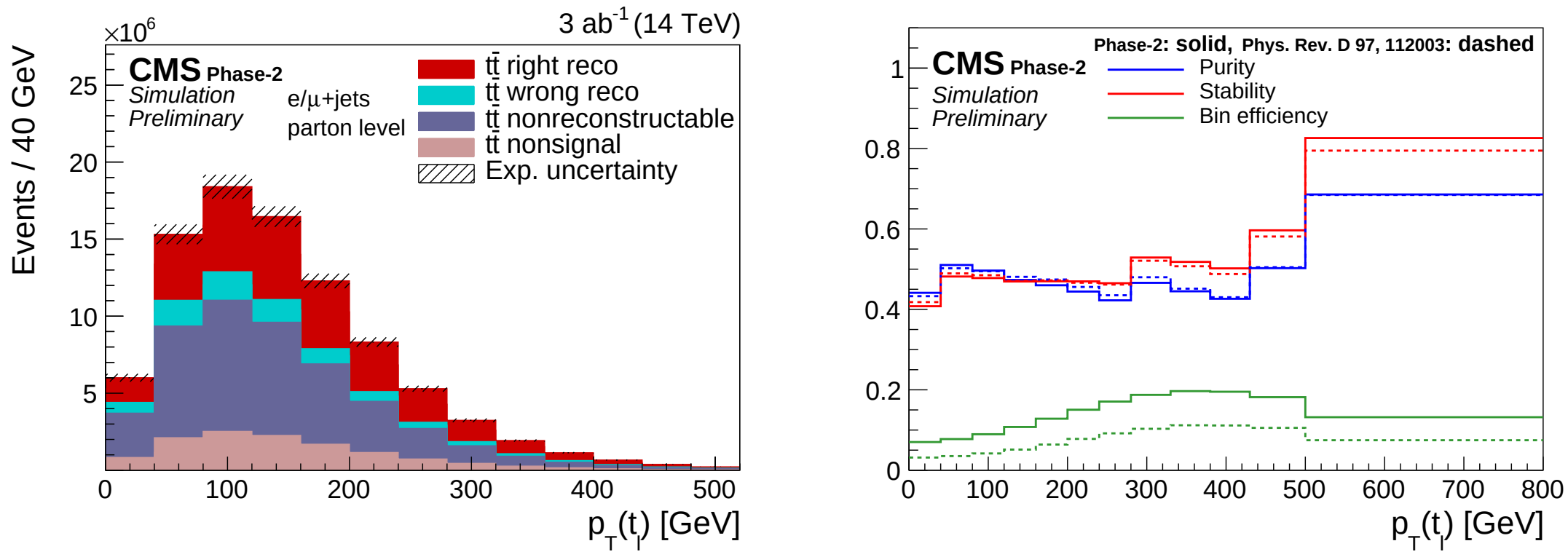

Figure 13: Expected event yields (left) and properties of the migration matrix (right) for the measurement of $p_{\mathrm{T}}(t_\ell)$. For comparison the properties are also shown for the 2016 analysis [2].

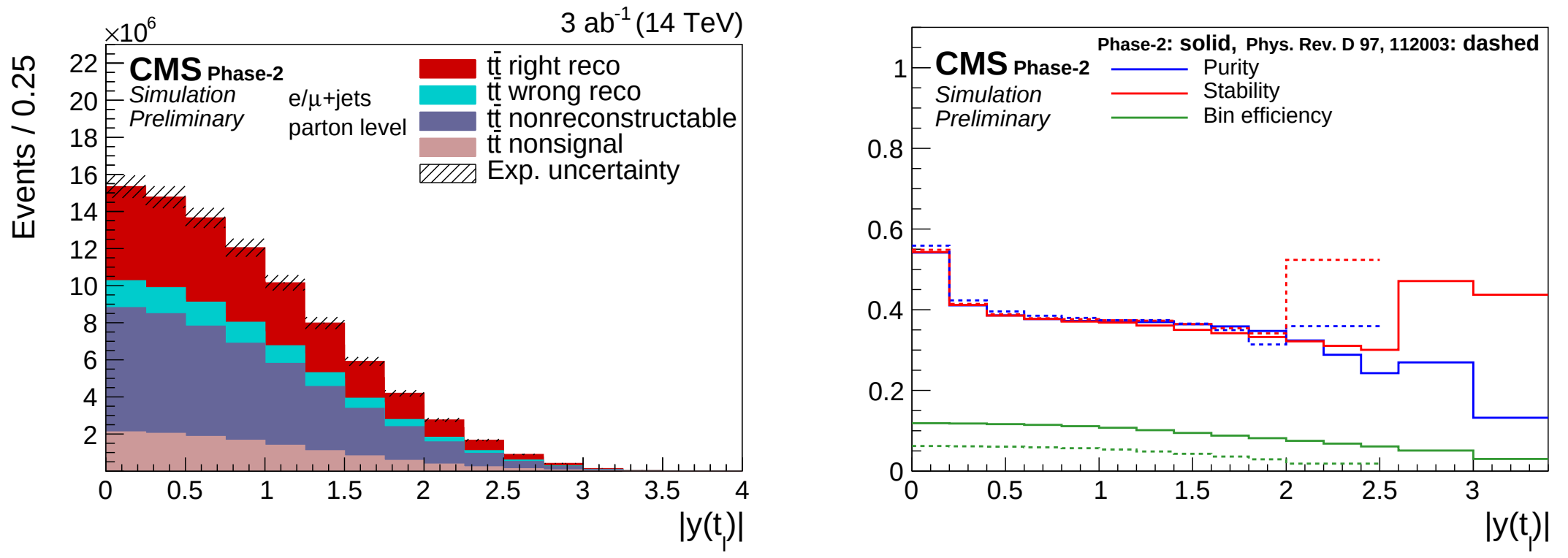

Figure 14: Expected event yields (left) and properties of the migration matrix (right) for the measurement of $|y(t_\ell)|$. For comparison the properties are also shown for the 2016 analysis [2].

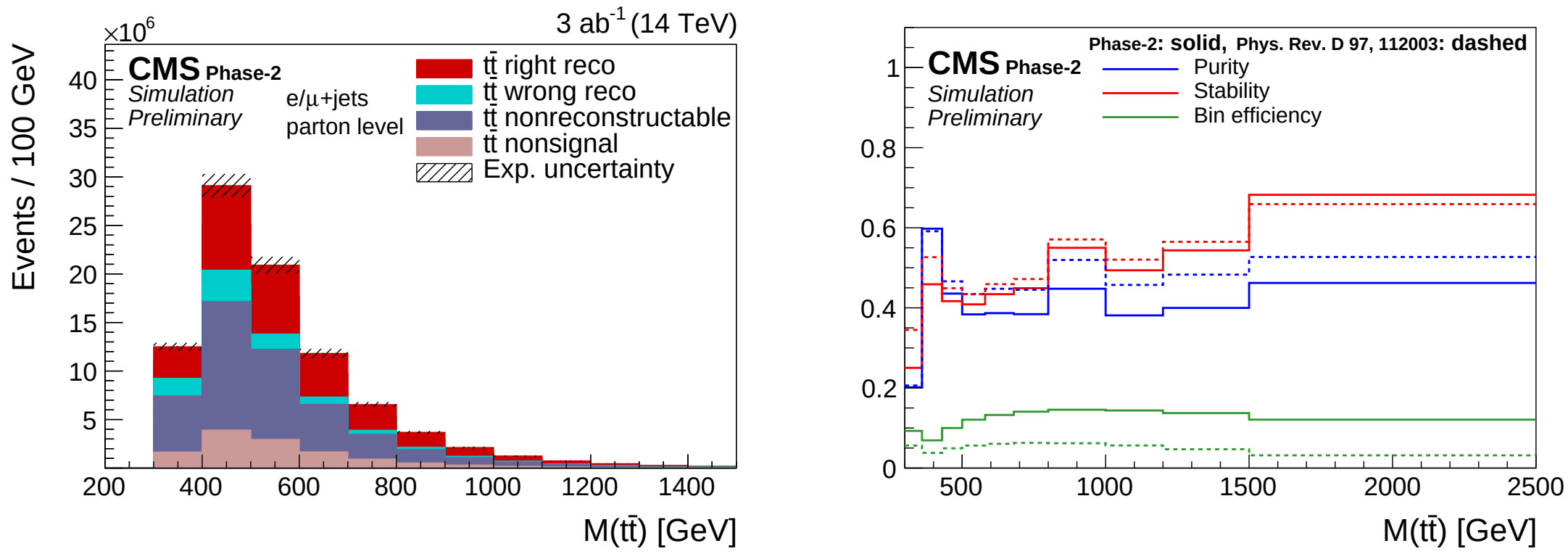

Figure 15: Expected event yields (left) and properties of the migration matrix (right) for the measurement of $M(\mathrm{t\bar{t}})$. For comparison the properties are also shown for the 2016 analysis [2].

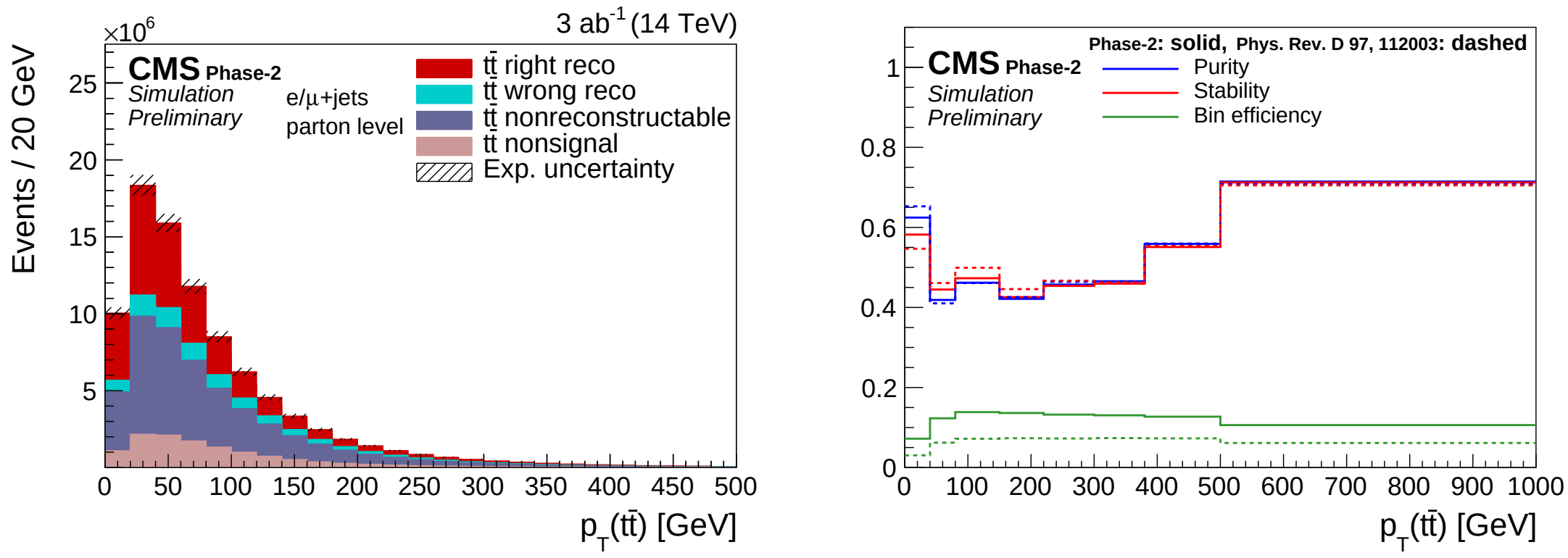

Figure 16: Expected event yields (left) and properties of the migration matrix (right) for the measurement of $p_\mathrm{T}(\mathrm{t\bar{t}})$. For comparison the properties are also shown for the 2016 analysis [2].

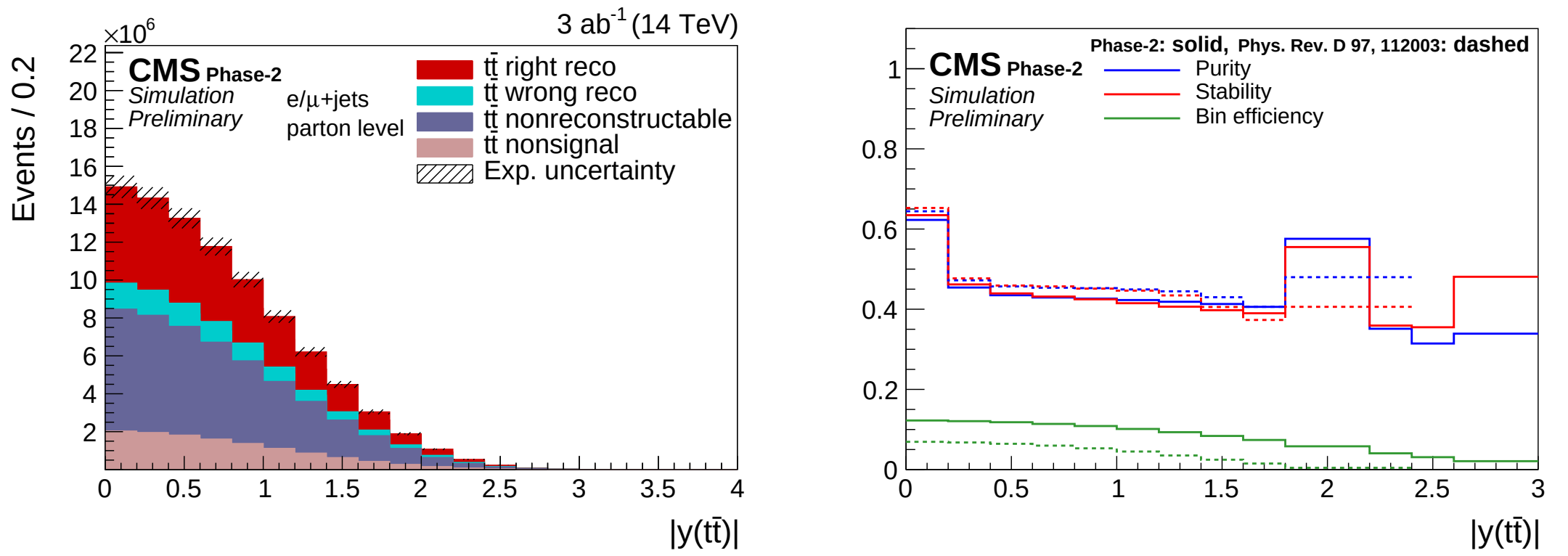

Figure 17: Expected event yields (left) and properties of the migration matrix (right) for the measurement of $|y(\mathrm{t\bar{t}})|$. For comparison the properties are also shown for the 2016 analysis [2].

# B Additional differential cross sections

In Figs. 18–20 the differential cross sections are shown normalized to unity within the measured range of each distribution. The absolute double differential cross section as a function of $M(\mathrm{t\bar{t}})$ *vs.* $|y(\mathrm{t\bar{t}})|$ is shown in Figs. 21 and 22.

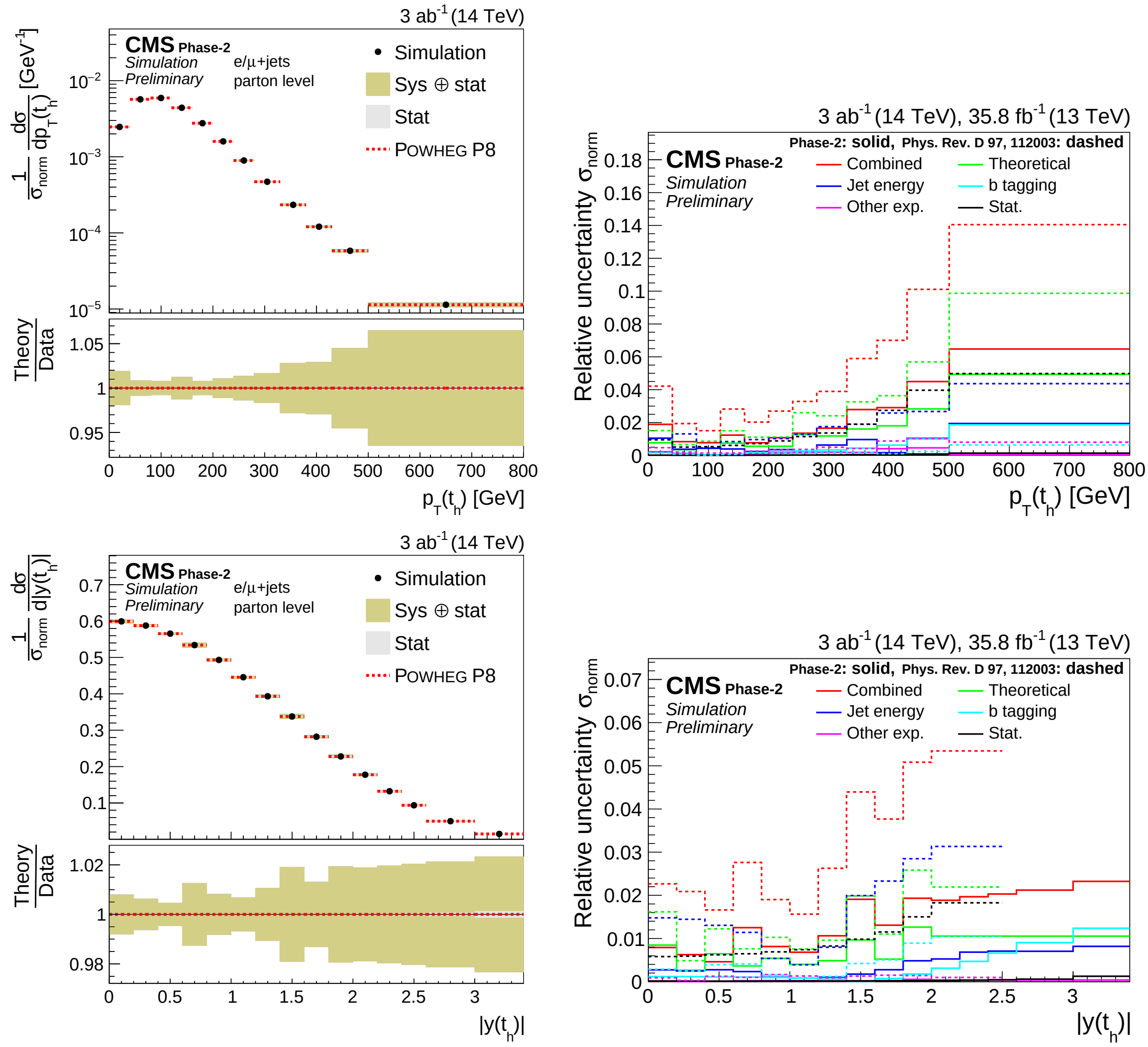

Figure 18: Normalized differential cross sections (left) as a function of $p_{\mathrm{T}}(\mathrm{t_h})$ (upper) and $|y(\mathrm{t_h})|$ (lower). The corresponding relative uncertainties (right) in the Phase-2 projections are compared to the uncertainties in the 2016 measurements [2].

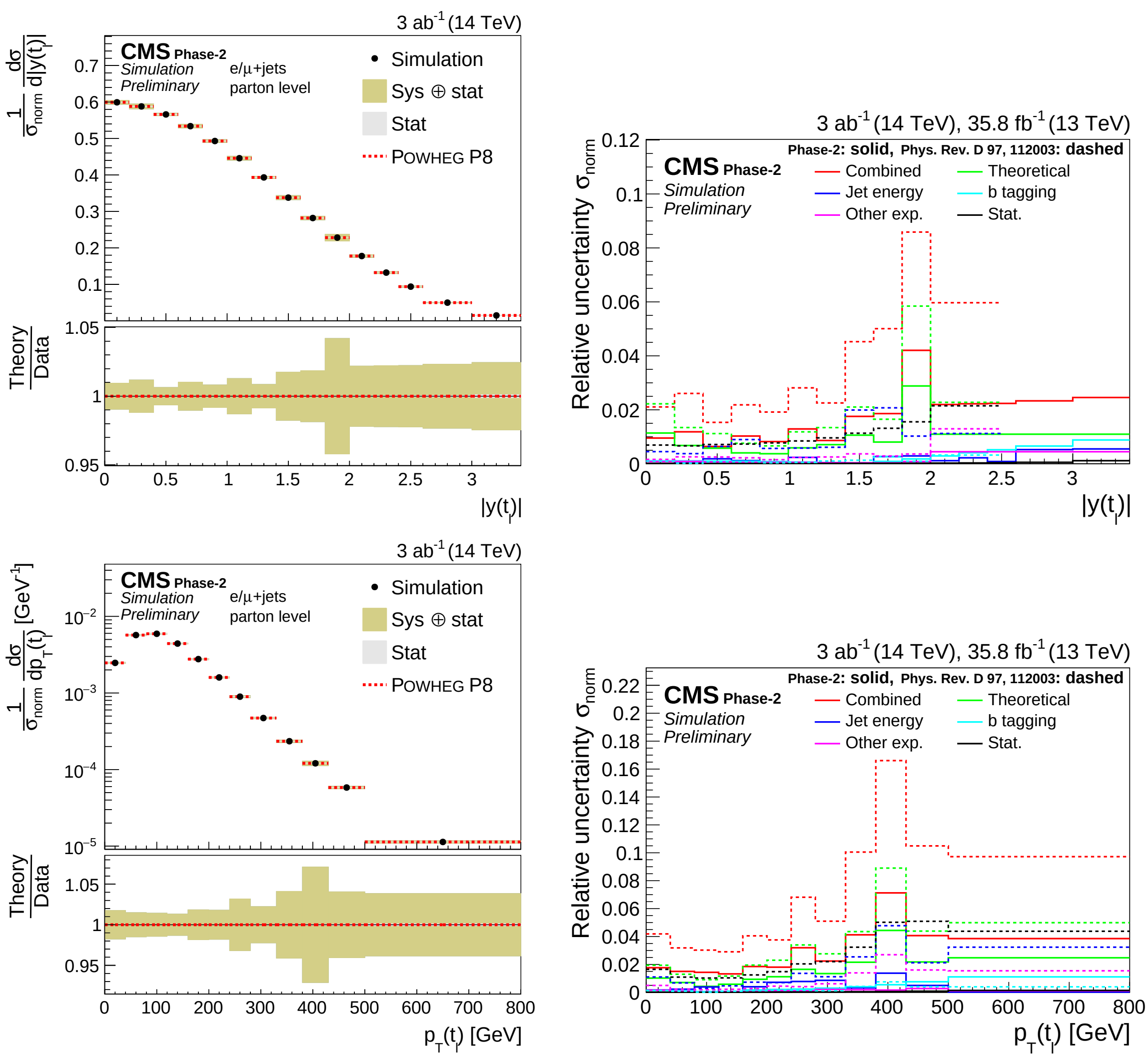

Figure 19: Normalized differential cross section as a function of $p_T(t_\ell)$ (upper) and $|y(t_\ell)|$ (lower). The corresponding relative uncertainties (right) in the Phase-2 projections are compared to the uncertainties in the 2016 measurements [2].

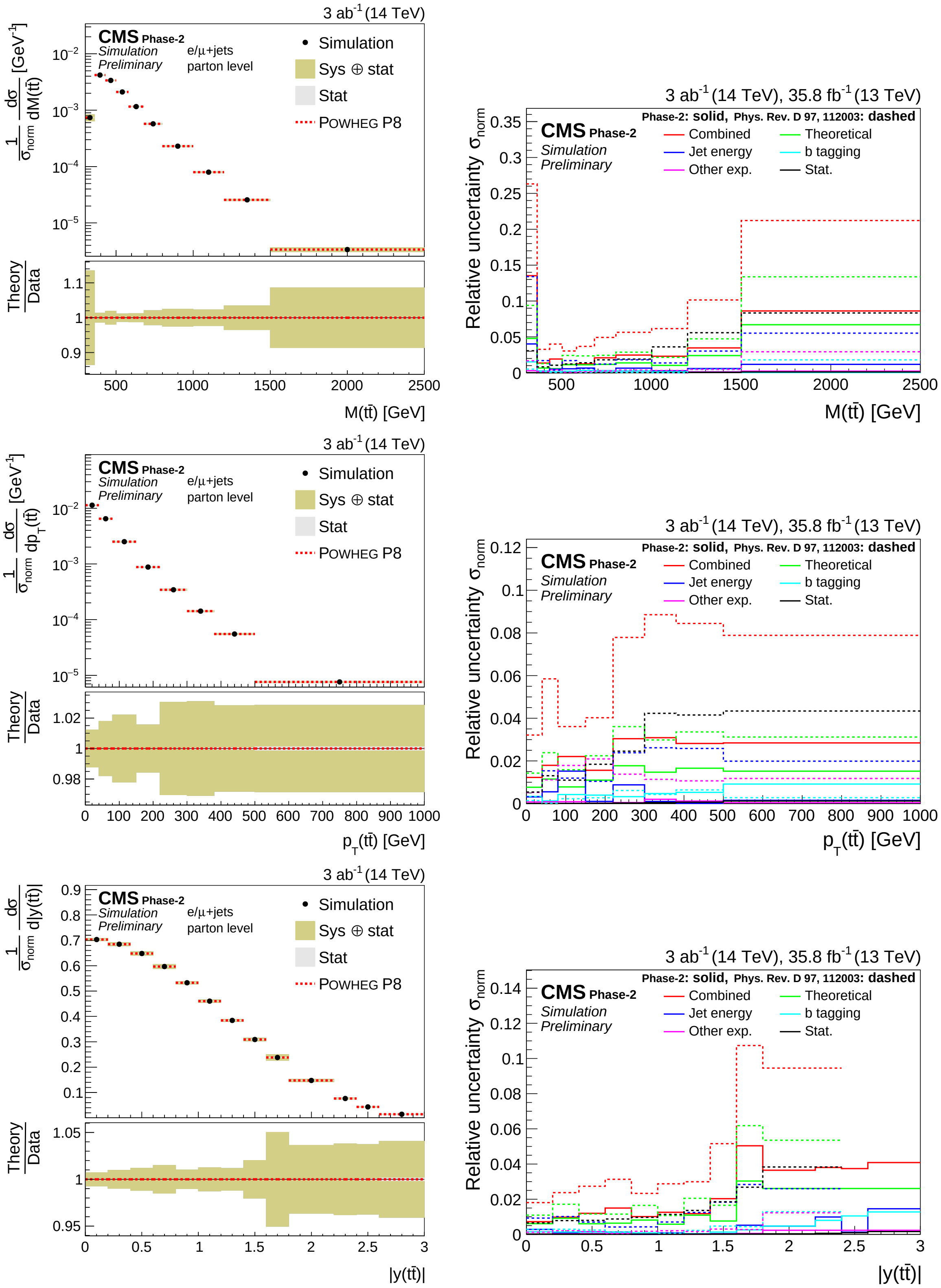

Figure 20: Normalized differential cross section as a function of $M(\mathrm{t\bar{t}})$ (upper), $p_{\mathrm{T}}(\mathrm{t\bar{t}})$ (middle), and $|y(\mathrm{t\bar{t}})|$ (lower). The corresponding relative uncertainties (right) in the Phase-2 projections are compared to the uncertainties in the 2016 measurements [2].

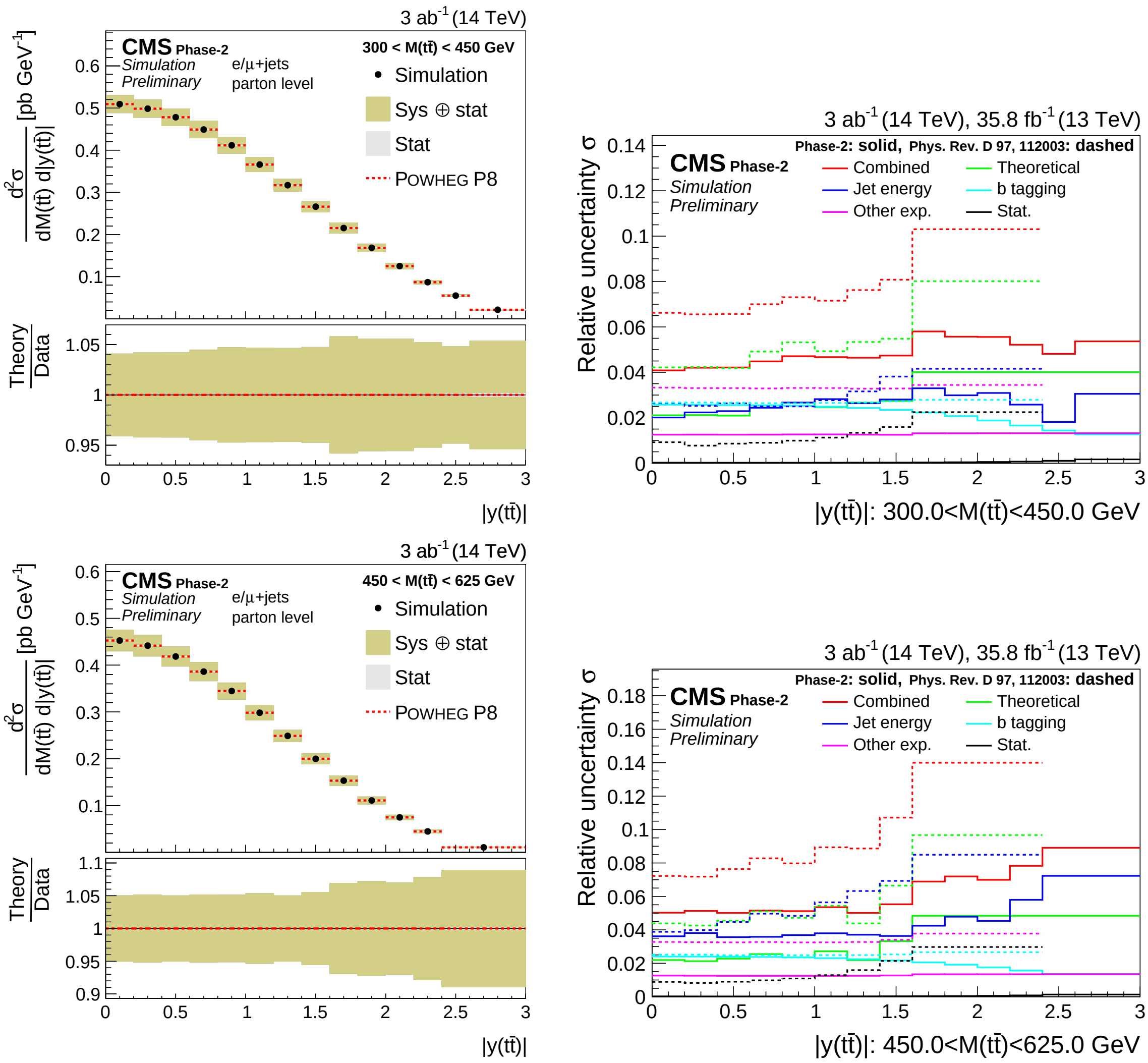

Figure 21: Projections of the double-differential cross section as a function of $M(\mathrm{t\bar{t}})$ *vs.* $|y(\mathrm{t\bar{t}})|$ (left). The corresponding relative uncertainties (right) in the Phase-2 projections are compared to the uncertainties in the 2016 measurements [2].

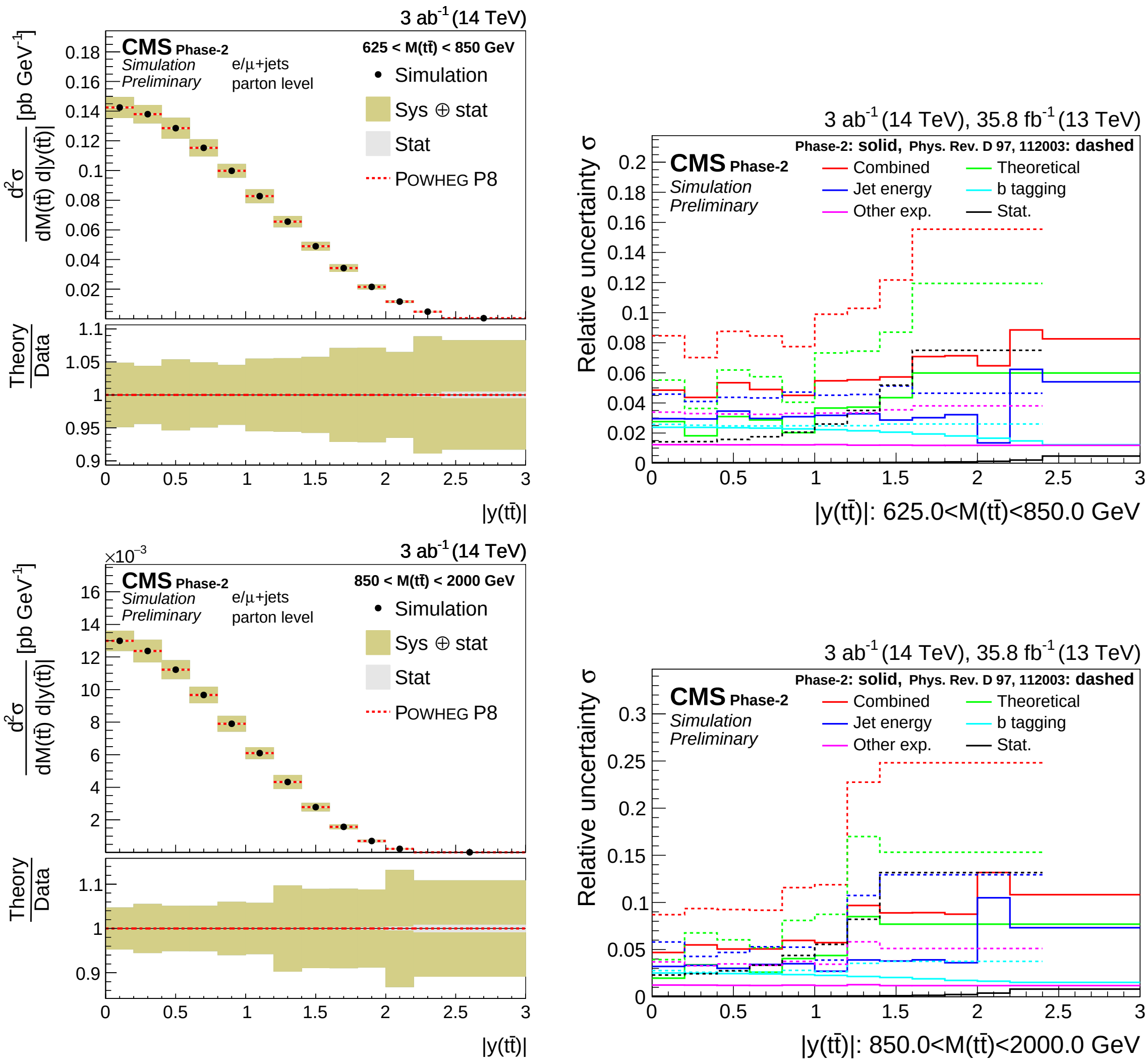

Figure 22: Projections of the double-differential cross section as a function of $M(\mathrm{t\bar{t}})$ *vs.* $|y(\mathrm{t\bar{t}})|$ (left). The corresponding relative uncertainties (right) in the Phase-2 projections are compared to the uncertainties in the 2016 measurements [2].

# ATLAS PUB Note

ATL-PHYS-PUB-2018-049

17th December 2018

# Prospects for the measurement of $t\bar{t}\gamma$ with the upgraded ATLAS detector at the High-Luminosity LHC

The ATLAS Collaboration

Measurements of $t\bar{t}\gamma$ production are studied in leptonic final states at the HL-LHC, where a data set with an integrated luminosity of 3 ab$^{-1}$ at a center-of-mass energy of 14 TeV is expected to be collected by the upgraded ATLAS detector. The expected precisions for the measurements of both fiducial and differential cross-sections are presented. The fiducial cross-section measurement can reach an uncertainty as low as 6% (3%) in the channel with one (two) lepton(s) and requiring a photon candidate with transverse momentum larger than 20 GeV. This uncertainty increases to 8% (12%) for photons with transverse momentum above 500 GeV. The uncertainty of differential cross-section measurements, normalised to unity, for several typical observables is in general below 5%.

# 1 Introduction

Measurements of top-quark properties play an important role in testing the Standard Model (SM) and its possible extensions. Studies of the production and kinematic properties of a top-quark pair ($t\bar{t}$) in association with a photon ($t\bar{t}\gamma$) probe the $t\gamma$ electroweak coupling. For instance, deviations in the transverse momentum ($p_{\mathrm{T}}$) spectrum of the photon from the SM prediction could point to new physics through anomalous dipole moments of the top quark [1–3]. A precision measurement of the $t\bar{t}\gamma$ production cross-section could effectively constrain some of the Wilson coefficients in top-quark effective field theories [4]. Furthermore, differential distributions of photon production in $t\bar{t}$ events can provide insight on the $t\bar{t}$ production mechanism, in particular about the $t\bar{t}$ spin correlation and the production charge asymmetry [5].

Evidence for the production of $t\bar{t}$ in association with an energetic, isolated photon was found in proton-antiproton ($p\bar{p}$) collisions at the Tevatron collider at a centre-of-mass energy of $\sqrt{s}$ = 1.96 TeV by the CDF Collaboration [6]. Observation of the $t\bar{t}\gamma$ process was reported by the ATLAS Collaboration in proton-proton ($pp$) collisions at $\sqrt{s}$ = 7 TeV [7]. Both the ATLAS and CMS Collaborations measured the $t\bar{t}\gamma$ cross-section at $\sqrt{s}$ = 8 TeV [8, 9]. In the ATLAS measurement, the differential cross-sections with respect to the transverse momentum $p_T$ and absolute pseudorapidity $|\eta|$ of the photon were reported. In the CMS measurement, the ratio of the $t\bar{t}\gamma$ fiducial cross-section to the $t\bar{t}$ total cross-section was measured. The ATLAS Collaboration also measured the $t\bar{t}\gamma$ cross-section at $\sqrt{s}$ = 13 TeV [10].

The study in this note is performed using the same framework and strategy as in the 13 TeV ATLAS measurement. This note investigates the precision of the $t\bar{t}\gamma$ measurement that can be achieved at the High-Luminosity LHC (HL-LHC) [11], using simulated data corresponding to an integrated luminosity of 3 ab$^{-1}$ at a centre-of-mass energy of 14 TeV, which is expected to be collected by the upgraded ATLAS detector during the full run of the HL-LHC. To cope with the harsh environment at the HL-LHC, the current ATLAS detector will be upgraded significantly [12, 13]: e.g. the inner tracker is being completely rebuilt and the TRT is removed in favor of an all-new all-silicon tracker.

The study is performed in final states with one or two leptons (electron or muon). The photon can originate not only from a top quark, but also from its charged decay products, including a charged fermion (quark or lepton) from the decay of the $W$-boson. In addition, it can be radiated from a charged incoming parton. In the study, no attempt is made to separate these different sources of photons, but an event selection is applied to suppress those radiated from top-quark decay products. The expected uncertainties are studied for the measurements of the fiducial cross-section and differential cross-sections, normalised to unity, for several typical observables.

# 2 Simulated event samples

Particle-level samples are generated at a centre-of-mass energy of 14 TeV without detector simulation. Leptons, photons, jets and missing transverse energy in the samples are smeared [14] to simulate the effect of object reconstruction and identification efficiencies as well as their momentum or energy resolutions in the upgraded ATLAS detector at the HL-LHC. Comparing with the current ATLAS detector, improvements are expected for the upgraded ATLAS detector: e.g. stronger fake electron suppression can be achieved for the same electron identification efficiency and the muon momentum resolution will be significantly improved. In the smearing, a pile-up condition corresponding to an average interaction per bunch-crossing of 200 is used.

The $t\bar{t}\gamma$ signal sample is simulated as a $2 \rightarrow 7$ process for the semileptonic and dileptonic decay channels of the $t\bar{t}$ system at leading order (LO) by MadGraph5_aMC@NLO v2.33 [15] (denoted as MG5_aMC in the following) interfaced with Pythia v8.212 [16], using the A14 tune [17] and the NNPDF2.3LO PDF set [18]. The photon can be radiated from an initial charged parton, an intermediate top quark, or any of the charged final state particles. The top-quark mass, top-quark decay width, $W$-boson decay width, and fine structure constant are set to 172.5 GeV, 1.320 GeV, 2.085 GeV, and 1/137, respectively. The five-flavour scheme is used where all the quark masses are set to zero, except for the top quark. Renormalisation and factorisation scales are set dynamically, corresponding to half the sum of transverse energies over all the particles generated from the matrix element, where the transverse energy of a particle of mass $m$ is defined as $E_T = \sqrt{m^2 + p_{\mathrm{T}}^2}$. The photon is requested to have $p_{\mathrm{T}} > 15$ GeV and $|\eta| < 5$. At least one lepton with $p_{\mathrm{T}} > 15$ GeV is required, with all the leptons satisfying $|\eta| < 5$. The $\Delta R$[1] between the photon and any of the charged particles among the seven final-state particles must be greater than 0.2. The resulting total cross-section of the process defined in this way is calculated to be 5.43 pb. Next-to-leading order (NLO) $k$-factors used in the 13 TeV $t\bar{t}\gamma$ measurement are applied to correct the fiducial cross-sections and acceptances to NLO.

The $t\bar{t}$ sample [19], where at least one of the top quarks decays leptonically, is generated with Powheg-Box v2 [20] using the NNPDF3.0NLO PDF set [21], and interfaced with Pythia v8.210 using the A14 tune and the NNPDF2.3LO PDF set. The $h_{\mathrm{damp}}$ parameter, which controls the $p_{\mathrm{T}}$ of the first additional emission beyond Born level in Powheg-Box, is set to 1.5 times the top-quark mass. The $t\bar{t}$ sample is normalised to the next-to-next-to-leading-order cross-section plus next-to-next-to-leading-logarithm corrections (NNLO+NNLL) [22], where the NNLL corrections correspond to resummation of soft gluon contributions. The $tW$ sample [19] is produced with Powheg-Box v1 using the CT10NLO PDF set [23], interfaced with Pythia v6.428 using the Perugia 2012 tune [24] and the CTEQ6L1 PDF set [18]. The $t$-channel single top sample is produced with Powheg-Box using the NNPDF2.3NLO PDF set, interfaced with Pythia v8.210 using the A14 tune and the NNPDF2.3LO PDF set. The production of $W$+jets is simulated with MG5_aMC v2.3.2 using the NNPDF3.0NLO PDF set, interfaced with Pythia v8.210 using the A14 tune and the NNPDF2.3LO PDF set. The production of $Z$+jets is simulated with Powheg-Box, interfaced with Pythia v8.210 using the AZNLO tune [25] and the CTEQ6L1 PDF set. The diboson ($WW$, $WZ$ and $ZZ$) samples [26, 27] are generated using Sherpa v2.2, using the NNPDF3.0NNLO PDF set. Different parton multiplicities are used for different production mechanisms of the diboson samples and with different precisions. For all these samples, the photon radiation is handled by the corresponding parton shower. The EvtGen program [28] is used to simulate the decay of bottom and charm hadrons, except for the Sherpa samples. A summary of all the simulation samples generated for this study is given in Table 1. In the table, the cross-sections of each simulation sample are given, for some of which higher order $k$-factors are applied on top of the cross-sections predicted by the generators.

The $t\bar{t}$ sample also contains $t\bar{t}\gamma$ events as the parton shower will add photon radiation to the $t\bar{t}$ events. To avoid the overlap between the $t\bar{t}$ and the $t\bar{t}\gamma$ samples, events in the $t\bar{t}$ sample are removed if they have a photon passing the prompt photon selection as defined in Section 3.

[1] ATLAS uses a right-handed coordinate system with its origin at the nominal interaction point (IP) in the centre of the detector and the $z$-axis along the beam pipe. The $x$-axis points from the IP to the centre of the LHC ring, and the $y$-axis points upward. Cylindrical coordinates $(r, \phi)$ are used in the transverse plane, $\phi$ being the azimuthal angle around the $z$-axis. The pseudorapidity is defined in terms of the polar angle $\theta$ as $\eta = -\ln\tan(\theta/2)$. The $\Delta R$ between two objects is defined as $\Delta R = \sqrt{\Delta\phi^2 + \Delta\eta^2}$.

Table 1: Summary of the simulation samples generated for this study. The $t\bar{t}\gamma$ sample has different $k$-factors applied to the single-lepton (1.30) and dilepton (1.44) fiducial regions.

| Sample | Cross-section | Generator | Parton shower | $k$-factor |
|---|---|---|---|---|
| $t\bar{t}\gamma$ | 5.43 pb (LO) | MG5_aMC | Pythia8 | NLO, 1.30 (1.44) |
| $t\bar{t}$ | 470 pb (NLO) | POWHEG | Pythia8 | NNLO+NNLL, 1.14 |
| $tW$ | 80.5 pb (NLO) | POWHEG | Pythia6 | - |
| $t$-channel single top | 67.5 pb (NLO) | POWHEG | Pythia8 | - |
| $W$+jets | 65200 pb (LO) | MC5_aMC | Pythia8 | - |
| $Z$+jets | 4120 pb (NLO) | POWHEG | Pythia8 | - |
| $WW/WZ/ZZ$ | 363 pb | Sherpa | | - |

The fake lepton background contribution is estimated from the data-driven background estimate in the 13 TeV analysis, scaled up by a factor of 83, the ratio of integrated luminosities between the 13 TeV (36 fb$^{-1}$) and the HL-LHC data samples. The extrapolation does not take into account the increase of the cross-section of the underlying physical processes which contribute to the fake-lepton background, due to the increase in centre-of-mass energy from 13 TeV to 14 TeV. However, the conservative systematics assigned to this background in Section 8 cover the possible difference. Moreover, this background is only a small background to the single-lepton channel.

The size of the simulated $Z\gamma$ sample (which is taken from the $Z$+jets sample by requiring the presence of a prompt photon) is insufficient to determine the background in the dilepton channel after the full event selection. As this background is expected to be dominant in the $ee$ and $\mu\mu$ channels, the estimate from the 13 TeV analysis is extrapolated to HL-LHC by scaling it up by a factor of 83, multiplied by a further factor of 1.08 to account for the increase in $Z\gamma$ cross section from 13 to 14 TeV. The 13 TeV estimate is based on a $Z\gamma$ MC sample, which is simulated with Sherpa v2.2.2. The possible change in shape of the distributions of the observables to be unfolded, when increasing the centre-of-mass energy from 13 TeV to 14 TeV, is checked using the $t\bar{t}\gamma$ samples and found to be negligible.

Due to the increased pile-up activity at the HL-LHC, more stringent isolation criteria are necessary to suppress jets being misidentified as photons and consequently the combined photon reconstruction and identification efficiency is expected to be smaller than in the 13 TeV analysis by around 30% for low-$p_T$ photons, while they are similar for high-$p_T$ photons [14]. This difference is taken into account in the extrapolation.

## 3 Object selection

Object and event selection closely follow Ref. [10].

Electron candidates are required to have a smeared $p_T > 25$ GeV and an absolute pseudorapidity $|\eta| < 2.47$, excluding the transition region between the barrel and endcap ($|\eta| \notin [1.37, 1.52]$). The parametrised "Medium" identification criteria is applied. Muon candidates are required to have a smeared $p_T > 25$ GeV and $|\eta| < 2.5$. The parametrised "Tight" identification criteria is applied. In order to apply the parameterisations, the leptons are required not to come from hadron decay. The potential increase in event yield by taking advantage of the improved $|\eta|$ acceptance of the upgraded ATLAS detector was studied, by accepting identified electrons and muons out to $|\eta| < 4$. This leads to a percent level gain in the signal yield, which is however accompanied by a similar increase in background.

Photon candidates must have a smeared $p_T > 20$ GeV and $|\eta_{\text{cluster}}| < 2.37$, excluding the transition region between the barrel and endcap. The "Tight" identification criteria is applied. The photon is required not to be from hadron decay. Both electrons and jets can be misidentified as photons, leading to electron-fake photons and hadronic-fake photons, respectively. The functions used to parametrise the fake rates are summarised in Table 2. Moreover, a set of $p_T$-dependent weights is applied to hadronic-fake photons to scale their contribution down to a similar level as in the 13 TeV analysis.

Jets are reconstructed using the anti-$k_t$ algorithm [29] with a distance parameter of $R = 0.4$ from stable final state particles after the parton shower. They are required to have a smeared $p_T > 25$ GeV and $|\eta| < 2.5$. Jets from pile-up are added randomly to the event with a certain probability. Jets containing $b$-hadrons ($b$-jets) are identified with a ghost-matching procedure [30] and are assigned $p_T$- and $\eta$-dependent weights to reproduce a 70% $b$-tagging efficiency. A simple overlap removal procedure is applied: jets which are within a $\Delta R < 0.4$ cone of a lepton or photon are removed.

The missing transverse momentum $E_T^{\text{miss}}$ is computed from all neutrinos.

Table 2 summarises the above object selection.

Table 2: Summary of the object selection.

| Object | Selection |
|---|---|
| Prompt photon | not from hadron decay<br>$p_T > 20$ GeV, $|\eta| < 2.37$, excluding transition region |
| Fake photon | hadronic fake ($j \to \gamma$) rate parametrised with Crystal-Ball or sigmoid (pile-up jet)<br>electron fake ($e \to \gamma$) rate 2% (5%) in barrel (endcap)<br>same $p_T/\eta$ as prompt photon |
| Prompt Electron | not from hadron decay<br>$p_T > 25$ GeV, $|\eta| < 2.47$, excluding transition region |
| Prompt Muon | not from hadron decay<br>$p_T > 25$ GeV, $|\eta| < 2.5$ |
| Jet | $p_T > 25$ GeV, $|\eta| < 2.5$<br>removed, if $\Delta R < 0.4$ wrt. lepton/photon |
| $b$-jet | 70% efficiency |
| $E_T^{\text{miss}}$ | non-interacting particles |

## 4 Event selection

Events are categorised into the $e$+jets or $\mu$+jets channel if their final state contains exactly one electron or one muon selected as above, and the two channels are referred to together as the single-lepton channel. Events containing exactly two electrons or two muons, or one electron and one muon, all of which must pass the above selection and be of opposite charge, are categorised into the $ee$ or $\mu\mu$ or $e\mu$ channel, and the three channels are referred to as the dilepton channel. The lepton $p_T$ thresholds of 25 GeV are high enough that the events can be efficiently triggered using single-lepton triggers.

The selected events must have at least four (two) jets in the single-lepton (dilepton) channel, at least one of which is $b$-tagged, and exactly one photon. A $Z$-boson veto is applied in the single electron channel by excluding events with invariant mass of the system of the electron and the photon around the $Z$-boson mass, i.e. by requiring $|m(e, \gamma) - m(Z)| > 5$ GeV, where $m(Z) = 91.188$ GeV. In the $ee$ and $\mu\mu$ same-flavour dilepton channels, events are excluded if the dilepton invariant mass or the invariant mass of the system of the two leptons and the photon is between 85 and 95 GeV, and $E_{\mathrm{T}}^{\mathrm{miss}}$ is required to be larger than 30 GeV. The dilepton invariant mass is required to be higher than 15 GeV to suppress low-mass Drell-Yan events. It is experimentally difficult to separate $t\bar{t}\gamma$ events where the photon is radiated from a top quark (i.e. those sensitive to the top-photon coupling) from other sources of photons in a $t\bar{t}\gamma$ event. But it is possible to suppress photons radiated from particles other than top quarks. To suppress photons radiated from the lepton(s) of top quark leptonic decay(s), the $\Delta R$ between the selected photon and lepton(s) must be greater than 1.0. This cut could be tightened in the HL-LHC analysis to increase its suppression power while still retaining a reasonable number of signal events. The event selection is summarised in Table 3.

Table 3: Summary of event selection. "OS" means the charges of the two leptons must have opposite signs.

| Channel | $e$+jets | $\mu$+jets | $ee$ | $\mu\mu$ | $e\mu$ |
|---|---|---|---|---|---|
| n($\ell$) | 1 $e$ | 1 $\mu$ | 2 $e$, OS | 2 $\mu$, OS | 1 $e$ + 1 $\mu$, OS |
| | | | m($\ell, \ell$) > 15 GeV | | |
| n($\gamma$) | 1 $\gamma$ | | | | |
| n(jet) | $\geq$ 4 jets | | $\geq$ 2 jets | | |
| n($b$-jet) | $\geq$ 1 $b$-jet | | | | |
| m(e,$\gamma$) veto | $\|m(e, \gamma)$ - $m(Z)\|$ > 5 GeV | - | | | |
| m($\ell, \ell$) veto | - | | $m(\ell, \ell) \notin$ [85,95] GeV | | - |
| m($\ell, \ell, \gamma$) veto | - | | $m(\ell, \ell, \gamma) \notin$ [85,95] GeV | | - |
| $E_{\mathrm{T}}^{\mathrm{miss}}$ | - | | $E_{\mathrm{T}}^{\mathrm{miss}} > 30$ GeV | | - |
| $\Delta R(\gamma, \ell)$ | $\Delta R(\gamma, \ell) > 1.0$ | | | | |

After the event selection, there are four types of backgrounds to the selected $t\bar{t}\gamma$ candidates, three of which are events with a misidentified object. Events with the selected photon being a misidentified jet or a non-prompt photon from hadron decays are referred to as hadronic-fake background and events with the selected photon being a misidentified electron are referred to as electron-fake background. Events with the selected lepton being a misidentified jet or non-prompt lepton from heavy-flavour decays are referred to as fake-lepton background. Finally, events with a prompt photon (excluding the $t\bar{t}\gamma$ signal) are referred to as prompt-photon background. Contributions from electron-fake and fake-lepton backgrounds to the dilepton channel were found to be very small in the 13 TeV analysis and are neglected here.

## 5 Event yields and distributions

The expected event yields of signal and backgrounds after event selection in each channel are summarised in Table 4. Statistical uncertainties due to the size of the simulation samples are shown. The differences between the $e$+jets and $\mu$+jets channels and between the $ee$ and $\mu\mu$ channels are due to the different reconstruction and identification efficiencies for electrons and muons.

The photon $p_{\mathrm{T}}$ distributions after event selection are shown in Figure 1 for all the channels. The error band represents the total statistical uncertainty due to the limited size of the samples.

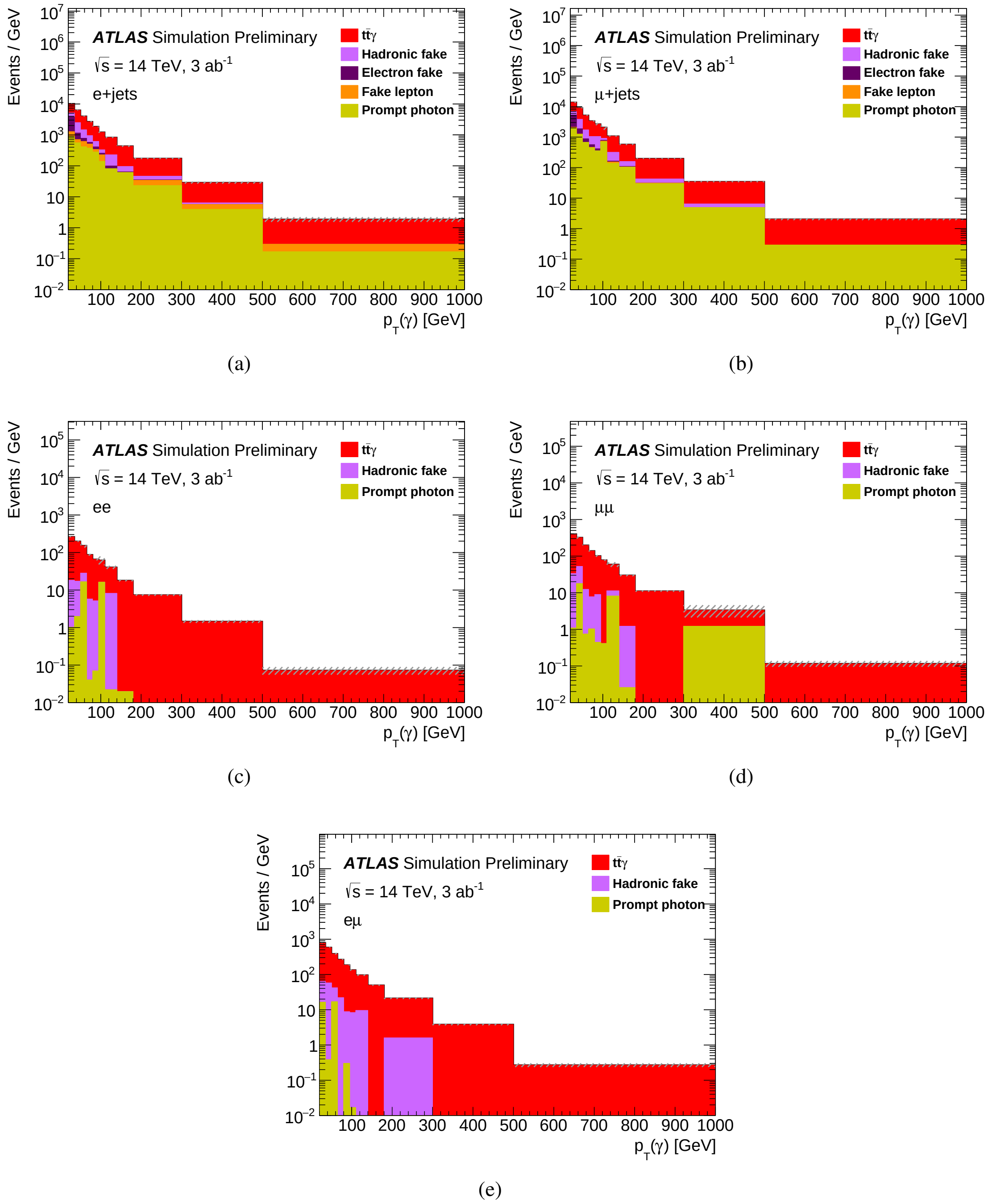

Figure 1: Distributions of the photon $p_{\mathrm{T}}$ in the five channels. The error band represents the total statistical uncertainty due to the limited size of the samples.

Table 4: Event yields of signal and background processes after the event selection. Statistical uncertainties due to the size of the samples are shown.

| Process | Signal | Hadronic fake | Electron fake | Prompt photon | Fake lepton |
|---|---|---|---|---|---|
| $e$+jets | 281100 ± 690 | 66500 ± 2000 | 50500 ± 2000 | 66300 ± 4200 | 17200 ± 2400 |
| $\mu$+jets | 376530 ± 800 | 91300 ± 4200 | 65000 ± 2300 | 104000 ± 11000 | 3300 ± 1400 |
| $ee$ | 13950 ± 160 | 1070 ± 160 | - | 2090 ± 430 | - |
| $e\mu$ | 39960 ± 270 | 3010 ± 240 | - | 530 ± 340 | - |
| $\mu\mu$ | 21240 ± 200 | 1550 ± 160 | - | 4700 ± 1100 | - |

# 6 Fiducial region

The fiducial region of the analysis is defined at particle level in a way that mimics the event selection in Section 4. Leptons (electron or muon, including those from $\tau$ decay) must have $p_{\mathrm{T}} > 25$ GeV and $|\eta| < 2.5$, and must not originate from hadron decays. Photons not from hadron decays and in a cone of radius R=0.1 around a lepton are added to the lepton before the lepton selection. Photons are required to have $p_{\mathrm{T}} > 20$ GeV and $|\eta| < 2.37$, and must not originate from hadron decays or be within $\Delta R$=0.1 of a lepton. The photon isolation computed from the ratio of the scalar sum of all charged stable particles' $p_{\mathrm{T}}$ around the photon over its transverse momentum must be smaller than 0.1. Jets are clustered using the anti-$k_t$ algorithm with $R = 0.4$, using all final state particles excluding non-interacting particles and muons that are not from hadron decay. Jets must have $p_{\mathrm{T}} > 25$ GeV and $|\eta| < 2.5$. A ghost matching method is used to determine the flavour of the jets, with those matched to $b$-hadrons tagged as $b$-jets. A simple overlap removal is performed: jets within $\Delta R < 0.4$ of a selected lepton or photon are removed. For events in the single-lepton (dilepton) channel, exactly one photon and exactly one lepton (two leptons) are required. At least four (two) jets are required with at least one of them being $b$-tagged. Events are rejected if there is any lepton and photon pair satisfying $\Delta R(\gamma, \ell) < 1.0$.

# 7 Normalised differential cross-section

In addition to the measurement of the absolute cross-section in the fiducial region defined above, normalised differential cross-section measurements are studied in this note. These measurements focus on the shape of the observables, while the overall rate is given by the absolute fiducial cross-section measurement.

The differential cross-section is given by

$$\frac{d\sigma}{dX_k} = \frac{1}{L_{\mathrm{int}} \cdot \Delta X_k} \cdot \frac{1}{\epsilon_k} \cdot \sum_j M_{jk}^{-1} \cdot (N_j^{\mathrm{obs}} - N_j^b) \cdot (1 - f_{\mathrm{out},j}) \,. \quad (1)$$

The indices $j$ and $k$ indicate the bin of the observable at detector and particle levels, respectively. The $X_k$ and $\Delta X_k$ are the observable and bin width of bin $k$. The $L_{\mathrm{int}}$ is the integrated luminosity. The $N_j^{\mathrm{obs}}$ and $N_j^b$ are the number of observed events and the number of estimated background events in bin $j$ at detector level, respectively. The efficiency $\epsilon_k$ is the fraction of signal events generated at particle level in bin $k$ of the fiducial region that are reconstructed and selected at detector level and have the objects, that are used to define the observable to be unfolded, matched between reconstruction and particle-levels with $\Delta R < 0.1$.

The migration matrix $M_{kj}$ expresses the probability for an event in bin $k$ at particle level to end up in bin $j$ at detector level, calculated from events passing both the fiducial region selection and the event selection, as well as the above matching procedure. The outside-migration fraction $f_{\text{out},j}$ is the fraction of signal events generated outside the fiducial region, but reconstructed and selected in bin $j$ at detector level or events failing the above matching. The normalised differential cross-section is calculated as

$$\frac{d\sigma^{\text{norm}}}{dX_k} = \frac{1}{\sum_k d\sigma/dX_k}\frac{d\sigma}{dX_k}, \tag{2}$$

where the sum is over all the bins of the observable.

The chosen observables to unfold are the photon $p_{\text{T}}$ and $|\eta|$, and the $\Delta R$ between the photon and the closest lepton for both single-lepton and dilepton channels, and the $\Delta\phi$ and $|\Delta\eta|$ between the two leptons for the dilepton channel. The kinematic properties of the photon are sensitive to the $t\gamma$ coupling, while the dilepton $\Delta\phi$ is sensitive to the $t\bar{t}$ spin correlation.

The signal sample is used to determine $\epsilon_k$, $f_{\text{out},j}$ and $M_{kj}$, which are shown in Figure 2 for the photon $p_{\text{T}}$ in the single-lepton channel. These efficiencies and migration matrices are quite similar to those in the 13 TeV analysis [31].

The inversion of the migration matrix $M_{kj}$ is performed using the iterative Bayesian method [32] implemented in the RooUnfold package [33]. The method relies on the Bayesian probability formula to invert the migration matrix, starting from a given prior of the particle-level distribution, and iteratively updating it with the posterior distribution. The binning choices for the unfolded distributions and the choice of three iterations for the unfolding are the same as in the 13 TeV analysis, except for the photon $p_{\text{T}}$, which has two additional bins of [300,500] GeV and [500,1000] GeV.

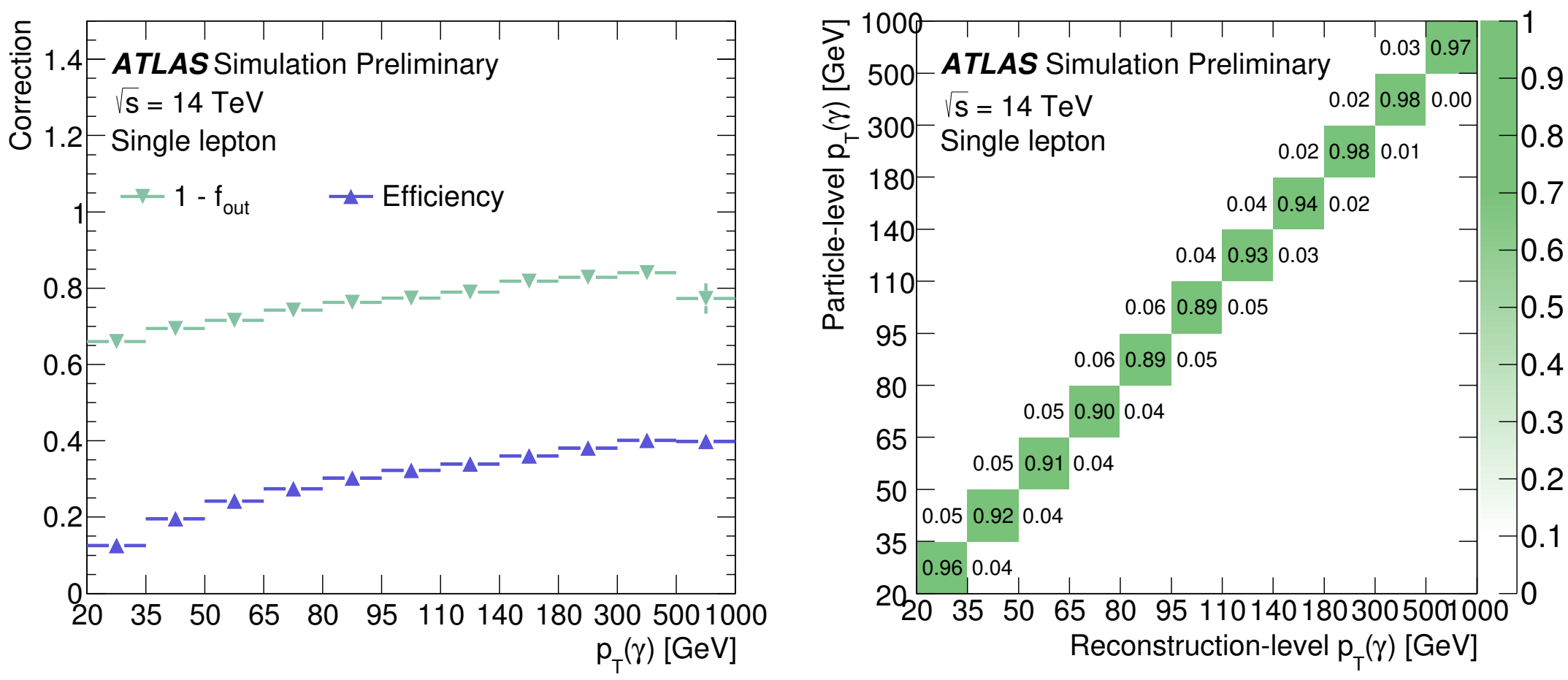

(a) (b)

Figure 2: The (a) efficiency and outside fraction and (b) migration matrix for the photon $p_{\text{T}}$ in the single-lepton channel.

# 8 Systematic uncertainties

The systematic uncertainties are extrapolated from the 13 TeV analysis [10]. For simplicity, only the most important sources of systematic uncertainty are considered, that contribute 90% of the total systematic uncertainty in the 13 TeV analysis. Uncertainties related to theoretical predictions, including event generators, are reduced by a factor of two compared to those used in the 13 TeV analysis to account for anticipated advancements in theoretical predictions and tools. The relative experimental uncertainties are in general kept at the same level as in the 13 TeV analysis.

For the systematic uncertainty of the modelling of the signal efficiency, the uncertainty of parton shower derived from a comparison between Pythia 8 and Herwig 7 [34] is considered and reduced by a factor of two, giving 1% for both the single-lepton and dilepton channels. For the $t\bar{t}$ modelling systematics, which affects the estimation of hadronic-fake background and the shape of electron-fake background, the uncertainties of initial-/final-state radiation (ISR/FSR) and the choice of event generator are considered. The former is derived from a comparison between the nominal $t\bar{t}$ sample and alternative ones with enhanced or suppressed ISR/FSR. The latter is derived from a comparison between Powheg-Box + Pythia and Sherpa. Both uncertainties are reduced by factors of two, giving 13% (14%) and 4% (5%) for the single-lepton (dilepton) channel, respectively.

The uncertainty of the hadronic-fake background estimation in the 13 TeV analysis is composed of the $t\bar{t}$ modelling uncertainty and the statistical uncertainties of the relevant control regions which will become negligible at the HL-LHC. For the electron-fake background systematics, the statistical uncertainty of the data-driven method used to estimate the background is ignored and the remaining uncertainty is 9%, which is from the choice of the templates used to do side-band fit in the method. For the fake-lepton background systematics, the uncertainty is taken to be the same as the 13 TeV analysis, giving 50%. In the 13 TeV analysis, the normalisation of the $W\gamma$ background was constrained with a relative precision of 13% using a template fit method, and this uncertainty is applied to the $W\gamma$ normalisation for the HL-LHC analysis. For the uncertainty of $Z\gamma$ background in the dilepton channel, 30% was assigned for QCD scale variation by varying the renormalisation and factorisation scale factors up and down by a factor or two independently and simultaneously in the 13 TeV analysis, which is reduced by half to 15% at the HL-LHC. For each of the prompt-photon backgrounds, except for the $W\gamma$ in the single-lepton channel, an additional 50% normalisation uncertainty was assigned in the 13 TeV analysis, which is reduced by half to 25% at the HL-LHC.

Experimental sources of uncertainty include the largest and second largest components of the jet energy scale (JES) uncertainty, which are called "JES NP 1" (NP means nuisance parameter) and "JES Rho Topology", and amount to 2% and 1% (1% and 1%), respectively in the single-lepton (dilepton) channel. The "JES Rho Topology" represents the uncertainty of a parameter "Rho", which is to quantify the transverse pile-up energy density of the event and used to subtract pile-up energy from the $p_{\mathrm{T}}$ of the jet. The uncertainties of the event selection efficiency or normalisation for all the simulated processes due to the uncertainties of the integrated luminosity, photon efficiency and pile-up are 1%, 1% and 2%, respectively. Among these uncertainties, the JES Rho Topology uncertainty is reduced by half based on the latest studies with respect to the 13 TeV analysis, while the others stay the same. The pile-up uncertainty has been increased by a factor of two with respect to the 13 TeV analysis to account for the significantly increased pile-up effect at the HL-LHC.

Simulation statistical uncertainties on the generated signal and background samples are expected to be negligible, assuming sufficiently large samples can be generated.

Table 5 summarises these uncertainties. For simplicity, the single-lepton channel systematics are applied to both the $e$+jets and $\mu$+jets channels and the dilepton channel systematics to the $ee$, $e\mu$ and $\mu\mu$ channels.

Table 5: Summary of systematic uncertainties. The last column indicates whether the shape effect of the corresponding uncertainties is considered in the differential cross-section measurement.

| Channel | Efficiency or normalisation | | Impact on | Shape |
|---|---|---|---|---|
| | Single lepton | Dilepton | | |
| Parton shower of $t\bar{t}\gamma$ | 1% | | Signal | Yes |
| ISR/FSR of $t\bar{t}$ | 13% | 14% | Hadronic fake and shape of electron fake | Yes |
| Generator of $t\bar{t}$ | 4% | 5% | | |
| Estimation of electron fake | 9% | - | Electron fake | Yes |
| Estimation of fake lepton | 50% | - | Fake lepton | Yes |
| Normalisation of $W\gamma$ in single-lepton | 13% | - | $W\gamma$ in single-lepton | No |
| QCD scale of $W\gamma$ in single-lepton | - | - | $W\gamma$ in single-lepton | Yes |
| Generator of $Z\gamma$ in dilepton | - | - | $Z\gamma$ in dilepton | No |
| QCD scale of $Z\gamma$ in dilepton | - | 15% | $Z\gamma$ in dilepton | Yes |
| Normalisation of prompt photon | 25% | | Prompt photon (except for $W\gamma$ in single-lepton) | No |
| Luminosity | 1% | | All except fake lepton | No |
| JES NP1 | 2% | 1% | | Yes |
| JES rho topology | 1% | | | |
| Pile-up | 2% | | | |
| Photon efficiency | 1% | | | |

In the differential cross-section measurement, in addition to the normalisation uncertainties considered above, shape effects are considered for the uncertainties of the $t\bar{t}\gamma$ and $t\bar{t}$ modelling and of the QCD scale choice of the $W\gamma$ ($Z\gamma$) modelling in the single-lepton (dilepton) channel. These uncertainties are reduced by factors of two in the same way as for the fiducial measurement. The shape effect of the experimental uncertainties are also considered, except for the uncertainty of integrated luminosity which only affects the normalisation. The background and experimental uncertainties are evaluated by varying the input distributions, unfolding them with corrections based on the nominal signal sample, and comparing the resulting unfolded distributions to the nominal one. The systematic uncertainty due to signal modelling is evaluated by varying the signal corrections, with which the nominal input distributions are unfolded, and comparing the resulting unfolded distributions to the nominal one.

# 9 Results

The expected precision of the fiducial cross-section measurement is studied by fitting an Asimov dataset to a likelihood function built in the fiducial region, which is the product of a single Poisson describing the total number of observed events and a group of Gaussians constraining the nuisance parameters used to parametrise each systematic uncertainty. The resulting total relative uncertainty of the signal strength, which is defined as the fitted number of signal over the predicted one, together with its decomposition into statistical and systematic uncertainties, are shown for each channel in Table 6 and compared with the corresponding results in the 13 TeV analysis, which are derived from fitting the output of a neural network using data. These numbers are also illustrated in Figure 3. The statistical uncertainties at the HL-LHC are in general below 1% and are much smaller than in the 13 TeV analysis, especially for the dilepton channels. The systematic uncertainties at the HL-LHC are also reduced, as a result of the reduction of the theoretical uncertainties by a factor of two. The fiducial measurement in the $e\mu$ channel has the smallest uncertainty of 3%. The $\mu\mu$ channel is less precise than the $ee$ channel due to the larger background contamination from $Z\gamma$ events.

Table 6: The relative uncertainty of the fiducial cross-section measurement expected at the HL-LHC in each channel. The results from the 13 TeV analysis are shown for comparison.

| Channel | | $e$+jets | $\mu$+jets | $ee$ | $e\mu$ | $\mu\mu$ |
|---|---|---|---|---|---|---|
| Stat. | HL-LHC | 0.2% | 0.2% | 0.9% | 0.5% | 0.8% |
| | Run-2 (36 fb$^{-1}$) | 2.8% | 3.0% | 8.1% | 4.6% | 10.0% |
| Sys. | HL-LHC | 6.8% | 6.3% | 5.0% | 3.3% | 6.6% |
| | Run-2 (36 fb$^{-1}$) | 7.9% | 8.4% | 8.9% | 5.7% | 9.8% |
| Total | HL-LHC | 6.8% | 6.3% | 5.0% | 3.4% | 6.6% |
| | Run-2 (36 fb$^{-1}$) | 8.4% | 8.9% | 12.0% | 7.3% | 14.0% |

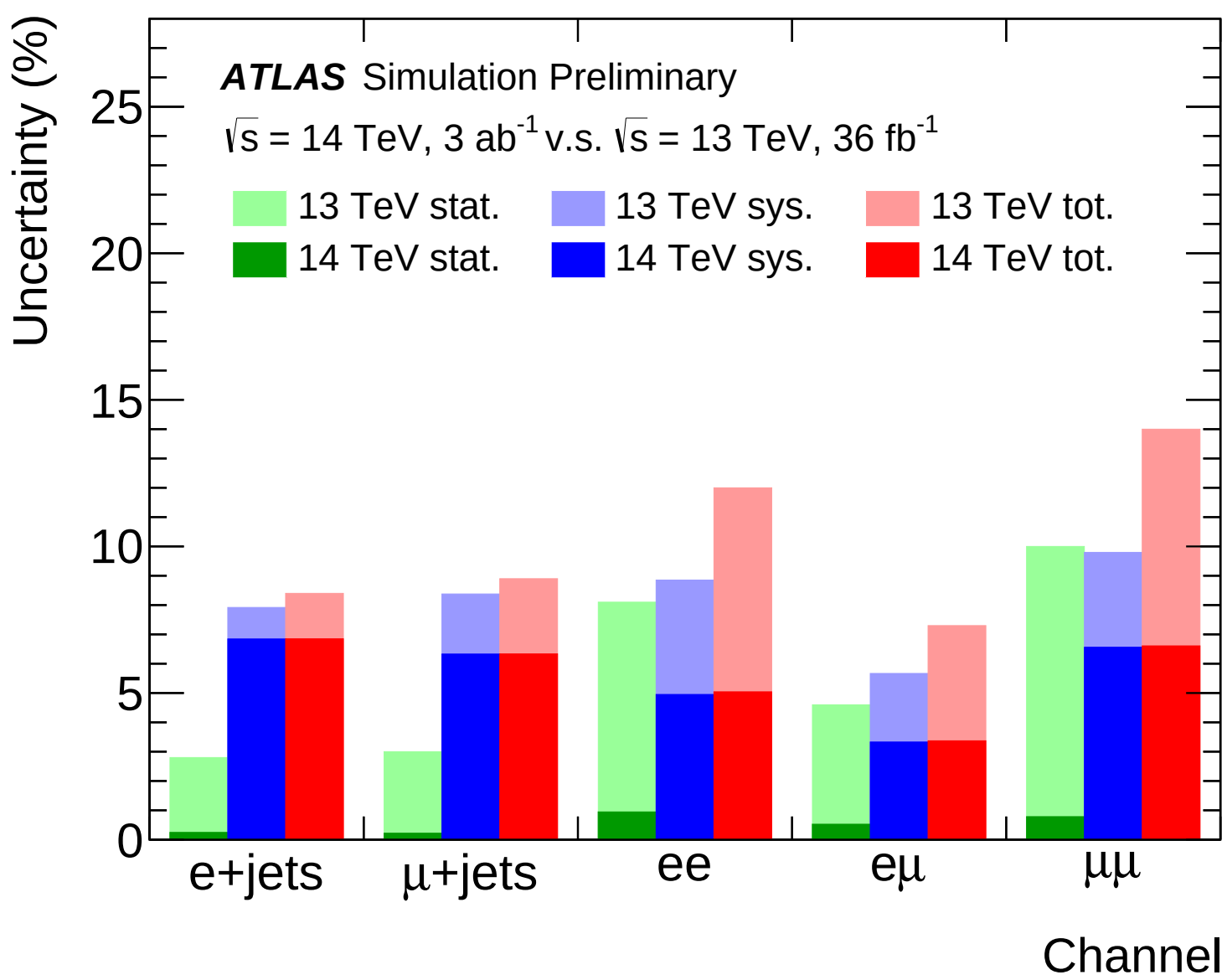

Figure 3: The relative statistical/systematic/total uncertainties of the measured fiducial cross-section in each channel for the HL-LHC and the 13 TeV analysis.

The contributions of different systematic uncertainties to the total uncertainty on the signal strength are summarised in Table 7. In the single-lepton channels, the uncertainties of $t\bar{t}$ ISR/FSR modelling, pile-up, JES and $W\gamma$ background estimation are the leading sources. If the lepton is an electron, the uncertainty of fake-lepton background estimation is also important. In the $ee$ and $\mu\mu$ channels, the uncertainties of $Z\gamma$ background estimation and pile-up are the dominant systematics, In the $e\mu$ channel, the pile-up uncertainty is the most important uncertainty.

Table 7: Decomposition of expected systematic uncertainties on the fiducial $t\bar{t}\gamma$ cross-section measurement. The "Total systematics" is the quadratic sum of all the individual uncertainties, ignoring their correlations.

| Source | $e$+jets | $\mu$+jets | $ee$ | $e\mu$ | $\mu\mu$ |
|---|---|---|---|---|---|
| $t\bar{t}\gamma$ PY8 vs H7 | 1.0% | 1.0% | 1.0% | 1.0% | 1.0% |
| $t\bar{t}$ ISR/FSR | 3.1% | 3.4% | 1.1% | 1.1% | 1.0% |
| $t\bar{t}$ MG5 vs Sherpa | 1.0% | 1.0% | 0.4% | 0.4% | 0.4% |
| $W\gamma$ norm. | 1.6% | 2.7% | | | |
| $Z\gamma$ norm. | 1.7% | 0.7% | 2.8% | <0.1% | 4.7% |
| $Z\gamma$ QCD scale | | | 1.7% | <0.1% | 2.8% |
| Single top norm. | 1.1% | 1.3% | 0.9% | 0.3% | 0.6% |
| Diboson norm. | <0.1% | <0.1% | 0.1% | | 0.1% |
| Fake-lep norm. | 3.0% | 0.5% | | | |
| e-fake norm. 1 | 1.5% | 1.5% | | | |
| e-fake norm. 2 | 0.7% | 0.8% | | | |
| JES NP 1 | 2.2% | 2.2% | 1.2% | 1.0% | 1.2% |
| JES Rho topo. | 1.1% | 1.1% | 1.2% | 1.0% | 1.2% |
| Photon eff. | 1.1% | 1.1% | 1.2% | 1.0% | 1.2% |
| Pile-up | 2.2% | 2.2% | 2.3% | 2.0% | 2.4% |
| Luminosity | 1.1% | 1.1% | 1.2% | 1.0% | 1.2% |
| Total systematics | 6.6% | 6.2% | 4.9% | 3.3% | 6.7% |

In the region of photon $p_T > 500$ GeV, the statistical/systematic uncertainties are 3.7%/7.4% and 11.5%/3.4% for the single-lepton and $e\mu$ channels, respectively. The statistical uncertainty of the $e\mu$ channel is large and one might consider combining it with the $ee$ and $e\mu$ channels. But it is found that combining these dilepton channels doesn't help to reduce the total uncertainty, due to the large $Z\gamma$ background and its large associated uncertainty in the $ee$ and $\mu\mu$ channels. Table 8 gives more details about these uncertainties.

Table 8: The relative statistical/systematic/total uncertainties for a fiducial cross-section measurement using photons with $p_T$ larger than 500 GeV at the HL-LHC, in the single-lepton, $e\mu$ and combined dilepton channels.

| Uncertainty | Stat. | Sys. | Total. |
|---|---|---|---|
| single-lepton | 3.7% | 7.4% | 8.2% |
| $e\mu$ | 11.5% | 3.4% | 12.0% |
| dilepton | 10.0% | 17.9% | 20.5% |

The differential cross-sections are unfolded using an Asimov dataset. The $e$+jets and $\mu$+jets channels are combined into the single-lepton channel. The $e\mu$ channel is unfolded by itself without combining with the $ee$ or $\mu\mu$ channel, since the latter two channels bring significantly more background contamination, making the results of the combined dilepton channels worse in most cases. The resulting uncertainties are shown in Figure 4 and 5 for the single-lepton and $e\mu$ channels, respectively. Statistical uncertainties are in general below 1% (2%) for the single-lepton ($e\mu$) channel, except for the high-$p_T$ bins: e.g. for photons above 500 GeV, the statistical uncertainty reaches 4% (12%). Systematic uncertainties are in general below 5% for both channels, with the background modelling uncertainties being the leading systematic uncertainty. The measurement in the single-lepton channel is mainly limited by systematic uncertainties, while that of the $e\mu$ channel by both statistical and systematical uncertainties. Overall, a 5% precision can be achieved for the differential measurement, except for the dilepton channel with a photon of $p_T$ larger than 500 GeV.

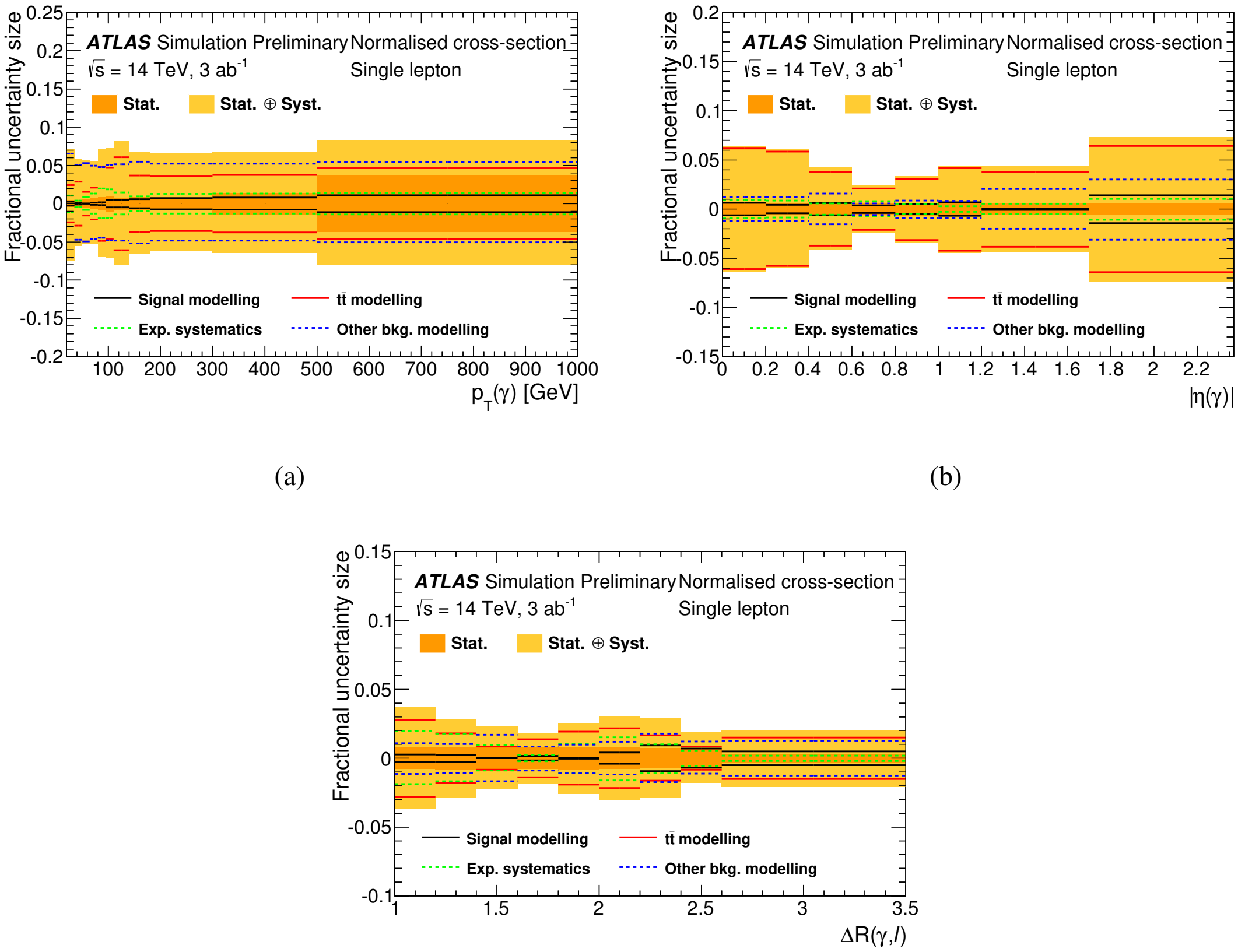

(a) (b) (c)

Figure 4: The systematic uncertainties for the normalised differential cross-sections as a function of the (a) photon $p_T$, (b) photon $|\eta|$ and (c) $\Delta R(\gamma, \ell)$ in the single-lepton channel.

The uncertainties of differential measurements for the HL-LHC and the 13 TeV analysis are compared in Figure 6 for the photon $p_T$ in the single-lepton and $e\mu$ channels, as well as for the $\Delta\phi(\ell, \ell)$ in the $e\mu$ channel. Both the statistical and systematic uncertainties are reduced significantly at the HL-LHC. The HL-LHC allows measurement of the photon $p_T$ spectrum up to 1000 GeV rather than the 300 GeV limit at Run 2.

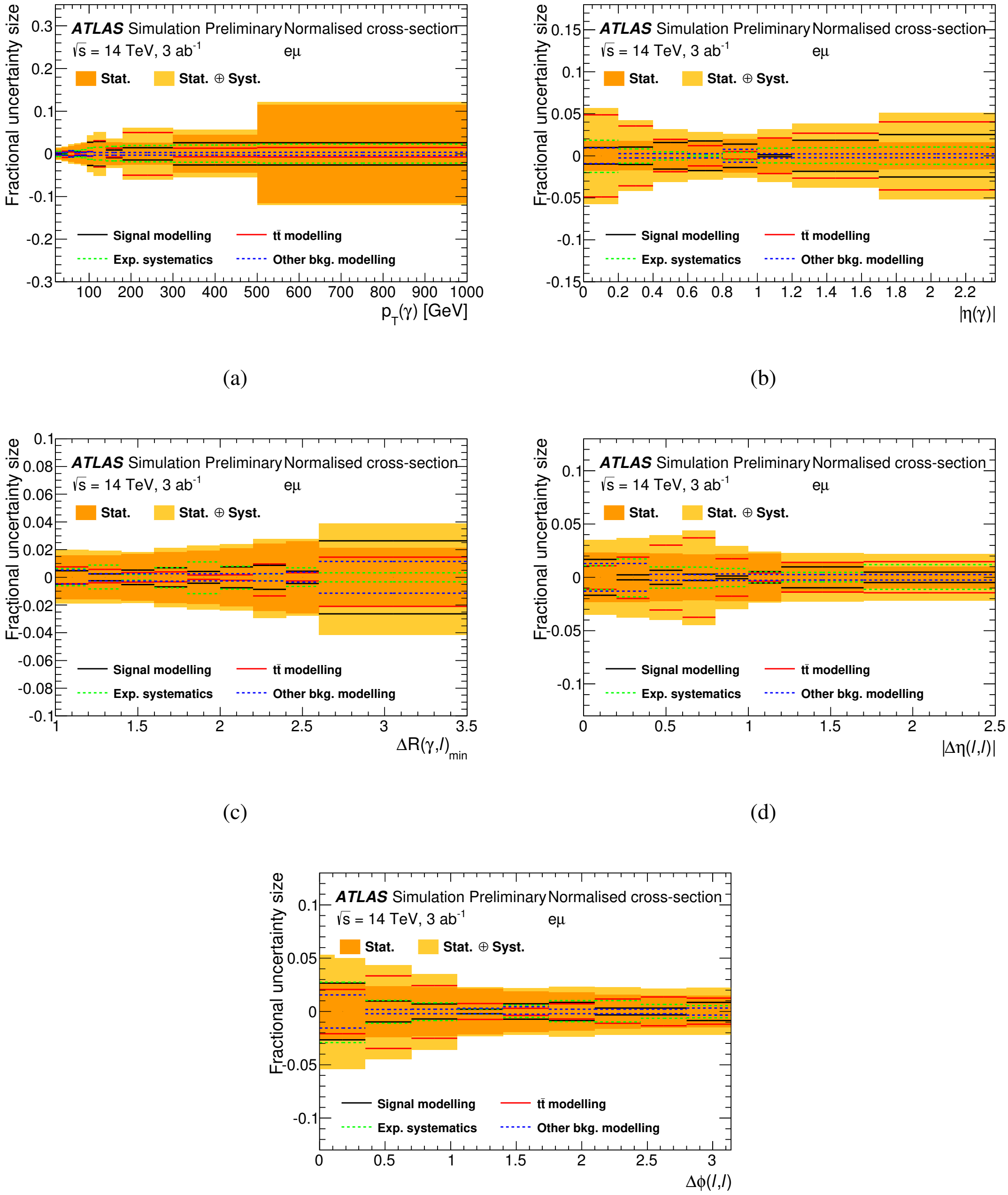

(a) (b) (c) (d) (e)

Figure 5: The systematic uncertainties for the normalised differential cross-sections as a function of the (a) photon $p_{\mathrm{T}}$, (b) photon $|\eta|$, (c) minimum $\Delta R(\gamma, \ell)$, (d) $|\Delta\eta(\ell, \ell)|$ and (e) $\Delta\phi(\ell, \ell)$ in the $e\mu$ channel.

15

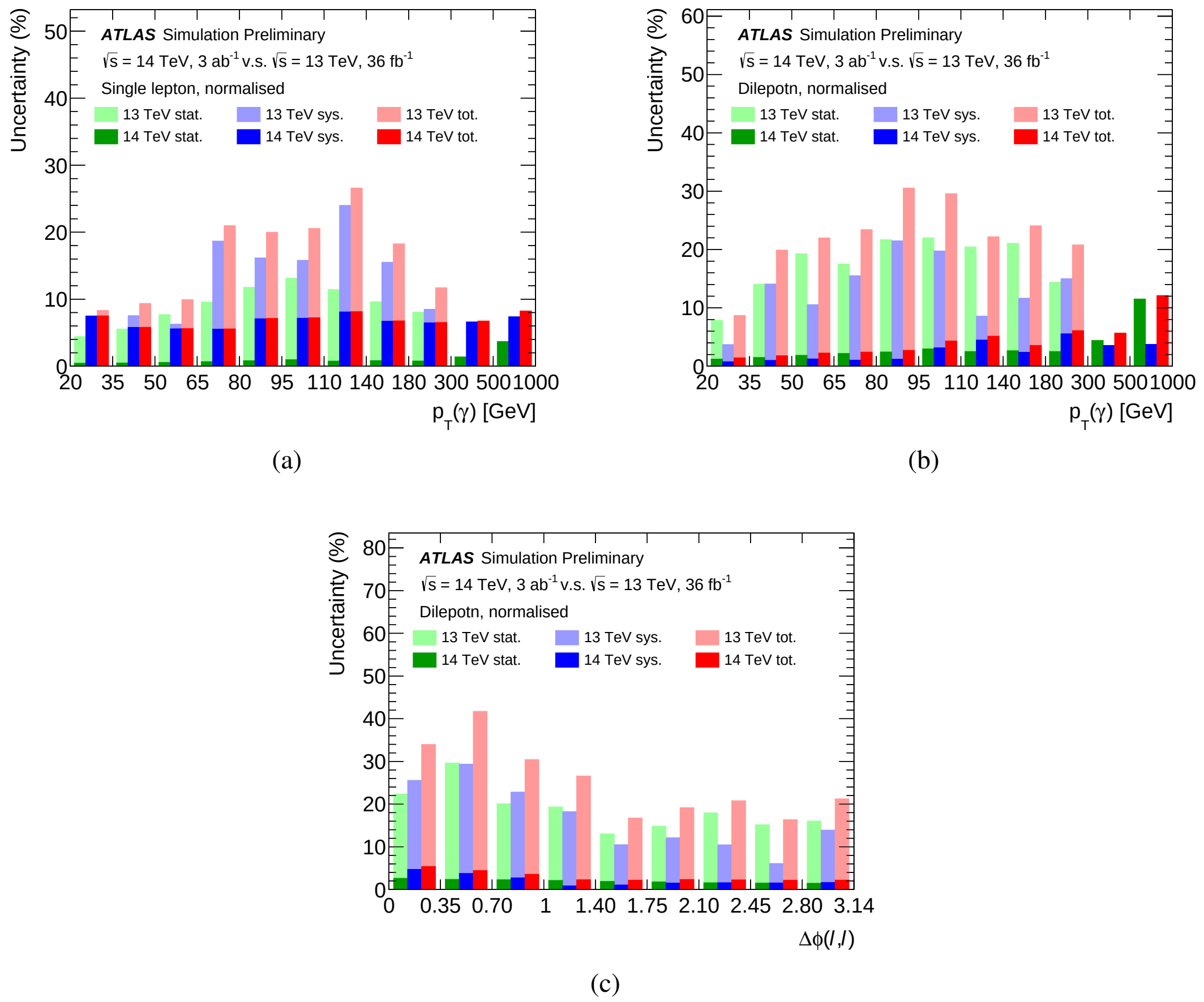

Figure 6: Comparison of statistical/systematic/total uncertainties for the normalised differential cross-sections as a function of the photon $p_{\mathrm{T}}$ in the (a) single-lepton channel and the (b) $e\mu$ channel, and the (c) $\Delta\phi(\ell, \ell)$ in the $e\mu$ channel, between the HL-LHC and the 13 TeV analysis.

# 10 Conclusion

The expected precision of fiducial and differential cross-section measurements of top-quark pair production in association with a photon are studied in the leptonic $t\bar{t}$ final states, using a simulated dataset corresponding to 3 ab$^{-1}$ of 14 TeV $pp$ collision data that is expected to be collected by the upgraded ATLAS detector at the HL-LHC. The differential cross-sections, normalised to unity, are measured as a function of the photon $p_{\mathrm{T}}$ and $|\eta|$, and the $\Delta R$ between the photon and the closest lepton for both channels, and the $|\Delta\eta|$ and $\Delta\phi$ between the two leptons for the dilepton channel. The best precision is achieved in the $e\mu$ channel with a 3% uncertainty for the measurement of fiducial cross-section with a photon $p_{\mathrm{T}}$ threshold at 20 GeV. The single-lepton channel provides the most precise measurement with an 8% uncertainty for photons with $p_{\mathrm{T}}$ above 500 GeV. The expected uncertainties of differential cross-section measurements are in general below 5%.

# CMS Physics Analysis Summary

Contact: cms-phys-conveners-ftr@cern.ch 

## Anomalous couplings in the tt+Z final state at the HL-LHC

The CMS Collaboration

### Abstract

The electroweak couplings of the top quark provide a crucial window to physics beyond the standard model and can be put to stringent tests with the CERN High-Luminosity LHC (HL-LHC). The expected sensitivity of the CMS detector for anomalous electroweak top quark interactions based on differential cross section measurements of the $\mathrm{t\bar{t}Z}$ process in the three lepton final state is provided for a HL-LHC scenario with 3000 $\mathrm{fb}^{-1}$ of proton-proton collision data at a centre-of-mass energy of 14 TeV.

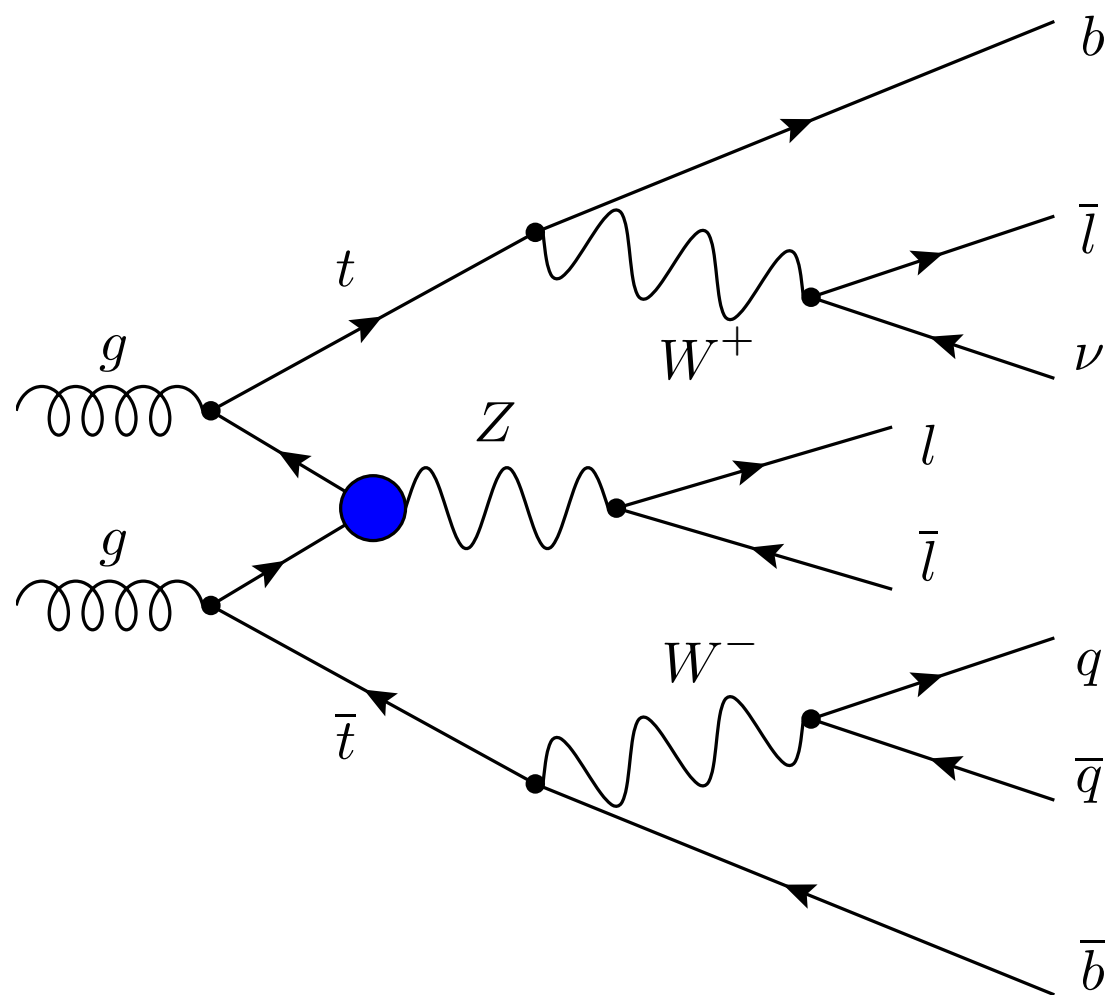

Figure 1: Representative Feynman diagram for the $\mathrm{t\bar{t}Z}$ process.

# 1 Introduction

Owing to the special role of the standard model (SM) top quark, many beyond standard model (BSM) predictions include anomalous couplings of the top quark to the electroweak gauge bosons [1–7]. Direct measurements of processes sensitive to the neutral-current interaction of the top quark have so far been limited by the amount of collision data available at the LHC [8]. With the data sample expected for the HL-LHC, it will be possible to measure the electroweak dipole moments of the top quark, as well as the (axial-)vector couplings of the top quark to the Z boson [9]. In this study we simulate differential cross section measurements of the $\mathrm{pp} \to \mathrm{t\bar{t}Z}$ process in the CMS Phase-2 detector using 3000 $\mathrm{fb}^{-1}$ of data, in events with three leptons (electrons or muons), where two are consistent with the Z boson mass hypothesis. A representative Feynman diagram is shown in Fig. 1. We use a DELPHES [10] detector simulation and consider an HL-LHC scenario with a centre-of-mass energy of 14 TeV.

We interpret the result in terms of the SM effective field theory (SM-EFT) [11]. In SM-EFT at mass dimension-6, there are 59 independent Wilson Coefficients [12] that form the so called Warsaw basis. Among them, 15 are relevant for top quark interactions [13]. In the Warsaw basis, several operators contribute both to the anomalous charged current interaction (the Wtb vertex) and the neutral current interactions (the ttZ and tt$\gamma$ vertex), albeit in different linear combinations. The parametrization used here allows a modification of the neutral top quark interactions while leaving the Wtb vertex unchanged [11].

# 2 CMS Phase-2 detector

The CMS detector [14] will be substantially upgraded in order to fully exploit the physics potential offered by the increase in luminosity at the HL-LHC [15], and to cope with the demanding operational conditions at the HL-LHC [16–20]. The upgrade of the first level hardware trigger (L1) will allow for an increase of L1 rate and latency to about 750 kHz and 12.5 $\mu$s, respectively, and the high-level software trigger (HLT) is expected to reduce the rate by about a factor of 100 to 7.5 kHz. The entire pixel and strip tracker detectors will be replaced to increase the granularity, reduce the material budget in the tracking volume, improve the radiation hardness, and extend the geometrical coverage and provide efficient tracking up to pseudorapidities of about $|\eta| = 4$. The performance of the muon system will be improved by upgrading the electronics of the existing cathode strip chambers (CSC), resistive plate chambers (RPC) and drift tubes (DT). New muon detectors based on improved RPC and gas electron multiplier (GEM) technologies will be installed to add redundancy, increase the geometrical coverage up to about $|\eta| = 2.8$, and improve the trigger and reconstruction performance in the forward region. The barrel electromagnetic calorimeter (ECAL) will feature the upgraded front-end electronics that will be able to exploit the information from single crystals at the L1 trigger level, to accommodate trigger latency and bandwidth requirements, and to provide 160 MHz sampling that allows high precision timing capability for photons. The hadronic calorimeter (HCAL), consisting in the barrel region of brass absorber plates and plastic scintillator layers, will be read out by silicon photomultipliers (SiPMs). The endcap electromagnetic and hadron calorimeters will be replaced with a new combined sampling calorimeter (HGCal) that will provide highly-segmented spatial information in both transverse and longitudinal directions, as well as high-precision timing information. Finally, the addition of a new timing detector for minimum ionizing particles (MTD) in both barrel and endcap region is envisaged to provide capability for 4-dimensional reconstruction of interaction vertices that will allow to significantly offset the CMS performance degradation due to high rate of simultaneous interactions per bunch crossing (pileup, PU).

The generated signal and background events are processed with the fast-simulation package DELPHES in order to simulate the expected response of the upgraded CMS detector. The object reconstruction and identification efficiencies, as well as the detector response and resolution, are parametrized in DELPHES using the detailed simulation of the upgraded CMS detector based on GEANT4 package [21, 22].

# 3 Event simulation

## 3.1 Generating weighted signal samples

While many BSM scenarios modify the $\mathrm{t\bar{t}Z}$ cross section, most have a large impact on other processes as well. Anomalous interactions between the top quark and the gluon (chromomagnetic and chromoelectric dipole moment interactions) are tightly constrained by the $\mathrm{t\bar{t}}$+jets final state [23]. Similarly, the modification of the Wtb vertex is best constrained by measurements of the W helicity fractions in top quark pair production [24] and in t-channel single top quark production [25]. The operators inducing anomalous interactions of the top quark with the remaining neutral gauge bosons, the Z boson and the photon, have Wilson coefficients $C_{\mathrm{tZ}}$, $C_{\mathrm{tZ}}^{[\mathrm{Im}]}$, $C_{\phi t}$ and $C_{\phi Q}^{-}$ [11]. The former two induce electroweak dipole moments while the latter two induce anomalous neutral current interactions. These Wilson coefficients are the main focus of this work. They amount to the linear combinations

$$C_{\mathrm{tZ}} = \mathrm{Re}\left(-\sin\theta_{\mathrm{W}} C_{\mathrm{uB}}^{(33)} + \cos\theta_{\mathrm{W}} C_{\mathrm{uW}}^{(33)}\right) \quad (1)$$

$$C_{\mathrm{tZ}}^{[\mathrm{Im}]} = \mathrm{Im}\left(-\sin\theta_{\mathrm{W}} C_{\mathrm{uB}}^{(33)} + \cos\theta_{\mathrm{W}} C_{\mathrm{uW}}^{(33)}\right) \quad (2)$$

$$C_{\phi t} = C_{\phi \mathrm{t}} = C_{\phi \mathrm{u}}^{(33)} \quad (3)$$

$$C_{\phi Q}^{-} = C_{\phi \mathrm{Q}} = C_{\phi \mathrm{q}}^{1(33)} - C_{\phi \mathrm{q}}^{3(33)} \quad (4)$$

where $\theta_W$ is the weak mixing angle and the Wilson coefficients in the Warsaw basis are denoted by $C_{\mathrm{uB}}^{(33)}$, $C_{\mathrm{uW}}^{(33)}$, $C_{\phi \mathrm{u}}^{(33)}$, $C_{\phi \mathrm{q}}^{1(33)}$, and $C_{\phi \mathrm{q}}^{3(33)}$ as defined in Ref. [11]. As we only consider $C_{\mathrm{tZ}}$, $C_{\mathrm{tZ}}^{[\mathrm{Im}]}$, $C_{\phi t}$ and $C_{\phi Q}^{-}$ in this analysis, we set other Wilson coefficients to zero. The constraints $C_{\phi \mathrm{q}}^{3(33)} = 0$ and $C_{\mathrm{uW}}^{(33)} = 0$ ensure a Wtb vertex according to the SM.

When scanning the BSM parameter space during event simulation, even a moderate number of four independent Wilson coefficients is prohibitive or severely restricts the achievable granularity. We eschew this limitation by a strategy taken from Ref. [26]. First, a sample of events is processed with MadGraph [27] at a reference parameter point. Then, for each event the compiled matrix element is reevaluated using MadWeight [28] at base points in the parameter space spanned by the Wilson coefficients $C_i$. By this procedure, event weights can be calculated at the parameter base points. Because the generic structure of a matrix element with operator insertions is polynomial in the Wilson coefficients, we can evaluate the matrix element at a sufficient number of parameter points and obtain a polynomial parametrization of the event weight in the full parameter space [29].

### 3.2 Simulated event samples

The t$\bar{\text{t}}$Z process is generated at the parton level using MADGRAPH5_aMC@NLO v2.3.3 [27] at leading order (LO), and decayed using MadSpin [30, 31] in order to preserve the spin correlation in the decays of the top quarks. It contains a small non-resonant t$\bar{\text{t}}\ell\bar{\ell}$ contribution. Parton showering and hadronization are generated using PYTHIA 8.2 [32, 33]. Fast detector simulation was performed using DELPHES, with the CMS reconstruction efficiency parametrization for the Phase-2 upgrade. The mean number of PU interactions per bunch crossing is varied from 0 to 200. Jets are reconstructed with the FastJet package [34], using the anti-$k_T$ algorithm [35], with a cone size $R = 0.4$.

The production of a Z boson in association with a top quark pair provides an ideal testbed for the t$\bar{\text{t}}$Z interaction. However, not all contributing Feynman diagrams contain a t$\bar{\text{t}}$Z vertex. Because of interference, the events with a boson originating from a top quark can not be perfectly separated. Therefore we single out the contribution from events where the Z boson originates at generator level from a W boson, a lepton (including $\tau$ leptons), (b-)jets, or an initial state quark. These 'non-informative' contributions do contain the t$\bar{\text{t}}$Z vertex and are therefore not affected by the Wilson coefficients considered here..

Important backgrounds to the t$\bar{\text{t}}$Z process in final states including leptons from the top quark decays include WZ production and single top quarks produced in association with a Z boson (tZq). In addition, we simulate background contributions for single top events in association with two bosons (tWZ) and for the t$\bar{\text{t}}\gamma$ process. The WZ, tZq, tWZ, t$\bar{\text{t}}\gamma$ and t$\bar{\text{t}}$Z processes are normalized to cross sections calculated up to next-to-leading order (NLO) in perturbative QCD with MADGRAPH5_aMC@NLO. The generated samples are summarized in Table 1.

As $C_{\text{tZ}}$ modifies the coupling of the Z to the top quark, the SM gauge symmetry requires a similar modification of the t$\bar{\text{t}}\gamma$ coupling. However in this work, we only consider affected BSM couplings in the t$\bar{\text{t}}$Z process. The effect of the considered Wilson coefficients $C_{\text{tZ}}$, $C_{\text{tZ}}^{[\text{Im}]}$, $C_{\phi t}$ and $C_{\phi Q}^{-}$ on the total yield due to modified couplings in the processes tZq, tWZ and t$\bar{\text{t}}\gamma$ is found to be negligible. Because the neglected variations generally affect the predicted yields with the same sign as the t$\bar{\text{t}}$Z process, this is a conservative choice.

## 4 Event selection

From the SM branching ratios of W and Z bosons as well as the 7, 8 and 13 TeV results on the inclusive t$\bar{\text{t}}$Z cross section from the ATLAS collaboration [36, 37] and the CMS collaboration [8, 38–40] it follows that the three lepton channel is the most sensitive search channel. Here, the Z boson decays to an opposite-sign same-flavor pair of electrons or muons, and one of the W bosons originating from a top quark decays to a lepton and neutrino. The other W boson can decay either leptonically or hadronically. We thus require exactly three reconstructed leptons (e or $\mu$) with $p_{\mathrm{T}}$ ($\ell$) thresholds of 10, 20, and 40 GeV, and $\lvert\eta(\ell)\rvert < 3.0$. We furthermore require that there is among them a pair of opposite-sign same-flavor leptons consistent with the Z boson by requiring $\lvert m(\ell\ell) - m_{\mathrm{Z}}\rvert < 10$ GeV. Here, and throughout the event selection, we remove reconstructed leptons within a cone of $\Delta R < 0.3$ to any reconstructed jet satisfying $p_{\mathrm{T}} > 30$ GeV. Furthermore, at least 3 jets with $p_{\mathrm{T}}$ (j)$>$ 30 GeV and $\lvert\eta(\mathrm{j})\rvert < 4.0$, where one of the jets has been identified as a b-tag jet according to the DELPHES specification (medium working point), are required. Because $\mathrm{t\bar{t}Z}$ is a process with very high invariant mass, the final state objects are typically produced centrally in the detector. A further increase of geometric acceptance of jets or leptons does therefore not increase the analysis sensitivity. The event selection is summarized in Table 2.

Table 1: Simulated processes with a Monte-Carlo sample size of one million events, the cross section for $\sqrt{s} = 13$ TeV and the scale factor for $\sqrt{s} = 14$ TeV. Here, $\ell = e, \mu, \tau$ and $\nu_\ell = \nu_e, \nu_\mu, \nu_\tau$.

| Process | | $\sigma_{13\,\text{TeV}}$ (pb) | $\sigma_{14\,\text{TeV}}/\sigma_{13\,\text{TeV}}$ |
|---|---|---|---|
| $\mathrm{t\bar{t}Z}$ | $\mathrm{pp} \to \mathrm{t\bar{t}}\ell\bar{\ell}$ | 0.0915 | 1.16 |
| WZ | $\mathrm{pp} \to \bar{\ell}\nu_\ell \ell\bar{\ell} + \mathrm{pp} \to \ell\bar{\nu}_\ell \ell\bar{\ell}$ | 4.666 | 1.16 |
| tZq | $\mathrm{pp} \to \mathrm{t}\ell\bar{\ell}\mathrm{q} + \mathrm{pp} \to \bar{\mathrm{t}}\ell\bar{\ell}\mathrm{q}$ | 0.0758 | 1.12 |
| tWZ | $\mathrm{pp} \to \mathrm{tW}\ell\bar{\ell} + \mathrm{pp} \to \bar{\mathrm{t}}\mathrm{W}\ell\bar{\ell}$ | 0.01123 | 1.12 |
| $\mathrm{t\bar{t}}\gamma$ | $\mathrm{pp} \to \mathrm{t\bar{t}}\gamma$ | 3.697 | 1.03 |

Table 2: Event selection and object level thresholds for the $\mathrm{t\bar{t}Z}$ selection.

| Observable | Selection |
|---|---|
| $N_{\text{lep}}$ | 3 |
| $N_{\text{jets}}$ | $\geq 3$ |
| $N_{\text{b-tag}}$ | $\geq 1$ |
| $p_{\mathrm{T}}$ ($\ell$) (GeV) | $> 10/20/40$ |
| $\lvert\eta(\ell)\rvert$ | $< 3.0$ |
| $p_{\mathrm{T}}$ (j) (GeV) | $> 30$ |
| $\lvert\eta(\mathrm{j})\rvert$ | $< 4.0$ |
| $\lvert m(\ell\ell) - m_{\mathrm{Z}}\rvert$ (GeV) | $\leq 10$ |

# 5 Signal regions

Because the dimension-6 operators introduce new momentum dependent tensor structures in the Lagrangian, the most sensitive observable is the Z boson transverse momentum $p_{\mathrm{T}}(\mathrm{Z})$ [41]. We consider its distributions in equally sized bins of 100 GeV. The second important observable is $\cos\theta^*_{\mathrm{Z}}$, the relative angle of the negatively charged lepton to the Z boson direction of flight in the rest frame of the Z boson. The differential cross sections for $\mathrm{t\bar{t}Z}$ with respect to $p_{\mathrm{T}}(\mathrm{Z})$ and $\cos\theta^*_{\mathrm{Z}}$ in SM and BSM scenarios are shown in Fig. 2. The distribution of $p_{\mathrm{T}}(\mathrm{Z})$ is more sensitive to BSM effects than $\cos\theta^*_{\mathrm{Z}}$, and the latter contributes approximately 10% to the sensitivity. We show the differential distributions for $C_{\mathrm{tZ}} = 2\ (\Lambda/\ \mathrm{TeV})^2$ and $C_{\mathrm{tZ}}^{[\mathrm{Im}]} = 2\ (\Lambda/\ \mathrm{TeV})^2$, corresponding to a signal hypothesis within the currently most stringent 95%

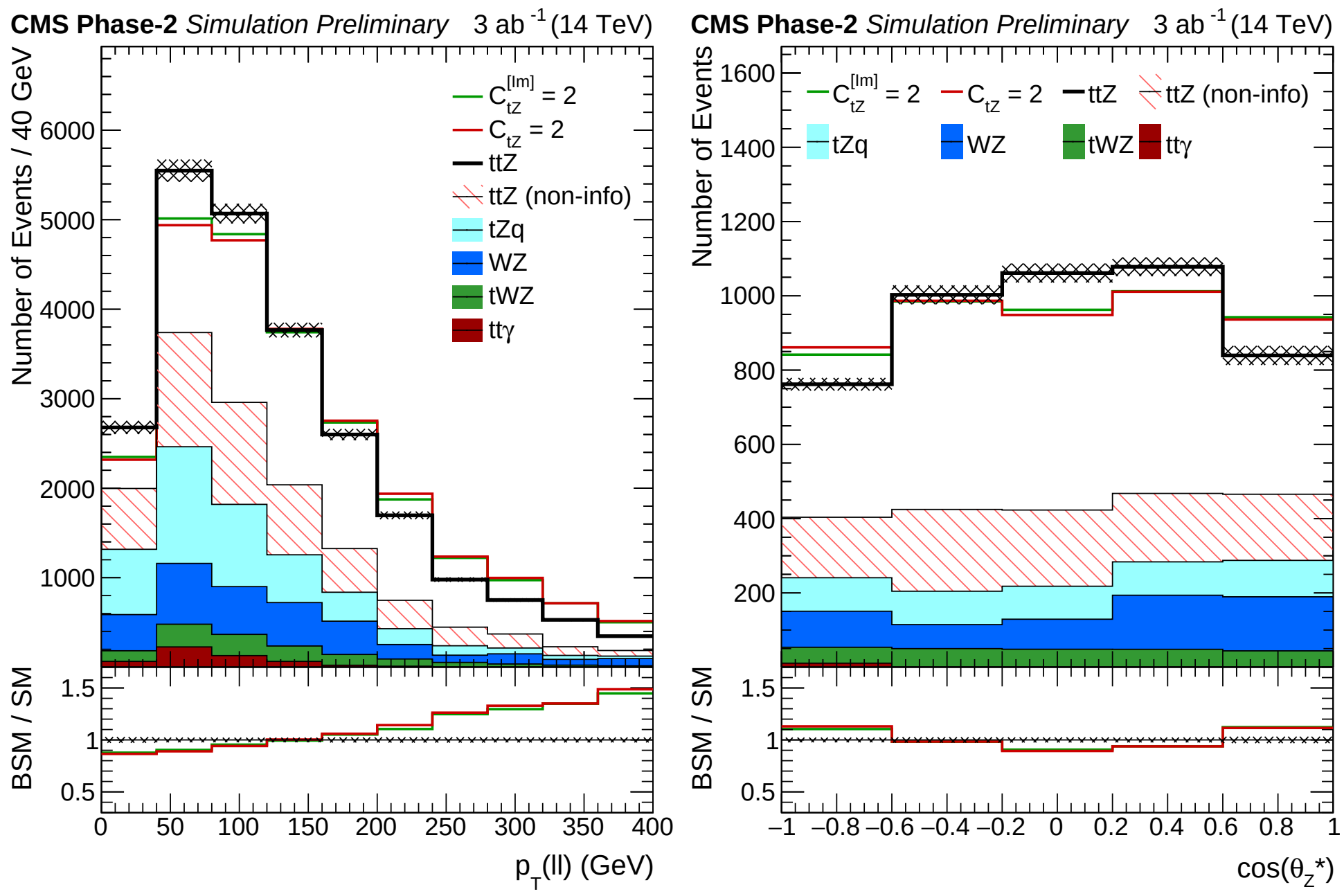

Figure 2: Differential cross sections with respect to $p_T(Z)$ (left) and $\cos\theta^*_Z$ (right) in the $t\bar{t}Z$ ($N_{lep}$=3) channel as specified in Table 2 and for the Phase-2 scenario. For $\cos\theta^*_Z$, an additional requirement of $p_T(Z) > 200$ GeV is applied. The SM distributions are shown in black with systematic uncertainties, while colored lines show hypotheses for $C_{tZ} = 2\ (\Lambda/\text{TeV})^2$ and $C^{[Im]}_{tZ} = 2\ (\Lambda/\text{TeV})^2$, with yields that are area-normalized to the SM distribution. The non-informative contribution to $t\bar{t}Z$ is described in Sec. 4 and shown hatched. Backgrounds are shown in solid colors.

CL limits [8, 42]. We normalize the BSM distributions to the SM yield in the plots to visualize the discriminating features of the parameters. The contribution from the $t\bar{t}Z$ process which does not contain information on the Wilson coefficients is shown hatched. A small background from non-prompt leptons is taken from Ref. [8] and scaled to 3 ab$^{-1}$. The choice of signal regions in $p_T(Z)$ and $\cos\theta^*_Z$ is shown in Fig. 3.

Table 3: Definition of the $t\bar{t}Z$ signal regions.

| $p_T(Z)$ (GeV) | $-1 \le \cos\theta^*_Z < -0.6$ | $-0.6 \le \cos\theta^*_Z < 0.6$ | $0.6 \le \cos\theta^*_Z$ |
|---|---|---|---|
| 0–100 | SR1 | SR2 | SR3 |
| 100–200 | SR4 | SR5 | SR6 |
| 200–400 | SR7 | SR8 | SR9 |
| $\ge 400$ | SR10 | SR11 | SR12 |

# 6 Systematic uncertainties

Experimental uncertainties are estimated based on the expected performance of the Phase-2 CMS detector. This scenario assumes that there will be further advances in both experimental methods and theoretical descriptions of relevant physics effects. Theoretical uncertainties are assumed to be reduced by a factor two with respect to the ones in the reference Run 2 analysis [43]. For experimental systematic uncertainties, it is assumed that those will be reduced by the square root of the integrated luminosity until they reach a defined lower limit based on estimates of the achievable accuracy with the upgraded detector [44].

Typical values for uncertainties are listed in Table 4. To propagate these uncertainties, events are either reweighted or the momenta of the respective objects are rescaled in order to follow the desired variation. The modified yields are subsequently compared to nominal ones and the resulting differences are taken as systematic uncertainties.

Table 4: The sources of systematic uncertainty grouped in experimental systematic uncertainties (exp.) and theoretical uncertainties (theo.) as well as their impacts on reconstructed objects and event yields.

| | Source | Affected processes | Unc. on pred. yield | Unc. obj. level |
|---|---|---|---|---|
| exp. | b-Tagging b-jets | all MC | 0.1 - 5.2 % | 1.0 - 4.6 % |
| | b-Tagging mis-tag | all MC | 0.1 - 4.5 % | 10 % |
| | Muon ID | all MC | 0.4 - 1.5 % | 0.5 % |
| | Electron ID | all MC | 0.4 - 1.5 % | 0.5 % |
| | Jet energy scale | all MC | 0.1 - 2.0 % | |
| | Integrated luminosity | all MC | 1.0 % | |
| | Trigger efficiency | all MC | 1.0 % | |
| | Non-prompt estimate | non-prompt background | 15.0 % | |
| theo. | Scale uncertainty | all MC | 0.2 - 1.7 % | |
| | PDF choice | all MC | 0.5 - 2.6 % | |
| | Parton shower | ttZ | 0.5 - 2.0 % | |
| | WZ cross section | WZ | 5.0 % | |
| | ttX cross section | tZq, tWZ, tt$\gamma$ | 5.5 % | |

# 7 Results

The predicted yields are estimated for the 3 ab$^{-1}$ HL-LHC scenario at $\sqrt{s} = 13$ TeV, scaled to 14 TeV, and are shown in Fig. 3. With the uncertainties described in Sec. 6, a binned likelihood function $L(\theta)$ is constructed where $\theta$ labels the set of nuisance parameters. We perform a profiled maximum likelihood fit of $L(\theta)$ and consider $q(r) = -2\log(L(\hat{\theta})/L(\hat{\theta}_{\mathrm{SM}}))$, where $\hat{\theta}$ and $\hat{\theta}_{\mathrm{SM}}$ are the set of nuisance parameters maximizing the likelihood function at the BSM and SM point, respectively. The largest contributions among the experimental uncertainties originate from the imperfect knowledge on the luminosity and trigger efficiencies. Uncertainties on the PDF and the cross section of the WZ process contribute significantly to the theoretical uncertainties.

In Fig. 4, the likelihood scan for the $\mathrm{t\bar{t}Z}$ process is shown, where we consider one non-zero Wilson coefficient at a time, and all others are set to zero. The corresponding 68% and 95% CL intervals are summarized in Table 5.

In Fig. 5, likelihood ratios for two pairs of Wilson coefficients corresponding to modified neutral current interactions ($C_{\phi t}$ and $C^{-}_{\phi Q}$) and dipole moment interactions ($C_{\mathrm{tZ}}$ and $C^{[\mathrm{Im}]}_{\mathrm{tZ}}$) are considered. The Wilson coefficient not shown on the $x$ axis is included in the profiling of nuisance parameters. The corresponding 68% and 95% CL intervals are summarized in Table 6.

In Fig. 6 (left), the likelihood scan for the $\mathrm{t\bar{t}Z}$ process is shown under the SM hypothesis in the $C^{-}_{\phi Q}/C_{\phi t}$ parameter plane of the Warsaw basis. The likelihood scan of the dipole moment parameters $C_{\mathrm{tZ}}/C^{[\mathrm{Im}]}_{\mathrm{tZ}}$ is shown in Fig. 6 (right). The green (red) lines show the 68% (95%) CL contour line and the SM parameter point corresponds to $C_{\phi t} = C^{-}_{\phi Q} = 0$ and $C_{tZ} = C^{[Im]}_{tZ} = 0$. For the neutral current interactions, the two-dimensional scan reveals that the sensitivity to $C_{\phi t}$ and $C^{-}_{\phi Q}$ is significantly correlated.

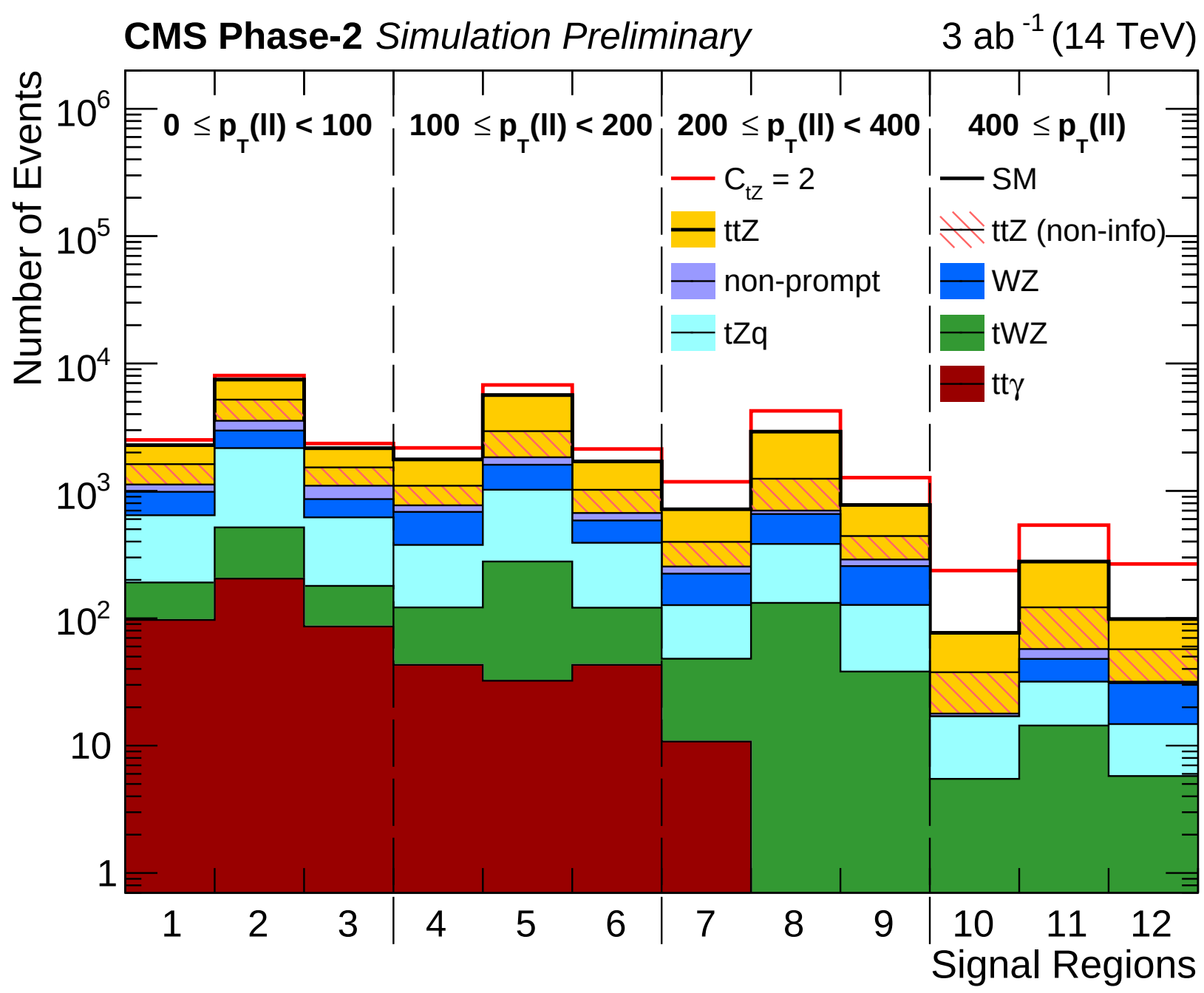

Figure 3: Signal region yields from simulation for SM processes (colored histograms). The yields are estimated for an integrated luminosity of 3/ab, the cross section is scaled to 14 TeV. The total SM yield is shown with the black line, the dashed red line reflects the total expected yield assuming modified couplings, with the chosen value $C_{tZ} = 2\ (\Lambda/\ \text{TeV})^2$. The hatched area represents the non-informative contribution to $t\bar{t}Z$ as described in Sec. 4.

Table 5: Expected 68 % and 95 % CL intervals, where one Wilson coefficient at a time is considered non-zero.

| Wilson coefficient | 68 % CL $(\Lambda/\text{TeV})^2$ | 95 % CL $(\Lambda/\text{TeV})^2$ |
|---|---|---|
| $C_{\phi t}$ | [-0.47, 0.47] | [-0.89, 0.89] |
| $C_{\phi Q}$ | [-0.38, 0.38] | [-0.75, 0.73] |
| $C_{tZ}$ | [-0.37, 0.36] | [-0.52, 0.51] |
| $C^{[Im]}_{tZ}$ | [-0.38, 0.36] | [-0.54, 0.51] |

Table 6: Expected 68 % and 95 % CL intervals for the selected Wilson coefficients in a profiled scan over the 2D parameter planes $C^{-}_{\phi Q}/C_{\phi t}$ and $C_{tZ}/C^{[Im]}_{tZ}$. The respective second parameter of the scan is left free.

| Wilson coefficient | 68 % CL $(\Lambda/\text{TeV})^2$ | 95 % CL $(\Lambda/\text{TeV})^2$ |
|---|---|---|
| $C_{\phi t}$ | [-1.65, 3.37] | [-2.89, 6.76] |
| $C_{\phi Q}$ | [-1.35, 2.92] | [-2.33, 6.69] |
| $C_{tZ}$ | [-0.37, 0.36] | [-0.52, 0.51] |
| $C^{[Im]}_{tZ}$ | [-0.38, 0.36] | [-0.54, 0.51] |

## 8 Summary

The CMS sensitivity to anomalous interactions using $t\bar{t}Z$ measurements in the HL-LHC era corresponding to a simulated data set of 3 ab $^{-1}$ of integrated luminosity has been been estimated in the context of SM-EFT. The considered scenario assumed advances in both experimental methods and theoretical descriptions of the relevant physics effects. With the reduced theoretical and experimental uncertainties, tight constraints are expected in two planes spanned by a total of four Wilson coefficients and in one dimensional log-likelihood scans.

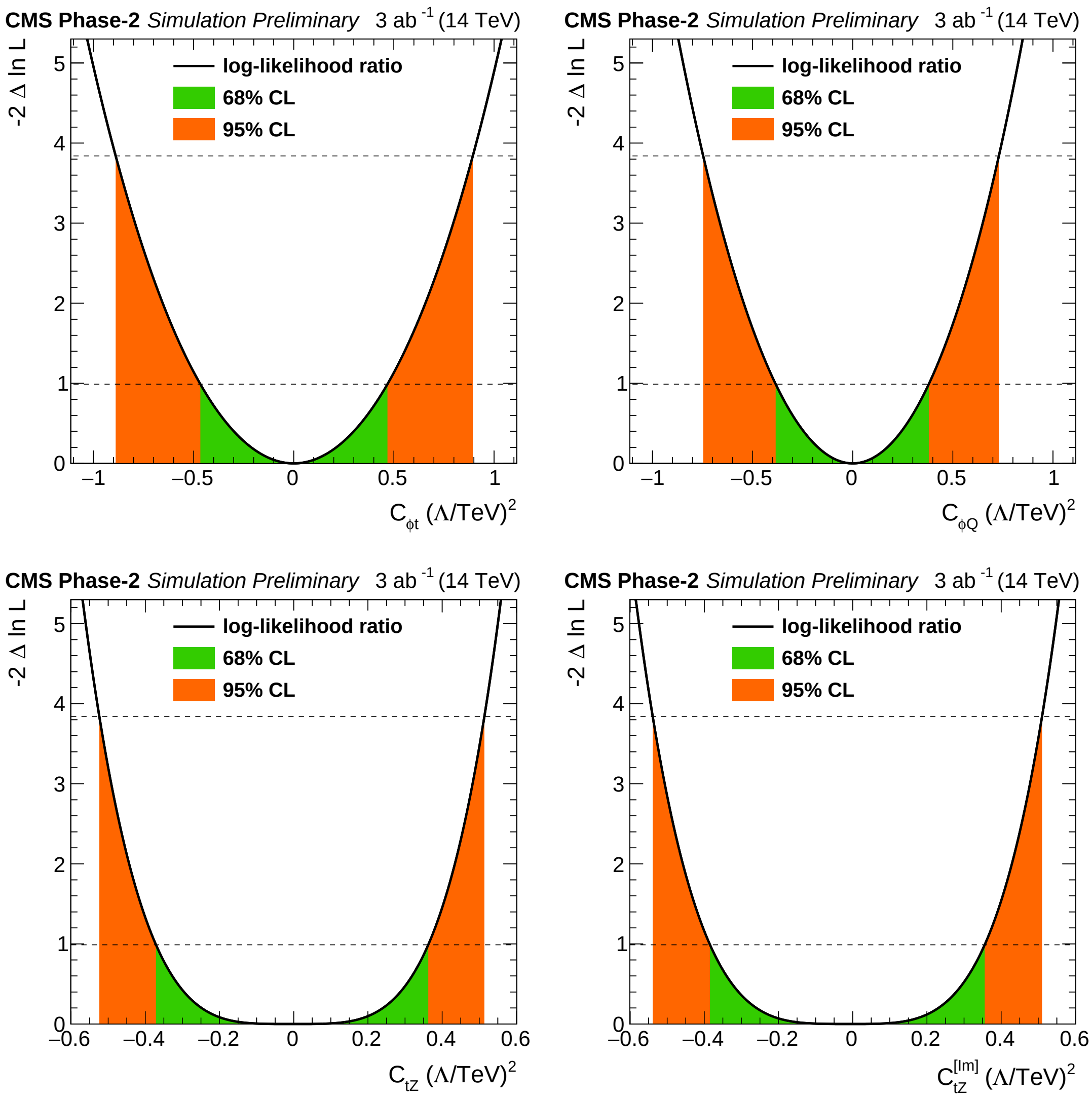

Figure 4: Individual likelihood ratio for the Wilson coefficients cpt and cpQM (top) and ctZ and ctZI (bottom) for the ttZ process. Here, only one Wilson coefficient at a time is considered non-zero. The 68% (95%) CL intervals are given in green (red).

**10**

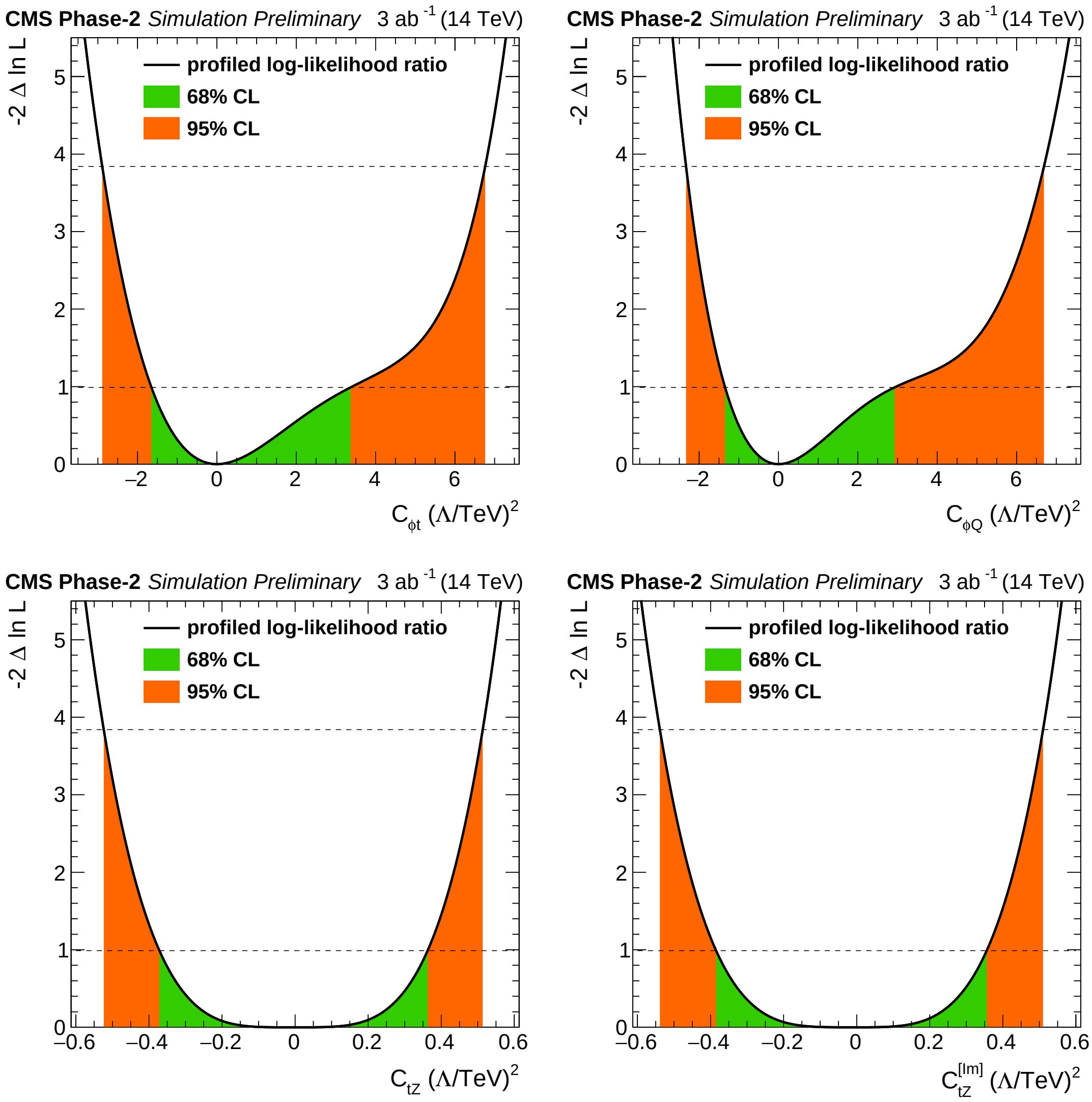

Figure 5: Individual profiled likelihood ratio for the Wilson coefficients $C_{\phi t}$ and $C^{-}_{\phi Q}$ (top) and $C_{tZ}$ and $C^{[Im]}_{tZ}$ (bottom) for the $t\bar{t}Z$ process under the SM hypothesis. The 68% (95%) CL intervals are given in green (red).

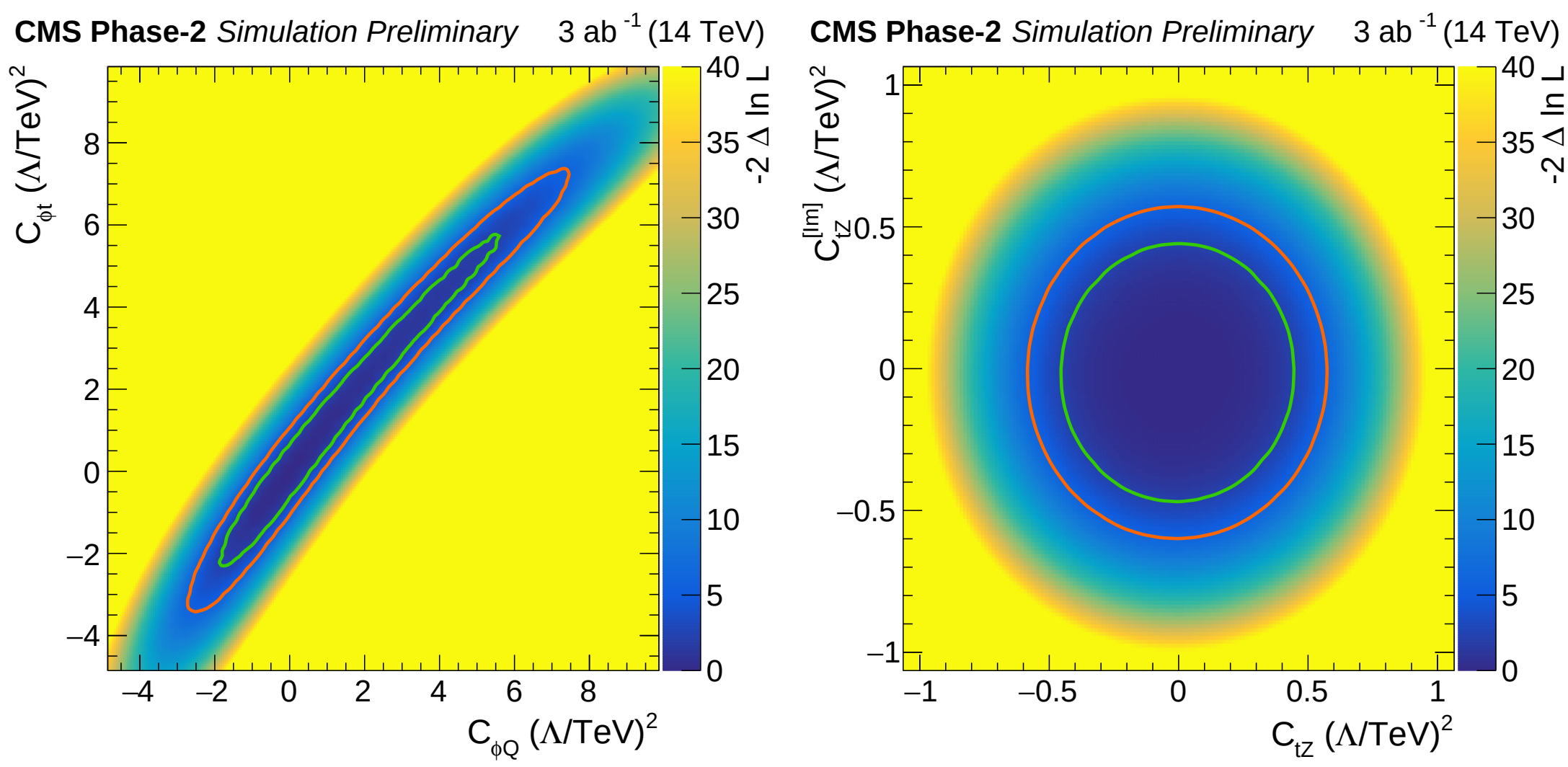

Figure 6: Scan of the negative likelihood in the $C^{-}_{\phi Q}/C_{\phi t}$ (left) and $C_{\mathrm{tZ}}/C^{[\mathrm{Im}]}_{\mathrm{tZ}}$ parameter planes (right) for the $\mathrm{t\bar{t}Z}$ process under the SM hypothesis. The 68% (95%) CL contour lines are given in green (red).

**12**

# CMS Physics Analysis Summary

Contact: cms-future-conveners@cern.ch

## Expected sensitivities for $t\bar{t}t\bar{t}$ production at HL-LHC and HE-LHC

The CMS Collaboration

### Abstract

The CMS searches for the production of four top quarks ($t\bar{t}t\bar{t}$) are used to provide projections for the High-Luminosity LHC and High-Energy LHC. Final states with same sign leptons or three or more leptons as well as multiple b-tagged jets are used in these projections. Several different scenarios for the systematic uncertainties are considered. For proton-proton collisions at $\sqrt{s} = 14$ TeV, the existing analysis strategies are expected to become dominated by systematic uncertainties. Evidence for $t\bar{t}t\bar{t}$ in a single analysis will become possible with around 300 fb$^{-1}$ of High-Luminosity LHC data at 14 TeV center-of-mass energy. With these datasets the uncertainty on the measured cross section will be of the order of 33 to 43%, depending on the systematic uncertainty. With 3 ab$^{-1}$ of High-Luminosity LHC data, the cross section can be constrained to 9% statistical uncertainty and 18 to 28% total uncertainty. At High-Energy LHC it would be possible to constrain the $t\bar{t}t\bar{t}$ cross section to within 1 to 2% statistical uncertainty.

The production of four top quarks ($t\bar{t}t\bar{t}$) is one of the rare processes in top quark physics that has large sensitivity to variety of New Physics effects that could be studied through direct searches, effective filed theory approaches or top quark-Higgs boson anomalous couplings, while at the same time it is interesting in the standard model context as a complex QCD process. The cross section is about one order of magnitude smaller than $t\bar{t}H$ production, with several precision calculations predicting values of $\sigma_{t\bar{t}t\bar{t}} = 9.2^{+2.9}_{-2.4}$ fb (NLO) and $\sigma_{t\bar{t}t\bar{t}} = 11.97^{+2.15}_{-2.51}$ fb (NLO+EWK)[1–3] for proton collisions at $\sqrt{s} = 13\,\text{TeV}$. The former value is used as a starting point in this study, as this was the value used in the experimental literature up to now.

CMS has published three analyses setting limits on $t\bar{t}t\bar{t}$ production [4–6] in the context of a search specifically designed for the standard-model signature, and both ATLAS and CMS have published multiple papers where limits on $t\bar{t}t\bar{t}$ production were derived as a side product of searches, typically coming from searches for vector-like quarks ($pp \to T\bar{T}/B\bar{B} \to t\bar{t}W^+W^-$) or MSSM ($pp \to \tilde{g}\tilde{g} \to \tilde{t}\bar{t}t\bar{t} \to t\bar{t}t\bar{t} + p_T^{miss}$) signatures [7–12]. The $t\bar{t}t\bar{t}$ process has not yet been observed, and the most sensitive CMS collaboration result sets 95% confidence level (CL) upper limits on the production cross section value of $20.8^{+11.2}_{-6.9}$ fb, which is equivalent to an excess with an expected significance of 1.0 standard deviations above the background-only hypothesis.

The production of $t\bar{t}t\bar{t}$ is one of the rare standard Model (SM) processes that is expected to be discovered and studied by future LHC runs, including the High-Luminosity LHC (HL-LHC) and the High-Energy LHC (HE-LHC). The increase in collision energy is important for $t\bar{t}t\bar{t}$ production because the cross section is still heavily dependent on the gluon parton density function (PDF) at pp collisions at $\sqrt{s} = 13\,\text{TeV}$, leading to a substantial improvement in the signal-to-background ratio when the collision energy of the LHC is increased. Investigations of the expected increase in cross section using the MadGraph5_aMC@NLO generator [1], indicated that the $t\bar{t}t\bar{t}$ cross section increases by a factor of approximately 1.3 when increasing the collision energy from 13 to 14 TeV, and by a factor of approximately 12.8 when increasing the collision energy from 13 to 27 TeV.

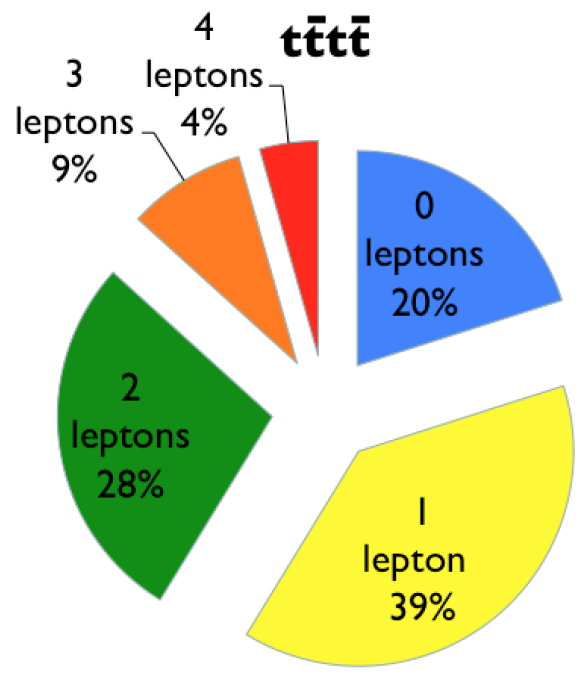

Figure 1: Summary of the branching fractions of $t\bar{t}t\bar{t}$ production.

At the LHC, $t\bar{t}t\bar{t}$ provides a particularly rich set of experimental signatures. In the standard model the four W bosons from the top quark decays can create striking leptonic signatures with four b quark jets and in association with many jets. Figure 1 summarises the branching fractions of the $t\bar{t}t\bar{t}$ process, where the largest fraction of events creates single charged lepton or dilepton signatures. The main backgrounds for $t\bar{t}t\bar{t}$ searches depend on the final state, but for the majority of the decay modes are originating from $t\bar{t}$ plus additional radiation that can include on-shell objects, such as $t\bar{t}$+vector bosons, Higgs bosons, and jets. Backgrounds with misidentified charged leptons are an another important source of backgrounds in decay channels with signatures containing many charged leptons, these originate from the production of one or more vector bosons and $t\bar{t}$+two vector bosons. Many additional jets are required be-yond the $t\bar{t} + b\bar{b}$ final state, which is the reason why many of these backgrounds can be further suppressed. The majority of the backgrounds are well understood processes which can be estimated from MC simulation with additional corrections from control regions or misidentification studies.

In this note, a simple rescaling of the results of Ref. [4] is presented. This paper considers the dataset collected at $\sqrt{s} = 13$ TeV using an integrated luminosity equivalent to 36 $\text{fb}^{-1}$. The final states sensitive to $t\bar{t}t\bar{t}$ production that are considered are those with two same-charge leptons or more than two charged leptons. The following sections will shortly summarize the analysis with a focus on the approach to statistical and systematic uncertainties as well signal isolation and background determination. The analysis is used unchanged, and quantitative information (e.g. selection efficiencies) on the objects used in the analysis can be found in [4].

### Search for $t\bar{t}t\bar{t}$ in same-sign dilepton and multilepton final states

The same-sign dilepton and multilepton search for $t\bar{t}t\bar{t}$ production [4] relies on a consolidated strategy in low-background searches that has been established by the CMS collaboration. Control Regions (CR) populated by events from specific background process are defined, and these CR are included in the maximum likelihood fit to determine the $t\bar{t}t\bar{t}$ signal strength. The dominant backgrounds determined using CR are $t\bar{t}W^{\pm}$ and $t\bar{t}Z/\gamma*$ production, while backgrounds from other rare processes, dominated by $t\bar{t}H$, are based on SM simulation predictions and are assigned large ($\pm$50%) normalisation uncertainties.

The experimental backgrounds from charge misidentification of leptons and non-prompt charged leptons are determined using data-driven methods in side bands. The invariant mass region around the Z boson resonance is rejected. Eight signal regions are defined, based on the number of charged leptons ($e, \mu$), number of jets, and number of b-tagged jets. Charged leptons are selected to pass well-established purity and efficiency criteria [13, 14] and $p_T > 20$ GeV, while b-tagged jets have a $p_T > 25$ GeV requirements. Untagged jets are subject to a more tight $p_T > 40$ GeV requirement. Jets are tagged as originating from b-quarks with the CMS DeepCSV algorithm [15]. A summary of the various signal regions that also gives an impression of the contribution of various backgrounds, including the expected yields, is listed in Tab. 1.

Using 35.9 $\text{fb}^{-1}$ of 13 TeV proton-proton collision data, this analysis is still dominated by statistical uncertainties. The systematic uncertainties are listed in Tab. 2.

Table 1: Definitions and expected yields with total uncertainties of the eight signal regions and the two control regions for $t\bar{t}W$ (CRW) and $t\bar{t}Z$ (CRZ), for a dataset of 35.9 $\text{fb}^{-1}$ at 13 TeV centre-of-mass. Adapted from Ref. [4].

| $N_{\text{leptons}}$ | $N_{\text{bjets}}$ | $N_{\text{jets}}$ | Region | SM background | $t\bar{t}t\bar{t}$ | Total |
|---|---|---|---|---|---|---|
| 2 | 2 | ≤5 | CRW | 83.7 ± 8.8 | 1.9 ± 1.2 | 85.6 ± 8.6 |
| | | 6 | SR1 | 7.7 ± 1.2 | 0.9 ± 0.6 | 8.6 ± 1.2 |
| | | 7 | SR2 | 2.6 ± 0.5 | 0.6 ± 0.4 | 3.2 ± 0.6 |
| | | ≥8 | SR3 | 0.5 ± 0.3 | 0.4 ± 0.2 | 0.8 ± 0.4 |
| | 3 | 5, 6 | SR4 | 4.0 ± 0.7 | 1.4 ± 0.9 | 5.4 ± 0.9 |
| | | ≥7 | SR5 | 0.7 ± 0.2 | 0.9 ± 0.6 | 1.6 ± 0.6 |
| | ≥4 | ≥5 | SR6 | 0.7 ± 0.2 | 1.0 ± 0.6 | 1.7 ± 0.6 |
| ≥3 | 2 | ≥5 | SR7 | 2.3 ± 0.5 | 0.6 ± 0.4 | 2.9 ± 0.6 |
| | ≥3 | ≥4 | SR8 | 1.2 ± 0.3 | 0.9 ± 0.6 | 2.1 ± 0.6 |
| Inverted Z veto | | | CRZ | 31.7 ± 4.6 | 0.4 ± 0.3 | 32.1 ± 4.6 |

Table 2: Summary of the sources of uncertainty in the Run 2 (dataset collected in 2016) analysis, and their effect on signal and background yields. The first group lists experimental and theoretical uncertainties in simulated signal and background processes. The second group lists normalisation uncertainties of the estimated backgrounds. As reported in Tab. 3 from [4].

| Source | Uncertainty (%) |
|---|---|
| Integrated luminosity | 2.5 |
| Pileup | 0–6 |
| Trigger efficiency | 2 |
| Lepton selection | 4–10 |
| Jet energy scale | 1–15 |
| Jet energy resolution | 1–5 |
| b tagging | 1–15 |
| Size of simulated sample | 1–10 |
| Scale and PDF variations | 10–15 |
| ISR/FSR (signal) | 5–15 |
| $t\bar{t}H$ (normalization) | 50 |
| Rare, $X\gamma$, $t\bar{t}VV$ (norm.) | 50 |
| $t\bar{t}Z/\gamma*$, $t\bar{t}W$ (normalization) | 40 |
| Charge misidentification | 20 |
| Nonprompt leptons | 30–60 |

## Treatment of systematic uncertainties and background cross sections

Scenarios for the evolution of systematic uncertainties are listed in Tab. 3, which are equivalent to the so-called *Run 2*, *YR18* and *YR18+* scenarios used in other CMS upgrade studies. A scenario with purely statistical uncertainties is also included for comparison. These scenarios are defined as follows:

- The *Stat. only* scenario only considers statistical uncertainties on the data. Uncertainties due to statistical fluctuations of data in control regions are substantially smaller than the statistical uncertainty on the yield in the signal regions, so are negligible in large datasets when derived from a data-driven method.
- The *Run 2* scenario considers the case where the systematic uncertainties remain unchanged. This means that all systematic uncertainties for the analysis are assumed to be unchanged with respect to the published analysis. Statistical uncertainties scale as expected by the increase in integrated luminosity, meaning as $1/\sqrt{L/L_{ref}}$, where $L_{ref}$ is the integrated luminosity with which the original analysis was performed.
- The *YR18* scenario considers the case where the theory and experimental systematic uncertainties improve over time. In this scenario the experimental systematic uncertainties that are sensitive to the size of the dataset are also reduced as $1/\sqrt{L/L_{ref}}$. As these systematic uncertainties will never completely be negligible, a limit to this reduction is set to 50% of the currently achieved uncertainty. Theoretical uncertainties on the background are of course also expected to improve due to developments in the calculations, techniques and orders considered. So systematic uncertainties on the simulation originating from theoretical sources are scaled by 50%.
- The *YR18+* scenario is identical to the *YR18* scenario except that, for the experimental systematic uncertainties, no floor values are assumed.

The fractional changes to the yields of the dominant background predictions, $t\bar{t}$ plus jets, W, Z or H bosons, were determined at next-to-leading order (NLO) for LHC collisions at 13, 14 and 27 TeV using the using the MadGraph5_aMC@NLO generator [1]. The extremely rare backgrounds (i.e., $t\bar{t}$WW, $t\bar{t}t$W production and similar) with negligible contributions were assumed to be sensitive to the same parton luminosity increase as $t\bar{t}$H production. Data-driven background estimates were increased by the ratio of the $t\bar{t}$ cross section increase as a function of integrated luminosity, since this is a process of similar $q^2$ and Bjorken-$x$ values as these backgrounds after preselection of multiple jets and charged leptons.

Table 3: Considered systematic uncertainty scenarios, described in detail in the text. The table reports the scale factor multiplied to the uncertainties taken from the published CMS analysis as reported in Tab. 2.

| Source uncert. | *Stat. only* | *Run 2* | *YR18* | *YR18+* |
|---|---|---|---|---|
| Statistical | $(L/L_{ref})^{-0.5}$ | $(L/L_{ref})^{-0.5}$ | $(L/L_{ref})^{-0.5}$ | $(L/L_{ref})^{-0.5}$ |
| Experimental | None | Original | $\max(0.5, (L/L_{ref})^{-0.5})$ | $(L/L_{ref})^{-0.5}$ |
| Int. Luminosity | None | Original | 0.4 | 0.4 |
| Data-driven bckgrnd | None | Original | $\max(0.5, (L/L_{ref})^{-0.5})$ | $(L/L_{ref})^{-0.5}$ |
| Theory (shapes) | None | Original | 0.5 | 0.5 |
| Bckgrnd cross section | None | Original | 0.5 | 0.5 |
| Signal cross section | None | Original | 0.5 | 0.5 |

## Results

Table 4: The expected significance of $t\bar{t}t\bar{t}$ signal over a background-only hypothesis in standard deviations (s.d.), is given for various CMS upgrade scenarios for sqrt(s)=14 TeV.

| Int. Luminosity | *Stat. only* | *Run 2* | *YR18* | *YR18+* |
|---|---|---|---|---|
| 300 $\text{fb}^{-1}$ | 4.09 | 2.71 | 2.85 | 2.93 |
| 3 $\text{ab}^{-1}$ | 12.9 | 3.22 | 4.26 | 4.49 |

For the high luminosity LHC, the *stat-only*, *S1+*, *S2+* and *S2NF+* scenarios are considered at a collision energy of 14 TeV. The assumed integrated luminosity is 3 $\text{ab}^{-1}$. For reference, a 300 $\text{fb}^{-1}$ scenario is also considered. The results are listed in Tab. 4 and Fig. 2, and show that for the statistics-only case and for the optimistic scenarios (*S2+* and *S2NF+*), the evidence for the $t\bar{t}t\bar{t}$ signal may be reached already with 300 $\text{fb}^{-1}$, while even with 3 $\text{ab}^{-1}$ of integrated luminosity will be a challenge to perform an observation with a single analysis. Alternatively, combining the analysis with complementary final states should be sufficient for observation. The reinterpreted analysis relies on small backgrounds which can be estimated with relatively large uncertainties. This means that the result becomes dominated by systematic uncertainties in large datasets, suggesting that a modified analysis strategy focusing on reduction of these uncertainties will greatly improve the sensitivity.

Considering the sensitivity of $t\bar{t}t\bar{t}$ production to new physics scenarios in the top quark and scalar sector, it is useful to consider how accurately the cross section can be measured with the analyses, once sufficient integrated luminosity has been collected. Of course in the future analysis techniques are also expected to improve, and dedicated analyses will surely improve this sensitivity, but this is beyond the scope of this study. It is, however, important to keep in mind that such a sensitivity study is less sensitive to systematic uncertainties on the background determination, while being more sensitive to the signal modelling uncertainties and overall branching fraction and acceptance of the selection. The expected sensitivity on the $t\bar{t}t\bar{t}$ cross section is listed in Tab. 5, and shows that measurements with 30% accuracy are possible at the start of HL-LHC which can be reduced to the order 20% at the end of the HL-LHC data taking, with a statistical uncertainty of 10% or less.

It is also possible to look further into the future, to the High-Energy LHC. At this point it is a valid question to ask if any of the systematic uncertainty scenarios are reasonable, but the statistical uncertainty should definitely still be possible to be assessed. At these time scales changes in analysis strategy might allow analysis improvements that focus on the optimization of the interplay between the statistical or systematic uncertainty. The process should at this point already be observed, so Tab. 5 only lists the expected sensitivity to measure the $t\bar{t}t\bar{t}$ cross section using $3 - 15$ $\text{ab}^{-1}$ of pp collision data at $\sqrt{s} = 27$ TeV.

### EFT interpretation

The expected sensitivity on the $t\bar{t}t\bar{t}$ cross section as listed in Tab. 5 can be interpreted in an effective field theory approach [16, 17]. The order-6 Effective-Field-Theory (EFT) Lagrangian reads

$$\mathcal{L}_{\text{EFT}} = \mathcal{L}_{\text{SM}}^{(4)} + \frac{1}{\Lambda}\sum_k C_k^{(5)}\mathcal{O}_k^{(5)} + \frac{1}{\Lambda^2}\sum_k C_k^{(6)}\mathcal{O}_k^{(6)} + o\left(\frac{1}{\Lambda^2}\right), \tag{1}$$

where $\mathcal{L}_{\text{SM}}^{(4)}$ is the renormalizable standard model Lagrangian, $\mathcal{O}_k^{(n)}$ denotes dim-$n$ composite operators, while $C_k^{(n)}$ are corresponding coupling parameters, which are called Wilson coefficients. Each term in the sum is suppressed by $\Lambda^{d-4}$ constant, where $d$ is the scaling dimension

Table 5: Expected sensitivity for the production cross section of $t\bar{t}t\bar{t}$ production, in percent, at 68% confidence level. The fractional uncertainty on the cross section signal strength is given for various LHC upgrade scenarios. Cross sections are corrected for the changes expected by $\sqrt{s}$. For the 15 ab$^{-1}$ 27 TeV scenario, the systematic uncertainty extrapolation is no longer valid, so only the statistical uncertainty is provided.

| Int. Luminosity | $\sqrt{s}$ | *Stat. only* (%) | *Run 2* (%) | *YR18* (%) | *YR18+* (%) |
|---|---|---|---|---|---|
| 300 fb$^{-1}$ | 14 TeV | $^{+30}_{-28}$ | $^{+43}_{-39}$ | $^{+36}_{-34}$ | $^{+36}_{-33}$ |
| 3 ab$^{-1}$ | 14 TeV | $\pm 9$ | $^{+28}_{-24}$ | $^{+20}_{-19}$ | $\pm 18$ |
| 3 ab$^{-1}$ | 27 TeV | $\pm 2$ | $^{+15}_{-12}$ | $^{+9}_{-8}$ | $^{+8}_{-7}$ |
| 15 ab$^{-1}$ | 27 TeV | $\pm 1$ | | | |

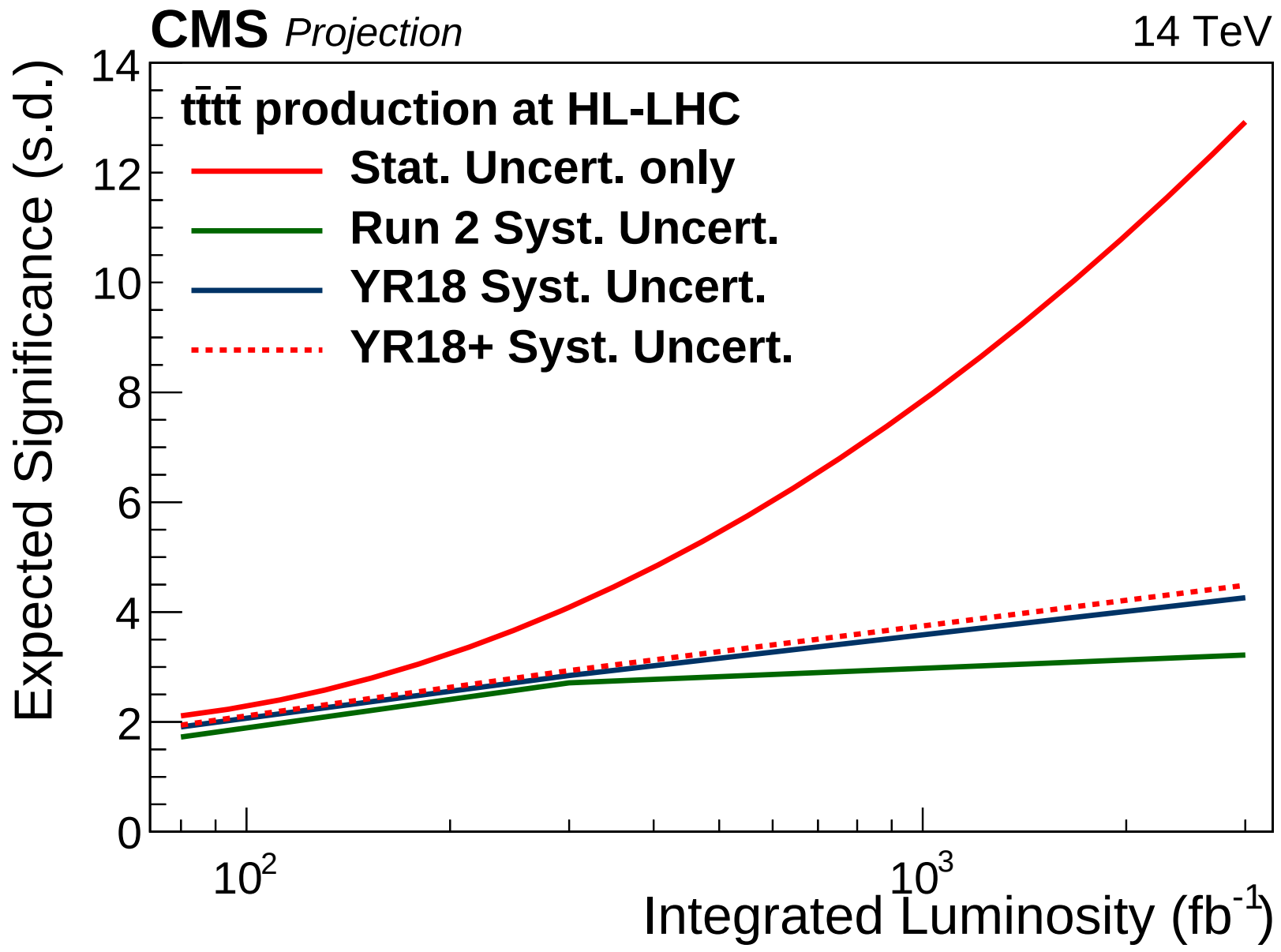

Figure 2: Expected significance of a search for $t\bar{t}t\bar{t}$ production with CMS at HL-LHC. The expected significance of $t\bar{t}t\bar{t}$ signal over a background-only hypothesis in standard deviations (s.d.) is given for various HL-LHC systematic uncertainty scenarios.

of relevant operators and $\Lambda$ is an effective energy cut-off of the model.

A minimal basis of composite dim-6 operators contributing in Eq. 1 was derived in [17]. Only a small subset of these operators can contribute to four top production. For the interpretation of the limits on $pp \to t\bar{t}t\bar{t}$ cross section, a different basis, proposed in [18, 19], is convenient. The list of contributing terms includes only following four-fermion operators

$$\mathcal{O}_R = (\bar{t}_R \gamma^\mu t_R)(\bar{t}_R \gamma_\mu t_R) \tag{2}$$

$$\mathcal{O}_L^{(1)} = (\bar{Q}_L \gamma^\mu Q_L)(\bar{Q}_L \gamma_\mu Q_L) \tag{3}$$

$$\mathcal{O}_B^{(1)} = (\bar{Q}_L \gamma_\mu Q_L)(\bar{t}_R \gamma_\mu t_R) \tag{4}$$

$$\mathcal{O}_B^{(8)} = \left(\bar{Q}_L \gamma_\mu T^A Q_L\right)\left(\bar{t}_R \gamma_\mu T^A t_R\right). \tag{5}$$

Since the data is sensitive only to the ratios $c_k \equiv C_k^{(6)}/\Lambda^2$, leading-order predictions for $pp \to t\bar{t}t\bar{t}$ cross section can be parametrised using new variables as

$$\sigma_{t\bar{t}t\bar{t}} = \sigma_{t\bar{t}t\bar{t}}^{\mathrm{SM}} + \sum_k c_k \sigma_k^{(1)} + \sum_{j \le k} c_j c_k \sigma_{j,k}^{(2)}, \tag{6}$$

where linear terms, $c_k\sigma_k^{(1)}$, represent interference of the SM production with dim-6 EFT contribution, while $c_jc_k\sigma_{j,k}^{(2)}$ terms correspond to insertion of two EFT operators. Arranging $c_k$ in a column-vector, $\vec{\mathbf{c}}$, the Eq. 6 can be expressed in the matrix form

$$\sigma_{t\bar{t}t\bar{t}}(\vec{\mathbf{c}}) = \sigma_{t\bar{t}t\bar{t}}^{\mathrm{SM}} + \vec{\mathbf{c}}^{\mathbf{T}} \cdot \vec{\boldsymbol{\sigma}}^{(1)} + \vec{\mathbf{c}}^{\mathbf{T}}\boldsymbol{\Sigma}^{(2)}\vec{\mathbf{c}}, \tag{7}$$

In order to find $\vec{\boldsymbol{\sigma}}^{(1)}$ and $\boldsymbol{\Sigma}^{(2)}$, a system of linear equations has to be solved, which is obtained by substituting linearly-independent vectors $\vec{\mathbf{c}}$ into Eq. 7. In the cross section calculation, the EFT interactions are implemented in the FEYNRULES [20] model and interfaced to MG5_aMC@NLO [1]. Coefficients of the Eq. 6 for $\sqrt{(s)} = 13, 14, 27$ TeV were determined independently. The NNPDF3.0LO [21] PDF set with $\alpha_S(M_{Z^0}) = 0.130$ were used and the high energy cut-off assumed the value $\Lambda = 1$ TeV.

The obtained combined experimental limit on $t\bar{t}t\bar{t}$ production can be utilized to provide constraints on effective field theory operators. The one standard-deviation uncertainties from from Tab. 5 can be used to constrain deviations from the standard model EFT when the $\sigma^{\mathrm{EFT}}/\sigma^{\mathrm{SM}}$ is larger than the uncertainty on the measurement of $\sigma^{\mathrm{SM}}$. Independent limits were obtained for the statistical uncertainties only, under the assumption that only two operator contribute to $t\bar{t}t\bar{t}$ cross section, while Wilson coefficients of the other operators were set to 0. The limits in two-dimensional space are shown in Figs. 3 and 4, where it is important to be aware that the reason why the HE-LHC ellipses behave differently is due to the drastically different PDF contribution of gluons vs quarks at increased collision energy. The resulting one-dimensional intervals are summarized in Table 6.

## Conclusion

The production of four top quarks has large sensitivity to new new physics effects and is interesting as a standard model QCD process. This note describes the reinterpretation of an analysis using 2016 data focusing on four top quark production using the same-sign dilepton and multilepton final states [4]. Multiple evaluation scenarios for the systematic uncertainties are considered. Evidence for $t\bar{t}t\bar{t}$ production will become possible with around 300 fb$^{-1}$ of HL-LHC data at $\sqrt{s} = 14$ TeV , at which point the statistical uncertainty on the measured cross section will be of the order of 30% and the measurement will have a total uncertainty of around 33-43%, depending on the systematic uncertainty scenario considered. For larger datasets at HL-LHC, all scenarios considered become dominated by systematic uncertainties. With 3 ab$^{-1}$ the cross section can be constrained to 9% statistical uncertainty, and the total uncertainty of a measurement ranges between 18% and 28% depending on the considered systematic uncertainties. At HE-LHC the $t\bar{t}t\bar{t}$ cross section is expected to be constrained to within a 1-2% statistical uncertainty.

Table 6: One-dimensional one-standard-deviation intervals for 14 and 27 TeV scenarios, for different fractional uncertainties on the measurement of $\sigma_{t\bar{t}t\bar{t}}$. Only total uncertainties are considered, allowing comparison to the uncertainties in Tab. 5.

| Wilson Coefficient | $\sqrt{s}$ | frac.unc. on $\sigma_{t\bar{t}t\bar{t}}$ (%) | Operator values | |
|---|---|---|---|---|
| $C_{O_R}$ | 14 TeV | 30 | 0.63 | -0.75 |
| | | 9 | 0.32 | -0.44 |
| | | 1 | 0.08 | -0.2 |
| $C_{O_R}$ | 27 TeV | 30 | 0.50 | -0.55 |
| | | 9 | 0.27 | -0.31 |
| | | 1 | 0.08 | -0.12 |
| $C_{O_{L1}}$ | 14 TeV | 30 | 0.63 | -0.75 |
| | | 9 | 0.32 | -0.44 |
| | | 1 | 0.08 | -0.2 |
| $C_{O_{L1}}$ | 27 TeV | 30 | 0.50 | -0.56 |
| | | 9 | 0.26 | -0.32 |
| | | 1 | 0.07 | -0.13 |
| $C_{B1}$ | 14 TeV | 30 | 1.21 | -1.22 |
| | | 9 | 0.66 | -0.67 |
| | | 1 | 0.22 | -0.23 |
| $C_{B1}$ | 27 TeV | 30 | 0.91 | -0.92 |
| | | 9 | 0.49 | -0.51 |
| | | 1 | 0.16 | -0.17 |
| $C_{B8}$ | 14 TeV | 30 | 2.04 | -2.64 |
| | | 9 | 1.00 | -1.61 |
| | | 1 | 0.22 | -0.82 |
| $C_{B8}$ | 27 TeV | 30 | 1.75 | -2.10 |
| | | 9 | 0.89 | -1.24 |
| | | 1 | 0.22 | -0.57 |

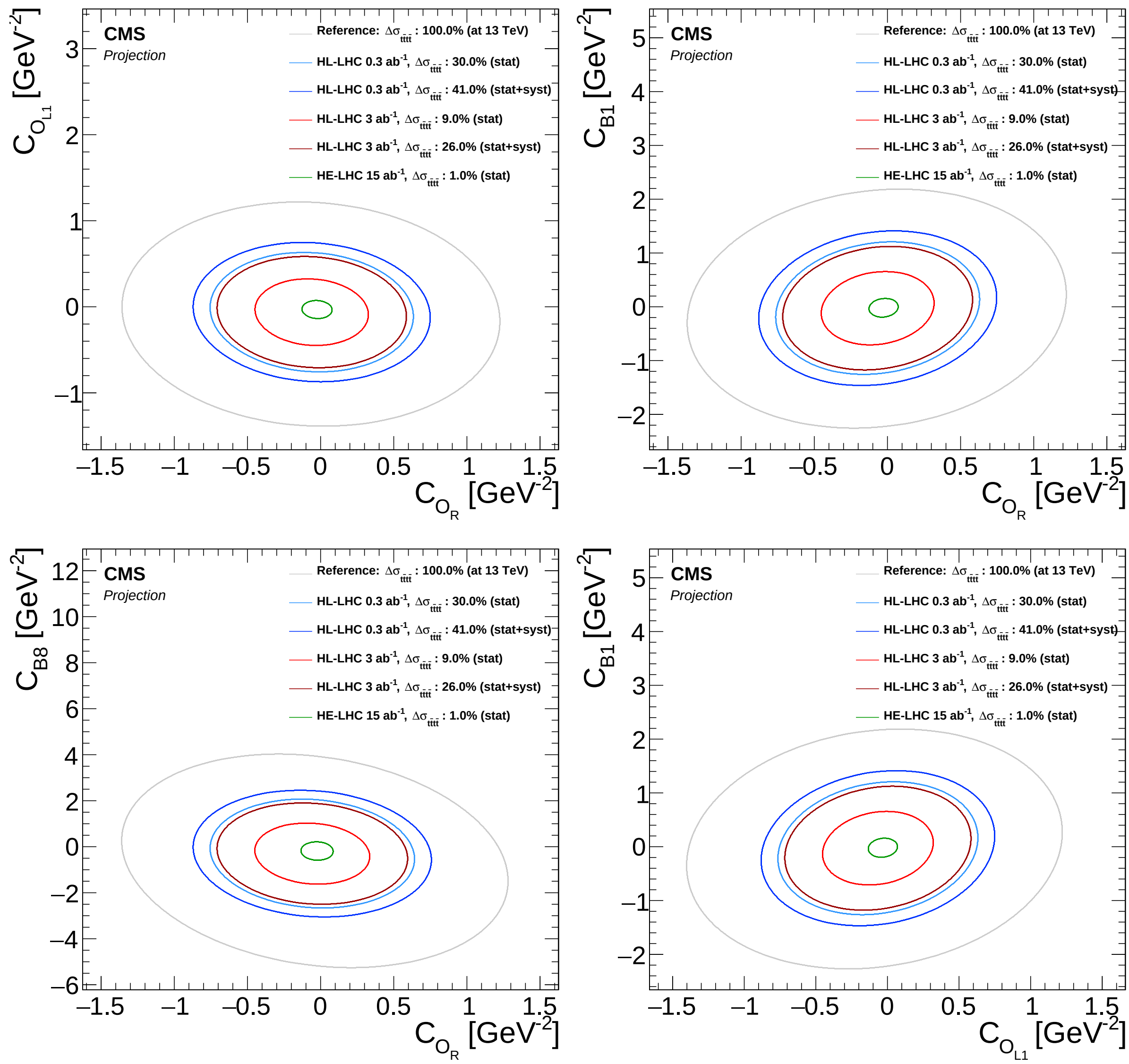

Figure 3: EFT interpretation plots in two dimensions. The shown ellipses are equivalent to the $t\bar{t}t\bar{t}$ cross section changing by one standard deviation of its statistical uncertainty from the projection. For reference, a curve with 100% expected uncertainty determined at $\sqrt{s} = 13$ TeV is shown. Only (expected) statistical uncertainties are considered unless explicitly mentioned..

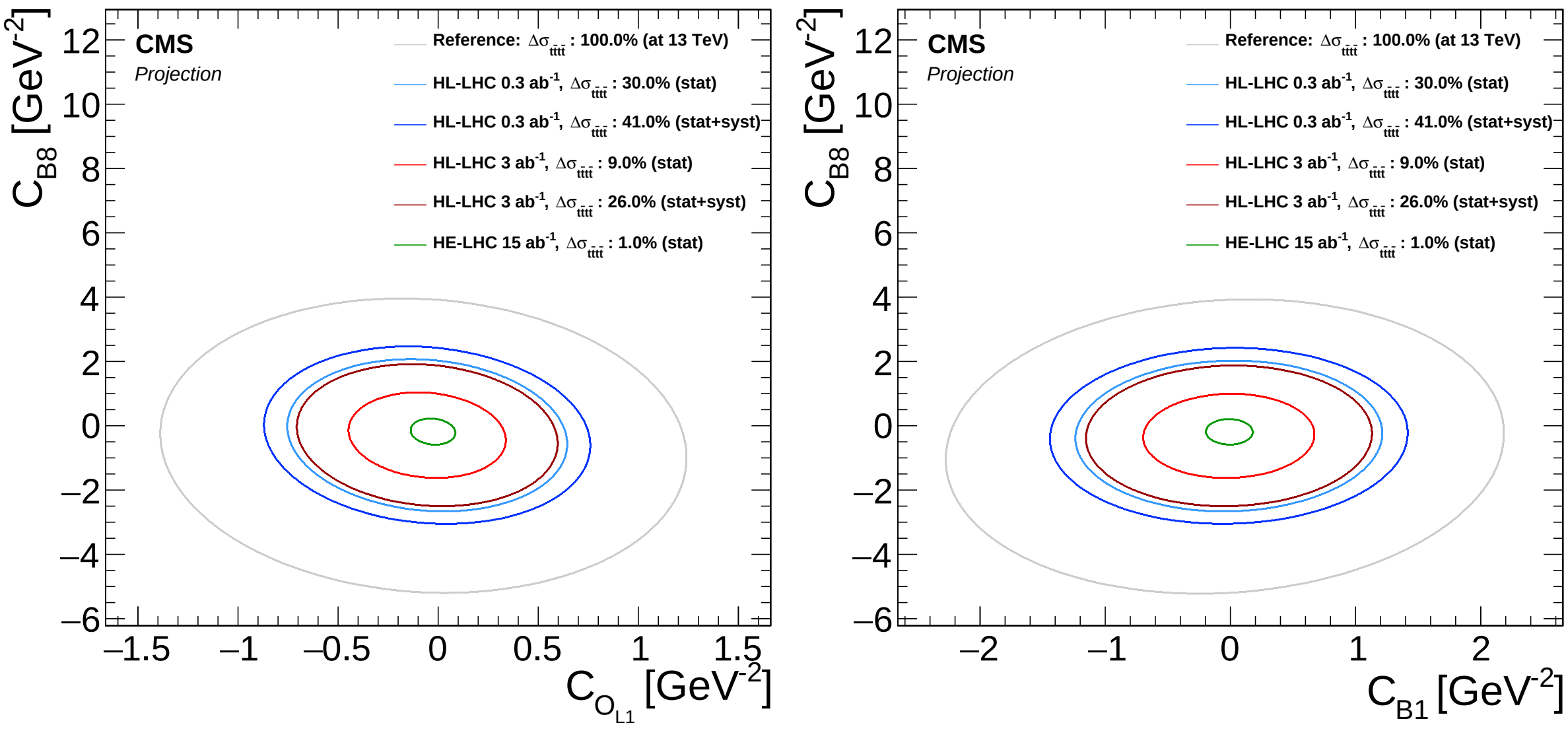

Figure 4: EFT interpretation plots in two dimensions. The shown ellipses are equivalent to the $t\bar{t}t\bar{t}$ cross section changing by one standard deviation of its statistical uncertainty from the projection. For reference, a curve with 100% expected uncertainty determined for $\sqrt{s} = 13$ TeV is shown. Only (expected) statistical uncertainties are considered unless explicitly mentioned.

**12**

**ATLAS PUB Note**

ATL-PHYS-PUB-2018-047

13th December 2018

# HL-LHC prospects for the measurement of the Standard Model four-top-quark production cross-section

The ATLAS Collaboration

This note presents projections for the measurement of the Standard Model four-top-quark production cross-section in the context of the High-Luminosity LHC with 3000 fb$^{-1}$ in proton-proton collisions at $\sqrt{s}$ = 14 TeV with the ATLAS experiment. The final states considered contain two same-charge leptons or at least three leptons. A precision of 11% on the four-top-quark production cross-section is expected to be achieved.

# 1 Introduction

In the Standard Model (SM) the production of four top quarks ($t\bar{t}t\bar{t}$) is a very rare process with an expected cross-section of $\sigma(pp \rightarrow t\bar{t}t\bar{t}) = 15.83^{+18\%}_{-21\%}$ fb at 14 TeV [1]. This process has not been observed. Many theories beyond the SM predict an enhancement of the $t\bar{t}t\bar{t}$ cross-section; examples include gluino pair production in supersymmetric models [2], pair production of scalar gluons [3, 4], and production of a heavy pseudoscalar or scalar boson in association with a $t\bar{t}$ pair in Type II two-Higgs-doublet models (2HDM) [5, 6]. In the context of Effective Field Theories, the $t\bar{t}t\bar{t}$ cross-section uniquely constrains the four-top-quark effective operators [7].

Four top quark production has been searched for at $\sqrt{s}$ = 13 TeV by the ATLAS and CMS experiments in the final states where one top quark decays leptonically (meaning to an electron or muon) [8, 9] and where at least two top quarks decay leptonically [10, 11]. The latter final state has a lower branching fraction but has lower background contamination when the two leptons have the same charge.

After the end of Run 3, the LHC will be upgraded to the High-Luminosity LHC (HL-LHC), significantly increasing its instantaneous luminosity. Upgrades of the ATLAS detector will be necessary to maintain its performance in the higher luminosity environment and to mitigate the impact of radiation damage and detector aging [12]. A new inner tracking system, extending the tracking region from pseudorapidity[1] $|\eta| < 2.7$ up to $|\eta| < 4$, will provide the ability to reconstruct forward charged particle tracks, which can be matched to calorimeter clusters for forward electron reconstruction, or associated to forward jets. The inner tracker extension also enables muon identification at high pseudorapidities if additional detectors are installed in the region $2.7 < |\eta| < 4$.

This note presents the prospect for measuring the SM $t\bar{t}t\bar{t}$ cross-section in the context of the HL-LHC with 3000 fb$^{-1}$ of proton-proton collisions at $\sqrt{s}$ = 14 TeV with the ATLAS experiment. Events with two same-charge leptons, and at least three leptons with at least five jets among which at least two are identified as originating from the hadronization of a $b$-quark ($b$-jet) are analyzed.

# 2 Simulated Samples

Samples that can give rise to two leptons with a same charge or at least three leptons are used. Monte Carlo (MC) samples for $t\bar{t}t\bar{t}$, $t\bar{t}$, single-top quarks (both $Wt$ and t-channels), a vector boson ($W$ or $Z$) or a Higgs boson in association with $t\bar{t}$, and multiboson production are processed. They are generated at $\sqrt{s}$ = 14 TeV and normalised to their theoretical cross-sections and to a luminosity of 3000 fb$^{-1}$. Only for the $t\bar{t}H$ process, the sample generated at $\sqrt{s}$ = 13 TeV is used while it is still normalised to the SM 14 TeV cross-section.

The $t\bar{t}t\bar{t}$ sample was generated at leading order in QCD with MG5_aMC@NLO v2.2.3 [13] and Pythia 8 (v8.186) [14] using the NNPDF2.3LO PDF set [15]. A sample of $t\bar{t}$ events was generated using Powheg-Box [16] and Pythia 8 using the NNPDF3.0NLO PDF set. Single-top quarks events were generated using Powheg-Box and Pythia 6 (v6.428). The $t\bar{t}Z/W$ samples were generated using MG5_aMC@NLO and Pythia 8 while multiboson events were generated using Sherpa [17]. The $t\bar{t}$ and single-top quarks

[1] ATLAS uses a right-handed coordinate system with its origin at the nominal interaction point (IP) in the centre of the detector and the $z$-axis along the beam pipe. The $x$-axis points from the IP to the centre of the LHC ring, and the $y$-axis points upward. Cylindrical coordinates $(r, \phi)$ are used in the transverse plane, $\phi$ being the azimuthal angle around the $z$-axis. The pseudorapidity is defined in terms of the polar angle $\theta$ as $\eta = -\ln\tan(\theta/2)$, and the rapidity $y$ is defined as $y = \frac{1}{2}\ln\frac{E+p_z}{E-p_z}$.

MC samples are used to model the background with fake, non-prompt or charge mis-identified leptons. This background is difficult to simulate, and it is usually evaluated with data-driven methods. For the purposes of this study, the normalisation of these samples is scaled based on the observed fake/non-prompt background fraction in the published Run 2 analysis of Ref. [10], as will be described in Section 4.

## 3 Object reconstruction and event selection

After the event generation step, a fast simulation of the trigger and detector effects is added with the dedicated ATLAS software framework [18]. The trigger, reconstruction and identification efficiencies, as well as the momentum/energy resolution of leptons and jets, are computed as function of their $\eta$ and $p_{\mathrm{T}}$ using simulation assuming an upgraded ATLAS detector [12], and are tabulated with functions which provide parameterised estimates of the ATLAS performance at the HL-LHC. These functions are then applied to the particle-level quantities. The functions assume the HL-LHC conditions of an instantaneous luminosity of $\mathcal{L} = 10^{34}$ cm$^{-2}$ s$^{-1}$ which implies an average number of additional collisions per bunch-crossing 200. More details on the object smearing and the corresponding performance can be found in Ref. [19].

Electrons and muons are reconstructed in the fiducial region of transverse momentum $p_{\mathrm{T}} > 25$ GeV and $|\eta| < 2.5$. Because of the $t\bar{t}t\bar{t}$ signal topology, no significant gain is obtained by extending the $\eta$ range for the leptons. Jets are selected with $p_{\mathrm{T}} > 25$ GeV, in the range $|\eta| < 4$. Leptons are required to be isolated, using the sum of the transverse energies of the charged and neutral truth particles within $\Delta R = \sqrt{(\Delta\eta)^2 + (\Delta\phi)^2} < 0.2$ around the lepton, denoted by $etcone20$. This amount of energy divided by the lepton $p_{\mathrm{T}}$ is required to be < 0.23(0.11) for electron (muon) candidates. The collection of selected jets includes simulated pileup effects. A track-based pileup jet rejection technique is simulated, assuming 2% efficiency to select a pileup jet as a hard-scatter jet. Jets containing $b$-hadrons are identified with a ghost-matching procedure [20] and are assigned $p_{\mathrm{T}}$- and $\eta$-dependent weights to reproduce a 70% $b$-tagging efficiency. The missing transverse momentum ($E_{\mathrm{T}}^{\mathrm{miss}}$) is simulated with parameterized contributions to soft terms, which are calculated with tracks matched to the primary vertex that are not associated with any reconstructed objects.

Events are selected if they contain at least two leptons with the same charge (2l) or at least three leptons (3l). At least five jets among which at least two are $b$-tagged are required. The selected leptons are required to have $\Delta R > 0.2$ with respect to any selected jets. In case an event contains a pair of same-flavour opposite-charge leptons, the invariant mass of these two leptons is required to satisfy: $|m_{ll} - 91| > 10$ GeV. In addition the scalar sum of the $p_{\mathrm{T}}$ of all selected jets and leptons ($H_T$) is required to be >500 GeV and $E_{\mathrm{T}}^{\mathrm{miss}}$ is required to be >40 GeV.

The distributions of $E_{\mathrm{T}}^{\mathrm{miss}}$ and $H_T$ are shown in Figure 1 after requiring two same-charge leptons or three leptons, at least 5 jets and at least 2 $b$-jets as preselection.

## 4 Analysis strategy

To extract the $t\bar{t}t\bar{t}$ cross-section a fit is performed to the $H_T$ distributions in several signal regions defined according to the jet and $b$-jet multiplicities. The analysis is fitting the $t\bar{t}t\bar{t}$ cross-section normalised to the prediction from the SM ($\mu$). The definition of the signal regions is provided in Table 1.

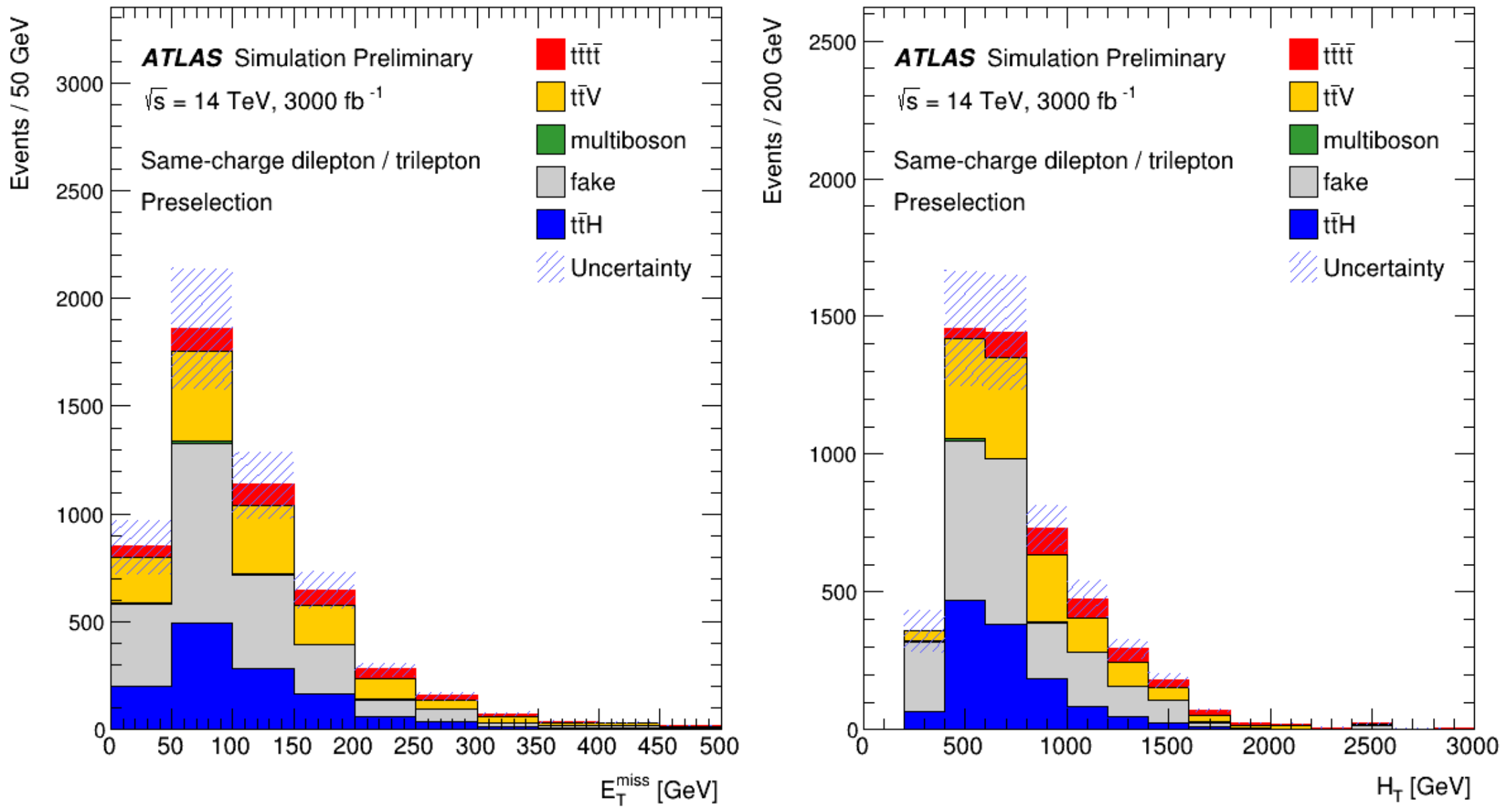

(a) (b)

Figure 1: Expected distributions of (a) $E_{\mathrm{T}}^{\mathrm{miss}}$ and (b) $H_T$ after the preselection requirements of two same-charge leptons or three leptons, at least 5 jets and at least 2 $b$-jets. The last bin contains overflows.

| | SR2l-6j3b | SR2l-6j4b | SR3l-6j3b | SR3l-6j4b |
|---|---|---|---|---|
| lepton requirement | 2l | 2l | 3l | 3l |
| jet requirement | $\geq 6$ | $\geq 6$ | $\geq 6$ | $\geq 6$ |
| $b$-jet requirement | $= 3$ | $\geq 4$ | $= 3$ | $\geq 4$ |

Table 1: Summary of selection requirements used to define the signal regions considered.

The rate of the instrumental background (electron with mis-identified charge, fake lepton or non-prompt lepton) is difficult to estimate using MC, but it has been shown in the published Run 2 analysis of Ref. [10] that it mostly comes from $t\bar{t}$ events. Therefore the fraction of instrumental background in the relevant regions with different lepton and $b$-tagged jet multiplicities observed in the Run 2analysis of Ref. [10] is used to scale the sum of $t\bar{t}$ and single-top MC events in the current analysis. These fractions are assumed to be independent of the requirement on the number of jets while varying with the lepton and $b$-jet multiplicities. To increase the statistics used to build $H_T$ template distributions for the fake background, $t\bar{t}$ and single-top MC events are selected with a relaxed isolation criteria: $etcone20/p_{\mathrm{T}} < 1.0$ for both electrons and muons. If no $t\bar{t}$ or single top MC events survive the selection, the $H_T$ distribution from the next lower $b$-jet multiplicity region is used. These $H_T$ templates are then scaled so that the fractions of fakes over the total background yield are: 44% for SR2l-6j3b, 32% for SR2l-6j4b, and 1.5% for SR3l-6j3b and SR3l-6j4b.

The number of events selected in the different signal regions are shown in Table 2 and in Figure 2.

The signal over background ratio in the different regions as well as pie charts representing the background composition are shown in Figures 3 and 4, respectively.

| | SR2l-6j3b | SR2l-6j4b | SR3l-6j3b | SR3l-6j4b |
|---|---|---|---|---|
| $t\bar{t}V$ | 20 ± 1 | 3 ± 1 | 13 ± 1 | 1.9 ± 0.8 |
| multiboson | <0.1 | <0.1 | <0.1 | <0.1 |
| fake | 54 ± 16 | 4 ± 1 | 0.4 ± 0.1 | 0.06 ± 0.02 |
| $t\bar{t}H$ | 48 ± 1 | 7 ± 3 | 13 ± 1 | 2.1 ± 0.9 |
| $t\bar{t}t\bar{t}$ | 78 ± 8 | 32 ± 3 | 61 ± 6 | 23 ± 2 |
| Total | 200 ± 18 | 46 ± 5 | 87 ± 6 | 27 ± 3 |

Table 2: Event yields of signal and background processes in the different signal regions used to extract the $t\bar{t}t\bar{t}$ cross section for an integrated luminosity of 3000 fb$^{-1}$. The uncertainties include the systematic sources described in Section 5.

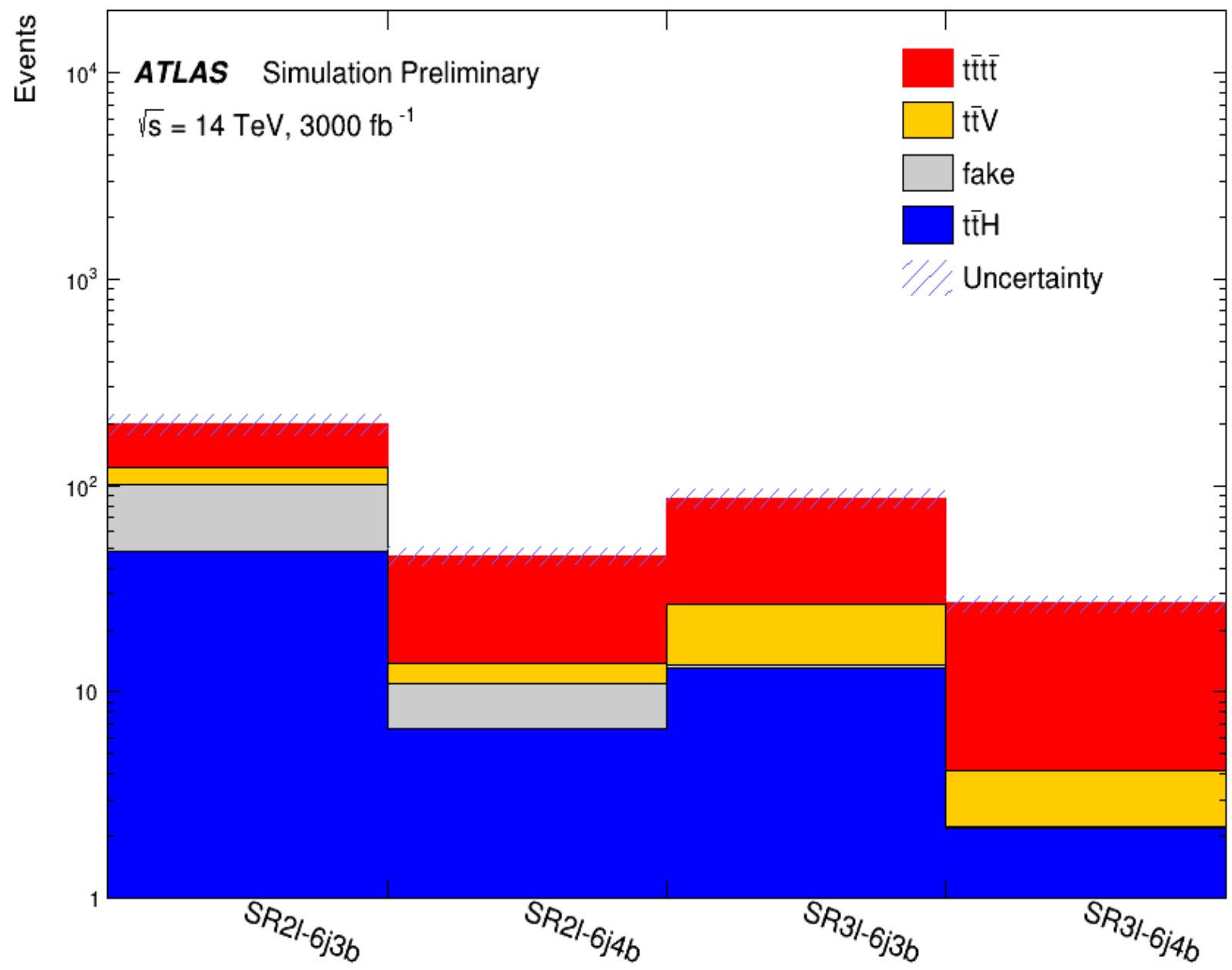

Figure 2: Number of selected events in the different signal regions. The hashed regions correspond to the systematic uncertainties described in the text.

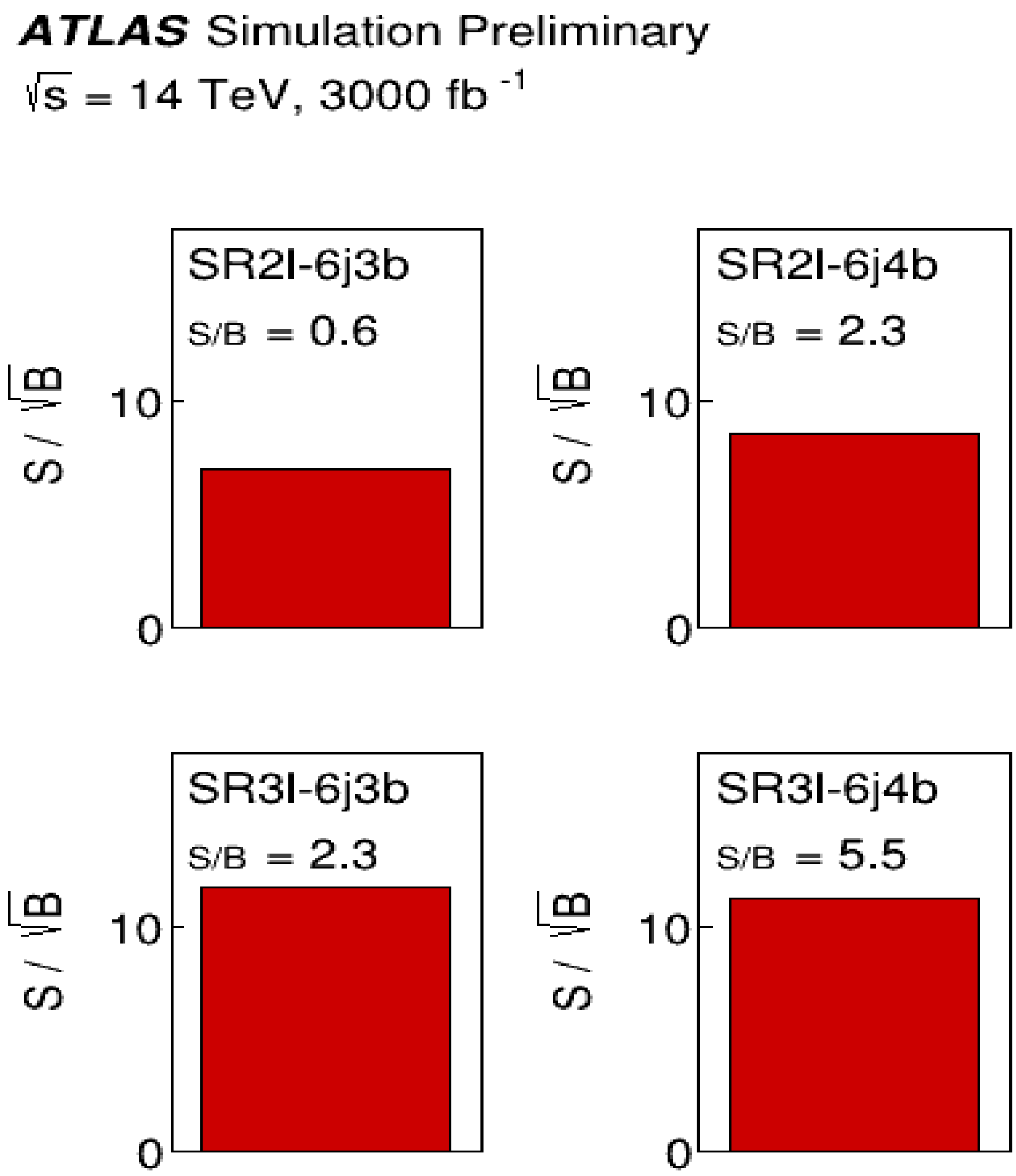

Figure 3: Signal over square root of background ratio in the different signal regions.

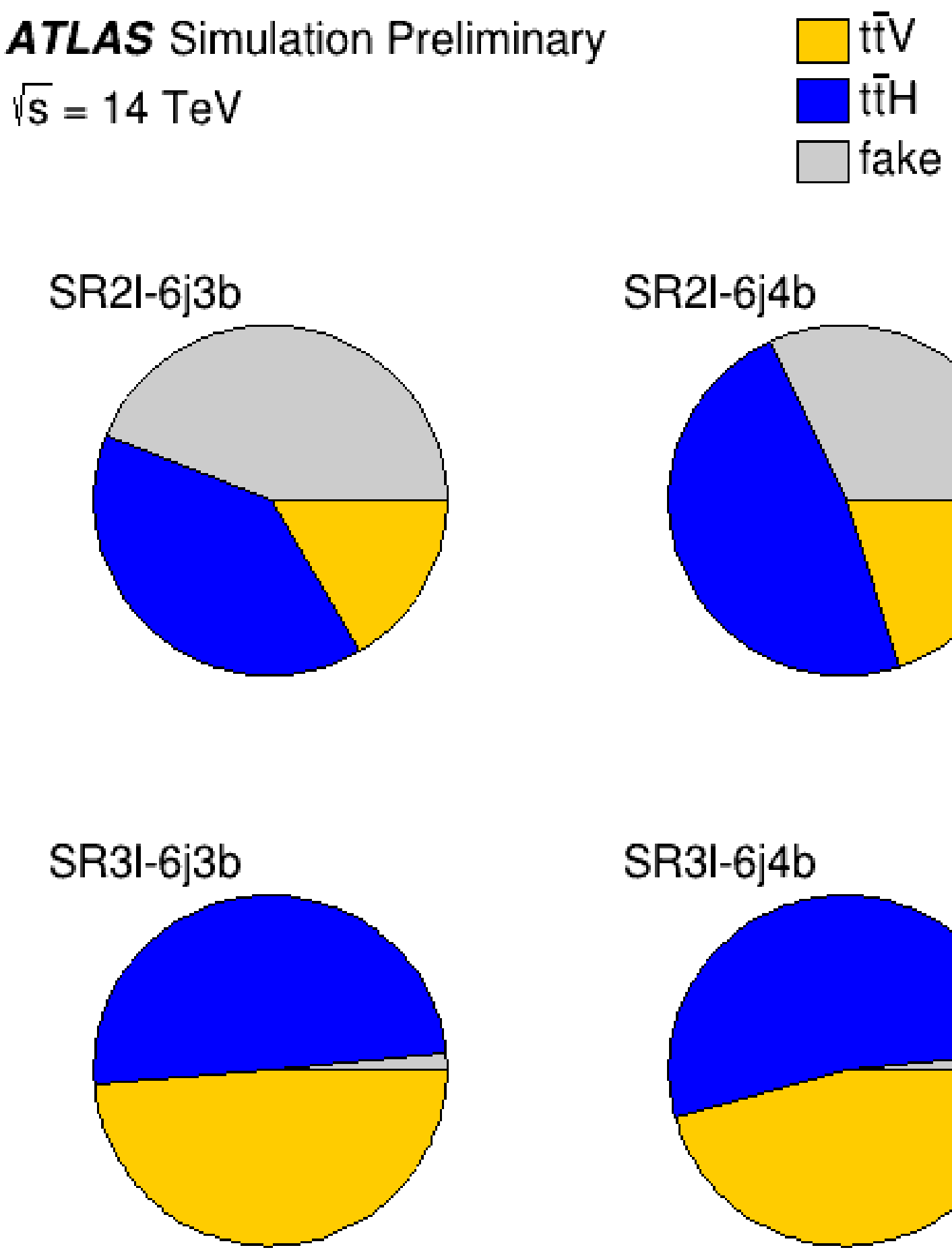

Figure 4: Fractional contributions of the various backgrounds to the total background prediction in each signal region.

# 5 Systematic uncertainties

In the the published Run 2 analysis of Ref. [10], the main sources of systematic uncertainties were found to be the uncertainties on the fake lepton background and on the SM background cross-section normalisation. Therefore, in this study experimental uncertainties (e.g. on jet energy scale or $b$-tagging efficiency) have been neglected. The theoretical uncertainty on the predicted signal cross section enters the determination of $\mu$ but not the uncertainty on the measured $t\bar{t}t\bar{t}$ cross section. The following systematic uncertainties are taken into account in this analysis assuming increasing uncertainty with the jet and $b$-jet multiplicity. A 15% (7%) overall normalisation uncertainty is assigned on the $t\bar{t}V$ ($t\bar{t}H$) backgrounds. In addition increasing uncertainties are assigned in the different signal regions as the relevant backgrounds come from $t\bar{t}V$ and $t\bar{t}H$ events with increasing number of jets: a 30% additional uncertainty is added on the $t\bar{t}V$ and $t\bar{t}H$ backgrounds in SR2l-6j3b, 40% in SR2l-6j4b and SR3l-6j3b and 50% in SR3l-6j4b. An overall shape uncertainty is added coming from scale variations, generator and parton shower variations corresponding to a 20% (10%) linear variation of the $H_T$ distributions for $t\bar{t}V$ ($t\bar{t}H$) events. As the instrumental background is estimated using scaling, a 30% uncertainty per signal region is assigned to it based on Run 2 analyses [10, 21]. A shape uncertainty of 20% per signal region is also assigned. No uncertainties related to the statistics of the MC samples used for the background estimation are applied. Finally a 10% uncertainty on the signal normalisation is assumed.

The systematic uncertainties are treated as nuisance parameters as described in Section 6. The input $H_T$ distributions used in the fit to extract the $t\bar{t}t\bar{t}$ cross-section are shown in Figure 5. The binning of these distributions is automatically determined to avoid bins with very low statistics while still keeping good significance.

# 6 Results

A maximum-likelihood fit of the $H_T$ distributions is performed simultaneously in the six signal regions to extract the $t\bar{t}t\bar{t}$ signal cross-section normalised to the prediction from the SM. The statistical analysis implemented in the `RooFit` package [22] uses a binned likelihood function $L(\mu, \theta)$. The $H_T$ distribution is used as the final discriminant in the six signal regions. The impact of systematic uncertainties on the background expectations is described by nuisance parameters $\theta$.

As a result of the fit, the uncertainy on the best-fit value of $\mu$ is found to be $\pm 0.16$ corresponding to a 11% uncertainty on the measured $t\bar{t}t\bar{t}$ cross-section. The corresponding significance is well above 5 standard deviations. A significance of around 5 standard deviations should be achievable with a luminosity of 300 fb$^{-1}$ assuming a center of mass energy of $\sqrt{s} = 14$ TeV.

The ranking obtained for the nuisance parameters ordered according to the largest contribution to the uncertainty in the signal strength is shown in Figure 6. The largest impacts come from the normalisation of the $t\bar{t}t\bar{t}$ signal as well as on the $t\bar{t}V$ background in the SR3l-6j3b and on the instrumental background in the SR2l-6j3b region. The shape uncertainties do not significantly affect the result. Measuring the $t\bar{t}V$ and $t\bar{t}H$ backgrounds differentially as a function of jet multiplicity would decrease the impact of these systematic uncertainties. Overall the impact of the systematic uncertainties is however modest as a fit without systematic uncertainties leads to a precision of 9% on the extracted $t\bar{t}t\bar{t}$ cross-section.

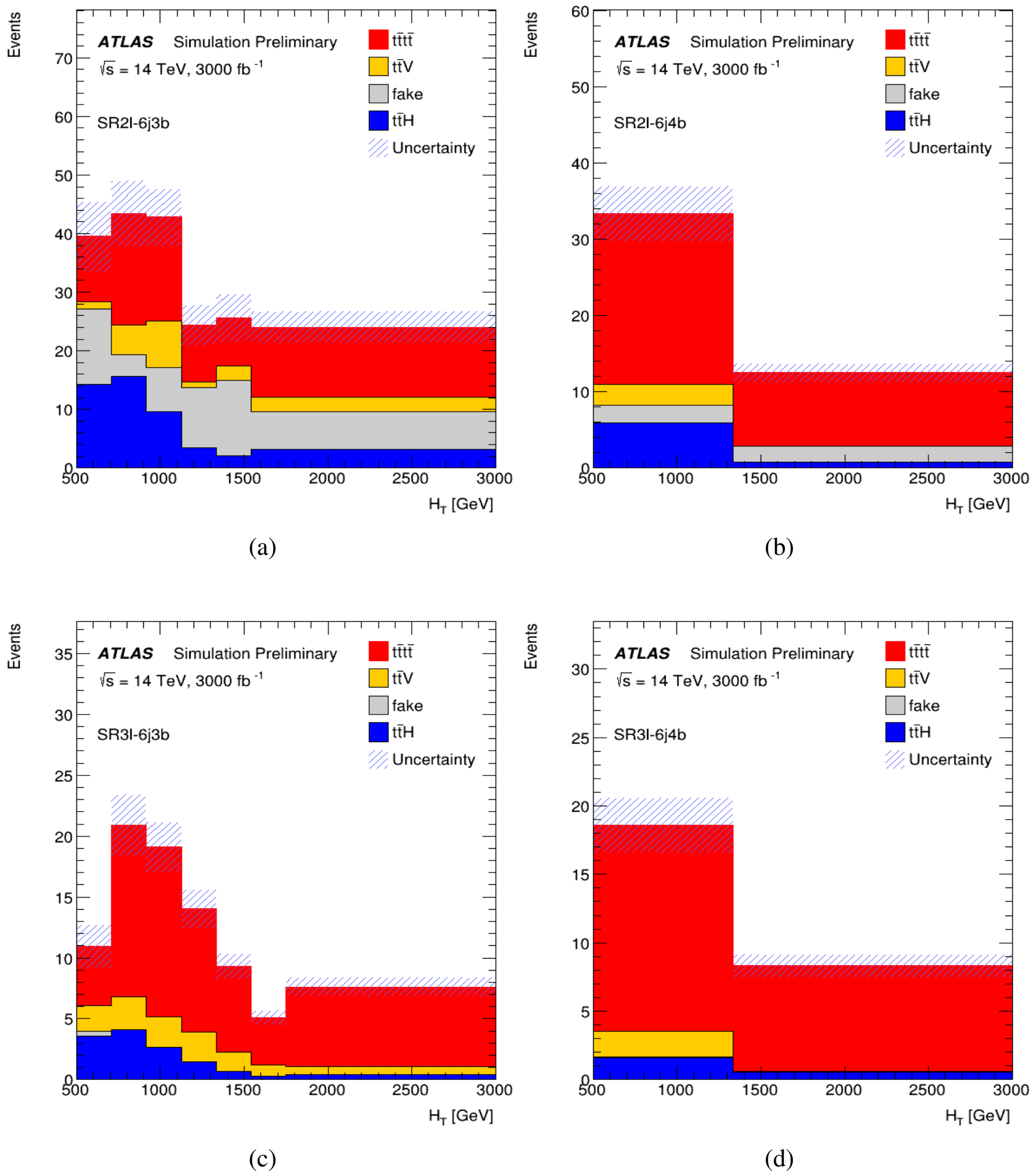

Figure 5: Expected $H_T$ distribution in the signal regions: (a) SR2l-6j3b, (b) SR2l-6j4b, (c) SR3l-6j3b and (d) SR3l-6j4b. The expected SM $t\bar{t}t\bar{t}$ signal (red histogram) is added on top of the background prediction.

9

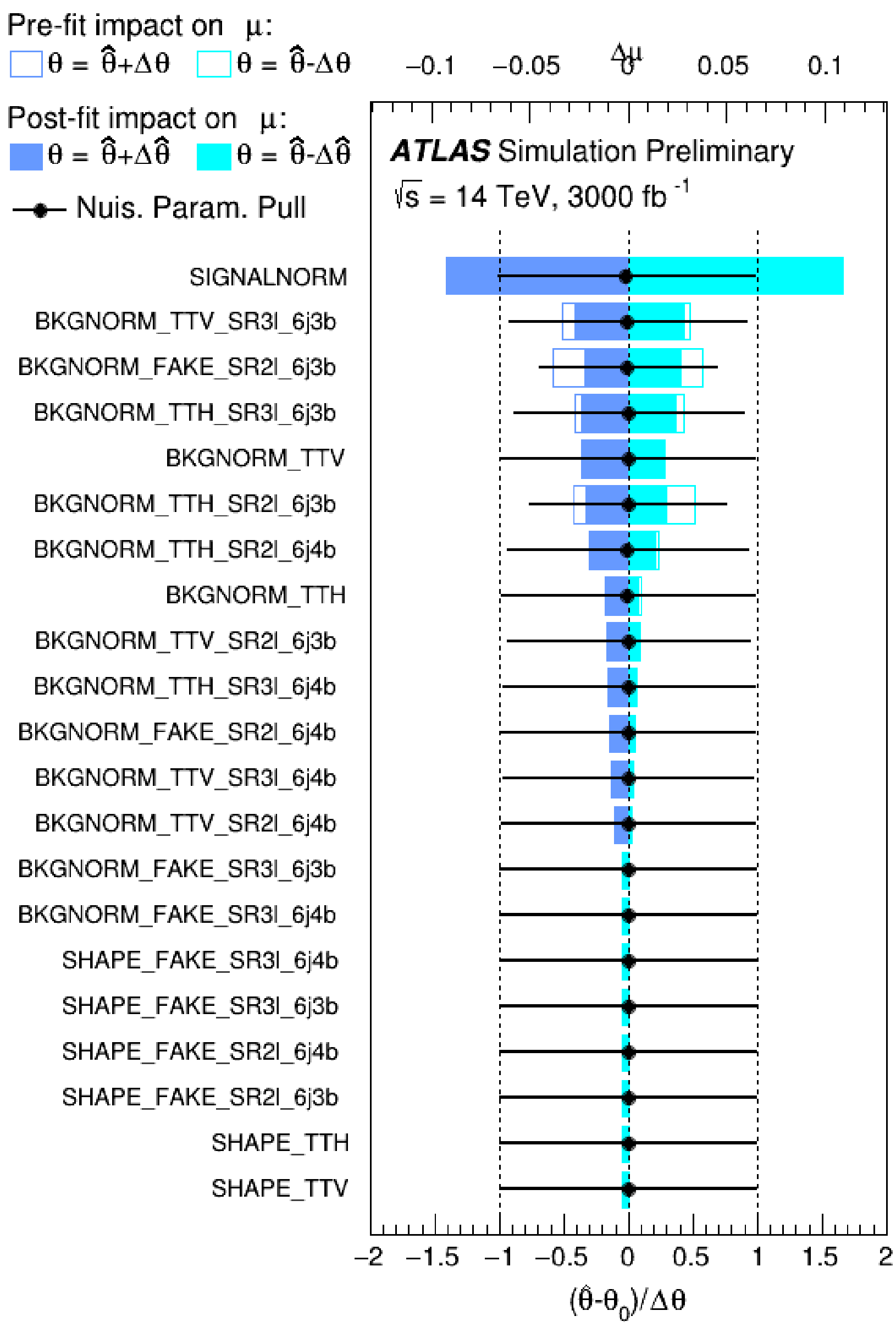

Figure 6: Fitted values of the nuisance parameters corresponding to the systematic uncertainties and their impact on the measured signal strength $\mu$. The fit is performed under the signal-plus-background hypothesis with the SM $t\bar{t}t\bar{t}$ production as signal. The black points, which are plotted according to the bottom horizontal scale, show the deviation of each of the fitted nuisance parameters, $\hat{\theta}$, in units of the pre-fit standard deviation $\Delta\theta$. The black error bars show the post-fit errors in units of pre-fit standard deviation. The nuisance parameters are sorted according to the post-fit effect of each of them on $\mu$ (solid dark and light blue areas, for positive and negative variations respectively), with those with the largest impact at the top. The pre-fit effect of each nuisance parameter is also shown (empty dark and light blue areas, for positive and negative variations respectively).

## 7 Conclusion

Projections for the measurement of the SM four-top-quark production cross-section in final states containing two same-charge leptons or at least three leptons, at least five jets and at least two $b$-jets at $\sqrt{s}$ = 14 TeV were performed in the context of the High-Luminosity LHC with 3000 fb$^{-1}$ of proton-proton collisions with the ATLAS experiment. An uncertainty on the $t\bar{t}t\bar{t}$ cross-section of 11% is expected with the precision being dominated by the statistical uncertainty. This corresponds to a significance to observe this yet-unmeasured signal well above 5 standard deviations. The current theoretical uncertainty on the computation of the four-top-quark production cross-section is roughly twice larger than the experimental projected uncertainty.

# CMS Physics Analysis Summary

Contact: cms-future-conveners@cern.ch

# Prospects for a search for gluon-mediated FCNC in top quark production using the CMS Phase-2 detector at the HL-LHC

The CMS Collaboration

## Abstract

Prospects are presented for a search for gluon-mediated flavour-changing neutral currents in the top quark production via tug and tcg vertices using the CMS Phase-2 detector at the HL-LHC. The analysis uses Monte Carlo samples of proton-proton collisions at $\sqrt{s} = 14$ TeV with a full simulation of the Phase-2 upgraded CMS detector assuming an average of 200 proton-proton interactions per bunch crossing. The final state signature of the signal is similar to that for the t-channel single top quark production in the $\mu/\mathrm{e}$ + jets final state. Bayesian and deep learning neural networks are used to discriminate the signal events against backgrounds. The 95% C.L. expected exclusion limits on the coupling strengths are $|\kappa_{\mathrm{tug}}|/\Lambda < 1.8 \times 10^{-3}\,(2.9 \times 10^{-3})\,\mathrm{TeV}^{-1}$ and $|\kappa_{\mathrm{tcg}}|/\Lambda < 5.2 \times 10^{-3}\,(9.1 \times 10^{-3})\,\mathrm{TeV}^{-1}$ for integrated luminosity of 3000 fb$^{-1}$ (300 fb$^{-1}$). The corresponding limits on branching fractions are $\mathcal{B}(\mathrm{t} \to \mathrm{ug}) < 3.8 \cdot 10^{-6}\,(9.8 \cdot 10^{-6})$ and $\mathcal{B}(\mathrm{t} \to \mathrm{cg}) < 32 \cdot 10^{-6}\,(99 \cdot 10^{-6})$ for integrated luminosity of 3000 fb$^{-1}$ (300 fb$^{-1}$). Therefore, the exploitation of the full HL-LHC data set with the upgraded CMS detector will allow to improve the current limits by an order of magnitude.

# 1 Introduction

Single top quark (t) production provides the opportunity to investigate aspects of top quark physics that cannot be studied with $t\bar{t}$ events [1]. Flavour-changing neutral currents (FCNC) are absent at lowest order in the SM, and are significantly suppressed through the Glashow–Iliopoulos–Maiani mechanism [2] at higher orders. Precise measurements of various rare decays of K, D, and B mesons, as well as of the oscillations in $K^0\overline{K}^0$, $D^0\overline{D}^0$, and $B^0\overline{B}^0$ systems, strongly constrain FCNC interactions involving the first two generations and the b quark [3]. The V–A structure of the charged current with light quarks is well established [3]. However, FCNC involving the top quark are significantly less constrained. In the SM, the FCNC couplings of the top quark are predicted to be very small ($\sim 10^{-10}$ [4]) and are not detectable at current experimental sensitivity. However, they can be significantly enhanced in various SM extensions, such as supersymmetry [4–6], and models with multiple Higgs boson doublets [7–9], extra quarks [10–12], or a composite top quark [13]. New vertices with top quarks are predicted, in particular, in models with light composite Higgs bosons [14, 15], extra-dimension models with warped geometry [16], or holographic structures [17]. Such possibilities can be encoded in an effective field theory through higher-dimensional gauge-invariant operators [18, 19]. Direct limits on top quark FCNC parameters have been established by the CDF [20], D0 [21], ATLAS [22], and CMS [23] Collaborations. Processes with FCNC vertices in the decay of the top quark are negligible since the current limits to the branching fractions are about $10^{-5}$, also the final states of such decays are difficult to distinguish from the backgrounds. This paper presents a search for FCNC interactions in the production of single top quarks. Models that have contributions from FCNC in the production of single top quarks can have sizable deviations relative to SM predictions, in particular those involving up quarks in the initial state as they profit from a large enhancement due to their parton distribution function (PDF). Also processes with charm quarks in the initial state benefit from a relative enhancement due to PDF with respect to processes initiated by bottom quarks, such as the background SM process of single top production in t channel. This is in contrast with searches for processes with FCNC vertices in the decay of the top quark where no such enhancement is present, and whose final states are difficult to distinguish from the backgrounds. The current limits on the branching ratios of the latter processes are about $10^{-5}$, and therefore this paper assumes negligible contribution of the FCNC decay modes to the total width of the top quark. The prospects for the search are estimated with a full simulation of the Phase-2 upgraded CMS detector with an average of 200 proton-proton interactions per bunch crossing. The Phase-2 upgrade of CMS detector is described in Technical Design Reports [24–29] and increases the angular coverage of the detector. The High Luminosity LHC regime with 3000 $fb^{-1}$ of integrated luminosity and $\sqrt{s} = 14$ TeV is assumed in this study.

# 2 Analysis strategy and simulation

There are two complementary strategies to search for FCNC in single top quark production. A search can be performed in the $s$ channel for resonance production through the fusion of a gluon (g) with an up (u) or charm (c) quark, as was the case in analyses by the CDF [20] and ATLAS [22] Collaborations. However, as pointed out by the D0 Collaboration, the $s$-channel production rate is proportional to the square of the FCNC coupling parameter and is therefore expected to be small [21]. On the other hand, the $t$-channel cross section and its corresponding kinematic properties have been measured accurately at the LHC [30–32], an important feature being that the $t$-channel signature contains a light-quark jet produced in association with the single top quark. This light-quark jet can be used to search for deviations from the SM prediction caused by FCNC in the top quark sector. This strategy was applied by the D0 Collaboration [21], as well as in the CMS Collaboration [23]. The FCNC tcg and tug interactions can be written in a model-independent form with the following effective Lagrangian [1]:

$$\mathfrak{L} = \frac{\kappa_{\mathrm{tqg}}}{\Lambda} g_s \bar{\mathrm{q}} \sigma^{\mu\nu} \frac{\lambda^{\mathrm{a}}}{2} \mathrm{t} G^{\mathrm{a}}_{\mu\nu}, \quad (1)$$

where $\Lambda$ is the scale of new physics ($\gtrsim$1 TeV), q refers to either the u or c quarks, $\kappa_{\mathrm{tqg}}$ defines the strength of the FCNC interactions in the tug or tcg vertices, $\lambda^{\mathrm{a}}/2$ are the generators of the SU(3) colour gauge group, $g_s$ is the coupling constant of the strong interaction, and $G^{\mathrm{a}}_{\mu\nu}$ is a gluon field strength tensor. The Lagrangian is assumed to be symmetric with respect to the left and right projectors. Single top quark production through FCNC interactions contains 48 subprocesses for both the tug and tcg channels, and the cross section is proportional to $(\kappa_{\mathrm{tqg}}/\Lambda)^2$. Representative Feynman diagrams for the FCNC processes are shown in Fig. 1. All these fea-tures are explicitly taken into account in the Single-Top Monte Carlo (MC) generator [33] based on the COMPHEP package [34], which was used to generate the signal events.

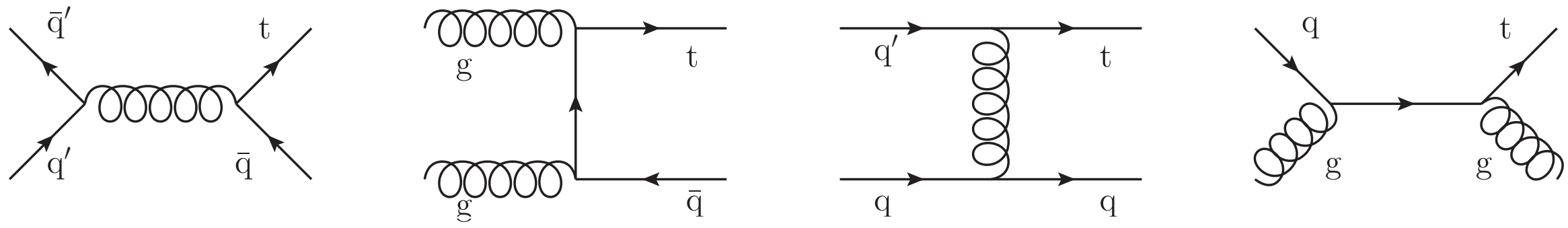

Figure 1: Representative Feynman diagrams for the FCNC processes with tqg interactions (q=u,c).

These signal samples as well as backgrounds from $\mathrm{t\bar{t}}$, single top, W+jets and Drell-Yan processes are estimated from full simulation of the CMS detector with realistic Phase-2 conditions, while the multijet QCD background is estimated with Run II data-driven template owing to the lack of statistics in the corresponding MC sample. The LO MADGRAPH 5.1 [35] generator is used to simulate W boson production with up to 4 additional jets in the matrix element, subdominant backgrounds from Drell–Yan in association with jets, and WW, WZ, and ZZ production. The POWHEG 1.0 NLO MC generator [36] provides a model for top quark pair and single production. Given the difficulty to reliably model QCD multijet events, this study makes use of a data-driven sample of 13 TeV data collected in 2016, with an anti-isolated selection. The resulting estimation of the QCD multijet background is rescaled to the appropriate luminosity and by the theoretical cross section ratio between 13 and 14 TeV, but other factors owing to differences in pileup, detector conditions, and some of the selection criteria are taken into account by a conservative normalization uncertainty.

## 3 Event selection and multivariate analysis

The particle-flow event algorithm [37] reconstructs and identifies each individual particle with an optimized combination of information from the various elements of the CMS detector. The energy of electrons is determined from a combination of the electron momentum at the primary interaction vertex as determined by the tracker, the energy of the corresponding ECAL cluster, and the energy sum of all bremsstrahlung photons spatially compatible with originating from the electron track. The energy of muons is obtained from the curvature of the corresponding track. The energy of charged hadrons is determined from a combination of their momentum measured in the tracker and the matching ECAL and HCAL energy deposits, corrected for zero-suppression effects and for the response function of the calorimeters to hadronic showers.

Finally, the energy of neutral hadrons is obtained from the corresponding corrected ECAL and HCAL energy. Jets are reconstructed offline from particle-flow candidates clustered by the anti-$k_T$ algorithm [38, 39]. More details are given in Section 9.4.1 of Ref. [29].

The final signature of the signal is selected by requiring to have only one isolated ($I^{\mu}_{rel} < 0.15$) muon or electron [40] with $p_T > 25\,GeV$ and $|\eta| < 2.8$. The relative isolation $I^{\mu}_{rel}$ is defined as the sum of the energy deposited by long-lived charged hadrons, neutral hadrons, and photons in a cone with radius $\Delta R = \sqrt{(\Delta\eta^2 + \Delta\phi^2)} = 0.4$, divided by the lepton $p_T$, where $\Delta\eta$ and $\Delta\phi$ are the differences in pseudorapidity and azimuthal angle (in radians), respectively, between the lepton and the other particle's directions. A similar definition is used for the electron isolation. Electrons in the overlap region $1.4 < |\eta| < 1.6$ are excluded from the analyses. Events with additional muons or electrons are rejected using a looser quality requirement of $p_T > 10\,GeV$, $|\eta| < 2.8$, and $I_{rel} < 0.25$. The event is required to have two or three PUPPI jets [41] reconstructed using the anti-$k_T$ algorithm with a distance parameter of $R = 0.4$, with $p_T > 30\,GeV$ and $|\eta| < 4.7$. We require at least one b tagged jet and at least one jet that fails the b tagging criteria. A high purity b tagging working point is used based on the cMVA [42] algorithm for jets with $|\eta| < 1.5$ and the DeepCSV [42] algorithm for jets with $1.5 < |\eta| < 3.5$. This high-purity working point corresponds to about 68% probability to identify jets from b-quarks and a misidentification probability of about 0.1% for the light-flavor jets.

The multijet QCD background is derived from the full single muon dataset collected in Run II in 2016 by CMS detector with 13 TeV center-of-mass energy. Owing to Run II detector conditions and the purpose to produce multijet-QCD-enriched sample, the event selection is modified to require one anti-isolated ($0.35 < I^{\mu}_{rel} < 1$) muon with $p_T > 26\,GeV$ and $|\eta| < 2.4$, without veto for additional low-$p_T$ leptons. The other requirements to select events with two or three jets are the same as in signal region described in previous paragraph. The b-tagging criteria are slightly different due to the limitation of $|\eta| < 2.4$ with the DeepCSV algorithm in Run II. Since the lepton is not isolated, we consider only jets outside a cone $\Delta R(\text{lepton}, \text{jet}) > 0.5$ to avoid a mismodelling in isolation-sensitive variables. The purity of the resulting QCD multijet sample is expected to be about 97% according to MC simulation in the Run II detector conditions. The normalization of the data-driven sample is obtained from the fit of multijet QCD template in 13 TeV CMS data and then rescaled to the expected luminosity of 3000 $fb^{-1}$ and by the theoretical cross section ratio of 1.09 between 13 and 14 TeV collision energy. Other factors related to little differences in event selection, pileup, detector conditions, and some of the selection criteria are taken into account by a conservative normalization uncertainty.

Several variables in the analysis require full kinematic reconstruction of the top quark and W boson candidates. For the kinematic reconstruction of the top quark, the W boson mass constraint is applied to extract the component of the neutrino momentum along the beam direction ($p_z$). This leads to a quadratic equation in $p_z$. When there are two real solutions of the equation, the smaller value of $p_z$ is used as the solution. For events with complex solutions, the imaginary components are eliminated by modifying $E_T^{miss}$ such that $m_T(W) \equiv \sqrt{2p_T(\mu)E_T^{miss}(1 - \cos[\Delta\phi(\mu, \vec{p}_T^{miss})])} = M_W$, where $M_W = 80.4$ [3].

The Bayesian Neural Network technique (BNN) and the slightly adapted FBM package [43, 44] are used to distinguish signal events from the standard model background. The input variables for each BNN are summarized in Table 1. Their choice is based on the difference in the structure of the Feynman diagrams contributing to the signal and background processes [45].

In the first step of the analysis one Bayesian Neural Network is trained to filter out multijet background events. A minimal set of the simplest and well-modeled variables to distinguish events with real W boson production from multijet QCD events are used and listed in Table 1. The input variables and the Multijet BNN discriminant distributions are shown in Fig. 2. The requirement on multijet BNN output discriminant to be greater than 0.7 rejects about 95% of multijet events and about 30% of signal events, as can be seen from Table 2. This requirement makes the multijet QCD background negligible and the uncertainty, in spite of being assigned a conservative value, has a much smaller impact than other uncertainties in the analysis. The events passing the multijet BNN requirement are passed to the next level of the analysis. At the next step two networks are trained, one for each type of signal processes, since the kinematics for the tug and tcg processes are slightly different due to the different initial states. The distributions of some of the BNN input variables are shown in Figs. 3. In addition to Bayesian Neural Networks, we prepare two fully connected Deep Learning Neural Networks (DNN) to compare DNN and BNN techniques. The input set of variables for DNN and BNN are the same. Five layers with about 100 nodes each are used for DNN architecture. The DNNs are built and trained using the Tensorflow [48] and Keras [49, 50] packages. The comparison of the BNN and DNN outputs are shown in Fig. 4 for the signal and background events. The back-ground is the properly weighted mixture of all SM processes. The comparison plots do not show a significant difference between BNN and DNN with respect to signal and background separation. However, in this analysis the BNN is used to obtain the limits for tug channel and DNN for tcg channel because of a slightly better performance in the corresponding channels.

Table 1: Input variables for the BNN/DNNs used in the analysis. The symbol X represents the variables used for each particular BNN/DNN. The notations "leading" and "next-to-leading" refer to the highest-$p_T$ and second-highest-$p_T$ jet, respectively. The notation "best" jet is used for the jet that gives a reconstructed mass of the top quark closest to the value of 172.5 GeV, which is used in the MC simulation.

| Variable | Description | Multijet BNN | tug FCNC BNN/DNN | tcg FCNC BNN/DNN |
|---|---|---|---|---|
| $p_T(j_1)$ | $p_T$ of the leading jet | | X | X |
| $p_T(j_2)$ | $p_T$ of the next-to-leading jet | | X | X |
| $p_T(j_1, j_2)$ | vector sum of the $p_T$ of the leading and the next-to-leading jet | | X | X |
| $p_T(j_L)$ | $p_T$ of the light-flavour jet (untagged jet with the highest value of $\|\eta\|$) | | X | X |
| $p_T(j_{\text{not best}})$ | $p_T$ of all jets without the one that best reconstructs the top quark | | X | X |
| $p_T(\text{lep})$ | $p_T$ of the lepton | X | X | X |
| $p_T(\text{top})_{b_1}$ | $p_T$ of the top quark reconstructed with leading c jet (the b-tagged jet with the highest $p_T$) | | X | X |
| $H_T(j)$ | scalar sum of the $p_T$ of the all jets | | X | X |
| $E_T^{\text{miss}}$ | missing transverse energy | X | | |
| $\eta(\text{lep})$ | $\eta$ of the lepton | | X | X |
| $\eta(j_L)$ | $\eta$ of the light-flavour jet | | X | X |
| $m_T(W)$ | transverse mass of the W boson | X | | |
| $m(j)$ | invariant mass of the all jets | | X | X |
| $m(j, W)$ | invariant mass of the W boson and all jets | | X | X |
| $m(\text{top})_{b_1}$ | invariant mass of the top quark reconstructed with leading b jet | | X | X |
| $N(j)$ | number of selected jets | | X | X |
| $\Delta\phi(\text{lep}, E_T^{\text{miss}})$ | azimuthal angle between the lepton and $\vec{p}_T^{\text{miss}}$ | X | | |
| $\cos(\theta_{\text{lep},j_L})\|_{\text{top}}$ | cosine of the angle between the lepton and the light-flavour jet in the top quark rest frame, for top quark reconstructed with the leading c jet [46] | | X | X |
| $\cos(\theta_{\text{lep},W})\|_W$ | cosine of the angle between the lepton momentum in the W boson rest frame and the direction of the W boson boost vector [47] | | X | X |
| $Q(\text{lep})$ | charge of the lepton | | X | |

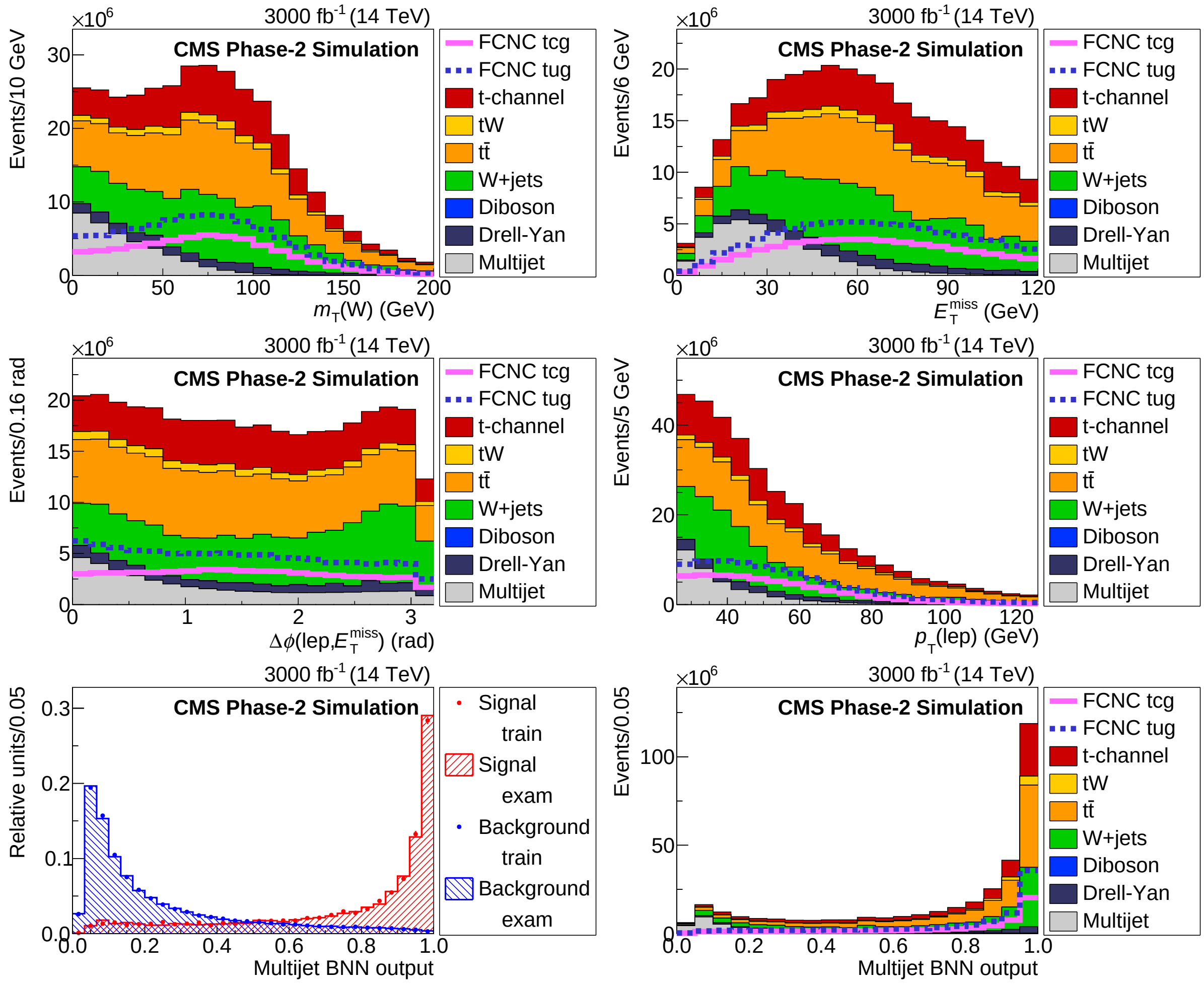

Figure 2: The Multijet BNN input variable distributions: $m_T(W)$ (top left), $E_T^{miss}$ (top right), $\Delta\phi(lep, E_T^{miss})$ (middle left) and $p_T(lep)$ (middle right). Comparison of distributions of the training and testing events of the Multijet BNN (bottom left) and resulting distribution of the Multijet BNN discriminant (bottom right). The solid and dashed lines give the expected distributions for FCNC tgu and tgc processes, respectively, assuming a coupling of $|\kappa_{tug}|/\Lambda = 0.09$ and $|\kappa_{tcg}|/\Lambda = 0.06\ \text{TeV}^{-1}$. Both muon and electron channels are presented on the plots.

The discriminant distributions of all SM and FCNC processes are shown in Fig. 5 for the BNN and in Fig. 6 for the DNN. All processes are normalized to their cross sections and selection efficiencies, and an integrated luminosity of 3000 fb$^{-1}$.

The shape of the neural networks discriminants are used in the statistical analysis to estimate the expected sensitivity to the contributions from FCNC.

# 4 Statistical analysis and expected limits

The statistical analysis is performed with the Theta package [51]. Bayesian inference is used to obtain the posterior probabilities based on an Asimov data set of the background-only model. We assume the same systematic scenario as in [24] and incorporate the following systematic uncertainties in the statistical model as nuisance parameters: luminosity measurement (1%), lepton identification and isolation (1% for electron and 0.5% for muon channel), jet energy scale (1%), b tagging (1% for b jets, 2% for c jets and 15% for light jets). The normalization of the $t\bar{t}$ contribution is varied by 6% [52], a prior normalization uncertainty for the multijet background is estimated conservatively to be 50% while the cross section of the remaining background sources is varied through their scale uncertainties as described in [53].

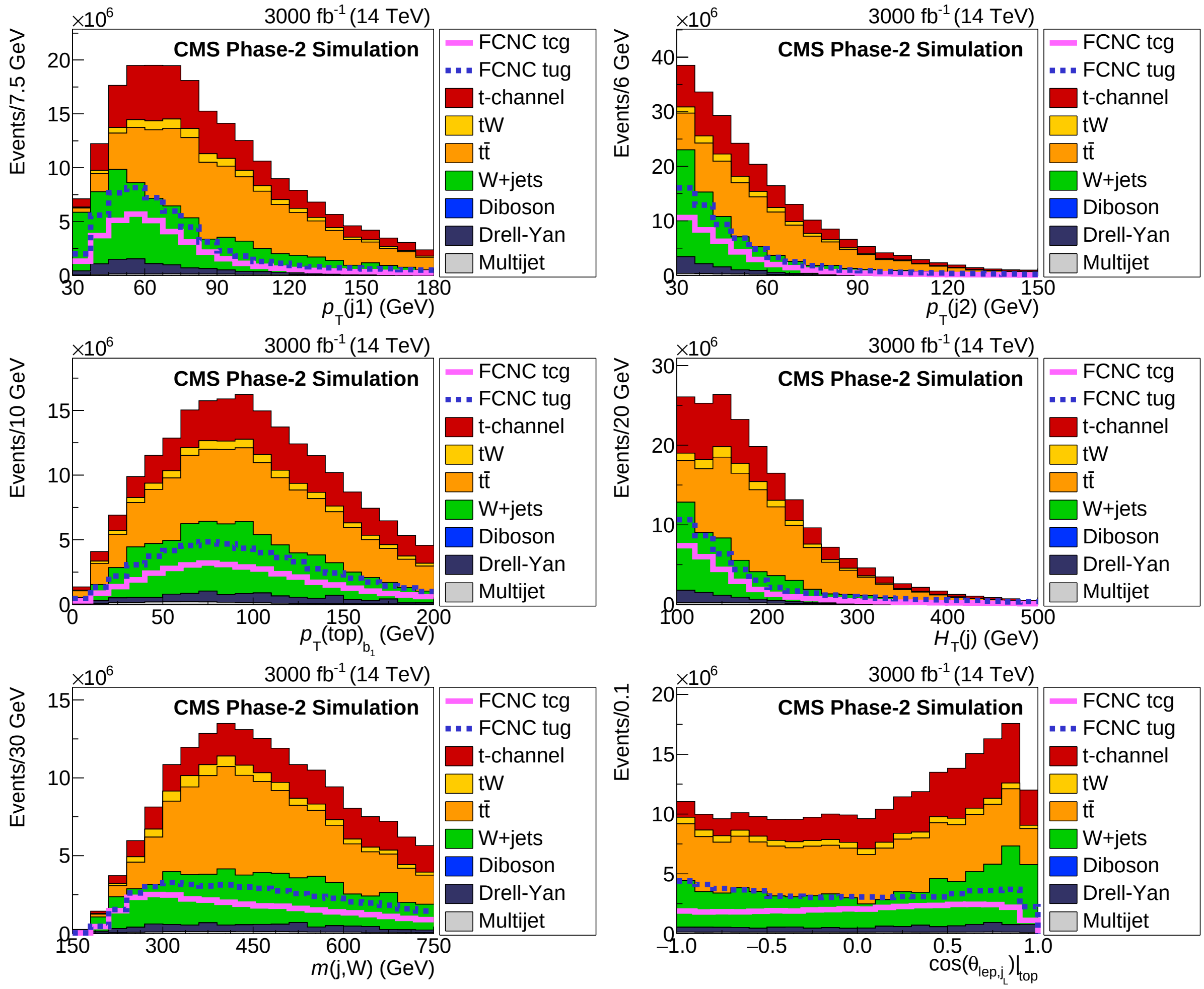

Figure 3: Comparison of FCNC tgc and tgu signal with the SM processes for the BNN input variables. The solid and dashed lines give the expected distributions for FCNC tgu and tgc processes, respectively, assuming the couplings $|\kappa_{tug}|/\Lambda = 0.06$ TeV$^{-1}$ and $|\kappa_{tcg}|/\Lambda = 0.09$ TeV$^{-1}$. The requirement of Multijet BNN > 0.7 is applied. The variables are described in Table 1.

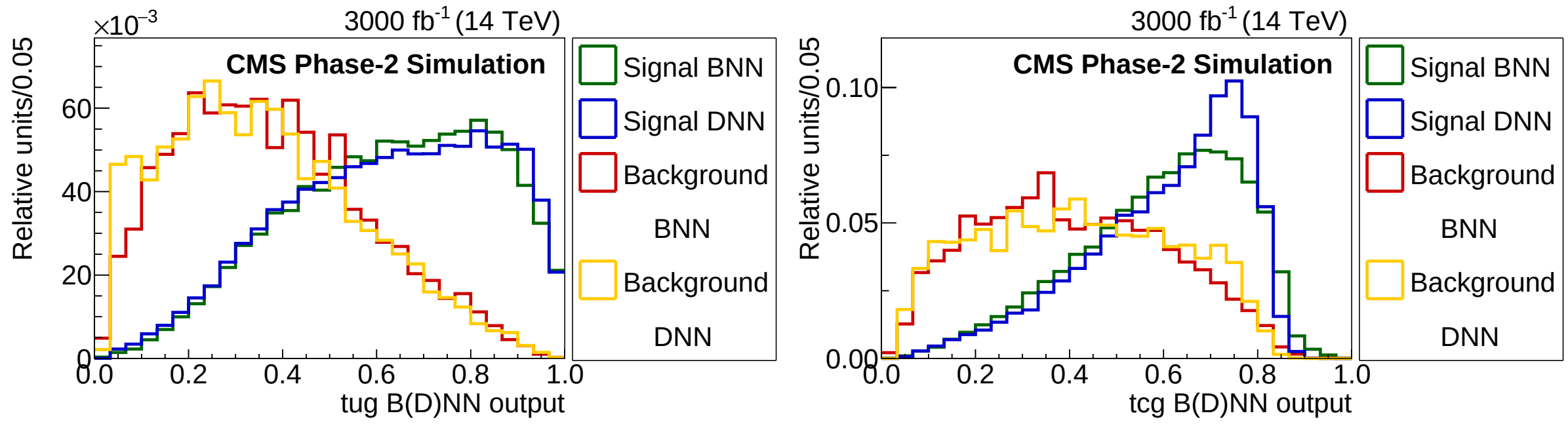

Figure 4: Comparison of the BNN and DNN FCNC discriminant distributions to distinguish FCNC tgu (left) and tgc (right) processes (signal) from the SM processes (background). The requirement of Multijet BNN > 0.7 is applied.

Table 2: The predicted event yields before and after the multijet BNN suppression for integrated luminosity of 3000 fb$^{-1}$. The estimations for tug and tcg processes assume coupling values of $|\kappa_{tug}|/\Lambda = 0.03$ and $|\kappa_{tcg}|/\Lambda = 0.03$ TeV$^{-1}$, respectively.

| Process | Basic selections | Multijet BNN > 0.7 |
|---|---|---|
| FCNC tcg | 646,000 | 434,000 |
| FCNC tug | 2,190,000 | 1,510,000 |
| $t$ channel | 7,420,000 | 5,270,000 |
| $tW$ channel | 1,190,000 | 846,000 |
| $t\bar{t}$ | 11,000,000 | 7,970,000 |
| W+jets | 9,690,000 | 6,380,000 |
| Dibosons | 97,500 | 58,000 |
| Drell–Yan | 1,600,000 | 870,000 |
| Multijets | 3,680,000 | 226,000 |

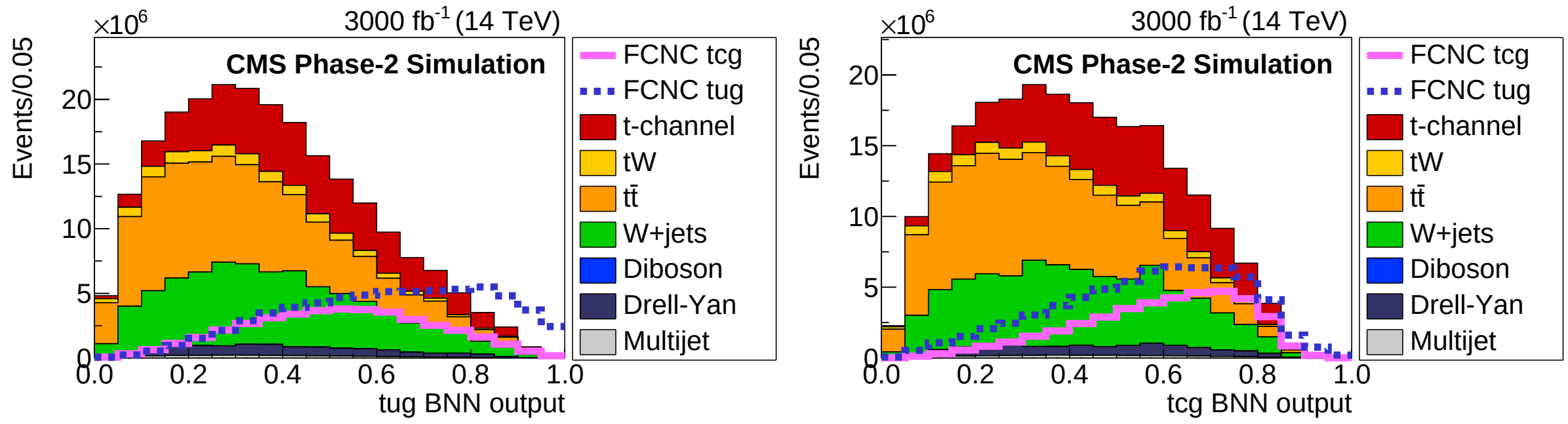

Figure 5: The FCNC BNN discriminant distributions to distinguish FCNC tgu (left) or tgc (right) processes from the SM contribution. The solid and dashed lines give the expected distributions for FCNC tgu and tgc processes, respectively, assuming the couplings to be $|\kappa_{tug}|/\Lambda = 0.06$ and $|\kappa_{tcg}|/\Lambda = 0.09$ TeV$^{-1}$. The requirement of Multijet BNN > 0.7 is applied.

The SM value for the top quark width is used in this analysis, since the influence of the FCNC parameters on the total top quark width is negligible for the allowed region of FCNC parameters. The COMPHEP package is used to simulate tug and tcg FCNC processes. The FCNC signal samples are normalized to the NLO cross sections using a K factors of 1.52 and 1.4 for $t \to ug$ and $t \to cg$ processes, respectively, for higher-order QCD corrections [54]. FCNC processes are kinematically different from any SM process. The posterior probability distributions of $|\kappa_{tug}|/\Lambda$ and $|\kappa_{tcg}|/\Lambda$ are obtained by fitting the histograms of BNN output in Fig. 5.

To obtain the individual exclusion limits on $|\kappa_{tug}|/\Lambda$ and $|\kappa_{tcg}|/\Lambda$ we assume the presence of only one corresponding FCNC parameter in the FCNC signal Monte Carlo model. These individual limits can be used to calculate the upper limits on the branching fractions $\mathcal{B}(t \to ug)$ and $\mathcal{B}(t \to cg)$ [55]. The expected exclusion limits at 95% C.L. on the FCNC couplings and the corresponding branching fractions are given in Table 3.

Table 3: The expected exclusion 1D limits at 95% C.L. on the FCNC couplings and the corresponding branching fractions for an integrated luminosity of 300 $fb^{-1}$ and 3000 $fb^{-1}$. In addition, a comparison with statistic-only uncertainties is shown.

| Integrated luminosity | $\mathcal{B}(t \to ug)$ | $\lvert\kappa_{tug}\rvert/\Lambda$ | $\mathcal{B}(t \to cg)$ | $\lvert\kappa_{tcg}\rvert/\Lambda$ |
|---|---|---|---|---|
| 300 $fb^{-1}$ | $9.8 \cdot 10^{-6}$ | 0.0029 $TeV^{-1}$ | $99 \cdot 10^{-6}$ | 0.0091 $TeV^{-1}$ |
| 3000 $fb^{-1}$ | $3.8 \cdot 10^{-6}$ | 0.0018 $TeV^{-1}$ | $32 \cdot 10^{-6}$ | 0.0052 $TeV^{-1}$ |
| 3000 $fb^{-1}$ Stat. only | $1.0 \cdot 10^{-6}$ | 0.0009 $TeV^{-1}$ | $4.9 \cdot 10^{-6}$ | 0.0020 $TeV^{-1}$ |

The dependence of the exclusion upper limits on integrated luminosity is shown in Fig. 7 with 1 and 2 $\sigma$ bands corresponding to 68% and 95% C.L. intervals of distributions of the limits. In addition the two-dimensional contours that reflect the possible simultaneous presence of both FCNC parameters are shown in Fig. 8. In this case both FCNC couplings are implemented in the FCNC signal Monte Carlo model. The expected limits can be compared with the recent CMS results [23] for the upper limits on the branching fractions of $2.0 \times 10^{-5}$ and $4.1 \times 10^{-4}$ for the decays $t \to ug$ and $t \to cg$, respectively.

The effect of each individual systematic uncertainty on parameter of interest is calculated by fixing the corresponding nuisance parameter at $\pm\,\sigma$ quantiles of the posterior distributions, and performing the Bayesian inference again. The impacts for the nuisance parameters are shown in Fig. 9. The biggest contribution for both signal channels come from the uncertainties of background cross sections and in particular from multijet QCD contribution and $t\bar{t}$ cross section uncertainties.

## 5 Conclusions

A direct search for model-independent FCNC $|\kappa_{tug}|/\Lambda$ and $|\kappa_{tcg}|/\Lambda$ couplings of the tug and tcg interactions has been projected for HL-LHC pp collisions at $\sqrt{s} = 14$ TeV based on full Monte Carlo simulation of the CMS experiment after the Phase II upgrades. The 95% C.L. expected exclusion limits on the coupling strengths are $|\kappa_{tug}|/\Lambda < 1.8 \times 10^{-3}$ $(2.9 \times 10^{-3})$ $TeV^{-1}$ and $|\kappa_{tcg}|/\Lambda < 5.2 \times 10^{-3}$ $(9.1 \times 10^{-3})$ $TeV^{-1}$ for the integrated luminosity of 3000 $fb^{-1}$ (300 $fb^{-1}$). The corresponding limits on branching fractions for the integrated luminosity of 3000 $fb^{-1}$ are $\mathcal{B}(t \to ug) < 3.8 \cdot 10^{-6}$ and $\mathcal{B}(t \to cg) < 32 \cdot 10^{-6}$. These results demonstrate that about one order of magnitude improvement can be achieved with respect to existing limits [23] on the branching fractions of rare FCNC top quark decays.

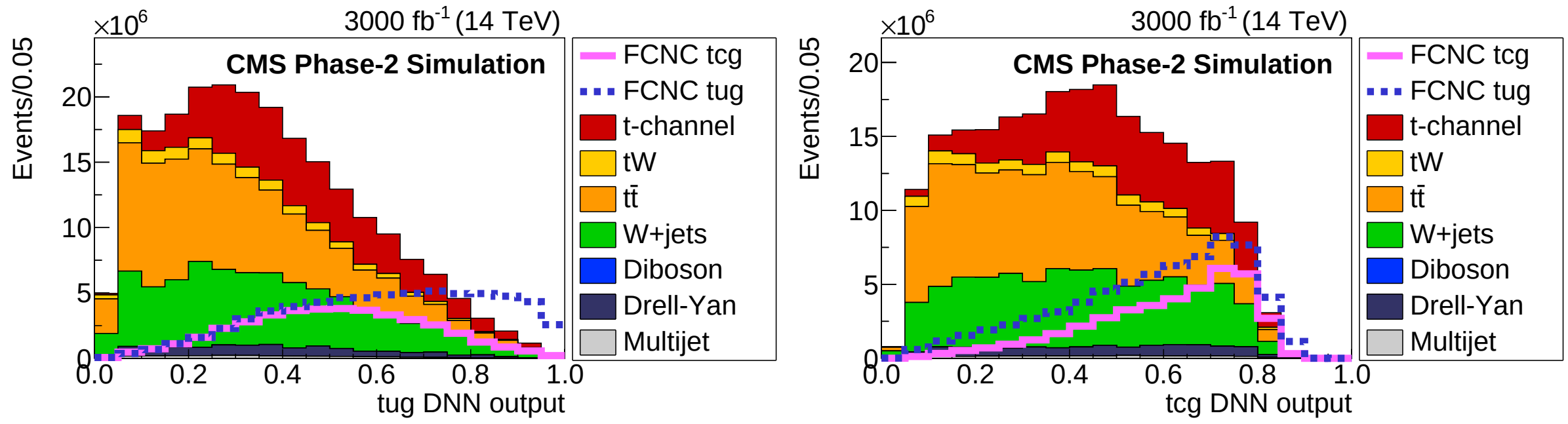

Figure 6: The FCNC DNN discriminant distributions when the DNN is trained to distinguish FCNC tgu (left) and tgc (right) processes from the SM processes. The solid and dashed lines give the expected distributions for FCNC tgu and tgc processes, respectively, assuming a coupling of $|\kappa_{\mathrm{tug}}|/\Lambda = 0.06$ and $|\kappa_{\mathrm{tcg}}|/\Lambda = 0.09\ \ \mathrm{TeV}^{-1}$ on the left (right) plots. The requirement of Multijet BNN $> 0.7$ is applied.

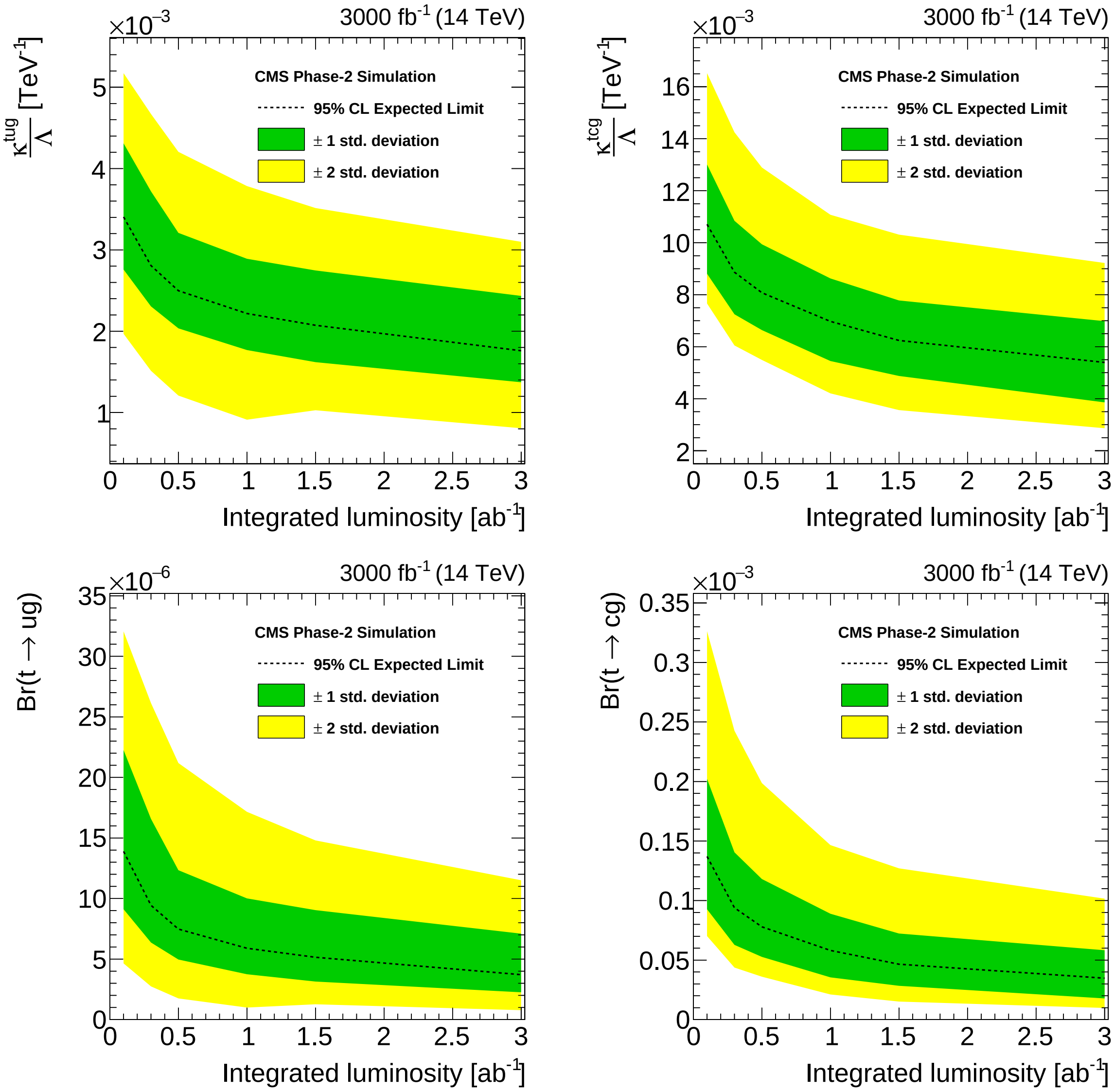

Figure 7: The expected exclusion limits at 95% C.L. on the FCNC couplings and the corresponding branching fractions as a function of integrated luminosity.

## 10

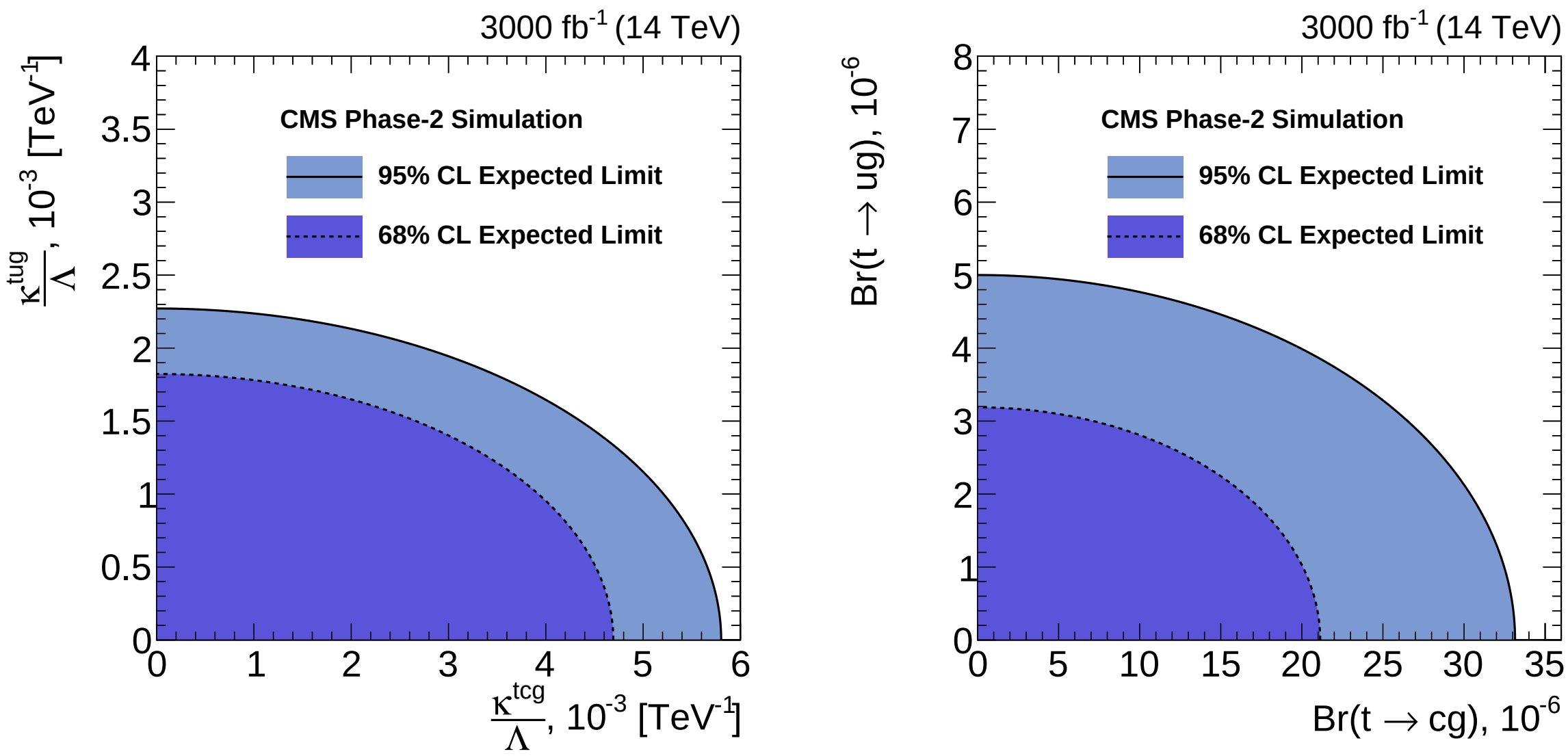

Figure 8: Two-dimensional expected limits on the FCNC couplings and the corresponding branching fractions at 68% and 95% C.L. for an integrated luminosity of 3000 $\mathrm{fb}^{-1}$.

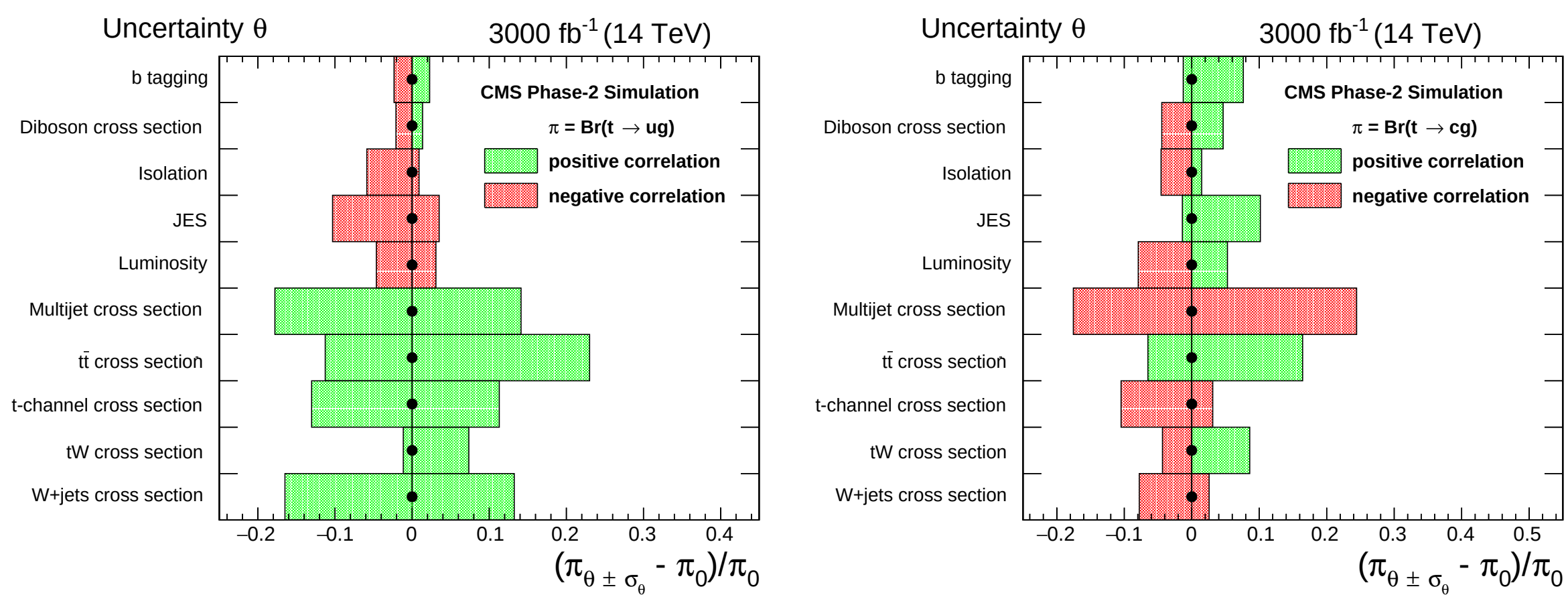

Figure 9: Effect of the systematic uncertainties on the expected exclusion limits on the branching fractions for $\mathcal{B}(\mathrm{t} \to \mathrm{ug})$ (left plot) and $\mathcal{B}(\mathrm{t} \to \mathrm{cg})$ (right plot).

**12**

# Sensitivity of searches for the flavour-changing neutral current decay $t \to qZ$ using the upgraded ATLAS experiment at the High Luminosity LHC

The ATLAS Collaboration

The sensitivity of the ATLAS experiment in the search for flavour-changing neutral-current top quark decays is presented. The study is performed in the context of the high luminosity phase of the Large Hadron Collider with a centre-of-mass energy of 14 TeV and an integrated luminosity of 3000 fb$^{-1}$. The three charged lepton final state of top-quark pair events is considered, in which one of the top quarks decays through the $t \to qZ$ ($q = u, c$) flavour-changing neutral-current channel and the other one decays to $bW$ ($t\bar{t} \to bWqZ \to b\ell\nu q\ell\ell$). An improvement by a factor of four is expected over the current Run-2 analysis results of $\mathcal{B}(t \to uZ) < 1.7 \times 10^{-4}$ and $\mathcal{B}(t \to cZ) < 2.4 \times 10^{-4}$ with 36.1 fb$^{-1}$ integrated luminosity. Obtained branching ratio limits are at the level of 4 to 5 $\times$ $10^{-5}$ depending on the considered scenarios for the systematic uncertainties.

# 1 Introduction

The top quark is the heaviest elementary particle known, with a mass of $m_t = 172.5 \pm 0.5$ GeV [1], and has such a small lifetime that it decays before hadronisation occurs. Flavour-changing neutral current (FCNC) decays such as $t \to qZ$ are forbidden at tree level. FCNC decays occur at one-loop level but are strongly suppressed by the GIM mechanism [2] with a suppression factor of 14 orders of magnitude relative to the dominant decay mode [3]. However, several SM extensions predict higher branching ratios for top-quark FCNC decays. Examples of such extensions are the quark-singlet model (QS) [4], the two-Higgs-doublet model with (FC 2HDM) or without (2HDM) flavour conservation [5], the Minimal Supersymmetric Standard Model (MSSM) [6], the MSSM with R-parity violation (RPV SUSY) [7], models with warped extra dimensions (RS) [8], or extended mirror fermion models (EMF) [9]. Reference [10] gives a comprehensive review of the various extensions of the SM that have been proposed. Table 1 provides the maximum values for the branching ratios $\mathcal{B}(t \to qZ)$ predicted by these models and compares them to the value predicted by the SM.

Table 1: Maximum allowed FCNC $t \to qZ$ ($q = u, c$) branching ratios predicted by several models [3–10].

| Model: | SM | QS | 2HDM | FC 2HDM | MSSM | RPV SUSY | RS | EMF |
|---|---|---|---|---|---|---|---|---|
| $\mathcal{B}(t \to qZ)$: | $10^{-14}$ | $10^{-4}$ | $10^{-6}$ | $10^{-10}$ | $10^{-7}$ | $10^{-6}$ | $10^{-5}$ | $10^{-6}$ |

Experimental limits on the FCNC branching ratio $\mathcal{B}(t \to qZ)$ were established by experiments at the Large Electron–Positron collider [11–15], HERA [16], the Tevatron [17, 18], and the Large Hadron Collider (LHC) [19–24]. The latest experimental limits are set by the CMS and ATLAS Collaborations. Limits of $\mathcal{B}(t \to uZ) < 2.4 \times 10^{-4}$ and $\mathcal{B}(t \to cZ) < 4.5 \times 10^{-4}$ at 95% confidence level (CL), are obtained by the CMS Collaboration using data collected at $\sqrt{s} = 13$ TeV [21]. For the same centre-of-mass energy, the ATLAS Collaboration derived the limits of $\mathcal{B}(t \to uZ) < 1.7 \times 10^{-4}$ and $\mathcal{B}(t \to cZ) < 2.4 \times 10^{-4}$ [24].

The High Luminosity upgrade of the Large Hadron Collider (LHC) (HL-LHC) is currently expected to begin operations in the second half of 2026 [25, 26], to achieve an ultimate luminosity of $7.5 \times 10^{34}$ cm$^{-2}$s$^{-1}$. The total integrated luminosity that is foreseen to be reached is 3000 fb$^{-1}$. This note presents a study of the sensitivity of the ATLAS experiment to top-quark decays via FCNC $t \to qZ$ ($q = u, c$ with $Z \to \ell^+\ell^-$). The top-quark–top-antiquark ($t\bar{t}$) events are studied, where one top quark decays through the FCNC mode and the other through the dominant SM mode ($t \to bW$). Only $Z$ boson decays into charged leptons and leptonic $W$ boson decays are considered. The final-state topology is thus characterized by the presence of three isolated charged leptons, at least two jets with exactly one being tagged as a jet containing a $b$-hadron, and missing transverse momentum from the undetected neutrino. The study is performed in the context of the LHC upgrade.

Based on the Run-1 search [23], the ATLAS detector sensitivity to FCNC $t \to qZ$ decays for the HL-LHC was studied and reported in Ref. [27], predicting a sensitivity of $(2.4 - 5.8) \times 10^{-5}$, when considering statistical uncertainties only, depending on the exact FCNC $t \to qZ$ modeling and $(8.3 - 41) \times 10^{-5}$, depending on the detailed assumptions for the systematic uncertainties. In the present analysis the description of the expected detector performance at the HL-LHC phase is improved and the analysis strategy closely follows the one of the Run-2 analysis [24] rather than the Run-1 search.

Since it is difficult to accurately estimate the relevant systematic uncertainties that will impact the analysis in the high luminosity environment, several scenarios are studied and compared.

## 2 Signal and background simulation samples

Particle-level samples are generated at a center-of-mass energy of 14 TeV without detailed detector simulation. To emulate the HL-LHC run conditions and detector response, physics objects defined in Section 3 are smeared using performance functions derived from MC events passed through a full GEANT4 simulation of the upgraded ATLAS detector [28–30]. The effect of objects reconstruction and identification efficiencies as well as their momentum or energy resolutions are parameterized assuming an average number of additional $pp$ collisions in the same or nearby bunch crossings (pile-up) of 200. In addition, pile-up jets are overlaid from a dedicated library.

In $pp$ collisions at a centre-of-mass energy of $\sqrt{s}$ = 14 TeV at the LHC, top quarks are produced according to the SM mainly in $t\bar{t}$ pairs with a predicted cross section of $\sigma_{t\bar{t}} = 0.98 \pm 0.06$ nb [31–36]. The uncertainty includes contributions from uncertainties in the factorisation and renormalization scales, the parton distribution functions (PDF), the strong coupling $\alpha_S$ and the top-quark mass. The cross section is calculated at next-to-next-to-leading order (NNLO) in QCD including resummation of next-to-next-to-leading logarithmic soft gluon terms with Top++ 2.0. The effects of PDF and $\alpha_S$ uncertainties are calculated using the PDF4LHC prescription [37] with the MSTW 2008 68% CL NNLO [38, 39], CT10 NNLO [40, 41] and NNPDF 2.3 5f FFN [42] PDF sets and are added in quadrature to those from the renormalization and factorisation scale uncertainties. These calculations are done for the top-quark mass value of 172.5 GeV used to simulate events as described in the following paragraphs.

The next-to-leading-order (NLO) simulation of signal events was performed with the event generator MG5_aMC@NLO [43] interfaced to Pythia8 [44] with the A14 [45] set of tuned parameters and the NNPDF30ME PDF set [42]. Top quark FCNC decay is done using the TopFCNC model [46, 47]. The effects of new physics at an energy scale $\Lambda$ were included by adding dimension-six effective terms to the SM Lagrangian. The Universal FeynRules Output (UFO) model [46, 47] is used for computation at NLO in QCD. No differences between the kinematical distributions from the $bWuZ$ and $bWcZ$ processes are observed. Due to the different $b$-tagging mistag rates for $u$- and $c$-quarks, the signal efficiencies differ after applying requirements on the $b$-tagged jet multiplicity. Hence limits on $\mathcal{B}(t \to qZ)$ are set separately for $q = u, c$. Only decays of the $W$ and $Z$ bosons with charged leptons were generated ($Z \to e^+e^-$, $\mu^+\mu^-$, or $\tau^+\tau^-$ and $W \to e\nu$, $\mu\nu$, or $\tau\nu$).

Several SM processes have final-state topologies similar to the signal, with at least three prompt[1] charged leptons, especially dibosons ($WZ$ and $ZZ$), but also $t\bar{t}Z$, $t\bar{t}W$, $t\bar{t}WW$, $tZ$ or $t\bar{t}t\bar{t}$ production. Events with non-prompt leptons, including the ones in which at least one jet is misidentified as a charged lepton, can also fulfil the event selection requirements. These events mainly consist of the $t\bar{t}$, $Z$+jets and $tW$ processes. Such background processes cannot be realistically estimated by the transfer function approach used for the HL-LHC studies. Therefore, these backgrounds are scaled to the same fraction of the total event yield as observed in the 13 TeV analysis [24]. All other background samples are normalized to their theory predictions.

[1] Prompt leptons are electrons or muons from the decay of $W$ or $Z$ bosons, either directly or through an intermediate $\tau \to \ell\nu\nu$ decay.

Table 2: Generators, parton shower simulation, parton distribution functions, and tune parameters used to produce particle-level samples without detailed detector simulation for this analysis. The acronyms ME and PS stand for matrix element and parton shower, respectively.

| Sample | Generator | Parton shower | ME PDF | PS PDF | Tune parameters |
|---|---|---|---|---|---|
| $t\bar{t} \to bWqZ$ | MG5_aMC@NLO [43] | Pythia8 [44] | NNPDF3.0NLO [48] | NNPDF2.3LO [42] | A14 [45] |
| $t\bar{t}Z$ | MG5_aMC@NLO | Pythia8 | NNPDF3.0NLO | NNPDF2.3LO | A14 |
| $ZZ, WZ$ | Sherpa v2 [49] | Sherpa v2 | NNPDF3.0NNLO | NNPDF3.0NNLO | Sherpa default |
| $tZ$ | MG5_aMC@NLO | Pythia8 | NNPDF3.0NLO | NNPDF2.3LO | A14 |
| $t\bar{t}WW$ | MG5_aMC@NLO | Pythia8 | NNPDF3.0NLO | NNPDF2.3LO | A14 |
| $Z$+jets | Powheg-Box v1 [50] | Pythia8 | CT10 [40] | CTEQ6L1 [51] | AZNLO [52] |
| $t\bar{t}$ | Powheg-Box v2 | Pythia8 | NNPDF3.0NLO | NNPDF2.3LO | A14 |
| $tW$ | Powheg-Box v1 | Pythia6 [53] | CT10f4 | CTEQ6L1 | Perugia2012 [54] |
| $t\bar{t}t\bar{t}$ | MG5_aMC@NLO | Pythia8 | NNPDF3.0NLO | NNPDF2.3LO | A14 |

Table 2 summarizes information about the generators, parton shower, and PDFs used to simulate the different event samples considered in this analysis.

# 3 Object reconstruction

Electrons and muons are required to have $p_T > 25$ GeV. This threshold is increased in relation to the Run 2 analysis [24] due to the expected higher yields of non-prompt lepton backgrounds. The single lepton trigger thresholds during the HL-LHC phase are expected to be 22 GeV for electrons and 20 GeV for muons [55], safely below the offline $p_T$ requirement of 25 GeV considered in this analysis. Therefore no significant efficiency loss is expected from trigger threshold effects.

Electrons are required to be outside the transition region between the barrel and endcap calorimeters with $1.37 < |\eta_{\text{cluster}}| < 1.52$. Electrons and muons with $|\eta| > 2.5$ are rejected. Reconstructed leptons within a cone of $\Delta R < 0.2$ of jets are removed. A truth-based isolation requirement is applied to the leptons, meaning that the sum of the transverse energies of stable[2] charged and neutral generator-level particles, with the exception of neutrinos, within a $\Delta R = 0.2$ cone around the lepton must be less than 23% (11%) of the electron (muon) candidate $p_T$. This requirement yields an efficiency of 95% for the prompt leptons and 37% (21%) efficiency for non-prompt electrons (muons) with $25 < p_T < 50$ GeV in the $t\bar{t}$ events.

The missing transverse momentum ($E_T^{\text{miss}}$) is defined at particle level as the transverse component of the vector sum of the final-state neutrino momenta. The $E_T^{\text{miss}}$ resolution is parameterized as a function of the overall event activity.

Jets are reconstructed using the anti-$k_t$ algorithm [58, 59] with a radius parameter $R = 0.4$. They are required to have $p_T > 30$ GeV and $|\eta| < 4.5$. Jets containing $b$-hadrons are randomly $b$-tagged to follow the 70% $b$-jet tagging efficiency working point of the MV2c10 algorithm [60]. The rejection rates for light-flavour jet and c-jet depend on the jet $p_T$ and can be found in Ref. [60].

[2] Particles in the MC event record with status code 1: a final-state particle, i.e. a particle that is not decayed further by the generator [56, 57].

4

# 4 Event selection and reconstruction

The selection requirements follow the ones from the Run 2 analysis. Events are required to have exactly three leptons (any combination of electrons and muons), at least two jets, with exactly one of them $b$-tagged, one pair of opposite charge and same flavour leptons with $|m_{\ell^+\ell^-} - 91.2\text{ GeV}| < 15\text{ GeV}$ and $E_{\mathrm{T}}^{\mathrm{miss}} > 20\text{ GeV}$. If more than one compatible lepton pair is found in the selection, the one with the reconstructed mass closest to 91.2 GeV is chosen as the $Z$ boson candidate. The selection is finalized with the kinematical requirements explained next. For each possible jet combination, the following $\chi^2$ function is minimized to derive the longitudinal momentum of the neutrino and, consequently, to reconstruct the top-quarks and the $W$ boson. The solution (among all possible jet combinations) that yields the minimum $\chi^2$ value is chosen.

$$\chi^2 = \frac{\left(m^{\mathrm{reco}}_{j_a\ell_a\ell_b} - m_{t_{\mathrm{FCNC}}}\right)^2}{\sigma^2_{t_{\mathrm{FCNC}}}} + \frac{\left(m^{\mathrm{reco}}_{j_b\ell_c\nu} - m_{t_{\mathrm{SM}}}\right)^2}{\sigma^2_{t_{\mathrm{SM}}}} + \frac{\left(m^{\mathrm{reco}}_{\ell_c\nu} - m_W\right)^2}{\sigma^2_W},$$

where $m^{\mathrm{reco}}_{j_a\ell_a\ell_b}$, $m^{\mathrm{reco}}_{j_b\ell_c\nu}$, and $m^{\mathrm{reco}}_{\ell_c\nu}$ are the reconstructed masses of the $qZ$, $bW$, and $\ell\nu$ systems, respectively, corresponding to the top-quarks and the $W$ boson, respectively. For each jet combination, $j_b$ corresponds to the $b$-tagged jet, while any non-$b$-tagged jet can be assigned to $j_a$. The central values of the masses and the widths of the top quarks and the $W$ boson are taken from simulated signal events. This is done by matching the particles in the simulated events to the reconstructed ones, setting the longitudinal momentum of the neutrino to the $p_z$ of the simulated neutrino, and then performing fitting to a Bukin function[3] [61] to the masses of the matched reconstructed top quarks and $W$ boson. The obtained values are $m_{t_{\mathrm{FCNC}}} = 171.4$ GeV, $\sigma_{t_{\mathrm{FCNC}}} = 13.1$ GeV, $m_{t_{\mathrm{SM}}} = 177.1$ GeV, $\sigma_{t_{\mathrm{SM}}} = 38.1$ GeV, $m_W = 85.7$ GeV and $\sigma_W = 30.2$ GeV. Values for the $\sigma_{t_{\mathrm{SM}}}$ and $\sigma_W$ are high due to the negative influence of the high pileup on the $E_{\mathrm{T}}^{\mathrm{miss}}$ resolution used to reconstruct the neutrino from the $t \to bW \to j\ell\nu$ decay. The events are then required to have $|m^{\mathrm{reco}}_{t_{\mathrm{FCNC}}} - 172.5\text{ GeV}| < 40$ GeV, $|m^{\mathrm{reco}}_{t_{\mathrm{SM}}} - 172.5\text{ GeV}| < 60$ GeV and $|m^{\mathrm{reco}}_W - 80.4\text{ GeV}| < 50$ GeV to remove outliers where the expected signal contribution is small. Note that the two last values were increased with respected to the 13 TeV analysis due to the worse resolutions shown here. The fractions of correct assignments between the reconstructed top quarks and the true simulated top quarks at parton level (evaluated as a match within a cone of size $\Delta R = 0.4$) are $\epsilon_{t_{\mathrm{FCNC}}} = 76\%$ and $\epsilon_{t_{\mathrm{SM}}} = 40\%$, where the difference comes from the fact that for the SM top-quark decay the match of the $E_{\mathrm{T}}^{\mathrm{miss}}$ with the simulated neutrino is less efficient.

Following the strategy of the Run-2 analysis, dedicated control regions (CR) are defined for the main background contributions to help constrain systematic uncertainties. Here only CR for $t\bar{t}Z$ and non-prompt leptons were defined. The $t\bar{t}Z$ CR requires exactly three leptons, two of them with the same flavour, opposite charge and reconstructed mass within 15 GeV of the $Z$ boson mass. Furthermore, the events are required to have at least four jets, two of which must be $b$-tagged, and $E_{\mathrm{T}}^{\mathrm{miss}} > 20$ GeV. The non-prompt lepton background CR requires three leptons with two of them having the same flavour, opposite charge and reconstructed mass outside 15 GeV of the $Z$ boson mass, at least two jets with one being $b$-tagged and $E_{\mathrm{T}}^{\mathrm{miss}} > 20$ GeV.

[3] These fits use a piecewise function with a Gaussian function in the centre and two asymmetric tails. Six parameters determine the overall normalization, the peak position, the width of the core, the asymmetry, the size of the lower tail, and the size of the higher tail. Of these, only the peak position and the width enter the $\chi^2$.

| Selection | Signal Region | $t\bar{t}Z$ CR | Non-prompt lepton CR |
|---|---|---|---|
| No. leptons | 3 | 3 | 3 |
| OSSF | Yes | Yes | Yes |
| $\lvert m_{\ell\ell}^{\text{reco}} - 91.2\text{ GeV}\rvert$ | $< 15$ GeV | $< 15$ GeV | $> 15$ GeV |
| No. jets | $\geq 2$ | $\geq 4$ | $\geq 2$ |
| No. $b$-tagged jets | 1 | 2 | 1 |
| $E_{\text{T}}^{\text{miss}}$ | $> 20$ GeV | $> 20$ GeV | $> 20$ GeV |
| $\lvert m_{\ell\nu}^{\text{reco}} - 80.4\text{ GeV}\rvert$ | $< 50$ GeV | - | - |
| $\lvert m_{j\ell\nu}^{\text{reco}} - 172.5\text{ GeV}\rvert$ | $< 60$ GeV | - | - |
| $\lvert m_{j\ell\ell}^{\text{reco}} - 172.5\text{ GeV}\rvert$ | $< 40$ GeV | - | - |

Table 3: The selection requirements applied for the signal and background control regions. OSSF refers to the presence of a pair of opposite-sign, same-flavour leptons.

Selection requirements applied in the signal and background control regions are summarized in Table 3. The expected distributions of relevant observables in the signal region are shown in Figure 1.

## 5 Systematic uncertainties

The background fit to the CRs, described in Section 6, reduces the systematic uncertainty from some sources, due to the constraints introduced by the Asimov simulated data. The main uncertainties, in both the background and signal estimations, are expected to come from theoretical normalization uncertainties and uncertainties in the modelling of background processes in the simulation. The effect of those uncertainties is estimated in the 13 TeV analysis, and then reduced by a factor of two, as recommended in Ref. [62], to account for expected improvements in theoretical predictions. The reduced uncertainty is then applied in this analysis. The uncertainties obtained before the combined fit are discussed below and are summarized in Table 4.

The cross section uncertainties of the $t\bar{t}Z$ and $tZ$ background processes are taken to be 6% and 15%, respectively. For diboson production, a 6% theoretical normalization uncertainty is considered as well as 24% uncertainty on the $WZ$ production in the SR due to the modelling in the simulation. In addition, a 12% uncertainty added in quadrature per jet is applied on the $WZ$ yield in each jet multiplicity bin to account for the imperfect knowledge of the jet multiplicity distribution in $WZ$ events. The $t\bar{t}$ production cross-section uncertainties from the independent variation of the factorisation and renormalization scales, the PDF choice, and $\alpha_{\text{S}}$ variations (see Refs. [36, 37] and references therein and Refs. [39, 41, 42]) give a 5% uncertainty in the signal normalization and 4% uncertainty on the total non-prompt leptons background in the SR. The 12% and 5% uncertainties due to the choice of NLO generator and amount of QCD radiation for the $t\bar{t}$ modelling are considered on the total non-prompt leptons background in the SR, while the uncertainty due to the choice of the parton shower algorithm is 1% in the SR and 19% in the non-prompt leptons CR. A 17% uncertainty is considered on the $Z$+jets normalization, which yields a 2.5% uncertainty on the total non-prompt leptons background in the SR. For the remaining small backgrounds, a 50% uncertainty is assumed.

For both the estimated signal and background event yields, experimental uncertainties resulting from detector effects are assumed to be same as in the 13 TeV analysis. The uncertainties on the lepton reconstruction, identification and trigger efficiencies, as well as lepton momentum scales and resolutions, are added in quadrature resulting in a 2.6% (1.9%) uncertainty on the total background (signal) event yield in the SR. The uncertainty due to the jet-energy scale and resolution is 9% (4%) on the total background (signal) event yield in the SR, while total $b$-tagging uncertainty, which includes the uncertainty of the $b$-, $c$-, mistagged- and $\tau$-jet scale factors, is 5% (3.4%). Uncertainties of the $E_{\mathrm{T}}^{\mathrm{miss}}$ scale and pile-up effects are 4% and 2.3% on the total background and signal yields in the SR, respectively.

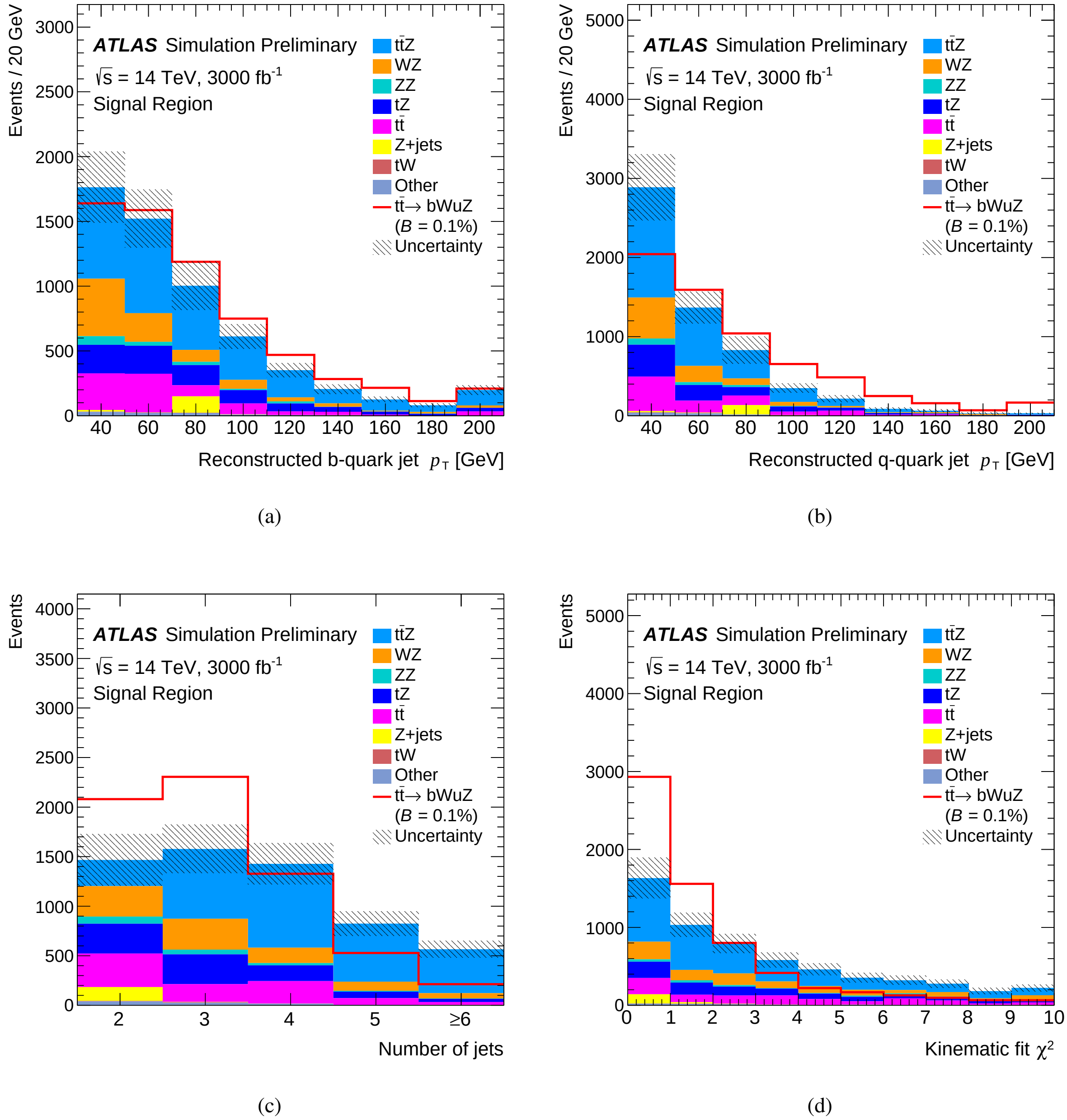

Figure 1: Expected distributions in the signal region for $p_T$ of the reconstructed a) $b$-quark jet from the $t \to bW$ decay and b) $q$-quark jet from the $t \to qZ$ decay, c) jet multiplicity and d) kinematic fit $\chi^2$. The signal is not shown stacked on top of the backgrounds, but is normalized separately to an arbitrary branching ratio of $\mathcal{B}(t \to qZ) = 0.1\%$. The dashed area represents the systematic uncertainty on the background prediction.

The total uncertainties of the leptons, jets, $b$-tagging, $E_{\mathrm{T}}^{\mathrm{miss}}$ and pile-up uncertainties on the total background/signal event yields are considered on each background/signal process as an input normalization uncertainty for the combined fit.

The shape uncertainties are not considered, assuming that their effect on the final results is not significant, as it is found in the 13 TeV analysis.

The MC statistical uncertainties are set to zero in the analysis, unless it is mentioned that they are considered, assuming that sufficiently large simulation samples will be available for the HL-LHC analysis.

| Source | Signal Region | | $t\bar{t}Z$ CR | Non-prompt CR |
|---|---|---|---|---|
| | S [%] | B [%] | B [%] | B [%] |
| Event modelling | 5 | 6 | 6 | 12 |
| Leptons | 1.9 | 2.6 | 2.1 | 2.9 |
| Jets | 4 | 9 | 6 | 4 |
| $b$-tagging | 3.4 | 5 | 7 | 3.0 |
| $E_{\mathrm{T}}^{\mathrm{miss}}$ | 1.4 | 5 | 0.4 | 0.8 |
| Pile-up | 2.3 | 4 | 5 | 1.8 |

Table 4: Summary of the relative impact of each type of uncertainty on the signal (S) and total background (B) yields in the signal region and on the total background yield in the background control regions before the combined fit.

## 6 Results

A simultaneous fit to the SR and the two CRs is used to search for a signal from FCNC decays of the top quark. A maximum-likelihood fit is performed to kinematic distributions in the signal and control regions to test for the presence of signal events. Contamination of the CRs by the signal is negligible. The kinematic distributions used in the fit are the $\chi^2$ of the kinematical reconstruction for the SR and the leading lepton $p_{\mathrm{T}}$ for the $t\bar{t}Z$ and non-prompt leptons CRs. The expected number of events in each region are shown in Table 5 with the total systematic uncertainties before (after) the combined fit under the background-only hypothesis, while the expected distributions are presented in Figures 2-4.

The statistical analysis to extract the signal is based on a binned likelihood function $L(\mu, \theta)$ as for the Run-2 search [24]. The $L(\mu, \theta)$ is constructed as a product of Poisson probability terms over all bins in each considered distribution, and Gaussian constraint terms for $\theta$, a set of nuisance parameters that parameterize effects of systematic uncertainties on the signal and background expectations. The parameter $\mu$ is a multiplicative factor for the number of signal events normalized to a branching ratio $\mathcal{B}_{\mathrm{ref}}(t \to qZ) = 0.1\%$. In the absence of FCNC signal, upper limits on $\mathcal{B}(t \to qZ)$ can be computed with the $\mathrm{CL_s}$ method [63, 64]. The expected 95% confidence level (CL) limit on $\mathcal{B}(t \to uZ)$ and on $\mathcal{B}(t \to cZ)$ are shown in Tables 6 and 7, which include the contribution from the statistical and systematic uncertainties. The latter one does not include contribution from the MC statistical uncertainty, which given the small size of some of the simulated event samples ($Z$+jets, for instance) is more realistic than the former one.

| Sample | Signal Region | $t\bar{t}Z$ CR | Non-prompt CR |
|---|---|---|---|
| $t\bar{t}Z$ | 2840 ± 400 (± 120) | 3330 ± 410 (± 90) | 1500 ± 160 (± 90) |
| $WZ$ | 920 ± 270 (± 150) | 210 ± 90 (± 60) | 660 ± 140 (± 90) |
| $ZZ$ | 156 ± 22 (± 12) | 20.6 ± 2.6 (± 1.6) | 154 ± 13 (± 11) |
| $tZ$ | 860 ± 170 (± 110) | 360 ± 70 (± 50) | 131 ± 21 (± 18) |
| Non-prompt leptons | 1000 ± 190 (± 90) | 257 ± 93 (± 25) | 4030 ± 900 (± 110) |
| Other | 90 ± 13 (± 8) | 70 ± 15 (± 13) | 1290 ± 130 (± 90) |
| Total bkg. | 5860 ± 810 (± 70) | 4240 ± 520 (± 60) | 7760 ± 1020 (± 90) |
| $t\bar{t} \to bWuZ$ | 299 ± 19 (± 8) | 6.77 ± 0.42 (± 0.19) | 17.7 ± 1.1 (± 0.5) |
| $t\bar{t} \to bWcZ$ | 331 ± 20 (± 9) | 11.64 ± 0.72 (± 0.32) | 23.5 ± 1.5 (± 0.7) |

Table 5: The expected event yields in the signal and background control regions. The number of signal events is normalized to the expected branching ratio limits of $\mathcal{B}(t \to uZ) = 4.6 \times 10^{-5}$ and $\mathcal{B}(t \to cZ) = 5.5 \times 10^{-5}$. Total systematic uncertainties are shown before (after) the combined fit under the background-only hypothesis. After the combined fit, the uncertainty on the total background is smaller than the uncertainty on some of the background contributions due to the negative correlations between some of the background sources.

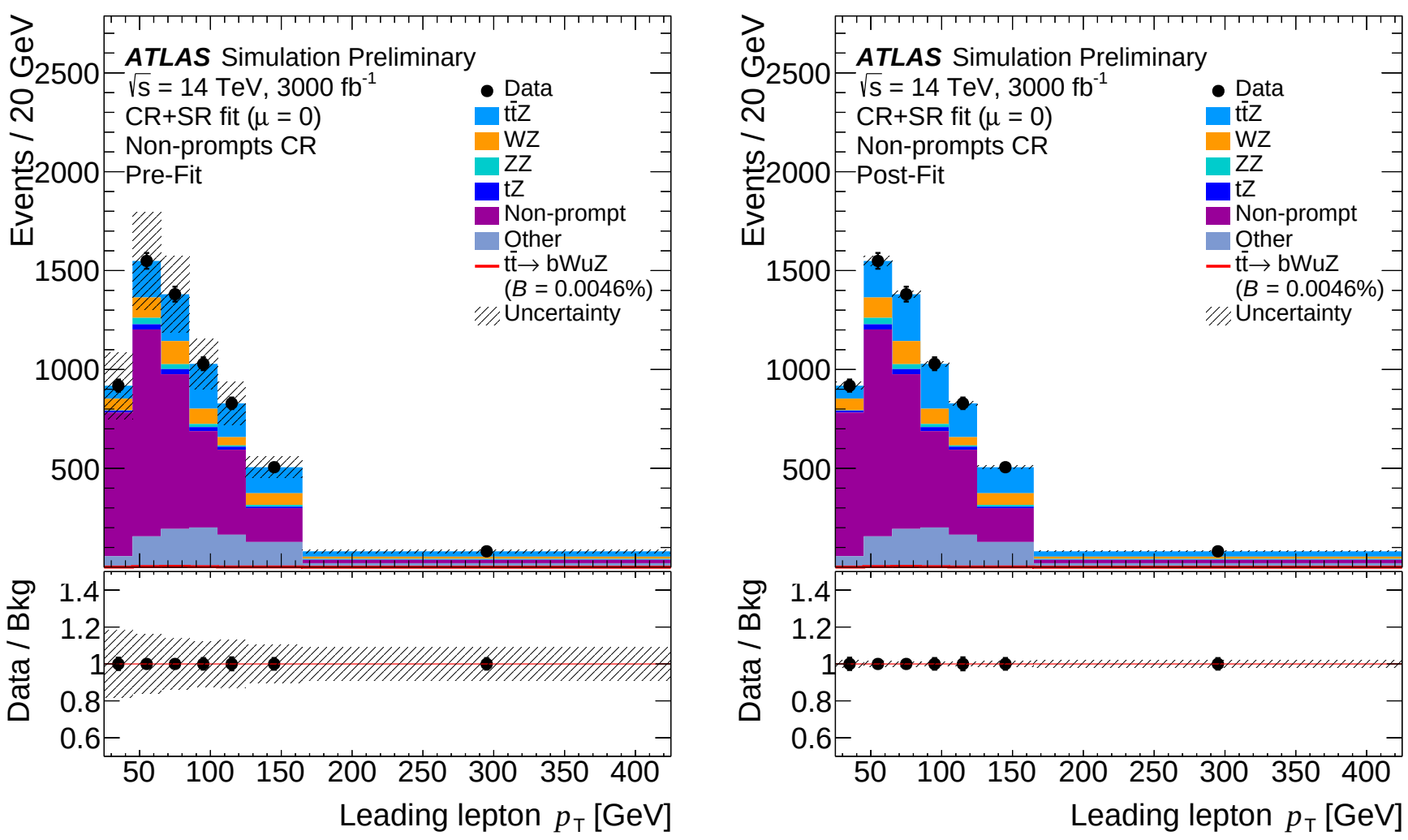

(a)

(b)

Figure 2: The distributions for the $p_{\mathrm{T}}$ of the leading lepton in the non-prompt leptons control region a) before and b) after the combined fit under the background-only hypothesis. The data points are from the "Asimov dataset", defined as a total expected pre-fit background. The number of signal events is normalized to the expected branching ratio limit of $\mathcal{B}(t \to uZ) = 4.6 \times 10^{-5}$. The dashed area represents the systematic uncertainty on the background prediction.

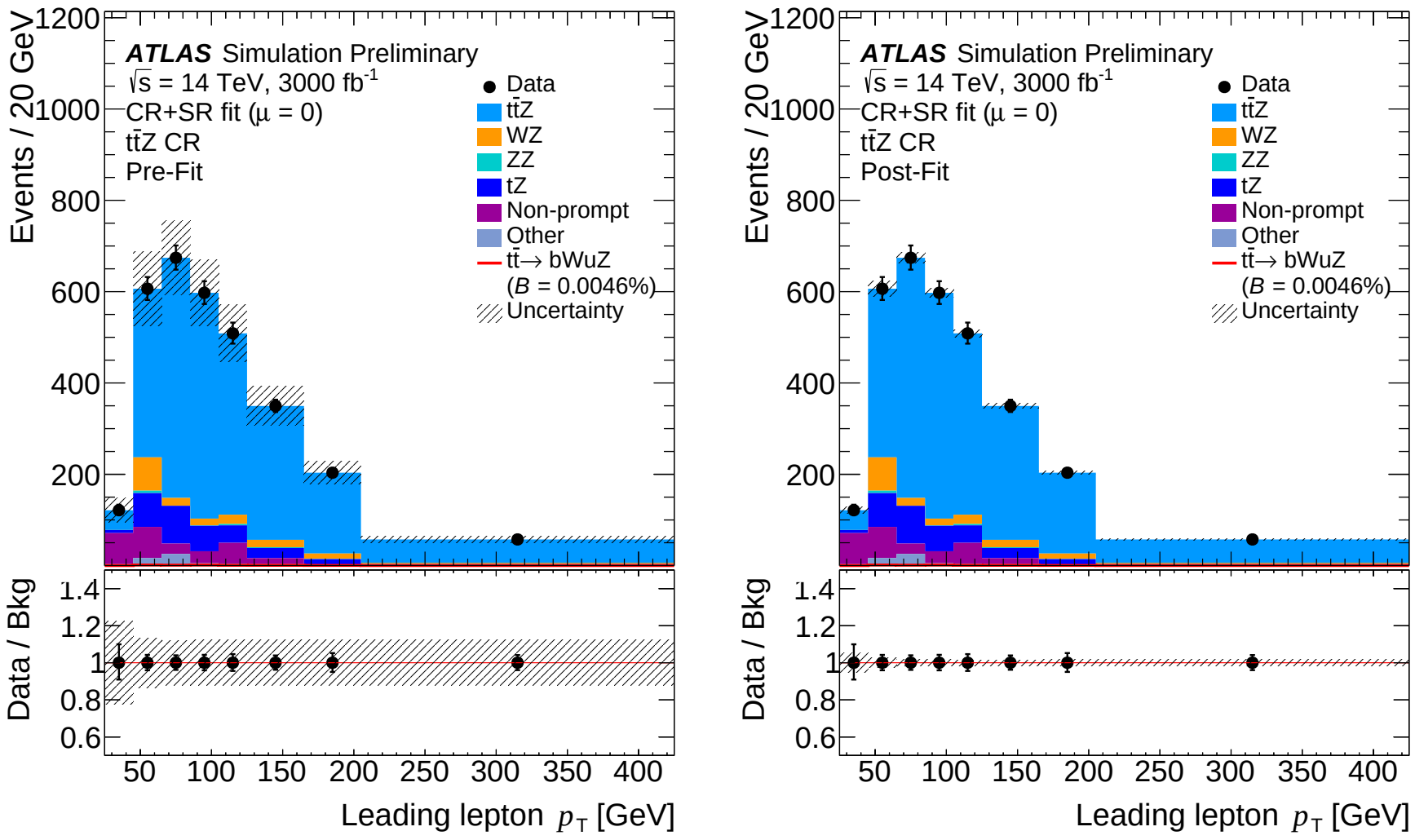

(a) (b)

Figure 3: The distributions for the $p_T$ of the leading lepton in the $t\bar{t}Z$ control region a) before and b) after the combined fit under the background-only hypothesis. The data points are from the "Asimov dataset", defined as a total expected pre-fit background. The number of signal events is normalized to the expected branching ratio limit of $\mathcal{B}(t \to uZ) = 4.6 \times 10^{-5}$. The dashed area represents the systematic uncertainty on the background prediction.

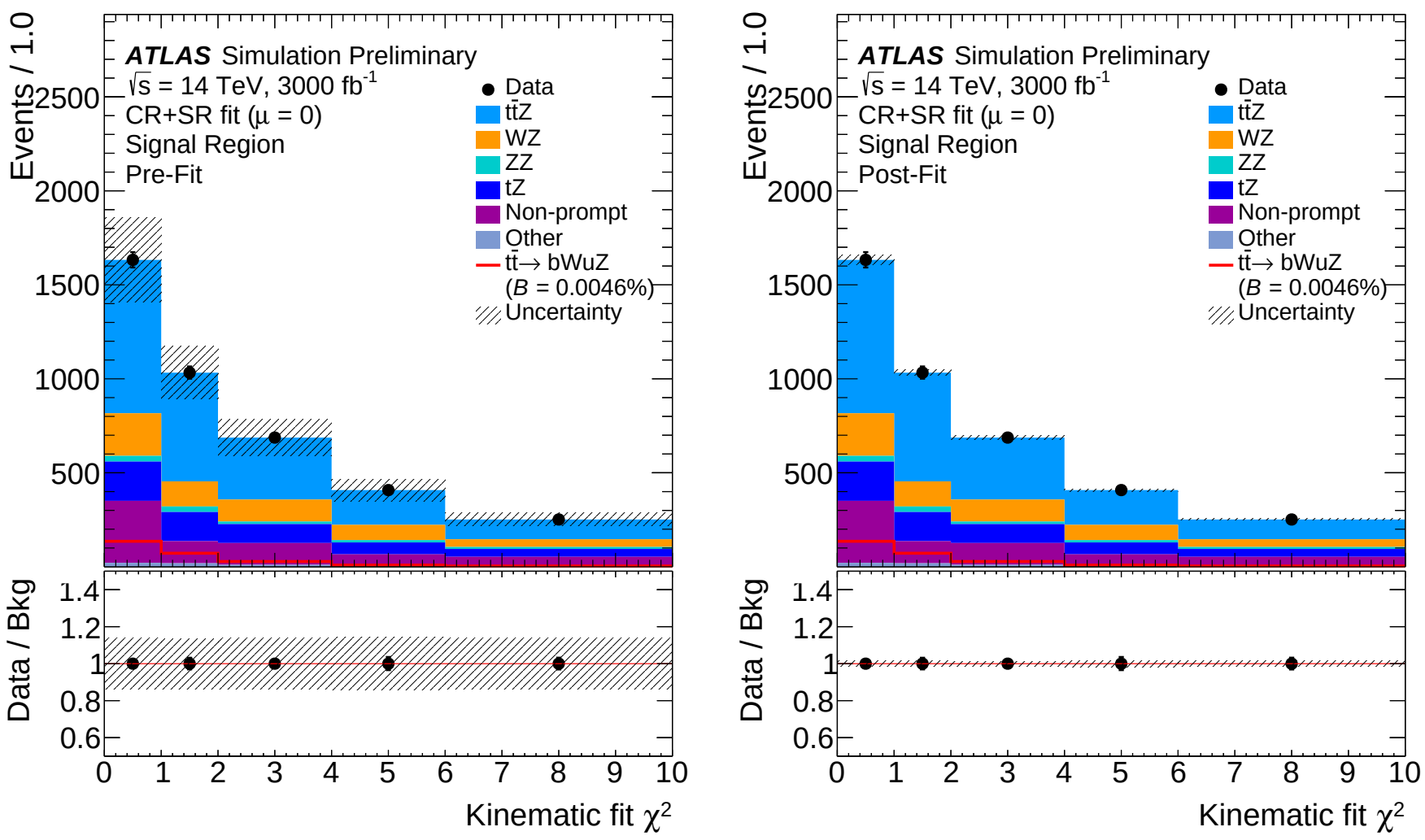

(a) (b)

Figure 4: The distributions for the $\chi^2$ after the event reconstruction in the signal region a) before and b) after the combined fit under the background-only hypothesis. The data points are from the "Asimov dataset", defined as a total expected pre-fit background. The number of signal events is normalized to the expected branching ratio limit of $\mathcal{B}(t \to uZ) = 4.6 \times 10^{-5}$. The dashed area represents the systematic uncertainty on the background prediction.

Inclusion of the CRs in the combined fit with the SR constrains backgrounds, reduces systematic uncertainties and thus improves the $\mathcal{B}(t \to qZ)$ limits. The limits obtained without inclusion of the CRs in the likelihood are about 13% worse compared to the results extracted from the CRs and SR combination. After the combined fit, the dominant contributions to systematic uncertainties come from $E_{\mathrm{T}}^{\mathrm{miss}}$ and jet reconstruction uncertainties. The effect of these uncertainties is estimated in the 13 TeV analysis and the same uncertainties are applied in the HL-LHC studies. If the expected improvements for these sources of systematic uncertainties are taken into account by reducing their effect by a factor of two [62], a further improvement of about 15% on the $\mathcal{B}(t \to qZ)$ limits is to be expected.

For comparison, an extrapolation of the 13 TeV analysis [24] is performed yielding the branching ratio limits of $\mathcal{B}(t \to uZ) < 1.0 \times 10^{-4}$ and $\mathcal{B}(t \to cZ) < 1.4 \times 10^{-4}$. These results are about factor two worse than the ones derived from the HL-LHC samples since the extrapolation approach does not incorporate other changes besides the cross-sections and integrated luminosity, such as the changes in the detector geometry or resolutions, or expected improvements in the estimation of uncertainties.

The limits on the branching ratios can be interpreted in the framework of an Effective Field Theory (EFT) approach, see for example Refs. [46, 47]. In this context limits can be set on the EFT coefficients. According to Ref. [47], the EFT operators to which the analysis is more sensitive are $C_{uB}^{(31)}$, $C_{uW}^{(31)}$, $C_{uB}^{(32)}$ and $C_{uW}^{(32)}$. Assuming a cut-off scale $\Lambda = 1$ TeV and that only one FCNC mode contributes, the branching ratio limits presented in Table 7 are converted to 95% CL upper limits on the moduli of the EFT coefficients. These are shown in Table 8. The results of this analysis should not depend of the handedness of the EFT couplings [65].

| | -1$\sigma$ | Expected | +1$\sigma$ |
|---|---|---|---|
| $\mathcal{B}(t \to uZ)$ | $4.9 \times 10^{-5}$ | $6.9 \times 10^{-5}$ | $9.7 \times 10^{-5}$ |
| $\mathcal{B}(t \to cZ)$ | $5.8 \times 10^{-5}$ | $8.1 \times 10^{-5}$ | $12 \times 10^{-5}$ |

Table 6: The expected 95% confidence level upper limits on the top-quark FCNC decay branching ratios are shown together with the $\pm 1\sigma$ bands, which include the contribution from the statistical and systematic uncertainties. Presented limits are extracted from "Asimov data" in the signal and background control regions, defined as the total expected pre-fit background. Systematic uncertainty from the MC statistical uncertainty is considered as well.

| | -1$\sigma$ | Expected | +1$\sigma$ |
|---|---|---|---|
| $\mathcal{B}(t \to uZ)$ | $3.3 \times 10^{-5}$ | $4.6 \times 10^{-5}$ | $6.5 \times 10^{-5}$ |
| $\mathcal{B}(t \to cZ)$ | $3.9 \times 10^{-5}$ | $5.5 \times 10^{-5}$ | $7.7 \times 10^{-5}$ |

Table 7: The expected 95% confidence level upper limits on the top-quark FCNC decay branching ratios are shown together with the $\pm 1\sigma$ bands, which include the contribution from the statistical and systematic uncertainties. Presented limits are extracted from "Asimov data" in the signal and background control regions, defined as the total expected pre-fit backgrounds. Systematic uncertainty from the MC statistical uncertainty is not considered.

| Operator | Expected limit |
|---|---|
| $\lvert C_{uB}^{(31)} \rvert$ | 0.13 |
| $\lvert C_{uW}^{(31)} \rvert$ | 0.13 |
| $\lvert C_{uB}^{(32)} \rvert$ | 0.14 |
| $\lvert C_{uW}^{(32)} \rvert$ | 0.14 |

Table 8: Expected 95% CL upper limits on the moduli of the operators contributing to the FCNC decays $t \to uZ$ and $t \to cZ$ within the TopFCNC model for a new-physics energy scale $\Lambda = 1$ TeV.

# 7 Conclusion

The sensitivity of the ATLAS experiment in the search for flavour-changing neutral-current top quark decays is presented. The study is performed in the context of the high luminosity phase of the Large Hadron Collider with a centre-of-mass energy of 14 TeV and an integrated luminosity of 3000 fb$^{-1}$. The three charged lepton final state of $t\bar{t}$ events is considered, in which one of the top quarks decays through the $t \to qZ$ ($q = u, c$) flavour-changing neutral-current channel and the other one decays to $bW$ ($t\bar{t} \to bWqZ \to b\ell\nu q\ell\ell$). An improvement by a factor of four is expected over the current Run-2 analysis results of $\mathcal{B}(t \to uZ) < 1.7 \times 10^{-4}$ and $\mathcal{B}(t \to cZ) < 2.4 \times 10^{-4}$ with 36.1 fb$^{-1}$ integrated luminosity. The branching ratio limits that are obtained are at the level of 4 to $5 \times 10^{-5}$ depending on the considered scenarios for the systematic uncertainties.

# ATLAS PUB Note

# Prospects for measurement of the top quark mass using $t\bar{t}$ events with $J/\psi \to \mu^+\mu^-$ decays with the upgraded ATLAS detector at the High Luminosity LHC

The ATLAS Collaboration

This document presents projections for the top quark mass measurement accuracy using $t\bar{t} \to$ lepton+jets events with $J/\psi \to \mu^+\mu^-$ in the final state at $\sqrt{s}$ = 14 TeV at the High-Luminosity LHC with 3000 fb$^{-1}$ of proton-proton collisions with the ATLAS experiment. A statistical uncertainty of 0.14 GeV is expected, with a systematic uncertainty of 0.48 GeV.

# 1 Introduction

A variety of alternative methods are exploited to supplement the top quark mass ($m_{\text{top}}$) measurements from direct mass reconstruction based on jet observables. In this note, one example is investigated, namely the measurement of $m_{\text{top}}$ from final states where one of the $b$-quarks hadronises into a $B$ hadron which decays into a $J/\psi$ meson, which then decays into a muon-antimuon pair. This approach relies on a template method exploiting the top quark mass sensitivity of the invariant mass $m(\ell\,\mu^+\mu^-)$ of the system made of the $J/\psi \to \mu^+\mu^-$ meson candidate and the isolated lepton coming from the associated $W$ boson decay. As this observable involves only three reconstructed leptons, the sensitivity to the light-jet and $b$-jet energy scale is expected to be reduced compared to the final states where the observables are based on jet reconstruction. One of the limiting factors of this approach is the small branching fraction, $\mathcal{B}(t\bar{t} \to (W^+b)(W^-b) \to (\ell\nu_\ell J/\psi(\to \mu^+\mu^-)X)(qq'b)) \sim 4.1 \times 10^{-4}$, where $\ell = e, \mu$. With this technique, the $b$-production and the $b$-fragmentation are expected to be among the dominating sources of systematic uncertainties. The use of different approaches and observables to measure $m_{\text{top}}$ should help to reduce the uncertainties in a combination of all measurements.

The use of $t\bar{t}$ events containing $J/\psi \to \mu^+\mu^-$ decays to measure the top quark mass has already been considered by both ATLAS and CMS [1–3]. These analyses are statistically limited, due to the low branching fraction, and thus will benefit from the High-Luminosity (HL) upgrade of the Large Hadron Collider (LHC) and the ATLAS detector [4].

Upgrades of the ATLAS detector [5] will be necessary to maintain its performance in the higher luminosity environment[1]. A new inner tracking system, extending the tracking region from $|\eta| < 2.7$ up to $|\eta| < 4$, will provide the ability to reconstruct charged particles in the forward region, which can be matched to calorimeter clusters for forward electron reconstruction, or associated to forward jets. The inner tracker extension also enables muon identification at high $\eta$ if additional detectors are installed in the region $2.7 < |\eta| < 4$.

This note presents projections for the accuracy of the top quark mass measurement at the HL-LHC, using the $J/\psi$ decay mode with the full expected luminosity of 3000 fb$^{-1}$ of proton-proton collisions at $\sqrt{s} = 14$ TeV. In a HL-LHC scenario, both the ATLAS detector and the analysis strategy are expected to change significantly from the Run-2 analysis and it is difficult to make reliable predictions for the systematic uncertainties relevant to the HL-LHC analysis. Following the existing recommendations for the HL-LHC studies [6, 7], a reduction of the $t\bar{t}$ modelling uncertainties by a factor of two and a reduction of some of the experimental uncertainties by up to a factor two are assumed. As such, the main result of this study is a statistical projection of the measurement. Where possible, the impact of typical sources of systematic uncertainty on the measurement are estimated.

The top quark mass determined in this method corresponds to the mass definition used in the MC simulation. Because of various steps in the event simulation, the mass measured in this way does not necessarily directly coincide with mass definitions within a given renormalization scheme, e.g. the top quark pole mass. Evaluating these differences is a topic of theoretical investigations [8–11].

[1] ATLAS uses a right-handed coordinate system with its origin at the nominal IP in the centre of the detector and the $z$-axis along the beam pipe. The $x$-axis points from the IP to the centre of the LHC ring, and the $y$-axis points upward. Cylindrical coordinates $(r, \phi)$ are used in the transverse plane, $\phi$ being the azimuthal angle around the $z$-axis. The pseudorapidity is defined in terms of the polar angle $\theta$ as $\eta = -\ln\tan(\theta/2)$. The transverse momentum and energy are defined as $p_{\text{T}} = p\sin\theta$ and $E_{\text{T}} = E\sin\theta$, respectively. The angular distance $\Delta R$ is defined as $\Delta R = \sqrt{(\Delta\eta)^2 + (\Delta\phi)^2}$.

## 2 Simulation samples

Samples of simulated events for signal and background processes are produced at 14 TeV centre-of-mass energy. They include the production of $t\bar{t}$ pairs, single-top quarks and $W/Z$ bosons in association with jets. Using their theoretically-predicted cross-sections, the MC samples are normalised to an integrated luminosity of 3000 fb$^{-1}$. The EvtGen (v1.2.0) [12] program is used to handle the decays of $b$- and $c$-flavoured hadrons in all samples.

The baseline $t\bar{t}$ simulation sample is produced using the next-to-leading order (NLO) Powheg-Box (r3026 v2) matrix-element (ME) event generator [13–16] and the NNPDF3.0 parton distribution function (PDF) set [17]. The parton shower, hadronisation and underlying event are simulated using Pythia 8 (v8.210) [18] with the A14 tune [19] using the NNPDF2.3 PDF set [20] in the parton shower. The number of generated $t\bar{t}$ events is $2 \times 10^7$ for the nominal sample corresponding to a an equivalent luminosity of 37.1 fb$^{-1}$.

Samples of single-top quarks corresponding to the $t$-channel and $s$-channel are generated with Powheg-Box (r3026 v2) and the NNPDF3.0 parton distribution function (PDF). The parton shower, hadronisation and underlying event are simulated using Pythia 8 (v8.210) with the A14 tune using the NNPDF2.3 PDF set in the parton shower. Samples for the $Wt$ production are generated with Powheg-Box v1 using the CT10 PDF set, interfaced to Pythia 6 with Perugia P2012C tunable parameters. The higher-order overlap with $t\bar{t}$ production is addressed using the "diagram removal" (DR) generation scheme [21].

In the aboved mentioned $t\bar{t}$ and single-top quarks simulation samples, the top quark mass is set to 172.5 GeV. Additional $t\bar{t}$ samples are generated for different assumed values of $m_{\text{top}}$, from 169 to 176 GeV, with otherwise unchanged parameters. Single-top quark production samples with different values of $m_{\text{top}}$ are not available for this analysis.

Detailed descriptions of other samples can be found in Ref. [22].

## 3 Event reconstruction and selection

After the event generation step, a fast simulation of the trigger and detector effects is added with the dedicated ATLAS software framework [23]. The trigger, reconstruction and identification efficiencies, the energy and transverse momentum resolution of leptons and jets are computed as functions of their $\eta$ and $p_{\text{T}}$ using full simulation studies assuming an upgraded ATLAS detector. They provide a parameterised estimate of the ATLAS performance at the HL-LHC. These smearing functions are applied to the particle-level quantities. The smearing functions assume the HL-LHC conditions of an instantaneous luminosity of $\mathcal{L} = 7.5 \times 10^{34}$ cm$^{-2}$ s$^{-1}$ and an average number of additional collisions per bunch-crossing of $< \mu >= 200$.

Details of the object selection and the corresponding assumed reconstruction efficiencies and resolutions can be found in Ref. [6].

Electrons and muons are reconstructed in the fiducial region of transverse momentum $p_{\text{T}} > 25$ GeV and pseudorapidity $|\eta| < 4$. Identification efficiencies [4] are applied to the lepton candidates to select which particles are identified as leptons. Similarly, isolation efficiencies are applied. The energy, the $p_{\text{T}}$ and the $\eta$ of the lepton candidates are smeared according to the detector resolution. This analysis makes use of additional muon candidates, selected with a $p_{\text{T}}$ threshold of 4 GeV, to reconstruct $J/\psi \rightarrow \mu^+\mu^-$ candidates within jets. These 'soft muons' are required to lie within $|\eta| < 4$ and do not have to satisfy isolation requirements.

Jets are reconstructed using the anti-$k_t$ algorithm [24, 25] implemented in the FastJet package [26], with a radius parameter of 0.4. Jets are accepted if they have $p_T > 25$ GeV and $|\eta| < 4.5$. Double counting of electrons as jets may arise from electron energy deposition in the calorimeter being clustered by the jet algorithm. To mitigate such effects, jets are removed if within $\Delta R = 0.2$ of a selected electron. After this step, electrons within $\Delta R = 0.4$ from a jet are rejected, since they are considered as decay products of the hadrons in the jet. For the same reason, isolated muons that are within $\Delta R = 0.04 + (10 \text{ GeV}/p_T)$ from a jet are also removed. A fraction of the particle-level jets are removed, according to the expected jet reconstruction efficiency [6]. The energy, $p_T$ and $\eta$ of remaining jets are smeared according to the detector resolution. Pile-up jets are rejected using tracking information.

Fake leptons are obtained from functions parametrising the expected level of light-jet to electron fake rate and the level of muon fake rate. Fake electrons are introduced according to the probability that a jet is misidentified as an electron as measured in full simulation, where the jet can come from the hard scattering vertex or from the pile-up. When a jet is reconstructed as an electron its energy is changed accordingly.

The event selection follows the analysis done at 8 TeV [1], except for the increased $\eta$ acceptance of leptons and jets. Events are required to have at least one charged isolated lepton with $p_T > 25$ GeV and $|\eta| < 4$ and at least 4 jets with $p_T > 25$ GeV and $|\eta| < 4.5$. No requirement is applied on the number of $b$-tagged jets. Figure 1 shows the distributions of the $p_T$ of the leading jet, the $p_T$ of the isolated lepton and the $p_T$ and $\eta$ of the soft muons.

$J/\psi$ candidates are reconstructed using all pairs of opposite charge sign soft muons. The two muon tracks are not refitted to a common vertex to form a $J/\psi$. Figure 2(a) shows the invariant mass distributions of the dimuon pairs with an invariant mass value below 20 GeV. A peak around the $J/\psi$ mass is clearly visible. More than $6 \times 10^6$ dimuon pairs remain at this level.

Only events with $J/\psi$ candidates in the mass range [3.0;3.2] GeV are retained. To reduce combinatoric background, only events with exactly one $J/\psi$ candidate in this mass range are kept. Finally, further selection criteria are applied. At least one of the two muons is required to be within a distance $\Delta R < 0.5$ from a jet. The angular distance between the two soft muons must be $\Delta R < 0.8$. The transverse momentum of the $J/\psi$ candidate must be $p_T > 8$ GeV. The transverse decay length of the $J/\psi$ candidate must be $L_{xy} > 0$ mm, with $L_{xy} = \frac{\vec{L}\vec{p}_T}{p_T}$ where $\vec{L}$ is the vector of the distance between the primary vertex and the extrapolated common vertex of the two soft muon candidates in the transverse plane, and $\vec{p}_T$ is the reconstructed transverse momentum vector of the dimuon candidate.

Figure 2(b) shows the distribution of the invariant mass of the $2 \times 10^5$ candidates, after the final selection, corresponding to an event rate of about 70 events per $\text{fb}^{-1}$ at 14 TeV. Due to the higher cross-section and despite possible losses due to trigger conditions and the increase of the pile-up, the amount of expected events is about 18% higher at 14 TeV than at 13 TeV. Furthermore, because of the increase of the $\eta$ coverage with the ATLAS HL-LHC detector to $|\eta| < 4$ for leptons an extra increase of about 10% of events is expected compared to the Run-2 analysis.

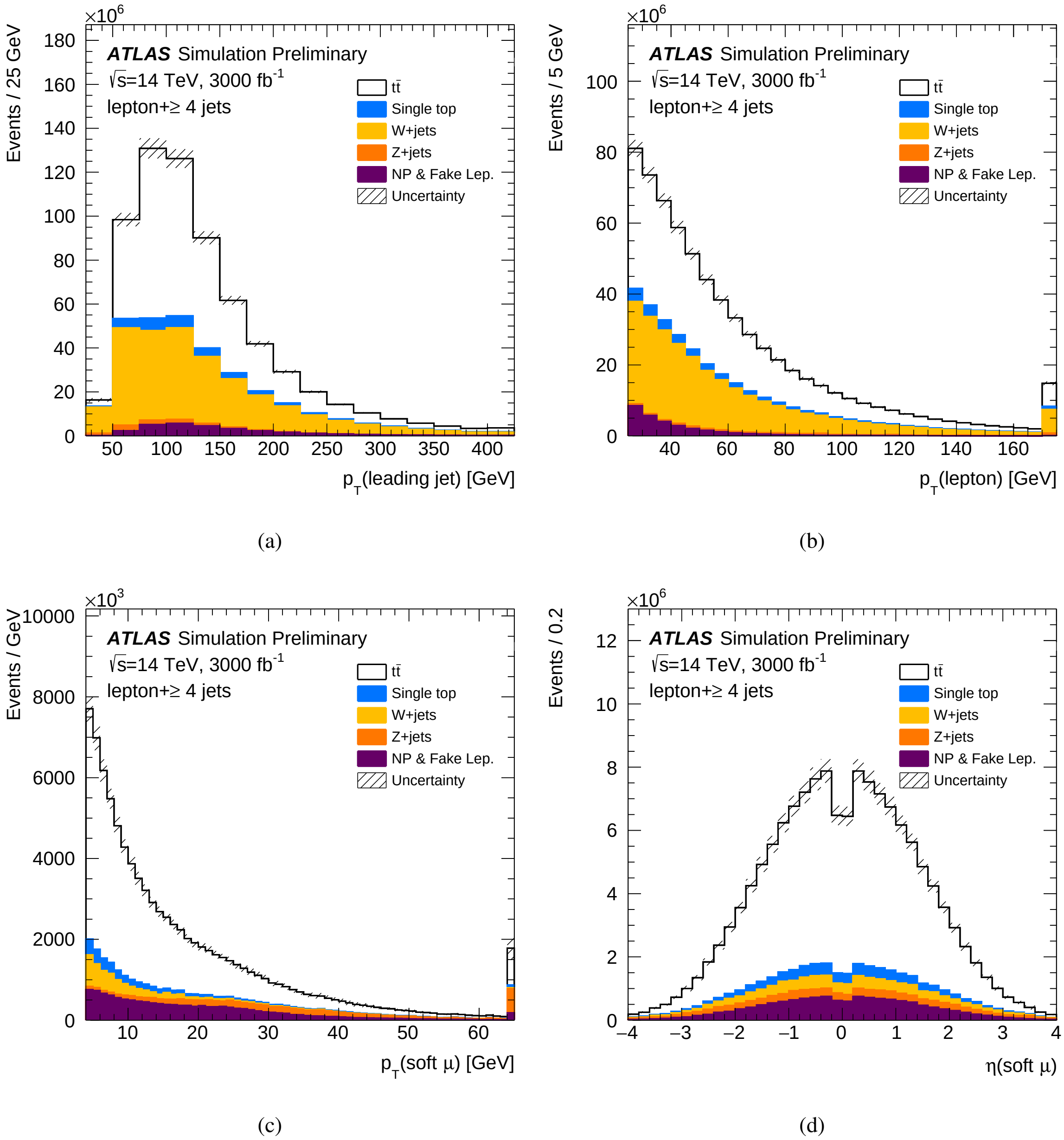

Figure 1: Distributions of (a) $p_\mathrm{T}$ of the leading jet and (b) $p_\mathrm{T}$ of the isolated lepton in events with one isolated lepton and at least four jets, (c) $p_\mathrm{T}$ and (d) $\eta$ of the soft muons in these events. Expectations are obtained from simulation, broken down into contributions from $t\bar{t}$, single-top, $W+$jets, $Z+$jets and events with non-prompt and fake leptons (referred to as 'NP & Fake Lep.'). The shaded area represents the combination of MC statistical uncertainties and systematic uncertainties on cross-sections. The rightmost bin contains all entries with values above the lower edge of this bin.

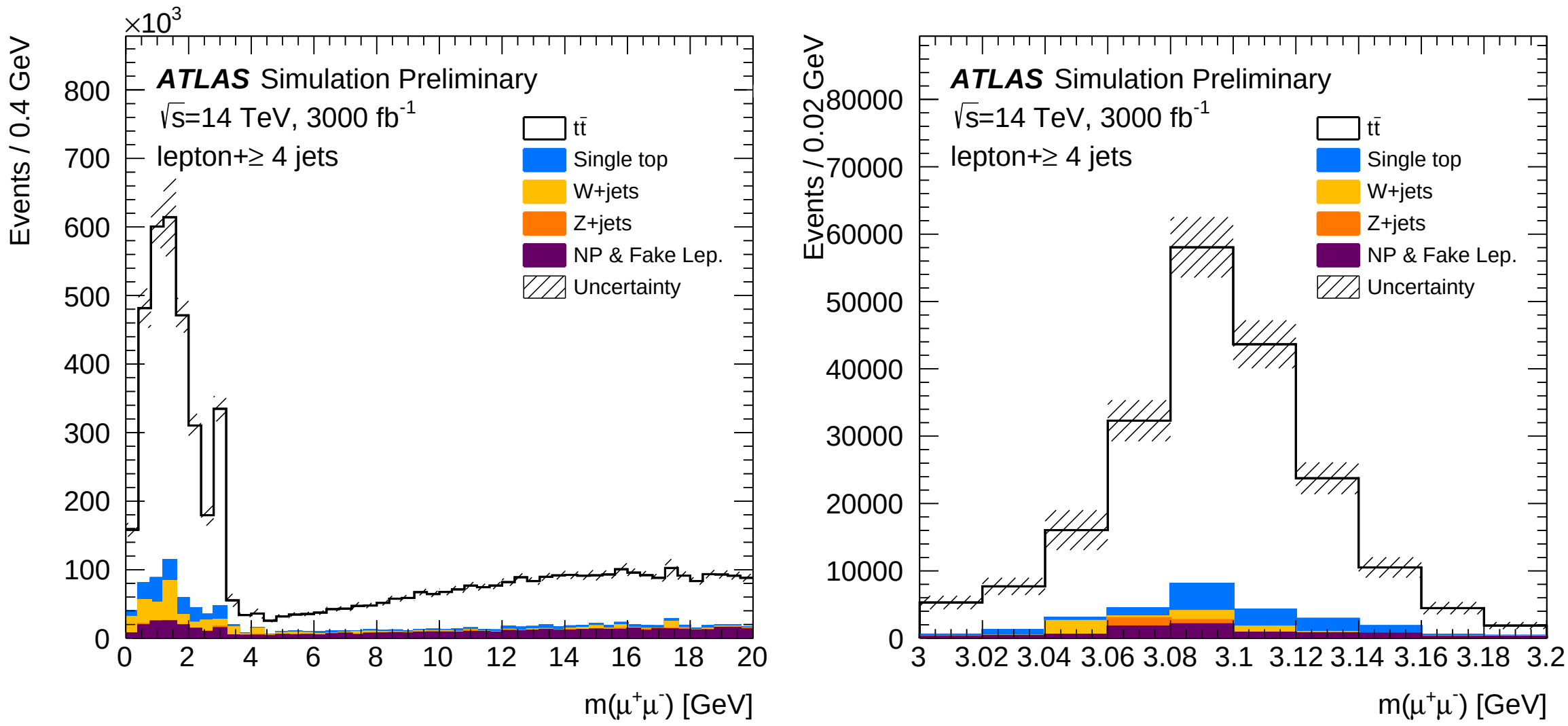

(a)

(b)

Figure 2: Distributions of the invariant mass of the dimuon candidates (a) in the mass range [0;20] GeV after the selection of soft muons and (b) in the mass range [3.0;3.2] GeV after the full selection. Expectations are obtained from simulation, broken down into contributions from $t\bar{t}$, single-top, $W+$jets, $Z+$jets and events with non-prompt and fake leptons (referred to as 'NP & Fake Lep.'). The shaded area represents the combination of MC statistical uncertainties and systematic uncertainties on cross-sections.

# 4 Measurement of the top quark mass

## 4.1 The template method

Figure 3 shows the distribution of the invariant mass $m(\ell\ \mu^+\mu^-)$ of the system made of the $J/\psi \to \mu^+\mu^-$ meson candidate and the isolated lepton. The top quark mass can be measured in the selected events by using the sensitivity of $m(\ell\ \mu^+\mu^-)$ to $m_{\text{top}}$. In the template method, probability density functions are constructed from all signal and background MC simulated samples. They are obtained for different $m_{\text{top}}$ values using $t\bar{t}$ samples generated at different top quark mass values: 169, 170, 171, 172, 172.25, 172.5, 172.75, 173, 174, 175 and 176 GeV. Appropriate functions are fitted to these templates, interpolating between different input $m_{\text{top}}$. The parameters of the functions are fixed by a simultaneous fit to all templates, imposing linear dependences of the parameters on $m_{\text{top}}$. The resulting template fit function has $m_{\text{top}}$ as the only free parameter and an unbinned likelihood maximisation gives the value of $m_{\text{top}}$ that best describes the data. The choosen analytical function is the sum of a Gaussian and a Gamma function. Figure 4 shows the templates obtained for different $m_{\text{top}}$ input values, overlaid with the corresponding probability density function from the fit. It shows the sensitivity of the templates to the input $m_{\text{top}}$ value, the knowledge of which is currently limited by the low number of MC simulated events.

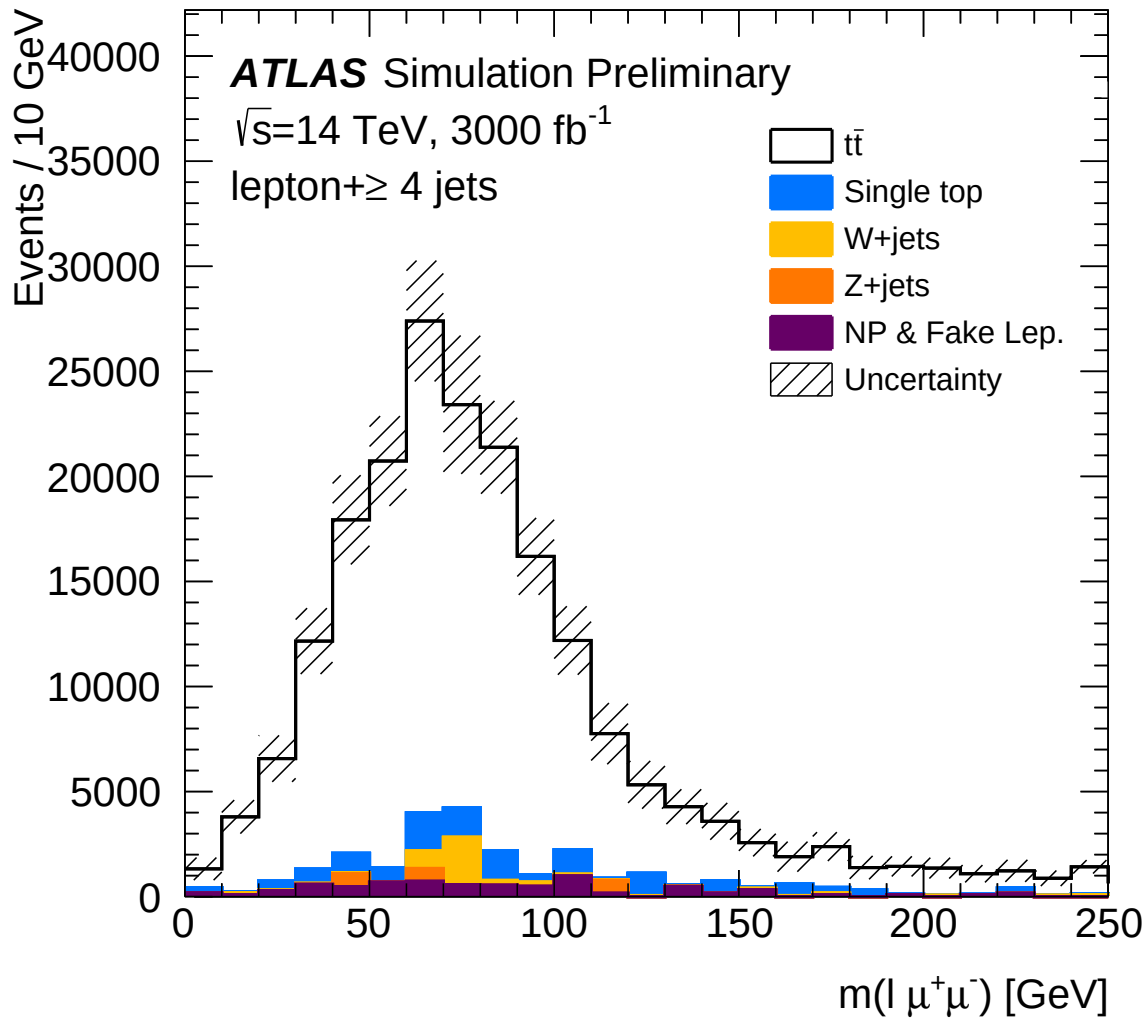

Figure 3: Distribution of the invariant mass $m(\ell\ \mu^+\mu^-)$ of the system made of the isolated lepton and the two soft muons, after the full selection, using dimuon candidates in the mass range [3.0;3.2] GeV. Expectations are obtained from simulation, broken down into contributions from $t\bar{t}$, single-top, $W+$jets, $Z+$jets and events with non-prompt and fake leptons (referred to as 'NP & Fake Lep.'). The shaded area represents the combination of MC statistical uncertainties and systematic uncertainties on cross-sections. The rightmost bin contains all entries with values above the lower edge of this bin.

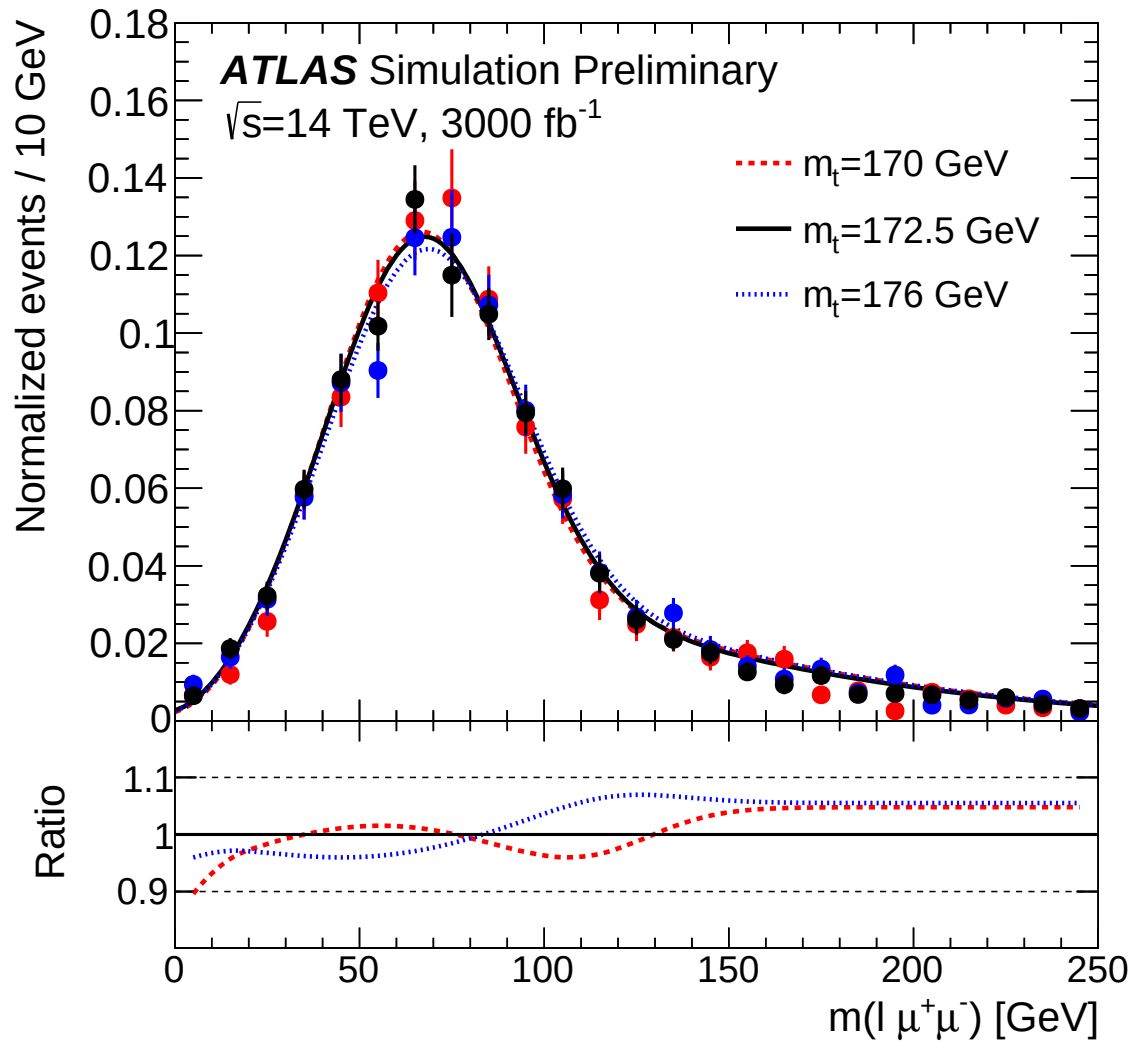

Figure 4: Template parametrization showing the sensitivity of $m(\ell\ \mu^+\mu^-)$ to the input $m_{\text{top}}$. Each template (shown as points with uncertainties corresponding to statistical uncertainties) is overlaid with the corresponding probability density function (shown as lines) from the fit to templates. In addition, the lower panel shows ratios of the three fitted functions and the fitted function for $m_{\text{top}}$ = 172.5 GeV.

### 4.2 Uncertainties affecting the $m_{top}$ determination

Due to the changes of the detector performance for the HL-LHC, it is difficult to estimate precisely the effects of systematic uncertainties. The sources of uncertainty are assumed to be the same as the current ones. Systematic uncertainties are estimated by varying each source of uncertainty and determining the impact on the mass measurement.

The residual difference between fitted and generated $m_{\text{top}}$ when analysing a template from a MC sample reflects the potential bias of the method. A constant is fitted to the observed $m_{\text{top}}$ residuals. This constant and its statistical uncertainty is assigned as the method uncertainty. This also covers effects from limited numbers of simulated events in the templates and potential deficiencies in the template parameterizations.

The signal modelling uncertainties of the $t\bar{t}$ physics processes concern the choices of the $t\bar{t}$ NLO generator, the parton shower and hadronisation model, the modelling of heavy-flavour production (hereafter $b$-production) in $t\bar{t}$ events, the $b$-fragmentation parameters, the modelling of initial- and final-state radiation, the underlying event and the colour reconnection effects. The $b$-production uncertainty originates from the effect of the uncertainties on the production fractions for different species of $b$-hadrons as well as the uncertainties on the branching fractions of the decays of $b$-hadrons to $J/\psi$, which likely will affect the kinematics of the $J/\psi$. The $b$-fragmentation uncertainty is assesed through a re-tune of the Pythia 8 parameters to describe the $b$-quark fragmentation function measured at LEP. All these uncertainties are evaluated from simulated events using different generators or tuning.

The background modelling uncertainties are related to the uncertainties on the shape and normalisation of the fake soft muons, the fake isolated leptons and other background processes.

The $t\bar{t}$ signal and background modelling uncertainties are measured in the preliminary Soft Muon Tagger analysis at 13 TeV, and are taken as suitable estimates for the $J/\psi$ analysis.

Experimental uncertainties arise from the modelling and calibration of the ATLAS detector response, affecting the performance of the event selection and the final state reconstruction. This measurement essentially relies on lepton reconstruction, especially on the reconstruction of muons, and is therefore susceptible to uncertainties on the lepton energy scales, resolution and reconstruction efficiencies. As $m(\ell\,\mu^+\mu^-)$ involves only three reconstructed leptons, the sensitivity to the light-jet and $b$-jet energy scale (JES/$b$-JES) as well as to the jet energy resolution (JER) is expected to be reduced compared to observables based on reconstructed jets.

### 4.3 Extrapolation scenario

The estimated Run-2 uncertainties are scaled to align with HL-LHC extrapolations developed by the ATLAS and CMS Collaborations and documented in Ref. [6, 7].

The theory modelling uncertainties are expected to be reduced by a factor two compared to existing values. The larger HL-LHC dataset will allow for generator tuning, as already started with Run-2 data [27], and smaller uncertainties based on the measurements of different kinematic distributions. The $b$-fragmentation parameters are expected to be measured directly in the LHC data, leading to a tuning of current generators to get a better match to data and smaller associated uncertainty.

The impact of the experimental systematic uncertainties will likely be reduced relative to their effect on the Run-2 analysis given the large datasets available, allowing precise performance studies to be conducted.

The jet reconstruction uncertainties on $m_{\text{top}}$ are expected to be reduced by a factor up to two, while uncertainties related to the reconstruction of electrons and muons remain the same as in Run-2.

## 4.4 Results

The projections for the accuracy of the top quark mass measurement using the $J/\psi$ decay mode with the full expected luminosity of 3000 fb$^{-1}$ of proton-proton collisions at $\sqrt{s}$ = 14 TeV are detailed in Table 1.

A statistical uncertainty of 0.14 GeV is expected, with a method uncertainty of 0.11 GeV.

| Source of uncertainty | $\sigma(m_{\text{top}})$ [GeV] |
|---|---|
| Statistical uncertainty | 0.14 |
| Method uncertainty | 0.11 |
| **Signal modelling uncertainties** | |
| $t\bar{t}$ NLO modelling | 0.06 |
| $t\bar{t}$ PS and hadronisation | 0.05 |
| $t\bar{t}$ $b$-production | 0.24 |
| $t\bar{t}$ $b$-fragmentation | 0.11 |
| Initial- and final-state radiation | 0.04 |
| Underlying event | 0.02 |
| Colour reconnection | 0.02 |
| Background modelling uncertainties | 0.10 |
| **Experimental uncertainties** | |
| Jet energy scale (JES) | 0.31 |
| $b$-jet energy scale ($b$-JES) | 0.06 |
| Jet energy resolution (JER) | 0.13 |
| Jet vertex fraction | 0.02 |
| Electrons | 0.03 |
| Muons | 0.09 |
| Pile-up | 0.04 |
| Total Systematic uncertainty | 0.48 |
| Total | 0.50 |

Table 1: The contributions of the various sources to the uncertainty on $m_{\text{top}}$ using $m(\ell\,\mu^+\mu^-)$ templates, extrapolated to an integrated luminosity of 3000 fb$^{-1}$ with a center-of-mass energy of 14 TeV. The last line refers to the sum in quadrature of the statistical and systematic uncertainties.

From preliminary studies using the Run-2 dataset, the dominant signal modelling uncertainties are related to the $t\bar{t}$ $b$-production and $t\bar{t}$ $b$-fragmentation. These uncertainties are expected to be respectively 0.24 and 0.11 GeV. Other $t\bar{t}$ modelling uncertainties are expected to be below 0.1 GeV each.

With the level of background being small, the background modelling uncertainties are expected to be 0.10 GeV.

The experimental uncertainties are expected to be 0.36 GeV. They are dominated by the uncertainties related to the JES (0.31 GeV) and the JER (0.13 GeV). Presently, the evaluation of both uncertainties still suffers from large statistical uncertainties. The uncertainties related to the muon reconstruction are expected to be 0.09 GeV.

Finally, the total systematic uncertainty is expected to be 0.48 GeV.

A total of 3000 $\text{fb}^{-1}$ of 14 TeV data would clearly decrease the statistical uncertainty in this analysis so that the precision would be limited by systematic effects. Therefore, the statistical precision could be traded in various ways for a reduced total systematic uncertainty by cutting into phase space regions where the systematic uncertainties are high.

# 5 Conclusion

The top quark mass measurement using $t\bar{t} \to$ lepton+jets events with $J/\psi \to \mu^+\mu^-$ in the final state is presented. Based on studies with a 13 TeV dataset, projections for the measurement accuracy at the High-Luminosity LHC using the full expected luminosity of 3000 $\text{fb}^{-1}$ of proton-proton collisions at $\sqrt{s} = 14$ TeV are derived. A statistical uncertainty of 0.14 GeV is expected, with a systematic uncertainty of 0.48 GeV.

ATLAS PUB Note

ATL-PHYS-PUB-2018-026

31st October 2018

# Prospects for the measurement of the $W$-boson mass at the HL- and HE-LHC

The ATLAS Collaboration

This note evaluates the potential of a dedicated dataset collected at low instantaneous luminosity and moderate pile-up for the measurement of the $W$-boson mass at the HL-LHC. The value of such data lies in the optimal reconstruction of missing transverse momentum allowed by the low detector occupancy, and in the extended pseudorapidity coverage of the upgraded ATLAS detector. Both effects allow a reduction of PDF uncertainties below what can be achieved using data from Run 1 and Run 2. The impact of a possible further increase in centre-of-mass energy, and of future deep-inelastic scattering data is also evaluated.

# 1 Introduction

Proton-proton collision data at low pile-up are of large interest for $W$ boson physics, as the low detector occupancy allows an optimal reconstruction of missing transverse momentum, and the $W$ production cross section is large enough to achieve small statistical uncertainties in a moderate running time. At $\sqrt{s}$ = 14 TeV and for an instantaneous luminosity of $\mathcal{L} \sim 5\times10^{32}$ cm$^{-2}$s$^{-1}$, corresponding to two collisions per bunch crossing on average at the LHC, about $2\times10^6$ W boson events can be collected in one week. Such a sample provides a statistical sensitivity at the permille level for cross section measurements, at the percent level for measurements of the $W$ boson transverse momentum distribution, and of about 10 MeV for a measurement of $m_W$.

Additional potential is provided by the new tracking detector, the ITk [1], which extends the coverage in pseudorapidity beyond $|\eta| < 2.5$ to $|\eta| < 4$. The increased acceptance allows $W$-boson measurements to probe a new region in Bjorken $x$ at $Q^2 \sim m_W^2$. This will in turn allow further constraints on the parton density functions (PDFs) from cross section measurements, and reduce PDF uncertainties in the measurement of $m_W$. A possible increase of the LHC centre-of-mass energy, such as the HE-LHC program with $\sqrt{s}$ = 27 TeV [2], could play a similar role.

This note presents a first quantitative study of this potential, focusing on the measurement of $m_W$ and restricting the discussion to statistical and PDF uncertainties. Experimental systematic uncertainties are not discussed in this note; their effect is largely of statistical nature for the moderate size, low pile-up samples considered here, and with adequate efforts and exploiting the full available data sample, their impact can be maintained at a level similar to the statistical uncertainty. Theoretical uncertainties in the modelling of $W$-boson production, like the description of the boson transverse momentum distribution, will also be constrained by measurements using these data. However, PDF uncertainties do not scale simply and become dominant once sufficient data are collected, which motivates the present study.

PDF uncertainties and their correlations are studied as a function of the $W$ boson charge, the decay lepton $|\eta_\ell|$, and $\sqrt{s}$; similar studies were presented in Refs. [3, 4]. Present-day PDF sets are considered as well as projected future sets representing the expected constraints from present and future $pp$ data, and from the LHeC deep inelastic scattering project [5]. Section 2 summarizes the analysis steps, including event generation, a procedure to incorporate detector effects, and the expected sample after event selection; Section 3 summarizes the methodology and results. Conclusions are presented in Section 4.

# 2 Simulation and event selection

Leptonic $W$ boson decays are characterized by an energetic, isolated electron or muon, and significant missing transverse momentum reflecting the decay neutrino. The hadronic recoil, $u_{\mathrm{T}}$, is defined from the vector sum of the transverse momenta of all reconstructed particles in the event excluding the charged lepton, and provides a measure of the $W$ boson transverse momentum. Lepton transverse momentum, $p_{\mathrm{T}}^{\ell}$, missing transverse momentum, $E_{\mathrm{T}}^{\mathrm{miss}}$, and the hadronic recoil are related through $\vec{E}_{\mathrm{T}}^{\mathrm{miss}} = -(\vec{p}_{\mathrm{T}}^{\ell} + \vec{u}_{\mathrm{T}})$. The $p_{\mathrm{T}}^{\ell}$ and $E_{\mathrm{T}}^{\mathrm{miss}}$ distributions have sharp peaks at $p_{\mathrm{T}}^{\ell} \sim E_{\mathrm{T}}^{\mathrm{miss}} \sim m_W/2$. The transverse mass $m_{\mathrm{T}}$, defined as $m_{\mathrm{T}} = \sqrt{2 p_{\mathrm{T}}^{\ell} E_{\mathrm{T}}^{\mathrm{miss}} \cos(\phi_\ell - \phi_{\mathrm{miss}})}$, peaks at $m_{\mathrm{T}} \sim m_W$.

Events are generated at $\sqrt{s}$ = 14 and 27 TeV using the W_EW_BMNNP process [6] of the Powheg event generator [7], with electroweak corrections switched off. The CT10 PDF set [8] is used, and parton shower effects are included using the Pythia event generator [9] with parameters set according to the AZNLO tune [10]. Final-state QED corrections are applied using Photos [11].

For a basic emulation of detector effects, the resolutions on the electron, muon and recoil reconstruction are parameterised as follows:

$$\sigma_e(E_\ell) = a(|\eta_\ell|)\sqrt{E_\ell} \oplus b(|\eta_\ell|) \oplus c(|\eta_\ell|) \cdot E_\ell, \quad (1)$$

$$\sigma_\mu(p_T^\ell) = r_0(|\eta_\ell|) \oplus r_1(|\eta_\ell|) \cdot p_T^\ell, \quad (2)$$

$$\sigma_{u_T}(p_T^W, s) = q_0 \cdot (s/s_0)^\alpha + q_1\sqrt{p_T^W}; \quad (3)$$

where $p_T^\ell$ and $p_T^W$ are the generator-level transverse momenta of the decay lepton and $W$ boson, $E_\ell$ the generator-level electron energy, and $s$ the centre-of-mass energy squared. The calorimeter resolution parameters $a$, $b$ and $c$, as well as the muon resolution parameters $r_{0,1}$ are functions of the lepton pseudorapidity and taken from Refs. [12, 13], which describe the expected performance of the upgraded ATLAS detector.

During four weeks in 2017 and 2018, ATLAS recorded $pp$ collision data at low pile-up and two centre-of-mass energies, $\sqrt{s}$ = 5 and 13 TeV. The recoil resolution parameters are determined for this study using Monte Carlo samples of $W$ boson events produced for the analysis of these data. The events were generated as described above, and passed through the full ATLAS simulation [14]. Pile-up is simulated using Pythia, and the average number of collisions per bunch crossing is set to $\langle\mu\rangle \sim 2$, matching the data taking conditions. The resolution parameters $q_0$ and $q_1$ are extracted from the behaviour of the recoil resolution as a function of $p_T^W$ at both energies. The first term reflects the contribution from the underlying event activity; for a reference centre-of-mass energy of $\sqrt{s_0}$ = 5 TeV, its parameters are found to be $q_0$ = 4.1 GeV and $\alpha = 0.40$, yielding a resolution in $u_T$ of about 6.0 GeV at $\sqrt{s}$ = 13 TeV, when $p_T^W = 0$. The second term is the contribution of the recoil jet, and its coefficient is determined to be $q_1 = 0.23$, independently of energy. The simulated and parameterised recoil resolutions agree well for 5 and 13 TeV, as shown in Figure 1. The curves for the HL- and HE-LHC are obtained extrapolating this parameterisation to $\sqrt{s}$ = 14 and 27 TeV, respectively.

Events are selected applying the following cuts to the object kinematics, after resolution corrections:

- $p_T^\ell > 25$ GeV, $E_T^{miss} > 25$ GeV, $m_T > 50$ GeV and $u_T < 15$ GeV;
- $|\eta_\ell| < 2.4$ or $2.4 < |\eta_\ell| < 4$.

The first set of cuts select the range of the kinematic peaks of the $W$ boson decay products, restricting to the region of small $p_T^W$ to maximize the sensitivity of the distributions to $m_W$. Two pseudorapidity ranges are considered, corresponding to the central region accessible with the current ATLAS detector, and to the forward region accessible in the electron channel with the ITk.

Signal cross sections, acceptance and the expected number of selected events are summarized in Table 1, accounting for typical electron and muon selection efficiencies. The cross sections are calculated at $O(\alpha_S)$ using Powheg; $O(\alpha_S^2)$ corrections would increase these numbers by 2 to 5% depending on the $W$ boson charge and $\sqrt{s}$. Small acceptance losses in the transition regions between the barrel, endcap and forward calorimeter systems are neglected.

The separation of $W^+$ and $W^-$ events for $2.4 < |\eta_\ell| < 4$ relies on a sufficiently accurate measurement of the electron charge. Given the small polar angle and limited magnetic field integral in this region, a rate of charge mis-identification of about 5% is estimated near the detector boundary. While significantly higher than in central region of the detector, this rate can be accurately controlled using the high-statistics $Z \to ee$ samples available from standard runs at the HL-LHC and does not significantly affect the PDF uncertainty estimates discussed in Section 3.

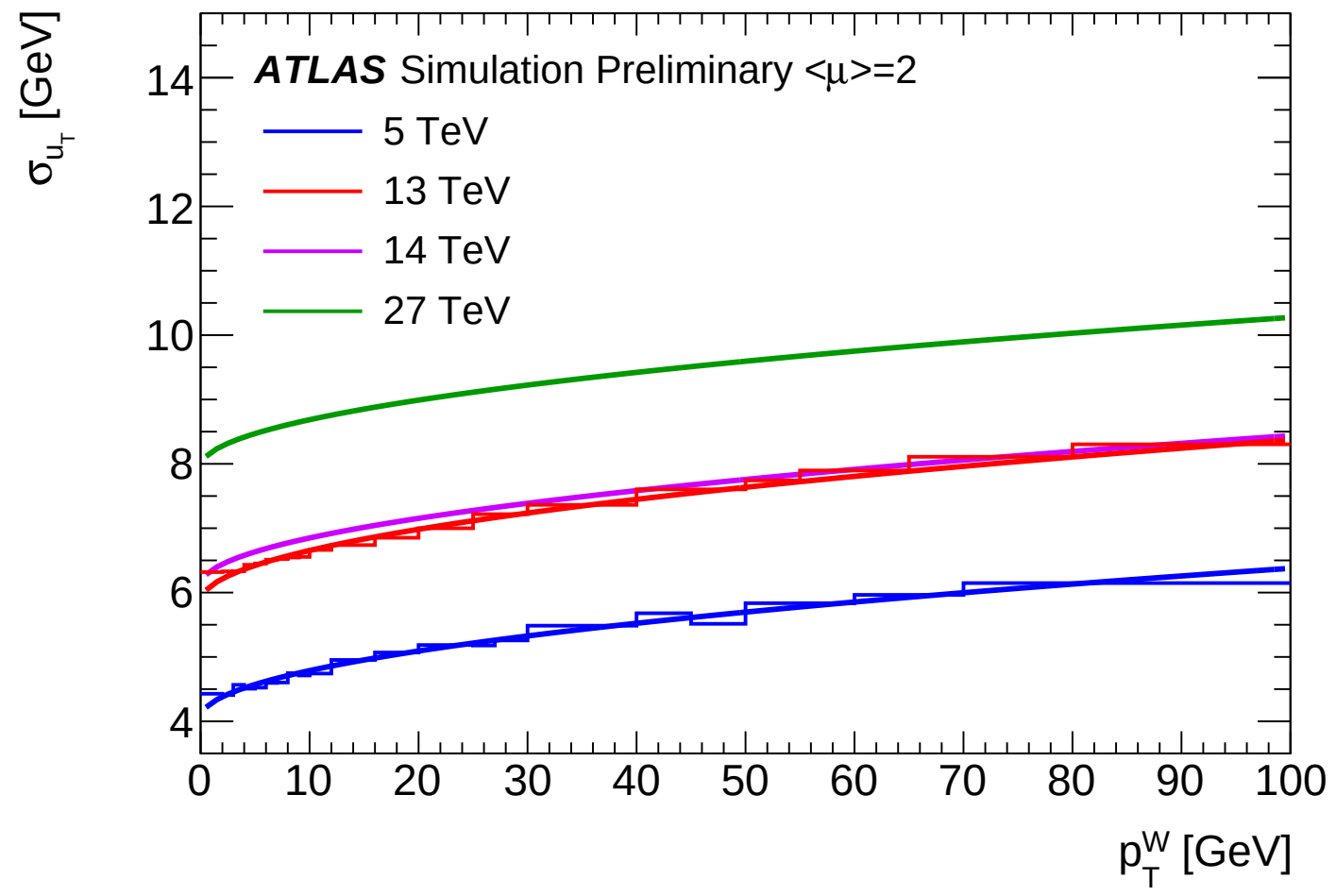

Figure 1: Recoil resolution as a function of $p_T^W$, for $\sqrt{s}$ = 5, 13, 14 and 27 TeV. The histograms correspond to the full detector simulation, and the curves to the parameterised resolution corrections described in the text.

| Process | $\sqrt{s}$ [TeV] | $\sigma$ [pb] | Acceptance | | $N_{\rm sel}$ (200 pb$^{-1}$) | |
|---|---|---|---|---|---|---|
| | | | $\lvert\eta_\ell\rvert < 2.4$ | $2.4 < \lvert\eta_\ell\rvert < 4$ | $\lvert\eta_\ell\rvert < 2.4$ | $2.4 < \lvert\eta_\ell\rvert < 4$ |
| $W^+ \to \ell^+\nu$ | 14 | 12160 | 0.24 | 0.14 | $1.2 \times 10^6$ | $3.5 \times 10^5$ |
| $W^- \to \ell^-\bar{\nu}$ | 14 | 8979 | 0.25 | 0.12 | $0.9 \times 10^6$ | $2.2 \times 10^5$ |
| $W^+ \to \ell^+\nu$ | 27 | 22926 | 0.17 | 0.11 | $1.6 \times 10^6$ | $5.0 \times 10^5$ |
| $W^- \to \ell^-\bar{\nu}$ | 27 | 17863 | 0.18 | 0.09 | $1.3 \times 10^6$ | $3.2 \times 10^5$ |

Table 1: Inclusive $W$ boson production cross sections, signal acceptance and expected number of selected events at $\sqrt{s}$ = 14 and 27 TeV. The cross sections correspond to a single decay lepton flavour. For $|\eta_\ell| < 2.4$, $N_{\rm sel}$ corresponds to the sum of the electron and muon decay channels; for $2.4 < |\eta_\ell| < 4$, only the electron channel is considered.

Distributions of the lepton transverse momentum and $W$ boson transverse mass are shown in Figure 2 for selected events. In both cases, the detector effects are dominated by the recoil resolution. The difference between the predictions of the full simulation and the parameterised resolution is a few percent of that between the detector-level and generator-level distributions, indicating that the present approach is adequate for this analysis.

## 3 PDF uncertainties in $m_W$

The Monte Carlo samples are produced using the CT10 PDF set, $m_W^{\rm ref}$ = 80.399 GeV, and the corresponding Standard Model prediction for $\Gamma_W$. Kinematic distributions for the different values of $m_W$ are obtained by

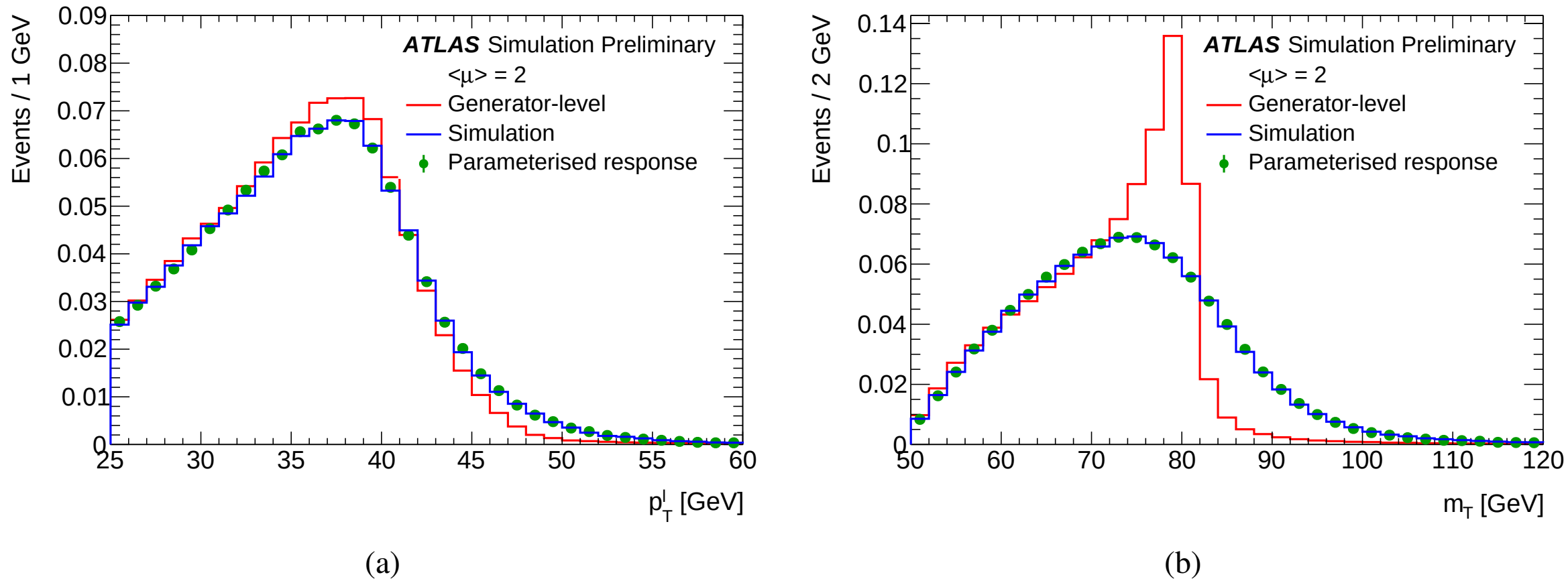

Figure 2: Generator- and detector-level $p_T^\ell$ (a) and $m_T$ (b) distributions for selected signal events. The detector-level distributions are shown as predicted by the full simulation and by the parameterised resolution corrections described in the text.

applying the following event weight to the reference samples:

$$w(m, m_W, m_W^{\text{ref}}) = \frac{(m^2 - m_W^{\text{ref}^2})^2 + m^4 \Gamma_W^{\text{ref}^2} / m_W^{\text{ref}^2}}{(m^2 - m_W^2)^2 + m^4 \Gamma_W^2 / m_W^2}, \tag{4}$$

which represents the ratio of the Breit–Wigner densities corresponding to $m_W$ and $m_W^{\text{ref}}$, for a given value of the final state invariant mass $m$.

A similar event weight, calculated internally by Powheg and corresponding to the ratio of the event cross sections predicted by CT10 and several alternate PDFs, is used to obtain final state distributions corresponding to the CT14 [15], MMHT2014 [16], HL-LHC [17] and LHeC [18] PDF sets and their associated uncertainties. Compared to current sets such as CT14 and MMHT2014, the HL-LHC set incorporates the expected constraints from present and future LHC data; it starts from the PDF4LHC convention [19] and comes in three scenarios corresponding to more or less optimistic projections of the experimental uncertainties. The LHeC PDF set represents the impact of a proposed future high-energy, high-luminosity $ep$ scattering experiment [5] on the uncertainties in the proton structure, using the theoretically best understood process for this purpose.

The shift in the measured value of $m_W$ resulting from a change in the assumed PDF set is estimated as follows. Considering a set of template distributions obtained for different values of $m_W$ and a given reference PDF set, and "pseudo-data" distributions obtained for $m_W = m_W^{\text{ref}}$ and an alternate set $i$ (representing, for example, uncertainty variations with respect to the reference set), the preferred value of $m_W$ for this set is determined by minimizing the $\chi^2$ between the pseudo-data and the templates. The preferred value is denoted $m_W^i$, and the corresponding bias is defined as $\delta m_W^i = m_W^i - m_W^{\text{ref}}$. The statistical uncertainty on the measurement is estimated from the half width of the $\chi^2$ function one unit above the minimum.

The present study considers measurements of $m_W$ in separate categories, corresponding to $W^+$ and $W^-$ events; five pseudorapidity bins, $|\eta_\ell| < 0.6$, $0.6 < |\eta_\ell| < 1.2$, $1.2 < |\eta_\ell| < 1.8$, $1.8 < |\eta_\ell| < 2.4$, and $2.4 < |\eta_\ell| < 4$; $p_T^\ell$ and $m_T$ distribution fits; and two centre-of-mass energies ($\sqrt{s}$ = 14 and 27 TeV). For

each category $\alpha$ and for the PDF sets considered here, the Hessian uncertainty corresponding to a given set is estimated as

$$\delta m_{W\alpha}^{+} = \left[ \sum_i \left( \delta m_{W\alpha}^{i} \right)^2 \right]^{1/2} \text{ if } \delta m_{W\alpha}^{i} > 0, \qquad \delta m_{W\alpha}^{-} = \left[ \sum_i \left( \delta m_{W\alpha}^{i} \right)^2 \right]^{1/2} \text{ if } \delta m_{W\alpha}^{i} < 0, \tag{5}$$

where $i$ runs over the uncertainty sets, and $\delta m_{W\alpha}^{i}$ is calculated with respect to the reference PDF set. For CT10 and CT14, the uncertainties are divided by a factor 1.645 to match the 68% CL. Only symmetrized uncertainties, $\delta m_{W\alpha} = (\delta m_{W\alpha}^{+} + \delta m_{W\alpha}^{-})/2$, are discussed below for simplicity. The correlation of PDF uncertainties between different measurement categories is calculated as

$$\rho_{\alpha\beta} = \frac{\sum_i \delta m_{W\alpha}^{i} \delta m_{W\beta}^{i}}{\delta m_{W\alpha} \delta m_{W\beta}}. \tag{6}$$

PDF variations applied as above generate correlated variations in the $p_{\mathrm{T}}^{W}$ and $p_{\mathrm{T}}^{Z}$ distributions, while the latter are strongly constrained by experimental data [10, 20]. These constraints were used in the ATLAS measurement of $m_W$ [21], bringing significant reduction in the PDF uncertainties. The uncertainties estimated here as thus conservative from this perspective, and partly account for uncertainties in the $p_{\mathrm{T}}^{W}$ distribution.

The overall measurement precision is evaluated by combining the results obtained in the different categories using the BLUE prescription [22]. Only statistical and PDF uncertainties are considered. The former are assigned assuming an integrated luminosity of 200 pb$^{-1}$, and normalizing the samples as summarized in Table 1. PDF uncertainties are estimated as described above.

The results of this procedure are summarized in Figures 3 and 4 and Table 2 for CT10. Figure 3 shows the PDF uncertainty correlations across the different measurement categories, for fits based on the $p_{\mathrm{T}}^{\ell}$ or $m_{\mathrm{T}}$ distributions. Moderate or negative correlations, which will lead to reduced combined uncertainties, are observed between categories of different $W$-boson charges, and between central and forward pseudorapidities, at given $\sqrt{s}$. On the other hand, PDF uncertainty correlations tend to be large and positive between $\sqrt{s}$ = 14 and 27 TeV, for a given boson charge and lepton pseudorapidity range. A similar behaviour is observed for CT14 and MMHT2014.

Table 2 and Figure 4 show the expected measurement uncertainties, and their statistical and PDF components. The numbers quoted for $0 < |\eta_\ell| < 2.4$ correspond to the combination of the four pseudorapidity bins in this range. As expected from the correlations, combining the central and forward pseudorapidity ranges brings significant reduction in the PDF uncertainties, both at 14 and 27 TeV. On the other hand, combining the 14 and 27 TeV samples mostly improves the statistical uncertainty. Statistical correlations between the fits to the $p_{\mathrm{T}}^{\ell}$ and $m_{\mathrm{T}}$ distributions were studied and found to be small, which explains the reduction in uncertainties obtained under their combination. With 200 pb$^{-1}$of data collected at each energy, a total uncertainty of about 11 MeV is obtained.

Figure 5 shows the evolution of the statistical and CT10 PDF uncertainties as a function of the size of the collected sample, for the combination of measurements at 14 TeV in all categories. The assumed integrated luminosities range from 200 pb$^{-1}$to 1 fb$^{-1}$, approximately corresponding to one to five weeks of machine time. The statistical uncertainty decreases as expected. The statistical sensitivity of the forward categories improves with increasing luminosity, which enhances their impact in the combination with the central categories and explains the slight decrease in PDF uncertainty.

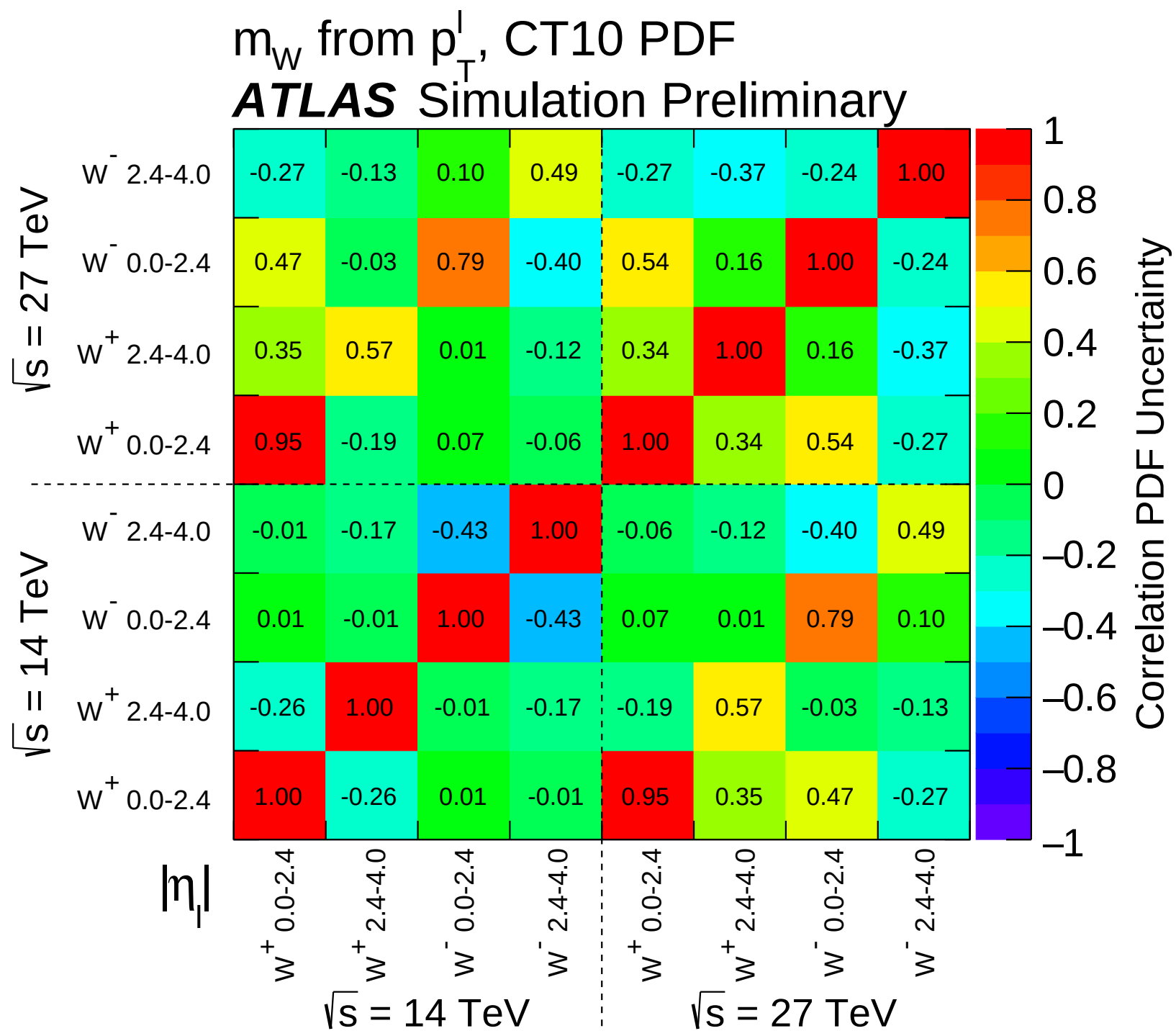

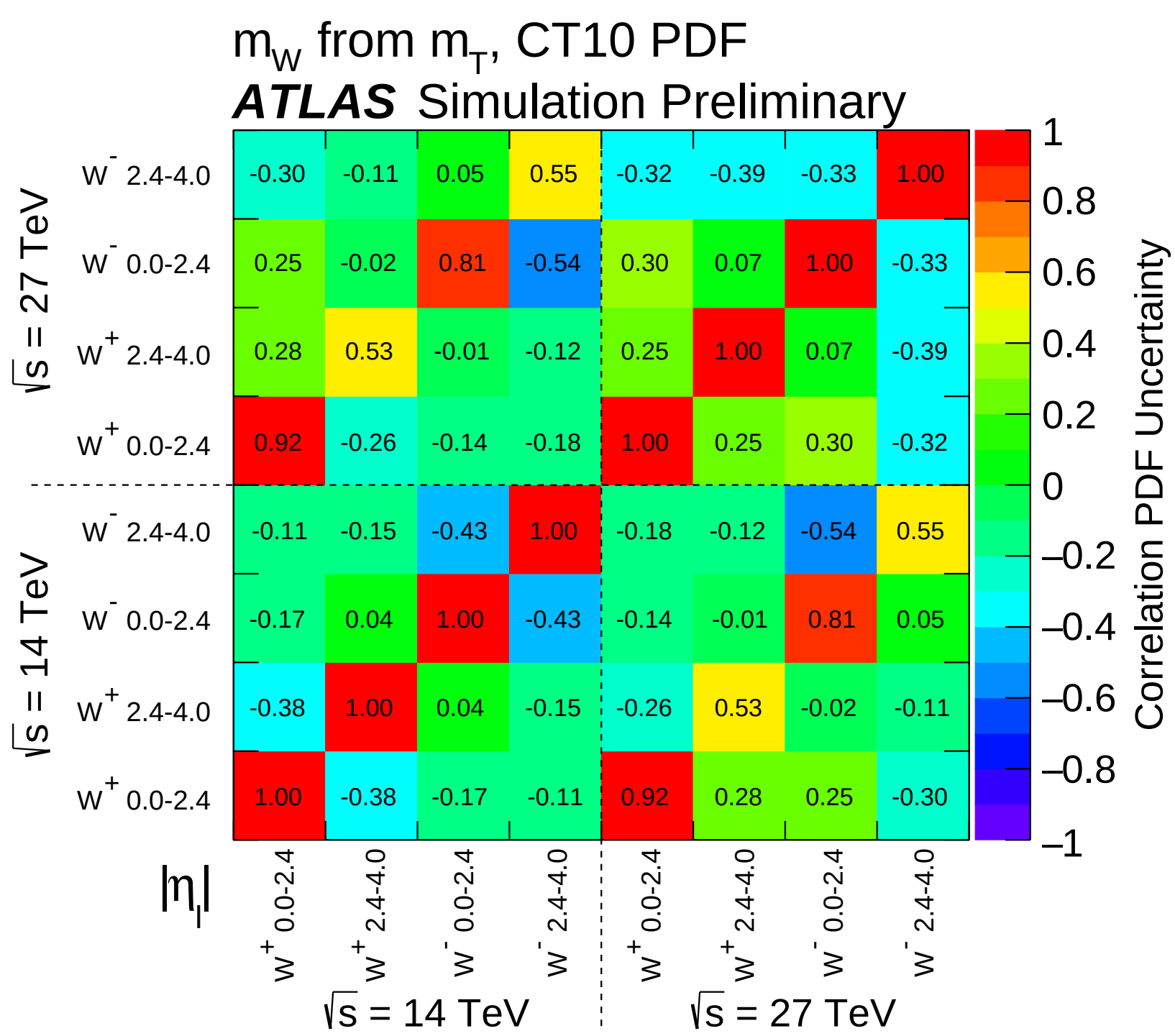

Figure 3: Correlation between the PDF uncertainties for the different measurement categories using the $p_T^\ell$ distribution (top) and the $m_T$ distribution (bottom), calculated using the CT10 PDF set. The numbers quoted for $0 < |\eta_\ell| < 2.4$ correspond to the combination of the four pseudorapidity bins in this range.

| $\sqrt{s}$ [TeV] | Lepton acceptance | Uncertainty in $m_W$ [MeV] | | |
|---|---|---|---|---|
| | | $p_T^\ell$ fits | $m_T$ fits | $p_T^\ell$, $m_T$ fits |
| 14 | $\lvert\eta_\ell\rvert < 2.4$ | 20.6 (14.8 ⊕ 14.4) | 18.0 (13.8 ⊕ 11.5) | 16.0 (10.6 ⊕ 12.0) |
| 14 | $\lvert\eta_\ell\rvert < 4$ | 15.6 (12.6 ⊕ 9.2) | 14.2 (12.0 ⊕ 7.6) | 11.9 (8.8 ⊕ 8.0) |
| 27 | $\lvert\eta_\ell\rvert < 2.4$ | 21.9 (13.5 ⊕ 17.2) | 20.0 (13.4 ⊕ 14.8) | 18.3 (10.2 ⊕ 15.1) |
| 27 | $\lvert\eta_\ell\rvert < 4$ | 14.8 (10.2 ⊕ 10.7) | 14.1 (10.4 ⊕ 9.5) | 12.3 (7.5 ⊕ 9.8) |
| 14+27 | $\lvert\eta_\ell\rvert < 4$ | 12.4 (8.4 ⊕ 9.1) | 11.3 (8.1 ⊕ 7.8) | 10.1 (6.3 ⊕ 7.9) |

Table 2: Measurement uncertainty for different lepton acceptance regions and centre-of-mass energies, using the $p_T^\ell$ and $m_T$ distributions and their combination in the fit, using the CT10 PDF set and for 200 pb$^{-1}$collected at each energy. The numbers quoted for $0 < |\eta_\ell| < 2.4$ correspond to the combination of the four pseudorapidity bins in this range. In each case, the first number corresponds to the sum of statistical and PDF uncertainties, and the numbers between parentheses are the statistical and PDF components, respectively.

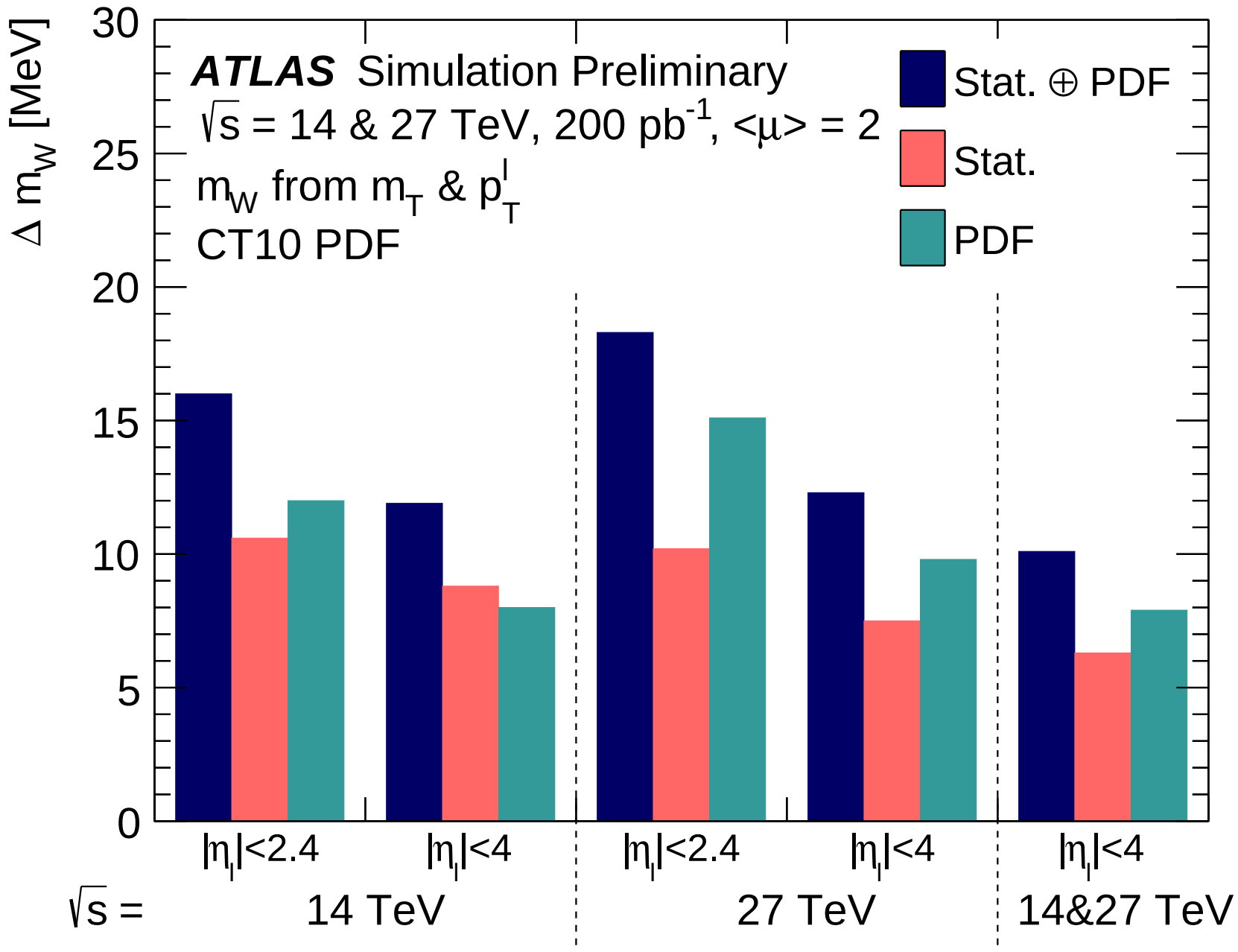

Figure 4: Measurement uncertainty for lepton acceptance regions and centre-of-mass energies, for combined fits to the $p_T^\ell$ and $m_T$ distributions, using the CT10 PDF set and for 200 pb$^{-1}$collected at each energy. The numbers quoted for $0 < |\eta_\ell| < 2.4$ correspond to the combination of the four pseudorapidity bins in this range.

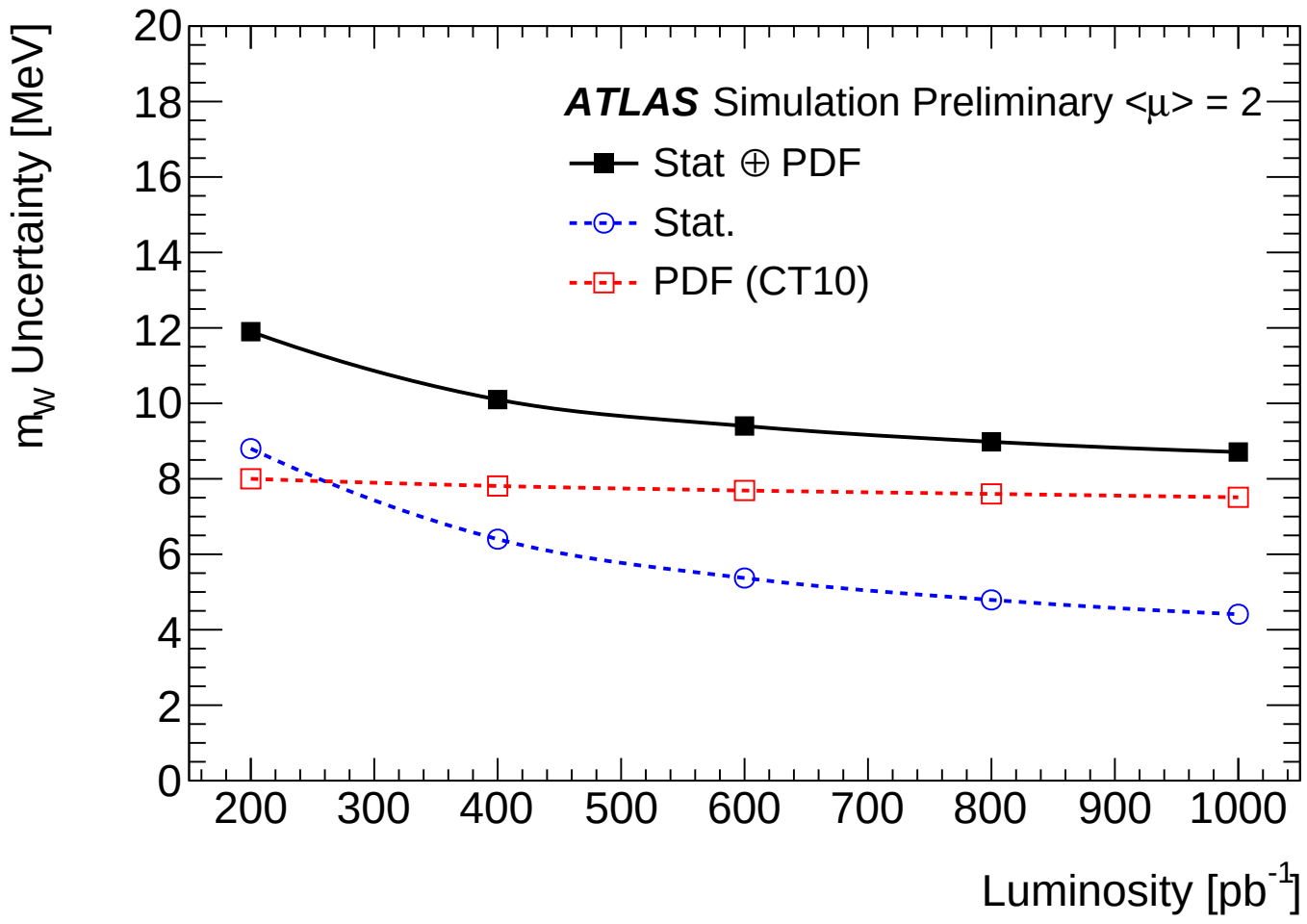

Figure 5: Statistical and PDF uncertainty components as a function of integrated luminosity, for fully combined measurements at $\sqrt{s}$ = 14 TeV. The CT10 PDF set is used.

Table 3 and Figure 6 compares the uncertainties obtained for different PDF sets. The CT10 and CT14 sets display similar uncertainty correlations, leading to similar improvement under combination of categories, and yielding comparable final uncertainties. The MMHT2014 uncertainties are about 30% lower. The three projected HL-LHC PDF sets give very similar uncertainties; scenario 2 is the most conservative and shown here. Compared to CT10 and CT14, a reduction in PDF uncertainty of about a factor of two is obtained in this case. Results for scenarios 1 and 3 are given in the appendix.

The LHeC projection results from a QCD fit to 1 ab$^{-1}$ of $ep$ scattering pseudodata, with $E_e$ = 60 GeV and $E_p$ = 7 TeV. Such a sample could be collected in about five years, synchronously with the HL-LHC operation. In this configuration, the neutral- and charged-current DIS samples are sufficient to disentangle the first and second generation parton densities without ambiguity, and reduce the PDF uncertainty below 2 MeV, a factor 5–6 compared to present knowledge. Also in this case the $m_W$ measurement will benefit from the large $W$ boson samples collected at the LHC, and from the anti-correlation between central and forward categories. In this context, PDF uncertainties would still be sub-leading with 1 fb$^{-1}$ of low pile-up data.

# 4 Conclusion

Given the high $W$ boson production cross section and the importance of an optimal reconstruction of missing transverse momentum in this channel, low-pile-up runs are an important tool for precision measurements of the $W$ boson properties. With $\langle\mu\rangle \sim 2$, a sample of 200 pb$^{-1}$can be collected in about one week, corresponding to about $2 \cdot 10^6$ $W$ boson events at $\sqrt{s}$ = 14 TeV, $3 \cdot 10^6$ events at $\sqrt{s}$ = 27 TeV, and a statistical sensitivity on $m_W$ below 10 MeV. If five to ten weeks can be spent collecting such data in the course of the HL- and HE-LHC, a statistical precision of about 3 MeV can be reached. Experimental systematic uncertainties are not discussed in this note, but their effect is largely of statistical nature; with adequate efforts and exploiting the full available data sample, their impact can be maintained at a level similar to the statistical uncertainty.

| $\sqrt{s}$ [TeV] | Lepton acceptance | Uncertainty in $m_W$ [MeV] | | |
|---|---|---|---|---|
| | | CT10 | CT14 | MMHT2014 |
| 14 | $\|\eta_\ell\| < 2.4$ | 16.0 (10.6 ⊕ 12.0) | 17.3 (11.4 ⊕ 13.0) | 15.4 (10.7 ⊕ 11.1) |
| 14 | $\|\eta_\ell\| < 4$ | 11.9 (8.8 ⊕ 8.0) | 12.4 (9.2 ⊕ 8.4) | 10.3 (9.0 ⊕ 5.1) |
| 27 | $\|\eta_\ell\| < 2.4$ | 18.3 (10.2 ⊕ 15.1) | 18.8 (10.5 ⊕ 15.5) | 16.5 (9.4 ⊕ 13.5) |
| 27 | $\|\eta_\ell\| < 4$ | 12.3 (7.5 ⊕ 9.8) | 12.7 (8.2 ⊕ 9.7) | 11.4 (7.9 ⊕ 8.3) |
| 14+27 | $\|\eta_\ell\| < 4$ | 10.1 (6.3 ⊕ 7.9) | 10.1 (6.9 ⊕ 7.4) | 8.6 (6.5 ⊕ 5.5) |

| $\sqrt{s}$ [TeV] | Lepton acceptance | Uncertainty in $m_W$ [MeV] | |
|---|---|---|---|
| | | HL-LHC | LHeC |
| 14 | $\|\eta_\ell\| < 2.4$ | 11.5 (10.0 ⊕ 5.8 ) | 10.2 (9.9 ⊕ 2.2) |
| 14 | $\|\eta_\ell\| < 4$ | 9.3 (8.6 ⊕ 3.7) | 8.7 (8.5 ⊕ 1.6) |

Table 3: Measurement uncertainty for different lepton acceptance regions, centre-of-mass energies and PDF sets, combined fits to the $p_T^\ell$ and $m_T$ distributions, and for 200 pb$^{-1}$collected at each energy. The numbers quoted for $0 < |\eta_\ell| < 2.4$ correspond to the combination of the four pseudorapidity bins in this range. In each case, the first number corresponds to the sum of statistical and PDF uncertainties, and the numbers between parentheses are the statistical and PDF components, respectively.

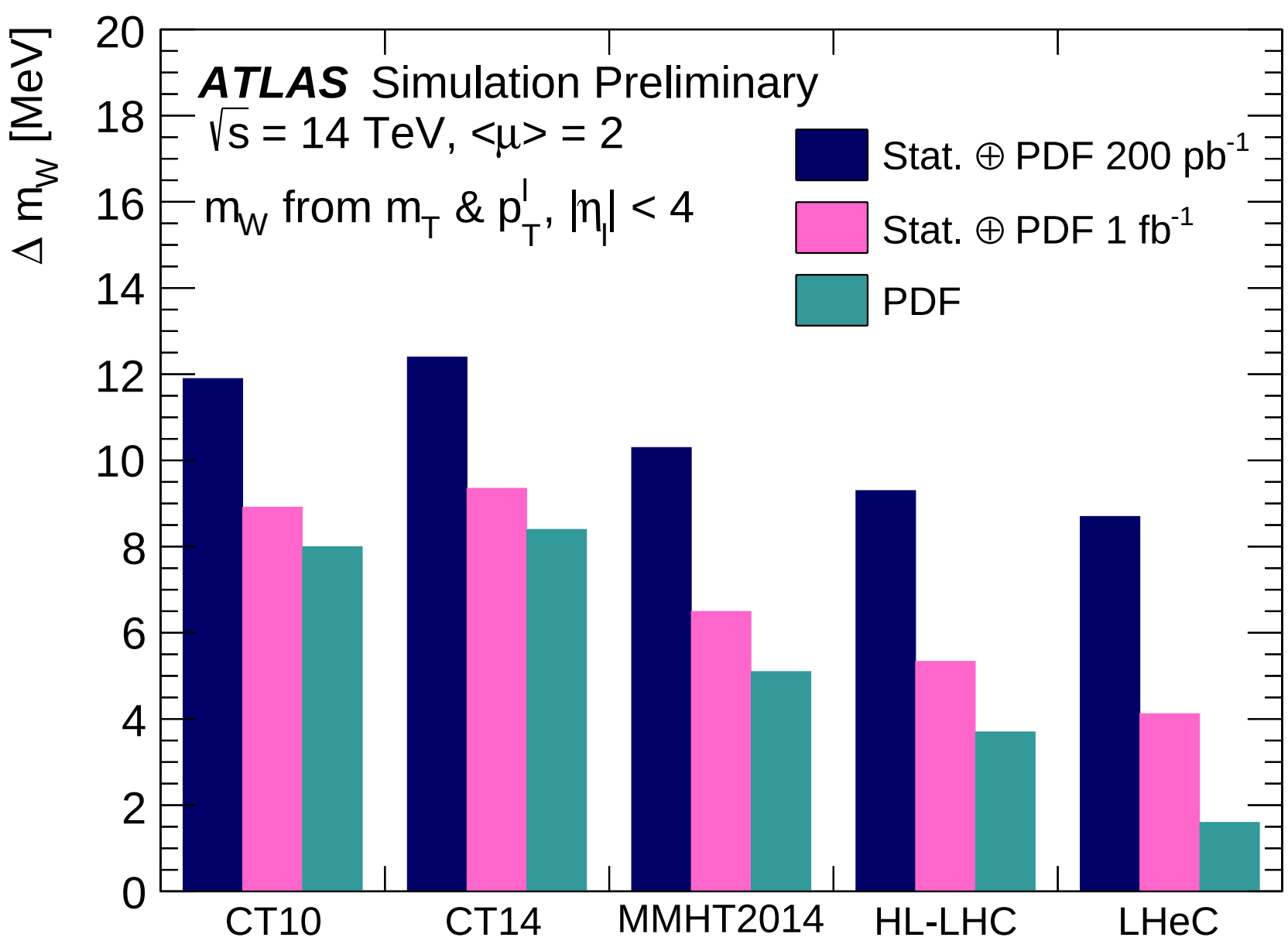

Figure 6: Measurement uncertainty for different PDF sets, combining $p_T^\ell$ and $m_T$ fits for $|\eta_\ell| < 4$, and for 200 pb$^{-1}$and 1 fb$^{-1}$collected at $\sqrt{s}$ = 14 TeV.

Using current PDF sets, PDF uncertainties and their impact on the $W$ boson transverse momentum distribution are typically 12–15 MeV when restricting to the current Inner Detector acceptance. They are reduced to about 5–8 MeV, for a range of current PDF sets, when exploiting the extended coverage allowed by the ITk. These uncertainties are further reduced to about 4 MeV when using the anticipated HL-LHC PDF set.

If the LHeC is built and runs synchronously with the HL-LHC, the combination of the large acceptance and excellent performance of the upgraded ATLAS detector and of the additional constraints on the proton structure from the theoretically clean DIS data reduces PDF uncertainties to less than 2 MeV.

# Appendix

Tables 4–6 show the PDF uncertainty correlations obtained for CT10, CT14 and MMHT2014. Figures 7, 8 summarize the same information.

Table 7 provides a comparison of the PDF uncertainties obtained for the HL-LHC PDF sets scenario 1, 2 and 3.

Figure 9 compares the expected measurement uncertainties for current PDF sets CT10, CT14 and MMHT2014. Figure 10 compares CT10 to the projected HL-LHC and LHeC sets.

| Channel | $\eta_\ell$ range | $\sqrt{s}$ [TeV] | | 1. | 2. | 3. | 4. | 5. | 6. | 7. | 8. |
|---|---|---|---|---|---|---|---|---|---|---|---|
| $W^+$ | 0–2.4 | 14 | 1. | 1 | -0.26 | 0.01 | -0.01 | 0.95 | 0.35 | 0.47 | -0.27 |
| $W^+$ | 2.4–4 | 14 | 2. | -0.26 | 1 | -0.01 | -0.17 | -0.19 | 0.57 | -0.03 | -0.13 |
| $W^-$ | 0–2.4 | 14 | 3. | 0.01 | -0.01 | 1 | -0.43 | 0.07 | 0.01 | 0.79 | 0.10 |
| $W^-$ | 2.4–4 | 14 | 4. | -0.01 | -0.17 | -0.43 | 1 | -0.06 | -0.12 | -0.40 | 0.49 |
| $W^+$ | 0–2.4 | 27 | 5. | 0.95 | -0.19 | 0.07 | -0.06 | 1 | 0.34 | 0.54 | -0.27 |
| $W^+$ | 2.4–4 | 27 | 6. | 0.35 | 0.57 | 0.01 | -0.12 | 0.34 | 1 | 0.16 | -0.37 |
| $W^-$ | 0–2.4 | 27 | 7. | 0.47 | -0.03 | 0.79 | -0.40 | 0.54 | 0.16 | 1 | -0.24 |
| $W^-$ | 2.4–4 | 27 | 8. | -0.27 | -0.13 | 0.10 | 0.49 | -0.27 | -0.37 | -0.24 | 1 |

| Channel | $\eta_\ell$ range | $\sqrt{s}$ [TeV] | | 1. | 2. | 3. | 4. | 5. | 6. | 7. | 8. |
|---|---|---|---|---|---|---|---|---|---|---|---|
| $W^+$ | 0–2.4 | 14 | 1. | 1 | -0.38 | -0.17 | -0.11 | 0.92 | 0.28 | 0.25 | -0.30 |
| $W^+$ | 2.4–4 | 14 | 2. | -0.38 | 1 | 0.04 | -0.15 | -0.26 | 0.53 | -0.02 | -0.11 |
| $W^-$ | 0–2.4 | 14 | 3. | -0.17 | 0.04 | 1 | -0.43 | -0.14 | -0.01 | 0.81 | 0.05 |
| $W^-$ | 2.4–4 | 14 | 4. | -0.11 | -0.15 | -0.43 | 1 | -0.18 | -0.12 | -0.54 | 0.55 |
| $W^+$ | 0–2.4 | 27 | 5. | 0.92 | -0.26 | -0.14 | -0.18 | 1 | 0.25 | 0.30 | -0.32 |
| $W^+$ | 2.4–4 | 27 | 6. | 0.28 | 0.53 | -0.01 | -0.12 | 0.25 | 1 | 0.07 | -0.39 |
| $W^-$ | 0–2.4 | 27 | 7. | 0.25 | -0.02 | 0.81 | -0.54 | 0.30 | 0.07 | 1 | -0.33 |
| $W^-$ | 2.4–4 | 27 | 8. | -0.30 | -0.11 | 0.05 | 0.55 | -0.32 | -0.39 | -0.33 | 1 |

Table 4: Correlation between the PDF uncertainties for the different measurement categories using the $p_T^\ell$ distribution (top) and the $m_T$ distribution (bottom), calculated using the CT10 PDF set. The numbers quoted for $0 < |\eta_\ell| < 2.4$ correspond to the combination of the four pseudorapidity bins in this range.

| Channel | $\eta_\ell$ range | $\sqrt{s}$ [TeV] | | 1. | 2. | 3. | 4. | 5. | 6. | 7. | 8. |
|---|---|---|---|---|---|---|---|---|---|---|---|
| $W^+$ | 0–2.4 | 14 | 1. | 1 | -0.38 | 0.52 | -0.08 | 0.94 | 0.18 | 0.67 | -0.07 |
| $W^+$ | 2.4–4 | 14 | 2. | -0.38 | 1 | -0.27 | -0.15 | -0.34 | 0.69 | -0.30 | -0.19 |
| $W^-$ | 0–2.4 | 14 | 3. | 0.52 | -0.27 | 1 | -0.36 | 0.58 | -0.13 | 0.93 | -0.15 |
| $W^-$ | 2.4–4 | 14 | 4. | -0.08 | -0.15 | -0.36 | 1 | -0.12 | -0.12 | -0.29 | 0.66 |
| $W^+$ | 0–2.4 | 27 | 5. | 0.94 | -0.34 | 0.58 | -0.12 | 1 | 0.14 | 0.75 | -0.07 |
| $W^+$ | 2.4–4 | 27 | 6. | 0.18 | 0.69 | -0.13 | -0.12 | 0.14 | 1 | -0.10 | -0.23 |
| $W^-$ | 0–2.4 | 27 | 7. | 0.67 | -0.30 | 0.93 | -0.29 | 0.75 | -0.10 | 1 | -0.23 |
| $W^-$ | 2.4–4 | 27 | 8. | -0.07 | -0.19 | -0.15 | 0.66 | -0.07 | -0.23 | -0.23 | 1 |

| Channel | $\eta_\ell$ range | $\sqrt{s}$ [TeV] | | 1. | 2. | 3. | 4. | 5. | 6. | 7. | 8. |
|---|---|---|---|---|---|---|---|---|---|---|---|
| $W^+$ | 0–2.4 | 14 | 1. | 1 | -0.49 | 0.42 | -0.18 | 0.91 | 0.09 | 0.58 | -0.20 |
| $W^+$ | 2.4–4 | 14 | 2. | -0.49 | 1 | -0.27 | -0.13 | -0.40 | 0.66 | -0.32 | -0.15 |
| $W^-$ | 0–2.4 | 14 | 3. | 0.42 | -0.27 | 1 | -0.44 | 0.44 | -0.17 | 0.92 | -0.32 |
| $W^-$ | 2.4–4 | 14 | 4. | -0.18 | -0.13 | -0.44 | 1 | -0.25 | -0.14 | -0.42 | 0.66 |
| $W^+$ | 0–2.4 | 27 | 5. | 0.91 | -0.40 | 0.44 | -0.25 | 1 | 0.04 | 0.65 | -0.21 |
| $W^+$ | 2.4–4 | 27 | 6. | 0.09 | 0.66 | -0.17 | -0.14 | 0.04 | 1 | -0.17 | -0.27 |
| $W^-$ | 0–2.4 | 27 | 7. | 0.58 | -0.32 | 0.92 | -0.42 | 0.65 | -0.17 | 1 | -0.44 |
| $W^-$ | 2.4–4 | 27 | 8. | -0.20 | -0.15 | -0.32 | 0.66 | -0.21 | -0.27 | -0.44 | 1 |

Table 5: Correlation between the PDF uncertainties for the different measurement categories using the $p_T^\ell$ distribution (top) and the $m_T$ distribution (bottom), calculated using the CT14 PDF set. The numbers quoted for $0 < |\eta_\ell| < 2.4$ correspond to the combination of the four pseudorapidity bins in this range.

| Channel | $\eta_\ell$ range | $\sqrt{s}$ [TeV] | | 1. | 2. | 3. | 4. | 5. | 6. | 7. | 8. |
|---|---|---|---|---|---|---|---|---|---|---|---|
| $W^+$ | 0–2.4 | 14 | 1. | 1 | -0.13 | 0.28 | -0.13 | 0.95 | 0.44 | 0.67 | -0.07 |
| $W^+$ | 2.4–4 | 14 | 2. | -0.13 | 1 | -0.40 | 0.10 | -0.07 | 0.58 | -0.18 | -0.26 |
| $W^-$ | 0–2.4 | 14 | 3. | 0.28 | -0.40 | 1 | -0.66 | 0.26 | -0.02 | 0.77 | 0.00 |
| $W^-$ | 2.4–4 | 14 | 4. | -0.13 | 0.10 | -0.66 | 1 | -0.11 | -0.09 | -0.44 | 0.52 |
| $W^+$ | 0–2.4 | 27 | 5. | 0.95 | -0.07 | 0.26 | -0.11 | 1 | 0.41 | 0.70 | -0.06 |
| $W^+$ | 2.4–4 | 27 | 6. | 0.44 | 0.58 | -0.02 | -0.09 | 0.41 | 1 | 0.20 | -0.26 |
| $W^-$ | 0–2.4 | 27 | 7. | 0.67 | -0.18 | 0.77 | -0.44 | 0.70 | 0.20 | 1 | -0.12 |
| $W^-$ | 2.4–4 | 27 | 8. | -0.07 | -0.26 | 0.00 | 0.52 | -0.06 | -0.26 | -0.12 | 1 |

| Channel | $\eta_\ell$ range | $\sqrt{s}$ [TeV] | | 1. | 2. | 3. | 4. | 5. | 6. | 7. | 8. |
|---|---|---|---|---|---|---|---|---|---|---|---|
| $W^+$ | 0–2.4 | 14 | 1. | 1 | -0.31 | 0.24 | -0.39 | 0.93 | 0.33 | 0.55 | -0.32 |
| $W^+$ | 2.4–4 | 14 | 2. | -0.31 | 1 | -0.46 | 0.18 | -0.18 | 0.56 | -0.35 | -0.15 |
| $W^-$ | 0–2.4 | 14 | 3. | 0.24 | -0.46 | 1 | -0.72 | 0.18 | -0.14 | 0.85 | -0.23 |
| $W^-$ | 2.4–4 | 14 | 4. | -0.39 | 0.18 | -0.72 | 1 | -0.38 | -0.10 | -0.77 | 0.62 |
| $W^+$ | 0–2.4 | 27 | 5. | 0.93 | -0.18 | 0.18 | -0.38 | 1 | 0.31 | 0.55 | -0.32 |
| $W^+$ | 2.4–4 | 27 | 6. | 0.33 | 0.56 | -0.14 | -0.10 | 0.31 | 1 | 0.01 | -0.35 |
| $W^-$ | 0–2.4 | 27 | 7. | 0.55 | -0.35 | 0.85 | -0.77 | 0.55 | 0.01 | 1 | -0.44 |
| $W^-$ | 2.4–4 | 27 | 8. | -0.32 | -0.15 | -0.23 | 0.62 | -0.32 | -0.35 | -0.44 | 1 |

Table 6: Correlation between the PDF uncertainties for the different measurement categories using the $p_T^\ell$ distribution (top) and the $m_T$ distribution (bottom), calculated using the MMHT2014 PDF set. The numbers quoted for $0 < |\eta_\ell| < 2.4$ correspond to the combination of the four pseudorapidity bins in this range.

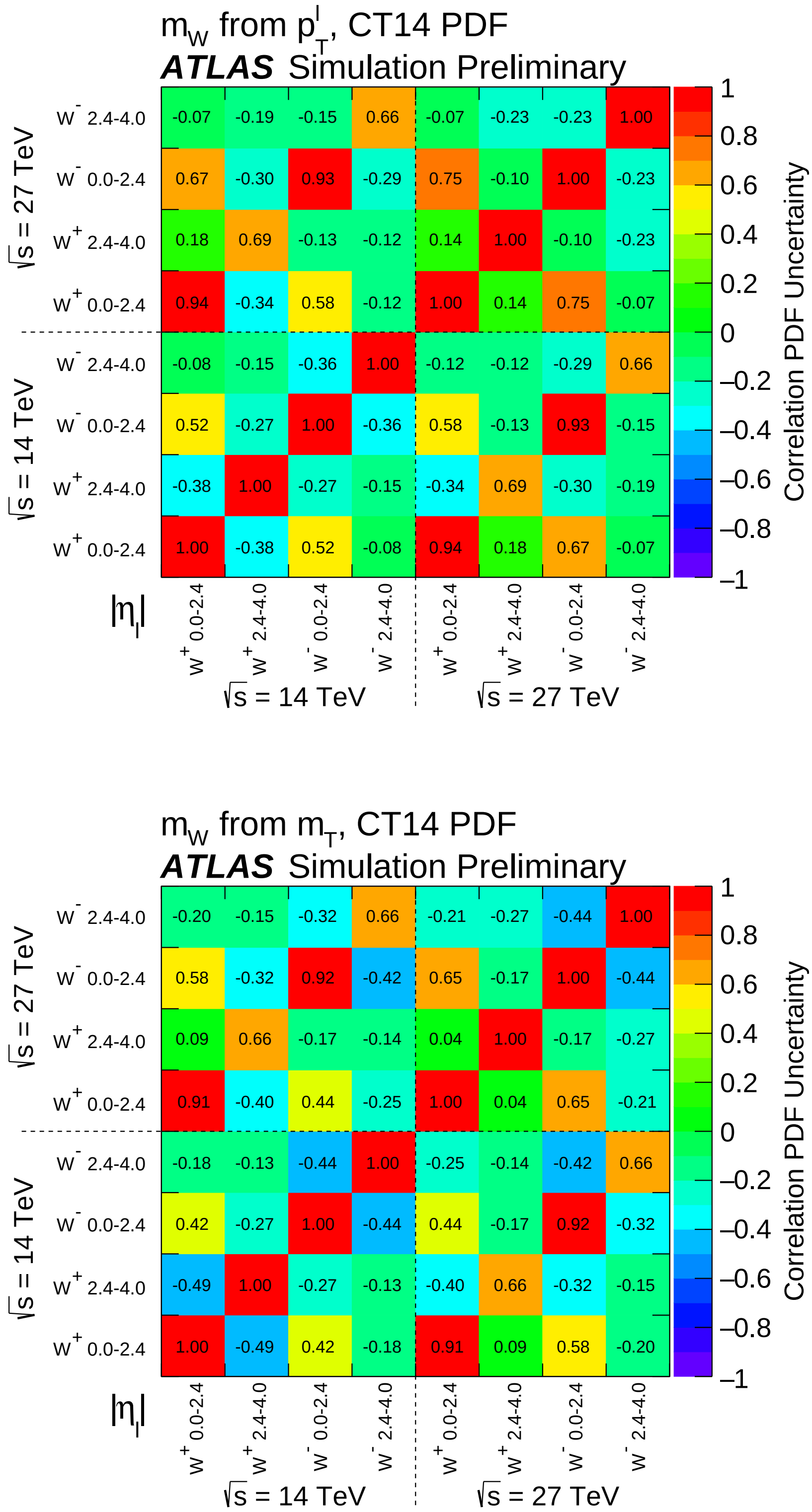

Figure 7: Correlation between the PDF uncertainties for the different measurement categories using the $p_{\mathrm{T}}^{\ell}$ distribution (top) and the $m_{\mathrm{T}}$ distribution (bottom), calculated using the CT14 PDF set. The numbers quoted for $0 < |\eta_\ell| < 2.4$ correspond to the combination of the four pseudorapidity bins in this range.

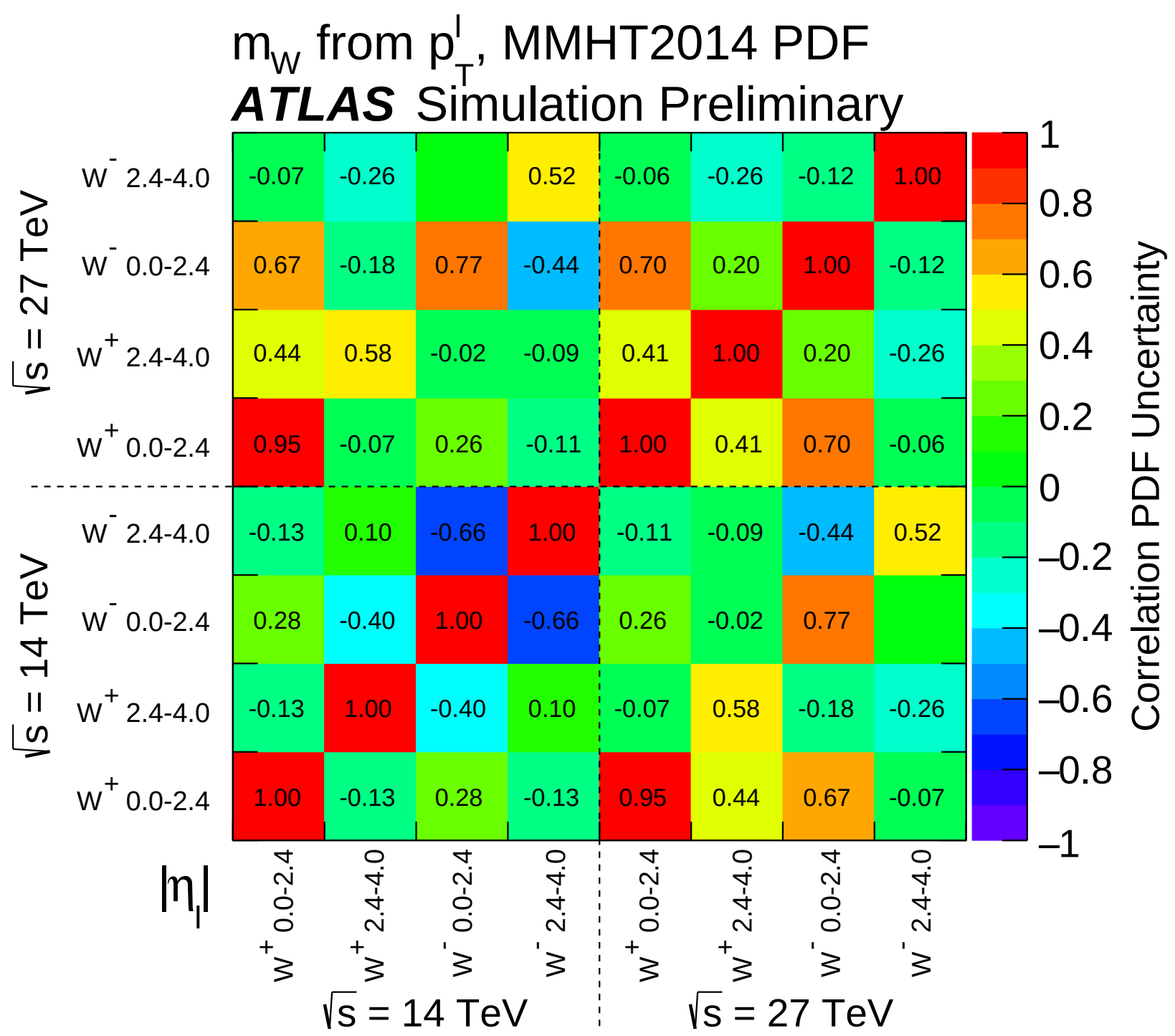

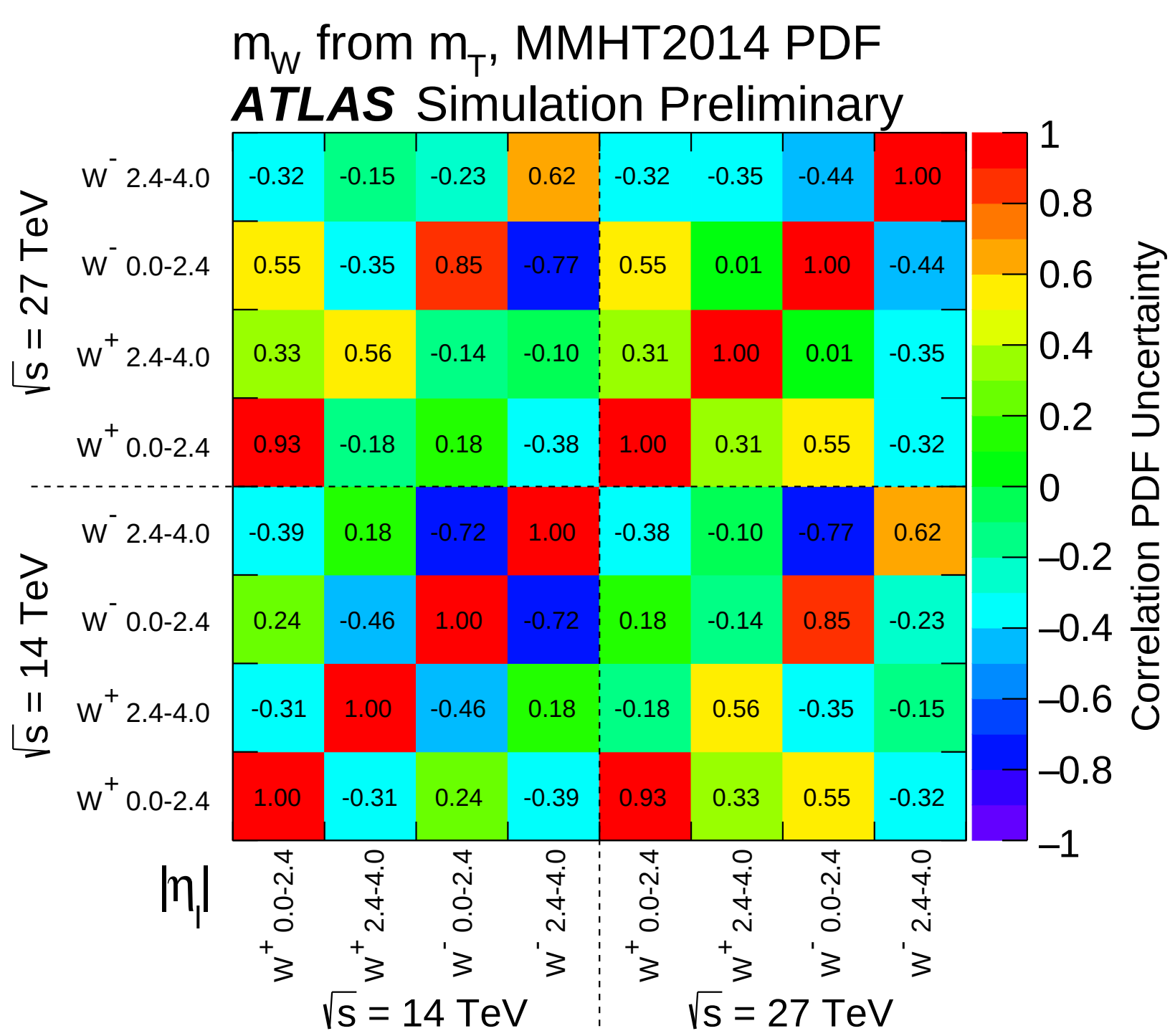

Figure 8: Correlation between the PDF uncertainties for the different measurement categories using the $p_T^\ell$ distribution (top) and the $m_T$ distribution (bottom), calculated using the MMHT2014 PDF set. The numbers quoted for $0 < |\eta_\ell| < 2.4$ correspond to the combination of the four pseudorapidity bins in this range.

| Measurement parameters | | Uncertainty in $m_W$ [MeV] | | |
|---|---|---|---|---|
| $\sqrt{s}$ [TeV] | $\eta_\ell$ range | Scenario 1 | Scenario 2 | Scenario 3 |
| 14 | $\lvert\eta_\ell\rvert < 2.4$ | 11.1 (9.9 ⊕ 5.0) | 11.5 (10.0 ⊕ 5.8 ) | 11.5 (10.0 ⊕ 5.7) |
| 14 | $\lvert\eta_\ell\rvert < 4$ | 9.1 (8.5 ⊕ 3.2) | 9.3 (8.6 ⊕ 3.7) | 9.2 (8.6 ⊕ 3.2) |

Table 7: Measurement uncertainty for different pseudorapidity ranges and scenarios of the HL-LHC PDF sets, for $\sqrt{s}$ = 13 TeV and 200 pb$^{-1}$, and combined $p_T^\ell$, $m_T$ fits. The numbers quoted for $0 < |\eta_\ell| < 2.4$ correspond to the combination of the four pseudorapidity bins in this range.

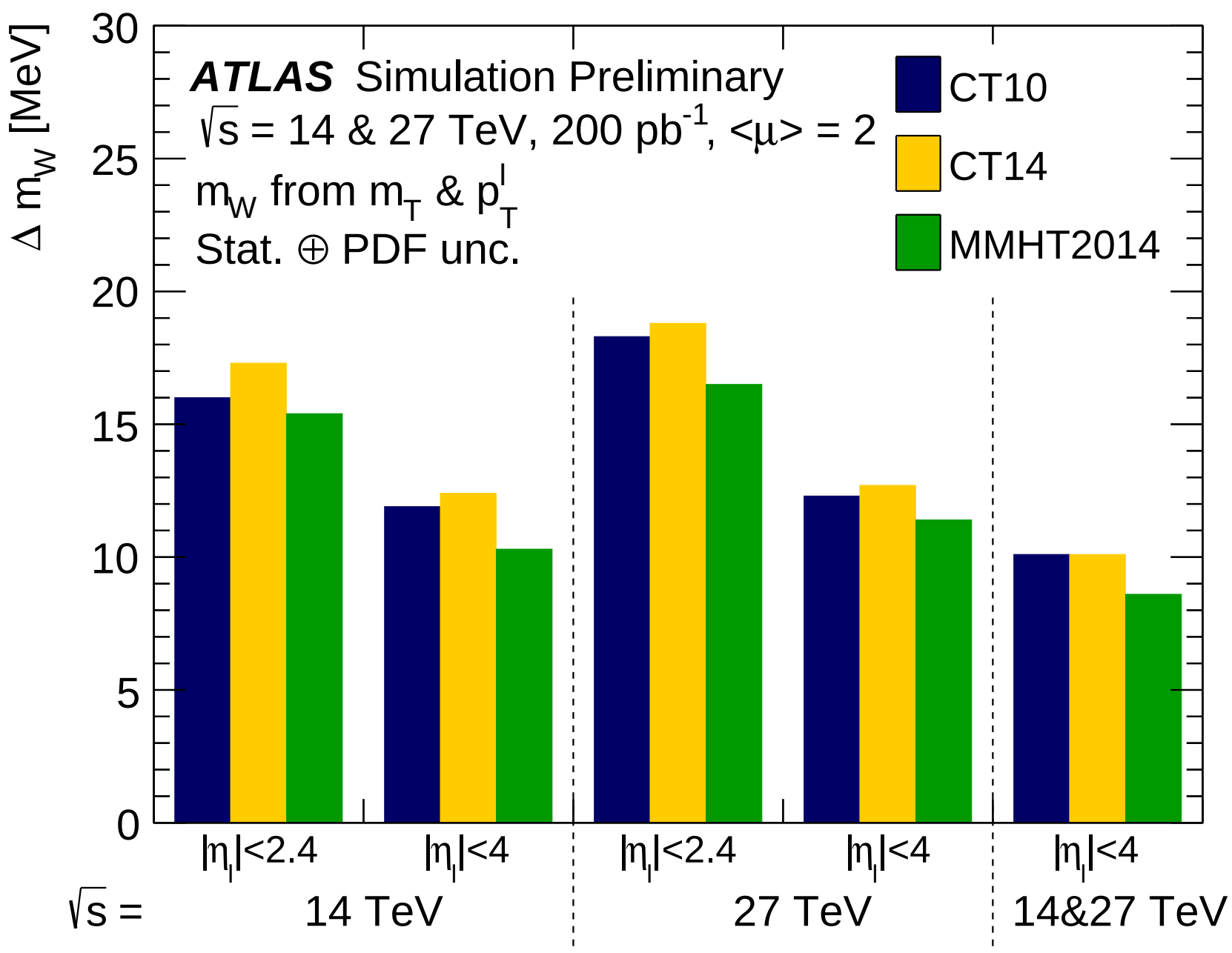

Figure 9: Measurement uncertainty for CT10, CT14 and MMHT2014, in different pseudorapidity ranges, for combined $p_T^\ell$ and $m_T$ fits and 200 pb$^{-1}$collected at $\sqrt{s}$ = 14 and 27 TeV.

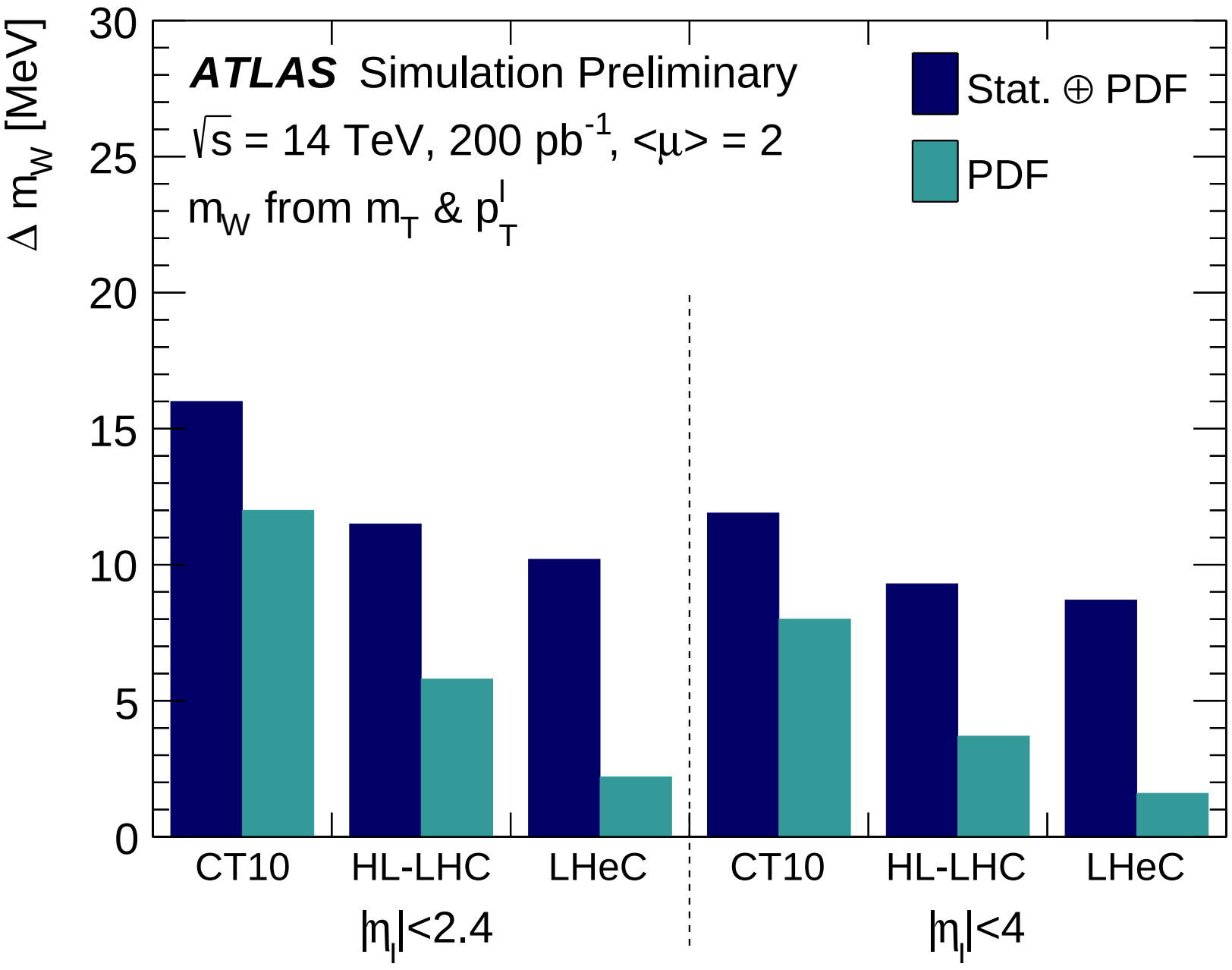

Figure 10: Measurement uncertainty for current and future PDF sets, for combined $p_T^\ell$ and $m_T$ fits within $|\eta_\ell| < 2.4$ and $|\eta_\ell| < 4$, and for 200 $\text{pb}^{-1}$collected at $\sqrt{s} = 14$ TeV.

# CMS Physics Analysis Summary

Contact: cms-future-conveners@cern.ch

# A proposal for the measurement of the weak mixing angle at the HL-LHC

The CMS Collaboration

## Abstract

A proposal is presented for measuring the weak mixing angle using the forward-backward asymmetry of Drell-Yan dimuon events in pp collisions at $\sqrt{s} = 14$ TeV with the CMS detector at the high luminosity LHC (HL-LHC). In addition to the increased luminosity, the upgraded part of the muon system extends the pseudorapidity coverage of the CMS experiment to $|\eta| < 2.8$ for muons. Since the measurement has higher sensitivity in this pseudorapidity region, both the statistical and systematic uncertainties will be significantly reduced. To estimate the increased potential for this measurement we use a Monte Carlo data sample of pp events corresponding to a luminosity of 3000 fb$^{-1}$ and compare to the recent CMS measurements at $\sqrt{s} = 8$ TeV.

## 1 Introduction

We report on a proposal for the measurement of the effective weak mixing angle using the forward-backward asymmetry, $A_{\mathrm{FB}}$, in Drell-Yan $\mu\mu$ events at the HL-LHC at CMS. The proposal is based on techniques used in Ref. [1] for such a measurement at $\sqrt{s} = 8\,\mathrm{TeV}$.

In leading order dilepton pairs are produced through the annihilation of a quark and antiquark via the exchange of a Z boson or a virtual photon: $\mathrm{q\bar{q}} \to Z/\gamma^* \to \ell^+\ell^-$. The definition of $A_{\mathrm{FB}}$ is based on the angle $\theta^*$ of the lepton ($\ell^-$) in the Collins-Soper [2] frame of the dilepton system:

$$A_{\mathrm{FB}} = \frac{\sigma_{\mathrm{F}} - \sigma_{\mathrm{B}}}{\sigma_{\mathrm{F}} + \sigma_{\mathrm{B}}}, \quad (1)$$

where $\sigma_{\mathrm{F}}$ and $\sigma_{\mathrm{B}}$ are the cross sections in the forward ($\cos\theta^* > 0$) and backward ($\cos\theta^* < 0$) hemispheres, respectively. In this frame the $\theta^*$ is the angle of the $\ell^-$ direction with respect to the axis that bisects the angle between the direction of the quark and opposite direction of the anti-quark. In pp collisions the direction of the quark is assumed to be in the boost direction of the dilepton pair. Here, $\cos\theta^*$ is calculated using laboratory-frame quantities as follows:

$$\cos\theta^* = \frac{2(p_1^+ p_2^- - p_1^- p_2^+)}{\sqrt{M^2(M^2 + P_{\mathrm{T}}^2)}} \times \frac{P_{\mathrm{z}}}{|P_{\mathrm{z}}|}, \quad (2)$$

where $M$, $P_{\mathrm{T}}$, and $P_{\mathrm{z}}$ are the mass, transverse momentum, and longitudinal momentum, respectively, of the dilepton system, and $p_1(p_2)$ are defined in terms of energy, $e_1(e_2)$, and longitudinal momentum, $p_{\mathrm{z},1}(p_{\mathrm{z},2})$, of the negatively (positively) charged lepton as $p_{\mathrm{i}}^{\pm} = (e_i \pm p_{\mathrm{z},i})/\sqrt{2}$ [2].

A non-zero $A_{\mathrm{FB}}$ in dilepton events arises from the vector and axial-vector couplings of electroweak bosons to fermions. At tree level, the vector $v_{\mathrm{f}}$ and axial-vector $a_{\mathrm{f}}$ couplings of Z bosons to fermions (f) are:

$$v_{\mathrm{f}} = T_3^{\mathrm{f}} - 2Q_{\mathrm{f}}\sin^2\theta_{\mathrm{W}}, \quad (3)$$

$$a_{\mathrm{f}} = T_3^{\mathrm{f}}, \quad (4)$$

where $T_3^{\mathrm{f}}$ and $Q_{\mathrm{f}}$ are the third component of the weak isospin and the charge of the fermion, respectively, and $\sin^2\theta_{\mathrm{W}}$ is the weak mixing angle, which is related to the masses of the W and Z bosons by the relation $\sin^2\theta_{\mathrm{W}} = 1 - M_{\mathrm{W}}^2/M_{\mathrm{Z}}^2$. Electroweak radiative corrections affect these leading-order relations. An effective weak mixing angle, $\sin^2\theta^{\mathrm{f}}_{\mathrm{eff}}$, is defined based on the relation between these couplings: $v_{\mathrm{f}}/a_{\mathrm{f}} = 1 - 4|Q_{\mathrm{f}}|\sin^2\theta^{\mathrm{f}}_{\mathrm{eff}}$, with $\sin^2\theta^{\mathrm{f}}_{\mathrm{eff}} = \kappa_{\mathrm{f}}\sin^2\theta_{\mathrm{W}}$, where flavor-dependent $\kappa_{\mathrm{f}}$ is determined by electroweak corrections. $A_{\mathrm{FB}}$ for dilepton events is primarily sensitive to the leptonic effective weak mixing angle ($\sin^2\theta^{\mathrm{lept}}_{\mathrm{eff}}$).

In this analysis we measure the leptonic effective weak mixing angle ($\sin^2\theta^{\mathrm{lept}}_{\mathrm{eff}}$) by fitting the mass and rapidity dependence of the observed $A_{\mathrm{FB}}$ in dilepton events. The most precise previous measurements of $\sin^2\theta^{\mathrm{lept}}_{\mathrm{eff}}$ were performed by the LEP and SLD experiments [3]. There is, however, a known tension of about 3 standard deviations between the two most precise measurements. Measurements of $\sin^2\theta^{\mathrm{lept}}_{\mathrm{eff}}$ were also reported by the LHC and Tevatron experiments [4–9]. The latest and the most precise LHC measurement was done by CMS [1], and its machinery is used in this analysis.

The analysis is based on samples of pp collisions simulated at $\sqrt{s} = 8$ and $14\,\mathrm{TeV}$ with next-to-leading order (NLO) matrix element implemented in the POWHEG event generator [10–13] using the NNPDF3.0 [14] PDFs and interfaced with PYTHIA 8 [15] with CUETP8M1* [16] underlying event tune for parton showering and hadronization and electromagnetic final-state radiation (FSR). The template variations for different values of $\sin^2\theta^{\text{lept}}_{\text{eff}}$ and PDFs are modeled using the POWHEG MC generator that provides matrix-element based event-by-event weights for each variation. The samples are normalized to the integrated luminosities of 19 $\text{fb}^{-1}$ for $\sqrt{s}$ = 8 TeV and to 10 – 3000 $\text{fb}^{-1}$ for $\sqrt{s}$ = 14 TeV samples. The analysis is done at the generator level, so the smearing due to detector effects is not taken into account, but comparison of 8 TeV predictions and measured values suggests that this effect is not significant. Moreover since the results are presented as comparison of 8 and 14 TeV measurements they can be directly applied to the measured 8 TeV results with real data.

The HL-LHC CMS detector extends the pseudorapidity, $\eta$, coverage for the muon reconstruction from current configuration of 2.4 to 2.8. In this analysis an event is selected if there are at least two muons with $|\eta| < 2.8$ and with the leading (i.e. having the largest transverse momentum $p_{\text{T}}$) muon $p_{\text{T}} > 25$ GeV and the second leading muon $p_{\text{T}} > 15$ GeV.

## 2 $\sin^2\theta^{\text{lept}}_{\text{eff}}$ extraction

We extract $\sin^2\theta^{\text{lept}}_{\text{eff}}$ by minimizing the $\chi^2$ value between the simulated data and template $A_{\text{FB}}$ distributions in 72 dilepton mass and rapidity bins. Figure 1 shows the $A_{\text{FB}}$ distributions in bins of dimuon mass and rapidity for different energies and pseudorapidity acceptances. As expected, at higher center-of-mass energies the observed $A_{\text{FB}}$ is smaller because the interacting partons have smaller $x$-values which results in a smaller fraction of dimuon events produced by the valence quarks, which also means more dilution. The simulated data are shown for $\sqrt{s}$ = 8 TeV and $\sqrt{s}$ = 14 TeV for two different selection requirements, $|\eta| < 2.4$ and 2.8. Extending the pseudorapidity acceptance significantly increases the coverage for larger $x$-values in the production and reduces both the statistical and PDF uncertainties, as shown below.

The observed $A_{\text{FB}}$ values depend on the size of the dilution effect, as well as on the relative contributions from u and d valence quarks to the total dilepton production cross section. Therefore, the PDF uncertainties translate into sizable variations in the observed $A_{\text{FB}}$ values. However, changes in PDFs affect the $A_{\text{FB}}(M_{\ell\ell}, Y_{\ell\ell})$ distribution in a different way from changes in $\sin^2\theta^{\text{lept}}_{\text{eff}}$. Changes in PDFs result in changes in $A_{\text{FB}}$'s in regions where the absolute values of $A_{\text{FB}}$ is large, i.e. at high and low dilepton masses. On the contrary, the effect of changes in $\sin^2\theta^{\text{lept}}_{\text{eff}}$ are largest near the Z-peak and are significantly smaller at high and low masses. Because of this behavior, which is illustrated in Fig. 2, we apply the Bayesian $\chi^2$ reweighting method to constrain the PDF uncertainties [17–19] and reduce the PDF errors in the extracted value of $\sin^2\theta^{\text{lept}}_{\text{eff}}$.

As a baseline, we use the NLO NNPDF3.0 set. In the Bayesian $\chi^2$ reweighting method, PDF replicas that better describe the observed $A_{\text{FB}}$ distribution are assigned larger weights, and PDF replicas that poorly describe the $A_{\text{FB}}$ are assigned smaller weights. Each weight factor is based on the best-fit $\chi^2$-value obtained with a given PDF replica $i$ used in the templates:

$$w_i = \frac{e^{-\frac{\chi^2_{\text{min}}}{2}}}{\frac{1}{N}\sum_{i=1}^{N} e^{-\frac{\chi^2_{\text{min}}}{2}}}, \tag{5}$$

where $N$ is the number of replicas in a PDF set. The final result is then calculated as a weighted average over the PDF replicas: $\sin^2\theta^{\text{lept}}_{\text{eff}} = \sum_{i=1}^{N} w_i s_i / N$, where $s_i$ is the best-fit $\sin^2\theta^{\text{lept}}_{\text{eff}}$ value

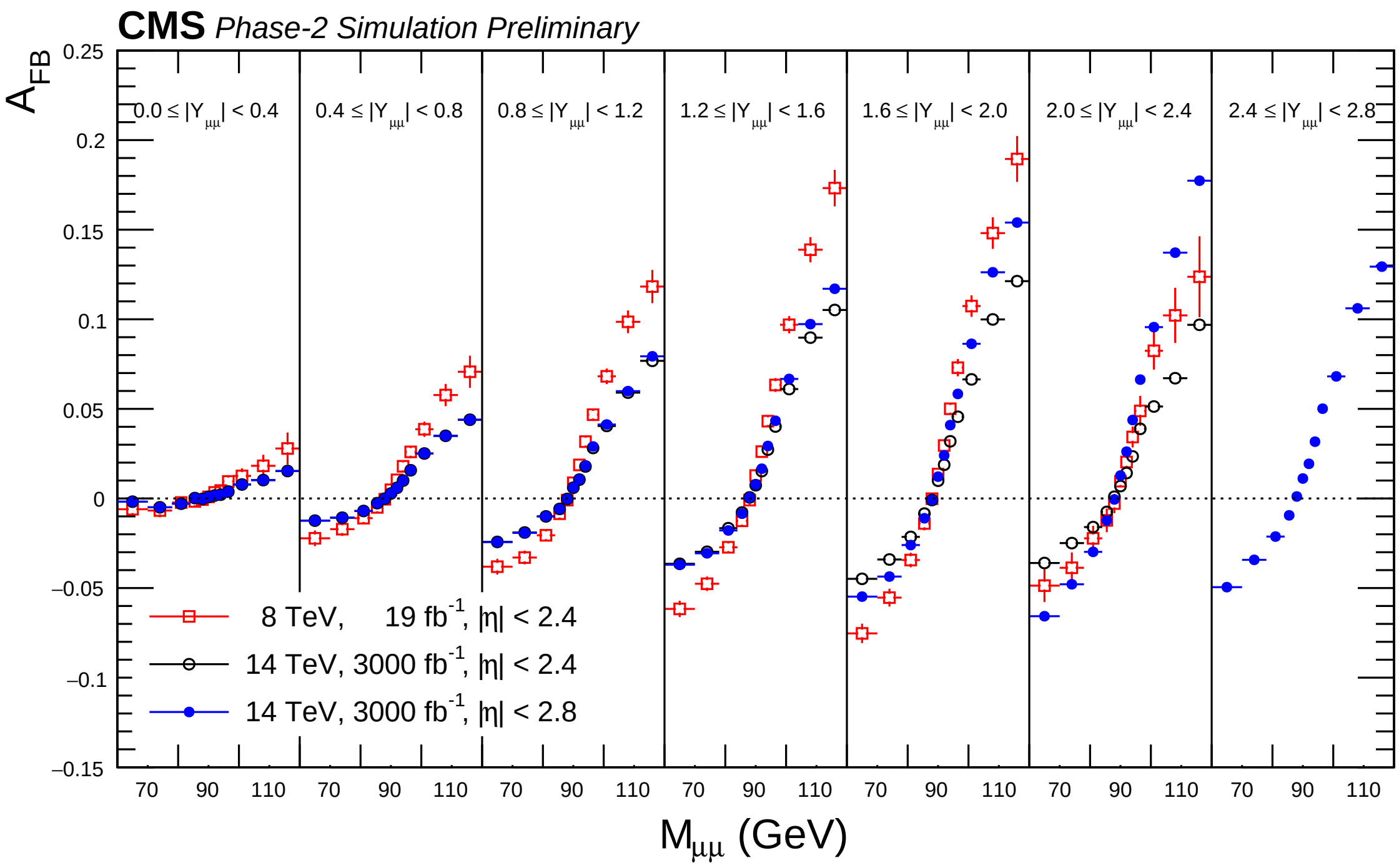

Figure 1: Forward-backward asymmetry distribution, $A_{\mathrm{FB}}(M_{\mu\mu}, Y_{\mu\mu})$, in dimuon events at $\sqrt{s} = 8\,\mathrm{TeV}$ and $14\,\mathrm{TeV}$. The distributions are made with POWHEG event generator using NNPDF3.0 PDFs and interfaced with PYTHIA 8 for parton-showering, QED final-state radiation (FSR) and hadronization. Following acceptance selections are applied to the generated muons after FSR: $|\eta| < 2.4$ (or $|\eta| < 2.8$), $p_{\mathrm{T}}^{\mathrm{lead}} > 25\,\mathrm{GeV}$, $p_{\mathrm{T}}^{\mathrm{trail}} > 15\,\mathrm{GeV}$. The error bars represent the statistical uncertainties for the integrated luminosities corresponding to $19\,\mathrm{fb}^{-1}$ at $\sqrt{s} = 8\,\mathrm{TeV}$ and $3000\,\mathrm{fb}^{-1}$ at $\sqrt{s} = 14\,\mathrm{TeV}$.

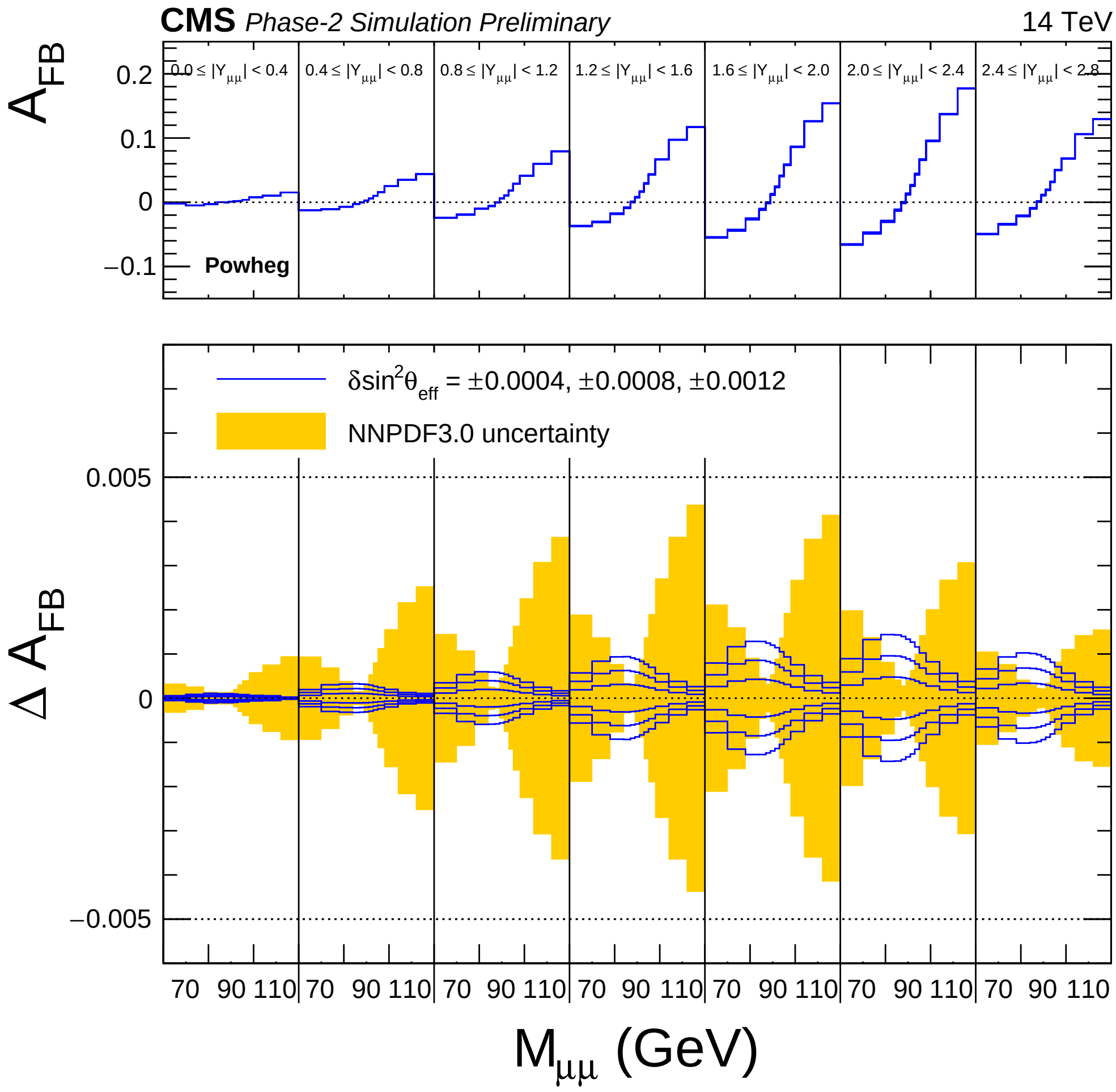

Figure 2: Forward-backward asymmetry distribution, $A_{\rm FB}(M_{\mu\mu}, Y_{\mu\mu})$, in dimuon events at $\sqrt{s} = 14\,\text{TeV}$. The distributions are made with POWHEG event generator using NNPDF3.0 PDFs and interfaced with PYTHIA 8 for parton-showering, QED final-state radiation (FSR) and hadronization. Following acceptance selections are applied to the generated muons after FSR: $|\eta| < 2.8$, $p_{\rm T}^{\rm lead} > 25\,\text{GeV}$, $p_{\rm T}^{\rm trail} > 15\,\text{GeV}$. The solid lines in the bottom panel correspond to six variations of $\sin^2\theta^{\rm lept}_{\rm eff}$ around the central value: $\pm0.0004$, $\pm0.0008$, and $\pm0.0012$. The shaded band shows the standard deviation over the 100 NNPDF3.0 replicas.

obtained for $i$-th PDF replica.

In the case of the 14 TeV analysis with large number of events ($> 200\ \mathrm{fb}^{-1}$), the pseudo-data are too precise to estimate the PDF uncertainties with the Bayesian reweighting approach because the replica distributions are too sparse compared to the statistical uncertainties. Therefore, the PDF uncertainties after the Bayesian reweighting is estimated by extrapolating from the lower values of integrated luminosities as illustrated in Fig. 3. The corresponding values for various luminosities are summarized in Table 1. One can see from the Table that with the extended pseudorapidity coverage of $|\eta| < 2.8$, the statistical uncertainties are reduced by about 30% and the PDF uncertainties are reduced by about 20%, compared to $|\eta| < 2.4$ regardless of the target integrated luminosity and for both nominal and constrained PDF uncertainties.

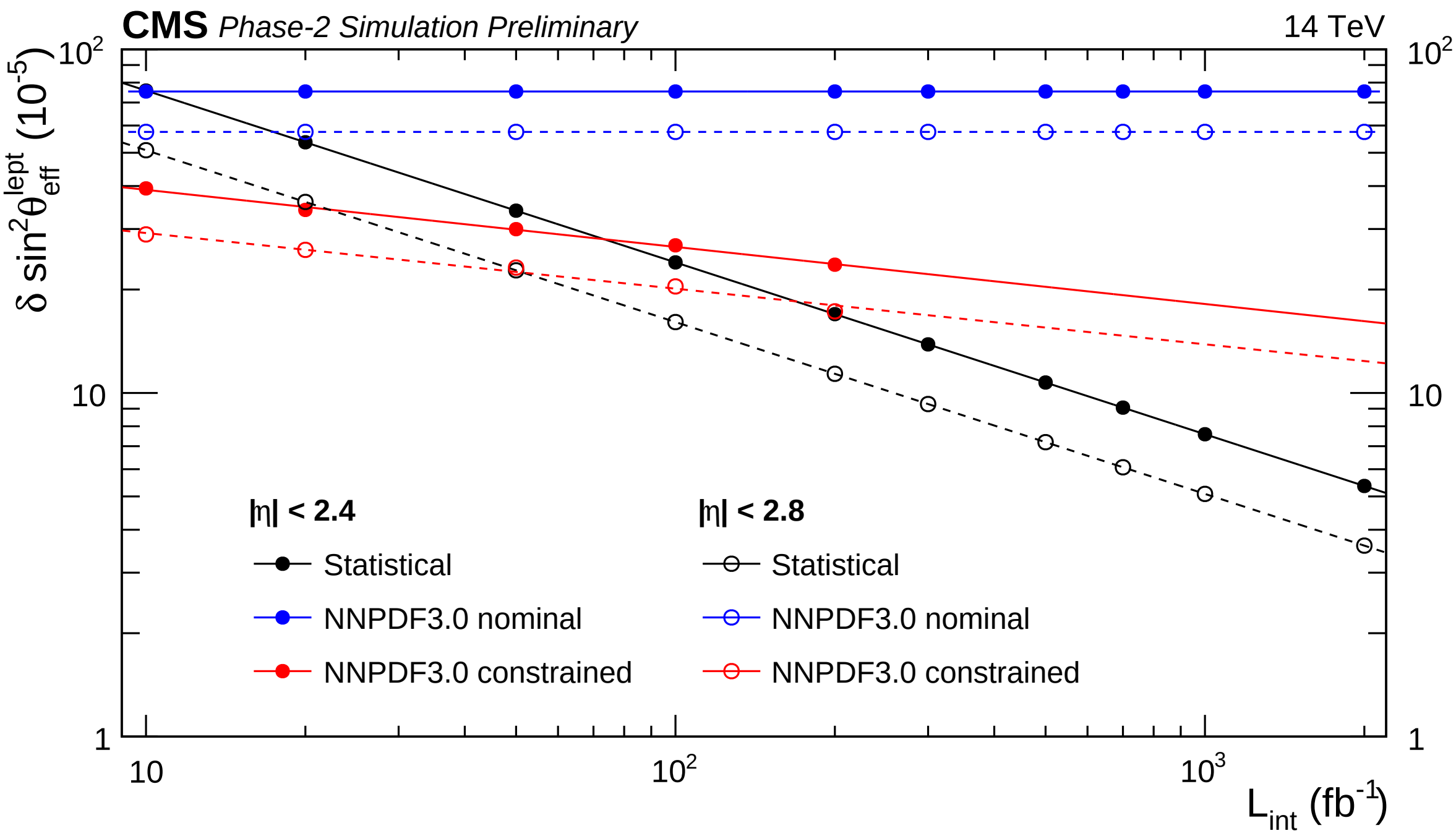

Figure 3: Projected statistical, nominal PDF and constrained PDF uncertainties in $\sin^2\theta^{\mathrm{lept}}_{\mathrm{eff}}$ extracted by fitting $A_{\mathrm{FB}}(m_{\mu\mu}, y_{\mu\mu})$ distributions at $\sqrt{s} = 14$ TeV with different values of integrated luminosities and for $|\eta| < 2.4$ and $|\eta| < 2.8$ acceptance selections for the muons. The nominal NNPDF3.0 uncertainty is calculated as a standard deviation of the extracted $\sin^2\theta^{\mathrm{lept}}_{\mathrm{eff}}$ over the 100 NNPDF3.0 replicas. To calculate the constrained NNPDF3.0 uncertainty, each replica is weighted by $\exp(-\chi^2_{\mathrm{min}}/2)$, where $\chi^2_{\mathrm{min}}$ is the best-fit $\chi^2$ obtained with this replica.

## 3 Summary

We presented prospects for precision $\sin^2\theta^{\mathrm{lept}}_{\mathrm{eff}}$ measurement with the upgraded CMS detector at the high-luminosity LHC. We find that extending the lepton acceptance from $|\eta| < 2.4$ to 2.8 decreases the statistical uncertainties by about 30% and PDF uncertainties by about 20% . We also find that starting from about $1000\,\mathrm{fb}^{-1}$, a single measurement would already have a negligible statistical uncertainty and the PDF uncertainty could be constrained to improve the precision of $\sin^2\theta^{\mathrm{lept}}_{\mathrm{eff}}$ measurement.

Table 1: Statistical, nominal NNPDF3.0, and constrained NNPDF3.0 uncertainties of extracted $\sin^2\theta_{\text{eff}}^{\text{lept}}$ at 14 TeV for muon acceptance of $|\eta| < 2.4$ and $|\eta| < 2.8$ and for the different values of integrated luminosity. For comparison, results of the 8 TeV estimate of this analysis are compared to the results obtained from 8 TeV measurement [1].

| $L_{int}$ | $\delta_{\text{stat}}[10^{-5}]$ | | $\delta^{\text{nominal}}_{\text{nnpdf3.0}}[10^{-5}]$ | | $\delta^{\text{constrained}}_{\text{nnpdf3.0}}[10^{-5}]$ | |
|---|---|---|---|---|---|---|
| ($\text{fb}^{-1}$) | $\|\eta\| < 2.4$ | $\|\eta\| < 2.8$ | $\|\eta\| < 2.4$ | $\|\eta\| < 2.8$ | $\|\eta\| < 2.4$ | $\|\eta\| < 2.8$ |
| 10 | 76 | 51 | 75 | 57 | 39 | 29 |
| 100 | 24 | 16 | 75 | 57 | 27 | 20 |
| 500 | 11 | 7 | 75 | 57 | 20 | 16 |
| 1000 | 8 | 5 | 75 | 57 | 18 | 14 |
| 3000 | 4 | 3 | 75 | 57 | 15 | 12 |
| 19 | 43 | | 49 | | 27 | |
| 19 (from [1]) | 44 | | 54 | | 32 | |

**ATLAS PUB Note**

ATL-PHYS-PUB-2018-037

28th November 2018

# Prospect for a measurement of the Weak Mixing Angle in $pp \to Z/\gamma^* \to e^+e^-$ events with the ATLAS detector at the High Luminosity Large Hadron Collider

The ATLAS Collaboration

This document describes a sensitivity study for the determination of the weak mixing angle from the measurement of the Z boson forward-backward asymmetry using 3000 fb$^{-1}$ of data to be collected by the ATLAS experiment with proton proton collisions at $\sqrt{s} = 14$ TeV at the High Luminosity Large Hadron Collider.

# 1 Introduction

In the Standard Model (SM), the $Z$ boson couplings differ for left- and right-handed fermions. The difference leads to an asymmetry in the angular distribution of positively and negatively charged leptons produced in $Z$ boson decays. This asymmetry depends on the weak mixing angle ($\sin^2\theta_W$) between the neutral states associated to the U(1) and SU(2) gauge groups, i.e. the relative coupling strengths between the photon and the $Z$ boson. The differential cross section for the decay of the $Z/\gamma^*$ to dilepton final state can be written at leading order as:

$$\frac{\mathrm{d}\sigma}{\mathrm{d}(\cos\theta)} = \frac{\alpha^2}{4s}\left[\frac{3}{8}A(1+\cos^2\theta) + B\cos\theta\right], \tag{1}$$

where $\sqrt{s}$ is the centre-of-mass energy of the quark and anti-quark, and $\theta$ is the angle between the negative lepton and the quark. The coefficients $A$ and $B$ depend on the charge of the fermions ($Q_f$) and are defined as [1]:

$$\begin{aligned} A &= Q_l^2 Q_q^2 - 2Q_l g_V^q g_V^l \chi_1 + ({g_A^q}^2 + {g_V^q}^2)({g_A^l}^2 + {g_V^l}^2)\chi_2\,, \\ B &= -4Q_l g_A^q g_A^l \chi_1 + 8 g_A^q g_V^q g_A^l g_V^l \chi_2\,, \end{aligned} \tag{2}$$

where $\chi_1$ is the interference between $Z$ and $\gamma^*$ contributions and $\chi_2$ is the $Z$ Breit-Wigner. The vector and axial-vector couplings of the fermions to the $Z$-boson are define respectively as $g_V^f = t_3^f - (2Q_f \times \sin^2\theta_W)$ and $g_A^f = t_3^f$. The vector coupling depends on the charge and on the weak isospin ($t_3^f$) of the fermions and on the weak mixing angle ($\theta_W$) . The coefficient $B$ introduces a forward-backward asymmetry in $\theta$ arising from the presence of both vector and axial-vector couplings.

Experimentally this asymmetry can be simply expressed as:

$$A_{\mathrm{FB}} = \frac{N(\cos\theta^* > 0) - N(\cos\theta^* < 0)}{N(\cos\theta^* > 0) + N(\cos\theta^* < 0)} = \frac{3}{8}\frac{B}{A}, \tag{3}$$

where $\theta^*$ is the angle between the negative lepton and the quark in the Collins-Soper frame [2] of the dilepton system and $N$ represents the number of forward decays ($\cos\theta^* > 0$) and the number of backward decays ($\cos\theta^* < 0$) . This forward-backward asymmetry is enhanced by the $Z/\gamma^*$ interference and exhibits significant dependence on the dilepton rapidity and invariant mass taking a different sign at high mass and at low mass. Since the asymmetry depends directly on the vector and axial-vector couplings, it is sensitive to the weak mixing angle which relates the two. In order to compare this studies with previous experimental determinations, a scheme is adopted in which the higher order corrections to the Z boson couplings are absorbed in effective couplings. The resulting effective parameter $\sin^2\theta_{eff}$ is defined [3], and is proportional to $\sin^2\theta_W$.

Several measurements of $\sin^2\theta_{eff}$ have been made at previous and current colliders, and the current world average is dominated by the combination of measurements at LEP and at SLD, which gives $\sin^2\theta_{eff} = 0.231530 \pm 16 \times 10^{-5}$. However, the two most precise measurements differ by over $3\sigma$ [3]. It is of great scientific interest the study of the weak mixing angle at the Large Hadron Collider (LHC) and at the High-Luminosity Large Hadron Collider (HL-LHC). Precision measurements may give insight into the tension between the previous precision measurements or may show signs of new physics.

Measurements of $A_{\mathrm{FB}}$ which at the LHC show the greatest sensitivity to $\sin^2\theta_{eff}$ are at high $Z$ rapidity when at least one lepton is present in the forward region [4]. At a rapidity of 0, the initial state is symmetric, with equal probability for the initial-state quark to be originated from either proton. Therefore, in such a kinematic configuration no forward-backward asymmetry is present due to a complete dilution of the parton-level asymmetry in the proton-proton collisions. For increasing $Z$ boson rapidities the momentum fraction of one parton reaches larger x where the valence quark Parton Density Function (PDF) dominate because the valence quarks typically carry more momentum than the antiquarks. Therefore, the $Z$ boson is more likely to be boosted in the quark direction and the incoming quark direction can be determined. Consequently a forward-backward asymmetry is visible, providing sensitivity to $\sin^2\theta_{eff}$, and it varies with the boson rapidity.

This note reports the projected sensitivity for the measurement of the $Z$ boson forward-backward asymmetry as a function of the dilepton invariant mass and rapidity, assuming 3000 fb$^{-1}$ of data at $\sqrt{s} = 14$ TeV to be collected with an upgraded ATLAS detector during the HL-LHC phase. Only $Z$ bosons decaying to electrons pairs are considered in this analysis since this final state provides the best experimental precision within the largest acceptance of the upgraded ATLAS detector.

## 2 The HL-LHC and the ATLAS detector

During the HL-LHC phase an instantaneous luminosity of around $5 - 7 \times 10^{34}\text{cm}^{-2}\text{s}^{-1}$ and an average number of collisions per bunch crossing (pile-up) of 200 are expected. To cope with the higher luminosity at the HL-LHC and its associated high pile-up and intense radiation environment several detector upgrades are foreseen for the ATLAS detector. A new inner tracking system (ITk) [5], extending the tracking region from $|\eta| \leq 2.5$ up to $|\eta| \leq 4.0$, will provide the ability to reconstruct forward charged particle tracks, which can be matched to calorimeter clusters for forward electron reconstruction, or associated to forward jets. A new detector, high granularity timing detector (HGTD) [6] designed to mitigate the pile-up is also foreseen. The other planned upgrades to the ATLAS detector are described in detail in Ref. [7].

## 3 Analysis

Monte Carlo (MC) simulated events with Powheg-BOX [8] generator interfaced to Pythia8 [9] are used to predict $pp \to Z/\gamma^* \to e^+e^-$ signal at $\sqrt{s} = 14$ TeV. The events are overlaid with additional inelastic $pp$ collisions per bunch-crossing simulated with Pythia. Parameterisations of the expected ATLAS detector performances during the HL-LHC runs have been derived [7] and applied on particle-level objects to emulate the detector response. Lepton trigger and identification efficiencies are derived as a function of $|\eta|$ and $p_T$ and used to estimate the likelihood of a given lepton to fulfil either the trigger or identification requirement, respectively. Electron energy resolutions and scale estimates are also parameterised as a function of $|\eta|$ and transverse energy for each electron.

Events are selected requiring at least one electron firing the single electron trigger except for events in which both electron candidates are in the forward region ($|\eta| > 2.5$) where a dielectron trigger is required. Only events with exactly two electron candidates of opposite charge and each having $p_T > 25$ GeV are futher selected. Each electron candidate must satisfy a set of *tight* selection criteria, which have been optimised for the level of pile-up expected at the HL-LHC [10]. The invariant mass of the electron pair is required to be loosely consistent with the $Z$ boson mass, $60 < m_{ee} < 200$ GeV. The fiducial acceptance of $pp \to Z/\gamma^* \to e^+e^-$ events is split into three orthogonal analysis channels depending on the electron pseudorapidity. Considering an integrated luminosity of 3000 fb$^{-1}$ at $\sqrt{s} = 14$ TeV, 540 million events are

3

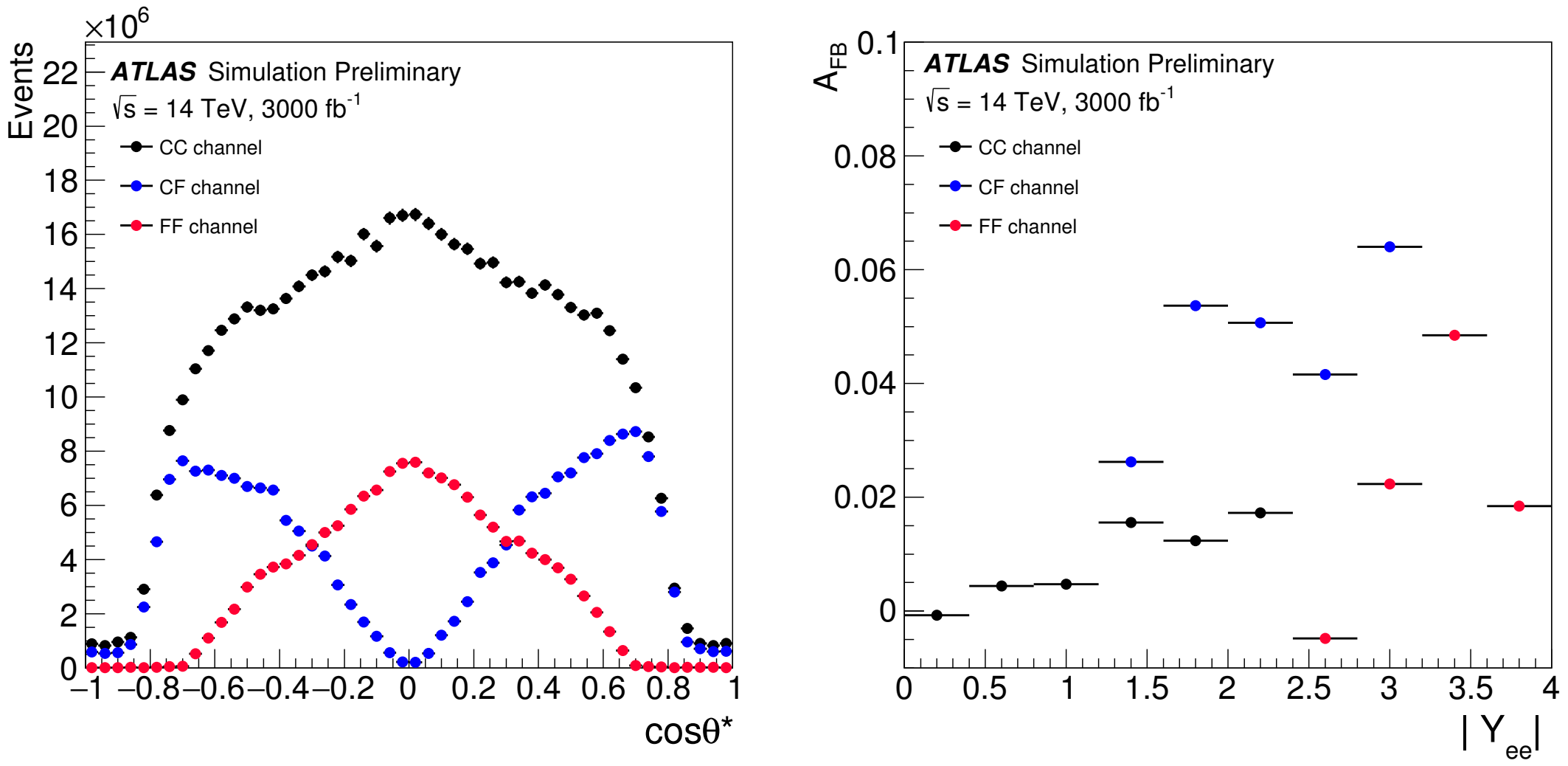

Figure 1: (left) The $\cos\theta^*$ distribution for CC, CF and FF channels for the selected $pp \to Z/\gamma^* \to e^+e^-$events expected for $3000\,\text{fb}^{-1}$ at $\sqrt{s} = 14$ TeV. (right) The $A_{\text{FB}}$ distribution at particle level in the fiducial volume as a function of the absolute dielectron rapidity for CC, CF and FF channels for $pp \to Z/\gamma^* \to e^+e^-$events expected for $3000\,\text{fb}^{-1}$ at $\sqrt{s} = 14$ TeV.

expected with a pair of electrons in the central region of the detector, $|\eta| < 2.47$, (CC channel), 210 million events are expected with a pair of electrons, where one electron is in the forward region ($2.5 < |\eta| < 4.2$) of the detector (CF channel) and 150 million events are expected with a pair of electrons in the forward region of the detector (FF channel).

The events in each channel are further categorised in 10 equal-size bins in absolute dilepton rapidity up to $|Y_{ee}| = 4.0$. On the left of Figure 1 the $\cos\theta^*$ distribution for the $pp \to Z/\gamma^* \to e^+e^-$candidate in the three different channels is shown for $3000\,\text{fb}^{-1}$ at $\sqrt{s} = 14$ TeV. The CF channel select event at high $\cos\theta^*$ value where the forward-backward asymmetry is more pronounced, therefore the sensitivity to the $A_{\text{FB}}$ and consequently to $\sin^2\theta_{eff}$ is higher in this channel.

The contribution of jets misidentified as electrons is suppressed using an additional track isolation requirement. In the forward region, the timing information provided by the HGTD is used to improve the electron isolation by rejecting additional tracks from interactions close in space, but separated in time from the hard-scatter vertex. The purity of the candidate sample is determined with simulation, and is found to be greater than 99% in the CC channel, between 90 and 98% in CF, and between 60 and 90% in the FF channel depending on the dilepton rapidity. The signal significance in the CF channel is improved up to 20% thanks to the enhance signal efficiency provided by the timing information of the HGTD.

The background-subtracted expected events in the three different channels are unfolded to forward and backward fiducial cross sections using the inverse of the response matrix to correct for detector effects and migrations between $m_{ee}$ and $|Y_{ee}|$ bins. In the CF and FF channels, migrations from one bin to another bin in $m_{ee}$ are up to 50 and 60% respectively.

Various sources of uncertainty are considered in the analysis and propagated via the unfolding procedure to the results. Significant uncertainties arise from the limited knowledge of the momentum scale and resolution of the electrons. Following the methodology in Ref. [11], in order to account for possible non-linearity in the energy scale of electrons reconstructed in the central (forward) region, a systematic of 0.5% (0.7%) is considered for electrons with $E_T$ < 55 GeV and up to 1.5% (2.1%) for $E_T$ > 100 GeV. Uncertainties associated with background are mostly relevant in the CF and FF channels. Given the lack of knowledge of the composition and the modelling of the misidentified electrons and the limited statistic of the MC samples used for the background determination, an overall normalisation systematic of 10% is considered for the background shape modelling. Since it was verified that this uncertainty has a negligible effect on the measurement, an additional 5% uncertainty, uncorrelated across bins, is considered on the background yield for all bin of invariant mass and rapidity .

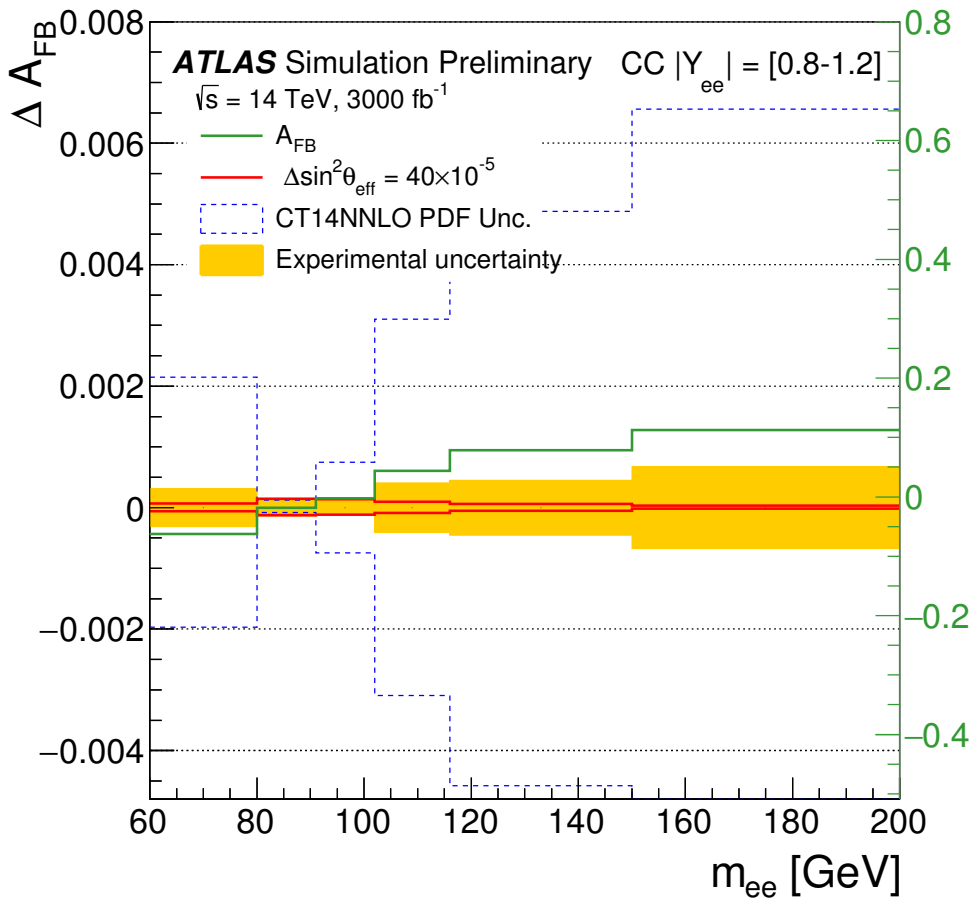

(a) Central-central (CC) channel

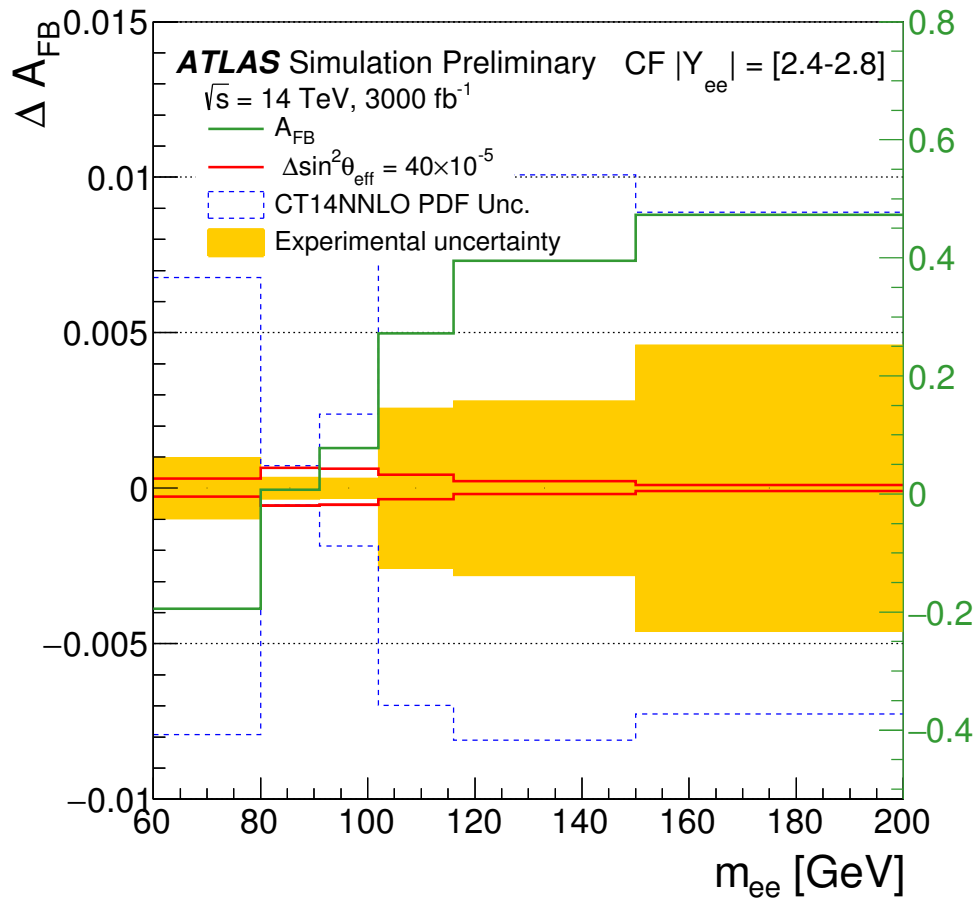

(b) Central-forward (CF) channel

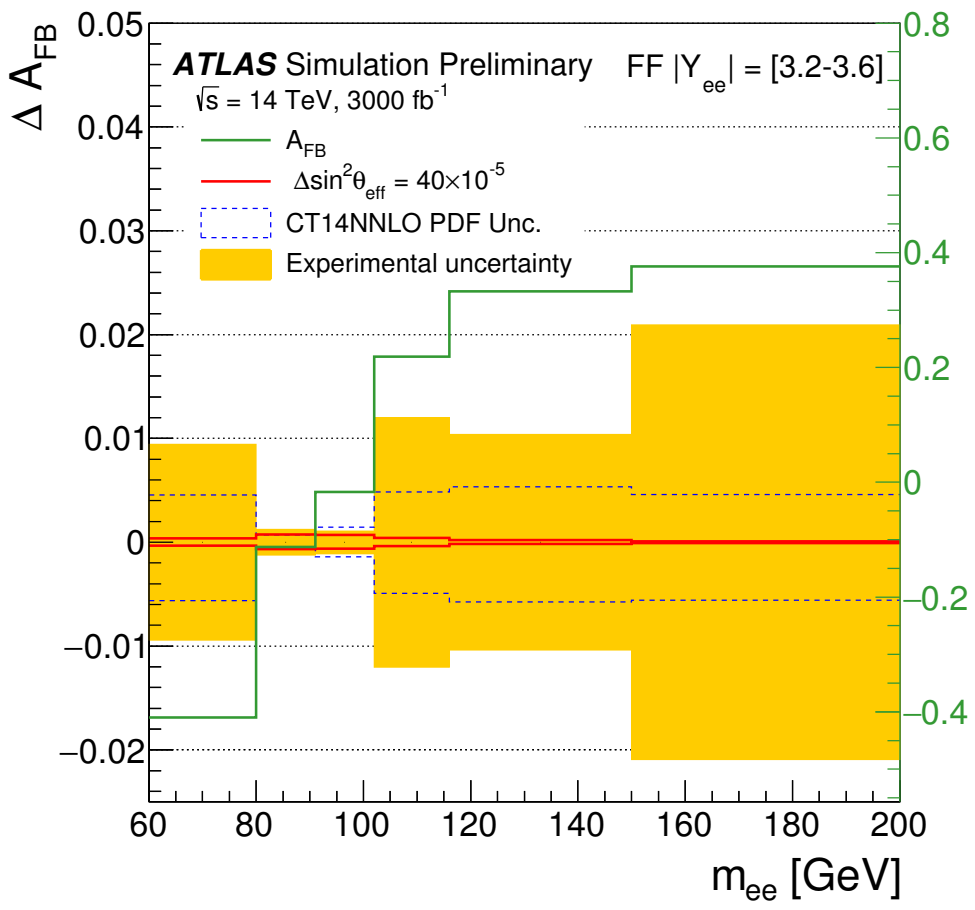

(c) Forward-forward (FF) channel

Figure 2: Distribution of the uncertainty on $A_{\mathrm{FB}}$ ($\Delta A_{\mathrm{FB}}$ ) as a function of dielectron mass for the CC, CF and FF channels. The filled bands correspond to the experimental uncertainty. The solid red lines correspond to a variations of $\sin^2\theta_{eff}$ corresponding to $40 \times 10^{-5}$. The dashed blue lines illustrate the total uncertainty of CT14 NNLO PDF set before in-situ profiling. Overlaid green line shows the particle-level $A_{\mathrm{FB}}$ distribution.

The $A_{\mathrm{FB}}$ for the CC, CF and the FF channel is calculated as function of $|Y_{ee}|$ and $m_{ee}$ following Equation 3 from projected measurements of particle level fiducial cross sections. On the right side of Figure 1 the expected amplitude of the $A_{\mathrm{FB}}$ as function of $|Y_{ee}|$ in the fiducial volume is shown for CC, CF and FF channels separately. The expected sensitivity to particle level $A_{\mathrm{FB}}$ as a function of $m_{ee}$ is also shown in Figure 2 for each channel for a chosen rapidity bin.

# 4 Results

The extraction of $\sin^2\theta_{eff}$ is done by minimising the $\chi^2$ value comparing particle-level $A_{\mathrm{FB}}$ distributions with different weak mixing angle hypotheses in invariant mass and rapidity bins combining CC, CF and FF channels. The default $A_{\mathrm{FB}}$ distributions are generated, at leading order (LO) in QCD, with DYTURBO an optimised version of DYRES/DYNNLO [12] with NNLO CT14 PDF [13] and the world average value for $\sin^2\theta_{eff}$= 0.23153. The same LO calculation is used to compute $A_{\mathrm{FB}}$ variations for different values of $\sin^2\theta_{eff}$ and PDFs. As shown in Figure 2, the imperfect knowledge of the PDF results in sizeable uncertainties in $A_{\mathrm{FB}}$, in particular in regions where the absolute value of the asymmetry is large, i.e. at high and low $m_{ee}$. On the contrary, near the $Z$ boson mass peak, the effect of varying $\sin^2\theta_{eff}$ is maximal, while being significantly smaller at high and low masses. Thus, a global fit is performed where $\sin^2\theta_{eff}$ is extracted while constraining at the same time the PDF uncertainties. The profiling procedure [14] used in this analysis to constraint the PDF uncertianies follows the one used in previous ATLAS publication [15] and is implemented in the *xFitter* package [16].

With this analysis, a significant reduction of the light quarks uncertainties at low $x$ is seen and the expected uncertainty on the extraction of $\sin^2\theta_{eff}$ are $25\times10^{-5}$, $21\times10^{-5}$ and $40\times10^{-5}$ for the CC, CF and FF channels respectively. Combining the three channels together, the measurement reaches a precision of $\Delta\sin^2\theta_{eff} = 18\times10^{-5}( \pm16\times10^{-5}$ (PDF) $\pm 9\times10^{-5}$ (exp.) ). The uncertainty of the results remains dominated by the currently limited knowledge of the PDFs. In Table 1 the $\sin^2\theta_{eff}$ sensitivity obtained with this analysis is compared with the latest published ATLAS measurement at $\sqrt{s}$ = 8 TeV [17]. When comparing results at 14 and 8 TeV, it should be noted that the forward-backward asymmetry in Z-events at $\sqrt{s}$ = 14 TeV is smaller than the asymmetry observed at $\sqrt{s}$ = 8 TeV, in addition the PDFs at lower $x$, more important at $\sqrt{s}$ = 14 TeV, are less known, resulting in an smaller overall sensitivity to extract $\sin^2\theta_{eff}$ . The loss of sensitivity is approximately compensated by the increased Z-boson production cross-section from 8 to 14 TeV (a factor of 1.8). Therefore, with the same integrated luminosity of 20 fb$^{-1}$, a simple extrapolation to $\sqrt{s}$ = 14 TeV centre-of-mass energy of the $\sin^2\theta_{eff}$ extracted at $\sqrt{s}$ = 8 TeV by the ATLAS experiment would result into the same statistical uncertainty and a 45% increase of the PDF uncertainty.

In the context of the Yellow Report for the HL-LHC, prospect PDF fits including HL-LHC pseudo-data of future PDF-sensitive measurements from ATLAS and CMS were performed [19]. Three prospect PDF scenarios were considered and compared with the reference PDF set PDF4LHC15 [20]. The expected sensitivity of the $\sin^2\theta_{eff}$ measurement with 3000 fb$^{-1}$ at $\sqrt{s}$ = 14 TeV is improved by 10−25% depending on the prospect PDFs scenario considered. In Table 1 the $\sin^2\theta_{eff}$ precision obtained with the "ultimate" HL-LHC PDF set is compared with the with the one obtained with CT14NNLO PDF set.

Finally the sensitivity of the analysis to $\sin^2\theta_{eff}$ extraction is also estimated with a prospect PDF set including expected data from LHeC collider [21]. In this case the PDF uncertainty is reduced by an additional factor of 5 with respect to the one obtained with the HL-LHC prospect PDFs.

Figure 3 compares the ATLAS sensitivity studies of $\sin^2\theta_{eff}$ presented in this note to previous measurements from the LHC experiments [17, 22–24], the combined legacy measurement from the CDF and D0 experiments at the Tevatron [25], and the most precise individual legacy measurements from LEP and SLD [26]. The accuracy of the measurement of the weak mixing angle obtained with an analysis of the $A_{\text{FB}}$ in Z-events with 3000 fb$^{-1}$ at $\sqrt{s}$ = 14 TeV with the ATLAS detector at HL-LHC exceeds the precision achieved in all previous single-experiment results to date and the measurement is dominated by PDF uncertainties. To explore the full potential of the HL-LHC data will be therefore essential to reduced PDF uncertainties. A moderate improvement of the sensitivity of this measurement is observed when using prospect PDF sets which include ancillary neutral current Drell-Yann measurements performed with the data collected during the high luminosity phase of the LHC, as included in PDF4LHC15$_{\text{HL-LHC}}$ set. Futher improvements may be achieved when using additional data on $W$ charge asymmetry and with the structure function data from the LHeC collider.

| | ATLAS $\sqrt{s}$ = 8 TeV | ATLAS $\sqrt{s}$ = 14 TeV | ATLAS $\sqrt{s}$ = 14 TeV |
|---|---|---|---|
| $\mathcal{L}$ [fb$^{-1}$] | 20 | 3000 | 3000 |
| PDF set | MMHT14 [18] | CT14 [13] | PDF4LHC15$_{\text{HL-LHC}}$ [19] |
| $\sin^2\theta_{eff}$ [$\times 10^{-5}$] | 23140 | 23153 | 23153 |
| Stat. | ± 21 | ± 4 | ± 4 |
| PDFs | ± 24 | ± 16 | ± 13 |
| Experimental Syst. | ± 9 | ± 8 | ± 6 |
| Other Syst. | ± 13 | - | - |
| Total | ± 36 | ± 18 | ± 15 |

Table 1: The value of $\sin^2\theta_{eff}$ with the breakdown of uncertainties from the ATLAS preliminary results at $\sqrt{s}$ = 8 TeV with 20 fb$^{-1}$ [17] is compared to the projected $\sin^2\theta_{eff}$ measurements with 3000 fb$^{-1}$ of data at $\sqrt{s}$ = 14 TeV for two PDF sets considered in this note. All the numbers values are given in units of $10^{-5}$. Note that other sources of systematic uncertainties, such as the impact of the MC statistical uncertainty, evaluated in Ref. [17] are not considered in this prospect analysis. For the HL-LHC prospect PDFs the "ultimate" scenario is chosen.

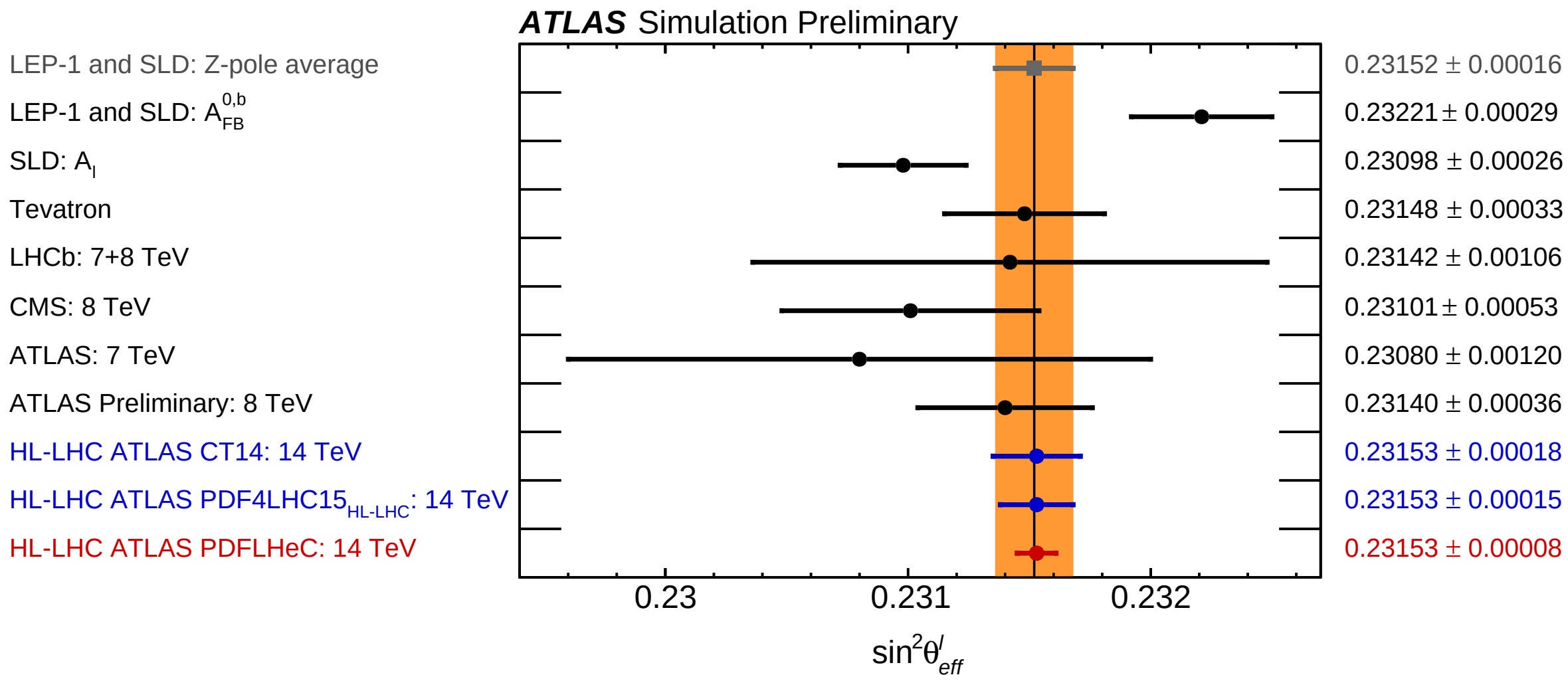

Figure 3: Comparison of the expected precision of the effective leptonic weak mixing angle presented in this note to previous measurements at LEP-1 and SLD [26], at the Tevatron [25], and at the LHC [17, 22–24]. The overall LEP-1 and SLD average [1] is represented together with its uncertainty as a vertical band. The ATLAS results from this analysis are shown with different PDF set senarios.

8

**ATLAS PUB Note**

# Prospect study of electroweak production of a $Z$ boson pair plus two jets at the HL-LHC

The ATLAS Collaboration

This note summarizes the prospect study for the electroweak production of a $Z$ boson pair plus two jets at the high-luminosity LHC in the four-lepton channel, using 3000 fb$^{-1}$ of simulated pp collisions at a centre-of-mass energy of 14 TeV, to be recorded by the ATLAS detector at the HL-LHC. Simulated events were produced at generator level and detector effects of lepton and jet reconstruction and identification were estimated by corrections, assuming the mean number of interactions per bunch crossing of 200. The expected significance of the electroweak production has been studied, as well as the precision of the expected measurements of differential cross sections as a function of the dijet or $ZZ$ invariant mass.

# 1 Introduction

The vector boson scattering (VBS) process is crucial for probing the mechanism of electroweak symmetry breaking in the Standard Model. At the LHC, the VBS process has been studied through the measurements of EW production of two vector bosons plus two jets (EW-VVjj). Evidence of the production of EW-VVjj processes at the LHC was seen by the ATLAS collaboration with 20.3 fb$^{-1}$ integrated luminosity of $\sqrt{s}$ = 8 TeV pp collision data [1], where a 3.6 $\sigma$ excess was observed in the data over the background-only prediction. A 2.0 $\sigma$ excess was observed by the CMS collaboration with 19.4 fb$^{-1}$ integrated luminosity of $\sqrt{s}$ = 8 TeV pp collision data [2]. In Run II, observations of the EW-VVjj processes have been reported by both the ATLAS and CMS collaborations in the same-sign $WW$ channel [3, 4]. In the $WZ$ channel, results were reported recently by both the ATLAS and CMS collaborations [5, 6]. With more data collected, evidence of the EW-VVjj processes with the $ZZ$ final states (EW-ZZjj) also becomes possible. A recent publication from the CMS collaboration with 35.9 fb$^{-1}$ of data at 13 TeV [7] reported an observed (expected) significance of 2.7 (1.6) $\sigma$ of the EW-ZZjj process in the $\ell\ell\ell\ell$ ($\ell = e, \mu$) final state. The $ZZ$ channel will benefit significantly from the increased luminosity at the high-luminosity LHC (HL-HLC).

In this note, a prospect study has been performed for the EW-ZZjj process at the HL-LHC in the $\ell\ell\ell\ell$ channel with the ATLAS detector. The VBS topology consists of two high-energy jets in the back and forward regions, with two vector bosons. Both the EW and QCD processes give the same final state, and the QCD VVjj process is the dominant background.

This study uses 3000 fb$^{-1}$ of simulated pp collisions at a centre-of-mass energy of 14 TeV, expected to be recorded by the ATLAS detector. Simulated events were produced at generator level. The detector effects of lepton and jet reconstruction and identification were estimated by corrections, assuming the mean number of interactions per bunch crossing of 200. The expected significance of the electroweak production has been studied, as well as the precision of the expected measurements of differential cross sections as a function of the dijet or $ZZ$ invariant mass.

# 2 The ATLAS detector at the HL-LHC

The new Inner Tracker (ITk) [8] will extend the ATLAS tracking capabilities to pseudorapidity ($|\eta|$) up to 4.0. The upgraded Muon Spectrometer [9] at the HL-LHC, where a forward muon tagger is included, will also provide lepton identification capabilities to $|\eta|$ up to 4.0. The new high granularity timing detector (HGTD) [10] designed to mitigate the pile-up (PU) effects is also foreseen in the forward region of $2.4 < |\eta| < 4.0$. The expected performance of the upgraded ATLAS detector has been studied at the HL-LHC as reported in Ref. [11].

## 3 Simulation

The analysis is performed using particle-level events of the signal and background processes. The samples are generated at 14 TeV and with a fast simulation based on the parametrization of the ATLAS detector at HL-LHC [11].

Both the EW-ZZjj and QCD-ZZjj processes with the $ZZ \to \ell\ell\ell\ell$ decays are modelled using Sherpa 2.2.2 [12] with the NNPDF3.0NNLO [13] PDF set. Those samples are generated inclusively with $\ell\ell\ell\ell$ plus two jets, where hadronic decay of vector bosons are included. The analysis selections are optimized to enhance the EW-ZZjj processes.

The signal sample is generated with two jets at Matrix Element (ME) level. The background process of $pp \to \ell\ell\ell\ell+$ n partons is modelled with next-to-leading order (NLO) QCD accuracy for events with up to one outgoing parton and with leading order (LO) accuracy for the case with two and three partons, in a phase space of $m_{\ell\ell} > 4$ GeV and at least two leptons with $p_{\mathrm{T}} >$ 5 GeV.

Other backgrounds, such as irreducible ones from triboson and contributions with misidentified or non-prompt leptons, have minor contributions to the $\ell\ell\ell\ell$ channel and therefore are not included in this analysis.

In addition to hard scattering events, pile-up collisions are included with a mean number of interactions per bunch crossing of 200, and the impact is included in the reconstruction for each physics object used in this analysis: electrons, muons and jets.

Signal and background yields are scaled to an integrated luminosity of 3000 fb$^{-1}$ for this study.

## 4 Event selections and phase space definition

The analysis selection is based on the Run-II analysis [14] and has been modified according to the expected changes in the ATLAS detector at the HL-LHC, which includes lepton identification in the forward region up to $|\eta| = 4.0$.

Candidate events are selected with exactly four leptons (electrons or muons), consistent with the decays from two on-shell $Z$ bosons. The VBS topology is enhanced by requiring at least two jets with large invariant mass and $\eta$ separation. Detailed selection criteria are summarized below:

- Exactly four leptons with $p_T^{\ell} >$ 20, 20, 10 and 7 GeV, and $|\eta| < 4.0$.
- Two $Z$ candidates, minimizing $|m_{Z_1} - m_{Z_{PDG}}| + |m_{Z_2} - m_{Z_{PDG}}|$, where $m_{Z_{PDG}}$ refers to the $Z$ mass value from PDG [15]. The candidate with the dilepton mass closest to $m_{Z_{PDG}}$ is labelled as $Z_1$, while the other one as $Z_2$.
- $60 < m_{Z_1} < 120$ GeV and $60 < m_{Z_2} < 120$ GeV.
- $m_{\ell^+\ell^-} > 10$ GeV for all the same-flavour opposite-sign lepton pairs.

- PU jet suppression is applied for all PU jets in the region of $|\eta| < 3.8$, based on the expected ATLAS detector performance at the HL-LHC.
- At lease two candidate jets satisfying $p_T > 30$ GeV with $|\eta| < 3.8$ or $p_T > 70$ GeV with $3.8 < |\eta| < 4.5$, and not overlapping with any of the four leptons within a cone in the $\phi - \eta$ space $\Delta R(\ell, j) < 0.2$.
- Two leading-$p_T$ jets (denoted as $j_1$ and $j_2$) satisfying $\eta^{j_1} \times \eta^{j_2} < 0$.
- $m_{jj} > 600$ GeV and $|\Delta\eta(jj)| > 2$.

In addition, a fiducial phase space is defined at generator level with the same kinematic selections as listed above, and it is used to study the expected precision of the cross-section measurements.

The numbers of selected signal and background events are summarized in Table 1, normalized to 3000 fb$^{-1}$ data. In addition to the baseline selection listed above, two alternative selections are also studied to compare different detector scenarios at the HL-LHC. The first one, with the lepton $|\eta|$ cut being limited to 2.7, is used to understand the improvement due to forward lepton reconstruction and identification with the upgraded ATLAS detector. The second one, with the PU jet suppression being only applicable up to $|\eta| = 2.4$, is deployed to see the benefit from extending the rejection range of PU jets at the HL-LHC. The extended tracking coverage improves the lepton detection efficiency and can increase the number of events with $\ell\ell\ell\ell$ and two jets in the final state by 15 to 30%, providing larger event yield for differential cross-section measurements. However, the overall significance of observing the EW-ZZjj process does not improve as much, due to larger increase of the QCD-ZZjj background contribution. This is an effect of the $ZZ$ system from EW processes being more centrally produced than the $ZZ$ system from QCD processes, as shown in Figure 1, where the distributions of the events as a function of the centrality variable is shown for both EW and QCD processes. The centrality of the $ZZ$ system is defined as

$$ZZ \text{ centrality} = \frac{|y_{ZZ} - (y_{j1} + y_{j2})/2|}{|y_{j1} - y_{j2}|} \tag{1}$$

The distributions of the dijet invariant mass ($m_{jj}$), the $ZZ$ invariant mass ($m_{ZZ}$) and the $\phi$ separation of two $Z$ bosons ($|\Delta\phi(ZZ)|$), after the event selection and normalized to 3000fb$^{-1}$, are also shown in Figure 1.

| Selection | $N_{\text{EW-ZZjj}}$ | $N_{\text{QCD-ZZjj}}$ | $N_{\text{EW-ZZjj}} / \sqrt{N_{\text{QCD-ZZjj}}}$ |
|---|---|---|---|
| Baseline | 432 ± 21 | 1402 ± 37 | 11.54 ± 0.58 |
| Leptons with $|\eta| < 2.7$ | 373 ± 19 | 1058 ± 33 | 11.46 ± 0.62 |
| PU jet suppression only in $|\eta| < 2.4$ | 536 ± 23 | 15470 ± 120 | 4.31 ± 0.19 |

Table 1: Comparison of event yields for signal ($N_{\text{EW-ZZjj}}$) and background ($N_{\text{QCD-ZZjj}}$) processes, and expected significance of EW-ZZjj processes, normalized to 3000 fb$^{-1}$ data at 14 TeV, with baseline and alternative selections. Uncertainties in the table refer to expected data statistical uncertainty at 14 TeV with 3000 fb$^{-1}$.

4

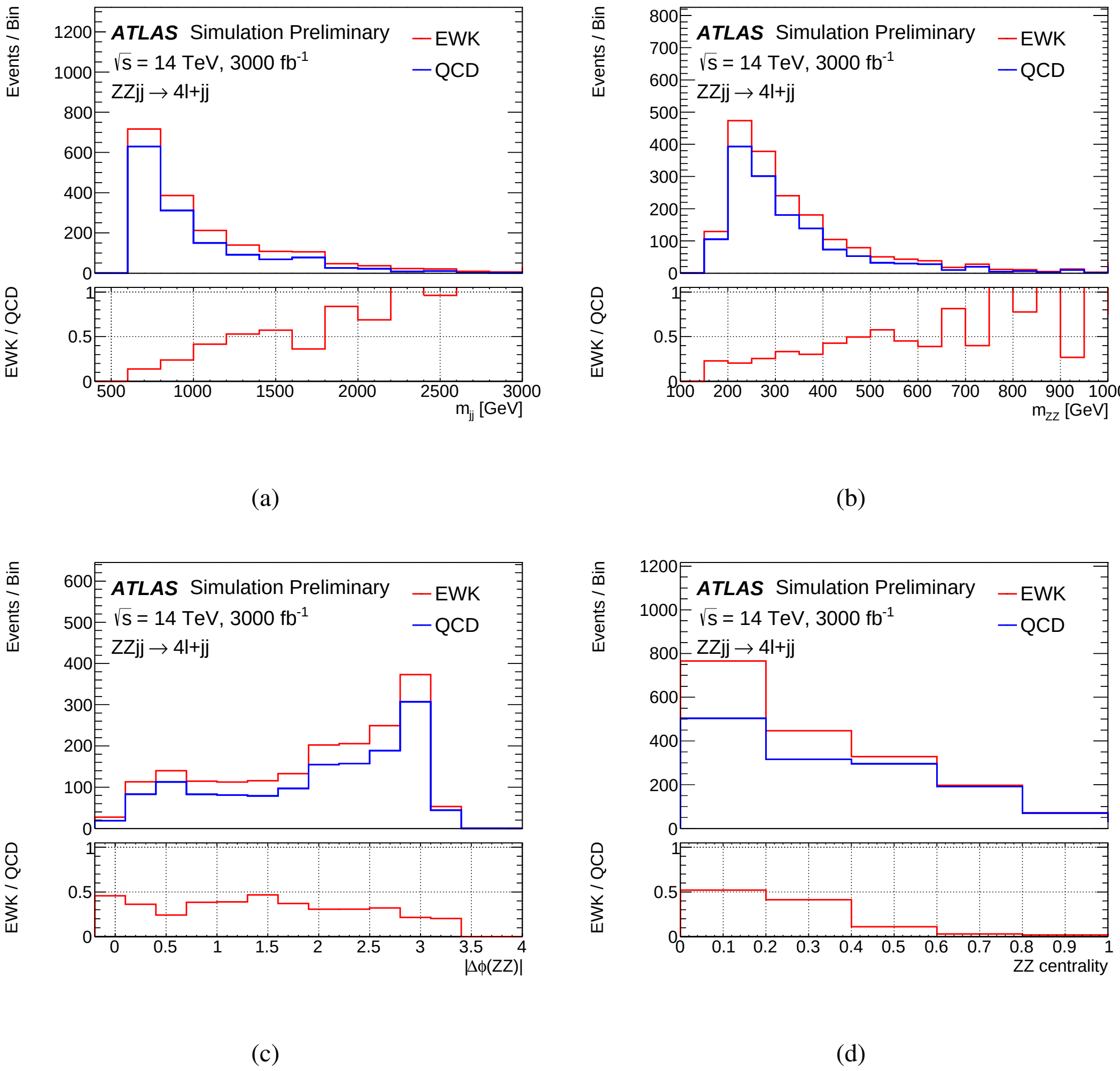

Figure 1: Detector-level distributions of EW and QCD processes after event selections for (a) $m_{jj}$, (b) $m_{ZZ}$, (c) $|\Delta\phi(ZZ)|$, (d) centrality of the $ZZ$ system, normalized to 3000 fb$^{-1}$.

5

## 5 Systematics

The dominant systematics for $\ell\ell\ell\ell$ channel are from theoretical modelling of the QCD-ZZjj background processes.

For theoretical sources, different sizes of systematics have been tested, at 5, 10 and 30% level on the background modelling. The 30% uncertainty is a conservative estimation from direct calculation by comparing different choices of PDF sets and QCD renormalization and factorization scales, following recommendation from PDF4LHC [16]. The uncertainty is driven by the QCD scale choices. The 5% one is an optimistic estimation where enough data events from QCD enriched control region at the HL-LHC could be used to provide constrain on the theoretical modelling of QCD-ZZjj processes.

For the experimental sources, the jet uncertainties have been checked following the studies in Ref. [11] and effect is within fluctuation of the simulated events, which is at 5% level. Thus, 5% uncertainty is used as a conservative estimate of the experimental uncertainties.

The final results largely rely on theoretical modelling of QCD-ZZjj background processes, thus results are presented under different conditions: the case with statistical uncertainty only, the case with statistical and experimental uncertainties, and the cases with additional theoretical uncertainties of 5, 10 and 30%. Statistical, experimental and theoretical uncertainties are treated as uncorrelated and summed up quadratically.

## 6 Results

The expected significance of EW-ZZjj production processes is calculated as

$$\text{Significance} = \frac{S}{\sqrt{\sigma(B)^2_{stat.} + \sigma(B)^2_{syst.}}}, \tag{2}$$

where $S$ denotes the number of signal events after selections, and $\sigma(B)_{stat.}$ and $\sigma(B)_{syst.}$ refer to the statistical and systematic uncertainties from background processes. Statistical uncertainty is estimated from expected data yield at 14 TeV with 3000 fb$^{-1}$.

A scan over different $m_{jj}$ cuts is performed and the result is shown in Figure 2 for luminosity of 3000 fb$^{-1}$. The scan is done from 600 GeV to 1.5 TeV, with a 50 GeV step.

The significance could be further improved with multivariate analysis techniques in the future.

In addition, the expected differential cross-section measurements of the EW-ZZjj processes at 14 TeV have been studied in the defined phase space, as a function of $m_{jj}$, and $m_{ZZ}$, as shown in Figure 3. The expected differential cross-section measurements are calculated bin by bin, following the equation

$$\sigma = \frac{N_{pseudo-data} - N_{QCD-ZZjj}}{L * C_{EW-ZZjj}}, \; C_{EW-ZZjj} = \frac{N^{det.}_{EW-ZZjj}}{N^{part.}_{EW-ZZjj}}, \tag{3}$$

where $N_{pseudo-data}$ is the expected number of data events with 3000 fb$^{-1}$ luminosity, and $N_{QCD-ZZjj}$ and $N_{EW-ZZjj}$ are the number of predicted events from QCD-ZZjj and EW-ZZjj processes, respectively. The $C_{EW-ZZjj}$ factor refers to the detector efficiency for EW-ZZjj processes, calculated as number of selected signal events at detector level ($N^{det.}_{EW-ZZjj}$), divided by number of selected events at particle level in the fiducial phase space ($N^{part.}_{EW-ZZjj}$). The interference between EW-ZZjj and QCD-ZZjj processes is at a few percent level and is ignored.

Both the statistical only case (statistical uncertainty is estimated from expected data yield at 14 TeV with 3000 fb$^{-1}$) and the ones with different sizes of theoretical uncertainties on the background modelling are shown in Figure 3.

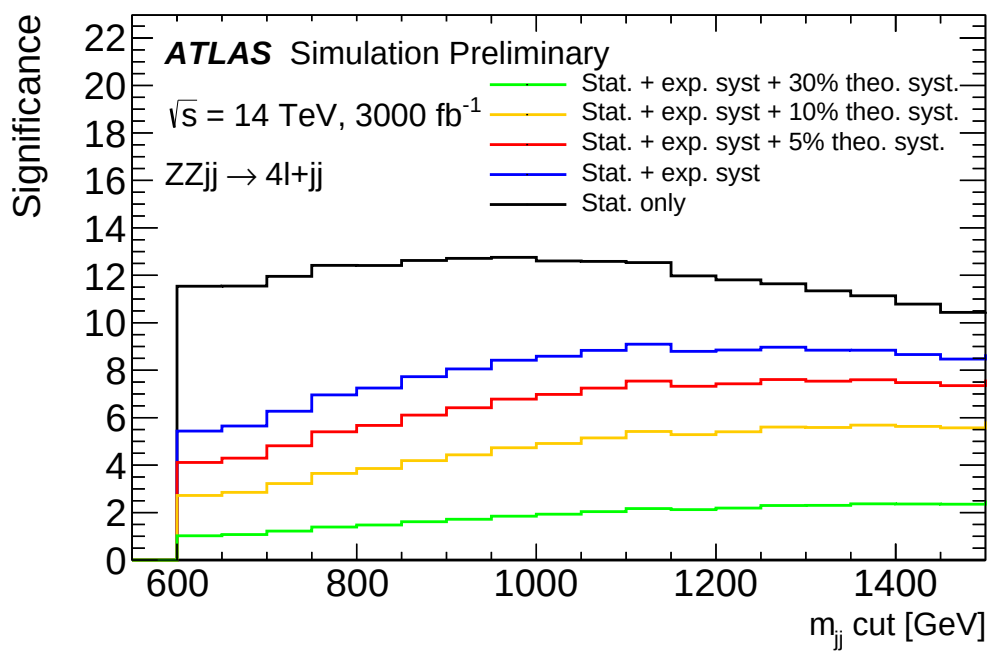

Figure 2: The expected significance of EW-ZZjj processes as a function of different $m_{jj}$ cut for 3000 fb$^{-1}$, under conditions of different sizes of theoretical uncertainties on the QCD-ZZjj background modelling. The statistical uncertainty is estimated from expected data yield at 14 TeV with 3000 fb$^{-1}$. Different uncertainties are summed up quadratically.

Table 2 shows the expected cross-section measurement in the phase space described at Section 4 for 3000 fb$^{-1}$, with the statistical only case, and the cases with different sizes of theoretical uncertainties. The statistical uncertainty is at 10% level and the integrated cross-section measurement becomes dominated by experimental and modelling uncertainty on the QCD-ZZjj background. For the possible extension of the HL-LHC run to 4000 fb$^{-1}$, the statistical uncertainty will be further reduced to 8% level.

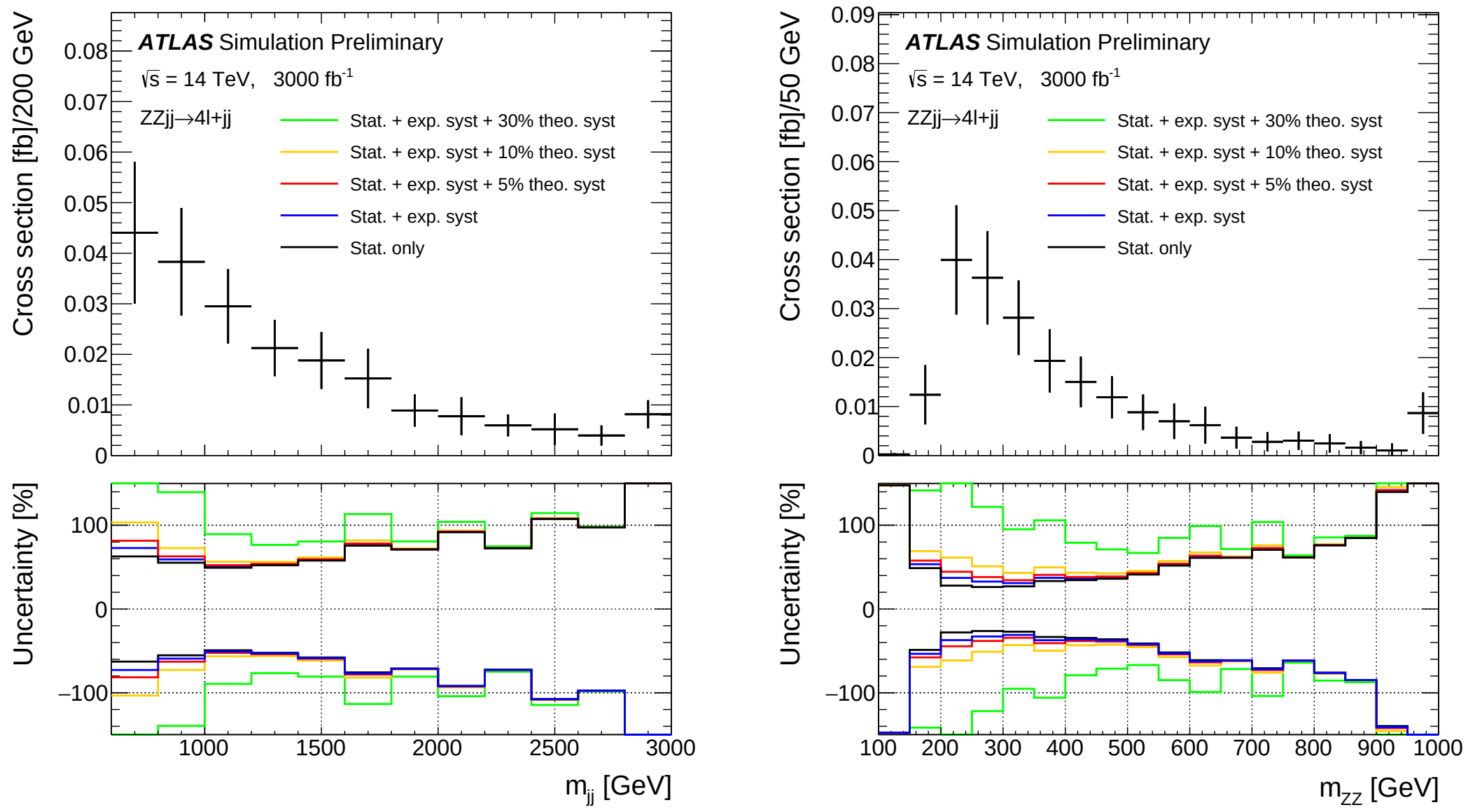

(a) (b)

Figure 3: The projected differential cross-sections at 14 TeV for the EW-ZZjj processes as a function of (a) $m_{jj}$ and (b) $m_{ZZ}$. The top panel shows measurement with statistical only case, where statistical uncertainty is estimated from expected data yield at 14 TeV with 3000 fb$^{-1}$. The bottom panel shows impact of different sizes of systematic uncertainties.

| | Cross section [fb] | Stat. only | Plus exp. | Plus 5% theo. | Plus 10% theo. | Plus 30% theo. |
|---|---|---|---|---|---|---|
| EW-ZZjj | 0.21 | ±0.02 | ±0.04 | ±0.05 | ±0.08 | ±0.21 |

Table 2: Summary of expected cross-section measurements with different theoretical uncertainties. The statistical uncertainty is estimated from expected data yield at 14 TeV with 3000 fb$^{-1}$. Different uncertainties are summed up quadratically.

# 7 Conclusion

The prospect study for the VBS $ZZ$ at the HL-LHC in the four-lepton channel, using 3000 fb$^{-1}$ simulated pp collision at a centre-of-mass energy of 14 TeV has been presented. With a simplified cut-based analysis, the expected significance of the electroweak production has been shown, as well as the precision of the expected measurements of the integrated cross section and differential cross sections as a function of dijet or $ZZ$ invariant mass. Under the assumption of theoretical uncertainty being constraint at 5% level for the QCD-ZZjj processes, the observation of the EW-ZZjj process can be reached with a significance of 7 $\sigma$. This is expected to be improved with multivariate analysis techniques in the future. For the integrated cross-section measurements of EW-ZZjj processes, the precision could reach at 20% level with the assumption of 5% theoretical uncertainty on the QCD-ZZjj processes. In the case of 30% theoretical uncertainty, the precision would be 100%.

8

# CMS Physics Analysis Summary

Contact: cms-phys-conveners-ftr@cern.ch 

# Study of $W^{\pm}W^{\pm}$ production via vector boson scattering at the HL-LHC with the upgraded CMS detector

The CMS Collaboration

## Abstract

The prospects for the study of $W^{\pm}W^{\pm}$jj final states, produced in proton-proton collisions at centre-of-mass energy of 14 TeV via vector boson scattering (VBS), with the upgraded CMS detector at the High-Luminosity LHC (HL-LHC) are presented. The W bosons are detected via their leptonic decays: W$\rightarrow$ $e\nu$ or $\mu\nu$. The results from a study using a full simulation of the upgraded detector along with an average number of 200 proton-proton interactions per bunch crossing are presented in terms of the precision of the cross section measurement as a function of the total integrated luminosity. The significance of the polarized cross section measurement is also discussed.

# 1 Introduction

The ultimate test of the Higgs mechanism in electroweak symmetry breaking lies in the vector boson scattering (VBS) process. In proton-proton collisions at the TeV energy scale at the LHC, the study of the scattering of a pair of weak gauge bosons, qq $\rightarrow$ qqVV, where V = W or Z, can potentially reveal the actual process responsible for the generation of mass of the W and Z bosons [1–3]. In particular, the study of longitudinally polarized vector boson scatterings would clearly indicate the presence of new interactions, if any, in the electroweak symmetry breaking sector.

Same-sign W-pair ($W^\pm W^\pm$) production is among the most promising channels to study this phenomenon, and involves several contributions at the leading order, including quartic gauge couplings due to the non-abelian nature of the weak gauge bosons [4, 5]. Figure 1 shows some of the representative Feynman diagrams for the $W^\pm W^\pm$ VBS process. The Higgs boson-mediated diagram cancels the divergence of the cross section from the other processes, thus restoring unitarity.

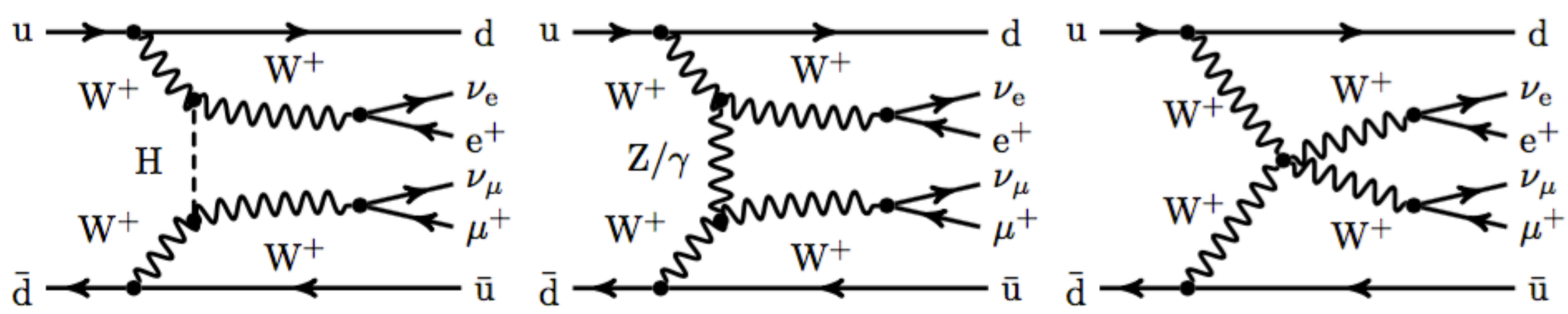

Figure 1: Representative Feynman diagrams for $W^\pm W^\pm$ electroweak production in proton-proton collisions: (left) t-channel Higgs boson exchange, (middle) t-channel $Z/\gamma$ exchange with triple gauge couplings, (right) quartic gauge coupling.

Electroweak $W^\pm W^\pm$ production at the LHC has been already observed by the CMS Collaboration with a significance of 5.5 standard deviations in proton-proton collisions corresponding to an integrated luminosity $\mathcal{L} = 35.9\,\text{fb}^{-1}$ at $\sqrt{s} = 13\,\text{TeV}$ [6]. The ATLAS Collaboration has also recently reported a similar measurement [7]. Only leptonic decays of the W bosons into electron or muon and a neutrino were considered, leading to final states with two same-sign leptons, two jets and missing transverse momentum from the neutrinos.

Since the $W^\pm W^\pm$ scattering process, followed by the leptonic decays of both Ws, has a very low cross section, $\mathcal{O}$(1 fb), any experimental study should strongly benefit from the large data volume to be delivered by the HL-LHC, corresponding to $\mathcal{L}$ up to about $3000\,\text{fb}^{-1}$ per experiment. The upgrades in the Phase 2 CMS detector, such as the tracking detector with extended acceptance, are expected to improve the experimental sensitivity in a complementary way. In particular, as the event topology is characterized by two jets in the forward and backward regions, the planned highly granular calorimeter in the mid-rapidity region should significantly enhance the capability to observe this signal.

This document describes the prospects for the study of VBS with $W^\pm W^\pm$ at $\sqrt{s} = 14\,\text{TeV}$ at the HL-LHC with the Phase 2 upgraded CMS detector. Results are presented for a range of $\mathcal{L}$, from $300\,\text{fb}^{-1}$ through $6000\,\text{fb}^{-1}$, where the first value corresponds to one year of proton-proton collision data, and the latter one to 10 years of combined data sets collected by the ATLAS and CMS experiments.

The data-taking conditions at the HL-LHC correspond to instantaneous luminosity of about 5–

$7.5\times10^{34}$ cm$^{-2}$ s$^{-1}$ and up to about 200 proton-proton interactions (pileup) per bunch crossing, on average. The $W^\pm W^\pm$ events are selected in the final state containing a same-sign lepton pair (ee/e$\mu$/$\mu\mu$) accompanied by a pair of jets consistent with the VBS process. We also explore the prospect of observing the longitudinally polarized component in W boson scattering.

## 2 The CMS detector and event simulation

The CMS detector [8] will be substantially upgraded in order to fully exploit the physics potential offered by the increase in luminosity at the HL-LHC [9], and to cope with the demanding operational conditions at the HL-LHC [10–14]. The upgrade of the first level hardware trigger (L1) will allow for an increase of L1 rate and latency to about 750 kHz and 12.5 $\mu$s, respectively; the high-level software trigger is expected to reduce the rate by about a factor of 100 to 7.5 kHz. The entire pixel and strip tracker detectors will be replaced to increase the granularity, reduce the material budget in the tracking volume, improve the radiation hardness, and extend the geometrical coverage and provide efficient tracking up to pseudorapidities of about $|\eta| = 4$. The muon system will be enhanced by upgrading the electronics of the existing cathode strip chambers, resistive plate chambers and drift tubes. New muon detectors based on improved resistive plate chamber and gas electron multiplier technologies will be installed to add redundancy, increase the geometrical coverage up to about $|\eta| = 2.8$, and improve the trigger and reconstruction performance in the forward region. The barrel electromagnetic calorimeter ($|\eta| < 1.44$) will feature upgraded front-end electronics that will be able to exploit the information from single crystals at the L1 trigger level, to accommodate trigger latency and bandwidth requirements, and to provide 160 MHz sampling, allowing high precision timing capability for photons. The hadronic calorimeter, consisting in the barrel region of brass absorber plates and plastic scintillator layers, will be read out by silicon photomultipliers. The endcap ($1.57 < |\eta| < 3.0$) electromagnetic and hadron calorimeters will be replaced with a new combined sampling calorimeter that will provide highly-segmented spatial information in both transverse and longitudinal directions, as well as high-precision timing information. Finally, the addition of a new timing detector for minimum ionizing particles in both barrel and endcap regions is envisaged to provide capability for 4-dimensional reconstruction of interaction vertices, which will help mitigate the performance degradation of the CMS experiment due to high pileup rates.

A detailed overview of the CMS detector upgrade program is presented in Refs. [10–14], while the expected performance of the reconstruction algorithms and the strategy to tackle the pileup issue is summarised in Ref. [15].

Signal samples for electroweak production, referred to as EWK $W^\pm W^\pm$, as well as the dominant background process, $W^\pm W^\pm$ + 2 jets produced through strong interactions, referred to as QCD $W^\pm W^\pm$, are generated using MADGRAPH v5.4.2 [16] and the leading order version of the parton distribution function set NNPDF v3.0 [17] with strong coupling constant $\alpha_s(m_Z) = 0.13$ and the four-flavor scheme. The information about the polarization of the individual W bosons in the signal process is extracted by generating a separate set of events using the v1.5.14 of the DECAY package of MADGRAPH. The other background processes considered in this analysis are $t\bar{t}$ + jets, single top, and single anti-top, which are simulated with POWHEG [18–21]; inclusive Drell–Yan, W$\gamma$, ZZ and WZ, which are simulated with MADGRAPH; triboson processes, including WW$\gamma$, WZ$\gamma$, WWW, WWZ, WZZ and ZZZ, which are simulated with MADGRAPH5_aMC@NLO [22, 23]; and QCD multijet production, which is simulated with PYTHIA v8.212 [24]. The PYTHIA package is used for parton showering, hadronization, and the underlying event simulation, with the parameter set of the CUETP9M1 tune [25] for all simulated samples. The generated events use a fully simulated description of the Phase 2 CMS detector, implemented using the GEANT4 package [26].

# 3 Physics object reconstruction and event selection

The physics objects are reconstructed with algorithms developed for the Phase 2 upgraded CMS detector proposed for the HL-LHC. The event selection strategy is similar to the recently published analysis using 13 TeV data [6].

Selected electrons or muons are required to pass identification criteria and have $p_{\mathrm{T}} > 20\,\mathrm{GeV}$ and $|\eta| \leq 3.0$. The relative isolation, which is defined as the ratio of the of the isolation variable to the $p_{\mathrm{T}}$ of the lepton candidate, is required to be less than 0.15 (0.21) for the barrel (endcap) region for electron candidates, and less than 0.15 for muon candidates. The isolation is defined as the sum of the $p_{\mathrm{T}}$ of tracks in a cone of radius $\Delta R = \sqrt{\Delta\eta^2 + \Delta\phi^2} = 0.3$ around the direction of the lepton candidate, and is corrected for the pile-up contribution. In addition, for electrons, no missing hits in the tracker subsystem are allowed and any electron overlapping with an isolated photon candidate within $\Delta R < 0.2$ is not considered as an electron candidate. The latter requirement significantly reduces the contribution from $\mathrm{W}\gamma$ events where the photon is misreconstructed as an electron.

Jets are reconstructed using FASTJET [27] with the anti-$k_{\mathrm{T}}$ clustering algorithm [28], using a distance parameter of 0.4. The pileup per particle identification [29] algorithm (PUPPI) is used to remove the contributions due to pileup from the resulting jets. The events are required to have at least two reconstructed jets with $p_{\mathrm{T}} > 50$ GeV and $|\eta| < 4.7$. The jets are not considered if they overlap with any of the isolated lepton or photon candidates, within $\Delta R = 0.4$. In the region $|\eta| > 3$ the probability for PU jets to pass the PUPPI requirements increases significantly. We expect this effect to be corrected at the HL-LHC. To keep the probability flat as a function of $\eta$ at a level of about 20%, only 20 randomly selected reconstructed jets in the region $|\eta| > 3$, out of 100, that do not match the generated jets originating from the hard interaction are accepted.

The shape of the invariant mass $m_{\mathrm{jj}}$ distribution of the two $p_{\mathrm{T}}$-leading jets is shown in Fig. 2 (left). Compared to the background processes, the EWK $\mathrm{W}^{\pm}\mathrm{W}^{\pm}$ signal distribution is harder, as expected. In all plots the symbol V corresponds to either a W or a Z boson, while "others" includes $\mathrm{t\bar{t}V}$, tV, tW, $\mathrm{\bar{t}W}$, tribosons, Drell–Yan and W + jets processes. The characteristic topology of the VBS process is even more evident in the distribution of the absolute value of the difference in pseudorapidity of the two leading jets, $|\Delta\eta_{\mathrm{jj}}|$, as presented in Fig. 2 (right). To select VBS-type events, the jets are required to satisfy $m_{\mathrm{jj}} \geq 500\,\mathrm{GeV}$ and $|\Delta\eta_{\mathrm{jj}}| \geq 2.5$.

The selected jets in the signal process are expected to originate from light quarks. In order to suppress background contribution from $\mathrm{t\bar{t}}$ production, events are rejected if there is a b tagged jet in the event with $|\eta| < 2.4$. The b tagging of a jet is performed with the Deep Combined Secondary Vertex discriminator [30], which based on a deep neural network [31].

The selected events are required to have exactly two isolated lepton candidates of the same charge. Any event with additional identified and isolated lepton candidates with $p_{\mathrm{T}} > 10\,\mathrm{GeV}$ is rejected. The invariant mass $m_{\ell\ell}$ of the two leptons is required to be above 20 GeV to avoid potential contributions from low-mass resonances. Furthermore, to reduce the background from $\mathrm{Z} \rightarrow \mathrm{ee}$ decays, where one of the electron charges is misidentified, the events with di-electron mass within 15 GeV from the nominal Z boson mass, 91.2 GeV, are excluded. Given that the charge misidentification probability for a muon is low, the estimated background from $\mathrm{Z} \rightarrow \mu\mu$ is negligible, and no dimuon invariant mass constraint is imposed.

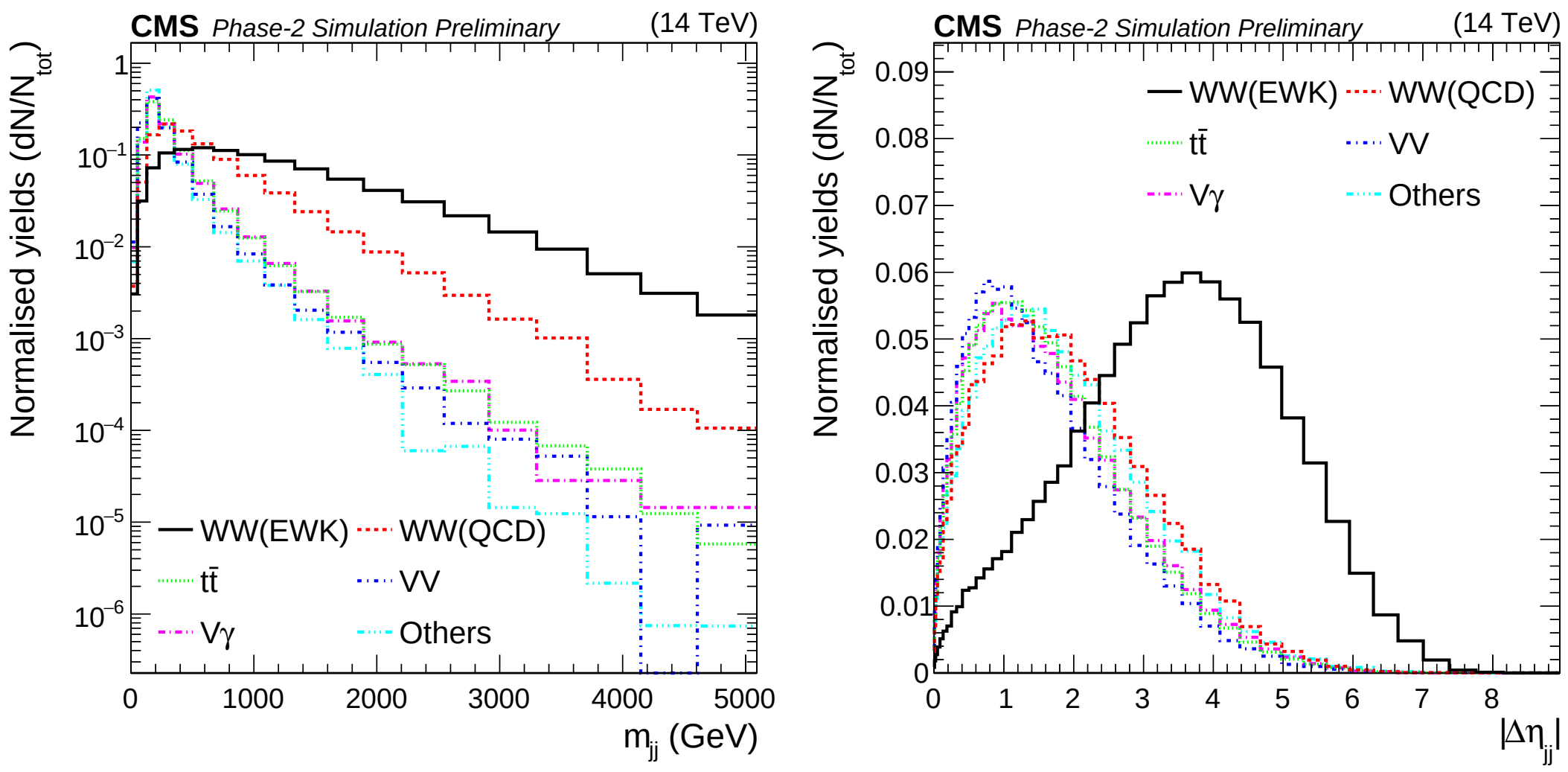

Figure 2: Shape comparisons for signal and background processes. Left: Invariant mass of the two leading jets. Right: The difference in pseudorapidity between them.

Since leptons in the EWK $W^\pm W^\pm$ process are expected to be located in the central region defined by the forward-backward jets, non-VBS background can be suppressed with the Zeppenfeld variable [32]. For a given lepton with pseudorapidity $\eta_\ell$, it is defined as

$$Z_\ell = \frac{[\eta_\ell - 0.5(\eta_1 + \eta_2)]}{|(\eta_1 - \eta_2)|},$$

where $\eta_1$ and $\eta_2$ refer to the pseudorapidities of the leading and subleading jets. The distribution of this variable is shown in Fig. 3 (left). The maximum value of this variable, $Z_{\mathrm{MAX}}$, for either of the leptons is required to be less than 0.75.

The missing transverse momentum for the signal events is presented in Fig. 3 (right) along with the distributions for the background processes. We further suppress background by requiring events to have missing transverse momentum above 40 GeV.

# 4 Results

## 4.1 Uncertainty in the VBS cross section measurement

The expected event yields are summarized in Table 1. The $m_{\mathrm{jj}}$ distribution after the full event selection for $\mathcal{L} = 3000\,\mathrm{fb}^{-1}$ is presented in Fig. 4 (left). At this stage the main background consists of inclusive $\mathrm{t\bar{t}}$ as well as WZ processes where the third lepton in the event was not reconstructed within the detector acceptance.

Since the current analysis is based on simulated events generated according to their corresponding standard model cross sections, the expected yields can only be used to estimate the uncertainty in the expected cross section measurement as a function of integrated luminosity. It is done by fitting the $m_{\mathrm{jj}}$ distribution using a binned maximum likelihood approach with all systematic uncertainties in the form of nuisance parameters with log-normal distributions. The correlations among different sources of uncertainties are taken into account while the different final states, ee, e$\mu$ and $\mu\mu$, are considered as independent channels in the fit.

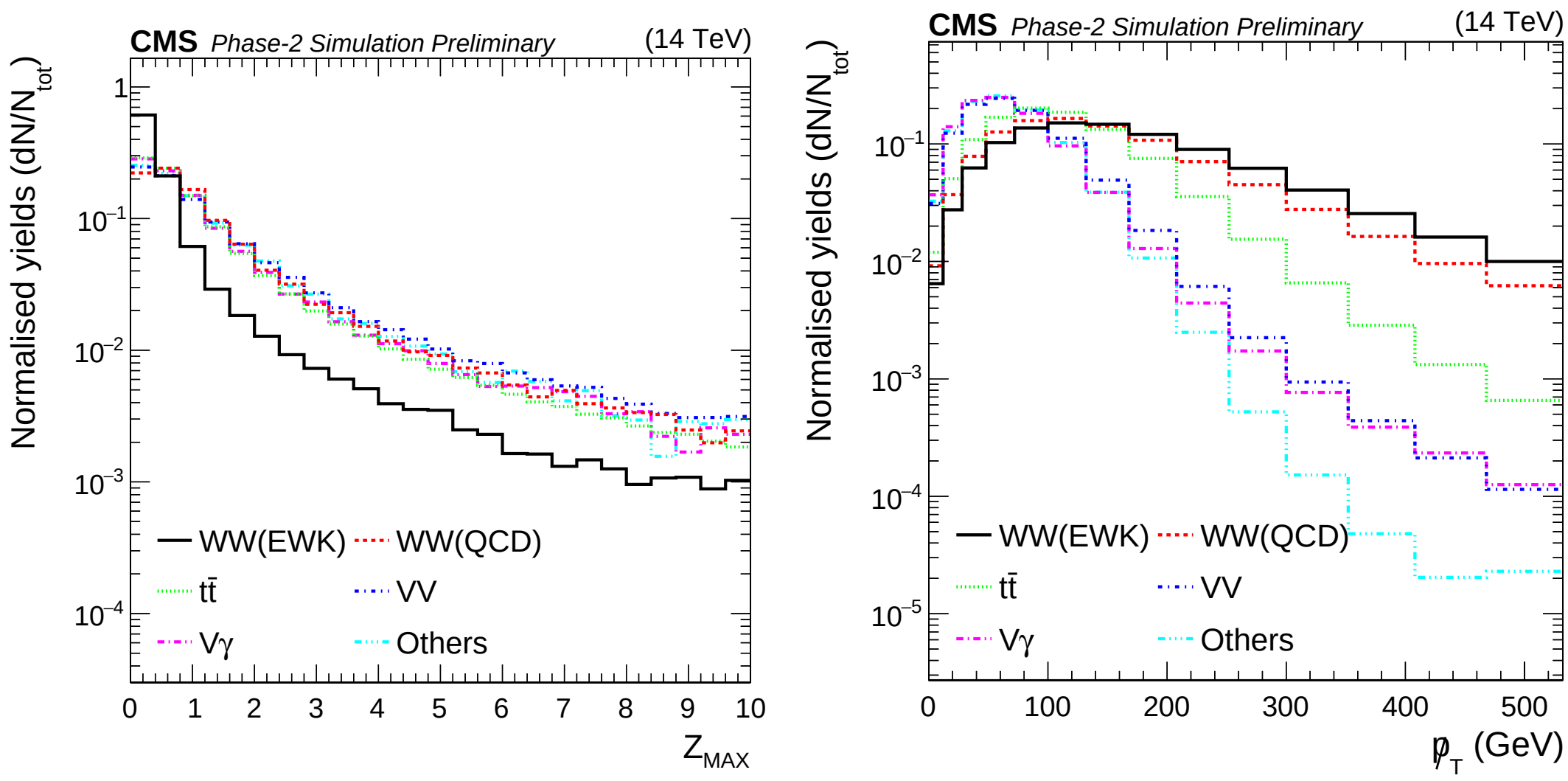

Figure 3: Shape comparisons for the signal and background processes. Left: The maximum of the Zeppenfeld variable for leptons. Right: Missing transverse momentum.

Table 1: Expected yields for signal and background contributions for $\mathcal{L} = 3000\,\text{fb}^{-1}$.

| Process | Expected yield, $\mathcal{L} = 3000\,\text{fb}^{-1}$ |
|---|---:|
| $W^{\pm}W^{\pm}$ EWK | 5368 |
| $t\bar{t}$ | 5515 |
| WZ | 1421 |
| W$\gamma$ | 406 |
| $W^{\pm}W^{\pm}$ QCD | 196 |
| Total background | 7538 |

The major sources of uncertainty considered in this analysis are referred to as the "YR18 systematic uncertainty scenario". Here theoretical uncertainties are reduced by a factor of two compared to the current situation, while experimental components are scaled down with the square root of the integrated luminosity until they reach a defined minimum value based on estimates of the achievable accuracy with the upgraded detector. The impact of the uncertainties on the signal strength, defined as the ratio of the observed cross section to the expected, are summarized in Table 2 for $\mathcal{L}$ = 300 and 3000 $fb^{-1}$, corresponding to 1 and 10 years of HL-LHC operation, respectively. The total uncertainty is presented in Fig. 4 (right) as a function of the integrated luminosity. The values of $\mathcal{L}$ exceeding 3000 $fb^{-1}$ are considered in the case where the measurements from CMS and ATLAS will be combined, effectively doubling the total integrated luminosity.

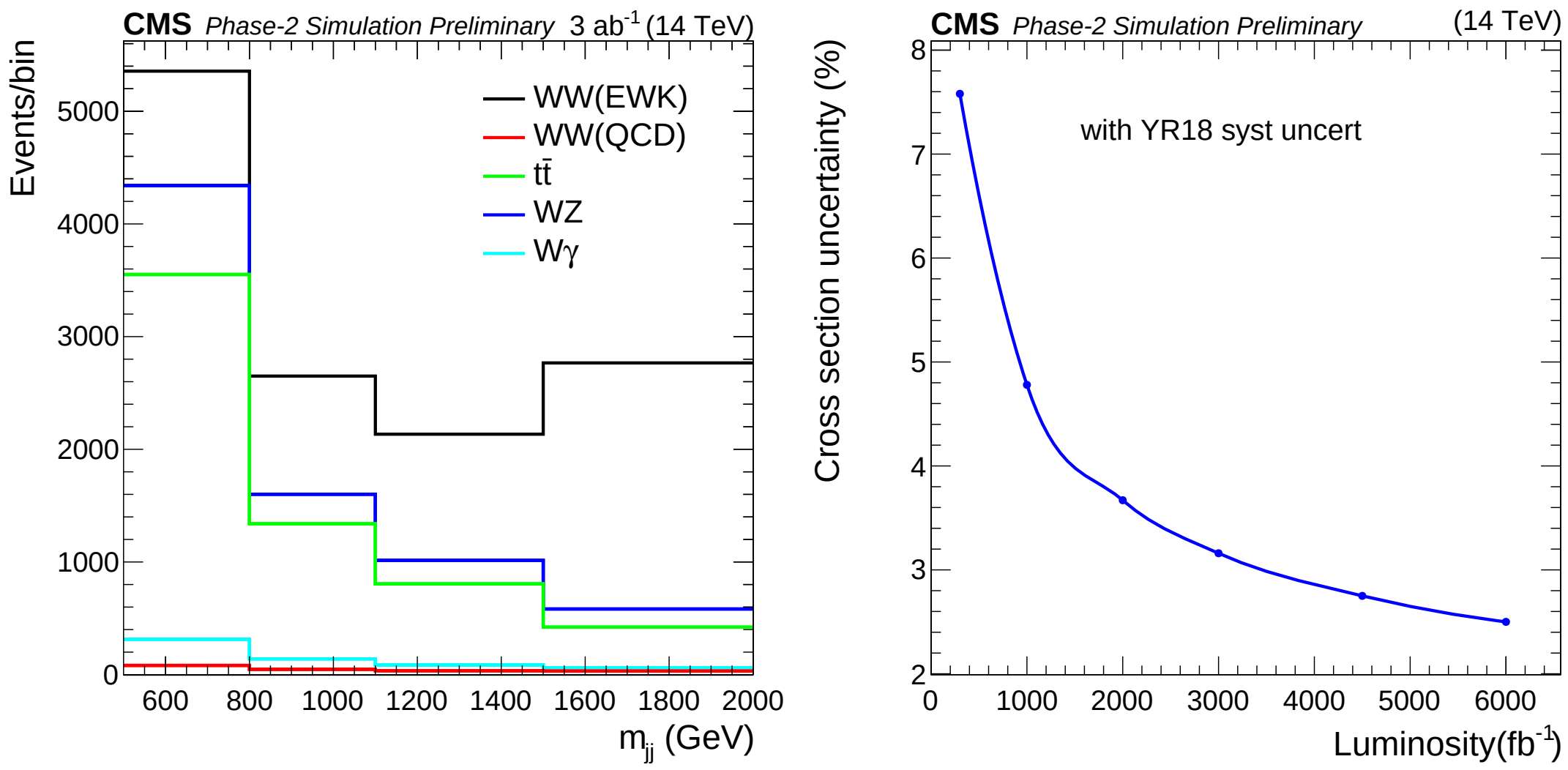

Figure 4: Left: The distribution of the invariant mass of the two leading jets after the selection requirements for an integrated luminosity of 3000 $fb^{-1}$. Right: The estimated uncertainty in the EWK $W^{\pm}W^{\pm}$ cross section measurement as a function of the integrated luminosity (with YR18 systematic uncertainties).

The input value for each source of the systematic uncertainty in Table 2 corresponds to the expected achievable precision at the HL-LHC; its effect on the signal strength is studied by varying the parameter of interest, keeping all others fixed. The uncertainty due to the b tag misidentification probability, the probability for a non-b jet to pass the b tag requirements, has the biggest impact, since it directly affects the number of selected signal events. Conversely, the b tag efficiency has very small impact, since the top contribution is constrained in the fit. It may be noted that the total experimental uncertainty decreases by more than a factor of two when moving from 300 to 3000 $fb^{-1}$. While none of the considered components of systematics has a statistical component, this arises from the better constraints on the nuisances from the fit with higher event yields.

### 4.2 Measurement of the longitudinally polarized $W^{\pm}W^{\pm}$ scattering

The total $W^{\pm}W^{\pm}$ VBS cross section can be decomposed into the polarized components based on the decays of the individual W bosons. Either or both can be longitudinally (L) or transversely (T) polarized, giving rise to final states of LL, TT and the mixed state of LT. The LL component is expected to be only about 6–7% of the total VBS cross section for jet $p_T > 50$ GeV. However, the shape of different kinematic variables can be used to extract the LL component of the total EWK $W^{\pm}W^{\pm}$ cross section.

Table 2: The systematic uncertainties considered in this analysis and their impact on the signal strength for two different integrated luminosities. For comparison, the expected statistical uncertainty is shown in the first row.

| Source of uncertainty | Input | 300 fb$^{-1}$ (1 year) | 3000 fb$^{-1}$ (10 years) |
|---|---|---|---|
| Statistical uncertainty | | 5.7% | 1.8% |
| Trigger efficiency (electron) | 1.0% | 0.5% | 0.2% |
| Trigger efficiency (muon) | 1.0% | 1.1% | 0.6% |
| Electron id + iso. efficiency | 1.0% | 0.6% | 0.3% |
| Muon id + iso. efficiency | 0.5% | 0.9% | 0.6% |
| Jet energy scale | 0.5–3.7% | 1.0% | 0.4% |
| b tag (stat. component) | 1.0% | 0.2% | 0.3% |
| b tag misidentification | 1–2% | 1.4% | 1.2% |
| Misidentified lepton from $t\bar{t}$ | 5–20% | 3.5% | 1.0% |
| Misidentified lepton from W$\gamma$ | 20% | 0.3% | 0.1% |
| Stat. accuracy of W$\gamma$ sample | 30% | 0.4% | 0.1% |
| Total (stat + experimental syst) | | 7.6% | 3.2% |
| Luminosity | 1.0% | 1.0% | 1.0% |
| Theoretical/QCD scale | 3.0% | 3.0% | 3.0% |
| Total (stat + syst + lumi + theory) | | 8.2% | 4.5% |

The difference in azimuthal angle between the two leading jets, $\Delta\phi_{jj}$, has the potential for discriminating the LL component of the VBS scattering from TT and LT contributions. Since the signal-to-background separation for the EWK $W^{\pm}W^{\pm}$ process improves with increasing $m_{jj}$ as shown in Fig. 4 (left), the $\Delta\phi_{jj}$ distributions are studied in two ranges of $m_{jj}$: for 500–1100 GeV and above 1100 GeV, as shown in Fig. 5. The plots present the relative contributions of different polarized states to the total EWK $W^{\pm}W^{\pm}$ cross section.

It is evident that the relative contribution of the LL component increases with $\Delta\phi_{jj}$. Figure 6 shows the combination of signal and background yields as a function of $\Delta\phi_{jj}$ in the two $m_{jj}$ regions. The signal, when both W bosons are longitudinally polarized (LL), is visible at high values of $\Delta\phi_{jj}$ above the contributions from transverse polarizations (LT and TT) and the dominant backgrounds.

Using these distributions and the same procedure as for the VBS cross section measurement, the significance for the observation of the LL process is estimated as a function of integrated luminosity. The significance is found to be 1.2 and 2.7 standard deviations for $\mathcal{L} = 300$ and 3000 fb$^{-1}$ respectively. The gradual improvement of signal significance as a function of integrated luminosity is shown in Fig. 7.

## 5 Summary

The prospects for the study of the $W^{\pm}W^{\pm}jj$ final states produced via vector boson scattering (VBS) in pp collisions at the HL-LHC have been presented. The signal and background events were generated with a full simulation of the response of the Phase 2 upgraded CMS detector. The W bosons are detected via their leptonic decays into $e\nu$ or $\mu\nu$. It is shown that the total ex-perimental uncertainty in the VBS $W^{\pm}W^{\pm}$ cross section measurement decreases by more than a factor of two when moving from a total integrated luminosity of 300 to 3000 $fb^{-1}$, down to about 3%, and can be decreased even further if the results from CMS and ATLAS experiments are combined. The analysis demonstrates the sensitivity for measuring the longitudinally polarized component of the $W^{\pm}W^{\pm}$ VBS production. The expected significance for an integrated luminosity of 3000 $fb^{-1}$ is estimated to be 2.7 standard deviations, and can exceed a value of 3 for a combination of the CMS and ATLAS measurements.

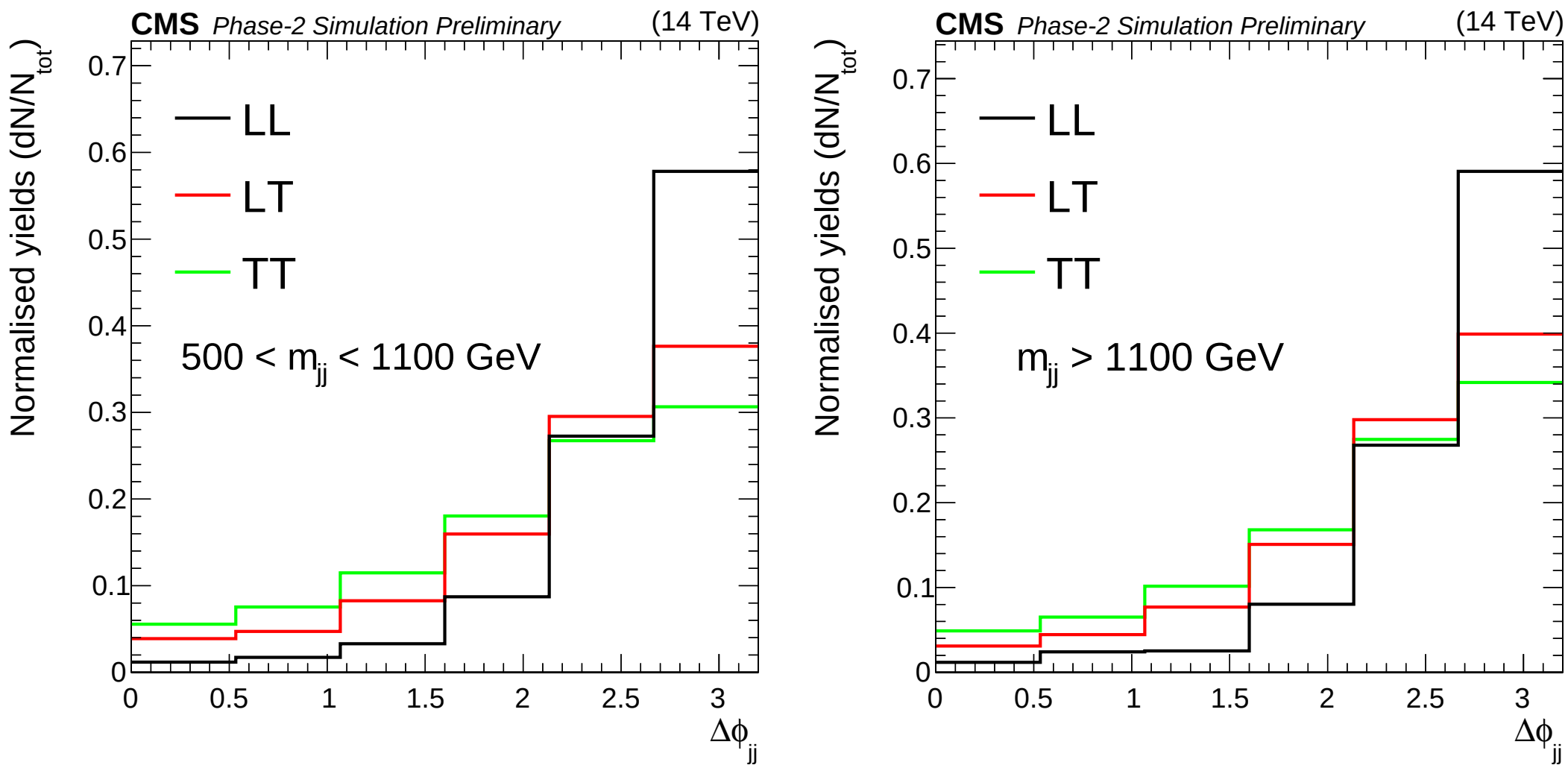

Figure 5: The shape comparison of the LL, LT and TT components in the distribution of the azimuthal angle difference between the two leading jets for the VBS $W^{\pm}W^{\pm}$ process for dijet invariant mass between 500 to 1100 GeV (left), and above 1100 GeV (right).

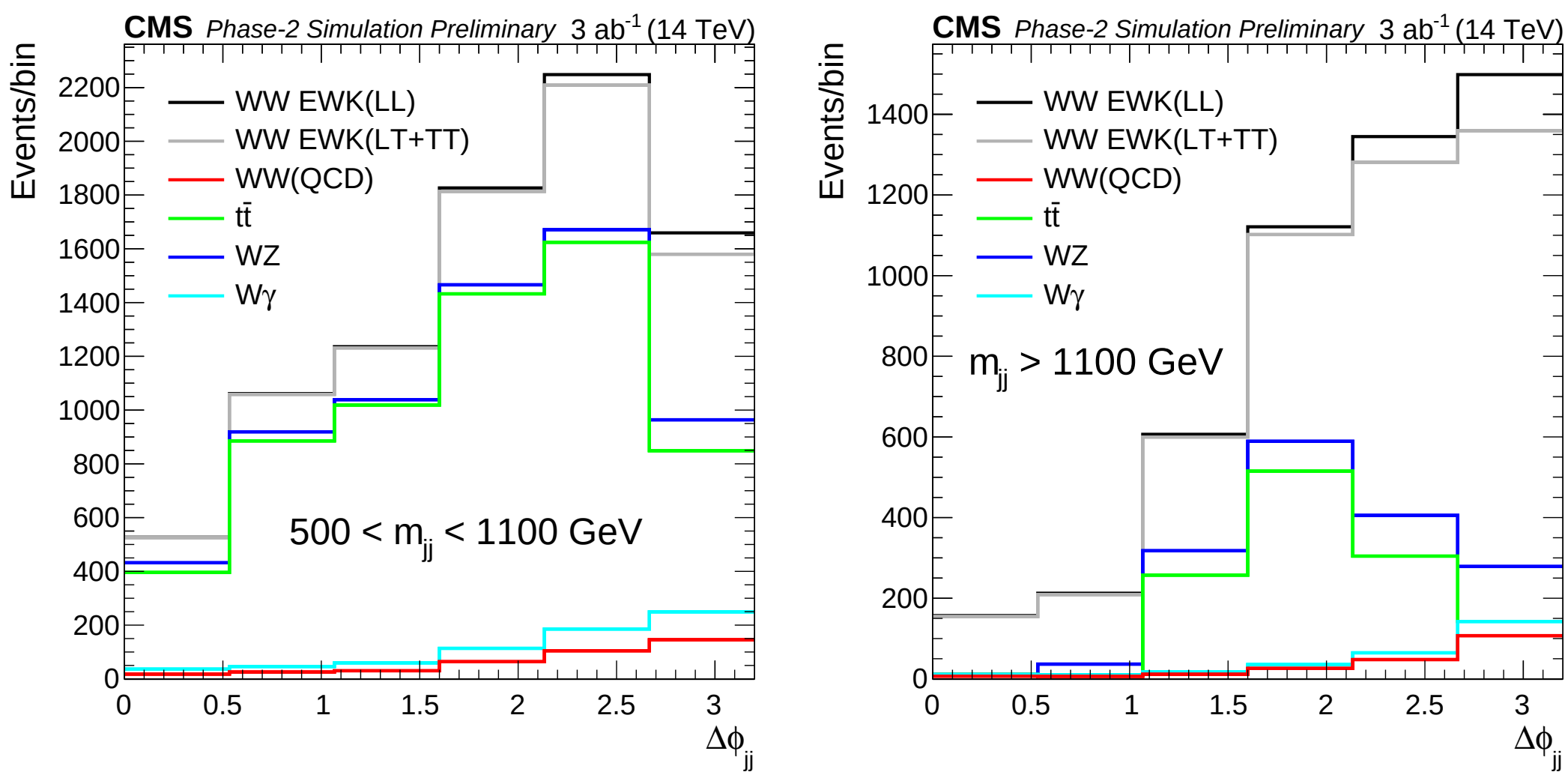

Figure 6: Distributions of the azimuthal angle difference between the two leading jets for dijet invariant mass in the range 500–1100 GeV (left) and above 1100 GeV (right). Stacked contributions from the signal and various backgrounds are shown.

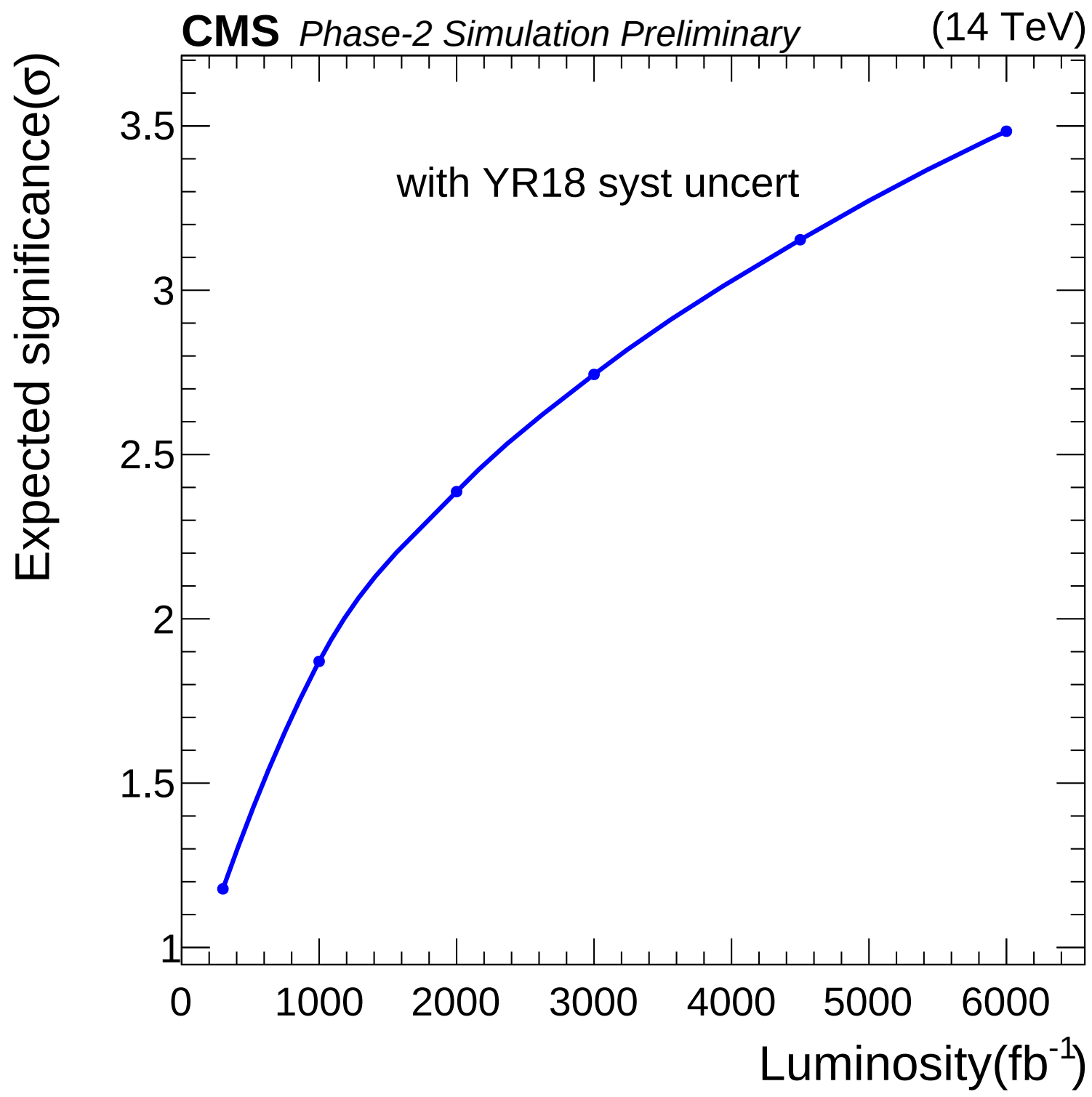

Figure 7: Significance of the observation of the scattering of a pair of longitudinally polarized W bosons as a function of the integrated luminosity (with YR18 systematic uncertainties).

**10**

# Prospects for the measurement of the $W^{\pm}W^{\pm}$ scattering cross section and extraction of the longitudinal scattering component in $pp$ collisions at the High-Luminosity LHC with the ATLAS Experiment

The ATLAS Collaboration

Prospects for measuring the $W^{\pm}W^{\pm}jj$ vector boson scattering process with 3000 fb$^{-1}$of proton-proton collisions at $\sqrt{s}$ = 14 TeV at the High-Luminosity Large Hadron Collider are studied. Events containing two same-sign leptons, missing transverse momentum, and at least two jets are analysed using simulated events parameterised to take into account the expected detector effects, and the typically data-driven backgrounds arising from charge misidentification and jets faking leptons are also included using the parameterisation functions. The cross section for the electroweak production process is extracted from a fit to the dijet invariant mass distribution, and an expected total uncertainty of 6% is achieved. We find the purely longitudinal scattering component can be extracted with an expected significance of 1.8$\sigma$ from a binned likelihood fit to the dijet azimuthal separation distribution.

# 1 Introduction

The study of the scattering of two massive vector bosons $V = W, Z$ (vector boson scattering, VBS) provides a key opportunity to probe the nature of the electroweak symmetry breaking (EWSB) mechanism as well as physics beyond the Standard Model (SM) [1, 2]. It is still unknown whether the discovered Higgs boson [3, 4] preserves unitarity of the longitudinal $VV$ scattering amplitude at all energies, or if other new physics processes are involved [5–9]. In the VBS topology, two incoming quarks radiate bosons which interact, yielding a final state of two jets from the outgoing quarks, and two massive bosons which decay into fermions. This final state can be the result of $VVjj$ electroweak (EW) production with and without a scattering topology, or of processes involving the strong interaction, shown in the Feynman diagrams in Figures 1, 2, and 3, respectively.

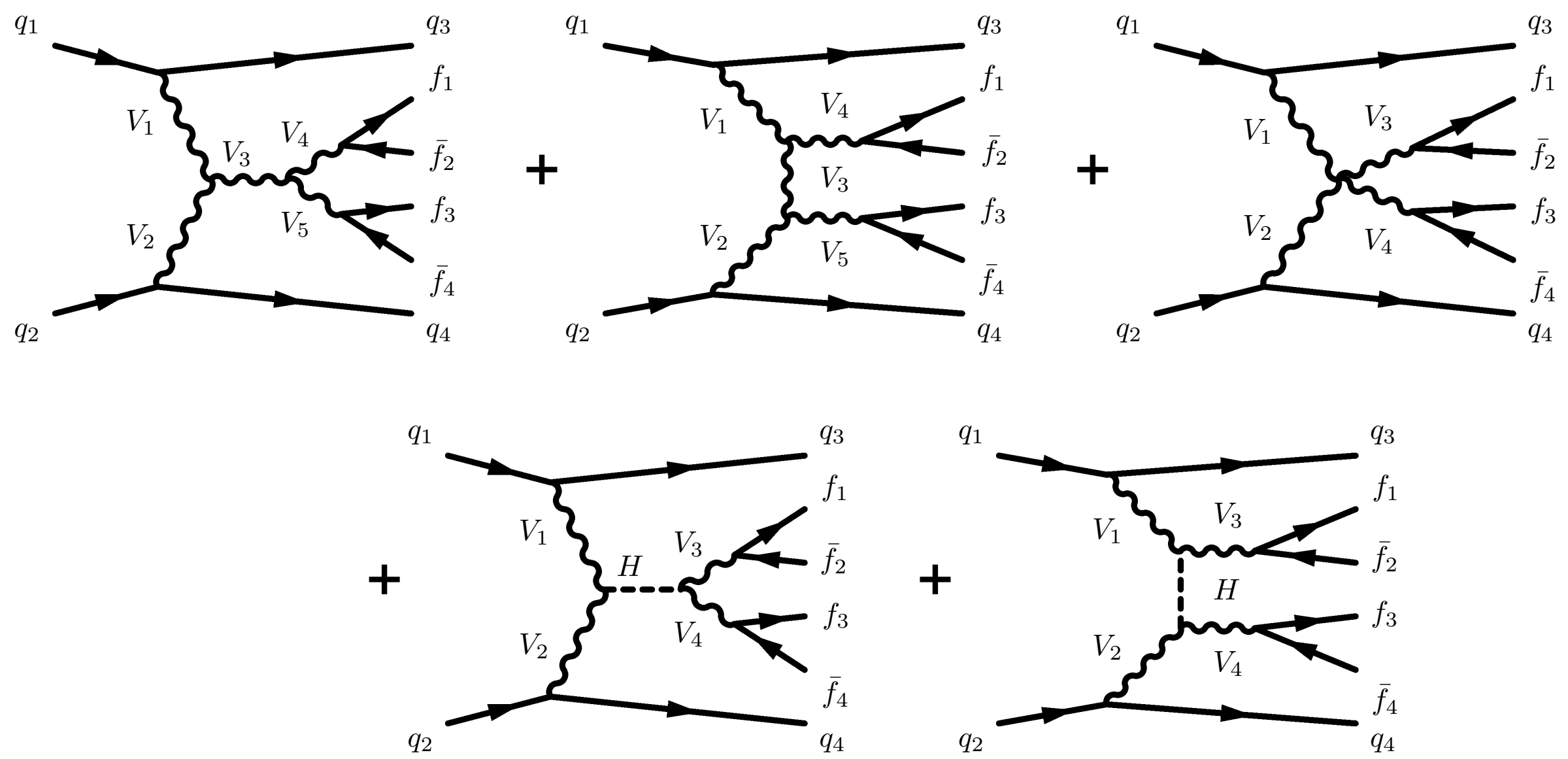

Figure 1: Representative Feynman diagrams for $VVjj$ EW production with a scattering topology (VBS) including either a triple gauge boson vertex with production of a boson in the $s$-channel (top left diagram), the $t$-channel exchange (top middle diagram), quartic gauge boson vertex (top right diagram), or the exchange of a Higgs boson in the $s$-channel (bottom left diagram) and $t$-channel (bottom right diagram). The lines are labeled by quarks ($q$), vector bosons ($V$), and fermions ($f$). Particles with different indices may or may not have the same flavour.

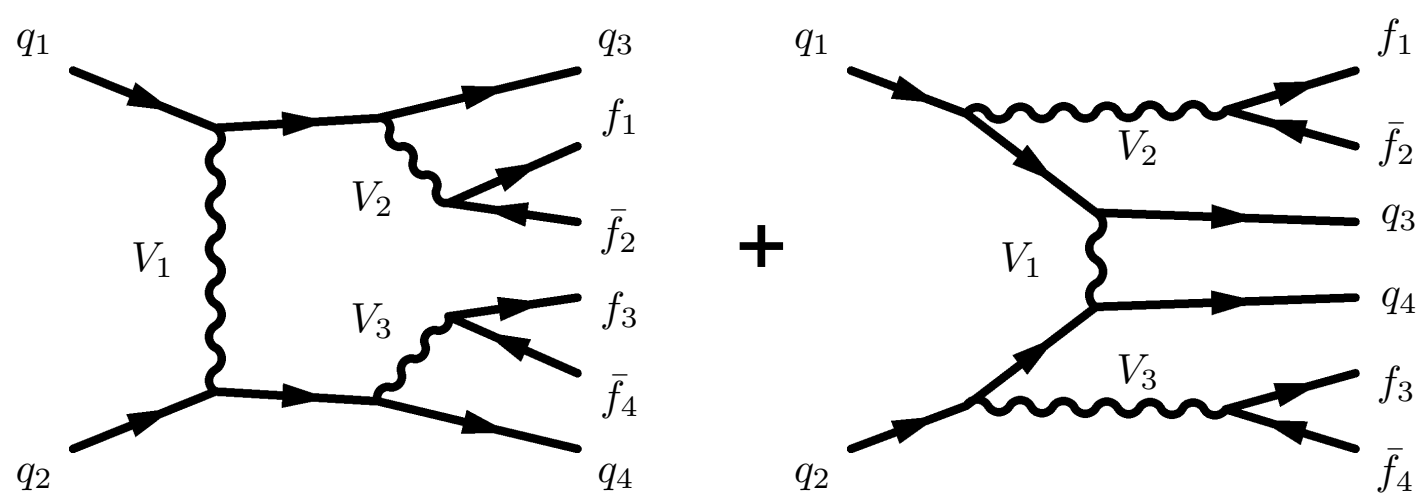

Figure 2: Representative Feynman diagrams for $VVjj$ EW production excluding the vector-boson scattering topology (non-VBS). The lines are labeled by quarks ($q$), vector bosons ($V$), and fermions ($f$). Particles with different indices may or may not have the same flavour.

With the largest cross-section ratio of electroweak to strong production [10, 11], events with $W^{\pm}W^{\pm}$ plus two jets ($W^{\pm}W^{\pm}jj$) provide one of the best opportunities to study the scattering of two vector bosons. The $W^{\pm}W^{\pm}jj$ VBS processes (Figure 1) include the contributions from triple gauge vertices in the $t$-channel, quartic gauge vertices, and the $t$-channel Higgs-mediated diagram. The non-VBS EW processes with the same final state (Figure 2) also contribute to the signal, but the signal region contributions can be suppressed through kinematic selections used in the analysis. Interactions involving at least two strong couplings ($W^{\pm}W^{\pm}jj$ QCD) are considered as background (Figure 3). We do not consider interference between the EW and QCD processes in this note. ATLAS and CMS have both observed the EW process at 13 TeV with significances 6.9 $\sigma$ and 5.5 $\sigma$, respectively [12, 13]. The focus for the High Luminosity LHC (HL-LHC) physics program will be the measurement of the scattering cross section of two longitudinally polarised $W$ bosons.

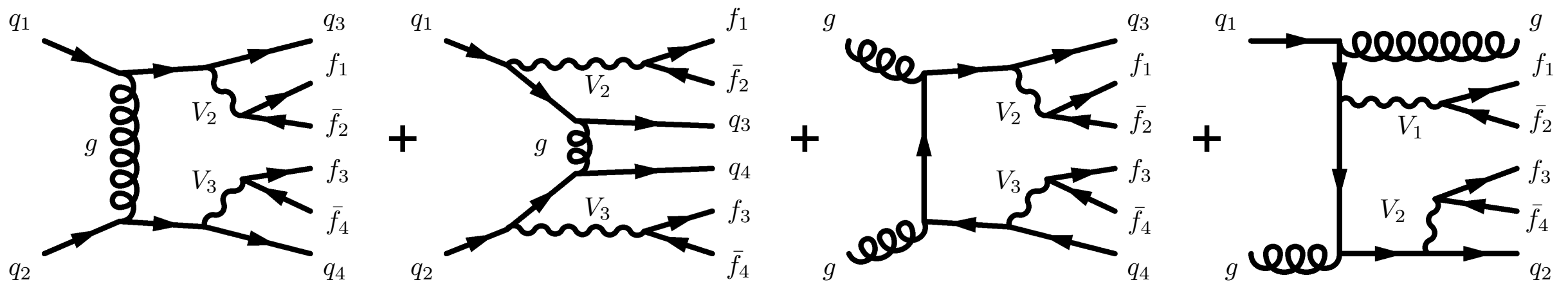

Figure 3: Representative Feynman diagrams for $VVjj$ QCD production. The lines are labeled by quarks ($q$), vector bosons ($V$), fermions ($f$), and gluons ($g$). Particles with different indices may or may not have the same flavour.

The ATLAS detector [14, 15] is a multi-purpose particle detector with a cylindrical geometry.[1] It consists of layers of inner tracking detectors surrounded by a superconducting solenoid, calorimeters, and a muon spectrometer, and will need several upgrades to cope with the expected higher luminosity at the HL-LHC and its associated high pileup and intense radiation environment. The primary motivation for the upgrade design studies is to maximise the potential of the experiment for searches and measurements despite these harsh conditions. A new inner tracking system, extending the tracking region from $|\eta| \leq 2.7$ up to $|\eta| \leq 4.0$, will provide the ability to reconstruct forward charged particle tracks, which can be matched to calorimeter clusters for forward electron reconstruction, or associated to forward jets. The inner tracker extension also enables muon identification at high pseudorapidities if additional detectors (such as micro-pattern gaseous or silicon pixel detectors) are installed between the endcap calorimeters and the New Small Wheel [16] in the region $2.7 < |\eta| \leq 4.0$. Despite being in an area without magnetic field, such detectors would increase the muon spectrometer acceptance and could be used to identify inner detector tracks in the forward region as muons, while relying entirely on the inner tracker for the momentum measurement. The impact of the addition of the High Granularity Timing Detector (HGTD) [17] is not considered.

In this note we study the $W^{\pm}W^{\pm}jj$ EW production in the context of these planned upgrades, presenting the prospects for the measurement of the inclusive EW, and purely longitudinal scattering cross sections. With a more realistic background and systematics estimation, this result supersedes the prediction of the inclusive $W^{\pm}W^{\pm}jj$ EW significance and measurement precision presented in [18].

[1] The ATLAS experiment uses a right-handed coordinate system with its origin at the nominal $pp$ interaction point at the centre of the detector. The positive $x$-axis is defined by the direction from the interaction point towards the centre of the LHC ring, with the positive $y$-axis pointing upwards, while the beam direction is along the $z$-axis. Cylindrical coordinates $(r, \phi)$ are used in the transverse $(x, y)$ plane, $\phi$ being the azimuthal angle around the beam direction. The pseudorapidity is defined in terms of the polar angle $\theta$ from the $z$-axis as $\eta = -\ln[\tan(\theta/2)]$. The distance in $\eta$–$\phi$ space between two objects is defined as $\Delta R \equiv \sqrt{(\Delta\eta)^2 + (\Delta\phi)^2}$. Transverse energy is computed as $E_{\mathrm{T}} = E \cdot \sin\theta$.

## 2 Monte Carlo Samples

Various processes aside from QCD and EW $W^{\pm}W^{\pm}jj$ contribute to producing an experimental signature of two prompt leptons with the same electric charge, two jets, and missing transverse momentum in the final state. The dominant prompt background process is $WZ$+jets production where both bosons decay leptonically and one lepton fails identification or falls outside of the detector acceptance. Other processes with two prompt leptons with the same electric charge in the final state include the $t\bar{t}V$ process, $ZZ$+jets production, and multiple parton interactions. These other processes contribute only a few percent to the total background. The non-prompt-lepton background arises from processes with one or two jets misreconstructed as leptons, or leptons from hadron decays (including $b$- and $c$-hadron decays), such as $W$+jets and top quark production. Finally, processes such as $t\bar{t}$ and Drell-Yan production contribute with two prompt leptons of opposite electric charge where one lepton's charge is misidentified. We do not consider background contributions from $W\gamma$ and $Z\gamma$ processes in this note.

Monte Carlo (MC) generators are used to model signal and background processes at a centre-of-mass energy of $\sqrt{s}$ =14 TeV, with the number of events scaled to an integrated luminosity of $\mathcal{L}$=3000 fb$^{-1}$ as expected for the full HL-LHC physics program. The signal (VBS and non-VBS EW) and background (QCD) production of $W^{\pm}W^{\pm}jj$ events are simulated using `Madgraph5_aMC@NLO` [19] with the `NNPDF3.0` PDF set [20], interfaced with `PYTHIA v8` [21] for parton showering, hadronisation and underlying event modelling. An additional `Madgraph5_aMC@NLO` sample with polarisation information is produced, to separate the longitudinal contribution (LL) to the $W^{\pm}W^{\pm}jj$ process from the transverse and mixed contributions (LT+TT). In this sample the $W$ bosons are decayed, assuming they are on-shell and have no spin correlations, using the `DECAY` routine provided with `MadGraph`.

The $WZ$ process is simulated using `SHERPA v2.2.0` [22–25] with next-to-leading order (NLO) accuracy in the strong coupling constant $\alpha_s$ for up to one associated parton, and leading order (LO) accuracy for two and three associated partons in the final state. Both QCD and EW production of $WZ$ processes are included in the $WZ$ background estimation. The generation of $ZZ$ events is done using `SHERPA v2.2.2` with up to two additional partons in the final state. For the generation of triboson events, matrix elements for all combinations of $pp \to VVV$ ($V = W, Z$) have been generated using `SHERPA v2.2.2` with up to two additional partons in the final state. Both fully leptonic decays, and processes in which one of the bosons decays hadronically, are considered. The generation of $W$+jets events is done for electron, muon, and tau final states using `Madgraph5_aMC@NLO` at LO and the `NNPDF3.0` PDF set, plus `PYTHIA v8` for fragmentation. They are simulated for up to one additional parton at NLO and up to two additional partons at LO. The generation of $Z$+jets events is done using the `POWHEG-BOX` event generator with the `CT10` PDF set [26], plus `PYTHIA v8` for parton showering and fragmentation. For the generation of top-quark pairs, the `POWHEG-BOX` v1 event generator with the `CT10` PDF set in the matrix element calculations is used. Electroweak $Wt$-channel single-top quark events are generated using the `POWHEG-BOX` v1 event generator. This event generator uses the four-flavour scheme for the NLO matrix-element calculations together with the fixed four-flavour PDF set `CT10f4`. For all top-quark processes, top-quark spin correlations are preserved (for t-channel, top-quarks are decayed using MadSpin). The parton shower, fragmentation, and underlying event are simulated using `PYTHIA v6` with the `CTEQ6L1` PDF set and the corresponding Perugia 2012 tune [27]. The top-quark mass is set to 172.5 GeV. The `EvtGen v1.2.0` program [28] is used to decay bottom and charm hadrons for the `POWHEG-BOX` samples.

Additional pileup interactions are generated with `PYTHIA v8` with the AU2 set of tuned parameters [29] and an average of 200 interactions per bunch crossing, and added event-by-event to the simulated samples.

# 3 Object and Event Selection

The analysis begins with particle-level objects, the detector simulation of which is estimated using smearing functions [30] derived from a full simulation of the ATLAS detector based on GEANT4 [31, 32], with the exception of pileup events which are fully simulated.

Lepton trigger and identification efficiencies are derived as a function of the lepton $\eta$ and $p_{\mathrm{T}}$ and used to estimate the likelihood of a given lepton passing either the trigger or identification requirement, respectively.[2] Muon transverse momentum and electron energy resolutions are also parameterised as a function of $\eta$ and either the transverse momentum (for muons) or transverse energy (for electrons). A tight isolation requirement is applied on the leptons. An estimation of the track- and calorimeter-based isolation is made, summing the momentum and energy, respectively, of stable generator-level particles with momenta greater than 1 GeV in a cone around each lepton. Only charged particles are taken into account for the track-based isolation, while the calorimeter-based isolation considers both charged and neutral particles, with the exception of neutrinos. The isolation criteria applied to each lepton are shown in Table 1, and are important for reducing the large contribution from non-prompt leptons originating from b-decays. An additional function for electrons[3] parameterises the likelihood of charge misidentification as a function of $p_{\mathrm{T}}$ and $\eta$.

| | Electron Isolation | Muon Isolation |
|---|---|---|
| Track-based isolation cone size | $\Delta R < 0.2$ | $\Delta R < 0.3$ |
| Track-based isolation requirement | $\Sigma p_T/p_T^e < 0.06$ | $\Sigma p_T/p_T^\mu < 0.04$ |
| Calorimeter-based isolation cone size | $\Delta R < 0.2$ | $\Delta R < 0.2$ |
| Calorimeter-based isolation requirement | $\Sigma E_T/p_T^e < 0.06$ | $\Sigma E_T/p_T^\mu < 0.15$ |

Table 1: Electron and muon isolation requirements.

The missing transverse momentum ($E_{\mathrm{T}}^{\mathrm{miss}}$) is defined at particle level as the transverse component of the vectorial sum of the final-state neutrino momenta. The $E_{\mathrm{T}}^{\mathrm{miss}}$ resolution is parameterised as a function of the overall event activity. Final-state particles with lifetime greater than 30 ps are clustered into jets (denoted as particle-level jets) using the anti-$k_t$ algorithm [33] with radius parameter $R = 0.4$. Final-state muons and neutrinos are not included in the jet clustering. To avoid double-counting jets associated with electrons, an overlap requirement is applied to exclude jets within a cone of $\Delta R_{e,j} < 0.05$ of any electron with transverse energy > 7 GeV. The particle-level jet momentum is smeared as a function of $p_{\mathrm{T}}$ and $\eta$ to account for detector effects.

The experimental signature of the $W^{\pm}W^{\pm}jj$ scattering process consists of two leptons (electrons or muons) with the same electric charge, two jets well-separated in rapidity, and moderate $E_{\mathrm{T}}^{\mathrm{miss}}$. Events are preselected by either a single-muon or single-electron trigger requiring transverse momentum $p_{\mathrm{T}} > 20$ and 25 GeV, respectively.

Jets with $p_{\mathrm{T}} > 30$ GeV and $|\eta| < 4.5$ are considered for the analysis, and a selection requirement on all jets with transverse momenta below 100 GeV is applied in order to distinguish between jets resulting from the hard scatter interaction and jets from the accompanying soft interactions (pileup jets). This jet vertex requirement is based on the $p_T$ fraction of the jet tracks originating from the hard scattering vertex[4]. The probability to misidentify a pileup jet as resulting from the hard scattering process is assumed to be 2%, with a resulting signal selection efficiency of 86% for $|\eta| < 1.5$ and 80% for $|\eta| > 2.9$ [34]. Jets either pass or fail this requirement, which is applied for jets with $|\eta| \leq 3.8$ where track information is available for the entire jet cone. To reduce the significant contamination from pileup jets, the $p_T$ threshold is increased from 30 to 70 GeV for jets outside the region where the vertex requirement is applied. The value of 70 GeV was chosen based on the $p_T$ and $\eta$ distributions of the fully-simulated pileup jets in the events, since most pileup jets have $p_T < 70$ GeV across the entire $\eta$ range. Jets passing the $p_T$ threshold and vertex requirements are preselected.

[2] The muon New Small Wheel Phase 1 upgrade will have extended the muon trigger to $|\eta| \leq 2.7$.
[3] The charge misidentification probability for muons is negligible.

Electrons and muons with transverse momenta $p_T > 7$ and 6 GeV, respectively, and with pseudorapidity $|\eta| \leq 4.0$ are preselected. An electron fake rate parameterisation allows for a fraction of jets to also be considered as preselected electrons. The fake electron function is based on the expected probability of a jet being misidentified as an electron, parameterised as a function of jet $p_T$ and $\eta$.

After all the object selection criteria have been applied, the two highest $p_T$ jets are defined as the leading and sub-leading *tag jets* in the event, and the two highest $p_T$ leptons are defined as the leading and subleading *signal leptons*.

| Selection requirement | Selection value |
|---|---|
| Signal lepton kinematics | $p_T > 25$ GeV and $|\eta| \leq 4.0$ |
| Tag jet kinematics | $p_T > 30$ GeV and $|\eta| \leq 4.5$ ($p_T > 70$ GeV for $|\eta| > 3.8$) |
| Dilepton separation and charge | Exactly two signal leptons with $\Delta R_{\ell,\ell} \geq 0.3$, $q_{\ell_1} \times q_{\ell_2} > 0$ |
| Dilepton mass | $m_{\ell\ell} > 20$ GeV |
| $Z_{ee}$ veto | $|m_{ee} - m_Z| > 10$ GeV |
| $E_T^{miss}$ | $E_T^{miss} > 40$ GeV |
| Jet selection and separation | at least two jets with $\Delta R_{\ell,j} > 0.3$ |
| Number of b-tagged jets | 0 |
| Dijet rapidity separation | $\Delta\eta_{j,j} > 2.5$ |
| Number of additional preselected leptons | 0 |
| Dijet mass | $m_{jj} > 500$ GeV |
| Lepton centrality[5] | $\zeta > 0$ |

Table 2: Default event selection criteria for $W^\pm W^\pm jj$ candidate events, with $\ell = e, \mu$ and $j$ as the leading or sub-leading jet.

The default event selection, as used in [34], is shown in Table 2, requiring two well-separated leptons of the same electric charge, and with $p_T > 25$ GeV. A minimum requirement on the dilepton mass reduces the contamination from low-mass Drell-Yan processes. Background contributions from $Z$ boson decays in the dielectron channel are significant, owing to a high likelihood of charge misidentification for electrons, so events with dilepton mass within 10 GeV of the $Z$ boson mass are excluded in the dielectron channel. In the dimuon channel we expect this contribution to be negligible. A requirement on $E_T^{miss}$ further reduces

[4] The vertex corresponding to the hard scatter interaction, typically the vertex with the highest $\Sigma p_T^2$.
[5] $\zeta = \min[\min(\eta_{\ell 1}, \eta_{\ell 2}) - \min(\eta_{j1}, \eta_{j2}), \max(\eta_{j1}, \eta_{j2}) - \max(\eta_{\ell 1}, \eta_{\ell 2})]$

the background from charge misidentified events. The *tag jets* are required to be well-separated in rapidity and to not overlap with the signal leptons, and events containing any b-tagged jets are vetoed. A veto on additional preselected leptons with $p_T > 7(6)$ GeV for electrons (muons) significantly reduces background from $WZ$ and $ZZ$ events. Finally, the *tag jets* are required to have a large invariant mass, and a requirement on the lepton centrality $\zeta$ is imposed to enhance the purity of the $W^{\pm}W^{\pm}jj$ electroweak signal.

| Selection requirement | Selection value |
|---|---|
| Signal lepton kinematics | **$p_T > 28$ GeV (leading lepton)** |
| | $p_T > 25$ GeV (subleading lepton) |
| Tag jet kinematics | **$p_T > 90$ GeV (leading jet)** |
| | **$p_T > 45$ GeV (subleading jet)** |
| Dilepton separation and charge | Exactly two signal leptons with $\Delta R_{\ell,\ell} \geq 0.3$, $q_{\ell_1} \times q_{\ell_2} > 0$ |
| Dilepton mass | **$m_{\ell\ell} > 28$ GeV** |
| $Z_{ee}$ veto | $\lvert m_{ee} - m_Z \rvert > 10$ GeV |
| $E_T^{\text{miss}}$ | $E_T^{\text{miss}} > 40$ GeV |
| Jet selection and separation | at least two jets with $\Delta R_{\ell,j} > 0.3$ |
| Number of b-tagged jets | 0 |
| Dijet rapidity separation | $\Delta\eta_{j,j} > 2.5$ |
| Number of additional preselected leptons | 0 |
| Dijet mass | **$m_{jj} > 520$ GeV** |
| Lepton centrality | **$\zeta > -0.5$** |

Table 3: Optimised event selection criteria for $W^{\pm}W^{\pm}jj$ candidate events, with $\ell = e, \mu$ and $j$ as the leading or sub-leading jet. Criteria that differ with respect to the default selection are shown in bold.

An optimisation of the object and event selection criteria was also performed using the random grid search (RGS) cut-based algorithm [35], to improve the signal selection against the total background, here considering only the longitudinal contribution to the $W^{\pm}W^{\pm}jj$ scattering as signal. The optimisation was performed over the lepton and jet $p_T$, dilepton and dijet invariant mass, and centrality requirements, with the additional requirement that there be more than 1000 signal events remaining post-optimisation. The optimised selection criteria are shown in Table 3. The increased lepton and jet $p_T$ requirements significantly reduce background contributions, and loosening the centrality requirement increases the number of the typically softer longitudinal $W^{\pm}W^{\pm}jj$ events passing the selection. In the next section the event yields and cross section measurement for both the default and optimised sets of event selection criteria are presented. For the extraction of the longitudinal scattering significance, only the optimised set is used.

## 4 Results

The total number of inclusive signal ($W^{\pm}W^{\pm}jj$ EW) and background events expected after the full event selection for an integrated luminosity of $\mathcal{L}$=3000 fb$^{-1}$ is shown in Table 4. Events with either a misidentified charge electron, or a jet faking an electron, are summed for all background samples and combined into a single entry titled "charge misidentification" or "jets faking leptons", respectively. The remaining $t\bar{t}$, single top, and $W$+jet events that pass the full event selection are listed as "Other non-prompt".

A total of 3490 signal events are expected, with 9888 background events. The relative fraction of signal to background events varies by final state, so the event yields for the separate channels, $\mu^{\pm}\mu^{\pm}$, $e^{\pm}e^{\pm}$, $\mu^{\pm}e^{\pm}$, and $e^{\pm}\mu^{\pm}$, are also shown. The number of expected signal and background events after the optimised full event selection are shown in Table 5. The dramatic reduction in the background yields is primarily a result of increasing the leading and subleading jet $p_{\mathrm{T}}$ requirements, which rejects most of the background from fake and misidentified charge contributions from $W$, $Z$+jet events.

| | All channels | $\mu^{\pm}\mu^{\pm}$ | $e^{\pm}e^{\pm}$ | $\mu^{\pm}e^{\pm}$ | $e^{\pm}\mu^{\pm}$ |
|---|---|---|---|---|---|
| $W^{\pm}W^{\pm}jj$ (QCD) | 206.4 | 91.1 | 22.8 | 38.4 | 54.1 |
| Charge Misidentification | 2300 | 0.0 | 2100 | 90 | 160 |
| Jets faking electrons | 5000 | 0.0 | 3400 | 1200 | 340 |
| $WZ + ZZ$ | 2040 | 500 | 438 | 423 | 680 |
| Tribosons | 115 | 47 | 15.4 | 21.6 | 31.2 |
| Other non-prompt | 210 | 110 | 20 | 60 | 27 |
| Total Background | 9900 | 750 | 6000 | 1900 | 1290 |
| Signal $W^{\pm}W^{\pm}jj$ (EW) | 3489 | 1435 | 432 | 679 | 944 |

Table 4: The expected signal and background event yields after the default full event selection for a corresponding integrated luminosity of $\mathcal{L}$=3000 fb$^{-1}$. Events tagged as either "charge misidentification" or "jets faking leptons" are summed for all background samples and combined into a single entry each in the table. Both QCD and EW production of $WZ$ processes are included in the diboson background.

| | All channels | $\mu^{\pm}\mu^{\pm}$ | $e^{\pm}e^{\pm}$ | $\mu^{\pm}e^{\pm}$ | $e^{\pm}\mu^{\pm}$ |
|---|---|---|---|---|---|
| $W^{\pm}W^{\pm}jj$ (QCD) | 168.7 | 74.6 | 19.7 | 32.2 | 42.2 |
| Charge Misidentification | 200 | 0.0 | 11 | 30 | 160 |
| Jets faking electrons | 460 | 0.0 | 130 | 260 | 70 |
| $WZ + ZZ$ | 1286 | 322 | 289 | 271 | 404 |
| Tribosons | 76 | 30.1 | 9.6 | 15.1 | 21.6 |
| Other non-prompt | 120 | 29 | 16.6 | 50 | 19 |
| Total Background | 2310 | 455 | 480 | 660 | 710 |
| Signal $W^{\pm}W^{\pm}jj$ (EW) | 2958 | 1228 | 380 | 589 | 761 |

Table 5: The expected signal and background event yields after the optimised full event selection for a corresponding integrated luminosity of $\mathcal{L}$=3000 fb$^{-1}$. Events tagged as either "charge misidentification" or "jets faking leptons" are summed for all background samples and combined into a single entry each in the table. Both QCD and EW production of $WZ$ processes are included in the diboson background.

The dijet invariant mass distributions for all events are shown in Figure 4 for the default and optimised event selections. Additionally for the optimised selection, the dilepton invariant mass and the integrated

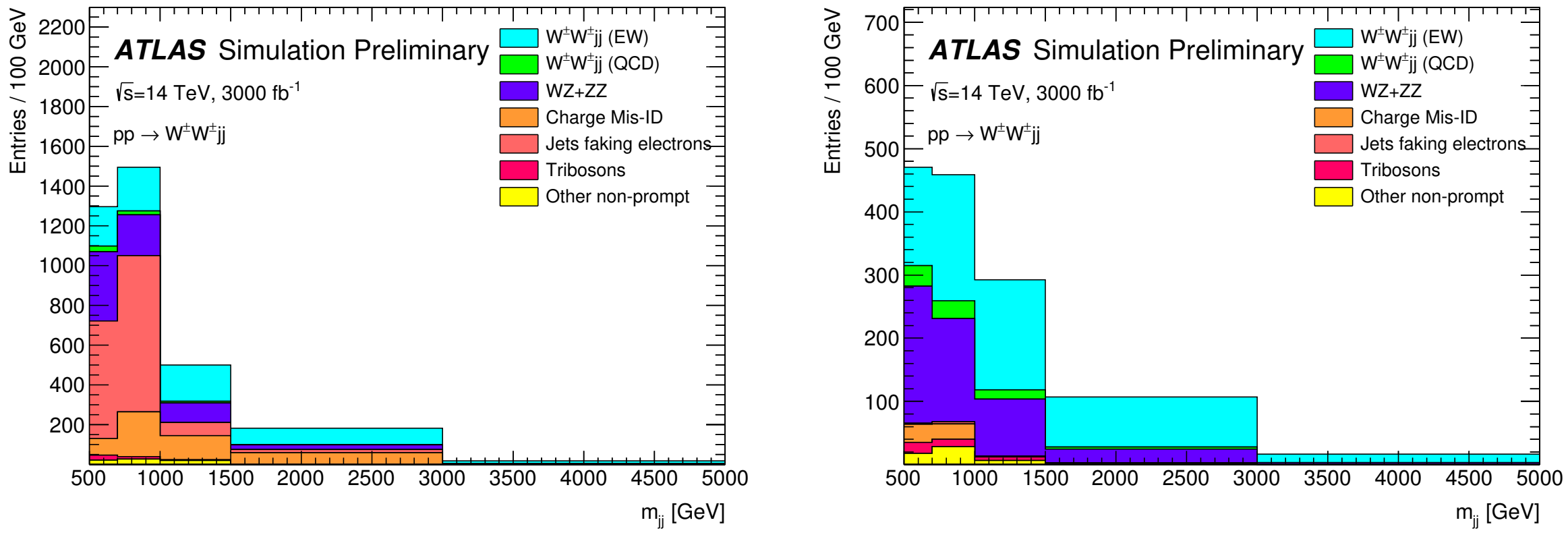

Figure 4: Dijet invariant mass distributions for events passing all selection criteria of the signal region, for the default (left) and optimised (right) event selections.

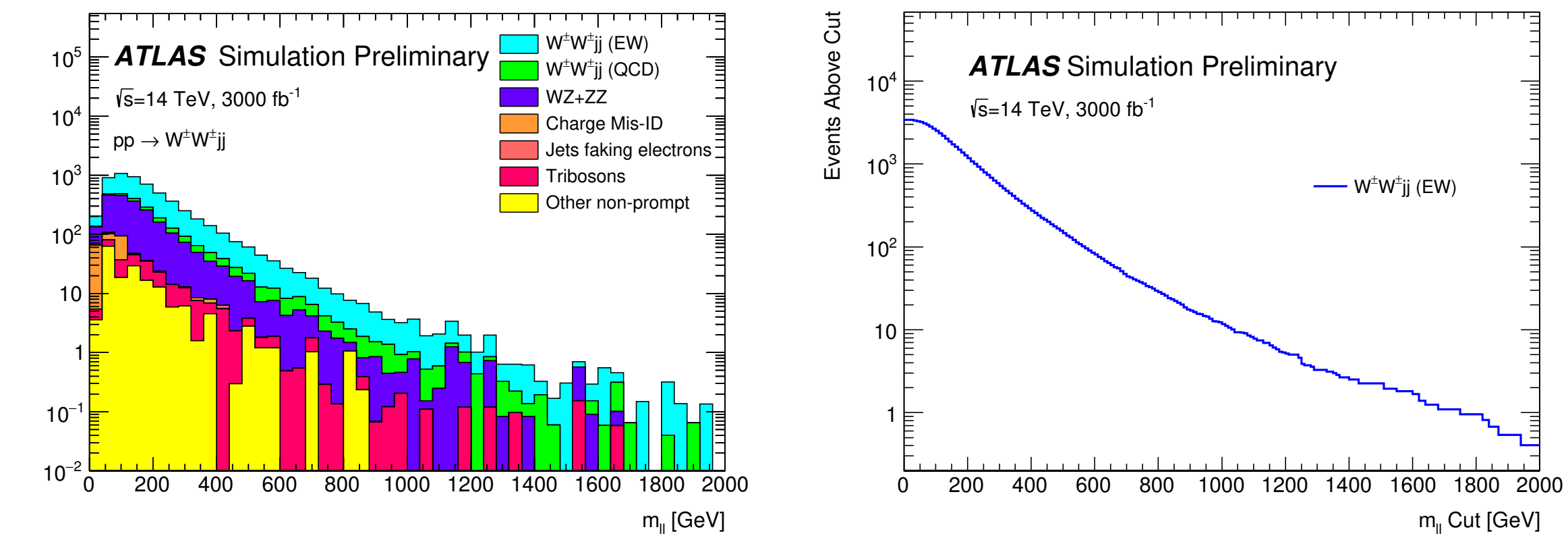

Figure 5: Dilepton invariant mass distribution (left), and integrated number of events as a function of dilepton invariant mass (right) for events passing all selection criteria of the signal region, for the optimised event selection.

9

| Source | Uncertainty (%) | |
|---|---|---|
| | Baseline | Optimistic |
| $W^{\pm}W^{\pm}jj$ (EW) | 3 | |
| Luminosity | 1 | |
| Trigger efficiency | 0.5 | |
| Lepton reconstruction and identification | 1.8 | |
| Jets | 2.3 | |
| Flavour tagging | 1.8 | |
| Jets faking electrons | 20 | |
| Charge mis-ID | 25 | |
| $W^{\pm}W^{\pm}jj$ (QCD) | 20 | 5 |
| Top | 15 | 10 |
| Diboson | 10 | 5 |
| Triboson | 15 | 10 |

Table 6: Expected experimental and rate uncertainties for an integrated luminosity of $\mathcal{L}$=3000 fb$^{-1}$.

number of signal and background events as a function of the dilepton invariant mass is shown in Figure 5. Categorisation by lepton flavour and charge increases the overall analysis sensitivity, so the $m_{jj}$ distribution is obtained for each of the eight channels[6] and combined in a profile likelihood fit to extract the $W^{\pm}W^{\pm}jj$ electroweak production cross section, using the same method as the ATLAS analysis that presents the observation of $W^{\pm}W^{\pm}jj$ (EW) using 13 TeV $pp$ collision data [12].

The uncertainties considered are given in Table 6. Experimental systematics on the trigger, leptons, jets, and flavour tagging are taken from the 13 TeV analysis unchanged, while for the baseline estimation, rate uncertainties on the backgrounds are halved. An "optimistic" set of uncertainties is also presented, where the uncertainties on the non-data-driven backgrounds are aggressively reduced.

With the default event selection and baseline set of uncertainties, the expected $W^{\pm}W^{\pm}jj$ cross section obtained from the fit is 16.89 ± 0.36 (stat) ± 0.53 (theory) ± 0.86 (sys) fb. With the optimised event selection the systematic uncertainty is reduced by 5%, with an expected cross section of 16.94 ± 0.36 (stat) ± 0.53 (theory) ± 0.78 (sys) fb. Figure 6 shows the projection of the expected total uncertainty on the cross section, as well as the individual components, as a function of integrated luminosity, for the optimised event selection. Additionally, Figure 6 shows the projections with the "optimistic" set of uncertainties, which results in a small effect overall.

In the SM, the Higgs boson unitarises the longitudinal $VV$ scattering amplitude completely. If, however, the SM is an effective theory of a more general one with an additional strongly-coupled sector, the unitarisation may be only partial, and new physics processes may be involved. In the context of $W^{\pm}W^{\pm}jj$ scattering, two quantities that are particular sensitive to the longitudinal scattering component are the dijet azimuthal separation $\Delta\phi(j, j)$ and leading lepton $p_{\mathrm{T}}$, with the longitudinal scattering preferentially occurring in the regions of large dijet separation and low leading lepton $p_{\mathrm{T}}$ [36, 37]. The shape comparisons of these distributions for the purely longitudinal scattering contribution (LL) and the combined mixed (LT) and transverse (TT) contributions are shown in Figure 7.

[6] $e^+e^+$, $e^-e^-$, $e^+\mu^+$, $e^-\mu^-$, $\mu^+e^+$, $\mu^-e^-$, $\mu^+\mu^+$, $\mu^-\mu^-$

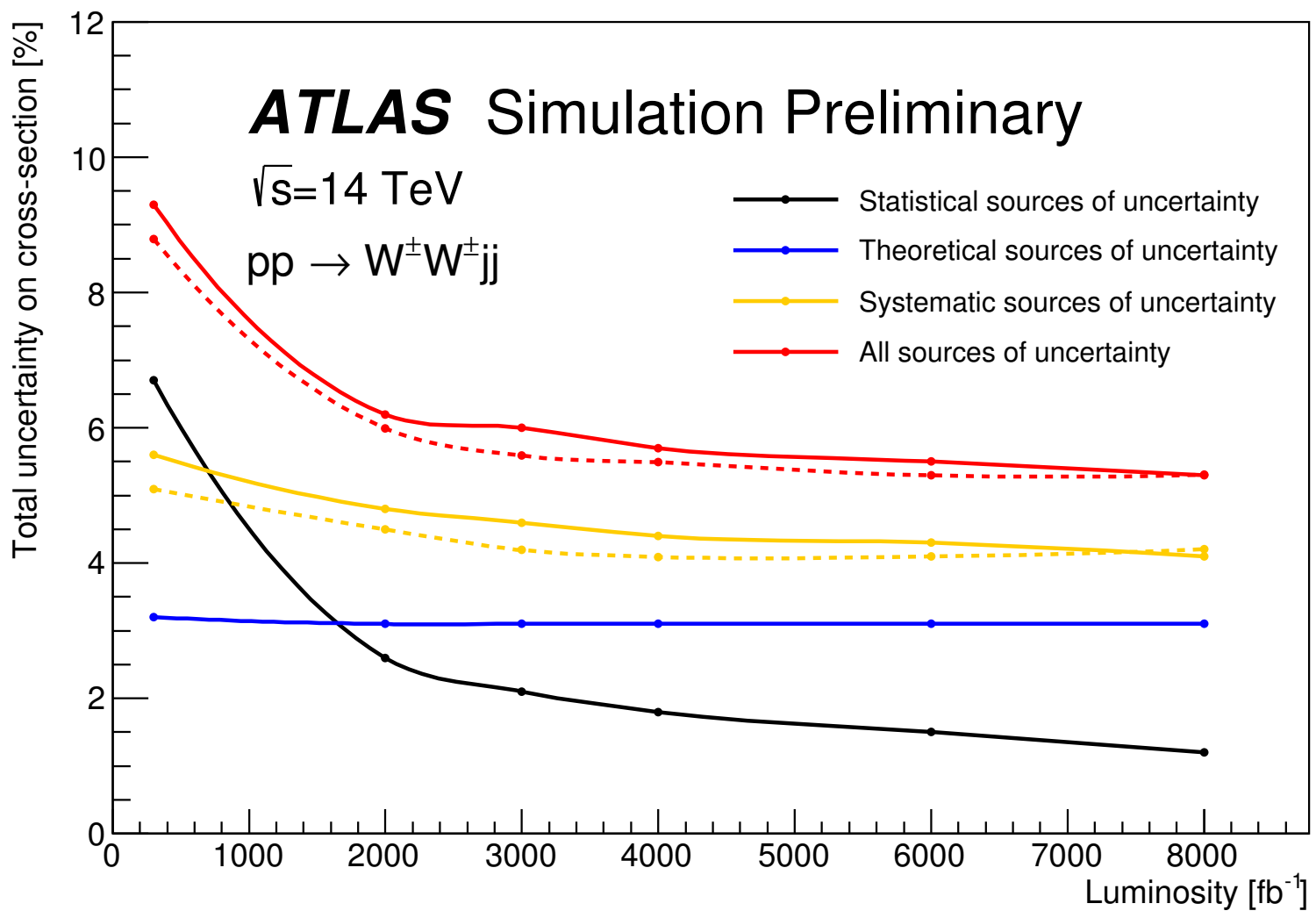

Figure 6: Projection of the statistical (black), theoretical (blue), systematic (yellow) and total (red) uncertainties on the cross section as a function of integrated luminosity, for the optimised event selection using the baseline scenario (solid lines). The dashed lines show the systematic and total uncertainties on the cross-section for the optimistic scenario (see Table 6). The theoretical uncertainty refers to the signal only.

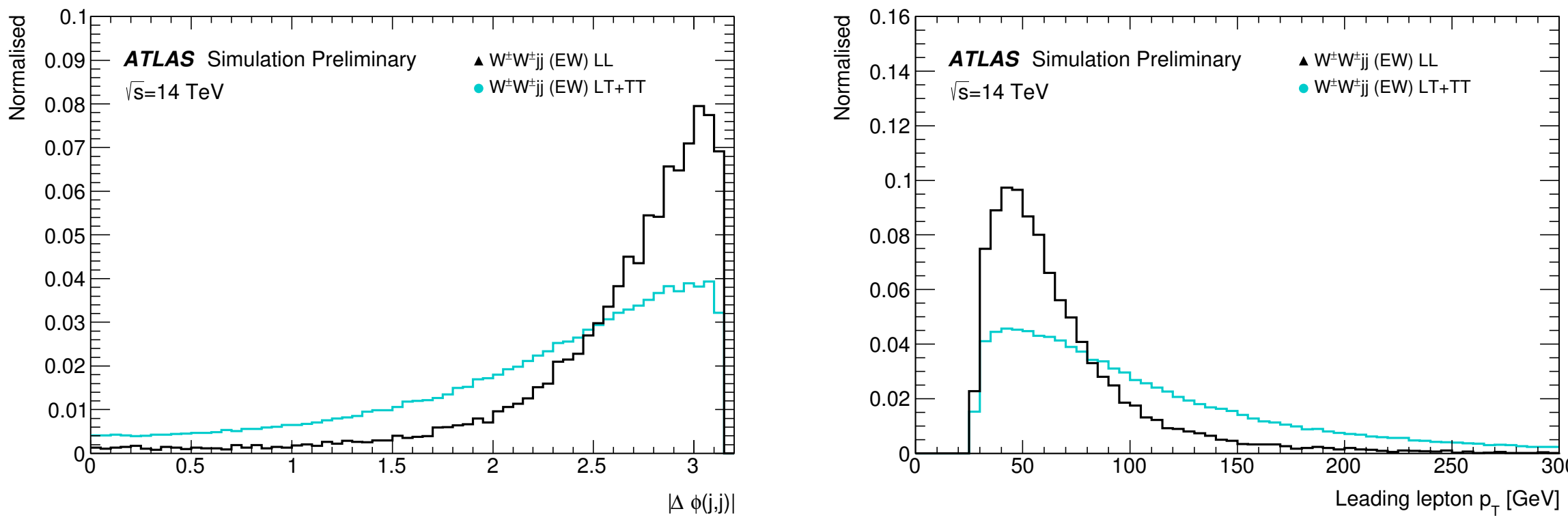

Figure 7: Shape comparisons for the dijet azimuthal separation $|\Delta\phi(j,j)|$ (left) and leading lepton $p_{\mathrm{T}}$ (right) distributions, for the purely longitudinal (LL) and combined mixed and transverse (LT+TT) $W^{\pm}W^{\pm}jj$ events.

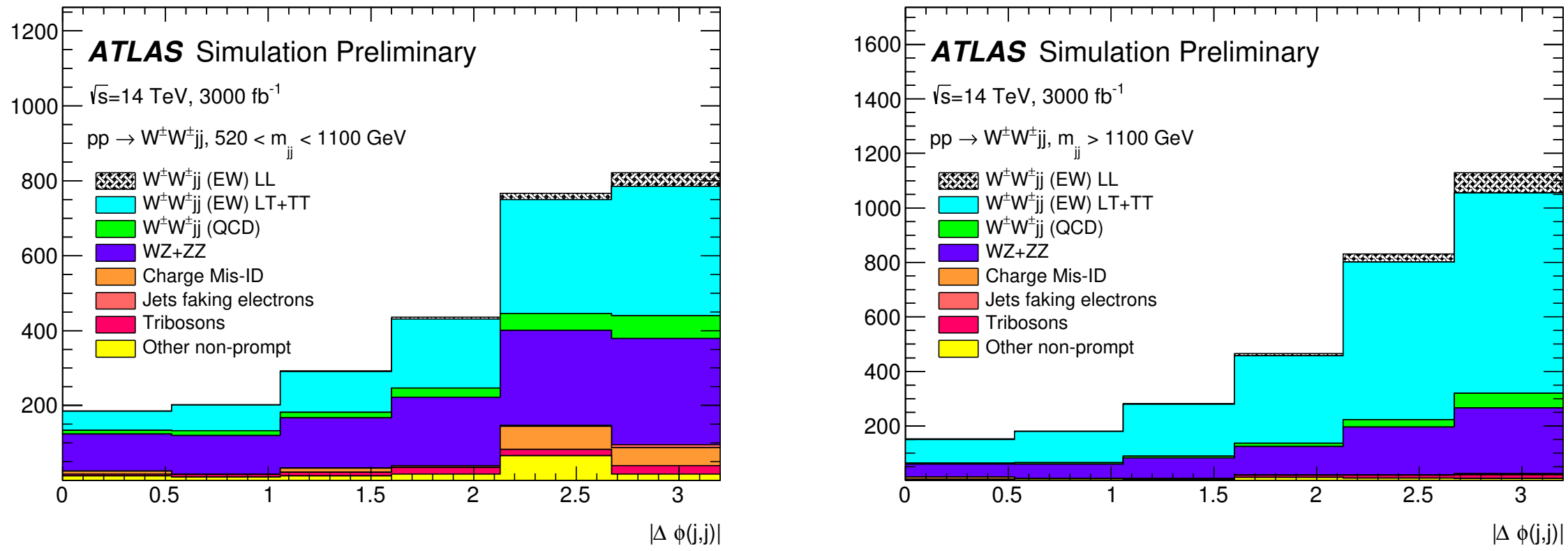

Figure 8: Dijet azimuthal separation ($|\Delta\phi(j, j)|$) for $520 < m_{jj} < 1100$ GeV (left) and $m_{jj} > 1100$ GeV (right). An additional requirement on the subleading lepton pseudorapidity ($|\eta| < 2.5$) is made to reduce the contributions from the fake and charge-misidentified backgrounds.

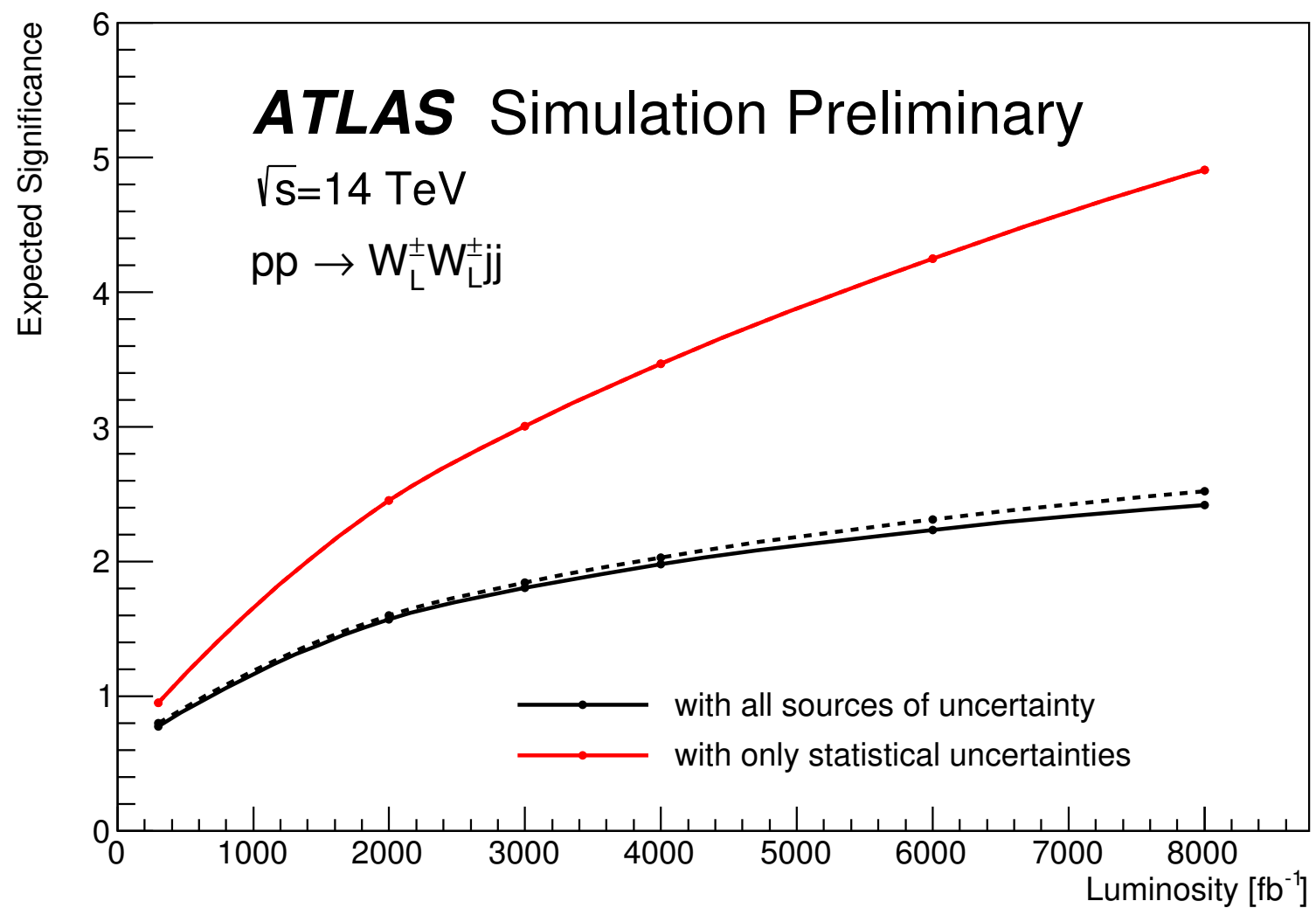

Figure 9: Projection of the expected significance of the observation of the $W_L^{\pm}W_L^{\pm}jj$ process as a function of integrated luminosity, for the optimised event selection using the baseline scenario, considering all the sources of uncertainty (black) or only the statistical uncertainty (red). The dashed lines show the expected significance for the optimistic scenario (see Table 6).

The $\Delta\phi(j, j)$, distribution is shown in Figure 8 for two regions of dijet mass, $520 < m_{jj} < 1100$ GeV and $m_{jj} > 1100$, with the contribution from the purely longitudinal scattering (LL) shown separately from that from the mixed and transverse contributions (LT+TT). An additional requirement restricting the pseudorapidity of the subleading lepton to the central region of the detector ($|\eta| < 2.5$) is made, which significantly reduces the contributions from the fake and charge-misidentified backgrounds. A simultaneous binned likelihood fit of the $\Delta\phi(j, j)$ distributions in the two regions of $m_{jj}$ is performed in four lepton flavour channels ($e^\pm e^\pm$, $e^\pm\mu^\pm$, $\mu^\pm e^\pm$, $\mu^\pm\mu^\pm$) to extract the longitudinal scattering significance, considering the LT+TT contributions as background. Due to the limited statistics, categorisation by lepton charge or leading lepton $p_{\mathrm{T}}$ is not done in this case. The expected significance of the observation of the $W_L^\pm W_L^\pm jj$ process obtained from the fit is $1.8\sigma$, with an expected precision of 47% on the measurement. Figure 9 shows the expected significance as a function of integrated luminosity.

Measuring VBS processes at a hadron collider is experimentally challenging due to small cross sections and the difficulty of separating longitudinal states from transverse ones. Recent studies [38] have shown that advances in machine learning can improve the prospects for the measurement of the $W_L^\pm W_L^\pm jj$ process. In addition, improvements in the lepton selection efficiency will also help to improve the measurement by increasing the statistics in the signal region.

# 5 Conclusion

The $W^\pm W^\pm jj$ channel is one of the best channels with which to measure the scattering of two vector bosons. Prospects for measuring the $W^\pm W^\pm jj$ vector boson scattering process in proton-proton collisions at $\sqrt{s}$ = 14 TeV at the High-Luminosity Large Hadron Collider have been studied using simulated events parameterised to take into account the expected detector effects in the high luminosity environment. With the optimised event selection a total of 2431 signal events are expected, for a background expectation of 1460 events. The cross section for the electroweak production process is extracted from a fit to the dijet invariant mass distribution, and an expected total uncertainty of 6% is achieved for an integrated luminosity of 3000fb$^{-1}$. Additionally, the purely longitudinal scattering component can be extracted with an expected significance of $1.8\sigma$ from a binned likelihood fit to the dijet azimuthal separation distribution.

# CMS Physics Analysis Summary

Contact: cms-phys-conveners-ftr@cern.ch

# Prospects for the measurement of electroweak and polarized WZ $\to 3\ell\nu$ production cross sections at the High-Luminosity LHC

The CMS Collaboration

## Abstract

Prospects for the measurement of WZ electroweak (EW) production in association with two jets at the High-Luminosity LHC are presented. The W and Z bosons are detected via their decays, W $\to$ e$\nu$, $\mu\nu$ and Z$\to$ ee, $\mu\mu$. The results are obtained by a projection of existing results at 13 TeV to 14 TeV. The expected uncertainty in the EW WZ cross section measurement and significance of the observation of the polarized portion of the EW WZ cross section are presented.

A study of the electroweak (EW) WZ production using 36.9 fb$^{-1}$ of proton-proton collisions at 13 TeV was recently reported by the CMS collaboration [1]. The existing EW results are strongly limited by the yields of the signal events, therefore the integrated luminosity expected at the end of the HL-LHC is mandatory to fully exploit this process via measurement of differential distributions and of the polarization of the final state bosons. The extension of these studies at the High-Luminosity LHC (HL-LHC) is of a great importance, since measurement of the polarized final states gives a direct access to the nature of the electroweak symmetry breaking via the exchange of a Higgs bosons in the t-channel. Any deviation from the standard model (SM) Higgs-gauge coupling could lead to a non-cancellation between gauge amplitudes and Higgs amplitudes, visible as an increase of the EW WZ cross section at large energies and can be measured by studying of the transverse and longitudinal portions of the WZ EW cross section. For such measurements the EW signal sample should be divided into three categories based on polarizations of the W and Z bosons: longitudinal-longitudinal (LL), with one boson transversely polarized (LT) and with both bosons transversely polarized (TT). In this note the precision of the EW WZ cross section measurement and significance of observation of the LL component are discussed. The results are presented as a function of integrated luminosity, where 300 fb$^{-1}$ corresponds to one year of CMS data taking, 3000 fb$^{-1}$ to ten years, as planned for the HL-LHC, and 6000 fb$^{-1}$ to the possible combination of the CMS and ATLAS results.

The future CMS measurements at the HL-LHC are expected to benefit from improvements in the detector and the event reconstruction, better accuracy in the luminosity measurements, and improved theoretical predictions that will become available at the HL-LHC. To perform the projection from existing data to the HL-LHC, the signal and background yields obtained from the Monte Carlo (MC) simulation and from data, for some backgrounds, at 13 TeV are scaled to 14 TeV using ratios of cross sections as predicted by the SM. These scaling factors vary for different processes. In general, the cross sections increase by 8 – 20% when changing from 13 to 14 TeV. For the EW production of WZ the increase is about 16%, for the QCD WW production about 8%. To justify that the 13 TeV MC samples can be used to describe the performance of the Phase-2 CMS detector at up to 200 additional pp interactions in the same and neighboring bunch crossings per event (pileup), the performance of lepton and jet identification algorithms at the HL-LHC is estimated using a Delphes simulation [2].

The CMS detector [3] will be upgraded to fully exploit the physics potential offered by the increase in luminosity, and to cope with the demanding operational conditions at the HL-LHC [4–8]. The upgrade of the first level hardware trigger (L1) will allow for an increase of L1 rate and latency to about 750 kHz and 12.5 $\mu$s, respectively, and the high-level software trigger (HLT) is expected to reduce the rate by about a factor of 100 to 7.5 kHz. The entire pixel and strip tracker detectors will be replaced to increase the granularity, reduce the material budget in the tracking volume, improve the radiation hardness, and extend the geometrical coverage and provide efficient tracking up to pseudorapidities of about $|\eta| = 4$. The muon system will be enhanced by upgrading the electronics of the existing cathode strip chambers (CSC), resistive plate chambers (RPC) and drift tubes (DT). New muon detectors based on an improved RPC design and gas electron multiplier (GEM) technologies will be installed to add redundancy, increase the geometrical coverage up to about $|\eta| = 2.8$, and improve the trigger and reconstruction performance in the forward region. The barrel electromagnetic calorimeter (ECAL) will feature the upgraded front-end electronics that will be able to exploit the information from single crystals at the L1 trigger level, accommodate trigger latency and bandwidth requirements, and provide 160 MHz sampling allowing high-precision timing capability for photons. The hadronic calorimeter (HCAL), consisting in the barrel region of brass absorber plates and plastic scintillator layers, will be read out by silicon photomultipliers (SiPMs). The endcap electromagnetic and hadron calorimeters will be replaced with a new high-granularity sampling calorimeter (HGCal) that will provide highly-segmented spatial information in both transverse and longitudinal directions, as well as high-precision timing information. Finally, the addition of a new timing detector for minimum ionizing particles (MTD) in both barrel and endcap regions is envisaged to provide the capability for 4-dimensional (3 space and 1 time) reconstruction of interaction vertices that will significantly offset the CMS performance degradation due to high PU rates.

A detailed overview of the CMS detector upgrade program is presented in Ref. [4–8], while the expected performance of the reconstruction algorithms and pile-up mitigation with the CMS detector is summarised in Ref. [9].

he CMS EW WZ measurement at 13 TeV [1] used single electron, single muon, double electron, double muon, and muon-electron triggers. For projection of these results to the HL-LHC conditions, we assume no changes in thresholds of the triggers; the trigger efficiencies are also assumed to be the same. The events are selected with exactly three leptons, two of them must have opposite charge, same flavor, and to be consistent with a Z boson. The four possible final states are labeled as eee, ee$\mu$, $\mu\mu$e, and $\mu\mu\mu$, where the first two leptons are the leptons associated with the Z boson, and the third lepton is associated with the W boson. The leading lepton from the Z boson must have $p_{\mathrm{T}} > 25$ GeV and the trailing lepton $p_{\mathrm{T}} > 15$ GeV. The third lepton, associated with the W boson, must have $p_{\mathrm{T}} > 20$ GeV. Events must have transverse missing momentum $P_{\mathrm{T}}^{\mathrm{miss}} > 30$ GeV to account for the presence of a neutrino from the W decay.

The events must additionally have at least two jets, reconstructed with anti-$k_T$ algorithm with distance parameter 0.4, with $p_{\mathrm{T}} > 50$ GeV and $|\eta| < 4.7$. The jets have to be separated from the lepton candidates by $\Delta R(\text{jet,lepton}) > 0.4$, to ensure distinct and isolated jets and leptons. The jet with the highest transverse momentum is chosen as leading jet and the jet with the second highest $p_{\mathrm{T}}$ as subleading jet. To exploit the unique signature of the EW process the two jets are required to have a high dijet mass $m_{jj} > 500$ GeV and a large pseudorapidity separation $|\Delta\eta_{jj}| > 2.5$. We further require $|\text{Zeppenfeld}(3\ell)| = \left|\eta_{3\ell} - \frac{1}{2}(\eta_{j_1} + \eta_{j_2})\right| < 2.5$, where $\eta_{3\ell}$ is the pseudorapidity of the trilepton system and $\eta_{j_1}, \eta_{j_2}$ are pseudorapidities of the leading and subleading jets. In addition, all lepton pairs must pass $m_{\ell\ell} > 4$ GeV to match the constraints, which are applied on the Monte Carlo samples. These constraints prevent problems with collinear emissions in theoretical calculations and suppress the contribution of low-mass resonances like J/$\psi$.

To reduce the $\mathrm{t\bar{t}}$ background the lepton pair associated with the Z boson must have an invariant mass between 76 GeV and 106 GeV (i.e. a 15 GeV window around the Z mass) and no jet with $p_{\mathrm{T}} > 30$ GeV passing the CSVv2 tight b tag working point [10] is allowed in the event. If the event contains more than one Z boson candidate, the lepton pair with an invariant mass closest to the Z mass is chosen. The trilepton mass has to be greater than 100 GeV to remove contributions from Z$\gamma$ events, where the photon radiated from the leptonic Z decay pair produces leptons. A detailed description of the identification and selection requirements can be found in Ref. [1] and is summarized in Table 1. The only difference with respect to the event selection in Ref. [1] is the extended pseudorapidity of the leptons reconstruction. In the Phase-2 CMS detector the electrons (muons) can be reconstructed in pseudorapidity range up to 3.0 (2.8), compared to 2.5 (2.4) in the existing CMS detector. The increase in the pseudopapidity coverage increases the yield for different decay channels by 5–8%. Extending the electron pseudorapidity coverage up to $|\eta| < 4$ is also under consideration, but this would require additional studies of systematic uncertainties and development of special reconstruc-tion algorithms. Since this should have a minor effect on the final results, this extension is not considered in this analysis.

Table 1: Summary of event selection requirements.

| Variable | Requirement |
|---|---|
| $p_T(\ell_{Z,1})$ [GeV] | $> 25$ |
| $p_T(\ell_{Z,2})$ [GeV] | $> 15$ |
| $p_T(\ell_W)$ [GeV] | $> 20$ |
| $\|\eta(\mu)\|$ | $< 2.8$ |
| $\|\eta(e)\|$ | $< 3.0$ |
| $\|m_Z - m_Z^{PDG}\|$ [GeV] | $< 15$ |
| $m_{3\ell}$ [GeV] | $> 100$ |
| $m_{\ell\ell}$ [GeV] | $> 4$ |
| $p_T^{miss}$ [GeV] | $> 30$ |
| $\|\eta(j)\|$ | $< 4.7$ |
| $p_T(j)$ [GeV] | $> 30$ |
| $p_T(j_{lead/subleading})$ [GeV] | $> 50$ |
| $\Delta R(j, \ell)$ | $> 0.4$ |
| $n_j$ | $\geq 2$ |
| $p_T(b)$ [GeV] | $> 30$ |
| $n_{b-jet}$ | $= 0$ |
| $m_{jj}$ | $> 500$ |
| $\Delta\eta(j_1, j_2)$ | $> 2.5$ |
| $\|\eta_{3\ell} - \frac{1}{2}(\eta_{j_1} + \eta_{j_2})\|$ | $< 2.5$ |

The expected signal and background yields after all selection requirements, corrected for the cross section increase from 13 to 14 TeV, detector acceptance improvements for the extension in pseudorapidity for leptons, and for integrated luminosity of 3000 fb$^{-1}$ are shown in Table 2. The nonprompt background is mainly caused by the production of $t\bar{t}$ and Drell–Yan events, where one or two jets are misidentified as leptons. This background is estimated from the 13 TeV data and scaled to 14 TeV. The scaling factor is based on simulation of the $t\bar{t}$ and Drell–Yan processes at 13 and 14 TeV.

As illustrated in Table 2, the major background to the EW production is QCD-induced production of WZjj events. Separating the EW and QCD-induced components requires exploiting the different kinematic signatures of the two processes. The relative fraction of EW process in WZjj production increases with increasing the dijet mass and angular separation of the leading jets. This motivates the use of a distribution of dijet mass in bins of angular separation,

Table 2: Expected signal and background yields for 3000 fb$^{-1}$, based on projection of corresponding yields from Ref. [1] and input systematic uncertainties as described in the text.

| Process | eee | ee$\mu$ | e$\mu\mu$ | $\mu\mu\mu$ | all |
|---|---|---|---|---|---|
| EW-WZjj | $380 \pm 8$ | $525 \pm 10$ | $763 \pm 14$ | $1089 \pm 20$ | $2757 \pm 28$ |
| QCD-WZjj | $476 \pm 16$ | $701 \pm 22$ | $927 \pm 28$ | $1383 \pm 43$ | $3486 \pm 58$ |
| t+V/VVV | $179 \pm 17$ | $264 \pm 9$ | $337 \pm 10$ | $594 \pm 19$ | $1374 \pm 24$ |
| Nonprompt | $19 \pm 2$ | $265 \pm 14$ | $665 \pm 41$ | $243 \pm 12$ | $1192 \pm 45$ |
| VV | $78 \pm 3$ | $49 \pm 2$ | $180 \pm 8$ | $92 \pm 4$ | $398 \pm 10$ |
| Z$\gamma$ | <1 | <1 | $296 \pm 37$ | <1 | $296 \pm 37$ |

**4**

$\Delta R_{jj} = \sqrt{(\Delta\eta_{jj})^2 + (\Delta\phi_{jj})^2}$, between jets for the extraction of the EW WZjj cross section as shown as a one-dimensional histogram in Fig. 1.

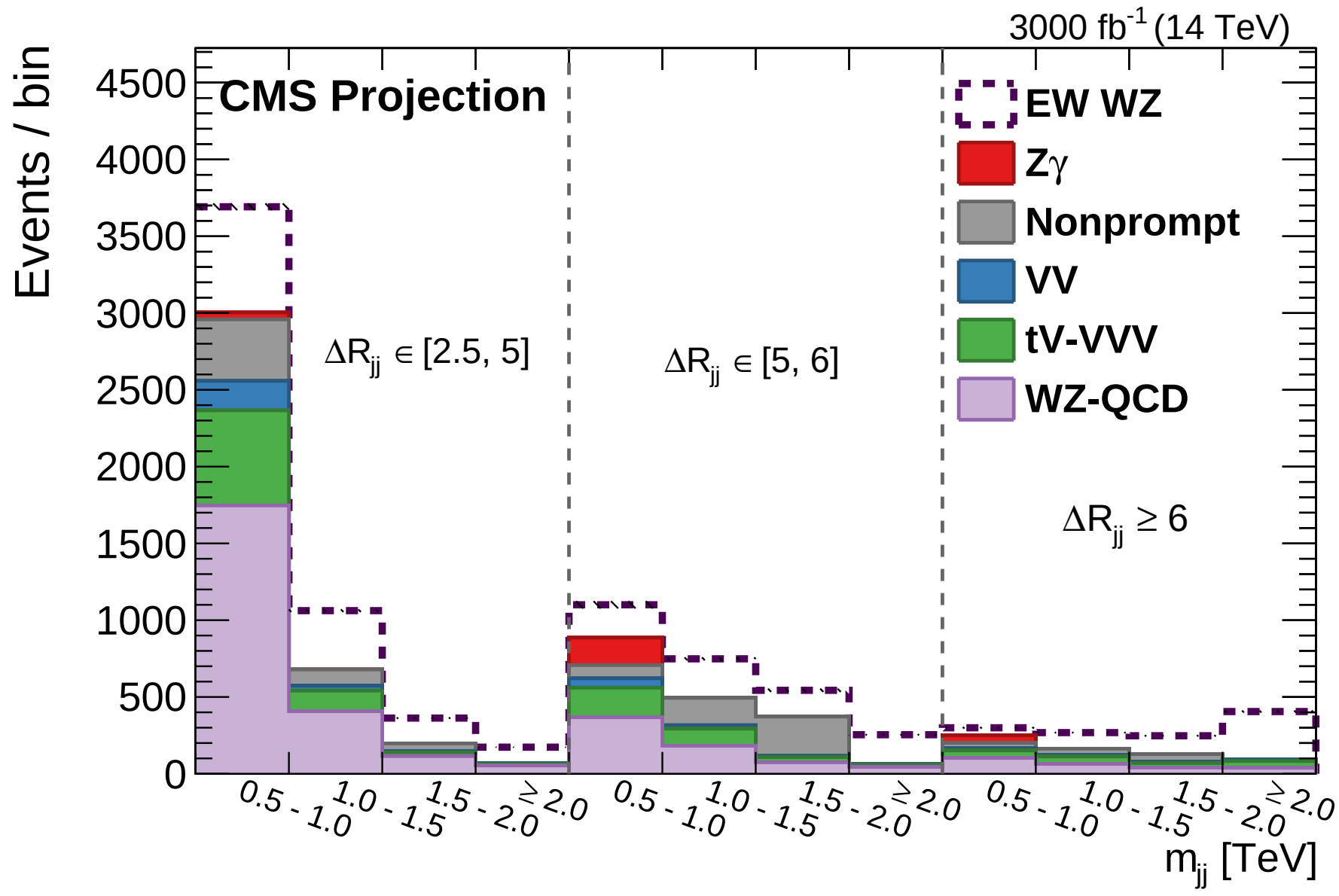

Figure 1: The $m_{jj}$ distributions in bins of $\Delta R_{jj}$ for 3000 $fb^{-1}$.

The measurement of the EW WZjj production cross section uses a maximum likelihood fit of this distribution performed simultaneously for four independent decay channels. The systematic uncertainties are represented by nuisance parameters in the fit and are allowed to vary according to their probability density functions. The correlations across bins, between different sources of uncertainty and decay channels are taken into account. The background contributions are allowed to vary within the estimated uncertainties.

Table 3: Input systematic uncertainty (%) for each nuisance parameter used in the fit.

| Systematic Source | Type | Amount, % |
|---|---|---|
| Integrated luminosity | Norm. | 1 |
| Nonprompt norm. | Norm. | 10 |
| b-tagging | Norm. | 1-3 |
| Electron scale and res. | Shape | 1 |
| Muon efficiency and res. | Shape | 0.5 |
| MET | Shape | 1-4 |
| Other background theory | Shape | 1-5 |
| QCD-WZjj PDF | Shape | 1 |
| QCD-WZjj Scale | Shape | 3-4 |
| EW-WZjj PDF | Shape | 1 |
| EW-WZjj Scale | Shape | 2-3 |
| Jet energy scale | Shape | 1-3 |
| Jet energy resolution | Shape | 1-4 |

The dominant sources of systematic uncertainty are summarized in Table 3. Since the cross section is measured using the differential distribution, the uncertainty can affect both the shapes and normalization of the distributions In the table, uncertainties of type ”Norm”, like luminosity uncertainty, only affect the yield of the events, while most of the uncertainties affect both normalization and shape of the distributions and are of the type ”Shape”. The uncertainties in the table represent input uncertainties to the fit. Most of them, such as lepton efficiency, PDF uncertainties, and other experimental and theoretical uncertainties are expected to decrease to the 1% level at the HL-LHC. The largest uncertainties are theoretical uncertainties from the renormalization and factorization scale choice ("QCD scale"), jet energy scale and resolution. The result of the fit gives an uncertainty in the EW WZ cross section measurement, which is plotted as a function of integrated luminosity in Fig. 2. The uncertainty is expected to decrease with integrated luminosity and approach 3-4% at 3000-6000 $\mathrm{fb}^{-1}$, where the systematic uncertainties will dominate the accuracy of the measured cross section.

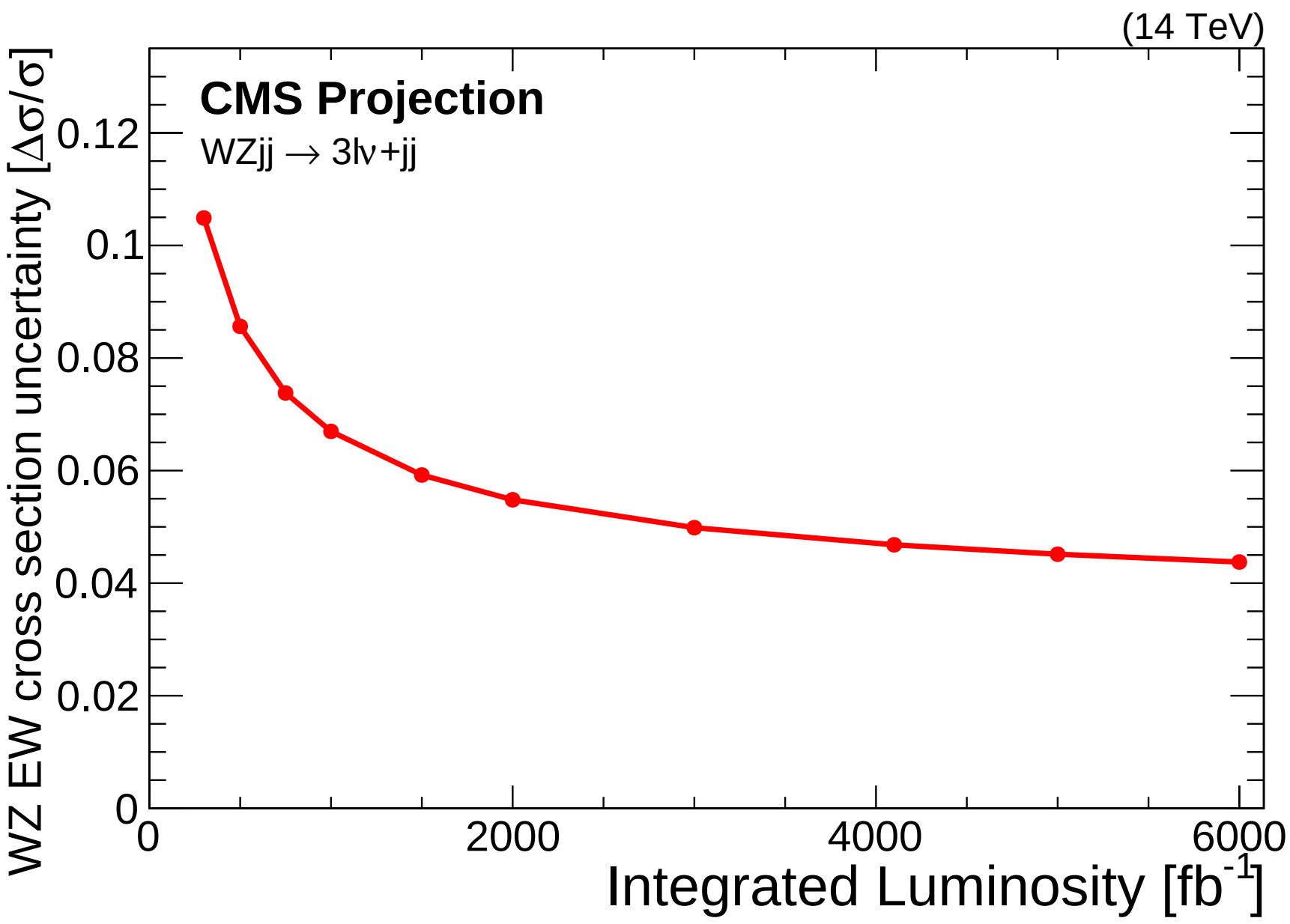

Figure 2: The uncertainty in the EW WZ cross section measurement as a function of integrated luminosity.

The polarized LL component of the EW WZ process is of the order of 5% of the total EW WZ cross section, but since it has a pronounced dependence on the angular separation between jets, one can extract the significance of the LL observation using the same fit procedure as described above. The portion of the LL cross section in the total EW cross section is shown in Fig. 3 left, the uncertainties are statistical only. The right plot shows the event yields for the LL polarized and non-LL polarized portions of the total EW cross section for 3000 $\mathrm{fb}^{-1}$. The LL contribution increases from 2-3% to 7-8% for high angular separation between jets and for high invariant mass of the dijet system. The distribution shown in the right plot is then used in the previously described fit instead of the total EW contribution. The LL is considered as a signal, the non-LL is considered as additional background together with other backgrounds shown in Fig. 1. The systematic uncertainties of the LL and non-LL portions within the EW cross section are considered as fully correlated. Since the LL yields are small and statistical uncertainties in each bin dominate, any additional systematic uncertainties that may change the LL to non-LL ratio are neglected. The significance of the LL observation as a function of integrated luminosity is shown in Fig. 4. The red curve presents the significance if only statistical uncertainties of the measurements are taken into account and the black line includes also systematic uncertainties. There are different possible improvements under discussion that may increase the sensitivity of this measurement in the future. The $\Delta\phi$ and $\Delta\eta$ separation between jets can be used separately, thus increasing the complexity of the fit to a three dimensional, multivariate approach with few variables may also increase significance of the LL measurement. Such studies will require significant increase in statistics of the MC samples and full simulation of the CMS detector response.

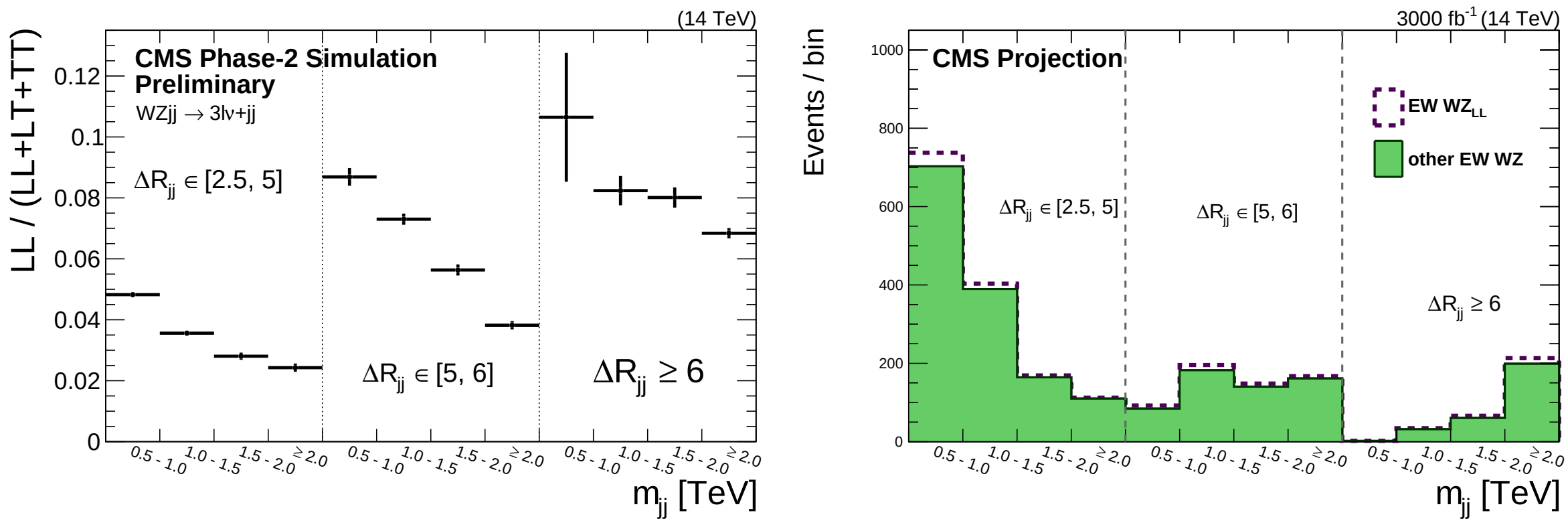

Figure 3: Graphs of unrolled 2D $\Delta R_{jj}; m_{jj}$ distribution of EW WZ. Left is a ratio of the LL portion of the EW WZ to the total sample and right is a stack plot of the LL portion (seen as signal) to the non-LL portion.

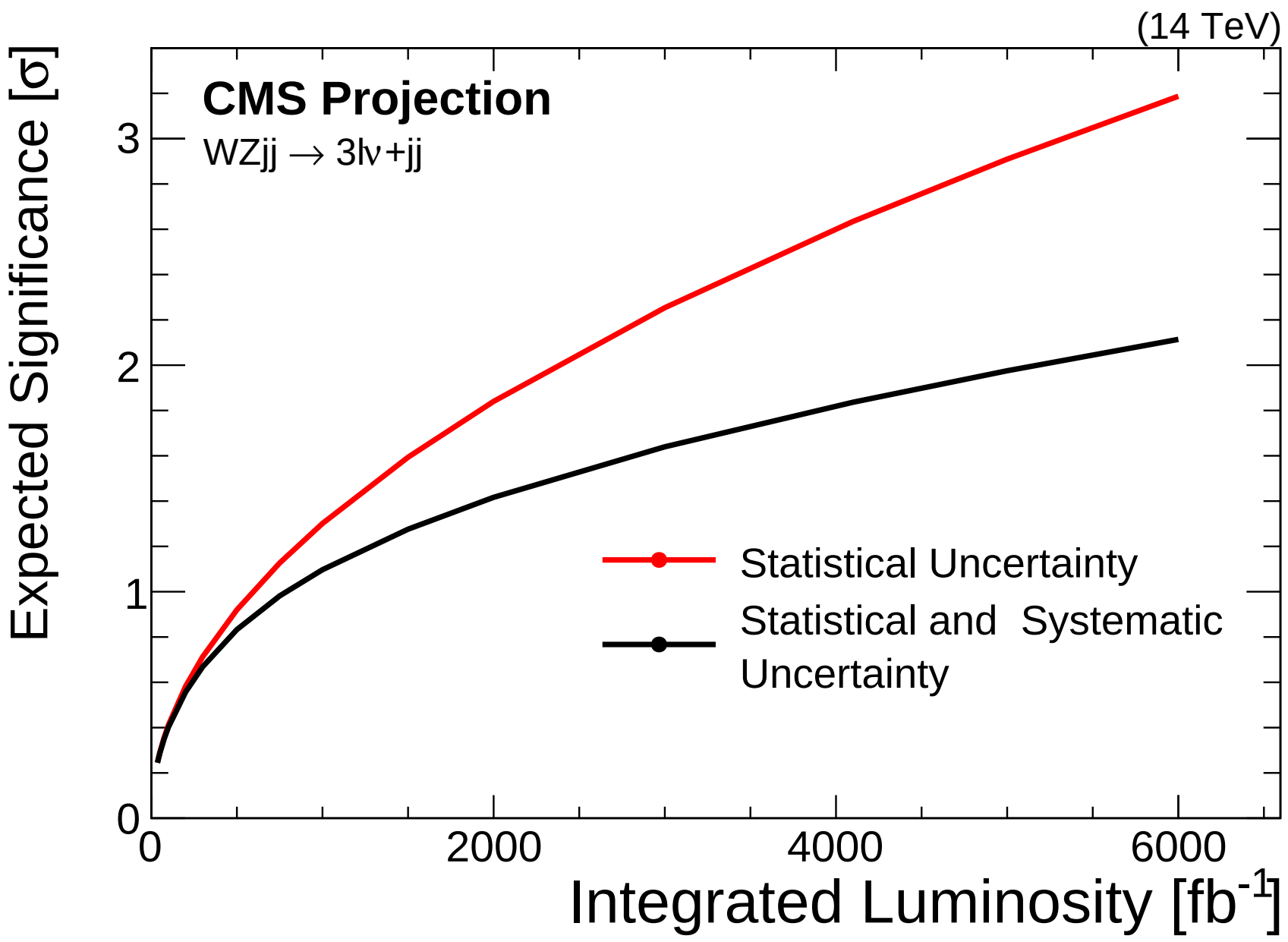

Figure 4: The expected significance to observe the LL portion of the EW WZ process as a function of luminosity.

The results presented in this note used the projection of existing Run 2 results at 13 TeV to estimate the uncertainty in the EW WZ measurement and to explore a possibility to measure the longitudinal part of the EW WZ cross section at the HL-LHC. The accuracy of the EW WZ cross section measurement is expected to significantly improve, down to 3–5% at 3000 fb$^{-1}$ of the integrated luminosity. The measurement of the LL polarized component of the EW WZ process will require improved analysis techniques, such as machine learning, or combining with results from additional WZ decay channels or other EW VBS measurements.

**ATLAS PUB Note**

ATL-PHYS-PUB-2018-023

October 29, 2018

# Prospective study of vector boson scattering in WZ fully leptonic final state at HL-LHC

The ATLAS Collaboration

Prospects are presented for measuring at HL-LHC in ATLAS phase II, the electroweak production of WZ in fully leptonic final state. Different detector setups are evaluated and compared to the phase 0 ATLAS detector; they include an extented tracker, a high granularity timing detector and a forward muon-tagger. In addition different conditions of pile-up event rejection are considered. Finally, studies on the polarisation of the vector bosons are reported. The results presented in this note were obtained with an integrated luminosity of 3000 fb$^{-1}$ at a centre of mass energy of 14 TeV.

# 1 Context of the study

The study of the electroweak Vector Boson Scattering (VBS) is an important goal of the LHC physics program, as it gives a direct access to the nature of the electroweak symmetry breaking mechanism, complementary to the study of the BEH boson properties. In this respect, of particular importance, is the study of the longitudinal states scattering of the vector bosons. Another relevant aspect lies in the probe of the non-abelian structure of the Standard Model via the sensitivity tests to triple and quartic gauge couplings.

In proton-proton collisions, VBS results from the interaction of two bosons radiated by the initial quarks and leading to a final state with two bosons and two jets and consists of purely electroweak processes which cannot be separated from other electroweak processes resulting in the same final state. An observation of the electroweak $WZ$ production exploiting the fully leptonic final states was first presented at the 2018 ICHEP conference [1] using 36.1 fb$^{-1}$ at 13 TeV. ATLAS has also set limits on anomalous quartic gauge couplings using $WZ$ final states with Run1 data [2, 3]. The results obtained so far are strongly limited by the available statistics: the increase of luminosity foreseen by the High-Luminosity upgrade of the LHC (HL-LHC) is mandatory to fully exploit the physics potential behind VBS.

The HL-LHC is currently expected to begin operation in the second half of 2026, at the energy of 14 TeV in the centre-of-mass and with a nominal levelled instantaneous luminosity of L = 5-7 $10^{34}$ cm$^{-2}$ s$^{-1}$ corresponding roughly to an average number of inelastic pp collisions $< \mu >$ of 140 to 200 for each beam crossing, and delivering an integrated luminosity of around 250-300 fb$^{-1}$ per year of operation. The design target is to collect 3000 fb$^{-1}$ in 10 years of operation[1].

To cope with the expected conditions, such as high pile-up, radiation doses and occupancies as well as large data transmission rates, the ATLAS detector will be upgraded. In particular the current Inner Detector will be replaced with an all-silicon Inner Tracker (ITk), which is described in [4] and will extend the tracking capabilities to larger $\eta$. A forward muon tagger is envisaged to extend the muon acceptance [5]. A new detector, a high granularity timing detector (HGTD) [6] designed to mitigate pile-up effects is also foreseen in the forward region. Finally, the online data acquisition system will be upgraded. This will allow the single lepton trigger threshold to be maintained similar to this of Run2 [7]. The other planned upgrades to the ATLAS detector are described in detail in the Scoping Document [8].

This note concentrates on VBS in $WZ$ final state with both bosons decaying in channels with electrons and muons. Decays of the $W$ or $Z$ via intermediate $\tau$ leptons decaying leptonically are included in the signal. The note presents the prospects for the measurements at a centre-of-mass energy of 14 TeV at the HL-LHC, of the cross section and the polarisation fractions with the planned upgraded ATLAS detector. In particular it is assumed that it will be possible to identify electrons and muons up to $|\eta^{\text{lep}}| = 4$ and likewise to associate jets to the hard-scattering vertex up to $|\eta^{\text{jet}}| = 3.8$. Since the topology of electroweakly produced di-bosons events consists of central bosons accompagnied with two high energy forward jets, the analysis benefits fully from this upgrade. The signal events analysed here, are included in the MC sample defined as $WZ - EW$, the main background $WZ - QCD$ is represented by events with the same final states but mediated by strong interactions and where the two gauge bosons are not the result of a scattering process.

[1] An 'ultimate' performance of L = 7.5 x $10^{34}$ cm$^{-2}$ s$^{-1}$ and 4000 fb$^{-1}$ is also under consideration.

The main challenge is to discriminate the signal from the $WZ - QCD$ background. This is achieved here with a standard cut-based event selection and can be improved or supplemented using a Boosted Decision Tree (BDT) [9], method which was used in [1].

## 2 Simulation

Signal and background processes are generated at a centre-of-mass energy of $\sqrt{s} = 14$ TeV and the number of events are scaled to an integrated luminosity of 3000 $\text{fb}^{-1}$ as expected for the nominal HL-LHC program. The signal $WZ - EW$ is simulated at LO using Sherpa 2.2.2 [10]. The dominant background $WZ - QCD$ as well as $ZZ - QCD$ are simulated at NLO, while $ZZ - EW$ is simulated at LO, all using Sherpa 2.2.0. The above samples are generated with the NNPDF30NNLO [11] probability density functions. The other backgrounds considered $t\bar{t}V$ and $tZ$ are generated using MadGraph5_aMC@NLO [12] interfaced with Pythia8 [13] and use respectively NNPDF23LO [14] and CTEQ6L1 [15] probability density functions.

The studies carried out in this note rely on a fast simulation based on the parametrisation of detector effects [16]. The trigger, reconstruction and identification efficiencies, the energy and transverse momentum resolution of leptons and jets are computed as function of their $\eta$ and $p_T$ using a full simulation of the ATLAS detector [17] based on GEANT4 [18] and are tabulated in performance functions. The following detector effects are implemented.

- The electron energy and the transverse momentum of muons[2] and jets are smeared using a gaussian shape with width values (depending on their $\eta$ and $p_T$ ) as returned by the performance functions. In addition, electrons and muons are dressed with prompt photons laying in a cone of $\Delta R = 0.1$ around the lepton direction, by adding their contributions to the lepton four-momentum.
- Final-state muons, electrons and jets are selected in order to reproduce statistically the reconstruction and identification efficiency measured in full simulation.
- Each final-state muons and electrons is flagged to reproduce statistically the single lepton trigger efficiency. The threshold is set at 24 GeV for electrons and muons.
- Jets from additionnal proton-proton interactions (PU) generated with Pythia8 and with $<\mu> = 200$ are added to the event record.
- Fake electrons are introduced according to the probability that a jet fakes an electron as measured in full simulation, where the jet can come from the hard scattering (HS) vertex or from the PU. When a jet is reconstructed as an electron its energy is changed accordingly.
- All jets below 100 GeV and $|\eta^{\text{jet}}| < 3.8$ are associated to the HS process depending on a probability based on the charged vertex fraction $R_{\text{pt}} = \Sigma_{\text{tracks}} p_T / p_T^{\text{jet}}$. In the following, this procedure is called track-confirmation (TC).
- The particle level missing energy is smeared according to its resolution measured in full simulation also taking into account the mean number of interactions in the event $<\mu>$.

Comparisons of the results of the fast simulation to the full analysis substantiate this approach. Several declinations of the effects listed above are considered:

[2] For muons the $p_T$ smearing is done in such a way that it allows for charge-flip.

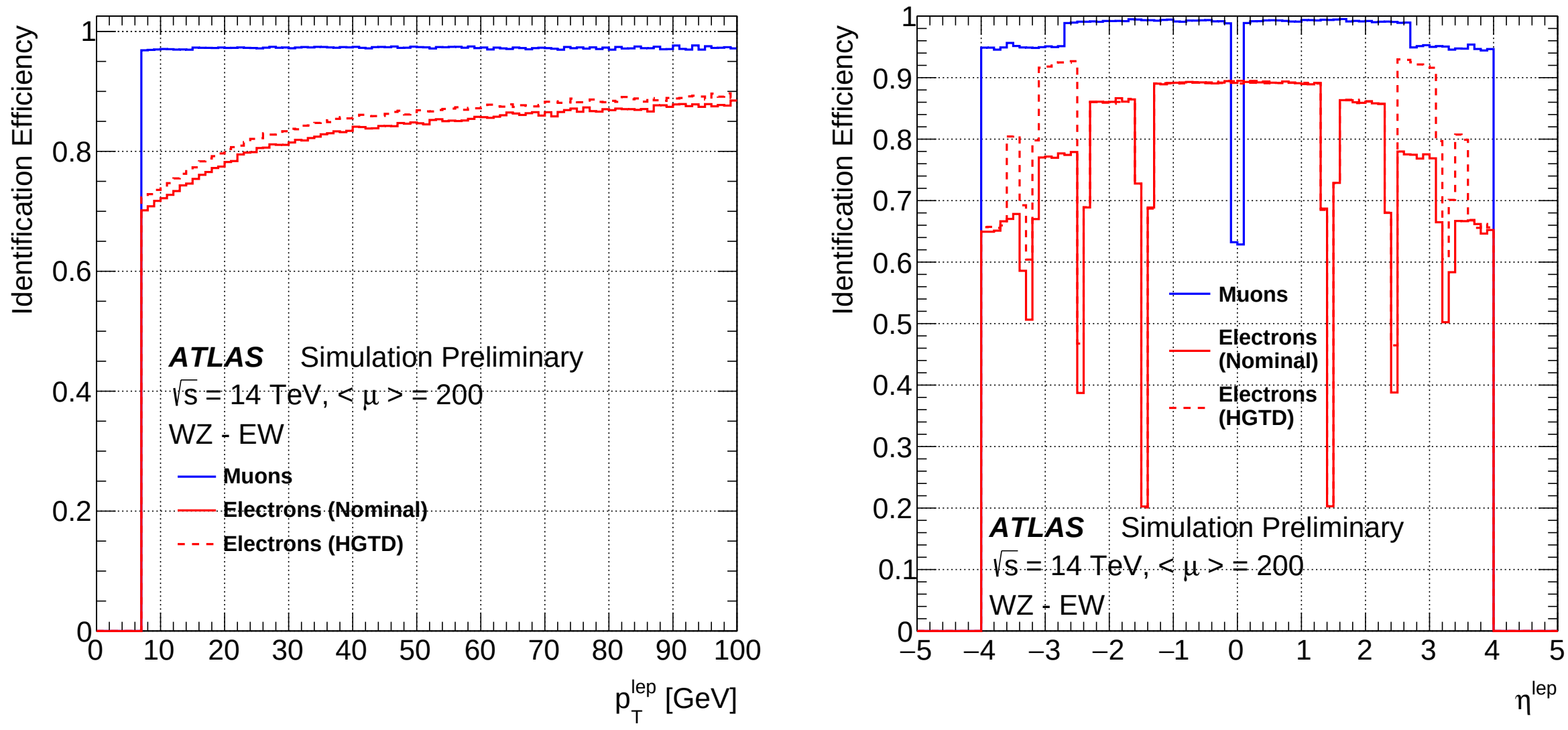

Figure 1: Left: Electrons and muons efficiencies versus $p_T^{lep}$. Right: versus $\eta^{lep}$ for leptons with $p_T^{lep}$ > 7 GeV and $|\eta^{lep}| < 4$. The improvement obtained by using the HGTD information is visible on the electron performance.

**[Nominal]** Leptons are identified up to $|\eta^{lep}| = 4$, and TC is used up to $|\eta^{jet}| = 3.8$. The probability to misidentify a PU jet as resulting from the HS process is 2%.

**[HGTD]** The simulation uses the same acceptance as above but the performance functions include the effect of the HGTD in the forward region ($3.8 < |\eta| < 2.4$). A 4 layer geometry is assumed.

**[High PU rejection]** The PU misidentifcation probability is set at 0.5% at the price of a less good HS jet reconstruction efficiency in the central region. This working point is also used in combination with the HGTD.

Figures 1 to 6 display for the effects described above, the results of different sets of performance functions. In Figures 1 displaying the lepton efficiencies versus $p_T$ and $\eta$, the expected improvement brought by the HGTD in the forward region on the electron efficiency is visible. Figures 2 indicate that the missing $E_T$ resolution is dominated by the PU effect as hardly no difference can be observed between $ZZ$ and $WZ$ final states. As shown in Figures 3, fake electrons represent less than 10% of electrons above a typical cut of 20 GeV used in the analysis; the expected improvement using the HGTD is also visible. The efficiency to associate a jet to the HS vertex is displayed in Figures 4: for $|\eta^{jet}| > 3.8$, all jets are associated to the HS vertex and the efficiency is 1. The results of the two PU rejection scenarios are also shown as well as the expected improvement brought by the HGTD in the forward region. Figures 5 represent the distribution of PU jets versus $p_T$ and $\eta$ associated to the HS vertex: PU jets with $p_T$ > 100 GeV or jets with $|\eta| > 3.8$ are all associated to the HS vertex. Finally, Figures 6 represent the fraction of HS jets in $WZ - EW$ and $WZ - QCD$ versus $p_T$ and $\eta$ before selection: jets in $WZ - QCD$ events are significantly more contaminated by PU jets.

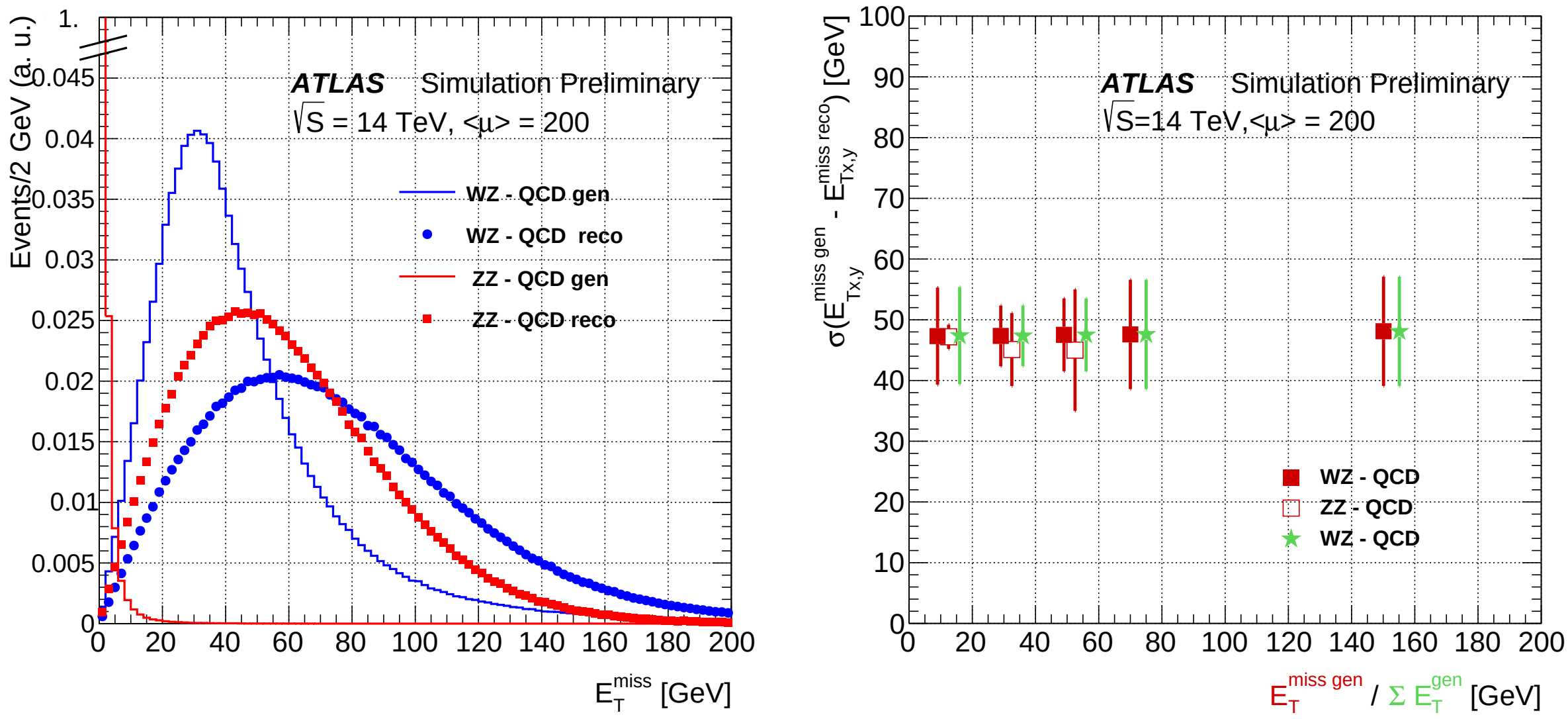

Figure 2: Left: Missing $E_T$ normalised distributions. Right: Missing $E_T$ resolutions in $WZ - QCD$ and $ZZ - QCD$ events.

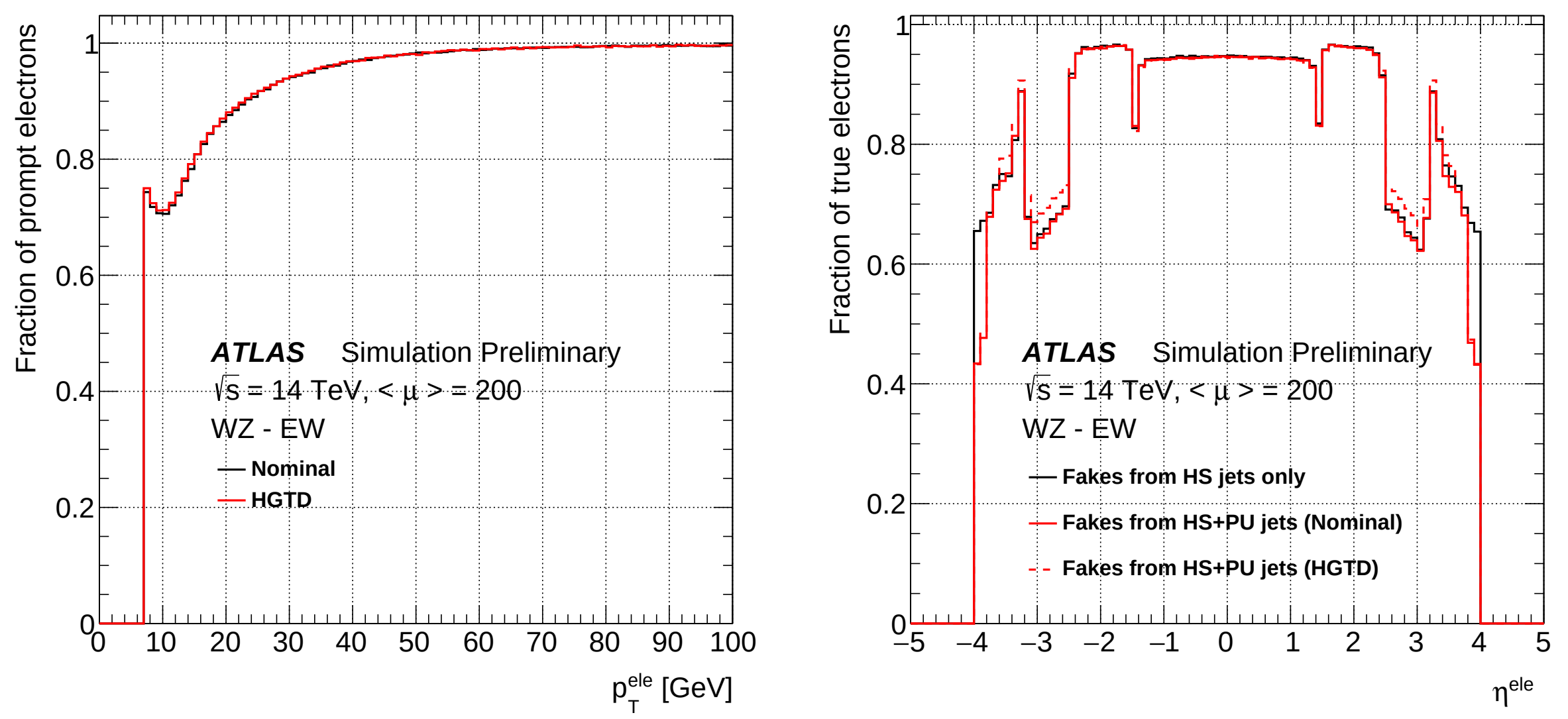

Figure 3: Left: Fraction of prompt electrons with respect to all - including fakes versus $p_T^{lep}$. Right: versus $\eta^{lep}$ for electrons with $p_T^{lep} > 7$ GeV and $|\eta^{lep}| < 4$ in signal events. Above a typical cut in $p_T^{lep}$ at 20 GeV, the fraction of fake electrons is less than 10%. An improvement due to using the HGTD information is visible in the $\eta^{lep}$ distribution in the forward regions.

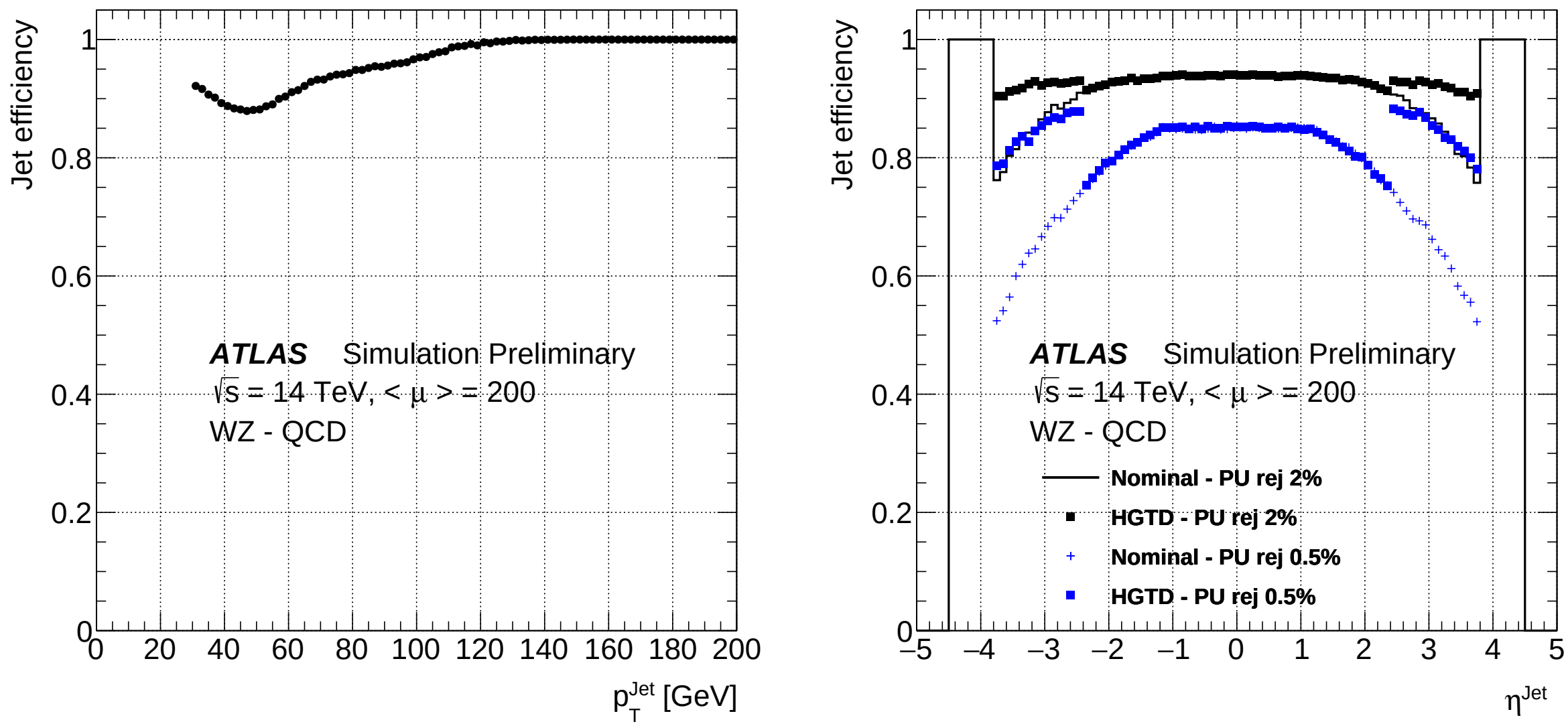

Figure 4: Left: Efficiency to associate a jet to the HS vertex versus $p_T^{jet}$. Right: versus $\eta^{jet}$ for jets with $p_T^{jet} > 30$ GeV and $|\eta^{jet}| < 4.5$. In the right figure, several working points mentioned in the Section 2 are represented; above $|\eta^{jet}| > 3.8$, every jet is considered as coming from the HS vertex.

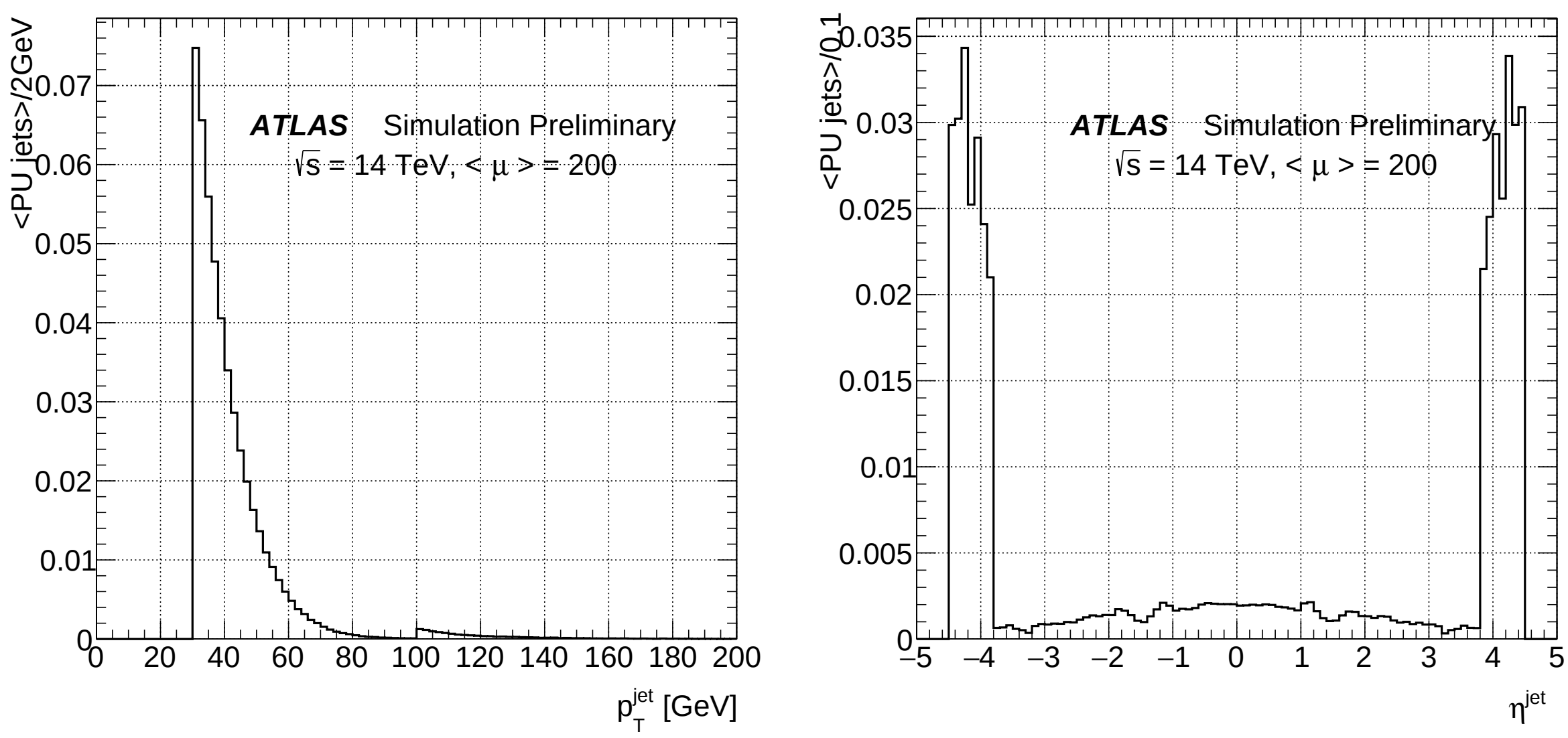

Figure 5: Left: $p_T$ normalised distributions. Right: $\eta$ normalised distributions of the mean number of PU jets for jets with $p_T^{jet} > 30$ GeV and $|\eta^{jet}| < 4.5$, after TC is applied.

6

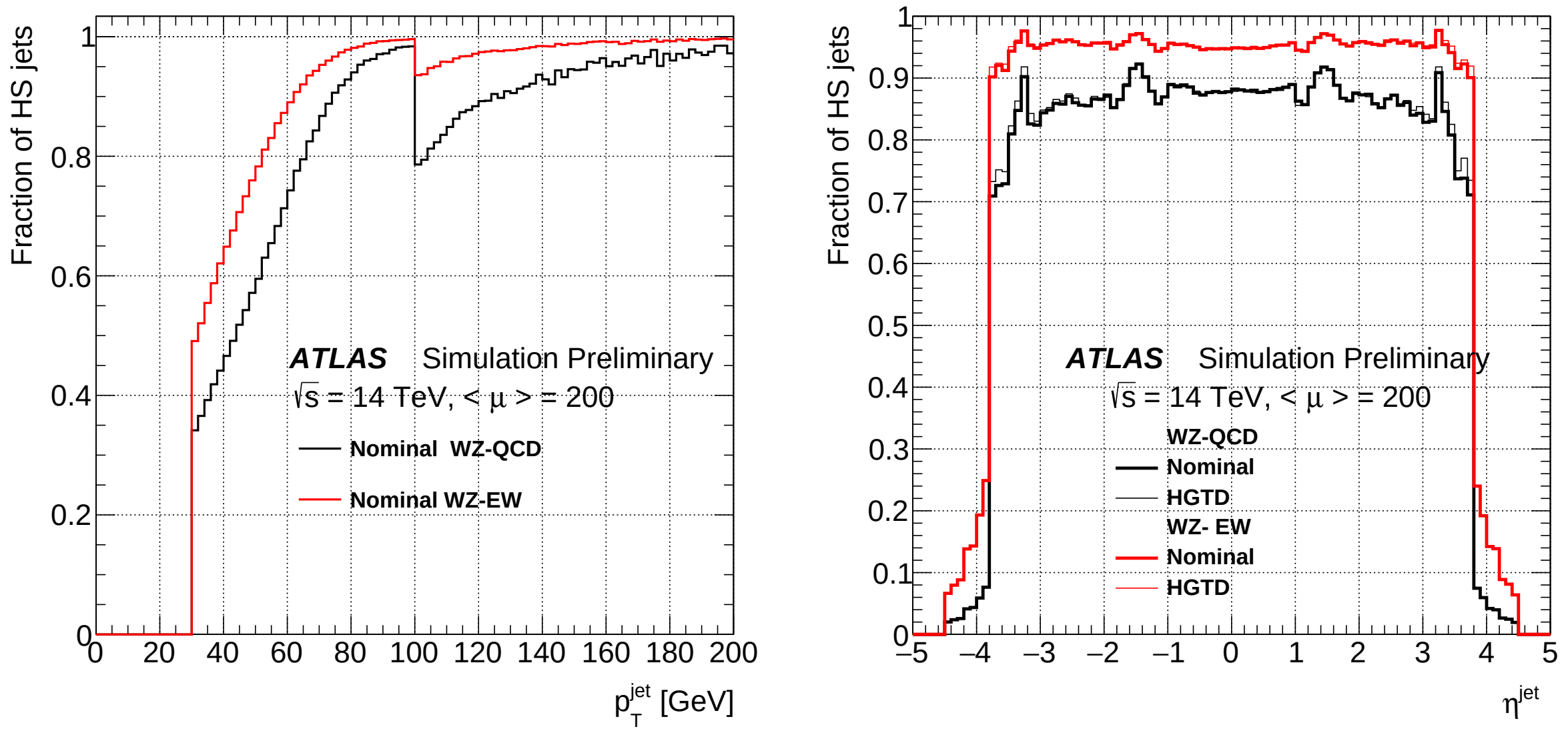

Figure 6: Left: Fraction of HS jets with respect to all jets - including PU jets - versus $p_T^{jet}$. Right: versus $\eta^{jet}$, for jets with $p_T^{jet}$ > 30 GeV and $|\eta^{jet}| < 4.5$ in $WZ - QCD$ and $WZ - EW$ events with and without the HGTD.

# 3 Event selection

The event selection follows the strategy developed for the Run2 analysis [1].

Events with three lepton candidates with $p_T^{lep}$ >15 GeV and $|\eta^{lep}| < 4$ are selected. At least one of the three leptons is required to have $p_T^{lep}$ > 25 GeV and it is checked in addition that at least one lepton passes the single lepton trigger. In order to suppress the background from $ZZ$ processes, events containing a fourth lepton with $p_T^{lep}$ > 7 GeV are discarded.

The event must have at least one pair of leptons of the same flavor and opposite charge (SFOC), with an invariant mass that is consistent within 10 GeV with the nominal $Z$ boson mass, $M_Z$ = 91.188 GeV. This pair is considered as a $Z$ boson candidate. If more than one pair is found, the pair whose invariant mass is closest to the nominal $Z$ boson mass is taken as the $Z$ boson candidate.

The third lepton is assigned to the $W$ boson. It is required to satisfy more stringent criteria than those required for the leptons attributed to the $Z$ boson: the $p_T$ threshold for this lepton is increased to 20 GeV.

Finally, the transverse mass of the $W$ candidate ($m_T^W$) computed using the missing energy of the event ($E_T^{miss}$) and the $p_T$ of the third lepton is required to be above 30 GeV.

The above selection will be referred as the $WZ$ inclusive selection. To select $WZ - EW$ events, additional criteria are applied.

At least, two jets with $p_T^{jet}$ > 30 GeV, laying into opposite detector hemisphere and with an invariant mass $M_{jj}$ greater than 200 GeV are required. This last cut is used to suppress the triboson events contribution present in the simulation of signal events. If there are more than one jet pair, the one with highest $p_T$ jets is chosen. This step will be referred as the VBS preselection.

| | ATLAS | YR |
|---|---|---|
| **Leptons** | 3 leptons | |
| $p_T^{lep}$ | > 15 GeV | |
| $\|\eta^{lep}\|$ | < 4.0 | < 3.0 |
| | At least one lepton with $p_T^{lep}$ > 25 GeV | |
| $ZZ$ Veto | No extra leptons with $p_T^{lep}$ > 7 GeV | |
| $Z$ boson | SFOC lepton pair<br>$\|m_{ll} - M_Z\|$ < 10 GeV | |
| $W$ boson | $p_T^{lW}$ > 20 GeV<br>$m_T^W$ > 30 GeV | |
| **Jets** | 2 jets | |
| | $p_T^{jet}$ > 30 GeV<br>$\|\eta^{jet}\|$ < 3.8 (see text)<br>opp. hemisphere<br>$M_{jj}$ > 200 GeV | $p_T^{jet}$ > 30 GeV for $\|\eta^{jet}\|$ < 3.8<br>$p_T^{jet}$ > 50 GeV for 3.8 < $\|\eta^{jet}\|$ < 4.5<br>$\|\delta_{jj}\|$ > 2.5<br>$M_{jj}$ > 200 GeV |
| **Final selection** | | |
| Benchmark | $M_{jj}$ > 500 GeV | |
| Optimised | $M_{jj}$ > 600 GeV<br>or BDT | |

Table 1: Event selection for the $WZ - EW$ signal.

The final selection requires an additional cut on the invariant mass $M_{jj}$ greater than typically 500 GeV. This is also the benchmark prescription for the Yellow Report (YR) on HE/HL-LHC physics (Section 4.3). Alternatively a BDT-based selection is developed and is presented in the section 5. The event selection is summarised in Table 1.

## 4 Comparitive studies between Run2 and HL-LHC setups

In order to be able to compare the performance of an upgraded detector with the current version of the ATLAS detector, the analysis was performed using the same event generation, the same level of PU and the expected luminosity for the HL-LHC phase but emulating the Run2 geometrical extension and detector set-up [19]. To draw this comparison, the standard cut on $M_{jj}$ > 500 GeV is applied for the final selection.

### 4.1 Emulating the ATLAS Run 2 setup

Electrons and muons are identified within $|\eta^{lep}| < 2.5$. Similarly, TC is valid up to $|\eta^{jet}| < 2.5$. The Run2 analysis uses jets up to $|\eta^{jet}| < 4.5$ but this selection is not optimised for the level of pile-up expected at HL-LHC [3]. Therefore, several optimisations varying the jet $\eta$ or $p_T$ cut were performed for a standard $M_{jj}$ cut in the range from 500 GeV to 600 GeV: the optimisation is performed to maximize the criterion $S/\sqrt{S+B}$. To illustrate the method, in Figures 7, the optimisation is performed versus a cut on $\eta$ of the jets. Curves resulting from different preselection cuts on $p_T$ and $M_{jj}$ are also shown: while changing the cut on $M_{jj}$ has a marginal effect, increasing the $p_T$ cut leads to a significant improvement. Keeping the full acceptance ($|\eta^{jet}| < 4.5$) , the optimised cut on $p_T$ is found to be high - 70 GeV - as shown in Figures 8. If the optimisation is done in 2 dimensions ($p_T^{jet}$ and $\eta^{jet}$) as shown in Figure 9 left : the $p_T$ cut obtained in this case is not significantly lower than the one found in the case mentionned above.

Since, one of the objective of HL-LHC is to measure and interpret differential distributions, it is felt that maximizing the phase space with keeping the $p_T$ cut as low as possible is preferable. Therefore a preferred optimisation consists in varying the $p_T$ cut for jets with $|\eta^{jet}| > 2.5$, keeping the $p_T$ cut at 30 GeV for jets with $|\eta^{jet}| < 2.5$. This latter option gives a good compromise with 4% more events than in the bare 2D optimisation for a $p_T$ cut raised at 75 GeV in the forward region (Figure 9 right ). Subsequent comparisons will be done with respect to this latest option.

### 4.2 Optimal fiducial phase-space for the ATLAS phase II setup.

In Table 2, the results of the optimised version of the analysis using the Run2 setup described in section 4.1 (column 1) are compared to those corresponding to the HL-LHC setup, where jets are selected up to $|\eta^{jet}| < 4.5$ (column 2). A gain of 60% in signal is observed at the expense of a large increase of $WZ-QCD$ background as it becomes likely to pick-up a PU jet. Consequently, $S/\sqrt{S+B}$ is significantly worse. This drawback can be mitigated by applying a tighter $p_T$ cut (80 GeV) for jets with $|\eta^{jet}| > 3.8$ (column 3): the signal gain is then reduced by 30% and is only marginally greater compared with restricting the jet acceptance to $|\eta^{jet}|$ less than 3.8 (last column).

| | Run2 optimised<br>$p_T > 30$ for $\|\eta\| < 2.5$<br>$p_T > 75$ GeV for $\|\eta\| > 2.5$ | HL-LHC<br>$p_T > 30$ GeV<br>$\|\eta\| < 4.5$ | <br>$p_T > 30$ for $\|\eta\| < 3.8$<br>$p_T > 80$ GeV for $\|\eta\| > 3.8$ | <br>$p_T > 30$ GeV<br>$\|\eta\| < 3.8$ |
|---|---|---|---|---|
| $WZ-EW$ | 3092 | 4942 | 3981 | 3889 |
| $WZ-QCD$ | 19618 | 93985 | 30613 | 29754 |
| $ZZ$ | 1671 | 6654 | 2029 | 1970 |
| $t\bar{t}V$ | 2830 | 4563 | 3189 | 3145 |
| $tZ$ | 1848 | 3190 | 2319 | 2221 |
| Total | 25967 | 108393 | 38150 | 37089 |
| $S/\sqrt{S+B}$ | 18.1 | 14.7 | 19.4 | 19.2 |
| S/(S+B) % | 11 | 4 | 9 | 9 |

Table 2: Expected number of $WZ-EW$ and background events correponding to an integrated luminosity of 3000 $fb^{-1}$. Comparison of the Run 2 optimised selection - with 2 $p_T$ cuts - in the first column with several options at HL-LHC (discussion in the text). The cut on $M_{jj}$ is 500 GeV in all cases.

[3] In the current Run2 analysis, cuts are also placed on the angle in space between the leptons and between jets and leptons; since they do not modify significantly the conclusions, they are not applied here for the sake of simplicity.

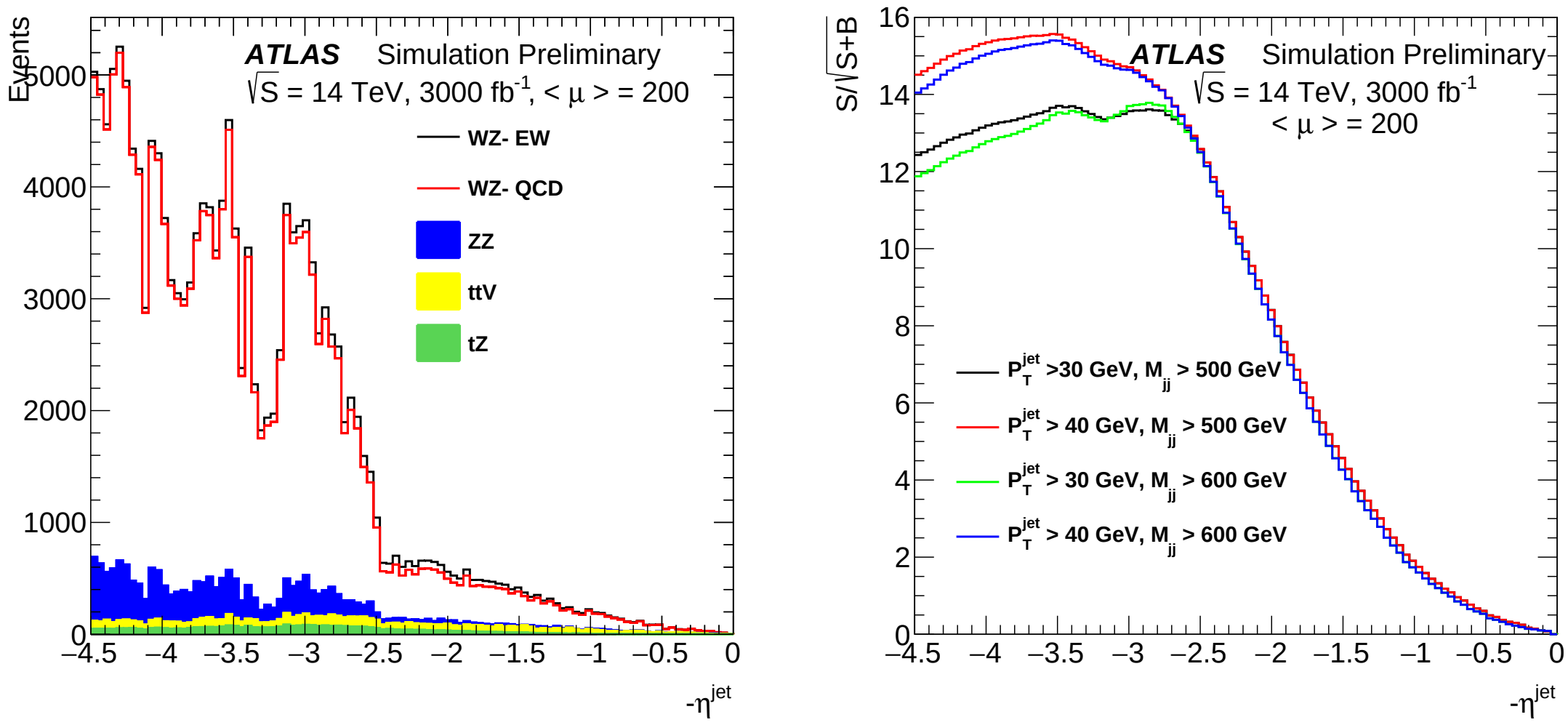

Figure 7: Left: Distribution of -$\eta^{\rm jet}$= -max($|\eta_{\rm jet1}|$, $|\eta_{\rm jet2}|$). Right: Evolution of $S/\sqrt{S+B}$ versus the cut on -$\eta^{\rm jet}$. The different curves correspond to variation of the jet $p_T$ or $M_{jj}$ cut as written in the legend.

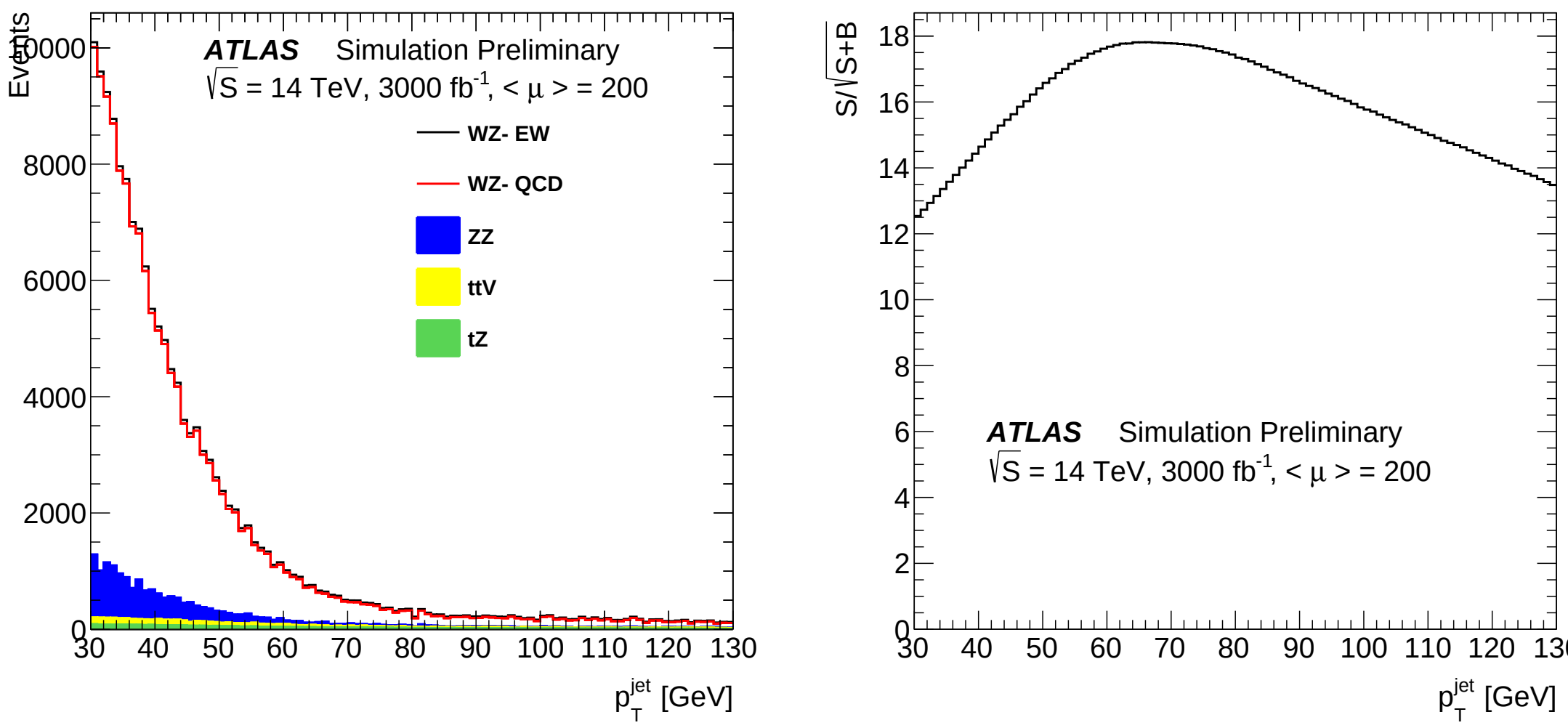

Figure 8: Left: Distribution of the subleading jet $p_T$. Right: Evolution of $S/\sqrt{S+B}$ versus the cut on the subleading of jet $p_T$.

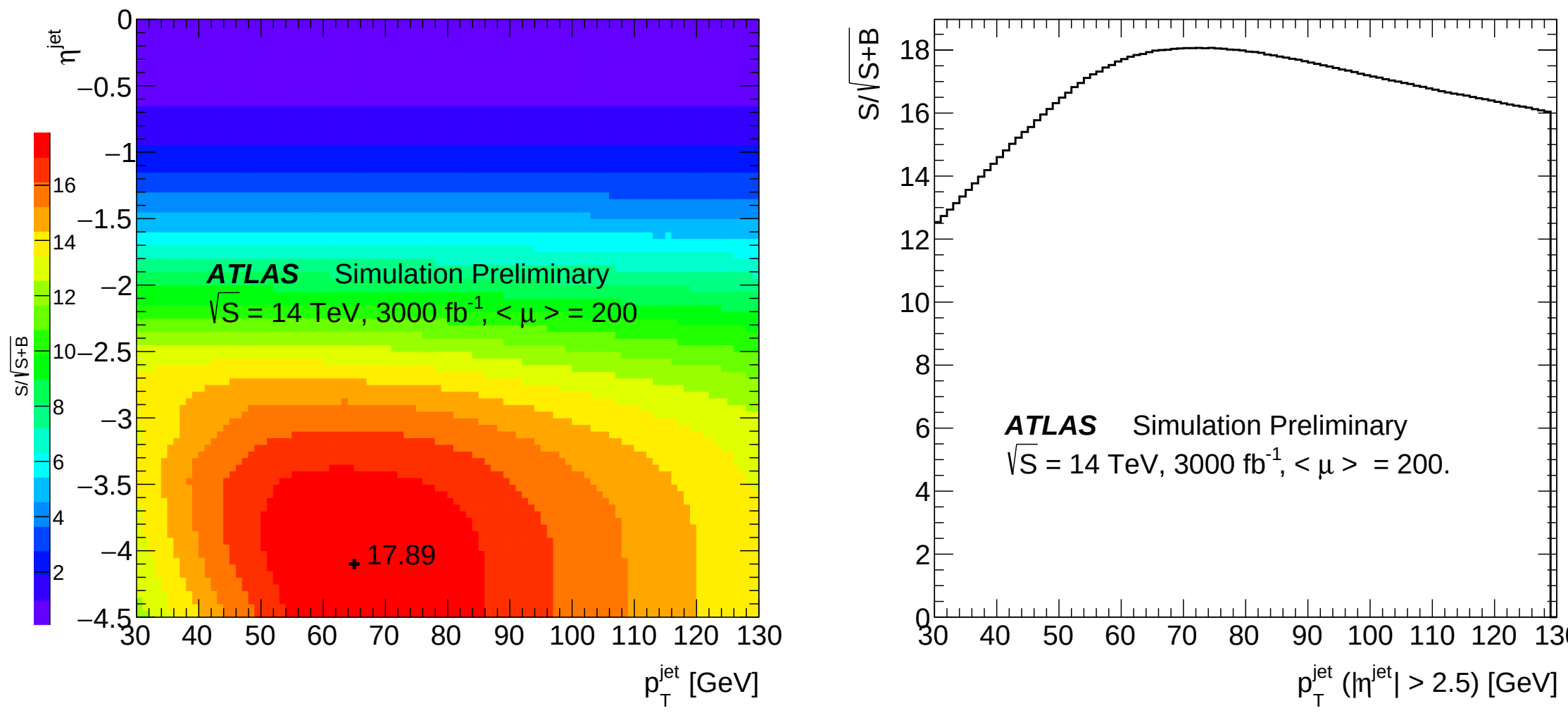

Figure 9: Left $S/\sqrt{S+B}$ in 2 dimensions: the position of the maximum is indicated. Right: Evolution of $S/\sqrt{S+B}$ versus the cut on $p_T^{\rm jet}$ for jets with $|\eta^{\rm jet}| > 2.5$ , while the $p_T$ of the jets with $|\eta^{\rm jet}| < 2.5$ is required to be greater than 30 GeV.

10

Therefore, the nominal fiducial phase-space in the rest of note is limited to $|\eta^{\text{jet}}| < 3.8$ and the results are presented in Table 2 (last column).

### 4.3 Yellow Report recommended selection

For the sake of comparisons with other channels and theoretical projections for the Yellow Report, common cuts were defined and are listed below and in Table 1. They modify the VBS preselection.

- Leptons: $|\eta^{\text{lep}}| < 3.0$
- Jets:

  $p_T^{\text{jet}} > 30$ GeV for $|\eta^{\text{jet}}| < 3.8$, $p_T^{\text{jet}} > 50$ GeV otherwise.

  $|\delta\eta_{jj}| > 2.5$

  $M_{jj} > 500$ GeV

The other cuts are kept identical. The resulting event yield of events are presented in the last two columns of Table 3 for the nominal version and for the option where the jet acceptance is restricted to $|\eta^{\text{jet}}| = 3.8$. It can be observed again that enlarging the jet acceptance increases the $WZ - QCD$ background as it becomes more likely to pick-up a PU jet.

### 4.4 Variations around the nominal ATLAS phase II layout

Along with an extended tracker, a HGTD, covering the high $\eta$ region is under study: the implications of this new detector for the $WZ - EW$ signal are presented in the subsection 4.4.1. In subsection 4.4.2, a restricted acceptance affecting the muon identification is also considered.

#### 4.4.1 HGTD and high PU rejection working point.

As the HGTD primary goal is to palliate the nuisance due to PU jets, its impact was estimated in the two scenarios mentioned in Section 2 corresponding to the different level of PU rejection in the nominal phase-space ($\eta^{\text{jet}} < 3.8$, $p_T^{\text{jet}} > 30$ GeV) and $M_{jj} > 500$ GeV. This is summarised in Table 3. The HGTD improves slightly $S/\sqrt{S+B}$ in the "nominal" PU configuration: the number of signal events is increased by 3.7% , while the number $WZ - QCD$ events is increased by 5.1%, roughly half of the gain being due to the gain in electron efficiency shown in Figure 1. However, considering the working point with a high PU rejection, the gain using the HGTD is more significant as the number of signal events is increased by 11% and the $WZ - QCD$ background by 13%. Additionally, $S/\sqrt{S+B}$ is improved.

Figures 10 enlighten where the gain in efficiency is localised in terms of $\eta^{\text{jet}}$ when the HGTD information is taken into account. It can then be noticed that the gain is provided by the sub-leading jet.

| | Nominal + $\|\eta_\mu\| < 2.7$ | Nominal | HGTD | Nominal + High PU rej | HGTD + High PU rej | YR Nominal | YR $\|\eta^{jet}\| < 3.8$ |
|---|---|---|---|---|---|---|---|
| $WZ - EW$ | 3661 | 3889 | 4033 | 3372 | 3744 | 3929 | 3492 |
| *Rel.* $\epsilon$ | 0.94 | 1. | 1.037 | 0.867 | 0.963 | 1.01 | 0.897 |
| *Rel.* $\epsilon$ | | | | 1. | 1.11 | | |
| *Rel.* $\epsilon$ | | | | | | 1. | 0.889 |
| $WZ - QCD$ | 26852 | 29754 | 31289 | 24261 | 27498 | 39537 | 25194 |
| *Rel.* $\epsilon$ | 0.90 | 1. | 1.051 | 0.815 | 0.924 | 1.33 | 0.847 |
| *Rel.* $\epsilon$ | | | | 1. | 1.13 | | |
| *Rel.* $\epsilon$ | | | | | | 1. | 0.637 |
| $ZZ$ | 2003 | 1970 | 1947 | 1508 | 1751 | 2924 | 1886 |
| $t\bar{t}V$ | 3209 | 3145 | 3209 | 2868 | 3104 | 3420 | 2741 |
| $tZ$ | 2142 | 2220 | 2292 | 1820 | 2062 | 2624 | 2185 |
| Total Back. | 34067 | 37089 | 38738 | 30456 | 34414 | 48495 | 32005 |
| $S/\sqrt{S+B}$ | 18.9 | 19.2 | 19.5 | 18.3 | 19.2 | 17.2 | 18.5 |
| S/(S+B) | 10 | 9 | 9 | 10 | 10 | 7 | 10 |

Table 3: Expected number of $WZ - EW$ signal and background events for an integrated luminosity of 3000 fb$^{-1}$ in different detector configurations, within the acceptance defined by $|\eta^{jet}| < 3.8$, $p_T^{jet} > 30$ GeV and $M_{jj} > 500$ GeV for the first five columns and within the YR report acceptance for the last two columns. The lines entitled *Rel.* $\epsilon$ give the relative efficiency of corresponding configurations relative to the column indicated by 1.

#### 4.4.2 Muon acceptance restricted to $|\eta| < 2.7$.

Considering that a muon tagger efficient up to $|\eta| = 4$ might not be available at the beginning of the HL-LHC phase, the implication of a muon identification acceptance limited with the New Small Wheel [20] to the region up to $|\eta| = 2.7$, was studied; the results are presented in the first column of the summary Table 3. In total, $S/\sqrt{S+B}$ is only marginally worse thanks to a sizeable reduction of the $WZ - QCD$ background. In addition, it was also found that $S/\sqrt{S+B}$ is maximum in the nominal case, ie when the lepton acceptance goes up to $|\eta| = 4$ for both electrons and muons.

### 4.5 Origin of jets

The final state of a VBS process is characterised by to two high $p_T$ quark jet preferentially in the forward regions. It is therefore important to differentiate them from gluon jets and from PU jets given the context of HL-LHC. Table 4 gives the fraction of events with at least one PU jets selected as one of the 2 jets after the final selection. For the signal and the main background, one can see the improvement brought by the extended tracker and the ability to associate jets to the HS vertex permitting to keep the jet $p_T$ cut as low as 30 GeV up to $|\eta^{jet}| < 3.8$.

Additionally, as shown in Figure 11, the parton origin of the sub-leading jet is for 45% of cases a gluon, while jets in signal events are issued preferentially from u or d quarks: an efficient quark/gluon tagger, primarily applied on the sub-leading jet could help to further reduce the $WZ - QCD$ background. This observation triggered several studies based on the full simulation - using HGTD , using multivariate

12

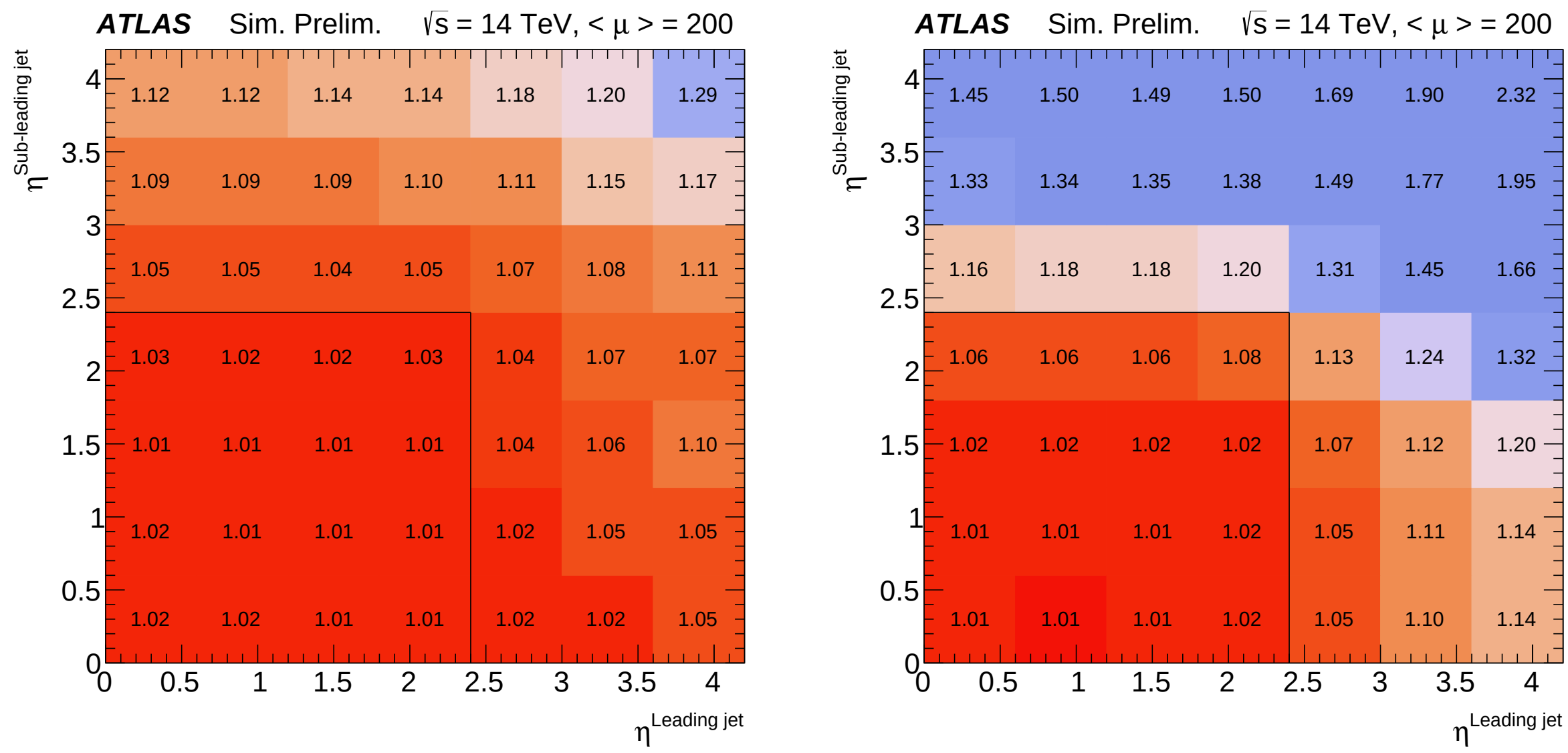

Figure 10: Left: Ratio of the number of events selected using HGTD information to the number of events selected without as function of $|\eta|$ of the leading and sub-leading jet for the nominal working point. Right: for the high PU rejection working point. The lines at $\eta^{\text{jet}} = 2.4$ materialise the boundaries of the HGTD.

| | $WZ - EW$ Frac PU | $WZ - QCD$ Frac PU |
|---|---|---|
| Run 2 | 2.77( 0.07) | 14.29 (0.47) |
| Nominal | 2.08 (0.05) | 11.20 (0.38) |
| Full acc. ($|\eta^{\text{jet}}| < 4.5$) | 17.74(0.14) | 69.37(0.61) |
| Full acc. + $p_T^{\text{jet}} > 80$ GeV | 2.21 (0.05) | 12.45 (0.39) |
| HGTD | 2.11 (0.05) | 10.64(0.37) |
| Nom. + High PU rej. | 1.30(0.05) | 5.52(0.30) |

Table 4: Fraction of events in % (and error) containing at least one PU jets after final selection.

technics - to improve the quark/gluon tagging and to evaluate its performance up to $|\eta^{\text{jet}}| < 4$ within the HL-LHC setup.

# 5 BDT based analysis and optimal $M_{jj}$ selection

A multivariate analysis (BDT) was developped to improve the signal and background separation. Only $WZ - QCD$ background was considered in the training as it is by far the largest contribution and only distinguishable by kinematics while an anti-b tagging can be used to reduce most of the other type of backgrounds.

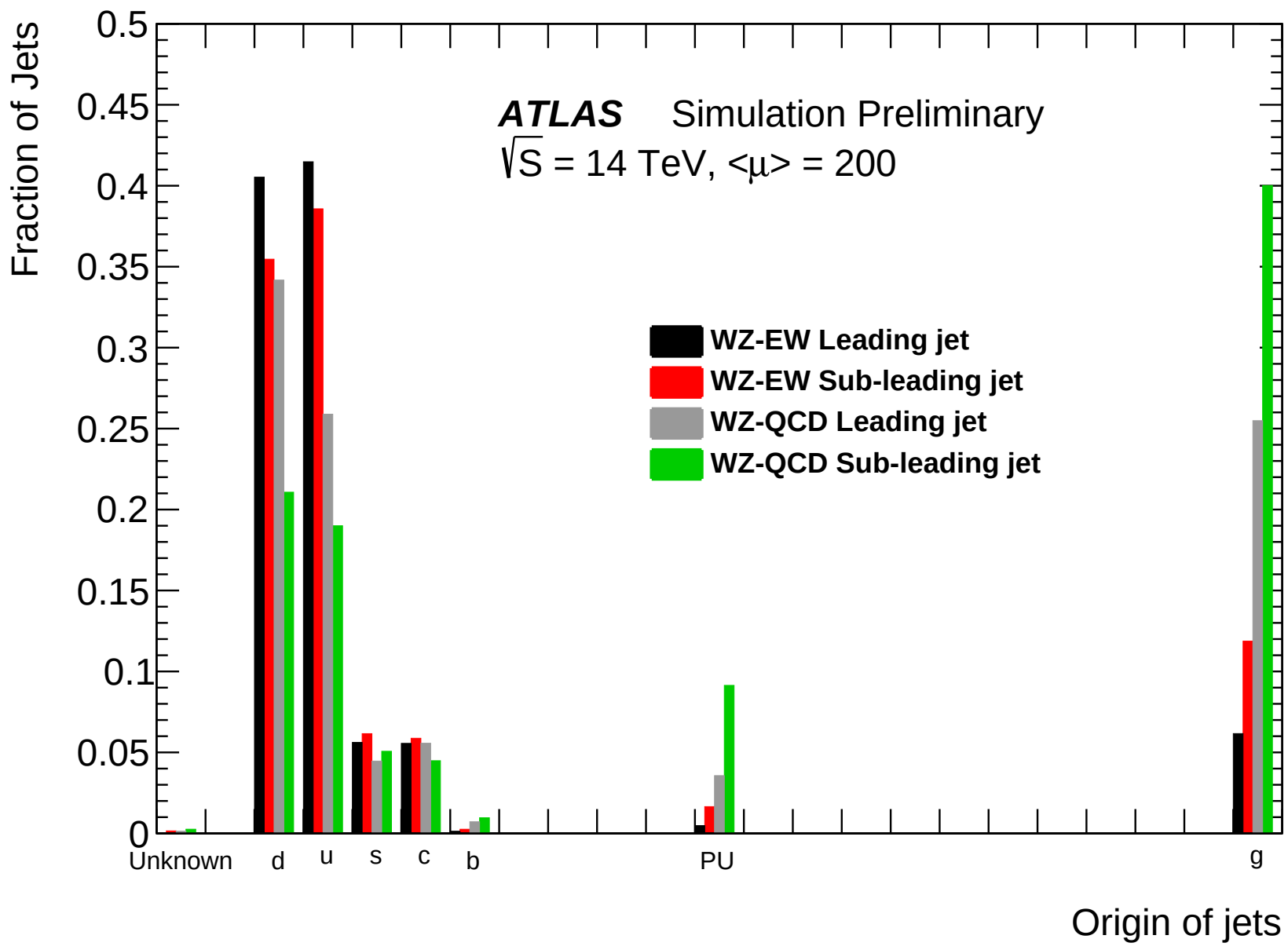

Figure 11: Parton origin of the leading and sub-leading jets in $WZ - EW$ and in $WZ - QCD$ events.

From a set of variables based on jets and leptons kinematics, the 25 best variables given by default by the TMVA package [9] are chosen and they are displayed in Appendix A. The improvement brought by an additionnal variable is not visible. For the training, 40000 signal events, with PU jets but without fake electrons are used and 20000 $WZ - QCD$ events, after the VBS preselection. Events are used unweighted. Two separate BDTs were trained on simulated events with and without HGTD.

Figures 12 compare the performance of a BDT cut with the classical $M_{jj}$ cut and in the right figure, it is quantified the relative gain in efficiency on the $WZ - EW$ signal using a BDT cut for a given $WZ - QCD$ background rejection: the gain in signal efficiency is larger for higher rejection working point.

Figure 13 left displays the distribution of $M_{jj}$ after the VBS preselection. In the Figure 13 right, the evolution of $S/\sqrt{S+B}$ is shown as well as this of $S/\sqrt{B}$. An equivalent content is displayed in Figures 14 for the BDT distribution.

The maximum of $S/\sqrt{S+B} = 19.4$ for the cut based analysis is obtained with a cut at $M_{jj}$ greater than 600 GeV while the maximum of $S/\sqrt{S+B} = 22.1$ for the BDT based analysis is obtained with a cut at 0.44. Relevant numbers are gathered in Table 5. Typically for the same $S/\sqrt{S+B}$, the BDT based analysis gives 40% more events and for the same signal efficiency, a signal purity 27% better.

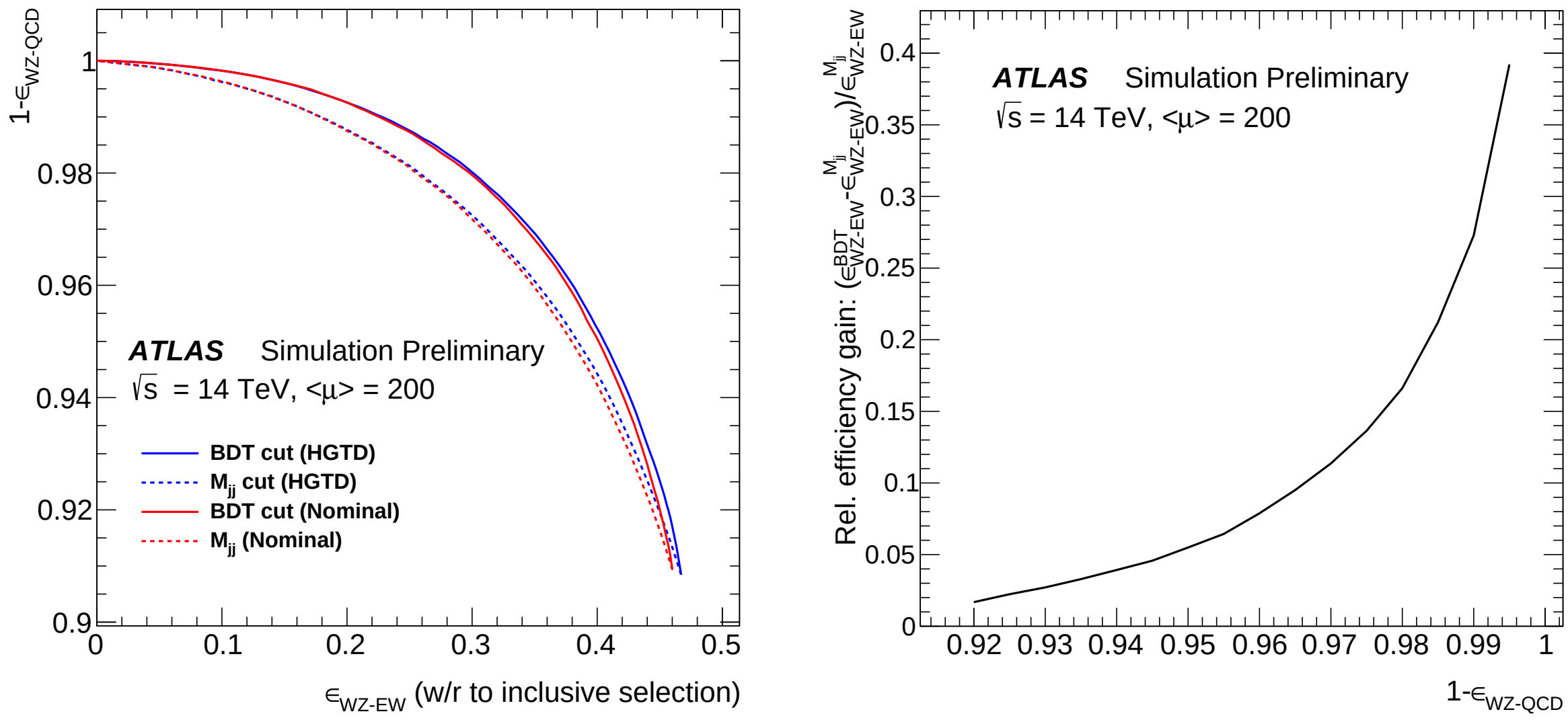

Figure 12: Left: Comparison in the plane $WZ - EW$ efficiency versus $WZ - QCD$ background rejection of the BDT cut with the classical $M_{jj}$ cut. Right: Gain in efficiency with a BDT cut relative to a $M_{jj}$ cut for the same background rejection.

| | $M_{jj}$ cut | $M_{jj}$ cut HGTD | BDT | BDT HGTD |
|---|---|---|---|---|
| $S/\sqrt{S+B}$ | 19.4 | 19.7 | 22.1 | 22.8 |
| N of events | 3422 | 3533 | 2651 | 2887 |
| S/(S+B) % | 11 | 11 | 19 | 18 |
| Same $S/\sqrt{S+B}$ | | | 19.4 | |
| Nb of events | | | 4740 | |
| S/(S+B) % | | | 8.0 | |
| $S/\sqrt{S+B}$ | | | 21.7 | |
| Same Nb of events | | | 3420 | |
| S/(S+B) % | | | 14 | |

Table 5: Summary of optimised performance with different detector options. The BDT based analysis is also compared with the cut based analysis in two additional scenarios: one where $S/\sqrt{S+B}$ is the same and the other one where the signal efficiency is the same.

15

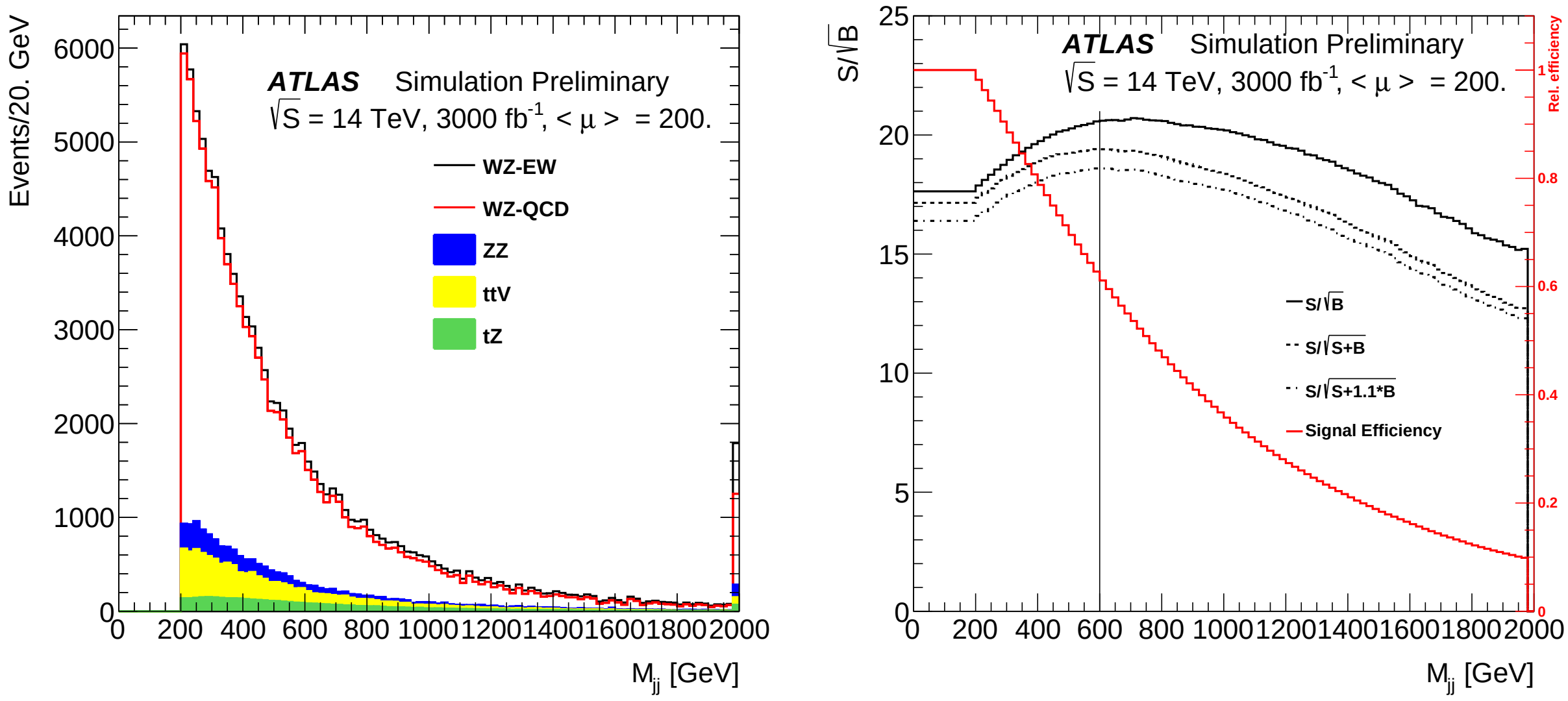

Figure 13: Left: $M_{jj}$ distribution. Right: $S/\sqrt{B}$ evolution with a $M_{jj}$ cut. The signal efficiency is also shown. The vertical line indicates the position where $S/\sqrt{S+B}$ is maximum. $S/\sqrt{S+1.1\times B}$ is also shown to illustrate that the position of the optimum is not very dependant on the knowledge of the level of background.

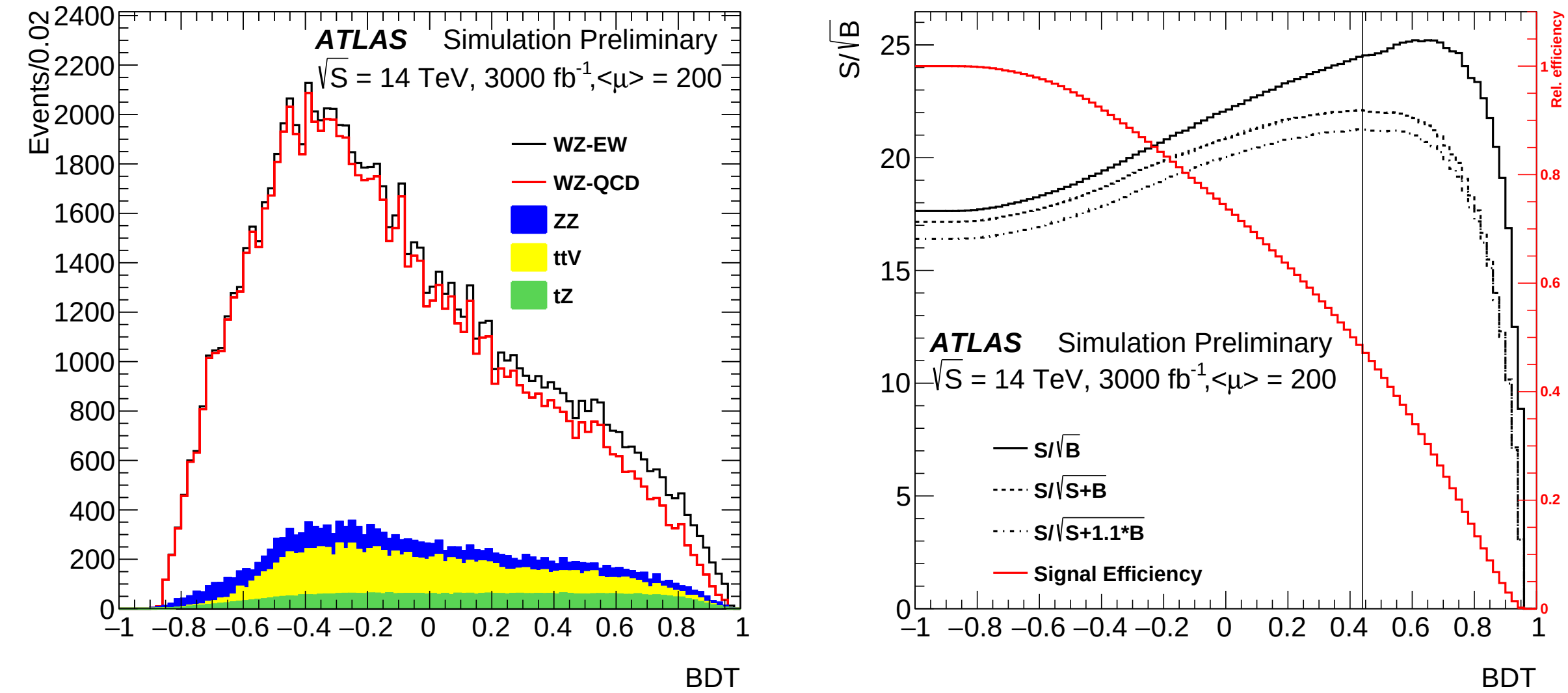

Figure 14: Left: BDT distribution. Right: $S/\sqrt{B}$ evolution with a BDT cut. The signal efficiency is also shown. The vertical line indicates the position where $S/\sqrt{S+B}$ is maximum. $S/\sqrt{S+1.1\times B}$ is also shown to illustrate that the position of the optimum is not very dependant on the knowledge of the level of background.

## 6 Systematics

Sources of systematics on the signal cross-section arise from the theoretical modelling of $WZ-QCD$ and $WZ-EW$ events and from experimental effects. In the 13 TeV VBS analysis [1], the detailed study of uncertainties shows that the former amounts in total to 7.3% ; the latter is by far dominated by the jet related uncertainties and amounts to 6.6%.

Following the recommandations given in [16] and [21], the jet energy scale was varied within 2.5% and jet energy resolution within 15%, leading respectively to a 2% and less that 1% systematic on the signal yield.

Given the signal over background ratio of the order of 1 to 10, the signal significance is limited by the uncertainty on the background as it is shown in Figure 15, and in particular by the theoretical uncertainty associated to the $WZ-QCD$ background of currently 5.8%. The projected experimental uncertainty from the jet related uncertainties amounts to 3.9%. However, it is expected that the background uncertainty can be controlled to a smaller value thanks to refined and diverse control regions, allowed by the larger number of background events. Figure 15 displays two scenarios in addition to the nominal one, where the signal significance is enhanced thanks to a better background rejection obtained either by the BDT explained in Section 5 or by applying a q/g discrimant on the jets: with an increased signal over background ratio, the systematic uncertainty on the background is less critical.

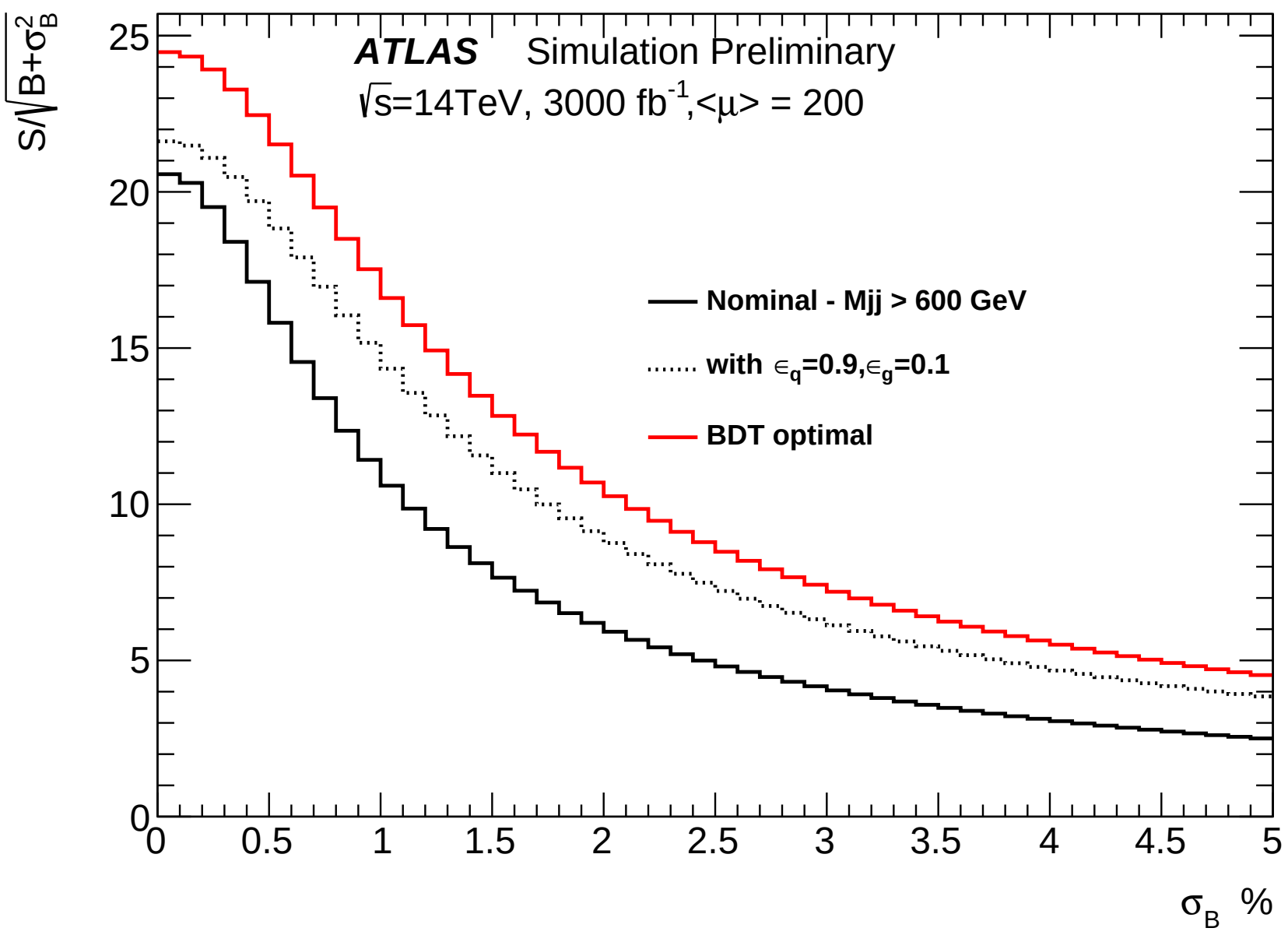

Figure 15: Evolution of the signal significance as a function of the uncertainty on the background.

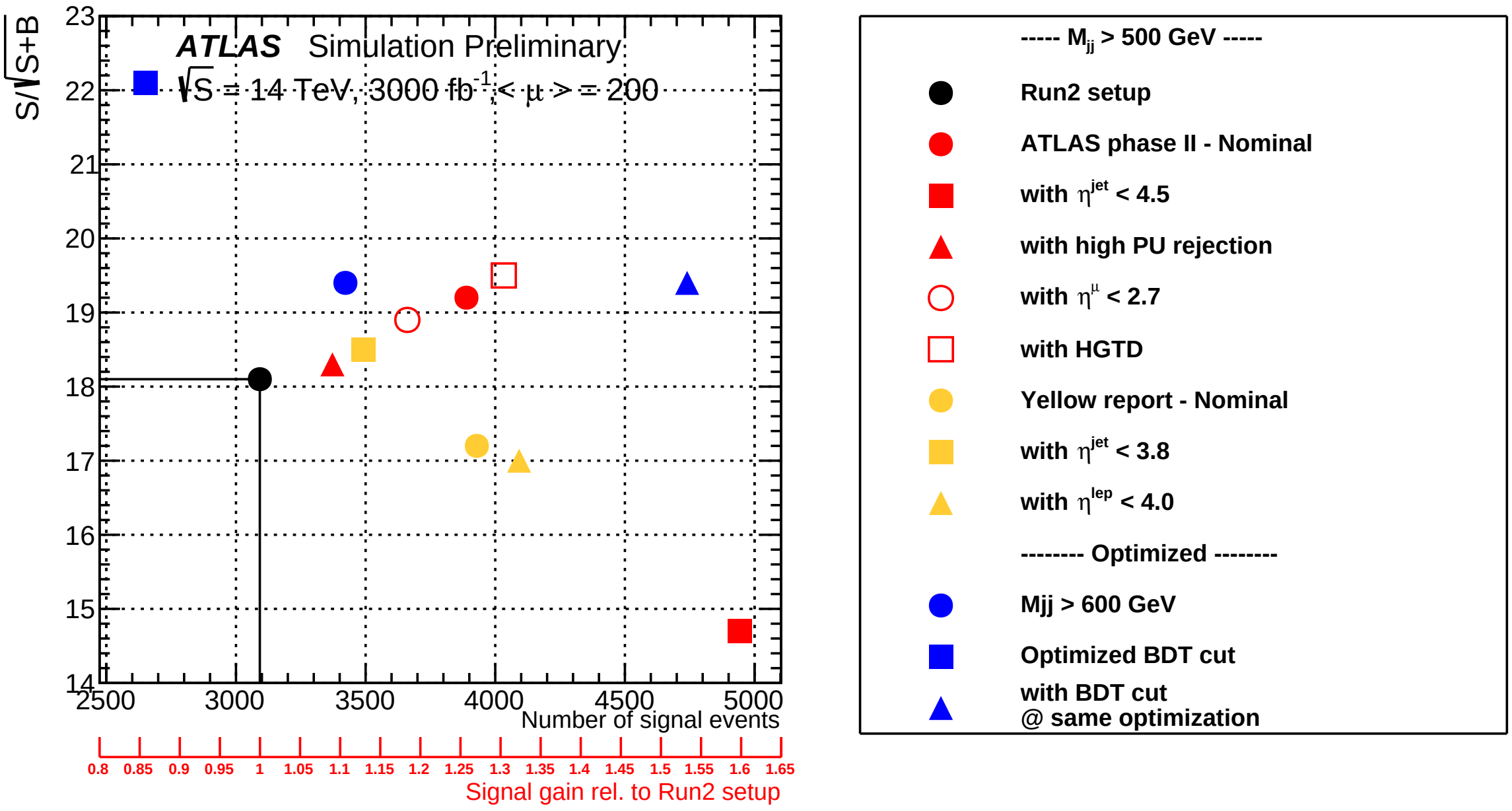

Figure 16: Event yield for a collection of HL-LHC setups and working points.

# 7 Recap of the main conclusions of the comparative studies.

Figure 16 summarises the event yield vs the optimisation criterion $S/\sqrt{S+B}$ for the different working points presented in the text.

The extended tracker setup brings 28% more signal events with an associated $S/\sqrt{S+B}$ more favorable. Within its acceptance, it also allows to maintain the jet $p_T$ cut as low as 30 GeV despite the large background from pile-up events. Though, the expected level of PU precludes considering jets in the full detector acceptance, unless a high $p_T^{jet}$ cut of the order of 80 GeV for $|\eta^{jet}| > 3.8$ is used. This in addition with a less extended $|\eta|$ range for leptons, makes the YR recommended selection non optimal for the ATLAS Phase II detector.

With the performance assumptions simulated by the HGTD performance function, an additional signal gain of 4% is brought by the HGTD detector. In the case where a high PU rejection is needed, the HGTD allows to recover the nominal signal efficiency, with a gain of 11%, the gain being brought essentially by the sub-leading jet.

A BDT based selection brings 40% more events for the same $S/\sqrt{S+B}$ or a purity 27% better for the same signal efficiency than an optimal $M_{jj}$ cut at 600 GeV. The optimal BDT cut gives an efficiency loss of 16% but increases the purity by 67% which, in fine, gives a signal significance greater than 5 if $\sigma_B < 5\%$. The $WZ - QCD$ background can also be mitigated by an efficient quark-gluon tagger since the sub-leading jet origin is a gluon in 45% of events.

# 8 Polarisation studies

In this section, the sensitivity to the measurement of the polarisation is investigated. The measurements of the longitudinal polarisation of the vector bosons, and especially to the doubly longitudinal production linked to the EWSB mechanism are of particular interest.

## 8.1 Methodology

The differential cross-section of a vector boson in the full phase space[4] can be expressed as the sum of the three polarisation states, left L, right R and longitudinal 0, according to the following formulae:

$$\frac{1}{\sigma}\frac{\mathrm{d}\sigma}{\mathrm{d}\cos\theta^*_{\mathrm{W}^\pm}} = \frac{3}{8}\mathrm{FL}(1 \mp \cos\theta^*_{\mathrm{W}})^2 + \frac{3}{8}\mathrm{FR}(1 \pm \cos\theta^*_{\mathrm{W}})^2 + \frac{3}{4}\mathrm{F0}(1 - \cos{\theta^*_{\mathrm{W}}}^2),$$

$$\frac{1}{\sigma}\frac{\mathrm{d}\sigma}{\mathrm{d}\cos\theta^*_{\mathrm{Z}}} = \frac{3}{8}\mathrm{FL}(1 + 2\mathrm{A}\cos\theta^*_{\mathrm{Z}} + \cos{\theta^*_{\mathrm{Z}}}^2) + \frac{3}{8}\mathrm{FR}(1 - 2\mathrm{A}\cos\theta^*_{\mathrm{Z}} + \cos{\theta^*_{\mathrm{Z}}}^2) + \frac{3}{4}\mathrm{F0}(1 - \cos{\theta^*_{\mathrm{Z}}}^2),$$

where $\cos\theta^*_{\mathrm{Z}}$ and $\cos\theta^*_{\mathrm{W}}$ represent the cosine of the decay angle of the lepton (or anti-lepton for $W^+$) as seen in the boson restframe with respect to the direction of the boson in the $WZ$ rest-frame. The term A is equal to $\frac{2c_v c_a}{c_v^2 c_a^2}$ where $c_v$ and $c_a$ represent the vector and axial coupling of the $Z$ boson to the leptons. The quantities FL, FR and F0 represent the fraction of each polarisation state and satisfy the relation FL+FR+F0 = 1.

Analytical fits are performed using only two variables, F0 and (FL-FR) thanks to the constraint mentioned above and for four different event distributions depending on the boson which is probed and on the charge of the $W$ present in the event. They are denoted $Z(W^+)$, $Z(W^-)$, $W^+$ and $W^-$. Results of the fit are illustrated in Figures 17 for the $Z$ and $W$ produced in $WZ - EW$ events in the full phase space at the generator level.

By weighting events according the analytical fit result, templates of each polarisation states are obtained at the reconstruction level and after the final event selection. An example of the templates normalised to unity are shown in Figures 18. By performing a binned profile-likelihood fit to the simulated distribution at the reconstruction level, the sensitivity to the measurement of F0 can be established.

[4] No kinematic or acceptance cuts are applied to the decay products of the bosons, except $|\mathrm{M}_{ll} - M_Z| < 10$ GeV.

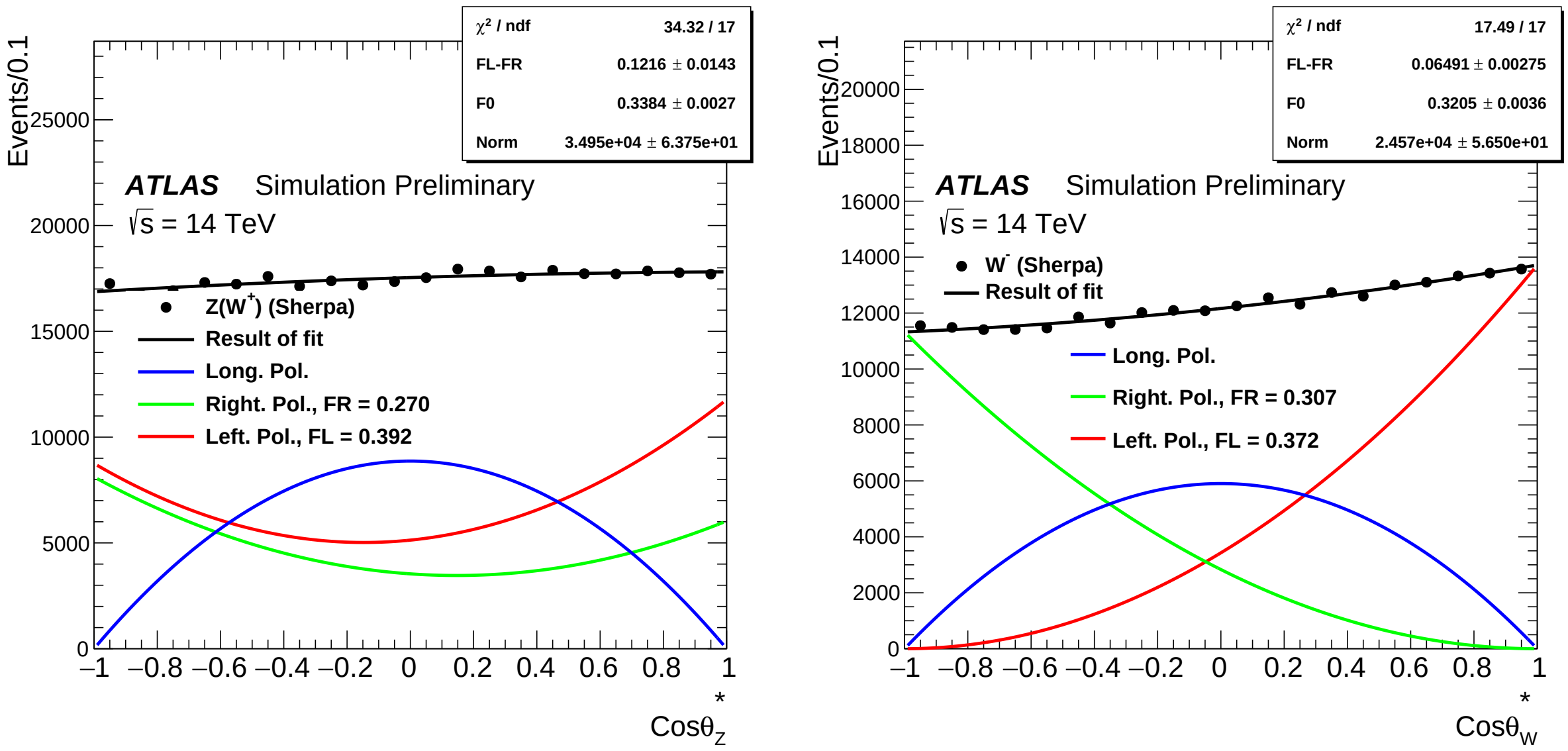

Figure 17: Left: Analytical fit to the three polarisation states at the generator level in the full phase space for $Z(W^+)$ in $WZ - EW$ events. Right: for $W^-$ in $WZ - EW$ events. As the constraint F0+FL+FR = 1 is used, the fitted parameters consist of F0, (FL-FR) and the normalisation.

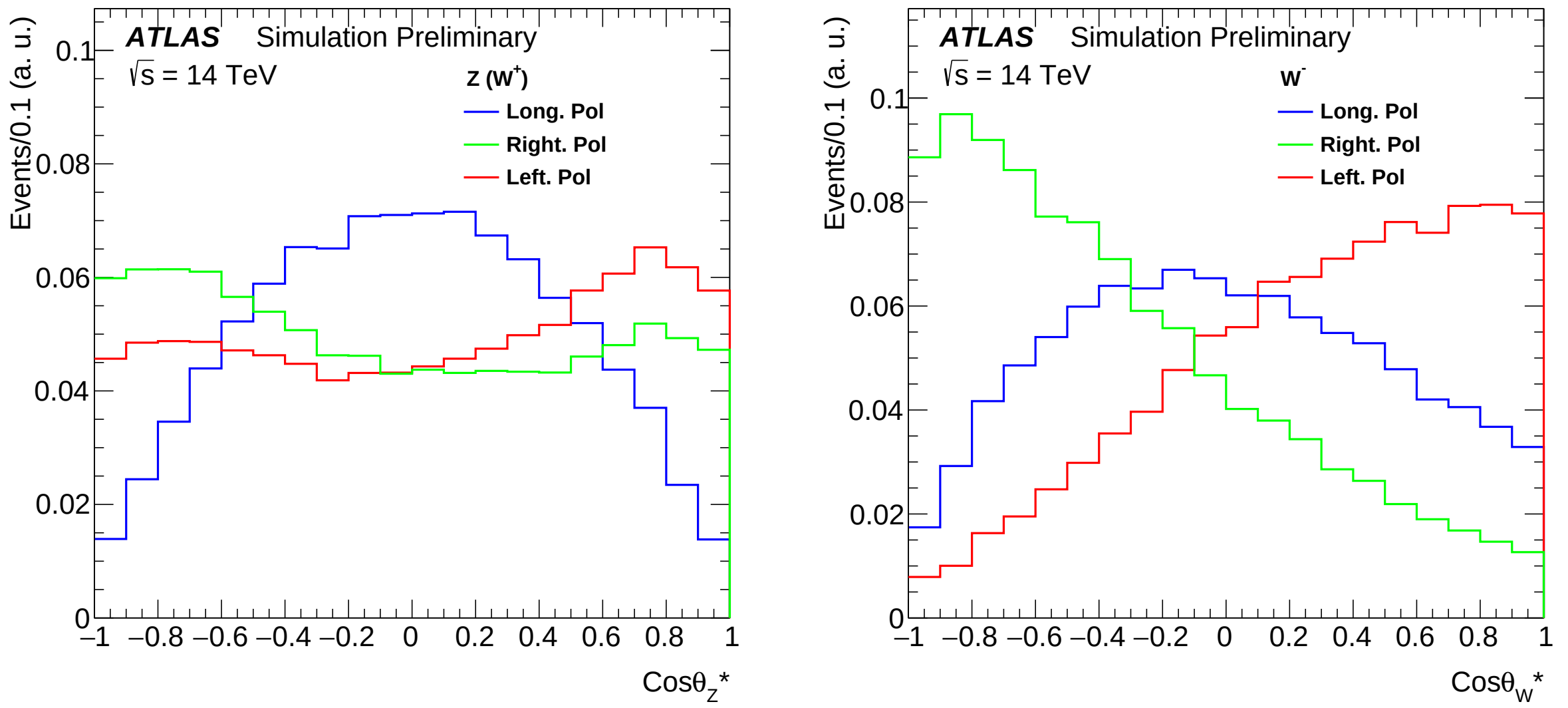

Figure 18: Left: Normalised templates at the reconstruction level and after the final event selection including $M_{jj}>$ 600 GeV of the three polarisation states for $Z(W^+)$. Right: for $W^-$.

## 8.2 Individual polarisation of $Z$ and $W$ bosons

In this section, each vector boson is treated independently within the nominal phase-space defined in Section 4.2.

### 8.2.1 Polarisation studies under no background conditions

Only $WZ - EW$ signal events are considered and the standard final selection with $M_{jj} > 600$ GeV is used. The distributions of $\cos\theta^*_Z$ and $\cos\theta^*_W$ are fitted with 3 parameters F0, (FL-FR) and the signal normalisation using the three polarisations templates histograms plus the background contribution coming from $WZ \to \tau X$[5]. This latter is normalised to the cross-section and the luminosity, with a 5% systematic error. The result of the fit is displayed in Figures 19 for two distributions. In Figure 20, the intrisic sensitivity to F0 of each channel can be appreciated: as expected the channels are affected either by a smaller statistic when a $W^-$ boson is involved or by a worse resolution due to the reconstruction of the neutrino when a $W$ boson is involved.

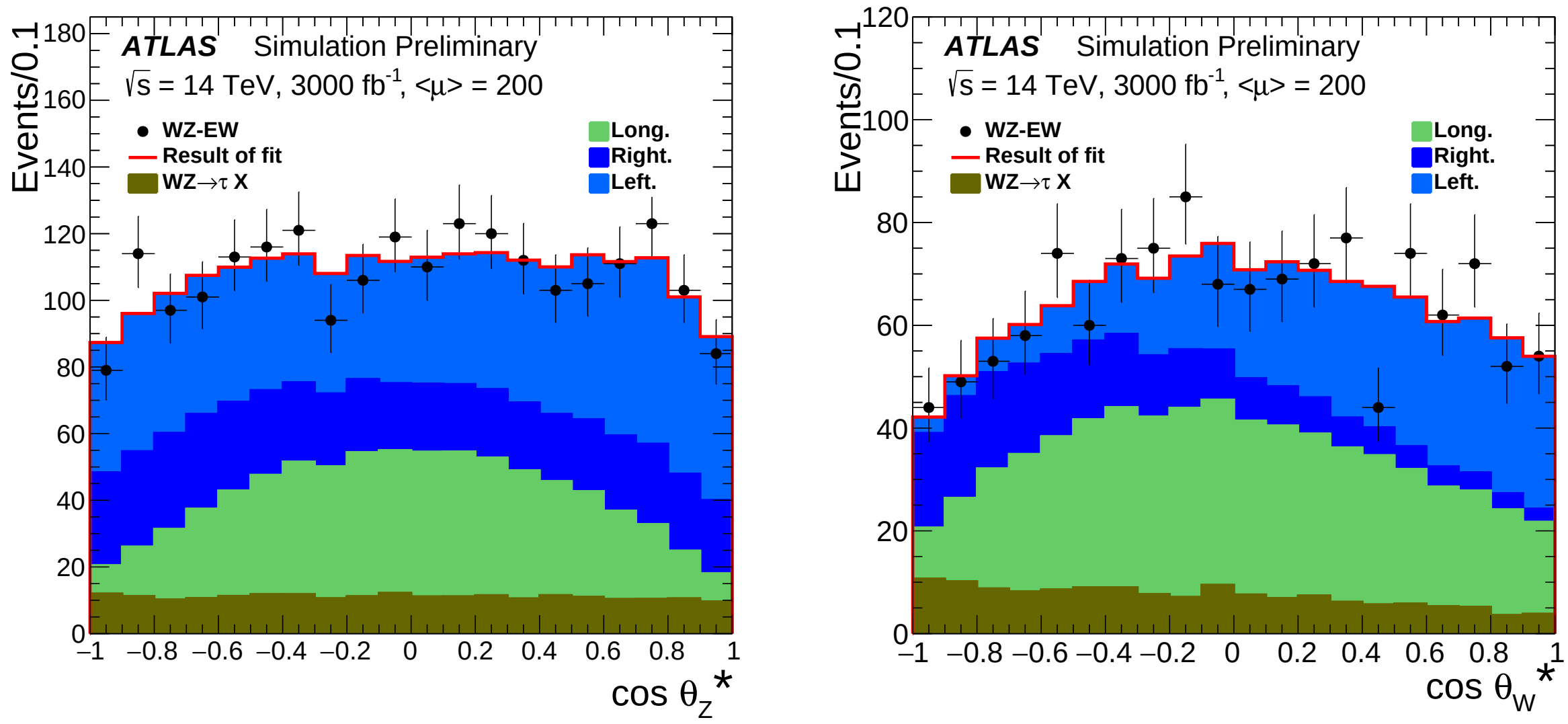

Figure 19: Left: Result of the fit for a set of pseudo data generated by varying the number of events in each bin within its statistical error for $Z(W^+)$. Right: for $W^-$. The dots represent the data normalised to a luminosity of 3000 $fb^{-1}$.

[5] $WZ \to \tau X$ events must be treated in this study as an additional background since the $\tau$ is, de facto, not reconstructed and consequently $\cos\theta^*_Z$ and $\cos\theta^*_W$ are not properly defined

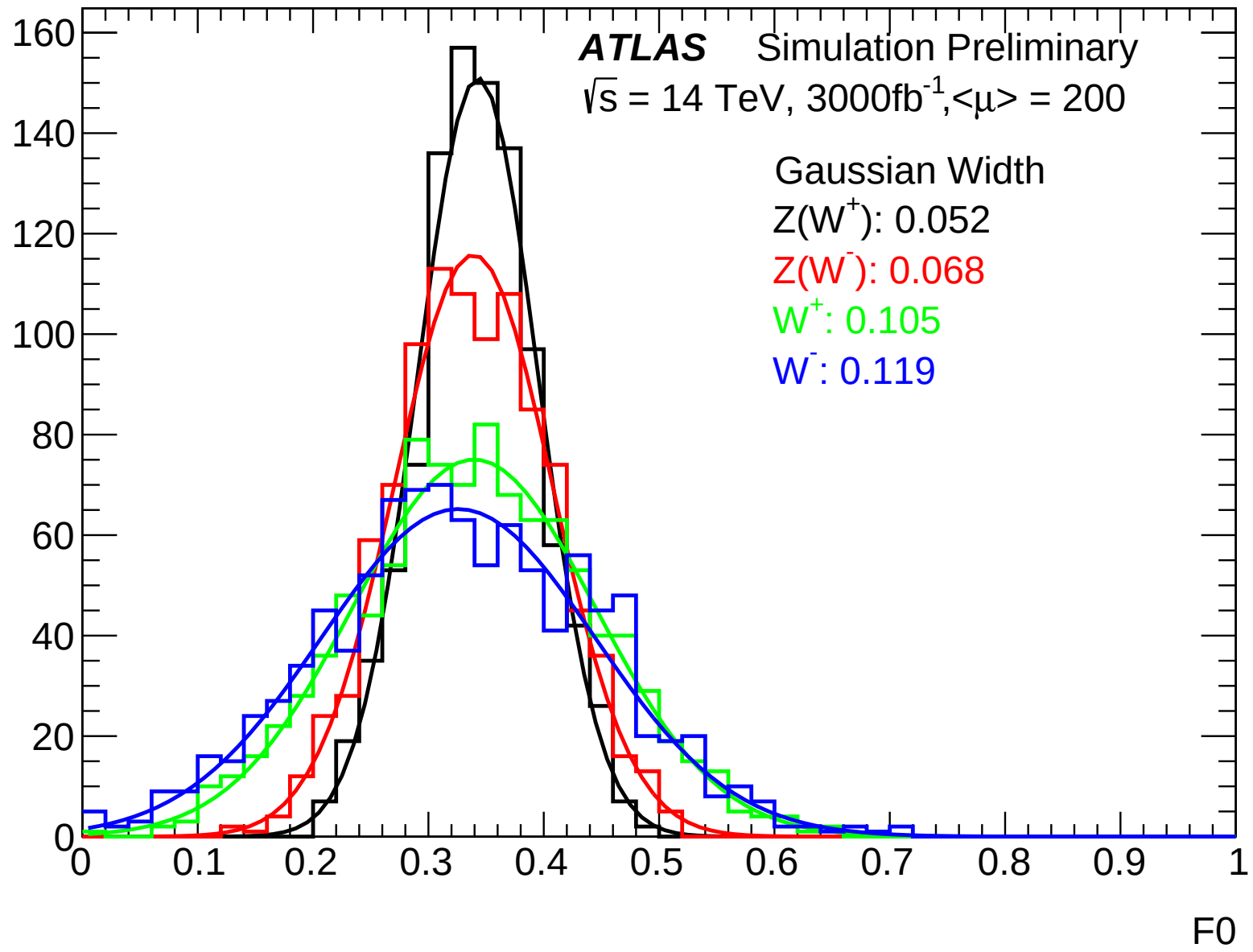

Figure 20: F0 distributions obtained from fits to 1000 sets of pseudo-data generated by varying the number of events in each bin within its statistical error for $Z(W^+)$,$Z(W^-)$, $W^-$and $W^+$.

### 8.2.2 Including the backgrounds

In this section, a more realistic approach is presented as the backgrounds are also considered. For the systematic associated to the total background normalisation, three cases were envisaged: 20%, 5% and 2.5%. For the best and worse channels, $Z(W^+)$ and $W^-$, a fit of F0 was performed for several hypothesis affecting the final selection:

- the $M_{jj}$ cut was varied from 500 GeV as the YR recommandations to the optimal cut of 600 GeV.
- the optimal BDT cut was applied as well as a more stringent cut
- the effect of q/g tagging on the subleading and leading jets was emulated. The current tagging performance given in [22] extrapolated to the HL-LHC acceptance, due to the large QCD background, are not efficient enough; the test was conducted with a bold scenario where $\epsilon_q$ is 0.9 while $\epsilon_g$ is 0.1.
- Finally the luminosity was doubled to emulate the combination of two experiments.

Significances for F0 ( = $\sqrt{-2\log L(0)}$ ) for the different hypothesis are presented in Table 6. The precision on the background is a key ingredient to improve the sensitivity on F0 as well as the signal purity for comparable signal significance. Adding a quark/gluon tagger, a gain of 10% on the significance can be achieved but in general, the improvement saturates when the signal statistics becomes too low. In the case of $W^-$, the conclusions are more critical as the sensitivity on F0 barely reaches a significance of 1 $\sigma$. Examples of the fitted distribution and of the loglikelihood profile are given in Figures 21 for the case enlightened in red in the table.

| | $M_{jj}$ > 500 GeV | $M_{jj}$ > 600 GeV | $M_{jj}$ >1000 GeV | BDT cut > 0.44 | BDT cut > 0.6 |
|---|---|---|---|---|---|
| **$Z(W^+)$** | | | | | |
| *Additionnal information:* | | | | | |
| *WZ-EW/ B* | *0.10* | *0.12* | *0.20* | *0.22* | *0.32* |
| *Exp. Numb. of long. pol. events* | *747* | *662* | *397* | *530* | *405* |
| Hyp. $\sigma_B$ = 20% | - | 1.12 | 1.77 | 1.82 | 2.19 |
| Hyp. $\sigma_B$ = 5% | 1.58 | 1.73 | 2.01 | 2.35 | 2.51 |
| Hyp. $\sigma_B$ = 2.5% | 1.88 | 1.97 | 2.07 | 2.48 | 2.56 |
| *$2 \times$ nominal luminosity:* | | | | | |
| Hyp. $\sigma_B$ = 20% | - | 1.50 | 2.44 | 2.45 | 3.00 |
| Hyp. $\sigma_B$ = 5% | 1.91 | 2.16 | 2.76 | 3.13 | 3.45 |
| Hyp. $\sigma_B$ = 2.5% | 2.39 | 2.58 | 2.88 | 3.40 | 3.57 |
| *q/g tagging (as $\epsilon_q$ = 90%, $\epsilon_g$ = 10% )* | | | | | |
| *WZ-EW/All B* | *0.14* | *0.16* | *0.24* | | |
| *Exp. Numb. of Long. pol. events* | *603* | *537* | *329* | | |
| Hyp. $\sigma_B$ = 20% | - | 1.28 | 1.91 | | |
| Hyp. $\sigma_B$ = 5% | 1.80 | 1.91 | 2.05 | | |
| Hyp. $\sigma_B$ = 2.5% | 2.03 | 2.08 | 2.08 | | |
| *q/g tagging + 2× luminosity* | | | | | |
| Hyp. $\sigma_B$ = 20% | 1.35 | 1.70 | 2.67 | | |
| Hyp. $\sigma_B$ = 5% | 2.23 | 2.45 | 2.85 | | |
| Hyp. $\sigma_B$ = 2.5% | 2.67 | 2.80 | 2.90 | | |
| **$W^-$** | | | | | |
| *Addtionnal information* | | | | | |
| *WZ-EW/B* | *0.10* | *0.11* | | *0.19* | |
| *Exp. Numb. of long. pol. events* | *424* | *368* | | *288* | |
| Hyp. $\sigma_B$ = 20% | - | 0.70 | | 0.91 | |
| Hyp. $\sigma_B$ = 5% | 0.76 | 0.78 | | 0.94 | |
| Hyp. $\sigma_B$ = 2.5% | 0.78 | 0.80 | | 0.95 | |
| *$2 \times$ nominal luminosity* | | | | | |
| Hyp. $\sigma_B$ = 20% | - | 0.97 | | 1.28 | |
| Hyp. $\sigma_B$ = 5% | 1.05 | 1.09 | | 1.33 | |
| Hyp. $\sigma_B$ = 2.5% | 1.09 | 1.12 | | 1.34 | |

Table 6: F0 significance (as defined in the text) for different selection hypothesis for the best channel $Z(W^+)$ (Top) and the worst channel $W^-$ (Bottom). The $WZ - EW$ signal purity and the expected number of longitudinally polarised events are also given as additionnal information. The case in red corresponds to Figures 21.

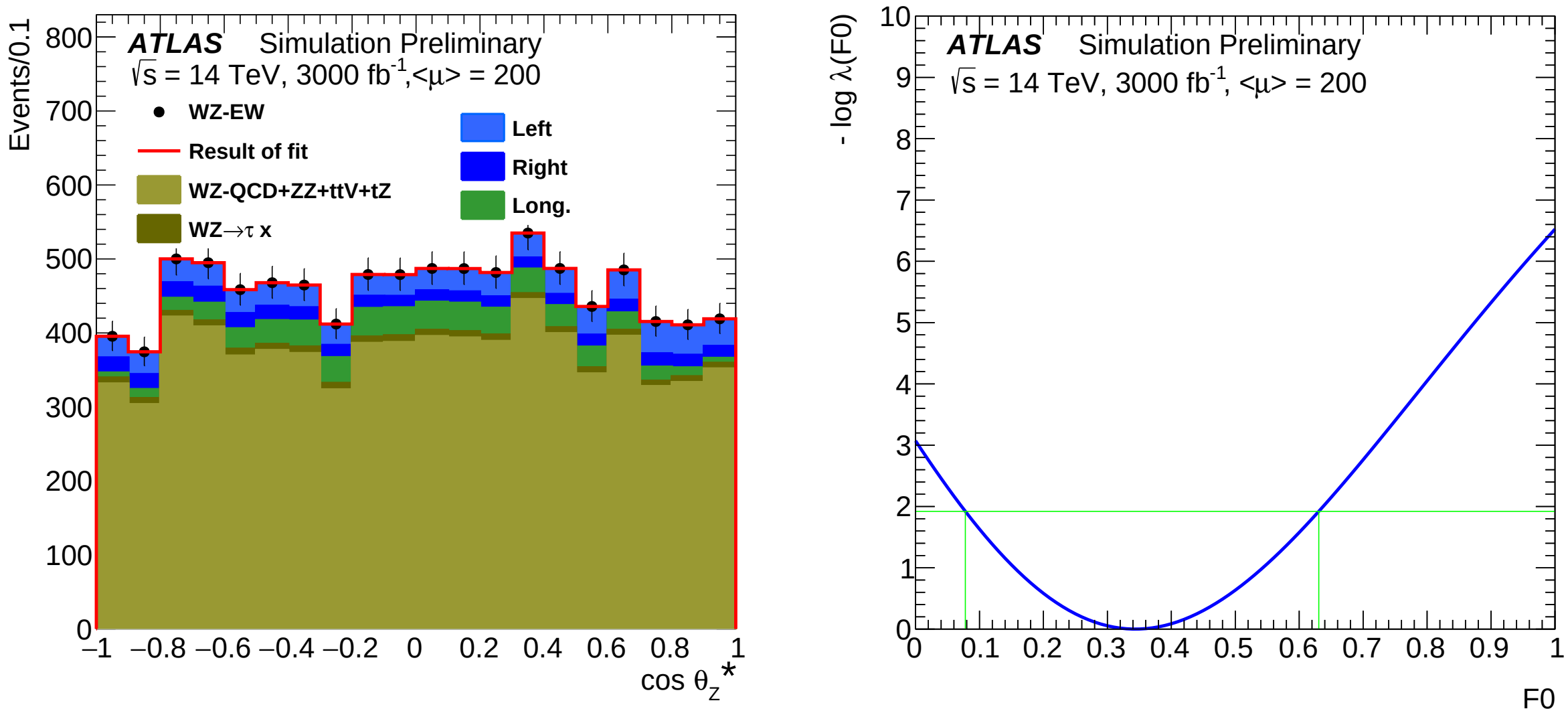

Figure 21: Left : Results of the fit of the F0 and FL-FR contributions on top of the $WZ - QCD$ background with an error of 2.5% and the $\tau$ background. Right: Profile of the negative log-likelihood vs F0 parameter.

### 8.3 Double polarisation of the $W$ and $Z$ bosons

Ultimately, one wants to measure the polarisation component of the $WZ$ final state as the sum of $W_0Z_0$, $W_0Z_T$, $W_TZ_0$ and $W_TZ_T$ components where T stands for transverse polarisation, the sum of the right and left fractions defined in the preceeding section.

The four corresponding templates are obtained in a similar way as in section 8.1, by combining the weights obtained from the analytical fits. Figures 22 illustrate the result of several fits which remain an exercice as the $WZ - QCD$ background was not taken into account. They show the results of a simultaneous fit of 2 data distributions to 4 data distributions consisting of the distributions of $\cos\theta^*_{\mathrm{Z}}$ and $\cos\theta^*_{\mathrm{W}}$ of the scalar sum of the lepton $\mathrm{p_T}$ from the Z boson and the scalar sum of the lepton $\mathrm{p_T}$ from the W and $\mathrm{E_T^{miss}}$. For $\mathrm{M_{jj}} > 500$ GeV, with 266 expected doubly longitudinally polarised events, the significance of the fraction of double longitudinally polarised final state stays below 1 $\sigma$. However, it should be noted that a 2-dimensional fit of $\cos\theta^*_{\mathrm{Z}}$ and $\cos\theta^*_{\mathrm{W}}$ gives a better sensitivity as shown in Figure 22 bottom left. Therefore, more sophisticated methods will be valuable to perform this measurement.

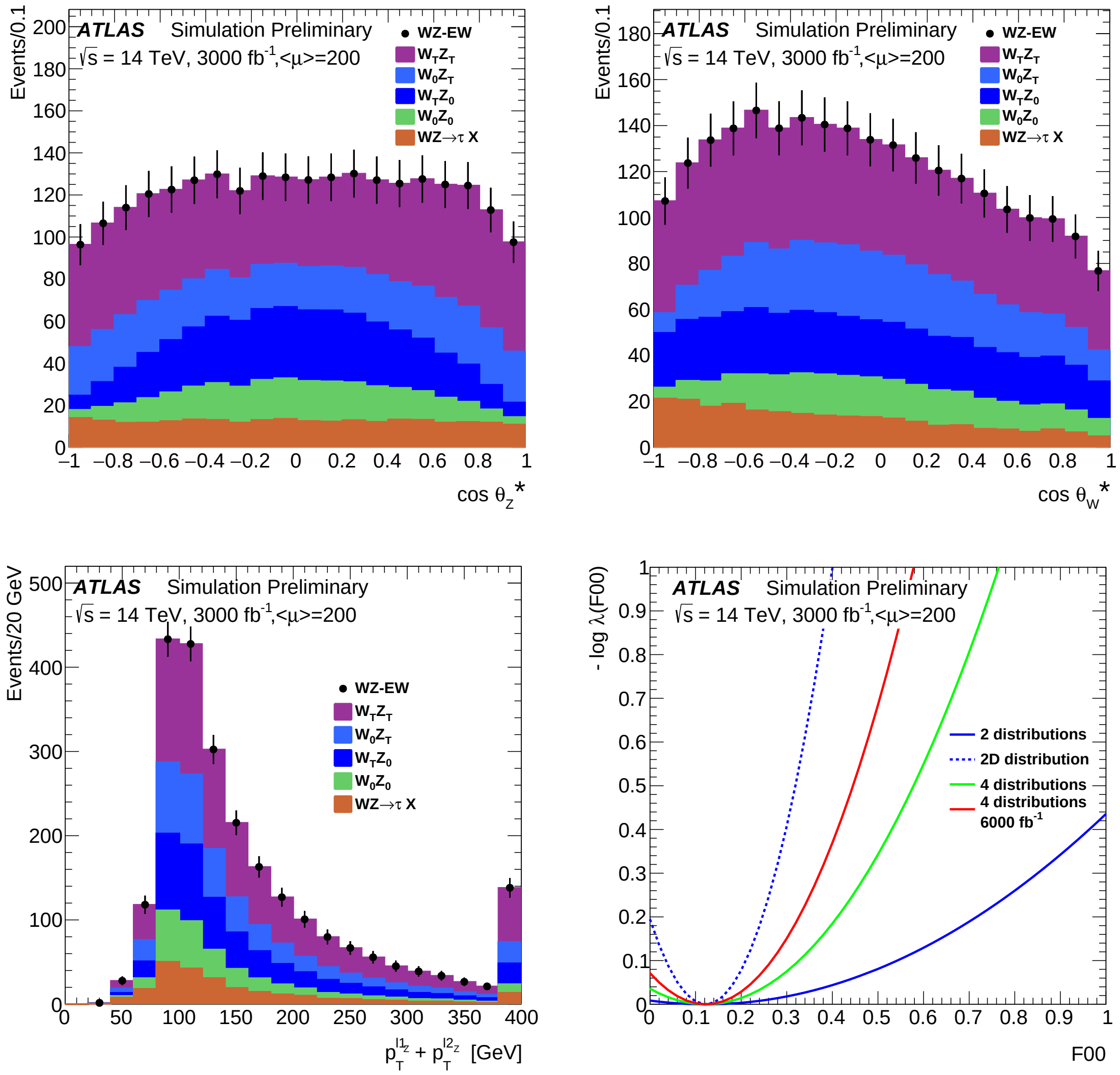

Figure 22: Results of the template fit for 3 distributions from the 4 used. Top Left: $\cos\theta^*_Z$ ,Top Right: $\cos\theta^*_W$ and Bottom Left: $p_T^{l1z}$ + $p_T^{l2z}$. Bottom Right: Negative log-likelihood profile vs F00 for different fits.

## 9 Kinematic distributions sensitive to aQGCs

In the reference [3], the variables $\sum|p_T^{lep}|$ and $|\Delta\phi(W,Z)|$ are shown to be sensitive on aQGC with the WZ final states. The distributions of these observables are shown in Figures 23 with 3000 fb$^{-1}$.

The sensitive region to aQGC lays at high $\sum|p_T^{lep}|$, typically above 500 GeV: extrapolating from the less than one signal event expected in [3] leads to less than 30 events at the end of Run3, while around 220 are expected with 3000 fb$^{-1}$. Similarly, from 2 events above $|\phi(W,Z)|$ equal 2.4, around 75 signal events are expected at the end of Run3, while about 950 are expected with 3000 fb$^{-1}$. Additionnal distributions are given in Appendix B.

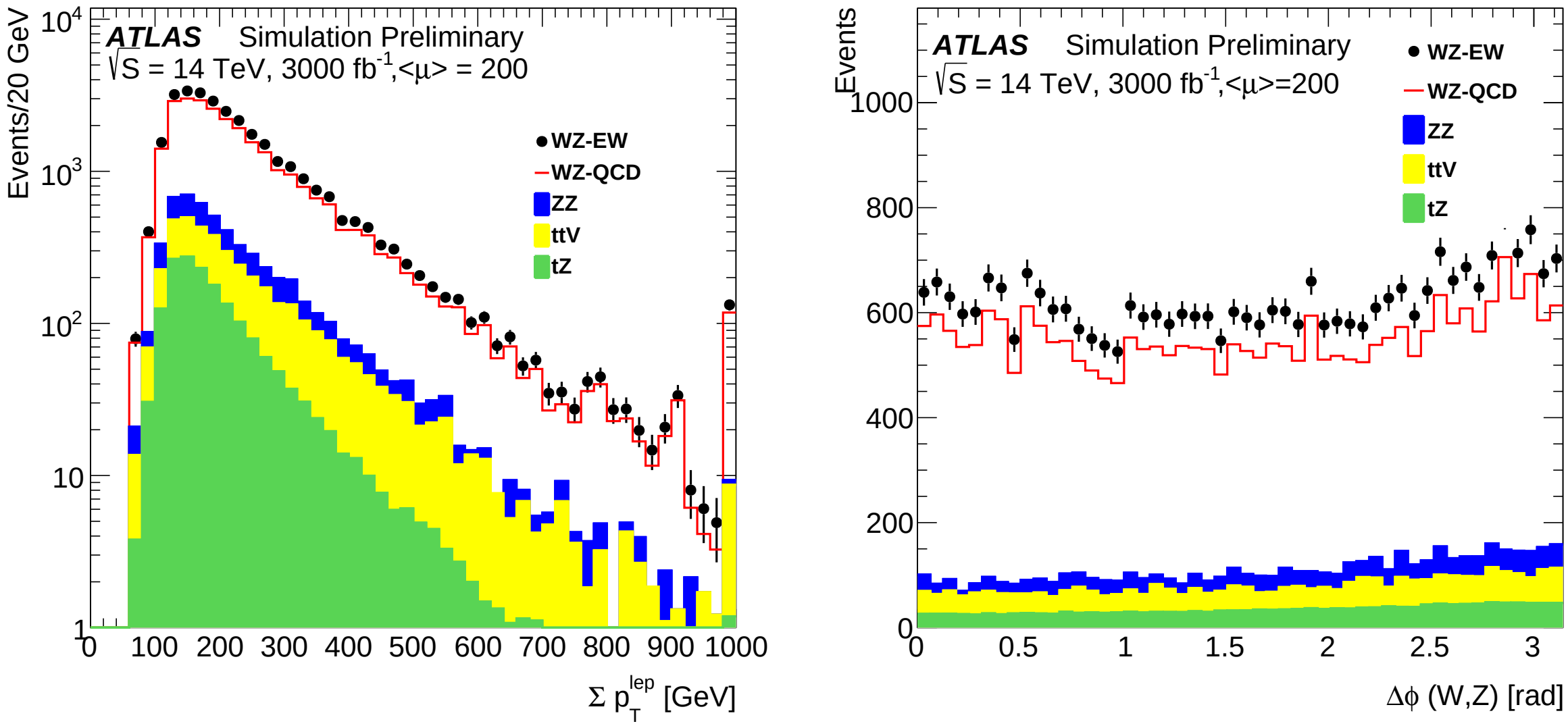

Figure 23: Left: Distribution of $\sum|p_T^{lep}|$. Right: Distribution of $|\Delta\phi(\mathrm{W,Z})|$ for an integrated luminosity of 3000 fb$^{-1}$.

# 10 Conclusion

Prospects for measuring vector boson scattering in the $WZ$ fully leptonic final state for HL-LHC are presented using a fast simulation based on the parametrisation of the ATLAS detector effects. Different detector setups and pile-up conditions are studied, extending or not the tracking acceptance with the implications of a lepton and a hard scatterring jet identification up to $|\eta| = 4$, including or not the high granularity timing detector and varying the working point for pile-up rejection.

The signal over background ratio is of the order of 1 to 10 and the main background (80%) comes from the strong production of $WZ$plus two jets final states. The extension of the tracking capabilities to large $\eta$ brings 28% more signal events and despite the expected level of PU, enables to maintain a jet $\mathrm{p_T}$ cut as low as 30 GeV within the tracker acceptance. Beyond, a more stringent cut on $\mathrm{p_T^{jet}}$, of the order of 80 GeV must be applied. Depending on the level of PU rejection, an additional signal gain, between 4% to 10% can be expected with the addition of HGTD. But for dealing with differential distributions, increasing the signal statistic or purity is important. Therefore, a multivariate analysis was also investigated to separate $WZ - EW$ events from the $WZ - QCD$ background. For the same $\mathrm{S}/\sqrt{\mathrm{S+B}}$, 40% more events are selected with the BDT. Additionnally, the $WZ - QCD$ background can be also mitigated by an efficient quark-gluon tagger in particular in the forward region.

Given the low signal over background ratio, the precision on the cross-sections measurement will be hampered by the systematic uncertainty associated to the $WZ - QCD$ background and in particular by the theoretical uncertainty. However, it is expected that the background uncertainty will be constrained by refined and diverse control regions, allowed thanks to the large amount of statistics.

Investigations were carried out to measure the polarisation fractions of the vector bosons. It was shown that the measurement of the longitudinal polarisation of the $Z$ bosons is reachable with an expected significance between 2 to 3 standard deviations. In general, the sensivity is increased with a larger signal purity. The measurement of the doubly longitudinal $W_0Z_0$ cross-section as predicted in the Standard Model will require more sophisticated methods.

# A Appendix

Figures 24 to 26 display the distributions of the 25 variables selected to build the BDT for the $WZ - EW$ signal and the $WZ - QCD$ background.

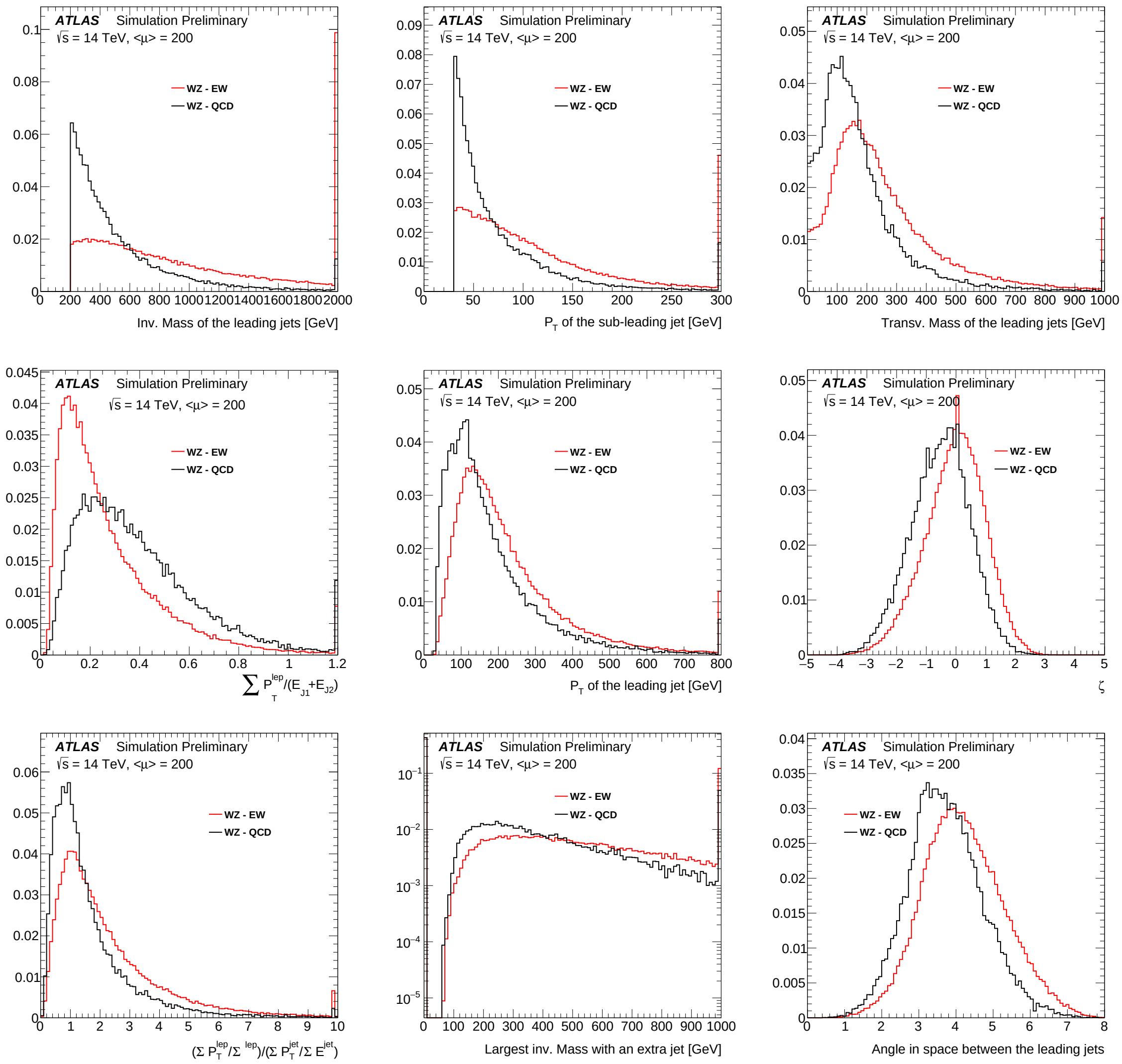

Figure 24: First 9 best ranked variables used to construct the BDT discriminant.

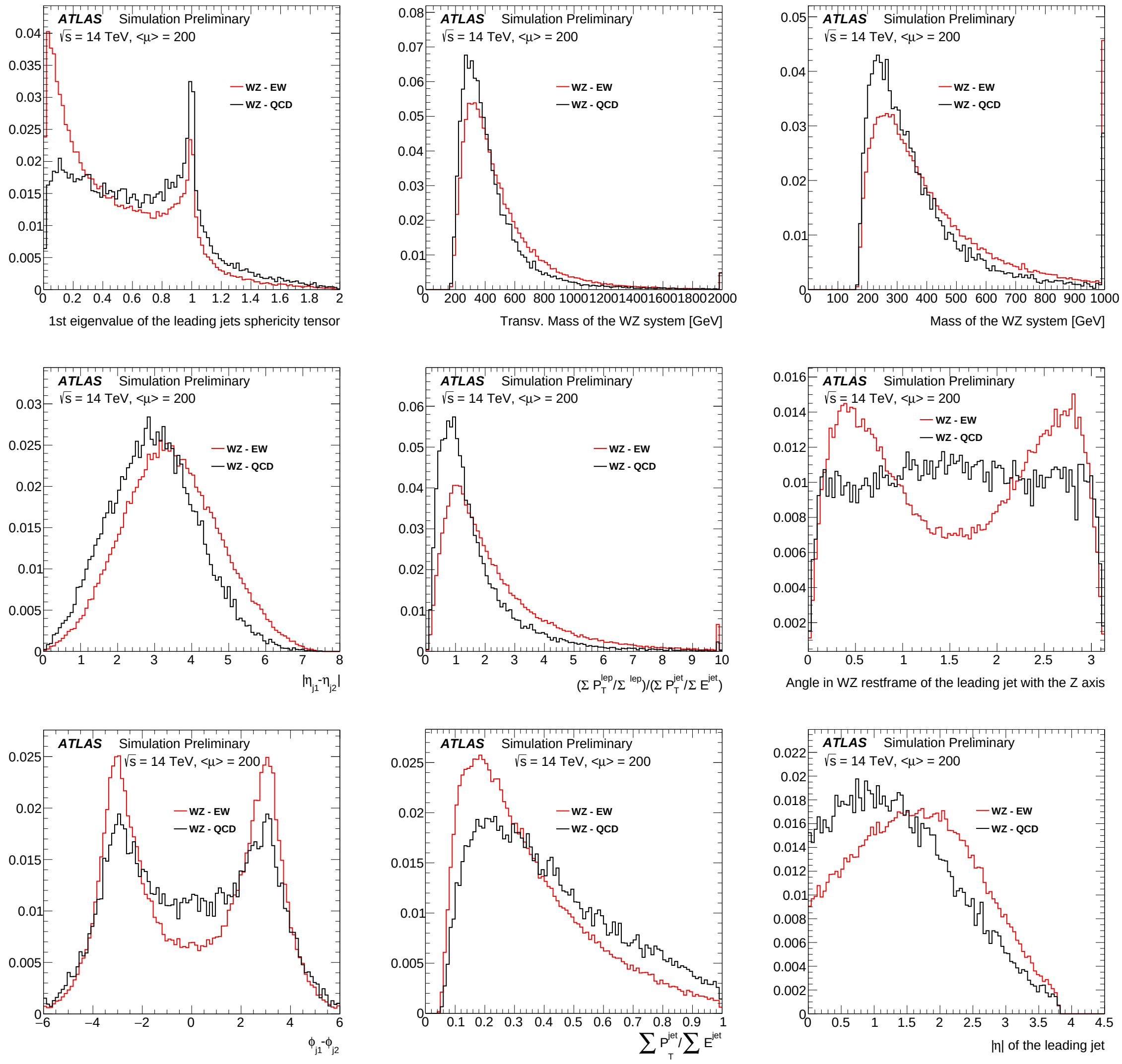

Figure 25: 10 to 18 ranked variables used to construct the BDT discriminant.

30

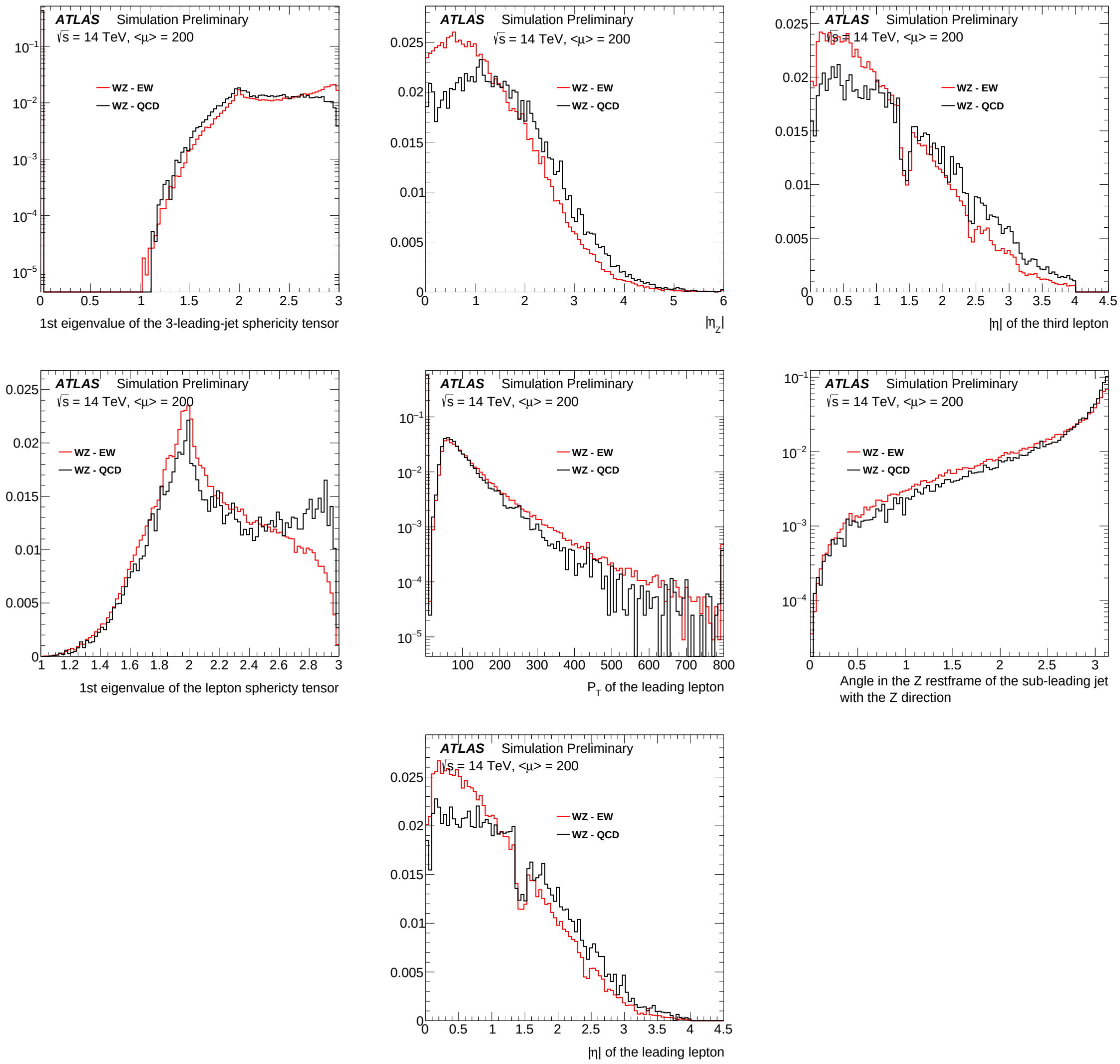

Figure 26: Last 7 ranked variables used to construct the BDT discriminant.

# B Appendix

Figures 27 represent the integrated number of events above $m_{WZ}^T$ or $m_{WZ}$ cuts versus the value of this cut. In Tables 7 and 8, the number of events in each bin for the signal and the sum of backgrounds is given as well as the MC statistical error as this type of cumulative distributions enlightens the tails of distributions.

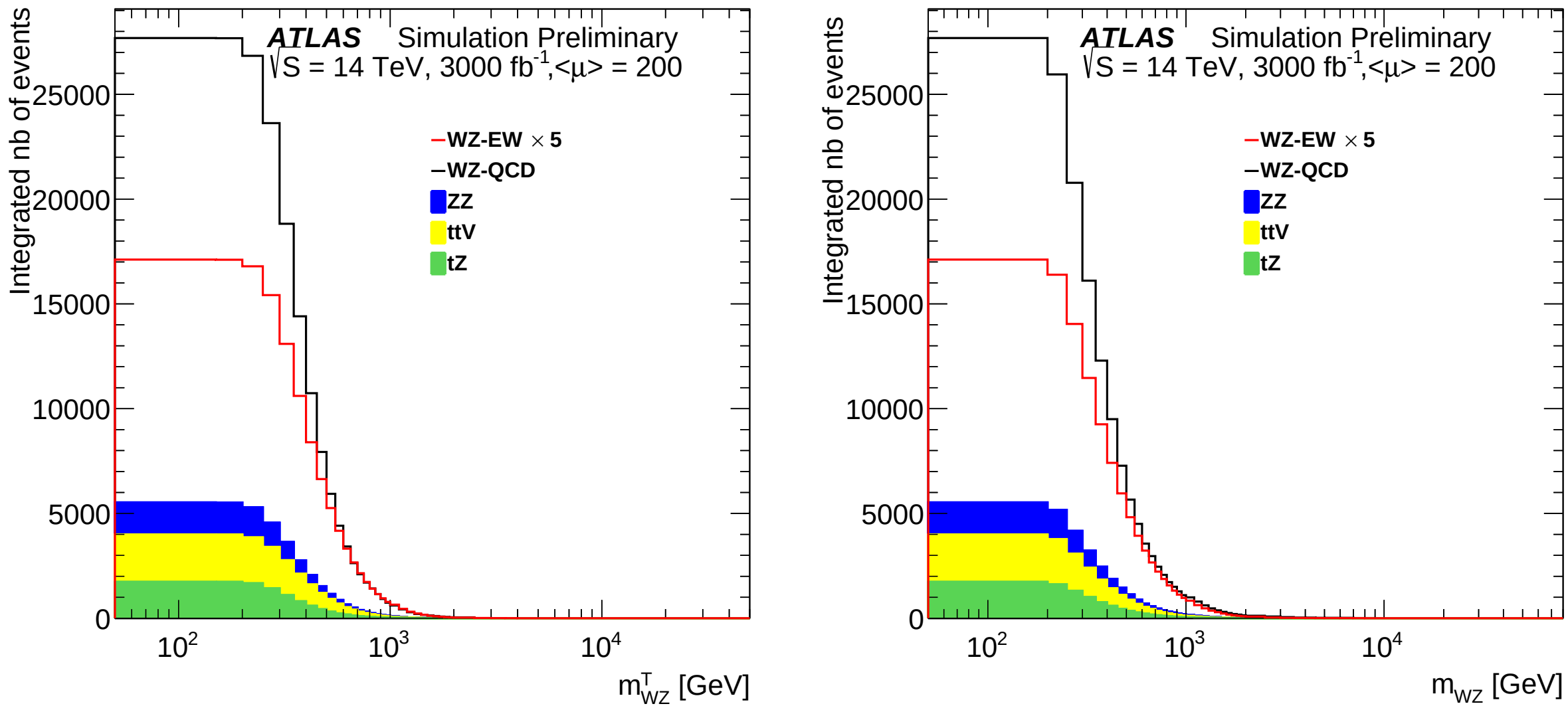

Figure 27: Left: Integrated number of events above $m_{WZ}^T$ cut vs $m_{WZ}^T$ cut. Right: Integrated number of events above $m_{WZ}$ cut vs $m_{WZ}$ cut.

| | Nb of signal events | MC statistical error | Nb of background events | MC statistical error |
|---|---|---|---|---|
| > 0 GeV | 3422.74 | 13.40 | 27686.18 | 272.24 |
| > 50 GeV | 3422.74 | 13.40 | 27686.18 | 272.24 |
| > 100 GeV | 3422.74 | 13.40 | 27686.18 | 272.24 |
| > 150 GeV | 3421.34 | 13.40 | 27677.83 | 272.18 |
| > 200 GeV | 3358.60 | 13.28 | 26837.26 | 268.83 |
| > 250 GeV | 3083.63 | 12.73 | 23625.10 | 252.20 |
| > 300 GeV | 2618.28 | 11.74 | 18822.81 | 222.12 |
| > 350 GeV | 2121.71 | 10.58 | 14407.57 | 194.66 |
| > 400 GeV | 1679.41 | 9.39 | 10738.51 | 166.73 |
| > 450 GeV | 1328.06 | 8.38 | 7934.36 | 142.88 |
| > 500 GeV | 1051.24 | 7.46 | 5936.06 | 123.23 |
| > 550 GeV | 834.45 | 6.65 | 4414.82 | 106.57 |
| > 600 GeV | 663.72 | 5.96 | 3423.02 | 95.17 |
| > 650 GeV | 532.40 | 5.36 | 2628.41 | 84.05 |
| > 700 GeV | 429.46 | 4.84 | 2098.40 | 76.39 |
| > 750 GeV | 346.17 | 4.37 | 1700.98 | 67.05 |
| > 800 GeV | 282.07 | 3.97 | 1419.26 | 62.07 |
| > 850 GeV | 230.39 | 3.61 | 1150.12 | 55.78 |
| > 900 GeV | 190.27 | 3.32 | 919.05 | 49.57 |
| > 950 GeV | 156.16 | 3.02 | 739.65 | 44.18 |
| > 1000 GeV | 129.65 | 2.78 | 607.95 | 40.43 |
| > 1100 GeV | 89.53 | 2.20 | 419.47 | 33.73 |
| > 1200 GeV | 62.23 | 1.86 | 285.62 | 27.92 |
| > 1300 GeV | 45.16 | 1.60 | 201.21 | 23.54 |
| > 1400 GeV | 32.82 | 1.37 | 165.87 | 21.97 |
| > 1500 GeV | 23.16 | 1.17 | 131.25 | 19.77 |
| > 1600 GeV | 18.20 | 1.06 | 104.84 | 18.05 |
| > 1700 GeV | 13.47 | 0.94 | 78.96 | 15.10 |
| > 1800 GeV | 10.06 | 0.77 | 60.74 | 13.83 |
| > 1900 GeV | 7.60 | 0.69 | 47.74 | 12.96 |
| > 2000 GeV | 5.95 | 0.63 | 40.26 | 12.36 |
| > 2500 GeV | 2.26 | 0.42 | 6.78 | 3.42 |
| > 3000 GeV | 1.05 | 0.34 | 4.70 | 2.83 |
| > 3500 GeV | 0.61 | 0.30 | 3.97 | 2.73 |
| > 4000 GeV | 0.56 | 0.30 | 2.04 | 1.93 |
| > 4500 GeV | 0.29 | 0.12 | 0.11 | 0.06 |
| > 5000 GeV | 0.14 | 0.08 | 0.11 | 0.06 |
| > 6000 GeV | 0.14 | 0.08 | 0.11 | 0.06 |
| > 7000 GeV | 0.14 | 0.08 | 0.11 | 0.06 |
| > 8000 GeV | 0.14 | 0.08 | 0.11 | 0.06 |
| > 9000 GeV | 0.14 | 0.08 | 0.07 | 0.05 |
| > 10000 GeV | 0.10 | 0.07 | 0.07 | 0.05 |
| > 20000 GeV | 0.05 | 0.05 | 0.07 | 0.05 |
| > 50000 GeV | 0.00 | 0.00 | 0.00 | 0.00 |

Table 7: Number of events for which the transvers mass $m_{WZ}^{T}$ is greater than the indicated value for $WZ-EW$ events and for the sum of backgrounds. The error due to the MC statistics is also quoted.

| $m_{WZ}$ | Nb of signal events | MC statistical error | Nb of background events | MC statistical error |
|---|---|---|---|---|
| > 0 GeV | 3422.76 | 13.40 | 27686.14 | 272.24 |
| > 50 GeV | 3422.76 | 13.40 | 27686.14 | 272.24 |
| > 100 GeV | 3422.76 | 13.40 | 27686.14 | 272.24 |
| > 150 GeV | 3422.76 | 13.40 | 27686.14 | 272.24 |
| > 200 GeV | 3278.35 | 13.12 | 25950.05 | 264.54 |
| > 250 GeV | 2808.34 | 12.16 | 20778.20 | 236.62 |
| > 300 GeV | 2293.04 | 11.02 | 16107.47 | 208.90 |
| > 350 GeV | 1850.80 | 9.95 | 12292.33 | 184.49 |
| > 400 GeV | 1482.64 | 8.92 | 9499.20 | 162.80 |
| > 450 GeV | 1191.53 | 8.03 | 7277.15 | 142.89 |
| > 500 GeV | 964.04 | 7.25 | 5660.48 | 125.56 |
| > 550 GeV | 786.99 | 6.58 | 4502.36 | 113.30 |
| > 600 GeV | 645.41 | 6.00 | 3560.94 | 100.56 |
| > 650 GeV | 532.03 | 5.50 | 2964.51 | 92.84 |
| > 700 GeV | 444.79 | 5.08 | 2458.29 | 85.70 |
| > 750 GeV | 374.50 | 4.71 | 2069.25 | 78.82 |
| > 800 GeV | 313.07 | 4.29 | 1722.52 | 71.66 |
| > 850 GeV | 263.34 | 3.90 | 1500.46 | 66.61 |
| > 900 GeV | 225.07 | 3.61 | 1281.54 | 62.29 |
| > 950 GeV | 193.75 | 3.37 | 1100.54 | 59.14 |
| > 1000 GeV | 167.71 | 3.16 | 995.30 | 56.84 |
| > 1100 GeV | 124.99 | 2.79 | 795.04 | 52.53 |
| > 1200 GeV | 94.47 | 2.46 | 615.52 | 46.94 |
| > 1300 GeV | 71.92 | 2.23 | 470.10 | 42.70 |
| > 1400 GeV | 57.71 | 2.05 | 376.19 | 38.77 |
| > 1500 GeV | 44.68 | 1.87 | 303.78 | 35.71 |
| > 1600 GeV | 34.17 | 1.41 | 245.97 | 31.95 |
| > 1700 GeV | 28.26 | 1.30 | 199.82 | 30.11 |
| > 1800 GeV | 22.87 | 1.19 | 172.09 | 29.24 |
| > 1900 GeV | 18.89 | 1.08 | 140.42 | 27.19 |
| > 2000 GeV | 15.44 | 0.94 | 117.22 | 26.01 |
| > 2500 GeV | 7.18 | 0.67 | 81.01 | 24.04 |
| > 3000 GeV | 3.54 | 0.50 | 52.79 | 22.44 |
| > 3500 GeV | 2.55 | 0.45 | 29.73 | 12.13 |
| > 4000 GeV | 1.87 | 0.41 | 28.37 | 11.75 |
| > 4500 GeV | 1.41 | 0.29 | 19.01 | 10.40 |
| > 5000 GeV | 1.07 | 0.26 | 18.89 | 10.40 |
| > 6000 GeV | 0.92 | 0.25 | 18.27 | 10.38 |
| > 7000 GeV | 0.63 | 0.17 | 8.84 | 4.41 |
| > 8000 GeV | 0.39 | 0.14 | 8.84 | 4.41 |
| > 9000 GeV | 0.34 | 0.13 | 4.10 | 2.80 |
| > 10000 GeV | 0.29 | 0.12 | 2.04 | 1.93 |
| > 30000 GeV | 0.00 | 0.00 | 0.04 | 0.04 |
| > 80000 GeV | 0.00 | 0.00 | 0.00 | 0.00 |

Table 8: Number of events for which the invariant mass $m_{WZ}$ is greater than the indicated value for $WZ - EW$ events and for the sum of backgrounds. The error due to the MC statistics is also quoted. In the reconstruction of the W boson, the longitudinal component of the neutrino momentum is chosen to be the smallest.

# HL-LHC prospects for diboson resonance searches and electroweak vector boson scattering in the $WW/WZ \to \ell\nu qq$ final state

The ATLAS Collaboration

This note presents the prospects of searches for new heavy resonances decaying to dibosons ($WW/WZ$) and measurements of electroweak $WW/WZ$ production via vector boson scattering in association with a high-mass dijet system in the $\ell\nu qq$ final states. The prospects are presented for an integrated luminosities of 300 and 3000 fb$^{-1}$ of proton-proton collisions at $\sqrt{s}$= 14 TeV to be recorded with the ATLAS detector at the high-luminosity LHC assuming the average number of $pp$ interactions per bunch crossing to be 200. The cross-section for electroweak $WW/WZ$ production in vector boson scattering processes is expected to have an observation greater than 5$\sigma$ at 300fb$^{-1}$ and to be measured to within 6.5% at 3000 fb$^{-1}$. The diboson resonance searches are interpreted for sensitivity to a heavy scalar singlet, a simplified phenomenological model with a heavy gauge boson and a Randall-Sundrum model with a spin-2 graviton. With 3000 fb$^{-1}$of $pp$ data, the exclusion limit reach for the new resonance is extended to 4.9 TeV in the heavy gauge boson model and 3.3 TeV in the Randall-Sundrum model. These are improvements of approximately 1 TeV in comparison to existing mass limits. With 3000 fb$^{-1}$of $pp$ data, the 5$\sigma$ discovery reach is 3.3 TeV for the heavy gauge boson model and 1.7 TeV for the Randall-Sundrum model.

# 1 Introduction

Searches for new heavy particles are an important part of the physics program at the Large Hadron Collider (LHC) and have been intensively performed over a broad range of final states to uncover physics beyond the Standard Model (SM). Many of these searches are motivated by models that aim to resolve the hierarchy problem, an unnaturally large difference in the strength between the electroweak and gravity forces, such as the Randall–Sundrum (RS) model with a warped extra dimension [1] or by models with composite Higgs bosons [2]. Possible extensions to the SM, such as extended Higgs sectors as in the two-Higgs-doublet model (2HDM) [3] or extended gauge sectors as in Grand Unified Theories [4–6], are also motivations for new heavy particle searches. No clear hint of new heavy particles has been observed to date at the LHC, placing strong constraints on the production of such new particles by both ATLAS and CMS [7–9].

The quest for new phenomena will therefore continue in future runs of the LHC and in future collider experiments. This note presents prospects for the search for resonances decaying to diboson ($WW$ or $WZ$, collectively called $VV$ where $V = W$ or $Z$) in the semileptonic channel where one $W$-boson decays leptonically and the other $W$ or $Z$-boson decays to quarks ($\ell\nu qq$ channel). The results include sensitivity for such new resonances based on an integrated luminosity of 300 or 3000 fb$^{-1}$ of $pp$ collisions at $\sqrt{s}$= 14 TeV using the ATLAS detector. An average number of 200 additional inelastic $pp$ collisions per bunch-crossing (denoted "pile-up" in the rest of the note) is assumed. Searches in other semileptonic and fully hadronic decay channels are expected to have similar sensitivities at high masses, as observed in the ATLAS searches with Run 2 data [7].

While the presence of resonances is the most dramatic signal for new phenomena, they may be too heavy or broad to be clearly seen at the LHC. The study of the high-energy scattering between the longitudinal components of the vector bosons (vector boson scattering or VBS) is a perfect case to search systematically for the presence of new particles or interactions behind the breaking of the EW symmetry. In fact the scattering amplitude of the VBS processes, in absence of the Higgs boson, would grow indefinitely with the center-of-mass energy, while it is finite if the Higgs boson is exactly the one predicted by the SM and its contributions are included. This important high-energy behavior still needs to be tested experimentally and it will be one of the main drivers of the physics program for the HL-LHC, the project upgrade of the LHC where the luminosity is expected to increase by a factor of 10 with respect to current conditions. ATLAS has recently presented results of VBS searches in the $W^{\pm}W^{\pm}$ channel [10] and $WZ$ channel [11] with 6.9$\sigma$ and 5.6$\sigma$ evidence respectively. The existing Run-2 VBS measurements have focused on channels involving the fully leptonic boson decays ($W(\ell\nu)$ and $Z(\ell\ell)$)[1] and photons. The semileptonic channels, i.e., $V(qq')Z(\nu\nu)$, $V(qq')W(\ell\nu)$ and $V(qq')Z(\ell\ell)$, can however offer some interesting advantages. The $V(qq')$ branching fractions are much larger than the leptonic branching fractions and the use of jet substructure techniques with large-radius jet reconstruction allows to reconstruct and identify the $V$-boson produced in the high-$p_{\mathrm{T}}$ region, which is the most sensitive to new physics effects. The sensitivity of the ATLAS experiment to VBS in the $V(qq')W(\ell\nu)$ final state, assuming an integrated luminosity of 300 or 3000 fb$^{-1}$ of $pp$ collisions at $\sqrt{s}$= 14 TeV, will also be studied. The analysis is based on event selection and classification similar to those used in the Run 1 and Run 2 ATLAS searches [8, 9].

[1] Unless otherwise noted, $\ell = e, \mu$ in this note.

## 2 Simulation Samples

Both exotic resonance and VBS analyses use generator-level samples of the main signal and background processes, combined with the parameterizations of the detector performance (muon and jet reconstruction and selection efficiencies and momentum resolutions) expected at the HL-LHC from fully simulated samples. The parametrized detector resolutions are used to smear the generator-level particle transverse momenta, while the parametrized efficiencies are used to reweigh the selected events [12]. All generated samples were produced at $\sqrt{s}$= 14 TeV and normalized to luminosities of 300 or 3000 fb$^{-1}$ when the results are presented.

### 2.1 Signal simulation

The prospect for resonance searches presented in this article are interpreted in the context of three different models: a heavy vector triplet (HVT) model [13, 14], a RS model [1] and a narrow heavy scalar resonance. The parameters of these models are chosen such that along the whole generated mass range, the resonance widths are less than 6% of the mass value, which is smaller than the detector resolution.

The HVT model [13, 14] provides a broad phenomenological framework to test a range of different scenarios involving new heavy gauge bosons and their couplings to SM fermions and bosons. In this model, a triplet $\mathcal{W}$ of colorless vector bosons is introduced with zero hypercharge. This leads to a set of nearly mass-degenerate charged ($W'^{\pm}$) and neutral ($Z'$) states, collectively denoted by $V'$. The masses of the $W'^{\pm}$ and $Z'$ bosons are taken to be the same in this prospect study. Two explicit HVT scenarios are used as benchmarks for interpretation of the results. The first scenario referred to as model A, reproduces the phenomenology of weakly coupled models based on an extended gauge symmetry [15]. The second DY scenario, referred to as model B, implements a strongly coupled model as in composite Higgs models [2].

The RS model postulates the existence of a warped extra dimension in which gravity propagates [1]. In the original “RS1" scenario only the effects of gravity are allowed to propogate in the extra dimension, while in the extended “bulk RS" scenario SM fields are also allowed to propagate in the extra dimension [16]. In both models [1, 16] the propagation in the extra dimension leads to the presence of a tower of Kaluza–Klein (KK) excitations of the graviton (denoted $G_{\mathrm{KK}}$). The bulk scenario avoids constraints on the orignial RS1 model from limits on flavour changing neutral currents and electroweak precision test by suppressing the graviton couplings to light fermions. This lead to the predominant decay modes of the bulk graviton to be to pairs of top-quarks, higgs bosons, and electroweak gauge bosons. The bulk KK gravitons are produced via both quark–antiquark annihilation and gluon–gluon fusion (ggF) processes, with the latter dominating due to suppressed couplings of the graviton to light fermions. The strength of the graviton interaction depends on the dimensionless coupling constant $k/\overline{M}_{\mathrm{Pl}}$, where $k$ corresponds to the curvature of the warped extra dimension and $\overline{M}_{\mathrm{Pl}} = 2.4 \times 10^{18}$ GeV is the effective four-dimensional Planck scale. This note assumes the value of $k/\overline{M}_{\mathrm{Pl}} = 1$.

The last model considered is an empirical model with a narrow heavy scalar resonance produced via the ggF and vector-boson-fusion (VBF) mechanisms and decaying directly into $VV$. The width of this new scalar is assumed to be negligible compared to the detector resolution. This benchmark model is used to explore sensitivity to extended Higgs sectors.

Signal events for the HVT and bulk RS models are generated with MadGraph5_aMC@NLO v2.2.2 [17] at leading order (LO) using the NNPDF23LO parton distribution function (PDF) set [18]. For the production of resonances in the HVT model, both the DY and VBF mechanisms are simulated. In the case of the heavy scalar model, signal events are generated at next-to-leading order (NLO) via the ggF and VBF mechanisms with Powheg-Box v1 [19, 20] and the CT10 PDF set [21].

For all signal models and production mechanisms, the generated events are interfaced to Pythia v8.186 [22] for parton showering, hadronization, and the underlying event. This Pythia interface includes the A14 set of tuned parameters [23] for events generated with MadGraph5_aMC@NLO at LO and the AZNLO set of tuned parameters [24] for events generated with Powheg-Box at NLO.

The electro-weak (EWK) $VVjj$ production is modeled using MadGraph5_aMC@NLO v2.3.3 [17], plus Pythia 8 [25] for fragmentation. The NNPDF30LO PDF set [18] is used. The EWK $VVjj$ samples are generated with two on-shell $V$ bosons, with one $V$ boson decaying leptonically ($Z \to \ell\ell$ with $\ell = e, \mu$, $W \to \ell\nu$ with $\ell = e, \mu, \tau$), and the other $V$ boson decaying hadronically. For each sample, all of the purely-electroweak tree-level diagrams (i.e. $O(\alpha_{EW}^6)$ diagrams) that contribute to the final state are included: VBS diagrams, non-VBS electroweak diagrams without $b$-quarks in the initial final states, and non-VBS electroweak diagrams with $b$-quarks in the initial final states. Example diagrams of these processes are shown in Figure 1. The non-VBS diagrams are suppressed by the analysis selction, e.g. diagrams including a $Wtb$ vertex, are suppressed by requiring that the tagging jets are not $b$-tagged ($b$-veto). (Section 4).

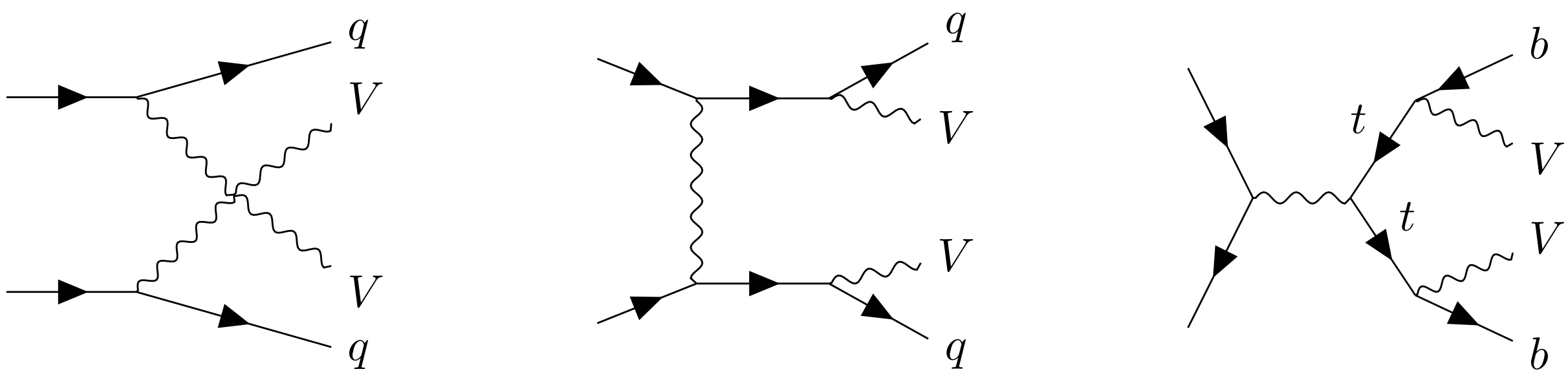

(a) (b) (c)

Figure 1: Example diagrams of purely-electroweak diagrams generated in the $VVjj$ process: VBS diagram (a), non-VBS diagram without $b$-quarks (b), and non-VBS diagrams with $b$-quarks (c).

Diagrams that contain a mixture of electroweak and QCD vertices (i.e. $O(\alpha_{EW}^4 \alpha_S^2)$ diagrams) are not included in these samples, and are not part of the signal definition. Such processes are accounted for by the background samples ($t\bar{t}$, single-top, and diboson).

### 2.2 Background simulation

Simulated background event samples are used to derive the main background estimates. The main background sources are $W$ bosons produced in association with jets ($W$+jets), with significant contributions from top-quark production (both $t\bar{t}$ pair and single-top), non-resonant vector-boson pair production ($ZZ$, $WZ$ and $WW$) and $Z$ bosons produced in association with jets ($Z$+jets). Background originating from multi-jet processes are expected to be negligible due to the event selection requirements described in Section 4. The list of simulated background samples and predicted generator cross-section values at 14 TeV are shown in Table 1.

| Background Process | Generator | Cross-section [pb] |
|---|---|---|
| $W$+jets | MadGraph5_aMC@NLO +Pythia 8 | $6.01\times10^{4}$ |
| $Z$+jets | Powheg-Box+Pythia 8 | $6.17\times10^{3}$ |
| $t\bar{t}$ | Powheg-Box+Pythia 6 | $8.24\times10^{2}$ |
| single-top $Wt$ channel | Powheg-Box+Pythia 6 | $8.05\times10^{1}$ |
| SM $WW/WZ \to \ell\nu qq$ | Powheg-Box+Pythia 6 | $5.99\times10^{1}$ |

Table 1: Summary of the simulated background samples considered in this analysis alongside the generator used and the predicted generator cross-section.

$W$+jets events are generated using MadGraph5_aMC@NLO v2.3.2 [17] at LO using the NNPDF30NLO PDF set [18], plus Pythia 8 [25] for fragmentation. They are simulated for up to one additional parton at NLO and up to two additional partons at LO. The generation of $Z$+jets events is done using Powheg-Box event generator plus PYTHIA8 [25] for showering and fragmentation. The generation use the CT10 PDF set [26] and the AZNLO CTEQL1 tune for Powheg-Box+Pythia [24]. $Z \to \tau\tau$ and $W \to \tau\nu$ events are included in the $Z$+jets and $W$+jets samples, respectively. Diboson processes with one of the bosons decaying hadronically and the other leptonically are simulated using Powheg-Box v2 [27] with the CT10 PDF set and showered/hadronized using Pythia 6 [28].

For the generation of top-quark pairs, the Powheg-Box event generator with the CT10 PDF set in the matrix element calculations is used. Electroweak $Wt$-channel single-top quark events are generated using the Powheg-Box v1 event generator [29–31]. This event generator uses the four-flavour scheme for the NLO matrix-element calculations together with the fixed four-flavour PDF set CT10f4 [26]. For all top-quark processes, top-quark spin correlations are preserved (for $t$-channel, top-quarks are decayed using MadSpin [32]). The parton shower, fragmentation, and underlying event are simulated using Pythia 6.428 [28] with the CTEQ6L1 [33] PDF set and the corresponding Perugia 2012 tune [34]. The top-quark mass is set to 172.5 GeV. The EvtGen v1.2.0 program [35] is used to decay bottom and charm hadrons for the Powheg-Box samples.

## 3 Object Selection

Studies on the performance of the upgraded phase-II ATLAS detector at the HL-LHC are documented in Ref [36, 37]. The performance studies assume a center of mass energy of $\sqrt{s}$ = 14 TeV with the average number of $pp$ interactions per bunch crossing of 200. The upgraded detector is fully simulated for the detector response. The results of these studies were used to derive functions that provide parameterized estimates of the detector performance for different physics objects. These functions are $p_{\mathrm{T}}$- and $\eta$-dependent[2] and are applied to the generator-level quantities to emulate energy resolution, efficiencies and mis-identification (fake) rates. Efficiency functions are available for the identification of electron, muon and heavy quark associated with jets ($b$-tagging). Functions to parameterize fake rates are also extracted for $b$-tagging (light flavor jets and jets from pile-up collisions can be mis-identified as $b$-jets). Energy resolution functions are applied to alter the four-momenta of generator-level objects by a random amount corresponding on average to the expected resolution of the detector. The object selection criteria are applied to the altered four-momenta.

[2] ATLAS uses a right-handed coordinate system with its origin at the nominal interaction point (IP) in the center of the detector and the $z$-axis along the beam pipe. The $x$-axis points from the IP to the center of the LHC ring, and the $y$-axis points upwards. Cylindrical coordinates $(r, \phi)$ are used in the transverse plane, $\phi$ being the azimuthal angle around the beam pipe. The pseudorapidity is defined in terms of the polar angle $\theta$ as $\eta = -\ln\tan(\theta/2)$. Angular distance is measured in units of $\Delta R \equiv \sqrt{(\Delta\eta)^2 + (\Delta\phi)^2}$.

The following objects are used in this analysis:

- Generator-level electrons or muons are required to be isolated by ensuring that the sum of the $p_{\mathrm{T}}$ of other final state charged particles within $\Delta R = 10\,\mathrm{GeV}/p_{\mathrm{T}}^{\ell}$ around the lepton is less than 6% of the lepton $p_{\mathrm{T}}$. Lepton identification efficiencies are applied to the selected isolated electrons and muons. Both electrons and muons are required to pass the tight identification criteria [36], which is the most effective to reduce mis-identified leptons . Once the identification efficiencies are applied, the energies of the remaining leptons are smeared according to the expected detector resolution. They are both required to have $p_{\mathrm{T}} > 27$ GeV and be within the acceptance of the inner detector ($|\eta| < 1.37$ or $1.52 < |\eta| < 2.47$ for electrons and $|\eta| < 2.5$ for muons). The $\eta$ acceptance cuts are chosen to match those used in previous searches even though the HL-LHC acceptances are expected to increase since the leptons in our signal models are expected to be mainly produced centrally in the detector.
- Small-$R$ jets (denoted by $j$ below): The anti-$k_t$ algorithm [38] with a radius parameter of R = 0.4 is used to reconstruct small-$R$ jets from final state generator-level particles. Pile-up jets are included from a pileup library built with the assumption of the average 200 proton-proton interactions per bunch crossing. The transverse momentum of the jets is smeared by 10-25% using an $\eta/p_{\mathrm{T}}$-dependent parameterization; the jets are requested to satisfy $|\eta| < 2.5$ and have a minimum $p_{\mathrm{T}}$ of 20 GeV. The identification of jets originating from $b$-quarks is done by finding jets with generator-level $b$-hadron within a cone of $\Delta R < 0.4$ around the jet direction. The $b$-tagging performance for these jets is then modeled by applying ($\eta/p_{\mathrm{T}}$-dependent) $b$-tagging efficiency function corresponding to a mean efficiency of 70% in simulated $t\bar{t}$ events. The $b$-tagging mis-identification rate is also applied to light flavor quark and gluon jets (including pile-up jets) to account for jets that do not contain $b$-hadrons but are accidentally identified as $b$-jets. A track confirmation algorithm is used to mitigate pile-up by selecting jets with generator-level charged hadrons that can be traced back to the primary vertex; the track confirmation efficiencies are applied to jets from hard-scattering and pile-up interactions.
- Large-$R$ jets (denoted by $J$ below): The anti-$k_t$ algorithm with a radius parameter of R = 1.0 is used to reconstruct large-$R$ jets. The large-$R$ jets are trimmed using the standard ATLAS trimming parameters of $p_{\mathrm{T}}$ fraction = 0.05 and R = 0.2 [39]. It is assumed that the performance of a future $W/Z$-boson tagger at the HL-LHC conditions will have similar, if not better, performance as existing boson taggers. To simulate the effect of Run-2 $W/Z$-boson tagging performance [40, 41] on local-calibrated topologically-clustered jets [42], events which contain a large-$R$ jet are scaled by the expected boson tagging efficiency for the $V \to qq$ with kinematics corresponding to the large-$R$ jet. The tagging efficiencies are calculated from fully-simulated 13 TeV Monte-Carlo (MC) samples as the fraction of events with a large-$R$ jet passing the tagger to the number of events with large-$R$ jets within $|m(J) - m(W/Z)| < 50$ GeV, where $m(J)$ is the large-R jet mass. The mass cut imposed by the tagger is always smaller then the 50 GeV window applied. The mass window cut is applied whenever the $W/Z$-boson tagger scale factors are applied and used to incorporate the shaping effect of a mass cut on the background distributions. Only the leading large-$R$ jet is considered for the tagging efficiency calculation. Separate efficiencies are calculated for each background and signal processes and applied as scale factors on the 14 TeV simulated samples. The $p_{\mathrm{T}}$ (mass) of the

large-$R$ jets is smeared using a Gaussian (Log-Normal) distribution with scale parameters derived as a function of $p_{\mathrm{T}}$ and $m(J)/p_{\mathrm{T}}$. The large-$R$ jets in events must have $p_{\mathrm{T}} > 200$ GeV and $|\eta| < 2.0$.

- Missing transverse momentum (with its magnitude $E_{\mathrm{T}}^{\mathrm{miss}}$, refered to as the missing transverse energy below): The missing transverse momentum is determined as the negative sum of the transverse momenta of all generator-level particles, except neutrinos, within the detector acceptance. The $x$ and $y$ components of the missing transverse momentum are smeared with the detector resolution and are used to obtain the total missing transverse energy in the event [43].
- Overlap Removal: Energy deposits in the calorimeter from electrons and muons can be reconstructed as jets. In this analysis, the jets are reconstructed from stable generator-level particles excluding muons and electrons from the decay of $W$, $Z$, Higgs and $\tau$ particles as well as neutrinos. To avoid double counting an electron as a jet, the nearest small-$R$ jet to an electron is removed from the list of jets if $\Delta R(e, \mathrm{jet}) < 0.2$. Additionally, small-$R$ jets within $\Delta R = 1.0$ of any large-$R$ jet in the event are removed from the list of jets.

# 4 Event Selection

Events are required to have exactly one lepton satisfying the selection criteria described above. It is assumed that the effect of trigger thresholds is negligible for the selected leptons with $p_T$ studied in this note. Events are further required to contain a hadronically-decaying $W/Z$ candidate, reconstructed either from two small-$R$ jets, defined as the resolved channel, or from one large-$R$ jet, designated the boosted channel (see below).

The missing transverse energy $E_{\mathrm{T}}^{\mathrm{miss}}$ has to be greater than 60 GeV, which suppresses multijet background to a negligible level. By constraining the $E_{\mathrm{T}}^{\mathrm{miss}}$ + lepton system to be consistent with the $W$ mass, the $z$ component of the neutrino ($\nu$) momentum can be reconstructed by solving a quadratic equation. The smallest solution is chosen and in the case where the solution is imaginary, only the real part is taken.

The analyses for the resonance search are detailed in Section 4.1 and the VBS search in Section 4.2. The event selections for the two analyses are considered separately and events can be selected by both analyses.

## 4.1 Resonance Search

The presence of narrow resonances is searched for in the distribution of reconstructed diboson mass using the signal shapes extracted from simulation of benchmark models. The invariant mass of the diboson system ($m(WV)$) is reconstructed from the leptonic $W$ candidate and hadronic $W/Z$ candidate, the latter of which is obtained from two small-$R$ jets in the resolved channel ($m(\ell\nu jj)$) or large-$R$ jet in the boosted channel ($m(\ell\nu J)$). The background shape and normalization are obtained from MC simulation with dedicated control regions to constrain systematic uncertainties of the background modeling and normalization. The following control regions are used in the final fit:

- If the event satisfies all the selection criteria except the $W/Z$-boson mass-window cut with no b-jets (b-veto) then the event is categorized as a $W$ control region event.

- If the event satisfies all the selection criteria and has additional b-jets then the event is categorized as a *top* control region event.

These regions are used to constrain the top and $W$+jets background normalization and shape uncertainties as described in Section 5.

The search is divided into two orthogonal categories to identify the ggF/q$\bar{\text{q}}$ and VBF production modes by identifying additional forward jets. If an event passes the VBF category selection for the additional forward jets (defined below) it is categorized as a VBF candidate event, otherise as a ggF/q$\bar{\text{q}}$ candidate event. Events are then processed by a merged-jet selection then a two resolved-jet selection if they fail the merged selection. This prioritization strategy provides the optimum signal sensitivity as it favors the merged selection which contains less background contribution. A summary of the selection in the resonance search is presented in Table 2. The final distributions in the signal regions for the resonance search can be seen in Figure 2. The acceptance curves for the resonance search as function of VBF and $(q\bar{q})$ produced HVT $W'$ signal masses can be found in Figure 4.

### 4.1.1 VBF Selection

Before any further selection, if an event contains two non-$b$-tagged small-$R$ jets with $\eta(j_1^{\text{tag}}) \cdot \eta(j_2^{\text{tag}}) < 0$, a pseudorapidity separation of $\Delta\eta = |\eta(j_1^{\text{tag}}) - \eta(j_2^{\text{tag}})| > 4.7$ and an invariant mass greater than 770 GeV, then the event is categorized as a VBF candidate event. If more than one such pairs of jets are found in the event, the one with the highest dijet invariant mass is chosen. These jets are not considered when searching for $V \to qq$ candidates.

### 4.1.2 Merged selection

In the merged selection, events are required to have at least one large-$R$ jets with $p_{\text{T}}(J) > 200$ GeV and $|\eta(J)| < 2$. If two or more large-$R$ jets are found, the one with the highest $p_{\text{T}}$ is chosen as a hadronically decaying $W/Z$ boson candidate. A mass window cut of $|m(J) - m(W/Z)| < 50$ GeV is then applied and the selected events are then scaled by the expected efficiency of a $W/Z$ boson tagger as described in Sec 3. If the selected large-$R$ jet contains a $b$-quark within $\Delta R < 1$ around the jet, then the event is rejected to reduce contribution from $t\bar{t}$ background.

Events are further required to pass the stricter $E_{\text{T}}^{\text{miss}}$ cut of $> 100$ GeV and have $p_{\text{T}}(\ell\nu) > 200$ GeV. If the lepton candidate is an electron, $E_{\text{T}}^{\text{miss}}/p_{\text{T}}(\ell\nu) > 0.2$ is required to reduce the multijet background contribution to a negligible level, as shown in the 13 TeV search [8]. Additional requirements that the $p_{\text{T}}$ of both hadronic and leptonic $V$ candidates are greater than 40% (30%) of $m(VV)$ for the ggF/q$\bar{\text{q}}$ (VBF) selection are also applied.

### 4.1.3 Resolved selection

Events which fail the merged selection and have at least two small-$R$ jets are then processed by the resolved selection. In the resolved selection, the hadronically decaying $W/Z$ candidate is reconstructed from the pair of small-$R$ jets with the mass $m(jj)$ closest to the $W/Z$ mass among all combinations of jets with $p_{\text{T}} > 20$ GeV. This jet pairing strategy is chosen as it gives the best signal sensitivity. The leading (sub-leading) signal jet is further required to have $p_{\text{T}} > 60(40)$ GeV after the jet pairs are selected to improve separation between the signal and the background. These signal jets are then required to be in either the $W$ mass window of $66 < m(jj) < 94$ GeV or the $Z$ mass window of $82 < m(jj) < 106$ GeV. To suppress backgrounds containing top quarks, events are vetoed if they contain any $b$-tagged jets not part of the hadronic $W/Z$-boson candidate. In addition, if the dijet mass is within the W mass window and both jets are $b$-tagged, the event is removed. This cut is not applied to the $Z$-boson candidate in order to retain signal with $Z \to bb$ decay.

The selected event is further required to fulfill the following angular conditions to enhance signal-like topology: $\Delta\phi(j_i, \ell) > 1$, $\Delta\phi(j_i, \nu) > 1$, $\Delta\phi(j_1, j_2) < 1$ and $\Delta\phi(\ell, \nu) < 1$. Furthermore, the leptonic $W$ candidate is required to have $p_{\mathrm{T}}(W) >$75 GeV and if the lepton candidate is an electron there is an additional requirement of $E_{\mathrm{T}}^{\mathrm{miss}}/p_{\mathrm{T}}(\ell\nu) > 0.2$. Similarly to the merged selection, both $V$ candidates are required to have $p_{\mathrm{T}}(\ell\nu/jj)/m(\ell\nu jj) > 0.35$ (0.3) for the ggF/q$\bar{\mathrm{q}}$ (VBF) selection.

## 4.2 Vector boson scattering search

Experimentally, VBS is characterized by the presence of a pair of vector bosons ($W$, $Z$, or $\gamma$) and two forward jets with a large separation in pseudorapidity and a large dijet invariant mass. Therefore the VBS search is required to have 2 additional forward VBS-topology tagging jets in the event in addition to jets associated with the boson decay, similar to the resonant VBF search.

The VBS tagging jets are required to be non-$b$-tagged, in order to suppress the contribution of diagrams with a $Wtb$ vertex (especially the electroweak $t\bar{t}$ production) in the electroweak $VVjj$ production. Tagging jets must be in the opposite hemispheres, $\eta(j_1^{\mathrm{tag}}) \cdot \eta(j_2^{\mathrm{tag}}) < 0$, and to have the highest dijet invariant mass among all pairs of jets remaining in the event after the $V \to jj$ jet selection. This jet prioritization scheme is the reverse of the resonant VBF scheme and offers better background rejection in the $p_{\mathrm{T}}$ regime of interest in the VBS search. After the tagging jet pair are selected, it is required that both tagging jets should have $p_{\mathrm{T}} >$30 GeV, in order to suppress the contribution from pile-up, and that the invariant mass of the two tagging jets system is greater than 400 GeV. In the VBS search the VBS-tagging jets are selected after the signal jets and are required to be $\Delta R > 1$ away from signal large-$R$ jets.

To optimize the signal sensitivity, Boosted Decision Trees (BDT) for the resolved and merged searches were trained on the background and signal MC samples in the respective regions.Four variables are included in the merged BDT: the invariant mass of the $lvJ$ system, the lepton $\eta$, the second tag jet $p_T$ and the boson centrality $\zeta_V$. The boson centrality is defined as $\zeta_V = min(\Delta\eta_+, \Delta\eta_-)$ where $\Delta\eta_+ = max(\eta(j_1^{\mathrm{tag}}), \eta(j_2^{\mathrm{tag}})) - max(\eta(\ell\nu), \eta(J))$ and $\Delta\eta_- = min(\eta(\ell\nu), \eta(J)) - min(\eta(j_1^{\mathrm{tag}}), \eta(j_2^{\mathrm{tag}}))$. In the resolved BDT, eight variables were used: the invariant mass of the $WVjj$ system , the lepton $\eta$, the $p_T$ of both VBS-tagging jets and sub-leading signal jet, the boson centrality defined similarly to above, the $\Delta\eta$ between signal jets, and the $\Delta R$ between the lepton and neutrino candidate. These variables were chosen as they are the minimal subset of variables with the greatest separation between the signal and background, that provide significant improvement when added during the training. The BDT were trained using a gradient descent BDT algorithm, maximizing the the Gini index, in the TMVA package [44]. The BDT are chosen as the discriminants and their distributions are used in the final fit for the VBS search. Similar to the resonance search, if any event fails either a mass-window cut or a $b$-veto but passes all other events then the event is categorized as a $W$ or top control region. The normalized BDT response and the invariant mass distribution of the diboson pair are displayed in Figure 3 for the VBS search. A summary of the selection in the VBS search is presented in Table 3.

### 4.2.1 Merged selection

In the merged selection, events are required to have at least one large-$R$ jet with $p_{\mathrm{T}}(J) > 200$ GeV and $|\eta(J)| < 2$. From those candidate large-$R$ jets, the one with the smallest $|m(J) - m(W/Z)|$ is selected as the signal large-$R$ jet. Mass window cuts and boson tagging efficiencies are applied in the same way as the resonance search.

To suppress backgrounds with top quarks, an event is rejected if any of the reconstructed jets outside the large R jet, is identified as containing a $b$-quark.

### 4.2.2 Resolved selection

If events fail the merged VBS selection, the resolved selection is then applied. Signal jets are chosen in the same way as the resolved resonance search, except that the additional $p_T$ requirements on the signal jets are not imposed. The signal jet pairs are then required to have $|m(jj) - m(W/Z)| < 15$ GeV. To suppress backgrounds with top quarks, an event is rejected if any of the reconstructed jets is identified as containing a $b$-quark.

<table>
<tr><th>Selection</th><th>Resonance Resolved</th><th>Resonance Merged</th></tr>
<tr><td rowspan="3">$W \to \ell\nu$</td><td colspan="2">1 isolated “tight" lepton<br>0 additional “loose" leptons</td></tr>
<tr><td>$E_{\mathrm{T}}^{\mathrm{miss}} > 60$ GeV<br>$p_{\mathrm{T}}(\ell\nu) > 75$ GeV</td><td>$E_{\mathrm{T}}^{\mathrm{miss}} > 100$ GeV<br>$p_{\mathrm{T}}(\ell\nu) > 200$ GeV</td></tr>
<tr><td colspan="2">$E_{\mathrm{T}}^{\mathrm{miss}}/p_{\mathrm{T}}(e\nu) > 0.2$</td></tr>
<tr><td>$V \to jj$</td><td>2 small-R jets<br>$min|m(jj) - m(W/Z)|$<br>$p_{\mathrm{T}}(j_1) > 60$ GeV, $p_{\mathrm{T}}(j_2) > 40$ GeV<br>$66 < m(jj) < 94$ GeV<br>or $82 < m(jj) < 106$ GeV</td><td>large-R jet<br>highest $p_T$<br>$p_{\mathrm{T}}(J) > 200$ GeV, $|\eta(J)| < 2$<br>$|m(J) - m(W/Z)| < 50$ GeV<br>Scale by $W/Z$-tagger efficiency</td></tr>
<tr><td>Tagged jets (VBF Category)</td><td colspan="2">Non-$b$-tagged<br>$\eta(j_1^{\mathrm{tag}}) \cdot \eta(j_2^{\mathrm{tag}}) < 0$, highest $m(jj)$<br>$p_{\mathrm{T}}(j_{1,2}^{\mathrm{tag}}) > 30$ GeV, $m(jj) > 770$ GeV, $\Delta\eta(j,j) > 4.7$</td></tr>
<tr><td>Topology</td><td>$p_{\mathrm{T}}(\ell\nu)/m(\ell\nu jj) > 0.35$ (0.3 for VBF)<br>$p_{\mathrm{T}}(jj)/m(\ell\nu jj) > 0.35$ (0.3 for VBF)<br>$\Delta\phi(j,\ell) > 1$, $\Delta\phi(j, E_{\mathrm{T}}^{\mathrm{miss}}) > 1$<br>$\Delta\phi(j,j) < 1$, $\Delta\phi(\ell, E_{\mathrm{T}}^{\mathrm{miss}}) < 1$</td><td>$p_{\mathrm{T}}(\ell\nu)/m(\ell\nu J) > 0.4$ (0.3 for VBF)<br>$p_{\mathrm{T}}(J)/m(\ell\nu J) > 0.4$ (0.3 for VBF)</td></tr>
<tr><td>b-veto</td><td colspan="2">No $b$-tagged jets in the event beside 1 (2) from $W(Z) \to jj$</td></tr>
</table>

Table 2: Summary of the event selection for the resonance search.

## 5 Systematics

Two types of systematic uncertainties are considered in the analysis: experimental uncertainties associated with the detector response and calibration of reconstructed objects, and uncertainties on the background modeling. Among those uncertainties, the most dominant systematic sources are considered for each type.

| Selection | VBS Resolved | VBS Merged |
|---|---|---|
| $W \to \ell\nu$ | 1 isolated "tight" lepton<br>0 additional "loose" leptons<br>$E_{\mathrm{T}}^{\mathrm{miss}} > 80$ GeV | |
| $V \to jj$ | 2 small-R jets<br>$min\|m(jj) - m(W/Z)\|$<br>$p_{\mathrm{T}}(j_1) >$40 GeV, $p_{\mathrm{T}}(j_2) >$20 GeV<br>$66 < m(j,j) < 106$ GeV | large-R jet<br>$min\|m(J) - m(W/Z)\|$<br>$p_{\mathrm{T}}(J) >$200 GeV, $\|\eta(J)\| < 2$<br>$\|m(J) - m(W/Z)\| < 50$ GeV<br>Scale by $W/Z$-tagger efficiency |
| Tagged jets | Non-$b$-tagged<br>$\eta(j_1^{\mathrm{tag}}) \cdot \eta(j_2^{\mathrm{tag}}) < 0$, highest $m(jj)$<br>$p_{\mathrm{T}}(j_{1,2}^{\mathrm{tag}}) > 30$ GeV, $m(jj)$>400 GeV | |
| | | $\Delta R(j_{1,2}^{tag}, J) > 1.0$ |
| b-veto | No $b$-tagged jets in the event | |

Table 3: Summary of the event selection for the VBS search.

For experimental sources, the jet energy resolution uncertainties are considered for the small (large)-R jets used in the resonance and VBS searches with the resolved (merged) selections. In addition, the large-R jet mass resolution uncertainties are considered for the merged selections. The resolution uncertainties are taken from the corresponding Run 2 search and reduced by 50% to account for the expected size at the time of the HL-LHC.

For theoretical uncertainties, the normalization uncertainties are considered for all backgrounds as well as the signal normalization uncertainty originating from choice of pdf set. The uncertainties on the background distribution shapes are taken into account for $W$+jets and $t\bar{t}$ processes by taking the variation of the $m(VV)$ and BDT distributions. The shape uncertainties for both $W$+jets and $t\bar{t}$ are based on generator comparisons with respect to Sherpa [45] and MadGraph respectively. The values of the normalization and shape uncertainties are taken to be half of those from the Run-2 search except for the $W$+jet which is taken to be reduced by a factor of 10. The different scaling of $W$+jet shape uncertainty is chosen since the Madgraph-Sherpa shape systematic is known to be overly conservative in the phase space of interest [46], which is expected to greatly improve with further data-MC comparisons. Likewise the $W$+jet normalization uncertainty is reduced by factor 10 to represent the expected increase in cross-section uncertainty for this process, which corresponds to $\sim 90\%$ of the expected background, with a factor of 10 or 100 increase in statistics with 300 and 3000 fb$^{-1}$. The uncertainties due to limited statistical accuracy of simulation predictions are not considered in the analysis.

The dominant uncertainties in the VBS search and low-mass regime of the Resonant search are diboson normalization and $W$+jet modeling uncertainties In the high-mass regime the $W$+jet and $t\bar{t}$ modeling uncertainties dominate.

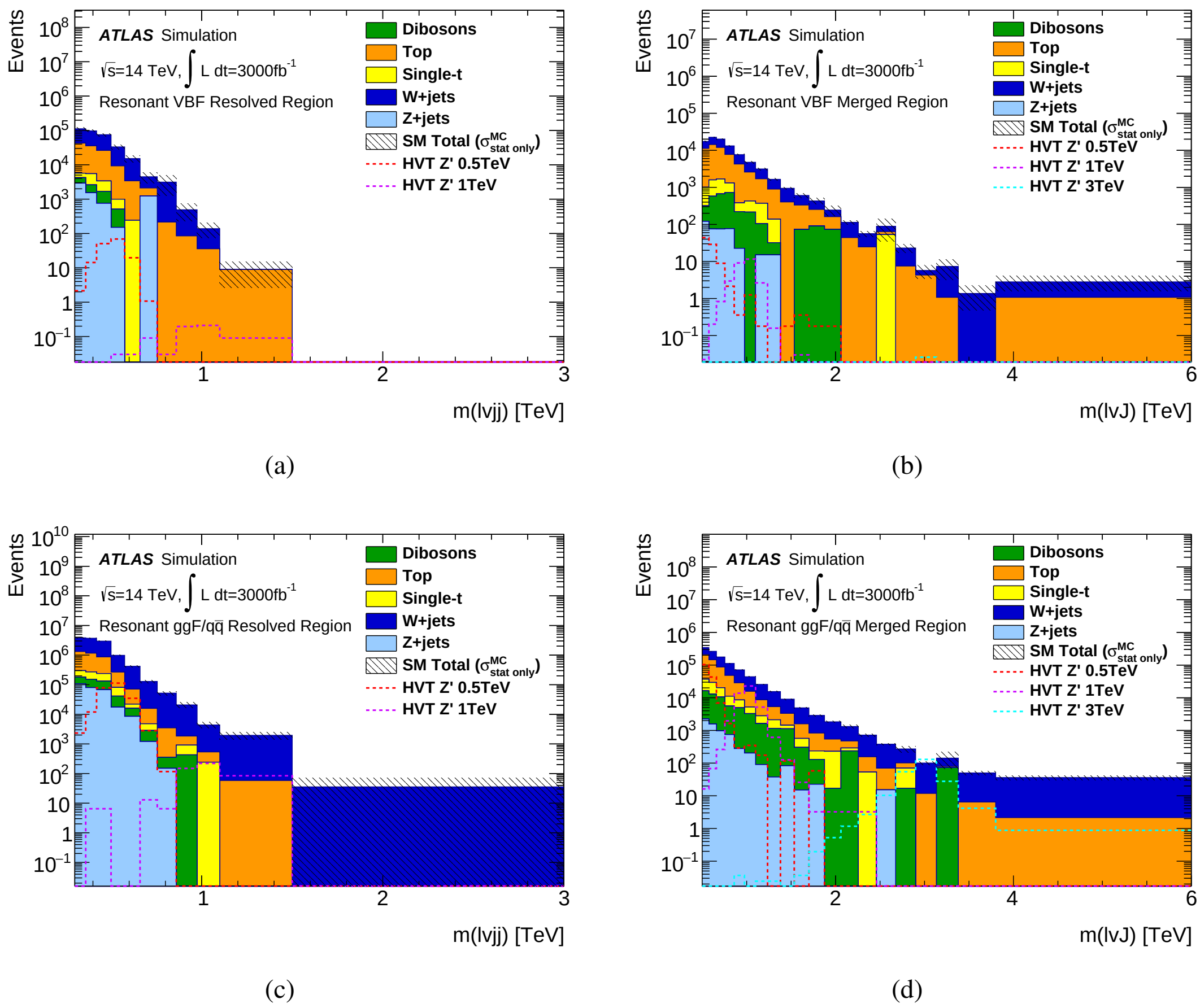

Figure 2: Final $m(\ell\nu jj)$ and $m(\ell\nu J)$ distributions in the resolved (left) and merged (right) signal regions respectively for the VBF resonance search (top) and ggF/q$\bar{q}$ resonance search (bottom). Background distributions are separated into production type. HVT signal for mass 0.5, 1, and 3 TeV are overlayed as dashed curves where appropriate.

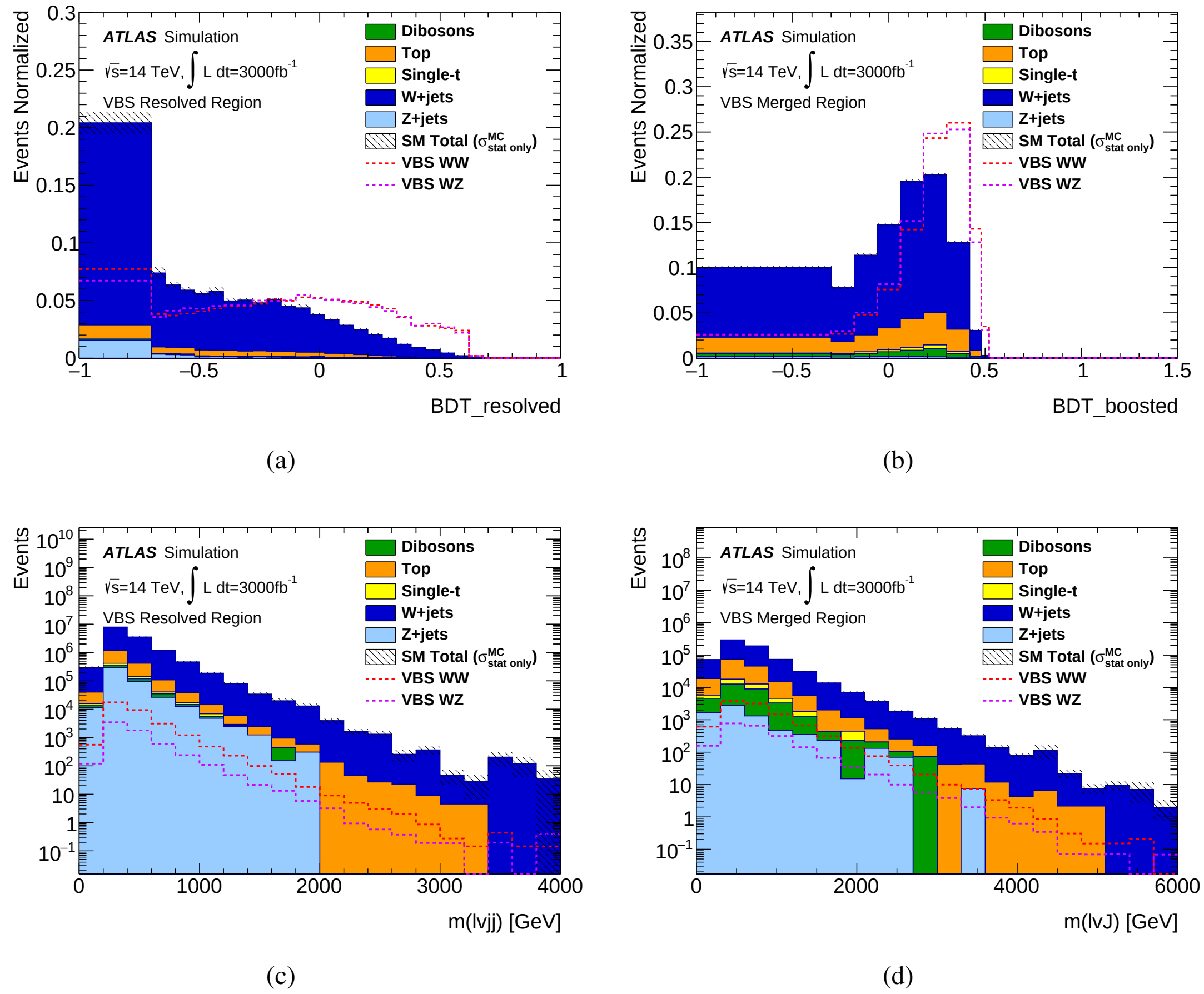

Figure 3: Final signal and background distributions for the VBS search in the respective resolved (left) and merged (right) signal regions for the normalized BDT response (top) and the reconstructed diboson invariant mass (bottom). Background distributions are separated into production type. VBS signals in $WW$ and $WZ$ mode are overlayed as dashed curves where appropriate. Both background and signal BDT distributions (top) are normalized to unity.

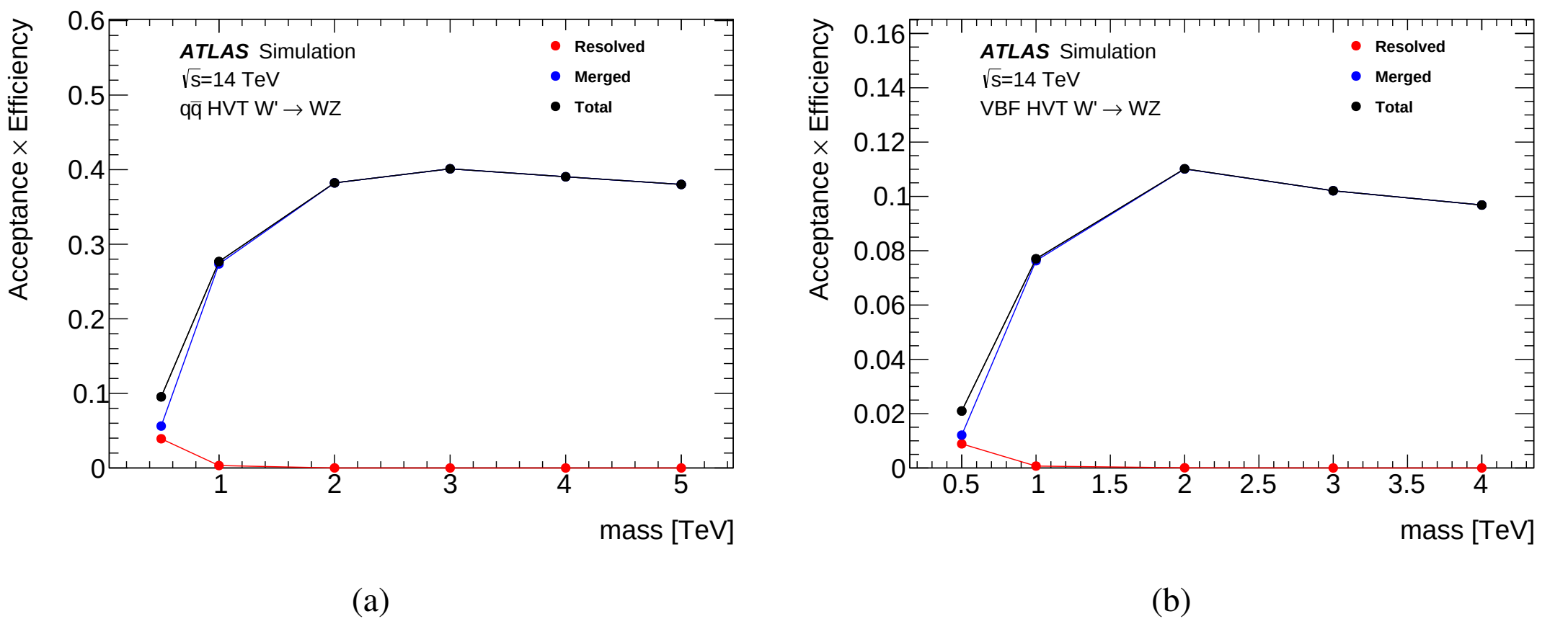

Figure 4: Acceptance times efficiency curves for the HVT $W'$ signal in the ggF/q$\bar{\text{q}}$ (left) and VBF (right) resonance selection as a function of signal mass. The q$\bar{\text{q}}$ produced signal is used for the ggF/q$\bar{\text{q}}$ plot and likewise for the VBF plot.

# 6 Results

## 6.1 Fit description

The results are extracted by performing a simultaneous binned maximum-likelihood fit to the $m(WV)$ distributions (BDT for the VBS search) in the signal regions and the $W$+jets and $t\bar{t}$ control regions. The $WW$ and $WZ$ channels for the resonance search are treated individually because they partially overlap. A test statistic based on the profile likelihood ratio [47] is used to test hypothesized values of the signal cross-section, separately for each model considered. The likelihood is defined as the product of the Poisson likelihoods for all signal and control regions for a given production mechanism category and channel. The fit includes five background contributions, corresponding to $W$+jets, $t\bar{t}$, single-top, $Z$+jets, and diboson.

Systematic uncertainties are taken into account as constrained nuisance parameters with Gaussian or log-normal distributions. For each source of systematic uncertainty, the correlations across bins of $m(WV)$ distributions and between different kinematic regions, as well as those between signal and background, are taken into account. The main background modelling systematics, namely the $W$+jets and $t\bar{t}$ shape uncertainties, are constrained by the corresponding control regions and are treated as uncorrelated among the resolved and merged signal regions.

The number of bins and the bin widths in each signal region are optimized according to the expected background event distribution and detector resolution. In all regions, the overflow events are included in the last bin.

## 6.2 Resonance search

For the resonance search, a statistical analysis is performed using the CLs [48] method to determine the expected upper limits that can be set on the signal cross section in the absence of signal. The expected upper limit set on the signal cross section is the greatest value of the signal cross-section that is not excluded with 95% confidence. This procedure is carried out for each signal mass.

The expected upper limits set on the signal cross section times branching ratio as a function of the signal mass are shown in Figure 5-6 for the ggF/q$\bar{\text{q}}$ and VBF categories at L =300 fb$^{-1}$ and L =3000 fb$^{-1}$ assuming pile-up conditions of 200 additional collisions per-crossing. A line showing the theoretical cross section for the HVT and Bulk Randall-Sundrum Graviton via ggF/q$\bar{\text{q}}$ production at each mass is overlayed and indicates the mass reach of the search.

For the HVT $W'$ and $Z'$ the limits are estimated to be 4.3 TeV with L =300 fb$^{-1}$ and and 4.9 TeV with L =3000 fb$^{-1}$ of $pp$ collisions, using the same detector configuration and pileup conditions. For the Bulk graviton the expected limits are estimated as 2.8 and 3.3 TeV at L =300 fb$^{-1}$ and L =3000 fb$^{-1}$. The values at L =3000 fb$^{-1}$ show an expected increase to the sensitivity of the search to the benchmark signals by $\sim$1 TeV with respect to existing limits in this channel [8].

In the circumstance that HL-LHC sees an excess, the expected sensitivity can also be characterized. The discovery significance is defined as the luminosity required to see a 5$\sigma$ effect of the signal. Here the signal significance is the quadratic sum of $s/\sqrt{s+b}$, for each bin of the final discriminant distribution at that luminosity, $s(b)$ representing the number of signal(background) events in the bin.

Figure 7 shows the expected discovery significance for the resonant search. In addition to the expected values, dashed curves shows the expected values for a future $W/Z$-tagger which has a 50% increase in signal efficiency and a further factor of 2 in background rejection. These values are representative of improvements seen in a recent diboson resonance search in the fully-hadronic $VV \to qqqq$ analysis[49] by using track-caloclusters[49] as opposed to locally-calibrated topologically-clustered calorimeter jets. Other possible improvements in $W/Z$-tagging in the HL-LHC era can originate from usage of more advanced machine-learning techniques to discriminate against the background contribution and better understanding of jet substructure variables with measurements at higher integrated luminosities.

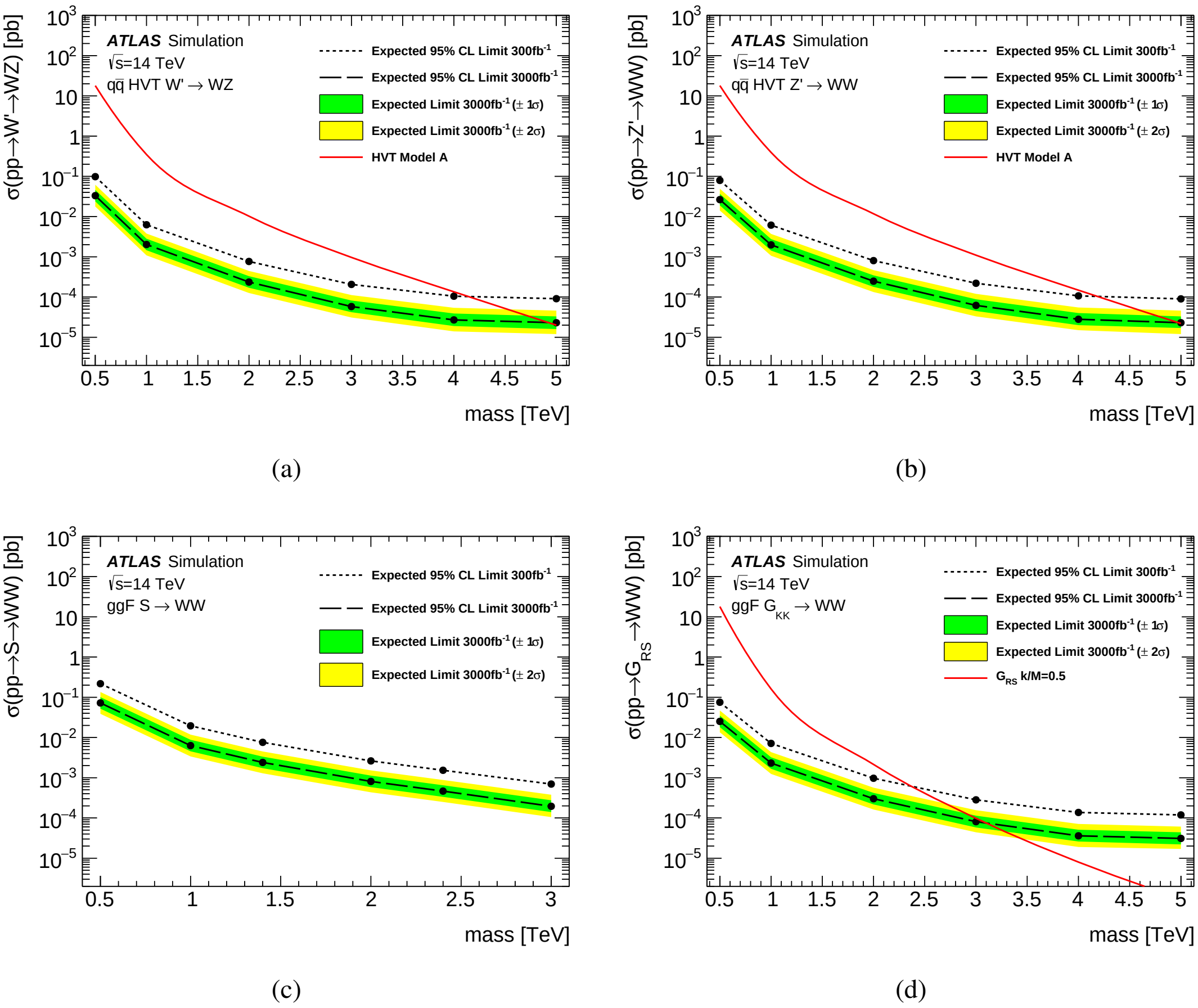

Figure 5: 95% Upper limit for the HVT $W'$ (top left), HVT $Z'$ (top right), Scalar (bottom left), and Graviton (bottom right) via ggF/$q\bar{q}$ production.

## 6.3 VBS search

For the VBS search, the statistical analysis is done on the signal strength of the SM VBS ($WW/WZ \to \ell\nu qq$) processes.

The expected significance for the SM VBS process is $5.7\sigma$ at 300 fb$^{-1}$. The expected cross-section uncertainties are 18% at 300 fb$^{-1}$ and 6.5% at 3000 fb$^{-1}$. The effects of unfolding were not considered for the cross-section estimates.

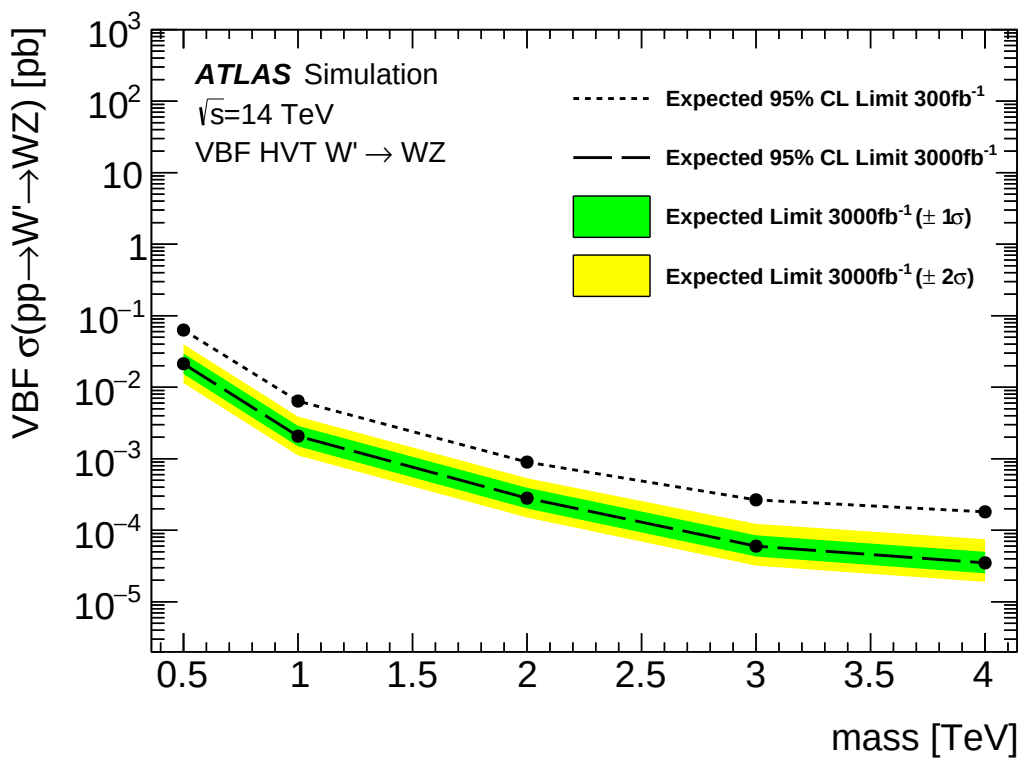

(a)

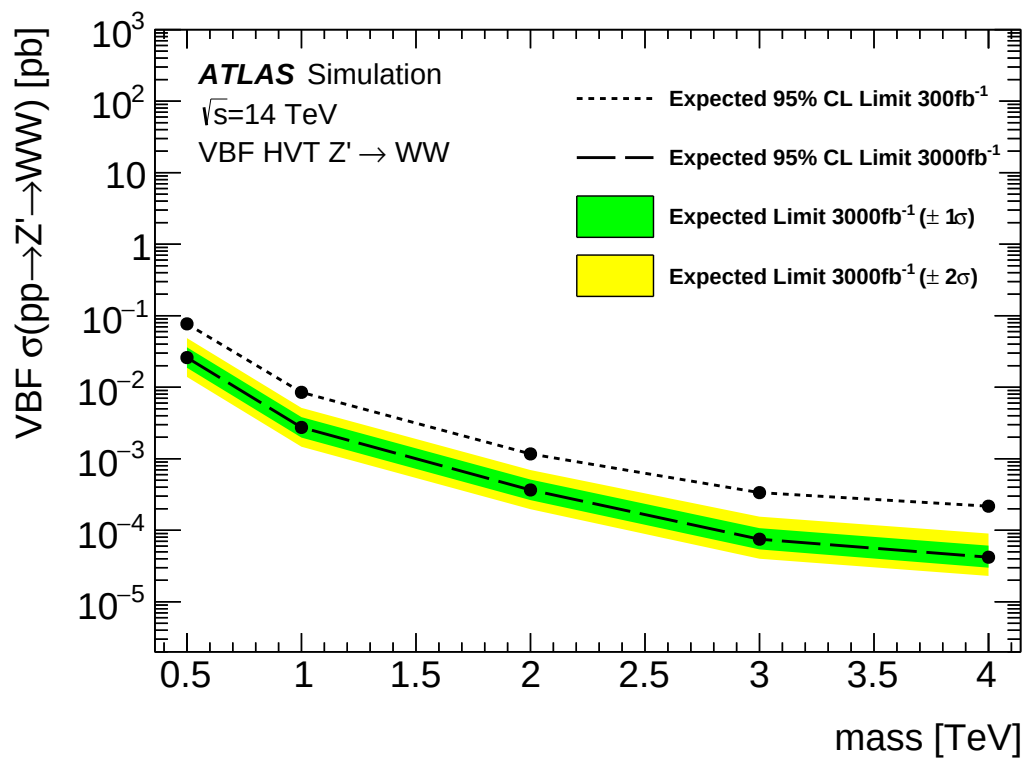

(b)

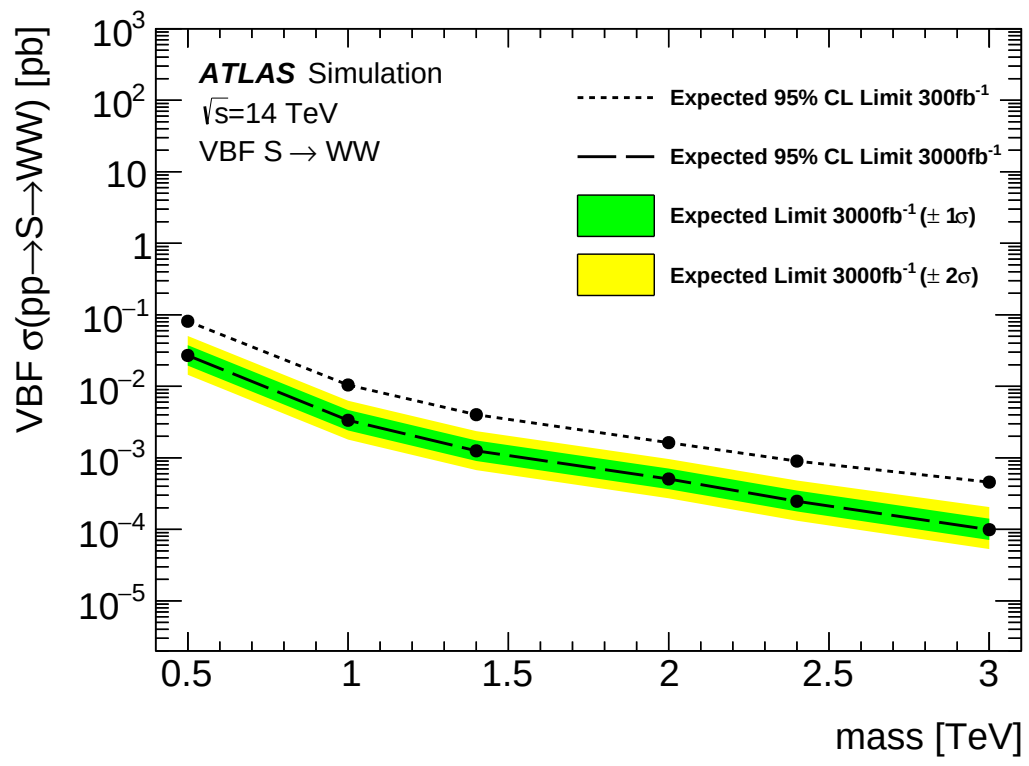

(c)

Figure 6: 95% Upper limit for the HVT $W'$ (left), HVT $Z'$ (right) and VBF Scalar(bottom) via VBF production.

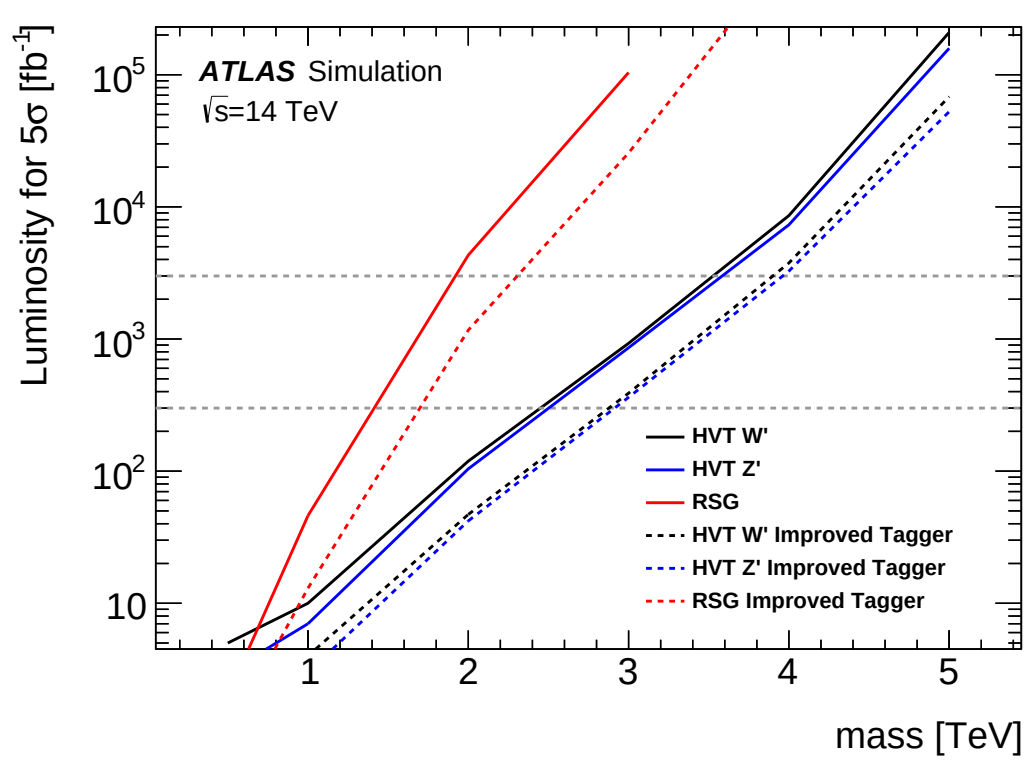

(a)

Figure 7: Expected luminosity required to observe a $5\sigma$ signal significance for the HVT $W'$ (black), HVT $Z'$ (blue) and $G_{RS}$ (red). The solid curves shows the sensitivity using the current $W/Z$-tagger and the dashed curves for a future tagger that has a 50% increased signal efficiency and a factor 2 increased rejection of background.

If control regions are not used to constrain the systematics the expected significance is reduced to $3.6\sigma$ at 300 fb$^{-1}$. Likewise the cross-section uncertainty are increased to 28% at 300 fb$^{-1}$ and 10% at 3000 fb$^{-1}$ when control regions are ignored.

Fig 8 shows the expected signal sensitivity and cross-section uncertainty as a function of integrated luminosity. In addition to the $\ell\nu qq$ channel, curves representing the estimated combined sensitivity including the other semi-leptonic channels, $\ell\ell qq$ and $\nu\nu qq$, are shown assuming they have equal sensitivity as the $\ell\nu qq$ channel. Here actual $p$-value calculations were done in comparison to Fig 7.

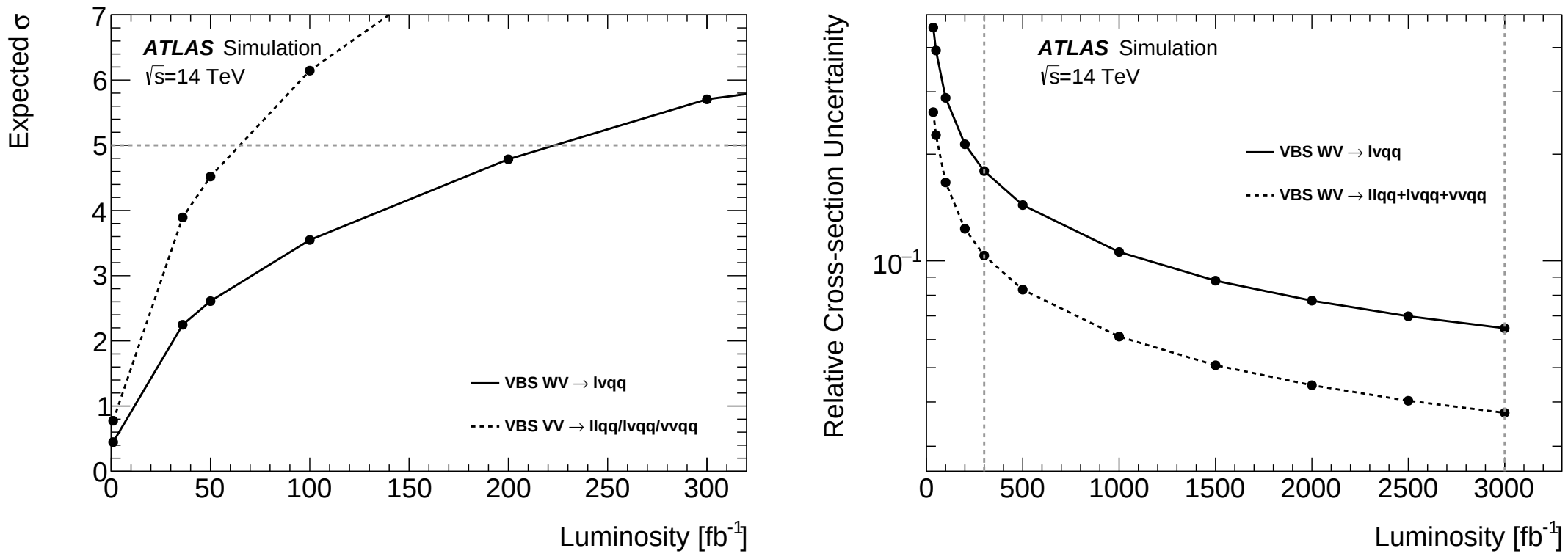

(a)

(b)

Figure 8: a) Expected signal significance as a function of integrated luminosity up to 300 fb$^{-1}$. The solid black curve is the significance from the $\ell\nu qq$ channel, while the black dashed curve shows the expected significance from all semi-leptonic channels assuming equal sensitivity. The grey dashed curve highlights the $5\sigma$ value. b) The expected cross-section uncertainty as function of integrated luminosity up to 3000 fb$^{-1}$. The solid black curve is the uncertainty from the $\ell\nu qq$ channel, while the dashed curve shows the expected uncertainty from all semi-leptonic channels assuming equal sensitivity. The grey dashed curve highlights the values at 300 fb$^{-1}$ and 3000 fb$^{-1}$. The effects of unfolding are not considered.

# 7 Conclusion

The prospects of searches for new heavy resonances decaying to diboson ($WW/WZ$) and measurements of electroweak $WW/WZ$ production via vector boson scattering in the semileptonic final states have been presented. The electroweak $WW/WZ$ production in vector boson scattering processes is expected to be observed with a significance of more than 5 standard deviations at 300 fb$^{-1}$and the expected cross-section measured to within 6.5% at 3000 fb$^{-1}$. The diboson resonance searches are interpreted for sensitivity to a heavy scalar singlet, a simplified phenomenological model with a heavy gauge boson and a Randall-Sundrum model with a spin-2 graviton. With 3000 fb$^{-1}$of $pp$ data, the mass limits for the new resonance is extended to 4.9 TeV for the HVT $W'/Z'$, and 3.4 TeV for the Bulk Graviton .

# CMS Physics Analysis Summary

Contact: cms-phys-conveners-ftr@cern.ch 

# Vector Boson Scattering prospective studies in the ZZ fully leptonic decay channel for the High-Luminosity and High-Energy LHC upgrades

The CMS Collaboration

## Abstract

Prospective studies for the vector boson scattering (VBS) in the ZZ channel at the HL-LHC are presented, where the Z bosons are identified and measured through their leptonic decays, $\ell = e, \mu$. The results obtained from the 2016 analysis with an integrated luminosity of 36 fb$^{-1}$ are projected to the HL-LHC luminosity of 3000 fb$^{-1}$ and center-of-mass energy of 14 TeV, taking into account the increased acceptance of the CMS detector. The projected uncertainty in the VBS ZZ cross section is 8.5–10.3% depending on the lepton $\eta$ coverage and assumptions made for the systematic uncertainties. A study is performed to separate the longitudinal polarization ($Z_L$) from the dominant transverse polarizations. The expected sensitivity for the VBS $Z_L Z_L$ fraction is 1.4 standard deviations. The foreseen upgrade coverage of up to $|\eta| = 3(2.8)$ for electrons (muons) leads to a 13% improvement in sensitivity compared to the Run 2 acceptance. Extending the coverage for electrons up to $|\eta| = 4$ would result in a modest increase in the sensitivity. Finally, the HE-LHC option would allow to bring the sensitivity at the $5\sigma$ level for this process.

# 1 Introduction

The high-luminosity LHC (HL-LHC) will operate at the center-of-mass (c.o.m) energy of 14 TeV and is expected to deliver to each experiment integrated luminosities of up to 3000 $\mathrm{fb}^{-1}$. It will provide a unique opportunity to search for rare physics processes. Weak vector boson scattering (VBS) is intimately related to the electroweak (EW) symmetry breaking mechanism (EWSB), the longitudinal mode of the gauge bosons being identified in the standard model (SM) with the Goldstone bosons of the EWSB. Unitarity restoration in the scattering of longitudinal weak bosons relies on the interference of the scattering amplitudes involving gauge couplings and couplings to the Higgs boson. While the studies of VBS have already been performed at the LHC Run 2 [1–4], the HL-LHC is expected to provide the first opportunity to study the longitudinal scattering of weak bosons.

Figure 1 shows some of the Feynman diagrams that contribute to EW production of the ZZjj signature, involving quartic (top left) and trilinear vertices (top right), as well as diagrams involving the Higgs boson (bottom left). The qq $\rightarrow$ ZZjj process can also be mediated through the strong interaction (bottom right in Fig. 1), which leads to the same final state as the VBS signal, and therefore constitutes an irreducible background.

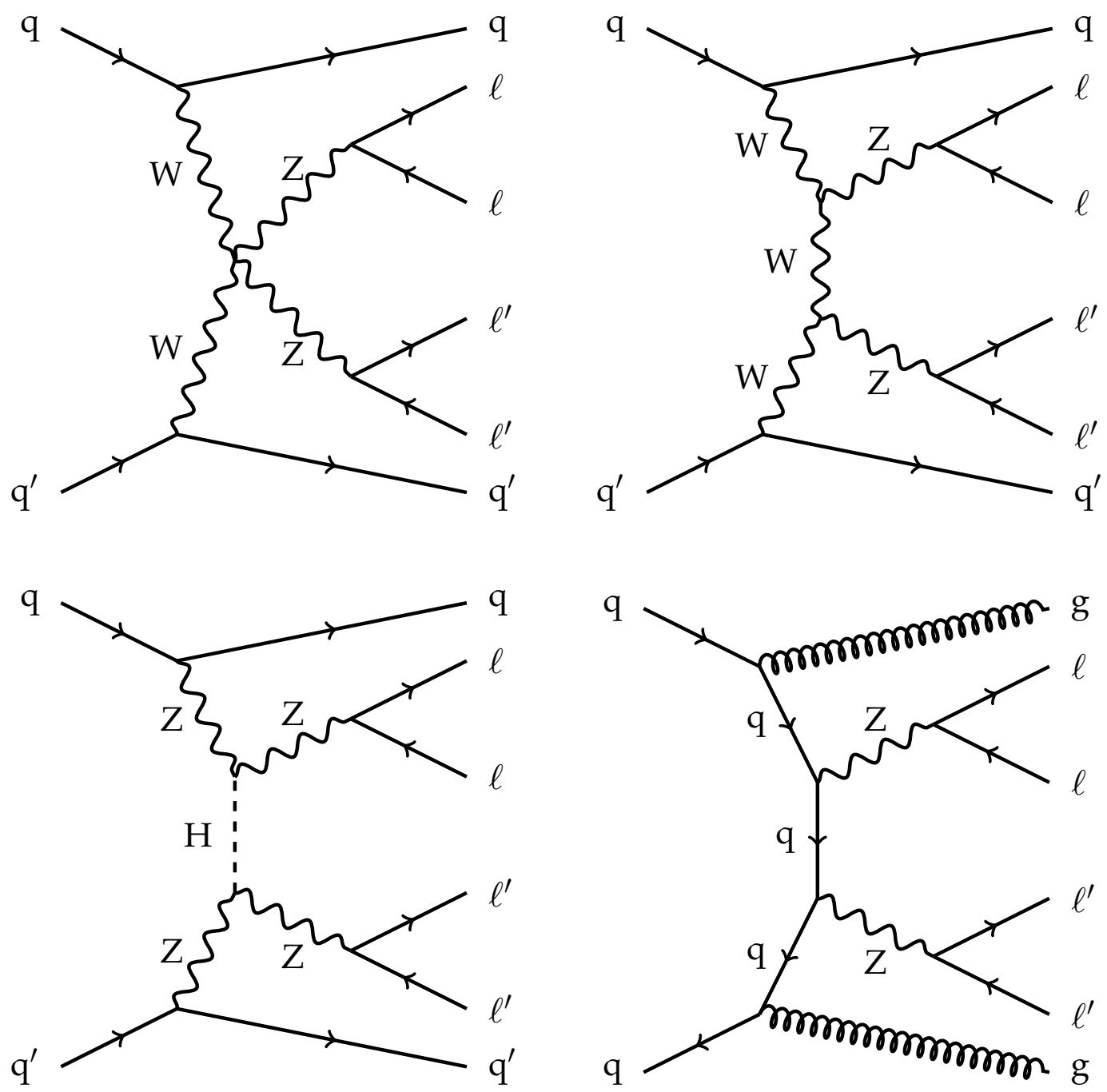

Figure 1: Representative Feynman diagrams for the EW- (top row and bottom left) and QCD-induced production (bottom right) of the ZZjj $\rightarrow \ell\ell\ell'\ell'$jj ($\ell$, $\ell' =$ e or $\mu$) final state. The scattering of massive gauge bosons as depicted in the top row is unitarized by the interference with amplitudes that feature the Higgs boson (bottom left).

This note presents prospective studies performed for VBS in the ZZ fully leptonic decay channel at the HL-LHC. It is based on the experimental investigation of VBS in the ZZ channel performed using data corresponding to an integrated luminosity of 36 $\mathrm{fb}^{-1}$ collected in 2016 and exploiting the fully leptonic final state, where both Z bosons decay into electrons or muons, $ZZ \rightarrow \ell\ell\ell'\ell'$ ($\ell$, $\ell' =$ e or $\mu$) [1]. Despite a low cross section, a small $Z \rightarrow \ell\ell$ branching fraction, and a large irreducible QCD background, this channel provides a favorable laboratory to study EWSB since all final-state particles are precisely reconstructed. In addition to a negligible reducible background, this channel provides a precise knowledge of the scattering energy through the measurement of $m_{4l}$. Furthermore, the measurement of the spin correlations of the final state fermions enables to identify the longitudinal contribution, which is the main interest of such studies. The longitudinal Z bosons ($Z_L$) are expected to be dominantly produced in the forward region [5], therefore a particular attention is payed on the lepton pseudorapidity coverage in the presented study.

The projected sensitivity for VBS ZZ is estimated by scaling the expected yields for the signal and the background processes, taking into account the increase in luminosity and scattering energy as well as the changes in acceptance and selection efficiencies between the Run 2 (13 TeV) and the Phase-2 (14 TeV) configurations. The Delphes simulation [6] is then used to assess the sensitivity to VBS $Z_LZ_L$.

After a brief reminder of the CMS detector Phase-2 upgrade in Section 2, the simulated samples used in this analysis are described in Section 3. The event selection and analysis strategy are then presented in Section 4. The effect of the increased acceptance and center-of-mass energy are discussed in Section 5 and the systematic uncertainties are addressed in Section 6. The sensitivity results for the VBS ZZ measurement at HL-LHC are presented in Section 7. The separation of the longitudinal component and results for the expected sensitivity and precision for the VBS $Z_LZ_L$ measurement are presented in Section 9. A summary of the analysis and results is given in Section 10.

## 2 CMS detector upgrade

The upgraded CERN High-Luminosity LHC is expected to deliver a peak instantaneous luminosity of up to $7.5 \times 10^{34}\,\mathrm{cm^{-2}\,s^{-1}}$ [7], which is an increase in instantaneous luminosity of about four times with respect to the LHC Run 2 performance. With this increase in instantaneous luminosity, the number of overlapping proton-proton interactions per bunch crossing, or pileup (PU), is expected to increase from its mean value of about 40 at the LHC to a mean value of up to 200 at the HL-LHC. Similarly, the levels of radiation are expected to significantly increase in all regions of the detector, in particular in its forward regions.

The CMS detector [8] will be substantially upgraded in order to fully exploit the physics potential offered by the increase in luminosity, and to cope with the demanding operational conditions at the HL-LHC [9–13]. In particular, in order to sustain the increased PU rate and associated increase in flux of particles, the upgrade will provide the detector with: higher granularity to reduce the average channel occupancy, increased bandwidth to accommodate the higher data rates, and improved trigger capability to keep the trigger rate at an acceptable level without compromising physics potential. The upgrade will also provide an improved radiation hardness to withstand the increased radiation levels.

The upgrade of the first level hardware trigger (L1) will allow for an increase of L1 rate and latency to about 750 kHz and 12.5 $\mu$s, respectively. The upgraded L1 will also feature inputs from the silicon strip tracker, allowing for real-time track fitting and particle-flow reconstruction [14] of objects at the trigger level. The high-level software trigger (HLT) is expected to reduce the rate by about a factor of 100 to 7.5 kHz.

The entire pixel and strip tracker detectors will be replaced to increase the granularity, reduce the material budget in the tracking volume, improve the radiation hardness, and extend the geometrical coverage and provide efficient tracking up to pseudorapidities of about $|\eta| = 4$.

In addition, the tracker will provide information on tracks above a configurable transverse momentum threshold to the L1 trigger, information presently only available at the HLT. It will also allow for including tracks with low momentum ( $\approx 2$ GeV).

The muon system will be enhanced by upgrading the electronics of the cathode strip chambers (CSC), resistive plate chambers (RPC) and drift tubes (DT). New muon detectors based on improved RPC and gas electron multiplier (GEM) technologies will be installed to add redundancy, increase the geometrical coverage up to about $|\eta| = 2.8$, and improve the trigger and reconstruction performance in the forward region.

The barrel electromagnetic calorimeter (ECAL) will be operated at lower temperatures to mitigate noise in avalanche photodiodes (APDs) due to radiation damage. Its upgraded front-end electronics will be able to exploit the information from single crystals at the L1 trigger level, to accommodate trigger latency and bandwidth requirements, and to provide an increased sampling rate of 160 MHz and high-precision timing capabilities. The hadronic calorimeter (HCAL), consisting in the barrel region of brass absorber plates and plastic scintillator layers, will be read out by silicon photomultipliers (SiPMs).

The endcap electromagnetic and hadron calorimeters will be replaced with a new combined sampling calorimeter (HGCal) that will provide coverage in pseudorapidity from about $|\eta| = 1.5$ up to $|\eta| = 3$. The new calorimeter will be based on a lead tungsten followed by stainless steel absorber with silicon sensors as the active material in the front section, and it will feature plastic scintillator tiles readout by SiPMs towards its back section at large distances from the beam. It will provide highly-segmented spatial information in both transverse and longitudinal directions, as well as 160 MHz sampling allowing high-precision timing capability for photons, which will allow for improved PU rejection and identification of electrons, photons, tau leptons, and jets.

Finally, the addition of a new precision timing detector for minimum ionizing particles (MTD) in both the barrel and endcap regions is envisaged to provide the capability for 4-dimensional reconstruction of interaction vertices that will significantly offset the CMS performance degradation due to high PU rates. The MTD is expected to achieve timing resolution of about 30 to 40 ps, and will provide coverage up to pseudorapidities of about $|\eta| = 3$.

A detailed overview of the CMS detector upgrade program is presented in Ref. [9–13]. The expected performance of the reconstruction algorithms and PU mitigation with the CMS detector is summarized in Ref. [15].

## 3 Monte Carlo samples and simulation

In addition to the samples used for the 13 TeV 2016 analysis and described in Ref. [1], simulated signal samples were produced for center-of-mass energies of $\sqrt{s} = 14$ and 13 TeV with polarization information on the outgoing Z bosons. Samples of simulated events for the main QCD background process were also produced for the center-of-mass energy of $\sqrt{s} = 14$ TeV.

The signal samples from purely electroweak VBS production, referred to as EW ZZ, are generated using MADGRAPH version 5.4.2 [16] and leading order (LO) version of PDFset NNPDF3.0 [17] with $\alpha_s = 0.13$ and using the 4-flavour scheme. The polarization information of individual Z bosons is kept by using the DECAY package from MADGRAPH5_aMC@NLO version 1.5.14 instead of MADSPIN. The 14 TeV sample is used to study the kinematics of polarized EW ZZ production and optimize the separation of the longitudinal component. The signal sample at 13 TeV is used to assess the effect of the change in center-of-mass energy.

Samples of events for the main irreducible QCD-induced pp $\rightarrow$ ZZjj process, referred to as QCD qqZZ, are produced at 14 TeV at next-to-leading-order (NLO) with up to two extra parton emissions with MadGraph5_aMCatNLO [18], and merged with parton showers using the FxFx scheme [19]. The jet multiplicities are simulated separately, in a similar way as was done in Ref. [1]. These samples are used to assess the effect of the change of center-of mass energy.

The PYTHIA v8.2 [20, 21] package is used for parton showering, hadronization and underlying event simulation. The fast-simulation package Delphes [6], with the CMS detector configuration corresponding to a number of pileup interactions of 200 (refered to as the 200PU configuration), is then used to simulate the expected response of the upgraded CMS detector.

# 4 Event selection and analysis

The analysis is based on the Run 2 investigation of VBS in the ZZ channel described in Ref. [1], with a data set corresponding to an integrated luminosity of 36 $\text{fb}^{-1}$. Run 2 results are projected into HL-LHC conditions, taking into account the effects of the increased lepton acceptance and center-of-mass energy in addition to the expected integrated luminosity.

The final state should contain at least two pairs of oppositely charged isolated leptons and at least two hadronic jets. The ZZ selection used is similar to that used in the CMS inclusive ZZ cross section measurement [22]. Events are required to contain at least two Z candidates, each formed from pairs of isolated and identified electrons or muons of opposite charges. Only reconstructed electrons (muons) with a $p_T > 7$ (5) GeV are considered. Among the four leptons, the highest $p_T$ lepton must have $p_T > 20$ GeV, and the second-highest $p_T$ lepton must have $p_T > 12$ (10) GeV if it is an electron (muon). Each pair of oppositely charged same-flavor leptons, is required to satisfy $60 < m_{\ell\ell} < 120$ GeV. At least two jets with $p_T > 30$ GeV and $|\eta| < 4.7$ are additionaly required. The two highest $p_T$ jets are referred to as the tagging jets and their invariant mass is required to be larger than 100 GeV. The above loose requirements defined the ZZjj selection used to extract the VBS signal.

The dominant background to the VBS search is the QCD-induced production of two Z bosons in association with jets. The yield and shape of the multivariate discriminant of this irreducible background is taken from simulation, but ultimately constrained by the data in the fit that extracts the EW signal. Reducible backgrounds arise from processes in which heavy-flavor jets produce secondary leptons or from processes in which jets are misidentified as leptons. The lepton identification and isolation and invariant mass requirements strongly suppress these backgrounds, which, after the selection, have a negligible impact on the results.

The determination of the signal strength for the EW production (ratio of the measured cross section to the SM expectation) employs a multivariate discriminant based on a boosted decision tree (BDT) to optimally separate the signal and the QCD background. Seven observables are used in the BDT, including $m_{jj}$, $|\Delta\eta_{jj}|$, $m_{ZZ}$, as well as the Zeppenfeld variables [23] $\eta^*_{Z_i} = \eta_{Z_i} - (\eta_{\text{jet}\,1} + \eta_{\text{jet}\,2})/2$ of the two Z bosons, and the ratio between the $p_T$ of the tagging jet system and the scalar $p_T$ sum of the tagging jets ($R(p_T)^{\text{jets}}$). The BDT also exploits the event balance $R(p_T)^{\text{hard}}$, which is defined as the transverse component of the vector sum of the Z bosons and tagging jets momenta, normalized to the scalar $p_T$ sum of the same objects [24]. The tunable hyper-parameters of the BDT training algorithm are optimized via a grid-search algorithm and the BDT performance was checked using a matrix element approach [25–27].

A maximum likelihood fit of the BDT distributions for signal and backgrounds is used to extract the signal strength. The shape and normalization of each distribution are allowed to vary within their respective uncertainties. The systematic uncertainties are treated as nuisance parameters in the fit and profiled.

## 5 Effect of the increased energy and acceptance

In addition to the luminosity scaling, a first effect comes from the difference in center-of-mass energy for the Run 2 (13 TeV) and the HL-LHC (14 TeV) configurations.

The cross sections are evaluated at LO with MADGRAPH (v5.4.2) [16] for the EW signal and the QCD qqZZ background, and with MCFM [28] for the QCD ggZZ background. The cross section ratio for the different processes are reported in Table 1. The signal cross section increases by about 15% while for the QCD qqZZ (ggZZ) background the increase is of about 17% (13%). The cross section ratios for the HE-LHC configuration (27 TeV) with respect to the HL-LHC configuration (14 TeV) is also reported.

Table 1: Cross section ratios $\sigma_{14\,\text{TeV}}$ / $\sigma_{13\,\text{TeV}}$ and $\sigma_{27\,\text{TeV}}$ / $\sigma_{14\,\text{TeV}}$ for the EW signal and the QCD background processes.

| | EW ZZ | QCD qqZZ | QCD ggZZ |
|---|---|---|---|
| $\sigma_{14\,\text{TeV}}$ / $\sigma_{13\,\text{TeV}}$ | 1.15 | 1.17 | 1.13 |
| $\sigma_{27\,\text{TeV}}$ / $\sigma_{14\,\text{TeV}}$ | 3.25 | 3.41 | 3.57 |

A second order effect arises from the difference in event acceptance between the two energies. It is estimated for each process at the reconstructed level with the 200PU configuration. The corrections are found to be small, up to $\sim$ 6%. It has been checked for all the observables used as input to the BDT that the shape differences induced by the change in center-of-mass energy are small.

The ratio of acceptances for various $\eta$ coverage configurations and for a center-of-mass energy of 13 TeV are reported in Table 2.

Table 2: Acceptance ratios for the Phase-2 detector with respect to Run 2 for various $\eta$ coverage configurations. The first number denotes the cut value for electrons while the number in parentheses denotes the cut value for muons. The numbers are for the center-of-mass energy of 13 TeV.

| | EW ZZ | QCD qqZZ | QCD ggZZ |
|---|---|---|---|
| $\|\eta\| < 3.0(2.8)/\|\eta\| < 2.5(2.4)$ | 1.13 | 1.18 | 1.12 |
| $\|\eta\| < 4.0(2.8)/\|\eta\| < 2.5(2.4)$ | 1.21 | 1.33 | 1.15 |

The increase in signal yield from the increased lepton acceptance for the Phase-2 detector is up to $\sim$ 20%. One can see also that an extension of up to $|\eta| < 4$ provides a sizeable increase in signal event yield, compared to $|\eta| < 3$. The event yield increase for the QCD qqZZ background is $\sim$ 10% higher than for the signal. The increase for the loop-induced ggZZ background is significantly lower, due to the Z production being more central for this process.

The shape differences induced on the variables used in the BDT by the change in detector acceptance at a given energy are found to be small. The most important difference appears for the Zeppenfeld variables as can be expected since these variables directly relate to the pseudo-rapidity of the final state Z bosons and therefore on the decay leptons. The change in $m_{jj}$ and $\Delta\eta_{jj}$, which weigh the most in the BDT discriminant, is very small.

# 6 Systematic uncertainties

In order to project the expected significance to HL-LHC configuration, two scenarios are considered for the systematic uncertainties. The first scenario ('Run 2 scenario') consists in using the same systematic uncertainties as that used for the Run 2 analysis, apart from the uncertainty in the QCD ggZZ background yield. In the second scenario ('YR18 scenario'), improved systematic uncertainties are assumed to be obtained from the more data and better understanding of the detector. In this scenario, the theory systematic uncertainties (PDF and QCD scales) are furthermore halved with respect to the Run 2 scenario.

Both shape and yield variations of the BDT output distributions for the signal and the various background are considered, in the same way as done for the Run 2 analysis [1].

For all processes apart from the sub-leading QCD ggZZ background, theoretical uncertainties were estimated by simultaneously varying the renormalization and factorization scales up and down by a factor of two with respect to the nominal value. As a VBS process the signal exhibits a weak dependence on the QCD scales choice and the size of the observed effect was found compatible with the NLO-LO comparison carried out in Ref. [29]. Uncertainties related to the choice of the PDF and strong coupling constant were evaluated following the PDF4LHC [30, 31] prescription and using the NNPDF [32] PDF sets. This procedure is also applied to the minor ttZ and WWZ backgrounds which have a negligible impact on the signal sensitivity.

The uncertainty associated to the QCD ggZZ background deserved a particular treatment. The gg $\rightarrow$ ZZ$jj$ loop-induced background, despite being suppressed by two order in $\alpha_S$ compared to the leading qq $\rightarrow$ ZZ$jj$, contributes significantly in the signal region. The kinematical distributions and in particular $m_{jj}$ appeared to be more signal-like. Being an $\alpha_S^4$ process at LO, this process is difficult to model and a flat uncertainty of 40% was assigned from the comparison of an MCFM simulation of gg $\rightarrow$ ZZ [28], therefore with the two extra jets from parton showers, and a MADGRAPH simulation of the QCD ggZZ background gg $\rightarrow$ ZZ$jj$.

The large uncertainty in the ggZZ loop-induced background yield has the highest impact on the significance, and is among the dominant uncertainties for the cross section measurement. Therefore, in addition to the values quoted in Table 3 for the YR18 scenario, the precision on the cross section measurement is also presented as a function of the uncertainty in the QCD ggZZ loop-induced background yield.

The experimental uncertainties are taken from the Run 2 analysis [1]. They include an uncertainty in the trigger efficiency, an uncertainty in the lepton selection efficiency (the numbers given in Table 3 stands for the 4e/2e2$\mu$/4$\mu$ final states, respectively) and an uncertainty in the pileup modeling estimated by varying the minimum bias cross section in the simulation by $\pm$4.6%. The jet energy scale (JES) uncertainty was estimated by varying the $p_T$ of the tagging jets by their respective uncertainty. The jet energy resolution (JER) in the simulation was corrected to match the distribution observed in the data and the uncertainty in the JER scaling factor is propagated to the simulated jets. The uncertainty in the data-driven reducible background, dominated by the statistic available in the control region, is sizeable but had a negligible effect on the sensitivity. The uncertainty in the luminosity is included as well.

The main source of systematic uncertainties and their effect on the signal and background yields are listed in Table 3. Other uncertainties in the minor ttZ and WWZ backgrounds are considered as well but are not listed in Table 3 as they have a negligible impact on the sensitivity.

Table 3: Effect of the systematic uncertainties on the signal and backgrounds yields for the two considered scenarios.

| Systematic source | Run 2 scenario | YR18 scenario |
|---|---|---|
| QCD scale, EW ZZ signal | 1–10% (shape) | 5% |
| PDF, EW ZZ signal | 8% (shape) | 4% |
| QCD scale, QCD qqZZ background | 8–12% (shape) | 6% |
| PDF, QCD qqZZ background | 3% (shape) | 1.5% |
| QCD ggZZ background | 10% | 10% or varied |
| Luminosity | 2.6% | 1% |
| Trigger efficiency | 2% | 1% |
| Lepton reco and selection efficiency | 6/4/2% | 2/1/0.5% |
| JES, EW ZZ signal | 1–5% (shape) | 1% |
| JER, EW ZZ signal | 1–2% (shape) | 1% |
| JES, QCD qqZZ background | 10–20% (shape) | 10% |
| JER, QCD qqZZ background | 3–6% (shape) | 1% |

For the cross section measurement, it is assumed that a fiducial cross section will be measured in a fiducial volume close to the detector level, such that the measurement is to first order insensitive to the theoretical uncertainties in the EW ZZ signal. Therefore, for this measurement, the nuisances corresponding to the EW ZZ signal uncertainties are frozen in the fit.

## 7 Results for VBS ZZ

The projected signal and background yields for the ZZjj selection defined in Section 5 and used in the statistical analysis, as well as for a VBS-enriched cut-based selection also requiring $m_{jj} > 400\,\text{GeV}$ and $|\Delta\eta_{jj}| > 2.4$, are reported in Table 4. The yields are given for an integrated luminosity of 3000 $\text{fb}^{-1}$. The reported uncertainties correspond to the Run 2 scenario, together with an uncertainty of 40% on the QCD ggZZ background yield as used for the Run 2 analysis. The reported event yields include the correction factors to account for the extended acceptance and the increase in center-of-mass energy as presented in Section 5. For the minor Z+X, ttZ and WWZ backgrounds, a correction factor similar to that evaluated for qqZZ is used. The corrections for the yields of these minor backgrounds lead to a change in projected significance of less than 1%.

A total of $\sim$ 705 events are expected for the VBS ZZ process in the fully leptonic final states for an integrated luminosity of 3000 $\text{fb}^{-1}$.

Figure 2 shows the scaled BDT output distribution for the signal and the various backgrounds for an integrated luminosity of 3000 $\text{fb}^{-1}$. The points represent pseudodata generated from the sum of the expected contributions of each process.

Figure 3 shows the projected significance for a 10% uncertainty in QCD ggZZ background yield, as a function of the integrated luminosity and for the two scenarios described in Section 6, as well as for a scenario with only the statistical uncertainty included. A sensitivity of $5\sigma$, where $\sigma$ stands for the standard deviation, is reached for an integrated luminosity of 225 $\text{fb}^{-1}$ if considering the statistical uncertainties only. It is reached for 280 (260) fb$^{-1}$ if considering the systematic uncertainties of the Run 2 (YR18) scenario.

Table 4: Signal and background yields projections for the ZZjj inclusive selection used in the statistical analysis and for a VBS cut-based selection also requiring $m_{jj} > 400\,\text{GeV}$ and $|\Delta\eta_{jj}| > 2.4$. Quoted uncertainties correspond to the systematic uncertainties for the Run 2 scenario together with a 40% uncertainty in the QCD ggZZ background yield, as used for the Run 2 analysis.

| Selection | $t\bar{t}Z$ and WWZ | QCD qqZZ + ggZZ | Total bkg. | EW ZZ signal | Total expected |
|---|---|---|---|---|---|
| ZZjj | $876 \pm 99$ | $11900 \pm 1700$ | $13600 \pm 1700$ | $706 \pm 79$ | $14300 \pm 1700$ |
| VBS cuts | $111 \pm 25$ | $2340 \pm 490$ | $2530 \pm 510$ | $456 \pm 57$ | $2990 \pm 480$ |

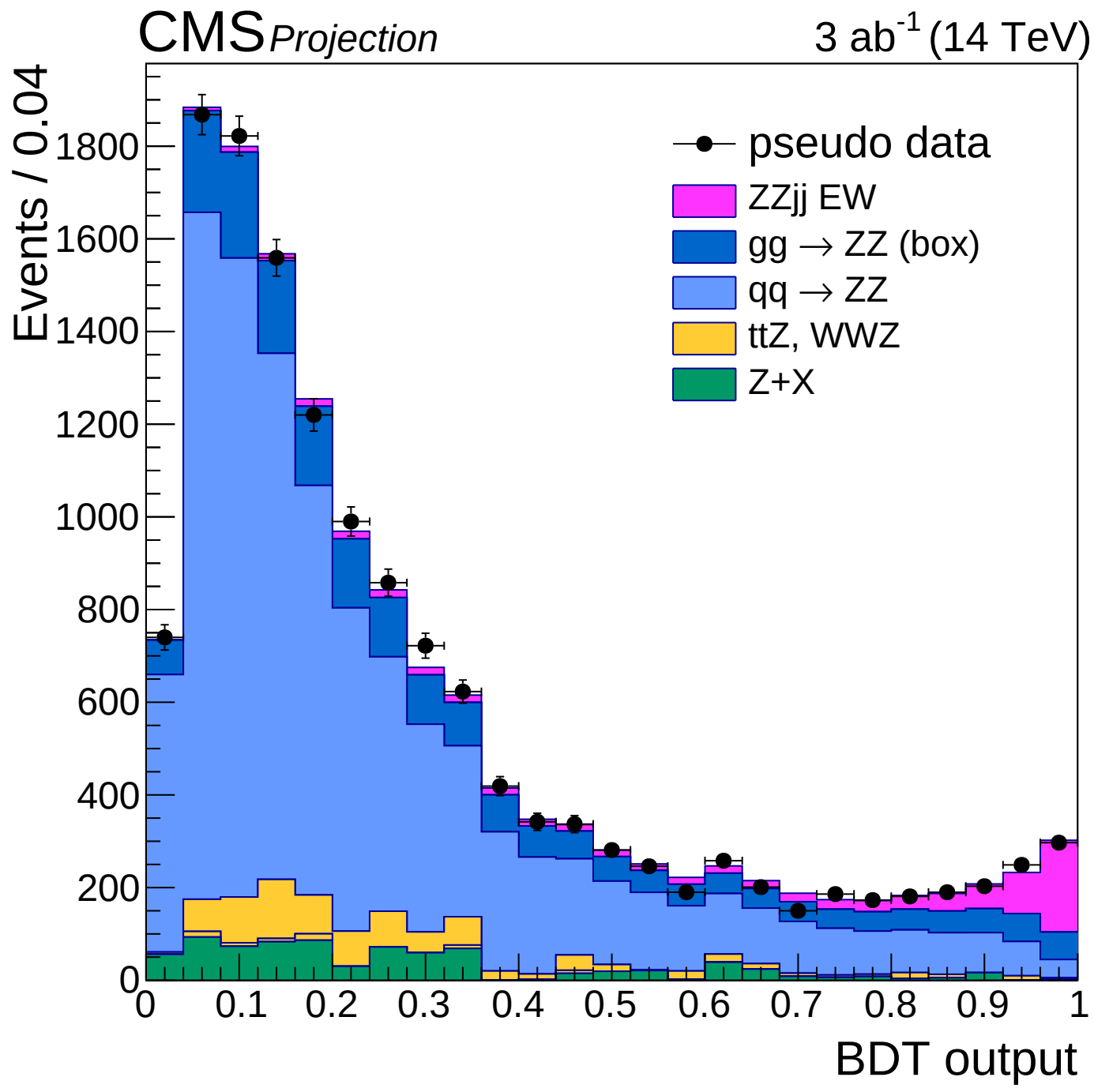

Figure 2: Expected distribution of the BDT output for 3000 fb$^{-1}$. The points represent pseudo data generated from the sum of the expected contributions for each process. The purple filled histogram represents the EW signal, the dark blue the QCD ggZZ background, the light blue the QCD qqZZ background, the yellow the ttZ plus WWZ backgrounds and the green the reducible background.

The expected significance for the Run 2 (YR18) scenario and for a 10% uncertainty in the QCD ggZZ background yield is 13.0 (13.6) for an integrated luminosity of 3000 fb$^{-1}$.

Figure 4 shows the projected relative uncertainty in the cross section measurement for 3000 fb$^{-1}$ as a function of the dominant systematic uncertainty, considering the YR18 scenario for the other uncertainties. Improving the uncertainty in the QCD ggZZ background from 40% to 5% leads to an improvement on the projected uncertainty in the cross section measurement of $\sim 13\%$.

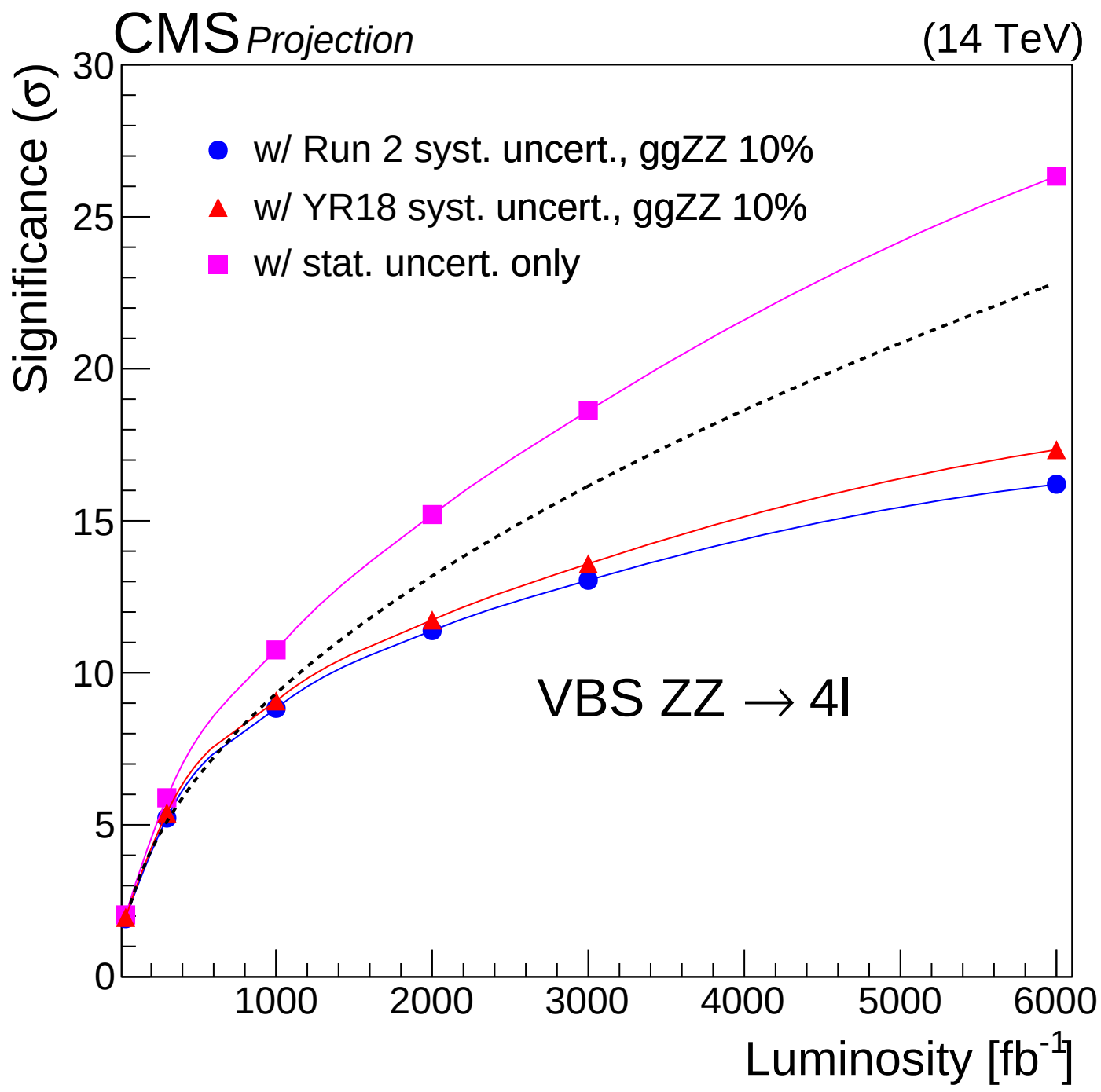

Figure 3: Projected significance for a 10% uncertainty in the QCD ggZZ background yield as a function of the integrated luminosity and for all other systematic uncertainties according to the Run 2 scenario (blue line and circles), and according to YR18 scenario (red line and triangles). The magenta line and squares show the results with only the statistical uncertainties included. The dashed line shows the projected significance as obtained scaling the 2016 result with statistical uncertainty only by the luminosity ratio.

Figure 5 shows the projected relative uncertainty in the cross section measurement as a function of the integrated luminosity and for the two scenarios described in Section 6, as well as for a scenario with only the statistical uncertainty included.

The projected measurement uncertainty is 9.5% (8.5%) for the Run 2 (YR18) scenario and for a 10% uncertainty in the QCD ggZZ background yield (blue filled circle and red filled triangle on Fig. 5), for an integrated luminosity of 3000 $fb^{-1}$. The projected measurement uncertainty is 10.3% (9.5%) for the Run 2 (YR18) scenario and for 3000 $fb^{-1}$ if considering only the luminosity increase. It is 9.8% (8.8%) if considering a pseudorapidity coverage of only up to $|\eta| = 3$ and 9.9% (9.0%) if considering a pseudorapidity coverage of only up to $|\eta| = 2.5$.

## 8 VBS $Z_LZ_L$ analysis

The decay angle $\cos\theta^*$ of the lepton direction in the Z decay rest frame with respect to the Z momentum direction in the laboratory frame is the most distinctive feature of the Z bosons polarisation states. The Z $p_T$ and $\eta$ distributions also carry information on the $Z_LZ_L$ production, in particular longitudinal Z bosons are produced with a lower $p_T$ and more forward, as compared to transverse polarizations ($Z_T$).

The distributions of $\cos\theta^*$, $p_T$ and $\eta$ of both Z bosons, together with the distributions of all observables previously used to separate the VBS process from the QCD backgrounds (see Section 4) are employed as input to a BDT to separate the VBS $Z_LZ_L$ signal from the VBS and QCD backgrounds. Reducible backgrounds are expected to be very small and are therefore neglected in this study.

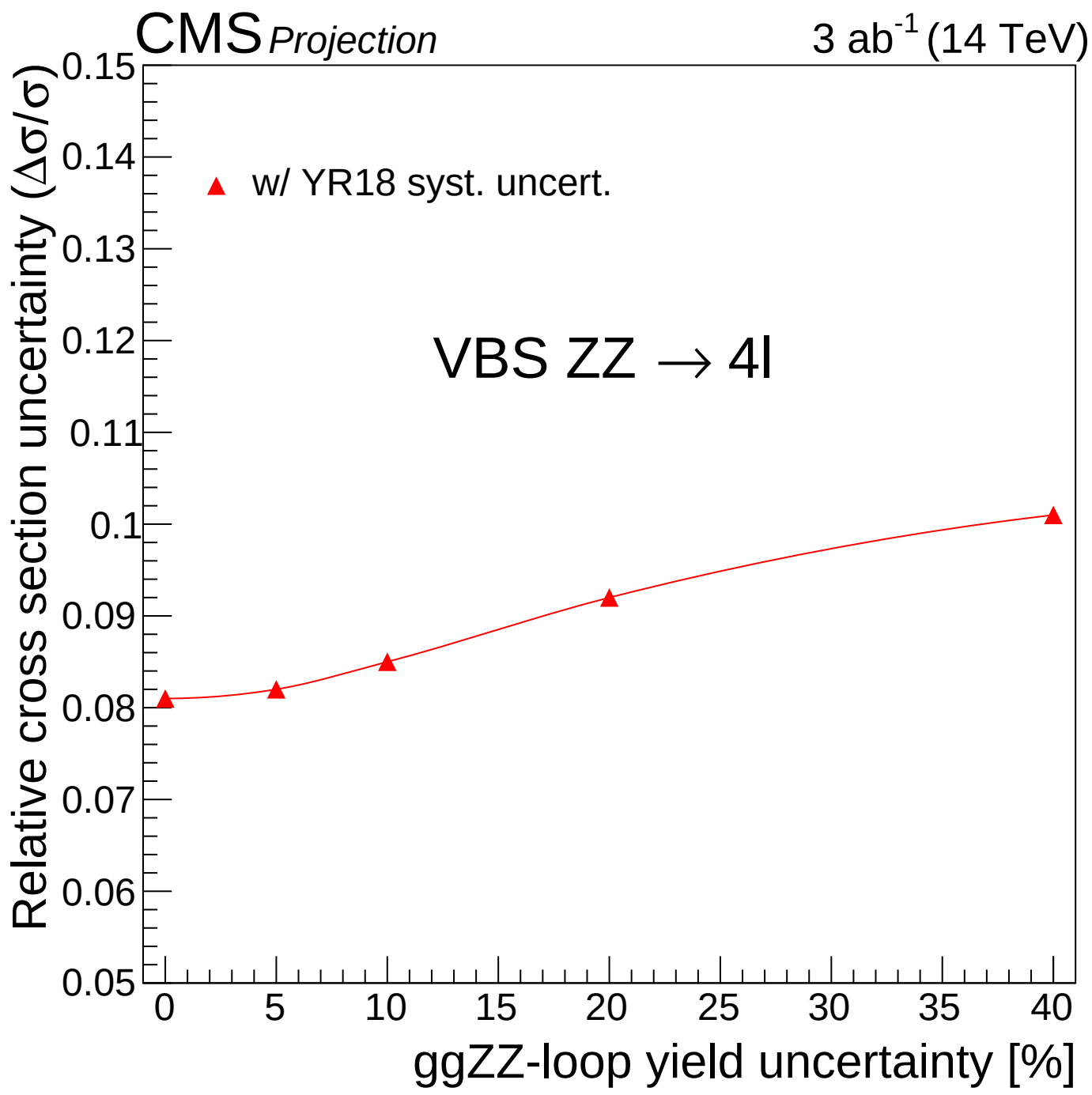

Figure 4: Projected relative uncertainty in the cross section for 3000 fb$^{-1}$ as a function of the uncertainty in the QCD ggZZ background yield (right). The YR18 scenario is used for the other systematic uncertainties.

Figure 6 presents the distributions of some of the discriminant variables used, as obtained from Delphes simulation with 200PU configuration. $Z_1$ ($Z_2$) refers to the $p_T$-leading ($p_T$-subleading) Z boson. The inclusive ZZjj selection that requires $m_{jj} > 100$ GeV is applied. The distributions are normalized to unity for shape comparison.

The BDT is trained separately to discriminate the LL signal from the QCD backgrounds (QCD BDT) and to discriminate the LL signal from the VBS background (VBS BDT). For the training of the QCD BDT a single background is considered, constituted by a weighted mixture of the QCD qqZZ and QCD ggZZ backgrounds.

Cut values are defined on the QCD BDT and on the VBS BDT ouput values, which maximize the overall significance estimator $S/\sqrt{B}$ for the selected events. The corresponding signal efficiency is 14.1% and the VBS, QCD qqZZ and QCD ggZZ background efficiencies are 1.6%, 0.03% and 0.05%, respectively.

It is assumed that the VBS $Z_LZ_L$ fraction, defined as VBS $Z_LZ_L$/VBS $(Z_LZ_L + Z_LZ_T + Z_TZ_T)$ will be measured, rather than the absolute VBS $Z_LZ_L$ cross section. In such ratio measurement, the systematic uncertainties from luminosity, and selection efficiency, as well as theoretical uncertainties on the VBS and VBS background cross section cancel out, such that among the sources of systematic uncertainties listed in Table 3 only the uncertainties in the QCD qqZZ and ggZZ background yields are considered.

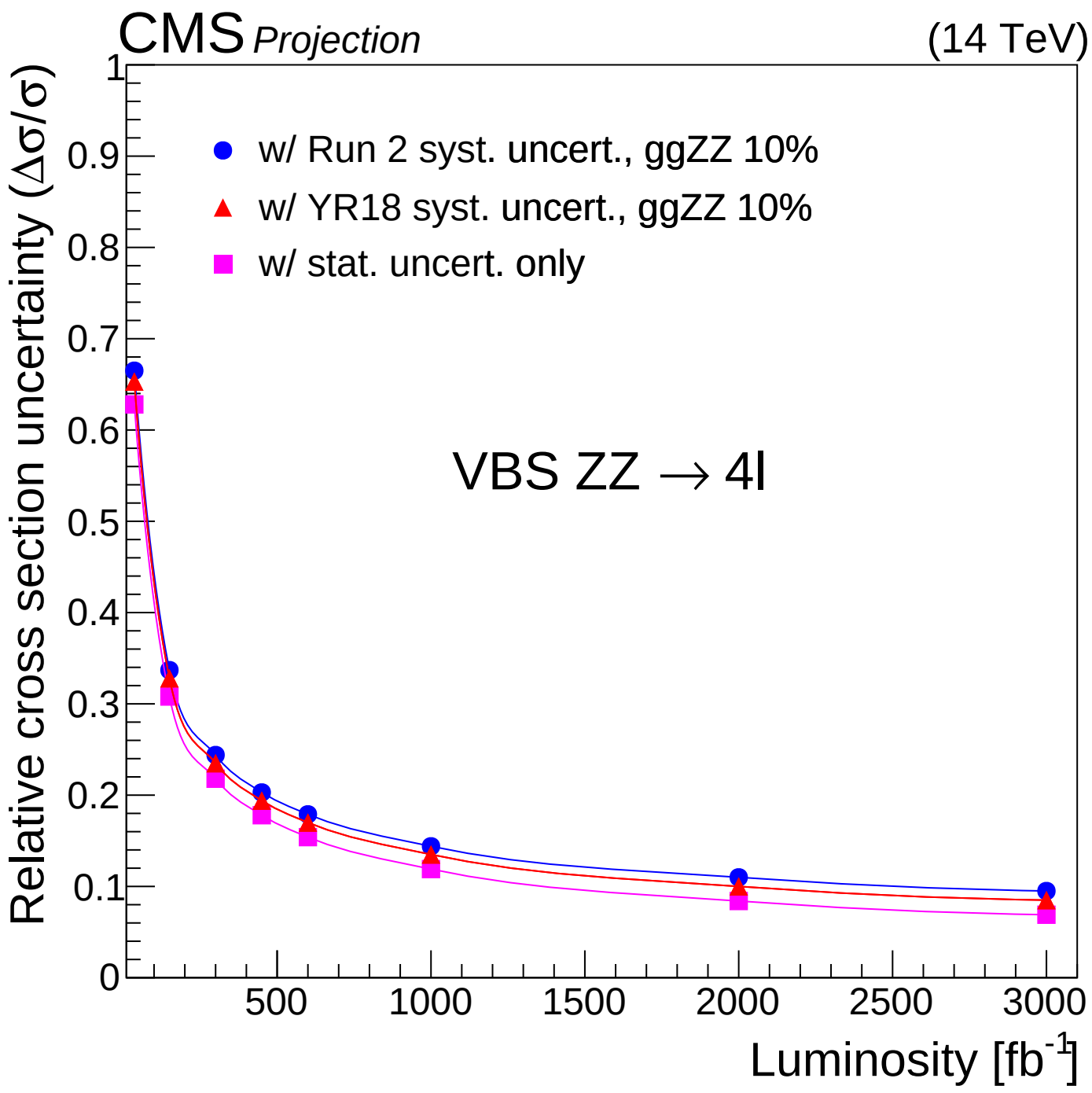

Figure 5: Projected relative uncertainty in the cross section as a function of the integrated luminosity and for all other systematic uncertainties according to the Run 2 scenario (blue line and circles), and according to the YR18 scenario (red line and triangles). Results are shown for 10% uncertainty uncertainty in the QCD ggZZ background yield. The magenta line and filled squares show the results with only the statistical uncertainties included.

## 9 Results for VBS $Z_L Z_L$

Figure 7 shows the expected significance for the VBS $Z_L Z_L$ fraction as a function of the integrated luminosity and for the scenarios described in Section 6 and for a 10% uncertainty in the ggZZ loop-induced background yield, as well as for a scenario with only the statistical uncertainty included. A significance of $1.4\sigma$ is reached for 3000 fb$^{-1}$. As expected from the ratio measurement, the effect of systematic uncertainties is very small. Results are also shown for an integrated luminosity of 6000 fb$^{-1}$, which would approximately correspond to combining ATLAS and CMS after 3000 fb$^{-1}$.

Figure 8 shows the expected relative uncertainty for the VBS $Z_L Z_L$ fraction measurement as a function of the integrated luminosity and for the YR18 scenario described in Section 6 with a 10% uncertainty in the ggZZ loop-induced background yield. The effect of systematic uncertainties is negligible. The result is also shown for an integrated luminosity of 6000 fb$^{-1}$, which would approximately correspond to combining ATLAS and CMS after 3000 fb$^{-1}$.

Table 5 presents the expected significance and relative uncertainty in the VBS $Z_L Z_L$ fraction for various $\eta$ coverage configurations. The foreseen coverage extension of up to $|\eta| = 3(2.8)$ for electrons (muons) leads to a significant improvement for the significance and uncertainty in the VBS $Z_L Z_L$ fraction. An extension of up to $|\eta| = 4$ for electrons would allow to further improve by $\sim 4\%$ both the significance and the cross section measurement uncertainty.

Finally, a simple scaling of the signal and background cross sections is performed to assess the sensitivity to the VBS $Z_L Z_L$ fraction at HE-LHC. An integrated luminosity of 15 ab$^{-1}$ is considered, together with a c.o.m energy of 27 TeV. The cross section ratios $\sigma_{27\,\text{TeV}}$ / $\sigma_{14\,\text{TeV}}$ are evaluated at LO with MADGRAPH (v5.4.2) [16] for the EW signal and the QCD qqZZ background, and with MCFM [28] for the QCD ggZZ background and reported in Table 1.

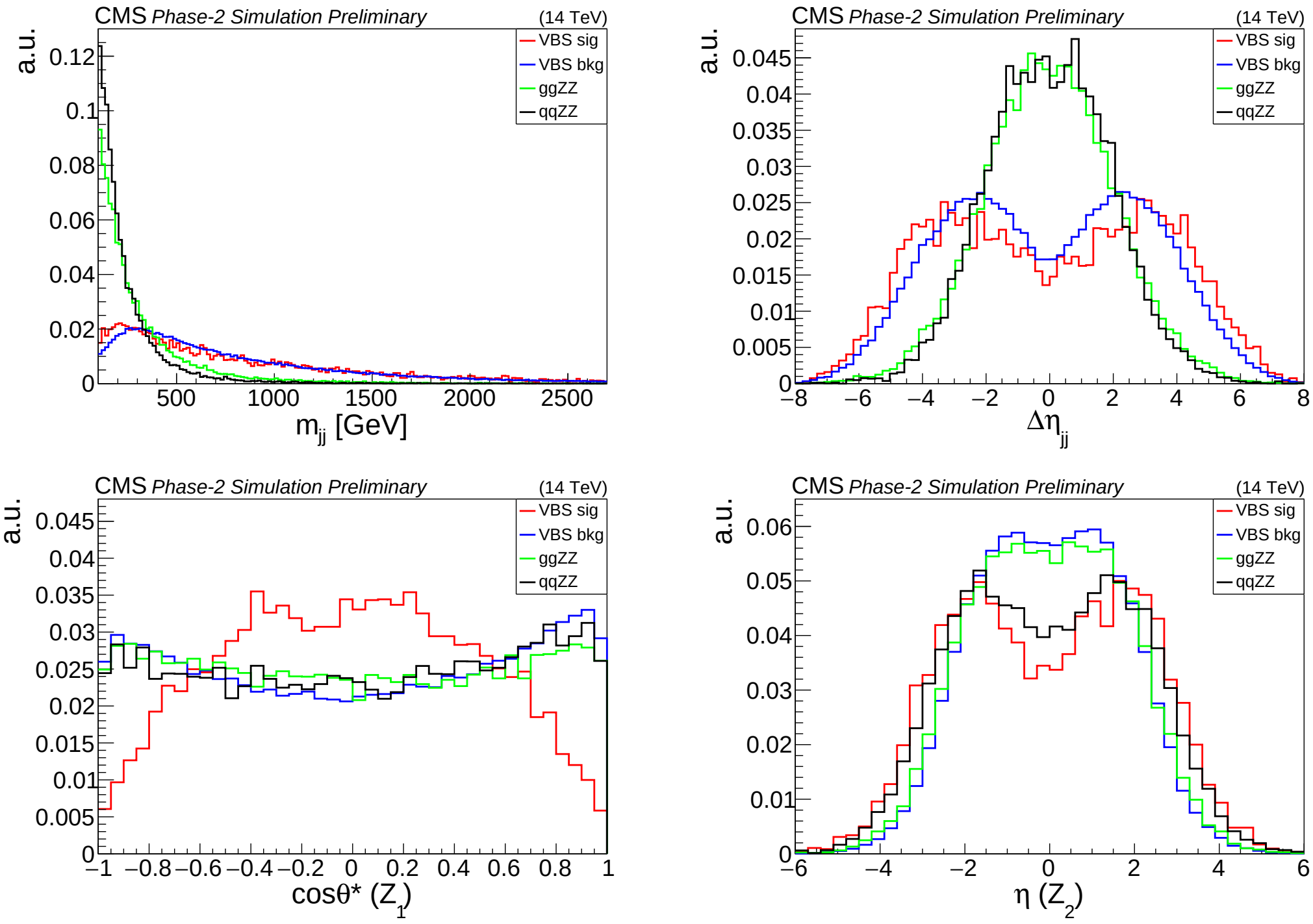

Figure 6: Distributions of some of the discriminant variables for the VBS $Z_LZ_L$ signal, the VBS $Z_LZ_T$ and $Z_TZ_T$ background and the QCD backgrounds from Delphes simulation and for the ZZjj inclusive selection that requires $m_{jj} > 100$ GeV. The distributions are normalized to unity for shape comparison.

Table 5: Significance and measurement uncertainty in the VBS $Z_LZ_L$ fraction for different lepton coverage configurations. The first configuration corresponds to the Run 2 configuration, the second to the Phase-2 upgrade and the third to an option for which the electron coverage would be extended up to $|\eta| = 4$. In the quoted $\eta$ coverages, the first number corresponds to electrons, while the number in parentheses corresponds to muons.

| $\eta$ coverage | significance | VBS $Z_LZ_L$ fraction uncertainty (%) |
|---|---|---|
| $\|\eta\| < 2.5(2.4)$ | $1.22\sigma$ | 88 |
| $\|\eta\| < 3.0(2.8)$ | $1.38\sigma$ | 78 |
| $\|\eta\| < 4.0(2.8)$ | $1.43\sigma$ | 75 |

Table 6 shows the expected significance and relative uncertainty for the measurement of the VBS $Z_LZ_L$ fraction at HE-LHC, compared to HL-LHC. The HE-LHC machine would allow to bring the sensitivity (uncertainty) for the measurement of the VBS $Z_LZ_L$ fraction at the level of $\sim 5\sigma$ ($\sim 20\%$).

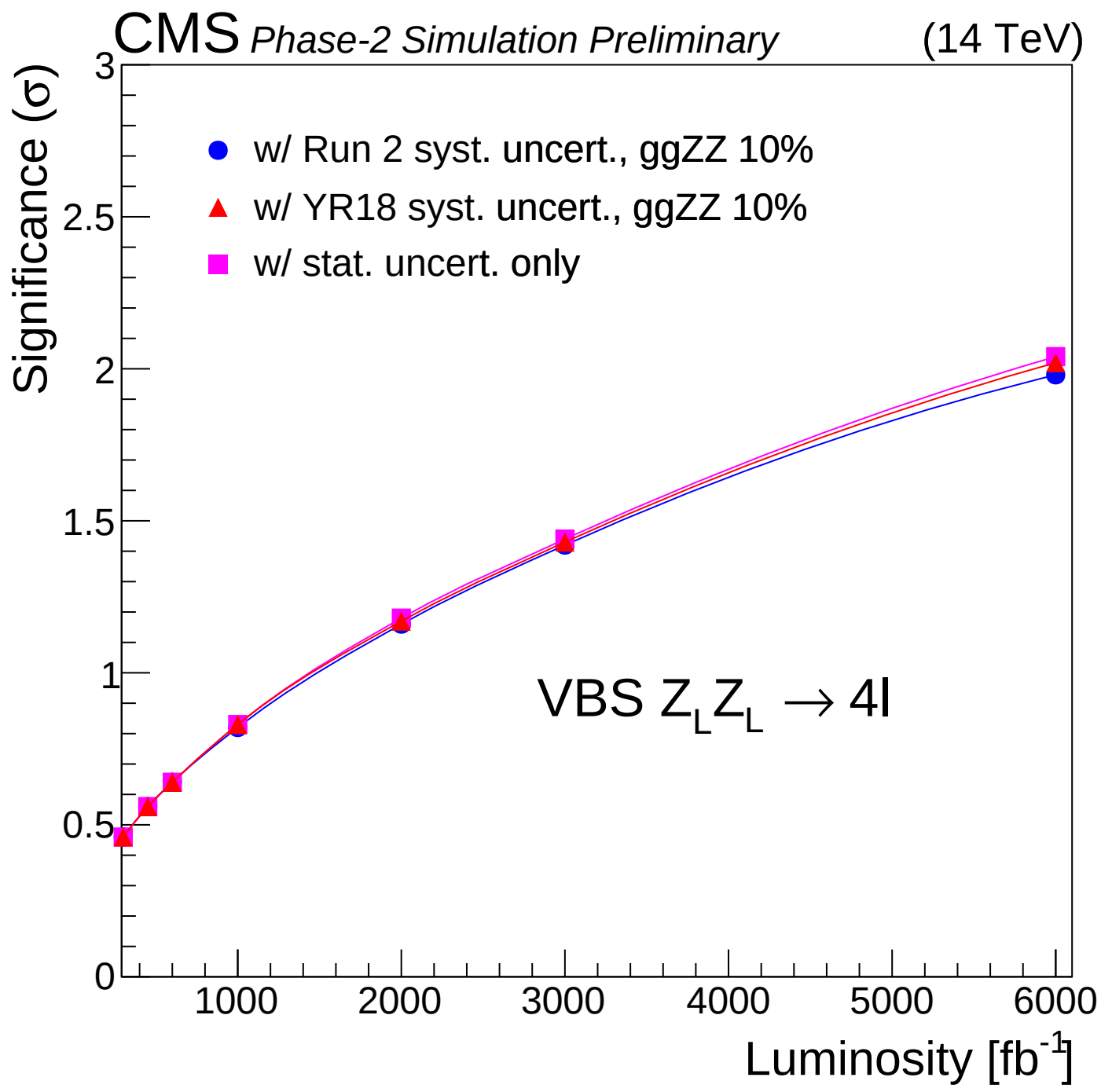

Figure 7: Expected significance for the VBS $Z_LZ_L$ fraction as a function of the integrated luminosity and for systematic uncertainties according to the Run 2 scenario (blue line and circles), and according to the YR18 scenario (red line and triangles). Results are shown for 10% uncertainty in the QCD ggZZ background yield. The magenta line and squares show the results with only the statistical uncertainties included.

Table 6: Expected significance and measurement uncertainty for the measurement of the VBS $Z_LZ_L$ fraction at HL-LHC and HE-LHC, with and without systematic uncertainties included.

| | significance | | VBS $Z_LZ_L$ fraction uncertainty (%) | |
|---|---|---|---|---|
| | w/ syst. uncert. | w/o syst. uncert. | w/ syst. uncert. | w/o syst. uncert.) |
| HL-LHC | $1.4\sigma$ | $1.4\sigma$ | 75% | 75% |
| HE-LHC | $5.2\sigma$ | $5.7\sigma$ | 20% | 19% |

## 10 Summary

We presented prospective studies for the vector boson scattering at the HL-LHC in the ZZ fully leptonic decay channel.

The analysis is based on the measurement performed using data recorded by the CMS experiment in 2016. The results previously obtained are projected to the expected integrated luminosity at HL-LHC of 3000 fb$^{-1}$ at the center-of-mass energy of 14 TeV, taking into account the increased acceptance of the new detector for the leptons and considering two scenario for the systematic uncertainties. The projected relative uncertainty in the VBS ZZ cross section measurement is 9.8% (8.8%) for the Run 2 (YR18) scenario and for a 10% uncertainty in the QCD ggZZ background yield, for an integrated luminosity of 3000 fb$^{-1}$ and a coverage of up to $|\eta| = 3$ for electrons. Extending the coverage up to $|\eta| = 4$ for electrons, the projected measurement uncertainty would be 9.5% and 8.5%, respectively.

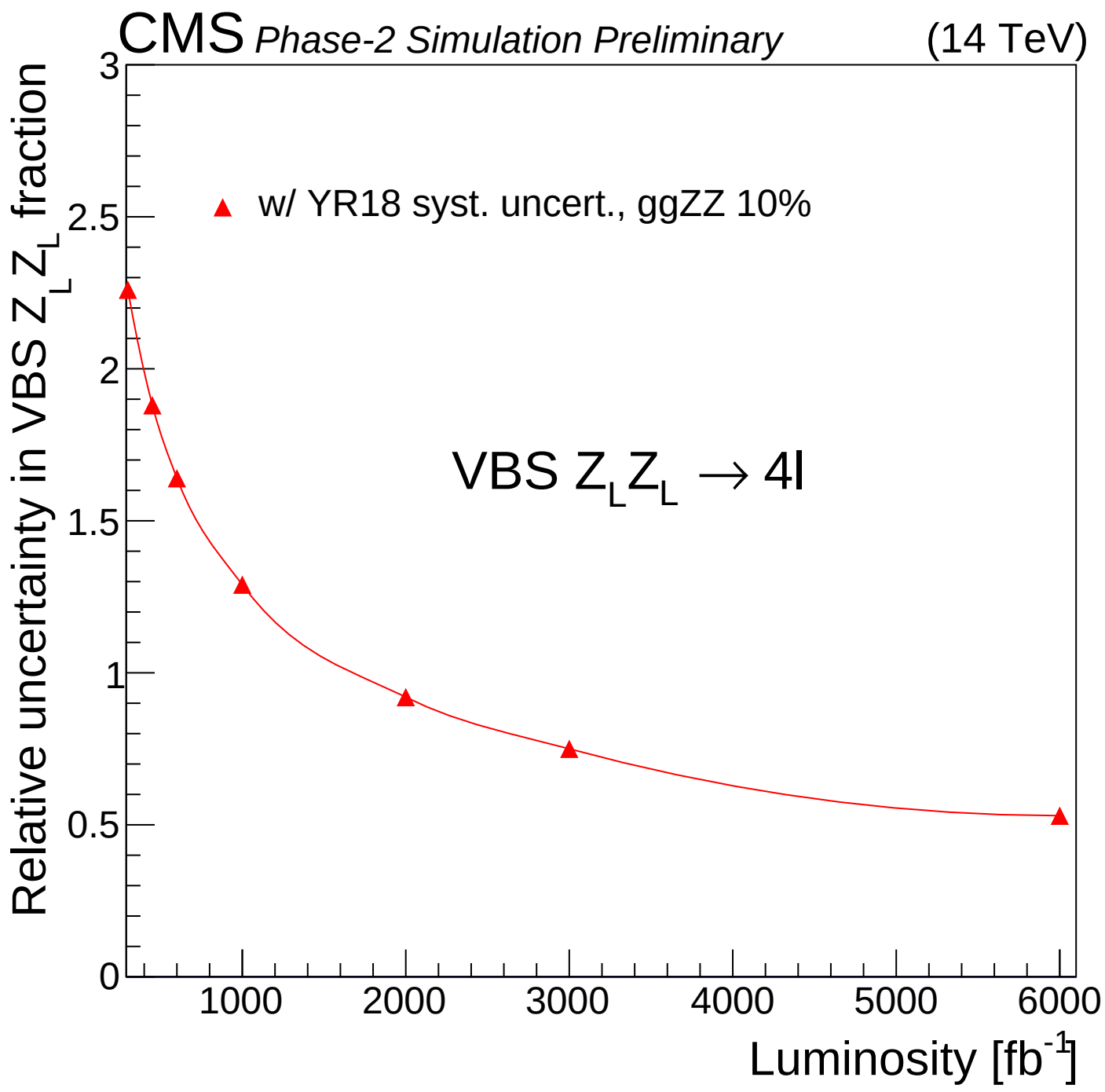

Figure 8: Expected relative uncertainty in the VBS $Z_LZ_L$ fraction as a function of the integrated luminosity and for systematic uncertainties according to the YR18 scenario. Results are shown for 10% uncertainty in the QCD ggZZ background yield.

The sensitivity for the longitudinal scattering VV $\rightarrow Z_LZ_L$ is assessed. The VBS $Z_LZ_L$ signal is separated from the VBS and QCD backgrounds by means of a multivariate discriminant that combines observables that discriminate VBS from QCD processes as well as observables that discriminate longitudinal from transverse Z boson polarizations. The expected significance for the VBS $Z_LZ_L$ fraction is $1.4\sigma$ for an integrated luminosity of 3000 fb$^{-1}$. With such integrated luminosity we enter measurement era for the VBS $Z_LZ_L$ fraction, with relative uncertainty below 100%. The measurement of such rare processes will of course benefit greatly of the highest luminosities. The lepton pseudorapidity coverage foreseen for the CMS detector upgrade leads to a significant improvement of the significance and cross section uncertainty for the VBS $Z_LZ_L$ process. Extending the coverage for electrons up to $|\eta| = 4$ would result in a modest improvement in the performance. Finally, the HE-LHC option would allow to bring the sensitivity at the $5\sigma$ level for this process.

## 16

# ATLAS PUB Note

ATL-PHYS-PUB-2018-030

November 14, 2018

# Prospect studies for the production of three massive vector bosons with the ATLAS detector at the High-Luminosity LHC

The ATLAS Collaboration

This document presents prospects for the expected ATLAS sensitivity to the production of three massive vector bosons $W^{\pm}W^{\pm}W^{\mp}$, $W^{\pm}W^{\mp}Z$ and $W^{\pm}ZZ$ at $\sqrt{s}$ = 14 TeV at the High-Luminosity Large Hadron Collider with 3000 fb$^{-1}$ and 4000 fb$^{-1}$. Final states with two same-sign, three, four or five leptons (electrons or muons) are considered. The results are presented in terms of expected signal significance and the estimated precision on the signal strength measurement.

# 1 Introduction

Measurements of the multi-boson production at the Large Hadron Collider (LHC) provide an excellent test of the electroweak sector of the Standard Model (SM). The triple gauge couplings (TGCs) and quartic gauge couplings (QGCs) that describe the strengths of the triple and quartic gauge boson self-interactions are completely determined by the non-Abelian nature of the electroweak $SU(2)_L \times U(1)_Y$ gauge structure in the SM. These interactions contribute directly to diboson and triboson production at colliders. Studies of triboson production can test these interactions and any possible observed deviation from the theoretical prediction would provide hints of new physics at a high energy scale. Triboson production also represent a source of irreducible background in many searches for physics beyond the SM (BSM).

The production of multiple heavy gauge bosons $V$ $(= W^{\pm}, Z)$ opens up a multitude of potential decay channels categorised according to the number of leptons and jets in the final state. Only charged leptons (electrons and muons) are considered in the studies. In this document we focus on the production of $W^{\pm}W^{\pm}W^{\mp}$, $W^{\pm}W^{\mp}Z$ or $W^{\pm}ZZ$ where at most one boson decays hadronically: $W^{\pm}W^{\pm}W^{\mp} \to \ell^{\pm}\nu\ell^{\pm}\nu\ell^{\mp}\nu$, $W^{\pm}W^{\pm}W^{\mp} \to \ell^{\pm}\nu\ell^{\pm}\nu jj$, $W^{\pm}W^{\mp}Z \to \ell^{\pm}\nu\ell^{\mp}\nu\ell^{+}\ell^{-}$, $W^{\pm}W^{\mp}Z \to \ell^{\pm}\nu jj\ell^{+}\ell^{-}$, $W^{\pm}ZZ \to \ell^{\pm}\nu\ell^{+}\ell^{-}\ell^{+}\ell^{-}$, $W^{\pm}ZZ \to \ell^{\pm}\nu\ell^{+}\ell^{-}\nu\nu$, $W^{\pm}ZZ \to jj\ell^{+}\ell^{-}\ell^{+}\ell^{-}$ and $W^{\pm}ZZ \to \ell^{\pm}\nu\ell^{+}\ell^{-}jj$, with $\ell = e$ or $\mu$. The $ZZZ$ channel is not included in these studies due to its very small cross-section. The leading order (LO) Feynman diagrams for the processes of interest are shown in Figure 1. Prospect studies have been performed, using a cut-based analysis, corresponding to an integrated luminosity of 3000 fb$^{-1}$ and 4000 fb$^{-1}$ of proton–proton collisions at a centre-of-mass energy of $\sqrt{s}$ = 14 TeV, expected to be collected by the ATLAS detector at the the High-Luminosity Large Hadron Collider (HL-LHC) [1, 2].

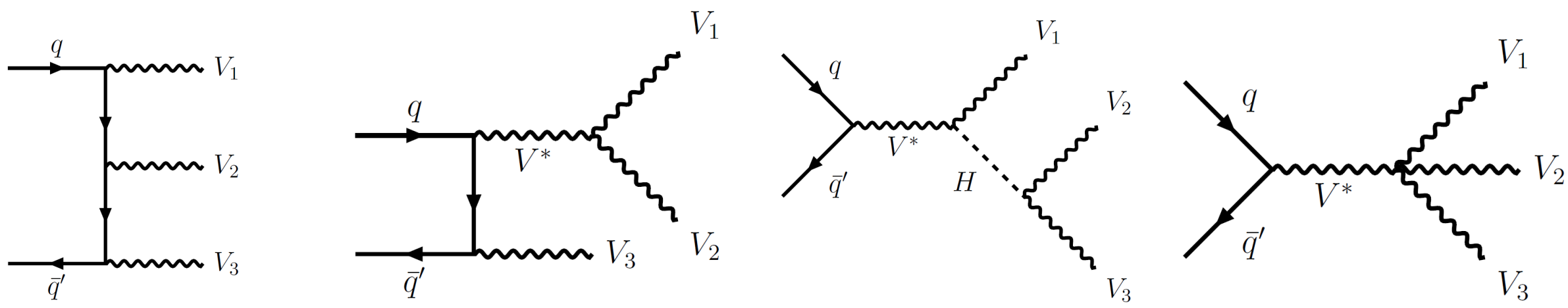

Figure 1: The LO Feynman diagrams for the production of three massive vector boson at the LHC.

# 2 The HL-LHC and the ATLAS detector

In the Run-2 data-taking period, the ATLAS detector collected ~ 140 fb$^{-1}$ of proton-proton collisions with a peak instantaneous luminosity of $2 \times 10^{34}$ cm$^{-2}$s$^{-1}$ and an average number of collisions per bunch crossing of $\langle\mu\rangle \sim 35$. A second long shutdown (LS2) will follow, after which the Run-3 will start. The data collected up to the next long shutdown (LS3) will amount to ~ 300 fb$^{-1}$ with an increase of the centre-of-mass energy to the designed value of 14 TeV. During LS3, the accelerator is foreseen to be upgraded to achieve instantaneous luminosities of 5–7$\times 10^{34}$ cm$^{-2}$s$^{-1}$ after which the HL-LHC phase will start. At the HL-LHC the average number of pile-up interactions per bunch crossing is expected to reach 200 and the data collected will amount to ~ 3000 fb$^{-1}$.

The ATLAS detector [3] is a multi-purpose particle detector with a cylindrical geometry.[1] It consists of layers of inner tracking detectors surrounded by a superconducting solenoid, calorimeters, and a muon spectrometer, and will need several upgrades to cope with the expected higher luminosity at the HL-LHC and its associated high pileup and intense radiation environment. The primary motivation for the upgrade design studies is to maximise the potential of the experiment for searches and measurements despite these harsh conditions. A new inner tracking system (ITk) [4], extending the tracking region up to $|\eta| \leq 4$, will provide the ability to reconstruct forward charged particle tracks, which can be matched to calorimeter clusters for forward electron reconstruction, or associated to forward jets. A new detector, the high granularity timing detector (HGTD) [5], designed to mitigate the pile-up, is also foreseen. The other planned upgrades to the ATLAS detector are described in detail in Ref. [6].

[1] ATLAS uses a right-handed coordinate system with its origin at the nominal interaction point in the centre of the detector.

## 3 Simulation samples

Monte Carlo (MC) simulated event samples are used to predict the background from SM processes and to model the multi-boson signal production. Simulation samples were generated at 14 TeV $pp$ centre-of-mass energy, with a number of events equivalent to at least 3000 fb$^{-1}$ integrated luminosity.

For the generation of triboson signal events, matrix elements for all combinations of $pp \rightarrow VVV$ have been generated using Sherpa v2.2.2 [7] with up to two additional partons in the final state, including next-to-leading-order calculations (NLO) in QCD [8–10] accuracy for the inclusive process, as described in Ref. [11]. All diagrams with three electroweak bosons are taken into account, including diagrams involving Higgs propagators, as in Figure 1. However, since these samples use factorised decays with on-shell vector bosons, the resonant contribution from those diagrams can not be reached from the 125 GeV Higgs boson. In order to account for the contribution coming from these diagrams, the corresponding production of $VH$ bosons is added to the signal. Electroweak NLO corrections to the signal production are not considered in this analysis. The diboson processes are generated with the Sherpa event generator following the approach described in Ref. [11]. For the simulation of the top-quark pair and the production of $VH$ bosons Powheg [12–14] + Pythia [15] was used as described in Ref. [16], while for $t\bar{t} + V$ and $t\bar{t} + H$, MadGraph5_aMC@NLO [17] interfaced to Pythia was used as in Ref. [18]. The top quark pair-production contribution is normalised to approximate NNLO+NNLL accuracy [19, 20].

## 4 Object and event selection

The generated events are overlaid with additional inelastic $pp$ collisions per bunch-crossing which are simulated with Pythia. Parameterisations of the expected performance of the ATLAS detector during the HL-LHC runs have been derived with dedicated studies based on the simulation of the ATLAS detector using the Geant4 program [21]. These parameterisations can be found in Ref. [22].

Lepton trigger and identification efficiencies are derived as a function of $\eta$ and $p_{\mathrm{T}}$ and used to estimate the likelihood of a given lepton to fulfil either the trigger or identification requirement, respectively. The muon transverse momentum and the electron energy resolutions are also parameterised as a function of $\eta$ and either the transverse momentum (for muons) or transverse energy (for electrons).

The positive $x$-axis is defined by the direction from the interaction point to the centre of the LHC ring, with the positive $y$-axis pointing upwards, while the beam direction defines the $z$-axis. Cylindrical coordinates $(r, \phi)$ are used in the transverse plane, $\phi$ being the azimuthal angle around the $z$-axis. The pseudorapidity $\eta$ is defined in terms of the polar angle $\theta$ by $\eta = -\ln\tan(\theta/2)$.

Final-state particles with lifetime greater than 10 ps are clustered into jets (denoted as particle-level jets) using the anti-$k_t$ jet clustering algorithm [23] with a radius parameter of 0.4. Final-state muons and neutrinos are not included in the truth jet clustering. To avoid double-counting of jets associated with electrons, an overlap requirement is applied to exclude jets within a cone of $\Delta R(e, j) < 0.2$ to any electron with transverse energy greater than 7 GeV. To account for detector effects, the truth jet momentum is smeared as a function of $p_T$ and $\eta$. To identify jets originating from the fragmentation of a $b$-quark ($b$-jets), an algorithm which has an efficiency of 70% was used. The corresponding rejection factor for jets originating from the fragmentation of a $c$ (light) quark is about 20 (750) [4].

The contribution of jets misidentified as electrons is derived as a function of $p_T$ and $\eta$.

The missing transverse momentum, $\boldsymbol{E}_T^{miss}$, with magnitude $E_T^{miss}$, is defined at particle level as the transverse component of the vectorial sum of the momenta of non-interacting particles in the event. The $E_T^{miss}$ resolution is parameterised as a function of the overall event activity.

The experimental signature of the triboson processes considered in this analysis consists of at least three charged leptons (or two leptons with the same electric charge in the case of the $W^\pm W^\pm W^\mp$ channel, where one $W$ boson decays hadronically), moderate $E_T^{miss}$ originating from the leptonic decay of $W$ bosons or neutrino decay of $Z$ bosons, and jets in case one of the vector bosons decays hadronically. Events are preselected by either a single-muon or single-electron trigger. The forward tracking is considered, allowing to select electron candidates up to $|\eta| \leq 4$ and muon candidates up to $|\eta| \leq 2.7$. In order to suppress contributions from the background processes and to increase the signal acceptance, the preselected electron or muon candidates are required to fulfil either *tight* or *loose* identification criteria described in Ref. [22]: for the $WWW$ channel two same-sign *tight* or three *tight* leptons are required, for the $WWZ$ two *tight* and one or two *loose* leptons, and for the $WZZ$ channel three, four or five *loose* leptons are required. Jets with $p_T$ >30 GeV and $|\eta| < 4.5$ are considered for the analysis if not stated otherwise.

The event selection criteria are based upon the one used in the ATLAS $W^\pm W^\pm W^\mp$ analysis at 8 TeV centre-of-mass energy, published in Ref. [24], but considers tighter selection criteria on the transverse momentum of the selected objets and missing transverse momentum of the event, in order to suppress higher pile-up contributions expected at the HL-LHC. Several kinematic variables have been considered in order to suppress events with non-prompt leptons and off-shell vector boson decays. The final selection requirements used to define the signal regions described in the following are obtained from an optimization to maximize the sensitivity to the $W^\pm W^\pm W^\mp$, $W^\pm W^\mp Z$ and $W^\pm ZZ$ processes and to reduce the contributions from SM background. In the case of the $W^\pm W^\pm W^\mp \to \ell^\pm \nu \ell^\pm \nu \ell^\mp \nu$ channel, three separate signal regions are defined based on the number of same-flavour opposite-sign (SFOS) lepton pairs in the event: 0SFOS ($e^\pm e^\pm \mu^\mp$, $\mu^\pm \mu^\pm e^\mp$), 1SFOS ($e^\pm e^\mp \mu^\pm$, $e^\pm e^\mp \mu^\mp$, $\mu^\pm \mu^\mp e^\pm$, $\mu^\pm \mu^\mp e^\mp$) and 2SFOS ($e^\pm e^\pm e^\mp$, $\mu^\pm \mu^\pm \mu^\mp$). Similarly, in the $W^\pm W^\mp Z \to \ell^\pm \nu \ell^\pm \nu \ell^+ \ell^-$ channel, two signal regions are defined based on the selection of SFOS or different-flavour opposite-sign (DFOS) lepton pairs events: SFOS ($e^\pm e^\mp \mu^\mp \mu^\pm$, $e^\pm e^\mp e^\pm e^\mp$, $\mu^\mp \mu^\pm \mu^\mp \mu^\pm$) and DFOS ($e^\pm e^\mp \mu^\mp e^\pm$, $\mu^\mp \mu^\pm \mu^\mp e^\pm$). To select $W^\pm W^\pm W^\mp \to \ell^\pm \nu \ell^\pm \nu jj$ candidates, events are required to have exactly two leptons with the same electric charge, and at least two jets. Three different final states are considered based on the lepton flavour, namely $e^\pm e^\pm$, $e^\pm \mu^\pm$ and $\mu^\pm \mu^\pm$. In the case of the $W^\pm ZZ$ process, a separate set of selection criteria is defined, depending on the decay of the bosons. In all channels, events are rejected if they have identified $b$-jets. This selection requirement suppresses background with top quarks, and it has a marginal impact on the signal efficiency. Tables 1 to 6 show the kinematic selection criteria used for the channels considered in this analysis.

| $W^\pm W^\pm W^\mp \to \ell^\pm\nu\ell^\pm\nu\ell^\mp\nu$ | 0 SFOS | 1 SFOS | 2 SFOS |
|---|---|---|---|
| Preselection | Exactly 3 *tight* leptons with $p_{\mathrm{T}} > 30$ GeV and $\|\eta\| < 4$ | | |
| $E_{\mathrm{T}}^{\mathrm{miss}}$ | - | $E_{\mathrm{T}}^{\mathrm{miss}} > 90$ GeV | $E_{\mathrm{T}}^{\mathrm{miss}} > 80$ GeV |
| SFOS dilepton mass | $m_{\ell\ell}^{\mathrm{SFOS}} > 20$ GeV | - | |
| Angle between the trilepton system and $\boldsymbol{E}_{\mathrm{T}}^{\mathrm{miss}}$ | $\|\phi^{3\ell} - \phi^{\boldsymbol{E}_{\mathrm{T}}^{\mathrm{miss}}}\| > 2.5$ | | |
| Z boson veto | $\|m_{ee} - m_Z\| > 15$ GeV | $m_Z - m_{\ell\ell}^{\mathrm{SFOS}} > 35$ GeV or $m_{\ell\ell}^{\mathrm{SFOS}} - m_Z > 20$ GeV | $\|m_{\ell\ell}^{\mathrm{SFOS}} - m_Z\| > 20$ GeV |
| Jet veto | At most one jet with $p_{\mathrm{T}} > 30$ GeV and $\|\eta\| < 2.5$ | | |
| *b*-jet veto | No identified *b*-jets with $p_{\mathrm{T}} > 30$ GeV | | |

Table 1: Event selection criteria for $WWW \to 3\ell\, 3\nu$ candidate events.

| $W^\pm W^\pm W^\mp \to \ell^\pm\nu\ell^\pm\nu jj$ | |
|---|---|
| Leptons | Exactly 2 same-charge *tight* leptons with $p_{\mathrm{T}} > 30$ GeV and $\|\eta\| < 4$ |
| Jets | At least 2 jets with $p_{\mathrm{T}} > 30$ GeV and $\|\eta\| < 2.5$ |
| $m_{\ell\ell}$ | $m_{\ell\ell} > 40$ GeV |
| $E_{\mathrm{T}}^{\mathrm{miss}}$ | $E_{\mathrm{T}}^{\mathrm{miss}} > 80$ GeV |
| $m_{jj}$ | 60 GeV $< m_{jj} <$ 120 GeV |
| $\Delta\eta_{jj}$ | $\Delta\eta_{jj} < 1.5$ |
| Z boson veto | $\|m_{ee} - 90\text{ GeV}\| > 10$ GeV |
| Third lepton veto | No third *loose* lepton with $p_{\mathrm{T}} > 7$ GeV and $\|\eta\| < 2.5$ |
| *b*-jet veto | No identified *b*-jets with $p_{\mathrm{T}} > 30$ GeV |

Table 2: Event selection criteria for $WWW \to 2\ell\, 2\nu\, 2j$ candidate events.

| $W^\pm W^\mp Z \to \ell^\pm\nu\ell^\pm\nu\ell^+\ell^-$ | SFOS | DFOS |
|---|---|---|
| Preselection | Exactly 4 *loose* (3rd and 4th *tight*) leptons with $\|\eta\| < 4$ and $p_{\mathrm{T}}(1,2) > 30$ GeV, $p_{\mathrm{T}}(3,4) > 25$ GeV | |
| SFOS dilepton mass | $\|m_{\ell\ell}^{\mathrm{SFOS}} - 91\text{ GeV}\| < 15$ GeV | |
| DFOS dilepton mass | – | $m_{\ell\ell}^{\mathrm{DFOS}} > 40$ GeV |
| Four-lepton mass | – | $m_{4\ell} > 250$ GeV |
| *b*-jet veto | No identified *b*-jets with $p_{\mathrm{T}} > 30$ GeV | |

Table 3: Event selection criteria for $WWZ \to 4\ell\, 2\nu$ candidate events. The four-lepton mass $m_{4\ell}$ is calculated as invariant mass of the four-lepton system.

| $W^{\pm}W^{\mp}Z \to \ell^{\pm}\nu jj\ell^{+}\ell^{-}$ | |
|---|---|
| Preselection | Exactly 3 *loose* leptons (2nd and 3rd *tight*) with $\|\eta\| < 4$ and $p_T(1) > 50$ GeV, $p_T(2) > 40$ GeV, $p_T(3) > 20$ GeV |
| SFOS dilepton mass | Exactly one $\|m_{\ell\ell}^{SFOS} - 91 \text{ GeV}\| < 15$ GeV |
| Jets | Exactly 2 jets with $p_T > 40$ GeV and $\|\eta\| < 3$ |
| Dijet mass | $50 < m_{jj} < 100$ GeV |
| Transverse mass | $m_T > 20$ GeV |
| *b*-jet veto | No identified *b*-jets with $p_T > 30$ GeV |

Table 4: Event selection criteria for $WWZ \to 3\ell\ 3\nu 2j$ candidate events. The transverse mass is calculated as $m_T = \sqrt{2p_T^{\ell}E_T^{miss}(1 - \cos(\varphi^{\ell} - \varphi^{\boldsymbol{E}_T^{miss}}))}$, where $\ell$ is the lepton that does not fulfil the SFOS dilepton mass requirement.

| $W^{\pm}ZZ \to$ | $\ell^{\pm}\nu\ell^{+}\ell^{-}\ell^{+}\ell^{-}$ | $\ell^{\pm}\nu\ell^{+}\ell^{-}\nu\nu$ |
|---|---|---|
| Preselection | Exactly 5 *loose* (4th and 5th *tight*) leptons with $\|\eta\| < 4$ and $p_T(1,2,3) > 30$ GeV, $p_T(4,5) > 25$ GeV | Exactly 3 *loose* leptons with $p_T > 20$ GeV and $\|\eta\| < 4$ |
| SFOS dilepton mass | $\|m_{\ell\ell}^{SFOS} - 91 \text{ GeV}\| < 15$ GeV | |
| $E_T^{miss}$ | - | $E_T^{miss} > 100$ GeV |
| Transverse mass | $m_T > 40$ GeV | $m_T > 50$ GeV |
| *b*-jet veto | No identified *b*-jets with $p_T > 30$ GeV | |

Table 5: Event selection criteria for $WZZ$ fully leptonic candidate events. For the $5\ell 1\nu$ channel two lepton pairs of the same flavour and opposite sign have to satisfy the dilepton mass selection requirement, while in $3\ell 3\nu$ exactly one pair is required to satisfy the given requirement. The transverse mass is calculated from the $E_T^{miss}$ and the lepton that does not fulfil dilepton mass requirement.

| $W^{\pm}ZZ \to$ | $jj\ell^{+}\ell^{-}\ell^{+}\ell^{-}$ | $\ell^{\pm}\nu\ell^{+}\ell^{-}jj$ |
|---|---|---|
| Lepton preselection | Exactly 4 *loose* leptons with $p_T > 20$ GeV and $\|\eta\| < 4$ | Exactly 3 *loose* leptons with $p_T > 20$ GeV and $\|\eta\| < 4$ |
| Jet preselection | At least two jets with $p_T > 30$ GeV and $\|\eta\| < 4.5$ | |
| SFOS dilepton mass | $\|m_{\ell\ell}^{SFOS} - 91 \text{ GeV}\| < 15$ GeV | |
| *b*-jet veto | No identified *b*-jets with $p_T > 30$ GeV | |

Table 6: Event selection criteria for $WZZ \to 4\ell 2j$ and $WZZ \to 3\ell 1\nu jj$ candidate events.

# 5 Results

The SM processes that mimic the multi-boson signal signatures by producing at least three prompt leptons or two prompt leptons with the same electric charge, can be grouped into the following categories:

- The $WZ$ and $ZZ$ processes, referred to as “diboson background”;
- The $WWW$, $WWZ$, $WZZ$, $ZZZ$ processes, excluding the signal process under study, referred to as “triboson background”;
- The $VH$ and $t\bar{t}H$ processes, excluding the processes which are added to the signal, referred to as “Higgs+X background”;
- The production of four top quarks, top quark associated with $WZ$ bosons or $t\bar{t}$ associated with $W$, $Z$, $WZ$ or $W^{\pm}W^{\mp}$ bosons, referred to as “top background”;
- Processes that have non-prompt leptons (electrons) originating from misidentified jets (referred to as “fake-lepton background”);
- Processes that produce prompt charged leptons, but the charge of one lepton is misidentified (referred to as “charge-flip background”).

The contributions from the $WW$ and $t\bar{t}$ processes are accounted for in the fake-lepton and charge-flip backgrounds. The diboson, triboson, Higgs+X and top background sources are estimated using simulated events, with the dominant irreducible background in most of the channels originating from the diboson processes. In some channels the contribution of the fake-lepton background, which is derived by applying the pre-defined ($p_{\mathrm{T}}$, $\eta$)-dependent likelihood as described in Section 4, becomes significant. The charge-flip background has been investigated and found to be negligible in all considered processes.

Regarding the $WWW \rightarrow 3\ell\,3\nu$ channel, one can observe that the 0SFOS signal region has the best signal-to-background ratio due to the requirement that suppresses $Z$ boson decays. Background is dominated by the diboson irreducible background and the fake-lepton contribution. The fake-lepton contribution mainly arises from $t\bar{t} \rightarrow \ell\ell$+jets events, with a jet misidentified as an electron. The contribution of signal events containing Higgs decays is at the level of 40%, and consistent between the three signal regions. In the $WWW \rightarrow 2\ell\,2\nu\,2j$ channel, the signal-to-background ratio is worse than in the leptonic one, with the dominant background from the $WZ$ diboson process. The total number of signal and background events expected after the full event selection in each channel, for an integrated luminosity of 3000 fb$^{-1}$ is shown in Table 7. Figure 2 shows the transverse mass distributions of the trilepton system, $m_{\mathrm{T}}^{3\ell}$, for the $WWW \rightarrow 3\ell\,3\nu$ channel after summing over the three signal regions and separately for 0SFOS events, and the distribution of the sum of the scalar $p_{\mathrm{T}}$ for all selected objects, $\Sigma p_{\mathrm{T}} = p_{\mathrm{T}}^{\ell_1} + p_{\mathrm{T}}^{\ell_2} + p_{\mathrm{T}}^{j_1} + p_{\mathrm{T}}^{j_2} + E_{\mathrm{T}}^{\mathrm{miss}}$ for the $WWW \rightarrow 2\ell\,2\nu\,2j$ channel.

In the $WWZ$ process the largest signal-to-background ratio is obtained in the four lepton channel, with two leptons being of different flavour, as this requirement suppresses a large fraction of the diboson background. The contribution of Higgs+X is smaller with respect to the $WWW \rightarrow 3\ell\,3\nu$ case due to the invariant mass requirement $m_{\ell\ell}^{\mathrm{DFOS}} > 40$ GeV. A very large fraction of the background arises from the production of top quarks in association with $Z$ and $W$ bosons. The selection requirement on high $E_{\mathrm{T}}^{\mathrm{miss}}$ has been checked and found to show no discrimination power between signal and background. The $WWZ \rightarrow 3\ell 1\nu 2j$ channel is dominated by the irreducible diboson background. The total number of signal and background events expected after the full event selection in each channel, for an integrated luminosity of 3000 fb$^{-1}$ is shown in Table 8. The expected signal and background distributions of the leading lepton

| $WWW \to$ | $3\ell\,3\nu$ 0SFOS | $3\ell\,3\nu$ 1SFOS | $3\ell\,3\nu$ 2SFOS | $2\ell\,2\nu\,2j$ $e^\pm e^\pm$ | $2\ell\,2\nu\,2j$ $e^\pm \mu^\pm$ | $2\ell\,2\nu\,2j$ $\mu^\pm \mu^\pm$ |
|---|---|---|---|---|---|---|
| Signal | | | | | | |
| $WWW$ | 191 | 146 | 85 | 133 | 413 | 282 |
| $WH$ | 121 | 98 | 56 | 53 | 198 | 139 |
| Total signal | 312 | 244 | 141 | 186 | 611 | 421 |
| Background | | | | | | |
| Diboson | 208 | 3 454 | 3 706 | 2 254 | 5 236 | 2 479 |
| Triboson | 37 | 37 | 26 | 18 | 43 | 23 |
| Higgs+X | 25 | 64 | 12 | 106 | 270 | 116 |
| Top | 60 | 48 | 59 | 148 | 314 | 174 |
| Fake-lepton | 97 | 163 | 58 | 285 | 257 | – |
| Total background | 427 | 3 766 | 3 861 | 2 811 | 6 120 | 2 792 |

Table 7: Expected number of signal and background events in the $WWW$ channels, after applying the selection criteria from Tables 1 and 2.

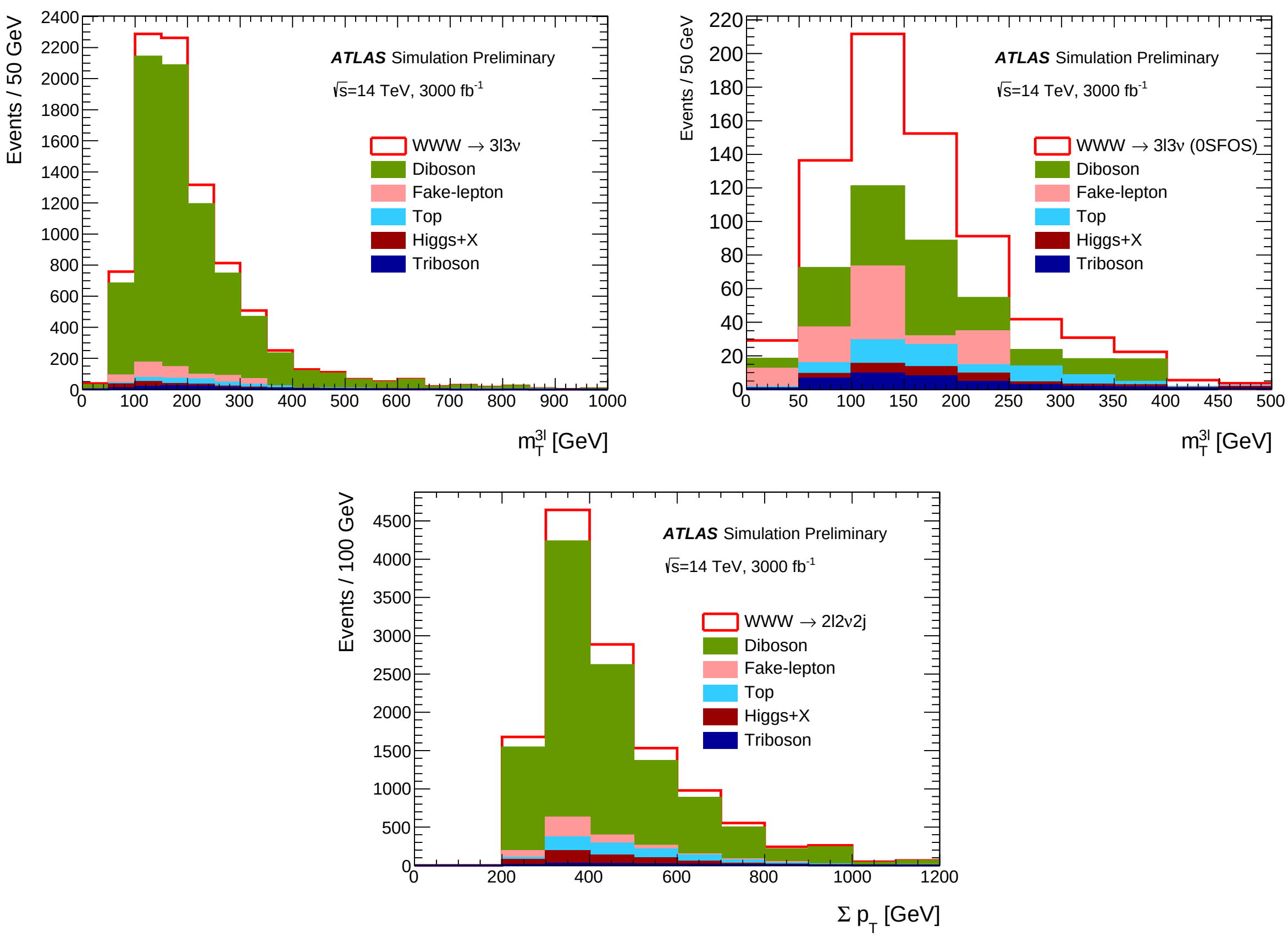

Figure 2: The distribution of $m_{\mathrm{T}}^{3\ell}$ for the $WWW \to 3\ell\,3\nu$ channel after summing over the three signal regions (top left) and separately for 0SFOS events (top right), and the distribution of $\Sigma p_{\mathrm{T}}$ for the $WWW \to 2\ell\,2\nu\,2j$ channel (bottom) as expected from the signal and background processes at 3000 fb$^{-1}$, after applying the selection criteria from Tables 1 and 2 respectively.

$p_T$, the two-lepton system transverse momenta $p_T^{\ell\ell}$ and the $\Sigma p_T = p_T^{\ell_1} + p_T^{\ell_2} + p_T^{\ell_3} + p_T^{\ell_4} + E_T^{miss}$ for the DFOS events in $WWZ \to 4\ell\, 2\nu$ channel, after applying the selection criteria from Table 3, are shown in Figure 3.

| $WWZ \to$ | $4\ell\, 2\nu$ SFOS | $4\ell\, 2\nu$ DFOS | $3\ell\, 3\nu 2j$ |
|---|---|---|---|
| Signal | | | |
| $WWZ$ | 197 | 146 | 267 |
| $ZH$ | 98 | 22 | 40 |
| Total signal | 295 | 168 | 307 |
| Background | | | |
| Diboson | 42 757 | 357 | 11 243 |
| Triboson | 78 | 11 | 77 |
| Higgs+X | 20 | 10 | 195 |
| Top | 650 | 390 | 807 |
| Fake-lepton | 18 | 16 | 8.1 |
| Total background | 43 523 | 784 | 12 330 |

Table 8: Expected number of signal and background events in the $WWZ$ channels, after applying the selection criteria from Tables 3 and 4.

| $WZZ \to$ | $5\ell 1\nu$ | $3\ell 3\nu$ | $4\ell 2j$ | $3\ell 1\nu 2j$ |
|---|---|---|---|---|
| Signal | | | | |
| $WZZ$ | 19 | 124 | 117 | 459 |
| $WH$ | 0.1 | 9.1 | 0.3 | 18 |
| Total signal | 19 | 133 | 117 | 477 |
| Background | | | | |
| Diboson | 4.0 | 97 766 | 18 481 | 166 270 |
| Triboson | 3.0 | 477 | 25 | 1 766 |
| Higgs+X | 0.3 | 668 | 30 | 1 196 |
| Top | 15 | 2 371 | 174 | 7 347 |
| Fake-lepton | 3.0 | 376 | 6.0 | 2 966 |
| Total background | 25 | 101 658 | 18 716 | 179 545 |

Table 9: Expected number of signal and background events in the $WZZ$ channels after applying the selection criteria from Tables 5 and 6.

Similarly, in the $WZZ$ channel the signal region with the largest signal-to-background ratio is the one with five charged leptons. In this case, the fake-lepton contribution becomes significant. The background is dominated by rare top production of $t\bar{t}WZ$. The other channels suffer from low signal-to-background ratio. The total number of signal and background events expected after the full event selection in each channel, for an integrated luminosity of 3000 fb$^{-1}$ is shown in Table 9. Figure 4 shows the distributions of the two-lepton invariant mass $m_{\ell\ell}$, selected to give the mass closest to the mass of the $Z$ boson, and of the $W$ boson decay lepton $p_T$ for the $WZZ \to 5\ell 1\nu$, as expected from the signal and background processes at 3000 fb$^{-1}$ after applying the selection criteria from Table 5.

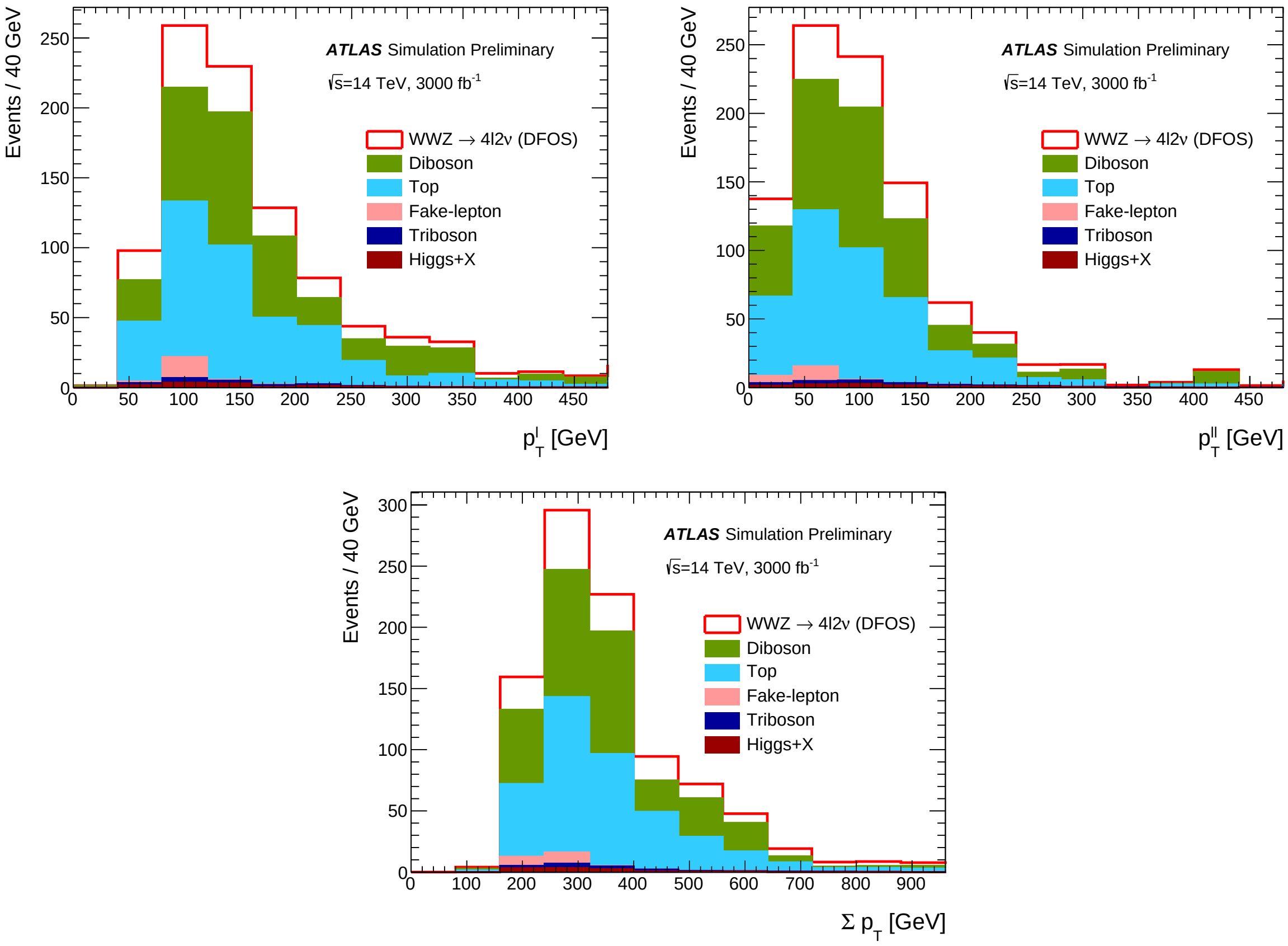

Figure 3: The distribution of the leading lepton transverse momentum $p_{\mathrm{T}}^{\ell}$ (top left), the distribution of the transverse momentum of the two-lepton system $p_{\mathrm{T}}^{\ell\ell}$ (top right) and the distribution of $\Sigma p_{\mathrm{T}}$ (bottom) for the DFOS events in $WWZ \rightarrow 4\ell\ 2\nu$ channel as expected from the signal and background processes at 3000 fb$^{-1}$ after applying the selection criteria from Table 3.

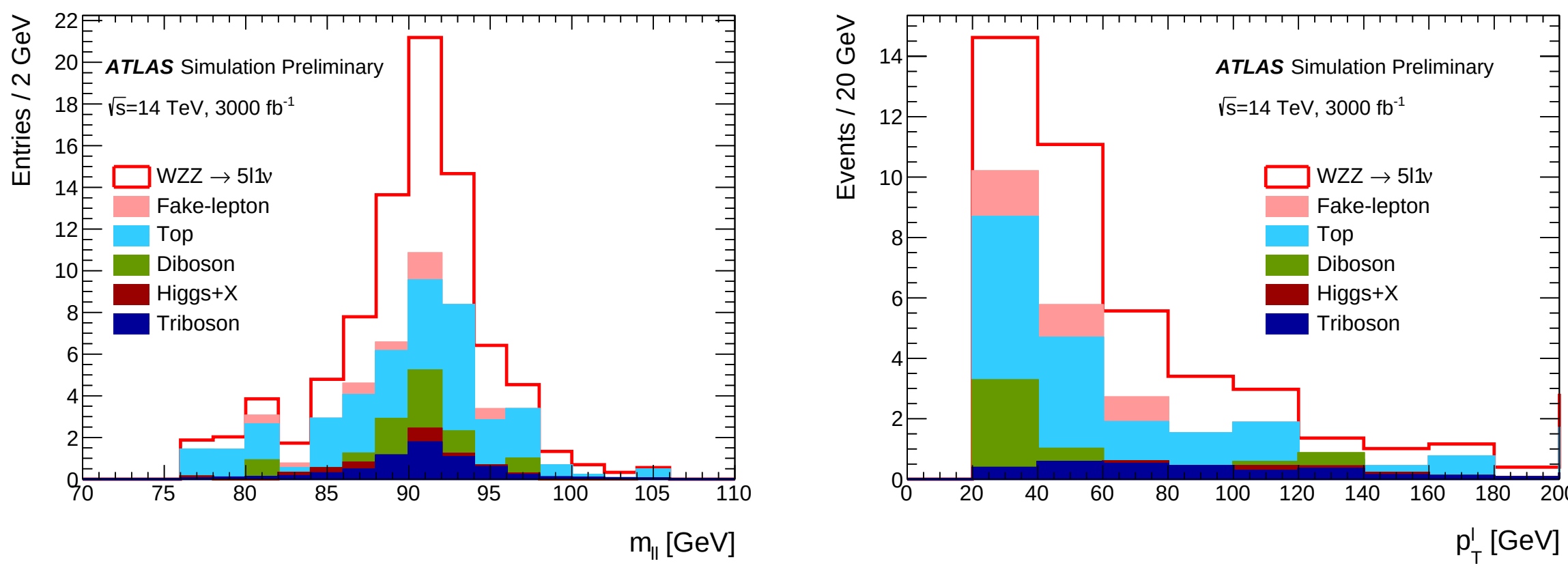

Figure 4: The distribution of the two-lepton invariant mass, $m_{\ell\ell}$, selected to give the mass closest to the mass of the $Z$ boson (left) and the distribution of the $W$ boson decay lepton $p_{\mathrm{T}}$ for $WZZ \rightarrow 5\ell 1\nu$, as expected from the signal and background processes at 3000 fb$^{-1}$, after applying the selection criteria from Table 5.

Systematic uncertainties in the signal and background predictions arise from the uncertainties in the measurement of the integrated luminosity, from the experimental modelling of the signal acceptance and detection efficiency, and from the background normalisation. With the much larger integrated luminosity and a sophisticated understanding of the detector performance and backgrounds at the HL-LHC, we expect experimental uncertainties related to the lepton reconstruction and identification efficiencies as well as lepton energy/momentum resolution and scale modelling of 1%, to the $E_{\mathrm{T}}^{\mathrm{miss}}$ modelling of 1%, to the jet energy scale and resolution of 1.5% and 5% in the fully leptonic and leptons+jets channels, respectively, to the luminosity measurement of 1% and to the expected pileup of 1% [22]. Based on the extrapolations of current ATLAS measurements and assuming a reduction of the uncertainty at the level of 15–80%, depending on the process and the origin of the systematics, the following systematic uncertainties on the cross-section normalisation for each of the background processes are assumed: 4% on $\sigma_{\mathrm{diboson}}$, 30% on $\sigma_{\mathrm{triboson}}$, 3% on $\sigma_{t\bar{t}}$, 20% on $\sigma_{t\bar{t}H}$, 6% on $\sigma_{t\bar{t}Z}$, and 11% on $\sigma_{t\bar{t}W}$. The uncertainty on the level of the fake-lepton background is estimated to be 10%. Taking these assumptions into account, we estimate the total systematic uncertainty on the background of 9%, 6% and 5% for the 0SFOS, 1SFOS and 2SFOS categories, respectively, in the $WWW \to 3\ell\, 3\nu$ channel, 8% for all categories in the $WWW \to 2\ell\, 2\nu\, 2j$ channel, 6% in the $WWZ \to 4\ell\, 2\nu$ channel, 8% in the $WWZ \to 3\ell\, 3\nu 2j$ channel, 9% in the $WZZ \to 5\ell 1\nu$ channel, 5% in the $WZZ \to 3\ell 3\nu$ channel and 7% in the $WZZ \to 4\ell 2j$ and $WZZ \to 3\ell 1\nu 2j$ channels.

Assuming that the number of signal events follows a Poisson distribution and taking into account an estimated total systematic uncertainty on the background ($\sigma_B$) explained above, the signal significance $Z_\sigma$ is calculated from the number of estimated signal and background events, indicated by $N_{\mathrm{sig}}$ and $N_{\mathrm{bkg}}$, respectively [25]:

$$Z_\sigma = \sqrt{2\left[\left(N_{\mathrm{sig}} + N_{\mathrm{bkg}}\right)\log\frac{N_{\mathrm{sig}} + N_{\mathrm{bkg}}}{B_0} + B_0 - N_{\mathrm{sig}} - N_{\mathrm{bkg}}\right] + \frac{\left(N_{\mathrm{bkg}} - B_0\right)^2}{\sigma_B^2}}, \tag{1}$$

where $B_0 = \frac{1}{2}\left(N_{\mathrm{bkg}} - \sigma_B^2 + \sqrt{\left(N_{\mathrm{bkg}} - \sigma_B^2\right)^2 + 4\left(N_{\mathrm{sig}} + N_{\mathrm{bkg}}\right)\sigma_B^2}\right)$.

The estimated precision on the signal strength measurement, $\frac{\Delta\mu}{\mu}$, is calculated as

$$\frac{\Delta\mu}{\mu} = \frac{\sqrt{N_{\mathrm{sig}} + N_{\mathrm{bkg}} + \sum_{i=0}^{\mathrm{b}}\left(N_i\sigma_i\right)^2 + \sum_{j=0}^{\mathrm{s}}\left(N_j\sigma_j\right)^2}}{N_{\mathrm{sig}}}, \tag{2}$$

where the SM backgrounds originating from the diboson, triboson, $t\bar{t}$, $t\bar{t}H$, $t\bar{t}W$, $t\bar{t}Z$ and fake-lepton are taken into account separately in the summation, assuming the systematic uncertainties on the cross-section normalisation listed above. Only experimental uncertainties are taken into account separately for the $VVV$ and $VH$ signal events. Uncertainties related to the limited number of simulated events are neglected. The expected signal significance is calculated separately for every channel and shown in Table 10. The expected precision on the signal strength measurement is calculated for integrated luminosities of 3000 fb$^{-1}$ and 4000 fb$^{-1}$ only for channels with $Z_\sigma > 3$, assuming the same values for the systematic uncertainties in both cases. The results are shown in Table 11.

The HL-LHC offers a large improvement to multi-boson production, as demonstrated by this simple cut-based analysis. Sensitivities larger than $3\sigma$ are expected in two channels and larger than $5\sigma$ in one channel. It should be noted that more mature analysis techniques such as MVA, would likely improve these results further. However, it is expected that precise estimations of the main backgrounds, i.e. from dibosons and fake-leptons, will be needed in HL-LHC data analysis, in order to obtain a high level of precision.

| Channel | $Z_\sigma$ at 3000 fb$^{-1}$ (4000 fb$^{-1}$) |
|---|---|
| $WWW \to 3\ell\ 3\nu$ | 0SFOS: 6.7 (7.0)<br>1SFOS: 1.0 (1.0)<br>2SFOS: 0.7 (0.7) |
| $WWW \to 2\ell\ 2\nu\ 2j$ | $e^\pm e^\pm$: 0.8 (0.8)<br>$e^\pm \mu^\pm$: 1.2 (1.2)<br>$\mu^\pm \mu^\pm$: 1.8 (1.8) |
| $WWZ \to 4\ell\ 2\nu$ | SFOS: 0.1 (0.1)<br>DFOS: 3.0 (3.1) |
| $WWZ \to 3\ell\ 3\nu 2j$ | 0.3 (0.3) |
| $WZZ \to 5\ell 1\nu$ | 3.0 (3.4) |
| $WZZ \to 4\ell 2j$ | 0.1 (0.1) |
| $WZZ \to 3\ell 3\nu$ | 0.03 (0.03) |
| $WZZ \to 3\ell 1\nu 2j$ | 0.04 (0.04) |

Table 10: Expected signal significance $Z_\sigma$ for an integrated luminosity of 3000 fb$^{-1}$ and 4000 fb$^{-1}$, calculated separately for each channel.

| Channel | $\frac{\Delta\mu}{\mu}$ (3000 fb$^{-1}$) | $\frac{\Delta\mu}{\mu}$ (4000 fb$^{-1}$) |
|---|---|---|
| $WWW \to 3\ell\ 3\nu$ (0SFOS) | 11% | 10% |
| $WWZ \to 4\ell\ 2\nu$ (DFOS) | 27% | 25% |
| $WZZ \to 5\ell 1\nu$ | 36% | 31% |

Table 11: Expected precision on the signal strength measurement $\frac{\Delta\mu}{\mu}$ for an integrated luminosity of 3000 fb$^{-1}$ and 4000 fb$^{-1}$, in the channels with $Z_\sigma > 3$.

# 6 Conclusion

Prospect studies of the searches for the production of three vector bosons, $W^\pm W^\pm W^\mp$, $W^\pm W^\mp Z$ or $W^\pm ZZ$, in fully leptonic and channels with jets are reported. Results correspond to an integrated luminosity of 3000 fb$^{-1}$ and 4000 fb$^{-1}$, expected to be collected at a centre-of-mass energy of 14 TeV by the ATLAS detector during the HL-LHC run. Signal regions have been defined for every channel separately, based on the selection requirements optimized to maximize the sensitivity to the $W^\pm W^\pm W^\mp$, $W^\pm W^\mp Z$ and $W^\pm ZZ$ processes and to reduce the contributions from SM background processes. Results in terms of the expected signal and background yields, the significance and the the signal strength measurement are given. Three channels, $W^\pm W^\pm W^\mp \to 3\ell\ 3\nu$ and $W^\pm W^\mp Z \to 4\ell\ 2\nu$ and $W^\pm ZZ \to 5\ell\ 1\nu$, are expected to provide sensitivities larger than $3\sigma$ with the precisions of the corresponding cross section measurements of 11%, 27% and 36%, respectively, at 3000 fb$^{-1}$.

# Higgs Physics

# Projections for measurements of Higgs boson cross sections, branching ratios, coupling parameters and mass with the ATLAS detector at the HL-LHC

The ATLAS Collaboration

A study is presented on the ATLAS experimental prospects for measuring Higgs boson cross sections, signal strengths and branching ratios, and determining couplings to individual fermions and bosons, at the High Luminosity LHC (HL-LHC) with 3000 fb$^{-1}$collected at 14 TeV. The results shown here are extrapolated from the Run 2 results using datasets of 36 fb$^{-1}$ and 80 fb$^{-1}$ collected at 13 TeV. The different analyses are combined to compute the expected precision of the measurement of the cross-sections for the main production processes of the Higgs boson and of the branching ratios into its main decay modes as well as their interpretation in terms of modifiers to Higgs boson couplings. It is assumed that the upgraded detector produces the same physics capabilities at HL-LHC as were achieved for Run 2, despite the increased pileup. The one exception for some of the measurements is muons, where an improved momentum resolution is anticipated. The baseline scenario for the expected systematic uncertainties assumes a reduction of many uncertainties compared to the extrapolated analyses. As a conservative benchmark, the results with Run 2 systematic uncertainty values are given as well. A similar exercise is also done for the Higgs boson mass measurement expected at HL-LHC.

# 1 Introduction

One of the main motivations for the planned high-luminosity upgrade of the LHC (HL-LHC), is to enable precise measurements of Higgs boson properties. In the Standard Model (SM), all properties of the Higgs boson are defined once its mass is known. However, this model leaves many open questions such as the hierarchy problem and the nature of dark matter. Many alternative theories addressing these issues make different predictions for the properties of one or more Higgs bosons. Precise measurements in the Higgs sector are therefore a high priority in the future programme of particle physics.

A comprehensive study of the projected precision of measurements of the Higgs boson production cross section times branching ratios and Higgs boson couplings was presented by the ATLAS Collaboration in 2014 [1]. It was based on the extrapolation of the Run 1 results to the 3000 $\text{fb}^{-1}$ integrated luminosity expected at HL-LHC.

The aim of this note is to update the expected measurement precision at HL-LHC of the above measurements taking advantage of the analysis improvements developed for Run 2 analyses. An extrapolation from the data samples collected in 2015 and 2016, the two first years of the LHC programme, corresponding to 36 $\text{fb}^{-1}$, was performed. This was implemented for the $WW$, $Z\gamma$, $t\bar{t}H$ and $\tau\tau$ final states. The extrapolation for the remaining channels ($\gamma\gamma$, $ZZ$, $VH$ with $H \to b\bar{b}$ and $\mu\mu$) used the latest results based on the data samples collected in 2015, 2016 and 2017 (80 $\text{fb}^{-1}$).

In addition to the increase in integrated luminosity, this extrapolation accounts for the increase in total cross sections from 13 to 14 TeV. The signal yields have been scaled by the ratio of Higgs boson production cross sections values at 13 and 14 TeV, reported in Ref. [2]. The background yields have been scaled according to the parton luminosity ratio between 13 and 14 TeV, reported in Ref. [3], taking into account whether the background process is predominantly quark pair or gluon pair initiated.

To simplify the extrapolation, object reconstruction efficiencies, resolutions and fake rates are assumed to be similar in the Run 2 and HL-LHC environments. This is based on the assumption that the improved performance of the ATLAS detector at HL-LHC will compensate for the degradation induced by higher pile up. The extrapolated measurement precision from Run 2 to HL-LHC integrated luminosity has been obtained by means of a likelihood fit to an Asimov data set based on the Run 2 model with the corrections described above applied.

Systematic uncertainties are separated into components for experimental uncertainties and theory uncertainties on signal and background processes. Their values in the baseline scenario considered for the HL-LHC (scenario S2) are reduced compared to those currently used in Run 2, reflecting the situation that is expected to be reached at the end of the HL-LHC programme. The correction factors, discussed with CMS and the theory community, are reported in Ref. [4]. By default, in scenario S2, all theory uncertainties for signal and background are halved. Exceptions are the PDF uncertainties for which reduction factors depend on the quark/gluon initial states and on energy scale values [5], and the uncertainties on the $t\bar{t}$+heavy flavour cross section, as explained in Section 2.6.

The uncertainty on the integrated luminosity is set to 1%. The uncertainties due to the limited size of simulation samples are assumed to be negligible. In the relevant analyses, the uncertainty on the modeling of the continuum background using a functional form description is also assumed to become negligible. To demonstrate the importance of the reduction of the systematic uncertainties in the HL-LHC era compared to Run 2, expected measurement uncertainties are provided for a scenario in which systematic uncertainties are kept at their current Run 2 values (called scenario S1). The exceptions are the uncertainties due to the size of simulation samples and the uncertainty on the modeling of the continuum background using a functional form description, which are assumed to be negligible, not only in the scenario S2, but also in scenario S1.

It should be noted that for this note, the breakdown of systematic uncertainties has been done in a sequential way, first separating the theory background uncertainties first from the total uncertainty, then theory signal uncertainties and finally the experimental uncertainties. In case of correlations between the experimental and theoretical uncertainty components, the quoted experimental and theory uncertainties in scenario S1 might differ significantly from those quoted in the corresponding Run 2 papers and conference notes, if a different procedure has been used there. In these cases, the different procedure will be highlighted in the relevant sections. Expected precision both of the measured signal strength $\mu$, defined as the ratio of the total Higgs boson signal yield to its SM prediction, and of the measured cross sections is reported for most channels.

All the extrapolated single-channel results are combined to compute the cross sections per production mode and the branching ratios. The measured production modes are gluon-gluon fusion (ggF), vector-boson fusion (VBF), Higgs boson production in association with a vector boson ( $VH$: sum of $WH$, $q\bar{q} \to ZH$, and $gg \to ZH$ processes), and Higgs boson production in association with a top-antitop quark pair ($t\bar{t}H$) or a single top quark ($tH$). The measured decays modes are the decays of the Higgs boson to $\gamma\gamma$, $ZZ$, $WW$, $\tau\tau$, $bb$, $\mu\mu$ and $Z\gamma$. These results are interpreted in a framework of multiplicative modifiers $\kappa$ applied to the SM values of Higgs boson couplings [3] and as limits on Beyond Standard Model (BSM) effects.

The last section of the note addresses the expected precision of the Higgs boson mass measurement at HL-LHC based on the $H \to ZZ^* \to 4\ell$ channel, which is the least sensitive to systematic uncertainties.

## 2 Single channels

### 2.1 $H \to \gamma\gamma$

The measurement of Higgs boson properties in the diphoton decay channel, using a dataset with an integrated luminosity of 80 fb$^{-1}$ from Run 2, has been recently released in Ref. [6]. The cross sections have been measured for several production modes of the Higgs boson: ggF, VBF, $VH$ and $t\bar{t}H + tH$ later labelled as 'top'.

Higgs boson production in association with a bottom–antibottom quark pair ($b\bar{b}H$) is merged with the ggF production mode. Simplified template cross sections (STXS) in the so-called strong merging scheme of Stage-1 have been measured as well. The analysis uses 29 event categories, that are optimised to be as pure as possible in STXS regions.

The signal in each category is parametrised using a double-sided Crystal Ball [7] function, whose parameters are fixed from a fit to the MC distributions. Its mean and width are allowed to vary within the systematic uncertainties on the photon energy scale (PES) and within the systematic uncertainties on the photon energy resolution (PER), respectively.

The background estimation uses fully data-driven techniques; an analytic function was chosen in each category to provide an accurate descriptions of the $m_{\gamma\gamma}$ shape in simulated background samples. This is the main source of systematic uncertainty, as described in Ref. [6] and labelled as 'spurious signal', and is often dominated by the uncertainty related to the limited size of the simulated samples. Since this latter uncertainty is assumed to be negligible at HL-LHC and moreover an improved modeling strategy of the $m_{\gamma\gamma}$ shape is expected, the spurious signal uncertainty is neglected in this extrapolation.
The other important systematic uncertainties are QCD scale uncertainties causing event migrations between categories, photon isolation efficiencies and jet energy uncertainties.

The luminosity and the signal cross sections have been scaled as described in Section 1. Due to the data-driven background estimation, no accurate studies have been performed on the exact background composition in each category. The relative fractions of each component are scaled within their uncertainties, but this is shown to have a negligible impact on the fitted event yield.

In the Run 2 analysis, a conservative uncertainty of 100% on the Higgs boson production in association with heavy flavour jets for the resonant background in $t\bar{t}H$ categories is used. Future measurements of the heavy flavour content of jets produced in association with the Higgs boson are expected to constrain these values. For scenario S2, the above uncertainty related to heavy flavour quark production modeling has been divided by two. In scenario S2, the photon isolation efficiency and the jet flavour composition uncertainties become the biggest source of systematics.

The results for the extrapolation of the production cross-section measurements times $H \to \gamma\gamma$ branching ratio at HL-LHC for scenario S2 and for Higgs-boson absolute rapidity $|y_H| < 2.5$ are :

$$\begin{aligned}
\sigma_{\text{ggF+b}\bar{\text{b}}\text{H}} \times \text{B}(H \to \gamma\gamma) &= 114.3^{+4.2}_{-4.0}\,\text{fb} = 114.3^{+2.0}_{-2.0}\,(\text{stat})\,^{+3.6}_{-3.4}\,(\text{exp})\,^{+0.7}_{-0.7}\,(\text{sig})\,\text{fb} \\
\sigma_{\text{VBF}} \times \text{B}(H \to \gamma\gamma) &= 9.03^{+0.86}_{-0.80}\,\text{fb} = 9.03^{+0.40}_{-0.39}\,(\text{stat})\,^{+0.57}_{-0.50}\,(\text{exp})\,^{+0.52}_{-0.48}\,(\text{sig})\,\text{fb} \\
\sigma_{\text{VH}} \times \text{B}(H \to \gamma\gamma) &= 5.01^{+0.47}_{-0.45}\,\text{fb} = 5.01^{+0.40}_{-0.40}\,(\text{stat})\,^{+0.19}_{-0.17}\,(\text{exp})\,^{+0.15}_{-0.14}\,(\text{sig})\,\text{fb} \\
\sigma_{\text{top}} \times \text{B}(H \to \gamma\gamma) &= 1.61^{+0.13}_{-0.12}\,\text{fb} = 1.61^{+0.08}_{-0.07}\,(\text{stat})\,^{+0.08}_{-0.07}\,(\text{exp})\,^{+0.07}_{-0.06}\,(\text{sig})\,\text{fb}
\end{aligned}$$

These results are compared to the SM prediction in Figure 1; the SM uncertainties on cross sections and branching ratio are divided by two compared to their current values, which approximately corresponds to the scaling expected for scenario S2. Table 1 shows the relative uncertainties on the cross-section measurements with Run 2 data and their expected values at HL-LHC with both systematic uncertainty scenarios. The last column displays the expected impact of signal theory uncertainties on the measured signal strength ($\Delta\mu_{\text{sig}}$). For the cross-section measurements, only signal theory uncertainties on the fiducial acceptance are considered, while for the signal strength measurements, in addition, also the uncertainty on the predicted SM production-mode cross sections times $H \to \gamma\gamma$ branching ratio is taken into account.

The expected ranking of the 10 systematic uncertainties for scenario S2 with the largest impact on the measured production-mode cross sections times $H \to \gamma\gamma$ branching ratio and on the measured production-mode signal strengths can be seen in Figures 2-3, respectively.

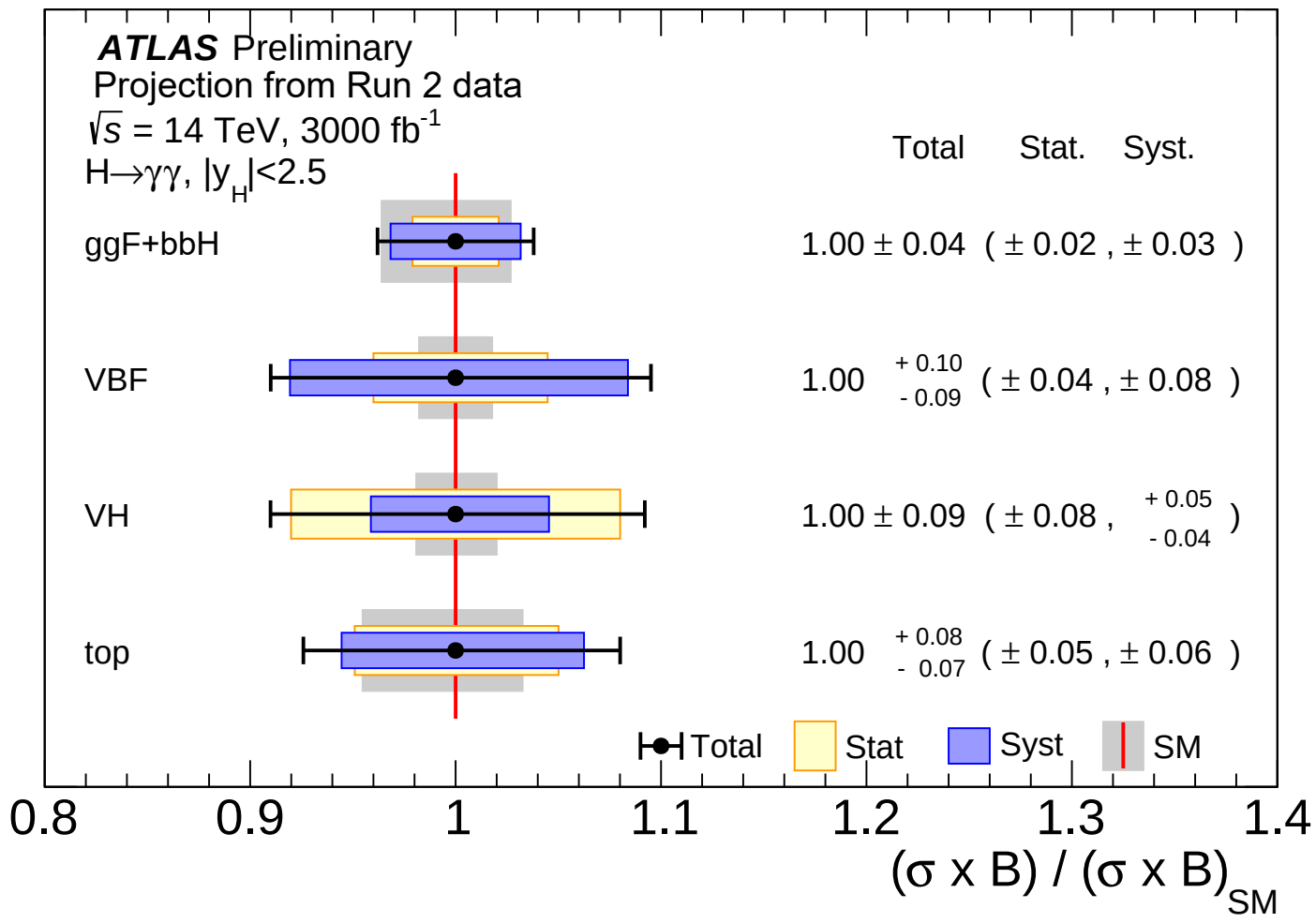

Figure 1: For each production mode, expected result for the measurement of the cross section times branching ratio normalised to their SM expectation for the $\gamma\gamma$ final state in scenario S2 at HL-LHC is shown for Higgs boson absolute rapidity $|y_H| < 2.5$. The statistical (yellow) and systematic (blue) uncertainty components are displayed as well as the theory uncertainty on the SM prediction (grey). The blue band of the systematic uncertainty includes both experimental and theory uncertainties.

| Prod. mode | Scenario | $\Delta_{\rm tot}/\sigma_{\rm SM}$ | $\Delta_{\rm stat}/\sigma_{\rm SM}$ | $\Delta_{\rm exp}/\sigma_{\rm SM}$ | $\Delta_{\rm sig}/\sigma_{\rm SM}$ | $\Delta\mu_{\rm sig}$ |
|---|---|---|---|---|---|---|
| ggF+$b\bar{b}H$ | Run 2, 80 fb$^{-1}$ | $^{+0.15}_{-0.14}$ | $^{+0.11}_{-0.11}$ | $^{+0.09}_{-0.08}$ | $^{+0.03}_{-0.02}$ | $^{+0.08}_{-0.06}$ |
| | HL-LHC S1 | $^{+0.06}_{-0.05}$ | $^{+0.02}_{-0.02}$ | $^{+0.05}_{-0.05}$ | $^{+0.01}_{-0.01}$ | $^{+0.07}_{-0.06}$ |
| | HL-LHC S2 | $^{+0.04}_{-0.03}$ | $^{+0.02}_{-0.02}$ | $^{+0.03}_{-0.03}$ | $^{+0.01}_{-0.01}$ | $^{+0.03}_{-0.03}$ |
| VBF | Run 2, 80 fb$^{-1}$ | $^{+0.36}_{-0.31}$ | $^{+0.30}_{-0.28}$ | $^{+0.16}_{-0.11}$ | $^{+0.13}_{-0.09}$ | $^{+0.15}_{-0.10}$ |
| | HL-LHC S1 | $^{+0.14}_{-0.13}$ | $^{+0.04}_{-0.04}$ | $^{+0.08}_{-0.07}$ | $^{+0.11}_{-0.10}$ | $^{+0.11}_{-0.10}$ |
| | HL-LHC S2 | $^{+0.10}_{-0.09}$ | $^{+0.04}_{-0.04}$ | $^{+0.06}_{-0.06}$ | $^{+0.06}_{-0.05}$ | $^{+0.06}_{-0.06}$ |
| $VH$ | Run 2, 80 fb$^{-1}$ | $^{+0.59}_{-0.54}$ | $^{+0.54}_{-0.50}$ | $^{+0.22}_{-0.20}$ | $^{+0.12}_{-0.09}$ | $^{+0.18}_{-0.11}$ |
| | HL-LHC S1 | $^{+0.11}_{-0.10}$ | $^{+0.08}_{-0.08}$ | $^{+0.06}_{-0.05}$ | $^{+0.05}_{-0.04}$ | $^{+0.09}_{-0.08}$ |
| | HL-LHC S2 | $^{+0.09}_{-0.09}$ | $^{+0.08}_{-0.08}$ | $^{+0.04}_{-0.03}$ | $^{+0.03}_{-0.03}$ | $^{+0.05}_{-0.05}$ |
| $top$ | Run 2, 80 fb$^{-1}$ | $^{+0.37}_{-0.32}$ | $^{+0.34}_{-0.30}$ | $^{+0.10}_{-0.07}$ | $^{+0.10}_{-0.07}$ | $^{+0.18}_{-0.11}$ |
| | HL-LHC S1 | $^{+0.11}_{-0.10}$ | $^{+0.05}_{-0.05}$ | $^{+0.07}_{-0.06}$ | $^{+0.07}_{-0.07}$ | $^{+0.13}_{-0.11}$ |
| | HL-LHC S2 | $^{+0.08}_{-0.08}$ | $^{+0.05}_{-0.05}$ | $^{+0.05}_{-0.04}$ | $^{+0.04}_{-0.04}$ | $^{+0.07}_{-0.06}$ |

Table 1: Expected precision of the production-mode cross-section measurements in the $H \to \gamma\gamma$ channel with 80 fb$^{-1}$ of Run 2 data and at HL-LHC. Uncertainties are reported relative to the SM cross section at the corresponding center-of-mass energy. Both scenarios S1 and S2 have been considered for the systematic uncertainties in the HL-LHC extrapolation. The spurious signal uncertainty is removed in both cases. The last column shows the signal theory uncertainty component when the measurement parameters are production-mode signal strengths instead of cross sections.

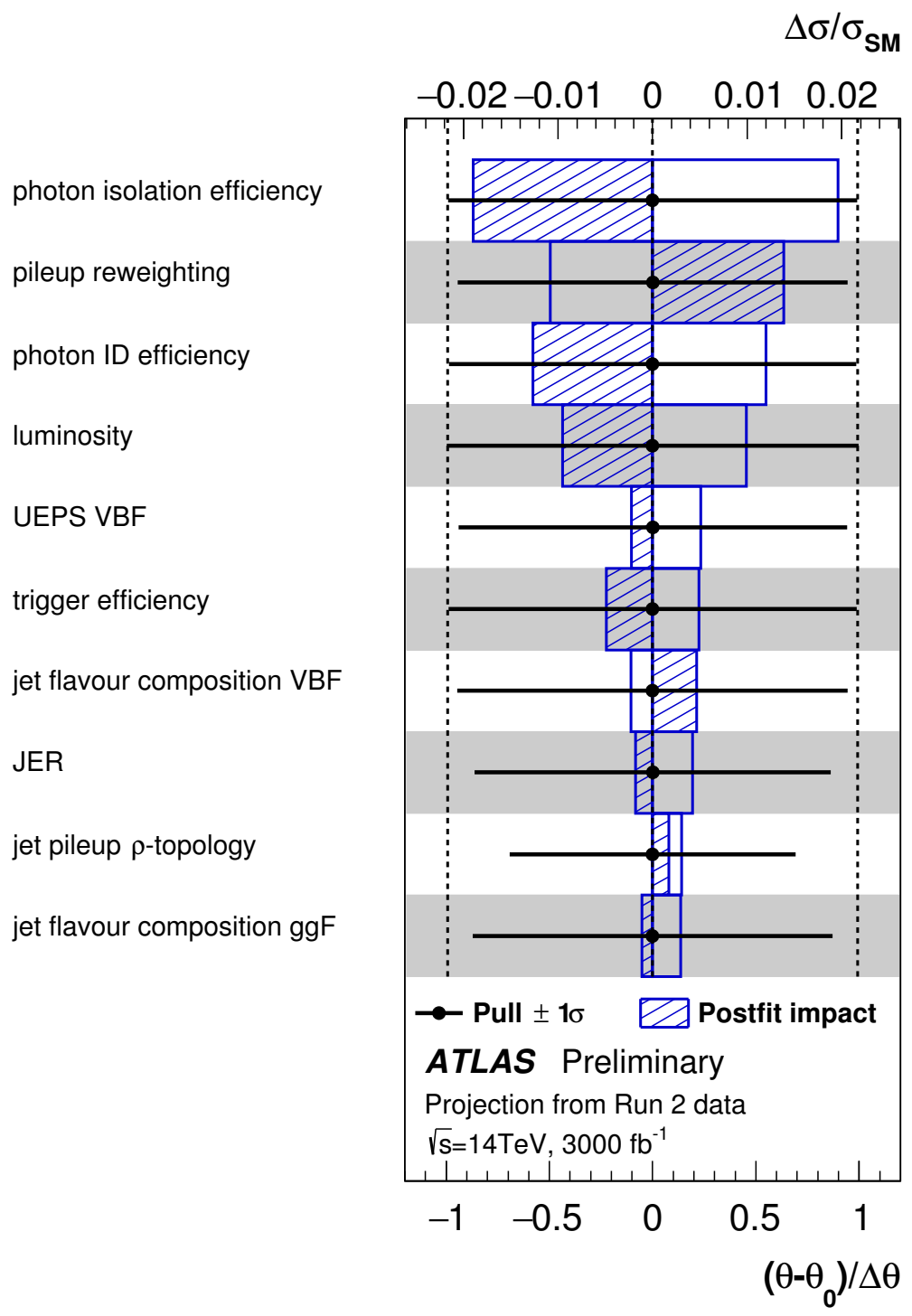

(a) $\sigma_{\mathrm{ggF+b\bar{b}H}}$

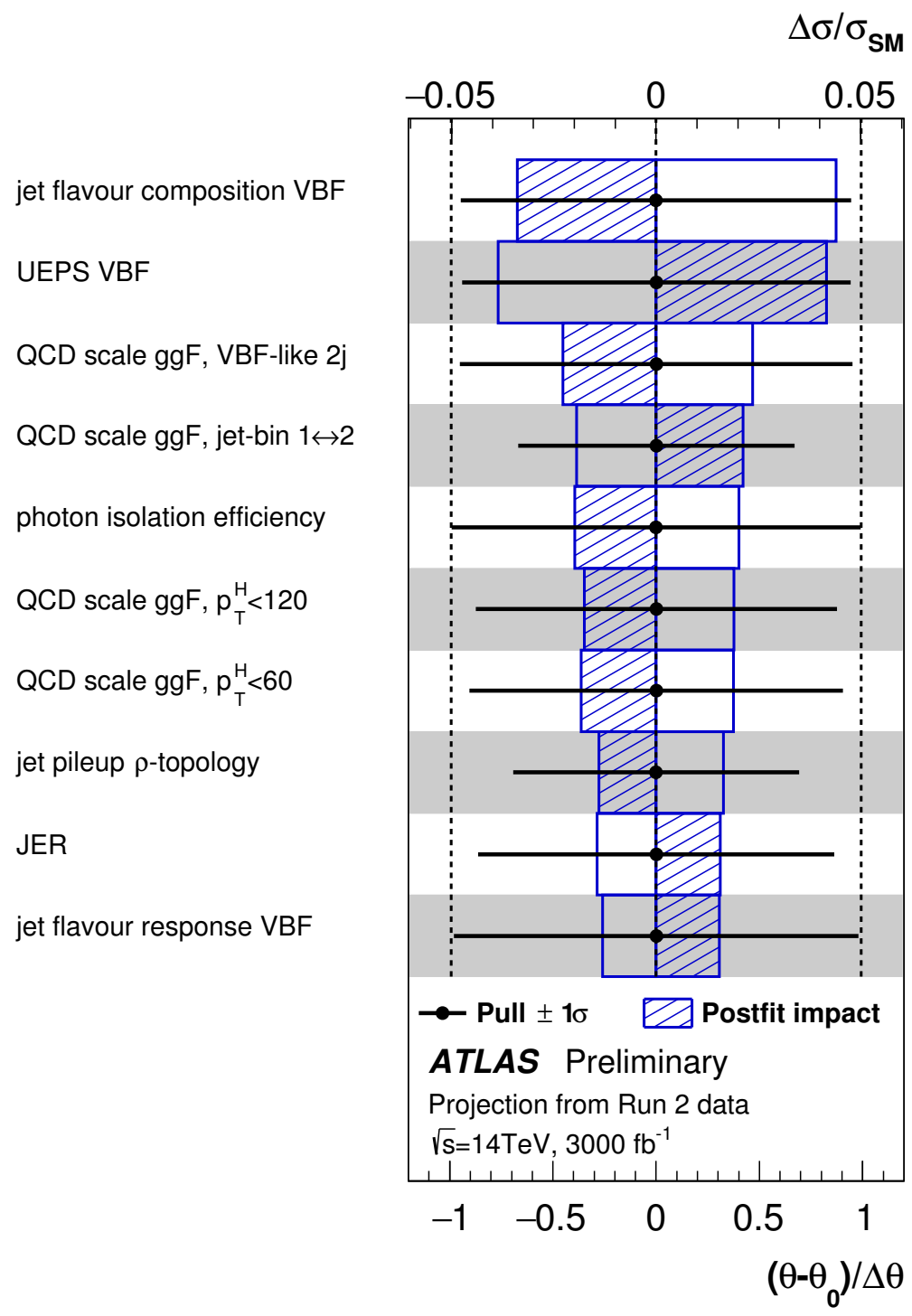

(b) $\sigma_{\mathrm{VBF}}$

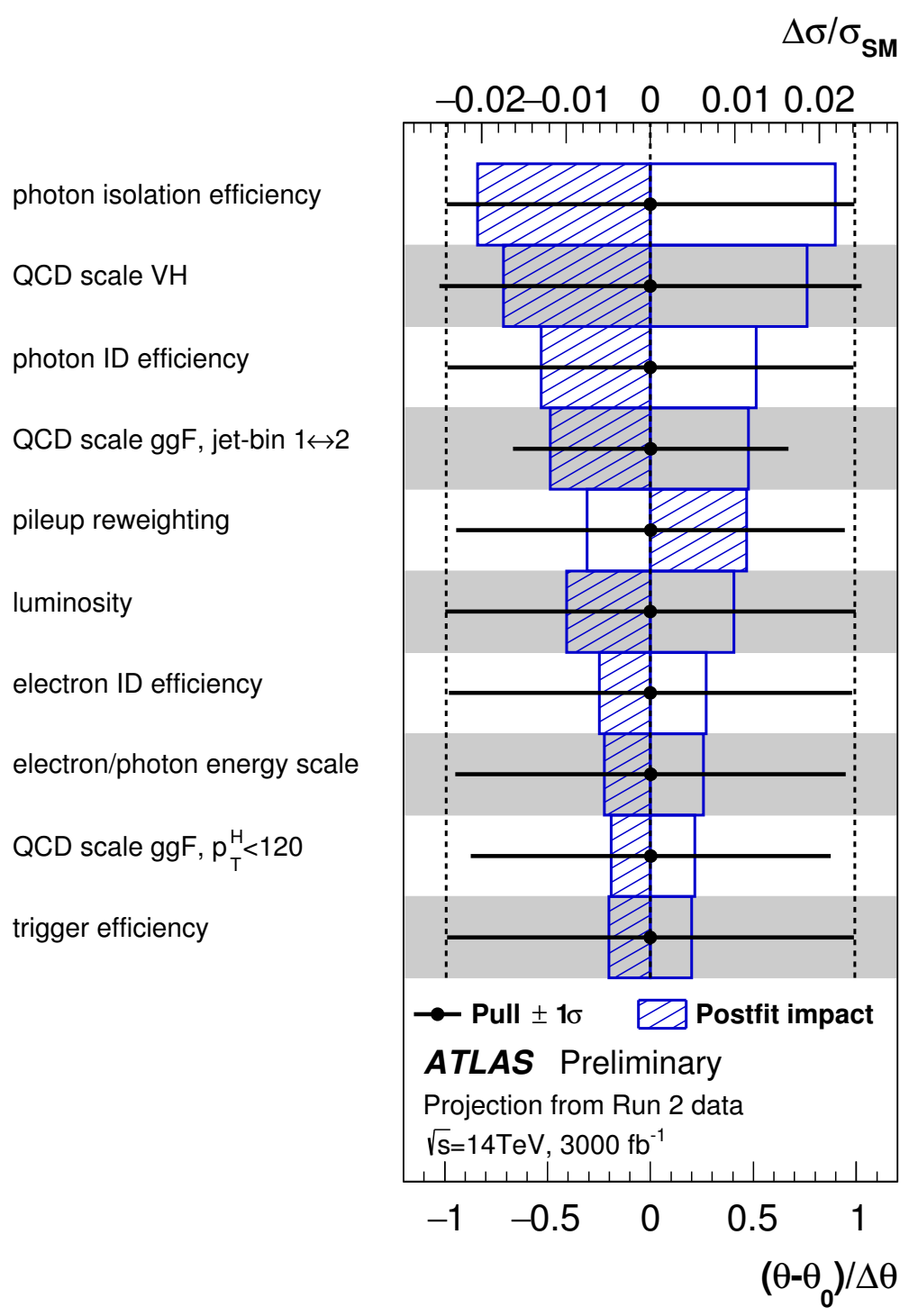

(c) $\sigma_{\mathrm{VH}}$

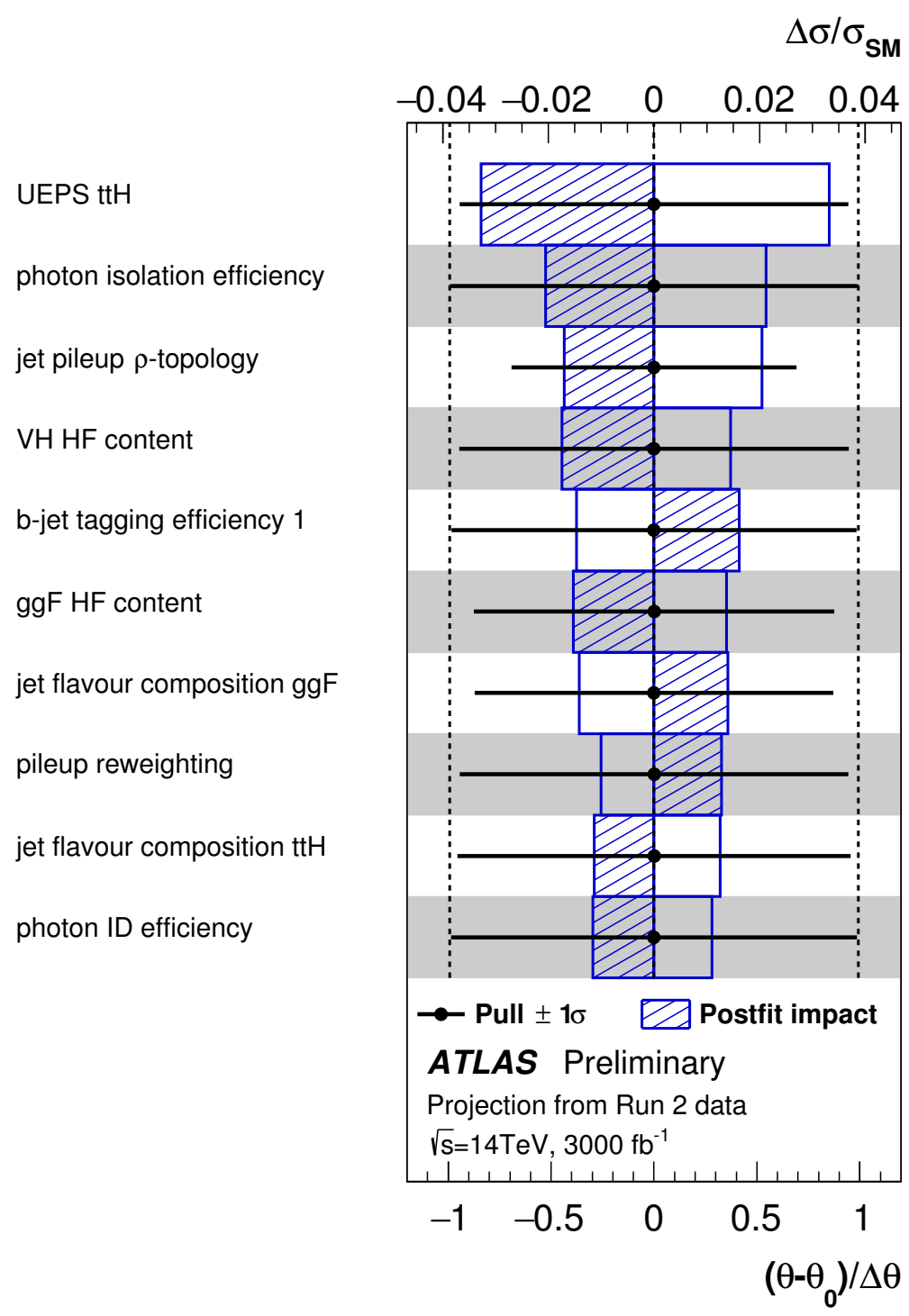

(d) $\sigma_{\mathrm{top}}$

Figure 2: Ranking of the 10 systematic uncertainties (scenario S2) with the largest impact on the expected cross-section times branching ratio measurement in the $H \to \gamma\gamma$ decay channel for the ggF+$b\bar{b}H$ (a), VBF (b), $VH$ (c) and top (d) production modes.

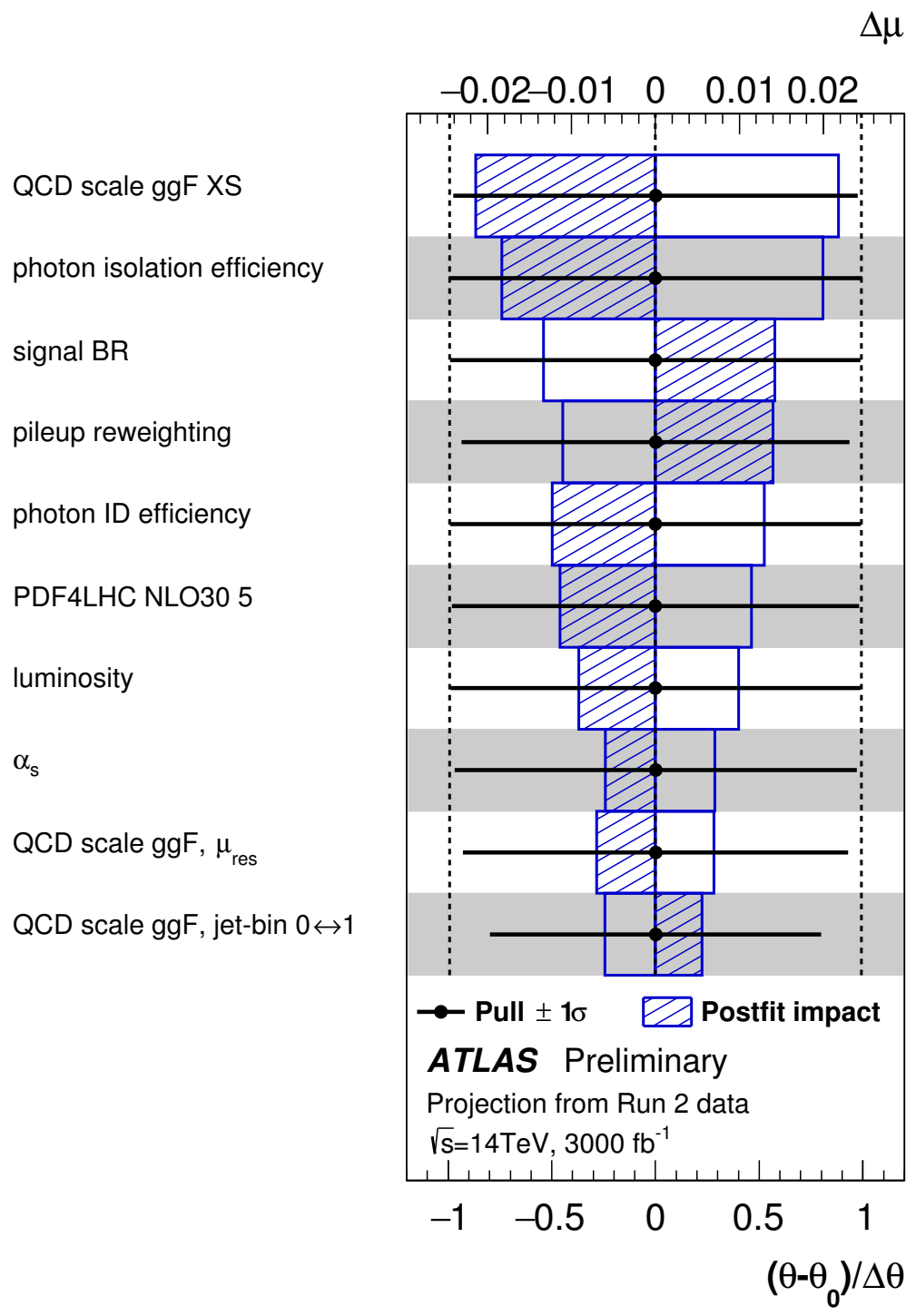

(a) $\mu_{\mathrm{ggF+b\bar{b}H}}$

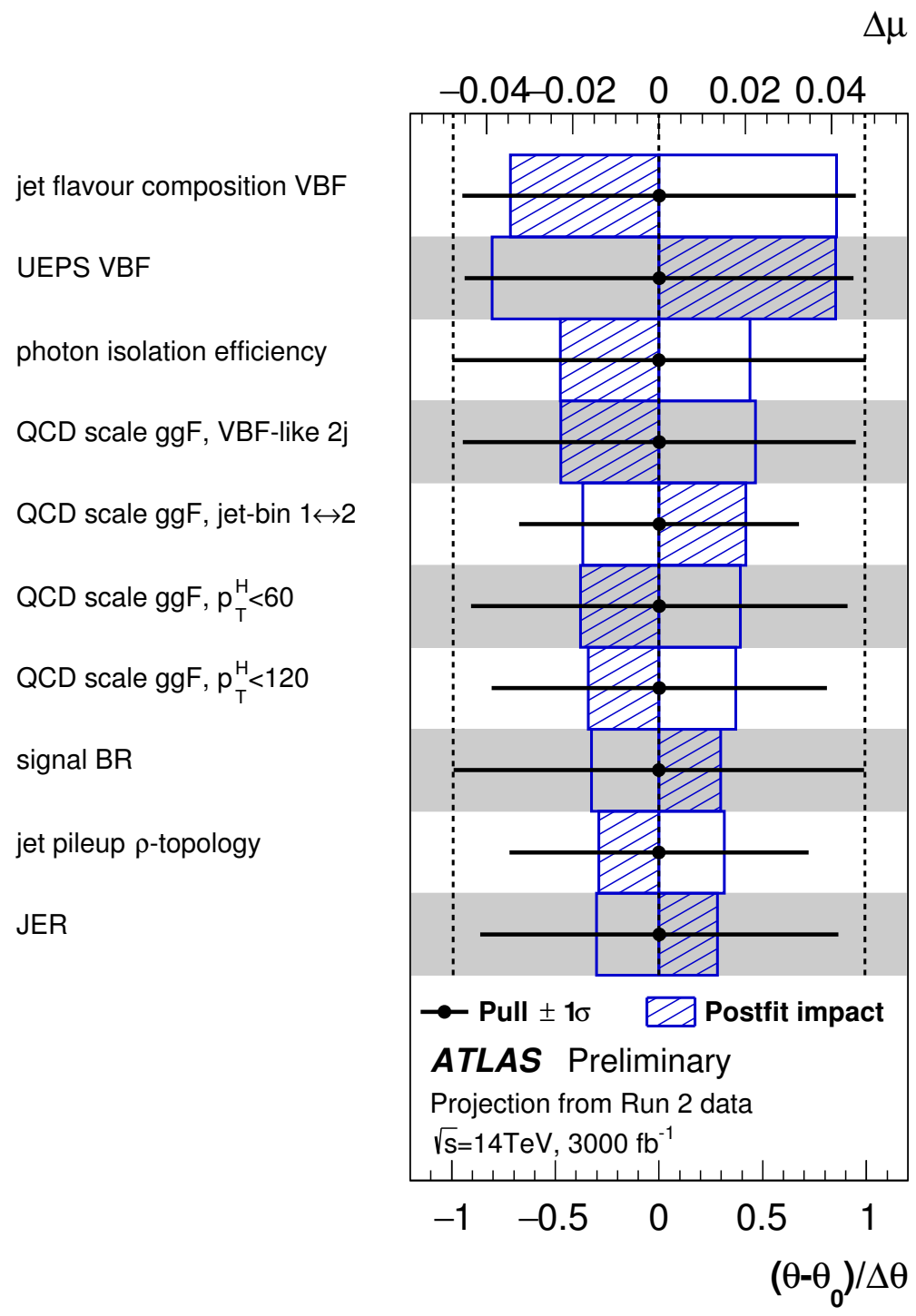

(b) $\mu_{\mathrm{VBF}}$

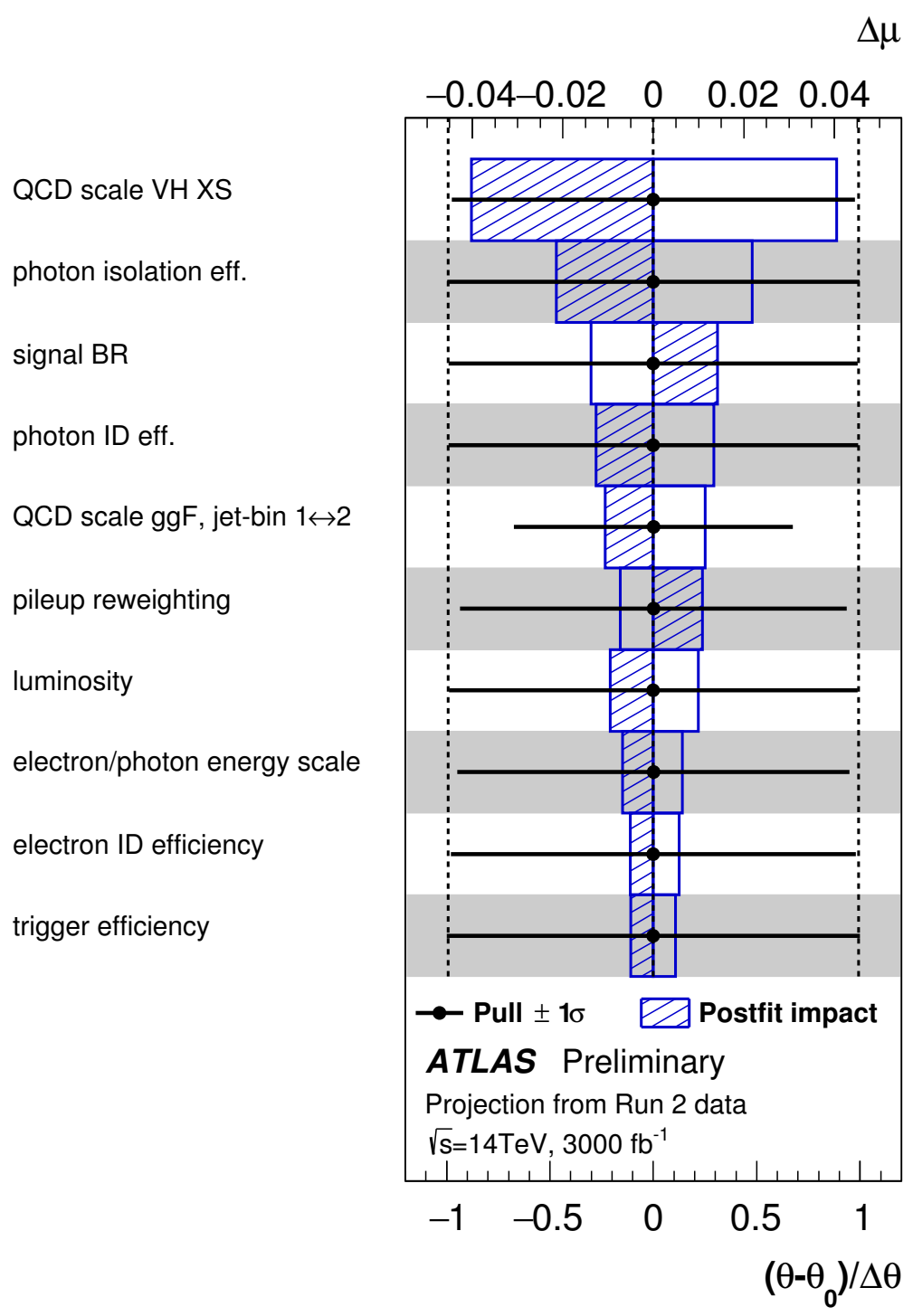

(c) $\mu_{\mathrm{VH}}$

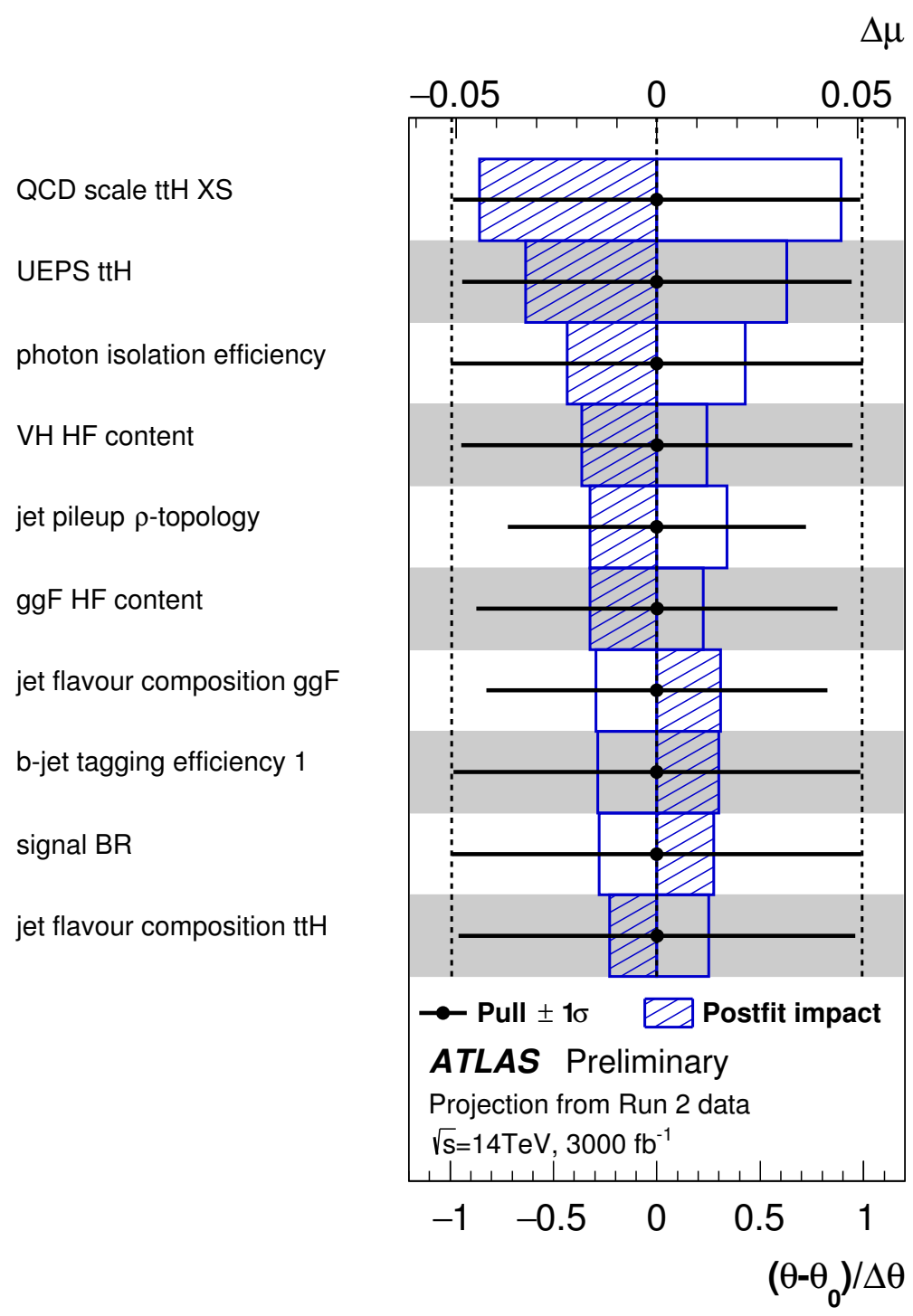

(d) $\mu_{\mathrm{top}}$

Figure 3: Ranking of the 10 systematic uncertainties (scenario S2) with the largest impact on the expected signal strength measurement in the $H \to \gamma\gamma$ decay channel for the ggF+$b\bar{b}H$ (a), VBF (b), $VH$ (c) and top (d) production modes.

## 2.2 $H \to ZZ^*$

Using a dataset with an integrated luminosity of 80 fb$^{-1}$ at Run 2, the measurement of the Higgs boson production cross section in the $H \to ZZ^* \to 4\ell$ decay channel [8] has been performed within the framework of the simplified template cross sections [2]. For the results in this note, the Stage 0 - Simplified Template Cross Sections framework (STXS Stage 0) has been used and the four main production modes are considered: ggF, VBF, $VH$ and $t\bar{t}H$.
In the Run 2 analysis [8], for the Stage-0 ggF bin, the dominant systematic uncertainties on the cross-section measurements come from electron/muon reconstruction and identification efficiency and pile-up modeling uncertainties. For VBF and $VH$, jet energy scale/resolution and QCD scale uncertainties related to bin migration are the largest uncertainties. In the $t\bar{t}H$ production bin, the largest impact comes from the theory uncertainties related to parton shower and to the heavy flavour quark production modeling for the ggF background contribution.

The expected results obtained for HL-LHC projections for scenario S1 (S2) and for Higgs boson absolute rapidity $|y_H| < 2.5$ are:

$$\begin{aligned}
\sigma_{\mathrm{ggF}} \times \mathrm{B}(H \to ZZ^*) &= 1.305^{+0.073}_{-0.072} \left(^{+0.057}_{-0.056}\right) \text{ pb} \\
&= 1.305^{+0.026}_{-0.026} \text{ (stat) } {}^{+0.055}_{-0.056} \left(^{+0.046}_{-0.046}\right) \text{ (exp) } {}^{+0.034}_{-0.031} \left(^{+0.021}_{-0.020}\right) \text{ (sig) } {}^{+0.009}_{-0.009} \left(^{+0.008}_{-0.008}\right) \text{ (bkg) pb} \\
\sigma_{\mathrm{VBF}} \times \mathrm{B}(H \to ZZ^*) &= 0.104^{+0.015}_{-0.014} \left(^{+0.013}_{-0.012}\right) \text{ pb} \\
&= 0.104^{+0.010}_{-0.010} \text{ (stat) } {}^{+0.006}_{-0.006} \left(^{+0.006}_{-0.005}\right) \text{ (exp) } {}^{+0.009}_{-0.008} \left(^{+0.005}_{-0.005}\right) \text{ (sig) } {}^{+0.001}_{-0.001} \left(^{+0.001}_{-0.001}\right) \text{ (bkg) pb} \\
\sigma_{\mathrm{VH}} \times \mathrm{B}(H \to ZZ^*) &= 0.058^{+0.012}_{-0.011} \left(^{+0.011}_{-0.010}\right) \text{ pb} \\
&= 0.058^{+0.010}_{-0.010} \text{ (stat) } {}^{+0.003}_{-0.002} \left(^{+0.002}_{-0.002}\right) \text{ (exp) } {}^{+0.005}_{-0.004} \left(^{+0.004}_{-0.003}\right) \text{ (sig) } {}^{+0.001}_{-0.001} \left(^{<0.001}_{<0.001}\right) \text{ (bkg) pb} \\
\sigma_{\mathrm{t\bar{t}H}} \times \mathrm{B}(H \to ZZ^*) &= 0.016^{+0.004}_{-0.003} \left(^{+0.004}_{-0.003}\right) \text{ pb} \\
&= 0.016^{+0.003}_{-0.003} \text{ (stat) } {}^{+0.001}_{-0.001} \left(^{+0.001}_{-0.001}\right) \text{ (exp) } {}^{+0.002}_{-0.001} \left(^{+0.001}_{-0.001}\right) \text{ (sig) } {}^{+0.001}_{-0.001} \left(^{<0.001}_{<0.001}\right) \text{ (bkg) pb}
\end{aligned}$$

These results are compared to the SM prediction in Figure 4; the SM uncertainties are divided by two compared to their current values, which approximately corresponds to the scaling expected from scenario S2.

To better compare the results of the Run 2 analysis and the extrapolated results with scenarios S1 and S2, Table 2 shows the uncertainties on the measured cross-section results divided by the predicted SM values.

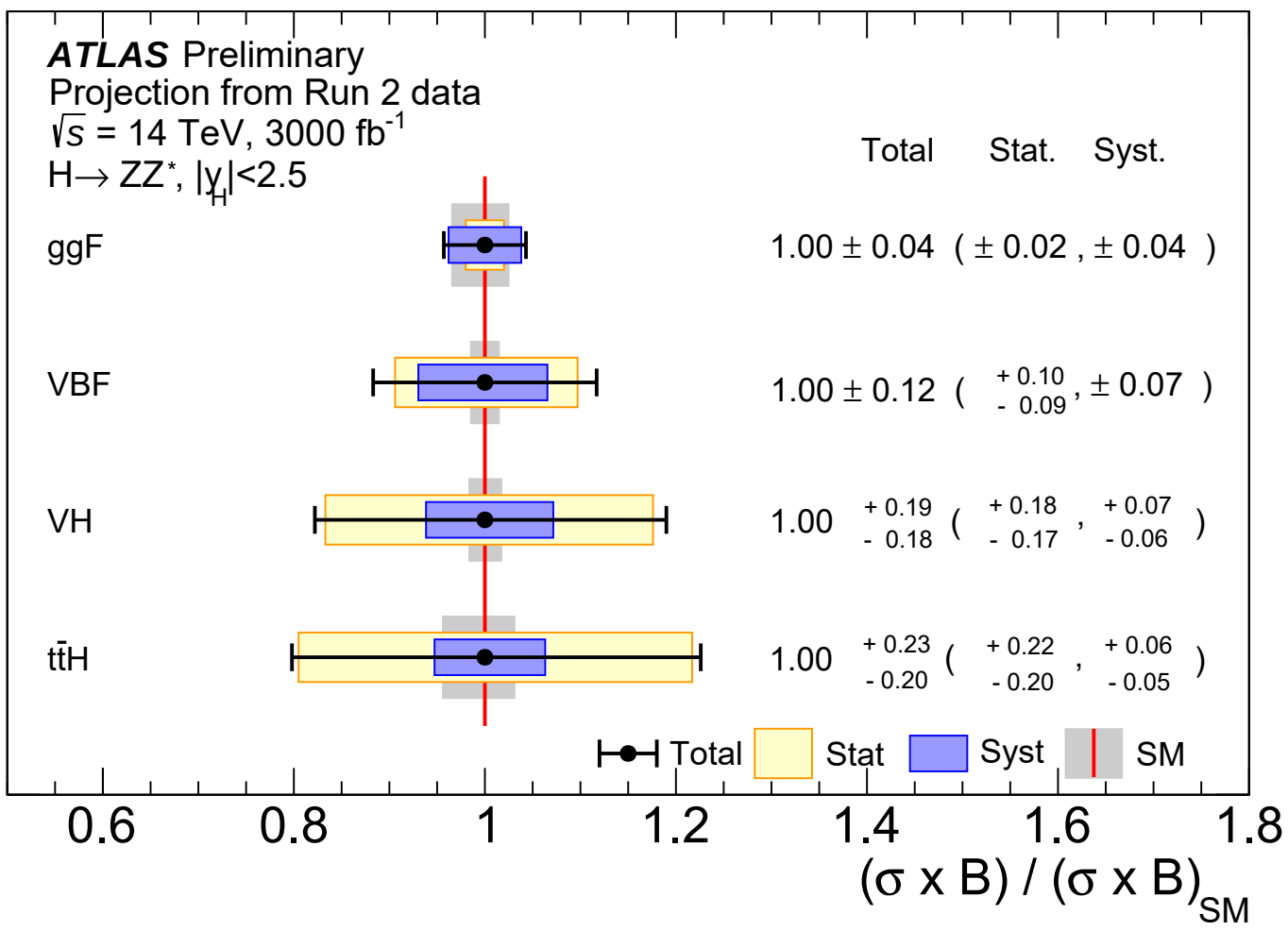

Figure 4: For each production mode, expected result for the measurement of the cross section times branching ratio normalised to their SM expectation for the $ZZ^*$ final state in scenario S2 at HL-LHC is shown for Higgs boson absolute rapidity $|y_H| < 2.5$. The statistical (yellow) and systematic (blue) uncertainty components are displayed as well as the theory uncertainty on the SM prediction (grey). The blue band of the systematic uncertainty includes both experimental and theory uncertainties.

| Prod. mode | Analysis | $\Delta_{tot}/\sigma_{SM}$ | $\Delta_{stat}/\sigma_{SM}$ | $\Delta_{exp}/\sigma_{SM}$ | $\Delta_{sig}/\sigma_{SM}$ | $\Delta_{bkg}/\sigma_{SM}$ | $\Delta\mu_{sig}$ |
|---|---|---|---|---|---|---|---|
| ggF | Run 2, 80 fb$^{-1}$ | $^{+0.160}_{-0.152}$ | $^{+0.143}_{-0.136}$ | $^{+0.053}_{-0.052}$ | $^{+0.043}_{-0.036}$ | $^{+0.011}_{-0.014}$ | $^{+0.070}_{-0.052}$ |
| | HL-LHC S1 | $^{+0.056}_{-0.055}$ | $^{+0.020}_{-0.020}$ | $^{+0.042}_{-0.043}$ | $^{+0.026}_{-0.024}$ | $^{+0.007}_{-0.007}$ | $^{+0.062}_{-0.056}$ |
| | HL-LHC S2 | $^{+0.043}_{-0.043}$ | $^{+0.020}_{-0.020}$ | $^{+0.035}_{-0.035}$ | $^{+0.016}_{-0.015}$ | $^{+0.006}_{-0.006}$ | $^{+0.030}_{-0.028}$ |
| VBF | Run 2, 80 fb$^{-1}$ | $^{+0.782}_{-0.598}$ | $^{+0.753}_{-0.583}$ | $^{+0.157}_{-0.095}$ | $^{+0.136}_{-0.074}$ | $^{+0.014}_{-0.029}$ | $^{+0.161}_{-0.101}$ |
| | HL-LHC S1 | $^{+0.147}_{-0.135}$ | $^{+0.097}_{-0.094}$ | $^{+0.059}_{-0.054}$ | $^{+0.088}_{-0.078}$ | $^{+0.007}_{-0.008}$ | $^{+0.087}_{-0.072}$ |
| | HL-LHC S2 | $^{+0.125}_{-0.117}$ | $^{+0.097}_{-0.094}$ | $^{+0.057}_{-0.052}$ | $^{+0.051}_{-0.047}$ | $^{+0.007}_{-0.006}$ | $^{+0.053}_{-0.050}$ |
| $VH$ | Run 2, 80 fb$^{-1}$ | $^{+1.410}_{-0.959}$ | $^{+1.381}_{-0.946}$ | $^{+0.155}_{-0.075}$ | $^{+0.228}_{-0.137}$ | $^{+0.012}_{-0.008}$ | $^{+0.283}_{-0.144}$ |
| | HL-LHC S1 | $^{+0.200}_{-0.185}$ | $^{+0.176}_{-0.167}$ | $^{+0.051}_{-0.042}$ | $^{+0.082}_{-0.070}$ | $^{+0.002}_{-0.001}$ | $^{+0.124}_{-0.084}$ |
| | HL-LHC S2 | $^{+0.190}_{-0.178}$ | $^{+0.176}_{-0.167}$ | $^{+0.043}_{-0.033}$ | $^{+0.064}_{-0.056}$ | $^{<0.001}_{<0.001}$ | $^{+0.077}_{-0.062}$ |
| $t\bar{t}H$ | Run 2, 80 fb$^{-1}$ | $< 5.75$ | | | - | | |
| | HL-LHC S1 | $^{+0.246}_{-0.213}$ | $^{+0.217}_{-0.195}$ | $^{+0.056}_{-0.042}$ | $^{+0.100}_{-0.074}$ | $^{+0.020}_{-0.026}$ | $^{+0.156}_{-0.095}$ |
| | HL-LHC S2 | $^{+0.226}_{-0.202}$ | $^{+0.217}_{-0.195}$ | $^{+0.042}_{-0.032}$ | $^{+0.047}_{-0.037}$ | $^{+0.010}_{-0.015}$ | $^{+0.074}_{-0.051}$ |

Table 2: Expected precision of the production-mode cross-section measurements in the $H \to ZZ^*$ channel with 80 fb$^{-1}$ of Run 2 data and at HL-LHC. Uncertainties are reported relative to the SM cross section at the corresponding center-of-mass energy. Both scenarios S1 and S2 have been considered for the systematic uncertainties in the HL-LHC extrapolation. The last column shows the theory uncertainty component when the measurement parameters are production mode signal strengths instead of cross sections. The value for the $t\bar{t}H$ channel with Run 2 data corresponds to the 95% CL limit.

For the ggF production mode, both scenarios S1 and S2 have larger systematic uncertainties than statistical uncertainties. For the other production modes, the statistical uncertainty dominates. For scenario S2, the signal theory (experimental) uncertainty is the largest systematic component for the $t\bar{t}H$ and $VH$ production modes (VBF production mode).

For scenario S2, the plots with the largest systematic uncertainties impacting the cross section and the signal strength respectively are shown in Figures 5 and 6 for the four production modes.

For the ggF production mode, the dominant systematic uncertainties for the cross-section measurement are the electron reconstruction and identification efficiencies, and the luminosity uncertainty. For the signal strength measurement the largest uncertainty is related to QCD scale uncertainty. The second largest source of signal theory uncertainty is related to the PDF uncertainties. These two theory uncertainties affect mostly the signal normalisation and therefore have a negligible impact on the ggF cross-section measurement accuracy.
For the VBF mode, the uncertainties related to the jet energy scale, the underlying event tune and parton shower and the QCD scale uncertainty for ggF events with jet bin migration passing the VBF selection, impact significantly both for the cross section and signal strength measurement.

For the $VH$ mode the QCD scale uncertainty for ggF events with jet bin migration passing the $VH$ event selection is the largest systematic uncertainty. The QCD scale uncertainty on the predicted $VH$ cross section has a large impact only on $\mu_{\mathrm{VH}}$.

For the $t\bar{t}H$ production mode, the uncertainty related to heavy flavour quark production modeling for the ggF background contribution has a large impact on the cross-section measurement. In addition to this last uncertainty, the QCD scale uncertainty on the total cross-section prediction also impacts the $t\bar{t}H$ signal strength measurement.

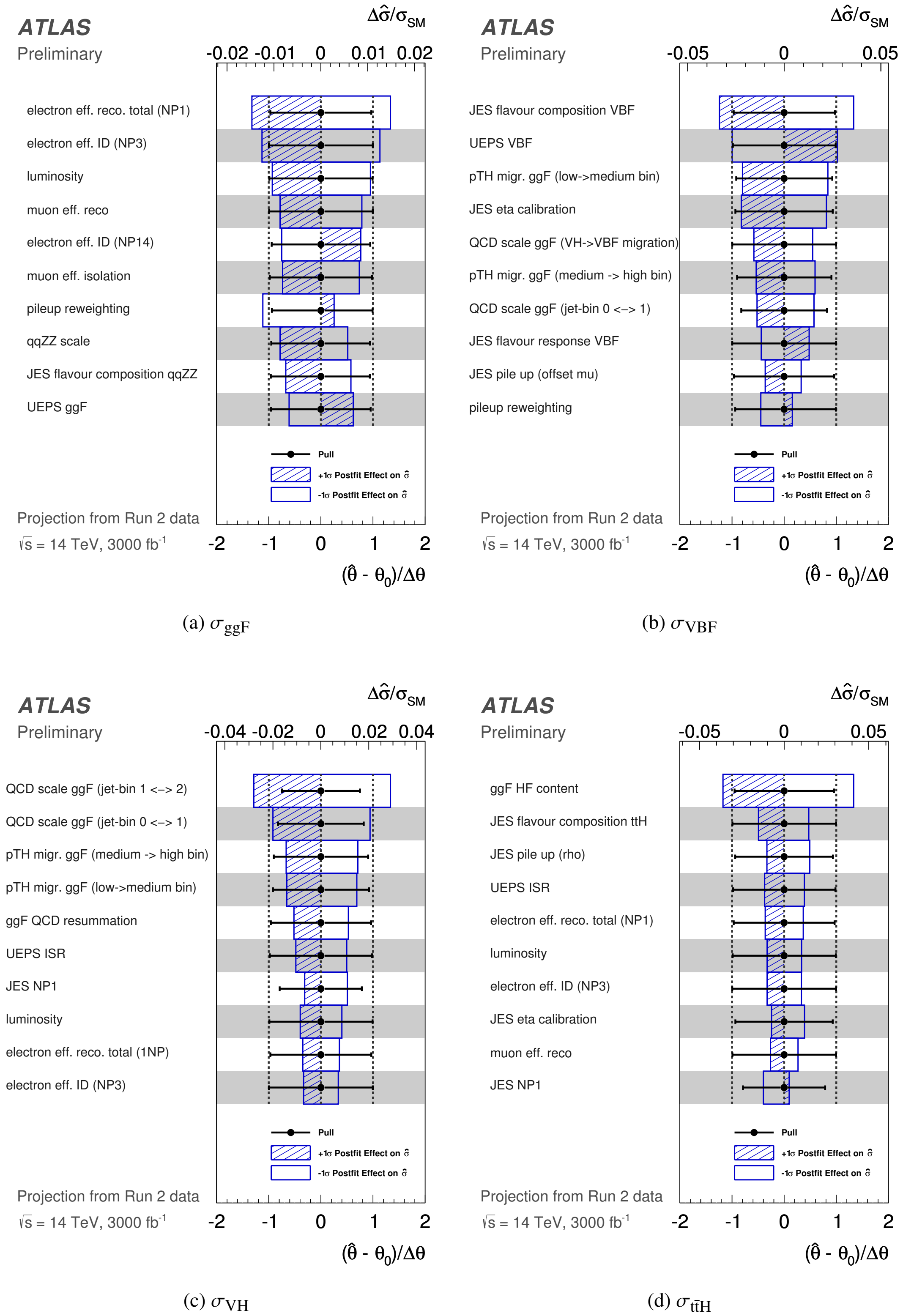

(a) $\sigma_{\rm ggF}$ (b) $\sigma_{\rm VBF}$ (c) $\sigma_{\rm VH}$ (d) $\sigma_{\rm t\bar{t}H}$

Figure 5: Ranking of the 10 systematic uncertainties (scenario S2) with the largest impact on the expected cross-section times branching ratio measurement of the $H \to ZZ^*$ decay channel for the ggF (a), VBF (b), $VH$ (c) and $t\bar{t}H$ (d) production modes.

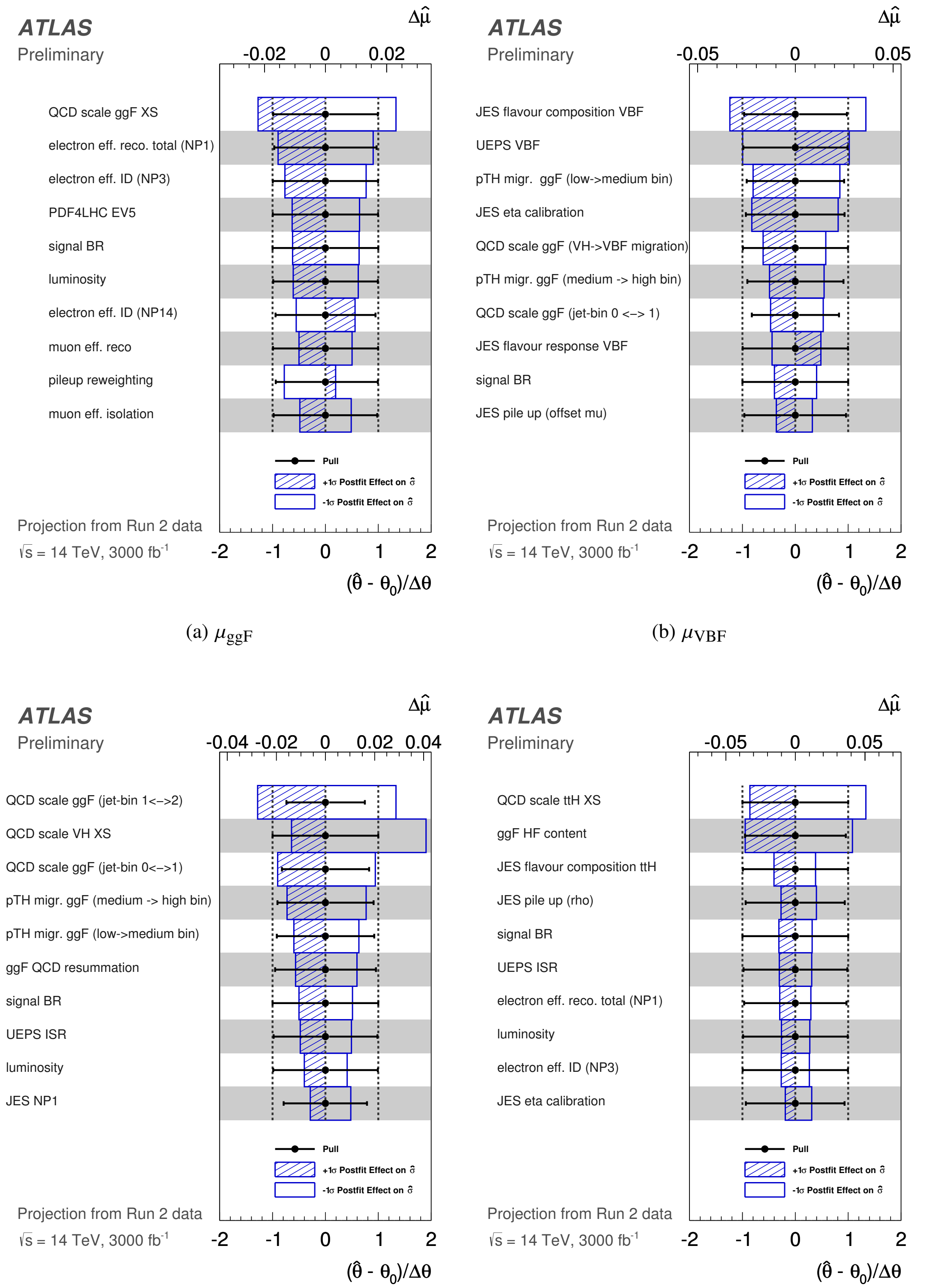

(a) $\mu_{\mathrm{ggF}}$ (b) $\mu_{\mathrm{VBF}}$ (c) $\mu_{\mathrm{VH}}$ (d) $\mu_{\mathrm{t\bar{t}H}}$

Figure 6: Ranking of the 10 systematic uncertainties (scenario S2) with the largest impact on the expected signal strength measurement of the $H \to ZZ^*$ decay channel for the ggF (a), VBF (b), $VH$ (c) and $t\bar{t}H$ (d) production modes.

## 2.3 $H \to WW^*$

Using a dataset corresponding to an integrated luminosity of 36.1 fb$^{-1}$ at Run 2, measurements of the cross sections times $H \to WW^*$ branching ratio have been reported in Ref. [9] for Higgs bosons produced via gluon-gluon fusion ggF and vector-boson fusion VBF and decaying via the $H \to WW^* \to e\nu\mu\nu$ channel. Signal regions for the measurement of gluon-gluon-fusion production are selected within events with less than 2 jets. A vector-boson fusion enriched category is defined using events with two or more jets. The normalisation of (non-resonant) $WW$, top ($t\bar{t}$ and $Wt$), and $Z \to \tau\tau$ backgrounds are set using dedicated control regions enriched with events from ggF and VBF processes, using similar selections as that used for the corresponding signal regions.

The expected results obtained for HL-LHC projections for scenarios S1(S2) are:

$$\sigma_{\text{ggF}} \times \text{B}(H \to WW^*) = 11.7^{+0.75}_{-0.76} \left(^{+0.54}_{-0.52}\right) \text{ pb}$$

$$= 11.7^{+0.12}_{-0.12} \text{ (stat) } ^{+0.43}_{-0.43} \left(^{+0.35}_{-0.34}\right) \text{ (exp) } ^{+0.39}_{-0.42} \left(^{+0.29}_{-0.29}\right) \text{ (sig) } ^{+0.47}_{-0.46} \left(^{+0.23}_{-0.27}\right) \text{ (bkg) pb}$$

$$\sigma_{\text{VBF}} \times \text{B}(H \to WW^*) = 0.916^{+0.099}_{-0.100} \left(^{+0.061}_{-0.060}\right) \text{ pb}$$

$$= 0.916^{+0.030}_{-0.030} \text{ (stat) } ^{+0.050}_{-0.044} \left(^{+0.027}_{-0.027}\right) \text{ (exp) } ^{+0.064}_{-0.061} \left(^{+0.035}_{-0.034}\right) \text{ (sig) } ^{+0.051}_{-0.059} \left(^{+0.029}_{-0.030}\right) \text{ (bkg) pb}$$

These results are compared to the SM prediction in Figure 7; the SM uncertainties are divided by two compared to their current values, which approximately corresponds to the scaling expected from scenario S2. The figure shows that, at HL-LHC, the ggF cross-section measurement will reach a level of precision which is similar to that of the theory predictions.

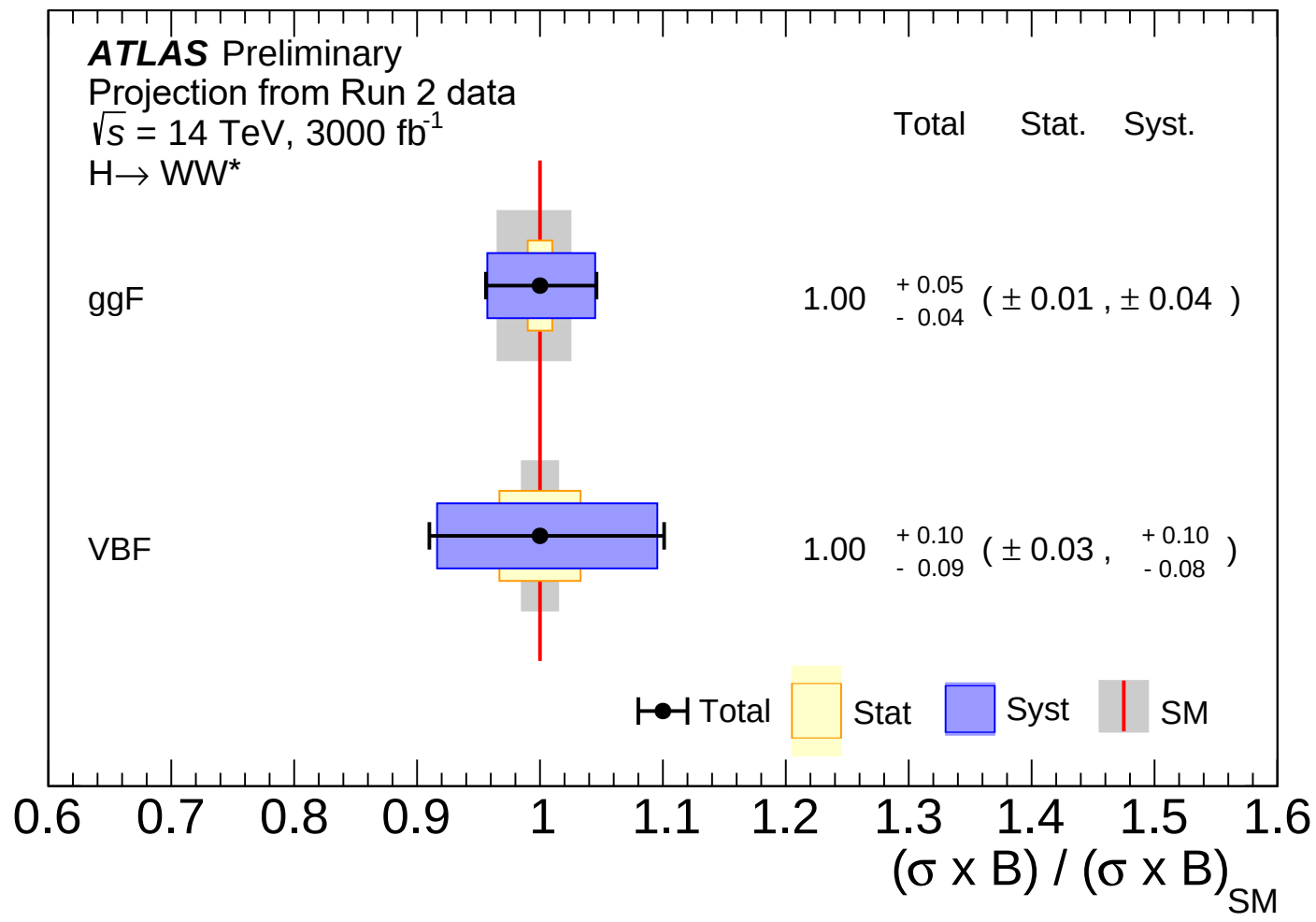

Figure 7: For each production mode, expected result for the measurement of the cross section times branching ratio normalised to their SM expectation for the $WW^*$ final state in scenario S2 at HL-LHC is shown. The statistical (yellow) and systematic (blue) uncertainty components are displayed as well as the theory uncertainty on the SM prediction (grey). The blue band of the systematic uncertainty includes both experimental and theory uncertainties.

Table 3 lists the total expected uncertainties on the cross section times branching ratio, as well as the contributions from statistical, experimental, signal and background theory uncertainties.

| Prod. mode | Scenario | $\Delta_{tot}/\sigma_{SM}$ | $\Delta_{stat}/\sigma_{SM}$ | $\Delta_{exp}/\sigma_{SM}$ | $\Delta_{sig}/\sigma_{SM}$ | $\Delta_{bkg}/\sigma_{SM}$ | $\Delta\mu_{sig}$ |
|---|---|---|---|---|---|---|---|
| ggF | Run 2, 36 fb$^{-1}$ | $^{+0.191}_{-0.189}$ | $^{+0.099}_{-0.098}$ | $^{+0.112}_{-0.110}$ | $^{+0.047}_{-0.036}$ | $^{+0.092}_{-0.096}$ | $^{+0.077}_{-0.058}$ |
| | HL-LHC S1 | $^{+0.064}_{-0.065}$ | $^{+0.010}_{-0.010}$ | $^{+0.037}_{-0.037}$ | $^{+0.040}_{-0.039}$ | $^{+0.033}_{-0.036}$ | $^{+0.068}_{-0.064}$ |
| | HL-LHC S2 | $^{+0.046}_{-0.044}$ | $^{+0.010}_{-0.010}$ | $^{+0.030}_{-0.029}$ | $^{+0.023}_{-0.020}$ | $^{+0.025}_{-0.025}$ | $^{+0.035}_{-0.033}$ |
| VBF | Run 2, 36 fb$^{-1}$ | $^{+0.391}_{-0.360}$ | $^{+0.332}_{-0.311}$ | $^{+0.122}_{-0.110}$ | $^{+0.115}_{-0.098}$ | $^{+0.106}_{-0.093}$ | $^{+0.119}_{-0.099}$ |
| | HL-LHC S1 | $^{+0.108}_{-0.109}$ | $^{+0.033}_{-0.033}$ | $^{+0.055}_{-0.048}$ | $^{+0.070}_{-0.067}$ | $^{+0.056}_{-0.064}$ | $^{+0.073}_{-0.070}$ |
| | HL-LHC S2 | $^{+0.067}_{-0.066}$ | $^{+0.033}_{-0.033}$ | $^{+0.029}_{-0.029}$ | $^{+0.038}_{-0.037}$ | $^{+0.032}_{-0.033}$ | $^{+0.039}_{-0.038}$ |

Table 3: Expected precision of the production-mode cross-section measurements in the $H \to WW^*$ channel with 36 fb$^{-1}$ of Run 2 data and at HL-LHC. Uncertainties are reported relative to the SM cross section at the corresponding center-of-mass energy. Both HL-LHC scenarios have been considered for the systematic uncertainties. The last column shows the theory uncertainty component when the measurement parameters are production-mode signal strengths instead of cross sections.

Figures 8 and 9 show the expected ranking of the leading sources of systematic uncertainties for the measured cross section times branching ratio and signal strength respectively, for the gluon-gluon fusion and vector-boson fusion processes.

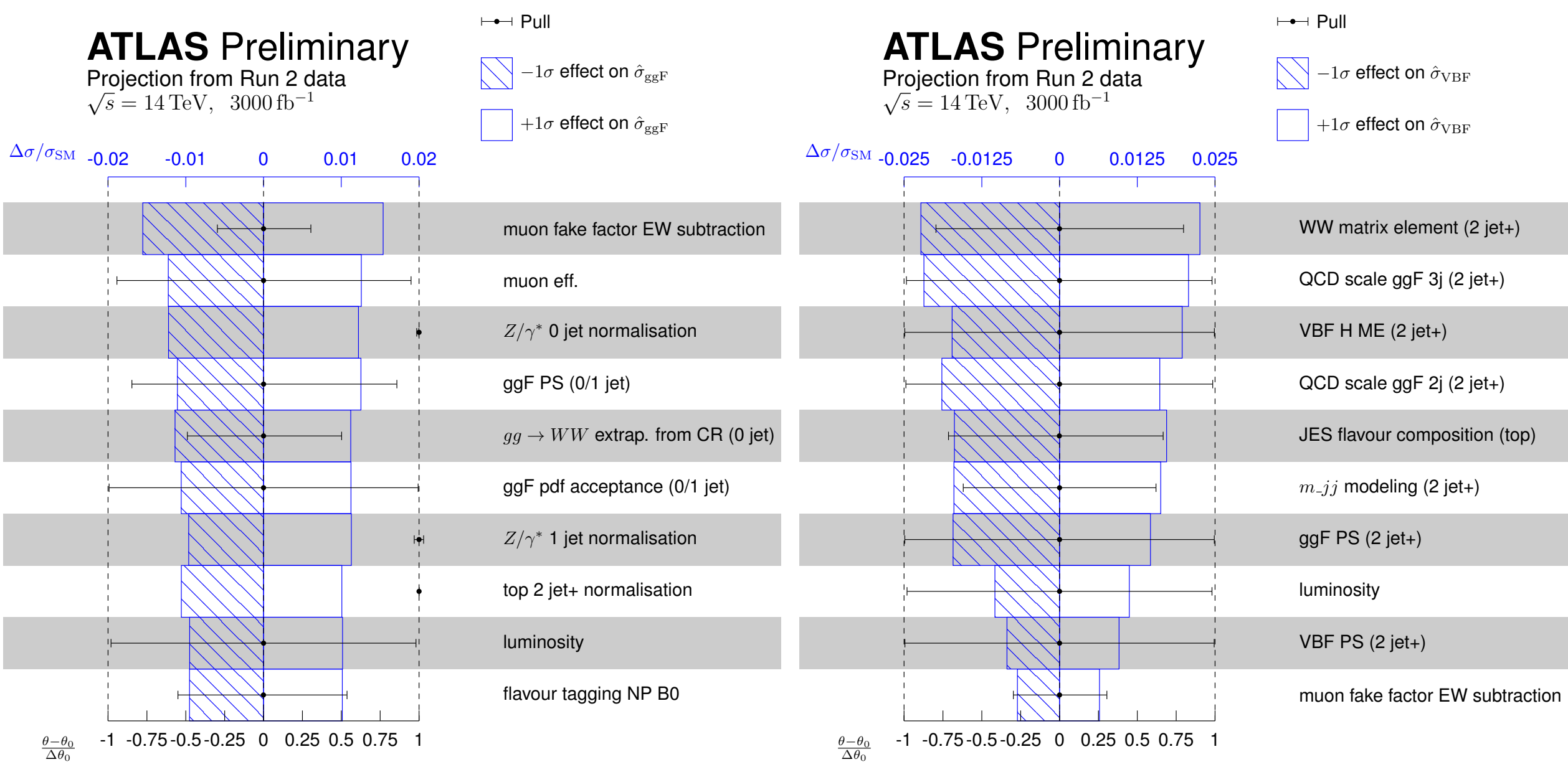

(a) $\sigma_{ggF}$ (b) $\sigma_{VBF}$

Figure 8: Ranking of the 10 systematic uncertainties (scenario S2) with the largest impact on the expected cross-section times branching ratio measurement of the $H \to WW^*$ decay channel for the ggF (a) and VBF (b) production modes. All +1$\sigma$ (-1$\sigma$ ) post-fit effects are displayed as having a positive (negative) values.

For the measured ggF cross section, the uncertainty related to the estimation of background contributions due to jets misidentified as muons, so called *fake muons*, and to the efficiency for actual muons are ranked higher in scenario S2, as these particular uncertainties are unchanged with respect to their Run 2 values. While the yield of $Z/\gamma^*$ events in the ggF enriched signal regions is small the corresponding $Z/\gamma^*$ enriched control regions contain a non-negligible number of events with fake leptons. The normalisation of the $Z/\gamma^*$ background can therefore indirectly affect the measured signal via fake lepton related nuisance parameters. Similarly the $t\bar{t} + Wt$ enriched control region with $\geq 2$ jets can provide additional constraints to the identification efficiency of jets originating from bottom quarks and thus on the $t\bar{t} + Wt$ event yield in the signal regions. Theory uncertainties on the parton shower modeling and PDFs also have a similar impact on the ggF cross-section measurement.

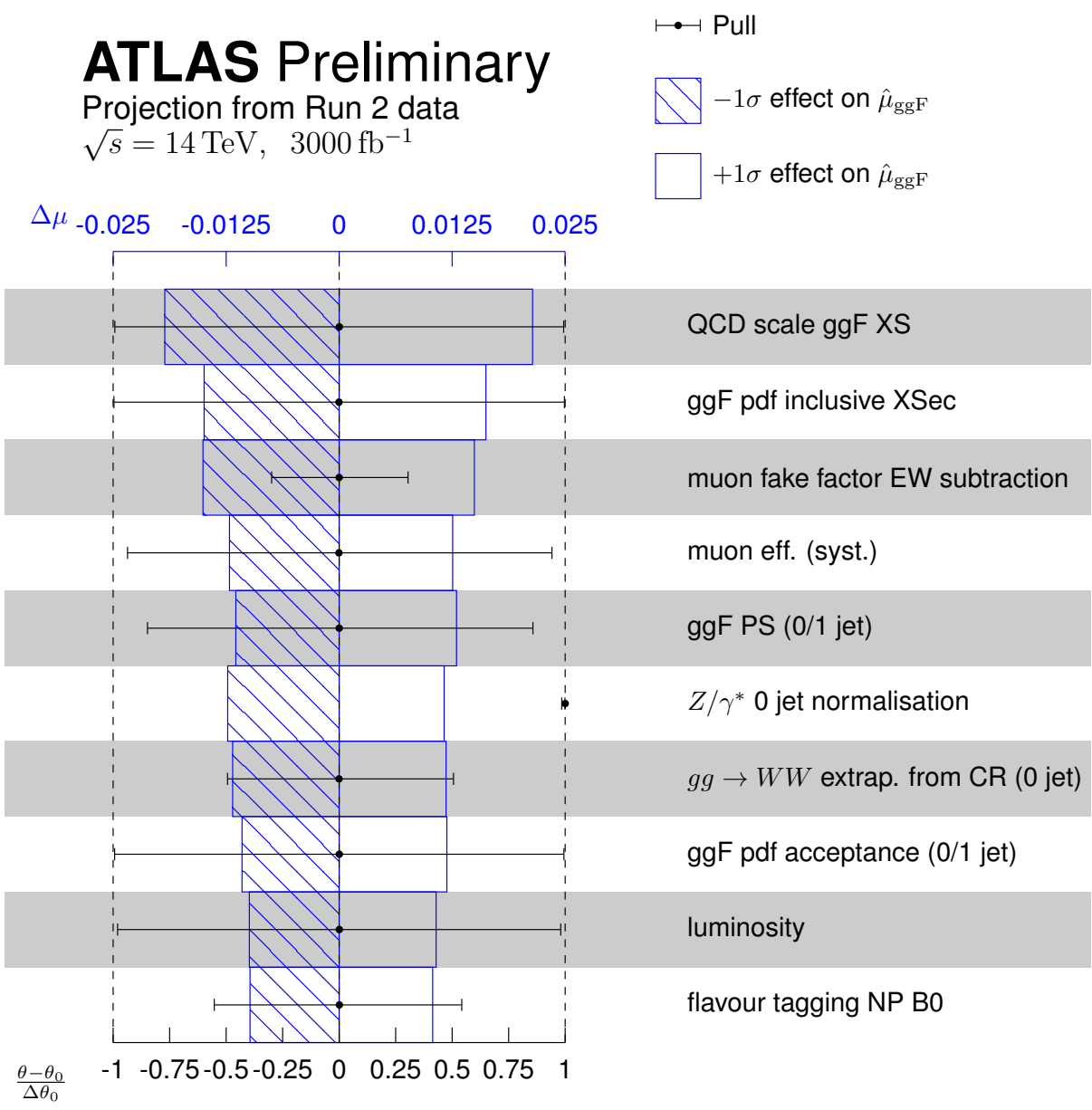

(a) $\mu_{\text{ggF}}$

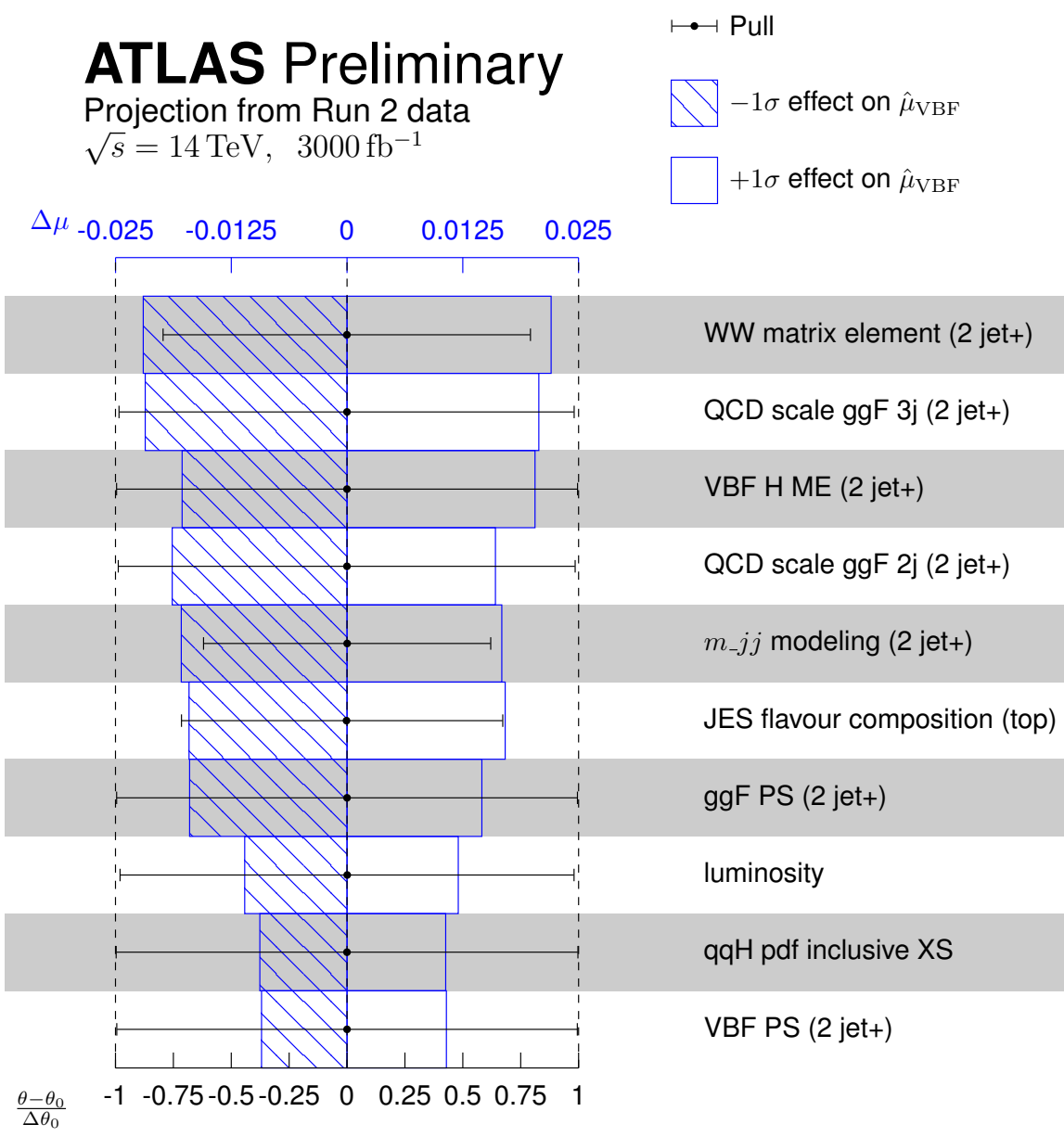

(b) $\mu_{\text{VBF}}$

Figure 9: Ranking of the 10 systematic uncertainties (scenario S2) with the largest impact on the expected signal strength measurement of the $H \to WW^*$ decay channel for the ggF (a) and VBF (b) production modes. All $+1\sigma$ ($-1\sigma$ ) post-fit effects are displayed as having a positive (negative) values.

The measurement in the vector-boson fusion channel shows similar rankings of systematic uncertainty sources for both the cross-section and signal strength measurements.

Besides the uncertainty due to the dependency of the jet energy scale on the jet flavour composition, the leading sources of uncertainties are mostly related to the acceptance of different signal and background processes in the VBF enriched phase space.

## 2.4 $H \to \tau\tau$

The measurement of the total cross section of the Higgs boson in the $H \to \tau\tau$ channel has been published in Ref. [10]. The analysis was performed on data collected in 2015 and 2016 corresponding to 36.1 fb$^{-1}$ at $\sqrt{s}$ = 13 TeV.
For the Higgs boson decay products all the leptonic ($\tau_{\mathrm{lep}}$) and hadronic ($\tau_{\mathrm{had}}$) decays of the $\tau$'s are considered. The analysis is done by splitting events into three categories depending on the possible $\tau\tau$ final states: ($\tau_{\mathrm{lep}}$, $\tau_{\mathrm{lep}}$), ($\tau_{\mathrm{lep}}$, $\tau_{\mathrm{had}}$) and ($\tau_{\mathrm{had}}$, $\tau_{\mathrm{had}}$).
The main backgrounds differ significantly between these three channels, as do the techniques used to determine them. For all final states, two categories targeting the VBF and ggF production modes are used. These two categories are further split in 13 exclusive signal regions, dependent on the decay channel.
The dominant uncertainties in Run 2 for the measured cross-section are theory uncertainties on the signal model, the uncertainty due to jets and $E_{\mathrm{T}}^{\mathrm{miss}}$ and the uncertainty due to background statistics, which include statistical uncertainties both of the simulated backgrounds and of the misidentified $\tau$ backgrounds, estimated using data.

The expected results for the cross sections times $H \to \tau\tau$ branching ratio of the two production modes at HL-LHC for the scenario S1 (S2) are:

$$\begin{aligned}
\sigma_{\mathrm{ggF}} \times \mathrm{B}(H \to \tau\tau) &= 3.43^{+0.79}_{-0.63} \left({}^{+0.42}_{-0.37}\right) \text{ pb} \\
&= 3.43^{+0.11}_{-0.11} \text{ (stat) } {}^{+0.21}_{-0.21} \left({}^{+0.14}_{-0.13}\right) \text{ (exp) } {}^{+0.70}_{-0.55} \left({}^{+0.36}_{-0.31}\right) \text{ (sig) } {}^{+0.27}_{-0.19} \left({}^{+0.11}_{-0.08}\right) \text{ (bkg) pb} \\
\sigma_{\mathrm{VBF}} \times \mathrm{B}(H \to \tau\tau) &= 0.268^{+0.025}_{-0.025} \left({}^{+0.021}_{-0.020}\right) \text{ pb} \\
&= 0.268^{+0.009}_{-0.009} \text{ (stat) } {}^{+0.014}_{-0.015} \left({}^{+0.013}_{-0.012}\right) \text{ (exp) } {}^{+0.017}_{-0.014} \left({}^{+0.007}_{-0.009}\right) \text{ (sig) } {}^{+0.009}_{-0.009} \left({}^{+0.010}_{-0.010}\right) \text{ (bkg) pb}
\end{aligned}$$

These results are compared to the SM prediction in Figure 10; the SM uncertainties are divided by two compared to their current values, which approximately corresponds to the scaling for the scenario S2.

Table 4 lists the total expected uncertainties on the cross section normalised to their SM values as well as the contributions from each uncertainty component. It's worthwhile to note that for this channel the breakdown of the systematics uncertainties for the scenarios S1 and S2 has been done differently from that used in the Run 2 analysis [10].

In Figure 11 the expected ranking of the systematic uncertainties for the scenario S2 with the largest impact on the measured inclusive ggF and VBF cross sections times $H \to \tau\tau$ branching ratio and signal strength is shown. Both measurements are largely affected by the QCD uncertainties related to the signal acceptance. The largest contribution to the total uncertainty comes from the theory signal uncertainty.

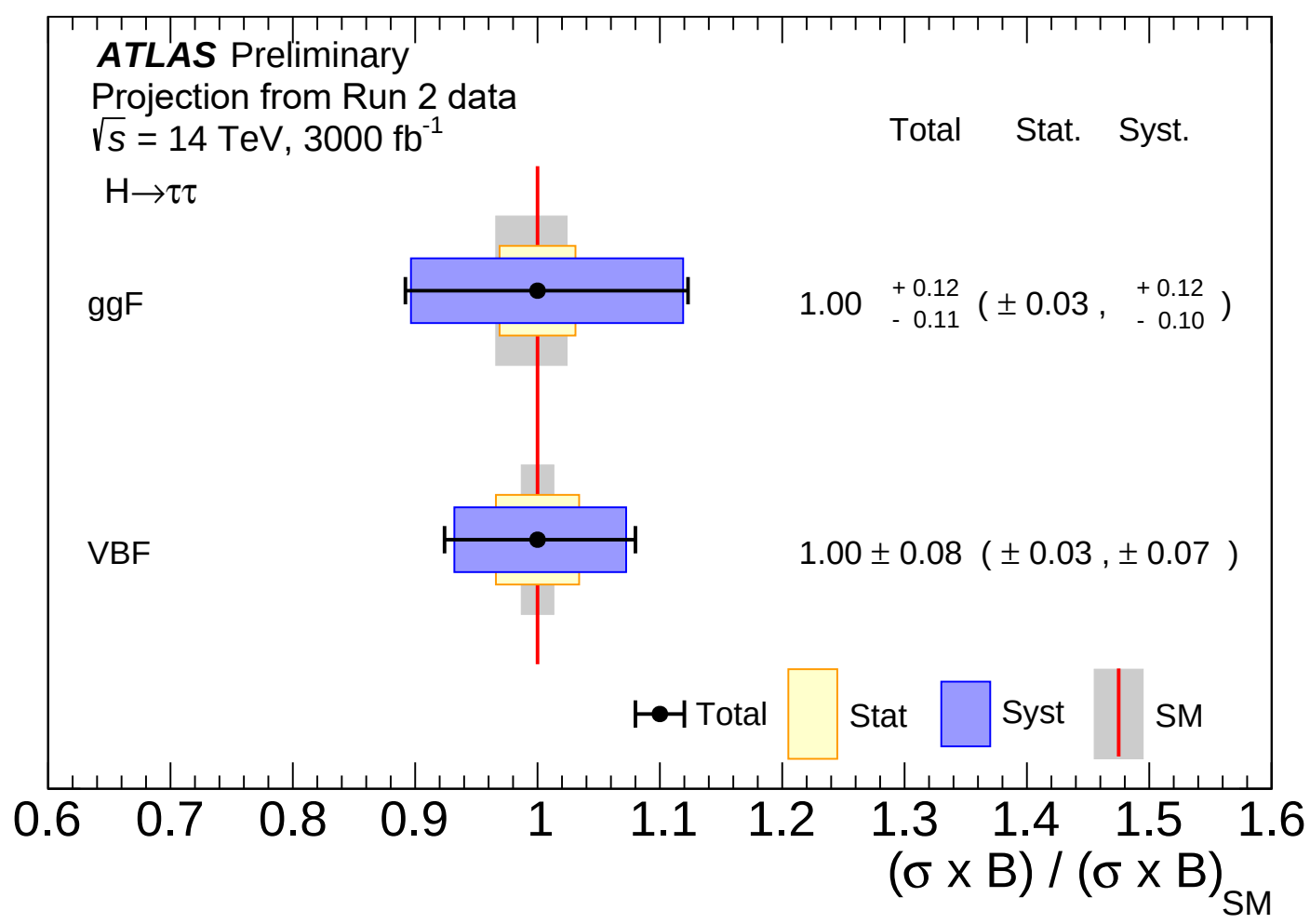

Figure 10: For each production mode, expected result for the measurement of the cross section times branching ratio normalised to their SM expectation for the $\tau\tau$ final state in scenario S2 at HL-LHC is shown. The statistical (yellow) and systematic (blue) uncertainty components are displayed as well as the theoretical uncertainty on the SM prediction (grey). The blue band of the systematic uncertainty includes both experimental and theory uncertainties.

| Prod. mode | Scenario | $\Delta_{\text{tot}}/\sigma_{\text{SM}}$ | $\Delta_{\text{stat}}/\sigma_{\text{SM}}$ | $\Delta_{\text{exp}}/\sigma_{\text{SM}}$ | $\Delta_{\text{sig}}/\sigma_{\text{SM}}$ | $\Delta_{\text{bkg}}/\sigma_{\text{SM}}$ | $\Delta\mu_{\text{sig}}$ |
|---|---|---|---|---|---|---|---|
| ggF | Run 2, 36 fb$^{-1}$ | $^{+0.629}_{-0.526}$ | $^{+0.337}_{-0.333}$ | $^{+0.365}_{-0.420}$ | $^{+0.364}_{-0.150}$ | $^{+0.139}_{-0.133}$ | $^{+0.360}_{-0.149}$ |
| | HL-LHC S1 | $^{+0.231}_{-0.185}$ | $^{+0.031}_{-0.031}$ | $^{+0.060}_{-0.062}$ | $^{+0.203}_{-0.160}$ | $^{+0.080}_{-0.055}$ | $^{+0.236}_{-0.185}$ |
| | HL-LHC S2 | $^{+0.123}_{-0.108}$ | $^{+0.031}_{-0.031}$ | $^{+0.041}_{-0.039}$ | $^{+0.104}_{-0.090}$ | $^{+0.031}_{-0.024}$ | $^{+0.123}_{-0.105}$ |
| VBF | Run 2, 36 fb$^{-1}$ | $^{+0.591}_{-0.538}$ | $^{+0.390}_{-0.373}$ | $^{+0.380}_{-0.389}$ | $^{+0.149}_{-0.078}$ | $^{+0.139}_{-0.110}$ | $^{+0.180}_{-0.091}$ |
| | HL-LHC S1 | $^{+0.093}_{-0.093}$ | $^{+0.034}_{-0.034}$ | $^{+0.052}_{-0.056}$ | $^{+0.063}_{-0.053}$ | $^{+0.034}_{-0.034}$ | $^{+0.081}_{-0.075}$ |
| | HL-LHC S2 | $^{+0.080}_{-0.076}$ | $^{+0.034}_{-0.034}$ | $^{+0.049}_{-0.045}$ | $^{+0.027}_{-0.033}$ | $^{+0.038}_{-0.038}$ | $^{+0.042}_{-0.042}$ |

Table 4: Expected precision of the production mode cross-section measurements in the $H \to \tau\tau$ channel with 36 fb$^{-1}$ of Run 2 data and at HL-LHC. Uncertainties are reported relative to the SM cross section at the corresponding center-of-mass energy. Both scenarios S1 and S2 have been considered for the systematic uncertainties in the HL-LHC extrapolation. The last column shows the theory uncertainty component when the measurement parameters are production mode signal strengths instead of cross sections.

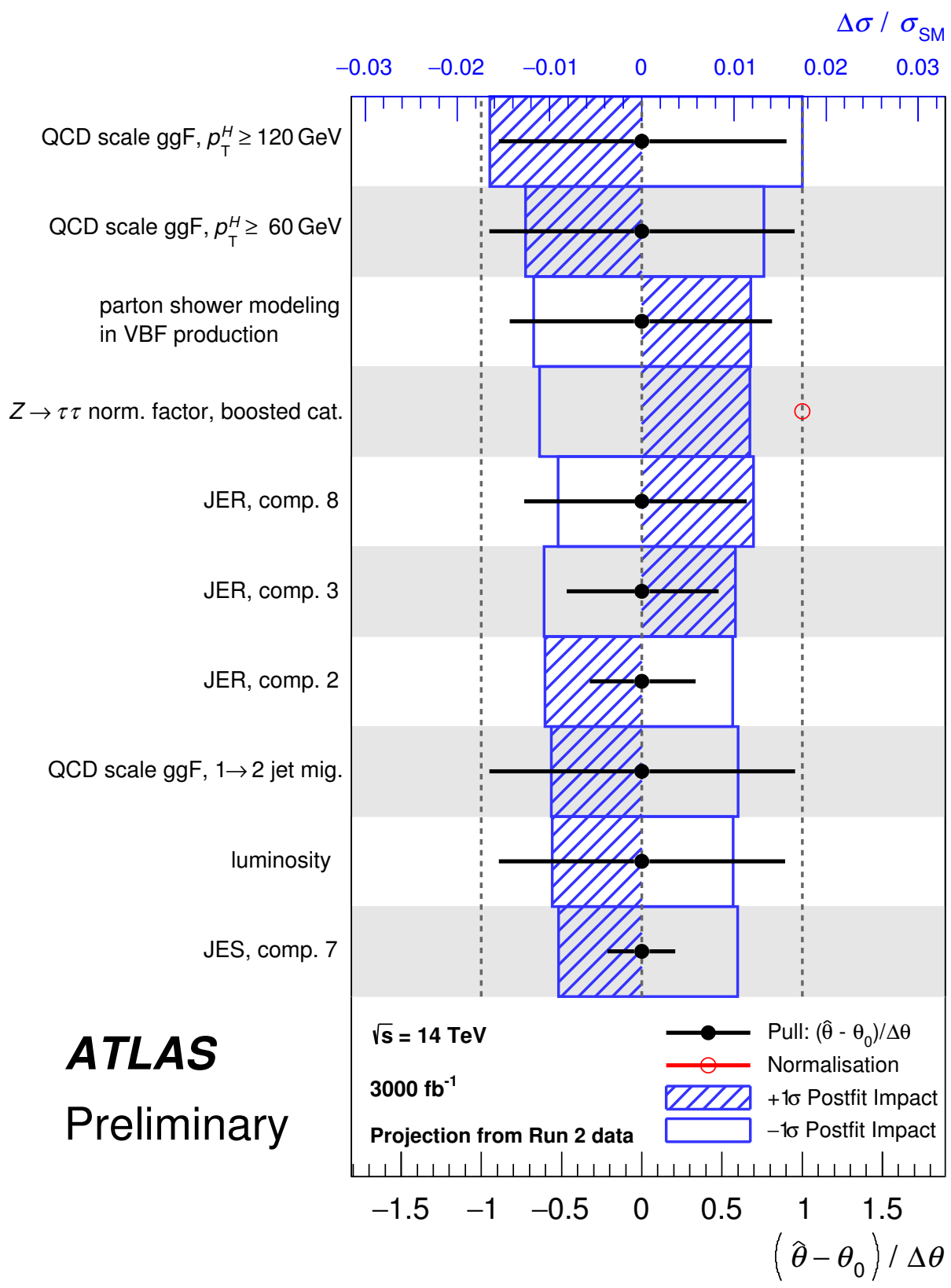

(a) $\sigma_{\text{ggF+VBF}}$

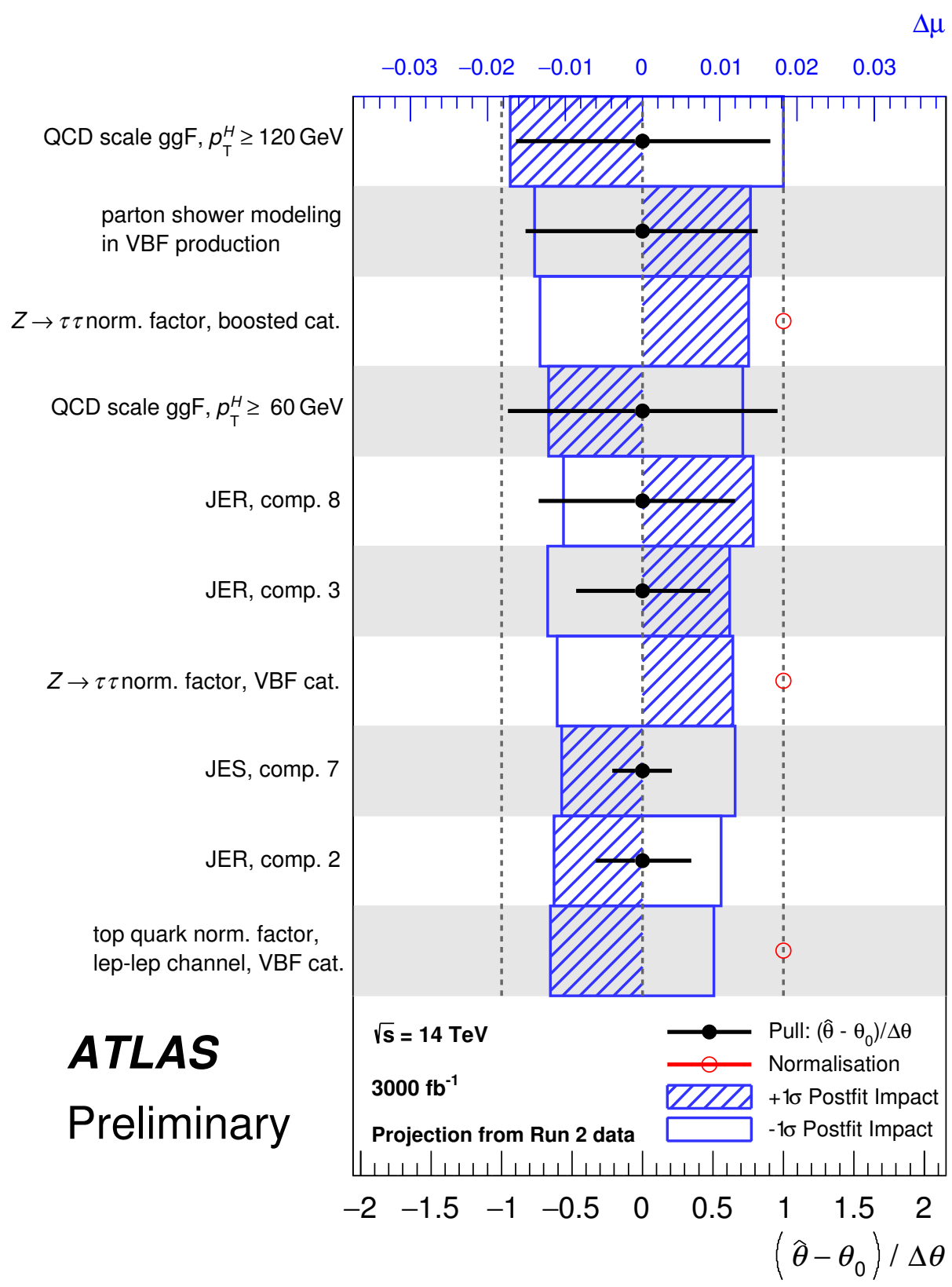

(b) $\mu_{\text{ggF+VBF}}$

Figure 11: Ranking of the 10 systematic uncertainties with the largest impact on the inclusive ggF and VBF cross sections times $H \to \tau\tau$ branching ratio (a) and signal strength (b) extrapolated at HL-LHC with the scenario S2.

## 2.5 $VH, H \to b\bar{b}$

The HL-LHC projections for the $VH$, $H \to b\bar{b}$ channel are performed using extrapolations based on the results of the analysis of 79.8 fb$^{-1}$ of pp collision data collected at $\sqrt{s}$ = 13 TeV [11]. The same statistical framework and analysis approach are used. In particular the same selection and event categories, for both signal and control regions are utilised.

The three cross sections for $WH$, $q\bar{q} \to ZH$ and $gg \to ZH$ production times the $H \to b\bar{b}$ branching ratio are determined from a fit where the three parameters are left free.

The measured product of the cross-section times the $H \to b\bar{b}$ branching ratio for each signal process, are the following, where the uncertainties outside (inside) the parentheses correspond to scenario S1 (S2):

$$\begin{aligned}\sigma(WH) \times \mathrm{B}(H \to b\bar{b}) &= 0.877\,^{+0.131}_{-0.121}\left(^{+0.091}_{-0.088}\right)\,\mathrm{pb}\\ &= 0.877\,^{+0.036}_{-0.036}\left(^{+0.036}_{-0.036}\right)\,(\mathrm{stat})\,^{+0.042}_{-0.041}\left(^{+0.039}_{-0.038}\right)\,(\mathrm{exp})\\ &\quad\ ^{+0.070}_{-0.061}\left(^{+0.040}_{-0.036}\right)\,(\mathrm{sig})\,^{+0.095}_{-0.088}\left(^{+0.063}_{-0.061}\right)\,(\mathrm{bkg})\,\mathrm{pb}\end{aligned}$$

$$\begin{aligned}\sigma(q\bar{q} \to ZH) \times \mathrm{B}(H \to b\bar{b}) &= 0.488\,^{+0.067}_{-0.064}\left(^{+0.059}_{-0.058}\right)\,\mathrm{pb}\\ &= 0.488\,^{+0.044}_{-0.043}\left(^{+0.044}_{-0.043}\right)\,(\mathrm{stat})\,^{+0.032}_{-0.031}\left(^{+0.028}_{-0.027}\right)\,(\mathrm{exp})\\ &\quad\ ^{+0.030}_{-0.027}\left(^{+0.015}_{-0.014}\right)\,(\mathrm{sig})\,^{+0.026}_{-0.023}\left(^{+0.023}_{-0.022}\right)\,(\mathrm{bkg})\,\mathrm{pb}\end{aligned}$$

$$\begin{aligned}\sigma(gg \to ZH) \times \mathrm{B}(H \to b\bar{b}) &= 0.084\,^{+0.042}_{-0.041}\left(^{+0.036}_{-0.036}\right)\,\mathrm{pb}\\ &= 0.084\,^{+0.028}_{-0.028}\left(^{+0.028}_{-0.028}\right)\,(\mathrm{stat})\,^{+0.021}_{-0.021}\left(^{+0.017}_{-0.017}\right)\,(\mathrm{exp})\\ &\quad\ ^{+0.015}_{-0.012}\left(^{+0.008}_{-0.007}\right)\,(\mathrm{sig})\,^{+0.017}_{-0.018}\left(^{+0.015}_{-0.015}\right)\,(\mathrm{bkg})\,\mathrm{pb}\end{aligned}$$

These numbers are translated in relative precision in Table 5. It's worthwhile to note that for this channel the breakdown of the systematics uncertainties for the scenarios S1 and S2 has been done differently from that used in the Run 2 analysis [11].

| Prod. mode | Scenario | $\Delta_{\text{tot}}/\sigma_{\text{SM}}$ | $\Delta_{\text{stat}}/\sigma_{\text{SM}}$ | $\Delta_{\text{exp}}/\sigma_{\text{SM}}$ | $\Delta_{\text{sig}}/\sigma_{\text{SM}}$ | $\Delta_{\text{bkg}}/\sigma_{\text{SM}}$ | $\Delta\mu_{\text{sig}}$ |
|---|---|---|---|---|---|---|---|
| $WH$ | Run 2, 80 fb$^{-1}$ | $^{+0.462}_{-0.425}$ | $^{+0.272}_{-0.265}$ | $^{+0.157}_{-0.127}$ | $^{+0.176}_{-0.075}$ | $^{+0.224}_{-0.213}$ | $^{+0.180}_{-0.077}$ |
| | HL-LHC S1 | $^{+0.149}_{-0.138}$ | $^{+0.041}_{-0.041}$ | $^{+0.048}_{-0.047}$ | $^{+0.080}_{-0.070}$ | $^{+0.108}_{-0.100}$ | $^{+0.085}_{-0.074}$ |
| | HL-LHC S2 | $^{+0.104}_{-0.100}$ | $^{+0.041}_{-0.041}$ | $^{+0.044}_{-0.043}$ | $^{+0.046}_{-0.041}$ | $^{+0.072}_{-0.069}$ | $^{+0.050}_{-0.045}$ |
| $q\bar{q} \to ZH$ | Run 2, 80 fb$^{-1}$ | $^{+0.667}_{-0.629}$ | $^{+0.578}_{-0.562}$ | $^{+0.129}_{-0.101}$ | $^{+0.175}_{-0.105}$ | $^{+0.143}_{-0.126}$ | $^{+0.180}_{-0.105}$ |
| | HL-LHC S1 | $^{+0.138}_{-0.132}$ | $^{+0.090}_{-0.089}$ | $^{+0.065}_{-0.063}$ | $^{+0.061}_{-0.055}$ | $^{+0.054}_{-0.048}$ | $^{+0.067}_{-0.059}$ |
| | HL-LHC S2 | $^{+0.121}_{-0.118}$ | $^{+0.090}_{-0.089}$ | $^{+0.057}_{-0.055}$ | $^{+0.031}_{-0.028}$ | $^{+0.048}_{-0.046}$ | $^{+0.037}_{-0.033}$ |
| $gg \to ZH$ | Run 2, 80 fb$^{-1}$ | $^{+2.629}_{-2.608}$ | $^{+2.105}_{-2.105}$ | $^{+0.606}_{-0.677}$ | $^{+0.658}_{-0.454}$ | $^{+1.012}_{-1.037}$ | $^{+1.269}_{-0.645}$ |
| | HL-LHC S1 | $^{+0.498}_{-0.490}$ | $^{+0.333}_{-0.333}$ | $^{+0.249}_{-0.250}$ | $^{+0.181}_{-0.140}$ | $^{+0.207}_{-0.218}$ | $^{+0.495}_{-0.209}$ |
| | HL-LHC S2 | $^{+0.432}_{-0.433}$ | $^{+0.333}_{-0.333}$ | $^{+0.208}_{-0.204}$ | $^{+0.096}_{-0.080}$ | $^{+0.177}_{-0.181}$ | $^{+0.222}_{-0.115}$ |

Table 5: Expected precision of the production-mode cross-section measurements in the $WH$, $q\bar{q} \to ZH$ and $gg \to ZH$ production modes for the $H \to b\bar{b}$ decay channel with 80 fb$^{-1}$ of Run 2 data and at HL-LHC. Uncertainties are reported relative to the SM cross section at the corresponding center-of-mass energy. Both HL-LHC scenarios have been considered for the systematic uncertainties. The last column shows the theory uncertainty component when the measurement parameters are production mode signal strengths instead of cross sections.

Figures 12 and 13 show the ranking, for each production mode, of the systematic uncertainties with the largest impact on the cross section times branching ratio and signal strength in the scenario S2, respectively.

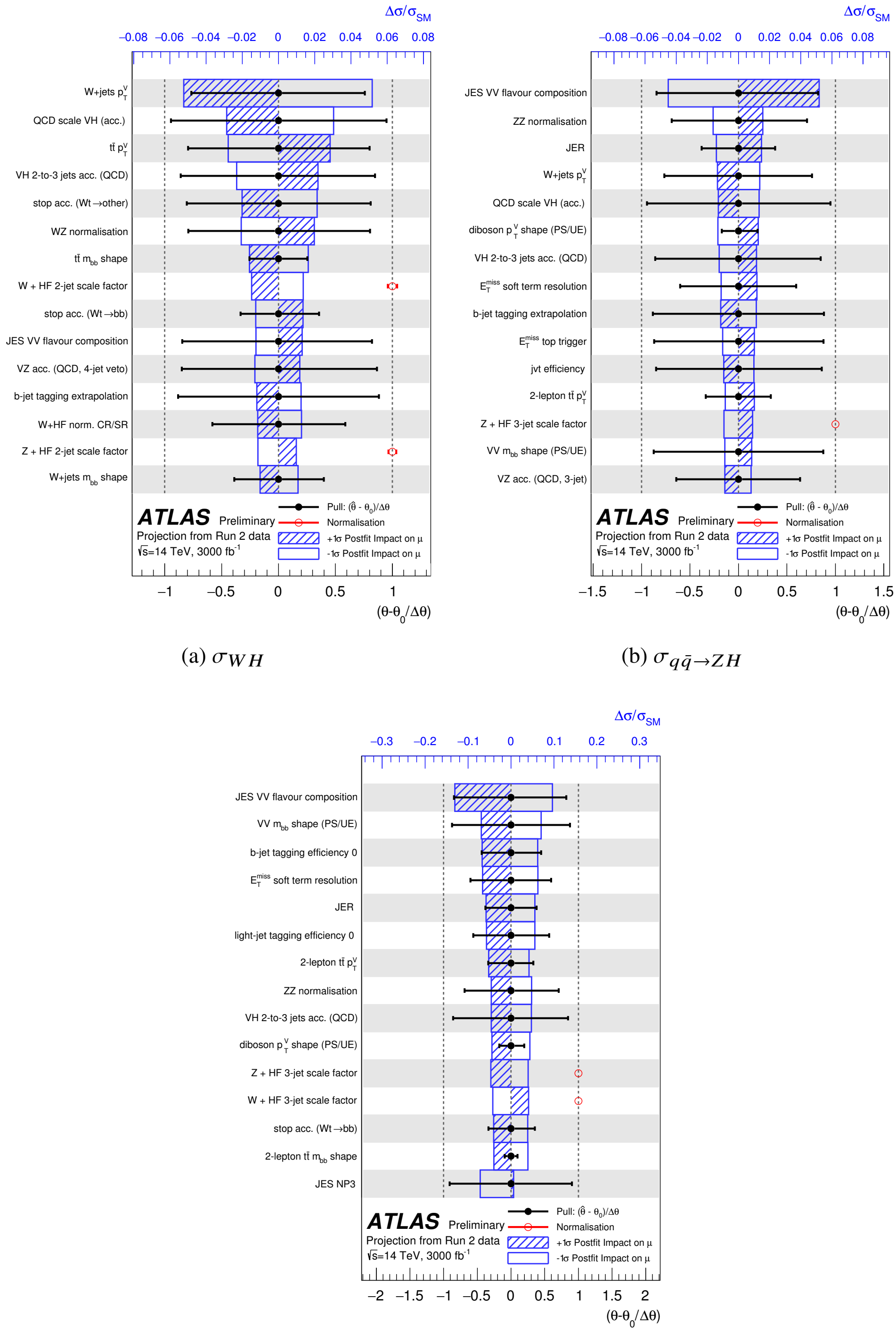

(a) $\sigma_{WH}$

(b) $\sigma_{q\bar{q}\to ZH}$

(c) $\sigma_{gg\to ZH}$

Figure 12: Ranking of the 15 systematic uncertainties (scenario S2) with the largest impact on the expected cross section times branching ratio measurement of the $H \to b\bar{b}$ decay channel for the $WH$ (a), $q\bar{q} \to ZH$ (b), $gg \to ZH$ (c) production modes.

Figure 14 summarises the expected precision of the measured cross sections for the three production modes. Figure 15 shows the expected precision of the measured cross sections when the $gg$ and $q\bar{q}$ to $ZH$ production modes are combined. It's worthwhile to note that in this latter fit, the uncertainty on the inclusive $ZH$ signal process is much smaller than the uncertainties on the single $q\bar{q} \to ZH$ and $gg \to ZH$ processes, due to correlations between their measurements.

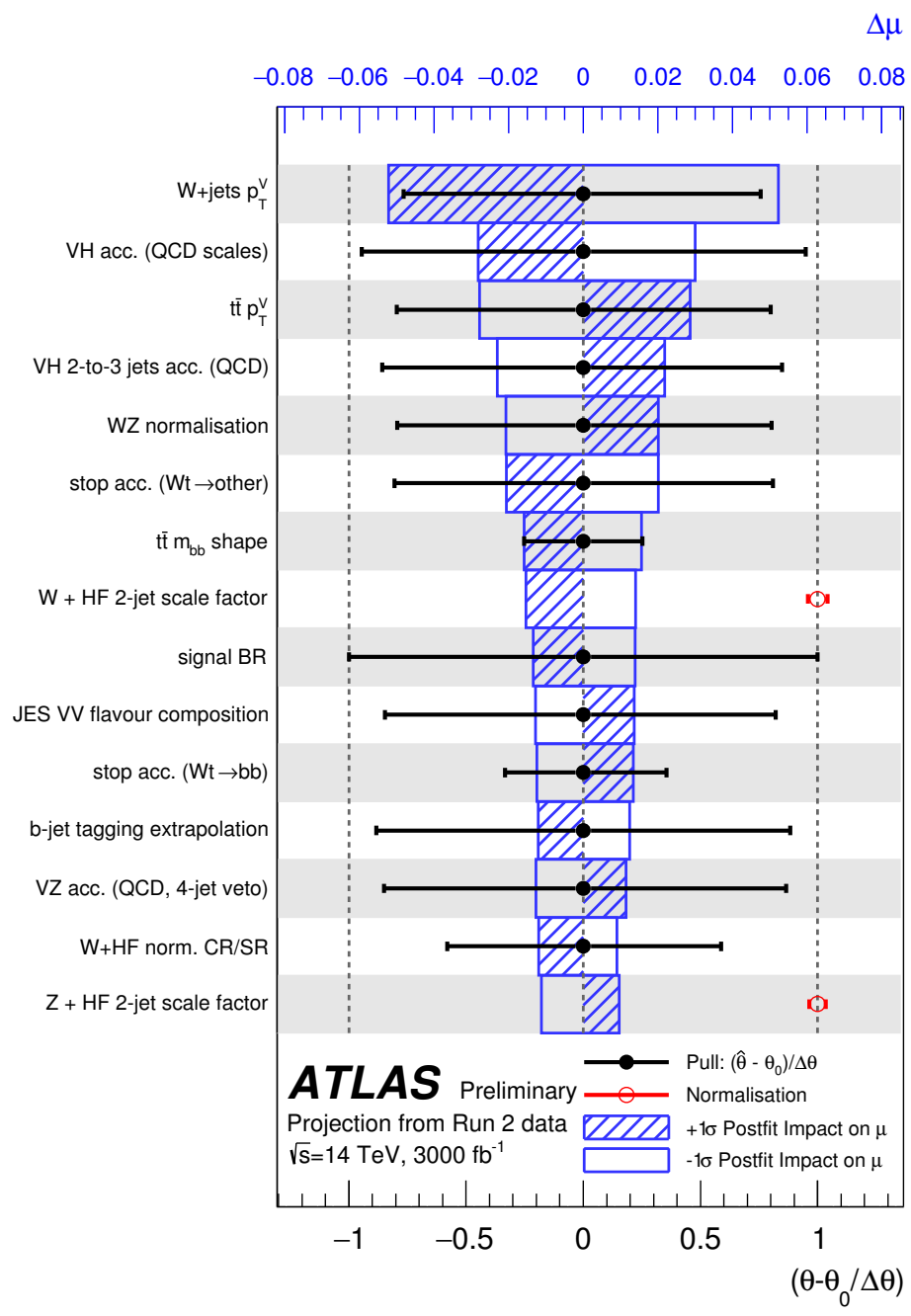

(a) $\mu_{WH}$

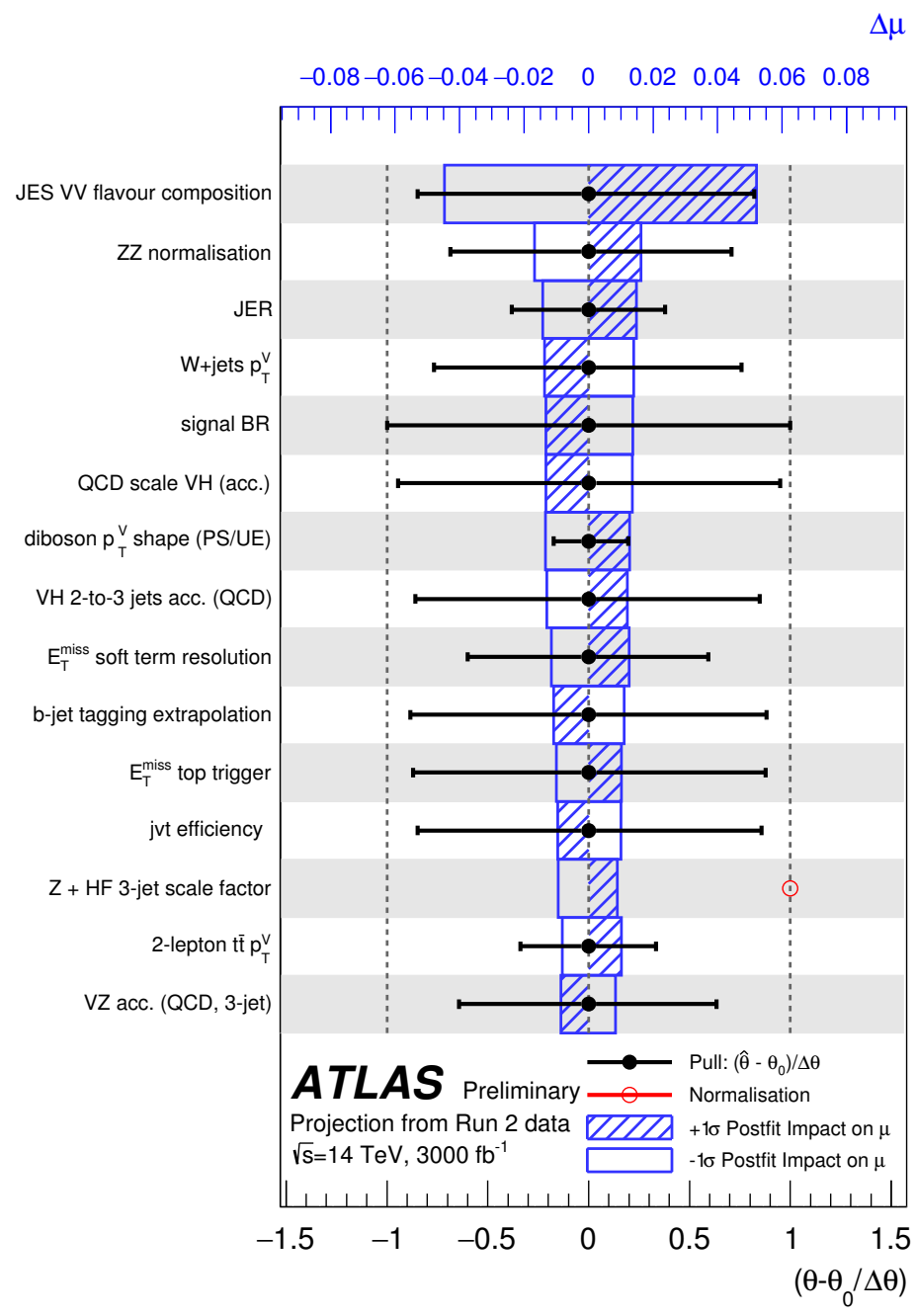

(b) $\mu_{q\bar{q}\to ZH}$

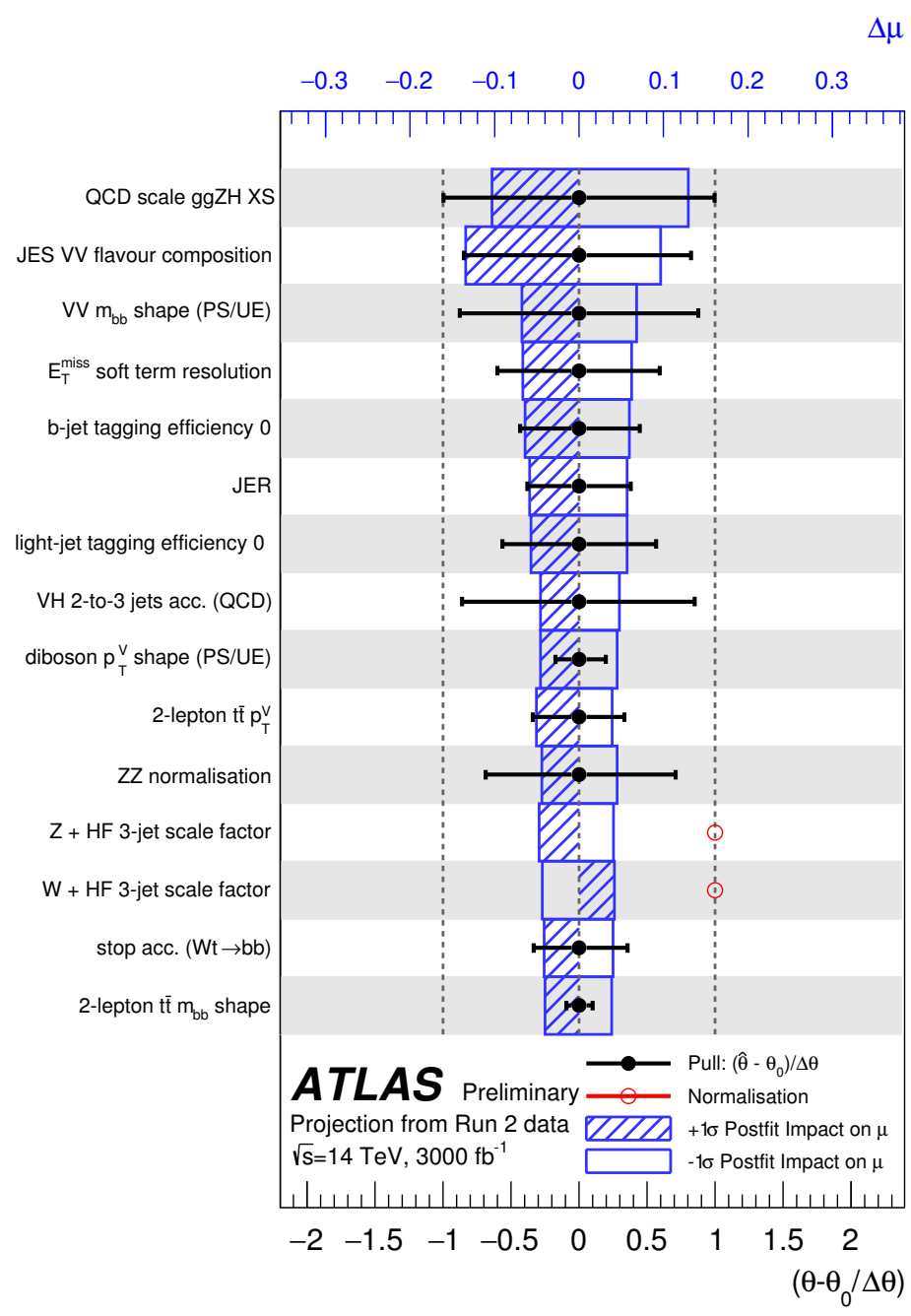

(c) $\mu_{gg\to ZH}$

Figure 13: Ranking of the 15 systematic uncertainties (scenario S2) with the largest impact on the expected signal strength measurement of the $H \to b\bar{b}$ decay channel for the $WH$ (a), $q\bar{q} \to ZH$ (b), $gg \to ZH$ (c) production modes.

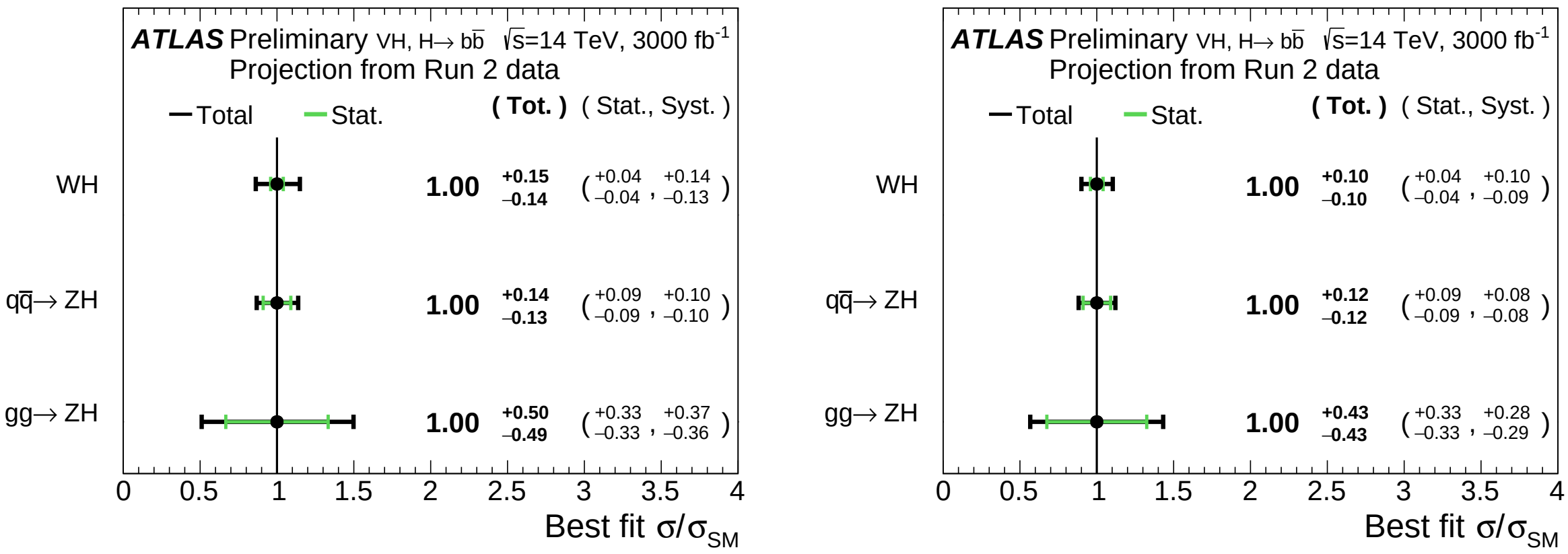

(a) Scenario S1 (b) Scenario S2

Figure 14: The fitted values of the Higgs boson cross section divided by their SM values for the $WH$, $q\bar{q} \to ZH$ and $gg \to ZH$ processes expected with 3000 fb$^{-1}$ at the HL-LHC in the (a) scenario S1 and (b) S2 extrapolations. The individual cross section values for the three processes are obtained from a simultaneous fit in which the cross section parameters for the $WH$, $q\bar{q} \to ZH$ and $gg \to ZH$ processes are floating independently.

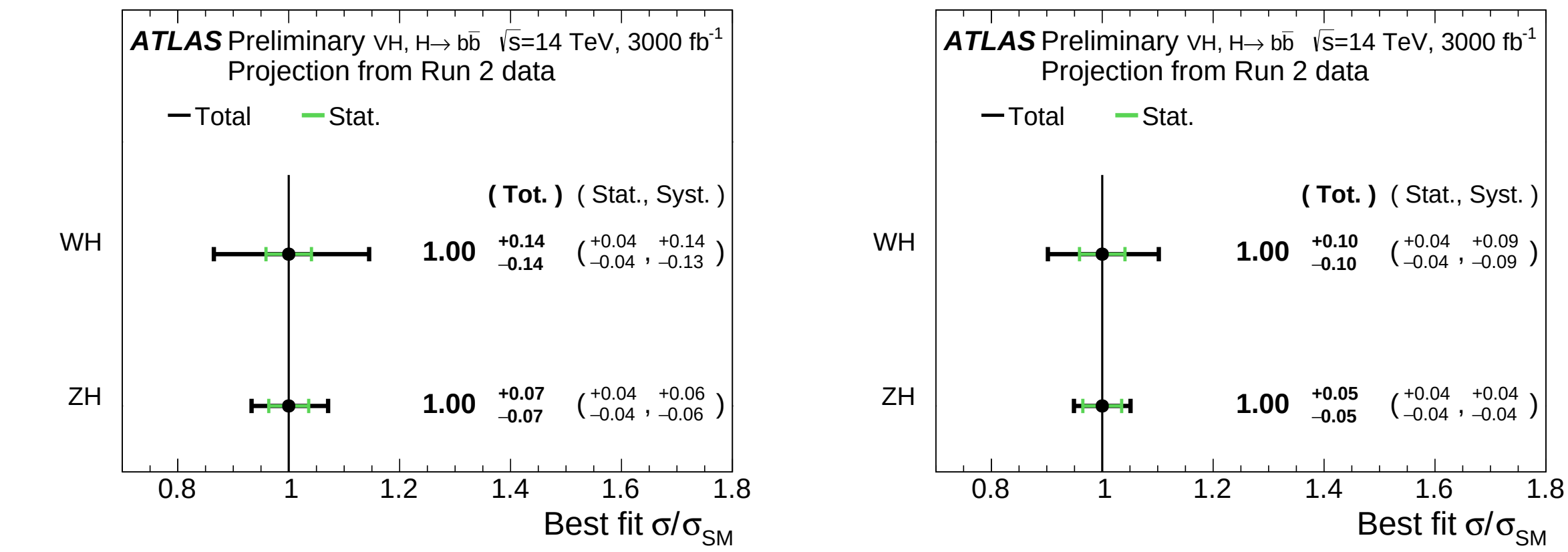

(a) Scenario S1 (b) Scenario S2

Figure 15: The fitted values of the Higgs boson cross section divided by their SM values for the $WH$ and $ZH$ processes expected with 3000 fb$^{-1}$ at the HL-LHC in the (a) scenario S1 and (b) S2 extrapolations. The individual cross section values for the two processes are obtained from a simultaneous fit in which the cross section parameters for the $WH$ and $ZH$ processes are floating independently.

## 2.6 $t\bar{t}H$

The ATLAS Collaboration has searched for the $t\bar{t}H$ production with LHC Run 2 data collected in 2015, 2016, and 2017, and observed Higgs boson production in association with a top-quark pair [12]. This analysis is sensitive to a large variety of final-state event topologies, $H \to WW^*$, $H \to ZZ^*$, $H \to \tau^+\tau^-$, $H \to b\bar{b}$ and $H \to \gamma\gamma$. They are all considered in what follows except the $H \to \gamma\gamma$ and $H \to ZZ^*$ final states which were covered in the Section 2.1 and 2.2, respectively.
The projection studies performed in this section are based on the extrapolation to HL-LHC of the published Run 2 results [13, 14] using the 2015-2016 dataset (36 fb$^{-1}$) for the $t\bar{t}H, H \to b\bar{b}$ and the $t\bar{t}H, H \to$ ML (multi-lepton final state) channels. This latter channel includes the $H \to WW^*$, $H \to ZZ^*$, $H \to \tau^+\tau^-$ final states, which are categorised according to the number of hadronically decaying $\tau$ leptons and the number of electrons or muons candidates in the event.

For the $t\bar{t}H, H \to$ML channel, the rankings of the top 10 nuisance parameters when fitting the cross section and the signal strength as a parameter of interest are reported in Figures 16 and 17 respectively, separating the selections which include (a) or exclude (b) hadronically decaying $\tau$ leptons.
In the $t\bar{t}H, H \to$ML including $\tau$ channel the dominant uncertainties are related to the treatment of the reconstruction of the $\tau$ leptons in the final state. In the $t\bar{t}H, H \to$ML excluding $\tau$ channel the modeling for the $t\bar{t}Z$ background is the largest systematic uncertainty. Leading uncertainties for both ML channels relate to the treatment of misidentified leptons in the control and signal regions. These uncertainties are not reduced in scenario S2 and some of them are constrained with 3000 fb$^{-1}$, as shown in Figures 16 and 17. It's worthwhile to note that those uncertainties are dominated by the statistics from control regions and therefore they are expected to decrease significantly at HL-LHC.
For the extrapolation of the $t\bar{t}H, H \to b\bar{b}$ channel, the large dataset of 3000 fb$^{-1}$ causes significant constraints on the background theoretical uncertainties, mostly those related to the description of the $t\bar{t}$+heavy flavour background component, as shown in Figures 16 and 17. This component is estimated in the Run 2 analysis as a large two-point systematics which encompasses the differences between several Monte Carlo simulations with respect to the nominal Powheg Pythia8 simulation and dominates the total uncertainty. The constraints on those uncertainties at 3000 fb$^{-1}$ induce a large reduction of the post-fit impact for the background theoretical uncertainties, which lead to a non realistic extrapolation. Therefore, the extrapolation of the $t\bar{t}H, H \to b\bar{b}$ channel has been performed injecting an additional uncertainty to get a reduction of a factor two (three) on the post-fit impact of the background theoretical uncertainties, respectively for scenario S1(S2). Such factors are consistent with the expected theory improvement on the uncertainty on the $t\bar{t}$+heavy flavour channel, as reported in Ref. [15]. The other uncertainties are treated as for the other channels in the note. The next largest experimental uncertainties in scenario S2 are related to flavour tagging and jet reconstruction.

Table 6 reports the value of the uncertainty on the cross section, normalised to the SM value, split in total uncertainty, statistical, experimental, and theory components, for both the $t\bar{t}H, H \to b\bar{b}$ and the $t\bar{t}H, H \to$ ML channels. The last column reports the theory signal uncertainty on the signal strength, which includes also the theoretical error on the predicted SM cross-section.

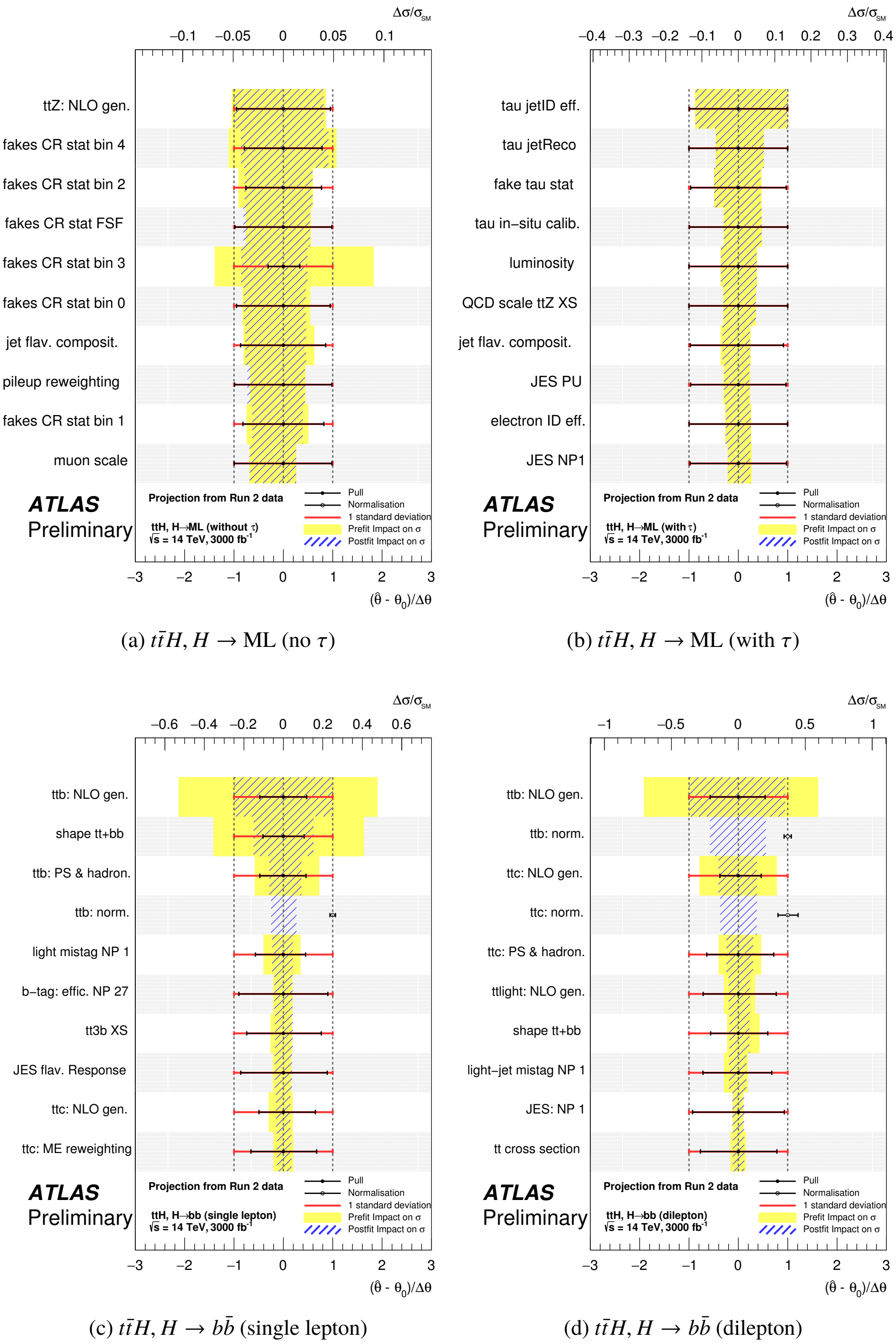

(a) $t\bar{t}H, H \to$ ML (no $\tau$)

(b) $t\bar{t}H, H \to$ ML (with $\tau$)

(c) $t\bar{t}H, H \to b\bar{b}$ (single lepton)

(d) $t\bar{t}H, H \to b\bar{b}$ (dilepton)

Figure 16: Ranking of the 10 systematic uncertainties with the largest impact on the $t\bar{t}H$ cross section at HL-LHC for scenario S2 for the $t\bar{t}H, H \to$ ML (no $\tau$) (a), $t\bar{t}H, H \to$ ML (with $\tau$) (b), $t\bar{t}H, H \to b\bar{b}$ (single lepton) (c) and $t\bar{t}H, H \to b\bar{b}$ (dilepton) (d) processes.

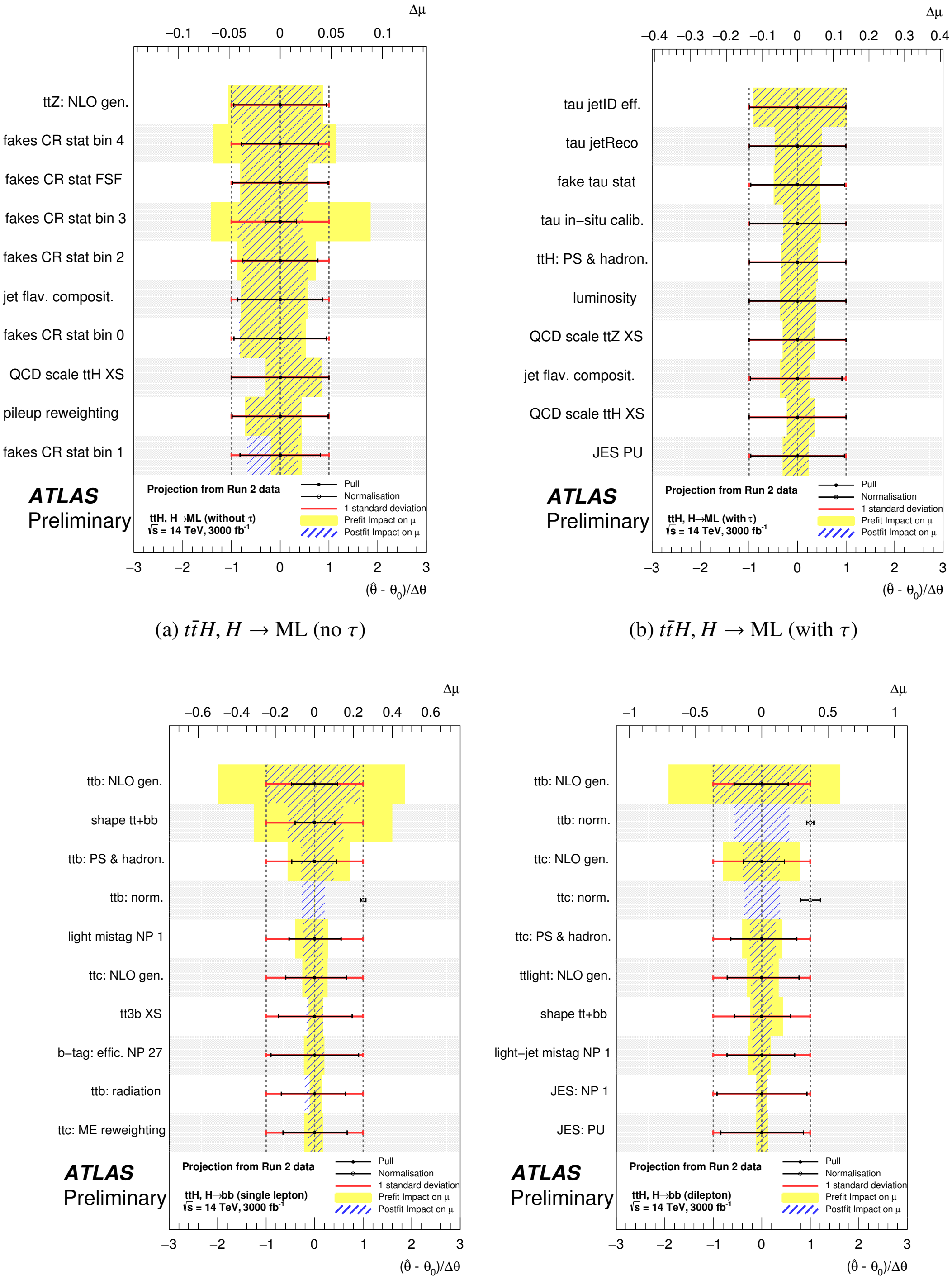

(a) $t\bar{t}H, H \to$ ML (no $\tau$)

(b) $t\bar{t}H, H \to$ ML (with $\tau$)

(c) $t\bar{t}H, H \to b\bar{b}$ (single lepton)

(d) $t\bar{t}H, H \to b\bar{b}$ (dilepton)

Figure 17: Ranking of the 10 systematic uncertainties with the largest impact on the $t\bar{t}H$ signal strength at HL-LHC for scenario S2 for the $t\bar{t}H, H \to$ ML (no $\tau$) (a), $t\bar{t}H, H \to$ ML (with $\tau$) (b), $t\bar{t}H, H \to b\bar{b}$ (single lepton) (c) and $t\bar{t}H, H \to b\bar{b}$ (dilepton) (d) processes.

| Final state | Scenario | $\Delta_{tot}/\sigma_{SM}$ | $\Delta_{stat}/\sigma_{SM}$ | $\Delta_{exp}/\sigma_{SM}$ | $\Delta_{sig}/\sigma_{SM}$ | $\Delta_{bkg}/\sigma_{SM}$ | $\Delta\mu_{sig}$ |
|---|---|---|---|---|---|---|---|
| $t\bar{t}H, H \to$ ML | Run 2, 36 fb$^{-1}$ | $^{+0.40}_{-0.40}$ | $^{+0.33}_{-0.34}$ | $^{+0.15}_{-0.15}$ | $^{+0.10}_{-0.10}$ | $^{+0.13}_{-0.13}$ | $^{+0.13}_{-0.13}$ |
| (no $\tau$) | HL-LHC S1 | $^{+0.18}_{-0.18}$ | $^{+0.04}_{-0.04}$ | $^{+0.13}_{-0.14}$ | $^{+0.08}_{-0.08}$ | $^{+0.12}_{-0.12}$ | $^{+0.11}_{-0.11}$ |
| | HL-LHC S2 | $^{+0.17}_{-0.17}$ | $^{+0.04}_{-0.04}$ | $^{+0.12}_{-0.13}$ | $^{+0.05}_{-0.05}$ | $^{+0.09}_{-0.09}$ | $^{+0.07}_{-0.07}$ |
| $t\bar{t}H, H \to$ ML | Run 2, 36 fb$^{-1}$ | $^{+0.64}_{-0.64}$ | $^{+0.54}_{-0.54}$ | $^{+0.29}_{-0.29}$ | $^{+0.10}_{-0.09}$ | $^{+0.14}_{-0.13}$ | $^{+0.13}_{-0.13}$ |
| (with $\tau$) | HL-LHC S1 | $^{+0.27}_{-0.28}$ | $^{+0.07}_{-0.07}$ | $^{+0.23}_{-0.23}$ | $^{+0.09}_{-0.08}$ | $^{+0.12}_{-0.12}$ | $^{+0.11}_{-0.11}$ |
| | HL-LHC S2 | $^{+0.25}_{-0.25}$ | $^{+0.07}_{-0.07}$ | $^{+0.22}_{-0.22}$ | $^{+0.05}_{-0.05}$ | $^{+0.07}_{-0.07}$ | $^{+0.07}_{-0.07}$ |
| $t\bar{t}H, H \to b\bar{b}$ | Run 2, 36 fb$^{-1}$ | $^{+0.61}_{-0.61}$ | $^{+0.22}_{-0.22}$ | $^{+0.27}_{-0.28}$ | $^{+0.10}_{-0.09}$ | $^{+0.47}_{-0.47}$ | $^{+0.15}_{-0.15}$ |
| (single lepton) | HL-LHC S1 | $^{+0.25}_{-0.20}$ | $^{+0.02}_{-0.02}$ | $^{+0.10}_{-0.10}$ | $^{+0.08}_{-0.06}$ | $^{+0.22}_{-0.17}$ | $^{+0.10}_{-0.11}$ |
| | HL-LHC S2 | $^{+0.18}_{-0.15}$ | $^{+0.02}_{-0.02}$ | $^{+0.09}_{-0.09}$ | $^{+0.06}_{-0.05}$ | $^{+0.14}_{-0.11}$ | $^{+0.08}_{-0.07}$ |
| $t\bar{t}H, H \to b\bar{b}$ | Run 2, 36 fb$^{-1}$ | $^{+1.06}_{-1.08}$ | $^{+0.51}_{-0.51}$ | $^{+0.32}_{-0.31}$ | $^{+0.11}_{-0.12}$ | $^{+0.90}_{-0.92}$ | $^{+0.14}_{-0.14}$ |
| (dilepton) | HL-LHC S1 | $^{+0.32}_{-0.26}$ | $^{+0.06}_{-0.06}$ | $^{+0.13}_{-0.13}$ | $^{+0.08}_{-0.07}$ | $^{+0.27}_{-0.21}$ | $^{+0.11}_{-0.09}$ |
| | HL-LHC S2 | $^{+0.23}_{-0.20}$ | $^{+0.06}_{-0.06}$ | $^{+0.11}_{-0.11}$ | $^{+0.06}_{-0.06}$ | $^{+0.17}_{-0.15}$ | $^{+0.08}_{-0.08}$ |

Table 6: Expected precision of the measurement of the $t\bar{t}H$ cross section for the $H \to$ML (first two row) and $H \to b\bar{b}$ (last two rows) decay channels with 36 fb$^{-1}$ of Run 2 data and at HL-LHC. Uncertainties are reported relative to the SM cross section at the corresponding center-of-mass energy. For the HL-LHC extrapolation, both scenarios S1 and S2 have been considered for the systematic uncertainties. The last column shows the theory uncertainty component when the measurement parameters are production-mode signal strengths instead of cross sections.

## 2.7 $H \to \mu\mu$

ATLAS has presented a search for the decay of the Higgs boson to a pair of muons using 79.8 fb$^{-1}$ of data collected at center-of-mass energy of $\sqrt{s}$ = 13 TeV [16]. The event selection is loose both at the muon object level and event level in order to retain as much signal as possible and the selected events are categorised to improve sensitivity. VBF-like events are first selected using a multivariate BDT classifier and split into two VBF-like categories of different purity. The remaining events are further split in 6 orthogonal categories based on signal purity and muon momentum resolution. The invariant mass distribution of the $m_{\mu\mu}$ signal is modelled using a weighted sum of a Gaussian and a Crystal Ball function, with parameter sets for each category based on the dimuon pseudorapidities and determined by a fit to simulated events. Background estimation, dominated by $Z/\gamma^* \to \mu\mu$, is fully data driven and is modelled using a weighted sum of a Breit-Wigner convolved by a Gaussian and an exponential function. The Run 2 analysis is limited by the statistical uncertainty, while the leading systematic uncertainty is the bias introduced by the choice of the function describing the background. This contribution is set to zero for the HL-LHC extrapolation. The other important uncertainties on the signal strength measurement include QCD scale, jet binning and Higgs boson $p_{\mathrm{T}}$ theoretical uncertainties.

This analysis is used as baseline to extrapolate the expected precision of the measurement at HL-LHC. In addition to the standard extrapolation procedure, the widths of the signal dimuon mass peaks are reduced in the Central and VBF categories by 30% and by 15% in the Forward categories to account for expected improvements in the muon $p_{\mathrm{T}}$ resolution in the new tracker system used at HL-LHC [17].

In the scenario S2 case, the most important reductions on systematic uncertainties are related to the signal modeling, namely Higgs boson $p_{\mathrm{T}}$, jet binning, and QCD scale, which are all reduced by 50%.

The results of the Run 2 analysis with 79.8 fb$^{-1}$ of data at $\sqrt{s}$ = 13 TeV are compared to the results at HL-LHC with 3000 fb$^{-1}$ of data at $\sqrt{s}$ = 14 TeV in Table 7.

| Scenario | $\Delta_{\mathrm{tot}}/\sigma_{\mathrm{SM}}$ | $\Delta_{\mathrm{stat}}/\sigma_{\mathrm{SM}}$ | $\Delta_{\mathrm{exp}}/\sigma_{\mathrm{SM}}$ | $\Delta_{\mu_{\mathrm{sig}}}/\sigma_{\mathrm{SM}}$ |
|---|---|---|---|---|
| Run 2, 79.8 fb$^{-1}$ | $^{+1.04}_{-1.06}$ | $^{+0.99}_{-1.03}$ | $^{+0.03}_{-0.03}$ | $^{+0.32}_{-0.27}$ |
| HL-LHC S1 | $^{+0.15}_{-0.14}$ | $^{+0.12}_{-0.13}$ | $^{+0.03}_{-0.03}$ | $^{+0.08}_{-0.05}$ |
| HL-LHC S2 | $^{+0.13}_{-0.13}$ | $^{+0.12}_{-0.13}$ | $^{+0.02}_{-0.02}$ | $^{+0.05}_{-0.04}$ |

Table 7: Expected precision of the measurement of the signal strength in the $H \to \mu\mu$ decay channel with 79.8 fb$^{-1}$ of Run 2 and 3000 fb$^{-1}$ of HL-LHC data. Both systematic uncertainty scenarios S1 and S2 for HL-LHC have been considered. For the HL-LHC extrapolation, the improved ITk resolution has been emulated.

Figure 18 shows the 10 systematic uncertainties with the largest impact on the signal strength measurement for scenario S2.

In both systematic uncertainty scenarios at HL-LHC, the expected precision of the measurement will be limited by the statistical uncertainty.

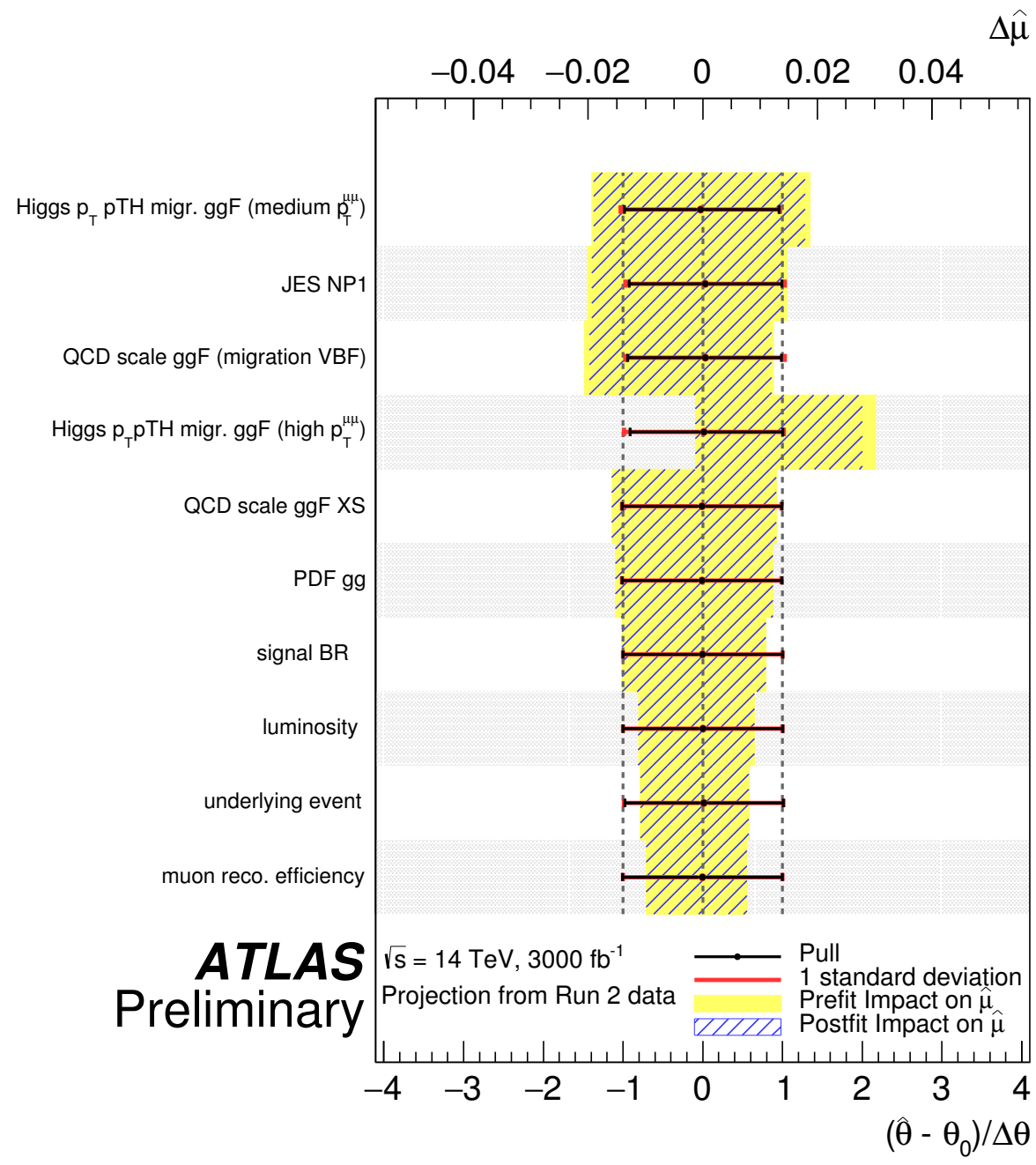

Figure 18: Ranking of the 10 systematic uncertainties (scenario S2) with the largest impact on the expected signal strength measurement in the $H \to \mu\mu$ decay channel.

## 2.8 $H \to Z\gamma$

The search for the $Z\gamma$ decay of the Higgs boson was performed using 36.1 fb$^{-1}$ at $\sqrt{s}$ = 13 TeV [18]. The observed (expected assuming SM Higgs hypothesis) upper limit on the production cross section times the branching ratio for $pp \to H \to Z\gamma$ is 6.6 (5.2) at the 95% confidence level for a Higgs boson mass of 125.09 GeV.

In this analysis, the event categorisation was optimised to improve the search sensitivity. VBF events were selected using a BDT. The analysis is strongly driven by the statistical uncertainty. The main systematic uncertainty comes from the bias induced by the choice of the background function.

The extrapolation to HL-LHC is performed with a simple scaling scenario. The modelings of the signal and background shape are kept as the above analysis. All the experimental and systematic uncertainties are also the same as before except the uncertainty from the background function choice which is taken to be negligible. With this scenario, the expected significance of the SM Higgs boson is 4.9 $\sigma$ with 3000 fb$^{-1}$. As summarised in Table 8 for the cross section times branching ratio for $pp \to H \to Z\gamma$, the precision for this measurement is expected to be 0.23 times the SM prediction. No number is provided for the current Run 2 integrated luminosity as the significance is too small. The expected measurement precision on the signal strength is evaluated to be 0.24.

| Scenario | $\Delta_{\mathrm{tot}}/\sigma_{\mathrm{SM}}$ | $\Delta_{\mathrm{stat}}/\sigma_{\mathrm{SM}}$ | $\Delta_{\mathrm{syst}}/\sigma_{\mathrm{SM}}$ |
|---|---|---|---|
| HL-LHC S1 | 0.23 | 0.20 | 0.11 |

Table 8: Expected precision of the measurement of the production cross section times the branching ratio for $pp \to H \to Z\gamma$ with 3000 fb$^{-1}$ of HL-LHC data.

Even with the whole HL-LHC dataset, the analysis sensitivity is strongly limited by the statistical component even with the conservative scenario S1. Among those different systematic uncertainty sources, the uncertainty from missing higher order corrections is the dominant one. To simplify the combination with other channels in Section 3, the systematic uncertainties for scenario S2 are assumed to be equal to the scenario S1 values.

# 3 Combination

## 3.1 Introduction

The results on the combination are obtained from a likelihood function defined as the product of the likelihoods of each input analysis. These are themselves products of likelihoods computed in mutually exclusive regions selected in the analysis, referred to as analysis categories. Since the input analyses are all based on scaling up expected results from the corresponding Run 2 versions, the correlation of nuisance parameters between channels is largely unchanged with respect to Ref. [19]. In particular, the branching ratio uncertainties have been broken down into different sources in order to take their correlations properly into account. The only differences are the following:

1. All the nuisance parameters corresponding to experimental uncertainties in the $t\bar{t}H$, $H \to b\bar{b}$ channel are uncorrelated from other channels due to the large constraints introduced at high luminosity in this channel.
2. All the nuisance parameters in the $H \to Z\gamma$ channel are uncorrelated from other channels. The impact from potential correlations is negligible, as this channel is limited by its statistical uncertainty.

Two sets of combinations, differing in signal theory uncertainty implementation and also in input channels, are performed.

1. **Cross-section combination**: the first combination is for measuring production mode cross sections and decay branching ratios. The exact configuration of theory uncertainties vary according to the assumptions involved in each measurement:

   - When measuring ratios of production-mode cross sections times decay branching ratios or the production-mode cross section in each decay channel, only theory uncertainties affecting the acceptance are considered. Neither QCD scale and PDF+$\alpha_S$ uncertainties on the cross sections, nor branching ratio uncertainties are included, since they affect only the signal normalisation.
   - When measuring production mode cross sections, in addition to the uncertainties related to the signal acceptance considered in the first case, the uncertainties on the assumed SM branching ratios are included.
   - When measuring decay branching ratios, in addition to the uncertainties related to the signal acceptance considered in the first case, uncertainties on the assumed SM cross sections, related to QCD scale and PDF+$\alpha_S$ uncertainties are included.

   The two rare decay channels, $H \to \mu\mu$ and $H \to Z\gamma$, are excluded from the combination of production mode measurements because their sensitivities are negligible. They are not used in the combination for the production mode cross sections per different decay channel measurement due to their negligible correlation with other channels but their results are added to the plots and tables for completeness.

2. **Signal strength combination**: the second combination is used to perform the combined measurement of the Higgs signal strength, as well as the interpretation of the measurements within the $\kappa$ model. The full theory uncertainties on the predicted cross sections and branching ratios are included. All the input channels, including $H \to \mu\mu$ and $H \to Z\gamma$, are included in the combination.

For each combination, the two systematic uncertainty scenarios S1 and S2 are considered. The same statistical methods as those used for the Run 2 combined measurements in Ref. [19] are used. Unless otherwise specified, projections discussed in the following sub-sections are based on an integrated luminosity of 3000 $\text{fb}^{-1}$.

### 3.2 Global signal strength

The global signal strength $\mu$ is defined as the ratio of the observed yields to their SM expectations. It corresponds to a global scaling of the expected Higgs boson yield in all categories by a single value. Its value depends on the SM predictions for each production-mode cross section and decay branching ratio.

The expected precision of the measurement of the global signal strength for the systematic scenario S1 (S2) is:

$$\begin{aligned}\mu &= 1.000^{+0.038}_{-0.037}\left(^{+0.025}_{-0.024}\right)\\ &= 1.000 \pm 0.006\ (\text{stat})\ \pm 0.016\ (0.013)\ (\text{exp})\ ^{+0.030}_{-0.028}\left(^{+0.017}_{-0.017}\right)(\text{sig})\ ^{+0.016}_{-0.015}\left(^{+0.010}_{-0.010}\right)(\text{bkg})\end{aligned}$$

### 3.3 Production cross sections

In this model, the measurement parameters are the cross sections for the five studied Higgs boson production modes: ggF, VBF, $WH$, $ZH$ and combination of $t\bar{t}H$ and $tH$ ($t\bar{t}H$+$tH$). The latter assumes their relative fractions to be as in the SM and for simplicity is labelled as $t\bar{t}H$ in the following for all the models. The small contribution from $b\bar{b}H$ is grouped with ggF. The $ZH$ process includes $ZH$ production with gluon-gluon initial state. The measurement is performed assuming SM values for its decay branching fractions. The expected production cross-section uncertainties obtained with 3000 $\text{fb}^{-1}$ are shown in Table 9 for scenarios S1 and S2.
The expected precision of the production cross-section measurement with only 300 $\text{fb}^{-1}$ at $\sqrt{s}$ = 14 TeV is shown in Table 10. Concerning the systematic uncertainties, the same scenarios S1 and S2 are considered. The expected production cross-section uncertainties obtained with 3000 $\text{fb}^{-1}$ are summarised in Figure 19. Figure 20 summarises the total uncertainties for both scenarios. Even for scenario S2, all production-mode cross-section measurements are limited by systematic uncertainties. All correlations are 0.08 or less, with the exception of that of $t\bar{t}H$ and ggF which is 0.16.

| POI | Scenario | $\Delta_{\text{tot}}$ | $\Delta_{\text{stat}}$ | $\Delta_{\text{exp}}$ | $\Delta_{\text{sig}}$ | $\Delta_{\text{bkg}}$ |
|---|---|---|---|---|---|---|
| $\sigma_{\text{ggF}}/\sigma^{\text{SM}}_{\text{ggF}}$ | HL-LHC S1 | $^{+0.035}_{-0.034}$ | $^{+0.008}_{-0.008}$ | $^{+0.021}_{-0.021}$ | $^{+0.022}_{-0.021}$ | $^{+0.016}_{-0.015}$ |
| | HL-LHC S2 | $^{+0.024}_{-0.024}$ | $^{+0.008}_{-0.008}$ | $^{+0.017}_{-0.017}$ | $^{+0.012}_{-0.012}$ | $^{+0.010}_{-0.010}$ |
| $\sigma_{\text{VBF}}/\sigma^{\text{SM}}_{\text{VBF}}$ | HL-LHC S1 | $^{+0.056}_{-0.054}$ | $^{+0.020}_{-0.020}$ | $^{+0.027}_{-0.026}$ | $^{+0.038}_{-0.037}$ | $^{+0.022}_{-0.020}$ |
| | HL-LHC S2 | $^{+0.042}_{-0.041}$ | $^{+0.020}_{-0.020}$ | $^{+0.023}_{-0.022}$ | $^{+0.023}_{-0.022}$ | $^{+0.018}_{-0.017}$ |
| $\sigma_{\text{WH}}/\sigma^{\text{SM}}_{\text{WH}}$ | HL-LHC S1 | $^{+0.095}_{-0.092}$ | $^{+0.041}_{-0.040}$ | $^{+0.041}_{-0.039}$ | $^{+0.053}_{-0.048}$ | $^{+0.055}_{-0.054}$ |
| | HL-LHC S2 | $^{+0.078}_{-0.076}$ | $^{+0.041}_{-0.040}$ | $^{+0.035}_{-0.034}$ | $^{+0.034}_{-0.031}$ | $^{+0.045}_{-0.045}$ |
| $\sigma_{\text{ZH}}/\sigma^{\text{SM}}_{\text{ZH}}$ | HL-LHC S1 | $^{+0.063}_{-0.061}$ | $^{+0.034}_{-0.034}$ | $^{+0.025}_{-0.024}$ | $^{+0.035}_{-0.033}$ | $^{+0.031}_{-0.030}$ |
| | HL-LHC S2 | $^{+0.049}_{-0.048}$ | $^{+0.034}_{-0.034}$ | $^{+0.018}_{-0.018}$ | $^{+0.020}_{-0.019}$ | $^{+0.022}_{-0.021}$ |
| $\sigma_{\text{t}\bar{\text{t}}\text{H}}/\sigma^{\text{SM}}_{\text{t}\bar{\text{t}}\text{H}}$ | HL-LHC S1 | $^{+0.069}_{-0.066}$ | $^{+0.019}_{-0.019}$ | $^{+0.032}_{-0.031}$ | $^{+0.038}_{-0.036}$ | $^{+0.044}_{-0.041}$ |
| | HL-LHC S2 | $^{+0.054}_{-0.052}$ | $^{+0.019}_{-0.019}$ | $^{+0.028}_{-0.027}$ | $^{+0.025}_{-0.023}$ | $^{+0.034}_{-0.033}$ |

Table 9: Expected uncertainty on the cross-section measurements for the ggF, VBF, $WH$, $ZH$ and $t\bar{t}H$ production modes normalised to their SM predictions for an integrated luminosity of 3000 fb$^{-1}$ for both systematic scenarios S1 and S2, assuming SM values for its decay branching fractions. The total uncertainties are decomposed into statistical uncertainties (stat), experimental systematic uncertainties (exp), and theory uncertainties in the modeling of the signal (sig) and background (bkg) processes.

| POI (300 fb$^{-1}$) | Scenario | $\Delta_{\text{tot}}$ | $\Delta_{\text{stat}}$ | $\Delta_{\text{exp}}$ | $\Delta_{\text{sig}}$ | $\Delta_{\text{bkg}}$ |
|---|---|---|---|---|---|---|
| $\sigma_{\text{ggF}}/\sigma^{\text{SM}}_{\text{ggF}}$ | HL-LHC S1 | $^{+0.059}_{-0.057}$ | $^{+0.027}_{-0.027}$ | $^{+0.037}_{-0.036}$ | $^{+0.023}_{-0.022}$ | $^{+0.028}_{-0.027}$ |
| | HL-LHC S2 | $^{+0.047}_{-0.046}$ | $^{+0.027}_{-0.027}$ | $^{+0.030}_{-0.029}$ | $^{+0.013}_{-0.012}$ | $^{+0.020}_{-0.020}$ |
| $\sigma_{\text{VBF}}/\sigma^{\text{SM}}_{\text{VBF}}$ | HL-LHC S1 | $^{+0.102}_{-0.098}$ | $^{+0.066}_{-0.066}$ | $^{+0.052}_{-0.048}$ | $^{+0.045}_{-0.044}$ | $^{+0.034}_{-0.031}$ |
| | HL-LHC S2 | $^{+0.097}_{-0.093}$ | $^{+0.067}_{-0.066}$ | $^{+0.043}_{-0.041}$ | $^{+0.049}_{-0.046}$ | $^{+0.025}_{-0.025}$ |
| $\sigma_{\text{WH}}/\sigma^{\text{SM}}_{\text{WH}}$ | HL-LHC S1 | $^{+0.220}_{-0.207}$ | $^{+0.129}_{-0.127}$ | $^{+0.100}_{-0.093}$ | $^{+0.089}_{-0.070}$ | $^{+0.118}_{-0.115}$ |
| | HL-LHC S2 | $^{+0.178}_{-0.172}$ | $^{+0.129}_{-0.127}$ | $^{+0.080}_{-0.076}$ | $^{+0.047}_{-0.037}$ | $^{+0.080}_{-0.078}$ |
| $\sigma_{\text{ZH}}/\sigma^{\text{SM}}_{\text{ZH}}$ | HL-LHC S1 | $^{+0.142}_{-0.138}$ | $^{+0.109}_{-0.108}$ | $^{+0.047}_{-0.042}$ | $^{+0.039}_{-0.032}$ | $^{+0.068}_{-0.067}$ |
| | HL-LHC S2 | $^{+0.123}_{-0.121}$ | $^{+0.109}_{-0.108}$ | $^{+0.035}_{-0.032}$ | $^{+0.019}_{-0.016}$ | $^{+0.041}_{-0.040}$ |
| $\sigma_{\text{t}\bar{\text{t}}\text{H}}/\sigma^{\text{SM}}_{\text{t}\bar{\text{t}}\text{H}}$ | HL-LHC S1 | $^{+0.112}_{-0.108}$ | $^{+0.061}_{-0.061}$ | $^{+0.059}_{-0.057}$ | $^{+0.033}_{-0.028}$ | $^{+0.065}_{-0.063}$ |
| | HL-LHC S2 | $^{+0.102}_{-0.099}$ | $^{+0.061}_{-0.061}$ | $^{+0.054}_{-0.052}$ | $^{+0.019}_{-0.016}$ | $^{+0.059}_{-0.057}$ |

Table 10: Expected uncertainty on the cross-section measurements for the ggF, VBF, $WH$, $ZH$ and $t\bar{t}H$ production modes normalised to their SM predictions for an integrated luminosity reduced to 300 fb$^{-1}$. Concerning the systematic uncertainties, the same scenarios S1 and S2 as those used for Table 9 are considered. The branching fractions are assumed to be as predicted by the SM. The total uncertainties are decomposed into statistical uncertainties (stat), experimental systematic uncertainties (exp), and theory uncertainties in the modeling of the signal (sig) and background (bkg) processes.

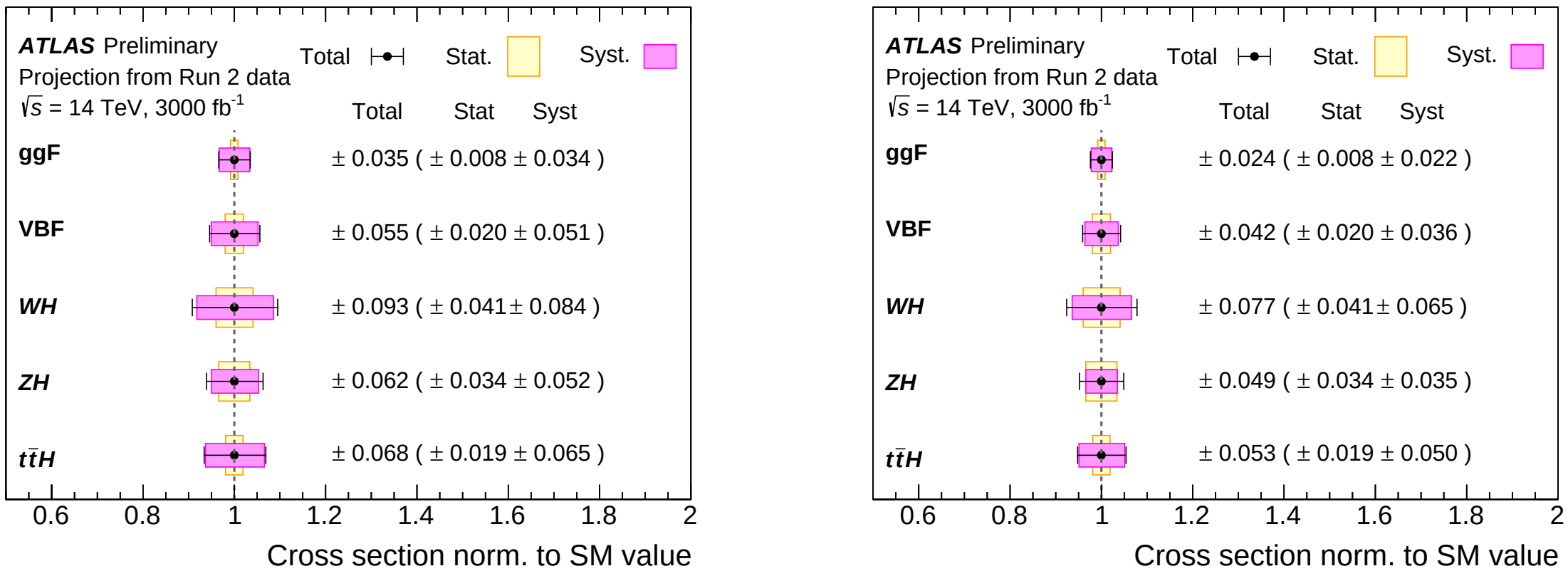

(a) Scenario S1 (b) Scenario S2

Figure 19: Expected result for the measured cross sections for the ggF, VBF, $WH$, $ZH$ and $t\bar{t}H$ production modes normalised to their SM predictions assuming SM branching fractions for scenarios S1 (a) and S2 (b). The black bars, yellow boxes and pink boxes show the total, statistical and systematic uncertainties respectively.

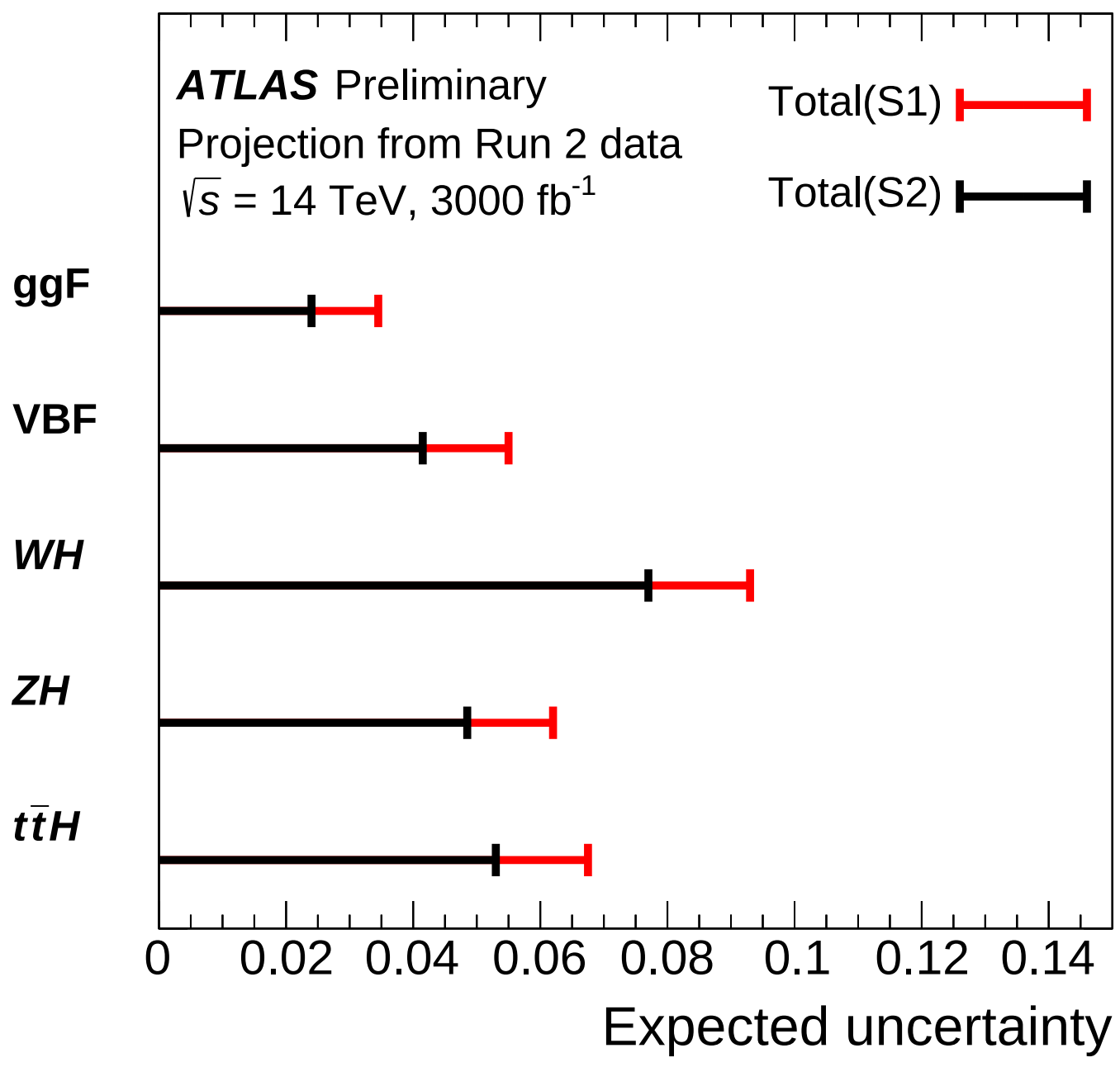

Figure 20: Expected uncertainty on the measurements of the cross sections for the ggF, VBF, $WH$, $ZH$ and $t\bar{t}H$ production modes normalised to their SM predictions assuming SM branching fractions for the scenarios S1 (red) and S2 (black).

### 3.4 Branching ratio

In this model, the POIs are the branching ratios for the seven Higgs boson decay channels to $\gamma\gamma$, $ZZ$, $WW$, $\tau\tau$, $bb$, $\mu\mu$ and $Z\gamma$. The measurement is performed assuming SM values for Higgs boson production cross sections.
The expected precision of branching ratio measurements for 3000 fb$^{-1}$ are shown in Table 11 for scenarios S1 and S2. Figure 21 summarises the expected results with the statistical and systematic components. In Figure 22, the total uncertainties for the branching ratios are summarised for scenarios S1 and S2.

| POI | Scenario | $\Delta_{tot}$ | $\Delta_{stat}$ | $\Delta_{exp}$ | $\Delta_{sig}$ | $\Delta_{bkg}$ |
|---|---|---|---|---|---|---|
| $BR_{bb}/BR_{bb,SM}$ | HL-LHC S1 | $^{+0.079}_{-0.072}$ | $^{+0.020}_{-0.020}$ | $^{+0.025}_{-0.024}$ | $^{+0.052}_{-0.047}$ | $^{+0.050}_{-0.045}$ |
| | HL-LHC S2 | $^{+0.052}_{-0.049}$ | $^{+0.020}_{-0.020}$ | $^{+0.020}_{-0.019}$ | $^{+0.029}_{-0.027}$ | $^{+0.032}_{-0.031}$ |
| $BR_{\tau\tau}$ / $BR_{\tau\tau,SM}$ | HL-LHC S1 | $^{+0.062}_{-0.058}$ | $^{+0.017}_{-0.017}$ | $^{+0.028}_{-0.027}$ | $^{+0.046}_{-0.043}$ | $^{+0.025}_{-0.023}$ |
| | HL-LHC S2 | $^{+0.045}_{-0.044}$ | $^{+0.017}_{-0.017}$ | $^{+0.025}_{-0.025}$ | $^{+0.029}_{-0.027}$ | $^{+0.018}_{-0.016}$ |
| $BR_{WW}/BR_{WW,SM}$ | HL-LHC S1 | $^{+0.059}_{-0.057}$ | $^{+0.010}_{-0.010}$ | $^{+0.028}_{-0.028}$ | $^{+0.044}_{-0.041}$ | $^{+0.026}_{-0.026}$ |
| | HL-LHC S2 | $^{+0.045}_{-0.043}$ | $^{+0.010}_{-0.010}$ | $^{+0.024}_{-0.024}$ | $^{+0.033}_{-0.031}$ | $^{+0.016}_{-0.016}$ |
| $BR_{\gamma\gamma}/BR_{\gamma\gamma,SM}$ | HL-LHC S1 | $^{+0.063}_{-0.058}$ | $^{+0.012}_{-0.012}$ | $^{+0.049}_{-0.045}$ | $^{+0.035}_{-0.032}$ | $^{+0.014}_{-0.013}$ |
| | HL-LHC S2 | $^{+0.038}_{-0.036}$ | $^{+0.012}_{-0.012}$ | $^{+0.030}_{-0.029}$ | $^{+0.018}_{-0.017}$ | $^{+0.007}_{-0.006}$ |
| $BR_{ZZ}/BR_{ZZ,SM}$ | HL-LHC S1 | $^{+0.053}_{-0.053}$ | $^{+0.016}_{-0.016}$ | $^{+0.027}_{-0.033}$ | $^{+0.039}_{-0.036}$ | $^{+0.018}_{-0.016}$ |
| | HL-LHC S2 | $^{+0.038}_{-0.038}$ | $^{+0.016}_{-0.016}$ | $^{+0.027}_{-0.027}$ | $^{+0.020}_{-0.018}$ | $^{+0.010}_{-0.010}$ |
| $BR_{\mu\mu}$ /$BR_{\mu\mu,SM}$ | HL-LHC S1 | $^{+0.157}_{-0.141}$ | $^{+0.127}_{-0.127}$ | $^{+0.036}_{-0.028}$ | $^{+0.084}_{-0.053}$ | $^{+0.000}_{-0.000}$ |
| | HL-LHC S2 | $^{+0.139}_{-0.134}$ | $^{+0.127}_{-0.127}$ | $^{+0.036}_{-0.027}$ | $^{+0.043}_{-0.031}$ | $^{+0.000}_{-0.000}$ |
| $BR_{Z\gamma}/BR_{Z\gamma,SM}$ | HL-LHC S1 | $^{+0.256}_{-0.228}$ | $^{+0.203}_{-0.203}$ | $^{+0.054}_{-0.036}$ | $^{+0.146}_{-0.099}$ | $^{+0.000}_{-0.000}$ |
| | HL-LHC S2 | $^{+0.256}_{-0.228}$ | $^{+0.203}_{-0.203}$ | $^{+0.054}_{-0.036}$ | $^{+0.146}_{-0.098}$ | $^{+0.000}_{-0.000}$ |

Table 11: Expected uncertainty on the measurements of the branching ratios of the Higgs boson normalised to their SM predictions, assuming SM values for its production cross section. The total uncertainties are decomposed into statistical uncertainties (stat), experimental systematic uncertainties (exp), and theory uncertainties in the modeling of the signal (sig) and background (bkg) processes. For the $BR_{Z\gamma}$ measurement, dominated by statistical uncertainties, the systematic uncertainties for scenario S2 are assumed to be equal to those used for the scenario S1.

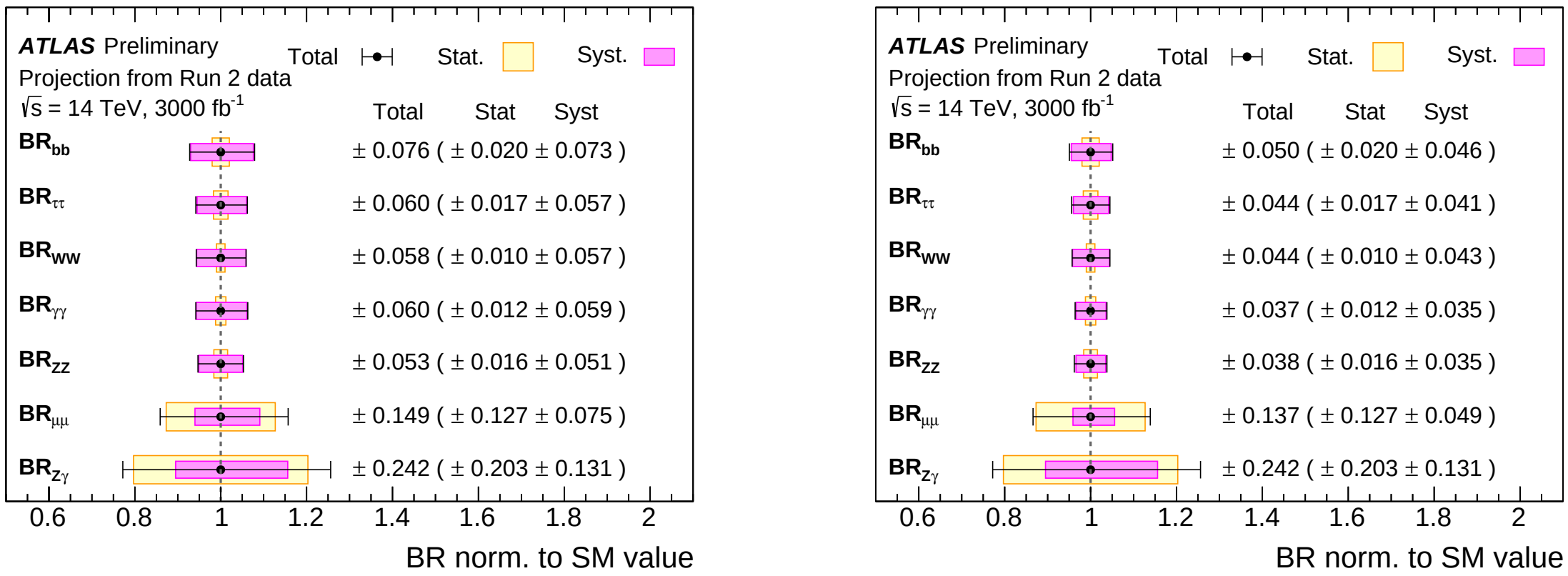

(a) Scenario S1 (b) Scenario S2

Figure 21: Expected result for the measured branching ratios of the $H \to \gamma\gamma$, $ZZ$, $WW$, $\tau\tau$, $bb$, $\mu\mu$ and $Z\gamma$ decay channels normalised to their SM predictions assuming SM production cross section for scenarios S1 (a) and S2 (b). The black bars, yellow boxes and pink boxes show the total, statistical and systematic uncertainties respectively. For the $\mathrm{BR}_{Z\gamma}$ measurement, dominated by statistical uncertainties, the systematic uncertainties for scenario S2 are assumed to be equal to those used for the scenario S1.

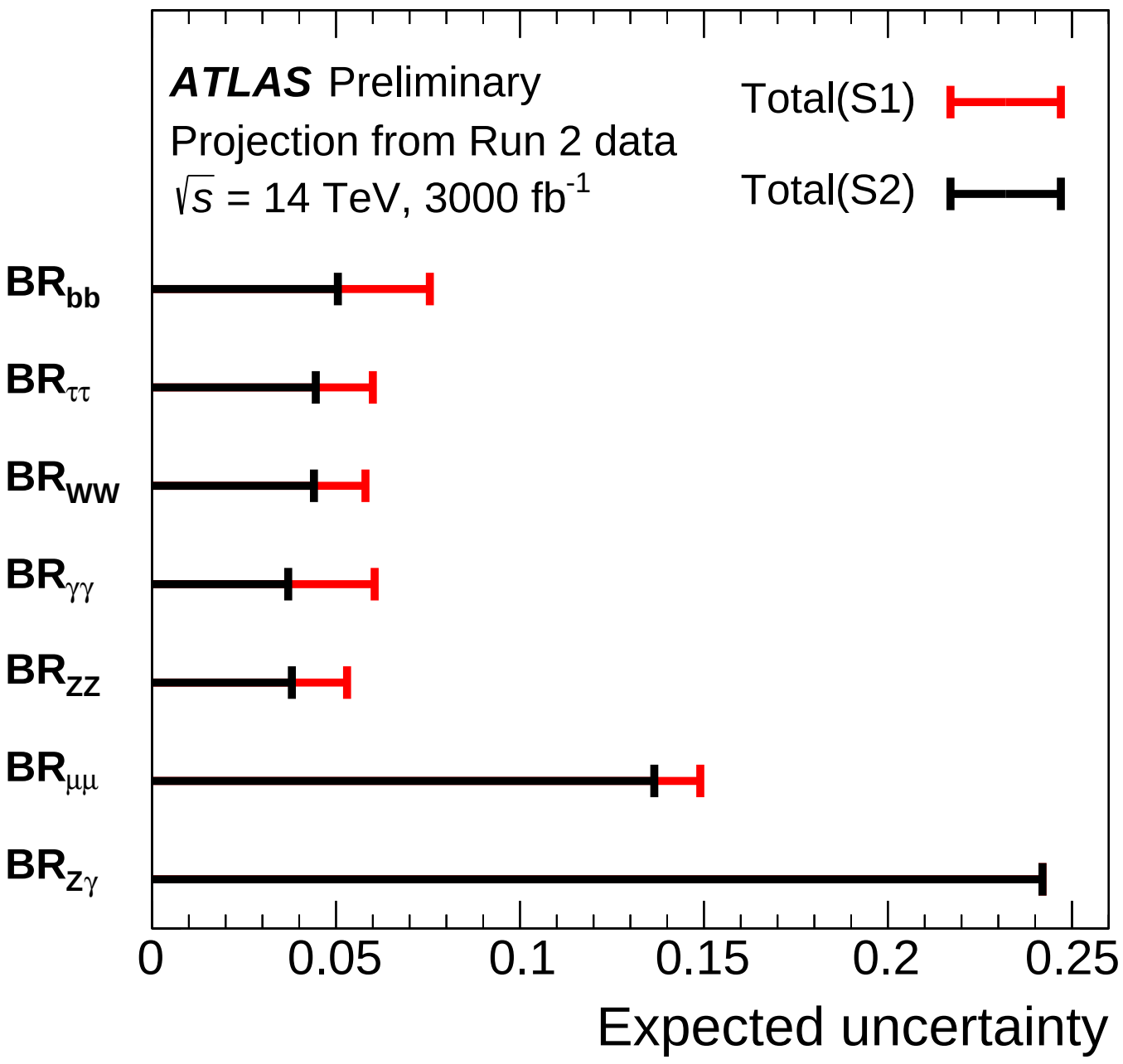

Figure 22: Expected uncertainty on the branching ratio measurements for the $\gamma\gamma$, $ZZ$, $WW$, $\tau\tau$, $bb$, $\mu\mu$ and $Z\gamma$ decay channels normalised to their SM predictions assuming SM production cross section for scenarios S1 (red) and S2 (black). For the $\mathrm{BR}_{Z\gamma}$ measurement, dominated by statistical uncertainties, the systematic uncertainties for scenario S2 are assumed to be equal to those used for the scenario S1.

## 3.5 Production-mode cross sections in different decay channels

In this model, the measured parameters are the combined production cross sections times branching fraction for ggF, VBF, $WH$, $ZH$ and $t\bar{t}H$ production in each relevant decay mode, normalised to their SM predictions. The results are obtained from a simultaneous fit to all decay channels, using as parameters

| POI | Scenario | $\Delta_{\text{tot}}$ | $\Delta_{\text{stat}}$ | $\Delta_{\text{exp}}$ | $\Delta_{\text{sig}}$ | $\Delta_{\text{bkg}}$ |
|---|---|---|---|---|---|---|
| $\sigma(\text{ggF}, H \to \gamma\gamma)/\sigma_{\text{SM}}$ | HL-LHC S1 | $^{+0.054}_{-0.050}$ | $^{+0.017}_{-0.017}$ | $^{+0.048}_{-0.045}$ | $^{+0.012}_{-0.011}$ | $^{+0.012}_{-0.011}$ |
| | HL-LHC S2 | $^{+0.037}_{-0.035}$ | $^{+0.017}_{-0.017}$ | $^{+0.031}_{-0.029}$ | $^{+0.009}_{-0.009}$ | $^{+0.006}_{-0.005}$ |
| $\sigma(\text{ggF}, H \to ZZ)/\sigma_{\text{SM}}$ | HL-LHC S1 | $^{+0.049}_{-0.049}$ | $^{+0.020}_{-0.020}$ | $^{+0.035}_{-0.038}$ | $^{+0.020}_{-0.016}$ | $^{+0.020}_{-0.018}$ |
| | HL-LHC S2 | $^{+0.039}_{-0.039}$ | $^{+0.020}_{-0.020}$ | $^{+0.030}_{-0.030}$ | $^{+0.011}_{-0.010}$ | $^{+0.010}_{-0.009}$ |
| $\sigma(\text{ggF}, H \to WW)/\sigma_{\text{SM}}$ | HL-LHC S1 | $^{+0.061}_{-0.059}$ | $^{+0.012}_{-0.012}$ | $^{+0.032}_{-0.031}$ | $^{+0.038}_{-0.036}$ | $^{+0.034}_{-0.033}$ |
| | HL-LHC S2 | $^{+0.044}_{-0.043}$ | $^{+0.012}_{-0.012}$ | $^{+0.027}_{-0.027}$ | $^{+0.021}_{-0.020}$ | $^{+0.024}_{-0.024}$ |
| $\sigma(\text{ggF}, H \to \tau\tau)/\sigma_{\text{SM}}$ | HL-LHC S1 | $^{+0.110}_{-0.101}$ | $^{+0.033}_{-0.033}$ | $^{+0.051}_{-0.049}$ | $^{+0.080}_{-0.070}$ | $^{+0.044}_{-0.044}$ |
| | HL-LHC S2 | $^{+0.085}_{-0.080}$ | $^{+0.033}_{-0.033}$ | $^{+0.045}_{-0.044}$ | $^{+0.058}_{-0.051}$ | $^{+0.028}_{-0.026}$ |
| $\sigma(\text{ggF}, H \to \mu\mu)/\sigma_{\text{SM}}$ | HL-LHC S1 | $^{+0.210}_{-0.188}$ | $^{+0.179}_{-0.179}$ | $^{+0.032}_{-0.024}$ | $^{+0.105}_{-0.054}$ | $^{+0.000}_{-0.000}$ |
| | HL-LHC S2 | $^{+0.187}_{-0.183}$ | $^{+0.179}_{-0.179}$ | $^{+0.031}_{-0.023}$ | $^{+0.047}_{-0.029}$ | $^{+0.000}_{-0.000}$ |
| $\sigma(\text{ggF}, H \to Z\gamma)/\sigma_{\text{SM}}$ | HL-LHC S1 | $^{+0.346}_{-0.320}$ | $^{+0.311}_{-0.311}$ | $^{+0.059}_{-0.039}$ | $^{+0.139}_{-0.062}$ | $^{+0.000}_{-0.000}$ |
| | HL-LHC S2 | $^{+0.346}_{-0.320}$ | $^{+0.311}_{-0.311}$ | $^{+0.059}_{-0.039}$ | $^{+0.139}_{-0.062}$ | $^{+0.000}_{-0.000}$ |
| $\sigma(\text{VBF}, H \to \gamma\gamma)/\sigma_{\text{SM}}$ | HL-LHC S1 | $^{+0.126}_{-0.114}$ | $^{+0.044}_{-0.044}$ | $^{+0.077}_{-0.070}$ | $^{+0.087}_{-0.077}$ | $^{+0.023}_{-0.019}$ |
| | HL-LHC S2 | $^{+0.093}_{-0.085}$ | $^{+0.044}_{-0.044}$ | $^{+0.058}_{-0.051}$ | $^{+0.056}_{-0.051}$ | $^{+0.009}_{-0.008}$ |
| $\sigma(\text{VBF}, H \to ZZ)/\sigma_{\text{SM}}$ | HL-LHC S1 | $^{+0.135}_{-0.126}$ | $^{+0.098}_{-0.094}$ | $^{+0.054}_{-0.049}$ | $^{+0.072}_{-0.065}$ | $^{+0.023}_{-0.020}$ |
| | HL-LHC S2 | $^{+0.122}_{-0.115}$ | $^{+0.098}_{-0.094}$ | $^{+0.053}_{-0.048}$ | $^{+0.047}_{-0.042}$ | $^{+0.014}_{-0.011}$ |
| $\sigma(\text{VBF}, H \to WW)/\sigma_{\text{SM}}$ | HL-LHC S1 | $^{+0.103}_{-0.103}$ | $^{+0.033}_{-0.033}$ | $^{+0.040}_{-0.037}$ | $^{+0.075}_{-0.078}$ | $^{+0.046}_{-0.044}$ |
| | HL-LHC S2 | $^{+0.066}_{-0.065}$ | $^{+0.033}_{-0.033}$ | $^{+0.029}_{-0.028}$ | $^{+0.040}_{-0.040}$ | $^{+0.028}_{-0.028}$ |
| $\sigma(\text{VBF}, H \to \tau\tau)/\sigma_{\text{SM}}$ | HL-LHC S1 | $^{+0.090}_{-0.085}$ | $^{+0.037}_{-0.037}$ | $^{+0.042}_{-0.041}$ | $^{+0.058}_{-0.053}$ | $^{+0.039}_{-0.036}$ |
| | HL-LHC S2 | $^{+0.079}_{-0.076}$ | $^{+0.037}_{-0.037}$ | $^{+0.050}_{-0.046}$ | $^{+0.033}_{-0.031}$ | $^{+0.037}_{-0.035}$ |
| $\sigma(\text{VBF}, H \to \mu\mu)/\sigma_{\text{SM}}$ | HL-LHC S1 | $^{+0.386}_{-0.388}$ | $^{+0.325}_{-0.325}$ | $^{+0.142}_{-0.092}$ | $^{+0.151}_{-0.192}$ | $^{+0.000}_{-0.000}$ |
| | HL-LHC S2 | $^{+0.370}_{-0.353}$ | $^{+0.325}_{-0.325}$ | $^{+0.142}_{-0.092}$ | $^{+0.104}_{-0.104}$ | $^{+0.000}_{-0.000}$ |
| $\sigma(\text{VBF}, H \to Z\gamma)/\sigma_{\text{SM}}$ | HL-LHC S1 | $^{+0.677}_{-0.688}$ | $^{+0.625}_{-0.619}$ | $^{+0.153}_{-0.065}$ | $^{+0.208}_{-0.293}$ | $^{+0.000}_{-0.000}$ |
| | HL-LHC S2 | $^{+0.677}_{-0.688}$ | $^{+0.625}_{-0.619}$ | $^{+0.153}_{-0.065}$ | $^{+0.208}_{-0.293}$ | $^{+0.000}_{-0.000}$ |

Table 12: Expected uncertainty on the measurements of the production cross section times branching fraction for the ggF and VBF production modes, normalised to their SM predictions, for scenarios S1 and S2. The total uncertainties are decomposed into statistical uncertainties (stat), experimental uncertainties (exp), and theory uncertainties in the modeling of the signal (sig) and background (bkg) processes. The values are obtained from a simultaneous fit to all decay channels. For the $Z\gamma$ measurement, dominated by statistical uncertainties, the systematic uncertainties for scenario S2 are assumed to be equal to those used for the scenario S1.

| POI | Scenario | $\Delta_{tot}$ | $\Delta_{stat}$ | $\Delta_{exp}$ | $\Delta_{sig}$ | $\Delta_{bkg}$ |
|---|---|---|---|---|---|---|
| $\sigma(\mathrm{WH},\ H \to \gamma\gamma)/\sigma_{\mathrm{SM}}$ | HL-LHC S1 | $^{+0.152}_{-0.144}$ | $^{+0.132}_{-0.130}$ | $^{+0.059}_{-0.046}$ | $^{+0.044}_{-0.037}$ | $^{+0.014}_{-0.012}$ |
| | HL-LHC S2 | $^{+0.141}_{-0.136}$ | $^{+0.132}_{-0.130}$ | $^{+0.037}_{-0.030}$ | $^{+0.030}_{-0.026}$ | $^{+0.007}_{-0.006}$ |
| $\sigma(\mathrm{WH},\ H \to bb)/\sigma_{\mathrm{SM}}$ | HL-LHC S1 | $^{+0.146}_{-0.135}$ | $^{+0.044}_{-0.043}$ | $^{+0.050}_{-0.048}$ | $^{+0.078}_{-0.068}$ | $^{+0.104}_{-0.097}$ |
| | HL-LHC S2 | $^{+0.102}_{-0.099}$ | $^{+0.044}_{-0.043}$ | $^{+0.042}_{-0.040}$ | $^{+0.044}_{-0.040}$ | $^{+0.070}_{-0.068}$ |
| $\sigma(\mathrm{ZH},\ H \to \gamma\gamma)/\sigma_{\mathrm{SM}}$ | HL-LHC S1 | $^{+0.177}_{-0.164}$ | $^{+0.151}_{-0.147}$ | $^{+0.059}_{-0.044}$ | $^{+0.070}_{-0.056}$ | $^{+0.014}_{-0.011}$ |
| | HL-LHC S2 | $^{+0.161}_{-0.153}$ | $^{+0.151}_{-0.147}$ | $^{+0.036}_{-0.028}$ | $^{+0.041}_{-0.034}$ | $^{+0.006}_{-0.005}$ |
| $\sigma(\mathrm{ZH},\ H \to bb)/\sigma_{\mathrm{SM}}$ | HL-LHC S1 | $^{+0.071}_{-0.068}$ | $^{+0.035}_{-0.035}$ | $^{+0.027}_{-0.026}$ | $^{+0.042}_{-0.038}$ | $^{+0.037}_{-0.035}$ |
| | HL-LHC S2 | $^{+0.052}_{-0.051}$ | $^{+0.035}_{-0.035}$ | $^{+0.020}_{-0.019}$ | $^{+0.022}_{-0.021}$ | $^{+0.024}_{-0.024}$ |
| $\sigma(\mathrm{VH},\ H \to ZZ)/\sigma_{\mathrm{SM}}$ | HL-LHC S1 | $^{+0.193}_{-0.181}$ | $^{+0.177}_{-0.168}$ | $^{+0.045}_{-0.039}$ | $^{+0.057}_{-0.050}$ | $^{+0.023}_{-0.021}$ |
| | HL-LHC S2 | $^{+0.187}_{-0.176}$ | $^{+0.177}_{-0.168}$ | $^{+0.037}_{-0.031}$ | $^{+0.043}_{-0.039}$ | $^{+0.018}_{-0.016}$ |
| $\sigma(t\bar{t}H,\ H \to \gamma\gamma)/\sigma_{\mathrm{SM}}$ | HL-LHC S1 | $^{+0.104}_{-0.096}$ | $^{+0.047}_{-0.046}$ | $^{+0.063}_{-0.056}$ | $^{+0.066}_{-0.061}$ | $^{+0.016}_{-0.014}$ |
| | HL-LHC S2 | $^{+0.076}_{-0.072}$ | $^{+0.047}_{-0.046}$ | $^{+0.043}_{-0.040}$ | $^{+0.041}_{-0.038}$ | $^{+0.005}_{-0.005}$ |
| $\sigma(t\bar{t}H,\ H \to WW, \tau\tau)/\sigma_{\mathrm{SM}}$ | HL-LHC S1 | $^{+0.235}_{-0.206}$ | $^{+0.063}_{-0.063}$ | $^{+0.193}_{-0.171}$ | $^{+0.085}_{-0.055}$ | $^{+0.084}_{-0.077}$ |
| | HL-LHC S2 | $^{+0.213}_{-0.191}$ | $^{+0.063}_{-0.063}$ | $^{+0.189}_{-0.170}$ | $^{+0.052}_{-0.034}$ | $^{+0.054}_{-0.048}$ |
| $\sigma(t\bar{t}H,\ H \to ZZ)/\sigma_{\mathrm{SM}}$ | HL-LHC S1 | $^{+0.219}_{-0.192}$ | $^{+0.196}_{-0.177}$ | $^{+0.046}_{-0.037}$ | $^{+0.085}_{-0.062}$ | $^{+0.018}_{-0.016}$ |
| | HL-LHC S2 | $^{+0.203}_{-0.183}$ | $^{+0.196}_{-0.177}$ | $^{+0.035}_{-0.026}$ | $^{+0.041}_{-0.035}$ | $^{+0.010}_{-0.009}$ |
| $\sigma(t\bar{t}H,\ H \to bb)/\sigma_{\mathrm{SM}}$ | HL-LHC S1 | $^{+0.218}_{-0.181}$ | $^{+0.032}_{-0.032}$ | $^{+0.042}_{-0.041}$ | $^{+0.078}_{-0.069}$ | $^{+0.197}_{-0.159}$ |
| | HL-LHC S2 | $^{+0.151}_{-0.133}$ | $^{+0.032}_{-0.032}$ | $^{+0.034}_{-0.033}$ | $^{+0.047}_{-0.041}$ | $^{+0.135}_{-0.118}$ |

Table 13: Expected uncertainty on the measurements of the production cross section times branching fraction for the $VH$ and $t\bar{t}H$ production modes, normalised to their SM predictions, for scenarios S1 and S2. The total uncertainties are decomposed into statistical uncertainties (stat), experimental uncertainties (exp), and theory uncertainties in the modeling of the signal (sig) and background (bkg) processes. The values are obtained from a simultaneous fit to all decay channels.

of interest the $(\sigma \times \mathrm{B})_{if}$ for each measured production mode $i$ and decay final state $f$.

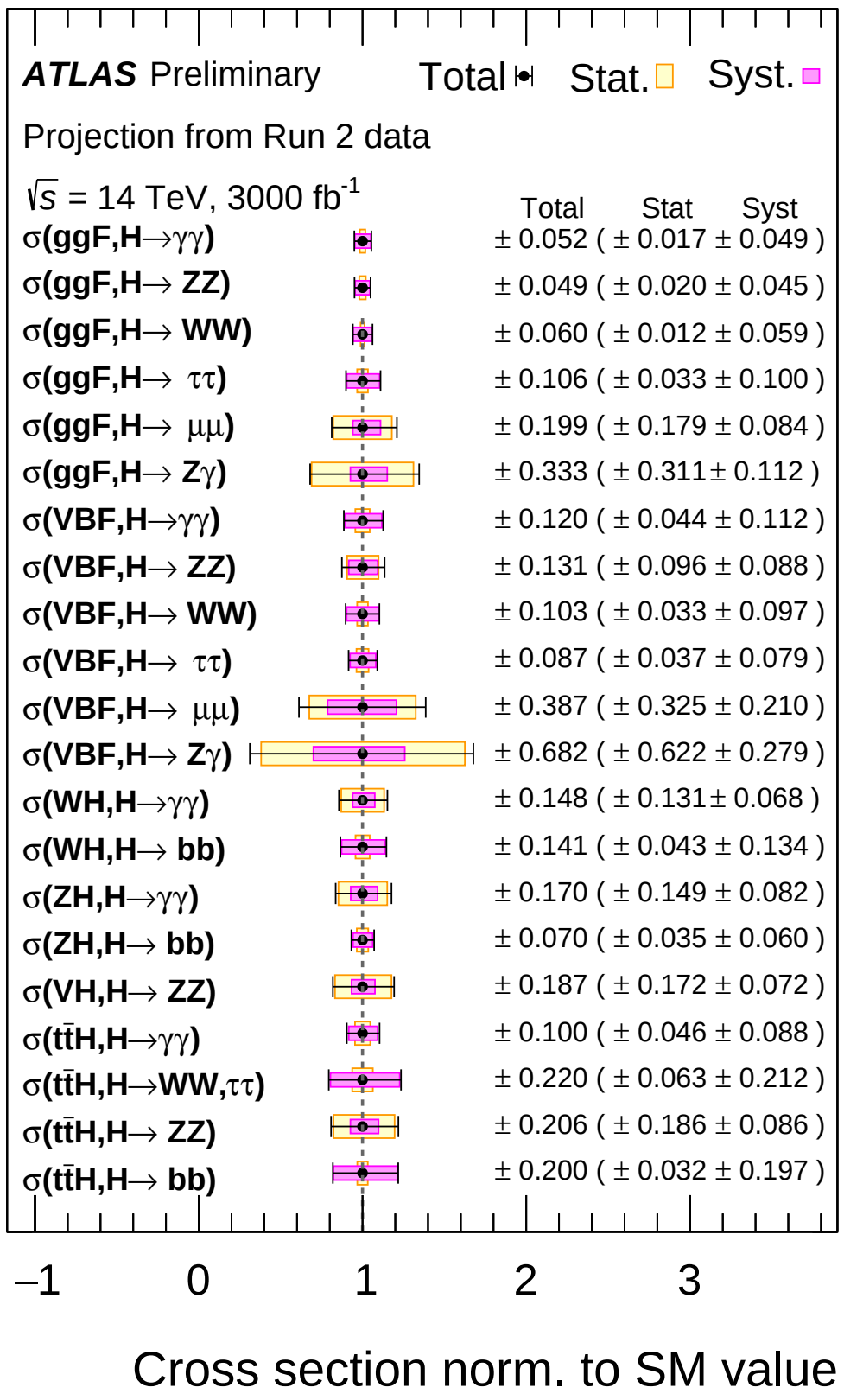

(a) Scenario S1

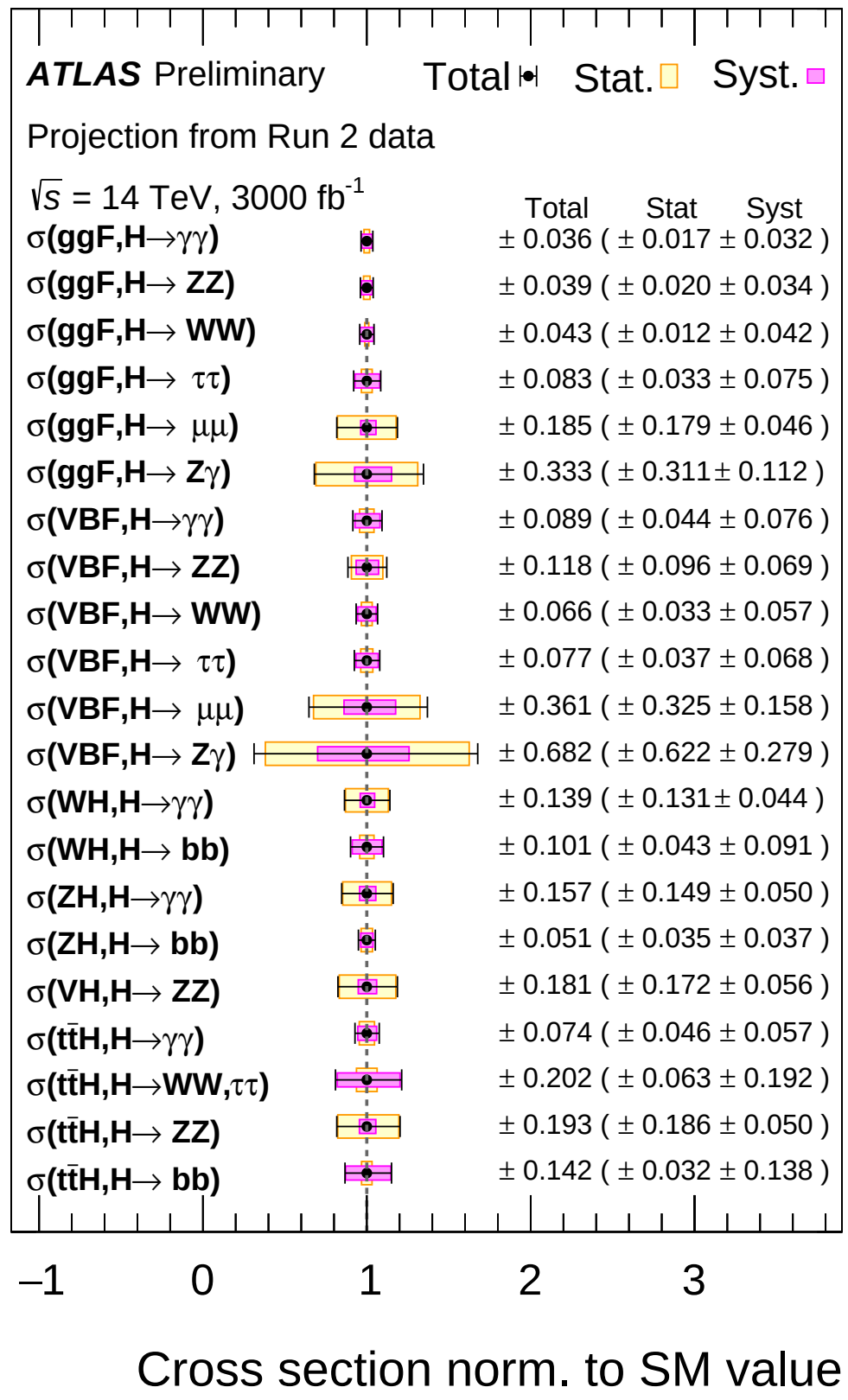

(b) Scenario S2

Figure 23: Expected results for the measured cross sections times branching fraction for the ggF, VBF, $WH$, $ZH$ and $t\bar{t}H$ production modes in each relevant decay channel, normalised to their SM predictions for scenarios S1 (a) and S2 (b). The values are obtained from a simultaneous fit to all decay channels. The black bars, yellow boxes and pink boxes show the total, statistical and systematic uncertainties respectively. For the $Z\gamma$ measurement, dominated by statistical uncertainties, the systematic uncertainties for scenario S2 are assumed to be equal to those used for the scenario S1.

The expected precision of the combined production cross sections times branching fraction for scenarios S1 and S2 is shown in Tables 12,13 and in Figure 23 as well. The total uncertainties for both scenarios are summarised in Figure 24.

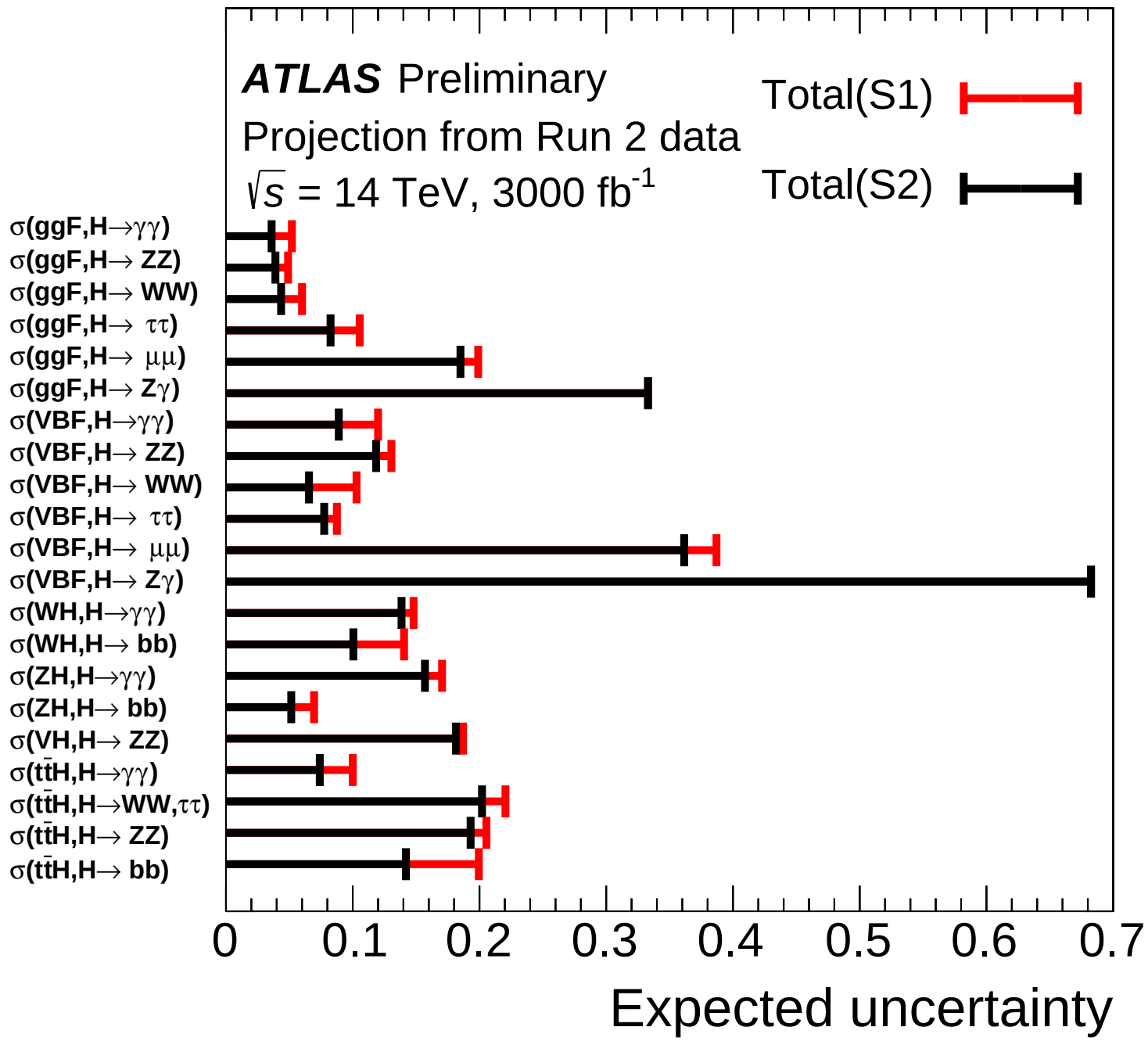

Figure 24: Expected uncertainty on the measurements of the cross sections times branching fraction for the ggF, VBF, $WH$, $ZH$ and $t\bar{t}H$ production modes in the different decay channels, normalised to their SM predictions for scenarios S1 (red) and S2 (black). The values are obtained from a simultaneous fit to all decay channels.

### 3.6 Ratios of cross sections and branching fractions

Ratios of cross sections and of branching fractions are measured using as reference the cross section of the $gg \to H \to ZZ^*$ process, $\sigma^{ZZ}_{ggF}$. The products $(\sigma \times \mathrm{B})_{if}$ of production cross sections in the process $i$ and branching fraction into the final state $f$ are expressed as

$$(\sigma \times \mathrm{B})_{\mathrm{if}} = \sigma^{\mathrm{ZZ}}_{\mathrm{ggF}} \cdot \left(\frac{\sigma_i}{\sigma_{\mathrm{ggF}}}\right) \cdot \left(\frac{\mathrm{B}_f}{\mathrm{B}_{ZZ}}\right), \tag{1}$$

in terms of the ratios of the production cross sections for VBF, $WH$, $ZH$ and $t\bar{t}H$ normalised to that of ggF and the ratios of the branching fractions into the $\gamma\gamma$, $WW^*$, $b\bar{b}$ and $\tau\tau$ final states normalised to that of $H \to ZZ^*$. The uncertainties on the measurements of these parameters for scenario S1 and S2 are shown in Table 14 and well as in Figure 25. The uncertainties for the two scenarios are summarized in Figure 26.

| POI | Scenario | $\Delta_{tot}/\sigma_{SM}$ | $\Delta_{stat}/\sigma_{SM}$ | $\Delta_{exp}/\sigma_{SM}$ | $\Delta_{sig}/\sigma_{SM}$ | $\Delta_{bkg}/\sigma_{SM}$ |
|---|---|---|---|---|---|---|
| $\sigma_{ggF}^{ZZ}$ | HL-LHC S1 | $^{+0.044}_{-0.044}$ | $^{+0.016}_{-0.016}$ | $^{+0.031}_{-0.034}$ | $^{+0.019}_{-0.017}$ | $^{+0.018}_{-0.016}$ |
| | HL-LHC S2 | $^{+0.034}_{-0.034}$ | $^{+0.016}_{-0.016}$ | $^{+0.027}_{-0.027}$ | $^{+0.010}_{-0.009}$ | $^{+0.010}_{-0.009}$ |
| $\sigma_{VBF}/\sigma_{ggF}$ | HL-LHC S1 | $^{+0.065}_{-0.062}$ | $^{+0.026}_{-0.026}$ | $^{+0.031}_{-0.029}$ | $^{+0.044}_{-0.043}$ | $^{+0.025}_{-0.023}$ |
| | HL-LHC S2 | $^{+0.050}_{-0.048}$ | $^{+0.026}_{-0.026}$ | $^{+0.026}_{-0.024}$ | $^{+0.026}_{-0.025}$ | $^{+0.022}_{-0.020}$ |
| $\sigma_{WH}/\sigma_{ggF}$ | HL-LHC S1 | $^{+0.102}_{-0.097}$ | $^{+0.054}_{-0.052}$ | $^{+0.047}_{-0.044}$ | $^{+0.054}_{-0.049}$ | $^{+0.050}_{-0.048}$ |
| | HL-LHC S2 | $^{+0.090}_{-0.086}$ | $^{+0.054}_{-0.052}$ | $^{+0.042}_{-0.040}$ | $^{+0.037}_{-0.034}$ | $^{+0.046}_{-0.045}$ |
| $\sigma_{ZH}/\sigma_{ggF}$ | HL-LHC S1 | $^{+0.106}_{-0.097}$ | $^{+0.051}_{-0.049}$ | $^{+0.043}_{-0.040}$ | $^{+0.051}_{-0.047}$ | $^{+0.064}_{-0.057}$ |
| | HL-LHC S2 | $^{+0.090}_{-0.084}$ | $^{+0.051}_{-0.049}$ | $^{+0.038}_{-0.036}$ | $^{+0.034}_{-0.032}$ | $^{+0.054}_{-0.049}$ |
| $\sigma_{t\bar{t}H}/\sigma_{ggF}$ | HL-LHC S1 | $^{+0.067}_{-0.064}$ | $^{+0.026}_{-0.026}$ | $^{+0.038}_{-0.037}$ | $^{+0.036}_{-0.034}$ | $^{+0.031}_{-0.030}$ |
| | HL-LHC S2 | $^{+0.055}_{-0.053}$ | $^{+0.026}_{-0.026}$ | $^{+0.036}_{-0.034}$ | $^{+0.023}_{-0.022}$ | $^{+0.022}_{-0.021}$ |
| $B_{\gamma\gamma}/B_{ZZ}$ | HL-LHC S1 | $^{+0.061}_{-0.057}$ | $^{+0.020}_{-0.019}$ | $^{+0.053}_{-0.049}$ | $^{+0.018}_{-0.017}$ | $^{+0.016}_{-0.014}$ |
| | HL-LHC S2 | $^{+0.045}_{-0.042}$ | $^{+0.020}_{-0.019}$ | $^{+0.037}_{-0.035}$ | $^{+0.011}_{-0.011}$ | $^{+0.010}_{-0.009}$ |
| $B_{WW}/B_{ZZ}$ | HL-LHC S1 | $^{+0.065}_{-0.061}$ | $^{+0.019}_{-0.018}$ | $^{+0.042}_{-0.038}$ | $^{+0.036}_{-0.034}$ | $^{+0.028}_{-0.027}$ |
| | HL-LHC S2 | $^{+0.049}_{-0.047}$ | $^{+0.019}_{-0.018}$ | $^{+0.036}_{-0.034}$ | $^{+0.020}_{-0.018}$ | $^{+0.019}_{-0.018}$ |
| $B_{\tau\tau}/B_{ZZ}$ | HL-LHC S1 | $^{+0.066}_{-0.062}$ | $^{+0.024}_{-0.024}$ | $^{+0.043}_{-0.038}$ | $^{+0.033}_{-0.033}$ | $^{+0.029}_{-0.026}$ |
| | HL-LHC S2 | $^{+0.053}_{-0.050}$ | $^{+0.024}_{-0.024}$ | $^{+0.037}_{-0.035}$ | $^{+0.023}_{-0.022}$ | $^{+0.019}_{-0.017}$ |
| $B_{bb}/B_{ZZ}$ | HL-LHC S1 | $^{+0.118}_{-0.105}$ | $^{+0.038}_{-0.037}$ | $^{+0.053}_{-0.048}$ | $^{+0.058}_{-0.052}$ | $^{+0.080}_{-0.069}$ |
| | HL-LHC S2 | $^{+0.092}_{-0.084}$ | $^{+0.038}_{-0.037}$ | $^{+0.046}_{-0.043}$ | $^{+0.036}_{-0.032}$ | $^{+0.061}_{-0.054}$ |

Table 14: Expected uncertainties on the measurements of $\sigma_{ggF}^{ZZ}$, of the ratios of production cross sections normalised to $\sigma_{ggF}$ and of the ratios of branching fractions normalised to $B_{ZZ}$ for both systematic scenarios S1 and S2. All measurements are normalised to their SM predictions. The total uncertainties are decomposed into statistical uncertainties (stat), experimental systematic uncertainties (exp), and theory uncertainties in the modeling of the signal (sig) and background (bkg) processes.

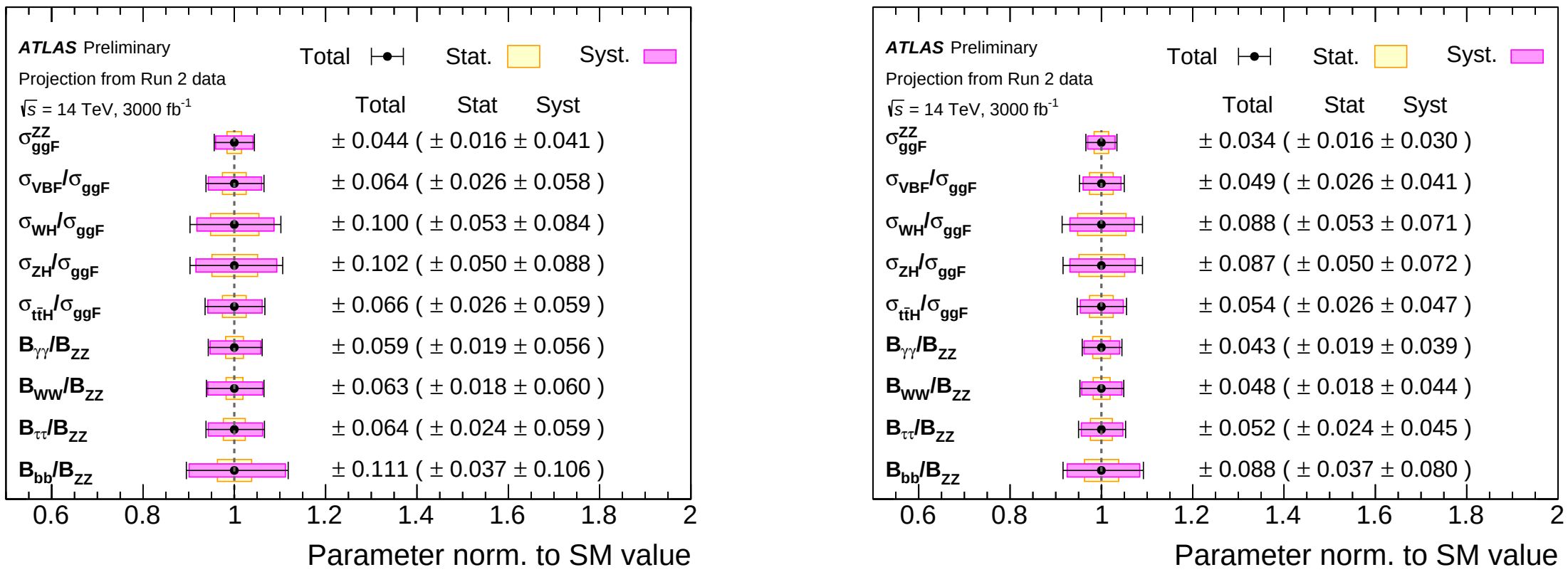

(a) Scenario S1 (b) Scenario S2

Figure 25: Expected result for the measurements of $\sigma^{ZZ}_{\text{ggF}}$, of the ratios of production cross sections normalised to $\sigma_{\text{ggF}}$ and of the ratios of branching fractions normalised to $B_{ZZ}$ for scenarios S1 (a) and S2 (b). The fit results are normalised to the SM predictions. The black error bars, yellow boxes and pink boxes show the total, statistical and systematic uncertainties, respectively.

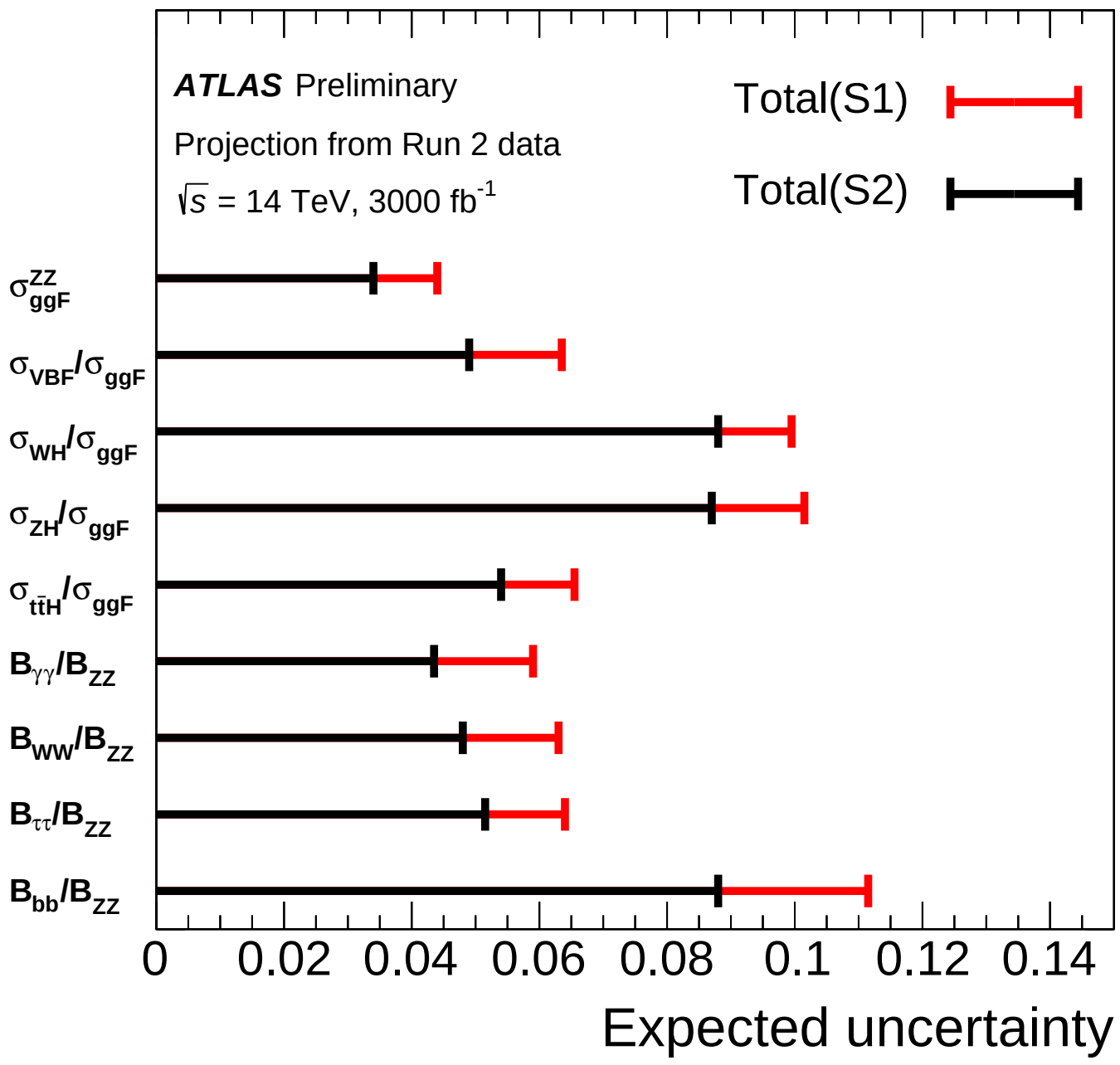

Figure 26: Expected uncertainty on the measurements of $\sigma^{ZZ}_{\text{ggF}}$, of the ratios of production cross sections normalised to $\sigma_{\text{ggF}}$ and of the ratios of branching fractions normalised to $B_{ZZ}$ for scenarios S1 (red) and S2 (black). The fit results are normalised to the SM predictions.

### 3.7 Measurement of coupling parameters in the $\kappa$ framework

This section discusses the measurements of coupling parameters in the "$\kappa$ framework" [3], as already included in the Run 1 combination [1]. In this framework, the $(\sigma \times \mathrm{B})$ for the various Higgs boson production and decay modes are expressed in the narrow-width approximation as

$$\sigma_i \times \mathrm{B}(H \to f) = \frac{\sigma_i \times \Gamma_f}{\Gamma_H} = \frac{\kappa_i^2 \kappa_f^2}{\kappa_H^2} \sigma_i^{\mathrm{SM}} \times \mathrm{B}^{\mathrm{SM}}(H \to f) \tag{2}$$

where $\kappa_i$ and $\kappa_f$ are multiplicative factors applied on the SM production and decay amplitudes respectively, and the factor $\kappa_H^2$ is applied on the total Higgs boson decay width $\Gamma_H$.

The $\kappa_i$, $\kappa_f$ are then expressed in terms of multiplicative factors applied to the Higgs boson couplings to SM particles, using expressions inspired by the leading-order Feynman diagrams for the corresponding processes. Factors related to fundamental couplings in the SM are $\kappa_W$, $\kappa_Z$, $\kappa_t$, $\kappa_b$, $\kappa_\tau$ and $\kappa_\mu$.

Couplings $\kappa_g$ for the ggF vertex, $\kappa_\gamma$ for the $H\gamma\gamma$ vertex and $\kappa_{Z\gamma}$ for the $HZ\gamma$ vertex are expressed either as a function of the more fundamental factors $\kappa_W$, $\kappa_Z$, $\kappa_t$, $\kappa_b$, $\kappa_\tau$ and $\kappa_\mu$ or kept as effective modifiers.

The $\kappa_H^2$ parameter can be expressed as

$$\kappa_H^2 = \frac{\sum_f \kappa_f^2 \mathrm{B}^{\mathrm{SM}}(H \to f)}{1 - \mathrm{B}_{\mathrm{BSM}}} \tag{3}$$

where $\mathrm{B}_{\mathrm{BSM}}$ includes both invisible decays and modifications to visible decays which are not measured in the analyses included in the combination.

The measurement of the $\kappa_j$ requires knowledge of $\kappa_H$: since on-shell Higgs boson $\sigma \times \mathrm{B}$ measurements only measure $\kappa_i/\kappa_H$, the $\kappa_i$ would otherwise be known only up to a common multiplicative factor. Since $\mathrm{B}_{\mathrm{BSM}}$ cannot be unambiguously measured at LHC, this requires specific assumptions. These assumptions can be any of the following:

- Assume $\mathrm{B}_{\mathrm{BSM}} = 0$, so that $\kappa_H$ can be expressed simply in terms of the measured $\kappa_j$.
- Include BSM contributions to the Higgs boson total width through the parameter $\mathrm{B}_{\mathrm{BSM}}$, constrained by assuming $\mathrm{B}_{\mathrm{BSM}} \geq 0$ and $\kappa_{W,Z} \leq 1$. The latter condition holds true for a broad class of extensions of the SM and disfavors large values of $\mathrm{B}_{\mathrm{BSM}}$ [3].
- Use off-shell measurements to constrain the Higgs boson total width and therefore $\kappa_H^2$, as was done in the ATLAS Run 1 combination [1].
- Probe ratios of coupling modifiers, which can be measured without any assumption on the total width of the Higgs boson

The formulas for each case are listed in Ref. [2].

### 3.7.1 Couplings to fermions vs. couplings to weak vector bosons

In this model, we assume a single coupling modifier for all fermions ($\kappa_F$) and for all weak vector bosons ($\kappa_V$). In addition, we assume only SM particles contribute to the total width of the Higgs boson so that $B_{BSM} = 0$. The effective couplings $\kappa_g$ and $\kappa_\gamma$ and the total width modifier $\kappa_H$ are expressed in terms of $\kappa_F$ and $\kappa_V$. The cross sections for the ggF and $t\bar{t}H$ production processes scale with $\kappa_F^2$, while those of VBF and $VH$ productions are proportional to $\kappa_V^2$. The $H \to ZZ^*$ and $H \to WW^*$ partial widths are proportional to $\kappa_V^2$, while those of $H \to b\bar{b}$ and $H \to \tau\tau$ scale with $\kappa_F^2$. The $H \to \gamma\gamma$ branching fraction depends on a combination of $\kappa_V^2$, $\kappa_F^2$, and $\kappa_V \cdot \kappa_F$ due to contributions from top-quark loops, $W$-boson loops and their interference to the decay process. The $\kappa_V$ parameter is assumed to be positive without loss of generality, and $\kappa_F$ is assumed to be positive since its negative range was excluded by previous measurements [20].

The expected contours at 68 and 95% CL in the $\kappa_V$ and $\kappa_F$ plane for scenarios S1 and S2 are shown in Figure 27.

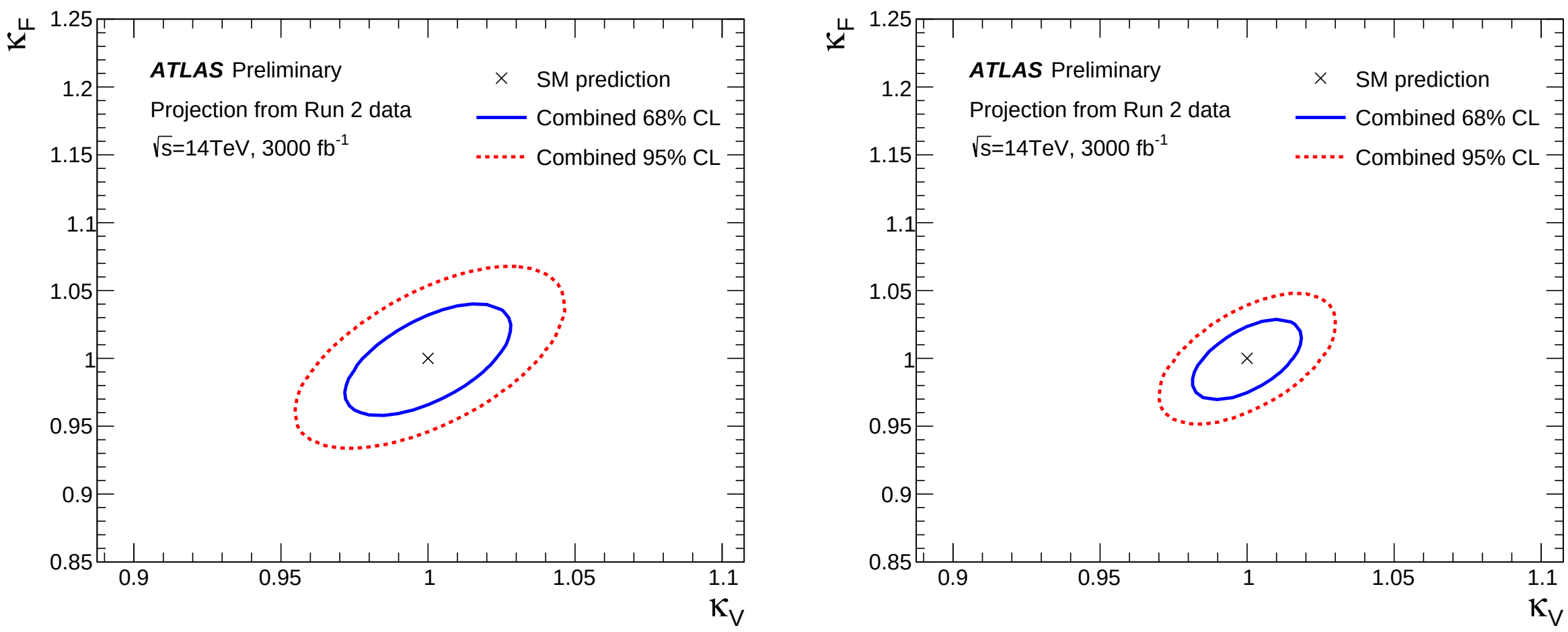

(a) Scenario S1 (b) Scenario S2

Figure 27: Expected contours at 68% and 95% CL in the ($\kappa_F$, $\kappa_V$) plane for scenarios S1 (a) and S2 (b) using a model with single coupling modifiers for fermions and weak vector bosons and assuming $B_{BSM} = 0$.

### 3.7.2 Probe BSM contributions to the production and decay loops

In this model, the coupling modifiers $\kappa_g$ for the ggF vertex and $\kappa_\gamma$ for the $H \to \gamma\gamma$ vertex are used as effective coupling modifiers.

The two modifiers are assumed to be positive without loss of generality. All other coupling modifiers related to SM particles are fixed to their SM values. No BSM contribution is included in the Higgs boson total width. Contours in the plane of $\kappa_\gamma$ and $\kappa_g$ for scenarios S1 and S2 are shown in Figure 28.

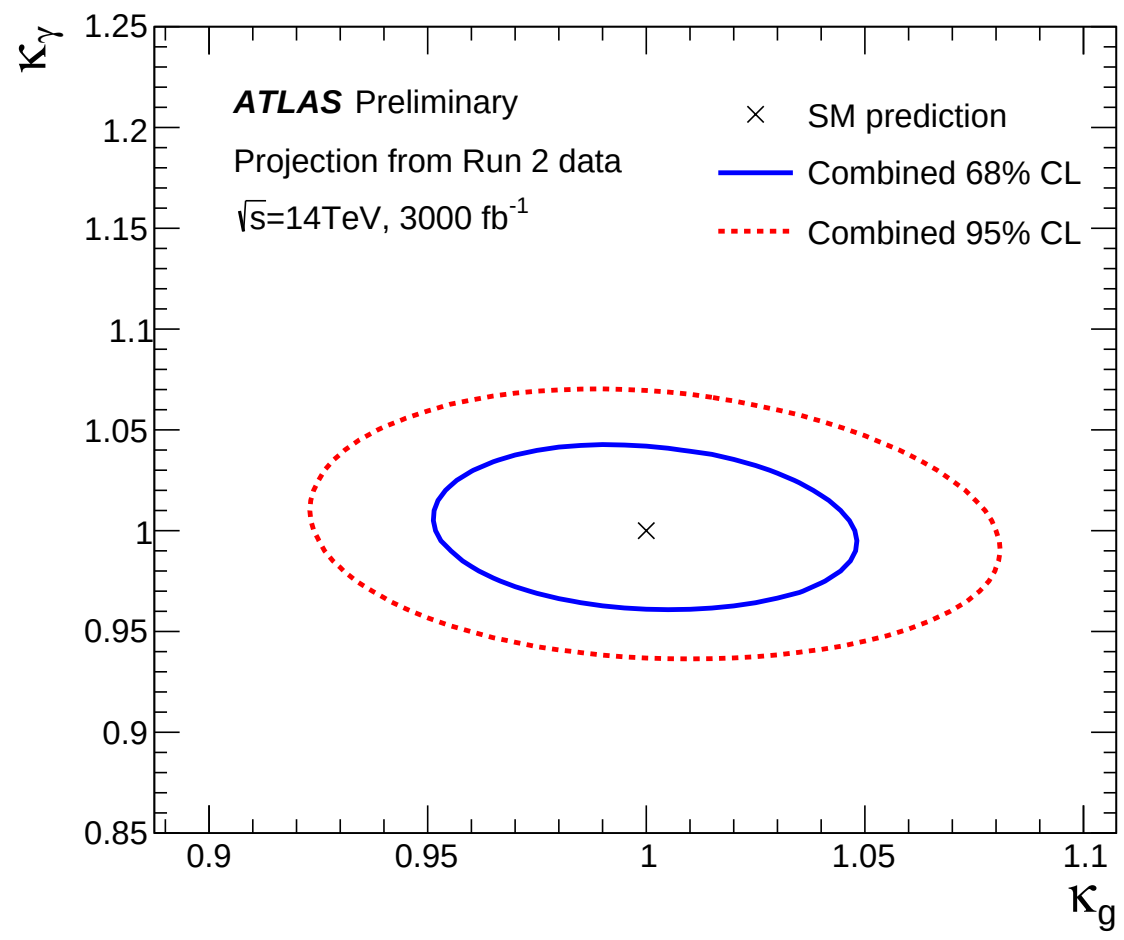

(a) Scenario S1

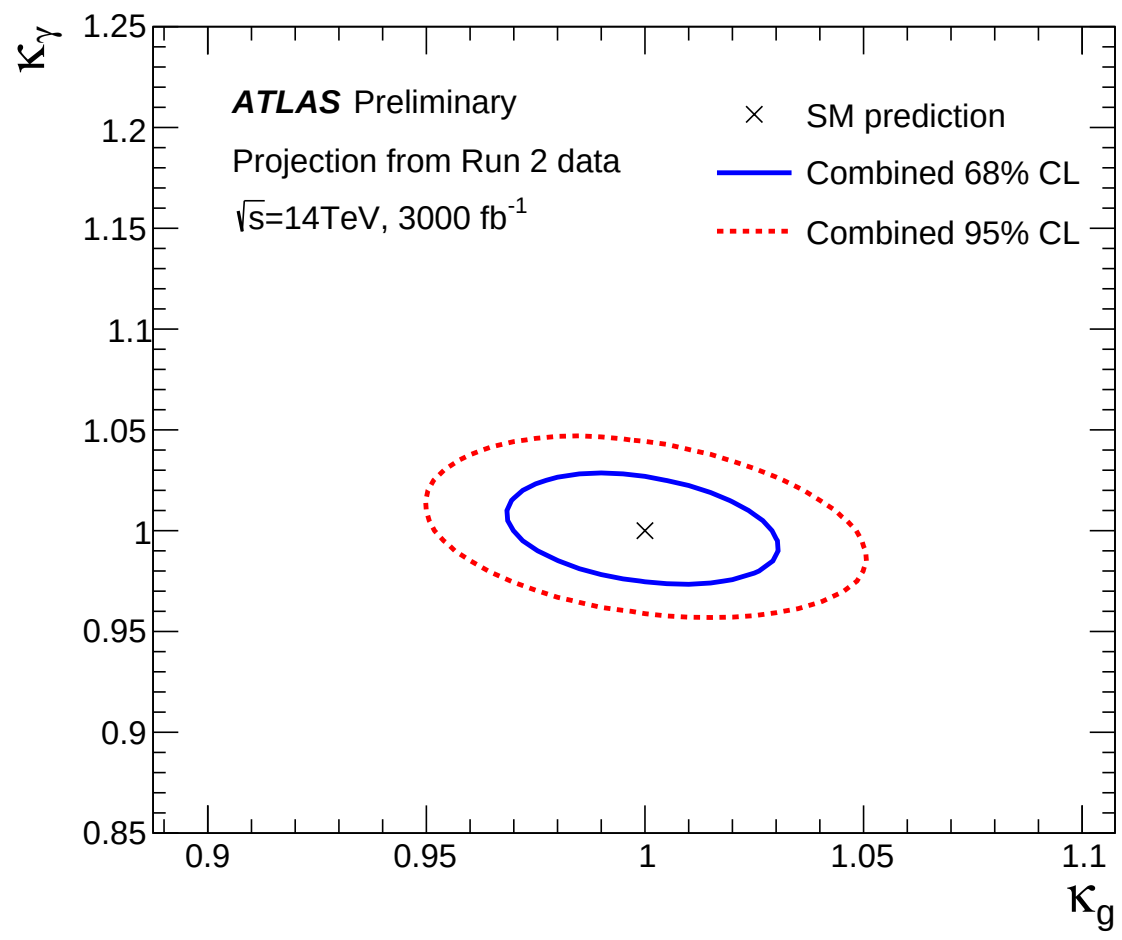

(b) Scenario S2

Figure 28: Expected contours at 68% and 95% CL in the $(\kappa_g, \kappa_\gamma)$ plane for scenarios S1 (a) and S2 (b) using a model with effective coupling modifiers for the ggF and $H \to \gamma\gamma$ loops, with other coupling modifiers fixed to their SM values, and assuming $\mathrm{B_{BSM}} = 0$.

### 3.7.3 Parametrisation assuming SM structure of the loops and no BSM contributions in decays

In this model, separate modifiers $\kappa_W$ and $\kappa_Z$ are considered for couplings to $W$ and $Z$ bosons, respectively. Separate couplings $\kappa_t$, $\kappa_b$, $\kappa_\tau$ and $\kappa_\mu$ are also introduced, respectively, for couplings to top and charm quarks, bottom and strange quarks, $\tau$ leptons, and muons. SM values are assumed for couplings to first-generation fermions and no BSM contribution to the Higgs boson total width is included. All couplings are assumed to be positive except $\kappa_W$ and $\kappa_Z$. The results are shown in Table 15.

Reduced coupling strength modifiers are defined for fermions ($F = t, b, \tau, \mu$) as $\kappa_F \frac{m_F}{v}$, and for gauge bosons ($V = W, Z$) as $\sqrt{\kappa_V}\frac{m_V}{v}$, where $\kappa_F$ ($\kappa_V$) is the coupling modifier, $m_F$ ($m_V$) is the mass of the fermion (boson), and $v = 246$ GeV is the vacuum expectation value of the Higgs field. The SM prediction is given by $m/v$ for both cases, where $m$ is the mass of the fermion or boson. Reduced couplings strengths are shown as a function of mass for the scenarios S1 and S2 in Figure 29.

| POI | Scenario | Precision |
|---|---|---|
| $\kappa_W$ | HL-LHC S1 | $^{+0.028}_{-0.027}$ |
| | HL-LHC S2 | $^{+0.019}_{-0.019}$ |
| $\kappa_Z$ | HL-LHC S1 | $^{+0.026}_{-0.025}$ |
| | HL-LHC S2 | $^{+0.017}_{-0.017}$ |
| $\kappa_t$ | HL-LHC S1 | $^{+0.043}_{-0.041}$ |
| | HL-LHC S2 | $^{+0.030}_{-0.029}$ |
| $\kappa_b$ | HL-LHC S1 | $^{+0.064}_{-0.060}$ |
| | HL-LHC S2 | $^{+0.044}_{-0.043}$ |
| $\kappa_\tau$ | HL-LHC S1 | $^{+0.038}_{-0.036}$ |
| | HL-LHC S2 | $^{+0.028}_{-0.027}$ |
| $\kappa_\mu$ | HL-LHC S1 | $^{+0.079}_{-0.076}$ |
| | HL-LHC S2 | $^{+0.070}_{-0.071}$ |

Table 15: Expected precision of the measurements of $\kappa_Z$, $\kappa_W$, $\kappa_b$, $\kappa_t$, $\kappa_\tau$ and $\kappa_\mu$ for scenarios S1 and S2 using a model where the couplings modifiers $\kappa_F$ and $\kappa_V$ are measured assuming no BSM contribution to the Higgs boson decays, and the SM structure of loop processes such as ggF, $H \to \gamma\gamma$ and $H \to gg$.

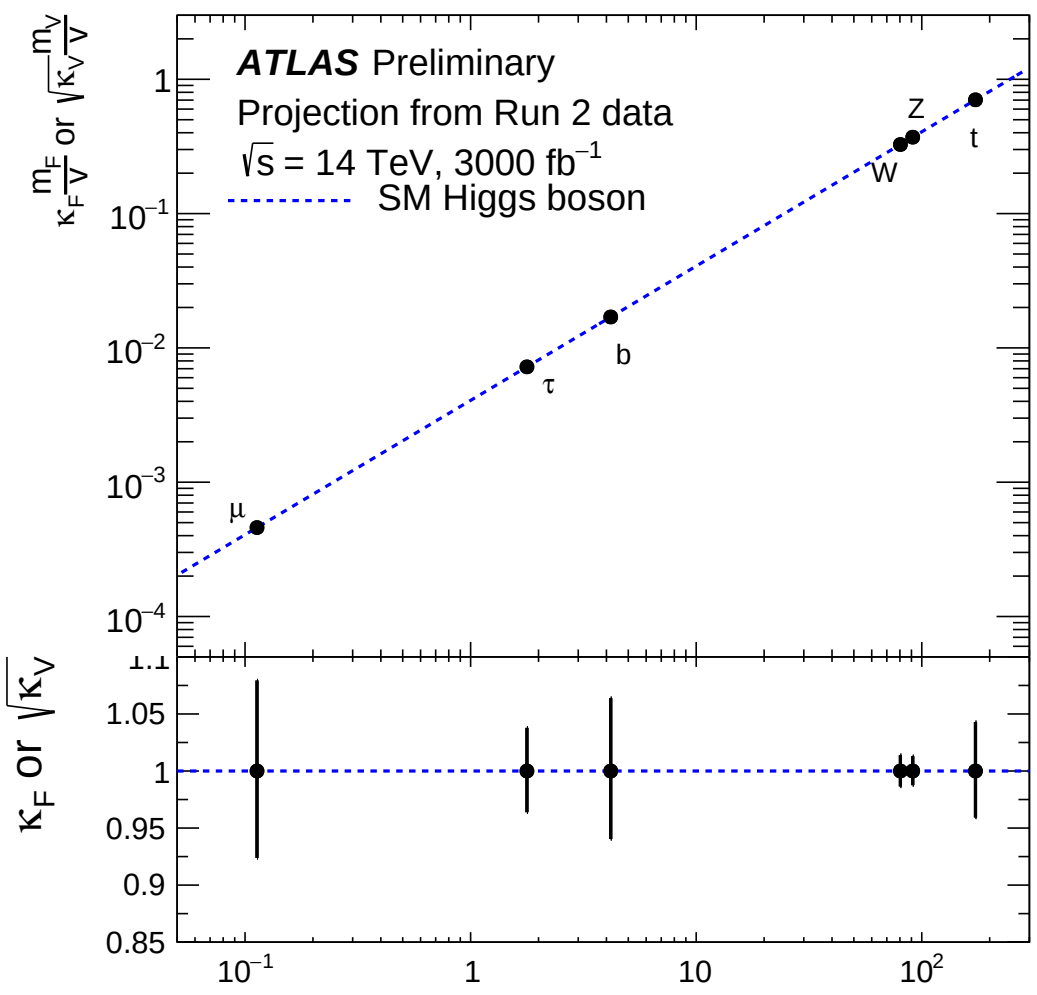

(a) Scenario S1

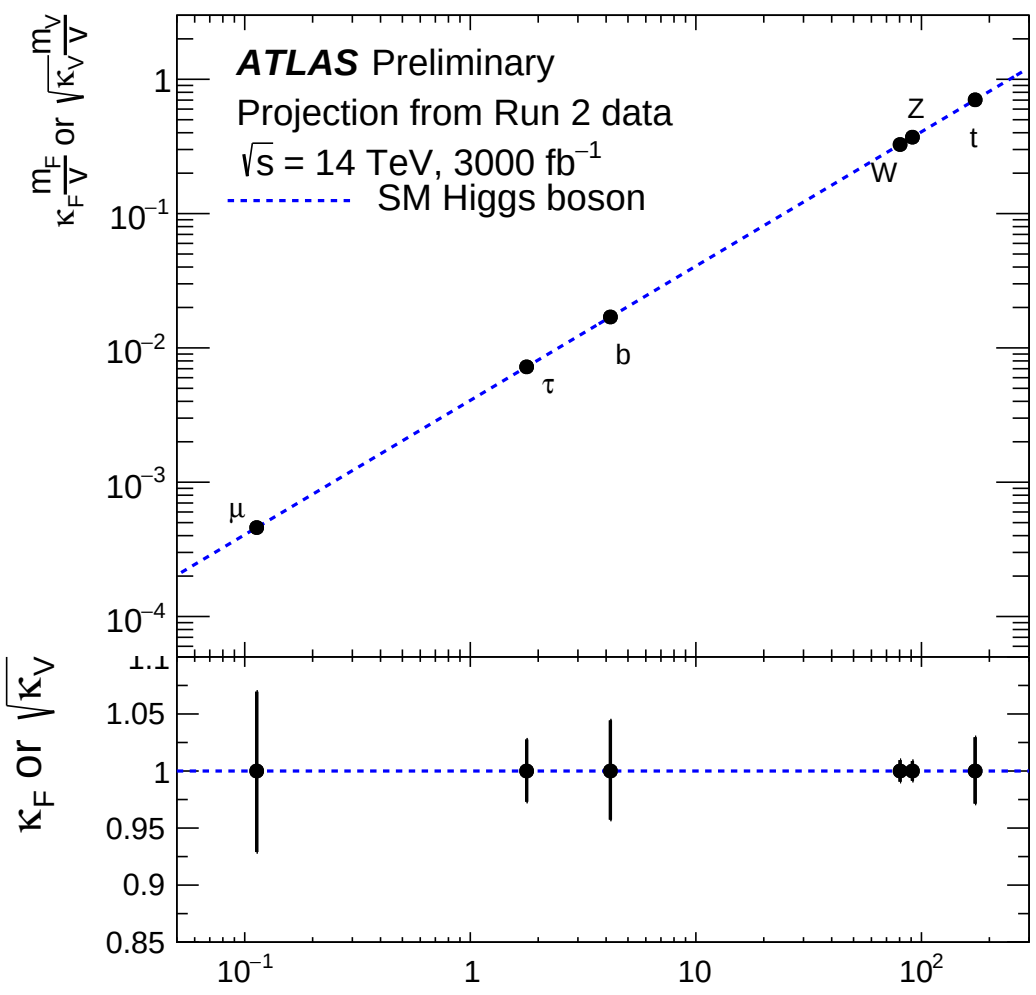

(b) Scenario S2

Figure 29: Reduced coupling strength modifiers $\kappa_F \frac{m_F}{v}$ for fermions ($F = t, b, \tau, \mu$) and $\sqrt{\kappa_V}\frac{m_V}{v}$ for weak gauge bosons ($V = W, Z$) as function of their masses $m_F$ and $m_V$, respectively, and the vacuum expectation value of the Higgs boson field $v = 246$ GeV. The SM prediction for both cases is also shown (dotted line). The uncertainties in the scenarios S1 and S2 are displayed in (a) and (b). The couplings modifiers $\kappa_F$ and $\kappa_V$ are measured assuming no BSM contribution to the Higgs boson decays, and the SM structure of loop processes such as ggF, $H \to \gamma\gamma$ and $H \to gg$.

### 3.7.4 Parametrisation including effective photon and gluon couplings with and without BSM contributions in decays

The two models considered in this section are based on the same parametrisation as the one in Section 3.7.3 but the ggF, $H \to gg$, $H \to \gamma\gamma$ and $H \to Z\gamma$ loop processes are parametrised using the $\kappa_g$, $\kappa_\gamma$ and $\kappa_{Z\gamma}$ modifiers in the same way as for the model of Section 3.7.2.

In the first model, no BSM contribution to the Higgs boson total width are considered ($B_{BSM} = 0$). The measured parameters are $\kappa_Z$, $\kappa_W$, $\kappa_b$, $\kappa_t$, $\kappa_\tau$, $\kappa_\gamma$, $\kappa_g$, $\kappa_\mu$ and $\kappa_{Z\gamma}$. The sign of $\kappa_t$ can be either positive or negative, while $\kappa_Z$ is assumed to be positive without loss of generality. The uncertainties of the first model without BSM contributions in the Higgs boson total width for scenario S1 and S2 are shown in Table 16.

| POI | Scenario | $\Delta_{tot}$ | $\Delta_{stat}$ | $\Delta_{exp}$ | $\Delta_{sig}$ | $\Delta_{bkg}$ |
|---|---|---|---|---|---|---|
| $\kappa_W$ | HL-LHC S1 | $^{+0.032}_{-0.031}$ | $^{+0.008}_{-0.008}$ | $^{+0.014}_{-0.013}$ | $^{+0.019}_{-0.019}$ | $^{+0.020}_{-0.019}$ |
| | HL-LHC S2 | $^{+0.022}_{-0.022}$ | $^{+0.008}_{-0.008}$ | $^{+0.012}_{-0.011}$ | $^{+0.012}_{-0.011}$ | $^{+0.013}_{-0.012}$ |
| $\kappa_Z$ | HL-LHC S1 | $^{+0.026}_{-0.025}$ | $^{+0.008}_{-0.009}$ | $^{+0.011}_{-0.011}$ | $^{+0.019}_{-0.017}$ | $^{+0.012}_{-0.012}$ |
| | HL-LHC S2 | $^{+0.018}_{-0.018}$ | $^{+0.008}_{-0.009}$ | $^{+0.009}_{-0.009}$ | $^{+0.010}_{-0.010}$ | $^{+0.008}_{-0.008}$ |
| $\kappa_t$ | HL-LHC S1 | $^{+0.068}_{-0.058}$ | $^{+0.011}_{-0.011}$ | $^{+0.016}_{-0.016}$ | $^{+0.056}_{-0.041}$ | $^{+0.033}_{-0.036}$ |
| | HL-LHC S2 | $^{+0.043}_{-0.040}$ | $^{+0.011}_{-0.011}$ | $^{+0.014}_{-0.014}$ | $^{+0.028}_{-0.024}$ | $^{+0.026}_{-0.027}$ |
| $\kappa_b$ | HL-LHC S1 | $^{+0.064}_{-0.060}$ | $^{+0.016}_{-0.016}$ | $^{+0.023}_{-0.022}$ | $^{+0.038}_{-0.036}$ | $^{+0.043}_{-0.040}$ |
| | HL-LHC S2 | $^{+0.044}_{-0.043}$ | $^{+0.016}_{-0.016}$ | $^{+0.020}_{-0.020}$ | $^{+0.022}_{-0.021}$ | $^{+0.029}_{-0.028}$ |
| $\kappa_\tau$ | HL-LHC S1 | $^{+0.038}_{-0.037}$ | $^{+0.011}_{-0.011}$ | $^{+0.017}_{-0.016}$ | $^{+0.026}_{-0.025}$ | $^{+0.019}_{-0.018}$ |
| | HL-LHC S2 | $^{+0.028}_{-0.027}$ | $^{+0.011}_{-0.011}$ | $^{+0.016}_{-0.016}$ | $^{+0.016}_{-0.015}$ | $^{+0.013}_{-0.012}$ |
| $\kappa_g$ | HL-LHC S1 | $^{+0.043}_{-0.041}$ | $^{+0.010}_{-0.010}$ | $^{+0.014}_{-0.014}$ | $^{+0.033}_{-0.031}$ | $^{+0.022}_{-0.021}$ |
| | HL-LHC S2 | $^{+0.032}_{-0.030}$ | $^{+0.010}_{-0.010}$ | $^{+0.012}_{-0.011}$ | $^{+0.022}_{-0.021}$ | $^{+0.016}_{-0.016}$ |
| $\kappa_\gamma$ | HL-LHC S1 | $^{+0.038}_{-0.036}$ | $^{+0.009}_{-0.009}$ | $^{+0.025}_{-0.024}$ | $^{+0.022}_{-0.021}$ | $^{+0.015}_{-0.014}$ |
| | HL-LHC S2 | $^{+0.024}_{-0.023}$ | $^{+0.009}_{-0.009}$ | $^{+0.017}_{-0.017}$ | $^{+0.011}_{-0.011}$ | $^{+0.009}_{-0.009}$ |
| $\kappa_\mu$ | HL-LHC S1 | $^{+0.079}_{-0.076}$ | $^{+0.062}_{-0.066}$ | $^{+0.021}_{-0.018}$ | $^{+0.041}_{-0.030}$ | $^{+0.015}_{-0.013}$ |
| | HL-LHC S2 | $^{+0.070}_{-0.071}$ | $^{+0.062}_{-0.066}$ | $^{+0.019}_{-0.016}$ | $^{+0.023}_{-0.018}$ | $^{+0.009}_{-0.009}$ |
| $\kappa_{Z\gamma}$ | HL-LHC S1 | $^{+0.128}_{-0.126}$ | $^{+0.097}_{-0.107}$ | $^{+0.028}_{-0.022}$ | $^{+0.077}_{-0.061}$ | $^{+0.015}_{-0.012}$ |
| | HL-LHC S2 | $^{+0.124}_{-0.123}$ | $^{+0.097}_{-0.107}$ | $^{+0.027}_{-0.022}$ | $^{+0.071}_{-0.056}$ | $^{+0.010}_{-0.008}$ |

Table 16: Expected uncertainties on each Higgs boson coupling modifier per particle type with effective photon, gluon and $Z\gamma$ couplings for scenarios S1 and S2. No BSM contribution to the Higgs boson total width is considered. The systematic uncertainties related to the $Z\gamma$ channel for scenario S2 are assumed to be equal to those used for the scenario S1, since the $Z\gamma$ measurement is dominated by statistical uncertainties.

The uncertainties of the second model which considers BSM in the Higgs boson total width are shown for scenarios S1 and S2 in Table 17. The restricted range of the $W$ and $Z$ couplings translates into asymmetric errors on parameters such as $\kappa_b$ which are largely measured in combination with vector bosons.

| POI | Scenario | $\Delta_{tot}$ | $\Delta_{stat}$ | $\Delta_{exp}$ | $\Delta_{sig}$ | $\Delta_{bkg}$ |
|---|---|---|---|---|---|---|
| $\kappa_W$ | HL-LHC S1 | $^{+0.000}_{-0.030}$ | $^{+0.000}_{-0.008}$ | $^{+0.000}_{-0.013}$ | $^{+0.000}_{-0.019}$ | $^{+0.000}_{-0.018}$ |
| | HL-LHC S2 | $^{+0.000}_{-0.022}$ | $^{+0.000}_{-0.008}$ | $^{+0.000}_{-0.011}$ | $^{+0.000}_{-0.011}$ | $^{+0.000}_{-0.012}$ |
| $\kappa_Z$ | HL-LHC S1 | $^{+0.000}_{-0.025}$ | $^{+0.000}_{-0.008}$ | $^{+0.000}_{-0.011}$ | $^{+0.000}_{-0.017}$ | $^{+0.000}_{-0.012}$ |
| | HL-LHC S2 | $^{+0.000}_{-0.017}$ | $^{+0.000}_{-0.008}$ | $^{+0.000}_{-0.009}$ | $^{+0.000}_{-0.010}$ | $^{+0.000}_{-0.007}$ |
| $\kappa_t$ | HL-LHC S1 | $^{+0.063}_{-0.058}$ | $^{+0.013}_{-0.011}$ | $^{+0.017}_{-0.016}$ | $^{+0.054}_{-0.041}$ | $^{+0.025}_{-0.036}$ |
| | HL-LHC S2 | $^{+0.039}_{-0.040}$ | $^{+0.013}_{-0.011}$ | $^{+0.015}_{-0.014}$ | $^{+0.027}_{-0.024}$ | $^{+0.020}_{-0.026}$ |
| $\kappa_b$ | HL-LHC S1 | $^{+0.043}_{-0.059}$ | $^{+0.013}_{-0.016}$ | $^{+0.018}_{-0.022}$ | $^{+0.028}_{-0.035}$ | $^{+0.023}_{-0.039}$ |
| | HL-LHC S2 | $^{+0.031}_{-0.042}$ | $^{+0.013}_{-0.016}$ | $^{+0.015}_{-0.020}$ | $^{+0.017}_{-0.020}$ | $^{+0.016}_{-0.027}$ |
| $\kappa_\tau$ | HL-LHC S1 | $^{+0.032}_{-0.036}$ | $^{+0.010}_{-0.011}$ | $^{+0.016}_{-0.016}$ | $^{+0.022}_{-0.025}$ | $^{+0.014}_{-0.017}$ |
| | HL-LHC S2 | $^{+0.024}_{-0.027}$ | $^{+0.010}_{-0.011}$ | $^{+0.015}_{-0.016}$ | $^{+0.014}_{-0.015}$ | $^{+0.009}_{-0.012}$ |
| $\kappa_g$ | HL-LHC S1 | $^{+0.042}_{-0.043}$ | $^{+0.012}_{-0.010}$ | $^{+0.013}_{-0.014}$ | $^{+0.036}_{-0.033}$ | $^{+0.013}_{-0.021}$ |
| | HL-LHC S2 | $^{+0.028}_{-0.030}$ | $^{+0.012}_{-0.010}$ | $^{+0.011}_{-0.011}$ | $^{+0.020}_{-0.021}$ | $^{+0.009}_{-0.016}$ |
| $\kappa_\gamma$ | HL-LHC S1 | $^{+0.029}_{-0.035}$ | $^{+0.008}_{-0.009}$ | $^{+0.024}_{-0.024}$ | $^{+0.013}_{-0.013}$ | $^{+0.005}_{-0.019}$ |
| | HL-LHC S2 | $^{+0.020}_{-0.023}$ | $^{+0.008}_{-0.009}$ | $^{+0.016}_{-0.017}$ | $^{+0.008}_{-0.010}$ | $^{+0.004}_{-0.009}$ |
| $\kappa_\mu$ | HL-LHC S1 | $^{+0.078}_{-0.076}$ | $^{+0.062}_{-0.066}$ | $^{+0.021}_{-0.018}$ | $^{+0.041}_{-0.031}$ | $^{+0.009}_{-0.012}$ |
| | HL-LHC S2 | $^{+0.069}_{-0.071}$ | $^{+0.062}_{-0.066}$ | $^{+0.019}_{-0.016}$ | $^{+0.022}_{-0.018}$ | $^{+0.005}_{-0.008}$ |
| $\kappa_{Z\gamma}$ | HL-LHC S1 | $^{+0.127}_{-0.126}$ | $^{+0.097}_{-0.107}$ | $^{+0.028}_{-0.022}$ | $^{+0.069}_{-0.061}$ | $^{+0.034}_{-0.011}$ |
| | HL-LHC S2 | $^{+0.123}_{-0.123}$ | $^{+0.096}_{-0.098}$ | $^{+0.031}_{-0.049}$ | $^{+0.070}_{-0.056}$ | $^{+0.005}_{-0.007}$ |
| $B_{BSM}$ | HL-LHC S1 | $^{+0.049}_{-0.000}$ | $^{+0.014}_{-0.000}$ | $^{+0.019}_{-0.000}$ | $^{+0.034}_{-0.000}$ | $^{+0.026}_{-0.000}$ |
| | HL-LHC S2 | $^{+0.033}_{-0.000}$ | $^{+0.015}_{-0.000}$ | $^{+0.015}_{-0.000}$ | $^{+0.019}_{-0.000}$ | $^{+0.017}_{-0.000}$ |

Table 17: Expected uncertainties on the measurement of each Higgs boson coupling modifier per particle type with effective photon, gluon and $Z\gamma$ couplings for scenarios S1 and S2. $B_{BSM}$ is included as a free parameter in the fit, assuming $B_{BSM} \geq 0$ and $\kappa_{W,Z} \leq 1$. The systematic uncertainties related to the $Z\gamma$ channel for scenario S2 are assumed to be equal to those used for the scenario S1, since the $Z\gamma$ measurement is dominated by statistical uncertainties.

The uncertainties for the models with and without BSM contributions to the Higgs boson total width for scenarios S1 and S2 are summarised in Table 18.

The results with the total, statistical and systematic uncertainties per $\kappa$ for a model not including BSM contributions to the Higgs boson total width are displayed in Figure 30 for scenarios S1 and S2. The same information for the model in which BSM contributions to the Higgs boson total width is provided in Figure 31. Figures 32 and 33 summarise the expected uncertainties for both models without and with BSM contributions. The results in the model with the BSM contributions can be translated into an upper limit for $B_{BSM}$: $B_{BSM} < 0.064$ (0.093) at 95 % CL for scenario S2 (S1).

| Scenario | Scenario S1 | | Scenario S2 | |
|---|---|---|---|---|
| Parameter | no BSM | with BSM | no BSM | with BSM |
| $\kappa_W$ | $^{+0.032}_{-0.031}$ | $-0.030$ | $^{+0.022}_{-0.022}$ | $-0.022$ |
| $\kappa_Z$ | $^{+0.026}_{-0.025}$ | $-0.025$ | $^{+0.018}_{-0.018}$ | $-0.017$ |
| $\kappa_t$ | $^{+0.068}_{-0.058}$ | $^{+0.063}_{-0.058}$ | $^{+0.043}_{-0.040}$ | $^{+0.039}_{-0.040}$ |
| $\kappa_b$ | $^{+0.064}_{-0.060}$ | $^{+0.043}_{-0.059}$ | $^{+0.044}_{-0.043}$ | $^{+0.031}_{-0.042}$ |
| $\kappa_\tau$ | $^{+0.038}_{-0.037}$ | $^{+0.032}_{-0.036}$ | $^{+0.028}_{-0.027}$ | $^{+0.024}_{-0.027}$ |
| $\kappa_g$ | $^{+0.043}_{-0.041}$ | $^{+0.042}_{-0.043}$ | $^{+0.032}_{-0.030}$ | $^{+0.028}_{-0.030}$ |
| $\kappa_\gamma$ | $^{+0.038}_{-0.036}$ | $^{+0.029}_{-0.035}$ | $^{+0.024}_{-0.023}$ | $^{+0.020}_{-0.023}$ |
| $\kappa_\mu$ | $^{+0.079}_{-0.076}$ | $^{+0.078}_{-0.076}$ | $^{+0.070}_{-0.071}$ | $^{+0.069}_{-0.071}$ |
| $\kappa_{Z\gamma}$ | $^{+0.128}_{-0.126}$ | $^{+0.127}_{-0.126}$ | $^{+0.124}_{-0.123}$ | $^{+0.123}_{-0.123}$ |
| $B_{BSM}$ | - | $+0.049$ | - | $+0.033$ |

Table 18: Expected uncertainties on the measurement of each Higgs boson coupling modifier per particle type with effective photon, gluon and $Z\gamma$ couplings for scenarios S1 and S2 either including $B_{BSM}$ as a free parameter or fixing it to zero. The SM corresponds to $B_{BSM}$=0 and all $\kappa$ parameters equal to unity. All parameters except $\kappa_t$ are assumed to be positive. In case BSM contributions are allowed, the conditions $\kappa_{W,Z} \leq 1$ are also applied. The systematic uncertainties related to the $Z\gamma$ channel for scenario S2 are assumed to be equal to those used for the scenario S1, since the $Z\gamma$ measurement is dominated by statistical uncertainties.

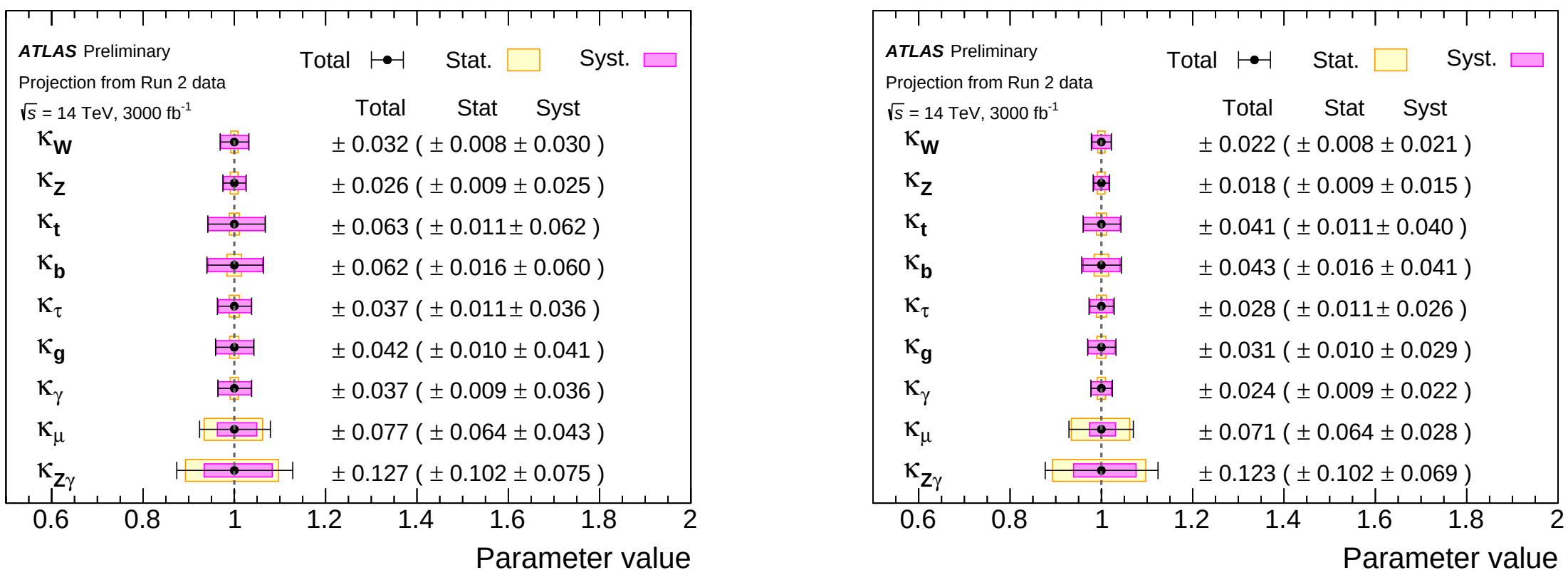

(a) Scenario S1

(b) Scenario S2

Figure 30: Expected result for the measurement of each Higgs boson coupling modifier per particle type with effective photon, gluon and $Z\gamma$ couplings, and without BSM contribution to the Higgs boson total width. The SM corresponds to all $\kappa$ parameters equal to unity. All parameters except $\kappa_t$ are assumed to be positive. Plot (a) corresponds to scenario S1 and (b) to scenario S2.

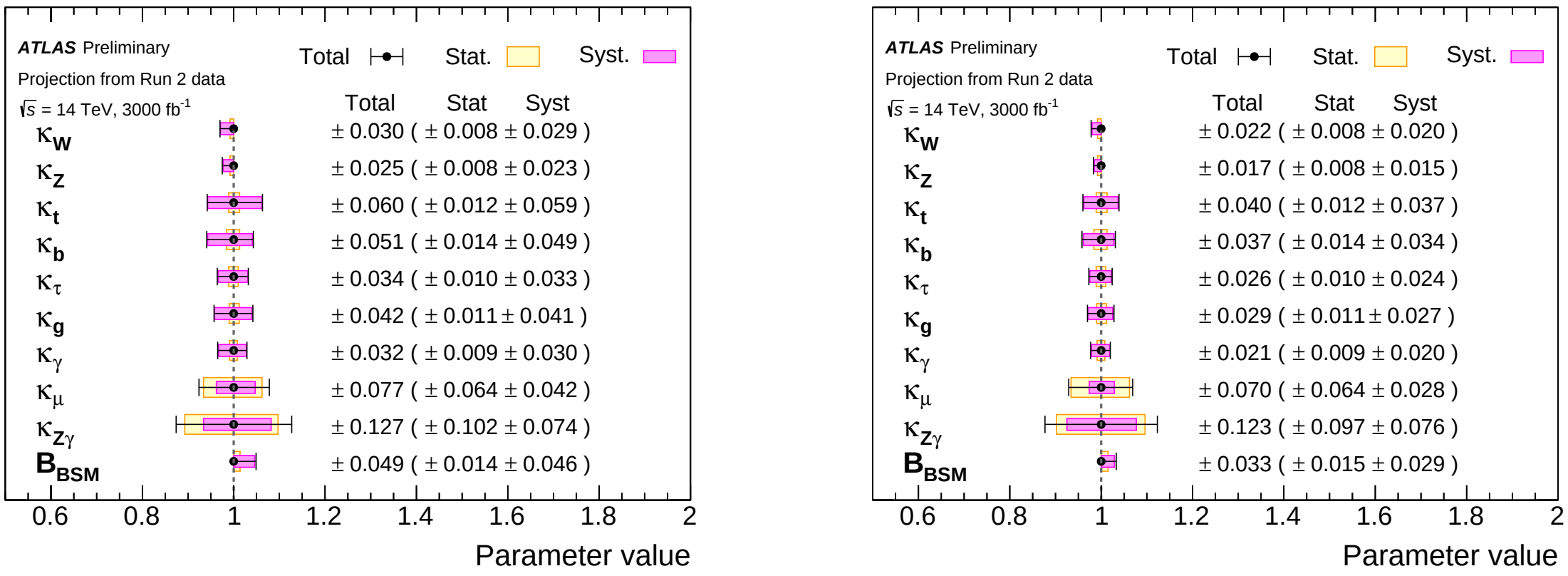

(a) Scenario S1 (b) Scenario S2

Figure 31: Expected result for the measurement of each Higgs boson coupling modifier per particle type with effective photon, gluon and $Z\gamma$ couplings, including BSM contribution to the Higgs boson total width. All parameters except $\kappa_t$ are assumed to be positive. The conditions $\kappa_{W,Z} \leq 1$ are applied. The SM corresponds to $B_{BSM} = 0$ and all $\kappa$ parameters equal to unity. Plot (a) corresponds to scenario S1 and (b) to scenario S2.

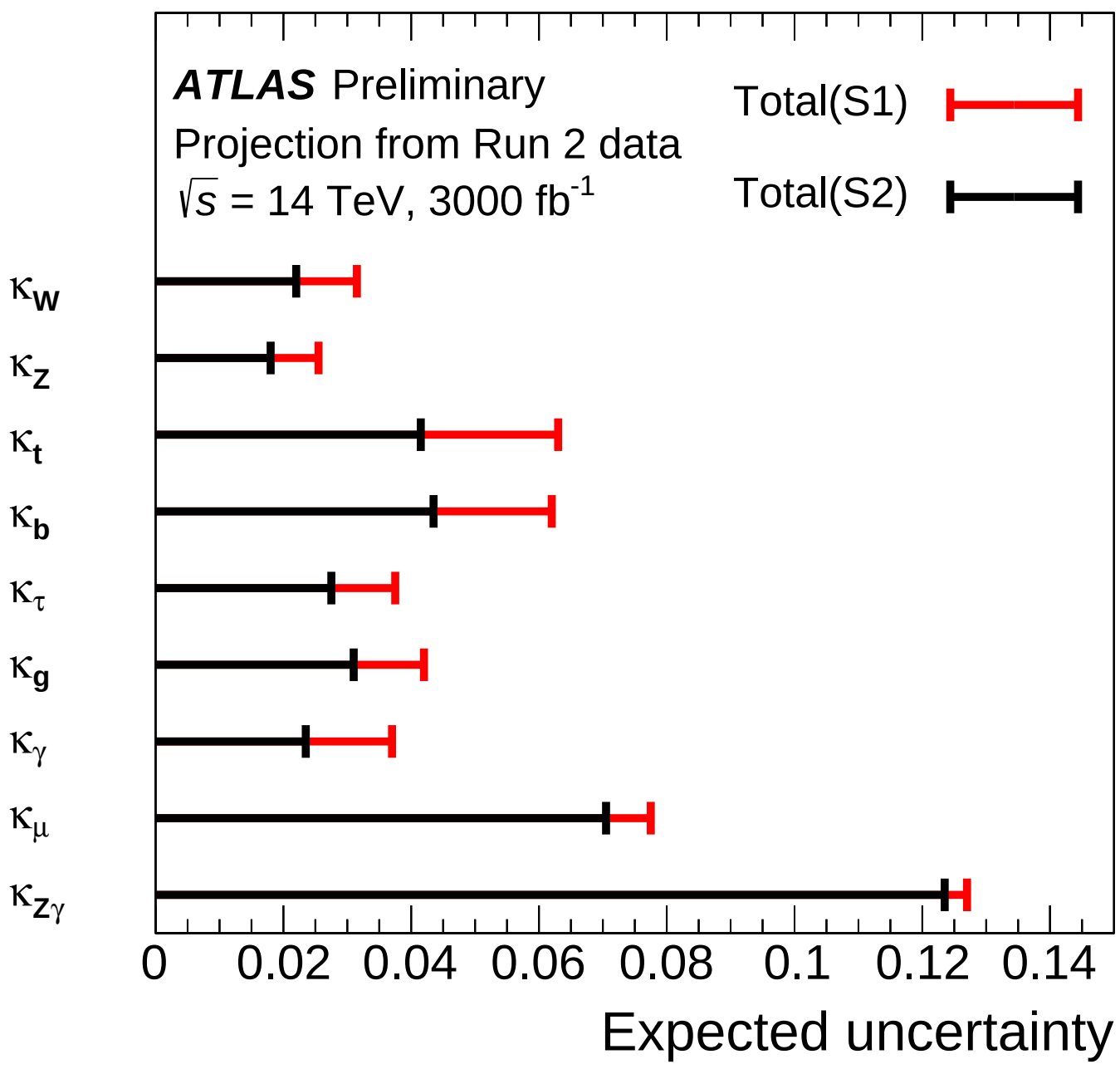

Figure 32: Expected uncertainty on the measurement of each Higgs boson coupling modifier per particle type with effective photon, gluon and $Z\gamma$ couplings, and without BSM contribution in the Higgs boson total width for scenarios S1 (red) and S2 (black). The SM corresponds to all $\kappa$ parameters equal to unity. All parameters except $\kappa_t$ are assumed to be positive.

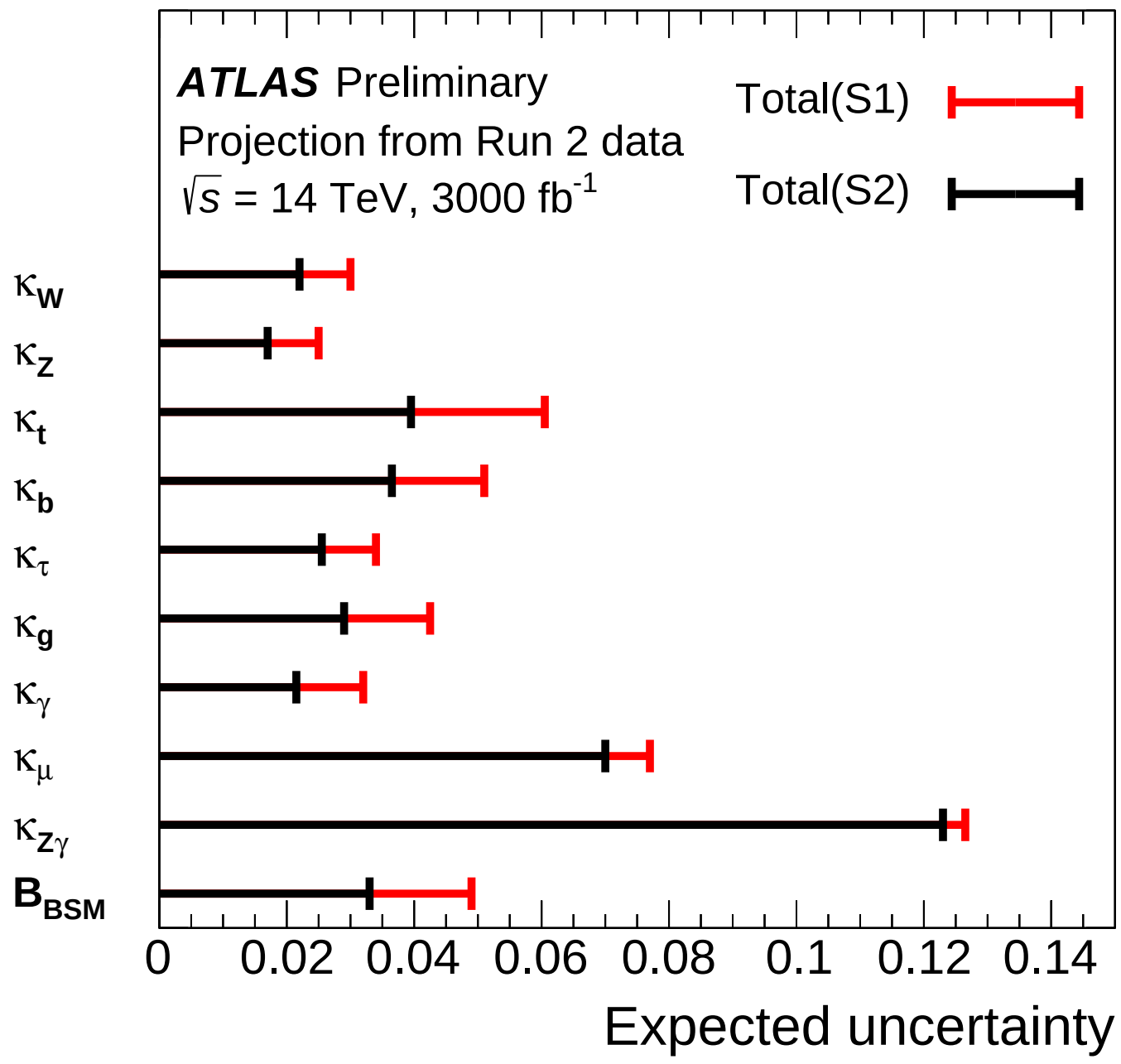

Figure 33: Expected uncertainty on the measurement of each Higgs boson coupling modifier per particle type with effective photon, gluon and $Z\gamma$ couplings, and with BSM contribution in the Higgs boson total width for scenarios S1 and S2. The conditions $\kappa_{W,Z} \leq 1$ are applied. The SM corresponds to $B_{BSM} = 0$ and all $\kappa$ parameters equal to unity.

### 3.7.5 Parametrisation using ratios of coupling modifiers

Finally, a model based on ratios of coupling modifiers is defined analogously to the cross-section ratio model of Section 3.4. The model parameters are the scaling factors defined in Table 19. The parametrisation requires no assumption on the Higgs boson total width. All parameters are assumed to be positive except $\lambda_{tg}$ and $\lambda_{WZ}$. The results for both scenarios S1 and S2 are summarised in Table 20 and Figures 34-35.

| Parameter | Definition in terms of $\kappa$ modifiers |
|---|---|
| $\kappa_{gZ}$ | $\kappa_g \kappa_Z / \kappa_H$ |
| $\lambda_{tg}$ | $\kappa_t / \kappa_g$ |
| $\lambda_{Zg}$ | $\kappa_Z / \kappa_g$ |
| $\lambda_{WZ}$ | $\kappa_W / \kappa_Z$ |
| $\lambda_{\gamma Z}$ | $\kappa_\gamma / \kappa_Z$ |
| $\lambda_{\tau Z}$ | $\kappa_\tau / \kappa_Z$ |
| $\lambda_{bZ}$ | $\kappa_b / \kappa_Z$ |
| $\lambda_{\mu Z}$ | $\kappa_\mu / \kappa_Z$ |
| $\lambda_{Z\gamma Z}$ | $\kappa_b / \kappa_Z$ |

Table 19: Definitions of ratios of coupling modifiers.

| POI | Scenario | $\Delta_{\mathrm{tot}}$ | $\Delta_{\mathrm{stat}}$ | $\Delta_{\mathrm{exp}}$ | $\Delta_{\mathrm{sig}}$ | $\Delta_{\mathrm{bkg}}$ |
|---|---|---|---|---|---|---|
| $\kappa_{gZ}$ | HL-LHC S1 | $^{+0.034}_{-0.033}$ | $^{+0.008}_{-0.008}$ | $^{+0.013}_{-0.016}$ | $^{+0.029}_{-0.027}$ | $^{+0.009}_{-0.009}$ |
| | HL-LHC S2 | $^{+0.022}_{-0.022}$ | $^{+0.008}_{-0.008}$ | $^{+0.013}_{-0.013}$ | $^{+0.015}_{-0.015}$ | $^{+0.005}_{-0.005}$ |
| $\lambda_{tg}$ | HL-LHC S1 | $^{+0.066}_{-0.057}$ | $^{+0.013}_{-0.013}$ | $^{+0.018}_{-0.018}$ | $^{+0.059}_{-0.048}$ | $^{+0.017}_{-0.020}$ |
| | HL-LHC S2 | $^{+0.040}_{-0.038}$ | $^{+0.013}_{-0.013}$ | $^{+0.017}_{-0.017}$ | $^{+0.031}_{-0.028}$ | $^{+0.013}_{-0.014}$ |
| $\lambda_{Zg}$ | HL-LHC S1 | $^{+0.046}_{-0.044}$ | $^{+0.013}_{-0.013}$ | $^{+0.016}_{-0.015}$ | $^{+0.038}_{-0.036}$ | $^{+0.017}_{-0.015}$ |
| | HL-LHC S2 | $^{+0.034}_{-0.033}$ | $^{+0.013}_{-0.013}$ | $^{+0.014}_{-0.014}$ | $^{+0.024}_{-0.024}$ | $^{+0.013}_{-0.013}$ |
| $\lambda_{WZ}$ | HL-LHC S1 | $^{+0.027}_{-0.026}$ | $^{+0.009}_{-0.009}$ | $^{+0.016}_{-0.014}$ | $^{+0.016}_{-0.015}$ | $^{+0.013}_{-0.013}$ |
| | HL-LHC S2 | $^{+0.022}_{-0.021}$ | $^{+0.009}_{-0.009}$ | $^{+0.014}_{-0.014}$ | $^{+0.011}_{-0.010}$ | $^{+0.010}_{-0.009}$ |
| $\lambda_{\gamma Z}$ | HL-LHC S1 | $^{+0.031}_{-0.030}$ | $^{+0.010}_{-0.010}$ | $^{+0.024}_{-0.023}$ | $^{+0.016}_{-0.014}$ | $^{+0.007}_{-0.006}$ |
| | HL-LHC S2 | $^{+0.022}_{-0.021}$ | $^{+0.010}_{-0.010}$ | $^{+0.018}_{-0.017}$ | $^{+0.008}_{-0.007}$ | $^{+0.005}_{-0.004}$ |
| $\lambda_{\tau Z}$ | HL-LHC S1 | $^{+0.035}_{-0.033}$ | $^{+0.012}_{-0.012}$ | $^{+0.019}_{-0.017}$ | $^{+0.023}_{-0.022}$ | $^{+0.014}_{-0.013}$ |
| | HL-LHC S2 | $^{+0.026}_{-0.026}$ | $^{+0.012}_{-0.012}$ | $^{+0.017}_{-0.016}$ | $^{+0.013}_{-0.012}$ | $^{+0.010}_{-0.009}$ |
| $\lambda_{bZ}$ | HL-LHC S1 | $^{+0.054}_{-0.051}$ | $^{+0.017}_{-0.016}$ | $^{+0.023}_{-0.021}$ | $^{+0.032}_{-0.030}$ | $^{+0.034}_{-0.032}$ |
| | HL-LHC S2 | $^{+0.040}_{-0.039}$ | $^{+0.017}_{-0.016}$ | $^{+0.021}_{-0.020}$ | $^{+0.019}_{-0.018}$ | $^{+0.024}_{-0.023}$ |
| $\lambda_{\mu Z}$ | HL-LHC S1 | $^{+0.079}_{-0.076}$ | $^{+0.062}_{-0.066}$ | $^{+0.023}_{-0.019}$ | $^{+0.041}_{-0.030}$ | $^{+0.009}_{-0.008}$ |
| | HL-LHC S2 | $^{+0.069}_{-0.071}$ | $^{+0.062}_{-0.066}$ | $^{+0.021}_{-0.017}$ | $^{+0.021}_{-0.017}$ | $^{+0.005}_{-0.005}$ |
| $\lambda_{Z\gamma Z}$ | HL-LHC S1 | $^{+0.128}_{-0.126}$ | $^{+0.097}_{-0.107}$ | $^{+0.029}_{-0.023}$ | $^{+0.077}_{-0.061}$ | $^{+0.010}_{-0.009}$ |
| | HL-LHC S2 | $^{+0.124}_{-0.123}$ | $^{+0.097}_{-0.107}$ | $^{+0.028}_{-0.022}$ | $^{+0.070}_{-0.056}$ | $^{+0.007}_{-0.003}$ |

Table 20: Expected uncertainty on the measurements of the ratios of Higgs boson coupling modifiers for scenarios S1 and S2. The systematic uncertainties related to the $Z\gamma$ channel for scenario S2 are assumed to be equal to those used for the scenario S1, since the $Z\gamma$ measurement is dominated by statistical uncertainties.

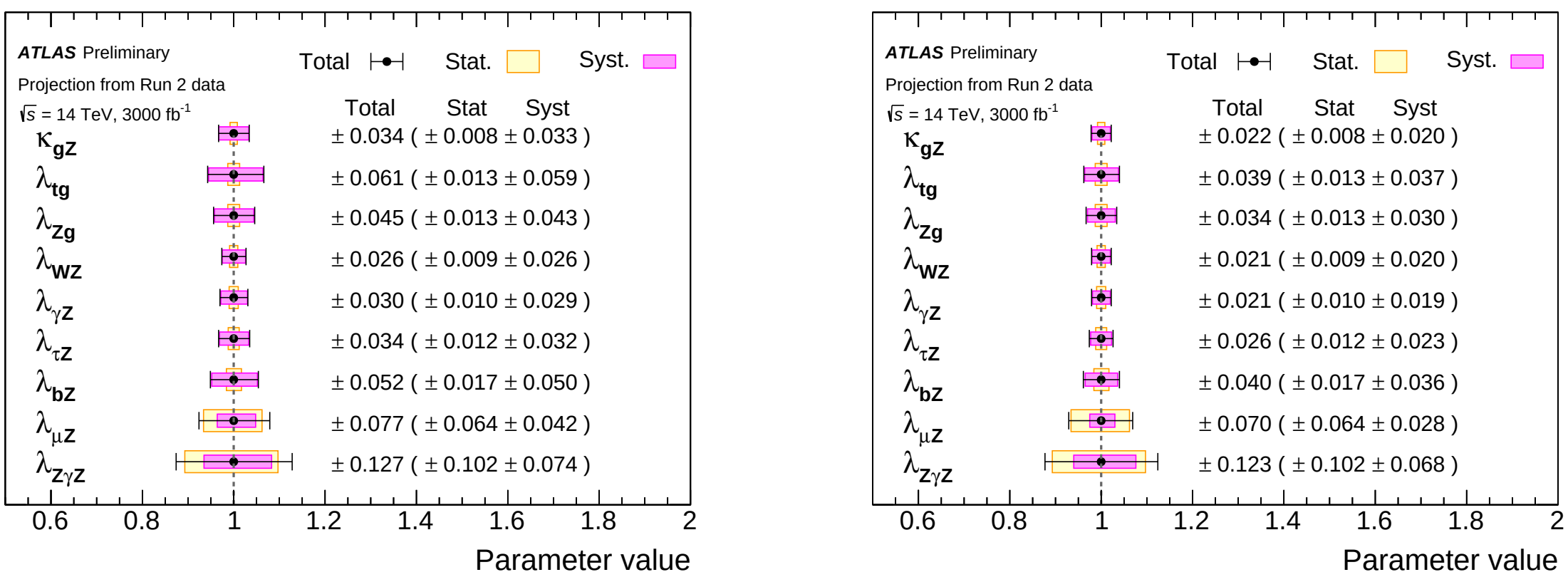

(a) Scenario S1

(b) Scenario S2

Figure 34: Expected result for the measurements of the coupling modifier ratios for scenarios S1 (a) and S2 (b).

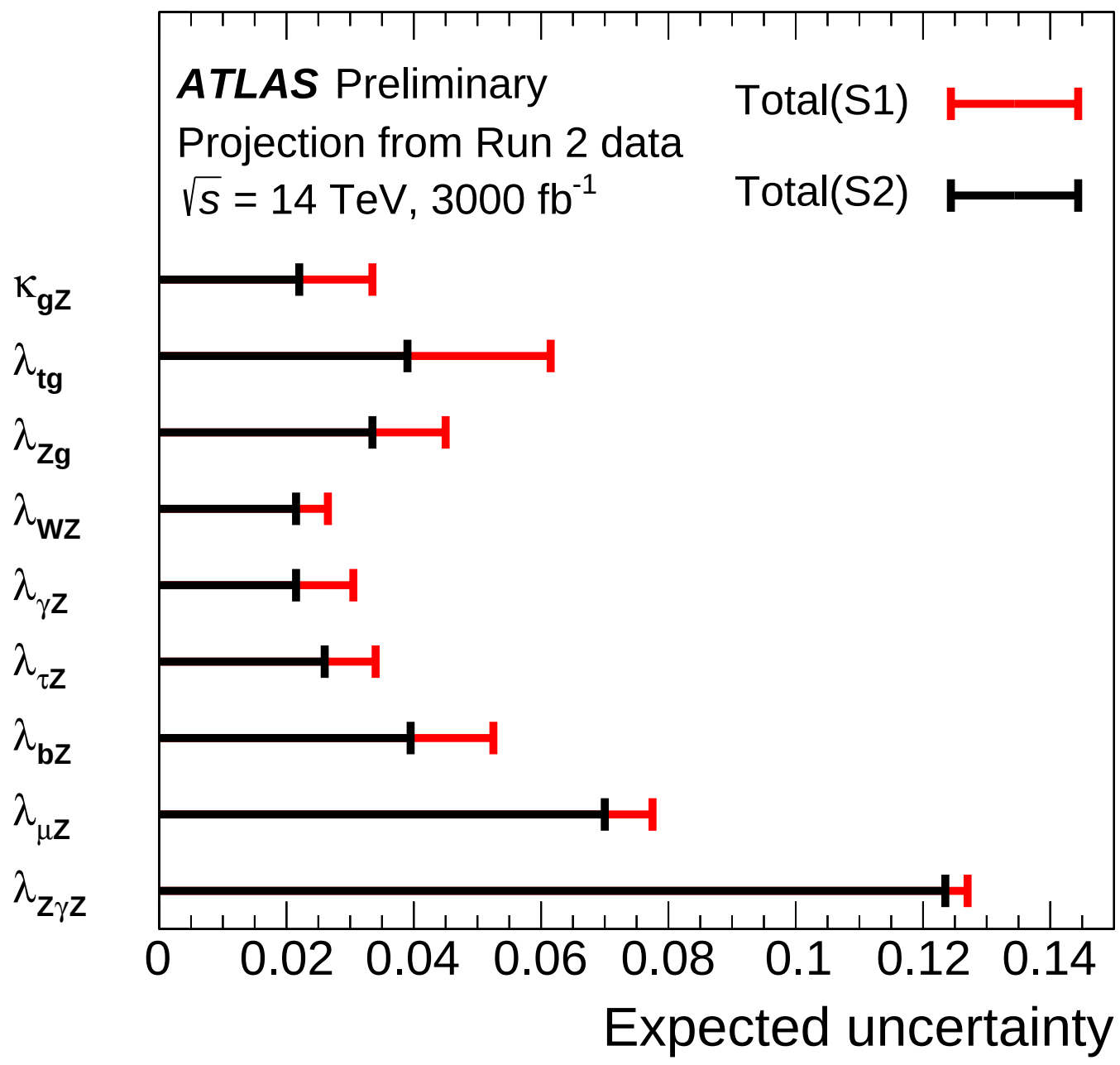

Figure 35: Expected uncertainty on the measurements of ratios of coupling modifiers for scenarios S1 (red) and S2 (black). The systematic uncertainties related to the $Z\gamma$ channel for scenario S2 are assumed to be equal to those used for the scenario S1, since the $Z\gamma$ measurement is dominated by statistical uncertainties.

# 4 Higgs boson mass with $ZZ^* \rightarrow 4$ leptons

The Higgs boson invariant mass has been measured with data collected in 2015 and 2016 (36 fb$^{-1}$) and published in Ref. [21]. The result is $m_H^{ZZ^*} = 124.79 \pm 0.36$ (stat.) $\pm$ 0.05 (syst.) GeV. The above analysis has been extrapolated to 3000 fb$^{-1}$ considering four scenarios. In the first scenario, the current systematic uncertainties and the current detector performance are assumed, as in the S1 scenario considered in this note. For the other scenarios, an improvement of 30 % in the transverse momentum resolution for muons of 45 GeV is considered, as expected thanks to the new tracking detector foreseen to be used at HL-LHC and whose performances are documented in the Technical Design Report for the ATLAS Inner Tracker Pixel Detector [17]. In the last two scenarios, a reduction of 50% and 80% on the muon transverse momentum uncertainty is assumed.

The total, statistical and systematic uncertainties are shown in Table 21.

| | $\Delta_{\text{tot}}$ (MeV) | $\Delta_{\text{stat}}$ (MeV) | $\Delta_{\text{syst}}$ (MeV) |
|---|---|---|---|
| Current Detector | 52 | 39 | 35 |
| $\mu$ momentum resolution improvement by 30% or similar | 47 | 30 | 37 |
| $\mu$ momentum resolution/scale improvement of 30% / 50% | 38 | 30 | 24 |
| $\mu$ momentum resolution/scale improvement 30% / 80% | 33 | 30 | 14 |

Table 21: Expected uncertainty on the measured mass of the Higgs boson for the S1 and upgraded detector scenarios with 3000 fb$^{-1}$ of HL-LHC data.

It should be noted that a detailed study of the calibration of the muons, electrons and photons with the very large HL-LHC sample has not been done. The large dataset available by the end of HL-LHC will give the opportunity to further optimise the analysis and to significantly reduce the systematic uncertainties on the muon transverse momentum scale.

# 5 Conclusion

The measurements of several Higgs boson properties have been extrapolated to the 3000 $fb^{-1}$ of integrated luminosity expected at HL-LHC. This large dataset will allow both to improve the measurement precision of the Higgs boson production and decay modes already observed, and to observe the currently unobserved decay modes $H \to \mu\mu$ and $H \to Z\gamma$.
In this extrapolation the expected improvements on the theory and experimental systematic uncertainties are taken into account. A precision at the level of few percents will be reached on all the production-mode cross sections and on the main decay channels. Only the measurement precision for the $W(Z)H, H \to \gamma\gamma$, $VH, H \to ZZ^*$ and $t\bar{t}H, H \to ZZ^*$ channels, beside the two aforementioned rare decays, will be limited by statistical uncertainty. These results highlight the importance of reaching, and eventually further improving, the expected theory and experimental systematic uncertainties in the next decade.
The projected measurement precision on the cross section times branching ratio is interpreted in terms of Higgs boson couplings to fermions and bosons in a variety of models. Projections of the measured ratios of couplings, branching ratios and cross sections are also reported. Finally, the Higgs boson mass is expected to be measured with a precision of the order of few tens of MeV.

# CMS Physics Analysis Summary

Contact: cms-phys-conveners-ftr@cern.ch

## Sensitivity projections for Higgs boson properties measurements at the HL-LHC

The CMS Collaboration

### Abstract

The expected sensitivities of Higgs boson measurements at the High-Luminosity LHC with integrated luminosities of up to 3000 $\text{fb}^{-1}$ are presented. These are determined by the extrapolation of analyses of 13 TeV collision data, amounting to 35.9 $\text{fb}^{-1}$, collected during Run 2 of the LHC. Projections are given for a combined measurement of coupling modifiers and signal strengths, with additional studies for ttH and VH production with $\text{H} \rightarrow \text{bb}$ decay, and for production in association with a single top quark. Projections are also given for the measurement of the Higgs boson transverse momentum differential cross section, and expected constraints on anomalous couplings and the total width are determined using on- and off-shell $\text{H} \rightarrow \text{ZZ}$ measurements.

# 1 Introduction

The discovery of the Higgs boson in 2012 by the ATLAS and CMS Collaborations [1–3] marked the beginning of a detailed programme to thoroughly measure its properties and to perform consistency tests with the predictions of the standard model (SM). The data delivered in Run 1 and Run 2 of the LHC has provided important measurements of Higgs boson properties including the mass, couplings to fermions and bosons, the tensor structure of the interaction with electroweak gauge bosons and differential production cross sections. So far no significant deviations from the SM predictions have been observed. However, a number of issues with the SM have motivated many theories beyond the SM (BSM) that can alter the properties of the Higgs boson. One such issue is the hierarchy problem, in which fine-tuning is required for the Higgs boson mass to be at the electroweak scale in the presence of large radiative corrections.

A significantly larger data set than currently available offers the possibility of new measurements, such as the Higgs trilinear self-coupling, as well as providing significant gains to existing measurements. A percent-level sensitivity to the couplings would allow discrimination between the SM predictions and many BSM theories. Detailed measurements of the rarer production processes, such as where the Higgs boson is produced in association with top quarks or a vector boson, are an important part of this programme.

The High-Luminosity LHC (HL-LHC) will provide such an opportunity. The instantaneous luminosity will increase substantially, resulting in a data set of 3000 $\text{fb}^{-1}$ by the end of the HL-LHC programme. This increase implies up to 200 pp interactions per bunch crossing, denoted as pileup (PU), and this constitutes a major experimental challenge. The HL-LHC is expected to operate with a centre-of-mass energy of 14 TeV for proton-proton collisions. The studies reported here assume the current 13 TeV energy of Run 2, though the increase in signal and background cross sections at 14 TeV is not expected to have a significant effect on the projected sensitivities.

The CMS detector [4] will be substantially upgraded in order to fully exploit the physics potential offered by the increase in luminosity at the HL-LHC [5], and to cope with the demanding operational conditions this will bring [6–10]. The upgrade of the first level hardware trigger (L1) will allow for an increase of L1 rate and latency to about 750 kHz and 12.5 $\mu$s, respectively, and the high-level software trigger (HLT) is expected to reduce the rate by about a factor of 100 to 7.5 kHz. The entire pixel and strip tracker detectors will be replaced to increase the granularity, reduce the material budget in the tracking volume, improve the radiation hardness, and extend the geometrical coverage and provide efficient tracking up to pseudorapidities of about $|\eta| = 4$. The muon system will be enhanced by upgrading the electronics of the existing cathode strip chambers (CSC), resistive plate chambers (RPC) and drift tubes (DT). New muon detectors based on improved RPC and gas electron multiplier (GEM) technologies will be installed to add redundancy, increase the geometrical coverage up to about $|\eta| = 2.8$, and improve the trigger and reconstruction performance in the forward region. The barrel electromagnetic calorimeter (ECAL) will feature upgraded front-end electronics that will be able to exploit the information from single crystals at the L1 trigger level, to accommodate trigger latency and bandwidth requirements, and to provide 160 MHz sampling allowing high precision timing capability for photons. The hadronic calorimeter (HCAL), consisting in the barrel region of brass absorber plates and plastic scintillator layers, will be read out by silicon photomultipliers (SiPMs). The endcap electromagnetic and hadron calorimeters will be replaced with a new combined sampling calorimeter (HGCal) that will provide highly-segmented spatial information in both transverse and longitudinal directions, as well as high-precision timing information. Finally, the addition of a new timing detector for minimum ionising particles (MTD) in both barrel and endcap regions is envisaged to provide the capability for 4-dimensional reconstruction of interaction vertices that will significantly offset the CMS performance degradation due to high PU rates.

A detailed overview of the CMS detector upgrade programme is presented in Refs. [6–10], while the expected performance of the reconstruction algorithms and pile-up mitigation with the CMS detector is summarised in Ref. [11].

The expected HL-LHC sensitivity for a range of Higgs boson measurements at CMS has previously been studied using both projections of existing analyses and dedicated simulation of the Phase-2 detector using DELPHES [12]. Previous projections have been based on 8 TeV measurements using Run 1 data [6, 13] and more recently on 13 TeV measurements using up to 12.9 $fb^{-1}$ of data [14]. The results presented here are produced in the context of an upcoming CERN Yellow Report (YR18) on the HL-LHC physics potential. They are based on analyses of up to 35.9 $fb^{-1}$ of 13 TeV data extrapolated to an integrated luminosity of 3000 $fb^{-1}$. They utilise the most recent analysis techniques and performance improvements, complementing and in some cases repeating the previous studies. Extrapolations to 300 $fb^{-1}$, the target integrated luminosity for the LHC, are also given for comparison.

The document is organized as follows. Section 2 contains a description of the systematic uncertainty extrapolation scenarios employed. Section 3 presents projections for Higgs boson production and decay rate measurements, as well as for coupling modifiers. These are based on analyses targeting production via the gluon fusion (ggH), vector boson fusion (VBF), vector boson associated (VH, V = W or Z), and top quark pair associated (ttH) modes, with decays into ZZ, WW, $\gamma\gamma$, $\tau\tau$, bb and $\mu\mu$ pairs. Here and throughout this note particles and antiparticles are not distinguished in the notation for particle pairs. Section 4 details the projection for a measurement of Higgs boson production in association with a single top quark (tH). The projected sensitivity to the distribution of the Higgs boson transverse momentum and the derivation of constraints on the couplings are reported in Section 5. Section 6 details the expected sensitivity to anomalous couplings and to the Higgs boson total width using on-shell and off-shell measurements in the $H \to ZZ$ channel.

## 2 Extrapolation procedure

In order to estimate the physics potential of the CMS detector by the end of LHC Run 3 and at the HL-LHC, projections are presented in different scenarios for the evolution of systematic uncertainties with increased data samples and improved theoretical predictions. The baseline scenarios assume that the CMS upgrades will provide the same level of detector and trigger performance as in the Run 2 data taking period [6–10]. Uncertainties due to the limited number of simulated events in the current analyses are neglected, under the assumption that sufficiently large samples of events will be available in future analyses. The two scenarios evaluated are:

- **“Run 2 systematic uncertainties” scenario (S1):** All systematic uncertainties are kept constant with integrated luminosity. The performance of the CMS detector is assumed to be unchanged with respect to the reference analysis;
- **“YR18 systematic uncertainties” scenario (S2):** Theoretical uncertainties are scaled down by a factor of two, while experimental systematic uncertainties are scaled down with the square root of the integrated luminosity until they reach a defined minimum value based on estimates of the achievable accuracy with the upgraded detector [11].

Table 1 summarises the Run 2 uncertainties for which a minimum value is set in S2. Systematic uncertainties in the identification and isolation efficiencies for electrons and muons are expected to be reduced to approximately 0.5%. The hadronic $\tau$ lepton ($\tau_h$) identification uncertainty is assumed to be reduced to approximately 2.5%. The uncertainty in the overall jet energy scale (JES) is expected to reach approximately 1% precision for jets with $p_T > 30$ GeV, driven primarily by improvements in the absolute scale and jet flavour calibrations. The missing transverse momentum uncertainty is obtained by propagating the JES uncertainties in its computation, yielding a reduction by up to a half of the Run 2 uncertainty. For the identification of b-tagged jets the uncertainty in the selection efficiency of b (c) quarks, and in misidentifying a light jet is expected to remain similar to the current level, with only the statistical component reducing with increasing integrated luminosity. The uncertainty in the integrated luminosity of the data sample could be reduced down to 1% thanks to a better understanding of the calibration and fit models employed in its determination, and making use of the finer granularity and improved electronics of the upgraded detectors.

Table 1: The sources of systematic uncertainty for which minimum values are applied in S2.

| Source | Component | Run 2 uncertainty | Projection minimum uncertainty |
|---|---|---|---|
| Muon ID | | 1–2% | 0.5% |
| Electron ID | | 1–2% | 0.5% |
| Photon ID | | 0.5–2% | 0.25–1% |
| Hadronic tau ID | | 6% | 2.5% |
| Jet energy scale | Absolute | 0.5% | 0.1–0.2% |
| | Relative | 0.1–3% | 0.1–0.5% |
| | Pileup | 0–2% | Same as Run 2 |
| | Method and sample | 0.5–5% | No limit |
| | Jet flavour | 1.5% | 0.75% |
| | Time stability | 0.2% | No limit |
| Jet energy res. | | Varies with $p_T$ and $\eta$ | Half of Run 2 |
| MET scale | | Varies with analysis selection | Half of Run 2 |
| b-Tagging | b-/c-jets (syst.) | Varies with $p_T$ and $\eta$ | Same as Run 2 |
| | light mis-tag (syst.) | Varies with $p_T$ and $\eta$ | Same as Run 2 |
| | b-/c-jets (stat.) | Varies with $p_T$ and $\eta$ | No limit |
| | light mis-tag (stat.) | Varies with $p_T$ and $\eta$ | No limit |
| Integrated lumi. | | 2.5% | 1% |

Theoretical uncertainties follow the prescriptions of the LHC Yellow Report 4 [15] in S1 and and are halved in S2 to account for future theoretical developments. In both scenarios the intrinsic statistical uncertainty on any measurement scales with $1/\sqrt{R_L}$, where $R_L$ is the projected integrated luminosity divided by that of the Run 2 analysis.

## 2.1 Statistical treatment

The results in this note are calculated using the standard statistical methods employed by the ATLAS and CMS Collaborations and described in detail in [16]. These are implemented in the ROOFIT [17] and ROOSTATS [18] software frameworks.

Expected uncertainties on parameters of interest (POIs), denoted $\vec{\alpha}$, are defined as the $1\sigma$ confidence level (CL) intervals determined using the profile likelihood ratio test statistic $q(\vec{\alpha})$ [19], in which experimental and theoretical uncertainties are incorporated via nuisance parameters $\vec{\theta}$:

$$q(\vec{\alpha}) = -2\ln\left(\frac{L(\vec{\alpha}, \hat{\hat{\theta}}_{\vec{\alpha}})}{L(\hat{\vec{\alpha}}, \hat{\vec{\theta}})}\right). \tag{1}$$

The quantities $\hat{\vec{\alpha}}$ and $\hat{\vec{\theta}}$ denote the unconditional maximum likelihood estimates of the parameter values, while $\hat{\hat{\theta}}_{\vec{\alpha}}$ denotes the conditional maximum likelihood estimate for fixed values of the parameters of interest $\vec{\alpha}$. The likelihood is evaluated for an Asimov data set [19] defined by the nominal model with the expected signal and background yields scaled to the projection integrated luminosity and with the SM expectation for the POIs. The $1\sigma$ CL interval for the measurement of each POI is determined as the interval for which $q(\vec{\alpha}) < 1$.

The uncertainties calculated with this method can also be decomposed into separate sources. To isolate the contribution from a group of systematic uncertainties, the corresponding nuisance parameters are first fixed to their maximum likelihood estimates, and the calculation of the interval is repeated but with only the remaining nuisance parameter values allowed to vary with $\alpha$. This results in a smaller uncertainty that is subtracted in quadrature from the total to yield the contribution of the chosen group. By extension, the statistical uncertainty on a measurement is defined as the uncertainty obtained when all nuisance parameters are fixed to their maximum likelihood values.

# 3 Production and decay rate signal strengths and coupling modifiers

The projections documented in this section are based on extrapolations of the following analyses:

- $H \to \gamma\gamma$, with ggH, VBF, VH and ttH production [20],
- $H \to ZZ^{(*)} \to 4\ell$, with ggH, VBF, VH and ttH production [21],
- $H \to WW^{(*)} \to \ell\nu\ell\nu$, with ggH, VBF and VH production [22],
- $H \to \tau\tau$, with ggH and VBF production [23],
- VH production with $H \to bb$ decay [24],
- Boosted H production with $H \to bb$ decay [25],
- ttH production with $H \to$ leptons [26],
- ttH production with $H \to bb$ [27, 28],
- $H \to \mu\mu$, with ggH and VBF production [29].

The projected results given in Section 3.1 are based on the combined measurement of these channels [30]. The projections for the ttH(bb) and VH(bb) measurements are studied further in Sections 3.2 and 3.3 respectively. A precise future measurement of ttH production is important as it offers the best direct probe of the top-Higgs Yukawa coupling with minimal model dependence. Measuring the bottom quark coupling precisely is also important due to the large $H \to bb$ branching fraction, around 58% in the SM, which also impacts the achievable precision of couplings from other decay channels. This is best measured using the VH production mode, where the overwhelming multijet background at the LHC is suppressed.

For the projected results at 1000 and 3000 $fb^{-1}$ the signal model in the $H \to \mu\mu$ channel is modified to account for the improved dimuon mass resolution in the Phase-2 tracker upgrade [7]. It is estimated that the reduced material budget and improved spatial resolution of the upgraded tracker will yield a 40% improvement in the relative dimuon mass resolution, for example a reduction from 1.1% to 0.65% for muons in the barrel region.

In the combined results in Section 3.1 the background theory uncertainty in the ttH(bb) analysis is modified such that in the maximum likelihood fit it is not reduced by more than a factor 2 (3) in S1 (S2) with respect to the current uncertainty in the 35.9 $fb^{-1}$ result. This is in order to reflect the expected theory improvements for the background cross-section uncertainty.

## 3.1 Combined measurements

The results in this section are presented under the two systematic uncertainty scenarios S1 and S2 as described in Section 2. Projections are made for three parametrisations of the signal. Two are based on signal strength modifiers $\mu$, defined as the ratio between the measured Higgs boson yield and its SM expectation. One set of parameters $\mu^f$, where $f$ = ZZ, WW, $\gamma\gamma$, $\tau\tau$, bb and $\mu\mu$, are introduced to scale the branching fractions of each decay mode independently, assuming the SM cross sections for the production modes. Another set, $\mu_i$, where $i$ = ggH, VBF, WH, ZH and ttH, scale each production cross section independently, assuming the SM values of the branching fractions. The third parametrisation is based on the coupling modifier, or $\kappa$-framework [31]. A set of coupling modifiers, $\vec{\kappa}$, is introduced to parametrise potential deviations from the SM predictions of the Higgs boson couplings to SM bosons and fermions. For a given production process or decay mode $j$, a coupling modifier $\kappa_j$ is defined such that,

$$\kappa_j^2 = \sigma_j/\sigma_j^{\mathrm{SM}} \quad \text{or} \quad \kappa_j^2 = \Gamma^j/\Gamma^j_{\mathrm{SM}}. \tag{2}$$

In the SM, all $\kappa_j$ values are positive and equal to unity. Six coupling modifiers corresponding to the tree-level Higgs boson couplings are defined: $\kappa_{\mathrm{W}}$, $\kappa_{\mathrm{Z}}$, $\kappa_{\mathrm{t}}$, $\kappa_{\mathrm{b}}$, $\kappa_\tau$ and $\kappa_\mu$. In addition, the effective coupling modifiers $\kappa_{\mathrm{g}}$ and $\kappa_\gamma$ are introduced to describe ggH production and $\mathrm{H} \to \gamma\gamma$ decay loop processes. The total width of the Higgs boson, relative to the SM prediction, varies with the coupling modifiers as $\Gamma_{\mathrm{H}}/\Gamma_{\mathrm{H}}^{\mathrm{SM}} = \sum_j \mathrm{B}^j_{\mathrm{SM}}\kappa_j^2/(1-\mathrm{B}_{\mathrm{BSM}})$, where $\mathrm{B}^j_{\mathrm{SM}}$ is the SM branching fraction for the $\mathrm{H} \to jj$ channel and $\mathrm{B}_{\mathrm{BSM}}$ is the Higgs boson branching fraction to BSM final states. In the results for the $\kappa_j$ parameters presented here $\mathrm{B}_{\mathrm{BSM}}$ is fixed to zero and only decays to SM particles are allowed. Projections are also given for the upper limit on $\mathrm{B}_{\mathrm{BSM}}$ when this restriction is relaxed, in which an additional constraint that $|\kappa_{\mathrm{V}}| < 1$ is imposed. A constraint on $\Gamma_{\mathrm{H}}/\Gamma_{\mathrm{H}}^{\mathrm{SM}}$ is also obtained in this model by treating it as a free parameter in place of one of the other $\kappa$ parameters.

### 3.1.1 Signal strength per-decay mode

The expected $\pm 1\sigma$ uncertainties on the per-decay-mode signal strength parameters in S1 and S2 for 300 $fb^{-1}$ and 3000 $fb^{-1}$ are summarised in Fig. 1 with numerical values given in Table 2. The table additionally gives the breakdown of the uncertainty into four components: statistical, signal theory, background theory and experimental. At 300 $fb^{-1}$ all S2 uncertainties are at or below 10%, with the exception of $\mu^{\mu\mu}$ at 42%. At 3000 $fb^{-1}$ the S2 uncertainties range from 3–4%, again with the exception of that on $\mu^{\mu\mu}$ at 10%. The S1 uncertainties are up to a factor of 1.5 larger than those in S2, reflecting the larger systematic component. The dominant uncertainty contribution is found to vary with the scenario and the integrated luminosity of the projection. At 300 $fb^{-1}$ the statistical, signal and experimental uncertainties tend to be of similar order in S1, whereas in S2 the latter two are reduced and the statistical becomes the largest component.

At 3000 $\mathrm{fb}^{-1}$ the systematic uncertainties generally dominate in both S1 and S2. In S2 the signal theory uncertainty is the largest, or joint-largest, component for all parameters except $\mu^{\mu\mu}$, which remains limited by statistics due to the small $\mathrm{H} \to \mu\mu$ branching fraction. The $\mu^{\mu\mu}$ uncertainty at 3000 $\mathrm{fb}^{-1}$ using the Run 2 dimuon mass resolution instead of the Phase-2 expectation is 14%.

Figures 18 and 19 in Appendix A give the evolution of the uncertainty components for each parameter in S1 and S2. This shows that for many parameters the experimental component reduces continuously with integrated luminosity. This is due to the expected data providing a stronger constraint on some of the systematic uncertainties than that which comes from the external measurements.

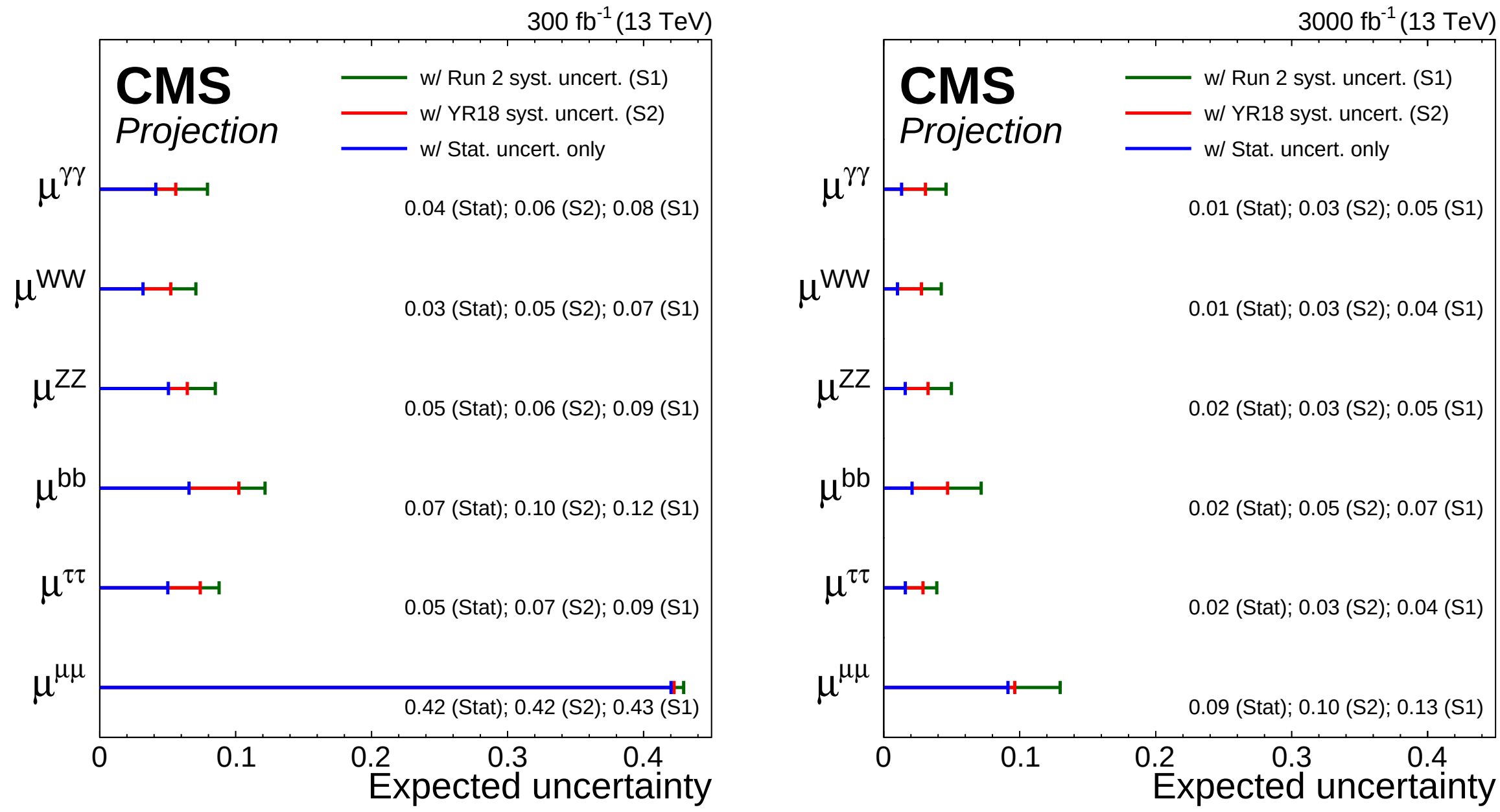

Figure 1: Summary plot showing the total expected $\pm1\sigma$ uncertainties in S1 (with Run 2 systematic uncertainties [30]) and S2 (with YR18 systematic uncertainties) on the per-decay-mode signal strength parameters for 300 $\mathrm{fb}^{-1}$ (left) and 3000 $\mathrm{fb}^{-1}$ (right). The statistical-only component of the uncertainty is also shown.

Another important aspect of the projected measurements is how the correlations between the measured parameters are expected to evolve. Correlations arise when analysis channels are sensitive to more than one production or decay mode and the chosen fit observables do not fully distinguish between these. In addition, correlations may arise when the same systematic uncertainties apply to multiple production or decay modes. Figure 2 shows the correlation coefficients between the signal strength parameters in S2 for 300 $\mathrm{fb}^{-1}$ and 3000 $\mathrm{fb}^{-1}$. At 300 $\mathrm{fb}^{-1}$ the correlations are small, at most +0.2, since the statistical uncertainties are relatively large and each decay channel is measured in dedicated analyses with low contamination from other final states. At 3000 $\mathrm{fb}^{-1}$ the correlations increase up to +0.44, and is largest between modes where the sensitivity is dominated by gluon-fusion production. This reflects the impact of the theory uncertainties affecting the SM prediction of the gluon-fusion production rate.

### 3.1.2 Signal strength per-production mode

The expected $\pm1\sigma$ uncertainties on the per-production-mode signal strength parameters in S1 and S2 for 300 $\mathrm{fb}^{-1}$ and 3000 $\mathrm{fb}^{-1}$ are summarised in Fig. 3 with numerical values given in

Table 2: The expected $\pm 1\sigma$ uncertainties, expressed as percentages, on the per-decay-mode signal strength parameters. Values are given for both S1 (with Run 2 systematic uncertainties [30]) and S2 (with YR18 systematic uncertainties). The total uncertainty is decomposed into four components: statistical (Stat), signal theory (SigTh), background theory (BkgTh) and experimental (Exp).

| | | 300 fb$^{-1}$ uncertainty [%] | | | | | 3000 fb$^{-1}$ uncertainty [%] | | | | |
|---|---|---|---|---|---|---|---|---|---|---|---|
| | | Total | Stat | SigTh | BkgTh | Exp | Total | Stat | SigTh | BkgTh | Exp |
| $\mu^{\gamma\gamma}$ | S1 | 7.9 | 4.1 | 4.8 | 0.3 | 4.8 | 4.6 | 1.3 | 3.5 | 0.3 | 2.6 |
| | S2 | 5.6 | 4.1 | 2.7 | 0.3 | 2.6 | 3.1 | 1.3 | 2.1 | 0.3 | 1.7 |
| $\mu^{WW}$ | S1 | 7.1 | 3.2 | 4.9 | 1.8 | 3.5 | 4.2 | 1.0 | 3.7 | 1.0 | 1.4 |
| | S2 | 5.2 | 3.2 | 2.7 | 1.4 | 2.8 | 2.8 | 1.0 | 2.2 | 0.9 | 1.1 |
| $\mu^{ZZ}$ | S1 | 8.5 | 5.1 | 5.1 | 0.4 | 4.5 | 5.0 | 1.6 | 3.5 | 1.9 | 2.5 |
| | S2 | 6.4 | 5.1 | 2.9 | 0.3 | 2.7 | 3.3 | 1.6 | 2.1 | 0.7 | 1.7 |
| $\mu^{bb}$ | S1 | 12.2 | 6.6 | 4.8 | 7.0 | 5.6 | 7.2 | 2.1 | 5.4 | 3.6 | 2.3 |
| | S2 | 10.2 | 6.6 | 2.4 | 5.6 | 4.9 | 4.7 | 2.1 | 2.5 | 2.9 | 1.7 |
| $\mu^{\tau\tau}$ | S1 | 8.8 | 5.0 | 5.1 | 0.9 | 5.0 | 3.9 | 1.6 | 2.6 | 1.5 | 1.9 |
| | S2 | 7.4 | 5.0 | 3.3 | 0.9 | 4.3 | 2.9 | 1.6 | 1.8 | 0.6 | 1.4 |
| $\mu^{\mu\mu}$ | S1 | 43.0 | 42.0 | 5.7 | 0.8 | 5.9 | 13.0 | 9.1 | 5.2 | 0.8 | 7.6 |
| | S2 | 42.2 | 42.0 | 3.0 | 0.8 | 2.6 | 9.6 | 9.1 | 2.6 | 0.8 | 1.7 |

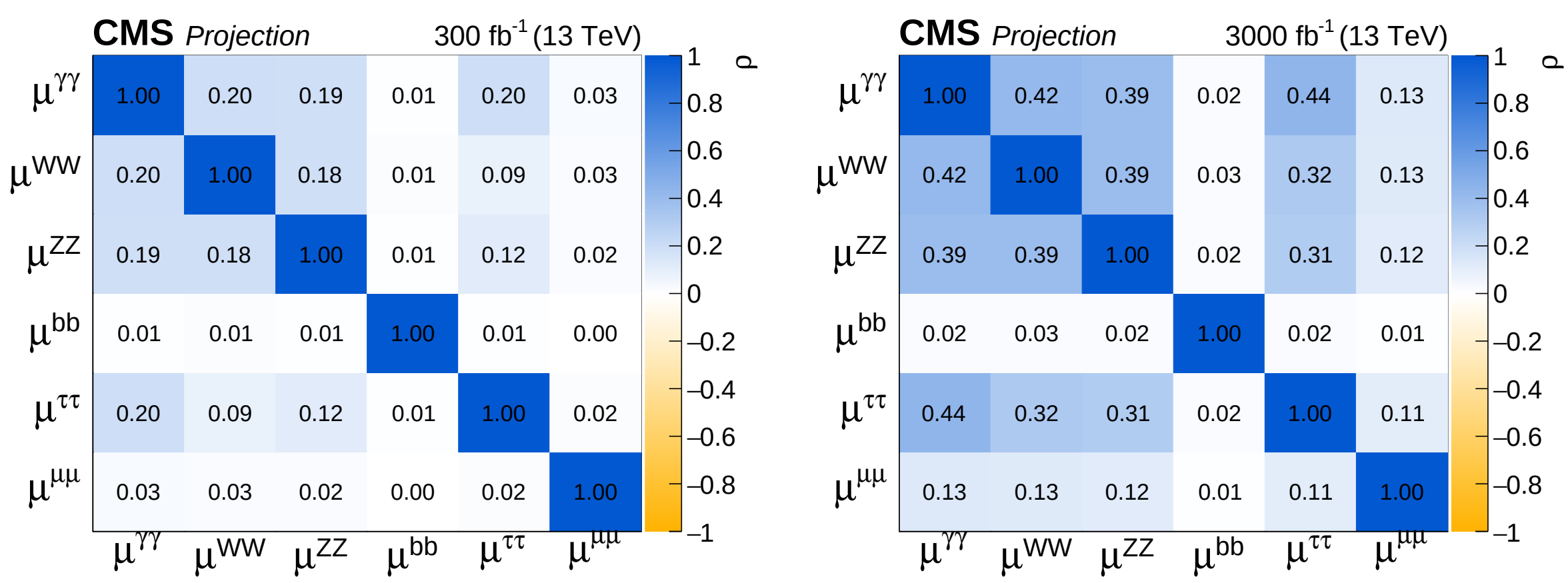

Figure 2: Correlation coefficients ($\rho$) between parameters in the signal strength per-decay-mode parametrisation for S2 (with YR18 systematic uncertainties) at 300 fb$^{-1}$ (left) and 3000 fb$^{-1}$ (right).

Table 3. The projections for 300 fb$^{-1}$ show that $\mu_{\text{ggH}}$ and $\mu_{\text{ttH}}$ will be limited by the signal theory uncertainty in S1. In S2, where this uncertainty is halved, it remains the largest component for $\mu_{\text{ggH}}$ whereas for $\mu_{\text{ttH}}$ it becomes the smallest, with the statistical, background theory and experimental contributions all at the 5–6% level. The other production modes are statistically limited in both scenarios. At 3000 fb$^{-1}$ in S1 the signal theory is the main contribution for all modes except WH which remains limited by statistics. In S2 $\mu_{\text{VBF}}$ and $\mu_{\text{WH}}$ are also statistically limited. Figures 20 and 21 in Appendix A show the evolution of the uncertainty components for each parameter in S1 and S2.

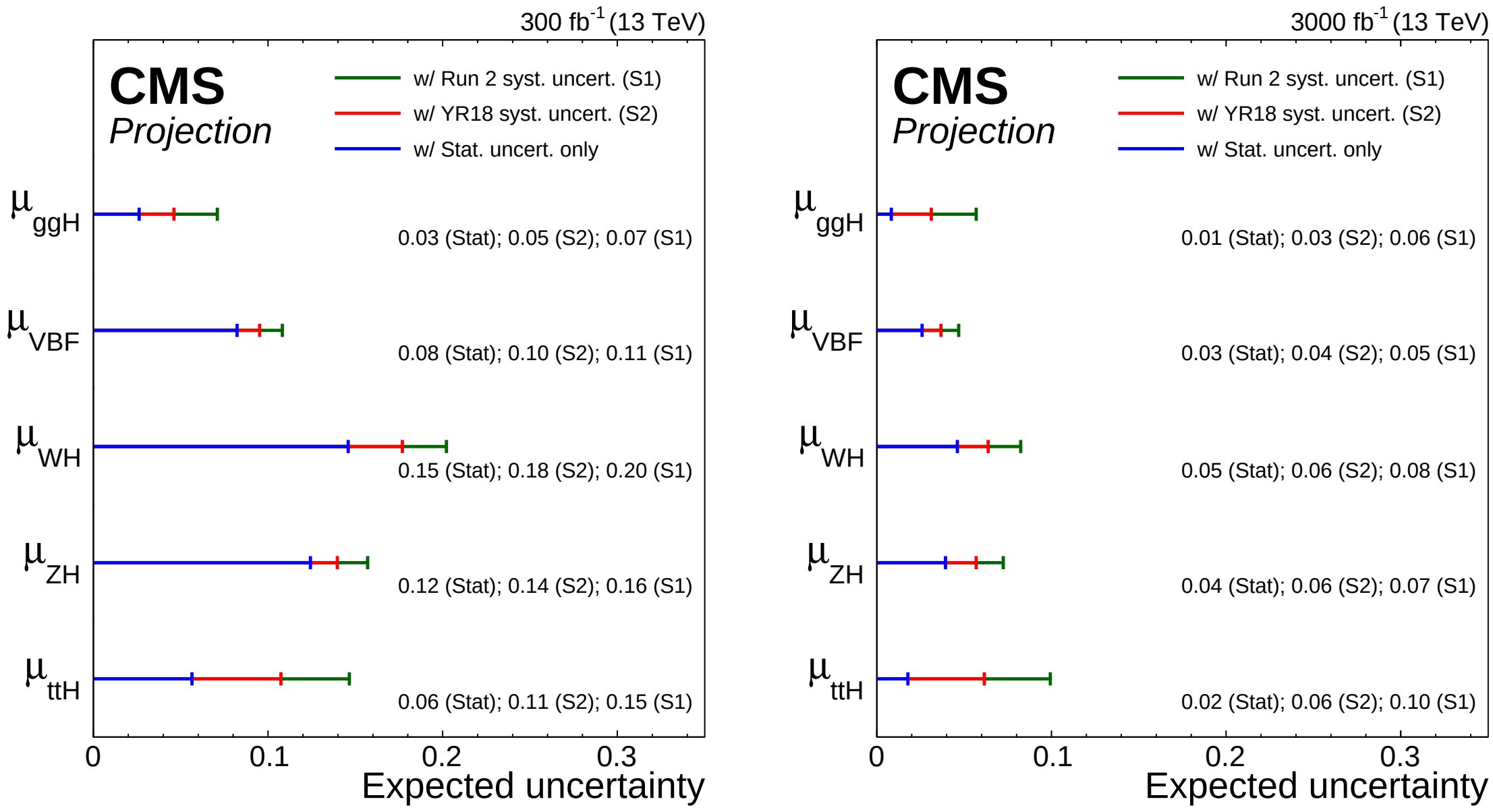

Figure 3: Summary plot showing the total expected $\pm 1\sigma$ uncertainties in S1 (with Run 2 systematic uncertainties [30]) and S2 (with YR18 systematic uncertainties) on the per-production-mode signal strength parameters for 300 fb$^{-1}$ (left) and 3000 fb$^{-1}$ (right). The statistical-only component of the uncertainty is also shown.

Figure 4 shows the correlation coefficients between the signal strength parameters in S2 for 300 fb$^{-1}$ and 3000 fb$^{-1}$. The correlations in this case are small compared to the per-decay measurements since production modes are generally well-isolated by independent analysis categories and the main theoretical uncertainties on the SM signal expectation are uncorrelated.

### 3.1.3 Coupling modifiers

The expected uncertainties for the coupling modifier parametrisation are summarised in Fig. 5 with numerical values given in Table 4. At 300 fb$^{-1}$ the total uncertainties on the $\kappa$ parameters, assuming $B_{\text{BSM}} = 0$, range from 4–10%, and at 3000 fb$^{-1}$ from 2–5%, with the exception of $\kappa_\mu$ which is 22% and 5% respectively in S2. The largest uncertainty component at 3000 fb$^{-1}$ is generally the signal theory in S1, whereas in S2 all four components contribute at a similar level for $\kappa_\gamma$, $\kappa_W$, $\kappa_Z$ and $\kappa_\tau$. The signal theory remains the main component for $\kappa_t$ and $\kappa_g$, and $\kappa_\mu$ is limited by statistics. Figures 22 and 23 in Appendix A show the evolution of the uncertainty components for each $\kappa$ parameter in S1 and S2.

Table 4 also gives the expected uncertainties on $B_{\text{BSM}}$ and $\Gamma_H/\Gamma_H^{\text{SM}}$ for the parametrisation with $B_{\text{BSM}} \geq 0$ and $|\kappa_V| \leq 1$. At 3000 fb$^{-1}$ the $1\sigma$ uncertainty on $B_{\text{BSM}}$ is 0.035 in S1 and 0.027 in S2, where in the latter case the statistical uncertainty is the largest component. The corresponding 95% CL expected upper limit is $B_{\text{BSM}} = 0.077(0.057)$ in S1 (S2) at 3000 fb$^{-1}$. The uncertainty on $\Gamma_H/\Gamma_H^{SM}$ is 0.05 in S1 and 0.04 in S2, equivalent to 0.16 and 0.21 MeV respectively, assuming the SM width of 4.1 MeV. The main contribution is the statistical uncertainty, followed by the experimental one.

Table 3: The expected $\pm1\sigma$ uncertainties, expressed as percentages, on the per-production-mode signal strength parameters. Values are given for both S1 (with Run 2 systematic uncertainties [30]) and S2 (with YR18 systematic uncertainties). The total uncertainty is decomposed into four components: statistical (Stat), signal theory (SigTh), background theory (BkgTh) and experimental (Exp).

| | | $300\ \text{fb}^{-1}$ uncertainty [%] | | | | | $3000\ \text{fb}^{-1}$ uncertainty [%] | | | | |
|---|---|---|---|---|---|---|---|---|---|---|---|
| | | Total | Stat | SigTh | BkgTh | Exp | Total | Stat | SigTh | BkgTh | Exp |
| $\mu_{\text{ggH}}$ | S1 | 7.1 | 2.6 | 5.8 | 1.4 | 2.8 | 5.7 | 0.8 | 5.4 | 0.9 | 1.2 |
| | S2 | 4.6 | 2.6 | 3.1 | 0.8 | 2.0 | 3.1 | 0.8 | 2.8 | 0.6 | 0.9 |
| $\mu_{\text{VBF}}$ | S1 | 10.8 | 8.2 | 4.8 | 1.2 | 4.9 | 4.7 | 2.6 | 3.0 | 1.3 | 2.1 |
| | S2 | 9.5 | 8.2 | 3.2 | 0.5 | 3.6 | 3.7 | 2.6 | 2.1 | 0.3 | 1.6 |
| $\mu_{\text{WH}}$ | S1 | 20.2 | 14.6 | 3.1 | 7.5 | 11.4 | 8.2 | 4.6 | 2.9 | 3.3 | 5.2 |
| | S2 | 17.7 | 14.6 | 2.3 | 4.4 | 8.7 | 6.4 | 4.6 | 1.4 | 2.7 | 3.2 |
| $\mu_{\text{ZH}}$ | S1 | 15.7 | 12.4 | 6.3 | 5.7 | 4.4 | 7.2 | 3.9 | 5.1 | 2.5 | 2.1 |
| | S2 | 14.0 | 12.4 | 3.3 | 3.7 | 4.1 | 5.7 | 3.9 | 3.0 | 2.3 | 1.7 |
| $\mu_{\text{ttH}}$ | S1 | 14.7 | 5.6 | 8.4 | 7.3 | 7.4 | 9.9 | 1.8 | 8.3 | 4.1 | 3.1 |
| | S2 | 10.7 | 5.6 | 4.1 | 5.6 | 5.9 | 6.2 | 1.8 | 4.2 | 3.4 | 2.4 |

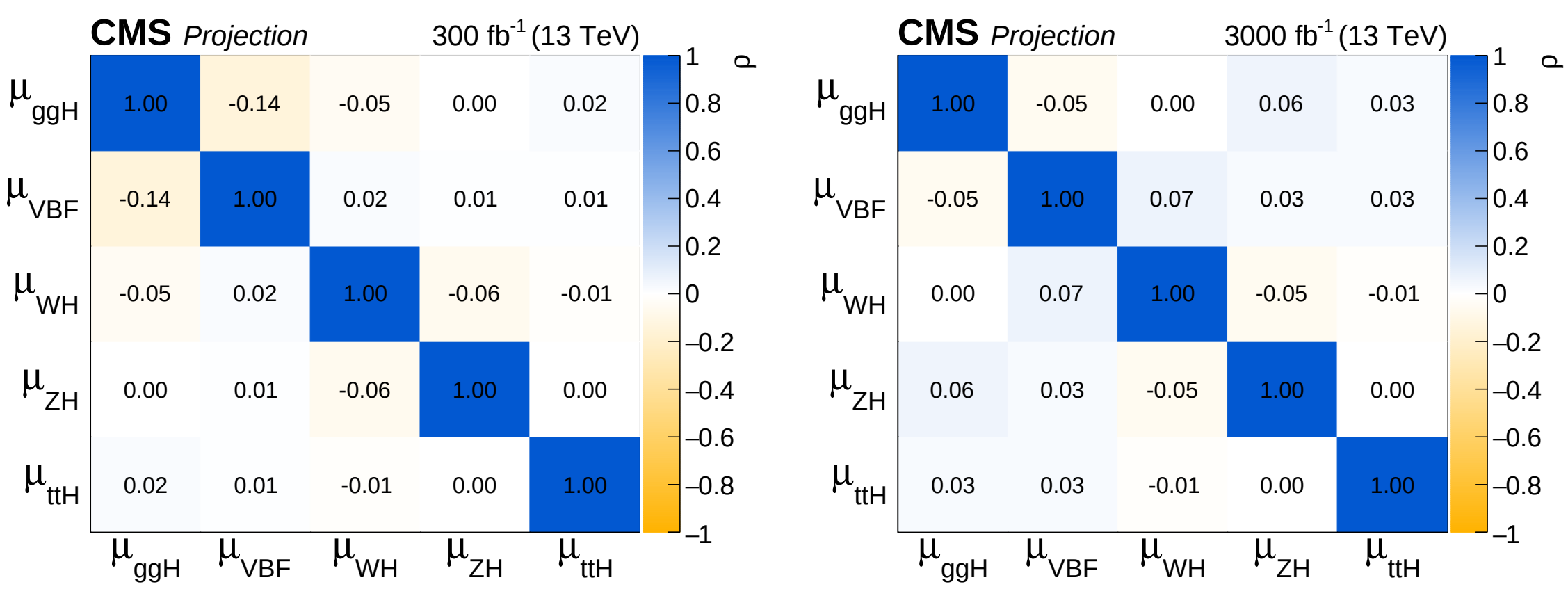

Figure 4: Correlation coefficients ($\rho$) between parameters in the signal strength per-production-mode parametrisation for S2 (with YR18 systematic uncertainties) at $300\,\text{fb}^{-1}$ (left) and $3000\,\text{fb}^{-1}$ (right).

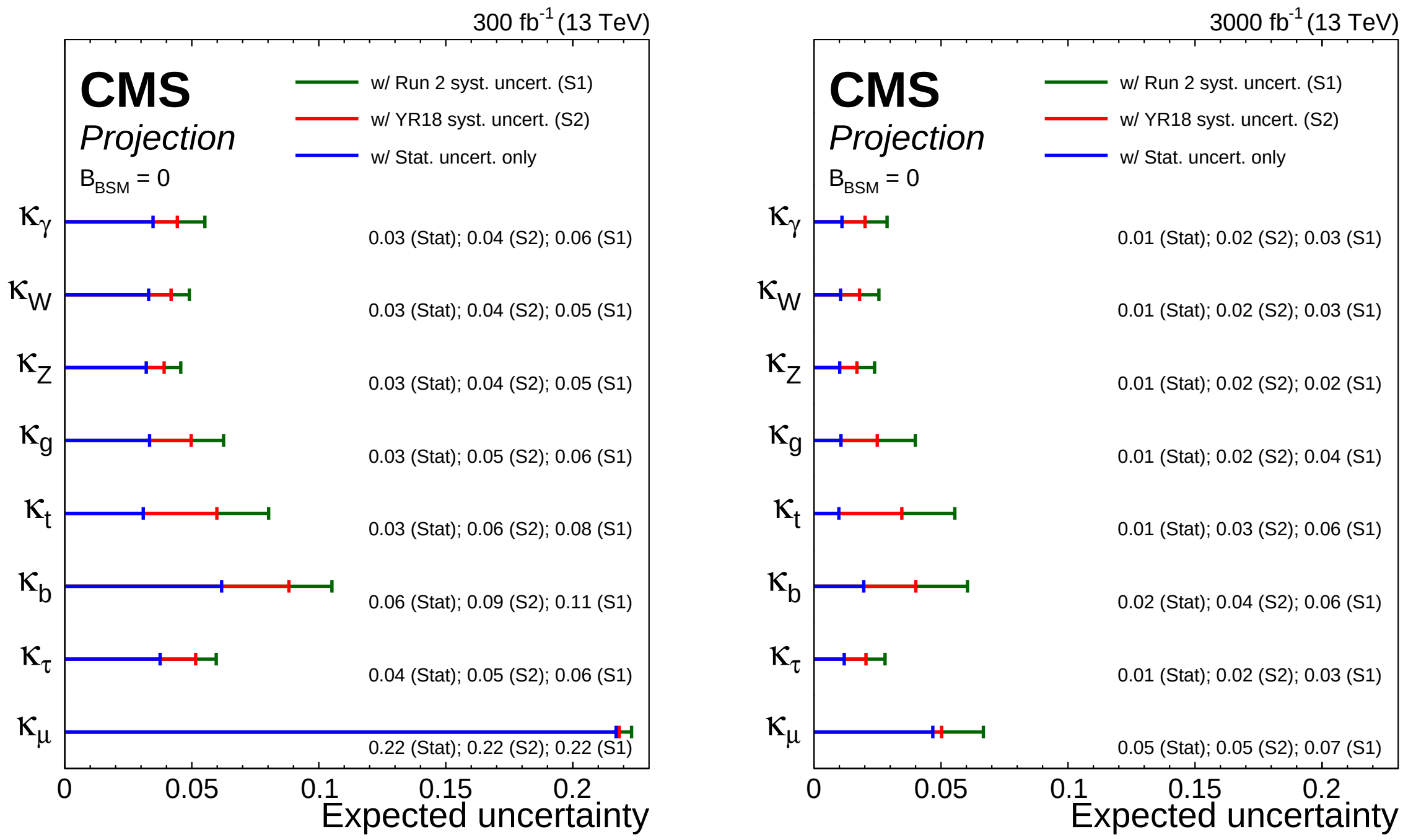

Figure 5: Summary plot showing the total expected $\pm 1\sigma$ uncertainties in S1 (with Run 2 systematic uncertainties [30]) and S2 (with YR18 systematic uncertainties) on the coupling modifier parameters for 300 fb$^{-1}$ (left) and 3000 fb$^{-1}$ (right). The statistical-only component of the uncertainty is also shown.

Figure 6 gives the correlation coefficients for the coupling modifiers for S2 at 300 fb$^{-1}$ and 3000 fb$^{-1}$. In contrast to the per-decay signal strength correlations in Fig. 2 the correlations here are larger, up to +0.74. One reason for this is that the normalisation of any signal process depends on the total width of the Higgs boson, which in turn depends on the values of the other coupling modifiers. The largest correlations involve $\kappa_b$, as this gives the largest contribution to the total width in the SM. Therefore improving the measurement of the H $\to$ bb process will improve the sensitivity of many of the other coupling modifiers at the HL-LHC.

Projections have also been determined for an alternative parametrisation, based on ratios of the coupling modifiers ($\lambda_{ij} = \kappa_i/\kappa_j$). A reference combined coupling modifier is defined which scales the yield of a specific production and decay process. This is chosen to be $\kappa_{gZ} = \kappa_g\kappa_Z/\kappa_H$, where $\kappa_H = \sum_j B^j_{SM}\kappa_j^2$. The results of this projection are given in Appendix B.

## 3.2 ttH production with H $\to$ bb

This section focuses on the analysis targeting ttH production with the H $\to$ bb decay channel and the single- and dilepton decay channels of the tt system using 35.9 fb$^{-1}$ of data collected at $\sqrt{s} = 13$ TeV [27]. In order to identify the signal against the background of tt+jets production, the analysis relies on dedicated multivariate techniques, including boosted decision trees and deep neural networks, that combine the information of several discriminating variables. The output of a matrix element method is also utilised. An excess of events above the background-only hypothesis with an observed (expected) significance of 1.6 (2.2) standard deviations is observed, corresponding to a best-fit signal strength of $\mu = 0.72 \pm 0.45$. This result is among the most sensitive to date and contributes significantly to the first observation of ttH production [32]. In this section, an extrapolation of the analysis to 3000 $fb^{-1}$ of data collected at the HL-LHC conditions is presented.

Table 4: The expected $\pm 1\sigma$ uncertainties, expressed as percentages, on the coupling modifier parameters, as well as $B_{BSM}$ and $\Gamma_H/\Gamma_H^{SM}$. The values for the $B_{BSM}$ parameter correspond to the $+1\sigma$ uncertainties only. Values are given for both S1 (with Run 2 systematic uncertainties [30]) and S2 (with YR18 systematic uncertainties). The total uncertainty is decomposed into four components: statistical (Stat), signal theory (SigTh), background theory (BkgTh) and experimental (Exp).

| | | 300 $fb^{-1}$ uncertainty [%] | | | | | 3000 $fb^{-1}$ uncertainty [%] | | | | |
|---|---|---|---|---|---|---|---|---|---|---|---|
| | | Total | Stat | SigTh | BkgTh | Exp | Total | Stat | SigTh | BkgTh | Exp |
| $B_{BSM} = 0$ | | | | | | | | | | | |
| $\kappa_\gamma$ | S1 | 5.5 | 3.5 | 2.0 | 1.8 | 3.3 | 2.9 | 1.1 | 1.8 | 1.0 | 1.7 |
| | S2 | 4.4 | 3.5 | 1.1 | 1.2 | 2.2 | 2.0 | 1.1 | 0.9 | 0.8 | 1.2 |
| $\kappa_W$ | S1 | 4.9 | 3.3 | 1.8 | 2.0 | 2.5 | 2.6 | 1.0 | 1.7 | 1.1 | 1.1 |
| | S2 | 4.2 | 3.3 | 1.0 | 1.3 | 2.0 | 1.8 | 1.0 | 0.9 | 0.8 | 0.8 |
| $\kappa_Z$ | S1 | 4.6 | 3.2 | 1.9 | 1.7 | 2.0 | 2.4 | 1.0 | 1.7 | 0.9 | 0.9 |
| | S2 | 3.9 | 3.2 | 1.0 | 1.1 | 1.7 | 1.7 | 1.0 | 0.9 | 0.7 | 0.7 |
| $\kappa_g$ | S1 | 6.3 | 3.3 | 3.6 | 2.5 | 3.0 | 4.0 | 1.1 | 3.4 | 1.3 | 1.2 |
| | S2 | 5.0 | 3.3 | 1.9 | 2.0 | 2.5 | 2.5 | 1.1 | 1.7 | 1.1 | 1.0 |
| $\kappa_t$ | S1 | 8.0 | 3.1 | 4.3 | 4.6 | 3.8 | 5.5 | 1.0 | 4.4 | 2.7 | 1.6 |
| | S2 | 6.0 | 3.1 | 2.2 | 3.5 | 3.0 | 3.5 | 1.0 | 2.2 | 2.1 | 1.2 |
| $\kappa_b$ | S1 | 10.5 | 6.2 | 3.9 | 5.2 | 5.4 | 6.0 | 2.0 | 4.3 | 2.9 | 2.3 |
| | S2 | 8.8 | 6.2 | 1.9 | 4.0 | 4.5 | 4.0 | 2.0 | 2.0 | 2.2 | 1.8 |
| $\kappa_\tau$ | S1 | 6.0 | 3.8 | 2.6 | 1.9 | 3.3 | 2.8 | 1.2 | 1.8 | 1.1 | 1.4 |
| | S2 | 5.2 | 3.8 | 1.7 | 1.4 | 2.8 | 2.0 | 1.2 | 1.0 | 0.9 | 1.0 |
| $\kappa_\mu$ | S1 | 22.3 | 21.7 | 2.7 | 1.8 | 3.6 | 6.7 | 4.7 | 2.5 | 1.0 | 3.9 |
| | S2 | 21.8 | 21.7 | 1.4 | 1.4 | 1.8 | 5.0 | 4.7 | 1.3 | 0.8 | 1.1 |
| $B_{BSM} \geq 0$, $\lvert\kappa_V\rvert \leq 1$ | | | | | | | | | | | |
| $B_{BSM}$ $(+1\sigma)$ | S1 | 8.2 | 6.0 | 2.7 | 3.1 | 3.7 | 3.8 | 1.9 | 2.4 | 1.5 | 1.7 |
| | S2 | 7.2 | 6.0 | 1.5 | 2.3 | 3.1 | 2.7 | 1.9 | 1.0 | 1.2 | 1.3 |
| $\Gamma/\Gamma_{SM}$ | S1 | 12.7 | 8.6 | 4.1 | 4.8 | 6.7 | 5.8 | 2.7 | 3.6 | 2.4 | 2.7 |
| | S2 | 11.2 | 8.6 | 2.3 | 3.9 | 5.5 | 4.3 | 2.7 | 1.9 | 1.8 | 2.1 |

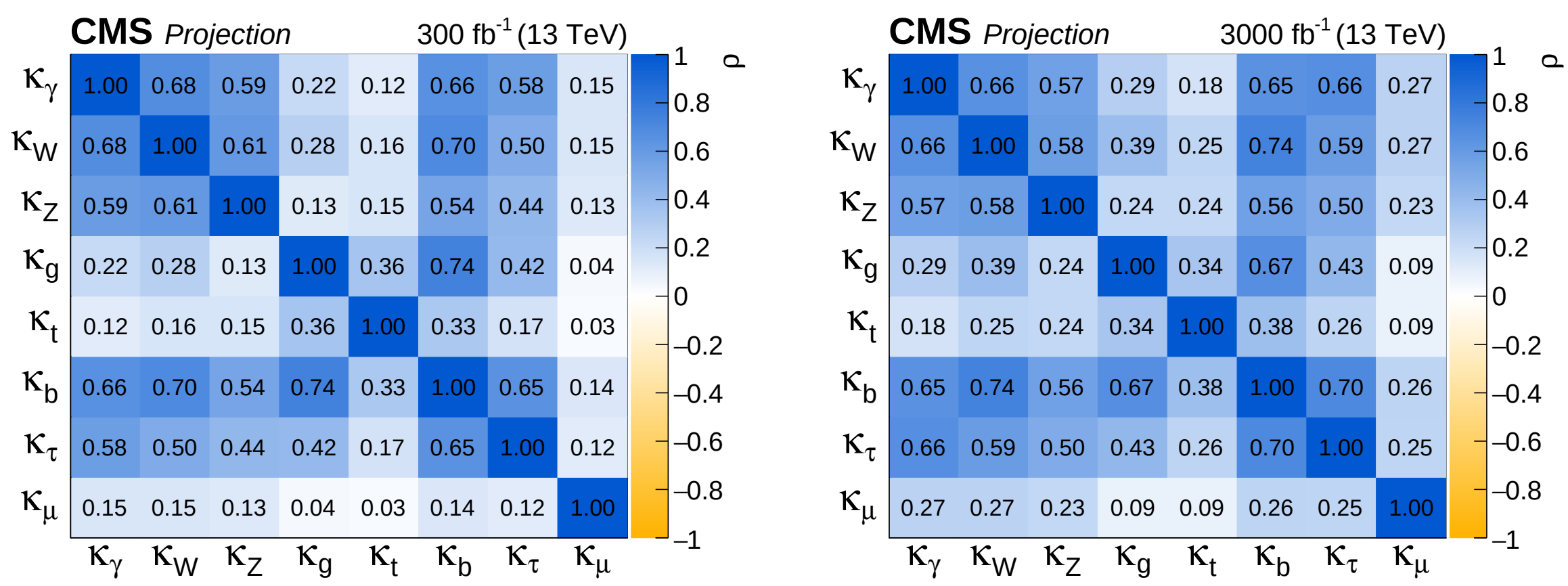

Figure 6: Correlation coefficients ($\rho$) between parameters in the coupling modifier parametrisation for S2 (with YR18 systematic uncertainties) at 300 $fb^{-1}$ (left) and 3000 $fb^{-1}$ (right).

The dominant tt+jets background is estimated from simulation and separated into five distinct processes based on the flavour of the additional jets in the event within acceptance that do not originate from the top quark decays. These are events with two (tt+bb) or one (tt+b) additional b-quark jets, events with one additional jet that contains two b quarks (tt+2b), events with additional c-quark jets and no additional b-quark jets (tt+cc), and all other events (tt + LF). The tt+bb, tt+b, tt+2b, and tt+cc processes are collectively referred to as tt + heavy-flavour (tt+HF) in the following and represent important sources of background. The final precision is limited to a large extent by the modelling of these background processes, which suffer from large theory uncertainties. Neither previous measurements of tt+HF production [33] nor higher-order theoretical calculations can currently constrain the normalisation of these contributions to better than 35% accuracy [34, 35]. Therefore, to account for differences in the phase space with respect to Ref. [33], a conservative extra 50% rate uncertainty is assigned to each of the four tt+HF processes in addition to the common theory uncertainties assigned to the inclusive tt + jets prediction. The statistical model is constructed such that these large conservative uncertainties can be constrained in the fit to data.

The expected precision of the signal strength measurement for the different integrated luminosities and scenarios is presented in Fig. 7 and also listed in Table 5. Shown are the total expected uncertainty as well as contributions from the following individual sources of uncertainties:

- **Stat:** the statistical uncertainty of the fit;
- **SigTh:** theoretical uncertainties related to generation of the ttH signal samples, including inclusive cross-section and PDF uncertainties;
- **BkgTh:** theoretical uncertainties related to the generation of the tt samples, including the pileup modelling, their inclusive cross-sections as well as parton shower modelling, and the additional 50% cross-section uncertainties for the tt+HF processes;
    - **Add. tt+HF XS:** subset of the BkgTh uncertainties containing only the additional 50% tt+HF cross-section uncertainties, motivated by their significant contribution to the analysis sensitivity;
- **Exp:** experimental uncertainties related to detector effects;
    - **Luminosity:** subset of the experimental uncertainties related to the integrated luminosity estimate;
    - **B tagging:** subset of the experimental uncertainties related to the b tagging;
    - **JES:** subset of the experimental uncertainties related to the jet energy scale measurement.

Additional sources of uncertainties considered in the analysis, including lepton identification, isolation, trigger, and jet energy resolution uncertainties, are not shown here due to their small impact on the final result.

Under the conservative S1 scenario, the expected total uncertainty on $\mu$ decreases from 0.49 at 35.9 $fb^{-1}$ to 0.20 and 0.11 at 300 $fb^{-1}$ and 3000 $fb^{-1}$, respectively. A rather sizeable reduction of the uncertainty with increased integrated luminosity is observed, even though the 35.9 $fb^{-1}$ result is dominated by systematic uncertainties. This is attributed to the increasing power of the profile likelihood fit to constrain the uncertainties due to the significantly increased number of events entering the fit. For example, relevant nuisance parameters, such as the ones describing the additional tt+HF cross-section uncertainties, are constrained to a few percent of their prior value.

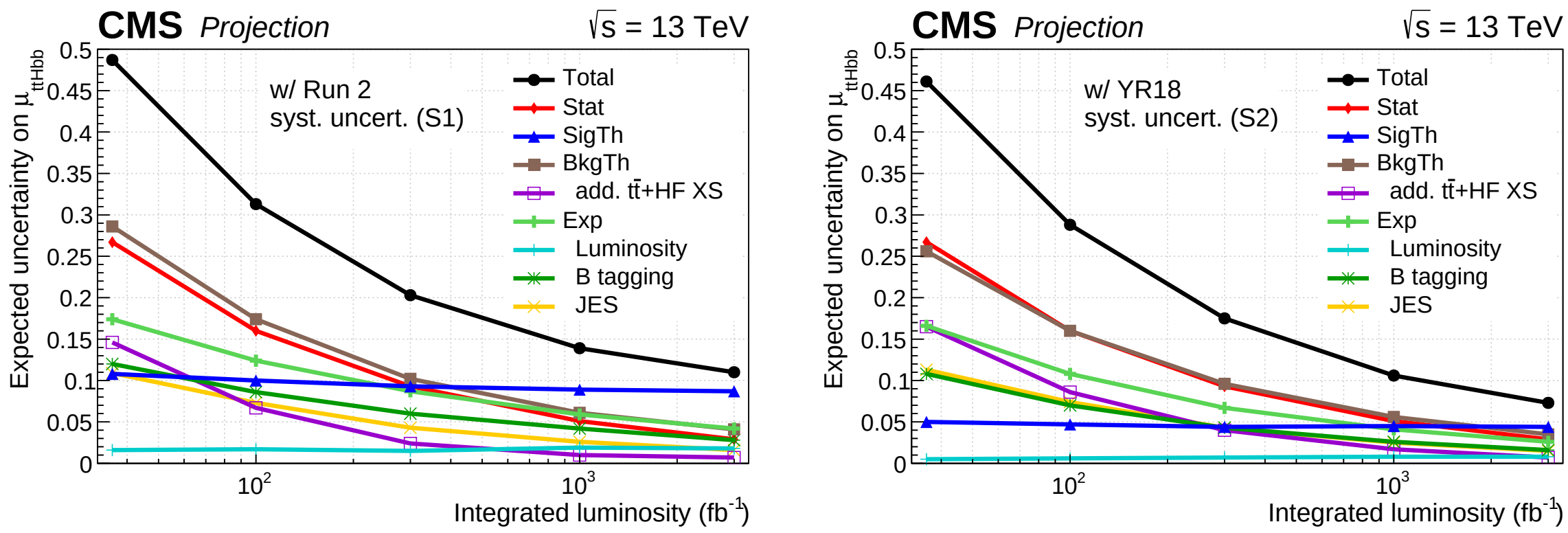

Figure 7: Expected uncertainties on the ttH signal strength as a function of the integrated luminosity under the S1 (left, with Run 2 systematic uncertainties [27]) and S2 (right, with YR18 systematic uncertainties) scenarios. Shown are the total uncertainty (black) and contributions of different groups of uncertainties. Results with 35.9 $\mathrm{fb}^{-1}$ are intended for comparison with the projections to higher luminosities and differ in parts from [27] for consistency with the projected results: uncertainties due to the limited number of MC events have been omitted and theory systematic uncertainties have been halved in case of the scenario S2.

Table 5: Breakdown of the contributions to the expected uncertainties on the ttH signal-strength $\mu$ at different luminosities for S1 (with Run 2 systematic uncertainties [27]) and S2 (with YR18 systematic uncertainties). The uncertainties are given in percent relative to $\mu = 1$. Results with 35.9 $\mathrm{fb}^{-1}$ are intended for comparison with the projections to higher luminosities and differ in parts from [27] for consistency with the projected results: uncertainties due to the limited number of MC events have been omitted and theory systematic uncertainties have been halved in case of the scenario S2.

| | | S1 | | | S2 | |
|---|---|---|---|---|---|---|
| Source | 35.9 $\mathrm{fb}^{-1}$ | 300 $\mathrm{fb}^{-1}$ | 3000 $\mathrm{fb}^{-1}$ | 35.9 $\mathrm{fb}^{-1}$ | 300 $\mathrm{fb}^{-1}$ | 3000 $\mathrm{fb}^{-1}$ |
| Total | 48.7 | 20.4 | 11.1 | 46.1 | 17.6 | 7.3 |
| Stat | 26.7 | 9.3 | 2.9 | 26.7 | 9.3 | 2.9 |
| SigTh | 10.8 | 9.3 | 8.7 | 5.0 | 4.5 | 4.4 |
| BkgTh | 28.6 | 10.3 | 4.1 | 25.6 | 9.6 | 3.5 |
| Add. tt+HF XS | 14.6 | 2.6 | 0.8 | 16.5 | 4.1 | 0.7 |
| Exp | 17.4 | 8.7 | 4.2 | 16.6 | 6.7 | 2.6 |
| Luminosity | 1.6 | 1.8 | 1.8 | 0.5 | 0.7 | 0.8 |
| B tagging | 12.0 | 6.1 | 2.8 | 10.8 | 4.4 | 1.6 |
| JES | 10.9 | 4.5 | 1.6 | 11.3 | 4.4 | 1.6 |

The background model has been developed to describe the data sufficiently well given the statistical precision obtained with 35.9 $fb^{-1}$. The results of this extrapolation illustrate that the background modelling will need to be refined at 3000 $fb^{-1}$, requiring improved simulations or more sophisticated control-region measurements. The observed constraints show that there will be enough data to obtain further information about the background beyond the current modelling, either by measuring the background directly or by sufficiently constraining the parameters of a model with more freedom. The predicted cross-section uncertainty of a few percent demonstrates the level of sensitivity at which the data will be able to distinguish different models. The results presented here have to be interpreted taking into account these considerations.

Under the more optimistic S2 scenario, the expected total uncertainties on $\mu$ are 0.46, 0.18, and 0.07 at 35.9 $fb^{-1}$, 300 $fb^{-1}$, and 3000 $fb^{-1}$, respectively, showing a clear improvement over the S1 scenario. The difference between S1 and S2 at 35.9 $fb^{-1}$ is due to the reduction of several theory uncertainties in scenario S2.

In both scenarios, the ttH(bb) process can be observed with a significance of approximately six standard deviations with 300 $fb^{-1}$ of data.

At each integrated luminosity value, the statistical uncertainties are the same in the two scenarios. A similar behaviour is observed for the uncertainties related to the jet energy scale, which indicates that the scaled components do not have a large impact on the overall sensitivity. Since several of the b tagging related uncertainties are reduced in scenario S2, their impact also decreases between scenarios S1 and S2. The total contribution of background modelling uncertainties (BkgTh) decreases between scenarios S1 and S2. As expected, the impact of the signal modelling uncertainty is reduced by about 50% between scenarios S1 and S2.

A key element of the background model are the large additional prior uncertainties on the normalisation of the tt+HF processes. By design these prior uncertainties are large, in order to provide the fit with enough freedom to cover potential differences between the data and the nominal tt background prediction. While the fit is able to constrain these uncertainties, they still have a major impact on the final result. To estimate the effect due to improvements in theory predictions, the extrapolation is also performed as a function of the tt+HF cross-section uncertainty, while all other uncertainties are set according to the S2 scenario. Already at 35.9 $fb^{-1}$, the gain in precision when reducing the prior tt+HF uncertainty is small: a factor 10 reduction of the prior uncertainty translates into only an 8% relative improvement of the total post-fit uncertainty. At 3000 $fb^{-1}$, the corresponding improvement is 3%. This is expected since the large prior uncertainties can be strongly constrained during the fit already with 35.9 $fb^{-1}$. The result supports the conclusion that a substantial improvement in analysis precision requires a further refinement of the background model beyond a mere reduction of the prior inclusive cross-section uncertainties.

The statistical power of the future data to extract further information on the background has been demonstrated with a modified background model, where the tt+bb normalisation is treated as a freely floating parameter in the final fit and the corresponding 50% prior rate uncertainty is omitted. Thus, the ttH signal-strength modifier $\mu$ and the tt+bb cross section were extracted simultaneously, while the other tt+HF processes remained constrained by a 50% prior rate uncertainty. In this case, the resulting statistical uncertainty after the fit was increased, since more information is extracted directly from the data, and the BkgTh uncertainty was reduced, while other uncertainties remained as in the nominal model. The resulting change in total uncertainty is negligible, and the expected overall significance at 300 $\text{fb}^{-1}$ is still close to six standard deviations in both scenarios. This example illustrates that the data at the HL-LHC will be sufficient to utilise a less constrained model at no substantial loss in sensitivity to the ttH signal.

## 3.3 VH production with $\mathrm{H} \rightarrow \mathrm{bb}$

The ATLAS and CMS Collaborations have both reported the observation of the $\mathrm{H} \rightarrow \mathrm{bb}$ decay [36, 37]. The studies presented here are performed based on a previous analysis, in which the CMS Collaboration reported evidence for the $\mathrm{H} \rightarrow \mathrm{bb}$ decay in the VH production mode using the 2016 proton-proton collision data set collected at $\sqrt{s} = 13$ TeV, which corresponds to an integrated luminosity of 35.9 $\text{fb}^{-1}$ [24]. This analysis makes use of leptonic decays of the vector boson which is produced in association with the Higgs boson. The final states of the VH system covered in this analysis always contain two b jets and either zero, one or two electrons or muons. Both leptons are required to have the same flavour in the two lepton selection. The b jets are identified using a combined multivariate (CMVA) tagging algorithm. The inputs include track impact parameter and secondary vertex information from the jet. Three thresholds on the CMVA discriminant are used in the analysis, denoted tight, medium and loose, which have efficiencies for tagging b jets ranging from 50–75% and for light quark or gluon jets from 0.15–3%.

Major backgrounds arising from SM production of vector boson plus heavy- or light-flavour jets, in addition to tt production, are controlled and constrained for each vector boson decay channel independently via dedicated control regions. Multivariate energy regression techniques are used to improve the b jet energy resolution, and a boosted decision tree is used to improve the discrimination between signal and background. The distribution of this multivariate discriminator is used as the discriminating variable in the signal extraction fit. The signal strength observed in this analysis is $\mu_{\mathrm{VHbb}} = 1.19^{+0.21}_{-0.20}\,(\text{stat})\,^{+0.34}_{-0.32}\,(\text{syst})$ . Here the projected uncertainty on the signal strength up to 3000 $\text{fb}^{-1}$ is reported, assuming $\mu_{\mathrm{VHbb}} = 1$.

Figure 8 shows the uncertainty on $\mu_{\mathrm{VHbb}}$ as a function of integrated luminosity, for scenario S1 (green points), scenario S2 (red points) and a scenario where all systematic uncertainties are ignored (blue points). In both scenarios S1 and S2 systematic uncertainties start to dominate very quickly, thus moving the projected uncertainty away from the statistical-only scaling curve.

Figure 9(a) shows the expected uncertainties on the signal strength at 300 and 3000 $\text{fb}^{-1}$ under the different uncertainty scenarios. Figure 9(b) shows the per-process and per-channel signal strength uncertainty at 3000 $\text{fb}^{-1}$, showing results for all three scenarios described above. The large improvement in the signal strength uncertainty for the 1-lepton channel, which is most sensitive to the WH production mode, is caused by the integrated luminosity scaling of an uncertainty in the modelling of the W boson $p_{\mathrm{T}}$ distribution. This uncertainty dominates this channel in scenario S1.

The contributions of different sources of uncertainty at 3000 $\text{fb}^{-1}$ in scenarios S1 and S2 are shown in Table 6. Both in scenario S1 and S2 the largest component of the systematic uncertainty is theoretical. Moving from S1 to S2 the total signal theoretical uncertainty reduces to half its size. This is expected as in scenario S2 the input uncertainties are scaled down to half the current size. In the case of the background theory, where the input uncertainties are also scaled to half their original size when going from scenario S1 to scenario S2, the total uncertainty due to this component is not halved. This is because at 3000 $\mathrm{fb}^{-1}$ some of the theoretical uncertainties on the backgrounds can be constrained in the fit. The same is true for the experimental uncertainties, which in some cases are already moderately constrained in the current analysis.

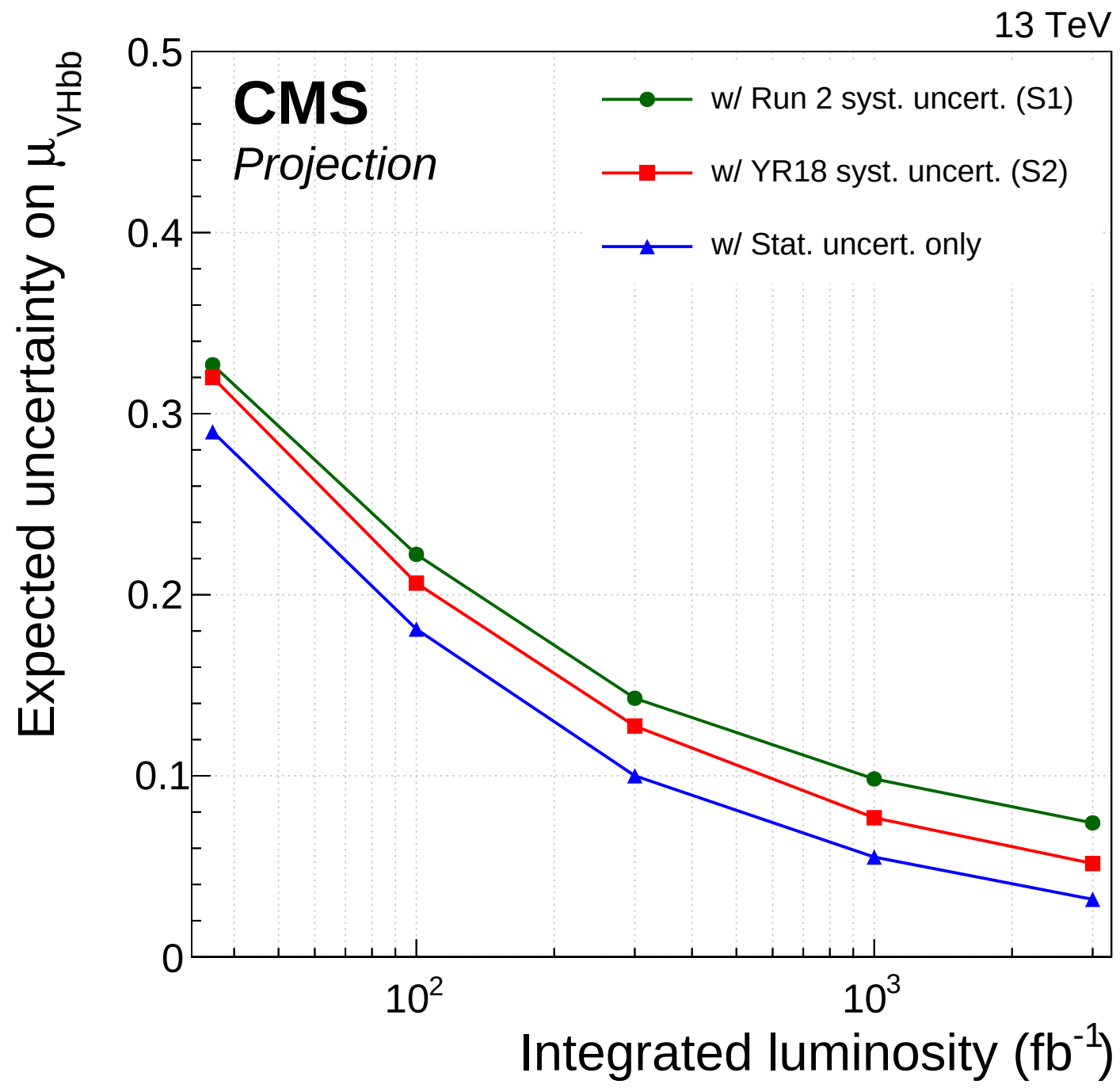

Figure 8: Uncertainty on the signal strength $\mu_{\mathrm{VHbb}}$ as a function of integrated luminosity for S1 (with Run 2 systematic uncertainties [24]) and S2 (with YR18 systematic uncertainties). Results with 35.9 $\mathrm{fb}^{-1}$ are intended for comparison with the projections to higher luminosities and differ in parts from [24] for consistency with the projected results: uncertainties due to the limited number of MC events have been omitted and theory systematic uncertainties have been halved in case of the scenario S2.

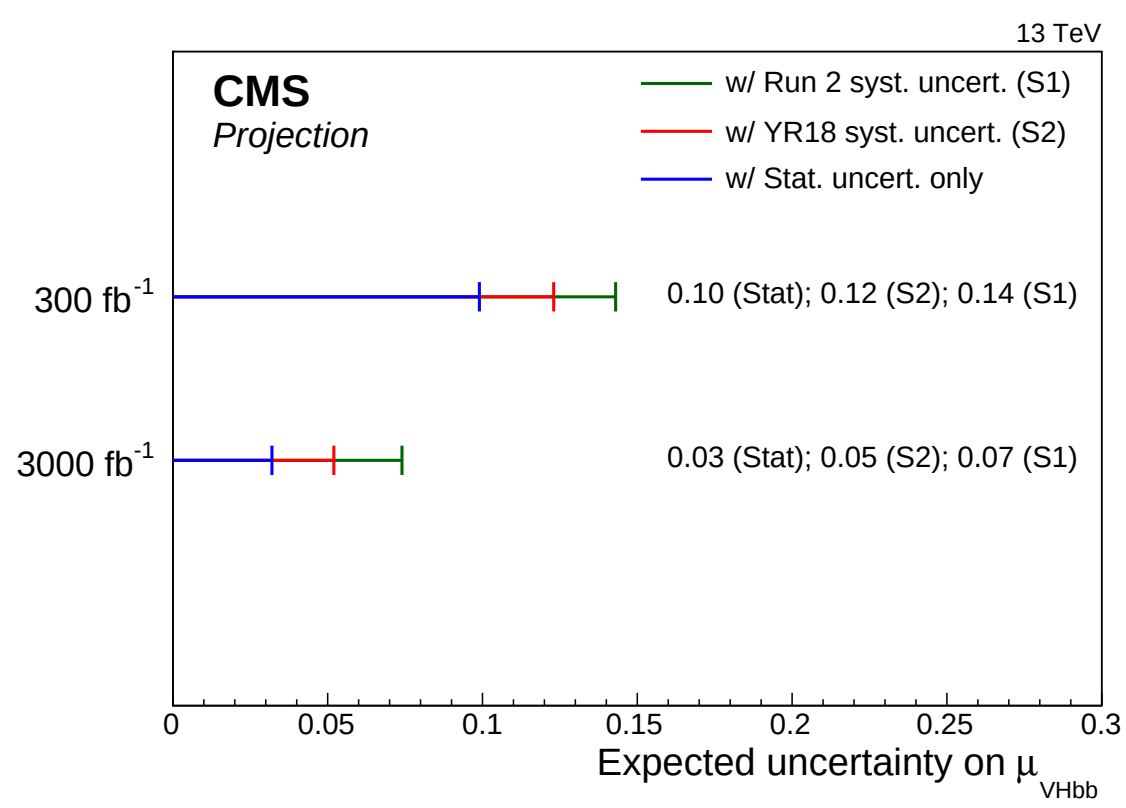

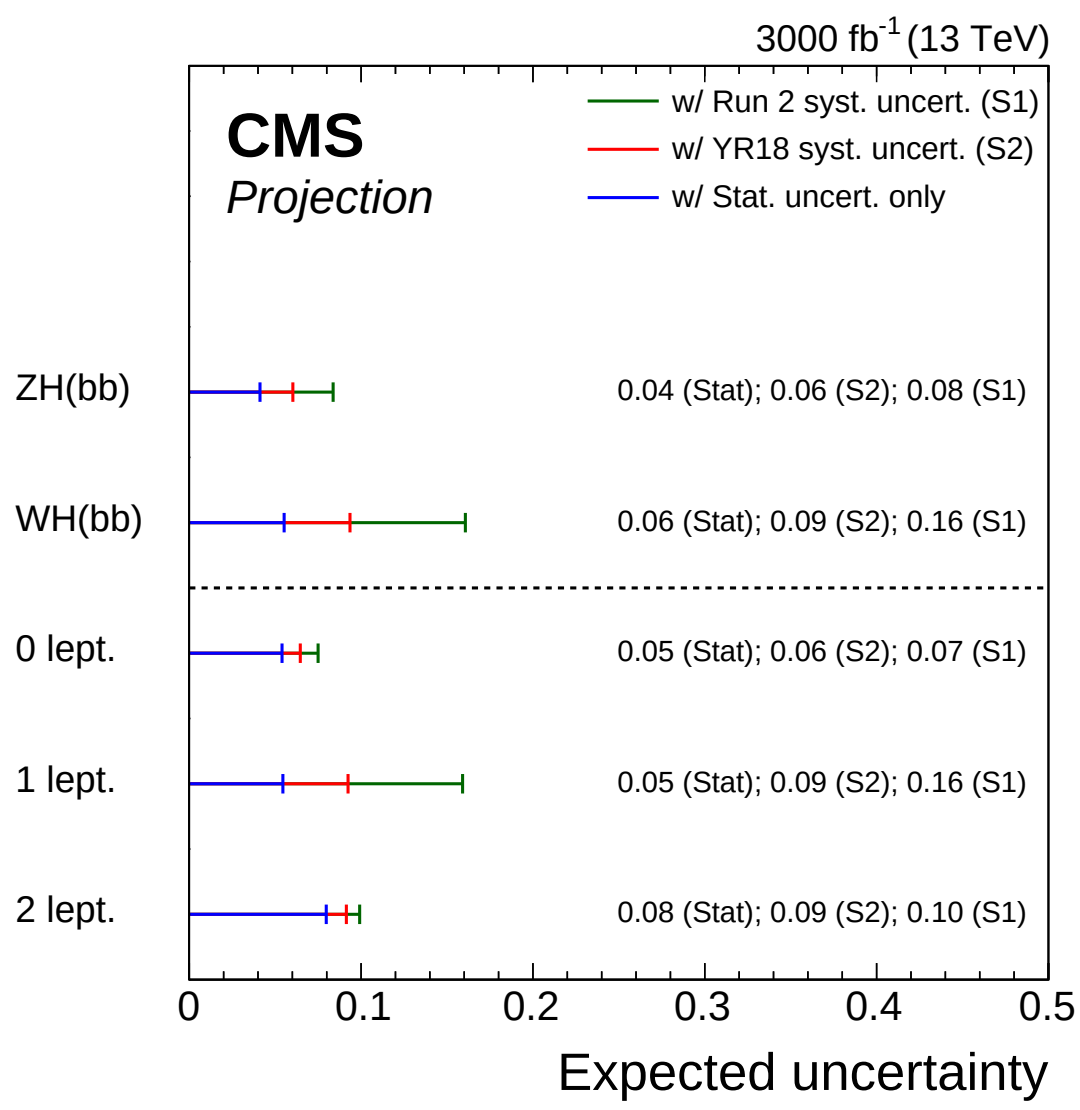

Figure 9: Uncertainties in the overall signal strength $\mu_{\mathrm{VHbb}}$ at 300 and 3000 $\mathrm{fb}^{-1}$ (left) and per-process and per-channel signal strengths at 3000 $\mathrm{fb}^{-1}$ (right). Values are given for the S1 (with Run 2 systematic uncertainties [24]) and S2 (with YR18 systematic uncertainties) scenarios, as well as a scenario in which all systematic uncertainties are removed.

Looking in more detail at the dominant signal theoretical uncertainties, the largest component in the uncertainty arises from the uncertainty in the gluon-induced ZH (ggZH) production cross section due to QCD scale variations. The ggZH process contributes a small fraction of the total ZH process. Despite this, the uncertainty in the production cross section for this process due to QCD scale variations becomes dominant because it is very large: 25% for the ggZH process, compared to approximately 4% for the ZH process [15]. The next most important uncertainties are category-acceptance uncertainties in the dominant Z+bb and W+bb backgrounds due to QCD scale variations, as well as the uncertainty in the ZH and WH production cross section due to QCD scale variations. In scenario S2 these four most important uncertainties contribute 1.6%, 1.5%, 1.3% and 1.2% (absolute) to the total uncertainty of 5.1%, respectively. To improve the precision of the measurement it is therefore important to improve these theoretical uncertainties.

Table 6: Contributions of particular groups of uncertainties, expressed as percentages, at an integrated luminosity of 3000 $\mathrm{fb}^{-1}$ in S1 (with Run 2 systematic uncertainties [24]) and S2 (with YR18 systematic uncertainties). The total uncertainty is decomposed into four components: signal theory, background theory, experimental and statistical. The signal theory uncertainty is further split into inclusive and acceptance parts, and the contributions of the b tagging and JES/JER uncertainties to the experimental component are also given.

| | S1 | S2 |
|---|---|---|
| Total uncertainty | 7.3% | 5.1% |
| Signal theory uncertainty | 5.4% | 2.6% |
| Inclusive | 4.6% | 2.2% |
| Acceptance | 2.7% | 1.3% |
| Background theory uncertainty | 2.8% | 2.3% |
| Experimental uncertainty | 2.6% | 2.2% |
| b-tagging | 2.2% | 2.0% |
| JES and JER | 0.7% | 0.6% |
| Statistical uncertainty | 3.2% | 3.2% |

In the future, and at the HL-LHC in particular, the b tagging efficiency may change. The conditions could worsen the efficiency, but at the same time new detectors and new techniques could also lead to an improvement in the b tagging efficiency. The effect of changes in b tagging efficiency on the overall signal strength uncertainty is evaluated. Changes in the b tagging efficiency are emulated by scaling the rates of processes with a single b tag by the change in b tagging efficiency, and scaling the rates of processes with two b tags by the change in b tagging efficiency squared. The modifications are applied only to the efficiency to select genuine b jets; the mistagging rates for light quark and gluon jets remain unchanged.

Figure 10 shows the results of the projections to 300 and 3000 $\mathrm{fb}^{-1}$, assuming various reductions and improvements in the b tagging efficiency relative to the performance of the three CMVA working points used in the analysis. A 10% improvement in the b tagging efficiency leads to a relative improvement in the signal strength uncertainty of up to 8.5% at 300 $\mathrm{fb}^{-1}$ and up to 6% at 3000 $\mathrm{fb}^{-1}$. The improvements on the signal strength precision are limited because the uncertainty is dominated by theoretical sources. When neglecting inclusive signal theory uncertainties this improvement becomes up to 9.5% at 300 $\mathrm{fb}^{-1}$ and up to 8% at 3000 $\mathrm{fb}^{-1}$.

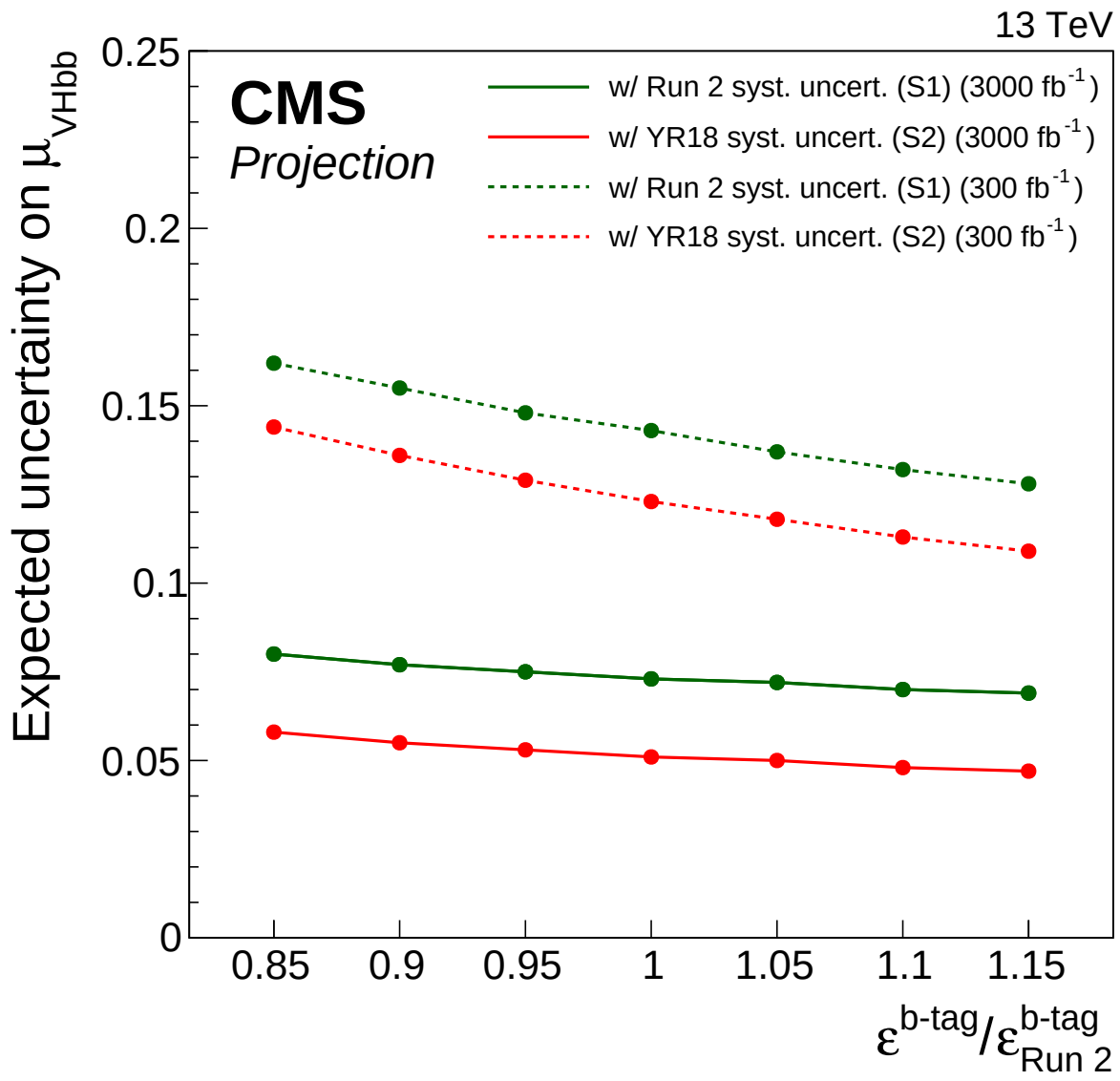

Figure 10: Effect of varying the b tagging efficiency ($\varepsilon^{\text{b-tag}}$) on the uncertainty in the signal strength measurement when considering all systematic uncertainties.

# 4 tH production

The strongest direct constraint on the absolute value of the top-Higgs Yukawa coupling ($y_t$) will be provided by the ttH measurement described in Section 3.2. As the ttH production cross section is only sensitive to the magnitude of this coupling further information is required to determine its sign. The production of the Higgs boson with a single top quark (tH) in association with either a W boson (tHW) or a light quark (tHq) provides unique information regarding the relative sign between $y_\mathrm{t}$ and the Higgs to vector boson coupling ($g_{\mathrm{HVV}}$). The dominant t-channel production of the tHq final state proceeds via two interfering leading-order processes involving these couplings, as shown in Fig. 11.

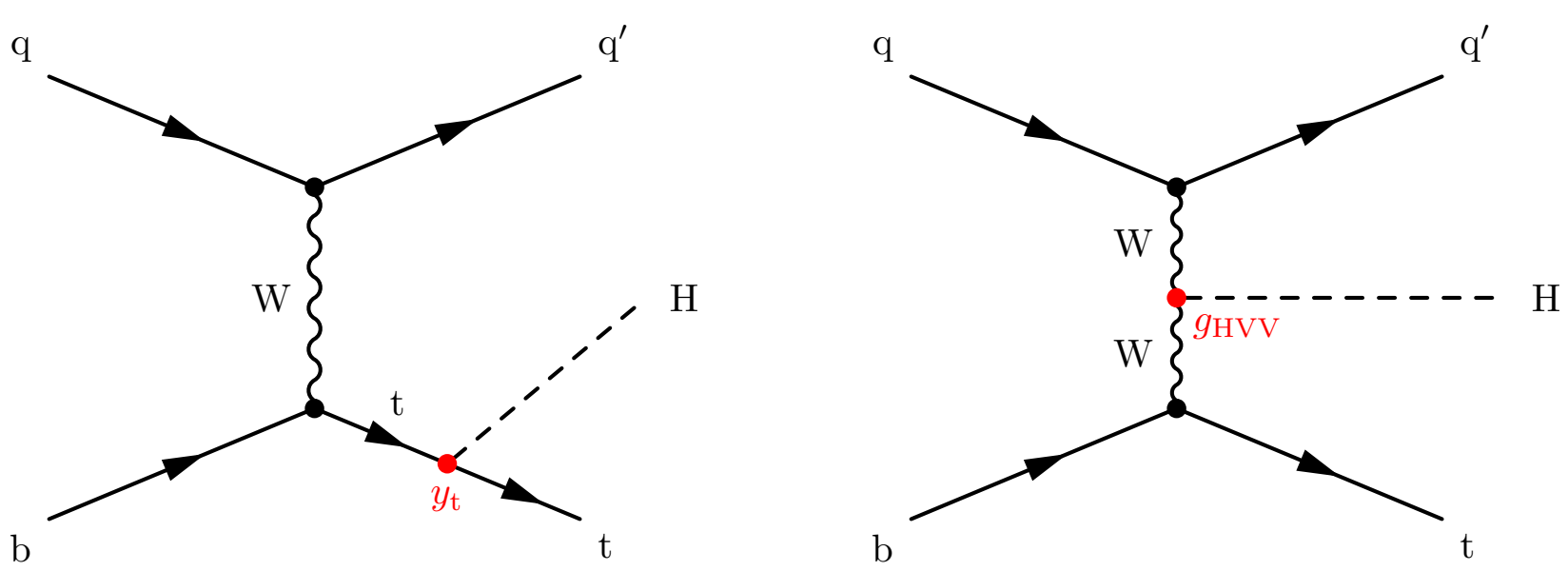

Figure 11: The representative leading-order diagrams for tHq production.

In the SM, the amplitudes of these processes interfere destructively leading to a very small cross section, approximately 71 fb at $\sqrt{s} = 13$ TeV, and therefore the observation of the tH process is not possible with the current data set. However, anomalous effects in the couplings leading to an opposite sign between $y_t$ and $g_{HVV}$ would cause constructive interference, enhancing the cross section by an order of magnitude or more, assuming the absolute values of the couplings remain close to the SM prediction. This would make the channel accessible at the LHC earlier with a smaller data set. This channel can therefore be used to determine the relative sign between the top-Higgs and the W-Higgs coupling modifiers with respect to the SM, denoted $\kappa_t$ and $\kappa_V$ respectively.

The sensitivity to the tH process at the HL-LHC is determined by extrapolating a combination of Run 2 analyses based on 35.9 fb$^{-1}$ of data at $\sqrt{s} = 13$ TeV [38]. Two of these analyses are dedicated searches for tHq: one targets a multi-lepton final state [39] in which the Higgs boson decays to WW, ZZ or $\tau\tau$ pairs, and the other targets the $H \to bb$ decay [40]. In both analyses the presence of at least one central b tagged jet and an isolated lepton from the top quark decay is required. Furthermore, the presence of a light quark jet at high pseudorapidity, a unique feature of the tHq production mode, is exploited. Both analyses also rely heavily on multivariate techniques to discriminate the signal against the large tt+jets background. The $\gamma\gamma$ final state is also utilised, via a reinterpretation of the inclusive $H \to \gamma\gamma$ analysis [20]. In this analysis the tHq and tHW processes primarily contribute to the"t$\bar{t}$H leptonic" and "t$\bar{t}$H hadronic" event categories, and these are included in the combination.

With the Run 2 analysis, values of $\kappa_t$ outside the ranges of approximately $[-0.9, -0.5]$ and $[1.0, 2.1]$ are excluded at the 95% CL, assuming $\kappa_V = 1$. To obtain the projected constraints on $\kappa_t$ a scan of the test statistic $q(\kappa_t)$ is performed, assuming $\kappa_V = 1$ and using an Asimov data set corresponding to the SM expectation ($\kappa_t = 1$, $\kappa_V = 1$). The expected contribution from the ttH process is also modified consistently with $\kappa_t$. Figure 12 gives the scan for the S1 and S2 projections, which shows that at 300 fb$^{-1}$ negative values of $\kappa_t$ are excluded at more than the 99% CL in the S1 scenario. The $\kappa_t$ ranges are even more constrained in S2. For 3000 fb$^{-1}$ a negative value of $\kappa_t$ is disfavoured with a significance larger than five standard deviations.

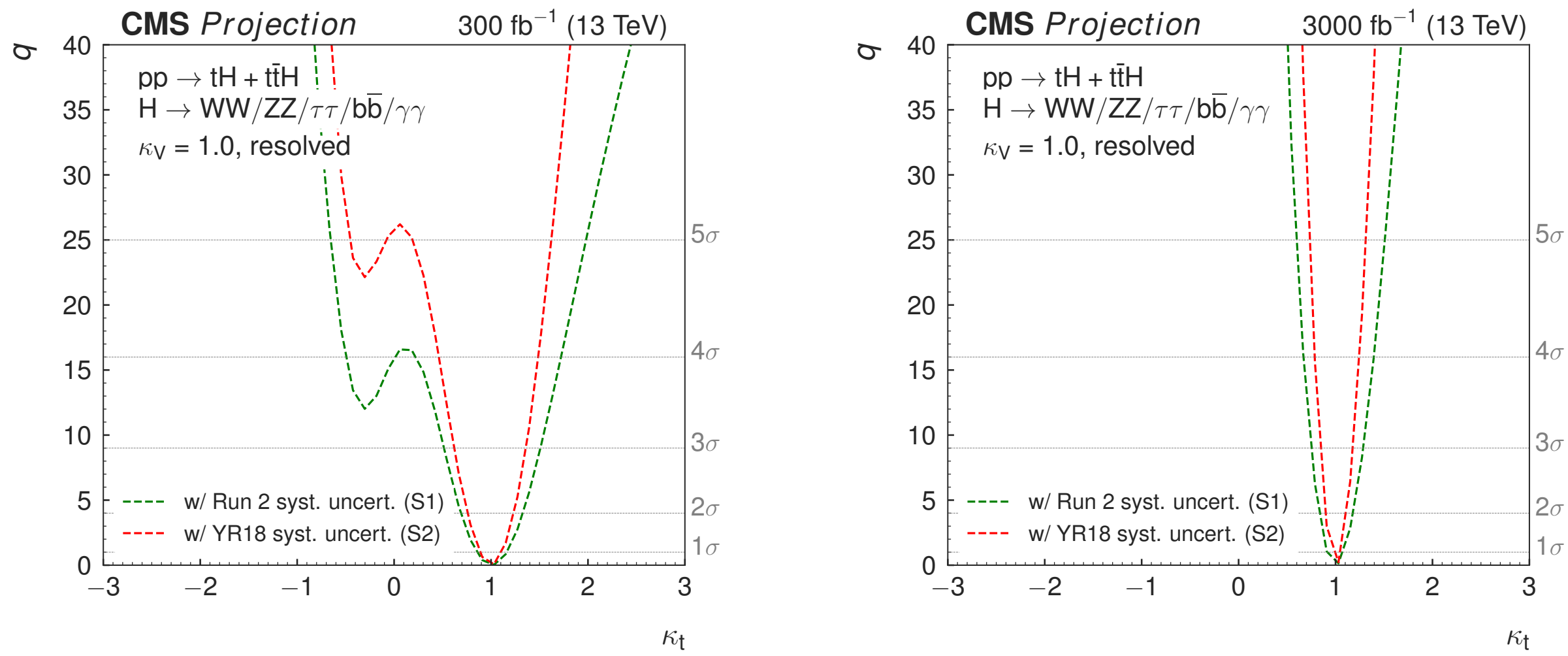

Figure 12: Scan of the test statistic $q$ versus $\kappa_t$ for the Asimov data sets corresponding to $\kappa_V = 1$ for the two integrated luminosity scenarios in S1 (with Run 2 systematic uncertainties [38]) and S2 (with YR18 systematic uncertainties).

With the larger HL-LHC data set it is therefore relevant to assess the sensitivity to the SM rate of tH production. In the Run 2 analysis with 35.9 fb$^{-1}$ of data, the observed upper limit on the tH signal strength relative to the SM ($\mu_{tH}$) is 25 compared to the median expected value of 12, assuming the ttH signal strength ($\mu_{\mathrm{ttH}}$) is fixed to the SM value.

In Fig. 13 the variation of the expected upper limits on $\mu_{\mathrm{tH}}$ is shown as a function of the integrated luminosity for the S1 and S2 scenarios. The limits are determined assuming a background-only hypothesis in which the ttH process is considered as following the SM epectation ($\mu_{\mathrm{ttH}} = 1$). In order to minimize further assumptions on the rate of ttH production, $\mu_{\mathrm{ttH}}$ is treated as a free parameter in the fit. In the S1 scenario the expected median upper limit on $\mu_{\mathrm{tH}}$ at 300 (3000) $\mathrm{fb}^{-1}$ is determined to be 5.60 (2.35). The corresponding value in S2 at 300 (3000) $\mathrm{fb}^{-1}$ is 3.98 (1.51). With the 3000 $\mathrm{fb}^{-1}$ data set and foreseen reduction in systematic uncertainties in S2, the expected upper limit on $\mu_{\mathrm{tH}}$ improves by about a factor of eight with respect to the current exclusion.

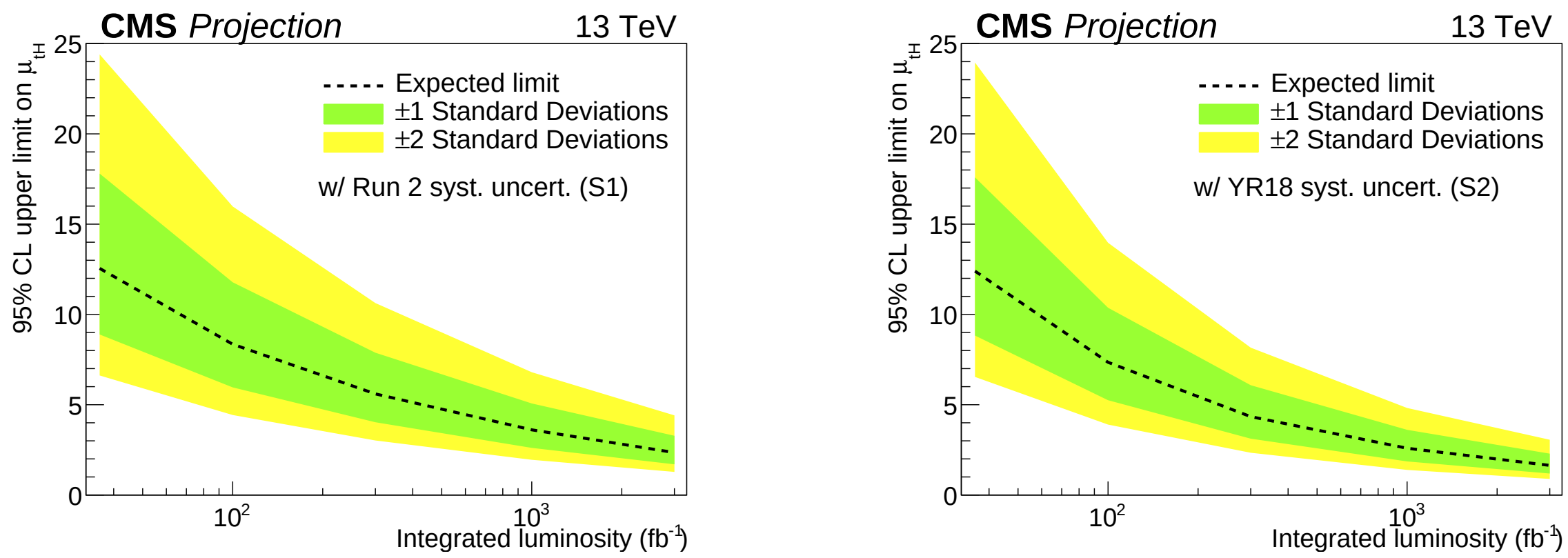

Figure 13: The variation of expected upper limit on $\mu_{\mathrm{tH}}$ with integrated luminosity for two projection scenarios S1 (with Run 2 systematic uncertainties [38]) and S2 (with YR18 systematic uncertainties).

The evolution of the expected uncertainty on the measurement of $\mu_{\mathrm{tH}}$, assuming the SM rate, is given in Table 7. Values are given for two assumptions of the ttH background: one in which $\mu_{\mathrm{ttH}}$ is unconstrained in the fit, and one in which it is fixed to the SM value of 1. In the latter case the uncertainties are reduced by around 10% at 3000 $\mathrm{fb}^{-1}$, indicating that a precise simultaneous measurement of the ttH signal strength will be needed to obtain the optimal sensitivity to the tH channel. In both cases it is found that the reduced systematic uncertainties in S2 improve the precision by up to 30%.

Table 7: The $\pm 1\sigma$ uncertainties on expected $\mu_{\mathrm{tH}}$=1 for scenarios S1 (with Run 2 systematic uncertainties [38]) and S2 (with YR18 systematic uncertainties) at all three luminosities, considering also the case when $\mu_{\mathrm{ttH}}$ is fixed at the SM value.

| | | $\mu_{\mathrm{ttH}}$ floating | $\mu_{\mathrm{ttH}}$ fixed |
|---|---|---|---|
| S1 | 35.9 $\mathrm{fb}^{-1}$ | $^{+6.2}_{-5.8}$ | $^{+5.8}_{-5.4}$ |
| | 300 $\mathrm{fb}^{-1}$ | $^{+2.9}_{-2.8}$ | $^{+2.5}_{-2.4}$ |
| | 3000 $\mathrm{fb}^{-1}$ | $^{+1.2}_{-1.2}$ | $^{+1.1}_{-1.0}$ |
| S2 | 35.9 $\mathrm{fb}^{-1}$ | $^{+6.2}_{-5.8}$ | $^{+5.8}_{-5.3}$ |
| | 300 $\mathrm{fb}^{-1}$ | $^{+2.2}_{-2.2}$ | $^{+2.0}_{-2.0}$ |
| | 3000 $\mathrm{fb}^{-1}$ | $^{+0.9}_{-0.9}$ | $^{+0.8}_{-0.8}$ |

# 5 Higgs boson $p_T$ distribution and coupling constraints

The measurement of Higgs boson differential cross sections can provide constraints on physical parameters that have a small effect on inclusive quantities such as total cross sections, but cause larger deviations in the distributions of certain observables. By varying the Higgs couplings, the strength with which quarks and other bosons couple to the Higgs boson, significant distortions in the shapes of differential cross sections appear, in particular for the transverse momentum distribution. The combination of differential cross sections from the $H \to \gamma\gamma$, $H \to ZZ$ and boosted gluon-fusion-induced $H \to bb$, along with an interpretation in the $\kappa$-framework [31], is documented in Ref. [41]. This section describes the projected results of the differential spectra and fits of the Higgs coupling modifiers. The projected differential cross section is computed as a function of $p_T(H)$, for each of the individual decay channels and for the combination. The cross sections in all bins are determined from a simultaneous maximum likelihood fit as described in section 2.1. The unfolding of the number of observed events in each bin to the particle level, to account for detector effects and bin-to-bin migrations, is performed within this fit.

## 5.1 The differential cross section for $p_T(H)$

The projection of the differential cross section for the $p_T(H)$ spectrum at 3000 fb$^{-1}$ is shown in Fig. 14, for both S1 and S2. The relative uncertainties for both scenarios are given in Tables 8 and 9. With respect to the uncertainties at the current integrated luminosity of 35.9 fb$^{-1}$, the uncertainties at 3000 fb$^{-1}$ in the higher $p_T(H)$ region are about a factor of ten smaller. This is expected, as the uncertainties in this region remain statistically dominated. The uncertainties in the lower $p_T(H)$ region are no longer statistically dominated however, as can been seen by comparing Table 8 with Table 9, where the reduced systematic uncertainties in S2 yield a reduction in the total uncertainty of up to 25% compared to S1.

| $p_T(H)$ (GeV) | 0-15 | 15-30 | 30-45 | 45-80 | 80-120 | 120-200 | 200-350 | 350-600 | 600-∞ |
|---|---|---|---|---|---|---|---|---|---|
| $H \to \gamma\gamma$ | 7.2% | 6.8% | 7.1% | 6.9% | 7.1% | 6.7% | 7.1% | 9.9% | 32.5% |
| $H \to ZZ$ | 6.2% | 5.7% | 5.0% | | 5.5% | | 9.6% | | |
| $H \to bb$ | *None* | | | | | | | 38.2% | 37.1% |
| Combination | 4.7% | 4.4% | 5.0% | 4.7% | 4.8% | 4.7% | 5.2% | 8.5% | 25.4% |

Table 8: Relative uncertainties on the projected $p_T(H)$ spectrum under S1 (with Run 2 systematic uncertainties [41]) at 3000 fb$^{-1}$.

| $p_T(H)$ (GeV) | 0-15 | 15-30 | 30-45 | 45-80 | 80-120 | 120-200 | 200-350 | 350-600 | 600-∞ |
|---|---|---|---|---|---|---|---|---|---|
| $H \to \gamma\gamma$ | 5.1% | 4.6% | 5.1% | 4.8% | 4.9% | 4.5% | 5.1% | 8.6% | 32.2% |
| $H \to ZZ$ | 5.4% | 4.8% | 4.1% | | 4.7% | | 9.1% | | |
| $H \to bb$ | *None* | | | | | | | 31.4% | 36.8% |
| Combination | 3.7% | 3.3% | 4.2% | 3.7% | 4.0% | 3.8% | 4.4% | 8.0% | 24.5% |

Table 9: Relative uncertainties on the projected $p_T(H)$ spectrum under S2 (with YR18 systematic uncertainties) at 3000 fb$^{-1}$.

## 5.2 Constraining Higgs coupling modifiers using the $p_T(H)$ spectrum

Theoretical predictions of differential distributions can be fit to data, and can subsequently be used to constrain physical parameters such as the couplings of the Higgs boson. Two sets of parametrisations are fit to the projected $p_T(H)$ distribution: Parametrisations dependent on $\kappa_b$ and $\kappa_c$, computed in Ref. [42], and parametrisations dependent on $\kappa_t$ and $c_g$, computed in

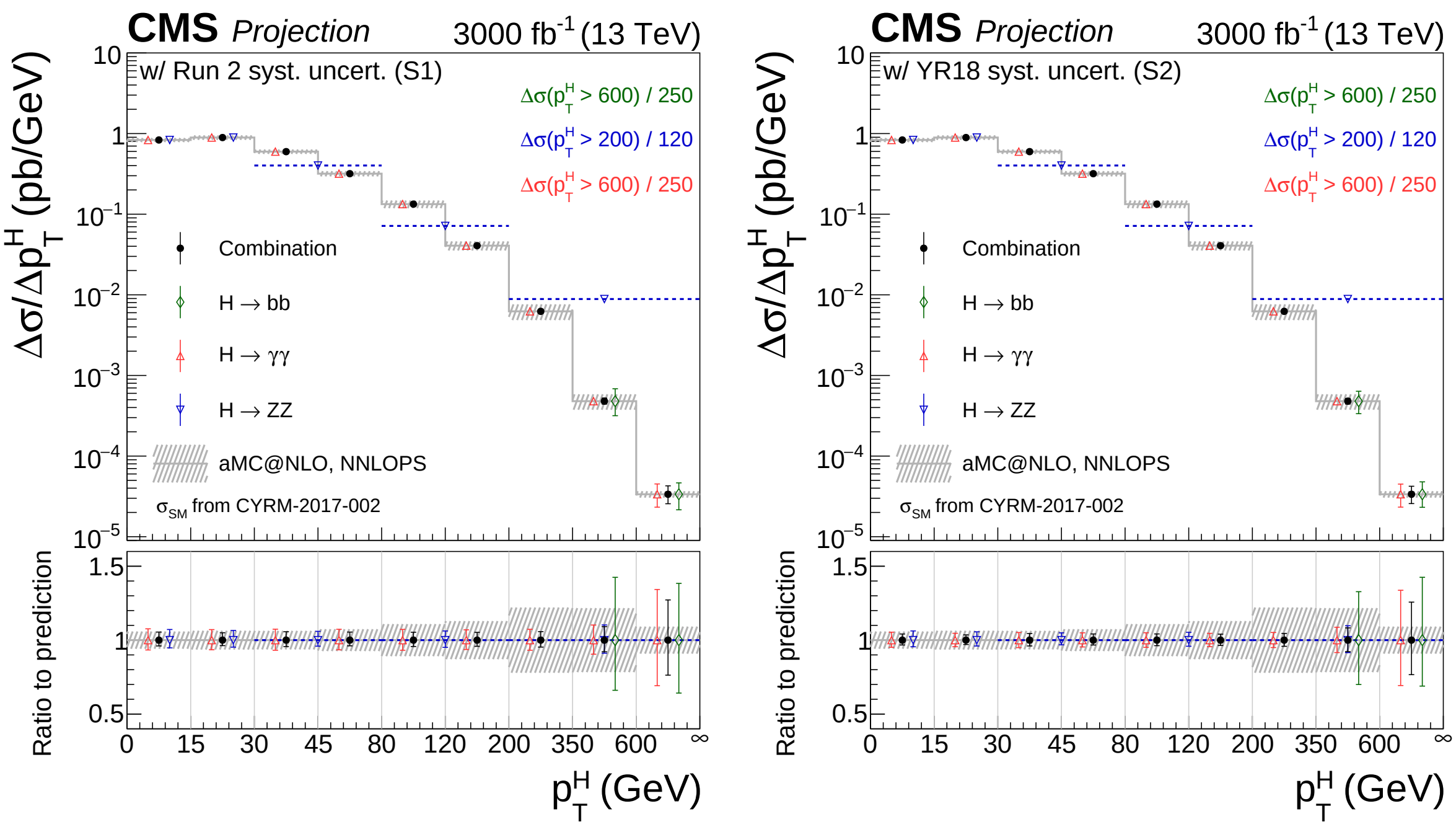

Figure 14: Projected differential cross section for the $p_T(H)$ spectrum at an integrated luminosity of 3000 fb$^{-1}$, under S1 (left, with Run 2 systematic uncertainties [41]) and S2 (right, with YR18 systematic uncertainties).

Ref. [43, 44]. The calculations of the former extend up to 120 GeV in $p_T(H)$, and therefore $H \to bb$ is not used in the fit. The results of the fits of $\kappa_b/\kappa_c$ and $\kappa_t/c_g$ are shown in Fig. 15 and Fig. 16 respectively. Additional theoretical uncertainties on the differential cross section predictions are included in the systematic uncertainties, and are not reduced with integrated luminosity. For this reason, the relative difference between S1 and S2 is expected not to be as pronounced as in the case of the differential $p_T(H)$ combination.

# 6 Constraints on anomalous HZZ couplings and the Higgs boson width using on-shell and off-shell measurements

Anomalous contributions in the spin-0 tensor structure of HVV interactions can be characterized by coefficients $a_2$, $a_3$, $\Lambda_1$, and $\Lambda_Q$ defined in Refs. [45, 46]. The contribution to the total cross section from these coefficients can be parametrised in terms of their fractional contribution to on-shell $H \to ZZ$ decays via the fractions $f_{ai}$ and phases $\phi_{ai}$ [45, 46]. Constraints on these anomalous contributions can further be improved by including off-shell Higgs boson production. An enhancement of signal events is expected in the presence of either anomalous HVV couplings or large Higgs boson total width, $\Gamma_H$ [15, 46, 47].

In this note, only the tensor structure proportional to $a_3$ is considered using either the combination of on-shell and off-shell events or with only on-shell events with $4\ell$ decay, following the techniques described in Refs. [15, 45, 47]. Constraints are placed in terms of $f_{ai}\cos(\phi_{ai})$ with the assumptions $\phi_{ai} = 0$ or $\pi$, $a_i^{ZZ} = a_i^{WW}$, and $\Gamma_H = \Gamma_H^{SM}$ (in the case of limits from the combined on-shell and off-shell likelihood parametrisation), and on $\Gamma_H$ with the assumption $f_{ai} = 0$.

The projections are shown in Fig. 17 and summarised in Table 10. Systematic uncertainties are found to have a negligible effect on the results for $f_{a3}\cos(\phi_{a3})$ using either on-shell and off-shell events combined or only on-shell events, so only scenario S1 is shown. In the case of $\Gamma_{\mathrm{H}}$ limits, theoretical systematic uncertainties are dominant over experimental ones. The dominant theoretical systematic effect comes from the uncertainty in the NLO EW correction on the $\mathrm{q\bar{q}} \to 4\ell$ simulation above the $2m_{\mathrm{Z}}$ threshold, but this uncertainty is also expected to be constrained from data with an integrated luminosity of 3000 $\mathrm{fb}^{-1}$. Limits on $\Gamma_{\mathrm{H}}$ are also given for an approximate S2 in which the experimental uncertainties are not reduced, while the theoretical uncertainties are halved with respect to S1. The 10% additional uncertainty applied on the QCD NNLO K factor on the gg background process is kept the same in this approximated S2 in order to remain conservative on the understanding of these corrections for this background component. It is also noted that the uncertainties on the signal and background QCD NNLO K factors are smaller in the Run 2 analysis [47] than in previous projections using Run 1 data [48].

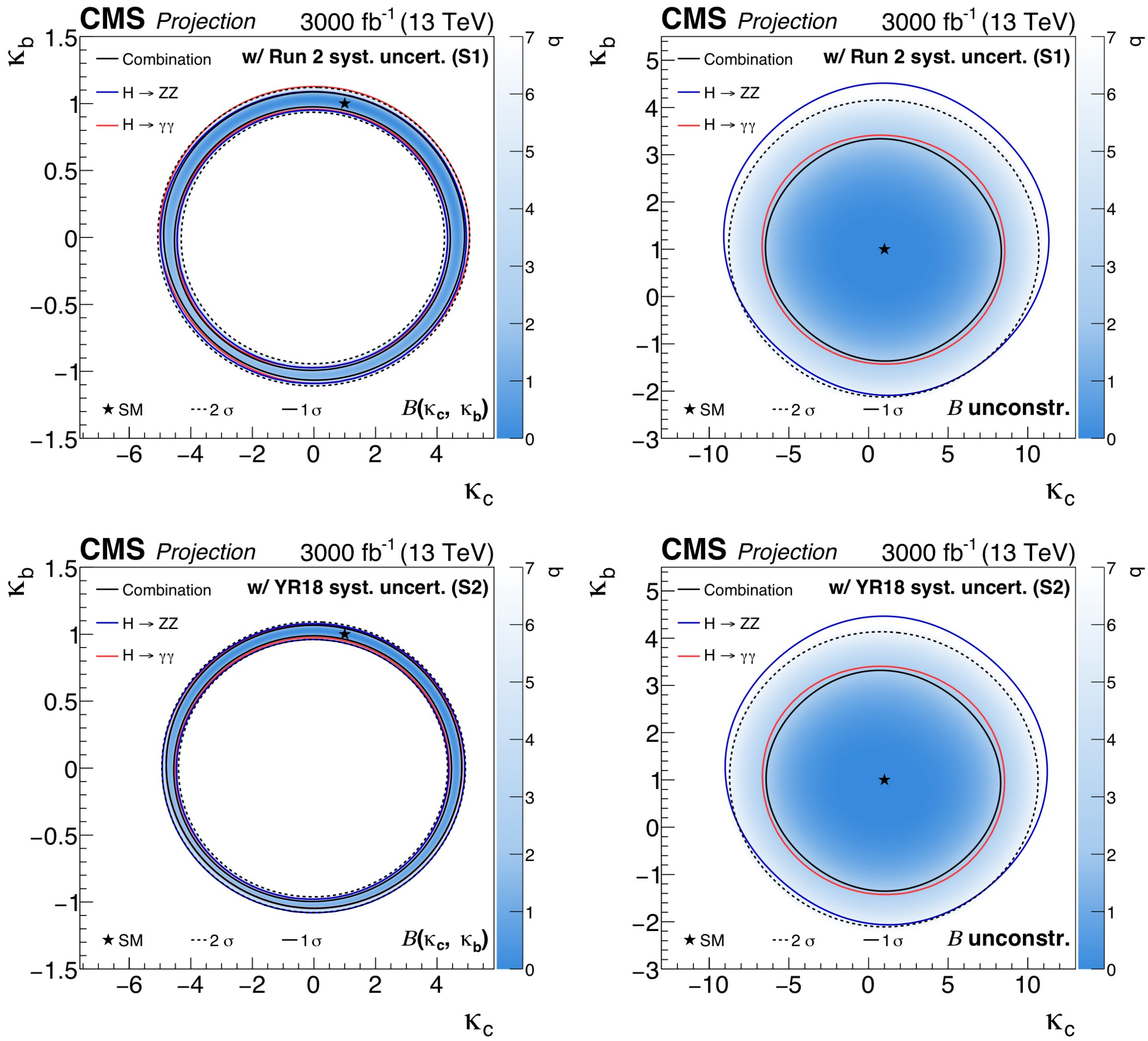

Figure 15: Projected simultaneous fit for $\kappa_b$ and $\kappa_c$, assuming the branching fractions to be determined by the couplings (left) and the branching fractions implemented as nuisance parameters with no prior constraint (right), under S1 (top) and S2 (bottom). The one standard deviation contour is drawn for the combination (H $\rightarrow \gamma\gamma$ and H $\rightarrow$ ZZ), the H $\rightarrow \gamma\gamma$ channel, and the H $\rightarrow$ ZZ channel in black, red, and blue, respectively. For the combination the two standard deviation contour is drawn as a black dashed line, and the shading indicates the negative log-likelihood, with the scale shown on the right hand side of the plots.

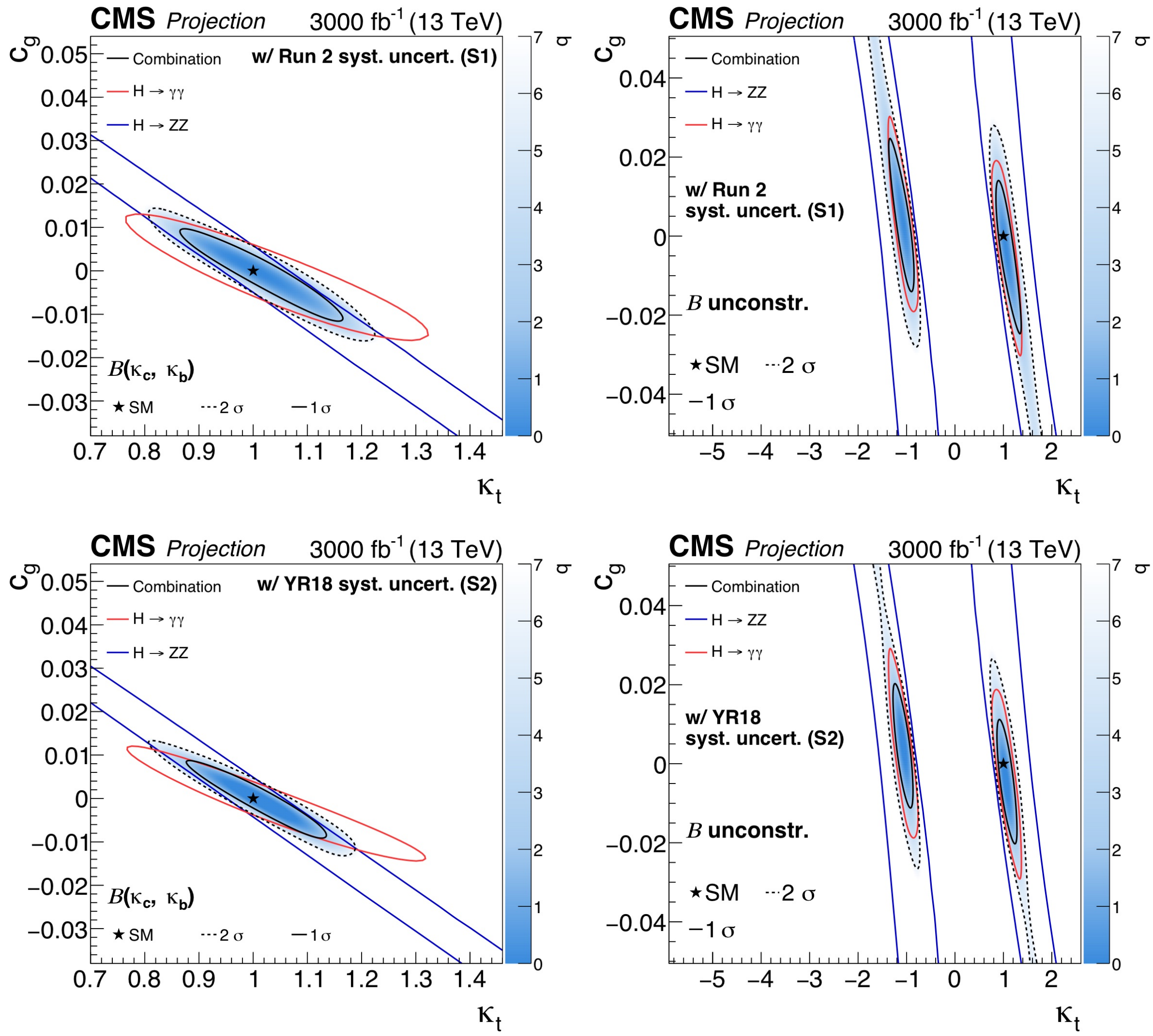

Figure 16: Projected simultaneous fit for $\kappa_t$ and $c_g$, assuming the branching fractions to be determined by the couplings (left) and the branching fractions implemented as nuisance parameters with no prior constraint (right), under S1 (top) and S2 (bottom). The one standard deviation contour is drawn for the combination ($H \to \gamma\gamma$, $H \to ZZ$, and $H \to bb$), the $H \to \gamma\gamma$ channel, and the $H \to ZZ$ channel in black, red, and blue, respectively. For the combination the two standard deviation contour is drawn as a black dashed line, and the shading indicates the negative log-likelihood, with the scale shown on the right hand side of the plots.

Table 10: Summary of the 95% CL intervals for $f_{a3}\cos(\phi_{a3})$, under the assumption $\Gamma_{\mathrm{H}} = \Gamma_{\mathrm{H}}^{\mathrm{SM}}$, and for $\Gamma_{\mathrm{H}}$ under the assumption $f_{ai} = 0$ for projections at $3000\,\mathrm{fb}^{-1}$. Constraints on $f_{a3}\cos(\phi_{a3})$ are multiplied by $10^4$. Values are given for scenarios S1 (with Run 2 systematic uncertainties [47]) and the approximate S2 scenario, as described in the text.

| Parameter | Scenario | Projected 95% CL interval |
|---|---|---|
| $f_{a3}\cos(\phi_{a3}) \times 10^4$ | S1, only on-shell | $[-1.8, 1.8]$ |
| $f_{a3}\cos(\phi_{a3}) \times 10^4$ | S1, on-shell and off-shell | $[-1.6, 1.6]$ |
| $\Gamma_{\mathrm{H}}$ (MeV) | S1 | $[2.0, 6.1]$ |
| $\Gamma_{\mathrm{H}}$ (MeV) | S2 | $[2.0, 6.0]$ |

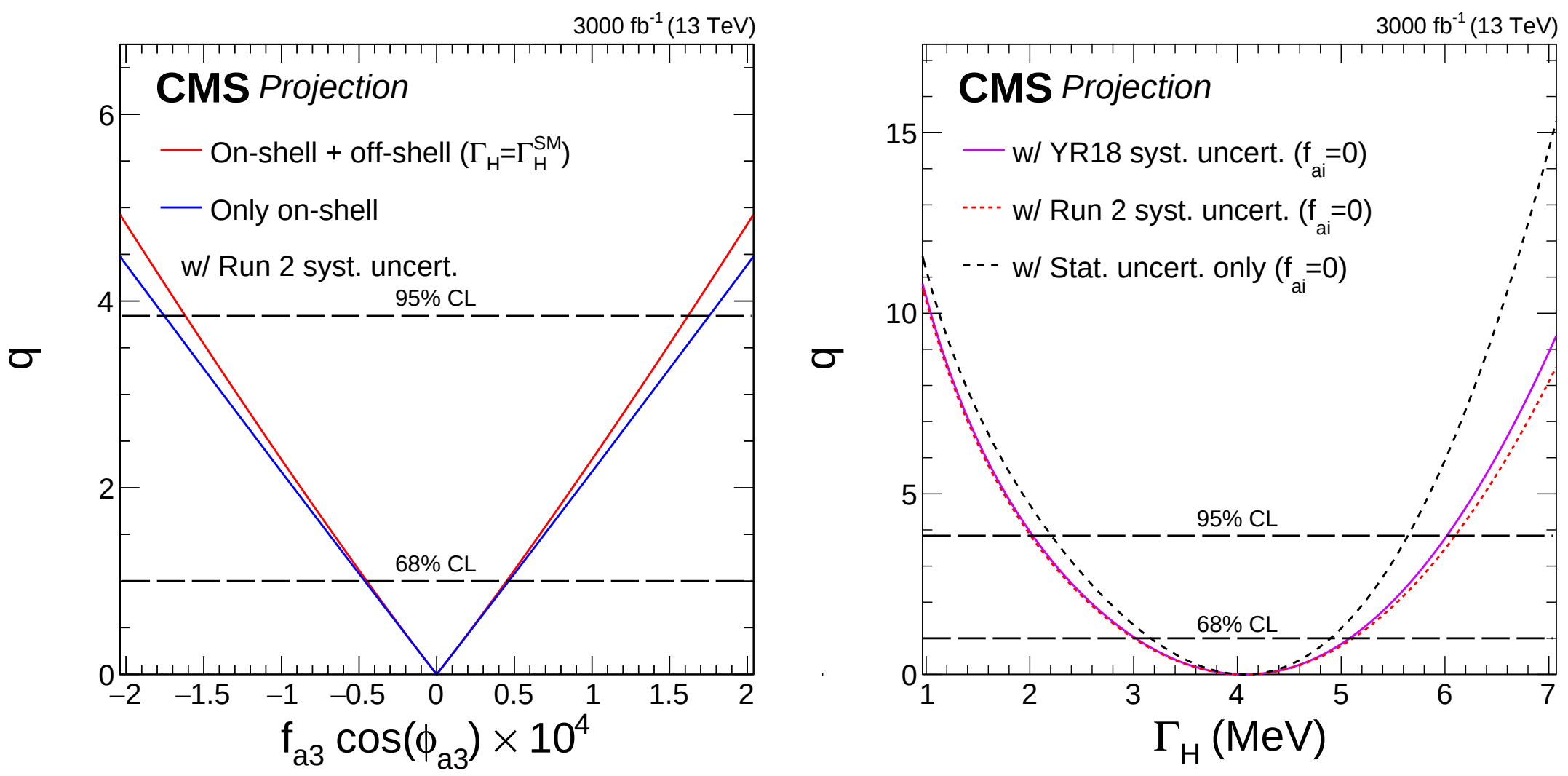

Figure 17: Likelihood scans for projections on $f_{a3}\cos(\phi_{a3})$ (left) and $\Gamma_{\mathrm{H}}$ (right) at 3000 $\mathrm{fb}^{-1}$. On the left plot, the scans are shown using either the combination of on-shell and off-shell events (red) or only on-shell events (blue). The dashed lines represent the effect of removing all systematic uncertainties. In the right plot, scenarios S2 (solid magenta) and S1 (dotted red) are compared to the case where all systematic uncertainties (dashed black) are removed. The dashed horizontal lines indicate the 68% and 95% CLs. The $f_{a3}\cos(\phi_{a3})$ scans assume $\Gamma_{\mathrm{H}} = \Gamma_{\mathrm{H}}^{\mathrm{SM}}$, and the $\Gamma_{\mathrm{H}}$ scans assume $f_{ai} = 0$.

# 7 Summary

The discovery of the Higgs boson opened a new era of precision measurements of the properties of the new particle, aimed to thoroughly test their consistency with the SM predictions. The present measurements of the Higgs boson couplings to fermions, bosons and of the tensor structure of the Higgs boson interaction with electroweak gauge bosons show no significant deviations with respect to the SM expectations. The HL-LHC will provide a unique environment in which to test the Higgs boson properties.

This summary describes the projected sensitivity to 300 and 3000 $fb^{-1}$ of several Higgs boson analyses performed on the 13 TeV data set collected in 2016. The projections are performed under different scenarios considering the systematic uncertainties under Run 2 and HL-LHC conditions. Results have been presented for a combined measurement of coupling modifiers and signal strengths, with additional studies for ttH and VH production with $H \to bb$ decay, and for production in association with a single top quark. Projections have also been given for the measurement of the Higgs boson transverse momentum differential cross section, and the expected constraints on anomalous couplings and the total width using on- and off-shell $H \to ZZ$ measurements.

# A Evolution of uncertainties with integrated luminosity

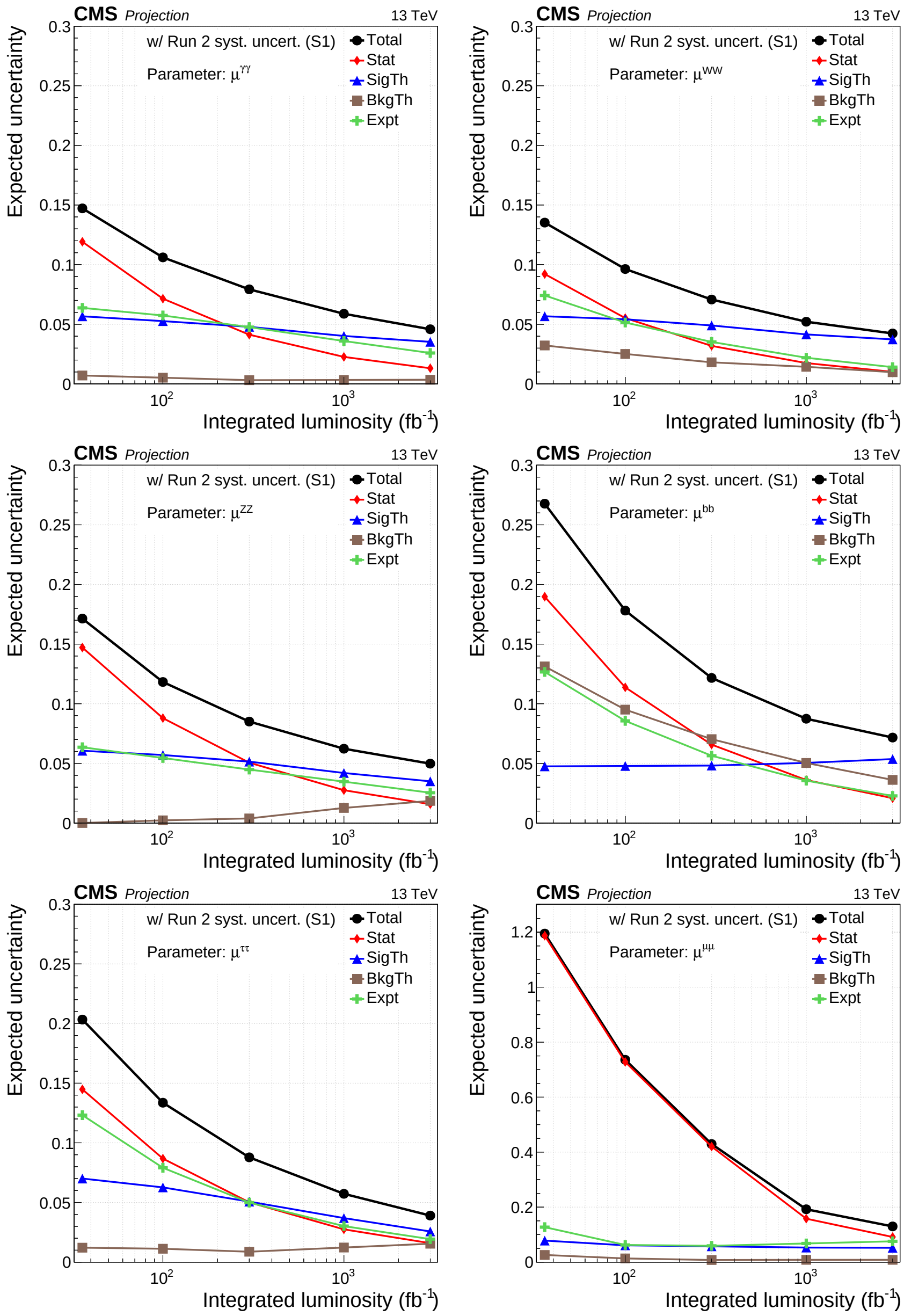

Figure 18: Evolution of the uncertainty contribution with integrated luminosity for the per-decay signal strength parameters in S1 (with Run 2 systematic uncertainties [30]).

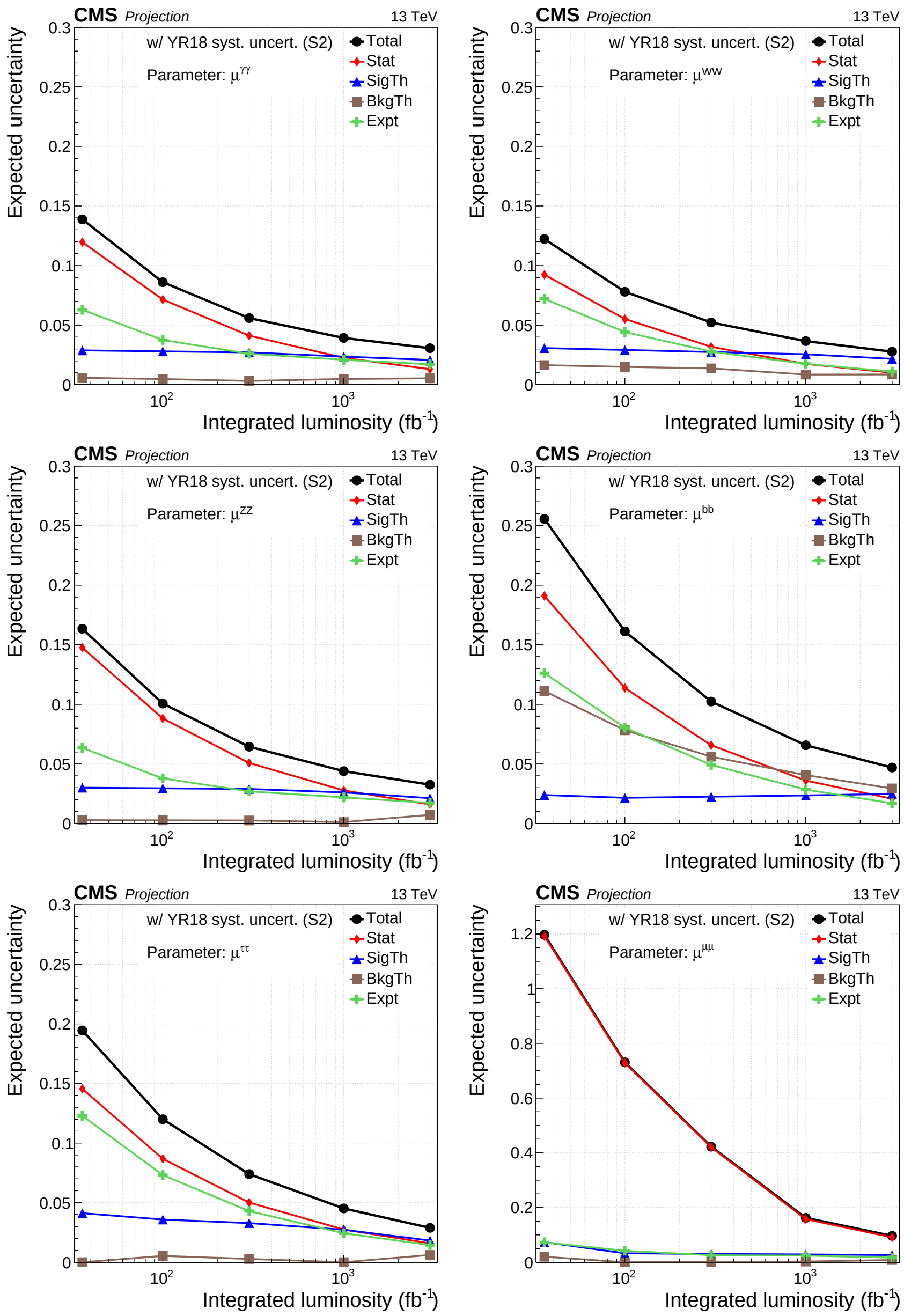

Figure 19: Evolution of the uncertainty contribution with integrated luminosity for the per-decay signal strength parameters in S2 (with YR18 systematic uncertainties).

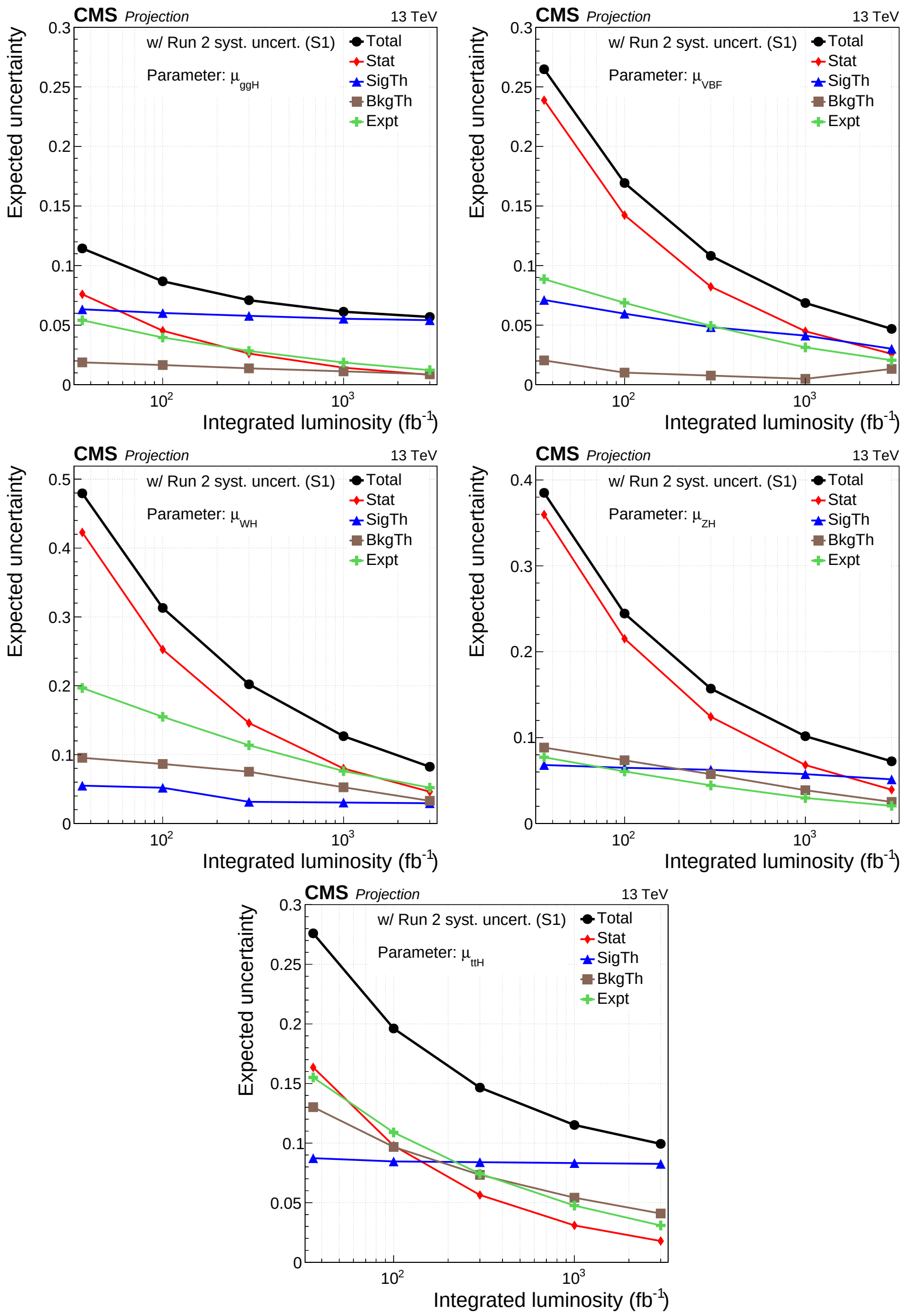

Figure 20: Evolution of the uncertainty contribution with integrated luminosity for the per-production signal strength parameters in S1 (with Run 2 systematic uncertainties [30]).

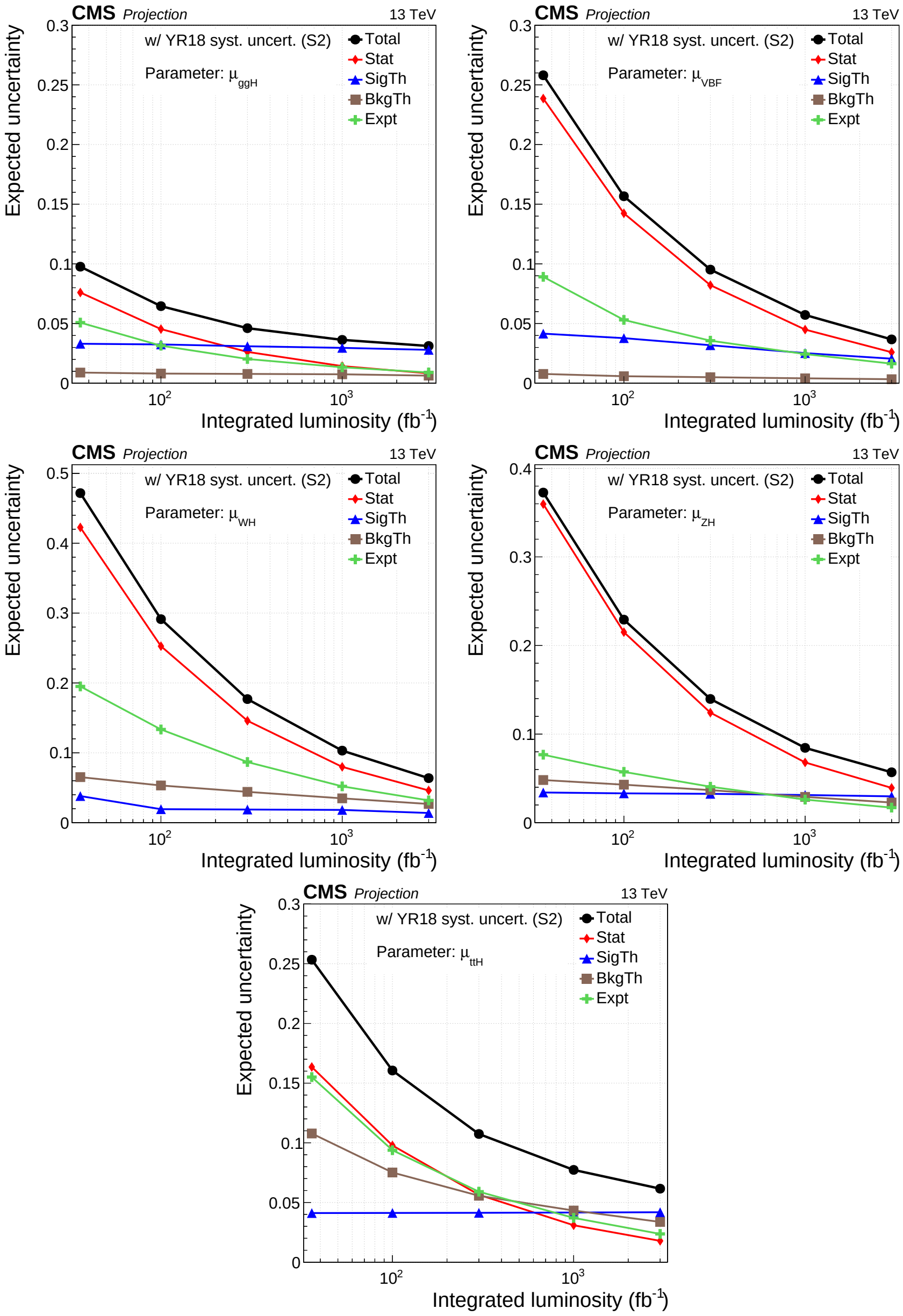

Figure 21: Evolution of the uncertainty contribution with integrated luminosity for the per-production signal strength parameters in S2 (with YR18 systematic uncertainties).

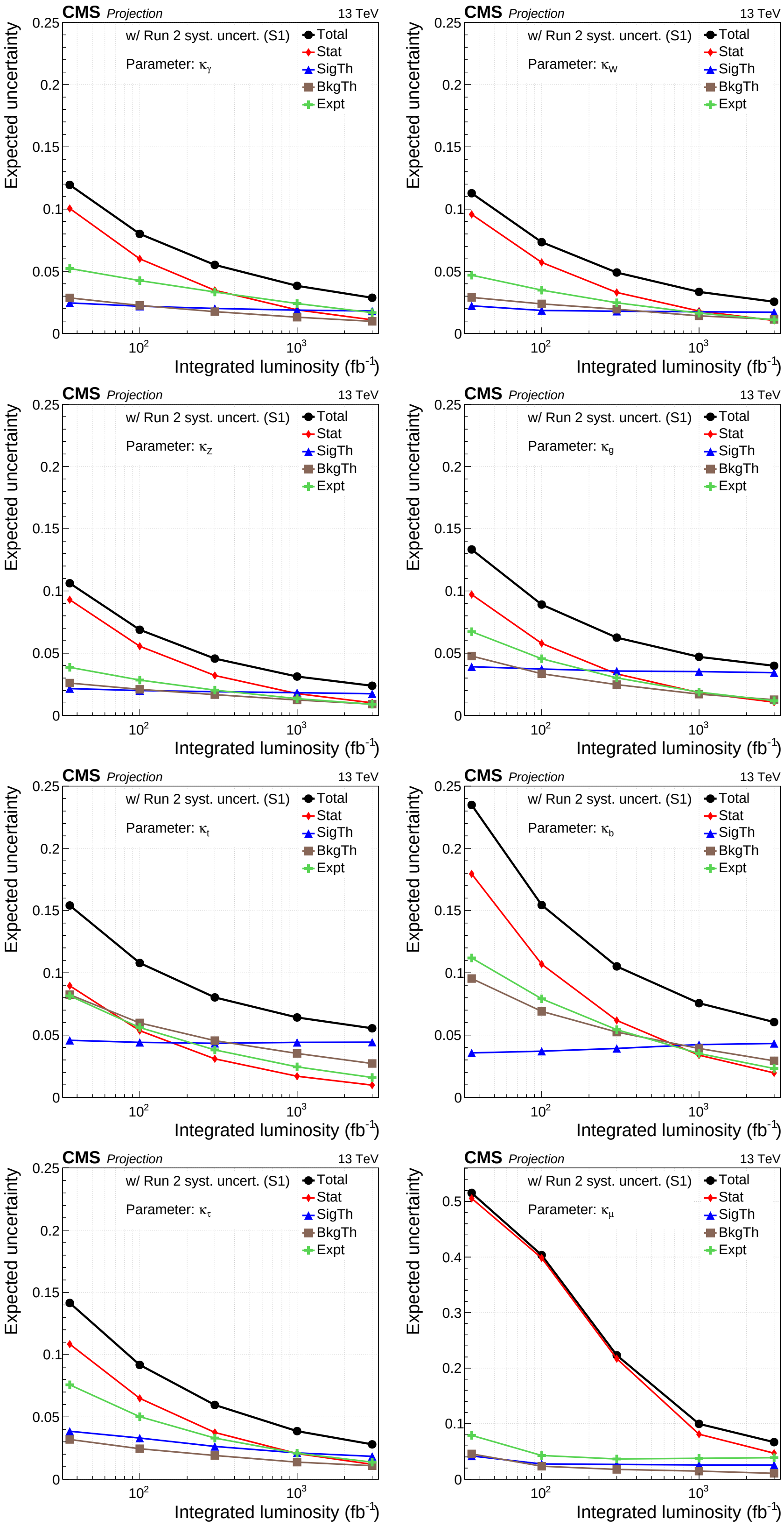

Figure 22: Evolution of the uncertainty contribution with integrated luminosity for the coupling modifier parameters in S1 (with Run 2 systematic uncertainties [30]).

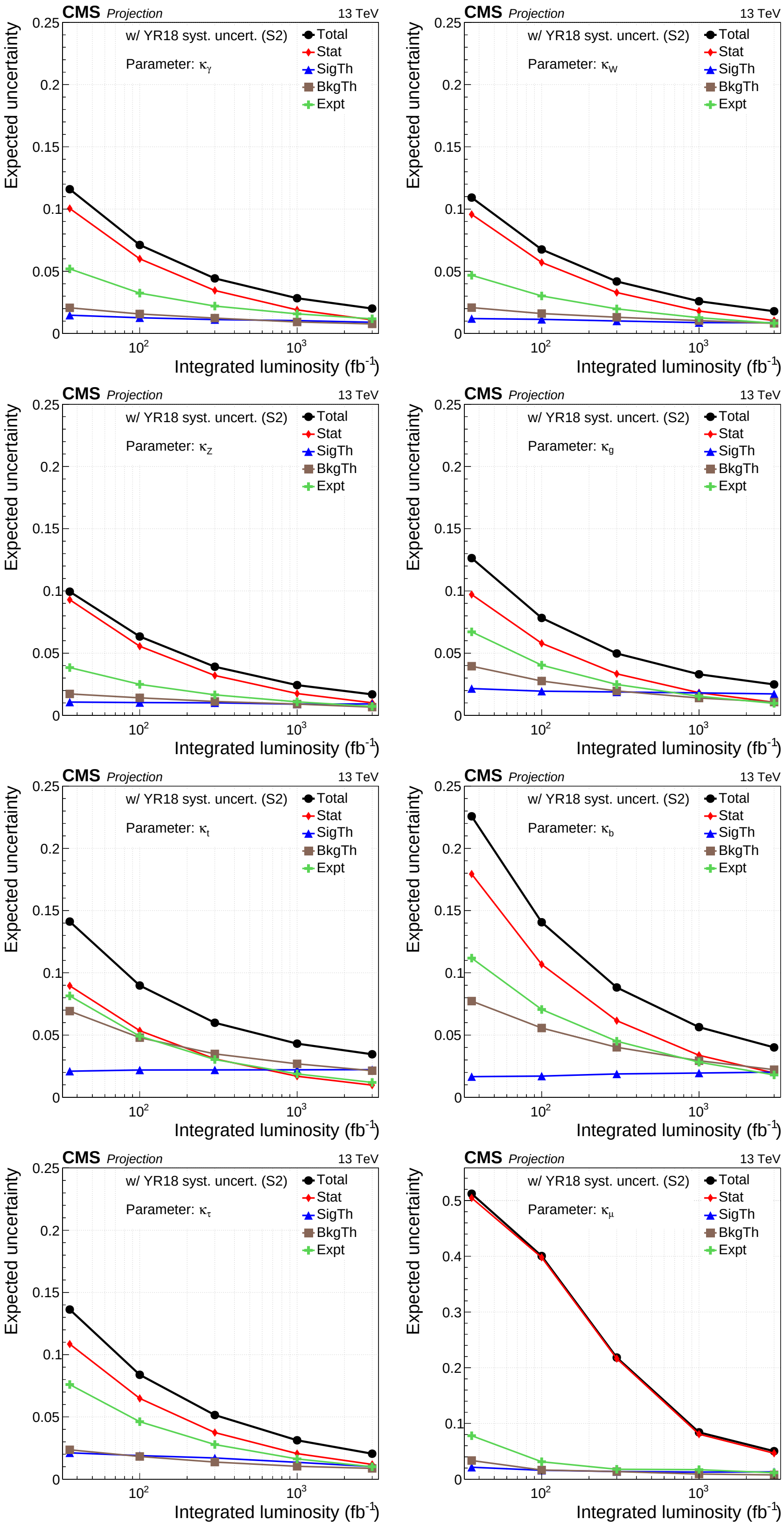

Figure 23: Evolution of the uncertainty contribution with integrated luminosity for the coupling modifier parameters in S2 (with YR18 systematic uncertainties).

# B Results for parametrisation with ratios of coupling modifiers

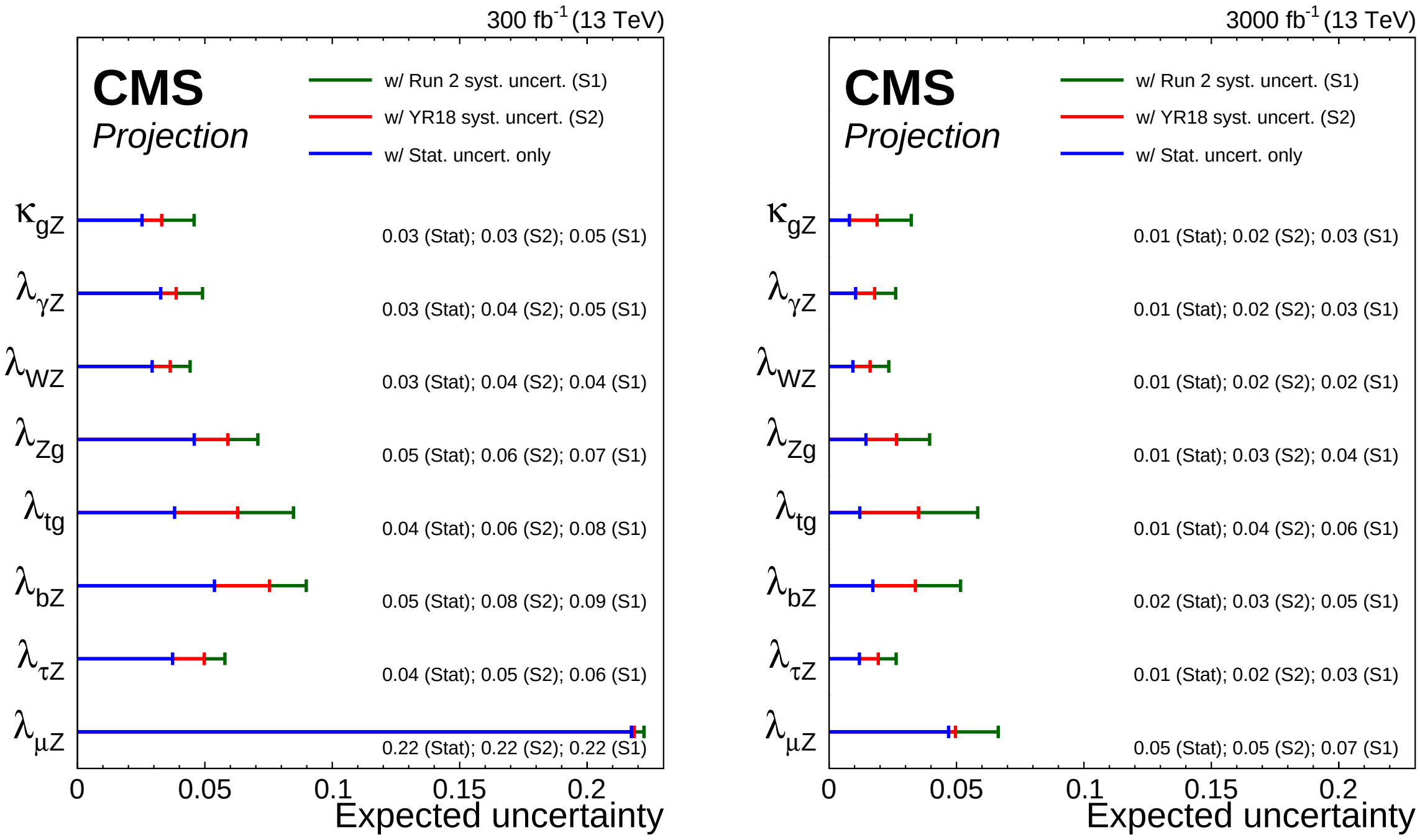

Figure 24: Summary plot showing the total expected $\pm 1\sigma$ uncertainties in S1 (with Run 2 systematic uncertainties [30]) and S2 (with YR18 systematic uncertainties) on the coupling modifier ratio parameters for 300 fb$^{-1}$ (left) and 3000 fb$^{-1}$ (right).

Table 11: The expected $\pm 1\sigma$ uncertainties, expressed as percentages, on the coupling modifier ratio parameters. Values are given for both S1 (with Run 2 systematic uncertainties [30]) and S2 (with YR18 systematic uncertainties). The total uncertainty is decomposed into four components: statistical (Stat), signal theory (SigTh), background theory (BkgTh) and experimental (Exp).

| | | 300 fb$^{-1}$ uncertainty [%] | | | | | 3000 fb$^{-1}$ uncertainty [%] | | | | |
|---|---|---|---|---|---|---|---|---|---|---|---|
| | | Total | Stat | SigTh | BkgTh | Exp | Total | Stat | SigTh | BkgTh | Exp |
| $\kappa_{gZ}$ | S1 | 4.6 | 2.5 | 3.1 | 0.4 | 2.2 | 3.2 | 0.8 | 2.7 | 0.9 | 1.2 |
| | S2 | 3.3 | 2.5 | 1.6 | 0.4 | 1.4 | 1.9 | 0.8 | 1.4 | 0.4 | 0.8 |
| $\lambda_{\gamma Z}$ | S1 | 4.9 | 3.3 | 1.6 | 0.4 | 3.3 | 2.6 | 1.0 | 1.1 | 1.0 | 1.8 |
| | S2 | 3.9 | 3.3 | 0.9 | 0.3 | 1.9 | 1.8 | 1.0 | 0.7 | 0.2 | 1.2 |
| $\lambda_{WZ}$ | S1 | 4.4 | 2.9 | 1.9 | 0.9 | 2.6 | 2.3 | 0.9 | 1.4 | 1.0 | 1.3 |
| | S2 | 3.6 | 2.9 | 1.1 | 0.7 | 1.8 | 1.6 | 0.9 | 0.8 | 0.5 | 0.9 |
| $\lambda_{Zg}$ | S1 | 7.1 | 4.6 | 3.8 | 1.7 | 3.4 | 3.9 | 1.4 | 3.2 | 1.1 | 1.4 |
| | S2 | 5.9 | 4.6 | 2.0 | 1.3 | 2.8 | 2.6 | 1.4 | 1.8 | 0.7 | 1.1 |
| $\lambda_{tg}$ | S1 | 8.5 | 3.8 | 5.2 | 2.8 | 4.7 | 5.8 | 1.2 | 5.0 | 1.8 | 1.9 |
| | S2 | 6.3 | 3.8 | 2.6 | 2.1 | 3.7 | 3.5 | 1.2 | 2.5 | 1.3 | 1.6 |
| $\lambda_{bZ}$ | S1 | 9.0 | 5.4 | 3.4 | 3.8 | 5.1 | 5.2 | 1.7 | 3.4 | 2.6 | 2.3 |
| | S2 | 7.5 | 5.4 | 1.8 | 3.0 | 3.9 | 3.4 | 1.7 | 1.7 | 1.7 | 1.7 |
| $\lambda_{\tau Z}$ | S1 | 5.8 | 3.7 | 2.4 | 0.9 | 3.6 | 2.6 | 1.2 | 1.2 | 1.2 | 1.6 |
| | S2 | 5.0 | 3.7 | 1.5 | 0.7 | 2.9 | 1.9 | 1.2 | 0.9 | 0.4 | 1.2 |
| $\lambda_{\mu Z}$ | S1 | 22.2 | 21.7 | 2.6 | 0.7 | 3.6 | 6.6 | 4.7 | 2.2 | 1.1 | 4.0 |
| | S2 | 21.8 | 21.7 | 1.7 | 0.6 | 1.4 | 5.0 | 4.7 | 1.1 | 0.4 | 1.2 |

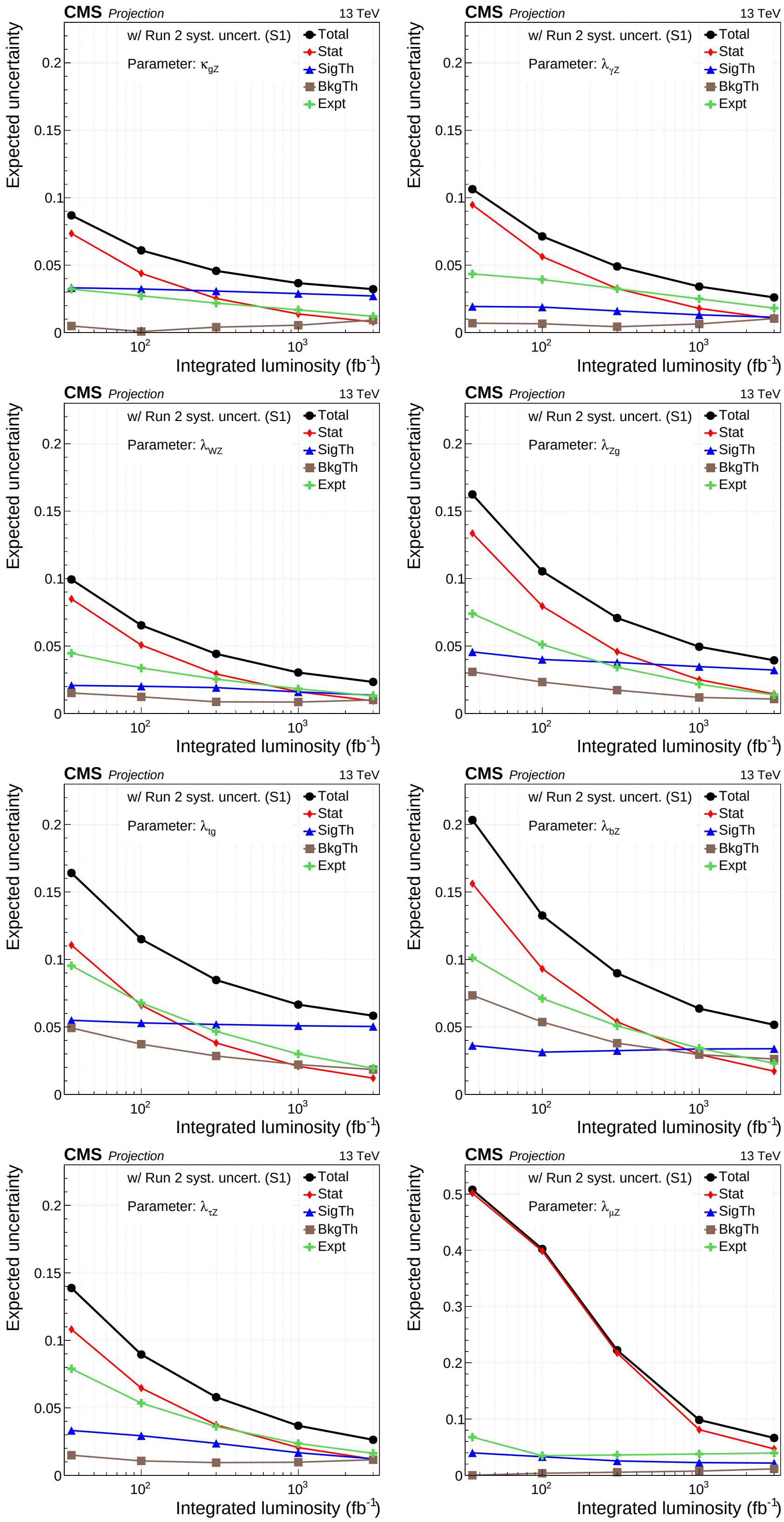

Figure 25: Evolution of the uncertainty contribution with integrated luminosity for the coupling modifier ratio parameters in S1 (with Run 2 systematic uncertainties [30]).

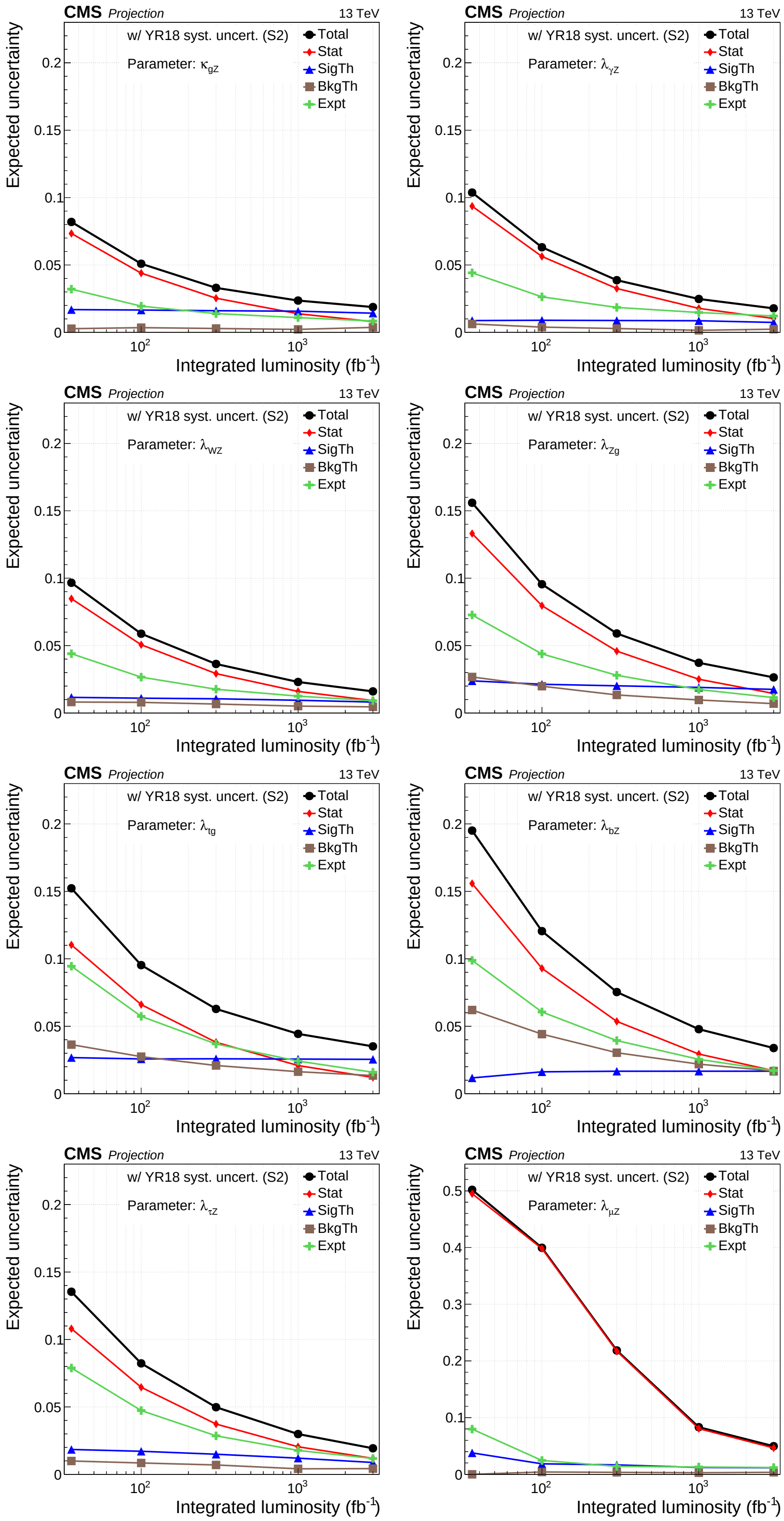

Figure 26: Evolution of the uncertainty contribution with integrated luminosity for the coupling modifier ratio parameters in S2 (with YR18 systematic uncertainties).

# C Cross section and branching fraction results

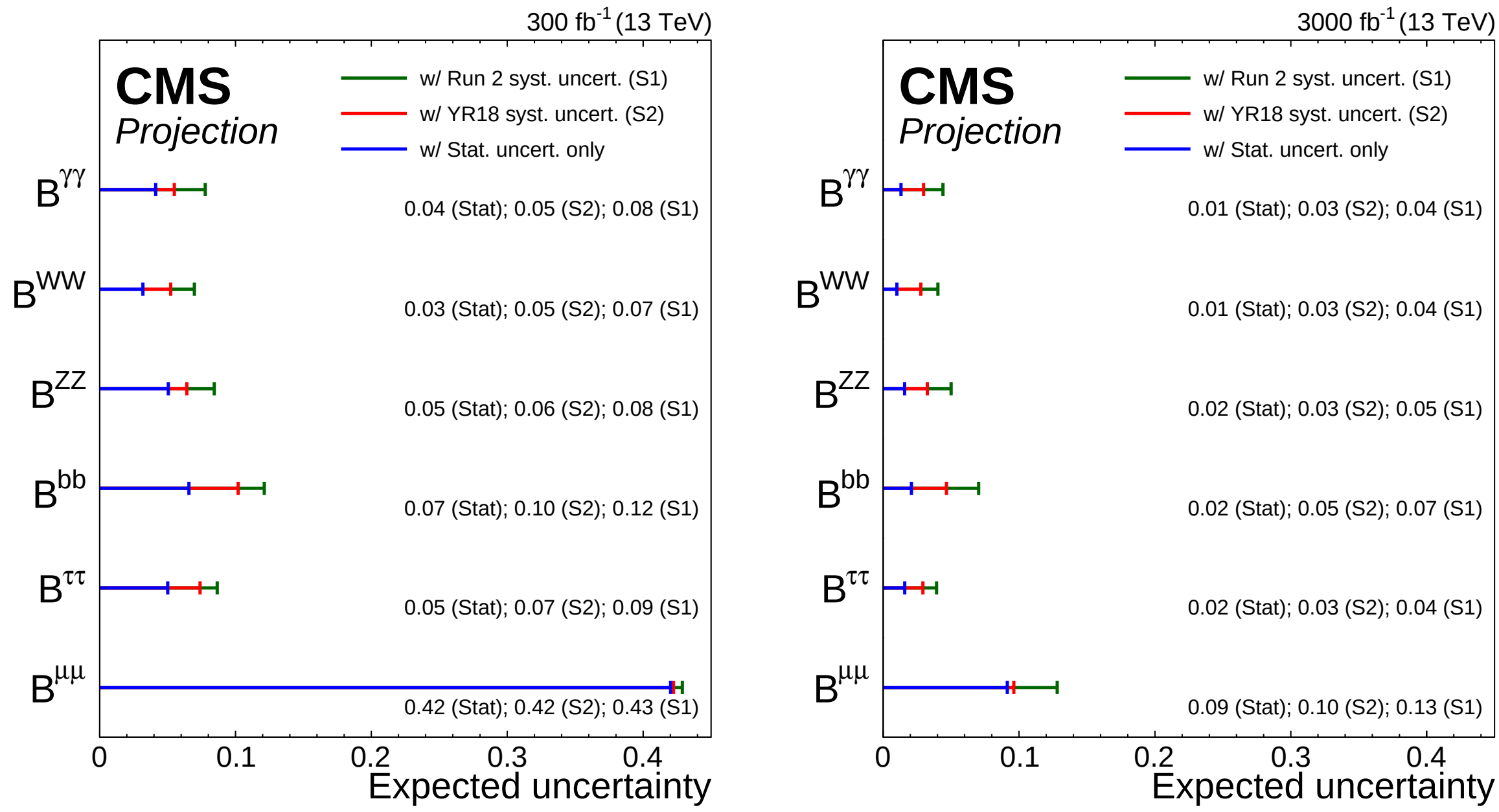

Figure 27: Summary plot showing the total expected $\pm 1\sigma$ uncertainties in S1 (with Run 2 systematic uncertainties [30]) and S2 (with YR18 systematic uncertainties) on the branching ratio measurements for $300\,\mathrm{fb}^{-1}$ (left) and $3000\,\mathrm{fb}^{-1}$ (right). The statistical-only component of the uncertainty is also shown.

Table 12: The expected $\pm 1\sigma$ uncertainties on the branching ratio measurements, expressed as percentages, and assuming the SM values for the production cross sections. Values are given for both S1 (with Run 2 systematic uncertainties [30]) and S2 (with YR18 systematic uncertainties). The total uncertainty is decomposed into four components: statistical (Stat), signal theory (SigTh), background theory (BkgTh) and experimental (Exp). The theory uncertainties on the branching ratios are neglected in these results.

| | | 300 $\mathrm{fb}^{-1}$ uncertainty [%] | | | | | 3000 $\mathrm{fb}^{-1}$ uncertainty [%] | | | | |
|---|---|---|---|---|---|---|---|---|---|---|---|
| | | Total | Stat | Exp | SigTh | BkgTh | Total | Stat | Exp | SigTh | BkgTh |
| $B^{\gamma\gamma}$ | S1 | 7.8 | 4.1 | 4.8 | 4.5 | 0.3 | 4.4 | 1.3 | 2.6 | 3.3 | 0.3 |
| | S2 | 5.5 | 4.1 | 2.6 | 2.5 | 0.3 | 3.0 | 1.3 | 1.7 | 1.9 | 0.3 |
| $B^{WW}$ | S1 | 7.0 | 3.2 | 3.5 | 4.8 | 1.8 | 4.0 | 1.0 | 1.4 | 3.5 | 1.0 |
| | S2 | 5.2 | 3.2 | 2.8 | 2.8 | 1.4 | 2.8 | 1.0 | 1.1 | 2.2 | 0.9 |
| $B^{ZZ}$ | S1 | 8.4 | 5.1 | 4.5 | 5.0 | 0.4 | 5.0 | 1.6 | 2.5 | 3.5 | 1.9 |
| | S2 | 6.4 | 5.1 | 2.7 | 2.8 | 0.3 | 3.2 | 1.6 | 1.7 | 2.1 | 0.7 |
| $B^{bb}$ | S1 | 12.1 | 6.6 | 5.6 | 4.7 | 7.0 | 7.0 | 2.1 | 2.3 | 5.2 | 3.6 |
| | S2 | 10.2 | 6.6 | 4.9 | 2.3 | 5.6 | 4.7 | 2.1 | 1.7 | 2.4 | 2.9 |
| $B^{\tau\tau}$ | S1 | 8.7 | 5.0 | 5.0 | 4.8 | 0.9 | 3.9 | 1.6 | 1.9 | 2.6 | 1.5 |
| | S2 | 7.4 | 5.0 | 4.3 | 3.3 | 0.9 | 2.9 | 1.6 | 1.4 | 1.9 | 0.6 |
| $B^{\mu\mu}$ | S1 | 42.9 | 42.0 | 5.9 | 5.3 | 0.8 | 12.8 | 9.1 | 7.6 | 4.7 | 0.8 |
| | S2 | 42.2 | 42.0 | 2.6 | 2.8 | 0.8 | 9.6 | 9.1 | 1.7 | 2.6 | 0.8 |

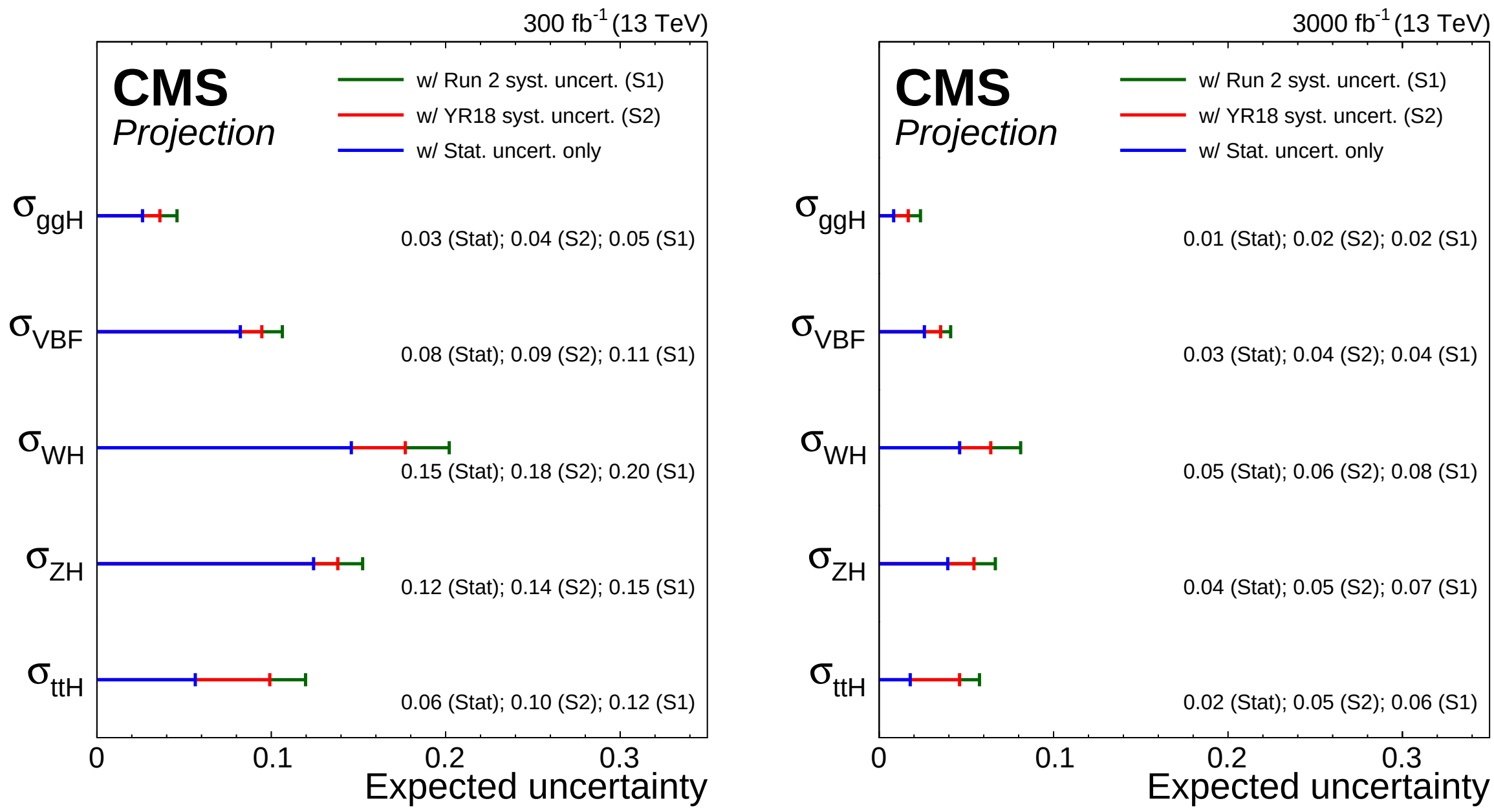

Figure 28: Summary plot showing the total expected $\pm 1\sigma$ uncertainties in S1 (with Run 2 systematic uncertainties [30]) and S2 (with YR18 systematic uncertainties) on the cross section measurements for 300 $\mathrm{fb}^{-1}$ (left) and 3000 $\mathrm{fb}^{-1}$ (right). The statistical-only component of the uncertainty is also shown.

Table 13: The expected $\pm 1\sigma$ uncertainties on the cross section measurements, expressed as percentages, and assuming the SM values for the branching fractions. Values are given for both S1 (with Run 2 systematic uncertainties [30]) and S2 (with YR18 systematic uncertainties). The total uncertainty is decomposed into four components: statistical (Stat), signal theory (SigTh), background theory (BkgTh) and experimental (Exp). The theory uncertainties on the production cross sections are neglected in these results.

| | | 300 fb$^{-1}$ uncertainty [%] | | | | | 3000 fb$^{-1}$ uncertainty [%] | | | | |
|---|---|---|---|---|---|---|---|---|---|---|---|
| | | Total | Stat | Exp | SigTh | BkgTh | Total | Stat | Exp | SigTh | BkgTh |
| $\sigma_{ggH}$ | S1 | 4.6 | 2.6 | 2.8 | 2.1 | 1.4 | 2.4 | 0.8 | 1.2 | 1.6 | 0.9 |
| | S2 | 3.6 | 2.6 | 2.0 | 1.2 | 0.8 | 1.7 | 0.8 | 0.9 | 0.9 | 0.6 |
| $\sigma_{VBF}$ | S1 | 10.6 | 8.2 | 4.9 | 4.4 | 1.2 | 4.1 | 2.6 | 2.1 | 2.0 | 1.3 |
| | S2 | 9.5 | 8.2 | 3.6 | 3.0 | 0.5 | 3.5 | 2.6 | 1.6 | 1.8 | 0.3 |
| $\sigma_{WH}$ | S1 | 20.2 | 14.6 | 11.4 | 3.1 | 7.5 | 8.1 | 4.6 | 5.2 | 2.6 | 3.3 |
| | S2 | 17.7 | 14.6 | 8.7 | 2.2 | 4.4 | 6.4 | 4.6 | 3.2 | 1.5 | 2.7 |
| $\sigma_{ZH}$ | S1 | 15.2 | 12.4 | 4.4 | 5.0 | 5.7 | 6.7 | 3.9 | 2.1 | 4.3 | 2.5 |
| | S2 | 13.8 | 12.4 | 4.1 | 2.5 | 3.7 | 5.4 | 3.9 | 1.7 | 2.4 | 2.3 |
| $\sigma_{ttH}$ | S1 | 12.0 | 5.6 | 7.4 | 1.5 | 7.3 | 5.8 | 1.8 | 3.1 | 1.9 | 4.1 |
| | S2 | 9.9 | 5.6 | 5.9 | 0.8 | 5.6 | 4.6 | 1.8 | 2.4 | 1.1 | 3.4 |

# Prospects for differential cross-section measurements of Higgs boson production measured in decays to ZZ and $\gamma\gamma$ with the ATLAS experiment at the High-Luminosity LHC

The ATLAS Collaboration

Prospects for Higgs boson differential cross section measurements for the full expected High-Luminosity LHC (HL-LHC) luminosity of 3 ab$^{-1}$ are performed using the $H \to \gamma\gamma$, $H \to ZZ^* \to 4\ell$ decay channels, as well as their combination. Differential cross section measurements are presented for the Higgs boson transverse momentum, Higgs boson rapidity, number of jets produced together with the Higgs boson, and the transverse momentum of the leading jet.

# 1 Introduction

Higgs boson differential cross section measurements are an important probe of the Standard Model (SM) and provide constraints on effects from physics beyond the SM. As almost model independent measurements, they are well suited to be used for a variety of interpretations (e.g. Ref. [1, 2]).
In this Note, the projections of Higgs boson differential cross-sections measured in the $H \to \gamma\gamma$ and $H \to ZZ^* \to 4\ell$ decay channels, as well as results obtained from combining the two decay channels, are shown for the full expected High-Luminosity LHC (HL-LHC [3]) integrated luminosity. Cross-sections are obtained from measured event yields by subtracting the backgrounds and taking into account detector efficiencies, resolutions, acceptances and branching fractions following the methods used in Ref. [4] for the $H \to \gamma\gamma$ decay channel, Ref. [5] and Ref. [6] for the $H \to ZZ^* \to 4\ell$, and Ref. [7] for their combination. The projections are obtained by scaling the signal and background Asimov datasets used for the Run2 analyses to the HL-LHC expected integrated luminosity of 3 ab$^{-1}$.
The performance of the future detector will be comparable to or better than the one in Run2. The higher pileup present in the HL-LHC is assumed to be compensated for by new detectors, reconstruction algorithms and analysis strategies to achieve the same performance as in Run2. Two different scenarios in the context of the HL-LHC are studied: in the first, the systematic uncertainties are considered to be the same as the Run2 analyses, while in the second expected improvements in systematic uncertainties are taken into account. In the latter, the current Run2 systematic uncertainties are scaled following Ref. [3].
The observables studied include the Higgs boson transverse momentum $p_T^H$ and rapidity $|y^H|$, the number of jets $N_{jets}$ with jet transverse momentum above 30 GeV, and the leading jet transverse momentum $p_T^{j1}$. For the $H \to \gamma\gamma$ channel, a different projection was performed in Ref. [8] for the $p_T^H$ using different pileup and systematic uncertainties scenarios, as well as a different estimation technique in the high $p_T^H$ regime. For the $H \to ZZ^* \to 4\ell$ channel, projections for the $|y^H|$ and $p_T^H$ were made taking into account improvements in ATLAS Muon Spectrometer [9]. In this note, the $p_T^H$ distribution is further studied by adding a new higher-$p_T$ bin with respect to the previous projections.
The methodology followed for this study is discussed in Section 2. Section 3 presents the results. The conclusions are reported in Section 4.

# 2 Analysis Strategy

This study aims to provide an estimated precision for the HL-LHC measurements of the $H \to \gamma\gamma$, $H \to ZZ^* \to 4\ell$, and combined differential cross sections for 3 ab$^{-1}$. The projections are performed using Asimov datasets matching the 2015-2017 luminosity, which are scaled to the expected integrated luminosity of 3 ab$^{-1}$ and center-of-mass energy of $\sqrt{s} = 14$ TeV [10]. The extrapolations are based on the Run2 analyses using an integrated luminosity of 36 fb$^{-1}$ [4, 5, 7], with the exception of the $p_T^H$ observable in $H \to ZZ^* \to 4\ell$ where the measurement based on 80 fb$^{-1}$ is used [6]. For the combination, since there are some observables where the binning choice is more granular in one channel than in the other, the bins with higher granularity are summed and combined with the corresponding bin of the other channel. Measurements are based on maximizing the profile-likelihood ratio

$$\Lambda(\vec{\sigma}) = \frac{\mathcal{L}(\vec{\sigma}, \hat{\hat{\theta}}(\vec{\sigma}))}{\mathcal{L}(\hat{\vec{\sigma}}, \hat{\theta})} \tag{1}$$

where $\vec{\sigma}$ is a vector whose elements correspond to the Higgs boson production cross section of each bin for a given observable, $\theta$ the nuisance parameters, which correspond to the systematic uncertainties considered, and $\mathcal{L}$ the likelihood function, which includes the signal extraction, correction to particle level, and extrapolation to the total phase space. The $\hat{\vec{\sigma}}$ and $\hat{\theta}$ terms denote the unconditional maximum-likelihood estimate of the parameters, and $\hat{\hat{\theta}}(\vec{\sigma})$ is the conditional maximum-likelihood estimate for the given parameter values. Under certain assumptions, the effects of foreseen changes to background measurements in each analysis are modeled. Previous measurements in the $H \rightarrow ZZ^* \rightarrow 4\ell$ channel have taken the normalization of the dominant background, non-resonant SM $ZZ^*$ production, from MC simulation: in the future, this normalization will likely be a parameter of the fit, constrained by the $m_{4\ell}$ sidebands. In the $H \rightarrow \gamma\gamma$ channel, the uncertainties on the background parameterization (spurious signal) are largely due to limited statistics in the MC and the data sidebands: for HL-LHC we expect to develop new methods which should render these uncertainties negligible. As with the Run2 results, the combined cross sections, as well as the cross sections in the individual channels, are extracted for the total phase space, and hence are slightly more model-dependent than the cross section measurements in the individual channels performed in Run2, which are extracted in fiducial phase spaces close to the selection criteria for reconstructed events in the detector. The center-of-mass energy increase is taken into account by applying a scale factor to the event yields, neglecting any change in the kinematic distributions, but taking into consideration the different production process composition for each observable bin. Among the four measured observables, the Higgs boson transverse momentum is of particular importance, since it can be used to test perturbative QCD calculations and it is also sensitive to the Lagrangian structure of the Higgs boson interactions [11]; therefore for this study the $p_{\mathrm{T}}^{H}$ measurement range was extended up to 1 TeV, with the highest-$p_{\mathrm{T}}^{H}$ bin defined to be $p_{\mathrm{T}}^{H} \in$ [350 GeV, 1 TeV]. Such a bin, while already present in the $H \rightarrow ZZ^* \rightarrow 4\ell$ 80 fb$^{-1}$ analysis, is not included in the $H \rightarrow \gamma\gamma$ Run2 measurement, which, as described in Ref. [4], gathered all the events beyond $p_{\mathrm{T}}^{H} = 350$ GeV in an overflow bin. For the projections, to construct a measurement for a bin $p_{\mathrm{T}}^{H} \in$ [350 GeV, 1000 GeV], the signal and background shapes of the overflow bin have been used, as well as the expected background and signal in this $p_{\mathrm{T}}^{H}$ range.

### 2.1 Systematic uncertainties

Systematic uncertainties are modeled as nuisance parameters in the fit. The impact of some systematic uncertainties is reduced to reflect expected improvements in the years leading up to the HL-LHC (Table 1) [3]. The theoretical and experimental uncertainties related to jet reconstruction and luminosity determination are expected to be ameliorated due to detector upgrades and measurements. The uncertainties related to the photon energy scale and resolution are scaled down, considering that the large amount of data expected and a better knowledge of the detector itself will allow for a more precise energy calibration. The uncertainties related to photon reconstruction, identification, and isolation are also scaled down, assuming that new methods will mitigate the higher pileup environment, which will allow the achievement of a better precision than the current one. Additionally, changes to the background estimation described in the previous section also affect the associated systematic uncertainties. For the $H \rightarrow \gamma\gamma$ analysis, the uncertainty related to the background modeling is set to zero regardless of whether the other systematic uncertainties are scaled. In the $H \rightarrow ZZ^* \rightarrow 4\ell$ channel, the uncertainties on the $ZZ$ irreducible background are scaled such that their sum in quadrature matches the normalization uncertainty expected from the new estimation procedure. Uncertainties affecting the shape of the $ZZ$ background are neglected.

Table 1: A summary list of the systematic uncertainties expected to be improved. The table briefly describes these uncertainties and specifies the amount by which they are scaled.

| Systematic Uncertainties | Scale Factor |
|---|---|
| Jet energy scale, forward region | Set to 0 |
| Jet energy scale, Jet punch-through | Set to 0 |
| High-$p_T$ jet energy scale | Set to 0 |
| $H \rightarrow \gamma\gamma$ background modeling | Set to 0 |
| $4\ell$ $m_H$ | Scaled by 0.25 |
| PDF | Scaled by 0.41 |
| Jet flavor | Scaled by 0.5 |
| Jet energy scale | Scaled by 0.5 |
| Pileup modelling | Scaled by 0.5 |
| QCD scale | Scaled by 0.5 |
| Underlying event and parton shower modeling | Scaled by 0.5 |
| Higgs branching ratios | Scaled by 0.5 |
| Photon energy scale and resolution | Scaled by 0.8 [1] |
| Photon reconstruction, ID, and isolation | Scaled by 0.8 |
| $qq \rightarrow ZZ$ irreducible background | Set to 2% |
| Luminosity | Set to 1% of expected integrated luminosity |

## 3 Results

The results of the HL-LHC projection of the differential cross section measurements are shown in Figure 1, for the $H \rightarrow \gamma\gamma$ decay channel, Figure 2, for the $H \rightarrow ZZ^* \rightarrow 4\ell$ decay channel, and Figure 3, for the combination of the two aforementioned decay channels. The dominant systematic uncertainties for the $N_{jets}$ and $|y^H|$ measurements are from luminosity and pileup, for $p_T^{j1}$ the jet flavour (quark or gluon) composition and the Higgs boson kinematics, and for $p_T^H$ the photon identification efficiency and the luminosity. A summary of the expected $p_T^H$ bins sensitivity is presented in Table 2 for the three channels. These results show that the highest $p_T^H$ bin will be sensitive to the quark top mass effects in the SM loop of the gluon fusion Higgs production according to theoretical predictions [12, 13].

[1] The impact of scaling down the photon energy scale and resolution uncertainty further down from a factor of 0.8 to 0.5 is found to be negligible.

Table 2: Summary of the expected relative uncertainty of the measurements in each $p_T^H$ bin for the (a) $H \to \gamma\gamma$ decay channel, (b) $H \to ZZ^* \to 4\ell$ decay channel, as well as (c) their combination.

| Bin [GeV] | Relative uncertainty [%] Without Sys. | Relative uncertainty [%] With Unscaled Syst. | Relative uncertainty [%] With Scaled Syst. |
|---|---|---|---|
| 0, 10 | 4.7 | 6.5 | 5.3 |
| 10, 20 | 3.9 | 5.9 | 4.6 |
| 20, 30 | 4.3 | 6.2 | 4.9 |
| 30, 45 | 4.1 | 6.0 | 4.7 |
| 45, 60 | 4.9 | 6.5 | 5.4 |
| 60, 80 | 5.0 | 6.7 | 5.7 |
| 80, 120 | 4.3 | 6.0 | 4.9 |
| 120, 200 | 3.4 | 5.4 | 4.2 |
| 200, 350 | 3.9 | 6.3 | 5.1 |
| 350, 1000 | 7.4 | 9.5 | 8.7 |

(a)

| Bin [GeV] | Relative uncertainty [%] Without Sys. | Relative uncertainty [%] With Unscaled Syst. | Relative uncertainty [%] With Scaled Syst. |
|---|---|---|---|
| 0, 10 | 5.5 | 9.0 | 8.3 |
| 10, 15 | 6.1 | 8.1 | 7.6 |
| 15, 20 | 6.2 | 8.9 | 8.3 |
| 20, 30 | 4.6 | 6.9 | 6.3 |
| 30, 45 | 4.3 | 6.3 | 5.7 |
| 45, 60 | 5.2 | 6.8 | 6.2 |
| 60, 80 | 5.4 | 6.8 | 6.3 |
| 80, 120 | 4.9 | 6.2 | 5.7 |
| 120, 200 | 5.6 | 6.7 | 6.4 |
| 200, 350 | 9.4 | 13.2 | 13.1 |
| 350, 1000 | 23 | 24 | 23 |

(b)

| Bin [GeV] | Relative uncertainty [%] Without Sys. | Relative uncertainty [%] With Unscaled Syst. | Relative uncertainty [%] With Scaled Syst. |
|---|---|---|---|
| 0, 10 | 3.2 | 5.5 | 4.5 |
| 10, 20 | 3.0 | 4.8 | 3.8 |
| 20, 30 | 2.8 | 5.0 | 3.9 |
| 30, 45 | 2.7 | 4.7 | 3.6 |
| 45, 60 | 3.2 | 5.0 | 4.1 |
| 60, 80 | 3.3 | 5.1 | 4.2 |
| 80, 120 | 2.9 | 4.6 | 3.7 |
| 120, 200 | 2.7 | 4.4 | 3.5 |
| 200, 350 | 3.4 | 5.4 | 4.5 |
| 350, 1000 | 6.8 | 8.7 | 8.2 |

(c)

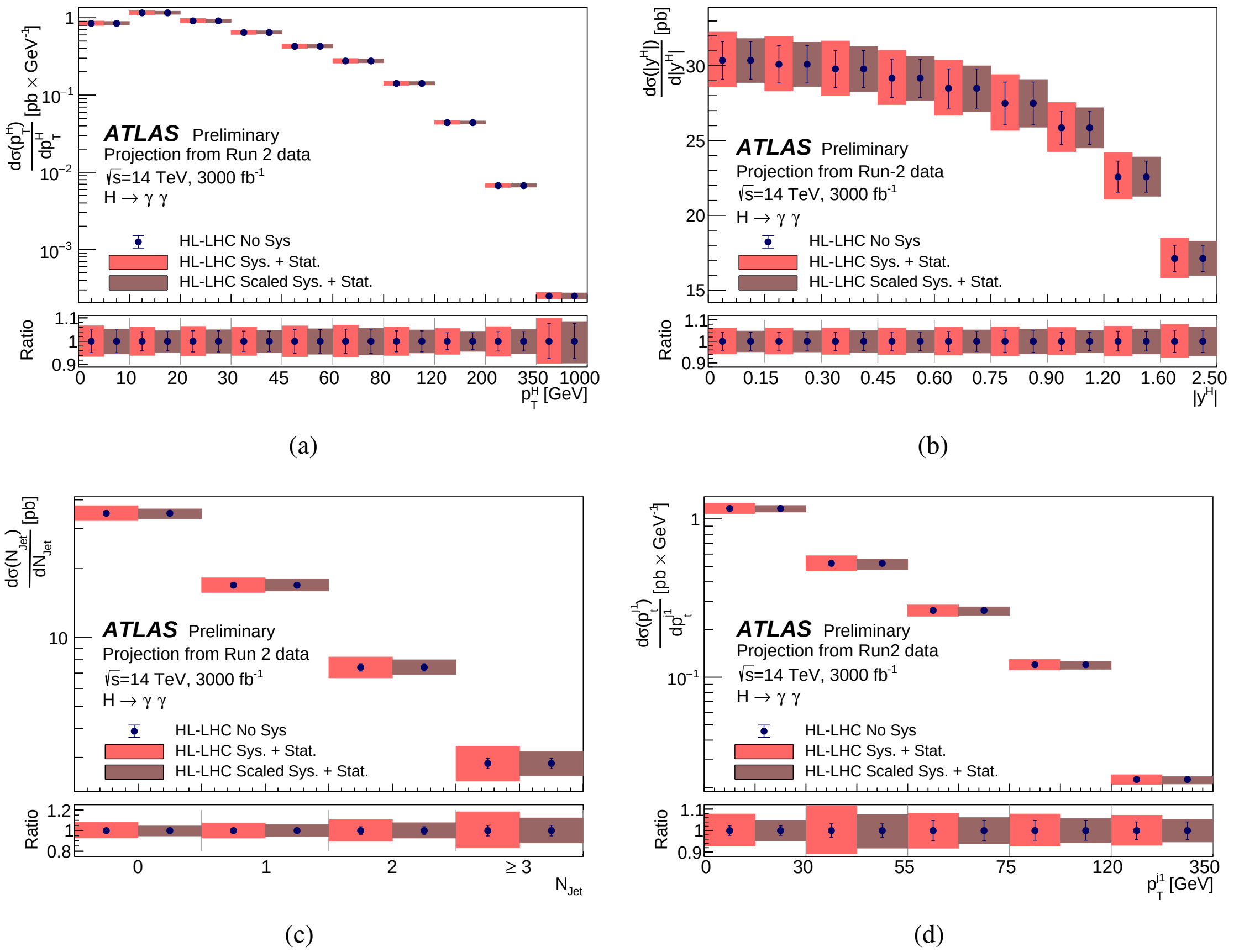

Figure 1: Differential cross section measurements in the total phase space extrapolated to the full HL-LHC luminosity for the $H \to \gamma\gamma$ decay channel for (a) Higgs boson transverse momentum $p_T^H$, (b) Higgs boson rapidity $|y^H|$, (c) number of jets $N_{jets}$ with $p_T > 30$ GeV, and (d) the transverse momentum of the leading jet $p_T^{j1}$. For each point both the statistical (error bar) and total (shaded area) uncertainties are shown. Two scenarios are shown: one with the current Run2 systematic uncertainties and one with scaled systematic uncertainties.

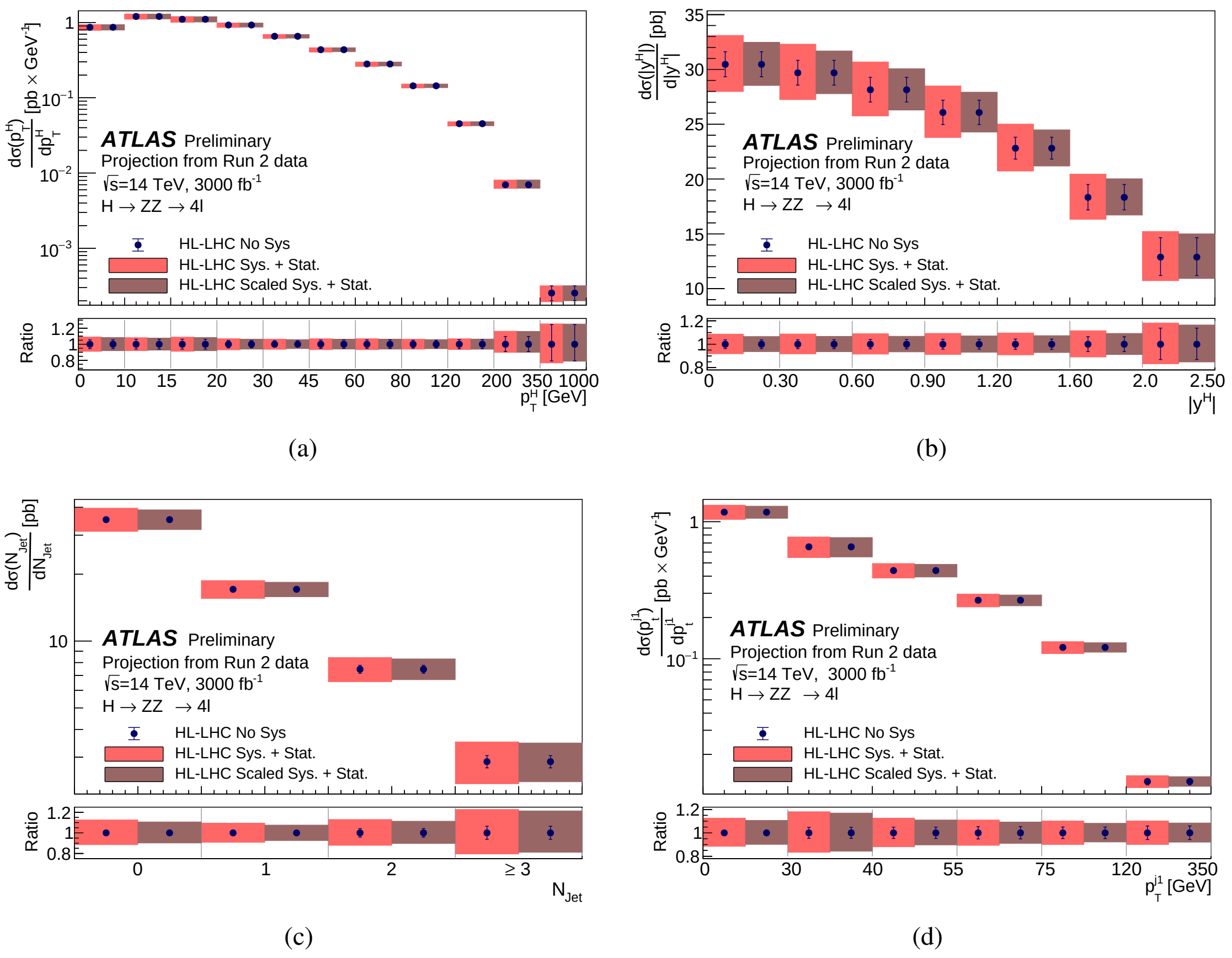

Figure 2: Differential cross section measurements in the total phase space extrapolated to the full HL-LHC luminosity for the $H \to ZZ^* \to 4\ell$ decay channel for (a) Higgs boson transverse momentum $p_{\mathrm{T}}^{H}$, (b) Higgs boson rapidity $|y^{H}|$, (c) number of jets $\mathrm{N_{jets}}$ with $p_{\mathrm{T}} > 30$ GeV, and (d) the transverse momentum of the leading jet $p_{\mathrm{T}}^{j1}$. For each point both the statistical (error bar) and total (shaded area) uncertainties are shown. Two scenarios are shown: one with the current Run2 systematic uncertainties and one with scaled systematic uncertainties.

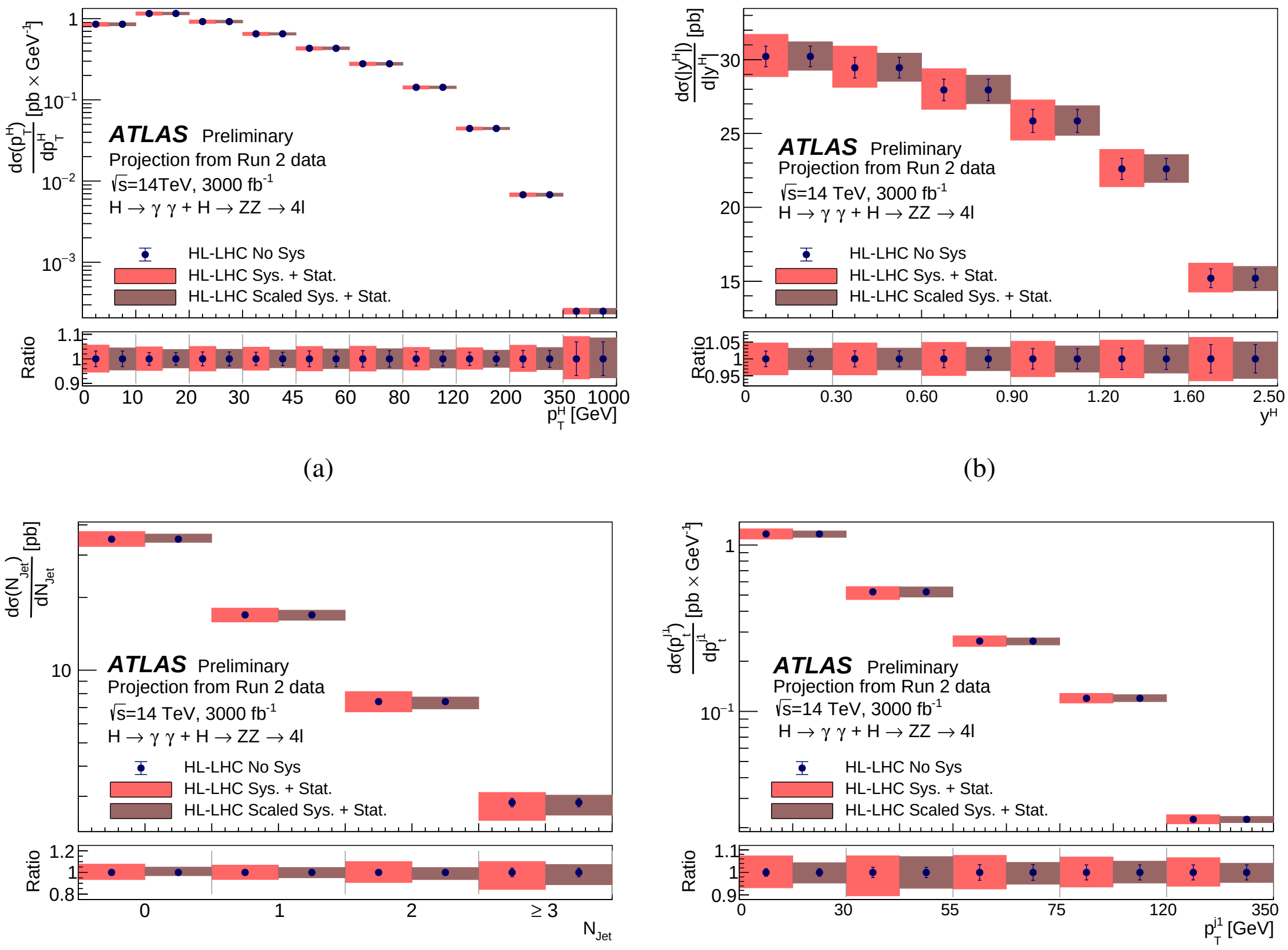

(a) (b)

(c) (d)

Figure 3: Differential cross sections in the total phase space extrapolated to the full HL-LHC luminosity for the combination of the $H \to \gamma\gamma$ and $H \to ZZ^* \to 4\ell$ decay channels for (a) Higgs boson transverse momentum $p_T^H$, (b) Higgs boson rapidity $|y^H|$, (c) number of jets $N_{jets}$ with $p_T > 30$ GeV, and (d) the transverse momentum of the leading jet $p_T^{j1}$. For each point both the statistical (error bar) and total (shaded area) uncertainties are shown. Two scenarios are shown: one with the current Run2 systematic uncertainties and one with scaled systematic uncertainties.

# 4 Conclusion

Projections for measurements of Higgs boson differential cross sections at the HL-LHC were presented. For a binning similar to what is used in current Run2 measurements, the uncertainties are expected to range between 3% and 20%. Furthermore, the high-$p_{\mathrm{T}}^{H}$ bin ($p_{\mathrm{T}}^{H} \in$ [350 GeV, 1 TeV]) will be accessible with a relative precision of about 8% after combining the $H \to \gamma\gamma$ and $H \to ZZ \to 4\ell$ decay channels, and it will be sensitive to the quark top mass effects in the SM loop of the gluon fusion Higgs production according to theoretical predictions [12, 13]. With the increased statistical precision, systematic uncertainties will play an important role for $|y^{H}|$, $\mathrm{N}_{\mathrm{jets}}$, $p_{\mathrm{T}}^{j1}$, and all $p_{\mathrm{T}}^{H}$ bins except the one with the highest $p_{\mathrm{T}}$, which will still be statistically limited. Such improved measurements will allow the single and combined measurements to further probe the SM with unprecedented precision and to test for indications of physics beyond the SM.

# ATLAS NOTE

# Prospects for the measurement of the rare Higgs boson decay $H \to \mu\mu$ with 3000 fb$^{-1}$ of $pp$ collisions collected at $\sqrt{s}$ = 14 TeV by the ATLAS experiment

The ATLAS Collaboration

## Abstract

This note presents a study of the prospects for the measurement of the rare Higgs boson decay $H \to \mu\mu$ using 3000 fb$^{-1}$ of proton-proton collisions at $\sqrt{s}$ = 14 TeV recorded with the ATLAS detector at the high-luminosity LHC. The studies assume an average number of interactions per bunch crossing $\langle\mu\rangle$ = 200. The $H \to \mu\mu$ signal from gluon fusion and vector-boson fusion is expected to be observed with a significance of more than 9 standard deviations. The uncertainty on the Higgs production cross section times the branching ratio to dimuons normalised by the Standard Model prediction is expected to be around 13%.

# 1. Introduction

In July 2012 a Higgs boson with mass near 125 GeV was discovered by ATLAS and CMS through its decays to di-boson final states, $H \to \gamma\gamma$, $H \to ZZ$ and $H \to WW$ [1, 2]. The analysis of the full Run 1 data collected by ATLAS and CMS [3] consolidated these results and also provided observation of decays to the leptonic final state $\tau\tau$ and evidence of the fermionic final state $b\bar{b}$. All the observed rates so far, even exploiting the 13 TeV $pp$ collisions collected by ATLAS and CMS in 2015 and 2016, are consistent with the Standard Model (SM) predictions. This is also the case for the measured $J^P$ properties of this particle, for which the SM $0^+$ hypothesis is favored over various alternatives, including $0^-$, $1^\pm$, and $2^+$ models [4, 5].

The final states and production modes observed so far are induced by the Higgs boson couplings to either the massive vector bosons ($W$, $Z$) or to third-generation fermions, either quarks ($b$, $t$) or leptons ($\tau$). To search for possible deviations from the SM predictions and also to expand the set of measurements of the properties of the discovered particle, it is thus natural to turn to the couplings of the Higgs boson with the other fermion families. In the SM, the largest among such couplings is that to the charm quark, but due to the huge QCD background this is not easily accessible at the LHC. The $H \to \mu^+\mu^-$ decay is an excellent candidate for such a search, with its small but not insignificant cross section times branching ratio (13.6 fb at $\sqrt{s}$ = 14 TeV [6]), smaller backgrounds and excellent invariant mass resolution that allows discrimination between signal and background. The limiting factor to the sensitivity of this search is the small expected signal yield and the very large contribution from the irreducible Drell-Yan background, $pp \to Z + X$, $Z \to \mu^+\mu^-$.

With the full dataset collected in Run 1, ATLAS and CMS were not able to find an evidence of a signal, and set upper limits on the product of the Higgs boson production cross section times muonic branching ratio of 7.0 and 7.4 times the SM expectation, respectively [7, 8]. With about 36.1 fb$^{-1}$ of $pp$ collisions collected by ATLAS at 13 TeV during the LHC Run 2, the observed (expected) upper limit on the cross section times branching ratio is 3.0 (3.1) times the SM prediction at the 95% confidence level for a Higgs boson mass of 125 GeV [9]. When combining together the ATLAS Run 1 and Run 2 results, the observed (expected) upper limit is 2.8 (2.9) times the SM prediction.

The sensitivity of the ATLAS experiment to this decay is studied here for a Higgs boson with a mass of $m_H$ = 125 GeV, assuming an integrated luminosity of 3000 fb$^{-1}$ of $pp$ collisions at $\sqrt{s}$ = 14 TeV. An average number of 200 additional inelastic $pp$ collisions per bunch-crossing (denoted "pile-up"in the rest of the note) is assumed. The analysis is based on a selection and event classification similar to that used for the Run 1 and Run 2 ATLAS searches, and updates the studies documented in Ref. [10], which were based on a simpler strategy without event categories, and assumed an average of 140 pile-up collisions per bunch-crossing ($\langle\mu\rangle$ = 140). The analysis uses generator-level samples of the main signal and background processes, combined with parametrisations of the detector performance (muon and jet reconstruction and selection efficiencies and momentum resolutions) obtained from fully simulated samples. The parametric resolutions are used to smear the generator-level particle transverse momenta, while the parametric efficiencies are used to reweigh the selected events. Compared to the previous study of Ref. [10], updated parametrizations of the ATLAS detector efficiencies and resolutions at the high-luminosity LHC – as described in the ATLAS Scoping Document [11] – have been used.

The three possible scenarios of detector upgrades presented in the ATLAS Scoping Document, defined by different costs and different performance and coverage, in term of pseudorapidity $\eta$ regions, are investigated by using the same event selection and classification. Results for the "reference" detector scenario, defined by an inner tracker coverage up to $|\eta| = 4.0$, and for a muon selection limited to $|\eta| < 2.5$, as in the ATLAS Run 1 and Run 2 analyses, are reported in the main text. The main results for the "low" and "middle"scenarios, characterised by reduced acceptance of the inner tracking system with respect to the reference scenario, and for the "reference" scenario with a muon selection extended to the range $|\eta| < 4.0$, are also summarised succintly in the main text (Sec. 6), while more details are provided in Appendix A.

## 2. MC Samples

The analysis is performed on generator-level samples of the main signal and background processes. The generated samples, including their theoretical cross sections for $\sqrt{s} = 14$ TeV, and their equivalent luminosities, used to normalise each sample to 3000 fb$^{-1}$, are summarised in Table 1.

Table 1: Signal and background processes and their production cross sections at the LHC for a centre-of-mass energy of $\sqrt{s} = 14$ TeV.

| **Process** | **Generator** | $\sigma_{\text{theory}}$ **[pb]** | **Order (in QCD) of the theory calculation** | $\int \mathcal{L}dt$ **[fb$^{-1}$] of generated events** |
|---|---|---|---|---|
| $H \to \mu^+\mu^-$, ggF | Powheg+Pythia | $11.9 \times 10^{-3}$ | N3LO | $4.20 \times 10^3$ |
| $H \to \mu^+\mu^-$, VBF | Powheg+Pythia | $0.93 \times 10^{-3}$ | approx NNLO | $5.36 \times 10^4$ |
| $Z(\to \mu^+\mu^-) + 0$ jets ($m_{\mu\mu} > 60$ GeV) | Alpgen+Herwig | 1516 | NNLO | 3.23 |
| $Z(\to \mu^+\mu^-) + 1$ jet ($m_{\mu\mu} > 60$ GeV) | Alpgen+Herwig | 389 | NNLO | 4.92 |
| $Z(\to \mu^+\mu^-) + 2$ jets ($m_{\mu\mu} > 60$ GeV) | Alpgen+Herwig | 140 | NNLO | 5.31 |
| $Z(\to \mu^+\mu^-) + 3$ jets ($m_{\mu\mu} > 60$ GeV) | Alpgen+Herwig | 47.4 | NNLO | 6.32 |
| $Z(\to \mu^+\mu^-) + 4$ jets ($m_{\mu\mu} > 60$ GeV) | Alpgen+Herwig | 15.1 | NNLO | 7.76 |
| $Z(\to \mu^+\mu^-) + 5$ jets ($m_{\mu\mu} > 60$ GeV) | Alpgen+Herwig | 5.8 | NNLO | 6.34 |
| $Z(\to \mu^+\mu^-) + 0$ jets ($m_{\mu\mu} > 100$ GeV) | Alpgen+Herwig | 57.4 | NNLO | 35 |
| $Z(\to \mu^+\mu^-) + 1$ jet ($m_{\mu\mu} > 100$ GeV) | Alpgen+Herwig | 17.7 | NNLO | 135 |
| $Z(\to \mu^+\mu^-) + 2$ jets ($m_{\mu\mu} > 100$ GeV) | Alpgen+Herwig | 6.70 | NNLO | 140 |
| $Z(\to \mu^+\mu^-) + 3$ jets ($m_{\mu\mu} > 100$ GeV) | Alpgen+Herwig | 2.38 | NNLO | 122 |
| $Z(\to \mu^+\mu^-) + 4$ jets ($m_{\mu\mu} > 100$ GeV) | Alpgen+Herwig | 0.796 | NNLO | 151 |
| $Z(\to \mu^+\mu^-) + 5$ jets ($m_{\mu\mu} > 100$ GeV) | Alpgen+Herwig | 0.314 | NNLO | 129 |
| $t\bar{t}$ | MC@NLO | 896 | NNLO+NNLL | 23.4 |
| $W^+W^- \to \ell_1^+\nu_1\ell_2^-\bar{\nu}_2$ | MC@NLO | 12.8 | NLO | 238 |

For the $H \to \mu^+\mu^-$ signal the Higgs boson mass is assumed to be $m_H = 125$ GeV. Only the dominant gluon fusion and VBF production modes are taken into account. The sum of their SM cross sections (taken from Ref. [6]), computed at N3LO (for ggF) or approximate NNLO (for VBF) in QCD with NLO electroweak corrections,

$$\sigma_{\text{ggF}} + \sigma_{\text{VBF}} = 58.95 \text{ pb}, \quad (1)$$

constitutes about 94% of the total SM Higgs boson production cross section (62.61 pb). The remaining 6% is due to $WH$ (2.4%), $ZH$ (1.6%), $ttH$ and $bbH$ (1% each), where the $WH$ and $ZH$ cross sections are computed at NNLO in QCD with NLO EW corrections included, while the $ttH$ and $bbH$ cross sections are computed at NLO in QCD. The theoretical values of the production cross sections, including their uncertainties, used for this study are listed in Table 2.

Both signal samples are generated with POWHEG [12, 13] interfaced to PYTHIA8 [14] for showering, hadronization and modeling of the underlying event. The Higgs boson is forced to decay to dimuons. The branching ratio predicted by the SM for $m_H = 125$ GeV is

$$BR_{\text{SM}}(H \to \mu^+\mu^-) = (2.18 \pm 0.05) \times 10^{-4} . \quad (2)$$

Table 2: Theoretical SM Higgs boson production cross sections at the LHC for a centre-of-mass energy of $\sqrt{s}$ = 14 TeV, for the dominant production modes.

| **Process** | $\sigma$ **[pb]** | **scale uncertainty [%]** | **PDF+$\alpha_s$ uncertainty [%]** |
|---|---|---|---|
| ggF | 54.7 | $^{+3.9}_{-3.9}$ | $^{+3.2}_{-3.2}$ |
| VBF | 4.28 | $^{+0.5}_{-0.3}$ | $^{+2.1}_{-2.1}$ |

Fifty thousand events are generated for both ggF and VBF production modes, corresponding to integrated luminosities of about 5 and 50 ab$^{-1}$, respectively.

The main backgrounds to the signal under study are due to processes that produce isolated muon pairs in the final state. They are, in decreasing order of relevance after the final selection:

- Drell-Yan events, $pp \to \gamma/Z^* X \to \mu^+\mu^- X$ (also denoted "$Z$+jets" events in the following);
- leptonic $t\bar{t}$ decays: $pp \to t\bar{t}X \to \mu^+\mu^- b\bar{b}\nu\bar{\nu}X$;
- leptonic $WW$ decays: $pp \to W^+W^-X \to \mu^+\mu^-\nu\bar{\nu}X$.

The production of $Z/\gamma^*$ in association with jets is generated with ALPGEN [15]. Up to five partons are produced in the hard scattering processes, with matrix elements implemented at LO. The $Z$ boson is forced to decay to dimuons. The production of $t\bar{t}$ and $WW$ pairs are generated with MC@NLO [16, 17]. Both ALPGEN and MC@NLO are interfaced to HERWIG [18] for parton showering, fragmentation into particles and to model the underlying event, using JIMMY [19] to generate multiple-parton interactions.

Two sets of $Z(\to \mu^+\mu^-)$+jet samples are produced. A first one, used to study events passing the final selection, which includes a dimuon invariant mass requirement $m_{\mu^+\mu^-} > 110$ GeV, is produced with a generator-level dimuon invariant mass, before final-state-radiation, greater than 100 GeV, and corresponds to an integrated luminosity of at least 35 fb$^{-1}$. A second set of $Z$+jet samples, used to study the background $m_{\mu^+\mu^-}$ distribution in a region around the $Z$ boson mass pole, is produced with a generator-level minimum dimuon invariant mass of 60 GeV and corresponds to an integrated luminosity of about 3 fb$^{-1}$. The samples are normalised by scaling the cross sections predicted by ALPGEN by an inclusive $k$-factor obtained as the ratio between the inclusive production cross section for $Z/\gamma^*(\to \mu^+\mu^-)$+jet events (2114 pb) from a calculation performed with MCFM [20] at NNLO in QCD for the study reported in Ref. [10], and the sum of the ALPGEN $Z(\to \mu^+\mu^-)$+n parton production cross sections (1822 pb) for the same $m_{\mu^+\mu^-}$>60 GeV requirement.

Nine $W^+W^-$ samples are produced correponding to the possible combinations of the different $W^\pm$ leptonic decays. The samples are normalised using an NLO calculation of the $W^+W^- \to \mu^+\nu_\mu\mu^-\bar{\nu}_\mu$ cross section (1.419 pb for each leptonic final state) performed with MCFM for the study reported in Ref. [10], and correspond to a luminosity of about 240 fb$^{-1}$.

Generated $t\bar{t}$ events are filtered by requiring that at least one top produces a $W$ boson that decays leptonically. The $t\bar{t}$ production cross section from a NNLL+approximate NNLO calculation is 896 pb (for $m_t$ = 173.3 GeV), with about 8.5% uncertainty. The generated number of $t\bar{t}$ events corresponds to an integrated luminosity of about 24 fb$^{-1}$.

4

# 3. Analysis strategy

The analysis proceeds through the following steps:

- an event selection aimed at keeping a high signal efficiency while suppressing as much as possible the backgrounds;
- an event classification to split the selected sample in subsets with different signal-to-background ratios ($S/B$) in order to improve the total sensitivity of the search;
- a maximum likelihood fit to the di-muon invariant mass distribution of the selected events, to estimate the signal yield and its uncertainty. The fit exploits the property that the invariant mass distribution of background events does not peak at the Higgs boson mass value, unlike the signal.

Systematic uncertainties are incorporated as nuisance parameters in the final fit, as described later. In the following, the three steps of the analysis are briefly reviewed.

## 3.1. Event selection

The following selection criteria are applied to the generated signal and background events after having applied the parametric efficiency and resolution corrections. The criteria are similar to those used in the Run 1 and Run 2 analyses, but the jet requirements are tighter to cope with the significantly larger number of pile-up jets.

- Events are required to have exactly one pair of reconstructed and identified opposite-sign muons with $p_T > 15$ GeV and $|\eta| < 2.5$. The muon reconstruction and identification efficiency is estimated to be 54% for $|\eta| < 0.1$ and 97% for $|\eta| > 0.1$. Muons overlapping with good jets within a cone of radius of $\Delta R = 0.4$ are ignored, where good jets are defined as all jets with $p_T > 30$ GeV passing the track-confirmation algorithm aimed at suppressing pile-up jets. This algorithm uses the tracking information by looking at the ratio $R_{p_T}$ of the scalar $p_T$ sum of the tracks that are associated with the jet and originate from the event primary vertex divided by the fully calibrated jet $p_T$. Small ratios typically arise from pile-up jets, while the ratio is larger for jets from the primary interaction. More details on this method, which in the following is referred to as track-confirmation (TC), and on its performance for different detector layouts are given in Ref. [11]. The track-confirmation requirements are applied to jets with $p_T < 100$ GeV and pseudorapidity $|\eta| < 3.8$.
- The leading-$p_T$ muon must have $p_T > 25$ GeV.
- The event should pass single or dimuon triggers, which have transverse momentum thresholds of 20 GeV and 11 GeV, respectively. The typical single muon trigger efficiency is around 96% for $|\eta| < 2.4$, with the exception of the region $|\eta| < 0.05$ where it drops to 77%, and is 0 for $|\eta| > 2.4$ due to the absence of fast muon chambers with triggering capability.
- For the extraction of the final results from a fit to the di-muon invariant mass, selected events are required to have $110 < m_{\mu\mu} < 160$ GeV.

The previous requirements define the nominal "signal region". A $Z$+jet "control region" is also defined by applying the same requirements with the exception of the di-muon invariant mass selection, which is changed to $87 < m_{\mu\mu} < 117$ GeV.

## 3.2. Event classification

Events are classified in seven orthogonal categories: one VBF-like category based on the presence of two forward jets with large rapidity gap and six categories based on the muon pseudorapidity and di-muon transverse momentum, $p_T^{\mu\mu}$. The Run 1 category definition is used, with the exception of the VBF-like category, whose definition has been reoptimised. Events considered as VBF candidates must still contain two high-$p_T$ jets, and the two leading-$p_T$ jets are required to be in opposite hemispheres ($\eta_{j1}\eta_{j2} < 0$), but the criteria applied on the minimum invariant mass and the minimum pseudorapidity separation between the leading-$p_T$ and subleading-$p_T$ jet have been tightened. Additional requirements are also applied to:

- the minimum value of the di-muon transverse momentum, since the muon pair from the Higgs boson in a VBF event is recoiling against the two quark-jets and is on average more boosted than muons from Higgs bosons produced through gluon fusion, and
- the maximum value of the quantity $\eta^* = |\eta_{\mu\mu} - \frac{\eta_{j1}+\eta_{j2}}{2}|$.

The optimisation is performed iteratively, by fixing the requirements on 3 of the four variables $m_{jj}$, $|\Delta\eta_{jj}|$, $p_T^{\mu\mu}$ and $\eta^*$ and finding the value of the cut on the fourth variable that maximises the VBF significance, computed as $N_{\text{VBF}}/\sqrt{N_{\text{VBF}} + N_{\text{ggF}} + B}$ in the VBF-like category. The requirement on the scanned variable is then fixed and the optimisation then proceeds to the next one. Only 2 iterations are performed, in the following order: $m_{jj} \to |\Delta\eta_{jj}| \to p_T^{\mu\mu} \to \eta^* \to |\Delta\eta_{jj}| \to m_{jj}$.

The final requirements for the VBF-like category are the following

- $m_{jj} > 650$ GeV ,
- $|\Delta\eta_{jj}| > 3.6$ ,
- $p_T^{\mu\mu} > 80$ GeV ,
- $\eta^* < 2.0$ .

The events not passing the VBF-like category are classified as either “central” or “non-central” whether both muons have $|\eta| < 1$ (and thus better $p_T$ resolution) or not. These two sub-samples are both split in three categories based on $p_T^{\mu\mu}$:

- low-$p_T^{\mu\mu}$: $p_T^{\mu\mu} < 15$ GeV,
- medium-$p_T^{\mu\mu}$: $15 < p_T^{\mu\mu} < 50$ GeV,
- high-$p_T^{\mu\mu}$: $p_T^{\mu\mu} > 50$ GeV,

to exploit the fact that signal events have on average larger di-muon transverse momentum than $Z/\gamma^*$ background events.

### 3.3. Signal extraction

The final results are obtained from a simultaneous fit to the invariant mass distribution of the seven event categories for events in the signal region. The signal yield $S$ and the background yield $B$ in each category, as well as the background shape parameters (except for the $Z$ lineshape ones), are floating in the fit.

Given the large number of events, a binned fit (with a sufficiently narrow binning compared to the excellent signal resolution) is performed, in order to reduce drastically the CPU time needed for the fit. The chosen binning is 0.1 GeV.

Two fits are performed:

1. an inclusive fit to all selected events, without categories, in which there is a single parameter of interest, the signal strength $\mu$
$$\mu = \frac{S}{S^{\mathrm{SM}}}, \tag{3}$$
defined as the ratio between the observed signal yield $S$ and its SM expectation, $S^{\mathrm{SM}}$;
2. a simultaneous fit to the selected events classified in the seven categories, using for each one its own signal and background model, with a single parameter of interest $\mu$.

Systematic uncertainties are incorporated in the likelihood function $L$ by introducing a nuisance parameter $\theta$ for each source of uncertainty, so that the signal and background expectations (yields or parameters of the model) become functions of $\theta$. A “penalty” or “constraint” term, which exploits the present estimate or guess of each systematic uncertainty, is then included as a multiplicative factor in the likelihood function. The nuisance parameters are then fitted (“profiled”) to the data, together with the parameter(s) of interest, when minimizing $-\log L$. The systematic uncertainties that are considered in this study are the theoretical uncertainties (currently around 5%) and the experimental luminosity uncertainty (expected to be around 3%). Other systematic uncertainties related to the background modelling and lepton identification efficiency are expected to be small and neglected in this note. Studies performed using 5000 fb$^{-1}$ of background MC samples produced at $\sqrt{s} = 13$ TeV for the analysis of the Run 2 data [9] show no evidence of a significant bias on the fitted signal yield induced by the choice of the background parametrization, and this effect is neglected. With such samples the statistical uncertainty on the estimated bias itself is around 30%. Much larger samples (around 50 ab$^{-1}$) are needed in order to reach a significantly smaller uncertainty on this estimated bias and verify that it is indeed negligible with respect to the other uncertainties considered in this document.

The likelihood function includes nuisance parameters for the theoretical uncertainties on the SM $H \to \mu^+\mu^-$ branching ratio and on the Higgs production cross sections, separately for gluon-fusion and VBF and separating effects due to the PDF uncertainties from those due to the choice of the renormalisation and factorisation scales in the fixed order calculations. The nominal uncertainties used in the fit are those given in Table 2 and in Eq. 2. A 3% uncertainty on the luminosity is also included. For all the uncertainties, Gaussian constraints are used, since the uncertainties are symmetric or only slightly skewed.

In order to estimate the median expected results and uncertainties, Asimov datasets are created for the signal and background, scaling the fitted signal and background models to the desired luminosity, and then combined.

### 3.3.1. Signal model

The signal di-muon invariant mass distribution is parametrised with the sum of a Crystal-Ball function and a Gaussian one. The peak position of the two functions are the same for all the seven categories. The parameters are extracted from a simultaneous unbinned maximum likelihood fit to the selected signal events, and then fixed in the final fit.

The signal model fits are illustrated in Fig. 1.

### 3.3.2. Background model

The background di-muon invariant mass distribution is parametrised with the sum of an exponential function and the convolution of a Breit-Wigner function (whose mass and width parameters are fixed to the nominal $Z$ pole ones) with a Gaussian resolution function.

The Gaussian resolution in each category is extracted through a simultaneous fit to the di-muon invariant mass distribution of simulated $Z$+jet events in the $Z$+jet control region selected in each category, with the model just described after removing the exponential component.

The nominal values of the parameters of the full background model (slope and relative normalisation of the exponential function) are extracted from unbinned maximum likelihood fits to the invariant mass distribution of the selected background events in the signal region, summing together the contribution from Drell-Yan, $t\bar{t}$ and $WW$, normalised to their respective SM cross sections. The parameters of the nominal background model will be used in the following to generate background distribution corresponding to high integrated luminosities, in order to overcome the limitations from the small (35–150 fb$^{-1}$) equivalent luminosity of the generated Drell-Yan MC samples.

In the final $S + B$ fit the slope and relative normalisation of the exponential function, as well as the background yields in each category, will be floating, in order to mimic the procedure adopted in data to extract the background parameters *in situ*, essentially exploiting the large background in the sidebands of the di-muon invariant mass distribution.

The $Z$ lineshape fits in the range $87 < m_{\mu^+\mu^-} < 117$ GeV, for an integrated luminosity of 3 fb$^{-1}$ (the equivalent luminosity of the Drell-Yan MC samples with a generator-level dimuon mass greater than 60 GeV) are illustrated in Fig. 2.

The background fits used to extract the nominal parameters are illustrated in Fig. 3.

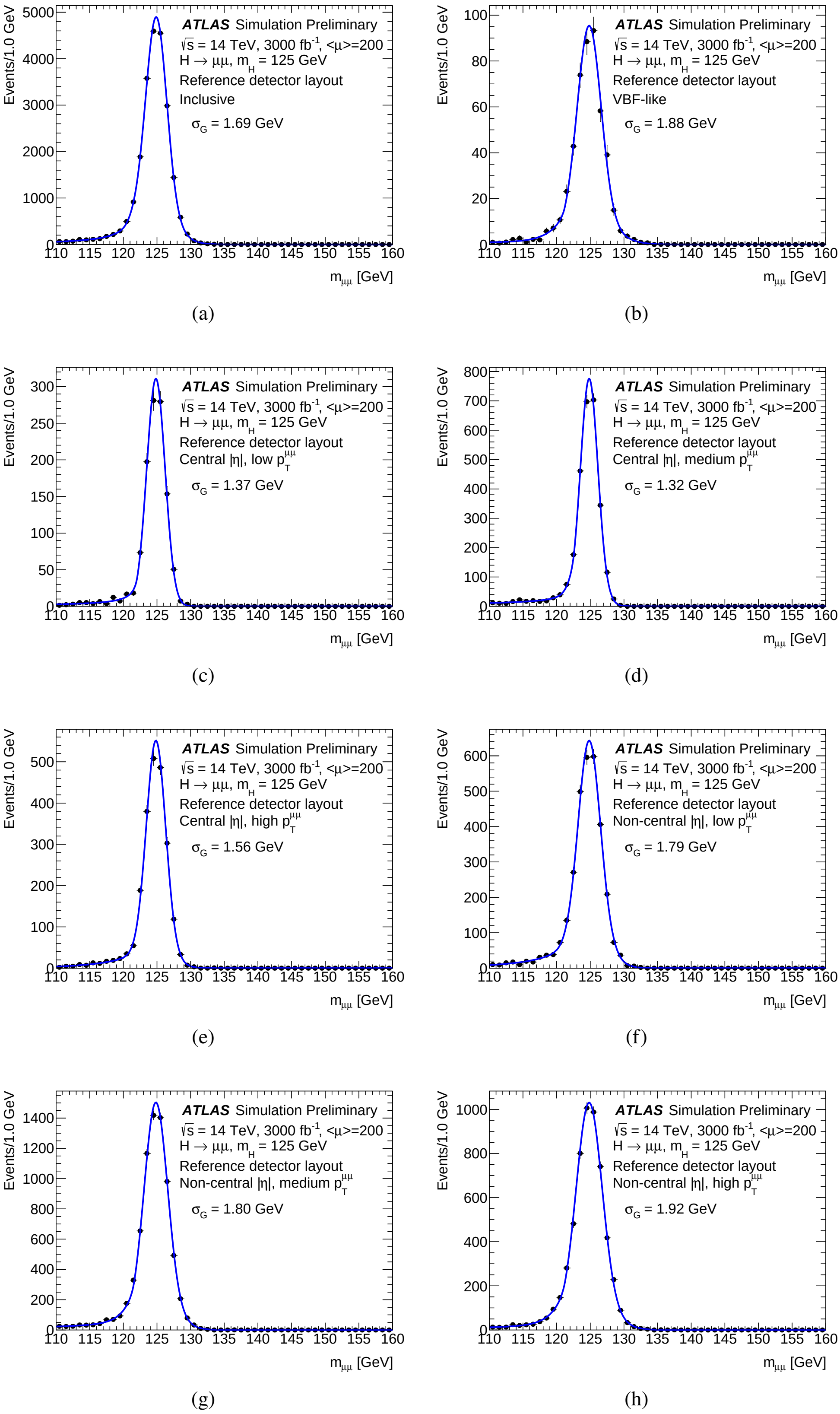

Figure 1: Invariant mass distribution of selected signal candidates, scaled to 3000 fb$^{-1}$, and result of the Crystal-Ball+Gaussian fit, for the reference detector scenario, assuming $\langle\mu\rangle = 200$. The resolution $\sigma_G$ of the Gaussian core of the distribution is overlaid. (a): inclusive sample. (b)-(h): categories.

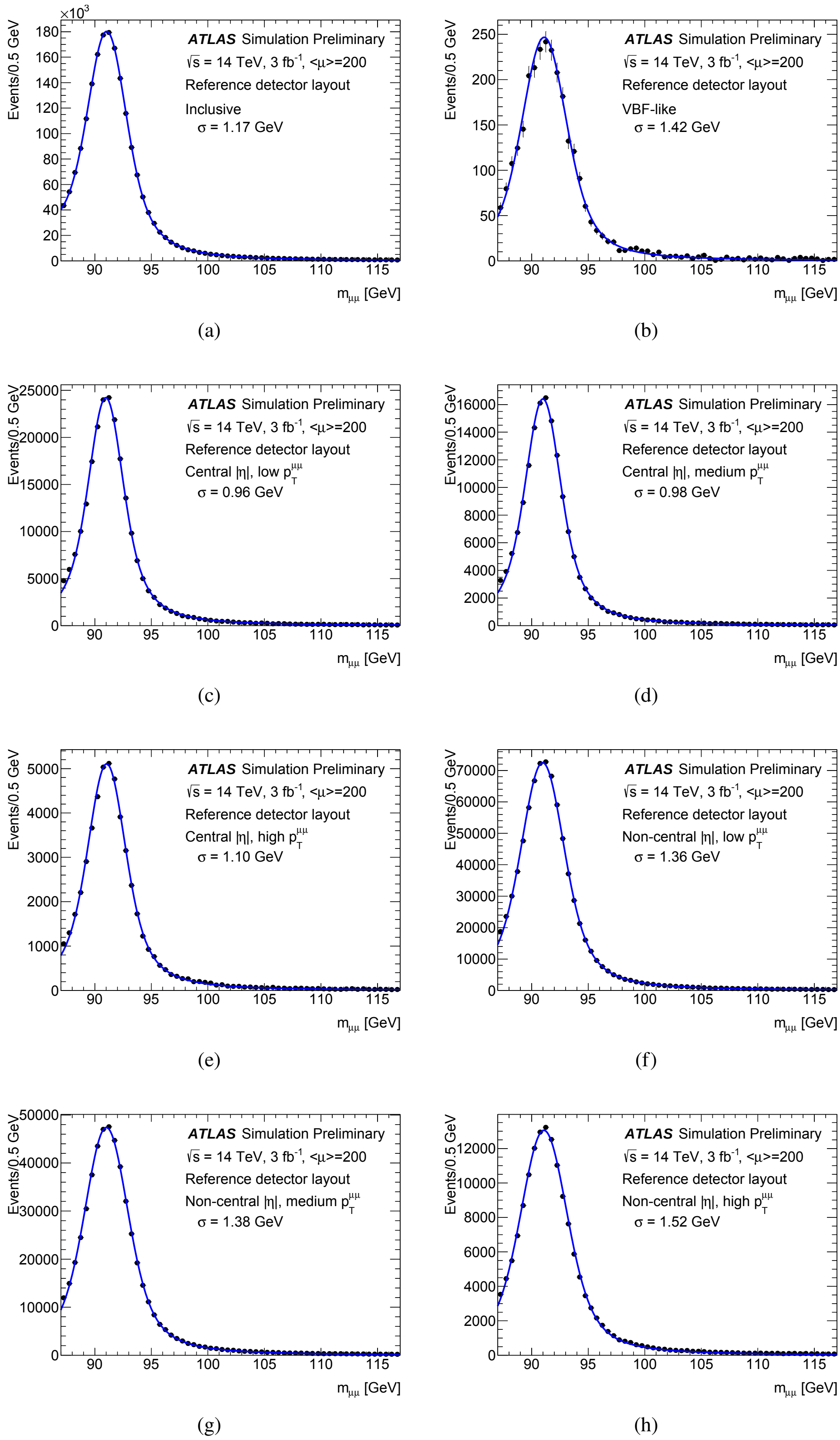

Figure 2: Invariant mass distribution of $Z \to \mu\mu$ background candidates, scaled to 3 fb$^{-1}$, and result of the fit with a Breit-Wigner convoluted with a Gaussian response function, for the reference detector scenario, assuming $\langle\mu\rangle = 200$. (a): inclusive sample. (b)-(h): categories.

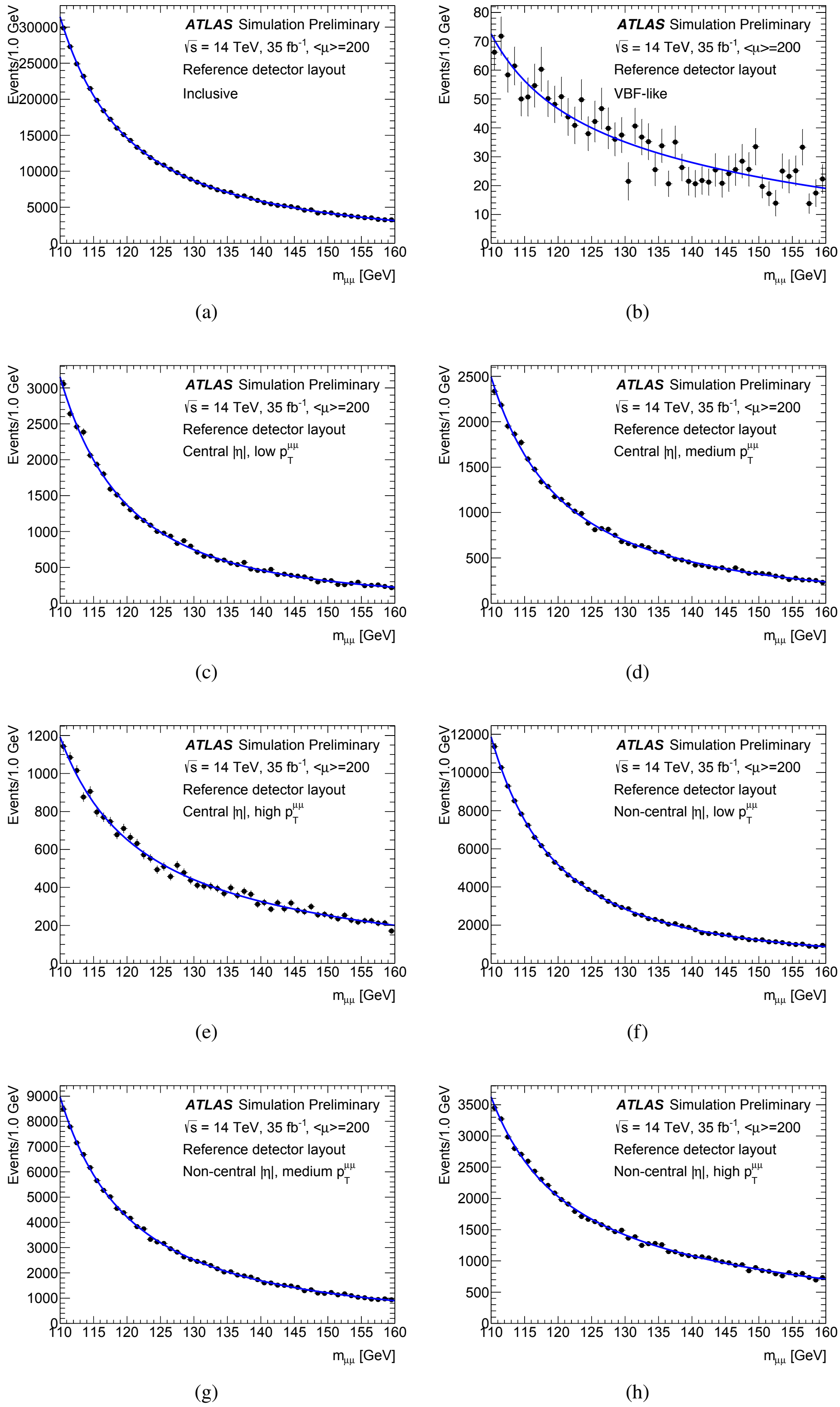

Figure 3: Invariant mass distribution of selected background candidates, scaled to 35 fb$^{-1}$, and result of the fit with the full background model, for the reference detector scenario, assuming $\langle \mu \rangle = 200$. (a): inclusive sample. (b)-(h): categories.

# 4. Results for the reference detector scenario

The expected signal and background yields after the various selection requirements are given in Table 3. The overall signal efficiency is 59% (59% for ggF events and 62% for VBF ones). The expected background overwhelms the signal: the $B/S$ ratio in the signal region is $\approx$ 1800, and it is $\approx$ 270 in a $\pm 1.5\sigma_G$ invariant-mass window around $m_{\mu\mu}$ = 125 GeV, where $\sigma_G$ is the resolution of the core of the invariant mass distribution of signal events.

Table 3: Expected signal and background yields in the signal region for 3000 fb$^{-1}$ in the reference detector scenario, for $\langle\mu\rangle = 200$.

| **Process** | **Expected yield** |
|---|---|
| ggF | 2.10e+04 |
| VBF | 1.74e+03 |
| ggF+VBF | 2.27e+04 |
| Z+jets | 3.67e+07 |
| top | 4.23e+06 |
| WW | 4.25e+05 |
| Total bkg | 4.14e+07 |

The expected transverse momentum and pseudorapidy distributions of the selected muons are shown in Fig. 4. The expected transverse momentum and pseudorapidy distributions of the leading and subleading jets in events with at least two good jets are shown in Fig. 5. The large increase of selected jets for $|\eta| > 3.8$ is due to pile-up jets not being vetoed by the track confirmation algorithm.

The expected signal and background dimuon invariant mass and transverse momentum distributions after the selection are illustrated in Figs. 6 (in the invariant mass range 70–270 GeV) and 7 (in the invariant mass range 110–160 GeV, used for the final fit).

The VBF-like category has an efficiency of about 14% for VBF signal events passing the selection. The VBF fraction of signal in the VBF-like category is around 51%; it is 1% in the low-$p_T^{\mu\mu}$ categories, 4% in the central-$p_T^{\mu\mu}$ and 13% in the high-$p_T^{\mu\mu}$ categories.

The expected yields in each category for $110 < m_{\mu\mu} < 160$ GeV are given in Table 4.

Table 4: Expected signal and background yields in each category for 3000 fb$^{-1}$ in the reference detector scenario, for $\langle\mu\rangle = 200$.

| **Process** | **VBF-like** | **low $p_T^{\mu\mu}$, central** | **med $p_T^{\mu\mu}$, central** | **hi $p_T^{\mu\mu}$, central** | **low $p_T^{\mu\mu}$, non central** | **med $p_T^{\mu\mu}$, non central** | **hi $p_T^{\mu\mu}$, non central** |
|---|---|---|---|---|---|---|---|
| ggF | 2.37e+02 | 1.12e+03 | 2.70e+03 | 1.93e+03 | 3.08e+03 | 7.11e+03 | 4.81e+03 |
| VBF | 2.47e+02 | 1.28e+01 | 1.07e+02 | 2.98e+02 | 3.55e+01 | 2.90e+02 | 7.46e+02 |
| ggF+VBF | 4.84e+02 | 1.13e+03 | 2.81e+03 | 2.23e+03 | 3.11e+03 | 7.40e+03 | 5.56e+03 |
| $Z$+jets | 8.45e+04 | 3.67e+06 | 2.95e+06 | 1.22e+06 | 1.39e+07 | 1.08e+07 | 4.10e+06 |
| top | 6.28e+04 | 3.37e+04 | 2.86e+05 | 7.13e+05 | 1.19e+05 | 9.62e+05 | 2.05e+06 |
| WW | 2.84e+03 | 7.37e+03 | 3.44e+04 | 3.09e+04 | 3.42e+04 | 1.71e+05 | 1.44e+05 |
| Total bkg | 1.50e+05 | 3.71e+06 | 3.27e+06 | 1.97e+06 | 1.41e+07 | 1.19e+07 | 6.30e+06 |

The expected signal and background yields and signal significance for each event category, in a $m_{\mu\mu}$ window of $\pm 1.5\sigma$ around 125 GeV, are given in Table 5. The total signal significance (from the sum in quadrature of $S/\sqrt{S+B}$ in a window of $\pm 1.5\sigma$ around 125 GeV) is $9.5\sigma$.

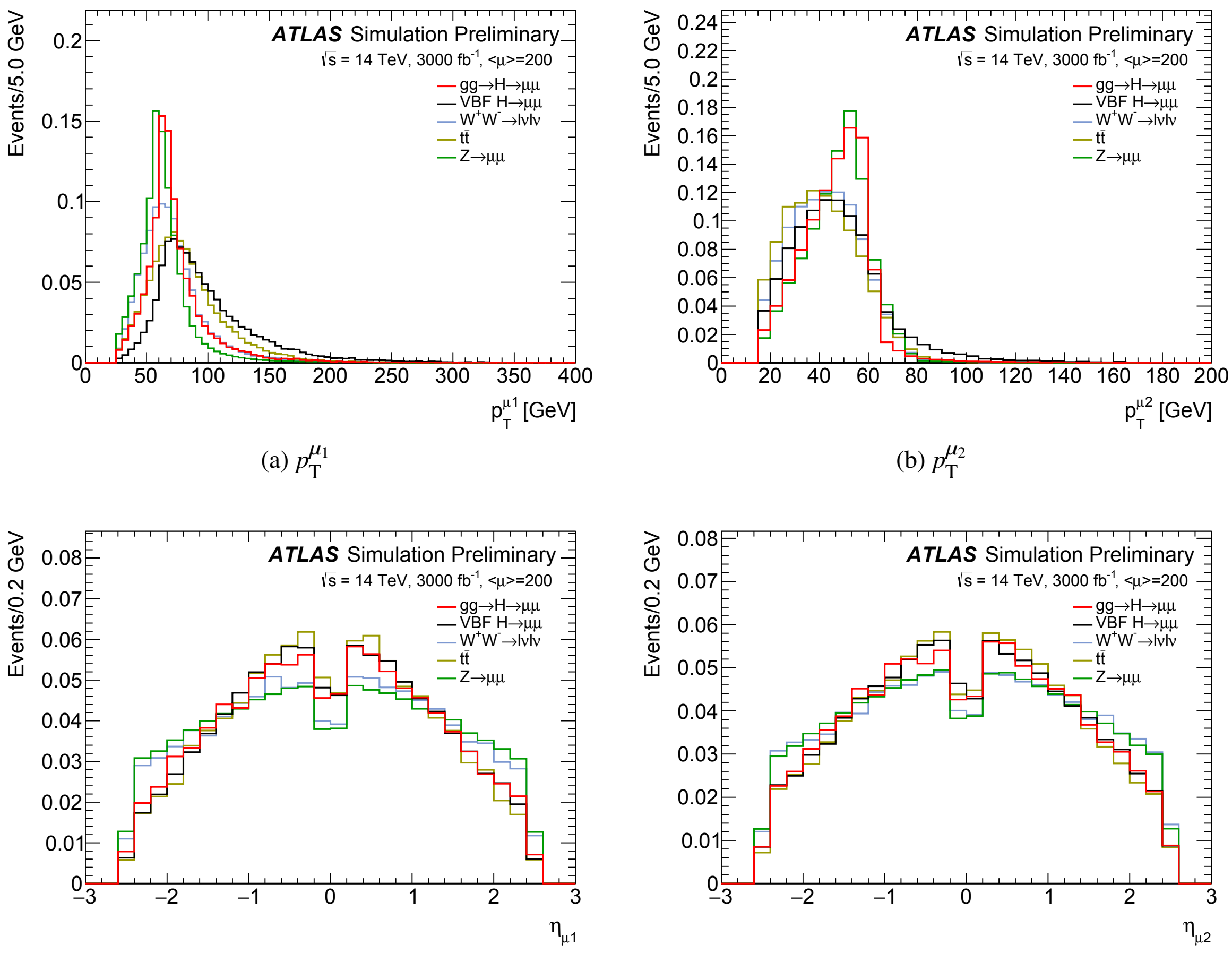

(a) $p_{\mathrm{T}}^{\mu_1}$ (b) $p_{\mathrm{T}}^{\mu_2}$ (c) $\eta_{\mu_1}$ (d) $\eta_{\mu_2}$

Figure 4: (a,b) Transverse momentum and (c,d) pseudorapidity distributions of selected (a,c) leading and (b,d) subleading muon candidates, normalised to unity, for the reference detector scenario, assuming $\langle\mu\rangle = 200$.

The result of the simultaneous fit to the seven event categories with one parameter of interest is illustrated in Fig. 8. Due to the huge background, the signal is only visible when subtracting the fitted background from the data, as illustrated in the bottom plot of the same figure.

The fitted signal strength in the inclusive fit is equal to $\mu = 1.00 \pm 0.15$.

The fitted signal strength in the simultaneous fit to the event categories is equal to $\mu = 1.00 \pm 0.13$. Compared to the inclusive fit the uncertainties are reduced by $\approx 12\%$.

Redoing the fits without sytematic uncertainties (*i.e.* without theory uncertainties and the luminosity uncertainty) yields:

- fitted signal strength in the fit to the inclusive sample: $\mu = 1.00 \pm 0.14$.
- fitted signal strength in the fit to the event categories: $\mu = 1.00 \pm 0.12$.

In the Appendix A the impact of extending the $|\eta|$ range for the muon selection from 2.5 to 4.0, profiting from the extended tracker coverage, is also evaluated.

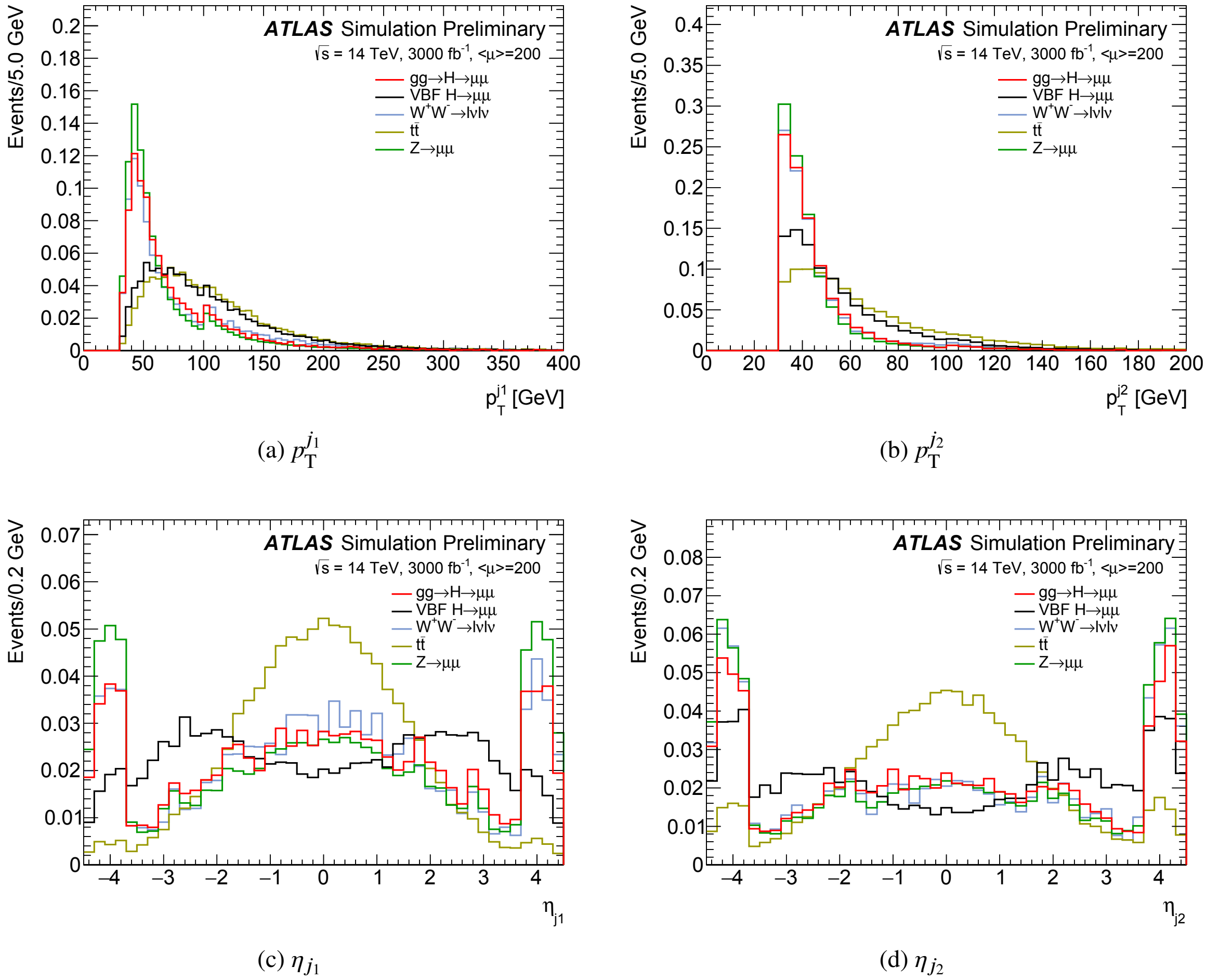

(a) $p_{\mathrm{T}}^{j_1}$ (b) $p_{\mathrm{T}}^{j_2}$

(c) $\eta_{j_1}$ (d) $\eta_{j_2}$

Figure 5: (a,b) transverse momentum and (c,d) pseudorapidity distributions of selected (a,c) leading and (b,d) subleading jet candidates, normalised to unity, for the reference detector scenario, assuming $\langle\mu\rangle = 200$. Only events with at least two good jets are considered.

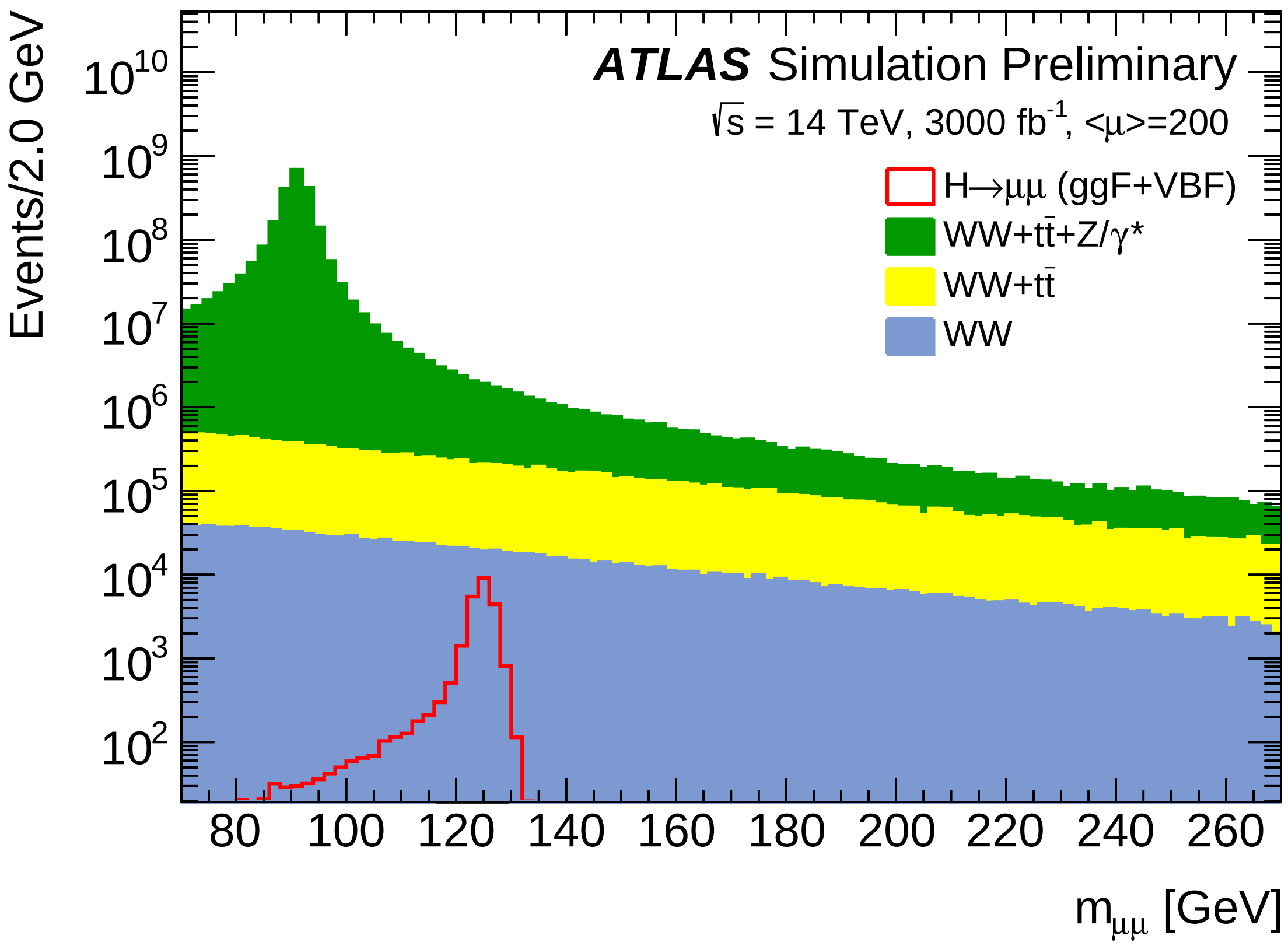

Figure 6: Invariant mass distribution of selected signal and background candidates, scaled to 3000 fb$^{-1}$, for the reference detector scenario, assuming $\langle\mu\rangle = 200$.

Table 5: Expected signal and background yields and signal significance in a $\pm 1.5\sigma_G$ invariant-mass window around $m_{\mu\mu}$ = 125 GeV for each category, where $\sigma_G$ is the resolution of the core of the invariant mass distribution of signal events. The last rows shows the total signal and background yields, the average invariant mass resolution, and the sum in quadrature of the significance of each category. The projections correspond to an integrated luminosity $\int \mathcal{L}dt$ = 3000 fb$^{-1}$ for a center-of-mass energy $\sqrt{s}$=14 TeV for the reference detector scenario.

| **Category** | $S$ | **VBF** | $B$ | **FWHM [GeV]** | $\sigma_G$ **[GeV]** | $S/\sqrt{S+B}$ |
|---|---|---|---|---|---|---|
| VBF-like | 386 | 197 | 19430 | 4.37 | 1.88 | 2.75 |
| low $p_T$, central | 921 | 11 | 350500 | 3.21 | 1.37 | 1.55 |
| med $p_T$, central | 2210 | 84 | 300500 | 3.08 | 1.32 | 4.01 |
| hi $p_T$, central | 1810 | 242 | 211800 | 3.50 | 1.56 | 3.91 |
| low $p_T$, non central | 2460 | 28 | 1740500 | 4.11 | 1.79 | 1.86 |
| med $p_T$, non central | 5860 | 230 | 1483600 | 4.24 | 1.80 | 4.80 |
| hi $p_T$, non central | 4380 | 588 | 829000 | 4.70 | 1.92 | 4.80 |
| Total | 18020 | 1380 | 4935500 | 3.93 | 1.69 | 9.53 |

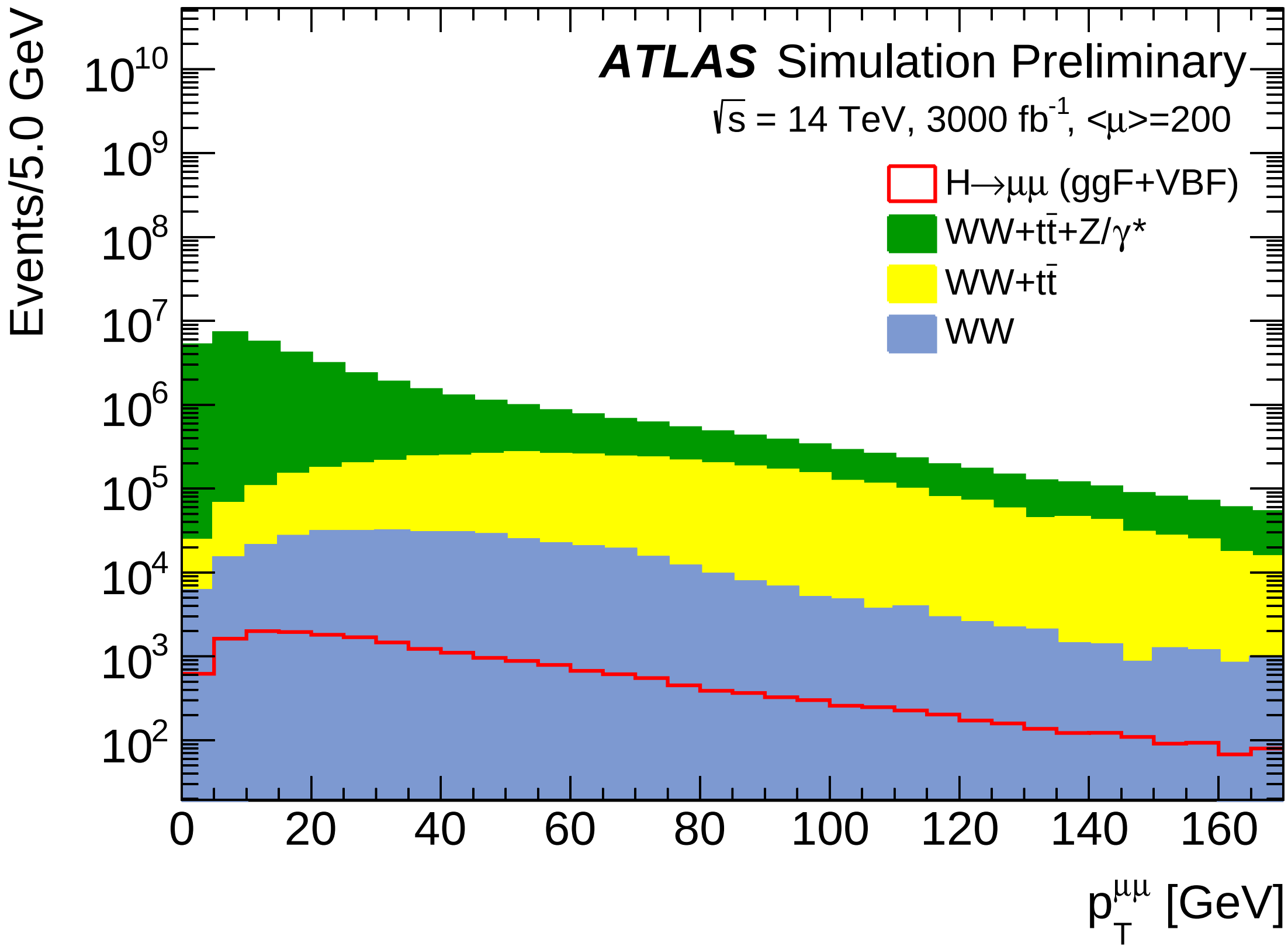

Figure 7: Dimuon transverse momentum distribution of selected signal and background candidates, scaled to 3000 fb$^{-1}$, for the reference detector scenario, assuming $\langle\mu\rangle = 200$.

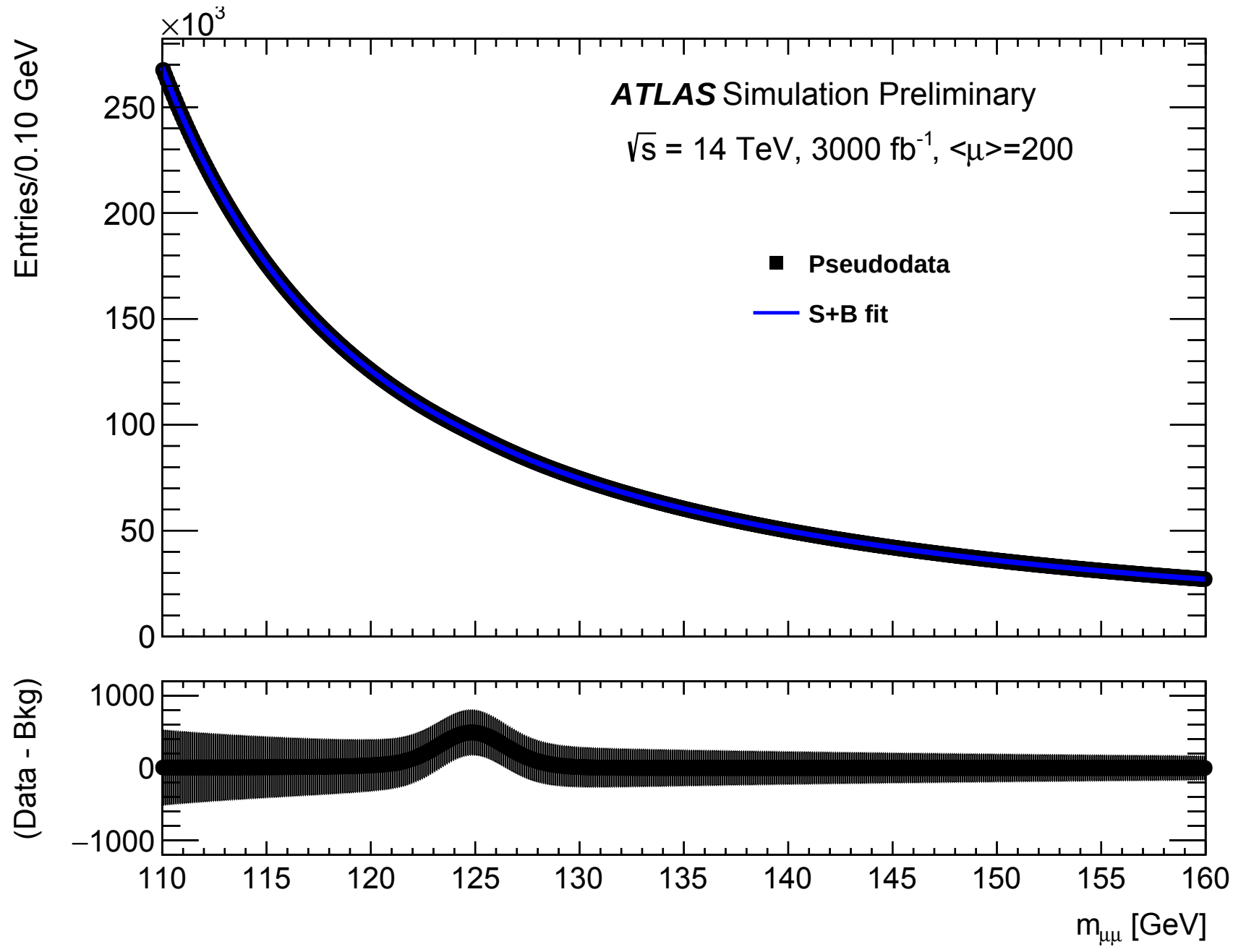

Figure 8: Top: invariant mass distribution of the signal+background Asimov dataset corresponding to 3000 fb$^{-1}$ of data (black dots with error bars) for the reference detector scenario, assuming $\langle\mu\rangle = 200$, and result of the simultaneous maximum likelihood fit, summed over the seven categories (blue line). Bottom: data–background-only fit residuals (black dots with error bars).

# 5. Validation with fully-simulated MC Samples

In this section the kinematic distributions of the selected muons and jets obtained by applying parametrised efficiency and resolution functions to generator-level events are compared to the same distributions obtained by applying the ATLAS reconstruction algorithms to a detailed GEANT4 simulation of the detector response. The comparison is limited to the signal samples, and provides a cross check of the validity of the simplified approach based on the parametrised efficiency and resolution functions. Only the shapes of the distributions are compared: in all the figures, the distributions are normalised to the same area.

The kinematic distributions ($p_{\mathrm{T}}$, $\eta$, $\phi$) of the selected muon candidates, as well as their ratio, are shown in Figure 9.

Figure 10 shows the invariant mass and transverse momentum distributions of the selected muon pairs.

Kinematic distributions of the selected jet pairs belonging to the VBF-dedicated category are shown in Figure 11. Jets are reconstructed with the anti-$k_t$ algorithm with a distance parameter of 0.4 [21] and fully calibrated. In the study based on full simulation samples, two alternative methods are adopted to suppress pile-up jets. The first one uses the same parametrisation of the efficiency of the track confirmation algorithm (described in the Section 3) applied in the analysis presented in the previous and following sections. The corresponding distributions in the figures are labelled as “Full Sim, with TC”. The second one uses an an explicit evaluation of the $R_{p_{\mathrm{T}}}$ variable, on which the track confirmation algorithm is based, using tracks and jets from the full simulation. The corresponding distributions in the figures are labelled as “Full Sim, with Rpt”.

Figure 12 shows the same distributions of Figure 11 but only for jets produced by the hard scattering (HS) process. A reconstructed jet is considered to be produced by the HS if a true HS jet is found inside a cone of radius 0.3 in $\eta - \phi$ centered on the axis of the reconstructed jet.

The shapes of the muon and jet kinematic distributions are in good agreement between the full simulation and the generator-level events corrected with parametrised efficiency and resolution functions.

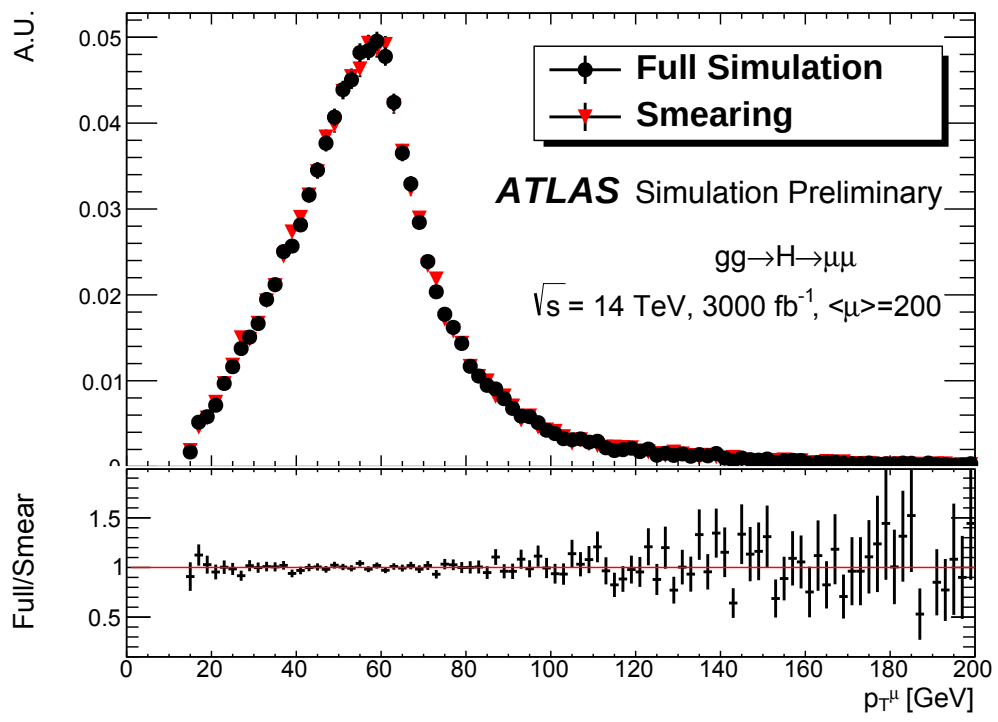

(a) ggF

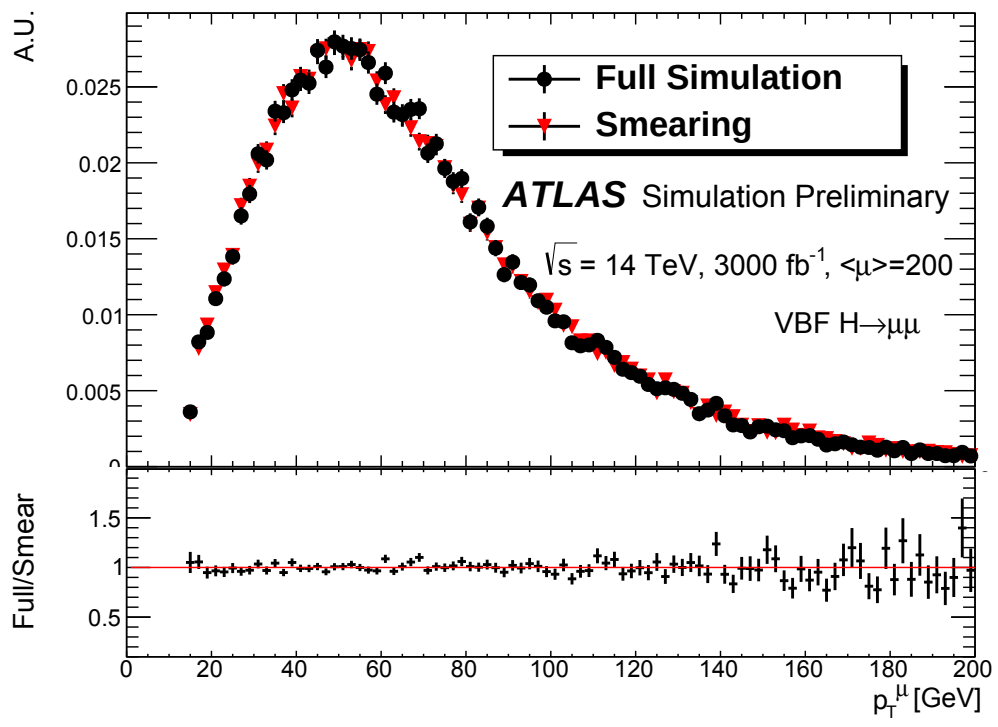

(b) VBF

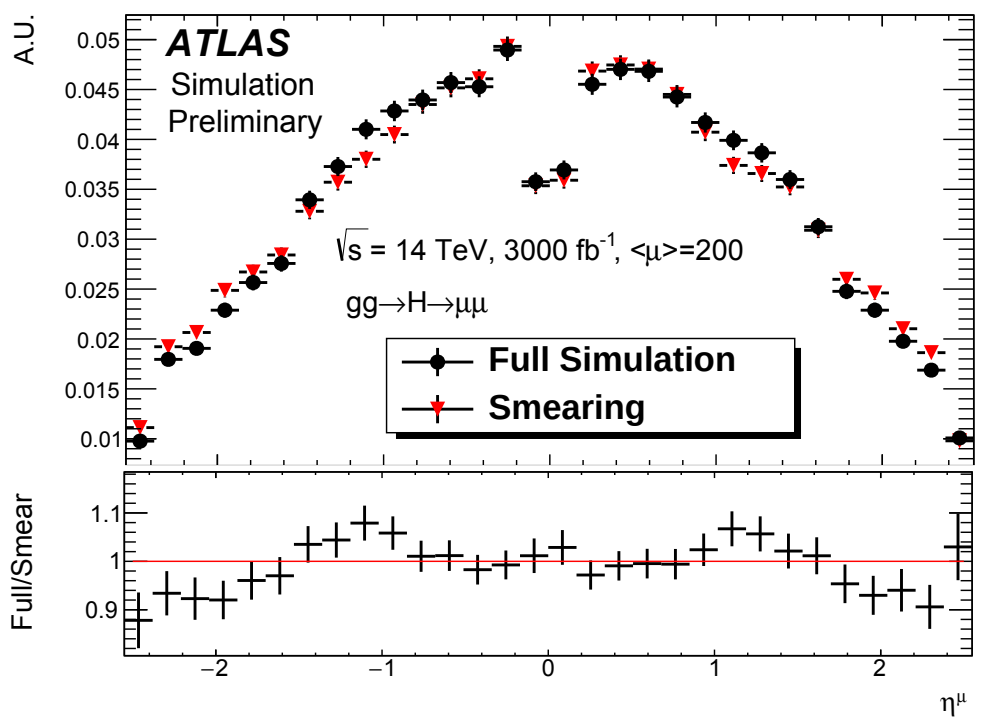

(c) ggF

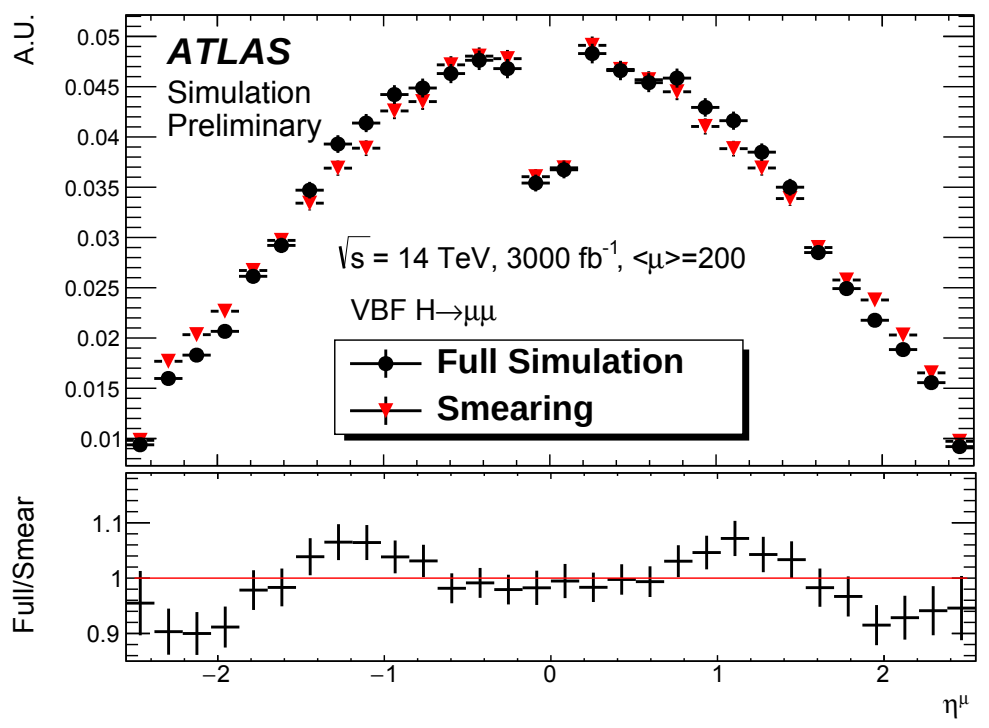

(d) VBF

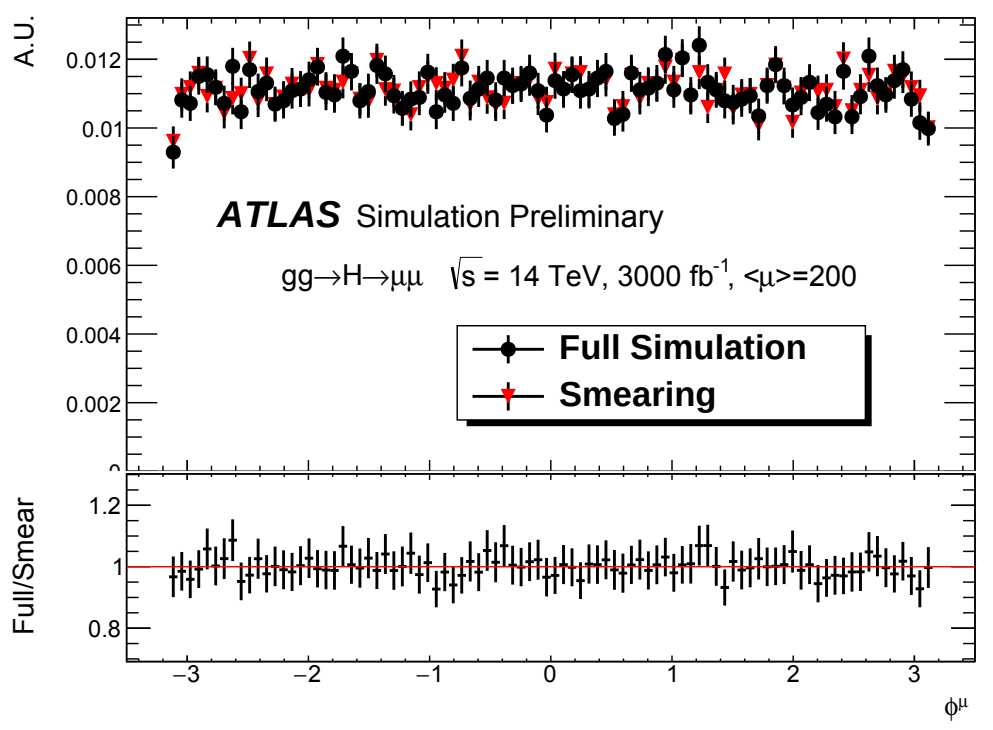

(e) ggF

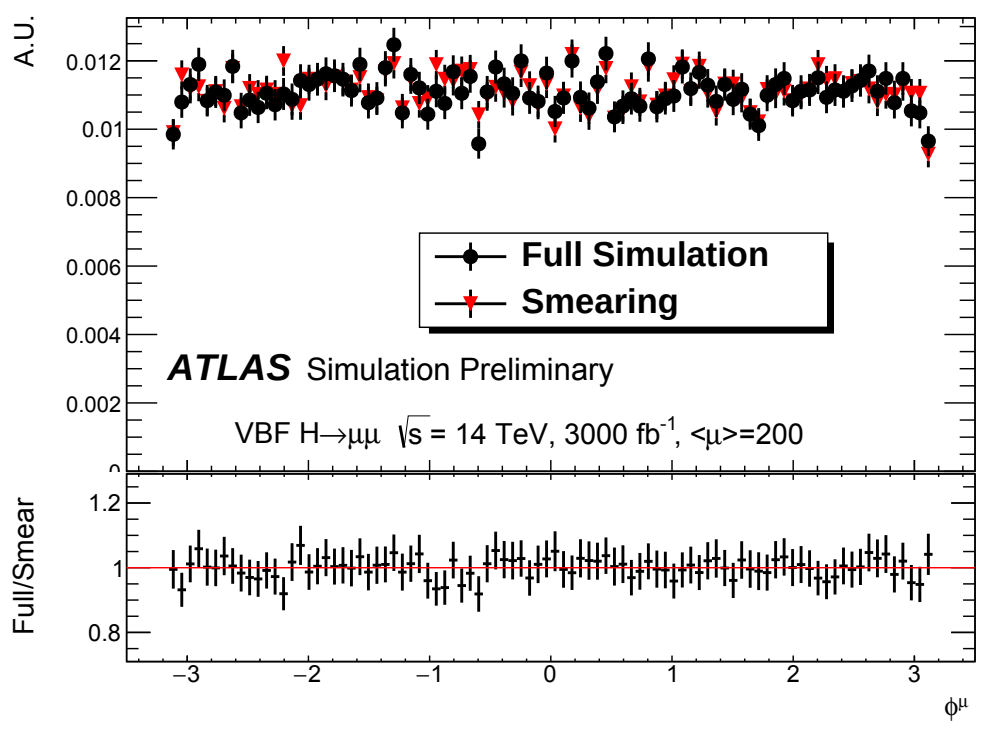

(f) VBF

Figure 9: Kinematic distributions of the selected muons, using either parametrised efficiency and resolution functions or a full detector simulation. The bottom panels show the ratio between the two distributions shown in the top panels. The distributions are normalised to the same area. (a) shows the muon $p_{\mathrm{T}}$ distribution for the ggF sample and (b) shows the $p_{\mathrm{T}}$ distribution for the VBF sample. (c) shows the $\eta$ distribution for the ggF sample and (d) shows the $\eta$ distribution for the VBF sample. (e) shows the $\phi$ distribution for the ggF sample and (f) shows the $\phi$ distribution for the VBF sample.

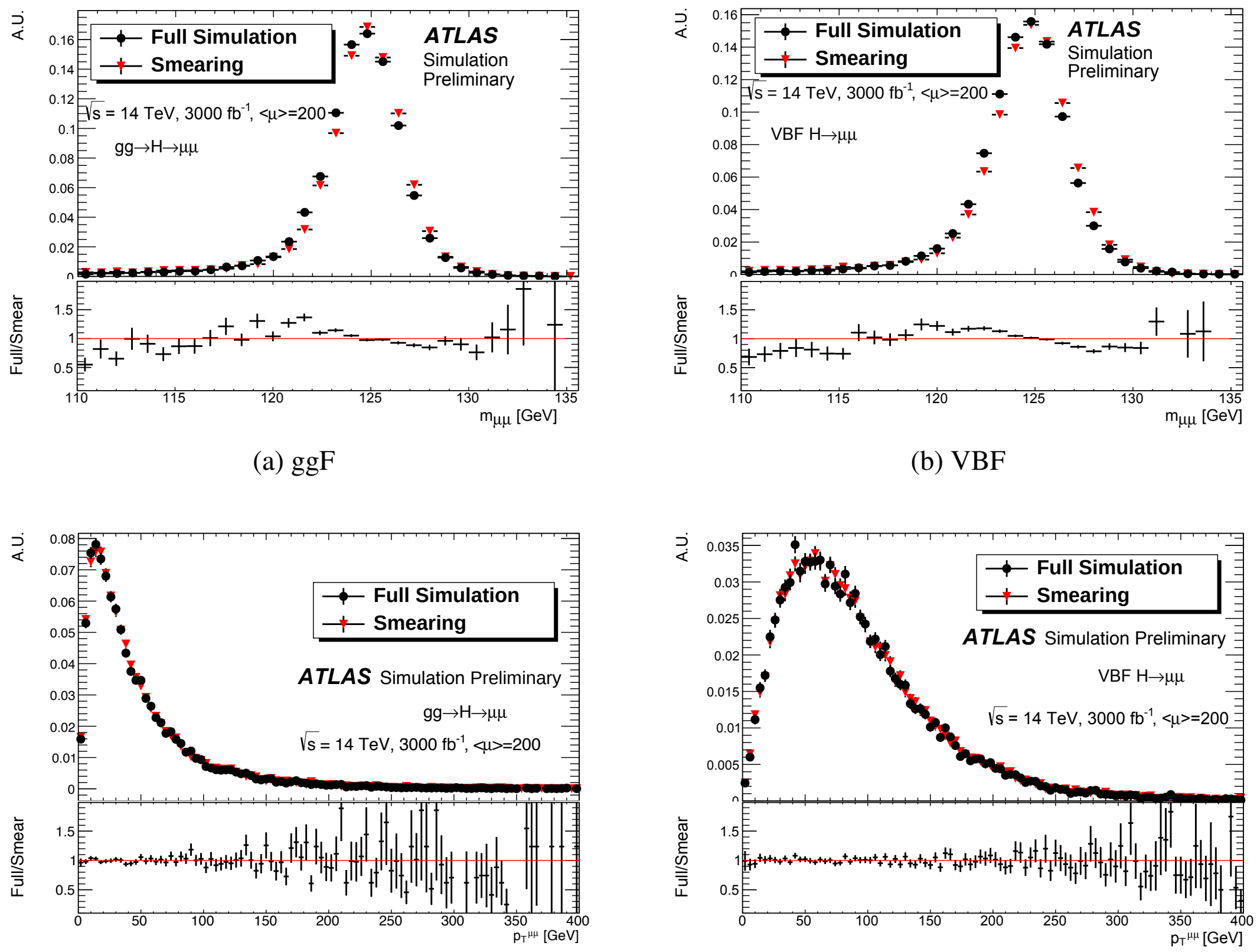

(a) ggF

(b) VBF

(c) ggF

(d) VBF

Figure 10: Kinematic distributions of the selected muon pairs, using either parametrised efficiency and resolution functions or a full detector simulation. The bottom panels show the ratio between the two distributions shown in the top panels. The distributions are normalised to the same area. (a) shows the di-muon invariant mass distribution for the ggF sample and (b) shows the di-muon invariant mass distribution for the VBF sample (c) shows the di-muon $p_T$ distribution for the ggF sample and (d) shows the di-muon $p_T$ distribution for the VBF sample.

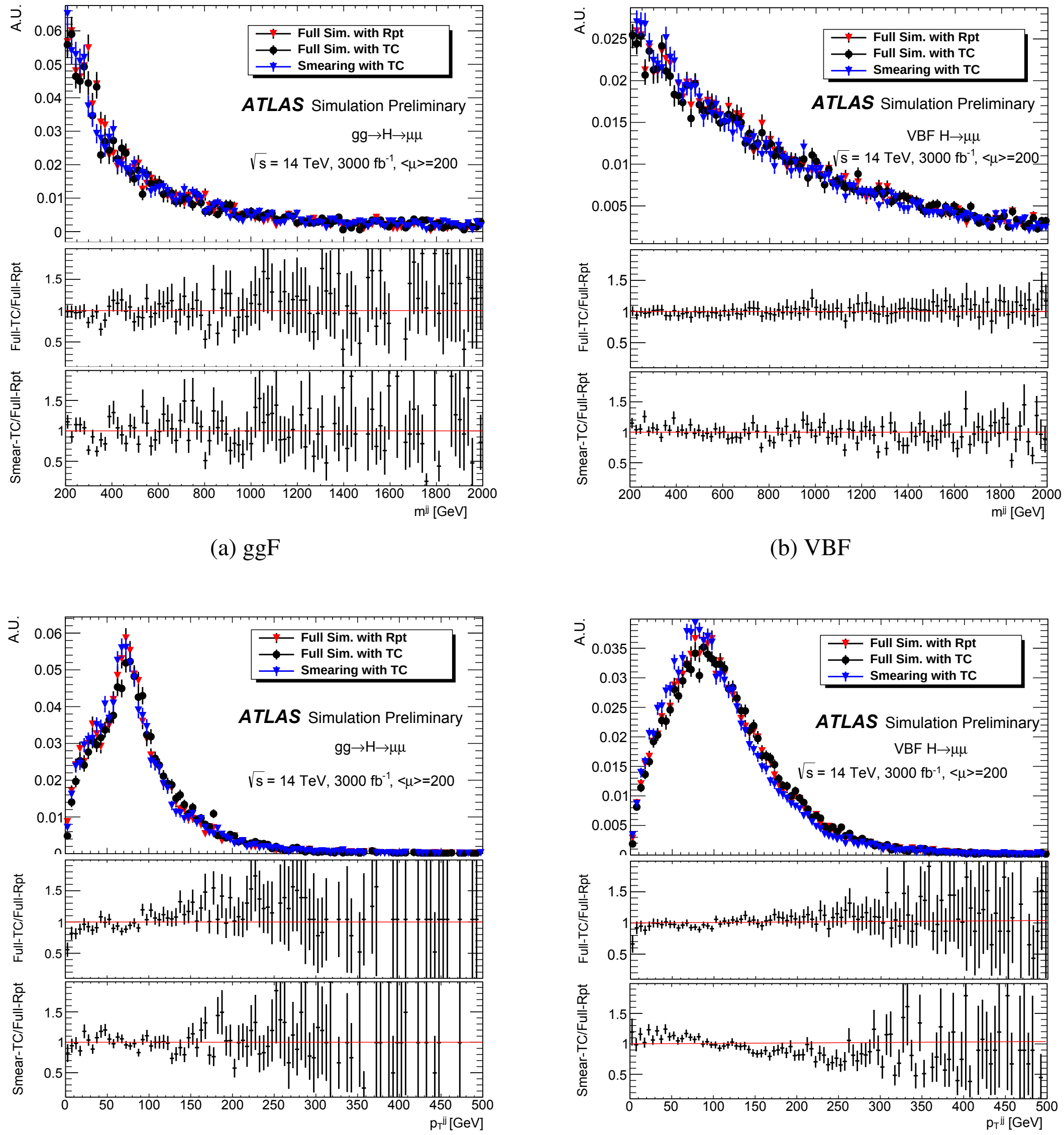

(a) ggF (b) VBF

(c) ggF (d) VBF

Figure 11: Kinematic distributions of the selected jet pairs, in events belonging to the VBF-dedicated category, using either parametrised efficiency and resolution functions or a full detector simulation. In the latter case two alternative methods to suppress pile-up jets are used (see text for more details). The two bottom panels show ratios between two of the three distributions shown in the top panels, either the two from the full simulation but with different pile-up suppression techniques (middle panel) or the distribution based on the parametrised efficiency and resolution functions and that from the full simulation using the $R_{p_T}$ computed from reconstructed tracks and jets explicitely. The distributions are normalised to the same area. (a) shows the di-jet invariant mass distribution for the ggF sample and (b) shows the di-jet invariant mass distribution for the VBF sample. (c) shows the di-jet $p_T$ distribution for the ggF sample and (d) shows the di-jet $p_T$ distribution for the VBF sample.

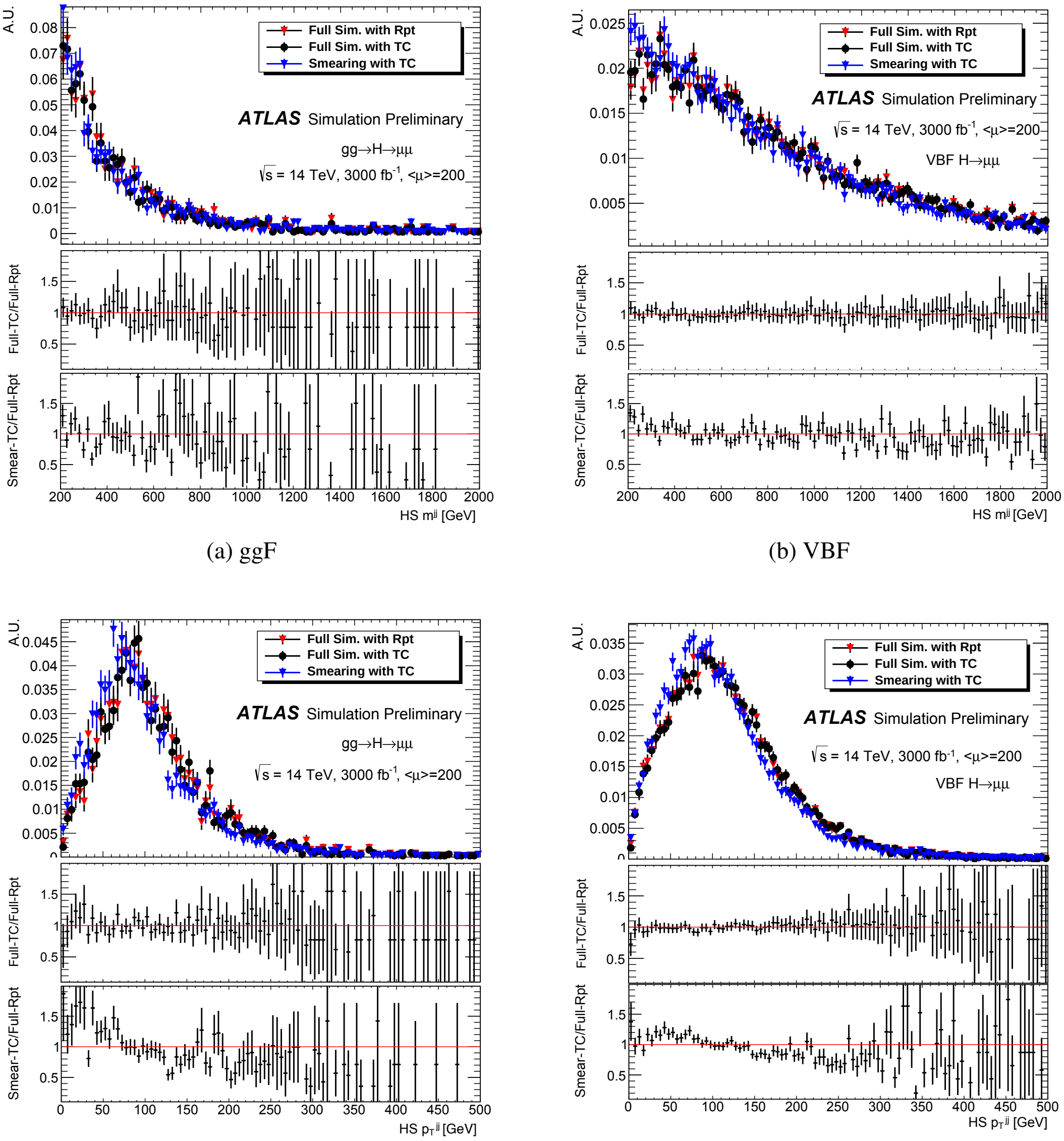

(a) ggF

(b) VBF

(c) ggF

(d) VBF

Figure 12: Kinematic distributions of the selected jet pairs, in events belonging to the VBF-dedicated category, using either parametrised efficiency and resolution functions or a full detector simulation. In the latter case two alternative methods to suppress pile-up jets are used (see text for more details). Only jets from the hard-scattering process are considered. The two bottom panels show ratios between two of the three distributions shown in the top panels, either the two from the full simulation but with different pile-up suppression techniques (middle panel) or the distribution based on the parametrised efficiency and resolution functions and that from the full simulation using the $R_{p_{\mathrm{T}}}$ computed from reconstructed tracks and jets explicitely. The distributions are normalised to the same area. (a) shows the di-jet invariant mass distribution for the ggF sample and (b) shows the di-jet invariant mass distribution for the VBF sample. (c) shows the di-jet $p_{\mathrm{T}}$ distribution for the ggF sample and (d) shows the di-jet $p_{\mathrm{T}}$ distribution for the VBF sample.

# 6. Summary

An updated study has been presented of the prospects for the measurement of the rare Higgs boson decay $H \to \mu\mu$ using 3000 fb$^{-1}$ of proton-proton collisions at $\sqrt{s}$ = 14 TeV recorded with the ATLAS detector at the high-luminosity LHC. The studies assume an average number of interactions per bunch crossing $\langle\mu\rangle$ = 200 and the latest performance assumptions for the various subdetectors, in three different upgrade scenarios.

The muonic Higgs boson decay has not been observed yet and constitutes the best way at the LHC to access the couplings of the Higgs boson to fermions of the second generation of matter particles.

The results for the estimated signal significance and the uncertainty on the signal strength are summarised in Table 6.

Table 6: The table compares the overall significance and signal strength uncertainty achievable with 3000 fb$^{-1}$ in the three different detector scenarios defined in the ATLAS Scoping Document, based on the event categories defined in the text.

| **Scoping Scenario** | $\langle\mu\rangle$ | **Overall significance** | $\Delta\mu$ **w/ syst. errors** | $\Delta\mu$ **w/o syst. errors** |
|---|---|---|---|---|
| reference | 200 | 9.5 | ±0.13 | ±0.12 |
| middle | 200 | 9.4 | ±0.14 | ±0.12 |
| low | 200 | 9.2 | ±0.14 | ±0.13 |

The muon efficiencies and resolutions between the three detector upgrade scenarios in the selected phase space are quite similar. This translates in significances that improve only slightly between the different scenarios. The overall signal strength uncertainties are very similar in all three scenarios.

No significant gain is found (without further re-optimisation) for the reference detector scenario when selecting muons with rapidities up to 4.0 instead of 2.5 as in the default analysis, due to the poor muon momentum resolution translating into a poor signal-vs-background separation in the distribution of the discriminating variable $m_{\mu\mu}$.

Finally, the comparison with the full simulation samples method based on parametrized efficiency and resolution functions.

# Appendix

## A. Results for alternative detector scenarios or selections

Alternative scenarios with reduced coverage of the inner tracking detector, or with an extended pseudorapidity selection for muons, have also been explored:

- “low” scenario: inner tracking detector acceptance reduced to $|\eta| < 2.5$,
- “middle” scenario: inner tracking detector acceptance reduced to $|\eta| < 3.2$,
- “extended” scenario: reference detector, with muon pseudorapidity selection extended to $|\eta| < 4.0$.

The main results about these alternative scenarios, in terms of expected significance and uncertainty on the signal strengths, have been given in Sec. 6. In this appendix a few more details are provided.

The reduction of the inner tracking detector acceptance has, as a major consequence, the reduction in pile-up jet rejection, which relies on tracking information. This is illustrated in Fig. 13. The purity of VBF signal events in the VBF category is thus worsened, as well as the signal efficiency for the “low” scenario, due to more muons with $|\eta|$ close to 2.5 being discarded by the jet-muon overlap algorithm with respect to the “reference” detector scenario.

In addition to the reduced inner detector acceptance, the “low” and “middle” scenarios are also characterised by lower muon trigger efficiencies and higher muon trigger $p_{\mathrm{T}}$ thresholds (25 GeV instead of 20 GeV for the single muon trigger and 15 GeV instead of 11 GeV for the dimuon trigger). The track momentum resolution of the “low” detector scenario is also worse than in the other cases due to a reduced number of layers in the inner tracking detector.

In the case of the extended scenario, the selection of muon candidates has been loosened to accept muon with absolute pseudorapidity up to 4, as shown in Fig. 14. This translates into a larger signal efficiency. However, the muon momentum resolution for $2.5 < |\eta| < 4.0$ is significantly worse than for $|\eta| < 2.5$, which implies a rather worse signal-to-background ratio for events with at least one forward muon ($2.5 < |\eta| < 4.0$), as the peaks from the Higgs boson (signal) and $Z$ boson (background) decays to dimuons are significantly smeared and their separation is decreased, as shown in Fig. 15. For this reason, in the case of the extended scenario, events with at least one forward muon are classified in three “forward” categories of their own, distinguished by the value of the dimuon transverse momentum in the same way as for the “central” and “non-central” ones. The signal model fits for the three forward categories are illustrated in Fig. 16. The signal invariant mass resolution for events with forward muons is at least two times worse than for events with both muon pseudorapidities within ±2.5.

The overall signal efficiency for the three alternative scenarios is:

- “low” scenario: 55% (55% for ggF events and 58% for VBF ones);
- “middle” scenario: 58% (58% for ggF events and 61% for VBF ones);
- “extended” scenario: 68% (68% for ggF events and 71% for VBF ones).

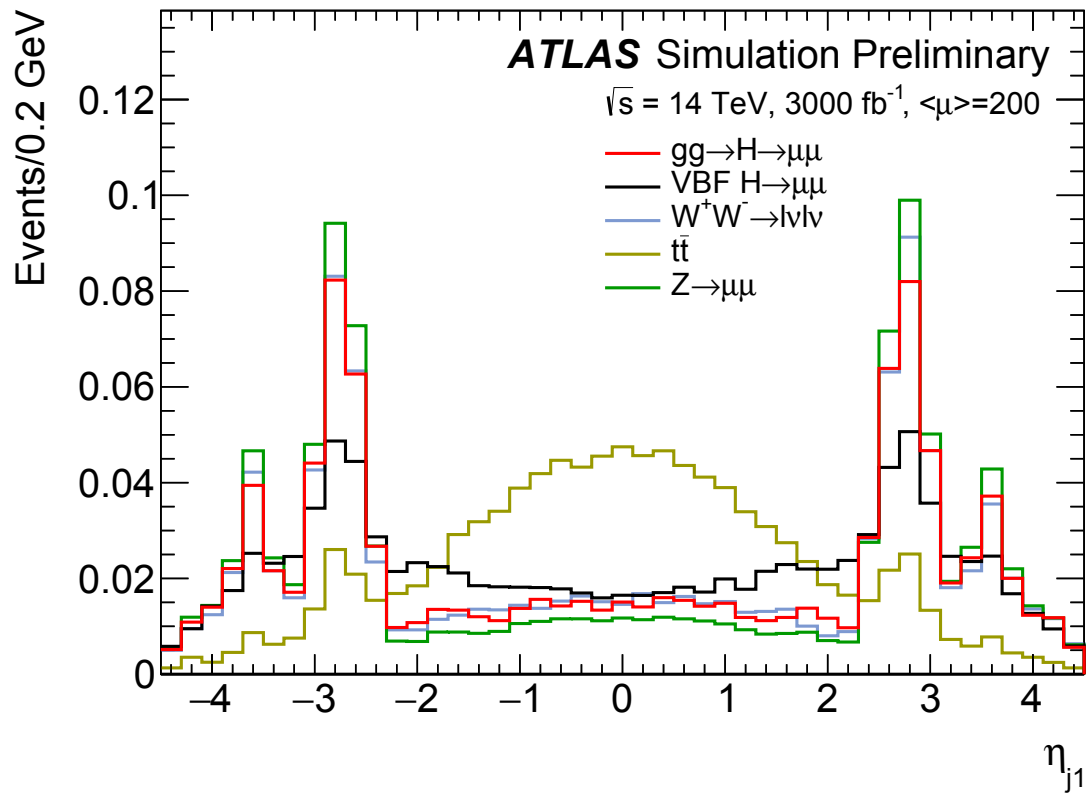

(a) $\eta_{j_1}$, “low” scenario

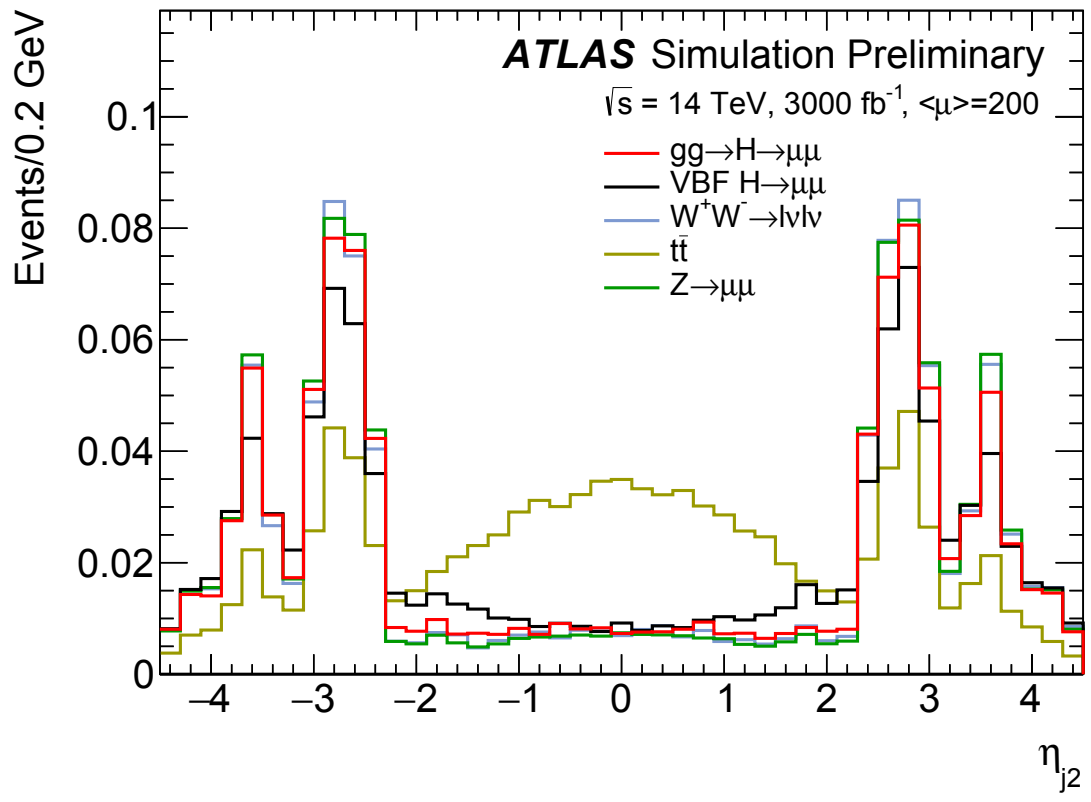

(b) $\eta_{j_2}$, “low” scenario

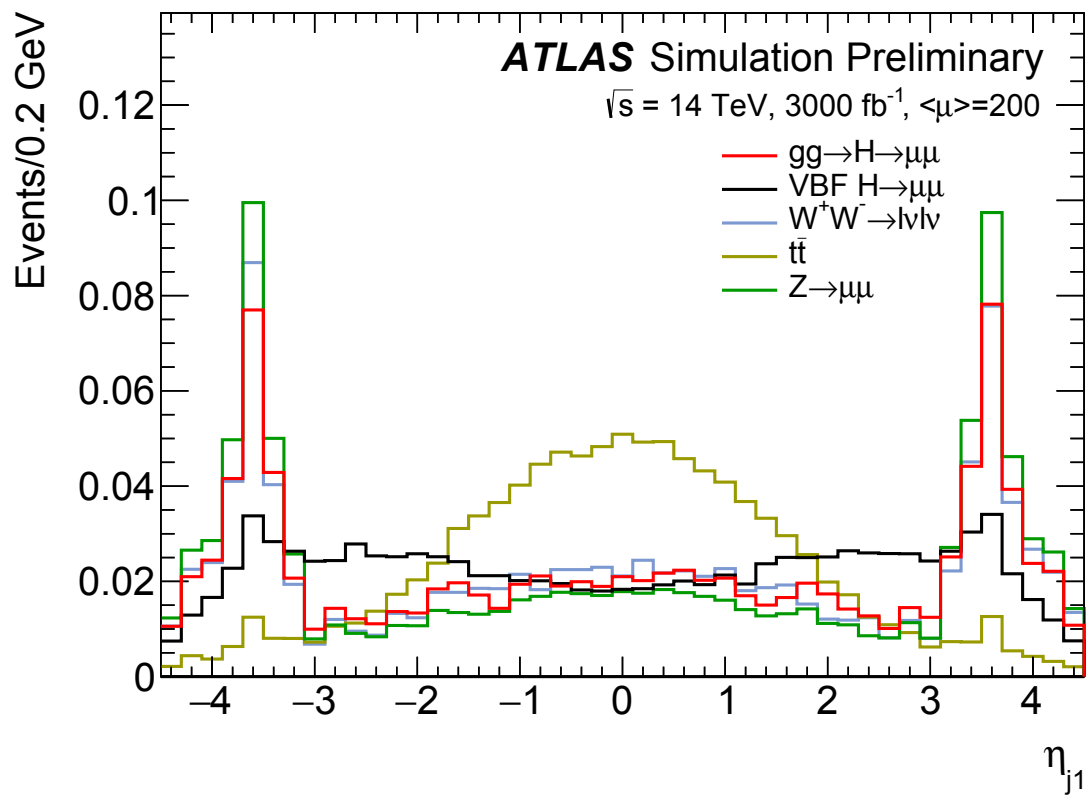

(c) $\eta_{j_1}$, “middle” scenario

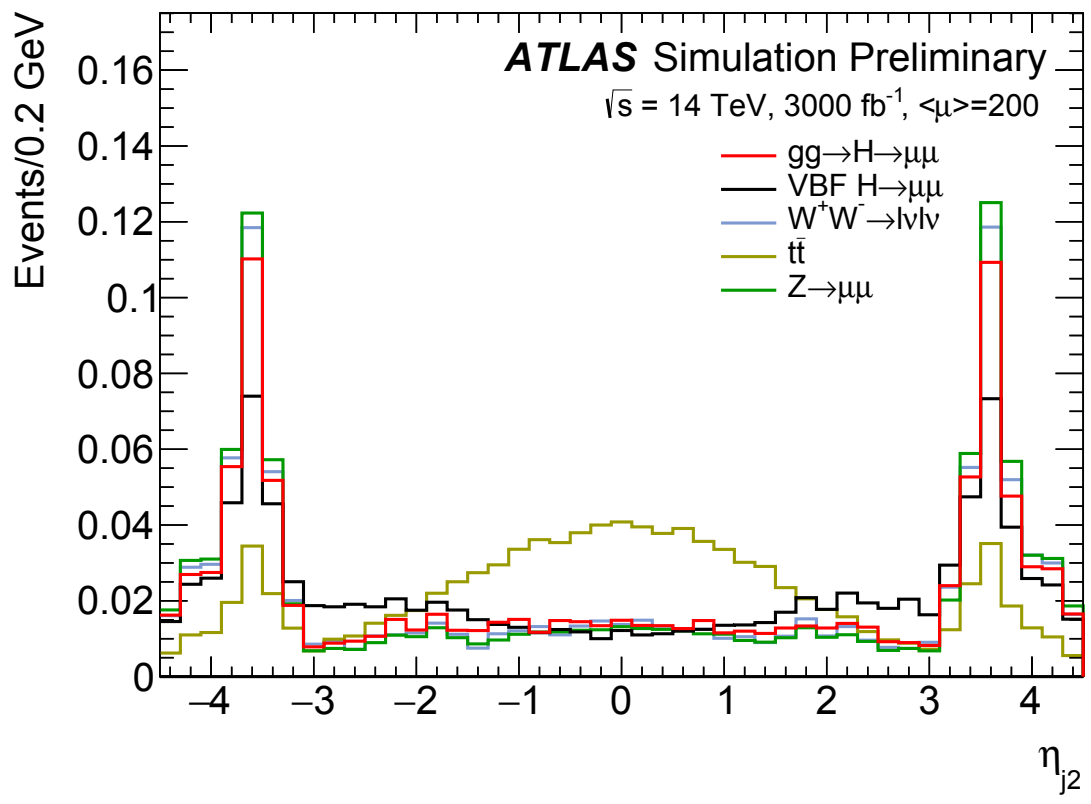

(d) $\eta_{j_2}$, “middle” scenario

Figure 13: Pseudorapidity distributions of selected (a,c) leading and (b,d) subleading jet candidates, normalised to unity, for the (a,b) “low” and (c,d) “middle” detector scenarios, assuming $\langle\mu\rangle = 200$. Only events with at least two good jets are considered.

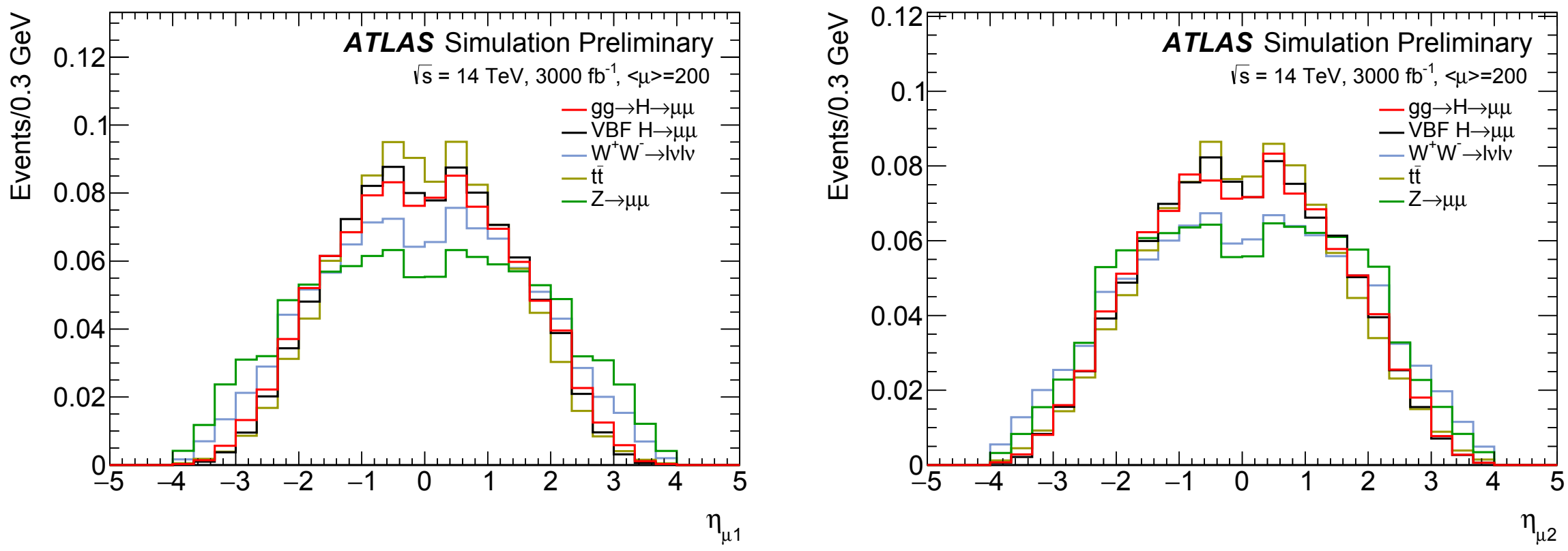

(a) $\eta_{\mu_1}$ (b) $\eta_{\mu_2}$

Figure 14: Pseudorapidity of selected (a) leading and (b) subleading muon candidates, scaled to 3000 fb$^{-1}$, for the extended detector scenario, assuming $\langle\mu\rangle$ = 200 and selecting muons with $|\eta|$ up to 4.0.

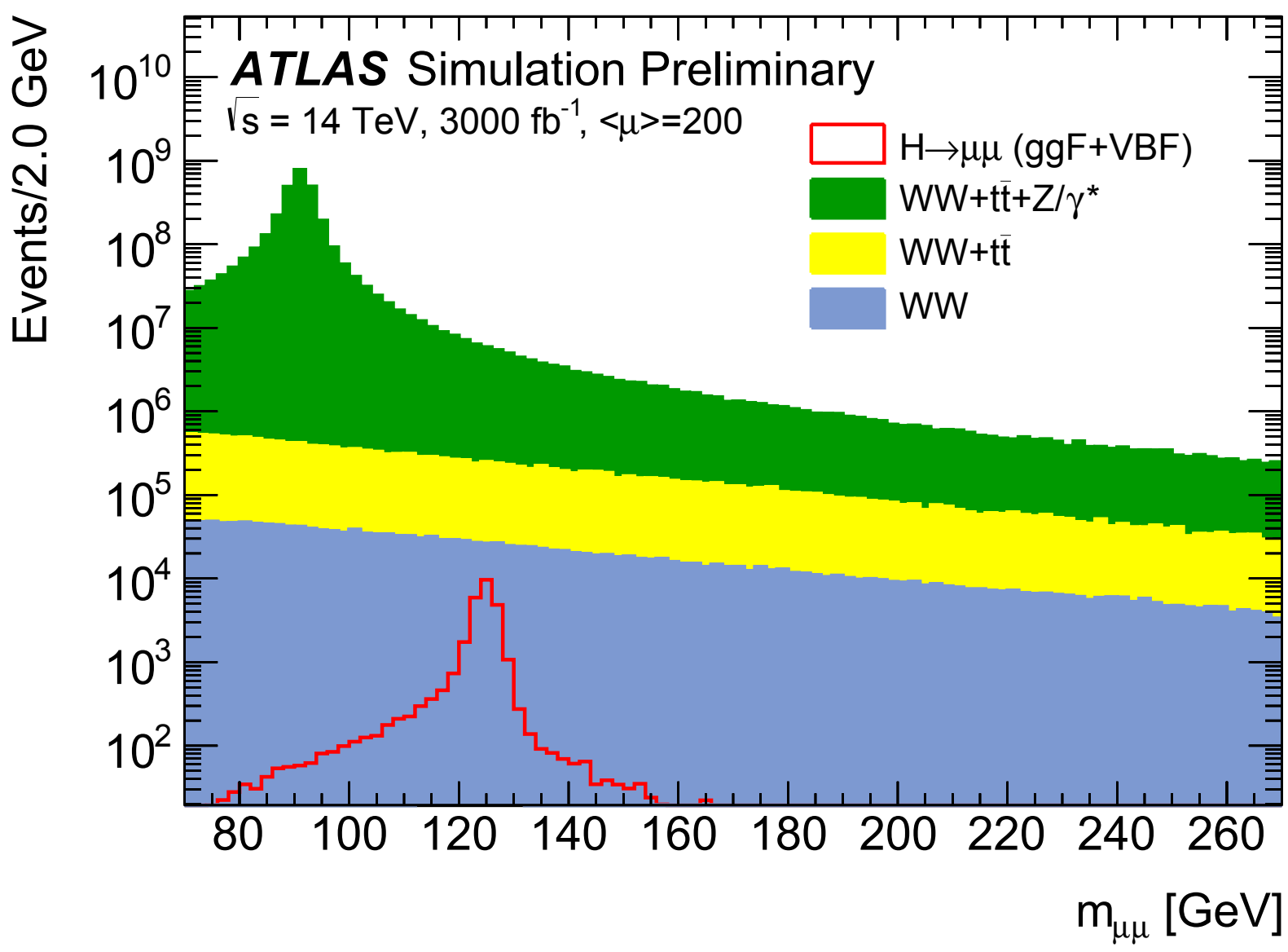

Figure 15: Invariant mass distribution of selected signal and background candidates, scaled to 3000 fb$^{-1}$, for the extended detector scenario, assuming $\langle\mu\rangle$ = 200 and selecting muons with $|\eta|$ up to 4.0.

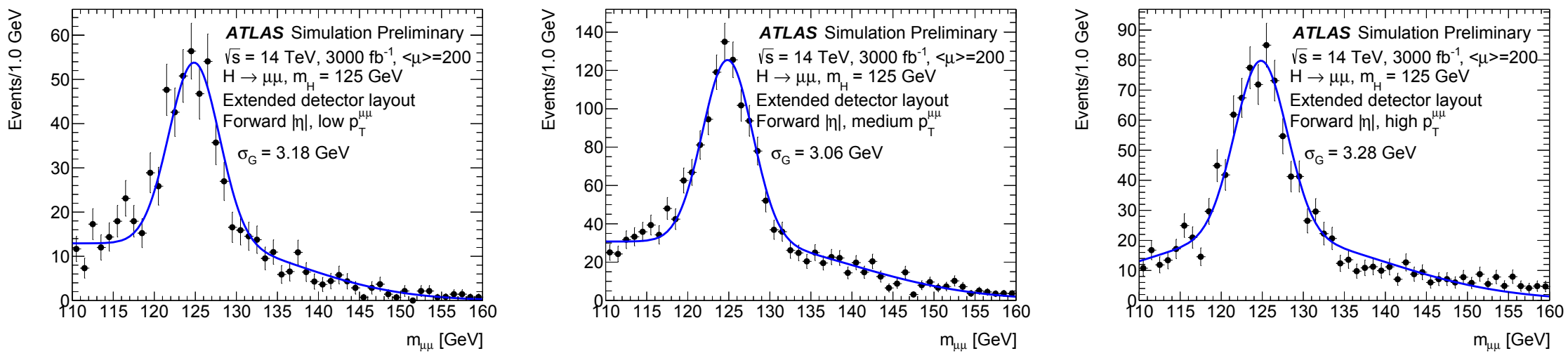

(a) forward, low $p_T^{\mu\mu}$ (b) forward, medium $p_T^{\mu\mu}$ (c) forward, high $p_T^{\mu\mu}$

Figure 16: Invariant mass distribution of selected signal candidates, scaled to 3000 fb$^{-1}$, and result of the Crystal-Ball+Gaussian fit, for the extended detector scenario, assuming $\langle\mu\rangle = 200$ and selecting muons with $|\eta|$ up to 4.0, for the three forward categories. The resolution $\sigma_G$ of the Gaussian core of the distribution is overlaid.

The expected $B/S$ ratio in a $\pm 1.5\sigma_G$ invariant-mass window around $m_{\mu\mu} = 125$ GeVafter the full selection is $\approx$ 270, 280 and 430 in the low, middle and extended scenarios, respectively.

The expected signal and background yields and signal significance for each event category for the low, middle and extended detector scenarios are given in Tables 7, 8, and 9, respectively.

Table 7: Expected signal and background yields and signal significance in a $\pm 1.5\sigma_G$ invariant-mass window around $m_{\mu\mu} = 125$ GeV for each category, where $\sigma_G$ is the resolution of the core of the invariant mass distribution of signal events. The last rows shows the total signal and background yields, the average invariant mass resolution, and the sum in quadrature of the significance of each category. The projections correspond to an integrated luminosity $\int \mathcal{L}dt = 3000$ fb$^{-1}$ for a center-of-mass energy $\sqrt{s}$=14 TeV for the "low" detector scenario.

| **Category** | $S$ | **VBF** | $B$ | **FWHM [GeV]** | $\sigma_G$ **[GeV]** | $S/\sqrt{S+B}$ |
|---|---|---|---|---|---|---|
| VBF-like | 550 | 226 | 34410 | 4.50 | 1.97 | 2.94 |
| low $p_T$, central | 820 | 10 | 304300 | 3.09 | 1.35 | 1.48 |
| med $p_T$, central | 1960 | 75 | 261900 | 3.07 | 1.31 | 3.82 |
| hi $p_T$, central | 1510 | 198 | 179200 | 3.47 | 1.54 | 3.56 |
| low $p_T$, non central | 2350 | 27 | 1685200 | 4.22 | 1.81 | 1.81 |
| med $p_T$, non central | 5600 | 219 | 1433100 | 4.31 | 1.83 | 4.67 |
| hi $p_T$, non central | 4010 | 526 | 780500 | 4.78 | 1.96 | 4.53 |
| Total | 16810 | 1280 | 4678700 | 4.03 | 1.71 | 9.15 |

Table 8: Expected signal and background yields and signal significance in a $\pm 1.5\sigma_G$ invariant-mass window around $m_{\mu\mu}$ = 125 GeV for each category, where $\sigma_G$ is the resolution of the core of the invariant mass distribution of signal events. The last rows shows the total signal and background yields, the average invariant mass resolution, and the sum in quadrature of the significance of each category. The projections correspond to an integrated luminosity $\int \mathcal{L}dt$ = 3000 fb$^{-1}$ for a center-of-mass energy $\sqrt{s}$=14 TeV for the "middle" detector scenario.

| **Category** | ***S*** | **VBF** | ***B*** | **FWHM [GeV]** | $\sigma_G$ **[GeV]** | $S/\sqrt{S+B}$ |
|---|---|---|---|---|---|---|
| VBF-like | 480 | 216 | 27300 | 4.50 | 1.91 | 2.88 |
| low $p_T$, central | 891 | 11 | 338400 | 3.16 | 1.37 | 1.53 |
| med $p_T$, central | 2130 | 82 | 291400 | 3.13 | 1.33 | 3.93 |
| hi $p_T$, central | 1690 | 223 | 199100 | 3.46 | 1.54 | 3.77 |
| low $p_T$, non central | 2440 | 28 | 1726600 | 4.12 | 1.79 | 1.85 |
| med $p_T$, non central | 5790 | 227 | 1465100 | 4.25 | 1.80 | 4.77 |
| hi $p_T$, non central | 4250 | 564 | 806000 | 4.69 | 1.93 | 4.72 |
| Total | 17660 | 1350 | 4854000 | 3.93 | 1.69 | 9.42 |

Table 9: Expected signal and background yields and signal significance in a $\pm 1.5\sigma_G$ invariant-mass window around $m_{\mu\mu}$ = 125 GeV for each category, where $\sigma_G$ is the resolution of the core of the invariant mass distribution of signal events. The last rows shows the total signal and background yields, the average invariant mass resolution, and the sum in quadrature of the significance of each category. The projections correspond to an integrated luminosity $\int \mathcal{L}dt$ = 3000 fb$^{-1}$ for a center-of-mass energy $\sqrt{s}$=14 TeV, in the reference detector scenario with muon pseudorapidity selection extended to $|\eta|$ = 4.0.

| **Category** | ***S*** | **VBF** | ***B*** | **FWHM [GeV]** | $\sigma_G$ **[GeV]** | $S/\sqrt{S+B}$ |
|---|---|---|---|---|---|---|
| VBF-like | 397 | 202 | 21300 | 4.60 | 1.90 | 2.70 |
| low $p_T$, central | 921 | 11 | 347000 | 3.21 | 1.37 | 1.56 |
| med $p_T$, central | 2210 | 84 | 300700 | 3.08 | 1.32 | 4.01 |
| hi $p_T$, central | 1810 | 242 | 211900 | 3.50 | 1.56 | 3.91 |
| low $p_T$, non central | 2460 | 28 | 1739700 | 4.11 | 1.79 | 1.86 |
| med $p_T$, non central | 5860 | 230 | 1483000 | 4.23 | 1.80 | 4.80 |
| hi $p_T$, non central | 4380 | 588 | 828000 | 4.68 | 1.92 | 4.80 |
| low $p_T$, forward | 398 | 4 | 1360900 | 8.88 | 3.18 | 0.34 |
| med $p_T$, forward | 907 | 36 | 1664100 | 8.88 | 3.06 | 0.70 |
| hi $p_T$, forward | 615 | 79 | 567000 | 9.34 | 3.28 | 0.82 |
| Total | 19950 | 1500 | 8523600 | 4.33 | 1.74 | 9.59 |

# Prospects for $H \to c\bar{c}$ using Charm Tagging with the ATLAS Experiment at the HL-LHC

The ATLAS Collaboration

The expected sensitivity of a search for decays of the Higgs boson into charm quarks is estimated for the ATLAS detector at the HL-LHC, by extrapolating the results obtained using a dataset with an integrated luminosity of 36.1 fb$^{-1}$ at $\sqrt{s}$ = 13 TeV. Associated production of Higgs and $Z$ bosons is targeted, where the $Z$ bosons decay to electrons or muons. Assuming an integrated luminosity of 3000 fb$^{-1}$ of $pp$ collision data at $\sqrt{s}$ = 14 TeV, an expected upper limit at the 95% confidence level of 6.3 times the Standard Model expectation for $\sigma(pp \to ZH) \times \mathcal{B}(H \to c\bar{c})$ is estimated, in the absence of systematic uncertainties. The impact of systematic uncertainties and possible improvements in the flavour tagging performance on the sensitivity are estimated.

# 1 Introduction

With a Standard Model (SM) branching ratio of 2.9% [1], decays of the Higgs boson to charm ($c$) quarks represent the fermionic decay mode with the largest branching ratio for which no experimental evidence exists. This decay mode also represents a promising window through which to probe the Yukawa couplings of the Higgs boson to the second generation quarks. Furthermore, given that the branching ratio for Higgs decays to first and second generation quarks is small in the SM, potential new physics affecting this sector could lead to notable modifications [2].

Measurements of the Yukawa couplings of the Higgs boson to second generation quarks are challenging at hadron colliders due to the large hadronic backgrounds, small branching ratios, and challenging jet flavour identification. However, exclusive searches for rare inclusive decays of the Higgs boson to $J/\psi\gamma$ [3], $\rho\gamma$, and $\phi\gamma$ [4] final states have been performed by the ATLAS Collaboration. Inclusive searches for decays of the Higgs boson into $c$ quarks are possible through its associated production with a vector boson, which provides a distinct final state and efficient trigger strategy. Recently, an inclusive Run 2 search based on a $\sqrt{s}$ = 13 TeV $pp$ collision dataset corresponding to an integrated luminosity of 36.1 fb$^{-1}$ collected by the ATLAS experiment [5] set a 95% CL$_s$ [6, 7] upper limit on $\sigma(pp \to ZH) \times \mathcal{B}(H \to c\bar{c})$ of 2.7 pb (110 times the SM expectation) [8].

This note describes the prospects for such an inclusive search at the upgraded HL-LHC ATLAS detector [9, 10], by extrapolating from the signal and background simulation used in the Run 2 analysis [8], to the expected integrated luminosity of 3000 fb$^{-1}$. The analysis strategy closely follows the Run 2 analysis [8], with any differences stated explicitly. However, in a HL-LHC scenario, both the ATLAS detector and the analysis strategy are expected to change significantly from the Run 2 analysis and it is difficult to make reliable predictions for the systematic uncertainties relevant to the HL-LHC analysis. As such, the main result of this study is a statistical projection of the limit, neglecting possible systematic uncertainties. Where possible, the impact of typical sources of systematic uncertainty on the limit are estimated, based on their Run 2 values. Finally, the effect of expected improvements in the ATLAS flavour tagging performance on the sensitivity of the analysis is explored.

# 2 Methods

This search targets the production of a Higgs boson in association with a $Z$ boson, in a final state where the $Z$ boson decays to a pair of oppositely charged electrons or muons and the Higgs boson decays to a pair of $c$ quarks. The selection considered for the HL-LHC is almost identical to the Run 2 analysis [8], with single lepton triggers used to collect events, followed by $Z$ and Higgs boson candidates being formed from pairs of leptons and one or two $c$-tagged jets, respectively.

The sensitivity of the search was quantified using a profile likelihood fit (referred to as the fit) to the invariant mass distributions (10 GeV bins between 50 and 200 GeV) of the two highest $p_T$ jets ($m_{c\bar{c}}$), in four event categories. These four event categories are defined as having either 1 or 2 of the Higgs boson candidate jets $c$-tagged, and $75 < p_T^Z < 150$ GeV or $p_T^Z > 150$ GeV, where $p_T^Z$ denotes the transverse momentum of the $Z$ boson candidate. The dominant background to this search is $Z$ + jets, with smaller contributions arising from diboson ($ZZ$, $ZW$), $t\bar{t}$, and $ZH(b\bar{b})$ production.

The higher expected signal yield in the HL-LHC scenario is exploited by considering a $c$-jet tagging working point with a greater $b$- and light-flavour jet rejection than that used by the Run 2 analysis.

The tighter working point exhibits efficiencies of 18%, 5% and 0.5% for $c$-, $b$-jet and light-flavour jets, respectively. This reduces the relative contribution from the kinematically irreducible $ZH(b\bar{b})$ background, and improves the overall background rejection by a factor of 5.3 for a 54% loss in signal, resulting in a 7% improvement to the expected limit.

The extrapolation from the Run 2 analysis to the expected HL-LHC result is performed by scaling the signal and background expectations from the Run 2 search [8] (using both the Run 2 shapes and normalisations) by process- and category-dependent scale factors (SF). These SFs account for modifications to the integrated luminosity, signal and background production cross sections, and expected $c$-jet tagging performance. Run 2 performance is assumed for all physics objects relevant to this analysis, with the exception of $c$-jet tagging, where the effect of performance improvements expected with the HL-LHC detector are investigated.

The increased integrated luminosity is accounted for by scaling the number of expected signal and background events by the ratio of the Run 2 (36.1 fb$^{-1}$) to expected HL-LHC (3000 fb$^{-1}$) integrated luminosities. The increase in the expected LHC center of mass energy $\sqrt{s}$ from 13 TeV to 14 TeV is accounted for with a $p_T^Z$-category dependent scaling of the number of events for the $ZH(c\bar{c})$ and $ZH(b\bar{b})$ processes. The scaling is based on the exclusive $pp \to ZH$ cross sections from [11], accounting for variations across the two $p_T^Z$-categories using a PYTHIA8-based generator-level Monte Carlo (MC) simulation [12]. The 13 TeV predictions for the $Z$ + jets and diboson backgrounds are scaled using the 14 TeV prediction assuming the ratio of parton luminosities for $q\bar{q}$ initiated processes, and the $t\bar{t}$ background is scaled assuming the ratio of parton luminosities for $gg$ initiated processes [11].

While the normalisation of the diboson and $t\bar{t}$ backgrounds are fixed to the SM expectation, the normalisation of the $Z$ + jets background is free to vary independently in the four fit categories. Given the large number of expected $Z$ + jets background events in the HL-LHC scenario, the normalisation of the $Z$ + jets background is estimated to be determined by the data with an uncertainty of less than 2% in all categories. As the normalisation of this background is determined in data, it is considered as a statistical uncertainty.

For the Run 2 analysis, the $ZH(b\bar{b})$ background normalisation was allowed to vary in the fit within the uncertainties associated with the SM expectation. For the HL-LHC extrapolation, the normalisation of the $ZH(b\bar{b})$ background is instead allowed to vary to within 14% of the SM value, corresponding to the expected uncertainty of the $VH(b\bar{b})$ signal strength (signal yield normalised to SM expectation) measurement for an integrated luminosity of 3000 fb$^{-1}$ at the HL-LHC [13]. This reflects the likely situation that the $VH(b\bar{b})$ normalisation will be determined in data, possibly in a combined $VH(b\bar{b})$ and $VH(c\bar{c})$ fit, hence the corresponding uncertainty is not considered as a systematic uncertainty associated to this analysis directly, but rather as statistical in nature.

## 3 Results

The expected 95% CL$_s$ [7] upper limit on the signal strength in the absence of systematic uncertainties is found to be $\mu_{ZH(c\bar{c})} < 6.3^{+2.5}_{-1.8}$, where the uncertainty corresponds to the $\pm 1\sigma$ interval of background-only pseudo-experiments. The best fit value for the $ZH(c\bar{c})$ expected signal strength is $\mu_{ZH(c\bar{c})} = 1.0 \pm 3.2$, and the $m_{c\bar{c}}$ distributions in the four analysis categories are shown in Figure 1. The expected yields for the signal and background processes in each category are shown in Table 1.

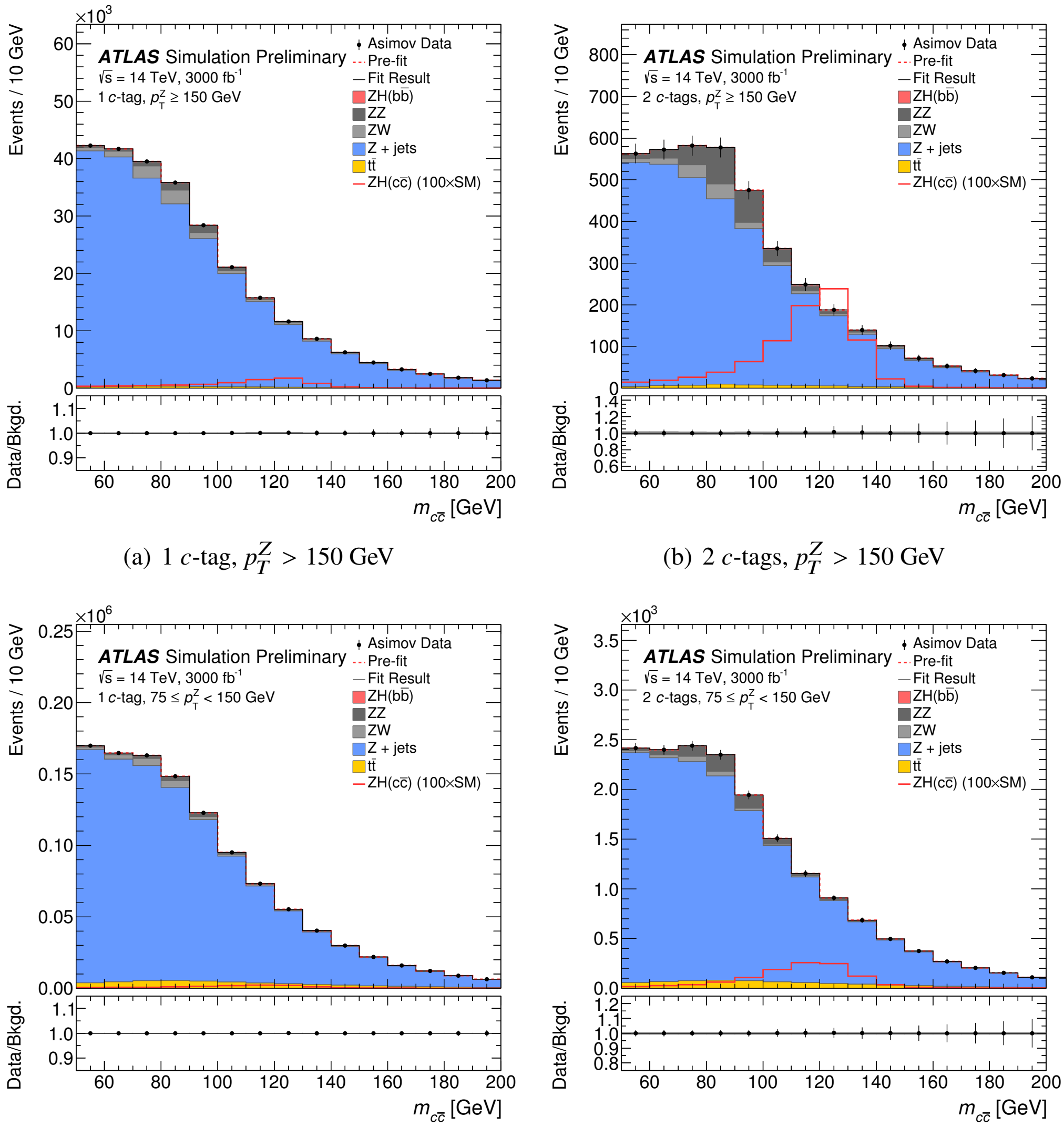

(a) 1 $c$-tag, $p_T^Z > 150$ GeV

(b) 2 $c$-tags, $p_T^Z > 150$ GeV

(c) 1 $c$-tag, $75 < p_T^Z < 150$ GeV

(d) 2 $c$-tags, $75 < p_T^Z < 150$ GeV

Figure 1: Post-fit $m_{c\bar{c}}$ distributions, for the four analysis categories used in the analysis. The expected signal is scaled by a factor of 100. The Asimov Data corresponds to the sum of expected signal and background events, while the stacked histogram corresponds only to the backgrounds. The error bars represent the statistical uncertainty in the expected number of data events.

Allowing the $Z$ + jets normalisation to float in the fit has an impact of +21% on the expected limit for an integrated luminosity of 3000 fb$^{-1}$, relative to the case where it is fixed to the nominal prediction in each category. The limit improves by only ~ 1% when the $ZH(b\bar{b})$ background normalisation is fixed to its SM expectation, showing that the expected $ZH(b\bar{b})$ signal strength measurement is sufficient to constrain this background.

In the context of this study, the sensitivity at the end of Run 3 was also evaluated under the assumption of 300 fb$^{-1}$ of 13 TeV $pp$ collision data. The analysis method and systematic uncertainties assumed are identical to those of the Run 2 analysis. At the end of Run 3 the 95% CL$_s$ expected upper limit on the $ZH(c\bar{c})$ signal strength is $\mu_{ZH(c\bar{c})} < 38^{+18}_{-10}$, estimated with a fit close to that of the Run 2 search [8].

4

| Sample | Yield | | | |
|---|---|---|---|---|
| | 1 $c$-tag | | 2 $c$-tags | |
| | $75 \leq p_{\mathrm{T}}^{Z} < 150$ GeV | $p_{\mathrm{T}}^{Z} > 150$ GeV | $75 \leq p_{\mathrm{T}}^{Z} < 150$ GeV | $p_{\mathrm{T}}^{Z} > 150$ GeV |
| $Z$ + jets | 271 000 ± 13 500 | 59 300 ± 2970 | 4350 ± 217 | 892 ± 44.6 |
| $WZ$ | 4080 ± 204 | 1700 ± 85.2 | 48.5 ± 2.42 | 29.6 ± 1.48 |
| $ZZ$ | 2570 ± 128 | 1020 ± 50.9 | 95.7 ± 4.79 | 49.7 ± 2.49 |
| $t\bar{t}$ | 16 000 ± 827 | 863 ± 44.6 | 241 ± 12.4 | 26.3 ± 1.36 |
| $ZH(b\bar{b})$ | 441 ± 16.8 | 327 ± 12.4 | 10.7 ± 0.407 | 9.38 ± 0.356 |
| $ZH(c\bar{c})$ | 74.4 ± 2.83 | 52.6 ± 2.00 | 8.54 ± 0.325 | 6.89 ± 0.262 |
| Total | 294 000 ± 13 600 | 63 300 ± 2970 | 4750 ± 218 | 1010 ± 44.7 |
| $S/\sqrt{S+B}$ | 0.137 ± 0.008 | 0.209 ± 0.013 | 0.124 ± 0.007 | 0.216 ± 0.013 |

Table 1: The expected yields for the signal and each background process in each signal region in the range $100 < m_{c\bar{c}} < 150$ GeV. The yields are estimated from MC simulation. The cross section uncertainties (not included in the fit) on the samples are shown. The final row shows the statistical significance of the $ZH(c\bar{c})$ signal ($S$), considered within the context of the sum of all background contributions ($B$).

# 4 Systematic Uncertainties

Systematic uncertainties affecting the Run 2 analysis are modelled as nuisance parameters in the fit. Due to the changes to the detector and the analysis strategy expected for the HL-LHC analysis, it is difficult to estimate the precise sources and effects of systematic uncertainties. However, the impact of some of the dominant sources of systematic uncertainty in the Run 2 analysis, estimated based on their impact in the Run 2 analysis, is studied to estimate the susceptibility of the HL-LHC sensitivity to pertinent systematic uncertainties. Individual nuisance parameters from the Run 2 analysis are assigned to broad groups (e.g. $c$-jet tagging and background shape) based on the nature of the systematic uncertainties they correspond to. The effect of each group of nuisance parameters on the 95% CL$_s$ expected upper limit on the $ZH(c\bar{c})$ signal strength is evaluated by repeating the fit with all of the nuisance parameters in a given group introduced to the fit. The impact of these groups of uncertainties on the sensitivity is summarised in Table 2. The uncertainties associated with the nuisance parameters can be constrained in the fit. The largest constraints occur for the nuisance parameters associated with the $Z$ + jets background shape, $c$-jet tagging efficiency and jet energy scale and resolution uncertainties, representing the ability to more precisely control these sources of uncertainty using the large amount of data at the HL-LHC. Compared to the Run 2 analysis, the HL-LHC analysis experiences a reduced exposure to the uncertainties associated with the tagging efficiency measurements, due to the reduced light-flavour jet component in the background as a result of the tighter operating point, and due to the uncertainty on the $c$-jet tagging efficiency being constrained in the fit.

The uncertainty in the shape of the $Z$ + jets background, due to the modelling of the underlying event and the parton shower, is likely to represent the dominant limitation to the sensitivity of the analysis, and will therefore require careful consideration in a HL-LHC analysis. However, the impact of the experimental systematics uncertainties (e.g. the $c$-jet tagging efficiency uncertainty) in a HL-LHC scenario will likely reduce relative to their effect on the Run 2 analysis given the large datasets available, allowing precise performance studies to be conducted. This effect is estimated in this study through the constraints on the associated nuisance parameters.

| Source of uncertainty | Change in limit |
|---|---|
| Background shape | +36% |
| Jet energy scale and resolution | +17% |
| Lepton reconstruction and identification | +12% |
| $c$-jet tagging efficiency | +11% |

Table 2: The increase in the nominal limit on the $ZH(c\bar{c})$ signal strength due to the introduction of typical systematic uncertainties, based on their effect on the Run 2 analysis. The "Background shape" uncertainties refer to the shape uncertainty in the $Z$ + jets, diboson, $t\bar{t}$ and $ZH(b\bar{b})$ backgrounds as estimated from MC generator comparisons in the Run 2 analysis. The "$c$-jet tagging efficiency" uncertainties refer to the uncertainty in the efficiencies of $c$-, $b$- and light-flavour jets in data, determined within the context of the Run 2 analysis.

## 5 Flavour Tagging Improvements

Preliminary studies into the $b$-jet tagging performance of ATLAS at the HL-LHC suggest an improvement of around a factor of 2.5 [14] in the light-flavour jet rejection. Assuming a factor of 2.5 improvement for the light-flavour jet rejection, for a fixed $b$-jet rejection and $c$-jet efficiency, an 8% improvement in the limit can be expected. Furthermore, $c$-jet tagging in a hadron collider environment is a very active area of research which is currently less mature than $b$-jet tagging. Significant improvements in the performance of $c$-jet tagging algorithms can be expected in coming years.

## 6 Potential Analysis Improvements

This limit likely represents an overestimation of the sensitivity of the analysis, due to the absence of systematic uncertainties. However, various improvements to the analysis strategy could increase the sensitivity significantly. In particular, other production channels, such as $Z(\nu\nu)H$ and $W(\ell\nu)H$, are known to exhibit comparable sensitivity to the $Z(\ell\ell)H$ channel in the analogous analysis for $H \to b\bar{b}$ decays [15]. Furthermore, the use of multivariate analysis (MVA) techniques (such as Boosted Decision Trees) was also shown to provide further sensitivity improvements. The inclusion of these additional production channels and an MVA-based analysis strategy could provide significantly enhanced sensitivity for the $H \to c\bar{c}$ analysis if adopted.

## 7 Conclusions

The expected sensitivity of a search for $pp \to Z(\ell^+\ell^-)H(c\bar{c})$ has been evaluated by extrapolating the results of a Run 2 search, performed by ATLAS using a $\sqrt{s}$ = 13 TeV $pp$ collision dataset corresponding to an integrated luminosity of 36.1 fb$^{-1}$. Assuming an integrated luminosity of 3000 fb$^{-1}$ of $\sqrt{s}$ = 14 TeV of $pp$ collision data at the HL-LHC, a 95% CL$_s$ upper limit on the $ZH(c\bar{c})$ signal strength of $\mu_{ZH(c\bar{c})} < 6.3$ can be expected, in the absence of systematic uncertainties. While, based on this projection, an observation is not expected, such a limit would provide strong constraints on new physics models, and provide a competitive direct limit on the Yukawa coupling of the Higgs boson to $c$ quarks .

**ATLAS PUB Note**

ATL-PHYS-PUB-2018-053

24th December 2018

# Measurement prospects of the pair production and self-coupling of the Higgs boson with the ATLAS experiment at the HL-LHC

The ATLAS Collaboration

A prospect study for the search for non-resonant Higgs-boson-pair production, using a statistical combination of the $b\bar{b}b\bar{b}$, $b\bar{b}\gamma\gamma$ and $b\bar{b}\tau^+\tau^-$ final states, is performed assuming 3000 fb$^{-1}$ of $pp$ collisions at a centre-of-mass energy of 14 TeV in the ATLAS detector at the High Luminosity LHC (HL-LHC). When systematic uncertainties are neglected, the expected signal significance is found to be 3.5$\sigma$ while the signal strength relative to the Standard Model prediction is expected to be measured with an accuracy of 31%. The Higgs boson self-coupling can be constrained to $-0.1 \leq \lambda_{HHH}/\lambda_{HHH}^{SM} \leq 2.7 \cup 5.5 \leq \lambda_{HHH}/\lambda_{HHH}SM \leq 6.9$, at 95% confidence level, and the measured value of $\lambda_{HHH}/\lambda_{HHH}^{SM}$ is expected to be $1.0^{+0.7}_{-0.6}$. If systematic uncertainties are included, the signal significance is found to be 3.0$\sigma$ while the signal strength relative to the Standard Model expectation is expected to be measured with an accuracy of 40%. In that case, the Higgs boson self-coupling can be constrained to $-0.4 \leq \lambda_{HHH}/\lambda_{HHH}^{SM} \leq 7.3$, at 95% confidence level, and the measured value of $\lambda_{HHH}/\lambda_{HHH}^{SM}$ is expected to be $1.0^{+0.9}_{-0.8}$. The $b\bar{b}\gamma\gamma$ and $b\bar{b}\tau^+\tau^-$ final states are further extrapolated to provide first estimates of the prospects at the High Energy LHC, assuming 15 ab$^{-1}$ of data at a centre-of-mass collision energy of 27 TeV.

# 1 Introduction

The Standard Model (SM) predicts the existence of a quantum field [1–4] that is responsible for the generation of masses of fundamental particles. This conjecture was tested and confirmed in 2012 with the discovery by the ATLAS and CMS Collaborations at the CERN Large Hadron Collider (LHC) of a particle associated with this field: the Higgs boson [5, 6].

After the observation, a number of measurements have been performed to quantify the properties of this particle, such as: mass, spin and parity quantum numbers, and couplings.

- Mass: The combination of the measurements of the Higgs boson mass performed by the ATLAS and CMS Collaborations using data collected in 2011-2012 (Run 1) yields a value of: $m_H = 125.09 \pm 0.21(\text{stat.}) \pm 0.11(\text{syst.})$ GeV [7]. This value is consistent with indirect constraints from precision electroweak data. Measurements based on a subset of the Run 2 data can be found in Refs. [8, 9].
- Spin and parity quantum numbers: The spin-parity of the observed particle agrees with the SM hypothesis (spin 0, $J^P = 0^+$): the ATLAS and CMS Collaborations exclude several alternative spin and parity hypotheses in favour of the SM hypothesis at more than 99.9% confidence level (CL) [10, 11].
- Couplings: Constraints on the coupling strength of the Higgs boson to vector bosons and fermions based on a combination of measurements by the ATLAS and CMS Collaborations are reported in Ref. [12]. Couplings to vector bosons are found to be compatible with those expected from the SM within an approximate 10% uncertainty, while in the case of the heavier SM fermions (top- and bottom-quarks, and $\tau$-leptons) the uncertainty is of the order of 15-20%. Those measurements are being significantly improved using Run 2 data [13, 14], and the couplings to fermions are now definitely established thanks to the recent observations of the ttH production and $H \to b\bar{b}$ decay mode [15–18]. The measurements of the coupling strengths can be used to constrain Beyond the Standard Model (BSM) scenarios.

In the SM, the physical Higgs field $H$ interacts with itself generating both mass and self-interaction terms. These terms arise from the Higgs potential in the perturbative expansion of the Higgs doublet $\phi$ around the electroweak symmetry breaking vacuum expectation value ($v$):

$$V(\phi^\dagger\phi) = \mu^2\phi^\dagger\phi + \lambda(\phi^\dagger\phi)^2 \tag{1}$$

$$\supset \lambda v^2 H^2 + \lambda v H^3 + \frac{\lambda}{4}H^4 \,. \tag{2}$$

The first term in the expansion describes the mass of the Higgs boson, $m_H = \sqrt{2\lambda v^2}$. The second and third terms describe the tri-linear and quartic self-interactions of the Higgs boson, with coupling strengths $\lambda_{HHH} = \kappa_{HHHH} = \lambda$. The vacuum expectation value of the Higgs field is related to the Fermi constant and its value can be inferred from measurements of the muon decay: $v \simeq 246$ GeV. The perturbative expansion described above implies the following relation between the coupling strengths associated with the Higgs self-interaction vertices, the mass of the Higgs boson and the vacuum expectation value:

$$\lambda_{HHH} = \kappa_{HHHH} = \frac{m_H^2}{2v^2} \,. \tag{3}$$

Measurements of the Higgs tri-linear and quartic interactions would provide constraints on the shape of the Higgs potential close to the minimum. Measurements of the strengths of the Higgs boson self-interactions and their comparison to SM predictions are necessary to verify the electroweak symmetry breaking mechanism of the SM. The existence of an extended scalar sector or the presence of new dynamics at higher scales could modify the Higgs boson self-couplings.

The parameters $\lambda_{HHH}$ and $\kappa_{HHHH}$ can be constrained via experimental studies of final states arising from the production of two and three Higgs bosons, respectively. However, the corresponding production cross-sections are significantly smaller than those for the production of single Higgs bosons. The production of $HH$ pairs through gluon-gluon fusion (ggF) has an expected cross-section of $36.69^{+2.1\%}_{-4.9\%}$ fb at 14 TeV [19]. Even with the higher LHC centre-of-mass energy of 14 TeV and the integrated luminosity expected to be reached by the end of the LHC era, a meaningful extraction of $\kappa_{HHHH}$ is impossible, but the observation of Higgs-boson-pair production and a determination of $\lambda_{HHH}$ appear to be feasible.

This note presents prospects for studies of Higgs-boson-pair production at the High Luminosity LHC (HL-LHC), assuming an integrated luminosity of 3000 fb$^{-1}$, and using three decay channels: $b\bar{b}b\bar{b}$, $b\bar{b}\gamma\gamma$ and $b\bar{b}\tau^+\tau^-$. The branching fractions of $H \to b\bar{b}$, $H \to \gamma\gamma$ and $H \to \tau^+\tau^-$ are 0.5824, $2.27 \times 10^{-3}$, and $6.272 \times 10^{-2}$, respectively [20]. These SM branching fractions are assumed for all the results presented. Only the dominant production mode via gluon-gluon fusion is examined. Scans are performed over a range of possible tri-linear self-coupling strengths, measured relative to the SM expectation and denoted by $\kappa_\lambda = \lambda_{HHH}/\lambda^{SM}_{HHH}$.

The current Run 2 measurements of the Higgs-boson-pair production are performed with approximately 36 fb$^{-1}$ of data [21, 22], combining different decay channels. The ATLAS collaboration reports the combined observed (expected) limit on the non-resonant Higgs-boson-pair production cross-section of 6.7 (10.4) times the SM expectation. The ratio of the Higgs boson self-coupling to its SM expectation is observed (expected) to be constrained at 95% CL to $-5.0 < \kappa_\lambda < 12.1$ ($-5.8 < \kappa_\lambda < 12.0$). The reported combined observed (expected) limit on the non-resonant Higgs-boson-pair production cross-section by the CMS collaboration is 22.2 (12.8) times the predicted Standard Model cross-section. The ratio of the Higgs boson self-coupling to its SM expectation is observed (expected) to be constrained at 95% CL to $-11.8 < \kappa_\lambda < 18.8$ ($-7.1 < \kappa_\lambda < 13.6$).

The expected performance for the $b\bar{b}b\bar{b}$ and $b\bar{b}\tau^+\tau^-$ channels at the HL-LHC is assessed through extrapolation of measurements [23, 24] performed by the ATLAS Collaboration using 24.3 fb$^{-1}$ and 36.1 fb$^{-1}$ of data, respectively, recorded at $\sqrt{s}$ = 13 TeV. The expected performance for the $b\bar{b}\gamma\gamma$ channel is assessed through the use of truth-level Monte Carlo (MC) samples. These MC samples have been adjusted with parameterised functions to estimate the response of the upgraded ATLAS detector at the HL-LHC. The average number of additional proton–proton interactions (pile-up) per bunch crossing is assumed to be $<\mu>$ = 200. An 8% improvement in $b$-tagging efficiency is expected for all channels as a result of improvements to the inner tracker (ITk) [25]. This improvement is factored into the $b\bar{b}b\bar{b}$ and $b\bar{b}\tau^+\tau^-$ extrapolations, and it is included in the smearing functions used in the $b\bar{b}\gamma\gamma$ analysis. In this note, the current Run 2 systematic uncertainties are scaled following Ref. [26].

## 2 $HH \to b\bar{b}b\bar{b}$

The projections for the $HH \to b\bar{b}b\bar{b}$ channel presented in this note are extrapolations of the recent results obtained by the ATLAS Collaboration [23] in which 27.5 fb$^{-1}$ of $pp$ collision data recorded at $\sqrt{s}$ = 13 TeV in 2015-2016 are used. These results set a 95% CL upper limit of 147 fb (234 fb expected) on the cross-section of SM-like Higgs-boson-pair production, where both Higgs bosons decay to $b\bar{b}$. This

3

corresponds to 13.0 times (20.7 times) the SM prediction. The CMS Collaboration recently set a 95% CL upper limit of 847 fb (419 fb expected), corresponding to 74.6 (36.9) times the SM prediction (under the same conditions but with an integrated luminosity of 35.9 $\text{fb}^{-1}$) [27].

The extrapolations are set using the same methodology as in Ref. [28], but the analysis upon which they are based is updated in line with the improved Run 2 analysis [23]. The assumption is made that the planned upgrades to the ATLAS detector [25, 29–33] and improvements to reconstruction algorithms will mitigate the effects of higher pile-up, resulting in the same jet reconstruction performance as that achieved in 2015-2016. The $b$-tagging efficiency is increased by 8% per jet, while the efficiency to incorrectly tag $c$- and light-jets remains the same, in order to reflect the expected improvements in $b$-tagging performance afforded by the ITk [25]. Jet transverse momentum (jet-$p_{\text{T}}$) thresholds, determined by the trigger requirement, will likely increase for HL-LHC running. The effect of raising jet thresholds is studied below. Furthermore, the assumption is made that the analysis will be unchanged in terms of selection and statistical interpretation – a rather pessimistic assumption given that the analysis will be improved to use new techniques and optimised to make best use of larger datasets (for example the boosted-decision-tree-based background reweighting approach of Ref. [34]).

The results presented here were included in the Pixel Detector TDR [25] and the TDAQ TDR [30].

## 2.1 Run 2 Analysis

The sensitivity of the $HH \to b\bar{b}b\bar{b}$ process is extrapolated from the analysis performed on 24.3 $\text{fb}^{-1}$ of data, collected during 2016 at a centre-of-mass energy of $\sqrt{s}$ = 13 TeV and documented in Ref. [23]. MC samples simulated at 13 TeV are used in this analysis to model the signal production and the background from $t\bar{t}$ events, as described in Ref. [23]. The SM $HH$ signal sample was generated using MadGraph5_aMC@NLO at next-to-leading order (NLO) in QCD, with the CT10 NLO [35] parton distribution function (PDF) set. Parton showers and hadronization were simulated with Herwig++ [36] using the UEEE5 set of tuned parameters (tune) [37]. The events were reweighted to reproduce the $m_{HH}$ spectrum obtained in Refs. [38, 39], which fully accounts for the finite top-quark mass. The dominant multijet background was modelled using the data-driven techniques described in Ref. [23]. This analysis reconstructs each $b$-quark from the Higgs boson decays as a distinct $R$ = 0.4 anti-$k_t$ jet [40] and has been optimised to search for low-mass and non-resonant Higgs-boson pairs. The mass of the two Higgs boson candidate system ($m_{HH}$) is used as the final discriminant. The remainder of this section briefly summarises this analysis, and the reader is directed to Ref. [23] for more details.

A combination of $b$-jet triggers is used to record events. Events are required to feature either one $b$-tagged jet with transverse momentum $p_{\text{T}} > 225$ GeV, or two $b$-tagged jets, either both satisfying $p_{\text{T}} > 35$ GeV or both satisfying $p_{\text{T}} > 55$ GeV, with different requirements on the $b$-tagging. Some triggers require additional non-$b$-tagged jets. This combination of triggers is 90% efficient for SM non-resonant signal, after the full offline selection described below.

Higgs boson kinematic properties are reconstructed using pairs of $R$ = 0.4 anti-$k_t$ jets built from topological clusters of energy deposits in calorimeters cells [41].

Jets containing $b$-hadrons are identified using a score value computed from a multivariate $b$-tagging algorithm (MV2c10 [42, 43]), which makes use of observables provided by an impact parameter algorithm, an inclusive secondary vertex finding algorithm and a multi-vertex finding algorithm. The $b$-tagging working point of 70% $b$-jet identification efficiency is chosen.

The selection begins with the requirement that the event contains at least four $b$-tagged jets with $p_T >$ 40 GeV and $|\eta| < 2.5$. The four jets with the highest $b$-tagging scores are paired to construct two Higgs boson candidates as described in Ref. [23].

Higgs-boson pairs are reconstructed over a wide range of masses, 200 GeV $\lesssim m_{HH} \lesssim$ 1300 GeV. Event selection criteria that vary as a function of the reconstructed mass are used to enhance the analysis sensitivity across this range. Mass-dependent requirements are made on the transverse momentum of the leading and sub-leading (in $p_T$) Higgs boson candidates. The absolute difference in pseudorapidity between the two Higgs boson candidates is then required to be less than 1.5.

A requirement on the masses of the Higgs boson candidates is made:

$$X_{HH} = \sqrt{\left(\frac{m_{2j}^{\text{lead}} - 120\,\text{GeV}}{0.1 m_{2j}^{\text{lead}}}\right)^2 + \left(\frac{m_{2j}^{\text{subl}} - 115\,\text{GeV}}{0.1 m_{2j}^{\text{subl}}}\right)^2} < 1.6 \quad (4)$$

where $m_{2j}^{\text{lead}}$ and $m_{2j}^{\text{subl}}$ are the masses of the leading and subleading Higgs boson candidates, respectively. The $0.1m_{2j}$ terms represent the expected experimental resolution of the Higgs boson candidate mass.

Finally, to reduce the $t\bar{t}$ background, hadronically decaying top-quark candidates are built from any three jets in the event, one of which must be a constituent of a Higgs boson candidate. The event is vetoed if a top-quark candidate is found where the reconstructed top-quark and $W$ boson masses are sufficiently close to their nominal values.

The acceptance times efficiency of the full event selection, including the trigger requirement, is 1.6%. The final analysis discriminant is the invariant mass of the selected four-jet system, $m_{HH}$, after a correction based on the known Higgs boson mass, where each Higgs boson candidate's four-momentum is multiplied by a correction factor $m_H/m_{2j}$.

After the full event selection, about 95% of the background consists of multijet events. It is difficult to model this background accurately using MC simulation, partly due to the complexity arising from the large number of multijet processes contributing to this background, but mainly due to the need for an extremely large number of events as a result of the large cross-section and high background rejection factor. As a result, the multijet background is modelled using a data-driven method, as described in Ref. [23]. Data featuring Higgs boson candidates reconstructed from exactly two $b$-tagged and two non-$b$-tagged jets (and satisfying all other selection criteria) are reweighted to represent the background in the signal region. The weights are derived in high-statistics control regions defined using mass sidebands in the $m_{2j}^{\text{lead}} - m_{2j}^{\text{subl}}$ plane, as shown in Figure 1.

The remaining 5% of the background originates from $t\bar{t}$ processes. The $m_{HH}$ spectrum is taken from MC simulation, while the event yield is normalised to data. There is negligible background from all other sources – including processes involving single Higgs bosons.

The uncertainties on the signal acceptance comprise: missing higher-order terms in the matrix elements and PDF set, as well as modelling of the underlying event, hadronic showers, initial- and final-state radiation. The total theoretical uncertainty is dominated by the uncertainties associated with the modelling of the initial- and final-state radiation.

The following detector modelling uncertainties are considered for the simulated signal and $t\bar{t}$ background: uncertainties in the jet energy scale (JES) and resolution (JER), uncertainties in the $b$-tagging efficiency and luminosity uncertainties.

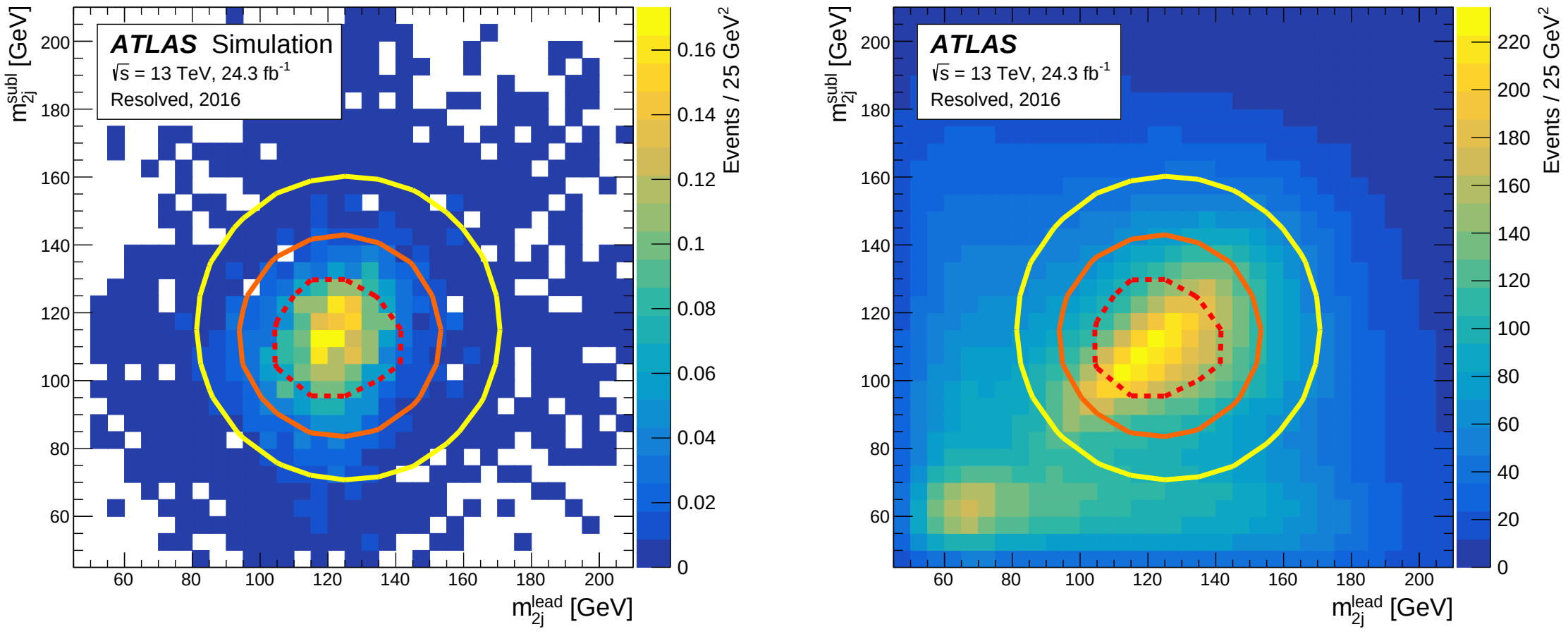

Figure 1: Higgs boson candidate mass-plane regions from the 2016 analysis. The signal region is inside the inner (red) dashed curve, and two control regions are defined outside the signal region: one outside the signal region and within the orange circle and the other between the orange and yellow circles. The left plot shows the distribution of events for the SM non-resonant $HH$ process, and the right plot shows the distribution of events for the estimated multijet background.

Uncertainties in the normalisation of the multijet and $t\bar{t}$ backgrounds are propagated from the fit which determines their yields. Shape uncertainties are assessed by deriving an alternative background model using the same procedure as in the nominal case, but from an independent control region. The differences between the baseline and the alternative models are used as a background-model shape uncertainty, with a two-sided uncertainty defined by symmetrising the difference about the baseline. The uncertainty is split into two components to allow two independent variations: a low-$H_{\mathrm{T}}$ and a high-$H_{\mathrm{T}}$ component, where $H_{\mathrm{T}}$ is the scalar sum of the $p_{\mathrm{T}}$ of the four jets constituting the pair of Higgs boson candidates. The boundary value is 300 GeV. The low-$H_{\mathrm{T}}$ shape uncertainty primarily affects the $m_{HH}$ spectrum below 400 GeV (close to the kinematic threshold) by up to 5%, and the high-$H_{\mathrm{T}}$ uncertainty mainly affects $m_{HH}$ above this by up to 30% relative to nominal. The size of these background normalisation and shape uncertainties are driven by the current statistical precision of the control regions and they were found to be the dominant systematic uncertainties in the analysis of the 2016 dataset.

## 2.2 Extrapolation Method

The statistical framework used to produce the Run 2 results documented in Ref. [23] is extended to assess the sensitivity of the analysis to non-resonant Higgs-boson-pair production with larger datasets. A test statistic based on the profile likelihood ratio [44] is used to test hypothesised values of the global signal strength factor, $\mu = \sigma_{HH}/\sigma_{HH}^{SM}$, for the SM non-resonant $HH$ signal model. Systematic uncertainties are treated using Gaussian or log-normal constraint terms in the definition of the likelihood function. The extended framework is used to produce $m_{HH}$ distributions for the signal and background, which have been modified to represent different integrated luminosities. These distributions are used to derive expected upper limits on the production cross-section for the signal process using a signal-plus-background fit to the background-only $m_{HH}$ distribution. Exclusion limits are based on the value of the statistic $CL_s$ [45], with a value of $\mu$ regarded as excluded at the 95% CL when $CL_s$ is less than 5%. The distributions can also be modified to investigate different assumptions and scenarios, for example assumptions related to the evolution of the systematic uncertainties.

The systematic uncertainties related to detector modelling – JES, JER, $b$-tagging, luminosity – are themselves largely set by the systematic uncertainties of the methods used to determine them. They could be reduced in the future through dedicated studies and development of new techniques, but since no reliable predictions of these improvements can be made, they are treated as constant in this extrapolation. Even with this conservative assumption, these uncertainties have a negligible impact on the analysis sensitivity.

The multijet and $t\bar{t}$ distributions have uncertainties associated with their normalisation and shape that are treated as nuisance parameters in the statistical analysis. In order to use the best models of these backgrounds in the dataset extrapolation process, a signal-plus-background fit is performed to the 2016 data (assuming $\mu = 1$) and the best-fit values of the nuisance parameters extracted. Background distributions are then generated to represent different integrated luminosities, using these best-fit nuisance parameter values. The statistical uncertainties on the data-driven multijet model are set to follow Poisson statistics corresponding to the dataset size, while the systematic uncertainties are left unchanged (additional constraints coming from the fit to 2016 data are ignored). Different assumptions regarding the evolution of the background uncertainties are explored in Section 2.4.

The signal model statistical uncertainty is dictated by MC resource limitations. As these will be substantially improved for the final HL-LHC analyses, the statistical uncertainty on the signal distributions is neglected.

All background distributions are corrected to account for the increase in collision energy from $\sqrt{s} = 13$ TeV to $\sqrt{s} = 14$ TeV. For simplicity this is done by scaling the number of expected events by 1.18, which accounts for the increase in cross-sections due to the change in gluon-luminosity. The signal distributions are normalised to the $\sqrt{s} = 14$ TeV prediction [19]. Possible effects on the $m_{HH}$ shape from the evolution of the PDFs are neglected for this study.

The $m_{HH}$ distributions extrapolated to 3000 fb$^{-1}$ are shown in Figure 2, and retain the same binning that was used in the Run 2 analysis. These scaled distributions are used in the statistical analysis along with those corresponding to systematic uncertainty variations. The signal-to-background ratio for all events is 0.06%, increasing to 0.19% if only events with $m_{HH} > 400$ GeV are considered. The low-$H_T$ uncertainty is constrained to 0.15 of the prior and the high-$H_T$ uncertainty is constrained to 0.03 of the prior. The uncertainty on the overall background yield, when the Run 2 uncertainty is extrapolated to 3000 fb$^{-1}$, is found to be 0.3%.

### 2.3 Results

The expected 95% CL upper limit on the global SM $HH$ signal strength as a function of the integrated luminosity is shown in Figure 3 for the best possible scenario where only statistical uncertainties are considered and for the conservative scenario where the pre-fit uncertainties remain as they were for the analysis of the 2016 dataset. The potential benefit of reducing the pre-fit systematic uncertainties is significant and becomes even more pronounced with larger datasets: the sensitivity with 3000 fb$^{-1}$ of data and the current systematic uncertainties is 2.5 times worse than in the scenario where systematic uncertainties are negligible. If systematic uncertainties were entirely eliminated, the excluded signal strength would be 1.4 with 3000 fb$^{-1}$ of data.

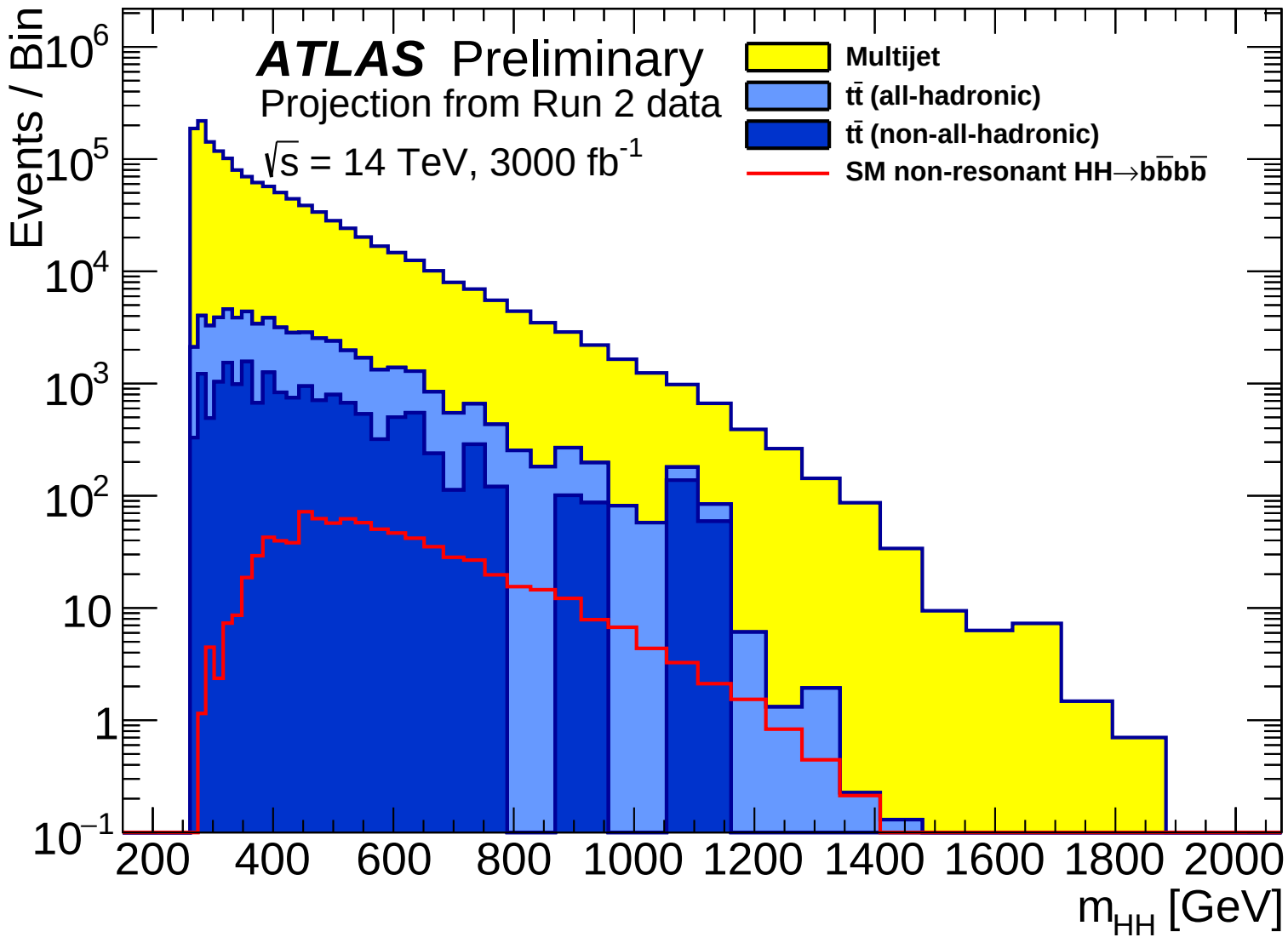

Figure 2: Stacked $m_{HH}$ histograms of the $t\bar{t}$ and multijet backgrounds extrapolated from 24.3 fb$^{-1}$ (the 2016 dataset) to 3000 fb$^{-1}$. The predicted SM non-resonant Higgs-boson-pair production signal is shown as the red line.

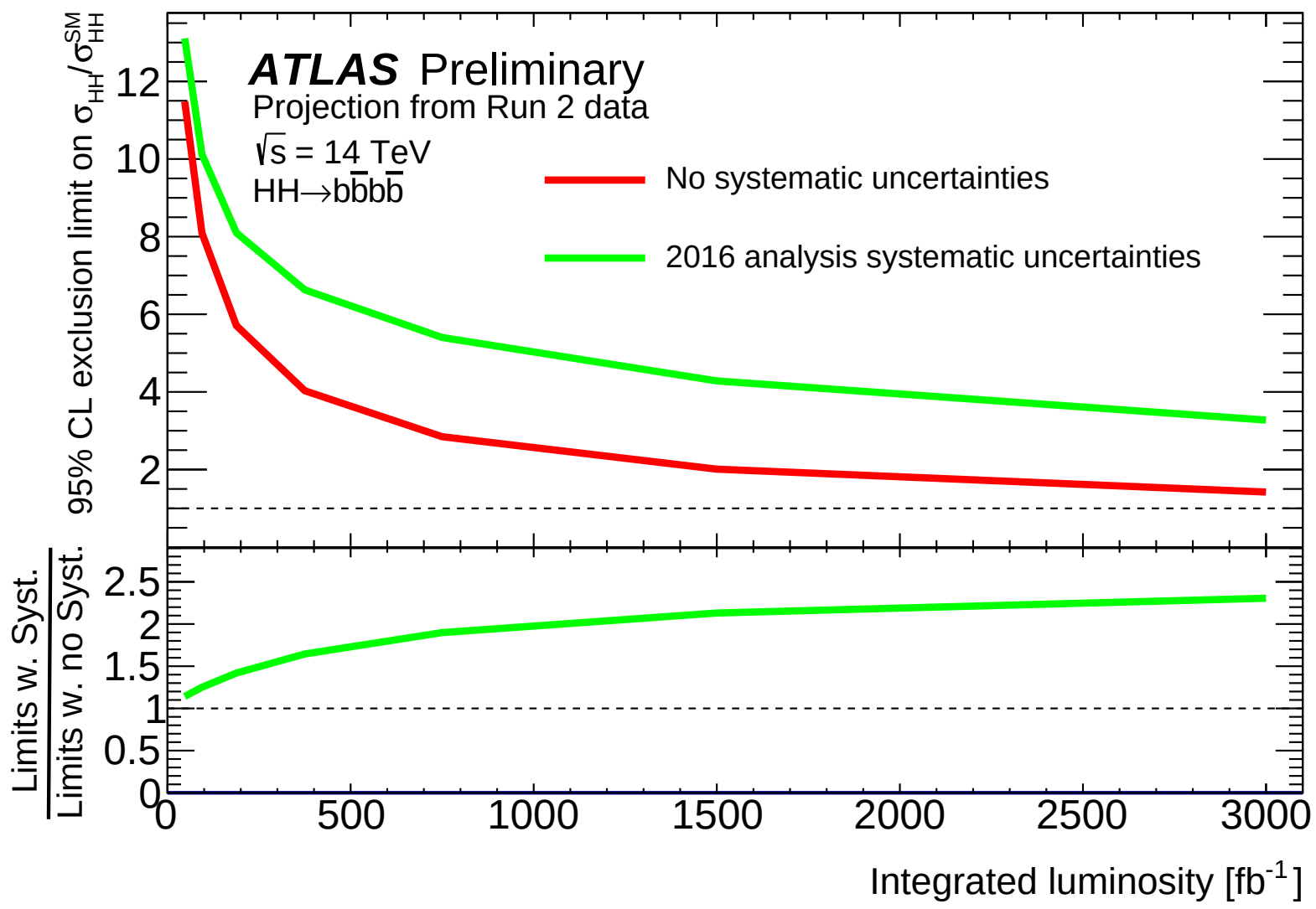

Figure 3: Expected 95% CL upper limit on $\sigma_{HH}/\sigma_{HH}^{SM}$, as a function of the integrated luminosity of the search between 47 and 3000 fb$^{-1}$. The red line shows the upper limit when evaluated without systematic uncertainties, while the green line assumes that the pre-fit systematic uncertainties remain as they were in 2016. The lower panel shows the ratio between these two limits. The extrapolated sensitivity is shown using a jet $p_{\mathrm{T}}$ threshold of 40 GeV.

Figure 4 shows a similar set of plots for the projected significance of the discrepancy from the background-only hypothesis. The potential benefit of reducing the systematic uncertainties is once again highlighted. If systematic uncertainties were entirely eliminated, the significance would be 1.4$\sigma$. The corresponding probability, $p_0$, to observe the predicted SM $HH$ signal because of fluctuations in the background is 0.0825. Keeping the current experimental systematic uncertainties, this becomes $0.62\sigma$, corresponding to a $p_0$-value of 0.269.

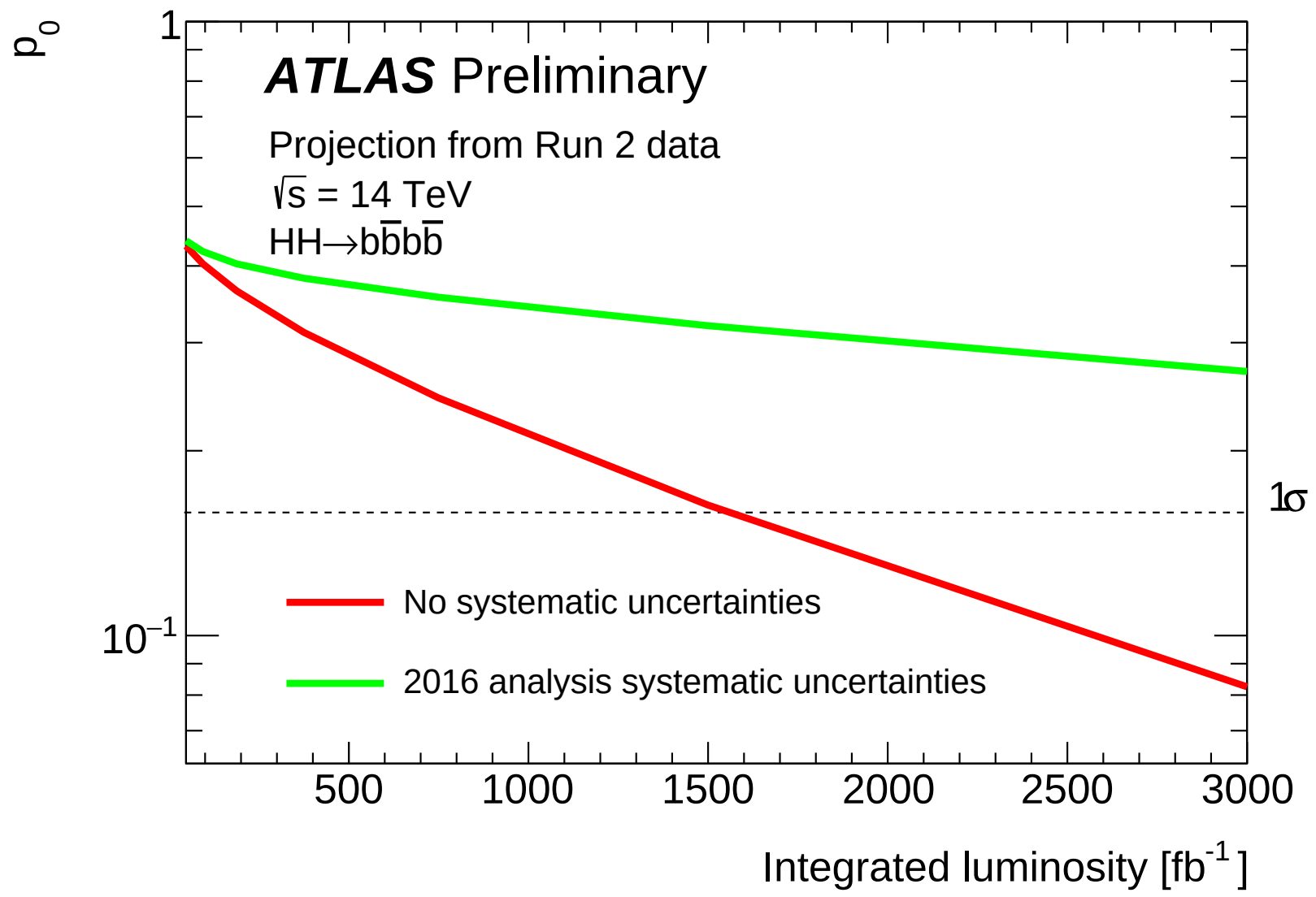

Figure 4: Expected significance, $p_0$, of the discrepancy from the background-only hypothesis when the predicted SM $HH$ signal is injected, as a function of the integrated luminosity of the search between 47 and 3000 fb$^{-1}$. The red line shows $p_0$ when evaluated without systematic uncertainties, while the green line assumes that the pre-fit systematic uncertainties remain as they were in 2016. The black dashed line shows the $p_0$ value corresponding to a signal significance of $1\sigma$. The extrapolated sensitivity is shown using a jet $p_T$ threshold of 40 GeV.

With 3000 fb$^{-1}$ of data, a degradation of the $b$-tagging efficiency to its Run 2 performance (assessed by removing the 8% improvement that was applied earlier) would result in a relative decrease by 10% (20%) of the significance for the case of current (no) systematic uncertainty.

To assess the sensitivity of the $HH \to b\bar{b}b\bar{b}$ analysis to $\kappa_\lambda$, samples for three different values of $\kappa_\lambda$ ($\kappa_\lambda$ = 0, 1 and 10) are produced using the leading-order (LO) version of MadGraph5_aMC@NLO [46] generator with the NNPDF 2.3 LO [47] PDF set and Pythia 8 [48] for the showering model. Using these three samples, the generator-level $m_{HH}$ distributions for $-20 \leq \kappa_\lambda \leq 20$ were then produced using the morphing technique documented in Ref. [49]. The binned ratios of $m_{HH}$ distributions for any given $\kappa_\lambda$ value to the SM case ($\kappa_\lambda = 1$) are computed at LO and then applied to the fully simulated SM NLO sample to reweight it to different $\kappa_\lambda$ values. In this way the SM NLO QCD and the finite top-quark mass corrections [38, 39] are applied for all $\kappa_\lambda$ signals, i.e. they are assumed to be independent of $\kappa_\lambda$.

An Asimov dataset with the predicted SM $HH$ signal was created, and maximum likelihood fits to this dataset were performed with different $\kappa_\lambda$ hypotheses. The negative natural logarithm of the ratio of the maximum likelihood for $\kappa_\lambda$ to that for the fit with $\kappa_\lambda = 1$ is shown in Figure 5. The $1\sigma$ and $2\sigma$ confidence interval (CI) constraints on $\kappa_\lambda$ from these curves, when considering only statistical uncertainties and when considering current pre-fit systematic uncertainties, are shown in Table 1.

The negative natural logarithm of the ratio of the maximum likelihood for $\kappa_\lambda$ to that for the fit with $\kappa_\lambda = 0$ is shown in Figure 6. The confidence intervals on $\kappa_\lambda$ are reported in Table 2

It can be seen in Figures 5 and 6 that there are typically two minima. The first minimum is located at $\kappa_\lambda = 1$ or 0, respectively, as the signal hypothesis used in the maximum likelihood fit corresponds to the signal used to create the Asimov dataset. The second minimum is observed at a $\kappa_\lambda$ value that corresponds to a similar fitted signal yield with respect to the $\kappa_\lambda$ point at the first minimum, which is a consequence of a higher cross-section, but lower acceptance and worse signal-to-background separation. The degeneracy is lifted because of the different shapes of the $m_{HH}$ distribution at each of the two minima, especially after including the systematic uncertainties.

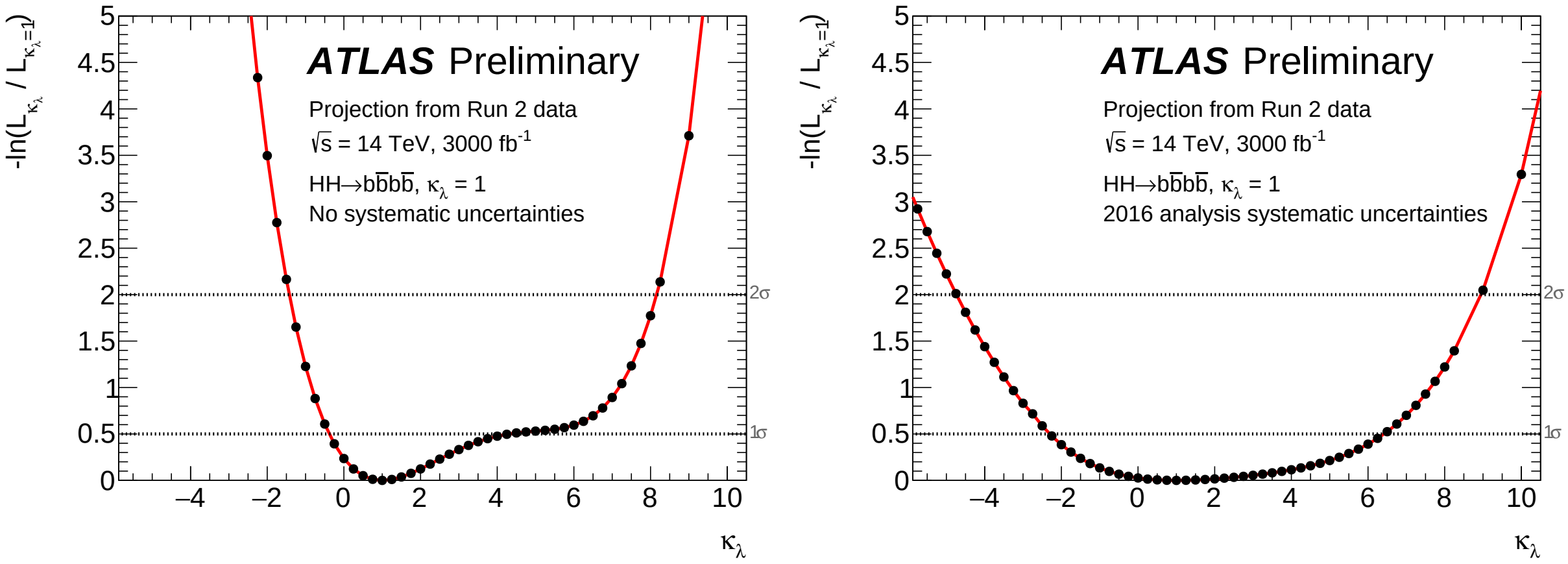

Figure 5: Negative natural logarithm of the ratio of the maximum likelihood for $\kappa_\lambda$ to the maximum likelihood for $\kappa_\lambda = 1$ for (left) the fits with only statistical uncertainties and (right) the fits with current systematic uncertainties as nuisance parameters. These extrapolated likelihood curves are produced with a jet $p_T$ threshold of 40 GeV. The dashed lines at $-\ln(L_{\kappa_\lambda}/L_{\kappa_\lambda=1}) = 0.5$ and 2.0 indicate the values corresponding to a $1\sigma$ and $2\sigma$ confidence interval, respectively (assuming an asymptotic $\chi^2$ distribution of the test statistic).

| Scenario | $1\sigma$ CI | $2\sigma$ CI |
|---|---|---|
| No systematic uncertainties | $-0.4 \le \kappa_\lambda \le 4.3$ | $-1.4 \le \kappa_\lambda \le 8.2$ |
| 2016 analysis systematic uncert. | $-2.3 \le \kappa_\lambda \le 6.4$ | $-4.7 \le \kappa_\lambda \le 8.9$ |

Table 1: Constraints on $\kappa_\lambda$ from the likelihood ratio test performed on the Asimov dataset created from the backgrounds and the SM $HH$ signal, as shown in Figure 5. Results are presented as a $1\sigma$ and $2\sigma$ CI on $\kappa_\lambda$ when considering only statistical uncertainties and when considering current pre-fit systematic uncertainties.

| Scenario | $1\sigma$ CI | $2\sigma$ CI |
|---|---|---|
| No systematic uncertainties | $-1.1 \le \kappa_\lambda \le 1.6$ | $-2.0 \le \kappa_\lambda \le 8.4$ |
| 2016 analysis systematic uncert. | $-2.8 \le \kappa_\lambda \le 6.1$ | $-5.4 \le \kappa_\lambda \le 9.0$ |

Table 2: Constraints on $\kappa_\lambda$ from the likelihood ratio test performed on the Asimov dataset created from the backgrounds and the $\kappa_\lambda = 0$ signal, as shown in Figure 6. Results are presented as a $1\sigma$ and $2\sigma$ CI on $\kappa_\lambda$ when considering only statistical uncertainties and when considering current pre-fit systematic uncertainties.

## 2.4 Impact of Reducing Background Modelling Uncertainties

Systematic uncertainties related to the data-driven background modelling are dominant, with the other sources (theoretical or detector modelling uncertainties) leading to a negligible change in the results.

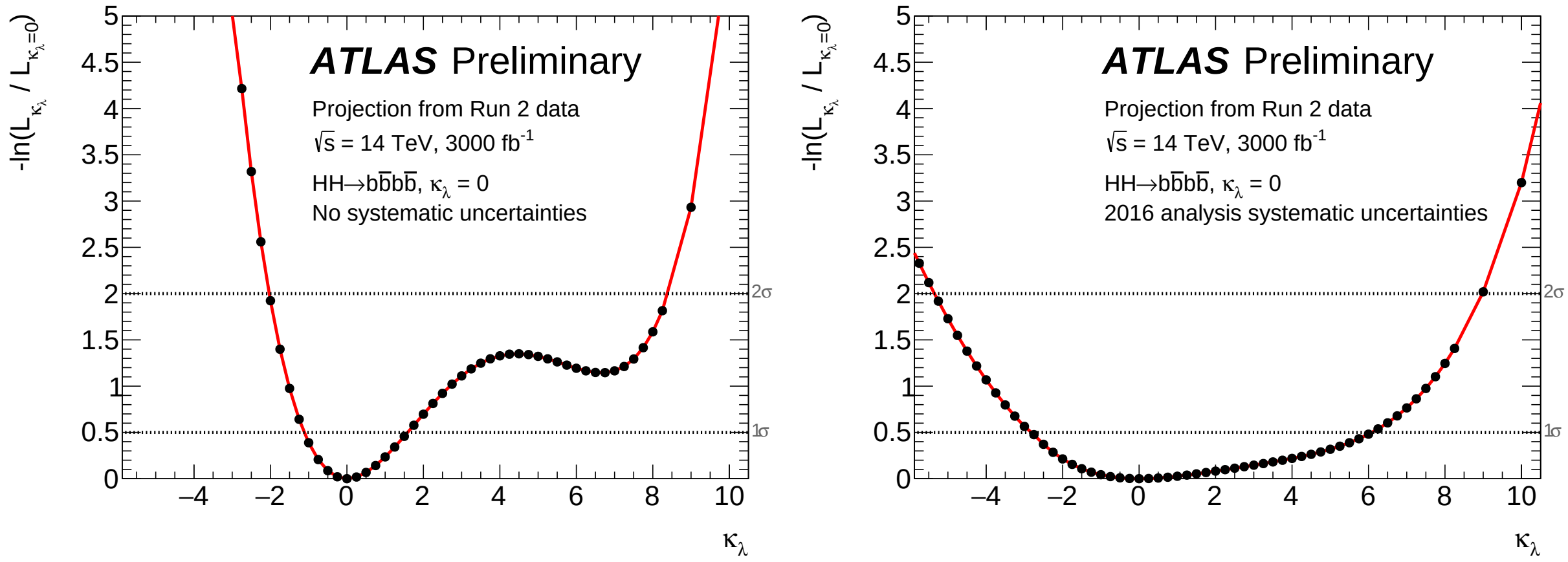

Figure 6: Negative natural logarithm of the ratio of the maximum likelihood for $\kappa_\lambda$ to the maximum likelihood for $\kappa_\lambda = 0$ for (left) the fits with only statistical uncertainties and (right) the fits with current systematic uncertainties as nuisance parameters. These extrapolated likelihood curves are produced with a jet $p_{\mathrm{T}}$ threshold of 40 GeV. The dashed lines at $-\ln(L_{\kappa_\lambda}/L_{\kappa_\lambda=0}) = 0.5$ and 2.0 indicate the values corresponding to a $1\sigma$ and $2\sigma$ confidence interval, respectively (assuming an asymptotic $\chi^2$ distribution of the test statistic).

The impact of potential reductions in the background modelling uncertainties is shown in Figure 7. The multijet background modelling uncertainties were determined in 2016 by examining the agreement between the background models derived in two different regions of the plane defined by the masses of the two Higgs boson candidates. The uncertainties were essentially limited by the statistical precision of these comparisons. As more data is accumulated, the statistical precision of these comparisons will increase and a reduction in the modelling uncertainties should be possible. Hence, the limit achievable for the case where the background uncertainties are scaled according to the integrated luminosity is also shown as the star in Figure 7.

## 2.5 Effect of Minimum Jet $p_{\mathrm{T}}$ Thresholds

The high number of pile-up events at HL-LHC cause difficulties in maintaining high acceptance when triggering on multijet final states. Jets produced in the pile-up events cause high trigger rates, potentially necessitating a rise in jet $p_{\mathrm{T}}$ thresholds, which is exacerbated by the deterioration in trigger $p_{\mathrm{T}}$ turn-on curves caused by the additional soft energy deposited in the calorimeters.

The impact of increasing the multijet trigger $p_{\mathrm{T}}$ thresholds has been examined by repeating the analysis using different minimum jet $p_{\mathrm{T}}$ requirements on the constituent jets of the Higgs boson candidates. The expected 95% CL upper limit on $\sigma_{HH}/\sigma_{HH}^{SM}$, as a function of the minimum jet $p_{\mathrm{T}}$ required is shown in Figure 8. Ref. [50] proposes a trigger menu with a multijet trigger that requires four jets, all satisfying a minimum $p_{\mathrm{T}}$ threshold equivalent to demanding $p_{\mathrm{T}} > 75$ GeV for jets reconstructed offline. As can be seen in Figure 8, this high minimum jet $p_{\mathrm{T}}$ threshold would have a significant negative impact on the sensitivity of the $HH \to b\bar{b}b\bar{b}$ analysis. In the scenario with no systematic uncertainties, this degrades the sensitivity by almost 50% relative to the current analysis threshold of $p_{\mathrm{T}} > 40$ GeV and is equivalent to halving the integrated luminosity of the final dataset. In the scenario with the current systematic uncertainties, the limits are 2.4 times higher, equivalent to reducing the dataset to only 200 fb$^{-1}$. The larger degradation in this case can be quantitatively explained by the background systematic uncertainty being currently dominated by the statistical precision of the comparison between the two models; so increasing the jet $p_{\mathrm{T}}$ threshold results in a loss of sensitivity from diminishing the precision of that comparison, as well as the expected hit from reduced signal region yields.

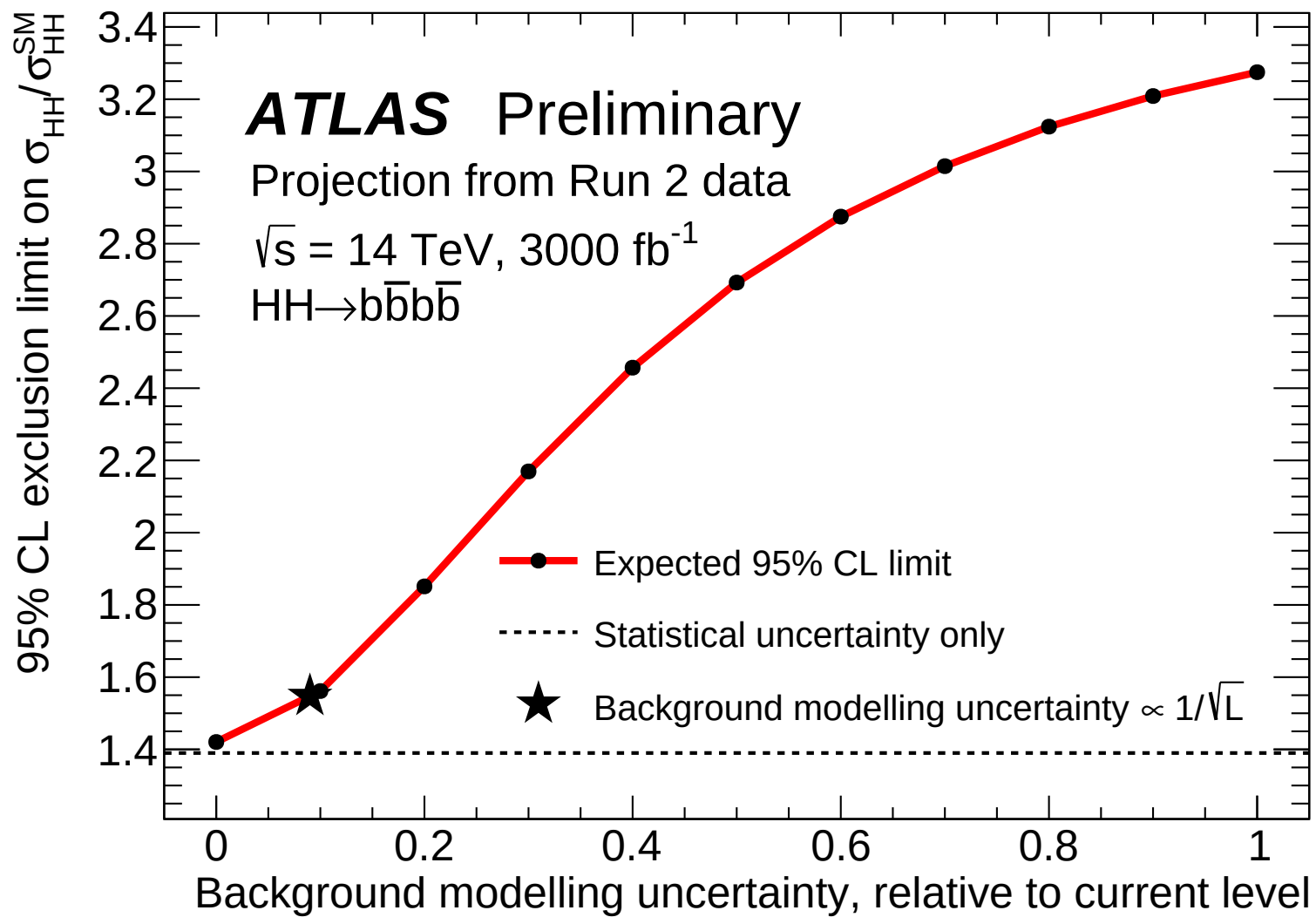

Figure 7: Expected 95% CL upper limit on $\sigma_{HH}/\sigma_{HH}^{SM}$, as a function of the pre-fit background modelling uncertainties, which are each scaled by a common, constant factor relative to the corresponding uncertainty in the Run 2 analysis (i.e. the uncertainties of the analysis of the 2016 dataset correspond to 1 here). The limit achievable assuming that the overall uncertainty scales with luminosity as $1/\sqrt{L}$ is shown by the star point. The limit obtained when considering only statistical uncertainties is shown as the dashed line. The extrapolated sensitivities are calculated assuming a jet $p_{\mathrm{T}}$ threshold of 40 GeV.

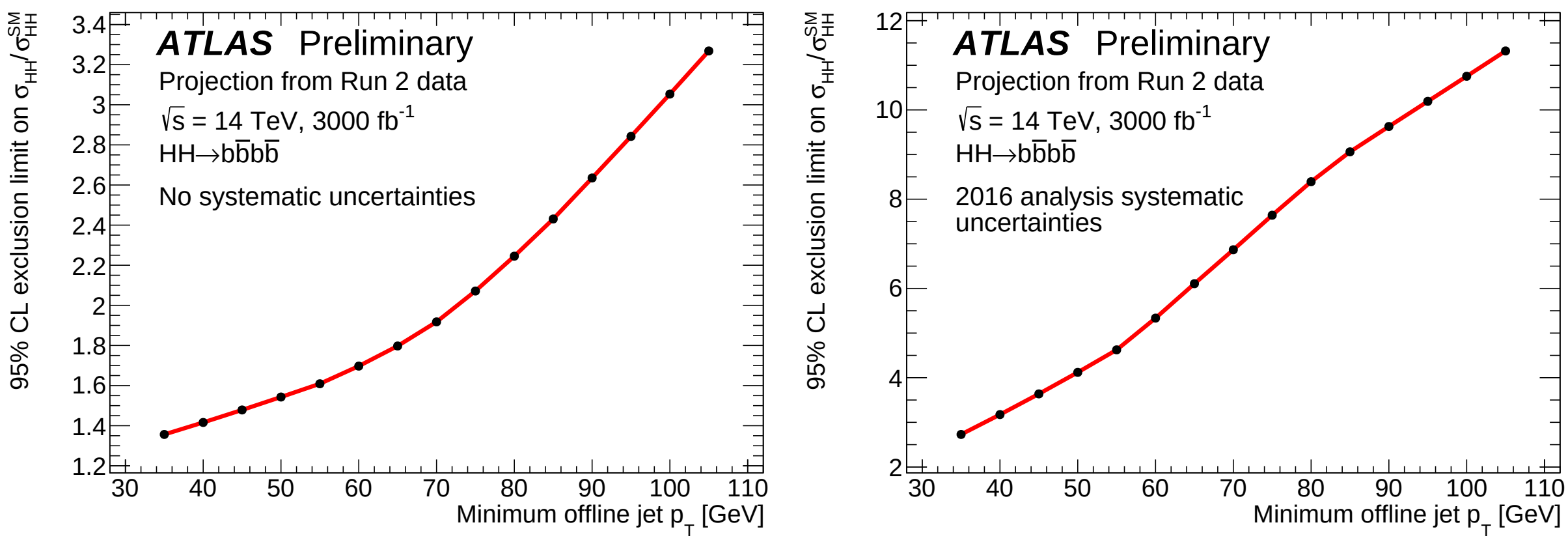

Figure 8: Expected 95% CL upper limit on $\sigma_{HH}/\sigma_{HH}^{SM}$, as a function of the minimum jet $p_{\mathrm{T}}$ required for the four Higgs boson candidate constituent jets. The left plot shows the case where only statistical uncertainties are considered, while the right plot includes the pre-fit systematic uncertainties as they were in 2016.

# 3 $HH \to b\bar{b}\tau^+\tau^-$

The ATLAS $HH \to b\bar{b}\tau^+\tau^-$ analysis [24] currently sets the world's strongest limit by a single channel on Higgs-boson-pair production. The results, obtained using 36.1 $\text{fb}^{-1}$ of $pp$ collision data at $\sqrt{s} = 13$ TeV collected during the LHC Run 2 in 2015 and 2016, constrain the production cross-section for non-resonant Higgs-boson-pair production to be less than 12.7 times the SM prediction, at 95% CL. The expected limit is 14.8 times the SM prediction. The CMS experiment excludes cross-sections greater than 31.4 times the SM prediction, while the expected limit is 25.1 times the SM prediction, under the same experimental conditions and for a similar integrated luminosity [51].

The results from the ATLAS Run 2 (2015+2016) analysis are extrapolated to $\sqrt{s} = 14$ TeV and the HL-LHC target integrated luminosity to obtain an estimate of the expected sensitivity of the $HH \to b\bar{b}\tau^+\tau^-$ channel to SM $HH$ production and $\lambda_{HHH}$ at the HL-LHC. The $\tau_{\text{lep}}\tau_{\text{had}}$ and $\tau_{\text{had}}\tau_{\text{had}}$ decay channels are considered, where the subscripts (lep = electron or muon, had = hadrons) indicate the decay mode of the $\tau$-lepton. The analysis strategy is briefly summarised below for convenience, Ref. [24] should be consulted for full details.

## 3.1 Run 2 Analysis

Events are characterised by the presence of either an electron or muon plus hadronic $\tau$-object ($\tau_{\text{had-vis}}$), or two $\tau_{\text{had-vis}}$, as well as two $b$-tagged jets, and $E_{\text{T}}^{\text{miss}}$ from neutrinos produced in the $\tau$-lepton decay. Events in the $\tau_{\text{lep}}\tau_{\text{had}}$ channel are required to pass a single-lepton trigger (SLT) or lepton-plus-$\tau_{\text{had-vis}}$ trigger (LTT). The electron or muon that passes the SLT condition is required to have $p_{\text{T}} > 26$ GeV at trigger level. Events that fail this requirement are considered for the LTT category if the electron (muon) has $p_{\text{T}} > 17$ (14) GeV. In the $\tau_{\text{had}}\tau_{\text{had}}$ channel events are required to pass either a single-$\tau_{\text{had-vis}}$ trigger (STT) with the leading $\tau_{\text{had-vis}}$ required to have $p_{\text{T}} > 80 - 160$ GeV (depending on the data-taking period of the Run 2 analysis) or a di-$\tau_{\text{had-vis}}$ trigger (DTT) with the leading (sub-leading) $\tau_{\text{had-vis}}$ required to have $p_{\text{T}} > 35$ (25) GeV. An additional jet with online $p_{\text{T}} > 25$ GeV is required at Level 1 for LTT and DTT events in order to keep manageable rates.

In the SLT and LTT channels, the electron or muon is required to have a minimum $p_{\text{T}}$ of at least 1 GeV more than the trigger threshold. In SLT events the $\tau_{\text{had-vis}}$ is required to have $p_{\text{T}} > 20$ GeV, while in the LTT channel the $\tau_{\text{had-vis}}$ is required to have $p_{\text{T}} > 30$ GeV. In the STT channel the leading $\tau_{\text{had-vis}}$ is required to have $p_{\text{T}} > 100 - 180$ GeV (depending on the data-taking period of the Run 2 analysis) and the sub-leading $\tau_{\text{had-vis}}$ is required to have $p_{\text{T}} > 20$ GeV. In the DTT channel the $p_{\text{T}}$ thresholds for the leading and sub-leading $\tau_{\text{had-vis}}$ are 40 GeV and 30 GeV, respectively. In all cases, the $\tau$-lepton decay products are required to have opposite-sign charge and events with additional electrons, muons or $\tau_{\text{had-vis}}$ are rejected. Signal region (SR) events are required to have two $b$-tagged jets, where the leading (sub-leading) jet $p_{\text{T}}$ is required to be at least 45 (20) GeV. In LTT and DTT channels, the leading jet $p_{\text{T}}$ threshold is raised to 80 GeV in order to be fully efficient with respect to the Level 1 requirements. The invariant mass of the di-$\tau$-lepton system, $m_{\tau\tau}^{\text{MMC}}$, is calculated using the Missing Mass Calculator [52], and is required to be greater than 60 GeV.

The dominant SM background processes are $t\bar{t}$, multijet and $Z$ bosons produced in association with two jets originating from heavy-flavour quarks ($bb$, $bc$, $cc$), subsequently referred to as $Z$+heavy-flavour[1]. The SM Higgs-boson production is also an important background process, in particular when the Higgs boson is produced in association with a $Z$ boson and the system subsequently decays into a $b\bar{b}\tau^+\tau^-$ final state. To model the SM $HH$ signal and background processes containing hadronic $\tau$-lepton decays MC samples with a fully simulated detector response are used. The SM $HH$ signal events were generated using MadGraph5_aMC@NLO at NLO in QCD, using the same setup used to generate the SM $HH \to b\bar{b}b\bar{b}$ signal events, which is described in Section 2.1. The events were reweighted to reproduce the $m_{HH}$ spectrum obtained in Refs. [38, 39], which fully accounts for the finite top-quark mass. The complete list of MC samples used in the analysis can be found in Ref. [24]. Contributions from processes in which a quark- or gluon-initiated jet is misidentified as a $\tau_{\text{had-vis}}$ candidate (fake-$\tau_{\text{had-vis}}$) are estimated using data-driven methods.

A Boosted Decision Tree (BDT) classification is used to separate the signal from background processes. In all channels the BDT uses the mass of the Higgs-boson pair, $m_{\tau\tau}^{\text{MMC}}$, the di-$b$-jet invariant mass, the angular distance $\Delta R \equiv \sqrt{(\Delta\eta)^2 + (\Delta\phi)^2}$ between the two visible $\tau$-lepton decay products, and the $\Delta R$ between the two $b$-jets. In the $\tau_{\text{lep}}\tau_{\text{had}}$ channel the transverse mass between the electron or muon and the $E_{\text{T}}^{\text{miss}}$ is also used. In the $\tau_{\text{had}}\tau_{\text{had}}$ channel the $E_{\text{T}}^{\text{miss}}\phi$ centrality that quantifies the angular position of the $E_{\text{T}}^{\text{miss}}$ relative to the visible $\tau$-lepton decay products in the transverse plane is used in the BDT as well. In the $\tau_{\text{lep}}\tau_{\text{had}}$ channel the BDT is trained against the dominant $t\bar{t}$ background, while in the $\tau_{\text{had}}\tau_{\text{had}}$ channel it is trained against $t\bar{t}$, $Z/\gamma^* \to \tau^+\tau^-$, and multijet processes.

## 3.2 Extrapolation Procedure

The expected HL-LHC sensitivity to a SM $HH$ signal in the $b\bar{b}\tau^+\tau^-$ final state is estimated by extrapolating the current Run 2 result [24] using the same statistical framework, based on the profile likelihood ratio [44]. This framework is used to produce modified BDT distributions for signal and background representing different target luminosities. These distributions are used to set upper limits on the signal strength, $\sigma_{HH}/\sigma_{HH}^{SM}$ (the $HH$ production cross-section relative to the SM prediction), using the $CL_{\text{s}}$ method [45]. The expected discovery significance assuming a SM $HH$ signal is also quoted. Various extrapolation assumptions are investigated.

The Run 2 BDT distributions for signal and background are scaled to luminosities up to 3000 fb$^{-1}$ by a single multiplicative factor, defined as the ratio of the target luminosity to the luminosity of the Run 2 result. In the baseline scenario, the uncertainty on the luminosity is assumed to be 1%. The performance of the HL-LHC detector is assumed to be broadly similar to that of the current detector, with the exception that the $b$-tagging efficiency is expected to improve by 8% for a given $c$- and light-jet rejections due to the upgraded inner tracker, after taking into account effects from high pile-up conditions at the HL-LHC. This is taken into account by conservatively assuming all backgrounds have two $b$-initiated jets and hence scaling the backgrounds and the signal by $(1.08)^2$. Although there is potential to reduce the systematic uncertainties related to modelling of the detector response by the methods employed, the Run 2 values are used in this study as they are currently limited by the methods rather than statistical precision.

The increase in centre-of-mass energy from $\sqrt{s}$ = 13 TeV to $\sqrt{s}$ = 14 TeV is accounted for by scaling the number of expected signal and background events (estimated using simulations at $\sqrt{s}$ = 13 TeV) by the ratio of the corresponding production cross-sections. In the case of the $HH$ signal and the single-Higgs-boson backgrounds, the latest LHC Higgs Cross-Section Working Group [19] recommendations are used; for other backgrounds a single factor of 1.18 is applied to account for the approximate cross-section increase arising from the enhanced gluon-luminosity. The effect on the shape of several kinematic and BDT input variables was checked at truth level for both signals and major backgrounds. Since no significant difference was found, the shape of the BDT distributions is assumed to be unchanged. The theoretical uncertainty on the signal cross-section is not taken into account for the presented results, but its potential impact on the expected upper limit on the signal strength was estimated and found to be negligible compared to other systematic uncertainties.

[1] Equivalently, $Z$ bosons produced in association with at least one light-flavour quark ($u$, $d$ or $s$) or gluon are referred to as $Z$+light-flavour.

In the Run 2 analysis, the binning used for the fit is determined by checking that the pre-fit relative background statistical uncertainty in each bin of the BDT output score is less than 0.5 (0.4) times the fraction of the signal in that bin for the $\tau_{\text{had}}\tau_{\text{had}}$ ($\tau_{\text{lep}}\tau_{\text{had}}$) channel. Additionally, the number of expected background events in each bin is required to be greater than 5 (10) for the $\tau_{\text{had}}\tau_{\text{had}}$ ($\tau_{\text{lep}}\tau_{\text{had}}$) channel. The binning criteria for the extrapolation is kept the same, but the binning itself changes due to the scaling of expected background events. Therefore, the statistical sensitivity improves beyond what is expected from just an increase in cross-section and luminosity.

In the Run 2 analysis, the normalisation of the simulated background samples from $t\bar{t}$ and $Z$+heavy-flavour production is allowed to freely float in the final profile-likelihood fit, as described in Ref. [24]. For simplicity, in the extrapolation the $Z$+heavy-flavour background is scaled by the normalisation factor of 1.34 obtained in the Run 2 analysis, while the $t\bar{t}$ normalisation is taken from simulation since its Run 2 normalisation factor is consistent with unity. The relative uncertainty on the normalisation is incorporated in the fit as a nuisance parameter for both backgrounds. Since the normalisations are obtained from a fit to data in a control region, their uncertainty is of statistical nature and therefore expected to be reduced. The extrapolation takes into account the effective increase in luminosity due to the larger dataset, the higher centre-of-mass energy, and the improvement in the $b$-jet identification described above. This increase leads to a reduction factor of approximately 10 in $t\bar{t}$ and $Z$+heavy-flavour normalisation uncertainty at 3000 fb$^{-1}$ for the baseline scenario. In the same scenario, the uncertainties accounting for acceptance differences between the control regions and the various signal regions are taken to be 50% of those derived in the Run 2 analysis, based on the assumption that the theoretical modelling of these processes will be improved.

Background processes with a quark- or gluon-initiated jet misidentified as a $\tau_{\text{had-vis}}$ were derived from data using fake-factor and fake-rate methods in the Run 2 analysis. In order to take into account the expected increase in statistics with higher luminosity, the statistical uncertainties on the data-driven fake-$\tau_{\text{had-vis}}$ background model in the baseline scenario are adjusted to follow Poisson statistics corresponding to the target dataset size; the systematic uncertainties are left unchanged. The binning of the BDT discriminant is conservatively always determined using the unscaled relative background MC statistical uncertainty.

In the Run 2 analysis, the uncertainty on the background due to single-Higgs-boson production in the $ZH$ and $t\bar{t}H$ modes was set to approximately the current measured ATLAS uncertainty on the signal strength in the relevant channel, i.e. to 28% and 30% respectively. In order to account for the expected improvement in these measurements with increased luminosity, the baseline scenario instead assumes 5% and 10% respectively, corresponding approximately to the recommended theory uncertainties [19]. The uncertainties on all other minor backgrounds are halved in this scenario. Finally, in the baseline scenario, the MC statistical uncertainties are neglected under the assumption that the size of the MC samples will increase significantly in line with the data luminosity.

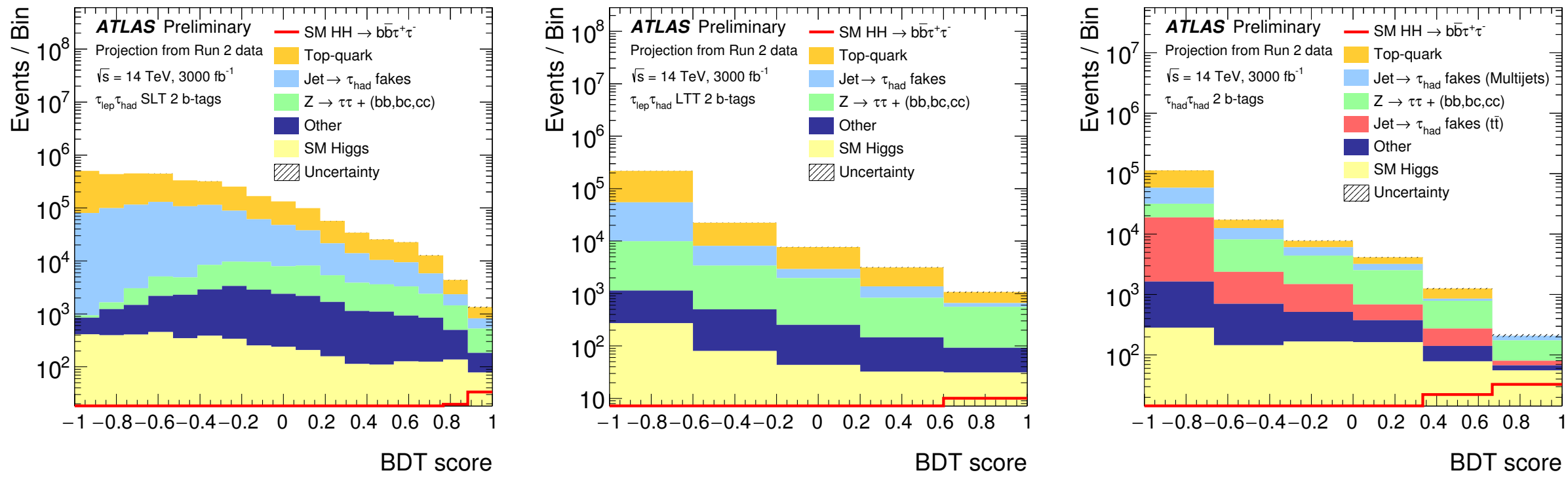

Figure 9: Distributions of the BDT score for the $\tau_{\text{lep}}\tau_{\text{had}}$ channel in the SLT category (left), $\tau_{\text{lep}}\tau_{\text{had}}$ channel in the LTT category (middle) and $\tau_{\text{had}}\tau_{\text{had}}$ channel (right). The background distributions are shown after the fit based on a background-only Asimov dataset and the signal is scaled to the SM prediction. The hatched bands represent the combined statistical and systematic uncertainty for the baseline scenario. These uncertainty bands are included in the plots for completeness but are very small.

In addition to the baseline scenario, an alternative conservative extrapolation is performed. Here, all systematic uncertainties are set to their Run 2 values unless otherwise stated. Analogously, the statistical uncertainties on the data-driven fake-$\tau_{\text{had-vis}}$ background are also set to their Run 2 value in this case. This extrapolation is split into two scenarios, one where the Run 2 MC statistical uncertainties are conservatively adopted and the other where they continue to be neglected. In all scenarios, additional constraints on the systematic uncertainties coming from the Run 2 fit are not taken into account. In order to show the ultimate limit of the expected performance, a final extrapolation is performed neglecting all systematic uncertainties, including the MC statistical uncertainty.

## 3.3 Systematics and Results

For each extrapolation scenario, a profile-likelihood fit is applied to the BDT score distributions shown in Figure 9 based on a background-only Asimov dataset. The fit is performed simultaneously in the three SRs to extract the signal cross-section. All sources of uncertainties are incorporated in the fit as nuisance parameters, as described in Section 3.2. Table 3 shows the number of events in each event category ($\tau_{\text{lep}}\tau_{\text{had}}$ SLT channel, $\tau_{\text{lep}}\tau_{\text{had}}$ LTT channel, $\tau_{\text{had}}\tau_{\text{had}}$ channel) in the baseline scenario, after applying the selection criteria described in Section 3.1. The numbers for the background are derived after the fit to the background-only Asimov dataset. The signal is estimated using a fit to an Asimov dataset with $\mu = 1$. The numbers are shown first for the entire SR, then for the last two bins of the BDT distribution where the BDT score is higher and finally only for the bin with the highest BDT score.

Figure 10 presents the upper limits on the $HH$ production cross-section normalised to the SM expectation as a function of the luminosity. The four extrapolation scenarios described above are shown: the scenario in which the systematic uncertainties remain the same as for the Run 2 analysis ("current systematic uncertainties"); the scenario with the current systematic uncertainties but neglected MC statistical uncertainties ("MC statistical uncertainty neglected"); the baseline scenario for the systematic uncertainties ("baseline"); and the scenario with no systematic uncertainties considered ("no systematic uncertainties"). In the absence of the SM $HH$ signal, the analysis is expected to set a 95% CL upper limit at 0.99 times the

16

| Full signal region | $\tau_{\text{lep}}\tau_{\text{had}}$ channel (SLT) | $\tau_{\text{lep}}\tau_{\text{had}}$ channel (LTT) | $\tau_{\text{had}}\tau_{\text{had}}$ channel |
|---|---|---|---|
| $t\bar{t}$ fake-$\tau_{\text{had-vis}}$ | - | - | 20400 ± 2200 |
| $t\bar{t}$ | 2218000 ± 13000 | 176000 ± 2300 | 57600 ± 2000 |
| Single top | 129200 ± 2800 | 8240 ± 230 | 4490 ± 150 |
| Multijet fake-$\tau_{\text{had-vis}}$ | - | - | 33500 ± 2100 |
| Fake-$\tau_{\text{had-vis}}$ | 867000 ± 13000 | 51100 ± 2300 | - |
| $Z \to \tau\tau + (bb, bc, cc)$ | 51800 ± 2100 | 14600 ± 600 | 23800 ± 1100 |
| Other | 24300 ± 1000 | 1710 ± 160 | 2550 ± 350 |
| SM Higgs boson | 4280 ± 360 | 460 ± 40 | 900 ± 60 |
| Total background | 3295300 ± 1800 | 252050 ± 500 | 143200 ± 400 |
| SM $HH$ | 107 ± 6 | 23.9 ± 1.3 | 81 ± 8 |

| Last two bins | $\tau_{\text{lep}}\tau_{\text{had}}$ channel (SLT) | $\tau_{\text{lep}}\tau_{\text{had}}$ channel (LTT) | $\tau_{\text{had}}\tau_{\text{had}}$ channel |
|---|---|---|---|
| $t\bar{t}$ fake-$\tau_{\text{had-vis}}$ | - | - | 146 ± 19 |
| $t\bar{t}$ | 1830 ± 40 | 1780 ± 30 | 370 ± 30 |
| Single top | 720 ± 20 | 420 ± 40 | 32.3 ± 2.8 |
| Multijet fake-$\tau_{\text{had-vis}}$ | - | - | 100 ± 20 |
| Fake-$\tau_{\text{had-vis}}$ | 640 ± 40 | - | 1210 ± 30 |
| $Z \to \tau\tau + (bb, bc, cc)$ | 1290 ± 70 | 1150 ± 70 | 610 ± 60 |
| Other | 460 ± 20 | 180 ± 20 | 80 ± 10 |
| SM Higgs boson | 220 ± 10 | 64 ± 3 | 134 ± 8 |
| Total background | 5730 ± 90 | 4230 ± 90 | 1470 ± 90 |
| SM $HH$ | 52 ± 3 | 16.2 ± 0.8 | 54 ± 5 |

| Last bin | $\tau_{\text{lep}}\tau_{\text{had}}$ channel (SLT) | $\tau_{\text{lep}}\tau_{\text{had}}$ channel (LTT) | $\tau_{\text{had}}\tau_{\text{had}}$ channel |
|---|---|---|---|
| $t\bar{t}$ fake-$\tau_{\text{had-vis}}$ | - | - | 12.9 ± 2.0 |
| $t\bar{t}$ | 235 ± 6 | 360 ± 30 | 0 |
| Single top | 283 ± 15 | 54 ± 3 | 0 |
| Multijet fake-$\tau_{\text{had-vis}}$ | - | - | 33.7 ± 7.2 |
| Fake-$\tau_{\text{had-vis}}$ | 300 ± 10 | 97 ± 9 | - |
| $Z \to \tau\tau + (bb, bc, cc)$ | 340 ± 20 | 470 ± 40 | 95 ± 16 |
| Other | 105 ± 5 | 61 ± 7 | 12.2 ± 2.1 |
| SM Higgs boson | 78 ± 4 | 31 ± 2 | 55 ± 3 |
| Total background | 1343 ± 25 | 1069 ± 55 | 209 ± 17 |
| SM $HH$ | 32.8 ± 1.6 | 9.8 ± 0.5 | 32 ± 3 |

Table 3: Post-fit expected number of events for the HL-LHC target integrated luminosity of 3000 fb$^{-1}$ after applying the selection criteria described in Section 3.1. The first part of the table shows the total number of events, the second shows the number of events in the last two bins of the BDT distribution and the third shows only the bin with the highest BDT score. In the $\tau_{\text{lep}}\tau_{\text{had}}$ channel, the "Fake-$\tau_{\text{had-vis}}$" background includes all processes ($t\bar{t}$, multijets and $W$+jets) in which a jet is misidentified as a $\tau_{\text{had-vis}}$, while in the $\tau_{\text{had}}\tau_{\text{had}}$ case the fake-$\tau_{\text{had-vis}}$ background from multijet processes ("Multijet Fake-$\tau_{\text{had-vis}}$") and $t\bar{t}$ production ("$t\bar{t}$ Fake-$\tau_{\text{had-vis}}$") are derived separately. The $t\bar{t}$ background includes events with true $\tau_{\text{had-vis}}$ and the very small contribution from leptons misidentified as $\tau_{\text{had-vis}}$. The 'Other' category includes contributions from $W$+jets (including fake $\tau_{\text{had-vis}}$ in the $\tau_{\text{had}}\tau_{\text{had}}$ channel), $Z \to \tau\tau$+light-flavour jets, $Z \to \ell\ell$+jets and diboson processes. The total background is not identical to the sum of the individual components since the latter are rounded for presentation, while the sum is calculated with the full precision before being subsequently rounded. Systematic uncertainties as defined in the baseline scenario are included in the fit. Due to the large correlations, individual uncertainties can be significantly larger than the total uncertainty.

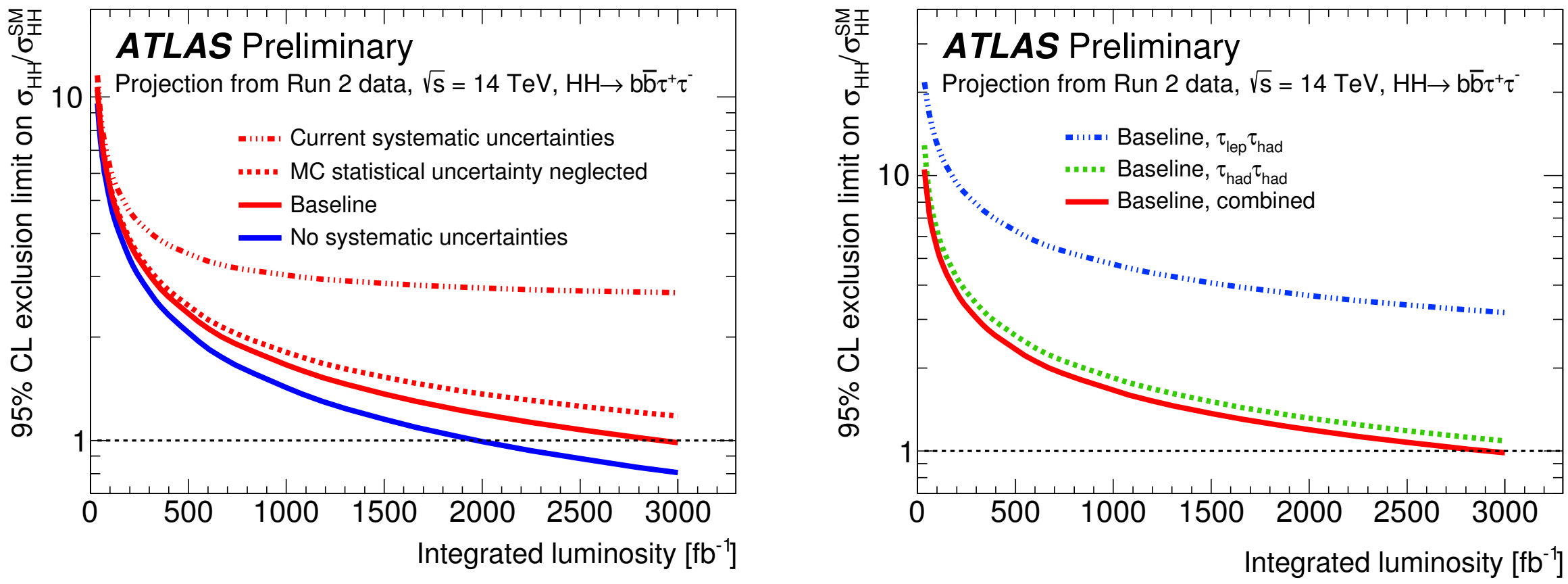

Figure 10: Expected 95% CL upper limit on $\sigma_{HH}/\sigma_{HH}^{SM}$, as a function of the integrated luminosity of the search between 36.1 and 3000 fb$^{-1}$. Results are shown for each extrapolation scenario where the $\tau_{\text{lep}}\tau_{\text{had}}$ and $\tau_{\text{had}}\tau_{\text{had}}$ channels are combined (left) and for the $\tau_{\text{lep}}\tau_{\text{had}}$ channel, the $\tau_{\text{had}}\tau_{\text{had}}$ channel and their combination separately for the baseline scenario (right).

SM expectation with 3000 fb$^{-1}$ of data for the baseline scenario. The $\tau_{\text{had}}\tau_{\text{had}}$ channel is most sensitive and can set an upper limit at 1.1 times the SM expectation. The expected significance, $p_0$, is shown in Figure 11 as a function of the integrated luminosity. The $2\sigma$ threshold can be reached for the baseline scenario and the full HL-LHC dataset. All results are summarised in Table 4.

| Scenario | $-1\sigma$ | Expected limit | $+1\sigma$ | Significance [$\sigma$] |
|---|---|---|---|---|
| No systematic uncert. | 0.58 | 0.80 | 1.12 | 2.5 |
| Baseline | 0.71 | 0.99 | 1.37 | 2.1 |
| MC statistical uncert. neglected | 0.8 | 1.2 | 1.6 | 1.7 |
| Current systematic uncert. | 1.9 | 2.7 | 3.7 | 0.65 |

Table 4: Expected 95% CL upper limit on $\sigma_{HH}/\sigma_{HH}^{SM}$, and significance at 3000 fb$^{-1}$ for the four extrapolation scenarios.

The effects of various categories of uncertainty in the baseline scenario are summarised in Table 5. The individual sources of uncertainty making up the categories listed in the table are grouped in the final fit to determine their correlated combined effect on the signal strength. The statistical uncertainty on the data is the dominant uncertainty in the fit.

The degradation of the $b$-tagging efficiency to its Run 2 values was assessed by removing the 8% improvement that was applied earlier. This would result in a relative decrease by up to 8% of the significance for 3000 fb$^{-1}$ of data.

### 3.4 Sensitivity to the Self-Coupling $\lambda_{HHH}$

Variations in the tri-linear Higgs boson self-coupling strength, $\kappa_\lambda = \lambda_{HHH}/\lambda_{HHH}^{SM}$, can be probed in addition to testing for the presence of a SM $HH$ signal. When probing $\lambda_{HHH} \neq \lambda_{HHH}^{SM}$, the top-quark

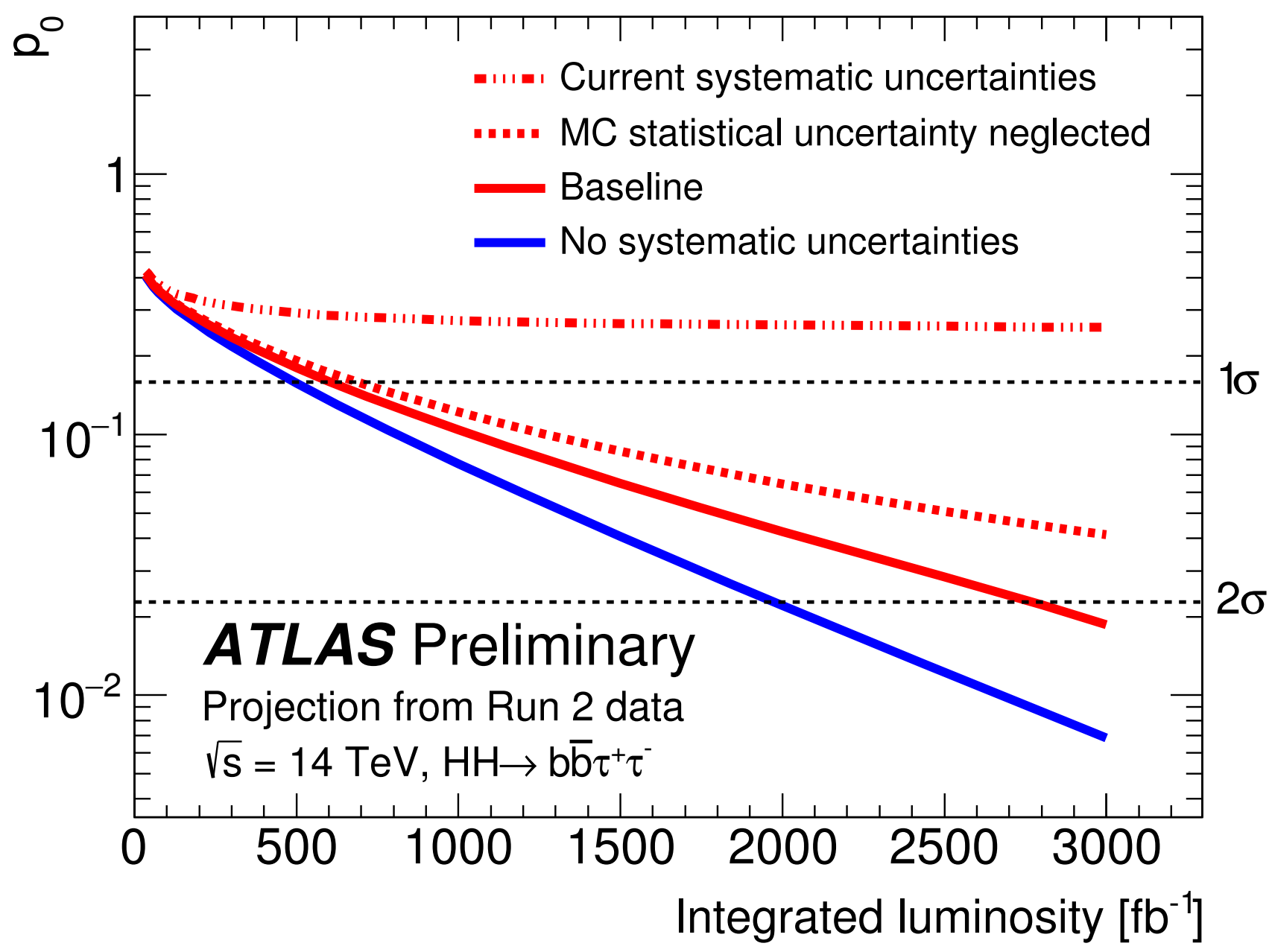

Figure 11: Expected significance, $p_0$, of the discrepancy from the background-only hypothesis when the predicted SM $HH$ signal is injected, as a function of the integrated luminosity of the search between 36.1 and 3000 fb$^{-1}$, for each extrapolation scenario with the $\tau_{\rm lep}\tau_{\rm had}$ and $\tau_{\rm had}\tau_{\rm had}$ channels combined.

Yukawa coupling is set to its SM value, while the effective Higgs boson self-coupling is changed by applying $\kappa_\lambda$ as a scale factor.

Samples for three different values of $\kappa_\lambda$ ($\kappa_\lambda$ = 0, 1 and 20) are produced at the generator-level using the LO version of MadGraph5_aMC@NLO [46] with the same setup as described in Section 2.3. Samples for any other $\kappa_\lambda$ value of interest are emulated by a linear combination of these three samples. Furthermore, following the procedure described in Section 2.3, the binned ratios of $m_{HH}$ distributions for any given $\kappa_\lambda$ value to the SM case ($\kappa_\lambda = 1$) are computed at LO and then applied to the fully simulated SM NLO sample to reweight it to different $\kappa_\lambda$ values, taking into account the SM NLO QCD and the finite top-quark mass corrections and assuming them to be independent of $\kappa_\lambda$.

The Run 2 results are extrapolated in order to estimate the sensitivity of the analysis to the Higgs boson self-coupling strength at the HL-LHC. Only the two most sensitive event categories are used for the extrapolation: the SLT category in the $\tau_{\rm lep}\tau_{\rm had}$ channel and the $\tau_{\rm had}\tau_{\rm had}$ channel. The default BDT trained with the SM $HH$ signal with $\kappa_\lambda = 1$ is replaced by another BDT trained with the BSM $HH$ signal generated with $\kappa_\lambda = 20$. The new BDT training increases the sensitivity of the analysis to the softer $m_{HH}$ spectrum and provides a similar performance as BDTs trained specifically for each $\kappa_\lambda$ value. The signal acceptance times efficiency varies significantly as a function of $m_{HH}$ as shown in Figure 12. The analysis sensitivity is driven by the high $m_{HH}$ region with a low background contamination. Therefore there is no significant gain from targeting a specific ($\kappa_\lambda$ dependent) $m_{HH}$ distribution.

The 95% CL expected upper limits on the $HH$ production cross-section as a function of $\kappa_\lambda$ are shown in Figure 13. The limits become significantly weaker for values of $\kappa_\lambda$ that correspond to a softer $m_{HH}$ spectrum, as expected from the decrease of the signal acceptance times efficiency and worse signal-to-background separation. All four scenarios for the systematic uncertainties discussed in Section 3.2 are considered.

| Source | Uncertainty (%) |
|---|---|
| Total | ±52 |
| Data statistics | ±43 |
| Simulation statistics | ±0 |
| Total systematic uncertainty | ±30 |
| **Experimental uncertainties** | |
| Luminosity | ±4.3 |
| Pile-up reweighting | ±7.0 |
| $\tau_{\text{had-vis}}$ | ±13 |
| Fake-$\tau_{\text{had-vis}}$ estimation | ±8.3 |
| $b$- tagging | ±8.1 |
| Jets and $E_{\text{T}}^{\text{miss}}$ | ±3.5 |
| Electron and muon | ±5.1 |
| Total experimental uncertainties | ±18 |
| **Theoretical and modelling uncertainties** | |
| Top | ±6.6 |
| Signal | ±8.6 |
| $Z/\gamma^* \rightarrow \tau^+\tau^-$ | ±11 |
| SM Higgs boson | ±8.5 |
| Other backgrounds | ±4.4 |
| Total theoretical and modelling uncertainties | ±17 |

Table 5: The percentage uncertainties on the simulated SM $HH$ signal strength, i.e. the simulated SM $HH$ yield assuming a cross-section times branching fraction equal to the 95% CL expected limit in the baseline scenario.

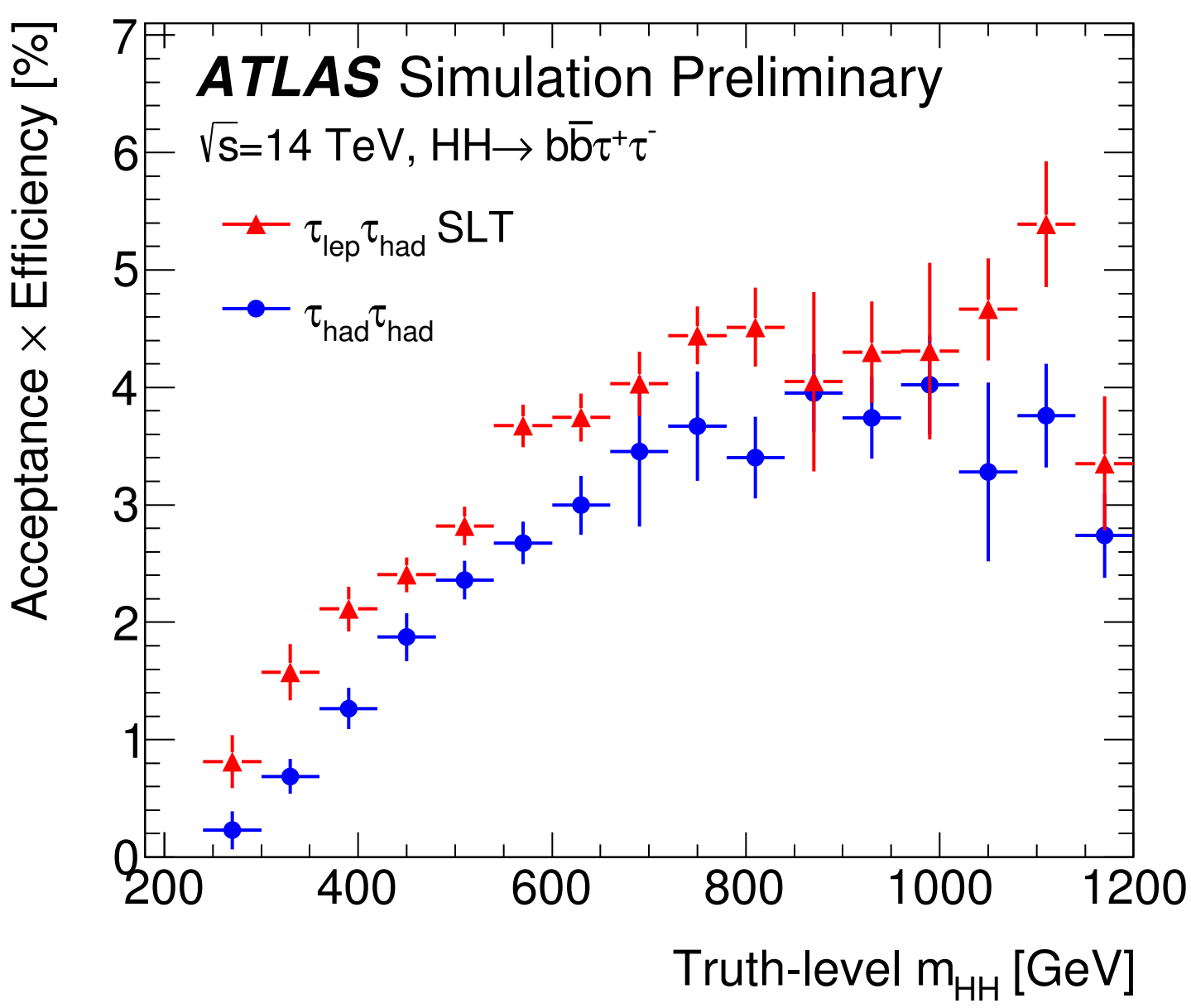

Figure 12: Signal acceptance times efficiency as a function of the truth-level $m_{HH}$ for the $\tau_{\text{lep}}\tau_{\text{had}}$ SLT (red) and $\tau_{\text{had}}\tau_{\text{had}}$ (blue) channels.

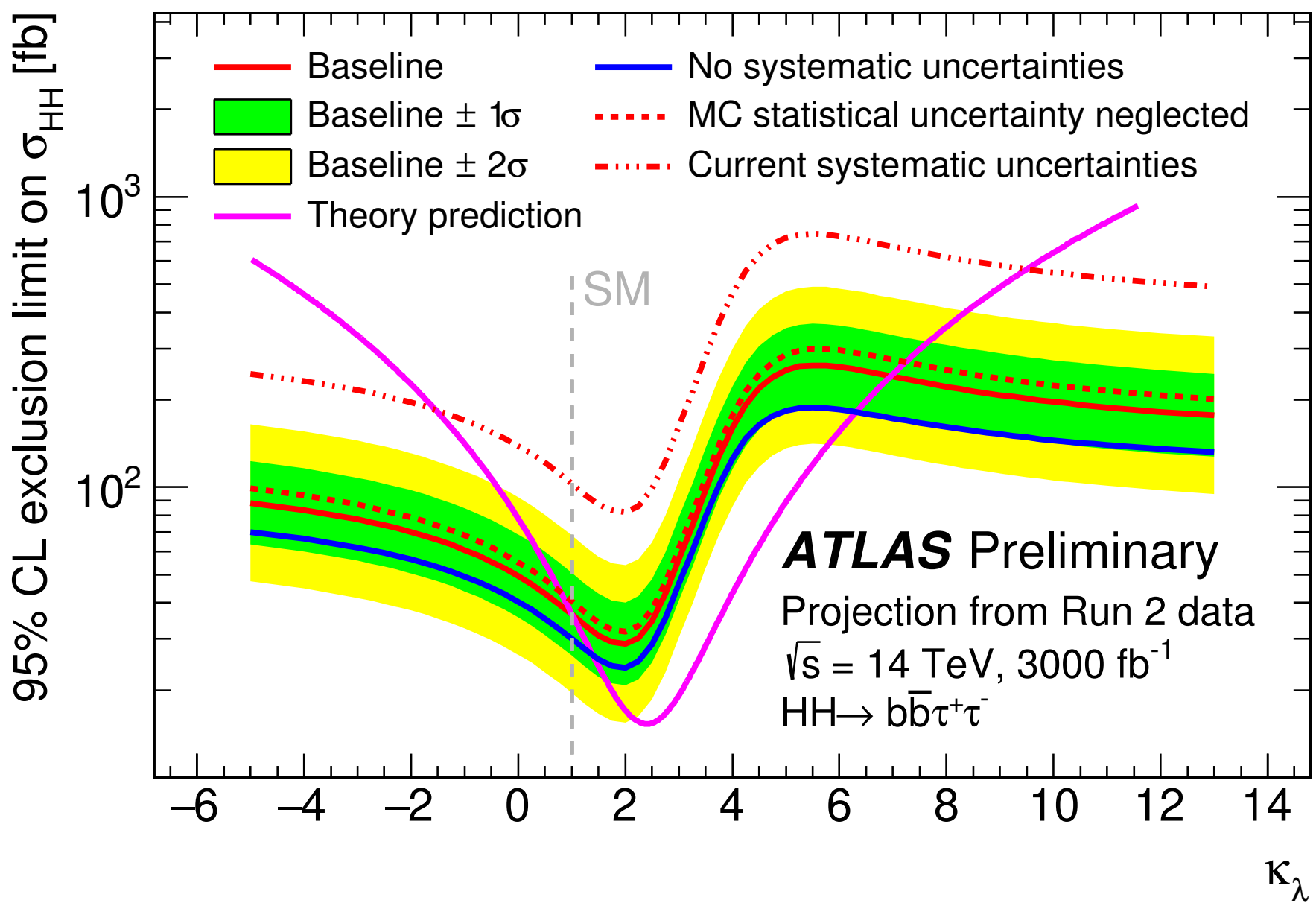

Figure 13: The 95% CL expected upper limits on the $HH$ production cross-section are shown as function of $\kappa_\lambda$. The value $\kappa_\lambda = 1$ corresponds to the SM prediction. The cross-section for $HH$ production predicted as a function of $\kappa_\lambda$ is superimposed on the limits. Four different scenarios for the systematic uncertainties are considered. From these results, the allowed $\kappa_\lambda$ interval at 95% CL is expected to be $1.0 < \kappa_\lambda < 7.0$ in the baseline scenario, and $1.4 < \kappa_\lambda < 6.3$ if systematic uncertainties are not considered.

To estimate the sensitivity of the analysis to the Higgs boson self-coupling, two different likelihood ratio tests are performed. In the first case the Asimov dataset is created from the backgrounds and the SM $HH$ signal. The maximum likelihood fits to this dataset are performed with different $\kappa_\lambda$ hypotheses. The ratio of the maximum likelihood for various $\kappa_\lambda$ values to that for the $\kappa_\lambda = 1$ value is used to set the $\kappa_\lambda$ confidence intervals assuming the SM $HH$ signal. The negative natural logarithm of this ratio is shown in Figure 14 on the left. Similarly, another Asimov dataset is built from the backgrounds and the $\kappa_\lambda = 0$ configuration. After performing the maximum likelihood fits to this dataset, the negative natural logarithm of the ratio of the maximum likelihood for various $\kappa_\lambda$ values to that for the $\kappa_\lambda = 0$ hypothesis is used to set the $\kappa_\lambda$ confidence intervals for the case where the $\kappa_\lambda = 0$ configuration, i.e. no Higgs boson self-coupling, is assumed. This is shown in Figure 14 on the right.

It can be seen from Figure 14 that in both cases two minima are observed. In the left (right) plot the first minimum is at $\kappa_\lambda = 1$ ($\kappa_\lambda = 0$) as the signal hypothesis used in the maximum likelihood fit corresponds to the signal used to create the Asimov dataset. The second minimum is observed at a $\kappa_\lambda$ value that corresponds to a higher cross-section, but lower acceptance and worse signal-to-background separation due to a softer $m_{HH}$ spectrum with respect to the $\kappa_\lambda$ point at the first minimum. This leads to a similar fitted signal yield between the two $\kappa_\lambda$ values and consequently a similar maximum likelihood result. For all other $\kappa_\lambda$ points the expected signal yield is either lower or higher, thus the likelihood ratio is greater. The confidence intervals on $\kappa_\lambda$ are given in Tables 6 and 7.

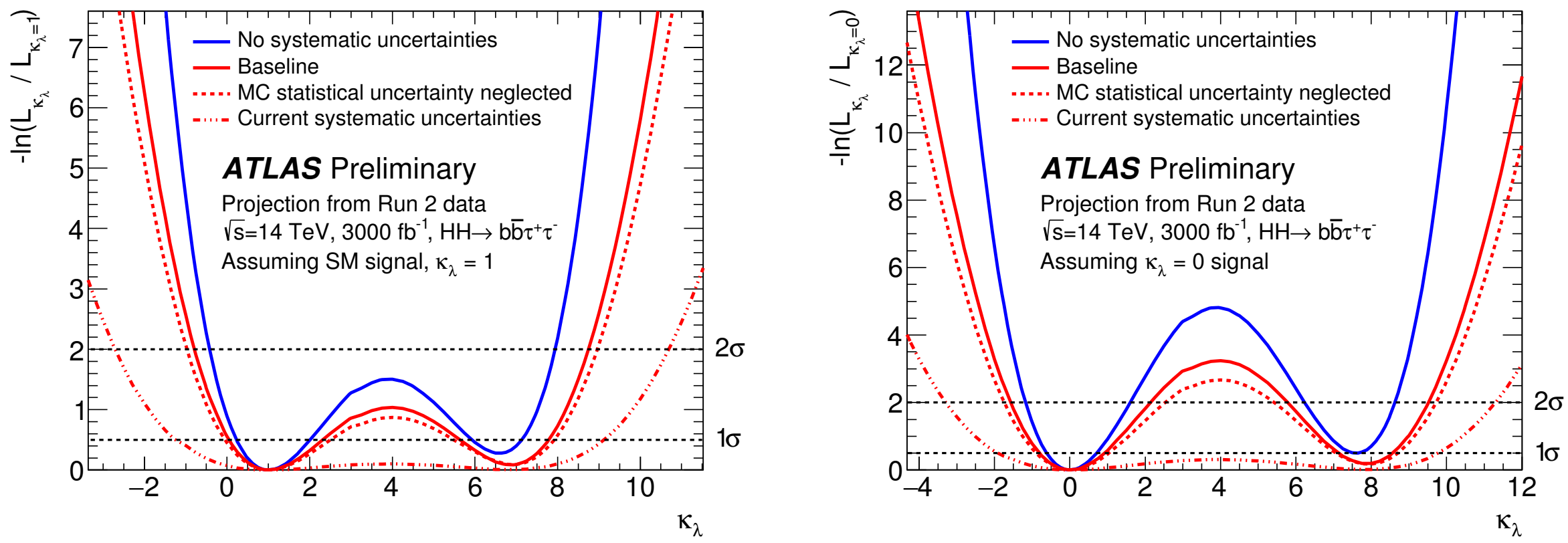

Figure 14: The left (right) plot shows the negative natural logarithm of the ratio of the maximum likelihood for $\kappa_\lambda$ to the maximum likelihood for $\kappa_\lambda = 1$ ($\kappa_\lambda = 0$), obtained from fits to the Asimov dataset that contains the $\kappa_\lambda = 1$ ($\kappa_\lambda = 0$) signal. The horizontal lines show the $1\sigma$ and $2\sigma$ confidence intervals (assuming an asymptotic $\chi^2$ distribution of the test statistic). Four different scenarios are considered for the systematic uncertainties.

| Scenario | $1\sigma$ CI | $2\sigma$ CI |
|---|---|---|
| No systematic uncert. | $0.2 < \kappa_\lambda < 2.0 \cup 5.9 < \kappa_\lambda < 7.2$ | $-0.4 < \kappa_\lambda < 7.9$ |
| Baseline | $0.1 < \kappa_\lambda < 2.3 \cup 5.7 < \kappa_\lambda < 7.8$ | $-0.8 < \kappa_\lambda < 8.8$ |
| MC statistical uncert. neglected | $-0.1 < \kappa_\lambda < 2.5 \cup 5.5 < \kappa_\lambda < 7.9$ | $-1.0 < \kappa_\lambda < 9.0$ |
| Current systematic uncert. | $-1.2 < \kappa_\lambda < 9.1$ | $-2.7 < \kappa_\lambda < 10.7$ |

Table 6: Constraints on $\kappa_\lambda$ from the likelihood ratio test performed on the Asimov dataset created from the backgrounds and the SM $HH$ signal, which is shown in Figure 14 on the left. Results are presented as a $1\sigma$ and $2\sigma$ CI on $\kappa_\lambda$ for the different scenarios for systematic uncertainties.

| Scenario | $1\sigma$ CI | $2\sigma$ CI |
|---|---|---|
| No systematic uncert. | $-0.6 < \kappa_\lambda < 0.7$ | $-1.2 < \kappa_\lambda < 1.6 \cup 6.2 < \kappa_\lambda < 8.6$ |
| Baseline | $-0.8 < \kappa_\lambda < 0.9 \cup 7.1 < \kappa_\lambda < 8.6$ | $-1.6 < \kappa_\lambda < 2.2 \cup 5.8 < \kappa_\lambda < 9.5$ |
| MC statistical uncert. neglected | $-0.9 < \kappa_\lambda < 1.0 \cup 7.1 < \kappa_\lambda < 8.7$ | $-1.7 < \kappa_\lambda < 2.5 \cup 5.4 < \kappa_\lambda < 9.7$ |
| Current systematic uncert. | $-1.9 < \kappa_\lambda < 9.8$ | $-3.3 < \kappa_\lambda < 11.3$ |

Table 7: Constraints on $\kappa_\lambda$ from the likelihood ratio test performed on the Asimov dataset created from the backgrounds and the configuration for $\kappa_\lambda = 0$, which is shown in Figure 14 on the right. Results are presented as a $1\sigma$ and $2\sigma$ CI on $\kappa_\lambda$ for the different scenarios for systematic uncertainties.

## 3.5 Impact of Trigger Thresholds

Maintaining high signal acceptance when triggering on hadronic final states at the HL-LHC will be difficult due to the high number of pile-up events. Triggering on final states with hadronically decaying $\tau$-leptons generally causes high trigger rates and it is therefore a limiting factor in the fully hadronic $b\bar{b}\tau^+\tau^-$ sub-channel. However, thanks to upgrades of several trigger systems, the expected performance of $\tau_{\text{had-vis}}$ triggers at the HL-LHC [30] should allow the analysis to maintain offline $p_{\mathrm{T}}$ thresholds of 40 (30) GeV on the leading (sub-leading) $\tau_{\text{had-vis}}$, with the additional advantage of no longer needing to require the presence of an additional jet at Level 1.

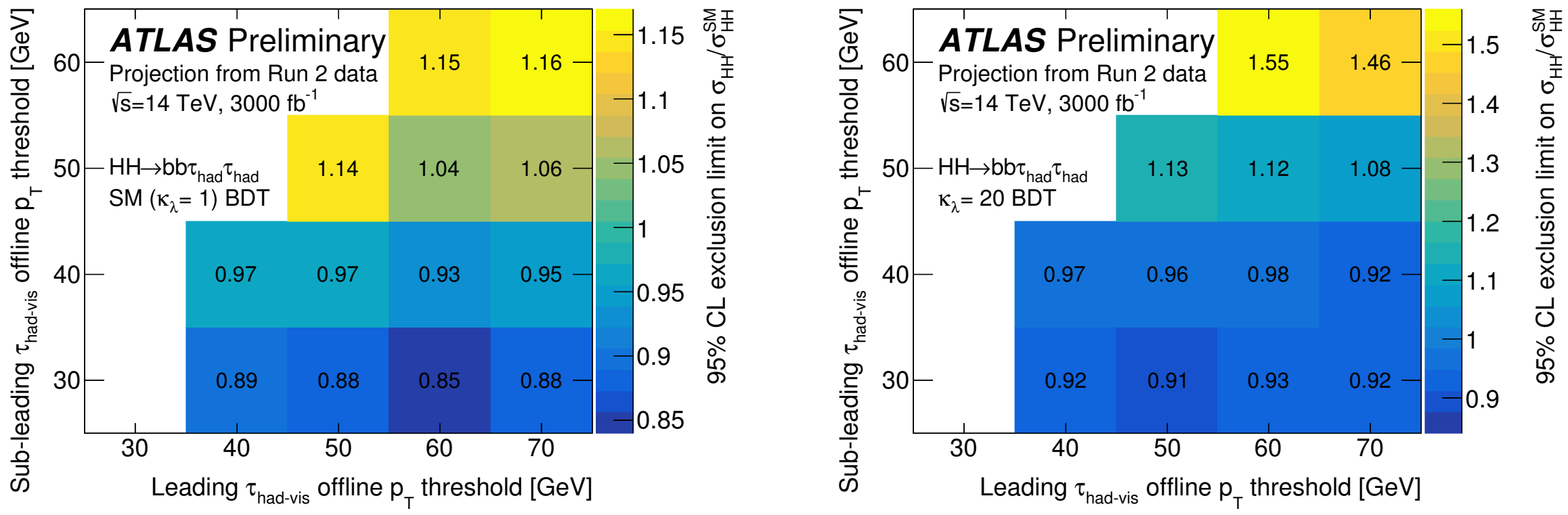

Figure 15: Expected 95% CL upper limit on $\sigma_{HH}/\sigma_{HH}^{SM}$, as a function of the minimum $p_{\mathrm{T}}$ thresholds for the leading and sub-leading $\tau_{\text{had-vis}}$. Systematic uncertainties are not taken into account and the results are shown for two different BDT classifiers, the nominal BDT classifier in the left plot and the BDT classifier trained on the $\kappa_\lambda = 20$ signal in the right plot.

The impact of increasing the leading and sub-leading $\tau_{\text{had-vis}}$ $p_{\mathrm{T}}$ thresholds has been studied by repeating the analysis for several variations of the minimum $\tau_{\text{had-vis}}$ $p_{\mathrm{T}}$ requirements, as illustrated in Figure 15. The requirement that DTT events contain an additional jet with offline $p_{\mathrm{T}} > 80$ GeV (as described in Section 3.1 and Ref. [24]) has to be maintained as otherwise the extrapolation of the estimated background would not be valid. The expected 95% CL upper limit on the SM Higgs-boson-pair production cross-section was extrapolated to 14 TeV and 3000 fb$^{-1}$, for each of the considered trigger scenarios, using the procedure previously described, without considering the impact of systematic uncertainties, as shown on the left plot of Figure 15. The sensitivity to the Higgs boson self-coupling is affected more by raising the $p_{\mathrm{T}}$ thresholds, compared to the overall SM Higgs-boson-pair production process. For this reason the study is repeated with a BDT classifier trained to be sensitive to the Higgs boson self-coupling production mode, as was used to perform the $\kappa_\lambda$ scan. The results are shown on the right plot of Figure 15.

In a pessimistic scenario where the leading and sub-leading minimum $\tau_{\text{had-vis}}$ $p_{\mathrm{T}}$ thresholds are required to be 70 and 60 GeV respectively (and the additional jet requirement is maintained), the upper limit on the $HH$ cross-section degrades by about 30% compared to the result obtained with the Run 2 thresholds. In the case where the BDT is trained on the $\kappa_\lambda = 20$ sample, this effect is even more pronounced and the limit degrades by about 60% compared to the Run 2 thresholds. It is important to note that the requirement of an additional jet with $p_{\mathrm{T}} > 80$ GeV to a large extent masks the effect of raising the leading $\tau_{\text{had-vis}}$ $p_{\mathrm{T}}$, and the degradation is expected to be even larger when comparing triggers that do not include the additional jet requirement.

# 4 $HH \to b\bar{b}\gamma\gamma$

One of the most promising channels to measure the Higgs boson self-coupling is the $b\bar{b}\gamma\gamma$ final state. It arises from the most probable Higgs boson decay, $H \to b\bar{b}$, in association with the $H \to \gamma\gamma$ decay, which provides a narrow mass peak and a very clean Higgs boson signal.

The current upper limits on the non-resonant Higgs-boson-pair production cross-section rely on data collected by the ATLAS and CMS experiments at a 13 TeV centre-of-mass energy with an integrated luminosity of 36.1 fb$^{-1}$. The ATLAS Collaboration reported an observed (expected) limit on the cross-section for non-resonant $HH$ production of 0.73 pb (0.93 pb) at 95% CL, corresponding to 22 (28) times the SM prediction, and constrained the Higgs boson self-coupling to $-8.2 < \kappa_\lambda < 13.2$ ($-8.3 < \kappa_\lambda < 13.2$) at 95% CL [53]. The CMS Collaboration reported an observed (expected) limit on the cross-section for non-resonant $HH$ production of 0.79 pb (0.63 pb) at 95% CL, corresponding to 24 (19) times the SM prediction, and constrained the Higgs boson self-coupling to $-11 < \kappa_\lambda < 17$ at 95% CL [54].

Unlike the $HH \to b\bar{b}b\bar{b}$ and $HH \to b\bar{b}\tau^+\tau^-$ analyses, which are based on extrapolating the Run 2 results to the HL-LHC, the $HH \to b\bar{b}\gamma\gamma$ analysis is fully based on simulations at $\sqrt{s}$ = 14 TeV. To emulate the upgraded ATLAS detector response, the final-state particles at truth level are smeared according to the expected detector resolutions assuming a pile-up scenario with 200 overlapping events ($<\mu>$ = 200) [55]. The expected identification efficiencies and fake rates for $b$-jets and photons are used accordingly. These smearing functions were obtained from fully simulated events using the detector layout described in the Pixel Detector TDR [25].

A previous study presented in the Pixel Detector TDR [25] resulted in an expected constraint on the Higgs boson self-coupling in the interval $0.2 < \kappa_\lambda < 6.9$ at 95% CL, with an expected signal significance of 1.5$\sigma$. This study improves the sensitivity by including a BDT method. This multivariate analysis exploits the full kinematic information to separate the $HH$ signal from the expected backgrounds. After selecting two isolated photons and two energetic $b$-tagged jets, and applying a BDT selection, the signal is extracted using a fit to the di-photon invariant mass distribution of the selected events. The variables used to train the BDT are carefully chosen to ensure that the $m_{\gamma\gamma}$ distribution is not sculpted. The signal appears as a narrow peak around $m_H$ superimposed on a resonant peak arising from single-Higgs-boson events and a smoothly falling continuum background. Unlike the Run-2 analysis the signal is extracted by selecting events in a narrow window of the $m_{\gamma\gamma}$ distribution around the Higgs boson mass, and then fitting the $m_{b\bar{b}\gamma\gamma}$ distribution divided in 8 bins.

## 4.1 Background and Signal Simulation

The expected signal and background processes are modelled using MC samples. This study reuses most of the samples of the analysis described in Ref. [56], which contains full details about their generation. The main backgrounds arise from various processes involving multiple jets and isolated photons that lead to two photons and two $b$-jets in the final state. The largest component of the background is the continuum coming from processes with multiple jet and photon production, such as $b\bar{b}\gamma\gamma$, $c\bar{c}\gamma\gamma$, $jj\gamma\gamma$, $b\bar{b}j\gamma$, $c\bar{c}j\gamma$, and $b\bar{b}jj$ events. It should be noted that the last three of these processes require at least one photon to be a misidentified jet. Other contributions come from $Z(b\bar{b})\gamma\gamma$, $t\bar{t}$ and $t\bar{t}\gamma$ production. A second component of the background comes from processes involving the production and subsequent decay of a single Higgs boson, such as $gg \to H(\to \gamma\gamma)$, $ZH(\to \gamma\gamma)$, $t\bar{t}H(\to \gamma\gamma)$, and $b\bar{b}H(\to \gamma\gamma)$. These processes are produced with MADGRAPH5_aMC@NLO [46, 57, 58] interfaced with PYTHIA [59] for the showering and hadronisation. These samples were generated inclusively (e.g. an additional jet in the tree-level matrix element is allowed). The list of signal and background samples considered is displayed in Table 8. As described below, in this analysis every possible combination of reconstructed candidates arising from a given truth-level event are considered, taking into account the probability of each combination. This procedure enhances the statistical power of the samples, thus compensating for the low equivalent integrated luminosity of some of the samples.

For the signal process $HH \to b\bar{b}\gamma\gamma$, only the dominant gluon-gluon fusion production mode is generated using MADGRAPH5_aMC@NLO at LO (with a finite top-quark mass) with PYTHIA 8 to model parton showering and hadronisation. The A14 tune [60] is used together with the NNPDF2.3LO PDF set [61]. The event yields are normalised to the next-to-next-to-leading order (NNLO) cross-sections of Refs. [62] and [63] (using the infinite top-quark mass approximation). Events were produced with self-coupling strengths $\kappa_\lambda = 0, \pm1, \pm2, \pm4, \pm6, \pm10$, with $\kappa_\lambda = 1$ corresponding to the SM expectation.

Overlaps between the different samples are taken into account at analysis level, for example $b\bar{b}\gamma\gamma$ events are excluded from the $b\bar{b}j\gamma$ samples and $t\bar{t}\gamma$ events from the $t\bar{t}$ sample.

## 4.2 Object Definitions

The event sample reconstruction uses the same approach as in Ref. [56]. The final-state particles at truth level are smeared according to the expected detector resolutions, and the impact of an average of 200 pile-up events is included in the reconstruction efficiency and energy smearing of the objects. Any truth jets that are matched to truth $e/\gamma$ objects are removed at this point.

The photon energy is smeared using the baseline photon resolution parameterisation described in Ref. [32], which is based on fully reconstructed simulation and provides a di-photon mass resolution comparable to the one obtained in Run 2. In this analysis, the efficiency to identify isolated photons is around 60% at $p_T = 50$ GeV and saturates at 85% above 150 GeV. The corresponding probability for a jet emerging from the primary interaction to be misidentified as a photon is at most $5 \times 10^{-4}$, which is also derived from fully reconstructed simulation. The probability that an electron fakes a photon is assumed to be 2% (5%) in the barrel (endcap) region.

The $b$-tagging efficiency and mistag rates have been updated according to the most recent ITk layout which is documented in Ref. [25]. The $b$-tagging working point of 70% identification efficiency was chosen. The $b$-tagging performance for the reconstructed jets is modelled by applying a $p_T$- and $\eta$-dependent efficiency or mistag rate function, as a function of the true jet flavour ($b/c$/light, where "light" refers to $u/d/s/g$). Such a parameterisation is derived by computing the responses of the ATLAS MV2 $b$-tagging algorithm [42], and the track confirmation algorithm, using fully simulated events.

## 4.3 Event Pre-Selection

The analysis in Ref. [56] typically processed events by first taking each truth-level particle and then examining all the possible object types that the detector might reconstruct it as. Each of these candidate reconstructed particles was assigned a probability based on the efficiency or fake rate, as appropriate, and then one was randomly chosen. In this way, each truth-level particle produced only one reconstructed object and each truth-level event produced a single reconstructed event. For this analysis, the statistical power is enhanced by considering all the possible object types that the truth-level particle could be reconstructed as, and examining every possible combination of reconstructed candidates arising from the set of truth-level particles. Each truth-level event produces multiple reconstructed events, where the weight of each combination of final-state particles is determined by multiplying together all the selection probabilities of the individual reconstructed objects. In order to reduce the number of final-state combinations to a manageable level, truth-level pile-up jets are only allowed to produce one reconstructed candidate. The pile-up jet may become a light jet, $b$-jet, fake photon, fake electron, or not be reconstructed at all – the choice is made randomly, based on the relative probabilities of each of the five possible reconstructed states. It should also be noted that, by construction, the sum of weights of all final-state combinations derived from a single truth-level event is unity.

| Process | Generator | $\sigma \cdot BR$ [fb] | Order QCD | Generated Events |
|---|---|---|---|---|
| $H(b\bar{b})H(\gamma\gamma)$, $\kappa_\lambda = 1$ | MADGRAPH5/PYTHIA 8 | 0.097 | NNLO | $6.0 \times 10^5$ |
| $H(b\bar{b})H(\gamma\gamma)$, $\kappa_\lambda = 0$ | MADGRAPH5/PYTHIA 8 | 0.204 | NNLO | $2.0 \times 10^5$ |
| $H(b\bar{b})H(\gamma\gamma)$, $\kappa_\lambda = 2$ | MADGRAPH5/PYTHIA 8 | 0.045 | NNLO | $2.0 \times 10^5$ |
| $H(b\bar{b})H(\gamma\gamma)$, $\kappa_\lambda = 4$ | MADGRAPH5/PYTHIA 8 | 0.112 | NNLO | $2.0 \times 10^5$ |
| $H(b\bar{b})H(\gamma\gamma)$, $\kappa_\lambda = 6$ | MADGRAPH5/PYTHIA 8 | 0.404 | NNLO | $2.0 \times 10^5$ |
| $H(b\bar{b})H(\gamma\gamma)$, $\kappa_\lambda = 10$ | MADGRAPH5/PYTHIA 8 | 1.662 | NNLO | $2.0 \times 10^5$ |
| $H(b\bar{b})H(\gamma\gamma)$, $\kappa_\lambda = -1$ | MADGRAPH5/PYTHIA 8 | 0.368 | NNLO | $2.0 \times 10^5$ |
| $H(b\bar{b})H(\gamma\gamma)$, $\kappa_\lambda = -2$ | MADGRAPH5/PYTHIA 8 | 0.588 | NNLO | $2.0 \times 10^5$ |
| $H(b\bar{b})H(\gamma\gamma)$, $\kappa_\lambda = -4$ | MADGRAPH5/PYTHIA 8 | 1.197 | NNLO | $2.0 \times 10^5$ |
| $H(b\bar{b})H(\gamma\gamma)$, $\kappa_\lambda = -6$ | MADGRAPH5/PYTHIA 8 | 2.030 | NNLO | $2.0 \times 10^5$ |
| $H(b\bar{b})H(\gamma\gamma)$, $\kappa_\lambda = -10$ | MADGRAPH5/PYTHIA 8 | 4.372 | NNLO | $2.0 \times 10^5$ |
| $gg \to H(\to \gamma\gamma)$ | POWHEG-BOX/PYTHIA 6 | $1.2 \times 10^2$ | NNNLO | $1.1 \times 10^7$ |
| $t\bar{t}H(\to \gamma\gamma)$ | PYTHIA 8 | 1.40 | NLO | $1.0 \times 10^5$ |
| $ZH(\to \gamma\gamma)$ | PYTHIA 8 | 2.24 | NLO | $1.0 \times 10^5$ |
| $b\bar{b}H(\to \gamma\gamma)$ | PYTHIA 8 | 1.26 | NLO | $5.0 \times 10^5$ |
| $b\bar{b}\gamma\gamma$ | MADGRAPH5/PYTHIA 8 | $1.4 \times 10^2$ | LO | $2.5 \times 10^6$ |
| $c\bar{c}\gamma\gamma$ | MADGRAPH5/PYTHIA 8 | $1.1 \times 10^3$ | LO | $3.5 \times 10^6$ |
| $jj\gamma\gamma$ | MADGRAPH5/PYTHIA 8 | $1.6 \times 10^4$ | LO | $3.8 \times 10^7$ |
| $b\bar{b}j\gamma$ | MADGRAPH5/PYTHIA 8 | $3.8 \times 10^5$ | LO | $3.5 \times 10^7$ |
| $c\bar{c}j\gamma$ | MADGRAPH5/PYTHIA 8 | $1.1 \times 10^6$ | LO | $3.0 \times 10^7$ |
| $b\bar{b}jj$ | MADGRAPH5/PYTHIA 8 | $4.6 \times 10^8$ | LO | $2.0 \times 10^6$ |
| $Z(\to b\bar{b})\gamma\gamma$ | MADGRAPH5/PYTHIA 8 | 5.07 | LO | $1.0 \times 10^5$ |
| $t\bar{t}(\geq 1$ lepton) | POWHEG-BOX/PYTHIA 6 | $5.3 \times 10^5$ | NNLO | $3.0 \times 10^8$ |
| $t\bar{t}\gamma(\geq 1$ lepton) | MADGRAPH5/PYTHIA 8 | $5.0 \times 10^3$ | NLO | $1.5 \times 10^6$ |

Table 8: List of the MC samples used in this analysis including the generators used for the matrix element and the parton showering. In addition, the production cross-section times branching ratio, the order in QCD of the cross-section calculation used (note that the order in QCD of the event generation can be lower) and the number of events generated are given.

The number of final-state combinations is further reduced by discarding any combination whose weight falls below a minimum threshold. This threshold is manually selected for each signal and background sample, in an effort to balance file sizes against statistical power. Typically, it has been possible to reduce the number of combinations by up to 80% with $< 0.1\%$ reduction in the integral of final-state weights.

Each final-state combination passing the minimum event weight threshold is first required to contain at least two photons, two $b$-jets, and less than ten jets in total. This loose initial selection criterion serves to reduce the processing time by identifying combinations that would fail later cuts, and removing them before the computationally expensive overlap removal procedure is performed. This selection criterion is not expected to have an impact on the number of combinations passing the event selection requirements. The event selection criteria are described in the following and summarised in Table 9.

Final-state combinations must be accepted by a di-photon trigger that requires at least two photons to have $p_{\mathrm{T}} > 25$ GeV within $|\eta| < 2.37$, excluding the crack region at $1.37 < |\eta| < 1.52$ where the photon energy resolution is poor. Offline, there must be at least two photons with $p_{\mathrm{T}} > 30$ GeV and located within the acceptance of the electromagnetic calorimeter. The leading photon must have $p_{\mathrm{T}} > 43$ GeV.

The following selection criteria are applied to leptons to facilitate overlap removal. Electrons are retained if they have $p_{\mathrm{T}} > 30$ GeV within $|\eta| < 2.37$ (excluding the crack region) and muons are retained if they have $p_{\mathrm{T}} > 25$ GeV and $0.1 < |\eta| < 2.5$ (driven by the acceptance of muon spectrometer and inner detector tracking). Muons separated from jets by $\Delta R_{\mu,jet} < 0.2$ are assumed to have originated from the jet, and combined with it. To ensure that particles are isolated, leptons separated from jets by $\Delta R_{\ell,jet} < 0.4$ are removed. Electrons separated from muons by $\Delta R_{e,\mu} < 0.2$ are also removed. To ensure that photons are isolated, photons within a cone $\Delta R_{\gamma,jet} < 0.2$ of a jet are removed.

Final-state combinations are required to have at most five jets with $p_{\mathrm{T}} > 30$ GeV located within $|\eta| < 2.5$. Of these five jets, at least two must be identified as $b$-jets with $p_{\mathrm{T}} > 35$ GeV. Finally, it is required that the final-state combination has no remaining isolated leptons.

A pair of Higgs boson candidates is then reconstructed by combining the two leading $b$-jets and the two leading photons. The Run 2 analysis [53] corresponds roughly to the pre-selection criteria adopted here, with some small exceptions such as the requirements on isolated leptons and the maximum number of jets.

## 4.4 Event Selection with a Boosted Decision Tree

The TMVA package [64] is used to perform a multivariate analysis with a BDT. The variables used for training the BDT are summarised in Table 10. It should be noted that most of the improvement with respect to the previous cut-based analysis [25] comes from the use of a BDT for event selection, not from the inclusion of additional kinematic variables.

| Event Selection Criteria |
|---|
| $\geq 2$ photons, with $p_T > 25$ GeV, $\lvert\eta\rvert < 1.37$ or $1.52 < \lvert\eta\rvert < 2.37$ (photon trigger) |
| $\geq 2$ isolated photons, with $p_T > 30$ GeV, $\lvert\eta\rvert < 1.37$ or $1.52 < \lvert\eta\rvert < 2.37$, leading $p_T > 43$ GeV |
| $\geq 2$ jets identified as $b$-jets with $p_T > 35$ GeV, $\lvert\eta\rvert < 2.5$ |
| $\leq 5$ jets with $p_T > 30$ GeV, $\lvert\eta\rvert < 2.5$ |
| No isolated electrons with $p_T > 30$ GeV, $\lvert\eta\rvert < 1.37$ or $1.52 < \lvert\eta\rvert < 2.37$<br>No isolated muons with $p_T > 25$ GeV, $0.1 < \lvert\eta\rvert < 2.5$ |

Table 9: Event pre-selection criteria applied in the analysis.

| Variable | Description |
|---|---|
| $\Delta R_{b\bar{b}\gamma\gamma}$ | Separation between Higgs boson candidates |
| $p_{Tb\bar{b}\gamma\gamma}$ | $p_T$ of Higgs-boson-pair candidate |
| $\Delta R_{b\bar{b}}$ | Separation between $b$-jets |
| $p_{Tb\bar{b}}$ | $p_T$ of $b\bar{b}$ Higgs boson candidate |
| $m_{b\bar{b}}$ | Invariant mass of $b\bar{b}$ Higgs candidate |
| $\Delta R_{\gamma\gamma}$ | Separation between photons |
| $p_{T\gamma\gamma}$ | $p_T$ of $\gamma\gamma$ Higgs candidate |
| $\eta_{\gamma\gamma}$ | $\eta$ of $\gamma\gamma$ Higgs candidate |
| $p_{Tb1}$ | $p_T$ of leading $b$-jet |
| $p_{Tb2}$ | $p_T$ of sub-leading $b$-jet |
| $p_{T\gamma 1}$ | $p_T$ of leading photon |
| $p_{T\gamma 2}$ | $p_T$ of sub-leading photon |
| $\cos(\theta^*)_{b\bar{b}}$ | Opening angle in $b\bar{b}$ frame |
| $\cos(\theta^*)_{\gamma\gamma}$ | Opening angle in $\gamma\gamma$ frame |
| $\cos(\theta^*)_{(b\bar{b})(\gamma\gamma)}$ | Opening angle in $b\bar{b}\gamma\gamma$ frame |
| $HT_{30}$ | Scalar sum of $p_T$ for all jets (before selections) with $p_T > 30$ GeV |
| $HT_{Central}$ | Scalar sum of $p_T$ for all jets (before selections) with $\lvert\eta\rvert < 2.37$ |
| $MHT_{30}$ | $\sqrt{\sum p_x^2 + \sum p_y^2}$ of all final-state objects with $p_T > 30$ GeV |
| massAllJets | Invariant mass of sum of four vectors of all jets in final-state combination |
| $n_J$ | Number of jets with $p_T > 20$ GeV and $\lvert\eta\rvert < 4.9$ |
| $n_B$ | Number of $b$-jets with $p_T > 20$ GeV and $\lvert\eta\rvert < 4.9$ |

Table 10: Summary of kinematic variables used when training the BDT.

The signal and background samples are split into two equal subsets with events selected at random – one subset is used to train the BDT, the other subset is used to evaluate the performance of the BDT following its creation. TMVA then evaluates the signal and background efficiencies as a function of the BDT response cut. The significance at each possible cut value is evaluated using the standard Asimov approximation [44], which provides the median significance $z_0$ in the hypothesis of $s$ signal and $b$ background events:

$$\text{median}[z_0|s+b] \approx \sqrt{2\left((s+b)\ln(1+s/b) - s\right)}. \tag{5}$$

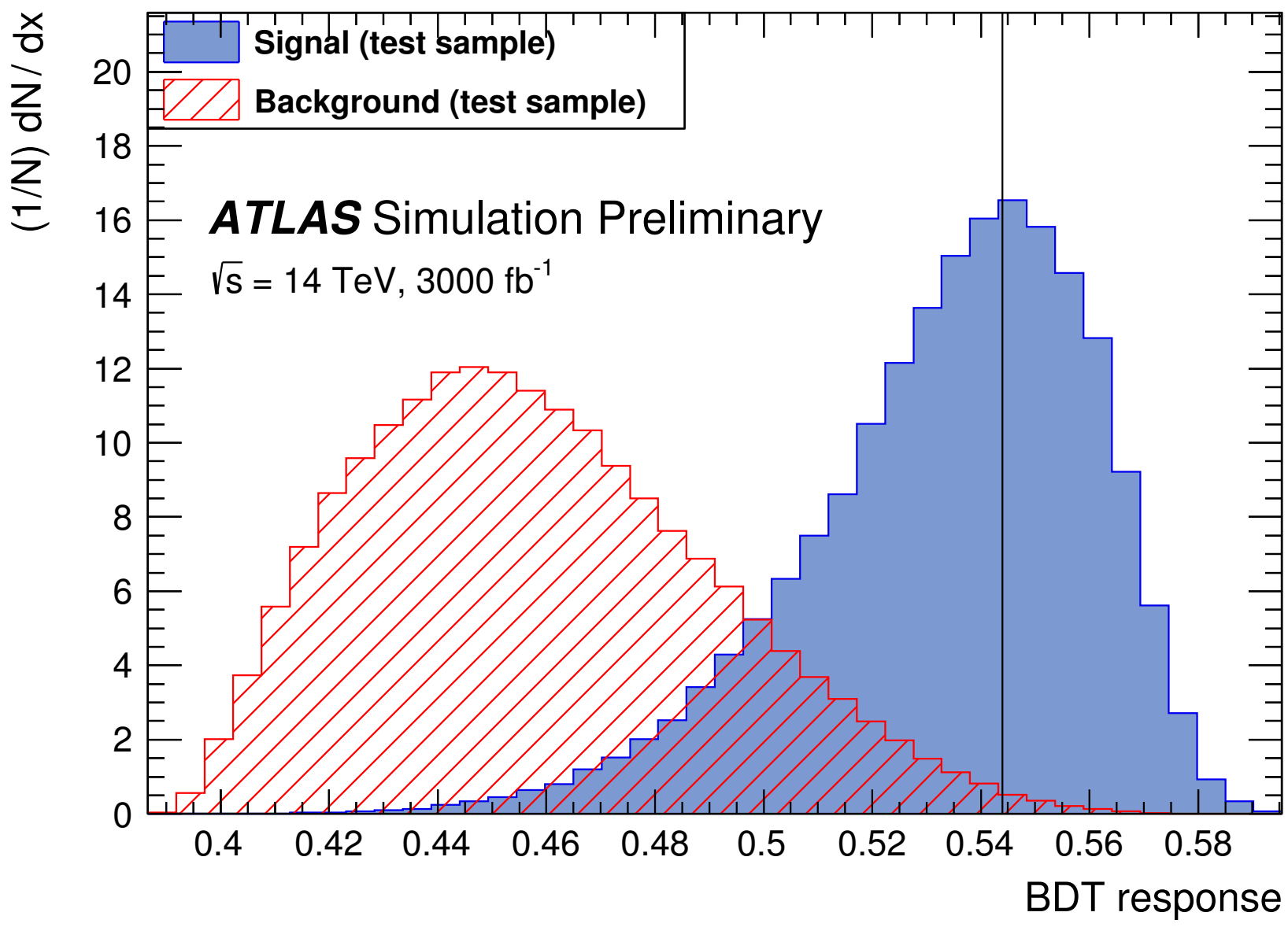

Figure 16: BDT response for signal and background test samples. The vertical line denotes the optimal cut on the BDT response that maximises the statistical-only significance, and is used in the remainder of the analysis.

The BDT response value that maximises the statistical-only significance is found to be 0.54, and is used in the following. The BDT response for the signal and background test samples is shown in Figure 16.

The response of this BDT is evaluated for every final-state combination in the full background samples, ignoring the earlier requirement of a minimum event weight. The distribution of $m_{\gamma\gamma}$ for signal and background events passing the BDT response cut is shown in Figure 17. The distribution of $m_{HH}$ for signal and background events passing the BDT response cut and a 123 GeV $< m_{\gamma\gamma} <$ 127 GeV mass cut is shown in Figure 18.

The number of events passing both the BDT response cut and the 123 GeV $< m_{\gamma\gamma} <$ 127 GeV mass cut are shown in Table 11. These two selections provide 6.46 signal events and a total of 6.8 background events. The relative MC statistical uncertainty on the selection efficiency of the pre-selection, the BDT response cut and the 123 GeV $< m_{\gamma\gamma} <$ 127 GeV mass cut is 1.2% for the sum of all single-Higgs-boson background processes and 1.9% for the sum of all continuum background processes.

Figure 19 shows the signal acceptance times selection efficiency as a function of $\kappa_\lambda$ for events passing the pre-selection criteria and the BDT response cut. The shape of this distribution is dominated by the acceptance, which is in turn affected by the interference between the box and triangle diagrams in the $HH$ production. Events are most boosted at $\kappa_\lambda \approx 2$ where the interference is greatest and the Higgs-boson-pair production cross-section is minimal.

### 4.5 Systematic Uncertainties

The systematic uncertainties considered here follow the recommendations for HL-LHC studies [26]. Experimental uncertainties are taken from Run 2 results [53] with scaling factors corresponding to the expected improvements. In the Run 2 analysis theory uncertainties are taken from Ref. [20], except for the QCD scale uncertainty assigned to the ggF production mode where a conservative uncertainty of 100% is considered, motivated by studies of heavy-flavour production in association with top-quark pairs and $W$ boson production in association with $b$-jets [53]. Those theory uncertainties are halved, following the prescriptions of Ref.[26]. The systematic uncertainties are summarised in Table 12.

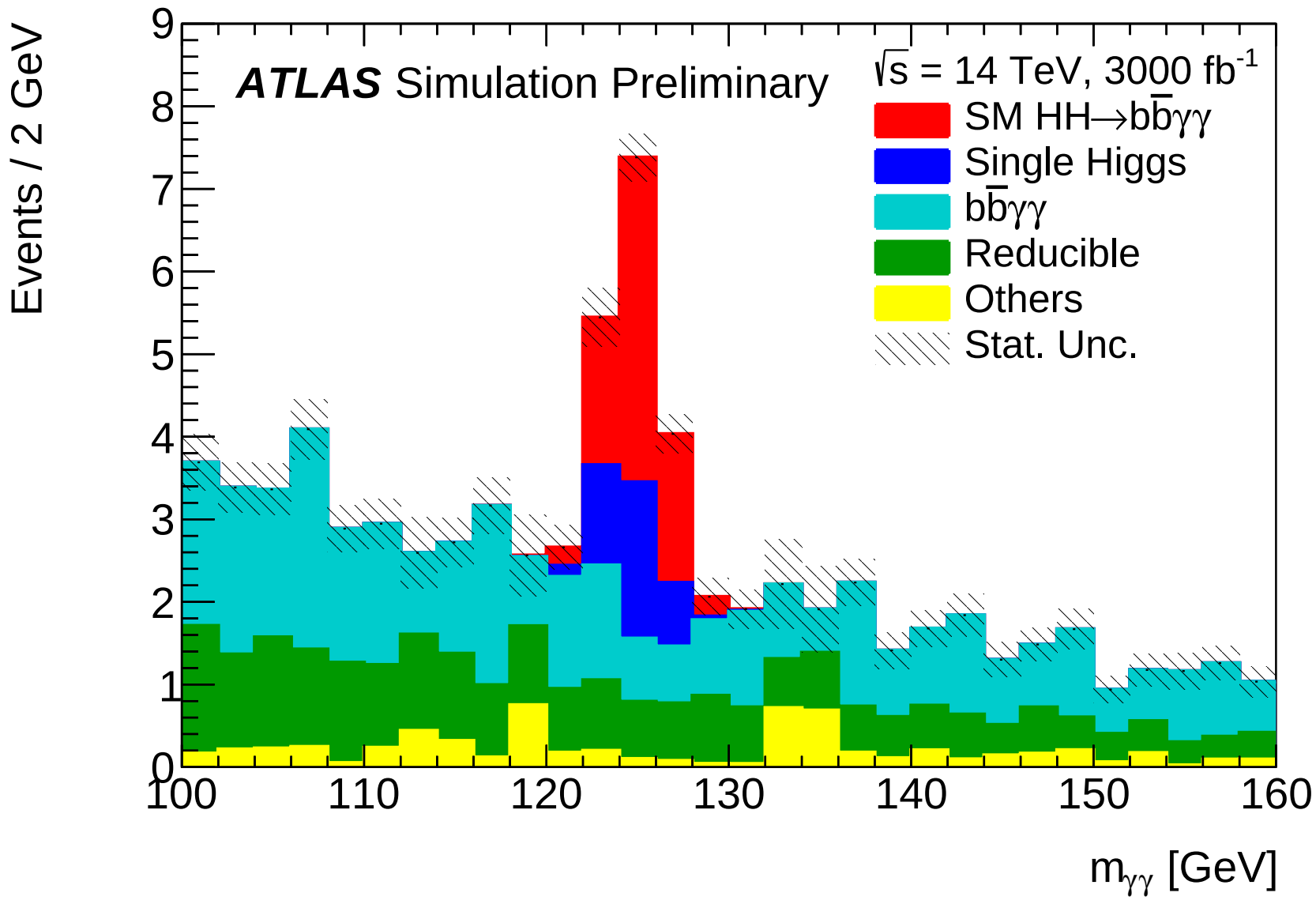

Figure 17: Distribution of $m_{\gamma\gamma}$ following the BDT response cut. The reducible background processes consist of $c\bar{c}\gamma\gamma$, $jj\gamma\gamma$, $b\bar{b}j\gamma$, $c\bar{c}j\gamma$, and $b\bar{b}jj$ events. Other background processes come from $Z(b\bar{b})\gamma\gamma$, $t\bar{t}$ and $t\bar{t}\gamma$.

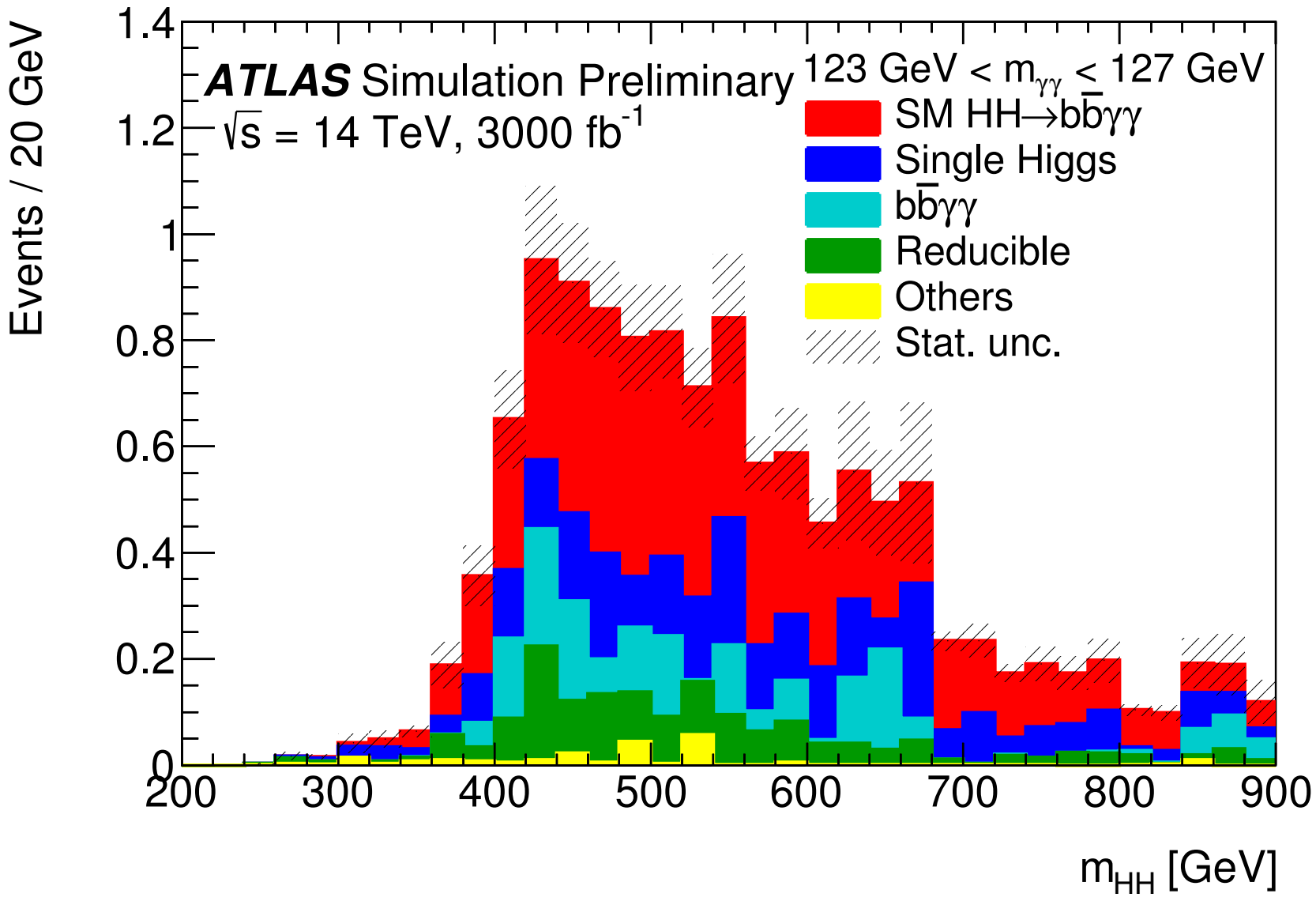

Figure 18: Distribution of $m_{HH}$ for the events passing the BDT selection and the requirement 123 GeV $< m_{\gamma\gamma} <$ 127 GeV. The reducible background processes consist of $c\bar{c}\gamma\gamma$, $jj\gamma\gamma$, $b\bar{b}j\gamma$, $c\bar{c}j\gamma$, and $b\bar{b}jj$ events. Other background processes come from $Z(b\bar{b})\gamma\gamma$, $t\bar{t}$ and $t\bar{t}\gamma$.

For the purpose of this prospect study, a cut-and-count analysis is performed on events within the

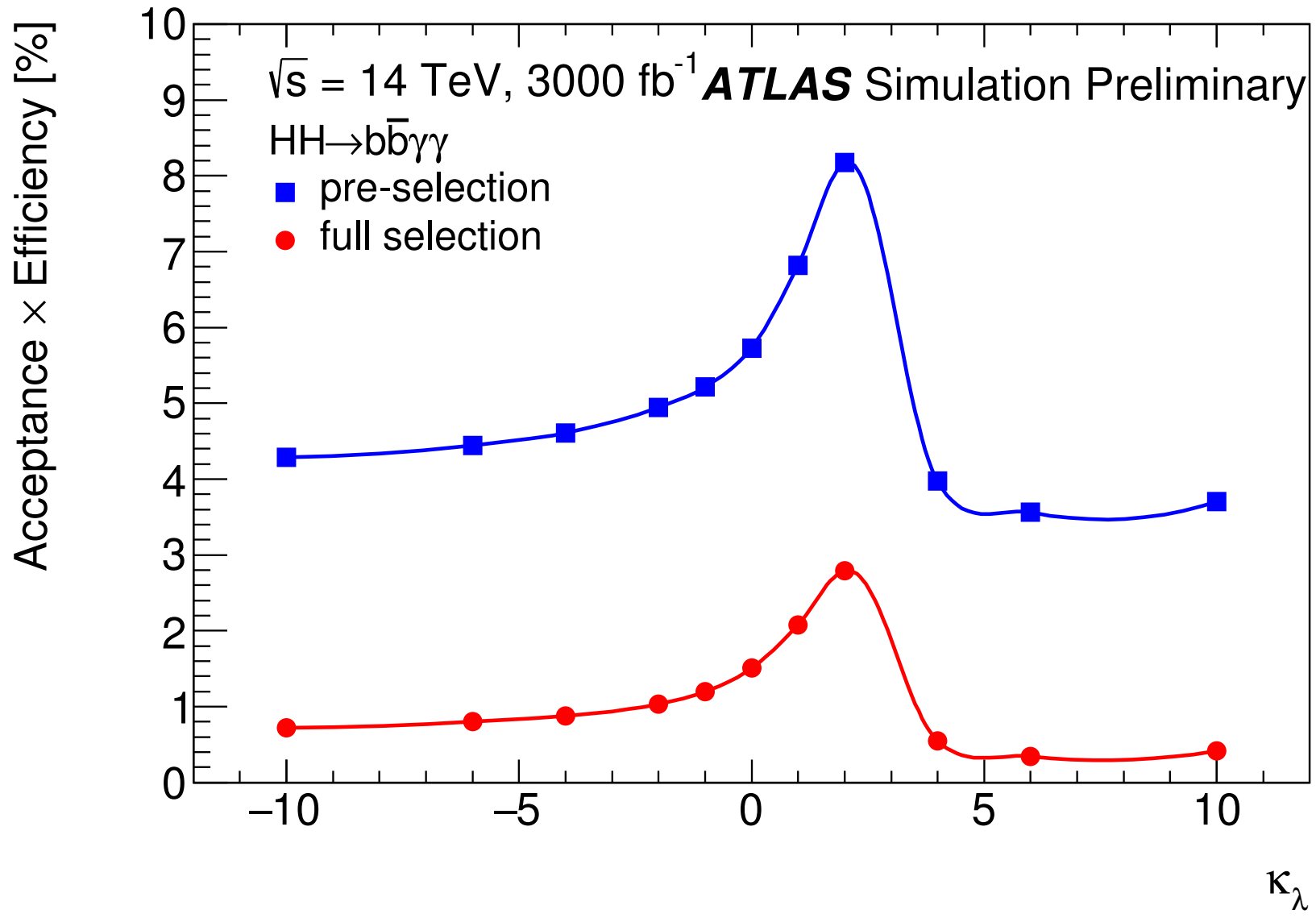

Figure 19: Signal acceptance times efficiency as a function of $\kappa_\lambda$.

123 GeV < $m_{\gamma\gamma}$ < 127 GeV mass region, and fitting the $m_{b\bar{b}\gamma\gamma}$ shape. However, with data the continuum background would be extracted from a fit to the $m_{\gamma\gamma}$ distribution. It is expected that the uncertainty introduced by such a fit will be sub-leading compared to the current statistical uncertainty, and therefore it is not considered here. For this reason no theory or experimental uncertainty has been considered on the expected yield for the continuum background passing the analysis and BDT selection, nor on the $m_{b\bar{b}\gamma\gamma}$ shape used later to compute the significance and constrain the value of $\kappa_\lambda$.

### 4.6 Results

The expected significance in this channel has been evaluated to be 2.0 (2.1) standard deviations with (without) systematic uncertainties. The degradation of the non-$b$-jet fake rates to its Run 2 performance (corresponding to the increase of the light- and $c$-jet fake rates by factors of 1.8 and 2.5, respectively) would result in a relative decrease by 5% of this significance. The expected 95% CL upper limit on the relative signal strength $\sigma_{HH}/\sigma_{HH}^{SM}$ is found to be 1.2 (1.1) with (without) systematic uncertainties.

The sensitivity of the $b\bar{b}\gamma\gamma$ analysis to $\kappa_\lambda$ is assessed as follows. Events passing the BDT response cut and having $m_{\gamma\gamma}$ falling in a narrow window of 123 GeV to 127 GeV are selected, and the resultant distribution of $m_{HH}$ is split into eight bins over the range 320 GeV < $m_{HH}$ < 900 GeV. The same morphing technique that was used in the $HH \to b\bar{b}b\bar{b}$ analysis is applied to generate a signal distribution at any arbitrary value of $\kappa_\lambda$. The distributions of $m_{HH}$ for various values of $\kappa_\lambda$ are shown in Figure 20. The $m_{HH}$ distributions were generated in the range $-20 \le \kappa_\lambda \le 20$.

Subsequently an Asimov dataset with the predicted SM $HH$ signal was created, and maximum likelihood fits to this dataset were performed for different $\kappa_\lambda$ hypotheses, combining all eight $m_{HH}$ bins. Figure 21 shows the resulting negative natural logarithm of the ratio of the maximum likelihood for $\kappa_\lambda$ to that for the fit with $\kappa_\lambda = 1$ for two scenarios: no systematic uncertainties and with systematic uncertainties. The $1\sigma$ and $2\sigma$ CIs are reported in Table 13. The negative natural logarithm of the ratio of the maximum likelihood for $\kappa_\lambda$ to that for the fit with $\kappa_\lambda = 0$ is shown in Figure 22 and the constraints on $\kappa_\lambda$ are reported in Table 14.

| Process | Events in sample | Events after pre-selection | Events passing BDT response | Events passing BDT response & 123 GeV $< m_{\gamma\gamma} <$ 127 GeV |
|---|---|---|---|---|
| $H(b\bar{b})H(\gamma\gamma)$, $\kappa_\lambda = 1$ | $3.0 \times 10^2$ | 20 | 8.0 | 6.46 |
| $gg \to H(\to \gamma\gamma)$ | $3.0 \times 10^5$ | 28 | 0.85 | 0.68 |
| $t\bar{t}H(\to \gamma\gamma)$ | $4.2 \times 10^3$ | 124 | 1.9 | 1.51 |
| $ZH(\to \gamma\gamma)$ | $6.7 \times 10^3$ | 26 | 1.33 | 0.93 |
| $b\bar{b}H(\to \gamma\gamma)$ | $3.8 \times 10^3$ | 3.7 | 0.028 | 0.025 |
| Single-Higgs-boson background | $3.2 \times 10^5$ | 182 | 4.1 | 3.2 |
| $b\bar{b}\gamma\gamma$ | $4.3 \times 10^5$ | 10100 | 92 | 1.9 |
| $c\bar{c}\gamma\gamma$ | $3.4 \times 10^6$ | 630 | 2.7 | 0.06 |
| $jj\gamma\gamma$ | $4.8 \times 10^7$ | 690 | 4.6 | 0.12 |
| $b\bar{b}j\gamma$ | $1.1 \times 10^9$ | 14000 | 130 | 1.16 |
| $c\bar{c}j\gamma$ | $3.3 \times 10^9$ | 480 | 2.5 | 0.021 |
| $b\bar{b}jj$ | $1.4 \times 10^{12}$ | 3600 | 26 | 0.16 |
| $Z(\to b\bar{b})\gamma\gamma$ | $1.5 \times 10^4$ | 230 | 4.5 | 0.10 |
| $t\bar{t}(\geq 1$ lepton) | $1.6 \times 10^9$ | 3530 | 11.3 | 0.05 |
| $t\bar{t}\gamma(\geq 1$ lepton) | $1.5 \times 10^7$ | 5600 | 23 | 0.07 |
| Continuum background | $1.4 \times 10^{12}$ | 38900 | 297 | 3.7 |
| Total background | $1.4 \times 10^{12}$ | 39100 | 301 | 6.8 |

Table 11: Number of events passing the pre-selection criteria, the BDT response cut, and passing the additional requirement of 123 GeV $< m_{\gamma\gamma} <$ 127 GeV. The number of background events was obtained by counting final-state combinations passing the selection criteria in samples that were generated using a single random seed for the smearing functions. All numbers are normalised to 3000 fb$^{-1}$. The totals appear inconsistent due to rounding.

It can be seen in Figures 21 and 22 that there are two minima. The first minimum is located at $\kappa_\lambda = 1$ or 0, respectively, as the signal hypothesis used in the maximum likelihood fit corresponds to the signal used to create the Asimov dataset. The second minimum is observed at the $\kappa_\lambda$ value for which the signal has a similar yield as that of the first minimum. The degeneracy of the two minima is lifted because of the different shapes of the $m_{HH}$ distribution at each of the two minima.

It should be noted that since the BDT was trained on the SM $HH$ signal only, the above constraints on $\kappa_\lambda$ are slightly pessimistic. Using separate BDTs trained on specific values of $\kappa_\lambda$ would bring slight improvements in the expected limit at high positive values of $\kappa_\lambda$, but have no impact at negative values of $\kappa_\lambda$.

# 5 Statistical Combination

Results based on the statistical combination of the $HH \to b\bar{b}b\bar{b}$, $HH \to b\bar{b}\tau^+\tau^-$ and $HH \to b\bar{b}\gamma\gamma$ channels are discussed in the following.

| **Experimental uncertainties** | | | | |
|---|---|---|---|---|
| Source | $HH$ signal | | Single-$H$ background | |
| *Yield* | | | | |
| Luminosity | ±1.0% | | ±1.0% | |
| Pile-up | ±1.6% | | ±1.0% | |
| Trigger | ±0.4% | | ±0.4% | |
| Photon identification | ±2.5% | | ±1.7% | |
| Photon isolation | ±0.8% | | ±0.8% | |
| Jet energy scale | ±0.8% | | ±1.5% | |
| Jet energy resolution | ±2.9% | | ±7.8% | |
| $b$-jet tagging | ±0.8% | | ±0.8% | |
| $c$-jet tagging | ±0.03% | | ±0.6% | |
| light-jet tagging | ±0.03% | | ±0.5% | |
| *Signal modelling* | | | | |
| Photon energy scale | ±0.6% | | ±0.6% | |
| Photon energy resolution | ±14% | | ±14% | |
| **Theoretical uncertainties** | | | | |
| *Cross-section* | | | | |
| Source | $ZH$ | ggF | $t\bar{t}H$ | $HH$ signal |
| QCD scale up | +3.0% | +50% | +1.9% | +2.5% |
| QCD scale down | −4.6% | −50% | −1.7% | −2.0% |
| PDF | ±1.4% | ±1.3% | ±0.64% | ±1.3% |
| *Branching ratios* | | | | |
| | $H \to \gamma\gamma$ | $H \to b\bar{b}$ | | |
| | 1.43% | 0.85% | | |

Table 12: In the top part, the dominant systematic uncertainties affecting the expected yield and the shape of the resonances are summarised. Sources not listed in the table are negligible by comparison. In the bottom part, theoretical uncertainties considered in the analysis are summarised. Uncertainties in both the Higgs-boson-pair signal and SM single-Higgs-boson background are presented.

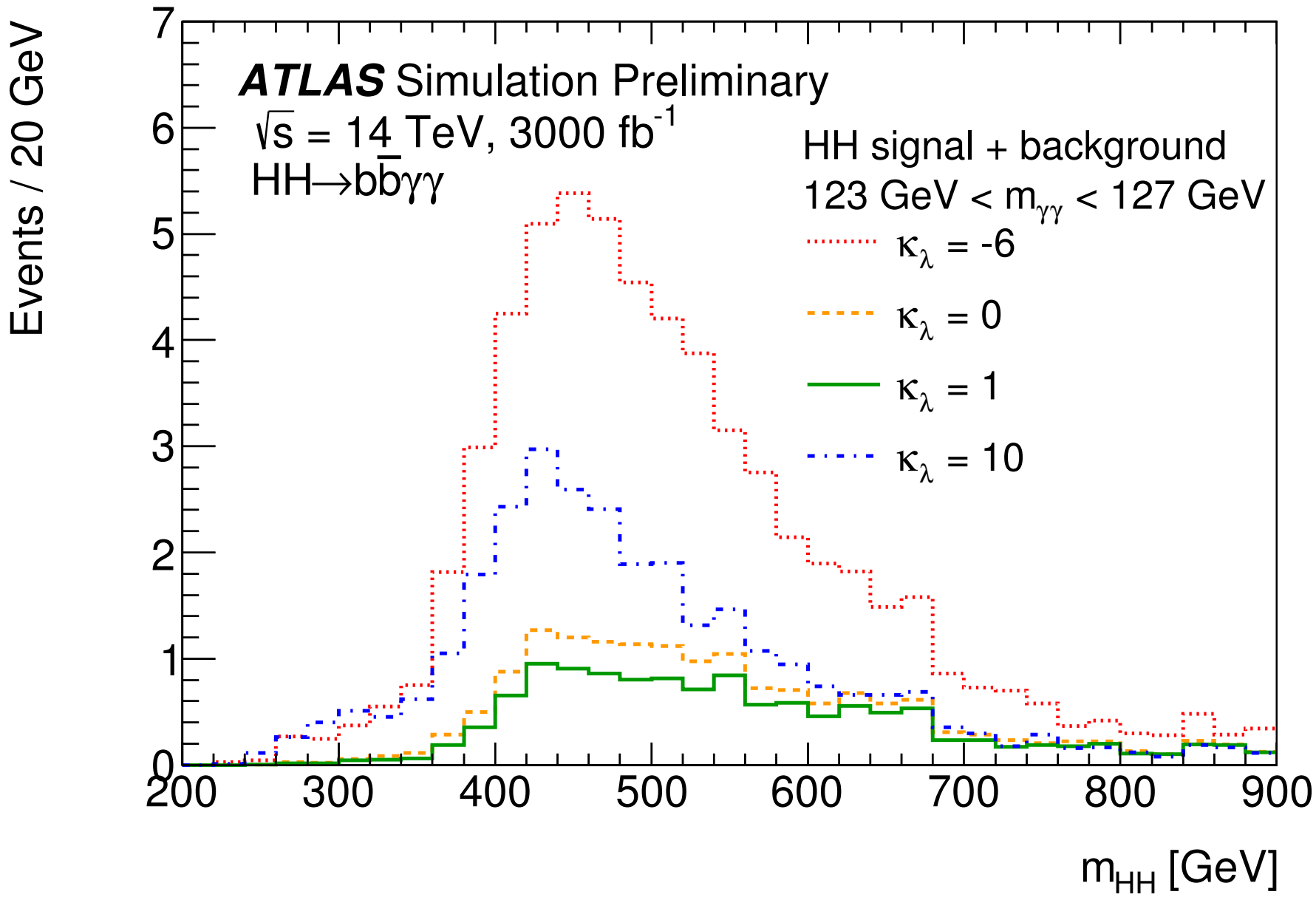

Figure 20: Distributions of $m_{HH}$ for combined signal and background events passing the BDT selection and the requirement 123 GeV < $m_{\gamma\gamma}$ < 127 GeV, for various values of $\kappa_\lambda$.

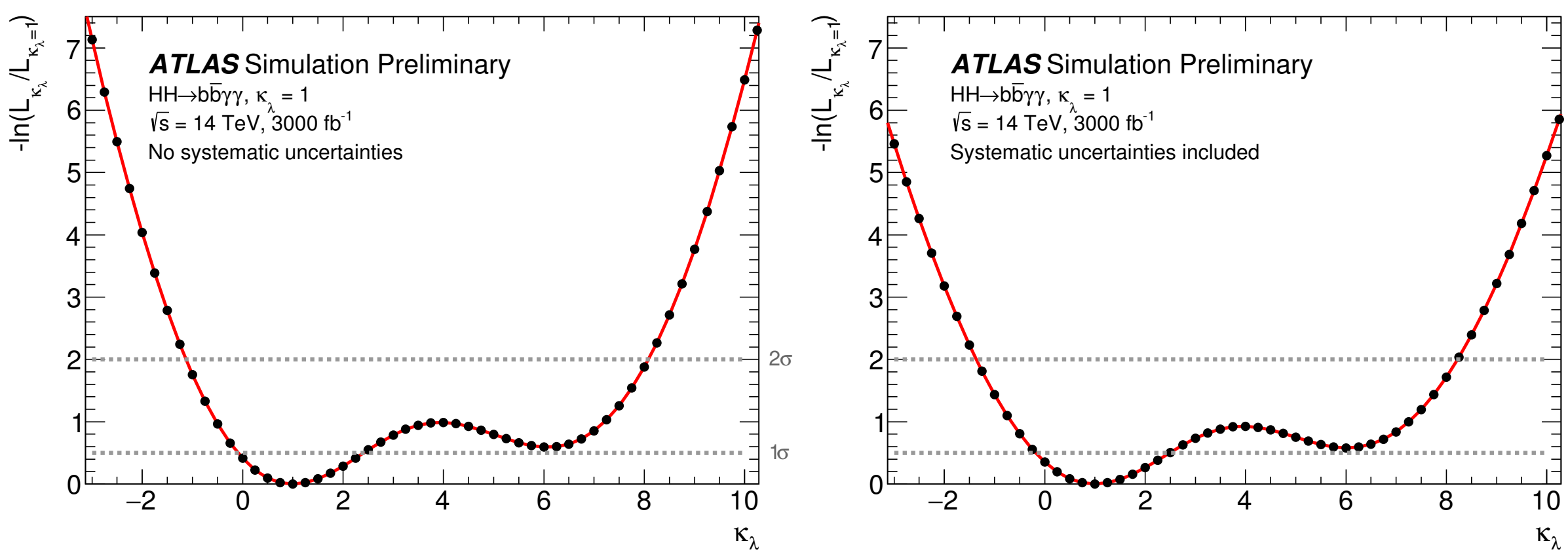

Figure 21: Negative natural logarithm of the ratio of the maximum likelihood for $\kappa_\lambda$ to the maximum likelihood for $\kappa_\lambda = 1$ for (left) the fits with only statistical uncertainties and (right) the fits with both statistical and systematic uncertainties. The dashed lines at $-\ln(L_{\kappa_\lambda}/L_{\kappa_\lambda=1}) = 0.5$ and 2.0 indicate the values corresponding to a $1\sigma$ and $2\sigma$ confidence interval, respectively (assuming an asymptotic $\chi^2$ distribution of the test statistic).

| Scenario | $1\sigma$ CI | $2\sigma$ CI |
|---|---|---|
| No systematic uncertainties | $-0.1 < \kappa_\lambda < 2.4$ | $-1.1 < \kappa_\lambda < 8.1$ |
| Systematic uncertainties included | $-0.2 < \kappa_\lambda < 2.5$ | $-1.4 < \kappa_\lambda < 8.2$ |

Table 13: Constraints on $\kappa_\lambda$ from the likelihood ratio test performed on the Asimov dataset created from the backgrounds and the SM $HH$ signal, as shown in Figure 21. Results are presented as a $1\sigma$ and $2\sigma$ CI on $\kappa_\lambda$ when considering only statistical uncertainties and when including systematic uncertainties.

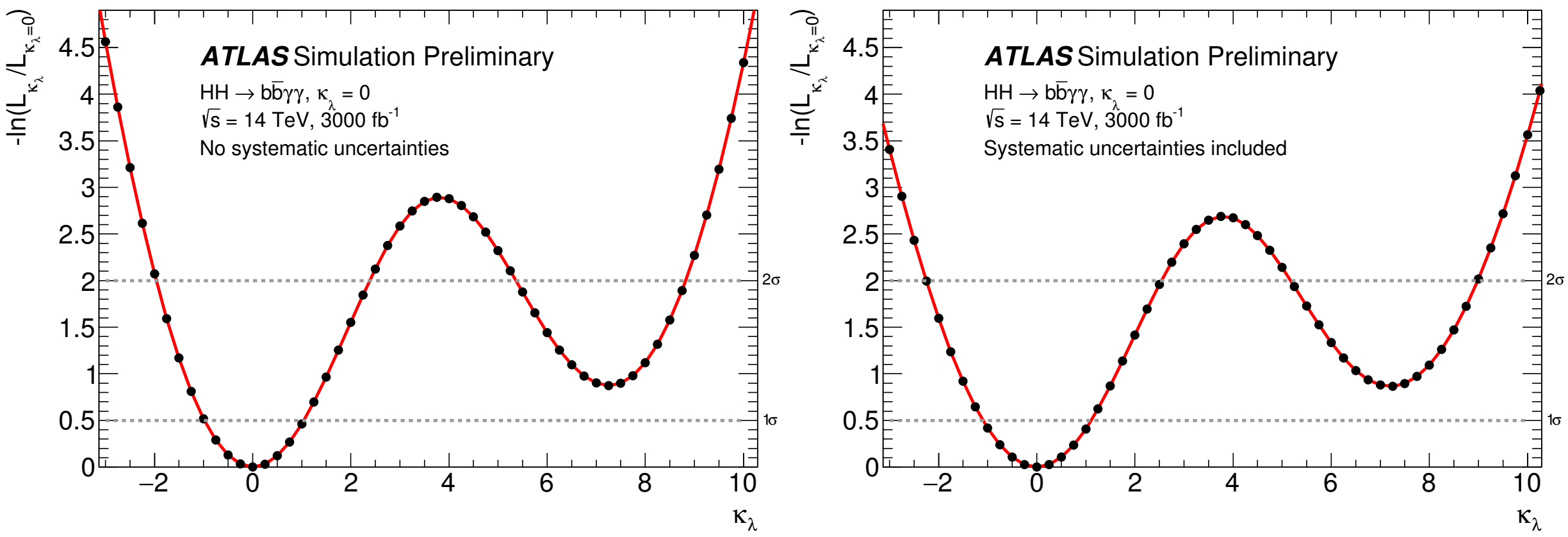

Figure 22: Negative natural logarithm of the ratio of the maximum likelihood for $\kappa_\lambda$ to the maximum likelihood for the $\kappa_\lambda = 0$ case for (left) the fits with only statistical uncertainties and (right) the fits with both statistical and systematic uncertainties. The dashed lines at $-\ln(L_{\kappa_\lambda}/L_{\kappa_\lambda=1}) = 0.5$ and 2.0 indicate the values corresponding to a $1\sigma$ and $2\sigma$ confidence interval, respectively (assuming an asymptotic $\chi^2$ distribution of the test statistic).

| Scenario | $1\sigma$ CI | $2\sigma$ CI |
|---|---|---|
| No systematic uncertainties | $-1.0 \leq \kappa_\lambda \leq 1.0$ | $-2.0 \leq \kappa_\lambda \leq 2.4 \cup 5.4 \leq \kappa_\lambda \leq 8.8$ |
| Systematic uncertainties included | $-1.1 \leq \kappa_\lambda \leq 1.1$ | $-2.3 \leq \kappa_\lambda \leq 2.5 \cup 5.2 \leq \kappa_\lambda \leq 9.0$ |

Table 14: Constraints on $\kappa_\lambda$ from the likelihood ratio test performed on the Asimov dataset created from the backgrounds and the $\kappa_\lambda = 0$ signal, as shown in Figure 22. Results are presented as a $1\sigma$ and $2\sigma$ CI on $\kappa_\lambda$ when considering only statistical uncertainties and when including systematic uncertainties.

The combination of various channels is realised by constructing a combined likelihood function that takes into account pseudo-data, signal and background models, and correlated systematic uncertainties from all channels. All the signal regions considered in the simultaneous fit either are orthogonal by construction or have negligible overlap due to different selection criteria.

Setting appropriate nuisance parameters to be correlated with one another induces a negligible change in the combination results compared to assuming all nuisance parameters are uncorrelated. Accordingly, only those nuisance parameters relating to Run 2 detector performance are correlated between the $HH \to b\bar{b}b\bar{b}$ and $HH \to b\bar{b}\tau^+\tau^-$ channels in the following results. There are 120 nuisance parameters included in the statistical analysis. No strong correlations between any of these nuisance parameters are found in the fit, with the exception of some correlation (up to about ±35%) between background-modelling nuisance parameters in the $HH \to b\bar{b}b\bar{b}$ analysis and between background-modelling nuisance parameters in the $HH \to b\bar{b}\tau^+\tau^-$ analysis. These correlations are also observed in the individual HL-LHC extrapolations. Theoretical uncertainties on the signal cross-section have negligible impact on the combined result.

Projected upper limits on the signal cross-section are set, using Asimov datasets that contain no signal. For the SM $HH$ signal, the 95% CL upper limit is 20.7 fb ($\mu = \sigma_{HH}/\sigma_{HH}^{SM} < 0.56$) when systematics are not included and 24.9 fb ($\mu < 0.68$) when they are. For the signal with $\kappa_\lambda = 0$, i.e. no self-coupling, the limits are 27.8 fb ($\mu < 0.36$) and 33.1 fb ($\mu < 0.42$), respectively.

Table 15 shows the expected median significance of the SM $HH$ signal relative to the background-only hypothesis for the individual channels and their combination. With 3000 fb$^{-1}$ of data, a SM $HH$ signal is expected to yield a significance of $3.0\sigma$ over the background-only expectation. The discrepancy when fitting the signal-plus-background model with $\kappa_\lambda = 0$ to pseudo-data generated with the SM $HH$ signal ($\kappa_\lambda = 1$) has a significance of $1.4\sigma$ ($1.8\sigma$) for the case with (without) systematic uncertainties.

| Channel | Statistical-only | Statistical + Systematic |
|---|---|---|
| $HH \to b\bar{b}b\bar{b}$ | 1.4 | 0.61 |
| $HH \to b\bar{b}\tau^+\tau^-$ | 2.5 | 2.1 |
| $HH \to b\bar{b}\gamma\gamma$ | 2.1 | 2.0 |
| Combined | 3.5 | 3.0 |

Table 15: Significance of the individual $HH \to b\bar{b}b\bar{b}$, $HH \to b\bar{b}\tau^+\tau^-$ and $HH \to b\bar{b}\gamma\gamma$ channels as well as their combination.

Table 16 shows the signal strength measured in the individual channels, as well as the combination, when the SM $HH$ signal is injected.

The combined sensitivity of the three analyses to $\kappa_\lambda$ is assessed by generating an Asimov dataset containing the backgrounds plus SM $HH$ signal. The negative natural logarithm of the ratio of the maximum likelihood fit for $\kappa_\lambda$ to that for the fit with $\kappa_\lambda = 1$ is shown in Figure 23. From these curves, the confidence intervals for $\kappa_\lambda$ reported in Table 17 are extracted. A second test of the sensitivity of the three analyses to $\kappa_\lambda$ was performed, in this case generating an Asimov dataset containing the backgrounds plus $\kappa_\lambda = 0$ signal (i.e. no Higgs boson self-coupling). The negative natural logarithm of the ratio of the maximum likelihood fit for $\kappa_\lambda$ to that for the fit with $\kappa_\lambda = 0$ is shown in Figure 24. From these curves, the confidence intervals for $\kappa_\lambda$ are extracted and documented in Table 18.

| Channel | Measured $\mu$ (Statistical-only) | Measured $\mu$ (Statistical + Systematic) |
|---|---|---|
| $HH \to b\bar{b}b\bar{b}$ | $1.0 \pm 0.6$ | $1.0 \pm 1.6$ |
| $HH \to b\bar{b}\tau^+\tau^-$ | $1.0 \pm 0.4$ | $1.0 \pm 0.5$ |
| $HH \to b\bar{b}\gamma\gamma$ | $1.0 \pm 0.6$ | $1.0 \pm 0.6$ |
| Combined | $1.00 \pm 0.31$ | $1.0 \pm 0.4$ |

Table 16: Signal strength measured in the individual channels and their combination using an Asimov dataset with SM $HH$ signal injected.

| Scenario | $1\sigma$ CI | $2\sigma$ CI |
|---|---|---|
| Statistical uncertainties only | $0.4 \leq \kappa_\lambda \leq 1.7$ | $-0.10 \leq \kappa_\lambda \leq 2.7 \cup 5.5 \leq \kappa_\lambda \leq 6.9$ |
| Systematic uncertainties | $0.25 \leq \kappa_\lambda \leq 1.9$ | $-0.4 \leq \kappa_\lambda \leq 7.3$ |

Table 17: Constraints on $\kappa_\lambda$ from the likelihood ratio test performed on the Asimov dataset created from the backgrounds and the SM $HH$ signal. Results are presented as $1\sigma$ and $2\sigma$ CI on $\kappa_\lambda$.

The significance with which the Higgs boson pair production would be observed is shown as a function of $\kappa_\lambda$ in Figure 25. The significance depends on the expected signal yield and therefore it is lower for those $\kappa_\lambda$ values for which the cross-section and the acceptance times efficiency is low.

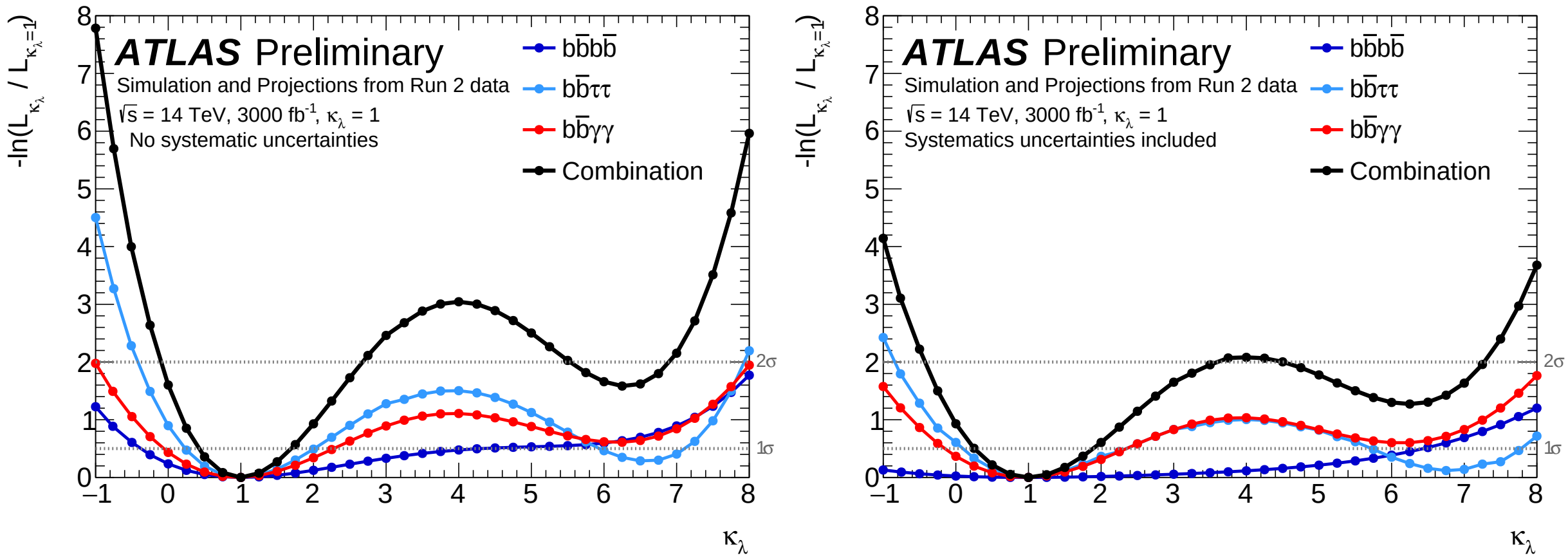

Figure 23: Negative natural logarithm of the ratio of the maximum likelihood for $\kappa_\lambda$ to the maximum likelihood for $\kappa_\lambda = 1$ for (left) the fits with only statistical uncertainties and (right) the fits with all systematic uncertainties as nuisance parameters. The black circles show the results for the combination, while the coloured markers show the values coming from the individual channels. The dashed lines at $-\ln\left(L_{\kappa_\lambda}/L_{\kappa_\lambda=1}\right) = 0.5$ and 2.0 indicate the values corresponding to the $1\sigma$ and $2\sigma$ confidence intervals, respectively (assuming an asymptotic $\chi^2$ distribution of the test statistic).

| Scenario | $1\sigma$ CI | $2\sigma$ CI |
|---|---|---|
| Statistical uncertainties only | $-0.5 \leq \kappa_\lambda \leq 0.5$ | $-0.9 \leq \kappa_\lambda \leq 1.1$ |
| Systematic uncertainties | $-0.6 \leq \kappa_\lambda \leq 0.7$ | $-1.3 \leq \kappa_\lambda \leq 1.5$ |

Table 18: Constraints on $\kappa_\lambda$ from the likelihood ratio test performed on the Asimov dataset created from the backgrounds and the $\kappa_\lambda = 0$ signal. Results are presented as $1\sigma$ and $2\sigma$ CI on $\kappa_\lambda$.

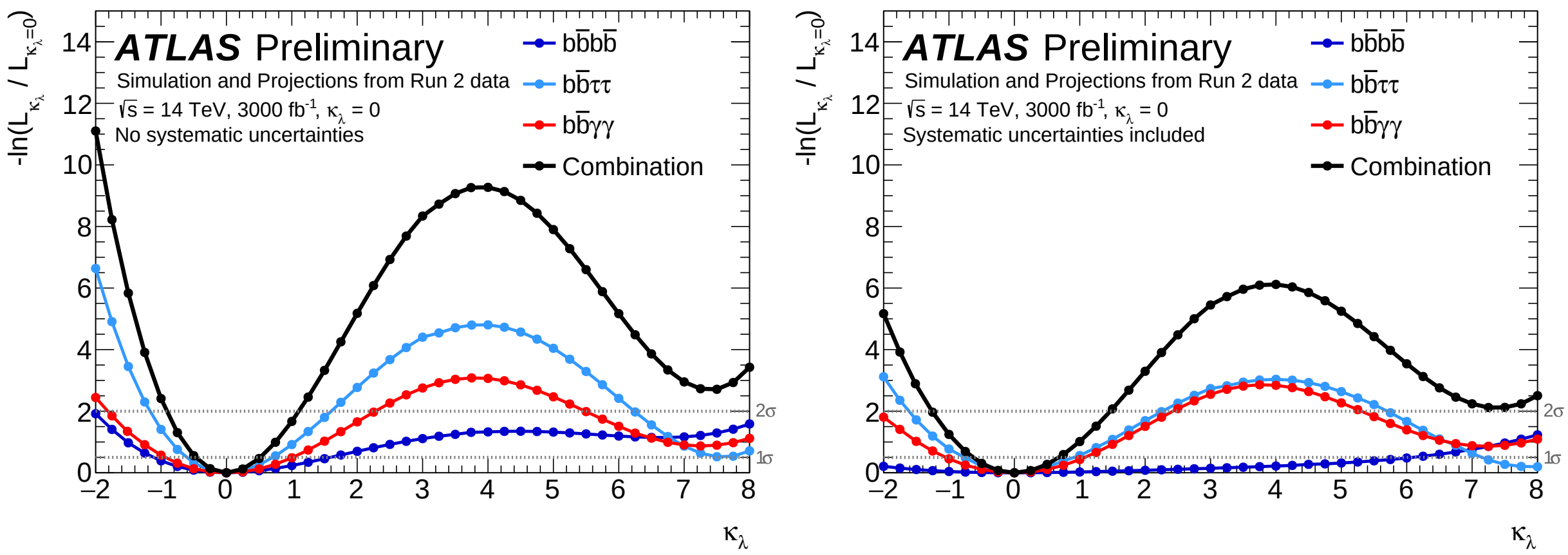

Figure 24: Negative natural logarithm of the ratio of the maximum likelihood for $\kappa_\lambda$ to the maximum likelihood for $\kappa_\lambda = 0$ for (left) the fits with only statistical uncertainties and (right) the fits with all systematic uncertainties as nuisance parameters. The black circles show the results for the combination, while the coloured markers show the values coming from the individual channels. The dashed lines at $-\ln\left(L_{\kappa_\lambda}/L_{\kappa_\lambda=0}\right) = 0.5$ and 2.0 indicate the values corresponding to the $1\sigma$ and $2\sigma$ confidence intervals, respectively (assuming an asymptotic $\chi^2$ distribution of the test statistic).

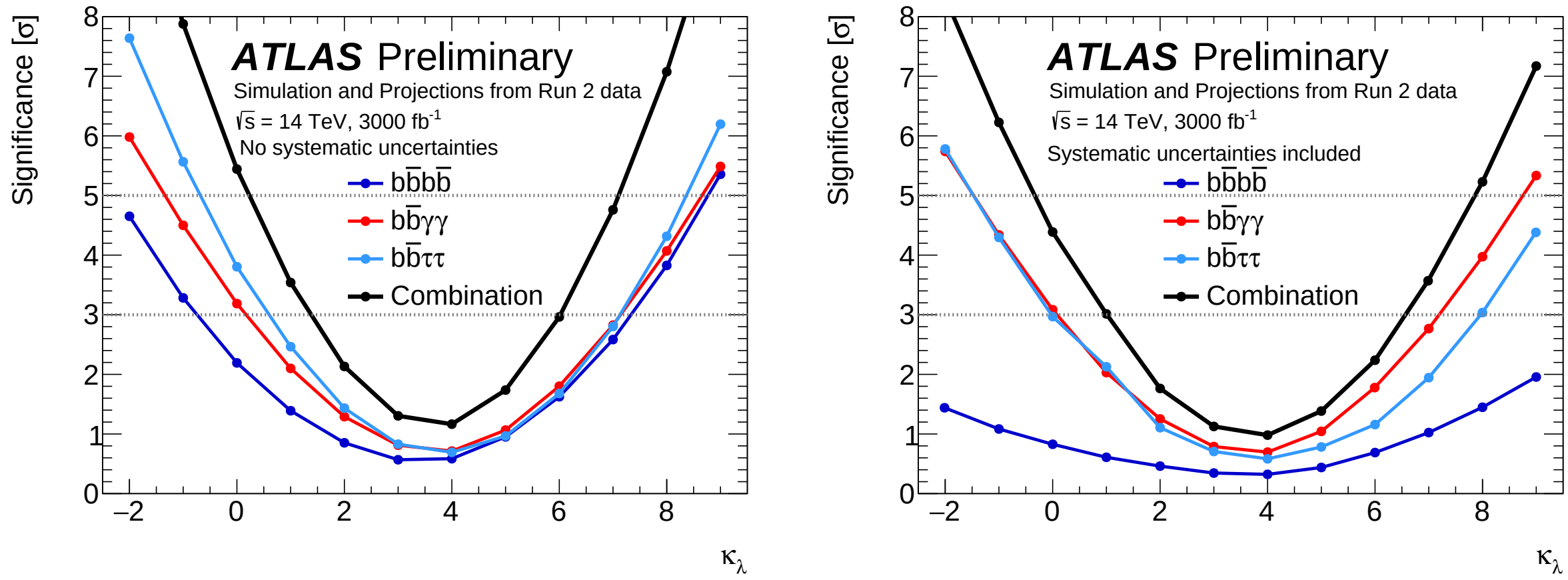

Figure 25: Expected significance of observing Higgs-boson-pair production for (left) the fits with only statistical uncertainties and (right) the fits with all systematic uncertainties as nuisance parameters. The two horizontal dashed lines show the $3\sigma$ and $5\sigma$ thresholds.

# 6 Projections for HE-LHC

The presented HL-LHC studies for the $b\bar{b}\gamma\gamma$ and $b\bar{b}\tau^+\tau^-$ final states were extended to provide first estimates of the prospects at the High Energy LHC (HE-LHC), assuming a centre-of-mass collision energy of 27 TeV and 15 ab$^{-1}$ of $pp$ collision data.

For the $b\bar{b}\gamma\gamma$ final state, the estimate is based on the results of the HL-LHC study (Section 4). Therefore, the event selection is based on the same BDT selection criteria as used in the HL-LHC analysis, which was optimised for achieving the highest significance for the SM $HH$ signal rather than the best sensitivity to BSM couplings. Furthermore, it is assumed that the detector performance will be the same as the HL-LHC ATLAS detector. This includes the same impact on the final-state particles from pile-up and on average the same number of jets arising from pile-up.

Comparisons between $\sqrt{s}$ = 14 TeV and $\sqrt{s}$ = 27 TeV simulations show that the kinematics at particle level of the $b$-jets and photons from the Higgs boson decay as well as the $m_{HH}$ distributions are similar. However, on average the photon and $b$-jet pairs from the Higgs boson decay tend to be more boosted and the $\eta$ spectrum of the decay particles tends to be pointing more frequently in the forward region, which will likely decrease the efficiency times acceptance by around 10%. This effect is not taken into account in the following, but it is not expected to have a significant impact on the results presented.

The event yields of each background sample, reported in Section 4.4, are scaled by the increase in luminosity and the ratio of the MC generator cross-sections at $\sqrt{s}$ = 27 TeV and $\sqrt{s}$ = 14 TeV. Such scaling is necessary since the background cross-section calculations that include the higher order corrections are not yet available for the $\sqrt{s}$ = 27 TeV centre-of-mass energy. The signal event yields are extrapolated in a similar way by increasing the luminosity and normalising the event yields to the SM $HH$ production cross-section of 139.9 fb, which corresponds to the latest LHC Higgs Cross-Section Working Group [19] recommendations. The yields are shown in Table 19.

Using the same method as in the $\sqrt{s}$ = 14 TeV analysis, the sensitivity of observing the SM $HH$ production and to various values of the Higgs boson self-coupling, $\kappa_\lambda$, is calculated. The sensitivity is extracted from the $m_{HH}$ distribution, which is constructed from the two leading $b$-jets and the two leading photons found in the event. The signal significance with respect to the background-only hypothesis is found to be 7.1$\sigma$ in the asymptotic approximation, assuming only statistical uncertainties. The systematic uncertainties are unknown for the HE-LHC. However, to get an idea of its impact, the significance would drop to 5.4$\sigma$ in case the same systematic uncertainties as for the HL-LHC were assumed. These uncertainties predominantly arise from the uncertainties in the photon energy scale. If these uncertainties were reduced by a factor of two a 5.9$\sigma$ significance would be achieved. Thus, $HH \to b\bar{b}\gamma\gamma$ production can be observed at the HE-LHC. The negative natural logarithm of the ratio of the maximum likelihood for $\kappa_\lambda$ to that for the fit with $\kappa_\lambda = 1$ is displayed in Figure 26. The value of $\kappa_\lambda$ is expected to be measured as $1 \pm 0.4$ if only statistical uncertainties are considered. Assuming the same systematic uncertainties (or photon energy resolution uncertainties reduced by a factor of two) as at the HL-LHC, the precision in measuring $\kappa_\lambda$ is around 0.50 (0.45). Although it is not possible to make a reliable prediction of the expected level of systematic uncertainties at HE-LHC, the high signal over background ratio close to 1 expected in the $HH \to b\bar{b}\gamma\gamma$ analysis will ensure a limited impact of background systematics on the signal significance. When only statistical uncertainties are considered, the $\kappa_\lambda = 0$ hypothesis can be rejected with a confidence level of 2.3$\sigma$.

Similarly, the expected HE-LHC sensitivity to a SM $HH$ signal in the $b\bar{b}\tau^+\tau^-$ final state is estimated by extrapolating the current Run 2 result [24] using the same statistical framework, based on the profile likelihood ratio [44]. Compared to the procedure described in Section 3.2, a more simplified extrapolation procedure is used. The Run 2 BDT distributions for signal and background are scaled to 15 $\text{ab}^{-1}$ by a single multiplicative factor, defined as the ratio of the target luminosity to the luminosity of the Run 2 result. The performance of the HE-LHC detector is assumed to be broadly similar to that of the current detector, with the exception that the efficiency to identify jets from b-quarks is expected to improve by 8% for a given light-jet rejection due to the upgraded inner tracker (similar to what has been assumed for the HL-LHC).

| Process | $\sigma$ [fb] | Number of events |
|---|---|---|
| $H(b\bar{b})H(\gamma\gamma)$, $\kappa_\lambda = 1$ | 0.37 | 122.4 |
| $gg \to H(\to \gamma\gamma)$ | $3.3 \times 10^2$ | 10.1 |
| $t\bar{t}H(\to \gamma\gamma)$ | 6.1 | 32.4 |
| $ZH(\to \gamma\gamma)$ | 5.3 | 13.7 |
| $b\bar{b}H(\to \gamma\gamma)$ | 2.9 | 0.4 |
| Single-Higgs-boson background | | 56.6 |
| $b\bar{b}\gamma\gamma$ | $4.0 \times 10^2$ | 26.6 |
| $c\bar{c}\gamma\gamma$ | $3.6 \times 10^3$ | 1.5 |
| $jj\gamma\gamma$ | $3.9 \times 10^4$ | 2.2 |
| $b\bar{b}j\gamma$ | $1.2 \times 10^6$ | 18.6 |
| $c\bar{c}j\gamma$ | $3.6 \times 10^6$ | 0.4 |
| $b\bar{b}jj$ | $1.2 \times 10^9$ | 1.7 |
| $Z(\to b\bar{b})\gamma\gamma$ | 10.3 | 1.0 |
| $t\bar{t}(\geq 1$ lepton) | $3.3 \times 10^6$ | 1.6 |
| $t\bar{t}\gamma(\geq 1$ lepton) | $3.1 \times 10^4$ | 2.1 |
| Continuum background | | 55.7 |

Table 19: Predicted production cross-sections and yields after event selection at the HE-LHC with 15 $ab^{-1}$ of data.

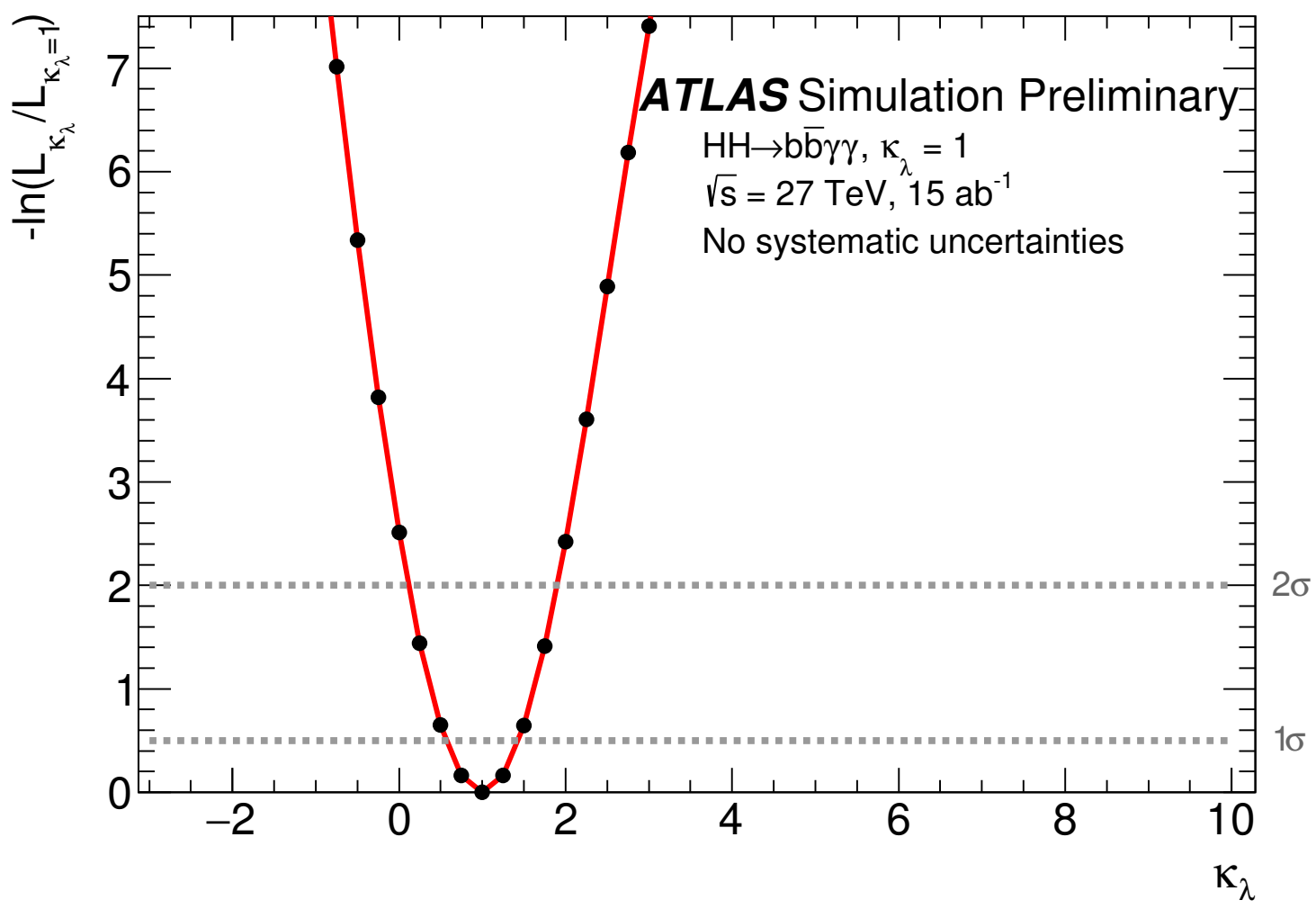

Figure 26: Negative natural logarithm of the ratio of the maximum likelihood for $\kappa_\lambda$ divided by the maximum likelihood for $\kappa_\lambda = 1$ for the fits assuming only statistical uncertainties. The dashed lines at $-\ln(L_{\kappa_\lambda}/L_{\kappa_\lambda=1}) = 0.5$ and 2.0 indicate the values corresponding to a $1\sigma$ and $2\sigma$ confidence interval, respectively (assuming an asymptotic $\chi^2$ distribution of the test statistic).

The increase in centre-of-mass energy from $\sqrt{s}$ = 13 TeV to $\sqrt{s}$ = 27 TeV is accounted for by scaling the number of expected background events by a factor of 4.5, to account for the approximate cross-section increase arising from the enhanced gluon-luminosity. In the case of the SM $HH$ signal, the signal yields are scaled to the cross-section of 139.9 fb [19]. In the extrapolation the $Z$+heavy-flavour background is scaled by the normalisation factor of 1.34 obtained in the Run 2 analysis, while the $t\bar{t}$ normalisation is taken from simulation since its Run 2 normalisation factor was consistent with unity.

The HE-LHC expected significance for the SM $HH$ process in the $b\bar{b}\tau^+\tau^-$ final state, based on the extrapolation procedure described above and without taking into account any systematic uncertainties, is $11\sigma$. In the same scenario, the $\kappa_\lambda = 0$ hypothesis can be rejected with a confidence level of $5.8\sigma$, while the Higgs-self coupling strength is expected to be measured as $\kappa_\lambda = 1.0 \pm 0.2$.

# 7 Conclusion

A prospect study for the search for non-resonant Higgs-boson-pair production at the HL-LHC has been performed, using the combination of the $b\bar{b}b\bar{b}$, $b\bar{b}\gamma\gamma$ and $b\bar{b}\tau^+\tau^-$ final states. The signal strength relative to the Standard Model is expected to be measured with an accuracy of 40% (31%) with (without) systematic uncertainties.

The Higgs boson self-coupling is constrained at 95% confidence level to $-0.4 \leq \kappa_\lambda \leq 7.3$ ($-0.1 \leq \kappa_\lambda \leq 2.7 \cup 5.5 \leq \kappa_\lambda \leq 6.9$), with (without) systematic uncertainties. The value of $\kappa_\lambda$ is expected to be measured as $1.0^{+0.9}_{-0.8}$ ($1.0^{+0.7}_{-0.6}$) with (without) systematic uncertainties. It may be possible to improve the sensitivity of the combination in the future through enhancements of the $b\bar{b}\tau^+\tau^-$ and $b\bar{b}\gamma\gamma$ analyses, such as explicitly training BDTs for specific values of $\kappa_\lambda$ or attempting to optimise event selection at low $m_{HH}$.

First estimates of the prospects at the HE-LHC, assuming 15 $\text{ab}^{-1}$ of data recorded at a centre-of-mass collision energy of 27 TeV, have also been obtained by further extrapolating the measurements in the $b\bar{b}\gamma\gamma$ and $b\bar{b}\tau^+\tau^-$ final states. For the $b\bar{b}\gamma\gamma$ final state, it is expected that the signal will be measured at a statistical-only significance of $7.1\sigma$ and $\kappa_\lambda$ will be measured as $1.0 \pm 0.4$. For the $b\bar{b}\tau^+\tau^-$ final state, it is expected that the signal will be measured at a statistical-only significance of $11\sigma$ and $\kappa_\lambda$ will be measured as $1.0 \pm 0.2$.

# CMS Physics Analysis Summary

Contact: cms-phys-conveners-ftr@cern.ch 

## Prospects for HH measurements at the HL-LHC

The CMS Collaboration

### Abstract

The prospects for the study of Higgs boson pair production at the High-Luminosity LHC with the CMS detector are presented. Five decay channels, bbbb, bbWW, bb$\tau\tau$, bb$\gamma\gamma$, and bbZZ, are studied. Analyses are developed using a parametric simulation of the upgraded detector response and optimised for a projected integrated luminosity of $3000\,\mathrm{fb}^{-1}$. The statistical combination of the five decay channels results in an expected significance for the standard model HH signal of $2.6\sigma$. Projections are also presented for the measurement of the Higgs boson self-coupling $\lambda_{\mathrm{HHH}}$. The expected 68 and 95% confidence level intervals for the coupling modifier $\kappa_{\lambda} = \lambda_{\mathrm{HHH}}/\lambda^{\mathrm{SM}}_{\mathrm{HHH}}$ are $[0.35, 1.9]$ and $[-0.18, 3.6]$, respectively.

# Contents

# 1 Introduction

The discovery of a scalar boson with a mass of about 125 GeV by the ATLAS and CMS Collaborations [1–3], and the determination of its properties so far consistent with those expected for the Higgs boson (H) of the standard model of particle physics (SM) [4], has stimulated interest for a detailed exploration and understanding of the Brout–Englert–Higgs (BEH) mechanism [5–7]. The SM predicts the existence of Higgs boson self-interactions, whose properties are directly determined by the structure of the BEH scalar potential. The study of the production of Higgs boson pairs (HH) represents therefore a crucial test of the SM since it gives experimental access to the Higgs boson self-coupling ($\lambda_{\mathrm{HHH}}$) and thus to the structure of the BEH potential itself. Moreover, HH production represents a unique way to probe the existence of physics beyond the standard model (BSM) that may manifest as a modification of $\lambda_{\mathrm{HHH}}$. In general, low energy scale effects of some high energy scale physics, as described in the context of an effective theory (EFT) [8], can result as contact interaction terms in the Lagrangian. Terms which can affect the double Higgs production are contact interactions between the HH pair and gluons (ggHH) and top quarks (ttHH), as well as a contact interaction between one Higgs boson and two gluons that we expect to be constrained in single Higgs boson measurements. The presence of these additional contributions will result in an anomalous HH production cross section and modified kinematic properties of the HH system.

In the SM, the dominant HH production mechanism in pp collisions is via gluon fusion, with an expected cross section of $31.05^{+2.2\%}_{-5.0\%}$ fb at $\sqrt{s} = 13$ TeV and $36.69^{+2.1\%}_{-4.9\%}$ at 14 TeV. These values were computed at the next-to-next-to-leading order (NNLO) of the perturbative quantum chromodynamics (QCD) calculation, including next-to-next-to-leading-logarithm (NNLL) corrections and finite top quark mass effects [9]. Because of the smallness of the cross section and the presence of backgrounds, searches for HH productions based on the current Run II dataset are not yet sensitive to SM HH production; large integrated luminosities are required for the experimental study of this very rare process.

The High-Luminosity LHC (HL-LHC) will provide a unique opportunity to study HH production as predicted in the SM and identify possible deviations induced by BSM physics in the signal cross section or properties. Upgrades of the LHC machine will increase the peak instantaneous luminosity to $5 - 7.5 \times 10^{34}\,\mathrm{cm}^{-2}\,\mathrm{s}^{-1}$ and the CMS experiment will collect more than 3000 $\mathrm{fb}^{-1}$ over a decade of operation. The high instantaneous luminosity will lead to 140 to 200 additional interactions per bunch crossing. This pileup will constitute a formidable challenge for the experiment both in terms of event reconstruction and radiation damage. A comprehensive detector upgrade program is under development to maintain and improve the detector performance under these challenging conditions.

The projected sensitivity to HH at the HL-LHC has so far been studied by the ATLAS [10–13] and CMS [14, 15] Collaborations either using a parametric simulation of the detector performance or with an extrapolation of the Run II analysis results. In both cases, the projected combined sensitivity to HH production is at the level of about $2\sigma$ or below. It is important to remark here that the usage of a parametric simulation requires a comprehensive knowledge of the expected upgraded detector performance. Very recent developments in the detector performance from CMS upgrade studies are not accounted in the referenced results. Similarly, the extrapolation of the current results cannot account for the optimisation of the analysis strategy to the large dataset collected at the HL-LHC.

Updated projections of the sensitivity have recently been developed by the CMS Collaboration to study the impact of the subsystem upgrades on the physics program, as described in the Technical Design Report (TDR) of the inner tracker [16], the barrel electromagnetic [17] and the endcap high-granularity [18] calorimeters.

The work described in this document improves and extends the previous projections to provide an updated and comprehensive study of the prospects for HH measurements at the HL-LHC. A parametric simulation, as detailed in Section 3, is used to model the upgraded detector response and simulate its performance considering the experience and understanding achieved in the preparation of the aforementioned TDRs. The five decay channels bbbb, bb$\tau\tau$, bbWW (WW $\to \ell\nu\ell'\nu'$ with $\ell, \ell' = \mathrm{e}, \mu$), bb$\gamma\gamma$, and bbZZ (ZZ $\to \ell\ell\ell'\ell'$ with $\ell, \ell' = \mathrm{e}, \mu$) are studied and dedicated analysis strategies are developed to exploit the HL-HLC dataset. The first four channels correspond to those expected to be the most sensitive to HH production at the HL-LHC based on the experience from Run II searches, while the very rare but clean bbZZ($\ell\ell\ell\ell$) final state is studied here for the first time. The corresponding branching fractions, computed considering a mass of the Higgs boson of 125 GeV [19], are summarised in Table 1. The event selection and analysis strategy of each channel are separately described in the following, and the sensitivity resulting from their statistical combination is discussed in Section 9.

Table 1: Branching fraction of the five decay channels considered in this document. The symbol $\ell$ denotes either a muon or an electron. In the bbWW decay channel, $\ell$ from the intermediate production of a $\tau$ lepton are also considered in the branching fraction.

| Channel | $\mathcal{B}$ [%] |
|---|---|
| bbbb | 33.6 |
| bb$\tau\tau$ | 7.3 |
| bbWW($\ell\nu\ell\nu$) | 1.7 |
| bb$\gamma\gamma$ | 0.26 |
| bbZZ($\ell\ell\ell\ell$) | 0.015 |

## 2 The CMS upgraded detector

The improvement of the performance of the CMS detector under HL-LHC conditions requires both increased radiation hardness to withstand over a decade of HL-LHC operations under high pileup and luminosity conditions, increased granularity to reduce particle occupancy and improve the object reconstruction, and increased bandwidth to accommodate higher data rates.

Both the hardware and software stages of the CMS trigger system, respectively denoted as the Level-1 (L1) and High Level Trigger (HLT), and the data acquisition system (DAQ) will undergo a substantial upgrade. The L1 trigger hardware will be replaced, allowing for an increase of its rate and latency to about 750 kHz and 12.5 $\mu$s respectively, while the HLT rate will be increased to 7.5 kHz. These values are to compare to the throughput of the current L1 and HLT systems of about 100 and 1 kHz respectively, and to a current L1 system latency of 3.8 $\mu$s. The L1 will also feature inputs from the silicon tracker, allowing real-time track fitting and particle-flow reconstruction of objects at the trigger level. The pixel and strip tracker detectors will be replaced to increase the granularity, reduce the material budget and extend the geometrical coverage. The front-end electronics of the barrel electromagnetic calorimeter (ECAL) will be upgraded to access the single-crystal information at the L1 trigger, as well as the electronics of the cathode strip chambers (CSC), resistive plate chambers (RPC) and drift tubes (DT) for muon detection. New muon detectors based on RPC and gas electron multiplier (GEM) technologies will also be installed to add redundancy, increase the geometrical coverage, and improve the trigger and reconstruction of muons in the forward region. The endcap electromagnetic and hadron calorimeters will be replaced with a new high-granularity sampling detector (HGCal) based on silicon pad sensor and will provide highly-segmented spatial information as well as timing information for a four-dimensional reconstruction of the interaction shower shapes. Finally, the addition of a new timing detector for minimum ionizing particles (MTD) is envisaged to provide additional capabilities, beyond spatial tracking algorithms, to correctly associate the reconstructed charged particles to the production vertex and thus suppress pileup effects.

A detailed overview of the upgrade program and of the CMS detector upgrades are presented in the Technical Proposal for the Phase-II upgrade [20]. The performance in object reconstruction is detailed and summarised in Ref. [21].

# 3 Signal and background modelling

## 3.1 Simulated physics processes

The signal and background processes in pp collisions at $\sqrt{s} = 14$ TeV are modelled using Monte Carlo (MC) event generators, that simulate the hard process, as well as the hadronisation and fragmentation effects. They are interfaced with the DELPHES [22] software to model the response of the upgraded CMS detector.

The HH signal is simulated using MADGRAPH5_aMC@NLO [23] at leading order (LO) accuracy. Five different samples of SM HH production are generated for the five final states considered here. The Higgs boson branching fractions prediction in the SM is used for the normalisation of the samples, resulting in the values previously reported in Table 1. Signals corresponding to anomalous values of the $\lambda_{\mathrm{HHH}}$ coupling are modelled by weighting the SM samples as a function of the invariant mass of the HH pair and of the angle of one of the two Higgs bosons with respect to the beam line in the reference of the HH system, following the procedure detailed in Ref. [24]. The same technique is also used to model the so-called shape benchmark signals, i.e. signals corresponding to specific combinations of Higgs boson effective couplings which kinematic properties may be used to approximately represent broader regions of the EFT parameter space. The definition of the shape benchmarks is given in [25].

Several background sources are considered for each decay channel, as detailed in the corresponding section of this documents.

Top quark processes are simulated using POWHEG [26–29] and multiparton interactions, parton shower, and hadronization effects are simulated with PYTHIA 8 [30]. The $\mathrm{t\bar{t}}$ production is normalised to the theoretical production cross section at $\sqrt{s} = 14$ TeV computed at NNLO+NNLL of $984.50^{+23.21}_{-34.69}(\mathrm{scale})^{+41.31}_{-41.31}(\mathrm{PDF} + \alpha_S)^{+27.14}_{-26.29}(\mathrm{mass})$ [31]. Single top quark production in the tW channel is normalised to the theoretical prediction at NNLO precision of $84.4 \pm 2.0(\mathrm{scale})^{+3.00}_{-4.80}(\mathrm{PDF})$ [32, 33]. The production of top quark pairs in association with Z bosons pairs is simulated with MADGRAPH5_aMC@NLO forcing the four lepton decay of the ZZ system.

Drell-Yan production of leptons pairs $Z/\gamma^* \to \ell\ell$ and of $W \to \ell\nu$ in association with jets are simulated at the LO precision using MADGRAPH5_aMC@NLO and normalised to the generator cross section at the same precision order. To increase the number of $Z/\gamma^*$ events that satisfy the event selections, inclusive samples are complemented by simulations in selected regions of the scalar sum of the transverse momentum of the partons emitted at matrix element level and of the invariant mass of the $\ell\ell$ system.

Multijet production from QCD interactions is simulated using MADGRAPH5_aMC@NLO at LO

accuracy. To increase the selection efficiency for the projection in the bbbb final state, the presence of at least one b quark emitted at matrix level is also required. Samples are generated in exclusive intervals of the scalar sum of the transverse momentum of the partons emitted at matrix element level and combined using the relative cross section. An inclusive sample, where generated events are required to contain a pair of partons with invariant mass larger than 1 TeV, is also used to improve the description for the phase space of highly boosted jets. The overall normalisation for the multijet background is obtained from a comparison between Run II data and MC produced with the same generator at $\sqrt{s} = 13$ TeV.

Single Higgs boson production in gluon (gg $\to$ H) and vector boson (VBF) fusion, and in associated production with top quarks ($t\bar{t}H$) and vector bosons (VH), is considered as a background for HH production. Decays of the Higgs boson are forced to exclusive final states to increase the acceptance for the HH decay channels studies, and different generators are used. For all the aforementioned processes where the Higgs boson decays to $ZZ^* \to \ell\ell\ell\ell$, the MADGRAPH5_aMC@NLO generator is used. For the decays of the Higgs boson to photon pairs, MADGRAPH5_aMC@NLO is used for the gg $\to$ H, VBF and VH processes, while POWHEG is used for $t\bar{t}H$. For the other channels, single Higgs samples for all the production modes are generated with POWHEG. Generated samples are normalised to the expected SM cross section as recommended in [19]. The gg $\to$ H production cross section is computed at the next-to-next-to-next-to-leading order ($N^3LO$) in perturbative QCD and at NLO in electroweak (EW) corrections. The VBF and WH processes cross sections are computed at NNLO QCD and NLO EW accuracies, and the $t\bar{t}H$ cross section is computed at NLO QCD and NLO EW accuracies. Finally, the ZH cross section is computed at NNLO QCD and NLO EW accuracy for quark-initiated contributions and at NLO QCD accuracy with NLL effects for the gluon fusion-induced component.

## 3.2 CMS detector response and object reconstruction

Both the signal and background samples are processed with the DELPHES fast parametric simulation software to simulate the response of the upgraded CMS detector and account for the pileup contributions by overlaying an average of 200 minimum bias interaction events simulated with PYTHIA 8. The DELPHES software simulates the performance of reconstruction and identification algorithms for electrons, muons, tau decays to hadrons ($\tau_h$) and a neutrino, photons, jets including those containing heavy flavour particles, and the missing transverse momentum vector $\vec{p}_T^{miss}$, defined as the projection onto the plane perpendicular to the beam axis of the negative vector sum of the momenta of all reconstructed particle-flow objects in an event. The resolution, energy and momentum scale, efficiencies, and misidentification rates for the various objects have been extensively compared and tuned to reproduce the performance obtained with a full simulation of the CMS detector based on GEANT4 [34] and the use of reconstruction algorithms tuned to the HL-LHC environment. In those cases where the reconstruction algorithms have not yet been developed or finalised, the parametrisation follows assumptions based on the Run II object performance and on the studies prepared for the CMS detector Technical Design Reports.

The simulation of the electron reconstruction is initiated by generator-level electrons, with a detection efficiency that parametrised in energy and pseudorapidity. The energy resolution of reconstructed electrons is a function of the ECAL and tracker resolutions. Similarly, the simulation of muon reconstruction is seeded by generator-level muons, applying a parametric reconstruction and identifications efficiency. The momentum of reconstructed muons is obtained via Gaussian smearing of the generator-level muon 4-momenta, with a resolution parametrised by transverse momentum and pseudorapidity.

Photons are reconstructed from neutral energy excess in a simplified version of the electromagnetic calorimeter. The efficiency for photons and jet backgrounds are parametrised based on studies of the upgraded CMS detector and correspond to the performance of a multivariate identification algorithm combined with an isolation selection.

Hadronic jets are reconstructed with a particle-flow algorithm run on simulated tracks and on the energy deposits in the ECAL, HCAL, and HGCal. The particle-flow candidates are clustered via sequential recombination of tracks and calorimeter deposits using the anti-$k_T$ algorithm [35, 36] operated with a distance parameters $R$ of 0.4. Pileup mitigation is performed using the "PileUp Per Particle Identification" (PUPPI) algorithm [37].

The missing transverse momentum $\vec{p}_T^{miss}$ is calculated for each event using particle-flow (PF) objects and corrected using PUPPI.

The efficiency of b jet identification and the corresponding misidentification rates for light flavour quarks and gluons jets are parametrised as functions of the jet $p_T$ and $\eta$, depending on the underlying jet flavour as determined by a geometric match with generated partons. The parametrised b tagging performance considers the improvements following the inclusion of the MTD detector measurements [38].

The efficiency of $\tau_h$ identification and hadron jet misidentification rates are parametrised as functions of the jet $p_T$ and $\eta$. Reconstructed jets are geometrically matched with $\tau_h$ objects at generator level to determine whether a jet candidate is a genuine $\tau_h$ or a quark or gluon jet. Parametrised probabilities corresponding to efficiency and misidentification rates, obtained from studies based on full CMS simulation, are consequently applied.

We do note that the aforementioned improvements in the b tag efficiency following the inclusion of the MTD detector information only correspond to the removal of spurious tracks in the reconstructed jets, effectively acting as suppression of the pileup effects, but that the track timing information is not used directly in a dedicated training of a b tag discriminant. Moreover, further improvements following the inclusion of the MTD detector are expected in the lepton, $\tau_h$ and photon isolation performance, as well as in the rejection of additional jets erroneously associated to the main production vertex. Improvements in the results described in this work may consequently be expected from a full, optimal use of the timing information.

## 3.3 Systematic uncertainties

Systematic uncertainties in the modelling of the signal and background processes due to theoretical (normalisation cross sections) and experimental effects (object reconstruction performance and background estimation) are considered. The assumption for the values reported below are discussed in detail in Ref. [21].

An uncertainty on the total integrated luminosity of 1% is considered. It is correlated in all the channels across all the processes that are assumed to be modelled with a MC simulation at the HL-LHC.

Uncertainties on the b tag efficiency are parametrised as a function of the jet $\eta$ and $p_T$, and amount to about 1% for genuine b jets with $p_T < 300\,\text{GeV}$ and range between 2 and 6% for larger $p_T$ values. An uncertainty of 1% is also assumed for the b tag efficiency of subjets identified within large-radius jets. The uncertainty on the scale of the reconstructed jets ranges between 0.2 and 2% depending on the source considered and is applied by varying the jet $p_T$ by the corresponding value and checking the changes in the processes yields.

Uncertainties in the electron identification and isolation efficiencies amount to 2.5% and 0.5%

for $p_T$ below and above 20 GeV, respectively. The muon identification and isolation efficiency uncertainty corresponds to 0.5% for all the $p_T$ values considered. For $\tau_h$ objects, this uncertainty amounts to 5% as in the Run II analyses. The uncertainties in the photon reconstruction and identification efficiency correspond to 1%, while their energy scale and resolution are assumed to be determined with an accuracy of 0.5 ad 5%, respectively.

Specific uncertainties in the modelling of the main physics processes are also considered and discussed in the context of the relevant analyses. Triggers are assumed to be fully efficient in the phase space considered, and the corresponding uncertainties are included in the object reconstruction and identification uncertainties.

The uncertainties in the theoretical cross sections used for the normalisation of simulated processes are assumed to be reduced by a factor of 1/2 with respect to the current value.

# 4 HH → bbbb

While characterised by the largest branching fraction among the HH final states, the bbbb decay channel suffers from a large contamination from the multijet background that makes it experimentally challenging. Two complementary strategies are explored here to identify the signal contribution. For those events where the four jets from the HH → bbbb decay can all be reconstructed separately, also referred to as the "resolved" topology, the use of multivariate methods is explored to efficiently identify the signal contribution in the overwhelming background. In cases where the invariant mass of the HH system, $m_{HH}$, is large, the high Lorentz boost of both Higgs bosons may results in a so-called "boosted" event topology where the two jets from a $H \rightarrow b\bar{b}$ decay overlap and are reconstructed as a single, large-area jet. Resolved topologies correspond to the large majority of SM HH events, giving the largest sensitivity on this signal. Boosted topologies help to suppress the multijet background and provide sensitivity to BSM scenarios where the differential HH production cross section is enhanced at high $m_{HH}$ by the presence of ggHH and ttHH effective contact interactions.

## 4.1 Event selection for resolved topologies

This work assumes the trigger efficiency to be 100% for the reconstructed object selections detailed below. This assumption is based on experience in the Run II analysis and is considered to be realistic considered the planned upgrades of the CMS trigger system, both at L1 and HLT, that will allow for an improvement and harmonisation of online and offline b tagging algorithms.

Events are preselected by requiring four jets with $p_T > 45$ GeV and $|\eta| < 3.5$ that satisfy the medium b tagging working point, corresponding to a b jet identification efficiency of approximately 70% for a light flavour and gluon jet misidentification rate of 1%. In case more than four jets are preselected, corresponding to less than 7% of the total signal events, the four highest $p_T$ candidates are selected. The efficiency of the jet preselection on the SM HH signal is of about 7%.

The four preselected jets are combined into the two Higgs boson candidates $H_1$ and $H_2$. Correct jet pairing is determined as the combination that minimises the difference in the invariant mass of the two jet pairs. This allows to explore the signature of two resonant $H \rightarrow b\bar{b}$ decays while minimising the bias induced in the multijet background selection, in particular suppressing jet combinations with both invariant masses close to $m_H$.

The signal region is defined by events that satisfy the following selection in the invariant mass of the two Higgs boson candidates:

$$\sqrt{(\mathrm{m_{H_1}} - 120\,\mathrm{GeV})^2 + (\mathrm{m_{H_2}} - 120\,\mathrm{GeV})^2} < 40\,\mathrm{GeV} \quad (1)$$

The selection has an efficiency of about 55% on the HH signal and rejects approximately 85% of the QCD multijet background. The low selection efficiency for signal events is due to the removal of events where at least one of the selected jets does not originate from the decay of a Higgs boson, and cases where jets have been incorrectly paired. For events correctly reconstructed the selection efficiency corresponds to about 90%. The expected event yields after the invariant mass selections are approximately 1370 for the HH signal and $1.1 \times 10^7$ for the background, mostly consisting of QCD multijet and $\mathrm{t\bar{t}}$ events. This difference of almost four orders of magnitude between the two processes calls for the use of multivariate methods that exploit the kinematic differences between the HH signal and the background processes.

A multivariate discriminant, consisting of a boosted decision tree (BDT), is built using the following kinematic variables:

- the invariant mass of the two Higgs candidates $\mathrm{H_1}$ and $\mathrm{H_2}$
- the transverse momentum of the two Higgs candidates $\mathrm{H_1}$ and $\mathrm{H_2}$
- the four-jet invariant mass $m_{\mathrm{HH}}$, and the reduced mass $M_{\mathrm{HH}} = m_{\mathrm{HH}} - (m_{\mathrm{H_1}} - 125\,\mathrm{GeV}) - (m_{\mathrm{H_2}} - 125\,\mathrm{GeV})$, the latter helping to remove part of the jet resolution effects by using the information on the Higgs boson invariant mass
- the minimal and max $\Delta\eta$ and $\Delta\varphi$ separation of the combinations of the four preselected jets
- the $\Delta\eta$, $\Delta\varphi$ and $\Delta R = \sqrt{(\Delta\eta)^2 + (\Delta\varphi)^2}$ separation of the jets that constitute $\mathrm{H_1}$ and $\mathrm{H_2}$
- the cosine of the angle formed by one of the Higgs candidates with respect to the beam line axis in the HH system rest frame

The BDT is trained with a gradient boosting algorithm and its parameters are optimised to ensure the best performance while verifying that no overtraining is introduced. The output of the BDT is used as the discriminant variable to look for the presence of a signal as an excess at high output values. The expected distribution of signal and background events is illustrated in Fig. 1. The binning of the distribution is optimised to maximise the sensitivity to the SM HH signal.

In addition to the systematic uncertainties described in Section 3.3, uncertainties in the BDT shape are studied. We expect that the huge dataset collected at the HL-LHC will allow to make a precise estimate of the multijet background in signal-depleted regions, defined for example by the inversion of the b tag or invariant mass criteria. We thus consider an uncertainty of 5% on the BDT bins with the highest S/B ratio. A quantitative study of the impact of the size of such uncertainty is presented at the end of this section.

## 4.2 Event selection for boosted topologies

The regions of phase space having large $m_{\mathrm{HH}}$ are best explored using dedicated physics object reconstruction and event selections aimed at identifying highly Lorentz-boosted Higgs bosons, improving substantially the efficiency and performance for $m_{\mathrm{HH}}$ values larger than about 1 TeV. The bbbb final state is experimentally favorable in this context given its large branching fraction. As boosted object reconstruction methods in this final state have been already successfully applied to HH searches [39], this projection aims at investigating their potential at HL-LHC.

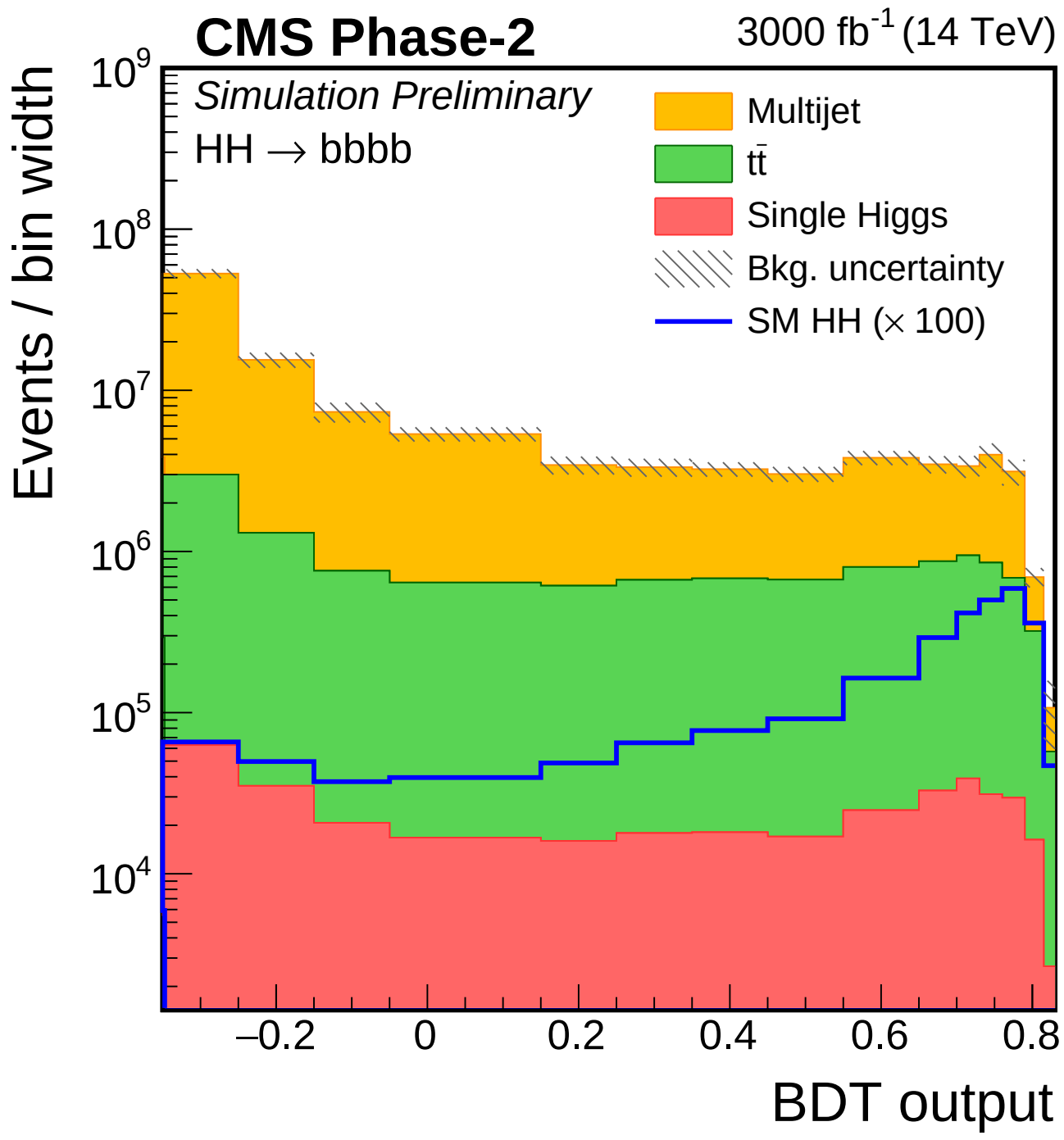

Figure 1: BDT output distribution for the signal and background processes considered in the bbbb resolved search.

Highly Lorentz-boosted $H \to b\bar{b}$ decays are experimentally reconstructed as a single, large area jet. The particle-flow candidates are clustered using the anti-$k_T$ algorithm with a distance parameter of 0.8 (AK8 jets). Contributions from pileup are mitigated using the pileup-per-particle (PUPPI) identification algorithm [40]. The vector sum of the clustered particle-flow candidates, weighted by their PUPPI weights, is assigned as the jet four-momentum. The jet energy is corrected to compensate for the nonlinear detector response to the energy deposited [41, 42].

The event selection aims to identify two boosted $H \to b\bar{b}$ decays, each associated with a single AK8 jet. The two leading-$p_T$ AK8 jets in the event, denoted as $J_1$ and $J_2$, are required to have $p_T > 300$ GeV and lie within $|\eta| < 3.0$. The soft-drop [43, 44] jet grooming algorithm is used to remove soft and collinear components of the jet and retain the two subjets associated with the showering and hadronization of the two b quarks from the $H \to b\bar{b}$ decay. The jets $J_1$ and $J_2$ are both required to have a soft-drop mass between 90–140 GeV, consistent with the observed mass of 125 GeV for the Higgs boson.

To further reduce backgrounds, the N-subjettiness algorithm [45] is used, which can differentiate between a jet containing an $N$ pronged decay from a jet containing a single hard parton. For a boosted $H \to b\bar{b}$ jet with a two-pronged structure, the N-subjettiness ratio $\tau_{21} \equiv \tau_2/\tau_1$ is much smaller than unity, while the background jets have larger values. Consequently, a requirement of $\tau_{21} < 0.6$ is made for both $J_1$ and $J_2$. Both the soft-drop and the $\tau_{21}$ requirements are optimized using $S/\sqrt{B}$ as figure of merit.

The soft-drop subjets are b-tagged using the DeepCSV algorithm [46] which uses machine learning techniques with inputs based on the tracks and displaced vertices associated to the jets. In this search the jet b tagging probability for light flavoured jets is required to be about 1%, corresponding to a probability of about 49% to correctly identify jets containing a b hadron. Events are classified as those having exactly three (3b category) or exactly four (4b category) b-tagged subjets, out of the four subjets belonging to $J_1$ and $J_2$.

The full event selection results in an efficiency for the SM signal of about 0.1%. For signals with a considerably harder $m_{HH}$ spectrum, such as those represented by the shape benchmark number 2 used here as a reference, the selection efficiency is 1.8%.

The main background is dijet production in QCD interactions. In the analysis of true collision data, it is expected that the background will be obtained from the data itself. Here, simulated samples of QCD dijet events are used. The background estimation follows closely the approach in Ref. [47]. The background obtained from simulations is scaled by a factor of 0.7, which has been derived comparing the LHC data at $\sqrt{s} = 13$ TeV selected as described in [47], and a MC simulation for the 13 TeV conditions based on the same generator used in this work.

After the event selection, the expected number of SM HH signal events is 96 in the 3b and 15 in the 4b event categories. The most sensitive shape benchmark using the boosted event selection is benchmark 2, that is typically associated with strong contact interactions and hence characterised by a large fraction of events at high $m_{HH}$. For this signal, the events yields are 537 in the 3b and 61 in the 4b event categories, assuming a cross section of 10 fb. The corresponding background yields are $1.08 \times 10^6$ and $1.40 \times 10^5$ in the 3b and 4b categories, respectively.

The main discriminating variable between the signal and the background is the invariant mass of $J_1$ and $J_2$ $m_{JJ}$, which is correlated with the HH invariant mass. Figure 2 shows the $m_{JJ}$ distributions of the signal and the background in the 3b and 4b event categories.

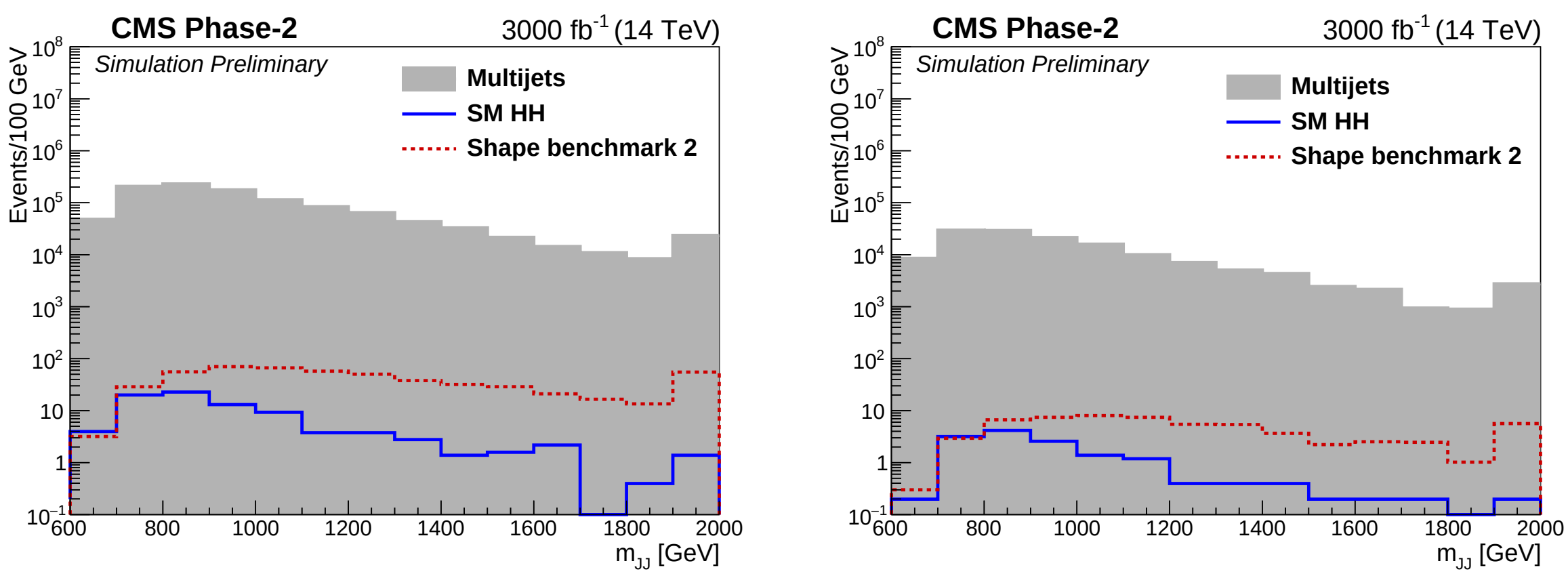

Figure 2: $m_{JJ}$ distributions for the estimated multijet background and the SM (blue) and shape benchmark 2 (red) signals. The distributions on the left are for the 3b and those on the right are for the 4b subjet b-tagged categories. Both signals are normalised to the SM HH production cross section for visualisation.

In addition to the uncertainties described in Section 3.3, dedicated uncertainties to the objects and variables used in this analysis are considered. The H jet mass scale and resolution uncertainties correspond to 1% each, the uncertainty in the selection efficiency correction on $\tau_{21}$ amounts to 13%, and another 3.5% uncertainty is assigned to the modelling of the parton shower and hadronization of the $H \to b\bar{b}$ decay within the H jets. These uncertainties are taken from [48] and reduced by a factor of 1/2.

## 4.3 Results

The resolved analysis is used in the search for the SM HH signal and for the study of the anomalous $\lambda_{HHH}$ couplings, while results on the expected constraint on anomalous HH production in the context of EFT models are derived for the boosted search.

Using the resolved bbbb search strategy, upper limits are computed at 95% CL given the projected signal and background distributions shown in Fig. 1. Considering the systematic uncertainties discussed above, an upper limit of 2.1 times the SM prediction is expected, corresponding to a local significance of the expected HH signal of $0.95\sigma$. If only statistical uncertainties are taken into account, the expected upper limit is 1.6 times the SM prediction and the significance is $1.2\sigma$.

Challenges towards achieving these sensitivities at the HL-LHC will be the capability to develop efficient triggers for the bbbb signal, and to precisely model the multijet background.

Triggering on multi jet signatures will be particularly challenging at the HL-LHC and, despite the upgrades at the L1 trigger and HLT systems, thresholds might be significantly higher than currently achieved in Run II collisions. A study of the change in the search sensitivity as a function of the minimal jet $p_T$ threshold is reported in Fig. 3. The study is realised by increasing the jet $p_T$ value applied at preselection and studying the resulting changes in the sensitivity with respect to the nominal $p_T$ threshold of 45 GeV discussed above. It has been verified that the loss of sensitivity does not arise from a reduced discrimination power of the BDT discriminant because of changes in the kinematic properties induced by the higher thresholds. Instead, the reduced sensitivity is a direct consequence of the reduced acceptance to HH → bbbb events, and an efficient trigger with low $p_T$ thresholds will be crucial at the HL-LHC.

Changes in the SM HH significance as a function of the uncertainty on the high S/B bins for the QCD multijet background are also shown in Fig. 3.

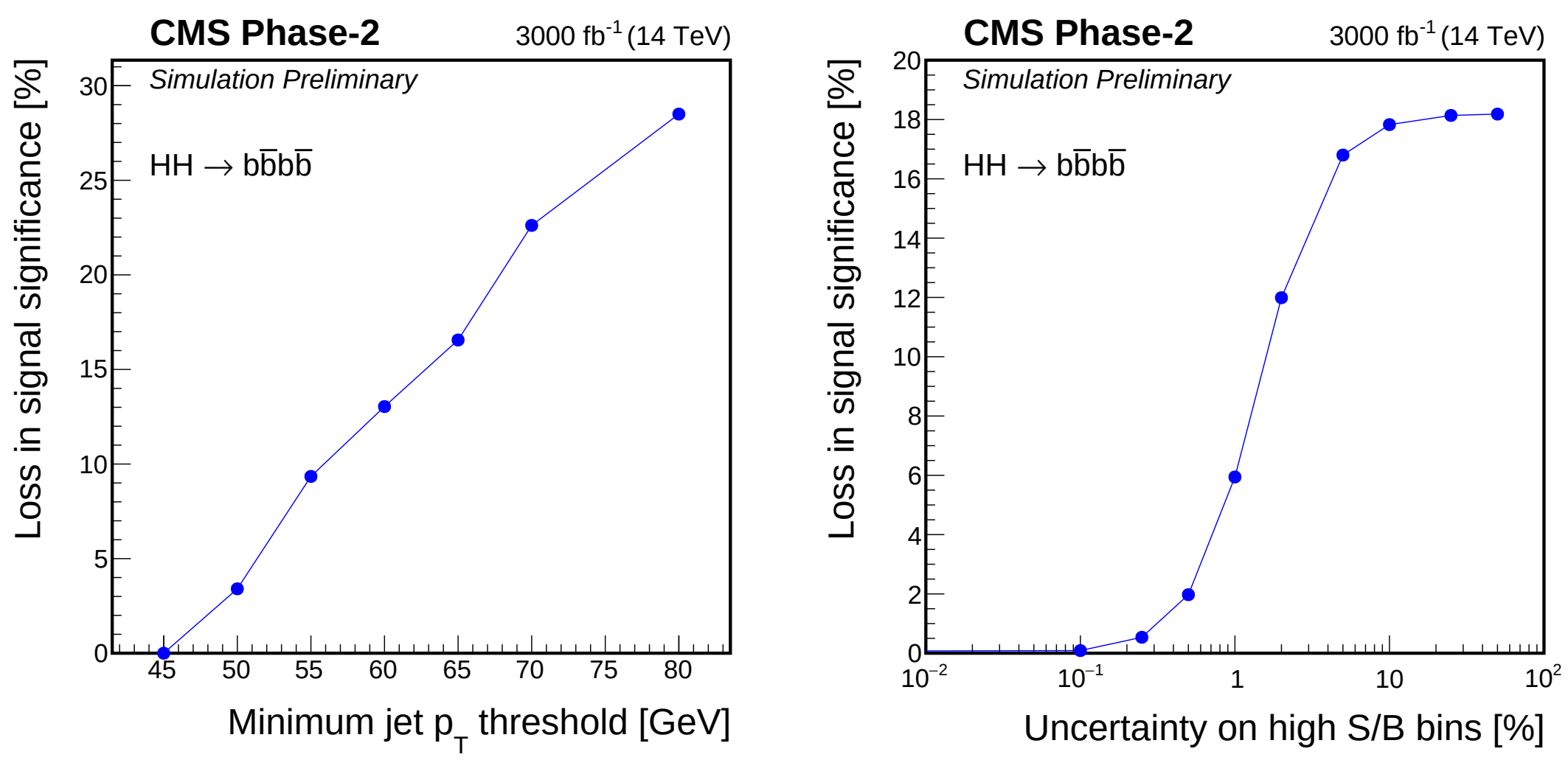

Figure 3: Loss of sensitivity of the HH → bbbb resolved search as a function of the minimal jet $p_T$ threshold (left) and as a function of the uncertainty assumed on high S/B bins for the QCD multijet background (right). In each curve, only the quantity shown on the horizontal axis is varied while the other are kept fixed to the nominal values assumed. The “loss” quantity plotted on the ordinate is defined $1 - Z/Z^0$, where $Z$ denotes the significance of the HH signal in the hypothesis considered and $Z^0$ the significance for the cases of a 45 GeV $p_T$ threshold (left) and of no uncertainty considered (right).

Using the event yields and distributions shown in Fig. 2 for the boosted search strategy, we calculate the 95% confidence level (CL) upper limits on the nonresonant HH productions in the SM and for other combinations of BSM couplings using the shape benchmark signals 1–12, as shown in Fig. 4.

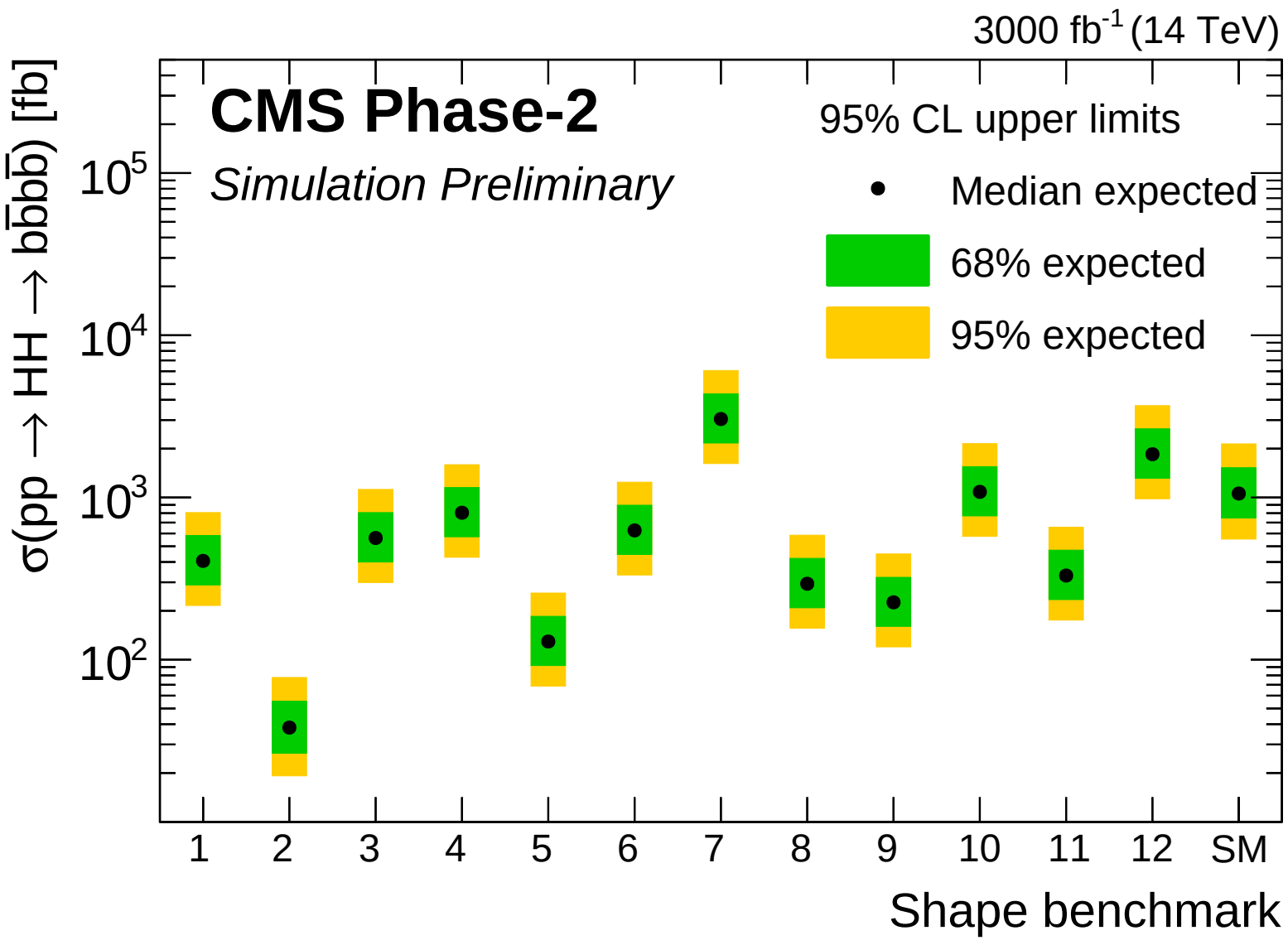

Figure 4: The expected upper limits for non-resonant HH production in the standard model and other shape benchmarks (1–12). The inner (green) and the outer (yellow) bands indicate the regions containing the 68 and 95%, respectively, of the distribution of limits expected under the background-only hypothesis.

## 5 HH → bb$\tau\tau$

The bb$\tau\tau$ final state is experimentally favourable thanks to its sizeable branching fraction of 7.3% and the moderate background contamination, mostly from irreducible processes such as $t\bar{t} \rightarrow b\bar{b}W^+W^-$ in a final state including a tau lepton and $Z\gamma^* + b\bar{b} \rightarrow \tau^+\tau^- + b\bar{b}$, as well as instrumental backgrounds where jets are misidentified as hadronically decaying taus, $\tau_h$. Furthermore, the presence of neutrinos in the final states represents a challenge to the signal identification as the final state is only partly reconstructed. State-of-the-art machine learning techniques are investigated here to study the search for a HH → bb$\tau\tau$ signal at the HL-LHC.

### 5.1 Event selection

We assume that events will be collected by using single-lepton, lepton-plus-$\tau_h$, and double-$\tau_h$ triggers with isolation criteria and transverse momentum thresholds similar to those used in Run II collisions. The assumptions appear reasonable considering the improved capabilities of the upgraded trigger system, the usage of track information in the L1 trigger to improve the $\tau_h$ reconstruction, and the possibility to develop more sophisticated kinematic triggers to specifically target the HH → bb$\tau\tau$ signal. This work assumes the trigger efficiency to be 100% for the reconstructed object selections detailed below.

Decays of the $\tau\tau$ system can result in six final states: e$\tau_h$, $\mu\tau_h$, $\tau_h\tau_h$, $\mu\mu$, e$\mu$, and ee. In this study, we only consider the three final states involving at least one $\tau_h$, that correspond to about 88% of the total decays of the $\tau\tau$ system and provide the largest sensitivity to the HH process.

Following the lepton and $\tau_h$ requirements defined in Tab. 2, events are exclusively selected into the three final state categories according to the following requirements:

Table 2: Kinematic requirements ($p_T$, $\eta$, and isolation) of electrons, muons, and hadronic taus. The hadronic tau requirements are listed according to the final states considered.

| Lepton | Min. $p_T$ [ GeV ] | Max. $\|\eta\|$ | Max. iso [ GeV ] |
|---|---|---|---|
| Primary $\mu$ | 23 | 2.1 | 0.15 |
| Primary e | 27 | 2.1 | 0.1 |
| Veto e/$\mu$ | 10 | 2.4 | 0.3 |
| **Hadronic tau** | **Min. $p_T$ [ GeV ]** | **Max. $\|\eta\|$** | |
| $\ell\tau_h$bb ($\ell = e, \mu$) | 20 | 2.3 | |
| $\tau_h\tau_h$bb | 45 | 2.1 | |

- $\mu\tau_h$: exactly one primary muon with no additional muons or electrons that satisfy looser veto selections, and at least one $\tau_h$ of opposite charge to the primary muon;
- e$\tau_h$: exactly one primary electron with no additional muons or electrons that satisfy looser veto selections, and at least one $\tau_h$ of opposite charge to the primary electron;
- $\tau_h\tau_h$: exactly zero veto muons or electrons and at least two $\tau_h$ of opposite charge to one another. In case of multiple possible choices of $\tau_h$ candidate, the highest $p_T$ ones are selected.

Events in all the three categories above are also required to contain at least two b-tagged jets with $p_T > 30$ GeV, $|\eta| < 2.4$. After these selections, about 100, 70 and 60 SM HH signal events are expected in the $\mu\tau_h$, e$\tau_h$, and $\tau_h\tau_h$ decay channels respectively. The corresponding total number of background events for the three channels are, respectively, $4.3 \times 10^6$, $2.9 \times 10^6$, and $1.25 \times 10^3$, mostly dominated by $t\bar{t}$ and Drell-Yan $\tau\tau$ production.

The selected visible $\tau\tau$ decay products, the b jets, and the missing transverse momentum are used to build the two Higgs boson candidates. The $H_{\tau\tau}$ candidate is defined as the sum of the four momenta of the selected lepton and $\tau_h$ in the $\mu\tau_h$ and e$\tau_h$ final states, and as the sum of the two $\tau_h$ four momenta in the $\tau_h\tau_h$ final state, plus the vector of missing momentum projected onto the plane transverse to the beam axis. Similarly, the $H_{b\bar{b}}$ candidate is defined as the sum of the four momenta of the two selected b jets. Finally, the four momentum of the HH system is computed as the vector sum of the selected object four momenta plus the vector of missing momentum projected onto the plane transverse to the beam axis.

## 5.2 DNN discriminant

### 5.2.1 Input variables to the neural network discriminant

Kinematic properties of the selected events in each of the three bb$\tau\tau$ final states are used to develop a neural network discriminant that is capable of separating the signal contribution from the background processes. A total of 52 input variables, also denoted as features, are used in this study. They are split into *basic* (27), *high-level/reconstructed* (21), and *high-level/global* (4) features. The specific choice of the input variables, as described below, was chosen since these proved to give the best discriminator performance, while other features could be implicitly computed by the network.

In what follows, $\tau_0$ is defined as the $\tau_h$ in $\ell\tau_h$bb events and as the highest $p_T$ $\tau_h$ candidate in $\tau_h\tau_h$bb events, while $\tau_1$ is defined as the other selected lepton or $\tau_h$. Additionally, we label the highest $p_T$ b jet as $b_0$ and the other as $b_1$.

**5.2.1.1 Basic features** 27 low-level final state features are calculated during the event reconstruction process: the 4-momenta in Cartesian coordinates (i.e. $(p_x, p_y, p_z, E)$), the magnitude of the 3-momenta, and mass of each selected final state ($\tau_0$, $\tau_1$, $b_0$, and $b_1$); and the magnitude of missing transverse momentum ($p_{\mathrm{T}}^{\mathrm{miss}}$), and its transverse components ($p_x^{\mathrm{miss}}$ and $p_y^{\mathrm{miss}}$).

**5.2.1.2 High level reconstructed features** 21 reconstructed features are calculated:

- The 4-momenta, magnitude of 3-momenta, and the invariant masses of the three systems which correspond to:
  1. the Higgs boson that decays to $\mathrm{b\bar{b}}$ ($\mathrm{H_{b\bar{b}}}$),
  2. the Higgs boson that decays to $\tau\tau$ ($\mathrm{H}_{\tau\tau}$), and
  3. the di-Higgs-boson system (HH);
- the stransverse mass ($M_{T2}$) of the system [49, 50];
- the transverse masses ($m_T$) of each tau, calculated according to Eq. 2:

$$m_T = \sqrt{2p_{T,\tau} \times p_{\mathrm{T}}^{\mathrm{miss}} \times \left(1 - \cos\Delta\phi_{\tau,p_{\mathrm{T}}^{\mathrm{miss}}}\right)}. \tag{2}$$

**5.2.1.3 High level global features** Four features help characterise the global event by returning the overall kinematics of physics objects and the (tagged) jet multiplicities:

- $s_T$, the scalar sum of muon $p_{\mathrm{T}}$, tau $p_{\mathrm{T}}$, b jet $p_{\mathrm{T}}$, and $p_{\mathrm{T}}^{\mathrm{miss}}$;
- the total number of jets (inclusive), b jets, and $\tau$ jets.

Example distributions of some of the features are shown in Fig. 5.

#### 5.2.2 Architecture and training summary

The dataset of simulated signal and background events, selected as described in the previous section, is divided into two equally sized subsamples and a pair of discriminators is trained, one on each half of the data. In Sec. 5.3, each discriminator is then used to classify the events in the subsample on which it was not trained.

Each discriminant consists of an ensemble of ten fully-connected deep neural networks (DNN), each consisting of three hidden layers of 100 neurons with SELU [51] activation functions. They are implemented in KERAS [52] using TENSORFLOW [53] as a back-end. The training and inference procedures make use of the improvements described in Ref. [54], however all improvements were reverified in this context using the approximate median significance (AMS), as defined in Ref. [55], as the optimisation metric. The various improvements were found to increase on the AMS by about 30% compared to the original model.

### 5.3 Signal inference

#### 5.3.1 Summary statistic construction

The expected discovery significances and cross section upper limits at 95% confidence are determined by considering the output of the DNN as a summary statistic of the signal and background events, and performing the hypothesis test of signal+background versus the null hypothesis of background-only in multiple regions of the statistic, spanning its full range (a shape analysis). This is achieved by parametrising the density of the signal and background events using histograms of variable bin width in order to better capture the shape of the distributions whilst not causing the statistical uncertainties in each bin to become too large. This process is done on a per decay-channel basis, resulting in three sets of distributions.

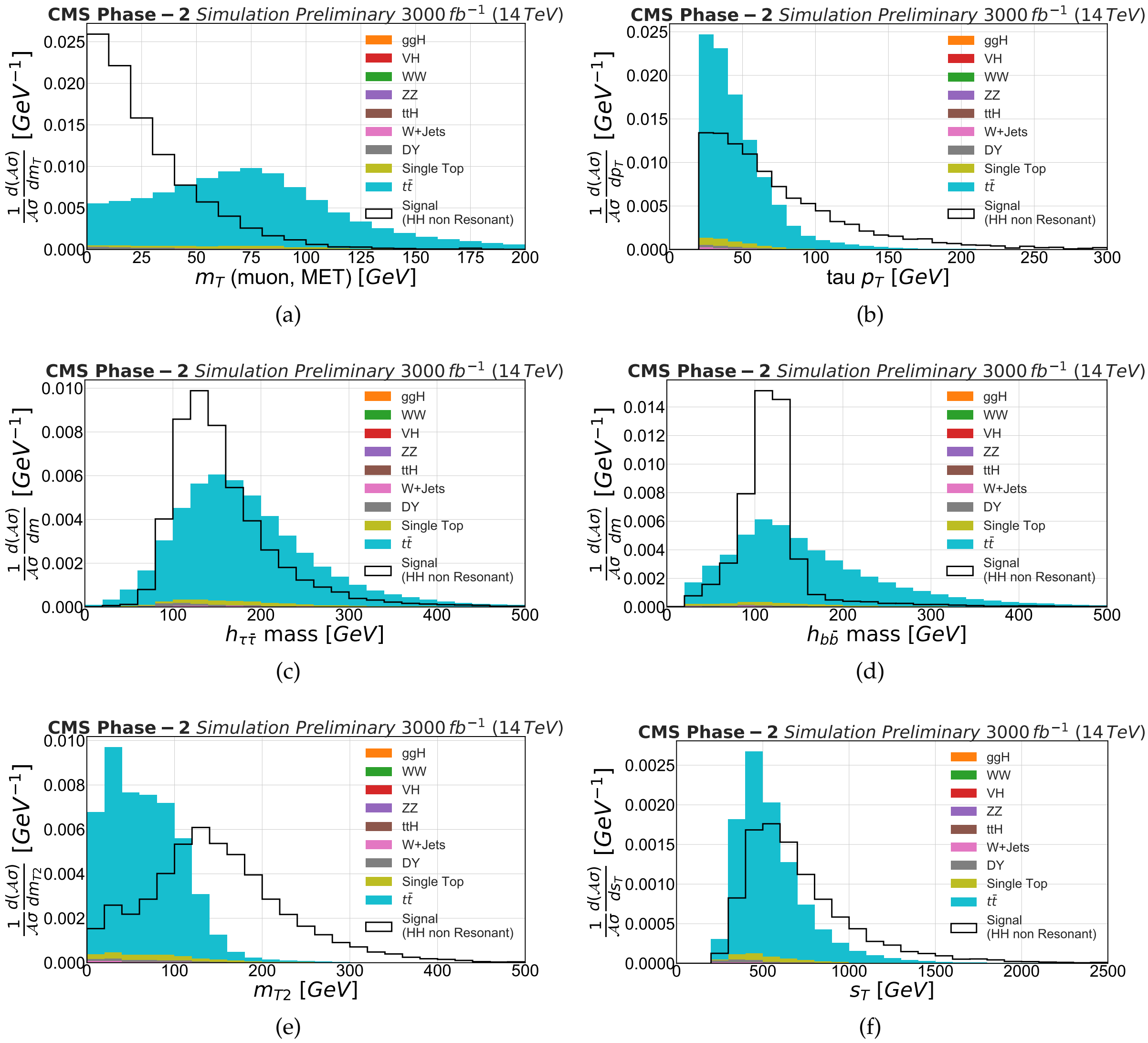

Figure 5: Example distributions for some of the features of the signal and background processes. Low-level features in the $\mu\tau_\mathrm{h}$ final-state:
(a) Transverse mass of the muon (i.e. $\tau_\mu$), as defined in Eq. 2,
(b) Transverse momentum of the $\tau_\mathrm{h}$.
Higgs-candidate masses for all final states together:
(c) $\mathrm{H}_{\tau\tau}$ mass,
(d) $\mathrm{H}_{\mathrm{b}\bar{\mathrm{b}}}$ mass.
High-level features for all final-states channels together:
(e) The stransverse mass $M_{T2}$,
(f) $s_T$ (defined as the scalar sum of lepton $p_\mathrm{T}$, $p_\mathrm{T}$ of both b-jets and $\tau_\mathrm{h}$, and $p_\mathrm{T}^{\mathrm{miss}}$).
Distributions are normalized to unit areas for signal and background, separately.

In order to make better use of the MC samples available, two DNN ensembles are trained, each on half of the samples. Each half of the data is then classified by the ensemble that was not trained on it. Since we make no fine-tuning of network hyper-parameters, and the architecture development was performed over cross-validation, each half of the data that was not used for training an ensemble represents an unseen sample of data for that ensemble. In doing this, we are able to classify the entire of our MC samples without the performance being biased due to overfitting.

Signal and background events are binned such that in a given bin, the statistical uncertainty on the population of each signal and background sample is less than 30%. However, only background samples with at least 100 MC events and a yield greater than 50 times that of the signal are considered when defining the bin edges, ensuring that the background samples considered have a sufficient number of simulated events to populate the distributions. These requirements aim to prevent our expected performance from being unduly limited by the sizes of the Monte Carlo samples currently available.

### 5.3.2 Hypothesis testing and results

A simultaneous fit is performed on the expected event distributions for the three final states considered, considering the systematic uncertainties described in Section 3.3. The uncertainty due to the number of MC events falling in each bin are neglected under the assumption that samples of sufficient size will be available for HL-LHC analyses.

With the assumed systematic uncertainties, an upper limit on the HH cross section times branching fraction of 1.4 times the SM prediction is obtained, corresponding to a significance of $1.4\sigma$. If only statistical uncertainties are considered, the upper limit amounts to 1.3 times the SM prediction for a significance of $1.6\sigma$.

## 6 HH $\rightarrow$ bbWW

We consider here HH final states containing two b jets and two neutrinos and two leptons, either electrons or muons. The decay channels involved are thus $\mathrm{H} \rightarrow \mathrm{b\bar{b}}$ in association with either a $\mathrm{H} \rightarrow \mathrm{Z}(\ell\ell)\mathrm{Z}(\nu\nu)$ or a $\mathrm{H} \rightarrow \mathrm{W}(\ell\nu)\mathrm{W}(\ell\nu)$ decay. While the analysis described in the following is optimised for HH $\rightarrow$ bbWW decays, that provide the largest branching fraction, the contribution of Higgs boson decays to both WW and ZZ, globally denoted as VV, is considered. Decays of the VV system to tau leptons subsequently decaying to electrons or muons with the associated neutrinos are also considered in the simulated signal samples. The corresponding branching fraction for the VV $\rightarrow \ell\nu\ell\nu$ decay is 1.73 % [19].

The dominant and subdominant background processes are $\mathrm{t\bar{t}}$ production in its fully leptonic decay mode, and Drell-Yan production of lepton pairs in association with jets. As both are irreducible background processes, i.e. they result in the same final state as the signal, the kinematic properties of the signal and background events are used and combined in an artificial Neural Network (NN) discriminant to enhance the sensitivity.

The single Higgs boson production backgrounds $\mathrm{t\bar{t}H}$ and $\mathrm{H} \rightarrow \mathrm{WW}(\ell\nu\ell\nu)$ were also considered but were found to have a negligible effect on the final result.

## 6.1 Event selection and background predictions

Events are selected based on the same selection criteria currently applied in the Run II CMS analysis of this final state. We assume that such events will be selected with dilepton triggers with transverse momentum thresholds similar to those deployed in Run II collisions, providing a 100% efficiency for the events that satisfy the selection described below.

Events are required to contain two leptons of opposite electric charge ($e^+e^-$, $\mu^+\mu^-$, $e^\pm\mu^\mp$), and with $p_T$ greater than 25 GeV and 15 GeV for ee events, 20 GeV and 10 GeV for $\mu\mu$ events, 25 GeV and 15 GeV for $\mu$e events, 25 GeV and 10 GeV for e$\mu$ events, for the higher and lower $p_T$ lepton, respectively. Electrons and muons in the pseudo-rapidity range $|\eta| < 2.8$ are considered, except the $1.444 < |\eta| < 1.5666$ being rejected for electrons. A dilepton mass requirement of $m_{\ell\ell} >$ 12 GeV is applied to all flavour combinations in order to suppress leptonia resonances.

Jets are required to have $p_T > 20$ GeV, $|\eta| < 2.8$, and be separated from identified leptons by a distance of $\Delta R = \sqrt{\Delta\phi^2 + \Delta\eta^2} > 0.3$. The magnitude of the negative vector sum of all PF candidates is referred to as $p_T^{miss}$. Selected jets must also satisfy the medium working point of the b tagging algorithm.

A summary of the object definitions and selections is shown in Table 3.

Table 3: Object definitions and event selections requirements.

| Object | ID and isolation requirements | Selection |
|---|---|---|
| Electrons | Isolation ($|\eta| > 1.5666$) $> 0.559$ (0.853) | $|\eta| < 1.444$ or $1.5666 < |\eta| < 2.8$<br>Leading $p_T > 25$ GeV<br>Sub-leading $p_T > 20$ GeV |
| Muons | Loose ID, Isolation $< 0.25$ | $|\eta| < 2.8$<br>Leading $p_T > 20$ GeV in $\mu\mu$ events<br>Leading $p_T > 25$ GeV in $\mu$e events<br>Sub-leading $p_T > 10$ GeV in $\mu\mu$ and e$\mu$ events |
| Jets | PUPPI Jet | $p_T > 20$ GeV, $|\eta| < 2.4$, $\Delta R_{lj} > 0.3$ |
| B-Jets | PUPPI medium MTD WP | $p_T > 20$ GeV, $|\eta| < 2.8$ |

Among all possible di-jets combinations fulfilling the previous criteria we select the two jets with the highest combined transverse momentum.

After the final object selection consisting of two opposite sign leptons and two b-tagged jets, a cut on $m_Z - m_{\ell\ell} > 15$ GeV is applied to remove the resonant Z peak and the high-$m_{\ell\ell}$ tail of the Drell-Yan+jets and $t\bar{t}$ background process.

The performance of the selections and object reconstruction was extensively checked looking at different kinematic distributions and comparing them to the ones obtained in the analysis of CMS data collected in the Run II.

After all the selection requirements described in this section, a total of about 50, 150, and 120 SM HH signal events is expected in the ee, e$\mu$, and $\mu\mu$ decay channels, respectively. The total numbers of background events in the three categories are $2.9 \times 10^6$, $8.4 \times 10^6$, and $7.5 \times 10^6$, respectively. The dominant background is $t\bar{t}$ production, with a sizeable contribution from Drell-Yan lepton pair production in the same-flavour dilepton channels.

### 6.2 Signal extraction

A neural network (NN) discriminant, based on the TMVA software [56], is used to improve the signal-to-background separation. In a phase space dominated by $t\bar{t}$ production, the NN utilizes information related to object kinematics. The variables provided as input to the NN exploit the presence of two Higgs bosons decaying into two b-jets on the one hand, and two leptons and two neutrinos on the other hand, which results in different kinematics for the di-lepton and di-jet systems between signal and background processes. The set of variables used as input is: $m_{\ell\ell}$, $m_{jj}$, $\Delta R_{\ell\ell}$, $\Delta R_{jj}$, $\Delta\phi_{\ell\ell,jj}$, defined as the $\Delta\phi$ between the di-jet and the di-lepton systems, $p_T^{\ell\ell}$, $p_T^{jj}$, $\min(\Delta R_{j,\ell})$, and $M_T$, defined as $M_T = \sqrt{2 p_T^{\ell\ell} p_T^{miss}(1-\cos(\Delta\phi(\ell\ell, p_T^{miss})))}$.

The output of the NN after selections in the $e^+e^-$, $\mu^+\mu^-$, $e^\pm\mu^\mp$ channels, is shown in Fig. 6. The Drell-Yan production of lepton pairs is modelled by interpolating separately in the three channels the simulated selected events with a third degree polynomial function, used to generate a smooth event distribution according to the expect event yields. The smoothing procedure has little impact on the projected sensitivity as discussed below.

As no shape uncertainty is considered in the $t\bar{t}$ background modelling and the total number of events for small NN output values is expected to reach tens of millions, the search may be subject to over-constraints of any systematic uncertainty that may be dependent or correlated with the $t\bar{t}$ shape. In order to mitigate this effect, the signal-depleted region with a NN output smaller than 0.5 will be excluded from the statistical inference.

### 6.3 Results

The results are derived as upper limits on the HH signal strength, defined as the ratio of the signal cross section times branching fraction to the SM expectation. A simultaneous fit is performed on the distributions of events shown in Fig. 6 in the $e^+e^-$, $\mu^+\mu^-$ and $e^\pm\mu^\mp$ event categories.

The expected upper limit at the 95% CL corresponds to 3.5 times the SM prediction when systematic uncertainties are considered, and to 3.3 times the same value if only statistical uncertainties are assumed. The corresponding significance of the HH signal is 0.56 and 0.59$\sigma$, respectively.

The impact of the assumed Drell-Yan contribution was shown to be small by scaling its expected yield by factors of 0.5 and 2 and verifying the changes in the sensitivity. Variations of 5% or below in the final result were observed, showing that the result is robust under different assumptions on the Drell-Yan background contamination.

## 7 $HH \rightarrow bb\gamma\gamma$

While characterized by a tiny branching fraction, the $bb\gamma\gamma$ final state is experimentally very clean and thus provides a large sensitivity. The main background processes are the continuum production of diphoton events and of single photons in association with a misidentified jet, as well as single Higgs boson production, where the Higgs boson decays to $\gamma\gamma$.

### 7.1 Selection

We assume that events selected as described below will be triggered with algorithms that require the presence of two photons and that are 100% efficient for selected events. This assumption is well verified in the current Run II search.

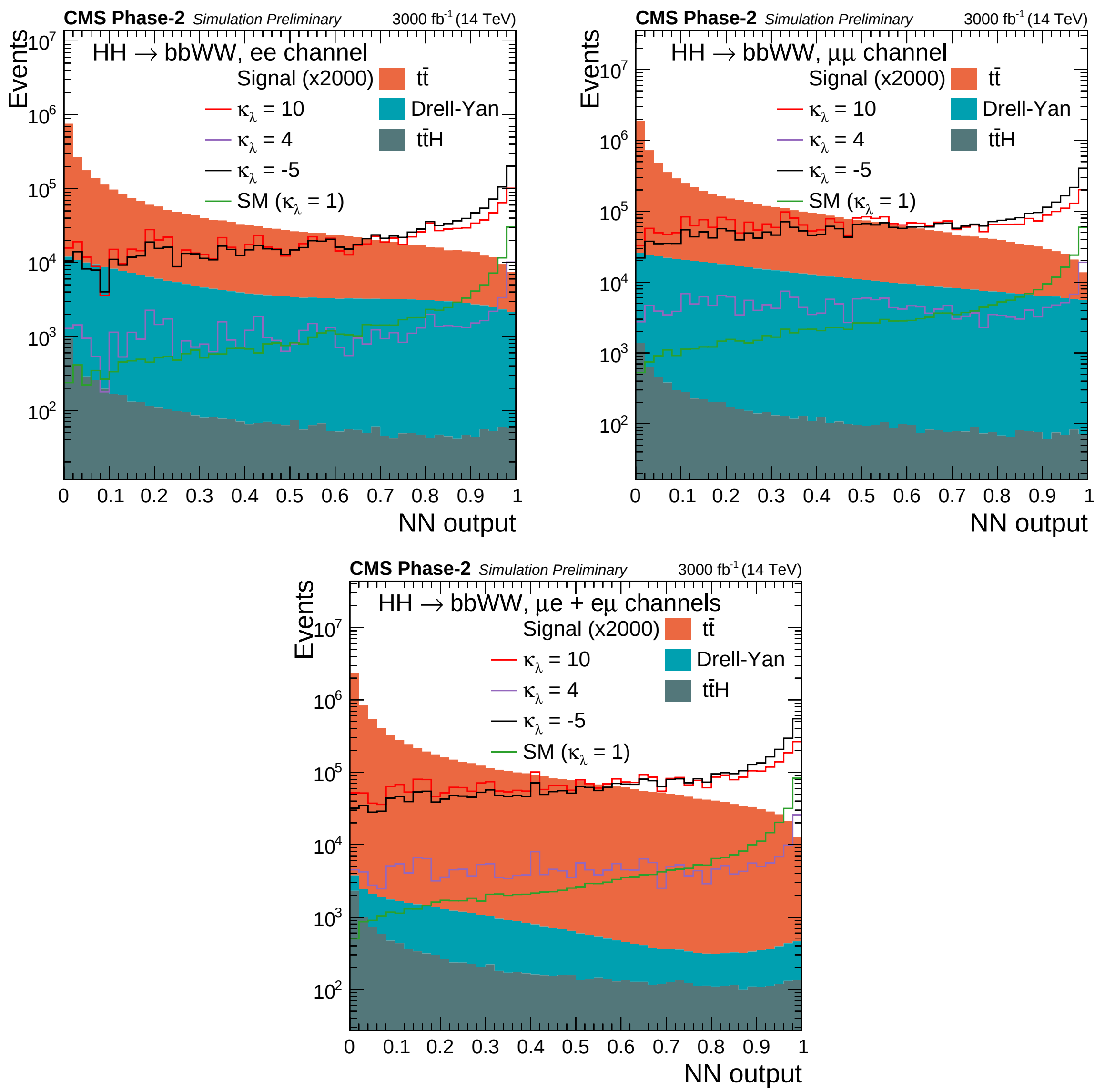

Figure 6: The output of the NN after the selections, evaluated in the $e^+e^-$ (top left) , $\mu^+\mu^-$ (top right), $e^\pm\mu^\mp$ (bottom) channels.

Photons satisfying the loose working point, corresponding to an efficiency of about 90% for a photon with $p_T > 30$ GeV and a jet misidentification rate of about 3%, are selected. The two photons with the highest $p_T$ that satisfy such requirements are considered and used to build the H → $\gamma\gamma$ candidate, and the kinematic selections reported in Table 4 are applied. While the acceptance of photons in the upgraded detector extends up to $\eta = 3$, the pseudorapidity is limited to 2.5 in order to increase the signal to background ratio. For events selected in the most sensitive categories, as described below, the signal photons are more central than background ones. An additional selection in $\eta$ is applied to exclude the transition region from the barrel electromagnetic to the endcap calorimeters.

The H → b$\bar{\text{b}}$ candidate is built from jets that satisfy the kinematic selection reported in Table 4. As discussed above, the $|\eta| < 2.5$ requirement is applied in order to increase the signal to background ratio. The angular distance $\Delta R_{\gamma j}$ between the jets and the selected photons is required to be larger than 0.4. In case more than two jets satisfy the kinematic requirements described above, candidates satisfying the highest b tagging criteria are preferred. In case ambiguities in the choice persist, the higher $p_T$ objects are selected. The background from light flavour jets is suppressed by requiring both jets to satisfy the loose working point of the b tagging algorithm, corresponding roughly to a 90% efficiency for a genuine b-jet and 10% of fake rate from jets initiated by light quarks or gluons. In addition, the dijet invariant mass $m_{jj}$ is required to be between 80 and 190 GeV.

Table 4: Photon and jet kinematic selections.

| Photon selections |
|---|
| $p_T/m_{\gamma\gamma} > 1/3$ (leading $\gamma$), $> 1/4$ (subleading $\gamma$) |
| $\|\eta\| < 1.44$ or $1.57 < \|\eta\| < 2.5$ |
| $100\,\text{GeV} < m_{\gamma\gamma} < 180\,\text{GeV}$ |
| **Jet selections** |
| $p_T > 25\,\text{GeV}$ |
| $\|\eta\| < 2.5$ |
| $\Delta R_{\gamma j} > 0.4$ |
| $80\,\text{GeV} < m_{jj} < 190\,\text{GeV}$ |
| At least 2 b-tagged jet (loose WP) |

The invariant mass of the $\gamma\gamma jj$ system is denoted as $m_{\gamma\gamma jj}$. The jet and photon resolution effects in $m_{\gamma\gamma jj}$ are mitigated by defining the variable $M_X$ as:

$$M_X = m_{\gamma\gamma jj} - m_{\gamma\gamma} - m_{jj} + 250\text{ GeV} \tag{3}$$

After such kinematic preselections, the main background contribution comes from nonresonant diphoton events. The main resonant backgrounds consist of Higgs boson production in association with two top quarks ($t\bar{t}H$), as expected because of their topology that is very similar to the signal. This background source is suppressed with the usage of a dedicated multivariate discriminant, consisting of a BDT trained to separate the HH and $t\bar{t}H$ processes. The discriminant combines twelve variables related to the presence of additional jets, electrons or muons, as well as the helicity angles of the HH and $b\bar{b}$ systems. A selection is applied on the output of the discriminant, rejecting approximately 75% of the $t\bar{t}H$ contamination for a signal efficiency of 90%. The selection on the BDT output was optimized based on the expected significance for the SM HH signal hypothesis.

## 7.2 Event categorization

A categorization based on a multivariate discriminant is performed. A BDT is trained to discriminate the signal from the sum of background processes. The $t\bar{t}H$ process is not considered because of the dedicated $t\bar{t}H$ discriminant described above. The variables used for the BDT training are:

- minimum angular distance between the selected jets and the selected photons. This variable is expected to reject collinear photon emission from a quark characteristic of QCD processes.
- angle between the diphoton object and the beam axis in the $\gamma\gamma jj$ rest-frame.
- angle between the leading selected jet and the beam axis in the dijet rest-frame.

- angle between the leading selected photon and the beam axis in the diphoton rest-frame.
- ratio of the transverse momentum to the mass, $p_T/M$, for the selected diphotons and dijets.
- b tag output for the two selected jets.
- photon energy resolution $\sigma_E/E$ for the two selected photons.
- angles on transverse plane between the direction of the missing momentum and the two selected jets.
- numbers of loose, medium, tight b-tagged jets in the event

It has been verified that the selection applied on the BDT output does not sculpt the diphoton and dijet invariant mass distributions for background events. Events with a low BDT classifier output are rejected, allowing to suppress approximately 90% of the background events including QCD events with light quarks. The events thus selected are divided in a medium and a high purity category based on the BDT output. The high purity (HP) category provides the best sensitivity and collects approximately 35% of the preselected events, while the medium purity (MP) category increases the overall acceptance to about 75% of the signal events, contributing to the overall sensitivity. Events are further divided into three categories depending on their $M_X$ value. For each category an optimization on the separation between MP and HP is performed to maximize the sensitivity to the SM HH production, and the SM HH signal is approximately equally shared among the resulting categories. The event categorization is summarized in Table 5.

In total, considering all the categories and a region centred on the $m_{\gamma\gamma}$ signal peak with a width of 2 times its resolution, about 40 SM HH signal events are expected to be recorded, for a total of about 190 resonant background events and 3600 nonresonant background events, approximately. High mass categories provide the best sensitivity to the SM HH signal, with about 35 signal events expected and a total of about 1600 background events. Low mass categories are sensitive to variations of the Higgs boson self coupling, that may enhance the production cross section at low $m_{HH}$ values.

Table 5: Categorisation applied for events selected in the bb$\gamma\gamma$ analysis. The symbols MP and HP denote, respectively, the medium and high purity categories based on the BDT output.

| MVA category | Classification on $M_X$ |
|---|---|
| 0: HP<br>1: MP | $250 < M_X < 350$ GeV: |
| 2: HP<br>3: MP | $350 < M_X < 380$ GeV |
| 4: HP<br>5: MP | $480 < M_X$ GeV |

## 7.3 Results

The background distributions are modelled by fitting the selected MC event distribution with an exponentially falling distribution, that was observed to describe well the simulated events. Results are obtained with a simultaneous fit of a pseudodataset (Asimov dataset) constructed from the modelled signal and background distributions in the categories defined above. An illustration of the expected distributions of events in the three high purity categories for the photon and jet pairs invariant mass is shown in Fig. 7. The events are obtained by generating random distribution with a Poissonian statistics from the Asimov dataset including SM signal and background.

The expected 95% CL upper limits on the HH signal is 1.09 times the HH cross section when systematic uncertainties are considered, and 1.11 when only statistical uncertainties are accounted for. The corresponding significance of the HH signal is 1.83 and 1.85$\sigma$, respectively.

## 8 HH $\rightarrow$ bbZZ

Up to now, the low signal rate leads to consider mostly final states with a sizable branching ratio. In view of HL-LHC, some rare but clean processes have been re-considered because of the increasing available statistics and the challenging conditions due to the enormous number of pile-up events. In this work, the sensitivity to the Higgs self-coupling for $m_H$ = 125 GeV is evaluated through the measurement of the non-resonant di-Higgs production final states in the bbZZ $(4\ell)$ decay mode, where $\ell = e, \mu$. Despite a small cross section ($\sigma_{b\bar{b}4\ell} = 5.3$ ab), the presence of four leptons associated with two b jets leads to a very clean final state topology allowing to maintain a rather good signal selection efficiency and to control the backgrounds. The main background processes are $t\bar{t}H(ZZ)$, $t\bar{t}Z$, ggH and ZH, followed by minor contributions such as WH and single Higgs production via vector boson fusion (VBF). The $t\bar{t}ZZ$ and $t\bar{t}H(WW)$ contributions are found to be negligible. Top quark pair production and Drell-Yan (DY) lepton pair production in association to jets are a reducible background for the analysis. As their contamination is due to hadrons misidentified as leptons (fake leptons) or to the selection of non-prompt leptons, large suppressions are expected with the selections used in this work. Nevertheless, their huge cross section, orders of magnitude larger than the HH signal, makes them a challenging background at the HL-LHC. The estimation of the $t\bar{t}$ and DY contribution in this work is difficult because of the limited number of MC events available, leading to very large uncertainties related to the few or zero events satisfying the selections. Moreover, the actual impact of the $t\bar{t}$ and DY backgrounds on the analysis largely depends on reconstruction techniques and performance in the rejection of fake and non-prompt leptons that are not fully optimized in the parametric simulation implemented in Delphes. We assume that dedicated techniques and optimized algorithms will be available by the HL-LHC operations to have negligible contamination, deeming this assumption reasonable from studies performed on the MC simulation. We also remark that the size of MC samples will not represent an issue at the HL-LHC, given the possibility to control the effective contaminations in data control regions. Nevertheless, the developments of such methods will represent a major challenge towards HL-LHC to maintains a high sensitivity in the bbZZ$(4\ell)$ channel. Signal and background samples are simulated as described in Section 3. In addition to the SM scenario ($\kappa_\lambda = 1$), samples with several other values of $\kappa_\lambda$, ranging from $\kappa_\lambda = -10$ to $\kappa_\lambda = 10$, are generated.

### 8.1 Event Selection

Events are required to have at least four identified and isolated (isolation $< 0.7$) muons (electrons) with $p_T > 5(7)$ GeV and $|\eta| < 2.8$, where muons (electrons) are selected if passing the loose (medium) working point identification. Z boson candidates are formed from pairs of opposite-charge, same flavour leptons ($\ell^+\ell^-$) requiring a minimum angular separation between two leptons of 0.02. At least two di-lepton pairs are required. The Z candidate with the invariant mass closest to the nominal Z mass is denoted as $Z_1$; then, among the other opposite-sign lepton pairs, the one with the highest $p_T$ is labelled as $Z_2$. In order to improve the sensitivity to the Higgs boson decay, Z candidates are required to have an invariant mass in the range

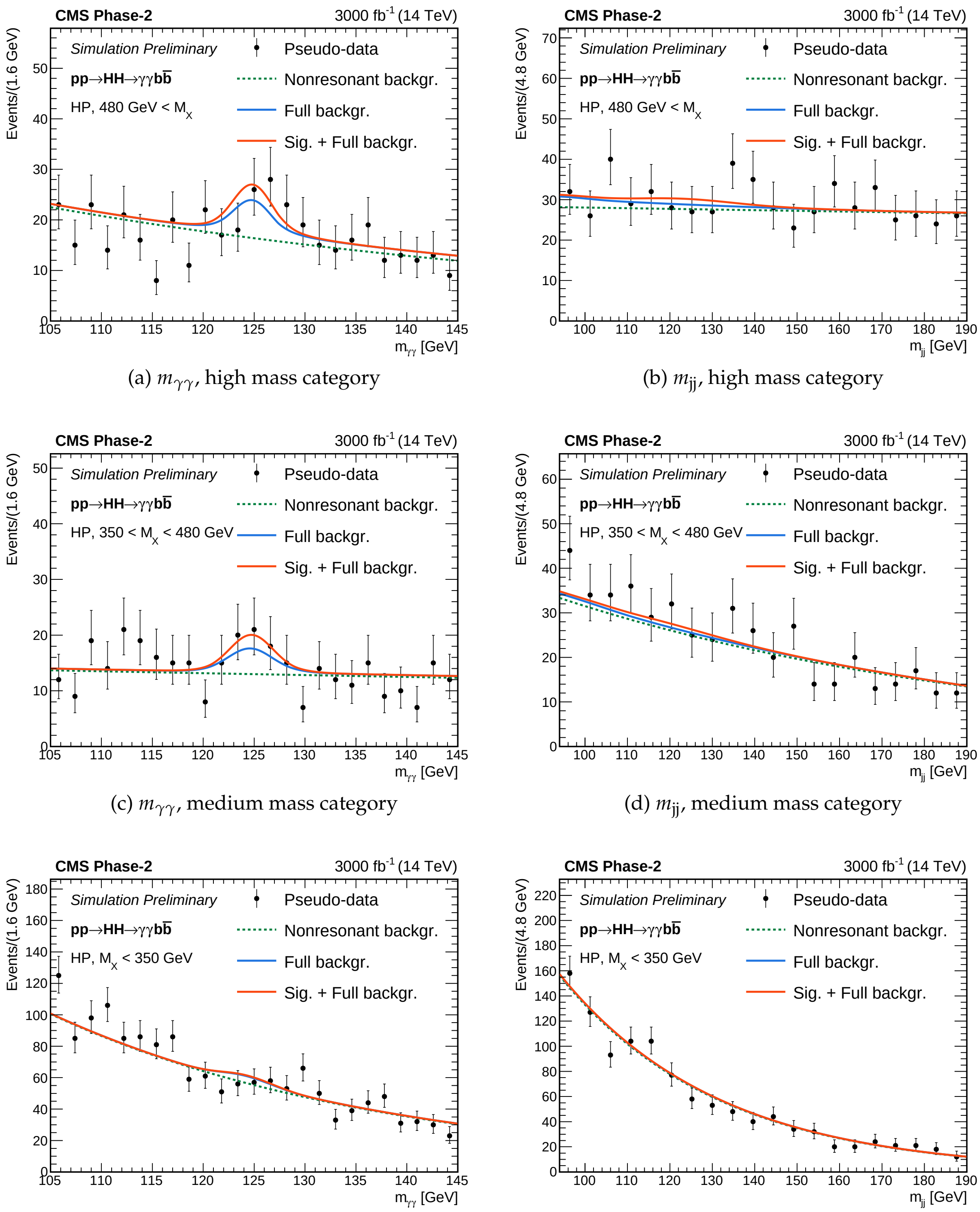

(a) $m_{\gamma\gamma}$, high mass category

(b) $m_{\mathrm{jj}}$, high mass category

(c) $m_{\gamma\gamma}$, medium mass category

(d) $m_{\mathrm{jj}}$, medium mass category

(e) $m_{\gamma\gamma}$, low mass category

(f) $m_{\mathrm{jj}}$, low mass category

Figure 7: Expected distribution of events in the photon (left column) and jet (right column) pair invariant mass. The full circles denote pseudo-data obtained from the expected events yields for the sum of the signal and background processes for 3000 fb$^{-1}$. Only the most sensitive high purity category is shown.

[50, 100] GeV ($Z_1$) and [12, 60] GeV ($Z_2$), respectively. At least one lepton is required to have $p_T > 20$ GeV and a second is required to have $p_T > 10$ GeV. The four leptons invariant mass, $m_{4\ell}$, is requested to be in the range [120, 130] GeV.

At least two (but not more than three) identified b jets, reconstructed with the anti-$k_T$ algorithm inside a cone of radius $R = 0.4$, are required; a b tag Medium working point is assumed. Their invariant mass, corrected assuming an improvement of 20% on the resolution on the $m_{b\bar{b}}$ peak, as expected for HL-LHC thanks to a proper b jet energy regression, is required to be in the range [90, 150] GeV. The angular distance between the two b jets has to be $0.5 < \Delta R_{b\bar{b}} < 2.3$; furthermore, the missing transverse energy of the event must be smaller than 150 GeV, and a selection on the angular distance between the two reconstructed Higgs is set to $\Delta R_{HH} \geq 2.0$.

## 8.2 Results

The invariant mass spectrum of the four leptons after the full event selection is shown in Figure 8. Considering the channels investigated, we expect to select 1 HH event for a total background yield of 6.8 in the inclusive $b\bar{b}4\ell$ ($\ell = e, \mu$) final state.

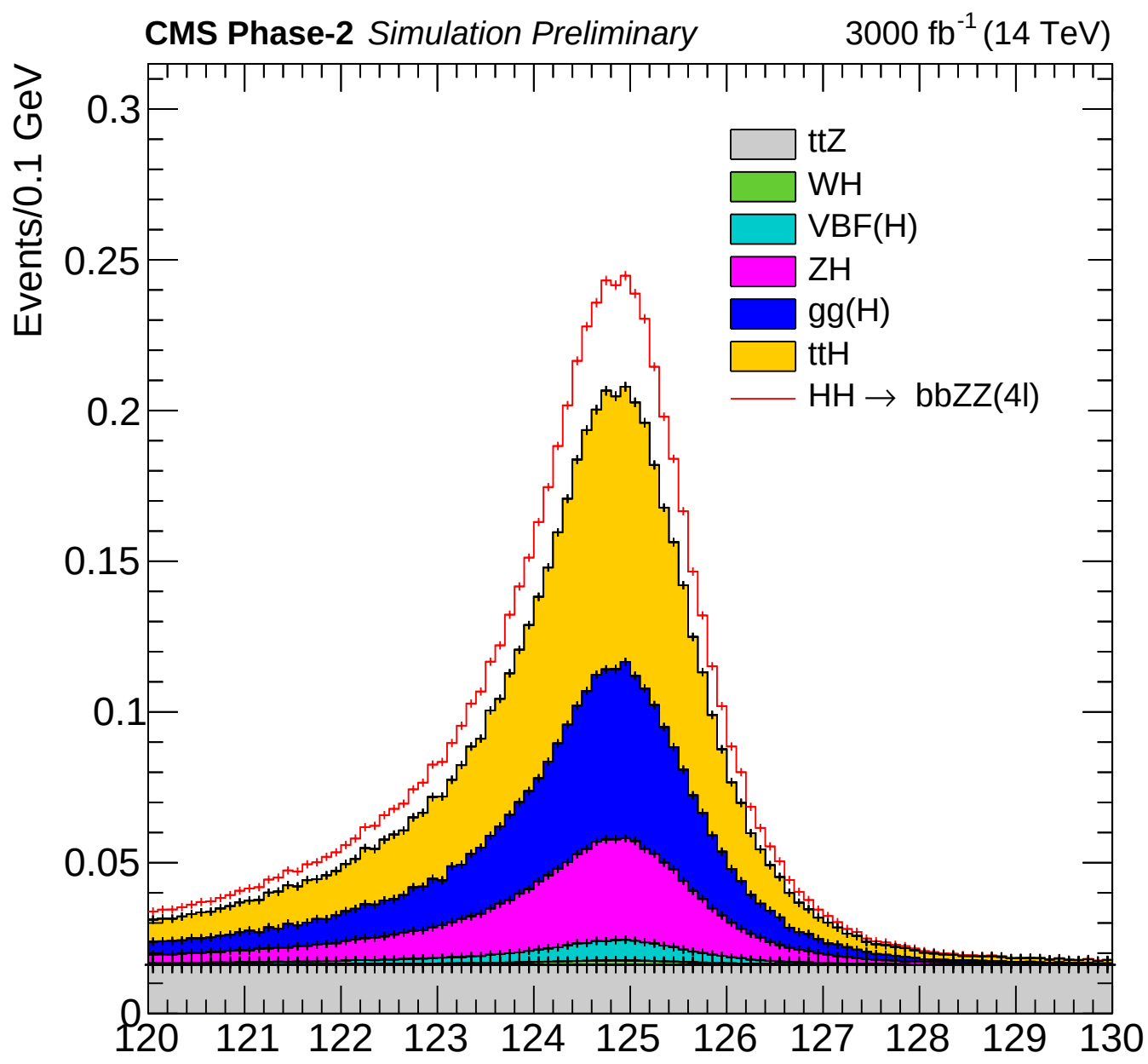

Figure 8: Invariant mass distribution of the four leptons selected at the end of the analysis for the $b\bar{b}4\ell$ final state.

The combined upper limit at the 95% CL on the HH cross section corresponds to 6.6 times the SM prediction, with a corresponding significance of $0.37\sigma$. The impact of the systematic uncertainties on the analysis is found to be almost negligible. The most sensitive channel is $b\bar{b}4\mu$, but a sizeable contribution to the sensitivity also comes from the $b\bar{b}2e2\mu$ and $b\bar{b}4e$ final states.

# 9 Decay channel combination and results

The results obtained in the five decay channels described above are combined statistically assuming the SM branching fractions for HH decays to the final states studied. The analyses of the five decay channels are designed to be orthogonal thanks to the mutually exclusive requirements in the objects used, or to have negligible overlap due to tight object identification criteria and the separation and the efficient separation achieved by the multivariate methods used. Systematic uncertainties associated to the same objects, such as the b tag efficiency uncertainties, and to the same processes, including common backgrounds and the HH signal, are correlated across the corresponding decay channels, while the others are left uncorrelated.

Table 6 summarises, for the five channels and their combination, the upper limit at the 95% confidence level (CL) and the significance for the SM HH signal. The combined 95% CL upper limit on the SM HH cross section amounts to 0.77 times the SM prediction, with a corresponding significance of the signal of $2.6\sigma$. These results significantly improve over previous projections thanks to the dedicated optimisation of the analysis strategies to the HL-LHC dataset. In comparison, the extrapolation to an integrated luminosity of $3000\,\mathrm{fb}^{-1}$ of the current Run II combination, obtained with a dataset of $35.9\,\mathrm{fb}^{-1}$ collected at $\sqrt{s} = 13\,\mathrm{TeV}$ [57], yields a projected SM HH significance of $1.8\sigma$ neglecting all the systematic uncertainties.

Table 6: Upper limit at the 95% confidence level, significance, projected measurement at 68% confidence level of the Higgs boson self coupling $\lambda_{\mathrm{HHH}}$ for the five channels studied and their combination. Systematic and statistical uncertainties are considered.

| Channel | Significance | | 95% CL limit on $\sigma_{\mathrm{HH}}/\sigma_{\mathrm{HH}}^{\mathrm{SM}}$ | |
|---|---|---|---|---|
| | Stat. + syst. | Stat. only | Stat. + syst. | Stat. only |
| bbbb | 0.95 | 1.2 | 2.1 | 1.6 |
| $\mathrm{bb}\tau\tau$ | 1.4 | 1.6 | 1.4 | 1.3 |
| $\mathrm{bbWW}(\ell\nu\ell\nu)$ | 0.56 | 0.59 | 3.5 | 3.3 |
| $\mathrm{bb}\gamma\gamma$ | 1.8 | 1.8 | 1.1 | 1.1 |
| $\mathrm{bbZZ}(\ell\ell\ell\ell)$ | 0.37 | 0.37 | 6.6 | 6.5 |
| Combination | 2.6 | 2.8 | 0.77 | 0.71 |

Prospects for the measurement of the $\lambda_{\mathrm{HHH}}$ coupling are also studied. Under the assumption that no HH signal exists, 95% CL upper limits on the SM HH production cross section are derived as a function $\kappa_\lambda = \lambda_{\mathrm{HHH}}/\lambda_{\mathrm{HHH}}^{\mathrm{SM}}$, where $\lambda_{\mathrm{HHH}}^{\mathrm{SM}}$ denotes the SM prediction. The result is illustrated in Fig. 9. A variation of the excluded cross section, directly related to changes in the HH kinematic properties, can be observed as a function of $\lambda_{\mathrm{HHH}}$. In the case of the bbWW analysis, these changes largely impact the DNN discriminant distribution that is optimised for the SM point. Parametrisation techniques, similar to those deployed in the Run II search, and further optimisations can be envisaged at the HL-LHC to mitigate this effect and improve the constraint on $\lambda_{\mathrm{HHH}}$.

Assuming instead that a HH signal exists with the properties predicted by the SM, prospects for the measurement of the $\lambda_{\mathrm{HHH}}$ are derived. The scan of the likelihood as a function of the $\kappa_\lambda$ coupling is shown in Fig. 10. The projected confidence interval on this coupling corresponds to $[0.35, 1.9]$ at the 68% CL and to $[-0.18, 3.6]$ at the 95% CL. The peculiar likelihood function structure, characterised by two local minimums, is related to the dependence of the total cross section and HH kinematic properties on $\kappa_\lambda$, while the relative height of the two minimums depends to the capability of the analyses to access differential $m_{\mathrm{HH}}$ information. The total HH cross section has a quadratic dependence on $\kappa_\lambda$ with a minimum at $\kappa_\lambda \approx 2.45$, while the kinematic differences for signals with $\kappa_\lambda$ values symmetric around this minimum are mostly relevant in the low region of the $m_{\mathrm{HH}}$ spectrum. Consequently, a partial degeneracy exists between the $\kappa_\lambda = 1$ value, that is assumed for the expected signal plus background modelling in the results of Fig 10, and a second $\kappa_\lambda$ value. The exact position of this second minimum depends on the interplay between the changes in the cross section and in the acceptance as a function of $\kappa_\lambda$. In analyses that retain sensitivity on the differential $m_{\mathrm{HH}}$ distribution, such as bbbb and bb$\tau\tau$ where this information is used as input to the multivariate methods, this degeneracy is partly removed. In the case of the bb$\gamma\gamma$ analysis, with a good acceptance and purity in the low $m_{\mathrm{HH}}$ region and a dedicated $m_{\mathrm{HH}}$ categorisation, a better discrimination of the second minimum is achieved. Further improvements can be envisaged in HL-LHC analyses by extending the $m_{\mathrm{HH}}$ categorisation to other channels beyond bb$\gamma\gamma$.

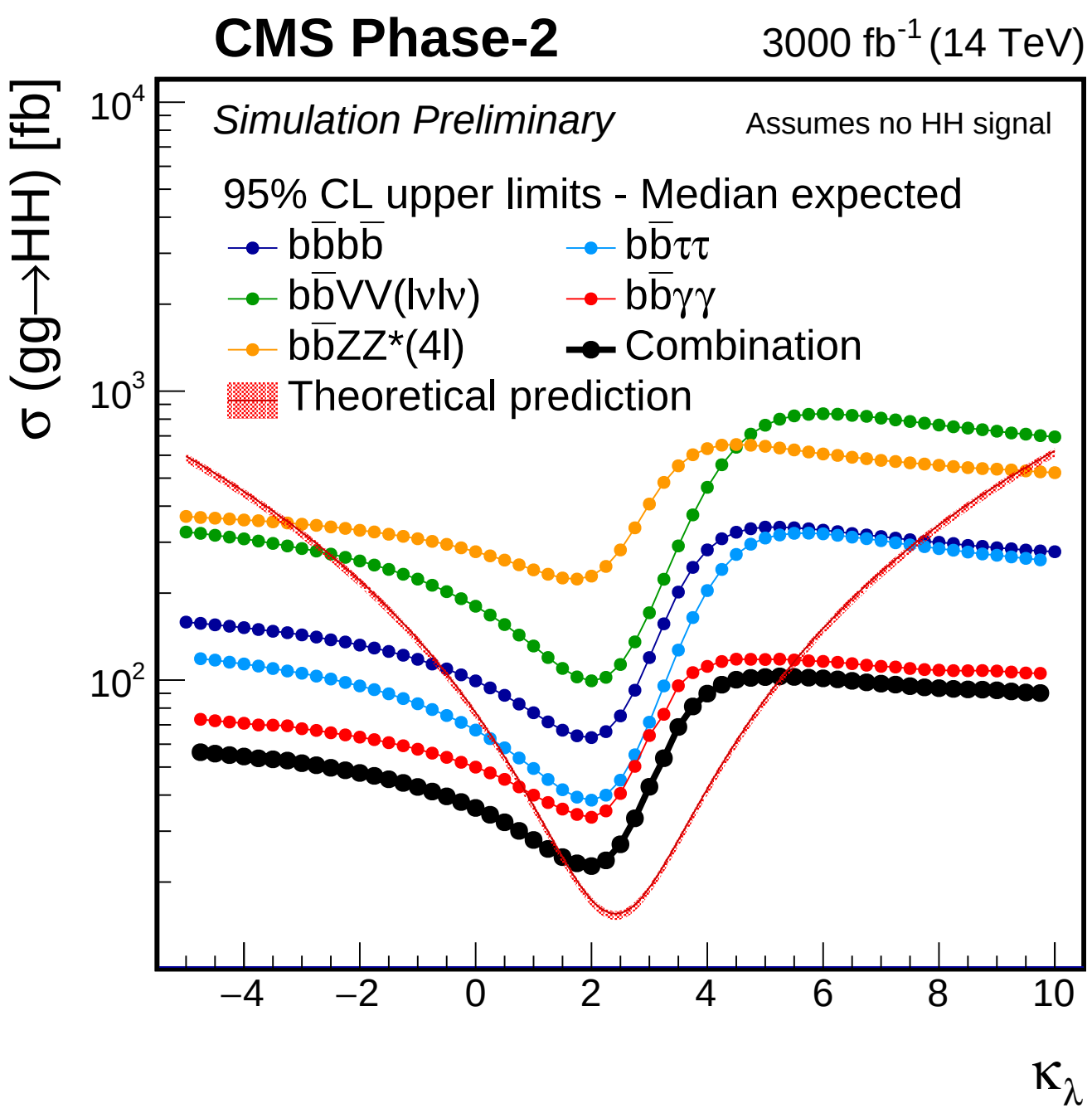

Figure 9: Upper limit at the 95% CL on the HH production cross section as a function of $\kappa_\lambda = \lambda_{\mathrm{HHH}}/\lambda^{\mathrm{SM}}_{\mathrm{HHH}}$ for the five decays channels investigated and their combination. The red band indicated the theoretical production cross section.

The combination of the five channels largely removes the degeneracy, and results in a plateau in the likelihood function for $\kappa_\lambda$ values between 4 and 6. Improvements in the combined sensitivity in this region have a large effect on the size of the 95% CL interval for the $\kappa_\lambda$ measurement.

## 10 Summary

Prospects for the search of Higgs boson pair (HH) production and for the measurement of the Higgs boson self-coupling ($\lambda_{\mathrm{HHH}}$) at the High-Luminosity LHC (HL-LHC) are presented. The study is performed using the five decay channels of the HH system to bbbb, bb$\tau\tau$, bbWW (with both W decaying leptonically), bb$\gamma\gamma$, and bbZZ (with both Z decaying to a pair of electrons or muons). The response of the upgraded CMS detector is studied with a parametric simulation that accounts for an average of 200 pp interactions per bunch crossing, and simulates the performance in the reconstruction and identification of physics objects. Assuming that no HH signal exists, a 95% confidence level (CL) upper limit on its cross section can be set to 0.77 times the SM prediction. Assuming that a HH signal exists with the properties predicted by the SM, we expect a combined significance of 2.6$\sigma$ and a determination of the $\lambda_{\mathrm{HHH}}$ coupling corresponding to the interval $[0.35, 1.9]$ at the 68% CL and to $[-0.18, 3.6]$ at the 95% CL.

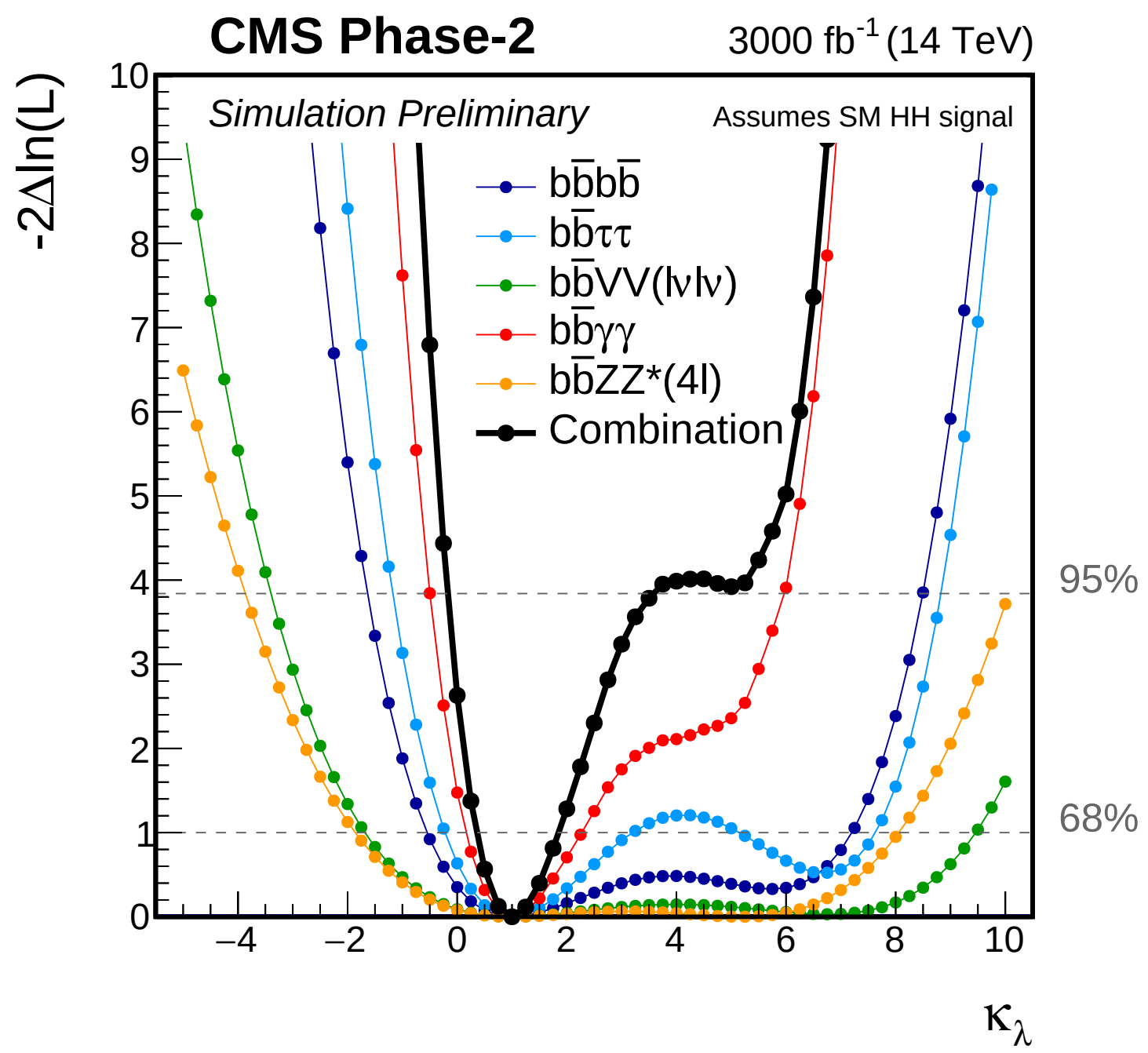

Figure 10: Expected likelihood scan as a function of $\kappa_\lambda = \lambda_{\mathrm{HHH}}/\lambda^{\mathrm{SM}}_{\mathrm{HHH}}$. The functions are shown separately for the five decay channels studied and for their combination.

# CMS Physics Analysis Summary

Contact: cms-phys-conveners-ftr@cern.ch 

# Constraints on the Higgs boson self-coupling from ttH+tH, H $\to \gamma\gamma$ differential measurements at the HL-LHC

The CMS Collaboration

## Abstract

This note details a study of prospects for ttH+tH, H $\to \gamma\gamma$ differential cross section measurements at the HL-LHC with the CMS Phase-2 detector. The study is performed using simulated proton-proton collisions at a centre-of-mass energy of $\sqrt{s} = 14$ TeV, corresponding to 3 ab$^{-1}$ of data. The expected performance of the upgraded CMS detector is used to model the object reconstruction efficiencies under HL-LHC conditions. The results are interpreted in terms of the expected sensitivity to deviations of the Higgs boson self-coupling, $\kappa_\lambda$, from beyond standard model effects. Using the HL-LHC data, the precision expected in ttH+tH, H $\to \gamma\gamma$ differential cross section measurements will constrain $\kappa_\lambda$ within the range $-4.1 < \kappa_\lambda < 14.1$, at the 95% confidence level, assuming all other Higgs boson couplings are fixed to standard model predictions. Moreover, it is possible to disentangle the effects of a modified Higgs boson self coupling from the presence of other anomalous couplings by using the differences in the shape of the measured spectrum. This separation is unique to differential cross section measurements. The ultimate sensitivity to the Higgs boson self coupling, achievable using differential cross section measurements, will result from a combination across Higgs boson production modes and decay channels.

*This document has been revised with respect to the version dated November 19, 2018.*

# 1 Introduction

In the standard model (SM) of particle physics [1–6], electroweak symmetry breaking (EWSB) is realised through the addition of a complex scalar doublet field, which, after EWSB, yields a physical, neutral, scalar particle, a Higgs boson (H). Since the discovery of the Higgs boson by the ATLAS and CMS Collaborations [7–9], several experimental measurements have been designed to test its compatibility with SM predictions. Despite the precision already achieved in measurements of the Higgs boson couplings to SM particles in the first two runs of the LHC [10–12], constraints on the Higgs boson self-coupling obtained from searches for double Higgs boson production [13, 14], remain limited.

An alternative approach to probing the Higgs boson self-coupling, exploiting radiative corrections to inclusive and differential Higgs boson production rates has been suggested in references [15–20]. At next-to-leading order (NLO), single-Higgs boson production includes processes with access to the Higgs boson trilinear coupling, $\lambda_3$, such as that shown in Fig. 1. The contributions from the Higgs boson self-coupling are sizeable for Higgs boson production in association with a pair of top quarks (ttH), a single top-quark (tH) or a massive vector boson (VH, V=W or Z). The effect is larger in these production modes due to the large mass of the V boson or top quark, providing a larger coupling to the virtual Higgs boson. Conversely, corrections to the dominant gluon-fusion (ggH) and vector-boson fusion (qqH) production modes are much smaller. Differential cross section measurements, in particular as a function of the Higgs boson transverse momentum $p_{\mathrm{T}}^{\mathrm{H}}$, allow one to disentangle the effects of modified Higgs boson self-coupling values from other effects such as the presence of anomalous top–Higgs couplings.

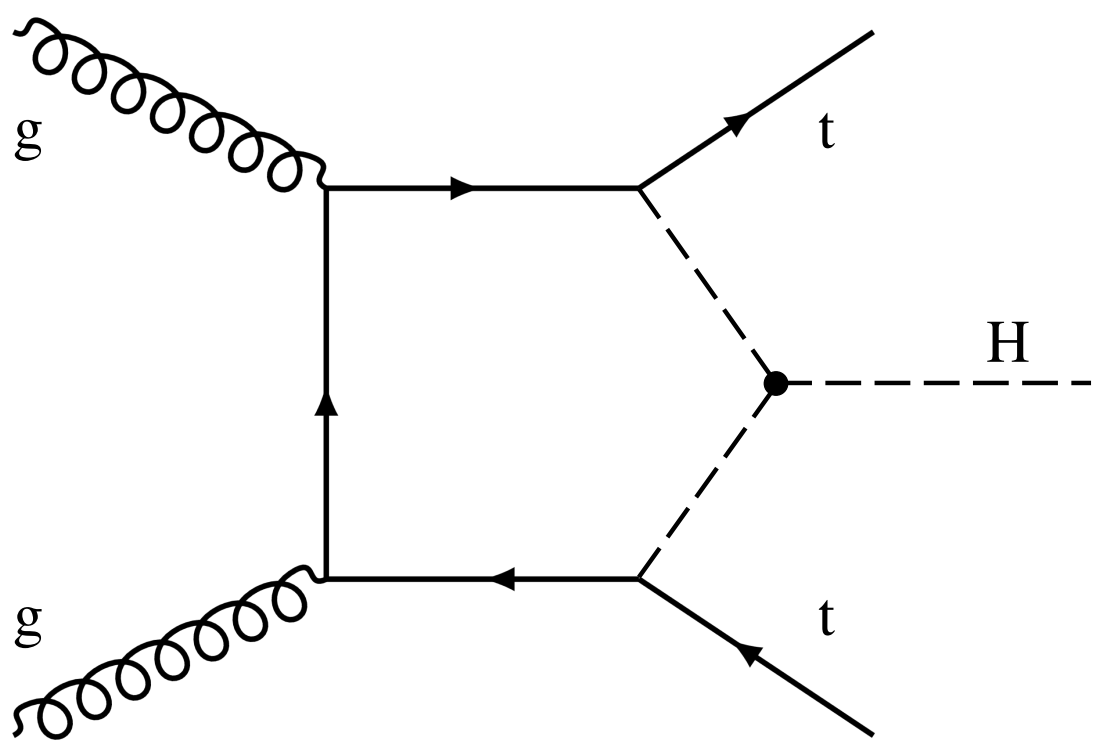

Figure 1: Example of a NLO Feynman diagram for ttH production which includes the Higgs boson self-coupling.

The dependence of the single-Higgs boson differential cross section is parameterised as a function of $\kappa_\lambda = \lambda_3/\lambda_3^{\mathrm{SM}}$, by considering NLO terms arising from the Higgs boson self-coupling such as the one in Fig. 1. This dependance is sensitive to both the production mode and kinematics of the Higgs boson. Scaling functions, $\mu_{ij}(\kappa_\lambda)$, are calculated using an electroweak reweighting tool [21] which determines the cross section, relative to the SM prediction, in a specific bin, $i$, of $p_{\mathrm{T}}^{\mathrm{H}}$, for each production mode, $j$. The $\kappa_\lambda$-dependent modifications are largest for ttH production, at threshold (low $p_{\mathrm{T}}^{\mathrm{H}}$). A 20% enhancement to the ttH production rate for $p_{\mathrm{T}}^{\mathrm{H}} \in [0,45]$ GeV, is predicted for $\kappa_\lambda \sim 10$. Further details on extracting $\mu_{ij}(\kappa_\lambda)$ relevant for this analysis, and the electroweak reweighting tool is provided in Section 3.

This note describes a strategy for measuring the $p_{\mathrm{T}}^{\mathrm{H}}$ differential cross section of a Higgs boson produced in association with at least one top quark and decaying to photons (ttH + tH, H $\rightarrow \gamma\gamma$), at the High-Luminosity LHC (HL-LHC) with the CMS Phase-2 detector, for a centre-of-mass energy of $\sqrt{s} = 14\,\text{TeV}$. The $\text{H} \rightarrow \gamma\gamma$ decay mode provides a clean final state in which the transverse momentum of the Higgs boson can be well reconstructed, owing to the excellent energy resolution of the CMS electromagnetic calorimeter for photons. Moreover, measurements of ttH production [22, 23] from a combination of decay channels obtain a significant contribution in sensitivity from the $\text{H} \rightarrow \gamma\gamma$ decay. The expected precision is determined based on simulated proton-proton (pp) events at the HL-LHC in which the conditions of the HL-LHC runs are accounted for in the object reconstruction performance of the upgraded CMS detector. The measurements are used to extract a constraint on the Higgs boson self-coupling which will be obtainable in this channel with the full HL-LHC dataset of $3\,\text{ab}^{-1}$.

This note is organised as follows: a summary of the upgraded CMS detector is provided in Section 2. Section 3 describes the simulated event samples and the parameterisation of the expected performance of the CMS Phase-2 detector. Section 4 describes the event selection optimised for the $\text{ttH} + \text{tH}$, $\text{H} \rightarrow \gamma\gamma$ differential cross section measurement. Section 5 describes the event categorisation and strategy for the signal and background modelling. Finally, the results are presented in Section 6.

## 2 The CMS Phase-2 detector

The CMS detector [24] will be substantially upgraded in order to fully exploit the physics potential offered by the increase in luminosity at the HL-LHC [25], and to cope with the demanding operational conditions at the HL-LHC [26–31].

The upgrade of the first level hardware trigger (L1) will allow for an increase of L1 rate and latency to about 750 kHz and 12.5 $\mu$s, respectively, and the high-level software trigger (HLT) is expected to reduce the rate by about a factor of 100 to 7.5 kHz.

The entire pixel and strip tracker detectors will be replaced to increase the granularity, reduce the material budget in the tracking volume, improve the radiation hardness, and extend the geometrical coverage and provide efficient tracking up to pseudorapidities of about $|\eta| = 4$. The muon system will be enhanced by upgrading the electronics of the existing cathode strip chambers (CSC), resistive plate chambers (RPC) and drift tubes (DT). New muon detectors based on improved RPC and gas electron multiplier (GEM) technologies will be installed to add redundancy, increase the geometrical coverage up to about $|\eta| = 2.8$, and improve the trigger and reconstruction performance in the forward region.

The barrel electromagnetic calorimeter (ECAL) will feature the upgraded front-end electronics that will be able to exploit the information from single crystals at the L1 trigger level, to accommodate trigger latency and bandwidth requirements, and to provide 160 MHz sampling allowing high precision timing capability for photons. The hadronic calorimeter (HCAL), consisting in the barrel region of brass absorber plates and plastic scintillator layers, will be read out by silicon photomultipliers (SiPMs). The endcap electromagnetic and hadron calorimeters will be replaced with a new combined sampling calorimeter (HGCal) that will provide highly-segmented spatial information in both transverse and longitudinal directions, as well as high-precision timing information.

Finally, the addition of a new timing detector for minimum ionizing particles (MTD) in both barrel and endcap region is envisaged to provide capability for 4-dimensional reconstruction of interaction vertices that will allow to significantly offset the CMS performance degradation due to high PU rates.

A detailed overview of the CMS detector upgrade program is presented in Ref. [26–31], while the expected performance of the reconstruction algorithms is summarised in Ref. [32].

# 3 Event generation and detector simulation

Simulated ttH events are produced using POWHEG v2.0 [33, 34] at NLO. Additional contributions from Higgs boson production via gluon-fusion (ggH), in association with a vector boson (VH), and in association with a single top (tH) and a quark jet (tHq) or a W boson (tHW) are generated using MADGRAPH5_aMC@NLO v2.2.2 [35], interfaced with PYTHIA v8.205 [36], at NLO. The inclusive cross sections of ttH, tH, and VH production are calculated to NLO precision in quantum chromodymanics (QCD) and electroweak (EW) theory [37], while the ggH cross section is calculated to next-to-next-to-next-to-leading order ($N^3$LO) precision in QCD and NLO precision in EW theory [38].

The irreducible background arises from top pair production in which two photons are radiated (tt + $\gamma\gamma$). Reducible backgrounds from top pair production in which one photon is radiated (tt + $\gamma$) and inclusive top pair production ($t\bar{t}$), where additional jets in the events are misidentified as isolated photons, contribute due to their larger cross sections. Simulated tt + $\gamma\gamma$ and tt + $\gamma$ events are generated using MADGRAPH5_aMC@NLO v2.2.2, while the inclusive $t\bar{t}$ sample is produced using POWHEG v2.0. In both cases, the events are interfaced with PYTHIA v8.205 for hadronisation and showering. Further backgrounds arise from events in which two isolated photons are produced ($\gamma - \gamma$) or in which one photon is reconstructed from a hadronic shower which has been misidentified as a photon ($\gamma -$ j). The $\gamma - \gamma$ sample is generated using SHERPA v2.2.5 [39] while the $\gamma -$ j sample is generated with PYTHIA v8.205.

The signal and background events are processed with DELPHES [40] to simulate the response of the CMS Phase-2 detector to showered particles. The object reconstruction and identification efficiencies, and the detector response and resolution are parameterised using events simulated with GEANT4 [41]. The mean number of simulated interactions per bunch crossing (pile-up) is set to 200 to model the expected conditions for pp collisions at the HL-LHC.

In DELPHES, photons are reconstructed as clusters of energy in the electromagnetic calorimeter, with no matching hits in the tracker. The photon reconstruction efficiency and fake rate are parametrised in accordance with a tight working point requirement on the photon identification score [42], available in the GEANT4 simulation. Generator-level muons and electrons from the interaction are reconstructed with some $p_T$ and $\eta$ dependent probability. These probabilities vanish outside of the tracker acceptance and below some energy threshold.

Hadronic jets are reconstructed using the CMS particle-flow (PF) algorithm [43], which uses information from the calorimeters and tracker in DELPHES. The output jet is a result of clustering the smeared particle-flow tracks and the particle-flow towers, using the common anti-$k_T$ algorithm [44], with a distance parameter of 0.4. For clustering, the FastJet package [45] is used and the PUPPI algorithm is employed to partially clean the effects of pile-up [46]. To identify jets originating from the hadronization of b quarks, a variable is constructed to match the performance of the DeepCSV algorithm [47], at different working points. This variable uses a parametric formula for the b tagging probability, which depends on the $p_T$, $\eta$ and truth-level parton flavour of the jet. A medium working point is used in this analysis, which corresponds to a 79% efficiency for true b jets with $p_T$ = 100 GeV, in the central region of the detector, and a misidentification probability of around 1.5% for jets originating from light quarks and gluons. The discriminant incorporates the expected improvements with the planned Phase-2 MTD [31].

The missing transverse momentum, $p_T^{miss}$, is taken as the negative vector $p_T$ sum of all recon-structed objects in DELPHES after employing the PUPPI algorithm to mitigate the effects of pile-up. The scalar $p_{\mathrm{T}}$ sum of all reconstructed objects, after the PUPPI algorithm corrections, is labelled as $S_T$.

To extract the signal scaling functions, $\mu_{ij}(\kappa_\lambda)$, leading order (LO) parton-level ttH, tH and VH events are generated using MADGRAPH5_aMC@NLO v2.5.5. These events are used as input to the electroweak reweighting tool, described in Ref. [21], which calculates $\lambda_3$-dependent corrections at NLO, $\mathcal{O}(\lambda_3)$, by reweighting events on an event-by-event basis. A diagram filter is applied to select only the relevant one-loop matrix elements which feature the trilinear coupling. The $\kappa_\lambda$ dependence is determined by taking the ratio of the $\mathcal{O}(\lambda_3)$ to LO contributions in bins of the generator-level $p_{\mathrm{T}}^{\mathrm{H}}$ spectrum, and feeding into the scaling equations provided in Ref. [16].

## 4 Event selection

To identify the H $\rightarrow \gamma\gamma$ final state, events are required to have two photons in the invariant mass range: $100 < m_{\gamma\gamma} < 180$ GeV, such that the leading (sub-leading) photon satisfies $p_{\mathrm{T}}^{\gamma}/m_{\gamma\gamma} > 1/3$ (1/4). The photons must lie within pseudorapidity, $|\eta^\gamma| < 2.4$, excluding the barrel-endcap transition region: $1.44 < |\eta^\gamma| < 1.57$. The two candidate photons are also required to be separated by $\Delta R_{\gamma\gamma} > 0.4$. Additionally, the photons must satisfy an isolation requirement such that the sum of charged transverse momentum in a cone of radius $\Delta R_\gamma = 0.4$, centred on the photon candidate, is less than $0.3\, p_{\mathrm{T}}^{\gamma}$. If more than two photons pass the above criteria, then the pair with $m_{\gamma\gamma}$ closest to 125 GeV is chosen.

Top quark decay products in the final state are used to select events consistent with originating from ttH or tH production. The top quark decays almost exclusively to a bottom quark and a W boson, hence, the selection requires all events to have at least one b tagged jet. Two orthogonal selection criteria are then imposed to distinguish between the possible final states of the W boson decay. The hadronic channel describes the situation in which all W bosons decay to a quark-antiquark pair, and the leptonic channel is designed to be enriched in events where at least one W boson decays leptonically, to the electron + neutrino ($e\nu$) or muon + neutrino ($\mu\nu$) final states.

In the hadronic channel, events must contain at least 3 jets, separated by $\Delta R > 0.4$ with respect to both photon candidates. The jets are required to have $p_{\mathrm{T}} > 25$ GeV, and to lie inside the region $|\eta| < 4$. Note, this pseudorapidity requirement incorporates the improved tracker coverage of the CMS Phase-2 detector. Additionally, a leptonic veto is applied to discard any events with at least one isolated electron or muon. Here, isolated leptons are required to have $p_{\mathrm{T}} > 20$ GeV and $|\eta| < 2.4$, excluding the barrel-endcap transition region for electrons. Muons must satisfy a similar isolation requirement to photons, such that the sum of all reconstructed particles $p_{\mathrm{T}}$, inside a cone of radius $\Delta R = 0.4$, excluding the muon itself, is less than 0.25 times the transverse momentum of the muon. In addition, for electrons, the invariant mass of the electron-photon pairs, $m_{e\gamma}$, is required to be greater than 5 GeV from the Z boson mass.

The selection criteria for the leptonic channel are defined to be completely orthogonal to the hadronic channel by inverting the lepton veto i.e. requiring at least one isolated lepton in the event. Finally, only 2 jets are required, satisfying the criteria discussed above.

To improve the signal-to-background ratio, two boosted decision tree (BDT) classifiers are trained independently in each channel. The classifiers aim to distinguish between signal-like and background-like events, using input variables related to the kinematics of the event con-stituents. Importantly, variables directly related to the external Higgs boson kinematics, such as diphoton rapidity, are avoided to minimise distortions to the $p_T^H$ spectrum. Variables related to the photon quality were included in the BDT classifiers for the ttH selection in the Run II $H \to \gamma\gamma$ analysis [48]. However, such variables, which are not available in DELPHES, are less effective at separating signal and background than the kinematicic ones. Therefore, the additional sensitivity gained by including photon quality variables is expected to be small. In training, ttH and tH are classified as signal, and ggH and VH events are included in the definition of background. The input variables for both the hadronic and leptonic BDTs are summarised in Table 1.

Table 1: Summary of the input variables for both the hadronic and leptonic BDT classifiers.

| Description | Hadronic | Leptonic |
|---|---|---|
| Leading and sub-leading photon variables | $p_T^{\gamma(1/2)}/m_{\gamma\gamma}$, $\eta^{\gamma(1/2)}$ | $p_T^{\gamma(1/2)}/m_{\gamma\gamma}$, $\eta^{\gamma(1/2)}$ |
| Leading and sub-leading photon isolation variables | $\sum_{\Delta R_\gamma<0.4} p_T^{\text{charged}}/p_T^\gamma$ | $\sum_{\Delta R_\gamma<0.4} p_T^{\text{charged}}/p_T^\gamma$ |
| Leading jet kinematics | $p_T^{j1/j2/j3/j4}$, $\eta^{j1/j2/j3/j4}$ | $p_T^{j1/j2/j3}$, $\eta^{j1/j2/j3}$ |
| Leading lepton kinematics | - | $p_T^\ell$, $\eta^\ell$ |
| Missing transverse momentum | $\lvert p_T^{\text{miss}}\rvert$ | $\lvert p_T^{\text{miss}}\rvert$ |
| Scalar sum of all energy, mitigating the effect of pile-up | $S_T$ | $S_T$ |
| Minimum difference in azimuthal angle between the diphoton system and object | Closest jet: $\Delta\phi_{\gamma\gamma,j}$ | Leading lepton: $\Delta\phi_{\gamma\gamma,\ell}$ |
| Global variables | $N_{\text{jets}}$, $N_{\text{b-jets}}$ | $N_{\text{jets}}$, $N_{\text{b-jets}}$, $N_{\text{leptons}}$ |

Figure 2 shows the respective BDT outputs for events in the hadronic and leptonic channels, after pre-selection. For the differential cross section measurement, it is necessary to select a relatively loose working point to maintain a high signal acceptance and thus obviate a large unfolding of the selection process. Events are required to have a BDT output value greater than 0.28 (0.13) in the hadronic (leptonic) selection. Tables 2 and 3 show the event yields at each stage of selection, for the hadronic and leptonic channels respectively. The yields are separated according to process, and all values are normalised to 3 ab$^{-1}$.

## 5 Event categorisation and signal and background modelling

Table 4 lists the bins in $p_T^H$ in which the ttH + tH differential cross sections are measured. This binning scenario is chosen to match the CMS + ATLAS agreed bin boundaries for inclusive $p_T^H$ differential measurements [49]. The simulated Higgs boson signal and background events which pass the selection are divided into bins of $p_T^{\gamma\gamma}$, whose boundaries correspond to those listed in Table 4. In the hadronic channel, all except the highest $p_T^{\gamma\gamma}$ bin, are further split into two categories according to the hadronic BDT output. The boundary is chosen at a BDT output value of 0.61, which effectively splits each bin into a low and high ttH purity category. This is not possible in the [350,∞] GeV bin due to limited data sample size.

For each event category, the Higgs boson signal events in that category are further divided into the contributions from different $p_T^H$ bins and according to their production mechanism. Due to the excellent photon energy resolution, each $p_T^{\gamma\gamma}$ category is dominated by events from the corresponding $p_T^H$ bin. The events are then fit using a parametric model, constructed using a sum of Gaussian probability density functions of the invariant diphoton mass $m_{\gamma\gamma}$. This model is sufficient to describe the mass resolution and peak position in each of the event categories while providing a smooth functional form. The simulated background events in each category are fit using a set of smoothly falling functions in $m_{\gamma\gamma}$. The normalisations for the signal and background models are defined as the sum of weights of the simulated signal (separated into $p_{\mathrm{T}}^{\mathrm{H}}$ bins and production processes) or background events which contribute.

Table 2: Number of events remaining at the subsequent stages of the ttH + tH hadronic selection. Also shown are the respective efficiencies of selection at each stage. The BDT efficiency, $\epsilon_{\mathrm{BDT}}$, is defined as the ratio of the number of events remaining after the cut on the BDT output, to the number of events remaining after pre-selection. All event yields are normalised to $3\,\mathrm{ab}^{-1}$.

| | Pre-selection | BDT (>0.28) | $\epsilon_{\mathrm{pre}}$ | $\epsilon_{\mathrm{BDT}}$ | $\epsilon_{\mathrm{tot}}$ |
|---|---|---|---|---|---|
| ttH | 820 | 650 | 20% | 79% | 16% |
| tH | 140 | 80 | 19% | 57% | 11% |
| ggH | 860 | 220 | 0.23% | 25% | 0.06% |
| $V$H | 170 | 43 | 1.1% | 25% | 0.27% |
| $\gamma-\gamma$ | $2.1\times10^{5}$ | $2.1\times10^{4}$ | 0.07% | 10% | $7.1\times10^{-5}$ |
| $\gamma-\mathrm{j}$ | $8.8\times10^{4}$ | 4100 | $2.8\times10^{-5}$ | 4.7% | $1.3\times10^{-6}$ |
| $\mathrm{tt}+\gamma\gamma$ | 340 | 244 | 0.56% | 71% | 0.40% |
| $\mathrm{tt}+\gamma$ | 5100 | 2700 | 0.08% | 54% | 0.04% |
| $\mathrm{t\bar{t}}$ | $3.1\times10^{4}$ | $1.3\times10^{4}$ | $1.2\times10^{-5}$ | 43% | $5.2\times10^{-6}$ |
| $\mathrm{t}+\gamma+\mathrm{j}$ | 3800 | 780 | 0.11% | 21% | 0.02% |
| Total Bkgd | $3.4\times10^{5}$ | $4.3\times10^{4}$ | $1.0\times10^{-4}$ | 13% | $1.3\times10^{-5}$ |

Table 3: Number of events remaining at the subsequent stages of the ttH + tH leptonic selection. Also shown are the respective efficiencies of selection at each stage. The BDT efficiency, $\epsilon_{\mathrm{BDT}}$, is defined as the ratio of the number of events remaining after the cut on the BDT output, to the number of events remaining after pre-selection. All event yields are normalised to $3\,\mathrm{ab}^{-1}$.

| | Pre-selection | BDT (>0.13) | $\epsilon_{\mathrm{pre}}$ | $\epsilon_{\mathrm{BDT}}$ | $\epsilon_{\mathrm{tot}}$ |
|---|---|---|---|---|---|
| ttH | 380 | 290 | 9.1% | 77% | 7.0% |
| tH | 45 | 32 | 6.1% | 72% | 4.4% |
| ggH | 18 | 2.0 | $4.7\times10^{-5}$ | 11% | $5.2\times10^{-6}$ |
| $V$H | 23 | 11 | 0.15% | 46% | 0.07% |
| $\gamma-\gamma$ | 6500 | 1400 | $2.3\times10^{-5}$ | 22% | $5.0\times10^{-6}$ |
| $\gamma-\mathrm{j}$ | 1100 | 157 | $3.5\times10^{-7}$ | 14% | $5.0\times10^{-8}$ |
| $\mathrm{tt}+\gamma\gamma$ | 630 | 390 | 1.0% | 62% | 0.63% |
| $\mathrm{tt}+\gamma$ | 2900 | 1100 | 0.08% | 40% | 0.03% |
| $\mathrm{t\bar{t}}$ | 8400 | 2500 | $9.0\times10^{-5}$ | 30% | $2.7\times10^{-5}$ |
| $\mathrm{t}+\gamma+\mathrm{j}$ | 780 | 100 | 0.02% | 13% | $2.6\times10^{-5}$ |
| Total Bkgd | $2.0\times10^{4}$ | 5700 | $6.2\times10^{-6}$ | 28% | $1.7\times10^{-6}$ |

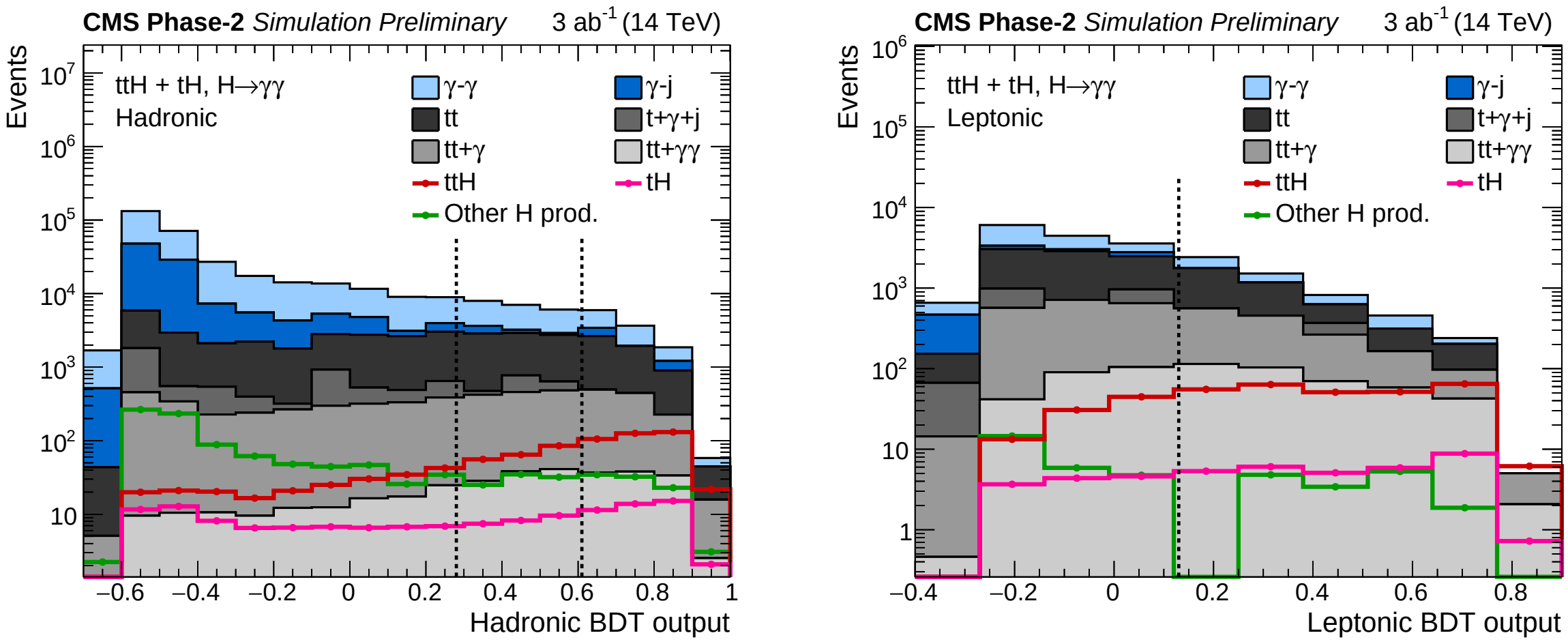

Figure 2: The BDT output distributions for the hadronic (left) and leptonic (right) channels, after pre-selection has been applied. Events with a BDT output value greater than 0.28 (0.13) are selected for the hadronic (leptonic) categories. This selection boundary is indicated by the leftmost (single) dashed line in the hadronic (leptonic) BDT output distribution. The second dashed line in the hadronic BDT output distribution shows the additional boundary at 0.61, which is used to further split the hadronic categories according to high and low ttH purity.

Table 4: The kinematic bins in which the differential cross sections are measured.

| Variable | bins [GeV] | | | | | | |
|---|---|---|---|---|---|---|---|
| $p_{\mathrm{T}}^{\mathrm{H}}$ | 0 | 45 | 80 | 120 | 200 | 350 | $\infty$ |

The sum of the fitted signal and background functions are shown for each category in Figs. 3, 4 and 5. The signal component due to ttH and tH only is also shown. A pseudo-data set is generated from the fitted distributions, and overlaid to illustrate a representative data sample which can be expected with 3 ab$^{-1}$ at the HL-LHC assuming SM Higgs boson production.

Since ttH + tH differential cross section measurements are statistically limited at the HL-LHC, only the dominant systematic uncertainties in such measurements are considered. These are incorporated into the signal models as nuisance parameters, $\vec{\theta}_s$, and are treated as log-normal variations in the yield of a particular $(i,j,k)$ signal combination, where $i$ labels the $p_{\mathrm{T}}^{\mathrm{H}}$ bin, $j$ labels the production mode, and $k$ corresponds to the reconstruction-level category. The systematic uncertainties considered are in line with the recommendations for HL-LHC projections [50], and are as listed below. All theoretical uncertainties are reduced by 50%, with respect to the Run II values, to represent improvements in theoretical predictions.

- *Integrated luminosity*: amounts to a flat 1% uncertainty in the yield in each $(i,j,k)$ combination.
- *Photon identification efficiency*: this uncertainty is expected to improve with respect to

**8**

the LHC Run II value. To accommodate this, a flat 0.5% uncertainty is used for all photons, irrespective of pseudorapidity. Assuming a uniform photon identification efficiency across all regions of the detector of 80%, and accounting for the fact that two photons are required in the selection, amounts to a 1.3% yield uncertainty for each ($i$,$j$,$k$) combination.

- *Jet energy scale*: the uncertainty in the jet energy scale depends on the measured transverse momentum of the jet and varies between 0.5-3%. The $p_T$ of all jets, in all signal samples, are scaled both up and down according to the uncertainty for the respective jet $p_T$. The scaled samples are subsequently propagated through the selection process, and the yield uncertainty is realised by comparing the scaled yield to the nominal yield in each ($i$,$j$,$k$) combination. This corresponds to an average yield uncertainty, across all combinations, between 2-3%.
- *b tagging efficiency*: a weight is applied to each event, according to the number of true $b$-jets at generator-level. The weight parametrises the increase/decrease in the signal yield realised at the $\pm 1\sigma$ bounds of the uncertainty in the b tagging efficiency, at the medium (with MTD) working point. The reweighted samples are propagated through the selection process, and the ($i$,$j$,$k$) yield uncertainty is determined by comparing the reweighted yields to the nominal yield. This provides an average yield uncertainty between 0-1%.
- *Theoretical uncertainties in the ggH yield*: a number of uncertainties are incorporated to account for the overall normalisation of the ggH production mode, as well as the migration of ggH events between ($i$,ggH,$k$) combinations. The yield variations are calculated according to the 2017 recommendations of the LHC Higgs Cross Section Working Group [37]. Events are reweighted, both up and down, according to uncertainties related to the QCD scale, the number of jets, the transverse momentum of the Higgs boson, and top mass effects. The reweighted events are propagated through selection, and the yield uncertainties are calculated as the ratio of the reweighted yields to the nominal yields in each ($i$,ggH,$k$) combination. The uncertainties, at most, provide a 15% variation in the yield for a given ($i$,ggH,$k$) combination.
- *Theoretical uncertainties in the inclusive ttH, tH and VH cross sections*: these uncertainties are implemented as yield uncertainties in all ($i$,$j$,$k$) combinations, for the ttH, tH and VH production modes. The values are taken directly from the LHC Higgs Cross Section Working Group, and are separated into the effects from the uncertainties in the factorisation and renormalisation scales, the parton distribution function (PDF), and the strong coupling constant, $\alpha_s$. The uncertainties in the QCD scales dominate, and after scaling down by 50%, yield variations are found to be less than 2.5% for VH, 5% for ttH and 7.5% for tH.
- *Theoretical uncertainties related to the shape of the $p_T^H$ spectrum*: the shape effects, originating from the uncertainty in the factorisation and renormalisation scales, are included for the ttH and tH production modes. Events are reweighted, on an event-by-event basis, for the situation where the renormalisation and factorisation scales are, independently, doubled and halved. These uncertainties predominanty account for the migration between $p_T^H$ bins and the efficiency of the cut on the BDT output values. In general, the shape uncertainties have a smaller impact than the overall normalisation uncertainties.

Systematic uncertainties in the photon scale and resolution, which modify the shape of the signal models, are expected to have a small effect on the final sensitivity, and are therefore ignored in this analysis.

# 6 Results

## 6.1 Differential cross section measurements

The differential ttH + tH production cross section is determined by defining a scaling parameter, $\mu_i$, for each bin, $i$, in $p_T^H$ such that the signal model, $S_k^{ij}$, scales according to,

$$S_k^{ij}(\mu_i, \vec{\theta}_s) = \mu_i \times S_k^{ij}(\vec{\theta}_s)\Big|_{\mu_i=1} \qquad \forall \text{ categories } k \text{ and } j = (\text{ttH}, \text{tH}). \tag{1}$$

A likelihood function is constructed in each category, using the signal and background models and an asimov dataset assuming that all $\mu_i = 1$. The product over all of these likelihoods is used to construct a profiled likelihood ratio test-statistic, to determine the best-fit values and uncertainties for each $\mu_i$, as described in Ref. [10]. The parameters of the background functions are profiled as nuisance parameters to model the statistical precision with 3 ab$^{-1}$ of data [48]. The fitted values of $\mu_i$ and their respective uncertainties are converted to fiducial cross sections times branching ratios, $\sigma_{\text{fid}}^{\text{ttH +tH}} \times \text{BR}(H \to \gamma\gamma)$, by correcting for the effects of the event selection. The fiducial region is defined by the following criteria:

- Higgs boson rapidity: $|Y^H| < 2.5$.
- Two photons from the Higgs boson decay: $p_T^{\gamma} > 20$ GeV and $|\eta^{\gamma}| < 2.5$.
- At least two jets: $p_T^{j} > 25$ GeV and $|\eta^{j}| < 4$.
- At least one of the jets, satisfying the above criteria, originates from a b quark.

Around 0.7% and 0.4% of the simulated ttH + tH events that pass the full event selection are not contained in the fiducial region, in the hadronic and leptonic categories, respectively. Although these events are included in the likelihood function as part of the signal component, they are subtracted when calculating the fiducial cross-section.

For each $p_T^H$ bin, $i$, we define selection efficiencies, $\epsilon_{\text{sel}}$, for the hadronic (H) and leptonic (L) channels as:

$$\epsilon_{\text{sel,H/L}}^{i} = \frac{N_{\text{obs,H/L}}^{i}}{N_{\text{fid}}^{i}} = \frac{\mu_i \times N_{\text{exp,H/L}}^{i}}{N_{\text{fid}}^{i}}, \tag{2}$$

where $N_{\text{fid}}^{i}$ is the number of simulated ttH + tH events passing the fiducial selection, in the $i^{\text{th}}$ $p_T^H$ bin. The value of $N_{\text{exp,H/L}}^{i}$ is determined as the number of ttH + tH events expected in the SM after the selection. The fiducial cross section times branching ratio in bin $i$ is given by,

$$\left[\sigma_{\text{fid}}^{\text{ttH+tH}} \times \text{BR}(H \to \gamma\gamma)\right]_i = \frac{N_{\text{fid}}^{i}}{\mathcal{L}_{\text{int}}}. \tag{3}$$

Figure 6 shows the expected ttH + tH differential cross sections times branching ratio, for the fiducial phase space defined above, in bins of $p_T^H$. The error bars represent the combined statistical and systematics uncertainties. With 3 ab$^{-1}$ of HL-LHC data, uncertainties of 20-40% in the differential cross sections are expected. To separate the contribution of the hadronic and leptonic channels, analogous likelihood scans are performed, using only the relevant categories. The hadronic channel is observed to provide, in general, greater sensitivity in terms of the differential cross section measurements due to a larger absolute signal yield after selection, compared to the leptonic channel. Additionally, the expected differential cross sections times branching ratio for anomalous values of the Higgs self coupling, $\kappa_\lambda$ = 10 and $\kappa_\lambda$ = -5 are shown.

### 6.2 Constraints on $\kappa_\lambda$

In order to extract the sensitivity to the Higgs boson self-coupling, we make the substitution $\mu_i \to \mu_{ij}(\kappa_\lambda)$. The parameterisations $\mu_{ij}(\kappa_\lambda)$ are determined using the electroweak reweighting tool provided in Ref. [21], which allows one to account for kinematic variations in the modifications to Higgs boson production due to non SM values of the Higgs boson self-coupling. For the contribution of ggH and to model the effect on the $\mathrm{H} \to \gamma\gamma$ decay rate, the scaling functions calculated for inclusive events provided in Ref. [15] are used directly.

A scan of the profiled likelihood, as a function of $\kappa_\lambda$, is shown in Fig. 7. In the scan, all other couplings are fixed to the SM predictions. The scan is performed in the region $\kappa_\lambda \in [-10, 20]$, beyond which, the physics model used here is no longer valid as next-to-next-to-leading order effects become important. Also shown are the results when only including the hadronic or leptonic categories in the scan, to demonstrate the relative contributions from each channel. It is observed that both channels contribute significantly towards the final sensitivity. For negative values of $\kappa_\lambda$, larger deviations in the $\mathrm{ttH} + \mathrm{tH}$ differential cross section are expected compared to positive values. The feature in the region around $5 < \kappa_\lambda < 15$ is a result of the degeneracy in the physics model. For the ttH production mode there exists a turning point in $\mu_{ij}(\kappa_\lambda)$, in this region, such that $\mu_{ij}(\kappa_\lambda)$ can take the same value for different $\kappa_\lambda$. This degeneracy is somewhat alleviated by the contamination of ggH in the signal model, which has a different scaling behaviour.

The individual contributions of the statistical and systematic uncertainties are determined by performing a likelihood scan with all systematic uncertainties removed. The only considerable deviation from the statistical-uncertainty-only curve, occurs in the $\kappa_\lambda \gtrsim 5$ region. This is a result of the theoretical uncertainties in the Higgs boson production yields, which have a larger effect in this region due to the degeneracy in the physics model. Using the profiled likelihood scan, values of $\kappa_\lambda$ outside of the range $-4.1 < \kappa_\lambda < 14.1$ can be expected to be excluded at the 95% confidence level with $3\,\mathrm{ab}^{-1}$ of data at the HL-LHC. The effect of additional fake photons, which may not be well modelled in DELPHES, is found to weaken the constraint at the 95% confidence level by around 10%.

Table 5 shows the 68% and 95% confidence level intervals for $\kappa_\lambda$, for different integrated luminosities recorded by the CMS Phase-2 detector at the HL-LHC, assuming constant detector performance. The intervals are extracted using the procedure described above, where the signal and background models are scaled, in each reconstruction level category, to different integrated luminosities. As the integrated luminosity increases, the constraint for positive values of $\kappa_\lambda$ improves more dramatically than for negative values.

Additionally, a two-dimensional likelihood scan is performed, in which an overall normalisation parameter for the Higgs boson signal processes, $\mu_{\mathrm{H}}$, is profiled. This incorporates other beyond-the-standard model effects, such as an anomalous top-Higgs coupling, which in general cause an inclusive shift across the whole $p_{\mathrm{T}}^{\mathrm{H}}$ spectrum. Figure 8 shows the results of the two-dimensional scan, in terms of the 68% and 95% confidence level contours. It can be seen that differential cross section measurements still provide sensitivity to $\kappa_\lambda$, without exploiting the overall normalisation of the $p_{\text{T}}^{\text{H}}$ spectrum. Table 6 shows the $1\sigma$ uncertainties in $\mu_H$ and the 95% confidence level intervals for $\kappa_\lambda$, when both profiling the other parameter and fixing the other parameter to the SM prediction.

Table 5: The 68% and 95% confidence level intervals for $\kappa_\lambda$ for different integrated luminosities recorded by the CMS Phase-2 detector at the HL-LHC, assuming constant detector performance. The 95% upper limit for $\mathcal{L}_{\text{int}} = 1\,\text{ab}^{-1}$ goes outside of the valid region, and is specified as 20+ in the table.

| $\mathcal{L}_{\text{int}}$ ( $\text{ab}^{-1}$ ) | 68% interval | 95% interval |
|---|---|---|
| 1 | [-3.1,10.9] | [-6.2,20+] |
| 2 | [-2.2,6.5] | [-4.6,17.0] |
| 3 | [-1.9,5.3] | [-4.1,14.1] |

Table 6: The $1\sigma$ uncertainties in $\mu_H$ and the 95% confidence level intervals for $\kappa_\lambda$, when the other parameter is profiled or fixed to the SM prediction.

| Other parameter | $\mu_H \pm \sigma_{\mu_H}$ | 95% interval on $\kappa_\lambda$ |
|---|---|---|
| Profile ($\kappa_\lambda$ or $\mu_H$) | $1.00^{+0.16}_{-0.11}$ | [-7.7,14.1] |
| Fix to SM ($\kappa_\lambda$ or $\mu_H$) | $1.00^{+0.08}_{-0.08}$ | [-4.1,14.1] |

As has been detailed in other studies, the combination of multiple production and decay channels will significantly improve the overall sensitivity to the Higgs boson self-coupling via single Higgs boson production measurements [15, 16, 19].

# 7 Conclusions

The precision of ttH + tH, H $\to \gamma\gamma$ differential cross section measurements, at the HL-LHC with the CMS Phase-2 detector, have been determined as a function of $p_T^H$. The analysis has been conducted using a simulated event sample corresponding to $3\,\text{ab}^{-1}$ of pp collision data under HL-LHC conditions. A combination of the hadronic and leptonic top decay channels is performed to maximise the sensitivity of the cross section measurements to the Higgs boson self-coupling. With the data expected by the end of the HL-LHC, the cross section in bins of $p_{\text{T}}^{\text{H}}$ can be measured within uncertainties of 20–40%, depending on the $p_{\text{T}}$ range. When deviations from the standard model prediction for the ttH + tH $p_{\text{T}}^{\text{H}}$ differential cross section are interpreted as modifications of the Higgs boson self-coupling, $\kappa_\lambda$, these measurements exclude values outside of the range $-4.1 < \kappa_\lambda < 14.1$, at the 95% confidence level. Furthermore, it has been shown such measurements still provide sensitivity to $\kappa_\lambda$, without exploiting the overall normalisation of the $p_{\text{T}}^{\text{H}}$ spectrum, thus allowing for other effects, such as the presence of anomalous top-Higgs couplings. This property is unique to differential cross section measurements.

This analysis indicates that additional sensitivity to the Higgs boson self-coupling is available through studies of the differential cross section of single Higgs boson production in association with top quarks. It should be noted that the ultimate sensitivity to the Higgs boson self-coupling, achievable at the HL-LHC, will result from a combination of analyses such as that described in this note with other Higgs decay channels and production modes, and with direct searches for double Higgs boson production.

**14**

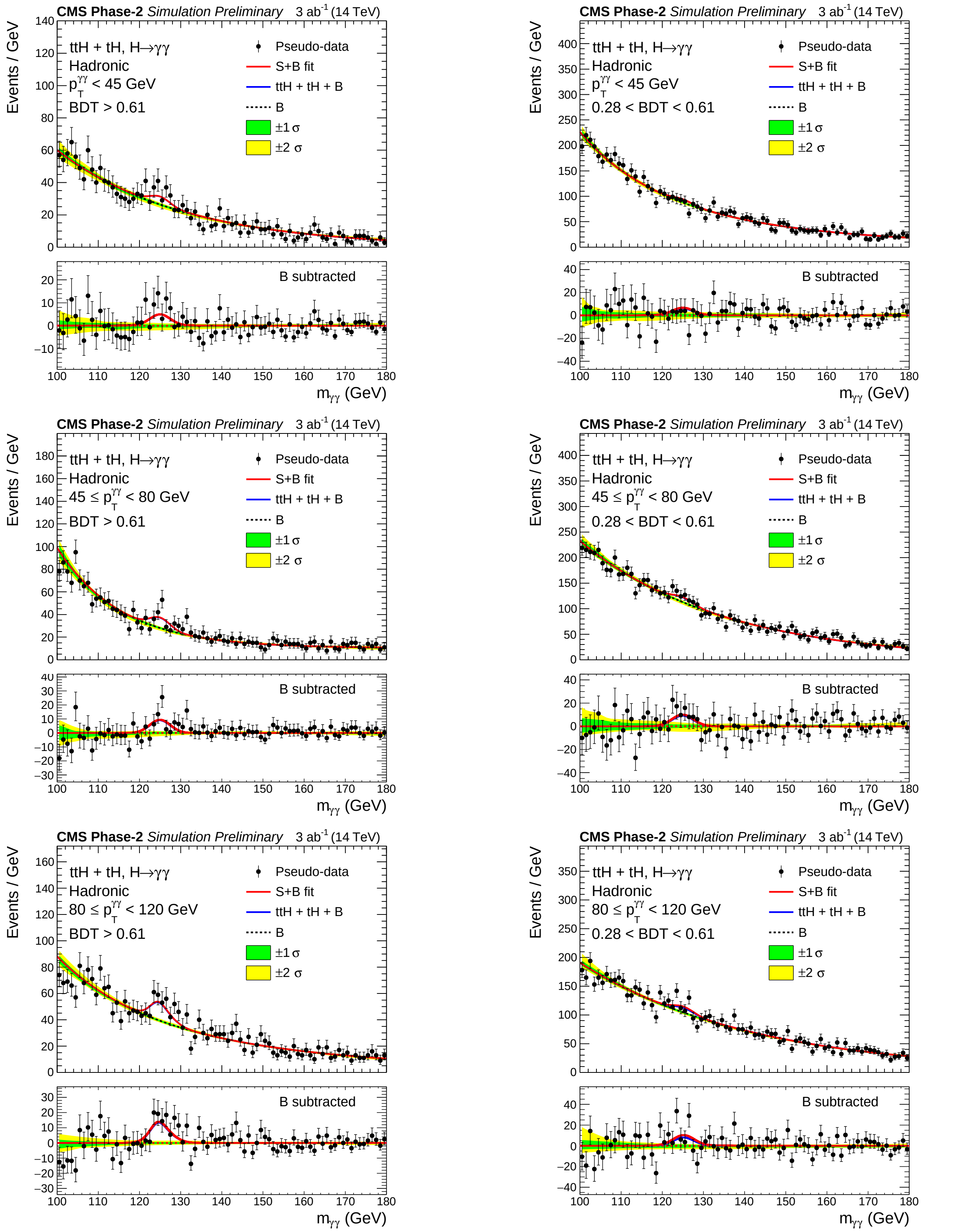

Figure 3: Best-fit signal (S) + background (B) models for the reconstruction-level categories in the ttH + tH hadronic channel, in the three lowest $p_T^{\gamma\gamma}$ bins: [0,45] GeV, [45,80] GeV and [80,120] GeV. A pseudo-data set is thrown from the best-fit functions, represented by the black points. The one (green) and two (yellow) standard deviation bands show the uncertainties in the background component of the fit. The residual plots, pseudo-data minus the background component, are shown in the lower panels.

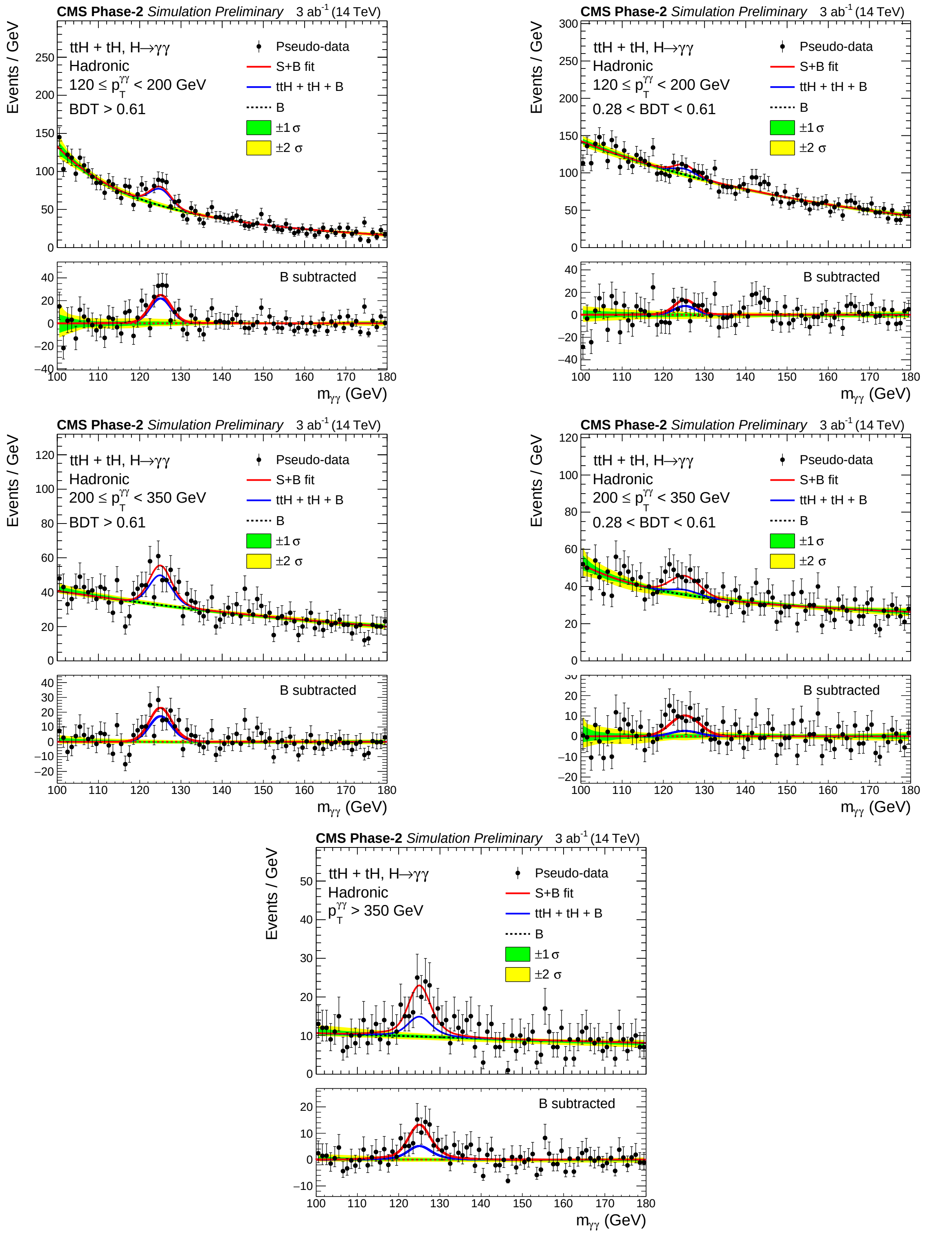

Figure 4: Best-fit signal (S) + background (B) models for the reconstruction-level categories in the ttH + tH hadronic channel, in the three highest $p_{\mathrm{T}}^{\gamma\gamma}$ bins: [120,200] GeV, [200,350] GeV and [350,∞] GeV. A pseudo-data set is thrown from the best-fit functions, represented by the black points. The one (green) and two (yellow) standard deviation bands show the uncertainties in the background component of the fit. The residual plots, pseudo-data minus the background component, are shown in the lower panels.

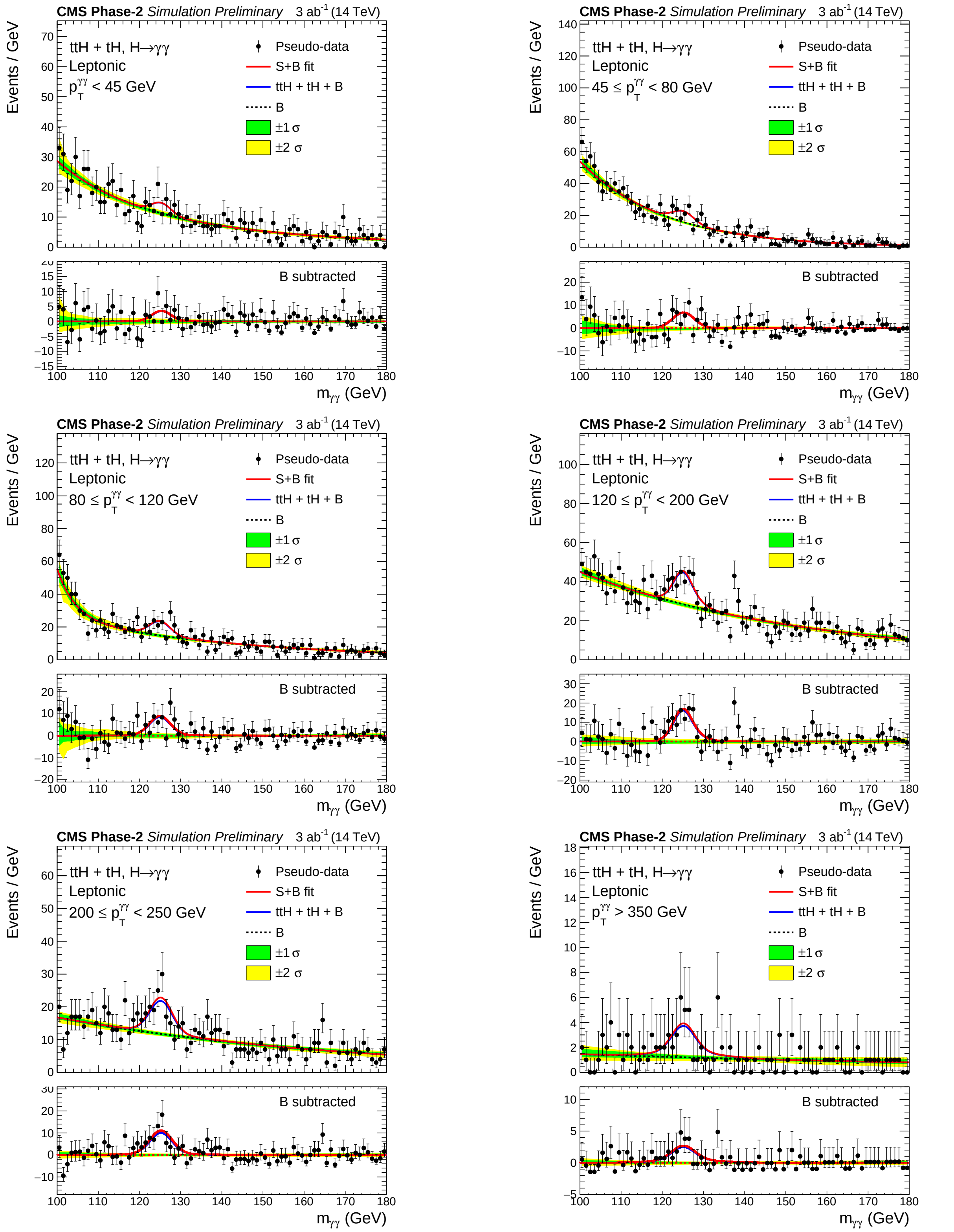

Figure 5: Best-fit signal (S) + background (B) models for each reconstruction-level category in the ttH + tH leptonic channel. A pseudo-data set is thrown from the best-fit functions, represented by the black points. The one (green) and two (yellow) standard deviation bands show the uncertainties in the background component of the fit. The residual plots, pseudo-data minus the background component, are shown in the lower panels.

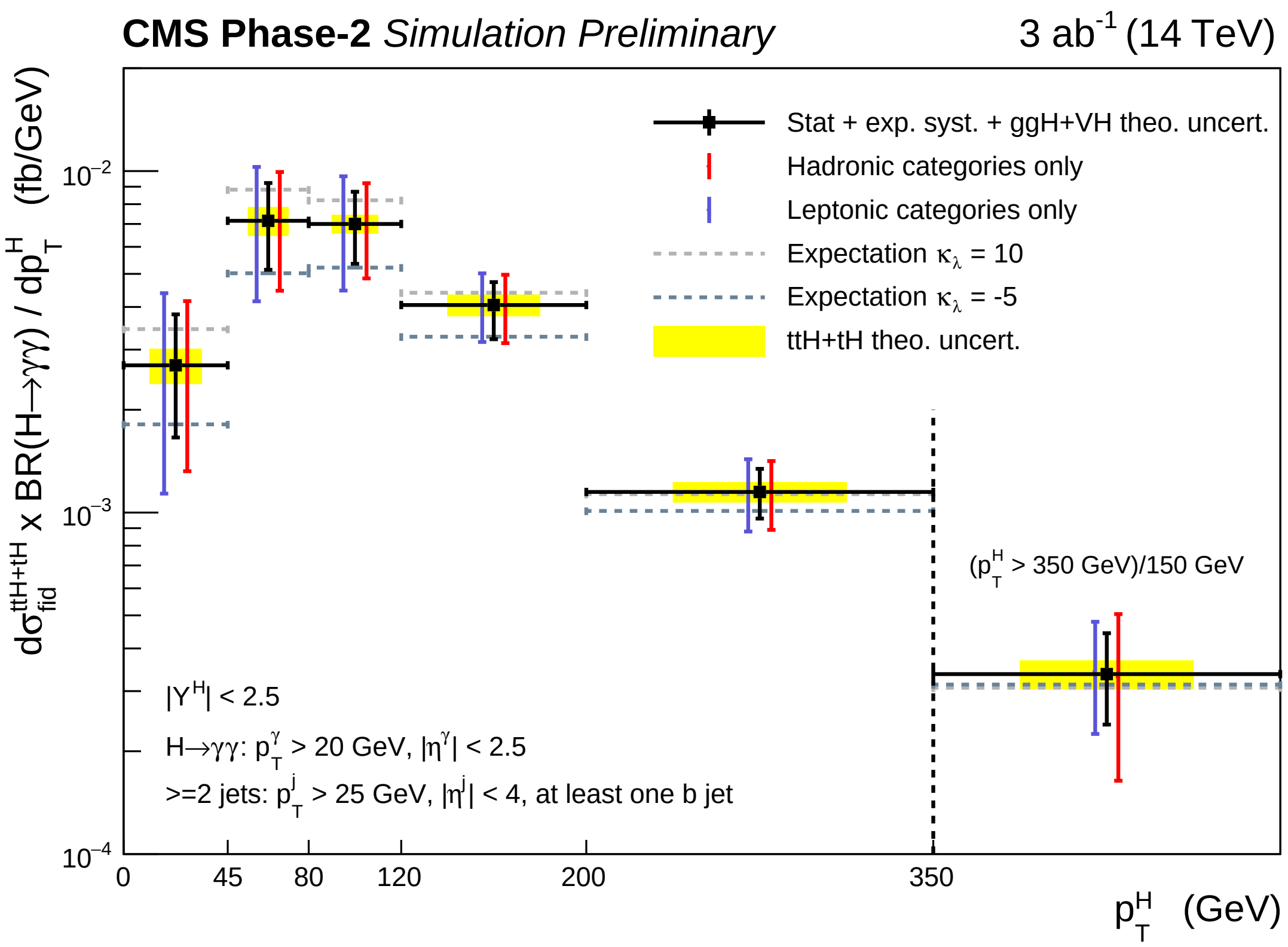

Figure 6: The expected differential ttH + tH cross sections times branching ratio, along with their respective uncertainties, in bins of $p_{\mathrm{T}}^{\mathrm{H}}$. These are for the fiducial region of phase space defined in the bottom left of the plot. The error bars on the black points include the statistical uncertainty, the experimental systematic uncertainties and the theoretical uncertainties related to the ggH and VH yields. The theoretical uncertainties in the inclusive ttH + tH cross section and those effecting the shape of the ttH + tH $p_T^H$ spectrum, originating from the uncertainty in the QCD scales, are shown by the shaded yellow regions. Contributions from the individual hadronic and leptonic channels are shown in red and purple respectively. The cross section for the $p_{\mathrm{T}}^{\mathrm{H}}$ = [350,∞] GeV bin is scaled by the width of the previous bin. Additionally, the expected differential ttH + tH cross sections for anomalous values of the Higgs boson self-coupling ($\kappa_\lambda$ = 10 and $\kappa_\lambda$ = -5) are shown by the horizontal dashed lines.

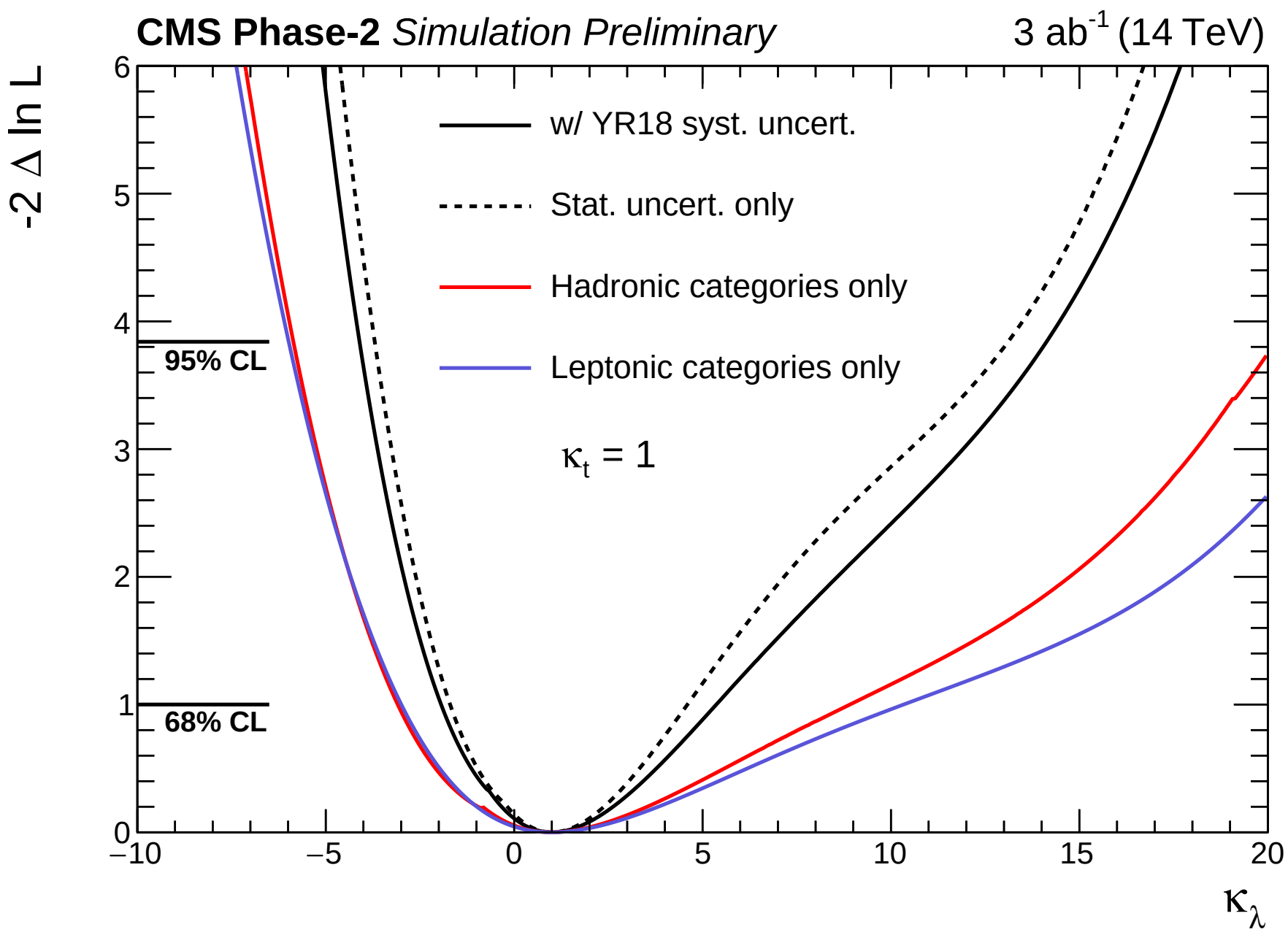

Figure 7: Results of the likelihood scan in $\kappa_\lambda$. The individual contributions of the statistical and systematic uncertainties are separated by performing a likelihood scan with all systematics removed. The observed deviation from the statistical uncertainty only curve is driven by the theoretical systematic uncertainties in the Higgs boson production yields. Additionally, the contributions from the hadronic and leptonic channels have been separated, shown in red and purple, respectively.

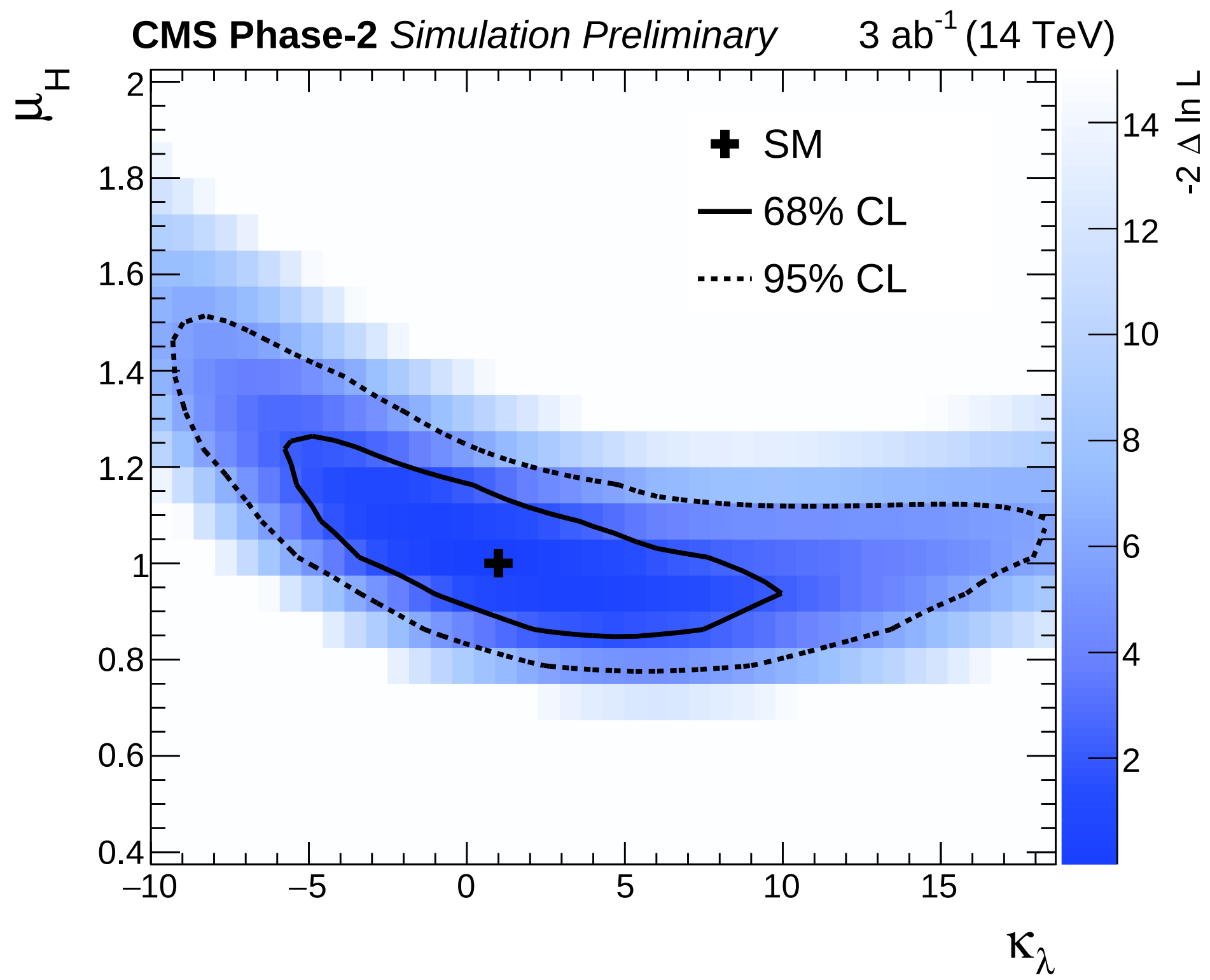

Figure 8: Results of the two-dimensional likelihood scan in $\kappa_\lambda$-vs-$\mu_\mathrm{H}$, where $\mu_\mathrm{H}$ allows all Higgs boson production modes to scale relative to the SM prediction. The 68% and 95% confidence level contours are shown by the solid and dashed lines respectively. The SM expectation is shown by the black cross.

# Prospects for the search for additional Higgs bosons in the ditau final state with the ATLAS detector at HL-LHC

The ATLAS Collaboration

Estimates of the sensitivity of the search for a heavy neutral Higgs boson in the $\tau\tau$ final state with the full High-Luminosity LHC dataset of 3000 fb$^{-1}$ proton–proton collisions at $\sqrt{s} = 14$ TeV are presented. These estimates are based on the extrapolation of current results obtained with the 36.1 fb$^{-1}$ ATLAS dataset collected in 2015–2016 at $\sqrt{s} = 13$ TeV. The expected 95% CL upper exclusion limits or, in alternative, the expected 5 $\sigma$ discovery reach are presented in terms of cross section times branching fraction of the gluon fusion production and $b$-associated production. In the hypothesis that no signal emerges, results are interpreted in the context of MSSM benchmark scenarios, e.g. in the hMSSM scenario $\tan\beta > 1$ is expected to be excluded for the mass range $250 < m_A < 350$ GeV. The parameter space with the expected 5 $\sigma$ discovery reach is also shown. The impact of the systematic uncertainties is also discussed.

# 1 Introduction

The discovery of a Standard Model (SM) like Higgs boson [1, 2] at the Large Hadron Collider (LHC) [3] has provided important insight into the mechanism of electroweak symmetry breaking. However, it remains possible that the discovered particle is part of an extended scalar sector, a scenario that is favored by a number of theoretical arguments [4, 5]. Searching for additional Higgs bosons is among the main goals of the High-Luminosity LHC (HL-LHC) programme [6]. The Minimal Supersymmetric Standard Model (MSSM) [4, 7, 8] is one of the well motivated extensions of the SM. Besides the SM-like Higgs boson, the MSSM requires two additional neutral Higgs bosons: one CP-odd ($A$) and one CP-even ($H$), which in the following are generically called $\phi$. At tree level, the MSSM Higgs sector depends on only two non-SM parameters, which can be chosen to be the mass of the CP-odd Higgs boson, $m_A$, and the ratio of the vacuum expectation values of the two Higgs doublets, $\tan\beta$. Beyond tree level, a number of additional parameters affect the Higgs sector, the choice of which defines various MSSM benchmark scenarios, such as $m_h^{\text{mod+}}$ [9] and hMSSM [10, 11]. The couplings of the additional MSSM Higgs bosons to down-type fermions are enhanced with respect to the SM Higgs boson for large $\tan\beta$ values, resulting in increased branching fractions to $\tau$-leptons and $b$-quarks, as well as a higher cross section for Higgs boson production in association with $b$-quarks.

The projections presented in this note are extrapolations of the recent results obtained by ATLAS using the 36.1 fb$^{-1}$ Run 2 dataset [12]. The MSSM Higgs boson with masses of 0.2–2.25 TeV and $\tan\beta$ of 1–58 is searched for in the $\tau_{\text{lep}}\tau_{\text{had}}$ and $\tau_{\text{had}}\tau_{\text{had}}$ decay modes, where $\tau_{\text{lep}}$ represents the leptonic decay of a $\tau$-lepton, whereas $\tau_{\text{had}}$ represents the hadronic decay. The main production modes are gluon–gluon fusion and in association with $b$-quarks. To exploit the different production modes, events containing at least one $b$-tagged jet enter the $b$-tag category, while events containing no $b$-tagged jets enter the $b$-veto category. The total transverse mass ($m_{\text{T}}^{\text{tot}}$), as defined in Ref. [12], is used as the final discriminant between the signal and the background.

In making these extrapolations, the assumption is made that the planned upgrades to the ATLAS detector and improvements to reconstruction algorithms will mitigate the effects of the higher pileup which can reach up to 200 in-time pileup interactions, leading to the overall reconstruction performance matching that of the current detector. Furthermore, the assumption is made that the analysis will be unchanged in terms of selection and statistical analysis technique, though the current analysis has not been re-optimised for the HL-LHC datasets.

# 2 Extrapolation method

To account for the integrated luminosity increase at HL-LHC, signal and background distributions are scaled by a factor of 3000/36.1. Furthermore, to account for the increase in collision energy from 13 TeV to 14 TeV, the background distributions are further scaled by a factor 1.18 which assumes the same parton-luminosity increase for quarks as that for gluons. The cross section of signals in various scenarios at 14 TeV are given in Ref. [13]. Possible effects on the kinematics and the $m_{\text{T}}^{\text{tot}}$ shape due to the collision energy increase are neglected for this study. The scaled $m_{\text{T}}^{\text{tot}}$ distributions for the four signal categories and one for the top control region are shown in Figures 1 and 2. These distributions are used in the statistical analysis.

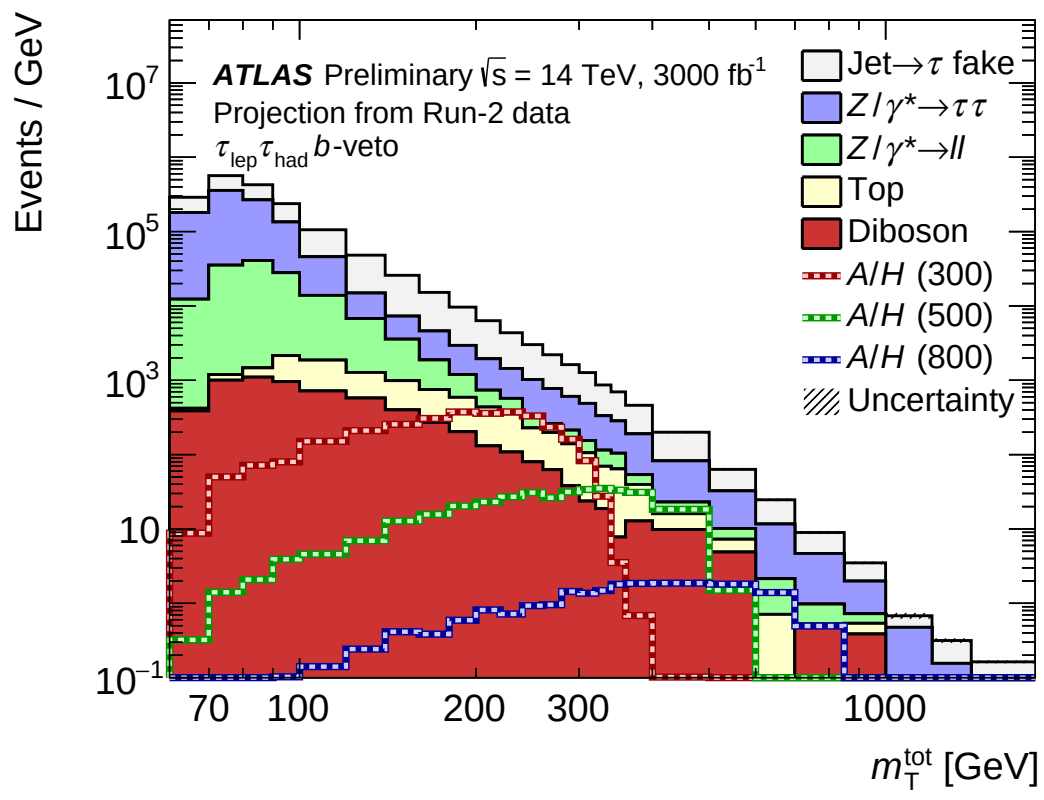

(a) $\tau_{\text{lep}}\tau_{\text{had}}$ $b$-veto category

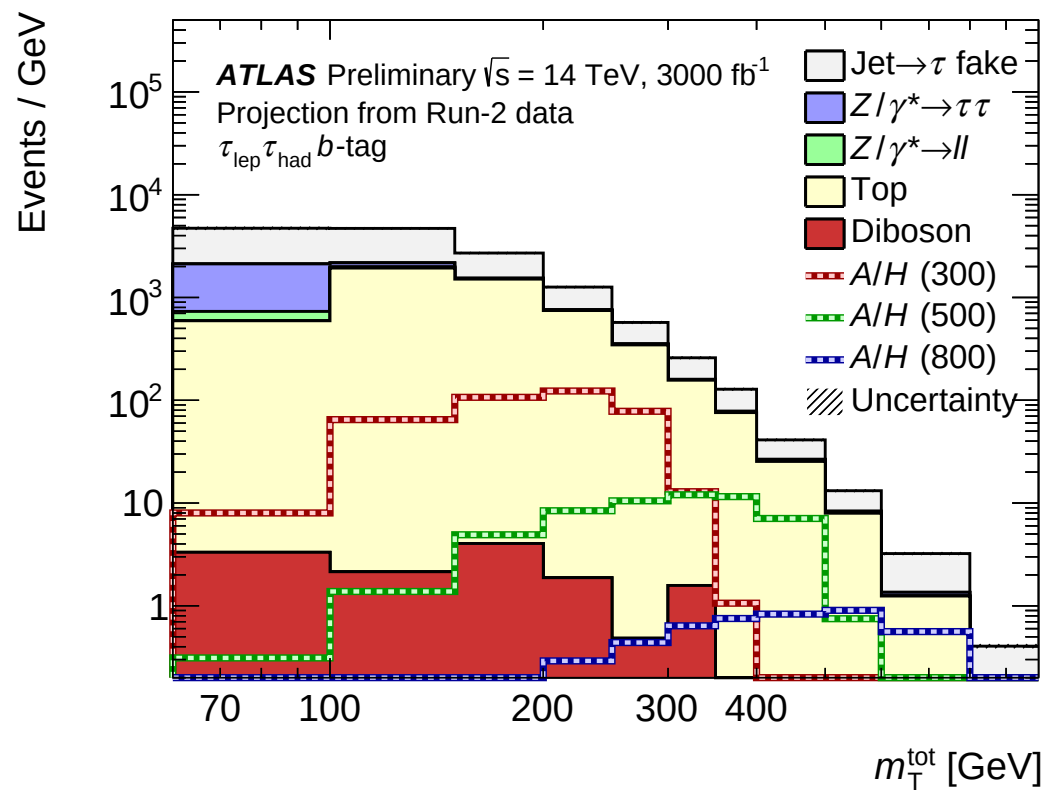

(b) $\tau_{\text{lep}}\tau_{\text{had}}$ $b$-tag category

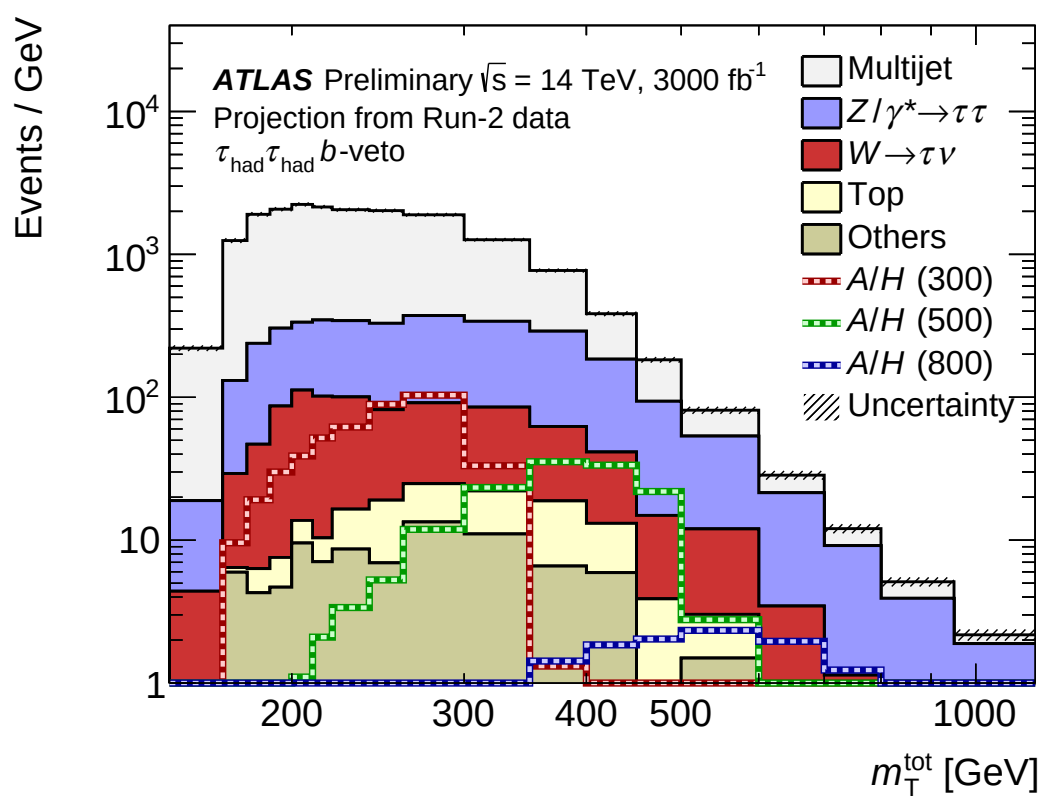

(c) $\tau_{\text{had}}\tau_{\text{had}}$ $b$-veto category

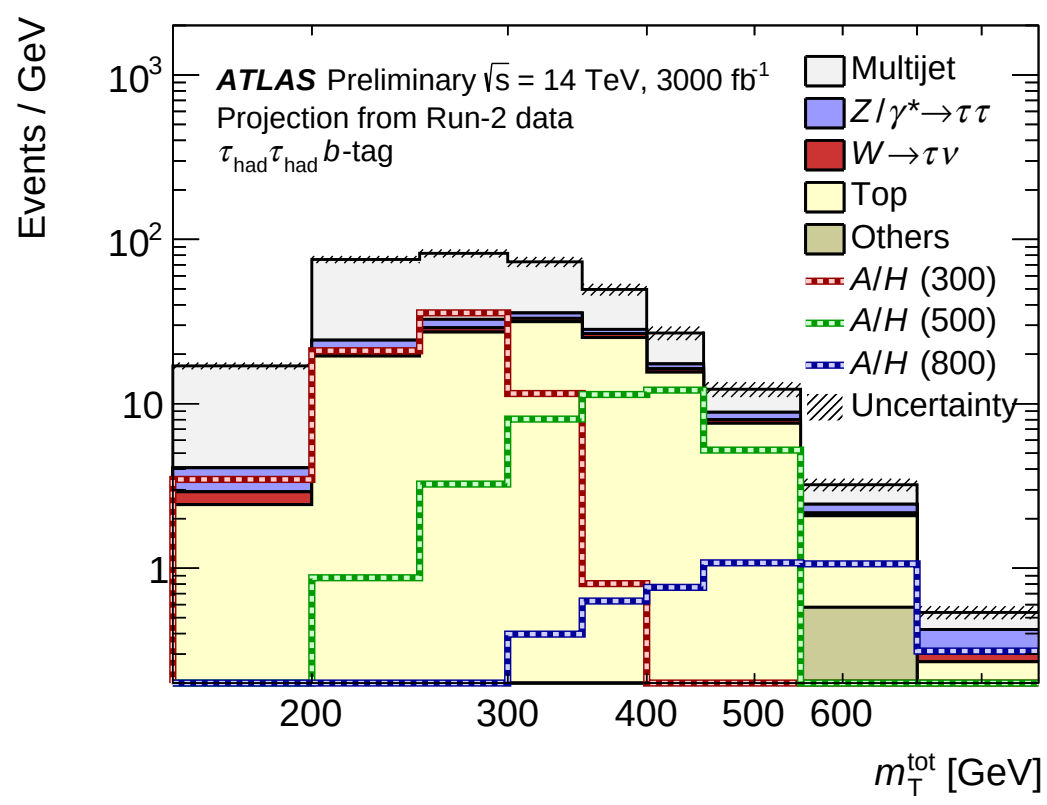

(d) $\tau_{\text{had}}\tau_{\text{had}}$ $b$-tag category

Figure 1: Distributions of $m_{\text{T}}^{\text{tot}}$ for each signal category. The predictions and uncertainties (including both statistical and systematic components) for the background processes are obtained from the fit under the hypothesis of no signal. The combined prediction for $A$ and $H$ bosons with masses of 300, 500 and 800 GeV and $\tan\beta = 10$ in the hMSSM scenario are superimposed.

The larger dataset at HL-LHC will give the opportunity to reduce the systematic uncertainties. The "Baseline" scenario for the systematic uncertainty reduction compared to current Run 2 values follows the recommendation of Ref. [14], according to which the systematic uncertainties associated with $b$-tagging, $\tau_{\text{had}}$ (hadronic $\tau$ decay) and theoretical uncertainties due to the missing higher order, the PDF uncertainty, etc., are reduced. The systematic uncertainties associated with the reconstruction and identification of the high-$p_{\text{T}}$ $\tau_{\text{had}}$ is reduced by a factor of 2 and becomes the leading systematic uncertainty for a heavy Higgs boson with mass $m_\phi > 1$ TeV. The systematic uncertainty associated with the modeling of the jet to $\tau_{\text{had}}$ fake background is assumed to be the same as in the current analysis. For the jet to $\tau_{\text{had}}$ fake background from multijet in $\tau_{\text{had}}\tau_{\text{had}}$ channel, the modeling uncertainty is mainly due to the limited data size in the control region and is reduced by a factor of 2. The statistical uncertainties on the predicted signal and background distributions, defined as the "template stat. uncertainty", is determined by the size of the MC samples and of the data sample in the control region where the $\tau_{\text{had}}$ fake factor is applied. The impact of the template stat. uncertainty is negligible in the Run 2 analysis. Assuming large enough MC samples will be generated for HL-LHC and sufficient data will be collected at HL-LHC, the uncertainties due to the sample size is ignored in this extrapolation study. To quantify the importance of the reduction of systematic uncertainties compared to current Run 2 values, results (labeled as "Unreduced") will also be given with current Run 2 values except for ignoring the template stat. uncertainty.

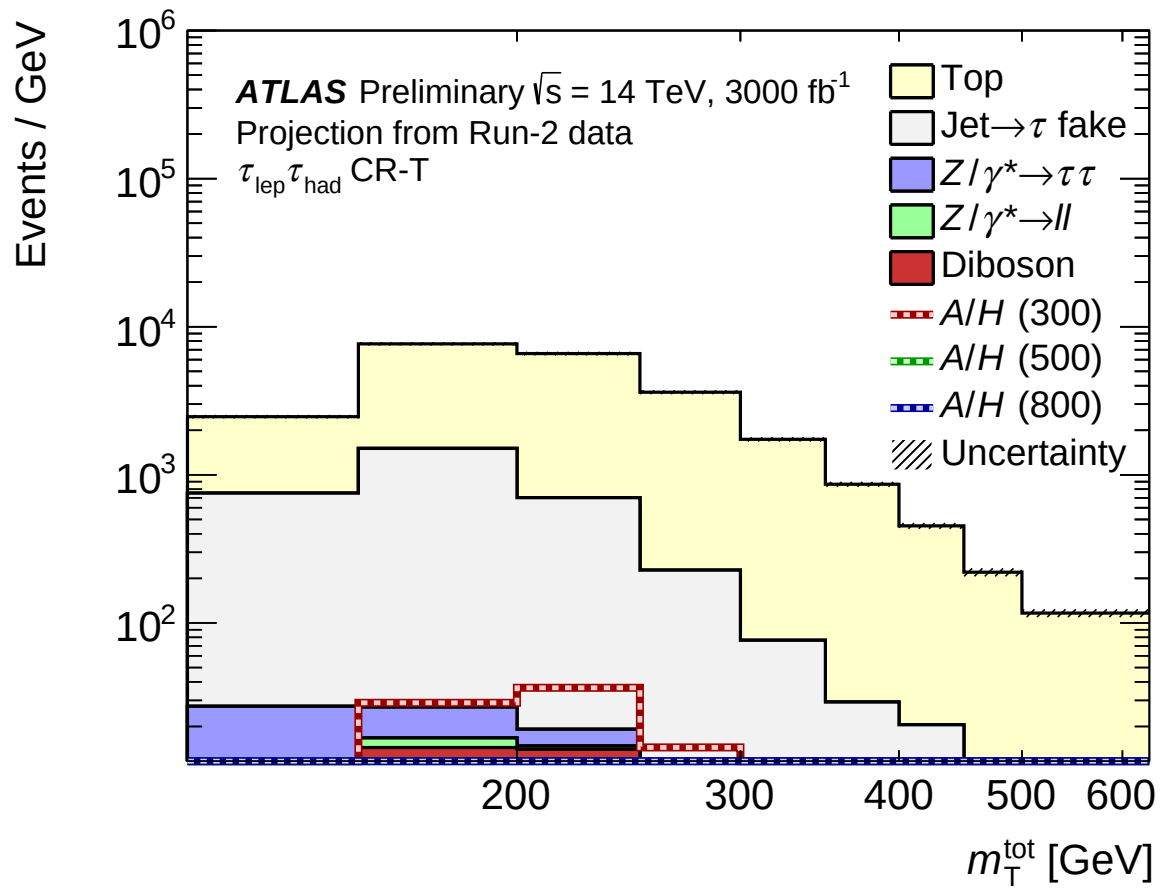

Figure 2: Distribution of $m_T^{tot}$ distributions in the top quark enriched control region of the $\tau_{lep}\tau_{had}$ channel.

# 3 Results

The $m_T^{tot}$ distributions from the $\tau_{lep}\tau_{had}$ (separately in the electron and muon channels) and $\tau_{had}\tau_{had}$ signal regions, as well as the top control region, are used in the final combined fit to extract the signal. The statistical framework used to produce the Run 2 results is documented in Ref. [12] and is adapted for this HL-LHC projection study. The results are given in terms of exclusion limits [15], as well as the 5 $\sigma$ discovery reach for gluon–gluon fusion and $b$-quarks association production modes.

## 3.1 Impact of systematic uncertainties

The impact of systematic uncertainties on the upper limit of the cross section times branching ratio ($\sigma \times BR(\phi \to \tau\tau)$) in Baseline scenario are calculated by comparing the expected 95% CL upper limit in case of no systematic uncertainties, $\mu_{stat}^{95}$, with a limit calculated by introducing a group of systematic uncertainties, $\mu_i^{95}$, as described in Ref. [12]. The systematic uncertainty impacts are shown in Figure 3(a) for gluon–gluon fusion production and Figure 3(b) for $b$-quarks association production as a function of the scalar boson mass. The major uncertainties are grouped according to their origin, while minor ones are collected as "Others" as detailed in Ref. [12].

The impact of systematic uncertainties is significant, as they degrade the expected limits by about 10–150 percent. In the low mass range, the leading uncertainties arise from the estimation of the dominant jet to $\tau_{had}$ fake background. At high masses, the leading uncertainty is from the reconstruction and identification of high-$p_T$ $\tau_{had}$. Because $\mu_{stat}^{95}$ is mainly determined by the data statistical uncertainty. In Figure 3(a) the impact of the $\tau_{had}$ related systematic uncertainties decreases after 1 TeV is due to the fact that the results at the higher mass regime are more limited by the data statistical uncertainty, while in Figure 3(b) the data statistical uncertainty in the $b$-tag category dominates in the high mass regime which leads the high-$p_T$ $\tau_{had}$ systematic uncertainty less outstanding.

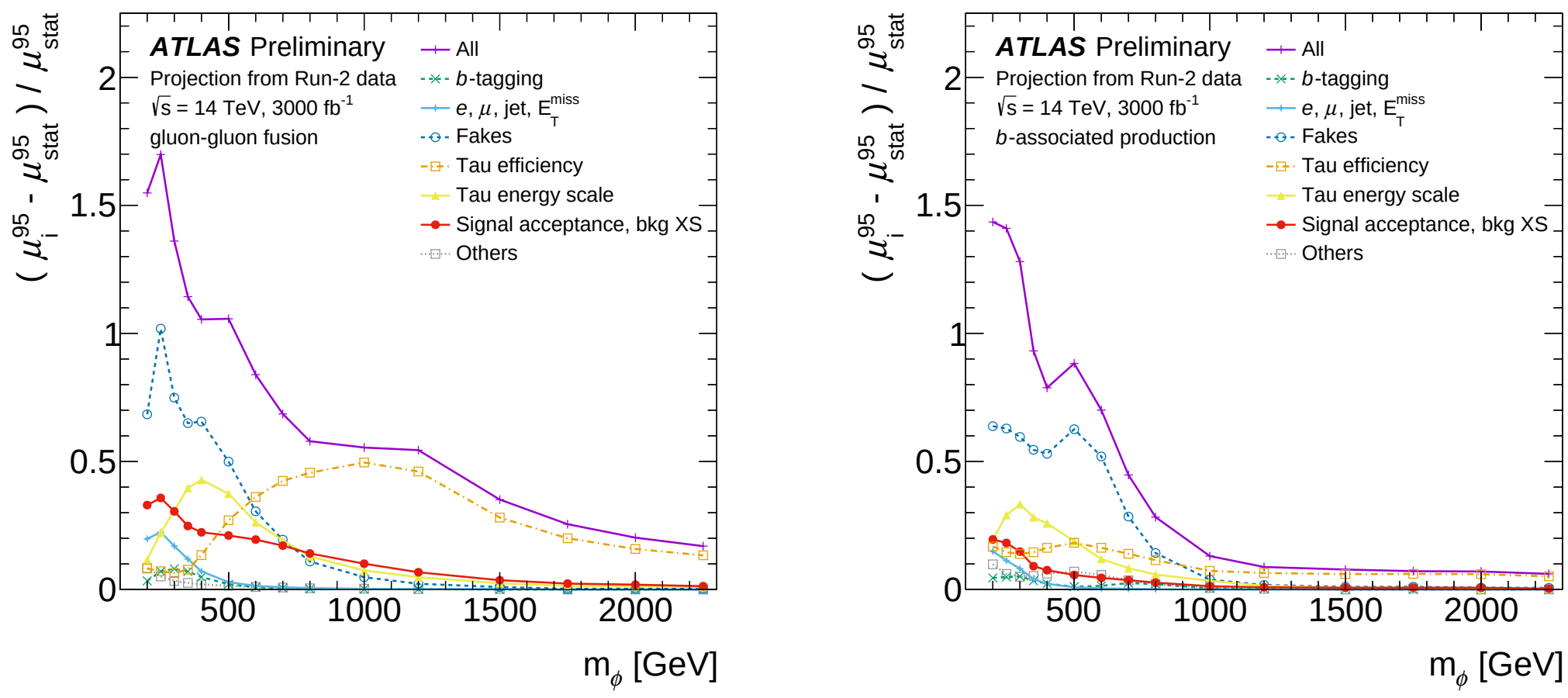

(a) gluon–gluon fusion production (b) $b$-associated production

Figure 3: Impact of major groups of systematic uncertainties (Baseline) on the $\phi \to \tau\tau$ 95% CL cross section upper limits as a function of the scalar boson mass, separately for the (a) gluon–gluon fusion and (b) $b$-associated production mechanisms.

### 3.2 Cross section limits and discovery reach

Figure 4 shows the upper limits on the gluon–gluon fusion and $b$-quark associated production cross section times the branching fraction for $\phi \to \tau\tau$. To demonstrate the impact of systematics, the expected exclusion limits with different systematic uncertainty scenarios are shown, as well as the Run 2 expected results [12]. The peaking structure around $m_\phi = 1$ TeV in figure 4(a) is due to the impact of the high-$p_T$ $\tau_{had}$ systematic uncertainty. The 5 $\sigma$ sensitivity line in the same figure illustrates the smallest values of the cross section times the branching fraction for which discovery level can be reached at HL-LHC: as clearly shown, the region where discovery is expected at HL-LHC extends significantly below the currently expected Run 2 exclusion region.

### 3.3 MSSM interpretation

Results are interpreted in terms of the MSSM. The cross section calculations follow the exact procedure used in Ref. [12], apart from the centre of mass energy is switched to 14 TeV. Figure 5 shows regions in the $m_A$–$\tan\beta$ plane excluded at 95% CL or discovered with 5 $\sigma$ significance in the hMSSM and $m_h^{\text{mod+}}$ scenarios. In the hMSSM scenario, $\tan\beta > 1.0$ for $250 < m_A < 350$ GeV and $\tan\beta > 10$ for $m_A = 1.5$ TeV could be excluded at 95% CL. When $m_A$ is above the $A/H \to t\bar{t}$ threshold, this additional decay mode reduces the sensitivity of the $A/H \to \tau\tau$ search for low $\tan\beta$. In the MSSM $m_h^{\text{mod+}}$ scenario, the expected 95% CL upper limits exclude $\tan\beta > 2$ for $250 < m_A < 350$ GeV and $\tan\beta > 20$ for $m_A = 1.5$ TeV.

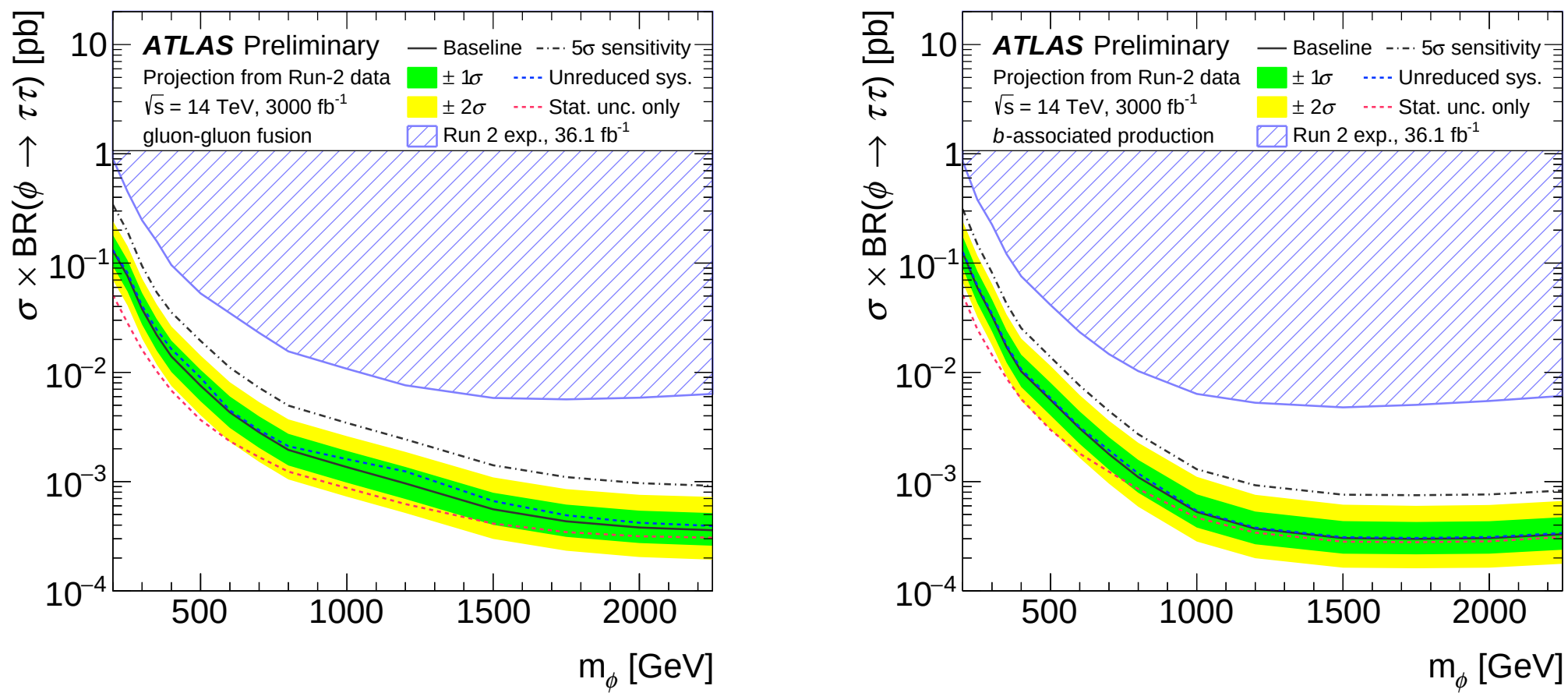

(a) gluon–gluon fusion production

(b) $b$-associated production

Figure 4: Projected 95% CL upper limits on the production cross section times the $\phi \to \tau\tau$ branching fraction for a scalar boson $\phi$ produced via (a) gluon–gluon fusion and (b) $b$-associated production, as a function of scalar boson mass. The limits are calculated from a statistical combination of the $\tau_e\tau_{\mathrm{had}}$, $\tau_\mu\tau_{\mathrm{had}}$ and $\tau_{\mathrm{had}}\tau_{\mathrm{had}}$ channels. "Baseline" uses the reduced systematic uncertainties scenario described in the text. "Unreduced sys." uses the same systematic uncertainties as the Run 2 analysis while ignoring the template stat. uncertainty. "Stat. unc. only" represents the expected limit without considering any systematic uncertainty. "5 $\sigma$ sensitivity" shows the region with the potential of 5 $\sigma$ significance in the Baseline scenario.

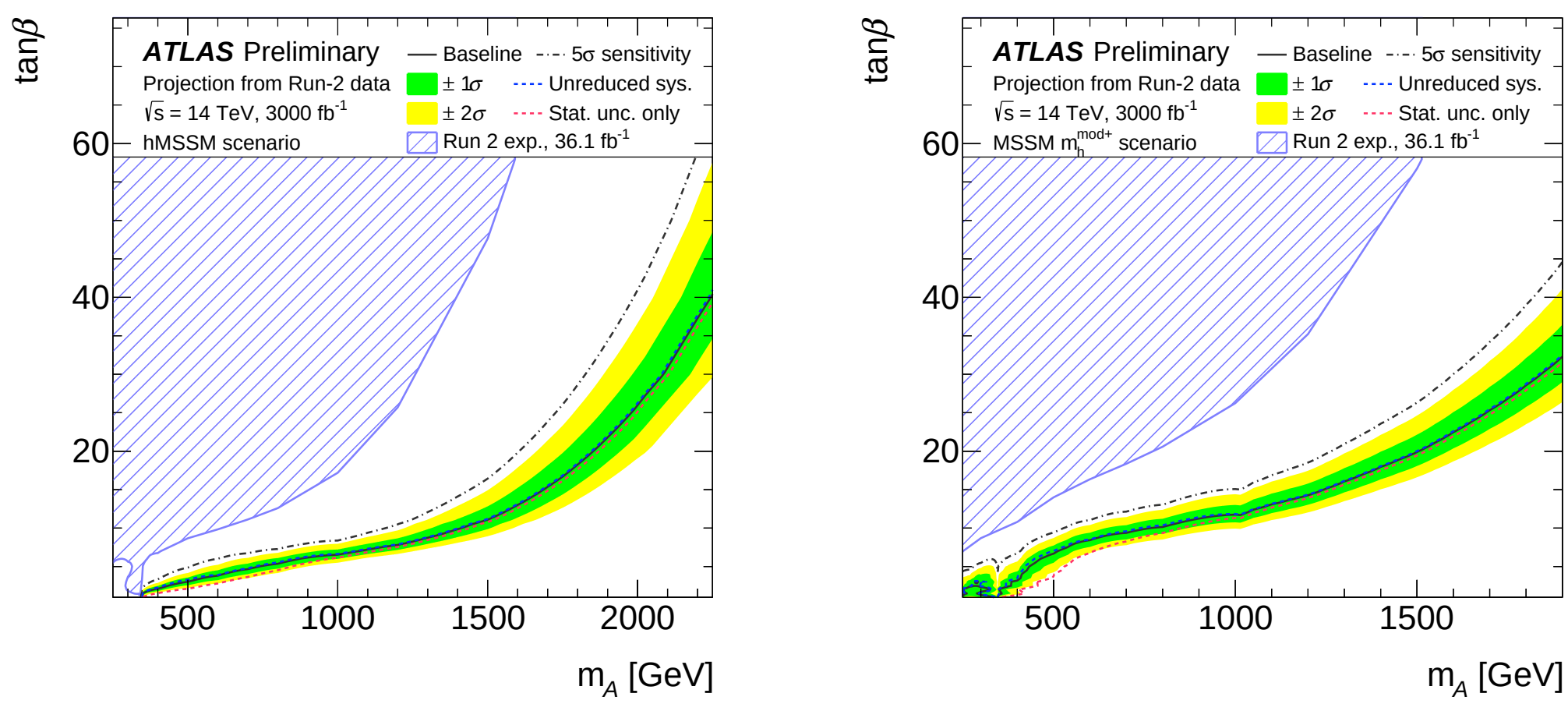

(a) hMSSM scenario

(b) $m_h^{\mathrm{mod+}}$ scenario

Figure 5: Projected 95% CL limits on tan $\beta$ as a function of $m_\phi$ in the MSSM (a) hMSSM and (b) $m_h^{\mathrm{mod+}}$ scenarios. The limits are calculated from a statistical combination of the $\tau_e\tau_{\mathrm{had}}$, $\tau_\mu\tau_{\mathrm{had}}$ and $\tau_{\mathrm{had}}\tau_{\mathrm{had}}$ channels. "Baseline" uses the reduced systematic uncertainties scenario described in the text. "Unreduced sys." uses the same systematic uncertainties as the Run 2 analysis while ignoring the template stat. uncertainty. "Stat. unc. only" represents the expected limit without considering any systematic uncertainty. "5 $\sigma$ sensitivity" shows the region with the potential of 5 $\sigma$ significance in the Baseline scenario.

## 4 Conclusion

The $H/A \to \tau\tau$ analysis documented in [12] has been extrapolated to estimate the sensitivity with 3000 fb$^{-1}$ of the HL-LHC dataset. The expected upper limits at 95% CL or, in alternative, the 5 $\sigma$ discovery reach in terms of cross section for the production of scalar bosons times the branching fraction to ditau final states have been estimated. The region with 5 $\sigma$ discovery potential at HL-LHC extends significantly below the currently expected Run 2 exclusion region. The expected limits are in the range 130–0.4 fb (130–0.3 fb) for gluon–gluon fusion ($b$-associated) production of scalar bosons with masses of 0.2–2.25 TeV. A factor of 6 to 18 increase in the sensitivity compared to the searches with the 36.1 fb$^{-1}$ Run 2 data [12] is projected. In the context of the hMSSM scenario, in the absence of a signal, the most stringent limits expected for the combined search exclude $\tan\beta > 1.0$ for $250 < m_A < 350$ GeV and $\tan\beta > 10$ for $m_A = 1.5$ TeV at 95% CL. The systematic uncertainties degrade the exclusion limit on $\sigma \times BR(\phi \to \tau\tau)$ by more than a factor of 2 for $m_\phi < 500$ GeV and about 10%–20% for $m_\phi = 2$ TeV. While the uncertainty on the estimate of fake $\tau_{\text{had}}$ dominates at low $m_\phi$, the uncertainty on high-$p_{\text{T}}$ $\tau_{\text{had}}$ reconstruction and identification is the leading systematic uncertainty at $m_\phi > 1.0$ TeV.

# CMS Physics Analysis Summary

Contact: cms-phys-conveners-ftr@cern.ch

# Projection of the Run 2 MSSM H $\to \tau\tau$ limits for the High-Luminosity LHC

The CMS Collaboration

## Abstract

A search for heavy Higgs bosons decaying to $\tau$ leptons was previously performed using data collected during Run 2 of the LHC, based on a data set of proton-proton collisions at $\sqrt{s} = 13$ TeV corresponding to an integrated luminosity of 35.9 fb$^{-1}$. A projection of these results to a High-Luminosity LHC data set of 3000 fb$^{-1}$ is described. For neutral Higgs boson masses above 1 TeV, an improvement by about one order of magnitude is expected in the 95% confidence level upper limits on the cross section. For the benchmark scenario $m_{\mathrm{h}}^{\mathrm{mod+}}$ of the minimal supersymmetric extension of the standard model, the expected lower limit on the mass of a heavy Higgs boson is extended from 1.25 to 2 TeV for $\tan\beta = 36$.

# 1 Introduction

Following the discovery of a Higgs boson by the ATLAS and CMS Collaborations in 2012 [1–3], a large number of measurements have established the compatibility of the new particle with standard model (SM) predictions. Nonetheless, there are many arguments in favor of theories that go beyond the SM. Many of these theories predict additional, heavy Higgs bosons. One such theory is supersymmetry [4, 5]. The minimal supersymmetric extension of the SM (MSSM) [6, 7] predicts two Higgs doublets, leading to five physical Higgs bosons: a light scalar (h), a heavy scalar (H), a pseudoscalar (A), and a charged pair ($H^{\pm}$).

Searches for MSSM Higgs bosons have been performed using the 2016 data from the LHC Run 2 [8–10]. So far, no significant evidence for physics beyond the SM has been found. However, the LHC to date has delivered only a small fraction of the integrated luminosity expected over its lifetime. Searches that are currently limited by statistical precision will see significant extensions in their reach as larger data sets are collected. Among the searches that will benefit are those for MSSM Higgs bosons.

Here, projections are presented for the reach that can be expected at higher luminosities in searches for heavy neutral Higgs bosons that decay to a pair of tau leptons. The projections are based on the most recent CMS publication for this search [10], performed using 35.9 $fb^{-1}$ of data collected during 2016 at a center-of-mass energy of 13 TeV. All the details of the analysis, including the simulated event samples, background estimation methods, systematic uncertainties, and different interpretations are described in Ref. [10]. Only details of direct relevance to the projection are presented here.

The analysis is a direct search for a neutral resonance decaying to two tau leptons. The following tau lepton decay mode combinations are considered: $\mu\tau_h$, $e\tau_h$, $\tau_h\tau_h$, and $e\mu$, where $\tau_h$ denotes a hadronically decaying tau lepton. In all these channels, events are separated into those that contain at least one b-tagged jet and those that do not contain any b-tagged jet. The goal of this categorization is to increase sensitivity to the dominant MSSM production modes: gluon fusion (ggH) and production in association with b quarks (bbH). The final discriminant is the total transverse mass, defined in Ref. [10]. The signal hypotheses considered consist of additional Higgs bosons in the mass range from 90 GeV to 3.2 TeV. The projection of the limits is performed by scaling all the signal and background processes to integrated luminosities of 300 and 3000 $fb^{-1}$, where the latter integrated luminosity corresponds to the total that is expected for the High-Luminosity LHC (HL-LHC).

The CMS detector [11] will be substantially upgraded in order to fully exploit the physics potential offered by the increase in luminosity at the HL-LHC [12], and to cope with the demanding operational conditions at the HL-LHC [13–17]. The upgrade of the first level hardware trigger (L1) will allow for an increase of L1 rate and latency to about 750 kHz and 12.5 $\mu$s, respectively, and the high-level software trigger (HLT) is expected to reduce the rate by about a factor of 100 to 7.5 kHz. The entire pixel and strip tracker detectors will be replaced to increase the granularity, reduce the material budget in the tracking volume, improve the radiation hardness, and extend the geometrical coverage and provide efficient tracking up to pseudorapidities of about $|\eta| = 4$. The muon system will be enhanced by upgrading the electronics of the existing cathode strip chambers (CSC), resistive plate chambers (RPC) and drift tubes (DT). New muon detectors based on improved RPC and gas electron multiplier (GEM) technologies will be installed to add redundancy, increase the geometrical coverage up to about $|\eta| = 2.8$, and improve the trigger and reconstruction performance in the forward region. The barrel electromagnetic calorimeter (ECAL) will feature the upgraded front-end electronics that will be able to exploit the information from single crystals at the L1 trigger level, to accommodate trigger latency and bandwidth requirements, and to provide 160 MHz sampling allowing high precision timing capability for photons. The hadronic calorimeter (HCAL), consisting in the barrel region of brass absorber plates and plastic scintillator layers, will be read out by silicon photomultipliers (SiPMs). The endcap electromagnetic and hadron calorimeters will be replaced with a new combined sampling calorimeter (HGCal) that will provide highly-segmented spatial information in both transverse and longitudinal directions, as well as high-precision timing information. Finally, the addition of a new timing detector for minimum ionizing particles (MTD) in both barrel and endocap region is envisaged to provide capability for 4-dimensional reconstruction of interaction vertices that will allow to significantly offset the CMS performance degradation due to high PU rates. A detailed overview of the CMS detector upgrade program is presented in Ref. [13–17], while the expected performance of the reconstruction algorithms and pile-up mitigation with the CMS detector is summarised in Ref. [18].

A previous CMS projection of the sensitivity for MSSM Higgs boson decays to a pair of tau leptons at the HL-LHC is reported in Ref. [19]. The results are presented in terms of model independent limits on a heavy resonance (either H or A, generically referred to as H below) decaying to two tau leptons, and are also interpreted in the context of MSSM benchmark scenarios.

## 2 Projection methodology

The projection assumes that the CMS experiment will have a similar level of detector and triggering performance during the HL-LHC operation as it provided during the LHC Run 2 period [13–17]. Three scenarios are considered for the projection of the size of systematic uncertainties to the HL-LHC:

- statistical uncertainties only: all systematic uncertainties are neglected;
- Run 2 systematic uncertainties: all systematic uncertainties are held constant with respect to luminosity, i.e., they are assumed to be the same as for the 2016 analysis;
- YR18 systematic uncertainties: systematic uncertainties are assumed to decrease with integrated luminosity following a set of assumptions described below.

In the YR18 scenario, selected systematic uncertainties decrease as a function of luminosity until they reach a certain minimum value. Specifically, all pre-fit uncertainties of an experimental nature (including statistical uncertainties in control regions and in simulated event samples) are scaled proportionally to the square root of the integrated luminosity. The following minimum values are assumed:

- muon efficiency: 25% of the 2016 value, corresponding to an average absolute uncertainty of about 0.5%;
- electron, $\tau_h$, and b-tagging efficiencies: 50% of the 2016 values, corresponding to average absolute uncertainties of about 0.5%, 2.5%, and 1.0%, respectively;
- jet energy scale: 1% precision for jets with $p_T > 30$ GeV
- estimate of the background due to jets misreconstructed as $\tau_h$ [20], for the components that are not statistical in nature: 50% of the 2016 values;
- luminosity uncertainty: 1%;
- theory uncertainties: 50% of the 2016 values, independent of the luminosity for all projections.

Note that for limits in which the Higgs boson mass is larger than about 1 TeV, the statistical uncertainties dominate and the difference between the systematic uncertainties found from the different methods has a negligible impact on the results.

The lightest Higgs boson h is excluded from the SM versus MSSM hypothesis test for the following reason: With increasing luminosity, the search will become sensitive to this boson. However, the current benchmark scenarios do not incorporate the properties of the h boson with the accuracy required at the time of the HL-LHC. Certainly the benchmark scenarios will evolve with time in this respect. Therefore the signal hypothesis includes only the heavy A and H bosons, to demonstrate the search potential only for these.

# 3 Projection results

## 3.1 Model independent limits

The model independent 95% confidence level (CL) upper limit on the cross sections for the ggH and bbH production modes, with the subsequent decays H $\to \tau\tau$, are shown in Figs. 1 and Fig. 2 for integrated luminosities of 300, 3000 and 6000 fb$^{-1}$. The 6000 fb$^{-1}$ limit is an approximation of the sensitivity with the complete HL-LHC dataset to be collected by the ATLAS and CMS experiments, corresponding to an integrated luminosity of 3000 fb$^{-1}$ each. The approximation assumes that the results of the two experiments are uncorrelated and that their sensitivity is similar. The first assumption is fulfilled to a high degree because the results are statistically limited; the validity of the second assumption is evident by comparing previous limits and projections.

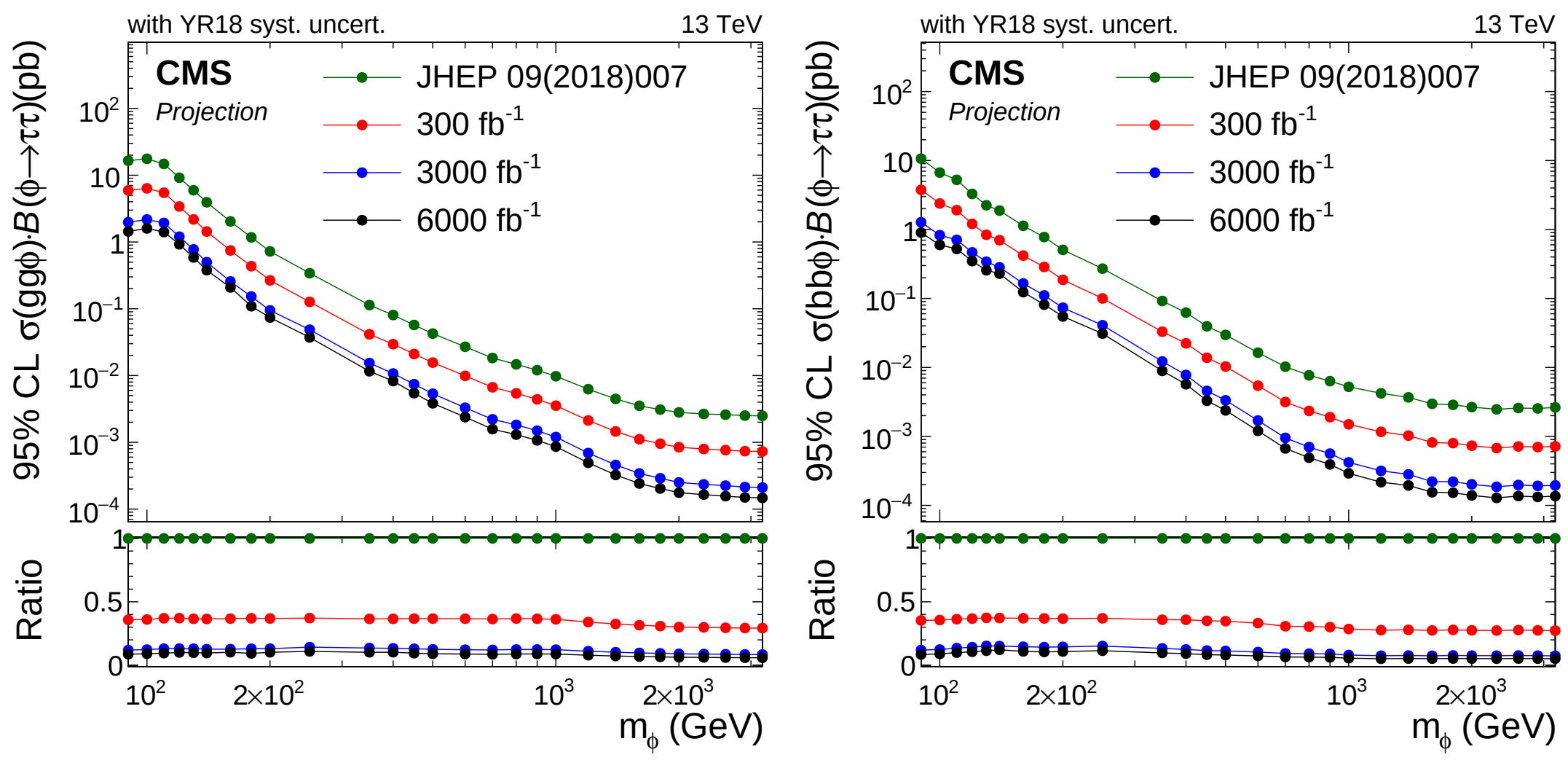

(a) ggH (b) bbH

Figure 1: Projection of expected model independent 95% CL upper limits based on 2016 CMS data [10] for ggH and bbH production with subsequent H $\to \tau\tau$ decays, with YR18 systematic uncertainties. The limit shown for 6000 fb$^{-1}$ is an approximation of the sensitivity with the complete HL-LHC dataset to be collected by the ATLAS and CMS experiments, corresponding to an integrated luminosity of 3000 fb$^{-1}$ each. The limits are compared to the CMS result using 2016 data [10].

For both production modes, the improvement in the limits at high mass values scales similarly to the square root of the integrated luminosity, as expected from the increase in statistical precision. The improvement at very low mass is almost entirely a consequence of reduced systematic uncertainties and not the additional data in the signal region. The difference between the Run 2 and YR18 scenarios is mostly because of the treatment of two kinds of systematic uncertainty of a statistical nature: the uncertainty related to the number of simulated events and that related to the number of events in the data control regions.

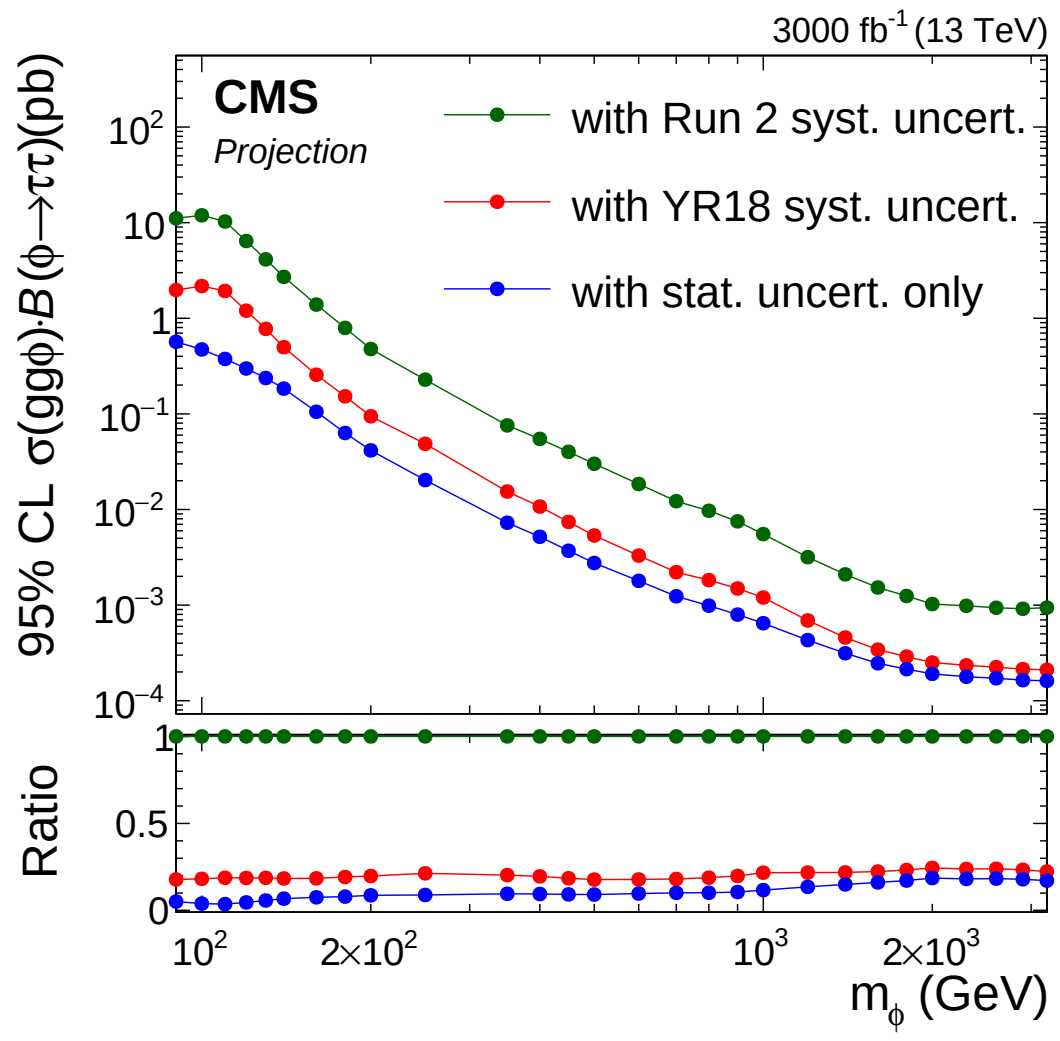

(a) ggH

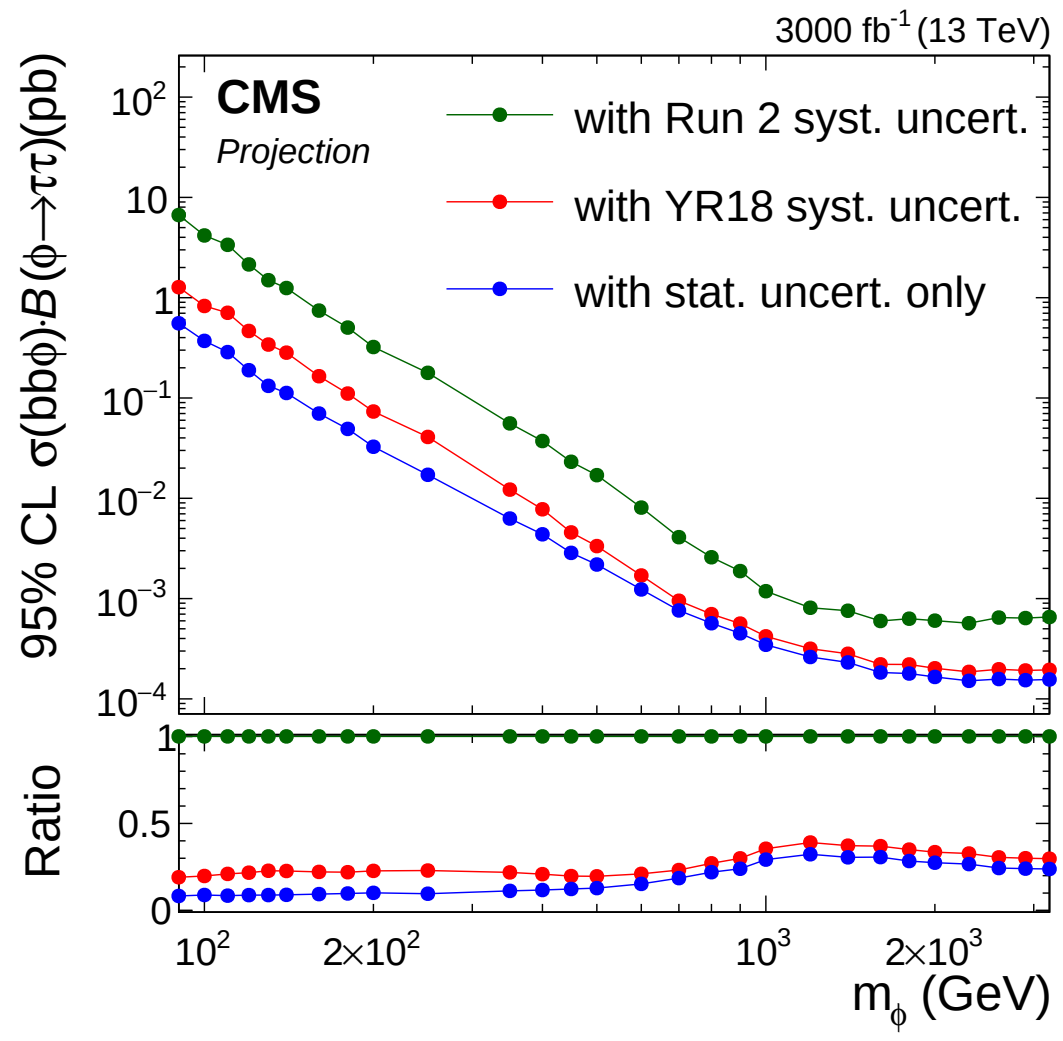

(b) bbH

Figure 2: Projection of expected model-independent limits based on 2016 CMS data [10] for ggH and bbH production with subsequent $\mathrm{H} \to \tau\tau$ decays, comparing different scenarios for systematic uncertainties for an integrated luminosity of 3000 fb$^{-1}$.

## 3.2 Model dependent limits

At the tree level, the Higgs sector of the MSSM can be specified by suitable choices for two variables, often chosen to be the mass $m_{\mathrm{A}}$ of the pseudoscalar Higgs boson and $\tan\beta$, the ratio of the vacuum expectation values of the two Higgs doublets. The typically large radiative corrections are fixed based on experimentally and phenomenologically sensible choices for the supersymmetric parameters, each choice defining a particular benchmark scenario [21]. Generally, MSSM scenarios assume that the 125 GeV Higgs boson is the lighter scalar h, an assumption that is compatible with the current experimental constraints for at least a significant portion of the $m_{\mathrm{A}}$–$\tan\beta$ parameter space. The di-tau lepton final state provides the most sensitive direct search for additional Higgs bosons predicted by the MSSM for intermediate and high values of $\tan\beta$, because of the enhanced coupling to down-type fermions.

The analysis results are interpreted in terms of these benchmark scenarios based on the profile likelihood ratio of the background-only and the tested signal-plus-background hypotheses. For this purpose, the predictions from both production modes and both heavy neutral Higgs bosons are combined. Figure 4 shows the results for three different benchmark scenarios: the $m_{\mathrm{h}}^{\mathrm{mod+}}$, the hMSSM, and the tau-phobic scenarios [10]. The sensitivity reaches up to Higgs boson masses of 2 TeV for values of $\tan\beta$ of 36, 26, and 28 for the $m_{\mathrm{h}}^{\mathrm{mod+}}$, the hMSSM, and the tau-phobic scenarios, respectively. Even at low mass, improvements are expected but in this case they are mostly a consequence of reduced systematic uncertainties and not the additional data in the signal region.

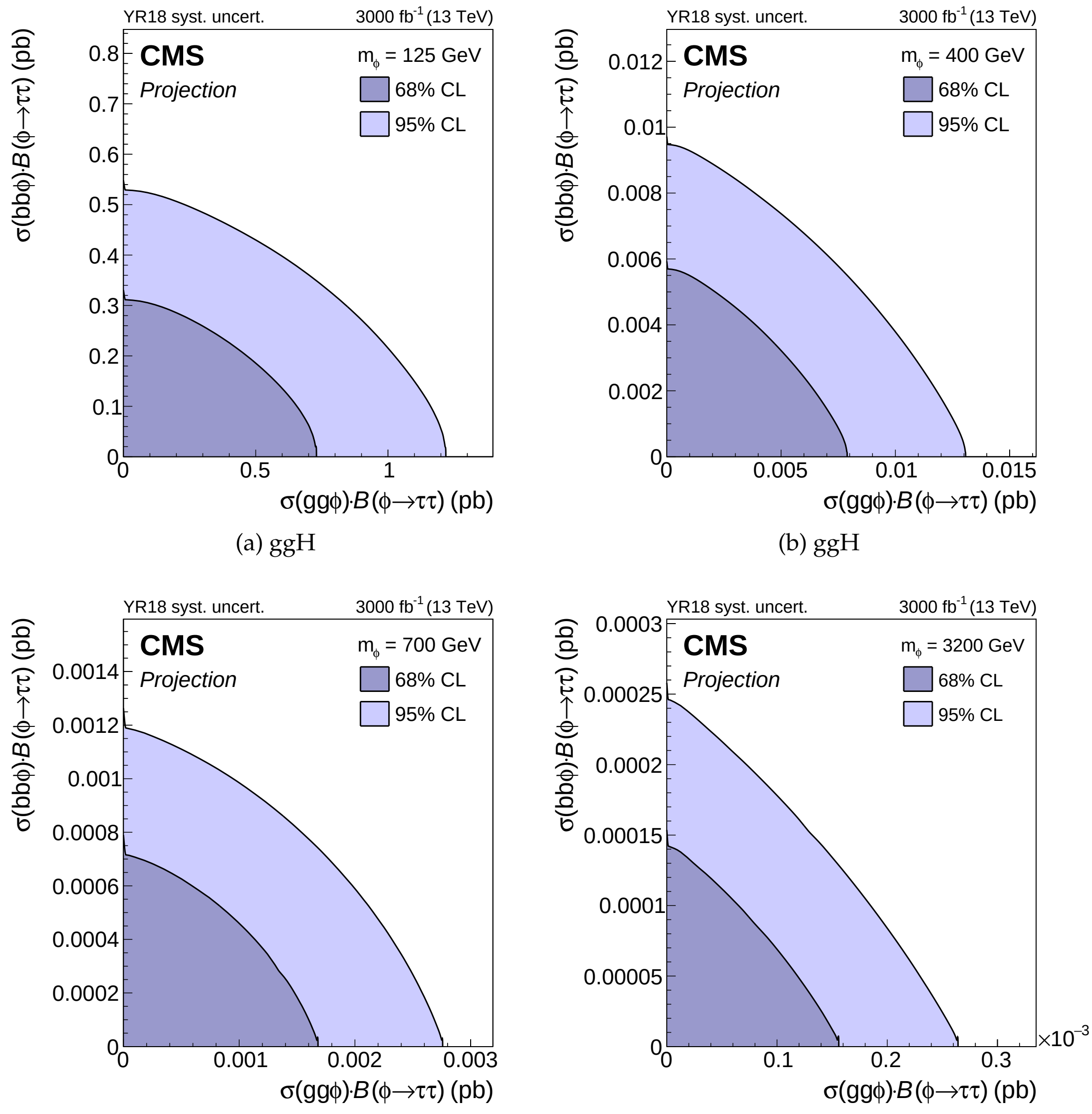

(a) ggH

(b) ggH

(c) ggH

(d) ggH

Figure 3: Projection of expected model-independent limits based on 2016 CMS data [10] for a simultaneous fit to the ggH and bbH production cross sections with subsequent $\mathrm{H} \to \tau\tau$ decays, for an integrated luminosity of $3000\,\mathrm{fb}^{-1}$ and with YR18 systematic uncertainties.

## 6

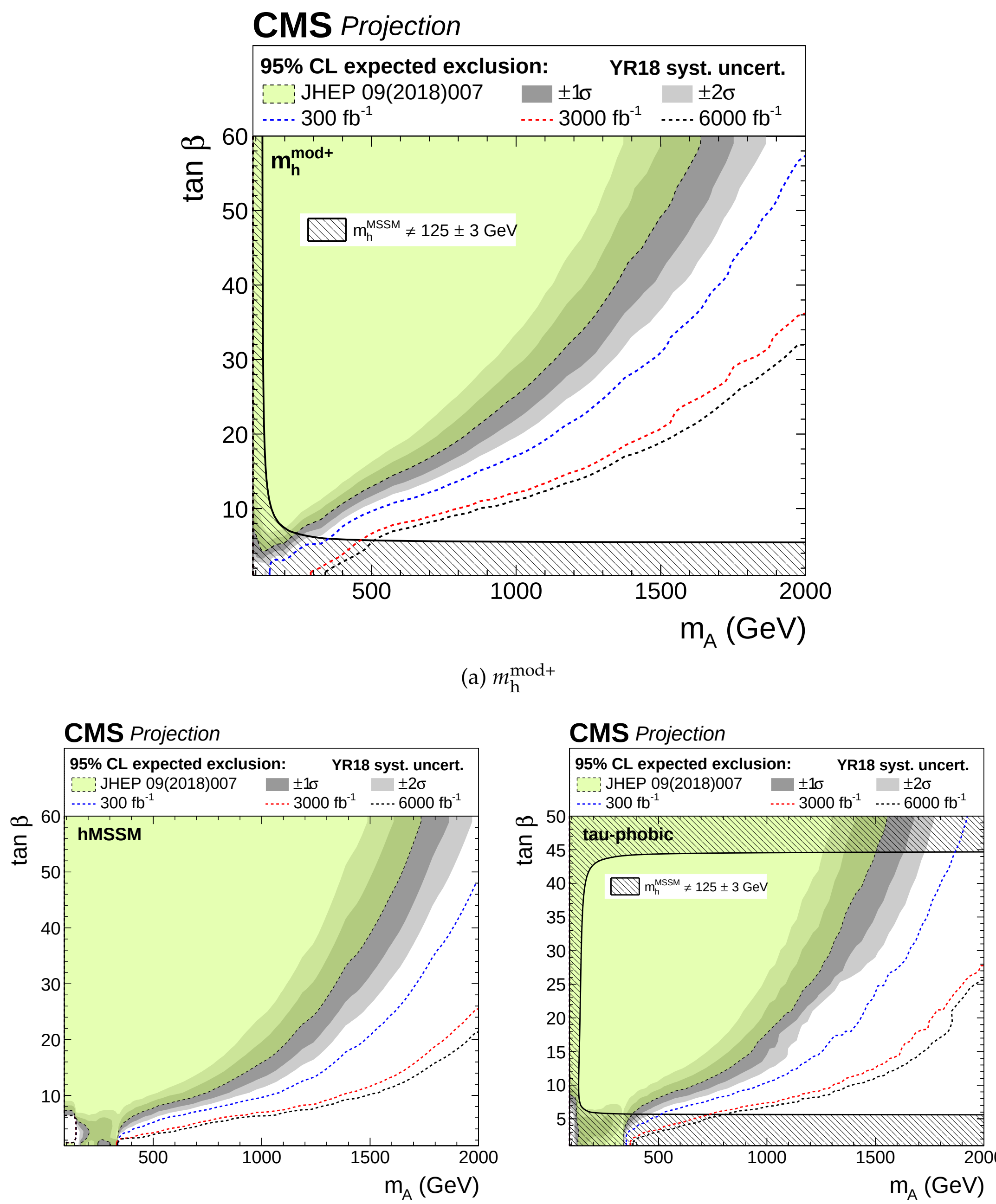

(a) $m_{\mathrm{h}}^{\mathrm{mod+}}$

(b) hMSSM

(c) tau-phobic

Figure 4: Projection of expected MSSM $\mathrm{H} \to \tau\tau$ 95% CL upper limits based on 2016 data [10] for different benchmark scenarios, with YR18 systematic uncertainties. The limit shown for $6000\,\mathrm{fb}^{-1}$ is an approximation of the sensitivity with the complete HL-LHC dataset to be collected by the ATLAS and CMS experiments, corresponding to an integrated luminosity of $3000\,\mathrm{fb}^{-1}$ each. The limits are compared to the CMS result using 2016 data [10]; for the tau-phobic scenario, it is a new interpretation of the information given in this reference.

# 4 Conclusions

The HL-LHC projections of the most recent results on searches for neutral MSSM Higgs bosons decaying to $\tau$ leptons have been shown, based on a data set of proton-proton collisions at $\sqrt{s} = 13$ TeV collected in 2016, corresponding to a total integrated luminosity of 35.9 fb$^{-1}$. The assumed integrated luminosity for the HL-LHC is 3000 fb$^{-1}$. In terms of cross section, an order-of-magnitude improvement in sensitivity is expected for neutral Higgs boson masses above 1 TeV since here the current analysis is statistically limited by the available integrated luminosity. For lower masses, an improvement of approximately a factor of five is expected for realistic assumptions on the evolution of the systematic uncertainties. For the MSSM benchmarks, the sensitivity will reach up to Higgs boson masses of 2 TeV for values of tan $\beta$ of 36, 26, and 28 for the $m_{\mathrm{h}}^{\mathrm{mod+}}$, the hMSSM, and the tau-phobic scenarios, respectively.

**8**

# CMS Physics Analysis Summary

Contact: cms-phys-conveners-ftr@cern.ch

# Search for invisible decays of a Higgs boson produced through vector boson fusion at the High-Luminosity LHC

The CMS Collaboration

## Abstract

The search for a Higgs boson decaying to invisible particles, produced through the vector boson fusion mode in the High-Luminosity LHC proton-proton collisions at $\sqrt{s} = 14$ TeV, is investigated based on simulation studies using Delphes, a fast-simulation package used to provide a parameterised response of the upgraded CMS detector. The event selection follows the existing CMS Run II data analysis, optimised for the High-Luminosity LHC conditions. The 95% confidence-level upper limits on the branching fraction of a standard-model-like Higgs boson decaying to invisible final states are studied with integrated luminosities of 300, 1000 and 3000 fb$^{-1}$ as a function of the thresholds applied on the transverse energy of the recoiling Higgs boson deposited in the detector.

# 1 Introduction

In the standard model (SM) of particle physics, the Higgs boson (H) is predicted to have a very small branching fraction to invisible particles (0.1% from $H \to 4\nu$ decays). While studies of the properties of the Higgs boson discovered by the ATLAS and CMS Collaborations [1–3] indicate a high level of compatibility with the SM [4], limits placed on invisible Higgs boson decays remain far less stringent than the SM prediction. In many theoretical extensions to the SM, for example Higgs-portal models [5–8], additional weakly-interacting particles interact with the Higgs boson and significantly increase its branching fraction to invisible final states.

Invisible decays of the Higgs boson have been searched for by the ATLAS and CMS Collaborations using data taken at $\sqrt{s} = 8$ and 13 TeV [9–14], considering different mechanisms for Higgs boson production. The signal is characterised by the presence of a transverse momentum imbalance, determined as the negative vectorial sum of the transverse momenta ($p_{\mathrm{T}}$) of all reconstructed particles $\vec{E}_{\mathrm{T}}^{\mathrm{miss}}$. The magnitude of $\vec{E}_{\mathrm{T}}^{\mathrm{miss}}$ is referred to as $E_{\mathrm{T}}^{\mathrm{miss}}$.

Gluon-gluon fusion (ggH) and vector-boson-associated production (VH, where V= W or Z) using hadronic decays of the W and Z bosons, are targeted by selecting events with a single high $p_{\mathrm{T}}$ jet and large $E_{\mathrm{T}}^{\mathrm{miss}}$. ZH production using leptonic decays of the Z boson is characterised by events with two well reconstructed electrons or muons and large $E_{\mathrm{T}}^{\mathrm{miss}}$. Lastly vector boson fusion production (VBFH) is targeted by searching for events containing a pair of jets with high invariant mass, large pseudo-rapidity separation, small separation in azimuthal angle, and large $E_{\mathrm{T}}^{\mathrm{miss}}$.

The combination of the searches for invisible decays of the Higgs boson in the ggH, VH (including ZH) and VBFH production modes using data taken in 2016 [10, 14], placed an observed (expected) upper limit on the branching fraction, $\mathcal{B}(H \to \text{inv.})$, of a SM-like Higgs boson of mass 125 GeV decaying to invisible particles at 0.26 (0.20) at the 95% confidence level (CL). The sensitivity of the combination is driven by the VBFH production channel, which alone excludes values smaller than 0.33 (0.25) at the 95% CL.

Given that VBFH production presents the best sensitivity, this channel is chosen to investigate the sensitivity of the search with the High-Luminosity LHC (HL-LHC). Previous sensitivity studies [15] indicated expected upper limits on $\mathcal{B}(H \to \text{inv.})$ of the order of 3 to 6% with an integrated luminosity of 3000 fb$^{-1}$, depending on the assumptions made for the experimental and theoretical uncertainties considered. In this document, a more thorough approach is considered by simulating proton-proton collisions at a centre-of-mass energy of 14 TeV, reconstructed using the upgraded CMS detector. A cut-and-count approach similar to the one described in the latest CMS search for invisible Higgs boson decays in the VBFH production mode [14] is used.

The CMS detector [16] will be substantially upgraded in order to fully exploit the physics potential offered by the increase in luminosity, and to cope with the demanding operational conditions at the HL-LHC [17–21]. The upgrade of the first level hardware trigger (L1) will allow for an increase of L1 rate and latency to about 750 kHz and 12.5 $\mu$s, respectively, and the high-level software trigger (HLT) is expected to reduce the rate by about a factor of 100 to 7.5 kHz. The entire pixel and strip tracker detectors will be replaced to increase the granularity, reduce the material budget in the tracking volume, improve the radiation hardness, and extend the geometrical coverage and provide efficient tracking up to pseudorapidities of about $|\eta| = 4$. The muon system will be enhanced by upgrading the electronics of the existing cathode strip chambers (CSC), resistive plate chambers (RPC) and drift tubes (DT). New muon detectors based on improved RPC and gas electron multiplier (GEM) technologies will be installed to add redundancy, increase the geometrical coverage up to about $|\eta| = 2.8$, and improve the trigger and reconstruction performance in the forward region. The barrel electromagnetic calorimeter (ECAL) will feature the upgraded front-end electronics that will be able to exploit the information from single crystals at the L1 trigger level, to accommodate trigger latency and bandwidth requirements, and to provide 160 MHz sampling allowing high precision timing capability for photons. The hadronic calorimeter (HCAL), consisting in the barrel region of brass absorber plates and plastic scintillator layers, will be read out by silicon photomultipliers (SiPMs). The endcap electromagnetic and hadron calorimeters will be replaced with a new combined sampling calorimeter (HGCal) that will provide highly-segmented spatial information in both transverse and longitudinal directions, as well as high-precision timing information. Finally, the addition of a new timing detector for minimum ionizing particles (MTD) in both barrel and endcap regions is envisaged to provide the capability for 4-dimensional reconstruction of interaction vertices that will significantly offset the CMS performance degradation due to high pileup (PU) rates. A detailed overview of the CMS detector upgrade program is presented in Ref. [17–21], while the expected performance of the reconstruction algorithms and PU mitigation with the CMS detector is summarised in Ref. [22].

The generated signal and background events are processed with the fast-simulation package Delphes [23] in order to simulate the expected response of the upgraded CMS detector. The object reconstruction and identification efficiencies, as well as the detector response and resolution, are parameterised in Delphes using the detailed simulation of the upgraded CMS detector based on GEANT4 package [24, 25].

## 2 Monte Carlo samples and objects

The VBFH signal samples are produced using POWHEG v2.0 [26, 27] at next-to-leading order (NLO) in quantum chromodynamics (QCD) theory, assuming $\mathcal{B}(\mathrm{H} \rightarrow \mathrm{inv.}) = 100\%$ and normalised using the inclusive Higgs boson production cross sections provided in Ref. [28]. Full-simulation samples produced at 13 TeV are used to derive the gluon-fusion contribution, applied as a fraction of the Delphes expected VBFH yields.

The main backgrounds are processes involving vector bosons (W and Z) produced in association with jets, either through QCD or electroweak (EWK) vertices. Monte Carlo samples for these backgrounds are generated at leading order (LO) using MADGRAPH5_aMC@NLO v2.2.2 [29] interfaced with PYTHIA v8.205 or higher. SM processes involving top quarks also contribute to the background, and are simulated using a combination of the POWHEG and MADGRAPH5_aMC@NLO generators. Backgrounds arising from QCD multijet events are simulated using MADGRAPH5_aMC@NLO interfaced with PYTHIA, imposing a minimum threshold of 1000 GeV on the dijet mass at parton level.

Electrons passing loose identification criteria, with a transverse momentum $p_{\mathrm{T}} > 10\,\mathrm{GeV}$, and pseudorapidity $|\eta| < 2.8$ are vetoed. Similarly, muons passing loose identification criteria with $p_{\mathrm{T}} > 20\,\mathrm{GeV}$ and $|\eta| < 3.0$ are vetoed. Taus passing loose identification criteria with $p_{\mathrm{T}} > 20\,\mathrm{GeV}$ and $|\eta| < 2.8$ are vetoed. Jets are reconstructed using the anti-$k_T$ algorithm [30, 31] with a parameter size of 0.4. The jets are required to have $p_{\mathrm{T}} > 30\,\mathrm{GeV}$ and $|\eta| < 5.0$, and are corrected for pileup effects using the "Puppi" algorithm [32].

A b-tagging algorithm is used to tag jets that originate from decays of B hadrons (b jets). The algorithm uses a combination of vertexing and timing information, and a working point with an efficiency of around 60% and a mis-tagging rate below 1% is defined to identify b jets. Events containing any identified b jets are vetoed.

The leading and sub-leading jets in the event are required to have $p_T > 80$ and 40 GeV, respectively, and be in opposite hemispheres of the detector. These two jets form the VBF dijet pair, and further requirements are applied on the invariant mass $M_{jj}$, and their separations in pseudorapidity $|\Delta\eta_{jj}|$ and azimuthal angle $|\Delta\phi_{jj}|$.

To reject the QCD multijet background, for which the transverse missing energy arises from jet mismeasurements, the $\vec{E}_T^{miss}$vector is required to not be aligned with a jet using min$\Delta\phi$(jet $\vec{p}_T >$ 30 GeV, $\vec{E}_T^{miss}$)$> 0.5$. The magnitude of the vectorial sum of the $p_T$ of all jets with $p_T > 30$ GeV is defined as $H_T^{miss}$.

The yields from the 14 TeV centre-of-mass energy Delphes samples are verified to be consistent with those expected from Ref. [14], with cross sections scaled to those predicted at 14 TeV.

The full-simulation and Delphes VBFH samples are compared in figure 1. The behaviour of $E_T^{miss}$ with 200 PU is not reproduced in Delphes, which shows a distribution with resolution similar to the one obtained in Ref. [14], whilst the current reconstruction algorithm of the upgraded detector in full simulation leads to a factor of 2 degradation of the $E_T^{miss}$ resolution. The $H_T^{miss}$variable is in much better agreement, indicating that the jet energy corrections with Puppi are performing adequately to mitigate the effect from the 200 PU environment. It should be noted that the current reconstruction algorithm, using the particle-flow approach [33], is not yet fully optimised for the upgraded detector design, and significant improvement in the $E_T^{miss}$ performance is expected once it has been tuned appropriately. Nevertheless, the impact of the observed degradation in the $E_T^{miss}$ resolution on the results presented here is considered in section 4.

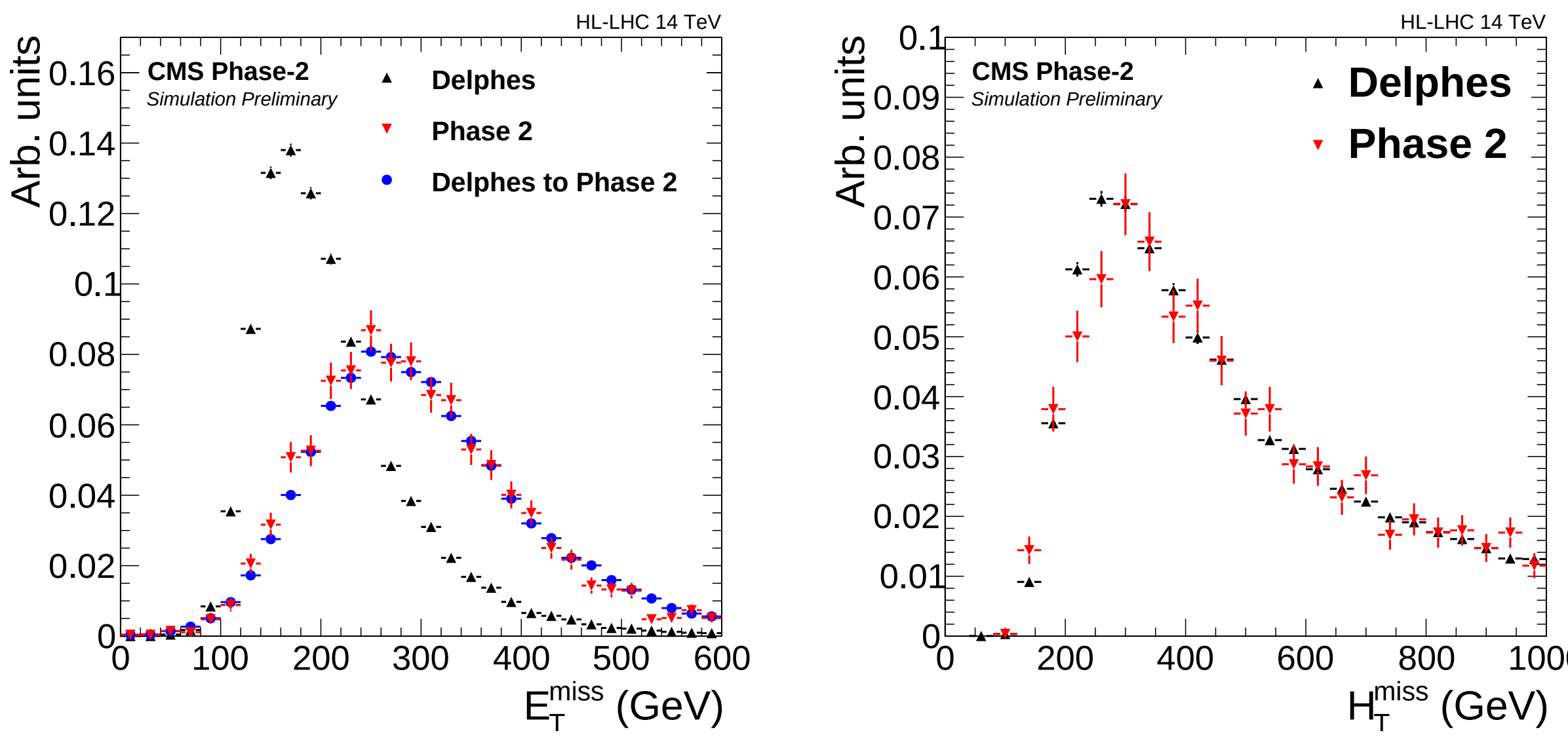

Figure 1: $E_T^{miss}$ (left) and $H_T^{miss}$(right) distributions in 200 PU VBFH signal samples, comparing full simulation (Phase 2) and Delphes. On the left, the distribution in Delphes is smeared as explained in the main text to reproduce the Phase 2 distribution.

# 3 Analysis strategy

The analysis uses five non-overlapping event regions: the signal region (SR) where events containing charged leptons ($\ell$, where $\ell = $ e or $\mu$) are vetoed, and four control regions (CR) with exactly one electron or muon (W $\rightarrow$ e$\nu$ CR and W $\rightarrow \mu\nu$ CR) or exactly two electrons or two muons (Z $\rightarrow$ ee CR and Z $\rightarrow \mu\mu$ CR). In the W $\rightarrow$ e$\nu$ and W $\rightarrow \mu\nu$ CRs, to further reject QCD multijet backgrounds, the transverse mass, defined as $\sqrt{2p_{\mathrm{T}}^{\ell}E_{\mathrm{T}}^{\mathrm{miss}}\left[1-\cos\Delta\phi(\ell,\vec{E}_{\mathrm{T}}^{\mathrm{miss}})\right]}$, where $p_{\mathrm{T}}^{\ell}$ is the transverse momentum of the lepton and $\Delta\phi(\ell,\vec{E}_{\mathrm{T}}^{\mathrm{miss}})$ is the azimuthal angle between the lepton momentum and $\vec{E}_{\mathrm{T}}^{\mathrm{miss}}$ vectors, is required to be less than 160 GeV. In the $\mathrm{W}\rightarrow\mathrm{e}\nu$ CR a selection on $E_{\mathrm{T}}^{\mathrm{miss}} > 60$ GeV is also applied due to the higher QCD multijet contamination than in the muon channel. In the $\mathrm{Z}\rightarrow\mathrm{ee}$ and $\mathrm{Z}\rightarrow\mu\mu$ CRs, the dilepton mass is required to be between 60 and 120 GeV. To account for the higher single-electron trigger thresholds that will be required at the HL-LHC , the leading electron $p_{\mathrm{T}}$ is required to be above 40 GeV, for both the $\mathrm{W}\rightarrow\mathrm{e}\nu$ and $\mathrm{Z}\rightarrow\mu\mu$ CRs.

The signal is separated from the backgrounds by using the characteristics of the VBF dijet pair. A first estimation of the best regions of interest is obtained by looking at the expected significance, $\sigma_A$, calculated using an Asimov dataset [34] assuming $\mathcal{B}(\mathrm{H}\rightarrow\mathrm{inv.}) = 100\%$, defined as,

$$\sigma_A = \sqrt{2\times((N_S+N_B)\times ln(1+N_S/N_B)-N_S)},$$

where $S$ is the VBFH production yield and $B$ is the sum of all background sample yields. Each selection criteria is varied in turn, one or two dimensions at a time, to identify the regions of phase-space with the highest expected significance. The highest $\sigma_A$ value is found to be for $\mathrm{M_{jj}} > 1000$ GeV, $|\Delta\eta_{\mathrm{jj}}| > 4$, $|\Delta\phi_{\mathrm{jj}}| < 1.8$, $E_{\mathrm{T}}^{\mathrm{miss}} > 130$ GeV. Further studies of the optimal region are done using the final output of the analysis, namely the 95% CL upper limits on $\mathcal{B}(\mathrm{H}\rightarrow\mathrm{inv.})$.

The lower threshold on the $E_{\mathrm{T}}^{\mathrm{miss}}$ is varied from 130 to 400 GeV in 10 to 50 GeV steps. Likewise, the lower threshold on $\mathrm{M_{jj}}$ is varied from 1000 to 4000 GeV in 100 GeV steps. The statistical uncertainty on the MC is considered to be negligible, assuming the available MC samples will have at least 10 times the integrated luminosity available in the data. For each ($E_{\mathrm{T}}^{\mathrm{miss}}$, $\mathrm{M_{jj}}$) selection, the yields are extracted in the four control regions and in the signal region, and a likelihood is constructed as the product of five Poisson terms, one per region. Upper limits on the Higgs boson production cross section times $\mathcal{B}(\mathrm{H}\rightarrow\mathrm{inv.})$ are placed at the 95% CL using the CLs criterion [35–37], with a profiled likelihood ratio as the test statistic in which systematic uncertainties are incorporated via nuisance parameters [3, 38]. Asymptotic formula are used to determine the distribution of the test statistic under signal and background hypotheses [34].

The scenario considered for the systematic uncertainties is described in table 1, together with the systematic uncertainties that were considered in Ref. [14], for comparison.

## 4 Results

The analysis is expected to be systematics dominated, with the dominant systematic uncertainties due to the muon and electron efficiencies, both in the control and signal regions, and the jet energy scale and trigger efficiencies. In Ref. [14], due to the limited size of the dilepton samples, the knowledge of the ratio of the cross sections of the W to Z boson production was used as a constraint between the two backgrounds, leading to an increased sensitivity. The theoretical uncertainty on this ratio is set at 12.5% from studies of missing higher order QCD and EWK corrections [14], for both QCD and EWK production. Once 300 $\mathrm{fb}^{-1}$ of data will be available, this constraint will play a smaller role. It is expected that improvements in theoretical calculations of the ratio will lead to half the current theoretical uncertainty, namely 7%. This uncertainty is expected to have an impact of at most 3–5% for the selection with the largest expected significance and is therefore neglected in the results presented herein. However, the uncertainty will be relevant when considering very tight selection criteria on $E_{\mathrm{T}}^{\mathrm{miss}}$ and $\mathrm{M_{jj}}$, i.e. when the statistical uncertainty in the CRs becomes dominant.

| Systematic | From Ref. [14] | This analysis |
|---|---|---|
| e-ID | 1%(gsf)⊕1%(idiso) | 1% |
| $\mu$-ID | 1%(reco)⊕1%(id)⊕0.5%(iso) | 0.5% |
| e-veto | 0.6%(gsf)⊕1.5%(idiso) | 1% |
| $\mu$-veto on QCD V+jets | 5%(reco)⊕5%(id)⊕2%(iso) | 2% |
| $\mu$-veto on EWK V+2jets | 10%(reco)⊕10%(id)⊕6%(iso) | 6% |
| $\tau$-veto | 1–1.5% for QCD–EWK | 0.5–0.75% |
| b-tag-veto | 0.1% (sig) 2% (top) | 0.05% (sig) 1% (top) |
| JES | 14%(sig) 2%(W/W) 1%(Z/Z) | 4.5%(sig) 0.5%(W/W) 0.2%(Z/Z) |
| Integrated luminosity | 2.5% | 1% |
| QCD multijet | 1.5% | 1.5% |
| Theory on W/Z ratio | 12.5% | 7% |
| ggH normalisation | 24% | 20% |

Table 1: Impact on the signal and background yields from the different sources of systematic uncertainty considered in Ref. [14] and for the HL-LHC setup considered in this analysis.

The most stringent upper limits are achieved in the regions with lower thresholds on $\mathrm{M_{jj}}$ and $E_{\mathrm{T}}^{\mathrm{miss}}$ of 2500 GeV and 190 GeV, respectively, for the 3000 fb$^{-1}$ scenario. The minimum is rather flat between $\mathrm{M_{jj}}$ values of 2300 and 3000 GeV, and between $E_{\mathrm{T}}^{\mathrm{miss}}$ values of 170 and 220 GeV, indicating limited impact from the size of the MC samples. The upper limits degrade steeply as the $E_{\mathrm{T}}^{\mathrm{miss}}$ threshold increases above 250 GeV. The behaviour is similar for the 300 and 1000 fb$^{-1}$ scenarios, with best thresholds found at lower values of $E_{\mathrm{T}}^{\mathrm{miss}}$ (170 GeV) and $\mathrm{M_{jj}}$ (1500 and 1800 GeV respectively) due to the interplay between the size of the control regions and the systematic uncertainties.

Distributions in $|\Delta\eta_{\mathrm{jj}}|$, $|\Delta\phi_{\mathrm{jj}}|$, $\mathrm{M_{jj}}$ for the leading jet pair, $E_{\mathrm{T}}^{\mathrm{miss}}$, $\mathrm{H_{T}^{miss}}$and min$\Delta\phi$(jet $p_{\mathrm{T}} > 30$ GeV, $E_{\mathrm{T}}^{\mathrm{miss}}$)), in the signal region are shown in figures 2, 3 and 4, for the 3000 fb$^{-1}$ scenario. The corresponding expected yields are shown in table 2. The uncertainties shown represent the statistical uncertainties due to the limited size of the Delphes samples and are not used in the calculations of the final limits.

The 95% CL upper limits for an integrated luminosity of 3000 fb$^{-1}$ are shown in figure 5, left, as a function of the thresholds applied on $E_{\mathrm{T}}^{\mathrm{miss}}$ assuming the MC statistical uncertainties are negligible, for the final selections described above. In the best case, the lowest 95% CL limit on B(H $\rightarrow$ inv.), assuming standard model production, is expected to be at 3.8%, for thresholds values of 2500 GeV (190 GeV) on the dijet mass ($E_{\mathrm{T}}^{\mathrm{miss}}$). If the $E_{\mathrm{T}}^{\mathrm{miss}}$ resolution was to be a factor of 2 worse, the re-optimisation of the selection leads to minimum thresholds of 1800 GeV (250 GeV) on the dijet mass ($E_{\mathrm{T}}^{\mathrm{miss}}$), but a similar 95% CL limit. The limits are shown for different integrated luminosities in figure 5, right.

# 5 Conclusion

The search for a Higgs boson decaying invisibly, produced in the vector-boson fusion mode, is investigated at the HL-LHC through simulation studies using a fast parametrisation of the upgraded CMS detector. The analysis follows the latest CMS publication, with an event selection optimised for the HL-LHC conditions. The expected 95% CL upper limits on the branching ratio of the standard model Higgs boson to invisible particles are presented as a function of the lower threshold applied on the transverse missing energy, for scenarios with integrated luminosities of 300, 1000 and 3000 fb$^{-1}$. The 95% CL upper limit on $B(H \to \text{inv.})$ assuming standard model production is expected to be at 3.8%, for thresholds values of 2500 GeV (190 GeV) on the dijet mass (missing transverse momentum). Even if the transverse missing energy resolution is degraded by a factor of two due to the high pileup conditions, a similar sensitivity is nevertheless achieved.

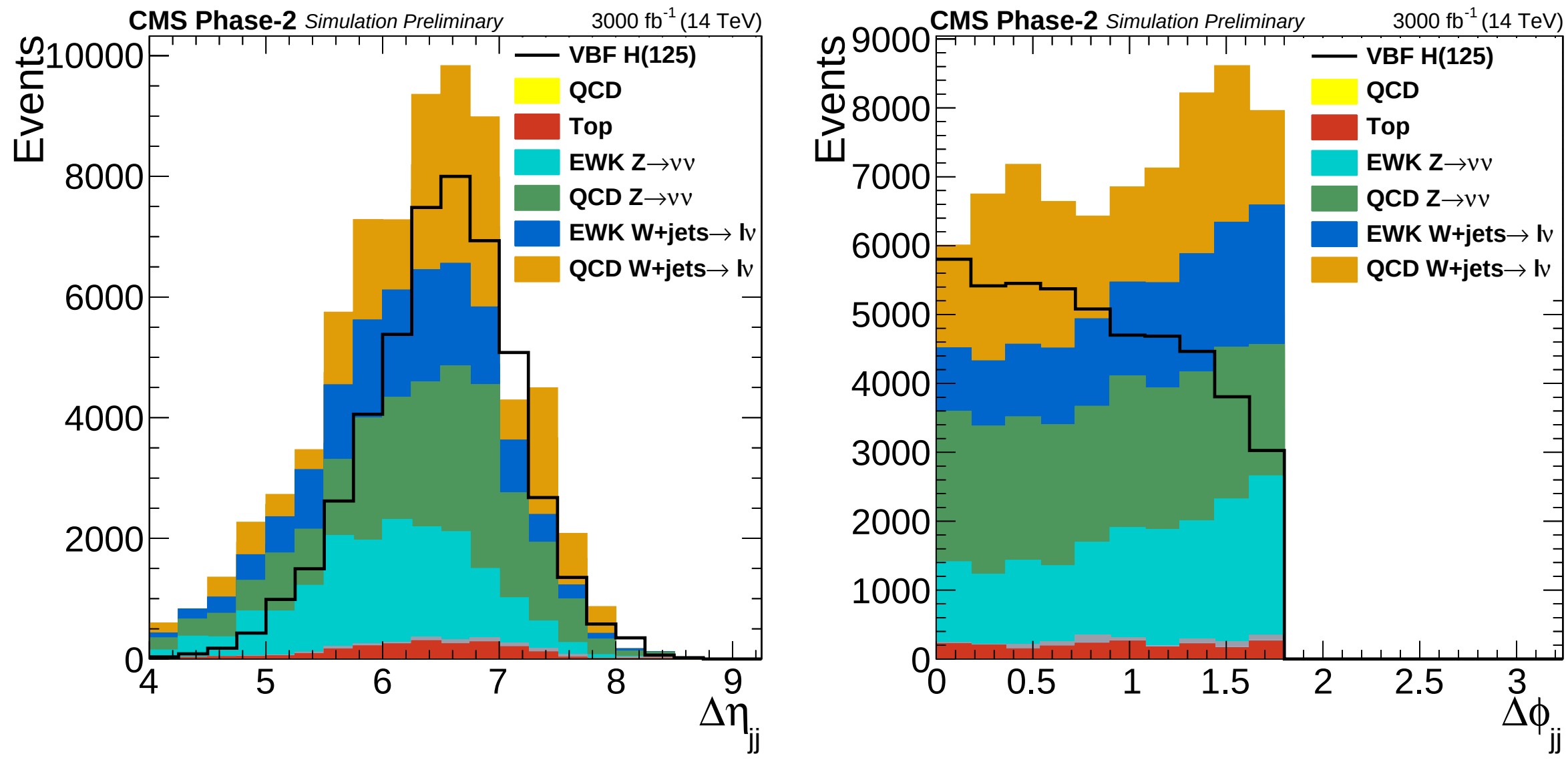

Figure 2: Distributions of $|\Delta\eta_{\mathrm{jj}}|$ and $|\Delta\phi_{\mathrm{jj}}|$ in the signal region for the final selection, $\mathrm{M_{jj}} > 2500\,\mathrm{GeV}$ and $E_{\mathrm{T}}^{\mathrm{miss}} > 190\,\mathrm{GeV}$.

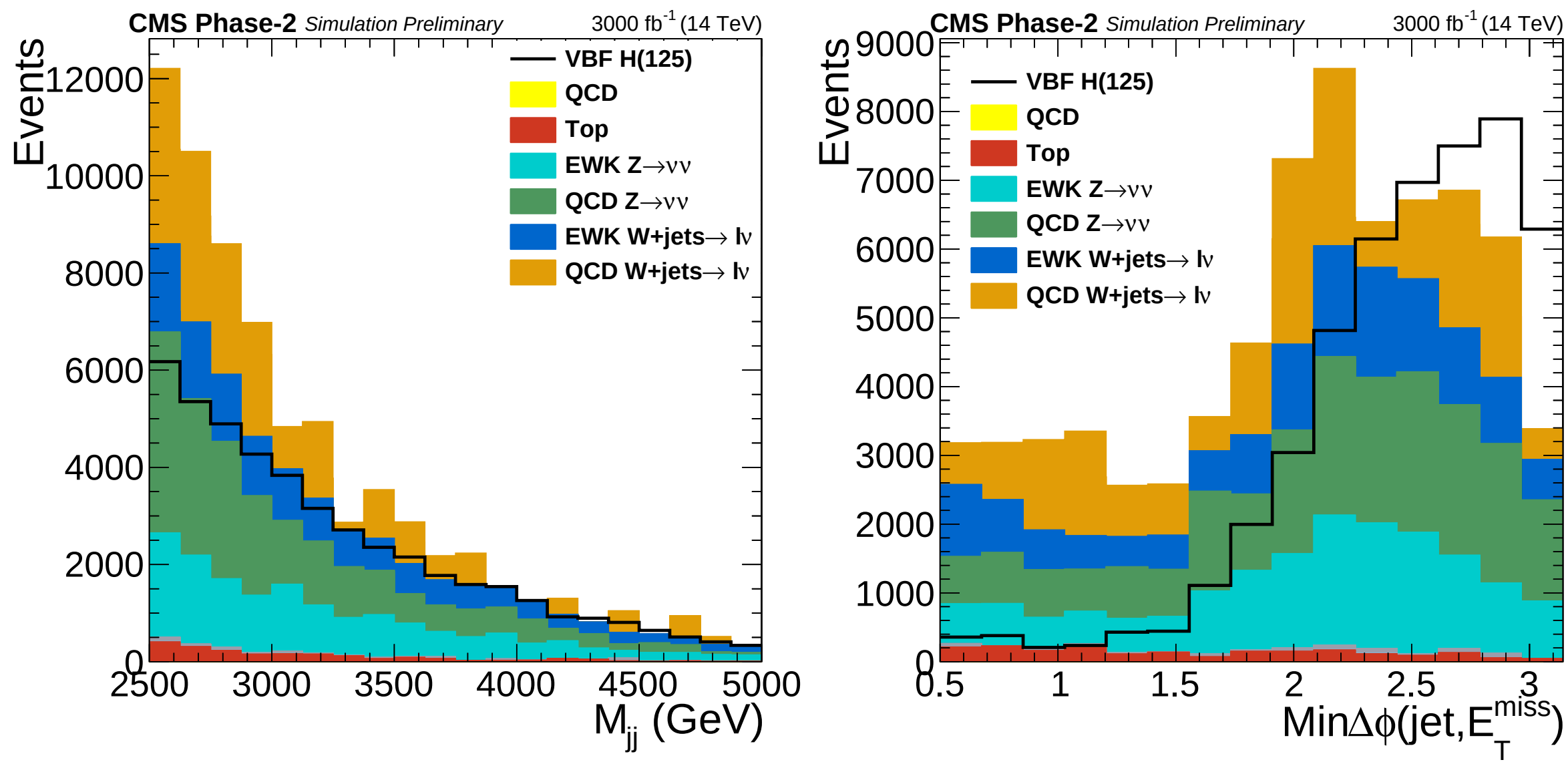

Figure 3: Distributions of $\mathrm{M_{jj}}$ and $\min\Delta\phi(\mathrm{jet}\ p_{\mathrm{T}} > 30\,\mathrm{GeV}, E_{\mathrm{T}}^{\mathrm{miss}})$ in the signal region for the final selection, $\mathrm{M_{jj}} > 2500\,\mathrm{GeV}$ and $E_{\mathrm{T}}^{\mathrm{miss}} > 190\,\mathrm{GeV}$.

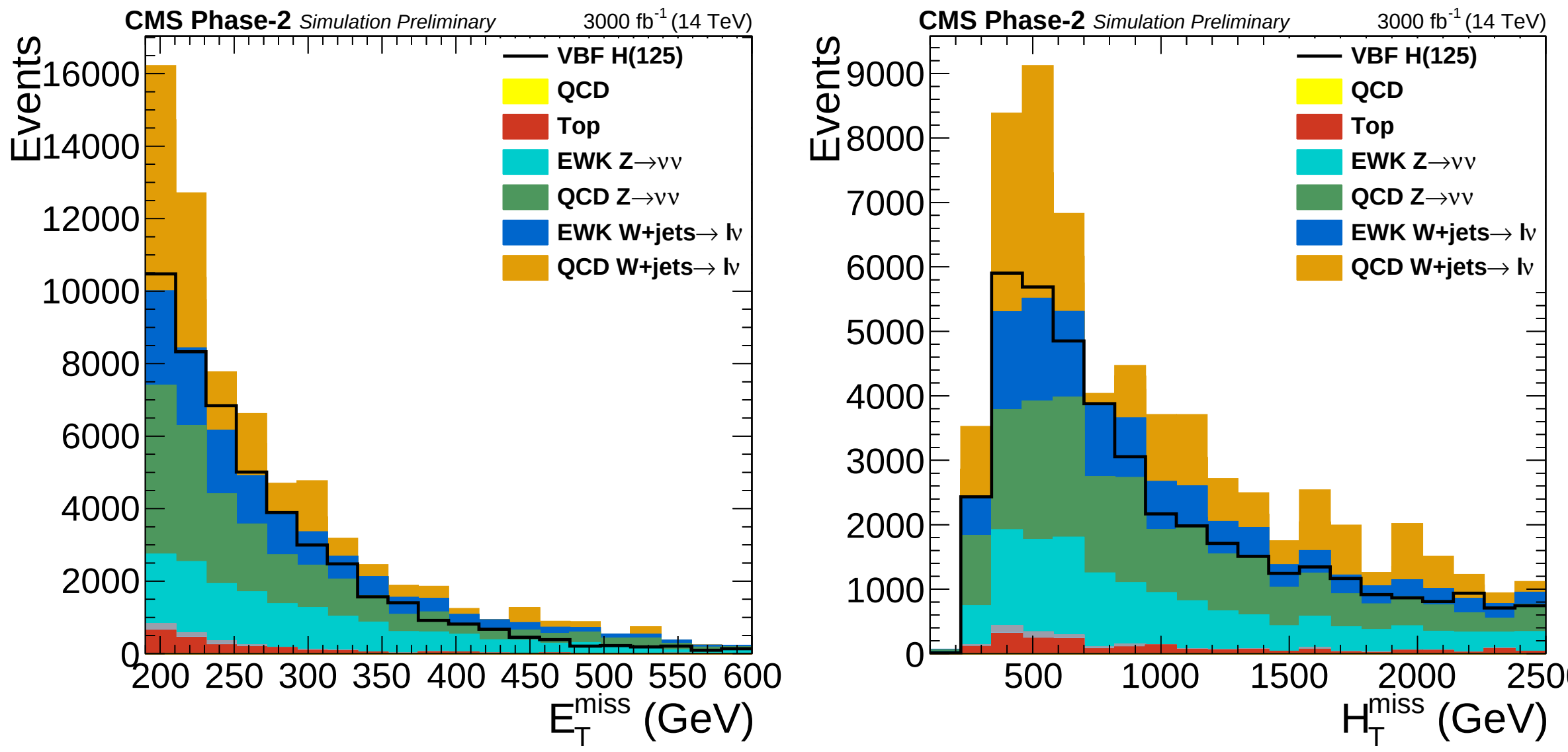

Figure 4: Distributions of $E_T^{miss}$ and $H_T^{miss}$in the signal region for the final selection, $M_{jj} >$ 2500 GeV and $E_T^{miss} > 190$ GeV.

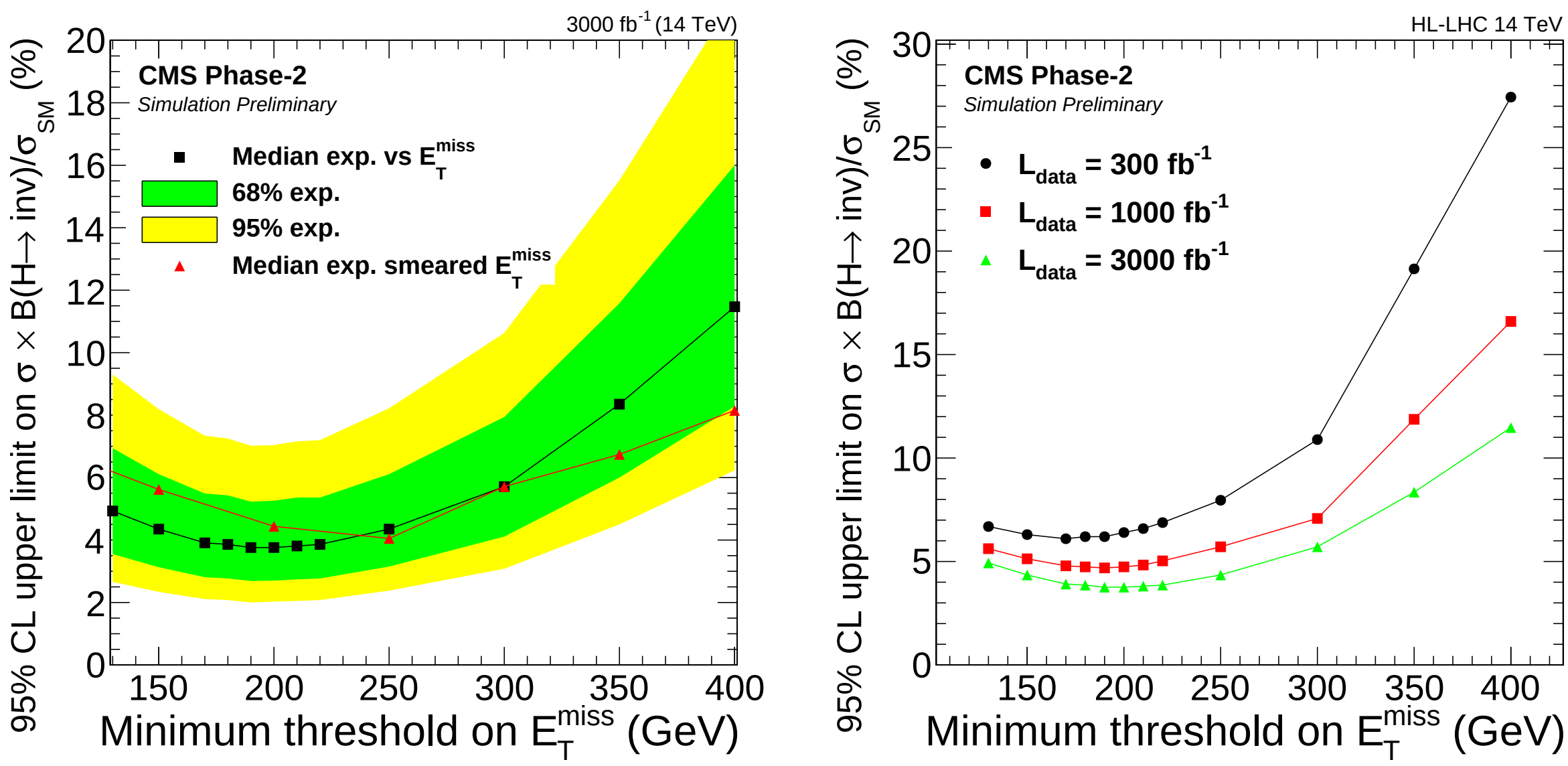

Figure 5: Left: 95% CL limits on $B(H \to \text{inv.})$ as a function of the minimum threshold on $E_T^{miss}$, for $M_{jj} > 2500$ GeV and an integrated luminosity of 3000 $fb^{-1}$. Right: 95% CL limits for scenarios with different integrated luminosities.

| Process | SR | $W \to e\nu$ CR | $W \to \mu\nu$ CR | $Z \to ee$ CR | $Z \to \mu\mu$ CR |
|---|---|---|---|---|---|
| VBFH | $47812 \pm 584$ | - | - | - | - |
| ggH | 972 | - | - | - | - |
| $Z \to \ell\ell$ (EWK) | $103 \pm 8$ | $398 \pm 16$ | $641 \pm 20$ | $1342 \pm 30$ | $1889 \pm 35$ |
| $Z \to \ell\ell$ (QCD) | $451 \pm 90$ | $944 \pm 126$ | $1048 \pm 116$ | $1347 \pm 118$ | $2297 \pm 158$ |
| $Z \to \nu\nu$ (EWK) | $15275 \pm 358$ | - | - | - | - |
| $Z \to \nu\nu$ (QCD) | $20968 \pm 599$ | - | - | - | - |
| $W \to e\nu$ (EWK) | $3358 \pm 62$ | $18986 \pm 146$ | $72 \pm 9$ | $33 \pm 6$ | - |
| $W \to \mu\nu$ (EWK) | $3426 \pm 62$ | $7 \pm 3$ | $29360 \pm 181$ | - | $17 \pm 4$ |
| $W \to \tau\nu$ (EWK) | $3595 \pm 64$ | $55 \pm 8$ | $87 \pm 10$ | - | - |
| $W \to e\nu$ (QCD) | $3994 \pm 999$ | $13376 \pm 1656$ | $170 \pm 168$ | - | - |
| $W \to \mu\nu$ (QCD) | $6891 \pm 1388$ | - | $23322 \pm 2096$ | - | - |
| $W \to \tau\nu$ (QCD) | $4308 \pm 938$ | - | - | - | - |
| Top | $2050 \pm 132$ | $2171 \pm 143$ | $3735 \pm 188$ | $107 \pm 36$ | $130 \pm 39$ |
| QCD | - | - | - | - | - |

Table 2: Number of events expected after the final selection, $M_{jj} > 2500$ GeV and $E_T^{miss} > 190$ GeV, with an integrated luminosity of 3000 fb$^{-1}$. The uncertainties are the statistical uncertainties from the Delphes samples.

# CMS Physics Analysis Summary

Contact: cms-phys-conveners-ftr@cern.ch

# Projection of searches for exotic Higgs boson decays to light pseudoscalars for the High-Luminosity LHC

The CMS Collaboration

## Abstract

This Physics Analysis Summary details the projections of two searches for exotic decays of the Higgs boson from LHC Run 2 to the High-Luminosity LHC. Decays to a pair of light pseudoscalar bosons are explored, in the final states of two $\tau$ leptons and two muons, or two $\tau$ leptons and two b quarks. The projections are based on analyses that use 35.9 $\mathrm{fb}^{-1}$ proton-proton collision data recorded at a center-of-mass energy of 13 TeV in 2016. Integrated luminosities of up to 3000 $\mathrm{fb}^{-1}$ are considered in the projections, with different scenarios for the extrapolation of the systematic uncertainties.

# 1 Introduction

After the discovery of the Higgs boson at the CERN LHC, extensive measurements have been performed to probe its consistency with the predictions of the standard model (SM). Indirect constraints on the branching fraction of the Higgs boson to beyond the SM (BSM) particles still leave a large room for potential exotic decays of the Higgs boson. Several direct searches for exotic Higgs boson decays have been performed with the CMS detector at 13 TeV center-of-mass energy in Run 2. With the increase of integrated luminosity expected at the High-Luminosity LHC (HL-LHC), the direct searches, which are currently mostly statistically dominated, are expected to play an increasingly important role in exploring the scalar sector.

This Physics Analysis Summary presents projections of two analyses searching for Higgs boson decays to two light pseudoscalars [1, 2]. Masses of the light pseudoscalars between 15 and 62.5 GeV are probed in the $2b2\tau$ and $2\mu2\tau$ final states. The projections assume that the same performance in the object reconstruction as in the Run 2 at the LHC can be achieved with the Phase-2 upgraded CMS detector at the HL-LHC. The event yields are scaled to integrated luminosities of up to 3000 $fb^{-1}$. The small difference in center-of-mass energy between the LHC and the HL-LHC, 13 and 14 TeV, respectively, is neglected.

The searches can be interpreted in the context of two-Higgs-doublet models extended with a scalar singlet (2HDM+S) [3]. Among the seven scalar and pseudoscalar particles predicted in 2HDM+S, one of the scalars (h) can be compatible with the discovered Higgs boson with a mass of 125 GeV and one of the pseudoscalars (a) can be light enough so that $h \to aa$ decays are allowed. One of the free parameters of these models is $\tan\beta$, defined as the ratio of the vacuum expectation values of the second and first doublets. Four types of 2HDM+S forbid flavor-changing neutral currents at tree level. In type I, all the fermions couple to the first doublet, and the branching fractions of the pseudoscalar a to fermions do not depend on $\tan\beta$. In type II, up-type quarks couple to the first doublet, whereas leptons and down-type quarks couple to the second doublet. This implies that pseudoscalar decays to b quarks and $\tau$ leptons are enhanced at high $\tan\beta$. The scalar sector of the next-to-minimal supersymmetric SM (NMSSM) is similar to that of 2HDM+S of type II. In type III, quarks couple to the first doublet, and leptons to the second one. This type is leptophilic at high $\tan\beta$. Finally, in type IV, leptons and up-type quarks couple to the first doublet, while down-type quarks couple to the second doublet.

The plans for the Phase-2 upgrade of the CMS detector are presented in Section 2, while the method used to project the results is described in Section 3. An overview of the two different searches using 35.9 $fb^{-1}$ of data collected in 2016 is given in Section 4, and the results of the projections to the HL-LHC conditions are detailed in Section 5.

# 2 Upgrade of the CMS detector

The CMS detector [4] will be substantially upgraded in order to fully exploit the physics potential offered by the increase in luminosity at the HL-LHC [5], and to cope with the demanding operational conditions at the HL-LHC [6–10]. The upgrade of the first level hardware trigger (L1) will allow an increase of L1 rate and latency to about 750 kHz and 12.5 $\mu$s, respectively, and the high-level software trigger (HLT) is expected to reduce the rate by about a factor of 100 to 7.5 kHz. The entire pixel and strip tracker detectors will be replaced to increase the granularity, reduce the material budget in the tracking volume, improve the radiation hardness, and extend the geometrical coverage and provide efficient tracking up to pseudorapidities of about $|\eta| = 4$. The muon system will be enhanced by upgrading the electronics of the existing cathode strip chambers (CSC), resistive plate chambers (RPC) and drift tubes (DT). New muon detectors based on improved RPC and gas electron multiplier (GEM) technologies will be installed to add redundancy, increase the geometrical coverage up to about $|\eta| = 2.8$, and improve the trigger and reconstruction performance in the forward region. The barrel electromagnetic calorimeter (ECAL) will feature the upgraded front-end electronics that will be able to exploit the information from single crystals at the L1 trigger level, to accommodate trigger latency and bandwidth requirements, and to provide 160 MHz sampling allowing high precision timing capability for photons. The hadronic calorimeter (HCAL), consisting in the barrel region of brass absorber plates and plastic scintillator layers, will be read out by silicon photomultipliers (SiPMs). The endcap electromagnetic and hadron calorimeters will be replaced with a new combined sampling calorimeter (HGCal) that will provide highly-segmented spatial information in both transverse and longitudinal directions, as well as high-precision timing information. Finally, the addition of a new timing detector for minimum ionizing particles (MTD) in both barrel and endcap region is envisaged to provide capability for 4-dimensional reconstruction of interaction vertices that will allow to significantly mitigate the CMS performance degradation related to high PU rates.

A detailed overview of the CMS detector upgrade program is presented in Ref. [6–10], while the expected performance of the reconstruction algorithms and pile-up mitigation with the CMS detector is summarised in Ref. [11].

# 3 Projection methodology

The extrapolations in this Physics Analysis Summary assume that the CMS experiment will have a similar level of detector and triggering performance during the HL-LHC operation as it provided during the LHC Run 2 period [6–10]. The results of extrapolations, hereafter named projections, are presented for different assumptions on the size of systematic uncertainties that will be achievable at the HL-LHC [11]:

- **"Run 2 systematic uncertainties" scenario:** This scenario assumes that the performance of the experimental methods at the HL-LHC will be unchanged with respect to the LHC Run 2 period, and there will be no significant improvement in the theoretical descriptions of relevant physics effects. All experimental and theoretical systematic uncertainties are assumed to be unchanged with respect to the ones in the reference Run 2 analyses, and kept constant with integrated luminosity.
- **"YR18 systematics uncertainties" scenario:** This scenario assumes that there will be further advances in both experimental methods and theoretical descriptions of relevant physics effects. Theoretical uncertainties are assumed to be reduced by a factor two with respect to the ones in the reference Run 2 analyses. For experimental systematic uncertainties, it is assumed that they will scale with the square root of the integrated luminosity until they reach a defined lower limit based on estimates of the achievable accuracy with the upgraded detector [11].

The "Run 2 systematic uncertainties" scenario allows for comparisons with current analyses, while the "YR18 systematics uncertainties" scenario is more realistic given the expected conditions for the HL-LHC.

In these scenarios, all the uncertainties related to the limited number of simulated events are neglected, under the assumption that sufficiently large simulation samples will be available by the time the HL-LHC becomes operational.

For all scenarios, the intrinsic statistical uncertainty in the measurement is reduced by a factor

$1/\sqrt{R_L}$, where $R_L$ is the projection integrated luminosity divided by that of the reference Run 2 analysis.

Table 1 summarises the Run 2 uncertainties for which a lower limit value is set in the "YR18 systematics uncertainties" scenario. Systematic uncertainties in the identification and isolation efficiencies for electrons and muons are expected to be reduced to around 0.5%. The hadronic $\tau$ lepton ($\tau_h$) performance is assumed to remain similar to the current level and therefore the associated uncertainties are not reduced in this scenario. The uncertainty in the overall jet energy scale (JES) is expected to reach around 1% precision for jets with $p_T > 30$ GeV, driven primarily by improvements for the absolute scale and jet flavour calibrations. The missing transverse momentum uncertainty is obtained by propagating the JES uncertainties in its computation, yielding a reduction by up to a half of the Run 2 uncertainty. For the identification of b-tagged jets, the uncertainty in the selection efficiency of b (c) quarks, and in misidentifying a light jet is expected to remain similar to the current level, with only the statistical component reducing with increasing integrated luminosity. The uncertainty in the integrated luminosity of the data sample could be reduced down to 1% by a better understanding of the calibration and fit models employed in its determination, and making use of the finer granularity and improved electronics of the upgraded detectors.

Table 1: The sources of systematic uncertainty for which limiting values are applied in the "YR18 systematics uncertainties" scenario. Systematic uncertainties of the reference Run 2 analyses are described in Refs. [1, 2].

| Source | Component | Run 2 unc. | Projection minimum unc. |
|---|---|---|---|
| Muon ID | | 1–2% | 0.5% |
| Electron ID | | 1–2% | 0.5% |
| Photon ID | | 0.5–2% | 0.25–1% |
| Hadronic $\tau$ ID | | 6% | Same as Run 2 |
| Jet energy scale | Absolute | 0.5% | 0.1–0.2% |
| | Relative | 0.1–3% | 0.1–0.5% |
| | Pileup | 0–2% | Same as Run 2 |
| | Method and sample | 0.5–5% | No limit |
| | Jet flavour | 1.5% | 0.75% |
| | Time stability | 0.2% | No limit |
| Jet energy resolution | | Varies with $p_T$ and $\eta$ | Half of Run 2 |
| $\vec{p}_T^{miss}$ scale | | Varies with analysis selection | Half of Run 2 |
| b-tagging | b-/c-jets (syst.) | Varies with $p_T$ and $\eta$ | Same as Run 2 |
| | light mis-tag (syst.) | Varies with $p_T$ and $\eta$ | Same as Run 2 |
| | b-/c-jets (stat.) | Varies with $p_T$ and $\eta$ | No limit |
| | light mis-tag (stat.) | Varies with $p_T$ and $\eta$ | No limit |
| Integrated luminosity | | 2.5% | 1% |
| Reducible bkg. ($h \to aa \to 2\mu 2\tau$) | | 20–40% | 4–8% |

## 4 Overview of the Run-2 analyses

In the $h \to aa \to 2\tau 2b$ search [1], the $\tau\tau$ pair is reconstructed as $e\mu$, $\mu\tau_h$, or $e\tau_h$, depending on the decay modes of the $\tau$ leptons. The symbol $\tau_h$ denotes a $\tau$ lepton decaying hadronically. Since the b jets originating from the pseudoscalar boson are typically soft, only one reconstructed b jet with $p_T > 20$ GeV is required. An improved signal sensitivity is obtained by dividing the events in four different categories depending on the visible invariant mass of the b jet and the $\tau$ candidates, denoted $m^{vis}_{b\tau\tau}$. The thresholds that define the categories depend on the final state. The categories with low $m^{vis}_{b\tau\tau}$ are enriched in signal events, while the categories with large $m^{vis}_{b\tau\tau}$ help to constrain the backgrounds. The results are extracted with a maximum likelihood fit of the visible $\tau\tau$ mass spectrum. The dominant backgrounds at low $m_{\mathrm{a}}$ are $\mathrm{t\bar{t}}$ production as well as events with jets misidentified as $\tau$ candidates, whereas the Drell–Yan background starts to contribute for $m_{\mathrm{a}} > 45\,\mathrm{GeV}$ values. This analysis is only sensitive to pseudoscalar masses above 15 GeV. The sensitivity of the analysis mostly comes from the low $m_{\mathrm{b}\tau\tau}^{\mathrm{vis}}$ category, which is statistically limited, and the statistical uncertainty strongly dominates the results.

In the $\mathrm{h} \to \mathrm{aa} \to 2\mu 2\tau$ search [2], the Higgs boson is reconstructed via its decay to a pair of pseudoscalar bosons in the final state with two $\tau$ leptons and two muons. Pseudoscalar masses between 15 and 62.5 GeV are investigated; in this mass range the decay products from the pseudoscalars are not collimated. Several $\tau\tau$ pair possibilities are considered: $\mathrm{e}\mu$, $\mathrm{e}\tau_{\mathrm{h}}$, $\mu\tau_{\mathrm{h}}$, and $\tau_{\mathrm{h}}\tau_{\mathrm{h}}$. In the case where there are 3 muons, the one with highest $p_{\mathrm{T}}$ is paired with the opposite-sign muon with the highest $p_{\mathrm{T}}$ among the other two, while the last muon is considered as originating from a $\tau$ lepton decay. To reduce the backgrounds from ZZ, Z+jets, and WZ+jets production, the invariant mass of the muon pair is required to be above the visible invariant mass of the $\tau\tau$ pair, and the visible invariant mass of the four objects is required to be less than 110–130 GeV depending on the final state. The limits are extracted with an unbinned maximum likelihood fit of the dimuon mass spectrum. The backgrounds are characterized by a rather flat dimuon mass spectrum, while the signal $\mathrm{h} \to \mathrm{aa} \to 2\mu 2\tau$ forms a narrow peak in the dimuon mass spectrum. The number of expected background events below the signal peak is almost zero, especially at low dimuon mass, and the analysis is strongly statistically dominated.

# 5 Results

For the $\mathrm{h} \to \mathrm{aa} \to 2\tau 2\mathrm{b}$ analysis, upper limits at 95% CL on $(\sigma(\mathrm{h})/\sigma_{\mathrm{SM}})\mathcal{B}(\mathrm{h} \to \mathrm{aa} \to 2\tau 2\mathrm{b})$ are shown in Fig. 1 for different integrated luminosities and systematic uncertainty scenarios. In this expression, $\sigma_{\mathrm{SM}}$ denotes the SM production cross section of the Higgs boson, whereas $\sigma(\mathrm{h})$ is the h production cross section. The limits improve proportionally to the square root of the integrated luminosity, as the analysis is statistically limited. For an integrated luminosity of $3000\,\mathrm{fb}^{-1}$, the difference between the limits in the systematic scenarios of Run 2 and YR18 is of the order of 5%, and the limits become another 5% better if all systematic uncertainties are neglected.

The limits of the $\mathrm{h} \to \mathrm{aa}$ analyses can be converted to limits on $\mathcal{B}(\mathrm{h} \to \mathrm{aa})$ in two-Higgs-doublet models extended with a scalar singlet (2HDM+S) [3], for a given type of model, $m_{\mathrm{a}}$, and $\tan\beta$. The limits in the four types of 2HDM+S are shown for $\mathrm{h} \to \mathrm{aa} \to 2\tau 2\mathrm{b}$ in Fig. 2, assuming $3000\,\mathrm{fb}^{-1}$ of data with YR18 systematic uncertainties. The color scale indicates the upper limits on $(\sigma(\mathrm{h})/\sigma_{\mathrm{SM}})\mathcal{B}(\mathrm{h} \to \mathrm{aa})$ that can be set assuming some values for $m_{\mathrm{a}}$ and $\tan\beta$.

The corresponding limits for the $\mathrm{h} \to \mathrm{aa} \to 2\mu 2\tau$ search are shown in Fig. 3. At low $m_{\mathrm{a}}$, the analysis is almost background-free, while, towards higher $m_{\mathrm{a}}$, backgrounds play an increasing role. For this reason the limit tends to scale inverse-proportionally to the luminosity at low $m_{\mathrm{a}}$ and to the square root of the luminosity at higher $m_{\mathrm{a}}$. The analysis is statistically limited, even with $3000\,\mathrm{fb}^{-1}$ of data. The difference between the Run 2 and YR18 systematic uncertainties in terms of upper limits is up to 5%, and is the largest at high $m_{\mathrm{a}}$. The limits in the four types of 2HDM+S are shown in Fig 4 for the $\mathrm{h} \to \mathrm{aa} \to 2\mu 2\tau$ analysis, assuming $3000\,\mathrm{fb}^{-1}$ of data in the "YR18 systematics uncertainties" scenario.

Since the branching fractions $\mathcal{B}(\mathrm{a} \to \mu\mu)$, $\mathcal{B}(\mathrm{a} \to \tau\tau)$, and $\mathcal{B}(\mathrm{a} \to \mathrm{bb})$ depend on the type of 2HDM+S and $\tan\beta$, the two analyses cover non overlapping parts of the parameter space.

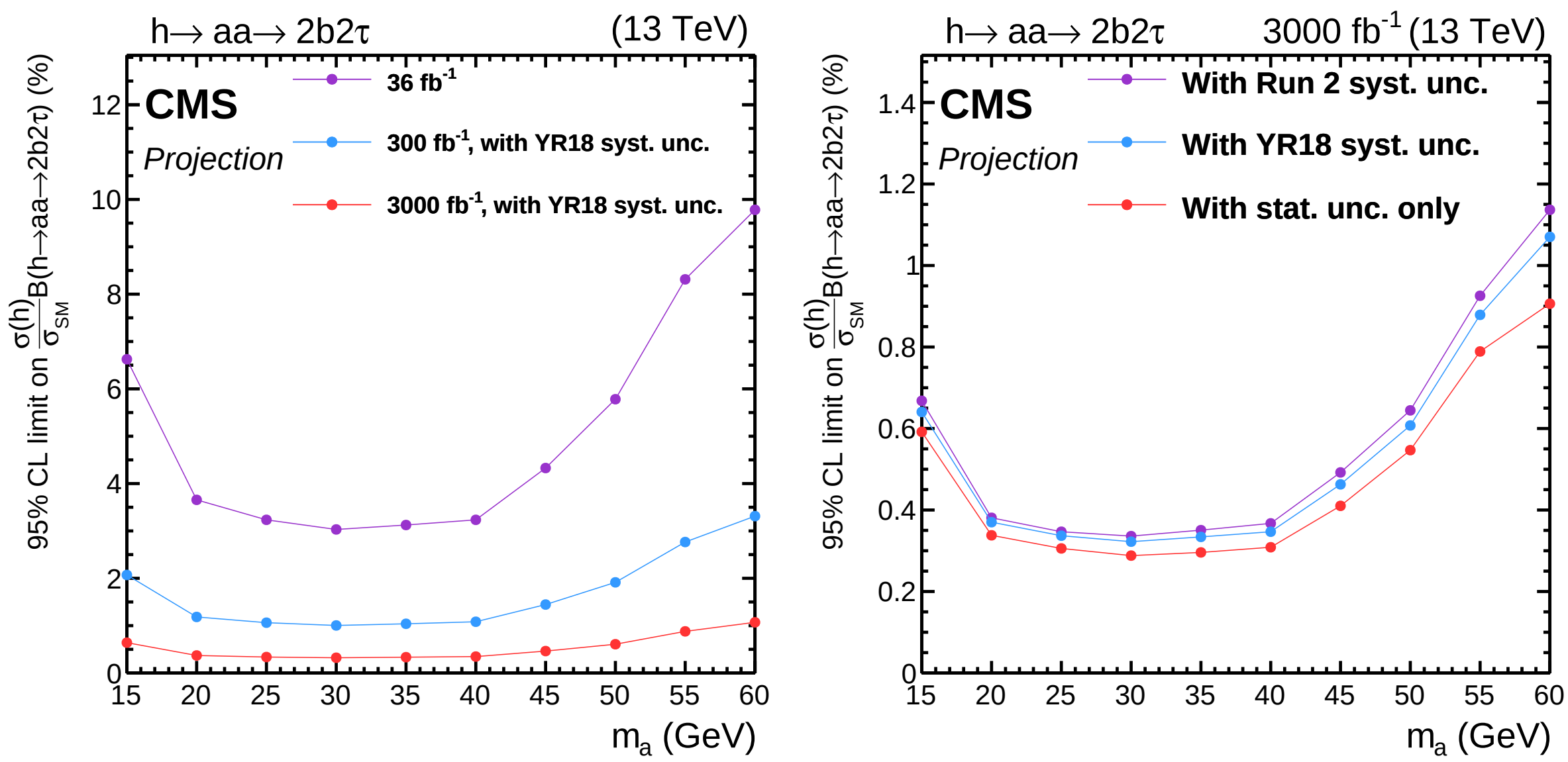

Figure 1: Left: Projected expected limits on $(\sigma(\mathrm{h})/\sigma_{\mathrm{SM}})$ times the branching fraction for $\mathrm{h} \to \mathrm{aa} \to 2\mathrm{b}2\tau$, for 36, 300, and 3000 $\mathrm{fb}^{-1}$. Right: Projected expected limits $(\sigma(\mathrm{h})/\sigma_{\mathrm{SM}})\mathcal{B}(\mathrm{h} \to \mathrm{aa} \to 2\tau 2\mathrm{b})$, comparing different scenarios for systematic uncertainties for an integrated luminosity of 3000 $\mathrm{fb}^{-1}$.

# 6 Summary

Recent searches for exotic decays of the Higgs boson performed with 35.9 $\mathrm{fb}^{-1}$ of data collected at 13 TeV [1, 2] have been projected to integrated luminosities of up to 3000 $\mathrm{fb}^{-1}$, achievable at the High-Luminosity LHC. They target decays of the Higgs boson to a pair of light pseudoscalars in the final states with two $\tau$ leptons and two muons, or two b quarks and two muons. The integrated luminosity of 3000 $\mathrm{fb}^{-1}$ will improve the sensitivity by about an order of magnitude for the search in the $2\mathrm{b}2\tau$ final state. The improvement is larger in the $2\mu 2\tau$ final state and scales almost linearly with the integrated luminosity for pseudoscalar masses close to 15 GeV.

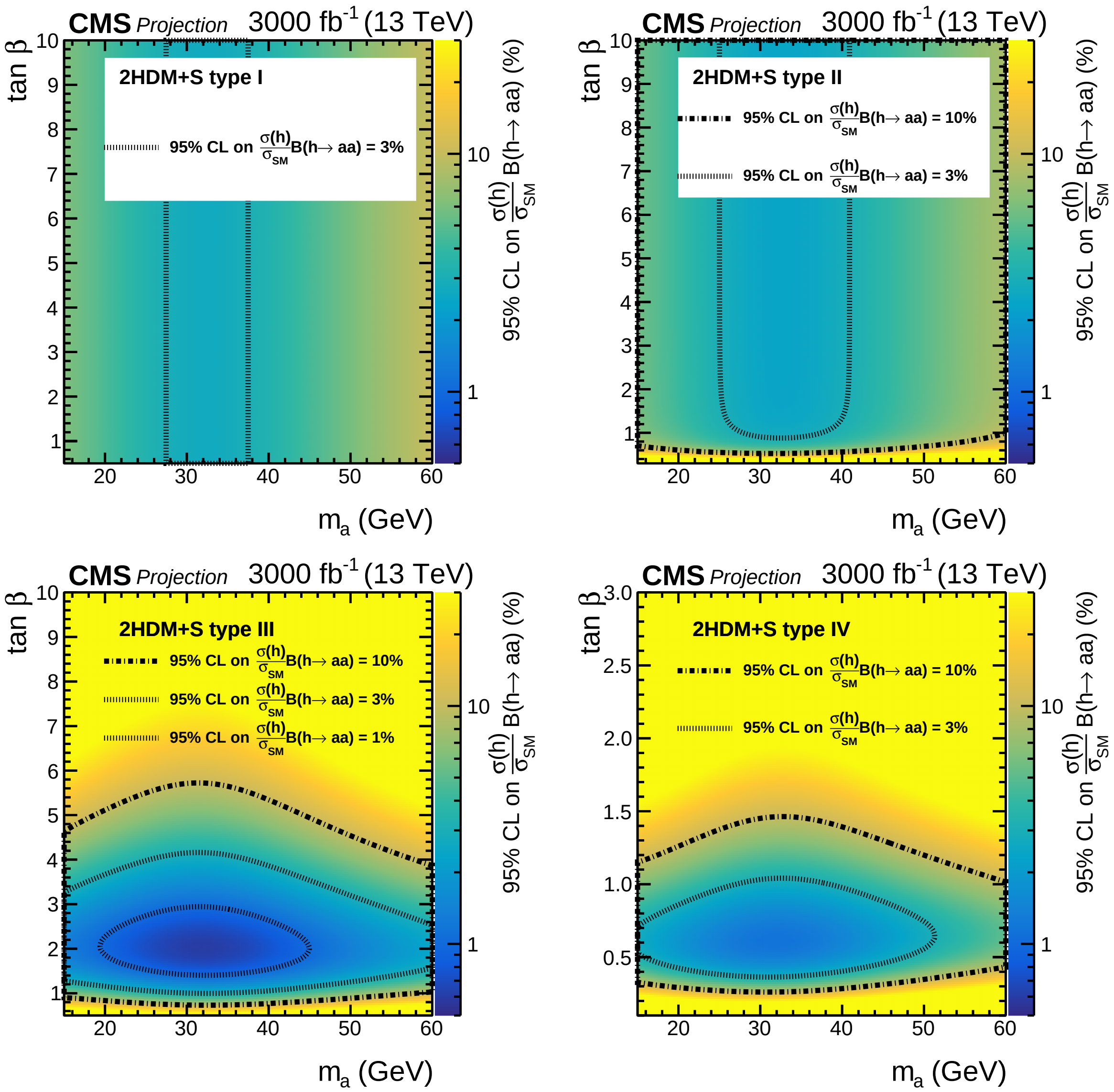

Figure 2: Expected upper limits on $(\sigma(\mathrm{h})/\sigma_{\mathrm{SM}})\mathcal{B}(\mathrm{h} \to \mathrm{aa})$ for 3000 fb$^{-1}$ of data with YR18 systematic uncertainties for the 2b2$\tau$ final state in 2HDM+S type-1 (top left), type-2 (top right), type-3 (bottom left), and type-4 (bottom right).

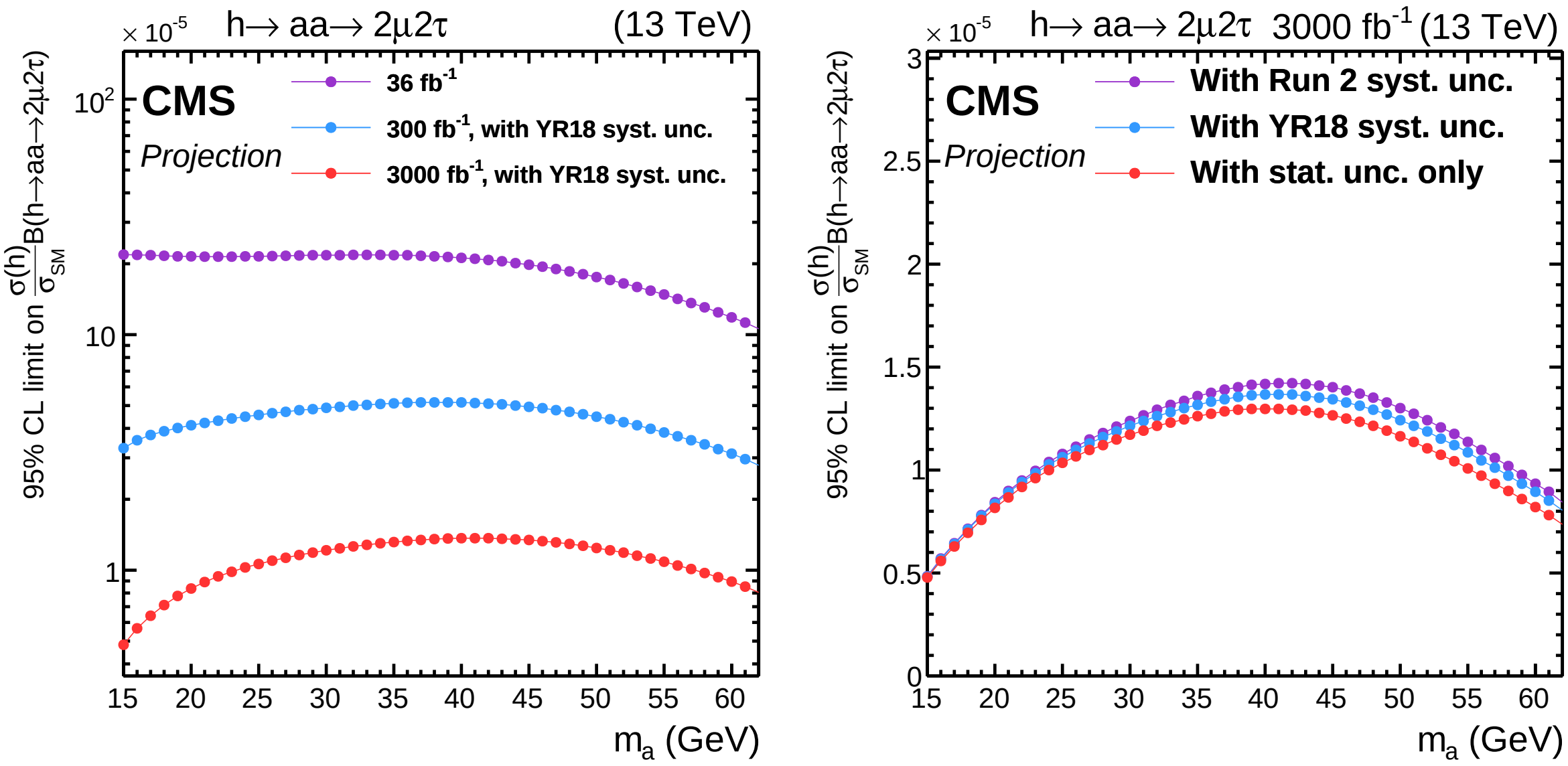

Figure 3: Left: Projected expected limits on $(\sigma(\mathrm{h})/\sigma_{\mathrm{SM}})\mathcal{B}(\mathrm{h} \to \mathrm{aa} \to 2\mu 2\tau)$, for 36, 300, and 3000 $\mathrm{fb}^{-1}$. Right: Projected expected limits on $(\sigma(\mathrm{h})/\sigma_{\mathrm{SM}})\mathcal{B}(\mathrm{h} \to \mathrm{aa} \to 2\mu 2\tau)$, comparing different scenarios for systematic uncertainties for an integrated luminosity of 3000 $\mathrm{fb}^{-1}$.

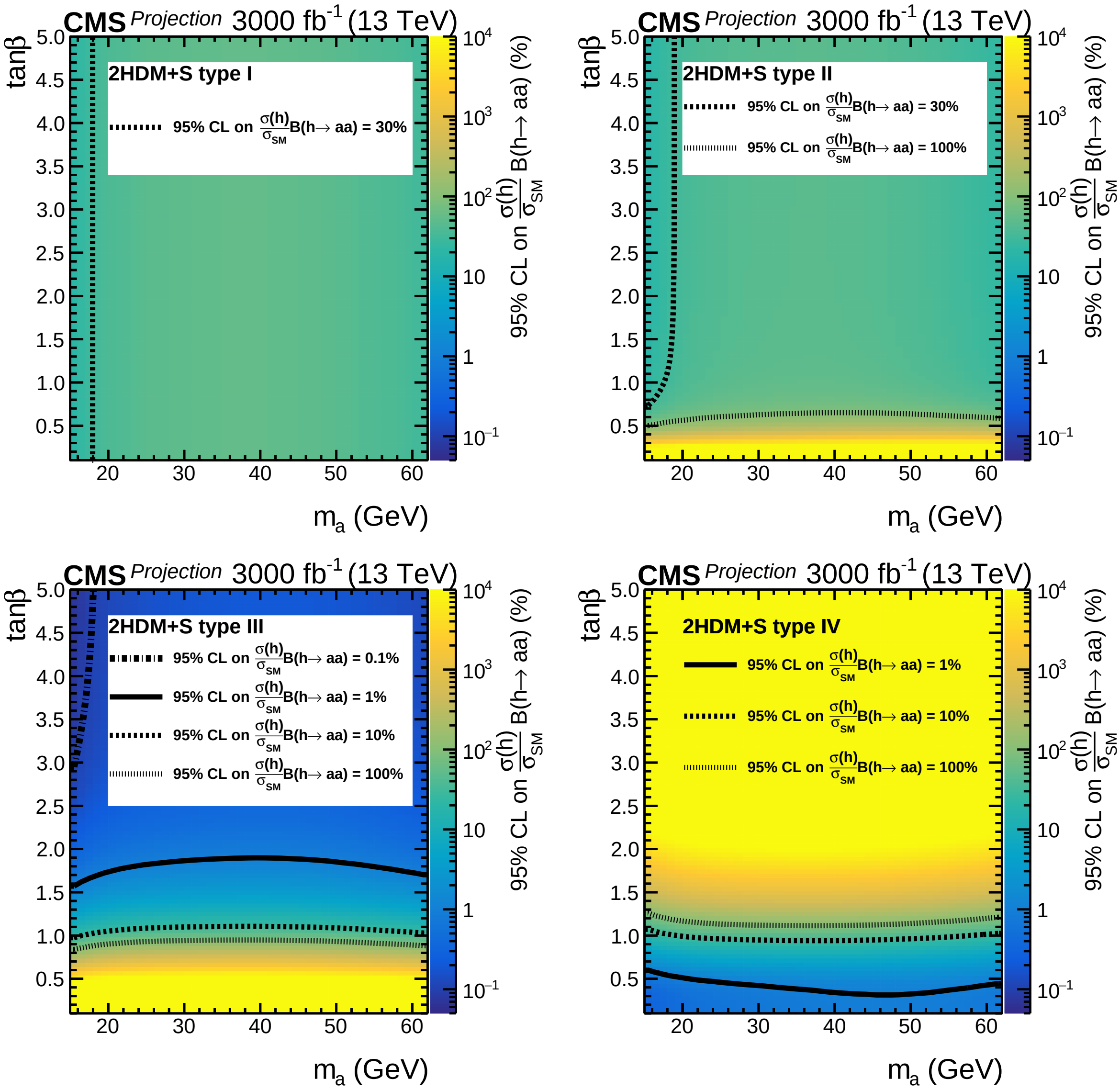

Figure 4: Expected upper limits on $(\sigma(\mathrm{h})/\sigma_{\mathrm{SM}})\mathcal{B}(\mathrm{h} \to \mathrm{aa})$ for $3000\,\mathrm{fb}^{-1}$ of data with YR18 systematic uncertainties for the $2\mu 2\tau$ final state in 2HDM+S type-1 (top left), type-2 (top right), type-3 (bottom left), and type-4 (bottom right).

# CMS Physics Analysis Summary

Contact: cms-phys-conveners-ftr@cern.ch

# Search for a new scalar resonance decaying to a pair of Z bosons at the High-Luminosity LHC

The CMS Collaboration

## Abstract

For a heavy resonance decaying into a pair of Z bosons, a projection of current CMS searches to the HL-LHC is presented. The study considers pp collisions for an integrated luminosity of 3000 $fb^{-1}$ and takes into account the Phase-2 upgrade of the CMS detector. The final state with two leptons and two quarks is used to search for heavy resonances in the mass range from 550 GeV to 3 TeV. The scalar particle X is assumed to have a decay width much narrower than the detector resolution. Upper limits on the cross sections for models predicting the production of this scalar resonance through gluon fusion and electroweak mechanisms are presented.

# 1 Introduction

The standard model (SM) of particle physics postulates the existence of a single Higgs boson as the manifestation of a scalar field responsible for electroweak (EW) symmetry breaking [1–7]. The ATLAS and CMS Collaborations have discovered a boson with a mass of 125 GeV [8–10] and with properties consistent with those expected for the SM Higgs boson [11–15].To-date there is no experimental evidence for the particles beyond the standard model. Nonetheless, searches for BSM physics are motivated by a number of phenomena such as the presence of dark matter or baryon asymmetry in the universe that are not explained by the SM. The BSM models that attempt to address these questions include two-Higgs-doublet models (2HDM) [16] as predicted by supersymmetry or other models predicting an extended Higgs-like EW singlet [17]. CMS and ATLAS collaborations have performed searches for a heavy scalar partner of the SM Higgs boson decaying into a pair of Z bosons [18, 19]. The ZZ decay has a sizable branching fraction for an SM-like Higgs boson of mass larger than the Z boson pair production threshold, $2m_{\mathrm{Z}}$, and is one of the main discovery channels for masses less than $2m_{\mathrm{Z}}$ [8–10]. The search for a new scalar boson X is performed over a range of masses from 550 GeV up to 3 TeV.

The CMS search for a heavy scalar partner of the SM Higgs boson using 35.9 $\mathrm{fb}^{-1}$ of pp collision data [19] will be referred to as Run-2 analysis throughout the article. In Run-2 analysis, the search for a scalar resonance X decaying to ZZ is performed over the mass range $130\,\mathrm{GeV} < m_{\mathrm{X}} < 3\,\mathrm{TeV}$, where three final states based on leptonic or hadronic decays of Z boson, $\mathrm{X} \to \mathrm{ZZ} \to 4\ell$, $2\ell 2\mathrm{q}$, and $2\ell 2\nu$ are combined. Because of the different resolutions, efficiencies, and branching fractions, each final state contributes differently depending on the signal mass hypothesis. The most sensitive final state for the mass range of 130–500 GeV is $4\ell$ due to its best mass resolution, whereas, for the intermediate region of 500–700 GeV, $2\ell 2\nu$ is most sensitive. For masses above 700 GeV $2\ell 2\mathrm{q}$ provides the best sensitivity. In this paper, we are particularly interested in the sensitivity in the high mass region, thus only $2\ell 2\mathrm{q}$ is used.

In the $2\ell 2\mathrm{q}$ final state, events are selected by combining leptonically and hadronically decaying Z candidates. The lepton pairs (electron or muon) of opposite sign and same flavor with invariant mass between 60 and 120 GeV are constructed. Hadronically decaying Z boson candidates ($\mathrm{Z}_{\mathrm{had}}$) are reconstructed using two distinct techniques, which are referred to as "resolved" and "merged". In the resolved case, the two quarks from the Z boson decay form two distinguishable narrow jets, while in the merged case a single wide jet with a large $p_{\mathrm{T}}$ is taken as a $\mathrm{Z}_{\mathrm{had}}$.

An arbitration procedure is used to rank multiple $\mathrm{Z}_{\mathrm{had}}$ candidates reconstructed in a single event: merged candidates have precedence over resolved candidates if they have $p_{\mathrm{T}} > 300\,\mathrm{GeV}$ and the accompanying leptonically decaying Z candidate has $p_{\mathrm{T}}(\mathrm{L}_{\mathrm{L}}) > 200\,\mathrm{GeV}$; resolved candidates have precedence otherwise. Within each selection category the candidate with the largest $p_{\mathrm{T}}$ has priority over the others.

The two dominant production mechanisms of a scalar boson are gluon fusion (ggF) and EW production, the latter dominated by vector boson fusion (VBF) with a small contribution of production in association with an EW boson ZH or WH (VH). We define the parameter $f_{VBF}$ as the fraction of the EW production cross section with respect to the total cross section. The results are given for two scenarios: $f_{VBF}$ floated, and $f_{VBF} = 1$. In the expected result, the two scenarios correspond to ggF and VBF production modes, respectively. To increase the sensitivity to the different production modes, events are categorized into VBF and inclusive types. Furthermore, since a large fraction of signal events is enriched with b quark jets due to the presence of $\mathrm{Z} \to \mathrm{b\bar{b}}$ decays, a dedicated category is defined. The definitions are as follows:

- **VBF-tagged** requires two additional and forward jets besides those from the hadronic decays of the Z boson candidate; a mass dependent criterion based on a dedicated discriminant defined for this category is applied;
- b **tagged** consists of the remaining events with two b tagged jets (in the resolved case) or two b tagged subjets from the hadronic Z boson candidate;
- **Untagged** consists of the remaining events.

The invariant mass of ZZ and a dedicated discriminant separating signal and background distributions are compared between observation and expected background to set limits on the production cross section.

Further details of the Run-2 analysis, including simulation samples, background estimation methods, systematic uncertainties, and different interpretations are described in Ref [19]. Only details of direct relevance to the projection of the Run-2 analysis are documented in the following.

A projection of this analysis is carried out by scaling all the signal and background processes to an integrated luminosity of 3000 $fb^{-1}$, expected to be collected at the high-luminosity LHC (HL-LHC). The projection does not account for the small cross section change due to the expected increase of the center of mass energy from Run-2 (13) TeV to HL-LHC (14 TeV). The upgrade and the expected performance of the CMS detector are described in the following section, and in detail in the Technical Proposal and the Technical Design Reports for the Phase-2 Upgrade of the CMS Detector [20–24]. Special care is taken to use realistic assumptions for the development of systematic uncertainties at high luminosity. The results are presented in terms of cross section limits on a heavy resonance decaying to a Z boson pair.

## 2 Upgraded CMS detector

The CMS detector [25] will be substantially upgraded in order to fully exploit the physics potential offered by the increase in luminosity at the HL-LHC [26], and to cope with the demanding operational conditions at the HL-LHC [20–24]. The upgrade of the first level hardware trigger (L1) will allow for an increase of L1 rate and latency to about 750 kHz and 12.5 $\mu$s, respectively, while in case of the high-level software trigger (HLT) its upgrade will allow the HLT rate to be increased to 7.5 kHz. The entire pixel and strip tracker detectors will be replaced to increase the granularity, reduce the material budget in the tracking volume, improve the radiation hardness, and extend the geometrical coverage and provide efficient tracking up to pseudorapidities of about $|\eta| = 4$. The muon system will be enhanced by upgrading the electronics of the existing cathode strip chambers (CSC), resistive plate chambers (RPC) and drift tubes (DT). New muon detectors based on improved RPC and gas electron multiplier (GEM) technologies will be installed to add redundancy, increase the geometrical coverage up to about $|\eta| = 2.8$, and improve the trigger and reconstruction performance in the forward region. The barrel electromagnetic calorimeter (ECAL) will feature the upgraded front-end electronics that will be able to exploit the information from single crystals at the L1 trigger level, to accommodate trigger latency and bandwidth requirements, and to provide an increased sampling rate of 160 MHz. The hadronic calorimeter (HCAL) consists in the barrel region of brass absorber plates and plastic scintillator layers, read out by hybrid photodiodes (HPDs), which will be replaced with silicon photomultipliers (SiPMs). The endcap electromagnetic and hadron calorimeters will be replaced with a new combined sampling calorimeter (HGCal) that will provide highly-segmented spatial information in both transverse and longitudinal directions, as well as high-precision timing information. Finally, the addition of a new timing detector for minimum ionizing particles (MTD) in both barrel and endocap region is envisaged to provide capability for 4-dimensional reconstruction of interaction vertices that will allow to significantly offset the CMS performance degradation due to high PU rates.

A detailed overview of the CMS detector upgrade program is presented in Ref. [20–24], while the expected performance of the reconstruction algorithms and pile-up mitigation with the CMS detector is summarised in Ref. [27].

# 3 Extrapolation procedure

This projection assumes that the CMS experiment will have a similar level of detector and triggering performance during the HL-LHC operation as it provided during the LHC Run 2 period [20–24]. The results of projection are presented for different assumptions based on the size of systematic uncertainties that is estimated for HL-LHC:

- **"Run 2 systematic uncertainties" scenario:** This scenario assumes that performance of the experimental methods at the HL-LHC will be unchanged with respect to the LHC Run 2 period, and there will be no significant improvement in the quantitative theoretical understanding of relevant physics effects. All experimental and theoretical systematic uncertainties are assumed to be unchanged with respect to the ones in the reference Run 2 analysis, and kept constant with integrated luminosity.
- **"YR18 systematics uncertainties" scenario:** This scenario assumes that there will be further advances in both experimental methods and theoretical descriptions of relevant physics effects. Theoretical uncertainties are assumed to be reduced by a factor two with respect to the ones in the reference Run 2 analysis. For experimental systematic uncertainties, it is assumed that those will be reduced by the square root of the integrated luminosity until they reach a defined lower limit based on estimates of the achievable accuracy with the upgraded detector [27].

In these scenarios, the statistical error from simulation is assumed to be negligible, under the assumption that sufficiently large simulation samples will be available by the time the HL-LHC becomes operational. For all scenarios, the intrinsic statistical uncertainty in the measurement is expected to scale by $1/\sqrt{\mathrm{L}}$, where L is the projection integrated luminosity divided by that of the reference Run 2 analysis.

Table 1 summarises the Run 2 uncertainties as well as the "YR18 systematics uncertainties" scenario. Systematic uncertainties in the identification and isolation efficiencies for electrons and muons are expected to be reduced to around 0.5%. The uncertainty in the overall jet energy scale (JES) is expected to reach around 1% precision for jets with $p_{\mathrm{T}} > 30\,\mathrm{GeV}$, driven primarily by improvements for the absolute scale and jet flavour calibrations. For the identification of b-tagged jets, the uncertainty in the selection efficiency of b(c) quarks, and in misidentifying a light jet is expected to reach around 1% precision. The uncertainty in the integrated luminosity of the data sample could be reduced down to 1% by a better understanding of the calibration and fit models employed in its determination, and making use of the finer granularity and improved electronics of the upgraded detectors.

Among other systematic uncertainties, the theoretical uncertainty from higher order QCD corrections on the ggZZ background and the signal is the most dominant for the ggF search. It is expected that theoretical description of these processes will be improved, thus the uncertainty is scaled by 0.5. The next important ones are the shape and yield uncertainties of the Z+jets background. They are determined from a data control region and are scaled with $1/\sqrt{L}$ in YR18 scenario. It is expected that at HL-LHC, the Z+jets background will have huge statistics, and the understanding of it will be at the percent level. Another important uncertainty is Z+jets fake rates. In the Run-2 analysis, they are derived from LO MC samples, and differences with repect to the NLO samples with limited statistics are assigned as systematic uncertainty. It is expected that larger statistics sample will be produced in the future or higher order description will be available to reduce this systematics uncertainty, thus it is scaled by 0.5 in YR18 scenario.

Table 1: The sources of systematic uncertainty where minimum values are applied in "YR18 systematics uncertainties" scenario. Systematic uncertainties of the reference Run 2 analysis are described in Ref. [19].

| Source | Run 2 uncertainty | Projection minimum uncertainty |
|---|---|---|
| Lepton selection efficiency | 4–8% | 0.5% |
| Lepton ID | 1–10% | 0.5% |
| Jet energy scale, resolution | 1–10% | 1% |
| b-tagging | 5–7% | 1% |
| Integrated luminosity | 2.5% | 1% |

## 4 Results

The $m_{ZZ}$ distribution of the events expected at 3000 $fb^{-1}$ is shown in Figure 1. Figure 2 shows upper limits at the 95% confidence level (CL) on the pp $\to$ X $\to$ ZZ cross section $\sigma_X \times \mathcal{B}_{X\to ZZ}$ as a function of $m_X$ for a narrow resonance whose $\Gamma_X$ is much smaller than the experimental resolution.

We follow the modified frequentist prescription [28–30] (CLs method), and an asymptotic approach with the profile likelihood ratio as the test statistic is used to estimate the upper limits at 95% confidence level. Systematic uncertainties are treated as nuisance parameters and profiled using log-normal priors.

The analysis uses $2\ell 2q$ final state to look for a scalar Higgs in the mass range of 550– 3000 GeV. It is the most sensitive channel above mass 700 GeV, while $2\ell 2\nu$ final state is the most sensitive for the intermediate region of 500–700 GeV. The exclusion limit for the cross section of the scalar decaying to a pair of Z bosons is 0.7–5 fb for the VBF production mode and 0.8–9 fb for the ggF production mode. This represents a factor of 10 improvement with respect to the results obtained using Run2 data. The differences between the two scenarios are minor and mostly present in the low mass region. It is because the search will still be limited by statistical uncertainties. The systematic uncertainties in this search have mild effects. If no $1/\sqrt{L}$ scaling is applied, the difference in the limit is 10% at low mass and almost none in the high mass region. The results for wide resonances are not given in this note for simplicity. The Run-2 result has shown that the excluded cross section for a 30% width resonance will be 40% higher at 1 TeV, compared to a narrow resonance assumption.

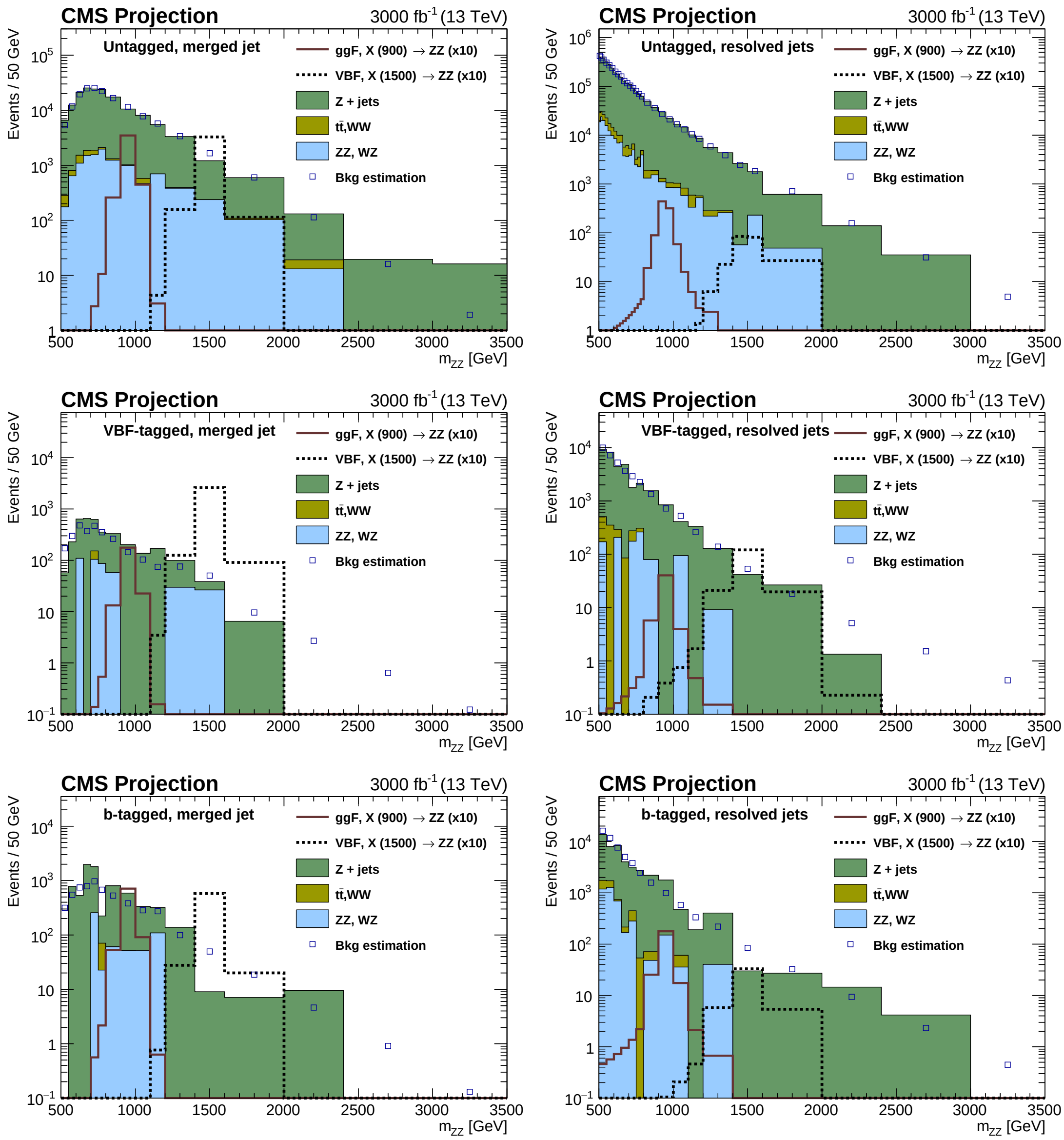

Figure 1: Distributions of the invariant mass $m_{ZZ}$ in the signal region expected at 3000 fb$^{-1}$, for the merged (left) and resolved (right) case in the different categories. The stacked histograms are the expected backgrounds from simulation. The blue points refer to the sum of background estimates derived from control samples. Examples of a 900 GeV ggF signal and a 1500 GeV VBF signal are given. The cross section corresponds to 10 times the excluded limit.

**6**

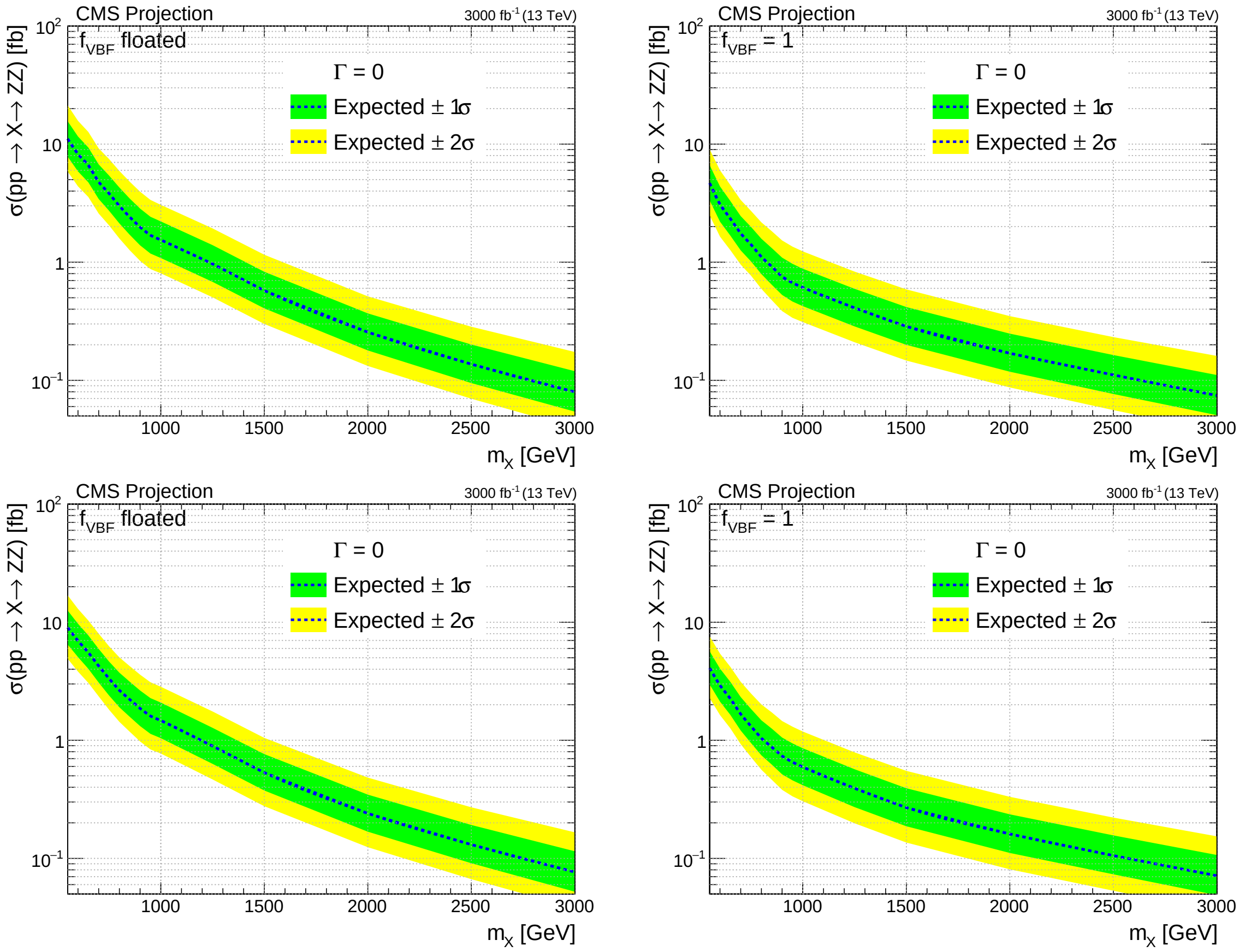

Figure 2: Expected upper limits at the 95% CL on the pp $\rightarrow$ X $\rightarrow$ ZZ cross section as a function of $m_X$, with $f_{\mathrm{VBF}}$ as a free parameter (left) and fixed to 1 (right). Scenario 1 (top) and scenario 2 (bottom) are shown. The scalar particle X is assumed to have a narrower decay width than the detector resolution. The results are shown for the $2\ell 2$q channel.

**8**

# CMS Physics Analysis Summary

Contact: cms-phys-conveners-ftr@cern.ch 

## First Level Track Jet Trigger for Displaced Jets at High Luminosity LHC

The CMS Collaboration

### Abstract

The CMS detector for the planned high luminosity LHC run will have a new first level (L1) hardware track trigger. The impact of the L1 track trigger is explored based on the potential increase of CMS sensitivity to signals beyond the standard model in final states with multiple jets and low total transverse energy. In particular, there is currently an blind spot for lifetimes of order 1 cm in searches for new long-lived scalars $\phi$ in Higgs decays, i.e. $\mathrm{H}(125) \to \phi\phi \to 4j$. It is found that a plausible extension of the L1 track trigger to tracks with an impact parameter of a few centimeters results in dramatic gains in the trigger efficiency. The gains are even larger for additional heavy SM-like Higgs bosons with the same decay.

# 1 Introduction

The high luminosity LHC program offers many exciting opportunities to search for rare processes. It is expected that the LHC will accumulate $3\,\mathrm{ab}^{-1}$ of proton-proton (pp) collisions at 14 TeV. The CMS detector will undergo major upgrades to all subsystems, including the tracker [1], the barrel [2] and endcap [3] calorimeters, the muon system [4], and the trigger [5].

The bandwidth limitations of the first level (L1) trigger are one of the main problems facing current searches for exotic Higgs boson decays, as well as many other signals beyond the standard model (BSM). The Higgs boson is assumed to be SM-like within this document with the same production cross-sections as the observed Higgs boson, but including rare unobserved decays with $\mathcal{B} = 10^{-5}$. The process where the Higgs boson decays to two new light scalars that in turn decay to jets, $\mathrm{H}(125) \to \phi\phi \to 4j$, is an important example. If the scalar $\phi$ has a macroscopic decay length, the offline analysis has no background from SM processes, but the majority of the signal events do not get recorded because they fail to be selected by the L1 trigger. The main obstacle is the high rate for low transverse momentum jets, which is made worse by additional extraneous pp collisions in the high luminosity environment.

In this note, we investigate the capabilities of L1 track finding [1] to increase the L1 trigger efficiency for such signals. We focus on small or moderate decay lengths of the new particles, 1–50 mm, and assume, as is demonstrated by many analyses [6–8], that the offline selection can remove all SM backgrounds with only a moderate loss of efficiency.

The investigation has two major thrusts. First, we propose a jet clustering algorithm that uses the L1 tracks found with a primary vertex constraint. Second, we consider the extension of the L1 track finder to off-pointing tracks, and develop a jet lifetime tag for tracks with $|\eta| < 1.0$. Off-pointing tracks do not point back to the primary collision point, but instead have a "kink" arising from the decay vertex of a long-lived decay. The kink is usually quantified in terms of the transverse impact parameter, $d_0$, which gives a measure of the smallest distance between the transverse projection of the track and the primary collision point. Future work will include: expanding the off-pointing track finding at L1 to the full acceptance of the outer tracker; matching the track jets with high transverse momentum ($p_\mathrm{T}$) deposits in the electromagnetic calorimeter; and finding new ways to evaluate track quality to suppress "fake" tracks that result from finding the wrong combination of track hits.

While in this study we focus on the specific Higgs boson decay to light scalars (see Ref. [9] for extensive review of physics motivations for such decays), the results and the proposed triggers are relevant for a broad spectrum of new physics searches, with or without macroscopic decay lengths.

# 2 Signal and background simulation

In these studies, the Phase-2 CMS detector is simulated using GEANT4 [10]. Event samples corresponding to 200 collisions per bunch crossing (pileup) [5] are used for the evaluation of trigger rates.

The following signal samples are considered:

1. Displaced single muons, generated with a uniform distribution of transverse momentum ($p_\mathrm{T}$) between 2 and 8 GeV, uniform in $\eta$ between -1 and 1, and with impact parameter $d_0$ distributed as a Gaussian with width $\sigma = 2\,\mathrm{cm}$.

2. The exotic decay of the SM-like Higgs boson $H(125) \rightarrow \phi\phi \rightarrow b\bar{b}b\bar{b}$, with $\phi$ masses of 15, 30, and 60 GeV, and $c\tau$ of 0, 1, and 5 cm. The production of the Higgs boson via gluon fusion is simulated by POWHEG v2.0 [11], while the hadronization and decay is performed by PYTHIA v8.205 [12].

3. The decay of a heavy SM-like Higgs boson with mass 250 GeV, $H(250) \rightarrow \phi\phi \rightarrow b\bar{b}b\bar{b}$, with $\phi$ masses of 15, 30, and 60 GeV, and $c\tau$ of 0, 1, and 5 cm. The production of the heavy SM-like Higgs boson via gluon fusion, its decay, and its hadronization are all simulated with PYTHIA v8.205 [12].

# 3 Track jets

The tracker is the most granular detector participating in the L1 decision, and therefore the most resilient to pileup. Track finding at L1 relies on selecting tracker hits that originate from high transverse momentum particles. This is achieved in the front-end electronics through use of the so-called $p_T$-modules, each consisting of two sensors separated by a few mm [1]. A particle crossing a tracker module produces a pair of hits in the two sensors. Such pairs form a "stub" if the azimuthal difference between the hits in the two sensors of a module is consistent with a prompt track with $p_T \gtrsim 2$ GeV.

In this section, we describe a simple jet clustering algorithm implementable in firmware, and compare it with anti-$k_T$ jets [13] with a size parameter of $R = 0.3$, as produced by FASTJET [14].

## 3.1 Algorithm description

A simplified algorithm for L1 track jets is used to facilitate the firmware implementation for the L1 trigger applications. L1 track jets are found by grouping tracks in bins of $z_0$, the point of closest approach to the $z$-axis, for the tracks. The bins are overlapping, staggered by half a bin, so that each track ends up in two bins, eliminating inefficiencies at bin edges. In each $z_0$ bin, the $p_T$ of the tracks are summed in bins of $\eta$ and azimuthal angle $\phi$ with bin size $0.2 \times 0.23$. A simplified nearest-neighbor clustering is performed, and the $\sum p_T^{trk}$ in the $z_0$ bin is calculated. The $z_0$ bin with the highest $\sum p_T^{trk}$ is chosen. Jets obtained through this algorithm are referred to as "TwoLayer Jets." For the studies below, $z_0$ bins with size 6 cm are used. Jets with $p_T > 50\,(100)$ GeV are required to have at least two (three) tracks.

## 3.2 Track selection

The track purity depends on the number of stubs in the track and the $\chi^2$ of the track fit. High-$p_T$ tracks are much less pure than low-$p_T$ tracks, with fake tracks distributed approximately uniformly in $1/p_T$ while real tracks are mostly low-$p_T$. To mitigate the effect of high-$p_T$ fake tracks, any track with a reconstructed $p_T$ above 200 GeV is assigned a $p_T$ of 200 GeV. The track quality selection used in this analysis is summarized in Table 1.

Table 1: Track selection for jet finding. The $\chi^2$ selections are per degree of freedom for a 4-parameter track fit.

| track $p_T$ | 4 stubs | 5 stubs | 6 stubs |
|---|---|---|---|
| 2–10 GeV | $\chi^2 < 15$ | $\chi^2 < 15$ | accept |
| 10–50 GeV | reject | $\chi^2 < 10$ | accept |
| >50 GeV | reject | $\chi^2 < 5$ | $\chi^2 < 5$ |

### 3.3 Comparison with FASTJET

We have verified that the TwoLayer trigger algorithm gives similar performance to a full jet clustering using the anti-$k_T$ algorithm with a size parameter $R = 0.3$, as implemented in FASTJET. Figure 1 shows the efficiency to reconstruct a track jet as function of the generator-level jet $p_T$. Figure 2 shows the calculated L1 trigger rates for an $H_T$ trigger, computed as the scalar sum of $p_T$ of jets. $H_T$ is computed from track jets with $p_T > 5\,\text{GeV}$. Figure 2 also shows the calculated L1 trigger rates for a quad-jet trigger with at least four track jets above a jet $p_T$ threshold.

The rates are computed based on a fixed number of colliding bunches. The trigger rate is computed as

$$\text{Rate} = \epsilon^{\text{L1 Thresh}} N_{\text{bunches}} f_{\text{LHC}},$$

where $N_{\text{bunches}} = 2750$ bunches for 25 ns bunch spacing operation, $f_{\text{LHC}} = 11246\,\text{Hz}$, and $\epsilon^{\text{L1 Thresh}}$ is the efficiency to pass a given L1 threshold as determined in simulation. For both the L1 trigger efficiency and rate, the performance of the TwoLayer hardware algorithm is compatible with the performance from the more sophisticated algorithm from FASTJET.

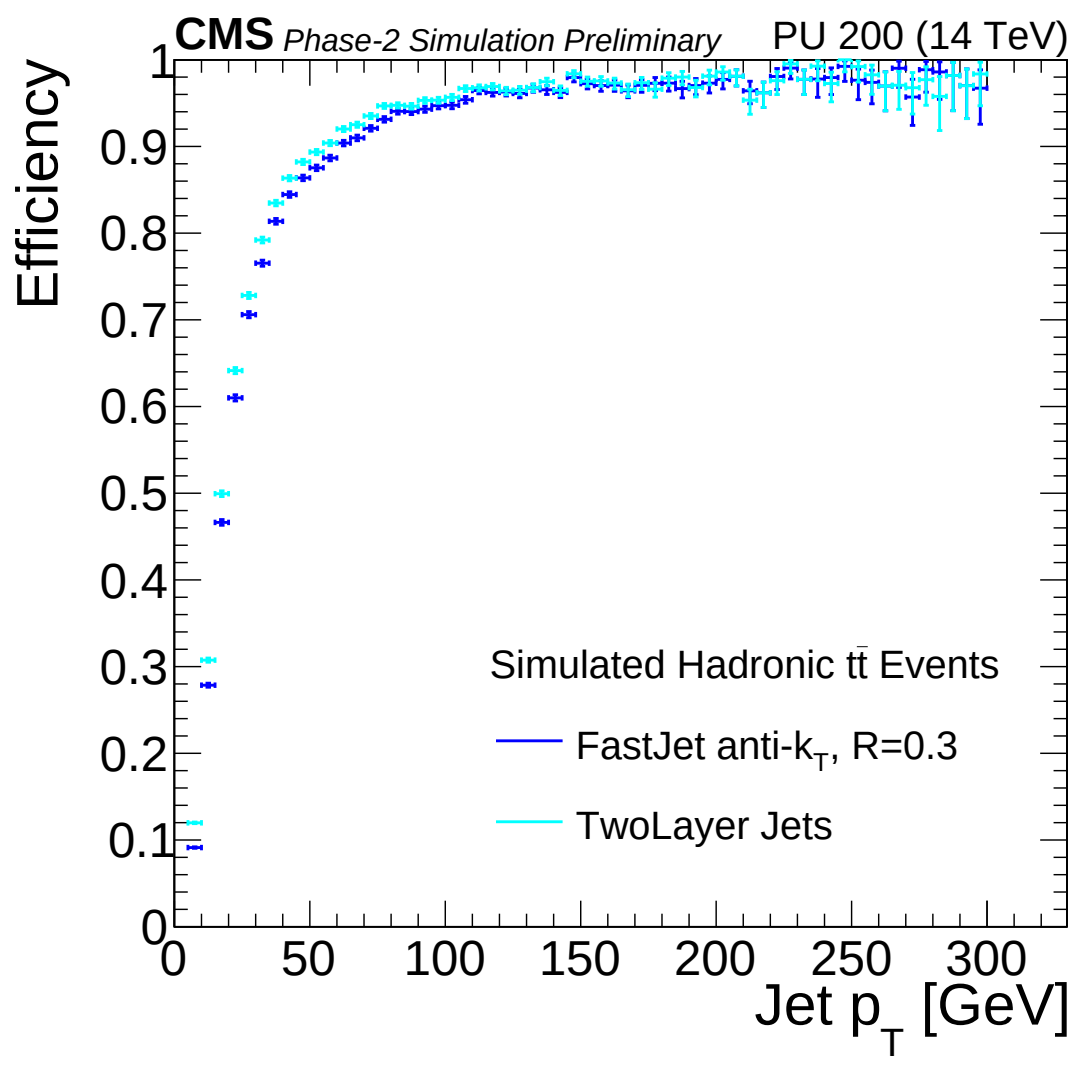

Figure 1: The efficiency for a jet to give rise to a L1 track jet as a function of the generator-level $p_T$ of the jet. The light and dark blue lines correspond to the trigger clustering (TwoLayer Jets) and anti-$k_T$ with $R = 0.3$ (FASTJET), respectively.

## 4 Displaced track finding

In this section, we briefly describe the performance of an algorithm for reconstruction of tracks with non-zero impact parameter. This approach extends the baseline L1 Track Trigger design to handle tracks with non-zero impact parameter and to include the impact parameter in the track fit. This enhanced design is feasible without greatly altering the track finding approach, but will require more FPGA computational power than the current proposal, which only considers only prompt tracks. Tracks passing the selection are clustered using the same algorithm as described in Section 3, and clusters containing tracks with high impact parameters are flagged as displaced jets. Though the baseline design of the L1 Track Trigger currently is optimized to find prompt tracks, these studies show that an enhanced L1 Track Trigger can extend the L1 trigger acceptance to include new BSM physics signals.

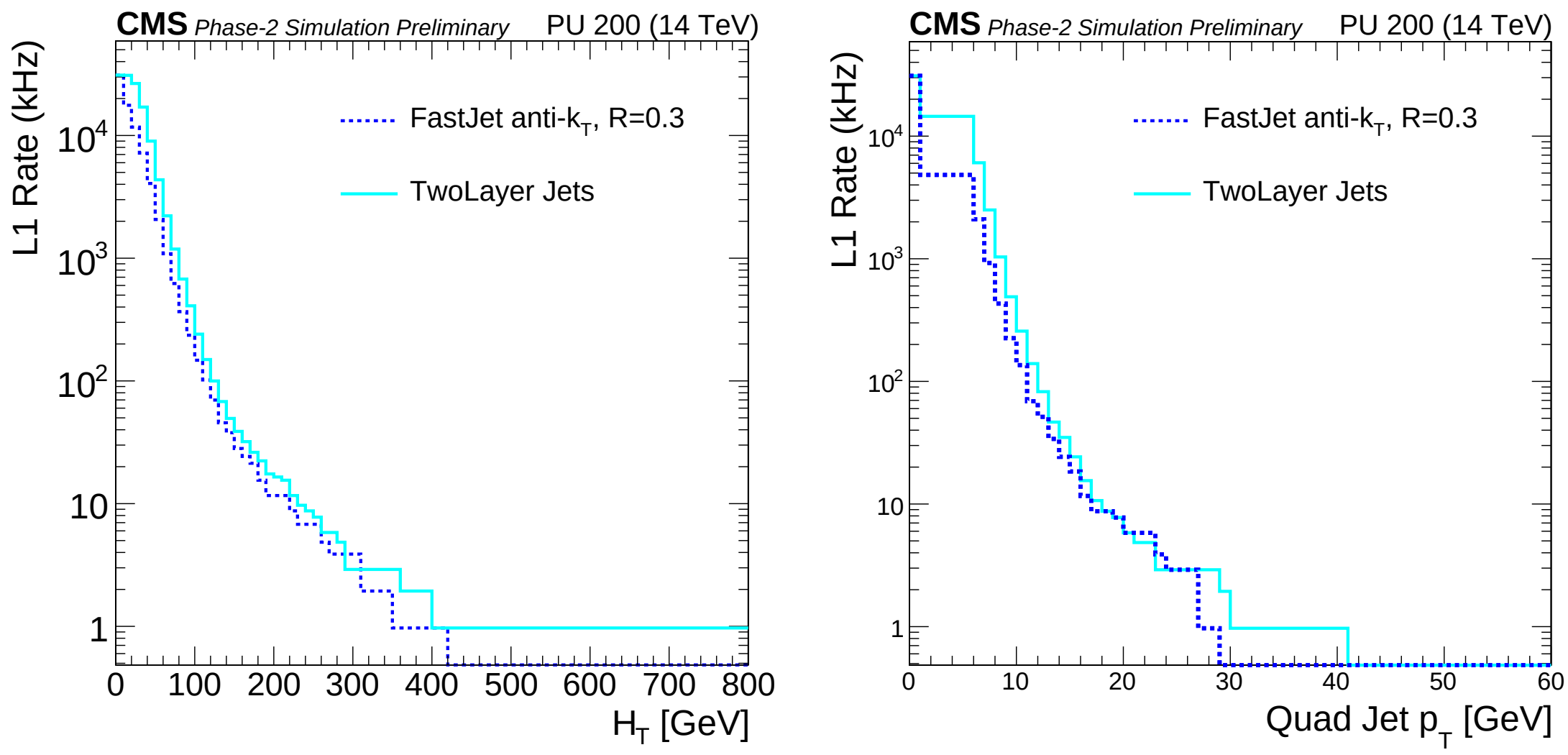

Figure 2: Calculated L1 trigger rates for track jet based $H_T$ (left) and quad-jet (right) triggers. The light and dark blue lines correspond to the trigger clustering (TwoLayer Jets) and anti-$k_T$ with $R = 0.3$ (FASTJET), respectively.

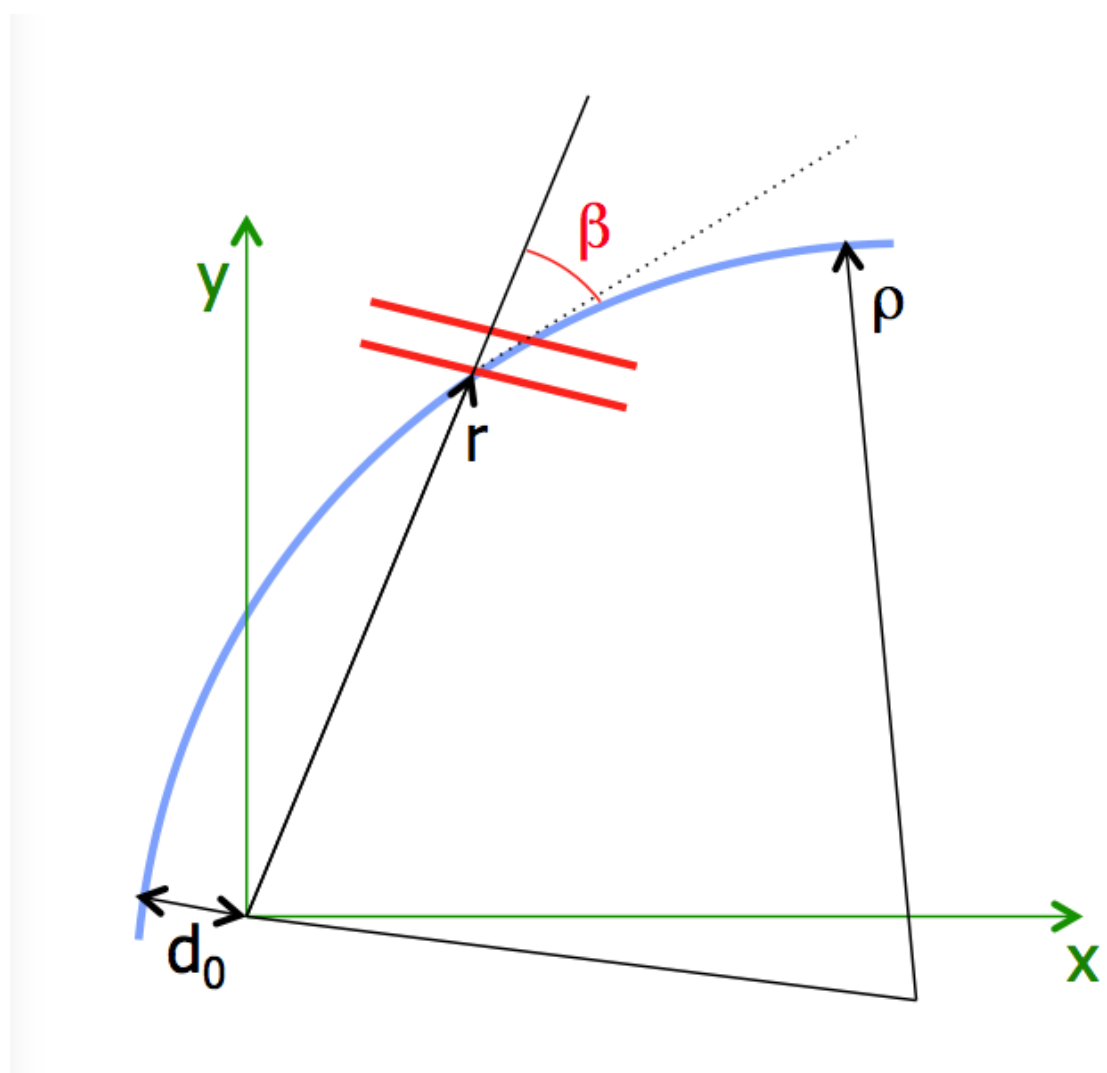

Figure 3: A sketch of a track crossing a $p_T$-module.

### 4.1 Stub efficiency

A track with a sufficiently small impact parameter can produce a stub. For tracks with large $p_T$ (i.e. large curvature radius $\rho$) and small $d_0$, the bending angle $\beta$ between the track and the prompt infinite momentum track, as shown in Fig. 3, is

$$\beta \approx \frac{r}{2\rho} - \frac{d_0}{r}.$$

Therefore, for a given $d_0$, one expects the stubs to be formed more efficiently as the radius of the module $r$ increases. Fig. 4 shows the efficiency for a displaced muon to produce a stub as a function of the signed transverse momentum and the impact parameter of the muon, as measured in the full GEANT4-based simulation of the Phase-2 detector. After the first layer of the tracker the stub reconstuction efficiency is high across a range of impact paramters. In the first layer of the tracker there is some inefficiency for impact parameters above 2 cm. Fig. 4 shows that for a range of impact paramters the stub reconstruction efficiency is large.

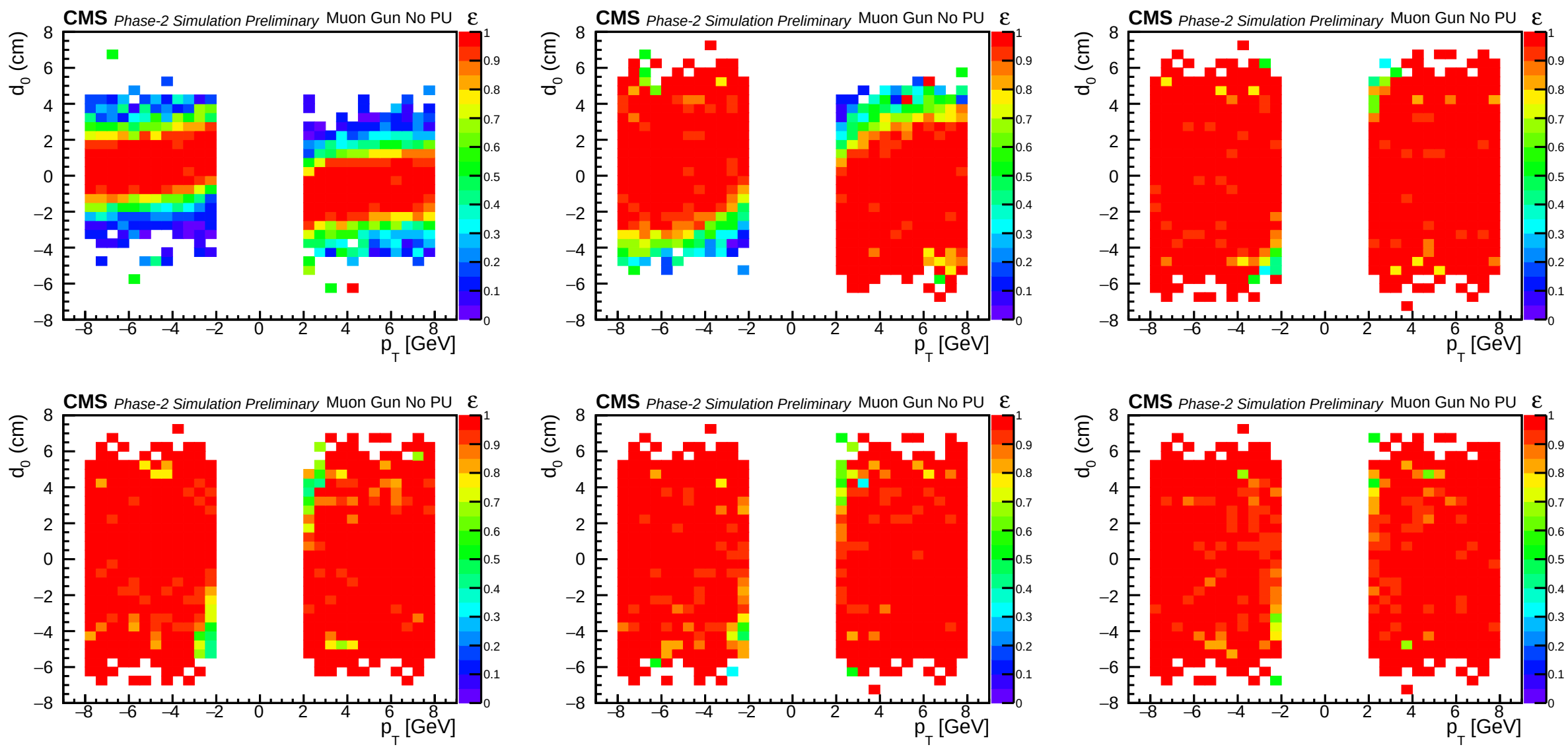

Figure 4: The efficiency for a displaced muon to form stubs in the six barrel layers of the Phase-2 tracker, as a function of the signed muon $p_T$ and impact parameter. The top row shows, from left to right, layers 1, 2, and 3; the bottom row shows layers 4, 5, and 6. The sample is comprised of 2000 muons generated with uniformly distributed transverse momentum between 2 and 8 GeV and pseudorapidity $|\eta| < 1$, and with the impact parameter $d_0$ distributed as a Gaussian with width of 2 cm.

### 4.2 Track finding efficiency

A special version of the tracklet algorithm [1] has been developed that is capable of reconstructing tracks with impact parameters of a few cm. For now, the reconstruction is limited to the barrel region ($|\eta| < 1.0$). Preliminary feasibility studies show that the algorithm will have similar performance in the entire outer tracker coverage.

Fig. 5 shows the track reconstruction efficiency requiring at least four and at least five stubs on the track. As expected, allowing only four stubs on a track gives a higher efficiency for high impact parameter tracks. The five stub efficiency is large at high momentum with impact parameter less than 3 cm. The five stub tracks will have larger purity than the 4 stub tracks, which motives the selection defined in the next section.

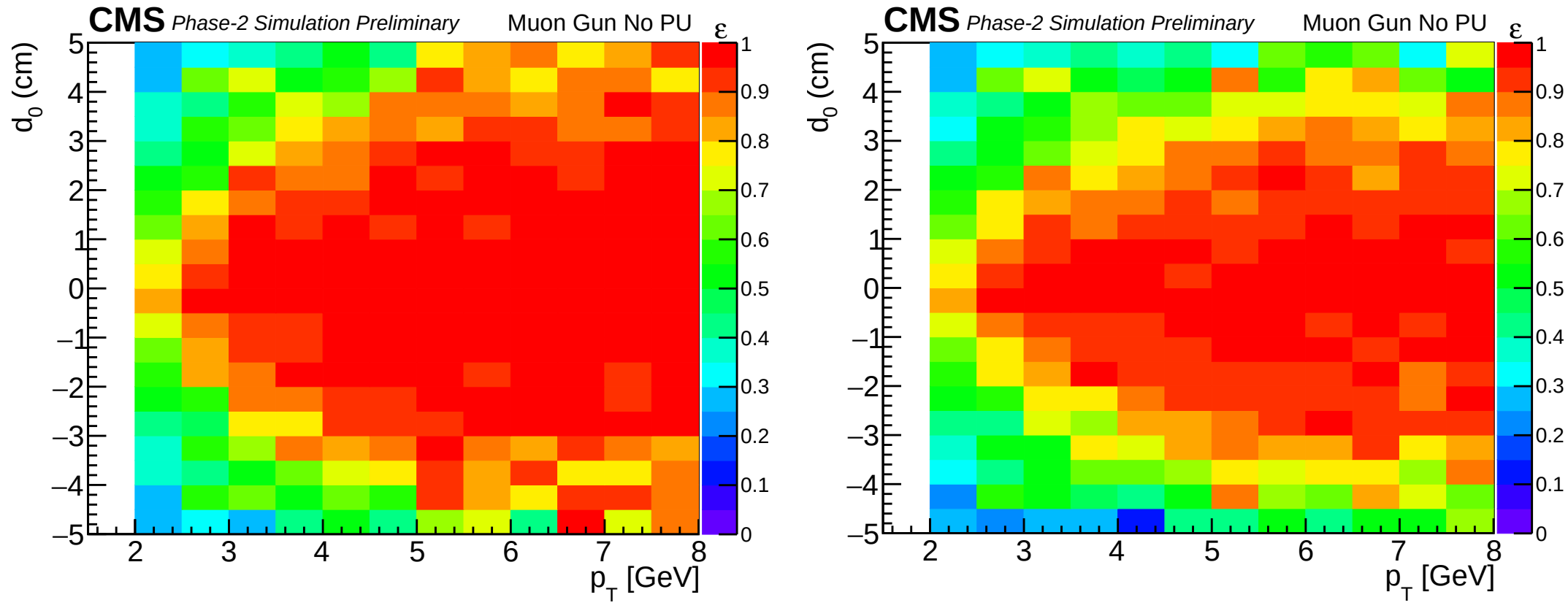

Figure 5: The efficiency for a displaced muon to be reconstructed as a track with at least four stubs (left) and at least five stubs (right).

### 4.3 Track selection

For the extended track finding algorithm, two track fits are performed: a 3-parameter $r\phi$ fit yielding $1/\rho$, $\phi_0$, and $d_0$, and a 2-parameter $rz$ fit yielding $t$ and $z_0$. The bend consistency variable is defined as

$$\text{consistency} = \frac{1}{N_{\text{stubs}}} \sum_{i=1}^{N_{\text{stubs}}} \left( \frac{\beta_i - \beta_i^{\text{exp}}}{\sigma_i} \right)^2 ,$$

where $N_{\text{stubs}}$ is the total number of stubs comprising the track, $\beta_i$ and $\beta_i^{\text{exp}}$ are the measured and expected bend angles for stub $i$, and $\sigma_i$ is the expected bend angle resolution.

Two track categories are defined, loose and tight. The selection is summarized in Table 2.

Table 2: Track selection criteria for jet finding with extended L1 track finding.

| | Loose | | | Tight | | |
|---|---|---|---|---|---|---|
| $N_{\text{stubs}}$ | $\chi^2_{r\phi}$ | $\chi^2_{rz}$ | consistency | $\chi^2_{r\phi}$ | $\chi^2_{rz}$ | consistency |
| 4 | <0.5 | <0.5 | <1.25 | | reject | |
| $\geq 5$ | <5.0 | <2.5 | <5.0 | <3.5 | <2.0 | <4.0 |

A jet is required to have at least two tracks passing the tight selection. If two or more tight tracks in a jet have $|d_0| > 0.1$ cm, the jet is tagged as a displaced jet.

## 5 Results

### 5.1 Track jets with prompt track reconstruction

Figure 6 shows the rate of the track jet $H_T$ trigger as a function of the efficiency of the heavy SM-like Higgs boson signal. While for prompt $\phi$ decays one can realistically achieve 20% efficiency at an L1 rate of 25 kHz, the efficiency quickly drops with the decay length, since the displaced tracks are not reconstructed for $d_0$ values above a few mm.

### 5.2 Track jets with a displaced tag

The rate for the $H_T$ trigger using the extended track finding is shown in Fig. 7, with and without a requirement of at least one jet with a displaced tag. The displaced tag requirement suppresses the rate by more than an order of magnitude. The displaced tracking and the trigger that requires a jet with a displaced tag make the signals with low $H_T$ accessible for displaced jets.

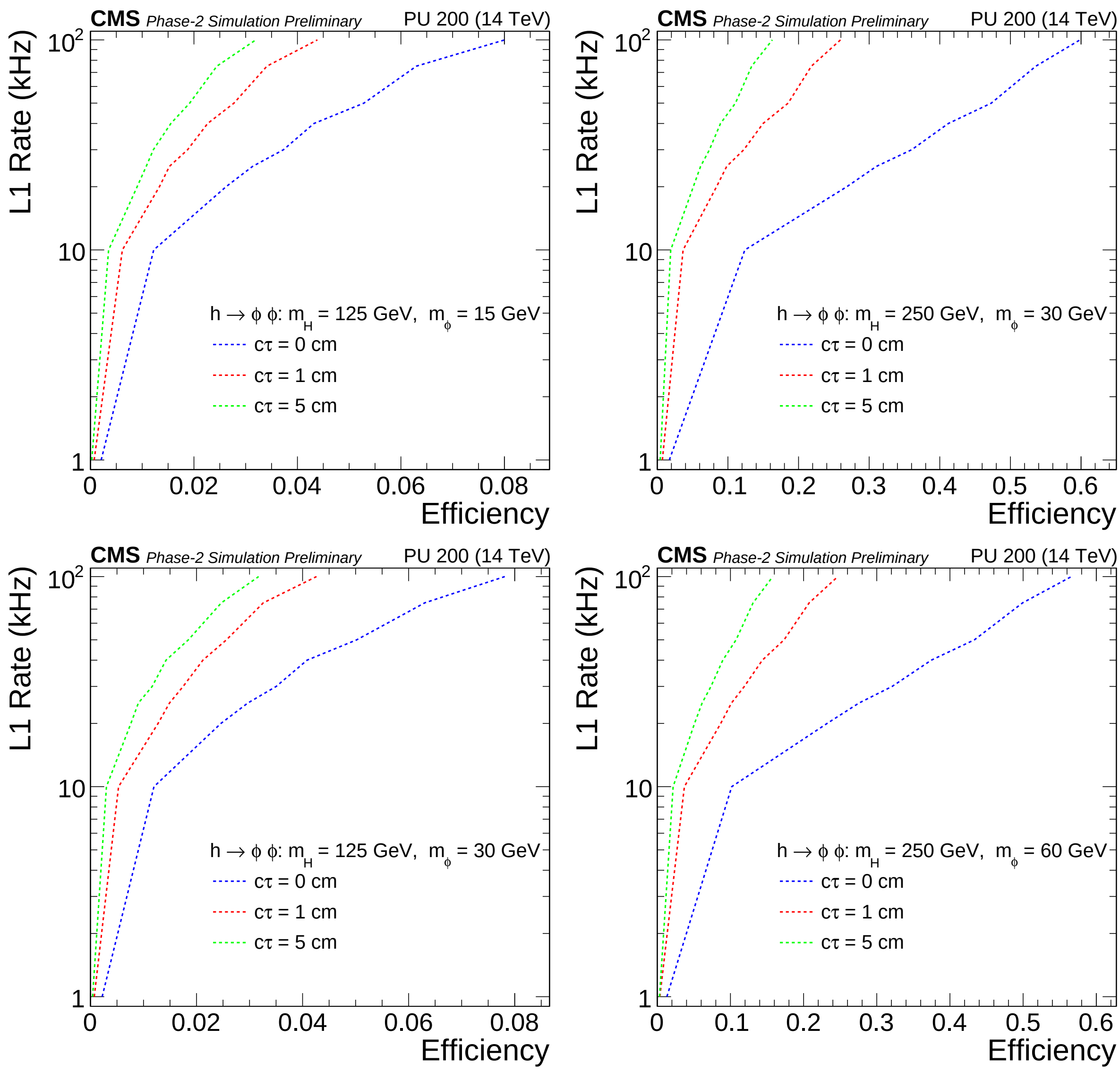

Figure 6: The rate of the track jet $H_T$ trigger as a function of signal efficiency for the SM-like Higgs boson (left) and the heavy SM-like Higgs boson (right) using prompt track finding.

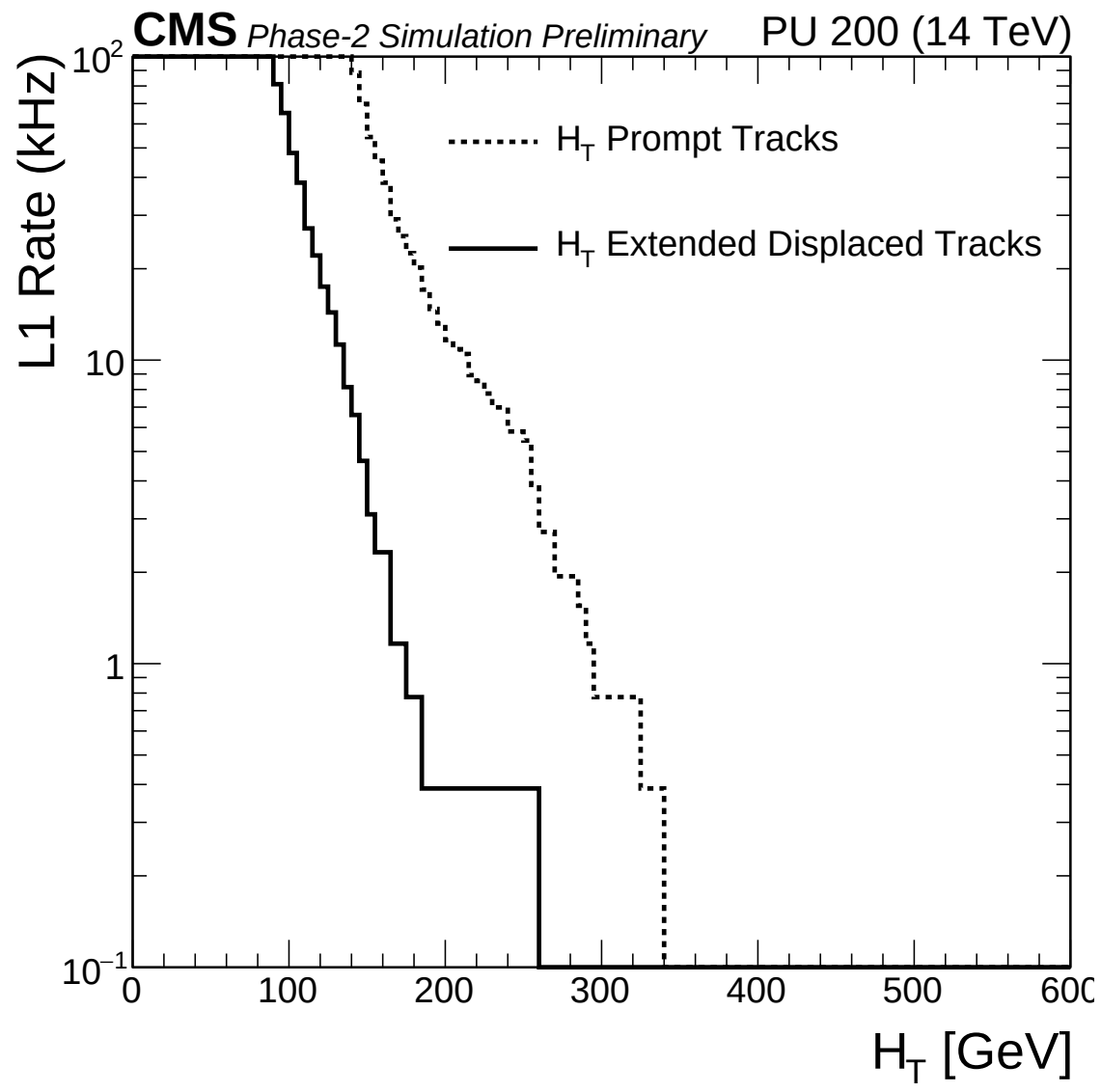

Figure 7: The rate of the track jet $H_T$ trigger using extended track finding with (solid line) and without (dashed line) a requirement of at least one jet with a displaced tag.

In order to compare the results with prompt and extended track reconstruction, one needs to make a correction for the rapidity coverage: prompt tracks are found in $|\eta| < 2.4$, while the extended track algorithm currently only reconstructs tracks in $|\eta| < 1.0$. For the feasible thresholds, the rate for $|\eta| < 0.8$ and $|\eta| < 2.4$ differ by a factor of five. To scale the efficiency for finding track jets to the full $|\eta| < 2.4$ range, we derive a scale factor (SF) based on efficiency in the full $\eta$ range and the central $\eta$ range. The signal efficiency SFs range from 4–6, which is comparable to the increase in the L1 rate. We have confirmed that such extrapolation works for the track jets clustered with prompt tracks. Figure 8 shows the expected trigger rate as a function of efficiency for the SM and the heavy SM-like Higgs bosons.

### 5.3 Expected event yields

The available bandwith for the triggers described above, if implemented, will be decided as a part of the full trigger menu optimization. Here, we consider two cases, 5 and 25 kHz. The expected event yield for triggers using extended and prompt tracking are shown in Fig. 9, assuming branching fraction $\mathcal{B}[\mathrm{H}(125) \to \phi\phi \to 4j] = 10^{-5}$ for the SM-like Higgs boson. For the heavy Higgs boson, the expected number of produced signal events is set to be the same as for the SM-like Higgs by requiring $\sigma_{\mathrm{pp}\to\mathrm{H}(250)}\mathcal{B}[\mathrm{H}(250) \to \phi\phi \to 4j] = 10^{-5}\sigma_{\mathrm{pp}\to\mathrm{H}(125)}$.

## 6 Conclusion

We have studied the upgraded CMS detector's ability to trigger on events with long lived particles decaying into jets. Currently, such events pass the L1 trigger only if the total transverse energy in the event is above a few hundred GeV. This is an important blind spot for searches, especially for the rare exotic Higgs boson decays like $\mathrm{H}(125) \to \phi\phi \to 4j$.

In this note, a new L1 trigger strategy based on the Phase-2 CMS detector's ability to find tracks at L1 is explored. Using L1 tracks for jet reconstruction significantly suppresses pile-up and

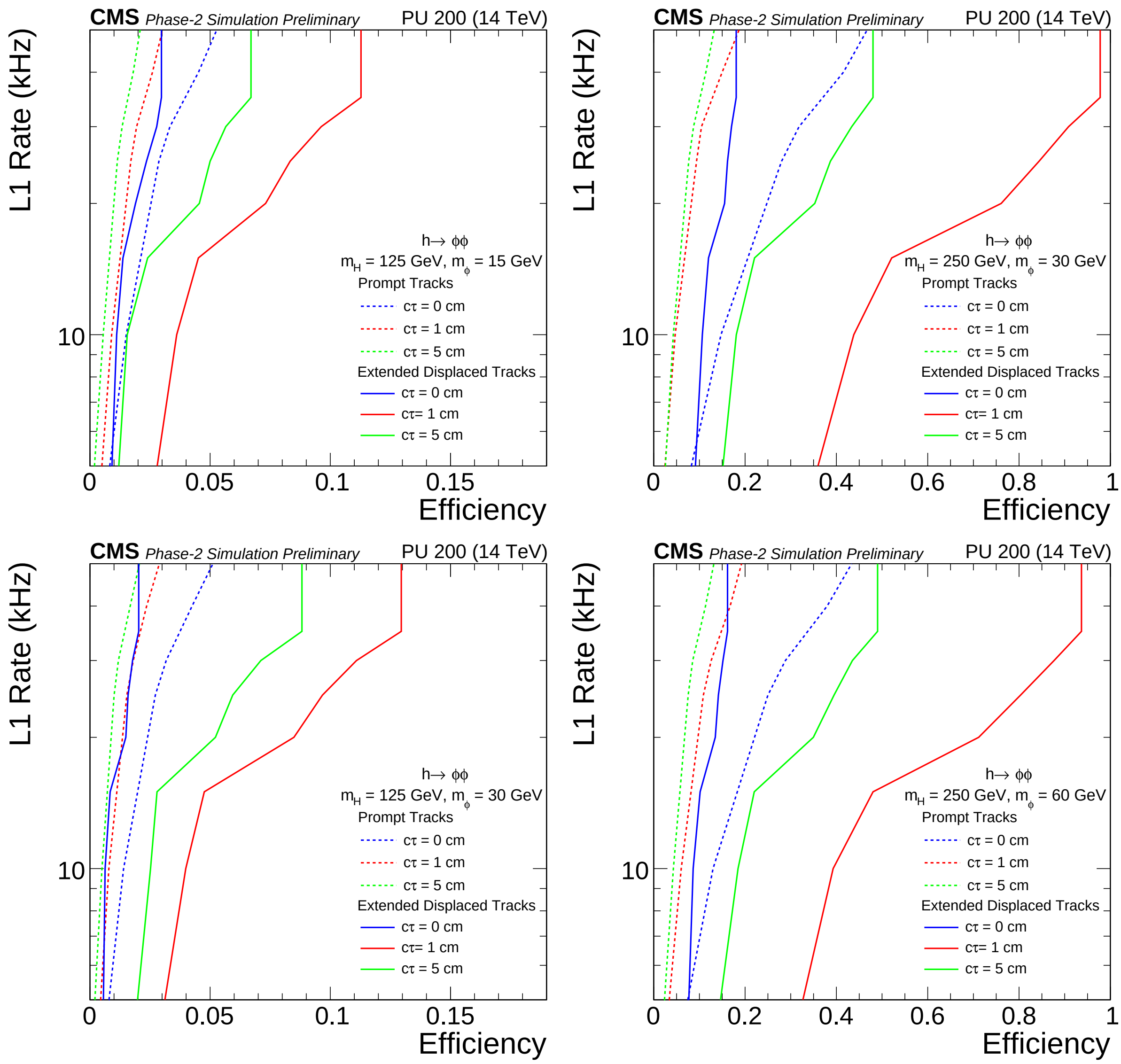

Figure 8: The rate of the track jet $H_T$ trigger as a function of signal efficiency using extended track finding for the SM-like Higgs (left) and the heavy SM-like Higgs (right). The extended track finding performance is extrapolated to the full outer tracker acceptance as described in text.

## 10

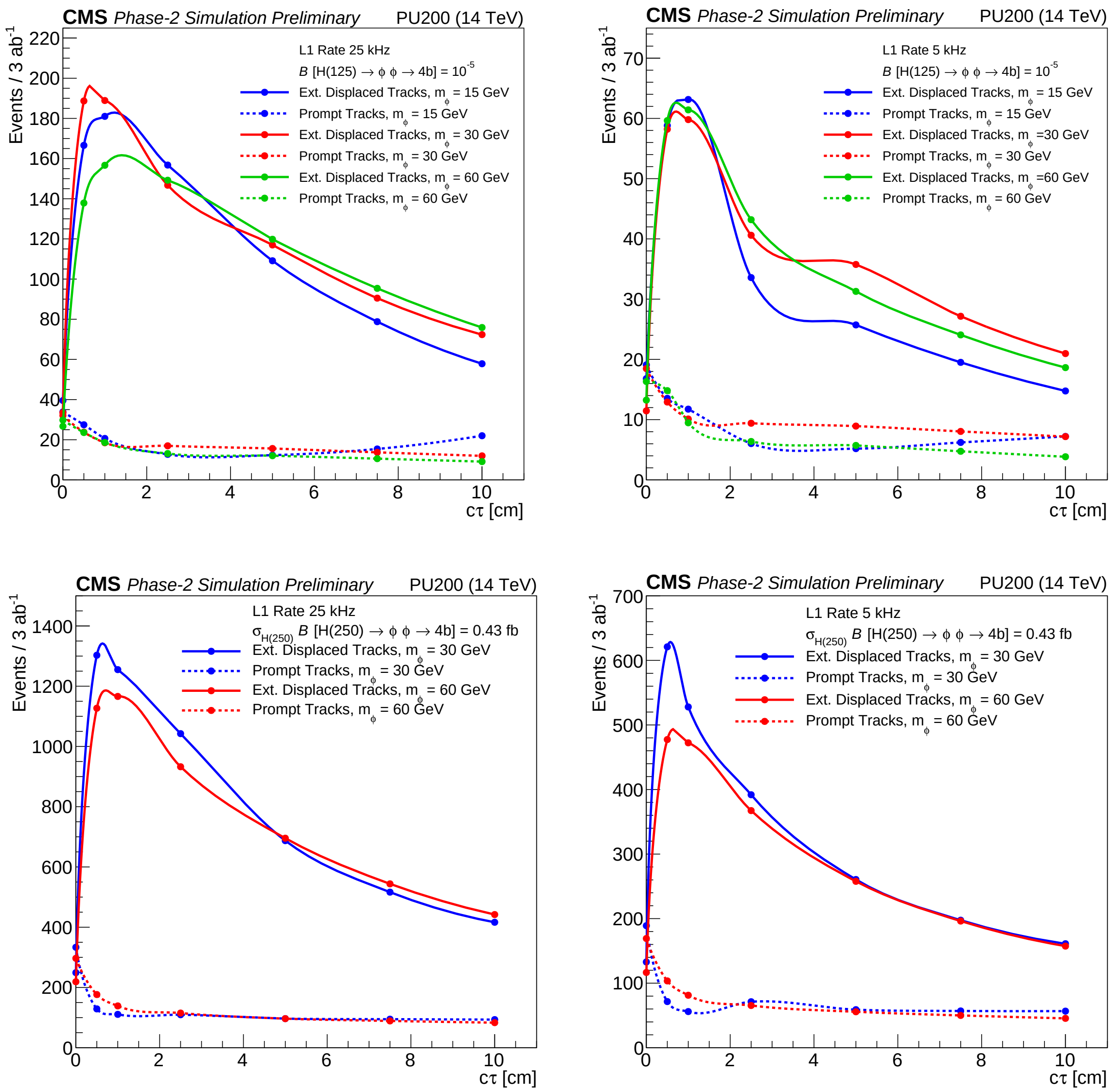

Figure 9: This plot shows the number of triggered Higgs events (assuming $\mathcal{B}[\mathrm{H}(125) \rightarrow \phi\phi \rightarrow 4j] = 10^{-5}$, corresponding to 1700 events) as a function of $c\tau$ for two choices for the trigger rates: 25 kHz (left), 5 kHz (right). Two triggers are compared: one based on prompt track finding (dotted lines) and another that is based on extended track finding with a displaced jet tag (solid lines).

allows to accept events with lower $H_T$. For the exotic Higgs decays considered, given the total Phase-2 dataset of $3\,ab^{-1}$ and branching fraction of $10^{-5}$, CMS would collect $\mathcal{O}(10)$ events, which should be sufficient for discovery. We also considered a plausible extension of the L1 track finder to consider tracks with impact parameters of a few cm. That approach improves the yield by more than an order of magnitude. The gains for the extended L1 track finding are even larger for the events with larger $H_T$, as demonstrated by the simulations of heavy Higgs boson decays.

# Beyond the Standard Model Physics

# CMS Physics Analysis Summary

Contact: cms-future-conveners@cern.ch

## Estimated Sensitivity for New Particle Searches at the HL-LHC

The CMS Collaboration

### Abstract

Sensitivity projections for new physics searches with 3000 $fb^{-1}$ of data anticipated at the high-luminosity LHC (HL-LHC) are presented. These results were obtained from dedicated studies performed for the ECFA 2016 upgrade workshop. Projections for heavy vector bosons ($Z'$ and $W'$) decays containing top quarks are obtained by extrapolating Run-2 results assuming scenarios with varying systematic uncertainties. Results for the dark matter and weak production of single vector-like quark searches are obtained by implementing detector performance specifications from the CMS Phase-2 technical proposal in the DELPHES simulation package.

# 1 Introduction

The High Luminosity LHC (HL-LHC) run, which is due to start in 2025, is expected to collect an integrated luminosity of approximately 3000 $\mathrm{fb}^{-1}$ at $\sqrt{s} = 14\,\mathrm{TeV}$. During the entire operation prior to this run, a dataset of only 10% of this size is expected, namely 300 $\mathrm{fb}^{-1}$. The discovery and study of physics beyond the standard model will remain one of the major goals of the CMS collaboration during the HL period. Such physics can yield exotic signatures, the observation of which will require high demands on the performance and capabilities of the detector. For a few selected physics models, we present studies of potential for new physics using CMS data during the HL-LHC run. The goal of these studies is to estimate the sensitivity for key channels at the HL-LHC either via a projection from $\sqrt{s} = 13\,\mathrm{TeV}$ analyses or via dedicated studies. This document summarizes new physics studies performed in preparation for the ECFA 2016 HL-LHC workshop in October 2016 [1] complementing similar high luminosity studies for the Higgs sector [2] and for standard model processes [3].

Two of these searches, $\mathrm{W}' \to \mathrm{tb}$ and $\mathrm{Z}' \to \mathrm{t\bar{t}}$, are projections extrapolated from current searches. These projections are based on present 2015/2016 $\sqrt{s}$=13 TeV analyses, referred to as Run-2 baseline analyses. Signal and background samples are taken from the corresponding 2015/2016 analysis and scaled to $\sqrt{s}$=14 TeV by the ratio of their production cross sections at 13 and 14 TeV. The sensitivity after accumulating 3000 $\mathrm{fb}^{-1}$ is estimated in terms of discovery potential and in terms of exclusion at the 95% confidence level (C.L.). The impact of systematic uncertainties is studied by considering different scenarios with the two extreme cases: (1) no improvement, keeping the systematic uncertainties at their Run-2 levels, (2) all systematics assumed to be negligible, corresponding to the detection limit. Systematic uncertainties have theoretical and experimental origins. For the former, improvements from higher order calculations are expected but hard to quantify at this time. On the experimental side, the Phase-2 detector will have better performance while pileup conditions will be much more severe. Uncertainties related to data-driven methods will decrease with larger datasets. Such considerations also apply to measurements of cross sections, lepton and trigger efficiencies, jet energy and tagging performances.

A second class of analyses simulates aspects of the upgraded CMS detector based on the documentation in the CMS Phase-2 Technical Proposal [4]. The studies target the physics reach with 3000 $\mathrm{fb}^{-1}$ but now include a parameterization of the expected detector performance. Different systematic scenarios are investigated using reasonable assumptions regarding their improvements in the future. More analyses of this type, including the Phase-2 detector performance, were performed previously [5], e.g. $\mathrm{W}' \to e\nu$ and $\mathrm{Z}' \to ee$, mono-W dark matter, heavy stable charged particles (HSCP) and long-lived signatures. The following two searches include detector performance in the parametrized detector simulation package DELPHES [6], dark matter (DM) in the jet+MET final state and single vector-like quarks (VLQ) in the $\mathrm{T} \to \mathrm{tH}$ channel.

This document is organized as follows. Sections 3 and 4 contain the projections for the heavy vector bosons ($\mathrm{Z}'$ and $\mathrm{W}'$) in decay channels with top quarks. Both projections are based on Run-2 results and take into account the impact of different scenarios for systematic uncertainties. Sections 5 and 6 follow with the summaries of upgrade analyses using the parametrized simulation package DELPHES with a performance parameterization according to the CMS Phase-2 Technical Proposal.

# 2 The CMS Phase-2 upgrade

In the Phase-2 CMS Technical Proposal [4], the performance of the CMS Phase-1 detector under the conditions of HL-LHC has been studied, considering the higher instantaneous luminosities leading to high pileup (PU) and high radiation levels. These studies show that the tracker and the endcap calorimetry will need to be replaced for Phase-2. With these changes, the performance issues due to high PU that are expected to be most pronounced in the inner and forward detector regions can be addressed. The new tracker and pixel vertex detector will have an extended forward acceptance. New endcap calorimeter detectors have higher segmentation and improve the energy resolution measurement. Additional improvements are foreseen for barrel calorimeters where the readout will be upgraded along with the electronics to handle higher event rates and larger trigger latencies which are necessary to accommodate the new track trigger. The forward muon system will be augmented with additional detectors in the region $|\eta| > 2.1$ which is the only region in the Phase-1 muon system without redundancy.

The performance parameters of this Phase-2 upgraded detector has been studied with simulations and is also documented in the "Technical Proposal for the Phase-2 Upgrade of the CMS Detector" [4]. Performance studies are not described in this document, which concentrates on physics sensitivity with the 3000 $\text{fb}^{-1}$ of HL-LHC data. These performance studies do provide the input for the parameterized detector simulation based on DELPHES (version 3.3.16) [6] which is used in the simulation for the mono-X and VLQ search projections.

The SM background samples are based on samples generated for Snowmass [7] with the generator information being reprocessed through the DELPHES version mentioned above. Dedicated trigger studies were not performed for these sensitivity estimates.

# 3 Sensitivity projections for W′ →tb in leptonic final states

Many SM extensions require additional heavy gauge bosons. For example, the sequential standard model (SSM) [8] predicts the existence of a new massive charged boson, W′, exhibiting the same couplings as the SM W boson with the additional decay channel W′ →tb opening up if the new boson is sufficiently massive. The benchmark analysis with maximum discovery sensitivity is the decay to a single lepton ($\ell = e, \mu$) and neutrino. In a scenario where a right-handed $\nu_R$ is heavier than a right-handed $W'_R$ boson, the decay to leptons is forbidden, leaving only the tb final state open for discovery.

The projection in this section is based on an analysis performed with 12.9 $\text{fb}^{-1}$ collected in 2016 at $\sqrt{s}$=13 TeV [9] and is referred to as the "baseline analysis". The analysis uses leptonic W boson decays, like $W'_R \to t(\to W(\ell\nu) + b)b$ with $\ell = e, \mu$. In this final state, we perform the search in four event categories in terms of lepton ($\ell = e, \mu$) and the number of b-tagged jets $N_{\text{b-tags}}$:

- electron plus one b-tagged jet, labeled "e+jets $\text{N}_{\text{b-tags}}$=1"
- electron plus two b-tagged jets, labeled "e+jets $\text{N}_{\text{b-tags}}$=2"
- muon plus one b-tagged jet, labeled "$\mu$+jets $\text{N}_{\text{b-tags}}$=1"
- muon plus two b-tagged jets, labeled "$\mu$+jets $\text{N}_{\text{b-tags}}$=2"

The simulated samples from the baseline analysis are scaled to the cross sections at 14 TeV. Details of the analysis strategy itself can be found in [9], while this note contains information relevant to the procedure used to extrapolate from 12.9 $\text{fb}^{-1}$ at $\sqrt{s}$=13 TeV to a projection for 3000 $\text{fb}^{-1}$ at $\sqrt{s}$=14 TeV.

## 3.1 Extrapolation details

The signal and background simulation is identical to the one used in the baseline analysis from 2016 [9] and scaled to $\sqrt{s}$=14 TeV by their cross section ratio. For signal samples, a ded-icated calculation of the 14 TeV cross sections for all signal masses of interest was performed using CompHEP (the same generator used for the 13 TeV samples). The resulting signal scaling factors are a function of the boson mass, and range from 1.16 for m($W'_R$)=1 TeV to 1.48 for m($W'_R$)=4 TeV. The lower limit of m($W'_R$) = 1 TeV is driven by the trigger thresholds. For the projection studies, additional mass points from 3.1 to 4 TeV in 100 GeV steps are included in order to better understand the analysis behavior in the region of interest for the projected luminosities. No correction is made for shape differences which may arise at 14 TeV from a slightly lower off-shell component.

The backgrounds are taken from simulation for this analysis, and then the modeling is checked in dedicated control regions enriched in the dominant background processes. For each background source extrapolations for the sample cross sections from 13 to 14 TeV are performed depending on the sample. All objects and efficiencies are similar to the baseline analysis.

### 3.2 Event selection and resulting distributions

The four search categories have been defined in Sec. 3.1. The discriminating variable is the invariant tb mass, M(tb), reconstructed the following way: we first reconstruct a W boson from the lepton and $E_{\mathrm{T}}^{\mathrm{miss}}$ in the event using the W mass to constrain the z-component of the neutrino momentum. Subsequently a top quark candidate is reconstructed using the jet in the event which gives a candidate mass closest to the top mass, and then combine the top quark candidate with the highest $p_T$ jet remaining in the event to give the $W'_R$ candidate and compute M(tb). The corresponding M(tb) distributions are shown in Fig. 1.

Here we briefly repeat the selection steps from the baseline analysis for 2016 data. It is expected that trigger thresholds and some specific selection steps will have to be adapted when performing this analysis at HL conditions where pileup is larger and the Phase-2 detector acceptance is larger. The triggers are the lowest unprescaled single lepton triggers, with trigger thresholds of 105 GeV and 45 GeV for electrons and muons, respectively. The following physics objects definitions are used in the analysis:

- **Electron** candidates are selected using a multivariate technique based on the shower-shape information, the quality of the track, the match between the track and electromagnetic cluster, the fraction of total cluster energy in the hadronic calorimeter, the amount of activity in the surrounding regions of the tracker and calorimeters and the probability of the electron originating from a converted photon.
- **Muon** candidates are required to be associated to a track with hits in the pixel and muon detectors, a good quality fit and transverse and longitudinal impact parameters close to the primary vertex.
- **Jets** are reconstructed within $|\eta| <2.4$ with the anti-$k_T$ algorithm [10, 11] with a size parameter of 0.4 and a transverse momentum requirement $p_T >25$ GeV. The b jets are identified with a b-tagging working point corresponding to 10% misidentification probability and 80% efficiency for b jets.

The kinematics selections in the analysis are:

- Events must contain one lepton with $p_T > 180$ GeV and excluded in the presence of an additional lepton with $p_T > 35$ GeV.
- Lepton and jet are required either to be well separated, quantified as $\Delta R$(lepton, closest jet)>0.4, or have a relative difference between the lepton and jet transverse momentum, $p_T$ (rel), above 60 (50) GeV for electrons (muons). The quantity $p_T$ (rel) is defined as the magnitude of the lepton momentum orthogonal to the jet axis.

- The leading jet is required to have $p_T$ greater than 350 (450) GeV for the electron (muon) channel with the subleading jet showing $p_T$ greater than 30 GeV.
- $E_T^{miss}$ has to be greater than 120 (50) GeV in the electron (muon) channel.
- $\Delta\phi$ between $E_T^{miss}$ and the electron has to be below 2.
- The vector sum of both jets is required to be $p_T$ (jet1 + jet2)>350 GeV and $p_T$ (top)>250 GeV.
- In the muon channel, the top mass is required to be between 100 and 250 GeV, in order to suppress background.

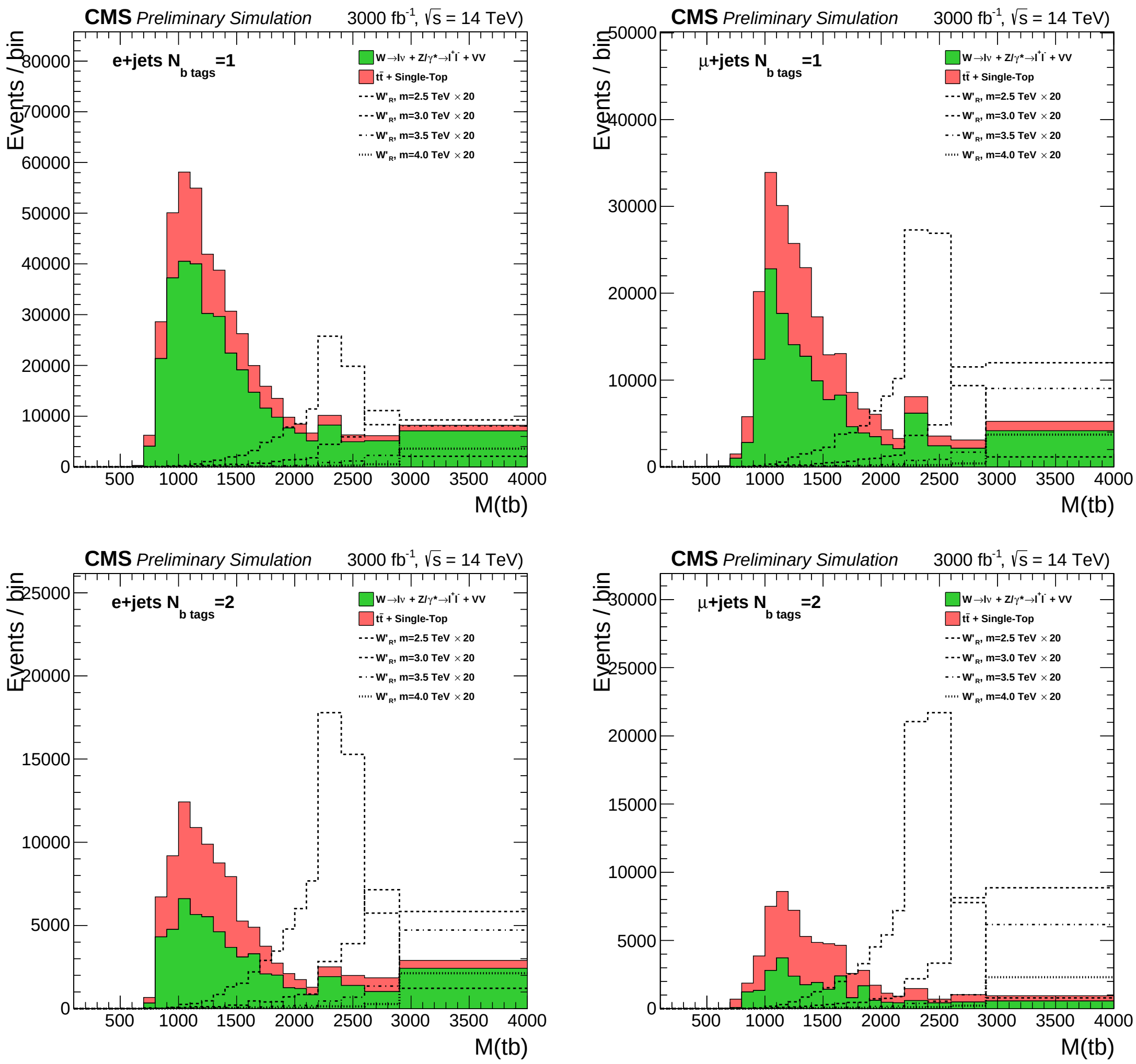

Figure 1: The reconstructed $tb$ invariant mass distributions in the 1 b-tag (top) and 2 b-tag (bottom) categories.

## 3.3 Systematic uncertainties

All systematic uncertainty estimates are taken from the baseline analysis. We disregard systematics affecting lepton efficiencies and photon identification that were specific to 2016 data. In addition a 15% (10%) uncertainty on the theoretical cross section of the top (bosonic) background is added. We then consider three scenarios for extrapolating systematics to 3000 fb$^{-1}$.

- **Current systematics** - We do not perform any adjustment to the magnitude of the systematics and keep the values from the Run-2 baseline analysis [9].

Table 1: Systematic uncertainties in two scenarios used for extrapolating from results using 12.9 $\mathrm{fb}^{-1}$ of data collected at $\sqrt{s}$ = 13 TeV [9]. The "current systematic" scenario assumes no change in systematics from their nominal values in the 12.9 $\mathrm{fb}^{-1}$ dataset used for projection. The "reduced systematic" scenario assumes a realistic reduction in the magnitude of systematic uncertainties from their nominal values, based on improvements in dataset size, detector performance, and theoretical accuracy among others. For systematics which affect the shape of the invariant mass distribution, the value quoted for the rate uncertainty is approximate.

| Source | Current systematics | Reduced systematics | Shape? |
|---|---|---|---|
| Luminosity | 6.2% | 1.5% | No |
| Trigger Efficiency ($e/\mu$) | 2%/5% | 1%/1% | No |
| Lepton ID Efficiency ($e/\mu$) | 5%/2% | 1%/1% | No |
| Jet Energy Scale | 3.8% | 1% | Yes |
| Jet Energy Resolution | 1% | 0.07% | Yes |
| $b/c$-tagging | 2.7% | 1% | Yes |
| light quark mis-tagging | 1.2% | 1.2% | Yes |
| W+jets Heavy Flavor Fraction | 2.3% | 1.1% | Yes |
| Top $p_T$ Reweighting | 18% | 6% | Yes |
| Pileup | 1.3% | 0.09% | Yes |
| PDF | 6.1% | 3% | Yes |
| Matrix element $Q^2$ scale | 18.9% | 9.5% | Yes |
| $t\bar{t}$ Parton matching $Q^2$ scale | 1.7% | 0.9% | Yes |
| Theoretical top cross section | 15% | 7.5% | No |
| Theoretical bosonic cross section | 10% | 5% | No |

- **Reduced systematics** - We scale theoretical cross section, PDF, and $Q^2$ scale uncertainties down by a factor of 2. The top $p_T$ uncertainty scaled down by a factor of 3 and the luminosity uncertainty is reduced to 1.5%. The magnitude of the jet energy scale uncertainty and the b-tag uncertainty is set to 1%. The mis-tag uncertainty stays unchanged. All other uncertainties are scaled down by a factor of $\sqrt{\mathcal{L}}$.
- **No systematics** - No systematics at all, corresponding to the best possible limit.

The systematic uncertainties and their sizes in the two scenarios in which systematics are considered are shown in Table 1. The leading uncertainties are not of experimental nature (e.g. $Q^2$, top and diboson cross sections, PDF) and should improve in the coming ten years when more data are recorded and theoretical calculations are refined, hence a factor two improvement is assumed in the scenario of "reduced systematics". Experimental uncertainties (e.g. efficiencies, scale factors, tagging efficiencies or luminosity) will be different with the real Phase-2 detector and expected to improve as well when this upgraded detector has been sufficiently studied. The last column in the Table indicates whether the source of systematics has an impact on the shape of the distribution. For these cases the value quoted for the rate uncertainty is approximate.

### 3.4 Projected exclusion reach

Exclusion limits for right-handed $W'_R$ bosons are shown in Fig. 2, with all four event categories combined. Theoretical W$'$ cross sections times branching ratios for two different theoretical assumptions on the right-handed neutrino mass are shown in red. On the top-left, the current scenario which assumes no change in systematics from their nominal values in the 12.9 $fb^{-1}$ dataset used for projection. $W'_R$ masses up to 4 TeV can be excluded. The reduced systematic scenario assumes a realistic reduction in the magnitude of systematic uncertainties from their nominal values based on improvements in dataset size, detector performance, and theoretical accuracy among others and is shown on the top-right. Not surprisingly, the sensitivity increases beyond 4 TeV. The selection was optimized for signal masses between 2-3 TeV corresponding to the reach of the 2016 baseline analysis. For masses beyond 4 TeV, where the off-shell part starts to become important, the selection should be re-optimized, which was not done for this projection. On the bottom-left in Fig. 2 the exclusion limit is displayed for the case without any systematics, exceeds significantly beyond 4 TeV.

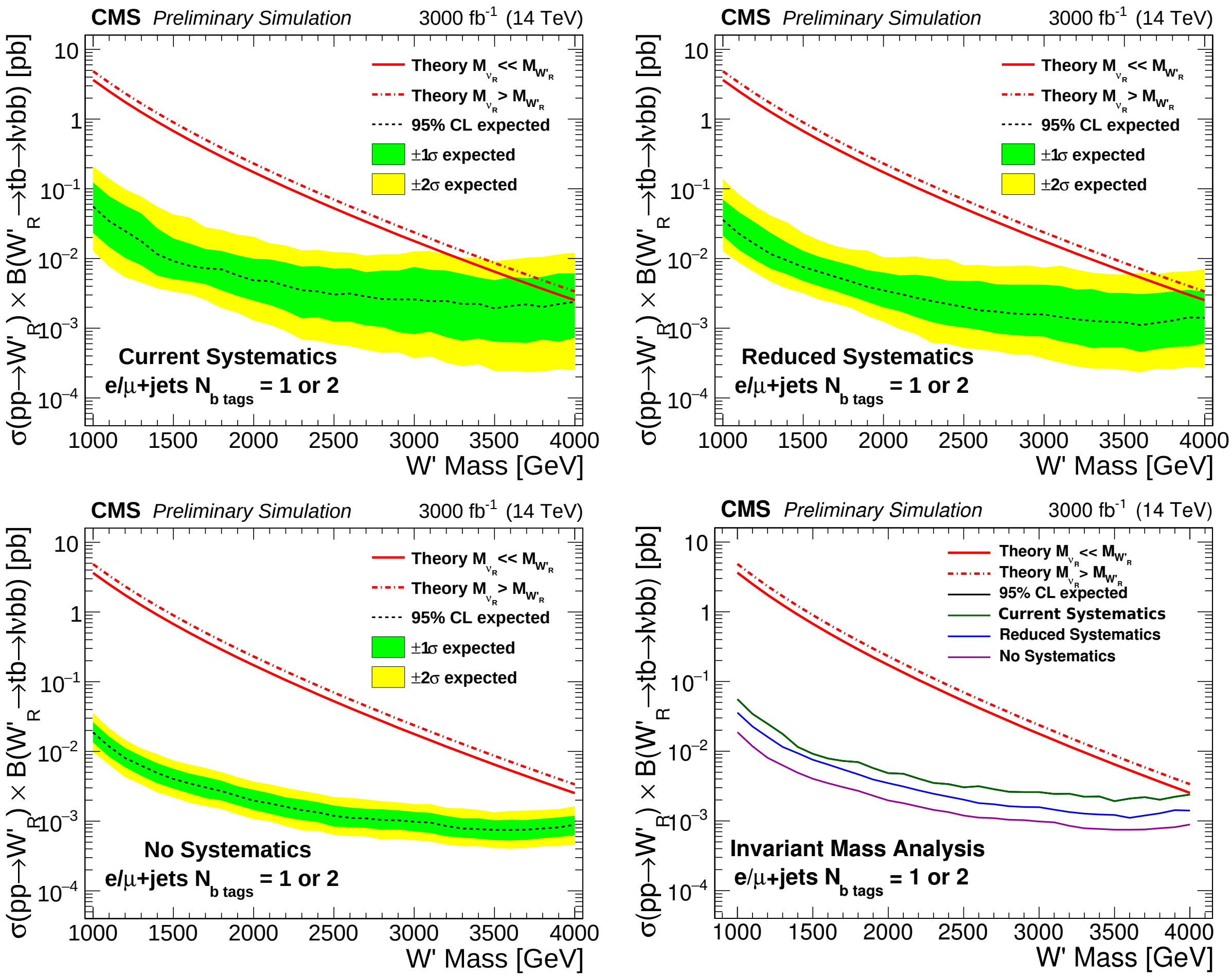

Figure 2: Projection of expected and observed Bayesian 95% C.L. upper limits on the production cross section times branching ratio of right-handed W′ bosons for an integrated luminosity of 3000 $fb^{-1}$. The projection combines electron/muon+jets channel and 1 or 2 b-tags. The ″current systematic″ scenario (top-left) assumes no change in systematics from the 12.9 $fb^{-1}$ dataset [9] used for projection. The ″reduced systematic″ scenario (top-right) assumes a realistic reduction from their nominal values. For the graph on the bottom-left, no systematic uncertainties are included. Theoretical W′ cross sections times branching ratios for two different theoretical assumptions on the right-handed neutrino mass are shown in red. Bottom-right: the three different uncertainty scenarios in the same figure.

### 3.5 Projected discovery reach

We also make projections for the discovery sensitivity for a range of signal masses and cross sections. A quasi-model-independent method is used where projections are performed for arbitrary cross sections and resonance mass. Toy datasets with different amounts of injected signal are studied. The p-values for these hypothesized datasets compared to the null-signal hypothesis yield significances which are reported in units of standard deviations in Fig. 3. Three exemplary values of $2\sigma$, $3\sigma$ (corresponding to "evidence") and $5\sigma$ (corresponding to discovery) are given. These projections are performed for the three systematic scenarios discussed previously.

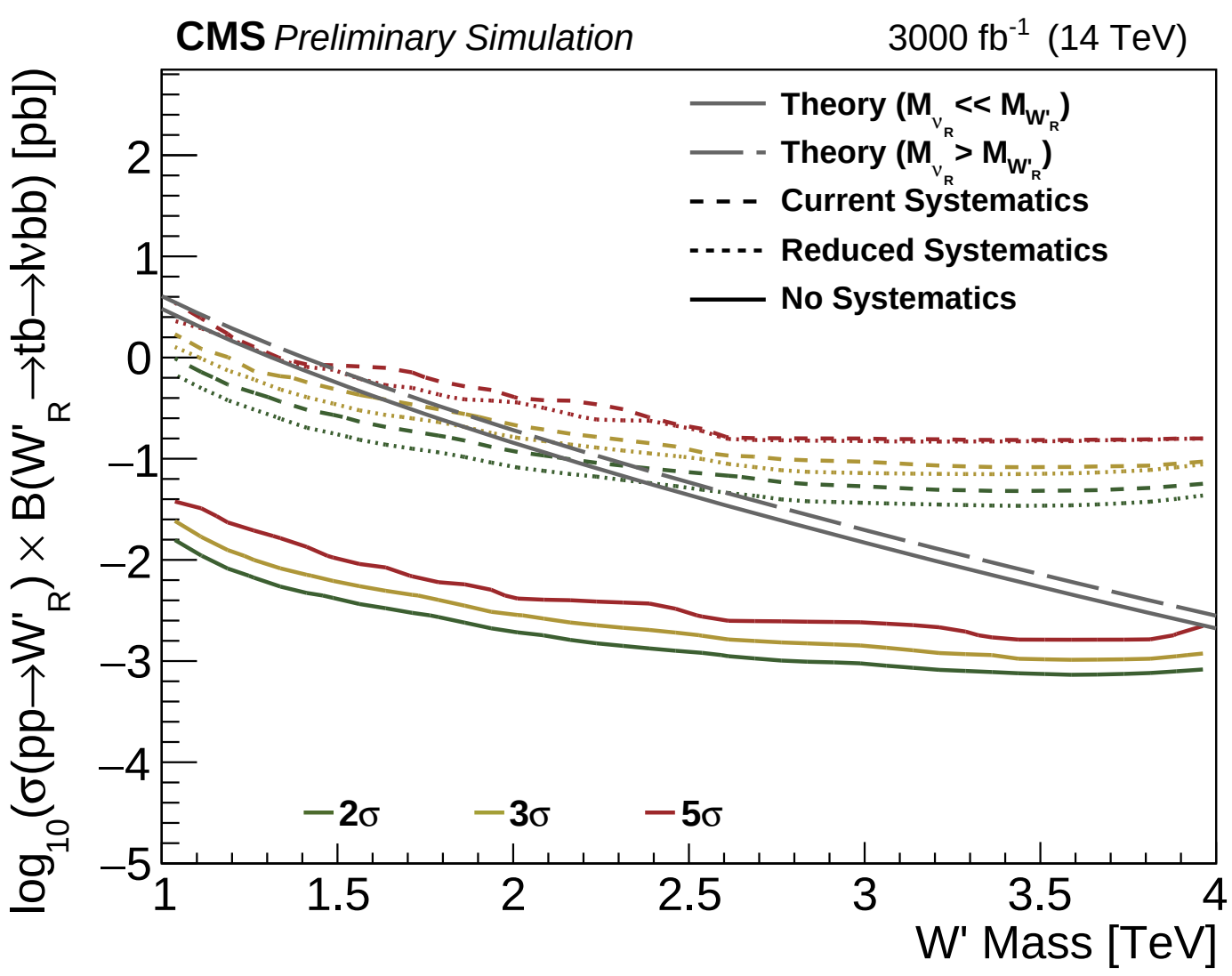

Figure 3: Expected discovery sensitivity for an integrated luminosity of 3000 $fb^{-1}$ as a function of the signal mass and the production cross section times branching ratio of right-handed $W'$ bosons in the combined electron/muon+jets channel, for combined 1 or 2 b-tags. Three scenarios for systematic are shown as explained in the legend. Theoretical $W'$ cross sections times branching ratios for two different theoretical assumptions on the right-handed neutrino mass are shown in grey (solid and dashed lines).

## 4 Sensitivity projection for $Z' \rightarrow t\bar{t}$

Additional neutral heavy vector bosons (denoted $Z'$) are also predicted. This section concentrates on the physics potential with 3000 $fb^{-1}$ in the decay channel $Z' \rightarrow t\bar{t}$. The $Z' \rightarrow t\bar{t}$ search comprises of two event categories:

- The **lepton+jets** channel as described in Ref. [12].
- The **all-hadronic** channel as described in Ref. [13].

The individual analyses use the 2015 LHC dataset corresponding to an integrated luminosity of 2.6 $fb^{-1}$. A combination of the baseline analyses is not publicly available, the results here are shown separately for the two event categories.

The projections are performed for two signal models: a narrow $Z'$ signal hypothesis [14], where the width of the resonance is set to 1% of the resonance mass, and a Randall-Sundrum Kaluza-Klein gluon resonance [15], where the resonance width is approximately 16% of the resonance mass. We use simulated events with masses up to, but not exceeding, 4 TeV, for both the $Z'$ and RS KK gluon signal models. Analysis above 4 TeV is challenging due to the large off-shell component important at high masses for wide-width signals.

## 4.1 Methodology of the extrapolation

The extrapolation is based on the analysis using 2.6 $fb^{-1}$ of 2015 LHC data, projecting to the planned 3000 $fb^{-1}$ of HL-LHC. The projection uses the existing Run-2 signal and background expectations, scaled by the 14-to-13 TeV luminosity ratio. The theta software framework [16] is used to compute expected cross section limits with these scaled templates. Table 2 lists the six all-hadronic and six lepton+jet channels that are considered. The $m_{t\bar{t}}$ distribution in each category is used for signal discrimination, as a peak on a falling background spectrum.

Table 2: Event categories used in the combination, from each of the two channels. $\Delta y$ represents the rapidity separation between the two top-tagged jets in the all-hadronic channel.

| Semileptonic Channel | All-Hadronic Channel |
|---|---|
| e + 0 b-tag + 0 top-tag | 0 subjet b-tag + $\lvert\Delta y\rvert < 1.0$ |
| e + 1 b-tag + 0 top-tag | 1 subjet b-tag + $\lvert\Delta y\rvert < 1.0$ |
| e + 1 top-tag | 2 subjet b-tag + $\lvert\Delta y\rvert < 1.0$ |
| $\mu$ + 0 b-tag + 0 top-tag | 0 subjet b-tag + $\lvert\Delta y\rvert > 1.0$ |
| $\mu$ + 1 b-tag + 0 top-tag | 1 subjet b-tag + $\lvert\Delta y\rvert > 1.0$ |
| $\mu$ + 1 top-tag | 2 subjet b-tag + $\lvert\Delta y\rvert > 1.0$ |

## 4.2 Systematic uncertainties

Two projections are made based on assumptions about the systematic uncertainties:

- **Current systematics** - same as in Run-2 baseline analysis, without scaling of the uncertainties.
- **Without any systematics** - only statistical uncertainties are included and scaled appropriately with the background and signal yield estimates.

The first projection uses the current uncertainties, with no improvements added. For example, the non-top multijet (NTMJ) background component for the all-hadronic channel is estimated using a data-driven approach, and improvements in the associated errors are expected when performing future analyses with larger datasets. Contributions to uncertainties from cross section measurements will also improve, as well as other contributions from components like jet energy scale, resolution, and lepton identification efficiency.

The dominant source of uncertainty in the all-hadronic channel is in the non-top multijet background. This is determined using a mistag rate which carries a momentum-dependent uncertainty of 5–100% depending on the b-tag content of the event. A corresponding mistag rate uncertainty of 19% is also applied in the semileptonic channel. Other important uncertainties include those applied to the simulated $t\bar{t}$ events, including uncertainties related to the choice of parton distribution functions as well as the scales used for the matrix element generation and parton shower evolution, which can be of order 10–20%. See the individual analysis documentation [12, 13] for further details on each of the uncertainty components.

In the second scenario, all systematic uncertainties are ignored, assuming only statistical uncertainties. This scenario yields the best possible limit with the existing analysis techniques. It assumes perfect knowledge of all the background components and associated modeling effects.

## 4.3 Projected exclusion reach

The projections in terms of 95% C.L. exclusion are shown in Fig. 4. In the first scenario of "current systematics", the expectation is to exclude the narrow $Z'$ model up to 3.3 TeV masses, and the RS KK Gluon model up to 4 TeV. For the second case where only statistical uncertainties are considered, signal models are excluded to well beyond the 4 TeV limit of this analysis. However, for the highest resonance masses, off-shell production of the $Z'$ becomes important, and the reconstructed $m_{t\bar{t}}$ does not peak at the resonance mass value. The analysis as presently designed will lose sensitivity quickly to the 5 TeV and higher-mass $Z'$ bosons. Therefore, a different analysis stategy should be designed and optimized for the off-shell decays of high mass resonances, which generally have less-boosted top quarks in the final state. Cross section limits of less than 1 fb and a few fb are obtained from the narrow $Z'$ and the RS KK gluon analysis, respectively.

## 4.4 Projected discovery reach

In addition to projections of exclusion limits, we also project expected discovery sensitivities in the possible presence of a new physics signal. The discovery sensitivities are estimated by using toy datasets with different amounts of injected signal. The p-values for these hypothesized datasets, compared to the null-signal hypothesis, are used to compute expected significances, reported as the number of standard deviations. The same two scenarios are examined regarding the systematic uncertainties, reusing the two channels. Figure 5 shows these results for the lepton+jets and all-hadronic channels. The significances are reported in the range of resonance cross section and resonance mass, for two width scenarios. This allows the estimation of sensitivities for arbitrary models with similar widths, if the mass and cross section are known.

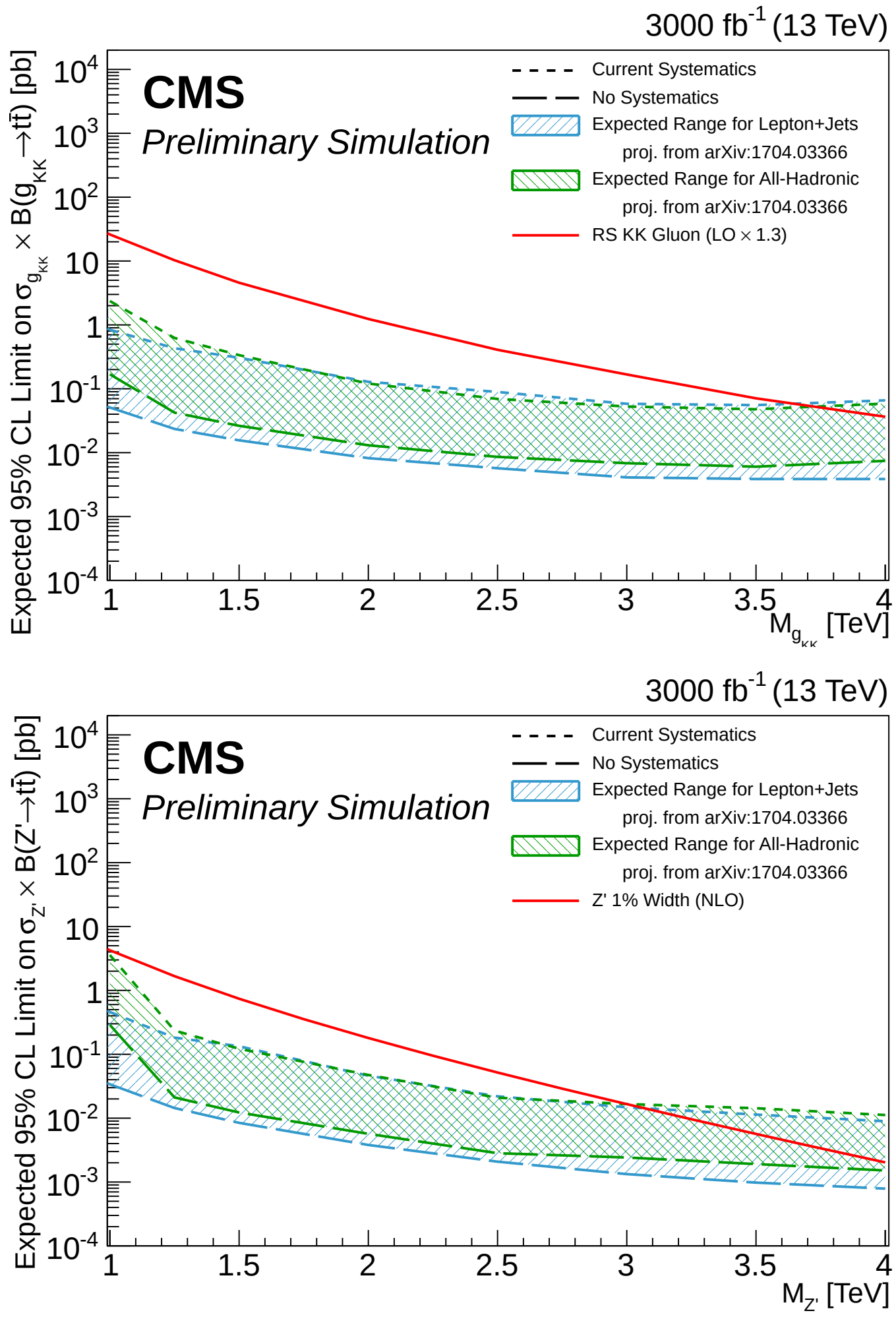

Figure 4: Projected ranges of cross section limits expected for 3000 fb$^{-1}$ of HL-LHC running, shown individually for the lepton+jets (blue) and all-hadronic event (green) categories. The short-dashed line shows the median expected limits using full systematics from the Run-2 analyses [17] assuming no improvements in systematic uncertainties. The long-dashed line shows the same when applying no systematic uncertainties.

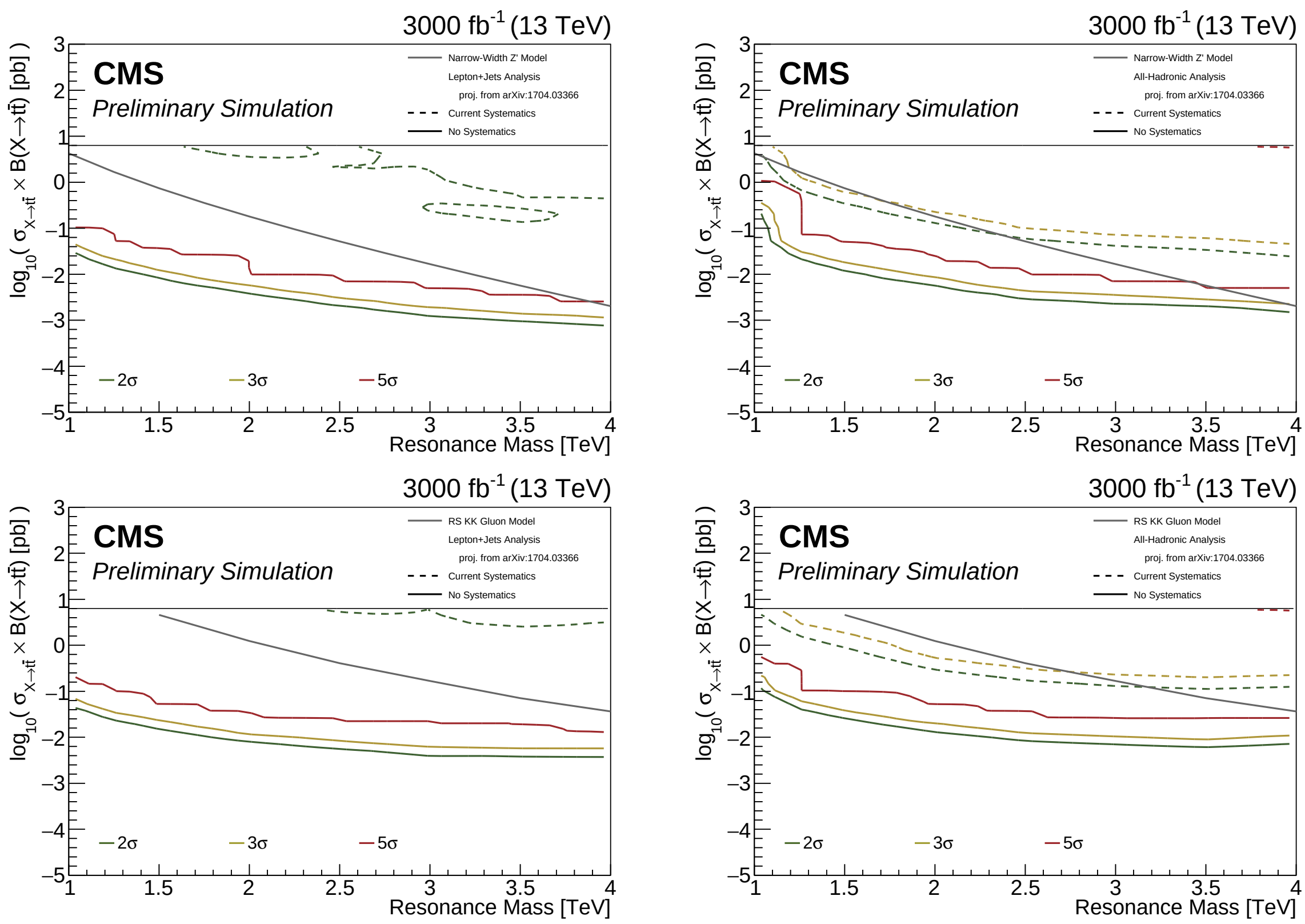

Figure 5: Discovery sensitivities for the lepton+jets channel (left column) and all-hadronic channel (right column), for 3000 fb$^{-1}$. The results are presented in the plane of the cross section versus the resonance mass, with the color contours representing the boundaries of areas with significances larger than 2, 3, or 5 standard deviations. The results are shown for the narrow-width signal hypothesis (top row) and RS KK gluon signal hypothesis (bottom row), with the "current systematic" uncertainties scenario from the Run-2 analysis [17] shown by the dashed lines and the "no systematic uncertainties" scenario shown by the solid lines.

# 5 Dark matter analysis

The search and/or characterization of dark matter (DM) in the form of Weakly Interacting Massive Particles (WIMPs) will be one of the top priorities of the HL-LHC. This section discusses the projected constraints on certain benchmark simplified dark matter models using the mono-jet search employing the signature of jets and missing transverse momentum.

This analysis uses DELPHES simulated signal and background samples and performs a full signal event selection which follows the actual Run-2 analysis described in Ref. [18] as closely as possible.

The simplified models of dark matter considered for these projections are the following with the corresponding Feynman diagrams for both processes in Fig. 6:

- s-channel DM pair production with an axial vector mediator with the couplings of the mediator to DM ($g_{DM}$) = 1.0 and to the SM ($g_{SM}$) =0.25.
- s-channel production via a pseudoscalar mediator with the couplings of the mediator to DM ($g_{DM}$) = 1.0 and to the SM ($g_{SM}$) = 1.0.

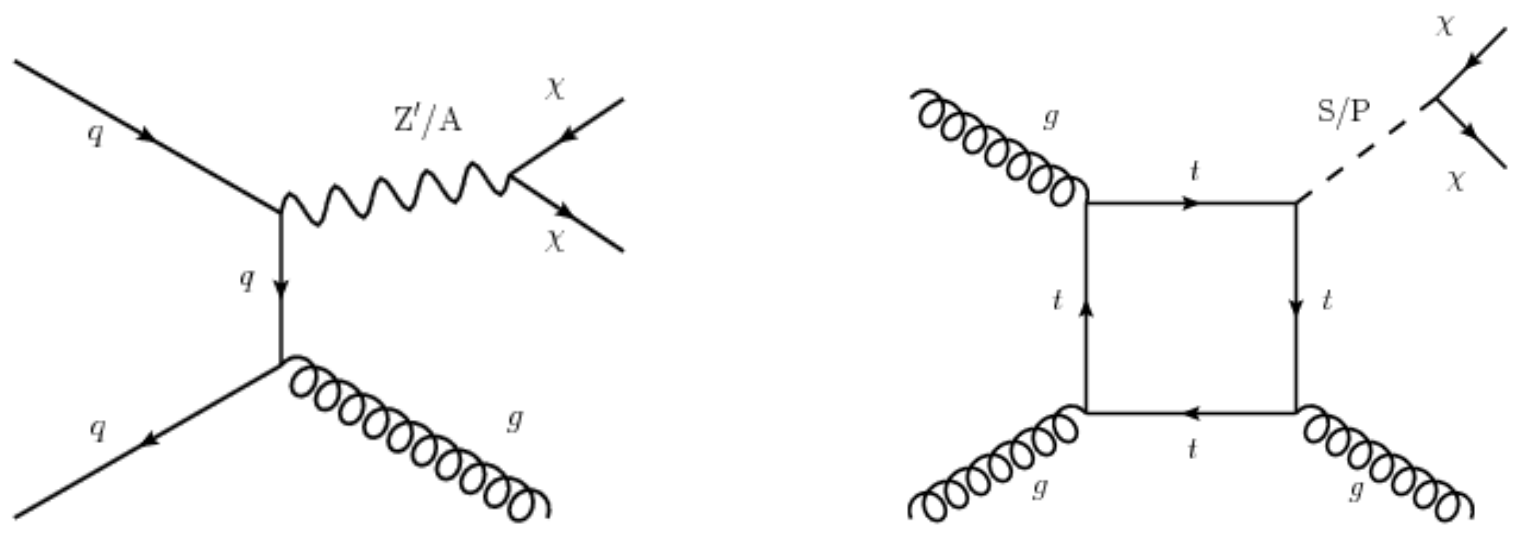

Figure 6: Feynman diagrams of DM pair production for an axial vector and pseudoscalar mediated interaction.

Constraints on the axial vector (AV) interaction can be translated to limits on spin-dependent DM-nucleon interactions and compared to those from the direct detection experiments. The results of searches for DM at the LHC so far have shown that colliders can place competitive constraints on spin-dependent interactions for this simplified model. For the pseudoscalar mediated model (PS) shown in Fig. 6, the LHC is uniquely placed to probe this interaction as it leads to velocity suppressed scattering cross sections for the direct detection experiments and is effectively inaccessible to them. Both models thus represent well-motivated benchmarks to study the projections of the HL-LHC.

## 5.1 Analysis strategy and event selection

Before the projection, the DELPHES implementation has been validated with respect to the Run-2 analysis with 13 $fb^{-1}$ at 13 TeV [18]. Details of the event selection for are presented in Tab. 3. The jet collection of AK4 jets is used for the validation study as well as for the ECFA projection in the 0 pileup (PU) scenario. Because the sensitivity of the analysis is dominated by events with very large MET, the effects from high pileup are not expected to cause a significant decrease in the expected sensitivity. Studies with the upgraded Phase-2 detector including the track trigger indicate that the Phase-2 trigger algorithms will allow to keep the thresholds around this value even in an environment with 200 PU events.

Signal samples are simulated with Powheg [19, 20] and subsequently passed through the DELPHES simulation with Phase-2 detector performance. A signal would manifest itself as an excess in the $E_T^{miss}$ distribution after requiring large $E_T^{miss}$ and a jet. This $E_T^{miss}$ distribution is the discriminating variable, displayed in Fig. 7 after the full event selection from Tab. 3. Also shown are signal examples for the scenario of an axial vector interaction for the example DM and mediator masses given in the legend. The signal-to-background ratio improves with increasing $E_T^{miss}$. The dominant background is due to $Z(\nu\nu)$ and $W(\ell\nu)$+j and is taken from simulation. It is labeled V+jets in Fig. 7.

Table 3: Summary of the event selection criteria used to select monojet events for this analysis.

| Event selection | |
|---|---|
| AK4 jets | $p_T\ (j_1) > 250$ for AV (200 for PS), $\lvert\eta\rvert < 2.5$ |
| $\Delta\phi$(jet, $E_T^{miss}$) | $\Delta\phi > 0.5$ |
| veto electrons | $p_T > 10, \lvert\eta\rvert < 2.4$ |
| veto muons | $p_T > 10, \lvert\eta\rvert < 2.5$ |
| veto taus | $p_T > 18, \lvert\eta\rvert < 2.3$ |
| b-jet veto | 'Loose', $p_T > 15, \lvert\eta\rvert < 2.5$ |
| $E_T^{miss}$ | $E_T^{miss} > 200$ GeV |

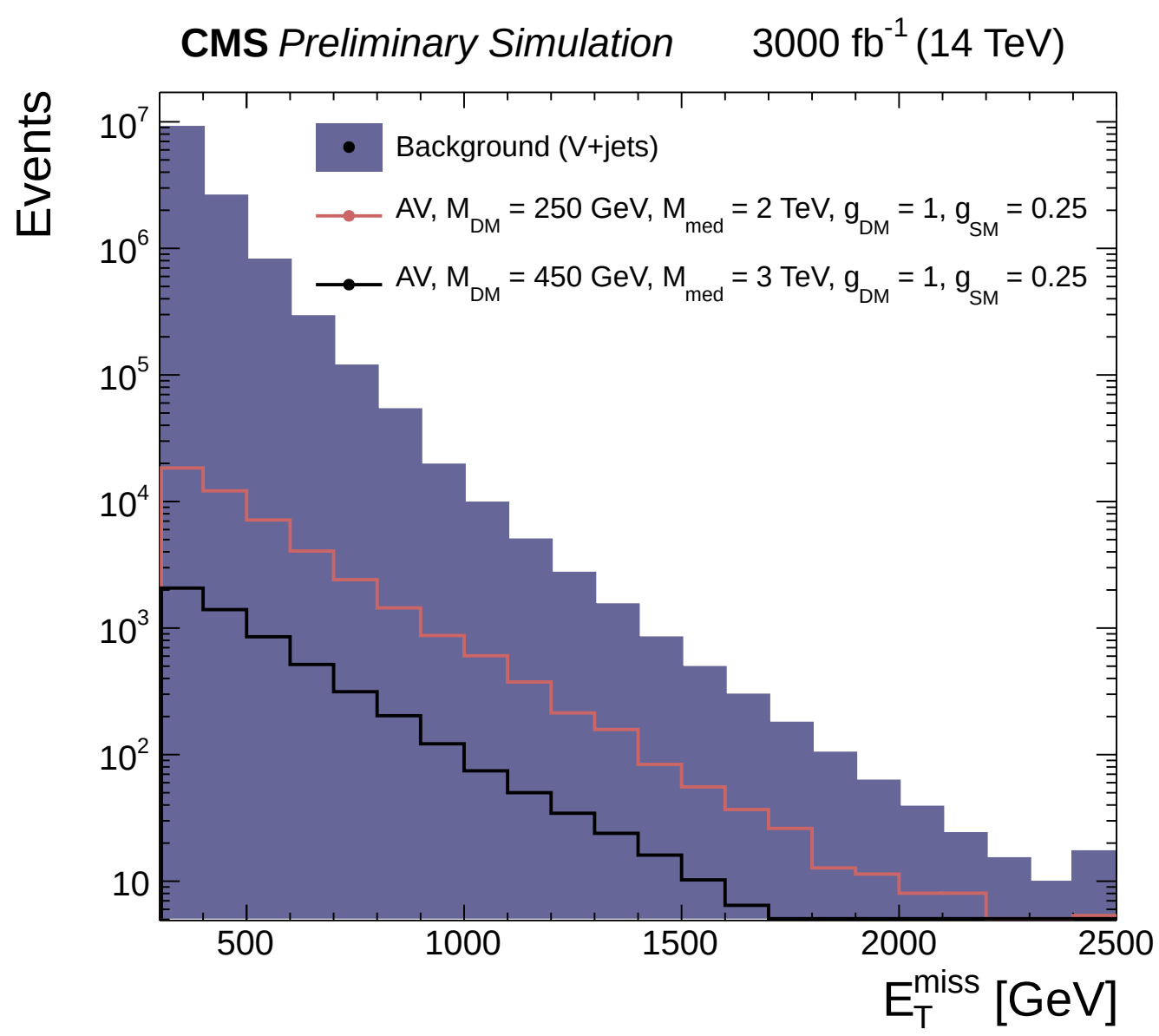

Figure 7: Distribution of the discriminating variable, $E_T^{miss}$, after full event selection. The V+jets background is taken from simulation. Two signal examples are shown for the axial vector model with the model parameters given in the legend.

## 5.2 Systematic uncertainties

The region of $E_T^{miss}$ dominating sensitivity to the two signal models chosen for the ECFA projections are different and hence the sources of systematic uncertainties. For the axial vector model, the tail of the $E_T^{miss}$ distribution plays the dominant role while for the pseudoscalar model it is bulk/low $E_T^{miss}$ region that provides the greatest sensitivity.

- For the axial vector model, the $E_T^{miss}$ range is extended to 2.4 TeV while presently the maximum $E_T^{miss}$ bin is at 1.2 TeV. The "current systematic" scenario is where the same systematic uncertainties on the $E_T^{miss}$ distribution in the current monojet analysis are used for the ECFA analysis with the extended $E_T^{miss}$ range, so the uncertainty in the last bin of the $E_{\mathrm{T}}^{\mathrm{miss}}$ is 10%. Other scenarios considered are, reducing the current uncertainties by a factor of 2 and a factor of 4.

- To address the variation in systematic error, the $E_{\mathrm{T}}^{\mathrm{miss}}$ distribution is divided into a low and high $E_{\mathrm{T}}^{\mathrm{miss}}$ region, where low $E_{\mathrm{T}}^{\mathrm{miss}}$ is $<500\,\mathrm{GeV}$ and high $E_{\mathrm{T}}^{\mathrm{miss}}$ is $>500\,\mathrm{GeV}$. The dominant systematic uncertainty in the low $E_{\mathrm{T}}^{\mathrm{miss}}$ region comes from the uncertainty on the lepton identification/isolation efficiency via the selection of the dilepton control sample which provides the dominant contribution to the estimation of the Z(vv) background and also the single muon control sample which predicts the W(lv) background. The low $E_{\mathrm{T}}^{\mathrm{miss}}$ region is hence systematics dominated and a systematic uncertainty of 1% per leg is taken for the ECFA $3000\,\mathrm{fb}^{-1}$ projection, compared to the current uncertainty of 2% per leg. The uncertainty in the high $E_{\mathrm{T}}^{\mathrm{miss}}$ region is dominated by the size of the control samples used for estimating the V+jets background. The uncertainty in this region is taken from the current monojet analysis and scaled by luminosity. The above-mentioned scenario is the "current systematics extrapolated to HL-LHC" scenario for the PS model. Other systematic scenarios studied are: the current systematics scaled down by a factor of 2, and, the uncertainties in the full $E_{\mathrm{T}}^{\mathrm{miss}}$ region taken from the CMS monojet analysis and scaled by luminosity.

### 5.3 Projected exclusion reach

Following the simplified model parametrization, the sensitivity is studied in terms of mediator mass, $M_{\mathrm{med}}$, and dark matter mass, $M_{\mathrm{DM}}$. The coupling values were given previously and are kept constant. The projected exclusion reach for $3000\,\mathrm{fb}^{-1}$ for both studied DM models is depicted in Fig. 8. The limits at 95% confidence level derived with the CLs method are shown for three systematic scenarios. They have a large impact on the reach in mediator mass. With the present knowledge one would reach 2.5 TeV for the AV-model and 600 GeV for the PS-model, while the limit with the best "scaled" uncertainty scenario corresponds to 3 TeV (AV) and 900 GeV (PS), respectively. The reach in DM mass improves accordingly for high mediator masses.

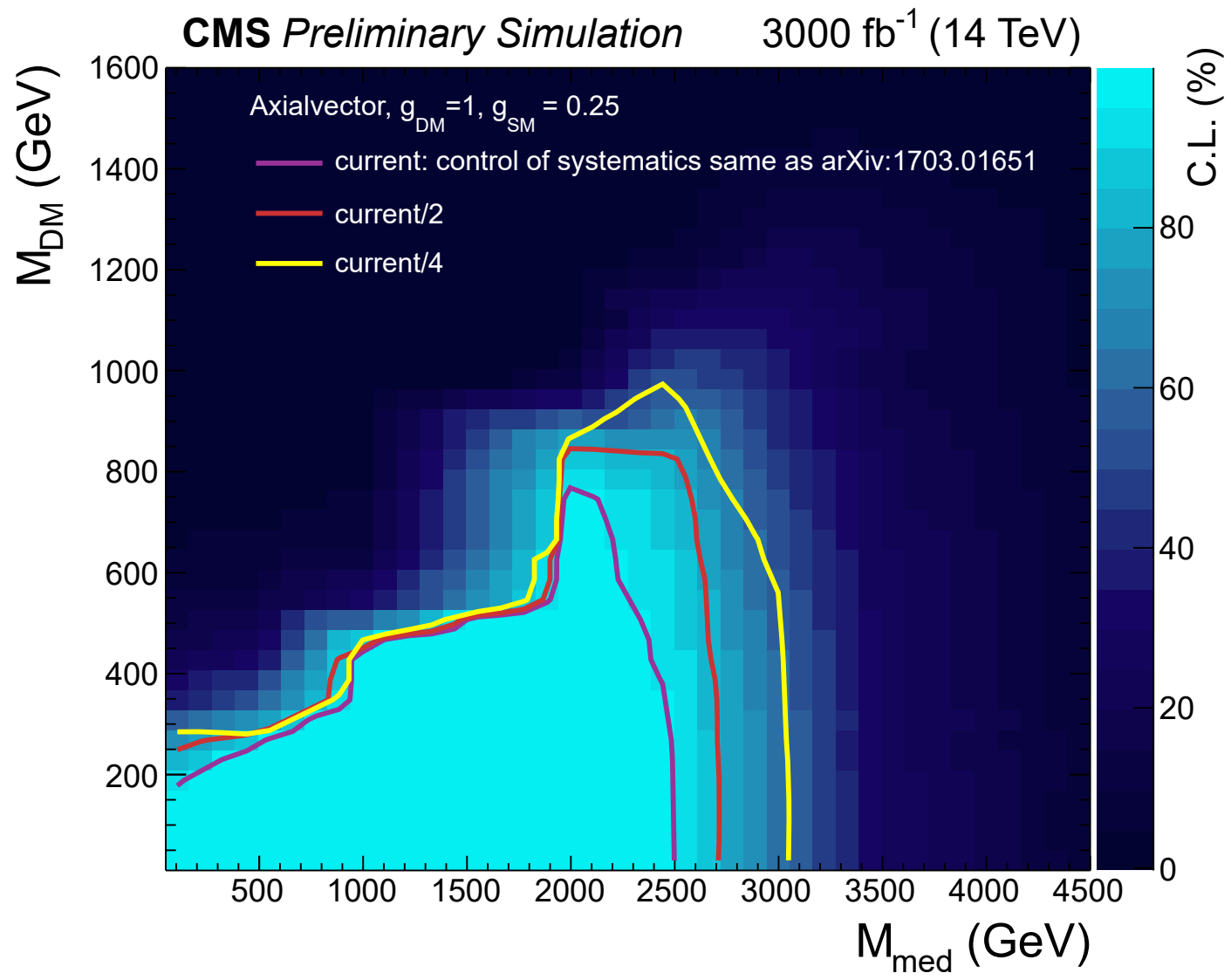

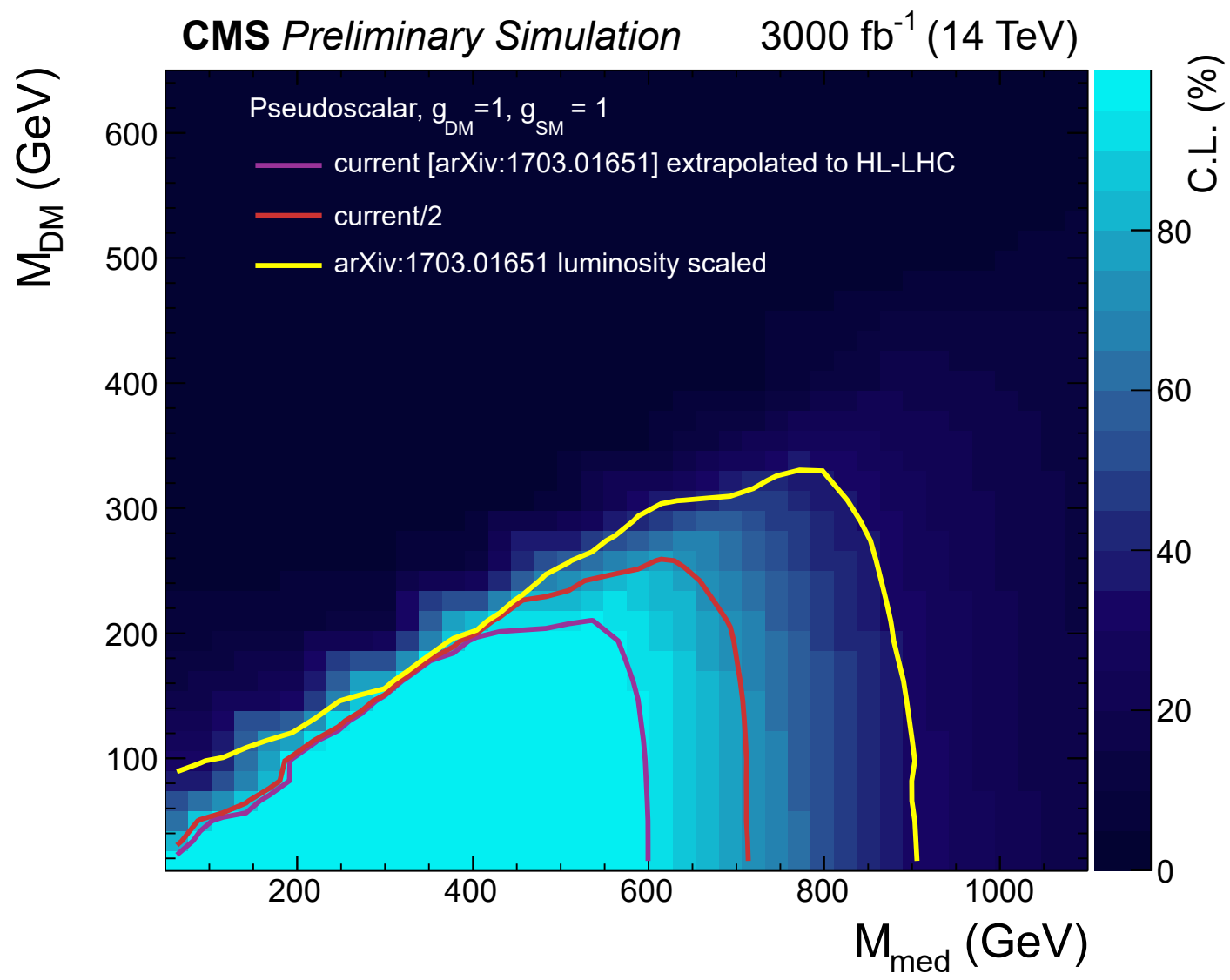

Figure 8: Projected exclusion limits at 95% C.L. for 3000 fb$^{-1}$ of HL-LHC statistics for two simplified dark matter models using the monojet analysis. On top the axial vector mediated simplified DM model ($g_{DM}$ =1, $g_{SM}$ =0.25), on the bottom the pseudoscalar mediated model ($g_{DM}$ =1, $g_{SM}$ =1). The limits are shown for three systematic scenarios. For the AV model: a "current" scenario assumes that the level of systematic control in the high $E_T^{miss}$ region is the same as the Run-2 analysis [18], while the "current/2" scenario scales it down by a factor of 2, and the "current/4" scenario by a factor of 4. For the PS model: a "current" scenario where the low $E_T^{miss}$ region is dominated by systematic uncertainties and the uncertainties in the high $E_T^{miss}$ region are taken from the Run-2 analysis and scaled by luminosity, the "current/2" scenario is the nominal systematics scaled down by a factor of 2, and the "luminosity scaled" scenario takes the uncertainties from the current analysis and scales by luminosity for the full $E_T^{miss}$ range.

# 6 Single vector-like quark T decaying to tH

Many SM extensions contain vector-like quarks (VLQ) which preferably mix with third generation quarks [21]. Such a particle could have a role in stabilizing the Higgs mass, and thus offers a potential solution to the hierarchy problem.

The analysis searches for the electroweak production of a vector-like partner of the top quark (T) decaying to a top quark and a Higgs boson ($T \to tH$) assuming 3000 fb$^{-1}$ of proton-proton collision data at 14 TeV. Much like the top quark itself, a vector-like top quark can be produced either in pairs dominantly through the strong interaction, or singly in association with additional quarks through the electroweak interaction via diagrams such as those depicted in Fig. 9. For pair production, lower limits at 95% C.L. on the mass between 720 and 920 GeV have been set depending on decay mode [22]. For very massive VLQs above TeV range, the pair-production cross section rapidly decreases as the phase space for producing two massive particles is limited. Hence, in this regime the single production via the electroweak process is expected to dominate over pair production [21]. In this search, the single T can be produced through the processes $qg \to T\bar{b}q'$ and $qg \to T\bar{t}q'$, and their charge conjugates. The production cross sections of single T quark and the branching fraction of $\mathcal{B}(T \to tH)$ depends on the strength of the electroweak couplings at the production vertex, i.e, $c^{bW}_{L/R}$ for charged and $c^{bZ}_{L/R}$ for neutral current interactions up to a factor of the electroweak coupling constant $g_W$. In this search we consider the simplest Simplified Model [23] for a singlet and a doublet T quark, where only the LH coupling $c^{bW}_{L}$ is allowed for the singlet case, and RH coupling $c^{bZ}_{R}$ for the doublet case. Therefore, we only focus on these two models in this search.

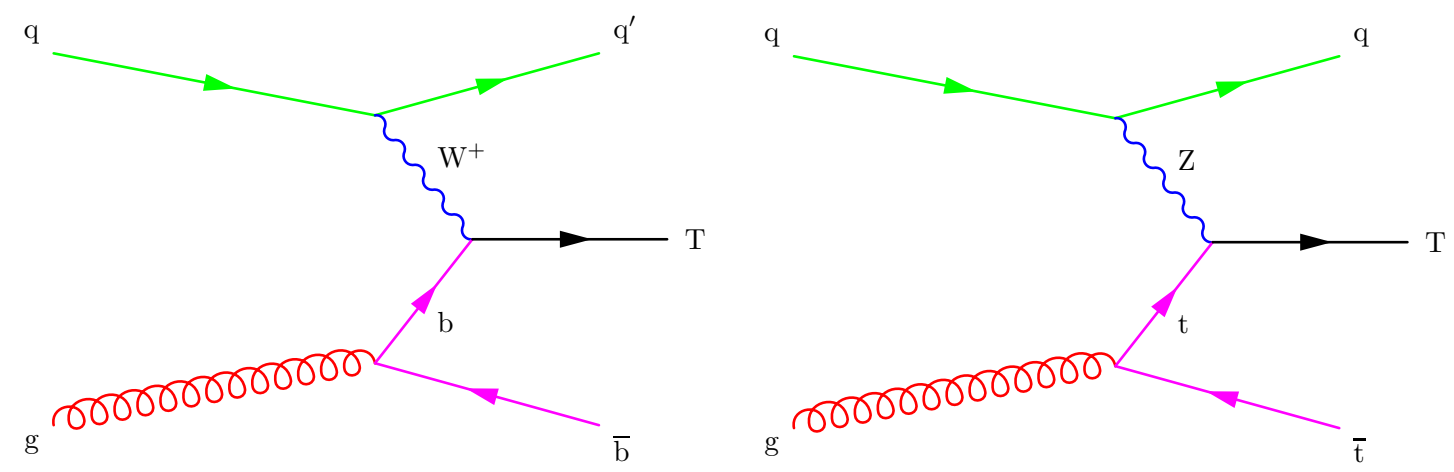

Figure 9: Example production diagrams. Charged-current (left) and neutral current (right).

The $qg \to T\bar{b}q'$ process, where a T decays into a semileptonically decaying top quark and Higgs boson, decaying into two b-quarks leading to the final state $qg \to (\ell\, \nu\, b)(b\bar{b})\bar{b}q'$ consisting of a lepton, missing energy from the neutrino, and possibly 4 b jets. The event signature has a very forward jet which can benefit from the plans to increase the acceptance of the tracker to $|\eta| = 4$. The forward tracking should distinguish primary vertices from a very high pileup of 200, hence reducing the fake background in forward region. For high values of the T mass, it is expected that the large boost from the decay, will lead to the decay products from the top quark, and the jets from the Higgs to become progressively more and more merged.

## 6.1 Analysis strategy

This study is a full analysis based on DELPHES using the Phase-2 performance from the technical proposal [4]. Samples for the process pp→T b, T →tH were generated using the leading order event generator MADGRAPH 5.2.3.30. [24]. The benchmark T quark masses used for the final result are 1, 1.5, 2.0, 2.5, and 3.0 TeV. The MADGRAPH samples are generated with an additional parton and interfaced with PYTHIA8 [25]. The NNPDF parton distribution function (PDF) was used [26]. The samples have the t decaying inclusively and H decaying 100% to $b\bar{b}$, with the mass of the Higgs set to 125 GeV. The mean pileup was set to 200 interactions per events. Separate samples were generated for left-handed (right-handed) couplings of the $c_L^{bW}$ ($c_R^{tZ}$ ) vertices for the $T \rightarrow tH$ decay using narrow width of 10 GeV.

The main SM backgrounds are: $t\bar{t}$ +jets, V+jets, single top and diboson events. The backgrounds are binned in $H_T$ and simulated using MADGRAPH interfaced with PYTHIA8. The backgrounds are normalized using the NLO cross sections, except for $t\bar{t}$ +jets and V+jets, where a k-factor of 1.68 and 1.23 is used respectively to normalize them to NNLO cross sections.

## 6.2 Event Selection

The event selection assumes the Phase-2 detector geometry including the increased acceptance. The object selection largely follows the present selection steps adapted to the Phase-2 detector performance. One requires events with one electron or muon with $p_T > 40$ GeV and $|\eta| < 4.0$. Jets and $E_T^{miss}$ are reconstructed with a new algorithm (denoted PUPPI) targeting high PU scenarios, which is an extension of the particle flow algorithm with charged hadron subtraction giving weights to particles based on the probability that they come from pileup or the primary vertex. Jets overlapping with leptons are removed if their separation within a cone $\Delta R(AK4, \ell)$ is greater than 0.4 and $\Delta p_T$ (rel) exceeds 40 GeV. This selection is assumed to suppress any QCD backgrounds in events. In addition at least one forward jet within the acceptance $2.4 < |\eta| < 5.0$ and $p_T > 30$ GeV is required. Also at least two central jets with $|\eta| < 2.4$, where the first leading jet has $p_T > 200$ GeV, the second leading jet has $p_T > 80$ GeV. At least one jet has to be identified as a b-jet with a tagging efficiency of around 70% and mistag rate of 5%. The $E_T^{miss}$ has to be greater than 20 GeV.

Higgs candidates are identified using boosted AK8 jets, where AK8 jets are defined as jets clustered within a cone size of radius 0.8. The AK8 jets with $p_T > 300$ GeV and $|\eta| < 2.4$ are first cleaned to the non-prompt leptons such as jets are rejected if they overlap with leptons within a cone $\Delta R(AK8, \ell) < 0.4$ radius, and $\Delta p_T$ (rel) $> 40$. The soft drop algorithm is used to identify subjets in a AK8 jet that are compatible with two body decay. Therefore to tag a Higgs boson, exactly two soft drop subjets are required with jet shape N-subjettiness variable $\tau_{21} < 0.6$ [27] and the soft drop jet mass within 90-160 GeV. Due to unavailability of the b-tagging on the subjets in DELPHES, no subjet b-tagging is applied to the subjets. To avoid the ambiguity between a Higgs candidate and a prompt lepton from a top quark decay of a T quark, the Higgs candidates are rejected if $\Delta R(H, \ell) < 1$.

The T mass is reconstructed by first identifying a top quark candidate and then combining it with a Higgs boson candidate using a $\chi^2$ minimization. To identify a top quark candidate decaying into a semileptonically decaying W boson and a b-quark, first the neutrino $p_Z$ solution is obtained by solving a quadratic equation using the following kinematic constraints $m(\ell\nu) = m(W) = 80.399$ GeV. Out of the two solutions, the smaller is kept. In case of solely one imaginary solution, its real part is used. Therefore using the neutrino and lepton four momentum, a top candidate is formed by combining them with one or two AK4 jets, and keeping the combination that results from the minimization the following $\chi^2$ function

$$\chi^2 = (\frac{M_{H,MC} - M_{H,rec}}{\sigma_{M_H,MC}})^2 + (\frac{M_{t,MC} - M_{t,rec}}{\sigma_{M_t,MC}})^2 + (\frac{dR(t,H)_{MC} - dR(t,H)_{rec}}{\sigma_{dR,MC}})^2.$$

Here $M_H$ is defined as Higgs mass, $M_t$ as top mass, and $\Delta R(t, H)$ as separation between a top quark and a Higgs boson candidate. The mean and widths for $\chi^2$ event are taken from generator level studies [28]. Only combinations that pass the requirement of $\Delta R$(AK4 jet, H) $> 1.0$ and $\Delta R$(t,H) $> 2$ are considered for the statistical analysis.

Figure 10 show the reconstructed mass distribution ($M_{T,\text{reco}}$) along with signal examples for various T masses as given in the legend. Signal efficiencies are about 4% for Tbq and about 3% for Ttq with only a light dependence on the T mass.

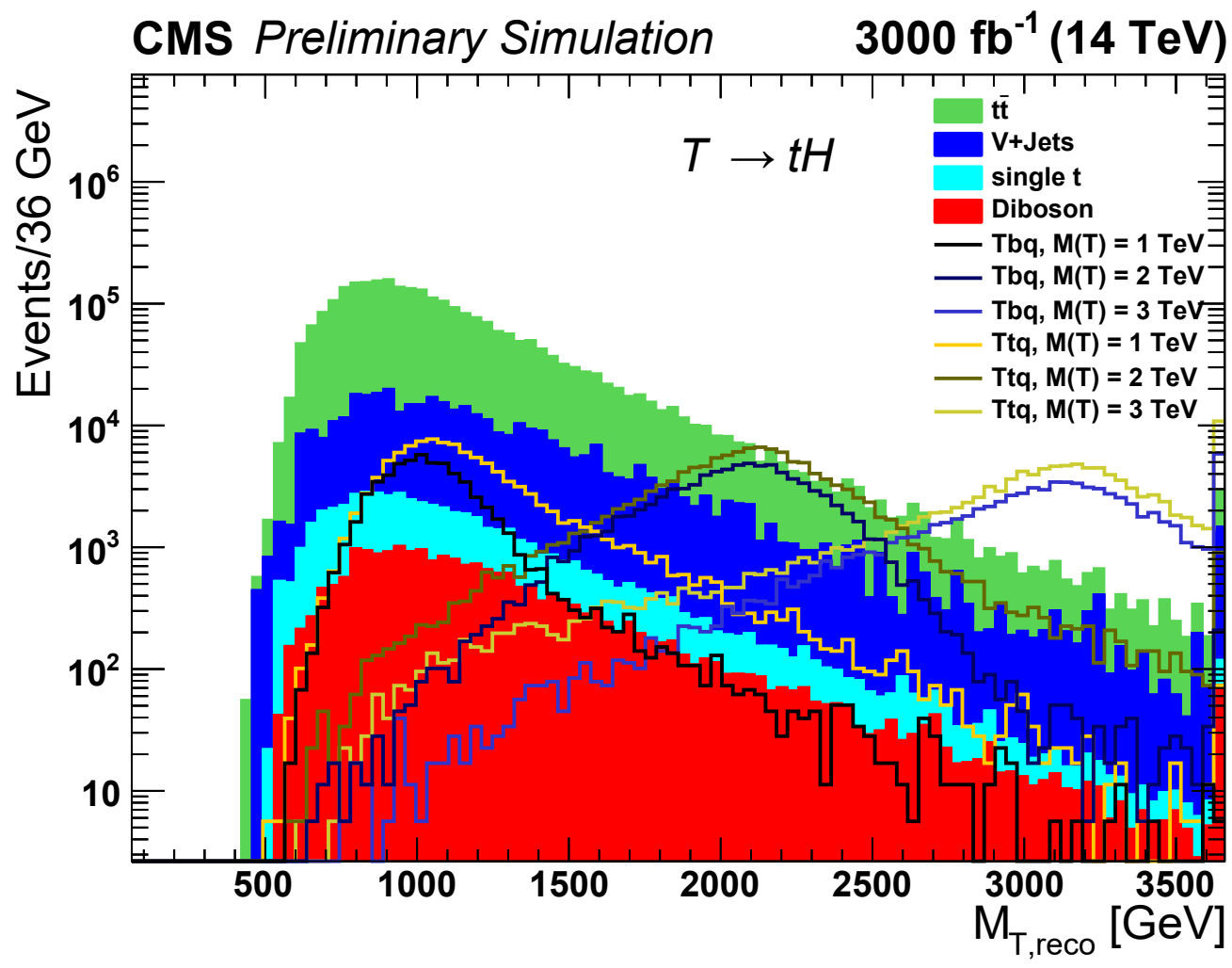

Figure 10: Distributions of the reconstructed mass of the T quark decaying into a top quark and a Higgs boson. The top quark further decays leptonically and the Higgs boson into a pair of $b\bar{b}$ quarks, leading to the final state of qg $\rightarrow (\ell\,\nu\,b)(b\bar{b})(\bar{b}q'/\bar{t}q')$. Selected signal samples of T masses of 1, 2, and 3 TeV from the processes pp $\rightarrow$ Tbq (Ttq), with left-handed (right-handed) couplings to the SM third generation quarks are overlaid on the total estimated background.

## 6.3 Systematic uncertainties

The main SM background in this search are $t\bar{t}$ + jets events, which are normalized to the NNLO cross section. We consider a total of 27% uncertainty on $t\bar{t}$ + jets normalization, which is 1/2 of the total uncertainty on $t\bar{t}$ theory cross section due to PDF and QCD scale, and top quark mass. We keep the same uncertainty on the single top quark background, since for the limit setting procedure, we combine the two samples and treat them as one template. Another large background is V + jets, where V can be a W or a Z boson. This background is combined with the smaller diboson backgrounds and a total of 20% conservative uncertainty is assigned on their normalization.

The largest shape uncertainties comes from b-tagging. To estimate them, we vary the nominal b-tagging SF at medium efficiency working point by scaling it up and down by 1% for b jets, 2% for c jets, and 5% for the light jet, and use the resultant shape template in our statistical analysis. For the jet energy scale an estimated flat uncertainty of 3.8% is applied. In Run-2 for Higgs-tagging, the measured uncertainty on jet mass and N-subjettiness selection scale factors are found to be 1.03±0.13, and we expect the uncertainty to improve with new tagging tools and hence do not apply any uncertainty due to these sources. Other uncertainties include 1% on jet energy resolution, 1.5% on luminosity, 1% on presumed trigger scale factor, 1% on lepton reconstruction and identification.

Without available QCD simulation, it is very difficult to estimate the QCD contribution, one does not expect QCD backgrounds in the signal region after the full event selection though. However, we checked the impact on expected limits by constraining the normalization of total background to 0%, 10%, 50% and 100% and found the median of the limits on cross section times branching ratio varies around 1%. Due to this observation and the analysis being shape based, we decided to drop any additional uncertainty due to QCD background.

## 6.4 Results

The Higgs combine package [29] has been used for the limit-setting procedure. A simultaneous fit of the background and the signal $M_T$ distributions is performed, with the systematic uncertainties treated as nuisance parameters with log-normal priors. A binned likelihood fit with Bayesian algorithm (asymptotic) is used to obtain a 95% C.L. upper limit on the signal strength. The expected limits for different mass hypotheses of the T quark are computed using the M(tH) distributions for the background and the signal. The results are shown in Fig. 11 and Tab. 4. The two scenarios are considered for signal interpretation. First is the singlet T quark production through pp $\rightarrow$ Tbq process, where only left-handed coupling ($c_L^{bW}$) of T quark to the third generation quark is allowed, and the second is production of doublet T quark through pp $\rightarrow$ Tbq process, where only right-handed coupling ($c_R^{tZ}$) is allowed. According to the equivalence theorem [30–32], a T quark can decay into bW, tZ or tH channels with the benchmark branching fraction of $\mathcal{B}(T \rightarrow bW) : \mathcal{B}(T \rightarrow tZ) : \mathcal{B}(T \rightarrow tH) = 0.5 : 0.25 : 0.25$. However, considering the simplest Simplified Model, this is only valid for a singlet T quark with $c_L^{bW}$ coupling. For the doublet T quark with $c_R^{tZ}$ coupling, we considered the benchmark branching fraction of $\mathcal{B}(T \rightarrow bW) : \mathcal{B}(T \rightarrow tZ) : \mathcal{B}(T \rightarrow tH) = 0.0 : 0.50 : 0.50$. The couplings $c_L^{bW}$ and $c_R^{tZ}$ are chosen as 0.5 due to the fact that signal simulations were performed with a fixed width of 10 GeV under the narrow width approximation, and the theoretical width of the VLQs is negligible compared to the experimental mass resolution for values, equal to or below 0.5. In future studies coupling values higher than 0.5 will be considered as signal simulations with wider width of around 20% and 30% of the T quark mass will be studied. In addition, with improved analysis methods such as the usage of subjet b-tagging in Higgs identification, the sensitivity to smaller couplings is expected to improve for T production through both the pp $\rightarrow$ Tbq and pp $\rightarrow$ Ttq processes.

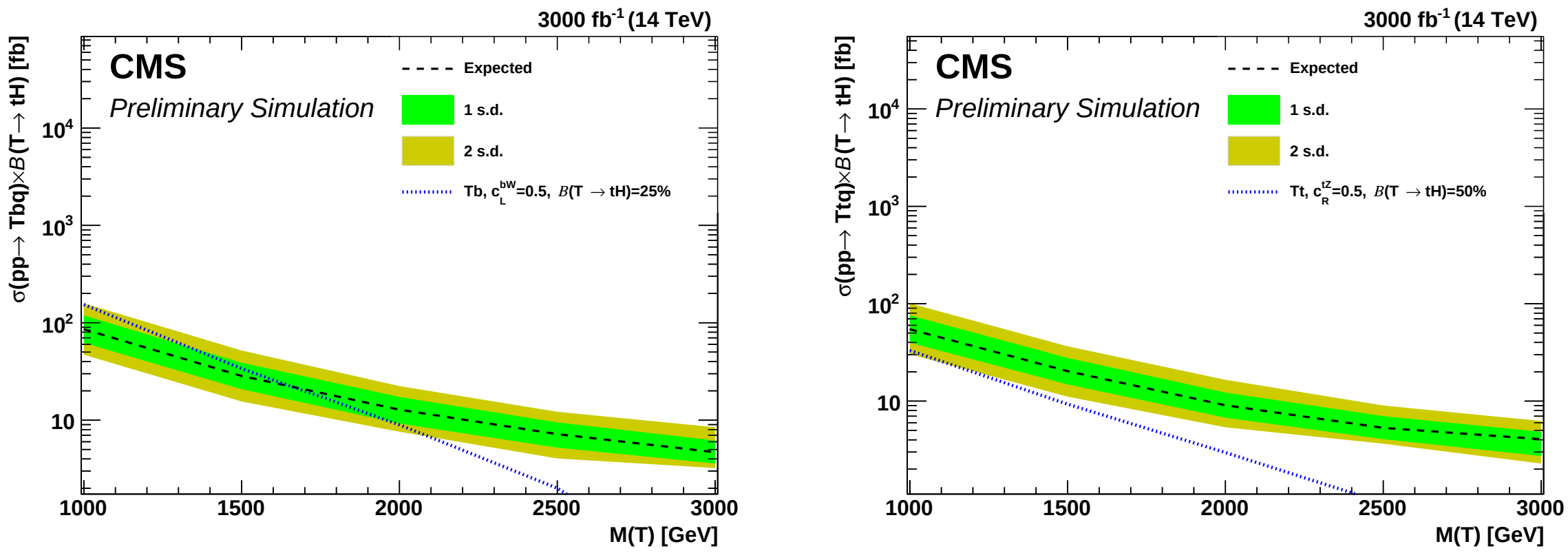

Figure 11: The expected limits at 95% C.L. on the $\sigma \times \mathcal{B}(T \rightarrow tH)$ of a T quark for different mass assumption of 1, 1.5, 2, 2.5, and 3 TeV, at an integrated luminosity of 3000 fb$^{-1}$. The left (right) plot shows the results for the process pp $\rightarrow$ Tbq (pp $\rightarrow$ Ttq) with left-handed (right-handed) coupling to the third generation SM quarks as described in models in Refs. [23, 33]. The dotted blue line in left (right) plot is the theory cross section assuming 0.5 coupling strength of the T quark to a W (Z) boson, and $\mathcal{B}$ (T $\rightarrow$ tH $=$ 0.25 (0.50)), and is obtained by scaling the NLO cross sections at 13 TeV to the k-factor obtained at 14 TeV with `CTEQ6L` PDF.

Table 4: The median expected upper limits at 95% C.L. on the cross section $\times\mathcal{B}(\mathrm{T} \rightarrow \mathrm{tH})$ of the T quark for the models pp $\rightarrow$ Tbq (Tbq) with $\mathcal{B}(\mathrm{T} \rightarrow \mathrm{tH} = 0.25)$ and pp $\rightarrow$ Ttq (Ttq) with $\mathcal{B}(\mathrm{T} \rightarrow \mathrm{tH} = 0.50)$ for different mass hypotheses and left-handed or right-handed couplings respectively. An integrated luminosity of 3000 fb$^{-1}$ at proton-proton collision at $\sqrt{s}$ = 14 TeV is assumed.

| Mass (GeV) | Expected cross section upper limit (fb) | |
|---|---|---|
| | Tbq (LH) | Ttq (RH) |
| 1000 | 85.9 | 54.7 |
| 1500 | 28.4 | 20.3 |
| 2000 | 12.8 | 9.06 |
| 2500 | 7.20 | 4.64 |
| 3000 | 4.69 | 4.69 |

# 7 Summary and conclusion

The physics reach with 3000 fb$^{-1}$ of HL-LHC data is studied in a number of searches for new physics. The projections described here demonstrate the gain from high-luminosity.

Discovering the nature of DM is an important problem in physics. The LHC is performing collider searches for dark matter and its mediators. One of the most common searches is the monojet search with j+MET in the final state. Based on the Run-2 event selection, the analysis for the two interesting couplings scenarios, namely the axial vector and pseudoscalar coupling, is performed. The precise knowledge of systematic uncertainties has a significant impact on the reach in mediator mass which corresponds to 2.5 TeV without any improvement of the present understanding of the systematic uncertainties. If, on the contrary, the systematics improves by 1/4 the sensitivity increases by 20% to 3 TeV.

Projections of searches are performed for new heavy vector bosons (Z′ and W′) at the HL-LHC. Analyses performed on the 13 TeV data are projected to the HL-LHC data set of 3000 fb$^{-1}$ to determine the maximum reach of excluded boson masses as well as the discovery reach. The projections are performed under different scenarios considering systematic uncertainties. A promising search is for a right-handed $W'_R$ in the decay channel to tb yielding final states of an electron or muon together with one or two b-tagged jets. The maximum reach in terms of boson mass is 4 TeV with present systematics and above with improved systematics. In addition, model independent discovery sensitivities are presented. The projected Z′ $\rightarrow\mathrm{t\bar{t}}$ exclusion limits are also around 3-4 TeV, depending on the widths of the new resonance and the knowledge of systematics. Two different widths are studied, $\Gamma$=1% for a SSM Z′ and 16% for a RS KK gluon. The expectation is to exclude the narrow Z′ model up to 3.3 TeV masses, and the RS KK Gluon model up to 4 TeV in the scenario of Run-2 systematics. In the best case scenario of only statistical uncertainties, the reach extends for both models beyond 4 TeV pushing into a region where the analysis strategy has to be adapted to accommodate the increasing off-shell component for high resonance masses.

The discovery of the Higgs boson provided theoretical constraints to physics beyond the SM and also opens new decay channels for searches. As one example, this document presents a search for weakly produced single vector like quarks (T) decaying to a t-quark and a Higgs boson. Since both particles decay further, this challenging analysis has to reconstruct them from the final state consisting of a lepton, $E_{\mathrm{T}}^{\mathrm{miss}}$ and up to four b-quarks. Due to the forward going jets an increased sensitivity is expected from the extended acceptance of the Phase-2 detector. Considering simplest Simplified Model for a singlet and a doublet T quark, two scenarios are considered for signal interpretation: LH coupling $c_L^{bW}$ with $\mathcal{B}(T \to bW) : \mathcal{B}(T \to tZ) : \mathcal{B}(T \to tH) = 0.5 : 0.25 : 0.25$ for the singlet T quark, and RH coupling $c_R^{tZ}$ with $\mathcal{B}(T \to bW) : \mathcal{B}(T \to tZ) : \mathcal{B}(T \to tH) = 0.0 : 0.5 : 0.5$ for the doublet T quark. The expected upper cross section limits range from 85.9 fb (54.7 fb) for a T mass of 1 TeV to 4.7 fb (4.1 fb) for a T mass of 3 TeV for the singlet (doublet) T quark.

# CMS Physics Analysis Summary

Contact: cms-phys-conveners-ftr@cern.ch

# Search for supersymmetry with direct stau production at the HL-LHC with the CMS Phase-2 detector

The CMS Collaboration

## Abstract

A search for the direct production of $\tau$ sleptons ($\widetilde{\tau}$) is developed assuming 3000 fb$^{-1}$ of proton-proton collision data produced by the HL-LHC at a center-of-mass energy of 14 TeV. Three final states are investigated: two $\tau$ leptons decaying hadronically, and one $\tau$ lepton decaying hadronically and the other one decaying to a muon or electron and neutrinos. The analysis is performed using the Delphes simulation of the CMS Phase-2 detector where the object reconstruction performance is tuned to the one achieved with CMS Phase-2 full simulation. In the mass-degenerate production scenario, $\widetilde{\tau}$ masses are excluded below 650 GeV, with the discovery contour of $\widetilde{\tau}$ masses reaching up to 470 GeV.

# 1 Introduction

Supersymmetry (SUSY) [1–8] is an attractive extension of the standard model (SM) of particle physics. It potentially provides solutions to some of the shortcomings affecting the SM, such as the need for fine tuning [9–14] to explain the observed value of the Higgs boson mass [15–20], and the absence of a dark matter (DM) candidate. Supersymmetric models are characterized by the presence of a superpartner for every SM particle with the same quantum numbers except that its spin differs from that of its SM counterpart by half a unit. The cancellation of quadratic divergences in quantum corrections to the Higgs boson mass from SM particles and their superpartners could resolve the fine-tuning problem. In SUSY models with $R$-parity conservation [21], the lightest supersymmetric particle (LSP) is stable [22, 23] and could be a DM candidate [24]. The superpartners of the electroweak gauge and Higgs bosons, namely the bino, winos, and Higgsinos, mix to form neutral and charged mass eigenstates, referred to as the neutralinos ($\widetilde{\chi}^0_i$) and charginos ($\widetilde{\chi}^{\pm}_i$), respectively. Here we assume $\widetilde{\chi}^0_1$, the lightest neutralino, to be the LSP.

The analysis reported in this note investigates the production of the hypothetical $\tau$ slepton (stau, denoted by $\widetilde{\tau}$), the superpartner of the $\tau$ lepton. Supersymmetric scenarios in which the $\widetilde{\tau}$ is light, lead to final states with one or more $\tau$ leptons. Coannihilation scenarios, characterized by a light $\widetilde{\tau}$ that has a small mass splitting with an almost pure bino-like LSP, lead to a DM relic density consistent with cosmological observations [25–30], making the search for new physics in these final states particularly interesting. In this analysis, we examine simplified SUSY models [31–34] in which the $\widetilde{\tau}$ can be produced directly through pair production and decays to a $\tau$ lepton and the LSP. The most sensitive searches for direct $\widetilde{\tau}$ pair production to date were performed at the CERN LEP collider [35–39]. At the CERN LHC, the ATLAS [40, 41] and CMS [42, 43] Collaborations have both performed searches for direct and indirect $\widetilde{\tau}$ production with 8 TeV LHC data. CMS has also investigated $\widetilde{\tau}$ production with 13 TeV data [44].

In many SUSY scenarios the $\widetilde{\tau}$ mass is lighter than the one of selectrons and smuons. The large data set expected at the HL-LHC provides an unprecedented opportunity to probe for the direct production of $\widetilde{\tau}$, which is a challenge due to the relatively small production cross section. For example, the cross section in the mass-degenerate scenario, where we assume that the left- and right-handed $\widetilde{\tau}$ have the same mass and add up their cross sections, for a $\widetilde{\tau}$ mass of 100 GeV is 0.41 pb, and for 300 GeV it is reduced to 0.0071 pb, while for a $\widetilde{\tau}$ mass of 500 GeV we expect only a cross section of 79 fb [45]. A search is therefore developed in events where both $\tau$ leptons decay either hadronically ("$\tau_h\tau_h$" analysis), and in events where one of the $\tau$ leptons decays hadronically (denoted in the following by $\tau_h$) and the other one to a muon or electron and neutrinos ("$\ell\tau_h$" analysis).

The simplified model used for the optimization of the search and the interpretation of the results is shown in Fig. 1. The search assumes $\widetilde{\tau}$ pair production in the mass-degenerate scenario. The cross sections have been computed for $\sqrt{s} = 14$ TeV at next-to-leading order (NLO) using the Prospino code [46]. Final values are calculated using the PDF4LHC recommendations for the two sets of cross sections following the prescriptions of the LHC SUSY Cross Section Working Group [45]. The branching ratio of the $\widetilde{\tau}$ into the $\tau$ lepton and the $\widetilde{\chi}^0_1$ is assumed to be 100%.

# 2 The upgraded CMS detector

The CMS detector [47] will be substantially upgraded in order to fully exploit the physics potential offered by the increase in luminosity at the HL-LHC [48], and to cope with the demand-ing operational conditions at the HL-LHC [49–53]. The upgrade of the first level hardware trigger (L1) will allow for an increase of L1 rate and latency to about 750 kHz and 12.5 $\mu$s, respectively, and the high-level software trigger (HLT) is expected to reduce the rate by about a factor of 100 to 7.5 kHz. The entire pixel and strip tracker detectors will be replaced to increase the granularity, reduce the material budget in the tracking volume, improve the radiation hardness, and extend the geometrical coverage and provide efficient tracking up to pseudorapidities of about $|\eta| = 4$. The muon system will be enhanced by upgrading the electronics of the existing cathode strip chambers (CSC), resistive plate chambers (RPC) and drift tubes (DT). New muon detectors based on improved RPC and gas electron multiplier (GEM) technologies will be installed to add redundancy, increase the geometrical coverage up to about $|\eta| = 2.8$, and improve the trigger and reconstruction performance in the forward region. The barrel electromagnetic calorimeter (ECAL) will feature the upgraded front-end electronics that will be able to exploit the information from single crystals at the L1 trigger level, to accommodate trigger latency and bandwidth requirements, and to provide 160 MHz sampling allowing high precision timing capability for photons. The hadronic calorimeter (HCAL), consisting in the barrel region of brass absorber plates and plastic scintillator layers, will be read out by silicon photomultipliers (SiPMs). The endcap electromagnetic and hadron calorimeters will be replaced with a new combined sampling calorimeter (HGCal) that will provide highly segmented spatial information in both transverse and longitudinal directions, as well as high-precision timing information. Finally, the addition of a new timing detector for minimum ionizing particles (MTD) in both barrel and endcap region is envisaged to provide capability for 4-dimensional reconstruction of interaction vertices that will allow to significantly offset the CMS performance degradation due to high PU rates.

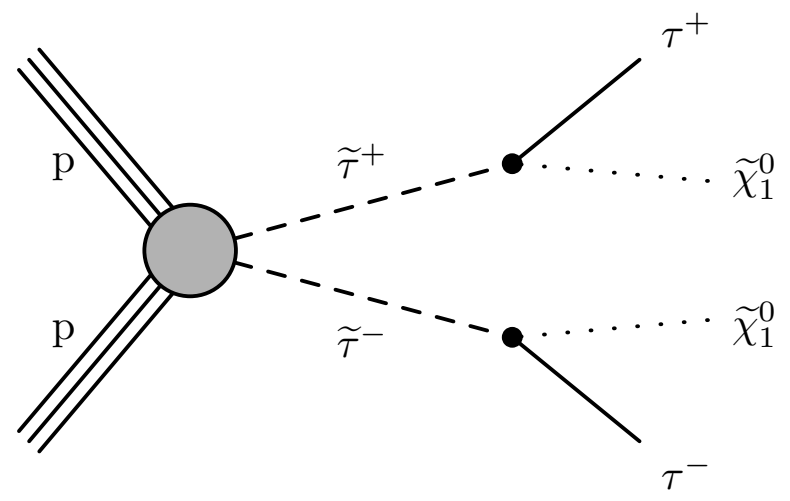

Figure 1: Diagram for the $\widetilde{\tau}$ pair production.

A detailed overview of the CMS detector upgrade program is presented in Ref. [49–53], while the expected performance of the reconstruction algorithms and the mitigation of pileup, i.e., additional proton-proton collisions within the same or neighboring bunch crossings, is summarized in Ref. [54].

## 3 Object reconstruction and simulated samples

The event reconstruction uses a particle-flow (PF) algorithm [55], combining information from the tracker, calorimeter, and muon systems to identify charged and neutral hadrons, photons, electrons, and muons in an event. Candidate events are expected to contain at least two leptons: either two $\tau_h$ candidates, or one $\tau_h$ and one muon or electron from $\tau$ lepton decays. In order to pass the selection, electrons (muons) are required to have a transverse momentum $p_T > 30$ GeV and pseudorapidity $|\eta| < 1.6(2.4)$. Dedicated lepton identification criteria are applied, providing 50% to 90% efficiency for muons and 25% to 80% efficiency for electrons, depending on the lepton $p_T$ and $\eta$. Both muons and electrons are required to be isolated. The isolation is calculated from the scalar sum of the $p_T$ of all particles within a cone of radius $R = \sqrt{(\Delta\eta)^2 + (\Delta\phi)^2} = 0.3$ around the lepton momentum vector, excluding the contribution of the lepton and applying an area-based correction to remove the contribution of particles from pileup [56]. The ratio $I_{rel}$ of the scalar sum of the $p_T$ in the cone to the transverse momentum of the lepton itself is required to be smaller than 0.05.

Jets are reconstructed using the anti-$k_T$ algorithm [57, 58], with a distance parameter of 0.4. For this study we use PUPPI jets [59] which are required to have $p_T > 30$ GeV and $|\eta| < 2.7$. Jets originating from b quarks are identified with the loose working point of the combined secondary vertex b tagging algorithm (CSVv2) [60], which corresponds to an efficiency of about 60–65%.

The $\tau_h$ candidates must satisfy $p_T > 40$ GeV in the $\ell\tau_h$ final states, while a slightly higher threshold of $p_T > 50$ GeV is required for the $\tau_h\tau_h$ final state, driven by the trigger thresholds foreseen for the HL-LHC. Since the main background in this analysis is due to events with jets misidentified as $\tau_h$ leptons, a tight working point with a small misidentification rate is chosen for $\tau_h$ identification. The $\tau_h$ reconstruction efficiency for this working point is about 30%, with a misidentification rate of about 0.08% assuming a multivariate analysis optimization. Overlaps between the two reconstructed leptons in the $\ell\tau_h$ final state are avoided by requiring them to have a minimum separation in $\Delta R$ of 0.3.

In order to ensure orthogonality between the different final states and suppress background, we reject events with additional electrons or muons beyond the two selected leptons that satisfy slightly less stringent selection criteria and transverse momentum of $p_T > 20$ GeV and $|\eta| < 2.7$.

The object selection requirements implemented in the analysis are summarized in Table 1.

Table 1: Summary of object selection requirements for the analysis.

| Selection requirement | $\ell\tau_h$ | $\tau_h\tau_h$ |
|---|---|---|
| Muon (electron) $p_T$ | $> 30$ GeV | — |
| Muon (electron) $p_T$ (veto) | $> 30$ GeV | $> 20$ GeV |
| Muon (electron) $\|\eta\|$ | $< 2.4(1.6)$ | — |
| Muon (electron) $\|\eta\|$ (veto) | $< 2.7$ | $< 2.7$ |
| $\tau_h$ $p_T$ | $> 40$ GeV | $> 50$ GeV |
| $\tau_h$ $\|\eta\|$ | $< 2.3$ | $< 2.3$ |
| $p_T$ ($\tau_h$ $\tau_h$) | — | $> 50$ GeV |
| jet $p_T$ (veto) | $> 30$ GeV | $> 30$ GeV |
| jet $\|\eta\|$ (veto) | $< 2.7$ | $< 2.7$ |
| b jet $p_T$ (veto) | $> 20$ GeV | $> 30$ GeV |

The MADGRAPH5_aMC@NLO 2.3.3 generator [61] is used to produce the parton-level background processes at leading order (LO), with the parton showering and hadronization provided by PYTHIA 8.212 [62, 63]. Signal models of direct $\widetilde{\tau}$ pair production are generated with MADGRAPH5_aMC@NLO at LO precision in perturbative quantum chromodynamics (QCD) up to the production of $\tau$ leptons, which are then decayed with PYTHIA 8.212. The NNPDF3.0LO set of parton distribution functions is used in the generation of all signal models.

The potential effect of pileup is estimated by overlaying the hard scatter event with minimum bias events drawn from a Poisson distribution with a mean of 200.

The generated signal and background events are processed with the fast-simulation package Delphes [64] in order to simulate the expected response of the upgraded CMS detector. The object reconstruction and identification efficiencies, as well as the detector response and resolution, are parameterized in Delphes using the detailed simulation of the upgraded CMS detector based on GEANT4 package [65, 66].

The detailed simulation of the upgraded CMS detector and objects performance at HL-LHC include the effects of aging in the barrel calorimeter that correspond to an integrated luminosity of 1000 $\text{fb}^{-1}$.

# 4 Event selection

The event selection for each final state requires the presence of exactly two reconstructed leptons with opposite charges, corresponding to the $\tau_h\tau_h$ or $\ell\tau_h$ final states. In order to suppress backgrounds with top quarks, we veto events containing any b-tagged jet in both final states. For the $\ell\tau_h$ analysis, the $p_T$ threshold for b-tagged jets is lowered to 20 GeV, as this allows to significantly reduce the background from W+jets events, where the W boson decays into an electron or muon and a neutrino, and a jet is misidentified as $\tau_h$.

The main background for the $\tau_h\tau_h$ final state after this selection consists of QCD multijet events, W+jets, DY+jets, and top quark events. Separating the background into prompt $\tau_h$ events, where both reconstructed $\tau$ leptons are matched to a generator $\tau_h$, and misidentified events, where one or more non-generator matched jets have been misidentified as prompt $\tau_h$, we find that the misidentified background dominates our search regions.

In the $\ell\tau_h$ final state, all events with at least one jet are rejected. Due to kinematical constraints in the signal, we reduce the background from QCD multijet events by requiring a maximum separation of the two leptons in $\Delta R$ of 3.5.

The baseline selection criteria described above are summarized in Table 2. The baseline events are then further selected using kinematic variables for each of the three final states to improve the sensitivity of the search to a range of sparticle masses.

Table 2: Summary of the baseline selection requirements in each final state.

| Selection requirement | $\ell\tau_h$ | $\tau_h\tau_h$ |
|---|---|---|
| $\Delta\phi(\ell_1, \ell_2)$ | $> 1.5$ | $> 1.5$ |
| $\Delta R(\ell_1, \ell_2)$ | $0.3 < \Delta R < 3.5$ | — |
| Veto of events with b-tagged jets | yes | yes |
| $N_{\text{jet}}$ | $= 0$ | — |

In order to further improve discrimination against the SM background, we take advantage of the expected presence of two $\tilde{\chi}_1^0$ in the final state for signal events, which would lead to missing transverse momentum, $\vec{p}_T^{\text{miss}}$. The missing transverse momentum vector $\vec{p}_T^{\text{miss}}$ is defined as the negative vector sum of all PF candidates with corresponding transverse momenta weighted through the PUPPI method. Its magnitude is referred to as $p_T^{\text{miss}}$.

In addition, mass observables that can be calculated from the reconstructed leptons and the $\vec{p}_T^{\text{miss}}$ provide strong discriminants between signal and background. For a mother particle decaying to a visible and an invisible particle, the transverse mass $M_T$ has a kinematic endpoint at the mass of the mother particle, and is calculated as follows:

$$M_T(\ell, \vec{p}_T^{\text{miss}}) \equiv \sqrt{2p_\ell p_T^{\text{miss}}(1 - \cos\Delta\phi(\vec{p}_\ell, \vec{p}_T^{\text{miss}}))}. \quad (1)$$

In addition, the scalar sum of the $M_T$ calculated with the first and second lepton and the missing transverse momentum, respectively, is used to further reduce the background events: $\Sigma M_T = M_T(\ell_1, \vec{p}_T^{\text{miss}}) + M_T(\ell_2, \vec{p}_T^{\text{miss}})$.

We also calculate the stransverse mass $M_{T2}$ [67, 68], defined as:

$$M_{T2}(m_s, \vec{s}, m_t, \vec{t}, \vec{p}_T^{\text{miss}}; \chi_1, \chi_2) = \min_{\substack{\vec{p}, \vec{q} \text{ s.t.} \\ \vec{p} + \vec{q} = \vec{p}_T^{\text{miss}}}} \left\{ \max \left[ M_T(m_s, \vec{s}, \chi_1, \vec{p}), M_T(m_t, \vec{t}, \chi_2, \vec{q}) \right] \right\} \quad (2)$$

where the transverse mass is given by

$$M_T(m, \vec{v}, \chi, \vec{p}) = \sqrt{m^2 + \chi^2 + 2\sqrt{m^2 + |\vec{v}|^2}\sqrt{\chi^2 + |\vec{p}|^2} - 2\vec{v} \cdot \vec{p}},$$

in which $\vec{s}, \vec{t}, \vec{p}, \vec{q}$, and $\vec{p}_T^{\text{miss}}$ are all real two-vectors, and the remaining quantities are real scalars which may all be assumed to be nonnegative as they only enter through their squares. As input for the visible particles ($\vec{s}$ and $\vec{t}$) we give the four-vectors of the two leptons, and we define the mass of the invisible particles $\chi_1 = \chi_2 = 0$. The $M_{T2}$ requirement reduces background from diboson production.

## 4.1 Search regions for the $\tau_h \tau_h$ analysis

The main variables that are used to define the search regions are $\Sigma M_T$ and $M_{T2}$, which are shown for the baseline selection in Fig. 2. All processes containing top quarks, i.e., $t\bar{t}$, single top quark, and $t\bar{t}$ +X production are combined and referred to "Top Quark" in the figure, while "Other SM" corresponds to background processes with low cross section that are combined, namely diboson and triboson production.

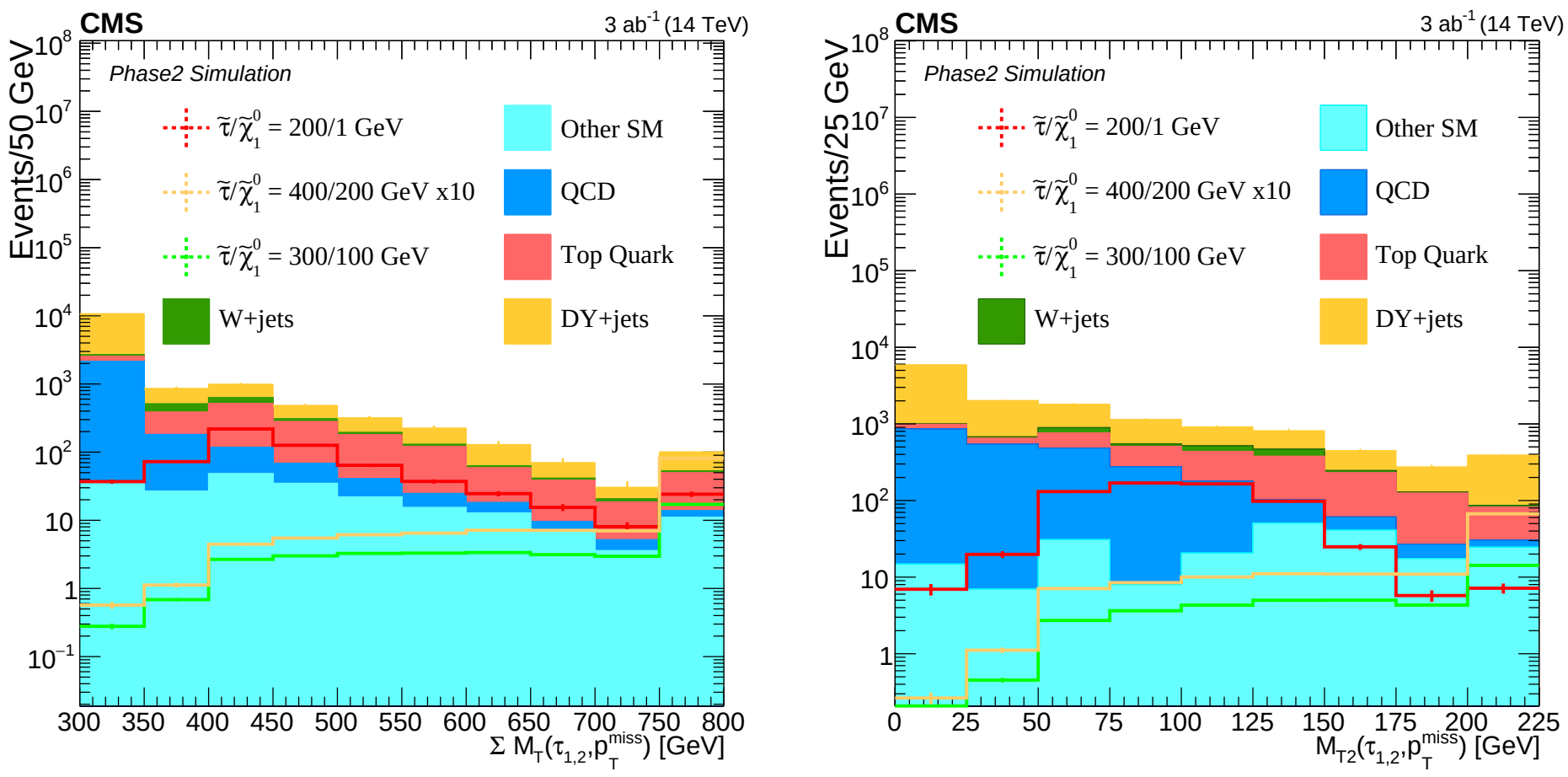

Figure 2: The main search variables for the $\tau_h\tau_h$ analysis, (left) $\Sigma M_T$ and (right) $M_{T2}$, both after the baseline selection. Scaled signal yields for direct $\tilde{\tau}$ production with the mass-degenerate cross section are shown for three separate scenarios of $\tilde{\tau}$ and LSP masses. All processes containing top quarks, i.e. $t\bar{t}$, single top quark, and $t\bar{t}$ +X production are combined and referred to "Top Quark" in the figure, while "Other SM" corresponds to background processes with a low number of events that are combined, diboson and triboson production.

While we apply a stringent requirement of at least 400 GeV for $\Sigma M_T$, we require $M_{T2}$ to be above 50 GeV.

The search regions, binned in $M_{T2}$, $\Sigma M_T$, and the number of jets $n_{jet}$, are summarized in Table 3. There are 24 regions in total.

Table 3: Definition of the search regions (SR) used in the $\tau_h\tau_h$ analysis. Signal depleted bins (low $\Sigma M_T$, high $M_{T2}$) are omitted. The full list of bins and background yields is presented in Table 6.

| Variable | Bin 0 | Bin 1 | Bin 2 | Bin 3 |
|---|---|---|---|---|
| $M_{T2}$ | $50 < M_{T2} < 100$ GeV | $100 < M_{T2} < 150$ GeV | $150 < M_{T2} < 200$ GeV | $M_{T2} > 200$ GeV |
| $\Sigma M_T$ | $400 < \Sigma M_T < 500$ GeV | $500 < \Sigma M_T < 600$ GeV | $\Sigma M_T > 600$ GeV | — |
| $n_{jet}$ | $= 0$ | $> 0$ | — | — |

### 4.2 Search regions for the $\ell\tau_h$ analysis

In the $\ell\tau_h$ final state, we require $M_T(\ell, \vec{p}_T^{miss}) > 120$ GeV, which reduces the W+jets background significantly. To further suppress the SM background in the leptonic final states, we require $p_T^{miss}$ to be at least 150 GeV, which mainly reduces QCD multijets and Drell-Yan events. Additional requirements on $M_{T2}$ and the $\tau_h$ $p_T$ are applied to define the search regions, as summarized in Table 4. Figures 3 and 4 show the distributions of $M_{T2}$, $M_T$, and $M_{T2}$ before the signal region selection for the $e\tau_h$ and $\mu\tau_h$ channel, respectively. In these figures, the ”Other SM” refers to processes with a low number of events after the baseline selection and includes diboson, triboson, $t\bar{t}$ and single top production.

Table 4: Search region requirements in the $\ell\tau_h$ analysis.

| Variable | Bin 0 | Bin 1 | Bin 2 | Bin 3 |
|---|---|---|---|---|
| $M_{T2}$ | $M_{T2} > 120$ GeV | $M_{T2} > 120$ GeV | $80 < M_{T2} < 120$ GeV | $80 < M_{T2} < 120$ GeV |
| $p_T(\tau_h)$ | $> 200$ GeV | $40 < p_T(\tau_h) < 200$ GeV | $> 200$ GeV | $40 < p_T(\tau_h) < 120$ GeV |

## 5 Systematic uncertainties

The dominant experimental uncertainties are those originating from jets being misidentified as $\tau_h$, the lepton efficiency, the jet energy scale and resolution, b tagging efficiency and integrated luminosity. These systematic uncertainties are correlated between the signal and the irreducible background yields. The sources of the systematic uncertainties and their values are reported in Table 5.

## 6 Results

The expected yields in the $\tau_h\tau_h$ final state after all selection requirements are given in Table 6.

The expected yields for the $e\tau_h$ and the $\mu\tau_h$ analysis are given in Tables 7 and 8, respectively, for all signal regions.

The expected upper limits at the 95% confidence level (CL), calculated using the asymptotic formulae [69] of the $CL_s$ criterion [70, 71], and the $5\sigma$ discovery potential are given in Fig. 5. The $\tau_h\tau_h$ analysis has been found to drive the sensitivity, but adding the $\ell\tau_h$ channel enlarges the exclusion bounds by about 60–80 GeV.

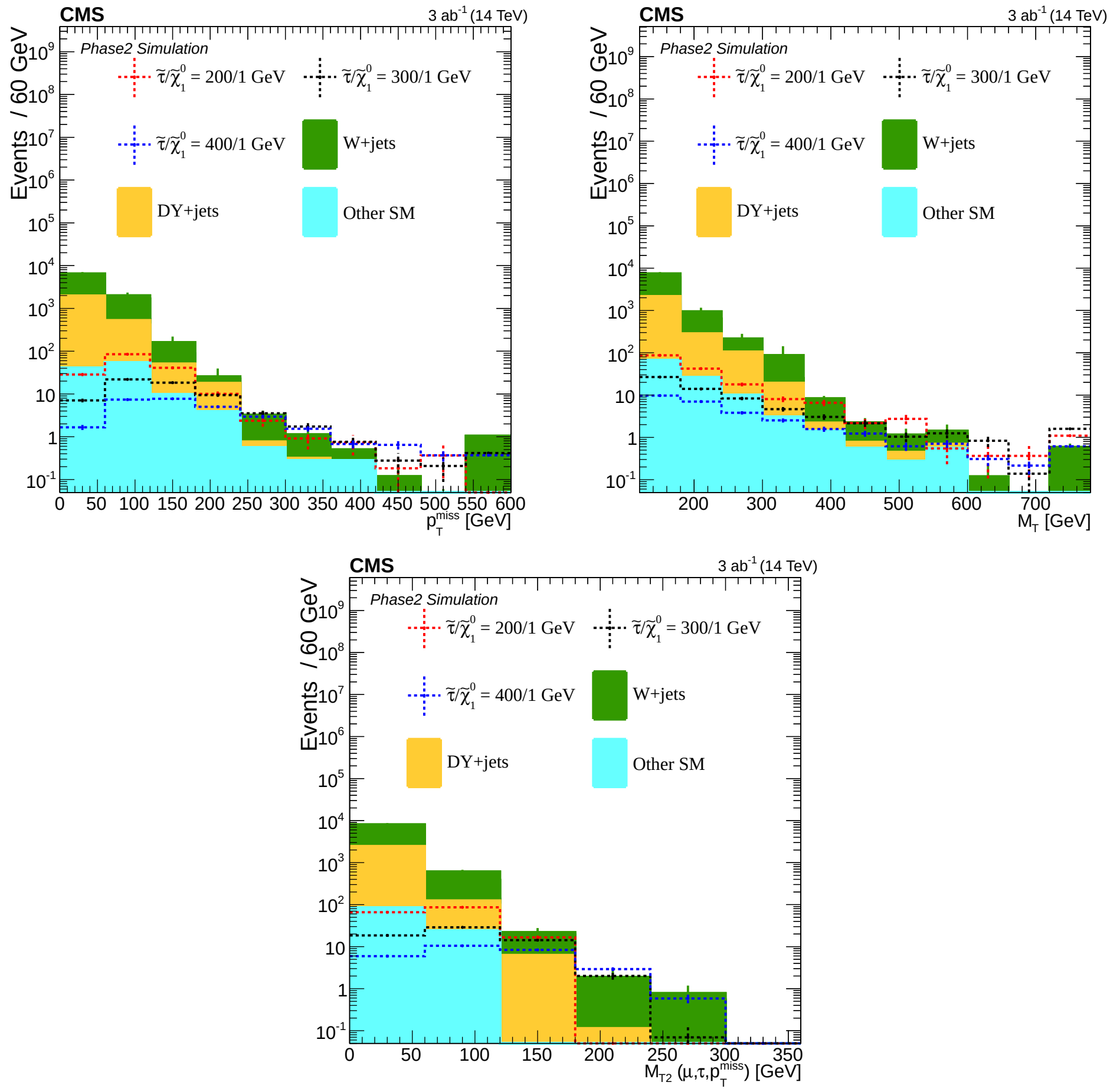

Figure 3: The variables used to determine the search regions in the $e\tau_h$ analysis after the baseline selection: (upper left) the $p_T^{miss}$ distribution, (upper right) the $M_T$ distribution, and (lower) the $M_{T2}$ distribution using $p_T^{miss}$ after the baseline selection. "Other SM" refers to processes with a low number of events after the baseline selection and includes diboson, triboson, $t\bar{t}$ and single top quark production.

Table 5: Summary of the experimental systematic uncertainties.

| Source of systematic uncertainty | Value |
|---|---|
| $\tau_h$ efficiency | 2.5% |
| $\tau_h$ misidentification rate | 15% |
| Muon efficiency | 0.5% |
| Electron efficiency | 1% |
| Jet energy scale | 1–3.5% |
| Jet energy resolution | 3–5% |
| b tagging | 1% |
| Integrated luminosity | 1% |

**8**

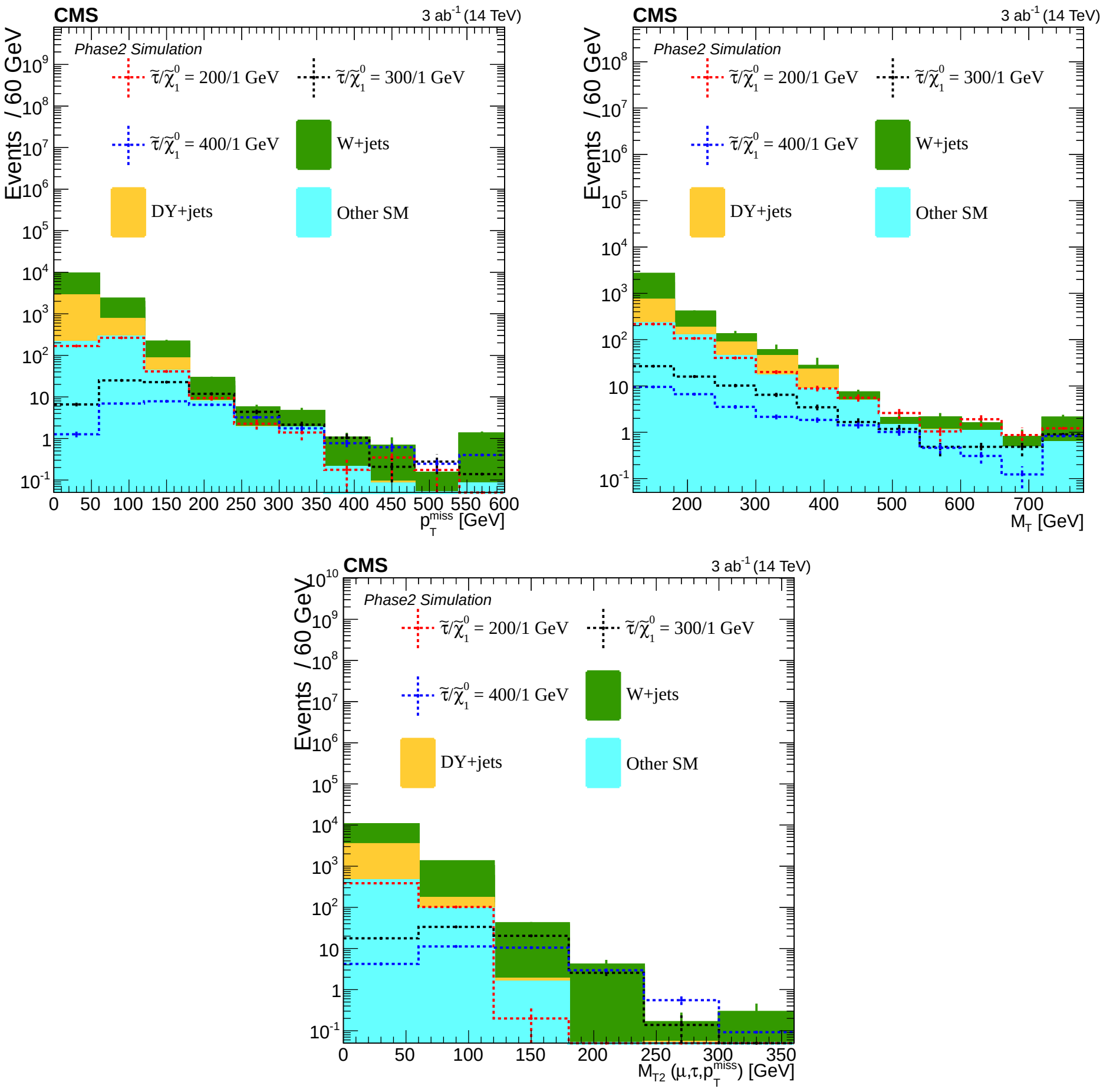

Figure 4: The variables used to determine the search regions in the $\mu\tau_h$ analysis after the baseline selection: (upper left) the $p_T^{miss}$ distribution, (upper right) the $M_T$ distribution, and (lower) the $M_{T2}$ distribution using $p_T^{miss}$ after the baseline selection. "Other SM" refers to processes with a low number of events after the baseline selection and includes diboson, triboson, $t\bar{t}$ and single top quark production.

Table 6: Signal region yields for for background and signal simulation in the $\tau_h\tau_h$ channel. The three rightmost columns show the signal predictions in the degenerate scenario, for masses given in the form of $(m_{\tilde{\tau}}/m_{\tilde{\chi}_1^0})$ in GeV.

| Bin | DY+jets | W+jets | $t\bar{t}$ | QCD | Other SM | Sum | (200/100) | (500/200) | (700/300) |
|---|---|---|---|---|---|---|---|---|---|
| SR-$\tau_h\tau_h$-$M_{T2}$_0_$M_T$_0_$N_j$_0 | 79.67 ± 32.14 | 58.80 ± 43.95 | 13.21 ± 3.86 | 5.41 ± 0.17 | 2.92 ± 2.35 | 160.00 ± 54.63 | 104.79 ± 4.62 | 1.19 ± 0.05 | 0.22 ± 0.01 |
| SR-$\tau_h\tau_h$-$M_{T2}$_0_$M_T$_0_$N_j$_1 | 57.76 ± 15.39 | 5.07 ± 0.52 | 104.54 ± 11.30 | 28.19 ± 0.33 | 8.78 ± 2.62 | 204.33 ± 19.28 | 56.96 ± 3.40 | 0.79 ± 0.04 | 0.16 ± 0.01 |
| SR-$\tau_h\tau_h$-$M_{T2}$_0_$M_T$_1_$N_j$_0 | 9.86 ± 6.28 | 3.96 ± 0.29 | 4.53 ± 2.24 | 1.26 ± 0.09 | 3.70 ± 1.54 | 23.31 ± 6.85 | 26.51 ± 2.32 | 0.72 ± 0.04 | 0.17 ± 0.01 |
| SR-$\tau_h\tau_h$-$M_{T2}$_0_$M_T$_1_$N_j$_1 | 1.36 ± 0.06 | 1.33 ± 0.13 | 31.25 ± 6.01 | 3.84 ± 0.11 | 2.79 ± 1.54 | 40.57 ± 6.21 | 18.99 ± 1.96 | 0.62 ± 0.04 | 0.14 ± 0.01 |
| SR-$\tau_h\tau_h$-$M_{T2}$_0_$M_T$_2_$N_j$_0 | 0.51 ± 0.04 | 2.85 ± 0.25 | 2.61 ± 1.79 | 0.38 ± 0.05 | - | 6.35 ± 1.81 | 21.82 ± 2.11 | 1.33 ± 0.06 | 0.40 ± 0.02 |
| SR-$\tau_h\tau_h$-$M_{T2}$_0_$M_T$_2_$N_j$_1 | 9.69 ± 6.28 | 0.86 ± 0.10 | 26.11 ± 5.56 | 0.86 ± 0.03 | 2.77 ± 1.54 | 40.28 ± 8.53 | 15.32 ± 1.76 | 1.11 ± 0.05 | 0.35 ± 0.02 |
| SR-$\tau_h\tau_h$-$M_{T2}$_1_$M_T$_0_$N_j$_0 | 32.60 ± 11.17 | 6.15 ± 0.53 | 16.36 ± 4.32 | 2.89 ± 0.18 | 4.99 ± 1.85 | 62.98 ± 12.13 | 83.71 ± 4.13 | 1.14 ± 0.05 | 0.19 ± 0.01 |
| SR-$\tau_h\tau_h$-$M_{T2}$_1_$M_T$_0_$N_j$_1 | 2.03 ± 0.10 | 1.34 ± 0.25 | 66.90 ± 8.74 | 18.17 ± 0.33 | 1.44 ± 1.62 | 89.89 ± 8.90 | 40.00 ± 2.84 | 0.74 ± 0.04 | 0.13 ± 0.01 |
| SR-$\tau_h\tau_h$-$M_{T2}$_1_$M_T$_1_$N_j$_0 | 19.59 ± 9.63 | 1.14 ± 0.20 | 3.96 ± 2.19 | 1.52 ± 0.11 | 0.56 ± 0.89 | 26.78 ± 9.92 | 25.73 ± 2.29 | 1.26 ± 0.05 | 0.25 ± 0.01 |
| SR-$\tau_h\tau_h$-$M_{T2}$_1_$M_T$_1_$N_j$_1 | 0.47 ± 0.05 | 0.56 ± 0.44 | 13.32 ± 3.91 | 5.19 ± 0.15 | 2.70 ± 1.36 | 22.24 ± 4.17 | 12.93 ± 1.62 | 0.91 ± 0.05 | 0.16 ± 0.01 |
| SR-$\tau_h\tau_h$-$M_{T2}$_1_$M_T$_2_$N_j$_0 | 9.08 ± 6.28 | 0.28 ± 0.06 | 0.05 ± 0.01 | 0.68 ± 0.07 | 2.11 ± 1.03 | 12.20 ± 6.37 | 10.83 ± 1.48 | 2.13 ± 0.07 | 0.57 ± 0.02 |
| SR-$\tau_h\tau_h$-$M_{T2}$_1_$M_T$_2_$N_j$_1 | 3.79 ± 2.51 | 0.06 ± 0.02 | 5.65 ± 2.53 | 1.37 ± 0.06 | 1.18 ± 1.03 | 12.05 ± 3.71 | 9.03 ± 1.35 | 1.78 ± 0.06 | 0.57 ± 0.02 |
| SR-$\tau_h\tau_h$-$M_{T2}$_2_$M_T$_1_$N_j$_0 | 0.17 ± 0.03 | 0.32 ± 0.10 | 0.05 ± 0.01 | 0.55 ± 0.08 | 1.03 ± 0.73 | 2.12 ± 0.74 | 2.69 ± 0.73 | 0.63 ± 0.04 | 0.11 ± 0.01 |
| SR-$\tau_h\tau_h$-$M_{T2}$_2_$M_T$_1_$N_j$_1 | 3.73 ± 2.51 | 0.22 ± 0.07 | 8.71 ± 3.13 | 1.84 ± 0.11 | 1.06 ± 0.73 | 15.57 ± 4.08 | 1.71 ± 0.58 | 0.39 ± 0.03 | 0.07 ± 0.01 |
| SR-$\tau_h\tau_h$-$M_{T2}$_2_$M_T$_2_$N_j$_0 | 0.23 ± 0.04 | 0.17 ± 0.05 | 0.04 ± 0.01 | 0.73 ± 0.07 | 0.02 ± 0.73 | 1.18 ± 0.73 | 2.48 ± 0.71 | 2.95 ± 0.08 | 0.80 ± 0.02 |
| SR-$\tau_h\tau_h$-$M_{T2}$_2_$M_T$_2_$N_j$_1 | 0.19 ± 0.02 | 0.04 ± 0.01 | 5.59 ± 2.53 | 1.51 ± 0.07 | 0.06 ± 0.73 | 7.38 ± 2.64 | 1.52 ± 0.54 | 2.19 ± 0.07 | 0.67 ± 0.02 |
| SR-$\tau_h\tau_c dh$-$M_{T2}$_3_$M_T$_2_$N_j$_0 | 53.02 ± 30.56 | 0.03 ± 0.02 | 0.02 ± 0.00 | 0.27 ± 0.03 | 0.03 ± 0.02 | 53.36 ± 30.56 | 0.24 ± 0.20 | 3.61 ± 0.09 | 1.74 ± 0.04 |
| SR-$\tau_h\tau_h$-$M_{T2}$_3_$M_T$_2_$N_j$_1 | 0.06 ± 0.01 | 0.02 ± 0.01 | 2.52 ± 1.59 | 0.50 ± 0.03 | 0.54 ± 0.51 | 3.66 ± 1.67 | 0.90 ± 0.41 | 3.17 ± 0.09 | 1.72 ± 0.04 |

Table 7: Signal region yields for background and signal simulation in the $e\tau_h$ channel. The three rightmost columns show the signal predictions in the degenerate scenario, for masses given in the form of $(m_{\tilde{\tau}}/m_{\tilde{\chi}_1^0})$ in GeV.

| SR name | DY | W+Jets | Other SM | Sum | (200/1) | (300/1) | (400/1) |
|---|---|---|---|---|---|---|---|
| SR-$e\tau_h$_1 | 0.18 ± 0.07 | 6.83 ± 1.45 | 0.03 ± 0.06 | 7.03 ± 1.45 | 3.13 ± 0.78 | 6.83 ± 0.71 | 2.54 ± 0.24 |
| SR-$e\tau_h$_2 | 0.44 ± 0.11 | 10.06 ± 1.52 | 0.98 ± 0.13 | 11.00 ± 1.53 | 8.60 ± 1.30 | 7.42 ± 0.74 | 2.36 ± 0.23 |
| SR-$e\tau_h$_3 | 0.15 ± 0.06 | 10.11 ± 1.41 | 0.62 ± 0.10 | 10.57 ± 1.41 | 5.86 ± 1.07 | 3.71 ± 0.52 | 1.30 ± 0.17 |
| SR-$e\tau_h$_4 | 0.10 ± 0.05 | 3.42 ± 0.87 | 0.38 ± 0.08 | 4.31 ± 0.97 | 4.10 ± 0.90 | 2.60 ± 0.44 | 0.58 ± 0.11 |

Table 8: Signal region yields for background and signal simulation in the $\mu\tau_h$ channel. The three rightmost columns show the signal predictions in the degenerate scenario, for masses given in the form of $(m_{\tilde{\tau}}/m_{\tilde{\chi}_1^0})$ in GeV.

| SR name | DY | W+Jets | Other SM | Sum | (200/1) | (300/1) | (400/1) |
|---|---|---|---|---|---|---|---|
| SR-$\mu\tau_h$_0 | 0.06 ± 0.02 | 7.82 ± 1.27 | 0.12 ± 0.13 | 7.94 ± 1.28 | 4.57 ± 0.91 | 9.50 ± 0.81 | 7.14 ± 0.47 |
| SR-$\mu\tau_h$_1 | 0.13 ± 0.04 | 20.51 ± 2.11 | 0.76 ± 0.29 | 21.62 ± 2.16 | 7.49 ± 1.17 | 9.43 ± 0.81 | 5.02 ± 0.39 |
| SR-$\mu\tau_h$_2 | 0.07 ± 0.03 | 12.02 ± 1.65 | 0.72 ± 0.19 | 12.53 ± 1.66 | 6.76 ± 1.11 | 6.03 ± 0.65 | 2.68 ± 0.29 |
| SR-$\mu\tau_h$_3 | 0.03 ± 0.02 | 3.19 ± 0.74 | 1.88 ± 0.31 | 4.86 ± 0.87 | 4.38 ± 0.89 | 1.25 ± 0.29 | 0.68 ± 0.14 |

**10**

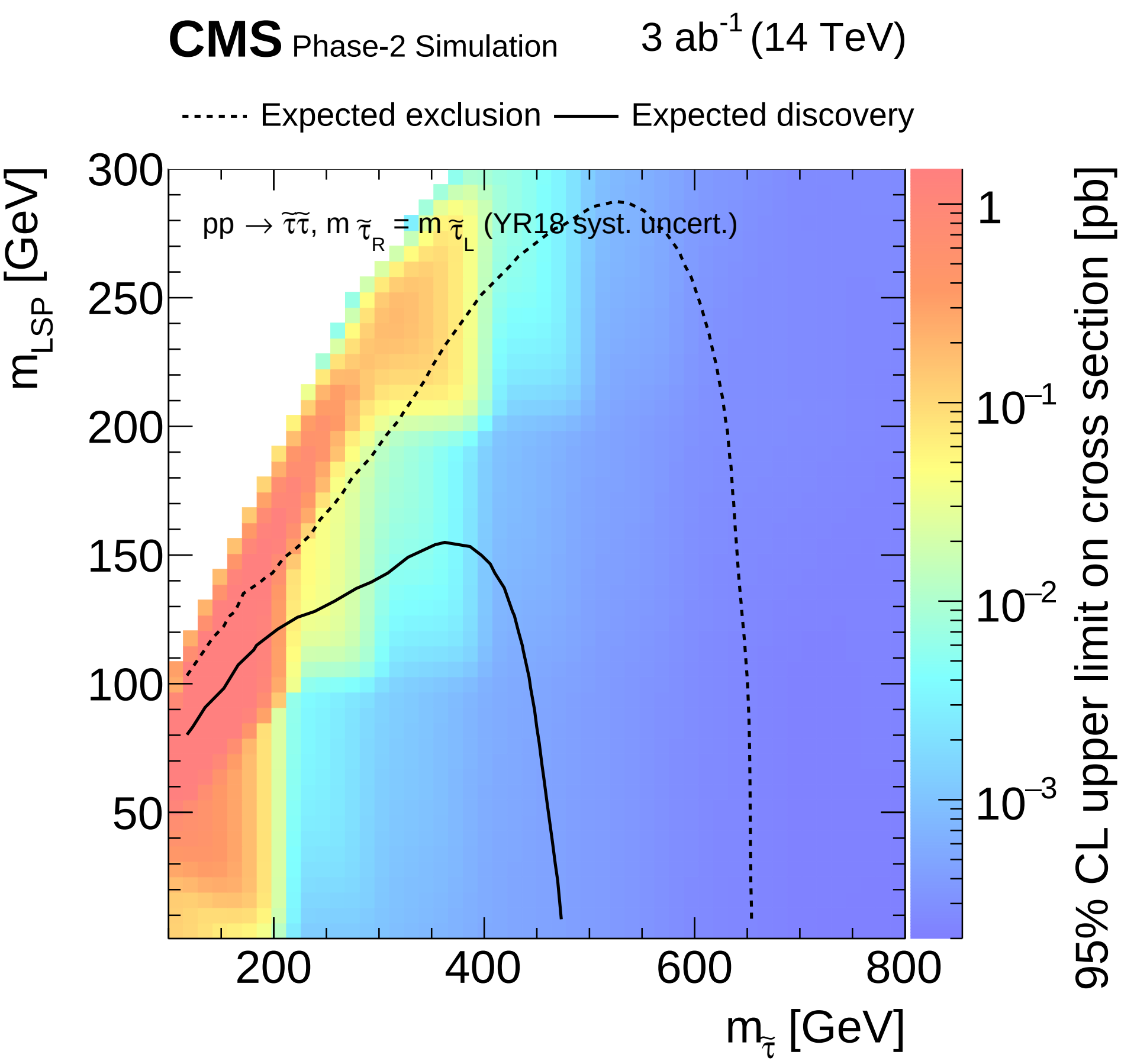

Figure 5: The expected upper limits at the 95% CL and the $5\sigma$ discovery potential for the combination of the results of the $\tau_h\tau_h$ and $\ell\tau_h$ channels.

# 7 Summary

A search for the direct production of $\tau$ sleptons has been presented, assuming $3000\,\mathrm{fb}^{-1}$ of proton-proton collision data produced by the HL-LHC at a center-of-mass energy of 14 TeV. Expected limits have been calculated for the final states that contain either two hadronically decaying $\tau$ leptons and missing transverse momentum, or one hadronically decaying $\tau$ lepton and one $\tau$ decaying to a muon or electron and neutrinos. The analysis is performed using the Delphes simulation of the CMS Phase-2 detector where the object reconstruction performance is tuned to the one achieved with CMS Phase-2 full simulation. In mass-degenerate scenarios, degenerate production of $\tau$ sleptons are excluded up to 650 GeV with the discovery contour reaching up to 470 GeV for a massless lightest neutralino.

**16**

# Prospects for searches for staus, charginos and neutralinos at the high luminosity LHC with the ATLAS Detector

The ATLAS Collaboration

The current searches at the LHC have yielded sensitivity to weakly-interacting supersymmetric particles in the hundreds of GeV mass range and the reach at the high-luminosity phase of the LHC is expected to significantly extend beyond the current limits. This document presents example benchmark studies for stau pair production ($\tilde{\tau}^+\tilde{\tau}^-$) using a final state with two hadronically decaying taus, chargino pair production ($\tilde{\chi}_1^+\tilde{\chi}_1^-$) using a final state with two leptons, and chargino-neutralino production ($\tilde{\chi}_1^\pm\tilde{\chi}_2^0$) using either a $\ell\ell\ell$ or $\ell bb$ final state. A parameterised simulation of the ATLAS detector at a centre-of-mass energy of 14 TeV is used. Expected results are shown for an integrated luminosity of 3000 fb$^{-1}$, where the discovery regions exceed the current limits on SUSY particle masses set at the LHC by hundreds of GeV. The discovery potential at the HL-LHC reaches stau masses of 530 GeV and chargino masses of 660 GeV, for $\tilde{\tau}^+\tilde{\tau}^-$ and $\tilde{\chi}_1^+\tilde{\chi}_1^-$ production, respectively. For $\tilde{\chi}_1^\pm\tilde{\chi}_2^0$ production, the discovery region reaches up to 920 GeV (1080 GeV) in $\tilde{\chi}_1^\pm$ and $\tilde{\chi}_2^0$ masses, where the $\tilde{\chi}_2^0$ decays via the Standard Model $Z$ ($h$) boson. The 95% CL expected exclusion potentials at the HL-LHC reach $\sim$ 200 GeV higher in mass than the discovery potentials for all cases considered.

# 1 Introduction

Supersymmetry (SUSY) [1–6] proposes that for every boson (fermion) of the Standard Model (SM) there exists a fermionic (bosonic) partner. The scalar superpartners of the SM fermions are called sfermions (comprising the charged sleptons, $\tilde{\ell}$, the sneutrinos, $\tilde{\nu}$, and the squarks, $\tilde{q}$), while the gluons have fermionic superpartners called gluinos ($\tilde{g}$). The bino, wino and higgsino fields are fermionic superpartners of the SU(2)×U(1) gauge fields of the SM, and the two complex scalar doublets of a minimally extended Higgs sector, respectively. Their mass eigenstates are referred to as charginos $\tilde{\chi}_i^\pm$ ($i = 1, 2$) and neutralinos $\tilde{\chi}_j^0$ ($j = 1, 2, 3, 4$), numbered in order of increasing mass. The direct production of charginos, neutralinos and sleptons through electroweak interactions may dominate the SUSY production at the LHC if the masses of the gluinos and squarks are large.

SUSY offers natural solutions to many of the problems with the SM. For example, SUSY particles with masses at the electroweak scale can cancel quadratic divergences to the Higgs mass corrections. SUSY can also accommodate the unification of the gauge interactions and a radiative breaking of the electroweak symmetry. Under the conservation of R-parity [7], the lightest SUSY particle (LSP) is stable and is a good candidate for the dark matter in the universe. Furthermore, the Minimal Supersymmetric Standard Model (MSSM) requires a Higgs boson with mass below $\sim$ 135 GeV which is consistent with the Higgs boson observed at the LHC.

The search for weak-scale SUSY is one of the highest physics priorities for the current and future LHC runs. The high luminosity upgrade of the LHC (HL-LHC) is expected to deliver proton-proton collisions at a centre-of-mass-energy of 14 TeV, with an integrated luminosity of around 3000 fb$^{-1}$. The large dataset expected at the end of HL-LHC offers an unprecedented discovery potential for heavy SUSY particles in the electroweak sector, of masses around or above a TeV. This note assesses the ATLAS sensitivity at the end of HL-LHC to direct production of various SUSY partners in the electroweak sector including the stau ($\tilde{\tau}$), chargino and neutralinos under the assumption of R-parity conservation.

# 2 The HL-LHC and the ATLAS detector

In the Run-2 data-taking period, the ATLAS experiment collected 149 fb$^{-1}$ of proton-proton collisions from the LHC at centre-of-mass energies of 13 TeV, with an average number of collisions per bunch crossing of $\langle\mu\rangle = 34$. A second long shutdown (LS2) will follow, during which the injection chain is foreseen to be modified and the accelerator will be able to achieve centre-of-mass-energies of 14 TeV. During LS3, the accelerator is foreseen to be upgraded to the HL–LHC, which is expected to deliver an integrated luminosity of about 3000 fb$^{-1}$, with an average number of pileup interactions per bunch crossing of $\langle\mu\rangle \sim 200$.

The ATLAS detector [8, 9] is a multi-purpose particle detector with a cylindrical geometry.[1] It consists of layers of inner tracking detectors surrounded by a superconducting solenoid, calorimeters, and a muon spectrometer, and will need several upgrades [10–15] to cope with the expected higher luminosity at the HL-LHC, the associated high pileup, and the intense radiation environment. The primary motivation for the upgrade design studies is to evaluate the potential of the experiment for searches and measurements despite these harsh conditions. A new inner tracking system, extending the tracking region from $|\eta| \leq 2.7$ up to $|\eta| \leq 4.0$, will provide the ability to reconstruct forward charged particle tracks, which can be matched to calorimeter clusters for forward electron reconstruction, or associated to forward jets. The inner tracker extension also enables muon identification at high pseudorapidities if additional detectors (such as micro-pattern gaseous or silicon pixel detectors) are installed between the endcap calorimeters and the New Small Wheel [16] in the region $2.7 < |\eta| \leq 4.0$. Despite being in an area without magnetic field, such detectors would increase the muon spectrometer acceptance and could be used to identify (tag) inner detector tracks in the forward region as muons, while relying entirely on the inner tracker for the momentum measurement.

[1] The ATLAS experiment uses a right-handed coordinate system with its origin at the nominal $pp$ interaction point at the centre of the detector. The positive $x$-axis is defined by the direction from the interaction point towards the centre of the LHC ring, with the positive $y$-axis pointing upwards, while the beam direction is along the $z$-axis. Cylindrical coordinates ($r$, $\phi$) are used in the transverse ($x$, $y$) plane, $\phi$ being the azimuthal angle around the beam direction. The pseudorapidity is defined in terms of the polar angle $\theta$ from the $z$-axis as $\eta = -\ln[\tan(\theta/2)]$. The distance in $y - \phi$ space between two objects is defined as $\Delta R = \sqrt{(\Delta y)^2 + (\Delta\phi)^2}$, where $y$ is the rapidity. Transverse energy is computed as $E_T = E \cdot \sin\theta$.

## 3 Electroweak SUSY searches at the HL–LHC

A broad range of electroweak SUSY scenarios and their experimental signatures are considered here, including the two-tau signature from $\tilde{\tau}^+\tilde{\tau}^-$ production in Section 4, the dilepton signature from $\tilde{\chi}_1^+\tilde{\chi}_1^-$ production in Section 5, and the three-lepton and $1\ell bb$ signatures from $\tilde{\chi}_1^\pm\tilde{\chi}_2^0$ production in Section 6. Hadronically decaying taus are used for the $\tilde{\tau}^+\tilde{\tau}^-$ search, while light leptons ($e$, $\mu$ only) are used for the $\tilde{\chi}_1^+\tilde{\chi}_1^-$ and $\tilde{\chi}_1^\pm\tilde{\chi}_2^0$ searches.

The individual analyses follow a coherent approach as much as possible, using the same parameterisations of the upgraded ATLAS detector configuration and the associated experimental uncertainties, the same Monte Carlo (MC) simulations for the common signal and background processes, and the same statistical framework for the interpretation of the results. The definitions of the physics objects follow similar strategies from either earlier publications using data or the previous studies for the HL-LHC. For the signal scenarios considered in this note, most of the final state particles are expected to be in the central region. Therefore the pseudorapidity selections for these final states physics objects remain mostly in the central regions. Signal regions (SR) are typically defined to target one or more regions in the signal model parameter space, using advanced kinematic variables including the output from multivariate methods. Event selections for the signal regions are usually optimised by maximising the expected sensitivity $Z_N$ [17], which takes into account the systematic uncertainties on the background.

The HistFitter [18] software framework is used for the statistical interpretation of the results. In order to quantify the probability for the background-only hypothesis to fluctuate to the observed number of events or higher, a one-sided $p_0$-value is calculated, where the profile likelihood ratio is used as a test statistic [19]. A signal model can be excluded at 95% confidence level (CL) if the $\mathrm{CL}_s$ [20] of the signal-plus-background hypothesis is below 0.05.

Experimental and theoretical uncertainties on the SUSY signal and SM background are accounted for in the exclusion fits. Experimental systematic uncertainties have been estimated based on the expected performance of the upgraded ATLAS detector as documented in Ref. [21]. The theoretical uncertainties, such as the overall cross-section and the modelling of the kinematic shapes, are halved compared to the state-of-art predictions found in Run-2 analyses. The systematic uncertainties arising from the statistics in the control region in data are assumed to scale with the inverse of the square-root of the integrated luminosity. MC-based, statistics-driven sources of uncertainty are considered negligible.

MC simulated event samples are used to predict the background from SM processes and to model the SUSY signal. The effects of an upgraded ATLAS detector are taken into account by applying energy smearing, efficiencies and fake rates to generator level quantities, following parameterisations based on detector performance studies with full simulation and HL-LHC conditions. The effect of the high pileup at the HL-LHC is incorporated by overlaying pileup jets onto the hard-scatter events. Jets from pileup are randomly selected as jets to be considered for analysis with ∼ 2% efficiency, based on the expected performance of a Jet Vertex Tagger at the HL-LHC [21]. The most relevant MC samples have equivalent luminosities (at $\sqrt{s}$ = 14 TeV) of at least 3000 fb$^{-1}$.

SUSY signal samples are generated at leading-order accuracy using MadGraph5_aMC@NLO [22] interfaced to Pythia 8 [23] with the A14 [24] tune for the modelling of the parton showering (PS), hadronisation and underlying event (UE). The matrix element (ME) calculation is performed at tree-level and includes the emission of up to two additional partons. The PDF set used for the generation is NNPDF23LO [25]. The ME–PS matching is done using the CKKW-L [26] prescription, with a matching scale set to one quarter of the mass of the pair produced particles. The cross-sections used to evaluate the signal yields are calculated to next-to-leading order in the strong coupling constant, adding the resummation of soft gluon emission at next-to-leading-logarithmic accuracy (NLO+NLL) [27, 28]. The nominal cross section and the uncertainty are taken from an envelope of cross section predictions using different PDF sets and factorisation and renormalisation scales, as described in Ref. [29].

Background samples were simulated using different MC generators depending on the process. The event generators, the accuracy of theoretical cross-sections, the underlying-event parameter tunes, and the PDF sets used for the background samples are summarised in Table 1. For all samples, except the ones generated using Sherpa [30], the Evtgen v1.2.0 [31] program was used to simulate the properties of the bottom- and charm-hadron decays.

| Process | Generator<br>+ fragmentation/hadronisation | Tune | PDF set | Cross-section<br>order |
|---|---|---|---|---|
| $W/Z$+jets | Powheg-Box v1 [32] + Pythia 8.186 [33]<br>Sherpa 2.2.1 [30] | AZNLO<br>Default | CTEQ6L1<br>NNPDF30NNLO [34] | NNLO<br>NNLO |
| $t\bar{t}$ | Powheg-Box v2 + Pythia 8.186 | A14 | NNPDF23LO [25] | NNLO+NNLL |
| Single top | Powheg-Box v1 or v2 + Pythia 6.428 [35] | Perugia2012 [36] | CT10 [37] | NNLO+NNLL |
| Diboson (fully leptonic)<br>(semi leptonic) | Sherpa 2.2.1<br>Powheg-Box v1 + Pythia 8.186 | Default<br>AZNLO [38] | NNPDF30NNLO<br>CTEQ6L1 | NLO<br>NLO |
| Triboson | Sherpa 2.2.2 | Default | NNPDF30NNLO | NLO |
| $t\bar{t} + X$ | MadGraph 2.2.2 [22] + Pythia 8.186 | A14 | NNPDF23LO | NLO |
| Higgs | Powheg-Box v2 + Pythia 8.186 | AZNLO | CTEQ6L1 | NNLO+NNLL |
| Multijet | Pythia 8.186 | AU2 [39] | CT10 | NLO |

Table 1: List of MC generators used for the SM background processes. Information is given about the underlying-event tunes, the PDF sets and the pQCD highest-order accuracy (LO, NLO, next-to-next-to-leading order, NNLO, and next-to- next-to-leading-log, NNLL) used for the normalization of the different samples. The Diboson process includes $WW$, $WZ$ and $ZZ$. The $t\bar{t}$+X process includes $t\bar{t}$+W, $t\bar{t}$+Z and $t\bar{t}$+WW. For the $W$+jets process, Sherpa was used in the ($1\ell bb$) final state, while Powheg+Pythia was used for the other final states. In the direct stau analysis, a combination of generators are used to model the $W$+jets events, where $W \to e/\mu\nu$ and $W \to \tau\nu$ are modelled with Powheg+Pythia and Sherpa, respectively.

# 4 Search for direct stau production

Searches for the direct production of light stau pairs at the HL-LHC are motivated by both experimental and theoretical considerations. As of today the most stringent exclusion limits are from LEP, making the search for direct staus a crucial "unturned stone" in the hunt for SUSY at the LHC. Staus are expected to be the lightest slepton flavor in models of GUT scale unification and the lighter stau is favoured to be mostly right-handed. Furthermore, in models with a light stau and lightest neutralino $\tilde{\chi}_1^0$ with a small mass difference, stau co-annihilation processes [40] in the early universe can reduce the $\tilde{\chi}_1^0$ relic density and make it consistent with observations from cosmological measurements [41].

A search for stau production is presented here, which uses a final state with two hadronically decaying $\tau$ leptons. Two simplified models describing the direct production of $\tilde{\tau}^+\tilde{\tau}^-$ are used in this document: one considers stau partners of the left-handed $\tau$ lepton ($\tilde{\tau}_{\mathrm{L}}$), and a second considers stau partners of the right-handed $\tau$ lepton ($\tilde{\tau}_{\mathrm{R}}$). In both models, the stau decays with a branching fraction of 100% to the SM $\tau$-lepton and the LSP, which is a common scenario in the phenomenological Minimal Supersymmetric SM [42] when $\tilde{\chi}_1^{\pm}$, $\tilde{\chi}_2^0$ and $\tilde{\nu}_\tau$ are heavier than the stau. The relevant diagram for this model can be seen in Figure 1.

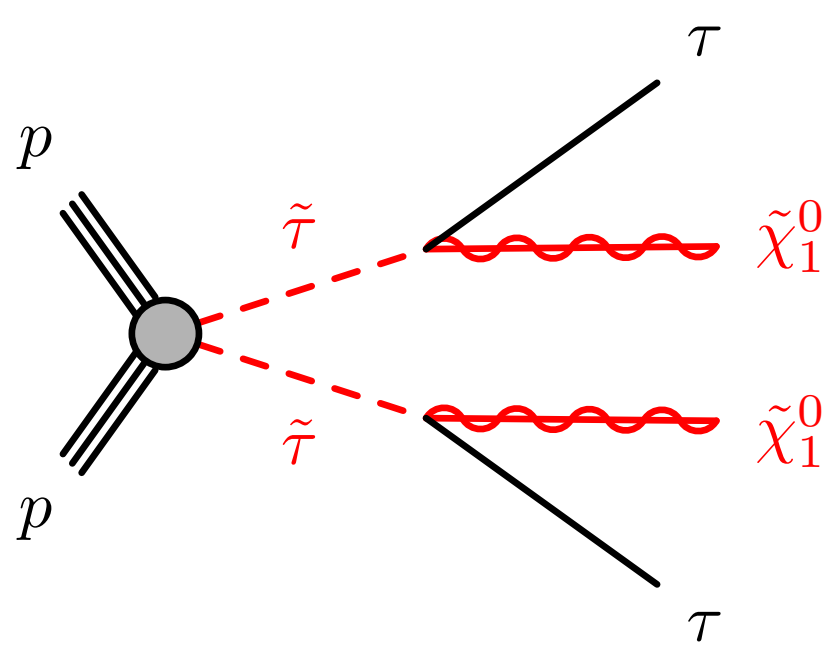

Figure 1: Diagram illustrating the signal scenario considered for the pair production of charged staus targeted by the two-tau final state.

The signature considered here is two hadronically decaying taus, low jet activity, and large missing transverse momentum ($E_{\mathrm{T}}^{\mathrm{miss}}$) from the $\tilde{\chi}_1^0$ and neutrinos. The SM background is dominated by $W/Z$+jets, multi-boson, multi-jet, and top pair production. In the ATLAS Run-1 search for combined $\tilde{\tau}_{\mathrm{L}}^+\tilde{\tau}_{\mathrm{L}}^-$ and $\tilde{\tau}_{\mathrm{R}}^+\tilde{\tau}_{\mathrm{R}}^-$ production [43], only a narrow range of stau masses ($m(\tilde{\tau}_{\mathrm{R}}, \tilde{\chi}_1^0) = (109, 0)$ GeV) was excluded due to the very small production cross section. Analysis of the 2015+2016 Run-2 data by CMS [44] did not further extend the sensitivity to direct stau production. Thus, this scenario is an interesting case to study at the HL-LHC, where significant gains in sensitivity could be made.

The event pre-selection is based on that of the previous 8 TeV analysis [43] and 13 TeV analysis [45]. Hadronically decaying taus are selected with $p_{\mathrm{T}} > 20$ GeV and $|\eta| < 4$, while electrons and muons are selected with $p_{\mathrm{T}} > 10$ GeV and $|\eta| < 2.47$ ($|\eta| < 2.5$ for muons). Jets are reconstructed with the anti-$k_t$ algorithm [46, 47] with a radius parameter of 0.4, with $p_{\mathrm{T}} > 20$ GeV and $|\eta| < 4$. Jets are identified as $b$-jets using the MV2c10 tagging algorithm, operating at an efficiency of 70% in $t\bar{t}$ simulation. To remove close-by objects from one another, an overlap removal based on $\Delta R$ is applied.

SM processes where one or more jet is mis-identified as a hadronically decaying tau (fake tau) contribute to the total background. To maximize the available MC statistics, these backgrounds are estimated by assigning a weight to each jet, where the weight corresponds to the tau fake rate in the HL-LHC detector performance parameterisation. The probability for an event to have one or two fake taus is assessed using all possible combinations of jets, and each event is then weighted by the probability it will contribute to the fake tau background. Cases with more than three fake taus are not considered due to the low probability (less than $10^{-6}$).

Before the optimization, pre-selection cuts are applied to suppress the SM background. Events are selected with exactly two tightly identified hadronic taus with $|\eta| < 2.5$, and the two taus must have opposite electric charge (OS). The tight tau algorithm correctly identifies one-prong (three-prong) taus with an efficiency of 60% (45%) and with a light-flavour jet misidentification probability of 0.06% (0.02%). Events with electrons, muons, $b$-jets or forward jets ($|\eta| > 2.5$) are vetoed. The effect of a di-tau trigger is considered by requiring the leading tau has $p_T > 50$ GeV and the sub-leading tau has $p_T > 40$ GeV, with an assumed trigger efficiency of 64%. To suppress the SM background, a loose jet veto is applied that rejects events containing jets with $|\eta| < 2.5$ and $p_T > 100$ GeV. Since the SUSY signal involves two undetected $\tilde{\chi}_1^0$, the resulting $E_T^{miss}$ spectrum tends to be harder than that for the major SM backgrounds, thus $E_T^{miss} > 200$ GeV is required to reject the multi-jet background. A $Z$ veto is imposed, where the invariant mass of the two taus, $m_{\tau\tau}$, is required to be larger than 100 GeV to suppress contributions from $Z/\gamma^*$ + jets production. To suppress the top quark and multi-jet backgrounds, the sum of the two-tau transverse mass[2] $m_{T\tau1} + m_{T\tau2}$, defined using the transverse momentum of the leading (next-to-leading) tau and $E_T^{miss}$, must be larger than 450 GeV.

The stransverse mass $m_{T2}$ [48, 49] is used to further discriminate SUSY events from SM processes. It can be shown to have a kinematic endpoint for events where two massive pair produced particles each decay to two objects, one of which is detected and the other escapes undetected. It is defined as

$$m_{T2} = \min_{\vec{q}_T}\{\max\left[m_T(\vec{p}_{T,1}, \vec{q}_T), m_T(\vec{p}_{T,2}, \vec{P}_T^{miss} - \vec{q}_T)\right]\}, \tag{1}$$

where $\vec{p}_{T,1}$ and $\vec{p}_{T,2}$ are the transverse momentum vectors of the two visible particles, $\vec{P}_T^{miss}$ is the missing transverse momentum, and $\vec{q}_T$ is the transverse vector that minimises the larger of the two transverse masses $m_T$. A requirement of $m_{T2} > 35$ GeV is applied to suppress the top, $W$+jets and $Z/\gamma^*$ + jets backgrounds.

Starting from this common pre-selection, a cut-and-count method is used to define various SRs. To target signal scenarios with different kinematics, three benchmark points are selected in the optimisation, based on the mass difference between the $\tilde{\tau}$ and $\tilde{\chi}_1^0$, $\Delta m \equiv m_{\tilde{\tau}} - m_{\tilde{\chi}_1^0}$:

- $\Delta m < 150$ GeV: $m(\tilde{\tau}, \tilde{\chi}_1^0) = (160, 40)$ GeV
- $\Delta m \in [150, 300]$ GeV: $m(\tilde{\tau}, \tilde{\chi}_1^0) = (400, 160)$ GeV
- $\Delta m \geq 300$ GeV: $m(\tilde{\tau}, \tilde{\chi}_1^0) = (500, 1)$ GeV

Finally, several kinematic variables that offer good discrimination power between signal and SM backgrounds are used to optimise the SR selection: the $p_T$ of the leading and next-to-leading tau, the event $E_T^{miss}$, the angular separation between the leading and next-to-leading tau $\Delta\phi(\tau1, \tau2)$ and $\Delta R(\tau1, \tau2)$, the jet veto $p_T$ threshold, along with $m_{T\tau1} + m_{T\tau2}$ and $m_{T2}$. The selection on these variables is optimized for high $Z_N$, assuming an uncertainty of 20% on the sum of all backgrounds. This uncertainty is a rough value of the total background uncertainty without the multi-jet uncertainty contributions from the Run-2 studies.

[2] The transverse mass is defined by $m_T = \sqrt{2p_{T,i}P_T^{miss}(1 - \cos\Delta\phi)}$, where $p_{T,i}$ is the transverse momentum vectors of the visible particle $i$, $P_T^{miss}$ is the missing transverse momentum, and $\Delta\phi$ is the angle between the particle and the $\vec{P}_T^{miss}$.

Three signal regions are defined to maximise model-independent discovery sensitivity based on the optimization for scenarios with low (SR-low), medium (SR-med) and high (SR-high) mass differences between the $\tilde{\tau}$ and $\tilde{\chi}_1^0$. Furthermore, another disjoint signal region binned in $m_{\mathrm{T2}}$ is defined to maximise model-dependent exclusion sensitivity based on the previous SR-high signal region with the jet veto threshold cut loosened to $p_{\mathrm{T}} > 100$ GeV. Each SR is identified by the range of the $m_{\mathrm{T2}}$, and is shown in Table 2. Figure 2 show the distributions of $m_{\mathrm{T2}}$ in these signal regions, applying all SR selections with the exception of $m_{\mathrm{T2}}$ itself.

Table 2: Summary of selection requirements for the direct stau signal regions.

| Common Selection | | | | |
|---|---|---|---|---|
| exactly two tight taus with opposite sign | | | | |
| $e/\mu$ veto, $b$-jet veto | | | | |
| $m_{\tau\tau} > 100$ GeV ($Z$-veto) | | | | |
| $E_{\mathrm{T}}^{\mathrm{miss}} > 200$ GeV | | | | |
| $p_{\mathrm{T}\tau 2} > 75$ GeV | | | | |
| $\Delta R(\tau 1, \tau 2) < 3$ | | | | |
| $\Delta\phi(\tau 1, \tau 2) > 2$ | | | | |
| **Selection** | **SR-low** | **SR-med** | **SR-high** | **SR-exclHigh** |
| jet veto threshold | $p_{\mathrm{T\,jet}} > 40$ GeV | $p_{\mathrm{T\,jet}} > 40$ GeV | $p_{\mathrm{T\,jet}} > 20$ GeV | $p_{\mathrm{T\,jet}} > 100$ GeV |
| $p_{\mathrm{T}\tau 1} >$ | 150 GeV | 200 GeV | 200 GeV | 200 GeV |
| $m_{\mathrm{T}\tau 1} + m_{\mathrm{T}\tau 2} >$ | 500 GeV | 700 GeV | 800 GeV | 800 GeV |
| $m_{\mathrm{T2}}(\tau 1, \tau 2)$ | $\in [80\,\mathrm{GeV}, \infty]$ | $\in [130\,\mathrm{GeV}, \infty]$ | $\in [130\,\mathrm{GeV}, \infty]$ | $\in [80\,\mathrm{GeV}, 130\,\mathrm{GeV}]$ |
| | | | | $\in [130\,\mathrm{GeV}, 180\,\mathrm{GeV}]$ |
| | | | | $\in [180\,\mathrm{GeV}, 230\,\mathrm{GeV}]$ |
| | | | | $\in [230\,\mathrm{GeV}, \infty]$ |

Tables 3 and 4 show the expected numbers of events for the SM backgrounds and three SUSY reference points in the ditau signal regions for an integrated luminosity of 3000 fb$^{-1}$. Only the statistical uncertainties for signal and backgrounds are shown in Table 3 and Table 4.

The systematic uncertainties are evaluated based on the SR-high systematic uncertainty in Ref. [45], where the dominant background experimental uncertainties in that study are the uncertainty on the multi-jet estimation ($\sim 33\%$), the tau energy scale in situ uncertainty ($\sim 8\%$), the tau energy scale uncertainty from modelling ($\sim 8\%$) and the detector ($\sim 13\%$), the tau ID efficiency uncertainty ($\sim 5\%$), the uncertainty from $E_{\mathrm{T}}^{\mathrm{miss}}$ reconstruction ($\sim 6\%$), and the uncertainty from the jet energy sale ($\sim 4\%$). The dominant signal uncertainties are the tau energy scale in situ uncertainty ($\sim 7\%$), the tau energy scale uncertainty from detector ($\sim 6\%$), the tau ID efficiency uncertainty ($\sim 13\%$), the MC/data related trigger systematics ($\sim 7\%$ in total), and the signal cross-section uncertainty ($\sim 9\%$).

A few of the experimental uncertainties are expected to be smaller at the HL-LHC compared to the 13 TeV studies, as described in Ref [21]. In particular, the tau energy scale insitu uncertainty is scaled

Table 3: Expected numbers of events for the SM background and the three benchmark signal models for combined $\tilde{\tau}^+_L\tilde{\tau}^-_L$ and $\tilde{\tau}^+_R\tilde{\tau}^-_R$ production in the signal regions SR-low, SR-med and SR-high. The "Other SM" contains contributions from the Top, Higgs boson and Multi-jet processes. Entries marked as '-' indicate negligible background contributions (less than 0.1). Uncertainties describe the MC statistical uncertainties only.

| | SR-low | SR-med | SR-high |
|---|---|---|---|
| $W$+jets | 8.8 ± 2.8 | 2.12 ± 0.56 | 1.00 ± 0.21 |
| Multi-boson | 2.6 ± 1.3 | 0.35 ± 0.18 | - |
| $Z/\gamma^*$ + jets | 1.4 ± 1.0 | - | - |
| Other SM | 0.98 ± 0.40 | - | - |
| SM total | 13.8 ± 3.3 | 2.57 ± 0.58 | 1.10 ± 0.21 |
| $m(\tilde{\tau}_L/\tilde{\tau}_R, \tilde{\chi}^0_1)$ = (160, 40) GeV | 34.9 ± 7.2 | 2.2 ± 1.6 | 0.63 ± 0.44 |
| $m(\tilde{\tau}_L/\tilde{\tau}_R, \tilde{\chi}^0_1)$ = (400, 160) GeV | 24.1 ± 1.6 | 13.8 ± 1.2 | 8.3 ± 1.0 |
| $m(\tilde{\tau}_L/\tilde{\tau}_R, \tilde{\chi}^0_1)$ = (500, 1) GeV | 19.4 ± 1.5 | 15.0 ± 1.3 | 11.6 ± 1.2 |

Table 4: Expected numbers of events for the SM background and the three benchmark signal points for combined $\tilde{\tau}^+_L\tilde{\tau}^-_L$ and $\tilde{\tau}^+_R\tilde{\tau}^-_R$ production in the exclusion ditau signal regions. The "Other SM" contains contributions from the Top, Higgs boson and Multi-jet processes. Entries marked as '-' indicate negligible background contributions (less than 0.1). Uncertainties describe the MC statistical uncertainties only.

| | SR-exclHigh | | | |
|---|---|---|---|---|
| $m_{T2}$ [ GeV ] | [80, 130] | [130, 180] | [180, 230] | [230, ∞] |
| $W$+jets | 2.42 ± 0.52 | 1.22 ± 0.26 | 1.10 ± 0.24 | 0.54 ± 0.12 |
| Multi-boson | 0.49 ± 0.10 | 0.08 ± 0.02 | 0.05 ± 0.01 | 0.04 ± 0.01 |
| $Z/\gamma^*$ + jets | - | - | - | - |
| Other SM | 0.14 ± 0.03 | 0.04 ± 0.01 | 0.03 ± 0.01 | 0.02 ± 0.00 |
| SM total | 3.06 ± 0.44 | 1.34 ± 0.18 | 1.19 ± 0.15 | 0.60 ± 0.06 |
| $m(\tilde{\tau}_L/\tilde{\tau}_R, \tilde{\chi}^0_1)$ = (160, 40) GeV | 0.96 ± 0.13 | 0.63 ± 0.09 | 0.00 ± 0.00 | 0.00 ± 0.00 |
| $m(\tilde{\tau}_L/\tilde{\tau}_R, \tilde{\chi}^0_1)$ = (400, 160) GeV | 4.79 ± 0.67 | 9.11 ± 1.28 | 6.43 ± 0.90 | 2.97 ± 0.42 |
| $m(\tilde{\tau}_L/\tilde{\tau}_R, \tilde{\chi}^0_1)$ = (500, 1) GeV | 1.84 ± 0.26 | 4.21 ± 0.59 | 5.99 ± 0.84 | 7.81 ± 1.10 |

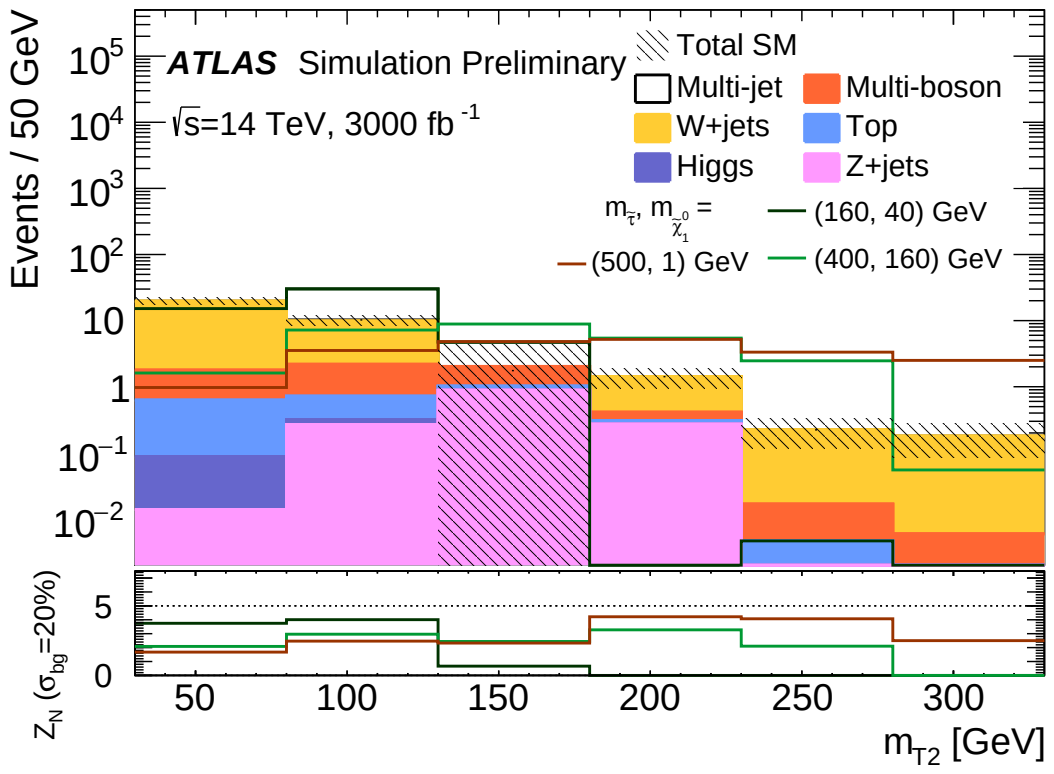

(a) SR-low $m_{T2}$

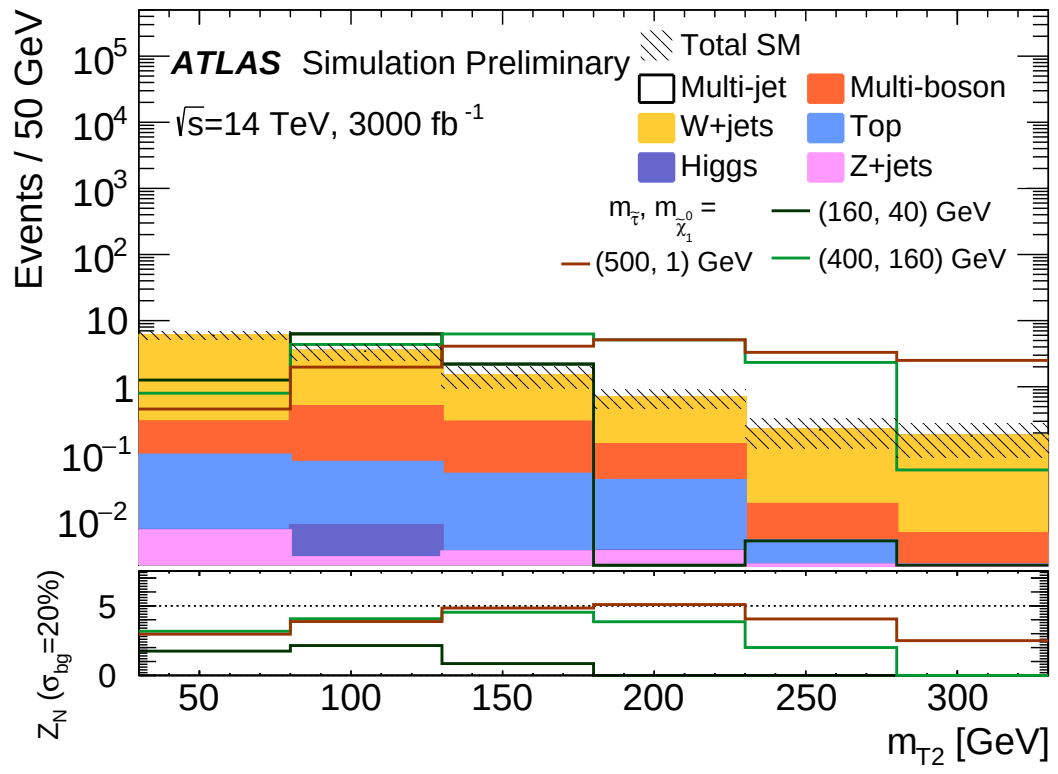

(b) SR-med $m_{T2}$

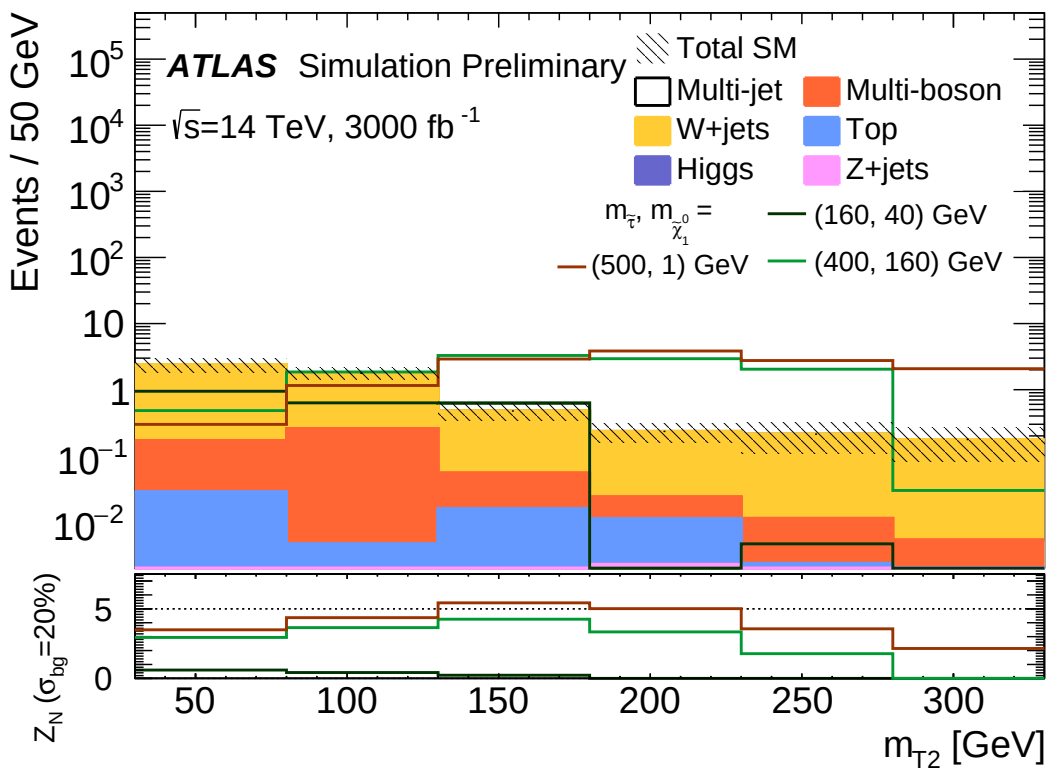

(c) SR-high $m_{T2}$

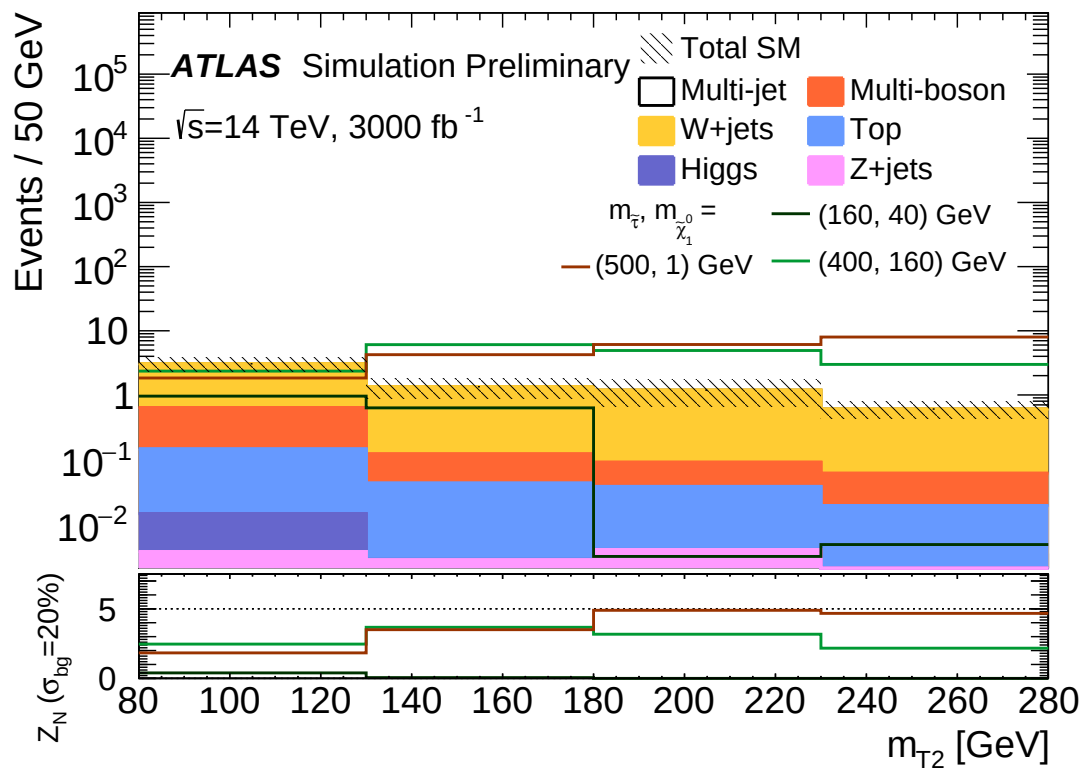

(d) SR-exclHigh $m_{T2}$

Figure 2: Distributions of each $m_{T2}$ variable in the SR-low, SR-med, SR-high and SR-exclHigh regions, applying all selections as specified in Table 2, with the exception of $m_{T2}$ itself. The stacked histograms show the expected SM backgrounds, while the hatched bands represent the statistical uncertainties on the total SM background. For illustration, the distributions of the SUSY reference points for combined $\tilde{\tau}_L^+\tilde{\tau}_L^-$ and $\tilde{\tau}_R^+\tilde{\tau}_R^-$ production are also shown as dashed lines. The last bin includes the overflow. The lower pad in each plot shows the significance, $Z_N$ using a background uncertainty of 20%, for the SUSY reference points. In (a), (b) and (c), $Z_N$ is shown for an $m_{T2}$ threshold, while for (d), $Z_N$ is shown in each $m_{T2}$ interval.

by a factor of 0.6 and the tau ID efficiency uncertainty is scaled by a factor of 0.45. The multi-jet uncertainties scale with the increased integrated luminosity, and the background theoretical uncertainties are halved. The theoretical cross-section uncertainty for direct stau production is taken as 10%, while the MC/data related systematics are considered negligible. All other uncertainties are assumed to be the same as the 13 TeV studies. In this Baseline Uncertainties assumption, the total background experimental uncertainty is $\sim$ 19%, with theoretical uncertainties on the Top, $Z/\gamma^*$ + jets and Higgs backgrounds of 13%, theoretical uncertainties on the $W$+jets and multi-jet backgrounds of 10%, and uncertainties on the multi-boson background of 8%. The total uncertainty on the SUSY signal is $\sim$ 14%.

A second scenario is also considered, where the expected uncertainties at the HL-LHC do not improve upon the 13 TeV studies for the SM background and signal. This results in a total background uncertainty of $\sim$ 38% and a signal uncertainty of $\sim$ 21% for the Run-2 Uncertainties scenario.

To calculate the discovery potential, SR-low, SR-med and SR-High defined in Table 2 are used, while for the model dependent exclusion limits the best expected signal region is used, considering SR-low, SR-med, SR-High, and the multi-bin SR-exclHigh. Experimental uncertainties are treated as correlated between signal and background and all uncertainties are treated as correlated across regions. The 95% CL exclusion and discovery potentials for combined $\tilde{\tau}_L^+\tilde{\tau}_L^-$ and $\tilde{\tau}_R^+\tilde{\tau}_R^-$ production, $\tilde{\tau}_L^+\tilde{\tau}_L^-$ production alone and $\tilde{\tau}_R^+\tilde{\tau}_R^-$ production alone under different uncertainty assumptions are shown in Figure 3. The $\pm 1\sigma_{\text{exp}}$ uncertainty band indicates the impact on the expected limit of the uncertainty included in the fit. For the Baseline Uncertainties scenario, the exclusion limit reaches 730 GeV in $\tilde{\tau}$ mass for the combined $\tilde{\tau}_L^+\tilde{\tau}_L^-$ and $\tilde{\tau}_R^+\tilde{\tau}_R^-$ production, and 680 GeV (420 GeV) for pure $\tilde{\tau}_L^+\tilde{\tau}_L^-$ (pure $\tilde{\tau}_R^+\tilde{\tau}_R^-$) production with a massless $\tilde{\chi}_1^0$. The discovery sensitivity reaches 110 – 530 GeV (110 – 500 GeV) in $\tilde{\tau}$ mass for the combined $\tilde{\tau}_L^+\tilde{\tau}_L^-$ and $\tilde{\tau}_R^+\tilde{\tau}_R^-$ (pure $\tilde{\tau}_L^+\tilde{\tau}_L^-$) production with a massless $\tilde{\chi}_1^0$. No discovery sensitivity is found for pure $\tilde{\tau}_R^+\tilde{\tau}_R^-$ production as the production cross section is very small.

For the Run-2 Uncertainties scenario, the exclusion limit is slightly reduced to 720 GeV in $\tilde{\tau}$ mass for the combined $\tilde{\tau}_L^+\tilde{\tau}_L^-$ and $\tilde{\tau}_R^+\tilde{\tau}_R^-$ production, and 670 GeV (390 GeV) for pure $\tilde{\tau}_L^+\tilde{\tau}_L^-$ (pure $\tilde{\tau}_R^+\tilde{\tau}_R^-$) production with a massless $\tilde{\chi}_1^0$. The discovery sensitivity is also slightly reduced, reaching 200 – 500 GeV (210 – 460 GeV) in $\tilde{\tau}$ mass for the combined $\tilde{\tau}_L^+\tilde{\tau}_L^-$ and $\tilde{\tau}_R^+\tilde{\tau}_R^-$ (pure $\tilde{\tau}_L^+\tilde{\tau}_L^-$) production with a massless $\tilde{\chi}_1^0$.

Based on the search channels and methods considered here, the HL-LHC is not expected to have discovery potential for the stau co-annhilitation scenario or for the production of light right-handed stau pairs, making these scenarios excellent benchmarks for further study at HL-LHC as well as for future collider-based experiments.

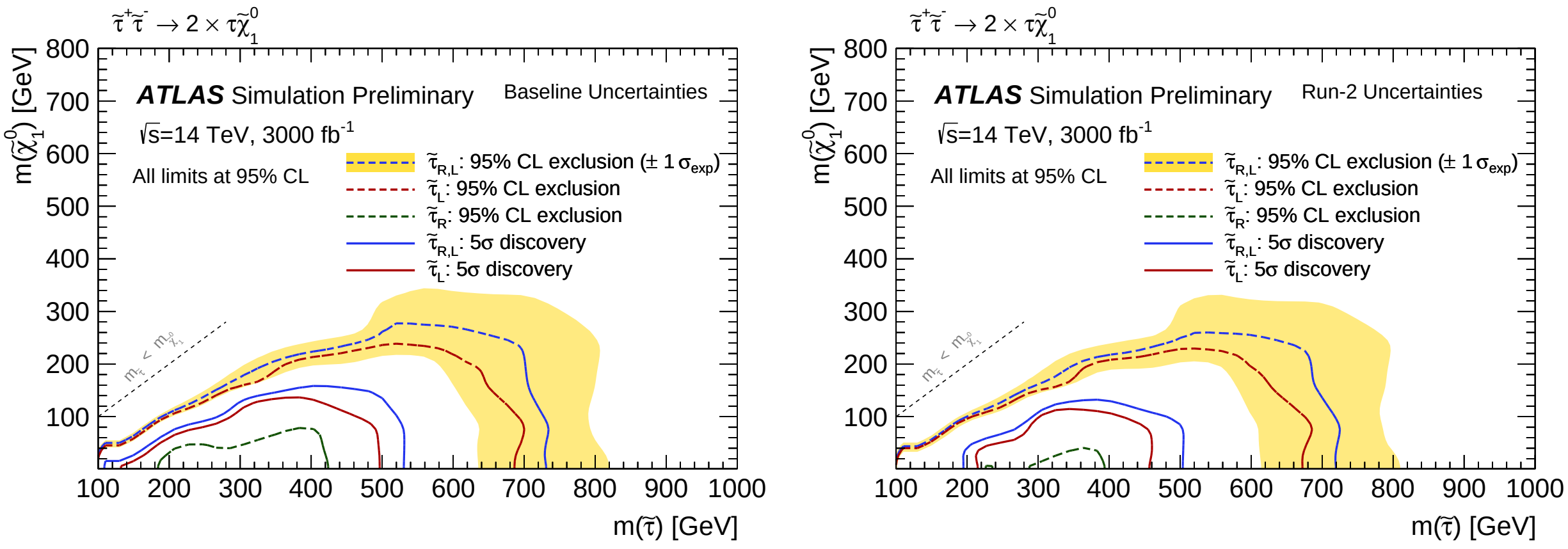

(a) Baseline Uncertainties

(b) Run-2 Uncertainties

Figure 3: The 95% CL exclusion and discovery potential for direct stau production at the HL-LHC (3000fb$^{-1}$ at $\sqrt{s}$ = 14 TeV), assuming $\tilde{\tau}_{\mathrm{L}}^{+}\tilde{\tau}_{\mathrm{L}}^{-} + \tilde{\tau}_{\mathrm{R}}^{+}\tilde{\tau}_{\mathrm{R}}^{-}$ production, $\tilde{\tau}_{\mathrm{L}}^{+}\tilde{\tau}_{\mathrm{L}}^{-}$ production, or $\tilde{\tau}_{\mathrm{R}}^{+}\tilde{\tau}_{\mathrm{R}}^{-}$ production, for (a) the Baseline Uncertainties scenario and (b) the Run-2 Uncertainties scenario.

# 5 Search for chargino pair production

In many SUSY models, the charged wino or higgsino states are light and decay via SM gauge bosons [50, 51]. A simplified model describing the direct production of $\tilde{\chi}_1^+\tilde{\chi}_1^-$ is studied here, where the $\tilde{\chi}_1^\pm$ is assumed to be pure wino, while the $\tilde{\chi}_1^0$ is the LSP and is assumed to be pure bino and stable. The $\tilde{\chi}_1^\pm$ decays with 100% branching fraction to $W^\pm$ and $\tilde{\chi}_1^0$, as seen in Figure 4. Only the leptonic decays of the $W$ are considered, resulting in final states with two opposite electric charge (OS) leptons and missing transverse momentum from the two undetected $\tilde{\chi}_1^0$.

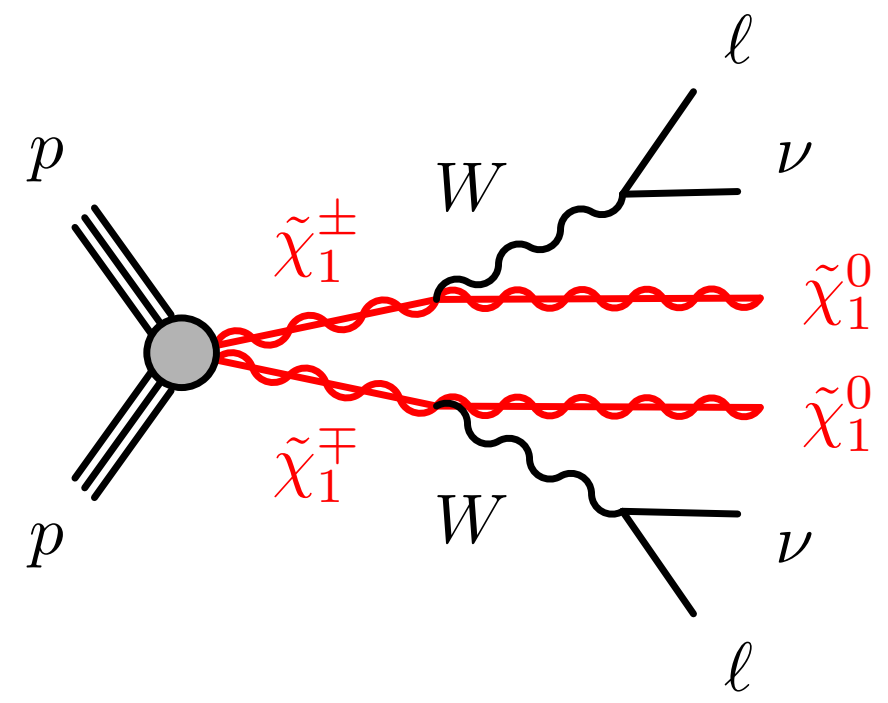

Figure 4: Diagram illustrating the signal scenario considered for the pair production of charginos targeted by the $2\ell$ final state.

The selection here closely follows the strategies adopted in the 8 TeV [52] and 13 TeV [53] searches. Events are required to contain exactly two leptons (electrons or muons) with $p_{\mathrm{T}} > 20$ GeV and $|\eta| < 2.5$ (2.47 for electrons). The lepton pair must satisfy $m_{\ell\ell} > 25$ GeV to remove contributions from low mass resonances. The two leptons must be OS, pass "tight" identification criteria [21], and be isolated (the scalar sum of the transverse momenta of charged particles with $p_{\mathrm{T}} > 1$ GeV within a cone of $\Delta R = 0.3$ around the lepton candidate, excluding the lepton candidate track itself, must be less than 15% of the lepton $p_T$). Jets are defined with $p_T > 30$ GeV and $|\eta| < 2.5$, and a $b$-tagging algorithm is used on those jets to correctly identify $b$-quark jets in simulated $t\bar{t}$ samples with an average efficiency of 85%, with a light-flavour jet misidentification probability of a few percent (parametrised as a function of jet $p_T$ and $\eta$). All leptons are required to be separated from each other and from jets. The latter requirement is imposed to suppress the background from semi-leptonic decays of heavy-flavour quarks, which is further suppressed by vetoing events having one or more $b$-tagged jets.

The selection strategy is shown in Table 5. The signal region is divided into two disjoint regions with a Same Flavour Opposite Sign (SFOS: $e^+e^-$, $\mu^+\mu^-$) or Different Flavour Opposite Sign (DFOS: $e^\pm\mu^\mp$) lepton pair to take advantage of the differing SM background composition for each flavour combination. The SFOS and DFOS regions are divided again into events with exactly zero jets or one jet, which target scenarios with large or small $\tilde{\chi}_1^\pm - \tilde{\chi}_1^0$ mass splittings, respectively. One lepton must have $p_T > 40$ GeV to suppress the SM background, and with $p_T^{\ell 1} > 40$ GeV and $p_T^{\ell 2} > 20$ GeV, either the single or double lepton triggers may be used to accept the event at the HL-LHC. Events with SFOS lepton pairs with an invariant mass within 30 GeV of the $Z$ boson mass are rejected to suppress the large $Z \to \ell\ell$ SM background. Large $E_T^{\text{miss}}$ and $E_T^{\text{miss}}$ significance ($E_T^{\text{miss}}$ significance = $E_T^{\text{miss}}/\sqrt{\sum p_T^{\text{leptons, jets}}}$) are chosen in accordance with the 13 TeV analysis [53] to suppress $Z$+jets events with poorly measured leptons.

The stransverse mass $m_{T2}$ defined in Equation 1 is calculated using the two leptons and $E_T^{\text{miss}}$, and used as the main discriminator in the SR selection to suppress the SM background. For $t\bar{t}$ or $WW$ decays, assuming an ideal detector with perfect momentum resolution, $m_{T2}(\ell, \ell, E_T^{\text{miss}})$ has a kinematic endpoint at the mass of the $W$ boson. Signal models with sufficient mass splittings between the $\tilde{\chi}_1^\pm$ and the $\tilde{\chi}_1^0$ feature $m_{T2}$ distributions that extend beyond this kinematic endpoint expected for the dominant SM backgrounds. Therefore, events in this search are required to have high $m_{T2}$ values. A set of disjoint signal regions "binned" in $m_{T2}$ are used to maximise model-dependent exclusion sensitivity. Each SR is identified by the lepton flavour combination (SFOS or DFOS), number of jets (-0J or -1J) and the range of the $m_{T2}$ interval, as seen in Table 5.

The stransverse mass $m_{T2}$ of SM and SUSY events in the signal regions is shown in Figure 5, for events passing $m_{T2} > 100$ GeV. Generally, the SM backgrounds drop off at lower $m_{T2}$ values (around the $W$ mass), while the SUSY signal and $2\ell$ diboson processes are seen to have long tails to high $m_{T2}$ values. In the 13 TeV analysis [53], long tails in $m_{T2}$ for $2\ell$ diboson processes were seen to be from the imperfect measurement of the leptons and $E_T^{\text{miss}}$ in $WW$, as well as $ZZ \to \ell^+\ell^-\nu\bar{\nu}$. Eleven high $m_{T2}$ intervals are defined to maximise the sensitivity to $\tilde{\chi}_1^+\tilde{\chi}_1^-$ production and the expected number of events from SM and SUSY processes in these signal regions are shown in Figure 5. After the application of the $Z$ veto, lepton $p_T$ thresholds and high $m_{T2}$, no $Z$+jets or $W$+jets events remain. The diboson processes are seen to dominate the total SM background across all signal regions. In the 13 TeV analysis this was seen to be mostly $WW$, due to its similarity with the SUSY signal.

To calculate the expected sensitivity to $\tilde{\chi}_1^+\tilde{\chi}_1^-$ production and decay via $W$ bosons, the expected uncertainties on the SM background are assessed. The level of accuracy achieved ($7 - 17\%$) in the 13 TeV analysis [53] was dominated by the normalisation of the $WW$ background (5%) and theoretical uncertainties on the $WW$ background ($\sim 5 - 10\%$), while the experimental uncertainties were $\sim 5\%$. The $t\bar{t}$ normalisation and theoretical uncertainties were similar to those for the $WW$ background. It is expected that the uncertainties from the normalisation of the $WW$ background will scale inversely with the increase in luminosity, and thus decrease to $\sim 1\%$, while a better understanding of $WW$ could halve the theoretical uncertainties to $\sim 2.5 - 5\%$. It is assumed that the experimental uncertainties will be understood to the same level, or better, than the 13 TeV analysis. Two scenarios for the uncertainties are considered for $\tilde{\chi}_1^+\tilde{\chi}_1^-$ production and decay via $W$ bosons at the HL-LHC, both assuming a 5% experimental uncertainty on the signal and SM background, and a 10% theoretical uncertainty on the signal. For the Run-2 Uncertainties scenario, the modelling uncertainty on the SM background is assumed to remain the same as for Run-2, at 10%. For the Baseline Uncertainties scenario, it is assumed the modelling of the $WW$ background can be understood to a better level, and the modelling uncertainty on the SM background halves to just 5%.

Table 5: Signal regions for the direct chargino pair production analysis.

| | Common | | | |
|---|---|---|---|---|
| $m_{\ell\ell}$ > [GeV] | 25 | | | |
| $p_T^{lep1}, p_T^{lep2}$ > [GeV] | 40, 20 | | | |
| number of $b$ jets | = 0 | | | |
| $E_T^{miss}$ > [GeV] | 110 | | | |
| $E_T^{miss}$ sig > [GeV$^{1/2}$] | 10 | | | |
| | SR-SFOS-0J | SR-SFOS-1J | SR-DFOS-0J | SR-DFOS-1J |
| lepton flavour/sign | SFOS | | DFOS | |
| $\lvert m_{SFOS} - m_Z \rvert$ > [GeV] | 30 | | – | |
| number of jets | = 0 | = 1 | = 0 | = 1 |
| $m_{T2}$ [GeV] | ∈ [100, 120] | | | |
| | ∈ [120, 140] | | | |
| | ∈ [140, 140] | | | |
| | ∈ [160, 180] | | | |
| | ∈ [180, 200] | | | |
| | ∈ [200, 250] | | | |
| | ∈ [250, 300] | | | |
| | ∈ [300, 350] | | | |
| | ∈ [350, 400] | | | |
| | ∈ [400, 500] | | | |
| | ∈ [500, ∞] | | | |

The statistical combination of all disjoint signal regions is used to set model-dependent exclusion limits. For each of the three uncertainties considered, half of the value is treated as correlated across signal regions, and the other half as uncorrelated. The exclusion potentials for $\tilde{\chi}_1^+\tilde{\chi}_1^-$ production and decay via $W$ bosons at the HL-LHC are shown in Figure 6. For the Run-2 Uncertainties scenario in the absence of an excess, $\tilde{\chi}_1^+\tilde{\chi}_1^-$ production may be excluded up to 840 GeV in $\tilde{\chi}_1^\pm$ mass. For the Baseline Uncertainties scenario, where the modelling uncertainty on the SM background halves from 10% to 5%, the expected exclusion potential increases by just a few GeV in $\tilde{\chi}_1^\pm$ mass and 20 GeV in $\tilde{\chi}_1^0$ mass. To calculate the discovery potential, eleven inclusive signal regions are defined with $m_{T2}$ larger than the lower bound of each $m_{T2}$ interval in Table 5, and the inclusive signal region with the best expected sensitivity is used. At the HL-LHC, the discovery potential reaches up to 610 GeV in $\tilde{\chi}_1^\pm$ mass for the Run-2 Uncertainties scenario, as seen in Figure 6(b). For the Baseline Uncertainties scenario, the discovery potential is extended by a further 50 GeV in $\tilde{\chi}_1^\pm$ mass and 80 GeV in $\tilde{\chi}_1^0$ mass.

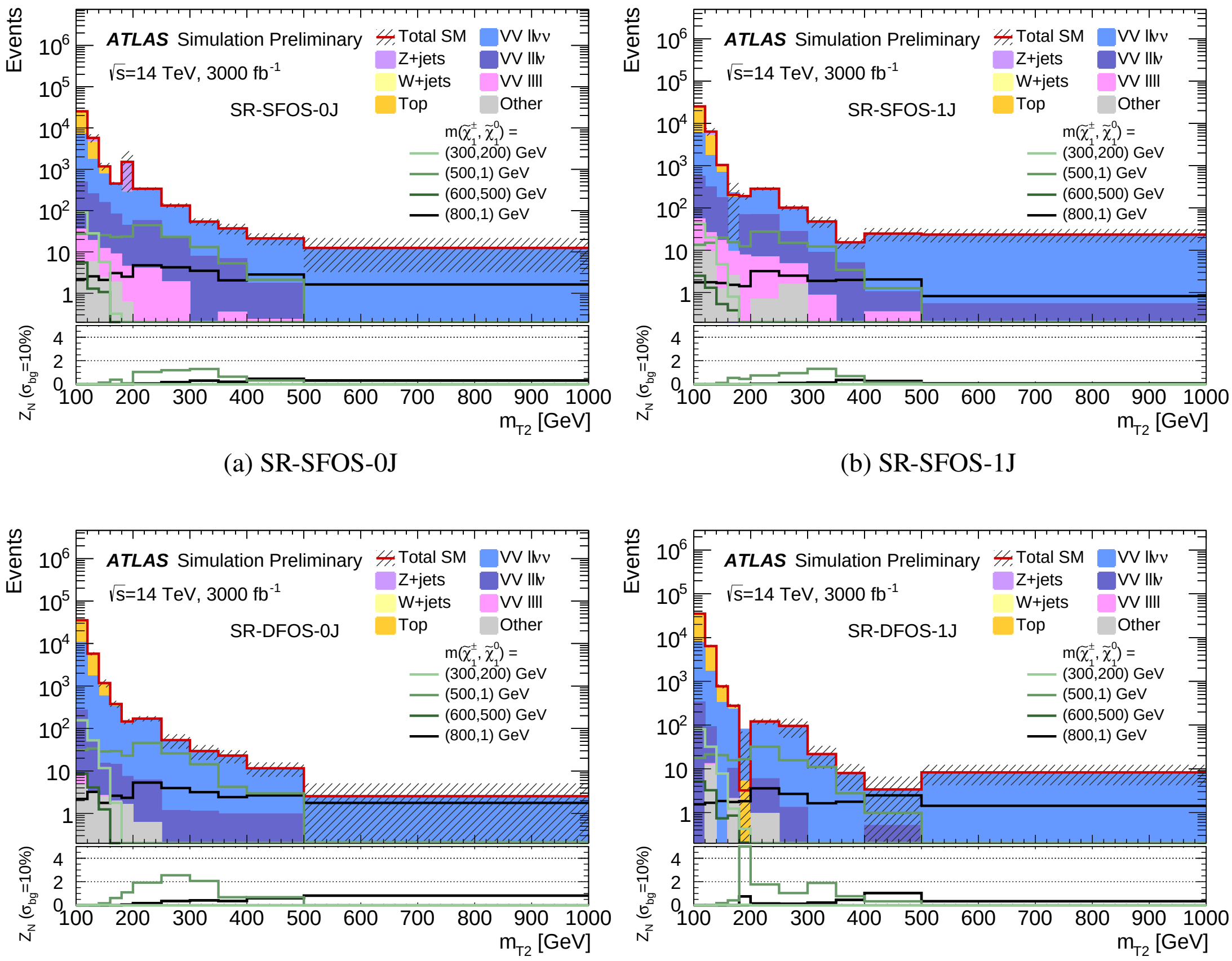

(a) SR-SFOS-0J

(b) SR-SFOS-1J

(c) SR-DFOS-0J

(d) SR-DFOS-1J

Figure 5: The expected number of events from SM and SUSY processes in the signal regions optimised for $\tilde{\chi}_1^+\tilde{\chi}_1^-$ production, for the HL-LHC. Uncertainties shown are the MC statistical uncertainties only. "Top" is the sum of the $t\bar{t}$ and single top backgrounds, while "Other" is the sum of the $t\bar{t}W$ and $t\bar{t}WW$ backgrounds. The last bin includes the overflow. The lower pad in each plot shows the significance, $Z_N$ using a background uncertainty of 10%, for a selection of SUSY scenarios in each $m_{T2}$ interval.

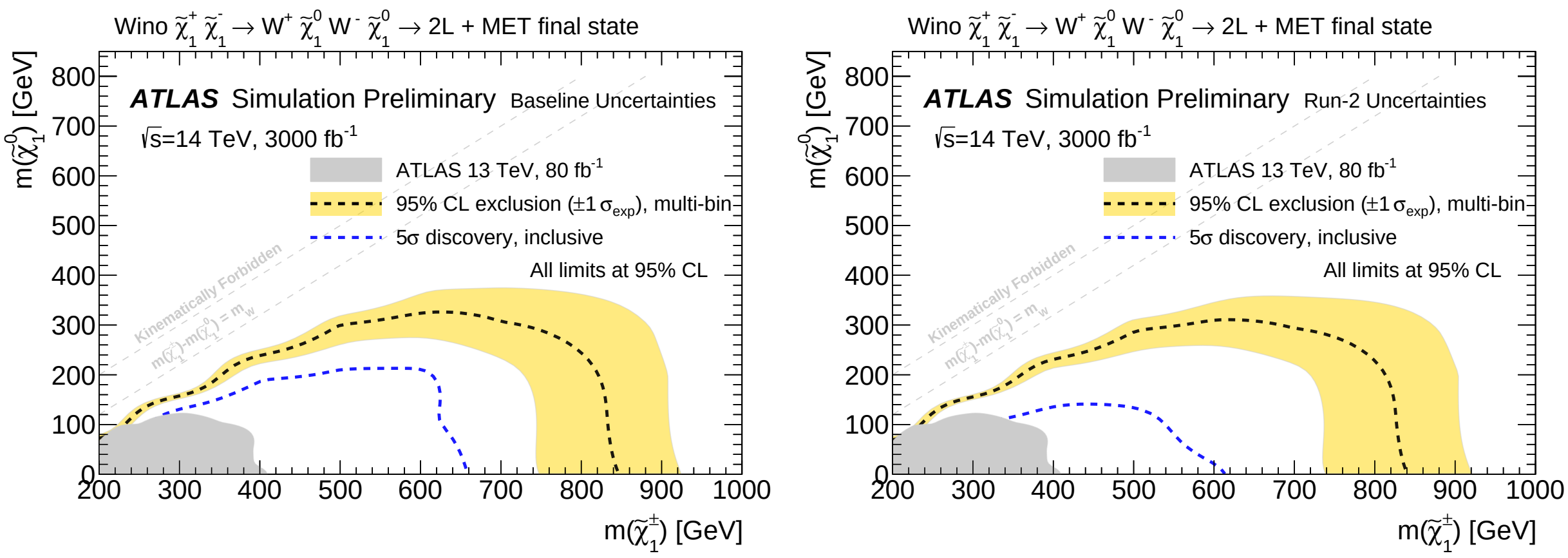

(a) Baseline Uncertainties, $\sigma_{\mathrm{bg}}^{\mathrm{modelling}} = 5\%$

(b) Run-2 Uncertainties, $\sigma_{\mathrm{bg}}^{\mathrm{modelling}} = 10\%$

Figure 6: The 95% CL exclusion and discovery potential for $\tilde{\chi}_1^+\tilde{\chi}_1^-$ production at the HL-LHC (3000fb$^{-1}$ at $\sqrt{s}$ = 14 TeV), assuming $\tilde{\chi}_1^\pm \rightarrow W\tilde{\chi}_1^0$ with a branching ratio of 100%, for an uncertainty on the modelling of the SM background of (a) 5% or (b) 10%. The observed limits from the analyses of 13 TeV data [53] are also shown.

# 6 Search for chargino-neutralino pair production

A simplified model describing the direct production of $\tilde{\chi}_1^\pm \tilde{\chi}_2^0$ is studied here, where the $\tilde{\chi}_1^\pm$ and $\tilde{\chi}_2^0$ are assumed to be pure wino and equal mass, while the $\tilde{\chi}_1^0$ is the LSP and is assumed to be pure bino and stable. The $\tilde{\chi}_1^\pm$ is assumed to decay with 100% branching fraction to $W^\pm$ and $\tilde{\chi}_1^0$, while two scenarios are considered for the $\tilde{\chi}_2^0$ decay, $\tilde{\chi}_2^0 \to Z\tilde{\chi}_1^0$ with 100% branching fraction as seen in Figure 7(a) or $\tilde{\chi}_2^0 \to h\tilde{\chi}_1^0$ with 100% branching fraction as seen in Figure 7(b). For $\tilde{\chi}_2^0 \to h\tilde{\chi}_1^0$, the light CP-even Higgs boson, $h$, of the MSSM Higgs sector is assumed to be practically identical to the SM Higgs boson [54], with the same mass and couplings as measured at the LHC [55–57]. A search for $\tilde{\chi}_1^\pm \tilde{\chi}_2^0 \to W\tilde{\chi}_1^0 Z\tilde{\chi}_1^0$ using the three lepton $(e, \mu)$ final state is described in Section 6.1, while the $1\ell bb$ final state is used for $\tilde{\chi}_1^\pm \tilde{\chi}_2^0 \to W\tilde{\chi}_1^0 h\tilde{\chi}_1^0$ and is described in Section 6.2.

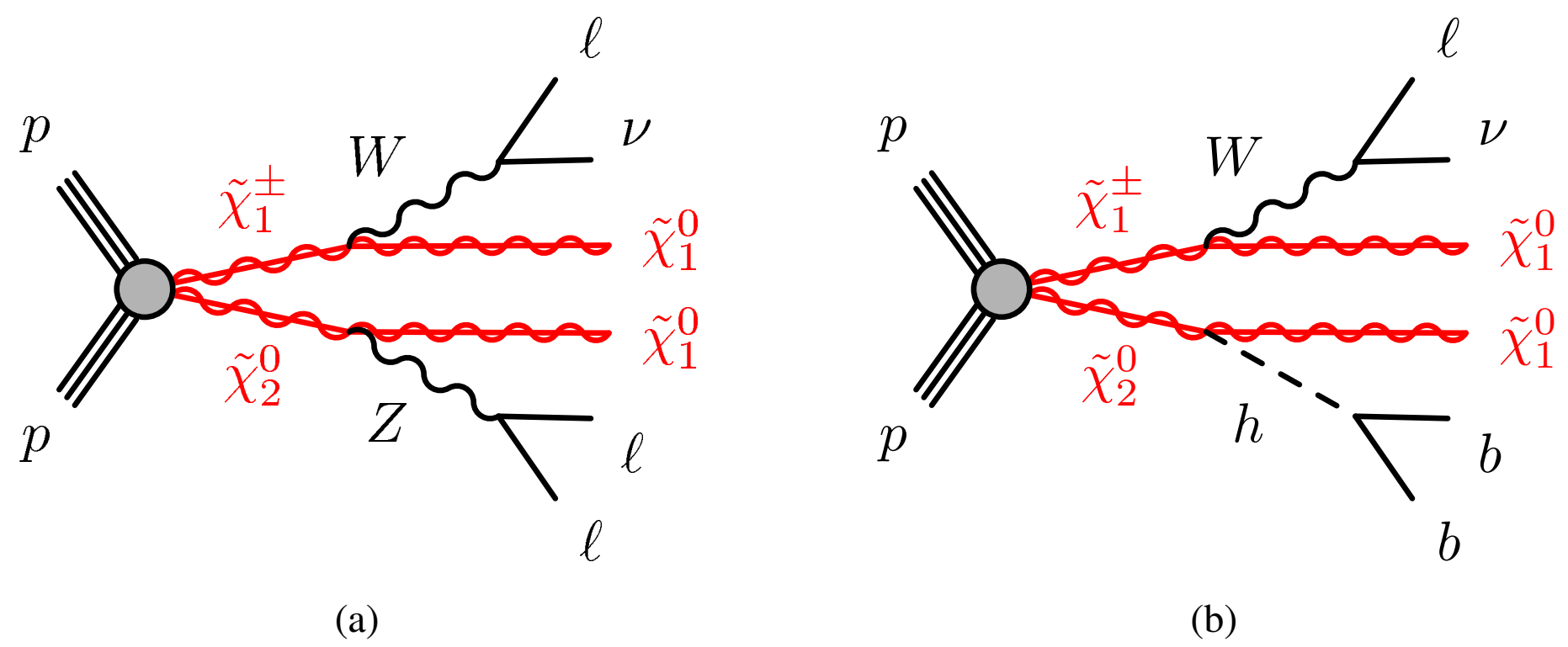

Figure 7: Diagrams illustrating the signal scenarios considered for the pair production of chargino and next-to-lightest neutralino which subsequently decay via (a) a $Z$ boson or (b) a Higgs boson $h$, targeted by the $3\ell$ and $1\ell b\bar{b}$ final states respectively.

## 6.1 Search for $\tilde{\chi}_1^\pm \tilde{\chi}_2^0 \to W\tilde{\chi}_1^0 Z\tilde{\chi}_1^0$ using three leptons

The selection for $\tilde{\chi}_1^\pm \tilde{\chi}_2^0 \to W\tilde{\chi}_1^0 Z\tilde{\chi}_1^0$ at the HL-LHC follows the strategy used in the 13 TeV search [58]. Events are selected with exactly three leptons (electrons or muons) with $p_T > 20$ GeV and $|\eta| < 2.5$, two of which must form an SFOS pair consistent with a $Z$ boson decay and have $|m_{\ell\ell} - m_Z| < 10$ GeV. To resolve ambiguities when multiple SFOS pairings are present, the transverse mass $m_T$ is calculated using the unpaired lepton for each possible SFOS pairing, and the combination that minimises the transverse mass, $m_T^{min}$, is chosen. The two leading leptons must have $p_T > 25$ GeV, and $m_{\ell\ell\ell}$ must be larger than 20 GeV to reject low mass SM decays. To suppress the $t\bar{t}$ background, events are vetoed if they contain $b$-tagged jets with $p_T > 30$ GeV and $|\eta| < 2.5$, while the $Z$+jets background is suppressed by requiring $E_T^{miss} > 50$ GeV. The chosen working point of the $b$-tagging algorithm correctly identifies $b$-quark jets in simulated $t\bar{t}$ samples with an average efficiency of 77%.

A set of disjoint signal regions binned in $m_T^{min}$ and $E_T^{miss}$ are used to maximise model-dependent exclusion sensitivity. Each SR is identified by the number of jets with $p_T > 30$ GeV and $|\eta| < 2.5$ (-0J or -1J), the range of the $E_T^{miss}$ interval and the range of the $m_T^{min}$ interval, as seen in Table 6.1. The SRs with at least one jet target signal scenarios in which the mass differences between the $\tilde{\chi}_1^+$ and $\tilde{\chi}_1^0$ is small. In such scenarios higher $E_{\mathrm{T}}^{\mathrm{miss}}$ in the event is expected when the $\tilde{\chi}_1^{\pm}\tilde{\chi}_2^0$ system recoils against the initial-state-radiation (ISR) jets. The distributions of $E_{\mathrm{T}}^{\mathrm{miss}}$ and $m_{\mathrm{T}}^{\mathrm{min}}$ in the 0-jet and 1-jet categories are shown in Figure 8 for events with $E_{\mathrm{T}}^{\mathrm{miss}} > 150$ GeV and $m_{\mathrm{T}}^{\mathrm{min}} > 150$ GeV.

Table 6: Signal regions for the search for chargino-neutralino pair production and decay to three leptons.

| | Common | | | | | |
|---|---|---|---|---|---|---|
| lepton flavour/sign | $e^+e^-\ell^{\pm}$ or $\mu^+\mu^-\ell^{\pm}$ | | | | | |
| $p_{\mathrm{T}}^{\ell 1}, p_{\mathrm{T}}^{\ell 2}, p_{\mathrm{T}}^{\ell 3}$ > [GeV] | 25, 25, 20 | | | | | |
| $\lvert m_{\mathrm{SFOS}} - m_Z \rvert$ < [GeV] | 10 | | | | | |
| $m_{\ell\ell\ell}$ > [GeV] | 20 | | | | | |
| number of $b$ jets | = 0 | | | | | |
| **SR** | **SR-0J** | | **SR-1J** | | **Inclusive** | |
| number of jets | = 0 | | $\geq 1$ | | $\geq 0$ | |
| [GeV] | $m_{\mathrm{T}}^{\mathrm{min}}$ | $E_{\mathrm{T}}^{\mathrm{miss}}$ | $m_{\mathrm{T}}^{\mathrm{min}}$ | $E_{\mathrm{T}}^{\mathrm{miss}}$ | $m_{\mathrm{T}}^{\mathrm{min}}$ | $E_{\mathrm{T}}^{\mathrm{miss}}$ |
| | $\in [150, 250]$ | $\in [200, 250]$ | $\in [150, 250]$ | $\in [200, 250]$ | $> 250$ | $> 200$ |
| | | $\in [250, 350]$ | | $\in [250, 350]$ | | $> 250$ |
| | | $\in [350, 450]$ | | $\in [350, 450]$ | | $> 350$ |
| | | $\in [450, \infty]$ | | $\in [450, 600]$ | | $> 450$ |
| | | | | $\in [600, \infty]$ | | $> 500$ |
| | | | | | | $> 600$ |
| | $\in [250, 400]$ | $\in [150, 250]$ | $\in [250, 400]$ | $\in [150, 250]$ | $> 400$ | $> 200$ |
| | | $\in [250, 350]$ | | $\in [250, 350]$ | | $> 250$ |
| | | $\in [350, 500]$ | | $\in [350, 500]$ | | $> 350$ |
| | | $\in [500, \infty]$ | | $\in [500, \infty]$ | | $> 450$ |
| | | | | | | $> 500$ |
| | | | | | | $> 600$ |
| | $\in [400, \infty]$ | $\in [150, 350]$ | $\in [400, \infty]$ | $\in [150, 350]$ | | |
| | | $\in [350, 450]$ | | $\in [350, 450]$ | | |
| | | $\in [450, 600]$ | | $\in [450, 600]$ | | |
| | | $\in [600, \infty]$ | | $\in [600, \infty]$ | | |

The expected number of events in the exclusive SRs, SR-0J and SR-1J, are summarised in Tables 7 and 8. In all regions the dominant background is $WZ$ production ($80 - 100\%$ of the total background), followed by the fakes from $t\bar{t}$ and $t\bar{t}V/\gamma$.

The exclusive SRs in the 13 TeV analysis [58] were dominated by statistical uncertainties on the background estimation ($5 - 30\%$), while uncertainties on the diboson modelling ($1 - 6\%$) and those on jet and $E_{\mathrm{T}}^{\mathrm{miss}}$ modelling ($2 - 7\%$) were also important. It is expected that the statistical uncertainties on the background estimation will scale with the inverse of the square root of the luminosity, and decrease to $1 - 5\%$. It is assumed that the experimental uncertainties will be understood to the same level as the 13 TeV analysis.

The statistical combination of all disjoint signal regions is used to set model-dependent exclusion limits. The expected sensitivity to ($\tilde{\chi}_1^{\pm}/\tilde{\chi}_2^0$) production is calculated considering 5% experimental uncertainties on the SM background and signal, a 10% theoretical uncertainty on the signal, and a 10% modelling uncertainty on the SM. For each of the three uncertainties considered, half of the value is treated as correlated across signal regions, and the other half as uncorrelated. With this Baseline Uncertainties scenario, Figure 9 shows the expected exclusion for $\tilde{\chi}_1^{\pm}\tilde{\chi}_2^0 \rightarrow W\tilde{\chi}_1^0 Z\tilde{\chi}_1^0$. In the absence of an excess, chargino and neutralino masses up to 1150 GeV may be excluded. The discovery potential is also shown in Figure 9, which reaches up to 920 GeV in chargino and neutralino masses. To calculate the discovery potential, twelve inclusive signal regions with $n_{\mathrm{jets}} > 0$ and lower thresholds on $E_{\mathrm{T}}^{\mathrm{miss}}$ and $m_{\mathrm{T}}^{\mathrm{min}}$ are defined, as shown in Table . The inclusive search region which gives the best expected sensitivity is used for the discovery potential calculation.

Table 7: The expected number of events from SM and SUSY processes in the three lepton 0J signal regions optimised for $\tilde{\chi}_1^{\pm}\tilde{\chi}_2^0 \rightarrow W\tilde{\chi}_1^0 Z\tilde{\chi}_1^0$ at the HL-LHC. Uncertainties shown describe the MC statistical uncertainties only. The event yields for two signal scenarios are also shown.

| $m_{\mathrm{T}}^{\mathrm{min}} \in [150, 250]$ GeV<br>$E_{\mathrm{T}}^{\mathrm{miss}}$ [GeV] | [200,250] | [250,350] | [350,450] | > 450 |
|---|---|---|---|---|
| Total SM | 190 ± 10 | 42 ± 6 | 5 ± 1 | 1.4 ± 0.7 |
| $m(\tilde{\chi}_1^{\pm}/\tilde{\chi}_2^0, \tilde{\chi}_1^0) = (1100, 0)$ GeV | 0.11 ± 0.03 | 0.44 ± 0.07 | 0.26 ± 0.05 | 0.88 ± 0.09 |
| $m(\tilde{\chi}_1^{\pm}/\tilde{\chi}_2^0, \tilde{\chi}_1^0) = (600, 400)$ GeV | 13 ± 2 | 11 ± 2 | 2.6 ± 0.8 | 1.2 ± 0.5 |
| $m_{\mathrm{T}}^{\mathrm{min}} \in [250, 400]$ GeV<br>$E_{\mathrm{T}}^{\mathrm{miss}}$ [GeV] | [150,250] | [250,350] | [350,450] | > 450 |
| Total SM | 34 ± 3 | 10 ± 5 | 4.1 ± 1.2 | 1.1 ± 0.7 |
| $m(\tilde{\chi}_1^{\pm}/\tilde{\chi}_2^0, \tilde{\chi}_1^0) = (1100, 0)$ GeV | 0.1 ± 0.03 | 0.36 ± 0.06 | 0.56 ± 0.07 | 1.4 ± 0.1 |
| $m(\tilde{\chi}_1^{\pm}/\tilde{\chi}_2^0, \tilde{\chi}_1^0) = (600, 400)$ GeV | 11 ± 1 | 8 ± 1 | 2.6 ± 0.8 | 0 ± 0 |
| $m_{\mathrm{T}}^{\mathrm{min}} > 400$ GeV<br>$E_{\mathrm{T}}^{\mathrm{miss}}$ [GeV] | [150,350] | [350,450] | [450,600] | > 600 |
| Total SM | 35 ± 3 | 8 ± 2 | 6 ± 1 | 2.1 ± 0.8 |
| $m(\tilde{\chi}_1^{\pm}/\tilde{\chi}_2^0, \tilde{\chi}_1^0) = (1100, 0)$ GeV | 0.41 ± 0.06 | 0.73 ± 0.09 | 1.3 ± 0.1 | 3.6 ± 0.2 |
| $m(\tilde{\chi}_1^{\pm}/\tilde{\chi}_2^0, \tilde{\chi}_1^0) = (600, 400)$ GeV | 4.2 ± 0.9 | 1.2 ± 0.5 | 0.23 ± 0.23 | 0 ± 0 |

Table 8: The expected number of events from SM and SUSY processes in the three lepton 1J signal regions optimised for $\tilde{\chi}_1^{\pm}\tilde{\chi}_2^0 \rightarrow W\tilde{\chi}_1^0 Z\tilde{\chi}_1^0$ at the HL-LHC. Uncertainties shown describe the MC statistical uncertainties only. The event yields for two signal scenarios are also shown.

| $m_{\mathrm{T}}^{\mathrm{min}} \in [150, 250]$ GeV<br>$E_{\mathrm{T}}^{\mathrm{miss}}$ [GeV] | [200,250] | [250,350] | [350,450] | [450,600] | > 600 |
|---|---|---|---|---|---|
| Total SM | 220 ± 15 | 74 ± 7 | 11 ± 2 | 6.2 ± 1.3 | 1.6 ± 0.9 |
| $m(\tilde{\chi}_1^{\pm}/\tilde{\chi}_2^0, \tilde{\chi}_1^0) = (1100, 0)$ GeV | 0.11 ± 0.03 | 0.38 ± 0.06 | 0.33 ± 0.06 | 0.48 ± 0.07 | 0.43 ± 0.07 |
| $m(\tilde{\chi}_1^{\pm}/\tilde{\chi}_2^0, \tilde{\chi}_1^0) = (600, 400)$ GeV | 9.4 ± 1.1 | 14 ± 2 | 4.2 ± 1.1 | 2.8 ± 0.8 | 1.4 ± 0.6 |
| $m_{\mathrm{T}}^{\mathrm{min}} \in [250, 400]$ GeV<br>$E_{\mathrm{T}}^{\mathrm{miss}}$ [GeV] | [150,250] | [250,350] | [350,500] | > 500 | |
| Total SM | 46 ± 8 | 21 ± 9 | 9.3 ± 2.4 | 2.5 ± 0.9 | |
| $m(\tilde{\chi}_1^{\pm}/\tilde{\chi}_2^0, \tilde{\chi}_1^0) = (1100, 0)$ GeV | 0.09 ± 0.03 | 0.36 ± 0.06 | 0.72 ± 0.09 | 1.2 ± 0.1 | |
| $m(\tilde{\chi}_1^{\pm}/\tilde{\chi}_2^0, \tilde{\chi}_1^0) = (600, 400)$ GeV | 7.3 ± 1.2 | 10 ± 1 | 4.2 ± 1.2 | 2.1 ± 0.7 | |
| $m_{\mathrm{T}}^{\mathrm{min}} > 400$ GeV<br>$E_{\mathrm{T}}^{\mathrm{miss}}$ [GeV] | [150,350] | [350,450] | [450,600] | > 600 | |
| Total SM | 31 ± 3 | 6.2 ± 1.4 | 4.3 ± 1.1 | 1.3 ± 0.7 | |
| $m(\tilde{\chi}_1^{\pm}/\tilde{\chi}_2^0, \tilde{\chi}_1^0) = (1100, 0)$ GeV | 0.44 ± 0.07 | 0.64 ± 0.08 | 1.3 ± 0.1 | 3.6 ± 0.2 | |
| $m(\tilde{\chi}_1^{\pm}/\tilde{\chi}_2^0, \tilde{\chi}_1^0) = (600, 400)$ GeV | 3.7 ± 0.9 | 1.4 ± 0.6 | 0.47 ± 0.33 | 0.47 ± 0.33 | |

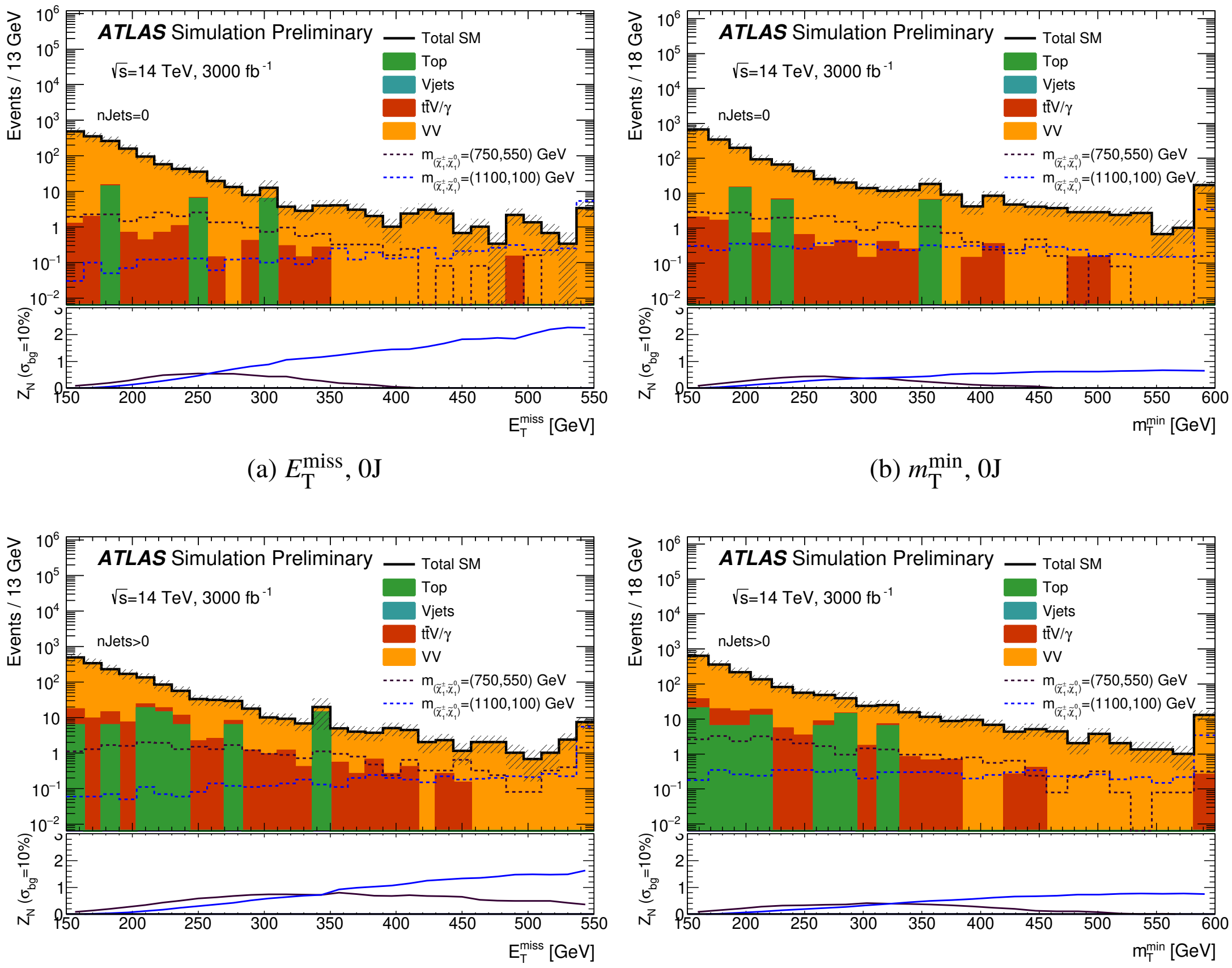

(a) $E_{\mathrm{T}}^{\mathrm{miss}}$, 0J

(b) $m_{\mathrm{T}}^{\mathrm{min}}$, 0J

(c) $E_{\mathrm{T}}^{\mathrm{miss}}$, 1J

(d) $m_{\mathrm{T}}^{\mathrm{min}}$, 1J

Figure 8: The distribution of $E_{\mathrm{T}}^{\mathrm{miss}}$ and $m_{\mathrm{T}}^{\mathrm{min}}$ in the events with zero jets (top) and the events with at least one jet (bottom). All common requirements along with $E_{\mathrm{T}}^{\mathrm{miss}} > 150$ GeV and $m_{\mathrm{T}}^{\mathrm{min}} > 150$ GeV are applied. The last bin includes the overflow. The lower pad in each plot shows the significance for the SUSY reference points, $Z_N$, as the threshold on the $x$-axis variable increases and assumes a background uncertainty of 10%.

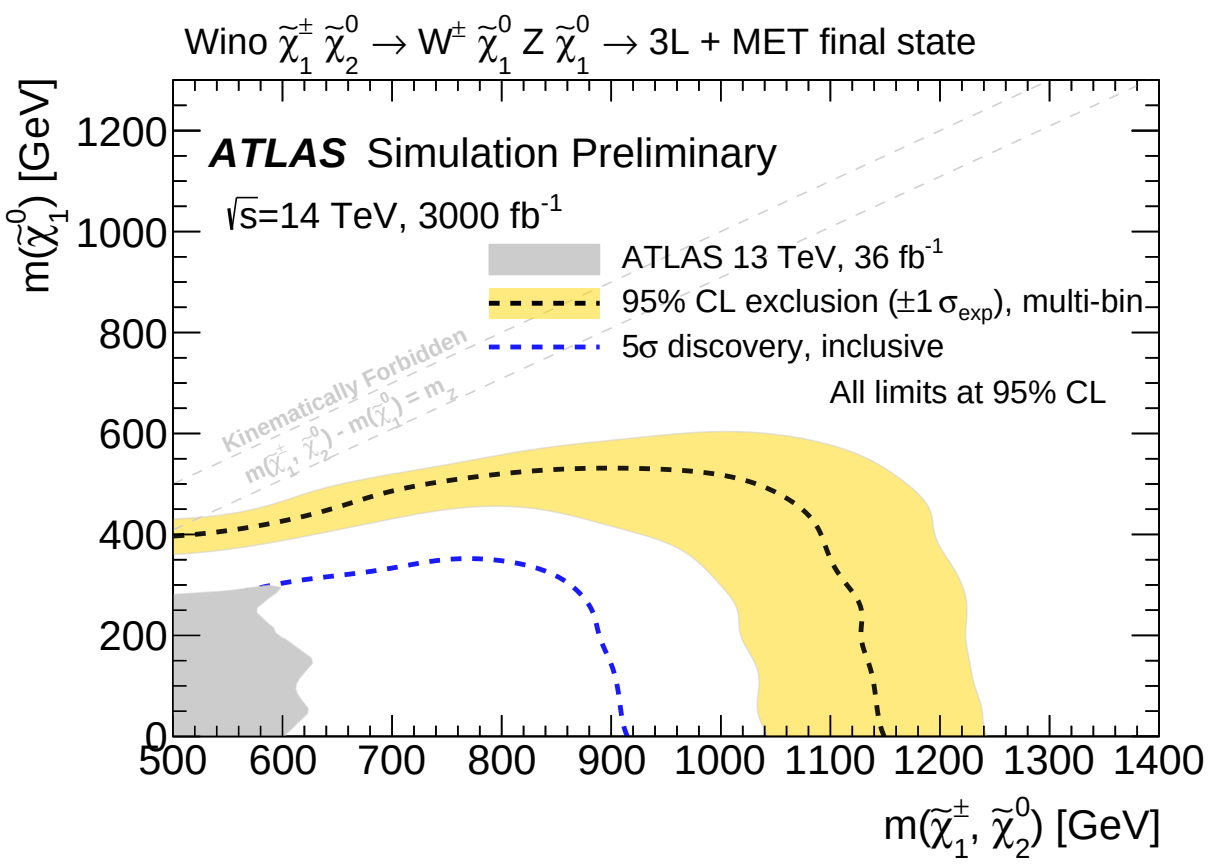

Figure 9: The 95% CL exclusion and discovery potential for $\tilde{\chi}_1^\pm \tilde{\chi}_2^0$ production at the HL-LHC (3000fb$^{-1}$ at $\sqrt{s}$ = 14 TeV), assuming $\tilde{\chi}_1^\pm \to W \tilde{\chi}_1^0$ and $\tilde{\chi}_2^0 \to Z \tilde{\chi}_1^0$ with a branching ratio of 100%. The observed limits from the analyses of 13 TeV data [52, 58–60] are also shown.

## 6.2 Search for $\tilde{\chi}_1^\pm \tilde{\chi}_2^0 \to W \tilde{\chi}_1^0 h \tilde{\chi}_1^0$ using $1\ell bb$

This analysis updates the previous studies in the same final states for the HL-LHC as in Ref. [61], using the latest parameterisations of the upgraded ATLAS detector configurations for the $\langle\mu\rangle \sim 200$ HL-LHC running conditions and the associated physics object systematic uncertainties, as well as a re-optimised multivariate based analysis method.

Signal models with $\tilde{\chi}_1^\pm$ and $\tilde{\chi}_2^0$ masses up to 1500 GeV are considered. The analysis is performed separately in three signal regions targetting signal models with different values of mass difference $\Delta m = m(\tilde{\chi}_1^\pm/\tilde{\chi}_2^0) - m(\tilde{\chi}_1^0)$, which leads to different kinematic shapes. In each region, one benchmark signal model is selected as a reference point for the optimisation of event selections and sensitivity estimations. The definitions of three regions and the corresponding benchmark signal models are:

- Low: $\Delta m < 300$ GeV, benchmark signal model $m(\tilde{\chi}_1^\pm/\tilde{\chi}_2^0, \tilde{\chi}_1^0) = (500, 310)$ GeV,
- Medium: $\Delta m \in [300, 600]$ GeV, benchmark signal model $m(\tilde{\chi}_1^\pm/\tilde{\chi}_2^0, \tilde{\chi}_1^0) = (800, 420)$ GeV,
- High: $\Delta m > 600$ GeV, benchmark signal model $m(\tilde{\chi}_1^\pm/\tilde{\chi}_2^0, \tilde{\chi}_1^0) = (1000, 1)$ GeV.

At each point in the signal model parameter space, the region with the best sensitivity is chosen for the estimate of the final analysis sensitivities, instead of a statistical combination of the regions. For this reason, event selections in different regions are not necessarily orthogonal.

The expected SM background is dominated by top quark pair-production $t\bar{t}$ and single top production, with smaller contributions from vector boson production $W$+jets, associated production of $t\bar{t}$ and a vector boson ($t\bar{t}V$) and dibosons.

### 6.2.1 Event selection

The event selection follows a similar strategy as in the previous studies documented in Ref. [61]. Candidate leptons (electrons or muons) are required to have $p_{\mathrm{T}} > 25$ GeV and $|\eta| < 2.47$ ($|\eta| < 2.7$ for muons), and pass "tight" and "medium" identification criteria for electrons and muons respectively. Candidate jets are reconstructed using the anti-$k_t$ algorithm with $R = 0.4$, are required to have $p_{\mathrm{T}} > 25$ GeV and $|\eta| < 2.5$. B-tagged jets are required to pass the jet requirements described previously, and pass the MV2c10 tagging algorithm operating at 77% b-jet tagging efficiency. Candidate jets and electrons are required to satisfy $\Delta R(e, \text{jet}) > 0.2$. Any leptons within $\Delta R = 0.4$ of the remaining jet are removed. The $E_{\mathrm{T}}^{\mathrm{miss}}$ at generator level is calculated as the vectorial sum of the momenta of neutral weakly-interacting particles, in this case neutrinos and neutralinos. The detector response is simulated using a set of parametrised functions as described in Section 3.

The impact of the trigger is not taken into account in this analysis. The planned upgrades to the detector, in particular an improved barrel muon coverage, are expected to allow lepton triggers that would have high efficiency for the studied scenarios with respect to the analysis selections.

Events containing exactly one lepton, and two or three jets passing the above object definitions are selected. Exactly two of the jets are required to be $b$-tagged with the criteria defined above. Four key variables are further used to discriminate signal from background:

- $m_{\mathrm{bb}}$ - the invariant mass of the two $b$-tagged jets
- $E_{\mathrm{T}}^{\mathrm{miss}}$ - the transverse momentum imbalance in the event
- $m_{\mathrm{T}}$ - the transverse mass constructed using the lepton $p_{\mathrm{T}}$ and the $E_{\mathrm{T}}^{\mathrm{miss}}$.
- $m_{\mathrm{CT}}$ - the contransverse mass, defined for the $b\bar{b}$ system as:

$$m_{\mathrm{CT}} = 2 p_{\mathrm{T}}^{b_1} p_{\mathrm{T}}^{b_2} (1 + \cos \Delta\phi_{bb}), \tag{2}$$

  where $p_{\mathrm{T}}^{b_1}$ and $p_{\mathrm{T}}^{b_2}$ are transverse momenta of the two leading $b$−jets and $\Delta\phi_{bb}$ is the azimuthal angle between them.

The $m_{\mathrm{bb}}$ is used to select events which have dijet masses within a window of the Higgs boson mass. The transverse mass variable $m_{\mathrm{T}}$ is effective at suppressing SM backgrounds containing $W$ bosons due to the expected kinematic endpoint around the $W$ boson mass assuming an ideal detector with perfect momentum resolution. The contransverse mass variable $m_{\mathrm{CT}}$ is an effective variable to select Higgs boson decays into $b$−quarks and to suppress the $t\bar{t}$ backgrounds [62, 63].

A set of common loose requirements, referred to as preselection, are applied first to suppress the fully hadronic multijet and $W$+jets backgrounds: $m_{\mathrm{T}} > 40$ GeV, $m_{\mathrm{bb}} > 50$ GeV, $E_{\mathrm{T}}^{\mathrm{miss}} > 200$ GeV. Figure 10 shows the distributions of the key discriminating variables after this selection, comparing the three benchmark signal models with the expected SM background.

To further distinguish between signal and background processes, a set of rectangular selections based on these kinematic observables is first studied to evaluate possible optimal selections and residual SM background. A multivariate method based on boosted decision trees (BDT) is then chosen for the optimal sensitivity. In this approach, three independent BDTs (referred to as M1, M2 and M3), are trained separately in each signal region for events passing the preselection and within the $m_{\mathrm{bb}}$ mass window of [105, 135] GeV. All the signal MC samples within a given signal region are combined to mimic the average kinematic shapes of the signal. Only the dominant $t\bar{t}$ background is considered in the training. In all regions, the following seven variables are used as inputs: $E_{\mathrm{T}}^{\mathrm{miss}}$, $m_{\mathrm{T}}$, $m_{\mathrm{CT}}$, the lepton $p_{\mathrm{T}}$, the leading and sub-leading b-jet $p_{\mathrm{T}}$, as well as the angular separation of the two $b$-tagged jets $\Delta R(b_1, b_2)$.

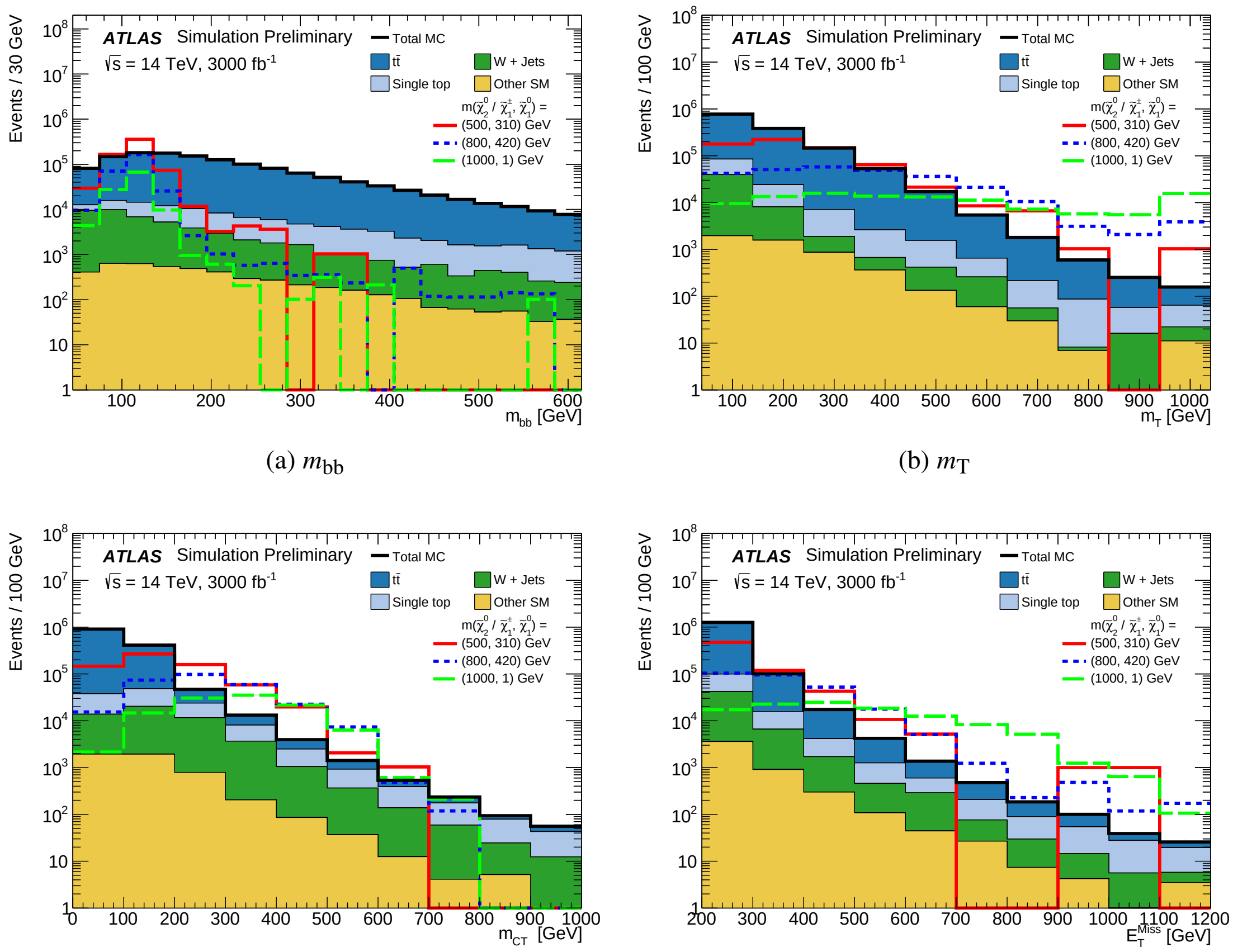

(a) $m_{\mathrm{bb}}$ (b) $m_{\mathrm{T}}$ (c) $m_{\mathrm{CT}}$ (d) $E_{\mathrm{T}}^{\mathrm{miss}}$

Figure 10: Distributions of the key discriminating variables at the preselection level. The contributions from all SM background are shown as stacked, and the expected distribution from the benchmark signal models are overlaid. The last bin does not include the overflow.

The BDT output distributions are then used to optimise signal regions maximising the expected significance $Z_N$ of the benchmark signal model. Examples of the BDT output distributions are shown in Figure 11. The resulting signal region definitions are shown in Table 9.

Table 9: Definitions of the signal regions with the benchmark signal model parameters used in the optimisation.

| SR | Benchmark signal model parameters $m(\tilde{\chi}_1^{\pm}/\tilde{\chi}_2^0, \tilde{\chi}_1^0)$ [GeV] | BDT range |
|---|---|---|
| SR-M1 | (500, 310) | > 0.25 |
| SR-M2 | (800, 420) | > 0.35 |
| SR-M3 | (1000, 1) | > 0.30 |

Table 10 shows the expected number of events for the SM background and three benchmark SUSY scenarios respectively. The SM background is dominated by the top backgrounds, including both the $t\bar{t}$ and single top processes.

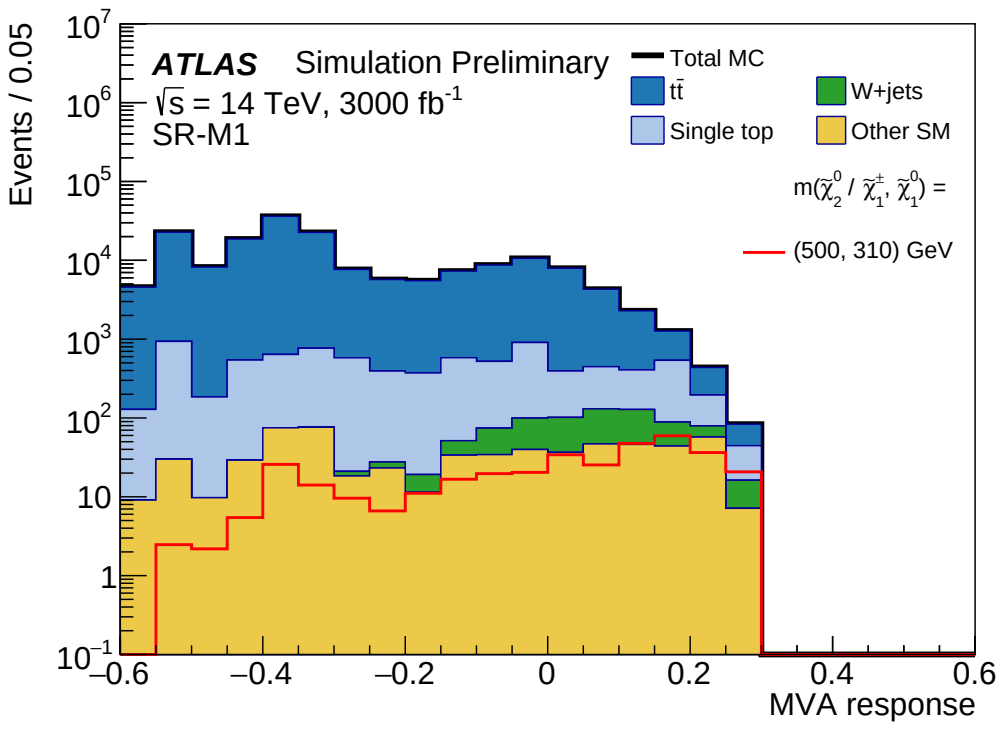

(a) M1

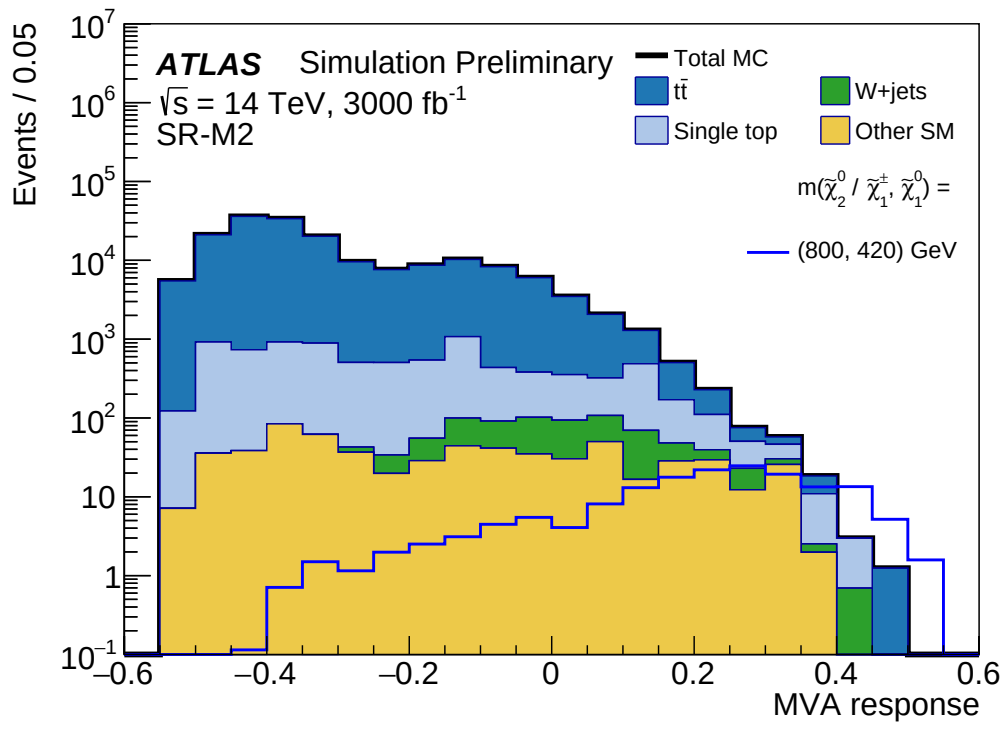

(b) M2

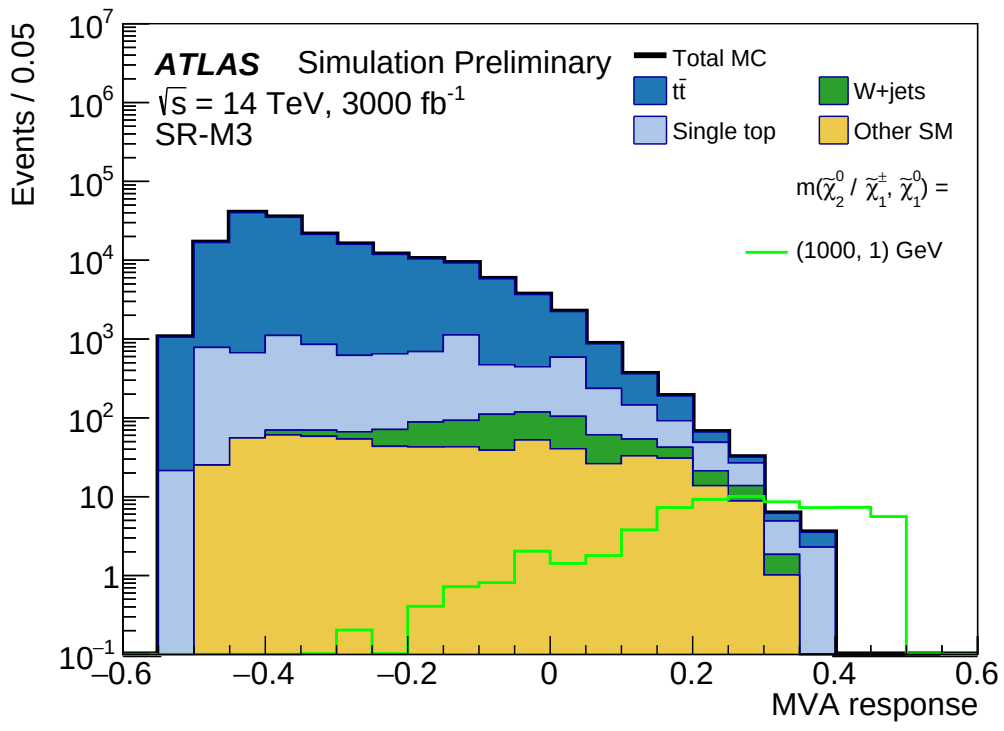

(c) M3

Figure 11: Distributions of the BDT responses in the three signal regions for the events that pass the preselection and are within $m_{\mathrm{bb}}$ mass window of [105, 135] GeV. The contributions from all SM background are shown as stacked, and the expected distribution from the benchmark signal models are overlaid.

The largest systematic uncertainties are from theoretical modelling of the irreducible backgrounds of $t\bar{t}$ and single top, mainly from the generator difference, renormalisation and factorisation scale variations and the interference between the $t\bar{t}$ and single top background. The total theoretical uncertainty is estimated to be about 7%. Experimental uncertainties are dominated by the jet energy scale (JES) and jet energy resolution (JER), on the order of 6%.

Figure 12 shows the expected 95% CL exclusion and $5\sigma$ discovery contours for the simplified models described earlier. In this model, masses of $\tilde{\chi}_1^{\pm}/\tilde{\chi}_2^0$ up to 1280 GeV could be excluded at 95% confidence level for a massless $\tilde{\chi}_1^0$. The discovery potential at $5\sigma$ could be extended up to 1080 GeV for a massless $\tilde{\chi}_1^0$. More mature analysis and reconstruction techniques such as performing a multi-bin shape fit, improving the training in the multivariate method by including other SM backgrounds, using jet substructure techniques in the boosted Higgs boson region, and performing a statistical combination of all signal regions would likely extend the sensitivity even further.

Table 10: Expected number of events for the SM background and the benchmark signal models in the $1\ell bb$ signal regions SR-M1, SR-M2, and SR-M3. The uncertainties describe the MC statistical uncertainties only.

| Processes | SR-M1 | SR-M2 | SR-M3 |
|---|---|---|---|
| $t\bar{t}$ | 38.9 ± 8.4 | 8.7 ± 3.3 | 2.5 ± 1.8 |
| single top | 28.3 ± 4.8 | 10.7 ± 3.2 | 5.4 ± 2.5 |
| W+jets | 22.2 ± 5.4 | 3.0 ± 2.0 | 2.0 ± 1.8 |
| $ttV$ | 5.1 ± 2.4 | 2.0 ± 1.4 | 1.0 ± 1.0 |
| Diboson | 2.0 ± 2.0 | - | - |
| total background | 97 ± 12 | 24.4± 5.2 | 10.9 ± 3.4 |
| $m(\tilde{\chi}_1^{\pm}/\tilde{\chi}_2^0, \tilde{\chi}_1^0) = (500, 310)$ GeV | 20.7 ± 4.8 | 4.6 ± 2.3 | 1.0 ± 1.0 |
| $m(\tilde{\chi}_1^{\pm}/\tilde{\chi}_2^0, \tilde{\chi}_1^0) = (800, 420)$ GeV | 44.3 ± 2.3 | 33.6 ± 2.0 | 21.2 ± 1.6 |
| $m(\tilde{\chi}_1^{\pm}/\tilde{\chi}_2^0, \tilde{\chi}_1^0) = (1000, 1)$ GeV | 32.2 ± 1.8 | 31.9 ± 1.8 | 28.9 ± 1.7 |

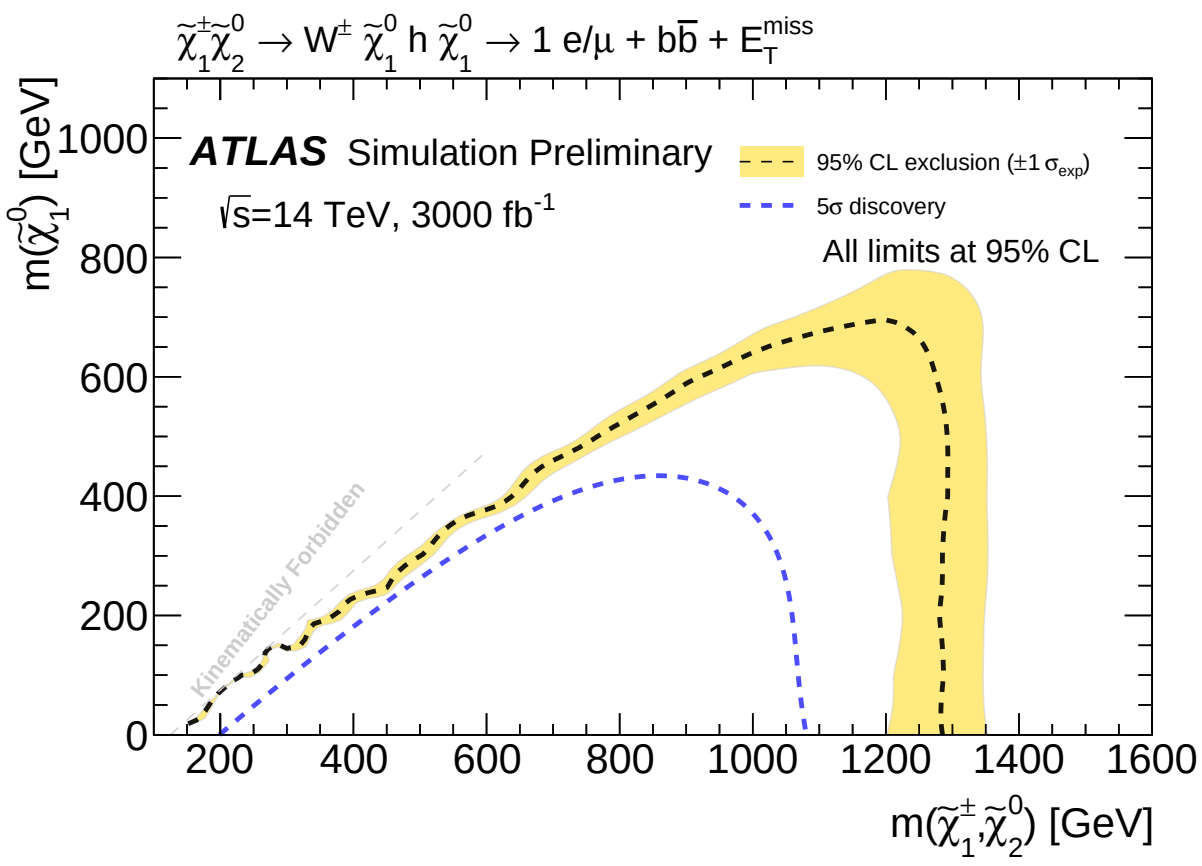

Figure 12: The 95% CL exclusion and discovery potential for $\tilde{\chi}_1^{\pm}\tilde{\chi}_2^0$ production at the HL-LHC (3000fb$^{-1}$ at $\sqrt{s} = 14$ TeV), assuming $\tilde{\chi}_1^{\pm} \rightarrow W\tilde{\chi}_1^0$ and $\tilde{\chi}_2^0 \rightarrow h\tilde{\chi}_1^0$ with a branching ratio of 100%.

# 7 Conclusion

The large dataset of around 3000 fb$^{-1}$ expected at the HL-LHC will significantly increase the ATLAS sensitivity to the productions of SUSY particles in the electroweak sector. This note summarises the expected sensitivities of direct productions of $\tilde{\tau}^+\tilde{\tau}^-$, $\tilde{\chi}_1^+\tilde{\chi}_1^-$, and $\tilde{\chi}_1^{\pm}\tilde{\chi}_2^0$ at the HL-LHC and the expected 95% exclusion regions and the 5$\sigma$ discovery regions are summarised in Table 11. The discovery sensitivities of the HL-LHC are still rather limited in the challenging compressed region where mass differences between the NLSP and LSP are small. In particular, there is no discovery potential for the theoretically favoured stau co-annhilitation with small mass differences ($\Delta m(\tilde{\tau}, \tilde{\chi}_1^0) < 100$ GeV) or for the production of $\tilde{\tau}_R$ pairs. These challenging scenarios serve as ideal benchmarks for further improvements in the detector performance, reconstruction techniques and analysis methods at the HL-LHC.

24

| SUSY particle | Final state | 95% CL exclusion region for $m(\tilde{\chi}_1^0) = 0$ GeV | $5\sigma$ discovery region for $m(\tilde{\chi}_1^0) = 0$ GeV |
|---|---|---|---|
| $\tilde{\tau}_L + \tilde{\tau}_R$ | $\tau\tau$ | $< 730$ GeV | [110, 530] GeV |
| $\tilde{\tau}_L$ | $\tau\tau$ | $< 680$ GeV | [110, 500] GeV |
| $\tilde{\tau}_R$ | $\tau\tau$ | $< 420$ GeV | |
| Wino $\tilde{\chi}_1^\pm$ | $WW$-mediated $2\ell$ | $< 840$ GeV | $< 660$ GeV |
| Wino $\tilde{\chi}_1^\pm, \tilde{\chi}_2^0$ | $WZ$-mediated $3\ell$ | $< 1150$ GeV | $< 920$ GeV |
| | $Wh$-mediated $1\ell bb$ | $< 1280$ GeV | $< 1080$ GeV |

Table 11: Summary of the 95% CL exclusion reach and the $5\sigma$ discovery reach at the end of HL-LHC for the direct productions of heavy SUSY partners in the electroweak sector, ssuming a massless $\tilde{\chi}_1^0$ LSP and baseline uncertainties. See text for the details of other assumptions of the signal models in each final state.

# CMS Physics Analysis Summary

Contact: cms-phys-conveners-ftr@cern.ch

# Searches for light higgsino-like charginos and neutralinos at the HL-LHC with the Phase-2 CMS detector

The CMS Collaboration

## Abstract

A search for the pair production of light higgsino-like charginos $\widetilde{\chi}_1^{\pm}$ and neutralinos $\widetilde{\chi}_2^0$ is presented, based on a simulation of 3000 fb$^{-1}$ of proton-proton collision data produced by the HL-LHC at a center-of-mass energy of 14 TeV. The Phase-2 CMS detector is simulated using Delphes. The $\widetilde{\chi}_1^{\pm}$ and $\widetilde{\chi}_2^0$ are assumed to be mass-degenerate, to be pair-produced ($\widetilde{\chi}_1^{\pm}\widetilde{\chi}_2^0$, $\widetilde{\chi}_2^0\widetilde{\chi}_1^0$), and to decay into the lightest stable superymmetric particle $\widetilde{\chi}_1^0$ via off-shell W and Z bosons. The $\widetilde{\chi}_1^0$ is also assumed to be higgsino-like. Candidate signal events are required to have two same-flavor, opposite-charge, low transverse momentum leptons (electrons or muons), one jet, and significant missing transverse momentum.

# 1 Introduction

Supersymmetry (SUSY) [1–5] is considered one of the most compelling theories of physics beyond the standard model (SM). It postulates the existence of new particles with spin differing by half a unit with respect to that of their SM partners. The linear superposition of the fermionic partners of the Higgs and electroweak gauge bosons, the higgsinos and gauginos respectively, are referred to as charginos $\tilde{\chi}^{\pm}_{1,2}$ and neutralinos $\tilde{\chi}^{0}_{1,2,3,4}$. If $R$-parity [6] is conserved, the lightest supersymmetric particle (LSP) $\tilde{\chi}^{0}_{1}$ is stable. The charginos and neutralinos are produced in pairs and decay into final states with SM particles and LSPs.

In scenarios of natural supersymmetry, the higgsinos may be the only low-mass supersymmetric states. The spectra will then be characterized by light higgsino-like $\tilde{\chi}^{\pm}_{1}$, $\tilde{\chi}^{0}_{2}$, and $\tilde{\chi}^{0}_{1}$ particles, while all other sparticles exhibit masses above the TeV scale. A search for higgsino-like charginos and neutralinos is therefore critical to probe for natural SUSY at the LHC and the High-Luminosity LHC (HL-LHC).

If $\tilde{\chi}^{\pm}_{1}$, $\tilde{\chi}^{0}_{2}$, and $\tilde{\chi}^{0}_{1}$ are higgsino-like, the mass splitting is driven by radiative corrections and acquires values up to a few GeV. As a result, pair-produced $\tilde{\chi}^{\pm}_{1}\tilde{\chi}^{0}_{2}$ or pair-produced $\tilde{\chi}^{0}_{2}\tilde{\chi}^{0}_{1}$ can decay promptly into $\tilde{\chi}^{0}_{1}$ only via off-shell W and Z bosons, leading to events with low transverse momentum ($p_{\mathrm{T}}$) SM particles. In leptonic decays of the Z boson, the events will contain one same-flavor, opposite-charge lepton pair, the invariant mass of which has a kinematic endpoint at $\Delta M(\tilde{\chi}^{0}_{2}, \tilde{\chi}^{0}_{1}) = m(\tilde{\chi}^{0}_{2}) - m(\tilde{\chi}^{0}_{1})$. Sensitivity to the signal is achieved by requiring at least one jet from initial-state radiation (ISR) that recoils against the two $\tilde{\chi}^{0}_{1}$ and produces significant missing transverse momentum ($p_{\mathrm{T}}^{\mathrm{miss}}$) in the event. Feynman diagrams for the signal processes are shown in Fig. 1. The ATLAS and CMS collaborations developed searches for higgsino-like $\tilde{\chi}^{\pm}_{1}$ and $\tilde{\chi}^{0}_{2}$ that used up to 36 fb$^{-1}$ of proton-proton collision data [7, 8] and started probing the parameter space beyond the LEP experiments' limits [9, 10]. By providing 3000 fb$^{-1}$ of proton-proton collision data at a center-of-mass energy of 14 TeV, the HL-LHC has the potential to significantly extend the experiments' sensitivity to higgsinos.

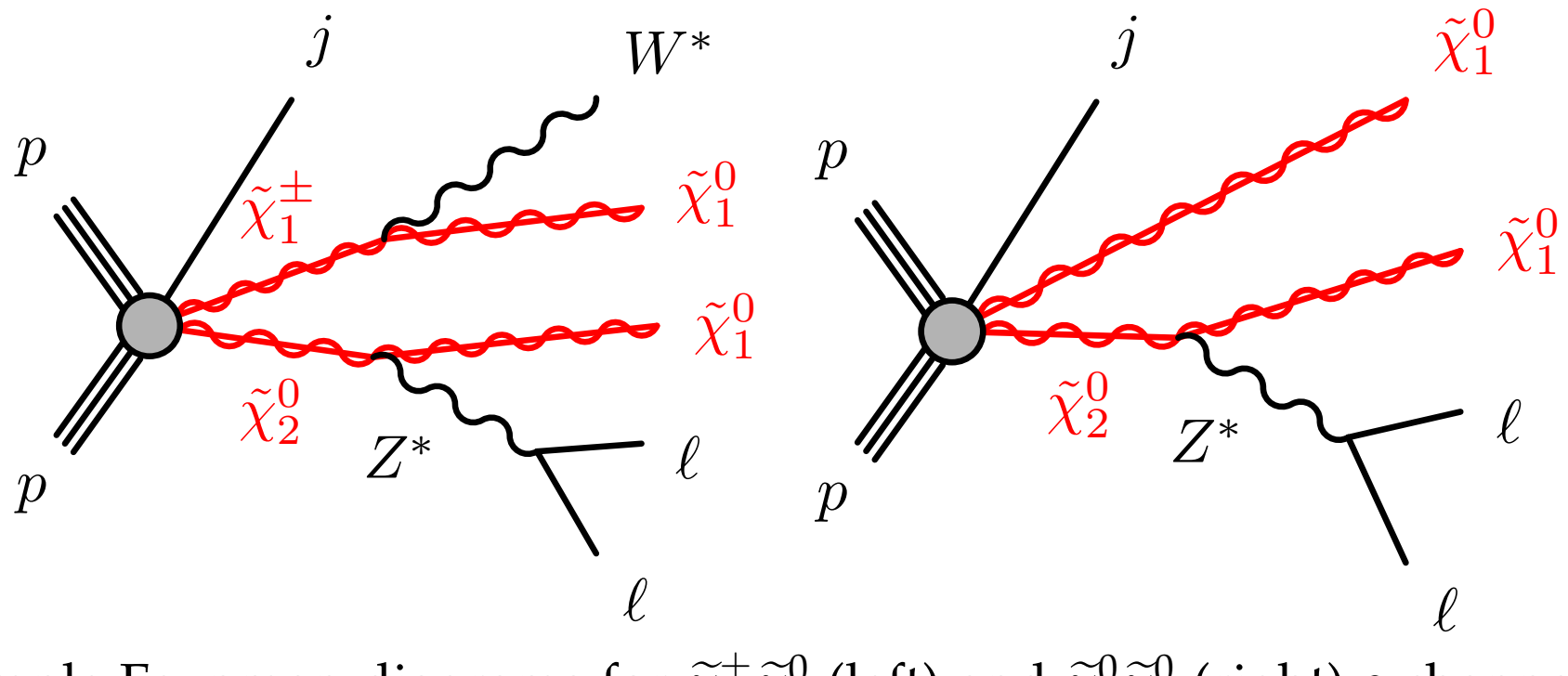

Figure 1: Example Feynman diagrams for $\tilde{\chi}^{\pm}_{1}\tilde{\chi}^{0}_{2}$ (left) and $\tilde{\chi}^{0}_{2}\tilde{\chi}^{0}_{1}$ (right) $s$-channel pair production, followed by the leptonic decay of the $\tilde{\chi}^{0}_{2}$.

The model used for the optimization of the search and its interpretation is a SUSY simplified model [11] where the higgsino-like $\tilde{\chi}^{\pm}_{1}$ and $\tilde{\chi}^{0}_{2}$ are assumed to be mass-degenerate and produced in pairs. The model thus contains both the $\tilde{\chi}^{\pm}_{1}\tilde{\chi}^{0}_{2}$ and the $\tilde{\chi}^{0}_{2}\tilde{\chi}^{0}_{1}$ production, where $\tilde{\chi}^{\pm}_{1}$ decays into $\mathrm{W}^{*}\tilde{\chi}^{0}_{1}$ and $\tilde{\chi}^{0}_{2}$ into $\mathrm{Z}^{*}\tilde{\chi}^{0}_{1}$, respectively, with a branching fraction of 100%. The region of parameter space explored by this analysis includes $m_{\tilde{\chi}^{\pm}_{1}} = m_{\tilde{\chi}^{0}_{2}}$ larger than 100 GeV to account for the lower limit set by the LEP experiments, as well as $5.0 \leq \Delta M(\tilde{\chi}^{0}_{2}, \tilde{\chi}^{0}_{1}) \leq 40$ GeV. The lower and upper bounds on $\Delta M(\tilde{\chi}^{0}_{2}, \tilde{\chi}^{0}_{1})$ are driven by the minimum lepton $p_{\mathrm{T}}$ requirement, the suppression of background events containing low mass resonances such as $J/\psi$, and the expected mass separation between the $\widetilde{\chi}^0_2$ and $\widetilde{\chi}^0_1$. In this region of parameter space, the $Z^* \rightarrow \ell\ell$ ($\ell$ = electron, muon) branching fraction depends only slightly on the $\widetilde{\chi}^0_2$ to $\widetilde{\chi}^0_1$ mass splitting (sub-percent impact); the branching fraction is therefore assumed to be the same as the branching fraction of the on-shell Z boson.

The cross sections are calculated for $\sqrt{s} = 14$ TeV at next-to-leading-order (NLO) plus next-to-leading-logarithmic (NLL) precision with the CTEQ 6.6 and MSTW 2008nlo90cl parton distribution functions (PDFs) using the RESUMMINO code [12, 13]. The signal samples are generated by the MADGRAPH5_aMC@NLO 2.2.2 [14] event generator up to two additional jets at leading order (LO) precision in perturbative QCD using the MLM merging scheme [15]. The supersymmetric particles are then decayed by the PYTHIA 8.212 package [16]. PYTHIA also provides parton showering and hadronization. The simplified models do not include any spin correlations in the decays. MADGRAPH5_aMC@NLO 2.2.2 is also used to produce selected parton-level background processes at LO (Drell-Yan, W+jets), with the parton showering and hadronization provided by PYTHIA. The $W^+W^-$ events are generated at NLO precision using the FxFx merging scheme [17] and scaled to the NLO cross-section [18]. The NNPDF3.0 [19] LO and NLO PDFs are used for the simulated samples generated at LO and NLO, respectively. Only the $t\bar{t}$ events are generated using MADGRAPH5 v1.5.10 and the CTEQ 6l1 PDF set [20].

All background events, except for the pair production of top quarks and of W bosons, are generated at $\sqrt{s} = 13$ TeV and the corresponding cross sections are scaled to 14 TeV. The $t\bar{t}$ and the $W^+W^-$ events are generated at a center-of-mass energy of 14 TeV. The effect of multiple interactions per bunch crossing (pileup) is estimated by overlaying the hard scatter event with minimum bias events drawn from a Poisson distribution with an average of 200.

The CMS detector [21] will be substantially upgraded in order to fully exploit the physics potential offered by the increase in luminosity at the HL-LHC [22], and to cope with the demanding operational conditions at the HL-LHC [23–27]. The upgrade of the first level hardware trigger (L1) will allow for an increase of L1 rate and latency to about 750 kHz and 12.5 $\mu$s, respectively, and the high-level software trigger is expected to reduce the rate by about a factor of 100 to 7.5 kHz. The entire pixel and strip tracker detectors will be replaced to increase the granularity, reduce the material budget in the tracking volume, improve the radiation hardness, extend the geometrical coverage, and provide efficient tracking up to pseudorapidities of about $|\eta| = 4$. The muon system will be enhanced by upgrading the electronics of the existing cathode strip chambers, resistive plate chambers (RPC), and drift tubes. New muon detectors based on improved RPC and gas electron multiplier technologies will be installed to add redundancy, increase the geometrical coverage up to about $|\eta| = 2.8$, and improve the trigger and reconstruction performance in the forward region. The barrel electromagnetic calorimeter will feature the upgraded front-end electronics that will be able to exploit the information from single crystals at the L1 trigger level, to accommodate trigger latency and bandwidth requirements, and to provide 160 MHz sampling allowing high precision timing capability for photons. The hadronic calorimeter, consisting in the barrel region of brass absorber plates and plastic scintillator layers, will be read out by silicon photomultipliers. The endcap electromagnetic and hadron calorimeters will be replaced with a new combined sampling calorimeter that will provide highly-segmented spatial information in both transverse and longitudinal directions, as well as high-precision timing information. Finally, the addition of a new timing detector for minimum ionizing particles in both barrel and endocap region is envisaged to provide capability for 4-dimensional reconstruction of interaction vertices that will allow to significantly offset the CMS performance degradation due to high pileup rates.

A detailed overview of the CMS detector upgrade program is presented in Refs. [23–27], while the expected performance of the reconstruction algorithms and pileup mitigation with the CMS detector is summarised in Ref. [28].

The generated signal and background events are processed with the fast-simulation package Delphes [29] in order to simulate the expected response of the upgraded CMS detector. The object reconstruction and identification efficiencies, as well as the detector response and resolution, are parameterised in Delphes using the detailed simulation of the upgraded CMS detector based on the GEANT4 package [30, 31].

# 2 Event reconstruction

In this analysis, the particle-flow (PF) [32] algorithm is used to attempt to reconstruct and identify each individual particle in the event. The algorithm considers information from all CMS sub-detectors to provide a global event description. In addition, the "Pileup Per Particle Identification" (PUPPI) [33] algorithm calculates the likelihood that each particle originates from a pileup interaction.

The $\tilde{\chi}_2^0$ is expected to decay into $\tilde{\chi}_1^0$ by emitting a low-mass $Z^*$ boson in the central region of the detector. Muons (electrons) are therefore selected with $5 \leq p_T \leq 30\,\text{GeV}$ and $|\eta| \leq 2.4\,(1.6)$. Dedicated identification criteria, including a requirement on the lepton impact parameter, are used to identify leptons. The identification efficiency for muons with $p_T$ of 5 GeV is 40% over the considered $|\eta|$ interval, while it ranges from 90 to 70% for muons with $p_T$ of 30 GeV. The efficiency for electrons with $p_T$ of 5 GeV is between 25 and 20% as $\eta$ increases from 0 to 1.6, and between 80 to 60% for electrons with $p_T$ of 30 GeV. Once identified, muons and electrons are considered candidate leptons if the scalar sum of track momenta in a cone around the lepton is less than 5 GeV and smaller than 50% of the lepton $p_T$. The cone's radius is defined to be $R = \sqrt{\Delta\phi(\ell, \text{tk})^2 + \Delta\eta(\ell, \text{tk})^2} = 0.3$, where $\ell$ refers to the lepton and tk to the tracks within the cone, and $\phi$ is the azimuthal angle. For electrons, the energy in the isolation cone is computed using the PUPPI algorithm. The isolation efficiency increases from 65% to 90% as the lepton $p_T$ increases from 5 to 30 GeV.

The anti-$k_T$ algorithm [34] with a size parameter of 0.4, implemented in the FastJet program [35], is adopted to reconstruct jets. In this analysis, candidate jets are jets with $p_T > 40\,\text{GeV}$ GeV and $|\eta| \leq 4.0$. Candidate jets consistent with the decay and hadronization of a B hadron are tagged as b jets with an efficiency of 74% [36]. Candidate jets with $p_T > 200\,\text{GeV}$ and $|\eta| \leq 2.4$ are referred to as ISR jets.

Candidate leptons and jets are required to be separated in space by $\Delta R = \sqrt{\Delta\phi(\ell, j)^2 + \Delta\eta(\ell, j)^2}$ greater than 0.4. $\Delta R$ is computed for each combination of a candidate lepton ($\ell$) and a jet ($j$).

The missing transverse momentum vector $\vec{p}_T^{\text{miss}}$ is defined as the negative vector sum of all PF objects in the event with their corresponding transverse momenta weighted through the PUPPI method. Its magnitude is referred to as $p_T^{\text{miss}}$.

# 3 Search strategy

This search targets the production of $\tilde{\chi}_1^{\pm}\tilde{\chi}_2^0$ and $\tilde{\chi}_2^0\tilde{\chi}_1^0$, followed by the decay of $\tilde{\chi}_2^0$ into $\tilde{\chi}_1^0$ via a low mass virtual Z boson. Events are therefore requested to contain at least two low-$p_T$, same-flavor, opposite-charge candidate leptons. In candidate signal events, the invariant mass of the candidate leptons will exhibit a kinematic end point at $m_{\ell_1,\ell_2} = \Delta M(\tilde{\chi}_2^0, \tilde{\chi}_1^0)$.

In this analysis, it is assumed that either a $p_{\mathrm{T}}^{\mathrm{miss}}$-based trigger reaching a plateau efficiency for $p_{\mathrm{T}}^{\mathrm{miss}} \geq 250\,\mathrm{GeV}$ (similar to that used in [8]) or a single jet trigger with $p_{\mathrm{T}} > 170\,\mathrm{GeV}$ [37] is adopted to select events. In order to ensure high trigger efficiency, the events are required to have $p_{\mathrm{T}}^{\mathrm{miss}} \geq 300\,\mathrm{GeV}$ and to contain at least one candidate ISR jet ($j_{\mathrm{ISR}}$) with $p_{\mathrm{T}}$ larger than 200 GeV. The ISR jet boosts the $\tilde{\chi}_1^{\pm}\tilde{\chi}_2^0$ or $\tilde{\chi}_2^0\tilde{\chi}_1^0$ system so that the outgoing $\tilde{\chi}_1^0$ particles are aligned, increasing the $p_{\mathrm{T}}^{\mathrm{miss}}$. To further exploit the boosted topology of the signal, events are accepted only if the $p_{\mathrm{T}}^{\mathrm{miss}}$ and the ISR candidate jet $p_{\mathrm{T}}$ satisfy $\Delta\phi(p_{\mathrm{T}}^{\mathrm{miss}}, p_{\mathrm{T}}(j_{\mathrm{ISR}})) \geq 2.0$. Since minor hadronic activity is expected from the electroweak production of $\tilde{\chi}_1^{\pm}$ and $\tilde{\chi}_2^0$, an upper bound of 4 is placed on the number of candidate jets $N_{\mathrm{jet}}$.

Several SM processes lead to events containing two same-flavor, opposite-charge candidate leptons, one ISR jet, and significant $p_{\mathrm{T}}^{\mathrm{miss}}$. One background category consists of prompt processes, where both candidate leptons originate from the prompt decay of W and Z bosons. Another category is misclassified processes, where at least one of the two candidate leptons originates from a semi-leptonic decay of a B hadron, a photon conversion, a decay in flight, or a misidentified quark or gluon. The prompt background is dominated by Drell-Yan (DY), diboson, and $\mathrm{t\bar{t}}$ production where both W bosons decay leptonically. The DY contribution is suppressed by requiring significant $p_{\mathrm{T}}^{\mathrm{miss}}$, while rejecting events with at least one b jet reduces the $\mathrm{t\bar{t}}$ background. The dominant misclassified processes are W and $\mathrm{t\bar{t}}$ production where one candidate lepton originates from the W boson decay and an additional misclassified lepton is selected in the event. Rejecting events with at least one b jet reduces both contributions. To further suppress the background contamination, events are accepted only if the angular separation between the two candidate leptons satisfies $\Delta R(\ell_1, \ell_2) \leq 2.0$, as expected of collinear leptons emerging from the decay of a boosted $\mathrm{Z}^*$ boson. Events satisfying the criteria described above, which are summarized in Table 1, form the baseline signal region.

Table 1: Definition of the baseline signal region. In the table below, $N_\ell$ is the number of lepton candidates; $\Delta R(\ell_1, \ell_2)$ is the angular separation between the two candidate leptons in the $\phi, \eta$ space; $N_{\text{b-jet}}$ is the number of b jets; $N_{\mathrm{jet}}$ is the number of candidate jets (including any ISR jet reconstructed in the event); $N_{\mathrm{ISR}}$ is the number of ISR jets; $\Delta\phi(p_{\mathrm{T}}^{\mathrm{miss}}, p_{\mathrm{T}}(j_{\mathrm{ISR}}))$ is the azimuthal distance between the $p_{\mathrm{T}}^{\mathrm{miss}}$ vector and the $j_{\mathrm{ISR}}$ $p_{\mathrm{T}}$ vector; and $m_{\ell_1,\ell_2}$ is the invariant mass of the two candidate leptons.

| Observable | Requirement |
|---|---|
| $N_\ell$ | $= 2$ (same flavor, opposite charge) |
| $\Delta R(\ell_1, \ell_2)$ | $\leq 2.0$ |
| $N_{\text{b-jet}}$ | $= 0$ |
| $N_{\mathrm{jet}}$ | $\leq 4$ |
| $N_{\mathrm{ISR}}$ | $\geq 1$ |
| $p_{\mathrm{T}}^{\mathrm{miss}}$ | $\geq 250\,\mathrm{GeV}$ |
| $\Delta\phi(p_{\mathrm{T}}^{\mathrm{miss}}, p_{\mathrm{T}}(j_{\mathrm{ISR}}))$ | $\geq 2.0$ |
| $m_{\ell_1,\ell_2}$ | $[5, 40]\,\mathrm{GeV}$ |

Figure 2 shows the $p_{\mathrm{T}}$ distributions of the candidate leptons with the highest and second-highest $p_{\mathrm{T}}$ in events satisfying the baseline signal region selection. In the signal models, the mean of the lepton $p_{\mathrm{T}}$ is correlated to the $\Delta M(\tilde{\chi}_2^0, \tilde{\chi}_1^0)$ mass difference. The $p_{\mathrm{T}}^{\mathrm{miss}}$ and $m_{\ell_1,\ell_2}$ distributions are presented in Fig. 3.

The missing transverse momentum, the invariant mass of the two candidate leptons, and the sub-leading lepton $p_{\mathrm{T}}(\ell_2)$ observables are found to provide the best discrimination between signal and background. Events in the baseline signal region are therefore classified in 60 cate-gories with $p_{\mathrm{T}}^{\mathrm{miss}}$ values in [250, 300, 350, 400, 500, ∞] GeV, $m_{\ell_1,\ell_2}$ values in [5, 10, 20, 30, 40] GeV, and $p_{\mathrm{T}}(\ell_2)$ in [5, 13, 21, 30] GeV. The categories are defined based on the $p_{\mathrm{T}}^{\mathrm{miss}}$ resolution and the expected kinematic endpoints of $m_{\ell_1,\ell_2}$ and $p_{\mathrm{T}}(\ell_2)$ in the signal models.

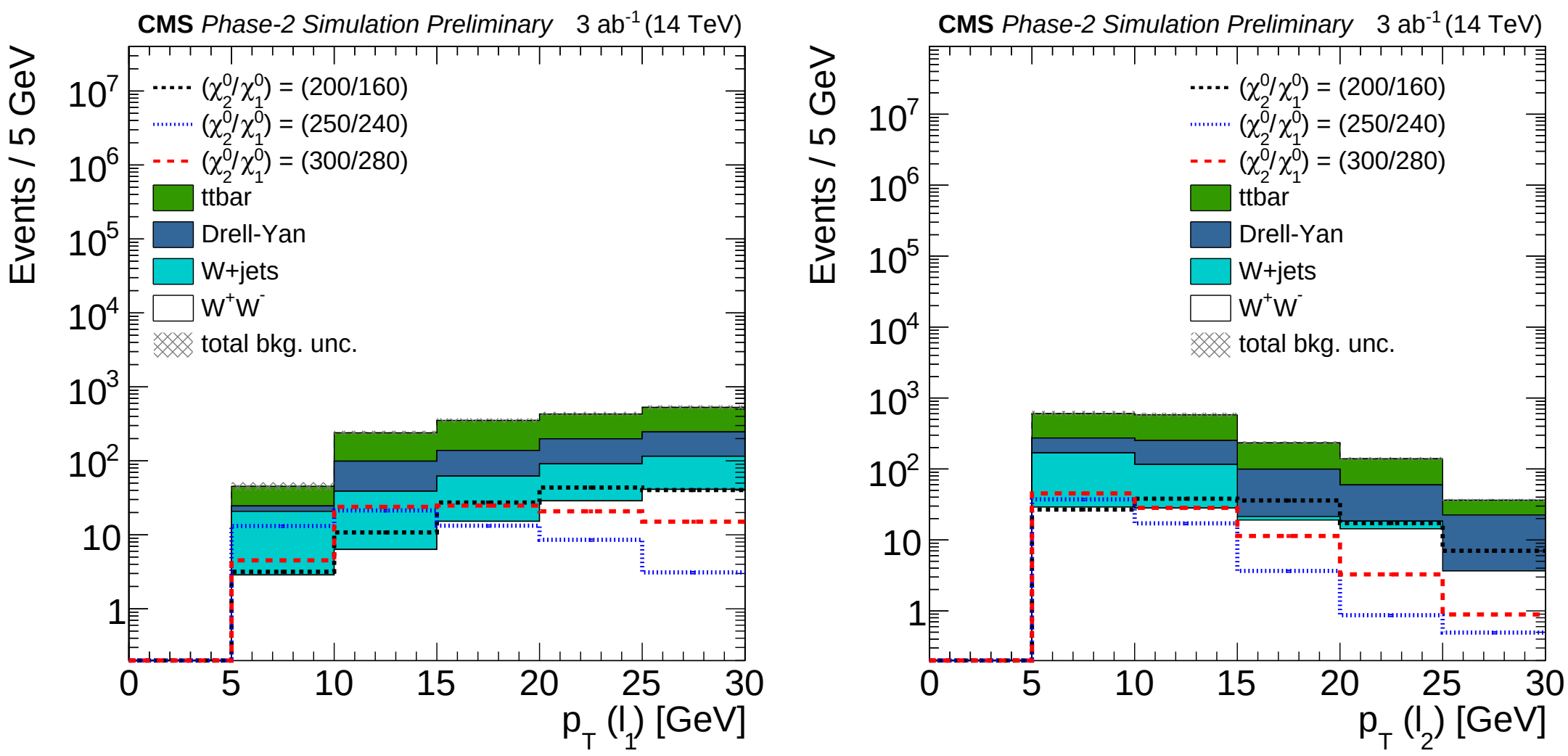

Figure 2: Distributions of the candidate lepton with the highest $p_T$ (left) and the second-highest $p_T$ (right) for background and signal events in the baseline signal region. Three selected $\widetilde{\chi}_1^{\pm}\widetilde{\chi}_2^0 + \widetilde{\chi}_2^0\widetilde{\chi}_1^0$ signal models are shown, where the first number corresponds to the mass of the $\widetilde{\chi}_2^0$ (and $\widetilde{\chi}_1^{\pm}$) and the second one to the mass of the $\widetilde{\chi}_1^0$. The uncertainty band represents systematical uncertainties.

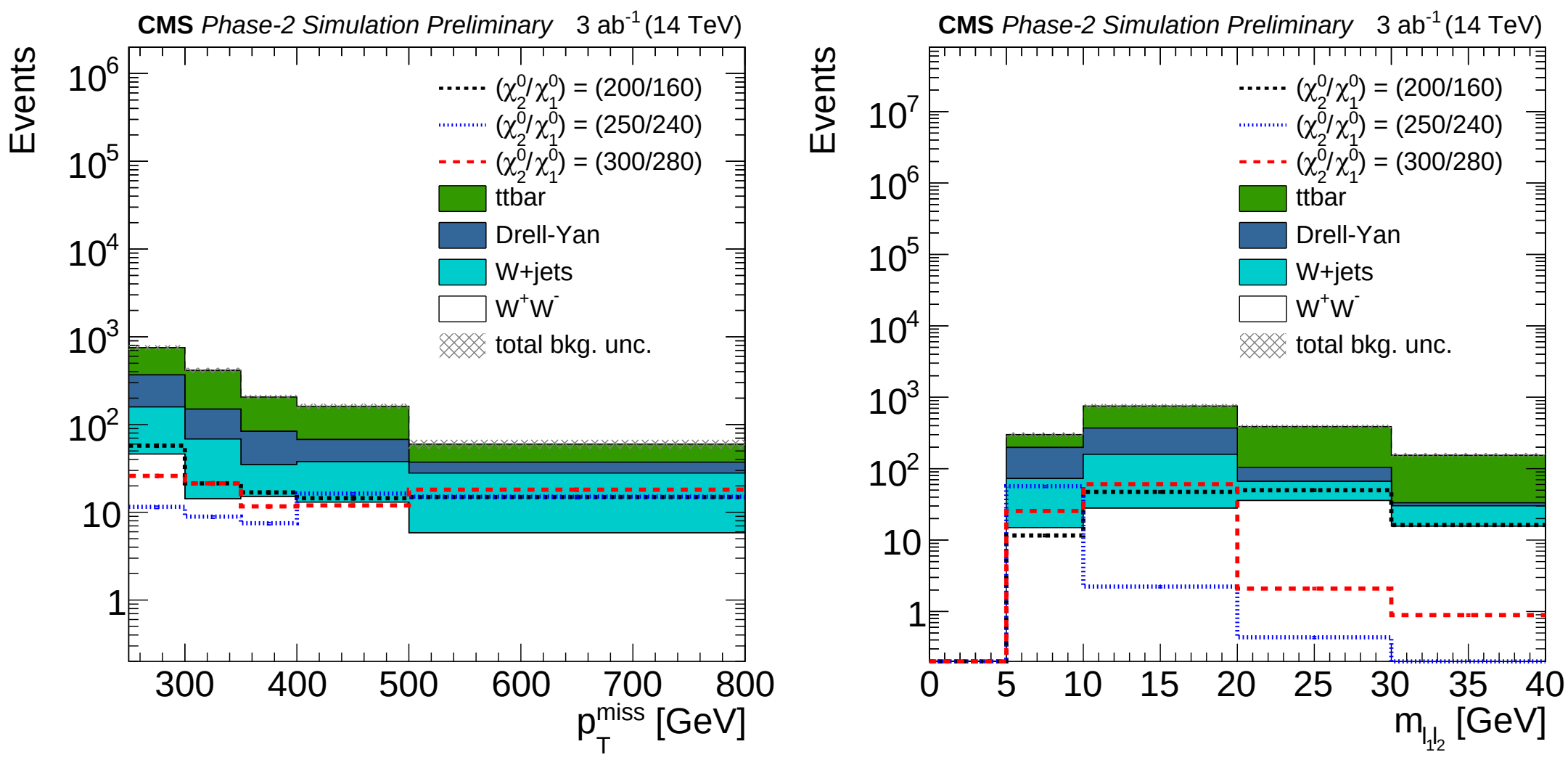

Figure 3: Distributions of the $p_T^{miss}$ (left) and $m_{\ell_1,\ell_2}$ (right) for backgrund and signal events in the baseline signal region. Three selected $\widetilde{\chi}_1^{\pm}\widetilde{\chi}_2^0 + \widetilde{\chi}_2^0\widetilde{\chi}_1^0$ signal models are shown, where the first number corresponds to the mass of $\widetilde{\chi}_2^0$ (and $\widetilde{\chi}_1^{\pm}$) and the second one to the mass of $\widetilde{\chi}_1^0$. The uncertainty band represents systematical uncertainties.

The search approach in this analysis differs in several ways from the one adopted in the Run-2 analysis presented in Ref. [8]. In this analysis, facilitated by the large size of the dataset expected at the HL-LHC, substantially more signal regions are used. In turn, the baseline selection is loosened, with no dedicated requirements to suppress the $\mathrm{Z} \to \tau\tau$ background.

# 4 Expected sensitivity

There are several systematic uncertainties in the yields of both the background and the signal processes. The dominant experimental uncertainties are those originating from the jet energy corrections, b-tagging efficiency, lepton identification and isolation efficiency (combined in Table 2), and integrated luminosity. The uncertainties values are derived from those estimated in the current Run-2 based analyses with proper scaling to account for the larger dataset expected at the HL-LHC. These systematic uncertainties are correlated among the prompt background processes and between the signal and the prompt background yields. The uncertainty values are reported per source in Table 2. The uncertainty in the total background also includes the uncertainty in the determination of the misclassified component. This is assumed to be 30% based on the prediction in Ref. [8] that uses observed data. It is assumed that the yields are not affected by the statistical uncertainty due to the limited number of generated event.

Table 2: Summary of the experimental systematic uncertainties assumed in the prediction of the yields for processes with prompt leptons.

| Source | Uncertainty |
|---|---|
| jet energy corrections | 1–2.5% |
| b-tagging | 1% |
| muon, electron efficiency | 0.5, 2.5% |
| integrated luminosity | 1% |

Theoretical uncertainties in the cross sections and in the acceptance from the choice of parton distribution functions are considered negligible and are not included. However, a systematic uncertainty of 10% in the signal acceptance, similar to the value from Ref. [8], is included to account for the modeling of the ISR jet. The systematic uncertainties are treated as nuisance parameters with log-normal probability density functions.

The search sensitivity is calculated within the modified frequentist framework using the asymptotic formulae and the $\mathrm{CL_s}$ criterion to compute the results [38–40]. The upper limit on the cross sections is computed at 95% confidence level (CL) and shown in Fig. 4. These contours correspond to the combination of $\tilde{\chi}_1^{\pm}\tilde{\chi}_2^0$ and $\tilde{\chi}_2^0\tilde{\chi}_1^0$ production. The signal and background yields for two representative event categories (out of the total 60) that are sensitive to high-mass $\tilde{\chi}_2^0$ signals are presented in Table 3. Higgsino-like mass-degenerate $\tilde{\chi}_1^{\pm}$ and $\tilde{\chi}_2^0$ are excluded for masses up to 360 GeV if the mass difference with respect to the lightest neutralino $\tilde{\chi}_1^0$ is 15 GeV, extending the sensitivity achieved in Ref. [8] by ≈210 GeV. Fig. 4 also shows the 5$\sigma$ discovery contour, computed using all signal regions without taking the look-elsewhere-effect into account. Under this assumption $\tilde{\chi}_1^{\pm}$ and $\tilde{\chi}_2^0$ can be discovered for masses as large as 250 GeV. These results demonstrate that the HL-LHC can significantly improve the sensitivity to natural SUSY.

Table 3: Signal and background yields in two representative event categories. SR1 is defined by $p_{\mathrm{T}}^{\mathrm{miss}} > 500\,\mathrm{GeV}$, $m_{\ell_1,\ell_2}$ in [10, 20] GeV, and $p_{\mathrm{T}}(\ell_2)$ in [13, 21] GeV. SR2 is defined by $p_{\mathrm{T}}^{\mathrm{miss}} > 500\,\mathrm{GeV}$, $m_{\ell_1,\ell_2}$ in [10, 20] GeV, and $p_{\mathrm{T}}(\ell_2)$ in [5, 13] GeV. The signal model considered here has $m_{\tilde{\chi}_1^\pm} = m_{\tilde{\chi}_2^0} = 300\,\mathrm{GeV}$ and $m_{\tilde{\chi}_1^0} = 280\,\mathrm{GeV}$. Only systematic uncertainties are given.

| Process | SR1 | SR2 |
|---|---|---|
| Signal | $3.3 \pm 0.2$ | $8.9 \pm 0.5$ |
| $\mathrm{t\bar{t}}$ | $1.7 \pm 0.1$ | $6.2 \pm 0.6$ |
| W+jets | $0.03 \pm 0.01$ | $15.8 \pm 4.8$ |
| $\mathrm{W^+W^-}$ | $0.7 \pm 0.04$ | $1.5 \pm 0.1$ |
| Drell-Yan | $0.9 \pm 0.1$ | $1.9 \pm 0.2$ |

A potential improvement to the analysis is the addition of final states with three leptons originating from $\tilde{\chi}_1^\pm \tilde{\chi}_2^0 \to \ell\nu\tilde{\chi}_1^0 \ell\ell\tilde{\chi}_1^0$ decays. The acceptance can also be increased with lower requirements on the minimum lepton $p_{\mathrm{T}}$ or on the minimum $m_{\ell_1,\ell_2}$. The latter improvement is expected to increase the sensitivity to models with mass splittings below 7.5 GeV, provided that the background from low-mass resonances is suppressed.

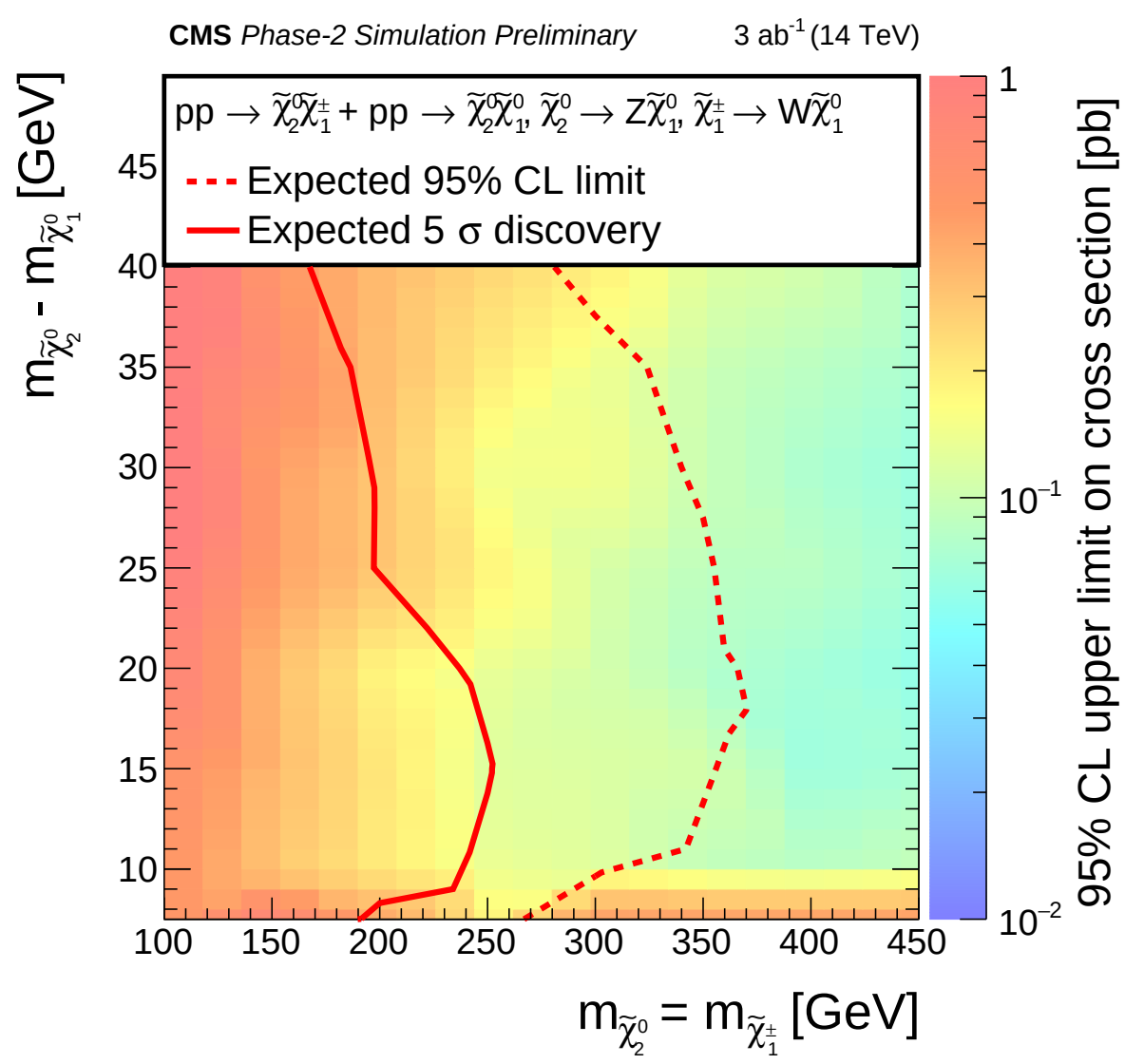

Figure 4: The $5\sigma$ discovery contours and expected 95% CL exclusion contours for the combined $\tilde{\chi}_1^\pm\,\tilde{\chi}_2^0$ and $\tilde{\chi}_2^0\,\tilde{\chi}_1^0$ production. Results are presented for $\Delta M(\tilde{\chi}_2^0, \tilde{\chi}_1^0) > 7.5\,\mathrm{GeV}$.

# 5 Summary

A search for the pair production of light higgsino-like charginos $\tilde{\chi}_1^\pm$ and neutralinos $\tilde{\chi}_2^0$ ($\tilde{\chi}_1^\pm\tilde{\chi}_2^0$, $\tilde{\chi}_2^0\tilde{\chi}_1^0$) is presented using $3000\,\mathrm{fb}^{-1}$ of simulated proton-proton collision data produced by the HL-LHC at 14 TeV. The $\tilde{\chi}_1^\pm$ and $\tilde{\chi}_2^0$ particles are assumed to be mass-degenerate, to be pair-produced, and to decay into the lightest stable superymmetric particle $\tilde{\chi}_1^0$ via off-shell W and Z bosons. The $\tilde{\chi}_1^0$ is also assumed to be higgsino-like. Higgsino-like mass-degenerate $\tilde{\chi}_1^\pm$ and $\tilde{\chi}_2^0$ particles with masses up to 250 GeV can be discovered for a mass difference of 15 GeV relative to the lightest neutralino $\tilde{\chi}_1^0$. For this mass splitting, $\tilde{\chi}_1^\pm$ and $\tilde{\chi}_2^0$ with masses up to 360 GeV can be excluded at 95% confidence level.

**10**

# ATLAS sensitivity to winos and higgsinos with a highly compressed mass spectrum at the HL-LHC

The ATLAS Collaboration

This note presents the prospects of two searches for direct chargino and neutralino production ($\tilde{\chi}_1^{\pm}\tilde{\chi}_1^{\pm}$, $\tilde{\chi}_1^{\mp}\tilde{\chi}_2^0$ and $\tilde{\chi}_1^{\mp}\tilde{\chi}_1^0$), in highly compressed mass scenarios, in proton-proton collisions at a centre-of-mass energy of 14 TeV. The first is a disappearing track search that investigates chargino production scenarios, where the chargino decays via a neutralino and a very soft pion, which is not reconstructed. The small mass splitting between the chargino and neutralino implies that the chargino has a significant lifetime and decays within the innermost tracking detector. The second is a two soft lepton search that investigates scenarios where the second lightest neutralino decays to the lightest neutralino and an off-shell Z-boson, giving rise to two soft leptons. The results of both searches are interpreted in pure-wino and pure-higgsino scenarios. The reach for these scenarios at the high-luminosity phase of the LHC with an assumed integrated luminosity of 3000 fb$^{-1}$ is presented and significantly extends beyond the current Run 2 limits of the respective analyses.

# 1 Introduction

Supersymmetry (SUSY) [1–6] is a space-time symmetry that relates fermions and bosons, predicting new particles that differ from their Standard Model (SM) partners by a half unit of spin. In the electroweak sector, SUSY partners of the Higgs, photon, $Z$, and $W$ are the spin $\frac{1}{2}$ higgsinos, photino, zino, and winos that further mix in neutralino ($\tilde{\chi}^0_{1,2,3,4}$) and chargino ($\tilde{\chi}^\pm_{1,2}$) states, also called the electroweakinos. In R-parity conserving scenarios [7], SUSY particles are produced in pairs and the lightest supersymmetric particle (LSP) is stable and is a dark matter candidate.

In anomaly-mediated supersymmetry breaking (AMSB) scenarios [8, 9], the supersymmetric partners of the SM $W$-bosons, the winos, are the lightest gaugino states. In this case, the lightest chargino ($\tilde{\chi}^\pm_1$) and the lightest neutralino ($\tilde{\chi}^0_1$), are both largely composed of the wino eigenstates and are nearly mass-degenerate. Due to this small mass difference, the $\tilde{\chi}^\pm_1$ can have a long enough lifetime such that it decays inside the detector. AMSB scenarios naturally predict a pure wino LSP, which is a dark-matter candidate. However other scenarios, which follow from naturalness arguments [10, 11], suggest that the absolute value of the Higgsino mass parameter $\mu$ is expected to be near the weak scale such that the higgsinos should be light, with masses below one TeV [12, 13], while the magnitude of the bino and wino mass parameters, $M_1$ and $M_2$ can be significantly larger, i.e. $|\mu| \ll |M_1|, |M_2|$. This results in the three lightest electroweakino states, $\tilde{\chi}^0_1$, $\tilde{\chi}^\pm_1$ and $\tilde{\chi}^0_2$ being dominated by the Higgsino component. In this scenario, the three lightest electroweakino masses are separated by hundreds of MeV to tens of GeV depending on the composition of these mass eigenstates, which is determined by the specific values of $M_1$ and $M_2$ [14]. Investigating either of these scenarios, with very small mass splitting between the lightest electroweakinos, is particularly challenging at hadron colliders, both due to the small cross-sections and due to the small transverse momenta ($p_{\mathrm{T}}$) of the final state particles.

The two searches presented in this note investigate electroweakino production in these two, nearly mass degenerate, scenarios assuming an integrated luminosity of 3000 fb$^{-1}$ collected with centre-of-mass energy $\sqrt{s}$ = 14 TeV proton-proton collisions at the High-Luminosity Large Hadron Collider (HL-LHC). The instantaneous luminosity of the HL-LHC is expected to be around $L = 7.5 \times 10^{34}$ cm$^{-2}$s$^{-1}$, leading to an average number of interactions per bunch crossing, $\langle\mu\rangle$, of approximately 200, and these conditions will provide a very challenging environment for physics analysis.

The disappearing track search investigates scenarios where the $\tilde{\chi}^\pm_1$, and $\tilde{\chi}^0_1$ are almost mass degenerate, leading to a long lifetime for the $\tilde{\chi}^\pm_1$ which decays after the first few layers of the inner detector, leaving a track in the innermost layers of the detector. The chargino decays as: $\tilde{\chi}^\pm_1 \rightarrow \pi^\pm \tilde{\chi}^0_1$, as shown in Figure 1 (left). The $\tilde{\chi}^0_1$ escapes the detector and the pion has a very low energy and is not reconstructed, leading to the disappearing track signature, as shown in Figure 2. The latest ATLAS results for this scenario [15] using the 2015–2016 dataset of the LHC $pp$ run at a centre-of-mass energy of $\sqrt{s}$ = 13 TeV, excluded wino masses below 430 GeV with a chargino lifetime, $\tau(\tilde{\chi}^\pm_1)$, of 0.2 ns.

The dilepton search investigates final states containing two soft muons and a large transverse momentum imbalance, which arise in scenarios where $\tilde{\chi}^0_2$ and $\tilde{\chi}^\pm_1$ are produced and decay via an off-shell $Z$ or $W$ boson respectively, $\tilde{\chi}^0_2 \rightarrow Z^* \tilde{\chi}^0_1$ and $\tilde{\chi}^\pm_1 \rightarrow W^* \tilde{\chi}^0_1$. Considering the $Z \rightarrow ee$ decay is beyond the scope of this note, but could further improve the sensitivity to these scenarios. Due to the very small mass splitting of the electroweakinos in this scenario, a jet arising from initial-state radiation (ISR) is required, to boost the sparticle system. Figure 1 (right) presents the scenario considered in this search. First constraints surpassing the earlier LEP limits [16] have recently been set by the ATLAS experiment [17], excluding mass splittings down to 2.5 GeV for $m(\tilde{\chi}_1^0) = 100$ GeV. Scenarios with direct $\tilde{\chi}_1^\pm$ pair production are also considered.

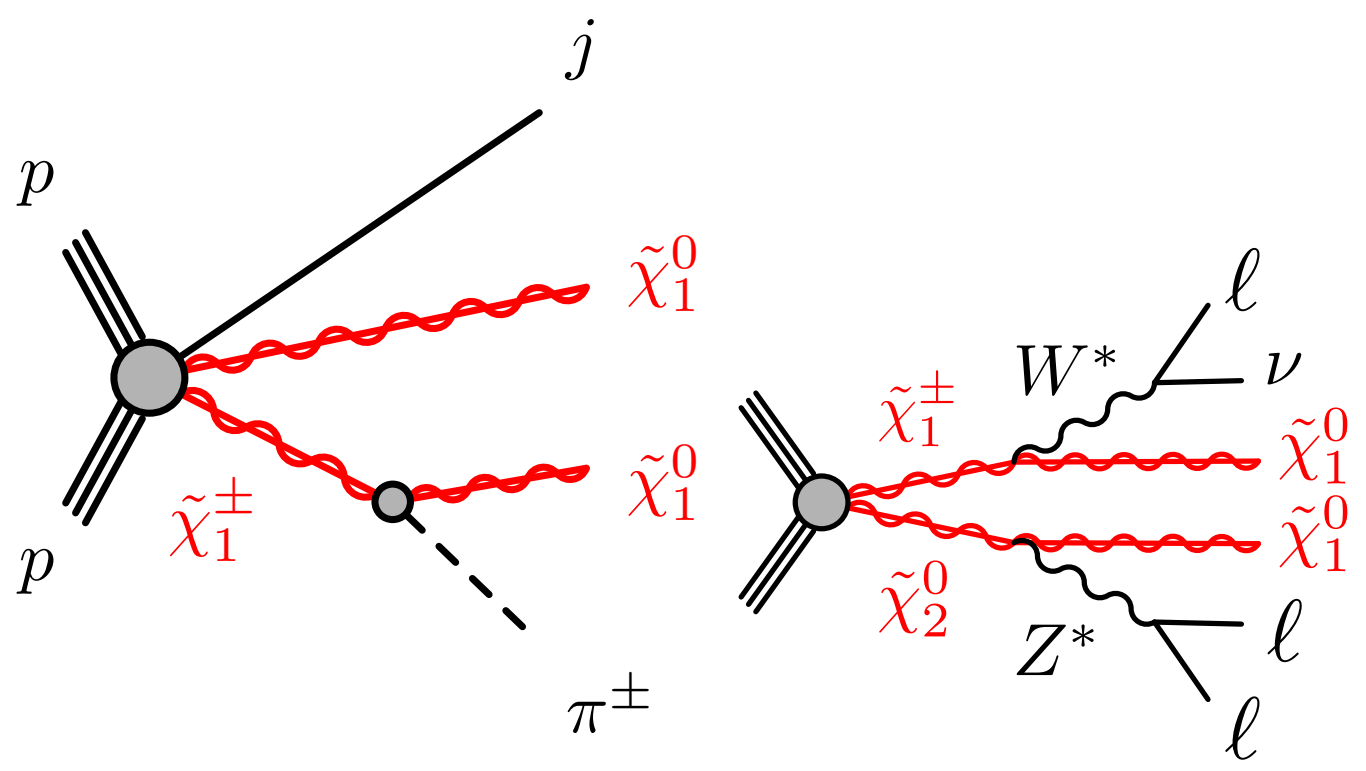

Figure 1: Diagrams depicting (left) $\tilde{\chi}_1^\pm\,\tilde{\chi}_1^0$ production and (right) $\tilde{\chi}_1^\pm\,\tilde{\chi}_2^0$ production.

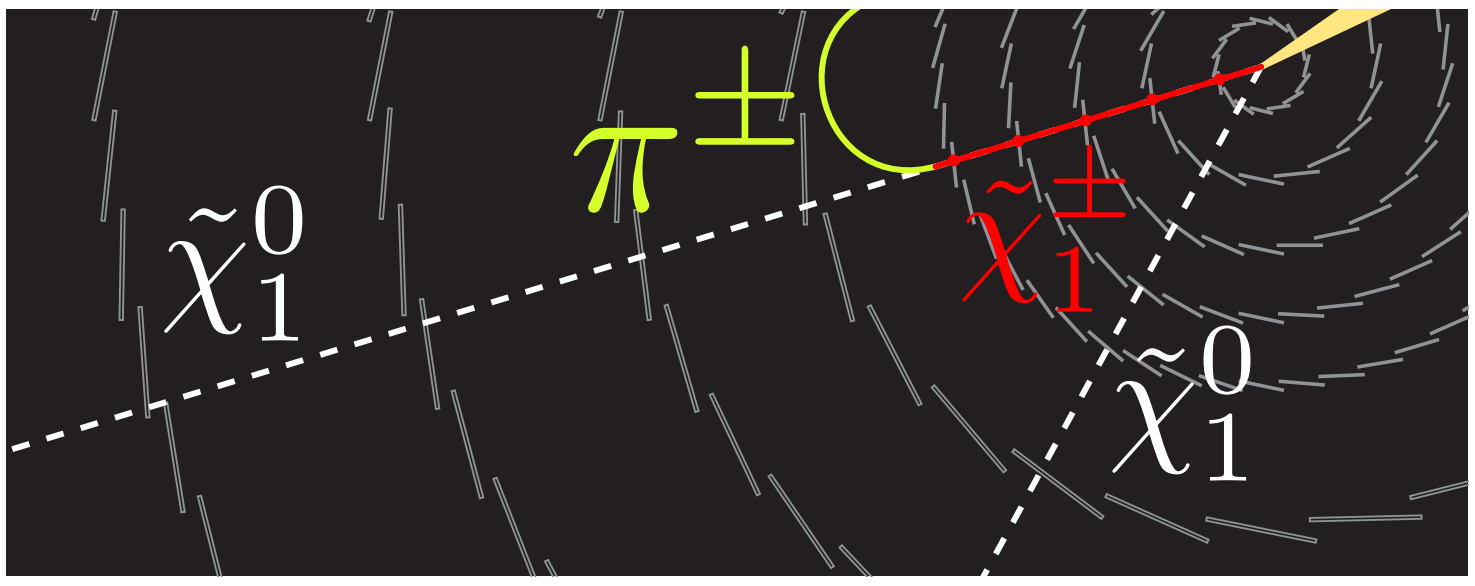

Figure 2: Schematic illustration of a $pp \to \tilde{\chi}_1^\pm\tilde{\chi}_1^0 +$ jet event in the HL-LHC ATLAS detector, with a long-lived chargino. Particles produced in pile-up $pp$ interactions are not shown. The $\tilde{\chi}_1^+$ decays into a low-momentum pion and a $\tilde{\chi}_1^0$ after leaving hits in the five pixel layers (indicated by red makers).

## 2 ATLAS Detector

The proposal for the upgraded ATLAS detector [18] which will operate at the HL-LHC includes a new all-silicon inner tracking detector, the Inner Tracker (ITk) [19, 20], composed of five layers of pixel detectors and four layers of double-sided silicon strip sensors, allowing for the reconstruction of the trajectory of charged particles within a pseudorapidity range $|\eta| < 4$. The inner tracker is immersed in a 2 T axial magnetic field, enabling the momentum measurement of the charged particles travelling through the inner tracker. The electromagnetic and hadronic calorimeters [21, 22] are located outside of the solenoid and provide high granularity energy measurements within $|\eta| < 4.9$. Beyond the calorimeters lies the muon spectrometer [23], consisting of a set of superconducting torodial magnets and three layers of gaseous chambers, allowing for the trajectories of muons to be measured up to $|\eta| < 2.7$. In addition to this fast detectors are also installed in the muon spectrometer for triggering purposes, with a range up to $|\eta| < 2.4$. The on-line trigger system will also be replaced by a new system, which will be capable of processing the rate increase anticipated at the HL-LHC. Both searches plan to use an $E_{\mathrm{T}}^{\mathrm{miss}}$ requirement to accepting events in the trigger, and the $E_{\mathrm{T}}^{\mathrm{miss}}$ selection used in the searches are expected to be in the $E_{\mathrm{T}}^{\mathrm{miss}}$ trigger plateau [24].

## 3 Monte-Carlo Samples

Monte Carlo (MC) samples are used to predict the expected backgrounds from SM processes and to model the SUSY signal scenarios under consideration.

The detector response is parameterised based on studies performed with Geant 4 [25] simulations of the upgraded detector in high luminosity conditions [18, 26]. All simulated events are overlaid with additional $pp$ interactions in the same and neighbouring bunch crossings (pile-up) simulated with the soft QCD processes of Pythia 8.186 using the A2 set of tuned parameters [27] and the MSTW2008LO [28] PDF set. The pile-up conditions of the HL-LHC are assumed to follow a Poisson distribution with an average of 200 additional interactions ($\langle\mu\rangle = 200$).

The disappearing track signals are generated assuming the minimal AMSB model with the ratio of the Higgs vacuum expectation values at the electroweak scale set to $\tan\beta = 5$, the sign of the higgsino mass term set to be positive, and the universal scalar mass set to $m_0 = 5$ TeV. The SUSY mass spectrum, the branching ratios and decay widths are calculated using ISASUSY version.7.80 [29]. The $m_{3/2}$ parameter is varied in these scenarios which allows the wino masses to vary and the LSP is expected to be a wino in these scenarios.

The higgsino simplified model used for the dilepton search includes the production of $\tilde{\chi}_2^0\tilde{\chi}_1^{\pm}$, $\tilde{\chi}_2^0\tilde{\chi}_1^0$ and $\tilde{\chi}_1^{\pm}\tilde{\chi}_1^{\mp}$. The $\tilde{\chi}_1^0$ and $\tilde{\chi}_2^0$ masses are varied, while the $\tilde{\chi}_1^{\pm}$ masses are set to $m(\tilde{\chi}_1^{\pm}) = \frac{1}{2}[m(\tilde{\chi}_1^0) + m(\tilde{\chi}_2^0)]$. It is expected that the mass splittings of pure higgsinos, generated by radiative corrections, are of the order of hundreds of MeV [30]. In the simplified models considered the masses of the $\tilde{\chi}_2^0$ and $\tilde{\chi}_1^0$ are varied independently while the $\tilde{\chi}_1^{\pm}$ masses are set to the average value of the $\tilde{\chi}_2^0$ and $\tilde{\chi}_1^0$. The production cross sections are calculated as if the states were pure higgsinos.

The signal samples are generated with up to two extra partons in the matrix element using MG5_aMC@NLO 2.3.3 [31] at leading order (LO) interfaced to Pythia 8.186 [32] for parton showering, hadronisation and SUSY particle decay. The NNPDF2.3LO [33] parton distribution function (PDF) set was used. Renormalisation and factorisation scales are determined by the default dynamic scale choice of MG5_aMC@NLO. The CKKW-L merging scheme [34] was applied to combine tree-level matrix elements containing multiple partons with parton showers. The scale parameter for merging was set to a quarter of the mass of the wino or higgsino depending upon the signal scenario. The A14 [35] set of tuned parameters with simultaneously optimised multiparton interaction and parton shower parameters was used for the underlying event together with the NNPDF2.3LO PDF set. The cross-sections for the disappearing track signals are calculated at next-to-leading order (NLO) in the strong coupling constant using Prospino2 [36]. The cross-sections for the higgsino signal samples are calculated to next-to-leading order (NLO) in the strong coupling, and next-to- leading-logarithm (NLL) accuracy for soft-gluon resummation using Resummino v2.1.0 [37–39]. For both signal scenarios the nominal cross-section and its uncertainty are taken from an envelope of cross-section predictions using different parton distribution function (PDF) sets and factorization and renormalization scales, as described in Ref. [40].

Events containing a $W$ or $Z$ boson with associated jets ($W$+jets and $Z/\gamma^*$+jets) are produced using the Sherpa v2.2.1 generator with the NNPDF30NNLO PDF set at $\sqrt{s} = 13$ TeVwith up to two extra partons in the matrix element. Diboson processes are generated at $\sqrt{s}$ = 13 TeV using Sherpa v2.2.2 with up to one extra parton in the matrix element. For the production of $t\bar{t}$ the Powheg-Box V2 generator interfaced to Pythia8 parton shower with the ATLAS A14 tune is used. The sample is generated at $\sqrt{s}$ = 14 TeV and a parton level filter requiring $E_{\mathrm{T}}^{\mathrm{miss}}$ >100 GeV is applied. The top-quark pair-production contribution is normalised to approximate next-to-next-to-leading-order calculations (NNLO) [41]. The NNLO FEWZ [42, 43] cross-sections are used for the normalisation of the inclusive $W$+jets and $Z$+jets samples. The expected diboson yields are normalised to the NLO cross-section from the generator. In addition, a series of dedicated "particle gun" samples are used for the disappearing track analysis to assess the probability of an isolated electron or hadron leaving a disappearing track in the detector.

# 4 Event Selection

The reconstruction of physics objects is performed at truth-level with parameterised detector functions [18]. Due to the different final states targeted by the disappearing track and dilepton searches, different requirements for the definition each object are applied.

The main signature of the disappearing track search is a short "tracklet" which is reconstructed in the inner layers of the detector and subsequently disappears. The tracklet reconstruction efficiency for signal charginos is estimated using fully simulated samples of $\tilde{\chi}_1^{\pm}$ pair production with $m(\tilde{\chi}_1^{\pm})$ = 600 GeV with the HL-LHC pile up conditions. Tracklet reconstruction is performed in two stages. Firstly "standard" tracks, hereafter referred to as tracks are reconstructed. Afterwards the track reconstruction is then rerun with looser criteria, requiring at least four pixel-detector hits. This second reconstruction uses only input hits which are not associated with tracks, referred to as "tracklets". The tracklets are then extrapolated to the strip detectors, and any compatible hits are assigned to the tracklet candidate. Tracklets are required to have $p_{\mathrm{T}}$ > 5 GeV and $|\eta|$ < 2.2. Candidate leptons, which are used only to veto events, are selected with $p_{\mathrm{T}}$ larger than 20 GeV and $|\eta| < 2.47$ (2.7) for electrons (muons).

The dilepton search targets scenarios that contain low $p_{\mathrm{T}}$ muons selected with $p_{\mathrm{T}}$ >3 GeV and $|\eta|$ < 2.5. Muons that originate from pile up interactions or from heavy flavour decays, referred as fake or non-prompt muons, are rejected by applying an isolation to the muon candidates. The main source of these muons is decays from heavy flavour mesons and baryons created in the quark hadronization process.

For both analyses candidate jets are reconstructed with the anti-$k_t$ algorithm with a radius parameter of 0.4. They are selected with $|\eta| < 2.8$ and the jet energy is smeared according to a Gaussian. Jets are tagged as originating from $b$-decays ($b$-tagged) using a parameterisation (versus the jet $p_{\mathrm{T}}$ and $\eta$) modelling the performance of the MV2c10 $b$-tagging algorithm [19]. In simulated $t\bar{t}$ events, the chosen working point identifies $b$-jets with an average efficiency of 70%, for a $c$-jet rejection factor of about 20 and a light-flavour jet rejection factor of about 750.

The magnitude of the missing transverse momentum ($E_{\mathrm{T}}^{\mathrm{miss}}$) is computed as the vectorial sum of the true momenta (at the particle level) of neutral weakly-interacting particles (neutrinos and neutralinos). It is then smeared, according to a Gaussian, to simulate the detector response, with a function parameterised as a function of the average number of interactions per bunch crossing $\mu$ and the scalar sum of energy in the calorimeter $\sum E_{\mathrm{T}}$. The method to resolve overlaps is as follows: candidate jets are required to be separated from candidate electrons by $\Delta R(e, \mathrm{jet})$ > 0.2; if a jet and electron are within 0.2, then the jet is removed and the electron is kept; after this step, leptons (both $e$ and $\mu$) are removed if they are within $\Delta R$ < 0.4 of a remaining jet.

For the disappearing-track search the final state contains zero leptons ($e$, $\mu$), large $E_{\mathrm{T}}^{\mathrm{miss}}$ and at least one tracklet, and events are reweighted by the expected efficiencies of tracklet reconstruction. The small mass splitting means events in which the $\tilde{\chi}_1^{\pm}$ and $\tilde{\chi}_1^0$ are produced back to back yield little $E_{\mathrm{T}}^{\mathrm{miss}}$, hence, it is necessary to select events where the system is boosted by recoiling against one or more jets from initial state radiation (ISR). The minimum azimuthal angular distance, $\min\{\Delta\phi(\mathrm{jet}_{1-4}, E_{\mathrm{T}}^{\mathrm{miss}})\}$, between the first four jets (ordered in $p_{\mathrm{T}}$) and the $E_{\mathrm{T}}^{\mathrm{miss}}$ is required to be greater than one, in order to reject events with mis-measured $E_{\mathrm{T}}^{\mathrm{miss}}$.

Table 1: Summary of the search selection criteria for the disappearing track SR.

| Variable | SR Selection |
|---|---|
| Lepton veto $p_{\mathrm{T}}$ [GeV] | >20 |
| $\min\{\Delta\phi(\mathrm{jet}_{1-4}, E_{\mathrm{T}}^{\mathrm{miss}})\}$ | $> 1$ |
| $E_{\mathrm{T}}^{\mathrm{miss}}$ [GeV] | $> 300$ |
| Leading jet $p_{\mathrm{T}}$ [GeV] | $> 300$ |
| Leading tracklet $p_{\mathrm{T}}$ [GeV] | $> 150$ |
| $\Delta\phi(E_{\mathrm{T}}^{\mathrm{miss}}, \mathrm{trk})$ | $< 0.5$ |

Different kinematic variables are used in the search to separate the signal from the SM background, such as the leading jet $p_{\mathrm{T}}$, the $E_{\mathrm{T}}^{\mathrm{miss}}$, the leading tracklet $p_{\mathrm{T}}$ and the azimuthal angle ($\Delta\phi(E_{\mathrm{T}}^{\mathrm{miss}}, \mathrm{trk})$) between the tracklet and the $E_{\mathrm{T}}^{\mathrm{miss}}$. The final SR selection is optimised using two signal models: one with $m(\tilde{\chi}_1^{\pm}) = 800$ GeV and a lifetime of 1 ns and the other with $m(\tilde{\chi}_1^{\pm}) = 200$ GeV and a lifetime of 0.04 ns. A minimum of three background events has been required in the optimisation process. Figure 3 shows the the $E_{\mathrm{T}}^{\mathrm{miss}}$ and the leading tracklet $p_{\mathrm{T}}$ distributions in events passing the full SR selection. The dominant background in the SR is found to be from fake tracklets, similarly to the Run 2 version of the search [17]. The full SR selection is presented in Table 1.

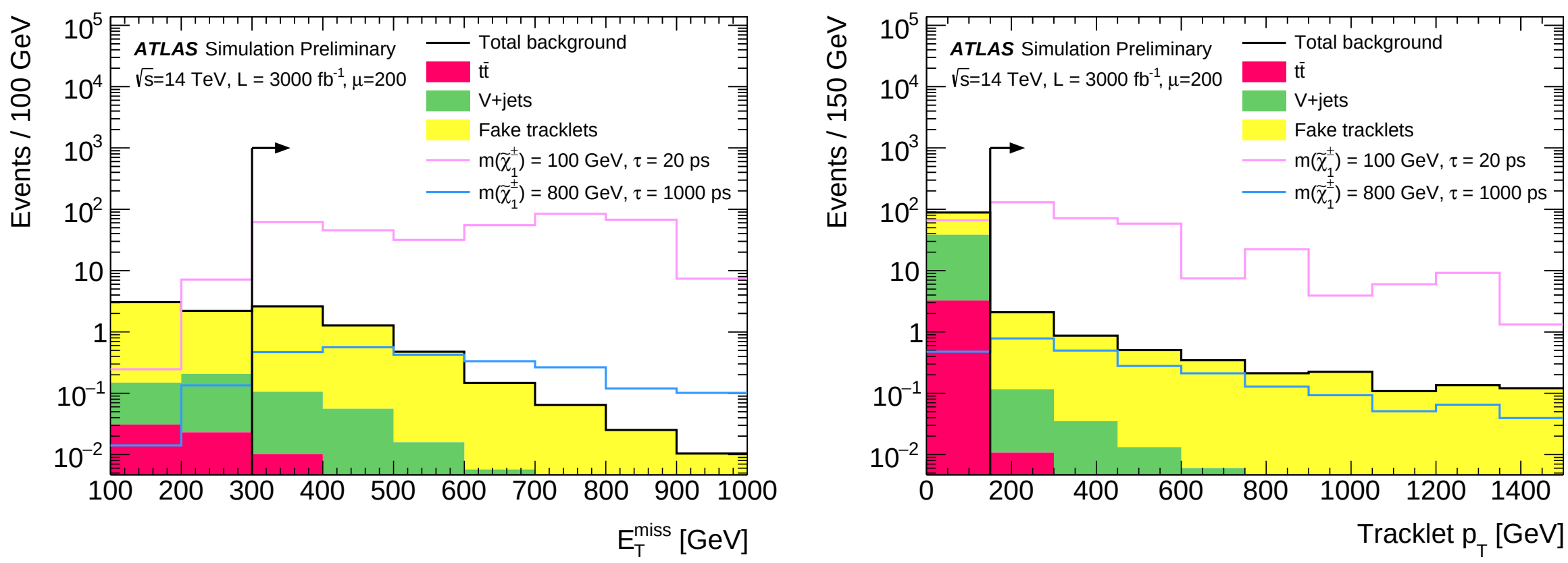

Figure 3: Kinematic variables for the disappearing track search: Left, $E_{\mathrm{T}}^{\mathrm{miss}}$; Right, tracklet $p_{\mathrm{T}}$. All figures contain events passing the full SR selection, aside from the selection on the variable under consideration. Two signal models are overlayed for reference, labelled by two numbers referring respectively to the chargino mass (in GeV) and lifetime (in ps).

For the dilepton search, only events with two opposite-sign muons are used in the final selection, as the muon reconstruction rate is not expected to fall dramatically and the muon fake rate is not expected to grow largely with increased pile-up. The final SR selections are summarised in Table 2. These requirements select events where the SUSY system recoils against a high-$p_T$ jet from ISR. This motivates additional requirements on the leading jet of $p_T(\text{jet}_1) > 100$ GeV and on the azimuthal separation $\Delta\phi(\text{jet}_1, E_T^{\text{miss}}) > 2.0$. In order to discriminate the signal from SM background processes, different kinematic variables are used such as the total number of muons in the event ($n_\mu$), the total number of jets ($n_{\text{jets}}$) and $b$-jets ($n_{b\text{-jets}}$) with $p_T > 30$ GeV. The missing transverse energy ($E_T^{\text{miss}}$), invariant mass of the dilepton system ($m_{\ell\ell}$), angular separation between the leptons ($\Delta R(\ell, \ell)$), invariant mass of two tau leptons ($m_{\tau\tau}$) (calculated as described in Ref. [17]) and ratio of the $E_T^{\text{miss}}$ to the scalar sum of the two leptons' $p_T$ ($E_T^{\text{miss}}/H_T^{\text{lep}}$) are also used. Figure 4 presents a selection of kinematic distributions after the full SR selection is applied, minus the selection on the variable under consideration. The final SR definitions split the $m_{\ell\ell}$ into six non-overlapping SRs, with $m_{\ell\ell}$ selections of [1, 3], [3.2, 5], [5, 10], [10, 20], [20, 30] and [30, 50] GeV.

Table 2: Summary of the SR selection requirements for the dilepton search.

| Variable | SR Selection ($m_{\ell\ell} < 20$ GeV) | SR Selection ($m_{\ell\ell} > 20$ GeV) |
|---|---|---|
| $n_\mu$ | $= 2$ | $= 2$ |
| $p_T(\mu_{1,2})$ [GeV] | $> 3$ | $> 8$ |
| $n_{\text{jets}}$ | $\geq 1$ | $\geq 1$ |
| $n_{\text{b-jets}}$ | $= 0$ | $= 0$ |
| $E_T^{\text{miss}}$ [GeV] | $> 500$ | $> 500$ |
| $\Delta R(\ell, \ell)$ | $< 2$ | $< 2$ |
| $m_{\ell\ell}$ [GeV] | [1, 20] excluding [3.0, 3.2] | [20, 50] |
| $p_T(\text{jet}_1)$ [GeV] | $> 100$ | $> 100$ |
| $\Delta\phi(j_1, E_T^{\text{miss}})$ | $> 2$ | $> 2$ |
| $\min(\Delta\phi(j, E_T^{\text{miss}}))$ | $> 0.4$ | $> 0.4$ |
| $m_{\tau\tau}$ [GeV] | $< 0$ or $> 160$ | $< 0$ or $> 160$ |
| $E_T^{\text{miss}}/H_T^{\text{lep}}$ | $> \max(5, 15 - 2m_{\ell\ell})$ | $> \max(10, 15 - 2m_{\ell\ell})$ |

The leading sources of background in the SR are from $t\bar{t}$, single-top, $WW$ + jets, and $Z/\gamma^*(\to \tau\tau)$ + jets. The dominant source of reducible background arises from processes where one or more leptons are fake or non-prompt, such as in $W$+jets production. The fake/non-prompt lepton background arises from jets misidentified as leptons, photon conversions, or semileptonic decays of heavy-flavor hadrons.

# 5 Background Estimation

MC simulated event samples are used to predict the backgrounds from SM processes in the SRs. Due to the differences in signal phenomenology between the disappearing track and dilepton searches, the main backgrounds and background estimation strategy differ between the searches.

For the disappearing track analysis, there are two main background contributions: SM particles that are reconstructed as tracklets, and events which contain fake tracklets. The SM particles reconstructed as tracklets are typically hadrons scattering in the detector material or electrons undergoing bremsstrahlung. The probability of an isolated electron or hadron leaving a disappearing track is calculated using samples of single electrons or pions passing through the current ATLAS detector layout, and is then scaled to take into account the ratio of material in the current ATLAS inner detector and the upgraded inner tracker. The second background contribution arises from events which contain "fake" tracklets (accidental alignment of hits in the inner detector). These events arise from $Z \to \nu\nu$ or $W \to \ell\nu$ (where the lepton is not reconstructed) and are scaled by the expected fake tracklet probability:

$$p_{\mathrm{fake,tight}}^{\mathrm{ITk}} = p_{\mathrm{fake,tight}}^{\mathrm{Run-2}} \times \frac{R_{\mathrm{fake,loose}}^{\mathrm{ITk}}}{R_{\mathrm{fake,loose}}^{\mathrm{Run-2}}} \times \frac{\epsilon_{z_0}^{\mathrm{ITk}}}{\epsilon_{z_0}^{\mathrm{Run-2}}}. \tag{1}$$

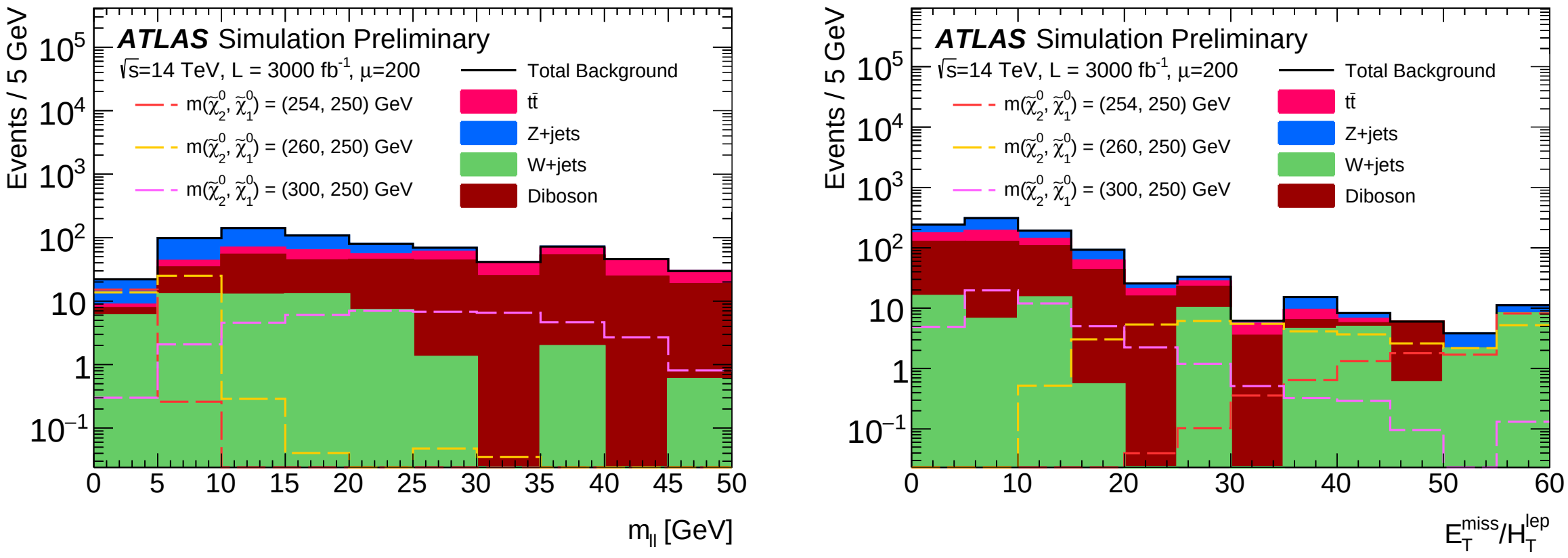

Figure 4: Distributions of a selection of kinematic variables used for the SR optimisation in the dilepton search. The variables are presented with the full SR selections implemented aside from the selection on the variable shown. Left, $m_{\ell\ell}$; Right, $E_{\mathrm{T}}^{\mathrm{miss}}/H_{\mathrm{T}}^{\mathrm{lep}}$. Three signal models with $m(\chi_1^0)$ = 250 GeV and different mass splittings $(\Delta m(\tilde{\chi}_2^0, \tilde{\chi}_1^0) = 4,\ 10,$ and $50\,\mathrm{GeV})$ are overlaid.

In this equation, $p_{\mathrm{fake,tight}}^{\mathrm{Run-2}}$ is the fake rate of the current Run-2 analysis [44], computed using a $d_0$ sideband for the track reconstruction, $R_{\mathrm{fake,loose}}^{\mathrm{ITk}}$ is the fake rate in the same $d_0$ sideband for ITk computed with a neutrino particle gun sample, such that all tracks are purely a result of pile-up interactions, $R_{\mathrm{fake,loose}}^{\mathrm{Run-2}}$ is the fake rate in the $d_0$ sideband for Run 2 computed on data, $\epsilon_{z_0}^{\mathrm{ITk}}$ is the selection efficiency of the tracklet $z_0$ selection in ITk, and $\epsilon_{z_0}^{\mathrm{Run-2}}$ is the selection efficiency of the tracklet $z_0$ selection in Run 2. The ratio of the fake rates in the ITk and Run 2 is found to be ≈200, as the fake rate depends strongly on pile-up, whilst the ratio of the tracklet selection efficiencies is ≈0.12, which takes into account the differences in tracklet selection between the search and the Run 2 analysis.

For the dilepton analysis the $t\bar{t}$ and $W$+jets background yields are calculated by fitting the shape of the background $E_{\mathrm{T}}^{\mathrm{miss}}$ distributions, in order to mitigate statistical fluctuations in the background estimations for these samples. The fit uses an exponential function after all of the SR selections are applied, aside from the $E_{\mathrm{T}}^{\mathrm{miss}}$ selection. As the final SR fit is performed on the $m_{\ell\ell}$ variable, the backgrounds are estimated in each $m_{\ell\ell}$ bin. The predicted values from the fit are found to be consistent with the yields estimated directly from MC, when MC statistics allow for a reasonable comparison to be made. The other SM background processes are estimated directly from MC.

## 6 Systematic Uncertainties

Systematic uncertainty projections for both searches have been determined starting from the systematic uncertainties studied in Ref. [17], and evolving them to a level which the ATLAS and CMS collaborations have agreed to consider as a sensible extrapolation to 3 ab$^{-1}$ of proton-proton collisions. Hence, the theory modelling uncertainties are expected to halve while the recommendations for detector-level and experimental uncertainties are dependent upon the systematic uncertainty under consideration and are scaled appropriately from the Run 2 analysis [18]. When setting exclusion limits, an additional systematic uncertainty of 15% is applied to the dilepton signal samples, whereas a 20% uncertainty is applied to the disappearing track analysis, to account for the theoretical systematic uncertainty on the models under consideration.

The dominant uncertainties in the disappearing track analysis arise from the modelling of the fake tracklet component, and the total uncertainty on the background yield is extrapolated to be 30%. In the dilepton Run 2 analysis the dominant uncertainties are due to the modelling of the fake and non-prompt lepton backgrounds and the experimental uncertainties related to the jet energy scale and flavour tagging. The total uncertainty for the dilepton search is also extrapolated to be 30%. The experimental uncertainty is assumed to be fully correlated between the background and the signal.

## 7 Results

The HistFitter framework [45], which utilises a profile-likelihood-ratio test statistic [46], is used to compute expected exclusion limits with the CL$_s$ prescription [47].

Table 3 presents the expected yields in the SR for the disappearing track search for each background source, corresponding to an integrated luminosity of 3000 fb$^{-1}$. As seen in the table the dominant background source is events with a “fake” tracklet, arising predominantly from $Z \to \nu\nu$ events with an ISR jet and high $E_{\mathrm{T}}^{\mathrm{miss}}$, which contain spurious hits that are reconstructed as a tracklet.

| | **SR** |
|---|---|
| Total SM | 4.6 ± 1.3 |
| $V$+jets events | 0.17 ± 0.05 |
| $t\bar{t}$ events | 0.02 ± 0.01 |
| Fake tracklets | 4.4 ± 1.3 |

Table 3: Yields are presented for the disappearing track SR selection with an integrated luminosity of 3000 fb$^{-1}$ at $\sqrt{s}$ = 14 TeV. The errors shown are the total statistical and systematic uncertainty.

Limits at 95% CL on the chargino lifetime are shown in Figure 5 as a function of the $\tilde{\chi}_1^{\pm}$ mass. The simplified models of chargino production considered include chargino pair production and chargino-neutralino production (both $\tilde{\chi}_1^{\pm}\tilde{\chi}_1^{0}$ and $\tilde{\chi}_1^{\pm}\tilde{\chi}_2^{0}$). The potential for the full HL-LHC dataset is expected to exclude at the 95% CL chargino lifetimes, assuming a wino-like (higgsino-like) LSP, of between 7 ps (10 ps) and 4 $\mu$s (1.5 $\mu$s) for light charginos with a mass of 100 GeV. Heavier wino-like (higgsino-like) charginos are expected to be excluded up to m($\tilde{\chi}_1^{\pm}$) = 1100 GeV (750 GeV) for lifetimes of 1 ns. The discovery potential of the analysis would allow for the discovery of wino-like (higgsino-like) charginos of mass 100 GeV with lifetimes between 20 ps and 700 ns (30 ps and 250 ns), or for a lifetime of 1 ns would allow the discovery of wino-like (higgsino-like) charginos of mass up to 800 GeV (600 GeV). Comparing the results to the theoretical prediction from Ref.[30], would allow for the exclusion at 95% CL of the theory with masses up to 850 GeV for the pure wino scenario and 250 GeV for the pure higgsino scenario. The discovery potential would be up to 450 GeV for the pure wino scenario and 150 GeV for the pure higgsino scenario.

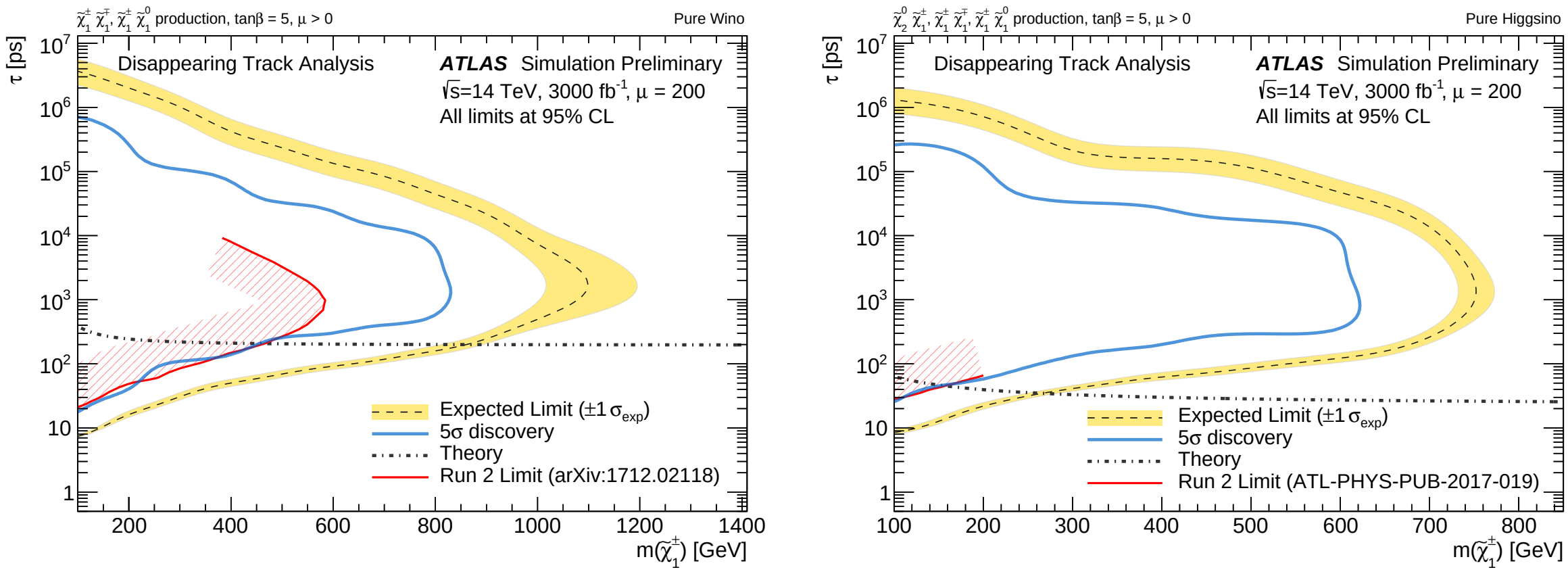

Figure 5: Expected exclusion limits at 95% CL from the disappearing track search using 3000 fb$^{-1}$of 14 TeV proton-proton collision data as a function of the $\tilde{\chi}_1^{\pm}$ mass and lifetime. Simplified models including both $\tilde{\chi}^{\pm}\tilde{\chi}^{\mp}$ and $\tilde{\chi}^{\pm}\tilde{\chi}^{0}$ are considered assuming pure-wino scenarios (left) and pure-higgsino scenarios (right). The yellow band shows the $1\sigma$ region of the distribution of the expected limits. The median of the expected limits is shown by a dashed line. The red line presents the current limits from the Run 2 analysis and the hashed region is used to show the direction of the exclusion. The expected limits with the upgraded ATLAS detector would extend these limits significantly. In the pure-wino scenario, the chargino lifetime as a function of the chargino mass calculated at the two loop level [48] is shown by the dashed grey line. In the pure-higgsino scenario the mass-lifetime relation is shown by the dashed grey line and is calculated at the one loop level [30]. The relationship between the masses of the chargino and the two lightest neutralinos in this scenario is $m(\tilde{\chi}_1^{\pm}) = \frac{1}{2}(m(\tilde{\chi}_1^0) + m(\tilde{\chi}_2^0))$.

The background yields for the dilepton SRs (split into the respective $m_{\ell\ell}$ intervals) are presented in Table 4. The main background in each SR is dependent upon the $m_{\ell\ell}$ interval under consideration, with $t\bar{t}$ the main background for the lowest $m_{\ell\ell}$ interval, the intermediate $m_{\ell\ell}$ selections dominated by $Z$+jets events, and the larger $m_{\ell\ell}$ intervals dominated by diboson production. The $t\bar{t}$ and diboson yields include the component from misidentified leptons. For the lowest $m_{\ell\ell}$ bin the component of $t\bar{t}$ from misidentified leptons is 40%, while it is 15% in the highest $m_{\ell\ell}$ bin.

Figure 6 shows the 95% CL expected exclusion limits in the m($\tilde{\chi}_2^0$), $\Delta m(\tilde{\chi}_2^0, \tilde{\chi}_1^0)$ plane. With 3000 fb$^{-1}$, $\tilde{\chi}_2^0$ masses up to 350 GeV could be excluded, as well as $\Delta m(\tilde{\chi}_2^0, \tilde{\chi}_1^0)$ between 2 and 20 GeV for m($\tilde{\chi}_2^0$) = 150 GeV. In the figure the blue curve presents the $5\sigma$ discovery potential of the search. To calculate the discovery potential a single-bin discovery test is performed by integrating over all of the $m_{\ell\ell}$ bins from 1 to the chosen $m_{\ell\ell}$ upper limit for a given SR selection (aside from $3 < m_{\ell\ell} < 3.2$ GeV).

Figure 7 presents the 95% expected exclusion limits in the $\tilde{\chi}_1^0, \Delta m(\tilde{\chi}_1^{\pm}, \tilde{\chi}_1^0)$ mass plane, from both the disappearing track and dilepton searches. The yellow contour shows the expected exclusion limit from the disappearing track search, with the possibility to exclude $m(\tilde{\chi}_1^{\pm})$ up to 600 GeV for $\Delta m(\tilde{\chi}_1^{\pm}, \tilde{\chi}_1^0) < 0.2$ GeV,

10

| | SR | | | | | |
|---|---|---|---|---|---|---|
| $m_{\ell\ell}$ bin [GeV] | [1, 3] | [3.2, 5] | [5, 10] | [10, 20] | [20, 30] | [30, 50] |
| Total SM | 2.5 ± 0.4 | 16.0 ± 2.5 | 62.4 ± 4.9 | 142.9 ± 10.7 | 102.6 ± 14.9 | 164 ± 20.2 |
| $t\bar{t}$ events | 1.7 ± 0.4 | 1.2 ± 0.2 | 7.5 ± 1.0 | 29.2 ± 4.2 | 21.8 ± 5.4 | 53.9 ± 9.2 |
| $VV$ events | 0.05 ± 0.01 | 0.8 ± 0.2 | 15.4 ± 2.1 | 40.5 ± 5.9 | 50.2 ± 12.3 | 104.8 ± 18.0 |
| $W$+jets events | 0.08 ± 0.02 | 0.9 ± 0.2 | 8.9 ± 1.2 | 25.7 ± 3.7 | 4.7 ± 1.2 | 3.5 ± 0.6 |
| $Z$+jets events | 0.7 ± 0.2 | 13.1 ± 2.5 | 30.6 ± 4.2 | 47.4 ± 6.9 | 25.9 ± 6.4 | 1.8 ± 0.3 |

Table 4: Yields are presented for the dilepton SR selection with an integrated luminosity of 3000 fb$^{-1}$ at $\sqrt{s}$ = 14 TeV. The errors shown are the total statistical and systematic uncertainty.

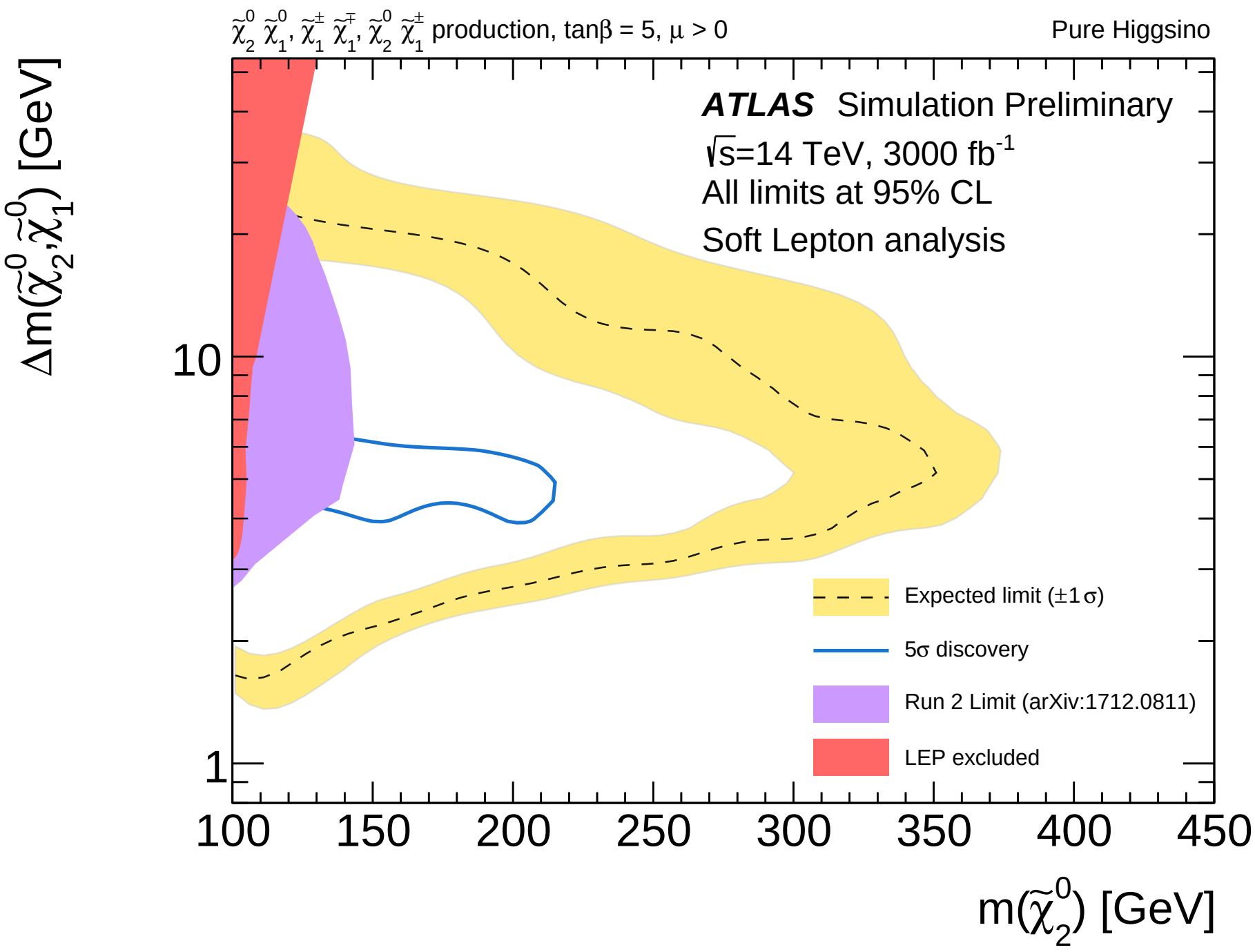

Figure 6: Expected exclusion limit (dashed line) in the $\Delta m(\tilde{\chi}_2^0, \tilde{\chi}_1^0)$, $m(\tilde{\chi}_2^0)$ mass plane, at 95% CL from the dilepton analysis with 3000 fb$^{-1}$of 14 TeV proton-proton collision data in the context of a pure Higgsino LSP with $\pm 1\sigma$ (yellow band) from the associated systematic uncertainties. The blue curve presents the $5\sigma$ discovery potential of the search. The purple contour is the observed exclusion limit from the Run 2 analysis.The figure also presents the limits on chargino production from LEP [16]. The relationship between the masses of the chargino and the two lightest neutralinos in this scenario is $m(\tilde{\chi}_1^{\pm}) = \frac{1}{2}(m(\tilde{\chi}_1^0) + m(\tilde{\chi}_2^0))$.

and could exclude up to $\Delta m(\tilde{\chi}_1^{\pm}, \tilde{\chi}_1^0) = 0.4$ GeV for $m(\tilde{\chi}_1^{\pm}) = 100$ GeV. The blue curve presents the expected exclusion limits from the dilepton search, which could exclude up to 350 GeV in $m(\tilde{\chi}_1^{\pm})$, and for a light chargino mass of 100 GeV would exclude mass differences between 2 and 15 GeV. Improvements that are expected with the upgraded detector, and search technique improvements may further enhance the sensitivity to these models. For example the sensitivity of the disappearing tracks search can be enhanced by optimising the tracking algorithms used for the upgraded ATLAS detector allowing for an increase in tracklet efficiency, the possibility of shorter tracklets produced requiring 3 or 4 hits, and further suppression of the fake tracklet component. The dilepton search sensitivity would be expected to improve by increasing the reconstruction efficiency for low $p_T$ leptons. The addition of the electron channel would also further enhance the search sensitivity.

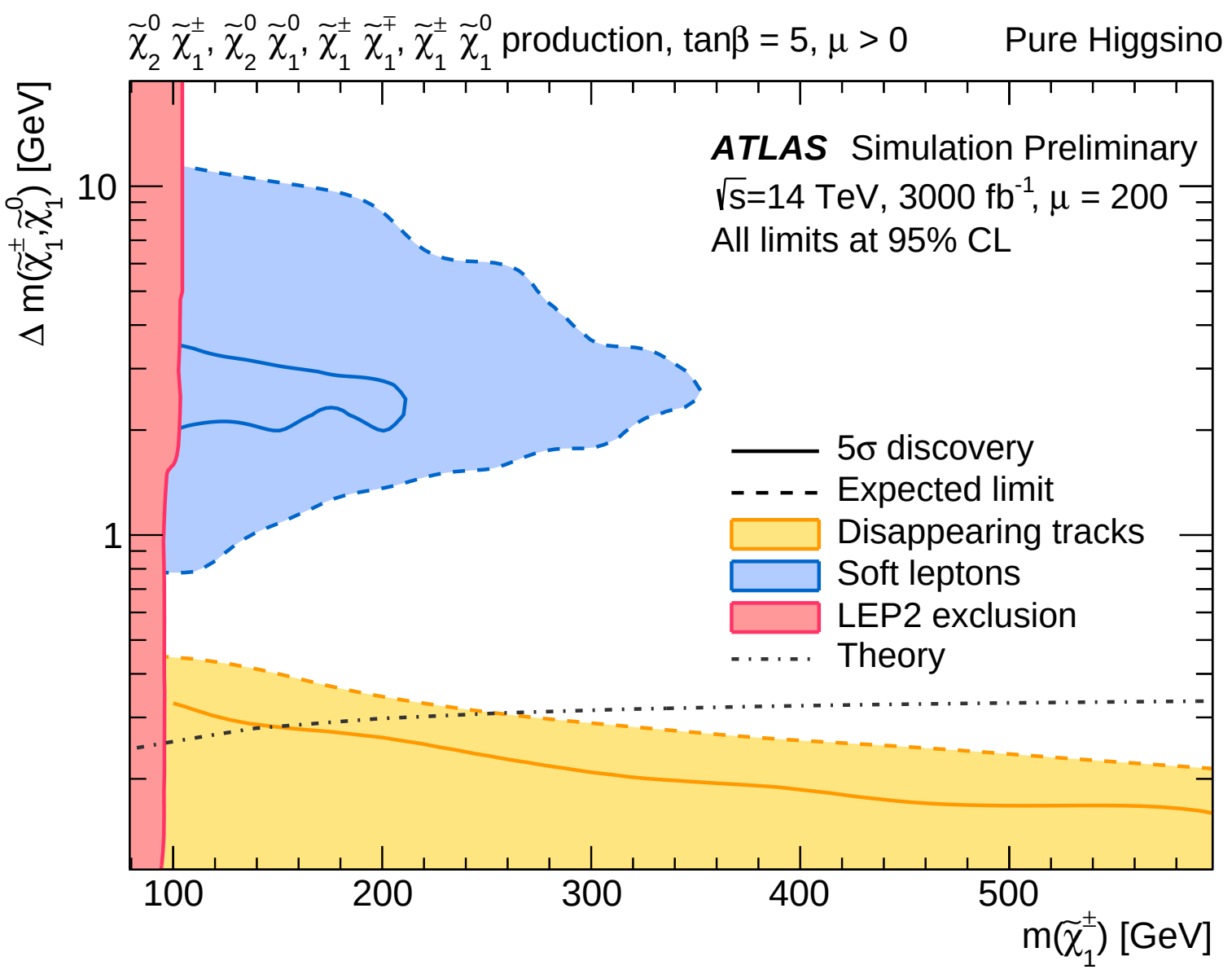

Figure 7: Expected exclusion at the 95% CL from the disappearing track and dilepton searches in the $\Delta m(\tilde{\chi}_1^{\pm}, \tilde{\chi}_1^0)$, $m(\tilde{\chi}_1^{\pm})$ mass plane. The blue curve presents the exclusion limits from the dilepton search. The yellow contour presents the exclusion limit from the disappearing track search. The figure also presents the limits on chargino production from LEP [16]. The relationship between the masses of the chargino and the two lightest neutralinos in this scenario is $m(\tilde{\chi}_1^{\pm}) = \frac{1}{2}(m(\tilde{\chi}_1^0) + m(\tilde{\chi}_2^0))$. The theory curve is a prediction from a pure higgsino scenario taken from Ref.[30].

## 8 Conclusion

This note presents studies performed to assess the sensitivity to electroweakino production with the HL-LHC and the upgraded ATLAS detector, using 3000 fb$^{-1}$ of $\sqrt{s}$ = 14 TeV data. Well motivated and natural SUSY scenarios predict a compressed electroweakinos sector. Two signatures with good discovery potential are considered in this prospect note (disappearing track and soft leptons). In a pure-Higgsino scenario, the former can discover up to 600 GeV charginos with 1 ns lifetime while the latter could discover the second lightest neutralino with mass up to 200 GeV. Improvements that could be expected with the upgraded detector will provide additional sensitivity for both of the searches presented.

# ATLAS sensitivity to top squark pair production at the HL-LHC

The ATLAS Collaboration

This document summarises the expected sensitivity of the ATLAS detector to top squarks with 3 ab$^{-1}$ of $\sqrt{s}$ = 14 TeV proton-proton collisions collected at the HL-LHC. The top squarks are pair produced and assumed to decay into a top quark and a neutralino. Prompt leptons are vetoed in the final state, which is only composed by jets and missing transverse momentum. A 5$\sigma$ discovery (95% CL exclusion) can be obtained for top squark masses up to 1.25 (1.7) TeV and small neutralino masses, assuming realistic projections of the systematic uncertainties. If the top squark mass equals the sum of the top quark and neutralino masses, then a 5$\sigma$ discovery (95% CL exclusion) can be achieved up to about 650 (850) GeV.

# 1 Introduction

The aim of this note is to assess the ATLAS sensitivity to top squark (stop in the following) pair production using the dataset expected to be collected by the upgraded detector during the Large Hadron Collider (LHC) high luminosity data-taking (HL-LHC in the following). The stops are the scalar supersymmetric [1–6] partners of the top quark fermionic degrees of freedom: for each of the two top chirality eigenstates $t_L$, $t_R$ the existence of partner scalar states $\tilde{t}_L$, $\tilde{t}_R$ is postulated. The two scalar states mix to form mass eigenstates $\tilde{t}_1$, $\tilde{t}_2$, where, by convention, $\tilde{t}_1$ is the lightest. Because of the large top quark Yukawa coupling, large stop masses tend to introduce large fine tuning [7, 8] in many supersymmetric models (and notably in the MSSM [9, 10]). Naturalness requirements normally set upper bounds for stop masses in the TeV range (although recent re-analyses of the fine tuning concept led to relax these requirements significantly [11]). These bounds may imply that the stops are within energetic reach of the LHC. This has triggered a lot of interest by the LHC collaborations (see, for example, Refs. [12–16]). Tight constraints have been set by both the ATLAS and CMS collaborations in many simplified and more realistic supersymmetric models.

This work aims to extend the analysis described in Ref. [17] and develop an event selection yielding optimal sensitivity to stop pair production with 3 ab$^{-1}$ of proton-proton collisions, expected to be collected by ATLAS by the end of the HL-LHC run. R-parity is assumed to be conserved [18]. The only supersymmetric particles assumed to have impact on the stop decay are the stop itself and the lightest supersymmetric particle (LSP), assumed to be a neutralino. With these assumptions, the stop decay is $\tilde{t}_1 \to t^{(*)}\tilde{\chi}_1^0$, where the star indicates that the top quark can possibly be off mass-shell, depending on the mass difference between the stop and the neutralino masses, $\Delta m(\tilde{t}_1, \tilde{\chi}_1^0)$. The final state considered is that where both top quarks decay hadronically: it is hence characterised by the presence of many jets and $b$-jets, and by missing transverse momentum $\mathbf{p}_T^{miss}$ (whose magnitude will be indicated by $E_T^{miss}$ in the following) stemming from the presence of the two $\tilde{\chi}_1^0$. The process is illustrated in Figure 1.

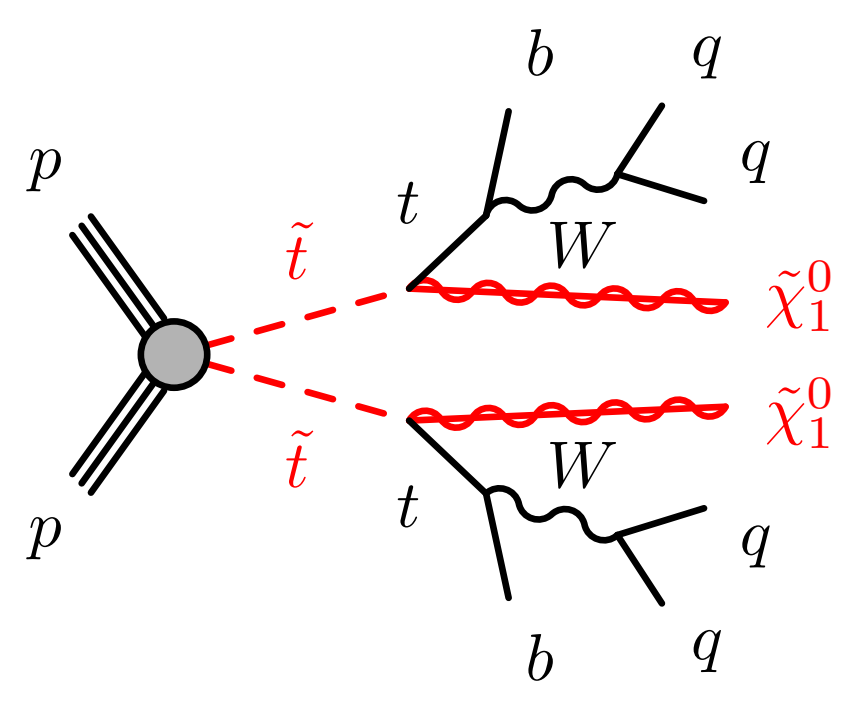

Figure 1: Signal processes considered in this analysis. The top quark can be either on or off mass-shell.

Two kinematic regimes are considered:

- If the difference between the stop and neutralino masses is large with respect to the top quark mass $\Delta m(\tilde{t}_1, \tilde{\chi}_1^0) \gg m_{top}$, then the top quarks emitted in the stop decay are produced on shell, and they have a boost in the laboratory frame proportional to $\Delta m(\tilde{t}_1, \tilde{\chi}_1^0)$. The final state is hence characterised by high $p_T$ jets and $b$-jets, and large $E_T^{miss}$. Typical analyses in this kinematic regimes have large signal acceptance, and the sensitivity is limited by the signal cross section, that decreases steeply

2

with increasing $m(\tilde{t}_1)$. The sensitivity to 3 ab$^{-1}$ of proton-proton collisions in this kinematic regime was already studied in Ref. [19]. This regime is the target of the "large $\Delta m$" analysis described in this document.

- If $\Delta m(\tilde{t}_1, \tilde{\chi}_1^0) \sim m_{\text{top}}$, then the extraction of the signal from the Standard Model (SM) background stemming from mainly $t\bar{t}$ production requires a focus on events where the stop pair system recoils against substantial initial-state hadronic activity (ISR). The upgrade sensitivity to this scenario has never been investigated before by ATLAS in final states with no leptons (see Ref. [20] for a study in final states with two leptons). It is the target of the "diagonal" analysis described in this document.

## 2 The ATLAS Detector

The predicted response of the ATLAS during HL-LHC is emulated by a set of smearing functions applied on top of the final-state particles, defined as those with a lifetime larger than $\tau = 30$ ps. A description of the emulation of the upgraded ATLAS detector is given in Ref. [21]. The smearing functions have been determined from a full Geant 4 [22] simulation of the upgraded ATLAS detector [23] assuming an average number of additional collisions per bunch-crossing $\langle\mu\rangle = 200$.

## 3 Event Simulation

The analysis is performed on datasets of SM background processes and supersymmetric signals simulated through different event generators. Signal models are all generated assuming a proton-proton collision centre-of-mass energy $\sqrt{s} = 14$ TeV with MadGraph5_aMC@NLO 2.2–2.4 [24] interfaced to Pythia 8 [25] for the parton showering (PS) and hadronisation and with EvtGen 1.2.0 [26] for the $b$- and $c$-hadron decays. The matrix element (ME) calculation is performed at tree level and includes the emission of up to two additional partons for all signal samples. In case of top quark off-shell decay, the MadSpin routine is used to preserve the correct spin correlations and phase space modelling. The parton distribution function (PDF) set used for the generation of the signal samples is NNPDF2.3LO [27] with the A14 [28] set of tuned underlying-event and shower parameters (UE tune). The ME–PS matching was performed with the CKKW-L [29] prescription, with a matching scale set to one quarter of the mass of the $\tilde{t}_1$. All signal cross sections were calculated at next-to-leading order in the strong coupling constant, adding the resummation of soft-gluon emission at next-to-leading-logarithm accuracy (NLO+NLL) [30–32]. They strongly depend on the stop mass: for example, the stop pair production cross section is 12.9 (0.14) fb for a stop mass of 900 (1600) GeV.

SM background samples were produced with different MC event generators depending on the process. The background sources of Z + jets and W+ jets events were generated with Sherpa 2.2.1 [33] using the NNPDF3.0NNLO [27] PDF set and the UE tune provided by Sherpa. Top-quark pair production where at least one of the top quarks decays semileptonically and single-top production were simulated with Powheg-Box 2 [34] and interfaced to Pythia 8 for PS and hadronization, with the CT10 [35] PDF set and using the Perugia2012 [36] set of tuned shower and underlying-event parameters. MadGraph5_aMC@NLO interfaced to Pythia 8 for PS and hadronization was used to generate the $t\bar{t}+V$ (where $V$ is a $W$ or $Z$ boson) samples at NLO with the NNPDF3.0NLO PDF set. The underlying-event tune used is A14 with the NNPDF2.3LO PDF set. Additional information can be found in Refs. [37–40]

Z + jets, W+ jets, $t\bar{t}$ +$V$, diboson and single top $s-$ and $t-$channel production events are all simulated assuming $\sqrt{s}$ = 13 TeV, and an event weight is assigned according to the ratio between the relevant PDF distributions to emulate $\sqrt{s}$ = 14 TeV events. $Wt$ and $t\bar{t}$ production events are generated directly assuming a centre-of-mass energy $\sqrt{s}$ = 14 TeV. The samples are normalised to the $\sqrt{s}$ = 14 TeV cross section at NNLO (for $t\bar{t}$ [41]) and NLO (for $Wt$ [42]). The values used are $\sigma_{t\bar{t}}$ = 984.5 pb and $\sigma_{Wt}$ = 84.4 pb.

# 4 Final State Object Definition

The event selection is based on variables constructed from the kinematics of particle-level objects, selected according to reconstruction-level quantities obtained from the emulation of the detector response for HL-LHC [21].

Electrons are defined with a `Loose` identification criterion with $p_{\mathrm{T}} > 7$ GeV and $|\eta| < 2.47$. Muons are defined with a `Loose` identification criterion, and are required to have $p_{\mathrm{T}} > 6$ GeV and $|\eta| < 2.7$. Baseline anti-$k_t$ $R = 0.4$ jets [43, 44] are required to have $p_{\mathrm{T}} > 20$ GeV and $|\eta| < 2.8$[1]. Jets arising from the fragmentation of $b$-hadrons are tagged with a nominal efficiency of 70%, computed on a $t\bar{t}$ sample simulated assuming $\langle\mu\rangle$. The corresponding rejection factor for jets originating from the fragmentation of a $c$ (light) quark is about 20 (750) [45]. Ambiguities between the reconstruction of leptons and jets are resolved following the same overlap-removal procedure outlined in Ref. [17].

Reclustered jets are created by applying the anti-$k_t$ algorithm with distance parameters $\Delta R = 0.8$ and $\Delta R = 1.2$ on signal jets, indicated in the following as anti-$k_t^{0.8}$ and anti-$k_t^{1.2}$ jet collections. A trimming procedure [46] is applied that removes $R = 0.4$ jets from the reclustered jets if their $p_{\mathrm{T}}$ is less than 5% of the $p_{\mathrm{T}}$ of the anti-$k_t^{0.8}$ or anti-$k_t^{1.2}$ jet $p_{\mathrm{T}}$.

# 5 Event Selection

Two different event selections are developed. They respectively target the signal parameter space where $\Delta m(\tilde{t}_1, \tilde{\chi}_1^0) \gg m_{\mathrm{top}}$ or $\Delta m(\tilde{t}_1, \tilde{\chi}_1^0) \sim m_{\mathrm{top}}$. They will be referred to as the "large $\Delta m$" and "diagonal" analyses respectively. In both cases, the selection follows closely that developed for the analysis of the dataset collected in 2015 and 2016, published in Ref. [17]: the same set of selection variables is used, although the thresholds are in some case modified to account for the higher integrated luminosity and the higher level of noise induced by pileup collisions at the HL-LHC.

## 5.1 Large $\Delta m$ Selection

The variables used in Ref. [17] for the event selection in the signal regions targeting large $\Delta m(\tilde{t}_1, \tilde{\chi}_1^0)$ are here briefly summarised:

- $\boldsymbol{N_{\mathrm{lep}}}$**:** The total number of baseline leptons in the event after overlap removal.

[1] Although the upgraded ATLAS detector will allow to efficiently suppress pileup up to large pseudorapidity, the final state objects produced by stop pair production tend to be central: it has been verified that increasing the pseudorapidity of the jet selection does not affect the final result of the analysis.

4

- $N_{\mathrm{jet}}$**:** The total number of signal jets.
- $N_{b-\mathrm{jet}}$**:** The total number of $b$-jets.
- $p_{\mathrm{T}}^{\mathrm{jeti}}$**:** The $p_{\mathrm{T}}$ of the $i$-th anti-$k_t^{0.4}$ jet (where the ordering is done in $p_{\mathrm{T}}$). The leading jet is labelled with $i = 1$.
- $m_i^{\text{anti-}k_t^R}$**:** The mass of the $i$-th (ordering done in mass) reclustered anti-$k_t$ jet reconstructed with distance parameter $R$. The leading jet is labelled with $i = 1$.
- $m_{\mathrm{Tb}}^{\mathrm{min}}$**:** The transverse mass $m_{\mathrm{T}}$[2] between the $\mathbf{p}_{\mathrm{T}}^{\mathrm{miss}}$ and the $b$-jet with the minimum $\Delta\phi$ to the $\mathbf{p}_{\mathrm{T}}^{\mathrm{miss}}$. This variable is known to have a kinematical end-point at the top quark mass for SM $t\bar{t}$ production.
- $\Delta R_{bb}$**:** The $\Delta R$ distance between the two $b$-jets in the event. If more than two $b$-jets are present, those with the highest $p_{\mathrm{T}}$ are considered.
- $m_{\mathrm{T}}^{\chi^2}$**:** Stransverse mass computed using the $\mathbf{p}_{\mathrm{T}}^{\mathrm{miss}}$ and the transverse momenta of the top candidates. They are defined by minimising (among all possible candidates) a $\chi^2$:

$$\chi^2 = \frac{\left(m_W^{\mathrm{cand1}} - m_W^{\mathrm{truth}}\right)^2}{m_W^{\mathrm{truth}}} + \frac{\left(m_W^{\mathrm{cand2}} - m_W^{\mathrm{truth}}\right)^2}{m_W^{\mathrm{truth}}} + \frac{\left(m_{\mathrm{top}}^{\mathrm{cand1}} - m_{\mathrm{top}}^{\mathrm{truth}}\right)^2}{m_{\mathrm{top}}^{\mathrm{truth}}} + \frac{\left(m_{\mathrm{top}}^{\mathrm{cand2}} - m_{\mathrm{top}}^{\mathrm{truth}}\right)^2}{m_{\mathrm{top}}^{\mathrm{truth}}}.$$

  The candidates $W$ are constructed using all possible combinations of one and two non-$b$-tagged jets. If more than two $b$-jets are present, the two with the highest $p_{\mathrm{T}}$ are considered.

A preselection is applied, which is summarised in Table 1. A lepton veto and a selection on the number of jets characterise the choice of focusing on fully hadronic events. The selection on $\Delta\phi\left(E_{\mathrm{T}}^{\mathrm{miss}}, \mathrm{jet}^{1,2}\right)$ is known to be extremely effective in suppressing multijet production events, where the $\mathbf{p}_{\mathrm{T}}^{\mathrm{miss}}$ vector tends to be aligned with one of the jets. The selection on $E_{\mathrm{T}}^{\mathrm{miss}}$ exploits the presence of the non-interacting neutralinos in the final state. The selections on the anti-$k_t^{1.2}$ and anti-$k_t^{0.8}$ jet masses exploit the potential presence of boosted top quarks and $W$-bosons in the final state. The selection on $m_{\mathrm{Tb}}^{\mathrm{min}}$ is effective in suppressing events from SM $t\bar{t}$ production.

For the evaluation of the final exclusion sensitivity, a set of mutually exclusive signal regions is defined. The background after preselection is dominated by $t\bar{t}$ and single top $Wt$ production. For both of these processes, the largest contribution comes from events where one of the two $W$ bosons decays hadronically and the other decays leptonically (there including $W \to \tau\nu$). The dominant background processes hence feature at most one hadronic top and/or $W$ decay, while the signal features two of them. The events are hence further classified in 30 different signal regions according to the number of identified $b$-jets, the value of $m_2^{\text{anti-}k_t^{1.2}}$ mass, and the value of the $E_{\mathrm{T}}^{\mathrm{miss}}$. In each $N_{b-\mathrm{jet}}$ bin, three bins are defined in $m_2^{\text{anti-}k_t^{1.2}}$. In order, it corresponds to having found a jet with mass: i) below that of the $W$ boson, ii) similar to that of the $W$ boson, iii) loosely consistent with that of the top quark. Finally, in each bin a set of $E_{\mathrm{T}}^{\mathrm{miss}}$ intervals are defined. In the bin with $N_{b-\mathrm{jet}} \geq 2$, there is no ambiguity[3] in the definition of the two expected $b$-jets from the stop decay, thus the $m_{\text{T}}^{\chi^2}$ and $\Delta R_{bb}$ are well defined. Additional selections on these variables are therefore applied. The full set of signal region selections is also presented in Table 1.

[2] The transverse mass between two vectors in the transverse plane $\mathbf{a}$ and $\mathbf{b}$ forming an angle $\theta$ between them is defined as $m_{\mathrm{T}} = \sqrt{2ab(1 - \cos\theta)}$.

[3] If $N_{b-\mathrm{jet}} > 2$ then the two $b$-jets with the highest $p_{\mathrm{T}}$ are used.

Table 1: Selection applied for the large $\Delta m$ analysis.

| Preselection | |
|---|---|
| $N_{\text{lep}} = 0$ | |
| $N_{\text{jet}} \geq 4$ | |
| $\Delta\phi\left(E_{\text{T}}^{\text{miss}}, \text{jet1}\right) > 0.4$, $\Delta\phi\left(E_{\text{T}}^{\text{miss}}, \text{jet2}\right) > 0.4$ | |
| $N_{b-\text{jet}} \geq 1$ | |
| $E_{\text{T}}^{\text{miss}} > 400$ GeV | |
| $p_{\text{T}}^{\text{jet1}} > 80$ GeV, $p_{\text{T}}^{\text{jet2}} > 80$ GeV | |
| $p_{\text{T}}^{\text{jet3}} > 40$ GeV, $p_{\text{T}}^{\text{jet4}} > 40$ GeV | |
| $m_1^{\text{anti-}k_t^{1.2}} > 120$ GeV | |
| $m_{\text{Tb}}^{\text{min}} > 250$ GeV | |
| $m_1^{\text{anti-}k_t^{0.8}} > 60$ GeV,$m_2^{\text{anti-}k_t^{0.8}} > 60$ GeV | |
| Signal region selection | |
| Number of $b$-tagged jets | Other selections |
| $N_{b-\text{jet}} = 1$ | $m_2^{\text{anti-}k_t^{1.2}} \in [0, 60), [60, 120), [120, \infty)$<br>$E_{\text{T}}^{\text{miss}} \in [400, 600), [600, 900), [900, 1200), [1200, 1600), [1600, \infty)$ |
| $N_{b-\text{jet}} > 1$ | $m_2^{\text{anti-}k_t^{1.2}} \in [0, 60), [60, 120), [120, \infty)$<br>$E_{\text{T}}^{\text{miss}} \in [400, 600), [600, 900), [900, 1200), [1200, 1600), [1600, \infty)$<br>$m_{\text{T}}^{\chi^2} > 400$, $\Delta R_{bb} \geq 1$ |

The $E_{\text{T}}^{\text{miss}}$ distributions for $N_{b-\text{jet}} \geq 2$ and for the two tightest bins in $m_2^{\text{anti-}k_t^{1.2}}$, that is, for the two bins that are most sensitive for large $\Delta m(\tilde{t}_1, \tilde{\chi}_1^0)$ values, are shown in Figure 2.

For the evaluation of the discovery sensitivity, a set of single bin cut-and-count signal regions is defined, which apply the full preselection, and then require $N_{b-\text{jet}} \geq 2$, $m_2^{\text{anti-}k_t^{1.2}} > 120$ GeV. Four different thresholds in $E_{\text{T}}^{\text{miss}}$ are then defined to achieve optimal sensitivity for a $5\sigma$ discovery: $E_{\text{T}}^{\text{miss}} > 400, 600, 800, 1000$ GeV. For each model considered, the signal region giving the lowest $p$-value against the background-only hypothesis in presence of the signal is used.

To avoid depending too much on the limited size of the background samples generated for this study, the background $E_{\text{T}}^{\text{miss}}$ distribution is parametrised independently for each $N_{b-\text{jet}}$, $m_2^{\text{anti-}k_t^{1.2}}$ bin and for each process with a simple exponential function. The function parameters are determined by fitting it to the MC predicted distribution in the range $E_{\text{T}}^{\text{miss}} > 400$ GeV.

## 5.2 Diagonal Selection

The selection for the region of the signal parameter space where $\Delta m(\tilde{t}_1, \tilde{\chi}_1^0) \sim m_{\text{top}}$ also follows closely that developed in Ref. [17]. The basic idea of the diagonal analysis arise from the fact that, given the mass relation between the stop and the neutralino, the stop decay products (the top quark and the neutralino) are produced nearly at rest in the stop reference frame. When looked at from the lab reference frame, the

6

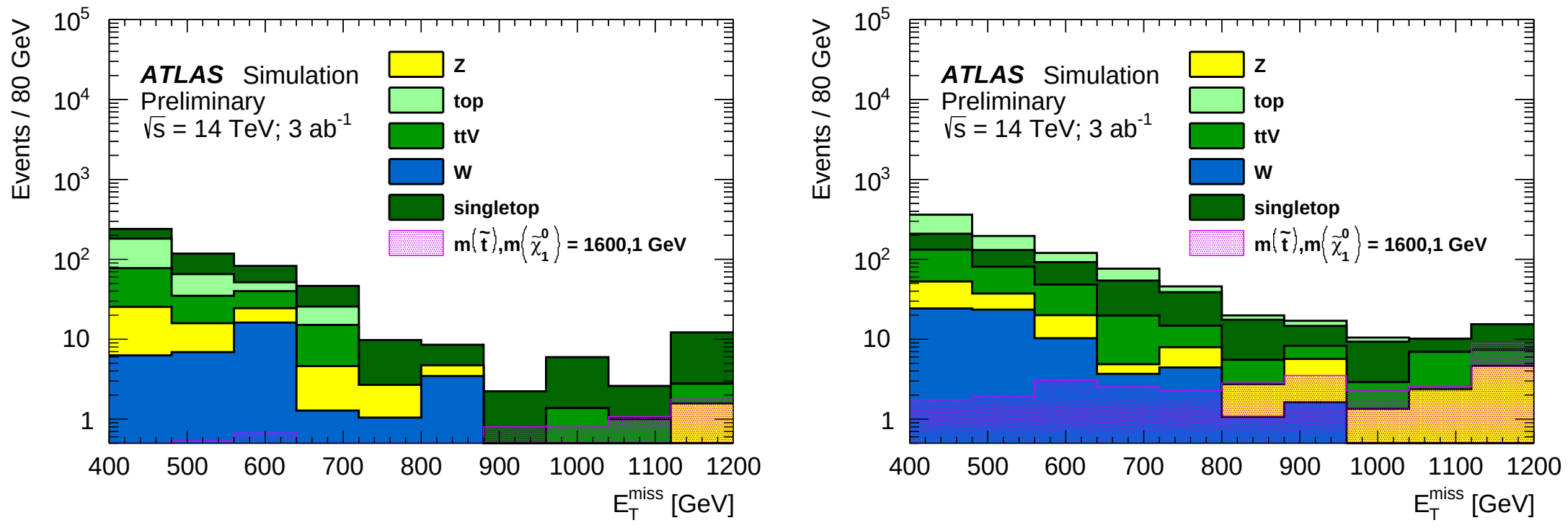

(a) $N_{b\text{-jet}} \geq 2$, 60 GeV < $m_2^{\text{anti-}k_t^{1.2}}$ $\leq$ 120 GeV

(b) $N_{b\text{-jet}} \geq 2$, $m_2^{\text{anti-}k_t^{1.2}}$ > 120 GeV

Figure 2: The $E_T^{\text{miss}}$ distributions for the two bins with the highest sensitivity to signals with large values of $\Delta m(\tilde{t}_1, \tilde{\chi}_1^0)$. The last bin includes overflow events.

transverse momentum acquired by the decay products will be proportional to their mass. If $p_T^{\text{ISR}}$ is the transverse momentum of everything that recoils against the stop pair, it can be shown that [47]

$$R_{\text{ISR}} = \frac{E_T^{\text{miss}}}{p_T^{\text{ISR}}} \sim \frac{m\left(\tilde{\chi}_1^0\right)}{m\left(\tilde{t}_1\right)} \tag{1}$$

Following this considerations, a recursive jigsaw [48] reconstruction is performed, which makes assumptions that allow the definition of a set of variables in different reference frames. In this specific case, it first defines the centre-of-mass of the primary proton-proton collision, or **CM** frame. In the **CM** frame, the sparticles frame **S** and the ISR system (**ISR**) are back-to-back to each other. One can then define the Visible (**V**) and Invisible (**I**) reference frames, composed respectively by the visible particles produced in the stop pair decay and the invisible particles produced in the stop pair decay. New variables are defined to exploit the relation suggested by Equation 1.

The following variables are used specifically for the diagonal analysis.

- $\boldsymbol{p_T^{\text{ISR}}}$ The total momentum of the **ISR** system in the **CM** frame.
- $\boldsymbol{R_{\text{ISR}}}$ This is defined as the ratio of the projection of total momentum of the invisible system in the **CM** frame on the total momentum of the **ISR** frame in the **CM** frame, to the the total momentum of the **ISR** frame in the **CM** frame. That is

$$R_{\text{ISR}} = \frac{|\vec{p}_{\mathbf{I}}^{\,\mathbf{CM}} \cdot \hat{p}_{\mathbf{ISR}}^{\mathbf{CM}}|}{|\vec{p}_{\mathbf{ISR}}^{\,\mathbf{CM}}|} \tag{2}$$

- $\boldsymbol{N_{b-\text{jet}}^{\text{S}}}$ The number of $b$-jets assigned to the frame **V**.
- $\boldsymbol{N_{\text{jet}}^{\text{S}}}$ The number of jets assigned to the frame **V**.
- $\boldsymbol{p_T^{4,\text{S}}}$ The transverse momentum of the 4th jet associated with the **V** frame.

- $\boldsymbol{\Delta\phi}(\mathbf{ISR}, \mathbf{p}_{\mathbf{T}}^{\mathbf{miss}})$ The distance in azimuthal angle between the **ISR** and **I** total momentum vectors in the **CM** frame.
- $\boldsymbol{m_{\mathrm{S}}}$ The mass of the **S** frame.
- $\boldsymbol{p_{\mathrm{T},b}^{0,\mathrm{S}}}$ The transverse momentum of the leading $b$-jet associated to the **V** frame.

The preselection applied for the diagonal analysis is summarised in Table 2.

Figure 3 shows the distribution of some of the key variables after the selection on $p_{\mathrm{T}}^{4,\mathrm{S}}$. The main features which were observed in Ref. [17] are still present: the $m_{\mathrm{S}}$ variable peaks at roughly twice the top mass for the signal, the position of the peak of the $R_{\mathrm{ISR}}$ variable increases with the ratio $m\left(\tilde{\chi}_1^0\right)/m\left(\tilde{t}_1\right)$, there is a strong peak at $\pi$ for the signal for $\Delta\phi(\mathrm{ISR}, \mathbf{p}_{\mathrm{T}}^{\mathrm{miss}})$. Already at this stage of the selection, the main background process is $t\bar{t}$.

Table 2: Selection applied for the diagonal analysis.

| Preselection | |
|---|---|
| $N_{\mathrm{lep}} = 0$ | |
| $N_{\mathrm{jet}} \geq 4$ | |
| $\Delta\phi\left(E_{\mathrm{T}}^{\mathrm{miss}}, \mathrm{jet1}\right) > 0.4$, $\Delta\phi\left(E_{\mathrm{T}}^{\mathrm{miss}}, \mathrm{jet2}\right) > 0.4$ | |
| $N_{b-\mathrm{jet}} \geq 1$ | |
| $E_{\mathrm{T}}^{\mathrm{miss}} > 400$ GeV | |
| $p_{\mathrm{T}}^{\mathrm{jet1}} > 80$ GeV, $p_{\mathrm{T}}^{\mathrm{jet2}} > 80$ GeV | |
| $p_{\mathrm{T}}^{\mathrm{jet3}} > 40$ GeV, $p_{\mathrm{T}}^{\mathrm{jet4}} > 40$ GeV | |
| $N_{b-\mathrm{jet}}^{\mathrm{S}} \geq 1$ | |
| $N_{\mathrm{jet}}^{\mathrm{S}} \geq 5$ | |
| $p_{\mathrm{T},b}^{0,\mathrm{S}} > 40$ GeV | |
| $m_{\mathrm{S}} > 300$ GeV | |
| $\Delta\phi(\mathrm{ISR}, \mathbf{p}_{\mathrm{T}}^{\mathrm{miss}}) > 3$ | |
| $p_{\mathrm{T}}^{\mathrm{ISR}} > 400$ GeV | |
| $p_{\mathrm{T}}^{4,\mathrm{S}} > 50$ GeV | |
| Signal region selection | |
| $R_{\mathrm{ISR}}$ selection | $E_{\mathrm{T}}^{\mathrm{miss}}$ selection |
| $0.5 < R_{\mathrm{ISR}} < 0.65$ | $E_{\mathrm{T}}^{\mathrm{miss}} \in [500, 700), [700, 1000), [1000, 1400), [1400, \infty)$ |
| $R_{\mathrm{ISR}} > 0.65$ | $E_{\mathrm{T}}^{\mathrm{miss}} \in [500, 700), [700, 1000), [1000, 1400), [1400, \infty)$ |

A possible strategy is suggested by figure 3(d): the $E_{\mathrm{T}}^{\mathrm{miss}}$ distribution shifts progressively higher values of $m\left(\tilde{t}_1\right)$. Hence a boost in sensitivity could be obtained by binning the signal regions in this variable. The final strategy for the assessment of exclusion sensitivity for the diagonal analysis is thus to use a set of mutually exclusive signal region defined in bins of $R_{\mathrm{ISR}}$ and $E_{\mathrm{T}}^{\mathrm{miss}}$. The final binning is shown in Table 2. Lower values of $R_{\mathrm{ISR}}$ are not considered given that the current analysis focuses mostly on the prospects for high $m(\tilde{t})$. For the evaluation of the discovery sensitivity, four cut-and-count signal regions are defined, which apply the full preselection, and then require $R_{\mathrm{ISR}} > 0.7$ and $E_{\mathrm{T}}^{\mathrm{miss}} > 500, 700, 900, 1100$ GeV. For each model considered, the signal region giving the lowest $p$-value against the SM hypothesis in presence of signal is used.

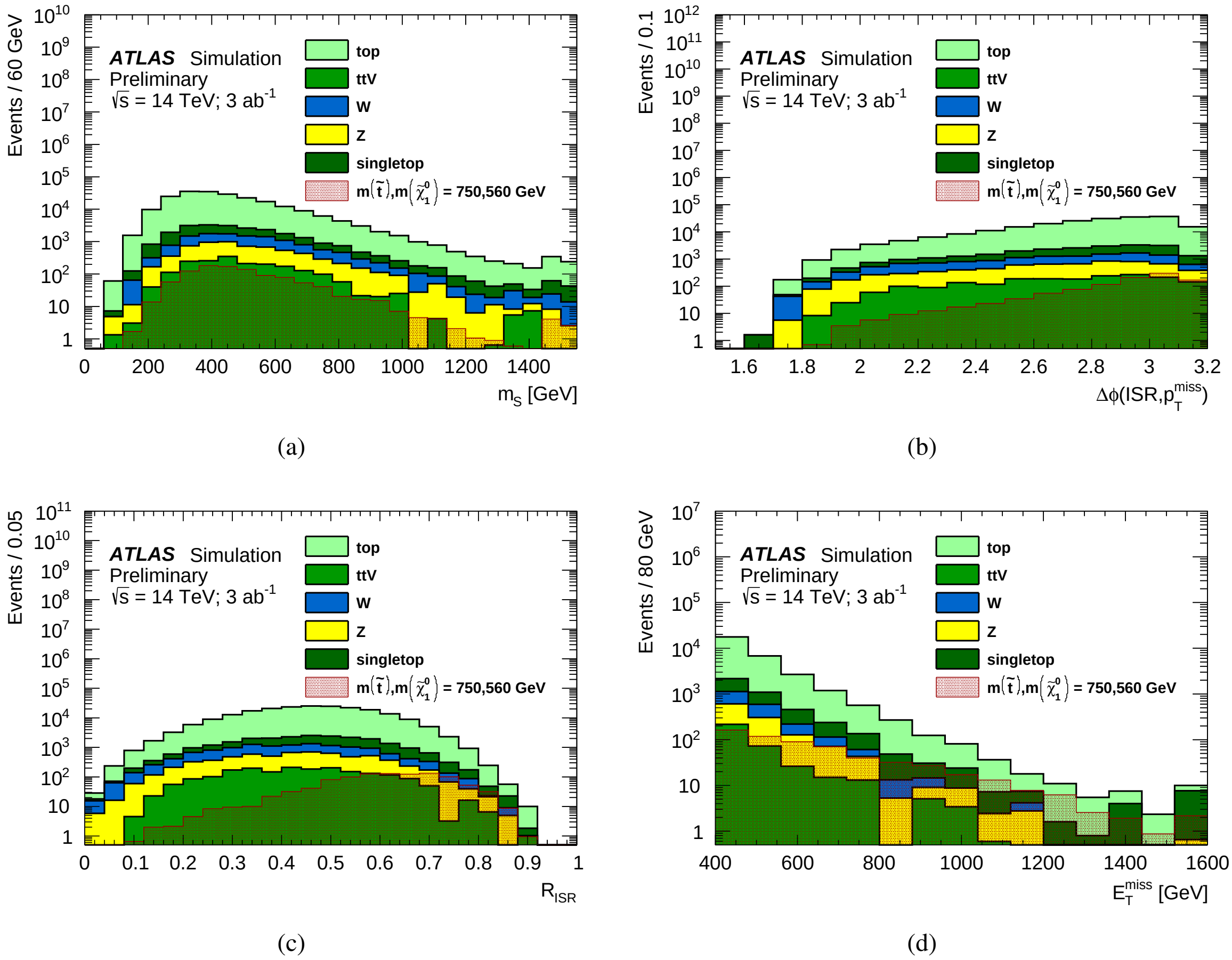

Figure 3: Distributions of the main variables of the diagonal selection, after all cuts up to that on $p_{\mathrm{T}}^{4,\mathrm{S}}$ have been applied. The signal and backgrounds are normalised to 3 ab$^{-1}$. The last bin includes overflow events.

Similarly to the large $\Delta m$ analysis, the background estimation used for the assessment of the analysis sensitivity stems from a parameterisation of the actual background MC. The background is parametrised in $E_{\mathrm{T}}^{\mathrm{miss}}$ in each bin of $R_{\mathrm{ISR}}$ and independently for each background process. The parametrisation is established for $E_{\mathrm{T}}^{\mathrm{miss}} > 500$ GeV, and it is done with a simple exponential function.

# 6 Systematic Uncertainties

Realistic and pessimistic uncertainty scenarios have been determined starting from the systematic uncertainties studied in Ref. [17], and extrapolating them to 3 ab$^{-1}$ following a common approach agreed upon by the ATLAS and CMS collaborations. Hence, the theory modelling uncertainties are expected to be reduced by a factor 2, while different recommendations have been provided for detector-level and experimental uncertainties.

With reference to Ref. [17]:

- For the large $\Delta m$ analysis, the total systematic uncertainty in the signal regions that targeted the same parameter space region as this analysis was evaluated to be 14–24%. The dominant uncertainties were due to jet energy scale (JES - 7%) and resolution (JER - 5–10%), $t\bar{t}$ and $Wt$ parton shower and generator uncertainties (5–12%). Owing to the reduced statistical uncertainty and a better understanding of the physics models, it is expected that JES, JER and top modelling uncertainties will all be reduced. It is assumed that they will all be halved by the end of the HL-LHC running. This leads to an estimate of the uncertainty for the large $\Delta m$ analysis of about 15% or less, depending on the phase space region. Given how relevant the $Wt$ background process is, special care will be needed to make sure that the treatment of interference terms between $t\bar{t}$ and $Wt$ and the corresponding uncertainty will be under control by the end of the HL-LHC.
- For the diagonal analysis, the estimated uncertainties in Ref. [17] were about 20%, with the exception of one region that was affected by large statistical uncertainty in the MC samples for $t\bar{t}$. The dominant uncertainty in all cases was connected with the modelling of ISR in $t\bar{t}$ events. For the uncertainty projection, it was decided to proceed as for the case of the large $\Delta m$ uncertainty to halve the $t\bar{t}$ modelling uncertainties. The result is a predicted uncertainty of 17% for the tightest $R_{\text{ISR}}$ bin.

In conclusion, a 15% uncertainty is retained as a baseline value of the expected uncertainty for both the large $\Delta m$ and the diagonal analysis to determine both the $5\sigma$ and the 95% CL exclusion reach of the analysis. For the case of the estimation of the 95% CL exclusion sensitivity, a further scenario with doubled uncertainty (30%) is also evaluated.

# 7 Results

The final $E_{\text{T}}^{\text{miss}}$ distribution in the bins with $m_2^{\text{anti-}k_t^{1.2}} > 120$ GeV, $N_{b\text{-jet}} \geq 2$ (for the large $\Delta m$ analysis) and $R_{\text{ISR}} > 0.65$ (for the diagonal analysis) are shown in Figure 4.

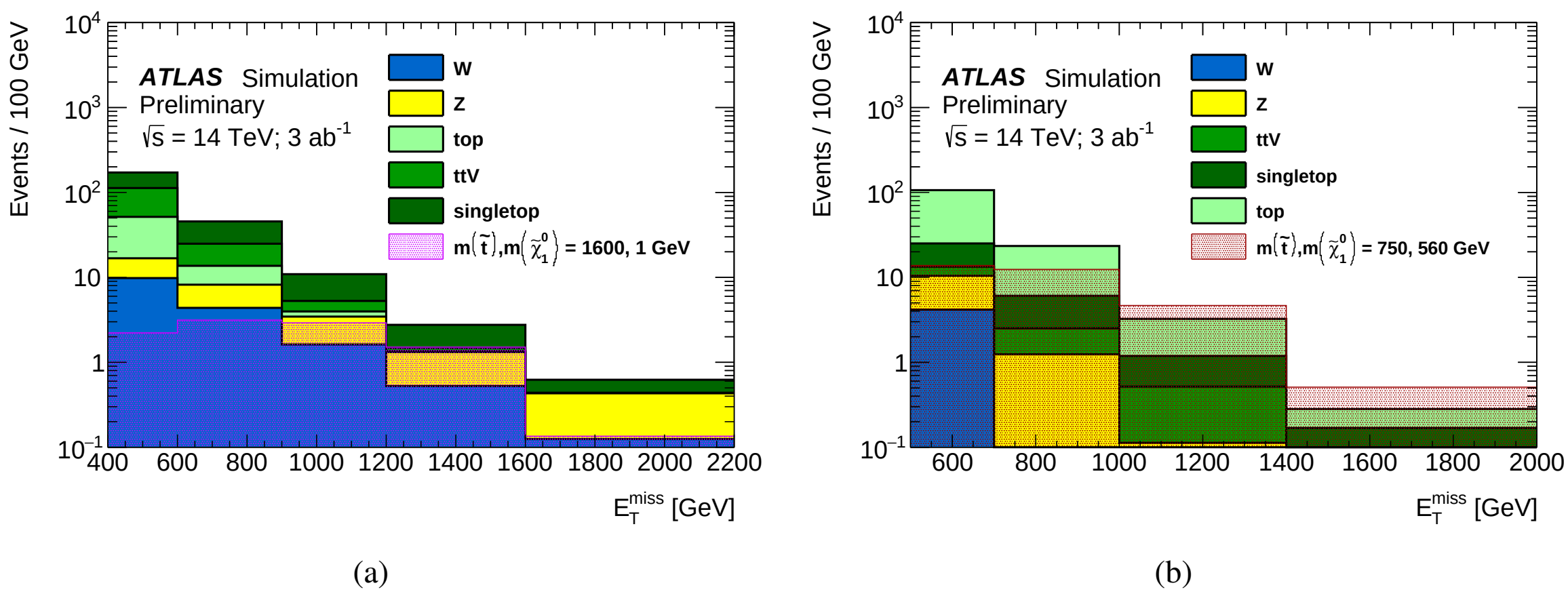

Figure 4: $E_{\text{T}}^{\text{miss}}$ distribution for (a) the $m_2^{\text{anti-}k_t^{1.2}} > 120$ GeV, $N_{b\text{-jet}} \geq 2$ bin of the large $\Delta m$ analysis and (b) $R_{\text{ISR}} > 0.65$ bin of the diagonal analysis. The last bin includes overflow events.

The final exclusion sensitivity evaluation is done by performing a profile-likelihood fit to a set of pseudo-data providing bin-by-bin yields corresponding to the background expectations. For each of the two analyses (large $\Delta m$ and diagonal), the likelihood is built as the product of poissonian terms, one for each of the considered bins. Systematic uncertainties are accounted for by introducing one independent nuisance parameter for each of the bins considered. The likelihood is modified introducing gaussian terms representing the assumed uncertainty. 95% CL exclusion contours on the masses of the supersymmetric particles are extracted using the CLs method [49]. For each mass of the stop and the neutralino, the analysis yielding the smallest CLs among the large $\Delta m$ and the diagonal is used.

The discovery sensitivity is obtained similarly from each of the single cut-and-count regions independently. For each signal point, the profile likelihood ratio fit is performed on pseudo-data corresponding to the sum of the expected background and the signal. The discovery contour corresponds to points expected to give a $5\sigma$ p-value against the background-only hypothesis. For each signal point, the discovery signal region yielding the smallest p-value is considered.

The final sensitivity of the analysis is summarised in Figure 5 assuming a 15% uncertainty for the $5\sigma$ discovery and 95% CL exclusion contour, and also assuming 30% uncertainty for the 95% CL exclusion contour.

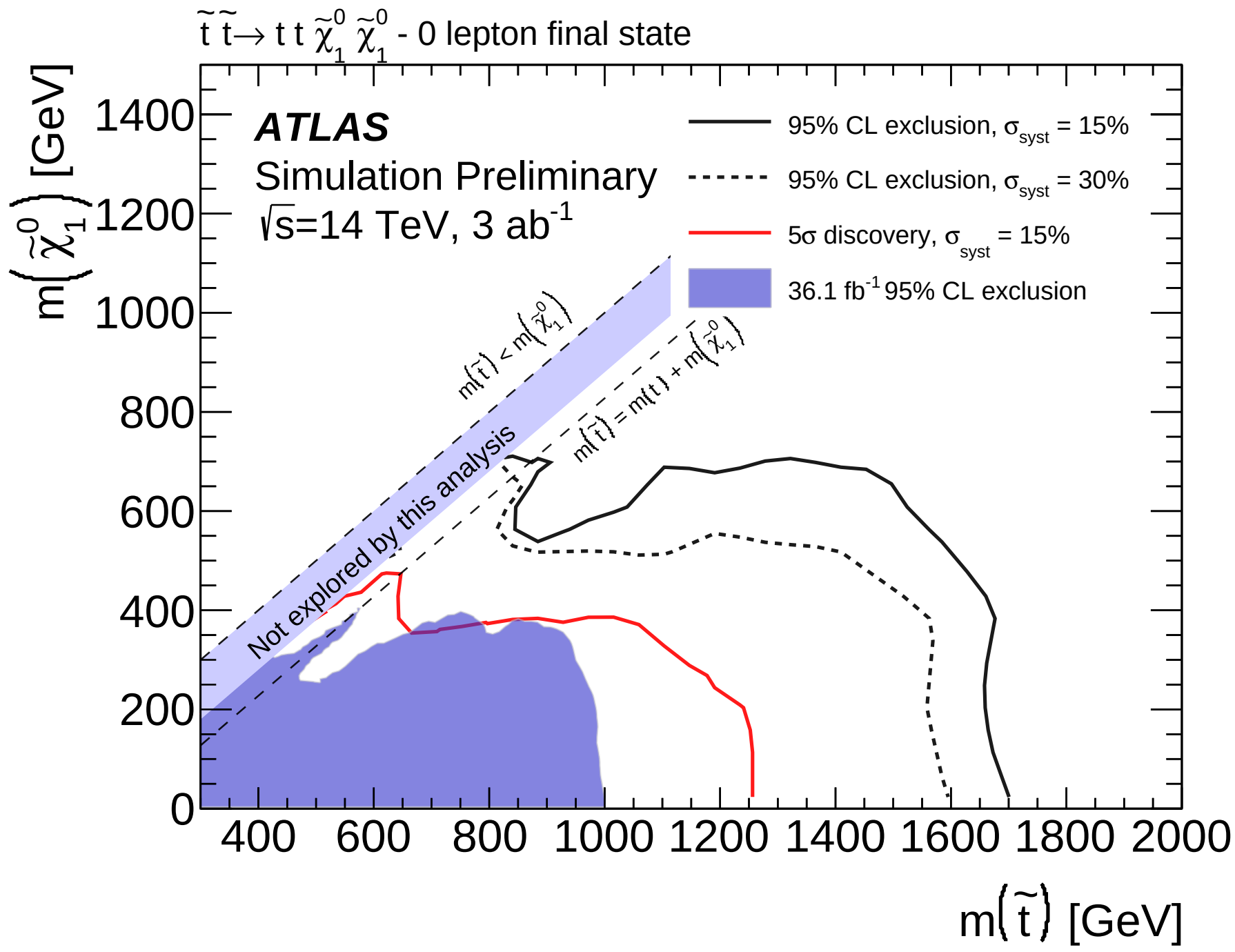

Figure 5: Final 95% CL exclusion reach and $5\sigma$ discovery contour corresponding to 3 ab$^{-1}$of proton-proton collisions collected by ATLAS at the HL-LHC.

# 8 Conclusions

The ATLAS sensitivity to stop pair production with 3 ab$^{-1}$ of proton-proton collisions and running conditions corresponding to those of the HL-LHC is estimated with an analysis that follows closely that published in Ref. [17]. The process of interest is $\tilde{t}_1 \rightarrow t^{(*)}\tilde{\chi}_1^0$. Event containing no leptons are retained, and two separate selections are developed targeting regions of the parameter space where $\Delta m(\tilde{t}_1, \tilde{\chi}_1^0) \gg m_{\rm top}$ or $\Delta m(\tilde{t}_1, \tilde{\chi}_1^0) \sim m_{\rm top}$. 95% CL exclusion and $5\sigma$ discovery contours are derived in the $\left(m\left(\tilde{t}_1\right), m\left(\tilde{\chi}_1^0\right)\right)$ plane for uncertainty assumptions which are either realistic or pessimistic extrapolations of the current uncertainties. Stops can be discovered (excluded) up to masses of 1.25 (1.7) TeV for $m\left(\tilde{\chi}_1^0\right) \sim 0$ under realistic uncertainty assumptions. The reach in stop mass degrades for larger neutralino masses. If $\Delta m(\tilde{t}_1, \tilde{\chi}_1^0) \sim m_{\rm top}$, then the discovery (exclusion) reach is 650 (850) GeV.

**ATLAS PUB Note**

# ATLAS sensitivity to Two-Higgs-Doublet models with an additional pseudoscalar exploiting four top quark signatures with 3 ab$^{-1}$ of $\sqrt{s}$ = 14 TeV proton-proton collisions

The ATLAS Collaboration

This document summarises the expected ATLAS sensitivity to four top-quark signatures in models involving two Higgs doublets and an additional pseudo-scalar mediator that couples to dark matter particles. The sensitivity is estimated using 3000 fb$^{-1}$ of proton-proton collisions in the context of the High Luminosity LHC update. The experimental signatures investigated include at least two charged leptons with the same electric charge or at least three leptons.

# 1 Introduction

A class of simplified models for dark matter (DM) searches at the LHC involving a two-Higgs-doublet extended sector together with an additional pseudo-scalar mediator to DM [1, 2] (denoted as 2HDM+$a$) are considered. This category of models represent one of the simplest ultra-violet (UV) complete and renormalisable frameworks to investigate the broad phenomenology predicted by spin-0 mediator-based DM models [2–16]. In these models, the 2HDM sector consists of a type-II coupling structure [17, 18]. Furthermore, the alignment ($\cos(\beta - \alpha) = 0$) and decoupling limit is assumed, such that the lightest CP-even state of the Higgs-sector, $h$, can be identified with the Standard Model (SM) Higgs boson and the electroweak vacuum-expectation-value (VEV), $v$, is set to 246 GeV. The additional pseudo-scalar mediator of the model, $a$, couples the DM particles to the SM and mixes with the pseudo-scalar partner of the SM Higgs boson, $A$. Following the prescriptions in Ref. [2], the masses of the heavy CP-even Higgs boson, $H$, and charged bosons, $H^{\pm}$, are set equal to the mass of the heavy CP-odd partner $m_A$, while the the three quartic couplings between the scalar doublets and the $a$ boson ($\lambda_{P1}, \lambda_{P2}$ and $\lambda_3$) are all set equal to 3, in order to simplify the phenomenology and evade the constraints from electroweak precision measurements. In addition, to reduce the multiplicity of the parameter space we consider unitary couplings between the $a$ and the DM particle $\chi$ ($y_\chi = 1$) and we fix the DM particle mass to 10 GeV. The ratio of the VEVs of the two Higgs-doublets, $\tan\beta$, is set to unity as well. This model is characterised by a rich phenomenology and can produce very different final states according to the production and decay modes for the various bosons composing the Higgs sector, which can decay both into dark matter or SM particles. The four-top signature is particularly interesting in this model if at least some of the neutral Higgs partners masses are kept above the $t\bar{t}$ threshold, since, when kinematically allowed, all four neutral bosons can contribute to this final state, as depicted in Fig. 1. The total four top-quark production cross-section is dominated by the light pseudo-scalar and the heavy scalar bosons. In order to highlight this interplay, four benchmark models will be considered, assuming different choices for the mass of the light CP-odd and heavy CP-even bosons and the mixing angle between the two CP-odd weak eigenstates ($\sin\theta$).

**Scenario 1** $m_a$ sensitivity scan assuming:

a) $m_H = 600$ GeV , $\sin\theta = 0.35$.

b) $m_H = 1$ TeV , $\sin\theta = 0.7$.

**Scenario 2** $\sin\theta$ sensitivity scan assuming:

a) $m_H = 600$ GeV , $m_a = 200$ GeV.

b) $m_H = 1$ TeV , $m_a = 350$ GeV.

Scenario 1a and Scenario 2 closely follow the DM Working Group recommendations and are therefore ideal benchmarks to compare the four top-quarks signature with other final states that characterise the phenomenology of this model [2]. The production cross section of the four benchmark scenarios are presented in Figure 2. It is found that for $m_H = 600$ GeV, the kinematic properties of the signal is relatively independent of $m_a$ (except for masses below the top-quark threshold) and of the mixing angle. Conversely, when the heavy scalar Higgs boson is chosen to be heavier ($m_H = 1$ TeV), the mass of the light pseudo-scalar and the mixing angle choice play an important role in determining the kinematic properties of the final state. This note considers four top-quarks final states involving at least two leptons with the same electric charge or at least three or more leptons. Final states with high jet multiplicity and one lepton are also very powerful to constrain these signature [19], but are not considered in this note.

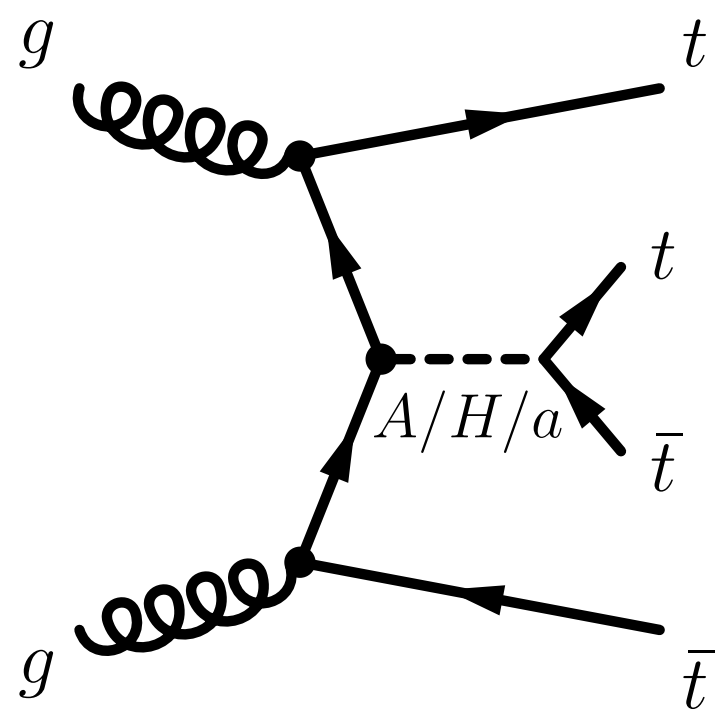

Figure 1: Schematic representation of the dominant production and decay modes for the 2HDM+$a$ model in the four top quarks final state.

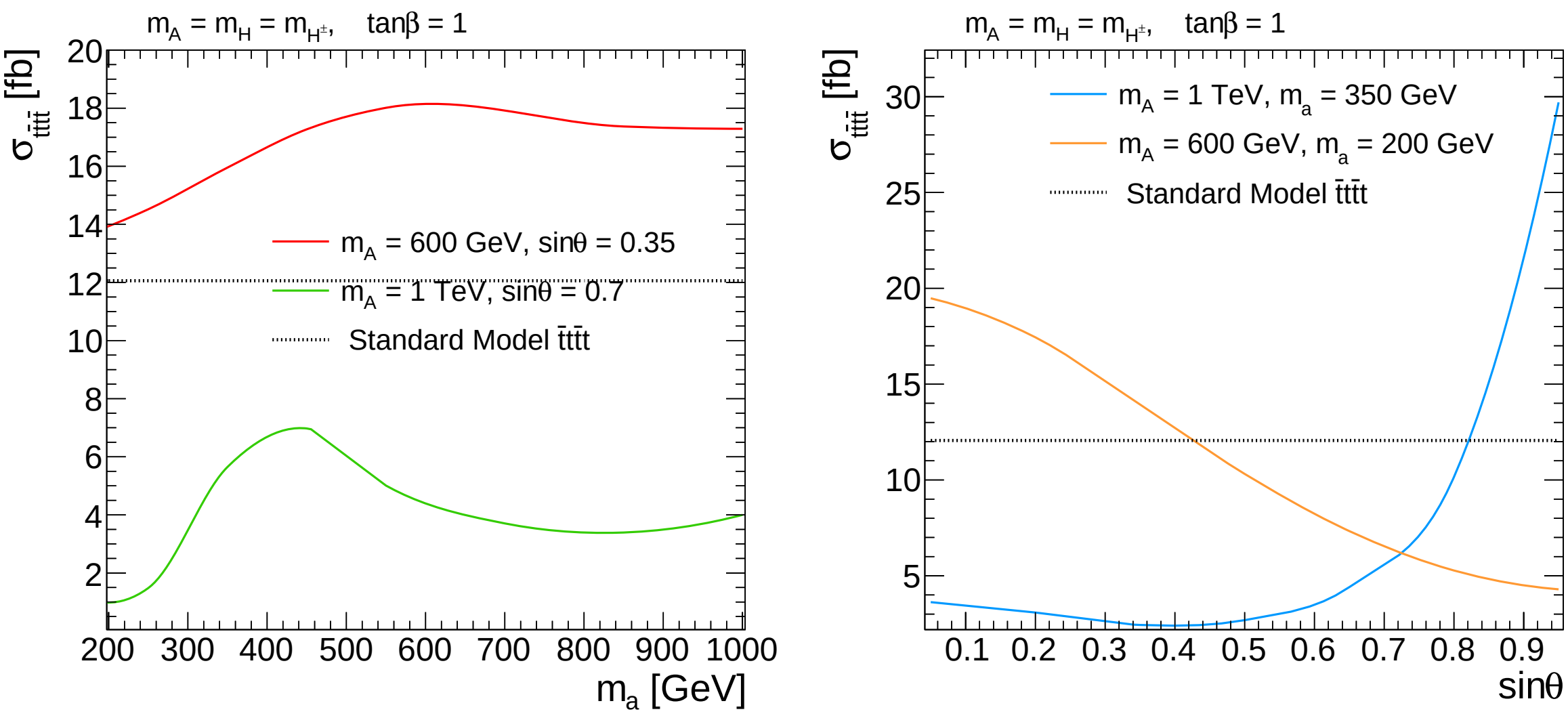

Figure 2: Production cross section for the four top quarks final states in the 2HDM+$a$ model for scenario 1 (left) and scenario 2 (right), as a function of light pseudoscalar mass and the mixing angle, respectively.

In the present data-taking period (Run-2), the LHC delivered $\sim$160 fb$^{-1}$ of proton-proton collisions with an instantaneous luminosity of $\sim 2\times10^{34}$ cm$^{-2}$s$^{-1}$ and an average number of collisions per bunch crossing of $\langle\mu\rangle \sim 35$. A second long shutdown (LS2) will follow, during which the injection chain is foreseen to be modified to allow for a higher instantaneous luminosity. The average number of proton-proton collisions per bunch crossing is expected to be $\langle\mu\rangle \sim 60$ and the data collected up to the next long shutdown (LS3) will amount to $\sim$300 fb$^{-1}$. An increase of the centre-of-mass-energy to 14 TeV is possible and is assumed to happen for this study. During LS3, the accelerator is foreseen to be upgraded to the High-Luminosity LHC (HL–LHC) which will be able to achieve luminosities of $\sim 7.5\times10^{34}$ cm$^{-2}$s$^{-1}$. The HL–LHC is expected to deliver an average number of pile up interactions per bunch crossing of $\langle\mu\rangle \sim 200$ and the data collected will amount to $\sim$3000 fb$^{-1}$.

## 2 ATLAS detector

The ATLAS experiment [20] is a multi-purpose particle detector with a forward-backward symmetric cylindrical geometry and nearly $4\pi$ coverage in solid angle[1]. The interaction point is surrounded by an inner detector (ID), a calorimeter system, and a muon spectrometer.

Upgrades to the detector and the triggering system are planned to adapt the experiment to the increasing instantaneous and integrated luminosities expected with the HL–LHC [21–27].

In the reference upgrade scenario, the ID will provide precision tracking of charged particles for pseudo-rapidities $|\eta| < 4.0$ and will be surrounded by a superconducting solenoid providing a 2 T axial magnetic field. It will consist of pixel and silicon-microstrip detectors.

In the pseudorapidity region $|\eta| < 3.2$, the currently installed high-granularity lead/liquid-argon (LAr) electromagnetic (EM) sampling calorimeters will be used. The current steel/scintillator tile calorimeter will measure hadron energies for $|\eta| < 1.7$. The end-cap and forward regions, spanning $1.5 < |\eta| < 4.9$, currently instrumented with LAr calorimeters for both the EM and hadronic energy measurements will be upgraded with a high granularity forward calorimeter in the $3.1 < |\eta| < 4.9$ range.

The muon spectrometer, consisting of three large superconducting toroids with eight coils each, a system of trigger and precision-tracking chambers, which provide triggering and tracking capabilities in the ranges $|\eta| < 2.4$ and $|\eta| < 2.7$, respectively, could be upgraded with the addition of a very forward muon tagger that will extend the trigger coverage up to $|\eta| = 4.0$.

A two-level trigger system will be used to select events, reducing the event rate to about 10 kHz. In the reference scenario, the bandwidth allocated to single lepton ($e$ or $\mu$) triggers will be of $\sim 2.2$ kHz each, with the respective thresholds set at $p_{\mathrm{T}} > 22$ GeV and $p_{\mathrm{T}} > 20$ GeV.

## 3 Dataset and simulated event samples

Monte Carlo (MC) simulated event samples are used to predict the background from SM processes and to model the 2HDM+$a$ signal. All samples have been generated assuming a proton-proton collision centre-of-mass energy $\sqrt{s} = 14$ TeV. The descriptions of these samples is summarised in Table 1. The detector response is taken into account using a set of parameterised response functions based on studies performed with Geant 4 simulations [28, 29] of the upgraded detector in high luminosity conditions [30]. The most relevant MC samples have equivalent luminosities of at least 3000 fb$^{-1}$.

Signal samples are generated from leading-order (LO) matrix elements, using the aMC@NLO v2.6.1 event generator interfaced to Pythia 8.230 with the A14 tune for the modelling of the parton showering, hadronisation and the description of the underlying event. Parton luminosities are provided by the NNPDF23LO Parton Distribution Function (PDF) set. The decay of the top quarks is modelled using MadSpin [52]. The interference of the signal processes with SM four top-quark production is neglected, as it amounts to less than 5% in the phase space considered in this analysis[2] [2]. In all cases, the mass of the top quark is fixed at 172.5 GeV.

[1] ATLAS uses a right-handed coordinate system with its origin at the nominal interaction point (IP) in the centre of the detector and the $z$-axis along the beam pipe. The $x$-axis points from the IP to the centre of the LHC ring, and the $y$-axis points upward. Cylindrical coordinates $(r, \phi)$ are used in the transverse plane, $\phi$ being the azimuthal angle around the beam pipe. The pseudorapidity is defined in terms of the polar angle $\theta$ as $\eta = -\ln\tan(\theta/2)$. Rapidity is defined as $y = 0.5 \ln[(E + p_z)/(E - p_z)]$ where $E$ denotes the energy and $p_z$ is the component of the momentum along the beam direction.

Table 1: Simulated signal and background event samples: the corresponding event generator, the parton shower, the cross-section normalisation, the PDF set and the underlying-event set of tuned parameters are shown.

| Physics process | Generator | Parton shower | Cross-section normalisation | PDF set | Tune |
|---|---|---|---|---|---|
| 2HDM+$a$ Signals | aMC@NLO 2.2.3 [31] | Pythia 8.186 [32] | LO [2] | NNPDF2.3LO [33] | A14 [34] |
| $t\bar{t}$ | powheg-box v2 [35] | Pythia 6.428 [36] | NNLO+NNLL [37–42] | NLO CT10 [43] | Perugia2012 [44] |
| Single-top ($t$-channel) | powheg-box v1 | Pythia 6.428 | NNLO+NNLL [45] | NLO CT10f4 | Perugia2012 |
| Single-top ($s$- and $Wt$-channel) | powheg-box v2 | Pythia 6.428 | NNLO+NNLL [46, 47] | NLO CT10 | Perugia2012 |
| $t\bar{t}W/Z/\gamma^*$ | aMC@NLO 2.2.2 | Pythia 8.186 | NLO [31] | NNPDF2.3LO | A14 |
| Diboson | Sherpa 2.2.1 [48] | Sherpa 2.2.1 | Generator NLO | CT10 [43] | Sherpa default |
| $t\bar{t}h$ | aMC@NLO 2.2.2 | Herwig 2.7.1 [49] | NLO [50] | CTEQ6L1 [51] | A14 |
| $Wh$, $Zh$ | aMC@NLO 2.2.2 | Pythia 8.186 | NLO [50] | NNPDF2.3LO | A14 |
| $t\bar{t}WW$, $t\bar{t}t\bar{t}$ | aMC@NLO 2.2.2 | Pythia 8.186 | NLO [31] | NNPDF2.3LO | A14 |
| $tZ$, $tWZ$, $t\bar{t}t$ | aMC@NLO 2.2.2 | Pythia 8.186 | LO | NNPDF2.3LO | A14 |
| Triboson | Sherpa 2.2.1 | Sherpa 2.2.1 | Generator LO, NLO | CT10 | Sherpa default |

Signal cross-sections are taken from the generator predictions and are between 1 and 30 fb. For comparison, the nominal cross-section of the largest irreducible background, the SM production of $t\bar{t}\,t\bar{t}$ is 12.06 fb [31].

# 4 Event selection

Candidate leptons are selected with $p_{\mathrm{T}}$ larger than 25 GeV to ensure that trigger efficiencies are fully efficient in the relevant phase space, and $|\eta| < 2.47$ (2.5) for electrons (muons). For this study, it was decided not to take advantage of the extended pseudorapidity coverage of the upgraded detector, since the lepton contribution at large values of $|\eta|$ for the targeted models has been found to be negligible. The electrons (muons) are required to pass "tight" ("medium") identification requirements [30]. For events with two electrons or one electron and one muon, electrons with $|\eta| > 1.37$ are not considered since such events are subject to backgrounds from electron charge misidentification, which has a substantially higher probability of occurring for electrons at high $|\eta|$. This background is subsequently assumed to be negligible [26]. They are also required to be isolated within the tracking volume: the scalar sum of the transverse momenta of charged tracks with $p_{\mathrm{T}} > 1$ GeV, not including the lepton track, within a cone of radius $\Delta R \equiv \sqrt{(\Delta\phi)^2 + (\Delta\eta)^2} = 0.2$ around the lepton candidate must be less than 15% of the lepton $p_{\mathrm{T}}$, where $\Delta\eta$ and $\Delta\phi$ are the separations in $\eta$ and $\phi$.

Candidate jets are reconstructed with the anti-$k_t$ algorithm [53, 54] with a radius parameter of 0.4. They are selected with $p_{\mathrm{T}} > 25$ GeV and $|\eta| < 3.8$ and requiring a tracking confirmation to reduce the contribution from pile-up. Signal events have been found to lie mostly in the central region of the detector, and the $\eta$ requirements for leptons and jets, tighter than the detector acceptance, have been optimised accordingly.

[2] The exact madgraph instruction used is: p p > t t t t / a z h1 QCD<=2

5

Jets are required to be separated from candidate electrons by $\Delta R(e, \text{jet}) > 0.2$. Leptons are removed if they are within $\Delta R = 0.4$ of a remaining jet. Jets are identified as originating from $b$-decays ($b$-tagged) using a parameterisation (as a function of the jet $p_{\text{T}}$ and $\eta$) modelling the performances of the MV2 $b$-tagging algorithm [55]. A requirement is chosen corresponding to a 77% average efficiency obtained for b-quark jets in simulated $t\bar{t}$ events. The rejection factors for light-quark and gluon jets, c-quark jets and $\tau \rightarrow$ hadrons+$\nu$ decays in simulated $t\bar{t}$ events are approximately 380, 12 and 54, respectively [30]. The $E_{\text{T}}^{\text{miss}}$ at generator level is computed as the vectorial sum of the momenta of neutral weakly-interacting particles. It is then smeared to simulate the detector response, with a function parameterised in the average number of interactions per bunch crossing $\mu$ and the scalar sum of energy in the calorimeter $\sum E_{\text{T}}$.

Events are accepted if they contain at least two electrons, two muons or one electron and one muon with the same electric charge or at least three leptons ($p_{\text{T}} > 25$ GeV). Furthermore, events are required to contain at least three $b$-jets. The up to four leading leptons and up to four leading $b$-jets in the event are grouped respectively in two systems, called $\mathcal{S}_\ell$ and $\mathcal{S}_b$. A signal system $\mathcal{S}$ is defined by $\mathcal{S} = \mathcal{S}_\ell \cup \mathcal{S}_b$.

Different discriminators and kinematic variables are used in the analysis to separate the signal from the SM background.

- $p_{\text{T}}(\mathcal{S}_\ell)$ : the vectorial sum of the lepton four momenta in $\mathcal{S}_\ell$;
- $\Delta R(\mathcal{S}_\ell, \mathcal{S}_b)$: the $\Delta R$ between the vectorial sum of the leptons in $\mathcal{S}_\ell$ and the vectorial sum of the $b$-jets in $\mathcal{S}_b$;
- m($\mathcal{S}$): the invariant mass of the signal system $\mathcal{S}$;

A common selection is applied to all events, before further categorisations. Events are required to have at least two jets with a $p_{\text{T}} > 50$ GeV. In events with exactly two (anti-)electrons, the contribution of SM processes including an on-shell $Z$ boson decaying leptonically with a lepton charge misidentification is reduced by vetoing events with 81.2 GeV $< m_{\ell\ell} <$ 101.2 GeV. Furthermore, low mass resonances are vetoed by requiring $m_{\ell\ell} > 15$ GeV. Figure 3 shows the $p_{\text{T}}(\mathcal{S}_\ell)$ and $\Delta R(\mathcal{S}_\ell, \mathcal{S}_b)$ distributions for events passing all the requirements described so far.

Two signal regions (SR) are defined selecting events with exactly two charged leptons with the same electric charge (denoted Same-Sign) or three or more charged leptons (denoted Multi-lep). A summary of the selections is presented in Table 2.

Figure 4 shows four key distributions ($\Delta R(\mathcal{S}_\ell, \mathcal{S}_b)$, $p_{\text{T}}(\mathcal{S}_\ell)$ , $b$-jet multiplicity, $m(\mathcal{S})$) for events passing one set of SRs requirements except for the requirement on the shown variable itself. The main backgrounds that survive the selections are the irreducible $t\bar{t}\,t\bar{t}$ and $t\bar{t}$ +V/h.

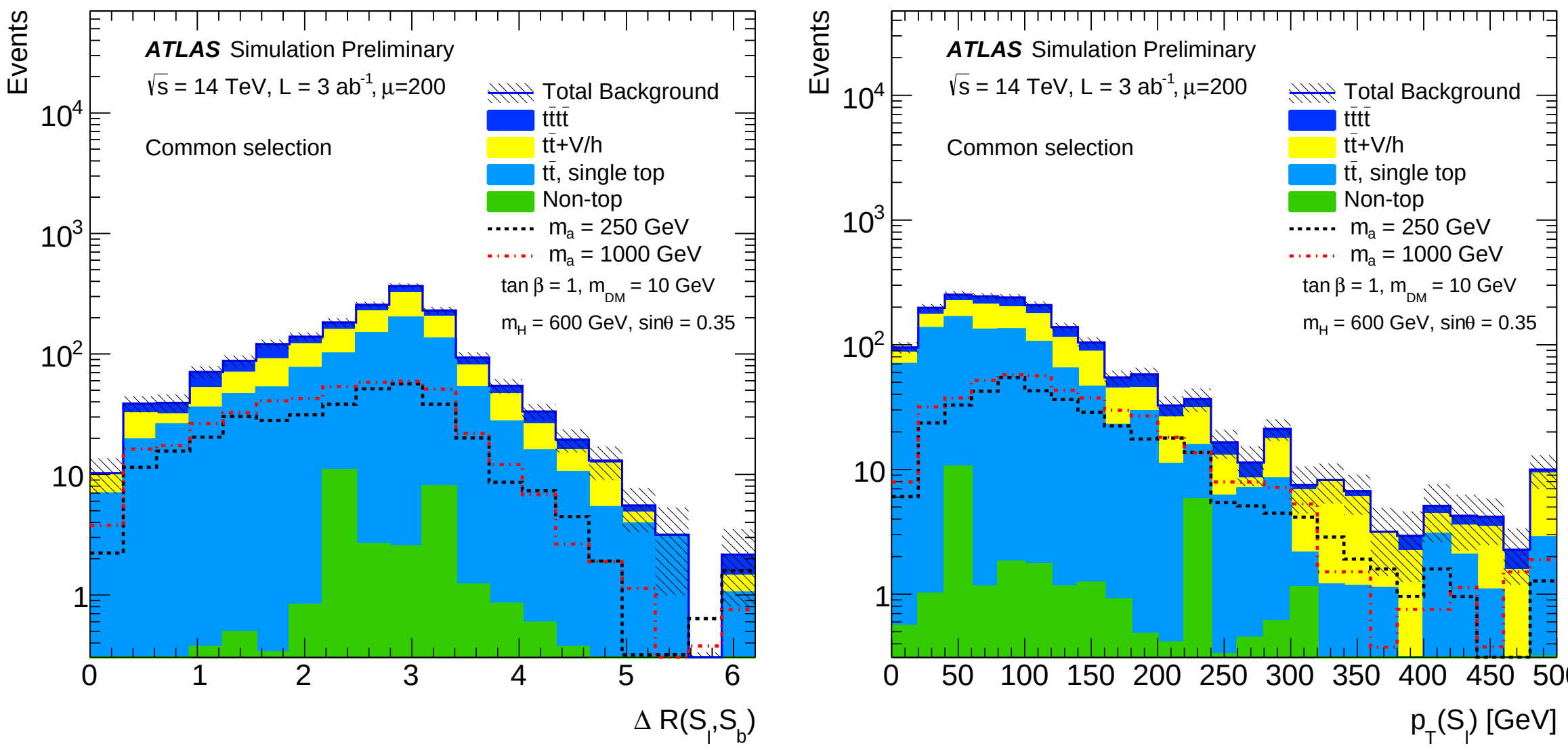

Figure 3: Distributions of $p_{\mathrm{T}}(\mathcal{S}_\ell)$ (left) and $\Delta R(\mathcal{S}_\ell, \mathcal{S}_b)$ (right) for events passing the $m_{\ell\ell}$ requirements described in Section 4. The contributions from all SM backgrounds are shown, and the hashed band represents the statistical uncertainty on the total SM background prediction. The expected distributions for signal models with $m_a = 250$ GeV, $m_H = 1000$ GeV, $\sin\theta = 0.35$ and $m_a = 550$ GeV, $m_H = 600$ GeV, $\sin\theta = 0.35$ are also shown as dashed lines for comparison.

Table 2: Summary of the analysis selection criteria (see text for details).

| | Same-Sign | Multi-lep |
|---|---|---|
| Lepton multiplicity ($p_{\mathrm{T}} > 25$ GeV) | exactly 2 (same charge) | $\geq 3$ |
| $m_{\ell\ell}$ allowed intervals [GeV] | ($> 15$ AND $< 81.2$) OR $> 101.2$ | – |
| $p_{\mathrm{T}}(\mathcal{S}_\ell)$ [GeV] | $> 50$ | $> 100$ |
| $b$-jet multiplicity ($p_{\mathrm{T}} > 25$ GeV) | $> 3$ | $> 2$ |
| $b$-jet multiplicity ($p_{\mathrm{T}} > 50$ GeV) | – | $> 1$ |
| Jet multiplicity ($p_{\mathrm{T}} > 50$ GeV) | $> 3$ | $> 1$ |
| $\Delta R(\mathcal{S}_\ell, \mathcal{S}_b)$ | $< 2.5$ | $< 2.5$ |
| $m(\mathcal{S})$ [GeV] | - | $> 550$ |

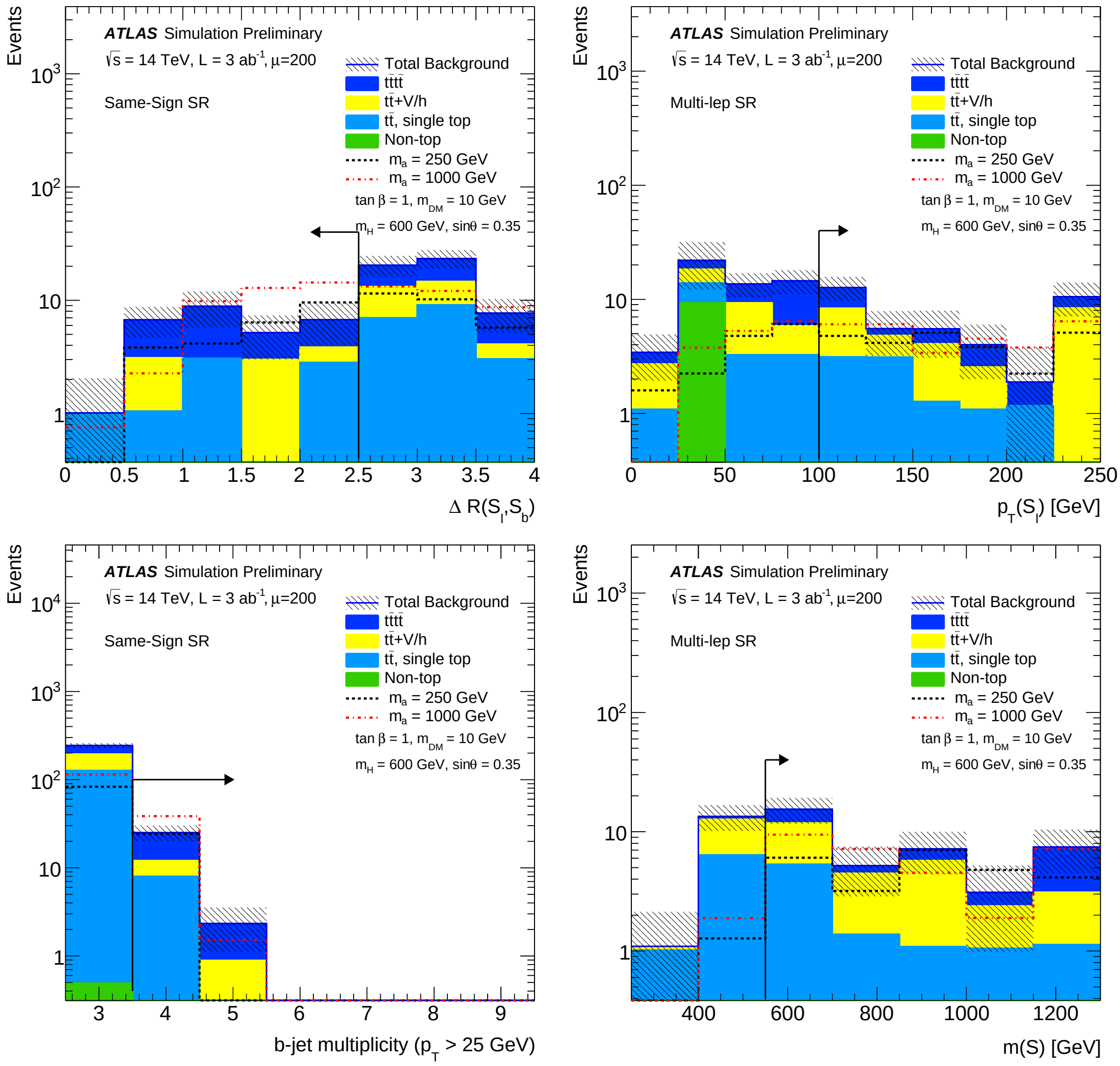

Figure 4: Key distributions for events passing all Same-Sign (left) or Multi-lep (right) selection requirements except that on the distribution itself. The contributions from all SM backgrounds are shown, and the hashed band represents the statistical uncertainty on the total SM background prediction. The expected distributions for signal models with $m_a$ = 150 GeV, $m_H$ = 1000 GeV, $\sin\theta$ = 0.7 and $m_a$ = 550 GeV, $m_H$ = 600 GeV, $\sin\theta$ = 0.35 are also shown as dashed lines for comparison.

# 5 Systematic uncertainties

The projection of systematic uncertainties determined starting from those studied in Ref [56] have been used. A sensible extrapolation has been developed by ATLAS and CMS Collaborations which is documented in Ref. [30]. The theory modelling uncertainties are expected to halve, while different recommendations have been provided for detector-level and experimental uncertainties.

In Ref. [56], the dominant uncertainties were due to theoretical modelling of the irreducible backgrounds (32%), the jet energy scale (JES - 6%) and resolution (JER - 6%), the b-tagging efficiency (7%) and the modelling of the fake and non-prompt lepton background. Owing to the reduced statistical uncertainty and a better understanding of the physics models, it is expected that JES, JER, b-tagging efficiency and irreducible background modelling uncertainties will all be reduced. It is assumed that they will all be halved by the end of the HL-LHC running. This leads to an estimate of the total background uncertainty of about 20%.

The resulting experimental uncertainty is assumed to be fully correlated between the background and the signal when setting 95% CL exclusion limits. Furthermore, an additional systematic of 5% is considered for the signal, in order to account for the theoretical systematic uncertainty on the model. For the expected discovery $p$-values values, only the uncertainty on the background is considered.

# 6 Results

Table 3 shows the expected yields in the SR for each background source, together with few benchmark signal models. The dominant contribution to the $t\bar{t}$ +V/h processes is $t\bar{t}$ +h for the Same-Sign SR and $t\bar{t}$ +$Z$ for the Multi-lep SR.

Table 3: Expected yields in the SRs together with their statistical uncertainties, for an integrated luminosity of 3000 fb$^{-1}$. The expected numbers of events for few signal samples are also reported.

| | Same-Sign SR | Multi-lep SR |
|---|---|---|
| Expected Standard Model | 27.4 ± 5.2 | 38.4 ± 6.4 |
| $t\bar{t}$ and single-top | 7.9 ± 1.5 | 9.1 ± 1.5 |
| $t\bar{t}$ +V/h | 4.94 ± 0.94 | 17.5 ± 2.9 |
| $t\bar{t}\,t\bar{t}$ | 14.5 ± 2.8 | 10.9 ± 1.8 |
| Others | 0.110 ± 0.020 | 0.86 ± 0.14 |
| ($m_a$ GeV, $m_H$ GeV, $\sin\theta$) | | |
| (200, 600, 0.05) | 31.7 ± 4.8 | 38.9 ± 5.8 |
| (200, 600, 0.50) | 17.4 ± 2.6 | 17.4 ± 2.6 |
| (200, 600, 0.95) | 6.06 ± 0.91 | 7.0 ± 1.1 |
| (250, 1000, 0.70) | 1.33 ± 0.20 | 3.32 ± 0.50 |
| (250, 600, 0.35) | 24.2 ± 3.6 | 25.2 ± 3.8 |
| (350, 1000, 0.05) | 9.5 ± 1.4 | 4.95 ± 0.74 |
| (350, 1000, 0.50) | 5.03 ± 0.75 | 5.63 ± 0.84 |
| (350, 1000, 0.95) | 32.9 ± 4.9 | 36.0 ± 5.4 |
| (550, 1000, 0.70) | 10.7 ± 1.6 | 11.4 ± 1.7 |
| (550, 600, 0.35) | 30.5 ± 4.6 | 34.9 ± 5.2 |

The HistFitter framework [57], which utilises a profile-likelihood-ratio test statistics [58], is used to compute expected discovery $p$-values. In case there is no excess seen in this channel, expected exclusion limits with the $CL_s$ prescription [59] are also calculated.

Scans of expected discovery $p$-values and expected exclusion limits at 95% CL are shown in Figures 5 and 6, respectively, as a function of $m_a$, for fixed $m_H$ and $\sin\theta$ and as a function of $\sin\theta$ for fixed $m_a$ and $m_H$. In all benchmarks, it is assumed that $\tan\beta = 1$ and $m_\chi = 10$ GeV.

For light pseudo-scalar masses above the $t\bar{t}$ decay threshold, a significance of about $3\sigma$ is expected if $m_H = 600$ GeV and $\sin\theta = 0.35$. The same benchmark is expected to be excluded for all light-pseudoscalar masses and for $\sin\theta < 0.35$ if $m_a = 200$ GeV. Mixing angles such that $\sin\theta > 0.95$ are also expected to be excluded for $m_a = 350$ GeV, $m_H = 1$ TeV and, under the same assumptions, an upper limit of about two times the theoretical cross section is set for $\sin\theta < 0.8$. Finally, $\sin\theta < 0.4$ is excluded for $m_H = 600$ GeV, $m_a = 200$ GeV.

In almost all cases the Same-Sign SR yields the strongest constraints on the parameter space considered in this work. However, the Multi-lep SR offers a complementary channel whose sensitivity is of the same order of magnitude. Possibly, exploiting dedicated techniques developed to suppress or better estimate the $t\bar{t} + V$ background that affects the Multi-lep SR, this signature can achieve sensitivity comparable to the Same-Sign selection.

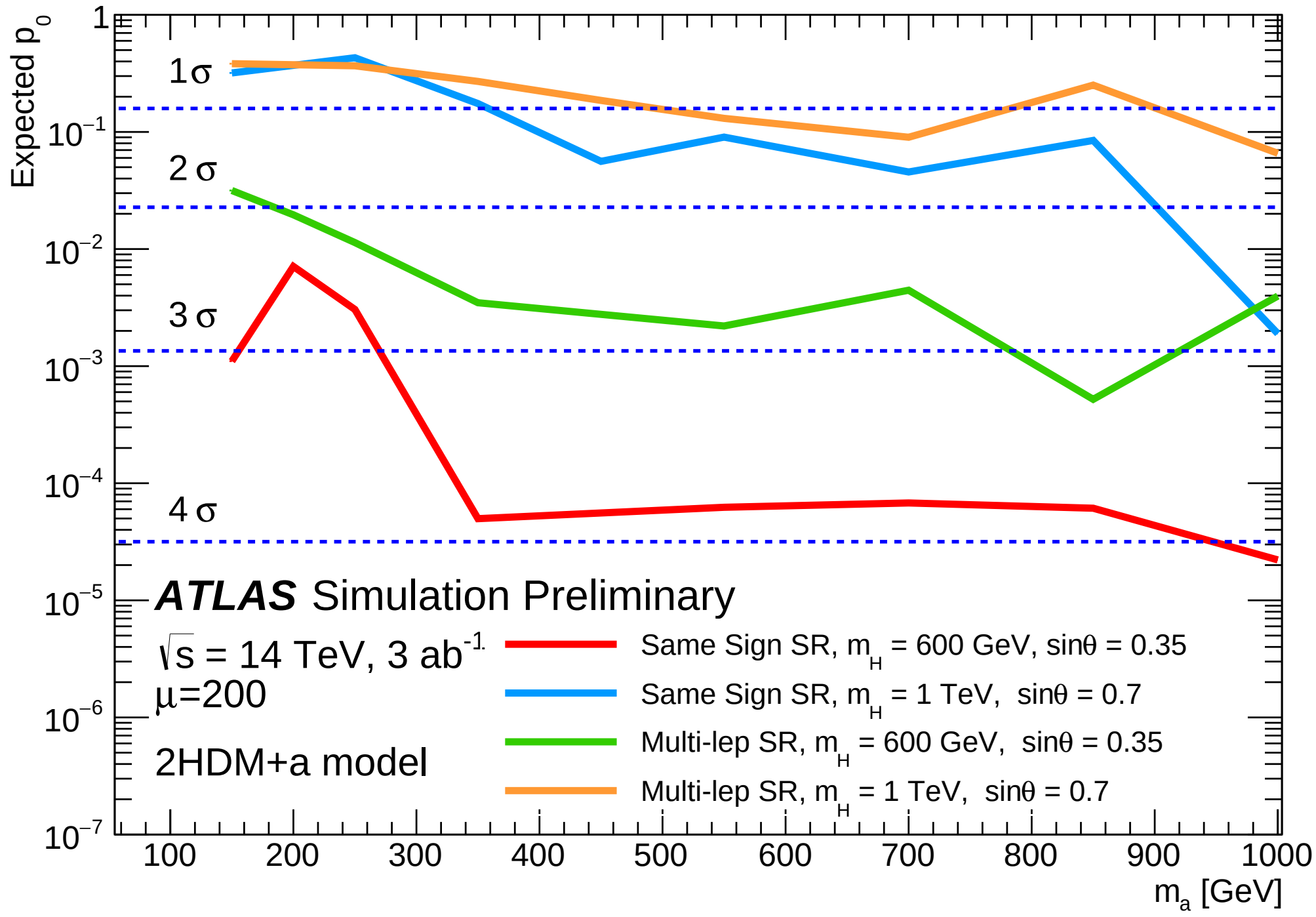

(a)

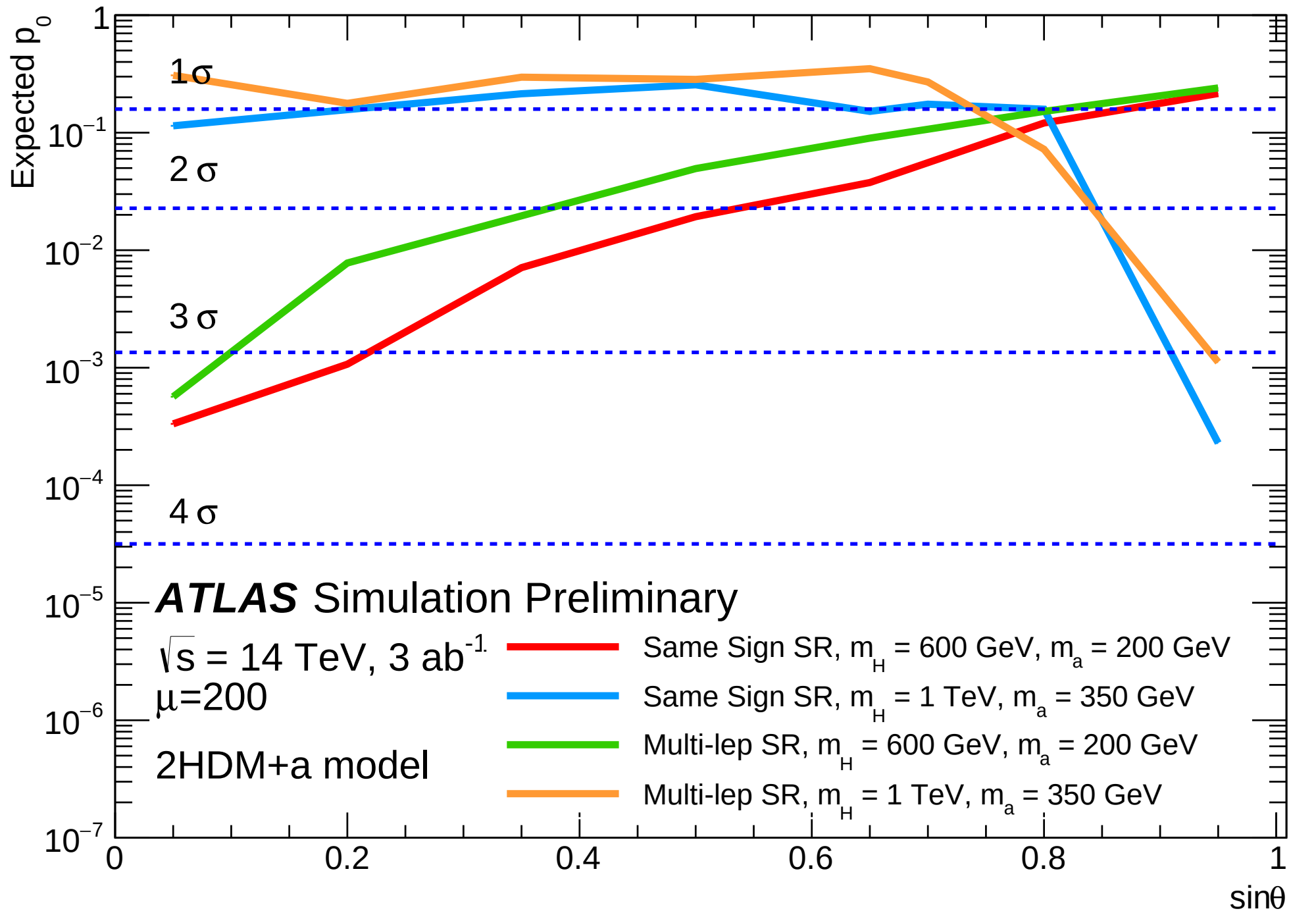

(b)

Figure 5: Discovery $p$-values for Same-Sign and Multi-lep SRs derived from the analysis of 3000 fb$^{-1}$ of 14 TeV proton-proton collision data as a function of $m_a$ (a) or as a function of $\sin\theta$ (b) for each parameter assumptions described in Sec. 1 and indicated in the legend.

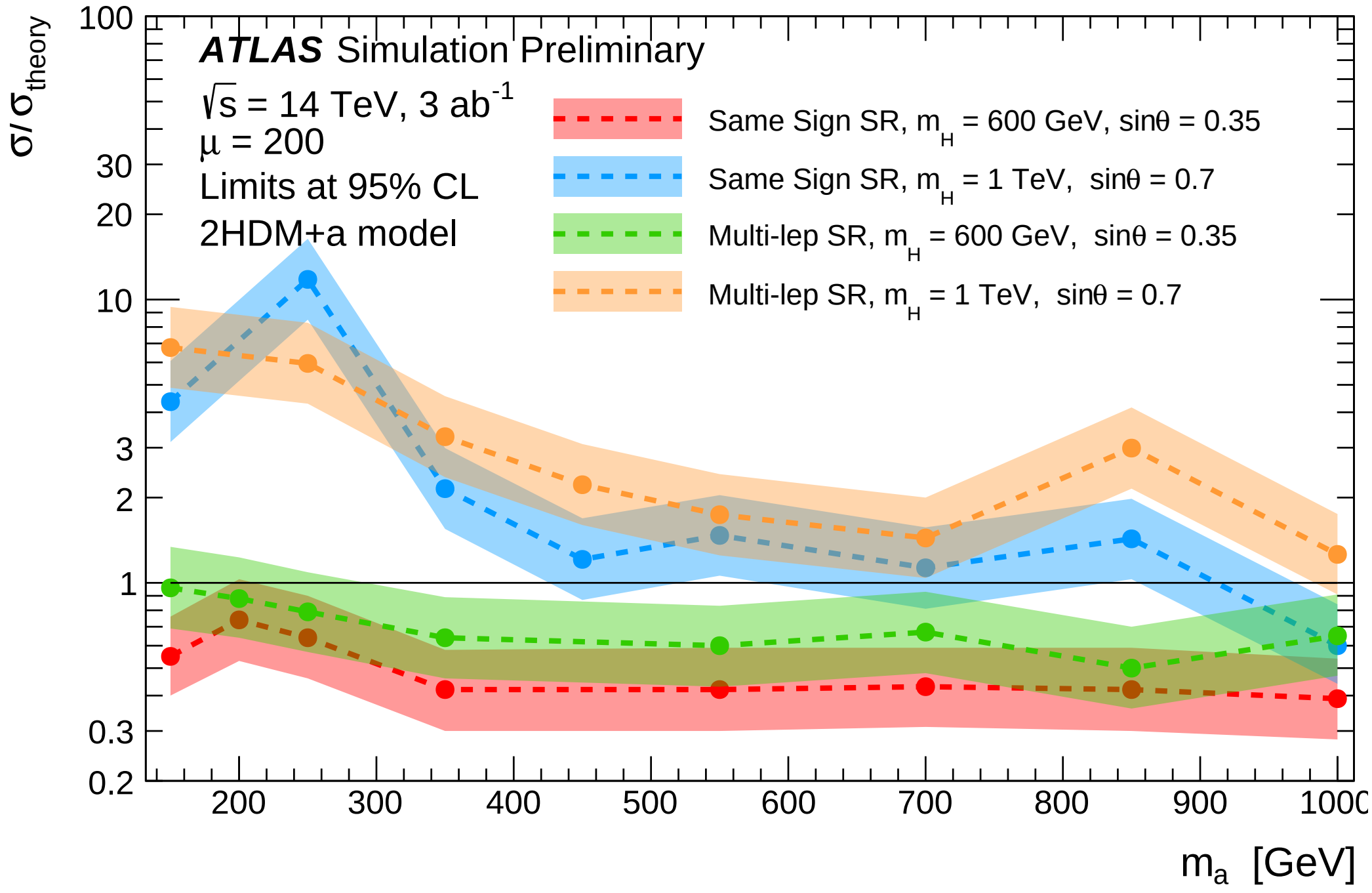

(a)

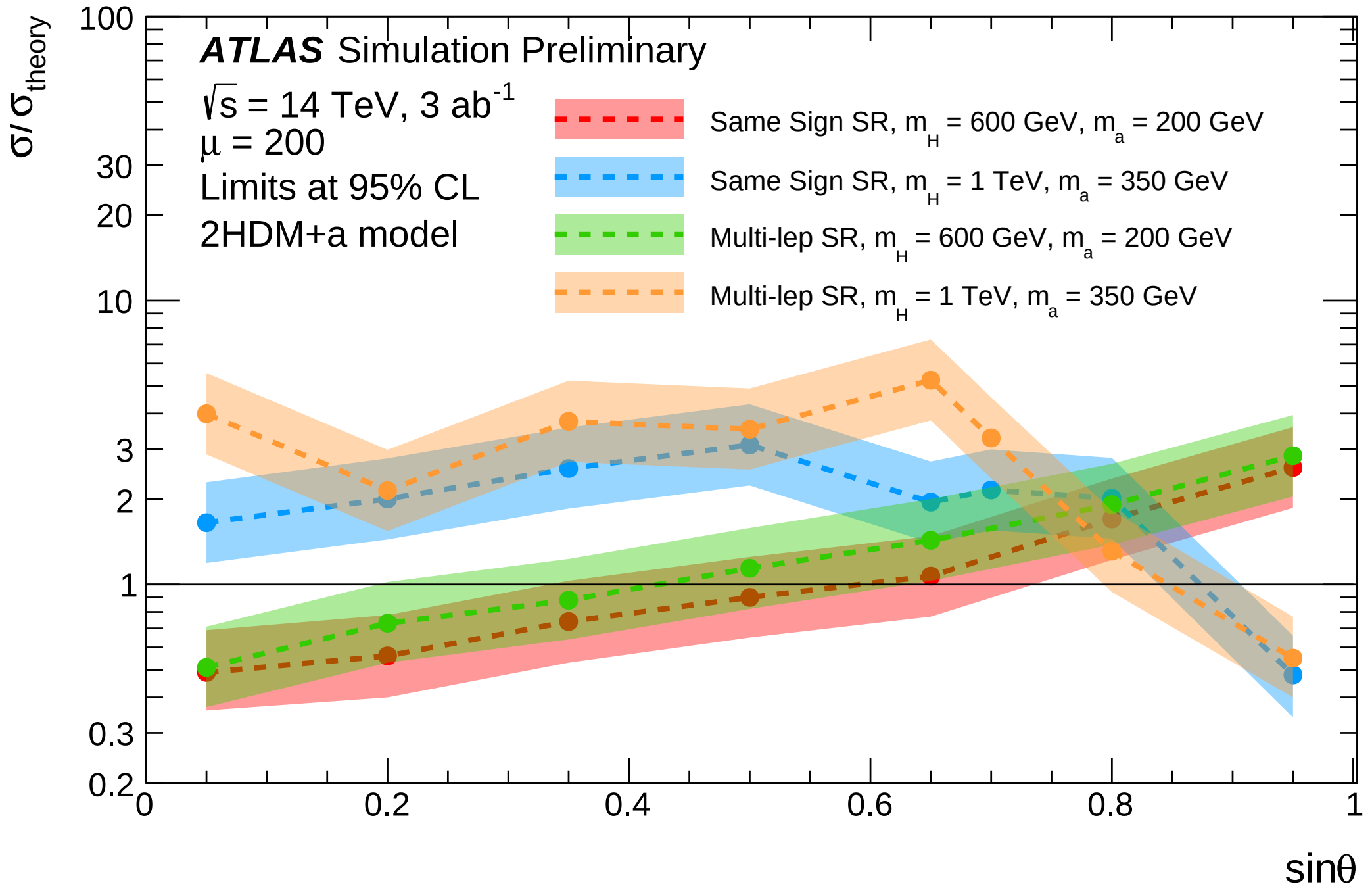

(b)

Figure 6: Exclusion limits at 95% CL for Same-Sign and Multi-lep SRs in terms of excluded cross-section ($\sigma$) over the cross-section predicted by the model ($\sigma_{theory}$). Limits are derived from the analysis of 3000 fb$^{-1}$ of 14 TeV proton-proton collision data as a function of $m_a$ (a) or as a function of $\sin\theta$ (b) for each parameter assumptions described in Sec. 1 and indicated in the legend. The $1\sigma$ variation of the total uncertainty on the limit is indicated as a band around each exclusion line.

# 7 Conclusion

The sensitivity to 2HDM+$a$ models is studied using simulated ATLAS data from proton-proton collisions at the LHC design centre-of-mass-energy of $\sqrt{s}$ = 14 TeV. Final states with with exactly two charged leptons with the same electric charge or three or more charged leptons have been considered. Emphasis has been put on four parameter scans: two scans of $m_a$, for $m_H$ = 1000 GeV, $\tan\beta = 1$, $\sin\theta = 0.7$ and for $m_H$ = 600 GeV, $\tan\beta = 1$, $\sin\theta = 0.35$ and two scans of $\sin\theta$, for $m_a$ = 350 GeV, $m_H$ = 1000 GeV, $\tan\beta = 1$ and for $m_a$ = 200 GeV, $m_H$ = 600 GeV, $\tan\beta = 1$. Searches are carried out with MC simulated events generated at $\sqrt{s}$ = 14 TeV, and with corrections accounting for the detector response applied to the generator-level particles. A dataset of 3000 fb$^{-1}$ extends the potential for evidence of a signal with $m_H$ = 600 GeV and mixing angles of $\sin\theta = 0.35$ assuming $m_a$ masses between 400 GeV and 1 TeV and the exclusion sensitivity for all $m_a$ between 200 GeV and 1 TeV. The addition of fully hadronic, semi-leptonic and di-leptonic final states with different electric charge channels, which have not been considered in this study, are likely to further extend this reach. Future improvements in the understanding of experimental and theoretical systematic uncertainties on the SM backgrounds would also provide additional gains in sensitivity.

# CMS Physics Analysis Summary

Contact: cms-phys-conveners-ftr@cern.ch

# HL-LHC searches for new physics in hadronic final states with boosted W bosons or top quarks using razor variables

The CMS Collaboration

## Abstract

We present High-Luminosity LHC (HL-LHC) projections of the Run 2 search for new physics in hadronic final states with boosted W bosons or top quarks using razor variables. Data event yields and signal/background cross sections from the 2016 analysis are scaled to obtain the HL-LHC sensitivity for center-of-mass energy of 14 TeV and integrated luminosity of 3 $ab^{-1}$. Different scenarios for systematic uncertainties are considered. The projection results are interpreted in terms of gluino pair production where each gluino decays to a top quark, an anti-top quark, and a neutralino; or to a top quark and a top squark; and direct top squark pair production where each top squark decays to a top quark and a neutralino.

# 1 Introduction

This note presents the projection of the CMS search for new physics with boosted W bosons or top quarks using the razor kinematic variables to the High-Luminosity LHC (HL-LHC) conditions of center-of-mass energy of 14 TeV and integrated luminosity of 3 $\mathrm{ab}^{-1}$. The projected search performed on the Run 2 2016 dataset is part of a larger inclusive new physics search with razor variables that includes an extensive set of hadronic and leptonic search regions, documented in [1].

The analysis targets final states consistent with supersymmetry (SUSY), and in particular, with a realization of it called natural SUSY [2, 3]. This specific scenario requires the existence of a light top squark, $\widetilde{\mathrm{t}}_1$, and a somewhat light gluino, $\widetilde{\mathrm{g}}$, which stabilize the Higgs field mass-squared term without excessive fine tuning. Observing light gluinos and top squarks at the LHC would provide a test for naturalness. The possibility that the top squark could be light has motivated several searches by the CMS and ATLAS collaborations for the direct production of top squarks. However, these searches tend to lose sensitivity in a few particular scenarios. One such scenario, called the compressed scenario, occurs when the mass of the $\widetilde{\mathrm{t}}_1$ approaches that of the lightest SUSY particle (LSP), assumed to be the lightest neutralino, $\widetilde{\chi}^0_1$. A second scenario, called the diagonal scenario, occurs when the mass difference between the top squark and the LSP is around the top quark mass, $\Delta m = m_{\widetilde{\mathrm{t}}_1} - m_{\widetilde{\chi}^0_1} \approx m_{\mathrm{t}}$. The diagonal scenario reduces the sensitivity of searches looking specifically for $\widetilde{\mathrm{t}}_1 \to \mathrm{t}\widetilde{\chi}^0_1$.

In the compressed scenario, the $\widetilde{\mathrm{t}}_1$ decays either through a 4-body decay to $\mathrm{b}\mathrm{f}\bar{\mathrm{f}}\widetilde{\chi}^0_1$, where f is any fermion, or through the loop-induced decay to $\mathrm{c}\widetilde{\chi}^0_1$. In both scenarios, the decay products of the top squark generally have very low transverse momentum ($p_{\mathrm{T}}$) and therefore are very hard to detect. In order to be sensitive to such cases, it is necessary to rely on another property of the events, often the presence of initial state radiation (ISR) jets. Instead, the search can target top squarks produced in a slightly longer decay chain. One possible assumption is that the heavy top squark $\widetilde{\mathrm{t}}_2$ is also accessible and decays to the $\widetilde{\mathrm{t}}_1$ via a Higgs or Z boson. Alternatively, one may postulate the existence of a gluino and search for top squarks from gluino decays.

This analysis targets gluino production, where the gluino decays to a top squark and a top quark. The Run 2 analysis excluded scenarios with a gluino mass around 2 TeV and a top squark mass of several hundred GeV; these limits are expected to increase significantly for the HL-LHC. Due to the significant mass gap between the gluino and the top squark, the top quark from the gluino decay receives a large boost. The top squark then decays, as in one of the scenarios explained above, to $\mathrm{c}\widetilde{\chi}^0_1$ for small $\Delta m$. The simplified model [4, 5] corresponding to this topology is called T5ttcc. In addition to these models, we also consider gluinos directly decaying to $\mathrm{t}\bar{\mathrm{t}}\widetilde{\chi}^0_1$, called T1tttt, and direct production of top squark pairs, where each top squark decays to a top quark and a neutralino, called T2tt. All of these models are illustrated by the diagrams in Fig. 1.

Boosted objects, which have high $p_{\mathrm{T}}$, are characterized by merged decay products separated by $\Delta R \sim 2m/p_{\mathrm{T}}$, where $m$ denotes the mass of the decaying massive particle. A top quark or W boson can be identified via boosted objects within a jet of size 0.8 if it has a momentum of $\gtrsim$430 GeV or $\gtrsim$200 GeV, respectively. As boosted objects become more accessible at the increased center-of-mass energies, they will be produced more frequently at the HL-LHC. Therefore this analysis is an interesting addition to the HL-LHC studies.

Figure 2 shows the generator-level $p_{\mathrm{T}}$ distributions for W bosons and top quarks from the gluino decay for several mass points of the T5ttcc simplified models, compared to the W boson and top quark $p_{\mathrm{T}}$ distributions from the standard model (SM) $\mathrm{t}\bar{\mathrm{t}}$+jets process. An initial

**2**

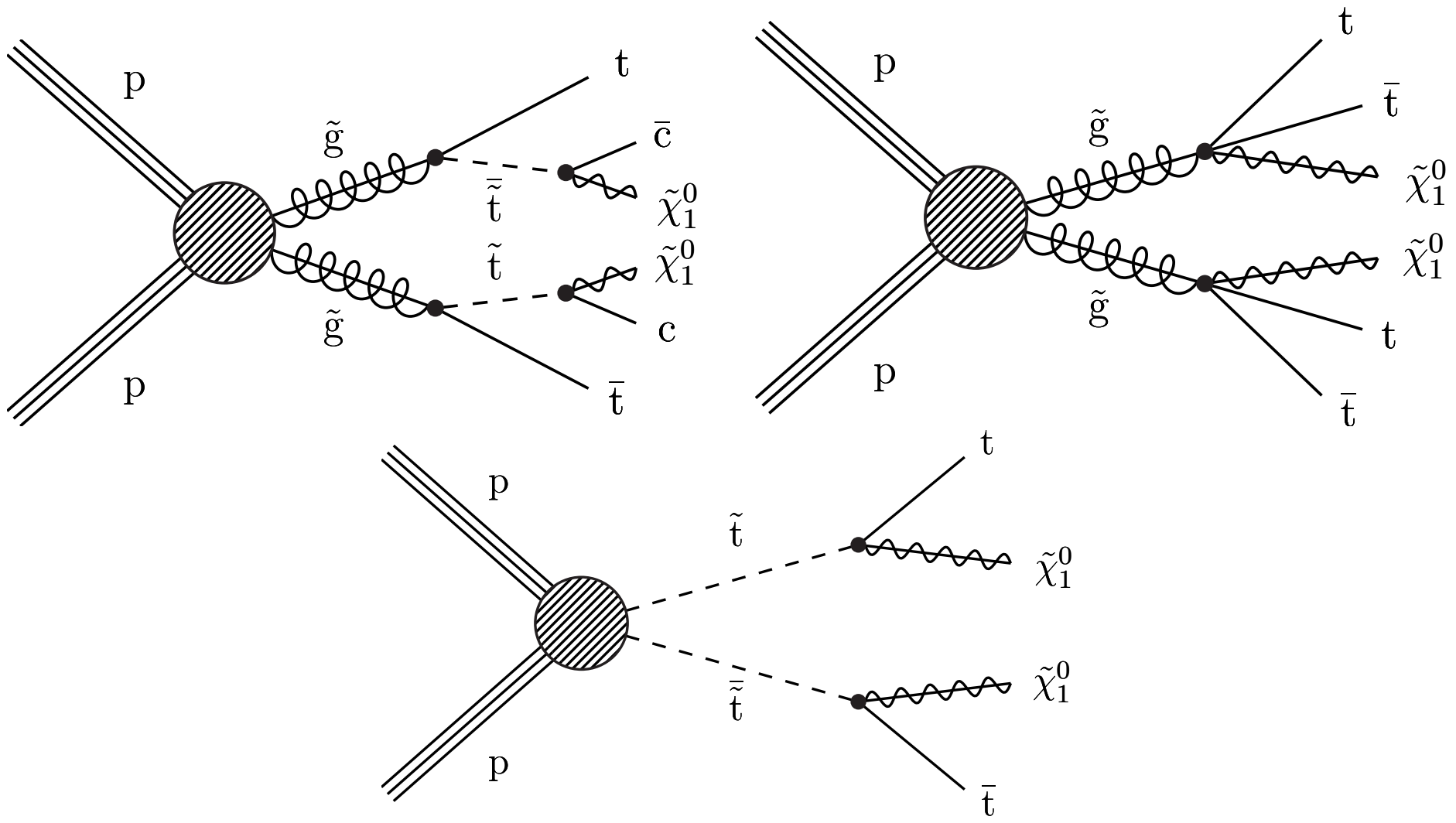

Figure 1: Signal models considered in this analysis: T5ttcc (top left), T1tttt (top right), and T2tt (bottom).

selection of a jet having size 0.8 with $p_{\mathrm{T}} > 200\,\mathrm{GeV}$ and razor variable $R^2 > 0.04$ has been applied to events shown in Fig. 2. From these distributions we can see that the W bosons and top quarks in the signal models have significantly higher momenta compared to the SM $\mathrm{t\bar{t}}$+jets process. This shows that the boosted top quarks and W bosons are a promising signature in new physics searches.

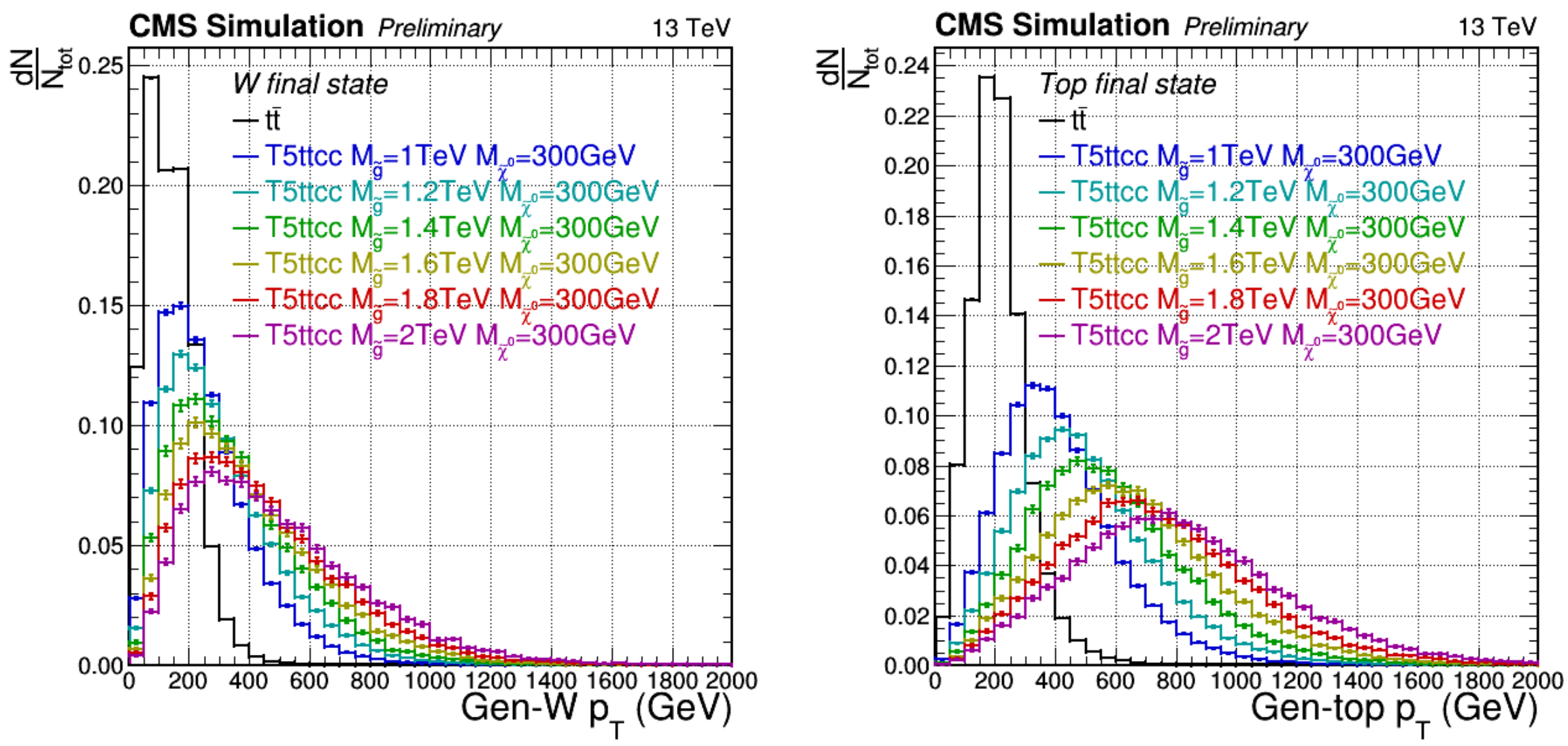

Figure 2: Generator-level W boson and top quark $p_{\mathrm{T}}$ distributions for several signal points from the T5ttcc simplified model, compared to the $\mathrm{t\bar{t}}$+jets background. Only a set of events selected with a requirement of a jet with size 0.8, $p_{\mathrm{T}} > 200\,\mathrm{GeV}$, and razor variable $R^2 > 0.04$, as explained in the text, are shown.

The analysis is performed in hadronic topologies with boosted top quarks, or boosted W bosons and b jets, using the razor kinematic variables (to be defined in Section 2), which are powerful tools that help to discriminate between SM processes and production of heavy new particles decaying to final states with massive invisible particles and massless visible particles. The analysis is performed in three signal search regions defined by selections on the razor variables. Boosted top quarks and W bosons are identified by finding massive jets that possess substructure, which can be identified with the n-subjettiness technique [6].

In this note, we will first introduce the razor variables in Section 2, followed by the HL-LHC and the upgraded CMS detector in Section 3. We will then explain the analysis methodology in Section 4, followed by the details of the projection of MC and data events in Section 5 and treatment of uncertainties in Section 6. Finally, we will present our results and their interpretation in Section 7, followed by the summary in Section 8.

## 2 The razor variables

The razor variables $M_R$ and $R^2$ map the event into a dijet topology [7]. They help to describe a signal coming from pair production of two heavy particles, each decaying to a massless visible particle and a massive invisible particle, as a peak over exponentially falling SM backgrounds. For this reason, the razor variables are robust discriminators for SUSY signals with pair-produced sparticles that subsequently decay to lighter SM particles and the invisible LSPs. For the simple case where the final topology has two visible particles, e.g., jets $j_1$ and $j_2$, the razor variables are defined using the 4-momenta of these two jets $(E^{j_i}, \vec{p}_{\mathrm{T}}^{\,j_i}, p_z^{j_i})$, where $i = 1, 2$, and the missing transverse momentum $\vec{p}_{\mathrm{T}}^{\mathrm{miss}}$, with magnitude $p_{\mathrm{T}}^{\mathrm{miss}}$, as

$$M_R \equiv \sqrt{\left(E^{j_1} + E^{j_2}\right)^2 - \left(p_z^{j_1} + p_z^{j_2}\right)^2} \tag{1}$$

$$M_{\mathrm{T}}^R \equiv \sqrt{\frac{p_{\mathrm{T}}^{\mathrm{miss}}\left(p_{\mathrm{T}}^{j_1} + p_{\mathrm{T}}^{j_2}\right) - \vec{p}_{\mathrm{T}}^{\mathrm{miss}} \cdot \left(\vec{p}_{\mathrm{T}}^{\,j_1} + \vec{p}_{\mathrm{T}}^{\,j_2}\right)}{2}}. \tag{2}$$

Given $M_R$ and $M_{\mathrm{T}}^R$, the razor dimensionless ratio is defined as

$$R \equiv \frac{M_{\mathrm{T}}^R}{M_R}. \tag{3}$$

However, if the decay chains are more complicated and there are multiple particles in the final state, we first form two “megajets” from the final state particles, such that each of the megajets contain the particles coming from one of the heavy pair-produced particles. $M_R$ and $R^2$ are then computed using the 4-momenta of these two megajets, where the megajet 4-momenta are computed as the vectorial sums of the 4-momenta of the jets contributing to each megajet. Of all the possible partitions of the jets into two megajets, we select the combination that minimizes the sum of the invariant masses of the two megajets. This choice will cluster together particles that are traveling in the same direction, and it has been found to perform well.

## 3 Upgraded CMS detector

The CMS detector [8] will be substantially upgraded in order to fully exploit the physics potential offered by the increase in luminosity, and to cope with the demanding operational conditions at the HL-LHC [9–13]. The upgrade of the first level hardware trigger (L1) will allow for an increase of the L1 rate and latency to about 750 kHz and 12.5 $\mu$s, respectively, and the high-level software trigger is expected to reduce the rate by about a factor of 100 to 7.5 kHz. The entire pixel and strip tracker detectors will be replaced to increase the granularity, reduce the material budget in the tracking volume, improve the radiation hardness, and extend the geometrical coverage and provide efficient tracking up to pseudorapidities of about $|\eta| = 4$. The muon system will be enhanced by upgrading the electronics of the existing cathode strip chambers, resistive plate chambers (RPC) and drift tubes . New muon detectors based on improved RPC and gas electron multiplier technologies will be installed to add redundancy, increase the geometrical coverage up to about $|\eta| = 2.8$, and improve the trigger and reconstruction performance in the forward region. The barrel electromagnetic calorimeter will feature the upgraded front-end electronics that will be able to exploit the information from single crystals at the L1 trigger level, to accommodate trigger latency and bandwidth requirements, and to provide 160 MHz sampling allowing high precision timing capability for photons. The hadronic calorimeter, consisting in the barrel region of brass absorber plates and plastic scintillator layers, will be read out by silicon photomultipliers. The endcap electromagnetic and hadron calorimeters will be replaced with a new combined sampling calorimeter that will provide highly-segmented spatial information in both the transverse and longitudinal directions, as well as high-precision timing information. Finally, the addition of a new timing detector for minimum ionizing particles in both the barrel and endcap regions is envisaged to provide the capability for 4-dimensional reconstruction of interaction vertices that will significantly offset the CMS performance degradation due to high pileup (PU) rates.

A detailed overview of the CMS detector upgrade program is presented in Refs. [9–13], while the expected performance of the reconstruction algorithms and pile-up mitigation with the CMS detector is summarised in Ref. [14].

# 4 Analysis methodology

The analysis is designed to look for an excess in events with high values of $M_R$ and $R^2$ in fully hadronic final states with at least one boosted W boson and a b jet, or one boosted top jet.

The 2016 analysis was performed using 35.9 $\text{fb}^{-1}$ of 13 TeV proton-proton collision data collected in 2016 [1]. The projection study presented here uses the same data and Monte Carlo (MC) events as in the 2016 analysis. It also follows exactly the same object selection, event selection, background estimation, systematic uncertainty calculation, and limit setting procedures as used in the 2016 analysis. As this is a projection study, event kinematics for individual processes are unchanged. The main differences introduced in the projection study are the scaling of event yields to higher cross sections and luminosities, which will be explained in Section 5, and the scaling of systematic uncertainties to the HL-LHC conditions [14], which will be detailed in Section 6. In the remainder of this section, we outline event selection and background estimation procedures which are directly adapted from the 2016 analysis by the HL-LHC projection study.

The 2016 analysis used data collected by triggers selecting events based on the $p_T$ of the leading jet and the scalar sum of the transverse momenta of all jets, $H_T$. These jets are reconstructed using the anti-$k_T$ algorithm [15, 16] with distance parameters of $R = 0.8$ (AK8) and $R = 0.4$ (AK4) for the $p_T$-based and $H_T$-based triggers, respectively. As these triggers were only ≈70% efficient for the $M_R$-$R^2$ selection, efficiencies were modeled as a function of jet $p_T$ and $H_T$ using orthogonal datasets. This trigger efficiency modeling is also applied in the projection, since data distributions are used in the control regions for background estimation. Detailed description of the objects used in the analysis are given in Ref. [1]. Boosted W bosons and top quarks are identified using the jet mass, the n-subjettiness variables $\tau_{2/1}$ and $\tau_{3/2}$ [6], and subjet b tagging.

Events in all signal, control, and validation regions in the analysis are required to have

- at least one good primary vertex

- at least four selected AK4 jets
- at least one AK8 jet with $p_T > 200$ GeV defining the boosted phase space; and
- $M_R > 800$ GeV and $R^2 > 0.08$, where the megajets are constructed from the selected AK4 jets. This selection, based on the kinematic properties of the target signals, provides an optimal balance between background suppression and signal acceptance.

The signal regions are required to have in addition:

- No leptons fulfilling the veto identification criteria
- Azimuthal distance between the two megajets, $\Delta\phi_{\text{megajets}} < 2.8$
- 3 categories based on boosted object and jet multiplicities are defined:
  - W boson categories: $\geq$1 AK4 b jet (identified with the medium tagger of the combined secondary vertex algorithm [17]) and $\geq$1 reconstructed AK8 W jet. Two bins of AK4 jet multiplicity:
    - W 4-5 jet: $4 \leq n_{\text{jet}} \leq 5$
    - W 6 jet: $n_{\text{jet}} \geq 6$
  - Top quark category (Top): $\geq$1 reconstructed AK8 top jet

The dominant SM backgrounds remaining in the signal regions originate from $t\bar{t}$+jets, single top quark production, quantum chromodynamics (QCD) multijet events that have jets produced through the strong interaction, and the $W(\ell\nu)$+jets and $Z(\nu\bar{\nu})$+jets processes. Because there are large uncertainties in the simulation modeling for these processes, data-driven methods are employed to estimate their contributions to the signal regions. The estimation method outlined below is directly taken from the 2016 analysis, and its complete details can be found in [1]. The procedure involves control regions that isolate a particular process to be estimated, or a process that can approximately mimic it. These control regions are generally defined by reversing or otherwise modifying one or more signal selection criteria, and are designed to be as similar as possible in kinematic properties to the signal regions, in order to reduce shape uncertainties. The projection study uses the control region definitions from the 2016 analysis as listed below:

- A multijet control region for the QCD multijet estimation obtained by inverting the $\Delta\phi_{\text{megajets}}$ selection, and by reversing the n-subjettiness criterion in the W and top tagging algorithms.
- A $t\bar{t}$+jets and single top control region for the $t\bar{t}$+jets and single top estimation, which requires exactly 1 lepton ($\ell = e$ or $\mu$), and transverse mass $m_T = \sqrt{2p_T^{\ell}p_T^{\text{miss}}(1-\cos\Delta\phi(\vec{p}_T^{\,\ell},\vec{p}_T^{\text{miss}}))} < 100$ GeV.
- A W+jets control region for the W+jets estimation, which requires exactly 1 lepton, 0 b jets, reversed subjet b tagging in the top tagging algorithm, and $30 < m_T < 100$ GeV.
- A $\gamma$+jets control region for the $Z(\nu\bar{\nu})$+jets estimation, with exactly 1 photon whose $\vec{p}_T$ is added to the $\vec{p}_T^{\text{miss}}$, 1 W- or top-tagged jet with only the jet mass requirement applied, and no requirement on b jets.
- A $Z(\ell^+\ell^-)$+jets control region with 2 same-flavor leptons (ee or $\mu\mu$) whose $\vec{p}_T$ are added to the $\vec{p}_T^{\text{miss}}$, 1 W- or top-tagged jet with only the jet mass requirement applied, and no requirement on b jets. This control region is used for correcting the primary $Z(\nu\bar{\nu})$+jets estimate, which uses the $\gamma$+jets control region defined above.
- A $W(\ell\nu)$+jets control region with exactly 1 lepton (e or $\mu$) whose $\vec{p}_T$ is added to the $\vec{p}_T^{\text{miss}}$, $30 < m_T < 100$ GeV, 1 W- or top-tagged jet with only the jet mass requirement applied, and no requirement on b jets. This control region is used to cross-check the $Z(\nu\overline{\nu})$+jets estimate and to derive a systematic uncertainty based on the difference with respect to the primary estimate from the $\gamma$+jets control region.

Data and simulation event yields in these regions are scaled to the HL-LHC cross sections, as described in Section 5. After scaling all distributions, background estimates in the signal regions are obtained by multiplying the observed data yields, binned in $M_R$ and $R^2$, by the simulation transfer factors computed as the ratios of the yields of background MC simulation events in the signal regions to the yields in control regions. Other SM processes that contribute less significantly, such as VV, VVV, and $\mathrm{t\bar{t}V}$, are estimated directly from the simulation, scaled to the HL-LHC cross sections and luminosities. The simulated events used for obtaining both the transfer factors and the direct estimates are corrected using various data-to-simulation correction factors and event weights. The uncertainties in these correction factors and weights are taken into account as systematic uncertainties (see Section 6). The validity of this background estimation procedure was established in the 2016 analysis by closure tests in two validation regions that resemble the topology and kinematic properties of the signal regions, but are background-dominated. These closure tests applied the full background estimation procedure to estimate the backgrounds in the validation regions and compared the estimated background yields to data counts, confirming their agreement.

Finally, these background estimates are used together with signal distributions obtained from MC simulation scaled to HL-LHC cross sections and luminosities to set exclusion limits on production cross sections and upper limits on gluino and top squark masses. Furthermore, projections of $5\sigma$ discovery sensitivity were computed.

# 5 Projection of MC and data event counts

The HL-LHC projections are performed on 2016 simulated and observed events. Simulated events are reweighted to better model data with trigger efficiency corrections, jet energy and resolution smearing, pileup corrections, W and top jet scale factors, b jet tagging scale factors, electron and muon identification and isolation scale factors, and various other corrections specific to the signal event generation and simulation modeling. After these corrections, the object and event selections that define the signal and control regions are applied to these events, as described in Section 4.

To do a projection to the HL-LHC conditions, we first take the $M_R$-$R^2$ distributions of the simulated events for each physics process $i$ and each selection region $j$, and scale the number of events as

$$N^{i,j}_{\text{HL-LHC}} = \left( \frac{\sigma^i_{\text{HL-LHC}}}{\sigma^i_{2016}} \frac{\mathcal{L}_{\text{HL-LHC}}}{\mathcal{L}_{2016}} \right) N^{i,j}_{2016} \tag{4}$$

$$= \kappa^i_{\frac{\text{HL-LHC}}{2016}} N^{i,j}_{2016}, \tag{5}$$

where $N^{i,j}_{2016}$ and $N^{i,j}_{\text{HL-LHC}}$ are the total number of events for a simulated process $i$ in search region $j$ for 2016 and HL-LHC; $\sigma^i_{2016}$ and $\sigma^i_{\text{HL-LHC}}$ are the cross sections for process $i$ for the 2016 and HL-LHC energies of 13 and 14 TeV; and $\mathcal{L}_{2016}$ and $\mathcal{L}_{\text{HL-LHC}}$ are the 2016 and HL-LHC integrated luminosities of 35.9 $\text{fb}^{-1}$ and 3 $\text{ab}^{-1}$.

The scaling applies to both the control and signal regions. Figure 3 shows the $\mathrm{pp} \to \widetilde{\mathrm{g}}\widetilde{\mathrm{g}}$ and $\mathrm{pp} \to \widetilde{\mathrm{t}}\widetilde{\mathrm{t}}$ production cross sections at the next-to-leading-order + next-to-leading-log (NLO+NLL) level versus the gluino or top squark masses, computed using PROSPINO and NLL-fast [18–22].

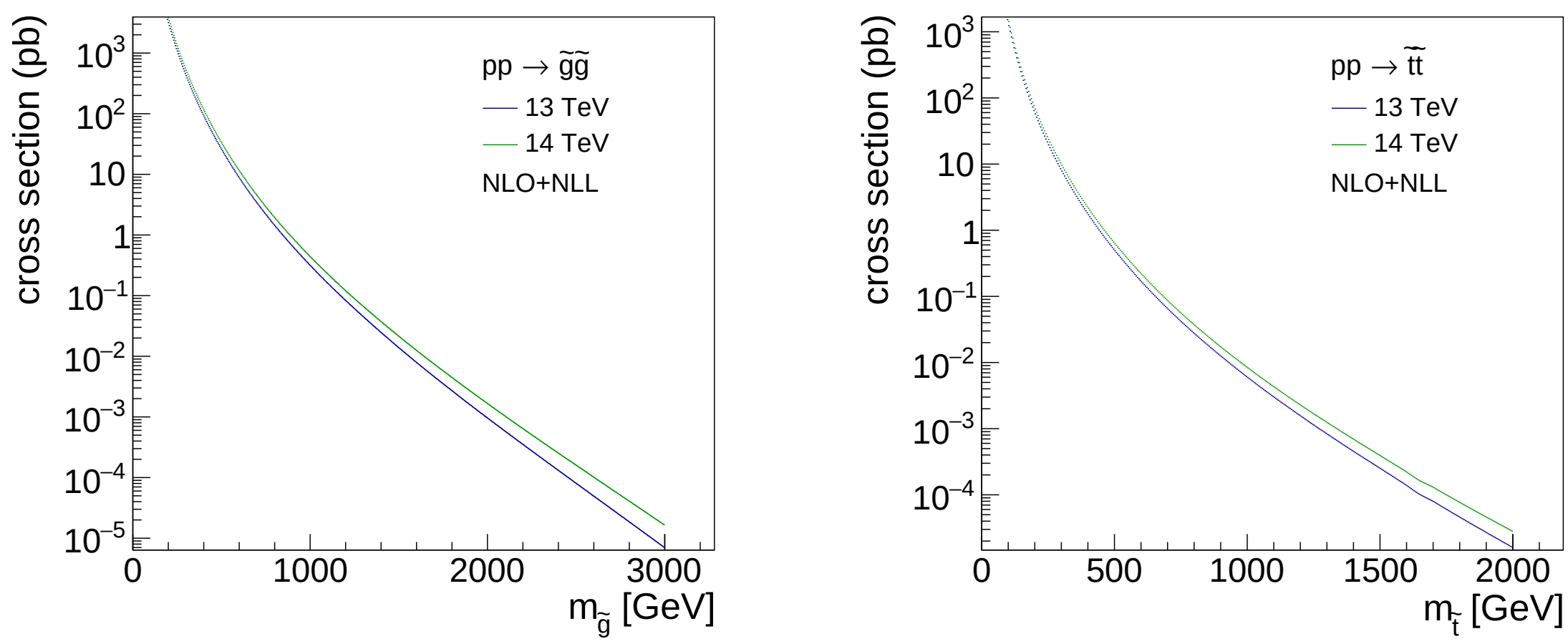

Figure 3: The pp → $\tilde{g}\tilde{g}$ (left) and pp → $\tilde{t}\tilde{t}$ (right) production cross sections at NLO+NLL precision versus the gluino and top squark masses, respectively.

Since some background estimates are based on the event yields measured in control regions in data, the correspondent data yields should be scaled to the HL-LHC conditions to deliver the proper projection of the backgrounds. The control regions represent a mixture of some dominant physics processes with minor contributions from additional backgrounds. To properly scale data yields, the simulated events are used. All background processes in MC are scaled to HL-LHC conditions separately and are mixed according to their cross sections to esimate the total event yield in the control region. This yield is compared to the total simulated event yields in the same control region without scaling. The ratio is used to project the existing data-based background estimates to the HL-LHC conditions. We compute this shape-dependent scaling on data distribution $D^{j,k}_{2016}$ in a control region $j$ for each $M_R$-$R^2$ bin $k$ as follows:

$$D^{j,k}_{\text{HL-LHC}} = \frac{\sum_i N^{i,j,k}_{\text{HL-LHC}}}{\sum_i N^{i,j,k}_{2016}} D^{j,k}_{2016} \tag{6}$$

$$= r^{j,k}_{\frac{\text{HL-LHC}}{2016}} D^{j,k}_{2016}, \tag{7}$$

where $N^{i,j,k}_{\text{HL-LHC}}$ and $N^{i,j,k}_{2016}$ are yields in bin $k$ of simulated distributions for each process $i$ for a control region $j$, and the resulting scaling factors $r^{j,k}_{\frac{\text{HL-LHC}}{2016}}$ vary depending on the bin $k$.

Once the data distributions $D^{j,k}_{\text{HL-LHC}}$ are obtained, a number of pseudo-data events $D^j = \sum_k D^{j,k}$ are produced from the distributions to match the expected yields from the HL-LHC. These pseudo-data event distributions and their statistical uncertainties are used to calculate the estimated backgrounds.

## 6 Treatment of uncertainties

During the HL-LHC runs, CMS will collect two orders of magnitude more data than it has collected so far in Run 2. This will significantly improve the precision of any analysis result. The large instantaneous luminosity will also cause an increase in the number of pileup events, which will introduce uncertainties in the results. However, improvements to the detector will help reduce various systematic uncertainties arising from detector inaccuracies and compensate for the pileup effects. The theoretical calculations are also expected to improve, providing more accurate and precise cross sections and event simulations. Furthermore, potential increases in computational speed and storage would help increase the number of simulated events produced and reduce MC-related statistical uncertainties. In this study, we use three scenarios to assess the effects of varying levels of the above-mentioned uncertainties, taken from conventions based on [14]. The integrated luminosities used in defining systematics are $\mathcal{L}_{\text{HL-LHC}} = 3\,\text{ab}^{-1}$ and $\mathcal{L}_{2016} = 35.9\,\text{fb}^{-1}$.

- Run 2 systematic uncertainties: This scenario is useful for direct comparison with the current analyses. Statistical uncertainties are scaled by $1/\sqrt{\mathcal{L}_{\text{HL-LHC}}/\mathcal{L}_{2016}} \equiv 1/\sqrt{\mathcal{L}}$. Systematic uncertainties (including experimental, theoretical and luminosity) are kept the same in relative terms as in the 2016 analysis.
- YR18 (CERN Yellow Report 2018) systematic uncertainties: This scenario reflects uncertainties that are considered achievable from today's perspective for the HL-LHC phase. Statistical uncertainties are scaled by $1/\sqrt{\mathcal{L}}$. Theoretical uncertainties are scaled down by 1/2. The remaining uncertainties, such as those on luminosity, jet energy scale and resolution, W, top, or b jet tagging scale factors, lepton scale factors, that are considered in the experimental systematic uncertainties category are scaled down based on the recommendations for the Yellow Report. While well-defined percent values were taken for some systematic uncertainties, such as $\pm 1\%$ for luminosity, for others, a fractional or luminosity-based scaling was done, except for the cases where the uncertainties are already small. Table 1 shows the list of uncertainties applied on background and signal processes and the corresponding scaling applied to these with respect to the current analysis.
- Stat-only: This scenario indicates the ultimate precision limit. Statistical uncertainties are scaled by $1/\sqrt{\mathcal{L}}$, while systematic uncertainties are neglected.

The effects of systematic uncertainties applied for the Run 2 and YR18 scenarios vary as a function of $M_R$ and $R^2$. The uncertainties in the 2016 analysis were dominated by statistical effects. Systematic uncertainties were relatively small for the final states of interest. For the YR18 scenario, some of the Run 2 uncertainties are small, such as those in the lepton reconstruction and identification scale factors, and thus not scaled down. Uncertainties arising from pileup are taken the same as in 2016. Even though pileup is expected to increase by about an order of magnitude, there will be large improvements in tracking, vertexing and $\eta$ coverage which will compensate for the increased effect. Figure 4 shows the average percentage contributions of the various systematic uncertainties to the overall background estimation as a function of $M_R$ and $R^2$ bins for the W 4-5 jet, W 6 jet, and Top categories for the Run 2 and YR18 scenarios. The most dominant systematic uncertainties affecting the simulated signal event yields come from W/top tagging ($\sim 8\%$), jet energy scale (JES) ($\sim 3\%$) and b tagging ($\sim 2\%$) variations.

Table 1: Summary of the scaling of uncertainties in the YR18 scenario for the background and signal processes for the HL-LHC projections. The "YR18 recommendation" treatment note specifies that the scaling of the uncertainty was done based on CMS recommendations for the Yellow Report, reflecting the potential upgrade performance of the CMS detector, summarised in Ref. [14].

| Uncertainty | Background | Signal | Treatment notes |
|---|---|---|---|
| | Statistical uncertainties | | |
| MC event yield | ignored | Run2 / $\sqrt{\mathcal{L}}$ | YR18 recommendation |
| Data event yield | Run2 / $\sqrt{\mathcal{L}}$ | – | YR18 recommendation |
| Extrapolation of background distributions in signal region | ignored | – | YR18 recommendation |
| | Theoretical systematic uncertainties | | |
| Scales (fact., renorm.) | Run2 $\times 1/2$ | Run2 $\times 1/2$ | YR18 recommendation |
| $\alpha_s$ | Run2 $\times 1/2$ | Run2 $\times 1/2$ | YR18 recommendation |
| Top $p_T$ reweighting | Run2 $\times 1/3$ | – | YR18 recommendation |
| ISR reweighting | – | Run2 $\times 1/2$ | YR18 recommendation |
| $Z \to \nu\nu$ modeling | Run2 $\times 1/2$ | Run2 $\times 1/2$ | YR18 recommendation |
| Multijet modeling | Run2 $\times 1/2$ | Run2 $\times 1/2$ | YR18 recommendation |
| | Experimental systematic uncertainties | | |
| Luminosity | $\pm 1.0\%$ | $\pm 1.0\%$ | YR18 recommendation |
| Pileup | Run2 | Run2 | Increased PU but better detector performance |
| Jet energy/mass scale | Run2 $\times 1/2$ | Run2 $\times 1/2$ | YR18 recommendation |
| Jet energy/mass resolution | Run2 $\times 1/2$ | Run2 $\times 1/2$ | YR18 recommendation |
| $p_T^{miss}$ | Run2 $\times 1/2$ | Run2 $\times 1/2$ | from JES, JER |
| Electron reconstruction | Run2 | Run2 | small |
| Electron identification | Run2 | Run2 | small |
| Muon tracking | Run2 | Run2 | small |
| Muon identification | Run2 | Run2 | small |
| Lost lepton shape | Run2 / $\sqrt{\mathcal{L}}$ | – | stat-dependent |
| b tag | $\pm 1\%$ | $\pm 1\%$ | YR18 recommendation |
| W/Top tag | Run2 | Run2 | YR18 recommendation |
| W/Top mistag | Run2 | Run2 | YR18 recommendation |
| W/Top masstag | Run2 | – | YR18 recommendation |
| W/Top antitag | Run2 | – | YR18 recommendation |
| Photon purity | Run2 $\sqrt{\mathcal{L}}$ | – | stat-dependent |
| Direct photon fraction | Run2 $\sqrt{\mathcal{L}}$ | – | stat-dependent |
| Z/$\gamma$ ratio | ignored | – | small |
| Z($\nu\bar{\nu}$) closure | Run2 / $\sqrt{\mathcal{L}}$ | – | stat-dependent |

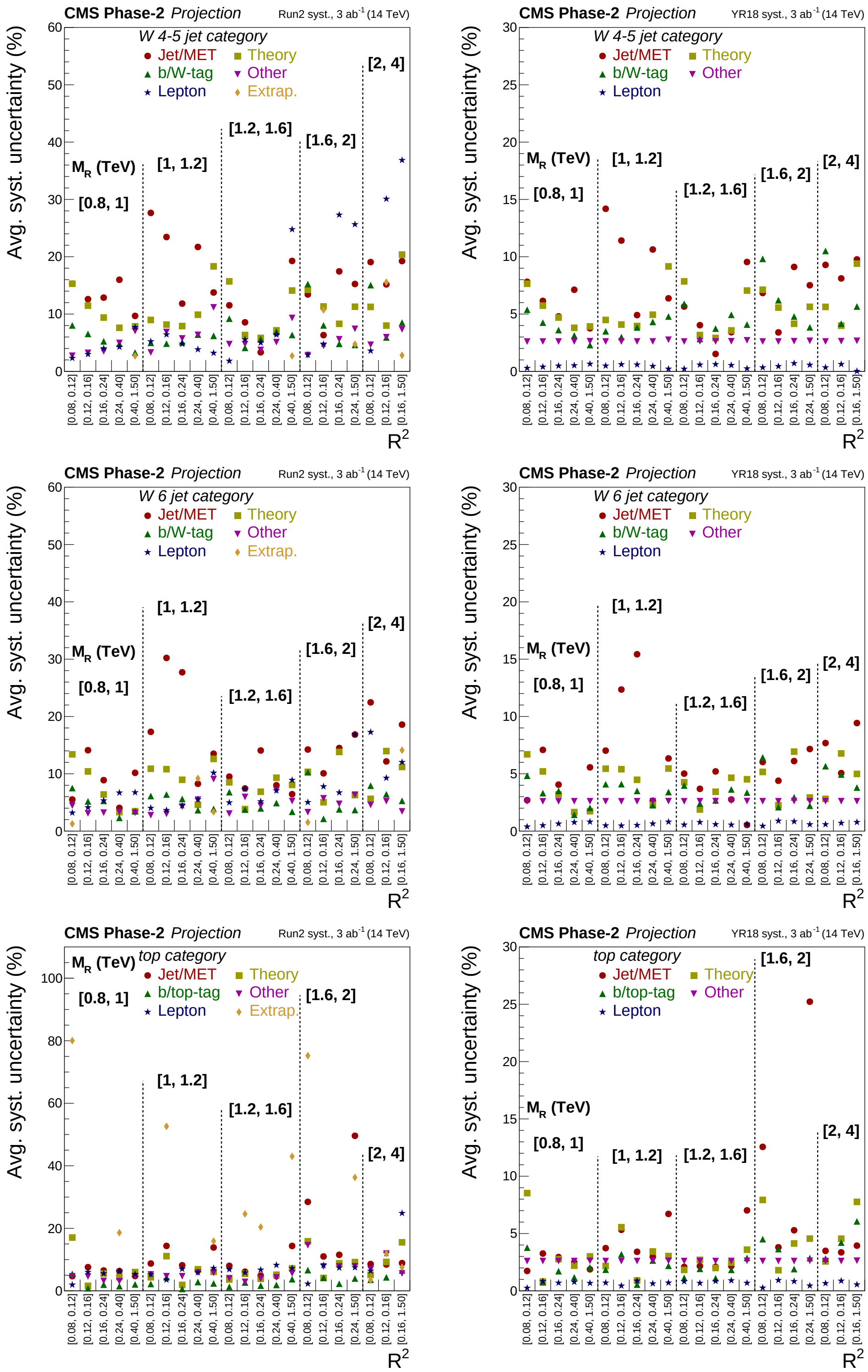

Figure 4: Average percentage contributions of various systematic uncertainties to the overall background estimation under the background-only assumption as a function of bins in $M_R$ and $R^2$ for the W 4-5 jet (top), W 6 jet (middle), and Top (bottom) categories for the Run 2 (left) and YR18 (right) scenarios.

# 7 Results and interpretation

We present the overall background estimation for the W 4-5 jet, W 6 jet, and Top categories along with distributions for several signal benchmark scenarios versus a one-dimensional representation of the bins in $M_R$ and $R^2$ in Fig. 5 for the HL-LHC. Statistical and systematic uncertainties are also shown for the YR18 case where systematic uncertainties are scaled down based on currently estimated projections of luminosity, detector conditions, and theoretical calculations.

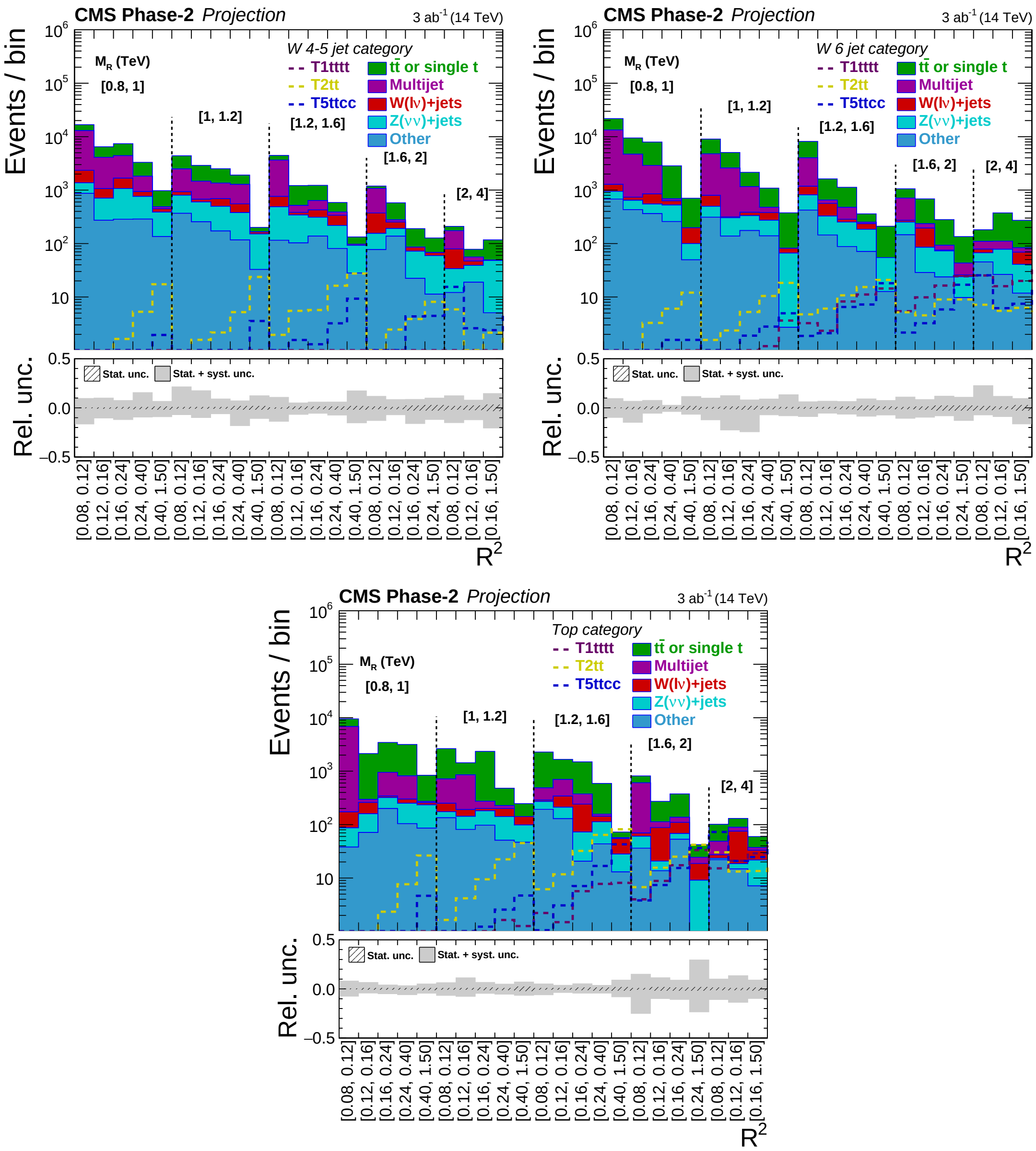

Figure 5: $M_R$-$R^2$ distributions shown in a one-dimensional representation for background predictions obtained for the W 4-5 jet (upper left), W 6 jet (upper right), and Top (lower) categories for the HL-LHC. Statistical and systematic uncertainties for the YR18 scenario are shown with the hatched and shaded error bars, respectively. Also shown are the signal benchmark models T5ttcc with $m_{\tilde{g}} = 2\,\text{TeV}$, $m_{\tilde{t}} = 320\,\text{GeV}$ and $m_{\tilde{\chi}_1^0} = 300\,\text{GeV}$; T1tttt with $m_{\tilde{g}} = 2\,\text{TeV}$ and $m_{\tilde{\chi}_1^0} = 300\,\text{GeV}$; and T2tt with $m_{\tilde{t}} = 1.2\,\text{TeV}$ and $m_{\tilde{\chi}_1^0} = 100\,\text{GeV}$.

The results are used to set expected upper limits on the production cross sections of various SUSY simplified models. We follow the LHC $CL_s$ procedure [23–25] by using the profile likelihood ratio test statistic and the asymptotic formula to evaluate the 95% confidence level (CL) expected limits on the production cross section. Systematic uncertainties are propagated by incorporating nuisance parameters that represent different sources of systematic uncertainty, which are profiled in the maximum likelihood fit. Fig. 6 shows the expected upper limits on the signal cross sections for the T5ttcc, T1tttt and T2tt simplified models for the combination of the W 4-5 jet, W 6 jet, and Top categories for the HL-LHC projection based on the YR18 scenario. Additionally, lower limits on gluino/top squark versus neutralino masses are shown for the cases of Run 2 systematic uncertainties, YR18 systematic uncertainties, and statistical-only scenarios. Gluino mass exclusion limit reaches over 2.6 TeV and 2.5 TeV for T5ttcc and T1tttt, respectively, and top squark mass limit reaches over 1.5 TeV for T2tt. For comparison, the figures also show the 2016 mass limits and the 300 $fb^{-1}$ limits for the Run 2 scenario.

Furthermore, projections of expected discovery sensitivity in the presence of a signal are computed. The p-values for the signal plus background and background-only hypotheses are used to obtain the expected significances in terms of number of standard deviations. Figure 7 shows the projected expected significance for the T5ttcc, T1tttt, and T2tt models based on the YR18 systematic uncertainties, along with the discovery upper bounds on the gluino/top squark versus neutralino masses for the three uncertainty scenarios for the HL-LHC. Discovery reach for gluino mass extend over 2.35 TeV and 2.3 TeV gluino mass for T5ttcc and T1tttt, and 1.4 TeV top squark mass for T2tt.

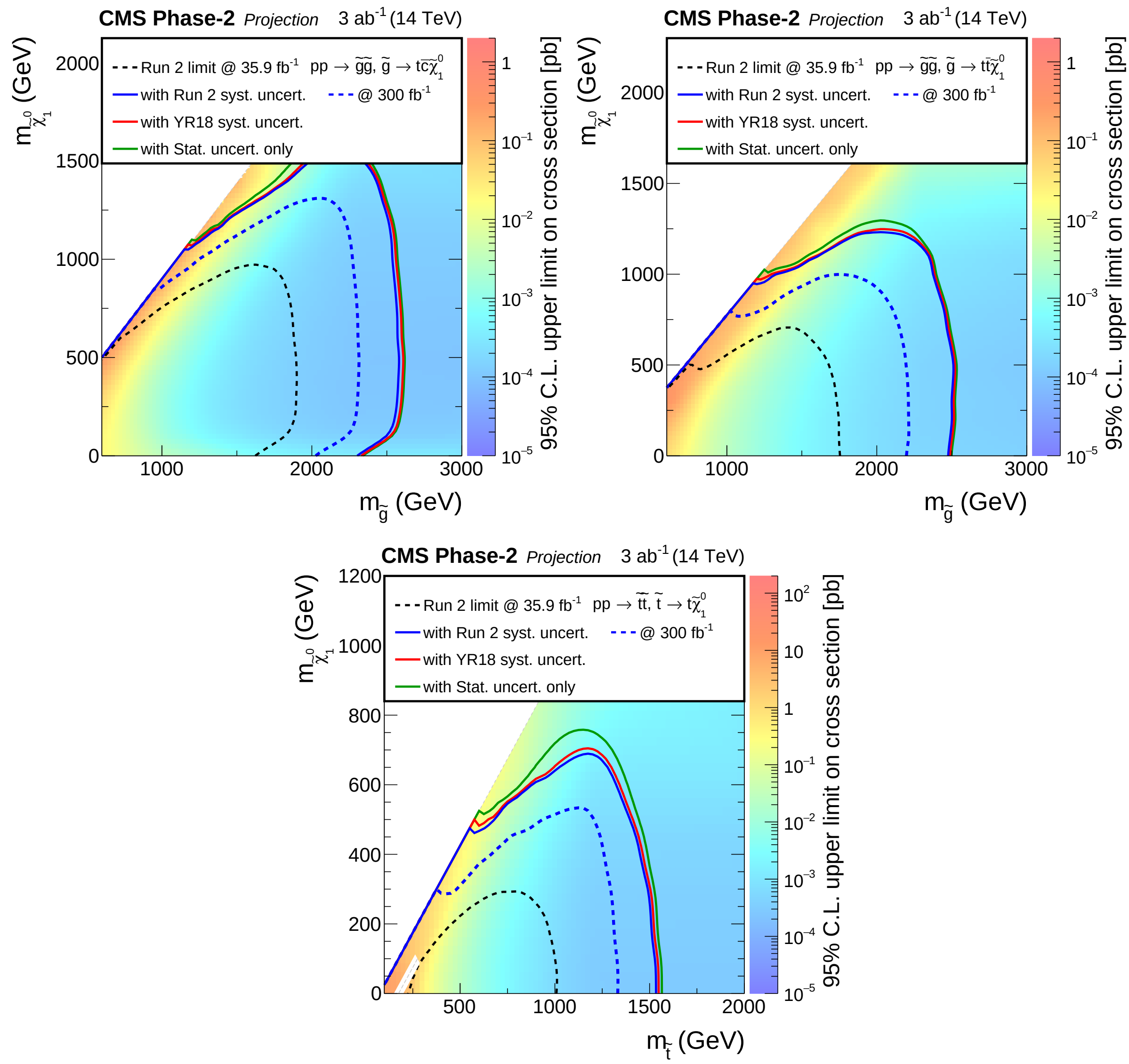

Figure 6: Projected expected upper limits on the signal cross sections for the HL-LHC using the asymptotic $CL_s$ method versus gluino/top squark and neutralino masses for the T5ttcc (top left), T1tttt (top right), and T2tt (bottom) models for the combined W 4-5 jet, W 6 jet, and Top categories for the YR18 scenario. The contours show the expected lower limits on the gluino/stop squark and neutralino masses based on the Run 2 systematic uncertainties, YR18 systematic uncertainties, and statistical-only scenarios, along with the 2016 analysis limit and the 300 $\text{fb}^{-1}$ limit for comparison. The lower left white diagonal band in the T2tt plot corresponds to the region $|m_{\tilde{t}} - m_t - m_{\tilde{\chi}_1^0}| < 25\,\text{GeV}$, where the mass difference between the $\tilde{t}$ and the $\tilde{\chi}_1^0$ is very close to the top quark mass. In this region, the signal acceptance depends strongly on the $\tilde{\chi}_1^0$ mass and is therefore difficult to model.

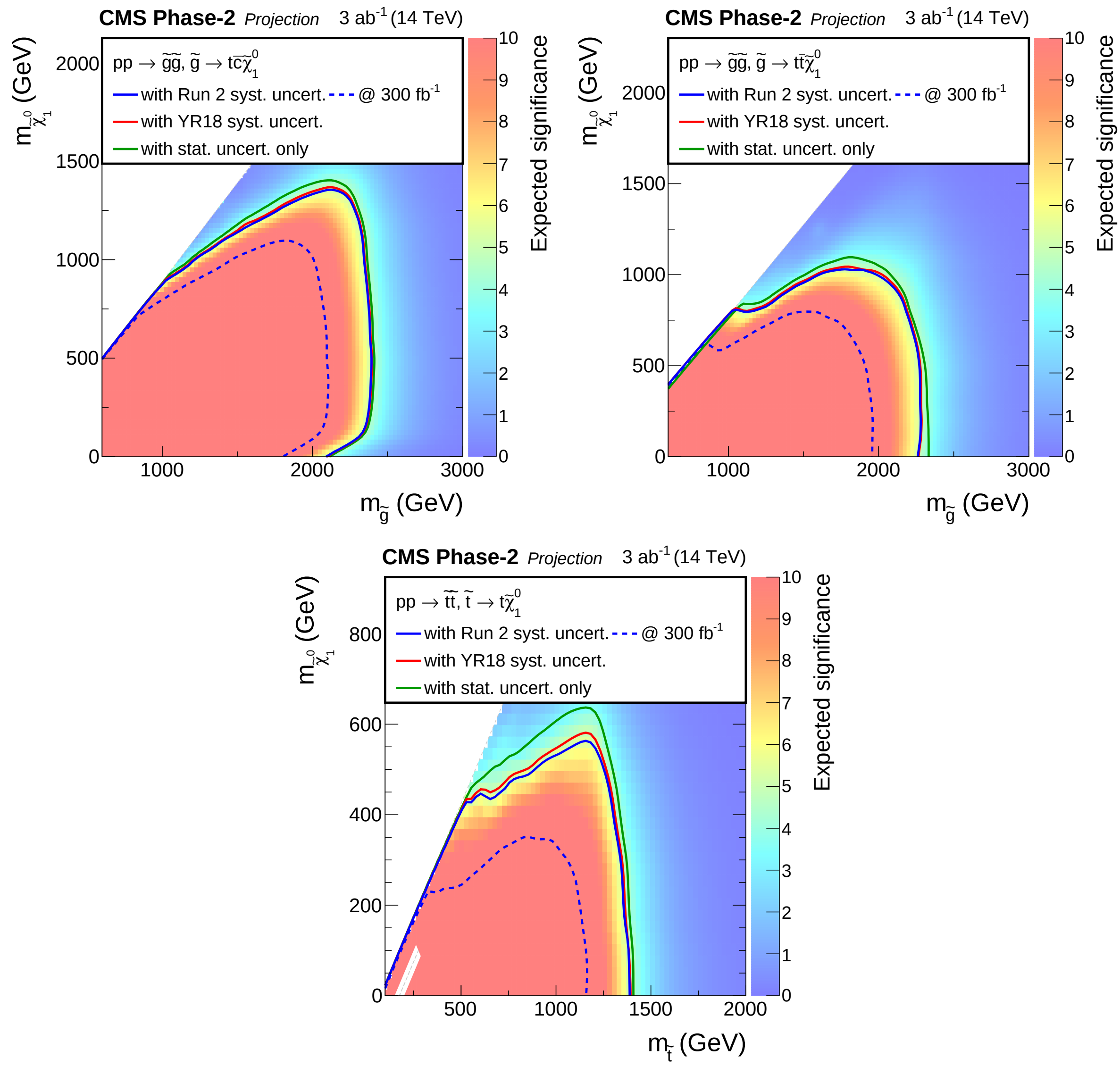

Figure 7: Projected expected significance for the HL-LHC versus gluino/stop and neutralino masses for the T5ttcc (top left), T2tttt (top right), and T2tt (bottom) models for the combined W 4-5 jet, W 6 jet, and Top categories for the YR18 scenario. The contours show the expected discovery bounds on the gluino/top squark and neutralino masses based on the Run 2 systematic uncertainties, YR18 systematic uncertainties, and statistical-only scenarios. The lower left white diagonal band in the T2tt plot corresponds to the region $|m_{\tilde{t}} - m_t - m_{\tilde{\chi}^0_1}| < 25\,\text{GeV}$, where the mass difference between the $\tilde{t}$ and the $\tilde{\chi}^0_1$ is very close to the top quark mass. In this region, the signal acceptance depends strongly on the $\tilde{\chi}^0_1$ mass and is therefore difficult to model.

# 8 Summary

We have presented the HL-LHC projection of the Run 2 search for new physics in hadronic final states with boosted W bosons or top quarks using the razor kinematic variables. Final states with boosted objects constitute an important search scenario, as they become more accessible at the increased center-of-mass energy at the HL-LHC. The projection study uses observed data yields and simulated signal and background events from the original analysis, which are scaled to obtain the HL-LHC sensitivity for center-of-mass energy of 14 TeV and integrated luminosity of $3\,\mathrm{ab}^{-1}$. The background estimation and limit setting procedures are fully adopted from the Run 2 analysis done using 2016 data. Different scenarios for systematic uncertainties, based on a common convention with other CMS analyses and ATLAS are considered. The projection results are interpreted in terms of gluino pair production where the gluinos decay into either a top quark, an anti-top quark, and a neutralino; or to a top quark and a top squark, and direct top squark pair production where top squarks decay into top quarks and neutralinos. The HL-LHC would exclude gluinos and top squarks up to 2.6 TeV and 1.5 TeV respectively, while making discovery possible for gluinos and top squarks up to masses of 2.35 TeV and 1.4 TeV, respectively, thus providing a very strong test of naturalness scenarios for supersymmetry.

# Extrapolation of $E_T^{miss}$ + jet search results to an integrated luminosity of 300 fb$^{-1}$ and 3000 fb$^{-1}$

The ATLAS Collaboration

This note presents a study of the impact of different systematic uncertainty scenarios on the sensitivity of the ATLAS search in the final state with at least one jet and large missing transverse momentum (monojet signature) to weakly-interacting massive particles (WIMPs). A simplified model in which WIMPs are pair-produced from the *s*-channel exchange of an axial-vector mediator is used as a benchmark for the sensitivity of the analysis to dark matter. The sensitivity prospectives are evaluated by extrapolating simulated results obtained by a recent ATLAS search based on 36 fb$^{-1}$ of *pp* collisions at a center-of-mass energy of 13 TeV to integrated luminosities of 300 fb$^{-1}$ and 3000 fb$^{-1}$.

# 1 Introduction

The final state with at least one jet and large missing transverse momentum ($E_{\mathrm{T}}^{\mathrm{miss}}$ + jet) is a key channel for the search for dark matter at the Large Hadron Collider (LHC) [1–3]. Different theoretical frameworks [4–6] beyond the Standard Model predict the production of pairs of Weakly Interacting Massive Particles (WIMPs) [7, 8] in association with hadronic jets, coming from initial state radiation, which can be used to discriminate signal from background events. This process is conveniently described with simplified benchmark models, which assume the existence of a massive mediator of the interaction between the initial state partons and the WIMPs. Depending on the spin-parity state of the mediator and on its couplings, different sensitivity is achieved at the LHC compared to direct and indirect detection experiments, with differing complementarity between LHC search channels, namely those with missing transverse momentum and those with resonant particles.

In the upcoming years, the LHC and the ATLAS experiment [9] will undergo significant upgrades. During the first phase, from 2021 to 2023 (Run-3), the ATLAS detector is expected to collect an integrated luminosity of 300 fb$^{-1}$ whereas in the second phase, from 2026 to 2036 (Run-4), the total amount of data corresponding of 3000 fb$^{-1}$ is foreseen. This unprecedented luminosity, which will be crucial for precision measurements in the Higgs sector and for increasing the discovery potential at the energy frontier, will be collected in two steps.

The goal of this study is to evaluate the impact of different assumed systematic uncertainty scenarios on the expected sensitivity to dark matter in the $E_{\mathrm{T}}^{\mathrm{miss}}$ + jet channel. The evaluation is based on the extrapolation to higher luminosity of the limits published by the ATLAS Collaboration with 36 fb$^{-1}$ of $pp$ collisions at a center-of-mass energy $\sqrt{s}$ = 13 TeV [10]; the same center-of-mass energy is assumed in all scenarios. A simplified benchmark model is chosen to illustrate the WIMP discovery potential, where Dirac fermion WIMPs, $\chi$, are pair-produced from the $s$-channel exchange of a spin-1 mediator, $Z_A$, with axial-vector couplings. The model is defined by four free parameters: the mediator mass ($m_{Z_A}$), the WIMP mass ($m_\chi$), the coupling strength of the flavour-universal mediator-quark interaction ($g_q$), and the coupling strength of the mediator-WIMP interaction ($g_\chi$). The width of the mediator is assumed to be minimal, as detailed in Refs. [5, 6].

This note is organized as follows. An overview of the ATLAS detector is provided in Sec. 2. Sec. 3 describes the simulation of Monte Carlo (MC) events used for these projections. Sec. 4 describes the event selection, while the methodology used to extrapolate limits to a luminosity of 300 fb$^{-1}$ and 3000 fb$^{-1}$ is discussed in Sec. 5. Results are discussed in Sec. 6, and conclusions are drawn in Sec. 7.

# 2 The ATLAS detector and the LHC programme

ATLAS [9] is a multi-purpose particle detector with a forward-backward symmetric cylindrical geometry. It covers almost the whole solid angle around the collision point with layers of tracking detectors immersed in a 2 T axial magnetic field produced by a solenoid, a calorimeter system which provides both the electromagnetic and hadronic energy measurements and a muon spectrometer which measures the deflection of muons in the magnetic field provided by large superconducting air-core toroidal magnets. During Run-2 (2015-2018) an integrated luminosity close to 150 fb$^{-1}$ has been collected with an instantaneous luminosity of $2.1 \times 10^{34}$ cm$^{-2}$s$^{-1}$ and an average number of collisions per bunch crossing $< \mu >\sim 35$.

After a long shutdown (2019-2020), collisions will restart in Run-3 at the instantaneous luminosity of $2.1 \times 10^{34}$ cm$^{-2}$s$^{-1}$ and $< \mu >\sim 60$. The integrated luminosity which is expected to be collected at the end of Run-3 corresponds to 300 fb$^{-1}$ at the centre-of-mass-energy to $\sqrt{s}$ = 14 TeV.

During the long shutdown between 2024-2026, the accelerator is foreseen to be upgraded to the High-Luminosity LHC (HL–LHC). It is currently expected to begin its operations in the second half of 2026, with a nominal levelled instantaneous luminosity of $7.5 \times 10^{34}$ cm$^{-2}$s$^{-1}$ at $\sqrt{s}$ = 14 TeV. This will lead to an average number of approximately 200 inelastic $pp$ collisions per bunch-crossing. This programme aims to provide a total integrated luminosity of 3000 fb$^{-1}$ by 2036. Upgrades to the ATLAS detector and the trigger system are planned in order to adapt the experiment to the new challenging conditions foreseen for LHC during Run-3 and Run-4 [11–21]. A detailed description of the expected performance of the ATLAS detector at the HL-LHC is addressed in Ref. [22].

# 3 Simulation

The same signal and background MC simulation setup as Ref. [10] is used in this study. Proton-proton collision events are simulated with $\sqrt{s}$ = 13 TeV; unless otherwise indicated, samples are processed with the full ATLAS detector simulation [23] based on Geant 4 [24]. Pile-up effects are taken into account by overlaying simulated minimum-bias events from Pythia 8.205 [25] onto the hard-scattering process, distributed according to the frequency in 36 fb$^{-1}$ data collected in 2015 and 2016. No correction for the different detector setup and pile-up conditions expected during Run-3 and the HL-LHC phase is applied assuming that similar performance to the ones gotten during Run-2 can be achieved in this channel.

## 3.1 Signal

Signal events are simulated in Powheg-Box v2 [26–28] (revision 3049) using the DMV model of WIMP pair production introduced in Ref. [29], at next-to-leading order (NLO) in the strong coupling constant $\alpha_S$. The mediator couplings are set to $g_q = 0.25$ and $g_\chi = 1$, and its propagator is described by a Breit-Wigner distribution. Events are generated using the NNPDF30 [30] parton distribution functions (PDFs) and interfaced to Pythia 8.205 with the A14 set of tuned parameters (tune) [31] for parton showering, hadronization and the underlying event.

A grid of samples is produced for WIMP masses ranging from 1 GeV to 1 TeV and mediator masses between 10 GeV and 10 TeV. For a few values of $m_{Z_A}$, additional samples with respect to those used in Ref. [10] are produced at generator-level only, in order to increase the granularity of the expected exclusion limit contours in the region where the mediator decay in WIMP pairs is on the mass shell. Full-simulation events have been used to validate the stability of acceptance and selection efficiency effects for on-shell events at fixed $m_{Z_A}$.

## 3.2 Backgrounds

$W/Z$+jets production processes are simulated using the Sherpa 2.2.1 [32] event generator. Matrix elements (ME) are calculated for up to two partons at NLO and up to four at LO using OpenLoops [33] and Comix [34], and merged with the Sherpa parton shower (PS) [35] using the ME+PS@NLO prescription [36] and the NNPDF3.0NNLO [30] PDF set.

A reweighting of the $W$+jets and $Z$+jets MC predictions is performed in order to account for higher-order QCD and electroweak corrections as described in Ref. [37], where parton-level predictions for $W/Z$+jets production, including NLO QCD corrections and NLO electroweak corrections supplemented by Sudakov logarithms at two loops, are provided as a function of the vector-boson $p_{\mathrm{T}}$. More details on the procedure are provided in Ref. [10].

The $t\bar{t}$ and single top quarks in the $Wt$-channel and $s$-channel processes are generated by Powheg-Box v2 [38] with CT10 [39] PDFs whereas electroweak $t$-channel single-top-quark events are generated using the Powheg-Box v1 event generator. The four-flavour scheme has been set to calculate NLO matrix elements, with the CT10 four-flavour PDF set. The parton shower, hadronization, and underlying event are simulated using Pythia 8.205 with the A14 tune. The top-quark mass is set to 172.5 GeV. In order to estimate the effects of the choice of matrix-element event generator and parton-shower algorithm, alternative sample based on MadGraph5_aMC@NLO v2.2.1 [40] interfaced to Herwig++ (v2.7.1) [41] are used.

Diboson processes ($WW$, $WZ$, and $ZZ$ production) are simulated by Sherpa 2.2.1 or Sherpa 2.1.1 with NNPDF3.0NNLO or CT10nlo PDFs, respectively, and normalized to NLO pQCD predictions [42]. Further diboson samples based on Powheg-Box [27] interfaced to Pythia 8.186 with CT10 PDFs are used for studies of systematic uncertainties.

## 4 Event reconstruction and event selection

Jets are reconstructed from energy deposits in the calorimeters, using the anti-$k_t$ algorithm[43, 44] with the radius parameter $R = 0.4$. Jets with $p_{\mathrm{T}} > 20$ GeV and $|\eta| < 2.8$ are considered in the analysis; for central jets ($|\eta| < 2.4$) with $p_{\mathrm{T}} < 50$ GeV, additional requirements based on the inner detector information[45] are applied to remove jets originating from pile-up collisions[1]. Additionally, jets with $p_{\mathrm{T}} > 30$ GeV and $|\eta| < 2.5$ are identified as $b$-jets if they pass the requirements of a multi-variate algorithm[46, 47] with 60% efficiency, as determined in a simulated sample of $t\bar{t}$ events.

Electrons are required to have $p_{\mathrm{T}} > 20$ GeV, $|\eta| < 2.47$, to satisfy the 'Loose' selection criteria described in Ref. [48] and to pass track-based isolation requirements. In case an electron overlaps within $\Delta R = \sqrt{(\Delta\eta)^2 + (\Delta\phi)^2} < 0.2$ with a jet with $p_{\mathrm{T}} > 30$ GeV, the electron is discarded (retained) if the jet is (not) $b$-tagged, while the other is removed. Electrons separated by $0.2 < \Delta R < 0.4$ from any remaining jet are removed.

Muons are reconstructed combining the information from the muon spectrometer and the inner detector. They are required to have $p_{\mathrm{T}} > 10$ GeV, $|\eta| < 2.5$ and to pass the 'Medium' identification requirements described in Ref. [49]. In case a muon overlaps within $\Delta R < 0.4$ with a jet with $p_{\mathrm{T}} > 30$ GeV, the jet (muon) is discarded if the number of tracks with $p_{\mathrm{T}} > 500$ MeV associated to the jet is less than (at least) three, while the other is removed.

The missing transverse momentum vector $\mathbf{p}_{\mathrm{T}}^{\mathrm{miss}}$ is reconstructed using all energy deposits in the calorimeter with $|\eta| < 4.9$; its magnitude is denoted as $E_{\mathrm{T}}^{\mathrm{miss}}$. Clusters associated with electrons, photons or jets with $p_{\mathrm{T}} > 20$ GeV make use of the corresponding calibrations. Softer jets and clusters not associated with electrons, photons or jets are calibrated using tracking information[50]. In this analysis, $E_{\mathrm{T}}^{\mathrm{miss}}$ is not corrected for the presence of muons.

[1] Similar performance in terms of jet identification and pile-up suppression reached during Run-2 are assumed in this context even in the challenging pile-up regimes expected in Run-3 and Run-4.

Events are pre-selected if they have at least one reconstructed primary vertex, with at least two associated tracks, and $E_{\mathrm{T}}^{\mathrm{miss}}$250 GeV. At least one jet with $p_{\mathrm{T}} > 250$ GeV and $|\eta| < 2.4$ must have been reconstructed, together with up to three additional jets with $p_{\mathrm{T}} > 30$ GeV and $|\eta| < 2.8$. In order to suppress background from multi-jet events, jets must have an azimuthal angle separation with respect to the missing transverse momentum direction $\Delta\phi(\text{jet}, \mathbf{p}_{\mathrm{T}}^{\mathrm{miss}}) > 0.4$. Jet quality criteria, based on quality requirements on calorimetric variables and on the jet charged particle fraction, are applied to suppress non-collision background and calorimeter noise.

Five regions are used in the analysis: a signal region, where only events without electrons or muons are selected, and four control regions enriched in $W \to \mu\nu$, $t\bar{t}$, $W \to e\nu$ and $Z \to \mu\mu$ events, respectively. Events containing exactly one muon and no electron, and with a transverse mass $m_{\mathrm{T}} = \sqrt{2p_{\mathrm{T}}^{\ell}p_{\mathrm{T}}^{\nu}[1 - \cos(\phi^{\ell} - \phi^{\nu})]}$ of the lepton-$\mathbf{p}_{\mathrm{T}}^{\mathrm{miss}}$ system between 30 and 100 GeV, are assigned to the $W \to \mu\nu$ ($t\bar{t}$) control region if they have zero (at least one) $b$-tagged jet. Similarly, events with exactly two muons, and with an invariant mass of the di-muon system between 66 and 116 GeV, are assigned to the $Z \to \mu\mu$ control region. Finally, events with exactly one isolated electron with $p_{\mathrm{T}} > 30$ GeV (reconstructed outside the transition region between the barrel and endcaps of the electromagnetic calorimeter, $1.37 < |\eta| < 1.52$), no other electron or muon, and with $30 < m_{\mathrm{T}} < 100$ GeV, are assigned to the $W \to e\nu$ region; in this region, the $E_{\mathrm{T}}^{\mathrm{miss}}$ calculation is corrected by subtracting from $\mathbf{p}_{\mathrm{T}}^{\mathrm{miss}}$ the contribution from the electron cluster in the calorimeter. Trigger requirements based on the missing transverse momentum, which make no use of muon information at trigger level, and fully efficient for $E_{\mathrm{T}}^{\mathrm{miss}} > 250$ GeV, are used for the signal region and for the $W \to \mu\nu$, $t\bar{t}$ and $Z \to \mu\mu$ control regions; dedicated single-electron triggers are used for the $W \to e\nu$-enriched control region. A summary of the signal region selection cuts is listed in Table 1.

Table 1: Summary of the cuts applied to define the signal region.

| Category | Selection criteria |
|---|---|
| trigger | fully efficient at $E_{\mathrm{T}}^{\mathrm{miss}} > 250$ GeV |
| vertex | $\geq 1$ vertex with $N_{\mathrm{trk}} \geq 2$ |
| pile-up suppression | JVT > 0.64 ($20 < p_{\mathrm{T}} < 50$GeV, $|\eta| < 2.4$) |
| jet cleaning | *Loose*, $p_{\mathrm{T}} > 30$GeV, $|\eta| < 2.8$ |
| electron veto | *Loose*, $p_{\mathrm{T}} > 20$GeV, $|\eta| < 2.47$ |
| muon veto | *Medium*, $p_{\mathrm{T}} > 10$GeV, $|\eta| < 2.5$ |
| jet multiplicity | $N_{\mathrm{jet}} \leq 4$ ($p_{\mathrm{T}} > 30$GeV, $|\eta| < 2.8$) |
| multi-jet suppression | $\Delta\phi(\text{jet}, E_{\mathrm{T}}^{\mathrm{miss}}) > 0.4$ |
| leading jet | *Tight*, $p_{\mathrm{T}} > 250$GeV, $|\eta| < 2.4$ |
| $E_{\mathrm{T}}^{\mathrm{miss}}$ | > 250GeV |

# 5 Methodology

The discriminating variable of the search is $E_{\mathrm{T}}^{\mathrm{miss}}$; 17 bins are used, as defined in Table 2, which corresponds to an improved coverage in $E_{\mathrm{T}}^{\mathrm{miss}}$ with respect to the 10 bin scenario of Ref. [10]. The lower end of the last $E_{\mathrm{T}}^{\mathrm{miss}}$ bin, 1.6 TeV, is chosen in order to keep a similar level of uncertainty on signal and control region yields for 300 fb$^{-1}$ and 3000 fb$^{-1}$ as in the Run-2 search.

Table 2: $E_{\mathrm{T}}^{\mathrm{miss}}$ bins used in the analysis.

| $E_{\mathrm{T}}^{\mathrm{miss}}$ bins | | |
|---|---|---|
| $250 < E_{\mathrm{T}}^{\mathrm{miss}} \leq 300$ GeV | $300 < E_{\mathrm{T}}^{\mathrm{miss}} \leq 350$ GeV | $350 < E_{\mathrm{T}}^{\mathrm{miss}} \leq 400$ GeV |
| $400 < E_{\mathrm{T}}^{\mathrm{miss}} \leq 450$ GeV | $450 < E_{\mathrm{T}}^{\mathrm{miss}} \leq 500$ GeV | $500 < E_{\mathrm{T}}^{\mathrm{miss}} \leq 550$ GeV |
| $550 < E_{\mathrm{T}}^{\mathrm{miss}} \leq 600$ GeV | $600 < E_{\mathrm{T}}^{\mathrm{miss}} \leq 650$ GeV | $650 < E_{\mathrm{T}}^{\mathrm{miss}} \leq 700$ GeV |
| $700 < E_{\mathrm{T}}^{\mathrm{miss}} \leq 800$ GeV | $800 < E_{\mathrm{T}}^{\mathrm{miss}} \leq 900$ GeV | $900 < E_{\mathrm{T}}^{\mathrm{miss}} \leq 1000$ GeV |
| $1000 < E_{\mathrm{T}}^{\mathrm{miss}} \leq 1100$ GeV | $1100 < E_{\mathrm{T}}^{\mathrm{miss}} \leq 1200$ GeV | $1200 < E_{\mathrm{T}}^{\mathrm{miss}} \leq 1400$ GeV |
| $1400 < E_{\mathrm{T}}^{\mathrm{miss}} \leq 1600$ GeV | $E_{\mathrm{T}}^{\mathrm{miss}} > 1600$ GeV | |

Differently from Ref. [10], the non-collision and multi-jet backgrounds, whose extrapolation to higher luminosity is non-trivial due to their strong dependence on the details of detector and LHC performance, and which nevertheless give a sub-leading, low-$E_{\mathrm{T}}^{\mathrm{miss}}$ contribution to the overall background in the signal region, are neglected in this analysis. Additionally, expected event yields for $W/Z$ + jets ($t\bar{t}$ and single-$t$) processes are rescaled by a factor 1.27 (1.06), in order to reflect the observation of the fitted normalisation factors obtained by performing a control regions based fit in an inclusive $E_{\mathrm{T}}^{\mathrm{miss}}$ bin selection [10].

A simultaneous, binned likelihood fit of a signal plus background model to the simulated $E_{\mathrm{T}}^{\mathrm{miss}}$ distributions of the five analysis regions is performed. The signal normalisation and two additional normalisation factors, one which rescales the prediction for processes containing $Z$ and $W$ bosons produced in association with jets, and one for $t\bar{t}$ and single-$t$ production, are free parameters of the fit. Nuisance parameters with gaussian constraints are used to describe the effect of systematic uncertainties on the signal and background $E_{\mathrm{T}}^{\mathrm{miss}}$ distributions. Correlations of systematic uncertainties across $E_{\mathrm{T}}^{\mathrm{miss}}$ bins are taken into account.

The different systematic uncertainty scenarios which are tested in this analysis are chosen to reflect the possible improvements in detector performance and in the theoretical modelling of signal and background processes, which could be achieved in the next years thanks to the foreseen detector upgrades and to progress in QCD and EW calculations. Table 3 illustrates the impact of the background uncertainties from Ref. [10], grouped in terms of different sources, on the total background yield in the signal and control regions, as determined from fits of the background-only model to the event yield in inclusive $E_{\mathrm{T}}^{\mathrm{miss}}$ bins of the four control regions. The systematic uncertainties which take into account limited MC-statistics are neglected in this study. Signal uncertainties affecting acceptance and cross-section are treated separately, and their impact before the fit is summarised in Table 4.

In this context the same systematic uncertainties considered in Ref. [10] are evaluated in the new binning scenarios. Three different scenarios are probed, which differ only by the assumed pre-fit value of uncertainties:

- **standard**: same uncertainties as in Ref. [10];
- **reduced by factor 2**: all pre-fit signal and background uncertainties are reduced by a factor two;
- **reduced by factor 4**: all pre-fit signal and background uncertainties are reduced by a factor four.

Table 3: Systematic uncertainties on the total signal region background yield, as determined after a background-only fit to simulated data in control regions. In order to illustrate the different impact of uncertainties across the $E_{\mathrm{T}}^{\mathrm{miss}}$ spectrum, the fit is performed using inclusive $E_{\mathrm{T}}^{\mathrm{miss}}$ bins.

| Source | Effect [%] $E_{\mathrm{T}}^{\mathrm{miss}} > 250$ GeV | Effect [%] $E_{\mathrm{T}}^{\mathrm{miss}} > 1000$ GeV |
|---|---|---|
| *Experimental* | | |
| Jet and $E_{\mathrm{T}}^{\mathrm{miss}}$ energy scales/resolutions | 0.5 | 5.3 |
| $b$-tagging efficiency | 0.9 | 0.5 |
| soft contributions to $E_{\mathrm{T}}^{\mathrm{miss}}$ | 0.4 | 1.7 |
| lepton identification, reconstruction, $E/p$ scale/resolution | 0.2 – 1.7 | 0.3 – 2.3 |
| *Theoretical* | | |
| $W/Z$ parton shower modelling, PDF | 0.8 | 0.7 |
| $W/Z$ QCD and EW corrections | 0.4 | 2 |
| $t$-quark parton shower modelling, ISR/FSR, MC generator choice | 0.3 | $\sim 0$ |
| diboson MC generator choice, NLO cross-section | 0.2 | 0.8 |

Table 4: Systematic uncertainties on the signal cross-section ($\sigma$) and acceptance ($A$), or on the signal yield ($N_{\mathrm{sig}}$), evaluated from MC simulation; see Ref. [10] for details.

| Source | Effect [%] |
|---|---|
| Jet and $E_{\mathrm{T}}^{\mathrm{miss}}$ energy scales/resolutions | 2 – 7 on $N_{\mathrm{sig}}$ ($E_{\mathrm{T}}^{\mathrm{miss}} > 250$ GeV)<br>3 – 9 on $N_{\mathrm{sig}}$ ($E_{\mathrm{T}}^{\mathrm{miss}} > 1000$ GeV) |
| ISR/FSR | 20 on $A$ |
| PDF | $< 20$ on $A$, $< 10$ on $\sigma$ |
| Renormalization/factorisation scales | $< 10$ on $A$, $< 25$ on $\sigma$ |
| Luminosity | 3.2 on $N_{\mathrm{sig}}$ |

# 6 Results

## 6.1 Exclusion sensitivity

The exclusion limits, obtained with the profile likelihood method [51], for each of the three systematic uncertainty scenarios are plotted in the $(m_\chi, m_{Z_A})$ mass plane in Fig. 1.

The choice of the new binning corresponds to an improvement of about 100 GeV in mediator mass reach with respect to the 10 bin scenario.

The plot on the left shows that with a integrated luminosity of 300 fb$^{-1}$, and assuming the same uncertainties as in Ref. [10], the 95% CL exclusion contour for $m_\chi = 1$ GeV can be extended up to $m_{Z_A} \sim 2.20$ TeV. The phase-space that can be probed by reducing by a factor two (four) all systematic uncertainties increases significantly, and the exclusion contour for low $m_\chi$ reaches mediator masses of about 2.34 (2.43) TeV.

Similarly, the plot on the right shows that with a luminosity of 3000 fb$^{-1}$, and assuming the same uncertainties as in Ref. [10], the 95% CL exclusion contour for $m_\chi = 1$ GeV can be extended up to $m_{Z_A} \sim 2.65$ TeV. The excluded region that can be obtained by reducing by a factor two (four) all systematic uncertainties reaches, for low $m_\chi$, mediator masses of about 2.77 (2.88) TeV.

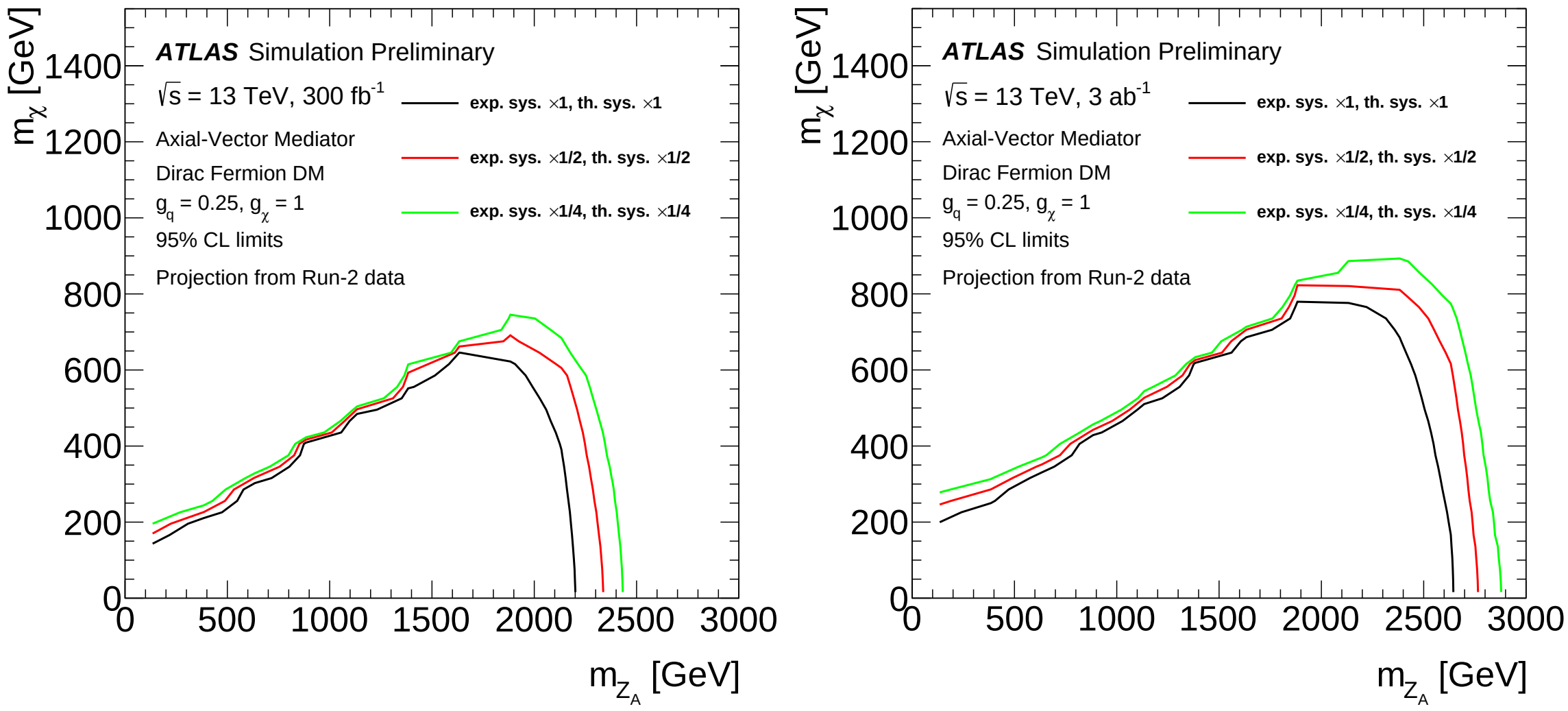

Figure 1: Expected 95% CL excluded regions on the $(m_\chi, m_{Z_A})$ mass plane for the axial-vector simplified model with couplings $g_\chi = 1$ and $g_q = 0.25$, for a luminosity of 300 fb$^{-1}$ (left) and 3000 fb$^{-1}$ (right). Three contours are shown in each plot, corresponding to the three different systematic uncertainty scenarios: standard (black), reduced by a factor 2 (red) and 4 (green). More details in the text.

Small differences between systematic uncertainty scenarios are observed when approaching the region where the decay of the mediator in two WIMPs is off the mass shell ($m_{Z_A} < 2m_\chi$), due to the decrease of the signal cross-section.

The contributions to sensitivity of experimental and theoretical uncertainties are investigated separately in Fig. 2 for 300 fb$^{-1}$ and in Fig. 3 for 3000 fb$^{-1}$. The left (right) plots show the effect of reducing by a factor two and four the effect of experimental (theoretical) systematic uncertainties on signal and backgrounds. The comparison of limit contours in these different systematic scenarios shows that the major impact to the sensitivity of the monojet search comes from theoretical uncertainties. Among these, $V$+jets and diboson uncertainties, as well as theoretical uncertainties on signal processes, are similar in size and give the leading contributions.

## 6.2 Discovery power

The discovery potential that can be reached with the integrated luminosities of 300 fb$^{-1}$ and 3000 fb$^{-1}$ is estimated in terms of the $p$-value of the background-only hypothesis, $p_0$, evaluated in the asymptotic approximation [51]. For each mass point, $p_0$ is evaluated after injecting the corresponding signal on top of the SM background. Results are shown in Fig. 4 for the different scenarios: contours corresponding to the $3\sigma$ evidence ($5\sigma$ discovery) are shown with solid (dashed) lines, for each of the tested systematic uncertainty scenarios.

With an integrated luminosity of 300 fb$^{-1}$ (3000 fb$^{-1}$) the existence of a dark matter signal described by the simplified model with an axial-vector mediator, $m_\chi = 1$ GeV and the coupling choice $g_\chi = 1$ and $g_q = 0.25$ would lead to a background incompatibility greater than $5\sigma$ at 1.81 (2.25) TeV, 1.94 (2.38) TeV and 2.02 (2.52) TeV assuming the same uncertainties as in Ref. [10], the scenario obtained by reducing by a factor two and by a factor four all the systematic uncertainties, respectively.

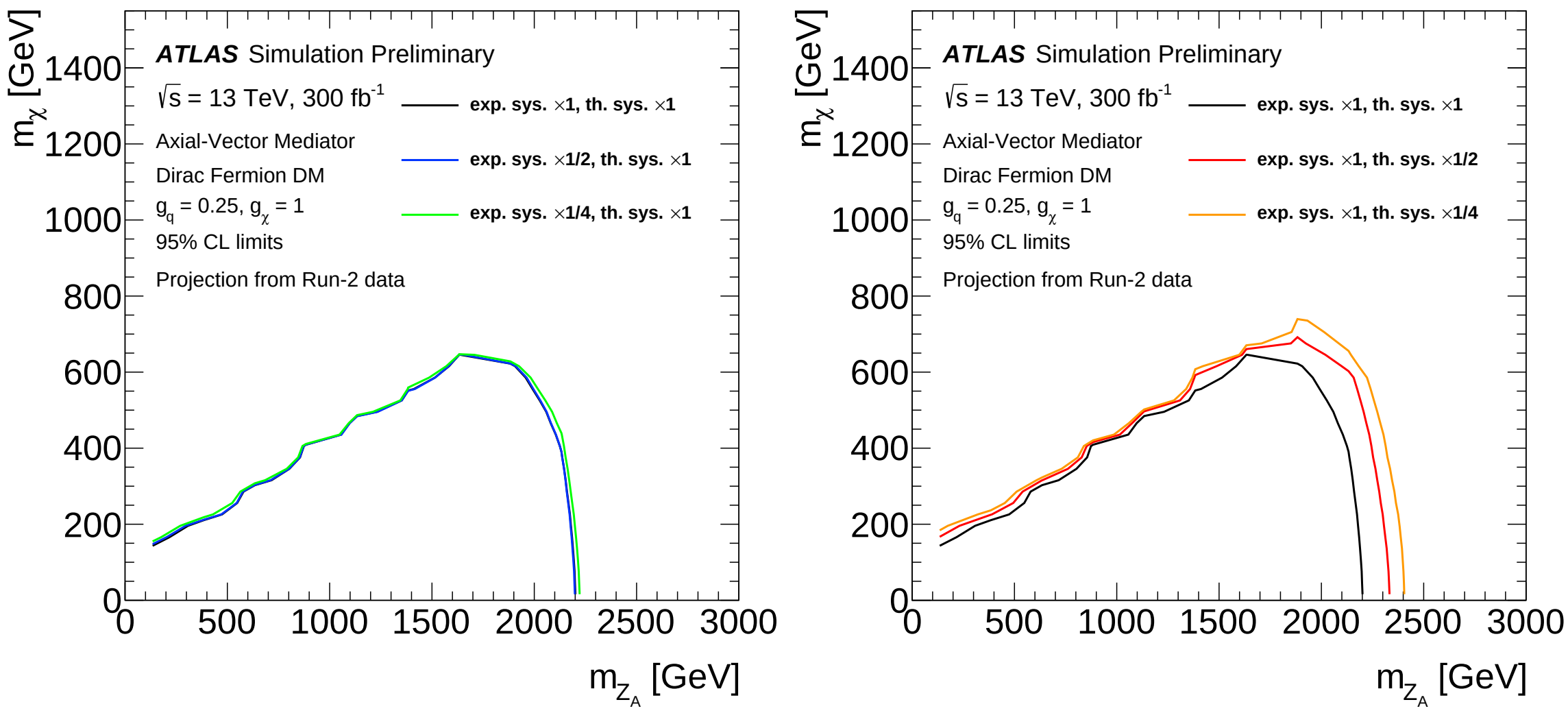

Figure 2: Expected 95% CL excluded regions on the $(m_\chi, m_{Z_A})$ mass plane for the axial-vector simplified model with couplings $g_\chi = 1$ and $g_q = 0.25$, for a luminosity of 300 fb$^{-1}$. The plot on the left (right) shows in black the expected contour assuming the same systematic uncertainties as in Ref. [10], while the scenarios with experimental (theoretical) uncertainties reduced by a factor two and four are shown in blue and green (red and orange), respectively.

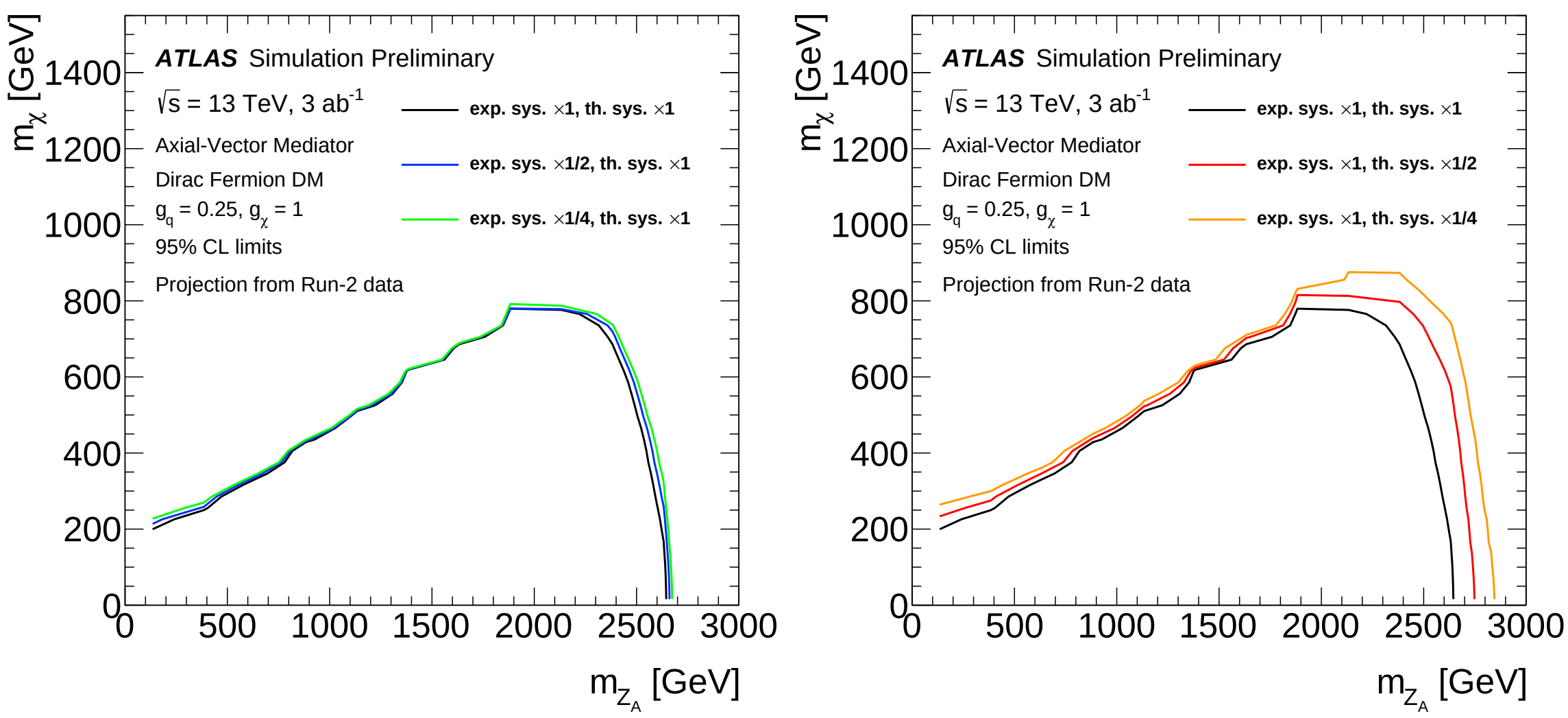

Figure 3: Expected 95% CL excluded regions on the $(m_\chi, m_{Z_A})$ mass plane for the axial-vector simplified model with couplings $g_\chi = 1$ and $g_q = 0.25$, for a luminosity of 3000 fb$^{-1}$. The plot on the left (right) shows in black the expected contour assuming the same systematic uncertainties as in Ref. [10], while the scenarios with experimental (theoretical) uncertainties reduced by a factor two and four are shown in blue and green (red and orange), respectively.

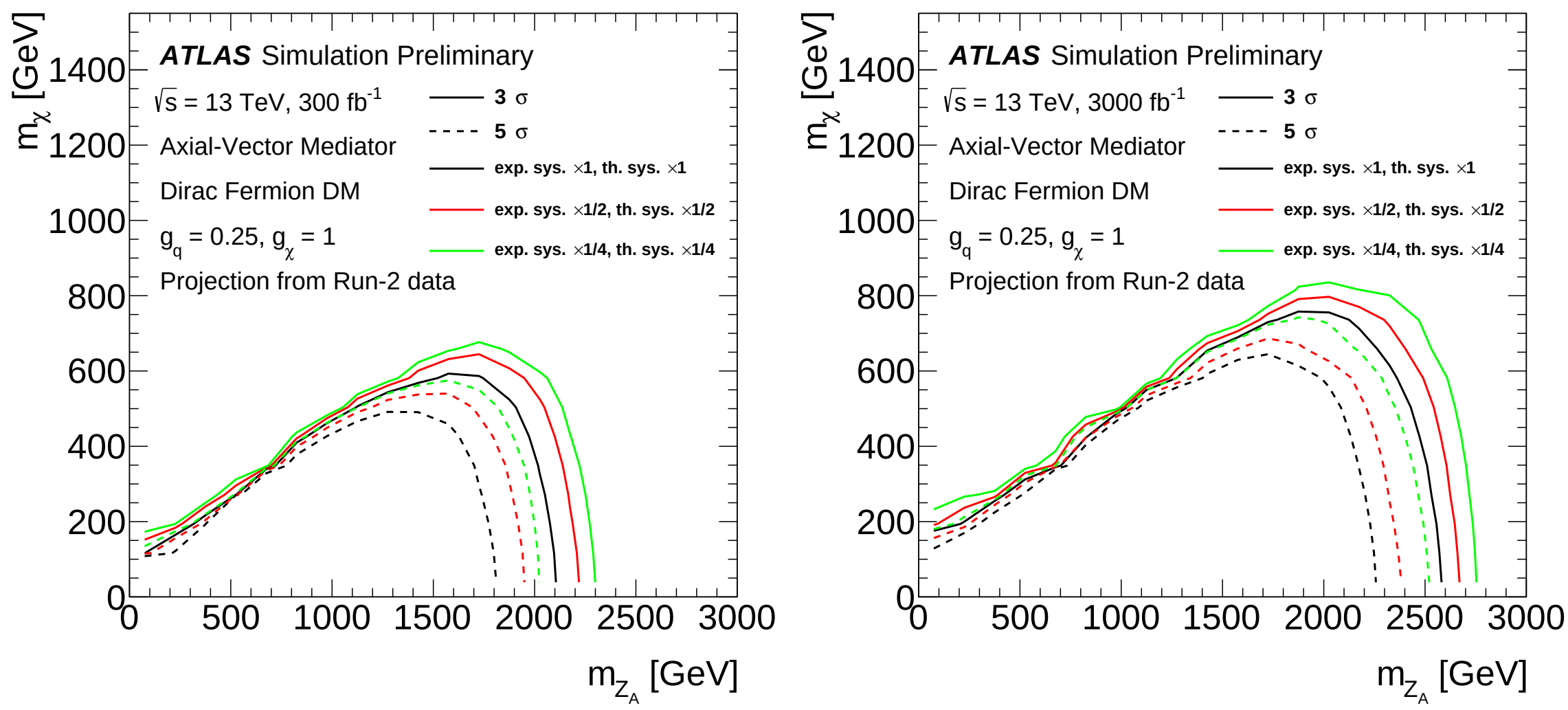

Figure 4: Expected $3\sigma$ (solid) and $5\sigma$ (dashed) discovery contours on the $(m_\chi, m_{Z_A})$ mass plane for the axial-vector simplified model with couplings $g_\chi = 1$ and $g_q = 0.25$, for a luminosity of 300 fb$^{-1}$ (left) and 3000 fb$^{-1}$ (right). Three contours are shown in each plot, corresponding to the three different systematic uncertainty scenarios: standard (black), reduced by a factor 2 (red) and 4 (green). More details in the text.

The increase of the center-of-mass energy from $\sqrt{s}$ = 13 TeV to $\sqrt{s}$ = 14 TeV, foreseen for the Run-3 and Run-4, will lead to an increase of the cross-sections of the dark matter signals considered in this context by 25-40%. On the other hand the overall cross-sections of the main $V$+jets background processes will increase by 10-15% leading to a slight increase of the $E_{\mathrm{T}}^{\mathrm{miss}}$ + jet channel sensitivity.

# 7 Conclusions

The impact of different systematic uncertainty scenarios on the sensitivity of the ATLAS search for dark matter in the final state with jets and large $E_{\mathrm{T}}^{\mathrm{miss}}$ has been estimated for the target luminosities of 300 fb$^{-1}$ and 3000 fb$^{-1}$, using a simplified model in which WIMP pairs are produced from the $s$-channel exchange of an axial-vector mediator coupling to quarks with a coupling strength of 0.25 and to WIMPs with unit coupling strength. Since the dominant contribution to sensitivity is given by the treatment and constraining of the systematic uncertainties, the same analysis strategy and simulated samples of the search with 36.1 fb$^{-1}$ of $pp$ collision data collected at $\sqrt{s}$ = 13 TeV [10] are used, and the same center-of-mass energy is assumed. Different scenarios in terms of experimental and theoretical systematic uncertainties on signal and backgrounds have been tested, where these uncertainties are improved by a factor two or four. For WIMP masses of 1 GeV, the expected 95% CL limit on the mediator mass for 300 fb$^{-1}$ (3000 fb$^{-1}$) increases from 2.20 (2.65) TeV to 2.43 (2.88) TeV if the total uncertainties are reduced by a factor four. An improvement of the systematic uncertainties related to the theoretical modelling of the signal and background processes is found to give the leading contribution to sensitivity. A discovery threshold of $5\sigma$ could be reached, for a signal with WIMP mass of 1 GeV and mediator mass of 1.81(2.25) TeV, with 300 fb$^{-1}$ (3000 fb$^{-1}$) and can be extended up to 2.02 (2.52) TeV by reducing the systematic uncertainties by a factor four. A further sensitivity improvement in this channel is expected with the increase of the center-of-mass energy from $\sqrt{s}$ = 13 TeV to $\sqrt{s}$ = 14 TeV, foreseen in the future LHC phases.

# Sensitivity of the ATLAS experiment to long-lived particles with a displaced vertex and $E_{\mathrm{T}}^{\mathrm{miss}}$ signature at the HL-LHC

The ATLAS Collaboration

This note presents the estimated sensitivity of a search for long-lived particles decaying within the tracking volume to multiple outgoing charged particles, using a signature of a displaced vertex and missing transverse momentum, in the upgraded ATLAS detector for the High-Luminosity LHC. The dataset is taken as 3000 fb$^{-1}$ of $pp$ collisions at $\sqrt{s}$ = 14 TeV. The replacement of the existing tracking detector with a full-silicon inner tracker (ITk) significantly changes the search sensitivity, which is estimated using extrapolations of the reconstruction capabilities for displaced tracks and displaced vertices from a combination of the performance of the current detector and a simulation of the proposed upgraded detector. Results are projected in the context of gluino $R$-hadron pair production. For long-lived gluinos which decay to SM quarks and a 100 GeV stable neutralino, ATLAS should have the sensitivity to discover $R$-hadrons with lifetimes from 0.1 to 10 ns with masses up to 2.8 TeV. In the absence of long-lived gluino production, the 95% CL upper limit on gluino masses will reach 3.4 TeV.

# 1 Introduction

One of the most intriguing scenarios of new physics to look for at the Large Hadron Collider (LHC) is new particles with measurably long lifetimes. Whether in the Standard Model (SM) or in new physics models, particles can acquire discernible lifetimes from a variety of mechanisms, including weak effective couplings to the final state due to heavy intermediate particles, which can arise in models with mass hierarchies, or due to small coupling constants, which can arise from conserved or nearly-conserved symmetries. Long lifetimes can also manifest due to decays that are phase-space suppressed or from new particles which interact only weakly with the SM particles via mediators.

If a new long-lived particle decays within the detector but at an observable distance from the proton-proton interaction point, and if its decay products include multiple charged particles reconstructed as tracks, it can produce a distinctive signature of an event containing at least one displaced vertex (DV). There are several recent papers at the LHC which have searched for displaced vertices, including Refs. [1–3].

This note presents a projection of the expected sensitivity of the analysis published in Ref. [2] in the context of the planned upgrades to the LHC and the ATLAS detector, presenting in detail and extending a study shown in Ref. [4]. The study assumes that the High-Luminosity LHC (HL-LHC) will deliver 3000 $\mathrm{fb}^{-1}$ of collisions at $\sqrt{s}$ = 14 TeV to the ATLAS experiment, and it explores the effect that the geometry of the new inner tracker (ITk) of the upgraded ATLAS detector will have in extending the acceptance for reconstructing DVs.

The search presented here requires at least one displaced vertex reconstructed within the ITk, and events are required to have at least moderate missing transverse momentum ($E_{\mathrm{T}}^{\mathrm{miss}}$), which serves as a discriminant against background as well as an object on which to trigger. The analysis sensitivity is projected for a benchmark supersymmetry model of pair production of long-lived gluinos, as shown in Figure 1. Each gluino hadronizes into an $R$-hadron and decays through a heavy virtual squark into a pair of SM quarks and a stable neutralino with a mass of 100 GeV. Results are presented for discovery and exclusion potential are explored for several different gluino lifetimes and for varying masses of the gluino.

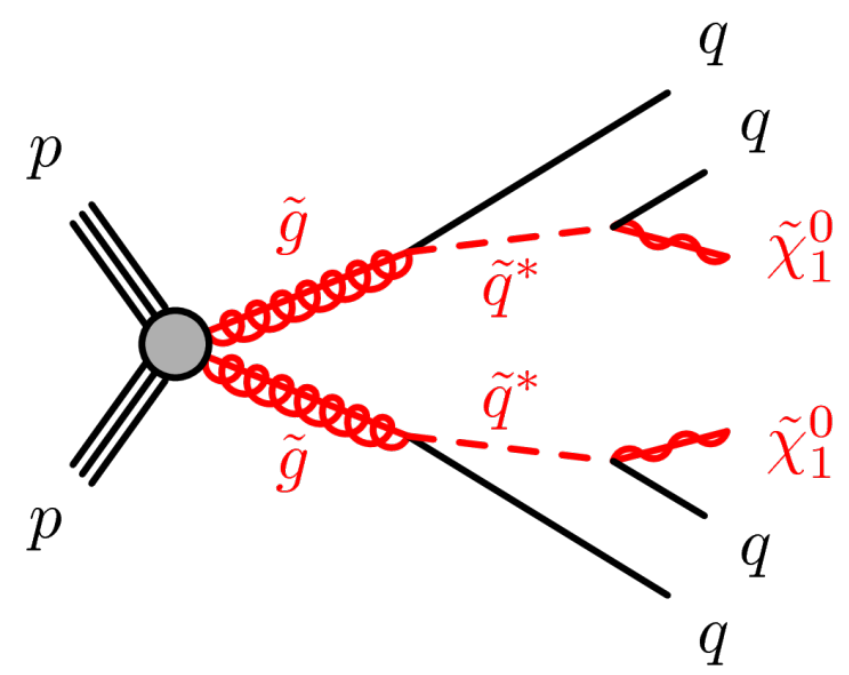

Figure 1: The benchmark model considered in this projection is a supersymmetric scenario with a long-lived gluino, which hadronizes after production into an $R$-hadron, and then decays through a virtual squark into a pair of SM quarks and a neutralino.

## 2 ATLAS Detector

ATLAS is a general-purpose detector[1] with a forward–backward symmetric cylindrical symmetry described in detail in Ref. [5]. For the purpose of this study, only the inner detector tracking system is described in detail below.

The current ATLAS tracker, referred to in this note as the Run 2 inner detector (ID), is composed of three detector systems organized in concentric layers, and covers the pseudorapidity range $|\eta| < 2.5$, immersed in a 2 T superconducting solenoid. The outermost layer, the transition radiation tracker (TRT), consists of densely packed proportional gas-filled straw tubes [6]. The semiconductor tracker (SCT) is equipped with silicon microstrip detectors and occupies the radial region from roughly 300 mm to 550 mm [7]. The innermost layer consists of a silicon pixel detector [8], which was upgraded in 2015 with an additional innermost pixel layer [9]. The pixel detector typically provides four precision measurements (hits) for each track at radial distances of 33 mm, 50 mm, 88 mm and 122 mm from the LHC beam line, while the SCT has four layers that typically provide eight hits in total, as each layer is composed of of double-sided stereo strip sensors.

The central barrel portion of the Phase II ATLAS tracker, the ITk, will be composed of four double-sided silicon strip detectors and five layers of silicon pixel sensors [4, 10]. There will be no gaseous detector to replace the TRT in the ITk; the new-generation silicon strip and pixel sensors will fill the entire volume inside the solenoid. The pixel coverage will be extended from $|\eta| < 2.5$ to $|\eta| < 4.0$, and the pixel layers in the central region will be located at approximately 39 mm, 99 mm, 160 mm, 220 mm, and 279 mm. The distance between one silicon layer and the next layer will increase relative to the ID. For charged particles coming from the primary vertex at $|\eta| \approx 0$, the typical number of silicon hits available for tracking will increase from twelve to thirteen, and will increase significantly more at higher $|\eta|$. While the exact configuration of the pixel layers in the endcap region is not yet finalized, this study used the Inclined Duals geometry described in Ref. [4]. The ID and ITk layouts are shown in Figure 2 and the expected performance are summarized in Ref. [11]. In addition to the new layout for ITk, the pixel size will decrease relative to the current detector.

## 3 Analysis overview

In this study, the event selection closely follows the requirements in the recent Run 2 search for a DV and $E_{\mathrm{T}}^{\mathrm{miss}}$ [2]. Events are required to have at least one DV in the ITk and at least five tracks must be associated to that vertex. To exclude hadronic interactions of SM particles, the vertex must not be located within a region of the detector filled with solid materials, and the invariant mass of the reconstructed vertex must be larger than 10 GeV. The event must pass the $E_{\mathrm{T}}^{\mathrm{miss}}$ trigger and have reconstructed $E_{\mathrm{T}}^{\mathrm{miss}} > 200$ GeV.

Estimating the future sensitivity to the distinctive signature of a DV in a new detector with new reconstruction tools requires creative methods of extrapolation and necessitates several assumptions about the future layout and performance of the detector. As the tracks comprising the signal DVs have fewer hits and much larger impact parameters than tracks from the primary vertex, the reconstruction algorithm needs to cover a much wider tracking phase space than tracking for prompt particles. To efficiently reconstruct signal tracks, a custom, extended configuration of the ATLAS tracking software is required. This custom configuration, hereafter referred to as *displaced tracking*, is optimized for the Run 2 detector and software [12]. Similarly, a custom algorithm is required for reconstruction of the displaced vertices. These customizations depend heavily on the geometry of the tracker, and neither a dedicated displaced tracking nor DV reconstruction setup has yet been developed for the Phase II detectors.

[1] ATLAS uses a right-handed coordinate system with its origin at the nominal interaction point (IP) in the centre of the detector and the $z$-axis along the beam pipe. The $x$-axis points from the IP to the centre of the LHC ring, and the $y$-axis points upward. Cylindrical coordinates $(r, \phi)$ are used in the transverse plane, $\phi$ being the azimuthal angle around the $z$-axis. The pseudorapidity is defined in terms of the polar angle $\theta$ as $\eta = -\ln\tan(\theta/2)$.

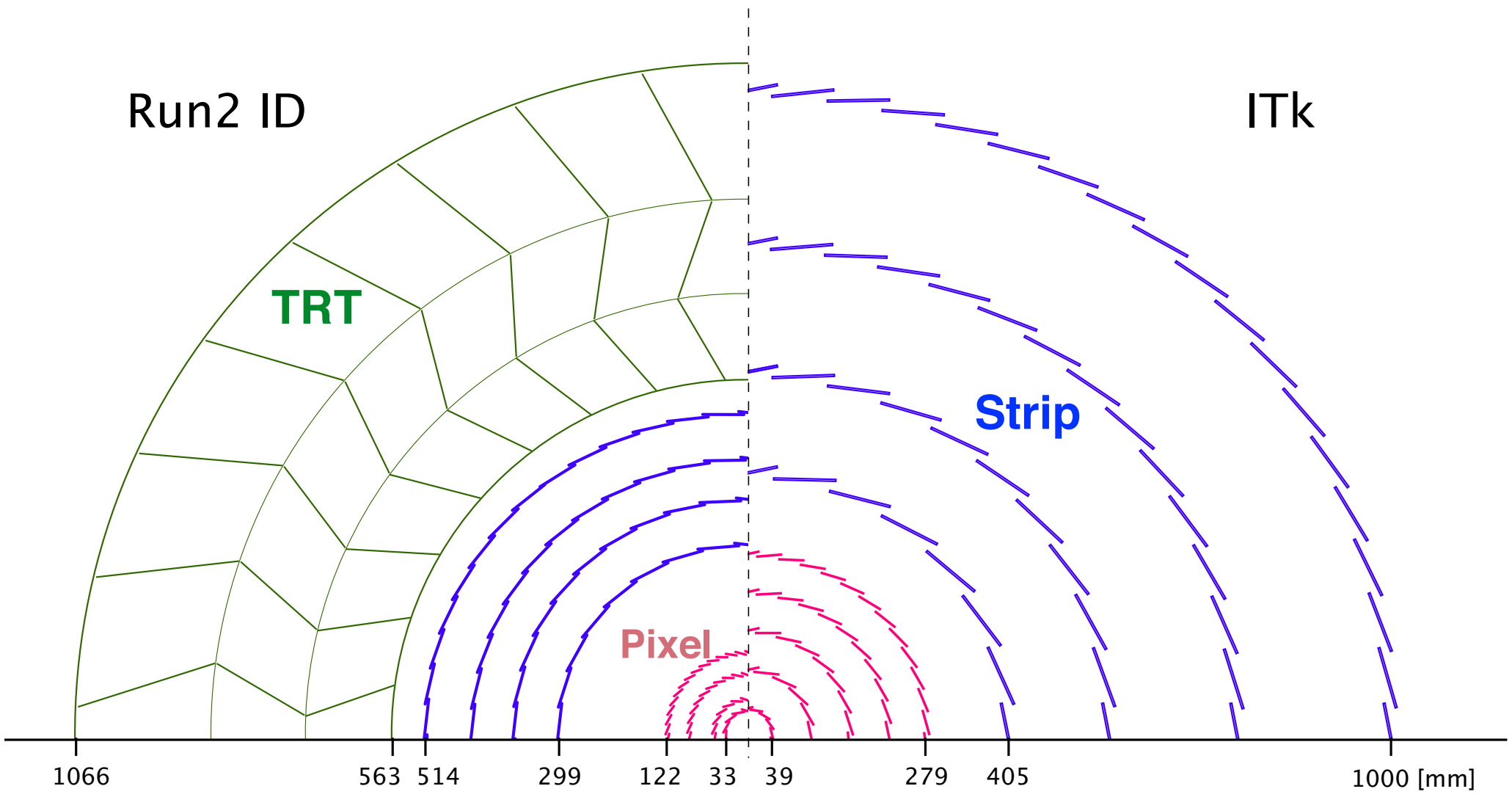

Figure 2: The radial layouts of the current ATLAS tracker and the proposed upgrade, the ITk.

Many ATLAS Phase II projection studies rely on simulated events which have been fully reconstructed by tracking software specially developed for finding prompt tracks in the Phase II detector. In other studies, generated Monte Carlo (MC) events are used to obtain particle-level kinematic properties, which are smeared by functions which estimate the future detector's performance. In this analysis, a hybrid approach is developed to estimate the prospective acceptance and efficiency for reconstructing and selecting displaced vertices. In this approach, particle-level Monte Carlo events are used to obtain kinematic distributions for the signal. The displaced tracking performance is estimated by factorizing the current displaced tracking performance into acceptance and efficiency terms, and assuming that the efficiency performance of the Run 2 algorithm will be reproduced for ITk for particles which pass the acceptance. The tracking acceptance is based on the number of hits left by a charged particle traversing the silicon sensors, and is calculated for the tracks of interest using a full simulation of the ITk geometry. The current DV performance is parameterized and extrapolated to the new detector geometry.

Other selections are either directly applied at particle-level to the signal events, or use the results of the Run 2 reinterpretation material which are publicly available [13] to estimate the efficiency. The background estimation is extrapolated from the Run 2 data-driven result, and systematic uncertainties are taken directly from the existing analysis.

## 4 Simulation samples

This study makes use of Monte Carlo simulation samples to obtain the kinematic properties of signal events, which are then used to estimate the efficiency for selecting signal events. The pair production of gluinos from proton-proton collisions at $\sqrt{s}$ = 13 TeV was simulated in Pythia 6.428 [14] at leading order with the AUET2B [15] set of tuned parameters for the underlying event and the CTEQ6L1 [16] parton distribution function (PDF) set. The gluino mass, $m_{\tilde{g}}$, ranges from 2.0 to 3.8 TeV. After production, the gluino hadronizes into an $R$-hadron and is propagated through the ATLAS detector by Geant4 [17, 18] until it decays. Pythia 6.428 is called to decay the gluino into a pair of SM quarks and a neutralino and models the three-body decay of the gluino, fragmentation of the remnants of the light-quark system, and hadronization of the decay products. The gluino lifetime ranges from 0.1 ns to 10 ns, and the neutralino mass is fixed to 100 GeV. In this study, only particle-level information about the $R$-hadron's decay products is used.

To normalize the expected number of signal events in the full HL-LHC dataset, the cross-sections for pair production of gluinos are calculated at next-to-leading order at $\sqrt{s}$ = 14 TeV using Prospino [19], assuming no contribution from the squark. The resummation of soft-gluon emission is taken into account at next-to-leading-logarithm accuracy (NLO+NLL) for cross-sections for gluino masses up to 3.5 TeV, beyond which a linear extrapolation of the correction for the NLL term is performed [20]. Uncertainties on the cross-section prediction are taken by varying the choice of PDF set and factorization and renormalization scales, as described in Ref. [21], with a reduction of 50% applied to the uncertainties to account for improvements by the time the analysis will be performed. This approach is also consistent with the expected improvements in PDF uncertainties described in Ref. [22].

## 5 Projection of displaced tracking performance

Both standard and displaced tracking reconstruction in the ATLAS ID require that each track have at least seven silicon-detector hits, where each side of the double-sided SCT strips counts as a hit. For a typical particle in the central part of the current ATLAS detector travelling away from the interaction point, the maximum radial distance at which a particle can originate and still have seven silicon hits is approximately 300 mm, the first layer of the SCT. Previous studies on reconstructing displaced tracks in the ID found that for simulated particles which do produce seven silicon hits, the probability of reconstructing a track is between 90–100% for a radius of production in the transverse plane, $r_{\mathrm{prod}}$, from 0 mm to 300 mm [12].

For this study, it is assumed that the tracking algorithms in Phase II will be able to match the current performance. Therefore, the total efficiency for reconstructing displaced tracks is factorized into an algorithmic efficiency, $\epsilon_{\mathrm{alg}}$, and a fiducial acceptance, $A_{\mathrm{fid}}$, where $\epsilon_{\mathrm{alg}}$ is the probability of reconstructing a track given a particle has deposited at least seven silicon hits, and $A_{\mathrm{fid}}$ is the fraction of particles which deposit at least seven hits. $\epsilon_{\mathrm{alg}}$ is taken to be 100% in this study. While this is consistent with the current performance, the increased instantaneous luminosity of the HL-LHC will make it more difficult to efficiently find tracks with large impact parameter. However, this study assumes that the following factors will work to mitigate the effect of high pileup: 1) the increase in the number of silicon layers is expected to increase the ability to separate tracks from actual prompt particles from combinatoric fakes, thereby mitigating the potential problem of fake prompt tracks claiming hits from displaced particles, 2) the increased resolution of the pixel detector will help to mitigate efficiency loss due to pileup, and 3) the collaboration will continue to adapt the tracking and vertexing algorithms until the era of the HL-LHC to deal with the more difficult conditions.

To estimate $A_{\mathrm{fid}}$ for the ITk layout, the particles from the $R$-hadron decay are propagated through a detailed simulation of the ITk Inclined Duals geometry using their generator-level radius of production and momentum vector. Only charged decay products with $p_{\mathrm{T}} > 1$ GeV are considered. Along each particle's trajectory, the number of active silicon layers traversed, $N_{\mathrm{Si}}$, is recorded, as well as the integrated amount of active and passive material traversed from one hit to the other in terms of the nuclear interaction length, $N_\lambda$. Each sensor's hit efficiency is assumed to be 100%. The effects of multiple scattering are not included. The probability that each particle reaches the next silicon strip layer of the tracker volume without undergoing a hadronic interaction is estimated as $e^{-N_\lambda}$. $A_{\mathrm{fid}}$ is taken as the ratio of the number of particles with $N_{\mathrm{Si}} \geq 7$ that do not undergo a hadronic interaction divided by the total number of particles.

In Figure 3, the total tracking efficiency calculated using this method is shown, for the decay products of $R$-hadrons with a mass of 2.0 TeV which decay with a mean proper lifetime of 1 ns. The tracking efficiency is shown as a function of the $R$-hadron decay position.[2] As $\epsilon_{\mathrm{alg}}$ is taken to be 100%, the total tracking efficiency is simply $A_{\mathrm{fid}}$, and is shown with and without the effects of hadronic interactions due to material. The tracking efficiency is shown for the ITk as well as the Run 2 ID geometry. The steep drop off in efficiency in the present ID at around 300 mm corresponds to the farthest radial extent of the first layer of the SCT, after which it is unlikely that a typical particle would traverse seven strip layers. In the ITk, the equivalent drop-off does not occur until after 400 mm due to the larger spacing between the silicon layers. Similar tracking efficiencies are observed within statistical uncertainty for the decay products of $R$-hadron with a range of of gluino masses, lifetimes, and neutralino masses.

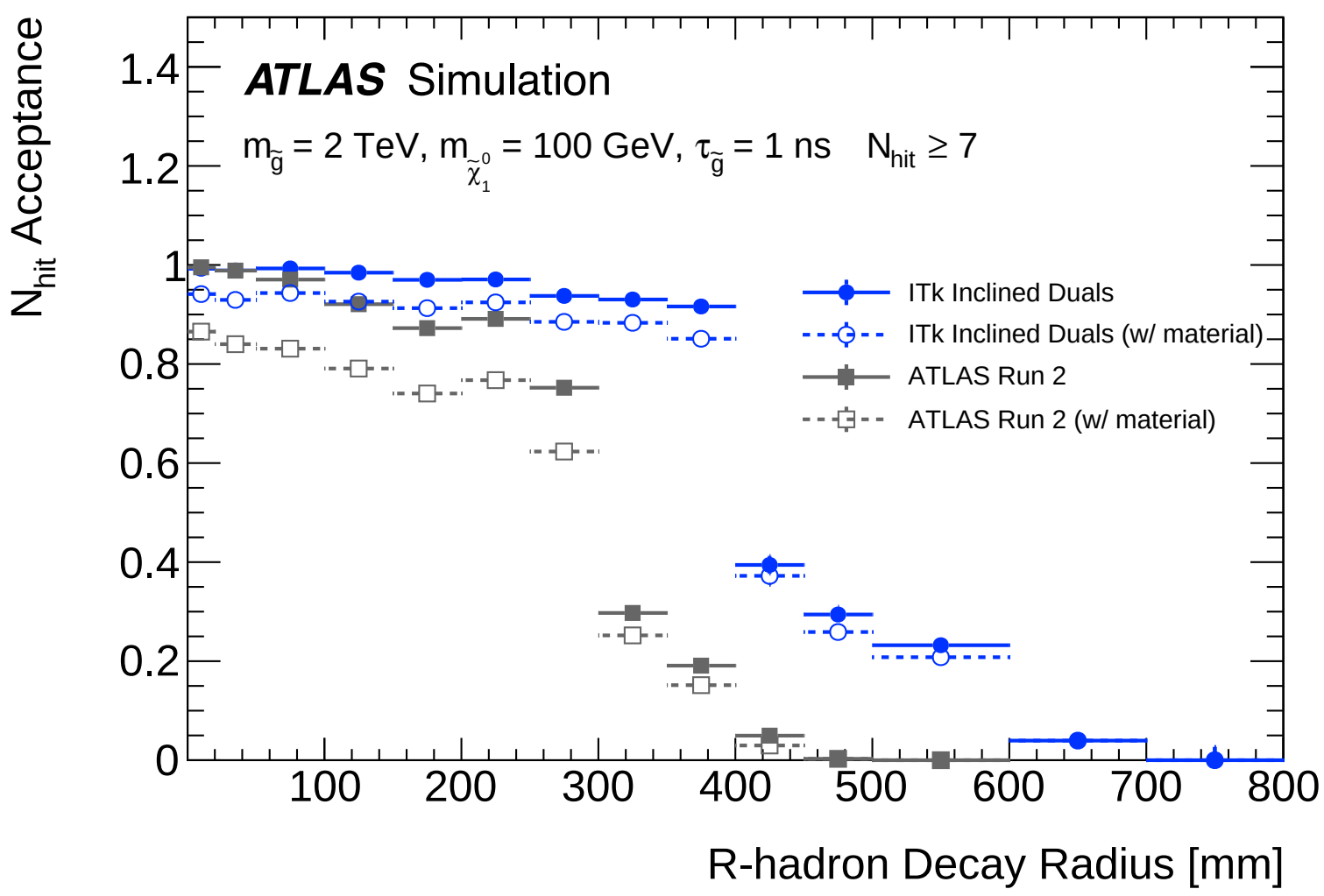

Figure 3: The probability that a charged particle, with $p_{\mathrm{T}} > 1$ GeV produced in the decay of a 2.0 TeV $R$-hadron with a lifetime of 1 ns, passes through at least seven silicon layers, as a function of the decay radius of the $R$-hadron, for both the Run 2 and ITk detector layouts [4]. The probability is shown with and without the simulated effect of material producing hadronic interactions.

[2] The values of the $R$-hadron decay radius, $r_{\mathrm{DV}}$, and the radius of the decay products production,$r_{\mathrm{prod}}$, are identical at the MC generator level for the $R$-hadron model used in this note, but $r_{\mathrm{DV}}$ is preferentially used when focusing on the vertex properties, while $r_{\mathrm{prod}}$ is used when focusing on the outgoing particles.

# 6 Projection of displaced vertexing performance

The configuration of the algorithm for reconstructing displaced vertices has recently been optimized within ATLAS for finding signals such as decaying $R$-hadrons. The efficiency of reconstructing a displaced vertex is measured in simulation as a function of the number of reconstructed tracks which belong to the vertex at generator-level, $n_{\mathrm{tracks}}$, and the radius of the vertex position in the transverse plane, $r_{\mathrm{DV}}$.

For this study, it is assumed that similar performance of the vertexing algorithms can be achieved in Phase II. However, given the different geometry of ITk relative to the current detector, the vertexing efficiency measured for Run 2 can not be applied directly to this study. The Run 2 vertex efficiency is parametrized and extrapolated as described below.

The Run 2 vertex efficiency is parameterized as a function of $r_{\mathrm{DV}}$ for bins of $n_{\mathrm{tracks}}$. One of the main reasons that the reconstruction efficiency depends on $n_{\mathrm{tracks}}$ is due to the fact that the number of trials of vertex forming using seed tracks increases with $n_{\mathrm{tracks}}$. The efficiency is fit with a function which smoothly combines an error function at low $r_{\mathrm{DV}}$ to model the initial rise in efficiency, a linear plateau, and an error function at high $r_{\mathrm{DV}}$ to model the falling off of efficiency near the beginning of the SCT. The fit values are compared across all bins of $n_{\mathrm{tracks}}$, and some smoothing of values is performed for bins which suffer from poor statistics.

To extrapolate from the Run 2 efficiency to the expected performance at ITk, the same fit values are used for each bin of $n_{\mathrm{tracks}}$. However, the mean of the error function used to model the turn-off is moved from 300 mm to 400 mm to reflect the change in the location of the inner silicon strip layer.

The vertex efficiency parameterized in this way in shown in Figure 4 for the case where $n_{\mathrm{tracks}} = 10$. The efficiency is shown as fit in the Run 2 measurement and as extrapolated to ITk. Also shown is the plateau of the efficiency as a function of $n_{\mathrm{tracks}}$ for the ITk extrapolation, assumed to be the same as that measured in Run 2.

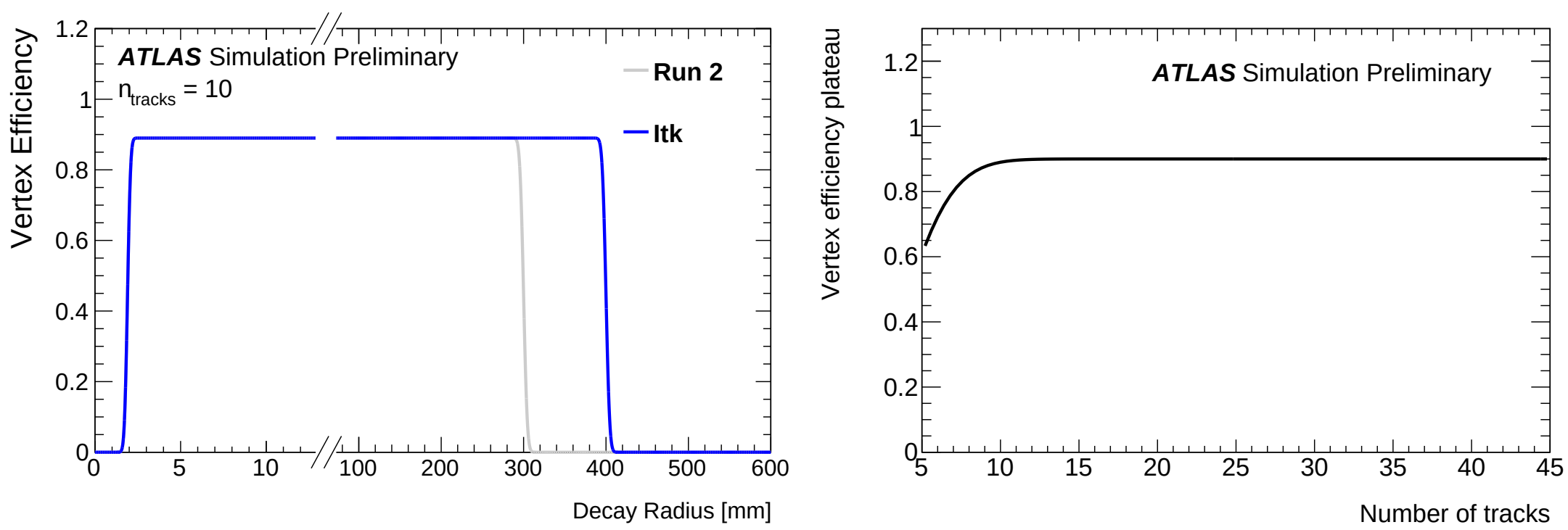

(a) (b)

Figure 4: (a) The parametrized efficiency for reconstructing a displaced vertex with $n_{\mathrm{tracks}} = 10$, as a function of the decay radius of the parent particle, as measured in Run 2 simulation and extrapolated to the ITk geometry. (b) The plateau of the vertex efficiency as a function of $n_{\mathrm{tracks}}$ for the ITk geometry.

## 7 Event selection

The event selection efficiency is estimated on signal events for the following selections:

- Event must pass $E_{\mathrm{T}}^{\mathrm{miss}}$ trigger and have offline $E_{\mathrm{T}}^{\mathrm{miss}} > 250$ GeV.
- Event must have at least one displaced vertex which passes the following requirements:
  - the vertex is not located in a region of the detector with material;
  - the vertex must contain at least 5 reconstructed tracks;
  - the invariant mass of the reconstructed tracks in the vertex, $m_{\mathrm{DV}}$, must be larger than 10 GeV.
- Each reconstructed track must satisfy:
  - 6 mm $< r_{\mathrm{prod}} <$ 400 mm;
  - $p_{\mathrm{T}} > 1$ GeV;
  - $|\eta| < 5$.

The efficiency of passing the $E_{\mathrm{T}}^{\mathrm{miss}}$ trigger and offline $E_{\mathrm{T}}^{\mathrm{miss}}$ requirements is taken from the Run 2 analysis, as parameterized in Ref. [13] as a function of the generator-level $E_{\mathrm{T}}^{\mathrm{miss}}$ and the decay radius of the $R$-hadron in the event which decays the farthest from the interaction point. The generator-level $E_{\mathrm{T}}^{\mathrm{miss}}$ is calculated as the magnitude of the transverse component of the vector sum of the momenta of the stable, weakly-interacting particles in the final state. For an event in which the farthest $R$-hadron decays before the calorimeter, the $E_{\mathrm{T}}^{\mathrm{miss}}$ efficiency plateaus near 100% for generator-level $E_{\mathrm{T}}^{\mathrm{miss}}$ of 500 GeV, while the plateau is between 80–90% for decays inside or after the calorimeter.

The efficiency of the material veto is estimated with a material map derived from a detailed geometry model of the ITk implemented in Geant4 [17, 18], which is used for the full Phase II MC simulations. All locations of passive and active material within the boundaries of 1200 mm in the radial ($r$) dimension and 3000 mm in longitudinal dimension ($z$) are stored in a binned $r$–$z$ projection of the ITk. For each $R$-hadron decay in an event, the position of the $R$-hadron decay is checked against the map. If the decay vertex is consistent with any material, the vertex is not considered further. For a $R$-hadron with a mass of 2.0 TeV and lifetime of 1 ns which decays into a 100 GeV neutralino, roughly 60% of the signal events pass the material veto.

For each $R$-hadron which passes the material selection, an estimation is made of which of its decay products would be reconstructed as a track with a random sampling of the displaced tracking efficiencies. It is parameterized as a function of the particle production radius, as calculated in Section 5. Only charged decay products with $p_{\mathrm{T}} > 1$ GeV, $|\eta| < 5$, and with 6 mm $< r_{\mathrm{prod}} <$ 400 mm are considered. The requirements on the track's production radius at truth level approximates a selection on the radius of a reconstructed vertex. If at least five charged particles of the decay are within the fiducial acceptance and modeled as reconstructed, then the invariant mass of those particles, $m_{\mathrm{DV}}$, is calculated, assuming all are pions.

If $m_{\mathrm{DV}} > 10$ GeV, the vertex is accepted based on a random-sampling of the vertex reconstruction efficiency computed in Section 6, which depends on the number of tracks and the vertex radius $r_{\mathrm{DV}}$. If at least one vertex is accepted in an event, the event is considered to have passed all selections, and an event weight is applied based on the event $E_{\mathrm{T}}^{\mathrm{miss}}$ efficiency.

## 8 Background

In the search for events with a displaced vertex and $E_{\mathrm{T}}^{\mathrm{miss}}$ in Run 2, the main source of background arises from hadronic interactions of SM particles with residual material not accounted for in the material map (such as gas molecules) and from low-mass vertices, such as from the decay of a SM hadron, which are either merged, or which are crossed by an unrelated, high-$p_{\mathrm{T}}$ track which promotes the vertex to higher $m_{\mathrm{DV}}$. The contribution from different background sources was estimated independently in control regions in data and tested in validation regions. The total expected background for 36 fb$^{-1}$ was estimated to be $0.02^{+0.02}_{-0.01}$ events.

Given that the background for this search is entirely instrumental in nature, it would not be reliable to perform a simulation-based projection of the expected background for a different detector and different reconstruction algorithms. Therefore, for the purpose of this projection, two different extrapolations of the size of current background are performed.

The default extrapolation assumes that the background will scale linearly with the size of the dataset, resulting in an expected background of $1.8^{+1.8}_{-0.9}$ events. However, several handles could be tightened in the analysis selection to continue to reject background without introducing appreciable signal efficiency loss. For example, additional requirements on the vertex goodness-of-fit or the compatibility of each track with the vertex could be imposed to further reduce backgrounds from low-mass vertices which are merged or crossed by an unrelated track. Therefore, a more optimistic scenario is also considered in which the total background is kept to the current level of $0.02^{+0.02}_{-0.01}$ events.

## 9 Results

The expected number of selected $R$-hadron events in the full HL-LHC dataset of 3000 fb$^{-1}$ is shown in Figure 5 for different gluino masses and lifetimes. This note focuses on the 0.1-10 ns lifetime range, which is expected to be the most sensitive range for this analysis. The sensitivity is expected to gradually drop off at lifetimes shorter than 0.1 ns, and below 0.01 ns searches for other signatures are expected to perform better.

To estimate the potential discovery and exclusion sensitivity, an estimate is needed of the uncertainty on the number of expected signal events. The signal selection uncertainties are taken to have the same relative size as in the existing Run 2 analysis. The uncertainties on the cross-sections for gluino pair production are described in Section 4. Finally, the uncertainties on the number of expected background events are taken to be the same relative amount as in the existing analysis, for both scenarios of the background size.

Using the number of expected signal and background events with their respective uncertainties, the expected exclusion limit at 95% confidence level (CL) on the gluino mass, as a function of lifetime, is calculated assuming no signal presence. In the case that signal is present, the $3\sigma$ evidence and $5\sigma$ discovery reaches are also calculated. Both the exclusion and discovery reach are estimated in the two background scenarios described in Section 8: linearly scaling it with integrated luminosity, and keeping the same value as in the Run 2 analysis. The exclusion and observation reach are calculated with pseudo-experiments following the $\mathrm{CL_s}$ prescription [23]. The results are shown in Figure 6 for both background scenarios.

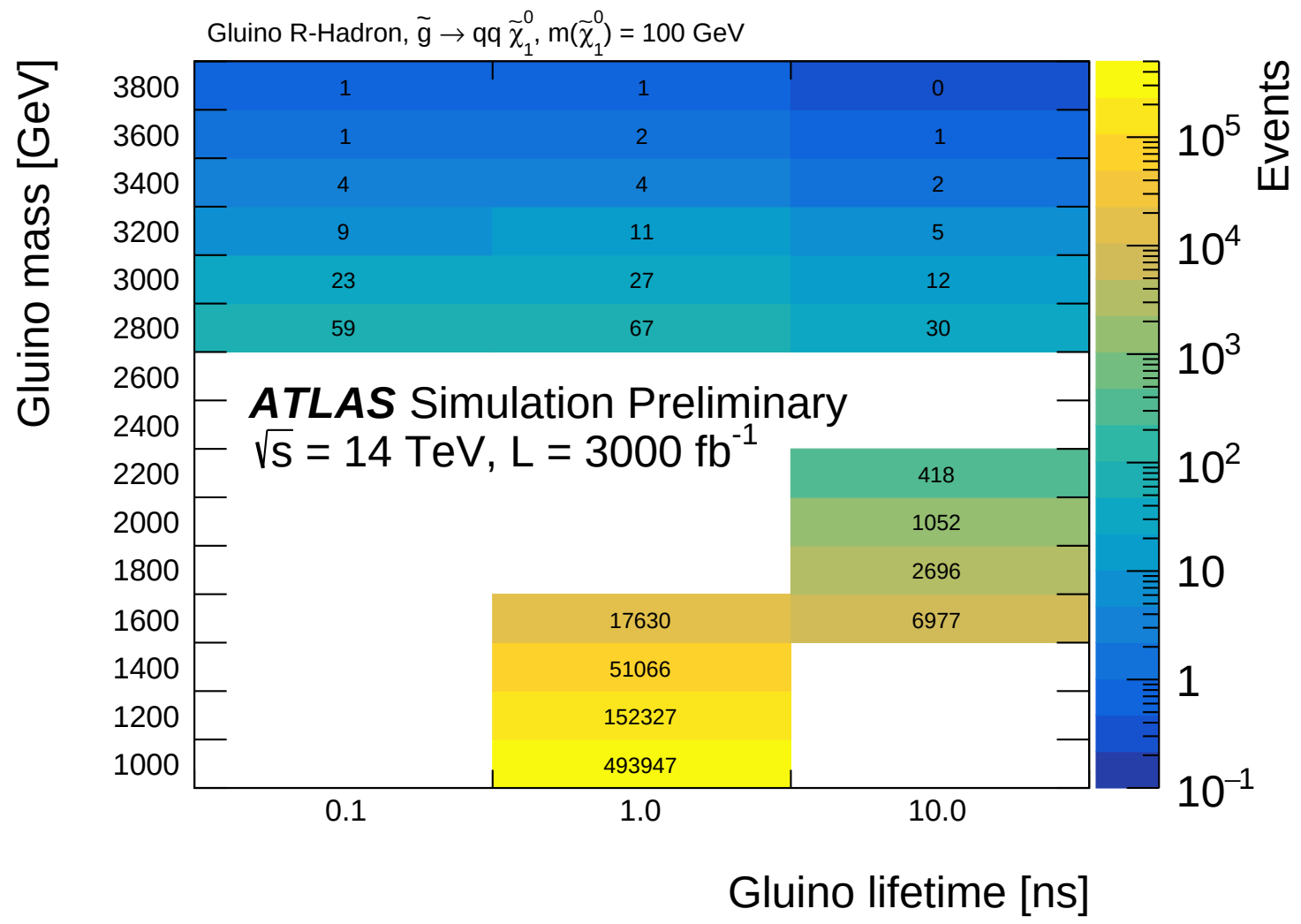

Figure 5: The number of expected $R$-hadron events selected for 3000 fb$^{-1}$ at $\sqrt{s}$ = 14 TeV as a function of the gluino lifetime and gluino mass for gluinos which decay to SM quarks and a neutralino with a mass of 100 GeV.

For an expected background of $1.8^{+1.8}_{-0.9}$ events, ATLAS should have sensitivity to a $5\sigma$ discovery of a gluino $R$-hadron decaying to SM quarks and a 100 GeV neutralino with masses up to 2.8 TeV over a range of lifetimes from 0.1 ns to above 10 ns. For the optimistic background scenario of $0.02^{+0.02}_{-0.01}$ background events expected, the projected discovery reach goes up to 2.9 TeV at a lifetime of 1 ns.

In the absence of long-lived gluino production, this analysis is projected to be able to exclude at the 95% CL $R$-hadrons with masses up to 3.3 TeV for lifetimes from 3 ns to above 10 ns, and up to 3.4 TeV for lifetimes between 0.1 ns and 3 ns, assuming the scaled background estimate. For the smaller background scenario, the 95% CL exclusion is above 3.4 TeV for the whole lifetime range, and up to 3.5 TeV around 1 ns. The significant increase in sensitivity relative to the ATLAS result with 33 fb$^{-1}$ at $\sqrt{s}$ = 13 TeV comes in part from the increase in collision energy and integrated luminosity. For longer lifetimes, a significant gain in selection efficiency and therefore reach is due to the larger volume of the silicon tracker, which allows displaced tracks and displaced vertices to be reconstructed at larger radii. This pushes the radius at which tracks from long-lived particles can be efficiently reconstructed from 300 to 400 mm, with corresponding gain in acceptance for longer lifetimes. While the existing analysis with the Run 2 ID loses sensitivity for $R$-hadrons with lifetimes of 10 ns or longer, it is expected that a similar analysis with the ITk will have sensitivity for lifetimes up to and beyond 10 ns.

While the results presented here were studied only for a fixed neutralino mass of 100 GeV, based on the results in Ref. [2], comparable sensitivity is expected over a large range of neutralino masses. As the neutralino mass increases for a fixed gluino mass, the multiplicity and momentum of the visible SM particles decreases, which in turn decreases the efficiency of the requirements on the track multiplicity, vertexing reconstruction, and vertex invariant mass as the difference between the neutralino mass and the gluino mass, $m_{DV}$, falls below 400 GeV. To efficiently select events with low $m_{DV}$, it is especially important to retain high $\epsilon_{\mathrm{alg}}$ and high vertexing efficiency in the HL-LHC conditions. Moreover, new background estimation techniques could allow a future analysis to relax the requirement on track multiplicity. While such studies require data and full reconstruction and are therefore beyond the scope of this note, future advances in reconstruction or analysis techniques could significantly improve the sensitivity to signals with fewer visible decay products due to a compressed mass spectrum or different signal models.

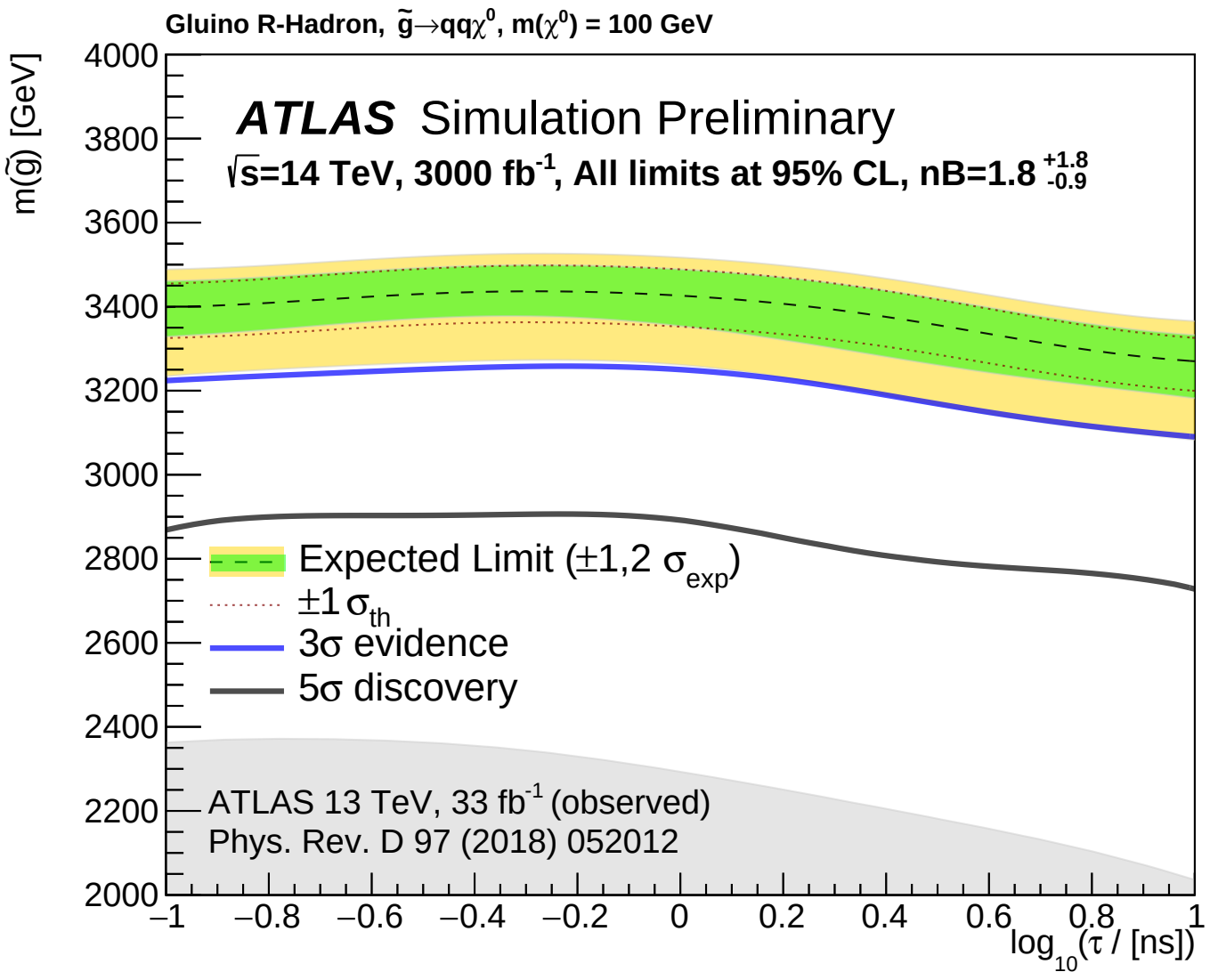

(a)

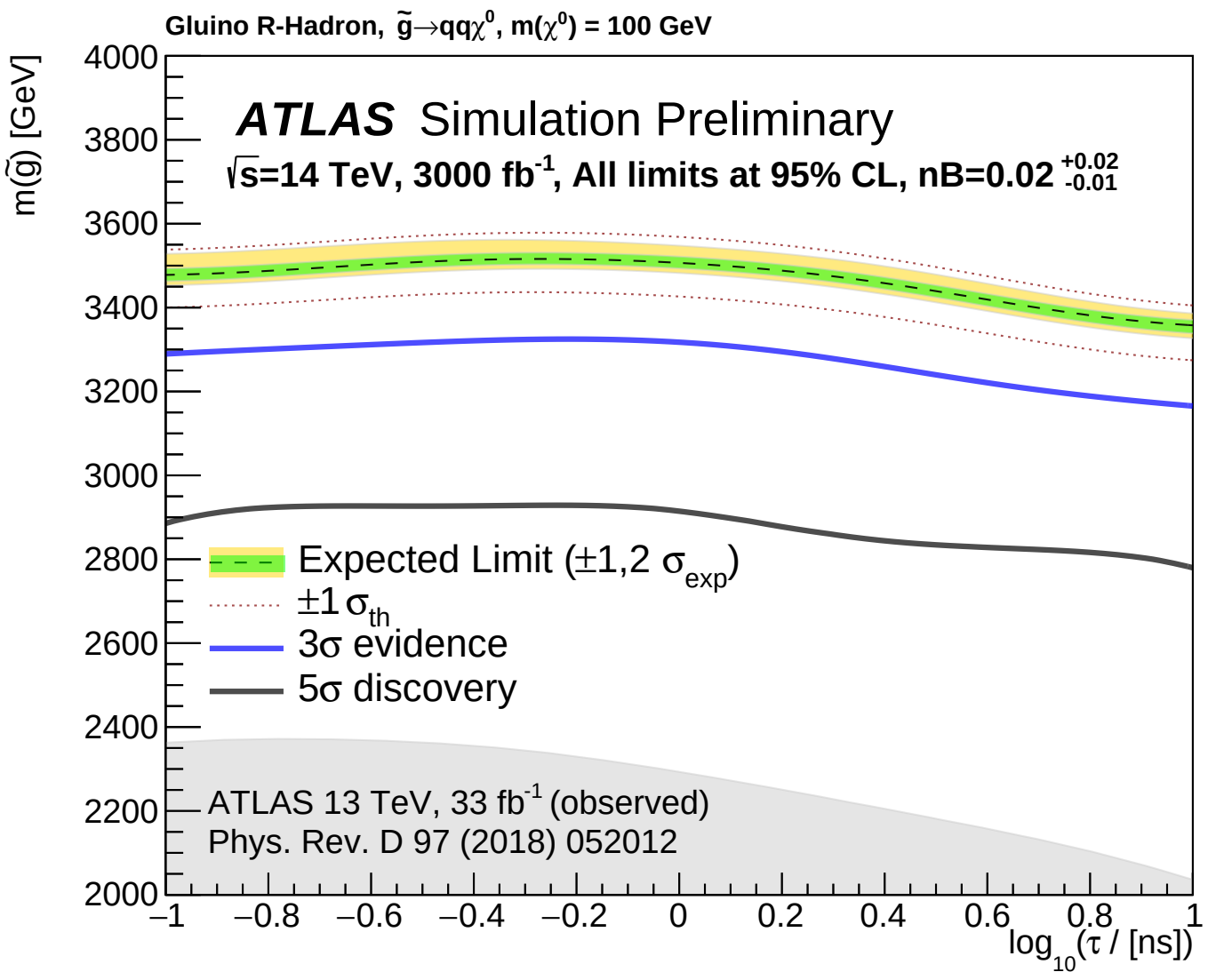

(b)

Figure 6: The projected sensitivity for the upper limit on the mass of a gluino $R$-hadron that can be observed with $3\sigma$ and $5\sigma$ confidence or excluded at 95% CL, as a function of the gluino lifetime, for (a) a background of $1.8^{+1.8}_{-0.9}$ events and (b) a background of $0.02^{+0.02}_{-0.01}$ events. These results are valid for a gluino which decays to SM quarks and a stable neutralino with a mass of 100 GeV. Results assume 3000 fb$^{-1}$ of collisions at $\sqrt{s}$ = 14 TeV collected with the upgraded ATLAS detector, and are compared to the observed ATLAS exclusion limits for a dataset of 33 fb$^{-1}$ at $\sqrt{s}$ = 13 TeV.

## 10 Conclusion

The upgraded energy and the larger dataset expected from the HL-LHC will increase significantly the sensitivity to new, heavy, long-lived particles. For models with a signature of a displaced vertex within the tracker and moderate $E_{\mathrm{T}}^{\mathrm{miss}}$, the larger silicon volume of the upgraded ATLAS detector will further extend the reach for lifetimes longer than 10 ns. For 3000 fb$^{-1}$ of $pp$ collisions at $\sqrt{s}$ = 14 TeV, ATLAS could discover gluino $R$-hadrons with masses up to 2.8 TeV and lifetimes from 0.1 ns to 10 ns, when the gluino decays to SM quarks and a 100 GeV stable neutralino. In the absence of long-lived gluino production, the 95% CL upper limit on gluino masses is projected to reach or exceed 3.4 TeV.

# CMS Physics Analysis Summary

Contact: cms-future-conveners@cern.ch

## Search sensitivity for dark photons decaying to displaced muons with CMS at the high-luminosity LHC

The CMS Collaboration

### Abstract

This note presents sensitivity studies for a search for pairs of displaced muons originating from long-lived dark photons using the Phase-2 CMS detector at the high-luminosity LHC with an integrated luminosity of 3000 $\mathrm{fb}^{-1}$. Projected sensitivities are obtained for broad ranges of dark photon masses ($1-30$ GeV) and lifetimes ($c\tau = 0.01 - 10$ m) in the context of Dark Supersymmetry models.

# 1 Introduction

A growing class of new physics models predicts long-lived particles (LLPs). A possible experimental signature of such particles at the LHC is an emergence of standard model (SM) particles at a large distance from the point of the primary proton-proton collision. Due to their low production cross section, LLPs are often beyond the sensitivity of the current searches. The high-luminosity LHC (HL-LHC) will provide proton-proton collisions at a center-of-mass energy of 14 TeV with an expected total integrated luminosity of 3000 $fb^{-1}$. Therefore, it is foreseen to be a powerful instrument to probe the production of LLPs, benefiting from both the increased center-of-mass energy, leading to larger cross-sections, and the significantly larger amount of data collected compared to the LHC.

We present a sensitivity study for a search for displaced muons that emerge from the decay of long-lived particles. The identification of such muons is challenging both at the trigger and final reconstruction level, especially if the LLPs decay outside the central tracking detector. Additional hits coming from the new CMS endcap muon stations [1], in combination with improved reconstruction algorithms, will allow one to extend triggering and efficient reconstruction of displaced muon tracks in the forward region.

Previous studies of displaced muon signatures at the HL-LHC, including a search sensitivity study in the context of Supersymmetry (SUSY) models with long-lived smuons, are presented in the Muon Upgrade TDR [2]. In this study, we explore another class of SUSY models containing an additional $U_D(1)$ symmetry [3, 4] and giving rise to massive long-lived bosons, so-called dark photons. If they have sufficiently long lifetimes, dark photons could yield signatures with pairs of displaced muons.

# 2 Muon upgrade for the CMS Phase-2 detector

The CMS detector will be upgraded in order to cope with the challenges during data taking at the HL-LHC. The upgraded existing detectors together with new detectors will allow CMS to maintain or even improve the trigger, reconstruction and identification capabilities for muons. The muon pseudorapidity acceptance range will be extended from $|\eta| < 2.4$ to $|\eta| < 2.8$, where $\eta = -\ln[\tan(\theta/2)]$ and $\theta$ denotes the polar angle with respect to the counterclockwise proton beam that is the $z$-axis of the CMS reference frame. This study assumes the geometry of the Phase-2 detector with the performance as documented in the recent TDRs. The analysis mainly depends on the capabilities of the muon system.

The current muon system consists of three detector types, namely drift tubes (DTs), resistive plate chambers (RPCs) and cathode strip chambers (CSCs) [5]. The muon forward region will be augmented with gas electron multipliers (GEMs) and improved RPCs for Phase-2 data taking while the electronics of the existing detectors will be upgraded. The foreseen upgrade of the CMS muon system is discussed in detail in Ref. [2].

# 3 Displaced muon reconstruction

This analysis relies on a dedicated muon reconstruction algorithm, the displaced standalone (DSA) algorithm, designed for highly displaced muons that potentially leave hits only in the muon system. In this algorithm, the muon reconstruction is performed using the Kalman-filter technique [6] without imposing the primary vertex constraint as it is done in the default standalone (SA) muon reconstruction algorithm. The DSA algorithm has a better reconstruction efficiency than the SA algorithm, for highly displaced muons (see Fig. 8.12 of the Muon TDR [2]). The DSA algorithm improves the transverse impact parameter ($d_0$) and the transverse momentum ($p_T$) resolution for displaced muons compared to the SA muon algorithm [7].

# 4 Signal model

In Dark SUSY models, in addition to supersymmetric fields, a dark sector of fermions and gauge fields is introduced. The gauge boson corresponding to the additional $U_D(1)$ symmetry is called the dark photon ($\gamma_D$) [3, 4], which can have a kinetic mixing with the SM photon. The dark photon acquires a mass after $U_D(1)$ symmetry breaking. In such models, the dark photon couples to SM charged particles in the same way as a photon, except that the couplings are scaled by a parameter $\epsilon$ that gives the strength of the kinetic mixing. The dark photon lifetime is proportional to $1/\epsilon^2$, and since $\epsilon$ can be very small, the dark photon lifetime can be long. If this is the case and if the dark photon has non-zero momentum, it can have a macroscopically long decay length.

Dark photons can be produced in cascade decays of the SM Higgs boson that would first decay to a pair of MSSM-like lightest neutralinos ($n_1$), each of which, in Dark SUSY models, can decay further to a dark sector neutralino ($n_D$) and the dark photon, as shown in Fig. 1.

For the branching fraction $\mathrm{BR}(H \to 2\gamma_D + \mathrm{X})$, where X denotes the particles produced in the decay of the SM Higgs boson apart from the dark photons, 20% is used. This value is in agreement with recent Run-2 studies [8] and taking into account the upper limit on invisible/non-conventional decays of the SM Higgs boson [9]. We assume neutralino masses $m(n_1) = 50$ GeV and $m(n_D) = 1$ GeV, and explore the search sensitivity for dark photon masses and lifetimes in the following ranges: $m(\gamma_D) = (1, 5, 10, 20, 30)$ GeV and $c\tau = (10, 10^2, 10^3, 5 \times 10^3, 10^4)$ mm. Final states with two and four muons are included in the analysis. In the former case, one dark photon decays to a pair of muons while the other dark photon decays to some other fermions (2-muon final state). In the latter case, both dark photons decay to muon pairs (4-muon final state). Both decay chains are shown in Fig. 1. The assumed Higgs production cross section via gluon-gluon fusion is 49.97 pb [10].

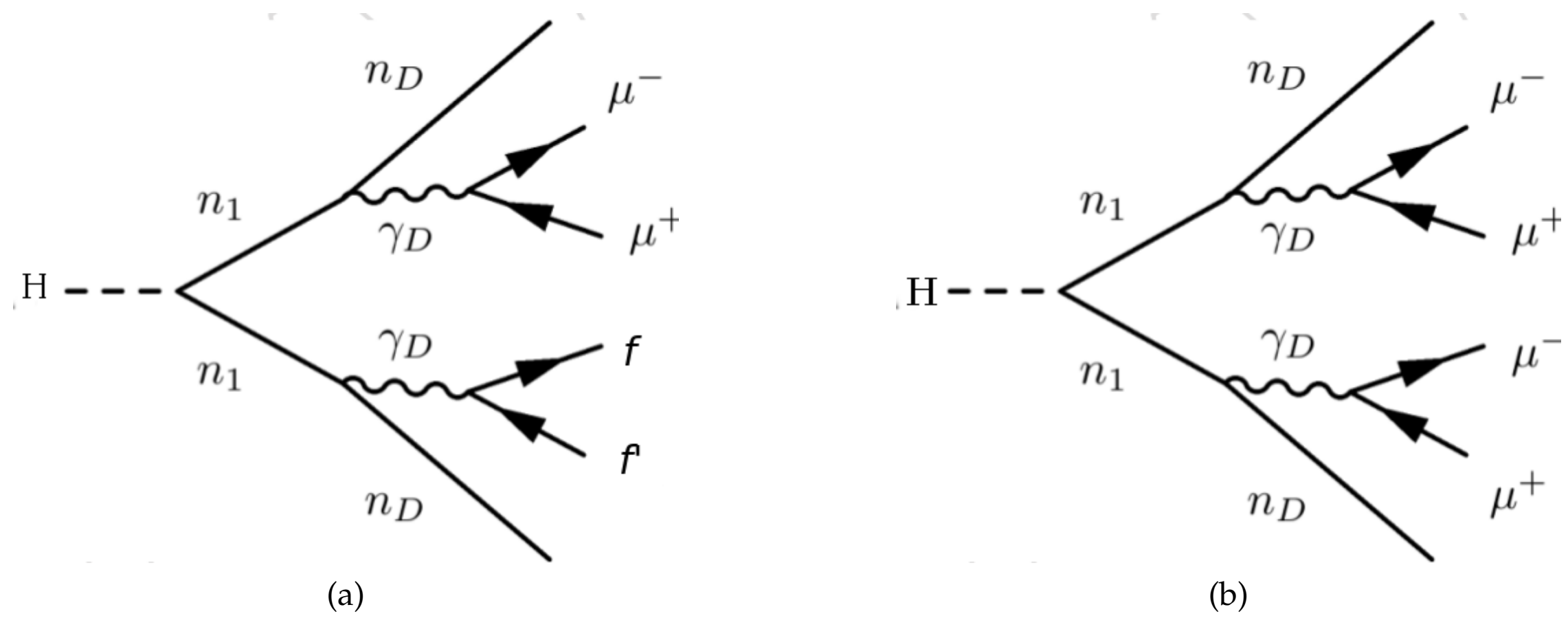

Figure 1: Feynman diagram of the decay of SM Higgs boson to a final state containing two or more muons in Dark SUSY models [11]. (a) Decay chain leading to a final state containing exactly two muons. (b) Decay chain leading to a final state containing exactly four muons.

The branching ratio of dark photons decaying to muons as a function of the dark photon mass is shown in Fig. 2. For dark photon masses close to masses of hadronic resonances such as $\rho$, $\omega$, $\phi$ and $\rho'$, the branching ratio to leptons is reduced. Narrow hadronic resonances (e.g. $\Upsilon$, $J/\psi$ and $\psi$(2S)) are not considered. Hence, in the vicinity of these narrow hadronic resonances in the range of the order of their natural widths, the analysis does not claim any sensitivity. For $m(\gamma_D) > 5$ GeV, the branching ratio to muons stays constant around 15% as shown in Fig. 2. In addition to muons, the final state contains missing transverse momentum ($p_{\mathrm{T}}^{\mathrm{miss}}$) originating from the dark neutralino in the $n_1 \to n_D + \gamma_D$ decay.

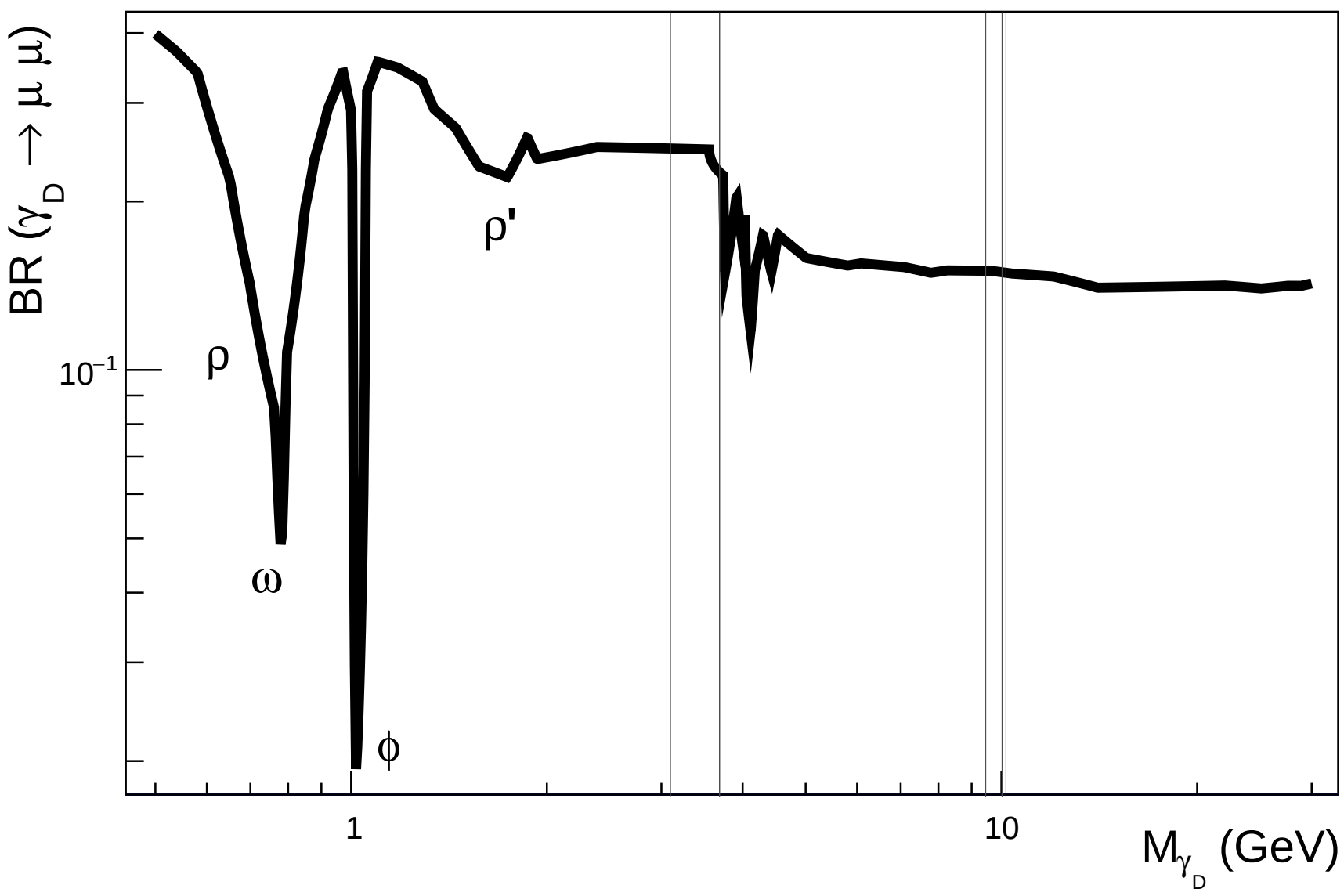

Figure 2: Branching ratio of dark photons to muons. The branching ratio calculations include the impact of hadronic resonances, such as $\rho$, $\omega$, $\phi$ and $\rho'$, as well as other leptonic decay modes of the dark photon. Narrow hadronic resonances (e.g. $\Upsilon$, $J/\psi$ and $\psi$(2S)), which are shown as gray areas, do not enter the branching ratio calculations.

## 5 Event simulation

The dark photon signal and the quantum chromodynamics (QCD) multijet background are both simulated with PYTHIA 8.212 [12, 13] at leading order. The Drell-Yan (DY) background is simulated with MADGRAPH5_aMC@NLO [14] and the $t\bar{t}$ background with POWHEG 2.0 [15–17], both with next-to-leading order cross sections. For hadronization, PYTHIA 8.212 is used with the underlying event tune CUETP8M1 [18]. The generated events are processed through a full simulation of the CMS Phase-2 detector based on GEANT4 [19].

Pileup interactions for the "PU200" scenario, with an average of 200 interactions per bunch crossing corresponding to expectations for the HL-LHC, are included in the simulation by overlaying additional simulated minimum bias events. Samples with no pileup and the CMS Phase-2 detector geometry are used to study effects from pileup. Beam halo muons are included in the simulation with the rate expected at HL. In addition, samples obtained with the Phase-1 detector performance are considered.

## 6 Backgrounds

A number of SM processes may yield the signal signature of two displaced muons and missing transverse momentum. The following three dominant processes are included in this study:

- The dominant background consists of QCD multijets events. Displaced muons can be produced in the decay of heavy quarks and neutrinos can be the source of missing transverse momentum.
- Similarly, $t\bar{t}$ production can lead to displaced muons and neutrinos that contribute to missing transverse momentum.
- Drell-Yan processes (DY $\to \mu\mu$) can appear as signal due to the misidentification of prompt muons as displaced. The missing transverse momentum can arise from instrumental effects.

Given the displaced signature, other sources of background besides the SM processes may contribute. However, they can be sufficiently suppressed, as described below.

- Beam halo: The protons of the LHC beam can collide with leftover molecules in the beam pipe. During these collisions, muons can be produced and can travel through the detector from one side to the other (see horizontal red lines in Fig. 3). These muons can have a large displacement from the primary interaction vertex. The amount of such beam halo muons scales with luminosity and exceeds the current conditions for the HL-LHC. However, these tracks can be identified by their very low transverse momentum (see Fig. 14 (b) in Ref. [20]). In 2% of the signal events simulated with PU200, a signature consistent with beam halo muons is observed before event selection. By selecting displaced muons with $p_T > 15$ GeV, muons from beam halo are completely suppressed.
- Cosmic ray muons: Cosmic ray muons traverse the detector usually from top to bottom and may be reconstructed as two displaced muon tracks. The contribution of cosmic ray muons is suppressed by rejecting displaced muon pairs which are back-to-back (Sec. 7). A suppression factor of $10^{-9}$ is calculated for a sample of cosmic ray muon events taken in 2017 with the active LHC clock while pp collisions are absent. As the rate of cosmic ray muons is independent of the collider conditions, this value is also valid for HL operation.

## 7 Event selection

In the context of the Phase-2 CMS detector and the HL-LHC, various studies have been performed to tackle the issue of triggering on displaced muons [2]. We use those results to set benchmark trigger scenarios in this analysis. We assume a dedicated displaced single-muon trigger with $p_T > 20$ GeV. For the Phase-2 upgraded CMS, such a trigger is expected to have 90% efficiency even for largely displaced muons [2].

For the offline selection, we require the DSA muons to have $p_T \geq 15$ GeV and $|\eta| \leq 2.8$. For the muon with the highest transverse momentum, $p_T \geq 20$ GeV is imposed to account for the displaced muon trigger threshold. To select muons of good quality, the fit of the hits in the muon chambers to build each muon track should meet the condition that the chi-squared divided by the number of degrees of freedom $\chi^2/\mathrm{ndof} \leq 2.0$. The corresponding track of the displaced muon has to have at least 17 muon hits for $|\eta| \leq 2.4$ and 6 hits in the region of the new ME0 station $2.4 \leq |\eta| \leq 2.8$ that are well associated with the track. Selected displaced muons should have a transverse impact parameter significance $|d_0|/\sigma(d_0) \geq 5.0$ (see Fig. 4).

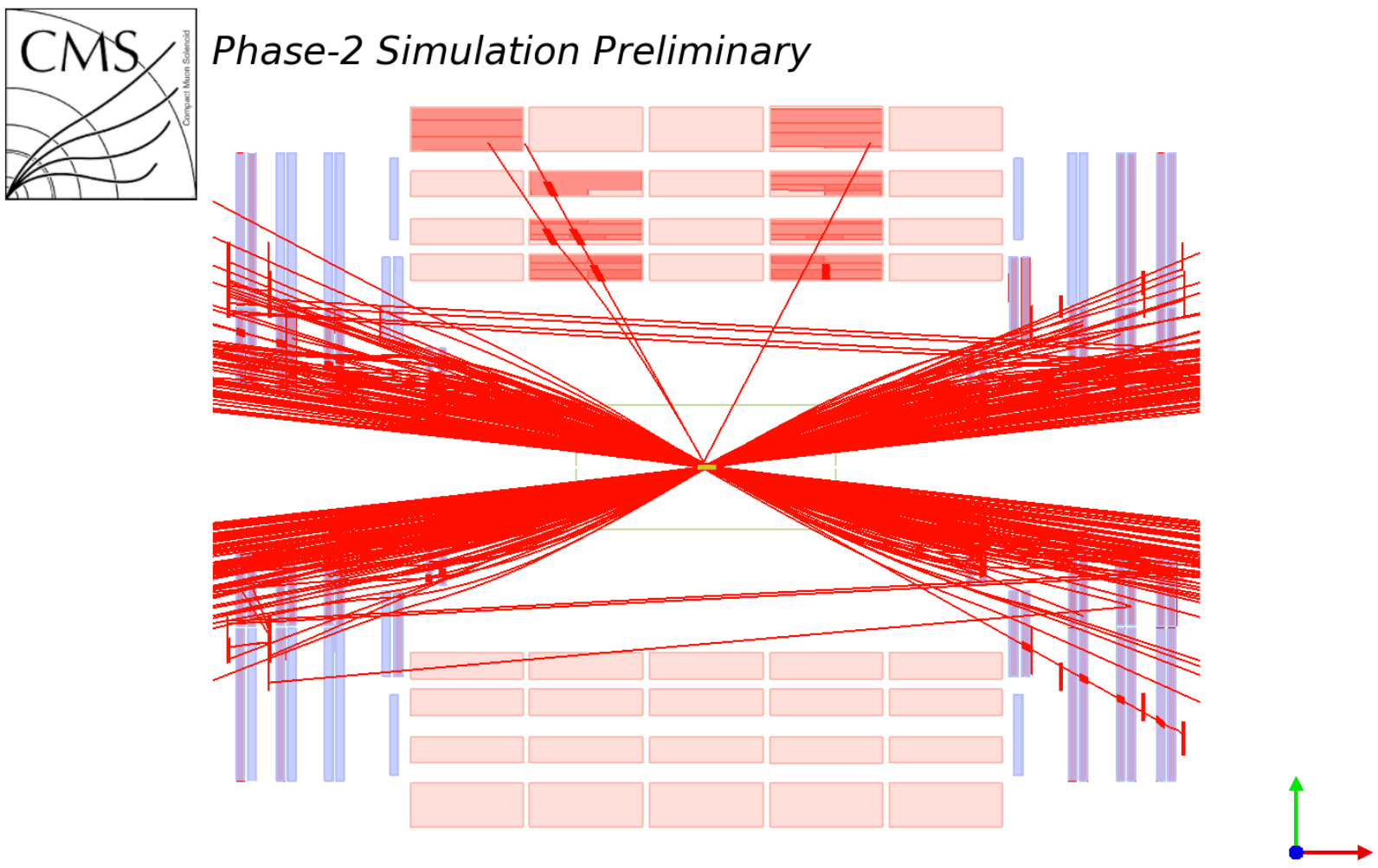

Figure 3: Event display of a $t\bar{t}$ event with high pileup. All reconstructed muons fulfilling $p_T > 1$ GeV are shown. Muons from pileup are predominantly in the forward region of the detector. The tracks going horizontally through detector with no origin at the center of the detector are muons from beam halo. Both types of muons, from pileup and beam halo, are very low-$p_T$ objects and are rejected by the muon $p_T$ criterion applied in the analysis.

Since requiring two muons to pass this criterion leaves very few events in the QCD background sample, we opt for assuming that the selection efficiency on two muons is factorizable and weight events following the procedure used in Ref. [21].

For each event, we require to have at least two DSA muons fulfilling the requirements mentioned above. If there are more than two selected muons, the ones with the highest $p_T$ are chosen. The two muons must have opposite charge ($q_{\mu,1} \cdot q_{\mu,2} = -1$) and must be separated by $\Delta R = \sqrt{\Delta\phi^2 + \Delta\eta^2} > 0.05$. The three-dimensional angle between the two displaced muons is required to be less than $\pi - 0.05$ (not back-to-back) in order to suppress cosmic ray backgrounds. Additionally, $p_{\mathrm{T}}^{\mathrm{miss}} \geq 50$ GeV is imposed to account for the dark neutralinos escaping the detector without leaving any signal.

In order to discriminate between background and signal, the three-dimensional distance from the primary vertex to the point of closest approach of the extrapolated displaced muon track, called $R_{\mathrm{Muon}}$, is used. A sketch illustrating $R_{\mathrm{Muon}}$ for the two selected muons is shown in Fig. 5. The event yield after full event selection of both selected muons as a function of $R_{\mathrm{Muon}-1}$ and $R_{\mathrm{Muon}-2}$ is used to search for the signal. Figure 6 shows $R_{\mathrm{Muon}-1}$ of the first selected muon for signal and background samples.

Dedicated search regions are defined using these distances symmetrically for both muons by summing up all events above a sliding lower threshold. With increasing threshold, the signal-to-background ratio improves due to the suppression of the backgrounds. The lower thresholds are optimized for every possible lifetime $c\tau$. By varying the lower bound, the sensitivity reaches its maximum at some point. This is taken as the predefined lower bound for the statistical interpretation of the results: 1 cm for $c\tau = 10$ mm, 10 cm for $c\tau = 100$ mm and 80 cm for $c\tau = 10^3, 5 \times 10^3, 10^4$ mm. The signal and background event yields for the different search regions after full selection are summarized in Tab. 1.

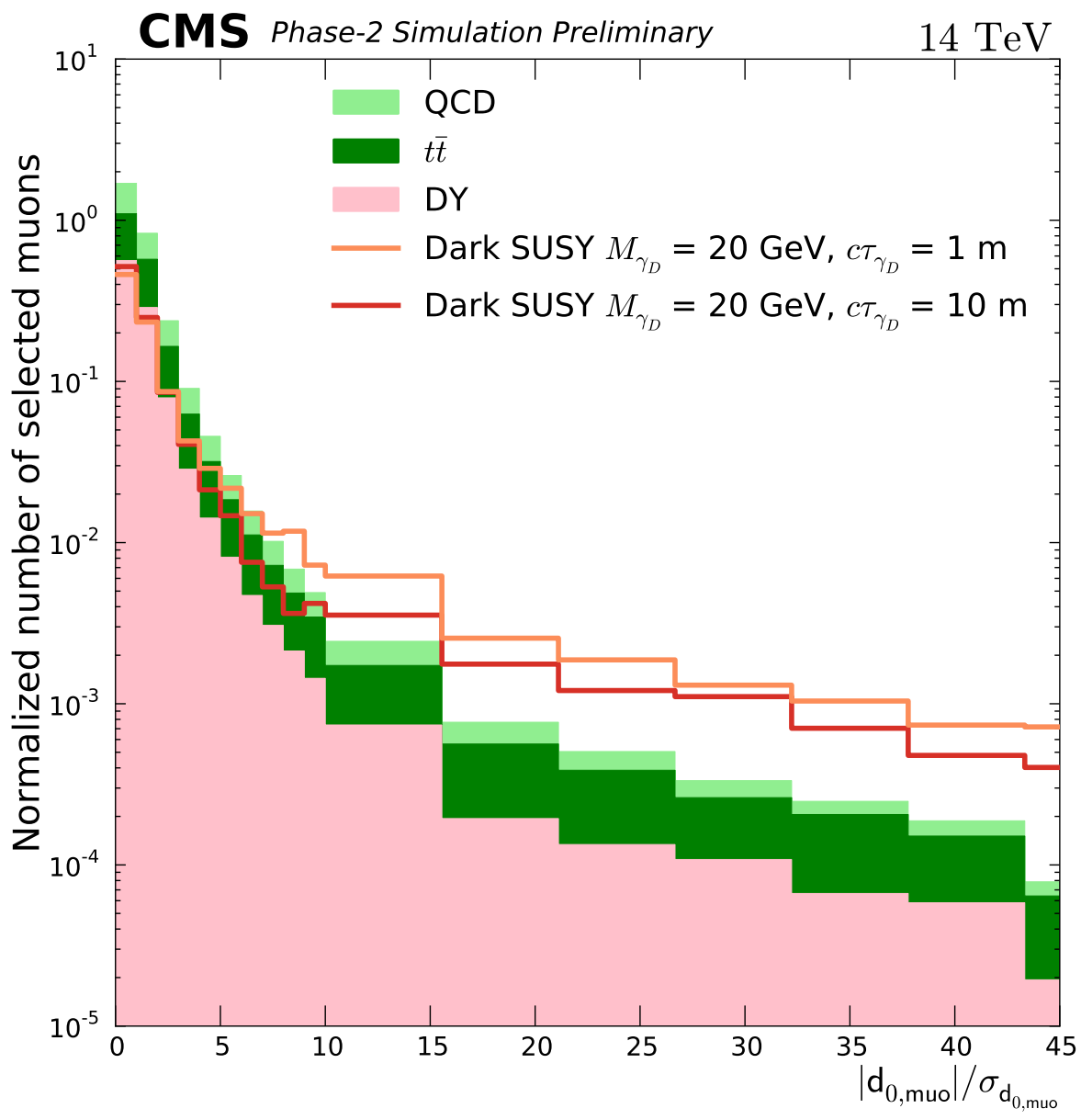

Figure 4: Distribution of the significance of the transverse impact parameter, $|d_0|/\sigma(d_0)$, for signal and background samples. Displaced standalone muons passing the kinematic selection ($p_T$ and $\eta$) of the object selection are shown.

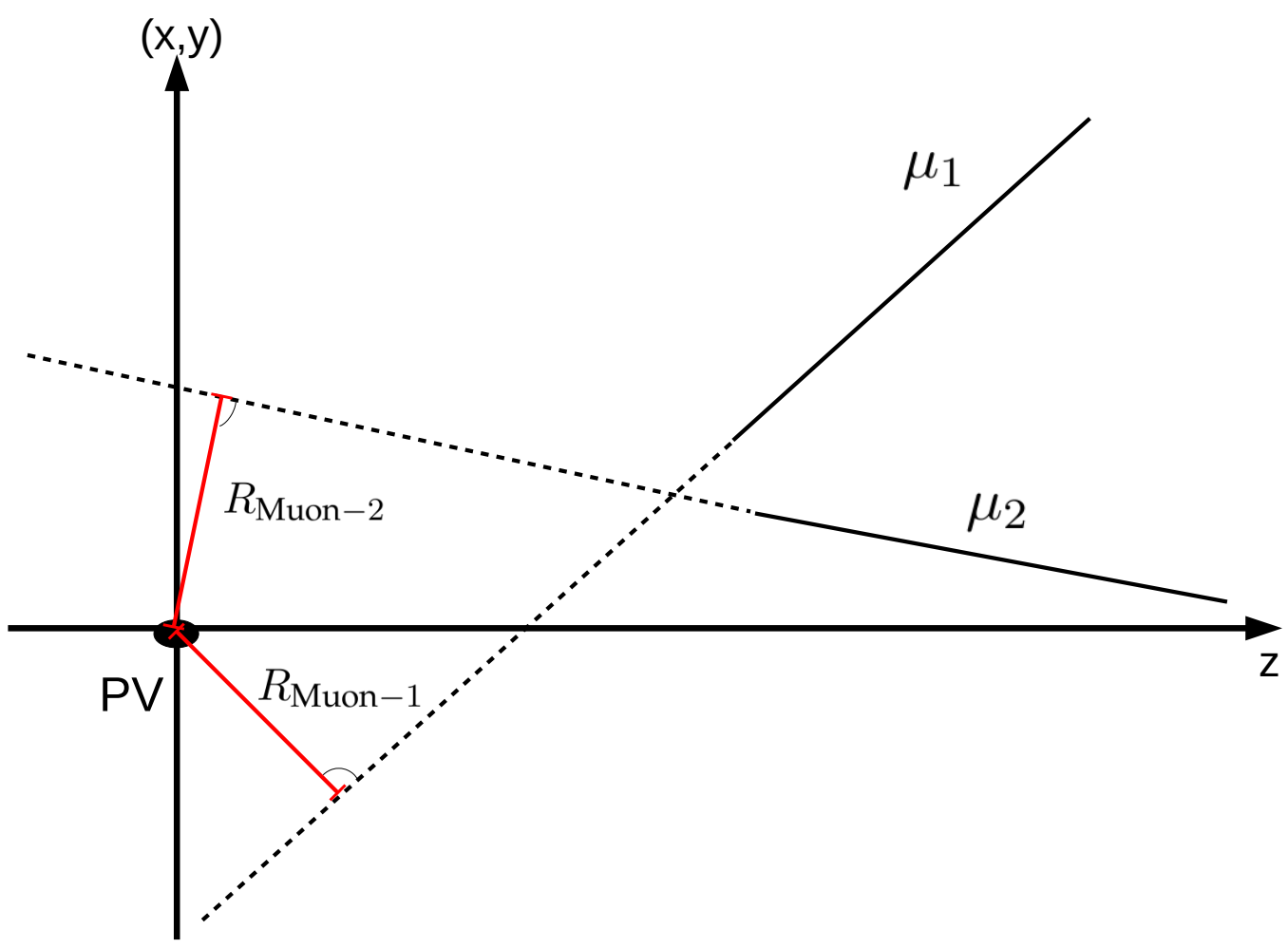

Figure 5: Sketch illustrating the three dimensional distances of the closest approach of the displaced muon track to the primary interaction vertex, $R_{\text{Muon-1}}$ and $R_{\text{Muon-2}}$, for the two selected muons. PV denotes the primary interaction vertex. $(x, y)$ illustrates the transverse plane and the $z$-axis is along the beam line.

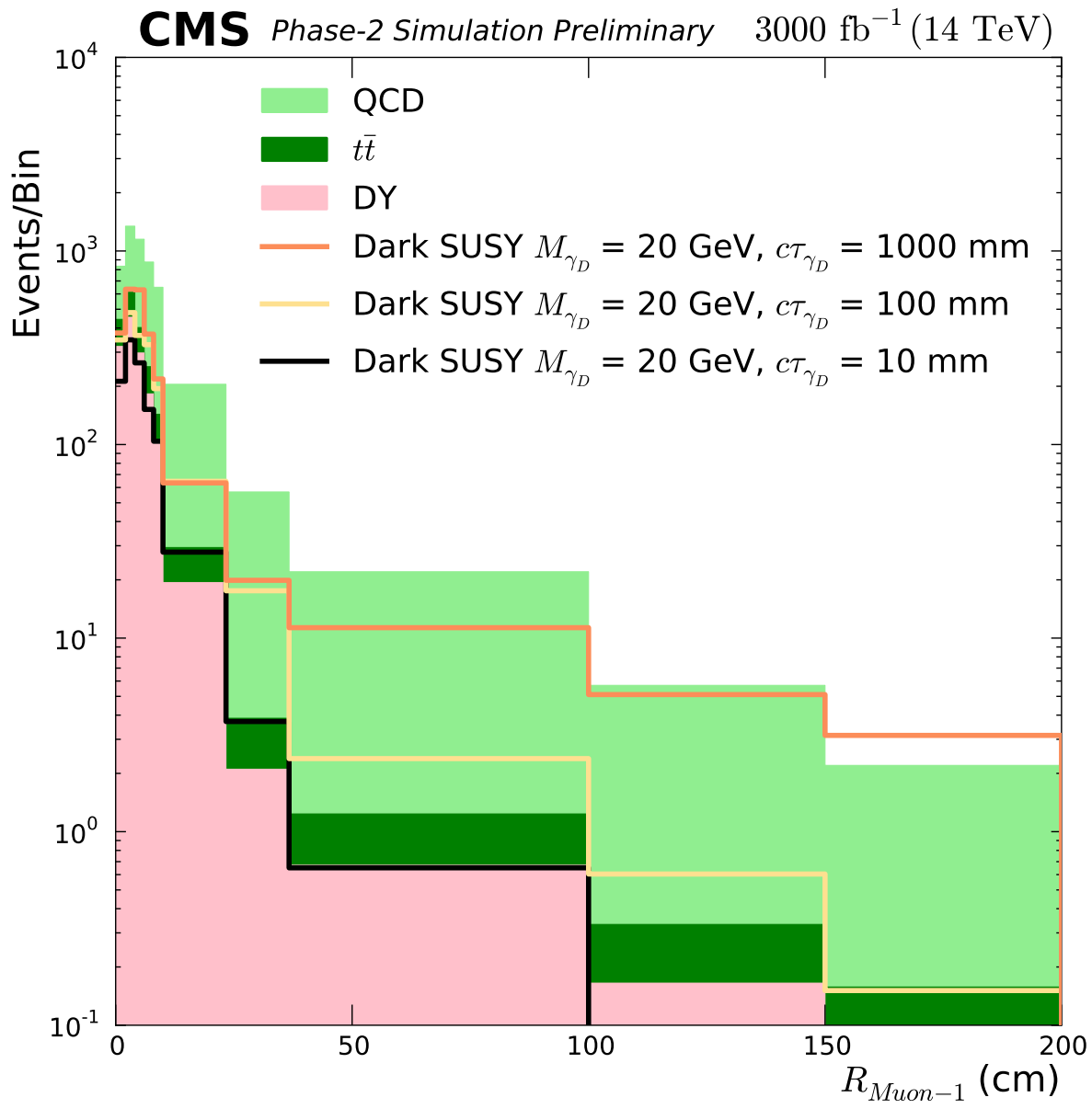

Figure 6: Distribution of the distance of the closest approach of the displaced muon track to the primary interaction vertex, $R_{\text{Muon}-1}$, for signal and background samples after the final event selection. The distance of the highest $p_T$ muon is shown.

Table 1: Signal and background event yields with statistical uncertainties in different search regions after the final event selection. The systematic uncertainties can be found in Sec. 8.

| Search Region [cm] | Event Yield | | | | | | | | |
|---|---|---|---|---|---|---|---|---|---|
| | Signal | | | | | | Background | | |
| | | $m_{\gamma_D}$[GeV] | | | | | | | |
| | $c\tau$[mm] | 1 | 5 | 10 | 20 | 30 | $t\bar{t}$ | Drell-Yan | QCD |
| >80 | 10000 | $0.00 \pm 0.00$ | $0.18 \pm 0.15$ | $0.22 \pm 0.20$ | $8.9 \pm 2.2$ | $29.8 \pm 4.8$ | $0.95 \pm 0.19$ | $2.06 \pm 0.47$ | $3.76 \pm 0.78$ |
| >80 | 5000 | $0.04 \pm 0.03$ | $0.83 \pm 0.37$ | $0.79 \pm 0.56$ | $35.3 \pm 6.3$ | $75.6 \pm 10.4$ | $0.95 \pm 0.19$ | $2.06 \pm 0.47$ | $3.76 \pm 0.78$ |
| >80 | 1000 | $0.06 \pm 0.03$ | $2.53 \pm 0.89$ | $12.8 \pm 3.7$ | $87.0 \pm 13.3$ | $132 \pm 16$ | $0.95 \pm 0.19$ | $2.06 \pm 0.47$ | $3.76 \pm 0.78$ |
| >10 | 100 | $0.96 \pm 0.14$ | $5.6 \pm 0.7$ | $11.7 \pm 1.7$ | $16.7 \pm 2.4$ | $12.9 \pm 1.7$ | $31.7 \pm 1.2$ | $215 \pm 5$ | $174 \pm 5$ |
| >1 | 10 | $4.02 \pm 0.25$ | $13.6 \pm 0.8$ | $10.5 \pm 0.5$ | $13.7 \pm 0.8$ | $9.3 \pm 0.6$ | $1020 \pm 6$ | $13320 \pm 30$ | $1150 \pm 10$ |

# 8 Systematic uncertainties

Since many systematic uncertainties can only be fully and finally determined using data, this study is left with establishing different scenarios to estimate uncertainties for the Phase-2 run period. The two scenarios considered in this study are taken from Ref. [22]. The first one assumes no change in systematic uncertainties with respect to 2016 Run-2 data taking, and the other one takes into account the larger dataset size, better detector performance, and higher theoretical accuracy for the HL-LHC, all of which lead to a reduction of systematic uncertainties compared to the nominal values of 2016.

For the Run-2 uncertainties, a 5% systematic uncertainty on the cross section of the different processes is applied, except for $t\bar{t}$ processes for which an uncertainty of 15% is considered. This is comparable to systematic uncertainties for the $t\bar{t}$ background applied for Run-2 searches [23, 24]. For the Higgs boson production cross section via gluon-gluon fusion, a 10% systematic uncertainty is assumed [10]. The instrumental uncertainties are taken from Ref. [25].

For Phase-2, the so-called "S2+" scenario, defined in Ref. [22], is considered. In this scenario, theory uncertainties are scaled by a factor 1/2, the muon identification uncertainty is taken at the expected floor value of 1% and, for luminosity, the anticipated 1% uncertainty is assumed. The efficiency of the displaced muon trigger varies with respect to the transverse displacement. This is covered by setting the uncertainty to 10% for the Phase-2 scenario. The systematic uncertainty on the trigger efficiency is taken from the efficiency evaluation in Fig. 3.4 of the Phase-2 Upgrade Level-1 Trigger Interim TDR [26].

# 9 Results

We present the search sensitivity results for the following three scenarios:

- Phase-2 scenario (DSA):
    - Integrated luminosity: 3000 $fb^{-1}$
    - Geometry: Phase-2 detector
    - Higher-efficiency trigger benchmark scenario (90%)
    - Pileup scenario: PU 200
    - Reconstruction: Dedicated displaced standalone algorithm
    - Systematic uncertainties: "S2+" scenario
- Phase-2 scenario (SA):
    - Same assumptions as Phase-2 scenario (DSA) except reconstruction: We assume the SA reconstruction efficiency is 1/3 of the dedicated DSA reconstruction efficiency (see Fig. 8.9 of the Muon TDR [2]).
- Phase-1 scenario:
    - Integrated luminosity: 300 $fb^{-1}$
    - Geometry: Phase-1 detector
    - Lower-efficiency trigger benchmark scenario: Since the trigger performance potentially decreases over the course of Phase-1 due to aging and increasing PU, an overall 60% efficiency is assumed here.
    - Pileup scenario: PU 200
    - Systematic uncertainties: Taken from 2016 Run-2 data taking

The search is performed using a simple counting experiment approach. In absence of a signal,

upper limits at 95% confidence level (CL) are obtained on a signal event yield with respect to the one expected for the considered model. We use the Bayesian method with a uniform prior for the signal event rate. The nuisance parameters associated with the systematic uncertainties are modeled with log-normal distributions. In presence of the expected signal, significance of the corresponding event excess over the expected background is assessed using the likelihood method.

The resulting limits for the Dark SUSY models are depicted in Fig. 7. While the results shown in Fig. 7 (a) are for a dark photon with a decay length of 1 m as a function of the dark photon mass, Fig. 7 (b) shows the results for a dark photon mass of 20 GeV as a function of the decay length.

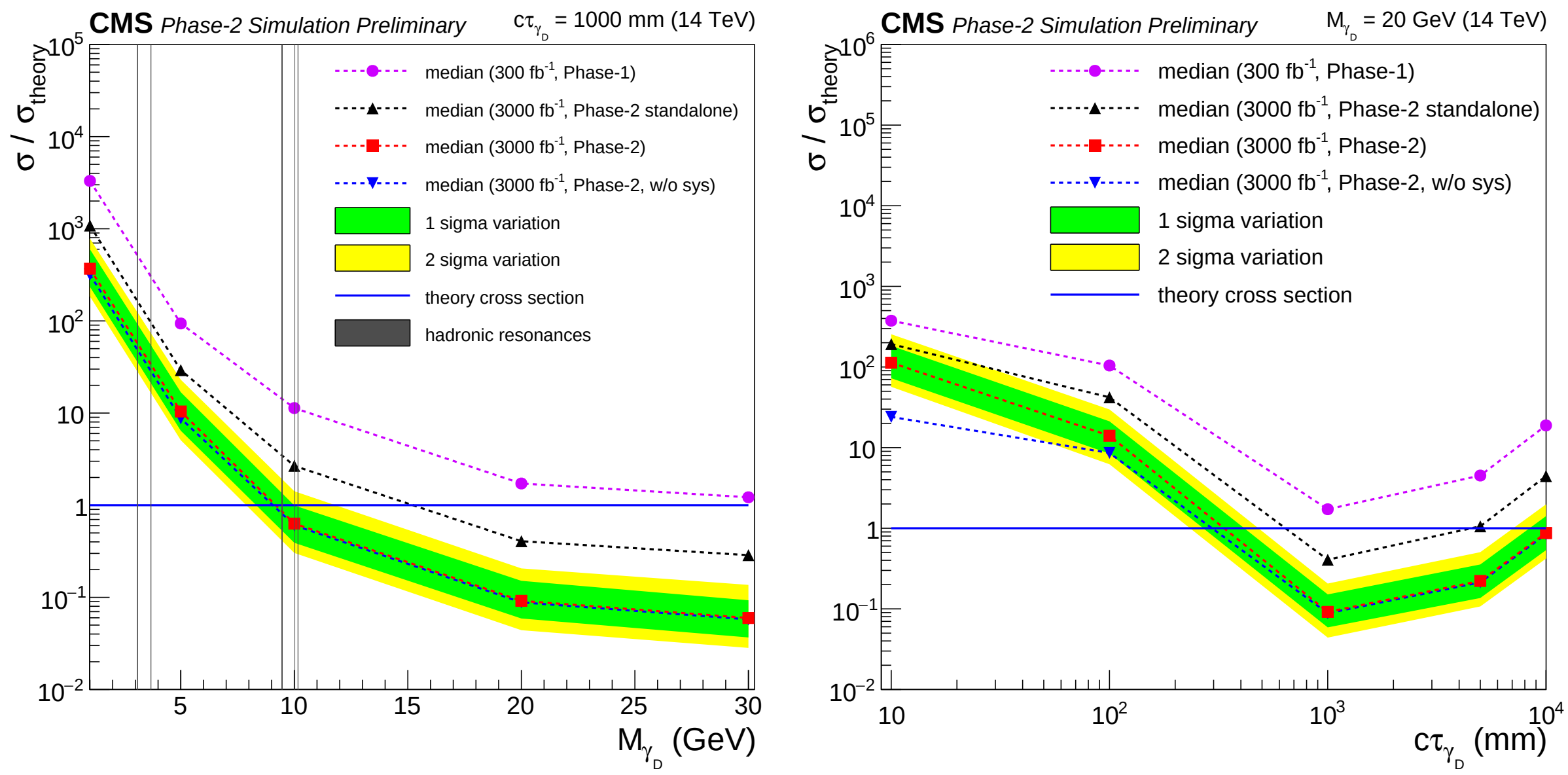

(a) (b)

Figure 7: 95% CL upper limits on production cross section $\sigma/\sigma_{\text{theory}}$ for various dark photon mass hypotheses and a fixed decay length of $c\tau = 1000$ mm (a) and a fixed mass of $M_{\gamma_D} = 20$ GeV as a function of the dark photon decay length (b). Green and yellow shaded bands show the one and two sigma range of variation of the expected 95% CL limits, respectively. Phase-2 results with 3000 fb$^{-1}$ (red) are compared to results obtained with 300 fb$^{-1}$ (violet) of integrated luminosity, which corresponds to the end of Phase-1 data taking. Another median of an excluded limit is shown here which represents the scenario with the reduced standalone reconstruction efficiency with 3000 fb$^{-1}$ (black) of integrated luminosity. Additionally, Phase-2 results with 3000 fb$^{-1}$ are determined without any systematic uncertainties (blue). The theoretical Dark SUSY cross section for 14 TeV is shown as a solid line. The gray lines indicate the regions of narrow hadronic resonances where the analysis does not claim any sensitivity (see Fig. 2).

Another presentation of the limits can be done in a parameter scan of the two-dimensional $\epsilon - m_{\gamma_D}$ plane. The closed area in Fig. 8 (b) shows the excluded region along with the region of discovery of dark photons compared to existing results (Fig. 8 (a)). Besides the searches at the LHC provided by ATLAS [27] and CMS [28] at a center-of-mass energy of $\sqrt{s} = 8(13)$ TeV, there are constraints from low-energy electron-positron colliders (KLOE [29], BaBar [30]), heavy ion colliders (PHENIX [31]) as well as from cosmological observations [32].

## 10

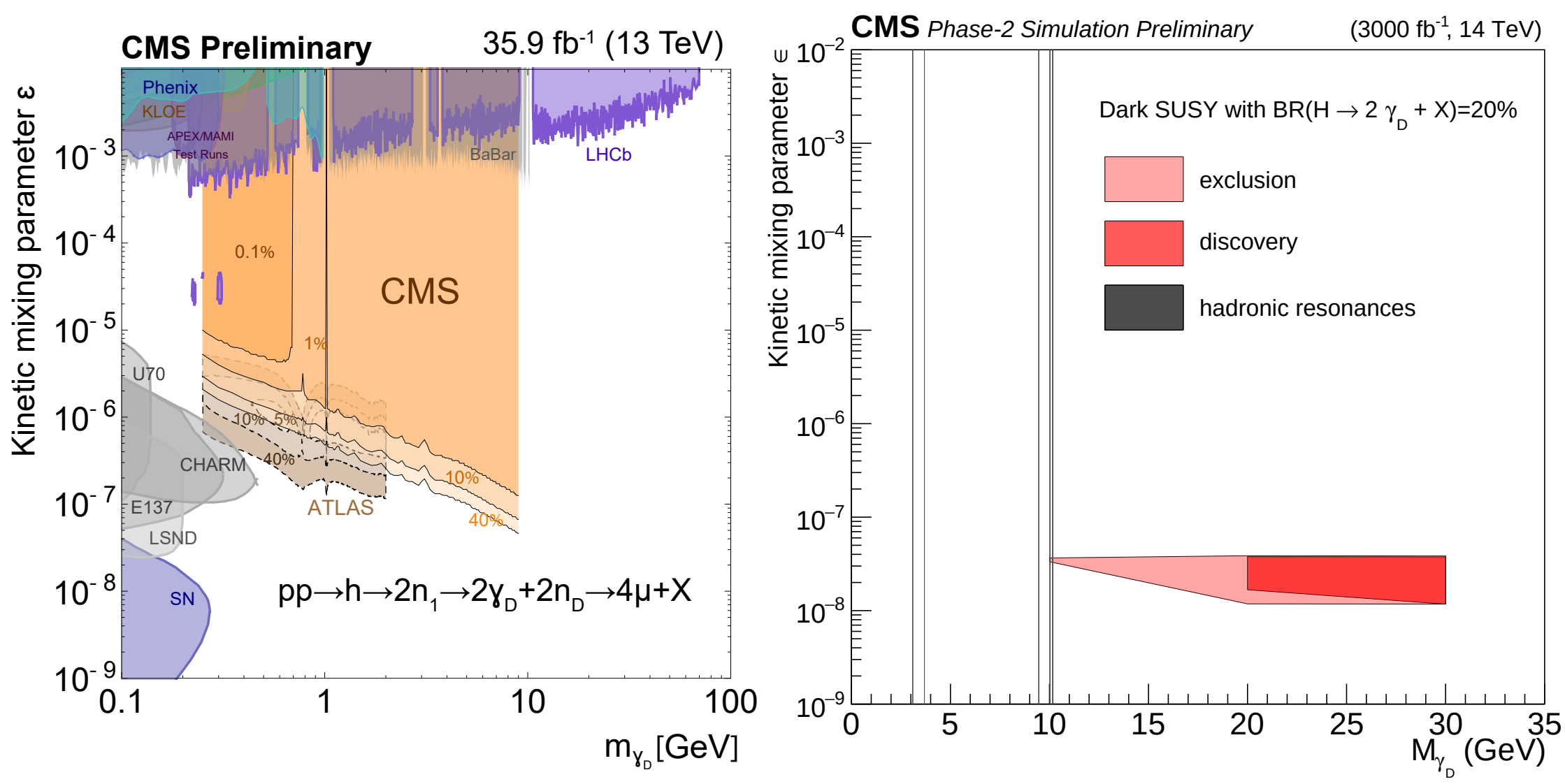

(a) (b)

Figure 8: Parameter scan in the $\epsilon - m_{\gamma_D}$ plane. (a) Collection of existing limits taken from Ref. [8]. (b) Results from this analysis for Phase-2 with 3000 fb$^{-1}$. The ranges with exclusion and discovery sensitivity are shown in light and dark red color, respectively. The gray lines indicate the regions of narrow hadronic resonances where the analysis does not claim any sensitivity (see Fig. 2).

Comparing the result to former CMS results at $\sqrt{s} = 8(13)$ TeV [28], Phase-2 searches will be sensitive to higher dark photon masses and lower values of the kinetic mixing parameter $\epsilon$ and, hence, longer lifetimes. The difference in the shape of the exclusion range has its origin in the usage of the dedicated displaced muon reconstruction algorithm instead of the standard muon reconstruction algorithm. By relaxing the constraint on the primary interaction vertex, the search becomes more sensitive to lower values of the kinetic mixing parameter $\epsilon$.

# 10 Summary

Present searches for displaced muons show no significant deviation with respect to the standard model expectation. However, there is quite a large phase-space which has not been explored yet and is unreachable with the current LHC conditions due to low signal cross sections and limited statistics. The high-luminosity LHC will provide a unique opportunity to search for new physics with a striking signature of highly displaced muons. This study presents the search sensitivity for pairs of displaced muons with an integrated luminosity of 3000 $fb^{-1}$. The transverse impact parameter significance and the three-dimensional distance between the extrapolated displaced muon track and the primary vertex are used as the discriminating variables in the search for two largely displaced muons that are reconstructed with a standalone algorithm using only muon chamber hits.

The search in this note is performed within a model belonging to a class of Dark SUSY models where dark photons decay into a pair of displaced muons. The study shows that searches at the high-luminosity LHC will be able to probe phase-space which has not been explored yet.

# 11 Appendix: Extension of Muon-TDR Analysis

This sensitivity study for displaced muons at the HL-LHC is an extension of a former study presented in the Muon Upgrade TDR [2] where displaced muons originate from smuons which serve as long-lived particles. In gauge-mediated SUSY breaking models, smuons can be (co-)NLSPs, i.e. the next-to-lightest SUSY particles, and almost degenerate in mass, decaying to a muon and a gravitino [33]. This decay can either be prompt, or the slepton can have a significant lifetime. In the latter case, the final state is given by two displaced opposite-sign muons and significant amount of missing transverse energy.

The study is focused on the process $q\bar{q} \to \widetilde{\mu}\widetilde{\mu} \to 2\mu 2\tilde{G}$ (see Fig. 9), where the two smuons decay far away from the primary interaction vertex with various decay length ranging from 10 mm up to 1 m and with masses from 200 GeV to 1.5 TeV. Distributions of the transverse impact parameter for different decay lengths can be found in Ref. [2].

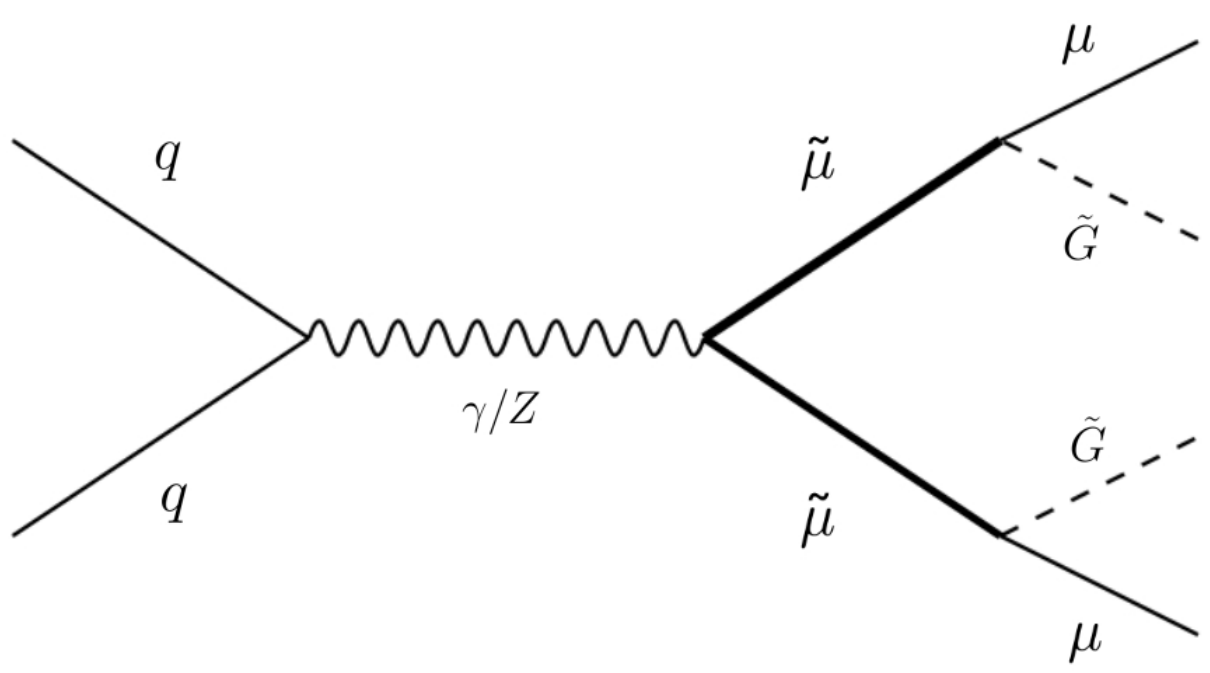

Figure 9: Feynman diagram of the process leading to smuon pair production at a hadron collider. The decay of the smuons leads to the final state including two muons.

In this section, additional results of this former study are presented. Figure 10 (a) shows the discovery sensitivity for smuons as a function of the decay length. The results are shown in terms of p-value and significance. A significance of $3\sigma$ corresponds to an evidence and a significance of $5\sigma$ to discovery. Figure 10 (b) shows the discovery sensitivity in the parameter space of mass and decay length. Smuons with masses up to 200 GeV could be discovered with the 3000 $\text{fb}^{-1}$ of the HL-LHC.

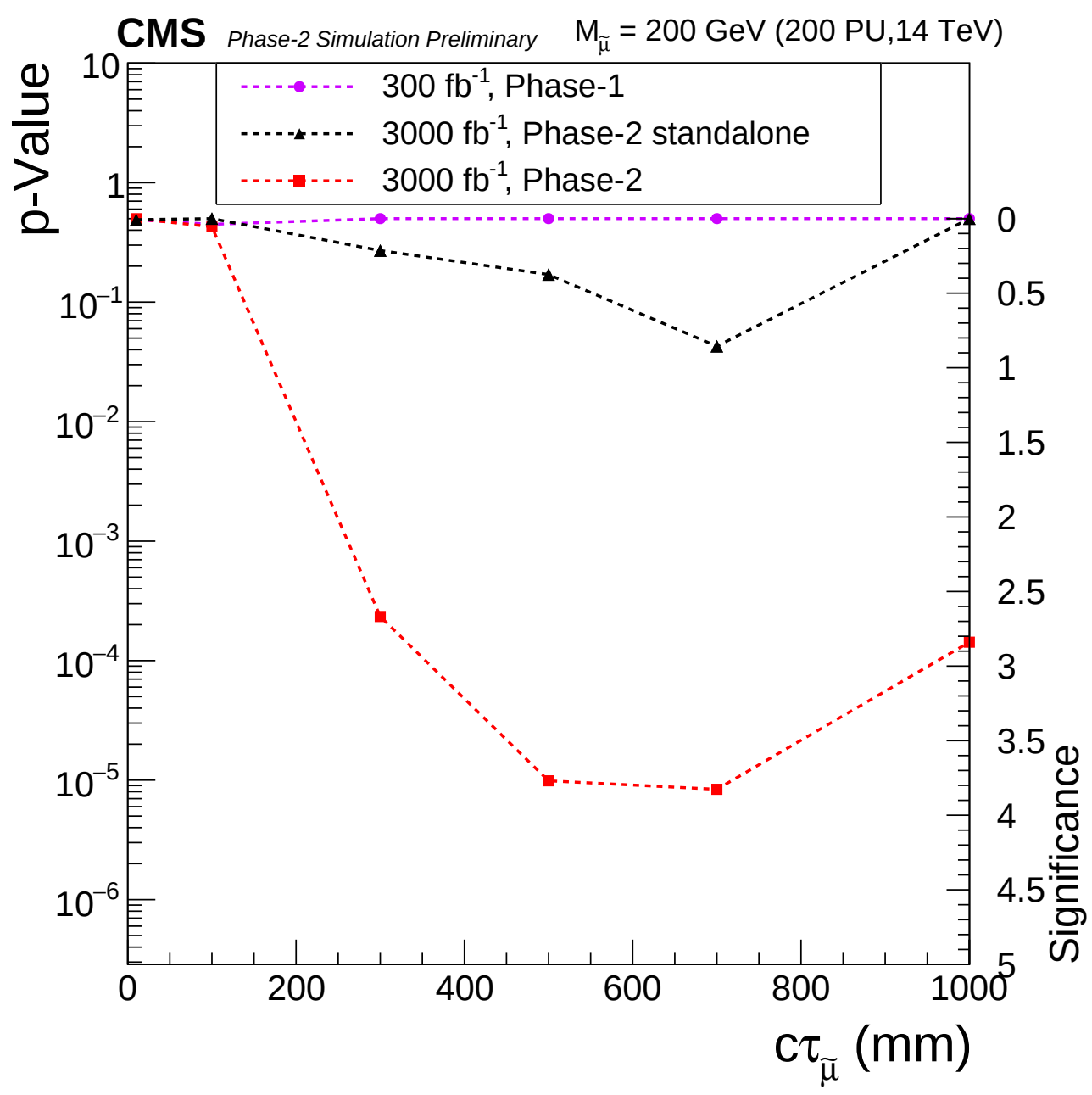

(a)

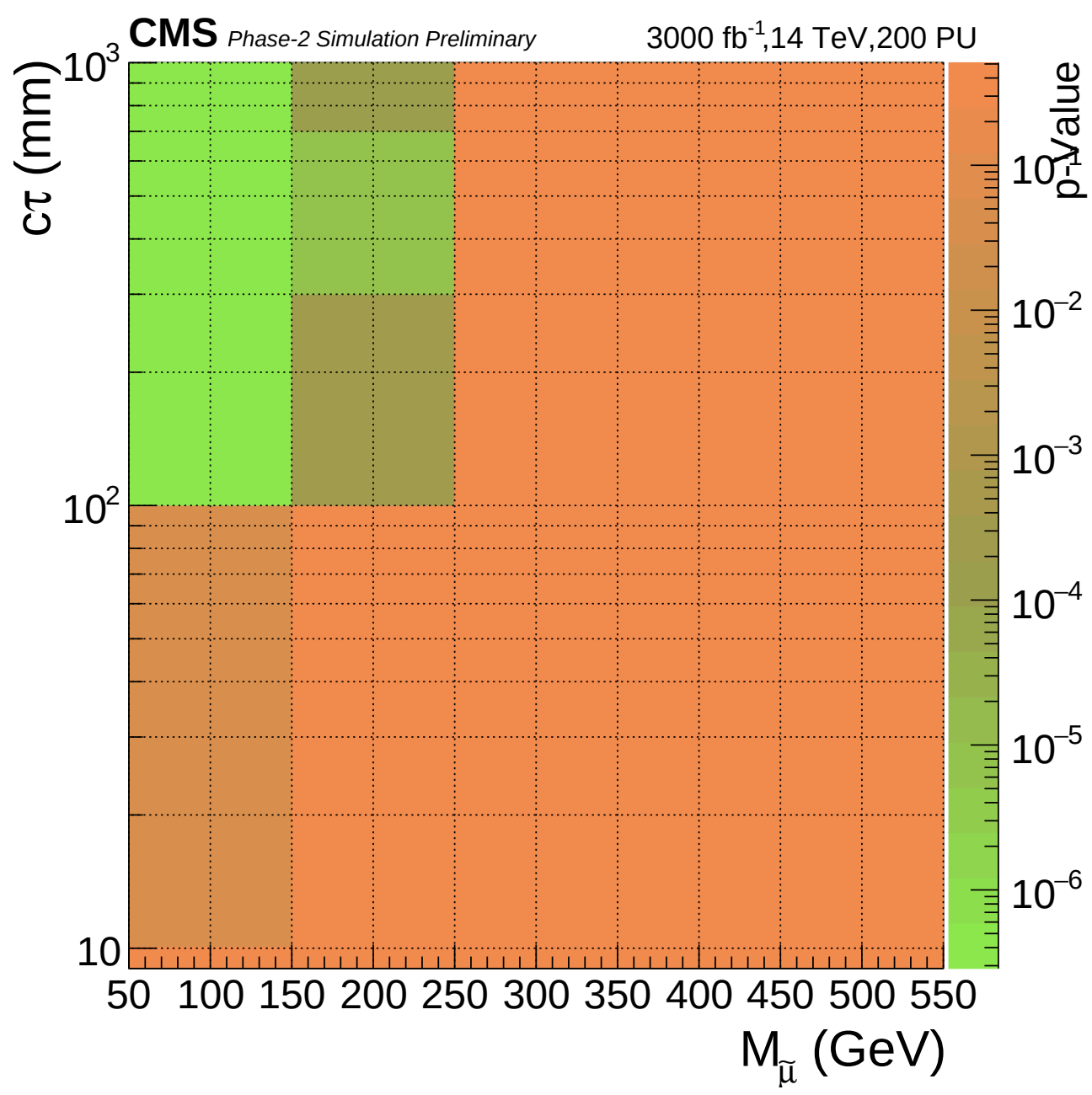

(b)

Figure 10: (a) Discovery significance and p-value for a fixed smuon mass of $M_{\tilde{\mu}} = 200$ GeV. (b) Discovery sensitivity in the parameter space of mass and decay length.

**16**

# ATLAS PUB Note

ATL-PHYS-PUB-2019-002

21st January 2019

# Search prospects for dark-photons decaying to displaced collimated jets of muons at HL-LHC

The ATLAS Collaboration

Several models of new physics beyond the Standard Model predict the existence of neutral particles that decay to pairs of leptons. These particles can also be long-lived with decay length comparable to, or even larger than, the LHC detectors dimensions. The triggering and the standalone tracking capabilities of the ATLAS muon spectrometer have been exploited to search for neutral long-lived particles decaying to pairs of muons in LHC Run-2 13 TeV data and set exclusion limits on their mass and lifetime. The enormous amount of data that will be collected by ATLAS during the Run-3 (300 $fb^{-1}$) and High-Luminosity (3000 $fb^{-1}$) 14 TeV LHC phase, and the updated ATLAS detector setup, will offer a unique opportunity to probe unexplored regions of phase space in the context of such searches. This note presents sensitivity prospects for Run-3 and High Luminosity LHC discussed in the context of a Hidden Sector model predicting the decay of the Higgs boson to two neutral long-lived particles subsequently decaying into a pair of displaced muons. Two new muon trigger algorithms are studied to improve the selection efficiency of displaced muon pairs.

# 1 Introduction

Long-lived particles (LLPs) arise in several new physics models that answer open questions in particle physics: dark matter, neutrino mass, matter–antimatter asymmetry and naturalness. Examples include: supersymmetric (SUSY) models such as mini-split SUSY [1, 2], gauge-mediated SUSY breaking [3], $R$-parity-violating (RPV) SUSY [4, 5] and Stealth SUSY [6, 7]; models addressing the hierarchy problem such as Neutral Naturalness [8–11] , Hidden Valleys [12, 13] and Hidden Sectors [14]; models addressing dark matter [15–19], and the matter–antimatter asymmetry of the universe [20]; and models that generate neutrino masses [21, 22]. Many of these theoretical models predict the existence of new neutral particles that can be long-lived, which may be produced in the proton–proton collisions of the LHC and decay back into Standard Model (SM) particles far from the interaction point (IP).

The Hidden Sector models predict the existence of a dark sector that is weakly coupled to the visible one. Depending on the structure of the dark sector and its coupling to the SM, some unstable dark states may be produced at colliders and decay back to SM particles with sizeable branching fractions (Br). An extensively studied case is one in which the two sectors couple through vector portals, i.e. a dark photon ($\gamma_d$) which mixes kinetically with the SM photon. If the dark photon cannot decay to a lighter dark fermion, it will decay to SM fermions. The kinetic mixing ($\epsilon$) can be small, resulting in dark photons with a non-negligible lifetime. Highly displaced decays of dark photons would produce unique signatures which may be overlooked by searches for promptly decaying particles, and thus require dedicated analyses that represent a challenge both for the trigger and for the reconstruction capabilities of the ATLAS detector. The triggering and standalone tracking capabilities of the ATLAS muon spectrometer (MS) have been usefully exploited in the searches for displaced decays of dark photons to muon pairs based on 7 TeV, 8 TeV and Run-2 13 TeV LHC $pp$ data [23–25], and exclusion limits have been set on the $\gamma_d$ mass and lifetime.

The standard ATLAS triggers [26] are designed assuming prompt production of particles at the interaction point (IP) and therefore are very inefficient in selecting the products of displaced decays. The searches for $\gamma_d$ decays are thus based on events selected by specialised triggers dedicated to the selection of events with displaced muon pairs [23, 25]. However these triggers are still far from optimal. If the dark photon is highly boosted, muons are collimated and the trigger efficiency is limited by the finite granularity of the current hardware trigger level. In terms of an interval of the azimuthal angle $\phi$ and pseudorapidity $\eta$ [1] , the granularity is $\Delta\eta \times \Delta\phi \simeq 0.2 \times 0.2$ (Region of Interest, RoI). If the dark photon is not boosted sufficiently, the out-going muons from a displaced decay are more open and may not point to the IP. The current hardware trigger level has a tight constraint on IP pointing resulting in non-optimal selection efficiency of displaced non-pointing muon tracks.

The new ATLAS detector setup [27] and the new Trigger & Data Acquisition system [28] for the High Luminosity LHC (HL-LHC) will offer the opportunity to develop new trigger algorithms overcoming the current limitations, both for collimated and non-pointing muon pairs. A study of two new trigger algorithms has been carried out using simulated Monte Carlo (MC) samples produced according to a Hidden Sector model predicting Higgs boson decays to dark photon pairs which in turn decay to a pair of muons. This model has been chosen as a benchmark also for the Run-2 13 TeV search for displaced dark photons decaying to collimated muon pairs [23]. The analysis sensitivity is studied here for Run-3 and

[1] ATLAS uses a right-handed coordinate system with its origin at the nominal interaction point in the centre of the detector and the $z$-axis coinciding with the beam pipe axis. The $x$-axis points from the interaction point to the centre of the LHC ring, and the $y$-axis points upward. Cylindrical coordinates $(r,\phi)$ are used in the transverse plane, $\phi$ being the azimuthal angle around the beam pipe. The pseudorapidity is defined in terms of the polar angle $\theta$ as $\eta = -\ln\tan(\theta/2)$.

HL-LHC conditions. The improvement introduced by adopting the new proposed trigger algorithm is also estimated.

The benchmark model and the simulated MC samples used for this study are presented in Section 2. The new ATLAS HL-LHC setup and updated detectors are described in Section 3. Section 4 presents a comparison of the trigger efficiency on MC signal samples simulated with the Run-2 and HL-LHC setup, and the new proposed triggers. Prospects of the Run-2 search for dark photon displaced decay to muons for Run-3 (300 fb$^{-1}$) and HL-LHC (3000 fb$^{-1}$) are presented in Section 5. Finally, the results of this study are summarised in Section 6.

## 2 Benchmark model and Monte Carlo samples

Among the numerous models predicting $\gamma_{\rm d}$, one class that is particularly interesting for the LHC features the hidden sector communicating with the SM through the Higgs portal. The benchmark model used in this analysis is the Falkowsky-Ruderman-Volansky-Zupan (FRVZ) vector portal model [29, 30]. In the FRVZ model a pair of dark fermions $f_{d2}$ is produced in the Higgs boson decay. As shown in Figure 1, the dark fermion decays in turn to a $\gamma_d$ and a lighter dark fermion assumed to be the Hidden Lightest Stable Particle (HLSP). The dark photon, assumed as vector mediator, mixes kinetically with the SM photon and decays to leptons or light hadrons. The branching fractions depend on its mass [29, 31, 32]. At the LHC, these dark photons would typically be produced with large boost, due to their small mass [33, 34], resulting in collimated structures containing pairs of leptons and/or light hadrons, known as lepton-jets (LJs). If produced away from the interaction point (IP), they are referred to as "displaced LJs". The mean lifetime $\tau$ of the $\gamma_{\rm d}$ is a free parameter of the model, and is related to the kinetic mixing parameter $\epsilon$ [35] by the relation:

$$\beta\gamma c\tau \;\propto\; \left(\frac{10^{-4}}{\epsilon}\right)^2 \left(\frac{100\ \mathrm{MeV}}{m_{\gamma_d}}\right)^2\ \mathrm{s}.$$

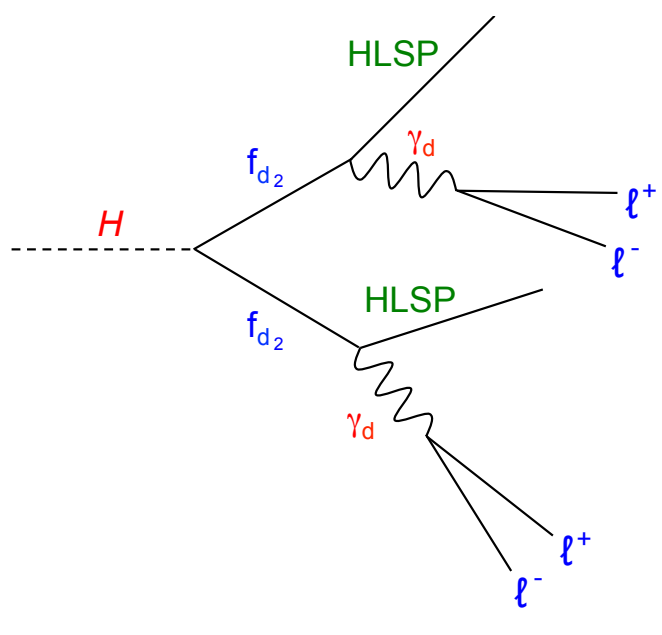

Figure 1: The Higgs boson decay to hidden particles according to the FRVZ model.

The analysis presented in this note focuses on displaced decays of dark photons into muon pairs, considering the expected $\gamma_{\rm d}$ decay BR of the FRVZ model [29]. The model assumes a gluon–gluon fusion (ggF) production mode H $\rightarrow\ 2\gamma_d + X$, thus the final results will be presented for different BR(H $\rightarrow\ 2\gamma_d + X$).

### 2.1 Monte Carlo samples

MC samples have been produced at 13 and 14 TeV center-of-mass energy for the FRVZ model and they are summarized in Table 1. Only the dominant ggF Higgs production mechanism is considered. The estimated cross section, calculated at next-to-next-to-leading order (NNLO) [36], in pp collisions at $\sqrt{s}$ = 13 TeV and $\sqrt{s}$ = 14 TeV are respectively $\sigma_{\mathrm{SM}}$ = 43.87 pb and $\sigma_{\mathrm{SM}}$ = 49.97 pb assuming $m_{\mathrm{H}}$ = 125 GeV. The mean lifetime $\tau$ and mass $m_{\gamma_{\mathrm{d}}}$ of the $\gamma_{\mathrm{d}}$ are free parameters of the model. In order to have boosted dark photons, two samples with light $\gamma_{\mathrm{d}}$ with a mass of $m_{\gamma_{\mathrm{d}}}$ = 400 MeV have been generated: a very displaced ('medium') sample with $c\tau$ = 49 mm, and a less displaced ('short') sample with $c\tau$ = 4.9 mm. A sample with unboosted dark photons has been generated considering a dark photon mass of 10 GeV. The 'medium' $m_{\gamma_{\mathrm{d}}}$ = 400 MeV sample and the $m_{\gamma_{\mathrm{d}}}$ = 10 GeV sample are used only for the trigger studies. The samples have been generated at leading order using MG5_aMC@NLO 2.2.3 [37] interfaced to the Pythia 8.210 [38] parton shower model. The A14 set of tuned parameters [39] has been used together with the NNPDF2.3LO parton distribution function (PDF) set [40].

| $\sqrt{s}$ [TeV] | $<\mu>$ | $m_H$ [GeV] | $m_{f_{d_2}}$ [GeV] | $m_{\mathrm{HLSP}}$ [GeV] | $m_{\gamma_{\mathrm{d}}}$ [GeV] | $c\tau_{\gamma_{\mathrm{d}}}$ [mm] |
|---|---|---|---|---|---|---|
| 13 | 25 | 125 | 5.0 | 2.0 | 0.4 | 49 |
| 13 | 25 | 125 | 5.0 | 2.0 | 0.4 | 4.9 |
| 13 | 25 | 125 | 30 | 10 | 10 | 856 |
| 14 | 200 | 125 | 5.0 | 2.0 | 0.4 | 49 |
| 14 | 200 | 125 | 5.0 | 2.0 | 0.4 | 4.9 |
| 14 | 200 | 125 | 30 | 10 | 10 | 856 |

Table 1: Parameters used for the Monte Carlo generation of the H $\rightarrow$ $2\gamma_d + X$ FRVZ benchmark samples.

One of the main SM backgrounds to the dark photon signal is multijet production. Samples of simulated 14 TeV multijet events are needed to compute scale factors to rescale the data-driven estimates at 13 TeV center-of-mass energy to 14 TeV. These samples are also used to evaluate the systematic uncertainties as discussed in Section 5. As shown in Table 2, the multijet MC samples have been generated with Pythia 8.210 using the A14 set of tuned parameters for parton showering and hadronisation, with the NNPDF2.3LO PDF set.

Simulated MC $Z \rightarrow \mu\mu$ events are needed for trigger and systematic uncertainties studies as discussed in Section 5. The MC samples have been generated with Powheg 1.2856 [41, 42] with Pythia 8.186 using the CT10 [43] PDF set and the AZNLO [44] tune. Four samples have been generated with different number of interactions per bunch crossing (pile-up) as shown in Table 2. The HL-LHC is expected to provide an increase of pile-up up to $\langle\mu\rangle$ = 200. These samples are used to study the effects of pile-up on the signal and background efficiency, discussed in Section 5.3.

Finally, a minimum bias sample has been generated for trigger rate evaluation for HL-LHC conditions. Pythia 8 with A2 [45] tune and MSTW2008LO [46] PDFs has been used for the generation of single proton-proton interactions.

For each MC sample, pile-up has been simulated with the soft strong-interaction processes of Pythia 8.210 using the A2 set of tuned parameters and the MSTW2008LO PDF set. Per-event weights were applied to the simulated events to correct for inaccuracies in the pileup simulation. All the generated MC samples have been processed through a full simulation of the ATLAS detector geometry and response [47] using the Geant4 [48] toolkit.

| Sample | $\sqrt{s}$ [TeV] | $< \mu >$ |
|---|---|---|
| QCD dijet | 13 | 25 |
| QCD dijet | 14 | 200 |
| $Z \rightarrow \mu\mu$ | 13 | 25 |
| $Z \rightarrow \mu\mu$ | 14 | 30 |
| $Z \rightarrow \mu\mu$ | 14 | 80 |
| $Z \rightarrow \mu\mu$ | 14 | 140 |
| $Z \rightarrow \mu\mu$ | 14 | 200 |
| Minimum bias | 14 | 200 |

Table 2: Parameters used for the Monte Carlo generation of the multijets, minimum bias and $Z \rightarrow \mu\mu$ samples.

## 3 ATLAS detector and trigger upgrades for the HL-LHC

The HL-LHC is expected to operate at a center-of-mass energy $\sqrt{s}$ = 14 TeV, providing a peak luminosity of $\sim 5 \times 10^{34}$ cm$^{-2}$s$^{-1}$ with a pile-up of $\langle\mu\rangle = 200$. A total integrated luminosity of 3000 fb$^{-1}$is expected at the end of the operations. The ATLAS collaboration has planned an extensive detector and trigger upgrade plan to cope with a luminosity ten times higher. The upgrade strategy is described in detail in the Phase-II Scoping Document [27]. For the purpose of this note, an overview of the upgrade plans of the ATLAS muon spectrometer in the barrel region ($|\eta| < 1$) [49] and of the muon trigger and data acquisition [28] will be presented in detail.

The ATLAS experiment plans to increase the maximum rate capability of the first trigger level (Level-0, L0) to 1 MHz at 10 $\mu$s latency. This requires new electronics for the MS. The replacement of the precision chamber read-out electronics will make it possible to include their data in the L0 decision and thus to increase the selectivity of the muon trigger. The acceptance of the present Resistive Plate Chamber (RPC) trigger system in the barrel region will be increased from 75% to 95% by the installation of additional thin-gap RPCs with a substantially increased high-rate capability compared to the current RPCs. In Figure 2, a transverse section of the barrel region is presented, showing the four layers of RPC chambers: the new RPC0 layer, also called Barrel Inner (BI), the two Barrel Middle (BM) RPC1-2 layers and the Barrel Outer (BO) RPC3 layer.

The new L0 muon trigger is designed to operate on the same principle as the Run-2 identification algorithm that runs at hardware level. The algorithm requires a coincidence of hits in the different RPC layers within a $\Delta\eta \times \Delta\phi$ window pointing to the IP. The width of the window is related to the transverse momentum ($p_{\mathrm{T}}$) threshold of the trigger. Different quality requirements are made on number of hits fully exploiting the four layers layout. The L0 will provide good efficiency at a moderate rate for low threshold single muon triggers. The expected RPC trigger system improvement in terms of acceptance $\times$ efficiency is shown in Figure 3 for prompt muons with $p_{\mathrm{T}}$ = 25 GeV. The efficiency of the current Run-2 setup, that requires a coincidence on all existing layers (BM-BM-BO) called "3/3 chambers" trigger, is shown by the red histogram. At the HL-LHC the additional RPC layer will be exploited requiring coincidence on 3 out of 4 layers ("3/4 chambers"). This trigger can be extended by requiring hits on both the innermost and outermost layers ("3/4 chambers + BI-BO"). The efficiencies of these two triggers are shown respectively in blue and green. The "3/4 chambers + BI-BO" selection will improve the trigger acceptance × efficiency of the Run-2 setup from 78% up to 96%.

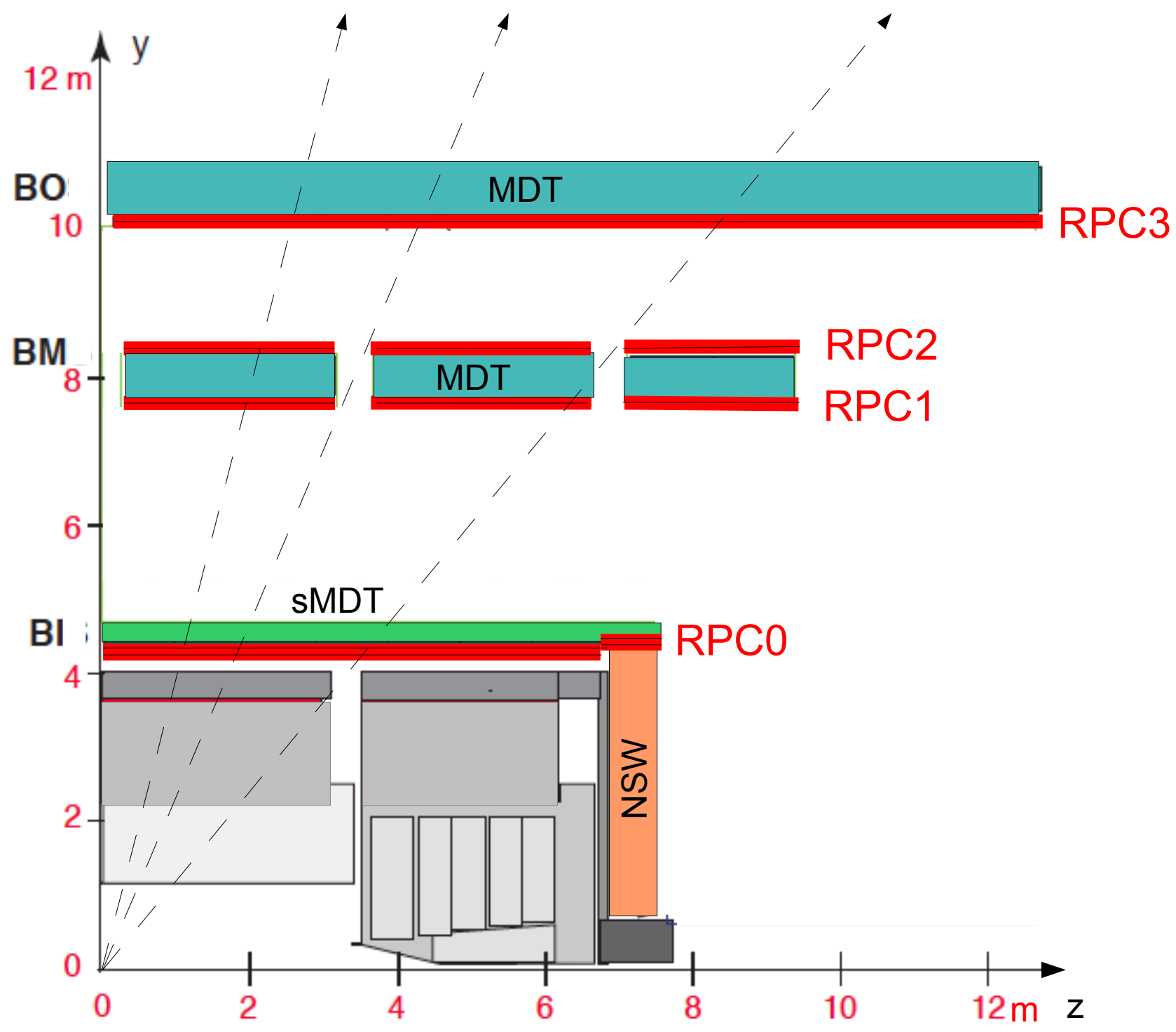

Figure 2: Transverse section of a small sector in the barrel region showing the four layers of RPC chambers: RPC0 in the barrel-inner (BI), RPC1-2 in the barrel-middle (BM), and RPC3 in the barrel-outer (BO) layers. The three dashed lines represent muon trajectories traversing two, three and four RPC chambers [28].

## 4 New proposed triggers

Hidden Sector scenarios can produce signatures which are not identified by the standard trigger system. The standard ATLAS triggers are optimised to select prompt events and are very inefficient in the selection of displaced objects. The upgraded detector will offer a great opportunity to implement dedicated triggers to improve the selection of displaced muons.

Two new trigger selections have been studied in this work: one dedicated to triggering on collimated LJs in boosted scenarios, based on requiring muons in a single RoI; a second one dedicated to triggering on non-boosted scenarios, loosening the pointing requirements applied in Run-2. With these new approaches it is possible to choose a lower single muon $p_T$ threshold as compared to the Run-2 configuration, improving the selection efficiency of events with displaced muon pairs without increasing significantly the trigger rate.

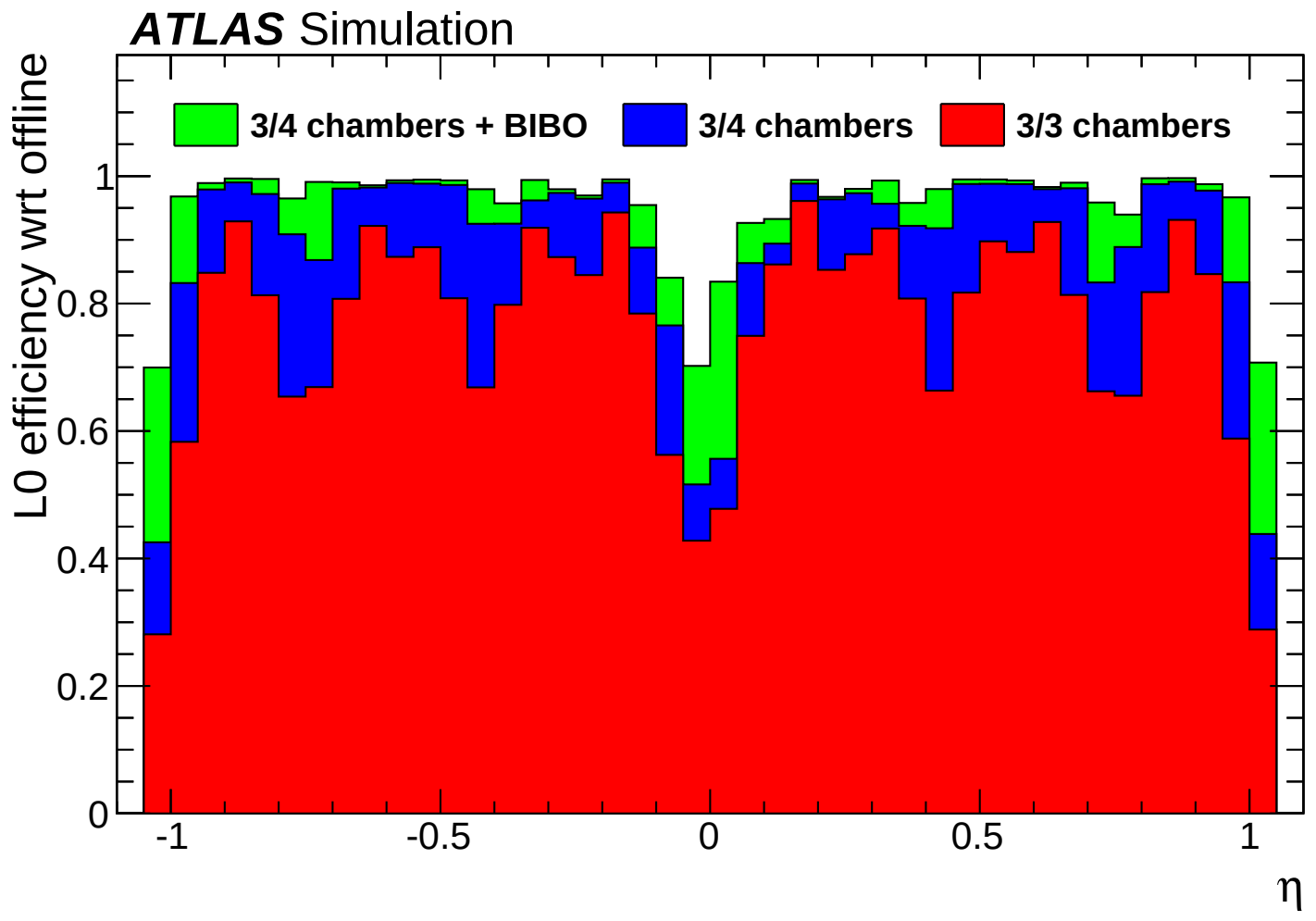

Figure 3: Acceptance × efficiency of the RPC trigger system as a function of $\eta$ for the Run-2 system "3/3 chambers" trigger (red), for the HL-LHC "3/4 chambers" trigger (blue) and for the HL-LHC "3/4 chambers + BI-BO" trigger (green). The efficiency is evaluated using Monte Carlo simulation of single muons with a fixed transverse momentum of 25 GeV [28].

### 4.1 Comparison between the Run-2 and the HL-LHC baseline setups

The efficiency of the low-level trigger of the Run-2 setup has been compared to the efficiency of the low-level trigger of the HL-LHC setup, using FRVZ 'medium' MC samples simulated for 13 TeV and 14 TeV conditions. The single muon trigger with $p_T$ = 20 GeV threshold has been used and only events with truth muons in the barrel region $|\eta| < 1.05$ have been considered. Figure 4 shows the $p_T$ = 20 GeV Run-2 (blue) and the L0 HL-LHC (red) low level muon trigger efficiency as a function of the truth transverse decay position (Lxy) of the $\gamma_d$. As expected, the HL-LHC L0 has a higher efficiency with respect to the Run-2 low level for decays that happen before the new BI RPC layer ($\sim$ 5 m). The two drops at $\sim$ 6 m and $\sim$ 7 m correspond to the $\gamma_d$ decaying after the BI and the BM RPC layer, respectively. The $p_T$ = 20 GeV threshold results in a reduced efficiency at small decay length that correspond to decays of low boosted dark photons.

### 4.2 L0 multi-muon scan trigger

In a scenario with highly boosted $\gamma_d$, the decay muons are close-by and likely fall in the same RoI. Figure 5 shows the opening $\Delta\phi$ angle of the two out-going muons of the dark photon decay as a function of the $p_T$ of the leading muon for the 'medium' MC FRVZ sample: most of the signal is between 10 and 20 GeV, and both muons fall in the same RoI. The Run-2 system is able to select only one muon candidate per RoI. Due to the high single muon trigger rate, it is not possible to go below the 20 GeV threshold, resulting in a major loss of events.

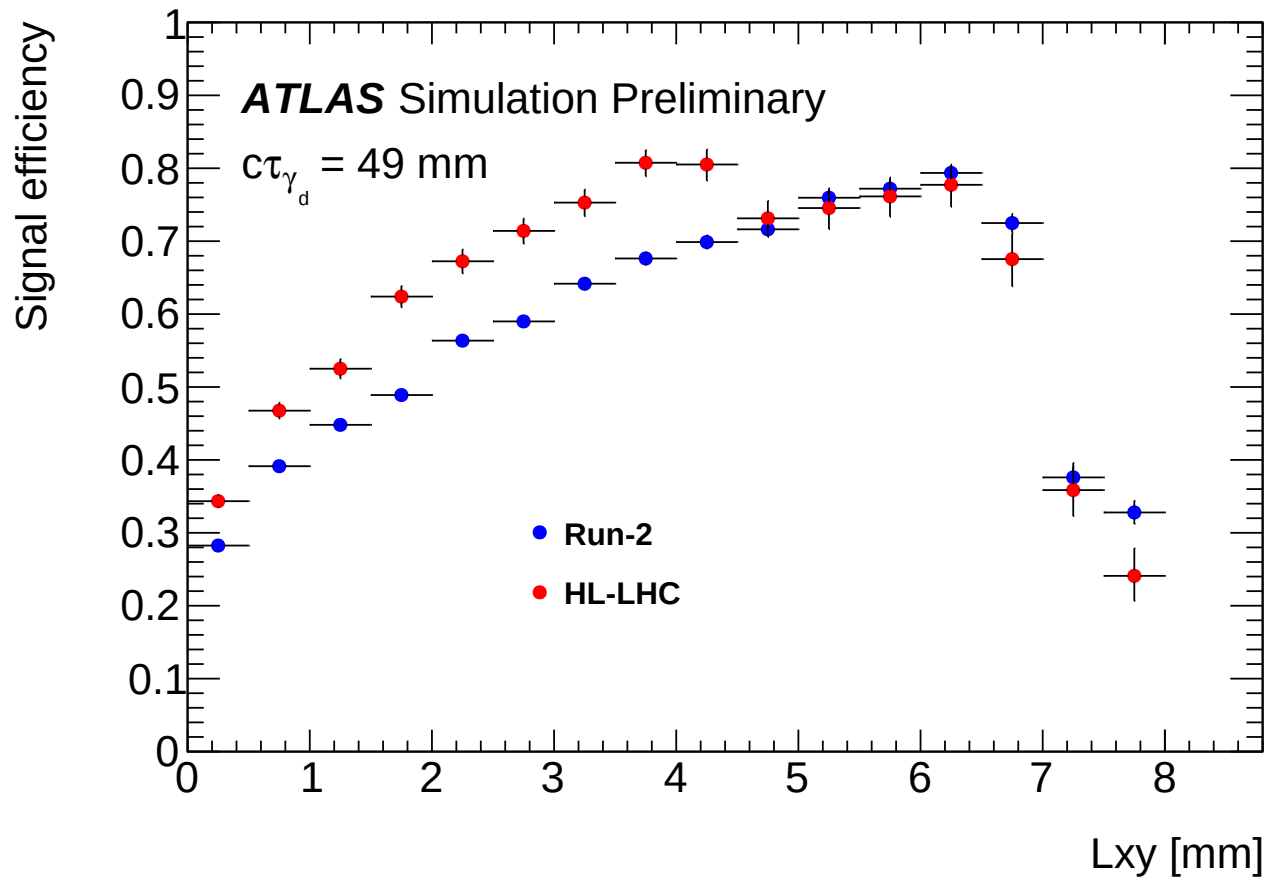

Figure 4: The muon trigger efficiency for $p_{\mathrm{T}} = 20$ GeV Run-2 (blue) and the HL-LHC (red) low level muon trigger as a function of the truth transverse decay position of the $\gamma_{\mathrm{d}}$.

A new approach is proposed to include in the sector logic multiple trigger candidates in the same RoI. This would allow the design of a new trigger selection called 'L0 multi-muon scan' with lower $p_{\mathrm{T}}$ thresholds resulting in a higher efficiency without increasing sensibly the trigger rate.

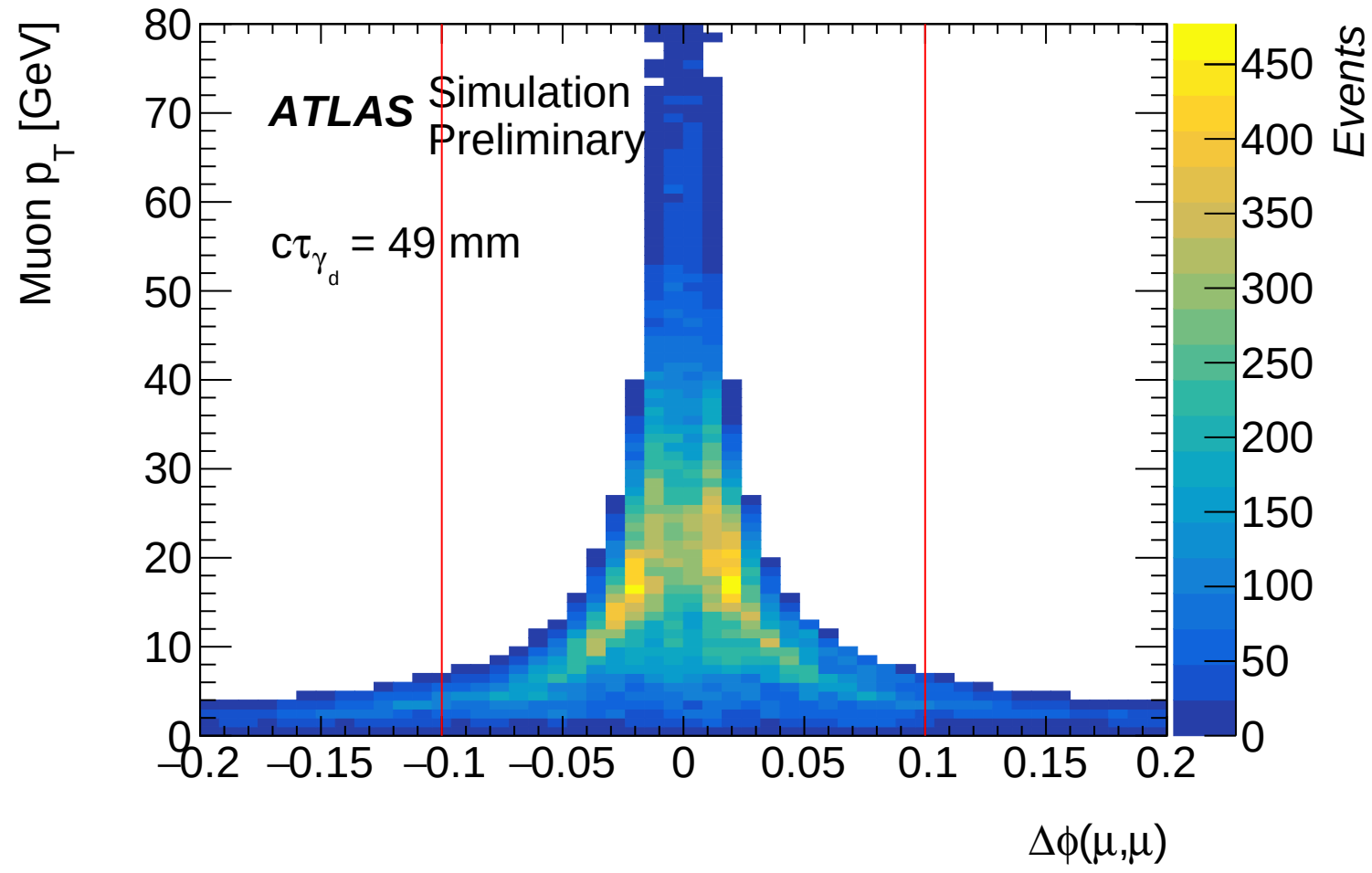

Figure 5: Truth transverse momentum distribution of the leading muon as a function of the opening angle in the $\phi$ plane of the two muons of the $\gamma_{\mathrm{d}}$ decay. Red lines show the RoI size. The 'medium' sample with average lifetime $c\tau = 49$ mm has been used.

The new trigger algorithm is designed to analyze hit patterns in the MS. As a first step, the algorithm searches for the pattern with the highest number of hit points, called best pattern, in the MS to form the primary L0 muon candidate. Then all the other possible hit patterns, not compatible with the best pattern, are searched for in the same RoI to form the secondary L0 muon candidates. A quality cut is applied to reduce the influence of noisy hits, requiring patterns with hits on at least three different RPC layers. Patterns are requested to not share RPC hits. If at least one secondary pattern is found, an additional L0 muon is assumed to be found in the RoI. The new L0 trigger algorithm is defined by the logical OR of a single muon L0 with $p_T$ = 20 GeV threshold and a multi-muon L0 with $p_T$ = 10 GeV threshold. The fake rate of the multi-muon trigger depends on the angular separation between the patterns found in the RoI. A minimal angular separation in the $\phi$ plane between the secondary pattern and the best one, $\Delta\phi_{RoI}$, is required to lower the fake rate. The angular separation $\Delta\phi_{RoI}$ is defined as a 'resolution parameter' and it is the only input to the algorithm. In order to fix the value of the resolution parameter, the efficiency of the proposed L0 trigger algorithm has been studied as a function of the resolution parameter for the 'medium' benchmark signal FRVZ sample. The fake rate of the trigger as a function of the resolution parameter has also been studied. The fake rate has been estimated using a sample of $Z \to \mu\mu$ decays simulated with HL-LHC conditions, by evaluating the trigger rate of events with single muons in the RoI. Figure 6 shows the trigger efficiency for the 'medium' signal FRVZ model and the fake rate as a function of the $\Delta\phi_{RoI}$ parameter. The efficiency is defined as the number of triggered events over the number of total events. The $\Delta\phi_{RoI}$ = 0.01 value for the resolution parameter has been adopted as basic selection (working point), this is the best compromise between efficiency and fake rate. The rate of the proposed L0 trigger is evaluated on the minimum bias MC sample and is estimated to be 13 kHz at 5 $\times$ $10^{34}$ cm$^{-2}$s$^{-1}$. From this estimate of the trigger rate we can assume the multijet background contamination to be small, a more complete study on multijet cannot be performed due to the limited samples statistics.

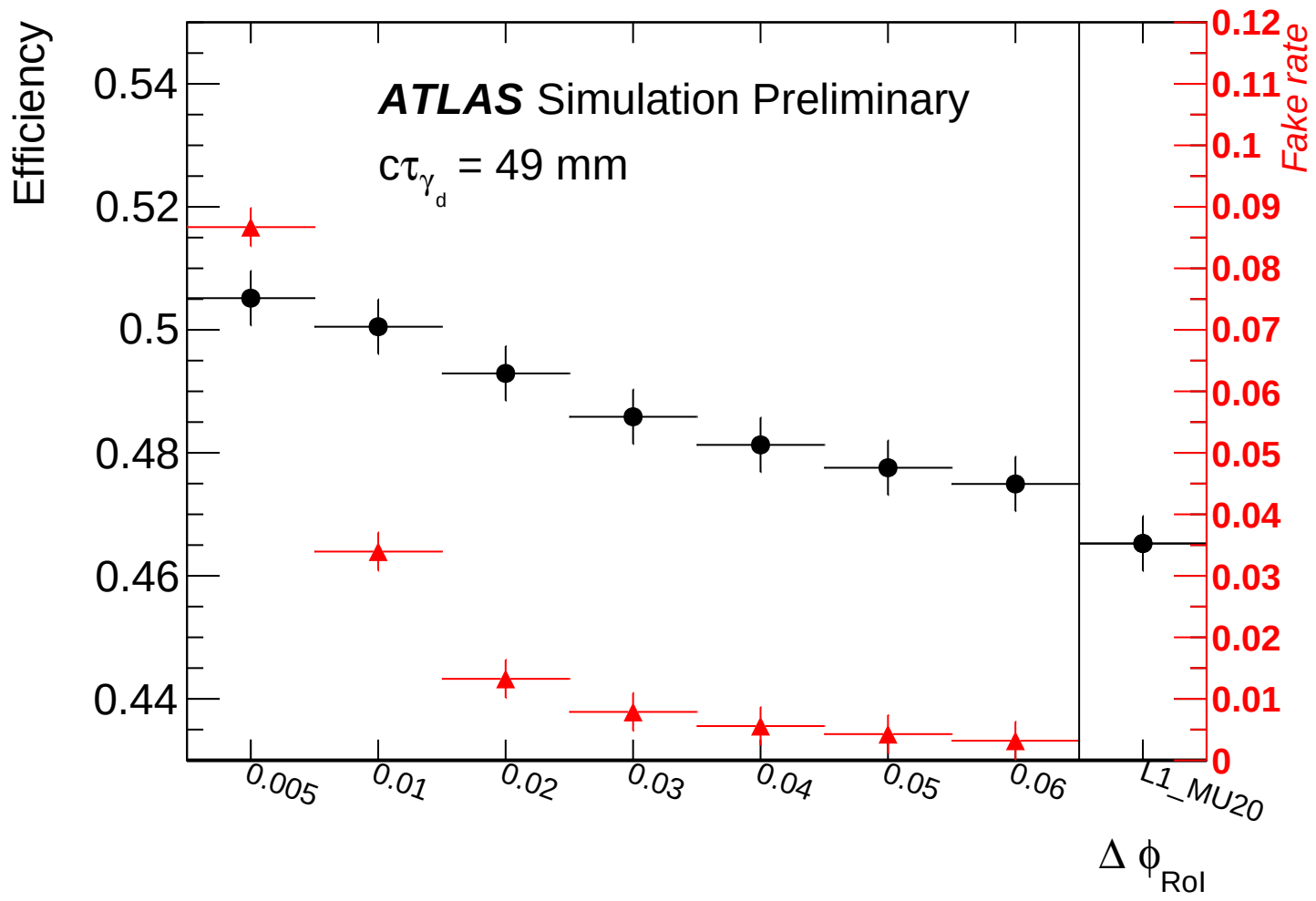

Figure 6: L0 multi-muon scan trigger efficiency for the 'medium' FRVZ signal sample in black (left axis) and fake rate in red (right axis) as a function of the resolution parameter $\Delta\phi_{RoI}$. In the separate box the Run-2 standard $p_T$ = 20 GeV (L1_MU20) trigger efficiency is shown for comparison.

Figure 7 shows the L0 multi-muon scan trigger efficiency as a function of the truth opening angle $\Delta\phi(\mu, \mu)$ between the two muons of the $\gamma_d$ decay. As reference, two single muon selections are shown with 10 (L0_MU10) and 20 (L0_MU20) GeV $p_T$ threshold. Moreover a preselection is made at truth level to select events with leading muon $p_T$ > 10 GeV and sub-leading muon $p_T$ > 5 GeV. The results are presented for both the 'short' and the 'medium' FRVZ MC samples. The opening angle $\Delta\phi(\mu, \mu)$ depends on both the decay distance and transverse momentum of the $\gamma_d$. For very small $\Delta\phi(\mu, \mu)$ the 'short' sample is expected to have on average larger $p_T$ of the $\gamma_d$ with respect to the 'medium' sample, and therefore the trigger efficiency is larger for this sample. At larger $\Delta\phi(\mu, \mu)$ both samples are expected to have the same trigger efficiency. An overall improvement up to 7% is achieved with respect to the baseline $p_T$ = 20 GeV selection.

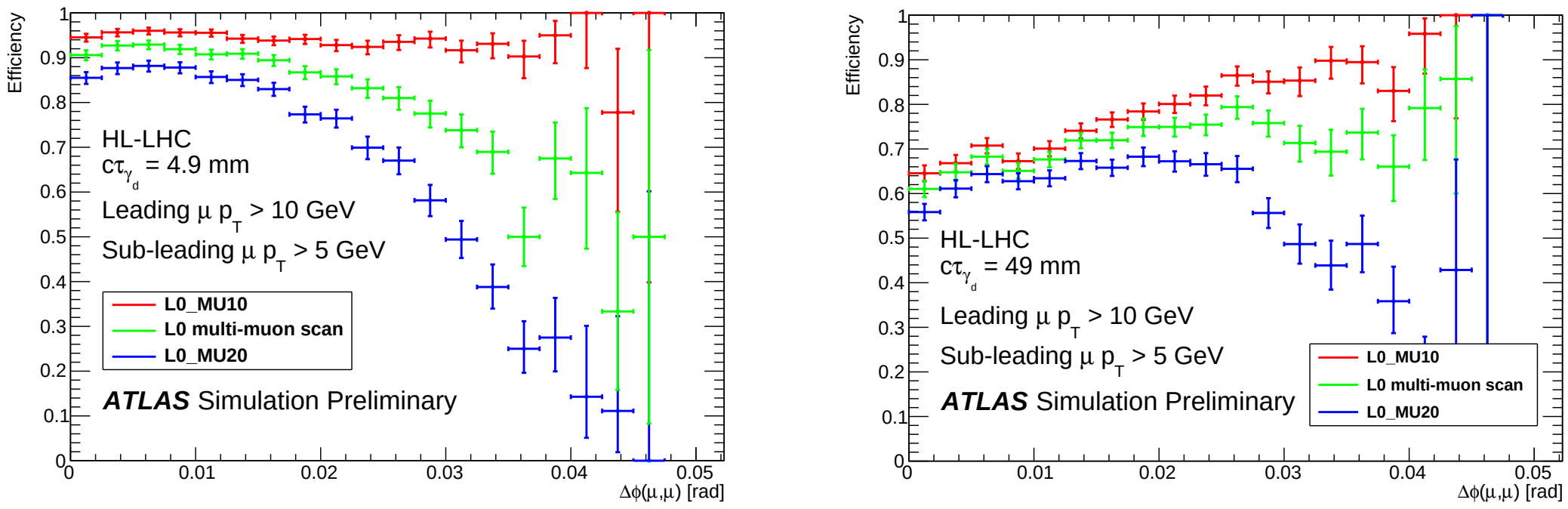

(a) 'short' sample, $c\tau = 4.9mm$ (b) 'medium' sample, $c\tau = 49mm$

Figure 7: Efficiency for different trigger selections as a function of the opening angle of the two muons of the $\gamma_\mathrm{d}$ decay. Single muon with 10 (L0_MU10) and 20 (L0_MU20) GeV $p_\mathrm{T}$ threshold are shown in red and blue, respectively. The L0 multi-muon scan trigger is shown in green.

## 4.3 L0 sagitta muon trigger

Considering a different scenario with unboosted $\gamma_\mathrm{d}$, the out-going muons may not be pointing to the IP. The L1 Run-2 trigger has a tight constraint on selecting only pointing muons resulting in non optimal selection of these exotic signatures.
The benchmark FRVZ sample with 10 GeV $\gamma_\mathrm{d}$ mass can be used to study a new trigger to select events with displaced non-pointing muons. In this sample the muons produced in the $\gamma_\mathrm{d}$ decay have a large track impact parameter $z_0$, defined as the minimum distance in the $z$ coordinate (along the beam axis) of the muon track extrapolated to the IP. Figure 8 shows the $z_0$ distribution as a function of the truth transverse momentum of the muons. The efficiency of the low level Run-2 muon triggers rapidly drops to zero for values of $|z_0| \geq$ 100 mm: the transverse momentum of the non-pointing muon is mis-reconstructed due to the pointing constraint to the IP, resulting in a underestimation of the true $p_\mathrm{T}$ value. As an example, a non-pointing muon that would have passed the $p_\mathrm{T}$ = 20 GeV trigger threshold is often only triggered by a 5 GeV threshold.
To recover this loss of efficiency, a new muon trigger, called 'L0 sagitta muon', and based on the sagitta method is proposed. The sagitta, defined as the vertical distance from the midpoint [2] of the chord [3] to the arc [4] of the muon trajectory itself, can be used to estimate the momentum of a charged particle travelling inside a magnetic field. The sagitta of a muon track can be computed at the L0 trigger level using $\eta - \phi$ measurement points in the BI, BM and BO RPC stations. The map between the inverse of the sagitta and the muon transverse momentum has been studied using a MC sample of single muons generated according to a uniform transverse momentum distribution. Figure 9 shows the distribution of the inverse of the sagitta as a function of the truth muon transverse momentum, the profile is also superimposed. The mean value of the inverse of the sagitta for $p_\mathrm{T}$ = 20 GeV pointing truth muon is $s^{-1} = 9 \times 10^{-6}$ mm$^{-1}$. High transverse momentum non-pointing muons can be thus selected using a L0 muon trigger with low $p_\mathrm{T}$ = 5 GeV threshold, computing the inverse of the sagitta and requesting a cut on $s^{-1} \leq 9 \times 10^{-6}$ mm$^{-1}$.

The performance of the L0 sagitta muon trigger has been studied with the FRVZ benchmark MC sample with $m_{\gamma_d}$ = 10 GeV. Figure 10 shows the efficiency, as a function of the muon transverse momentum, of the L0 $p_T$ = 20 GeV muon trigger (red), the L0 sagitta muon trigger (blue) and the logical OR of the two triggers (green). A ~20% improvement in efficiency is achieved by adding the new trigger.

[2] The midpoint is defined as the middle point of a segment
[3] The chord of a circle is a line segment that connects two points of the circle itself
[4] The arc is a portion of the circumference of a circle.

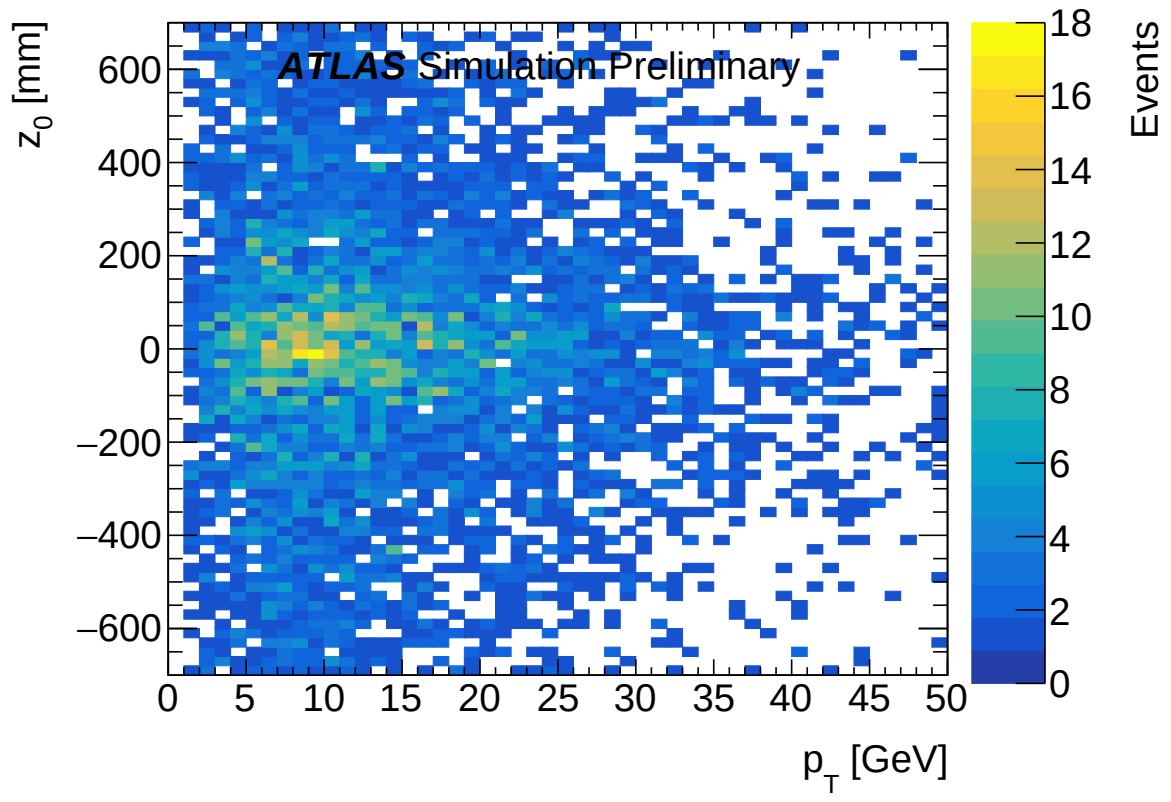

Figure 8: $z_0$ muon impact parameter as a function of the truth muon transverse momentum in the FRVZ MC sample with 10 GeV $\gamma_d$'s.

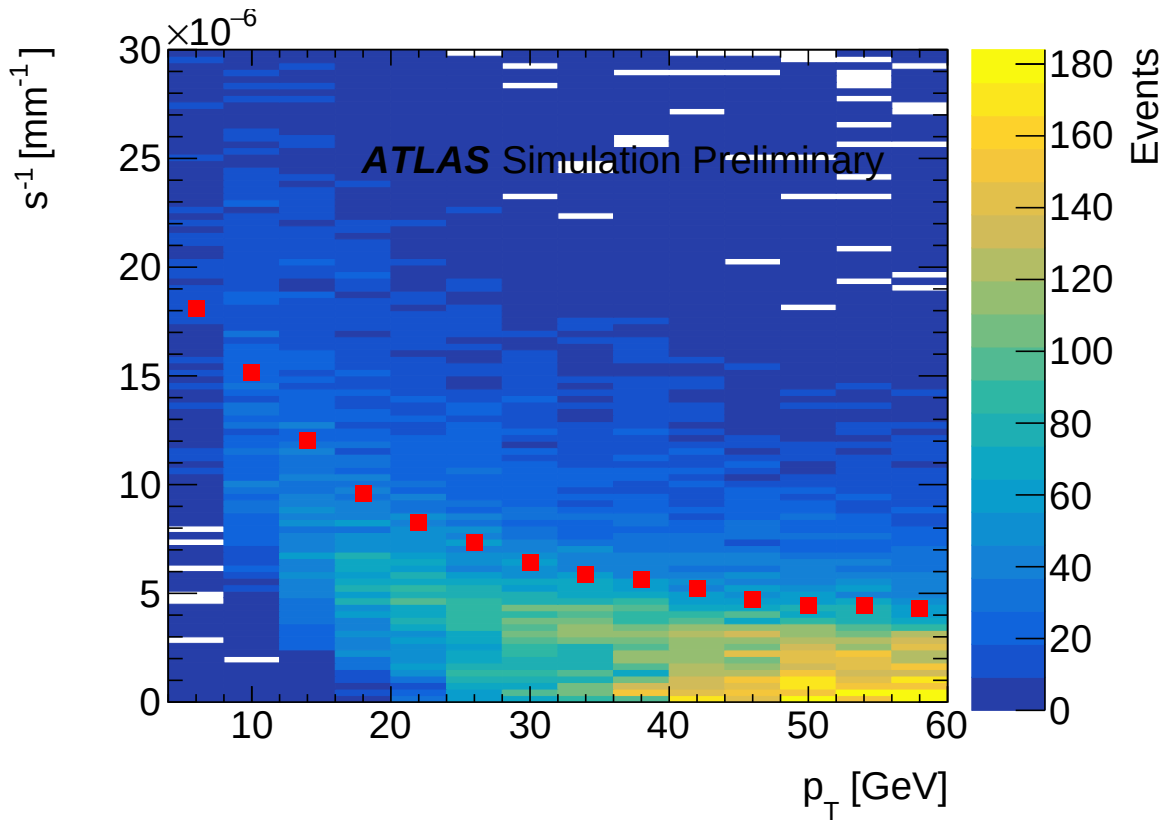

Figure 9: Inverse of the sagitta of pointing muons as a function of the muon truth transverse momentum. The profile of the inverse of the sagitta over the muon transverse momentum is overlaid in red.

The L0 sagitta muon trigger has been tested on single pointing muon events generated with a flat $p_T$ in the range 1-50 GeV. Figure 11 shows the standard 20 GeV muon trigger efficiency (red) and the L0 sagitta muon trigger efficiency when the $p_T$ = 20 GeV trigger has not fired (blue) as a function of the muon transverse momentum. The contamination from low $p_T$ muons is low. Furthermore, it can be further reduced via tuning of the sagitta threshold.

# 5 Prospects for Run-3 and the HL-LHC

The evaluation of the expected sensitivity of the displaced dark photon search after Run-3 and HL-LHC operations is based on the 2015+2016 Run-2 ATLAS analysis. This analysis, which is in the finalisation phase and uses 36 fb$^{-1}$ of 13 TeV data, improves the early Run-2 analysis based on 3.4 fb$^{-1}$ [23] by making use of multivariate techniques for signal discrimination against the backgrounds. The benchmark signal model used in the Run-2 search is a FRVZ model with 400 MeV $\gamma_d$ mass and lifetime $c\tau = 49$ mm. The branching fraction of the $\gamma_d$ decay to muons is 45%.

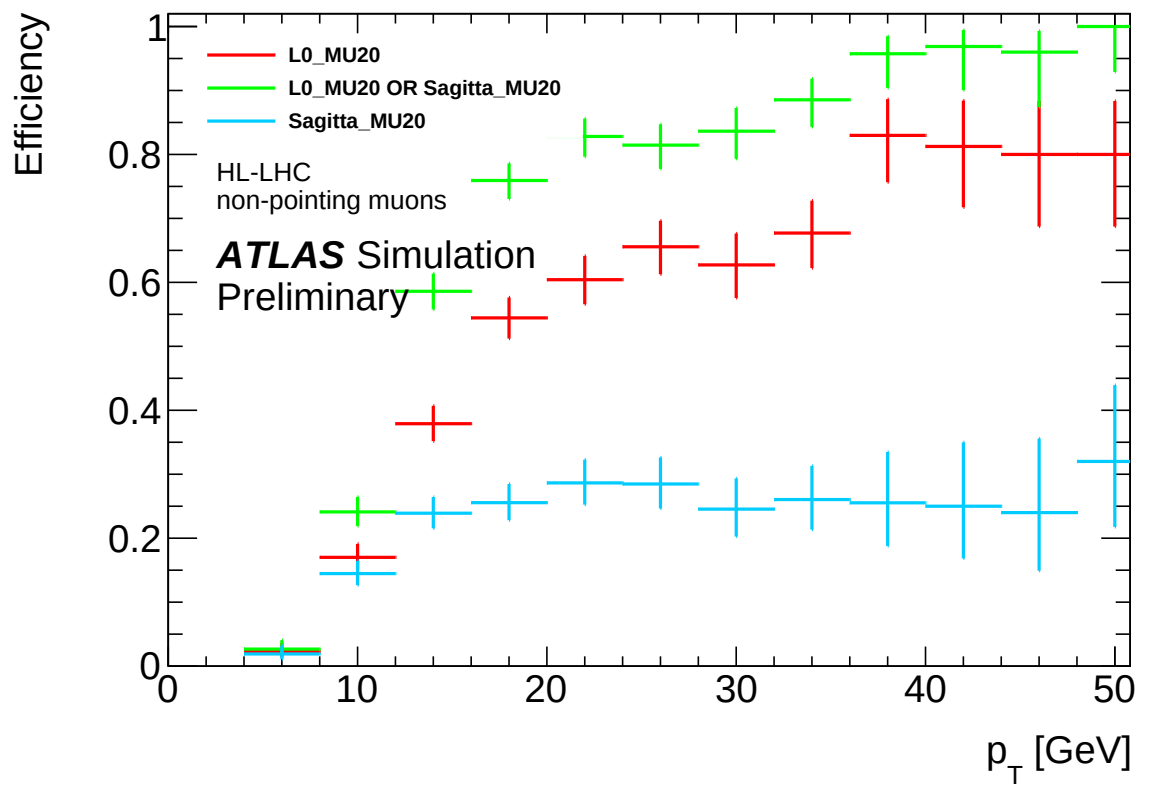

Figure 10: Trigger efficiency comparison for FRVZ sample with $m_{\gamma_d} = 10$ GeV: L0 $p_T = 20$ GeV threshold (red), L0 sagitta muon trigger (blue) and the OR of the two triggers (green).

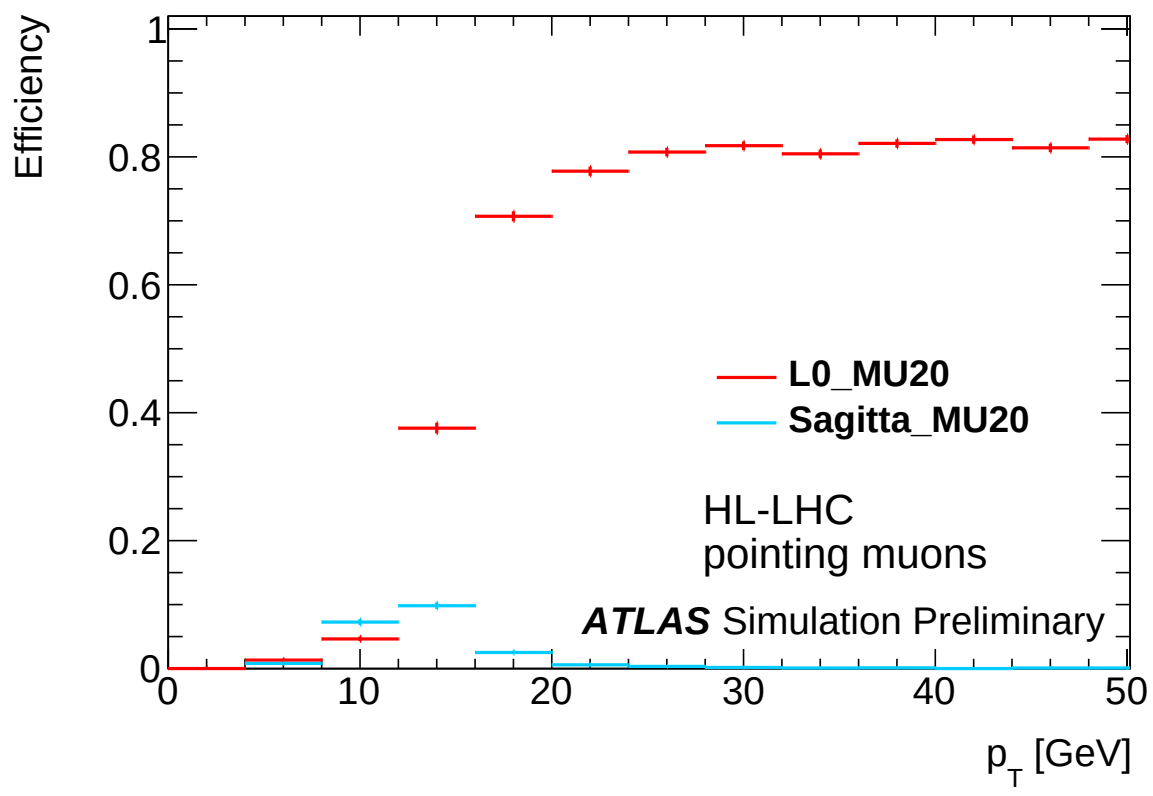

Figure 11: Muon trigger efficiency for MC single pointing muon events: standard 20 GeV threshold (red) and L0 sagitta muon trigger when the previous trigger is not fired (blue).

### 5.1 Event selection and background estimation in the Run-2 analysis

Only dark photons decaying to muons after the pixel detector and before the BM RPC trigger chambers are considered. Muons are reconstructed using information from the MS only (no match with an Inner Detector (ID) track is required). At least two muons reconstructed in a $\Delta R = 0.4$ cone, isolated with respect to calorimeter jets, identify a dark photon decay to muons (muonic LJ). The search is limited to a pseudorapidity interval $-2.4 \leq \eta \leq 2.4$, rejecting events in the barrel-endcap transition region $1.0 \leq |\eta| \leq 1.1$. Only events with two muonic lepton-jets are selected.

Multijet events and cosmics-rays are the sources of background to the muonic lepton-jet signal.In Run-2 analysis the secondary cosmics-ray background contributes around 7% of the total background and depends only on the duration time of data taking, therefore it is expected to be a marginal background for the Run-3 and HL-LHC prospects. The residual cosmics-ray background in the signal region is estimated by applying the analysis selection to empty bunch crossing data and then scaling the remnant events to the filled bunch crossing data.
The multijet background is reduced using track isolation around the muonic lepton-jet direction: displaced muonic lepton-jet are expected to be highly isolated in the inner tracker. The track isolation ($\sum p_{\mathrm{T}}$) is defined as the sum of the transverse momenta of the tracks reconstructed in the inner tracker and matched to the primary vertex of the event, in a $\Delta R = 0.4$ cone around the muonic lepton-jet direction. Residual multijet background has been estimated with a data-driven ABDC method, relying on the assumption that the background events distribution can be factorised in a plane of two uncorrelated variables in four sub-regions and expecting most of the signal events in only one of them. The two uncorrelated variables used in the ABCD methods are the maximum value of the $\sum p_{\mathrm{T}}$ of the two muonic lepton-jets and the opening angle of the two muonic lepton-jets in the azimuthal plane ($|\Delta\phi|$). The opening angle $|\Delta\phi|$ is expected to be large as the dark photons in the FRVZ model are produced almost back-to-back.

### 5.2 Extrapolations of Run-2 results to Run-3 and HL-LHC

Run-2 results have been extrapolated to Run-3 and HL-LHC assuming an integrated luminosity at the end of the operations respectively of 300 fb$^{-1}$ and 3000 fb$^{-1}$ at $\sqrt{s}$ = 14 TeV. Analysis selection and detector efficiency for Run-3 and HL-LHC are considered to be the same of the Run-2 analysis. The extrapolation procedure is described below.

For the extrapolation to Run-3 at 300 fb$^{-1}$ and $\sqrt{s}$ = 14 TeV, considering no change in pileup with respect to Run-2, signal events and multijet background events have been scaled according to the difference in integrated luminosity and centre-of-mass energy. The cosmics-ray background is assumed to scale with duration of data taking, a scale factor of 2.5 has been assumed.

For the extrapolation to HL-LHC, in addition to the difference in integrated luminosity, a scale factor has been considered to take into account the increase in centre-of-mass energy from 13 TeV to 14 TeV and the pileup conditions up to 200 interactions per bunch crossing. The scale factor is calculated directly from the comparison between the simulated MC samples at HL-LHC conditions and the simulated MC samples with Run-2 conditions. The resulting scaling factor are 1.25 for multijet events and 1.13 for the FRVZ signal model. The cosmics-ray background is assumed to scale with duration of data taking, comparing Run-2 to the expected 12 years duration of the HL-LHC data taking, a scale factor of 6 has been assumed . Moreover, since the Run-2 analysis is sensitive to $\gamma_{\mathrm{d}}$ with mass up to 2 GeV, we can assume an improvement in signal selection of 7 % by adopting the L0 multi-muon scan trigger selection discussed in Sec. 4.2.

The expected number of background and signal events after Run-3 and HL-LHC data taking are summarised in Table 3.

### 5.3 Uncertainties

Uncertainties have been extrapolated from the Run-2 reference analysis. The statistical sources of uncertainties have been scaled with the expected integrated luminosity, for both Run-3 and HL-LHC.
The systematic uncertainties for Run-3 have been assumed to be the same as in the Run-2 analysis.

| Muonic channel | $\sqrt{s}$ TeV | Expected background | Expected signal FRVZ model |
|---|---|---|---|
| Run-3 | 14 | 930 ± 12 (stat.) | 5325 ± 213 (stat.) |
| HL-LHC | 14 | 11685 ± 48 (stat.) | 65648 ± 2626 (stat.) |

Table 3: Expected number of background and FRVZ signal events after Run-3 and HL-LHC operations. Statistical errors only are presented and $BR(\gamma_d \to \mu\mu) = 45$ % is used. Cosmics-ray events are subtracted.

For the HL-LHC projection systematic uncertainties have been evaluated according to the specifications of the ATLAS collaboration for upgrade studies [50]. The upgraded ATLAS detector is assumed to perform at least as well as in Run-2, therefore analysis specific uncertainties (like reconstruction and trigger efficiency) have been considered to be the same as in the Run-2 analysis. The systematic uncertainty on the jet energy resolution has been taken to be the same as in the Run-2 analysis, whilst the uncertainty on the jet energy scale has been halved. The high pileup conditions during the HL-LHC operations will affect the efficiency of the track isolation variable used in the Run-2 analysis. The effect of the higher pile-up on the $\sum p_{\mathrm{T}}$ has been evaluated by computing the efficiency of $\sum p_{\mathrm{T}}$ selection for isolated muons from the $Z \to \mu\mu$ decay, using the MC samples generated with different pileup conditions. The distributions of the isolation efficiency as a function of the isolation variable $\sum p_{\mathrm{T}}$ for four different samples with an increasing number of interaction vertices are shown in Figure 12. The systematic uncertainty has been assumed to be 18%, corresponding to the maximum variation of the efficiency at $\sum p_{\mathrm{T}} = 4.5$ GeV which is the value that defines the signal region in the Run-2 analysis. Finally the uncertainty on the integrated luminosity of the full HL-LHC dataset has been assumed to be 1%. A summary of the systematic uncertainties is given in the Table 4.

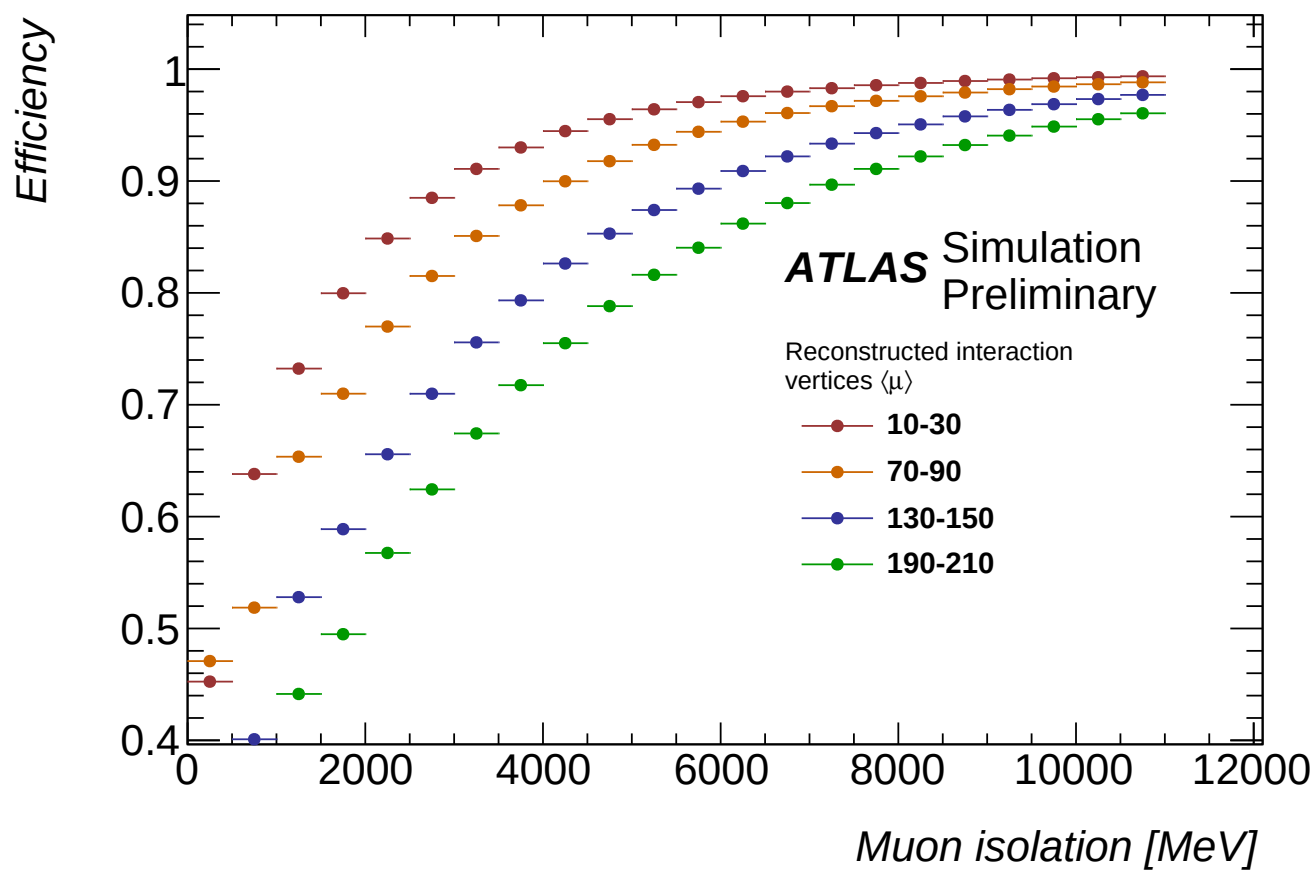

Figure 12: Isolation efficiency as a function of $\sum p_T$ for four intervals of the number of reconstructed interaction vertices per event in a $Z \to \mu\mu$ MC sample.

## 5.4 Results

The *CLs* method [51] has been used to set upper limits at 95% CL on the cross-section times branching fraction of H $\to$ $2\gamma_d + X$ as a function of the $\gamma_{\mathrm{d}}$ lifetime, considering a 45% dark photon branching ratio to muons.

| Systematic uncertainty (in %) | Run-3 | HL-LHC |
|---|---|---|
| Luminosity | 2.2 | 1.0 |
| Reconstruction efficiency $\gamma_d$ | 9.7 | 9.7 |
| Effect of pile-up on $\Sigma p_T$ | 10 | 18 |
| Reconstruction of the $p_T$ of the $\gamma_d$ | 5.1 | 5.1 |
| Pile-up | 2.0 | 2.0 |
| Jet energy scale | 5.0 | 2.5 |
| Jet energy resolution | 2.0 | 2.0 |

Table 4: Summary of the systematic uncertainties used for sensitivity extrapolation to Run-3 and HL-LHC.

| Excluded $c\tau$ [mm] muonic-muonic | Run-2 | Run-3 | HL-LHC | HL-LHC w/ L0 muon-scan |
|---|---|---|---|---|
| BR(H $\rightarrow 2\gamma_d + X$)=10 % | $2.2 \leq c\tau \leq 111$ | $1.15 \leq c\tau \leq 435$ | $0.97 \leq c\tau \leq 553$ | $0.97 \leq c\tau \leq 597$ |
| BR(H $\rightarrow 2\gamma_d + X$)=1 % | - | $2.76 \leq c\tau \leq 102$ | $2.18 \leq c\tau \leq 142$ | $2.13 \leq c\tau \leq 148$ |

Table 5: Ranges of $\gamma_d$ $c\tau$ excluded at 95 % CL for H $\rightarrow 2\gamma_d + X$ assuming BR(H $\rightarrow 2\gamma_d + X$) = 10 % and BR(H $\rightarrow 2\gamma_d + X$) = 1 %.

Results for the three different scenarios are presented in Figure 13: 300 fb$^{-1}$after Run-3, 3000 fb$^{-1}$after HL-LHC and 3000 fb$^{-1}$after HL-LHC including the multi-muon scan trigger improvement. Table 5 shows the excluded $c\tau$ ranges assuming BR(H $\rightarrow 2\gamma_d + X$) = 10 % and BR(H $\rightarrow 2\gamma_d + X$) = 1 %.

The exclusion limits are re-interpreted in the context of the vector portal model. The exclusion contour plot in the plane defined by the $\gamma_\mathrm{d}$ mass and the kinetic mixing parameter $\varepsilon$ is presented in Figure 14. Two different scenarios are shown assuming a Higgs decay branching fraction to the hidden sector of 1% [5]: 300 fb$^{-1}$after Run-3, 3000 fb$^{-1}$after HL-LHC including multi-muon scan trigger improvement.

[5] Results for 10% BR are visually very similar to the 1% ones in log-scale and are not shown in the figure.

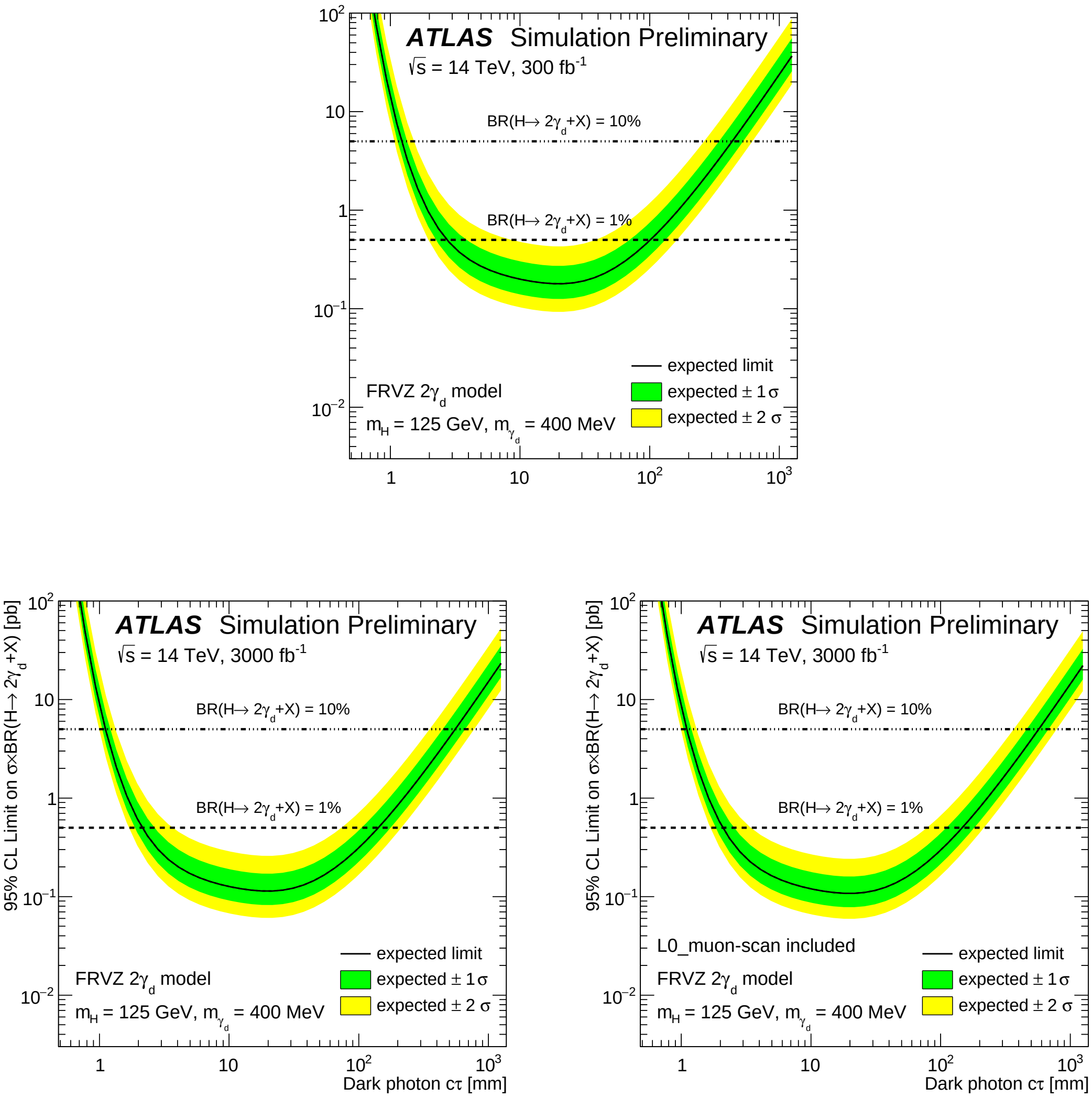

Figure 13: 95% CL upper limit on the cross-section times branching fraction of $H \rightarrow 2\gamma_d + X$ as a function of the $\gamma_d$ lifetime, considering 45% dark photon branching ratio to muons. Three different scenario are considered: 300 fb$^{-1}$after Run-3 (top), 3000 fb$^{-1}$after HL-LHC (right) and 3000 fb$^{-1}$after HL-LHC including multi-muon scan trigger improvement (left).

## 6 Conclusions

Two new muon trigger algorithms to improve the selection of displaced dark photons decaying to muons at the HL-LHC have been presented. The performance of the two triggers has been evaluated on MC simulated events based on a simplified model that predicts the Higgs boson decay to dark photons pairs. A first trigger, the L0 multi-muon scan trigger, has been designed to improve trigger efficiency for close-by muon pairs. Tests on the MC benchmark sample show a gain in efficiency of ~7% with respect to the baseline selection used in Run-2. A second trigger, the L0 sagitta muon trigger, has been designed to trigger on displaced non-pointing muons. An efficiency improvement of ~20% is achieved on the benchmark MC sample with respect to the Run-2 baseline selection.

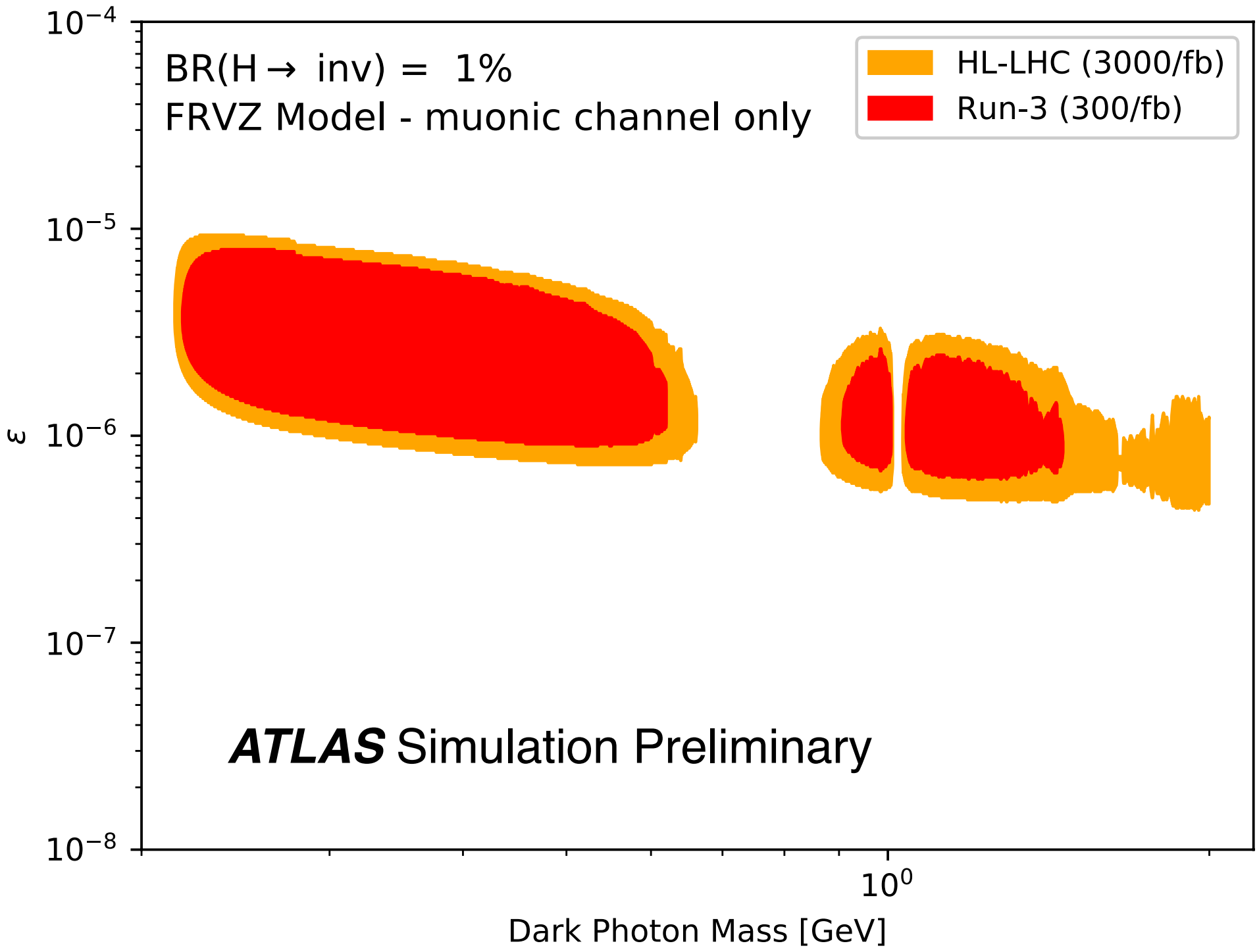

Figure 14: Exclusion contour plot in the plane defined by the $\gamma_{\rm d}$ mass and the kinetic mixing parameter $\epsilon$. Two different scenarios are shown assuming a Higgs decay branching fraction to the hidden sector of 1%: 300 fb$^{-1}$after Run-3 (red) ad 3000 fb$^{-1}$after HL-LHC including multi-muon scan trigger improvement (orange).

Sensitivity prospects of the ATLAS dark photon search for Run-3 and HL-LHC have been estimated at the expected integrated luminosity of 300 fb$^{-1}$and 3000 fb$^{-1}$respectively, extrapolating the results of the Run-2 search. The 95% CL exclusion limit on the dark photon average $c\tau$ is expected to improve, extending the lower bound down to 0.97 mm and the upper bound up to 597 mm, assuming a branching ratio of the Higgs boson decay to the Hidden sector of 10%. Moreover, the search at the HL-LHC is expected to probe BR(H $\rightarrow$ $2\gamma_d + X$) down to $\sim$ 1%, where the Run-2 analysis lacks of sensitivity.

# CMS Physics Analysis Summary

Contact: cms-phys-conveners-ftr@cern.ch

# Projection of the Mono-Z search for dark matter to the HL-LHC

The CMS Collaboration

## Abstract

A study of the expected discovery sensitivity and exclusion power of a search for new invisible particles in events with a Z boson and missing transverse momentum at the high-luminosity LHC is presented. Sensitivity estimates are derived from a CMS Run 2 result with the use of rescaling techniques. Different scenarios of integrated luminosity and systematic uncertainties are explored, and results are presented in the parameter space of a simplified model of dark matter production with a spin-1 mediator, as well as a simplified model with a pseudoscalar mediator and second Higgs doublet.

*This document has been revised with respect to the version dated October 22, 2018.*

# 1 Introduction

One of the most important open questions in physics today is the origin and nature of dark matter (DM). At the end of Run 2 of the LHC, no evidence has been found that DM particles are produced in proton-proton collisions. In this document, the discovery sensitivity and exclusion potential of a search for DM particles in collision events with a Z boson and missing transverse momentum $p_{\mathrm{T}}^{\mathrm{miss}}$ at the high-luminosity LHC (HL-LHC) is explored, where the Z boson decays to either a pair of electrons or muons.

The results are interpreted in the context of simplified models of DM production. As a general benchmark, a scenario with a vector mediator $Z'$ and Dirac fermion DM candidate $\chi$ is considered. In addition to the masses of these particles, $m_{\mathrm{med}}$ and $m_{\mathrm{DM}}$, the model has two free coupling parameters: $g_q$, the universal mediator-quark coupling, and $g_{DM}$, the mediator-DM coupling. Following the recommendations of the LHC DM forum [1], default coupling values of $g_q = 0.25$ and $g_{DM} = 1$ are chosen.

In addition, a model with a pseudoscalar mediator and a second Higgs doublet, referred to as a+2HDM, is considered [2]. The addition of a second Higgs doublet results in the presence of two new neutral and charged scalars H and $\mathrm{H}^{\pm}$, as well as an additional pseudoscalar A. The a and A bosons mix, resulting in the possibility of $\mathrm{H} \to \mathrm{aZ}$ decays. The a boson subsequently decays to DM particles, resulting in an overall Z+$p_{\mathrm{T}}^{\mathrm{miss}}$ signature. The DM candidate is assumed to be a Dirac fermion. Parameter choices follow the recommendations from the LHC DM working group [3] based on the work of Ref. [2]. To ensure compatibility with the measurements of the coupling strengths of the known h(125) boson, the coupling strength of the H boson to the SM gauge bosons $\cos(\beta - \alpha)$ is set to zero ("alignment limit"). Constraints from precision measurements of the properties of the W and Z bosons are evaded by setting the masses of the heavy bosons to be equal $m_{\mathrm{H}} = m_{\mathrm{H}^{\pm}} = m_{\mathrm{A}}$. The ratio of the vacuum expectation values of the two Higgs doublets is chosen to be $\tan(\beta) = 1$, and the mass of the DM candidate is $m_{\mathrm{DM}} = 10\,\mathrm{GeV}$. To ensure perturbativity of the model, the mixing angle of the a and A bosons is set to $\sin(\theta) = 0.35$, and the quartic couplings in the extended Higgs sector are chosen to be $\lambda_{P1} = \lambda_{P2} = \lambda_{P3} = 3$. All one-loop diagrams for $\mathrm{gg} \to \ell\ell\chi\chi$ are taken into account. For the parameter space covered in this analysis, the width of the heavy scalar H ranges between 5% at $m_{\mathrm{H}} = 500\,\mathrm{GeV}$ and 30% at $m_{\mathrm{H}} = 2\,\mathrm{TeV}$. Due to the large relative width at large $m_{\mathrm{H}}$, the H boson is dominantly produced off-shell with a mass much lower than the $m_{\mathrm{H}}$ parameter. This ensures that even at high values of $m_{\mathrm{H}}$, the Z boson is not overly boosted and the leptons from its decay are well separated.

The main kinematic differences between the two signal models are visible in the distributions of $p_{\mathrm{T}}(Z)$ and $p_{\mathrm{T}}^{\mathrm{miss}}$. While these distributions are strictly nonresonant in the case of the vector mediator scenario, there is a Jacobian peak present in the distributions of both variables for the a+2HDM model, which is caused by the resonant $\mathrm{H} \to \mathrm{aZ}$ decay. The peaking structure is generally attractive for this search as it facilitates the discrimination between the signal and the fully nonresonant backgrounds. Representative Feynman diagrams for the signal processes in both models are shown in Fig 1.

# 2 Method

The strategy and implementation of this analysis follows directly that of Ref. [4], which is a search for the production of beyond the standard model particles, such as dark matter particles, that show no detector interaction. A possible signal is constrained by considering the production of such invisible particles in association with a Z boson decaying to either a pair of electrons or muons.

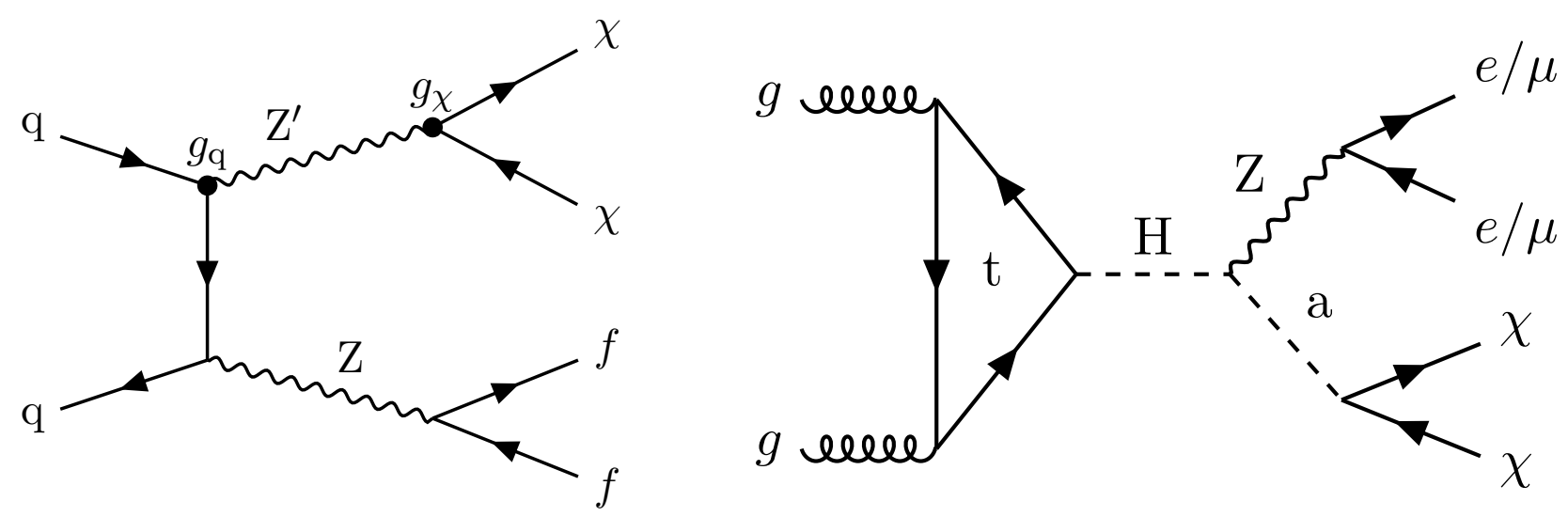

Figure 1: Representative leading-order Feynman diagrams of the signal processes in the simplified model with a vector mediator $Z'$ and DM candidate $\chi$ (left), as well as the a+2HDM model (right). In the case of the a+2HDM scenario, box diagrams without a heavy scalar particle are also taken into account, but give a sub-dominant contribution in the parameter range of interest.

The sought-after signal topology is a well-reconstructed Z boson and large missing transverse momentum. Missing transverse momentum is calculated as transverse component of the vector sum of all particle-flow particle momenta in an event [5]. In the following, this vector is referred to as $\vec{p}_{\mathrm{T}}^{\mathrm{miss}}$, while its magnitude is referred to as $p_{\mathrm{T}}^{\mathrm{miss}}$. Except for the effects of initial state radiation, the invisible particles and the Z boson are expected to be produced back-to-back in the laboratory frame, and the signal selection thus focuses on extracting events in a balanced topology. The exact selection criteria are listed in Tab. 1.

In addition to the signal region (SR), the following control regions are used:

- Opposite-flavour lepton region: Same selection as SR, but using $e\mu$ instead of $ee/\mu\mu$ events. This region is used to estimate nonresonant backgrounds such as $\mathrm{t\bar{t}}$ and WW production.
- Low-$p_{\mathrm{T}}^{\mathrm{miss}}$ region: Same selection as SR, except $50 < p_{\mathrm{T}}^{\mathrm{miss}} < 200\,\mathrm{GeV}$. This region is used to derive the normalization for the Drell-Yan process.
- 3-lepton/4-lepton regions: Same selection as SR except requiring 3 or 4 leptons instead of 2 and using emulated $p_{\mathrm{T}}^{\mathrm{miss}}$ instead of the standard reconstructed $p_{\mathrm{T}}^{\mathrm{miss}}$ to evaluate all selection criteria. From all opposite-sign same-flavour combinations of leptons, the pair with the invariant mass closest to the nominal mass of the Z boson is used to reconstruct the Z candidate. Emulated $p_{\mathrm{T}}^{\mathrm{miss}}$ is calculated by excluding the charged leptons that are not part of the Z candidate from the $p_{\mathrm{T}}^{\mathrm{miss}}$ calculation. The goal of this procedure is to mimic the effect of not reconstructing the lepton from the decay of the W boson (WZ) or substituting the Z boson decay mode from $\mathrm{Z} \to \ell\ell$ to $\nu\nu$ (ZZ). The emulated $p_{\mathrm{T}}^{\mathrm{miss}}$ in the control regions is then representative of the standard $p_{\mathrm{T}}^{\mathrm{miss}}$ in the signal region. These regions are used to constrain the WZ and ZZ processes.

The signal extraction is performed using a maximum-likelihood (ML) fit to the $p_{\mathrm{T}}^{\mathrm{miss}}$ spectrum in signal and control regions. The fit strategy of Ref. [4] is extended by introducing one freely floating normalization parameter for each (emulated) $p_{\mathrm{T}}^{\mathrm{miss}}$ bin. These parameters have a multiplicative effect on the yield of the WZ and ZZ background contributions and are correlated between both processes, as well as between signal and control regions, but uncorrelated between the (emulated) $p_{\mathrm{T}}^{\mathrm{miss}}$ bins. Effectively, this implementation allows the full shape of the background distribution of (emulated) $p_{\mathrm{T}}^{\mathrm{miss}}$ to be determined from data, while constraining the ratio between the WZ and ZZ contributions in signal and control regions to the values predicted in simulation. It is therefore well suited to the case of large integrated luminosities, where large event yields in signal and control regions are expected. An example of a successful application of this method is given in Ref. [6].

Simulated samples of the relevant background processes are used as inputs to the ML fit. The dominant ZZ and WZ, as well as the $t\bar{t}$ and WW contributions making up the nonresonant backgrounds, are simulated using POWHEG at next-to-leading order (NLO) in quantum chromodynamics (QCD) [7–9]. For WZ and ZZ, scale factors are applied to account for NNLO QCD and NLO electroweak corrections [10–14]. The Drell-Yan process is simulated using Madgraph5_aMC@NLO at NLO in QCD with up to two additional partons at the matrix element level [15, 16]. Signal samples for the scenario with a vector mediator are generated using the implementation of Ref. [17] with Madgraph5_aMC@NLO version 2.2.2 at NLO in QCD with up to one additional parton at the matrix element level. Events for the a+2HDM scenario are produced using Madgraph5_aMC@NLO version 2.4.2 at leading order in QCD without no additional partons using the implementation of Ref. [2]. The parton distribution functions (PDFs) in all samples are modeled using NNPDF3.0 NLO [18], except for the a+2HDM case, which uses NNPDF3.1 NNLO [19].

For all simulated samples, parton showering is applied using PYTHIA version 8.2 [20] with the CUETP8M1 tune for the underlying event description [21]. A full implementation of the CMS detector in GEANT4 [22] is used to simulate the detector response. The samples are processed with the CMS detector configuration and data taking conditions of the 2016 data taking period.

Systematic uncertainties are comprised of experimental and theoretical uncertainties. The main experimental uncertainties are related to the energy scale of identified objects (approximately 1% for leptons, less than 3% for jets), and lepton identification efficiencies (approximately 1% per lepton). Theoretical uncertainties related to the choice of parton distribution functions amount to approximately 2% for the background and signals alike. The uncertainty related to the choice of renormalization and factorization scales is up to 10% for the background and 5% for signals. The dominant theoretical uncertainty is related to missing mixed electroweak-QCD higher order corrections in the WZ and ZZ processes and is estimated as the product of the LO to NLO scale factors in both QCD and electroweak. This uncertainty ranges from approximately 10% for $p_{\mathrm{T}}^{\mathrm{miss}} = 200\,\mathrm{GeV}$ up to 30% for $p_{\mathrm{T}}^{\mathrm{miss}} > 800\,\mathrm{GeV}$.

Any relevant differences between the HL-LHC and Run 2 conditions are covered by event-by-event rescaling procedures. The following scaling procedures are applied:

- Center-of-mass energy: The change from $\sqrt{s} = 13\,\mathrm{TeV}$ to $\sqrt{s'} = 14\,\mathrm{TeV}$ is incorporated by recalculating the PDF weight for each event with shifted values of the Bjorken $x$ variables for the incoming partons 1, 2: $x_{1/2} \rightarrow x'_{1/2} = \frac{\sqrt{s}}{\sqrt{s'}} x_{1/2}$. The event-by-event weights are derived using the LHAPDF interface [26]. The method is validated on dedicated simulation samples produced at different values of $\sqrt{s}$ and a good agreement between the reweighted samples and the validation samples is observed.
- Experimental $p_{\mathrm{T}}^{\mathrm{miss}}$ performance: The large instantaneous luminosity at the HL-LHC will come at the cost of an increase in the number of collision events per bunch crossing, referred to as pileup (PU). For this analysis, an average number of PU events of 200 is considered. This may result in a degradation of the $p_{\mathrm{T}}^{\mathrm{miss}}$ resolution compared to the Run 2 case [27]. Based on studies of simulated events with both Run 2 and HL-LHC conditions with current reconstruction methods, the $p_{\mathrm{T}}^{\mathrm{miss}}$ resolution is expected to be worse at the HL-LHC by a factor of $1.5 - 2$, depending on the value of the generator-level $p_{\mathrm{T}}^{\mathrm{miss}}$. As a conservative choice, a resolution degradation by a constant factor of 2 is assumed for this analysis, which is incorporated in simulated events by artificially increasing the vectorial difference between the generated and reconstructed $\vec{p}_{\mathrm{T}}^{\mathrm{miss}}$ vectors by a factor of two,

$$(\vec{p}_{\mathrm{T}}^{\mathrm{miss}}(reco) - \vec{p}_{\mathrm{T}}^{\mathrm{miss}}(gen)) \quad \rightarrow \quad 2 \times (\vec{p}_{\mathrm{T}}^{\mathrm{miss}}(reco) - \vec{p}_{\mathrm{T}}^{\mathrm{miss}}(gen)).$$

Table 1: Requirements for the signal region selection. The requirements fall in three categories: Lepton selection, vetoes based on the multiplicities of hadronic objects, dilepton candidate selection, and high-$p_{\mathrm{T}}^{\mathrm{miss}}$ back-to-back topology requirements. The requirements are identical to those of Ref. [4], except for the $p_{\mathrm{T}}^{\mathrm{miss}}$ requirement, which has been increased to remove increased background contributions at low $p_{\mathrm{T}}^{\mathrm{miss}}$ due to degraded $p_{\mathrm{T}}^{\mathrm{miss}}$ resolution at high PU. Jets are clustered using the anti-$k_{\mathrm{T}}$ algorithm [23] implemented in the FastJet program [24] with a radius parameter of 0.4. Bottom quark jets are identified using the CSVv2 algorithm [25]. The dilepton angular separation is defined as $\Delta R = \sqrt{(\eta(\ell_1) - \eta(\ell_2))^2 + (\phi(\ell_1) - \phi(\ell_2))^2}$.

| Quantity | Requirement |
|---|---|
| Number of charged leptons | = 2, with opposite charge, same flavour |
| Muon $p_{\mathrm{T}}$ | $> 20\,\mathrm{GeV}$ |
| Leading (trailing) Electron $p_{\mathrm{T}}$ | $> 25(20)\,\mathrm{GeV}$ |
| | |
| Jet multiplicity | $\leq 1$ jet with $p_{\mathrm{T}} > 30\,\mathrm{GeV}$ |
| b Jet multiplicity | No b jet $p_{\mathrm{T}} > 20\,\mathrm{GeV}$ |
| Hadronic $\tau$ multiplicity | No $\tau$ with $p_{\mathrm{T}} > 18\,\mathrm{GeV}$ |
| | |
| Dilepton mass | $\lvert M(\ell\ell) - m_{\mathrm{Z}}\rvert < 15\,\mathrm{GeV}$ |
| Dilepton $p_{\mathrm{T}}$ | $> 60\,\mathrm{GeV}$ |
| Dilepton $\Delta R$ | $< 1.8$ |
| | |
| $p_{\mathrm{T}}^{\mathrm{miss}}$ | $> 200\,\mathrm{GeV}$ |
| $\Delta\phi(\vec{p}_{\mathrm{T}}^{\,\ell\ell}, \vec{p}_{\mathrm{T}}^{\,\mathrm{miss}})$ | $> 2.6$ |
| $\lvert p_{\mathrm{T}}^{\mathrm{miss}} - p_{\mathrm{T}}^{\ell\ell}\rvert / p_{\mathrm{T}}^{\ell\ell}$ | $< 0.4$ |
| $\Delta\phi(\vec{p}_{\mathrm{T}}^{\,j}, \vec{p}_{\mathrm{T}}^{\,\mathrm{miss}})$ | $> 0.5$ rad |

- Luminosity: The normalization of simulation samples is scaled to values of the integrated luminosity between 300 $\mathrm{fb}^{-1}$ and 3 $\mathrm{ab}^{-1}$.

# 3 Results

The distribution of $p_{\mathrm{T}}^{\mathrm{miss}}$ in the signal region is shown in Fig. 2. Signal significances and exclusion limits are calculated with the asymptotic approximation of the CLs method [28–31]. The significance is calculated as the quantile of a two-sided Gaussian distribution corresponding to the probability that an observed excess caused by a signal of unit signal strength could be the result of a statistical fluctuation of the standard model backgrounds. It is given in units of the standard deviation of the Gaussian distribution. For all interpretations, three systematic uncertainty scenarios are considered:

- **Run 2 syst. uncert.**: Systematic uncertainties are estimated to be of the same size as the Run 2 analysis [4]. This scenario does not consider any possible future improvements to the systematic uncertainties.
- **YR18 syst. uncert.**: The effect of expected future improvements in the control of systematic uncertainties is included according to the conventions of the 2018 CERN yellow report (YR18) [32]. Theoretical and experimental uncertainties are reduced by 50%, and the statistical uncertainties due to the finite size of the simulation samples are neglected. This scenario is the current best estimate of what can be achieved at the HL-LHC.
- **Stat. uncert. only**: Only statistical uncertainties are considered. This scenario demonstrates the maximal reach of the analysis strategy if systematic uncertainties are negligible.

The discovery significance and signal strength exclusion limits for a signal in the vector mediated simplified model are shown in Fig. 3. The sensitivity to a signal in this scenario does not depend strongly on the value of the DM candidate mass $m_{\mathrm{DM}}$ as long as $m_{\mathrm{med}} < m_{\mathrm{med}}/2$ and the results are thus shown for the representative case $m_{\mathrm{DM}} = 1\,\mathrm{GeV}$. At the lowest considered integrated luminosity $\mathcal{L}_{\mathrm{int}} = 300\ \mathrm{fb}^{-1}$, the search is statistically limited, and the result shows only a limited dependence on the choice of the systematic uncertainty scenario. Depending on the choice of mediator mass, the signal overlaps with different regions of the standard model $p_{\mathrm{T}}^{\mathrm{miss}}$ background spectrum. Accordingly, the effect of systematic uncertainties is largest for lower values of the mediator masses ($\approx 300\,\mathrm{GeV}$), where there is significant overlap of the signal and background distributions. With increasing mediator mass, the effect subsides, as the signal moves towards the tails of the background $p_{\mathrm{T}}^{\mathrm{miss}}$ distribution. Depending on the mass of the mediator, different values of $\mathcal{L}_{\mathrm{int}}$ are required to achieve a discovery. For the intermediate masses between 750 and 1000 GeV, a signal could have been small enough in Run 2 not to be detected, while still achieving discovery sensitivity over the run time of the HL-LHC. The case of $m_{\mathrm{med}} = 1\,\mathrm{TeV}$ highlights the need for improved systematic uncertainties: Depending on the assumed scenario for systematic uncertainties, the signal may either already be discovered with an integrated luminosity of 1 $ab^{-1}$, or it may remain below the discovery threshold even with 3 $ab^{-1}$. In addition to the discovery sensitivity, Fig. 4 shows the expected limits on the couplings in the vector-mediated DM scenario. The general dependence on luminosity and systematic uncertainty scenarios is similar to the case of the discovery significance. For the case of the quark coupling $g_q$, values of approximately $0.04 - 0.10$ will be testable at the end of the HL-LHC run. For the DM coupling $g_{DM}$, values between 0.15 and 0.45 will be accessible, depending on the mediator mass. The difference in exclusion reach in the two couplings is due to their different effects on the product of signal cross section and branching fraction: While a reduction of $g_q$ decreases the mediator production cross section, it increases the branching fraction of the mediator to DM particles and thus partly counteracts the first effect. In the case of the DM coupling, there is no effect compensating the reduction it induces in the branching fraction of the mediator to DM particles.

The exclusion in the two-dimensional $m_{\text{med}}$-$m_{\text{DM}}$ plane for $\mathcal{L}_{\text{int}} = 3\ \text{ab}^{-1}$ is shown in Fig. 5. Assuming the YR18 systematic uncertainty scenario, mediator masses up to approximately 1.5 TeV can be probed, which is an improvement over the Run 2 result by a factor of approximately 2.3. In the "stat. only" and "Run 2 syst. unc." scenarios, the exclusion is improved, respectively weakened, by slightly more than 100 GeV.

For the a+2HDM model, results are presented in terms of the two-dimensional exclusion reach with $\mathcal{L}_{\text{int}} = 3\ \text{ab}^{-1}$ in the plane of $m_{\text{a}}$ and $m_{\text{H}} = m_{\text{A}}$, which is shown in Fig. 6. In the YR18 scenario, the light pseudoscalar can be probed up to masses of approximately 600 GeV, with the maximum reach being achieved around $m_{\text{H}} = m_{\text{A}} = 1.3$ TeV. Again, the range of outcomes defined by the "stat. only" and "Run 2 syst. unc." scenarios spans $100 - 150$ GeV in the pseudoscalar mass. The maximal exclusion reach in the mass of the heavy bosons is approximately 1.9 TeV for low values of $m_{\text{a}} \approx 100$ GeV with a corresponding range of $100 - 200$ GeV for the different uncertainty scenarios.

## 4 Summary

A sensitivity study for a search for dark matter (DM) particles in events with a Z boson and missing transverse momentum at the HL-LHC has been presented. The effects of the increase in integrated luminosity and center-of-mass energy, as well as the impact of changing experimental conditions and expected future improvements in the size of systematic uncertainties are taken into account. Assuming an integrated luminosity of 3 $ab^{-1}$, it will be possible to probe vector-mediated DM production up to values of the mediator mass of approximately 1.5 TeV. In a simplified model with a second Higgs doublet and a pseudoscalar mediator, heavy scalars will be probed up to masses of 1.9 TeV, and the light pseudoscalar mediator will be accessible up to masses of 600 GeV. A comparison of different scenarios of systematic uncertainties shows that even moderate differences in the size of uncertainties can significantly affect the size of the dataset necessary for discovery. Independent of the details of the systematic uncertainty treatment, significant improvements in the mass and coupling reach over current results are to be expected.

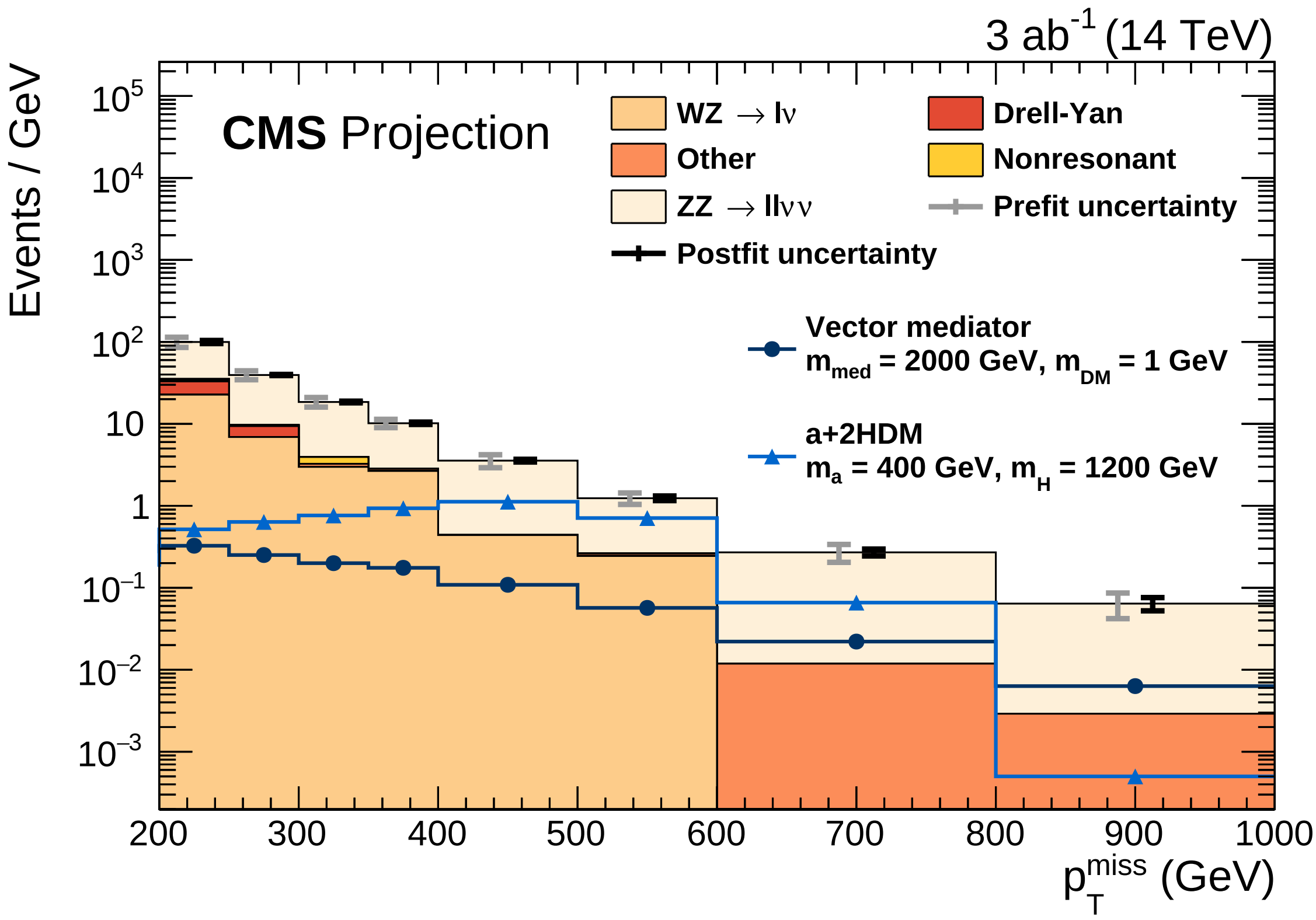

Figure 2: Spectrum of $p_{\mathrm{T}}^{\mathrm{miss}}$ in the signal region. The summed background spectrum is overlaid with the spectra for two signal hypotheses. The uncertainty bands for the background prediction correspond to the YR18 uncertainty scenario described in the text and are shown both before and after applying a background-only maximum-likelihood fit to the Asimov dataset in signal and control regions ("prefit" and "postfit", respectively).

**8**

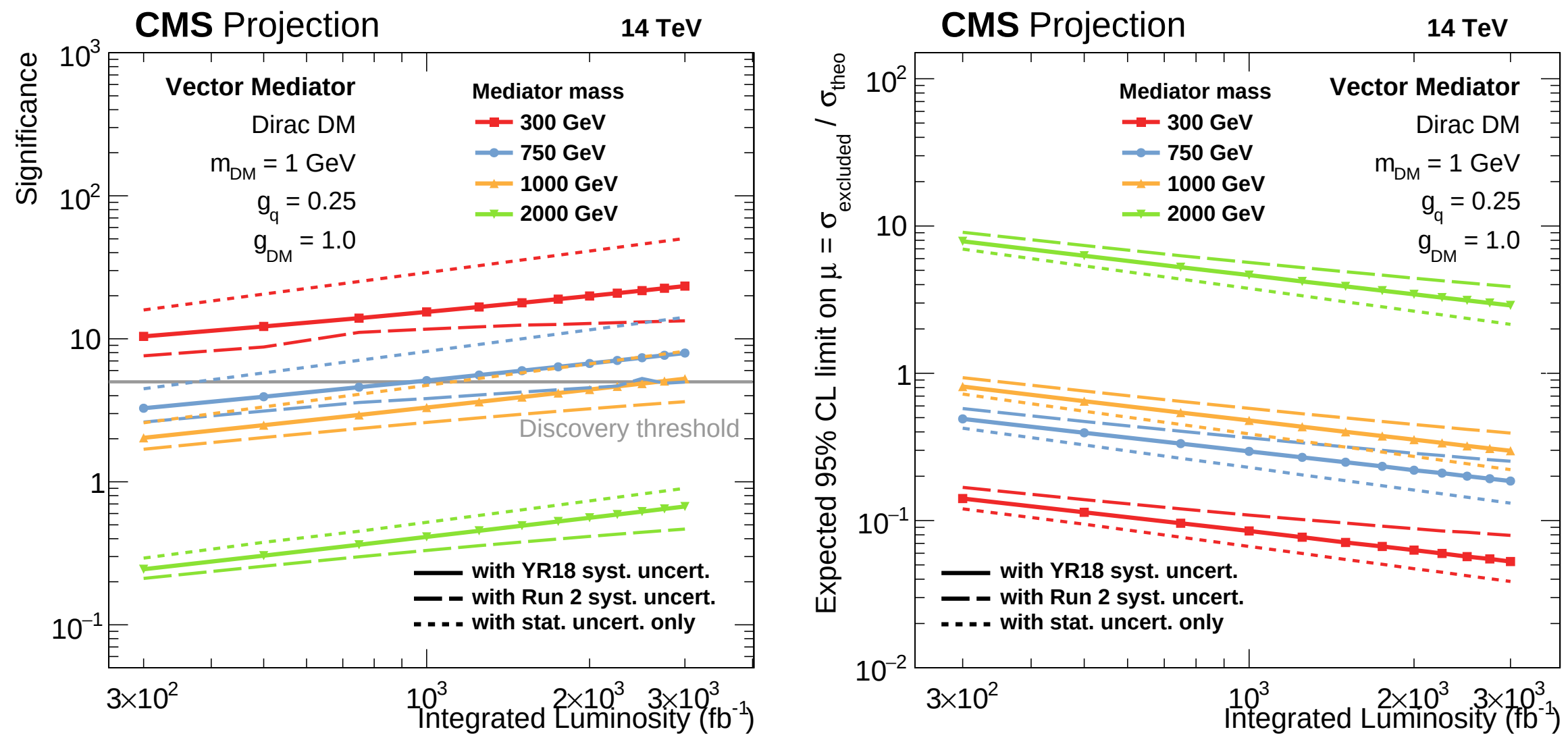

Figure 3: Expected discovery significance (left) and signal strength exclusion limits (right) for the vector-mediated DM signal as a function of $\mathcal{L}_{\text{int}}$ and for different values of the mediator mass. The results are shown for the three systematic uncertainty scenarios described in the text, with the scenario labeled as "Run 2" corresponding to Ref. [4]. The significance is calculated for unity signal strength.

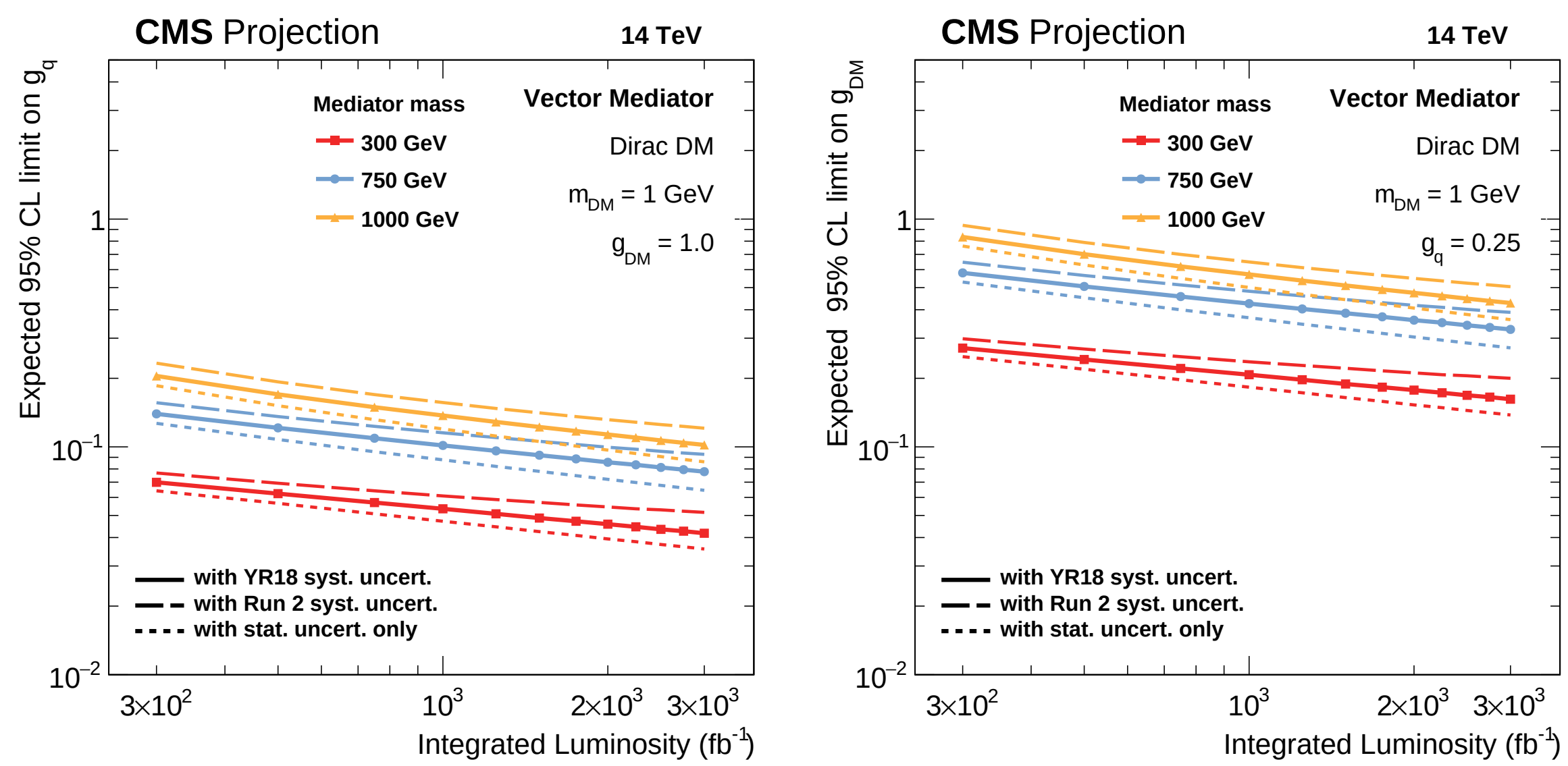

Figure 4: Exclusion sensitivity for the couplings $g_q$ (left) and $g_{DM}$ (right) in the vector-mediated DM scenario as a function of $\mathcal{L}_{\text{int}}$ and for different values of the mediator mass. The results are shown for the three systematic uncertainty scenarios described in the text, with the scenario labeled as "Run 2" corresponding to Ref. [4]. Note that no limit can be set if the sensitivity for a given point is too low. For increasing values of $g_q$ and $g_{DM}$, the product of cross section and branching fraction eventually reaches a plateau and does not increase further with an increase in one of the couplings. Due to this effect, no coupling limits can be set for $m_{\text{med}} = 2\,\text{TeV}$.

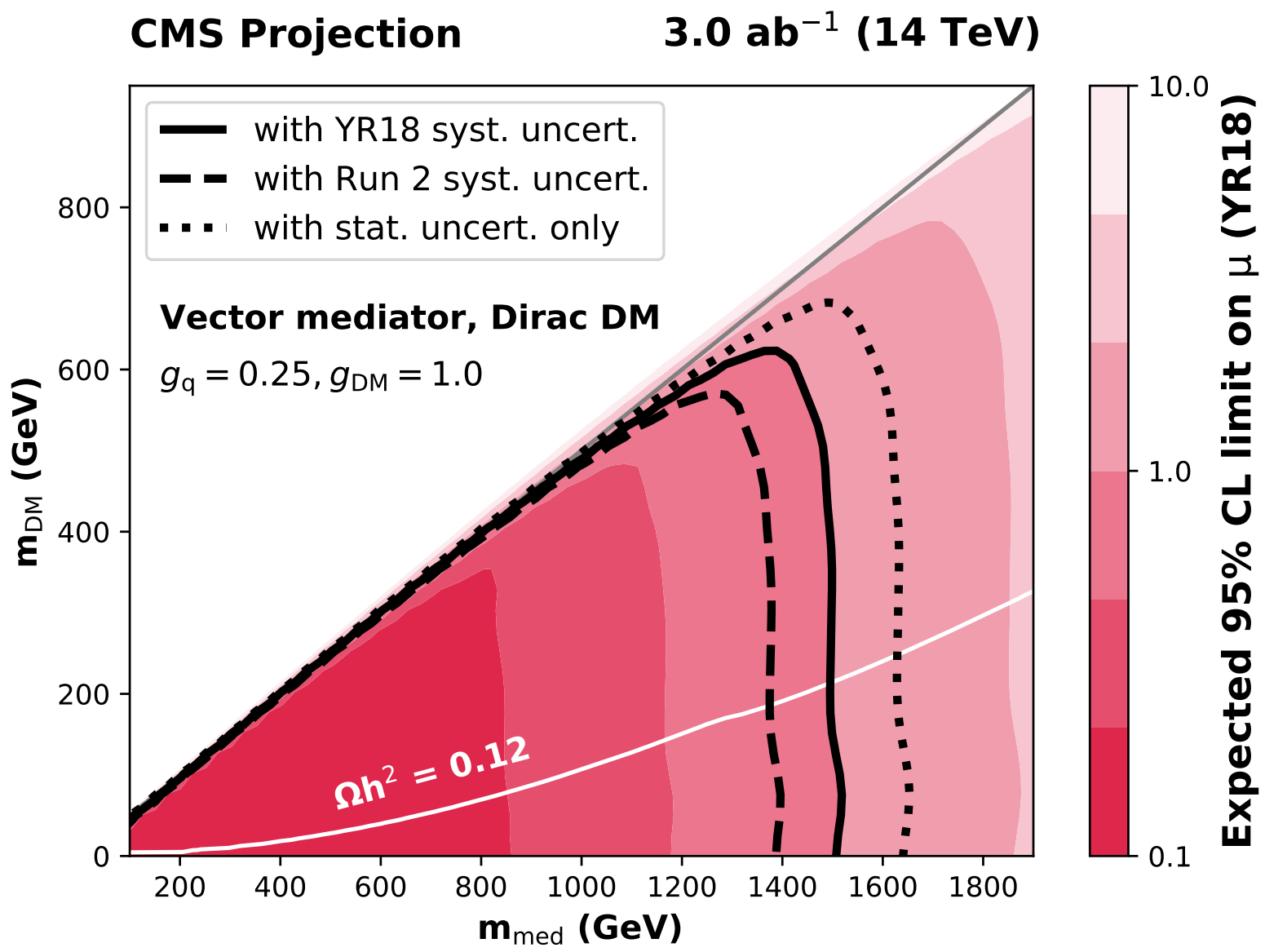

Figure 5: Expected 95% CL exclusion limits on the signal strength of vector-mediated DM production in the plane of mediator and dark matter masses. The results are shown for the three systematic uncertainty scenarios described in the text, with the scenario labeled as "Run 2" corresponding to Ref. [4]. The $m_{\mathrm{med}} = 2 \times m_{\mathrm{DM}}$ diagonal, which is the kinematic boundary for decay of an on-shell mediator to DM particles, is indicated as a grey line. The white line indicates parameter combinations for which the observed DM relic density in the universe can be reproduced [33]. Points below (above) this line have relic densities that are larger (smaller) than the observed value of $\Omega h^2 = 0.12$ [34].

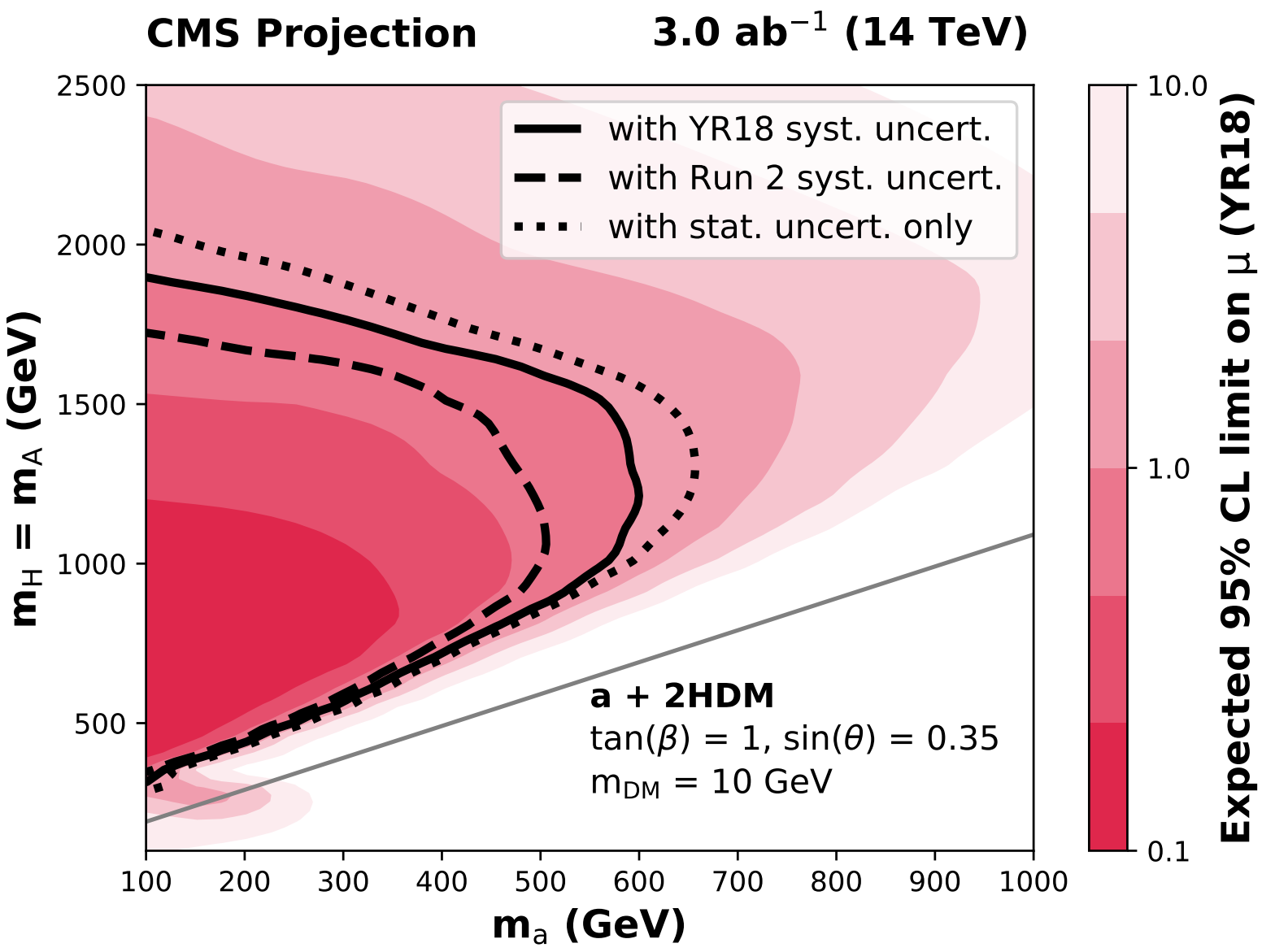

Figure 6: Expected 95% CL exclusion limits on the signal strength in the a+2HDM scenario as a function of the mass of the main DM mediator a and the masses of the H and A bosons $m_{\mathrm{H}} = m_{\mathrm{A}}$. The results are shown for the three systematic uncertainty scenarios described in the text, with the scenario labeled as "Run 2" corresponding to Ref. [4]. The grey line indicates the kinematic boundary $m_{\mathrm{H}} = m_{\mathrm{a}} + m_{\mathrm{Z}}$, below which the $\mathrm{H} \to \mathrm{aZ}$ decay is prohibited for an on-shell H and the sensitivity of this search is limited.

# Prospects for Dark Matter searches in mono-photon and VBF+$E_{\mathrm{T}}^{\mathrm{miss}}$ final states in ATLAS

The ATLAS Collaboration

This document presents a prospect study for dark matter searches with the ATLAS detector at luminosities as expected at HL-LHC. A scenario where the Standard Model is extended by the addition of an electroweak fermionic triplet with null hyper charge is considered. The lightest mass state of the triplet constitutes a weakly interacting massive particle dark matter candidate. This model is inspired by Supersymmetry and by the Minimal Dark Matter setup, and provides a benchmark in the spirit of simplified models. Projections for an integrated luminosity of 3000 fb$^{-1}$ are presented for the dark matter searches in the mono-photon and VBF+$E_{\mathrm{T}}^{\mathrm{miss}}$ final states, based on the run-2 analyses strategy. To illustrate the experimental challenges associated to a high pile-up environment due to the high luminosity, the VBF+$E_{\mathrm{T}}^{\mathrm{miss}}$ topology is considered and the effect of the increased pile-up on the VBF invisibly decaying Higgs boson is studied as a benchmark process.

# 1 Pure WIMP Dark Matter triplet

A fermionic triplet $\chi$ of the $SU(2)_L$ group with null hypercharge $(Y)$,

$$\chi = \begin{pmatrix} \chi^+ \\ \chi_0 \\ \chi^- \end{pmatrix} \tag{1}$$

is added to the SM with a Lagrangian:

$$\begin{aligned}\mathcal{L}_{MDM} &= \frac{1}{2}\bar{\chi}(i\not{D} + M)\chi \\ &= \frac{1}{2}\bar{\chi}_0(i\not{\partial} - M_{\chi^0})\chi_0 + \bar{\chi^+}(i\not{\partial} - M_{\chi^+})\chi^+ \\ &+ g(\bar{\chi^+}\gamma_\mu\chi^+(\sin\theta_W A_\mu + \cos\theta_W Z_\mu)) + \bar{\chi^+}\gamma_\mu\chi_0 W^-_\mu + \bar{\chi}_0\gamma_\mu\chi^+ W^+_\mu\end{aligned}$$

where $g$ is the $SU(2)$ gauge coupling; $M$ is the tree-level mass of the particle; $\sin\theta_W$ and $\cos\theta_W$ are the sine and cosine of the Weinberg angle; $A_\mu$, $Z_\mu$, $W_\mu$ are the SM boson fields. The lightest component of the triplet is stable if some extra symmetry is imposed, like lepton number, baryon minus lepton number (B-L) or a new symmetry under which $\chi$ is charged (e.g. R-parity in SUSY).

At tree level all the $\chi$ components have the same mass, but a mass splitting is induced by the electroweak corrections given by loops of SM gauge bosons between the charged and neutral components of $\chi$. These corrections make the charged components heavier than the neutral one. The neutral component ($\chi_0$) is therefore the lightest one and its mass differs by $\simeq$ 165 MeV [1] from the one of the charged components. Being neutral and stable, $\chi_0$ constitutes a potential DM candidate. If the thermal relic abundance is assumed, the mass of $\chi_0$ is $M_{\chi_0} \simeq 3$ TeV. However, if $\chi$ is not the only particle making the whole dark matter or if it is not thermally produced [2], its mass can be $M_{\chi_0}$ <3 TeV.

This model provides a benchmark of a typical WIMP DM candidate and its phenomenology recreates the one of supersymmetric models where the Wino is the lightest supersymmetric particle (LSP). For this reason this triplet is referred to as *Wino-like*. Together with providing a good DM candidate, this Wino-like triplet has other interesting features, for instance it modifies the running of the Higgs quartic coupling stabilizing the electroweak vacuum. It also changes the running of the gauge couplings helping with their unification. As studied in Ref. [3], treating $M$ as a free parameter, this triplet can be probed at the LHC in different ways. Once produced, the charged components of the triplet decay into the lightest neutral component $\chi_0$ plus very soft charged pions. $\chi_0$ is identified as $E_{\mathrm{T}}^{\mathrm{miss}}$ in the detector while the pions, because of the small mass splitting between the neutral and charged components, are so soft that they are lost and are not reconstructed. Therefore, the production of $\chi$ can be searched for by:

- mono-X searches, such as mono-jet (see also Ref. [4]) and mono-photon;
- VBF +$E_{\mathrm{T}}^{\mathrm{miss}}$ searches as $\chi$ can also be produced via VBF (see also Ref. [5]);
- disappearing tracks searches: the lifetime of $\chi^\pm$ is about $\tau \simeq 0.2$ ns and it corresponds to a decay length at rest $d = c\tau \simeq 6$ cm, this means that almost all the $\chi^\pm$ particles decay before reaching the detector. However, a small fraction of them can travel enough to leave a track in the detector. In this case, the signature of these events is characterized by high $p_T$ tracks (caused by $\chi^\pm$) which end inside the detector once they have decayed into $\chi_0$ and soft pions [6].

This document is focused on the VBF production mode and mono-photon final state. They constitute a necessary complement to the mono-jet and disappearing track searches, because of the very different dependencies on the model parameters like the EW representation and the value of the mass splitting. For example, the strong reach of disappearing track searches [7] would be significantly reduced for larger values of the mass splitting because the lifetime scales as the inverse of the third power of the mass splitting. On the other hand, the reach of VBF and mono-photon searches would not be substantially altered because the decay products other than $\chi^0$ would still be too soft for our vetos, see e.g. [8]. Furthermore, the mono-photon final state provides a complementary channel with respect to the mono-jet one as the mono-jet signature arises from initial state radiation while in case of the mono-photon the radiation can also be emitted from an intermediate $W$ boson or $\chi^{\pm}$ particle.

LEP limits exclude masses below ~90 GeV [9–11], therefore we focus on $M_{\chi_0}$ ≥90 GeV.

## 2 The LHC, HL-LHC and the ATLAS detector

The expected luminosity that will be collected at the end of run-2 is estimated to be ~150 fb$^{-1}$ (end of 2018) with an instantaneous luminosity of ~1.5×10$^{34}$ cm$^{-2}$s$^{-1}$ and an average number of collisions per bunch crossing $< \mu > \sim 30$. The run-2 will be followed by a long shutdown (2019-2020). Collisions will restart in early 2021, in the so-called run-3. During this run, the instantaneous luminosity will increase up to ~2×10$^{34}$ cm$^{-2}$s$^{-1}$ and $< \mu >$ is expected to be ~ 60. The amount of data which is expected at the end of run-3 corresponds to 300 fb$^{-1}$. An increase of the centre-of-mass-energy to $\sqrt{s}$ =14 TeV is foreseen.

The HL LHC will start after 2025. The centre-of-mass-energy will be $\sqrt{s}$ =14 TeV for an instantaneous luminosity of 7.5×10$^{34}$ cm$^{-2}$s$^{-1}$ and the aim is to achieve ~3000 fb$^{-1}$ of total integrated luminosity per experiment in 10 years of running time. The pile-up will increase substantially with an expected $< \mu >$ growing to ~ 200. This implies the necessity of developing techniques able to mitigate it in order to perform physics analysis.The ATLAS detector and the trigger system will undergo several upgrades to collect data during the HL-LHC [12, 13].

In this note projections corresponding to an integrated luminosity of 3 ab$^{-1}$ are presented for analyses performed with run-2 data [14, 15], this allows to exploit the full complexity of run-2 analyses. The upgrades of the detector are expected to lead to a better background rejection, while the different pile-up conditions will constitute a challenge, as shown in Sec. 5.2 for the VBF+$E_{\mathrm{T}}^{\mathrm{miss}}$ analysis.

## 3 Simulation setup

### 3.1 EW triplet DM signal generation for mono-photon and VBF plus $E_{\mathrm{T}}^{\mathrm{miss}}$ analyses

The model has been implemented in FeynRules 2.3.24 [16] and considers the electroweak triplet $\chi = (\chi^+, \chi_0, \chi^-)$ described in Sec. 1. This model implementation is the same that has also been used for the phenomenological studies in Ref. [3]. MadGraph [17–19] at Leading Order (LO) has been used to generate the hard-scatter process. Madgraph is then interfaced to Pythia8 [20], with the parameter values set according to the ATLAS tune A14 [21], for parton shower, hadronization and underlying event simulation. The parton distribution function (PDF) set used is NNPDF2.3 at leading order (LO) [22].

3

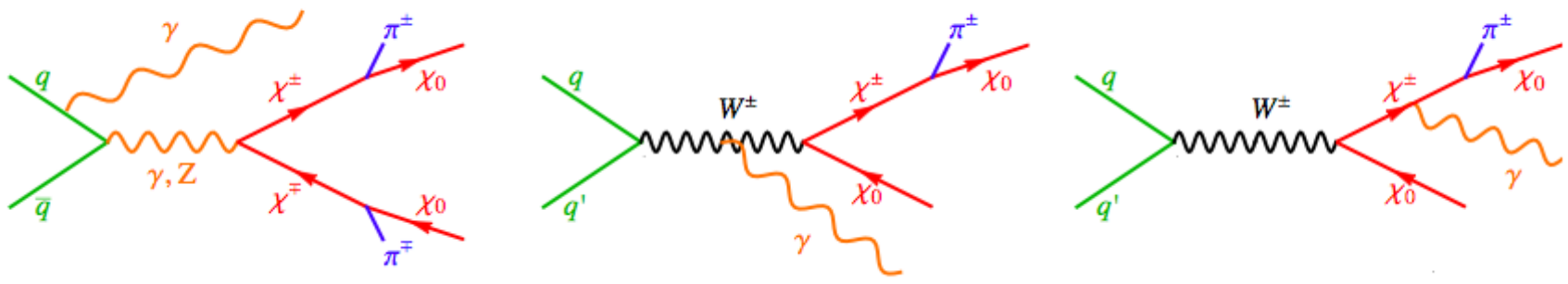

Figure 1: Some representative diagrams for the pure WIMP triplet in $\gamma$+$E_{\mathrm{T}}^{\mathrm{miss}}$ final states. The $\chi^{\pm}$ particles decay into the stable $\chi_0$ DM candidate and soft pions which are not reconstructed [3].

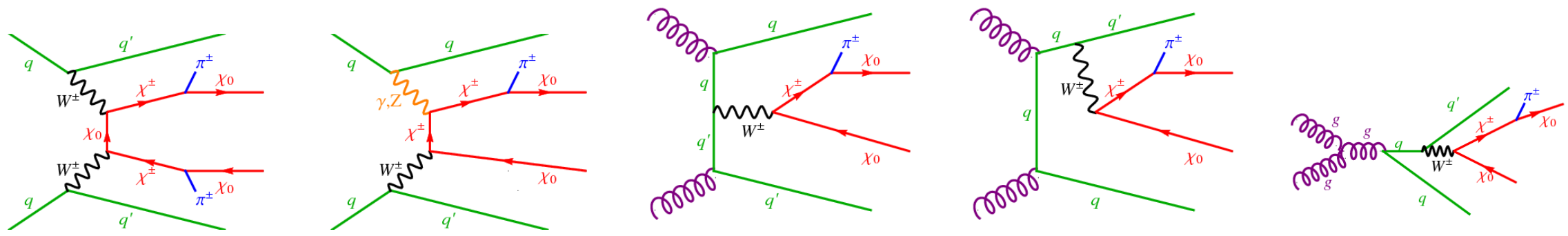

Figure 2: Some representative diagrams for the pure WIMP triplet produced via VBF. The $\chi^{\pm}$ particles decay into the stable $\chi_0$ DM candidate and soft pions which are not reconstructed [3].

For the mono-photon analysis, events with a pair of $\chi$ and one final state $\gamma$ with at least $E_T^{\gamma}$=130 GeV are generated. For the VBF+$E_{\mathrm{T}}^{\mathrm{miss}}$ analysis, events with a pair of $\chi$ and two final state partons with transverse momentum $p_T$ >40 GeV, pseudorapidity separation $\Delta\eta$ >3 and invariant mass of at least 500 GeV have been generated at the matrix-element level. Some Feynman diagrams for the two processes are shown in Figure 1 and Figure 2 respectively. Notice that, for the VBF+$E_{\mathrm{T}}^{\mathrm{miss}}$ analysis, also diagrams not properly originating from two vector bosons (in contrast to pure VBF processes) contribute to the signal (and also background) events as those diagrams produce a jets+$E_{\mathrm{T}}^{\mathrm{miss}}$ signature where the jets have large pseudorapidity separation. In particular, all the diagrams generated from processes with a final state pair of $\chi$ particles and two quarks, passing the generator level selections listed above, are included as part of the signal. Cross sections, computed at $\sqrt{s}$ =13 and 14 TeV for comparisons, for different values of the $\chi_0$ mass, are shown in table 1 and 2 for the mono-photon and VBF+$E_{\mathrm{T}}^{\mathrm{miss}}$ final states respectively. Only samples generated at $\sqrt{s}$ =13 TeV have been used in the analyses described in Sec. 4 and Sec. 5.1, the cross sections at $\sqrt{s}$ = 14 TeV are only listed for reference to show the relative change due to the increase in the center-of-mass energy.

Signal samples have been simulated for different values of $\chi_0$ mass with the official ATLASFAST-II simulation of the current detector [23] at $\sqrt{s}$ =13 TeV. For the mono-photon analysis Monte Carlo (MC) samples have been simulated for $\chi_0$ masses ranging from 90 GeV to 1 TeV, while for the VBF+$E_{\mathrm{T}}^{\mathrm{miss}}$ analysis, a scan of $\chi_0$ mass has been performed in the range 90-200 GeV.

### 3.2 Datasets for pile-up studies

To investigate the realistic conditions for the HL-LHC, a study using the VBF+$E_{\mathrm{T}}^{\mathrm{miss}}$ topology employs a full simulation of the upgraded ATLAS detector with $< \mu >= 200$ and at $\sqrt{s}$=14 TeV. The main background[1] for the VBF $H \to$ invisible search is the strong production of $Z$+jets, where the $Z$ decays to neutrinos. Event samples of VBF $H \to ZZ^* \to \nu\bar{\nu}\nu\bar{\nu}$ and strong production of $Z$+jets are generated at $\sqrt{s}$ =14 TeV with Powheg-Box v1_r2856 [24–28] using the CT10 [29] PDF set and interfaced with Pythia8 v8.186 [20, 30]. The invisible Higgs event sample further includes EvtGen (v1.2.0) [31] and uses the AZNLOCTEQL1 [32] set of tuned parameters for Pythia 8. The cross-section for this signal sample is 4.32 fb. In the run-2 analysis, there is also a non-negligible contribution of $ggH$ production (about 15%); this is ignored in the results presented in Sec. 5.2 as it does not qualitatively change the conclusions. The $Z$+jets event sample additionally includes PhotosPP [33] and uses the AU2 [34] set of tuned parameters for Pythia 8. Detector simulation including digitization uses a full simulation based on Geant 4 [35, 36]. The detector geometry includes the upgraded ATLAS inner tracker (ITk) [37, 38] with five barrel pixel layers and four barrel strip layers. Pythia 8 minimum bias events are overlaid on top of each hard-scatter event to simulate pile-up.

| $\chi^0$ Mass [GeV] | Cross-section at $\sqrt{s}$ = 13 TeV [fb] | Cross-section at $\sqrt{s}$ = 14 TeV [fb] |
|---|---|---|
| 90 | 5.55 ± 0.05 | 6.36 ± 0.06 |
| 100 | 4.67 ± 0.03 | 5.28 ± 0.03 |
| 200 | 1.164 ± 0.009 | 1.36 ± 0.01 |
| 300 | 0.399 ± 0.002 | 0.478 ± 0.003 |
| 500 | 7.54e-02 ± 0.05e-02 | 9.32e-02 ± 0.06e-02 |
| 750 | 1.369e-02 ± 0.009e-02 | 1.804e-02 ±0.012e-02 |
| 1000 | 3.169e-03 ± 0.021e-03 | 4.393e-03 ±0.029e-02 |

Table 1: Summary of cross sections for different $\chi_0$ mass hypotheses considered in the analysis for triplet $\chi$ generated with MadGraph at $\sqrt{s}$ = 13 TeV and $\sqrt{s}$ = 14 TeV after the generation level cuts described in Sec. 3.1 in final states with an energetic matrix element photon (mono-photon final state). Notice that only MC samples generated at $\sqrt{s}$ =13 TeV have been simulated and used in the analysis described in Sec. 4, the cross sections at $\sqrt{s}$ = 14 TeV are only listed for reference to show the relative change due to the increase in the center of mass energy.

| $\chi_0$ Mass [GeV] | Cross-section at $\sqrt{s}$ =13 TeV [fb] | Cross-section at $\sqrt{s}$ =14 TeV [fb] |
|---|---|---|
| **90** | **195.8 ± 0.6** | 238.1 ± 0.8 |
| 100 | 140.9 ± 0.4 | 172.9 ± 0.5 |
| **110** | **105.6 ± 0.4** | 129.2 ± 0.4 |
| 120 | 80.7 ± 0.2 | 99.7 ± 0.3 |
| **200** | **16.49 ± 0.04** | 20.56 ± 0.05 |
| 500 | 0.612 ± 1e-03 | 0.815 ± 2e-03 |
| 1000 | 2.132e-02 ± 8e-05 | 3.16e-02 ± 1e-04 |

Table 2: Summary of cross sections for different $\chi_0$ mass hypotheses for triplet $\chi$ generated with MadGraph at $\sqrt{s}$ = 13 TeV and $\sqrt{s}$ = 14 TeV after the generation level cuts described in Sec. 3.1 in final states with a pair of $\chi$ and two matrix element partons (VBF+$E_{\mathrm{T}}^{\mathrm{miss}}$ analysis). MC samples have been simulated and considered in the analysis only for masses shown in bold, the other values are shown for reference. Notice that only samples generated at $\sqrt{s}$ =13 TeV have been used in the analysis described in Sec. 5.1, the cross sections at $\sqrt{s}$ = 14 TeV are only listed for reference to show the relative change due to the increase in the center of mass energy.

[1] Strong production of $W$+jets is also a significant background for the current search. This background is reducible, resulting from leptons that fall out of acceptance. As discussed in Sec. 5.2, this background contribution is accounted for by simply doubling the $Z$+jets yield.

# 4 Mono-Photon final state

The mono-photon analysis is characterized by a relatively clean final state, containing a photon with a high transverse energy and large $E_{\mathrm{T}}^{\mathrm{miss}}$, which can be mimicked by few SM processes. The run-2 search for new phenomena performed in mono-photon events in pp collisions at $\sqrt{s}$ = 13 TeV at the LHC, using data collected by the ATLAS experiment in 2015 and 2016 corresponding to an integrated luminosity of 36.1 fb$^{-1}$[15], has shown no deviations from the SM expectations. The mono-photon search has been interpreted in terms of the pure WIMP triplet model described in Sec. 1 at high luminosity by keeping the same event selection and the same strategy for the background estimates to exploit the full complexity of the analysis.

The dominant backgrounds consist in processes with a Z or W boson produced in association with a photon, mainly Z($\rightarrow \nu\nu$) + $\gamma$ and processes containing a photon with associated jets, $\gamma$+jets. They are estimated by normalizing the MC prediction for those backgrounds with factors obtained from a simultaneous fitting technique, based on control regions built by reverting one or more cuts of the signal region such that one type of process becomes dominant in that region. Other subleading backgrounds, like W/Z + jet, top and diboson, in which electrons or jets can fake photons are estimated with data-driven techniques and their contribution obtained with 2015 and 2016 data, is rescaled to the high luminosity scenario.

Events passing the lowest unprescaled single photon trigger are selected requiring $E_{\mathrm{T}}^{\mathrm{miss}}$ > 150 GeV. The leading photon has to satisfy the "tight" identification criteria and is required to have $p_{\mathrm{T}}^{\gamma}$ >150 GeV, $|\eta|$ < 2.37 and to be isolated. The photon and $E_{\mathrm{T}}^{\mathrm{miss}}$ are required to be well separated, with $\Delta\phi(\gamma, E_{\mathrm{T}}^{\mathrm{miss}}) > 0.4$. Finally, events are required to have no electrons or muons and no more than one jet with $\Delta\phi(\mathrm{jet}, E_{\mathrm{T}}^{\mathrm{miss}}) >$ 0.4. Five signal regions (SRs) are defined corresponding to different $E_{\mathrm{T}}^{\mathrm{miss}}$ ranges. The run-2 analysis shows that the total background prediction uncertainty, including systematic and statistical contributions varies from 6.1% to 13.5% for the various SRs, dominated by the statistical uncertainty in the control regions which varies from approximately 4.3% to 10.4%. The largest systematic uncertainties are due to the uncertainty in the rate of fake photons from jets and to the uncertainty in the jet energy scale. Exclusion limits have been set on the production cross section of DM models using a one-sided profile likelihood ratio and the CLs technique [39, 40] with the asymptotic approximation[41]. In the run-2 analysis, a multiple-bin fit has been performed on the expected $E_{\mathrm{T}}^{\mathrm{miss}}$ distributions of the signal samples by combining the information from the three SRs with increasing $E_{\mathrm{T}}^{\mathrm{miss}}$ ranges. Model independent limits on the fiducial cross section, defined as $\sigma \times A$, where $A$ is the acceptance, have been obtained using inclusive SRs with increasing $E_{\mathrm{T}}^{\mathrm{miss}}$ thresholds in order to provide useful constraints on new physics which can be re-interpreted in terms of signal models not covered by this study.

| $\chi_0$ mass [GeV] | Signal Efficiency [%] |
|---|---|
| 90 | 36.6 ± 0.5 |
| 100 | 36.7 ± 0.5 |
| 200 | 38.2 ± 0.5 |
| 300 | 38.5 ± 0.5 |
| 500 | 39.8 ± 0.5 |
| 700 | 40.3 ± 0.5 |
| 1000 | 40.5 ± 0.5 |

Table 3: Summary of signal efficiencies for different values of $\chi^0$ masses for the mono-photon analysis in the SR.

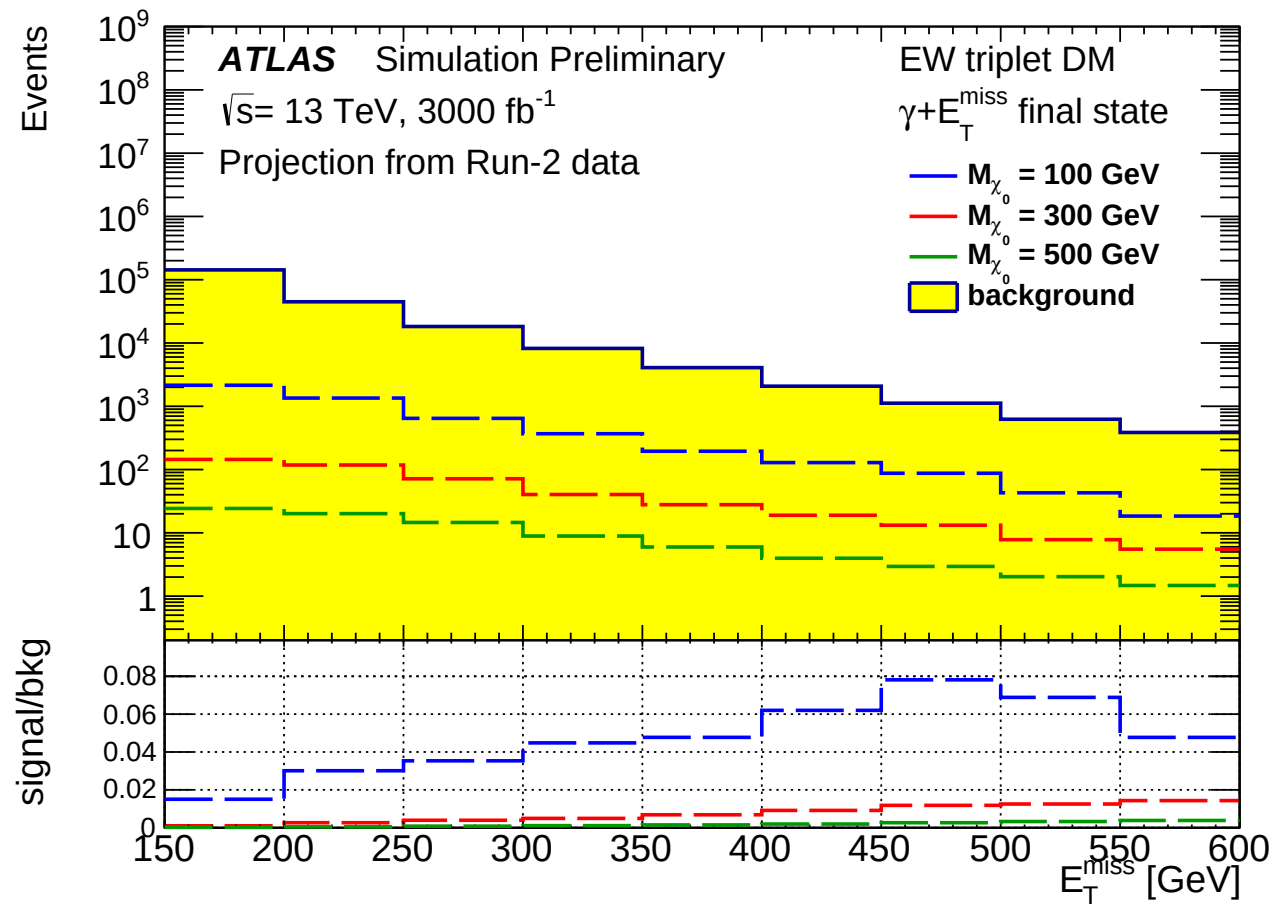

Figure 3: Pre-fit distribution of $E_{\mathrm{T}}^{\mathrm{miss}}$ for the dominant backgrounds W/Z+$\gamma$ and $\gamma$+jets and some signal samples in the SR corresponding to an integrated luminosity of 3000 fb$^{-1}$. Subleading backgrounds, estimated in run-2 analysis with data-driven techniques, are not included.

The reinterpretation of mono-photon search in the context of the pure WIMP triplet model uses the full simulated MC signal samples described in Sec. 3. The simultaneous fit is performed on the most inclusive SR, corresponding to $E_{\mathrm{T}}^{\mathrm{miss}}$ >150 GeV, that gives the best expected sensitivity as this model provides a medium-low $E_{\mathrm{T}}^{\mathrm{miss}}$ distribution for the various signal samples studied. All background samples, including fake photons estimated with data-driven techniques, have been included in the fit rescaling the run-2 estimates to the high luminosity scenario. For an integrated luminosity of 3000 fb$^{-1}$, the number of background events in the SR before any fit is 217400 ± 7600, where the error is statistical only. The pre-fit distribution of $E_{\mathrm{T}}^{\mathrm{miss}}$ for the dominant backgrounds and some signals is shown in Figure 3. All the systematic uncertainties on the MC background samples have been taken into account to obtain projections of the expected upper limits on the $\chi_0$ at 95% CL for an integrated luminosity of 3000 fb$^{-1}$and $\sqrt{s}$ = 13 TeV, that are shown in Figure 4. The expected limit is obtained from a fit to the so-called Asimov dataset [41], with the signal and all backgrounds scaled to their predicted values. Masses of $\chi_0$ below 310 GeV can be excluded at 95% CL by the analysis assuming the same systematic uncertainties adopted in Ref. [15]. For an integrated luminosity of 36.1 fb$^{-1}$, corresponding to the run-2 analysis, masses of $\chi_0$ below 50 GeV could be excluded. A sensitivity to masses above 90 GeV, which corresponds to the current limit by LEP, can be reached only with the increased luminosity provided by the HL-LHC scenario.

The impact of the systematic uncertainty on the sensitivity of the analysis has been checked considering that the analysis will no more be limited by the statistical uncertainty at high luminosity. In a scenario in which the current systematic uncertainties are halved, an exclusion of $\chi_0$ masses up to about 340 GeV could be reached. Thanks to the increased statistics, the analysis at high luminosity could be further optimized by performing a multiple-bin fit, thus on more bins in $E_{\mathrm{T}}^{\mathrm{miss}}$ improving the overall sensitivity of the analysis.

The increase in the $\sqrt{s}$ from 13 TeV to 14 TeV has not been taken into account: the cross section of the signals close to exclusion are expected to increase by ~20% (see table 1), while those of the main $Z \rightarrow \nu\nu + \gamma$ background are expected to increase by ~10-15% leading to a slight increase in the signal significance.

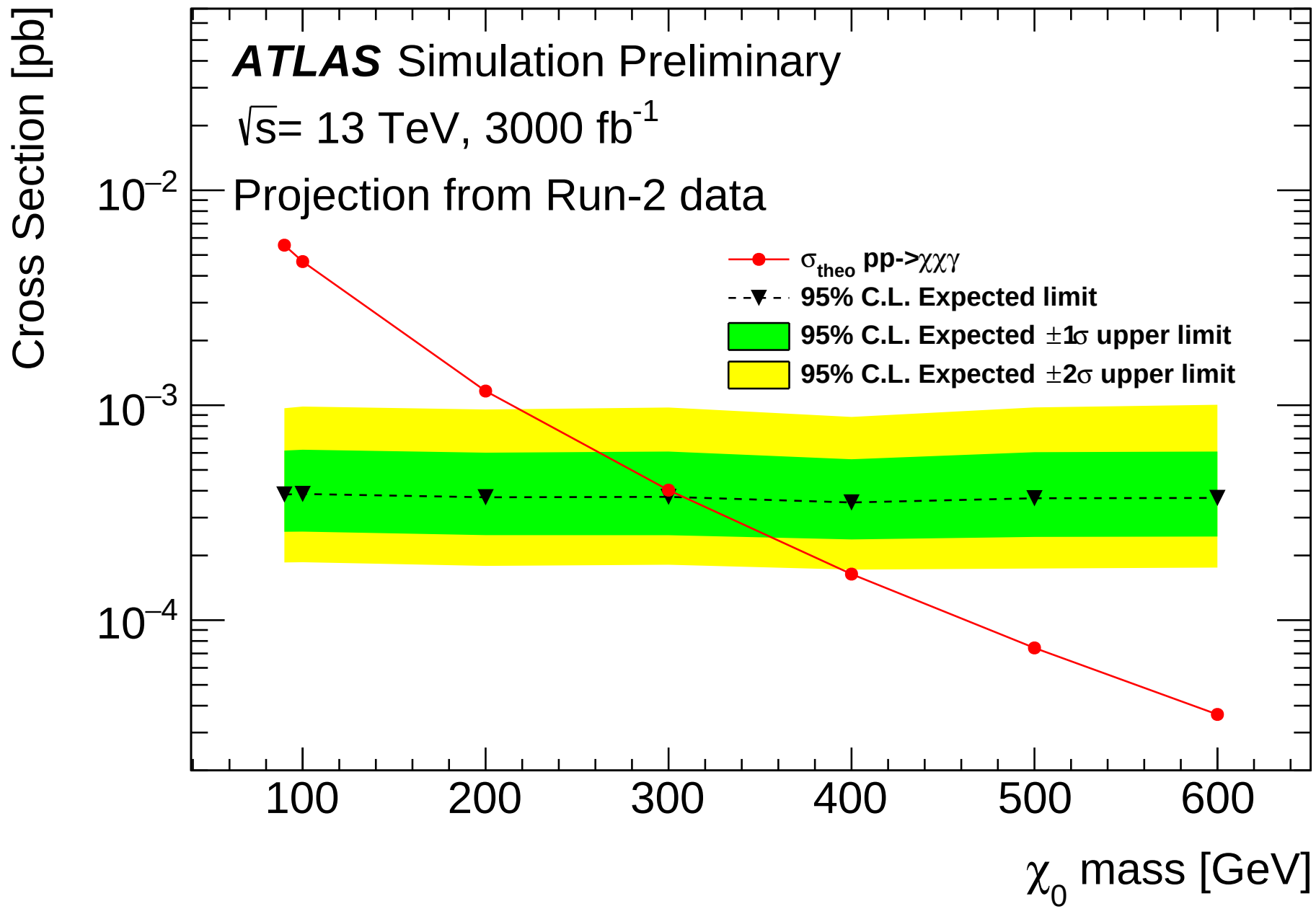

Figure 4: Expected upper limits at 95% CL on the production cross section of $\chi$ as a function of $\chi_0$ mass in mono-photon final state corresponding to an integrated luminosity of 3000 fb$^{-1}$and assuming the same systematic uncertainties adopted in Ref [15]. The red line shows the theoretical cross section.

# 5 VBF plus $E_{\mathrm{T}}^{\mathrm{miss}}$ final state

The VBF+$E_{\mathrm{T}}^{\mathrm{miss}}$ topology is characterized by two quark-initiated jets with a large separation in rapidity and $E_{\mathrm{T}}^{\mathrm{miss}}$. In analogy to the mono-photon result presented in the previous section, the sensitivity of the VBF+$E_{\mathrm{T}}^{\mathrm{miss}}$ analysis to the pure WIMP triplet model is presented in Sec. 5.1 as a reinterpretation of the run-2 results for the high luminosity scenario foreseen for the HL-LHC. As pile-up is a key experimental challenge for event reconstruction in the VBF topology at the HL-LHC, a dedicated study of its impact is shown in Sec. 5.2 using VBH $H \rightarrow$invisible as benchmark.

## 5.1 Projections at high luminosity for DM for EW triplet DM

As the mass splitting $\Delta M$ between $\chi^{\pm}$ and $\chi_0$ is small ($\Delta M \sim 165$ MeV), the $\chi^{\pm}$ particles decay into $\chi_0$ and very soft pions, which are not reconstructed and are lost in the detector. Therefore the signature of a triplet produced via VBF is defined by the presence of two energetic jets, largely separated in pseudorapidity, with $O(1)$ TeV invariant mass, and large $E_{\mathrm{T}}^{\mathrm{miss}}$ coming from the DM particles. A search for an invisibly decaying Higgs boson produced via VBF has been performed by ATLAS using a dataset corresponding to an integrated luminosity of 36 fb$^{-1}$ of $pp$ collision at $\sqrt{s}$ =13 TeV [14]. The final state is defined by the presence of two energetic jets, largely separated in pseudorapidity and with $O(1)$ TeV invariant mass, and large $E_{\mathrm{T}}^{\mathrm{miss}}$ coming from the invisible particles from the Higgs decay. This analysis set limits on the Branching Ratio ($\mathcal{B}$) of the Higgs boson decaying into invisible particles. The main backgrounds to this analysis arise from $Z \to \nu\nu$+jets and $W \to \ell\nu$+jets events. The contribution of $W/Z$ events is estimated with the following approach: dedicated regions (control regions) enriched in $W \to \ell\nu$ (where the lepton is found) and $Z \to \ell\ell$[2] events are used to normalize to data the MC estimates using a simultaneous fitting technique and to extrapolate them to the signal region. The multijet background comes from multijet events where large $E_{\mathrm{T}}^{\mathrm{miss}}$ is generated mainly by jet mismeasurements. This is highly reduced by a tight $E_{\mathrm{T}}^{\mathrm{miss}}$ cut and is estimated via data-driven methods resulting in less than 1% of the total background.

This analysis is reinterpreted in the context of the model described in Sec. 1 for an integrated luminosity of 3000 fb$^{-1}$: the same selections, background samples and analysis strategy are used to set limits on the cross section of the pure WIMP triplet produced via VBF at an integrated luminosity corresponding to the one which will be reached by HL-LHC. The only selection which has been changed is the separation in pseudorapidity between the two leading jets ($\Delta\eta(j_1, j_2)$) which has been relaxed from $\Delta\eta(j_1, j_2)$ >4.8 to $\Delta\eta(j_1, j_2)$ >3.5. A relaxed $\Delta\eta(j_1, j_2)$ selection increases the sensitivity to the model as, in addition to the pure VBF Feynman diagrams, also diagrams with strong production contribute to the signal. Therefore, a signal region is defined by selecting events passing the lowest unprescaled $E_{\mathrm{T}}^{\mathrm{miss}}$ trigger, containing no electron and muon, having exactly two jets with transverse momentum $p_T(j_1)$ >21 nov 80 GeV and $p_T(j_2) > 50$ GeV, which are not back to back in the transverse plane ($\Delta\Phi(j_1, j_2)$ <1.8) and which are separated in pseudorapidity ($\Delta\eta(j_1, j_2)$ >3.5). Events are required to have large $E_{\mathrm{T}}^{\mathrm{miss}}$ ($E_{\mathrm{T}}^{\mathrm{miss}}$ >180 GeV), the two leading jets are separated from the $E_{\mathrm{T}}^{\mathrm{miss}}$ ($\Delta\Phi(j_1,\ E_{\mathrm{T}}^{\mathrm{miss}})$ >1, $\Delta\Phi(j_2,\ E_{\mathrm{T}}^{\mathrm{miss}})$ >1), the vectorial sum of all the jets (including the pile-up ones) is required to be $H_{\mathrm{T}}^{\mathrm{miss}}$ >150 GeV and the invariant mass of the dijet system is required to be $M(j_1, j_2)$ >1 TeV. For all the details about jet, electron, muon and $E_{\mathrm{T}}^{\mathrm{miss}}$ reconstruction and details about the selection refer to Ref. [14]. As done in the run-2 analysis, the selected events are then split into three categories (*bins*) according to the invariant mass of the dijet system. In particular, the following bins are selected by requiring: 1 TeV< $M(j_1, j_2)$ <1.5 TeV; 1.5 TeV< $M(j_1, j_2)$ <2 TeV; $M(j_1, j_2)$ >2 TeV. Similarly, the control regions enriched in $W/Z$+ jets processes used in the analysis to constrain the backgrounds, are also split in the same three categories. The same background MC samples that have been used for the run-2 analysis, produced at $\sqrt{s}$ =13 TeV, have been used for this study by rescaling the luminosity to 3000 fb$^{-1}$, the signal samples used are the ones described in Sec. 3.1. Following the same run-2 strategy, a simultaneous fit in SR and CRs, using the three $M(j_1, j_2)$ bins to increase the signal sensitivity, is used for the $W/Z$+ jets background estimation and for the limit setting. The signal efficiencies, prior to any fits, for the three values of $\chi_0$ mass considered in the analysis, are shown in table 4 for an integrated luminosity of 3000 fb$^{-1}$. The total background, before any fits, in the three bins is respectively 215000±4900; 72000±3300; 48400±1900, where the error is statistical only. Pre-fit distributions of $\Delta\eta(j_1, j_2)$ and $M(j_1, j_2)$, for the total background and the signals, in the inclusive $M(j_1, j_2) > 1$ TeV SR bin, are shown in Fig. 5.

Exclusion limits are set on the production cross section of the model using a one-sided profile likelihood ratio and the CLs technique [39, 40] with the asymptotic approximation[41].

Experimental and theoretical systematic uncertainties have been taken into account and are included in the likelihood as Gaussian-distributed nuisance parameters. Experimental systematic uncertainties include the ones related to the jet energy scale (JES) and resolution (JER), $E_{\mathrm{T}}^{\mathrm{miss}}$ soft term and lepton measurements. The main experimental systematic uncertainties for the run-2 VBF+$E_{\mathrm{T}}^{\mathrm{miss}}$ analysis come from JES and JER [42]. At HL, some of the experimental systematic uncertainties are expected to be reduced or become negligible, therefore, for this projection, all the run-2 experimental systematic uncertainties

[2] With $\ell$ being electrons or muons.

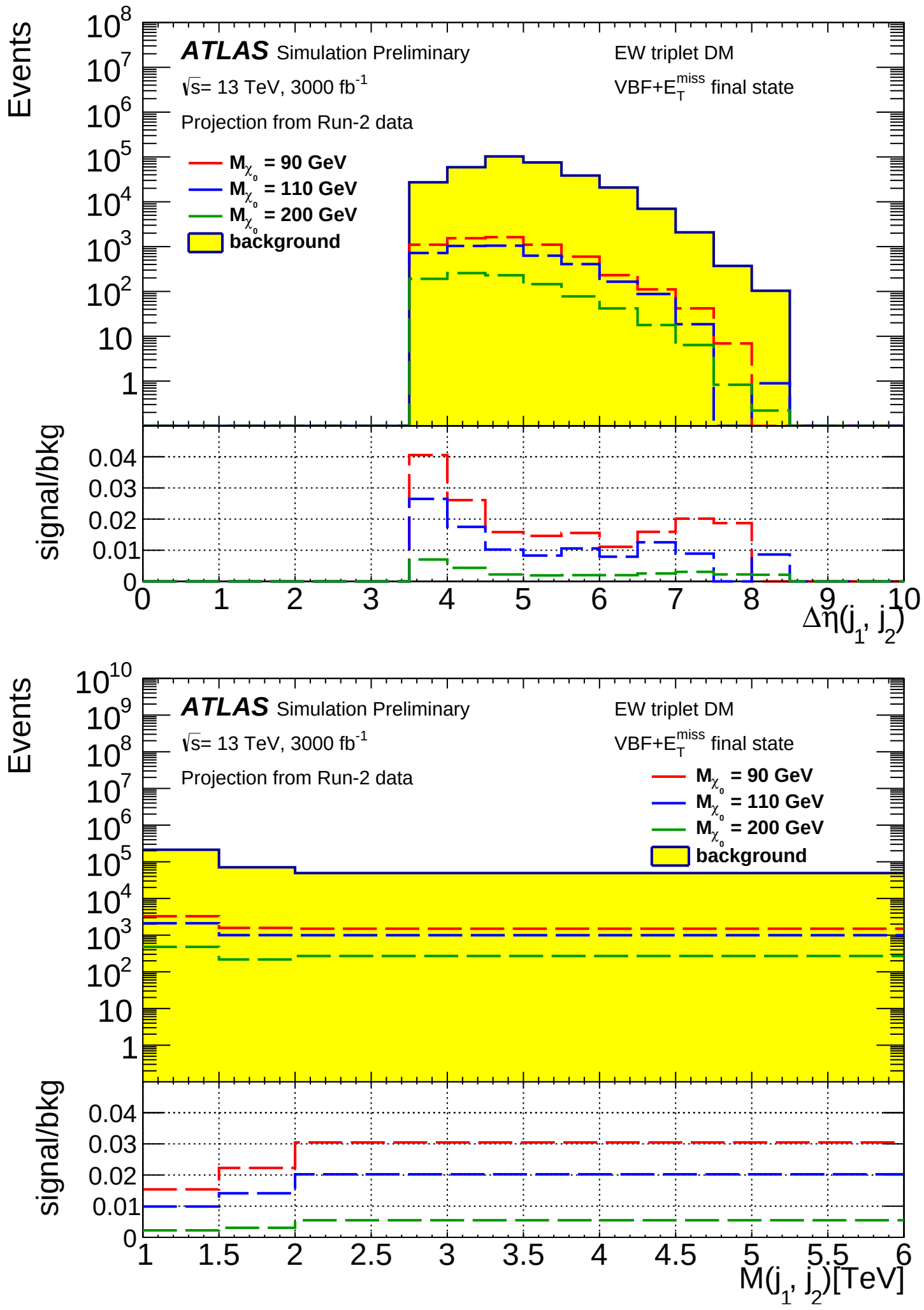

Figure 5: Pre-fit distribution of event yields for $\Delta\eta(j_1, j_2)$ (top) and $M(j_1, j_2)$ (bottom) in the signal region (inclusive $M(j_1, j_2) > 1$ TeV bin). The backgrounds ($W/Z$+jets, dibosons, top) and the three signal samples considered in the analysis are shown.

10

| | Signal Efficiency [%] | | |
|---|---|---|---|
| $\chi_0$ mass [GeV] | $M(j_1, j_2)$ bin1 | $M(j_1, j_2)$ bin2 | $M(j_1, j_2)$ bin3 |
| 90 | 0.56 ± 0.02 | 0.27 ± 0.01 | 0.26 ± 0.01 |
| 110 | 0.67 ± 0.02 | 0.32 ± 0.01 | 0.32 ± 0.01 |
| 200 | 0.97 ± 0.03 | 0.44 ± 0.02 | 0.55 ± 0.02 |

Table 4: Summary of signal efficiencies for different values of $\chi_0$ mass, for the VBF analysis, in the three $M(j_1, j_2)$ bins considered in the analysis.

have been rescaled according to the HL expectations which are discussed in Ref. [43]. In particular some of the main JES and JER systematic uncertainties are halved. The main theoretical sources of uncertainty for the run-2 analysis come from choices on the resummation, renormalization, factorization and CKKW matching scale for the $W/Z$+jets backgrounds processes. A significant improvement in these systematic uncertainties is expected. Therefore, the current run-2 theoretical systematic uncertainties on the $W/Z$+jets backgrounds have been rescaled down to reach the level of few % (5% of the run-2 theoretical systematic uncertainties is kept). This level of systematic uncertainties corresponds to the one employed by the run-2 ATLAS mono-jet search [44]. Such an improvement in theoretical systematics for the VBF final state may be reached using similar techniques and here it is assumed that these improvements on the theoretical side will be reached for the HL-LHC phase. The same correlation scheme that has been used in [14] is also used for the projections presented here. Uncertainties arising from the finite MC statistics of the samples used are assumed to be negligible.

The results obtained by rescaling the signals and backgrounds to an integrated luminosity of 3000 fb$^{-1}$ are shown in Fig. 6. The results indicate that with such a luminosity the lowest masses considered ($M_\chi$ ~110 GeV) can be excluded at 95% CL. This study is aimed at providing a first idea of the potential reach of such an analysis with an increased luminosity. Indeed, with an integrated luminosity corresponding to the one that will be reached at the end of the run-2, there is no sensitivity to masses above 90 GeV. This analysis is very sensitive to the systematic uncertainties and a further optimization of the selection cuts on this model, taking into account the higher luminosity, could help to achieve a better reach. Furthermore, the HL-LHC trigger scenario has not been considered in this note. VBF analyses will probably benefit from a combination of $E_{\mathrm{T}}^{\mathrm{miss}}$ and VBF jet triggers. However, even with $E_{\mathrm{T}}^{\mathrm{miss}}$ thresholds raised by 50-100 GeV with respect to the current ones, the analysis is still sensitive to this model for the masses considered. The increase in the center of mass energy $\sqrt{s}$ from 13 TeV to 14 TeV, which is expected at HL-LHC, has not been taken into account: the cross-section of the signals considered are expected to increase by ~20% at $\sqrt{s}$ =14 TeV, as shown in table 2, while the $W/Z$+jets cross sections will increase by ~8%. Therefore, a slight increase in the signal significance is expected.

### 5.2 The challenge of pile-up for VBF at the HL-LHC

In this study, jets are built from particle flow objects [45] using the anti-$k_t$ [46] algorithm with radius parameter $R = 0.4$ as implemented in FastJet [47]. Even though the current ATLAS standard jet collection is constructed from locally calibrated topological calorimeter-cell clusters [48], particle flow jets are used for this analysis because of their superior jet energy resolution and pile-up stability at low $p_{\mathrm{T}}$. Aside from a constant calibration to bring the jet response close to unity on average, jets are not further corrected for pile-up or calibrated for the detector response. Jets are only considered if $p_{\mathrm{T}} > 25$ GeV and $|\eta| < 4.5$.

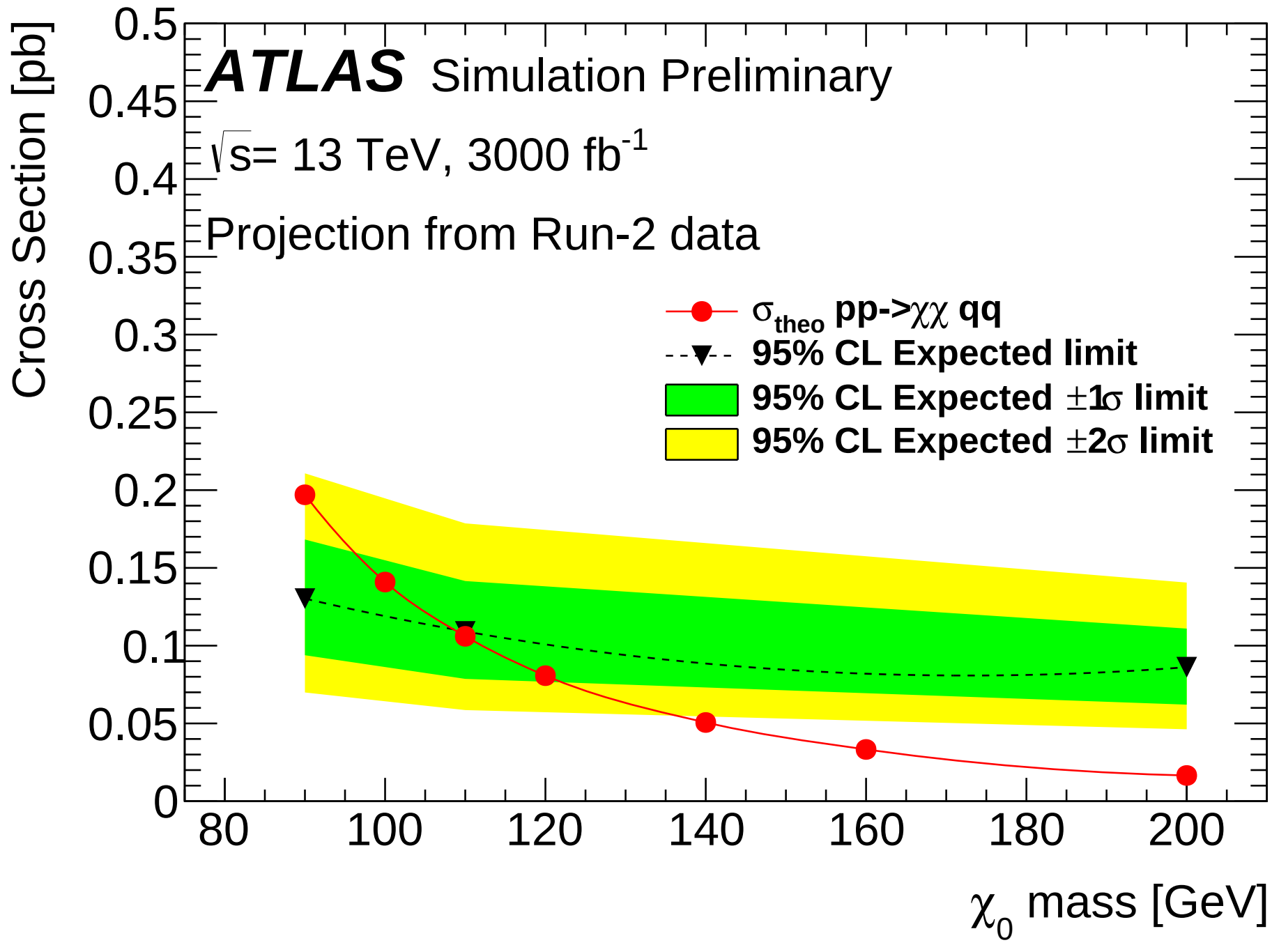

Figure 6: Expected upper limit on the production cross section of $\chi$ in VBF+$E_T^{miss}$ final state. Results are shown for an integrated luminosity of 3000 fb$^{-1}$.

Particle-level jets are built from detector-stable particles ($c\tau > 10$ mm) excluding muons and neutrinos as well as pile-up and any particles resulting from interactions with the detector. The same algorithm is used at particle level as is used at detector-level. A particle-level jet is matched to a detector-level jet if it is the highest $p_T$ particle-level jet satisfying $p_{T,part} > 10$ GeV and $\Delta R(\text{part.,det.}) < 0.3$. The parton label of a detector-level jet is the type of the highest energy parton ghost-associated [49] to the matched particle-level jet. Detector-level jets originating mostly from pile-up (henceforth, 'pile-up jets') will have no associated particle-level jet. There are ambiguous cases where a jet may have significant contributions from both the hard-scatter process as well as pile-up. The ambiguous case is assigned when there is a particle-level jet with $p_T > 4$ GeV close ($0.3 < \Delta R < 0.6$) to a given detector-level jet.

Figure 7 shows the average number of jets of each type in $H \to$ invisible and $Z$+jets events. Most Higgs events have two quark-initiated jets, but there is only 1 quark-initiated jet with $p_T > 50$ GeV on average. The average number of pile-up jets per event in the two processes is nearly the same and is about 0.2. In the absence of any pile-up jet mitigation, there is also a spike in the $\Delta\eta$ distribution at the edge of acceptance that is dominated by pile-up jets. Therefore, pile-up jet rejection is critical for the success of this analysis.

Charged particle tracks are constructed from hits in the upgraded inner tracker. No explicit track quality selection is applied in terms of the hit pattern in the ITk, though the baseline fake rate is expected to be significantly better than for the present detector. Tracks are ghost associated to the jets and required to have $p_T > 900$ MeV and $p_T < 40$ GeV (to suppress fake tracks). The difference between the primary vertex[3] $z$ position and the track $z_0$ (longitudinal impact parameter) must be less than $2\sigma$, where $\sigma$ is the sum in quadrature of the track $z_0$ and the vertex $z$ uncertainties. The left plot of Figure 8 shows the vertex $z$ resolution for the signal and background samples. The standard deviation of the central peak is about 0.1 mm. The reason for the long tails is from events where the selected vertex is not the primary vertex. The right plot of Figure 8 shows the vertex $z$ resolution extended over the entire beamspot size. A Gaussian fit away from the peak indicates that the wrong vertex is chosen about 25% of the time. Further implications of this mis-identification are discussed below.

[3] This is the vertex with the highest $\sum p_T^2$.

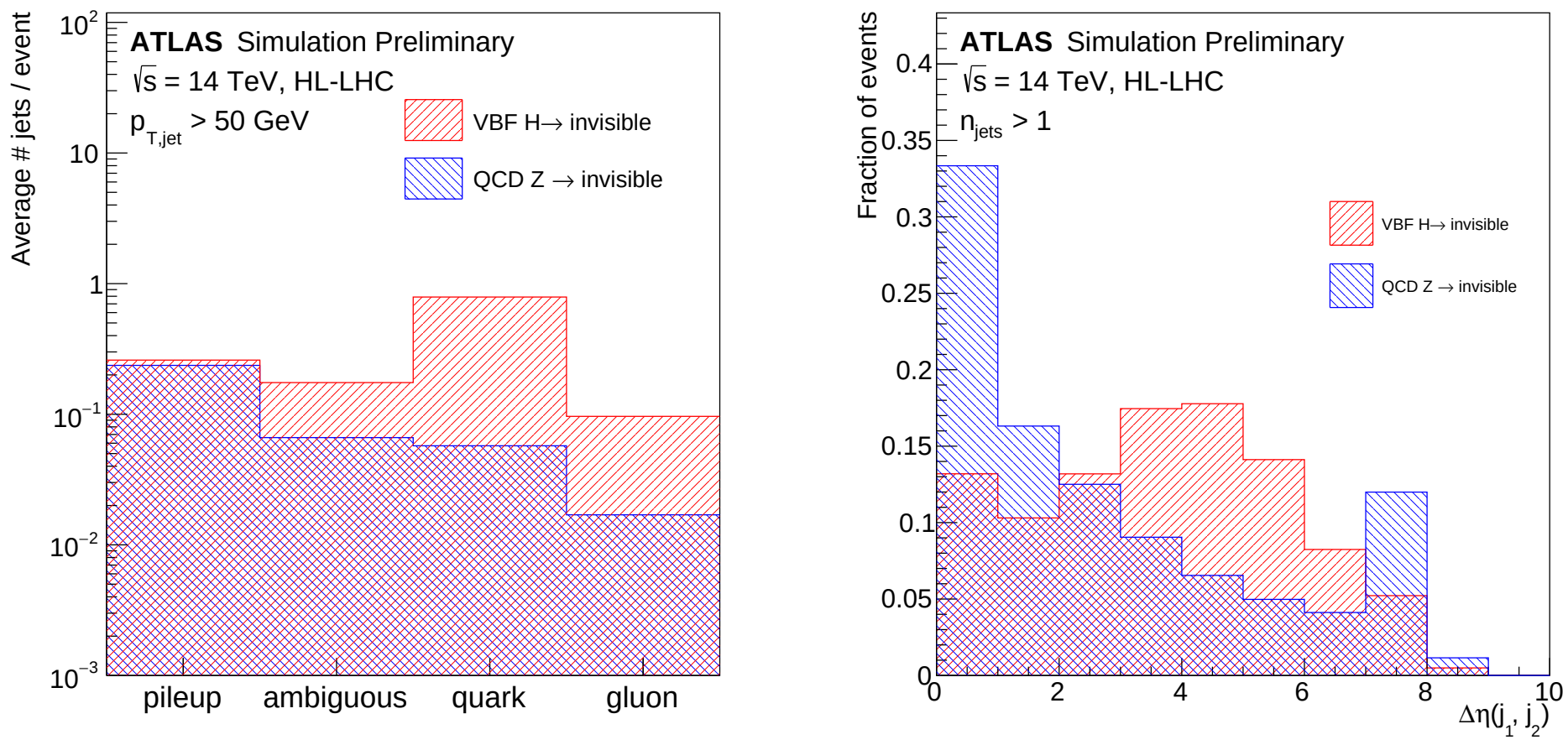

Figure 7: Left: The average number of jets in $H \to$ invisible and $Z$+jets events for the various jet types prior to any event selection for jets with $p_T > 50$ GeV. Right: The distribution of $\Delta\eta(j_1, j_2)$ of the leading two jets (requiring events to have at least two jets, each with $p_T > 50$ GeV). No pile-up jet rejection is applied.

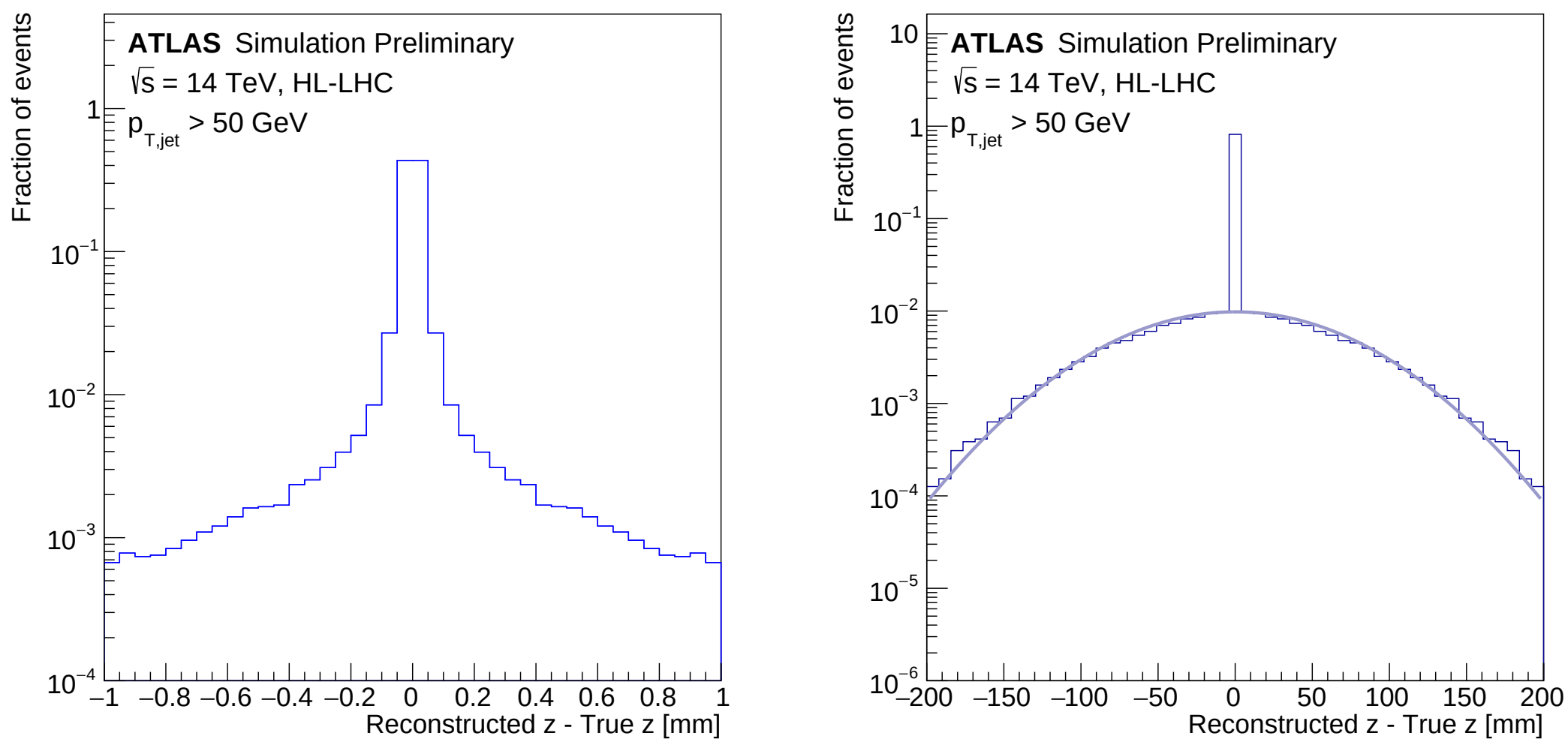

Figure 8: The $z$ resolution for the primary vertex. The left plot is simply a zoomed in version of the right plot. In the right plot, the range away from 10 mm in both tails is fit to a Gaussian. The area of the fitted region is about 25% of the total.

One of the key discriminating observables between pile-up jets and hard scatter jets is $R_{p_T}$ [50], which is the sum of the $p_T$ of the tracks associated to the jet normalized by the jet $p_T$. Only tracks with $\Delta R < 0.3$ are considered in the calculation of $R_{p_T}$. Figure 9 shows the distribution of $R_{p_T}$ for hard-scatter and pile-up jets which are relatively forward/central ($|\eta| > 1.2$) and cases where the primary vertex is or is not correctly identified. When the primary vertex is mis-identified, the distribution of $R_{p_{\mathrm{T}}}$ for hard-scatter jets is nearly the same as it was for pile-up jets when the vertex is correctly identified. For forward jets, the separation between the two classes is worse, due to the poorer vertex and track-to-vertex performance. As a baseline[4], jets are declared 'hard-scatter' if $R_{p_{\mathrm{T}}} > 0.05$ which corresponds to 85% hard-scatter efficiency and 2% pile-up jet efficiency when $|\eta| < 1.2$ and $|z^{\mathrm{reco}} - z^{\mathrm{true}}| < 0.1$.

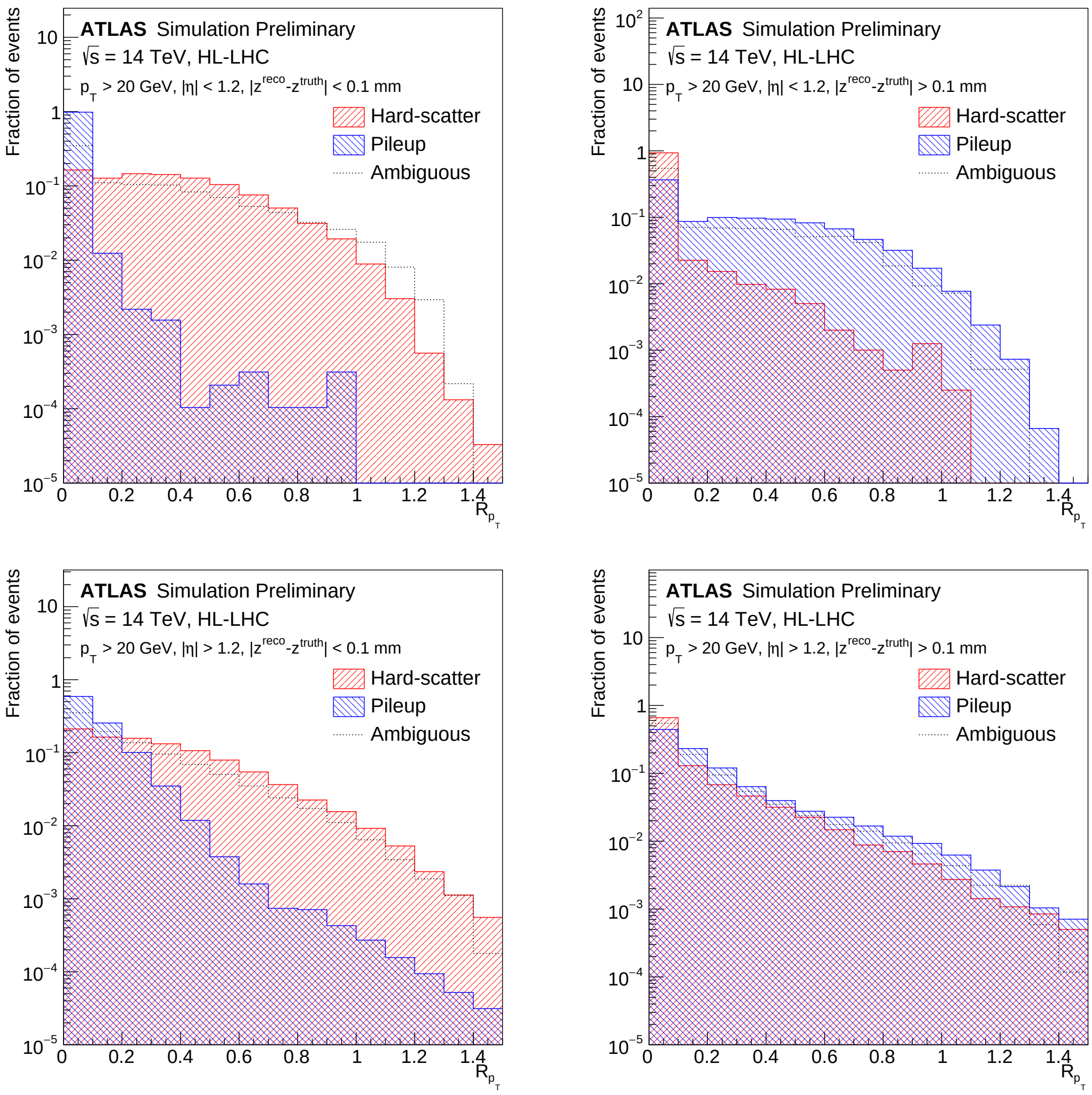

Figure 9: The distribution of $R_{p_{\mathrm{T}}}$ for $|\eta| < 1.2$ and $|z_0^{\mathrm{reco}} - z_0^{\mathrm{true}}| < 0.1$ mm (top left), $|\eta| < 1.2$ and $|z_0^{\mathrm{reco}} - z_0^{\mathrm{true}}| > 0.1$ mm (top right), $|\eta| > 1.2$ and $|z_0^{\mathrm{reco}} - z_0^{\mathrm{true}}| < 0.1$ mm (bottom left), and $|\eta| > 1.2$ and $|z_0^{\mathrm{reco}} - z_0^{\mathrm{true}}| > 0.1$ mm (bottom right). Summed over $H$ and $Z$ jets, about 14% are in the top left, 6% in the top right, 53% in the bottom left, and 27% in the bottom right.

A last input that is critical to the $H \to$ invisible search is the $E_{\mathrm{T}}^{\mathrm{miss}}$, a proxy for the Higgs $p_{\mathrm{T}}$. As it depends on all of the reconstructed objects, $E_{\mathrm{T}}^{\mathrm{miss}}$ reconstruction is a complex task. There has not been a detailed optimization of the $E_{\mathrm{T}}^{\mathrm{miss}}$ reconstruction for the upgraded ATLAS detector; in this analysis, the negative sum of the transverse momenta of all reconstructed jets is used (and indicated by the symbol $E_{\mathrm{T,jet}}^{\mathrm{miss}}$).

[4] Clearly, the optimal configuration will have an $|\eta|$-dependent definition.

14

An event selection based on the jets and $E_{\mathrm{T}}^{\mathrm{miss}}$ described above is modeled after the run-2 analysis [14] in order to assess the impact of various pile-up mitigation scenarios. It is likely that a dedicated optimization for the amount of data and detector conditions will improve the result, but this is beyond the scope of this illustrative study[5]. Due to limitations of MC statistics, a simplified version of the run-2 VBF $H \to$ invisible analysis as introduced in Sec. 5.1 is used. In particular, all of the angular requirements with jets are removed and there is no binning in $M(j_1, j_2)$. Additionally, $E_{\mathrm{T,jet}}^{\mathrm{miss}} > 150$ GeV[6].

For the run-2 analysis, the event selection efficiency for $Z \to \nu\bar{\nu}$ events is about $2 \times 10^{-6}$ and about 0.5% for the signal with a $\mathcal{B}(H \to \text{invisible}) = 100\%$ (which is about 85% from VBF). The background is nearly half QCD $Z \to \nu\bar{\nu}$ and half QCD $W$+jets. Since the following results are only based on $Z$+jets, branching ratio limits are computed by doubling the $Z$+jets background. It is likely that with the extended coverage of the ITk relative to the current tracker the lost leptons will be suppressed and thus the $W$+jets background will be less than the $Z$+jets rate so this approximation is conservative.

A simplified statistical analysis is performed to assess the impact of several scenarios on the $H \to$ invisible branching ratio limit with the full HL-LHC dataset. A one-bin statistical test with one overall source of systematic uncertainty is performed to determine if a particular signal yield is excluded. The signal yield is scanned to determine the largest branching ratio that would be not excluded at the 95% confidence level. Table 5 presents the corresponding limits, normalized to the one for the run-2 systematic uncertainties and truth-based pile-up tagging to show the relative gains and losses possible under various scenarios. The choice of showing only a normalized $\mathcal{B}$ is motivated by the fact that a simplified analysis has been performed with the aim to show the importance of having a pile-up tagger for this analysis and not to optimize the reach of an Higgs invisible search at HL-LHC. The three rows correspond to different strategies for removing pile-up jets. The first row corresponds to the case where pile-up jets are not actively removed, the second row indicates the performance when pile-up jets are identified using $R_{p_{\mathrm{T}}}$, and the last row represents the case where truth labels are used to reject pile-up. The four columns correspond to different assumptions on the systematic uncertainties at the time of the HL-LHC. A total systematic uncertainty of 10% is used[7], which is similar to the run-2 result; this scenario is indicated by the results in the first column. The second and third columns show results with systematic uncertainties that are half as large and the last column uses systematic uncertainties that are a factor of ten smaller than the present analysis. To give an indication of the potential of harsher event selections, the third column supposes that the HL-LHC analysis will have the same number of selected signal events as the current analysis. For this scenario, the background efficiency is assumed to be 10% of the signal efficiency.

With a realistic reduction in the systematic uncertainty and tighter selection criteria, it may be possible to significantly improve the sensitivity. The limit improves from including a simple $R_{p_{\mathrm{T}}}$-based pile-up jet rejection, though the gap with the truth-information-based tagger indicates that there is room (and reward) for developing a more sophisticated approach.

[5] Furthermore, the challenges of event triggering are ignored. With the many improvements to upgraded ATLAS triggering system, it is likely that despite a more challenging environment, triggering may be more effective at the HL-LHC than the present LHC as mentioned before.

[6] Reference [14] required the full $E_{\mathrm{T}}^{\mathrm{miss}} > 180$ GeV and the $H_{\mathrm{T}}^{\mathrm{miss}}$ to be above 150 GeV. The latter also includes pile-up jets. The requirement used here is a hybrid that uses the sum of jets (so like $H_{\mathrm{T}}^{\mathrm{miss}}$) but does not include pile-up jets (like $E_{\mathrm{T}}^{\mathrm{miss}}$) that compromises performance and the availability of MC statistics.

[7] Implemented as a single component uncertainty for a one-bin counting experiment.

| | | Systematic Uncertainties [% of nominal] | | | |
|---|---|---|---|---|---|
| | $\frac{\mathcal{B}(H\to\text{invs.})}{\mathcal{B}_{\text{Truth, Nominal}}(H\to\text{invs.})}$ | 100% | 50% | 50% + fixed efficiency | 10% |
| PU jet rejection | None | – | – | 0.31 | 0.59 |
| PU jet rejection | $R_{p_{\mathrm{T}}}$ | – | – | 0.28 | 0.48 |
| PU jet rejection | Truth | 1.0 | 0.48 | 0.07 | 0.10 |

Table 5: The limit on the $H \to$ invisible branching ratio using the full HL-LHC dataset (3 ab$^{-1}$) normalized to the one for the run-2 systematic uncertainties and truth-based pile-up tagging to show the relative gains and losses possible under various scenarios. A '–' indicates a value bigger than 1. In addition to rejecting more background that would have otherwise passed the $\Delta\eta$ and $M(j_1, j_2)$ requirements, a truth pile-up tagger has more signal as the third jet veto is also more efficient. The amount by which the truth pile-up tagger is more efficient for the signal for the third jet veto is about the same as the inefficiency on the background for the $\Delta\eta$ requirement.

# 6 Conclusion

A new $SU(2)$ fermionic triplet, added on top of the SM with an approach inspired by the Minimal Dark Matter model, provides a good dark matter candidate if a new symmetry is imposed (e.g. the SM B-L). This triplet has mass ~3 TeV if the relic abundance is matched, however smaller masses are also allowed in case of non-thermal production mechanisms or if it constitutes only a fraction of the DM abundance. Such a triplet can be produced at the LHC and it can be probed in different ways. This note presents results for the mono-photon and the VBF+$E_{\mathrm{T}}^{\mathrm{miss}}$ final states.

The run-2 mono-photon search has been reinterpreted in the context of this model assuming an integrated luminosity of 3000 fb$^{-1}$, which will be available with the HL-LHC. The result of this study shows that masses of $\chi_0$ below 310 GeV will be probed at 95% CL assuming the same systematic uncertainties adopted in the run-2 analysis.

The pure WIMP triplet can also be produced via VBF giving rise to a final state with jets largely separated in rapidity and $E_{\mathrm{T}}^{\mathrm{miss}}$. This final state is the same final state that has been defined to look for the invisible decay of the Higgs boson produced via VBF. The run-2 analysis, applying small changes in the selection, is used as basis, to test the DM triplet model at an integrated luminosity of 3000 fb$^{-1}$. Projections are presented for the HL-LHC scenario are presented showing that, with such a luminosity, it will be possible to test the lower masses of this model (up to ~110 GeV). A slight improvement in the signal significance from the increase of the center-of-mass energy to $\sqrt{s}$ =14 TeV foreseen for the HL-LHC is expected for both analyses. Complementary searches, such as mono-jet searches and disappearing track signatures, would also set stringent constraints to this model.

Many experimental aspects of the search in the VBF+$E_{\mathrm{T}}^{\mathrm{miss}}$ channel will be particularly challenging including the rejection of pile-up jets, the identification of the primary vertex, and the resolution of low $p_{\mathrm{T}}$ jets. With a combination of pile-up robustness studies, analysis optimization, and theory uncertainty reduction, perhaps in part from auxiliary SM measurements, then $H \to$ invisible and EW triplet DM searches at the HL-LHC may be significantly improved.

# Prospects for a search of invisible particles produced in association with single-top quarks with the ATLAS detector at the HL-LHC

The ATLAS Collaboration

The expected sensitivity of a search for events with one top quark and large missing transverse momentum is estimated using simulated proton–proton collisions at a centre-of-mass energy of 14 TeV with the ATLAS experiment at the HL-LHC. A non-resonant production of an exotic state decaying to a pair of invisible dark-matter particles in association with a right-handed top quark is considered. Only the topologies where the $W$ boson from the top quark decays into an electron or a muon and a neutrino are considered. Assuming an integrated luminosity of 3000 fb$^{-1}$, the expected exclusion limit (discovery reach) at 95% CL on the mass of the exotic state is 4.6 TeV (4.0 TeV) using a multivariate analysis based on a boosted decision tree.

# 1 Introduction

The discovery of a Standard Model (SM) Higgs-like boson in 2012 by the ATLAS [1] and CMS [2] Collaborations opened up new possibilities in searches for new physics. In fact, even with the existence of a Higgs boson confirmed, the SM cannot be considered a complete description of nature. For example, the theory does not explain the fermion generations and mass hierarchy, nor the origin of the matter–antimatter asymmetry in the Universe. Neither does it describe the existence of non light-emitting matter, usually referred to as dark matter (DM), nor describe gravitational interactions. The SM is therefore generally regarded as a low-energy approximation of a more fundamental theory with new degrees of freedom and symmetries that would become manifest at higher energy.

Despite the strong evidence from astrophysical measurements [3–5] which support the existence of DM, there is no evidence yet for non-gravitational interactions between DM and SM particles. DM particles are not expected to interact with the detector and therefore can not be directly detected but can be inferred through a large amount of missing transverse momentum. The specific search strategy depends on what type of particle or system is recoiling against the unseen DM. Both the ATLAS and CMS Collaborations have carried out searches for DM particles produced in association with jets [6–9], photons [10, 11], $W$ or $Z$ [7, 12, 13] and Higgs [14–17] bosons, significantly constraining the allowed parameter space for generic classes of models predicting DM candidates.

This note describes the expected sensitivity of a search for the non-resonant production of an exotic state decaying into a pair of invisible DM particle candidates in association with a right-handed top quark. Such final-state events, commonly referred to as "monotop" events, are expected to have a reasonably small background contribution from SM processes [18]. In this analysis only the topologies where the $W$ boson from the top quark decays into a lepton and a neutrino are considered.

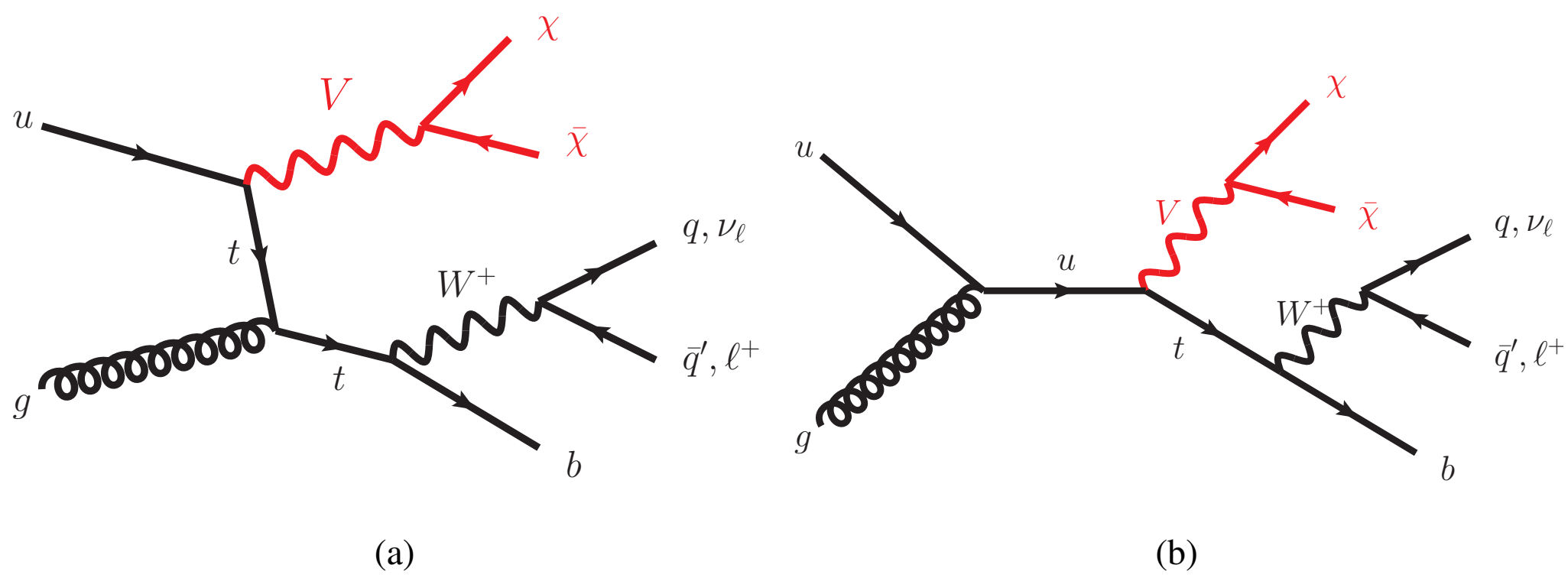

Figure 1: Representative leading-order Feynman diagrams corresponding to the monotop signals searched for non-resonant (a) $t$- and (b) $s$-channel DM production in association with a top quark.

The non-resonant monotop production via a flavour-changing neutral interaction is shown in Figure 1 where a top quark ($t$), a light-flavour up-type quark ($u$) and an exotic massive vector-like particle $V$ can be parametrised through a general Lagrangian [18, 19]:

$$\mathcal{L}_{\text{int}} = aV_\mu \bar{u}\gamma^\mu P_{\text{R}} t + g_\chi V_\mu \bar{\chi}\gamma^\mu \chi + \text{h.c.}, \quad (1)$$

where $V$ is coupled to a pair of DM particles (represented by Dirac fermions $\chi\bar{\chi}$) whose strength can be controlled through a parameter $g_\chi$ and where $P_{\text{R}}$ represents the right-handed chirality projector. The parameter $a$ stands for the coupling constant between the massive invisible vector boson $V$, and the $t$- and $u$-quarks, and $\gamma^{\mu}$ are the Dirac matrices. A detailed description of further assumptions present in the benchmark models can be found in Refs. [19, 20].

The study presented here is performed with simulated proton–proton ($pp$) collisions at a centre-of-mass energy of 14 TeV within the framework of the HL-LHC with an upgraded ATLAS detector [21, 22] and assuming an integrated luminosity of 3000 fb$^{-1}$. Similar searches for such topologies were previously done by the CDF Collaboration in proton–antiproton ($p\bar{p}$) collisions at $\sqrt{s}$ = 1.96 TeV at the Tevatron, using 7.7 fb$^{-1}$ [23], excluding the presence of such vector particles of masses of up to 150 GeV. Using $pp$ collisions at the LHC, the ATLAS Collaboration set a limit of 657 GeV using 20.3 fb$^{-1}$ of $pp$ collision data at $\sqrt{s}$ = 8 TeV [24] and the CMS Collaboration in a search using 36 fb$^{-1}$ at $\sqrt{s}$ = 13 TeV [25], excluded masses up to 2 TeV. This result superseded the previous search by CMS using 19.7 fb$^{-1}$ at $\sqrt{s}$ = 8 TeV [26].

## 2 Upgraded ATLAS detector at the HL-LHC

The HL-LHC is currently expected to begin its operations in the second half of 2026, with a nominal levelled instantaneous luminosity of $7.5 \times 10^{34}$ cm$^{-2}$ s$^{-1}$ at $\sqrt{s}$ = 14 TeV. This will lead to an average number of approximately 200 inelastic $pp$ collisions per bunch-crossing (pile-up). This will be significantly higher than the average pile-up of 50 during 2018 data-taking at $2.1 \times 10^{34}$ cm$^{-2}$ s$^{-1}$. This programme aims to provide a total integrated luminosity of 3000 fb$^{-1}$ by 2036. Upgrades of the ATLAS detector[1] will be necessary to maintain its performance in the higher luminosity environment and to mitigate the impact of radiation damage and detector ageing. The inner detector will be completely replaced for the HL-LHC, using an all-silicon design (referred to as “ITk”) with increased granularity, higher read-out bandwidth and reduced material budget [27, 28]. It will be extended to provide tracking in the region $|\eta| < 4$. The performance of the ITk will be as good, and in most cases better, than the existing inner detector in an environment with significantly higher overlapping events. All of the calorimeters except the forward calorimeters will maintain their current performance and they will not be replaced, although the readout electronics will be replaced to enable improved triggering [29, 30]. A new high-granularity timing detector (HGTD) will also be installed in the forward regions to reduce occupancy from $|\eta| < 2.4$ up to $|\eta| < 4.0$ in the high pile-up HL-LHC environment [31]. The muon detector will be upgraded [32] in order to: extend coverage for muon identification to $|\eta| < 4.0$; permit the use of precision tracking for early trigger decisions; reduce the fake trigger rate in the forward region while preserving high efficiency; and increase trigger acceptance to $|\eta| < 2.7$ by eliminating gaps. The trigger and data acquisition systems will be improved to preserve high signal acceptance in the high-rate and high-occupancy HL-LHC environment [33]. The improvements will include: higher bandwidth readout; using high granularity measurements and tracking information earlier in the trigger. The hardware-based first-level trigger accept rate is planned to be 400-1000 kHz, while the software-based high-level trigger accept rate will be 10 kHz, i.e. an increase of about a factor 10 compared to the high-level trigger at the current ATLAS detector. The $b$-jet efficiency and light-flavour-quark rejection of the projected ATLAS detector at the HL-LHC is expected to be similar to that of the current detector while the $c$-jet rejection is expected to be about a factor of two lower than that of the current detector [34].

[1] ATLAS uses a right-handed coordinate system with its origin at the nominal IP in the centre of the detector and the $z$-axis along the beam pipe. The $x$-axis points from the IP to the centre of the LHC ring, and the $y$-axis points upward. Cylindrical coordinates $(r, \phi)$ are used in the transverse plane, $\phi$ being the azimuthal angle around the $z$-axis. The pseudorapidity is defined in terms of the polar angle $\theta$ as $\eta = -\ln\tan(\theta/2)$. The transverse momentum and energy are defined as $p_{\mathrm{T}} = p \sin\theta$ and $E_{\mathrm{T}} = E \sin\theta$, respectively. The $\Delta R$ is the distance defined as $\Delta R = \sqrt{(\Delta\eta)^2 + (\Delta\phi)^2}$.

## 3 Signal and background simulation samples

Samples of events generated using Monte Carlo (MC) simulations were produced using different event generators interfaced to various parton showering (PS) and hadronisation generators. After the event generation step, a fast simulation of the trigger and detector effects was added with the dedicated ATLAS software framework [35]. The trigger, reconstruction and identification efficiencies, the energy and transverse momentum resolution of leptons and jets were computed as a function of their $\eta$ and $p_{\mathrm{T}}$ using full simulation studies assuming an upgrade ATLAS detector [22], and were tabulated in smearing functions which provide parameterised estimates of the ATLAS performance at the HL-LHC. These smearing functions were applied to the truth-level quantities, defined in Section 4. The smearing functions assume the HL-LHC conditions of an instantaneous luminosity of $7.5 \times 10^{34}$ cm$^{-2}$ s$^{-1}$ and the presence of 200 overlapping events in each bunch-crossing [36]. Detailed studies are shown in Refs. [37, 38].

All the signal and background processes involving top quarks were simulated assuming a top-quark mass of $m_t = 172.5$ GeV and a branching ratio (BR) of 100% for the decay $t \to Wb$. All samples are normalised using their corresponding theoretical production cross-sections.

### 3.1 Signal samples

For the matrix-element (ME) calculations, samples of signal events generated using the non-resonant monotop model were produced using the MadGraph5_aMC@NLO (v2.3.3) [39] generator at leading-order (LO) using the NNPDF3.0 LO [40] parton distribution function (PDF) set. The PS, hadronisation and the underlying event (UE) were handled by the Pythia 8 (v8.30) event generator [41] with the A14 [42] set of tuned parameters, using the NNPDF2.3 LO PDF set [43]. The EvtGen (v1.6.0) program [44] was used to describe the properties of the bottom and charmed hadron decays. All these MC simulation samples were generated for a range of the mediator masses between $m_V = 1.0$ and 5.0 TeV, in steps of 0.5 TeV. The benchmark DM particle masses are assumed to be $m_\chi = 1$ GeV (larger masses, up to around 100 GeV, can be considered since kinematic distributions predicted by the model do not change as shown in Ref. [25]). The values of the coupling parameter $a$ was set to 0.5 and $g_\chi$ was set to 1.0.

### 3.2 Background samples

Samples of simulated events for background processes include production of single-top quark, top-quark-antiquark pair ($t\bar{t}$), $W/Z$ boson in association with jets, vector-boson pairs, associated production of a $t\bar{t}$ pair and a $W/Z$ boson and single-top quark in association with a $Z$ boson.

Samples of simulated events for $t\bar{t}$ production and electroweak production of single-top quarks in the $s$-channel, associated $tW$ and $t$-channel were produced using the next-to-leading-order (NLO) Powheg-Box generator [45–47]. In the $t\bar{t}$ event generation the resummation damping factor[2] was set to $1.5 \times m_t$ and in the $t$-channel event generation the four-flavour scheme was used, treating the $b$-quark as massive. For $t\bar{t}$ and $s$-channel the NNPDF3.0 NLO PDF set was used in the ME generation, while NNPDF3.04f NLO PDF set was used for the $t$-channel, and CT10 [48] PDF set for the associated $tW$ process. All these simulation samples except the latter were interfaced to Pythia 8 for the PS, fragmentation and the UE simulation, using the A14 set of tuned parameters and the NNPDF2.3 LO PDF set. The associated $tW$ production sample was interfaced to Pythia 6 [49], using the CT10 PDF set and the corresponding Perugia 2012 tuneable parameters [50].

[2] The resummation damping factor is one of the parameters controlling the ME/PS matching in Powheg and effectively regulates the high-$p_{\mathrm{T}}$ gluon radiation.

The $W$ boson production in association with jets was produced using the MadGraph5_aMC@NLO generator at LO using the NNPDF3.0 NLO PDF set. These $W$+jets event samples were simulated for up to one additional parton at NLO and up to two additional partons at LO. The $Z$ boson production in association with jets ($Z$ + jets) was produced using the Powheg-Box generator at NLO in QCD with the CT10 PDF set and the AZNLO [51] set of tuned parameters of the UE are used. The final-state photon radiation was modelled by the Photos [52] MC simulation. Both productions were interfaced with Pythia 8 generator for the PS, fragmentation and UE, using the CT10 PDF set in the case of $W$+jets and CTEQ6L1 [53] PDF set in the case of $Z$ + jets.

Samples of vector-boson pairs events ($WW$, $ZZ$, $WZ$), containing up to three additional partons where at least one of the bosons decays leptonically, were produced using the Sherpa generator [54] with the NNPDF3.0 NNLO PDF set.

The associated productions of a $t\bar{t}$ pair and either a $W$ or $Z$ boson ($t\bar{t}W$, $t\bar{t}Z$) were generated using MadGraph5_aMC@NLO at NLO using the NNPDF3.0 NLO [40] PDF set. The generated events were then processed with Pythia 8 to perform the fragmentation and hadronisation, and to generate the UE, using the NNPDF2.3 LO PDF set and the A14 set of tuned parameters.

Samples of single-top quark production in association with a $Z$ boson events ($tZq$) were generated at LO in QCD using MadGraph5_aMC@NLO in the four-flavour scheme with the CTEQ6L1 LO PDF set. The $Z$ boson was simulated to be on-shell and off-shell $Z/\gamma^*$ contributions and their interference were not taken into account. The PS, hadronisation and the UE were generated by Pythia 8 with the A14 set of tuned parameters using the NNPDF2.3 LO PDF set.

In all background samples where Pythia 6 or Pythia 8 were used, the EvtGen program was also used to model bottom and charmed hadron decays.

# 4 Object definition

Particle-level definitions are used for electrons, muons, jets and missing transverse momentum, which are the final-state objects used by this analysis. These are constructed from stable particles of the MC event record with a lifetime larger than $0.3 \times 10^{-10}$ s within the observable pseudorapidity range.

Electrons and muons, hereafter referred to as leptons ($\ell$), need to originate from a $W$ boson, including from an intermediate tau decay. Leptons from hadron decays, either directly or via a tau decay, are rejected. Leptons are requested to have $|\eta| < 2.5$. The selected lepton four-momentum is calculated including photons within a cone of size of $\Delta R = 0.1$. In order to simulate the electron/muon track match requirement (i.e. the overlap removal between electrons and muons), events are rejected if a matching in $\phi$ and $\theta$ of 0.005 is found between these two particle-level objects. Identification efficiencies [22] are applied to the

5

lepton candidates to select which particles are identified as leptons. These have their energy, $p_{\mathrm{T}}$ and $\eta$ smeared according to the detector resolution.

Neutrinos are required, similarly to electrons and muons, not to originate from a hadron or quark decay. The missing transverse momentum, with magnitude $E_{\mathrm{T}}^{\mathrm{miss}}$, is calculated from the negative vector sum of true final-state particles within the detector acceptance. The contribution due to pile-up is taken into account before applying detector resolution effects.

Jets are reconstructed using the anti-$k_t$ algorithm [55] implemented in the FastJet [56] library, with a radius parameter of 0.4. All stable final-state particles are used to reconstruct the jets, except the selected neutrinos, leptons and the photons associated with these leptons. This implies that the $b$-jet energy is close to that of the $b$-quark before hadronisation and fragmentation. The $b$-tagging is performed if the jet is within $|\eta| < 2.5$ applying a tagging efficiency, function of the true flavour of the jet, $p_{\mathrm{T}}$ and $\eta$. Since the $b$-tagging is particularly sensitive to the contamination of pile-up tracks, tracks with large impact parameters are considered. Therefore tracks from nearby pile-up are likely to be selected in order to mitigate effects from pile-up. These efficiencies are evaluated considering the latest layout of the ITk detector [28] though not the HGTD, pile-up of 200 and using the MV2 $b$-tagging algorithm [34, 57, 58] at the 70% working point. Double counting of electrons as jets may arise from electron energy deposition in the calorimeter being clustered by the jet algorithm. To mitigate such effect jets are removed if within $\Delta R = 0.2$ from a selected electron. After this step, electrons within $\Delta R = 0.4$ from a jet are rejected, since they are considered as decay products of the hadrons in the jet. For the same reason, muons that are within $\Delta R(\mathrm{muon, jet}) = 0.04 + 10\ \mathrm{GeV}/p_{\mathrm{T}}(\mathrm{muon})$ from a jet are also removed. A fraction of the particle-level jets are removed, according to the expected mis-identification rate shown in Ref. [37]. Energy, $p_{\mathrm{T}}$ and $\eta$ of remaining jets are smeared according to the detector resolution. Pile-up jets are rejected using tracking information.

## 5 Event selection and analysis strategy

The experimental signature of the non-resonant monotop events with $W$ boson decaying leptonically is one lepton from the $W$-boson decay, large $E_{\mathrm{T}}^{\mathrm{miss}}$, and one jet identified as likely to be originated from a $b$-quark. The signal event candidates are selected by requiring exactly one lepton with $p_{\mathrm{T}} > 30$ GeV, exactly one jet with $p_{\mathrm{T}} > 30$ GeV identified as a $b$-jet and $E_{\mathrm{T}}^{\mathrm{miss}} > 100$ GeV. Since the considered monotop process favours final states with positive leptons, events with negative lepton charge are rejected. These criteria defines the base selection.

In order to maximise the sensitivity of the study, in addition to the base selection further discrimination is achieved by applying additional criteria according to the kinematic properties of the signal while rejecting background. Events entering the pre-selection region defined in Section 5.1 are used to train a boosted decision tree (BDT) algorithm. A selection on the BDT output is used to define the BDT-based signal region. A study was performed to optimise a cut-based analysis and signal region, but it was found to be less effective than a multivariate analysis approach. The results of this study are also described in this Section 5.2. To extract exclusion limits the $E_{\mathrm{T}}^{\mathrm{miss}}$ distribution is used as the discriminating variable when executing the statistical analysis.

### 5.1 BDT-based analysis

In addition to the base selection, further discrimination between the monotop signal events and background events is achieved by applying additional criteria. The transverse mass of the lepton–$E_{\mathrm{T}}^{\mathrm{miss}}$ system,

$$m_{\mathrm{T}}(\ell, E_{\mathrm{T}}^{\mathrm{miss}}) = \sqrt{2 p_{\mathrm{T}}(\ell) E_{\mathrm{T}}^{\mathrm{miss}} \left(1 - \cos\Delta\phi(\ell, E_{\mathrm{T}}^{\mathrm{miss}})\right)}, \tag{2}$$

where $p_{\mathrm{T}}(\ell)$ denotes the magnitude of the lepton transverse momentum and $\Delta\phi(\ell, E_{\mathrm{T}}^{\mathrm{miss}})$ is the azimuthal difference between the lepton momentum and the $E_{\mathrm{T}}^{\mathrm{miss}}$ directions, is required to be larger than 100 GeV in order to reduce the background contribution. In background events the spectrum of this quantity decreases rapidly for values higher than the $W$-boson mass. In signal events instead, the spectrum has a tail at higher mass values, as seen in the search performed by ATLAS at $\sqrt{s} = 8$ TeV [24]. When originating from the decay of a top quark, the lepton and the $b$-jet are close to each other. Therefore, events are required to have an azimuthal difference between the lepton momentum and the $b$-jet momentum directions ($|\Delta\phi(\ell, b\text{-jet})|$) of less than 2.0, which disfavours the $W+$jets and diboson backgrounds. Table 1 shows a summary of the previous criteria which defines the pre-selection region. Figure 2 shows the distributions of $|\Delta\phi(\ell, b\text{-jet})|$, the angular distance between the lepton and the $b$-jet ($\Delta R(\ell, b\text{-jet})$), and $m_{\mathrm{T}}(\ell, E_{\mathrm{T}}^{\mathrm{miss}})$.

| Variable | Requirement |
|---|---|
| Multiplicity (leptons) | 1 |
| $p_{\mathrm{T}}(\ell)$ [GeV] | $> 30$ |
| Lepton charge sign | $> 0$ |
| $p_{\mathrm{T}}(b\text{-jet})$ [GeV] | $> 30$ |
| $E_{\mathrm{T}}^{\mathrm{miss}}$ [GeV] | $> 100$ |
| Multiplicity ($b$-jets) | 1 |
| $m_{\mathrm{T}}(\ell, E_{\mathrm{T}}^{\mathrm{miss}})$ [GeV] | $> 100$ |
| $\|\Delta\phi(\ell, b\text{-jet})\|$ | $< 2.0$ |

Table 1: Overview of the pre-selection criteria used to define the pre-selection region.

Further selection is performed via a BDT algorithm provided by the Toolkit for Multivariate Analysis [59]. The BDT is trained to discriminate the monotop signal from the dominant $t\bar{t}$ background. For the training, since no significant difference is observed for the different mass values, the sample with $m_V = 2.5$ TeV is used. Half of the events of both signal and background samples are selected randomly and used to train the BDT. The other half is used to probe the BDT behaviour in order to avoid the presence of overtraining. The variables entering the BDT are selected from a pool of fundamental quantities, like $p_{\mathrm{T}}$ of jets and $b$-jets, and angular distances. The variables selected are the ones showing the best discriminating power. In particular, $|\Delta\phi(\ell, b\text{-jet})|$ and $m_{\mathrm{T}}(\ell, E_{\mathrm{T}}^{\mathrm{miss}})$ are found to be the most effective variables. A full list and description of the variables used in the BDT training is given in Table 2. Figure 3 shows the distribution of the BDT response in the pre-selection region. Only events with BDT response $> 0.9$ and $E_{\mathrm{T}}^{\mathrm{miss}} > 150$ GeV enter in the signal region and are used in the extraction of the result. This value is chosen because it maximises the significance while leaving sufficient statistics for the result to be meaningful.

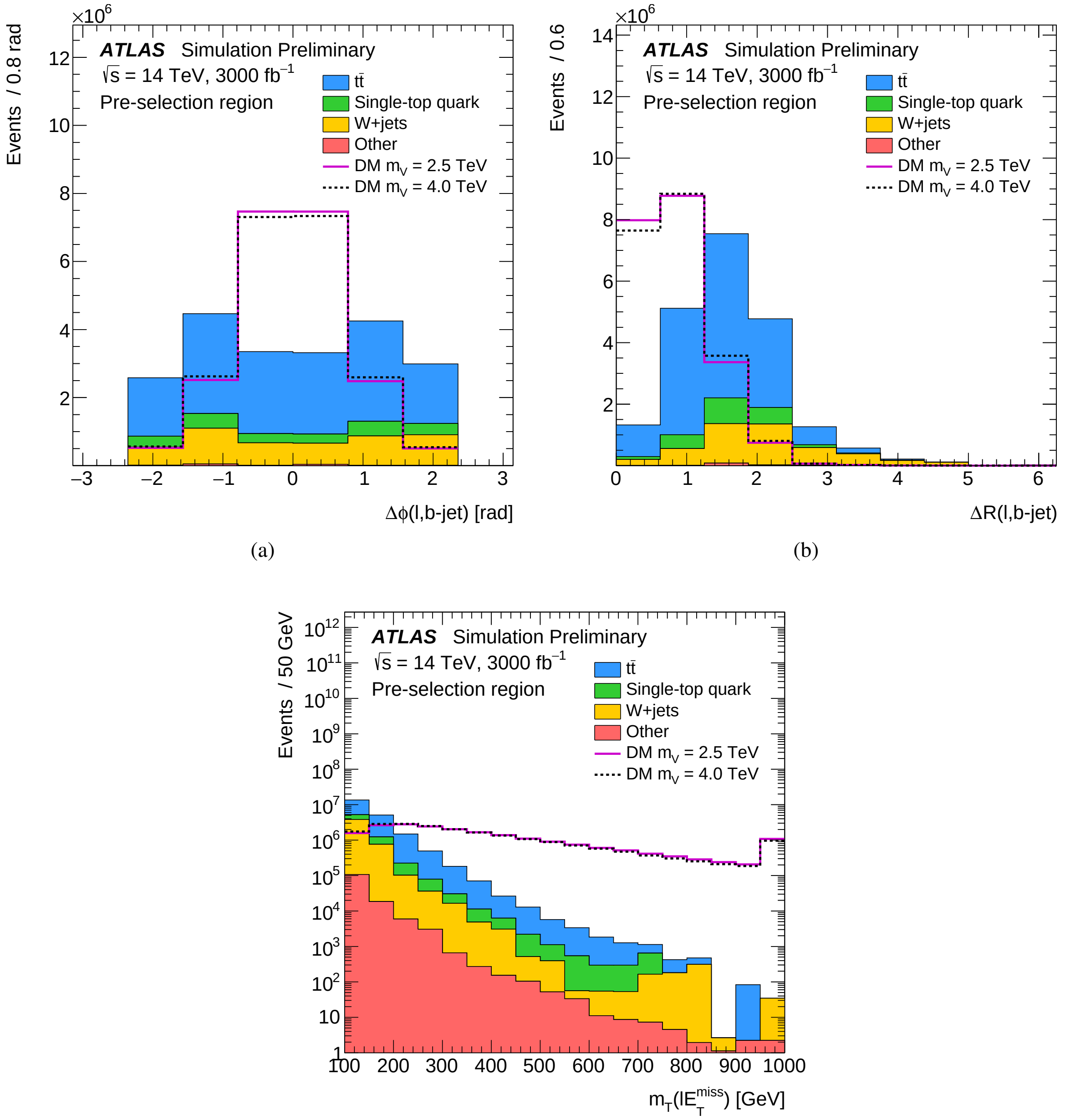

(a) (b) (c)

Figure 2: Distributions of (a) $\Delta\phi$ between the lepton and the $b$-jet, (b) $\Delta R$ between the lepton and the $b$-jet and (c) transverse mass of the lepton–$E_{\mathrm{T}}^{\mathrm{miss}}$ system. The stack distribution shows the background prediction which includes $t\bar{t}$, single-top quark, $W+$jets and Other (i.e. $Z+$jets, dibosons, $t\bar{t}W/Z$ and $tZq$). Solid and dashed lines represent the signal corresponding to a mediator mass of 2.5 and 4.0 TeV, respectively. The background event samples are normalised to their theoretical predictions and the signal event samples are normalised to the number of background events.

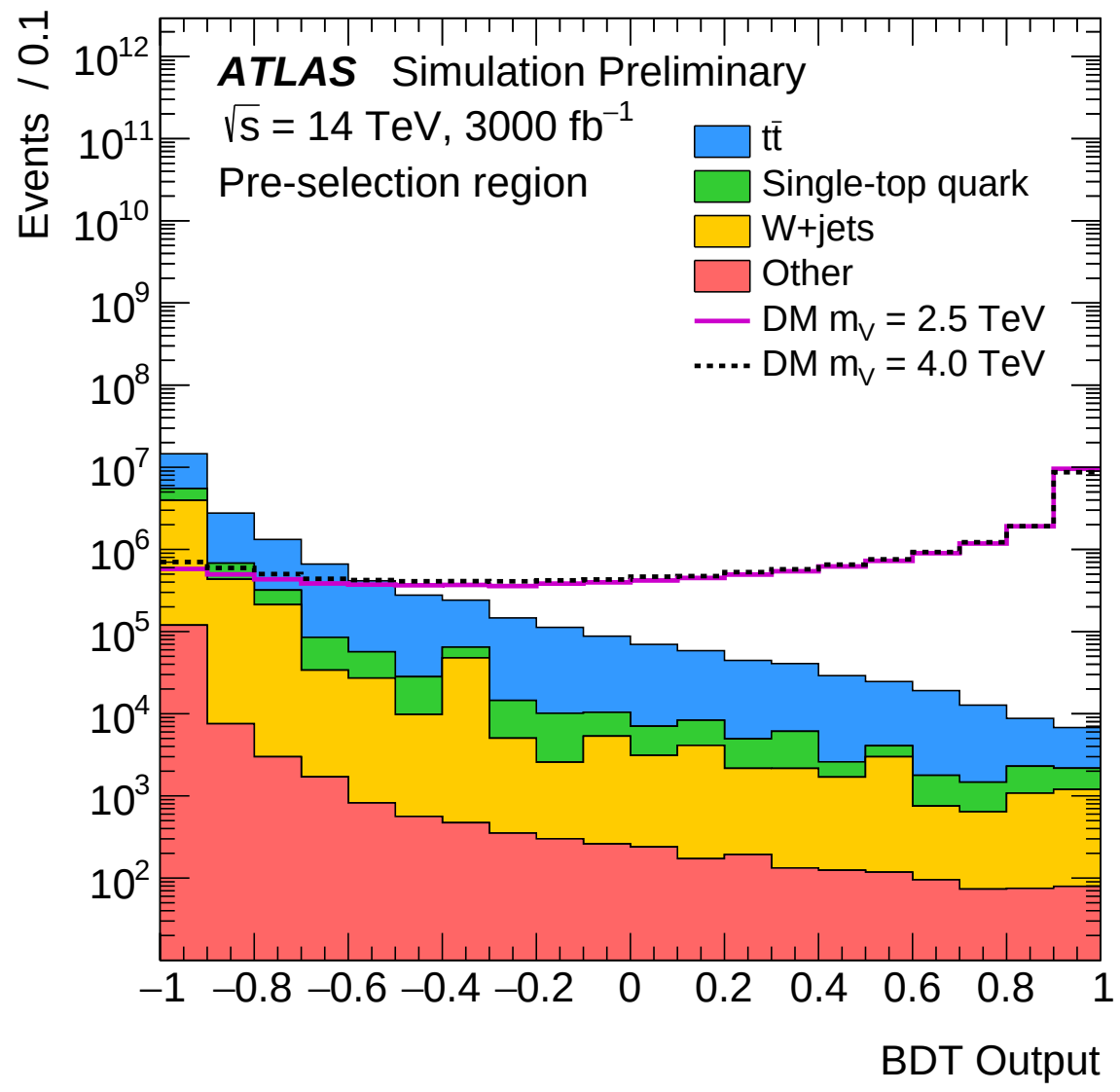

Figure 3: Response of the BDT algorithm for events in the pre-selection region. The stack distribution shows the background prediction which includes $t\bar{t}$, single-top quark, $W$+jets and Other (i.e. $Z$+jets, dibosons, $t\bar{t}W/Z$ and $tZq$). Solid and dashed lines represent the signal corresponding to a mediator mass of 2.5 and 4.0 TeV, respectively. The background event samples are normalised to their theoretical predictions and the signal event samples are normalised to the number of background events.

## 5.2 Cut-based analysis

Events used in this study are selected with the base selection together with additional requirements in three variables properly optimised. The optimisation is performed by varying systematically the thresholds of $|\Delta\phi(\ell, b\text{-jet})|$, $\Delta R(\ell, b\text{-jet})$ and $m_{\mathrm{T}}(\ell, E_{\mathrm{T}}^{\mathrm{miss}})$, and without taking into account systematics uncertainties. The $E_{\mathrm{T}}^{\mathrm{miss}}$ is used as discriminant variable in the likelihood fit. The tested selection on $m_{\mathrm{T}}(\ell, E_{\mathrm{T}}^{\mathrm{miss}})$ ranges between $> 50$ GeV and $> 300$ GeV in steps of 25 GeV. Selections on the angular variables range from $< 0.5$ to $< 2.9$, in steps of 0.2. The figure of merit used in this process is the excluded signal strength obtained from the likelihood fit. The fitting procedure is described in Section 6. The signal with $m_V =$ 2.5 TeV is used for this study. The optimal set of requirements is found to be the base-section criteria with $\Delta R(\ell, b\text{-jet}) < 1.2$ and $m_{\mathrm{T}}(\ell, E_{\mathrm{T}}^{\mathrm{miss}}) > 225$ GeV and with no requirements on $|\Delta\phi(\ell, b\text{-jet})|$. Additionally a cut on $E_{\mathrm{T}}^{\mathrm{miss}} > 150$ GeV is applied to further reduce background. These criteria define the signal region of the cut-based analysis.

Table 3 shows the predicted event yields in the pre-selection region and in the signal regions of the BDT- and cut-based analyses. Comparing the two signal regions, the former analysis has about two order of magnitude larger signal-to-background ratios than the latter analysis. In both analyses the dominant background is the $t\bar{t}$ production. In the BDT-based analysis, the $t\bar{t}$ background represents the 65% of the total background, followed by an important contribution of $W$+jets and single top-quark backgrounds. In the cut-based analysis, the $t\bar{t}$ background represents the 90% of the total background with minor contribution of single top-quark production and negligible contribution of the rest.

| Variable name | Description |
|---|---|
| Kinematic variables | |
| $E_{\mathrm{T}}^{\mathrm{miss}}$ | Magnitude of the missing transverse momentum |
| $p_{\mathrm{T}}(b\text{-jet})$ | Transverse momentum of the $b$-jet |
| $p_{\mathrm{T}}(\text{leading-jet})$ | Transverse momentum of the leading jet |
| Lepton $p_{\mathrm{T}}$ | Transverse momentum of the lepton |
| $m_{\mathrm{T}}(\ell, E_{\mathrm{T}}^{\mathrm{miss}})$ | Transverse mass of lepton–$E_{\mathrm{T}}^{\mathrm{miss}}$ system |
| Azimuthal differences | |
| $\lvert\Delta\phi(\ell, \text{leading-jet})\rvert$ | $\Delta\phi$ between the lepton and the leading jet |
| $\lvert\Delta\phi(\ell, b\text{-jet})\rvert$ | $\Delta\phi$ between the lepton and the $b$-jet |
| $\Delta\phi(\ell, E_{\mathrm{T}}^{\mathrm{miss}})$ | $\Delta\phi$ between the lepton and $E_{\mathrm{T}}^{\mathrm{miss}}$ |
| Angular distance differences | |
| $\Delta R(\ell, \text{leading-jet})$ | $\Delta R$ between the lepton and the leading jet |
| $\Delta R(\ell, b\text{-jet})$ | $\Delta R$ between the lepton and $b$-jet |
| Masses | |
| Leading-jet mass | Mass of the leading jet |

Table 2: List of variables entering the BDT and their definitions.

| Process | Pre-selection region | Signal region (BDT-based) | Signal region (Cut-based) |
|---|---|---|---|
| $m_V$ = 1.0 TeV | 183100 ± 400 | 58900 ± 200 | 100300 ± 300 |
| $m_V$ = 1.5 TeV | 33700 ± 180 | 13000 ± 110 | 19800 ± 140 |
| $m_V$ = 2.0 TeV | 8400 ± 90 | 3530 ± 60 | 5110 ± 70 |
| $m_V$ = 2.5 TeV | 2540 ± 50 | 1100 ± 30 | 1560 ± 40 |
| $m_V$ = 3.0 TeV | 890 ± 30 | 380 ± 19 | 540 ± 20 |
| $m_V$ = 3.5 TeV | 360 ± 19 | 150 ± 12 | 220 ± 15 |
| $m_V$ = 4.0 TeV | 160 ± 13 | 64 ± 8 | 97 ± 10 |
| $m_V$ = 4.5 TeV | 83 ± 9 | 31 ± 6 | 48 ± 7 |
| $m_V$ = 5.0 TeV | 47 ± 7 | 17 ± 4 | 27 ± 5 |
| Single-top quark | 2058000 ± 1400 | 490 ± 20 | 32600 ± 180 |
| $t\bar{t}$ | 14146000 ± 4000 | 2270 ± 50 | 407500 ± 600 |
| $W$+jets | 4617000 ± 2000 | 710 ± 30 | 16900 ± 130 |
| Other | 136000 ± 400 | 57 ± 8 | 1260 ± 40 |
| Total background | 20957000 ± 5000 | 3520 ± 60 | 458300 ± 700 |

Table 3: Predicted pre-fit event yields for the merged electron and muon channels in the pre-selection region and for the signal regions of the BDT- and cut-based analyses. The signal and backgrounds are normalised to their theoretical predictions. The uncertainties shown are statistical only.

## 6 Results

The BDT-based approach is selected given the significantly better results obtained compared to the cut-based analysis. Thus, unless explicitly stated, the content on this section refers to the BDT-analysis.

For the statistical analysis all backgrounds except the $t\bar{t}$ production are merged in a non-$t\bar{t}$ background to avoid problems of poor statistics in the signal region. This allows to use a binned likelihood fit. The shape of the $E_{\mathrm{T}}^{\mathrm{miss}}$ distribution is used in the statistical analysis, as it is expected to be the most sensitive variable to the presence of new physics. The binning of this distribution is optimised for the sensitivity of the analysis in the signal region while ensuring the stability of the fit. This results in a non-equidistant binning which exhibits wider bins in regions with a large signal contribution, while preserving a sufficiently large number of background events in each bin. Figure 4 shows the post-fit $E_{\mathrm{T}}^{\mathrm{miss}}$ distribution in the signal region. The result does not include MC statistical uncertainties but incorporates effects of systematic uncertainties. The theoretical modelling of signal and background has the largest prior, 15%. The second largest source of uncertainty is the one relative to the $E_{\mathrm{T}}^{\mathrm{miss}}$ reconstruction, with 6% prior. Jet energy scale (JES) and jet energy resolution (JER) contribute with a total of 5%. The uncertainty on the requirements for pile-up jets rejection is 5%. The ones on lepton identification and $b$-tagging efficiencies are 1.2% and 2.5%, respectively. The uncertainty on the expected luminosity is also taken into account, with a 1% effect.

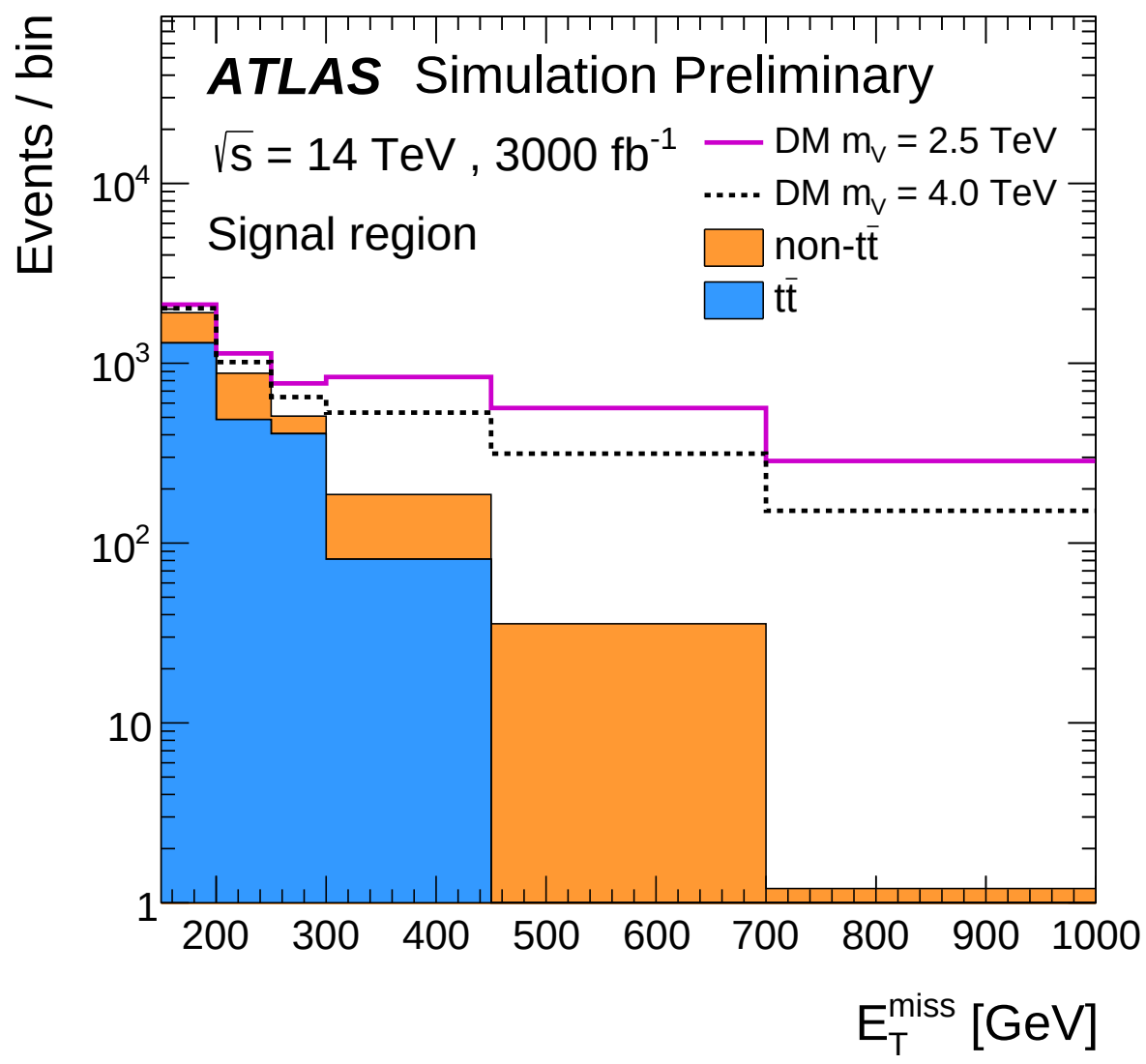

Figure 4: Expected post-fit $E_{\mathrm{T}}^{\mathrm{miss}}$ distribution in the signal region. The stack distribution shows the $t\bar{t}$ and non-$t\bar{t}$ background predictions. Solid and dashed lines represent the signal corresponding to a mediator mass of 2.5 and 4.0 TeV, respectively. The signal event samples are normalised to the number of background events. The binning is the same as the optimised, non-equidistant binning used in the fit. Last bin includes overflow events.

Hypothesis testing is performed using a frequentist approach which uses the asymptotic approximation described in Ref. [60]. Figure 5 shows the expected 95% confidence level (CL) upper limits as a function of the mediator mass for the non-resonant model assuming $m_\chi$ = 1 GeV, $a$ = 0.5 and $g_\chi$ = 1. After the fit, the largest impact on the result is coming from the uncertainty on the $E_{\mathrm{T}}^{\mathrm{miss}}$ reconstruction. This is expected since the $E_{\mathrm{T}}^{\mathrm{miss}}$ is the final discriminant in the analysis. The second largest contribution is coming from background and signal modelling. The other contributions are, in order of importance: pile-up jet rejection requirements, JES and JER, lepton reconstruction efficiency and $b$-tagging efficiency. The uncertainty on the expected luminosity is found to have the smallest effect. The expected mass limit at 95% CL is 4.6 TeV while the discovery reach (based on $5\sigma$ significance) is 4.0 TeV. For the current analysis the effect of possible improvements in the systematic uncertainties is estimated by reducing by half the uncertainties. This has the effect of increasing the exclusion limit (discovery reach) by 80 (50) GeV.

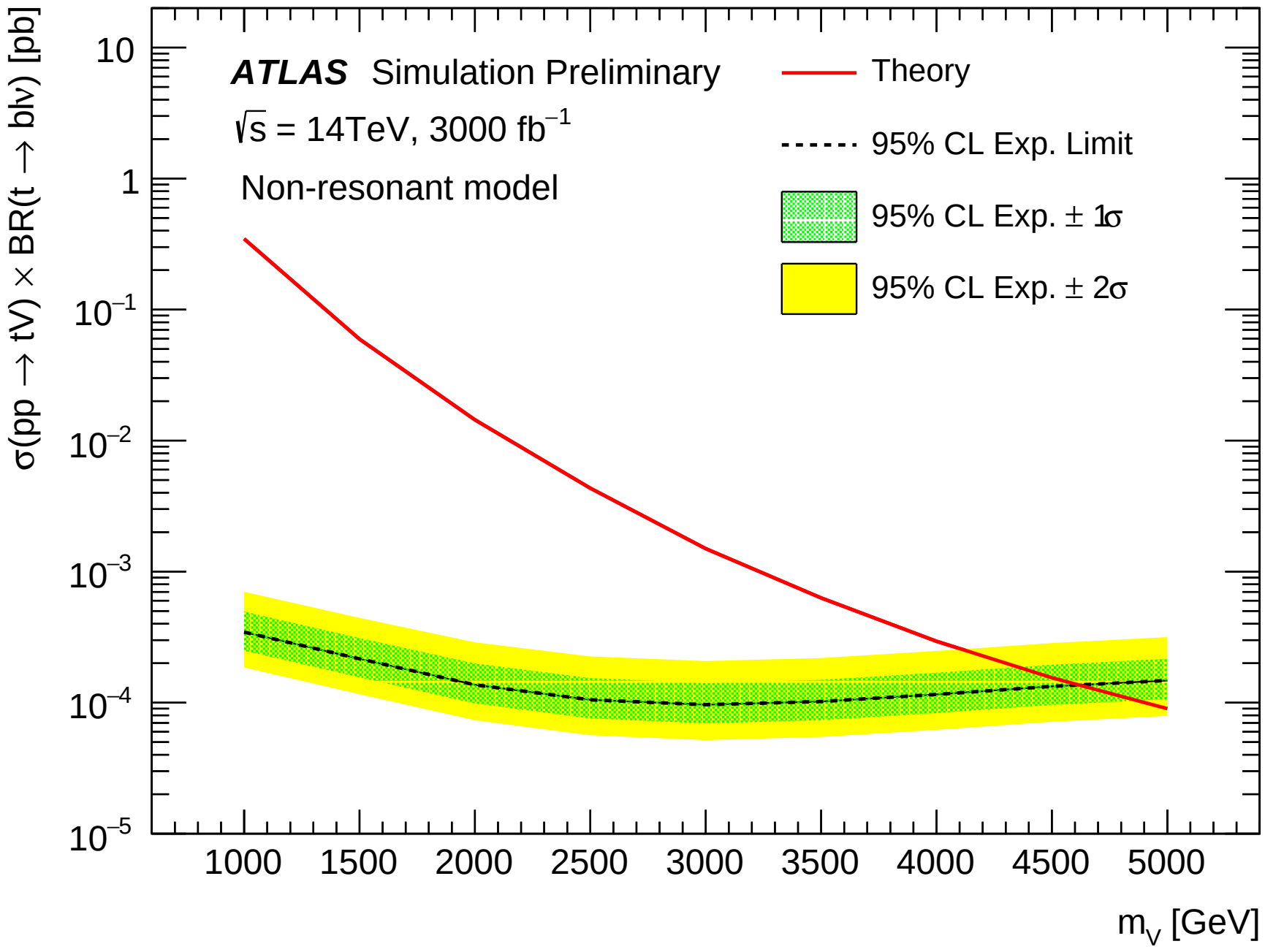

Figure 5: Expected 95% CLs upper limits on the signal cross-section as a function of the mass of the mediator for the non-resonant model assuming $m_\chi$ = 1 GeV, $a$ = 0.5 and $g_\chi$ = 1 using a BDT analysis. The MC statistical uncertainty is not considered but the full set of systematics, extrapolated from the 13 TeVanalysis is considered.

The expectations for the equivalent of Run-3 integrated luminosity (300 fb$^{-1}$) is checked, obtaining an exclusion limit (discovery reach) of 3.7 TeV (3.2 TeV).

The expected mass limit at 95% CL obtained with the cut-based analysis, assuming an integrated luminosity of 3000 fb$^{-1}$ and including same systematic uncertainties, is 3.2 TeV. As anticipated at the beginning of the section, this limit is significantly lower than what is obtained with the BDT-based analysis.

## 7 Conclusion

The expected sensitivity of a search for events with one top quark and large missing transverse momentum is estimated in $pp$ collisions at a centre-of-mass energy of $\sqrt{s}$ = 14 TeV with the ATLAS detector at the HL-LHC. A non-resonant production of an exotic state $V$, decaying to a pair of invisible dark-matter particles $\chi\bar{\chi}$, in association with a right-handed top quark is considered. Only the topologies where the $W$ boson from the top quark decays into an electron or a muon and a neutrino are considered. The number of signal and background events are estimated from simulated truth particle-level information after applying smearing functions to mimic an upgraded ATLAS detector response in the HL-LHC environment. The expected exclusion limit at 95% CL on the mass of the exotic state $V$ is, in the absence of MC statistical uncertainty but considering systematic uncertainties, 4.6 TeV using a multivariate analysis based on a BDT and assuming an integrated luminosity of 3000 $fb^{-1}$. The discovery reach obtained is 4.0 TeV. If improvements in systematics would be translated in to a reduction of the uncertainties by a factor 2, the expected exclusion (discovery) would increase by 80 (50) GeV. Expected exclusion for Run-3 equivalent integrated luminosity (300 $fb^{-1}$) including systematics is 3.7 TeV, while the discovery reach is 3.2 TeV.

# ATLAS sensitivity to dark matter produced in association with heavy quarks at the HL-LHC

The ATLAS Collaboration

This note presents the prospects of a search for weakly interacting dark matter produced in association with heavy flavour quarks at the HL–LHC. The search is performed assuming 3000 $\text{fb}^{-1}$ of proton-proton collisions collected by the ATLAS detector at a centre of mass energy of 14 TeV. Two experimental signatures are investigated, characterised by missing transverse momentum and either a pair of bottom quarks or two opposite-charge leptons (electrons or muons) resulting from the decay of a top quark pair or a top quark and a $W$-boson. The results are interpreted within the framework of Simplified Models which couple the dark and Standard Model sectors via the exchange of colour-neutral spin-0 mediators, assuming unitary couplings and a dark matter mass of 1 GeV. Compared to a previous search conducted with 36.1 $\text{fb}^{-1}$ of data at $\sqrt{s}$ = 13 TeV, the reach achievable for dark matter detection in events with bottom quarks is extended by a factor of 3–8.7 following the increased luminosity and centre of mass energy expected for the HL–LHC final dataset, along with the upgrades to the ATLAS detector. For events with top quarks in the final state, the expected sensitivity to scalar mediator production extents by a factor of 5, and exclusion of pseudoscalar mediator masses up to 385 GeV becomes possible.

# 1 Introduction

While the existence of dark matter (DM) is supported by a plethora of astrophysical observations [1–4], its particle nature remains largely unexplained. The Weakly Interacting Massive Particle (WIMP) [5] is a well-motivated candidate for the bulk of dark matter, possessing the requisite properties and appearing in many Beyond the Standard Model (SM) theories. WIMPs created at colliders escape detection, resulting in a signature characterised by missing transverse momentum. Searches for WIMP dark matter, observable by the presence of an accompanying SM particle(s), have been performed extensively at the Large Hadron Collider (LHC) [6–12]. This note presents the prospects of a search for dark matter produced in association with heavy flavour (bottom or top) quarks at the High Luminosity LHC (HL–LHC).

Signatures involving heavy flavour quarks are expected to be the most sensitive to models where the dark and SM sectors couple via the exchange of a spin-0 mediator [13]. This study therefore focuses on two simplified models, defined by either a scalar, $\phi$, or pseudoscalar, $a$, mediator [14–16]. In both cases, the mediating particle is taken to be colour-neutral and the dark matter candidate is assumed to be a weakly interacting Dirac fermion, $\chi$, uncharged under the SM. The models are described by five common parameters: the dark matter mass, $m(\chi)$, the mediator mass, $m(\phi)$ or $m(a)$, the dark matter–mediator coupling, $g_\chi$, the flavour-universal SM–mediator coupling, $g_v$, and the decay width of the mediator, $\Gamma(\phi)$ or $\Gamma(a)$. For simplicity, an assumption of $g_\chi = g_v = g$ is made and the mediator width is taken to be the minimal width described in Ref. [15]. For this scenario, $\chi\bar{\chi}$ production in association with top-quarks is expected to dominate at the HL–LHC. Two signatures featuring top quarks in the final state are therefore considered. The first signature, denoted DM+$t\bar{t}$, is characterised by two tops decaying di-leptonically as shown in Figure 1(a). The second signature, DM+$Wt$, involves a single top produced in tandem with a $W$-boson, both of which decay leptonically as shown in Figures 1(b) and 1(c).

While the couplings of the mediator to the up- and down-type quarks are assumed to be indistinguishable in this study, the condition is by no means a necessary one. In the event that coupling to up-type quarks is suppressed - a possibility in UV completions of the aforementioned models - production of dark matter in association with bottom quarks becomes relevant. The DM+$b\bar{b}$ final state is also well motivated as an avenue for probing the parameter space of two-Higgs doublet models (2HDM). In the context of the 2HDM+$a$ model [17, 18] for example, the rate for $pp \to b\bar{b} + a$ is enhanced by the ratio of the Higgs doublet vacuum expectation values, $\tan\beta$, if a Yukawa sector of type-II is realised. Constraints on $\tan\beta$ can be extracted via a straightforward recasting of exclusion limits on the simplified pseudoscalar mediator model (see Appendix A in Ref. [18]). Consequently, a search for DM+$b\bar{b}$ targeting the latter model is optimised to also set bounds on $\tan\beta$. DM+$t\bar{t}/Wt$, mono-$X$, and di-top searches can be exploited in a similar manner, however the DM+$b\bar{b}$ channel is uniquely situated to probe the high $\tan\beta$ region, a region not currently well constrained in two-Higgs doublet models. Dark matter production in association with $b$-quarks is therefore also considered in this note, an example diagram for which is shown in Figure 1(d) where the model parameters are as defined in the previous paragraph.

A search targeting the DM+$b\bar{b}$ and DM+$t\bar{t}$ signatures was performed at the LHC using 36.1 fb$^{-1}$ of data collected in 2015 and 2016 at a centre of mass energy of 13 TeV [13]. No evidence of physics beyond the SM was found and constraints were placed on the ratio of the measurable cross-section to the theoretically predicted cross-section, $\sigma/\sigma(g = 1.0)$, as a function of the mediator mass in the range 10-500 GeV. Likewise, constraints on spin-0 mediator production with a DM+$Wt$ signature were projected for 35 fb$^{-1}$ and 300 fb$^{-1}$ of data in Ref. [19]. This note presents the prospects for further constraining these models with HL-LHC data and is divided into two independent analyses. The first is optimised for the DM+$b\bar{b}$ final state and serves to additionally quantify the expected gain in performance potential for HL–LHC searches involving flavour tagged jets and large missing transverse momentum. Similarly, the second analysis, which is optimised for the DM+$t\bar{t}$ and DM+$Wt$ final states, also serves to showcase the gain for searches featuring a combination of flavour tagged objects, missing transverse momentum, and leptons.

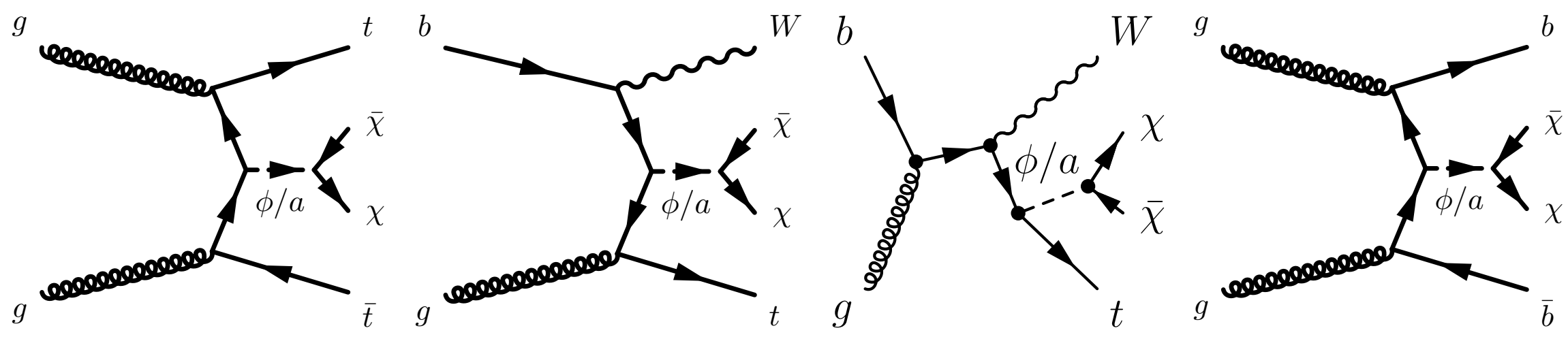

Figure 1: Representative tree-level diagrams for the production of dark matter ($\chi$) in association with (a-c) top quarks and (d) bottom quarks following the exchange of either a colour-neutral scalar ($\phi$) or pseudoscalar ($a$) particle.

## 2 The LHC and HL–LHC

In the present data-taking period, the LHC delivered ~150 fb$^{-1}$ of proton-proton collisions with a peak instantaneous luminosity of 2×10$^{34}$ cm$^{-2}$s$^{-1}$ and an average number of collisions per bunch crossing of $\langle\mu\rangle \sim 35$. A long shutdown (LS2) will follow, during which the injection chain is foreseen to be modified to allow for instantaneous luminosities up to ~2.5×10$^{34}$ cm$^{-2}$s$^{-1}$. The data collected up to the next long shutdown (LS3) will amount to ~300 fb$^{-1}$. An increase of the centre-of-mass-energy to 14 TeV is possible and is assumed to happen for this study. An upgrade of the accelerator to the HL–LHC is planned to take place during LS3, enabling luminosities of ~7×10$^{34}$ cm$^{-2}$s$^{-1}$ to be achieved. The HL–LHC is expected to deliver an average number of pile up interactions per bunch crossing of $\langle\mu\rangle \sim 200$ during its operation with the total data collected amounting to ~3000 fb$^{-1}$.

## 3 The ATLAS Detector

The ATLAS experiment [20] is a multi-purpose particle detector with a forward-backward symmetric cylindrical geometry and nearly 4$\pi$ coverage in solid angle[1]. The interaction point is surrounded by an inner detector (ID), a calorimeter system, and a muon spectrometer.

Upgrades to the detector and the triggering system are planned to adapt the experiment to the increasing instantaneous and integrated luminosities expected with the HL–LHC [21–27].

[1] ATLAS uses a right-handed coordinate system with its origin at the nominal interaction point (IP) in the centre of the detector and the $z$-axis along the beam pipe. The $x$-axis points from the IP to the centre of the LHC ring, and the $y$-axis points upward. Cylindrical coordinates ($r$, $\phi$) are used in the transverse plane, $\phi$ being the azimuthal angle around the beam pipe. The pseudorapidity is defined in terms of the polar angle $\theta$ as $\eta = -\ln\tan(\theta/2)$. Rapidity is defined as $y = 0.5\ln[(E + p_z)/(E - p_z)]$ where $E$ denotes the energy and $p_z$ is the component of the momentum along the beam direction.

In the reference upgrade scenario, the ID will provide precision tracking of charged particles for pseudorapidities $|\eta| < 4.0$ and will be surrounded by a superconducting solenoid providing a 2 T axial magnetic field. It will consist of silicon pixel and microstrip detectors.

In the pseudorapidity region $|\eta| < 3.2$, the currently installed high-granularity lead/liquid-argon (LAr) electromagnetic (EM) sampling calorimeters will be used. The current steel/scintillator tile calorimeter will be used, although the readout electronics will be replaced to enable improved triggering [23, 24]. A new high-granularity timing detector (HGTD) will also be installed in the forward regions to reduce occupancy from $|\eta| < 2.4$ up to $|\eta| < 4.0$ in the high pile-up HL-LHC environment [27].

The muon spectrometer, consisting of three large superconducting toroids with eight coils each, and a system of trigger and precision-tracking chambers, which provide triggering and tracking capabilities in the ranges $|\eta| < 2.4$ and $|\eta| < 2.7$ respectively, could be upgraded with the addition of a very forward muon tagger that would extend the trigger coverage up to $|\eta| = 4.0$ [22].

A two-level trigger system will be used to select events, reducing the event rate to about 10 kHz. In the reference scenario, the bandwidth allocated to di-lepton ($ee$, $\mu\mu$, $e\mu$) triggers is expected to be 0.2 kHz per trigger, where an offline selection of $p_{\mathrm{T}} > 10$ GeV for each lepton ensures full efficiency. For the missing transverse energy ($E_{\mathrm{T}}^{\mathrm{miss}}$) trigger, the bandwidth allocated is $\sim 0.4$ kHz, with $> 210$ GeV representing the offline $E_{\mathrm{T}}^{\mathrm{miss}}$ above which a typical analysis would use the data according to the Technical Design Report for the Phase-II upgrade of the ATLAS TDAQ system [25].

## 4 Monte Carlo Samples

Monte Carlo (MC) simulated event samples are used to predict the background from SM processes and to model the dark matter signal. The most relevant MC samples have equivalent luminosities (at 14 TeV) of at least 3000 fb$^{-1}$. The technical implementation of these samples is summarised in Table 1, including the packages used to perform matrix element generation and parton showering. The order at which the cross-section is computed for a given process is also shown, along with the specific choice of Parton Distribution Function (PDF) and tune. The dark-matter $t\bar{t}$ and $b\bar{b}$ signal samples are generated following the prescriptions in Ref. [15] and the $Wt$ signal samples ($Wt + \phi/a$) are generated following Ref [19]. For the $Wt + \phi/a$ model the production cross-section is computed at leading-order (LO) accuracy in the strong coupling constant $\alpha_{\mathrm{S}}$. For the $t\bar{t} + \phi/a$ and $b\bar{b} + \phi/a$ models the production cross-section is computed at next-to-LO (NLO) accuracy. Lastly, the $Z/\gamma^{*}$+jets and $W$+jets samples are generated with $\sqrt{s} = 13$ TeV. A collision energy of 14 TeV is replicated by applying an event weight based on the momentum fraction carried by the colliding partons and the ratio of PDF distributions for the different beam energies.

To emulate the Phase-II run conditions and detector response, the signal and SM background samples are smeared using performance functions derived from MC events passed through a full Geant 4 simulation of the upgraded ATLAS detector [28–30]. Specifically, smearing is applied to the resolution and reconstruction efficiencies of the physics objects discussed in Section 5 using parameterisations made with $\langle\mu\rangle = 200$. The contribution from pileup is emulated by overlaying jets from a dedicated library.

4

Table 1: Summary of the simulated signal and SM background event samples used in this analysis, including the event generator, parton shower package, cross-section normalisation, PDF set, and underlying event parameter tune.

| Physics process | Generator | Parton shower | Cross-section normalisation | PDF set | Tune |
|---|---|---|---|---|---|
| $t\bar{t}+\phi/a$ Signal<br>$b\bar{b}+\phi/a$ Signal | aMC@NLO 2.3.3 | Pythia 8.212 | NLO | NNPDF30NLO | A14 [31] |
| $Wt+\phi/a$ Signal | aMC@NLO 2.4.3 | Pythia 8.212 | LO | NNPDF23LO | A14 |
| $t\bar{t}$ | Powheg-Box v2 [32–34] | Pythia 8.186 [35] | NNLO | NNLO CT10 [36] | A14 |
| Single-top ($t$-channel) | powheg-box v1 | Pythia 6.428 | NNLO+NNLL [37] | NLO CT10f4 | Perugia2012 |
| Single-top ($s$- and $Wt$-channel) | powheg-box v2 | Pythia 6.428 | NNLO+NNLL [38, 39] | NLO CT10 | Perugia2012 |
| $t\bar{t}W/Z/\gamma^*$ | aMC@NLO 2.2.2 | Pythia 8.186 | NLO [40] | NNPDF2.3LO | A14 |
| Diboson | Sherpa 2.2.1 [41] | Sherpa 2.2.1 | Generator NLO | CT10 [36] | Sherpa default |
| $t\bar{t}h$ | aMC@NLO 2.2.2 | Herwig 2.7.1 [42] | NLO [43] | CTEQ6L1 [44] | A14 |
| $Wh$, $Zh$ | aMC@NLO 2.2.2 | Pythia 8.186 | NLO [43] | NNPDF2.3LO | A14 |
| $t\bar{t}WW$, $t\bar{t}t\bar{t}$ | aMC@NLO 2.2.2 | Pythia 8.186 | NLO [40] | NNPDF2.3LO | A14 |
| $tZ$, $tWZ$, $t\bar{t}t$ | aMC@NLO 2.2.2 | Pythia 8.186 | LO | NNPDF2.3LO | A14 |
| Triboson | Sherpa 2.2.1 | Sherpa 2.2.1 | Generator LO, NLO | CT10 | Sherpa default |
| $Z/\gamma^*$+jets | Sherpa 2.2.1 [45] | Sherpa 2.2.1 [45] | NNLO [46] | NLO CT10 [36] | Sherpa default |
| $W$+jets | Sherpa 2.2.1 [45] | Sherpa 2.2.1 [45] | NNLO [46] | NLO CT10 [36] | Sherpa default |

# 5 Final State Object Selections

Selecting events consistent with the production of dark matter in a final state with either bottom or top quarks requires the reconstruction of jets, muons, electrons, and missing transverse momentum, $\vec{p}_{\mathrm{T}}^{\mathrm{miss}}$, where $E_{\mathrm{T}}^{\mathrm{miss}} = |\vec{p}_{\mathrm{T}}^{\mathrm{miss}}|$. This section describes the object definitions and kinematic variables used to discriminate signal from SM background processes in the two search channels.

In the previous search, $\vec{p}_{\mathrm{T}}^{\mathrm{miss}}$ is calculated as the negative vector sum of the transverse momenta of all identified physics objects [13]. A soft term constructed from all tracks unmatched to any physics object and originating from the primary vertex is also added. For the studies performed in this note however, the $\vec{p}_{\mathrm{T}}^{\mathrm{miss}}$ is computed at generator-level as the vectorial sum of the momenta of all neutral weakly-interacting particles in an event, including neutrinos and the dark matter candidate. This quantity is then smeared based on the $\vec{p}_{\mathrm{T}}^{\mathrm{miss}}$ resolution associated with the smeared sum of energies of interacting particles in the event.

Jet candidates are reconstructed using the anti-$k_t$ jet clustering algorithm with a radius parameter R = 0.4 [47] and are required to have transverse momentum $p_{\mathrm{T}} > 20$ GeV and $|\eta| < 3.8$. Tracking confirmation is applied to all jets to reduce the contribution from particle decays originating from pile-up interactions [48].

Decays from $b$-quarks are identified ($b$-tagged) using parametrisations that model the performance of the Run-2 multivariate $b$-tagging algorithm MV2c10 [49–51] as a function of jet $p_{\mathrm{T}}$ and $\eta$. Candidate $b$-jets must pass an identification requirement corresponding to an efficiency of 70% for jets containing $b$-hadrons in simulated $t\bar{t}$ events. This requirement represents the tightest set of restrictions on $b$-jets for which flavour-tagging performance functions are available. The corresponding rejection factor for jets originating from the fragmentation of a c (light) quark is ~20 (750) [26].

Baseline electron candidates with $p_{\mathrm{T}} > 7$ GeV are reconstructed in the region $|\eta| < 4.0$ and required to pass the "loose" likelihood-based identification requirements [52, 53]. Similarly, muon candidates with $p_{\mathrm{T}} > 6$ GeV and $|\eta| < 2.7$ are required to pass the "medium" identification criteria [54, 55]. Signal leptons in the DM+$t\bar{t}/Wt$ channels are further required to have $p_{\mathrm{T}} > 20$ GeV, to ensure constant trigger efficiencies in the relevant phase, and $|\eta| < 2.47$ (2.5) for electrons (muons). The reduced pseudorapidity range compared with the DM+$b\bar{b}$ channel is motivated by the topologies of DM+$t\bar{t}/Wt$ events, which are characterised by central leptons.

To resolve reconstruction ambiguities, an overlap removal algorithm is applied to baseline leptons and jets. Where a baseline electron is found to lie within $\Delta R = \sqrt{\Delta\phi^2 + \Delta\eta^2} = 0.2$ of a candidate jet, the jet is removed if it fails to pass $b$-jet identification criteria corresponding to an efficiency of 85%. The same is applied to jets in the DM+$b\bar{b}$ (DM+$t\bar{t}/Wt$) channel which lie within $\Delta R = 0.2$ (0.4) of a selected muon and which are not true $b$-jets. To avoid rejecting events featuring leptonic $c$- or $b$-hadron decays, electrons (muons) are discarded if they are found within a cone of $\Delta R = 0.4$ ($\Delta R = \max(0.4, 0.04 + (10\ \mathrm{GeV})/\mathrm{p}_T^{\mu})$) of any surviving jet.

# 6 Signatures with $b$-quarks and $E_{\mathrm{T}}^{\mathrm{miss}}$

To isolate the event topology of the DM+$b\bar{b}$ final state, events are required to have at least two $b$-tagged jets. The contribution from SM background processes is suppressed via the application selection criteria based on that of the 13 TeV analysis and updated to align with HL–LHC design considerations.

To reduce the contribution from leptonic and semi-leptonic $t\bar{t}$ decays and from leptonic decays of W and Z bosons, events containing at least one baseline lepton ($N_l^B$) are vetoed. A further requirement of no more than 2 or 3 jets is imposed in order to control the large background from hadronic $t\bar{t}$ decays which are typically characterised by high jet multiplicities. A minimum requirement on the azimuthal separation between each jet, $j$, and the missing transverse momentum, $\Delta\phi(j, \vec{p}_{\mathrm{T}}^{\mathrm{miss}}) > 0.4$, is also imposed in accordance with the treatment used at 13 TeV to suppress fake $E_{\mathrm{T}}^{\mathrm{miss}}$ in multi-jet events.

To reduce the contribution from the dominant Z+jets background, several variables exploiting the difference in spin between the scalar and pseudoscalar particles and the Z boson are defined. These variables make use of the pseudorapidity and azimuthal separations between jets, $b$-jets and the missing transverse momentum and include:

- The azimuthal correlation variables:

$$\delta^- = \Delta\phi(j, \vec{p}_{\mathrm{T}}^{\mathrm{miss}}) - \Delta\phi(b, b)$$
$$\delta^+ = |\Delta\phi(j, \vec{p}_{\mathrm{T}}^{\mathrm{miss}}) + \Delta\phi(b, b) - \pi|$$

  where $\Delta\phi(j, \vec{p}_{\mathrm{T}}^{\mathrm{miss}}) = \phi(j) - \phi(\vec{p}_{\mathrm{T}}^{\mathrm{miss}})$ is the azimuthal separation between any jet in an event and $\vec{p}_{\mathrm{T}}^{\mathrm{miss}}$, and $\Delta\phi(b, b) = \phi(b_1) - \phi(b_2)$ is the azimuthal separation between the leading $b$-jet ($b_1$) and sub-leading $b$-jet ($b_2$).

6

- The momentum imbalance between the leading and sub-leading $b$-jets:

$$\mathrm{Imb}(b,b) = \frac{p_T(b_1) - p_T(b_2)}{p_T(b_1) + p_T(b_2)}$$

- The cosine of $\pi - \Delta\phi(b,b)$:

$$\cos(\pi - \Delta\phi(b,b))$$

- The hyperbolic tangent of the pseudorapidity separation between the leading and sub-leading $b$-jet, $\Delta\eta(b,b) = \eta(b_1) - \eta(b_2)$:

$$\cos\theta^*_{bb} = \left|\tanh\left(\frac{|\Delta\eta(b,b)|}{2}\right)\right|$$

$b$-jets produced in association with vector particles are expected to yield a reasonably flat $\cos\theta^*_{bb}$ distribution. For $b$-jets accompanying the production of a heavy scalar or pseudoscalar mediator however, $\cos\theta^*_{bb}$ is expected to peak around 1. The distribution of this variable, along with those of other key discriminants, is shown in Figure 2.

To reduce the contribution from processes where spin anti-correlations are not present or easily exploited, events are required to pass a cut on $H_{\mathrm{T}}^{\mathrm{ratio}}$, the ratio of the leading jet transverse momentum, $p_T(j_1)$, to the scalar sum of the transverse momenta of all jets in the event, $H_{\mathrm{T}}$. Also used is the hyperbolic tangent of the $\eta$ separation between the leading ($j_1$) and third-leading jet ($j_3$):

$$\cos\theta^*_{j_1 j_3} = \left|\tanh\left(\frac{\Delta\eta(j_1, j_3)}{2}\right)\right|$$

In signal events with 3 jets the first and third jet are largely produced back-to-back, leading to a peak at approximately 1 in the $\cos\theta^*_{j_1 j_3}$ distribution. In contrast, $j_1$ and $j_3$ in events from SM background processes – in particular, from $t\bar{t}$ decays – exhibit strong collinearity, leading to dominance in the region below 0.5 as shown in Figure 2(b). Note that a cut on $\cos\theta^*_{j_1 j_3}$ is only applied to events with 3 jets.

As seen from Figure 2(c), the shape of the $\cos\theta^*_{bb}$ distribution for the scalar and pseudoscalar signals can depend strongly on the mass of the mediating particle. Consequently, separate selections are derived for $m(\phi/a) < 100$ GeV and $m(\phi/a) \geq 100$ GeV. The resulting signal regions, denoted by $\mathrm{SR}_{b,\mathrm{low}}$ and $\mathrm{SR}_{b,\mathrm{high}}$ respectively, are defined in Table 2.

The $\cos\theta^*_{bb}$ variable provides the best discrimination between signal and background events. As such, $\mathrm{SR}_{b,\mathrm{low}}$ and $\mathrm{SR}_{b,\mathrm{high}}$ are divided into four equal-width exclusive bins in $\cos\theta^*_{bb}$, reflecting the configuration used in Run 2. The bins are denoted by the labels $\mathrm{SR}_{b,X}$-bin1 through $\mathrm{SR}_{b,X}$-bin4 where $X = \{\mathrm{low}, \mathrm{high}\}$.

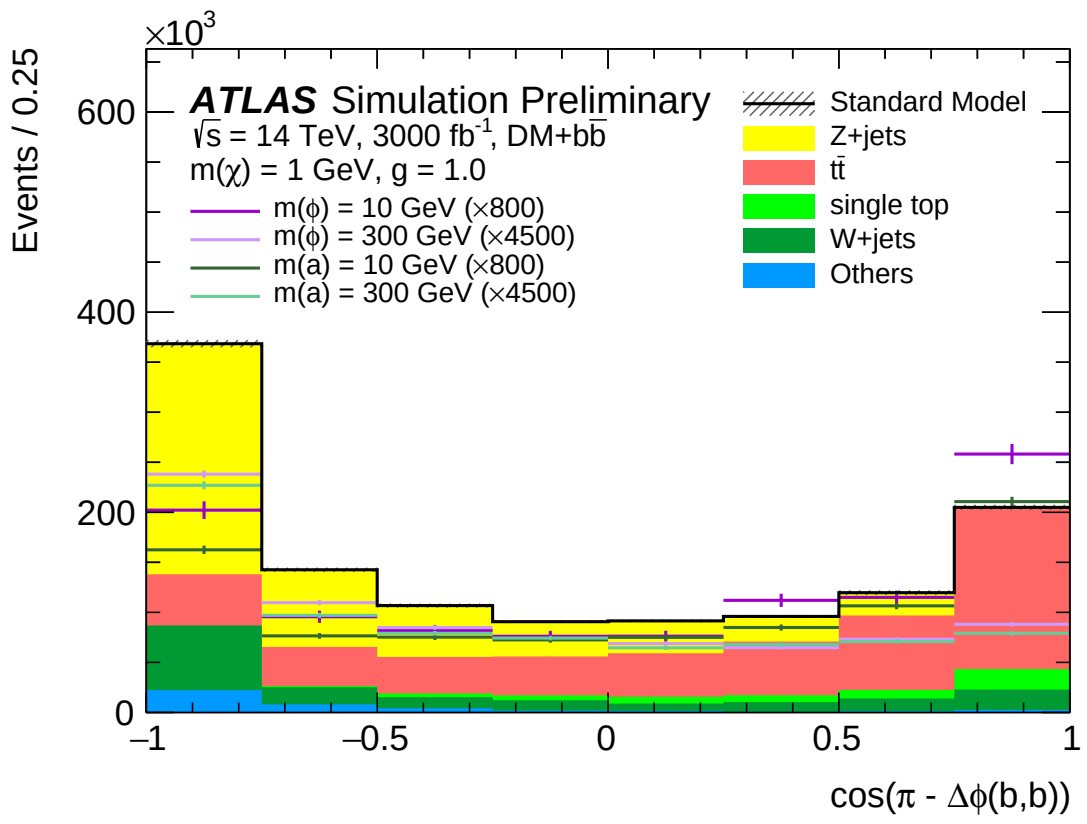

(a)

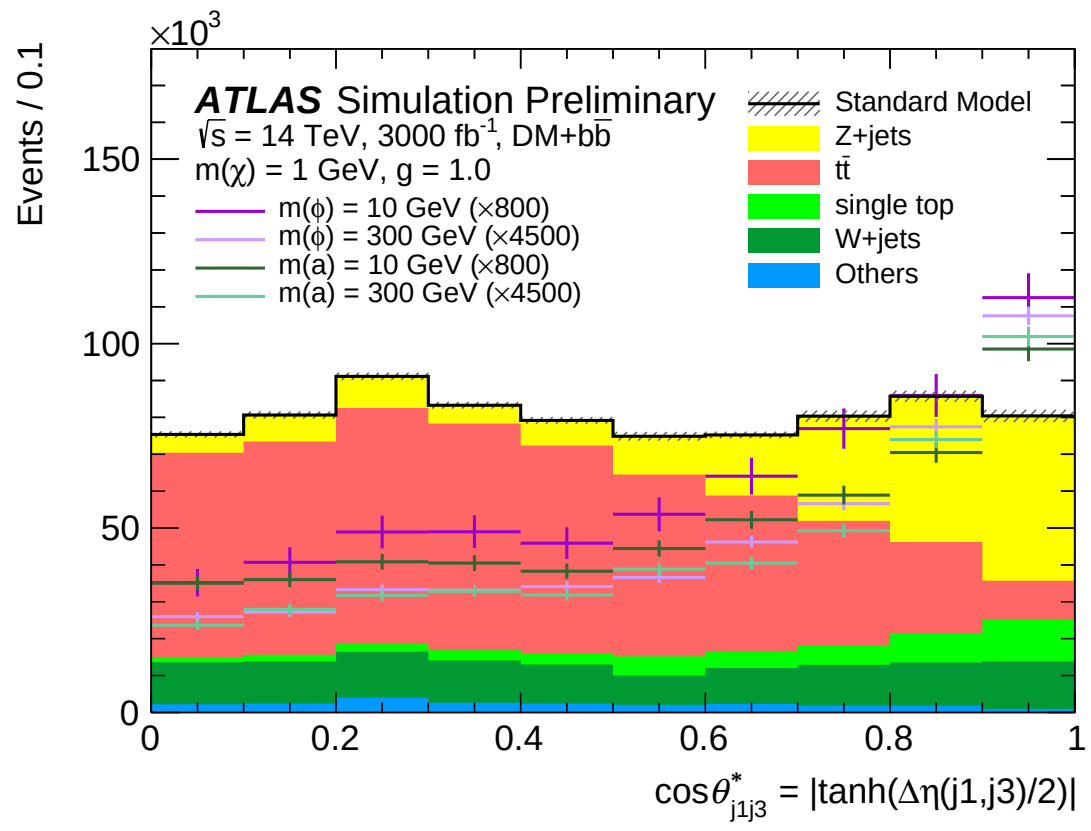

(b)

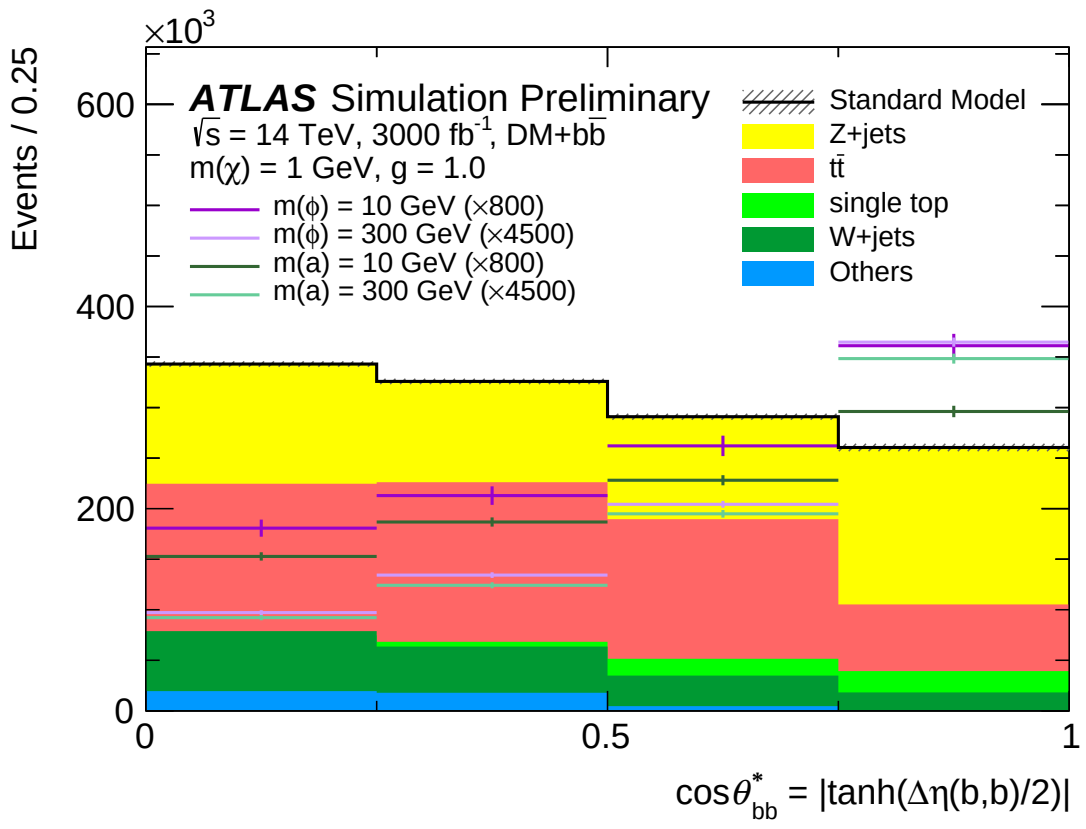

(c)

Figure 2: Distributions of several key discriminants in the DM+$b\bar{b}$ analysis following the requirement of $E_{\mathrm{T}}^{\mathrm{miss}}$ >210 GeV, no leptons, 2 or 3 jets, and at least two identified $b$-jets. The hatched bands and error bars represent the statistical uncertainty on the total SM background and signal yields respectively.

# 7 Signatures with top quarks and $E_{\mathrm{T}}^{\mathrm{miss}}$

In order to target dark matter produced in association with one (DM+$Wt$) or two (DM+$t\bar{t}$) top quarks, one signal region is defined, and it is denoted denoted $\mathrm{SR}_{2\ell}$. Events are required to have exactly two opposite electric charge leptons, electrons or muons, either same- or different-flavour with an invariant mass (regardless of the flavours of the leptons in the pair), $m_{\ell\ell}$, being larger than 100 GeV in order to reduce the $t\bar{t}$ background. Furthermore, candidate signal events are required to have at least one identified $b$-jet. Different discriminators and kinematic variables have been used to further separate the $t\bar{t} + \phi/a$ and $Wt + \phi/a$ signal from the SM background.

- $m_{b2\ell}^{\mathrm{min}}$ is the smallest invariant mass computed between the leading $p_{\mathrm{T}}$ $b$-tagged jet and each of the two leptons in the event. In events with two top quarks decaying dileptonically, at least one of the two mass combinations must be bounded from above by $m_{b2\ell}^{\mathrm{min}} < \sqrt{m_t^2 - m_W^2}$.
- $\mathbf{p}_{\mathrm{T,boost}}^{\ell\ell}$: defined as the vector

8

$$\mathbf{p}_{\mathrm{T,boost}}^{\ell\ell} = \mathbf{p}_{\mathrm{T}}^{\mathrm{miss}} + \mathbf{p}_{\mathrm{T}}(\ell_1) + \mathbf{p}_{\mathrm{T}}(\ell_2).$$

The $\mathbf{p}_{\mathrm{T,boost}}^{\ell\ell}$ variable, with magnitude $p_{\mathrm{T,boost}}^{\ell\ell}$, can be interpreted as the opposite of the vector sum of all the transverse hadronic activity in the event.

- $\Delta\phi_{\mathrm{boost}}$: the azimuthal angle between the $\mathbf{p}_{\mathrm{T}}^{\mathrm{miss}}$ vector and the $\mathbf{p}_{\mathrm{T,boost}}^{\ell\ell}$ vector [56].
- $m_{\mathrm{T2}}$: lepton-based stransverse mass. The stransverse mass [57, 58] is a kinematic variable used to bound the masses of a pair of intermediate particles which are presumed to each have decayed semi-invisibly into one visible and one invisible particle. The stransverse mass is defined as

$$m_{\mathrm{T2}}(\mathbf{p}_{\mathrm{T},1}, \mathbf{p}_{\mathrm{T},2}, \mathbf{q}_{\mathrm{T}}) = \min_{\mathbf{q}_{\mathrm{T},1}+\mathbf{q}_{\mathrm{T},2}=\mathbf{q}_{\mathrm{T}}} \left\{\max[\, m_{\mathrm{T}}(\mathbf{p}_{\mathrm{T},1}, \mathbf{q}_{\mathrm{T},1}), m_{\mathrm{T}}(\mathbf{p}_{\mathrm{T},2}, \mathbf{q}_{\mathrm{T},2})\,]\right\},$$

where $m_{\mathrm{T}}$ indicates the transverse mass[2], $\mathbf{p}_{\mathrm{T},1}$ and $\mathbf{p}_{\mathrm{T},2}$ are the transverse momentum vectors of the two particles (assumed to be massless), and $\mathbf{q}_{\mathrm{T},1}$ and $\mathbf{q}_{\mathrm{T},2}$ are the unknown transverse momentum vectors of the invisible particles, with $\mathbf{q}_{\mathrm{T}} = \mathbf{q}_{\mathrm{T},1} + \mathbf{q}_{\mathrm{T},2}$. The minimisation is performed over all the possible decompositions of $\mathbf{q}_{\mathrm{T}}$. For $t\bar{t}$ or $WW$ events, where the transverse momenta of the two leptons in each event are taken as $\mathbf{p}_{\mathrm{T},1}$ and $\mathbf{p}_{\mathrm{T},2}$, and $p_{\mathrm{T}}^{\mathrm{miss}}$ as $\mathbf{q}_{\mathrm{T}}$, $m_{\mathrm{T2}}(\ell_1, \ell_2, E_{\mathrm{T}}^{\mathrm{miss}})$ is bounded sharply from above by the mass of the $W$ boson [59, 60], while signal events do not respect this bound because of the additional $E_{\mathrm{T}}^{\mathrm{miss}}$ coming from the undetected DM particles.

A summary of the analysis selections of $\mathrm{SR}_{2\ell}$ is presented in Table 2. For reference, the distribution of the $m_{\mathrm{T2}}$ variable for events passing all of the $\mathrm{SR}_{2\ell}$ requirements except that on $m_{\mathrm{T2}}$ is shown in Figure 3.

For the exclusion limits presented in Section 9, the $m_{\mathrm{T2}}$ distribution is divided into five exclusive bins between ([200,220],[220,240],[240,260],[260,280],[>280]) GeV following an approach similar to that used for $\mathrm{SR}_{b,\mathrm{low}}$ and $\mathrm{SR}_{b,\mathrm{high}}$. The bins are denoted by the labels $\mathrm{SR}_{2\ell}$-bin1 through $\mathrm{SR}_{2\ell}$-bin5 and span the range of $m_{\mathrm{T2}}$ between 200 GeV and 300 GeV. The last bin also includes events with $m_{\mathrm{T2}} > 300$ GeV.

[2] The transverse mass is defined as $m_{\mathrm{T}} = \sqrt{2|\mathbf{p}_{\mathrm{T},1}||\mathbf{p}_{\mathrm{T},2}|(1-\cos(\Delta\phi))}$, where $\Delta\phi$ is the angle between the particles with transverse momenta $\mathbf{p}_{\mathrm{T},1}$ and $\mathbf{p}_{\mathrm{T},2}$ in the plane perpendicular to the beam axis.

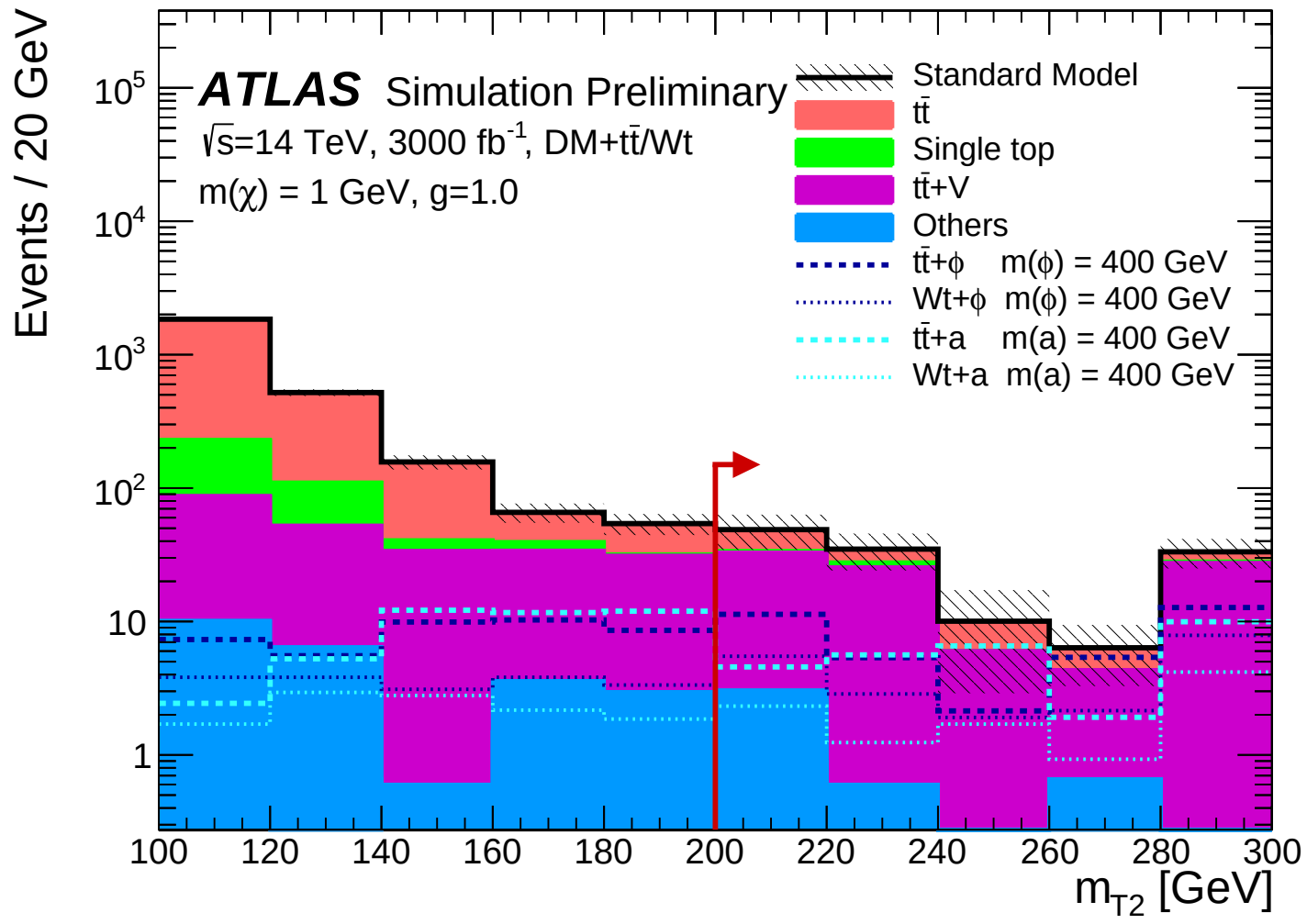

Figure 3: Distribution of $m_{\mathrm{T2}}$ for events satisfying the SR criteria except that on $m_{\mathrm{T2}}$. The contributions from all SM backgrounds are shown; the hatched bands represent the systematic uncertainty. The rightmost bin includes overflow events.

Table 2: Summary of the analysis selection criteria (see text for details).

| | $\mathrm{SR}_{b,\mathrm{low}}$ | $\mathrm{SR}_{b,\mathrm{high}}$ | $\mathrm{SR}_{2\ell}$ |
|---|---|---|---|
| $N_\ell$ | 0 | 0 | 2 |
| $N_{\mathrm{jets}}$ | 2 or 3 | 2 or 3 | $\geq 1$ |
| $N_{b\text{-jets}}$ | $\geq 2$ | $\geq 2$ | $\geq 1$ |
| $\lvert\eta_j\rvert$ | $< 3.0$ | $< 3.8$ | $< 2.5$ |
| $m_{\ell\ell}$ | - | - | $> 100$ GeV |
| $E_{\mathrm{T}}^{\mathrm{miss}}$ | $> 210$ GeV | $> 300$ GeV | $> 300$ GeV |
| $m_{b2\ell}^{\mathrm{min}}$ | - | - | $< 150$ GeV |
| $\Delta\phi_{\mathrm{boost}}$ | - | - | $< 1.5$ |
| $p_T(j_1)$ | $> 130$ GeV | $> 200$ GeV | $> 100$ GeV |
| $p_T(j_3)$ | $< 50$ GeV | $< 90$ GeV | - |
| $H_{\mathrm{T}}^{\mathrm{ratio}}$ | $> 0.75$ | $> 0.4$ | - |
| $\delta^-$ [rad] | - | $< 0.5$ | - |
| $\delta^+$ [rad] | - | $< 1.0$ | - |
| Imb($b$,$b$) | - | $< 0.6$ | - |
| $\cos(\pi - \Delta\phi(b, b))$ | $> 0.75$ | - | - |
| $\cos\theta^*_{j_1 j_3}$ | $> 0.8$ | $> 0.75$ | - |
| $m_{\mathrm{T2}}$ | - | - | $> 200$ GeV |

## 8 Systematic Uncertainties

Systematic uncertainties based on those in Ref. [13] are applied to signal and SM background processes. The uncertainties are scaled to align with HL–LHC extrapolations developed by the ATLAS and CMS Collaborations and documented in Ref. [30]. During Phase-II operation, the theory modelling uncertainties are expected to halve, while the degree of reduction of experimental uncertainties like, for example, the jet energy scale and $b$-jet mis-identification depend on the signal region. This results in a total expected systematic uncertainty on the SM background of 13.42% for $\mathrm{SR}_{b,\mathrm{low}}/\mathrm{SR}_{b,\mathrm{high}}$ and 13% for $\mathrm{SR}_{2\ell}$, corresponding to a reduction of ∼15% and ∼54% respectively compared to the 13 TeV analysis. Statistical uncertainties due to the limited size of the Monte Carlo samples used for the modelling of signal and SM background processes are neglected.

Two types of hypothesis tests are performed in order to extract expected discovery $p$-values and 95% CL exclusion limits. For the $p$-values, a cut-and-count experiment is employed assuming uncertainty only on the SM background yield. The value of this uncertainty is set to the total background uncertainty reported above. For the exclusion limits, the sensitivity is evaluated by performing a profile-likelihood fit to pseudo-data corresponding to the expected background and signal yields in each multi-bin signal region. The likelihood is built as the product of Poissonian terms, one for each of the bins considered. Systematic uncertainties are incorporated as Gaussian distributed nuisance parameters affecting both the overall normalisation of the fit variable and the individual bin yields. For the former, the uncertainty on the SM background contribution is modelled by a nuisance parameter with value equal to the total background uncertainty. For the signal contribution, experimental uncertainty is accounted for by a nuisance parameter with a value of 10% (9.4%) for $\mathrm{SR}_{b,\mathrm{low}}/\mathrm{SR}_{b,\mathrm{high}}$ ($\mathrm{SR}_{2\ell}$), corresponding to the HL–LHC extrapolation of the Run 2 detector- and reconstruction-based uncertainties. A separate nuisance parameter with a value of 5% is also included to account for theoretical uncertainties on the signal models. The nuisance parameters affecting the individual bin yields account for potential inaccuracies in the extrapolated theory and experimental uncertainties. Such inaccuracies may result from, for example, the difference in selection criteria between a HL–LHC search and the reference Run 2 search.

## 9 Results

The predicted yields in the $\mathrm{SR}_{b,\mathrm{low}}$, $\mathrm{SR}_{b,\mathrm{high}}$ and $\mathrm{SR}_{2\ell}$ signal regions are reported in Tables 3, 4 and 5 respectively. For both $\mathrm{SR}_{b,\mathrm{low}}$ and $\mathrm{SR}_{b,\mathrm{high}}$, the main background consists of $Z/\gamma^*$+jets events followed by hadronic decays of $t\bar{t}$. A significant contribution also comes from single top quark processes and events featuring a $W$-boson produced in association with jets (“$W$+jets”). Note that the minor background from di-boson, tri-boson, and $t\bar{t} + Z/W$ , $t\bar{t} + WW/ZZ/WZ$ processes in $\mathrm{SR}_{b,\mathrm{low}}$ and $\mathrm{SR}_{b,\mathrm{high}}$ is referred to collectively as “Others”.

In $\mathrm{SR}_{2\ell}$, the dominant background consists of di-leptonic decays of $t\bar{t}$ and $t\bar{t}Z$ with $Z \to \nu\nu$. As with $\mathrm{SR}_{b,\mathrm{low}}$ and $\mathrm{SR}_{b,\mathrm{high}}$, the SM processes that make a minor contribution are merged into an “Others” category. In $\mathrm{SR}_{2\ell}$, this category contains the background from di-/tri-boson, $Z/\gamma^*$+jets, $t\bar{t}\,t\bar{t}$, and $t\bar{t} + WW$ processes.

| | $SR_{b,\text{low}}$-bin1 | $SR_{b,\text{low}}$-bin2 | $SR_{b,\text{low}}$-bin3 | $SR_{b,\text{low}}$-bin4 |
|---|---|---|---|---|
| SM events | 2542 ± 75 | 2436 ± 92 | 2861 ± 103 | 2585 ± 138 |
| $Z/\gamma^*$+jets events | 1337 ± 64 | 1410 ± 82 | 1885 ± 96 | 2030 ± 136 |
| $t\bar{t}$ events | 785 ± 37 | 708 ± 41 | 685 ± 35 | 384 ± 26 |
| Single top quark events | 166.8 ± 8.0 | 137.6 ± 8.0 | 143.8 ± 7.3 | 146.0 ± 9.8 |
| W+jets events | 252.9 ± 12.1 | 151.0 ± 8.8 | 108.1 ± 5.5 | 24.7 ± 1.7 |
| Others events | 0.81 ± 0.04 | 28.2 ± 1.6 | 39.1 ± 2.0 | −− |
| $b\bar{b} + \phi$ (10 GeV) | 12.21 ± 0.83 | 11.43 ± 0.86 | 14.55 ± 0.98 | 15.0 ± 1.2 |
| $b\bar{b} + a$ (10 GeV) | 10.48 ± 0.50 | 14.62 ± 0.85 | 14.73 ± 0.75 | 13.11 ± 0.88 |

Table 3: Expected yields in $SR_{b,\text{low}}$ for SM background processes and a selection of signal masses for an integrated luminosity of 3000 fb$^{-1}$at $\sqrt{s}$ = 14 TeV. The uncertainties on the quoted numbers correspond to the MC statistical uncertainty.

| | $SR_{b,\text{high}}$-bin1 | $SR_{b,\text{high}}$-bin2 | $SR_{b,\text{high}}$-bin3 | $SR_{b,\text{high}}$-bin4 |
|---|---|---|---|---|
| SM events | 1130 ± 54 | 1208 ± 47 | 1218 ± 52 | 1054 ± 52 |
| $Z/\gamma^*$+jets events | 572 ± 45 | 594 ± 39 | 665 ± 46 | 698 ± 49 |
| $t\bar{t}$ events | 346 ± 27 | 343 ± 23 | 312 ± 22 | 214 ± 15 |
| Single top quark events | 101.4 ± 8.0 | 110.3 ± 7.3 | 108.2 ± 7.5 | 101.6 ± 7.1 |
| $W$+jets events | 40.4 ± 3.2 | 108.1 ± 7.1 | 104.6 ± 7.2 | 40.1 ± 2.8 |
| Others events | 69.8 ± 5.5 | 53.2 ± 3.5 | 28.2 ± 2.0 | −− |
| $b\bar{b} + \phi$ (300 GeV) | 0.70 ± 0.05 | 0.76 ± 0.045 | 0.92 ± 0.06 | 1.22 ± 0.07 |
| $b\bar{b} + a$ (300 GeV) | 0.51 ± 0.04 | 0.68 ± 0.04 | 0.66 ± 0.05 | 0.94 ± 0.07 |

Table 4: Expected yields in $SR_{b,\text{high}}$ for SM background processes and a selection of signal masses for an integrated luminosity of 3000 fb$^{-1}$at $\sqrt{s}$ = 14 TeV. The uncertainties on the quoted numbers correspond to the MC statistical uncertainty.

The results are translated into constraints on the scalar and pseudoscalar models using the HistFitter package [61], which employs a profile-likelihood-ratio test statistic to perform hypothesis testing [62]. The package is used to compute the expected discovery $p$-values for the scalar and pseudoscalar mediator models. For a flavour-universal coupling between $\phi/a$ and the SM quarks, only $SR_{2\ell}$ is sensitive to dark matter production. Scans of the expected discovery significance in this signal region are shown in Figure 4 as a function of the mediator mass. The $5\sigma$ discovery potential for the full HL–LHC dataset is expected to extend up to $m(\phi)$ = 105 GeV and $m(a)$ = 150 GeV for the DM+$t\bar{t}$ channel. Addition of the DM+$Wt$ channel further extends the discovery potential to $m(\phi)$ = 155 GeV and $m(a)$ = 250 GeV.

The HistFitter package is also used to compute the expected exclusion limits for each model with the $CL_s$ prescription [63] and assuming no excess in the observed data. The limits are shown in Figures 5 and 6 for $\phi/a \to \chi\bar{\chi}$ production in association with bottom quarks and top quarks respectively for $\mathcal{L}$ = 3000 fb$^{-1}$ at $\sqrt{s}$ = 14 TeV. The contours correspond to the 95% CL upper limit on the ratio of the measurable cross-section with respect to the theoretically predicted cross-section for $g$ = 1.0. Also shown for comparison are the corresponding limits at $\sqrt{s}$ = 13 TeV with 36.1 fb$^{-1}$ of data [13].

For the DM+$b\bar{b}$ channel, cross-sections ~45–100 times the theoretically predicted for $g$ = 1.0 and $m(\phi/a) < 100$ GeV are excluded with the anticipated HL–LHC dataset. This corresponds to a factor of 3–3.5 (3–4.3) improvement with respect to the previous reach achievable for the scalar (pseudoscalar) mediator model. Similarly, for $m(\phi/a) \geq 100$ GeV, the extended coverage in pseudorapidity afforded by the upgrade to the ATLAS Inner Tracker allows for better exploitation of anti-correlations in jet and $b$-jet spin-sensitive variables like $\cos\theta^*_{bb}$. This results in a larger gain in the exclusion potential for the high-mass region, equivalent to a factor of 5.8–8.7 (3–5) increase with respect to the 13 TeV limit for scalar (pseudoscalar) masses in the range 100–500 GeV.

| | $SR_{2\ell}$ | $SR_{2\ell}$-bin1 | $SR_{2\ell}$-bin2 | $SR_{2\ell}$-bin3 | $SR_{2\ell}$-bin4 | $SR_{2\ell}$-bin5 |
|---|---|---|---|---|---|---|
| | 133 ± 21 | 49 ± 14 | 35 ± 10 | 10.0 ± 7.1 | 6.4 ± 3.0 | 33.3 ± 8.3 |
| $t\bar{t}$ events | 33.3 ± 5.3 | 15.1 ± 4.5 | 7.1 ± 2.2 | 4.05 ± 2.9 | 2.0 ± 1.0 | 5.0 ± 1.3 |
| $t\bar{t}+V$ events | 92 ± 15 | 29.9 ± 8.8 | 24.9 ± 7.7 | 6.00 ± 4.3 | 3.7 ± 1.8 | 27.4 ± 6.7 |
| Single top quark events | 3.82 ± 0.61 | 0.76 ± 0.23 | 2.30 ± 0.70 | −− | −− | 0.76 ± 0.20 |
| Others events | 4.30 ± 0.43 | 3.00 ± 0.70 | 0.60 ± 0.18 | −− | 0.66 ± 0.32 | −− |
| $t\bar{t}/Wt+a$ (50 GeV) | 235 ± 18 | 62.9 ± 9.6 | 42.6 ± 7.3 | 45.1 ± 8.6 | 19.6 ± 4.3 | 64.5 ± 8.1 |
| $t\bar{t}/Wt+\phi$ (50 GeV) | 219 ± 33 | 61 ± 17 | 58 ± 16 | 10.6 ± 4.4 | 17.0 ± 9.8 | 71 ± 22 |
| $t\bar{t}/Wt+a$ (400 GeV) | 39.0 ± 4.9 | 6.9 ± 1.8 | 6.8 ± 1.8 | 8.3 ± 3.3 | 2.8 ± 1.0 | 14.1 ± 2.6 |
| $t\bar{t}/Wt+\phi$ (400 GeV) | 57.2 ± 6.6 | 16.8 ± 3.7 | 8.2 ± 2.0 | 4.0 ± 2.0 | 7.5 ± 2.0 | 20.6 ± 3.4 |

Table 5: Expected yields in $SR_{2\ell}$ for SM background processes and a selection of signal masses for an integrated luminosity of 3000 fb$^{-1}$at $\sqrt{s}$ = 14 TeV. The uncertainties on the quoted numbers correspond to the MC statistical uncertainty.

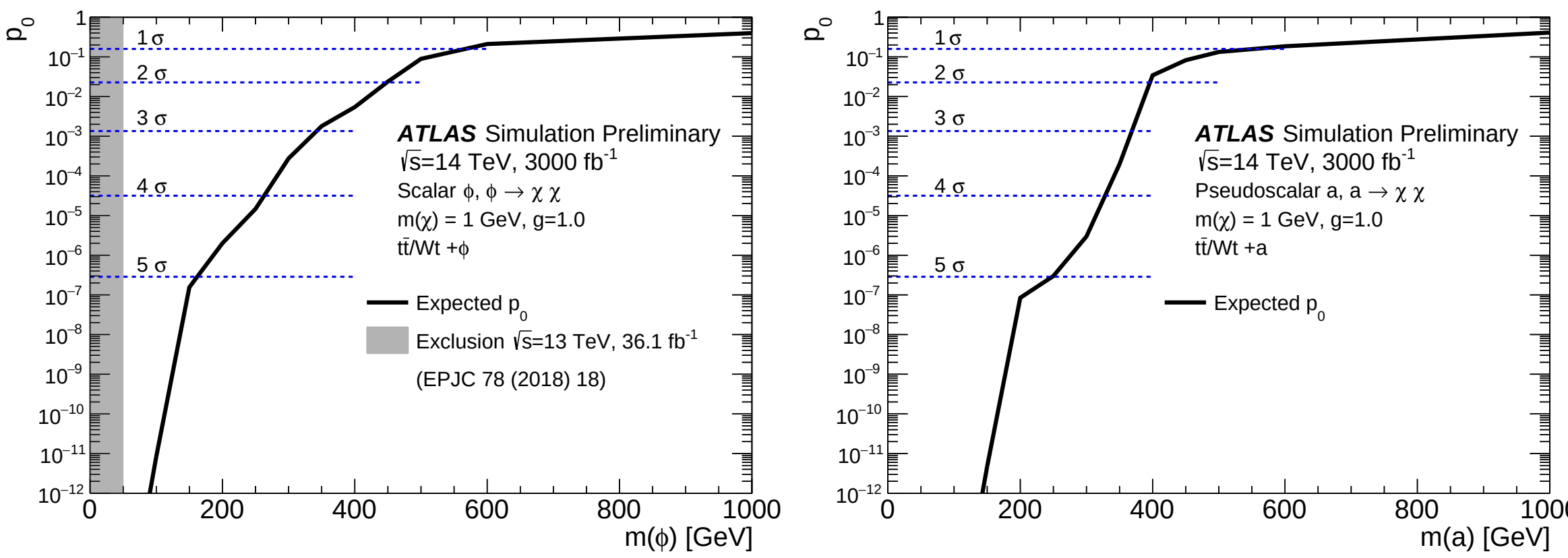

(a) (b)

Figure 4: Expected compatibility, represented by the $p$-value $p_0$, of the background-only hypothesis with the production of a colour-neutral (left) scalar or (right) pseudoscalar mediator in association with one or two top quarks for 3000 fb$^{-1}$ of 14 TeV proton-proton collision data. The compatibility is given as a function of the mediator mass assuming $\phi/a \to \chi\bar{\chi}$ and $g = 1.0$.

As mentioned previously, the DM+$b\bar{b}$ channel is better motivated within the context of the 2HDM+$a$ model, offering appealing prospects for constraints on $\tan\beta$. Using the same 2HDM+$a$ model as in Ref. [18] and assuming a large mass splitting between the two pseudoscalar states ($A$ and $a$ with $m(A) >$

$m(a)$), an upper bound on $\tan\beta$ can be approximated by the formula [17]:

$$\tan\beta \simeq \left[\frac{g_\chi g_\nu}{y_\chi \sin\theta}\left(\frac{\sigma}{\sigma(g=1.0)}\right)\right]^{1/2} = \left[\frac{1}{y_\chi \sin\theta}\left(\frac{\sigma}{\sigma(g=1.0)}\right)\right]^{1/2}$$

where $\sigma/\sigma(g = 1.0)$ corresponds to the value of the exclusion limit for $pp \to a + b\bar{b}$ in the context of the simplified pseudoscalar mediator model. For $\sin\theta = 0.35$ and $y_\chi = 1$ (a common choice of parameter values), expected bounds on $\tan\beta$ achievable at the HL–LHC range from ~19 for $m(a) = 10$ GeV to ~100 for $m(a) = 500$ GeV, significantly extending the current phase space coverage.

The exclusion limits in Figure 6 include the contributions from both the DM+$t\bar{t}$ and DM+$Wt$ final states. Considering only the DM+$t\bar{t}$ channel, the limit is expected to extend up to $m(\phi) = 405$ GeV and $m(a) = 385$ GeV. In the case of the scalar mediator model, this represents a factor of 5 improvement with respect to the 13 TeV result. The statistical precision of the signal models has been found to be the main limiting factor in assessing the sensitivity of the DM+$t\bar{t}/Wt$ channel. If a single-bin signal region defined by the inclusive selection ($SR_{2\ell}$) is considered in place of the multi-bin selection, the exclusion limits in Figure 6 are reduced by a maximum of 20%, with scalar mediator masses below 100 GeV predominantly affected. For reference, the statistical uncertainties on the signal and SM background yields are provided in Table 5. Statistical uncertainties are also provided for the DM+$b\bar{b}$ channel in Tables 3 and 4.

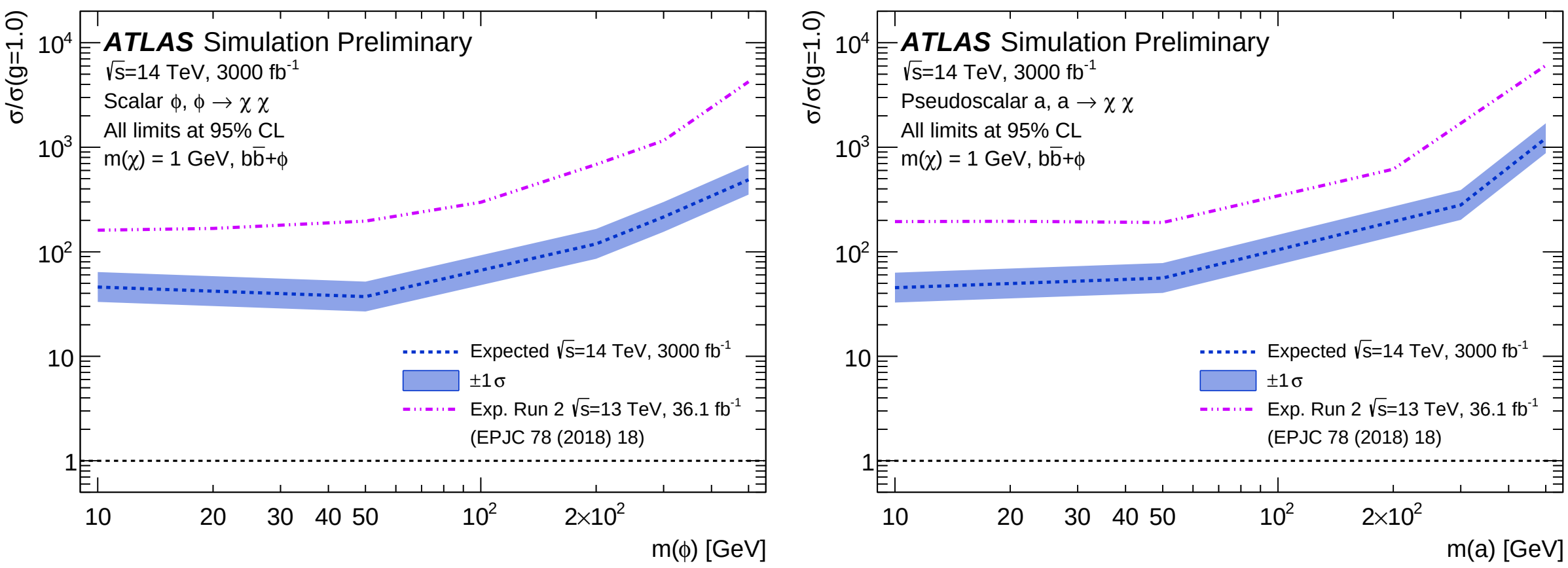

(a) (b)

Figure 5: Exclusion limits for the production of a colour-neutral (left) scalar or (right) pseudoscalar mediator in association with bottom quarks and decaying to a pair of dark matter particles with mass 1 GeV. The limits, calculated at 95% CL, are given as a function of the mediator mass and represent the ratio of the excluded cross-section to the theoretically predicted cross-section for a coupling, $g = 1.0$, and for 3000 fb$^{-1}$ of 14 TeV proton-proton collision data. The solid bands correspond to the expected limit $\pm 1\sigma$. Also shown for comparison is the expected limit for 36.1 fb$^{-1}$ of 13 TeV proton-proton collision data taken from the previous analysis [13] (pink).

For each dark matter and mediator mass pair, the exclusion limit on the production cross-section of colour-neutral scalar mediator particles can be converted into a limit on the spin-independent DM–nucleon scattering cross-section using the procedure described in Ref. [64]. Figure 7 shows the resulting constraints in the plane defined by the dark-matter mass and the scattering cross-section, which are derived considering only the contribution from the $t\bar{t} + \phi$ model. The maximum value of the DM–nucleon scattering cross-section shown in the plot corresponds to the value of the cross section for a mediator mass of 10 GeV. The red contour is the exclusion limit at 90% CL. The green contour indicates the $5\sigma$ discovery potential. The lower horizontal line in the green (red) contour corresponds to the value of the cross section for $m(\phi) = 105$ GeV ($m(\phi) = 430$ GeV). Overlaid for comparison are the most stringent direct detection limits to date from the LUX [65, 66], CRESST-III [67], XENON1T [68], PandaX [69] and DarkSide-50 [70] Collaborations.

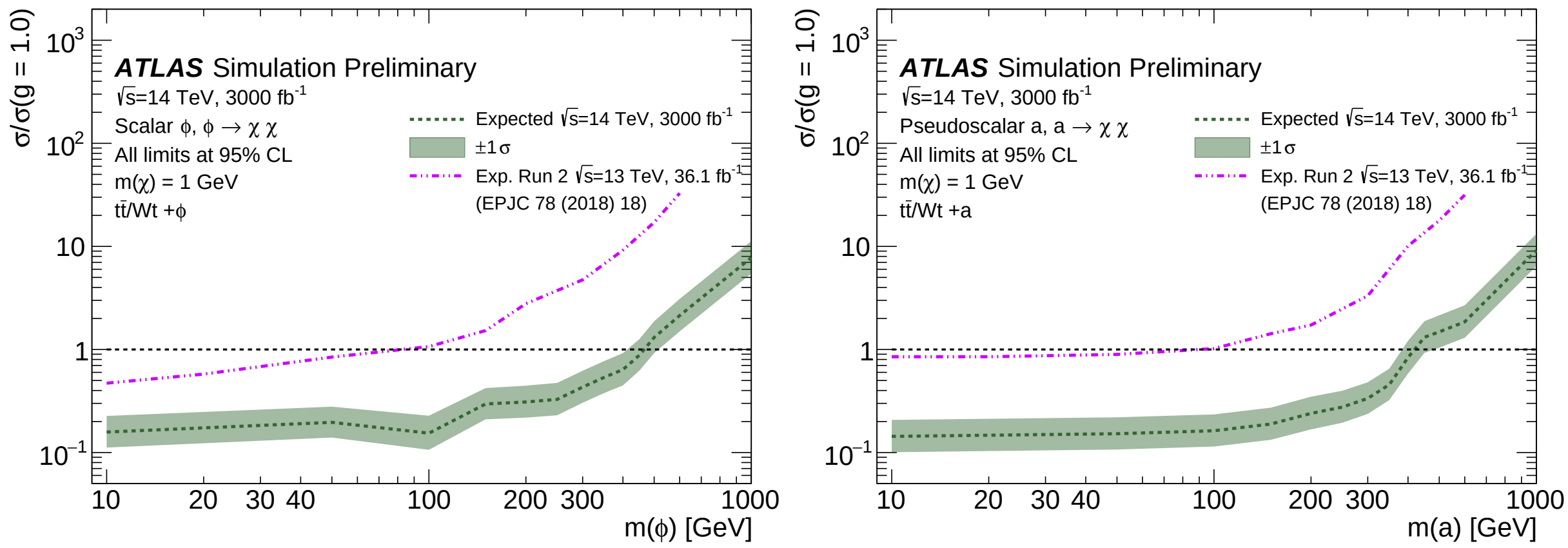

(a) (b)

Figure 6: Exclusion limits for the production of a colour-neutral (left) scalar or (right) pseudoscalar mediator in association with one or two top quarks and decaying to a pair of dark matter particles with mass 1 GeV. The limits, calculated at 95% CL, are given as a function of the mediator mass and represent the ratio of the excluded cross-section to the theoretically predicted cross-section for a coupling, $g = 1.0$, and for 3000 $\text{fb}^{-1}$ of 14 TeV proton-proton collision data. The solid bands correspond to the expected limit $\pm 1\sigma$. Also shown for comparison is the expected limit for 36.1 $\text{fb}^{-1}$ of 13 TeV proton-proton collision data taken from the previous analysis [13] (pink).

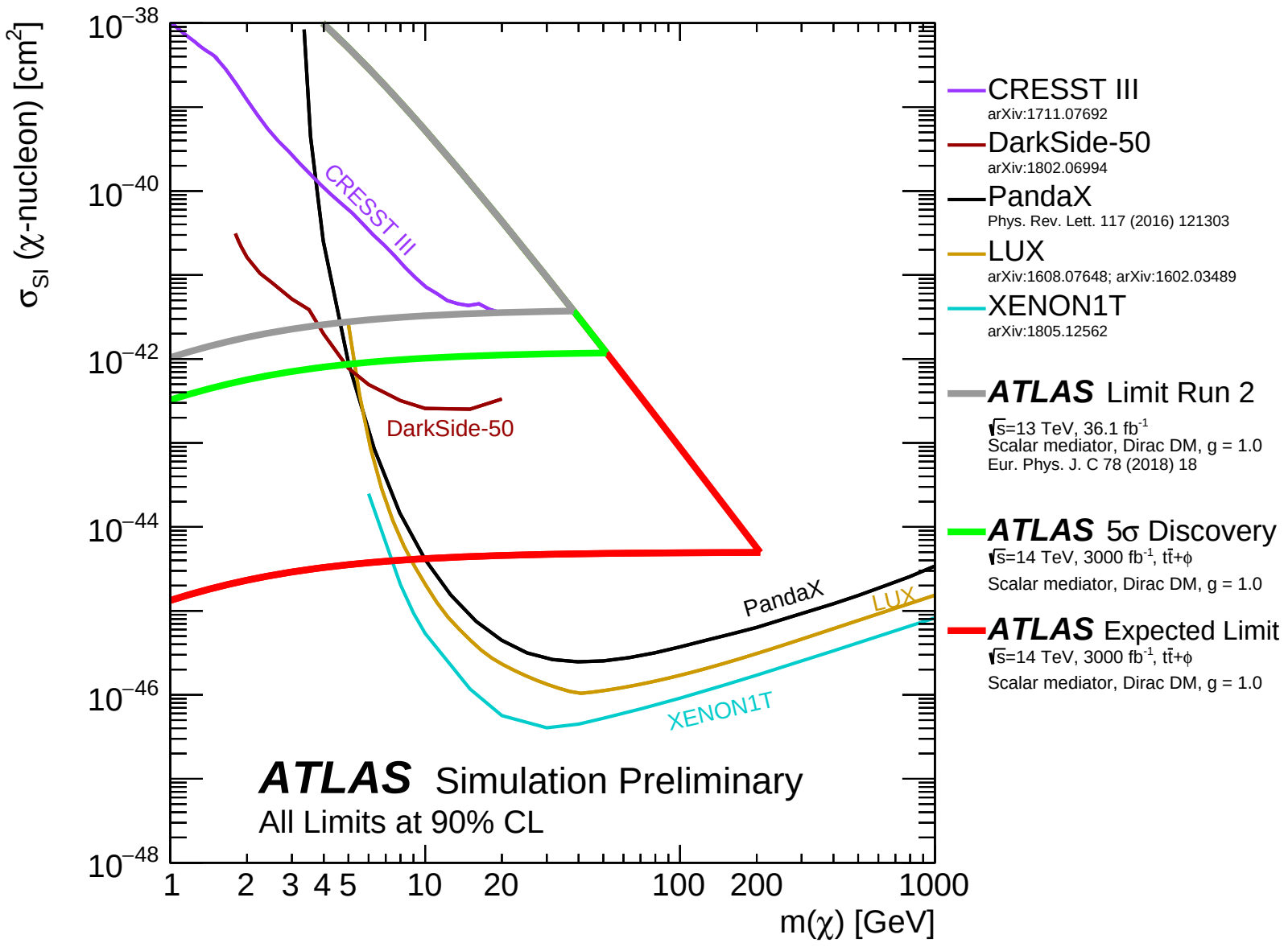

Figure 7: Comparison of the 90% CL limits on the spin-independent DM-nucleon cross-section as a function of DM mass between these results and the direct-detection experiments, in the context of the colour-neutral simplified model with scalar mediator. The green contour indicates the $5\sigma$ discovery potential at HL–LHC. The lower horizontal line of the DM–nucleon scattering cross-section for the red (green) contour corresponds to value of the cross section for $m(\phi)$ = 430 GeV ($m(\phi)$ = 105 GeV). The grey contour indicates the exclusion derived from the observed limits for 36.1 fb$^{-1}$ at 13 TeV taken from Ref. [13]. The results are compared with limits from direct detection experiments.

## 10 Conclusion

This note presents an estimate of the ATLAS sensitivity to dark matter production in association with heavy flavour quarks with 3000 fb$^{-1}$ of proton-proton collisions at $\sqrt{s}$ = 14 TeV. Feasibility studies are carried out for two simplified models in which the dark and SM sectors are assumed to couple via the exchange of a spin-0 mediator. The studies are divided into two categories: dark matter production in association with a pair of bottom quarks, and dark matter production in association with one or two top quarks. Parametrisations derived from performance studies are used to emulate the response of the upgraded ATLAS detector with 200 interactions per bunch crossing. Exclusion limits are derived at 95% CL for mediator masses in the range 10–500 GeV assuming systematic uncertainties consistent with current forecasts. In comparison to results obtained with 36 fb$^{-1}$ in Run 2, the exclusion potential at the HL–LHC is found to improve by a factor of ∼3–8.7 for scalar and pseudoscalar masses produced in association with bottom quarks, assuming unitary couplings and a dark matter mass of 1 GeV. In final states with one or two leptonically-decaying top quarks, the mass range for which a colour-neutral scalar mediator is expected to be excluded extends from 80 GeV to 405 GeV. Similarly, exclusion of pseudoscalar masses up to 385 GeV is expected.

# CMS Physics Analysis Summary

Contact: cms-phys-conveners-ftr@cern.ch 

# Search for excited leptons in $\ell\ell\gamma$ final states in proton-proton collisions at the HL-LHC

The CMS Collaboration

## Abstract

A search for excited leptons (electrons and muons) is presented, using simulation of the upgraded CMS detector at the High-Luminosity LHC (HL-LHC). Excited leptons are predicted by many theories beyond the standard model (SM) where quarks and leptons are not elementary but instead are themselves composite objects. Excited leptons ($\ell^* = e^*, \mu^*$) in $\ell\ell\gamma$ ($\ell = e, \mu$) final states where the excited lepton decays to a SM lepton and a photon ($\ell^* \to \ell\gamma$) are studied. The main background is Drell-Yan production in association with a photon. The HL-LHC environment (a centre of mass energy of 14 TeV and an integrated luminosity of 3000 fb$^{-1}$) allows for an extension of the discovery potential for excited leptons. The analysis is optimised for HL-LHC conditions, and indicates that $5\sigma$ discovery of excited leptons is possible for masses up to 5.1 TeV. If no significant discrepancies are seen in the data, excited leptons with masses below 5.8 TeV could be excluded at 95% confidence level.

*This document has been revised with respect to the version dated December 17, 2018.*

# 1 Introduction

The standard model (SM) provides a very precise description of many phenomena in particle physics observed over the last half century. Notwithstanding its huge success, it does not explain the origin of the mass hierarchy and the existence of three generations of quarks and leptons. As an attempt to answer such fundamental questions, compositeness of quarks and leptons is introduced in many models [1–10]. These compositeness models suggest that quarks and leptons are made of more fundamental constituents that are bound by a new strong interaction with a characteristic energy scale $\Lambda$ (called the compositeness scale).

Compositeness models predict the existence of excited states of quarks and leptons. In proton-proton (pp) collisions, excited fermions could be produced via contact interactions (CI) governed and decay either through SM gauge interactions or via CI to SM fermions. The contact interaction can be described by an effective Lagrangian:

$$\mathcal{L}_{CI} = \frac{g^{*2}}{2\Lambda_m^2} j^\mu j_\mu \quad (1)$$

where $g^{*2}$ is chosen to be $4\pi$, $j^\mu$ is the fermion current and $\Lambda_m$ is the energy scale of the substructure, assumed to be equal to or larger than the excited lepton mass. An illustration of the production decay mode is shown in Fig. 1.

This analysis presents a search for excited leptons ($\ell^* = e^*, \mu^*$) in $\ell\ell\gamma$ ($\ell = e, \mu$) final states where the excited lepton decays to a SM lepton and a photon ($\ell^* \to \ell\gamma$) in an upgraded CMS detector at the High-Luminosity LHC (HL-LHC).

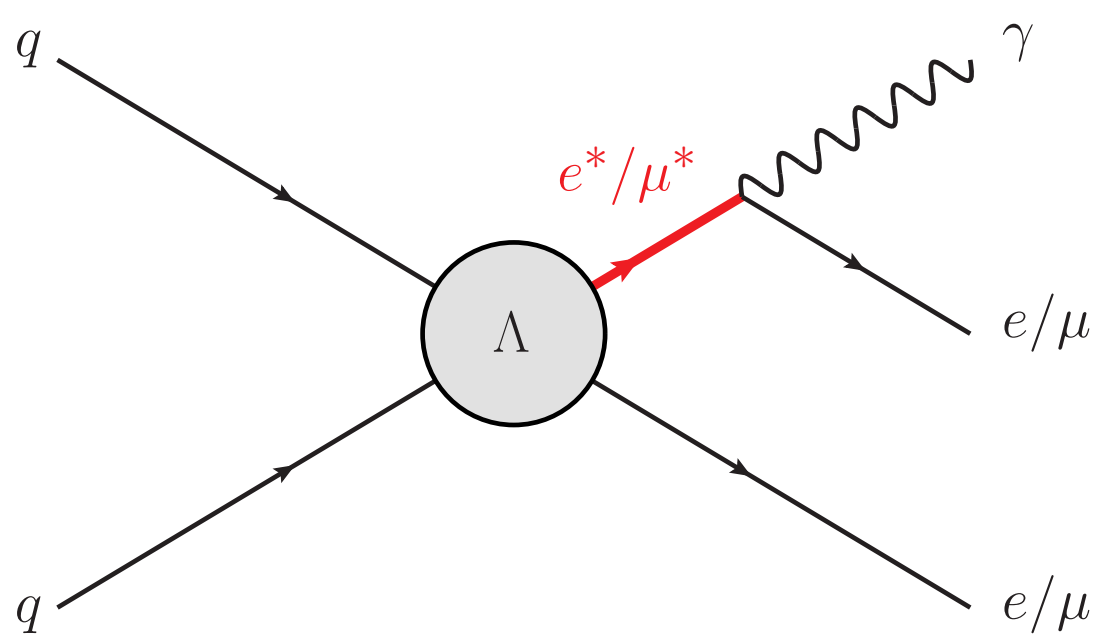

Figure 1: The Feynman diagram of the production of excited leptons in $\ell\ell\gamma$ final states.

The upgraded CERN HL-LHC is expected to deliver a peak instantaneous luminosity of up to $7.5 \times 10^{34}\,\mathrm{cm}^{-2}\,\mathrm{s}^{-1}$ [11], which is an increase in instantaneous luminosity of about four times with respect to the LHC Run 2 conditions. With this increase in instantaneous luminosity, the number of overlapping proton-proton interactions per bunch crossing, or pileup (PU), is expected to increase from its mean value of about 40 at the LHC to a mean value of up to 200 at the HL-LHC. Similarly, the levels of radiation are expected to significantly increase in all regions of the detector, in particular in its forward regions.

The CMS detector will be substantially upgraded in order to fully exploit the physics potential offered by the increase in luminosity, and to cope with the demanding operational conditions at the HL-LHC [12–16]. In particular, in order to sustain the increased PU rate and associated increase in flux of particles, the upgrade will provide the detector with: higher granularity to reduce the average channel occupancy, increased bandwidth to accommodate the higher data rates, and improved trigger capability to keep the trigger rate at an acceptable level without compromising physics potential. The upgrade will also provide an improved radiation hardness to withstand the increased radiation levels.

A detailed overview of the CMS detector upgrade program, known as 'Phase-2' is presented in Ref. [12–16]. The expected performance of the reconstruction algorithms and PU mitigation with the CMS detector is summarised in Ref. [17].

In this search, a clear signature of an opposite-sign same-flavour (SF) lepton pair and a photon allows highly efficient signal selection to assess the CMS upgrade physics reach. However, an ambiguity between the lepton from the excited lepton decay and the lepton from CI makes it challenging to identify the reconstructed mass of the excited lepton between invariant masses of two possible pairings of a lepton and the photon. For this search, information of both invariant masses is used to discriminate the excited lepton signal from SM background processes. We consider a benchmark model based on the formalism described in Ref. [8]. Because the excited lepton is produced in association with a SM lepton, there are two leptons in the final state.

Searches for excited leptons have been previously performed by the ATLAS [18–20] and CMS [21–23] Collaborations, the LEP [24–27], HERA [28], and Tevatron [29–32] colliders. No evidence for their existence was found in the searches so far and excited electrons (muons) are excluded for $m_{\ell^*} < 3.8(3.9)$ TeV by the CMS 13 TeV results [23].

For this HL-LHC sensitivity projection, we use an analysis strategy and procedure similar to the previous CMS data analysis [23]. The following scenario is considered: a centre-of-mass energy 14 TeV, an integrated luminosity of 3000 fb$^{-1}$ accumulated at the end of the HL-LHC program, and an increased average PU of 200 under the Phase-2 CMS detector upgrade.

## 2 Simulated samples and event selection

The signal samples are generated with PYTHIA 8.205 [33] at $\Lambda = 10$ TeV for $\ell^*$ masses ranging from 3.5 TeV to 6.5 TeV in steps of 250 GeV. The simulated signal samples are generated at leading order (LO) in perturbative quantum chromodynamics and corrected by using a mass dependent k-factor for next-to-leading-order (NLO) normalisation, ranging from 1.28 to 1.46. The main background is the SM Z$\gamma$ process, which is generated at NLO using the MADGRAPH5_aMC@NLO 2.3.3 [34]. The generated signal and background samples are interfaced to DELPHES [35], which is a parametric simulation of the CMS Phase-2 detector at the particle level. All simulated samples used in this analysis include a simulation with 200 average PU.

The signature of the signal event in this search has a SF lepton pair and a photon. The leptons and photon from the signal event are produced centrally for $m_{\ell^*} > 3.5$ TeV and therefore they are separated from the PU which is more significant in the high $\eta$ region. We select events having two isolated electrons or muons and a photon with requirements as follows. Electron and photon candidates are required to have pseudorapidity $|\eta| < 2.5$ and transverse momentum $p_T > 35$ GeV, and they are excluded in the electromagnetic calorimeter barrel-endcap transition region ($1.44 < |\eta| < 1.57$). Muon candidates should be isolated with $|\eta| < 2.4$ and $p_T > 35$ GeV. The leptons are required to have opposite charge and the selected electrons and muons must be separated from the photon by $\Delta R = \sqrt{\Delta\eta^2 + \Delta\phi^2} > 0.7$. In addition, the invariant mass of the two SF leptons $m_{\ell\ell}$ is required to be larger than 116 GeV in order to suppress the dominant background contribution from real Z boson production (Z resonance veto criteria). The detailed criteria are based on the definitions used in the 2016 $\ell^*$ search [23].

# 3 Background and signal modelling

The main SM background after the event selection is Drell-Yan production associated with a photon ($Z\gamma$), which has the same signature as the final state of the signal, when the Z boson decays into two leptons. This background is significantly suppressed by the Z boson veto requirement. Contributions of other SM processes like diboson and top quark pair production in association with a photon ($t\bar{t} + \gamma$) are relatively small, in particular in the signal search region of excited lepton masses above 2 TeV. Simulated $t\bar{t} + \gamma$ events are studied in this analysis, however the background events are imperfectly estimated due to insufficient sample size. Hence, we only consider the dominant $Z\gamma$ background in this search, and additional background contributions are considered as systematic uncertainties on the total background estimate. The photon misidentification rate under the HL-LHC conditions is studied in Ref. [36] using PHASE-2 DELPHES samples. The photon misidentification rate is expected to be about 1% when the photon $p_T$ is on the order of 100 GeV, which is compatible with the 2016 result. In the previous CMS Run 2 $\ell^*$ search [23], we observed 20% and 5% contribution from $t\bar{t} + \gamma$ and misidentified photon backgrounds, respectively, at $m_{\ell^*} > 1$ TeV; therefore a 25% systematic uncertainty is assigned for the missing background contributions.

To distinguish between signal and background events, a two-dimensional distribution of the two invariant masses $m_{\ell\gamma}^{\rm min}$ and $m_{\ell\gamma}^{\rm max}$ is used. A search window is set in the two-dimensional distribution of $m_{\ell\gamma}^{\rm max}$ versus $m_{\ell\gamma}^{\rm min}$. For $\ell^*$ events, either $m_{\ell\gamma}^{\rm min}$ or $m_{\ell\gamma}^{\rm max}$ corresponds to the reconstructed invariant mass of $\ell^*$. Therefore, the mass resonance of the signal is concentrated in an "L" shape [21, 22]. On the other hand, background events have no such correlation in $m_{\ell\gamma}^{\rm min}$ and $m_{\ell\gamma}^{\rm max}$ and are scattered around at low masses below about 2 TeV. This distinction between signal and background events in the distribution of $m_{\ell\gamma}^{\rm max}$ versus $m_{\ell\gamma}^{\rm min}$ is used to set the search window. We only set the lower $m_{\ell\gamma}^{\rm max}$ bound at 2 TeV in the two-dimensional distribution as the search window in order to maximise the signal yields. The reason why the L-shaped search window is not applied in this analysis is due to the insufficient MC statistics, therefore the results of the limits are conservative. The distributions of the dominant $Z\gamma$ background and signal are shown in Fig. 2.

The product of signal acceptance and efficiency ($A \times \varepsilon_{\rm sig}$) is obtained using the simulated DELPHES signal samples. The result as a function of $m_{\ell^*}$ is shown in Fig. 3 and is approximately 43% (55%) in the electron (muon) channel.

# 4 Systematic uncertainties

Systematic uncertainties for the performance of the lepton (0.5%) and photon (2.0%) reconstruction and identification, and the integrated luminosity (1.0%) follow the recommendation for upgrade analyses [37]. The theoretical systematic uncertainty is reduced by a scale of 1/2 with respect to the 2016 result. A statistical uncertainty in the entire signal region is dominant in this analysis. The missing background contribution is considered to be the main systematic uncertainty in the background estimation, as discussed in Sec. 3. The systematic uncertainties in the signal and background yields are summarised in Table 1.

# 5 Results and their interpretations

We explore the discovery potential for excited electrons and muons under the HL-LHC scenario, based on an integrated luminosity of 3000 fb$^{-1}$. We set 95% confidence level (CL) upper limits on the $\ell^*$ production cross sections, which are computed with the modified frequentist $CL_s$ method [38, 39], with a likelihood ratio used as a test statistic. The systematic uncertainties are treated as nuisance parameters with log-normal priors.

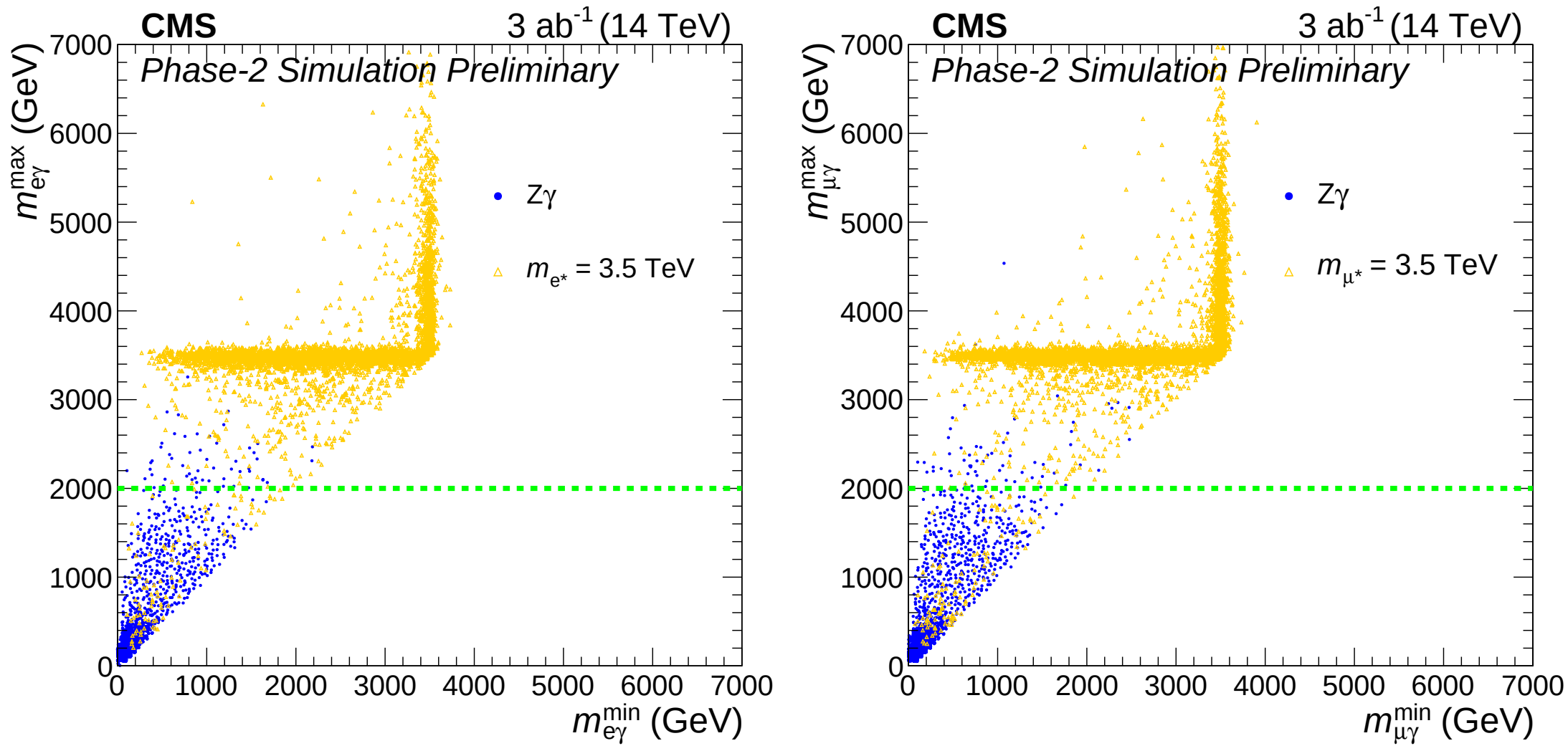

Figure 2: The 2D $m_{\ell\gamma}^{max}$ versus $m_{\ell\gamma}^{min}$ distribution after full selection with $Z\gamma$ background (blue circles) and signal $e^*$ and $\mu^*$ samples of $m_{\ell^*}$ = 3.5 TeV (orange triangles) in the electron channel (left) and muon channel (right). The dashed green line indicates the lower bound of the search window.

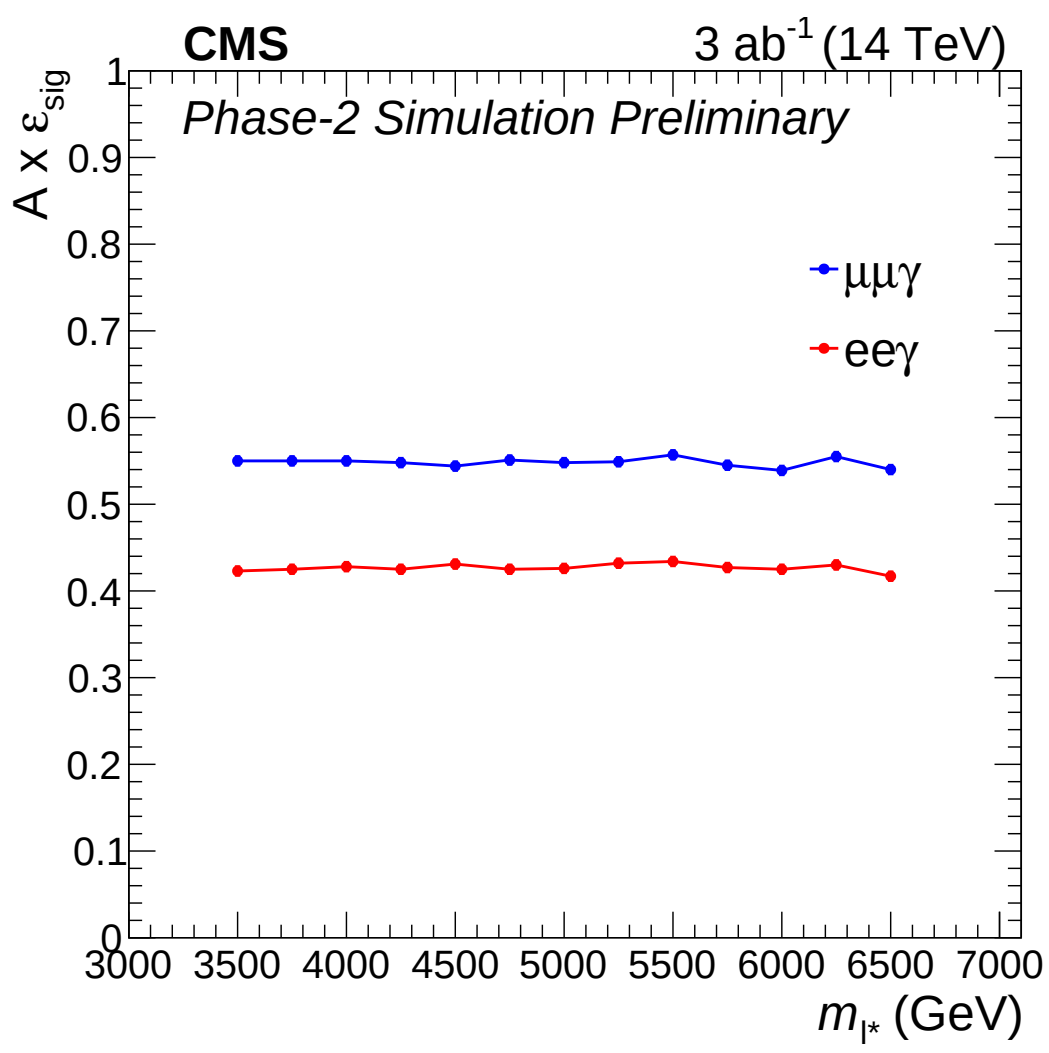

Figure 3: The product of signal acceptance and efficiency as a function of the generated resonance mass for the ee$\gamma$ (lower) and $\mu\mu\gamma$ (upper) channels. Each marker denotes the value measured from the simulated signal sample at a given mass point.

| | Electron channel | | Muon channel | |
|---|---|---|---|---|
| | Signal | Bkg | Signal | Bkg |
| Integrated luminosity | 1.0% | 1.0% | 1.0% | 1.0% |
| Lepton efficiency | 0.5% | 0.5% | 0.5% | 0.5% |
| Photon efficiency | 2.0% | 2.0% | 2.0% | 2.0% |
| PDF & scales | 1.0% | 5.0% | 1.0% | 5.0% |
| Summary of all backgrounds | - | 25.0% | - | 25.0% |

Table 1: Summary of the systematic uncertainties.

The discovery potential as a function of excited lepton mass shown in Fig. 4, indicates that $3\sigma$ evidence ($5\sigma$ discovery) is possible for both excited electrons and muons with masses up to 5.5 (5.1) TeV.

Fig. 5 shows the result of the expected upper limits for $e^*$ (left) and $\mu^*$ (right). The expected exclusion of the excited leptons is $m_{\ell^*} < 5.8$ TeV for both $e^*$ and $\mu^*$ in the case where $m_{\ell^*} = \Lambda$. While the electron channel has a lower signal yield than the muon channel, it also has lower background, and the net result is that the excluded cross sections differ only by about 10%, producing a similar exclusion limit on the excited lepton mass.

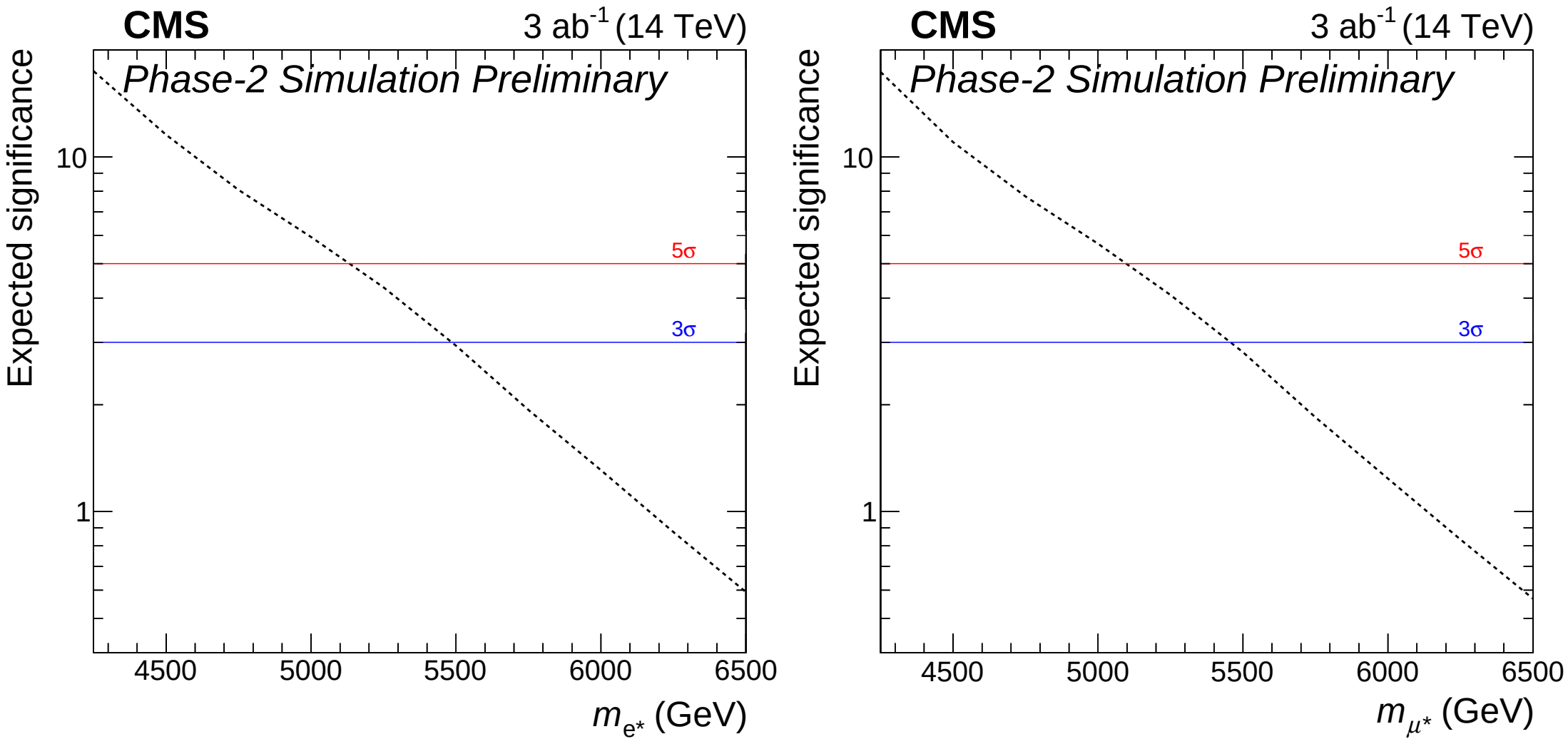

Figure 4: Discovery significance for excited electrons (left) and muons (right) with 3000 $fb^{-1}$ at the HL-LHC.

# 6 Summary

The search for excited leptons in final states with two leptons and a photon in proton-proton collisions at the High-Luminosity LHC (HL-LHC) was studied. The HL-LHC environment (a centre of mass energy of 14 TeV and an integrated luminosity of 3 $ab^{-1}$) allows for an extension of the discovery potential for excited leptons. The results were optimised for HL-LHC conditions, and it was found that excited leptons masses up to 5.5 (5.1) TeV can be excluded (discovered), for both excited electrons and excited muon states. Excited leptons could be excluded for masses below 5.8 TeV, at 95% confidence level.

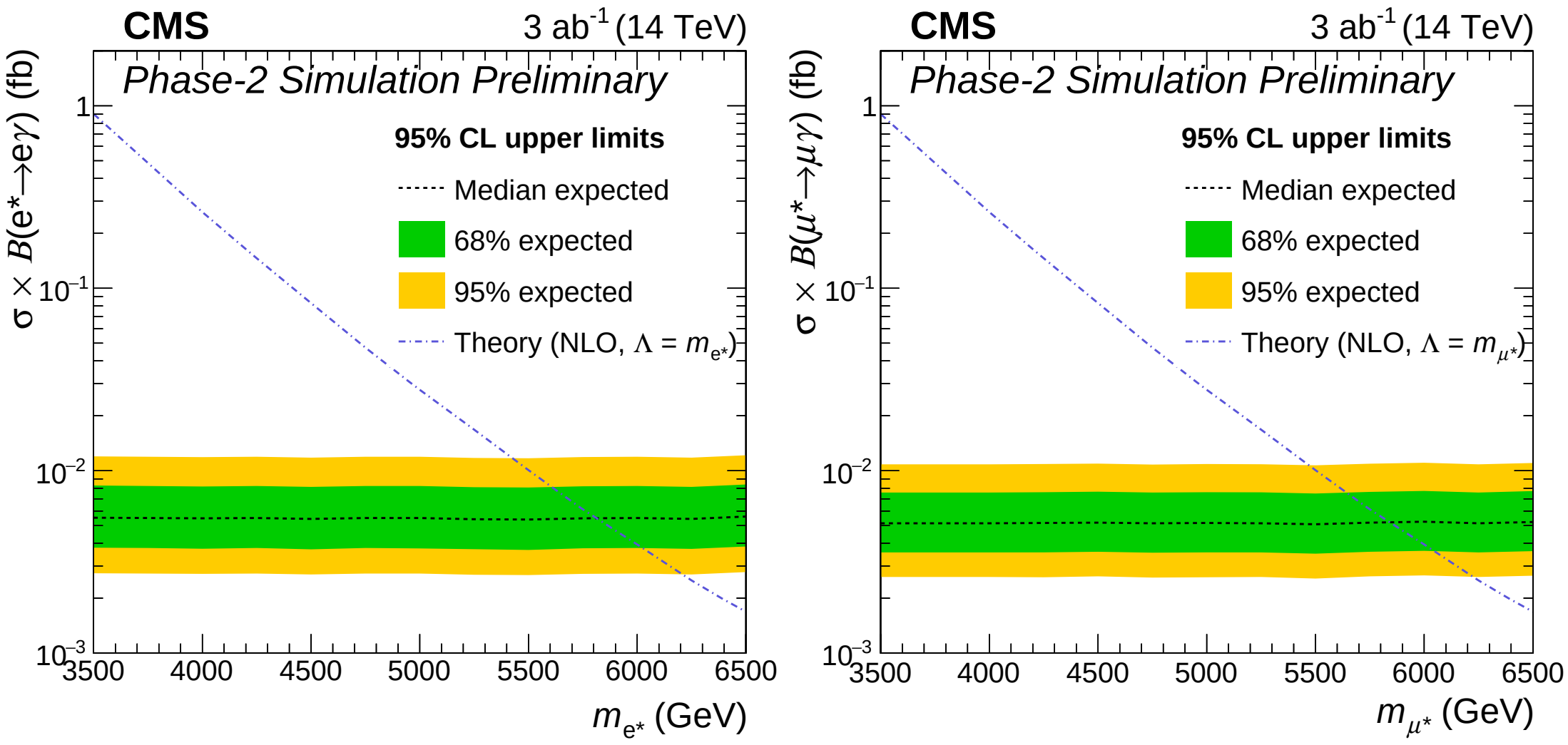

Figure 5: Exclusion limits for excited electrons (left) and muons (right) on the product of cross section and branching fraction.

**8**

# CMS Physics Analysis Summary

Contact: cms-phys-conveners-ftr@cern.ch

## Search for heavy composite Majorana neutrinos at the High-Luminosity and the High-Energy LHC

The CMS Collaboration

### Abstract

The sensitivity of a search for heavy Majorana neutrinos with the CMS Phase-2 detector in a final state with two leptons and at least one large-radius jet is investigated. Such new particles arise in theories beyond the standard model with compositeness. The study is based on searches previously performed with Run 2 CMS data, where no evidence for a signal was found. The High-Luminosity LHC (HL-LHC) with a centre-of-mass energy of 14 TeV and an integrated luminosity of 3 $ab^{-1}$ will allow an extension of the sensitivity to cross sections of order of a few ab for heavy neutrino masses $M(N_\ell)$ ranging from 3 to 9 TeV for the $\ell\ell q\bar{q}'$ channel, where $\ell$ is an electron or a muon and $q$ is a quark. For the compositeness scale $\Lambda = M(N_\ell)$, the existence of a heavy Majorana neutrino could be excluded for masses up to 8 TeV at the 95% confidence level. The projection of the study to the High-Energy LHC (HE-LHC) scenario, with a centre-of-mass energy of 27 TeV, is also presented here.

# 1 Introduction

While the standard model (SM) of fundamental interactions continues to be experimentally verified, it fails to solve several open problems. Whether it is from a purely theoretical approach, looking, for instance, at the quadratic sensitivity of the Higgs field to high scale, or from experimental evidence like the discovery of neutrino oscillations and matter-antimatter asymmetry, there is clearly a need to extend the SM in some way.

Many scenarios beyond the SM (BSM) suggest that the dynamics of new physics at high energies may be represented by an effective theory of higher-dimensional operators, with new degrees of freedom. Compositeness of ordinary fermions is one possible BSM scenario that may lead to a solution of the hierarchy problem or to the explanation of the proliferation of ordinary fermions [1–6]. Typically in a composite scenario, SM quarks and leptons are assumed to have an internal substructure that should become manifest at some sufficiently high energy scale, the compositeness scale $\Lambda$. Ordinary fermions are thought to be bound states of some as–yet–unobserved fundamental constituents generically referred to as *preons*. Two model–independent features [4, 7–9] are relevant in a composite scenario: (i) the existence of excited states of quarks and leptons with masses lower than or equal to the compositeness scale $\Lambda$, interacting via magnetic type gauge couplings with the ordinary SM fermions; (ii) contact interactions, which are supposed be residual interactions due to the as–yet–unknown preon dynamics. The heavy composite Majorana neutrino $N_\ell$ would be a particular case of such excited states. Early literature investigated the production at hadron colliders of heavy composite Majorana neutrinos [10] as well as their effects in low energy reactions such as neutrinoless double $\beta$ decay [11, 12]. Recently, excited quarks and leptons have been searched for at the LHC [13, 14], boosting previous limits on their mass and on the value of the compositeness scale $\Lambda$ [15–20]. From the theoretical point of view, the phenomenology of excited leptons and quarks has also been the object of renewed interest with respect to the LHC phenomenology [21–25] as well as with respect to possible connections of heavy composite neutrinos to leptogenesis [26].

For this study of a heavy Majorana neutrino, the Lagrangian density for gauge mediated interactions, which are of the magnetic type for current conservation, reads [9, 27]

$$\mathcal{L}_G = \frac{1}{2\Lambda} \overline{L^*_{\mathrm{R}}} \sigma^{\mu\nu} \left( gf \frac{\overrightarrow{\tau}}{2} \cdot \overrightarrow{W}_{\mu\nu} + g' f' Y B_{\mu\nu} \right) L_{\mathrm{L}} + h.c. \tag{1}$$

where $L^*_R$ and $L_L$ are respectively the right handed doublet of the excited fermions and the left handed doublet of the standard model, $g$ and $g'$ are the $SU(2)_L$ and $U(1)_Y$ gauge couplings, and $f$ and $f'$ are dimensionless couplings, which are expected to be of order unity [9] and henceforth simply assumed to be 1.

Contact interactions can be thought of as an effective field theory description of the effects of the unknown internal dynamics of compositeness. The corresponding Lagrangian describes the four-fermion contact interactions by a dimension-6 effective operator, and hence two inverse powers of the compositeness scale $\Lambda$ appear:

$$\mathcal{L}_C = \frac{g_*^2}{\Lambda^2} \frac{1}{2} j^\mu j_\mu \tag{2}$$

with

$$j_\mu = \eta_L \bar{\psi}_L \gamma_\mu \psi_L + \eta'_L \bar{\psi}^*_L \gamma_\mu \psi^*_L + \eta''_L \bar{\psi}^*_L \gamma_\mu \psi_L + h.c. + (L \to R), \tag{3}$$

where $g_*^2 = 4\pi$, the $\eta$ factors that define the chiral structure are usually set equal to 1, and $\psi$ and $\psi^*$ are the SM and excited fermion fields [9]. Thus, the total interaction, shown diagram-matically in Fig. 1, is the sum of the gauge interaction, as described in Eq. 1, and the contact interaction, as described in Eq. 3.

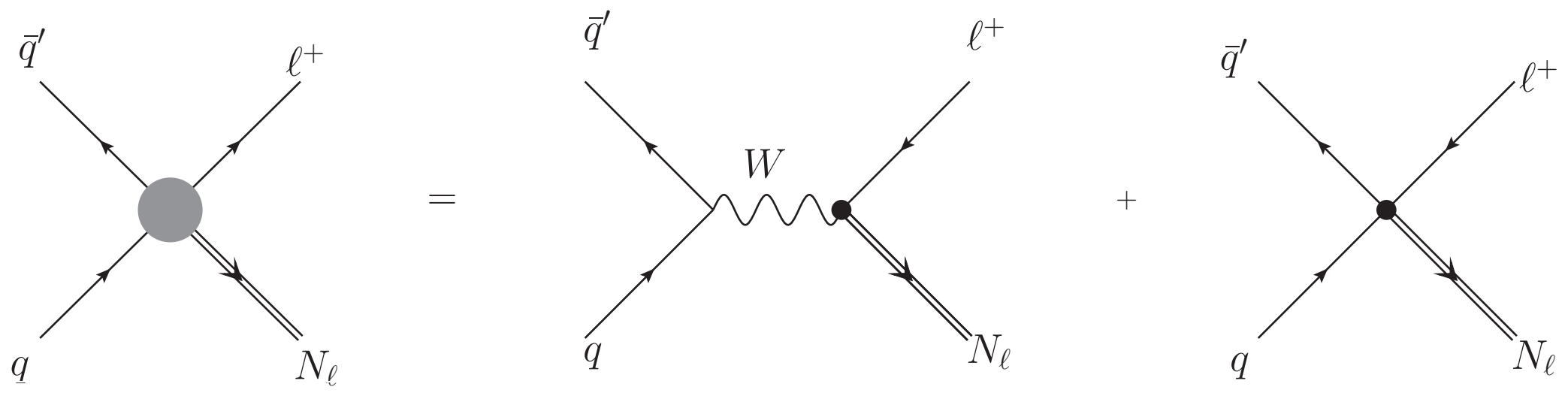

Figure 1: The production of a heavy composite Majorana neutrino via the fermion interaction discussed in the text as a sum of the gauge and contact contributions.

The heavy composite Majorana neutrino can be produced in association with a lepton, in pp collisions, via quark–antiquark annihilation ($q\bar{q}' \to \ell N_\ell$). Being its own antiparticle, the heavy composite Majorana neutrino can decay either as a neutrino or an anti-neutrino. This implies for instance, that in gauge mediated interactions the reactions $N_\ell \to \ell^+ W^-$ and $N_\ell \to \ell^- W^+$ occur, to lowest order, with equal probability (50%). The production and decay processes can occur via both gauge and contact interactions, although the former is dominated by the contact interaction mechanism [28] for all values of the compositeness scale $\Lambda$ and of the mass of $N_\ell$ relevant in this analysis, as shown in Fig. 2. The dominant interaction in the decay mechanism

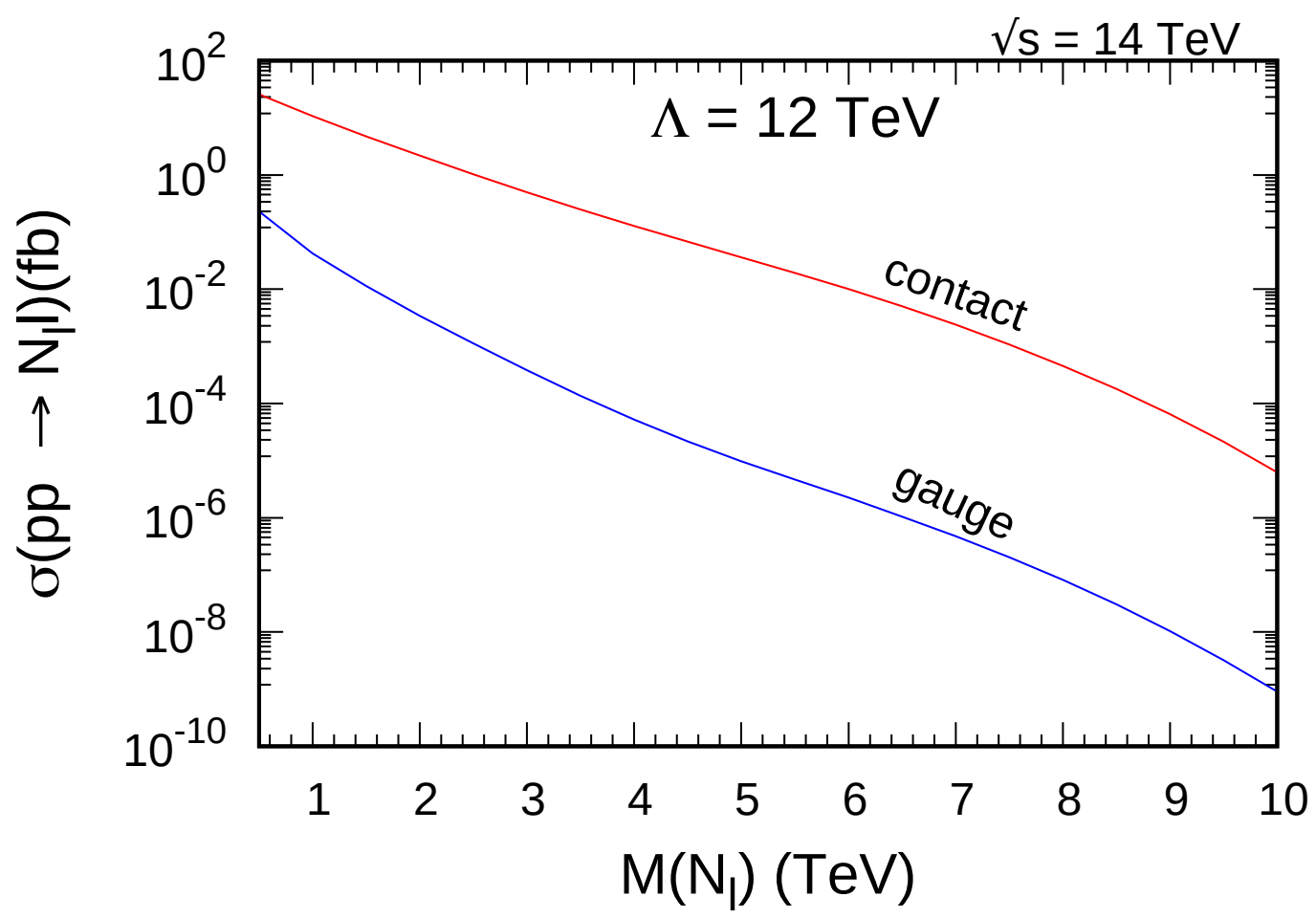

Figure 2: Production cross sections of the heavy composite Majorana neutrino for gauge and contact interactions at $\Lambda = 12$ TeV for pp collisions at $\sqrt{s} = 14$ TeV, obtained with CalcHEP (v3.6) [29].

depends on $\Lambda$ and the mass of $N_\ell$, as shown in Fig. 3. The possible decays are:

$$N_\ell \to \ell q \bar{q}' \qquad N_\ell \to \ell^+ \ell^- \nu(\bar{\nu}) \qquad N_\ell \to \nu(\bar{\nu}) q \bar{q}'.$$

This implies that the allowed final states are:

$$\ell\ell q\bar{q}' \qquad \ell\ell\ell\nu(\bar{\nu}) \qquad \ell\nu(\bar{\nu})q\bar{q}'.$$

In this work, the final state $\ell\ell q\bar{q}'$ is considered, as in the first phenomenological study of the model [23]. We focus on the cases in which $\ell$ is either an electron or a muon, giving rise to the final state signatures $eeq\bar{q}'$ and $\mu\mu q\bar{q}'$. In Fig. 4, the Feynman diagram of the entire process $pp \to \ell N_\ell \to \ell\ell q\bar{q}'$ is shown. We recall that, since particle flavor is conserved within the heavy sector of the model, we do not foresee interference between the electron and muon channels, and thus they are considered separately.

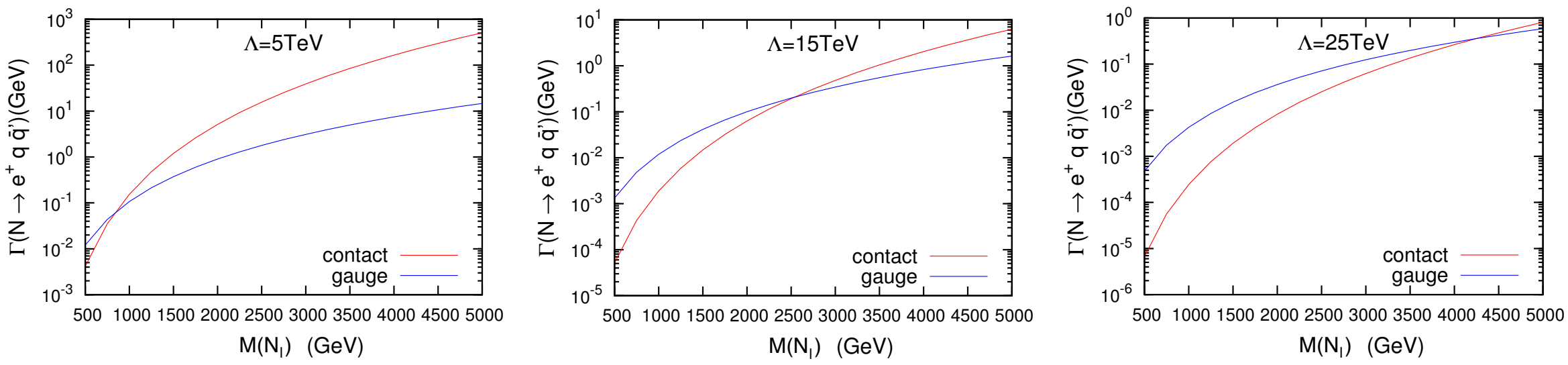

Figure 3: Decay amplitude of the heavy composite Majorana neutrino to a lepton and two quarks, for $\Lambda = 5$ TeV (left), $\Lambda = 15$ TeV (centre), and $\Lambda = 25$ TeV (right), as a function of its mass, obtained with CalcHEP (v3.6) [29]. The x-axis range has been restricted to emphasize the interplay.

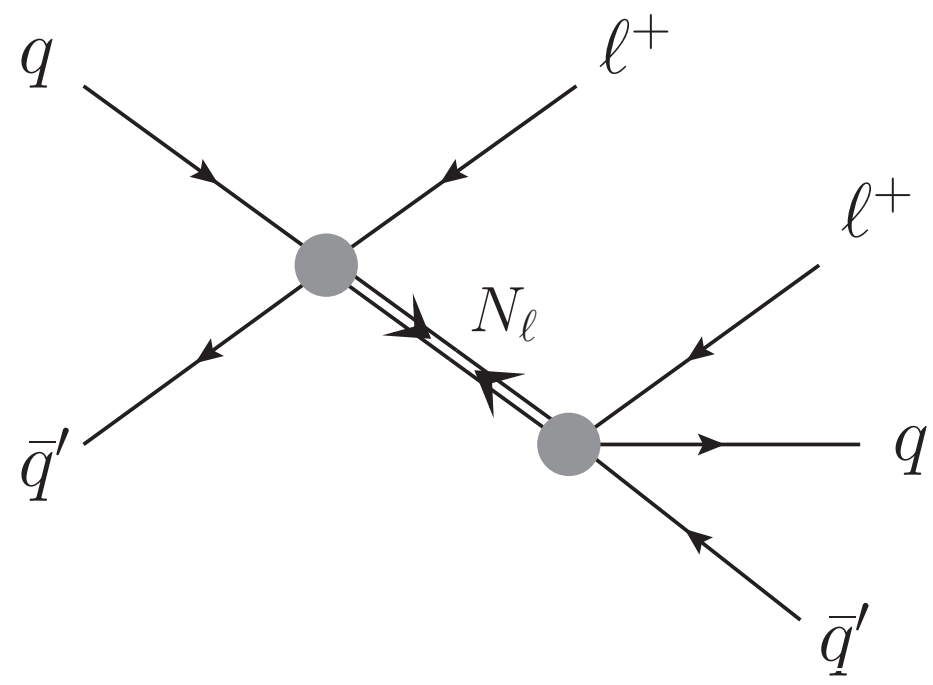

Figure 4: The Feynman diagram of the process for the production and decay of a heavy composite Majorana neutrino, according to the decay chain $pp \to \ell N_\ell \to \ell\ell q\bar{q}'$.

The direct search for a heavy Majorana neutrino within the same framework has been performed by the CMS Collaboration, measuring the final state with two leptons and at least one large-radius jet, with data from pp collisions at $\sqrt{s} = 13$ TeV and with an integrated luminosity of 2.3 fb$^{-1}$ [30]. Good agreement between the data and the SM expectations was observed in the search, and the heavy composite Majorana neutrino is excluded for masses up to 4.60 TeV in the electron channel and 4.70 TeV in the muon channel, for the representative case $\Lambda = M(N_\ell)$.

In the present work, we apply the selection inspired by the Run 2 data analysis within the same theoretical scenario at $\sqrt{s} = 14$ TeV assuming the Phase-2 CMS detector response and a pileup scenario that corresponds to the collection of 3 ab$^{-1}$ of integrated luminosity in ten years of data taking.

## 2 The CMS Phase-2 detector

The CMS detector [31] will be substantially upgraded in order to fully exploit the physics potential offered by the increase in luminosity at the HL-LHC [32], and to cope with the demanding operational conditions at the HL-LHC [33–37]. The upgrade of the first level hardware trigger (L1) will allow for an increase of L1 rate and latency to about 750 kHz and 12.5 $\mu$s, re-spectively, and the high-level software trigger (HLT) is expected to reduce the rate by about a factor of 100 to 7.5 kHz. The entire pixel and strip tracker detectors will be replaced to increase the granularity, reduce the material budget in the tracking volume, improve the radiation hardness, and extend the geometrical coverage and provide efficient tracking up to pseudorapidities of about $|\eta| = 4$. The muon system will be enhanced by upgrading the electronics of the existing cathode strip chambers (CSC), resistive plate chambers (RPC) and drift tubes (DT). New muon detectors based on improved RPC and gas electron multiplier (GEM) technologies will be installed to add redundancy, increase the geometrical coverage up to about $|\eta| = 2.8$, and improve the trigger and reconstruction performance in the forward region. The barrel electromagnetic calorimeter (ECAL) will feature the upgraded front-end electronics that will be able to exploit the information from single crystals at the L1 trigger level, to accommodate trigger latency and bandwidth requirements, and to provide 160 MHz sampling allowing high precision timing capability for photons. The hadronic calorimeter (HCAL), consisting in the barrel region of brass absorber plates and plastic scintillator layers, will be read out by silicon photomultipliers (SiPMs). The endcap electromagnetic and hadron calorimeters will be replaced with a new combined sampling calorimeter (HGCal) that will provide highly-segmented spatial information in both transverse and longitudinal directions, as well as high-precision timing information. Finally, the addition of a new timing detector for minimum ionizing particles (MTD) in both barrel and endcap region is envisaged to provide capability for 4-dimensional reconstruction of interaction vertices that will allow to significantly offset the CMS performance degradation due to high PU rates.

A detailed overview of the CMS detector upgrade program is presented in Ref. [33–37], while the expected performance of the reconstruction algorithms and pile-up mitigation with the CMS detector is summarised in Ref. [38].

## 3 Simulated samples

We use Monte Carlo (MC) samples for the signal and the SM backgrounds. The MC samples for the signal are generated at Leading Order (LO) with CalcHEP (v3.6) [29] for $\sqrt{s} = 14$ TeV proton-proton collisions, using the NNPDF3.0 LO parton distribution functions with the four-flavor scheme [39], for the $\Lambda$ and mass values given in Table 1.

| $\Lambda$ [$TeV$] | $M(N_\ell)$ [$TeV$] | | | |
|---|---|---|---|---|
| 6 | - | - | 6 | - |
| 9 | - | 3 | 6 | 9 |
| 12 | 0.5 | 3 | 6 | 9 |
| 15 | 0.5 | 3 | 6 | - |
| 18 | 0.5 | 3 | - | - |
| 21 | 0.5 | - | - | - |

Table 1: Parameters used in the HL-LHC analysis, $\ell = e, \mu$.

The background samples considered are top quark pair production ($t\bar{t}$), single top quark production (tW), the Drell-Yan (DY) process, W+jets and diboson production (WW, WZ, ZZ), and are generated with MADGRAPH5_aMC@NLO [40] using the CTEQ6L1 PDF set [41]. Following the notation of the Run 2 analysis [42], the $t\bar{t}$ and tW are considered together and called "TTtW", while the W+jets and the diboson production, being a small contribution ($\sim 5\%$ of the total), is referred to as "Other".

For all of the MC samples the hadronization of partons is simulated with PYTHIA 8 [43] and the expected response of the upgraded CMS detector with the fast-simulation package DELPHES [44]. The object reconstruction and identification efficiencies, as well as the detector response and resolution, are parameterized in DELPHES using the detailed simulation of the upgraded CMS detector based on the GEANT4 package [45, 46]. The contribution from 200 additional pileup events has been included in the simulation as well.

# 4 Event selection

Final state objects are reconstructed by the particle flow (PF) algorithm [47]. The PF algorithm combines information from all CMS subdetectors and reconstructs individual particles in the event such as electrons, muons, photons, neutral hadrons and charged hadrons. The event selection of the present analysis is based on exploiting some specific kinematic features of the leptons and of the large-radius jet in the signal samples in order to minimize the contamination from the SM backgrounds.

The transverse momentum $p_T$ of the leading lepton is required to be greater than 110 GeV, while the $p_T$ of the subleading lepton must be greater than 40 GeV. All lepton candidates are required to be in the pseudorapidity range $|\eta| < 2.4$. Restricting to the high-mass region given by $M(\ell, \ell) > 300$ GeV, where $M(\ell, \ell)$ is the dilepton invariant mass, allows reducing the DY background and part of the TTtW background, without affecting the signal acceptance. The large-radius jets, i.e. clustered with a size parameter $R = 0.8$ ("AK8 jets"), are reconstructed using the anti-k$_T$ algorithm [48], implemented in the FASTJET package [49]. The large-radius jets are analyzed using the Pileup-per-particle-identification (PUPPI) mitigation algorithm [50]. This algorithm is designed to remove PU using event information both at the global and local level, identifying pileup at the particle level. The AK8 jets are required to have a minimum $p_T$ of 200 GeV, to be within a pseudorapidity region with $|\eta| < 2.4$ and to be separated from leptons by a distance $\Delta R = \sqrt{(\Delta\eta)^2 + (\Delta\phi)^2} > 0.8$. The process under consideration has a fairly central distribution of its final state particles. A study of the effect of increasing the pseudo-rapidity acceptance for the final state objects, made possible by the Phase-2 CMS detector upgrade, has shown that the increased background contribution would spoil the advantage given by the larger efficiency, lowering the overall sensitivity. Hence the more central selection, $|\eta| < 2.4$, has been kept for both leptons and large-radius jets. Requiring one or more large-radius jets is suitable regardless of whether $N_\ell$ decays through gauge or contact interactions. In fact, for gauge mediated decays of the heavy composite neutrino, the two quarks are expected to overlap and thus form a large-radius jet, while in the case of contact-mediated decays, the two quarks are well separated, but form two large-radius jets because of the overlap with final state radiation. The signal region is therefore defined by requiring two same-flavor isolated leptons (electrons or muons) and at least one large-radius jet. With this selection, the total efficiency for the signal is about 55% in the $eeq\bar{q}'$ channel and 65% in the $\mu\mu q\bar{q}'$ channel, for heavy neutrinos with masses greater than 3 TeV.

A shape-based analysis is performed investigating the invariant mass, $M(\ell\ell J)$, of the two leptons and the leading large-radius jet. As shown in Fig. 5, this variable provides a good discrimination between the signal and the SM backgrounds.

# 5 Results

Figure 5 shows the distribution of the variable $M(\ell\ell J)$ for both the SM backgrounds and the signal for a particular benchmark choice of the model parameters, namely $\Lambda = M(N_\ell) = 6$

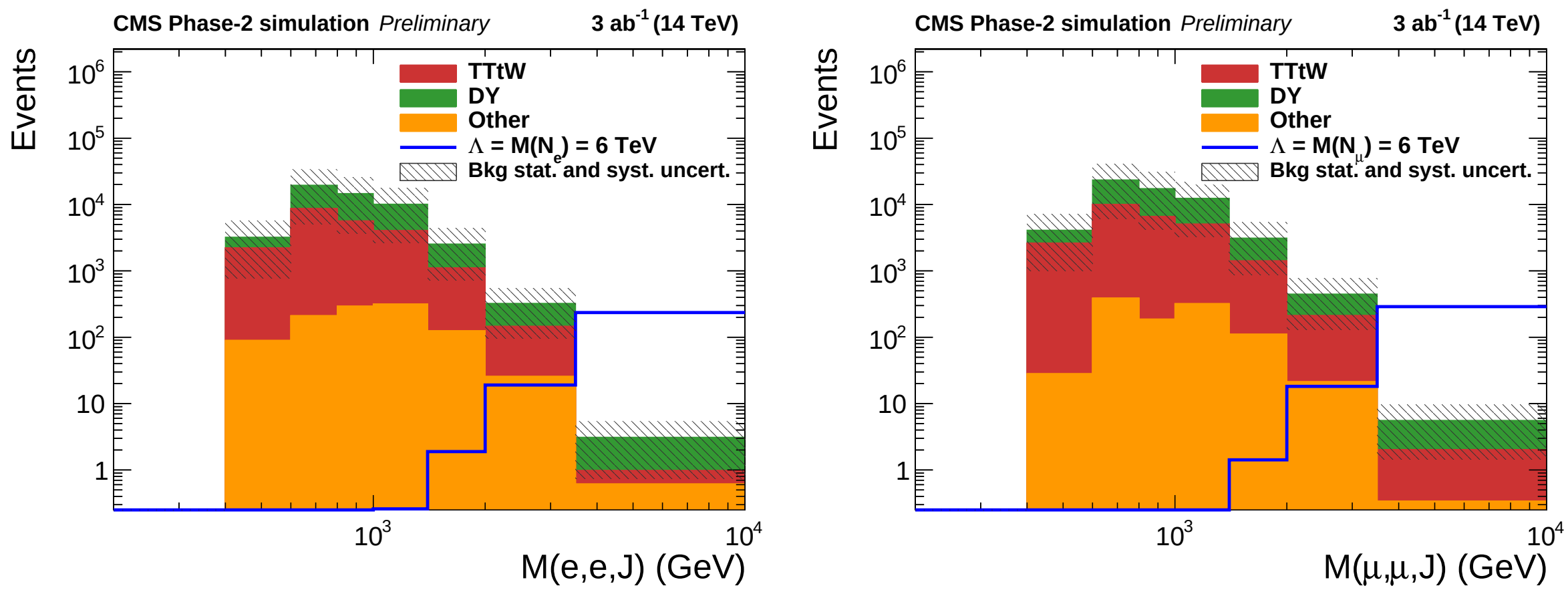

Figure 5: Distribution of the variable $M(\ell\ell J)$ of backgrounds (stacked plots) and expected signal (lines) in the signal region, considering the model parameters $\Lambda = M(N_\ell) = 6$ TeV, for the $eeq\bar{q}'$ channel (left) and for the $\mu\mu q\bar{q}'$ channel (right). The background statistical and systematic uncertainties have been combined.

TeV. The expected discovery sensitivity of a heavy composite Majorana neutrino, produced in association with a lepton, and decaying into a same-flavor lepton and two jets, is shown in Figure 6. The CMS Phase-2 detector will be able to find evidence for a composite neutrino with mass below $M(N_\ell) = 7.6$ TeV.

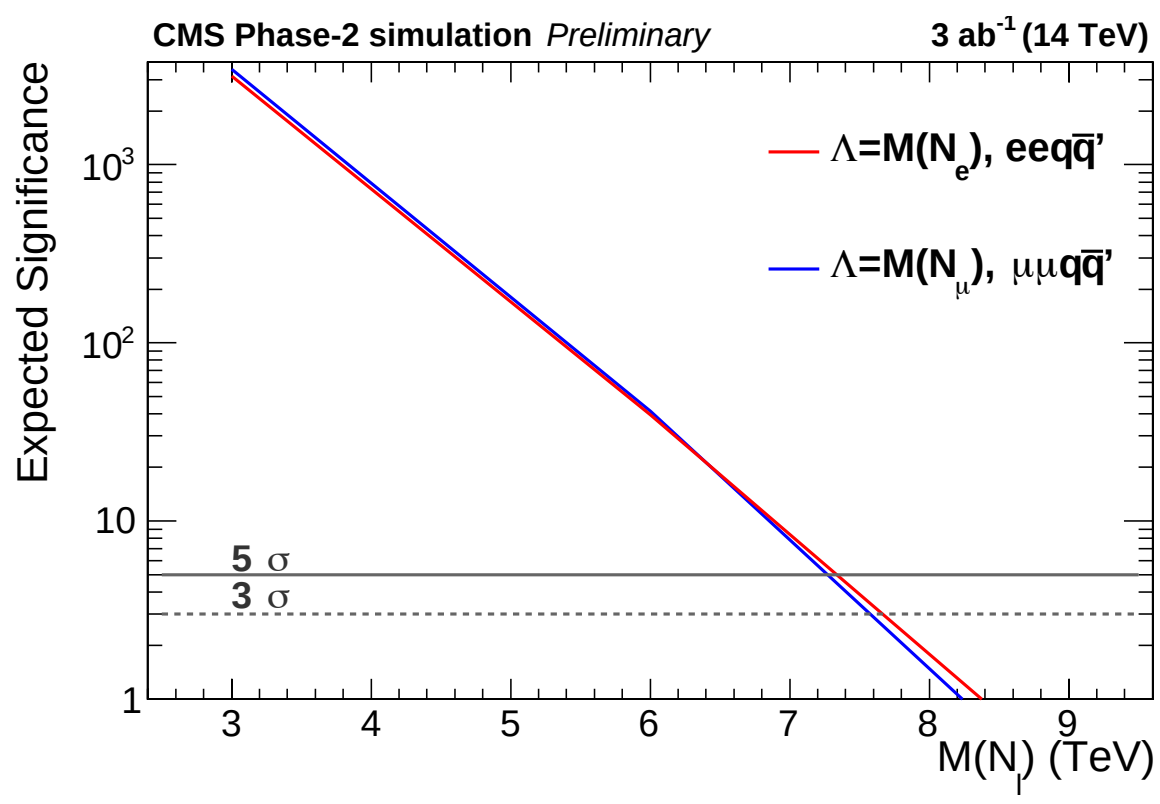

Figure 6: The expected statistical significance for both the $eeq\bar{q}'$ (red line) and $\mu\mu q\bar{q}'$ (blue line) channel for the case $\Lambda = M(N_\ell)$. The gray solid (dotted) line represents 5 (3) standard deviations, respectively.

Expected exclusion limits on the mass of the heavy neutrino are also evaluated. An asymptotic $CL_s$ criterion [51, 52] is used to set an upper limit at 95% confidence level on the cross section of the heavy composite Majorana neutrino produced in association with a lepton times its branching fraction to a same-flavor lepton and two quarks, $\sigma(pp \to \ell N_\ell) \times \mathcal{B}(N_\ell \to \ell q\bar{q}')$. The $M(\ell\ell J)$ distributions from MC simulations of signal and SM backgrounds are used as input in the limit computation together with the systematic uncertainties, as discussed in Ref. [30]. The systematic uncertainties, listed in Table 2, are evaluated in accordance with the most recent recommendations [53] and assumed to be independent of mass. The results are presented in Fig. 7 for the $eeq\bar{q}'$ channel and for the $\mu\mu q\bar{q}'$ channel for different values of the compositeness scale: $\Lambda = 12, 24, 35$ TeV and $\Lambda = M(N_\ell)$. Figure 8 displays the corresponding upper limits on the $(\Lambda, M(N_\ell))$ plane for both of the final states considered. The extrapolation is similar for the two channels. We see that the sensitivity to $\Lambda$ is higher at low neutrino masses, reaching the exclusion of $\Lambda = 35$ TeV at $M(N_\ell) = 0.5$ TeV and decreases at higher $N_\ell$ masses. For the representative case $\Lambda = M(N_\ell)$, while in the analysis of Run 2 data it was possible to exclude heavy neutrino masses up to 4.60 (4.70) TeV in the $eeq\bar{q}'$ ($\mu\mu q\bar{q}'$) channel [30], under planned HL-LHC conditions this limit would be extended to 8 TeV.

# 6 Prospects for the HE-LHC

In this section we extend the study presented above to the HE-LHC, which is projected to reach a centre-of-mass energy of 27 TeV and an integrated luminosity of 15 ab$^{-1}$.

The MC simulation samples for the signal are generated at Leading Order (LO) with CalcHEP (v3.6) [29] for $\sqrt{s} = 27$ TeV proton-proton collisions, using the NNPDF3.0 LO parton distribution functions with the four-flavor scheme [39]. The signal samples are generated for both the $eeq\bar{q}'$ and $\mu\mu q\bar{q}'$ channels for the following choice of parameters: $\Lambda = 10, 100$ TeV for $M(N_\ell) = 0.5, 5, 10$ TeV, $\Lambda = 25, 50$ TeV for $M(N_\ell) = 0.5, 5, 10, 12.5, 15$ TeV, and $\Lambda = M(N_\ell) \in [3, 18]$ TeV. The main background sources considered are the top quark pair production, Drell-Yan (DY) process, vector boson+jet and diboson production, and are provided by the Future Circular Collider group [54]. For both signal and background samples the parton hadronization is treated with PYTHIA 8 and the response of the detector (assumed to perform the same as the CMS Phase-2 detector) with DELPHES.

In the spirit of a projection study, the event selection criteria remain identical to those of the HL study described in Section 4.

The expected discovery sensitivity and the exclusion limits are extracted using the $M(\ell\ell J)$ variable shown in Fig. 9. According to the prescriptions given in Ref. [53], the systematic uncertainties are those listed in Table 2. Figure 10 shows that with the HE-LHC we could find evidence for a composite Majorana neutrino with mass below $M(N_\ell) = 12$ TeV, for $\Lambda = M(N_\ell)$. Upper limits on the production cross section times the branching fraction $\sigma(pp \to \ell N_\ell) \times \mathcal{B}(N_\ell \to \ell q\bar{q}')$ are shown in Fig. 11 for some benchmark choices of the parameters. The projection on the exclusion limits is also presented in Fig. 12 for the $(\Lambda, M(N_\ell))$ plane. We can conclude that, given the model condition $\Lambda = M(N_\ell)$, the HE-LHC could exclude a heavy composite Majorana neutrino with mass up to 12.5 TeV in both $eeq\bar{q}'$ and $\mu\mu q\bar{q}'$ channels.

Table 2: List of systematic uncertainties used in the present analysis.

| Source | Value |
|---|---|
| Integrated luminosity | 1% |
| Pileup | 2% |
| Electron ID | 0.5% |
| Electron scale | 0.5% |
| Muon ID | 0.5% |
| Muon scale | 0.5% |
| Jet energy scale | 1% |
| Jet energy resolution | 1% |
| Background | 0.3% |
| Drell-Yan (theory) | 4% |

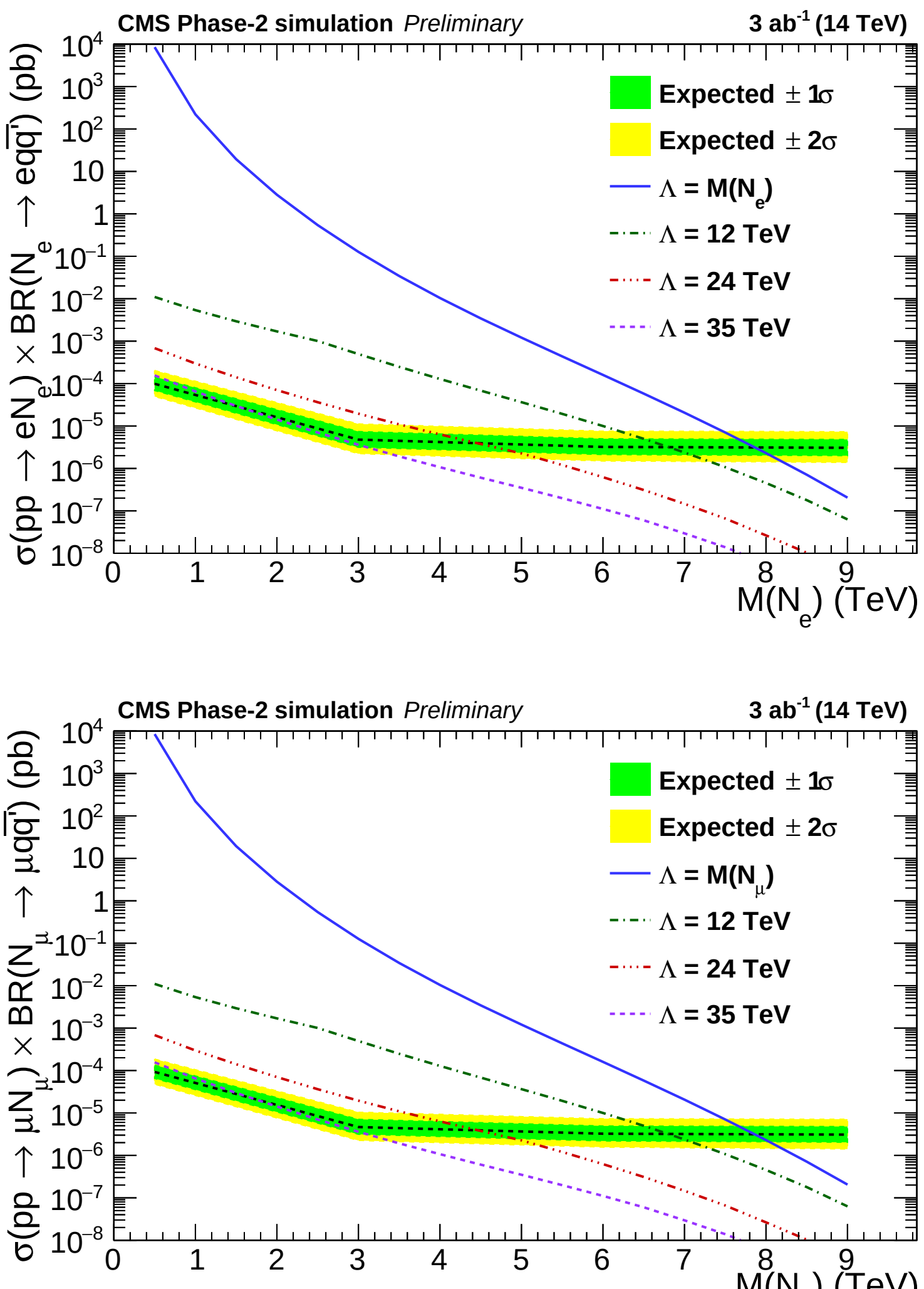

Figure 7: The expected 95% CL upper limits (black dotted lines) on $\sigma(pp \to \ell N_\ell) \times \mathcal{B}(N_\ell \to \ell q\bar{q}')$, obtained in the analysis of the $eeq\bar{q}'$ (top) and the $\mu\mu q\bar{q}'$ (bottom) final states, as a function of the mass of the heavy composite Majorana neutrino. The corresponding green and yellow bands represent the expected variation of the limit to one and two standard deviation(s). The solid blue curve indicates the theoretical prediction of $\Lambda = M(N_\ell)$. The textured curves give the theoretical predictions for $\Lambda$ values ranging from 12 to 35 TeV.

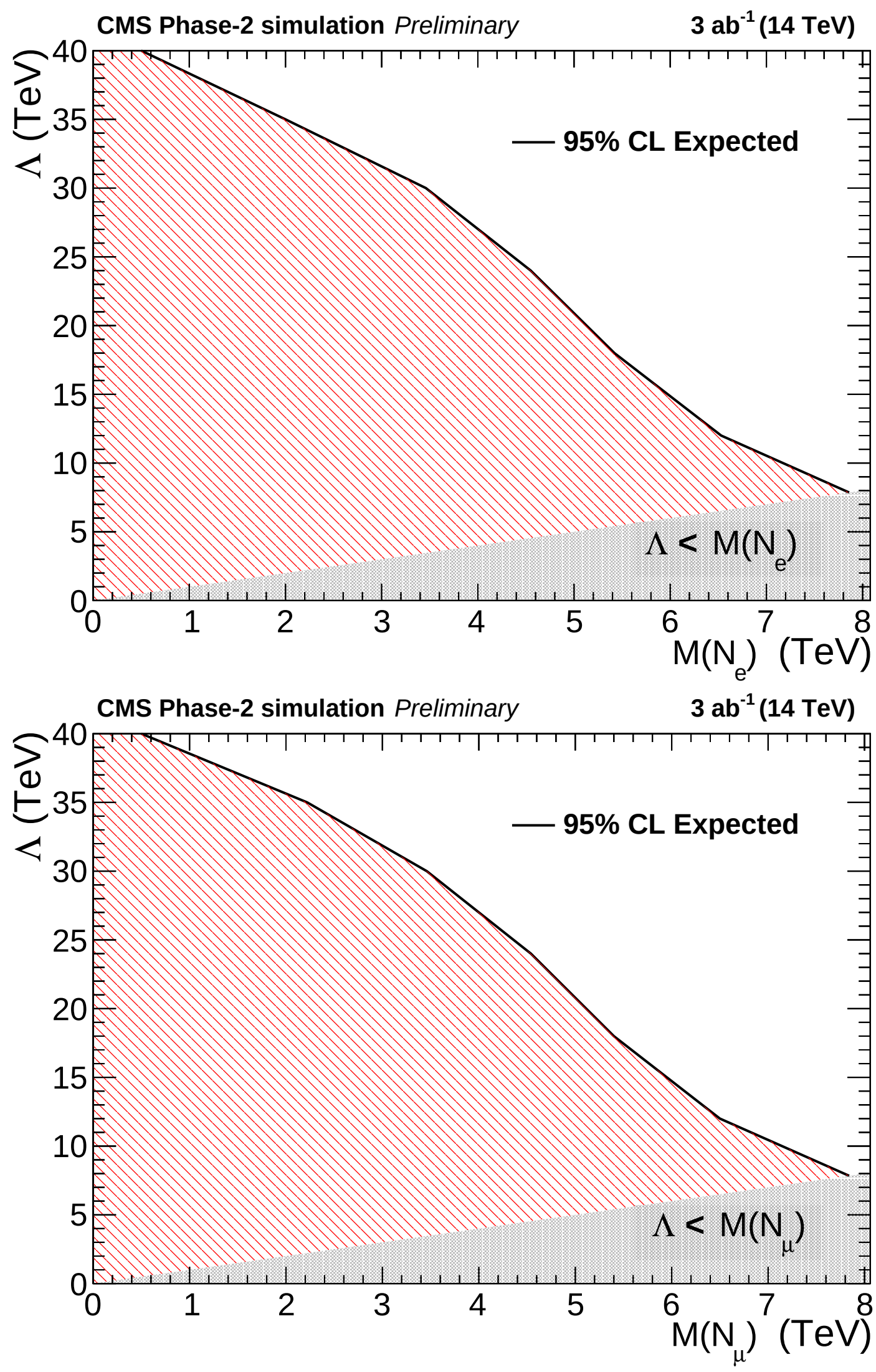

Figure 8: The expected 95% CL lower limits (black lines) on the compositeness scale $\Lambda$, obtained in the analysis of the $eeq\bar{q}'$ (top) and the $\mu\mu q\bar{q}'$ (bottom) final states, as a function of the mass of the heavy composite Majorana neutrino. The gray zone corresponds to the phase space $\Lambda < M(N_\ell)$ not allowed by the model.

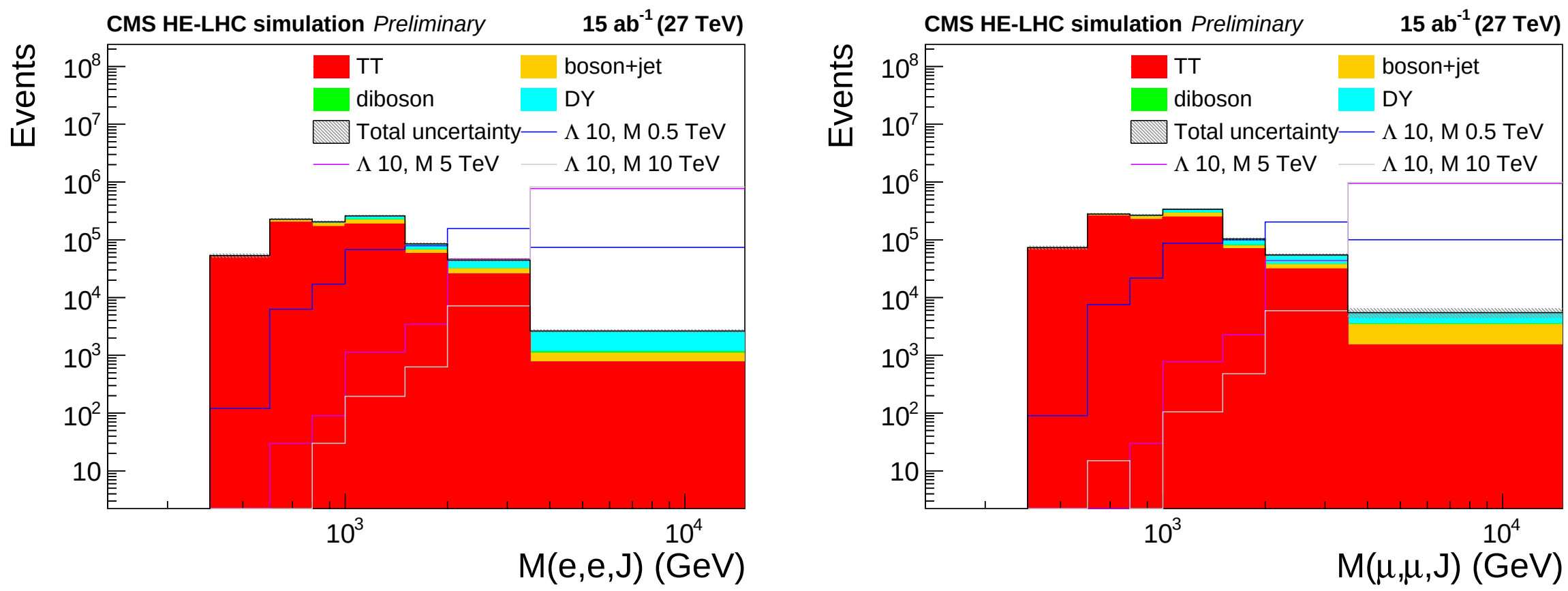

Figure 9: Distribution of the invariant mass of two leptons and leading large-radius jet of backgrounds and three signal samples for the $eeq\bar{q}'$ channel (left) and $\mu\mu q\bar{q}'$ channel (right) for the HE-LHC. The error bands are given by the combination of systematic and statistical uncertainties.

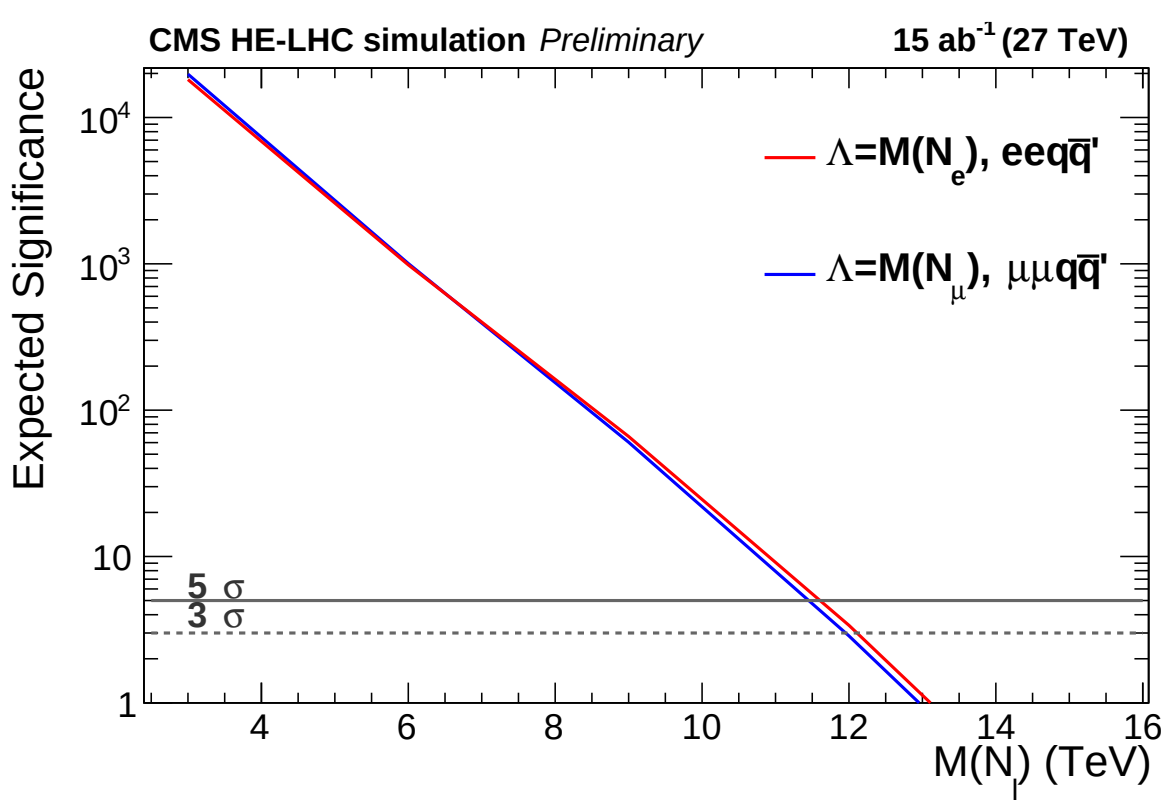

Figure 10: The expected statistical significance for the HE-LHC projection of the $eeq\bar{q}'$ (red line) and the $\mu\mu q\bar{q}'$ (blue line) channel for the case $\Lambda = M(N_\ell)$. The gray solid (dotted) line represents 5 (3) standard deviations, respectively.

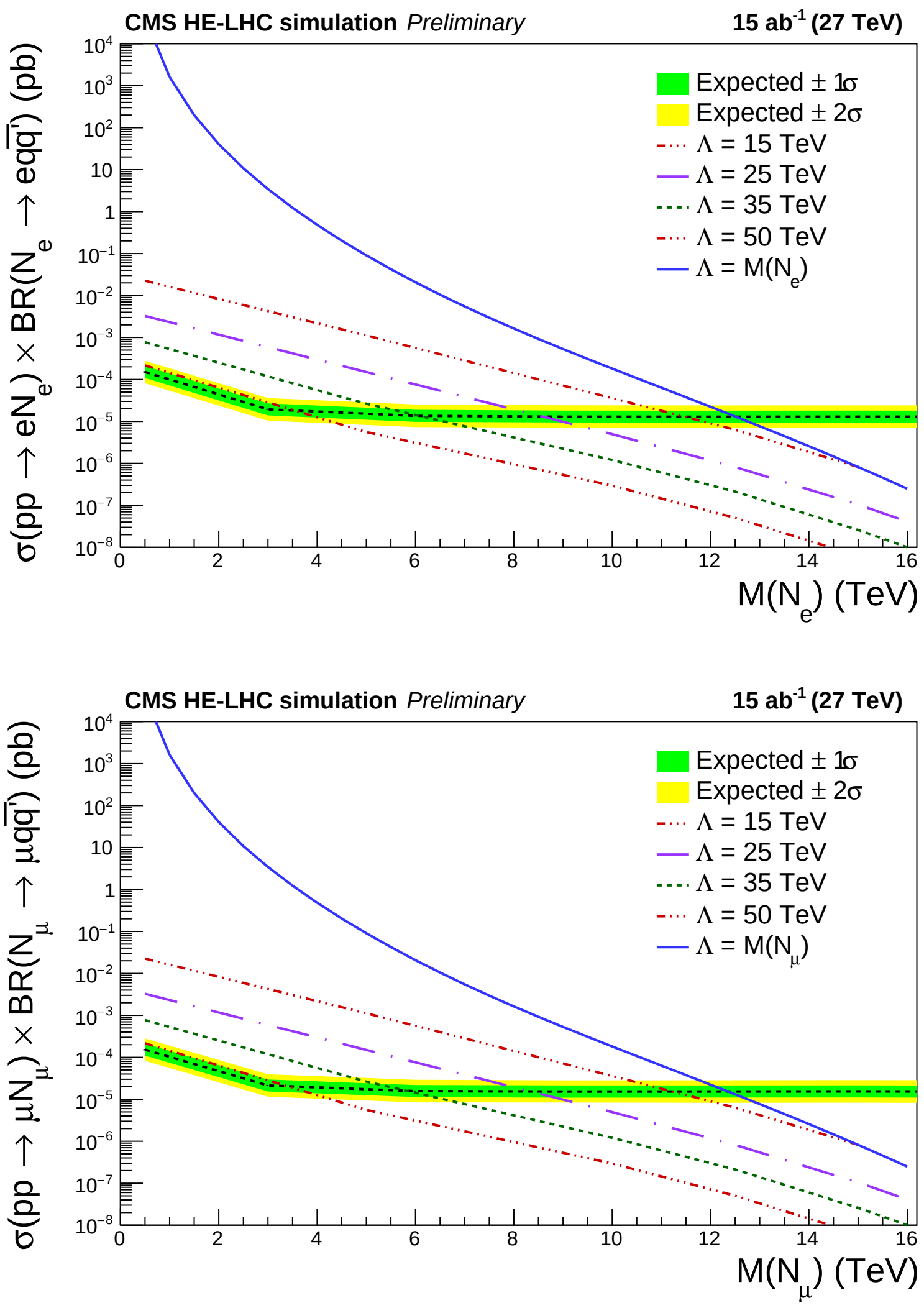

Figure 11: The expected 95% CL upper limits for the HE-LHC projection of the $eeq\bar{q}'$ channel (top) and the $\mu\mu q\bar{q}'$ channel (bottom). The cross section limits are higher in the HE case because of the much larger background expectation at 27 TeV.

**12**

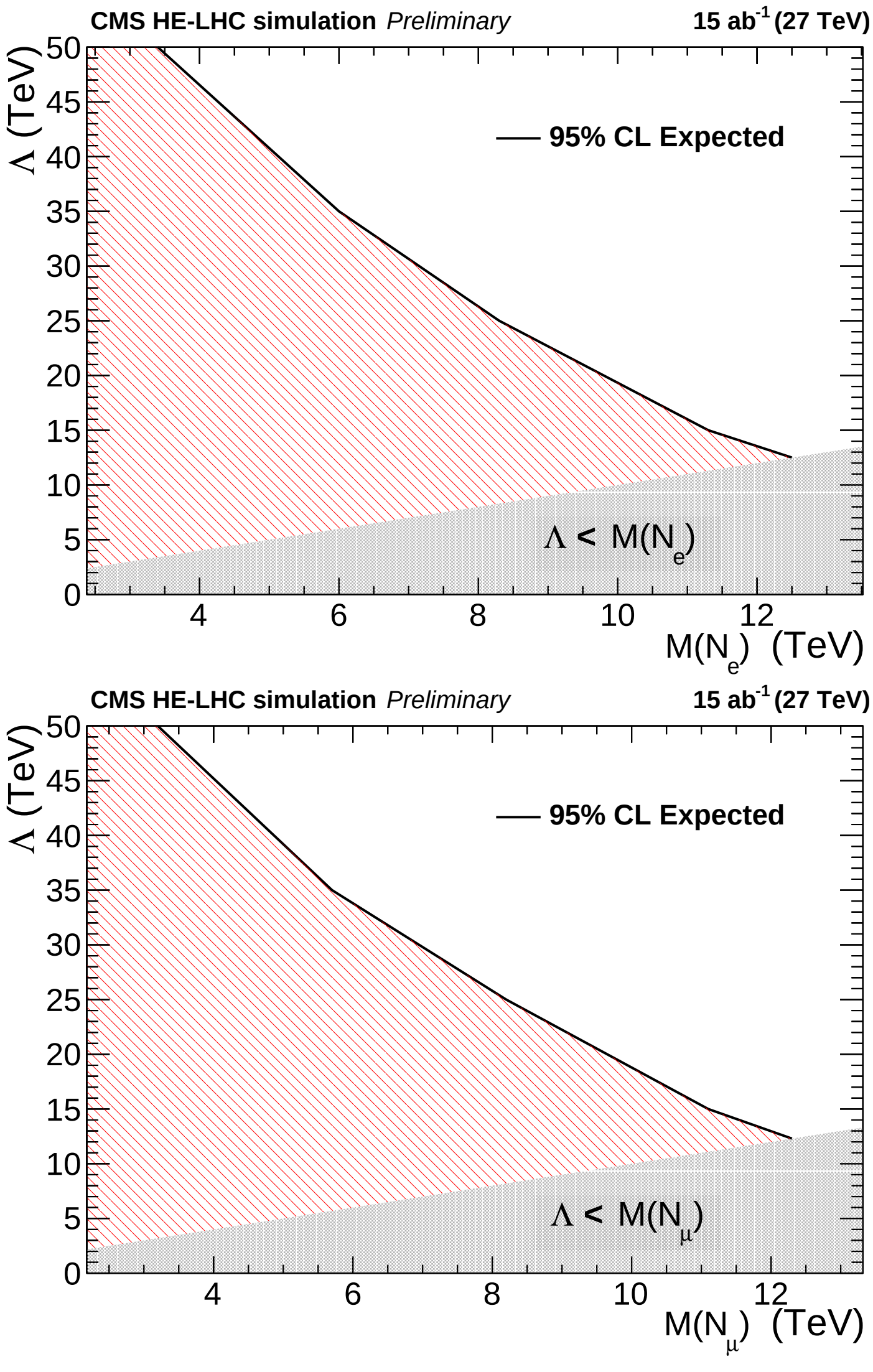

Figure 12: The expected 95% CL lower limits (black lines) on the compositeness scale Λ, obtained in the analysis of the $eeq\bar{q}'$ (top) and the $\mu\mu q\bar{q}'$ (bottom) final states, as a function of the mass of the heavy composite Majorana neutrino for the HE-LHC projection. The red shaded zone highlights the excluded parameter space. The gray zones are not allowed by the model.

# 7 Summary

Studies have been conducted of the expected performance at the HL-LHC and at the HE-LHC of a search for a composite Majorana neutrino. The study has been carried out considering a heavy composite Majorana neutrino produced in association with a lepton and decaying into a same-flavor lepton plus two quarks, with the requirement of two leptons and at least one large-radius jet in the signal region. The HL-LHC running conditions and Phase-2 detector allow a significant extension of the parameter space that can be probed.

# CMS Physics Analysis Summary

Contact: cms-phys-conveners-ftr@cern.ch 

# Search for $\mathrm{t\bar{t}}$ resonances at the HL-LHC and HE-LHC with the Phase-2 CMS detector

The CMS Collaboration

## Abstract

A search for a heavy resonance decaying into a $\mathrm{t\bar{t}}$ pair is presented using the upgraded Phase-2 CMS detector design at the High-Luminosity LHC (HL-LHC) and High-Energy LHC (HE-LHC), with center-of-mass energies of 14 and 27 TeV, respectively, and integrated luminosities of 3 and 15 $\mathrm{ab^{-1}}$. Two distinct final states with either a single lepton or no leptons are considered. Jet substructure techniques and top quark identification algorithms are used for the object reconstruction. At the HL-LHC (HE-LHC), the production of a Randall–Sundrum gluon can be excluded at 95% confidence level with a mass up to 6.6 (10.7) TeV or can be discovered at $5\sigma$ significance with a mass up to 5.7 (9.4) TeV.

# 1 Introduction

Many models of new physics predict heavy resonances with enhanced couplings to the third generation of fermions in the standard model (SM) [1–9]. Thus, the study of the top quark can provide important insights into the validity of such models. This analysis considers top quark pair production to search for the presence of heavy resonances. In particular, we focus on the production of a Randall–Sundrum Kaluza–Klein gluon (RSG) [8].

This note presents projections for the search for resonant $t\bar{t}$ production with simulated events at center-of-mass energies of 14 and 27 TeV and the upgraded Phase-2 CMS detector in the all-jets and lepton-plus-jets final states. The average number of proton-proton (pp) interactions per bunch crossing (pileup) is assumed to be 200. In the high mass ranges accessible at the High-Luminosity LHC (HL-LHC) with $\sqrt{s} = 14$ TeV and the High-Energy LHC (HE-LHC) with $\sqrt{s} = 27$ TeV, reconstructing the event topology of $t\bar{t}$ production requires special techniques. Jet substructure variables and top quark identification algorithms are used to handle the case where the hadronic decay products of the top quark are fully merged into a single jet. This is likely to occur if a hypothetical RSG resonance has a mass larger than 1 TeV. Figure 1 shows a schematic representation of a fully hadronic $t\bar{t}$ event, where each top quark decays as $t \to Wb$.

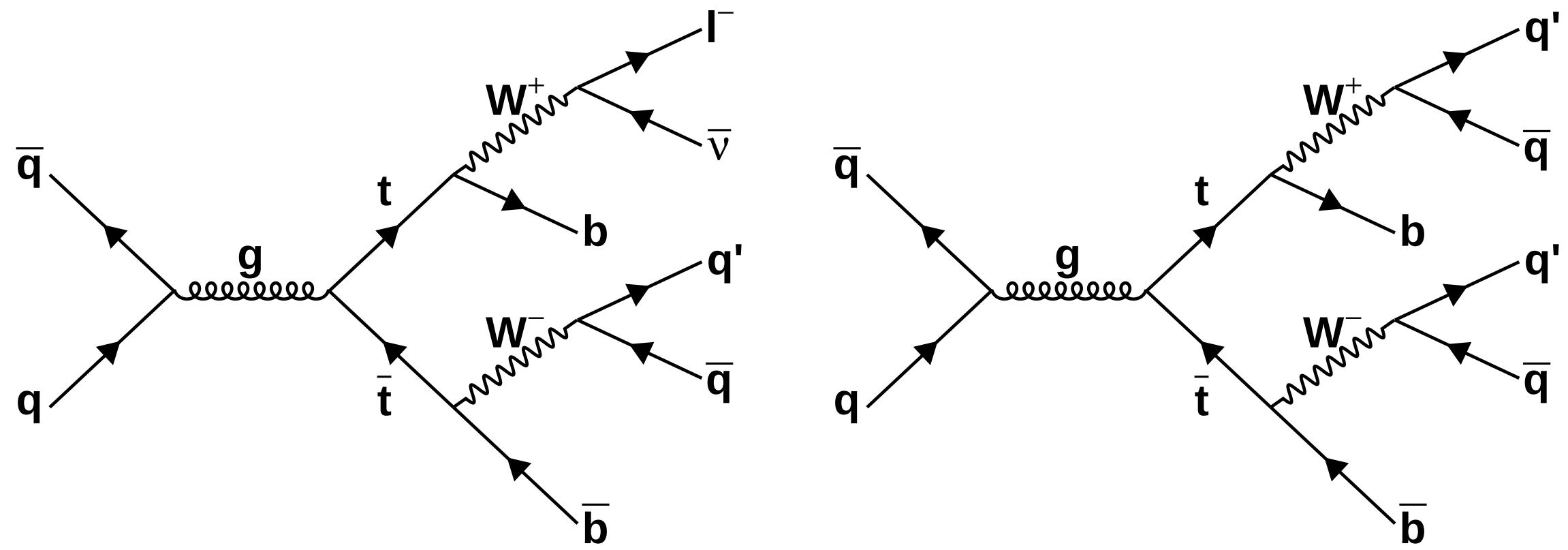

Figure 1: Leading order Feynman diagrams showing pair production and decays of top quark with (left) single-lepton and (right) fully hadronic final states.

The results of the search will be presented as a combination of the all-jets and single-lepton plus jets final states with boosted topologies. The single-lepton final state considers a single electron or a single muon. A search for $t\bar{t}$ resonances in all-hadronic final states was previously performed by CMS at $\sqrt{s} = 7$ TeV [10], 8 TeV [11], and 13 TeV [12].

The dominant background for the all-jets final state, given that dijet events are selected, is quantum chromodynamics (QCD) multijet production. In analyses that use observed data, this contribution is determined by a data-based method, called the modified mass procedure [12], in which the top-tagging misidentification rate is derived in a QCD-enriched sideband. However, for this study the contribution will be estimated from simulated events. The systematic uncertainties associated to this background, nevertheless, will be estimated assuming that the modified mass procedure is used. Background contributions from standard model $t\bar{t}$ production, the most dominant background source in the single-lepton final state, are also determined from simulation. The second most dominant background for the single-lepton final state is represented by production of a W boson in association with one or more jets. Additional background sources are due to Drell-Yan processes, diboson production, and single top quark production.

## 2 The CMS detector

The CMS detector [13] will be substantially upgraded for the HL-LHC in order to fully exploit the physics potential offered by the increase in luminosity and to cope with the demanding operational conditions [14–18].

The upgrade of the first level hardware trigger (L1) will allow for an increase of the rate and latency to about 750 kHz and 12.5 $\mu$s, respectively. The high-level software trigger is expected to reduce the rate by about a factor of 100 to 7.5 kHz.

The entire pixel and strip tracker detectors will be replaced to increase the granularity, to reduce the material budget in the tracking volume, to improve the radiation hardness, and to extend the geometrical coverage with efficient tracking up to pseudorapidities of about $|\eta| = 4$.

The muon system will be enhanced by upgrading the electronics of the existing cathode strip chambers, resistive plate chambers (RPC), and drift tubes. New muon detectors based on improved RPC and gas electron multiplier technologies will be installed to add redundancy, to increase the geometrical coverage up to about $|\eta| = 2.8$, and to improve the trigger and reconstruction performance in the forward region.

The barrel electromagnetic calorimeter will feature the upgraded front-end electronics that will be able to exploit the information from single crystals at the L1 trigger level, to accommodate trigger latency and bandwidth requirements, and to provide 160 MHz sampling to allow high-precision time measurements for photons. The barrel hadronic calorimeter, which consists of brass absorber plates and plastic scintillator layers, will be read out by silicon photomultipliers. The endcap electromagnetic and hadron calorimeters will be replaced with a new combined high granularity sampling calorimeter that will provide highly-segmented spatial information in both transverse and longitudinal directions, as well as high-precision time information.

Finally, the addition of new timing detectors for minimum ionizing particles in both the barrel and endcap regions is envisaged to provide the capability for 4-dimensional reconstruction of interaction vertices that will significantly mitigate the CMS performance degradation due to high pileup rates.

A detailed overview of the CMS detector upgrade program is presented in Refs. [14–18], while the expected performance of the reconstruction algorithms and pileup mitigation with the CMS detector is summarised in Ref. [19].

## 3 Simulation

The RSG signal processes are generated using PYTHIA 8.212 [20] at leading order (LO), assuming a decay width of 17%, for a mass of the resonance ranging from 2 to 12 TeV. A variety of event generators are used for the Monte Carlo (MC) simulation of the background processes. The POWHEG 2.0 [21–24] event generator is used to generate $\mathrm{t\bar{t}}$ and single top quark events in the $t$-channel and tW channel to next-to-LO (NLO) accuracy. The single top quark events in the $s$-channel, Z+jets, and W+jets are simulated using MADGRAPH5_aMC@NLO 2.2.2 [25]. The Z+jets and W+jets events are generated at LO using the MLM matching scheme [26]. The single top quark events in the $s$-channel are generated at NLO using the FxFx matching scheme [27]. The PYTHIA event generator is used to simulate the QCD multijet and WW events at NLO.

Parton showering, hadronization, and the underlying event are simulated with PYTHIA, using NNPDF 3.0 parton distribution functions (PDFs) and the CUETP8M1 [28, 29] tune for all processes, except for the $\mathrm{t\bar{t}}$ sample which is produced with the CUETP8M2T4 [30] tune. The Phase-2 CMS detector simulation and the reconstruction of physics objects are performed with the Delphes software package [31].

# 4 Object reconstruction

The particle flow (PF) algorithm [32] is used together with the pileup per particle identification (PUPPI) [33] method to reconstruct the final state objects such as electrons, muons, jets, and missing transverse momentum. The leptons, small-radius jets, and missing tranverse momentum are used only for the single-lepton final state. The large-radius jets are used for both final states.

The events are required to have at least one primary reconstructed vertex. The reconstructed vertex with the largest value of summed physics-object $p_{\mathrm{T}}^2$ is taken to be the primary pp interaction vertex. The physics objects are the jets, clustered using the jet finding algorithm [34, 35] with the tracks assigned to the vertex as inputs, and the associated missing transverse momentum, taken as the negative vector sum of the $p_{\mathrm{T}}$ of those jets.

The "medium" working point for electron identification criteria (ID) and the "tight" working point for muon ID are used. The electrons are selected if they have $p_{\mathrm{T}} > 80\,\mathrm{GeV}$ and $|\eta| < 3$. The muons are required to have $p_{\mathrm{T}} > 55\,\mathrm{GeV}$ and $|\eta| < 3$. Exactly one lepton is required in the single-lepton final state. The selected leptons are not required to be isolated because of the boosted final state that often places them near other particles.

Jets are clustered from the reconstructed PF candidates using the anti-$k_{\mathrm{T}}$ algorithm [34] with the Fastjet 3.1 software package [35] with a size parameter of 0.4 (AK4). Jets which overlap with the selected lepton have the lepton energy subtracted. We consider only AK4 jets that have $p_{\mathrm{T}} > 30\,\mathrm{GeV}$ and $|\eta| < 4$. The single-lepton final state requires at least two AK4 jets. The jets are ordered by their $p_{\mathrm{T}}$ values; the first jet is required to have $p_{\mathrm{T}} > 185\,(150)\,\mathrm{GeV}$ and the second jet is required to have $p_{\mathrm{T}} > 50\,(50)\,\mathrm{GeV}$ in the electron (muon) channel.

The missing transverse momentum ($\vec{p}_{\mathrm{T}}^{\,\mathrm{miss}}$) is defined as the negative of the vector sum of the $p_{\mathrm{T}}$ of all reconstructed PF candidates in an event and its magnitude is denoted as $p_{\mathrm{T}}^{\mathrm{miss}}$. In the single-lepton final state, $p_{\mathrm{T}}^{\mathrm{miss}}$ is required to be greater than 120 (50) GeV in the electron (muon) channel.

Both final states use large-radius anti-$k_{\mathrm{T}}$ jets with a size parameter of 0.8 (AK8). PF candidates weighted by the PUPPI algorithm are used as input for the AK8 jet clustering. The AK8 jet mass is computed from the jet components remaining after the soft-drop grooming procedure [36] is applied with $\beta = 0$ and $z_{\mathrm{cut}} = 0.1$. For this choice of parameters, the soft-drop algorithm is identical to the modified mass-drop procedure from [37]. This is called the soft-drop mass ($m_{\mathrm{SD}}$). Additionally, the $N$-subjettiness ($\tau_N$) jet substructure variables [38] are computed. In particular, the ratio $\tau_3/\tau_2$ provides the best discrimination between jets from top quarks and jets from light-flavored quarks (u, d, s, c) or gluons (g). The implementation of these algorithms as provided in the Delphes package is used. In the single-lepton final state, the AK8 jets are required to be separated from the selected lepton by $\Delta R(\text{lepton}, \text{AK8 jet}) > 0.8$, where $\Delta R = \sqrt{(\Delta\phi)^2 + (\Delta\eta)^2}$ and $\phi$ is the azimuthal angle. In the fully hadronic final state, the soft-drop subjets are tagged as originating from the production of a b quark by using the deep combined secondary vertex (DeepCSV) algorithm [39]. The efficiency for tagging true b jets is around 49% and probability for mis-tagging light quarks is roughly 1%. The subjet b tagging is used to categorize events in the fully hadronic final state.

# 5 Event selection and reconstruction

For top quarks with a large boost, $p_{\mathrm{T}} > 400\,\mathrm{GeV}$, an algorithm based on the soft-drop mass $m_{\mathrm{SD}}$, the $N$-subjettiness ratio $\tau_3/\tau_2$, and subjet b-tagging is used to identify the decay of the top quark with no leptons. Figure 2 shows the distributions of generated and reconstructed RSG mass in each final state after the full event selection. Because of the off-shell production at high mass, the RSG signals display broader spectra as the invariant mass increases. As shown in Table 1, this distorts the mass distribution toward lower masses.

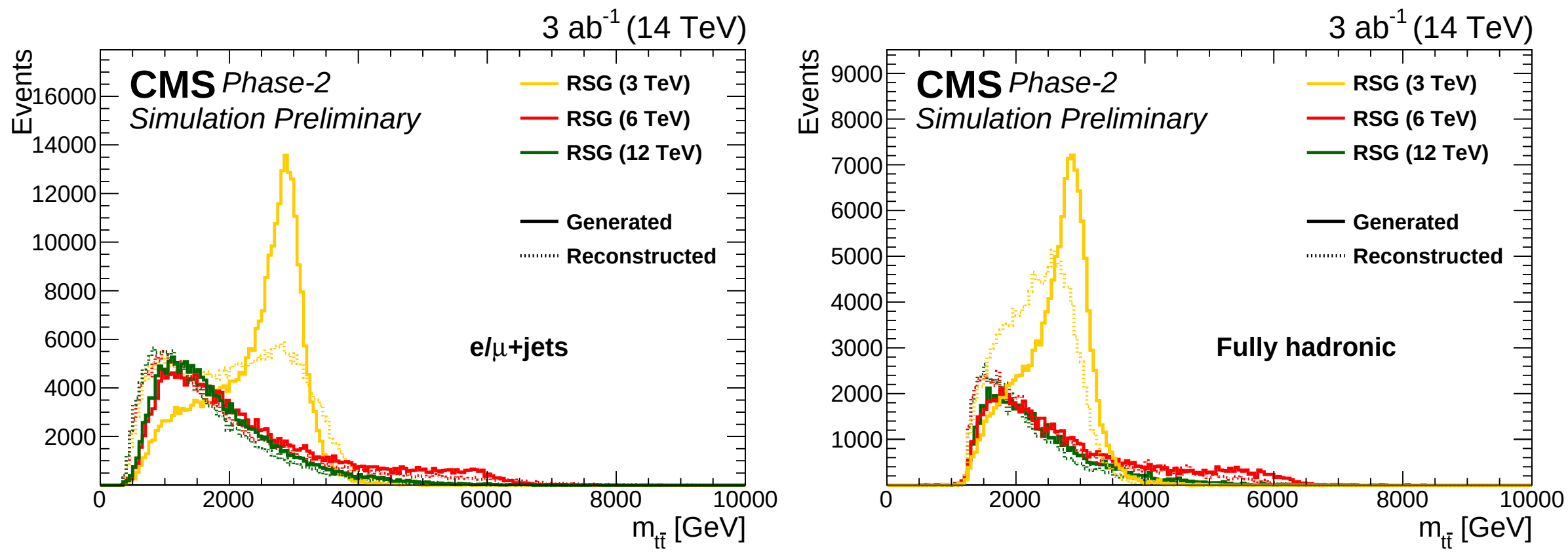

Figure 2: The generated and reconstructed RSG mass distributions in the (left) single-lepton and (right) fully hadronic final states. The distributions are shown after the full event selection in each final state, as described in Sections 5.1 and 5.2.

Table 1: Summary of the selection efficiencies for the two final states for the signal hypotheses considered.

| RSG mass [TeV] | Single-lepton efficiency [%] | Fully hadronic efficiency [%] |
|---|---|---|
| 2 | 9.6 | 4.2 |
| 3 | 10 | 4.6 |
| 4 | 8.8 | 3.8 |
| 5 | 7.7 | 3.0 |
| 6 | 6.9 | 2.5 |
| 8 | 6.2 | 2.1 |
| 10 | 6.1 | 2.0 |
| 12 | 6.0 | 1.9 |

## 5.1 Single-lepton final state

The single-lepton final state consists of events where one of the top quarks in the event decays leptonically and the other decays hadronically. The event selection and reconstruction from Ref. [12] is implemented. In addition to the selections described in Section 4, the single-muon events are required to have $H_{\mathrm{T}}^{\mathrm{lep}} > 150\,\mathrm{GeV}$, where $H_{\mathrm{T}}^{\mathrm{lep}} = p_{\mathrm{T}}^{\mathrm{miss}} + p_{\mathrm{T}}^{\mathrm{lep}}$.

The large-radius AK8 jets are also used in the single-lepton final state to identify hadronically decaying top quarks. In addition to the requirements in Section 4, we select the AK8 jets that are separated from the lepton by $\Delta R(\text{lepton, AK8 jet}) > 0.8$. The events with two or more AK8 jets satisfying these criteria are rejected in order to avoid any overlap with the selection for the fully hadronic final state.

The mass of the $\mathrm{t\bar{t}}$ system provides strong discrimination between signal and background events. We follow the methodology used in Ref. [12] for the single-lepton final state to re-construct the mass of the $\mathrm{t\bar{t}}$ system. We first find the $z$-component of the neutrino momentum using a W boson mass constraint, assuming the W boson is produced on-shell. All solutions are considered in the mass reconstruction. In the cases where there are only complex solutions, the real part is taken to be the only solution. The mass reconstruction procedure is then split into two cases:

- Events with zero t-tagged AK8 jets: all possible assignments of the selected AK4 jets to either a leptonically decaying top quark, a hadronically decaying top quark, or neither case are considered.
- Events with exactly one t-tagged AK8 jet: the tagged AK8 jet is taken as the hadronically decaying top quark, and all possible assignments of the selected AK4 jets to a leptonically decaying top quark or no top quark are considered. In this case, for an AK4 jet to be assigned to the leptonic decay, it is required to be separated from the AK8 jet by $\Delta R(\mathrm{AK8}, \mathrm{AK4}) > 1.2$.

Among all possible hypotheses built as described above, we choose the hypothesis that gives the smallest $\chi^2$, defined by

$$\chi^2 = \chi^2_{\mathrm{lep}} + \chi^2_{\mathrm{had}} = \left[\frac{M_{\mathrm{lep}} - \overline{M}_{\mathrm{lep}}}{\sigma_{M_{\mathrm{lep}}}}\right]^2 + \left[\frac{M_{\mathrm{had}} - \overline{M}_{\mathrm{had}}}{\sigma_{M_{\mathrm{had}}}}\right]^2.$$

Here, for events with zero (one) t-tagged AK8 jets, $\overline{M}_{\mathrm{lep}} = 175\,(175)\,\mathrm{GeV}$, $\overline{M}_{\mathrm{had}} = 177\,(173)\,\mathrm{GeV}$, $\sigma_{M_{\mathrm{lep}}} = 19\,(19)\,\mathrm{GeV}$, and $\sigma_{M_{\mathrm{had}}} = 16\,(15)\,\mathrm{GeV}$. Further requirements are applied based on the reconstructed top quarks: $\chi^2_{\mathrm{lep}} + \chi^2_{\mathrm{had}} < 30$ and $\Delta R(\mathrm{t_{lep}}, \mathrm{t_{had}}) > 1$, where $\mathrm{t_{lep}}$ ($\mathrm{t_{had}}$) and $\chi^2_{\mathrm{lep}}$ ($\chi^2_{\mathrm{had}}$) are the reconstructed leptonically (hadronically) decaying top quark and the corresponding $\chi^2$ value, respectively.

In order to improve the sensitivity, the events are categorized based on the lepton flavor (electron or muon) and the number of t-tagged (zero or one) AK8 jets. In total, there are four analysis categories. Figure 3 shows the reconstructed mass distributions of the $\mathrm{t\bar{t}}$ system ($m_{\mathrm{t\bar{t}}}$) in each analysis category, and these correspond to the templates used for the statistical interpretation of the analysis.

## 5.2 Fully hadronic final state

The following selection is used in the all-hadronic final state. The first two AK8 jets must have $p_{\mathrm{T}} > 400\,\mathrm{GeV}$, $|\eta| < 4$, and $m_{\mathrm{SD}} > 50\,\mathrm{GeV}$. To obtain the final templates, stricter selection criteria are used, in addition to the above. The first two AK8 jets must have $105 < m_{\mathrm{SD}} < 210\,\mathrm{GeV}$, $\tau_3/\tau_2 < 0.65$, $H_{\mathrm{T}} > 1.2\,\mathrm{TeV}$, where $H_{\mathrm{T}}$ is the scalar $p_{\mathrm{T}}$ sum of the two AK8 jets, and $\Delta\phi > 2.1$. These are the same selection criteria used in Ref. [12], except for the extended pseudorapidity range. The pseudorapidity selection follows the recommendations for objects for the Yellow Report [40].

The following categories are considered, as in Ref. [12]:

- 0 b-tagged jets, $|\Delta y(j_1, j_2)| < 1$
- 1 b-tagged jets, $|\Delta y(j_1, j_2)| < 1$
- 2 b-tagged jets, $|\Delta y(j_1, j_2)| < 1$
- 0 b-tagged jets, $|\Delta y(j_1, j_2)| > 1$
- 1 b-tagged jets, $|\Delta y(j_1, j_2)| > 1$

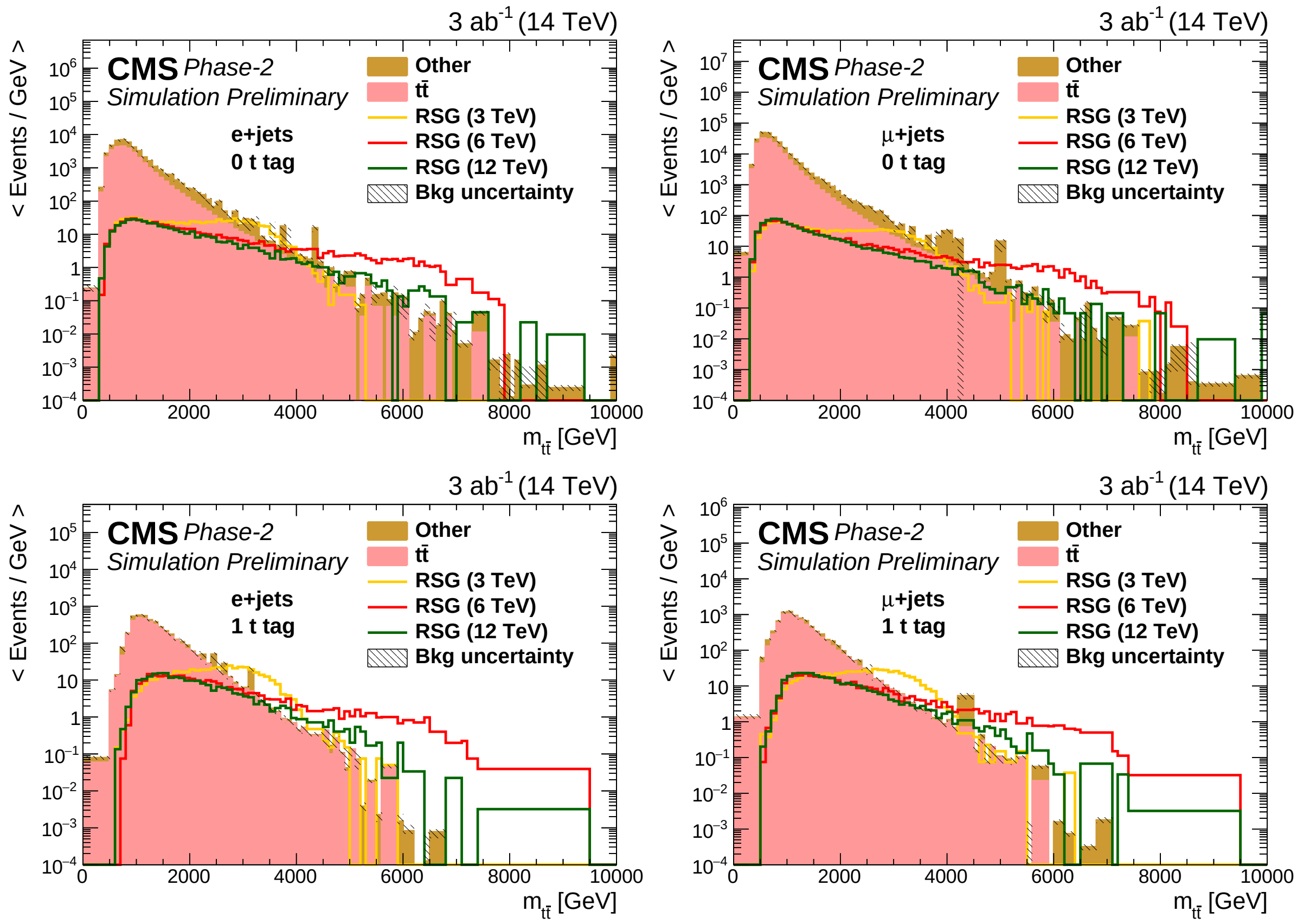

Figure 3: The distributions of $m_{t\bar{t}}$ in events with (top) zero or (bottom) one t-tagged jets for (left) single-electron or (right) single-muon samples. The statistical uncertainties are scaled down by the square root of the projected luminosity. Variable sized bins are used for each category so that the statistical uncertainty on the total background in each bin is less than 10%. The bin contents of the distributions are divided by their bin width. The overflow events are added to the last bin and its content is also divided by the width of the last bin.

- 2 b-tagged jets, $|\Delta y(j_1, j_2)| > 1$

Figure 4 shows the $m_{t\bar{t}}$ distributions for each category after the final selection, which are used as templates for the statistical interpretation of the results.

# 6 Systematic uncertainties

Several systematic uncertainties that affect the distributions of reconstructed mass of the $t\bar{t}$ system ($m_{t\bar{t}}$) in each final state are considered, while the statistical uncertainties in the simulated samples are not included. The uncertainties in electron and muon identification are taken to be 1 and 0.5%, respectively. In addition, several measurement uncertainties are considered, including the jet energy scale (0.5–4%, dependent on jet $p_T$), the jet energy resolution (3%), and the luminosity measurement (1%). Uncertainties are also included to account for the expected data-to-simulation differences in the t- and b-tagging efficiencies.

The uncertainties in the parton distribution functions (PDFs) and the QCD renormalization and factorization scales amount to 2.4% and 3–4%, where the latter is based on Ref. [12], scaled by the projected luminosity. These theory uncertainties are included only for the background MC processes. The QCD multijet background in the fully hadronic final state is expected to be derived from data and therefore additional uncertainties are applied for modified mass procedure and its closure test. These uncertainties are derived from Ref. [12] and scaled by the projected luminosity.

Finally, the theoretical uncertainties in the cross section of each background process are also considered. A summary of all systematic uncertainties is listed in Table 2.

Table 2: Summary of the sources of systematic uncertainties.

| Source | Uncertainty | Single-lepton | Fully hadronic |
|---|---|---|---|
| Integrated luminosity | 1% | ✓ | ✓ |
| Electron identification | 1% | ✓ | |
| Muon identification | 0.5% | ✓ | |
| Jet energy scale | 0.5–4% | ✓ | ✓ |
| Jet energy resolution | 3% | ✓ | ✓ |
| b tagging (subjet) | 1% | | ✓ |
| Light quark mistagging (subjet) | 10% | | ✓ |
| t tagging | 5% | ✓ | ✓ |
| PDF | 2.4% | ✓ | ✓ |
| QCD renorm./fact. scale | 3–4% | ✓ | ✓ |
| $t\bar{t}$ cross section | 3% | ✓ | ✓ |
| Single top quark cross section | 6% | ✓ | |
| W+jets cross section | 3% | ✓ | |
| Z+jets cross section | 6% | ✓ | |
| WW cross section | 6% | ✓ | |
| QCD multijet cross section | 6% | ✓ | |
| QCD multijet modified mass procedure | 2% | | ✓ |
| QCD multijet estimate closure test | 5% | | ✓ |

# 7 Projections at the HE-LHC

We also present projections for the sensitivity to $t\bar{t}$ resonances at a center-of-mass energy of 27 TeV, accessible by the HE-LHC. The signal processes are generated for a resonance mass

**8**

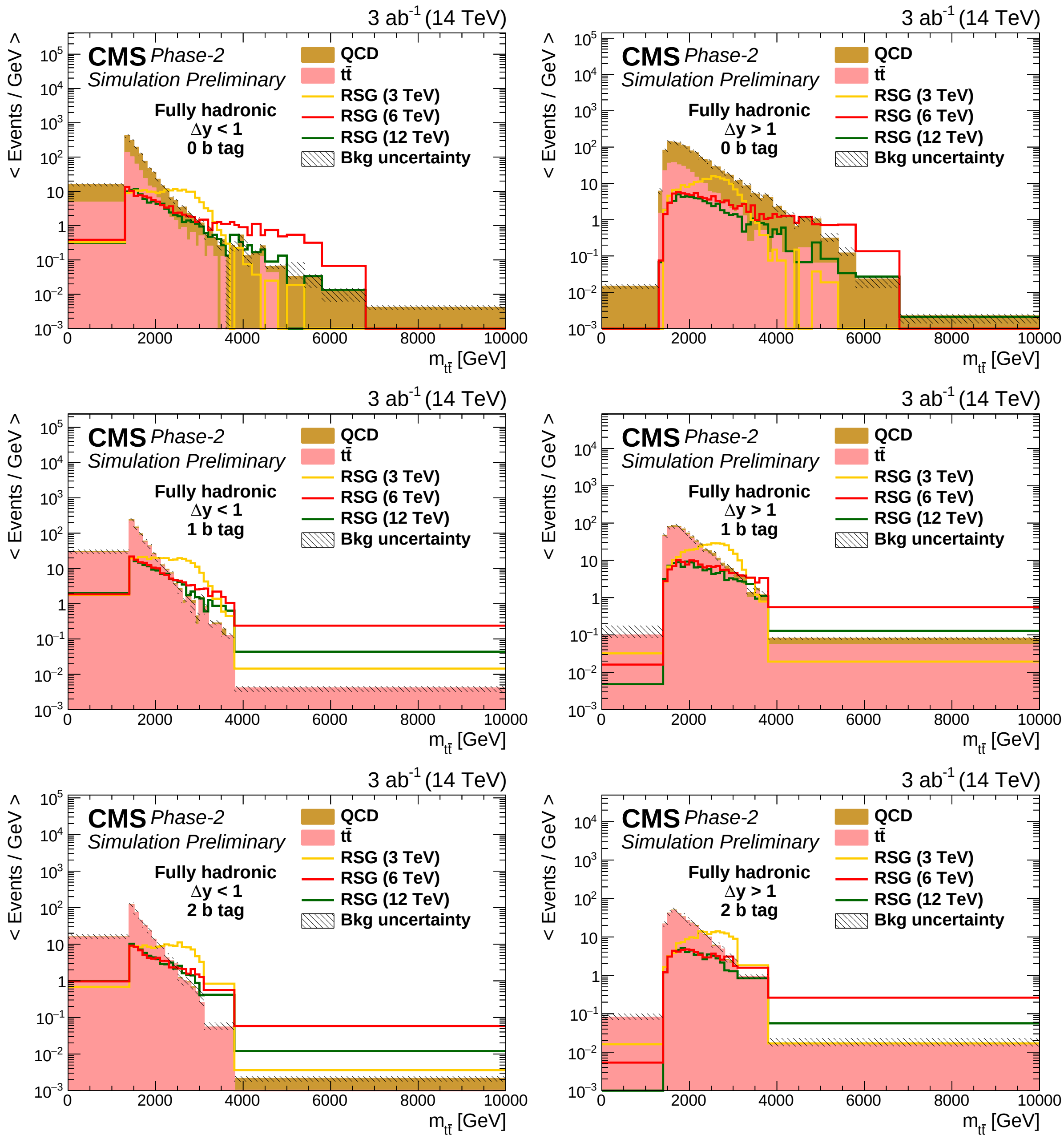

Figure 4: The distribution of the RSG candidate mass for (left) $\Delta y < 1$ or (right) $\Delta y > 1$ in the (top) zero, (middle) one, or (bottom) two b-tag event categories in the fully hadronic final state. All variables presented after full selection. Along with the RSG signal, the two main backgrounds are shown: $t\bar{t}$ and QCD multijets. The statistical uncertainties are scaled down by the square root of the projected luminosity. Variable sized bins are used for each category so that the statistical uncertainty on the total background in each bin is less than 10%. The bin contents of the distributions are divided by their bin width. The overflow events are added to the last bin and its content is also divided by the width of the last bin.

ranging from 4 to 12 TeV, in steps of 2 TeV. The same background processes are also considered for the HE-LHC projections in both final states. All simulated backgrounds are generated at $\sqrt{s} = 27$ TeV, assuming the same number of pileup interactions as the HL-LHC. The reconstruction of physics-level objects are simulated with the Delphes software package with the Phase-2 CMS detector design.

We follow the same analysis strategy described before for HL-LHC projections. The event selection and reconstruction methods are kept the same, while the requirements on the $p_T$ of the first and second AK4 jets are increased to 200 and 100 GeV, respectively, in both the single-electron and single-muon final states, and $H_T^{lep}$ is required to be greater than 200 GeV in the single-muon final state. Because of the low number of events in the QCD multijet background simulation in the fully hadronic final state, the events are categorized based on only the number of b-tagged AK8 subjets. The $m_{t\bar{t}}$ distributions after applying the final selection criteria are shown in Fig. 5 for the single-lepton final state categories and in Fig. 6 for the fully hadronic final state categories.

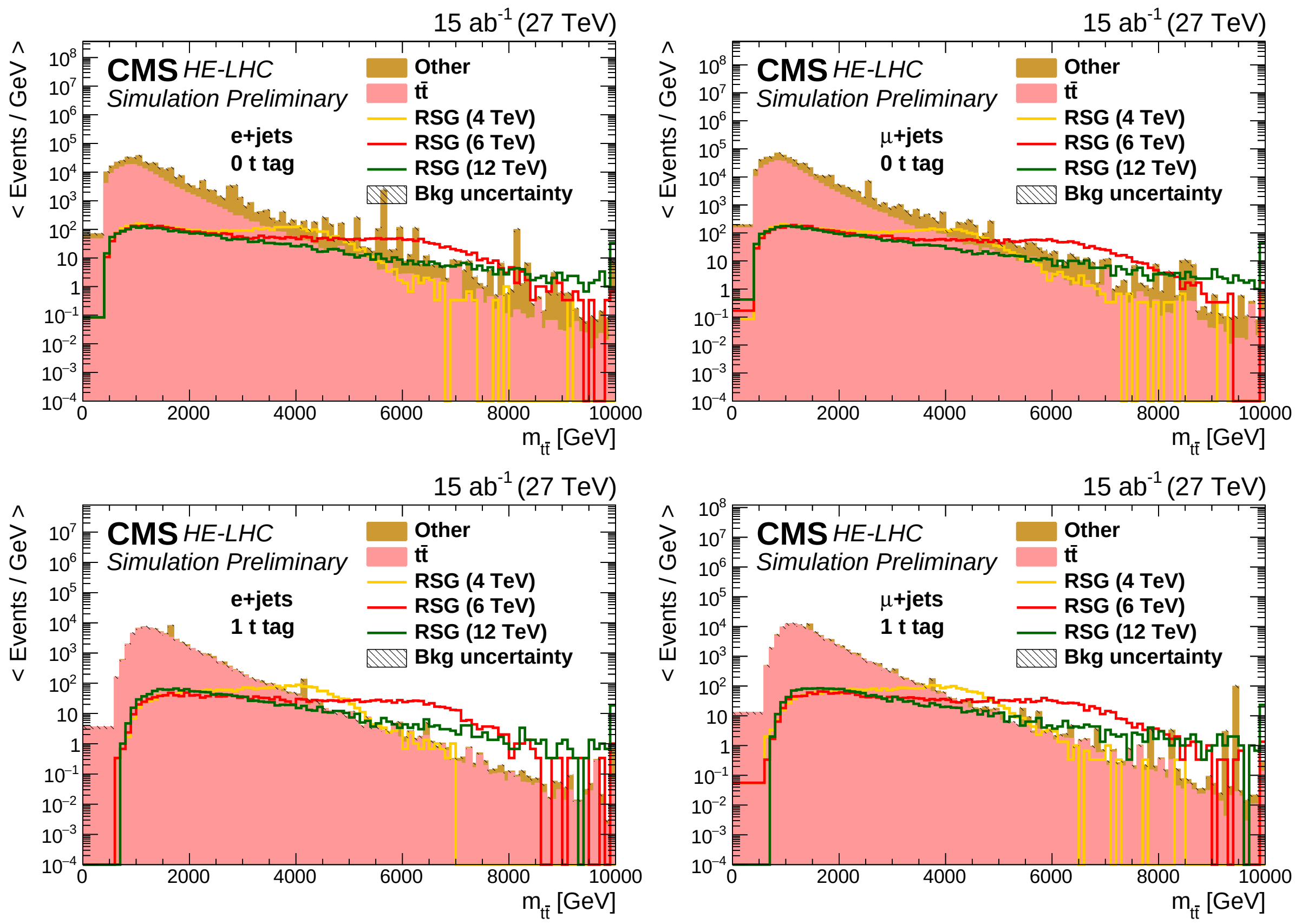

Figure 5: Distributions of $m_{t\bar{t}}$ in events with (top) zero or (bttom) one t-tagged jets for (left) the single-electron or (right) single-muon final state. The statistical uncertainties are scaled down by the square root of the projected luminosity. Variable sized bins are used for each category so that the statistical uncertainty on the total background in each bin is less than 10%. The bin contents of the distributions are divided by their bin width. The overflow events are added to the last bin and its content is also divided by the width of the last bin.

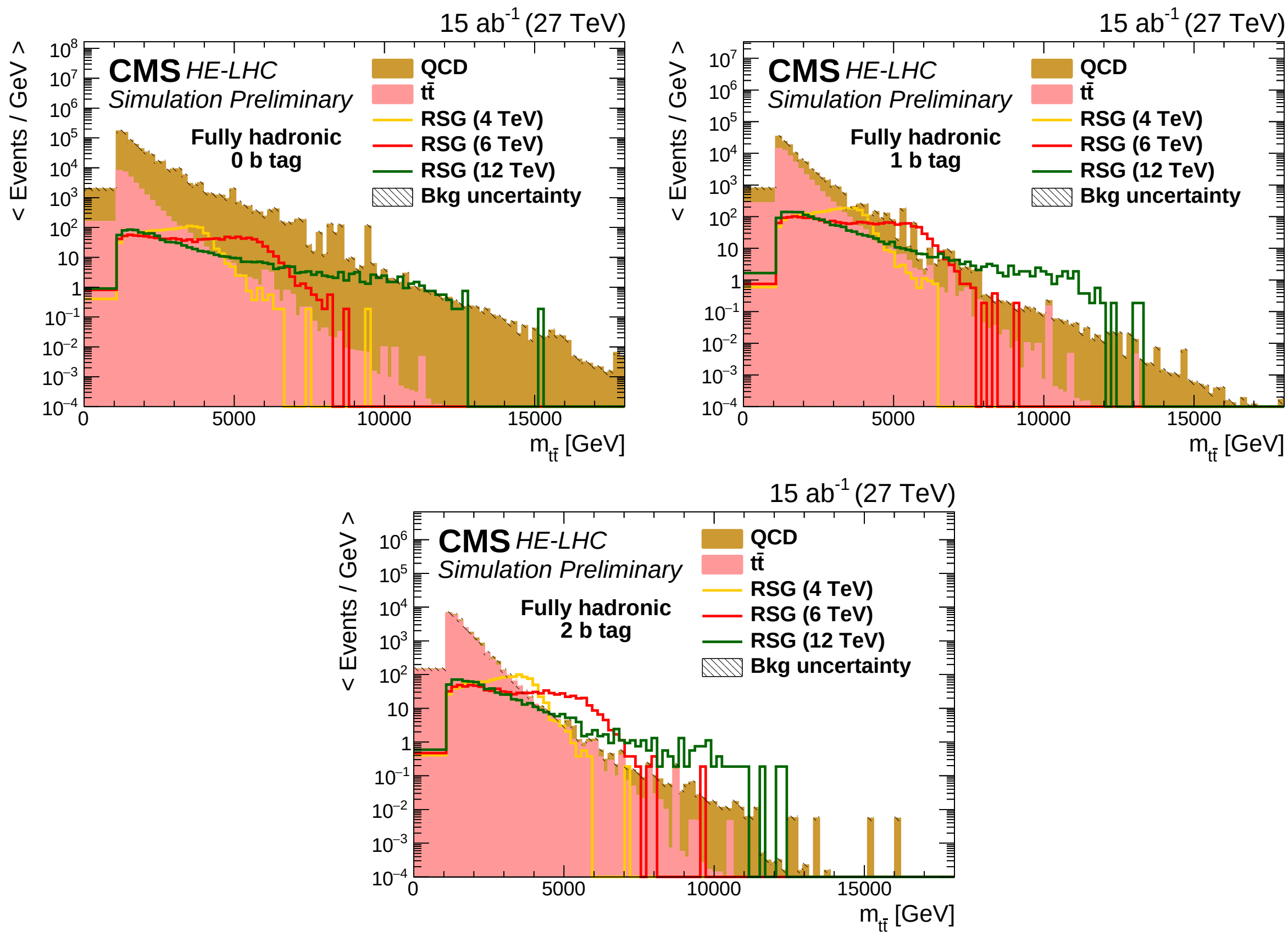

Figure 6: Distributions of $m_{t\bar{t}}$ for (top left) zero, (top right) one, or (bottom) two b tag event categories in the fully hadronic final state. The statistical uncertainties are scaled down by the square root of the projected luminosity. Variable sized bins are used for each category so that the statistical uncertainty on the total background in each bin is less than 10%. The bin contents of the distributions are divided by their bin width. The overflow events are added to the last bin and its content is also divided by the width of the last bin.

# 8 Results

We use the Theta package [41] to derive the expected cross section limits at 95% confidence level (CL) on the production of an RSG decaying to $\mathrm{t\bar{t}}$. The limits are computed using the asymptotic $\mathrm{CL_s}$ approach. A binned likelihood fit to the distributions of the reconstructed $m_{\mathrm{t\bar{t}}}$ is performed in both the single-lepton and fully hadronic final states. The systematic uncertainties are included as nuisance parameters with log-normal probability density functions. The results are limited by the statistical uncertainties in the background estimates. These uncertainties are scaled down by the projected integrated luminosity and are treated using the Barlow–Beeston light method [42, 43]. The expected limits at 95% CL and the discovery potential at $3\sigma$ and $5\sigma$ significance for resonance masses from 2 to 12 TeV and two different projected integrated luminosities for the combined single-lepton and fully hadronic final states are listed in Table 3. The production of an RSG with a mass up to 6.6 TeV is excluded at 95% CL for a projected integrated luminosity of 3 $\mathrm{ab^{-1}}$, as shown in Fig. 7. An RSG with a mass up to 5.7 TeV could be discovered at $5\sigma$ significance.

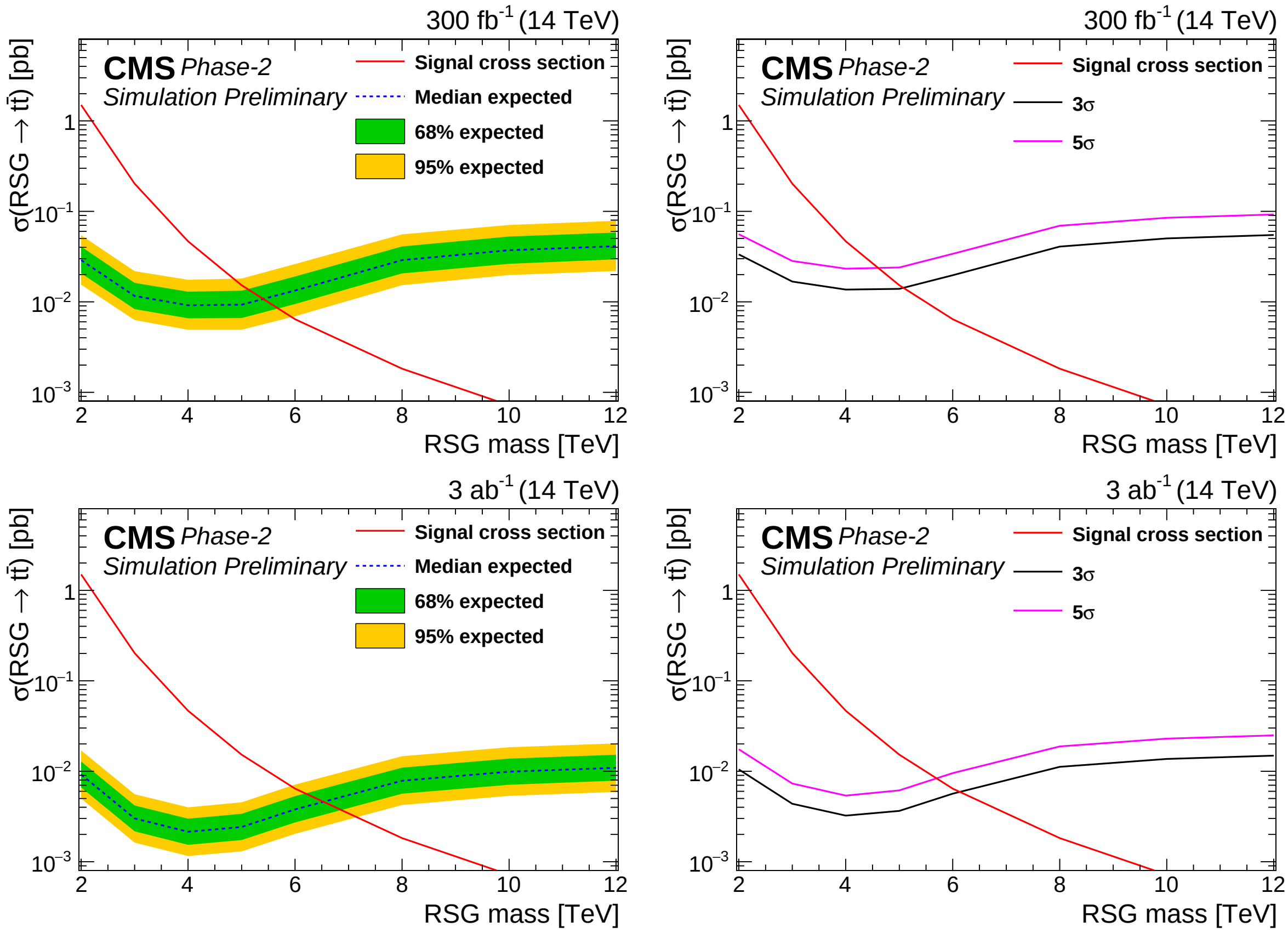

Figure 7: 95% CL expected upper limits (left) and $3\sigma$ and $5\sigma$ discovery reaches (right) for an RSG decaying to $\mathrm{t\bar{t}}$ at 300 $\mathrm{fb^{-1}}$ (top) and 3 $\mathrm{ab^{-1}}$ (bottom) for the combined single-lepton and fully hadronic final states. The LO signal theory cross sections are scaled to NLO using a $k$ factor of 1.3 [44].

Figure 8 shows a comparison of the expected limits for RSG with corresponding results using exclusively the statistical uncertainties. Figure 8 also shows a comparison of the expected sensitivity contribution from each final state.

The expected limits at 95% CL and the discovery potential at $\sqrt{s} = 27$ TeV for resonance masses from 4 to 12 TeV and a projected integrated luminosity of 15 $\mathrm{ab^{-1}}$ for the combined single-

**12**

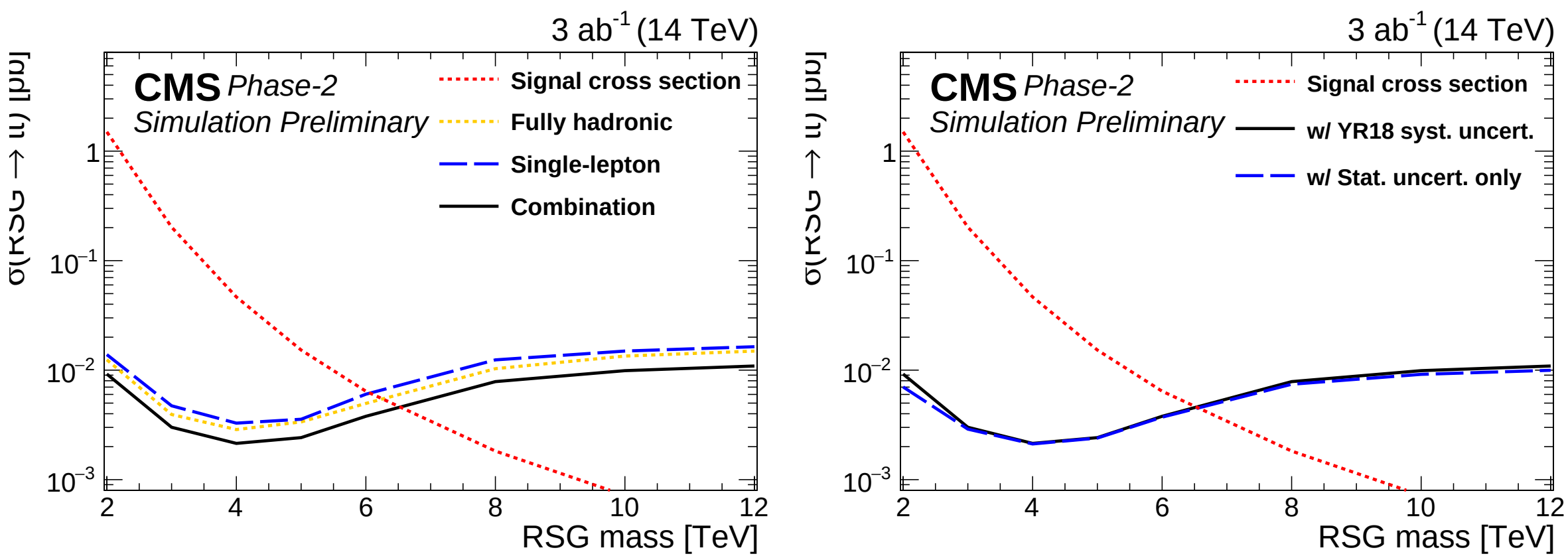

Figure 8: 95% CL expected cross section limits for 3 ab$^{-1}$ projection. Comparisons of the contributions from each final state to the combination is shown on the left. The effect of different systematic uncertainty scenarios on the combined limits is shown on the right. The LO signal theory cross sections are scaled to NLO using a $k$ factor of 1.3 [44].

Table 3: Expected cross section limits at 95% CL and discovery reaches at 3 and 5$\sigma$ in the combined single-lepton and fully hadronic final states for an RSG decaying to $t\bar{t}$. The LO signal theory cross sections are scaled to NLO using a $k$ factor of 1.3 [44].

| Mass [TeV] | Theory [pb] | Median exp. [pb] | 68% exp. [pb] | 95% exp. [pb] | 3$\sigma$ [pb] | 5$\sigma$ [pb] |
|---|---|---|---|---|---|---|
| $\sqrt{s} = 14$ TeV, $\mathcal{L} = 300$ fb$^{-1}$ | | | | | | |
| 2 | 1.4989 | 0.029 | [0.021, 0.041] | [0.015, 0.054] | 0.033 | 0.056 |
| 3 | 0.2023 | 0.012 | [0.008, 0.016] | [0.006, 0.022] | 0.017 | 0.028 |
| 4 | 0.0466 | 0.009 | [0.007, 0.013] | [0.005, 0.018] | 0.014 | 0.023 |
| 5 | 0.0153 | 0.009 | [0.007, 0.013] | [0.005, 0.018] | 0.014 | 0.024 |
| 6 | 0.0064 | 0.013 | [0.009, 0.019] | [0.007, 0.026] | 0.020 | 0.034 |
| 8 | 0.0018 | 0.029 | [0.021, 0.041] | [0.015, 0.056] | 0.041 | 0.069 |
| 10 | 0.0007 | 0.037 | [0.026, 0.053] | [0.020, 0.071] | 0.050 | 0.085 |
| 12 | 0.0003 | 0.041 | [0.030, 0.058] | [0.022, 0.079] | 0.055 | 0.092 |
| $\sqrt{s} = 14$ TeV, $\mathcal{L} = 3$ ab$^{-1}$ | | | | | | |
| 2 | 1.4989 | 0.009 | [0.007, 0.013] | [0.005, 0.017] | 0.010 | 0.017 |
| 3 | 0.2023 | 0.003 | [0.002, 0.004] | [0.002, 0.006] | 0.004 | 0.007 |
| 4 | 0.0466 | 0.002 | [0.002, 0.003] | [0.001, 0.004] | 0.003 | 0.005 |
| 5 | 0.0153 | 0.002 | [0.002, 0.003] | [0.001, 0.005] | 0.004 | 0.006 |
| 6 | 0.0064 | 0.004 | [0.003, 0.005] | [0.002, 0.007] | 0.006 | 0.010 |
| 8 | 0.0018 | 0.008 | [0.006, 0.011] | [0.004, 0.015] | 0.011 | 0.019 |
| 10 | 0.0007 | 0.010 | [0.007, 0.014] | [0.005, 0.018] | 0.014 | 0.023 |
| 12 | 0.0003 | 0.011 | [0.008, 0.015] | [0.006, 0.020] | 0.015 | 0.025 |

lepton and fully hadronic final states are shown in Fig. 9 and Table 4. The same systematic uncertainties listed in Table 2 are considered. The off-shell production of the resonance causes the cross section limits to increase at very high masses.

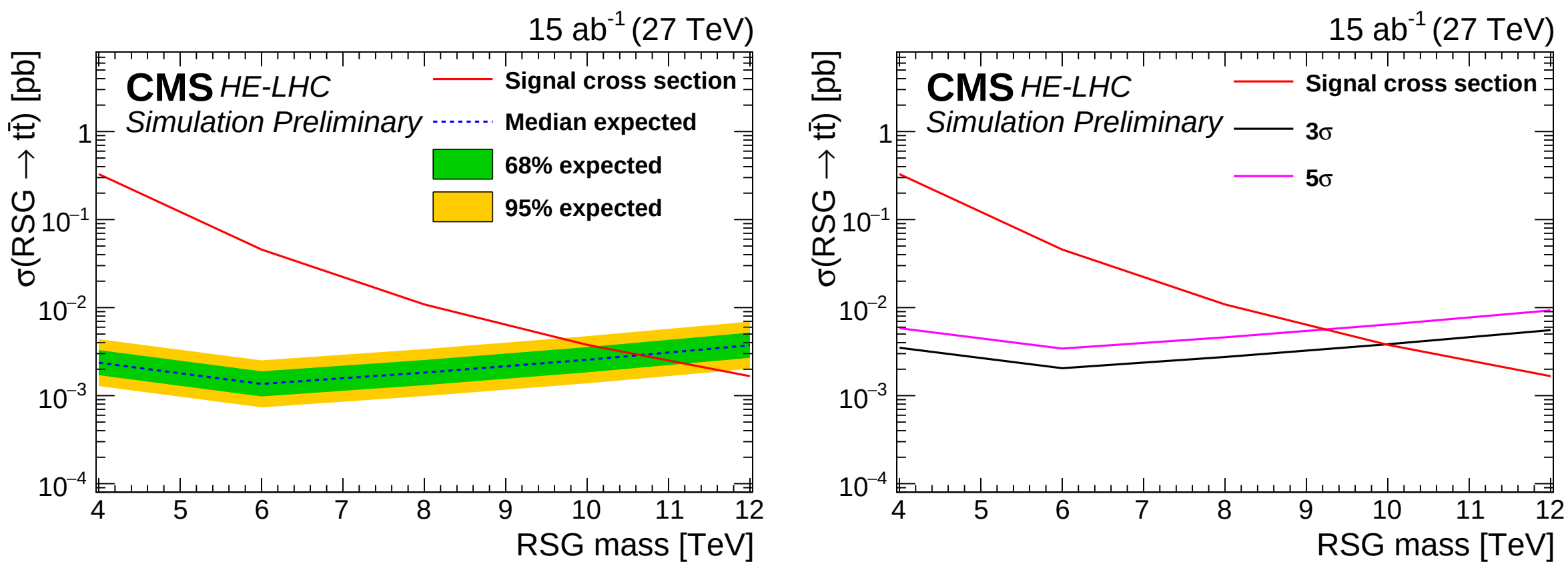

Figure 9: 95% CL expected upper limits (left) and $3\sigma$ and $5\sigma$ discovery potential (right) for an RSG decaying to $t\bar{t}$ with 15 $ab^{-1}$ from the HE-LHC for the combined single-lepton and fully hadronic final states. The LO signal theory cross sections are scaled to NLO using a $k$ factor of 1.3 [44].

Table 4: Expected cross section limits at 95% CL and discovery potential at $3\sigma$ and $5\sigma$ in the combined single-lepton and fully hadronic final states for an RSG decaying to $t\bar{t}$ with 15 $ab^{-1}$ from the HE-LHC. The LO signal theory cross sections are scaled to NLO using a $k$ factor of 1.3 [44].

| Mass [ TeV ] | Theory [fb] | Median exp. [fb] | 68% exp. [fb] | 95% exp. [fb] | $3\sigma$ [fb] | $5\sigma$ [fb] |
|---|---|---|---|---|---|---|
| $\sqrt{s} = 27$ TeV, $\mathcal{L} = 15$ $ab^{-1}$ | | | | | | |
| 4 | 328.77 | 2.37 | [1.71, 3.29] | [1.28, 4.38] | 3.50 | 5.84 |
| 6 | 45.68 | 1.36 | [0.98, 1.89] | [0.74, 2.52] | 2.05 | 3.43 |
| 8 | 10.87 | 1.83 | [1.32, 2.54] | [0.99, 3.38] | 2.75 | 4.60 |
| 10 | 3.78 | 2.55 | [1.84, 3.56] | [1.38, 4.74] | 3.84 | 6.43 |
| 12 | 1.66 | 3.73 | [2.69, 5.20] | [2.02, 6.94] | 5.55 | 9.29 |

# 9 Summary

We have presented a sensitivity projection for heavy resonant $t\bar{t}$ pair production using the upgraded Phase-2 CMS detector design at the High-Luminosity LHC (HL-LHC) and High-Energy LHC (HE-LHC), with center-of-mass energies of 14 and 27 TeV and integrated luminosities of 3 and 15 $ab^{-1}$. Two distinct final states, single-lepton or fully hadronic, are considered. We set limits on the production cross sections of a Randall–Sundrum gluon and exclude masses up to 6.6 (10.7) TeV at 95% confidence level at the HL-LHC (HE-LHC). The Randall–Sundrum gluon with a mass up to 5.7 (9.4) TeV can be discovered with $5\sigma$ significance at the HL-LHC (HE-LHC).

# CMS Physics Analysis Summary

Contact: cms-future-conveners@cern.ch

# Search for vector boson fusion production of a massive resonance decaying to a pair of Higgs bosons in the four b quark final state at the HL-LHC using the CMS Phase 2 detector

The CMS Collaboration

## Abstract

The prospects of a search for a massive resonance produced through vector boson fusion and decaying to a pair of standard model Higgs bosons at the high luminosity LHC at CERN is explored. Simulated events from proton-proton collisions at a centre-of-mass energy of 14 TeV collected by the upgraded CMS detector are used. Both the Higgs bosons are assumed to decay to a b quark-antiquark pair, each. For a high mass resonance, the Higgs bosons are highly Lorentz-boosted and are each reconstructed as a large-area jet. The signal also contains two energetic jets in the forward regions of the detector. The expected signal significances for a bulk graviton in warped extradimensional models, having a mass between 1500 and 3000 GeV and a narrow width compared to its mass, is presented, assuming a cross section of 1 fb, for a data set corresponding to an integrated luminosity of 3.0 $ab^{-1}$.

# 1 Introduction

The search for new physics resonances decaying to a pair of Higgs bosons (H) [1–3] is motivated by several beyond standard model (BSM) scenarios. Such models include warped extra dimensions (WED) [4] having particles such as the spin-0 radion [5–7] and the spin-2 first Kaluza–Klein (KK) excitation of the graviton [8–10]. Others, such as the two-Higgs doublet models [11] (particularly, the minimal supersymmetric model [12]) and the Georgi-Machacek model [13] also contain spin-0 resonances. These resonances may have a sizeable branching fraction to a H pair.

Searches for a new particle X in the HH decay channel have been performed by the ATLAS [14–16] and CMS [17–21] Collaborations in proton-proton (pp) collisions at $\sqrt{s} = 7$ and 8 TeV. The ATLAS Collaboration published limits on the production of a KK bulk graviton decaying to HH in the final state with a pair of b quark and antiquark ($b\bar{b}b\bar{b}$), using pp collision data at $\sqrt{s} = 13$ TeV [22–24]. More recently, the CMS Collaboration has published limits on the production of a KK bulk graviton and a radion, decaying to HH, in the $b\bar{b}b\bar{b}$ final state, using pp collision data at $\sqrt{s} = 13$ TeV, corresponding to an integrated luminosity of 35.9 $\mathrm{fb}^{-1}$ [25, 26]. Overall, the searches from ATLAS and CMS set a limit on the production cross sections and the branching fractions $\sigma(\mathrm{pp} \to \mathrm{X})\mathcal{B}(\mathrm{X} \to \mathrm{HH} \to b\bar{b}b\bar{b})$ for masses of X ($m_{\mathrm{X}}$) up to 3000 GeV.

The above searches looked at the $s$-channel production of a narrow resonance X produced from the standard model (SM) quark-antiquark or gluon-gluon interactions. The WED models are used in the interpretations of the results. In these models, the extra spatial dimension is compactified between two branes (called the bulk) via an exponential metric $\kappa l$, where $\kappa$ is the warp factor and $l$ the coordinate of the extra spatial dimension [27]. The fundamental scale is the reduced Planck scale ($\overline{M_{\mathrm{Pl}}} \equiv M_{\mathrm{Pl}}/8\pi$, $M_{\mathrm{Pl}}$ being the Planck scale) and the ultraviolet cutoff of the theory $\Lambda_{\mathrm{R}} \equiv \sqrt{6}\mathrm{e}^{-\kappa l}\overline{M_{\mathrm{Pl}}}$ [5]. Assuming $\Lambda_{\mathrm{R}} = 3$ TeV, a spin-0 radion of mass below 1400 GeV is excluded at a 95% confidence level [25]. The cross section limit of a bulk graviton decaying to $\mathrm{HH} \to b\bar{b}b\bar{b}$ is between 4 and 1.4 fb for masses between 1400 and 3000 GeV.

In addition to the $s$-channel production of X, there is the vector boson fusion (VBF) production mode, as depicted in Fig. 1. The $s$-channel production cross section of a bulk graviton, assuming $\kappa/\overline{M_{\mathrm{Pl}}} = 0.5$, is in the range 5–0.05 fb for masses between 1500 and 3000 GeV and the VBF production mode is expected to have a cross section an order of magnitude smaller [28]. On the other hand, the absence of a signal from the $s$-channel process may point to highly suppressed couplings of X with the SM quarks and gluons. In such a case, the VBF production mode may be the most dominant production process for X in pp collisions.

The VBF production process $\mathrm{pp} \to \mathrm{X}q\bar{q}$ should be observable with the data from the high luminosity LHC (HL-LHC), which is expected to collide protons on protons at a centre-of-mass energy of 14 TeV and deliver a total integrated luminosity of 3 $\mathrm{ab}^{-1}$.

In this analysis, the prospects of a search for a massive resonance produced through VBF and decaying to HH using pp collisions at a centre-of-mass energy of 14 TeV at the HL-LHC, assuming a data set corresponding to an integrated luminosity of 3 $\mathrm{ab}^{-1}$ collected by the upgraded CMS detector, is explored. For a high mass resonance, the Higgs bosons are highly Lorentz-boosted and are each reconstructed as a large-area jet (Higgs jet). In addition, a signal event is characterized by two energetic jets at large pseudorapidity $\eta$ ($\equiv -\ln[\tan(\theta/2)]$), $\theta$ being the polar angle of the jet measured in the CMS detector coordinate system, typical of the VBF production mode.

This analysis summary is organized as follows: In Section 2 the CMS experimental apparatus and simulations are described, followed by the event selection in Section 3 and the estimation of the backgrounds in Section 4. The projections of the search sensitivity are presented in Section 5 followed by the summary in Section 6.

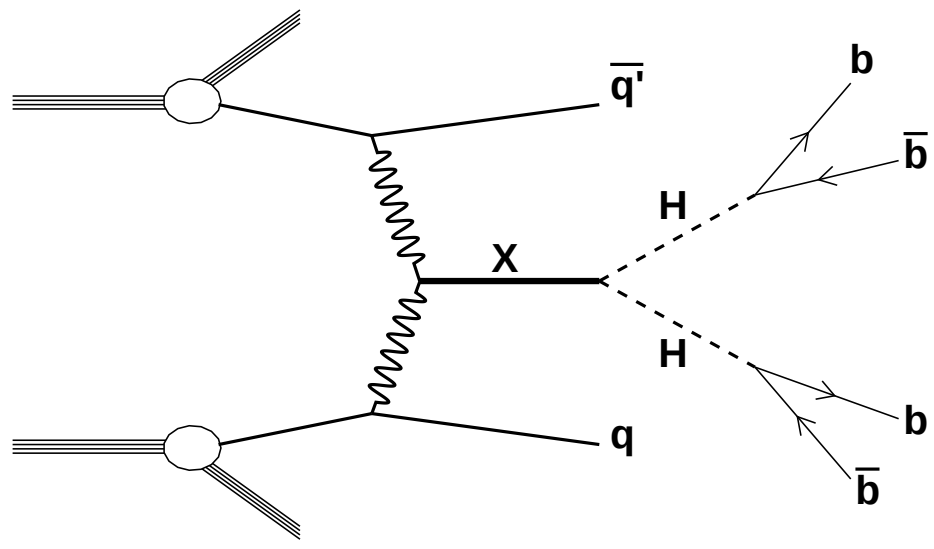

Figure 1: The vector boson fusion mode of production of a resonance X decaying to a pair of Higgs bosons H, with both Higgs bosons decaying to $b\bar{b}$ pairs.

## 2 The CMS detector and simulations

A detailed description of the CMS detector with the associated coordinate system and relevant kinematic variables can be found in Ref. [29]. The CMS experiment will be upgraded [30–33] (Phase 2) in order to cope with the challenges during data taking at HL-LHC, which primarily includes a large number of simultaneous pp collisions (pileup), up to 200, in the detector. In order to maintain or even improve trigger, reconstruction and identification capabilities, several new detector technologies will be used to upgrade the currently used detector subsystems. A simulation of the upgraded Phase 2 CMS detector was used for this study.

Signal events for bulk gravitons were simulated using at leading order MADGRAPH5_aMC@NLO 2.4.2 [34] event generator for masses in the range 1500-3000 GeV and for a width of 1% of the mass. The NNPDF3.0 leading order parton distribution functions (PDFs) [35], taken from the LHAPDF6 PDF set [36–39], with the four-flavour scheme, were used. The main background are events comprised uniquely of jets arising from the SM strong interaction (multijet events). This background was simulated using PYTHIA 8.212 [40], for events containing two hard partons, with the invariant mass of the two partons required to be greater than 1000 GeV.

For both the signal and the background processes, the showering and hadronization of partons was simulated with PYTHIA 8. The pileup events contribute to the overall event activity in the detector, the effect of which was included in the simulations assuming a pileup distribution averaging to 200, as anticipated at the HL-LHC beam conditions. All generated samples were processed through a GEANT4-based [41, 42] simulation of the upgraded CMS detector.

## 3 Event selection

The simulated particle hits in the CMS detector elements are reconstructed using the particle-flow (PF) algorithm [43] into physics objects (charged and neutral hadrons, electrons, muons, and photons), which are used for further reconstruction and analysis.

Among the many collision vertices in an event, the primary interaction vertex for pp collisions is taken to be the one with the highest $\sum p_{\mathrm{T}}^2$ of the associated clusters of physics objects. The physics objects are the jets, clustered using the anti-$k_{\mathrm{T}}$ jet finding algorithm [44, 45], with a distance parameter of 0.4, having the tracks assigned to the vertex as inputs, and the associated missing transverse momentum, taken as the negative vector sum of the $p_{\mathrm{T}}$ of those jets. The other interaction vertices are considered to be pileup vertices.

The contribution of pileup collisions in the event is mitigated using the pileup per particle identification (PUPPI) algorithm [46]. This algorithm removes charged particles originating from pileup vertices, while retaining those from the primary vertex. Neutral particles are assigned a weight between zero or one, with a higher value indicating a higher likelihood of the particle to be from the primary vertex. Particles from the PF algorithm are clustered into jets using the anti-$k_{\mathrm{T}}$ algorithm with a distance parameter of 0.8 (AK8 jets) or 0.4 (AK4 jets). The vector sum of the momenta of all clustered particles, weighted by their PUPPI weights, is taken to be the jet momentum. Jet energy corrections are applied as a function of jet $\eta$ and $p_{\mathrm{T}}$ [47, 48] to compensate for the nonlinear response of the detector to the collected energy.

The two leading-$p_{\mathrm{T}}$ AK8 jets, $J_1$ and $J_2$, respectively, in the event are required to have transverse momenta $p_{\mathrm{T}} > 300\,\mathrm{GeV}$ and a pseudorapidity range $|\eta| < 3.0$. To identify the two leading-$p_{\mathrm{T}}$ AK8 jets with the boosted $\mathrm{H} \to \mathrm{b\bar{b}}$ candidates from the $\mathrm{X} \to \mathrm{HH}$ decay (H tagging), these jets are first groomed [49] to remove soft and wide-angle radiation using the modified mass drop algorithm [50, 51], with the soft radiation fraction parameter $z$ set to 0.1 and the angular exponent parameter $\beta$ set to 0, also known as the soft-drop algorithm [52, 53]. The soft-drop algorithm gives two subjets each, for $J_1$ and $J_2$, by undoing the last stage of the jet clustering. The invariant mass of the two subjets is the soft-drop mass of each AK8 jet, which has a distribution with a peak near the Higgs boson mass $m_{\mathrm{H}} = 125\,\mathrm{GeV}$ [54, 55], and a width of about 10%. The soft-drop mass window selection was optimized using a figure of merit of $S/\sqrt{B}$ and is required to be in the range 90–140 GeV for both $J_1$ and $J_2$.

The N-subjettiness algorithm identifies substructures arising from hard partons inside a jet to distinguish a two-pronged $\mathrm{H} \to \mathrm{b\bar{b}}$ decay from the background of jets arising from a single quark or a gluon, using inclusive jet shape variables $\tau_1$ and $\tau_2$ [53, 56, 57]. The ratio $\tau_{21} \equiv \tau_2/\tau_1$ has a value much smaller than unity for a jet with two subjets, and hence, for signal selection, $J_1$ and $J_2$ is required to have $\tau_{21} < 0.6$ following an optimization of the above figure of merit.

The H tagging of $J_1$ and $J_2$ further requires identifying the subjet pairs from each of $J_1$ and $J_2$ to be b tagged using the DeepCSV algorithm [58], which combines information from tracks and secondary vertices associated to the subjets into a multivariate discriminator using deep machine learning techniques. The output of the DeepCSV algorithm can be interpreted as the probability of a jet to belong to one of five flavour categories, defined by whether the jet contains exactly one or two b hadrons, exactly one or two c hadrons in the absence of any b hadrons, or no b or c hadrons [58]. In this search, b-tagged subjets are required to have a probability of about 49% to contain at least one b hadron, and a corresponding probability of about 1% of having no b or c hadrons. Events are classified into two categories: those having exactly three out of the four b-tagged subjets (3b category), and those that have all four subjets b-tagged (4b category).

An event is required to have at least two AK4 jets ($j_1$ and $j_2$), which are separated from the H jets by $\Delta R > 1.2$, in the pseudorapidity-azimuthal angle plane, with $p_{\mathrm{T}} > 50\,\mathrm{GeV}$ and $|\eta| < 5$. To pass the VBF selections, these jets must lie in opposite $\eta$ regions of the detector, and the pseudorapidity separation between them $|\Delta\eta(j_1, j_2)| > 5$. The invariant mass $m_{\mathrm{jj}}$ reconstructed using these AK4 jets is required to pass $m_{\mathrm{jj}} > 300\,\mathrm{GeV}$.

Table 1: Event yields and efficiencies for the signal and multijet background for an average pileup of 200. The product of the cross sections and branching fractions of the signals $\sigma(\mathrm{pp} \to \mathrm{Xjj} \to \mathrm{HHjj})$ is assumed to be 1 fb. Owing to the large sample sizes of the simulated events, the statistical uncertainties are small.

| Process | 3b category | | 4b category | |
|---|---|---|---|---|
| | Events | Efficiency (%) | Events | Efficiency (%) |
| Multijets | 4755 | $1.6 \times 10^{-3}$ | 438 | $1.5 \times 10^{-4}$ |
| BG ($m_X$ =1500 GeV) | 326 | 11 | 95.2 | 3.2 |
| BG ($m_X$ =2000 GeV) | 316 | 11 | 81.2 | 2.7 |
| BG ($m_X$ =3000 GeV) | 231 | 7.7 | 41.4 | 1.4 |

The bulk graviton invariant mass $m_{JJ}$ is reconstructed from the 4-momenta of the two Higgs jets in events passing the above mentioned full selection criteria. The main multijet background is smoothly falling, above which the signal is searched for, as a localized excess of events, for a narrow resonance $X$.

## 4 Background estimation

It is expected that the multijet background component in a true search at the HL-LHC will rely on the data for a precise estimate. Methods such as those described in Ref. [25] are known to provide an accurate prediction of the multijet background $m_{JJ}$ shape as well as the yield. Here, the expected background yields based on simulations are described.

The simulated multijet background sample consists of $\sim$4 million events none of which survive the full selection. To estimate the background, the subjet b-tagging efficiency is determined using a loose set of selections which require events to have $J_1$ and $J_2$ passing only the soft-drop mass and $\tau_{21}$ requirements. The b-tagging efficiency is obtained for the different subjet flavours and as a function of $p_T$ and $\eta$.

Multijet events passing all selections but the subjet b tagging are then reweighted according to the subjet efficiencies to obtain the probability of the event to pass the three of four subjet b-tagging categories. The $m_{JJ}$ distributions for the multijet background after the full selection are then obtained from the weighted events in these two categories.

From the analysis of current LHC data at $\sqrt{s} = 13$ TeV, it was found that the multijet backgrounds in simulations is overestimated by a factor of 0.7 compared to the yields in the data. Accordingly, the multijet background yield has been corrected by this factor, assuming this to hold also for the simulations of the multijet processes at $\sqrt{s} = 14$ TeV . The $m_{JJ}$ of the backgrounds thus obtained and the signals are shown in Fig. 2. The event yields after full selection are given in Table 1.

The efficiency of events to pass the VBF jet selection depends strongly on pileup due to the combinatorial backgrounds from pileup jets, which affects both the signal and the background selection. Moreover, the VBF selection efficiency for multijets grows faster than the signal efficiency with pileup, since the latter has true VBF jets which already pass the selection in the absence of pileup. Hence, in the present search, the requirement of additional VBF jets does not result in any appreciable gain in the signal sensitivity. It is anticipated that developments in the rejection of pileup jets in the high $\eta$ region will eventually help suppress the multijets background and improve the signal sensitivity further.

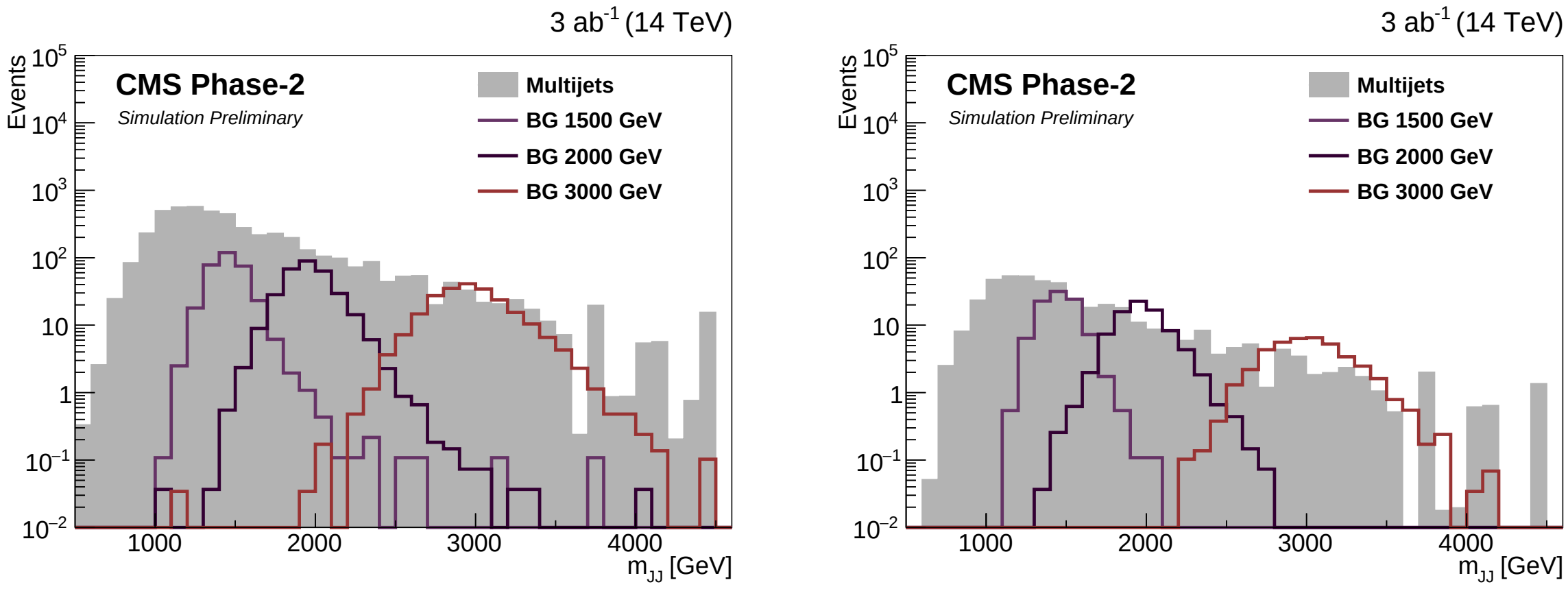

Figure 2: The estimated multijet background and the signal $m_{jj}$ distributions for bulk gravitons (BG) of masses 1500, 2000, and 3000 GeV, assuming a signal cross section of 1 fb. The distributions on the left are for the 3b and those on the right are for the 4b subjet b-tagged categories and for an average pileup of 200.

# 5 Projections

The expected significance of the signal, assuming a production cross section of 1 fb is estimated.

Several systematic uncertainties are considered. The uncertainty in the jet energy scale amounts to 1%. The uncertainty in the subjet b-tagging efficiency difference between the data and simulations is taken to be 1%. An uncertainty of 1% is assigned to the integrated luminosity measurement. These uncertainties are based on the projected values for the full data set at the HL-LHC.

In addition, several measurement uncertainties are considered based on the 2016 search for a resonance decaying to a pair of boosted Higgs bosons [25], scaled by 0.5. The H jet selection uncertainties include the uncertainties in the H jet mass scale and resolution (1%), the uncertainty in the data to simulation difference in the selection on $\tau_{21}$ (13%), and the uncertainty in the showering and hadronization model for the H jet (3.5%). The uncertainties in the signal acceptance because of the parton distribution functions (1%) and the simulation of the pileup (1%) are also taken into account

The expected signal significance of a bulk graviton of mass between 1500 and 3000 GeV, produced through vector boson fusion, and decaying into a pair of Higgs bosons, each of which decays to a $b\overline{b}$ pair, is given in Fig. 3 for an integrated luminosity of 3 $ab^{-1}$.

# 6 Summary

The vector boson fusion production mode for diboson resonances is extremely challenging to probe using the current data because of its small cross section. The search for these processes are however feasible at the high luminosity LHC, and we present here the search for a massive spin-2 bulk graviton decaying to two Higgs bosons. The bulk gravitons are predicted in various new physics scenarios like the warped extradimensional models, which aim to explain the so-called hierarchy problem of the standard model. The search focuses on the final state where both the Higgs bosons decay to b quark-antiquark pairs that are boosted, thus forming Higgs jets. Assuming a signal production cross section of 1 fb, with a data set corresponding to an integrated luminosity of 3 $ab^{-1}$ in proton-proton collisions at the at a centre-of-mass of 14 TeV, the CMS experiment should be able to find the evidence for the presence of a bulk graviton of mass between 1500 and 3000 GeV. It is expected that future advances in the event reconstruction and physics object identification techniques, spurred on by the Phase 2 CMS detector design, will help improve these projections even further.

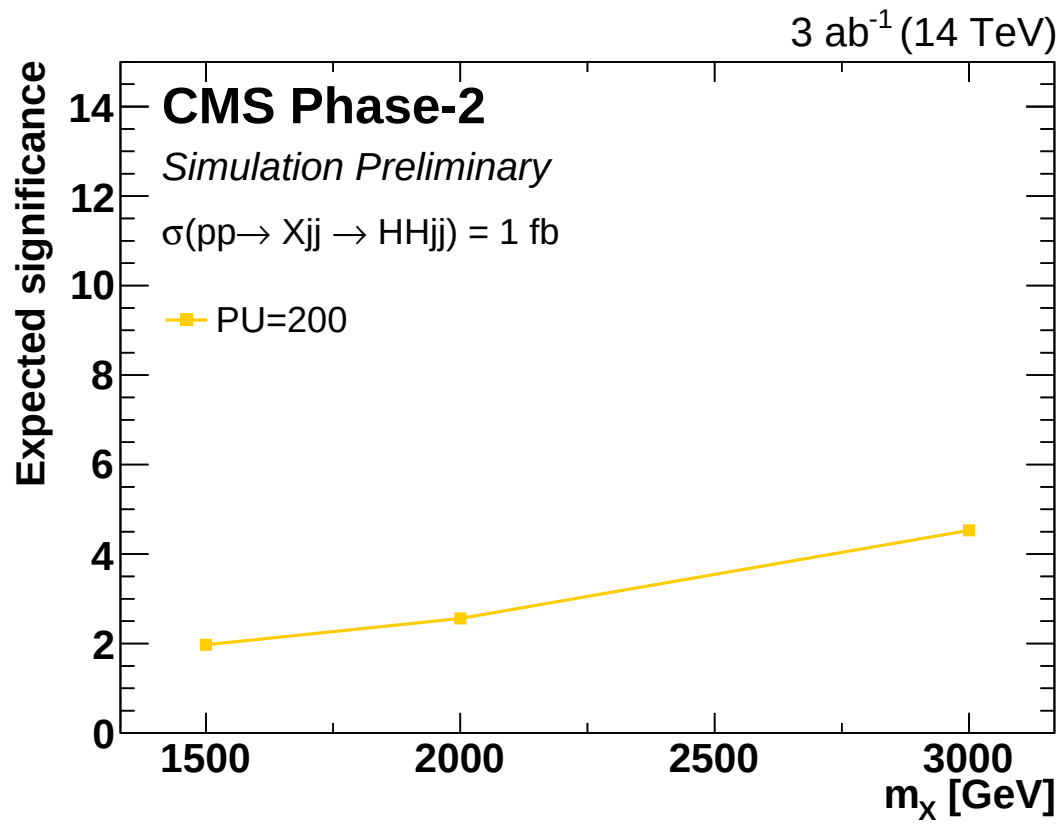

Figure 3: The expected signal significance for Bulk Gravitons of masses 1500, 2000, and 3000 GeV, assuming a production cross section of 1 fb. The data set corresponds to an integrated luminosity of 3 ab$^{-1}$ and with a pileup of 200.

**8**

## 10

# Search prospects for resonant Higgs boson pair production in the $b\bar{b}b\bar{b}$ final state with large-radius jets from $pp$ collisions at the HL-LHC

The ATLAS Collaboration

This note presents a study of resonant Higgs boson pair production in the $b\bar{b}b\bar{b}$ final state with 3000 fb$^{-1}$ of $\sqrt{s} = 14$ TeV $pp$ collisions at the HL-LHC. Results are based on a combination of extrapolation from a search carried out with $\sqrt{s} = 13$ TeV data and smearing of particle-truth level Monte Carlo simulation for the Kaluza–Klein graviton signal and the dominant multijet background. The truth-based analysis emulates the “boosted” analysis in which each of the Higgs bosons decaying into a $b\bar{b}$ system is reconstructed as a large-radius jet. Upper limits on the cross section for $pp \rightarrow G_{\mathrm{KK}} \rightarrow HH \rightarrow b\bar{b}b\bar{b}$ are estimated to be in the range between 1.44 fb at a resonance mass of 1.0 TeV and 0.025 fb at 3.0 TeV, at the 95% confidence level.

# 1 Introduction

The discovery of the Higgs boson ($H$) with a relatively low mass of 125 GeV is reaffirming the need to search for natural solutions to the hierarchy problem. Many new physics models predict rates of Higgs boson pair production significantly higher than the Standard Model (SM) rate [1–3]. For example, TeV-scale resonances such as the first Kaluza-Klein (KK) excitation of the graviton, $G_{\mathrm{KK}}$, predicted in the bulk Randall–Sundrum (RS) model [4, 5] or the heavy neutral scalar of two-Higgs-doublet models (2HDM) [6] can decay into pairs of Higgs bosons, $HH$.

The projection study presented here assumes an integrated luminosity of 3000 fb$^{-1}$ collected with proton–proton collisions at the High-Luminosity Large Hadron Collider (HL-LHC) [7] operating at a center-of-mass energy of $\sqrt{s}$ = 14 TeV and with a mean number of $pp$ interactions per bunch crossing $\mu$ = 200. The projection uses the search for high-mass spin-2 KK gravitons decaying into $HH$ as a benchmark, with each of the Higgs bosons decaying to $b\bar{b}$, thereby yielding a final state with two highly boosted $b\bar{b}$ systems, which are reconstructed as two large-radius jets. The most recent results of searches for $HH$ resonances in the $b\bar{b}b\bar{b}$ final state have been released by the ATLAS and CMS collaborations [8, 9] and correspond to the analysis of about 36 fb$^{-1}$ of $pp$ collision data at $\sqrt{s}$ = 13 TeV. Based on these results, a lower mass limit of 1.36 TeV is set at the 95% confidence level (CL) for coupling parameter $k/\overline{M}_{\mathrm{Pl}}$ = 1.0 [8].

An upgraded ATLAS detector [10] will operate at the HL-LHC. The upgraded detector includes a new all-silicon inner tracking detector [11] consisting of five layers of pixel detectors and four layers of double-sided silicon strip detectors to reconstruct the trajectory of charged particles within the pseudorapidity range $|\eta| < 4.0$. A superconducting solenoid with a 2 T axial magnetic field surrounds the inner tracker to enable the momentum of charged particles to be measured. Electromagnetic and hadronic calorimeters [12, 13] located outside of the solenoid provide energy measurements with high granularity within $|\eta| < 4.9$. The outer most muon spectrometer [14] consists of a set of superconducting toroidal magnets and three layers of gas chambers to measure the trajectories of muons up to $|\eta| = 2.7$. It also incorporates fast detectors for triggering purposes up to $|\eta| = 2.4$. In addition to upgrades to the detector, the on-line trigger system will be replaced by a new system [15] capable of processing the ten-fold rate increase anticipated at the HL-LHC.

The following strategy is followed to obtain sensitivity estimates at the HL-LHC: (i) signal and background mass distributions for the pair of candidate Higgs bosons in the event are taken from the most recent ATLAS data analysis at $\sqrt{s}$ = 13 TeV [8] and scaled to 3000 fb$^{-1}$; (ii) simulated signal and background mass distributions are used to derive mass-dependent scaling functions to extrapolate the distributions from $\sqrt{s}$ = 13 TeV to 14 TeV; (iii) simulated signal and background mass distributions are used for further scaling of the distributions to reproduce the impact of additional selection criteria not included in Ref. [8]. Systematic uncertainties are scaled down by factors of two or more (where applicable) relative to the values from the 13 TeV data analysis to account for the increased precision available with 3000 fb$^{-1}$ at the HL-LHC.

# 2 Signal model

The projection is based on an interpretation in terms of $G_{\mathrm{KK}}$ production in the bulk RS model. This model is used as a benchmark against which the improvement in sensitivity at the HL-LHC can be assessed. The RS model postulates the existence of a warped extra dimension in which only gravity propagates as in the original "RS1" scenario [16] or in which both gravity and all SM fields propagate as in the "bulk RS" scenario [17]. Propagation in the extra dimension leads to a tower of Kaluza–Klein excitations of gravitons (denoted $G_{\mathrm{KK}}$) and SM fields. In the bulk RS model considered here, KK gravitons are produced via both quark–antiquark annihilation and gluon–gluon fusion, with the latter dominating due to suppressed couplings to light fermions. The strength of the coupling depends on $k/\overline{M}_{\mathrm{Pl}}$, where $k$ corresponds to the curvature of the warped extra dimension and $\overline{M}_{\mathrm{Pl}} = 2.4 \times 10^{18}$ GeV is the effective four-dimensional Planck scale. Both the production cross section and decay width of the KK graviton scale as the square of $k/\overline{M}_{\mathrm{Pl}}$. For the values $k/\overline{M}_{\mathrm{Pl}} = 0.5$ and 1.0 used in the interpretation, the relative $G_{\mathrm{KK}}$ resonance widths ($\Gamma/m$) are approximately 1.5% and 6%, respectively. In both cases, the width of the reconstructed signal mass peaks is dominated by the experimental resolution. The $G_{\mathrm{KK}}$ branching fraction to $HH$ is approximately 10%; other final states are $t\bar{t}$ (60%), $WW$ (20%), and $ZZ$ (10%).

# 3 Monte Carlo simulation

## 3.1 Signal samples

Samples for the signal process are generated for $\sqrt{s} = 13$ and 14 TeV with the MadGraph5_aMC@NLO Monte Carlo (MC) event generator [18] using the implementation of the RS bulk model from Ref. [19] and cross sections computed at higher order from Ref. [20]. The coupling parameter for the warped extra dimension is set to $k/\overline{M}_{\mathrm{Pl}} = 1.0$. Predictions for $k/\overline{M}_{\mathrm{Pl}} = 0.5$ use the $k/\overline{M}_{\mathrm{Pl}} = 1.0$ samples scaled down by a factor of four to account for the decrease in cross section. The output of the event generation is interfaced to Pythia8 [21] for the decay of Higgs bosons to $b\bar{b}$ pairs, as well as for parton showering, hadronization, and underlying-event simulation. The Higgs boson mass is set to 125 GeV. The NNPDF2.3 [22] leading-order (LO) parton distribution functions (PDFs) are used throughout for signal production.

## 3.2 Background samples

The dominant background for the 13 TeV data analysis stems from multijet production. This background source is estimated directly from data in that analysis and represents about 80%, 90%, and 95% of the total background in the signal region for events with 2, 3, and 4 $b$-tags (the classification of events based on the number of $b$-tags is described below). The remaining source of background originates almost completely from $t\bar{t}$ production. The shape of the dijet mass distribution for $t\bar{t}$ events is taken from MC samples of events generated with Powheg-Box v1 [23, 24] and the CT10 PDF set [25]. These events are interfaced with Pythia6 [26] for simulation of the parton shower, hadronization, and the underlying event. The normalization of the $t\bar{t}$ background in Ref. [8] is extracted from a fit to the leading large-$R$ jet mass distribution in the 13 TeV data [27].

Samples of simulated multijet background events are generated to derive scaling functions to be applied to the background predictions from the 13 TeV data analysis. These samples are generated at both $\sqrt{s} = 13$ TeV and $\sqrt{s} = 14$ TeV. Two different sets of MC samples are used to study the impact of differences in jet flavor composition on the multijet background scaling functions. The first set of events corresponds to the $2 \rightarrow 2$ processes $pp \rightarrow jj$ (with $j = g$ or $q$) generated with Pythia8 with truth jet $p_{\mathrm{T}}$ in the range between 400 and 2500 GeV. The second set of events corresponds to the $2 \rightarrow 4$ processes $pp \rightarrow bbbb$ generated with MadGraph5_aMC@NLO requiring $b$-quarks to have $p_{\mathrm{T}}$ above 100 GeV and at least one $b$-quark with $p_{\mathrm{T}}$ above 200 GeV. For both sets, the events are interfaced with PYTHIA8 to simulate the parton shower, hadronization, and the underlying event.

## 4 Event reconstruction and selection

The truth-level events produced by the event generators with parton shower, hadronization, and underlying event as described in the previous section are processed with a reconstruction that builds a series of different jet collections. Large-radius jets are built from generated particles with the anti-$k_t$ algorithm [28] operating with a radius parameter $R = 1.0$, then trimmed [29] by reclustering the jet with the $k_t$ algorithm into $R = 0.2$ subjets and removing those subjets with $p_{\mathrm{T}}^{\mathrm{subjet}}/p_{\mathrm{T}}^{\mathrm{jet}} < 0.05$, where $p_{\mathrm{T}}^{\mathrm{subjet}}$ is the transverse momentum of the subjet and $p_{\mathrm{T}}^{\mathrm{jet}}$ that of the original jet. Previous studies indicate that the trimming effectively removes the impact of pileup up to $\mu = 300$ [30]. Small-radius jets are built from generated charged particles with the anti-$k_t$ algorithm and a radius of $R = 0.2$. Only charged particles with $p_{\mathrm{T}} > 0.5$ GeV are used in the clustering to emulate the track jets [31] used in the 13 TeV data analysis.

The detector response for the truth-level analysis is simulated with a set of functions providing a parameterized response derived from fully-simulated event samples. This includes efficiency maps for jets of different quark flavors to satisfy the $b$-tagging requirements of the `MV2c10` algorithm [32] at the 70% working point (i.e. the $b$-tagging efficiency for $b$-quarks in a sample of simulated $t\bar{t}$ events is equal to 70%) for both the 13 TeV data and future HL-LHC detector conditions, the latter being evaluated with an average pileup of $\mu = 200$. For the same $b$-tagging 70% working point, the rejection of charm and light jets improves by a factor of ~2 and ~2.5, respectively, for the HL-LHC detector relative to the Run 2 detector relevant to the 13 TeV data analysis. More information about the $b$-tagging performance can be found in Section 3.2.2 of Ref. [11].

The event selection applied to the truth-level analysis proceeds similarly to that used for the 13 TeV data analysis and is summarized below:

- **Trigger:** Events are required to have at least one large-radius jet reconstructed with the anti-$k_t$ algorithm with a radius $R = 1.0$ and with $p_{\mathrm{T}} > 420$ GeV. The trigger requirement is 100% efficient for signal events passing the selection described below. Trigger studies for the HL-LHC on-line system indicate that it will be possible to maintain the $p_{\mathrm{T}}$ threshold for large-$R$ jets at values as low or lower than that for the 13 TeV data taking (see Section 6.11 in Ref. [15]).
- **2 large-*R* jets:** The event must contain two large-$R$ jets with $p_{\mathrm{T}} > 250$ GeV, mass $m_{\mathrm{J}} > 50$ GeV, and $|\eta| < 2.0$. The leading jet must have $p_{\mathrm{T}} > 450$ GeV to guarantee 100% trigger efficiency.
- **Rapidity separation:** The pseudorapidity separation between the leading and subleading large-$R$ jets must satisfy $|\Delta\eta| < 1.7$ to suppress multijet background processes.
- **Track jets:** The leading and subleading large-$R$ jets must contain at least one track jet each to be considered as Higgs boson candidates. The track jets are associated with the large-$R$ jets via ghost association [33, 34].
- ***b*-tagging**: Events are classified by the number of $b$-tagged track jets associated with the leading and subleading large-$R$ jets, other large-$R$ jets are not considered in the event selection. Two-tag events are characterized by the presence of exactly one $b$-tagged track jet in each of the two large-$R$ jets. Three-tag events have exactly one $b$-tagged track jet in one large-$R$ jet and two in the other

4

large-$R$ jet. Four-tag events have exactly two $b$-tagged track jets in each of the two large-$R$ jets. It should be noted that the $b$-tagging requirement is applied via weights that correspond to the probability for a given large-$R$ jet to contain zero, one, two, or more $b$-tagged track jets. The individual track jet $b$-tagging probability is given by the corresponding $b$-tagging efficiency, which depends on the true quark flavor of each track jet.

- **Higgs mass:** The masses of the leading large-$R$ jet ($m_{\mathrm{J}}^{\mathrm{lead}}$) and subleading large-$R$ jet ($m_{\mathrm{J}}^{\mathrm{subl}}$) must be consistent with the Higgs boson mass: $X_{HH} < 1.6$, to optimize the search sensitivity, see Eq. (1) below:
$$X_{HH} = \sqrt{\left(\frac{m_{\mathrm{J}}^{\mathrm{lead}} - 125\,\mathrm{GeV}}{0.1\, m_{\mathrm{J}}^{\mathrm{lead}}}\right)^2 + \left(\frac{m_{\mathrm{J}}^{\mathrm{subl}} - 120\,\mathrm{GeV}}{0.1\, m_{\mathrm{J}}^{\mathrm{subl}}}\right)^2}\,. \tag{1}$$
The peak values are set to be 125 GeV and 120 GeV for the leading and subleading large-$R$ jets, respectively, in the simulated samples to match the peak positions in the truth-level analysis (this differs slightly from the full-simulation analysis). A lower peak value is set for the subleading large-$R$ jets due primarily to energy carried away by neutrinos from semileptonic decays.

Three regions in the plane formed by the leading large-$R$ jet mass and the subleading large-$R$ jet mass are used in the analysis. The signal region (SR) is defined by the requirement $X_{HH} < 1.6$ and the control region (CR) is defined by the requirements $R_{HH} = \sqrt{\left(m_{\mathrm{J}}^{\mathrm{lead}} - 125\,\mathrm{GeV}\right)^2 + \left(m_{\mathrm{J}}^{\mathrm{subl}} - 120\,\mathrm{GeV}\right)^2} < 33$ GeV and $X_{HH} \geq 1.6$. The sideband region (SB) is defined by the requirements $R_{HH} \geq 33$ GeV and $\sqrt{\left(m_{\mathrm{J}}^{\mathrm{lead}} - 135\,\mathrm{GeV}\right)^2 + \left(m_{\mathrm{J}}^{\mathrm{subl}} - 130\,\mathrm{GeV}\right)^2} < 58$ GeV. The choices made to define the control and sideband regions are driven by the need to select events that are kinematically similar to those in the signal region while providing sufficiently large samples to derive the background estimate from the sideband region and validate it in the control region. These regions are depicted in Figure 1 which shows a two-dimensional distribution in the leading–subleading large-$R$ jet mass plane for the dijet MC sample at $\sqrt{s} = 13$ TeV.

For the 13 TeV data analysis, the dominant multijet background is estimated with a data-driven approach which utilizes events with a smaller number of $b$-tags in the sideband region. The events used for this estimation are required to have the same track-jet topology as in the event categories they are used to model the background as illustrated in Figure 2. Events with 1 $b$-tag are used to model the background in the 2-tag category. Likewise, events with 2 $b$-tags are used to model the background in the 3- and 4-tag categories.

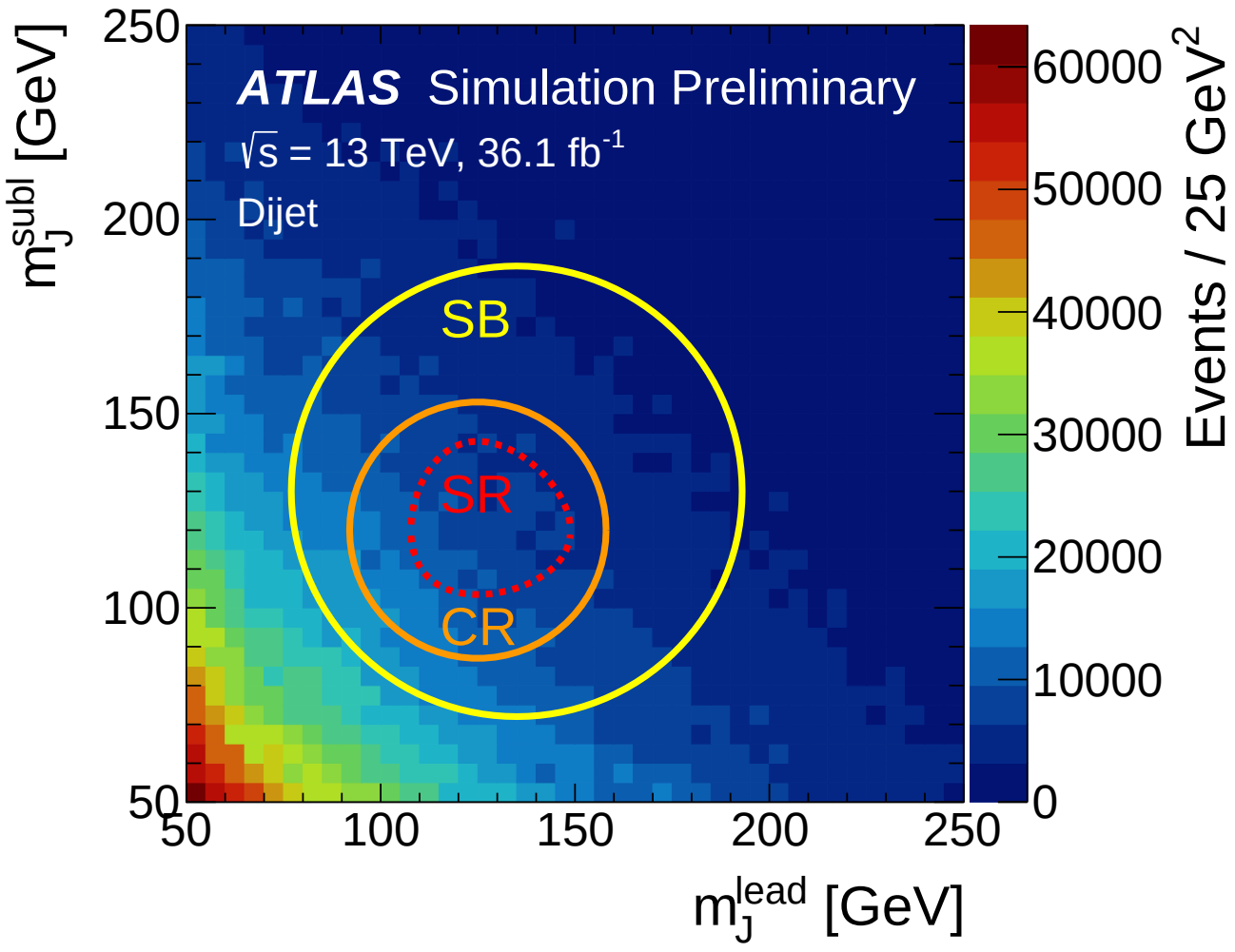

Figure 1: Leading–subleading large-$R$ jet mass plane distribution for the dijet MC sample at $\sqrt{s}$ = 13 TeV for events passing the trigger, large-$R$ jets, and rapidity separation requirements in the truth-level analysis. The inner-most contour delineates the signal region (SR). The control region (CR) lies between the inner and middle contours whereas the sideband region (SB) lies between the middle and outer contours.

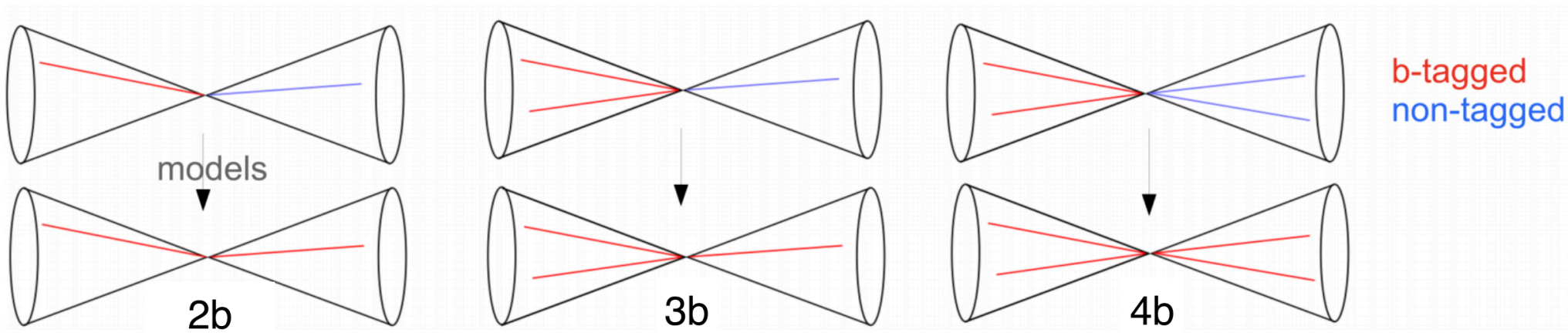

Figure 2: Sketch of event topologies used in the data-driven multijet background estimation. The upper row shows the track-jet topologies used to model the background in the events passing the event selection as indicated in the bottom row. Events in the upper row have fewer $b$-tags.

Table 1: Cumulative acceptance times efficiency at each selection stage for the signal MC samples, as well as for the 13 TeV data analysis from Ref. [8]. The acceptance times efficiency is only available for the broader 2-tag category that includes events in which both $b$-tagged track jets are associated with the same large-$R$ jet in the case of the 13 TeV data analysis.

| Requirement | signal ($m$ = 2.0 TeV) | | signal ($m$ = 3.0 TeV) | |
|---|---|---|---|---|
| | truth analysis | 13 TeV analysis | truth analysis | 13 TeV analysis |
| Trigger | 1.00 | 1.00 | 1.00 | 1.00 |
| Two large-$R$ jets | 0.85 | 0.80 | 0.87 | 0.86 |
| Large-$R$ jet $\Delta\eta$ | 0.81 | 0.76 | 0.83 | 0.82 |
| At least 2 $b$-tagged track jets | 0.66 | 0.59 | 0.46 | 0.43 |
| $X_{HH} < 1.6$ (SR) | 0.23 | 0.26 | 0.14 | 0.17 |
| 2-tag SR (incl. same jet) | 0.086 | 0.090 | 0.083 | 0.116 |
| 2-tag SR | 0.057 | – | 0.057 | – |
| 3-tag SR | 0.097 | 0.113 | 0.042 | 0.038 |
| 4-tag SR | 0.041 | 0.036 | 0.009 | 0.005 |

## 5 Validation against 13 TeV data analysis

To validate the truth-level analysis described above, the acceptance times efficiency $A \times \epsilon$ is compared with results from the 13 TeV data analysis [8]. Table 1 presents $A \times \epsilon$ at successive stages of event selection (see Section 4) for the signal samples. It is found that the truth-level analysis emulates the 13 TeV data analysis fairly well for the $m$ = 2.0 TeV and 3.0 TeV signal samples used as a test. Perfect agreement is not required since the truth-level analysis is only used to derive scaling functions based on relative changes in normalization or shape of the dijet mass distributions rather than to derive absolute predictions for signal and background event yields.

The invariant mass distribution for the combination of leading and subleading large-$R$ jets is shown in Figure 3 for signal events with $m$ = 2.0 and 3.0 TeV in the 2-tag, 3-tag, and 4-tag categories. Distributions for the truth-level analysis are normalized to those for the 13 TeV data analysis. A slight offset is observed which is due to a different treatment of large-$R$ jets in the full detector simulation and reconstruction and in the truth-level analysis. However, this does not affect the scaling of signal events used for the HL-LHC projection appreciably.

The 13 TeV data analysis can be further utilized to check the multijet background simulation used here for scaling purposes. For this check, the $t\bar{t}$ background shape and normalization are taken from the 13 TeV data analysis while the multijet background shapes are taken from the truth-level analysis with a normalization provided by the 13 TeV data analysis such that the sum of the multijet and $t\bar{t}$ backgrounds matches the number of data events in the combined signal and control regions (see Figure 4). There is good agreement in the shapes of the dijet mass distributions for both the dijet and $4b$ multijet MC samples described in Section 3.2, although the former does suffer from a limited number of events in the 4-tag category.[1]

[1] A wider region including the sideband region is used to enlarge the number of events and derive the scaling functions for the HL-LHC projection described in Section 6.

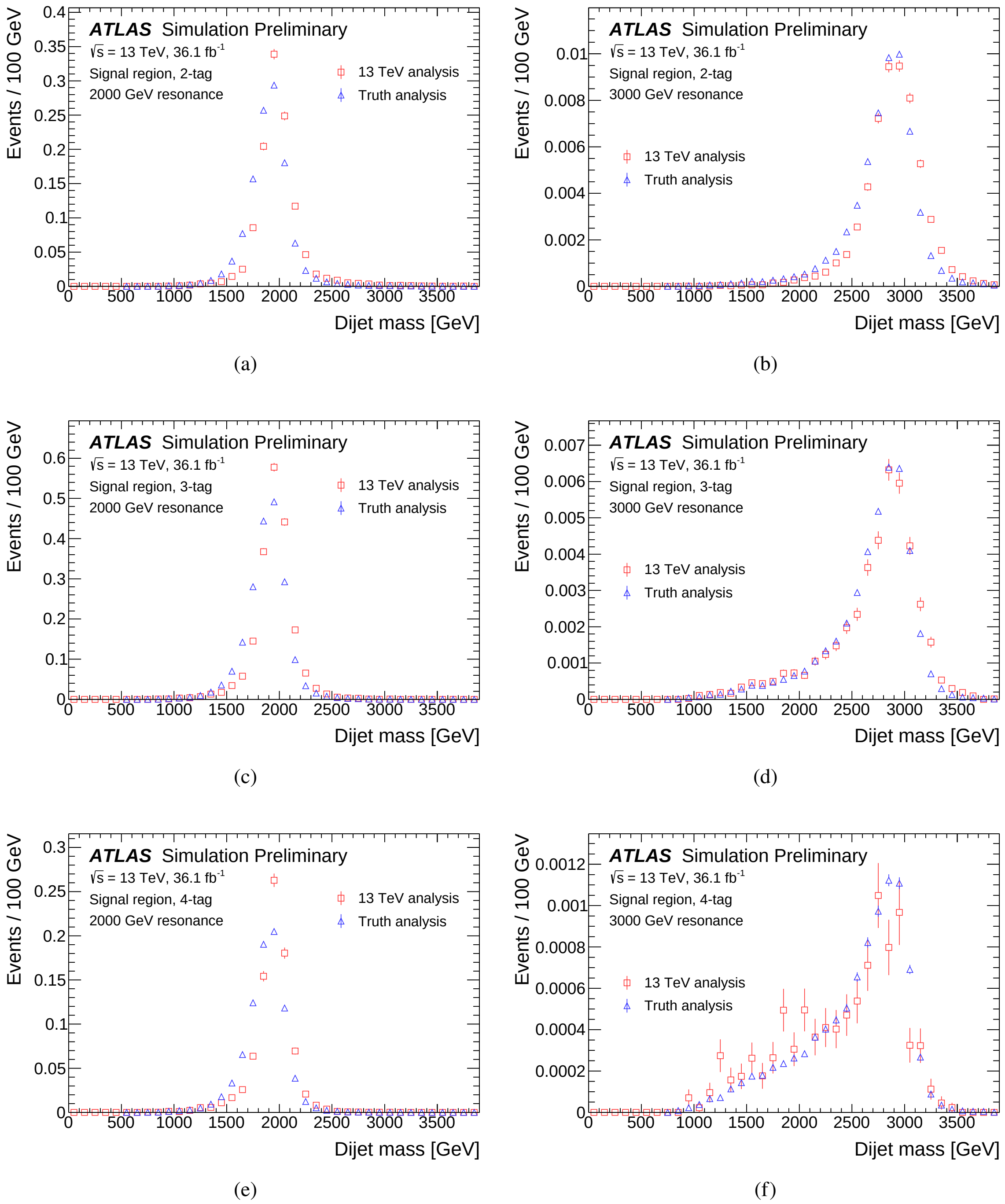

Figure 3: Dijet mass distributions for 2-tag (top), 3-tag (middle), and 4-tag (bottom) signal events in the signal region for $G_{\mathrm{KK}}$ resonances with masses of 2 TeV (left) and 3 TeV (right) in the bulk RS model from the 13 TeV data analysis in Ref. [8] and the truth-level analysis. The histograms for the truth-level analysis are normalized to those from Ref. [8].

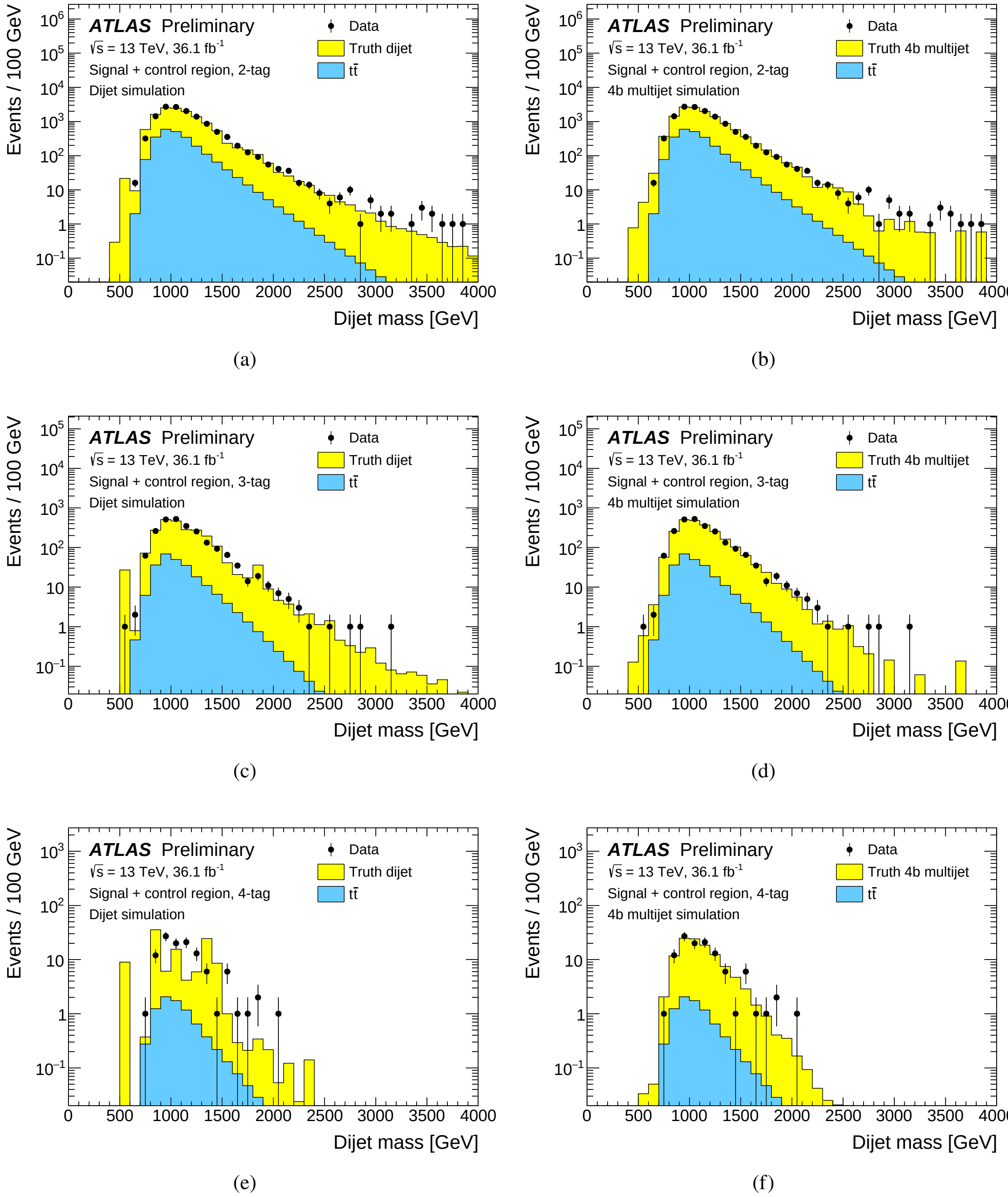

Figure 4: Dijet mass distributions for 2-tag (top), 3-tag (middle), and 4-tag (bottom) events in the combined signal and control regions for data (points) and the expected background (histograms). The $t\bar{t}$ background comes from the 13 TeV data analysis in Ref. [8]. The multijet background comes from the truth-level analysis and corresponds to the $pp \to jj$ process generated with Pythia8 (left) or $pp \to bbbb$ process generated with MadGraph5_aMC@NLO (right) normalized such that the total background matches the number of events observed in the data.

Figure 5 presents the dijet mass distributions in the signal and control regions combined for the multijet background predicted by either the "dijet" multijet background generated with Pythia8 or the "4$b$" multijet background generated with MadGraph5_aMC@NLO as described in Section 3.2. The distributions are normalized to the number of events estimated with the 13 TeV data. There is good agreement in the shape of the dijet mass distributions produced with either event generator. This indicates that either multijet simulation can be used reliably to predict the shape of the dijet mass distributions. However, differences in the flavor content of the large-$R$ jets in the dijet and 4$b$ multijet MC samples do affect the predicted background yields in the following study and the two samples are used to extract a range of projections at the HL-LHC.

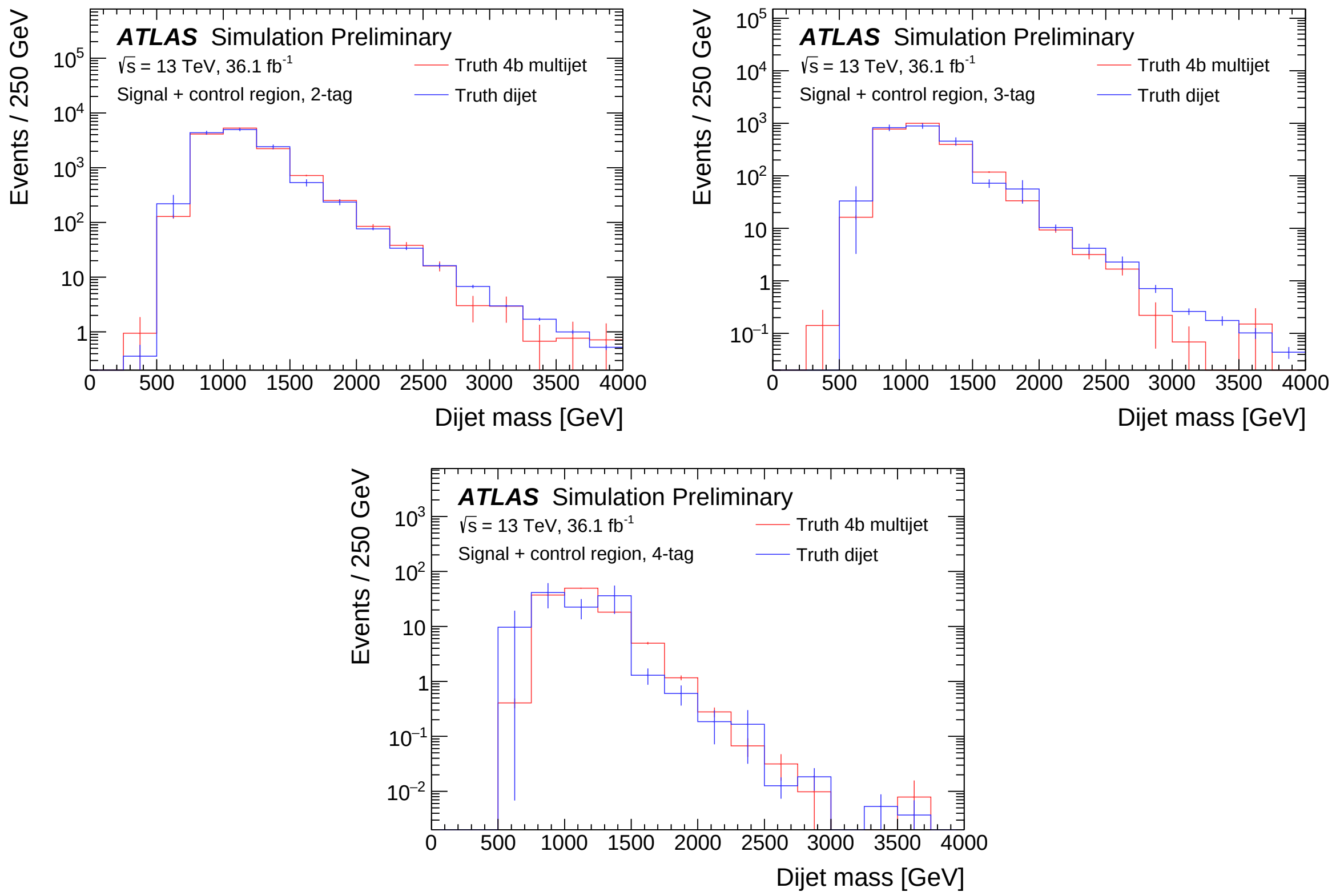

Figure 5: Dijet mass distributions for 2-tag, 3-tag, and 4-tag events in the combined signal and control region for the dijet and 4$b$ multijet MC samples. The distributions are extracted from the truth-level analysis and normalized to the predicted multijet event yields from the 13 TeV data analysis.

# 6 Projection for HL-LHC

The projection for the HL-LHC proceeds as follows:

1. Dijet mass distributions for signal and both multijet and $t\bar{t}$ background events from the 13 TeV data analysis in Ref. [8] are scaled from 36.1 fb$^{-1}$ to 3000 fb$^{-1}$.

2. Background distributions are further scaled with mass-dependent functions to extrapolate from $\sqrt{s}$ = 13 TeV to 14 TeV. These functions take into account both increases in cross section and changes in detector performance from the Run 2 ATLAS detector (as applicable to Ref. [8]) to the future upgraded detector at the HL-LHC. The functions are obtained separately for the 2-, 3-, and 4-tag categories in two steps. First, the dijet mass distributions are extracted from the truth-level analysis of multijet MC events with the Run 2 detector at $\sqrt{s}$ = 13 TeV and with the HL-LHC detector at $\sqrt{s}$ = 14 TeV, normalized using the same integrated luminosity. An example of such distributions is shown in Figure 6 for the 3-tag category. Lower event yields are obtained for the 14 TeV dijet MC sample in Figure 6(a) due to the stronger rejection of charm- and light-jets with the HL-LHC detector, despite a higher cross section relative to 13 TeV. Second, the ratio between the $\sqrt{s}$ = 14 TeV and 13 TeV distributions is fit with a low-order polynomial that serves as the scaling function. The differences in cross section values amount to a ~25% average increase in the kinematic region relevant to this study.

   Signal distributions are scaled by the ratio of calculated cross sections at $\sqrt{s}$ = 14 TeV and 13 TeV for each resonance mass value employed in the statistical analysis. The signal cross section increases linearly as a function of the resonance mass, yielding an increase growing from 26% at 1 TeV to 58% at 3 TeV [20].

3. Further mass-dependent scaling is applied to all signal and background distributions to reflect improvements in the reconstruction of highly boosted jets, as obtained by using variable-radius track jets [35] instead of fixed-radius ($R = 0.2$) track jets, or in the background suppression by applying a requirement on the maximum number of charged particles associated with each large-$R$ jet. Different scalings are applied depending on the number of $b$-tags. These additional scalings are discussed in more detail below.

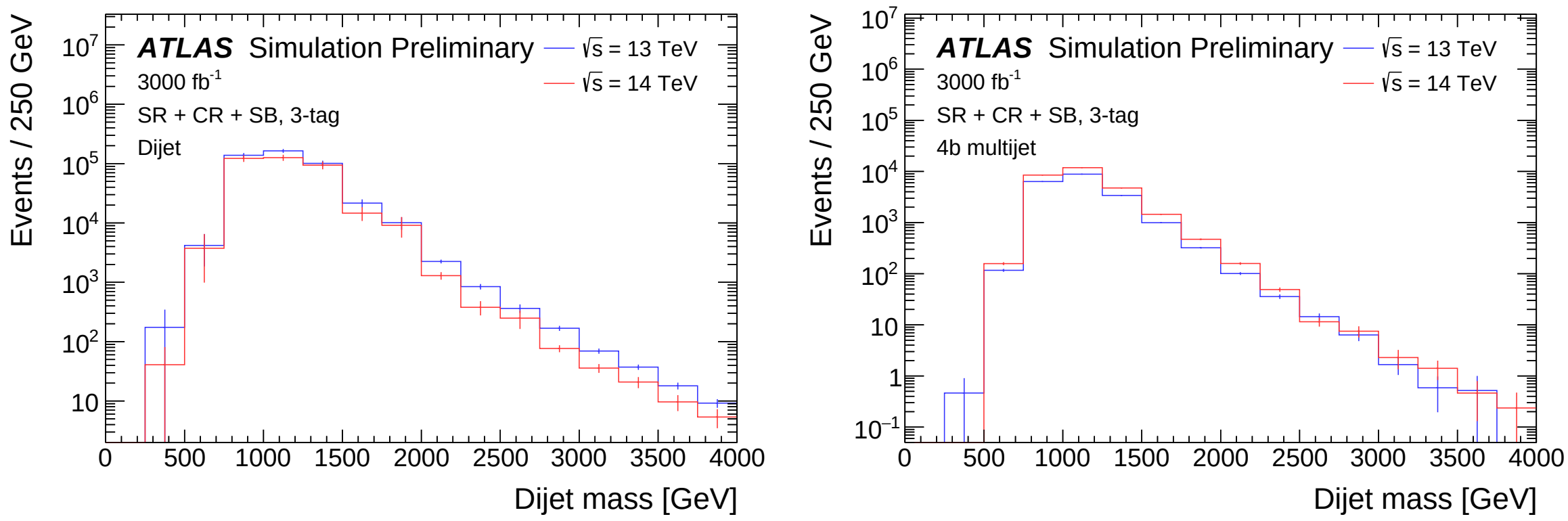

(a) (b)

Figure 6: Dijet mass distributions for 3-tag events at $\sqrt{s}$ = 13 TeV and 14 TeV in the combined signal+control+sideband region for (a) the dijet and (b) the $4b$ multijet MC samples. The distributions are extracted from the truth-level analysis and normalized to the predicted multijet event yields obtained using the cross sections computed by the event generators. The distributions are only used to derive scaling functions from fits to the ratio of the two distributions.

Scaling functions derived from the truth-level analysis of the multijet MC samples are applied equally to multijet and $t\bar{t}$ dijet mass distributions from the 13 TeV data analysis in Ref. [8]. Events in a combined signal+control+sideband region are used to enlarge the number of MC events available to derive these scaling functions.

The first improvement relative to the 13 TeV data analysis arises from the use of variable-radius track jets. This circumvents the problem that $R = 0.2$ track jets from the $H \to b\bar{b}$ decay start merging for Higgs boson $p_{\mathrm{T}}$ values larger than approximately $2m_H/R = 1250$ GeV. As an illustration, the dijet mass distributions for 3-tag events in the dijet and $4b$ multijet MC samples are shown in Figure 7 for each choice of track jet algorithm. Distributions for signal MC events with $m(G_{\mathrm{KK}}) = 3.0$ TeV are shown in Figure 8(a). A consequence of using variable-radius track jets is that $A \times \epsilon$ remains essentially unchanged for 2-tag events but increases appreciably for 3-tag and especially 4-tag events as the resonance mass increases, as shown in Figure 8(b).

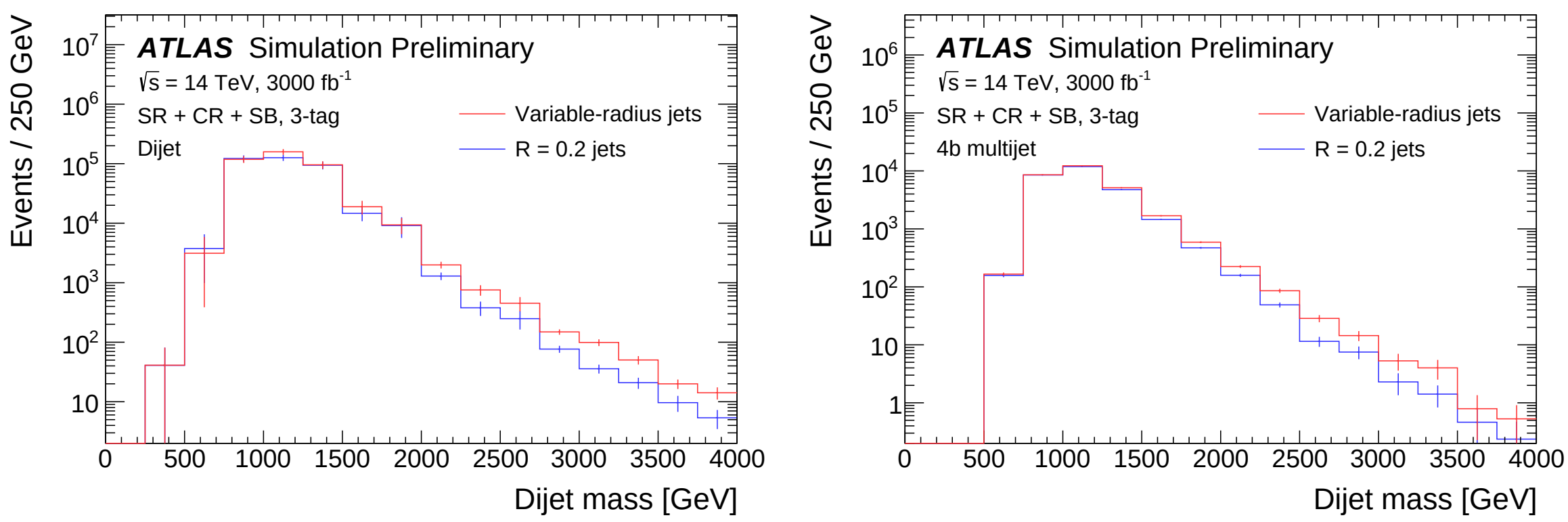

(a) (b)

Figure 7: Dijet mass distributions from the truth-level analysis for 3-tag events with fixed-radius ($R = 0.2$) and variable-radius track jets in the combined signal+control+sideband region for (a) the dijet and (b) the $4b$ multijet MC samples. The distributions are extracted from the truth-level analysis and normalized to the predicted multijet event yields obtained using the cross sections computed by the event generators. The distributions are only used to derive scaling functions from fits to the ratio of the two distributions.

Dijet mass distributions from the 13 TeV data analysis for signal and background events are scaled with mass-dependent functions to account for the change in $A \times \epsilon$ due to the use of variable-radius track jets. Separate signal and background functions are derived for the different numbers of $b$-tags. Background functions are derived from the $4b$ multijet MC sample and found to agree statistically with those derived from the dijet MC sample, indicating weak dependence on the flavor composition of the jets. Both signal and background event yields show increases for 3-tag and 4-tag events, although the increase is more pronounced for signal events and thus the search sensitivity improves by reconstructing track jets with a variable radius.

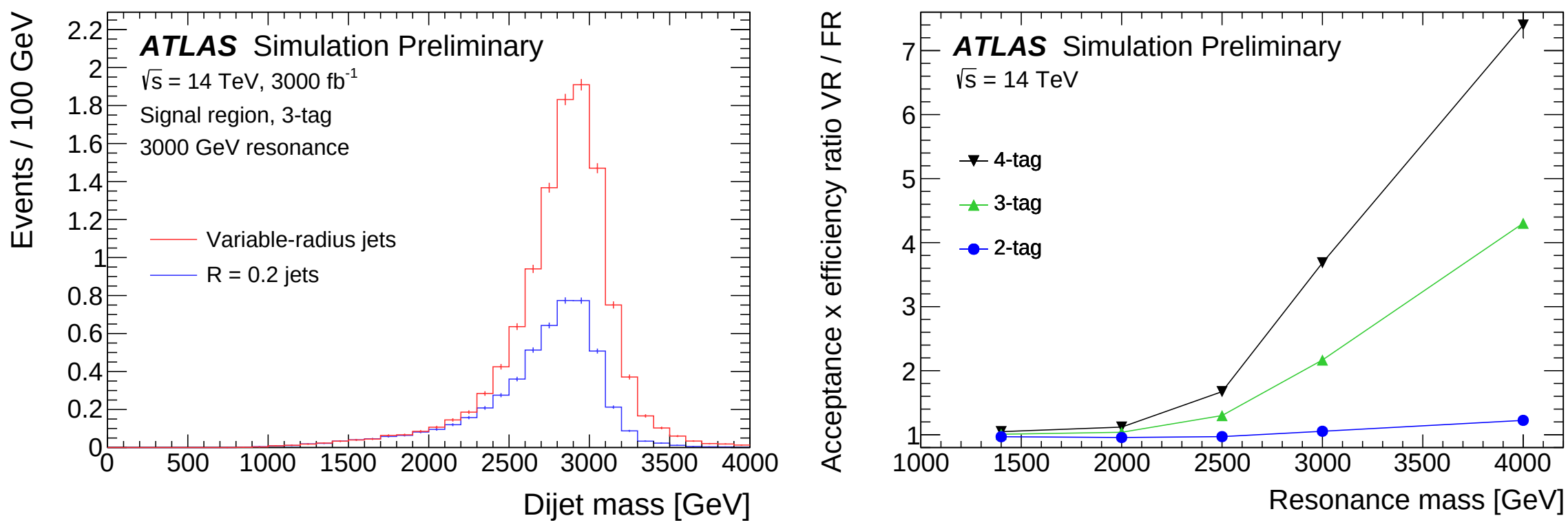

(a) (b)

Figure 8: (a) Dijet mass distributions from the truth-level analysis for 3-tag events with fixed-radius ($R = 0.2$) and variable-radius track jets in the signal region for the bulk RS model $m(G_{\mathrm{KK}}) = 3.0$ TeV MC sample. (b) Ratio of acceptance times efficiency values for variable-radius track jets relative to fixed-radius track jets for 2-tag, 3-tag, and 4-tag events as a function of $G_{\mathrm{KK}}$ mass in the bulk RS model MC samples at $\sqrt{s} = 14$ TeV.

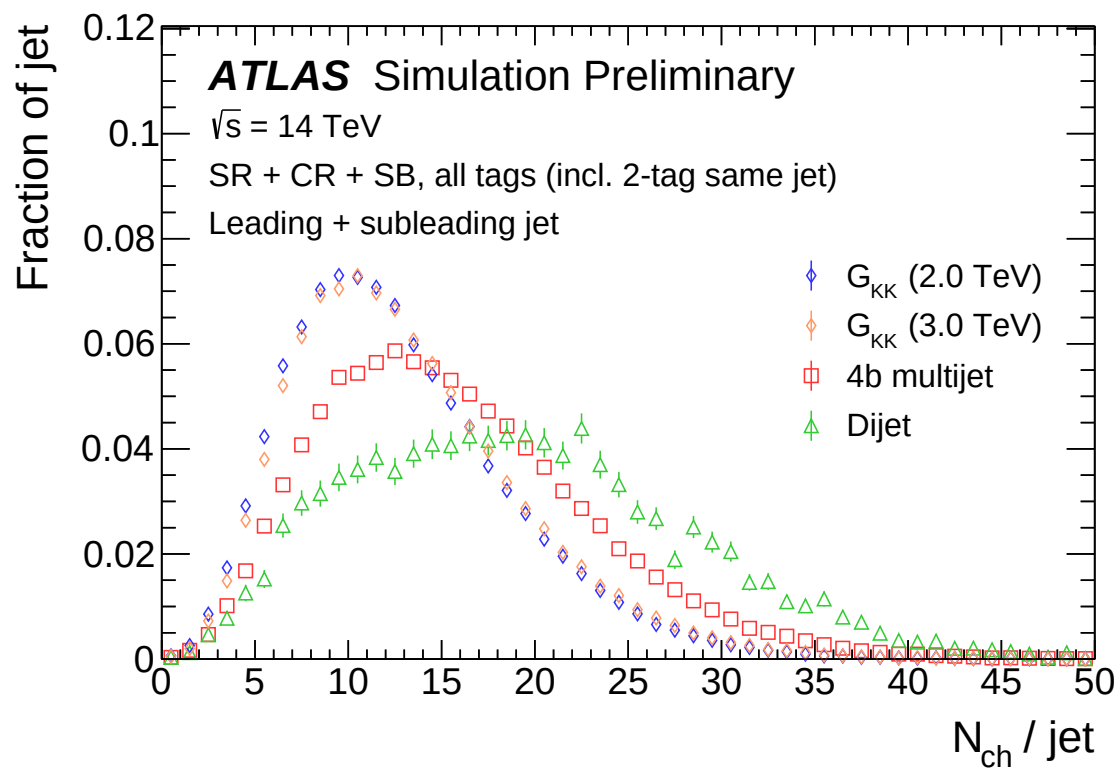

Figure 9: Large-$R$ jet charged particle multiplicity distributions from the truth-level analysis for events with variable-radius track jets in the combined signal+control+sideband region for signal events with $m(G_{\mathrm{KK}}) = 2.0$ TeV and $m(G_{\mathrm{KK}}) = 3.0$ TeV, as well as dijet and $4b$ multijet MC samples, at $\sqrt{s} = 14$ TeV. Charged particles are required with have $p_{\mathrm{T}} > 1$ GeV and be within $\Delta R < 0.6$ of the large-$R$ jet.

The second improvement relative to the 13 TeV data analysis is the requirement of a maximum number of charged particles associated with large-$R$ jets to exploit differences between quark- and gluon-initiated jets, the latter being an important component of the multijet background. A distribution of the number of associated charged particles shows a clear separation between signal and background contributions (see Figure 9). The impact of pileup at $\mu = 200$ has been studied for charged particle tracks with $p_{\mathrm{T}} > 1$ GeV associated with the primary vertex (see Section 3 of Ref. [11]). In the case of $t\bar{t}$ events, the average number of tracks associated with the primary vertex increases by about 15% due to pileup. Further pileup suppression is possible with additional requirements on the longitudinal impact parameter or track-vertex association probability.

Both leading and subleading large-$R$ jets are required to have fewer than 20 charged particles with $p_T > 1$ GeV and $\Delta R < 0.6$ with respect to the jet axis. This additional requirement retains approximately 80% of the signal events in the signal region, with little dependence on the number of $b$-tags or the resonance mass. The impact of this requirement is considerably larger on multijet background events than it is on signal events, as illustrated in Figures 10 and 11, which show the dijet mass distributions for background and signal events with 3 $b$-tags. The reduction in background level increases with dijet mass. The mass-dependent scaling functions derived from the 4$b$ multijet MC sample provide a smaller degree of improvement in sensitivity of the analysis since the impact of the charged multiplicity requirement is more pronounced in the case of the dijet MC sample.

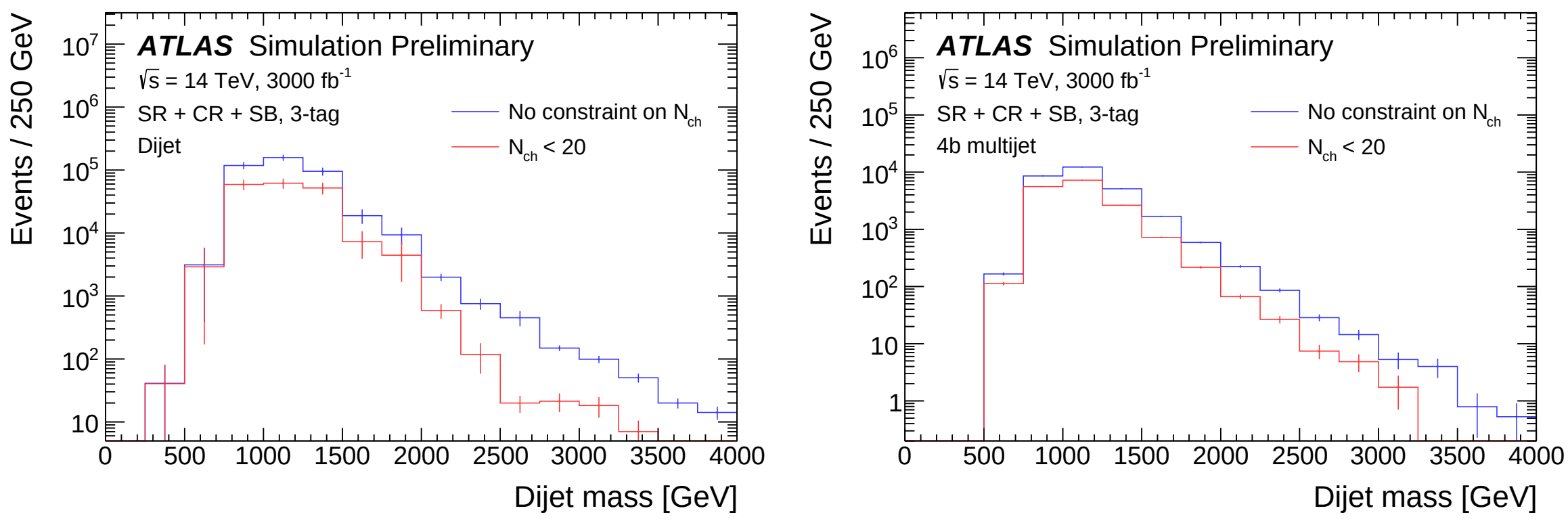

(a) (b)

Figure 10: Dijet mass distributions from the truth-level analysis for 3-tag events before and after the requirement on number of charged particles associated with the large-$R$ jets for (a) the dijet and (b) the 4$b$ multijet MC samples. The distributions are normalized to the predicted multijet event yields obtained using the cross sections computed by the event generators. The distributions are only used to derive scaling functions from fits to the ratio of the two distributions.

The dijet mass distributions at $\sqrt{s} = 14$ TeV with 3000 fb$^{-1}$ resulting from the scaling procedure described above, including variable-radius track jets and the requirement on the maximum number of charged particles per large-$R$ jet, are shown in Figure 12.

## 7 Statistical analysis and results

The statistical analysis uses a test statistic based on the profile likelihood ratio [36] with a signal strength parameter equal to $\sigma/\sigma_{\text{model}}$. If no discovery is made, exclusion limits are set with the asymptotic assumption and following the CL$_s$ method [37]. Signal strength values are excluded at the 95% confidence level if the CL$_s$ value is smaller than 0.05. The 2-tag, 3-tag, and 4-tag categories are combined in the analysis.

The full set of systematic uncertainties from the 13 TeV data analysis is included in the statistical analysis. These comprise theoretical uncertainties in the signal acceptance as well as experimental uncertainties affecting the large-$R$ jet reconstruction (both scale and resolution uncertainties in the jet energy and mass) and the $b$-tagging efficiencies. Also included are uncertainties in the shape and normalization of the multijet and $t\bar{t}$ backgrounds. Further details are available in Ref. [8]. The dominant uncertainties in the signal event yields arise from $b$-tagging and large-$R$ jet mass resolution, whereas for the background event yields they arise from the data-driven estimate.

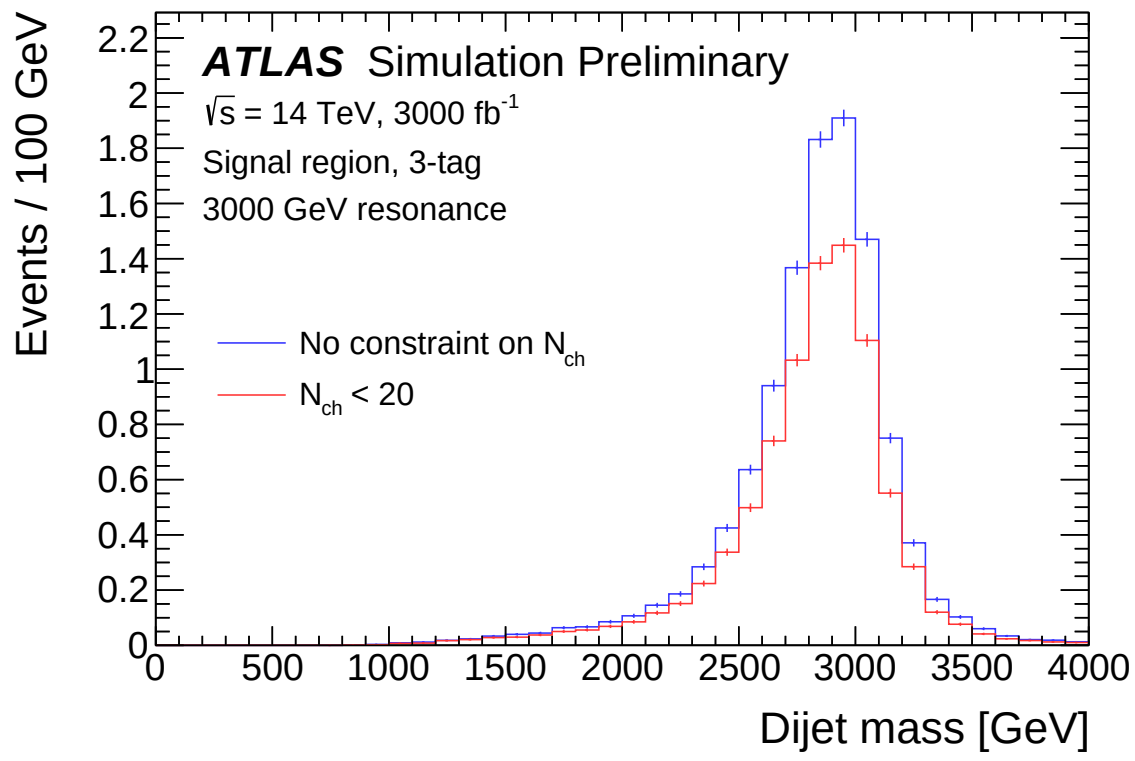

Figure 11: Dijet mass distributions for 3-tag events with variable-radius track jets for signal events in the signal region for the bulk RS model $m(G_{\mathrm{KK}}) = 3.0$ TeV MC sample before and after the requirement on the number of charged particles associated with the large-$R$ jets.

For the signal, the HL-LHC extrapolation assumes the following reduction in systematic uncertainties: the $b$-tagging efficiency and large-$R$ jet mass resolution uncertainties are reduced by factors of 3 and 2, respectively, relative to the 13 TeV data analysis. A 1% uncertainty is assumed for the measurement of the integrated luminosity, which has negligible impact on the sensitivity. A reduction by a factor of 3 is assumed for all other sources of uncertainty—the most important sub-dominant sources of uncertainty are the jet energy scale and resolution as well as the jet mass scale. For the background, the HL-LHC extrapolation assumes that the background estimate uncertainties are fully driven by the statistical uncertainties in the samples used in the estimate and thus scale according to $1/\sqrt{N}$.

The expected 95% CL cross section upper limits for the 13 TeV data analysis with 36.1 fb$^{-1}$ and the HL-LHC projection at $\sqrt{s}$ = 14 TeV with 3000 fb$^{-1}$ are shown in Figure 13. The limits for the 13 TeV data analysis with 36.1 fb$^{-1}$ presented here are not identical to those from Ref. [8] since the latter also include the resolved $b\bar{b}b\bar{b}$ channel which contributes the most at lower resonance mass. For resonance masses above 1 TeV, the merged channel considered here dominates the sensitivity of the search.

Upper limits on $\sigma \times \mathcal{B}$ at $\sqrt{s}$ = 14 TeV range from 1.44 fb (1.82 fb) at a mass of 1.0 TeV to 0.025 fb (0.040 fb) at a mass of 3.0 TeV when dijet (4$b$) scaling and the variable-radius track jets with a maximum requirement on the number of charged particles are applied. The benefit from the use of variable-radius track jets becomes significant at the highest resonance masses considered here, with an improvement in the upper limits of at least 24% (depending on the choice of scaling) at 3.0 TeV. Addition of a requirement on the maximum number of charged particles further improves the upper limits by factors of about 20% and 45% at masses of 2.0 and 3.0 TeV, respectively. Systematic uncertainties have a modest impact on the limits with an effect of at most 20% at 1.0 TeV and decreasing to $\sim$ 5% at high mass. A hypothetical relative improvement of 10% in the $b$-tagging efficiency for true $b$-jets with unchanged charm- and light-jet rejection yields improvements in the limits of at most 12% when 4$b$ scaling is assumed, with more modest gains when dijet scaling is assumed.

The lower mass limits on KK gravitons improve from the current 1.36 TeV at $\sqrt{s}$ = 13 TeV to 2.95 TeV (2.75 TeV) at the HL-LHC in the scenario with coupling $k/\overline{M}_{\mathrm{Pl}} = 1.0$. In the case of $k/\overline{M}_{\mathrm{Pl}} = 0.5$, no limit can be set with the 13 TeV data analysis but a lower limit of 2.15 TeV (2.00 TeV) can be set at the HL-LHC given a multijet background scaling modeled with the dijet (4$b$ multijet) MC samples. These values are summarized in Table 2. Higher sensitivity is obtained for projections with scaling functions derived from the dijet MC samples due to the mix of different quark flavors in multijet events and the significant improvement in charm- and light-jet rejection expected with the upgraded HL-LHC detector. The projection obtained with the dijet MC sample is taken as the primary estimate of the analysis sensitivity at the HL-LHC, as this sample better approximates the relative fractions of 2-, 3-, and 4-tag events in the 13 TeV data analysis. This is likely due to the mix of different quark flavors in that sample. The projection obtained with the 4$b$ multijet MC sample, which assumes the multijet background to be of irreducible four $b$-quark composition with no charm- or light-flavor contribution, provides a more conservative estimate of the analysis sensitivity.

Table 2: Expected 95% CL lower limits on $G_{\mathrm{KK}}$ mass for the 13 TeV data analysis and the extrapolation to the HL-LHC for $k/\overline{M}_{\mathrm{Pl}} = 0.5$ and 1.0 in the bulk RS model. Different extrapolations are provided based on modeling of the changes in multijet background using either the dijet or the 4$b$ multijet MC samples.

| Model | $\sqrt{s}$ = 13 TeV, 36.1 fb$^{-1}$ as in Ref. [8] | $\sqrt{s}$ = 14 TeV, 3000 fb$^{-1}$ dijet scaling | 4$b$ scaling |
|---|---|---|---|
| $k/\overline{M}_{\mathrm{Pl}} = 0.5$ | no limit | 2.15 TeV | 2.00 TeV |
| $k/\overline{M}_{\mathrm{Pl}} = 1.0$ | 1.36 TeV | 2.95 TeV | 2.75 TeV |

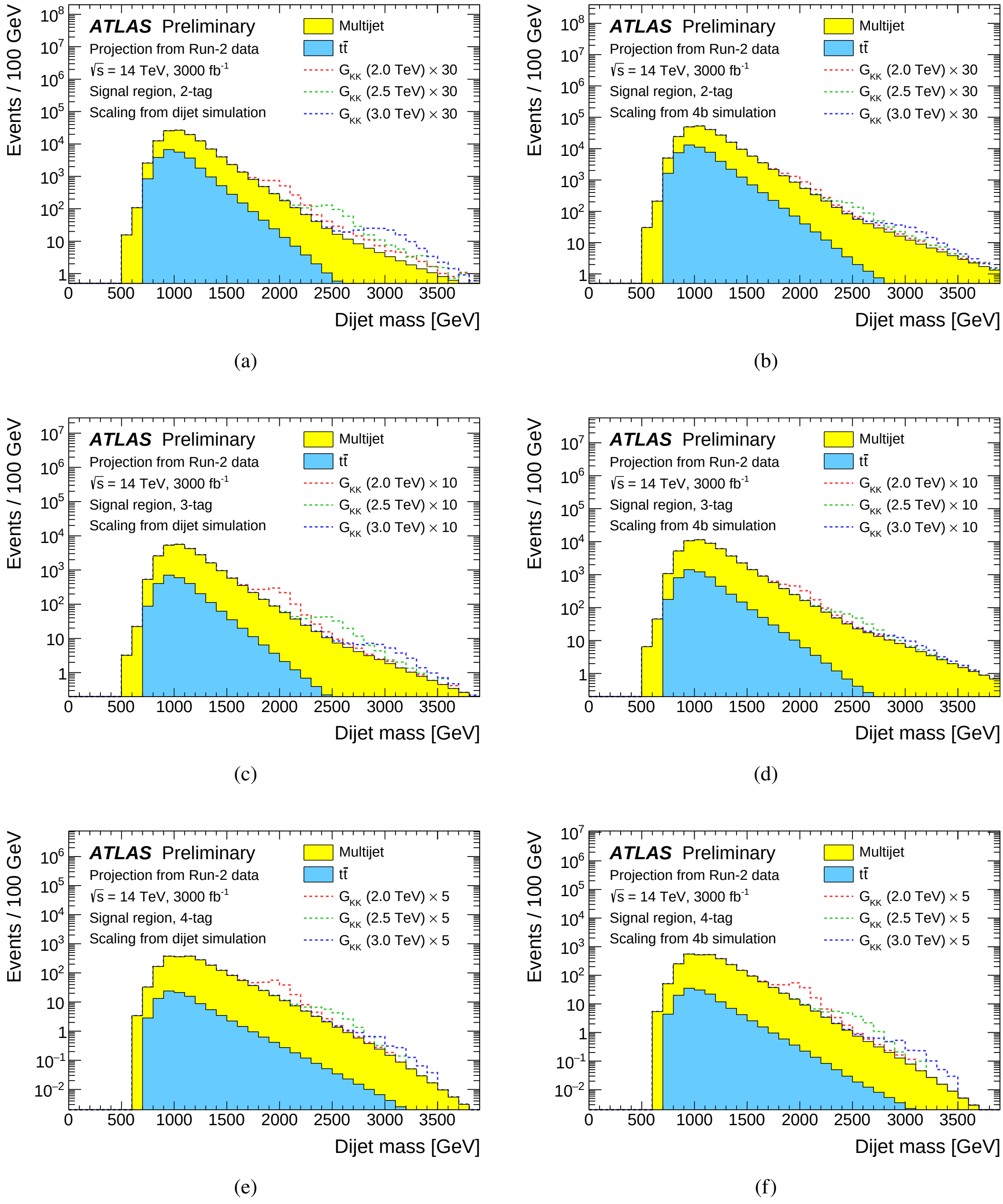

Figure 12: Dijet mass distributions from the truth-level analysis for 2-tag, 3-tag, and 4-tag events in the signal region for the expected background and signals at the HL-LHC. The multijet background is scaled using either the dijet (left) or the 4$b$ multijet (right) MC samples. The event yields for signal events at $G_{\mathrm{KK}}$ masses of 2.0, 2.5, and 3.0 TeV are scaled up for visibility.

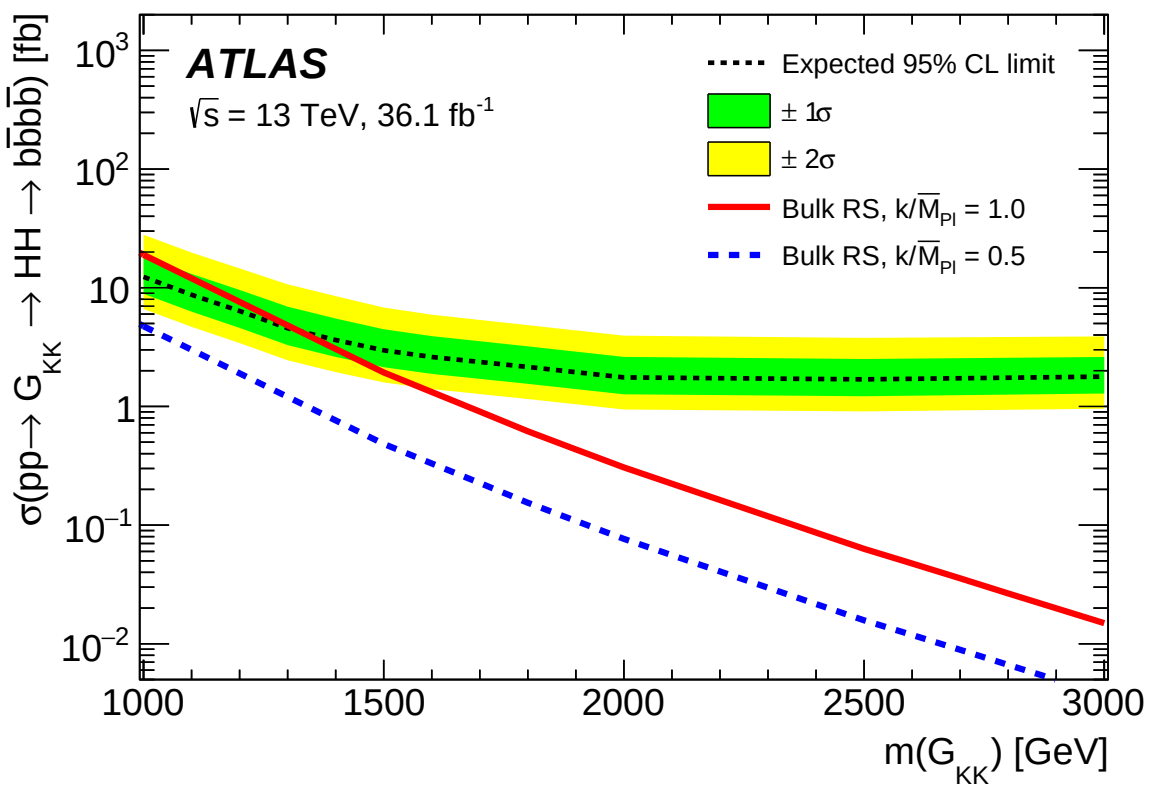

(a)

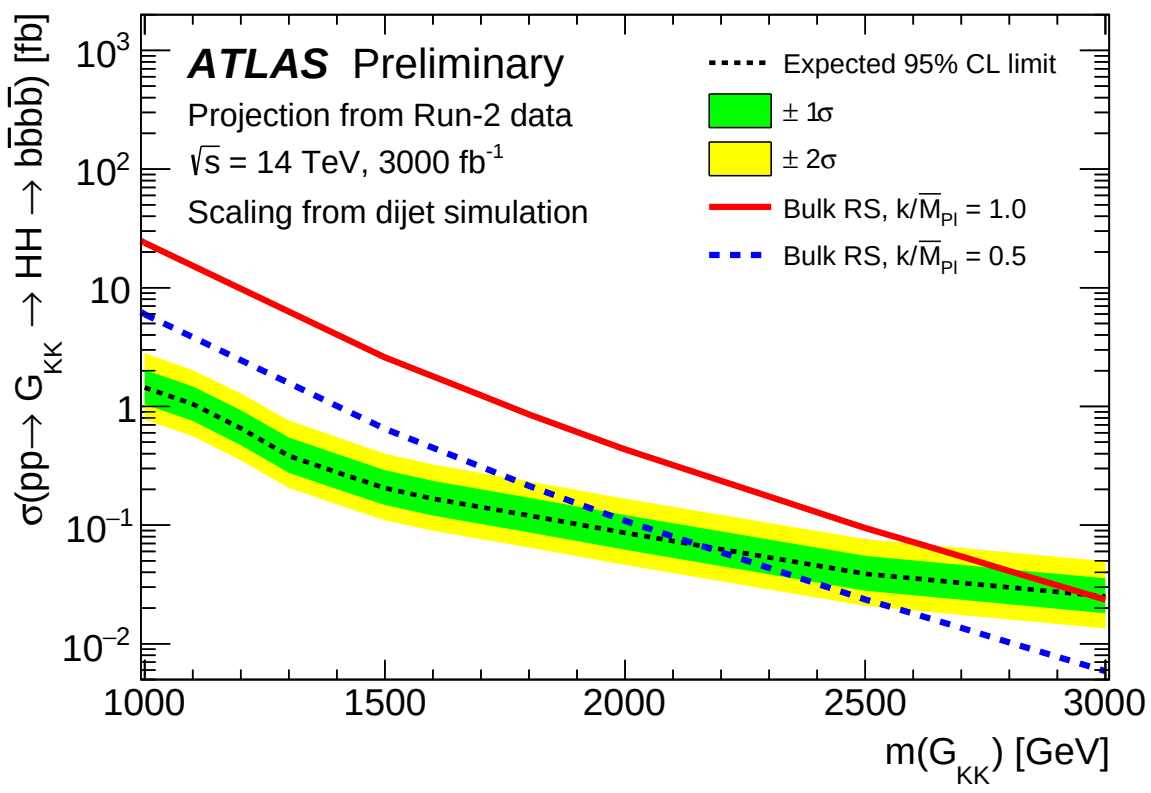

(b)

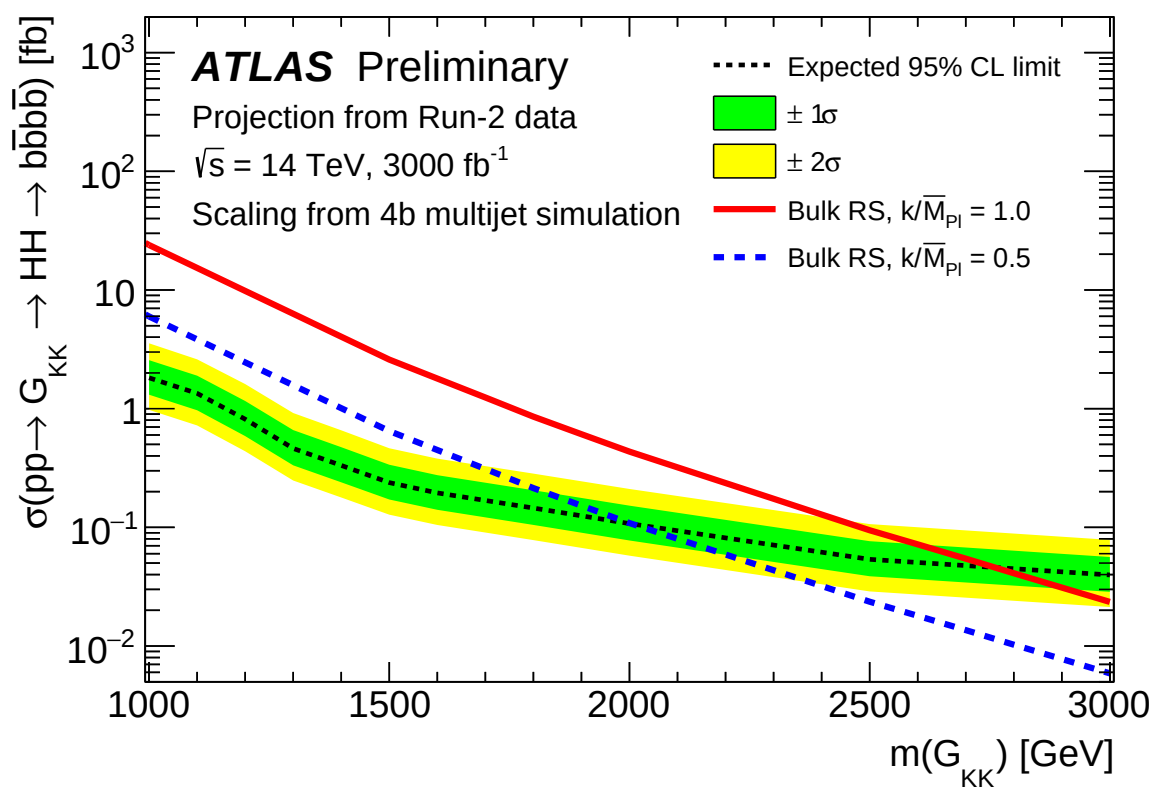

(c)

Figure 13: Expected upper limits on the $G_{\rm KK}$ production cross section times branching fraction to $b\bar{b}b\bar{b}$ at the 95% CL for (a) the 13 TeV data analysis published in Ref. [8] and the extrapolated HL-LHC analysis with analysis improvements to track jet reconstruction and event selection included using (b) the dijet MC samples or (c) the 4$b$ multijet MC samples to model the changes in the multijet background relative to the background predictions from the 13 TeV data analysis.

# 8 Conclusions

A study exploring the search potential for high-mass resonances decaying into a pair of Higgs bosons at the high-luminosity LHC with 3000 $\mathrm{fb}^{-1}$ is presented. The search exploits the large $H \to b\bar{b}$ branching fraction and the reconstruction of highly boosted Higgs bosons with large-radius jets and $b$-tagged track subjets to discriminate the signal from the dominant multijet background. The prospects for the HL-LHC are obtained by extrapolating the results of the analysis of 36.1 $\mathrm{fb}^{-1}$ of $pp$ collisions at $\sqrt{s}$ = 13 TeV with scaling functions derived from a truth-level analysis. Cross section upper limits as low as 0.025 fb are projected at resonance masses near 3.0 TeV, with lower mass limits on the KK graviton as high as 2.15 TeV and 2.95 TeV in the benchmark bulk Randall–Sundrum model with a warped extra dimension and $k/\overline{M}_{\mathrm{Pl}} = 0.5$ and 1.0, respectively. All limits are evaluated at the 95% confidence level. The study is limited to the resonance mass range between 1.0 and 3.0 TeV due to the extrapolation procedure followed here. Future searches will investigate the higher resonance mass region as well.

# Prospects for searches for heavy $Z'$ and $W'$ bosons in fermionic final states with the ATLAS experiment at the HL-LHC

The ATLAS Collaboration

This note presents the prospects for new heavy $W'$ and $Z'$ bosons at the high luminosity LHC with 3000 fb$^{-1}$ of $\sqrt{s}$ = 14 TeV proton–proton collision data. These analyses are based on generator-level information with parameterised estimates applied to the final state particles to simulate the response of the upgraded ATLAS detector and pile-up collisions. $W'$ bosons are searched for in final states with a bottom and a top quark or with a lepton plus missing transverse energy. Assuming a right-handed $W'$ boson, $W'$ masses up to 4.9 TeV can be excluded with the first search, while the second search excludes $W'$ bosons predicted in the Sequential Standard Model for masses up to 7.9 TeV in the lepton plus missing transverse energy final state. Results are presented for a search for $Z'$ bosons in $t\bar{t}$ and dilepton final states. Masses below 6.5 TeV for Sequential Standard Model $Z'$ bosons can be excluded.

# 1 Introduction

Extensions to the Standard Model (SM) may include heavy gauge bosons, called $W'$ and $Z'$ bosons in the following, which are heavier versions of the SM $W$ and $Z$ bosons. New heavy gauge bosons appear if the gauge group of the SM is extended, as in Grand Unified Theories, Little-Higgs models [1, 2], left-right symmetric models [3–5]. As an example, the Sequential Standard Model (SSM) [6] posits $W'_{\text{SSM}}$ and $Z'_{\text{SSM}}$ bosons with couplings to fermions that are identical to those of the SM $W$ or $Z$ boson. This model is often used in analyses as a benchmark and is useful for comparing the sensitivity of different experiments. An example inspired by Grand Unified Theories are the $\text{E}_6$-motivated [7, 8] theories which predict as a special case $Z'_\psi$ bosons, which are considered in this note. In addition, $W'$ and $Z'$ bosons are forecast in theories with universal extra dimension, like Kaluza-Klein excitation of the SM $W$ and $Z$ bosons [9–11], and in theories that with a strongly-coupled sector, such as Topcolour-assisted Technicolar [12] and composite-Higgs [13, 14] theories.

New heavy gauge bosons are searched for in various final states in ATLAS using the $\sqrt{s} = 13$ TeV data. In this note, the prospects for $W'$ bosons searches in the $W' \to t\bar{b} \to \ell\nu b\bar{b}$ and $W' \to \ell\nu$ decay channel, where the lepton is an electron or a muon[1], are summarised. $Z'$ bosons are searched for in the final states $Z' \to \ell\ell$ and $Z' \to t\bar{t}$. The ATLAS search for $W'$ bosons in the $W' \to t\bar{b} \to \ell\nu b\bar{b}$ final state excluded at the 95% confidence limit (CL) right-handed $W'$ bosons ($W'_\text{R}$), with a coupling to the SM particles equal to the SM weak coupling constant, with masses up to 3.15 TeV using 36 fb$^{-1}$ of data [15]. The search for $W' \to e\nu$ and $W' \to \mu\nu$ signals [16] used 79.8 fb$^{-1}$ of data taken at $\sqrt{s} = 13$ TeV. This analysis used a SSM $W'$ boson as benchmark and reported a lower exclusion limit at 95% CL on the $W'_{\text{SSM}}$ pole mass of 5.6 TeV. The $Z' \to \ell\ell$ analysis [17] based on 36.1 fb$^{-1}$ of data 95% CL reports lower mass limits of 4.5 TeV for $Z'_{\text{SSM}}$ bosons and 3.8 TeV for $Z'_\psi$ bosons. This analysis also provides lower mass limits on several other $Z'$ boson models. The search for $Z'$ bosons in the $t\bar{t}$ final state [18] using 36.1 fb$^{-1}$ of data allowed to exclude at 95% CL the $Z'$ bosons as predicted by Topcolour-assisted Technicolour model ($Z'_{\text{TC2}}$), using a width of 1% for the $Z'_{\text{TC2}}$ boson, upper limits on the production cross-sections vary from 25 pb to 0.02 pb for masses from 0.4 TeV to 5 TeV.

Searches for new heavy gauge bosons are an important part of the physics programme at the HL-LHC as they profit significantly from the expected increase in luminosity to 3000 fb$^{-1}$ of $pp$ collisions to be taken at $\sqrt{s} = 14$ TeV. In case of a discovery, the search for such particles in different decay channels (not only the ones reported here) will help to shed light on the underlying theory. Some of the above mentioned analyses were studied in the past for the HL-LHC running for ATLAS. The $W' \to \mu\nu$ prospect studies shown in Ref. [19] only looked at the improvements of the foreseen muon trigger. A first prospective High Luminosity (HL) LHC ATLAS analysis for $Z'_{\text{SSM}}$ boson decaying leptonically is summarised in Ref. [20]. Compared to the previous analysis, this study uses the latest layout of the upgraded ATLAS detector, and considers a higher value for average pile-up $\langle\mu\rangle$. In addition, it was found that the signal cross-sections used in the previous analysis were too high. A first update of these results was presented in Ref. [21] while a prospects study for a search for $Z'$ bosons from CMS can be found in Ref. [22].

The results of the analyses reported in this note are based on Monte Carlo (MC) simulations with an upgraded detector [19, 21, 23–26]. As discussed in more detail in Section 3, parameterised estimates are applied to the final state particles to simulate the response of the upgraded ATLAS detector and pile-up collisions. The prospects for a discovery or exclusion of such new heavy bosons has been studied in several decay channels. To allow comparisons with the published results using 36.1 fb$^{-1}$ and 79.8 fb$^{-1}$ of data, the same benchmark scenarios are used as in the respective results.

[1] In the following, the term lepton ($\ell$) is used to refer to an electron or a muon.

Section 4 reports the prospects of a search for $W'$ bosons in final states with a top and a bottom quark. $W'$ bosons are also searched for in final states with an electron or muon and large missing transverse energy ($E_{\mathrm{T}}^{\mathrm{miss}}$) in Section 5. Section 6 reports the results for $Z'$ bosons decaying leptonically into two electrons or muons. The exclusion and discovery mass limits are also computed for a hypothetic running at $\sqrt{s}$ = 13 TeV and $\sqrt{s}$ = 15 TeV as well as running at the high energy (HE) LHC assuming 15 ab$^{-1}$ of $pp$ collisions taken at $\sqrt{s}$ = 27 TeV. The HE-LHC is a possible future accelerator. The detectors are not yet designed but a similar physics performance is envisaged as for the HL-LHC detectors. Therefore, these studies will use the same parameterised estimates for the object performances and assume the same detector coverage as for the HL-LHC case. Finally Section 7 summarises previous results [21, 27] on a search for $Z'$ bosons decaying into a top anti-top quark pair in the $t\bar{t} \to WbWb \to \ell\nu bqq'b$ final state.

## 2 Monte Carlo samples

For each of the analyses presented in this note dedicated MC samples for $\sqrt{s}$ = 14 TeV $pp$ collisions were produced. The samples used for each of the analyses are summarised in Tables 1 – 3.

Table 1: Event generators used for the simulation of the signal and background processes used in the $W' \to t\bar{b}$ search. The PS/Had column describes the program used for parton shower and hadronisation.

| Process | Generator | PS/Had | MC Tune | PDF set |
|---|---|---|---|---|
| $W'_{\mathrm{R}}$ | MadGraph _aMC@NLO | Pythia8 | A14 [28] | NNPDF23LO [29] |
| $t\bar{t}$ | Powheg-Box | Pythia6 | Perugia 2012 [30] | CTEQ6L1 [31] |
| Single-top $t$-ch | Powheg-Box | Pythia8 | A14 | NNPDF30NLO [32] |
| Single-top $W$+$t$ | Powheg-Box | Pythia6 | Perugia 2012 | CT10 [33] |
| Single-top $s$-ch | Powheg-Box | Pythia8 | A14 | NNPDF30NLO |
| $W$ + jets | MadGraph _aMC@NLO | Pythia8 | A14 | NNPDF30NLO |
| $Z$ + jets | Powheg | Pythia8 | AU2 [34] | CT10 |
| $WW$, $WZ$, $ZZ$ | Powheg-Box | Pythia8 | AZNLO [35] | CTEQ6L1 |

Table 2: Event generators used for the simulation of the signal and background processes used in the $W' \to \ell\nu$ search. The PS/Had column describes the program used for parton shower and hadronisation. These samples were produced in several bin in $m_{\mathrm{T}}$.

| Process | Generator | PS/Had | MC tune | PDF set |
|---|---|---|---|---|
| $W' \to \ell\nu, \ell = e, \mu$ | Pythia8 | Pythia8 | A14 | NNPDF23LO |
| $W \to \ell\nu, \ell = e, \mu, \tau$ | Powheg-Box | Pythia8 | AZNLO | CTEQ6L1 |
| $Z/\gamma \to \ell\ell, \ell = e, \mu, \tau$ | Powheg-Box | Pythia8 | AZNLO | CTEQ6L1 |
| $t\bar{t} \to \ell\nu\ell\nu$ | Powheg-Box | Pythia6 | A14 | NNPDF30NNLO |
| $WW \to \ell\nu\ell\nu$ | Sherpa | Sherpa | default Sherpa tune | CT10 |
| $WZ \to \ell\nu\nu\nu$ | Sherpa | Sherpa | default Sherpa tune | CT10 |

Table 3: Event generators used for the simulation of the signal and background processes used in the $Z' \to \ell\ell$ search. The PS/Had column describes the program used for parton shower and hadronisation. These samples were produced in several bin in $m_{\ell\ell}$.

| Process | Generator | PS/Had | MC tune | PDF set |
|---|---|---|---|---|
| $Z' \to \ell\ell, \ell = e, \mu$ | Pythia8 | Pythia8 | A14 | NNPDF23LO |
| $Z/\gamma \to \ell\ell, \ell = e, \mu$ | Powheg-Box | Pythia8 | AZNLO | CTEQ6L1 |

The $W'_R \to t\bar{b}$ search is performed in the channel where the $W'_R$ decays into a top quark and a $\bar{b}$-quark, the top quark decays into a $W$ boson and a $b$-quark, and the $W$ boson decays in turn into a lepton and a neutrino, which is undetected and results in large $E_T^{miss}$. The productions of $t\bar{t}$ pairs and $W$+jets are the dominant background processes. Single top quarks ($t$-channel, $Wt$ and $s$-channel), $Z$+jets and dibosons ($WW$, $WZ$, and $ZZ$) production are also present in the signal regions. The $W'_R$ signals were generated at leading order (LO) by MadGraph5_aMC@NLO v2.2.3 [36–39]. Samples of simulated signal events were rescaled to next-to-leading-order (NLO), calculated with ZTOP [40] and generated for signal masses between 1 and 7 TeV in steps of 1 TeV in mass. Simulated top-quark pair and single-top-quark processes were produced using the NLO Powheg-Box [41, 42] programme with the parton distribution function (PDF) set CT10 [43]. The parton shower and the underlying event were added using Pythia6 or Pythia8 [44]. The background contributions from $W$ boson production in association with jets were simulated using the MadGraph generator while the $Z$ boson process was generated with Powheg. The production of vector-boson pairs ($WW$, $WZ$ or $ZZ$) with at least one charged lepton in the final state was simulated by the Powheg-Box generator in combination with Pythia8.

The main SM background in the $W'_{SSM} \to \ell\nu$ search arises from processes with at least one prompt final-state lepton. Background MC samples were produced for the charged- and neutral current Drell-Yan (DY) production, as well as top-quark pair ($t\bar{t}$) and diboson production. The background samples were simulated for different ranges of the boson transverse mass ($m_T$) to ensure that a large number of MC events is available across the entire $m_T$ region probed. The SSM signal $W' \to e\nu$ and $W' \to \mu\nu$ samples were generated at LO in QCD using the Pythia 8 event generator and includes off-shell production. Mass-dependent correction factors are applied to normalise the samples to the same mass-dependent NNLO pQCD calculation as used for the $W$ background. Compared to the LO prediction using NNPDF2.3 LO, the corrections increase the cross-section by about 40% around a boson invariant mass of 1–2 TeV, and by about 10% at 5 TeV. Further EW corrections beyond QED final-state radiation (FSR) are not considered for the signal. The sample was produced providing a flat spectrum in $W'$ masses which is reweighted in the analysis to produce $W'$ signals for given masses. The backgrounds from $W \to \ell\nu$, $Z/\gamma^* \to \ell\ell$, $W \to \tau\nu$, and $Z/\gamma^* \to \tau\tau$ were simulated using the Powheg-Box v2 matrix-element calculation up to NLO in perturbative quantum chromodynamics (pQCD). The final-state photon radiation (QED FSR) was modelled by the Photos [45] MC simulation. The samples are normalised using the same prescription as used in the Run 2 $\sqrt{s}$ = 13 TeV analysis [16] assuming higher order corrections to the cross-section at $\sqrt{s}$ = 14 TeV are the same as at $\sqrt{s}$ = 13 TeV. The normalisation is done as a function of the boson invariant mass to a next-to-next-to-leading order (NNLO) pQCD calculation using the numerical programme VRAP which is based on Ref. [46] and the CT14NNLO PDF set [47]. In addition to the modelling of QED FSR, a fixed-order electroweak (EW) correction to NLO is calculated as a function of the boson mass with the Mcsanc [48, 49] event generator at LO in pQCD.

In the $Z' \to \ell\ell$ search the signal and background $Z/\gamma^* \to \ell\ell$ event generation and normalisation to higher order cross-section predictions is done using the same methodology and assumptions as the $W'_{SSM} \to \ell\nu$ search. The signal events were generated in LO with PYTHIA 8.186 and also includes off-shell production. $Z/\gamma^* \to \ell\ell$ production is generated with POWHEG-BOX v2. The only difference with the samples used in the $W'_{\mathrm{SSM}} \to \ell\nu$ search is that the samples were produced in bins of $m_{\ell\ell}$.

For the comparisons with running at $\sqrt{s} = 13$ TeV, the generator-level particles of the Run 2 simulated datasets were used and the parameterisations for the detector response are applied. These datasets use the same MC generator setups as the $\sqrt{s} = 14$ TeV datasets. The $\sqrt{s} = 15$ TeV projections use the $\sqrt{s} = 14$ TeV datasets which are renormalised by the cross-section ratio to the higher centre of mass energy. Signal and background events were also produced at $\sqrt{s} = 27$ TeV for the HE-LHC projections. As higher order corrections are not yet available, the LO predictions are used for the signal and NLO generator cross-sections for the DY background.

## 3 Event reconstruction

The detector response is emulated with a set of functions providing a parameterised response derived from fully-simulated event samples. These fully simulated samples were produced using the detector layout described in the ATLAS Technical Design Reports [19, 21, 23–26] and assume an instantaneous luminosity of $7.5 \times 10^{34}\mathrm{cm}^{-2}\mathrm{s}^{-1}$, which corresponds to a pile-up scenario of 200 overlapping events ($<\mu>= 200$) [50] in each bunch crossing. The results of these studies were used to derive $\eta$ and $p_{\mathrm{T}}$-dependent functions which are applied to the generator-level quantities to emulate the transverse energy and momentum resolutions as well as efficiencies and fake rates and also include a dedicated pile-up overlay library to add pile-up jets to a hard-scatter truth event. Functions are available for leptons, jets and $E_{\mathrm{T}}^{\mathrm{miss}}$. Parameterisations are also derived for the identification and misidentification for $b$-jets. Details on the performance are summarised in Ref. [51]. In general, the performance is found to be as good or even better as obtained with the current detector.

Similar to the analyses using 36.1 fb$^{-1}$ and 79.8 fb$^{-1}$ of data, overlaps between jets and leptons are being taken care of in the following way. If the jet closest to an electron candidate is within a cone of size $\Delta R = \sqrt{(\Delta\eta)^2 + (\Delta\phi)^2} = 0.2$, that jet is removed from the event, as they most probably correspond to the same reconstructed object. If a remaining jet is found close to an electron within a cone of size $\Delta R = 0.4$, then the electron candidate is discarded. Selected muon candidates near jets that satisfy $\Delta R(\text{muon, jet}) < 0.04 + 10\ \text{GeV}/p_{\mathrm{T}}^{\mu}$ are rejected if the jet has at least three tracks originating from the primary vertex. Any jets with less than three tracks that overlap with a muon are rejected.

## 4 Search for $W'$ bosons decaying into a top and a bottom quark

### 4.1 Analysis strategy

The $\mathrm{W}'_{\mathrm{R}}$ boson search is performed in the channel where the $\mathrm{W}'_{\mathrm{R}}$ decays into a top quark and a $\bar{b}$-quark, the top quark decays into a $W$ boson and a $b$-quark, and the $W$ boson decays in turn into a lepton and a neutrino, which is undetected and results in missing transverse momentum, $\mathrm{E}_{\mathrm{T}}^{\mathrm{miss}}$. The final-state signature consists of two $b$-quarks, one charged lepton (electron or muon) and $\mathrm{E}_{\mathrm{T}}^{\mathrm{miss}}$.

This analysis uses single lepton triggers. Events are selected containing at least one electron with $p_{\mathrm{T}} > 22$ GeV in $|\eta| < 2.5$ or at least one muon with $p_{\mathrm{T}} > 20$ GeV and $|\eta| < 2.65$. Selected electrons candidates satisfy the *tight* working point [52] requirements and are required to have $p_{\mathrm{T}} > 25$ GeV and a pseudorapidity, $|\eta|$, smaller than 2.47. If the electron candidates fall in the calorimeter barrel–endcap transition region, $1.37 < |\eta_{\mathrm{cluster}}| < 1.52$, the event is rejected. Similarly, muon candidates must meet the *tight* identification working point [53] requirements and have a transverse momentum $p_{\mathrm{T}} > 25$ GeV, with $|\eta| < 2.65$.

The dominant background processes are the production of $t\bar{t}$ pairs and $W$+jets. Single top quarks ($t$-channel, $Wt$ and $s$-channel), $Z$+jets and dibosons ($WW$, $WZ$, and $ZZ$) production also contribute to the background in the signal regions. They all modelled with MC simulated events. Instrumental background coming from misindentified electrons, referred to as the multijet background, is also present but it is very small and further suppressed by applying dedicated selections, and it will be neglected in the following. In the muon channel there is a requirement that $\mathrm{E}_{\mathrm{T}}^{\mathrm{miss}}$>30 GeV and also on the sum of $m_{\mathrm{T}}^{W}$ and $\mathrm{E}_{\mathrm{T}}^{\mathrm{miss}}$: $m_{\mathrm{T}}^{W}$ + $\mathrm{E}_{\mathrm{T}}^{\mathrm{miss}} > 100$ GeV. In the electron channel the same requirement is applied and in addition the $\mathrm{E}_{\mathrm{T}}^{\mathrm{miss}}$ threshold is raised to 80 GeV to further suppress the multijet background.

Starting from the baseline objects, $W$ boson and top-quark candidates are reconstructed, and are used to define signal regions. The reconstruction of the W′ candidate starts with the calculation of the $z$ component of the neutrino momentum from the invariant mass of the lepton–$\mathrm{E}_{\mathrm{T}}^{\mathrm{miss}}$ system with the constraint that the mass of $W$ boson is 80.4 GeV. The constraint yields a quadratic equation and in the case of two real solutions, the smallest $|p_z|$ solution is chosen. Two imaginary solutions are obtained if the transverse mass, $m_{\mathrm{T}}^{W}$, of the reconstructed W boson is larger than the mass of the $W$-boson used in the constraint. This can happen due to the resolution of the $E_{\mathrm{T}}^{\mathrm{miss}}$. For these cases, the $E_{\mathrm{T}}^{\mathrm{miss}}$ components are adjusted to satisfy $m_{\mathrm{T}}^{W} = m_W$, yielding a single real solution. The top-quark candidate is reconstructed by first choosing the jet among all selected jets in the event, which yields the invariant mass closest to the top-quark mass ($m_{\mathrm{top}} = 172.5$ GeV), and than adding the four-momenta of that jet and $W$-boson candidate. This jet is referred to as "$b_{\mathrm{top}}$", and may not be actually $b$-tagged. Finally, the four-momentum of the candidate W′ boson is reconstructed by adding the four-momentum of the reconstructed top-quark candidate and the four-momentum of the highest-$p_{\mathrm{T}}$ remaining jet (referred to as "$b_1$"). The invariant mass of the reconstructed $W' \to t\bar{b}$ system ($m_{t\bar{b}}$) is the discriminating variable of this search. The signal is searched for in the range from 0.5 TeV to 8 TeV. An event selection common to all signal regions is defined as: lepton $p_{\mathrm{T}} > 50$ GeV, $p_{\mathrm{T}}(b_1) > 200$ GeV, $p_{\mathrm{T}}(\mathrm{top}) > 200$ GeV. As the signal events are expected to be boosted, there is a requirement on the angular separation of the lepton and $b_{\mathrm{top}}$: $\Delta R(\ell, b_{\mathrm{top}}) < 1.0$.

The phase space is divided into eight signal regions (SR) defined by the number of jets and $b$-tagged jets, and are labelled as "$X$-jet $Y$-tag" where $X = 2, 3$ and $Y = 1, 2$, that are further separated into electron and muon channels. Figure 1 shows the signal selection acceptance × efficiency (defined as the number of events passing all selection requirements divided by the total number of simulated simulated $W' \to t\bar{b} \to \ell\nu b\bar{b}$ events) as a function of the $W'_{\mathrm{R}}$ mass. The selection acceptance × efficiency curves are shown separately for electron and muon channel, and also for their combination. The maximum signal acceptance × efficiency is around 2 TeV and it decreases above that mass value due to decrease in the $b$-tagging efficiency and as the events become more boosted. The muon channel outperforms the electron channel due to overlap removal requirements, as they are relaxed by using a variable $\Delta R$ cone size. The variable $\Delta R$ cone size is not used for electrons because of the possible double counting of the energies of electron and jet.

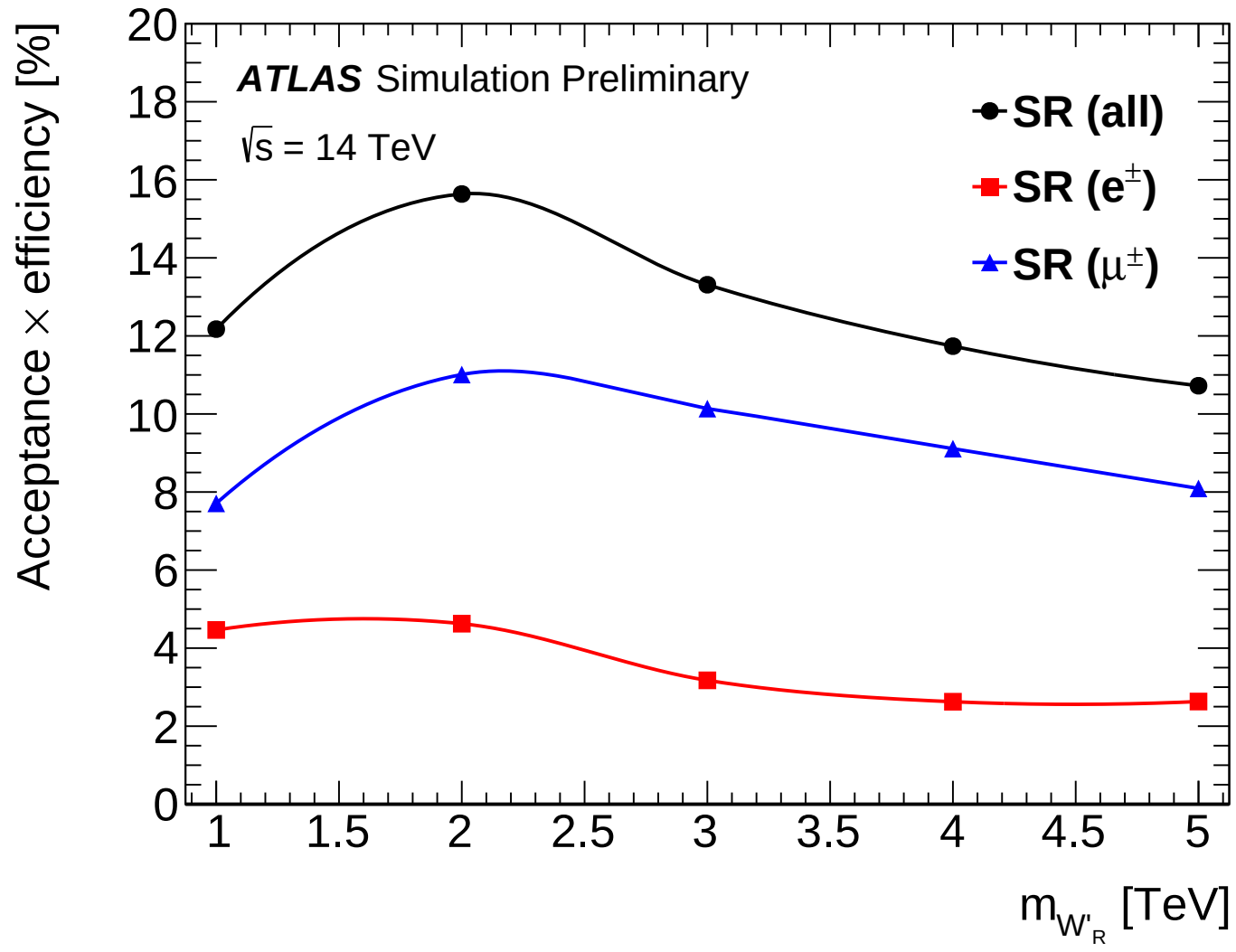

Figure 1: Signal selection acceptance × efficiency in the signal regions as a function of the $W'_{\mathrm{R}}$ mass. It is defined as the number of events passing all selections divided by the total number of simulated $W' \to t\bar{b} \to \ell\nu b\bar{b}$ events. Acceptances × efficiencies are shown for all SRs combined (full circle), electron channel (full square) and muon channel (full triangle).

## 4.2 Systematic uncertainties

Systematic uncertainties are evaluated following the prescriptions used in the analysis [54] and then scaled according to the recommendations [51]. The uncertainty on the luminosity (1%) and on the theory cross sections (5% for diboson, 10% for $Z$+jets, and 3% for single top) are included in all the plots and for the calculations of expected limits and significances. The $b$-tagging and the modelling uncertainties (which are the dominant uncertainties on the shape of the discriminating variable from the previous analysis) are also included.

## 4.3 Results

The presence of a massive resonance is tested by simultaneously fitting the $m_{tb}$ templates of the signal and background simulated event samples using a binned maximum–likelihood approach (ML) based on the RooStats framework [55–57]. Each signal region is treated as an independent search channel.

The normalisations of the $t\bar{t}$ and $W$+jets backgrounds were found to be different than one in the analysis of 36.1 fb$^{-1}$, therefore they are free parameters in the fit. They are constrained by Asimov dataset to one by construction. The other background normalisations are assigned Gaussian priors based on their respective normalisation uncertainties. The systematic uncertainties described in Section 4.1 are incorporated in the fit as nuisance parameters with correlations across regions and processes taken into account. The signal normalisation is a free parameter in the fit.

The expected event yields after the ML fit are shown in Tables 4 to 7 and correspond to an integrated luminosity of 3000 fb$^{-1}$.

The $m_{tb}$ distributions for the SR after the ML fit are shown in Figures 2 and 3. An expected signal contribution corresponding to a $W'_{\mathrm{R}}$ boson with a mass of 2.0 TeV is shown as a dashed histogram overlay.

Table 4: The numbers of signal and background events are shown in the 2-jet 1-tag signal regions. For signal, the values correspond to expected event yields and quoted uncertainties account for the statistical uncertainty of the number of events in the simulated samples. The number of background events is obtained following a ML fit to the Asimov dataset and uncertainties contain systematic uncertainties.

| | 2-jet 1-tag ($e^{\pm}$) | | | 2-jet 1-tag ($\mu^{\pm}$) | | |
|---|---|---|---|---|---|---|
| $W'_{\mathrm{R}}$ (1.0 TeV) | 179 000 | ± | 51 000 | 291 000 | ± | 65 000 |
| $W'_{\mathrm{R}}$ (2.0 TeV) | 8700 | ± | 2500 | 19 300 | ± | 3700 |
| $W'_{\mathrm{R}}$ (3.0 TeV) | 680 | ± | 220 | 1800 | ± | 370 |
| $W'_{\mathrm{R}}$ (4.0 TeV) | 79 | ± | 28 | 226 | ± | 48 |
| $W'_{\mathrm{R}}$ (5.0 TeV) | 14 | ± | 5 | 36 | ± | 8 |
| $t\bar{t}$ | 504 180 | ± | 770 | 8879 | ± | 1700 |
| Single-top | 56 840 | ± | 500 | 102 650 | ± | 930 |
| $W$+jets | 169 490 | ± | 990 | 331 900 | ± | 2100 |
| $Z$+jets, diboson | 1000 | ± | 51 | 5200 | ± | 180 |
| Total background | 731 500 | ± | 1300 | 1 327 600 | ± | 2700 |

Table 5: The numbers of signal and background events are shown in the 3-jet 1-tag signal regions. For signal, the values correspond to expected event yields and quoted uncertainties account for the statistical uncertainty of the number of events in the simulated samples. The number of background events is obtained following a ML fit to the Asimov dataset and uncertainties contain systematic uncertainties.

| | 3-jet 1-tag ($e^{\pm}$) | | | 3-jet 1-tag ($\mu^{\pm}$) | | |
|---|---|---|---|---|---|---|
| $W'_{\mathrm{R}}$ (1.0 TeV) | 137 000 | ± | 45 000 | 247 000 | ± | 60 000 |
| $W'_{\mathrm{R}}$ (2.0 TeV) | 10 400 | ± | 2700 | 24 600 | ± | 4200 |
| $W'_{\mathrm{R}}$ (3.0 TeV) | 950 | ± | 260 | 2640 | ± | 440 |
| $W'_{\mathrm{R}}$ (4.0 TeV) | 113 | ± | 34 | 350 | ± | 59 |
| $W'_{\mathrm{R}}$ (5.0 TeV) | 20 | ± | 6 | 55 | ± | 10 |
| $t\bar{t}$ | 1 271 600 | ± | 1300 | 2 167 500 | ± | 2500 |
| Single-top | 50 950 | ± | 440 | 88 600 | ± | 800 |
| $W$+jets | 171 300 | ± | 1100 | 388 900 | ± | 2300 |
| $Z$+jets, diboson | 731 | ± | 26 | 3100 | ± | 160 |
| Total background | 1 494 600 | ± | 1300 | 2 648 200 | ± | 3500 |

Table 6: The numbers of signal and background events are shown in the 2-jet 2-tag and 3-jet 2-tag signal regions. For signal, the values correspond to expected event yields and quoted uncertainties account for the statistical uncertainty of the number of events in the simulated samples. The number of background events is obtained following a ML fit to the Asimov dataset and uncertainties contain systematic uncertainties.

| | 2-jet 2-tag ($e^{\pm}$) | | | 2-jet 2-tag ($\mu^{\pm}$) | | |
|---|---|---|---|---|---|---|
| $W'_{\mathrm{R}}$ (1.0 TeV) | 184 000 | ± | 52 000 | 320 000 | ± | 68 000 |
| $W'_{\mathrm{R}}$ (2.0 TeV) | 6000 | ± | 2100 | 14 700 | ± | 3200 |
| $W'_{\mathrm{R}}$ (3.0 TeV) | 270 | ± | 140 | 1170 | ± | 290 |
| $W'_{\mathrm{R}}$ (4.0 TeV) | 28 | ± | 17 | 135 | ± | 37 |
| $W'_{\mathrm{R}}$ (5.0 TeV) | 6 | ± | 3 | 22 | ± | 6 |
| $t\bar{t}$ | 225 600 | ± | 220 | 418 090 | ± | 510 |
| Single-top | 16 090 | ± | 200 | 28 200 | ± | 300 |
| $W$+jets | 6760 | ± | 230 | 11 810 | ± | 130 |
| $Z$+jets, diboson | 0 | ± | | 0 | ± | |
| Total background | 248 450 | ± | 320 | 458 100 | ± | 520 |

Table 7: The numbers of signal and background events are shown in the 2-jet 2-tag and 3-jet 2-tag signal regions. For signal, the values correspond to expected event yields and quoted uncertainties account for the statistical uncertainty of the number of events in the simulated samples. The number of background events is obtained following a ML fit to the Asimov dataset and uncertainties contain systematic uncertainties.

| | 3-jet 2-tag ($e^{\pm}$) | | | 3-jet 2-tag ($\mu^{\pm}$) | | |
|---|---|---|---|---|---|---|
| $W'_{\mathrm{R}}$ (1.0 TeV) | 153 000 | ± | 47 000 | 270 000 | ± | 63 000 |
| $W'_{\mathrm{R}}$ (2.0 TeV) | 77 400 | ± | 2300 | 19 400 | ± | 3700 |
| $W'_{\mathrm{R}}$ (3.0 TeV) | 410 | ± | 170 | 1760 | ± | 360 |
| $W'_{\mathrm{R}}$ (4.0 TeV) | 43 | ± | 21 | 200 | ± | 45 |
| $W'_{\mathrm{R}}$ (5.0 TeV) | 8 | ± | 4 | 34 | ± | 8 |
| $t\bar{t}$ | 842 140 | ± | 470 | 1 482 430 | ± | 860 |
| Single-top | 23 530 | ± | 230 | 43 510 | ± | 440 |
| $W$+jets | 21 100 | ± | 240 | 39 930 | ± | 240 |
| $Z$+jets, diboson | 5.7 | ± | 0.2 | 1370 | ± | 100 |
| Total background | 900 920 | ± | 550 | 1 567 250 | ± | 900 |

The binning of the $m_{tb}$ distribution is chosen to optimise the search sensitivity while minimising statistical fluctuations. Requirements are imposed on the expected number of background events per bin, and the bin width is adapted to a resolution function that represents the width of the reconstructed mass peak for each studied $W'_{\mathrm{R}}$ boson signal sample. This results in different number of bins in each region.

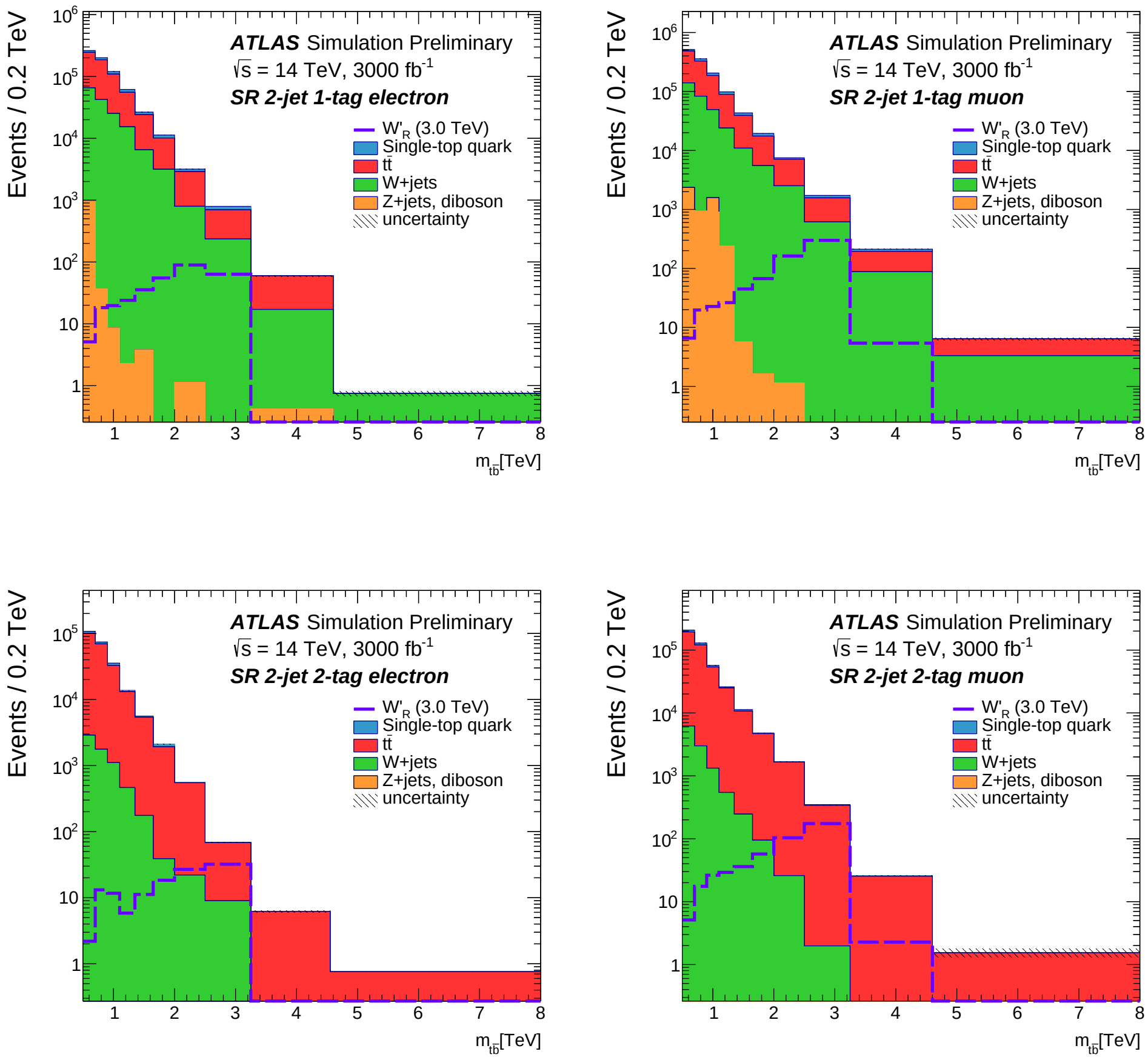

Figure 2: Post-fit distributions of the reconstructed mass of the $W'_{\mathrm{R}}$ boson candidate in the (top) 2-jet 1-tag and (bottom) 2-jet 2-tag signal regions, for (left) electron and (right) muon channels. An expected signal contribution corresponding to a $W'_{\mathrm{R}}$ boson mass of 3 TeV is shown. Uncertainty bands include all the systematic uncertainties.

The limits are evaluated assuming a modified frequentist method known as $\mathrm{CL_s}$ [58] with a profile-likelihood-ratio test statistic [59] and using the asymptotic approximation.

The 95% confidence level (CL) upper limits on the production cross section multiplied by the branching fraction for $W'_{\mathrm{R}} \to t\bar{b}$ are shown in Figure 4 as a function of the resonance mass. The expected exclusion limits range between 0.04 pb and $2.3 \times 10^{-3}$ pb for $W'_{\mathrm{R}}$ boson masses from 1 TeV to 5 TeV. The existence of $W'_{\mathrm{R}}$ bosons with masses $m_{W'_{\mathrm{R}}} < 4.9$ TeV is expected to be excluded for the benchmark model for $W'_{\mathrm{R}}$, assuming that the $W'_{\mathrm{R}}$ coupling $g'$ is equal to the SM weak coupling constant $g$. This would increase the current limit reported in Ref. [54] by 1.8 TeV.

The expected discovery significances is calculated using the profile likelihood test statistic for different mass hypotheses of the $W'_{\mathrm{R}}$ benchmark model for luminosity of 3000 $\mathrm{fb}^{-1}$ in asymptotic approximation.

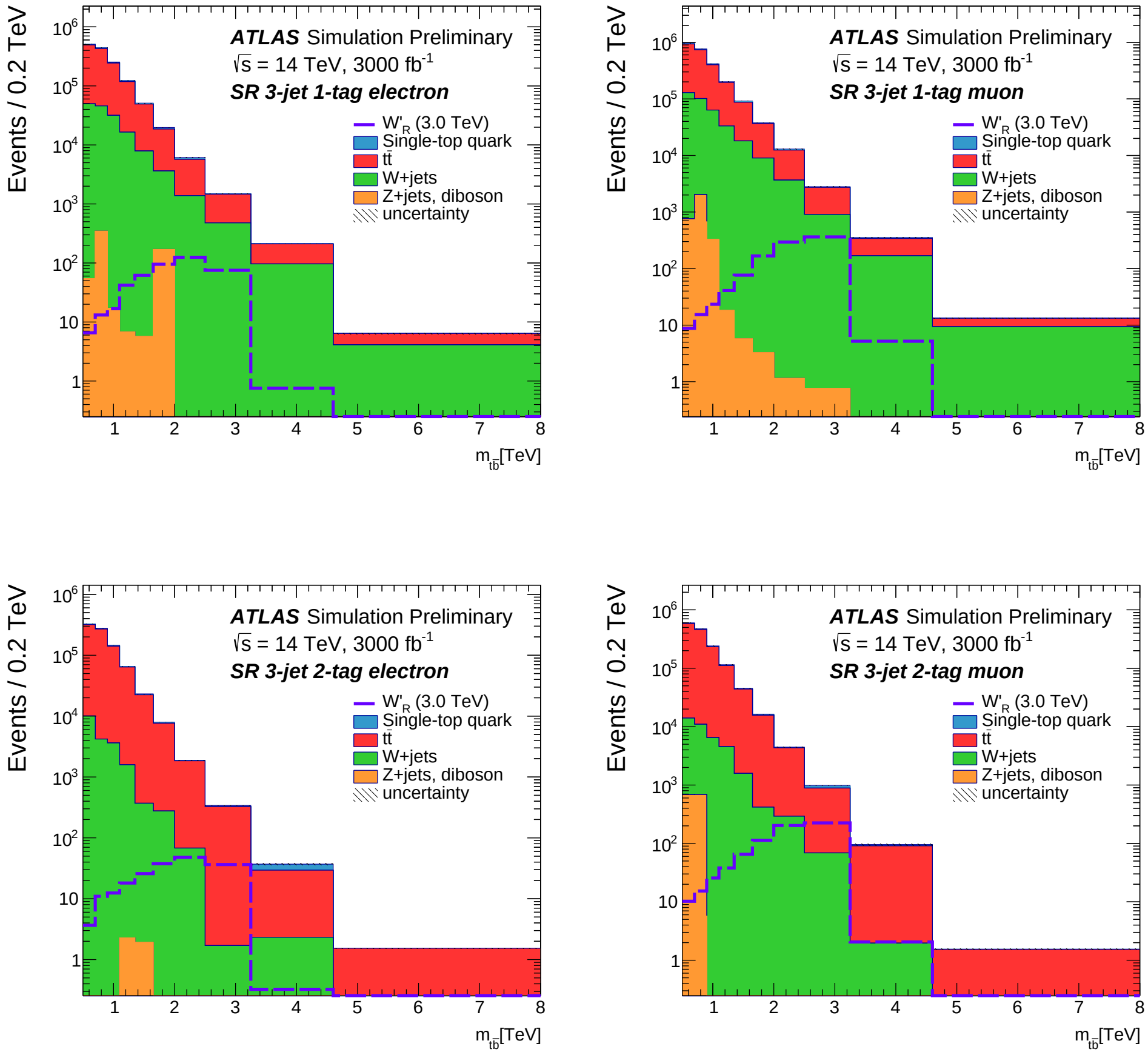

Figure 3: Post-fit distributions of the reconstructed mass of the $W'_{\mathrm{R}}$ boson candidate in the (top) 3-jet 1-tag and (bottom) 3-jet 2-tag signal regions, for (left) electron and (right) muon channels. An expected signal contribution corresponding to a $W'_{\mathrm{R}}$ boson mass of 3 TeV is shown. Uncertainty bands include all the systematic uncertainties.

Based on $5\sigma$ significance, it is found that $W'_{\mathrm{R}}$ with masses up to 4.3 TeV can be discovered.

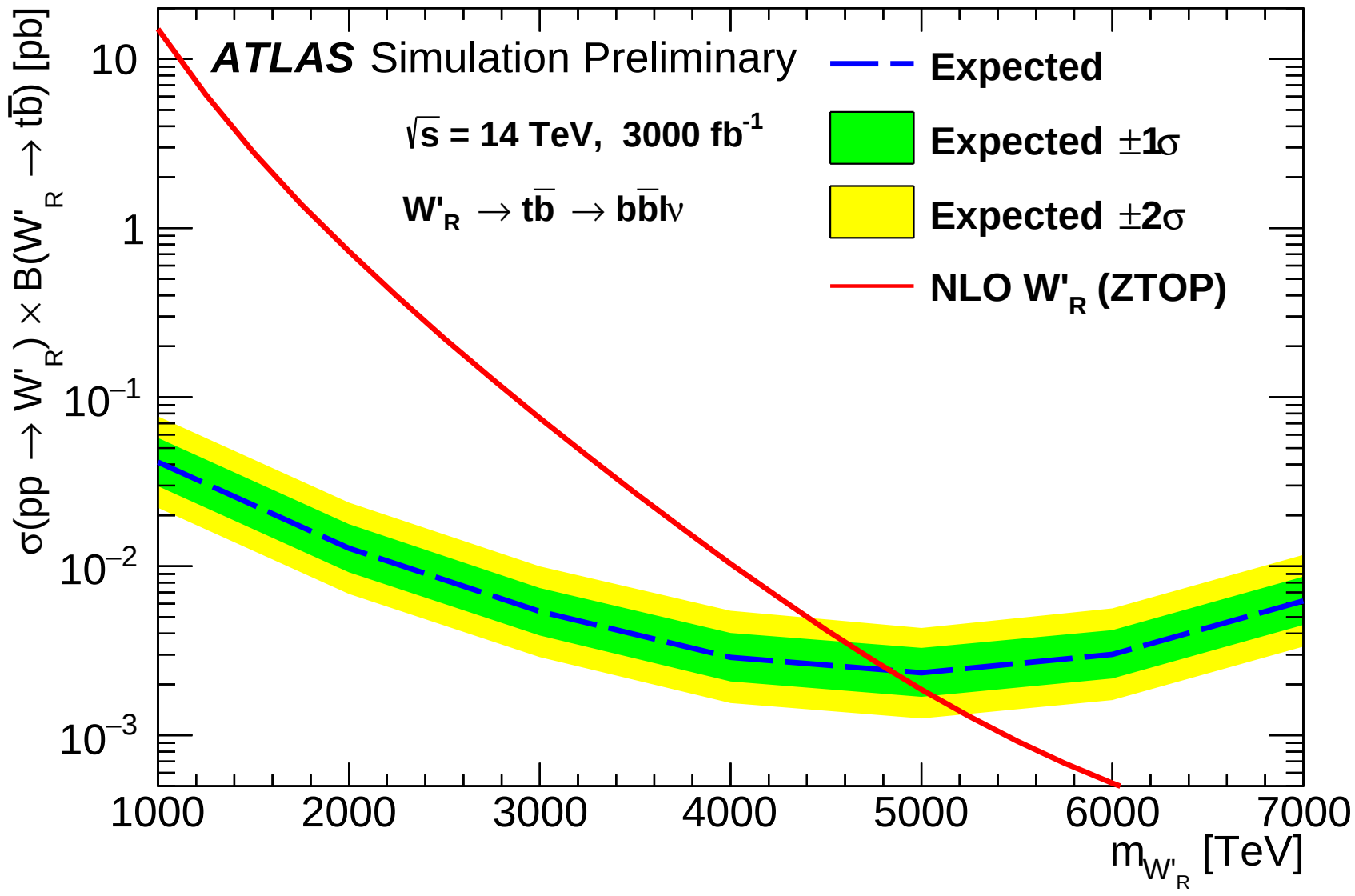

Figure 4: Upper limits at the 95% CL on the $W'_{\mathrm{R}}$ production cross section times branching fraction as a function of resonance mass. The dashed curve and shaded bands correspond to the limit expected in the absence of signal and the regions enclosing one/two standard deviation ($\sigma$) fluctuations of the expected limit. The theory prediction is also shown.

# 5 Search for $W'$ bosons decaying into a lepton and missing transverse momentum

## 5.1 Analysis Strategy

HL-LHC projections of a search for a new heavy, charged $W'$ boson decaying into an electron or muon and a neutrino are presented in this section. A high mass $W'$ signal would appear as an excess of events above the SM background at high $m_{\mathrm{T}}$. The SM background mainly arises from processes with at least one prompt final-state lepton, with the largest source being the charged-current DY $W$ boson production, where the $W$ boson decays into an electron or muon and a neutrino. Other non-negligible contributions are from top-quark pair ($t\bar{t}$) and single-top-quark production, neutral-current DY ($Z/\gamma^{*}$) process, diboson production, and from events in which one final-state jet or photon satisfies the lepton selection criteria. This last component of the background, referred to in the following as the multijet background, receives contributions from multijet, heavy-flavour quarks and $\gamma$ + jet production and is one of the smallest backgrounds in this analysis. It is evaluated in a data-driven way in the Run 2 analysis and cannot be yet reliably estimated from MC samples and is therefore not considered in this analysis. It was found to be negligible in the muon channel at $m_{\mathrm{T}}$> 3 TeV in the Run 2 analysis [16] based on 79.8 fb$^{-1}$ of $pp$ collisions. In the electron channel, the contribution constitues around 10% of the overall cross section at $m_{\mathrm{T}}\approx$ 3 TeV and mainly arises from jets misidentified as electrons.

Events are selected in a very similar way to the data analysis reported in Ref. [16]. Only events passing the single electron or muon triggers are considered. The single electron trigger selects events containing at least one electron with $p_{\mathrm{T}}$ > 22 GeV in $|\eta|$ < 2.5, while the single muon trigger requires a muon with $p_{\mathrm{T}}$ > 20 GeV in $|\eta|$ < 2.65. Events are required to contain exactly one lepton which can be either an electron or a muon. In the muon channel, muons with $p_T > 55$ GeV within $|\eta| < 2.65$ are selected which fulfil the criteria of the *high-$p_T$* working point [51]. In the electron channel, the electron must satisfy the *tight* identification criteria, and the electron has $p_T > 65$ GeV and has to lie in the rapidity region for precision measurements $|\eta| < 1.37$ or $1.52 < |\eta| < 2.47$. These $p_T$ thresholds are the same as in the Run 2 analysis and are motivated by the triggers which select events containing leptons with loose identification criteria and without isolation requirements. Though not applied in this analysis, such events will be needed for the data-driven background subtraction methods, as employed in Run 2, to work. The $p_T$ thresholds for these "looser" triggers are not yet available and therefore in the following it is assumed the $p_T$ thresholds will be similar to those used in Run 2. The magnitude of $E_T^{miss}$ must exceed 55 GeV (65 GeV) in the electron (muon) channel. Events in both channels are vetoed if they contain additional leptons satisfying loosened selection criteria, namely electrons with $p_T > 20$ GeV satisfying the *medium* identification criteria or muons with $p_T > 20$ GeV passing the *loose* muon selection.

The total acceptance times efficiency as a function of the $W'_{SSM}$ pole mass is presented in Figure 5 for both the electron and muon channel. A value 83% (68%) is reached around a pole mass of 2.5 TeV in the electron (muon) channel. For higher pole masses the acceptance times efficiency is falling again due to lower parton luminosities and reaches 65% (60%) for pole masses between 6 and 10 TeV. The resulting $m_T$ distribution is shown in Figure 6 for the expected background as well as for a possible SSM $W'$ boson with a mass of 6.5 TeV. The $m_T$ bins used in this figure is the binning used to compute the exclusion limits and the discovery reach in the following. The expected number of events in various $m_T$ bins are shown also in Table 8 for both the overall background and several $W'_{SSM}$ signals.

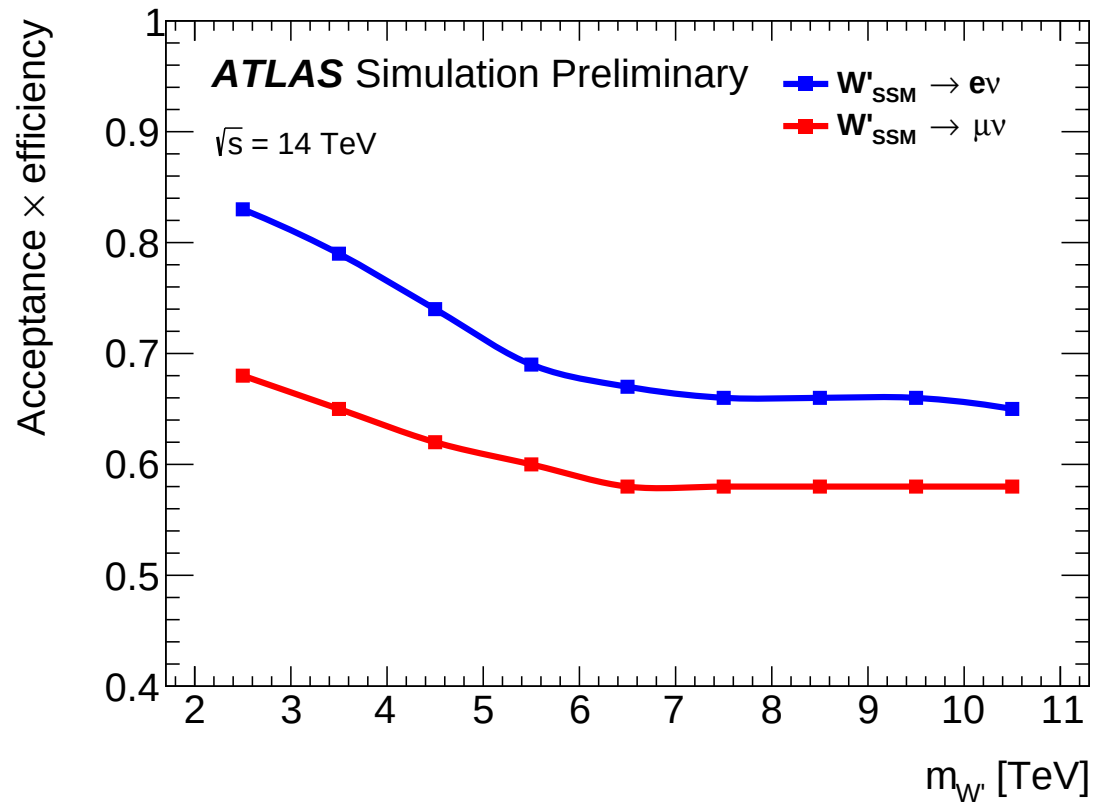

Figure 5: Total signal acceptance times efficiency versus SSM $W'$ pole mass for the SSM $W'$ model in the electron and muon channel of the $W'_{SSM} \to \ell\nu$ search.

### 5.2 Systematics

The systematic uncertainties arise from both experimental and theoretical sources. The uncertainties found in the Run 2 analysis [16] increase for larger values of $m_T$. Therefore, in order to describe this increase as a function of $m_T$, the uncertainties are expressed as a percentage value multiplied by the value of $m_T$ given in TeV in the following. The values are derived from the uncertainties found in the TeV range in the Run 2 analysis. They are then scaled to more realistic values to be expected by the end of the HL-LHC following the recommendations given in Ref. [51].

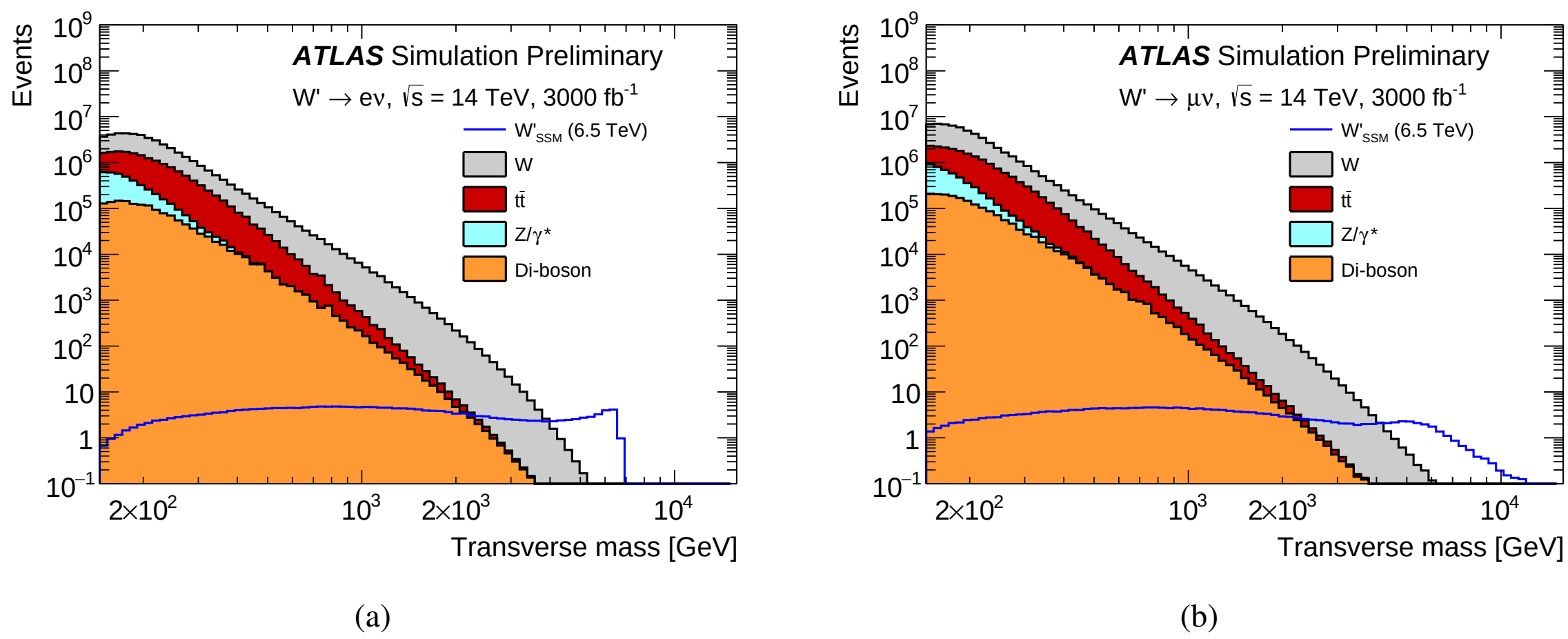

Figure 6: Transverse mass distributions for events satisfying all selection criteria in the (a) electron and (b) muon channels of the $W'_{\mathrm{SSM}} \rightarrow \ell\nu$ search. The distributions in data are compared to the stacked sum of the main expected backgrounds. As an example, the expected signal distributions for a SSM $W'$ boson with a mass of 6.5 TeV is shown. The bin width is constant in log $m_{\mathrm{T}}$.

Table 8: Expected event yields and their statistical uncertainties in the electron and muon channel in different $m_{\mathrm{T}}$ intervals. For presentational purposes several of the $m_{\mathrm{T}}$ bins used to compute the exclusion limits and the discovery reach are merged in larger bins in this table. The yields are given for the SM background (arising from charged and neutral current and top-quark production) and for $W'_{\mathrm{SSM}}$ bosons for three values of the pole mass.

Electron channel

| $m_T$ [GeV] | 60–240 | 240–950 | 950–3000 | 3000–5000 | 5000–15000 |
|---|---|---|---|---|---|
| Total SM | 44787000 ± 7000 | 10100500 ± 3200 | 29680 ± 170 | 63 ± 8 | 0.33 ± 0.58 |
| $W'_{\mathrm{SSM}}$ (3.5 TeV) | 225.1 ± 2.3 | 2009 ± 24 | 7920 ± 80 | 5590 ± 80 | 1.167 ± 0.012 |
| $W'_{\mathrm{SSM}}$ (6.5 TeV) | 16.57 ± 0.11 | 101.23 ± 0.31 | 76.88 ± 0.26 | 24.13 ± 0.12 | 18.05 ± 0.13 |
| $W'_{\mathrm{SSM}}$ (7.5 TeV) | 9.18 ± 0.06 | 55.49 ± 0.17 | 38.12 ± 0.13 | 6.557 ± 0.029 | 3.754 ± 0.018 |

Muon channel

| $m_T$ [GeV] | 60–240 | 240–950 | 950–3000 | 3000–5000 | 5000–15000 |
|---|---|---|---|---|---|
| Total SM | 77526000 ± 9000 | 9962400 ± 3200 | 25240 ± 160 | 63 ± 8 | 1.3 ± 1.2 |
| $W'_{\mathrm{SSM}}$ (3.5 TeV) | 312.9 ± 3.2 | 1940 ± 40 | 6590 ± 120 | 4240 ± 110 | 63 ± 12 |
| $W'_{\mathrm{SSM}}$ (6.5 TeV) | 24.39 ± 0.14 | 100.14 ± 0.32 | 68.50 ± 0.25 | 20.81 ± 0.12 | 14.52 ± 0.12 |
| $W'_{\mathrm{SSM}}$ (7.5 TeV) | 13.26 ± 0.35 | 55.7 ± 0.8 | 34.2 ± 0.6 | 5.67 ± 0.13 | 2.98 ± 0.07 |

The experimental systematic uncertainties due to the reconstruction, identification and isolation of muons result in a value of 2.5% $\times m_{\mathrm{T}}$ [TeV], while these uncertainties are negligible for electrons. Systematic uncertainties due to the energy resolution and scale are set to 2.5% $\times m_{\mathrm{T}}$ [TeV]. The main systematic uncertainties in the $E_{\mathrm{T}}^{\mathrm{miss}}$ calculation and on the jet energy scale are found to be negligible already in Run 2 and are therefore not considered in this analysis.

Theoretical uncertainties are related to the production cross-sections estimated from MC simulation. The effects when propagated to the total background estimate are significant for $W$ and $Z/\gamma^{*}$ production, and to some extent for top-quark production, but are negligible for diboson production. No theoretical uncertainties are considered for the $W'$ boson signal in the statistical analysis.

The largest uncertainties arise from the PDF uncertainty for the DY background, which is obtained from the 90% CL CT14NNLO PDF uncertainty set using VRAP to calculate the NNLO cross-section as a function of the boson mass. Rather than using the original 28 CT14 uncertainty eigenvectors, a re-diagonalised set of seven PDF eigenvectors, as provided by the authors of the CT14 PDF using MP4LHC [60, 61], is used. This sum is referred to as "PDF variation" in the following. An additional uncertainty arises from the choice of the nominal PDF set used. In Ref. [16], the central values of the CT14NNLO PDF set were compared to the MMHT2014 [62] and NNPDF3.0 [32] PDF sets. A comparison between these PDF sets showed that the central value for NNPDF3.0 falls outside the "PDF variation" uncertainty at large $m_{\mathrm{T}}$. Thus, an envelope of the "PDF variation" and the NNPDF3.0 central value was formed, where the former was subtracted in quadrature from this envelope, and the remaining part, which is non-zero only when the NNPDF3.0 central value is outside the "PDF variation" uncertainty, is referred to as "PDF choice". In the electron channel one of the largest sources of uncertainty arises from the multijet background estimation. These uncertainties amount to around 2.5% $\times m_{\mathrm{T}}$ [TeV] for the PDF choice and 5% $\times m_{\mathrm{T}}$ [TeV] for the PDF variations. The uncertainties of the multijet background in the electron channel is assumed to be 2.5% $\times m_{\mathrm{T}}$ [TeV]. Rounding up all of these systematics values gives an estimate of 7% $\times m_{\mathrm{T}}$ [TeV] for each signal and channel. It should be noted that this search is limited by the statistic uncertainties as this search looks for a signal in the very high $m_{\mathrm{T}}$ tail.

## 5.3 Results

A statistical analysis is performed for the search for a $W'_{\mathrm{SSM}}$ boson using the $m_T(\ell\nu)$ distribution in the electron and muon channels as the discriminant. Pole masses are tested in 1 TeV intervals throughout the region of interest ranging from 2.5 TeV to 11.5 TeV. The exclusion limits and discovery reach is interpolated between the two nearest mass points where necessary. This interpolation has a negligible effect on the accuracy of the result. For calculating the exclusion limits the same methodology is used as in the Run 2 analysis, where the limits are calculated in a Bayesian analysis [63]. The same statistical model implementation is used in the following for both the calculation of the exclusion limits and the discovery reach, the latter based on a profile likelihood ratio test assuming an asymptotic test statistic distribution [59]. The systematics are taken into account with a Gaussian prior, and the prior of the signal parameter of interest is Log Normal.

Assuming no signal is observed, exclusion upper limits on the cross-section for producing a $W'_{\mathrm{SSM}}$ boson times its branching ratio to only one lepton generation ($\sigma \times BR$) are computed at the 95% CL as a function of the $W'_{\mathrm{SSM}}$ boson mass. The exclusion limits use a uniform positive prior probability distribution for $\sigma \times BR$. The expected upper limits are extracted using $W'_{\mathrm{SSM}}$ templates binned in $m_{\mathrm{T}}$. The expected limits are derived from pseudo-experiments obtained from the estimated background distributions. The median of the distribution of the limits from the pseudo-experiments is taken as the expected limit, and $1\sigma$ and $2\sigma$ bands are defined as the ranges containing respectively 68% and 95% of the limits obtained with the pseudo-experiments.

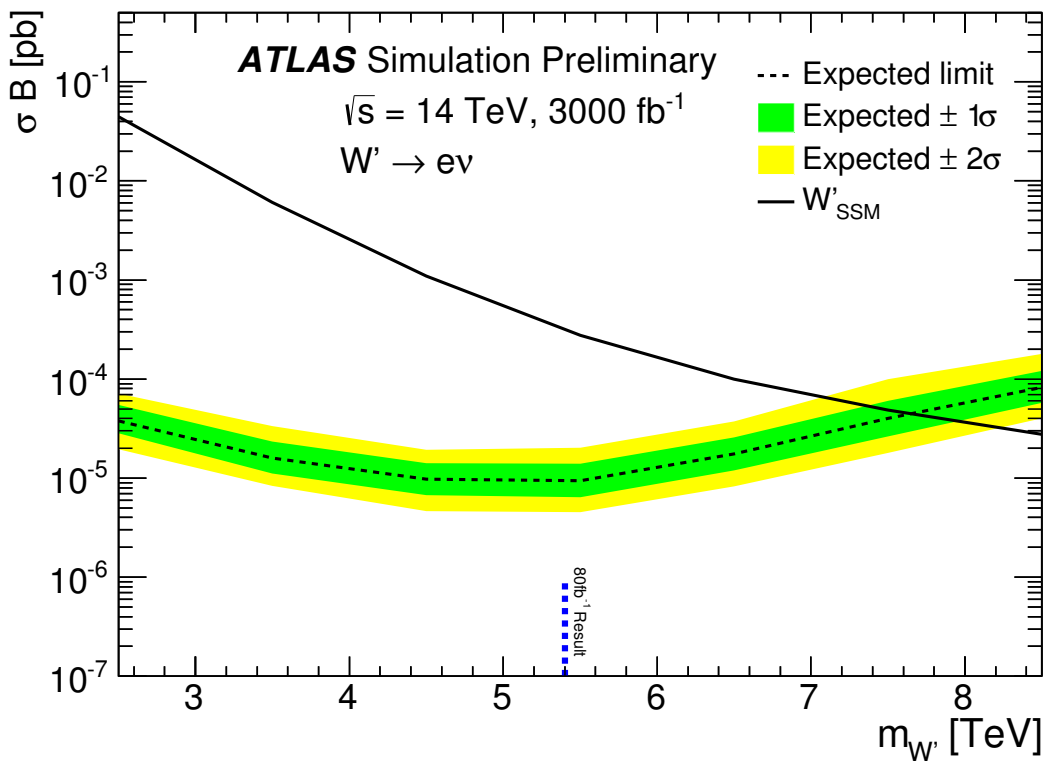

(a)

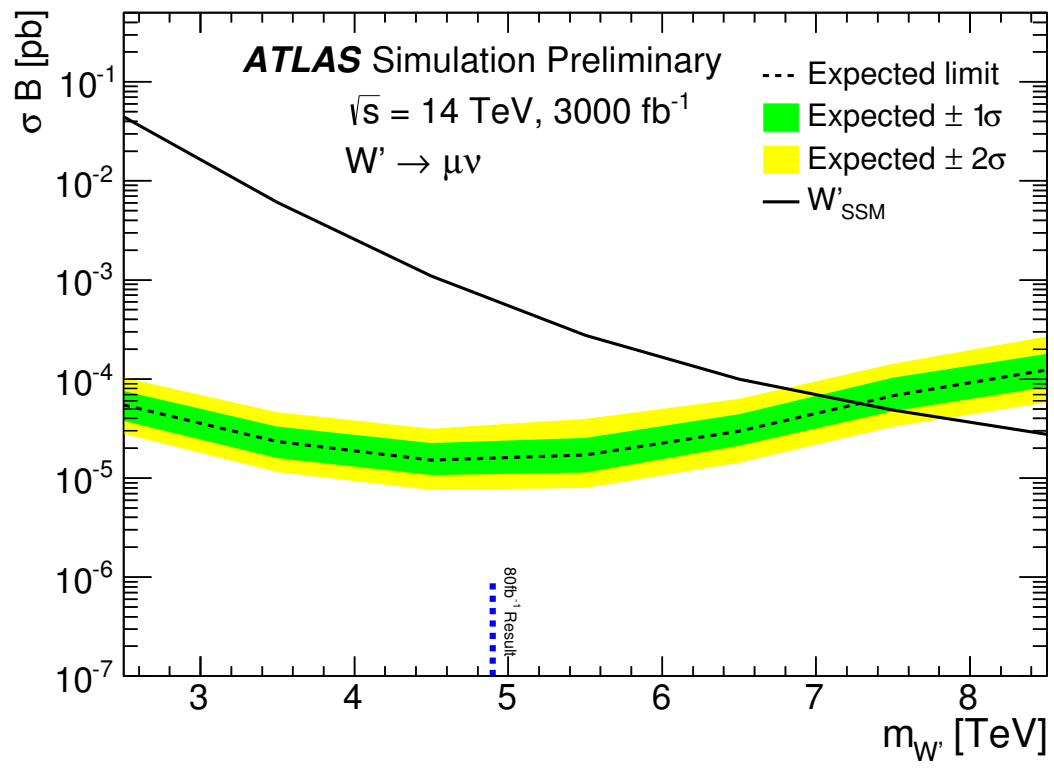

(b)

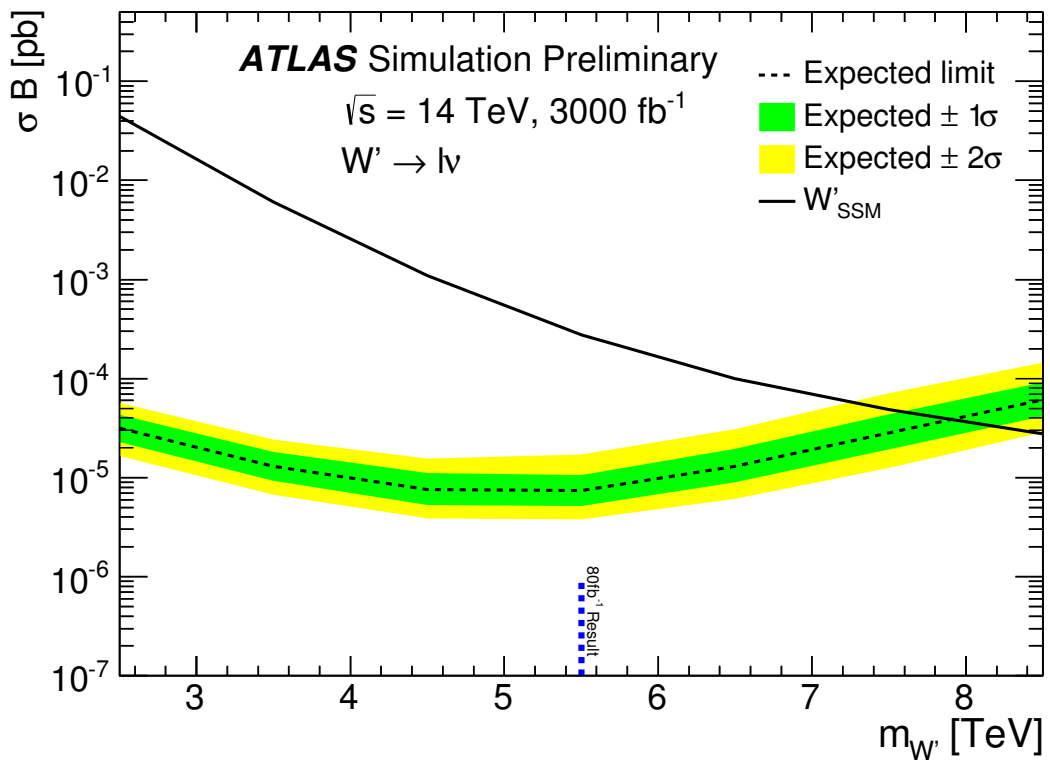

(c)

Figure 7: Expected (dashed black line) upper limits on cross-section times branching ratio ($\sigma \times BR$) as a function of the SSM $W'$ boson mass in the (a) electron, (b) muon and (c) combined electron and muon channels of the $W'_{\mathrm{SSM}} \to \ell\nu$ search assuming 3000 fb$^{-1}$ of data. The $1\sigma$ (green) and $2\sigma$ (yellow) expected limit bands are also shown. The predicted $\sigma \times BR$ for SSM $W'$ production is shown as a black line. These limits are based on the theory NNLO cross-section including off-shell production. The blue marker shows the current limits obtained with the latest Run 2 analysis based on 79.8 fb$^{-1}$ of data.

The 95% CL upper limits on $\sigma \times BR$ as a function of the $W'_{\mathrm{SSM}}$ mass are shown in Figure 7 for an integrated luminosity of 3000 fb$^{-1}$. The upper limits on $\sigma \times BR$ for $W'$ bosons start to weaken above a pole mass of 5 TeV, which is mainly caused the combined effect of a rapidly falling signal cross-section towards the kinematic limit and the increasing proportion of the signal being produced off-shell and falling in the low-$m_{\mathrm{T}}$ tail. It can be seen that $W'_{\mathrm{SSM}}$ bosons can be excluded up to masses of 7.6 (7.3) TeV in the electron (muon) channel. The limits in the electron channel are better as the calorimeter resolution is much better than the muon spectrometer one for very high-$p_{\mathrm{T}}$ leptons. The combination of these two channels increases the limits to just over 7.9 TeV. This is an improvement of more than 2 TeV with respect to the current exclusion limits using 79.8 fb$^{-1}$ of $\sqrt{s}$ = 13 TeV of data. For comparison, assuming the performance of the upgraded ATLAS detector and a luminosity of 300 fb$^{-1}$, the combined $W'_{\mathrm{SSM}}$ boson masses up to 6.7 TeV can be excluded. Though the detector resolutions for the upgraded detector at the HL-LHC are applied, this is a good approximation of the reach with the current detector at the end of LHC Run 3.

The discovery reach is based on a 5$\sigma$ significance. It is found that SSM $W'$ bosons can be discovered up to masses of 7.7 TeV. The discovery reach is shown in Table 9 together with the exclusion limits discussed above. As can be seen, the discovery reach typically is only few hundred GeV lower than the mass limits obtained with a background-only hypothesis. The similarity of the values for the discovery reach and the exclusion limit is expected, as in the high-$m_{\mathrm{T}}$ tail the background contribution approaches zero, while the number of signal events is around three. The expected reach with 300 fb$^{-1}$ of data will be 1.2 TeV lower assuming the same detector performance.

Table 9: Expected 95% CL lower limit on the $W'_{\mathrm{SSM}}$ mass in TeV in the electron and muon channels and their combination of the $W'_{\mathrm{SSM}} \to \ell\nu$ search assuming 3000 fb$^{-1}$ of data. In addition, the discovery reach for finding such new heavy particles is shown.

| Decay | Exclusion [TeV] | Discovery [TeV] |
|---|---|---|
| $W'_{\mathrm{SSM}} \to e\nu$ | 7.6 | 7.5 |
| $W'_{\mathrm{SSM}} \to \mu\nu$ | 7.3 | 7.1 |
| $W'_{\mathrm{SSM}} \to \ell\nu$ | 7.9 | 7.7 |

# 6 Search for $Z'$ bosons decaying to dilepton pairs

## 6.1 Analysis Strategy

This section reports on HL-LHC projections of a search for a $Z'$ boson which would manifest as a narrow resonance through its decay, in the dielectron and/or dimuon mass spectrum. Exclusion mass limits and discovery reach in mass will be presented in the following. Besides looking at the mass limits using $\sqrt{s}$ = 14 TeV $pp$ collisions at the HL-LHC, results will be also given at $\sqrt{s}$ = 13 TeV and $\sqrt{s}$ = 15 TeV. This will give an idea on the impact of a 1 TeV change in centre of mass energy on the exclusion limits and the discovery reach. Furthermore the mass reach at the HE-LHC running at $\sqrt{s}$ = 27 TeV and collecting 15 ab$^{-1}$ of data assuming the same physics performance and same detector setup as the upgraded ATLAS detector is presented. This scenario is only considered for the dielectron channel, as the performance in the dimuon channel is not representative with the upgraded ATLAS detector as the inner tracking detector for a $\sqrt{s}$ = 27 TeV accelerator would be imbedded in a larger magnetic field to ensure a good momentum resolution for very high-$p_{\mathrm{T}}$ tracks.

A $Z'$ signal would appear as an excess of events above the SM background at high dilepton invariant masses. This analysis considers two $Z'$ banchmark models: $Z'_{\mathrm{SSM}}$ bosons predicted by the SSM and $Z'_{\psi}$ bosons motivated by the models based on the E6 gauge group. In comparison to the benchmark $Z'_{\mathrm{SSM}}$ boson, which has a width of approximately 3% of its mass, the E6 models predict narrower $Z'$ signals with the $Z'_{\psi}$ boson having the smallest width which is 0.5% of its mass. The dominant source of background arises from DY production. Background from diboson ($WW$, $WZ$, $ZZ$) and top-quark production is not considered as their contribution to the overall SM background is negligible for dilepton invariant masses ($m_{\ell\ell}$) exceeding 2 TeV. This background is more pronounced at lower masses and was found to amount to around 10% (20%) of the total background for an invariant mass of 1 TeV (300 GeV). In the dielectron channel, additional background arises from $W$+jets and multi-jet events in which at most one real electron is produced and one or more jets satisfy the electron selection criteria. While this background is negligible in the muon channel, it amounts to around 15% for $m_{\ell\ell} > 1$ TeV after the event selection criteria in the Run 2 analysis [64] in the electron channel. The omission of this background and the tiny contribution from top-quark and diboson processes for large masses is covered by the systematic uncertainties and does not impact on the exclusion limits and the discovery reach.

Similar event selection criteria are applied as in the Run 2 analysis. Only events accepted by the single electron or muon trigger are considered. The single electron trigger requires at least one electron with transverse momentum $p_{\mathrm{T}} > 22$ GeV in $|\eta| < 2.5$ or one muon with $p_{\mathrm{T}} > 20$ GeV in $|\eta| < 2.65$ to be present in the event. In the offline analysis, exactly two same-flavoured leptons with $p_{\mathrm{T}} > 25$ GeV within $|\eta| < 2.65$ ($|\eta| < 2.47$ excluding $1.37 < |\eta| < 1.52$) for muons (electrons) are required. The electrons and muons have to fulfil the *tight* and *high-$p_T$* identification working points, respectively.

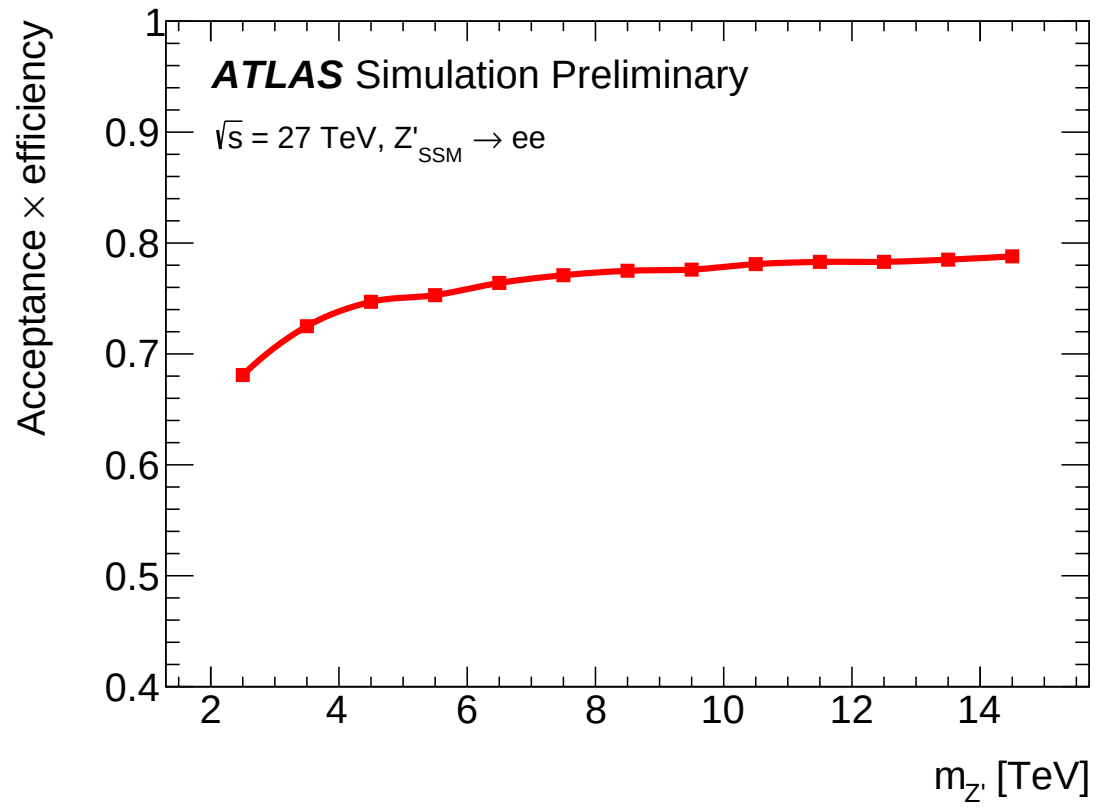

Figure 8: Total signal acceptance times efficiency versus SSM $Z'$ pole mass obtained for the $Z'_{\mathrm{SSM}} \rightarrow \ell\ell$ search assuming collisions taken at $\sqrt{s} = 27$ TeV.

The acceptance times efficiency after these selection cuts as a function of the $Z'_{\mathrm{SSM}}$ pole mass for collisions at the HL-LHC is very similar to the findings from Run 2 [17]. For running at the HE-LHC, the acceptance × efficiency curve for the electron channel is displayed in Figure 8. The resulting dilepton invariant mass spectrum (using the binning which is used to compute the exclusion limits and the discovery reach) is shown in Figure 9 for the DY background as well as for an example $Z'_{\mathrm{SSM}}$ boson with a mass of 5 TeV at the HL-LHC. The differences in the shapes of the reconstructed $Z'_{\mathrm{SSM}}$ signals in the electron and muon channels arise from the effects of the momentum resolution. Figure 10 shows the expected dielectron mass spectrum for the HE-LHC scenario. The differences in shape with respect to Figure 9 arise from the kinematics of the leptons from $Z'$ boson decay due to differences in the rapidity distribution. The expected event yields and their statistical uncertainties, in bins of invariant mass, are given in Table 10 for the dielectron and dimuon channel. The yields are shown separately for the SM DY background and three $Z'_{\mathrm{SSM}}$ signals.

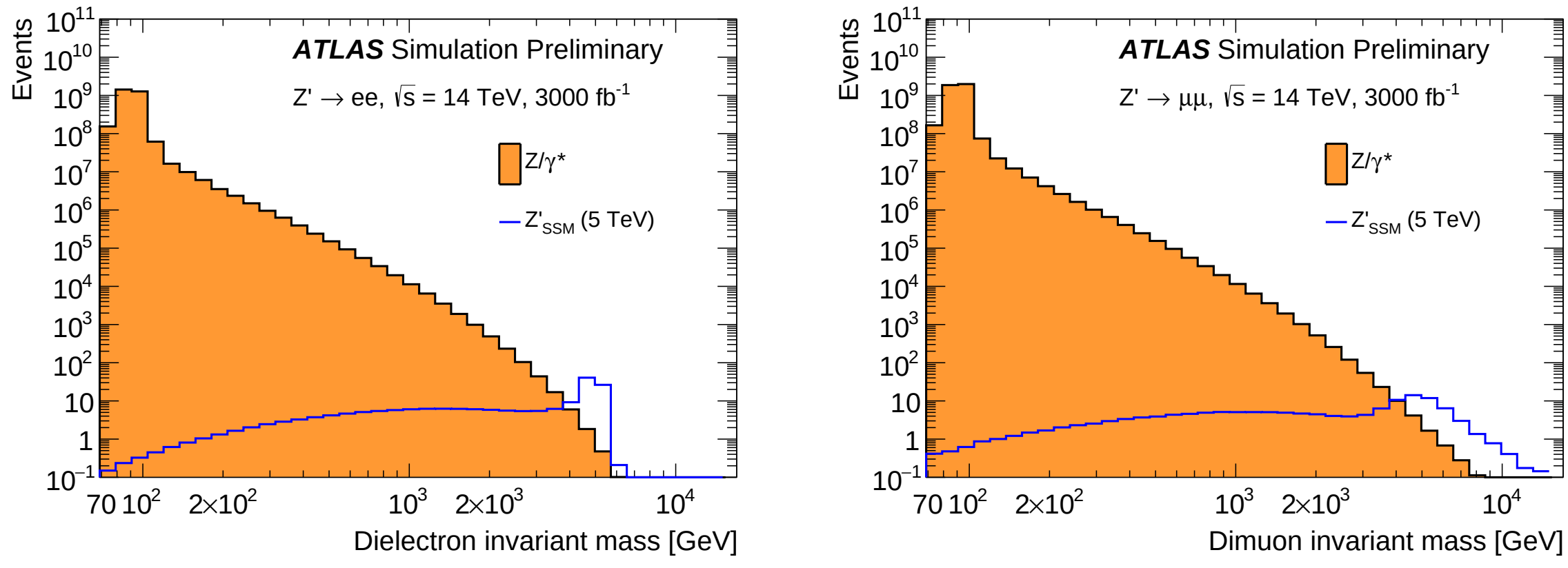

(a) (b)

Figure 9: Invariant mass distributions for events satisfying all selection criteria in the (a) dielectron and (b) dimuon channel. The distribution is shown for the DY background and as an example a SSM $Z'$ boson with a mass of 5 TeV.

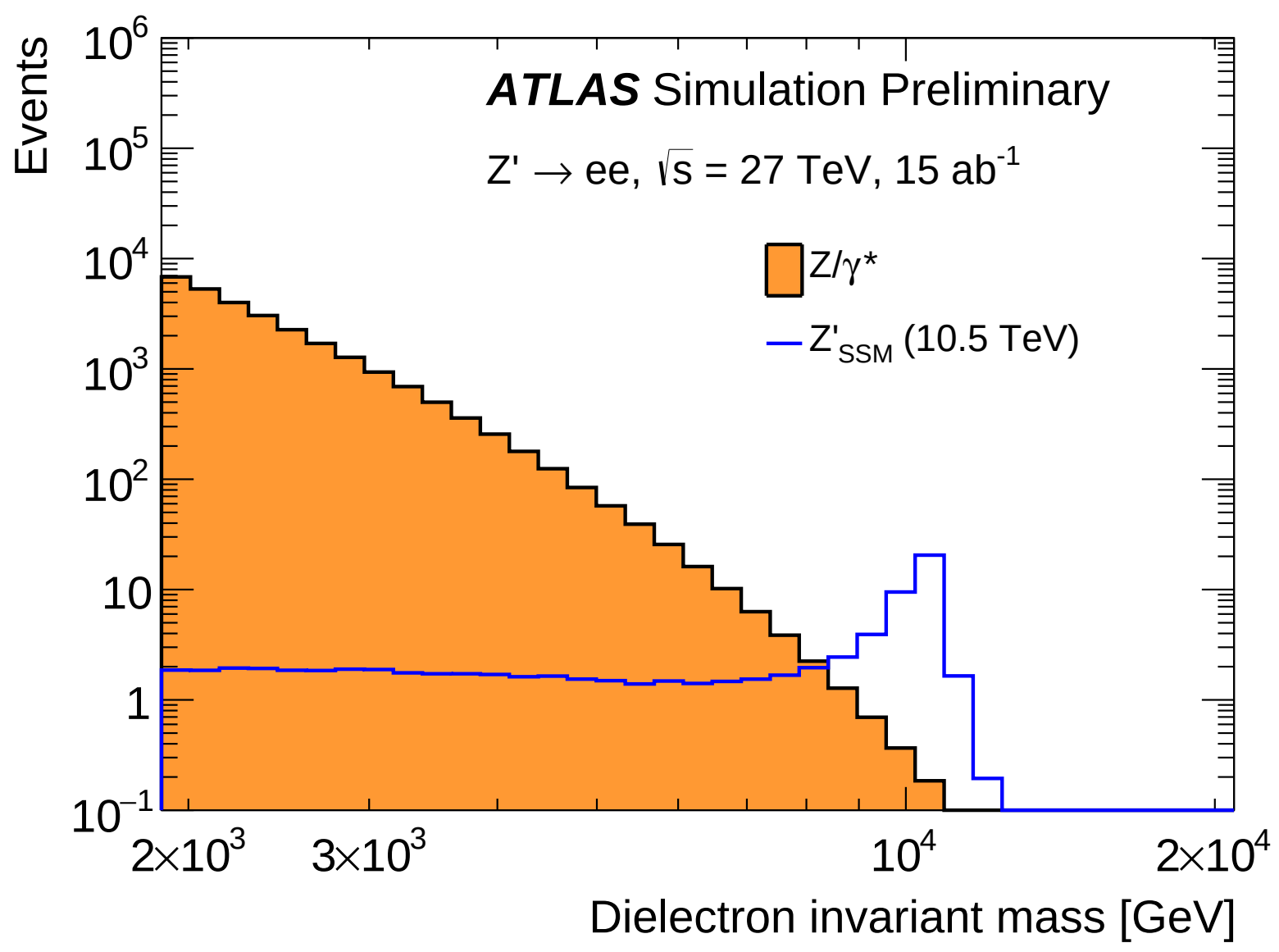

Figure 10: Invariant mass distributions for events satisfying all selection criteria in the dielectron channel for running at $\sqrt{s}$ = 27 TeV. The distribution is shown for the DY background and as an example a SSM $Z'$ boson with a mass of 10.5 TeV.

Table 10: Expected event yields for 3000 fb$^{-1}$ of $\sqrt{s}$ = 14 TeV collisions and their statistical uncertainties in the electron and muon channel in different mass intervals. For presentational purposes several of the mass bins used to compute the exclusion limits and the discovery reach are merged in larger bins in this table. The yields are given for the DY background and for $Z'_{\mathrm{SSM}}$ bosons for three values of the pole mass.

Electron channel

| $m_{\ell\ell}$ [GeV] | 60–240 | 240–950 | 950–3000 | 3000–5000 | 5000–15000 |
|---|---|---|---|---|---|
| Total SM | 3011040000 ± 50000 | 4065000 ± 2000 | 25100 ± 160 | 69 ± 8 | 0.6 ± 0.8 |
| $Z'_{\mathrm{SSM}}$ (3.5 TeV) | 26.74 ± 0.12 | 171.7 ± 0.7 | 427.3 ± 2.2 | 1524 ± 7 | 0.3547 ± 0.0013 |
| $Z'_{\mathrm{SSM}}$ (5.5 TeV) | 4.034 ± 0.018 | 24.34 ± 0.05 | 27.2 ± 0.04 | 11.58 ± 0.04 | 20.15 ± 0.11 |
| $Z'_{\mathrm{SSM}}$ (6.5 TeV) | 2.437 ± 0.011 | 14.263 ± 0.028 | 14.741 ± 0.020 | 3.098 ± 0.011 | 3.41 ± 0.04 |

Muon channel

| $m_{\ell\ell}$ [GeV] | 60–240 | 240–950 | 950–3000 | 3000–5000 | 5000–15000 |
|---|---|---|---|---|---|
| Total SM | 4215910000 ± 60000 | 4317000 ± 2100 | 25500 ± 160 | 92 ± 10 | 2.8 ± 1.7 |
| $Z'_{\mathrm{SSM}}$ (3.5 TeV) | 38.5 ± 0.4 | 155.0 ± 0.9 | 365 ± 5 | 973 ± 13 | 34.7 ± 2.2 |
| $Z'_{\mathrm{SSM}}$ (5.5 TeV) | 6.85 ± 0.7 | 26.18 ± 0.13 | 24.2 ± 0.9 | 11.59 ± 0.14 | 12.24 ± 0.23 |
| $Z'_{\mathrm{SSM}}$ (6.5 TeV) | 3.610 ± 0.035 | 13.58 ± 0.07 | 11.76 ± 0.05 | 2.380 ± 0.029 | 1.93 ± 0.06 |

## 6.2 Systematics

The experimental and theoretical uncertainties assumed in this analysis are estimated from the Run 2 results [64]. These uncertainties will be smaller by the end of the HL-LHC running and the Run 2 values are scaled to more realistic values following the recommendations given in Ref. [51]. In the following, only the largest sources of uncertainties are considered. As the uncertainties vary with $m_{\ell\ell}$, the uncertainties are expressed as a percentage value multiplied by the value of $m_{\ell\ell}$ given in TeV, using the uncertainties reported in the TeV range in the Run 2 analysis.

The experimental systematic uncertainties due to the reconstruction, identification and isolation of muons sums up to approximately $2.5\% \times m_{\ell\ell}$ [TeV]. Systematic uncertainties due to the energy resolution and scale are set to $1.5\% \times m_{\ell\ell}$ [TeV]. The uncertainties due to the resolution and reconstruction of the leptons will be added in quadrature to the dominant sources of theoretical uncertainties due to the choice of the PDF and the variation on the PDF. The uncertainties due to the choice of the PDF on the signal and the DY background are taken as $2.5\% \times m_{\ell\ell}$ [TeV] and the uncertainties due to any variation in PDF are assumed to be $5\% \times m_{\ell\ell}$ [TeV].

Overall these uncertainties add up to $6.5\% \times m_{\ell\ell}$ [TeV]. This search is looking for an excess in the high $m_{\ell\ell}$ tail and therefore, as the $W'$ searches, it is limited by the statistic uncertainties.

## 6.3 Results

The statistical analysis performed for this search uses the same Bayesian analysis is employed as in the $W' \to \ell\nu$ search presented in Section 5.3 and as used in the Run 2. As in the $W' \to \ell\nu$ search, pole masses are tested in 1 TeV intervals in the mass region between 2.5 TeV and 11.5 TeV and the exclusion limits and discovery reach are interpolated between the two nearest mass points where necessary.

In case no signal is observed at the LHC or the HL-LHC, 95% CL exclusion upper limits can be set for producing a $Z'$ boson times its branching ratio to only one lepton generation ($\sigma \times BR$). The exclusion is performed for the $Z'$ boson models $Z'_{\mathrm{SSM}}$ and $Z'_{\psi}$. Exclusion limits are shown in Figure 11 assuming a

20

$Z'_{\psi}$ boson as benchmark. The expected mass limits for the two $Z'$ models are also summarised in Table

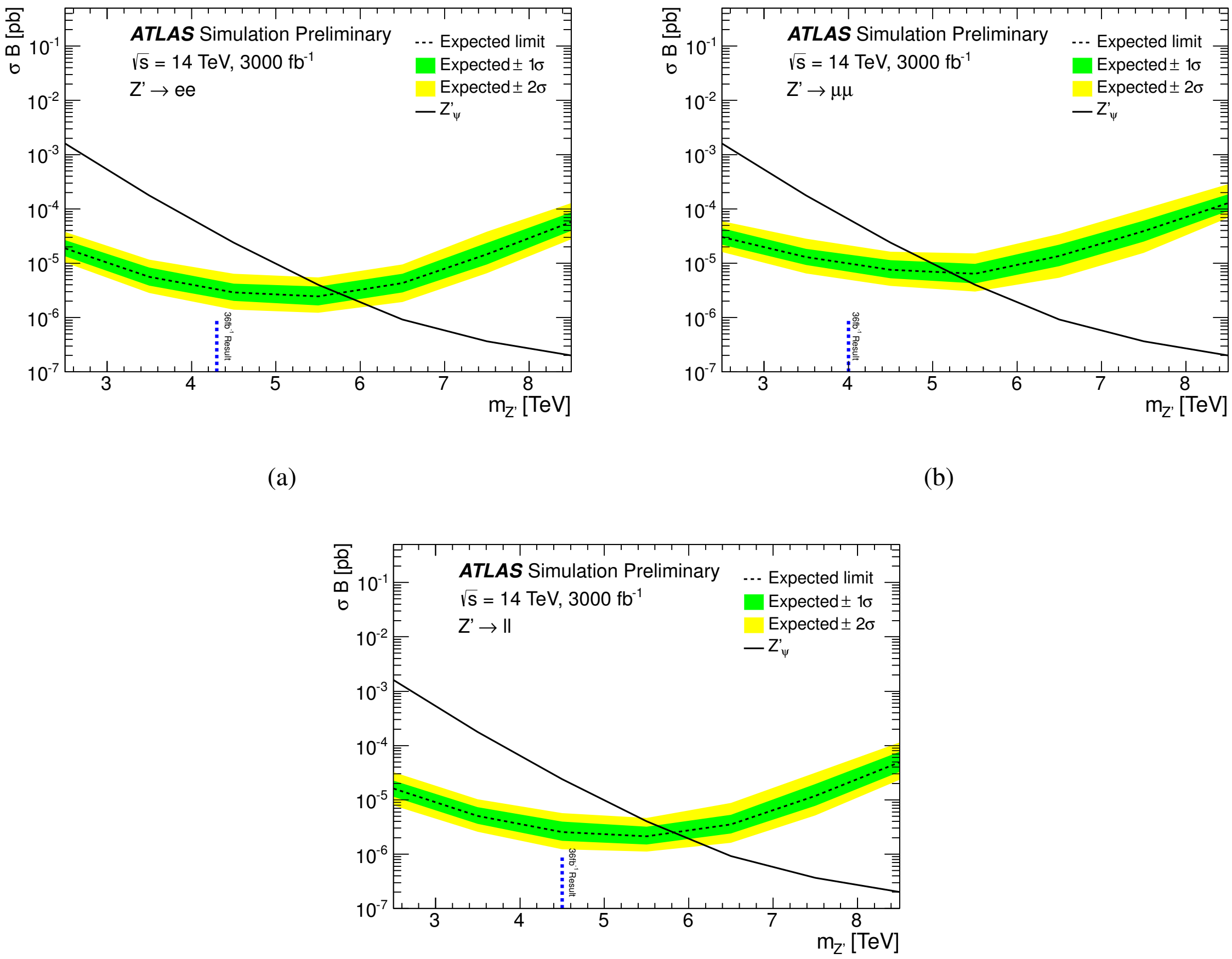

(a)

(b)

(c)

Figure 11: Expected (dashed black line) upper limits on cross-section times branching ratio ($\sigma \times BR$) as a function of the $Z'_{\psi}$ boson mass in the (a) dielectron, (b) dimuon and (c) combined electron and muon channels for $\sqrt{s}$ = 14 TeV collisions and an integrated luminosity value of 3000 fb$^{-1}$. The $1\sigma$ (green) and $2\sigma$ (yellow) expected limit bands are also shown. The predicted $\sigma \times BR$ for $Z'_{\psi}$ production is shown as a black line. These limits are based on the theory NNLO cross-section including off-shell production. The blue marker shows the current limits obtained with the Run 2 analysis which is based on 36 fb$^{-1}$ of data.

11 and visualised in Figure 12. These exclusion limits will extend the current $Z'_{\mathrm{SSM}}$ ($Z'_{\psi}$) mass limit of 4.5 (3.8) TeV obtained using 36.1 fb$^{-1}$ of data taken at $\sqrt{s}$ = 13 TeV to 6.7 (6.1) TeV. Table 11 does not only show these expected lower limits on the pole mass for the two $Z'$ scenarios at the HL-LHC with $\sqrt{s}$ = 14 TeV collisions, but also show the impact on these results if the collision energy would vary by 1 TeV. As can be seen, the lower limits on the pole mass would differ by 200–300 GeV. These exclusion limits are driven by the performance of the dielectron channel as the calorimeter resolution is much better than the muon spectrometer one for very high $p_{\mathrm{T}}$ leptons. In order to compare these findings with the expectations at the end of Run 3, the exclusion limits are also extracted for a luminosity of 300 fb$^{-1}$ of $\sqrt{s}$ = 14 TeV collisions. Though the detector resolutions for the upgraded detector at the HL-LHC are applied, this is a reasonable approximation of the expectations with the current detector at the end of the LHC data-taking. At 95% CL $Z'_{\mathrm{SSM}}$ ($Z'_{\psi}$) boson masses up to 5.4 TeV (4.8 TeV) can be excluded.

Table 11: Expected 95% CL lower limit on the $Z'$ mass in TeV in the dielectron and dimuon channels and their combination for two benchmark $Z'$ models for different centre of mass energies and 3000 fb$^{-1}$ of data. In addition, the discovery reach for finding such new heavy particles is shown.

| | $\sqrt{s}$ = 13 TeV | | $\sqrt{s}$ = 14 TeV | | $\sqrt{s}$ = 15 TeV | |
|---|---|---|---|---|---|---|
| Decay | Exclusion | Discovery | Exclusion | Discovery | Exclusion | Discovery |
| $Z'_{\mathrm{SSM}} \rightarrow ee$ | 6.0 TeV | 5.9 TeV | 6.4 TeV | 6.3 TeV | 6.7 TeV | 6.6 TeV |
| $Z'_{\mathrm{SSM}} \rightarrow \mu\mu$ | 5.5 TeV | 5.4 TeV | 5.8 TeV | 5.7 TeV | 6.0 TeV | 5.9 TeV |
| $Z'_{\mathrm{SSM}} \rightarrow \ell\ell$ | 6.1 TeV | 6.1 TeV | 6.5 TeV | 6.4 TeV | 6.7 TeV | 6.7 TeV |
| $Z'_{\psi} \rightarrow ee$ | 5.3 TeV | 5.3 TeV | 5.7 TeV | 5.6 TeV | 6.1 TeV | 6.0 TeV |
| $Z'_{\psi} \rightarrow \mu\mu$ | 4.9 TeV | 4.6 TeV | 5.2 TeV | 5.0 TeV | 5.5 TeV | 5.2 TeV |
| $Z'_{\psi} \rightarrow \ell\ell$ | 5.4 TeV | 5.4 TeV | 5.8 TeV | 5.7 TeV | 6.1 TeV | 6.1 TeV |

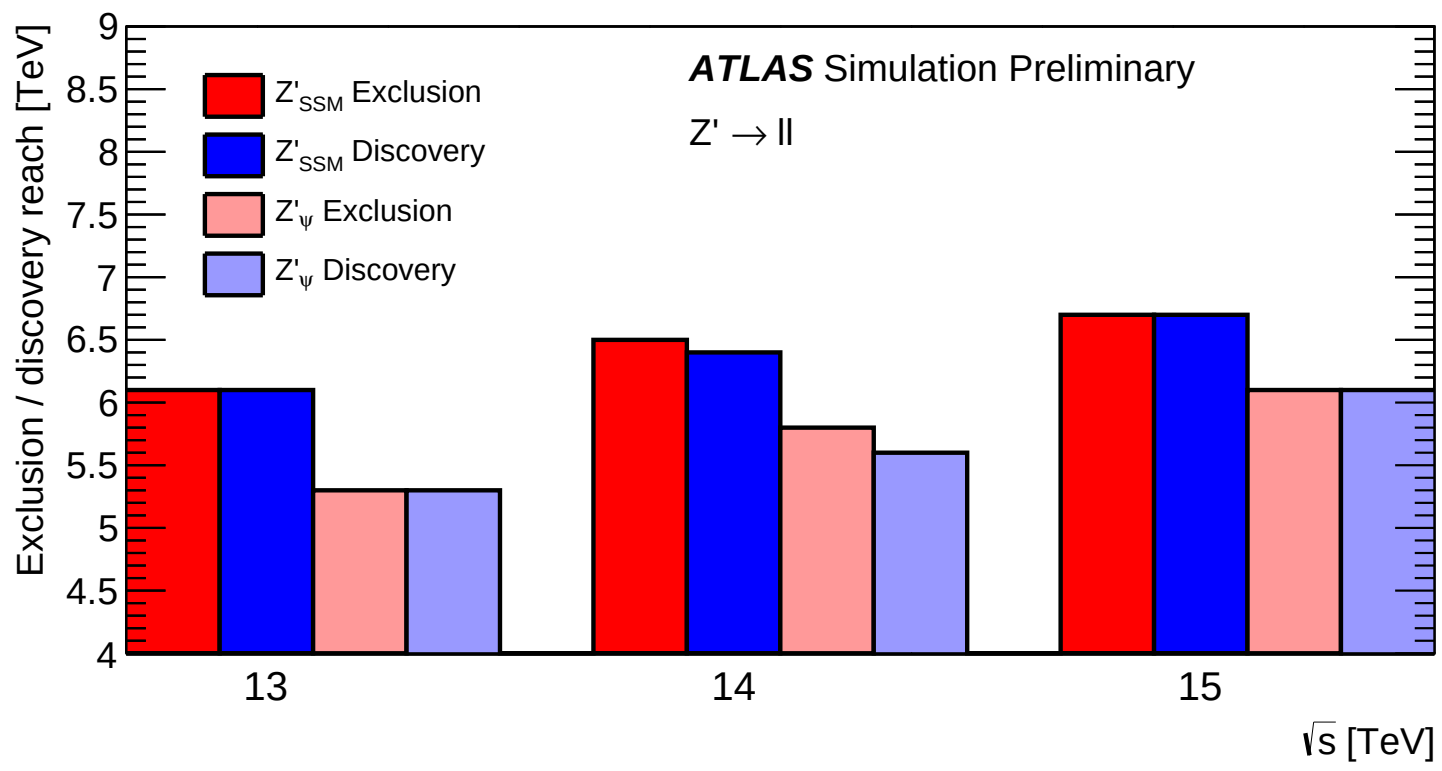

Figure 12: Expected (blue bars) upper limits on the cross-section times branching ratio ($\sigma \times BR$) as well as the discovery reach (red bars) for the search for $Z'_{\mathrm{SSM}}$and $Z'_{\psi}$ bosons as a function of the centre of mass energy. The limits given correspond to the results after combining the results of the electron and muon channel.

Table 11 also shows the discovery reach for finding $Z'_{\mathrm{SSM}}$ and $Z'_{\psi}$ bosons. As can be seen the $Z'$ discovery reach and exclusion mass limit differ by at most few hundred GeV. In some cases the difference is less than 100 GeV and the reach given in the Table is the same as the exclusions limits as the numbers are rounded to a precision of 100 GeV. The similarity of these values is expected, as in the high-mass tail the background contribution approaches zero, while the number of signal events is around three. Compared to the results presented in Ref. [21] the discovery reach reported here is higher due to a change in how the reach is calculated. This is based on the shape of the signal and background $m_{\ell\ell}$ distribution here while the $5\sigma$ significance was calculated in a mass range between $m(Z')/2$ to infinity in Ref. [21].

The discovery reach and lower exclusion limits at 95% CL in mass are also calculated for a detector at the HE-LHC in the dielectron channel. This is done assuming the same physics performance as for the ATLAS detector at the HL-LHC. The exclusion limits and the discovery reach are summarised in Table 12. At the HE-LHC $Z'_{\mathrm{SSM}}$ and $Z'_{\psi}$ bosons can be discovered up to 12.8 TeV and 11.2 TeV, respectively, thus increasing their discovery reach by 6.5 TeV compared the HL-LHC. In case $Z'$ bosons are not discovered yet, the HE-LHC will be able to further rule out $Z'_{\mathrm{SSM}}$ and $Z'_{\psi}$ bosons up to 12.8 TeV and 11.4 TeV, respectively. This corresponds to an increase of the discovery potential by a factor of two compared to the expectations at the HL-LHC.

Table 12: 95% CL lower limits and discovery reach on the $Z'_{\mathrm{SSM}}$ and $Z'_{\psi}$ boson mass in the dielectron channel assuming 15 ab$^{-1}$ of $pp$ data to be taken at the HE-LHC with $\sqrt{s}$ = 27 TeV.

| Decay | Exclusion [TeV] | Discovery [TeV] |
|---|---|---|
| $Z'_{\mathrm{SSM}} \to ee$ | 12.8 | 12.8 |
| $Z'_{\psi} \to ee$ | 11.4 | 11.2 |

# 7 Search for $Z'$ bosons decaying to a $t\bar{t}$ pair

Besides these studies, the prospects for $Z'$ bosons in the $t\bar{t}$ final state [27] were studied. Updated results using a more recent parameterisation of the $b$-tagging efficiencies and misidentification rates were shown in Ref. [21]. These studies used the event selection criteria and systematic uncertainty based on the Run 1 analysis [65] and a summary of the results of these studies without any further updates are given below.

The analysis looks for a narrow width $Z'$ boson searched for in a final state in which one of the $W$ bosons from the top quark decays to two jets and the other decays to a lepton (electron or muon) and a neutrino ($t\bar{t} \to WbWb \to \ell\nu bqq'b$). Events are required to contain exactly one lepton, several jets and at least a moderate amount of $E_{\mathrm{T}}^{\mathrm{miss}}$ must be present. Events are separated into boosted and resolved channels with most of the signal events falling in the former category. In the resolved channel the decay products of the hadronic top-quark decay are reconstructed as three separate jets and in total events must contain at least four jets. In the boosted channel, the hadronic top-quark decay products are highly boosted and end up in one broad so-called large-R jet. Events are selected if at least one large-R jet and one jet (from the other top-quark decay) is present. Subsequently $m_{t\bar{t}}$ is reconstructed based on the reconstruction of the $W$ bosons and b-jets in the event. Using $m_{t\bar{t}}$ as discriminant, upper limits were set on the signal cross-section times branching ratio as a function of the $Z'$ boson mass. Using as benchmark a Topcolour-assisted Technicolour $Z'_{\mathrm{TC2}}$ boson with a narrow width of 1.2%, $Z'_{\mathrm{TC2}}$ bosons can be excluded up to $\simeq$ 4 TeV with 3000 fb$^{-1}$ of pp collisions [21]. This mass limit is pessimistic, which is coming partly from the treatment of the systematic uncertainties which were taken from the Run 1 analysis [65]. These systematics are already smaller in the Run 2 analysis [66] and will be further reduced at the time of the HL-LHC. In particular the systematic uncertainty in the boosted channel are now reduced due to the significant improvements of the performance of boosted jets in Run 2 (in particular using more tracking information to look for sub-jets within the large-R jets). This gain in performance also improves the signal over background ratio. In addition, the usage of the top-tagger algorithm will help to further reject background.

# 8 Conclusion

This note summarises the prospects from four different searches for new heavy $W'$ and $Z'$ bosons at the HL-LHC, which is expected to run at $\sqrt{s}$ = 14 TeV and collect 3000 fb$^{-1}$of data. These studies are based on MC simulations. To simulate the response of the upgraded ATLAS detector and pile-up collisions the MC truth information is convoluted with parameterised estimates to emulate the response of the upgraded ATLAS detector and pile-up collisions.

The first search uses as benchmark a right-handed $W'$ boson and looks at the $W'_{\mathrm{R}} \to t\bar{b} \to \ell\nu b\bar{b}$ final state. In case such particles are not discovered, they can be excluded up to masses of 4.9 TeV at the HL-LHC. This increases the current limits based on 36 fb$^{-1}$of LHC data by 2.5 TeV. Another search uses the $W'_{\mathrm{SSM}}$ boson as benchmark and studies the $W' \to \ell\nu$ final state. The lower exclusion limit at 95% CL on the $W'_{\mathrm{SSM}}$ pole mass is expected to improve from 5.6 TeV currently to 7.9 TeV. If such particles exist they can be discovered up to $W'_{\mathrm{SSM}}$ masses of 7.7 TeV. Though different benchmark models are used in these analyses, in case of a discovery the search for signals in different final states will help to unravel the underlying theory.

$Z'$ bosons are searched for in the dilepton final states as well as their decays into a $t\bar{t}$ pair. $Z'_{\mathrm{SSM}}$ and $Z'_{\psi}$ bosons are expected to be discovered at the HL-LHC up to masses of 6.4 TeV and 5.7 TeV. In case no signs for such particles are found, they can be excluded up to $Z'_{\mathrm{SSM}}$ ($Z'_{\psi}$) masses of 6.5 TeV (5.8 TeV). In case the HL-LHC would run at $\sqrt{s}$ = 13 TeV or $\sqrt{s}$ = 15 TeV these numbers would decrease or increase by few hundred GeV. Searches for a $Z'$ predicted by a Topcolour-assisted Technicolour model in the $t\bar{t} \to WbWb \to \ell\nu bqq'b$ final state were studied in Ref. [27]. If no $t\bar{t}$ resonances are found such particles can be excluded up to masses of around 4 TeV. This study based on the Run 1 results does not include the improvements in boosted jet reconstruction and the top tagger already available in Run 2 and these results therefore represent a very conservative estimate of the expected performance.

# CMS Physics Analysis Summary

Contact: cms-future-conveners@cern.ch

# Projection of searches for pair production of scalar leptoquarks decaying to a top quark and a charged lepton at the HL-LHC

The CMS Collaboration

## Abstract

Projections for searches for pair-produced scalar leptoquarks decaying into top quarks and muons or tau leptons for high integrated luminosities of up to 3000 $fb^{-1}$ at the high luminosity LHC are presented. This study is based on published analysis results of data recorded in the year 2016. It uses scaled signal and background templates to estimate the reach in terms of discovery potential and upper limits on the leptoquark pair production cross section. Two different ways of treating systematic uncertainties are studied. The mass reach for a $5\sigma$ discovery or a 95% confidence level exclusion is expected to increase by 400 to 500 GeV with respect to the 2016 results. In the case of mixed decays between these two channels, the mass expected to be in reach for a $5\sigma$ discovery and the expected 95% confidence level limit on excluded leptoquark masses ranges from 1200 to 1700 GeV and from 1400 to 1900 GeV depending on the value of the branching fraction, respectively.

# 1 Introduction

Leptoquarks (LQs) are hypothetical particles that carry both baryon and lepton quantum numbers. They are color-triplets and carry fractional electric charge. Their possible quantum numbers can be restricted by the assumption that their interactions with standard model (SM) fermions are renormalizable and gauge invariant [1]. The spin of an LQ state is either 0 (scalar LQ) or 1 (vector LQ). LQs appear in theories beyond the SM such as grand unified theories [2–4], technicolor models [5, 6], or compositeness scenarios [7, 8]. Models [9–20] proposing the existence of LQs as an explanation for the tension between the SM prediction and experimental data in flavor observables, such as the ratios $R_{D^{(*)}}$ [21–28] and $R_{K^{(*)}}$ [29–32], or the muon anomalous magnetic moment $a_\mu$ [33, 34] favor large couplings of the LQ to third-generation quarks and masses at the TeV scale.

At the CERN LHC, LQs can be produced in pairs via gluon-gluon fusion and quark-antiquark annihilation. The pair production cross section depends on the mass of the LQ. For scalar LQs, it is known at next-to-leading (NLO) order precision in perturbative quantum chromodynamics [35]. The production of a single LQ coupled exclusively to top quarks is suppressed, as it requires a top quark in the initial state.

A study is presented of the expected sensitivity of searches for pair-produced scalar LQs decaying into top quarks and charged leptons with the high luminosity LHC (HL-LHC) [36]. The analysis is based on published CMS results of the $t + \mu$ [37] and $t + \tau$ [38] LQ decay channels, which have been carried out using $35.9\,fb^{-1}$ of proton-proton collision data collected in the year 2016 with a center-of-mass energy of $\sqrt{s} = 13\,TeV$. The analysis strategies are kept unchanged with respect to the ones in Refs. [37, 38], only different total integrated luminosities, the higher center-of-mass energy of 14 TeV, and different scenarios of systematic uncertainties are considered.

# 2 Summary of the analyses of the 2016 dataset

The analysis strategies of Refs. [37, 38] are briefly summarized in the following. The results from these searches are used for estimating the expected sensitivity with the HL-LHC by scaling the predictions for the SM backgrounds, the LQ signals, and the corresponding uncertainties to higher integrated luminosities. Feynman diagrams of the two signal processes under consideration are shown in Fig. 1.

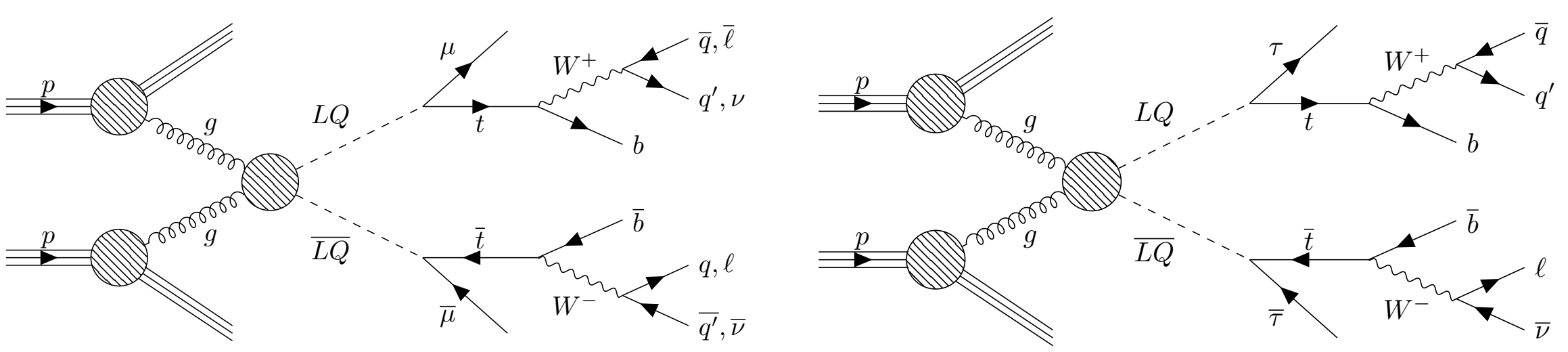

Figure 1: Feynman diagrams of the gluon-induced production and the subsequent decay of a pair of LQs into top quarks and muons (left) and into top quarks and $\tau$ leptons with a subsequent decay of the two top quarks into the $\ell$+jets final state (right), where $\ell$ denotes an electron or muon.

### 2.1 Search in the LQ $\rightarrow$ t$\mu$ decay channel

The search for pair-produced LQs decaying exclusively into top quarks and muons [37] is carried out in the final state with at least two muons and at least two jets. The signal region is split into two orthogonal categories. Events in category A are required to have at least two muons and one additional electron or muon, where at least one pair of muons must have opposite-sign electric charge. Category B contains all remaining events in the signal region.

The LQ system is reconstructed in category A under the assumption that the two top quarks, decay products of the LQs, decay into the $\ell$+jets final state ($\ell$ denoting an electron or muon), producing jets, the missing transverse momentum ($p_{\mathrm{T}}^{\mathrm{miss}}$), and either one electron or muon in addition to the two prompt muons from the LQ decays. First, top quark candidates are built from permutations of the seven jets with the highest transverse momentum ($p_{\mathrm{T}}$), the additional electron or muon, and $p_{\mathrm{T}}^{\mathrm{miss}}$. Second, the LQ candidates are constructed from top quark candidates and muons that have not already been used for the leptonically decaying top quark. As there are multiple LQ candidates in an event, the pair of LQ candidates is chosen based on a $\chi^2$ variable.

Since the reconstructed LQ mass shows strong discrimination power between signal and SM background, its distribution is used for the final statistical interpretation in category A. In category B, where the LQ mass is not reconstructed, the spectrum of $S_{\mathrm{T}}$, defined as the scalar sum of the $p_{\mathrm{T}}$ of all selected leptons, jets, and $p_{\mathrm{T}}^{\mathrm{miss}}$, is used for this purpose.

The main SM background in category A is top quark pair production with smaller contributions from DY+jets, diboson, $\mathrm{t\bar{t}}$ +V, where V denotes a heavy gauge boson, and single t production. The background prediction here is taken purely from simulation and has been corrected for different electron/muon misidentification rates in data and simulated events. In category B, the production of $\mathrm{t\bar{t}}$ and DY+jets constitutes the major backgrounds, while single t, diboson, and $\mathrm{t\bar{t}}$ +V production contribute to a minor degree. The major backgrounds are estimated with a data-driven technique, leading to a strongly reduced impact of systematic uncertainties on category B.

No significant deviation from the SM prediction is observed in either category in the 2016 dataset [37]. A binned-likelihood fit and a Bayesian method [39–41] are used to set upper limits on the pair-production cross section of LQs exclusively decaying to top quarks and muons, excluding scalar LQs with masses below 1420 GeV.

### 2.2 Search in the LQ $\rightarrow$ t$\tau$ decay channel

The search for pair-produced scalar LQs decaying exclusively into top quarks and $\tau$ leptons [38] is performed in the final state with at least one electron or muon, at least one $\tau_{\mathrm{h}}$ lepton, where the subscript h indicates the hadronic decay of the $\tau$ lepton, and at least three jets. Two exclusive categories of events are defined based on the number of $\tau_{\mathrm{h}}$ leptons. In category A, events contain exactly one $\tau_{\mathrm{h}}$ lepton while at least two $\tau_{\mathrm{h}}$ leptons are required in category B. Events in category A are further sorted into one out of four subcategories. These are defined by the electric charges of the electron or muon and the $\tau_{\mathrm{h}}$ lepton, which can be either same-sign or opposite-sign, and the value of $S_{\mathrm{T}}$. Considering final states with an electron or a muon, a total of ten orthogonal search regions is used in the analysis. Different event selections are applied in these regions in order to maximize the expected significance of a hypothetical LQ signal.

In category A, the four-momentum of the top quark decaying into the hadronic final state (hadronic top quark) is reconstructed and the distribution of its transverse momentum ($p_{\mathrm{T}}^{\mathrm{t}}$) is used for a shape analysis. Top quarks produced in the decay of an LQ are expected to carry a larger $p_T$ compared to those originating from SM background processes, hence analyzing the $p_T^t$ spectrum provides discrimination power between signal and background. A counting experiment is performed in category B due to the small number of selected events.

Events containing a jet misidentified as a $\tau_h$ lepton are a considerable source of background in all categories. A data-driven technique is employed to estimate the contribution of $t\bar{t}$ events with such misidentified $\tau_h$ leptons in both categories. In addition, the contribution of events with misidentified $\tau_h$ leptons from W+jets production in category A is estimated from data in a background-enriched control region, defined by inverting the $\tau_h$ isolation.

The data are found to be in agreement with the SM prediction in all search regions in the 2016 dataset [38]. The results from all channels are combined in a binned-likelihood fit and a Bayesian method is used to set upper limits on the pair-production cross section of LQs decaying exclusively into top quarks and $\tau$ leptons. Such scalar LQs with masses below 900 GeV are excluded.

### 2.3 Combination of searches

A statistical combination of the previously introduced two analyses allows to set limits on the LQ pair production cross section for varying branching fractions $\mathcal{B}$ between the two decay modes, where we assume $\mathcal{B}(\mathrm{LQ} \to \mathrm{t}\mu) = 1 - \mathcal{B}(\mathrm{LQ} \to \mathrm{t}\tau)$. In this way, pair-produced LQs could be excluded up to masses of 900 GeV for all values of $\mathcal{B}(\mathrm{LQ} \to \mathrm{t}\mu)$ using the 2016 dataset of CMS [37].

## 3 The upgraded CMS detector

The CMS detector [42] will be substantially upgraded in order to fully exploit the physics potential offered by the increase in luminosity at the HL-LHC, and to cope with the demanding operational conditions at the HL-LHC [43–47]. The upgrade of the first level hardware trigger (L1) will allow for an increase of L1 rate and latency to about 750 kHz and 12.5 $\mu$s, respectively, and the high-level software trigger (HLT) is expected to reduce the rate by about a factor of 100 to 7.5 kHz. The entire pixel and strip tracker detectors will be replaced to increase the granularity, reduce the material budget in the tracking volume, improve the radiation hardness, and extend the geometrical coverage and provide efficient tracking up to pseudorapidities of about $|\eta| = 4$. The muon system will be enhanced by upgrading the electronics of the existing cathode strip chambers (CSC), resistive plate chambers (RPC) and drift tubes (DT). New muon detectors based on improved RPC and gas electron multiplier (GEM) technologies will be installed to add redundancy, increase the geometrical coverage up to about $|\eta| = 2.8$, and improve the trigger and reconstruction performance in the forward region. The barrel electromagnetic calorimeter (ECAL) will feature the upgraded front-end electronics that will be able to exploit the information from single crystals at the L1 trigger level, to accommodate trigger latency and bandwidth requirements, and to provide 160 MHz sampling allowing high precision timing capability for photons. The hadronic calorimeter (HCAL), consisting in the barrel region of brass absorber plates and plastic scintillator layers, will be read out by silicon photomultipliers (SiPMs). The endcap electromagnetic and hadron calorimeters will be replaced with a new combined sampling calorimeter (HGCal) that will provide highly-segmented spatial information in both transverse and longitudinal directions, as well as high-precision timing information. Finally, the addition of a new timing detector for minimum ionizing particles (MTD) in both barrel and endocap region is envisaged to provide capability for 4-dimensional reconstruction of interaction vertices that will allow to significantly offset the CMS performance degradation due to high PU rates.

| Uncertainty | Value at 3000 $\mathrm{fb}^{-1}$ [%] |
|---|---|
| Luminosity | 1 |
| SM production cross sections | 2.8–12.5 |
| b-tagging (b/c) | 1 |
| b-tagging (light) | 5 |
| JES | 1–2.5 |
| JER | 3–6 |
| e, $\mu$ efficiencies | 1 |
| e, $\mu$ misidentification | 1–16 |
| $\tau$ identification | 2.5 |
| $\tau$ energy scale | 3 |
| $\tau$ charge misidentification | 2 |
| Background extrapolation | LQ $\to$ t$\mu$: 1.2–3.6 |
| | LQ $\to$ t$\tau$: $\leq 1$ |

Table 1: Scaled relative systematic uncertainties in the "YR18 syst." scenario at $\mathcal{L}_{\mathrm{int}}^{\mathrm{target}} = 3000\,\mathrm{fb}^{-1}$.

A detailed overview of the CMS detector upgrade program is presented in Ref. [43–47], while the expected performance of the reconstruction algorithms and pile-up mitigation with the CMS detector is summarised in Ref. [48].

## 4 Projection to higher integrated luminosities

The projection of expected limits and significances of the individual analyses discussed in the previous section is performed by scaling the expected distributions of signal and background in each region, which are used in the final limit setting procedure of each of the analyses, to a higher value of integrated luminosity. The scaling factor $f$ of the bin contents of each histogram is given by $f = \mathcal{L}_{\mathrm{int}}^{\mathrm{target}}/35.9\,\mathrm{fb}^{-1}$. Here, $\mathcal{L}_{\mathrm{int}}^{\mathrm{target}}$ is the target value of integrated luminosity under consideration and takes values of up to $3000\,\mathrm{fb}^{-1}$, which corresponds to the total integrated luminosity the HL-HLC could deliver. The relative statistical uncertainties in both simulated and data-driven histogram templates are reduced by a factor of $1/\sqrt{f}$. In this projection, two different scenarios of scaling systematic uncertainties are considered in addition.

In the first scenario (denoted "YR18 syst."), the relative experimental uncertainties are scaled by a factor of $1/\sqrt{f}$ until they reach a defined lower limit based on estimates of the achievable accuracy with the upgraded detector [48]. The relative experimental uncertainties considered at $\mathcal{L}_{\mathrm{int}}^{\mathrm{target}} = 3000\,\mathrm{fb}^{-1}$ are listed in Table 1. The jet energy scale (resolution) uncertainty is referred to by "JES (JER)" and the uncertainty in the (mis)identification efficiency for charged leptons is denoted "(mis)identification". In the LQ $\to$ t$\mu$ search, the uncertainties in b-tagging and e/$\mu$ efficiencies have the largest impact on the final sensitivity, while in the LQ $\to$ t$\tau$ analysis the uncertainty in the $\tau$ lepton identification is dominant. The relative theoretical uncertainties, which in these analyses are the uncertainties due to missing higher orders, estimated by varying the renormalization and factorization scales $\mu_r$ and $\mu_f$, as well as the PDF uncertainty, are halved at $\mathcal{L}_{\mathrm{int}}^{\mathrm{target}} = 3000\,\mathrm{fb}^{-1}$. The uncertainties on the predicted LQ pair production cross section are not scaled. In the second scenario (denoted "stat. only"), no systematic uncertainties are considered.

The increased center-of-mass energy of $\sqrt{s} = 14\,\mathrm{TeV}$ at the HL-LHC with respect to $\sqrt{s} = 13\,\mathrm{TeV}$ in 2016 is accounted for by reweighting on an event-by-event basis, taking into account the shift in the momentum fractions of the initial-state partons and the resulting change in the differential production cross section of each simulated sample. As a result, the production cross sections of the backgrounds and, in particular, the LQ signal increase, especially for events with high momentum transfer, leading to an overall gain in sensitivity of $\mathcal{O}(5\%)$ compared to $\sqrt{s} = 13$ TeV. The object reconstruction under HL-LHC conditions in combination with the upgraded CMS detector is assumed to be sufficiently robust against additional interactions in the same bunch crossing in order not to introduce additional systematic uncertainties.

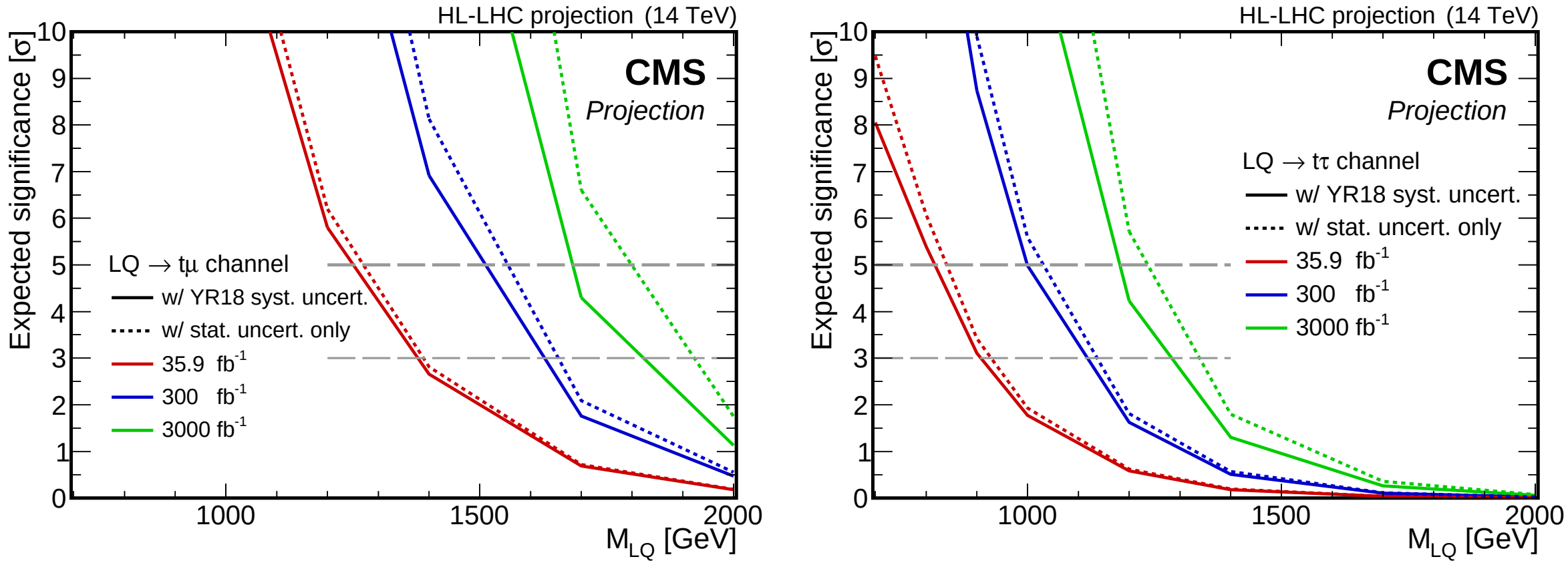

Figure 2: Expected significances for an LQ decaying exclusively to top quarks and muons (left) or $\tau$ leptons (right) as a function of the LQ mass and for different integrated luminosities in the "YR18 syst." (solid) and "stat. only" (dotted) scenarios. All results were obtained with templates for $\sqrt{s} = 13$ TeV that were scaled to $\sqrt{s} = 14$ TeV.

Figure 2 presents the expected significances of the analyses as a function of the LQ mass scaled to different assumed integrated luminosities in the "YR18 syst." and "stat. only" scenarios. The significances were computed from a log-likelihood ratio with the THETA [39] package, based on the final distributions of the LQ $\rightarrow$ t$\mu$ (left) and LQ $\rightarrow$ t$\tau$ (right) analyses, respectively, that are also used for setting expected limits. Increasing the target integrated luminosity from $\mathcal{L}_{\text{int}}^{\text{target}} = 35.9\,\text{fb}^{-1}$ up to $\mathcal{L}_{\text{int}}^{\text{target}} = 3000\,\text{fb}^{-1}$ greatly increases the discovery potential of both analyses. The LQ mass corresponding to a discovery at $5\sigma$ significance with a dataset corresponding to $3000\,\text{fb}^{-1}$ increases by more than 500 GeV, from about 1200 GeV to roughly 1700 GeV, in the LQ $\rightarrow$ t$\mu$ decay channel. For LQs decaying exclusively to top quarks and $\tau$ leptons, a gain of 400 GeV is expected, pushing the LQ mass in reach for a $5\sigma$ discovery from 800 GeV to 1200 GeV.

In Fig. 3, the expected projected limits are shown for the "YR18 syst." and the "stat. only" scenarios. They were obtained from the final distributions of the LQ $\rightarrow$ t$\mu$ (left) and LQ $\rightarrow$ t$\tau$ (right) analyses, respectively. The kink in the "YR18 syst." scenarios at high integrated luminosities in Fig. 3 (left) is related to category A in the LQ $\rightarrow$ t$\mu$ analysis becoming limited by systematic uncertainties. Leptoquarks decaying only to top quarks and muons are expected to be excluded below masses of 1900 GeV at $3000\,\text{fb}^{-1}$, which is a gain of 500 GeV compared to the limit of 1420 GeV obtained in the published analysis of the 2016 dataset [37] with data corresponding to $35.9\,\text{fb}^{-1}$. The mass below which LQs decaying exclusively to top quarks and $\tau$ leptons are expected to be excluded increases by 500 GeV, from 900 GeV to approximately 1400 GeV.

Figure 4 shows the expected significances and upper limits on the pair-production cross section of scalar LQs allowed to decay to top quarks and muons or $\tau$ leptons at the 95% CL as a function of the LQ mass and a variable branching fraction $\mathcal{B}(\text{LQ} \rightarrow \text{t}\mu) = 1 - \mathcal{B}(\text{LQ} \rightarrow \text{t}\tau)$ for an integrated luminosity of $3000\,\text{fb}^{-1}$ in the two different scenarios. These results were obtained from the combination of the LQ $\rightarrow$ t$\mu$ and LQ $\rightarrow$ t$\tau$ analyses. For all values of $\mathcal{B}$, LQ masses up to approximately 1200 GeV and 1400 GeV are expected to be in reach for a discovery at the $5\sigma$ level and a 95% CL exclusion, respectively.

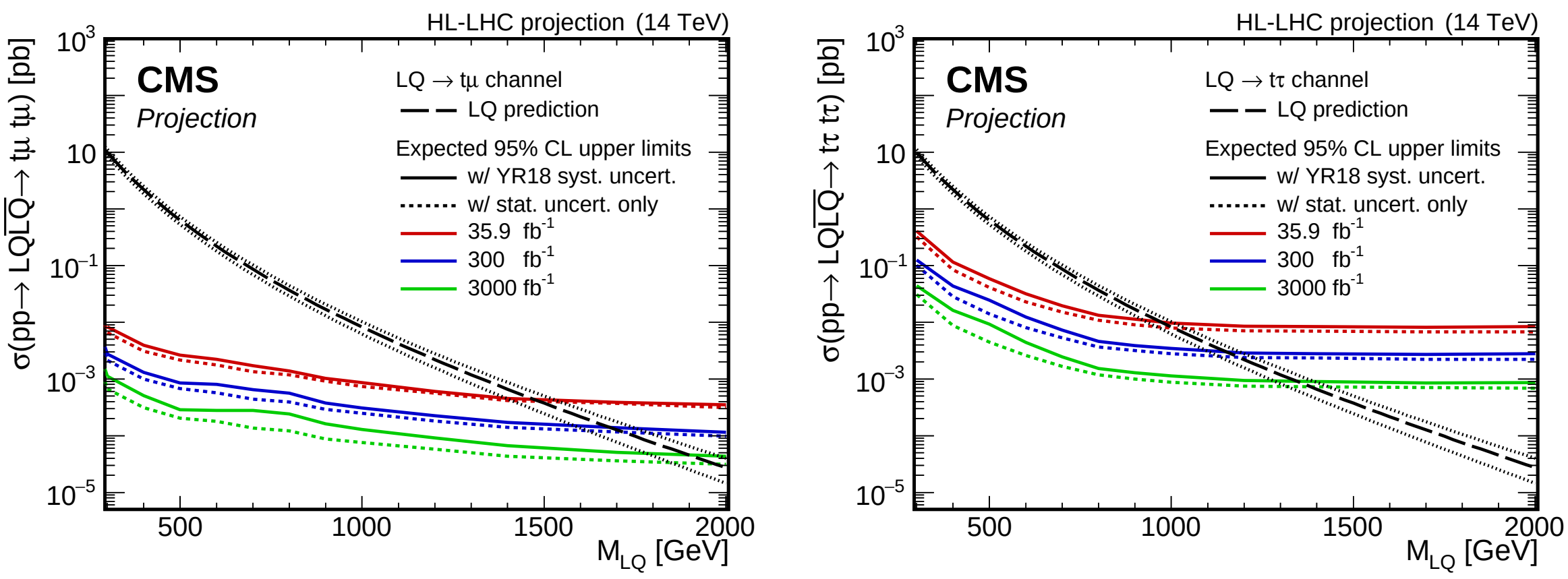

Figure 3: Expected upper limits on the LQ pair production cross section at the 95% CL for an LQ decaying exclusively to top quarks and muons (left) or $\tau$ leptons (right) as a function of the LQ mass and for different integrated luminosities in the "YR18 syst." (solid) and "stat. only" (dotted) scenarios. All results were obtained with templates for $\sqrt{s} = 13\,\text{TeV}$ that were scaled to $\sqrt{s} = 14\,\text{TeV}$. The LQ pair production cross section was calculated at NLO [35], its uncertainty takes into account PDF and scale variations.

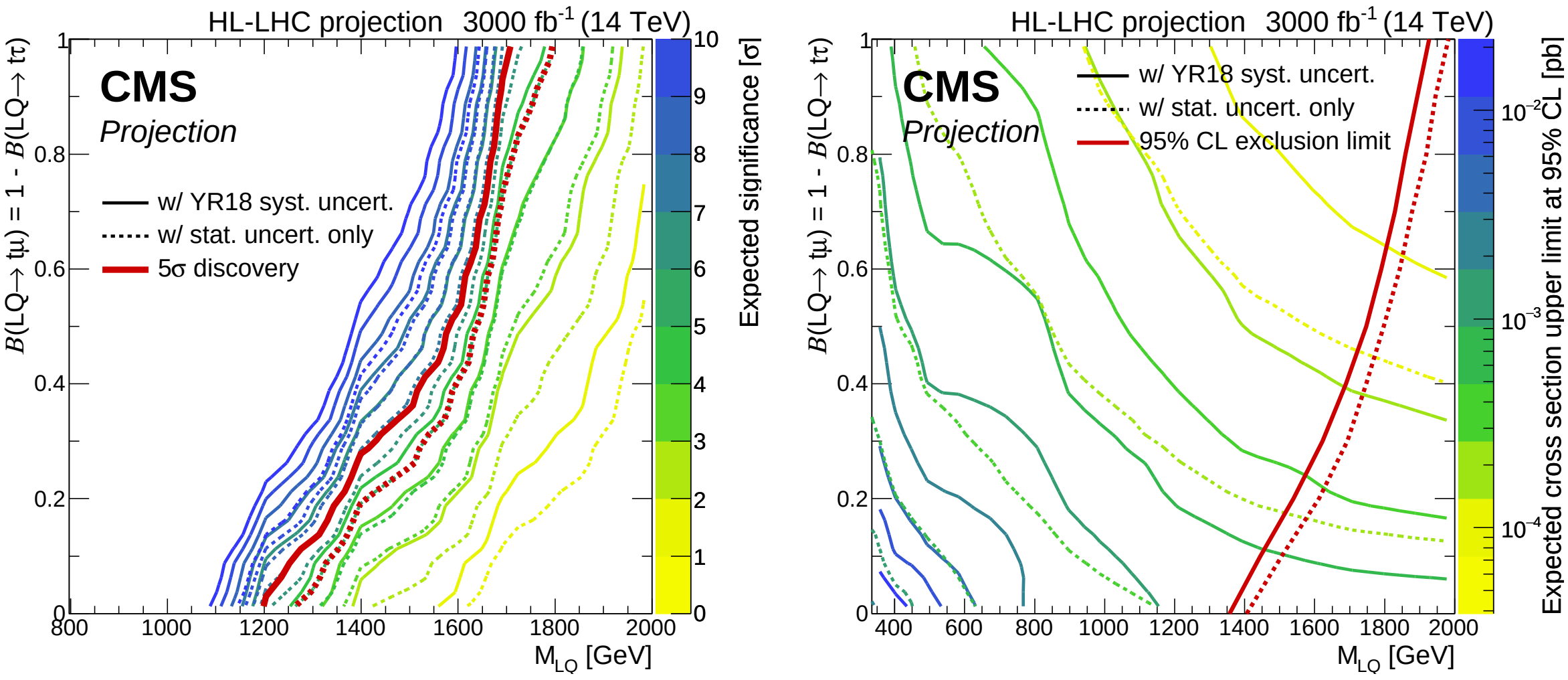

Figure 4: Expected significances (left) and expected upper limits on the LQ pair-production cross section at the 95% CL (right) as a function of the LQ mass and the branching fraction at $3000\,\text{fb}^{-1}$ in the "YR18 syst." and the "stat. only" scenarios. Color-coded lines represent lines of a constant expected significance or cross section limit, respectively. The red lines indicate the $5\sigma$ discovery level (left) and the mass exclusion limit (right).

## 5 Summary

Projections for searches for pair production of scalar leptoquarks decaying into top quarks and muons or $\tau$ leptons at the high-luminosity LHC have been presented. They are based on published analyses of the dataset recorded by the CMS experiment in 2016. The effect of an increased center-of-mass energy of $\sqrt{s} = 14\,\text{TeV}$ and the impact of reduced systematic uncertainties have been taken into account. The results of the analyses of the 2016 dataset are expected to be improved significantly with an integrated luminosity of $3000\,\text{fb}^{-1}$.

# CMS Physics Analysis Summary

Contact: cms-phys-conveners-ftr@cern.ch

# Prospects for exclusion or discovery of a third generation leptoquark decaying into a $\tau$ lepton and a b quark with the upgraded CMS detector at the HL-LHC

The CMS Collaboration

## Abstract

Projections for searches for production of third generation leptoquarks (LQs) in proton-proton collisions at a center-of-mass energy of 14 TeV are presented. The projections use simulated data samples corresponding to integrated luminosities of 300 and 3000 fb$^{-1}$. The analysis utilizes the DELPHES simulation package for the upgraded CMS detector at the High-Luminosity LHC, and considers both the single production channel, with a final state consisting of one b quark and two $\tau$ leptons, and the pair production channel, producing two b quarks and two $\tau$ leptons. In both cases, only $\tau$ leptons that decay hadronically are considered. Assuming a Yukawa coupling of one for the LQ-b-$\tau$ vertex, the expected 95% confidence level mass limit is 732 and 1130 GeV for integrated luminosities of 300 and 3000 fb$^{-1}$ for singly-produced LQs. The corresponding limits for pair production are 1249 (1518) GeV. Discovery sensitivity ($5\sigma$) is expected to be reached for masses below 800 (1000) GeV for the single production channel and 1200 (1500) GeV for the pair production channel in the 300 (3000) fb$^{-1}$ scenario. Limits are calculated both assuming negligible systematic uncertainties and utilizing ones extrapolated from searches at 13 TeV.

# 1 Introduction

Leptoquarks (LQs) are hypothetical color-triplet bosons which carry both baryon and lepton quantum numbers and have fractional electric charge. They are predicted by many extensions of the standard model (SM) of particle physics, such as theories invoking grand unification [1–8], technicolor [9–11], or compositeness [12]. To satisfy experimental constraints on flavour changing neutral currents and other rare processes [13, 14], it is generally assumed that there are three types of LQs, each type coupling only to leptons and quarks of a single generation.

Third-generation scalar LQs have recently received considerable interest from the theory community, as the existence of leptoquarks with large couplings can explain the anomaly in the $\overline{\mathrm{B}} \to \mathrm{D}\tau\overline{\nu}$ and $\overline{\mathrm{B}} \to \mathrm{D}^{*}\tau\overline{\nu}$ decay rates reported by the BaBar [15, 16], Belle [17–22], and LHCb [23] Collaborations. These decay rates collectively deviate from the SM predictions by about four standard deviations [24]. Such LQs could also provide a consistent explanation for other anomalies in B physics reported by LHCb [25–30] and Belle [31].

The production cross sections and decay widths of LQs in proton-proton (pp) collisions are determined by the LQ mass $m_{LQ}$, its branching fraction $\beta$ to a charged lepton and a quark, and the Yukawa coupling $\lambda$ of the LQ-lepton-quark vertex. Leptoquarks can be produced in pairs via gluon fusion or quark-antiquark annihilation, and singly via quark-gluon fusion. The LQ pair production cross section does not depend on $\lambda$, while that for single production does, and thus the sensitivity of searches for singly produced LQs depends on $\lambda$. For $\lambda = 1$, at lower masses, the cross section for pair production is greater than that of single production. However, the single-LQ production cross section decreases more slowly with increasing $m_{LQ}$, eventually exceeding that of pair production. For a third-generation LQ to explain the observed B physics anomalies, $\lambda$ has to be large ($\lambda \sim m_{LQ}$ measured in TeV). For such $\lambda$, the single production cross section is larger than the pair production cross section when $m_{LQ}$ is greater than 1.0-1.5 TeV [32]. Feynman diagrams of the signal processes at leading order (LO) are shown in Fig. 1.

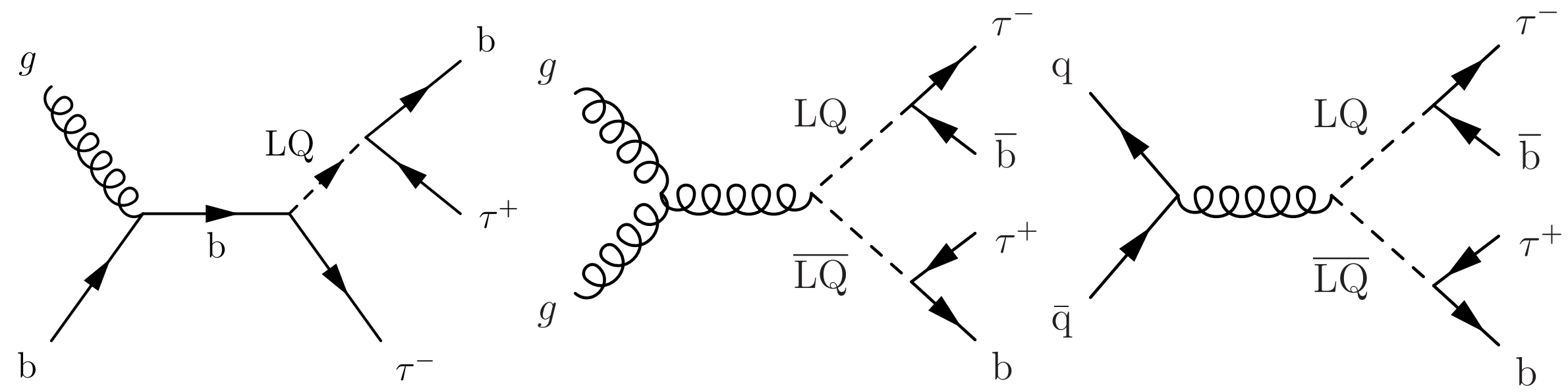

Figure 1: Leading order Feynman diagrams for the production of a third-generation LQ in the single production s-channel (left) and the pair production channel via gluon fusion (center) and quark fusion (right).

The most stringent limits on the production cross section of a third-generation LQ decaying to a $\tau$ lepton and a bottom quark come from a search by the CMS Collaboration, in which a scalar LQ with mass below 1 TeV is excluded at 95% confidence level (CL) in a search for LQ pair production in the $\tau\tau$bb final state [33]. Limits on the LQ mass are set at 740 GeV for the single production channel [34]. Another type of third-generation scalar LQ decaying to a $\tau$ lepton and a top quark is excluded by the CMS Collaboration for masses up to 900 GeV [35].

The analysis described in this document calculates the future discovery and exclusion prospects for singly and pair produced third-generation scalar LQs, each decaying to a hadronically decaying $\tau$ (65% of the $\tau$ decays [36]), denoted as $\tau_h$, and a bottom quark. Signal is separated from background using the distribution of the scalar sum of the transverse momenta of jets and taus in the $\tau\tau$b and $\tau\tau$bb final states. The analysis uses event samples of simulated pp collisions at a center-of-mass energy of 14 TeV, corresponding to integrated luminosities of 300 and 3000 fb$^{-1}$. Additional pp interactions (pileup or PU) within the same or adjacent bunch crossings are included, with an average of 200 interactions per event.

## 2 The upgraded CMS detector

The CMS detector [37] will be substantially upgraded in order to fully exploit the physics potential offered by the increase in luminosity [38], and to cope with the demanding operational conditions at the HL-LHC [39–43]. The upgrade of the first level hardware trigger (L1) will allow for an increase of L1 rate and latency to about 750 kHz and 12.5 $\mu$s, respectively, and the output rate of the high-level software trigger (HLT) will also be increased to 7.5 kHz. The entire pixel and strip tracker detectors will be replaced to increase the granularity, reduce the material budget in the tracking volume, improve the radiation hardness, and extend the geometrical coverage and provide efficient tracking up to pseudorapidities of about $|\eta| = 4$. The muon system will be enhanced by upgrading the electronics of the existing cathode strip chambers (CSC), resistive plate chambers (RPC) and drift tubes (DT). New muon detectors based on improved RPC and gas electron multiplier (GEM) technologies will be installed to add redundancy, increase the geometrical coverage up to about $|\eta| = 2.8$, and improve the trigger and reconstruction performance in the forward region.

The barrel electromagnetic calorimeter (ECAL) will feature upgraded front-end electronics that will be able to exploit the information from single crystals at the L1 trigger level, to accommodate trigger latency and bandwidth requirements, and to provide 160 MHz sampling allowing high precision timing capability for photons. The hadronic calorimeter (HCAL), consisting in the barrel region of brass absorber plates and plastic scintillator layers, will be read out by silicon photomultipliers (SiPMs). The endcap electromagnetic and hadron calorimeters will be replaced with a new combined sampling calorimeter (HGCal) that will provide highly-segmented spatial information in both transverse and longitudinal directions, as well as high-precision timing information. Finally, the addition of a new timing detector for minimum ionizing particles (MTD) in both barrel and endcap region is envisaged to provide capability for 4-dimensional reconstruction of interaction vertices that will allow to significantly offset the CMS performance degradation due to high PU rates.

A detailed overview of the CMS detector upgrade program is presented in Ref. [39–43], while the expected performance of the reconstruction algorithms and pileup mitigation with the CMS detector is summarized in Ref. [44].

## 3 Simulated samples

The LQ signals for both single and pair LQ production are generated at leading order (LO) using version 2.6.0 of MADGRAPH5_aMC@NLO [45] for $m_{LQ}$ =500, 1000, 1500, and 2000 GeV. The branching fraction is assumed to be $\beta = 1$, i.e. the LQ always decays to a $\tau$ lepton and a bottom quark. The Yukawa coupling of the LQ to a $\tau$ lepton and a bottom quark is set to $\lambda = 1$. The width $\Gamma$ is calculated using $\Gamma = m_{LQ}\lambda^2/(16\pi)$ [46], and is less than 10% of the LQ mass for most of the considered search range. The signal samples are normalized to the cross section calculated at LO, multiplied by a $K$ factor to account for higher order contributions [47].

The main background arises from pair production of top quarks ($\mathrm{t\bar{t}}$). Other significant contributions to the background include Drell-Yan (DY) production, multijet production via quantum chromodynamics (QCD), single top quark production, Z or W boson+jets and diboson production (WW, WZ, ZZ). The generated signal and background events are processed with the fast-simulation package DELPHES [48] in order to simulate the expected response of the upgraded CMS detector.

# 4 Event selection

Similar event selections are used in both the singly and pair produced LQ searches, except for the requirement on the number of jets. In the search for single production, the presence of at least one reconstructed jet is required, while at least two are required in the search for pair production. Jets are reconstructed using FASTJET [49], using the anti-$k_T$ algorithm [50], with a distance parameter of 0.4.

Since no precise $\tau_h$ identification efficiency or hadron misidentification rate is predefined in the DELPHES software package, a parameterization is used to emulate the $\tau$ identification efficiency. Jets that match generator-level $\tau$ are selected as reconstructed taus with an efficiency of 61%, independent of jet $p_T$. Reconstructed jets that do not match generator-level $\tau$ can be misidentified as $\tau$ jets, with a misidentification probability that depends on the jet $p_T$. The average misidentification rate is 1%, ranging from 1.9% for a $p_T$ of 50 GeV to 0.5% for a $p_T$ of 150 GeV.

In both channels, two reconstructed $\tau$ leptons with opposite sign are required, each with transverse momentum $p_{T,\tau} > 50$ GeV and a maximum pseudorapidity $|\eta_\tau| < 2.3$. We utilize reconstructed jets with $p_{T,\text{jet}} > 50$ GeV and $|\eta_{\text{jet}}| < 2.4$. We require at least one such jet for the $\tau\tau$b channel and 2 for the $\tau\tau$bb channel. These jets are required to be neither generator-level $\tau$ nor jets misidentified as $\tau$.

To reduce background due to Drell-Yan (particularly Z$\to \tau\tau$) events, the di-$\tau$ invariant mass of the two $\tau$ leptons $m_{\tau\tau}$ is required to be $> 95$ GeV.

We require that at least one of the previously selected jets is b-tagged, with $p_T > 50$ GeV and not considering jets labelled as $\tau$ as eligible, neither if they come from generator-level $\tau$ leptons or if they are misidentified jets. Finally, we reject any events with identified and isolated electrons (muons), with $p_T > 10$ GeV and $|\eta| < 2.4$ (2.5).

After applying these selections, and considering the branching ratio for a $\tau$ lepton to decay hadronically, a signal efficiency of 4.9% (11%) is obtained for the single (pair) production.

# 5 Systematic uncertainties

The systematic uncertainties considered in the study are summarized in Table 1. They are calculated by scaling the current experimental uncertainties. For uncertainties limited by statistics, including the uncertainty on the DY and QCD cross sections, a scale factor of $1/\sqrt{L/35.9}$ is applied, where $L$ is the integrated luminosity in fb$^{-1}$. For uncertainties coming from theoretical calculations, a scale factor of 1/2 is applied with respect to current uncertainties, as is the case for the uncertainties on the cross sections for top or diboson events.

Other experimental systematic uncertainties are scaled by the square root of the integrated luminosity until the uncertainty reaches a minimum value based on estimates of the achievable accuracy with the upgraded detector [44]. Uncertainties on the integrated luminosity, $\tau$ identification and b tagging/misidentification are examples of experimental systematic uncertainties, which are expected to reach the minimum value for both luminosity scenarios. The uncertainty of 5% on the $\tau$ identification efficiency arises from the sum of the uncertainties of each of the two $\tau$ leptons considered in the selection.

| Uncertainty | | LQ | $t\bar{t}$ - single top | DY | QCD | Diboson |
|---|---|---|---|---|---|---|
| Integrated luminosity | | 1% | | | | |
| $\tau$ identification | | 5% | | | | |
| b tagging | | 1% | | - | - | |
| b misidentification | | - | - | 5% | | |
| $\sigma_{top}$ | | - | 2.75% | - | - | - |
| $\sigma_{DY}$ | 300 $fb^{-1}$ | - | - | 10.4% | - | - |
| | 3000 $fb^{-1}$ | - | - | 3.3% | - | - |
| $\sigma_{QCD}$ | 300 $fb^{-1}$ | - | - | - | 10.4% | - |
| | 3000 $fb^{-1}$ | - | - | - | 3.3% | - |
| $\sigma_{diboson}$ | | - | - | - | - | 3% |

Table 1: Summary of the main systematic uncertainties, where $\sigma_{bkg}$ represents the uncertainty in the cross section of the background bkg. Uncertainty in b misidentification refers to the tagging of light jets as b jets.

# 6 Results

Signal extraction is based on a binned maximum likelihood fit to the distribution of the scalar $p_T$ sum $S_T$. This variable is defined as the sum of the transverse momenta of the two selected $\tau$ leptons and either the highest-$p_T$ jet in the case of single LQ production, or the two highest-$p_T$ jets in the case of LQ pair production. The two versions of this variable are shown in Fig. 2, for the HL-LHC 3000 $fb^{-1}$ scenario.

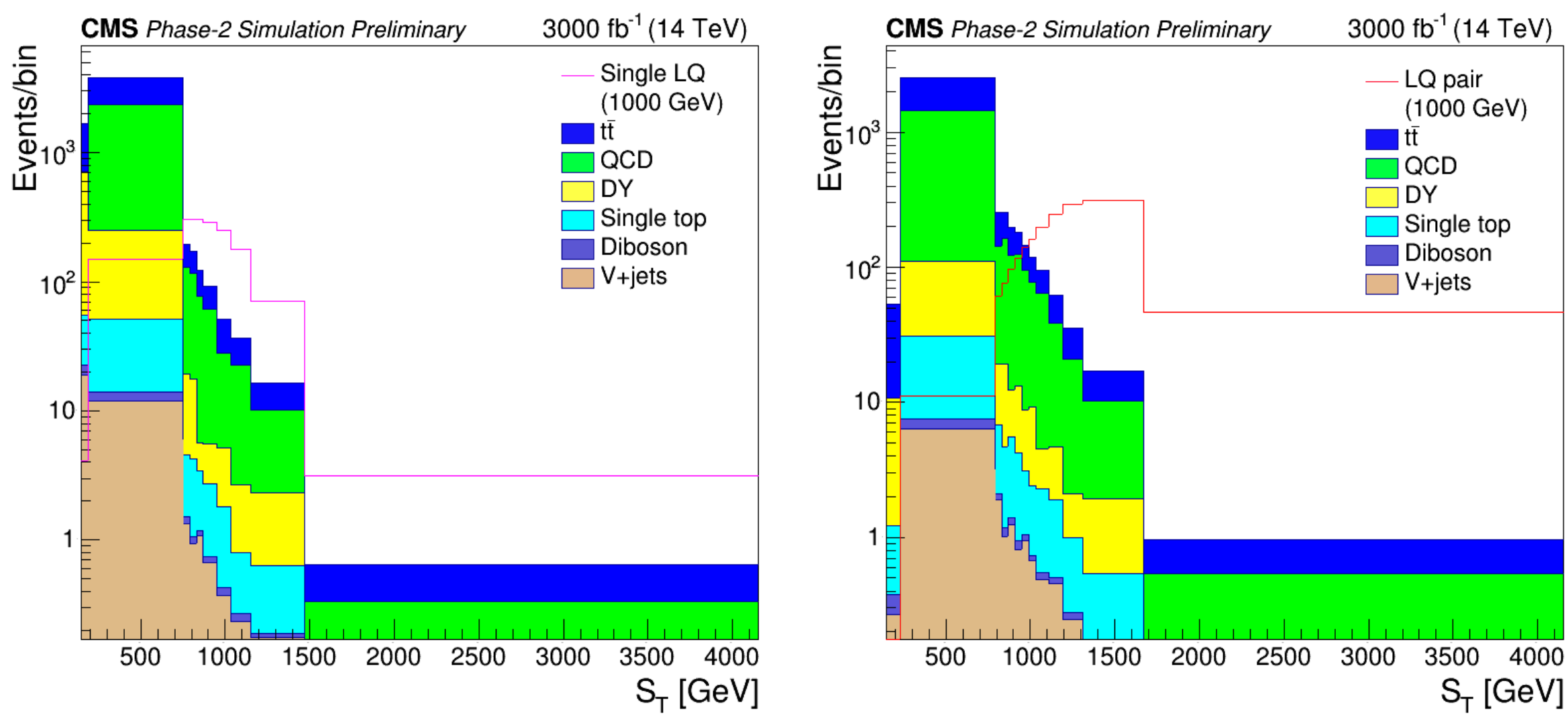

Figure 2: (left) Scalar sum of the $p_T$ of the two selected $\tau$ leptons and the highest-$p_T$ jet in the single LQ selected sample. (right) Scalar sum of the $p_T$ of the two selected $\tau$ leptons and the two highest-$p_T$ jets in the LQ pair selected sample. The considered backgrounds are shown as stacked histograms, while empty histograms for signals for the single LQ and LQ pair channels (for $m_{LQ} = 1000$ GeV) are overlaid to illustrate the sensitivity. Both signal and backround are normalized to a luminosity of 3000 $fb^{-1}$.

The uncertainties described in Table 1 are represented by nuisance parameters in the fit. We set an upper limit at 95% confidence level (CL) on the cross section times branching fraction $\beta$ as a function of $m_{LQ}$ by using the asymptotic $CL_s$ modified frequentist criterion [51–54]. Upper limits are calculated considering two different scenarios. The first one, hereafter abbreviated as ”stat. only”, considers only statistical uncertainties, to observe how the results are affected by the increase of the integrated luminosity. The second scenario, hereafter abbreviated as ”stat. + syst.”, also includes the best estimate of the systematic uncertainties at the HL-LHC, as defined in Table 1. Two projections are calculated, one for an integrated luminosity of 300 $fb^{-1}$ (Run 3) and another one for 3000 $fb^{-1}$ (HL-LHC). The limits are shown in Fig. 3 for both single LQ (left) and LQ pair production (right) channels, together with the theoretical prediction for the cross section [47] assuming $\lambda = 1$ and $\beta = 1$, for both the stat. only and stat. + syst. scenarios.

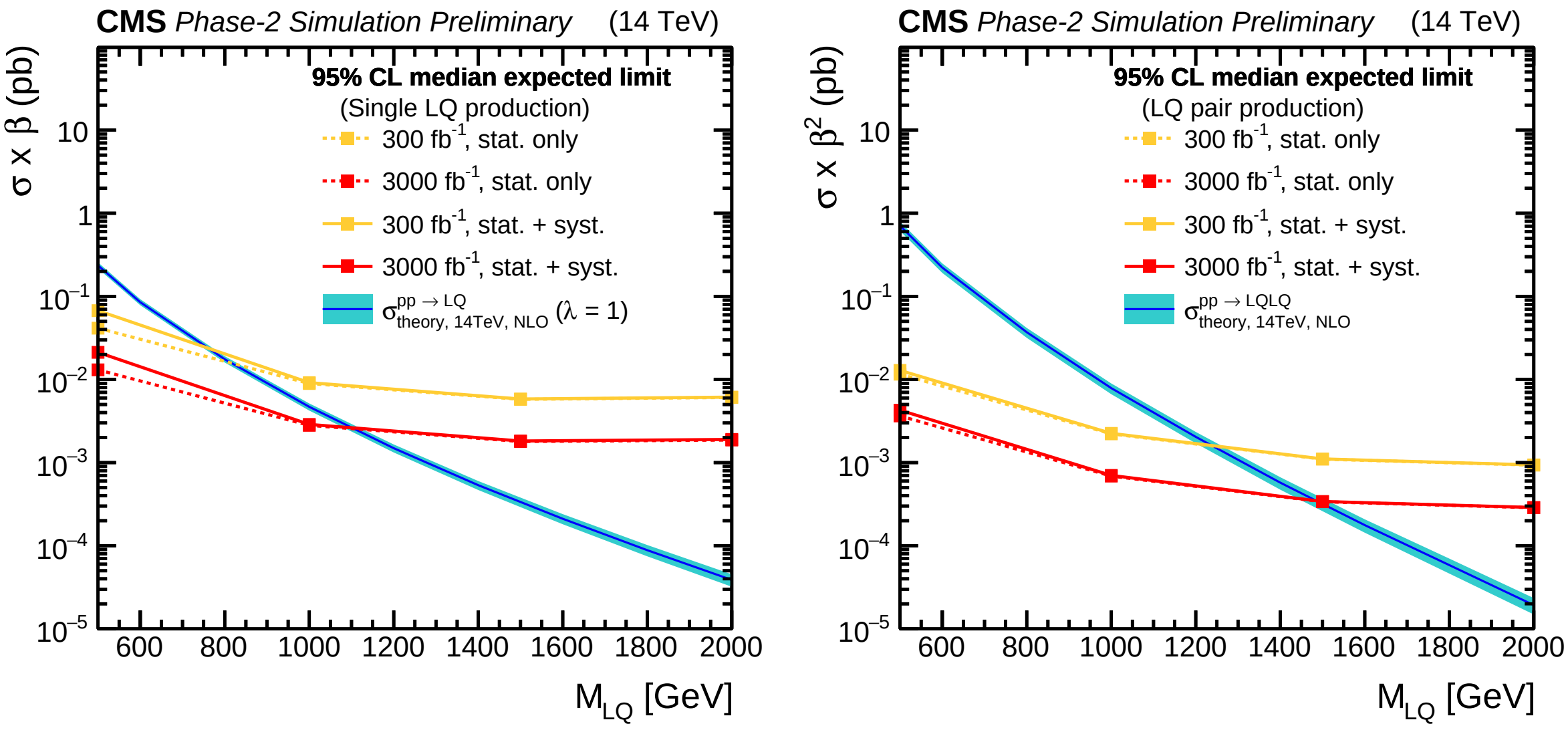

Figure 3: Expected limits at 95% CL on the product of the cross section $\sigma$ and the branching fraction $\beta$ for the single (left) and pair (right) LQ production channels. Note that, in the case of pair LQ production, the limit is calculated for $\sigma \times \beta^2$. Limits are calculated as a function of the LQ mass, for the two high luminosity projections, 300 $fb^{-1}$ (red) and 3000 $fb^{-1}$ (orange), for both the stat. only (dashed lines) and the stat. +syst. scenarios (solid lines). This is shown in conjunction with the theoretical predictions at NLO [47], in cyan.

Comparing these expected limits with the theoretical predictions, projected limits on the LQ mass are calculated, as shown in Table 2.

| Production channel | Uncertainty scenario | LQ mass [GeV] | |
|---|---|---|---|
| | | 300 $fb^{-1}$ | 3000 $fb^{-1}$ |
| single LQ | stat. only | 784 | 1135 |
| | stat. + syst. | 732 | 1130 |
| LQ pair | stat. only | 1253 | 1520 |
| | stat. + syst. | 1249 | 1518 |

Table 2: Lower limits on the LQ mass for each considered production mechanism, uncertainty scenario, and integrated luminosity hypothesis considered in the analysis.

Since the single-LQ signal cross section scales with $\lambda^2$, it is straightforward to recast the results presented in Fig. 3 in terms of expected upper limits on $\lambda$ as a function of $m_{LQ}$, as shown in Fig. 4. Values of $\lambda$ up to 3 are considered, so that the width of the LQ signal stays narrow compared to detector resulution and constraints from electroweak precision measurements are satisfied [55]. We have used the assumption that the shape of the $S_T$ distribution does not change as a function of $\lambda$, which has been verified based on the simulation for the $\lambda$ range used in this analysis [34]. The blue band shows the parameter space (95% CL) for the scalar LQ preferred by the B physics anomalies: $\lambda = (0.95 \pm 0.50) m_{LQ}(\text{TeV})$ [32]. The pair production limits are clearly stronger than those of the single production channel. Results under the two integrated luminosity scenarios are compatible with the latest third generation LQ searches in CMS for both the single [34] and pair production [35]. For the luminosity scenario of 300 (3000) fb$^{-1}$, the leptoquark pair production channel is more sensitive if $\lambda < 2.7$ $(2.3)$, while the single leptoquark production dominates otherwise.

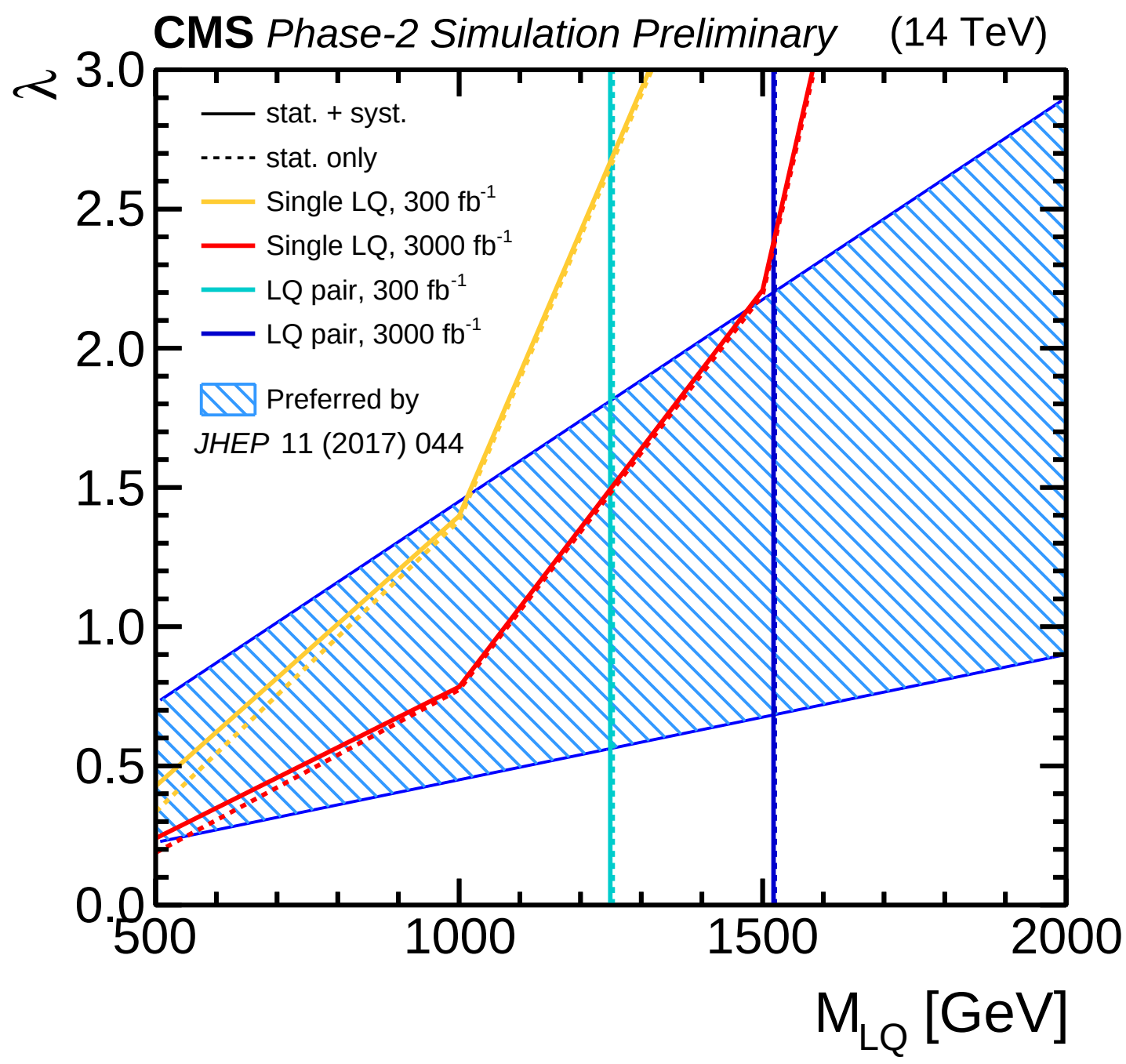

Figure 4: Expected exclusion limits at 95% CL on the Yukawa coupling $\lambda$ at the LQ-lepton-quark vertex, as a function of the LQ mass. A unit branching fraction $\beta$ of the LQ to a $\tau$ lepton and a bottom quark is assumed. Future projections for 300 and 3000 fb$^{-1}$ are shown in red and blue respectively, for both the stat. only and stat. + syst. scenarios, shown as dashed and filled lines respectively, and for both the single LQ and LQ pair production, where the latter corresponds to the vertical line (since it does not depend on $\lambda$). The left hand side of the lines represents the exclusion region for each of the projections, whereas the region with diagonal blue hatching shows the parameter space preferred by one of the models proposed to explain anomalies observed in B physics [32].

Using the predicted cross section [47] of the signal, it is possible to estimate the maximal LQ mass expected to be in reach for a $5\sigma$ discovery sensitivity for both high luminosity scenarios. Figure 5 shows the expected local significance of a signal-like excess as a function of the LQ mass hypothesis. Discovery significance of $5\sigma$ is projected for LQ masses below 800 (1000) GeV for the single production channels and 1200 (1500) GeV for the pair production channel in the 300 (3000) fb$^{-1}$ scenario. However, given the limited number of mass points, limits on the LQ mass must be regarded as an estimation, as the extrapolation between each mass point is not precise. The discovery sensitivity is found to be approximately the same if the stat. only or the stat. + syst. uncertainty scenarios are studied, with the exception of the $5\sigma$ significance for the single production channel for the 300 fb$^{-1}$ scenario, found to be at 900 GeV.

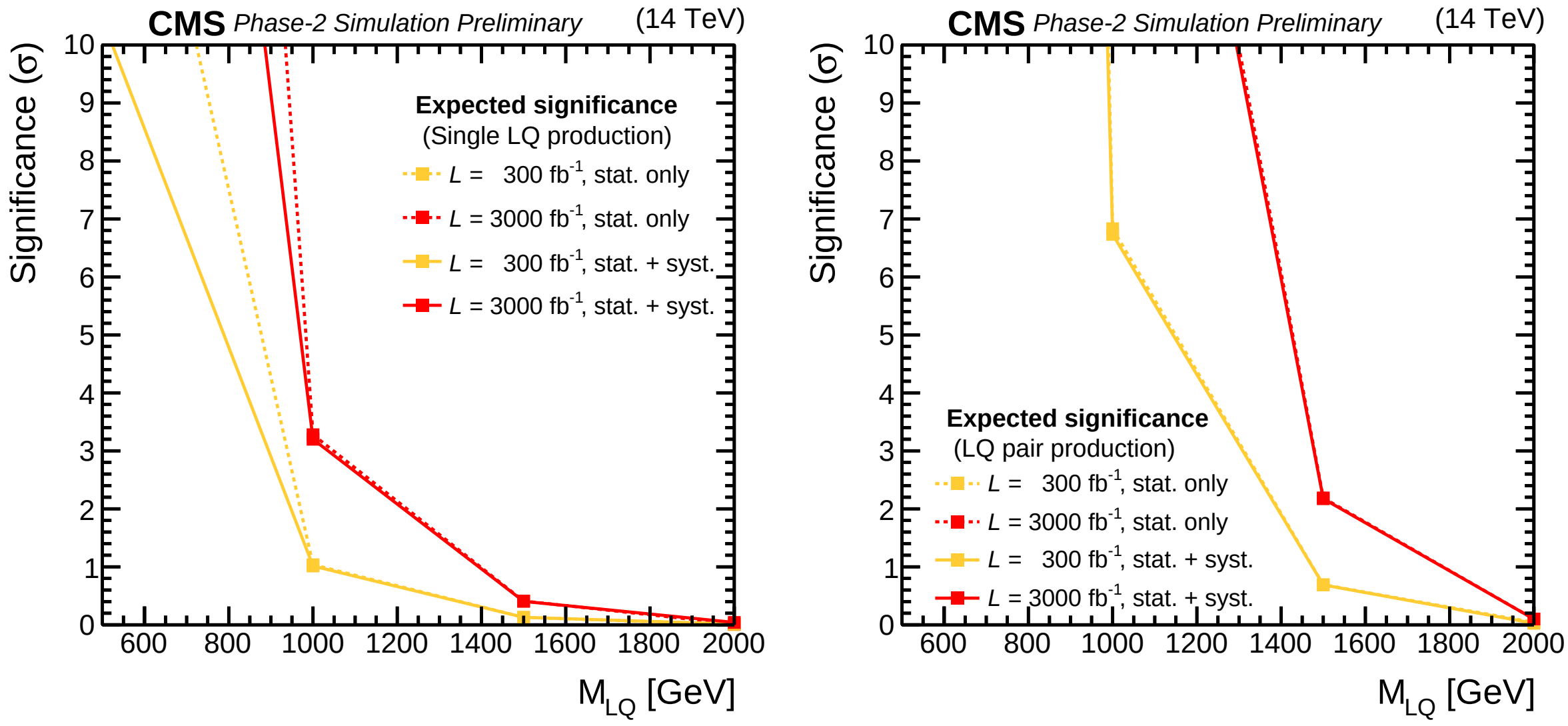

Figure 5: Expected local significance of a signal-like excess as a function of the LQ mass, for the two high luminosity projections, 300 fb$^{-1}$ (red) and 3000 fb$^{-1}$ (orange), assuming the theoretical prediction for the LQ cross section at NLO [47], calculated with $\lambda = 1$ and $\beta = 1$. Projections are calculated for both single LQ (left) and LQ pair production (right).

# 7 Summary

Expected limits on the cross section for singly and pair produced third-generation scalar leptoquarks (LQ), each of which decays to a $\tau$ lepton and a bottom quark, have been presented as a function of the LQ mass. Projections have been made using DELPHES simulated samples at 14 TeV, for two luminosity scenarios at 300 fb$^{-1}$ and 3000 fb$^{-1}$.

Comparing the limits with theoretical predictions assuming unit Yukawa coupling $\lambda = 1$, third-generation scalar leptoquarks are expected to be excluded at 95% confidence level for LQ masses below 732 and 1130 GeV for the single LQ production channel for the 300 and 3000 fb$^{-1}$ scenarios, considering both statistical and systematic uncertainties. The corresponding limits for LQ pair production are 1249 GeV and 1518 GeV.

Limits on $\lambda$ are also placed as a function of the leptoquark mass. For the 300 (3000) fb$^{-1}$ luminosity scenario, the leptoquark pair production channel is more sensitive if $\lambda < 2.7$ (2.3), while the single leptoquark production dominant otherwise. These results show that future LQ searches under higher luminosity conditions are promising, as they are expected to greatly increase the reach of the search. They also show that the pair production channel is expected to be the most sensitive. A significance of $5\sigma$ is projected for LQ masses below 800 (1000) GeV for the single production channels and 1200 (1500) GeV for the pair production channel in the 300 (3000) fb$^{-1}$ scenario.

**8**

# CMS Physics Analysis Summary

# Sensitivity study for a heavy gauge boson W′ in the decay channel with a tau lepton and a neutrino at the High-Luminosity LHC

The CMS Collaboration

## Abstract

A sensitivity study for the discovery or exclusion of a heavy vector boson W′ in the final state with a tau lepton and a neutrino is presented. Event samples are simulated for the Phase-2 CMS detector at the High-Luminosity LHC (corresponding to an integrated luminosity of 3 $ab^{-1}$), using the parameterized detector simulation program DELPHES. A signal would appear as an excess of events with high transverse mass of the hadronic tau and missing transverse momentum, compared to the standard model background. With the high integrated luminosity during Phase-2, a W′ boson with SM-like couplings could be observed with a significance exceeding five standard deviations with a mass up to 6.0 TeV. In case of no observation, the results are interpreted as lower limits on the mass of the W′ boson in the context of the sequential standard model. In addition, variations in the coupling strength are studied, and a model-independent cross section limit is provided.

# 1 Introduction

New heavy gauge bosons are predicted by various extensions of the standard model (SM). The charged version of such heavy gauge bosons is generally referred to as W′. This note describes a sensitivity study for a W′ boson decaying to a tau lepton ($\tau$) and a neutrino ($\nu_\tau$) at the High-Luminosity LHC (HL-LHC) [1] with 3000 fb$^{-1}$ of expected data at a proton-proton (pp) center-of-mass energy of 14 TeV. The performance of the upgraded Phase-2 CMS detector is simulated with DELPHES [2] following the recently established performance parameters summarised in Ref. [3].

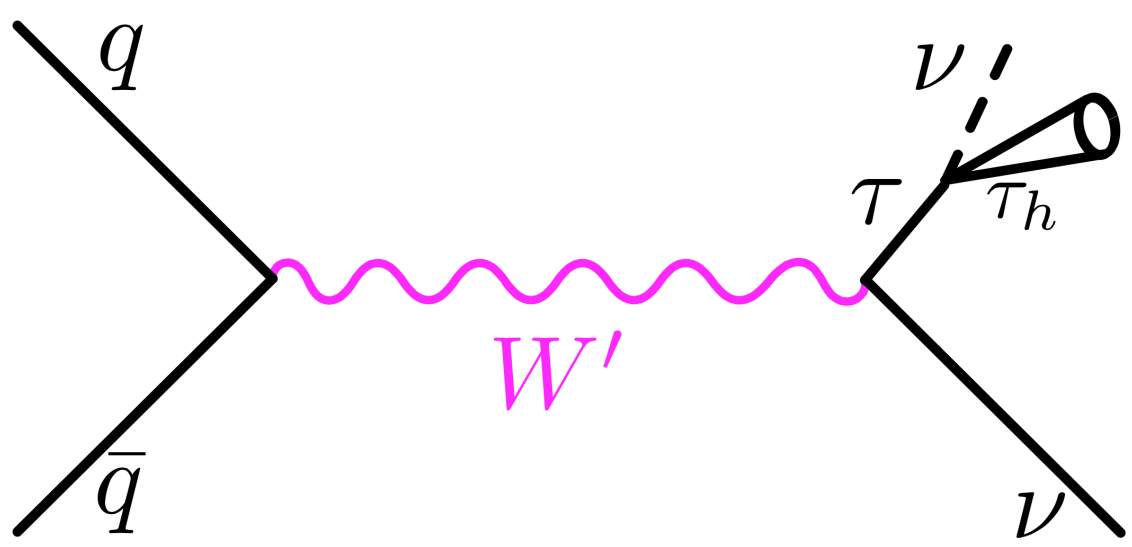

Figure 1: Illustration of the production and decay of the W′ boson with the subsequent hadronic decay of tau ($\tau_h$).

The signature of a W′ boson is similar to a high mass W boson, yielding in the decay $W \to \tau\nu_\tau$ a single tau lepton, of which we consider the hadronic decay ($\tau_h$), and missing energy due to the neutrinos. The hadronic decay of the tau lepton gives rise to tau-jets, which are experimentally distinctive because of their low charged hadron multiplicity, unlike quantum chromodynamics (QCD) multi-jets, which have high charged hadron multiplicity, or other leptonic W′ boson decays, which yield no jet in the decay.

This Phase-2 study follows closely the recently published Run 2 result [4], which used proton-proton collision data collected by the CMS experiment at the LHC at a center-of-mass energy $\sqrt{s} = 13$ TeV, corresponding to an integrated luminosity of 35.9 fb$^{-1}$. The results are interpreted in the context of the sequential standard model (SSM) [5]. In addition, variations in the coupling strength are studied, and a model-independent cross section limit is provided.

# 2 The upgraded CMS detector

The CMS detector [6] will be substantially upgraded in order to fully exploit the physics potential offered by the increase in luminosity at the HL-LHC, and to cope with the demanding operational conditions at the HL-LHC [7–11]. The upgrade of the first level hardware trigger (L1) will allow for an increase of L1 rate and latency to about 750 kHz and 12.5 $\mu$s, respectively, and the high-level software trigger is expected to reduce the rate by about a factor of 100 to 7.5 kHz. The entire pixel and strip tracker detectors will be replaced to increase the granularity, reduce the material budget in the tracking volume, improve the radiation hardness, and extend the geometrical coverage and provide efficient tracking up to pseudorapidities of about $|\eta| = 4$. The muon system will be enhanced by upgrading the electronics of the existing cathode strip chambers, resistive plate chambers (RPC), and drift tubes. New muon detectors based on improved RPC and gas electron multiplier technologies will be installed to add redundancy, increase the geometrical coverage up to about $|\eta| = 2.8$, and improve the trigger and reconstruction performance in the forward region. The barrel electromagnetic calorimeter will feature the upgraded front-end electronics that will be able to exploit the information from single crystals in the L1 trigger system, to accommodate trigger latency and bandwidth requirements, and to provide 160 MHz sampling allowing high precision timing capability for photons. The hadronic calorimeter, consisting in the barrel region of brass absorber plates and plastic scintillator layers, will be read out by silicon photomultipliers. The endcap electromagnetic and hadron calorimeters will be replaced with a new combined sampling calorimeter that will provide highly-segmented spatial information in both transverse and longitudinal directions, as well as high-precision timing information. Finally, the addition of a new timing detector for minimum ionizing particles in both barrel and endcap region is envisaged to provide capability for 4-dimensional reconstruction of interaction vertices that will allow to significantly offset the CMS performance degradation due to the large number of pp interactions per bunch crossing (pileup, PU).

A detailed overview of the CMS detector upgrade program is presented in Refs. [7–11], while the expected performance of the reconstruction algorithms and pileup mitigation with the CMS detector is summarised in Ref. [3].

# 3 Physics model and signal simulation

The presence of a W′ boson signal over the SM background could be observed in the distribution of the transverse mass ($m_T$) of the transverse momentum of the $\tau_h$ ($p_T^\tau$) and the missing transverse momentum ($p_{\mathrm{T}}^{\mathrm{miss}}$):

$$m_T = \sqrt{2 p_T^\tau p_{\mathrm{T}}^{\mathrm{miss}} (1 - \cos \Delta\phi(\vec{p}_{\mathrm{T}}^{\tau}, \vec{p}_{\mathrm{T}}^{\mathrm{miss}}))}. \quad (1)$$

Unlike the leptonic search channels, the signal shape of W′ boson with hadronically decaying tau leptons does not show a Jacobian peak structure, because of the presence of two neutrinos in the final state. Despite the multi-particle final state, the decay appears as a typical two-body one. The axis of the hadronic tau jet is back to back with $\vec{p}_{\mathrm{T}}^{\mathrm{miss}}$ and the magnitude of both is comparable such that their ratio is about unity.

The SSM is a benchmark model used as a reference point for experimental searches of W′ bosons for more than two decades. In the SSM, the W′ boson, as shown in Fig. 1, is considered to be a heavy analogue of the SM W boson, with similar decay modes and branching fractions. These are modified by the presence of the $\mathrm{t\bar{b}}$ decay channel, which is accessible for W′ boson masses above 180 GeV. The resulting branching fraction for the tau channel is 8.5%, and the width of a 1 TeV W′ boson would be about 33 GeV.

The SSM W′ signal was simulated with MADGRAPH5_aMC@NLO [12] at leading order (LO) and hadronized using Pythia 8.212 [13] with the underlying event tune CUETP8M1 [14]. The detector simulation was performed with DELPHES Version 3.4.1. Samples for eight values of the W′ boson mass were simulated at intervals of 1 TeV, ranging from masses of 1 TeV up to 8 TeV with a coupling as suggested by the SSM.

In addition, a range of weaker couplings was also simulated and studied. The W′ boson coupling strength, $g_{\mathrm{W}'}$, is given in terms of the SM weak coupling strength $g_{\mathrm{W}} = e/\sin^2\theta_{\mathrm{W}} \approx 0.65$. Here, $\theta_{\mathrm{W}}$ is the weak mixing angle. If the W′ boson is a heavier copy of the SM W boson, its coupling ratio is $g_{\mathrm{W}'}/g_{\mathrm{W}} = 1$ and the SSM W′ boson theoretical cross sections, signal shapes, and widths apply. However, different couplings are possible. Because of the dependence of the width of a particle on its coupling, and the consequent effect on the $m_T$ distribution, a limit can also be set on the coupling strength. Samples for a range of values for the ratio of the couplings $g_{W'}/g_W$ from 0.01 to 3 were simulated with MADGRAPH. These signals exhibit different widths as well as different cross sections. They were reweighted to take into account the decay width dependence, thus providing the appropriate reconstructed $m_T$ distributions for $g_{W'}/g_W \neq 1$.

# 4 Background simulation

The dominant background appears in the off-shell tail of the $m_T$ distribution of the SM W boson. This background is generated at LO using MADGRAPH5_aMC@NLO including a dedicated sample of high mass ($m(\tau+\nu) > 400$ GeV) events to sufficiently model the background in the signal region. Subleading background contributions arise from $t\bar{t}$ and QCD multijet events. The number of background events are reduced by the event selection. These backgrounds primarily arise as a consequence of jets misidentified as $\tau_h$ candidates and populate the lower transverse masses while the signal exhibits an excess of events at high $m_T$. Multijet (QCD) background is simulated with PYTHIA in nine bins of $p_T$ ranging from 50 GeV to infinity. Other background processes are: $Z/\gamma* \to \ell\ell$ generated with MADGRAPH5_aMC@NLO, diboson processes generated with PYTHIA 8.212, and top quark processes generated with POWHEG 2.0 [15–20] and MADGRAPH5_aMC@NLO.

As for the signal, the detector performance is simulated with DELPHES. For all the processes, parton fragmentation and hadronization are performed with PYTHIA 8.212 and the underlying event tune CUETP8M1. All simulated event samples are normalized to the expected luminosity of 3000 $\text{fb}^{-1}$, using the theoretical cross section values. Additional pp collisions during the same bunch crossing (pileup) are taken into account by superimposing simulated minimum bias interactions onto all events. The average pileup value at the HL-LHC is expected to be 200.

# 5 Object reconstruction and event selection

The strategy of this analysis is to select a heavy charged boson candidate decaying almost at rest to a hadronic jet consistent with a $\tau_h$ candidate and neutrinos, the latter manifesting themselves as $p_T^{miss}$. Hadronically decaying tau leptons are selected since the corresponding branching fraction, about 60%, is the largest among all $\tau$ lepton decays.

Since no precise $\tau_h$ identification efficiency or hadron misidentification rate is predefined in DELPHES, a parameterization is used to emulate the $\tau_h$ identification efficiency. In a first step, jets are reconstructed using the anti-$k_T$ algorithm [21, 22], with a distance parameter of 0.4. This study uses PUPPI jets [23] with $p_T > 30$ GeV and $|\eta| < 2.7$. Jets matching generator-level hadronically decaying tau leptons are selected with an efficiency of the medium working point with 45% (independent of $p_T$). This working point has been tuned such that the misidentification rate is about a factor two within the Run 2 values [24]. The fraction of jets misidentified as $\tau_h$ is $p_T$ dependent, for example 0.16% for $p_T = 100$ GeV and 0.1% for $p_T > 190$ GeV, nearly independent of $\eta$. If, after these selections, zero or more than one $\tau_h$ candidates are found in the event, the event is discarded. To avoid overlaps with possible W$'$ boson searches in the electron or muon channel, events are rejected if they contain a loosely identified electron or muon.

The following event selection is identical to the Run 2 analysis [4]. Events with one hadronically decaying $\tau$ lepton and $p_T^{miss}$ are selected if the ratio of $p_T^{\tau}$ to $p_T^{miss}$ satisfies $0.7 < p_T^{\tau}/p_T^{miss} < 1.3$ and the angle $\Delta\phi(\vec{p}_T^{\,\tau}, \vec{p}_T^{\,miss})$ is greater than 2.4 radians. The key distributions and achievable sensitivity of the analysis were compared to the Run 2 performance and found to be comparable.

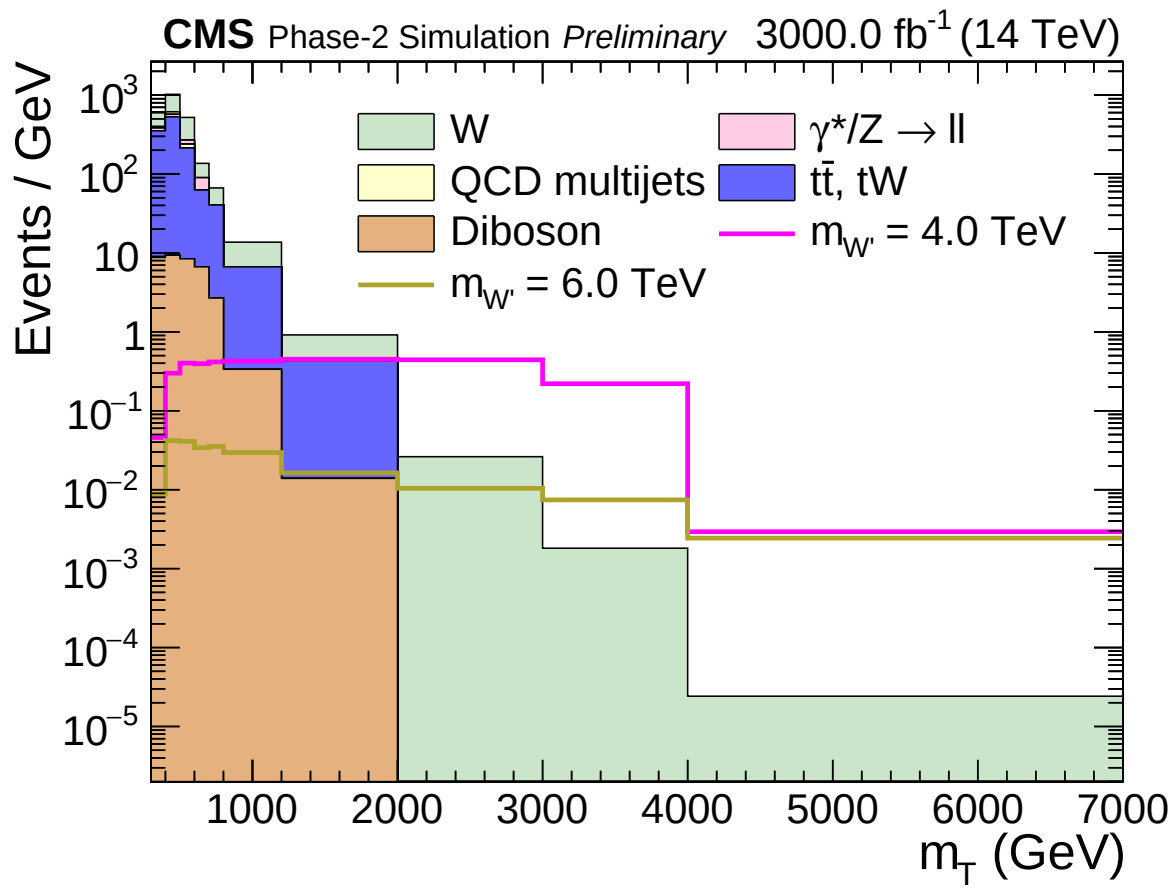

Figure 2: Distribution of $m_T$, after all selections for HL-LHC conditions of 3000 fb$^{-1}$ and 200 PU. The relevant SM backgrounds are shown according to the labels in the legend. Signal examples for values of the W′ boson mass of $m_{W'}$ =4 TeV and 6 TeV are scaled to their SSM LO cross section and 3000 fb$^{-1}$ integrated luminosity.

The discriminating variable is the $m_T$ distribution. The expected background distribution after applying all selection criteria is shown in Fig. 2, along with predicted signal distributions for different values of the mass of the W′ boson. The product of the signal efficiency and acceptance for SSM W′ $\rightarrow \tau\nu$ events depends on the W′ boson mass. It reaches about 18% for values of the W′ boson mass in the range of 3-4 TeV, and decreases to about 11% for higher and lower values of 8 TeV and 1 TeV, respectively. The signal efficiency decreases for higher W′ boson masses, as off-shell production increases yielding more events in the low $m_T$ region (similar to lower masses), where the kinematic cuts apply. Overall, this signal efficiency is about 5% lower than the Run 2 signal efficiency due to the less efficient DELPHES tau identification. Due to this and much higher pileup, more events are expected in this final state, mainly in the low $m_T$ region.

## 6 Results

The sensitivity at a center-of-mass energy of 14 TeV at HL-LHC is studied based on the $m_T$ distribution in Fig. 2. Upper limits on the product of the production cross section and branching fraction, $\sigma(pp \rightarrow W') \times \mathcal{B}(W' \rightarrow \tau\nu)$, are determined using a Bayesian method [25] with a uniform positive prior probability distribution for the signal cross section. All limits presented here are at 95% confidence level (CL). For every bin the signal expectation is compared to the sum of all background processes thus considering the full shape information of the $m_T$ distribution. This procedure is performed for different values of parameters of each signal, to obtain limits in terms on these parameters, such as the W′ boson mass. Signal events are expected to be particularly prominent at the upper end of the $m_T$ distribution, where the expected SM background is low.

The nuisance parameters associated with the systematic uncertainties are modeled through log-normal distributions for uncertainties in the normalization. Systematic uncertainties related to object performance follow the recommendation for upgrade analyses [3], with the uncertainty values for tau identification (2.5%), tau energy scale (3%), and for jet and $p_T^{miss}$ energy scale (2.5%) and resolution (3%), respectively. Uncertainties on the SM background cross sections are reduced by a factor 1/2 with respect to Run 2. The uncertainty on the integrated luminosity is expected to be 1%. In the high-mass region, the expected number of background events is consistent with zero, the effect of the systematic uncertainty on the exclusion limits is negligi-ble.

With 3000 $\mathrm{fb}^{-1}$ of the integrated luminosity during Phase-2, the W′ mass reach for a potential observation increases to 6.9 TeV and 6.4 TeV for evidence with a significance exceeding three standard deviations (3 $\sigma$) and discovery with 5 $\sigma$, respectively, as shown in Fig. 3 (left). The sensitivity is shown for 3000 $\mathrm{fb}^{-1}$ and 200 PU as expected during the HL operation, along with the reach for 300 $\mathrm{fb}^{-1}$ corresponding to the LHC Phase-1 operation. In the absence of a signal in the data, the existence of SSM W′ bosons with a mass up to 7.0 TeV can be excluded at 95% confidence level (CL) as depicted in Fig. 3 (right), improving significantly the present sensitivity [4], which excludes SSM W′ bosons decaying to tau and $p_{\mathrm{T}}^{\mathrm{miss}}$ up to 4.0 TeV in mass.

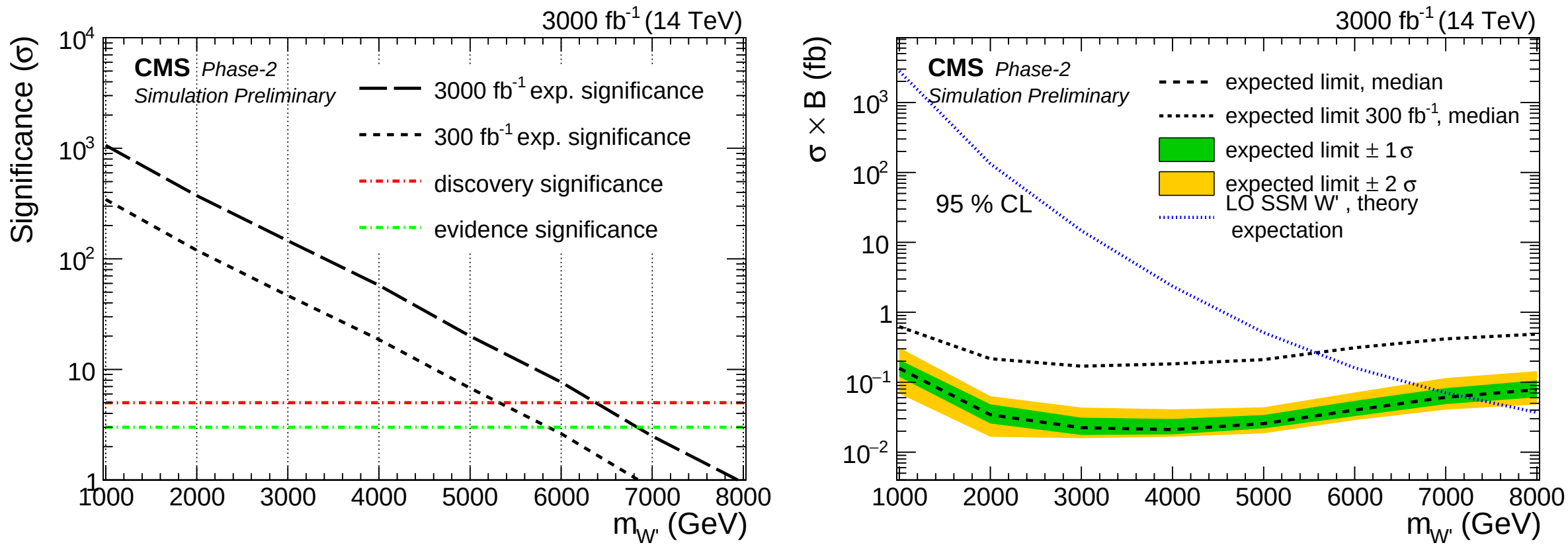

Figure 3: Sensitivity for a SSM W′ boson for 300 $\mathrm{fb}^{-1}$ and 3000 $\mathrm{fb}^{-1}$. Discovery significance (left) and expected exclusion limit on the SSM W′ boson mass at 95% CL (right).

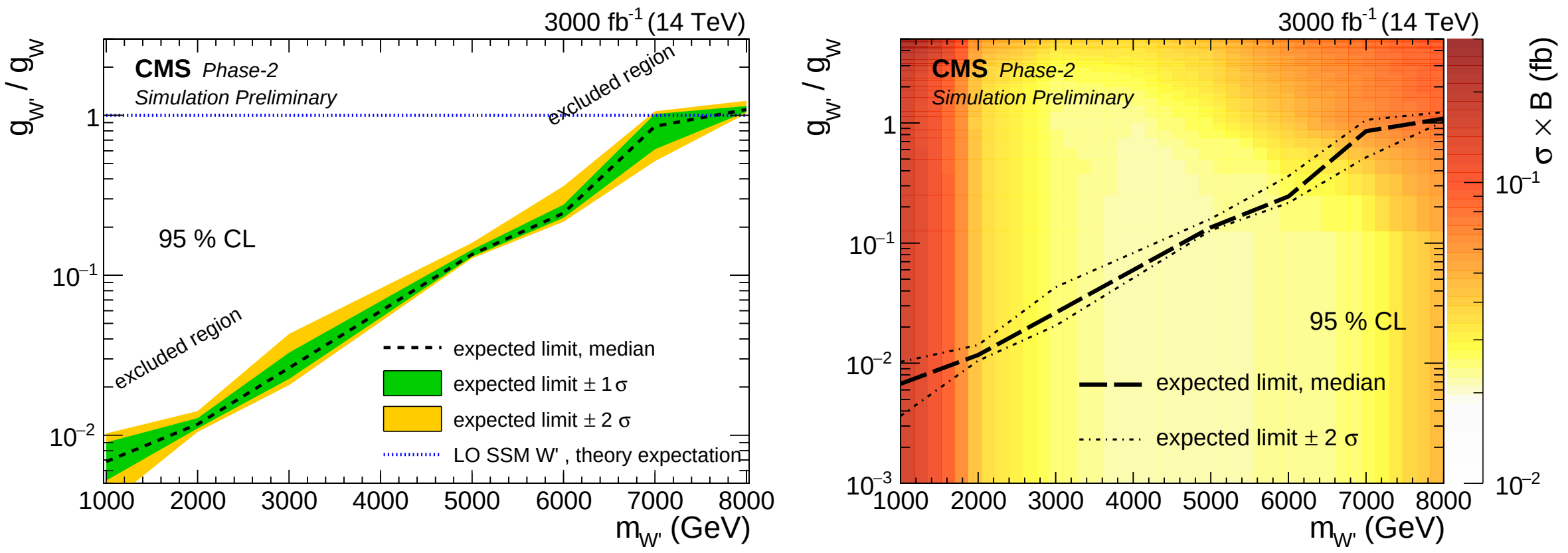

Figure 4: Sensitivity to the coupling ratio $g_{\mathrm{W'}}/g_{\mathrm{W}}$ of a W′ boson using 3000 $\mathrm{fb}^{-1}$ of integrated luminosity at the HL-LHC. On the left, the coupling ratio $g_{\mathrm{W'}}/g_{\mathrm{W}}$ is shown as a function of the W′ boson mass. The theory line of the SSM W′ boson is shown in blue. The 2D graph on the right includes additionally the limit on the cross section represented by the color code.

While the SSM assumes SM-like couplings of the fermions, the couplings could be weaker if further decay channels occur. The HL-LHC has a good sensitivity to study the coupling ratio $g_{\mathrm{W'}}/g_{\mathrm{W}}$. The sensitivity to smaller values for the couplings extends significantly, as shown in Fig. 4 (left) as a function of the W′ boson mass. In Fig. 4 (right) additionally the limit on the cross section is represented by the color code.

To allow further interpretations, a model-independent cross section limit is determined. A major difference with respect to the SSM limit is the fact that this limit has to be calculated as a single bin ranging from a lower threshold $m_{\mathrm{T}}^{\mathrm{min}}$ to infinity. For this reason, fluctuations in the number of events at very high $m_T$ have a larger impact on the limit. The resulting exclusion sensitivity for any new physics model with a $\tau_h$ and substantial $p_T^{miss}$ is shown in Fig. 5. Compared to the present result [4] it improves by an order of magnitude over the entire $m_T$ mass range and extends to even higher $m_T$.

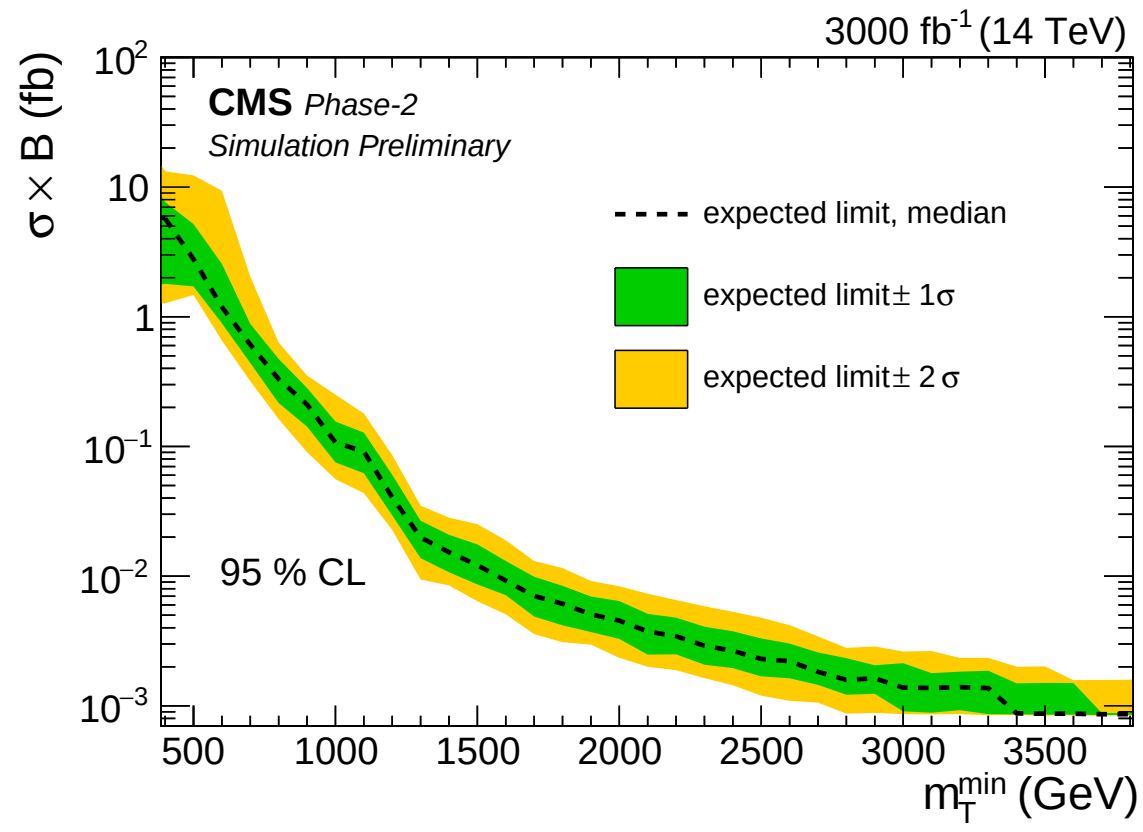

Figure 5: Model-independent cross section limit scaled to 3000 fb$^{-1}$. For this, a single-bin limit is calculated considering events above a lower threshold $m_T^{min}$ while keeping the signal yield constant in order to avoid including any signal shape information on this limit calculation.

# 7 Summary

Taking guidance from the published Run 2 analysis based on proton-proton collisions corresponding to an integrated luminosity of 35.9 fb$^{-1}$ at $\sqrt{s} = 13$ TeV [4], the physics reach at the High-Luminosity LHC with 3000 fb$^{-1}$ at $\sqrt{s} = 14$ TeV with the upgraded CMS detector was studied. The final state consists of a hadronically decaying tau lepton and $p_T^{miss}$ caused by neutrinos. The interpretation was performed in the benchmark sequential standard model (SSM) with an additional charged gauge boson W′. With the high luminosity, the sensitivity can be substantially improved. The discovery at a significance level of 3(5) standard deviations is possible for W′ boson masses of 6.9(6.4) TeV, respectively. In case of no observation, SSM W′ boson masses up to 7.0 TeV can be excluded.

While the SSM assumes standard-model-like couplings, weaker couplings are possible. Depending on the value of the W′ boson mass, the high-luminosity data will allow to study couplings down to nearly $10^{-2}$. To allow interpretations in other models, a model-independent limit on the cross section was provided.

**8**

# Flavour Physics

# Prospects for the $\mathcal{B}(B^0_{(s)} \to \mu^+\mu^-)$ measurements with the ATLAS detector in the Run 2 and HL-LHC data campaigns

The ATLAS Collaboration

This note estimates the ATLAS detector performance in measuring the branching fractions of the very rare decays $B^0_s \to \mu^+\mu^-$ and $B^0 \to \mu^+\mu^-$ using data collected during the whole LHC Run 2 campaign and during the whole HL-LHC campaign. The estimation is obtained by means of pseudo-MC experiments based on the measurement of the two processes performed by the ATLAS experiment using the full integrated luminosity collected during the Run 1 data taking campaign.

# 1 Introduction

This note documents a simulation-based study of the confidence regions for the search for very rare decays $B^0 \to \mu^+\mu^-$ and $B^0_s \to \mu^+\mu^-$ using the statistics to be collected in Run 2 and HL-LHC with the ATLAS detector [1]. Both processes are flavour-changing neutral-current (FCNC) mediated and therefore highly suppressed in the SM, nevertheless their branching fractions have been accurately predicted [2]: $\mathcal{B}(B^0_s \to \mu^+\mu^-) = (3.65 \pm 0.23) \times 10^{-9}$ and $\mathcal{B}(B^0 \to \mu^+\mu^-) = (1.06 \pm 0.09) \times 10^{-10}$. The measurement of the two branching fractions is relevant to indirect searches for physics beyond the Standard Model [3–11], therefore these processes are a flagship in B-physics analyses.
The present studies are based on the latest result from the ATLAS collaboration regarding this search, which is based on the data collected during Run 1 [12].

# 2 Extrapolation Procedure

The analysis sensitivity is expressed with expected $BR(B_d \to \mu\mu)$ - $BR(B_s \to \mu\mu)$ contour plots centred at the SM expected value for these processes. These contour plots are obtained predicting the expected statistics relative to the Run 1 analysis, and then running the same fitting and toy-MC machinery used for the Run 1 result. The yield extrapolation is performed for the total expected Run 2 yield, and then for 3 ab$^{-1}$ of HL-LHC integrated luminosity.

All sensitivities are obtained assuming as central value the SM expectation [2]. Offline analysis selections and reconstruction efficiencies are assumed to be the same as the published Run 1 result, while yields are corrected for different center of mass energy, trigger selections and integrated luminosities. The S/B ratio is assumed to be the same as the Run 1 analysis, since the analysis selection variables make it very robust against pile-up (thanks e.g. to vertexing quality and pointing angle requirements), and the dominant source of background after the final analysis selection is from heavy flavour decays. The HL-LHC ATLAS tracker upgrades [13, 14] entail improvements in vertex and mass determination, partly enhanced by the increased muon trigger momentum thresholds. While the decay length resolution improvements are neglected in favor of a conservative S/B assumption, the signal mass resolution variations are taken into account when assessing the analysis sensitivity. This variation is assessed using HL-LHC simulations: figure 1 ([13]) shows the comparison of signal width obtained from Run 2 simulations to the one obtained simulating HL-LHC detector and collision conditions. The substantial improvement in mass resolution reflects in an improved signal statistical significance and separation of $B_s \to \mu\mu$ and $B_d \to \mu\mu$ decays, as demonstrated in section 6.2.

The expected Run 2 confidence level bands need to be derived (due to the low-statistics regime of this analysis) using a profiled likelihood ratio Neyman belt construction. As the sample statistics increases, the fit likelihood maximum approaches asymptotic Gaussian behaviour, closely reproducing the two-dimensional Neyman construction contours. This approximation is exploited to reduce the computational burden of the results.

The HL-LHC extrapolations are expected to be sufficiently well approximated with likelihood contours. The full Run 2 extrapolation is a pivotal point in this study, justifying the asymptotic approach used in confidence band extraction for the HL projections.

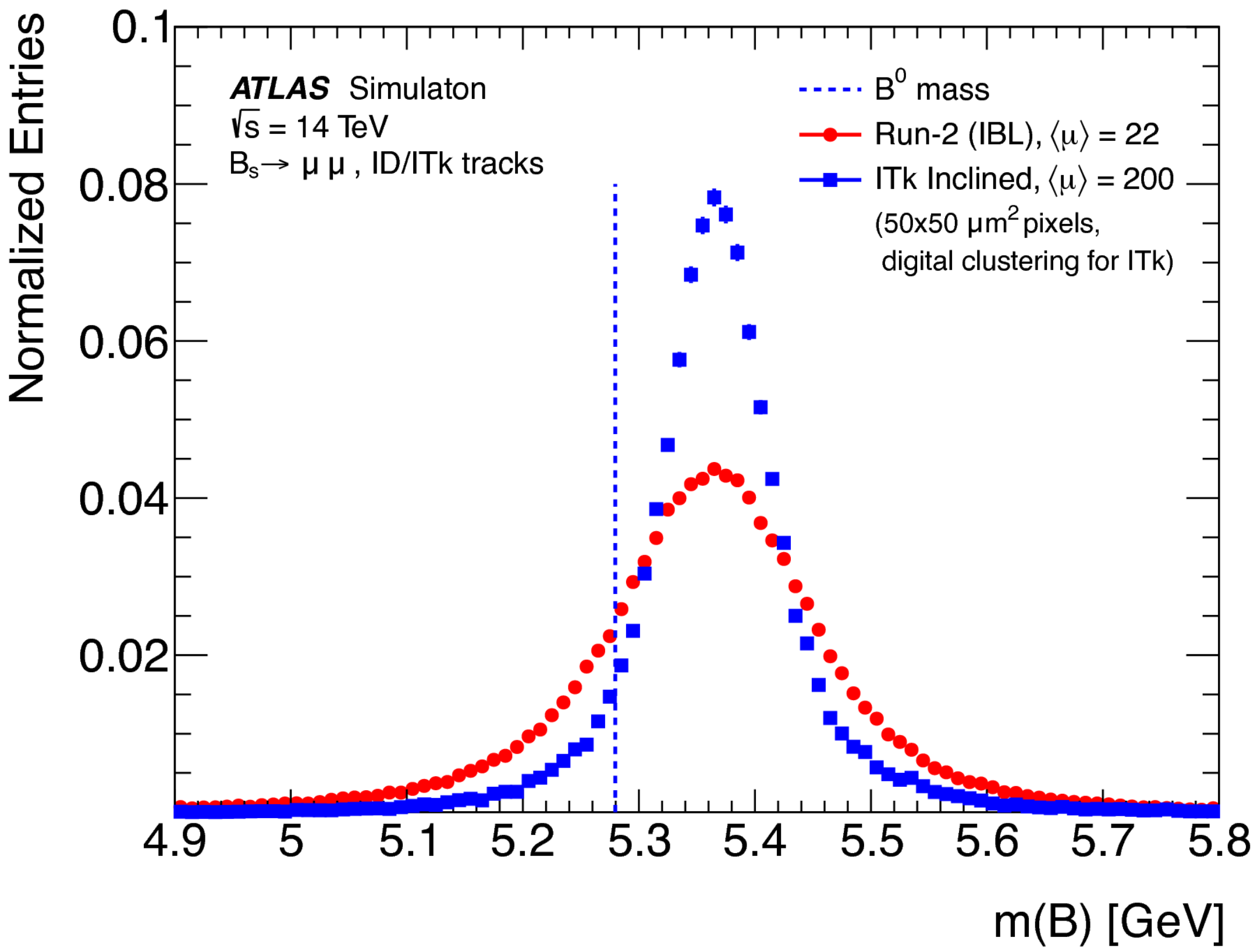

Figure 1: Reconstructed $B_s \to \mu\mu$ mass spectrum for muons with $|\eta| < 2.5$. For reference, the $B_d$ mass value [15] is shown as a dotted line. HL-LHC and Run 2 are compared. The figure shows the mass resolution using only the ID/ITk track parameter measurement, evaluated at the fitted B-vertex.

In order to determine the $BR(B_d \to \mu\mu)$ - $BR(B_s \to \mu\mu)$ confidence regions, the expected signal yield is obtained through the following expression:

$$\begin{aligned}\text{(projected signal statistics)} = \text{(Run 1 signal statistics)} \times \frac{\text{(projected B cross-section)}}{\text{(Run 1 B cross-section)}} \\ \times \frac{\text{(projected Luminosity)}}{\text{(Run 1 Luminosity)}} \times \frac{\text{(projected trigger efficiency)}}{\text{(Run 1 trigger efficiency)}}.\end{aligned} \quad (1)$$

Where the B production cross-section scale factor we use assumes, conservatively, that Run 1 data have been all collected at a center of mass energy of 8 TeV.
Each of the ratios in equation 1 are quantified in sections 4 and 5.

### 2.1 Systematic uncertainties

The Run 1 analysis parametrizes two classes of systematic uncertainties: the ones coming from external inputs (e.g. the $f_s/f_d$ ratio and the $B^+ \to J/\psi[\to \mu\mu]K^\pm$ branching ratio) and those depending on internal analysis effects (invariant mass fit shapes, efficiencies, etc.). The latter category is parameterized as a function of the signal yields where dependencies are found to be significant.

This study extrapolates the same systematic uncertainties including the same signal yield parameterization found in the Run 1 studies. As for the external sources of systematic uncertainties, it is plausible to expect that these will be reduced with other measurements and could optimistically scale for the most part like statistical uncertainties. This study however conservatively assumes their values to be those used in the Run 1 analysis.

## 3 Run1 cross-check

Since this study re-implements in the Run 2 $B \to \mu\mu$ analysis framework the same techniques employed in the run 1 analysis, the ATLAS Run 1 analysis is reproduced first. Like for the Run 1 publication, contour plots are obtained as profiled likelihood contours, with a central fitted value consistent with the Run 1 analysis. To this effect, a mock dataset similar to the real Run 1 data sample is generated using toy simulations and then fit with the same fitting technique used in Run 1. The resulting contours are compatible with the ones obtained in Run1 with real data.

## 4 Extrapolation to full Run 2 statistics

The three main ingredients of equation 1 are computed as follows:

1. **B production cross section with respect to Run 1**:
due to the Run 2 increased center of mass energy a value ~ 1.7 times higher is expected , according to studies performed using FONLL [16].
2. **The expected collected luminosity for the selected triggers in Run 2**:
we assume 130fb$^{-1}$ as the total Run 2 integrated luminosity. Similar trigger conditions as the ones present in 2017 are projected to 2018.
3. **The efficiency of the dimuon triggers available in Run 2 with respect to the Run 1 triggers**:
these have been calculated exploiting $B^0_s \to \mu^+\mu^-$ MC simulations for the Run 2 data-taking conditions.

In order to compute the relative Run 2/Run 1 trigger efficiency, simulated signal events were fully reconstructed and the same preselections as the Run 1 analysis [12] were applied. All efficiencies are normalized to the 4 GeV dimuon (2MU4) trigger selection, with the Run 2 dataset assumed to be collected with an admixture of triggers extrapolating to 2018 data-taking the same trigger selections and prescales implemented in 2017.
These ingredients predict a 7-fold increase with respect to Run 1 statistics.

## 5 Extrapolation to HL-LHC statistics

The prediction of the HL-LHC $B \to \mu\mu$ signal yield is based on equation 1, taking into account 3 ab$^{-1}$ of expected integrated luminosity and the triggers foreseen by [17]:

1. **B production cross section with respect to Run 1**,
   any further increase in the enter of mass energy relative to Run 2 is neglected, conservatively assuming the same $\times 1.7$ factor as for the Run 2 extrapolation.
2. **expected collected luminosity for the selected triggers at HL-LHC**,
   an integrated luminosity of $\approx 3\text{ab}^{-1}$ ($3000\text{fb}^{-1}$) is assumed to be collected during the whole HL-LHC data taking period;
3. **efficiency of the dimuon triggers available at HL-LHC with respect to the Run 1 triggers**,
   the HL-LHC signal trigger efficiencies have been calculated exploiting the same MC simulations used in section 4. The pile-up conditions in these simulations differ from what expected for HL-LHC, but this is irrelevant when assessing truth-matched signal efficiencies. Based on the prospective triggers foreseen in [17], we explore different dimuon transverse momentum thresholds ($p_{\mathrm{T}}^{\mu_1}$,$p_{\mathrm{T}}^{\mu_2}$): (6 GeV, 6 GeV), (6 GeV, 10 GeV) and (10 GeV, 10 GeV).

Depending on these dimuon trigger thresholds, three working points are inferred:

- Conservative: $\times 15$ Run 1 statistics;
- Intermediate: $\times 60$ Run 1 statistics;
- High-yield: $\times 75$ Run 1 statistics.

# 6 Results

## 6.1 Run 2 Contours

A two-dimensional Neyman construction [18] based on likelihood ratio ranking is used to identify the 68.3%, 95.5% and 99.7% confidence level regions for the combined measurement of $\mathcal{B}(B_s^0 \to \mu^+\mu^-)$ and $\mathcal{B}(B^0 \to \mu^+\mu^-)$. Pseudo-MC experiments are used in the Neyman construction procedure and to verify the coverage.
Figure 2 compares the Neyman-construction based Run 1 and Run 2 confidence regions and the Standard Model theoretical prediction with its uncertainty [2]. Systematic uncertainties are included in both sets of contours.
As introduced in section 2, at the level of full Run 2 statistics the behavior of the likelihood function is expected to be almost asymptotic. This is shown in figure 3, where 68.3%, 95.5% and 99.7% (stat+syst) confidence regions constructed with Neyman belt and and the asymptotic likelihood approaches are compared: consistency is very good for the 68.3% contour, and progressively deviates from gaussianity (with inconsistencies at the level of 15-20%) at larger $\Delta \log L$. This is due to the progressively less adequate Gaussian approximation of the Likelihood maximum. These residual non-gaussianities decrease at the higher statistics projected for HL-LHC, therefore allowing the $\Delta \log L$ contour approach to be sufficient for HL-LHC extrapolations.
The effect of systematic uncertainties is illustrated in Figure 4. This figure shows the 2D Neyman belt based Run 2 extrapolated confidence regions and the Standard Model theoretical prediction with its uncertainty. The dominant systematic uncertainty is coming from the $f_s/f_d$ ratio and fit shape uncertainties, affecting predominantly or exclusively $BR(B_s \to \mu\mu)$, and therefore resulting in a 'rotation' of the measurement ellipsoid to be more parallel to the x-axis.

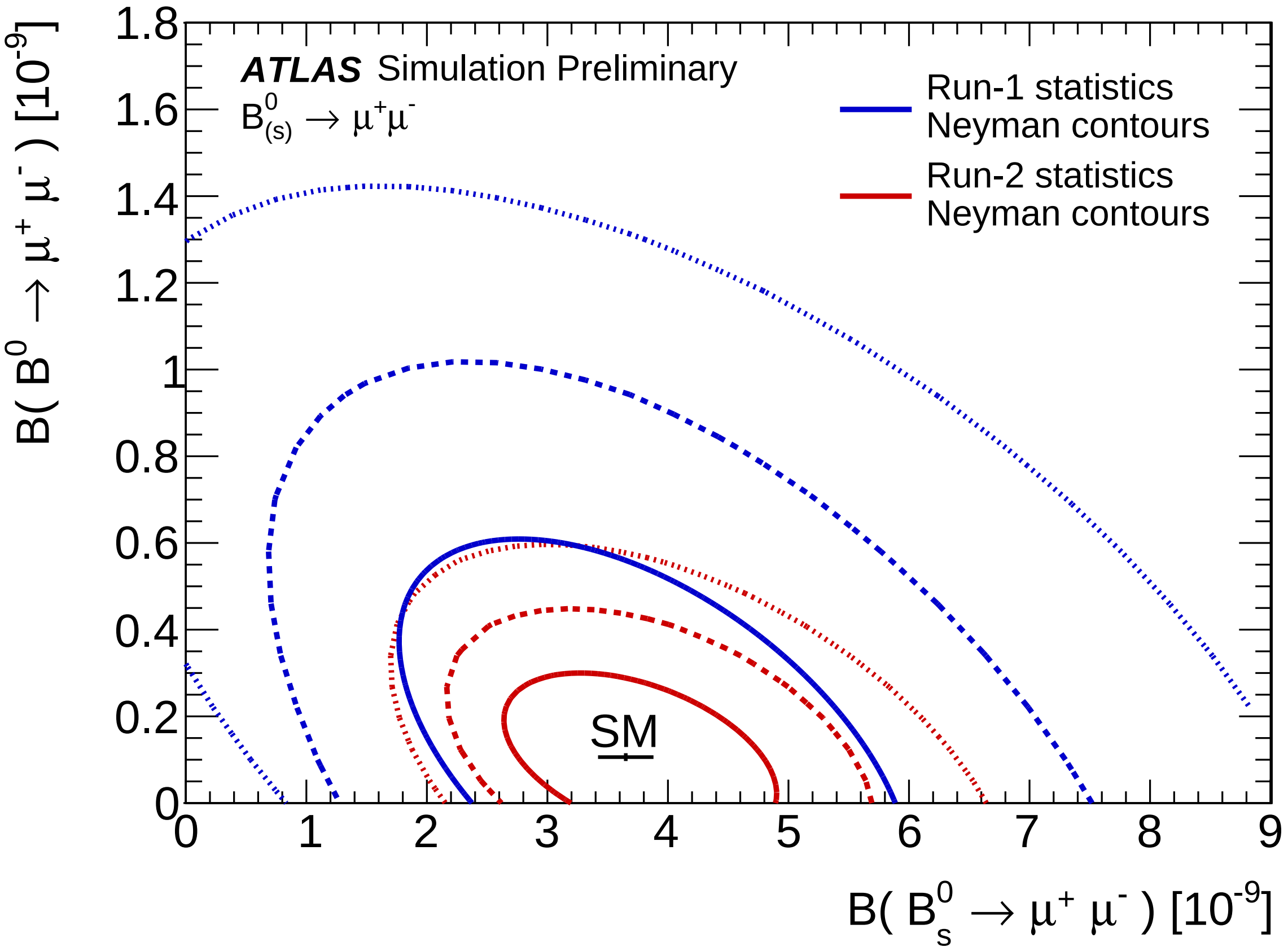

Figure 2: Comparison of the 68.3% (solid), 95.5% (dashed) and 99.7% (dotted) stat.+syst. confidence regions for the Run 1 statistics (blue, outermost) and the extrapolated Run 2 statistics (red, innermost). The confidence regions are obtained with the 2D Neyman belt construction, based on pseudo-MC experiments and the Run 1 analysis likelihood. The Run 2 pseudo-MCs reproduce the expected signal mass resolution and have been scaled with respect to their Run 1 counterpart according to the triggers available in Run 2, the different integrated luminosity and the different B production cross section. The black point shows the SM theoretical prediction and its uncertainty [2].

## 6.2 HL-LHC Contours

Section 6.1 indicates that with the statistics expected in the HL scenarios considered the profiled likelihood contours sufficiently well approximate the expected analysis performance.

Two different sets of likelihood ratio contours corresponding to 68.3%, 95.5% and 99.7% probability are obtained, with and without systematic uncertainties. All working points described in section 5 are considered and result in the contours illustrated in figures 5, 6 and 7.
In Table 1 we compare the profiled likelihood uncertainties separately for the $\mathcal{B}(B^0_s \to \mu^+\mu^-)$ and $\mathcal{B}(B^0 \to \mu^+\mu^-)$ measurements, at the various data taking points discussed in the previous paragraphs. Statistical uncertainty uniformly decreases with statistics as expected, while systematic uncertainties show a distinct behaviour for $B^0_s$ ($f_s/f_d$ dominated) and $B^0$ yields.

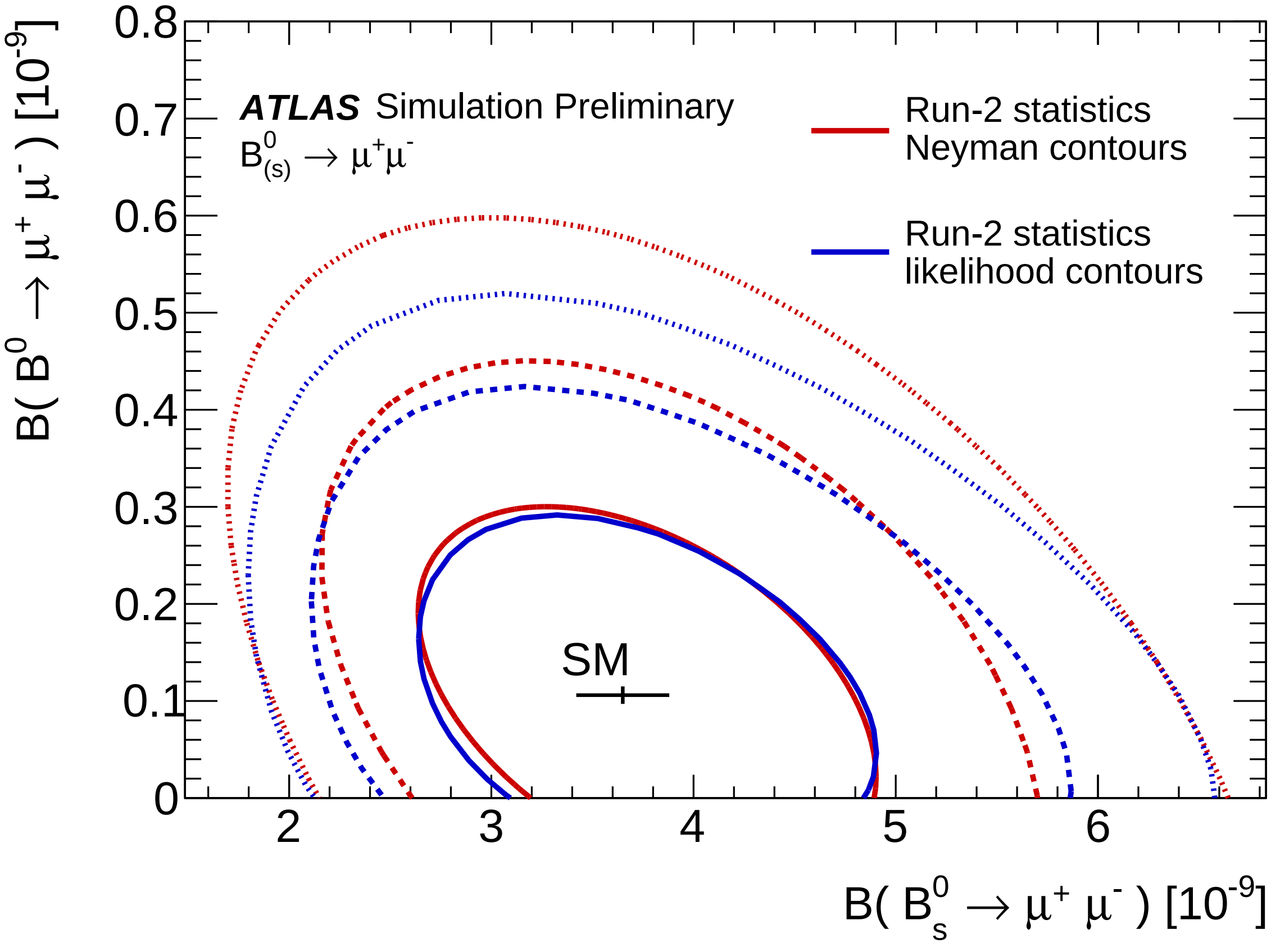

Figure 3: Comparison of the 68.3% (solid), 95.5% (dashed) and 99.7% (dotted) stat.+syst. confidence regions for the extrapolated Run 2 statistics. Red contours are obtained exploiting the 2D Neyman belt construction based on pseudo-MC experiments, while blue contours are drawn at constant $\Delta \log L$ in the gaussian maximum approximation. The Run 2 pseudo-MCs reproduce the expected signal mass resolution and have been scaled with respect to their Run 1 counterpart according to the triggers available in Run 2, the different integrated luminosity and the different B production cross section. The black point shows the SM theoretical prediction and its uncertainty [2].

## 7 Conclusions

A detailed study of th ATLAS experiment reach in the search for rare decays of $B^0_s$ and $B^0$ into oppositely charged muons is presented. The study is based on the results of the analysis performed on the data collected during Run 1 of LHC, and takes into account several aspects of the extrapolation such as e.g. trigger selections and efficiencies, detector performance effects, luminosity and collision energy conditions.

Systematic uncertainties are extrapolated from the Run 1 analysis, without assumptions on the evolution of external sources of systematic uncertainties such as $f_s/f_d$ and the BR of the reference channel. Extrapolations of the statistics available in the datasets to be collected during Run 2 and HL-LHC are performed and the confidence regions for Run 1, Run 2 and HL-LHC statistics are obtained. The extrapolations take into account different mass resolution performances of the configurations considered as well as the effect of different trigger selections and signal and background yields dependencies on the collision center of mass energy.

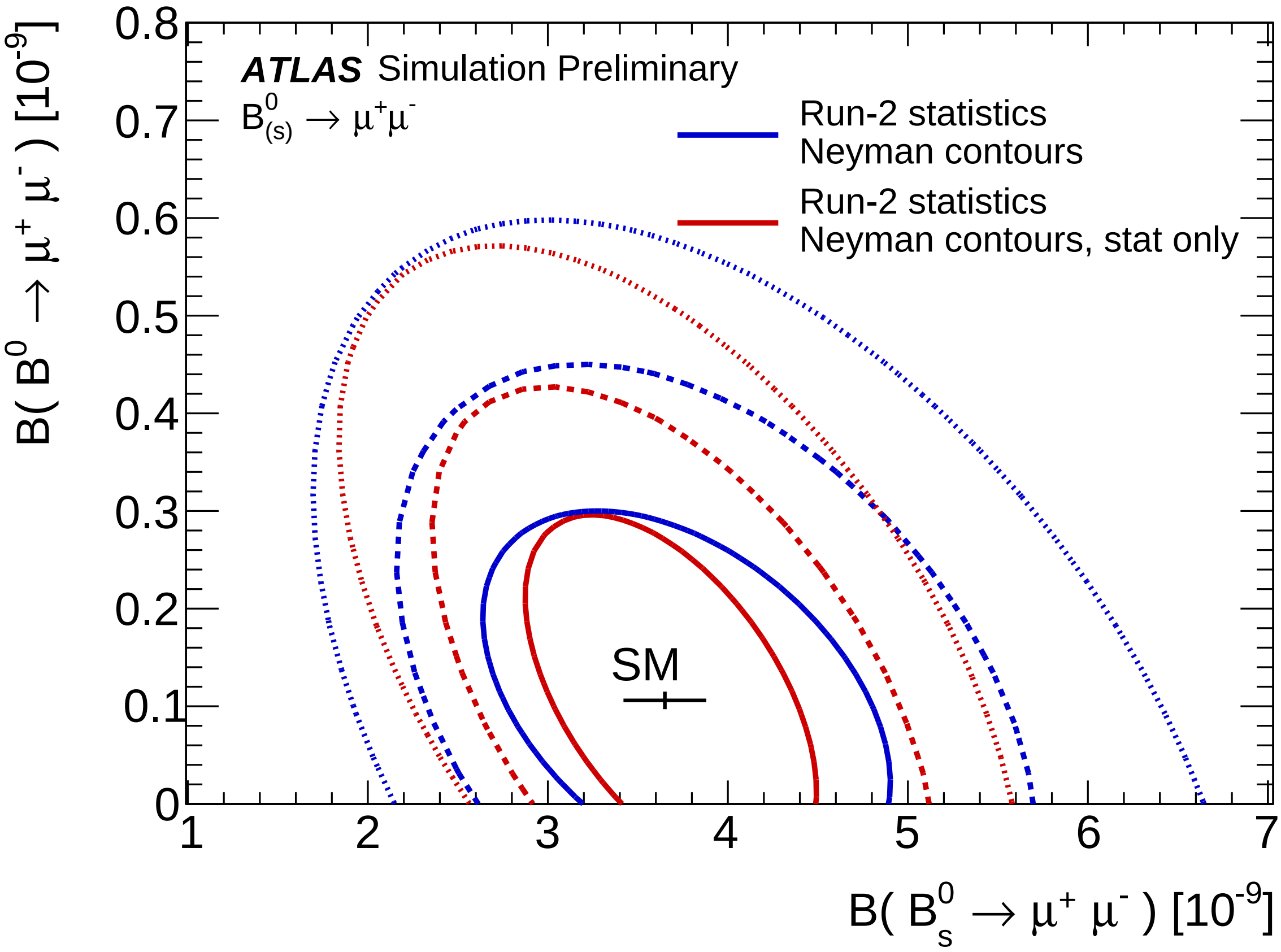

Figure 4: Comparison of 68.3% (solid), 95.5% (dashed) and 99.7% (dotted) confidence level contours obtained exploiting the 2D Neyman belt construction for the Run 2 case. Red contours are statistical only; blue contours include systematics uncertainties from the ATLAS Run1 analysis [12] extrapolated to Run 2 statistics. The Run 2 pseudo-MCs reproduce the expected signal mass resolution and have been scaled with respect to their Run 1 counterpart according to the triggers available in Run 2, the different integrated luminosity and the different B production cross section. The black point shows the SM theoretical prediction and its uncertainty [2].

This study reports also in table 1 the projected uncertainties on the individual $B_s$ and $B_d$ branching ratios for the scenarios considered.

| | $\mathcal{B}(B^0_s \to \mu^+\mu^-)$ | | $\mathcal{B}(B^0 \to \mu^+\mu^-)$ | |
|---|---|---|---|---|
| | stat [$10^{-10}$] | stat + syst [$10^{-10}$] | stat [$10^{-10}$] | stat + syst [$10^{-10}$] |
| Run 2 | 7.0 | 8.3 | 1.42 | 1.43 |
| HL-LHC: Conservative | 3.2 | 5.5 | 0.53 | 0.54 |
| HL-LHC: Intermediate | 1.9 | 4.7 | 0.30 | 0.31 |
| HL-LHC: High-yield | 1.8 | 4.6 | 0.27 | 0.28 |

Table 1: Uncertainty on $\mathcal{B}(B^0_s \to \mu^+\mu^-)$ and $\mathcal{B}(B^0 \to \mu^+\mu^-)$ as reported by the fitting procedure applied to the toy simulations. The results are centred on the SM theoretical prediction [2]. For each extrapolation performed, statistical and statistical+systematic uncertainties are reported in units of $10^{-10}$. These values can be compared with the combined Run 1 measurement of CMS and LHCb [19] $\mathcal{B}(B^0_s \to \mu^+\mu^-) = (2.8^{+0.7}_{-0.6}) \times 10^{-9}$, $\mathcal{B}(B^0 \to \mu^+\mu^-) = (3.9^{+1.6}_{-1.4}) \times 10^{-10}$ and the latest 2015+2016-data LHCb result [20] $\mathcal{B}(B^0_s \to \mu^+\mu^-) = (3.0 \pm 0.6^{+0.3}_{-0.2}) \times 10^{-9}$. The table reports a sufficient number of significant digits to highlight the difference between statistical+systematics and systematics-only uncertainties.

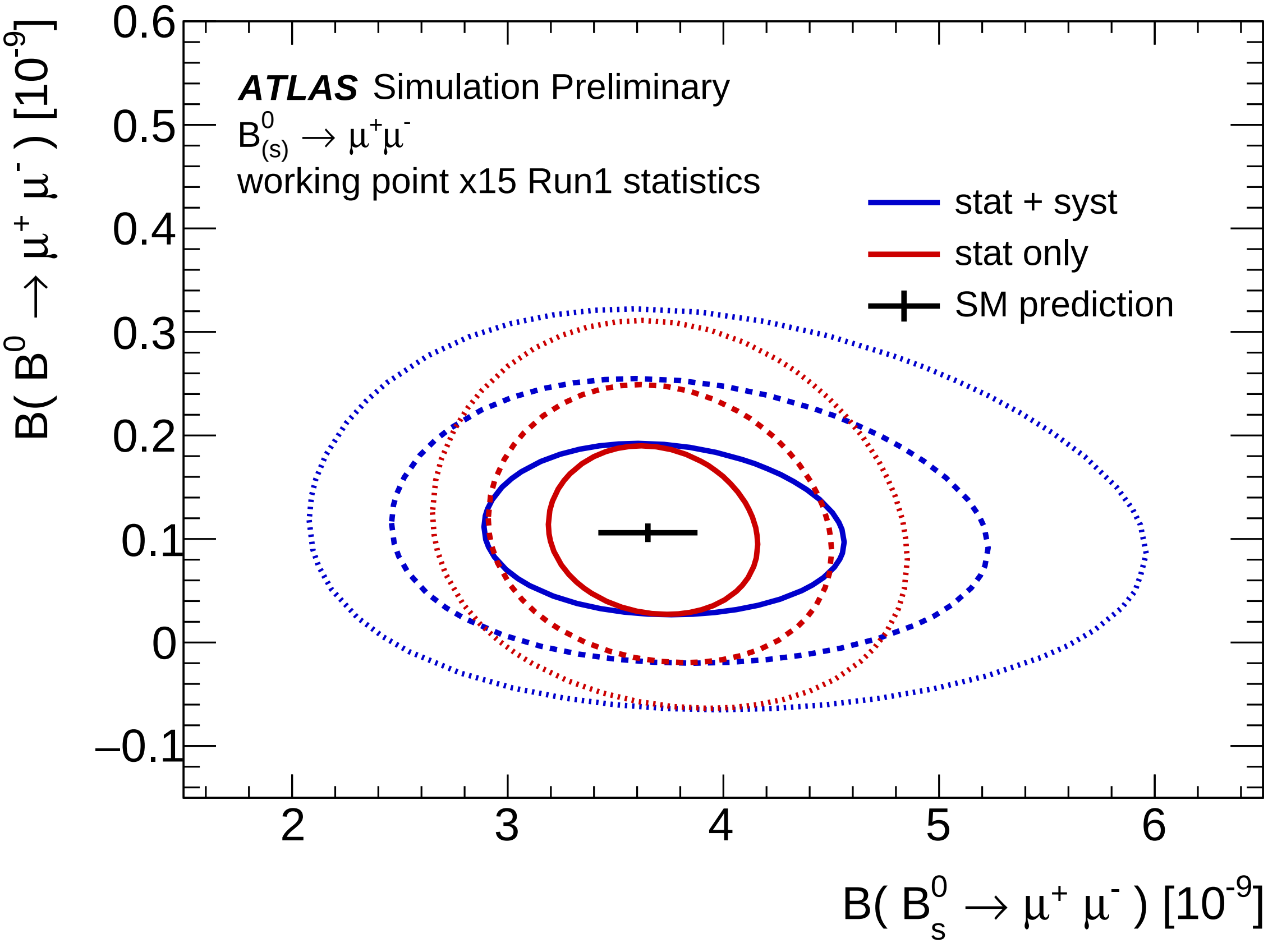

Figure 5: Comparison of 68.3% (solid), 95.5% (dashed) and 99.7% (dotted) confidence level profiled likelihood ratio contours for the working point at ×15 Run 1 statistics. This corresponds to the 'conservative' HL-LHC extrapolation, based on yield projections for the (10GeV, 10GeV) dimuon trigger. Red contours do not include the systematic uncertainties, which are then included in the blue ellipsoids. The black point shows the SM theoretical prediction and its uncertainty [2].

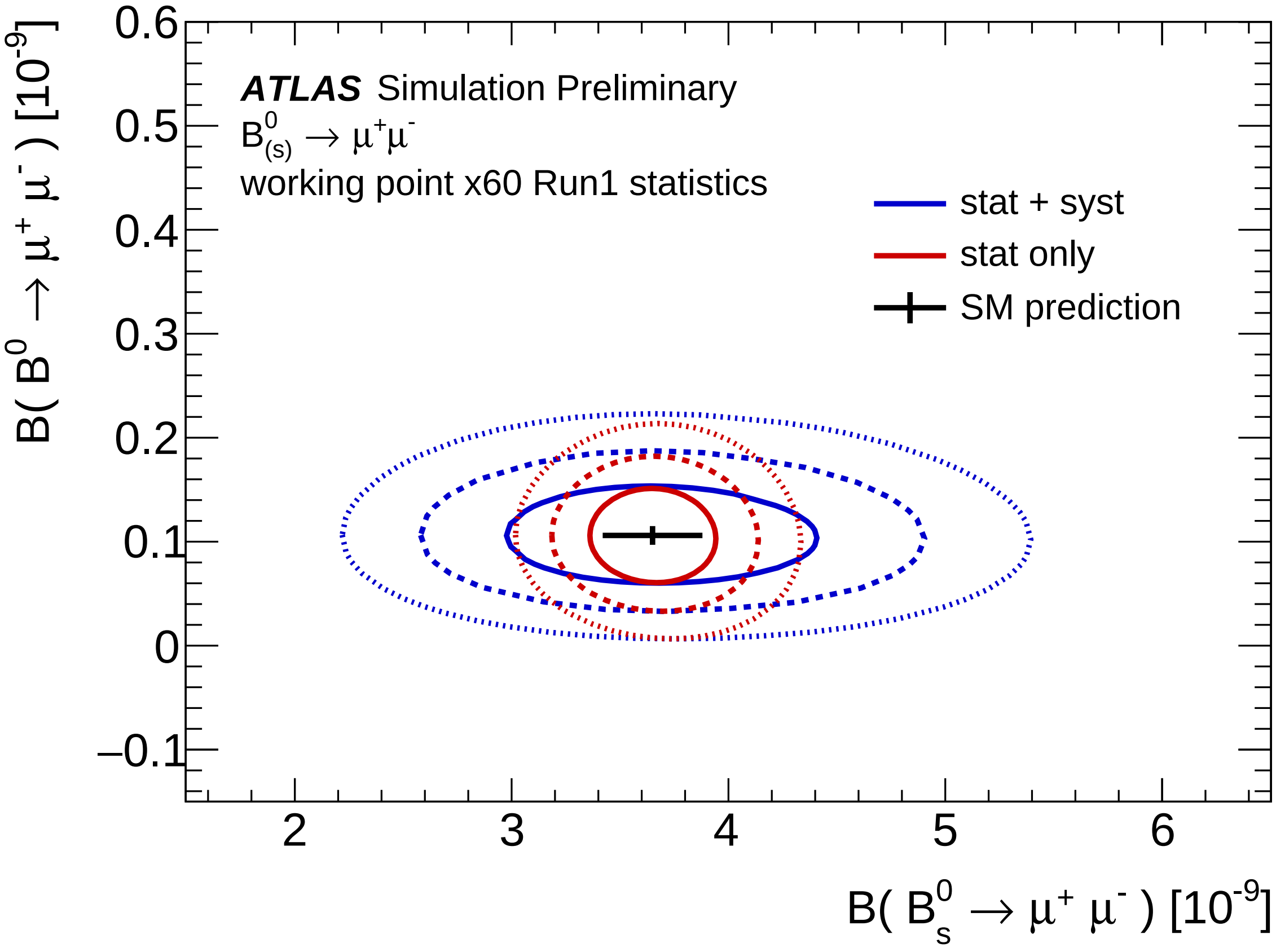

Figure 6: Comparison of 68.3% (solid), 95.5% (dashed) and 99.7% (dotted) confidence level profiled likelihood ratio contours for the working point at ×60 Run 1 statistics. This corresponds to the 'intermediate' HL-LHC extrapolation, based on yield projections for the (6GeV, 10GeV) dimuon trigger. Red contours do not include the systematic uncertainties, which are then included in the blue ellipsoids. The black point shows the SM theoretical prediction and its uncertainty [2].

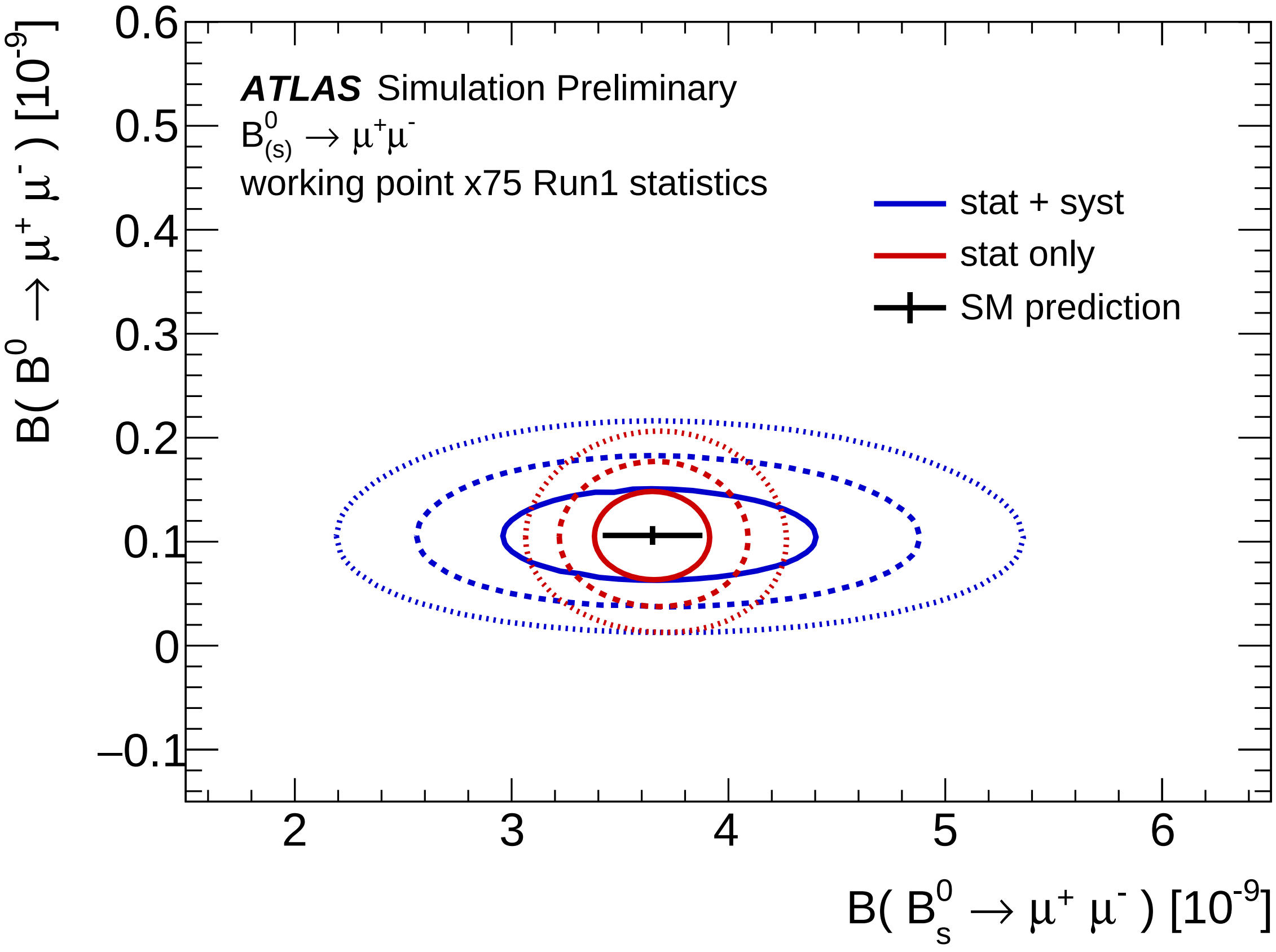

Figure 7: Comparison of 68.3% (solid), 95.5% (dashed) and 99.7% (dotted) confidence level profiled likelihood ratio contours for the working point at ×75 Run 1 statistics. This corresponds to the 'high-yield' HL-LHC extrapolation, based on yield projections for the (6GeV, 6GeV) dimuon trigger. Red contours do not include the systematic uncertainties, which are then included in the blue ellipsoids. The black point shows the SM theoretical prediction and its uncertainty [2].

# CMS Physics Analysis Summary

# Measurement of rare $B \to \mu^+\mu^-$ decays with the Phase-2 upgraded CMS detector at the HL-LHC

The CMS Collaboration

## Abstract

The sensitivity of the upgraded CMS detector for measuring the rare decays $B^0_s \to \mu^+\mu^-$ and $B^0 \to \mu^+\mu^-$ in the HL-LHC scenario is studied. The upgraded detector, especially with its improved momentum resolution, and the foreseen total integrated luminosity of 3000 fb$^{-1}$ are expected to enable high precision measurements of the branching fractions of $B^0_s \to \mu^+\mu^-$ and the effective lifetime of the $B^0_s \to \mu^+\mu^-$ decay with reduced systematic and statistical uncertainties. At 3000 fb$^{-1}$, it will also be possible to observe the $B^0 \to \mu^+\mu^-$ decay with more than $5\sigma$ significance.

# 1 Introduction

The decays $B^0_s \to \mu^+\mu^-$ and $B^0 \to \mu^+\mu^-$ are flavor changing neutral current (FCNC) transitions $b \to s(d)$ that cannot proceed at the tree level but can occur at the one-loop level via electroweak penguin and box diagrams in the standard model (SM) [1]. The decays are furthermore helicity suppressed by a factor of $m(\mu)^2/m(B)^2$, where $m(\mu)$ and $m(B)$ are the masses of the muon and $B^0_s$ meson, respectively. The SM predictions [2, 3] for the branching fractions are $\mathcal{B}(B^0_s \to \mu^+\mu^-) = (3.57 \pm 0.17) \times 10^{-9}$ and $\mathcal{B}(B^0 \to \mu^+\mu^-) = (1.06 \pm 0.09) \times 10^{-10}$. These are time-integrated branching fractions where the decay width differences of the heavy and light states of the $B^0_s$ meson are taken into account.

Moreover, new physics (NP) models [4, 5] also predict enhancements to the branching ratios for these rare decays and therefore studies of rare $B$ decays provide excellent opportunities to search for NP.

In the $B^0_s - \overline{B^0_s}$ mixing, there is a sizable difference $\Delta\Gamma_s = \Gamma_L - \Gamma_H$ between the decay widths of the $B^0_s$ light and heavy mass eigenstates which was experimentally measured [6]. From this measured decay width difference, it is possible to define the parameter $y_s \equiv \tau_{B^0_s}\Delta\Gamma_s/2$ = $0.062 \pm 0.006$, where $\tau_{B^0_s} = 1.510 \pm 0.005$ ps is the $B^0_s$ mean lifetime. The parameter $A_{\Delta\Gamma}$ is defined as $A_{\Delta\Gamma} = -2\Re(\lambda)/(1 + |\lambda|^2)$, with $\lambda = (q/p)(A(\overline{B^0_s} \to \mu^+\mu^-)/A(B^0_s \to \mu^+\mu^-)$. The complex coefficients $q$ and $p$ define the mass eigenstates of the $B^0_s - \overline{B^0_s}$ system in terms of the flavour eigenstates. Within the SM, $A_{\Delta\Gamma}$ is expected to be +1, i.e. that the decay occurs mostly through the heavier $B^0_s$ eigenstate with an effective lifetime defined by

$$\tau_{\mu^+\mu^-} \equiv \frac{\int_0^\infty t\langle(\Gamma(B^0_s(t)) \to \mu^+\mu^-)\rangle}{\int_0^\infty \langle(\Gamma(B^0_s(t)) \to \mu^+\mu^-)\rangle}, \quad (1)$$

where t is the proper decay time. The effective lifetime is related to the $B^0_s$ mean lifetime through the relation

$$\tau_{\mu^+\mu^-} = \frac{\tau_{B^0_s}}{1 - y_s^2}\left(\frac{1 + 2A_{\Delta\Gamma}y_s + y_s^2}{1 + A_{\Delta\Gamma}y_s}\right). \quad (2)$$

However, the $B^0_s \to \mu^+\mu^-$decay could receive contributions beyond the SM, whose current bounds do not actually exclude any $A_{\Delta\Gamma}$ value in the whole range [-1, +1].

CMS [7], ATLAS [8] and LHCb [9] have measured the $B^0_s \to \mu^+\mu^-$ branching fraction using data collected during the LHC Run-I. The measured branching fractions are $(3.0^{+1.0}_{-0.9}) \times 10^{-9}$, $(0.9^{+1.1}_{-0.8}) \times 10^{-9}$ and $(2.9^{+1.1}_{-1.0}) \times 10^{-9}$ from CMS, ATLAS and LHCb, respectively. The combination of the CMS and LHCb results gives $\mathcal{B}(B^0_s \to \mu^+\mu^-) = (2.8^{+0.7}_{-0.6}) \times 10^{-9}$ [10], in good agreement with the SM expectation. The CMS analysis is based on an integrated luminosity of 25 $\text{fb}^{-1}$ and 5 $\text{fb}^{-1}$ collected at the center-of-mass energy of $\sqrt{s} = 7$ TeV and 20 $\text{fb}^{-1}$ at 8 TeV. The uncertainty of the measured branching ratio is dominated by the statistical uncertainty, hence it is expected to decrease with the accumulation of more data. Similarly, all experiments reported upper limits on the $\mathcal{B}(B^0 \to \mu^+\mu^-)$ (1.1 $\times 10^{-9}$ for CMS, $4.2 \times 10^{-10}$ for ATLAS and $7.4 \times 10^{-10}$ for LHCb, at 95% confidence level) and again a larger sample size should improve the result. In the meantime, the ATLAS [11] and LHCb [12] updated their analyses using partial Run 2 data where no evidence of the decay $B^0 \to \mu^+\mu^-$is reported.

In this study, we focus on the mass resolution improvements with the Phase-2 CMS detector.

The new tracker detector will feature 4 pixel barrel layers and 5 disks on either endcap. The outer tracker material budget will diminish by roughly a factor of 2 in the central region ( $|\eta| < 1$) and about a factor of 3 in the intermediate region around $1.2 < |\eta| < 1.5$ [13]. This, combined with a smaller silicon sensors pitch, will improve the momentum resolution especially in the barrel region ($|\eta| < 1.4$) and will help to separate the $B^0$ signal from the tail of the $B^0_s$ signal, which now becomes a background to the $B^0$ measurement. Therefore, we investigate the contamination of the $B^0$ signal region from the $B^0_s$ candidates and also from the background candidates of the rare semileptonic $B$ decays (e.g. from $B^0 \to \pi^-\mu^+\nu$), where a hadron is misidentified as a muon. Then, we focus on the estimation of the sensitivity of branching fraction and decay time measurements using pseudo-experiments. The studies are based on the detailed Run-2 and Phase-2 full detector simulations using the GEANT4-based simulation package [14]. Monte-Carlo (MC) samples of the $B^0_s \to \mu^+\mu^-$ and $B^0 \to \mu^+\mu^-$ signal and the $B^0 \to \pi^-\mu^+\nu$ and $B^0 \to K^+\pi^-$ background with the simulation of the CMS detector for Run-2 and Phase-2 are used for the following studies.

## 2 Analysis Strategy

This study is based on the Run-2 analysis strategy that is applied to the upgraded CMS tracking system.

We measure rare leptonic neutral B decays, in particular $B^0_s \to \mu^+\mu^-$and $B^0 \to \mu^+\mu^-$. The $B$ candidate reconstruction starts with two opposite sign "global muons" [15] that are combined in a common displaced vertex fit to form a dimuon candidate. We require for the muon track $p_{\mathrm{T}} > 4$ GeV and $|\eta| <1.4$. We keep candidates with an invariant mass $4.5 < m_{\mu^+\mu^-} <6.5$ GeV. The fitted dimuon candidate is required to fulfill $p_T > 6.5$ GeV. Some useful variables for rejecting the backgrounds are

- $l_{3D}/\sigma(l_{3D})$: The flight length significance of the B candidate (the distance between the secondary and primary vertex, divided by its uncertainty);
- $\delta_{3D}/\sigma(\delta_{3D})$: The significance of the 3D impact parameter of the B candidate with respect to the selected primary vertex;
- $\alpha_{3D}$: The pointing angle of the B candidate;
- $d^0_{\mathrm{ca}}$: The minimum distance of closest approach to the B candidate vertex of a track (not belonging to the B candidate) in the event;
- $\chi^2/\mathrm{ndf}$: The vertex fit $\chi^2$ of the dimuon vertex;
- $N^{\mathrm{close}}_{\mathrm{trk}}$: The number of tracks in the vicinity of the B decay vertex;
- Isolation variables:
  - $I \equiv p_{\perp_B}/(p_{\perp_B} + \sum_{\mathrm{trk}} p_\perp)$: The isolation of the B candidate while $p_\perp >0.9$ GeV, $\Delta R < 0.7$ and $d^0_{\mathrm{ca}} < 0.05$ cm.
  - $I_\mu \equiv p_{\perp_\mu}/(p_{\perp_\mu} + \sum_{\mathrm{trk}} p_\perp)$: The isolation of the muon candidate while $p_\perp >0.5$ GeV, $\Delta R < 0.5$ and $d^0_{\mathrm{ca}} < 0.1$ cm.

The background is composed of several sources. Combinatorial background arises from two uncorrelated semileptonic $B$ decays that result in a random combination of muons. Rare semileptonic $B$ decays, such as $B^0 \to h\mu + \nu$ where a hadron is misidentified as a muon and where the neutrino carries away only a small amount of energy. There is also a background component from two-body hadronic decays, "peaking" background (e.g. from $B^0 \to K^+\pi^-$), when both hadrons from the decay are misidentified as muons.

The peaking background has significantly higher branching fraction than the signal branching fractions, thus an advanced muon identification algorithm was developed based on boosted decision tree (BDT), muon BDT, to separate the genuine muons from the hadrons that are misidentified as muons.

A second BDT, based on different properties of events, is used to separate signal events from other backgrounds. The variables used for this BDT are basically the same as in the CMS Run-I analysis [7], which includes those listed above. The signal-to-background ratio $(S/B)$, which depends on momentum resolution, is best in the barrel region and degrades if one or both muons are detected in the forward region. Therefore, the analysis is performed in two different regions defined by the pseudorapidity of the most forward muon $|\eta_f|$: Channel 0 is defined as $|\eta_f| < 0.7$ and channel 1 is given by $0.7 < |\eta_f| < 1.4$.

To extract the signal yield, an unbinned maximum likelihood fit to the dimuon invariant mass distribution is performed in bins of the discriminant variable of the second BDT. For the determination of the $B^0_s \to \mu^+\mu^-$ branching fraction, a normalization decay channel $B^+ \to J/\psi K^+$ is used. The latter decay has a well measured branching fraction and has similar topology/kinematics as the signals, so that their trigger and selection efficiencies do not differ significantly. Therefore, the $\mathcal{B}(B^0_s \to \mu^+\mu^-)$ is expressed as a function of the number of signal events ($N_{(B^0_s \to \mu^+\mu^-)}$) normalized to the number of $B^+ \to J/\psi K^+$events. This approach eliminates uncertainties related to the $b$-quark production cross section and the integrated luminosity and reduces the systematic uncertainties because of partial cancellation between the signal and normalization.

The formula for the branching fraction is:

$$\mathcal{B}(B^0_s \to \mu^+\mu^-) = \frac{N_{\text{sig}}}{N_{\text{norm}}} \times f_u/f_s \times \frac{\varepsilon_{\text{norm}}}{\varepsilon_{\text{sig}}} \times \mathcal{B}(B^+ \to J/\psi K^+)\,, \tag{3}$$

where $N_{\text{norm}}$ is the number of reconstructed $B^+ \to J/\psi K^+$ decays, $\mathcal{B}(B^+ \to J/\psi K^+) = (1.010 \pm 0.029) \times 10^{-3}$ [16], $\varepsilon_{\text{norm}}$ is the total efficiency for the normalization channel, $N_{\text{sig}}$ is the number of signal candidates, $\varepsilon_{\text{sig}}$ is the total signal efficiency, and $f_u/f_s$ is the fragmentation function, which is $0.250 \pm 0.012$ [16].

For the measurement of the $B^0_s \to \mu^+\mu^-$ effective lifetime the following procedure is applied. First an unbinned maximum likelihood fit to the dimuon invariant mass distribution is performed. Based on the fit results, a projection along the proper decay time distribution for the $B^0_s$ signal events is built with the sPlot [17] technique. To make it more clear, for each event, we have the information of the reconstructed invariant mass, mass resolution and decay time. In this analysis, we perform an unbinned maximum likelihood (UML) fit to mass and mass resolution to extract the "weight" to be a signal (sWeight) for each event with the sPlot method. By applying these weights to the decay time for each event, one obtains the background-subtracted decay time distribution. Subsequently, a binned maximum likelihood fit to the signal proper time distribution obtained from sPlot is carried out to extract the effective lifetime of the $B^0_s$ meson. The model used in the lifetime fit is formed by an exponential function, convolved with a Gaussian function that describes the expected decay time resolution, multiplied by an efficiency function that accounts for reconstruction and selection effects on the shape of the proper time distribution.

## 3 Results

The results of this analysis are split into two parts. The first one describes the improvements in the invariant mass reconstruction of the dimuon system due to the new inner tracking system with improved granularity. The second part includes the sensitivities of the effective lifetime and branching fraction measurements using pseudo-experiments generated with a MC technique. The baseline of the pseudo-experiments is the Run-2 probability density functions (PDF) for the signal and background components, which is modified with the improved mass resolutions obtained from the full simulation of the HL-LHC detector.

Compared to the CMS Run-I analysis [7], the Run-2 analysis improvements include a more advanced UML fit, an improved muon identification algorithm, and most importantly the determination of the $B^0_s \to \mu^+\mu^-$ effective lifetime.

We have estimated the evolution of the systematic uncertainties from their Run-II values to the HL-LHC era as follows: The major sources of systematic uncertainties are from external physics parameters (e.g. $f_u/f_s$ ratio and SM branching fractions of $B \to \mu^+\mu^-$ and $B^+ \to J/\psi K^+$) and those depending on internal analysis effects (e.g. individual signal and background yields, efficiencies, etc.)

The uncertainty on the muon identification (ID) efficiency ratio is determined by the difference of data and MC efficiency ratio from $[B^+ \to J/\psi K^+]/[B^0_s \to J/\psi\phi]$ and is assumed to diminish to 1% for the Phase-2 case. In this study, it is expected that the dimuon trigger for the signal and normalization channels will remain the same.

The major systematics shown in Table 1 are implemented in the PDFs as nuisance parameters. The external input to the UML, the value of $f_u/f_s$ relative uncertainty is currently known as 5.8% [18], as it is the same value used in Run-2 analysis. This uncertainty is assumed to be 3.5% for 14 TeV which is dominated by systematic uncertainties from form-factor ratios and branching fraction measurements.

The systematic uncertainty on the normalization yield enters the branching ratio formula (Eq. 3) directly. We determine this uncertainty from the yield difference between the results of the fit to the unconstrained $B^+$ and the fit to the $J/\psi$ mass-constrained $B^+$ invariant mass distribution. The Belle II collaboration is expected to be able to improve this measurement and a residual total systematic uncertainty of 1.4% seems reasonable for the Phase-2 scenario. The uncertainty due to the peaking and semileptonic backgrounds is currently dominated by the uncertainty on the hadron misidentification probability (proton, charged pion or kaon). To determine the systematic uncertainty on the muon misidentification probability for pions, kaons and protons, we calculate the bin-averaged misidentification probability for data and MC simulation. From these values and their errors, we calculate the error-weighted average and its uncertainty. The relative uncertainty of this average is used as systematic uncertainty. We assume the uncertainty on the hadron misidentification probability for kaons and pions to be 10%. As a result of this, the relative uncertainties on the yield of peaking background and semileptonic background are 20% and 15%, respectively, during the Run-2 era. As a reasonable assumption, these two uncertainties are expected to reduce by a factor of 2, resulting in estimates of 10% and 7.5%, respectively, for the Phase-2 scenario.

The selection efficiency depends on the effective lifetime of $B^0_s$ and its uncertainty, assumed to be 2% during Phase-2 era. The systematic uncertainty for the determination of the $B^0_s \to \mu^+\mu^-$ effective lifetime will be limited by the knowledge of the trigger efficiency as a function of the $B^0_s \to \mu^+\mu^-$ decay time. We expect that this uncertainty can be well measured with $B^+ \to J/\psi K^+$ decays using similar triggers (except for the dimuon mass range). The systematic uncertainty in the trigger efficiency is estimated by the difference between the dimuon trigger for signal and the dimuon trigger for $J/\psi$, which is assumed to diminish to 1.5% during the Phase-2 era.

Table 1: Input sources of systematic uncertainties and the propagated uncertainties on the $B \to \mu^+\mu^-$ branching fractions, $\delta\mathcal{B}(B^0_s \to \mu^+\mu^-)$ and $\delta\mathcal{B}(B^0 \to \mu^+\mu^-)$.

| Source | Input uncertainties | $\delta\mathcal{B}(B^0_s \to \mu^+\mu^-)$ | $\delta\mathcal{B}(B^0 \to \mu^+\mu^-)$ |
|---|---|---|---|
| Muon ID efficiency ratio | 1% | 1% | 1% |
| $B^+$ normalization yield | 1.4% | 1.4% | 1.4% |
| $f_u/f_s$ ratio | 3.5% | 3.5% | - |
| Effective lifetime | 2% | 2% | - |
| Trigger efficiency | 1.5% | 1.5% | 1.5% |
| Other sources | 3% | 3% | 3% |
| Peaking background yield | 10% | 0.5% | 2.7% |
| Semileptonic background yield | 7.5% | | |

Other sources of the systematics, e.g. acceptance, analysis selection, kaon track efficiency etc., add up to 6% (for the Run-2 case) and their impact on the final results has been studied by reducing them by a factor of two. It has been observed that the sensitivities for the significance of the $B^0$ observation and its branching fraction are not significantly affected, whereas there is a $\sim$15% improvement on the sensitivity for branching fraction of $B^0_s$.

### 3.1 Study of dimuon mass resolution

The sensitivity of the analysis to the signal is determined not only by the relative signal and background yields, but also by the mass resolution of the dimuon system. The background contribution to the $B^0 \to \mu^+\mu^-$ signal yield from the $B^0_s \to \mu^+\mu^-$ and $B^0 \to \pi^-\mu^+\nu$ decays are studied.

Both cases are strongly affected by the mass resolutions. For this study, the mass resolutions of the $B^0_s \to \mu^+\mu^-$ and $B^0 \to \mu^+\mu^-$ signal and $B^0 \to \pi^-\mu^+\nu$ backgrounds are obtained from detailed MC simulation for the Phase-2 and Run-2 scenarios. In the simulation, $pp$ collisions are generated using PYTHIA 8.212 [19] with the configuration of $2 \to 2$ QCD process. Decays of hadrons are described by EVTGEN [20] and final state radiation is generated using PHOTOS [21]. The interaction of the generated particles with the detector and its response are implemented using the GEANT4 [14] toolkit.

The mass resolutions are determined by fitting the dimuon mass with a single Gaussian function in the region $\pm2\sigma$ around the most probable mass value. The resolutions for signal $B^0_s \to \mu^+\mu^-$ and $B^0 \to \mu^+\mu^-$ processes are compared in the Phase-2 and Run-2 reconstructions in Table 2 and Fig. 1. They show significant improvements of the order 40-50% over the Run-2 scenario.

We use $|\eta_f|$, the pseudorapidity of the most forward muon (of the candidate), to visualize the results and compare the performance of Phase-2 against Run-2. In Fig. 1, the signal mass distributions for $|\eta_f| < 1.4$ are overlayed. The improved separation between $B^0 \to \mu^+\mu^-$ and $B^0_s \to \mu^+\mu^-$ in Phase-2 is evident. This will help to separate the $B^0$ signal from the tails of $B^0_s$ signal, which now becomes a background for the $B^0$ measurement. It is crucial to separate the $B^0$ and $B^0_s$ peak in the determination of the significance of the $B^0 \to \mu^+\mu^-$ observation. In Fig. 2, the improvement on the mass resolutions is shown with the $B^0_s \to \mu^+\mu^-$decay; left, the mass distributions for Run-2 and Phase-2 are overlayed within $|\eta_f| < 1.4$ and right, the resolution as a function of $\eta$ is illustrated.

Table 2: Mass resolutions for $B_s^0 \to \mu^+\mu^-$ and $B^0 \to \mu^+\mu^-$, obtained from Gaussian fits to the core of the respective mass distributions (see text for details). The last column shows the ratio between the Run-2 and Phase-2 resolutions.

| Category | Run-2 [Mev] | Phase-2 [Mev] | Ratio |
|---|---|---|---|
| $B_s^0 \to \mu^+\mu^-$, channel 0 | 37 | 26 | 1.4 |
| $B_s^0 \to \mu^+\mu^-$, channel 1 | 56 | 37 | 1.5 |
| $B^0 \to \mu^+\mu^-$, channel 0 | 37 | 26 | 1.4 |
| $B^0 \to \mu^+\mu^-$, channel 1 | 56 | 37 | 1.5 |

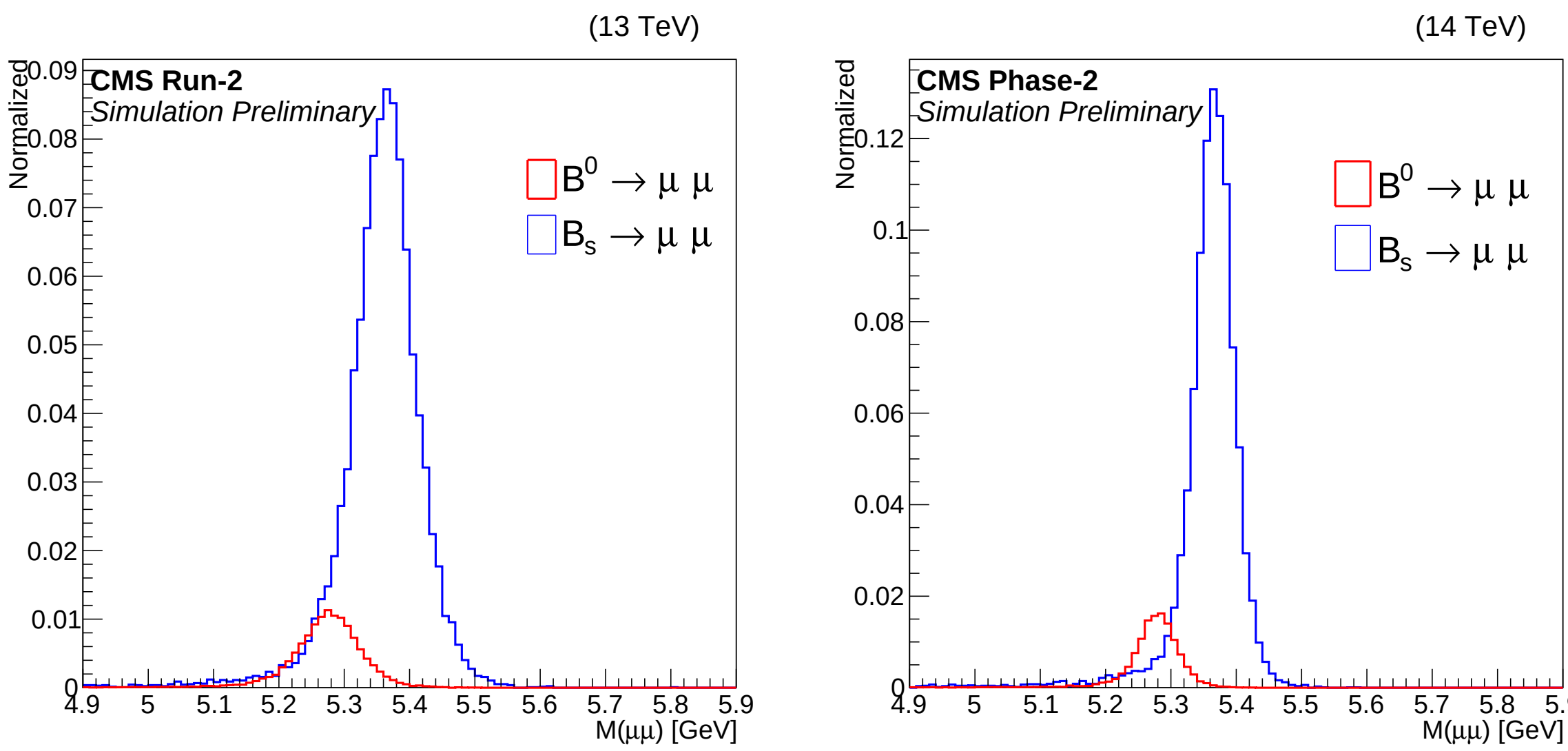

Figure 1: The left plot shows the $B_s^0$ and $B^0$invariant mass distributions in the Run-2 scenario. The right plot shows the $B_s^0$ and $B^0$ invariant mass distributions for Phase-2. The $B_s^0$ distribution is normalized to unity and the $B^0$ distribution is normalized according to the SM expectation.

The effect of the improved mass resolution is also visible in the invariant mass distribution of $B^0 \to \pi^-\mu^+\nu$ decays where the pion is misidentified as a muon. This background source is a limitation on the sensitivity of the $B^0 \to \mu^+\mu^-$ search. Even though it is not as significant in the case of $B_s^0 \to \mu^+\mu^-$, it also contributes to the $B_s^0$ signal region. In the Run-2 analysis, the $\Lambda_b \to p\mu\nu$ decay is not a large component of the semileptonic background anymore. The decay is now simulated with form factors, calculated in QCD sum rules on the light-cone (LCSR) [22], instead of phase space and its contribution is less then 10% of the total semileptonic background. With improved mass resolution of CMS Phase-2 tracker, it is expected to lower the contribution of $B^0 \to \pi^-\mu^+\nu$ background into the $B^0$ signal region by ∼30% in the mass interval $5.2 < m < 5.3$ GeV as shown in Fig 3.

### 3.2 Pile-up effects

To check the effects of the pile-up on offline reconstruction, we study the distributions of the $B_s^0$ candidate isolation discriminant in the Phase-2 scenario, respectively with no simulated pile-up or with average pile-up of 200 events per bunch crossing. Isolation parameters are important in terms of their sensitivity to pile-up, e.g. the more pile-up, the more tracks will be present in the vicinity of the B candidate. Since these parameters are not affected by high pile-up, we may conclude the signal efficiency will remain the same during the Phase-2 era. Apart from this fact, it is also one of the most important input variables of the BDT in order to separate signal

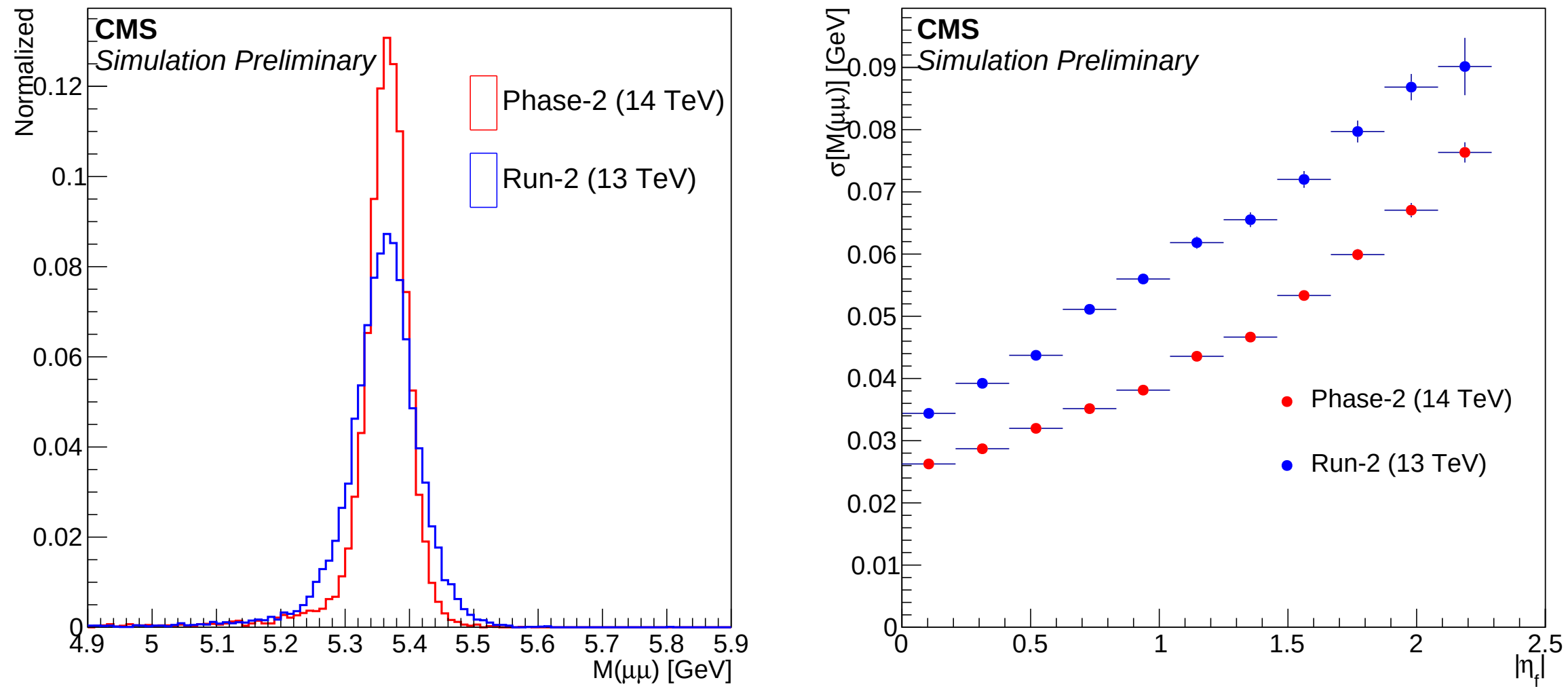

Figure 2: (left)Mass distributions for $B^0_s \to \mu^+\mu^-$ in the Run-2 and Phase-2 scenarios for $|\eta_f| < 1.4$. A single Gaussian is fit to the core of the mass distribution (see text for details). (right) Mass resolution as a function of $|\eta_f|$.

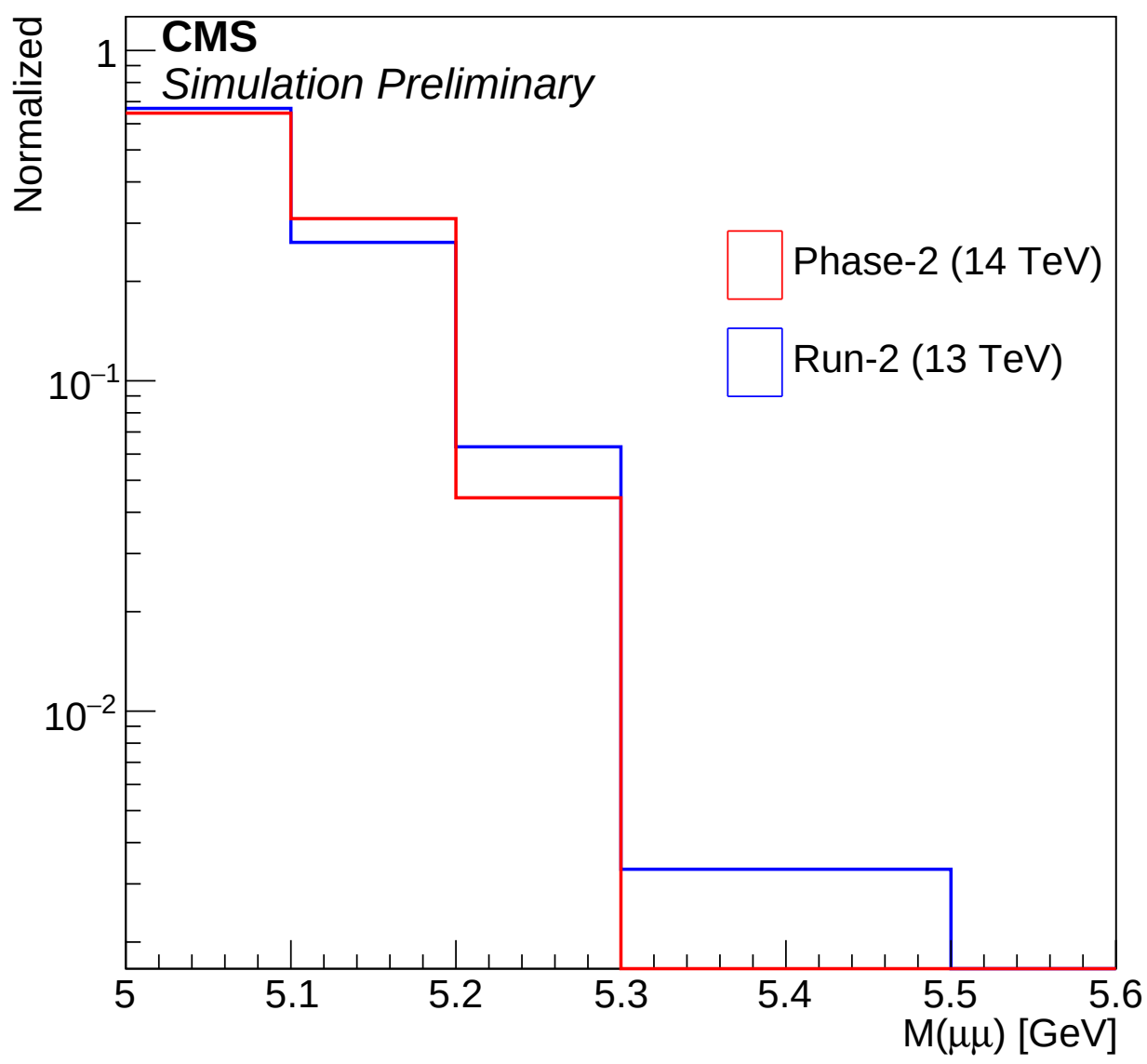

Figure 3: Contribution of $B^0 \to \pi^-\mu^+\nu$ background events (with the pion misidentified as a muon) into the signal regions. The ratio of number of $B^0 \to \pi^-\mu^+\nu$ events for Phase-2 to Run-2 is 5/19 in the mass interval $5.2 < m < 5.3$ GeV of the $B^0$ signal region.

events from background.

The definition of the B-candidate isolation is the same as in Run-2:

$$I = \frac{p_T(B)}{p_T(B) + \sum_{\text{trk}} p_T}, \tag{4}$$

where $p_{\mathrm{T}}(B)$ is the B-candidate transverse momentum. The $\sum_{\mathrm{trk}} p_T$ is extended over all the other tracks in a cone with $\Delta R = 0.7$, having $p_T > 0.9$ GeV and that are not associated to any PV but have the distance of the closest approach to the B decay vertex $d^0_{\mathrm{ca}} < 0.05$cm. The last requirement is made to minimize the pile-up dependence of the variable.

The normalized distribution of the isolation variable for the two pile-up scenarios are shown in Fig. 4. Although the PU-200 MC sample has factor of 5 less statistics than the no-PU MC sample, we have observed no significant changes on the isolation variable distributions for Phase-2.

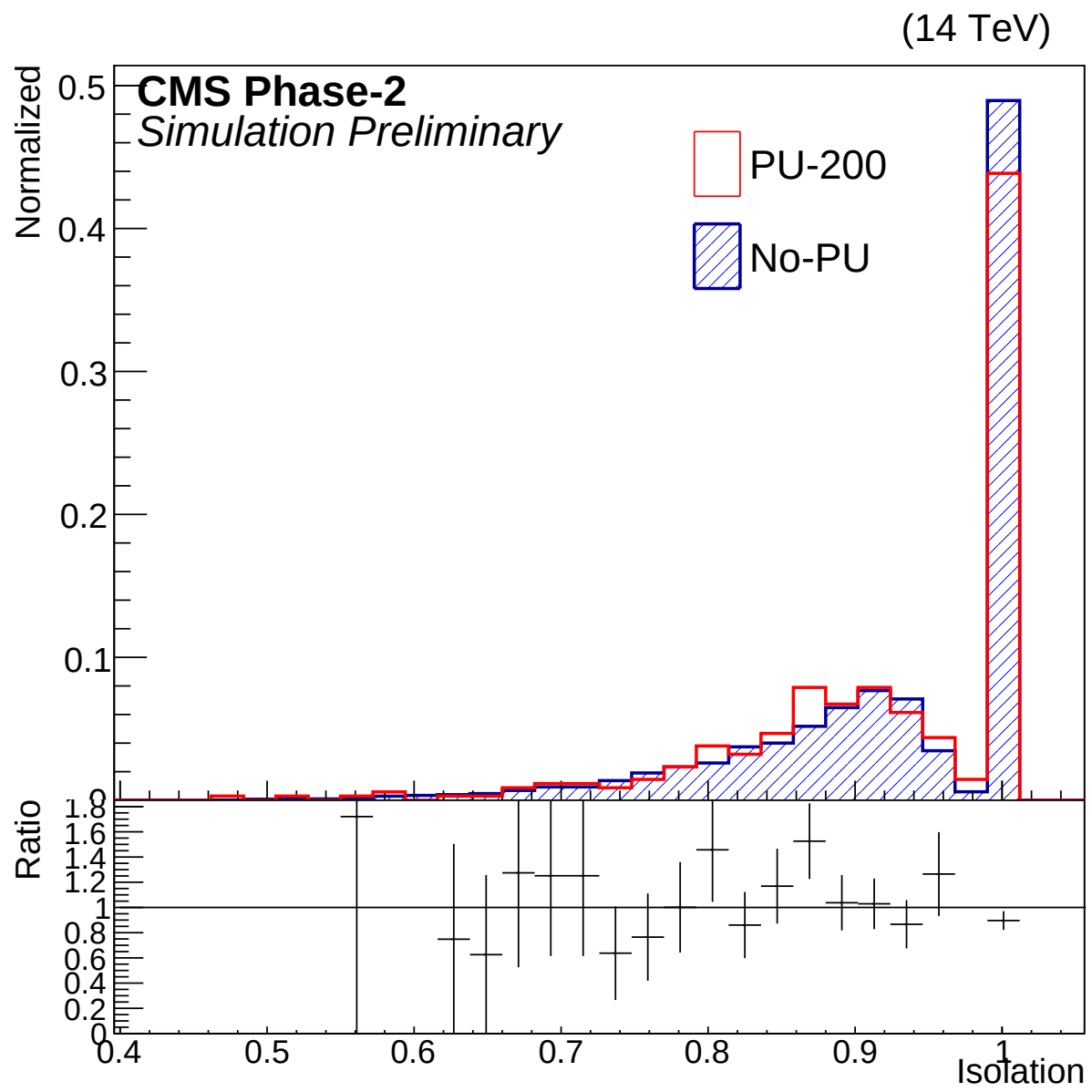

Figure 4: Normalized isolation variable distributions for the $B^0_s$ signal for the two pile-up scenarios is shown. The blue distribution represents the case with no pile-up while the red one is for average pile-up of 200 interactions per bunch crossing. In the bottom, the ratio between the PU=0 and the PU=200 distributions is also shown.

### 3.3 Sensitivity of branching fraction and decay time measurements

The expected performance of the analysis is estimated with pseudo-experiments based on toy MC, which provides a proper estimate of the statistical uncertainties.

The full simulated events are used to study the expected detector resolutions at Phase-2. This information is then used to construct the PDF models in the UML. The pseudo-experiments are carried out by generating the toy MC events based on the complete model and the expected yields. The corresponding systematic uncertainties are included as nuisance parameters (with either Gaussian or Lognormal constraints) in the likelihood fit. In the fits to each toy MC, the statistical and systematic uncertainties are evaluated together within MINUIT. The numerical values given as results are from the resulting uncertainty distributions of the fits (we are reporting median of the distribution here), and including both statistical and systematic effects.

Fig. 5 shows the invariant mass fit projections corresponding to an integrated luminosity of 3000 fb$^{-1}$ for both channels. In Fig. 6 the corresponding decay time distribution is shown with the fit projection overlayed. The expected variations from the pseudo-experiments divided by its uncertainties (pull) agree with a standard normal distribution.

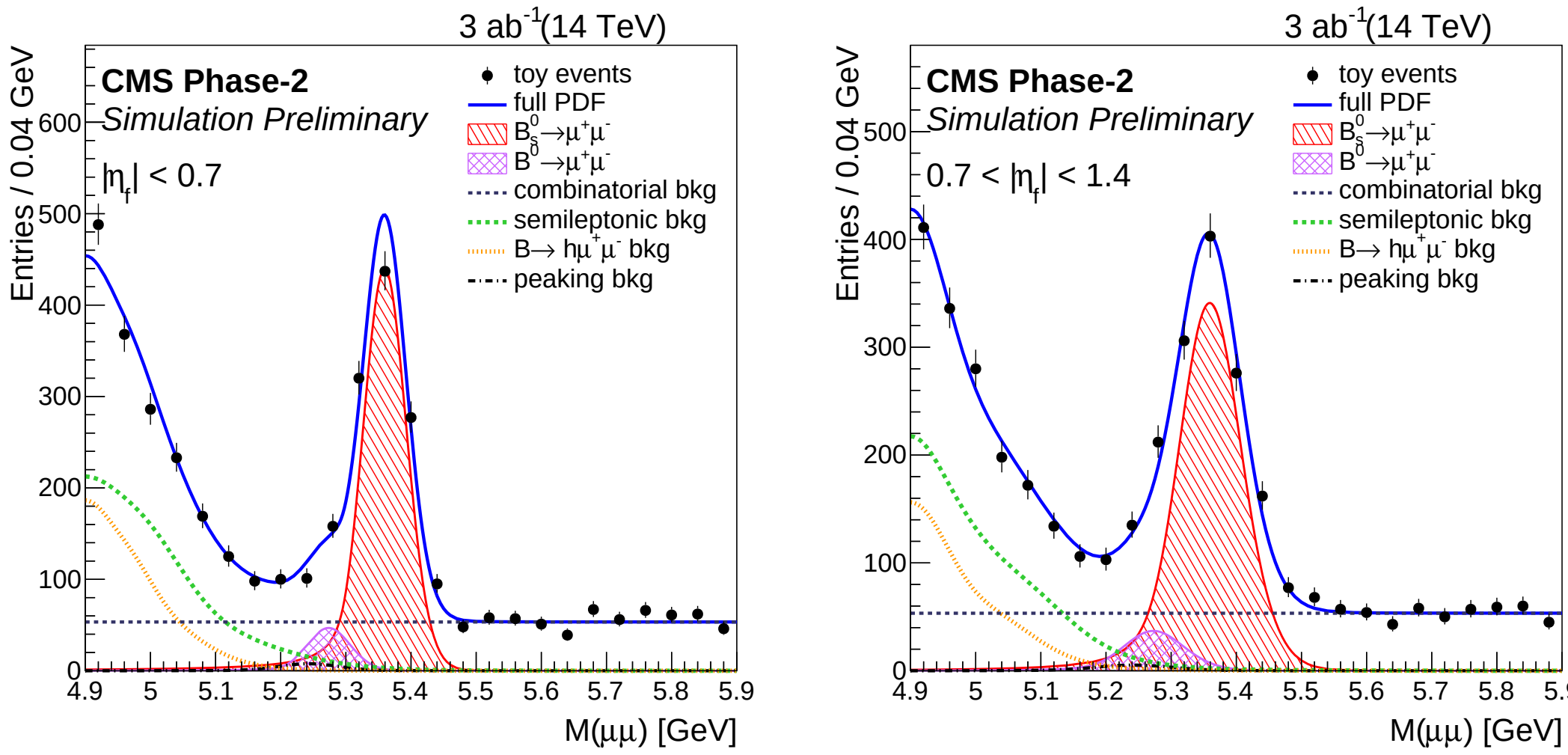

Figure 5: Invariant mass distributions with the fit projection overlayed, corresponding to an integrated luminosity of 3000 $\text{fb}^{-1}$. The left plot shows the central barrel region, $|\eta_f| < 0.7$ and the right plot is for $0.7 < |\eta_f| < 1.4$.

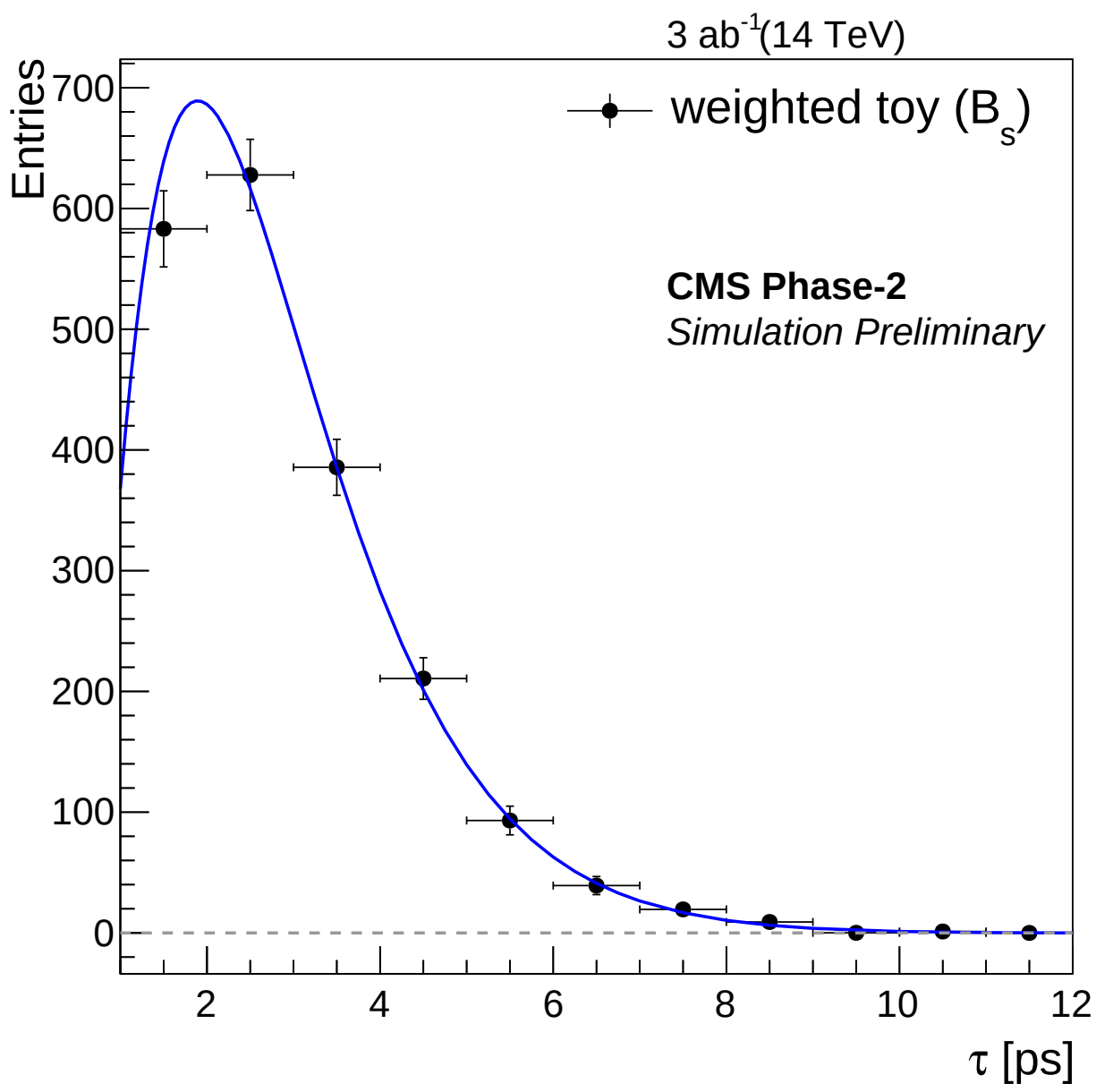

Figure 6: The binned maximum likelihood fit to the background-subtracted decay time distribution for the Phase-2 scenario. The effective lifetime from the fit is 1.61 ±0.05 ps.

We provide the sensitivities of the measurement for the $B^0_s \to \mu^+\mu^-$ effective lifetime and the branching fractions of the rare decays of $B^0_s$ and $B^0$ mesons to dimuons in Table 3. In the table, the total relative uncertainties on the branching fractions of the $B^0_s \to \mu^+\mu^-$ and $B^0 \to \mu^+\mu^-$ include both systematics and statistical uncertainties, while the absolute uncertainty on the $B^0_s$ effective lifetime is the statistical only. Based on the Run-2 analysis, it can be noted that the total uncertainty on the $B^0_s$ effective lifetime is currently dominated by the statistical uncertainty.

We have also repeated the pseudo-experiments without any systematics included. The results show that the sensitivities of the $B^0$ branching ratios and of the range of the significance of $B^0$ observation do not change significantly. Therefore, it can be concluded that they are dominated by the statistics of the total uncertainties. On the contrary, the sensitivity of the $B^0_s$ branching ratio reduces significantly that it is mostly driven by the systematic (∼75%) uncertainties.

As an additional test to investigate the effect of the improved mass resolutions on the final results, we have performed the pseudo-experiments assuming the Run-2 mass resolutions. The studies show that there is a ∼20% improvement in the sensitivity of the $B^0$ branching fraction and the significance of its observation has a ∼25% gain due to the upgraded Phase-2 CMS tracker system.

Table 3: Estimated analysis sensitivity for different integrated luminosities. Columns in the table, from left to right: the total integrated luminosity, the median expected number of reconstructed $B^0_s$ and $B^0$ mesons, the total uncertainties on the $B^0_s \to \mu^+\mu^-$ and $B^0 \to \mu^+\mu^-$ branching fractions, the range of the significance of $B^0$ observation (the range indicates the $\pm 1\sigma$ of the distribution of significance) and the statistical uncertainty on the $B^0_s \to \mu^+\mu^-$ effective lifetime.

| $\mathcal{L}$ (fb$^{-1}$) | $N(B_s)$ | $N(B^0)$ | $\delta\mathcal{B}(B_s \to \mu\mu)$ | $\delta\mathcal{B}(B^0 \to \mu\mu)$ | $\sigma(B^0 \to \mu\mu)$ | $\delta[\tau(B_s)]$(stat-only) |
|---|---|---|---|---|---|---|
| 300 | 205 | 21 | 12% | 46% | $1.4 - 3.5\sigma$ | 0.15 ps |
| 3000 | 2048 | 215 | 7% | 16% | $6.3 - 8.3\sigma$ | 0.05 ps |

## 4 Conclusions

The inner tracker of the Phase-2 detector provides an order of 40-50% improvement on the mass resolutions over the Run-2 case that will allow precise measurements of the $B^0_s \to \mu^+\mu^-$ and $B^0 \to \mu^+\mu^-$ rare decays. The semileptonic background contribution into the signal regions will be reduced substantially and the improved separation of the $B^0_s$ and $B^0$ yields will lower the signal cross feed contamination, which is crucial for the $B^0$ observation. With an integrated luminosity of 3000 fb$^{-1}$, CMS will have the capability to measure the $B^0_s \to \mu^+\mu^-$ effective lifetime with an error of about 0.05 ps and to observe the $B^0 \to \mu^+\mu^-$ decay with more than 5 standard deviation significance.

**12**

# CP-violation measurement prospects in the $B_s^0 \to J/\psi\phi$ channel with the upgraded ATLAS detector at the HL-LHC

The ATLAS Collaboration

This note estimates the ATLAS detector performance in measuring the CP-violating phase $\phi_s$ during the whole HL-LHC campaign. The study is based on simulations of the upgraded detector and projections of the collected signal data-sample.

# 1 Introduction

New phenomena beyond the predictions of the Standard Model (SM) may alter a $CP$-symmetry violation ($CPV$) in $b$-hadron decays. $CPV$ in $B_s^0 \to J/\psi\phi$ occurs due to interference between direct decay and $B_s^0$ mixing. The oscillation frequency of $B_s^0$ mixing is characterized by the mass difference $\Delta m_s$ of the heavy ($B_H$) and light ($B_L$) mass eigenstates. Other physical quantities involved in $B_s^0$ mixing are the decay width $\Gamma_s = (\Gamma_L + \Gamma_H)/2$ and the width difference $\Delta\Gamma_s = \Gamma_L - \Gamma_H$, where $\Gamma_L$ and $\Gamma_H$ are the decay widths of the $B_L$ and $B_H$ states. The $CPV$ phase $\phi_s$ is induced by the weak phase difference between the $B_s^0$ mixing and the $b \to c\bar{c}s$ decay amplitudes. In the SM $\phi_s$ is small and can be related to CKM quark mixing matrix elements via the relation $\phi_s \simeq -2\beta_s$, with $\beta_s = arg[-(V_{ts}V_{tb}^*)/(V_{cs}V_{cb}^*)]$. Assuming no BSM contributions a value of $-2\beta_s = -0.0370 \pm 0.0006$ rad can be predicted by combining beauty and kaon physics observables, see [1]. The value of $\phi_s$ is potentially expected to be sensitive to the physics phenomena Beyond SM (BSM), hence the experimental measurement of the $CPV$ phase $\phi_s$ is a viable probe of BSM effects in Heavy Flavours.

Currently all LHC $\phi_s$ measurements are consistent with the SM, see Figure 1. The combined Run 1 LHC value is $\phi_s = -0.021 \pm 0.031$ rad, taken from [2, 3]. As reported in that publication, the most precise LHCb result is obtained from the $B_s^0 \to J/\psi\phi$ channel, with some improvements deriving from the combination with additional channels. It is clear that the SM precision in predicting $\phi_s$ is still much better than the current experimental reach. Some improvements are expected from Run 2 and Run 3, however the search for New Physics will flourish with the HL-LHC upgrade.

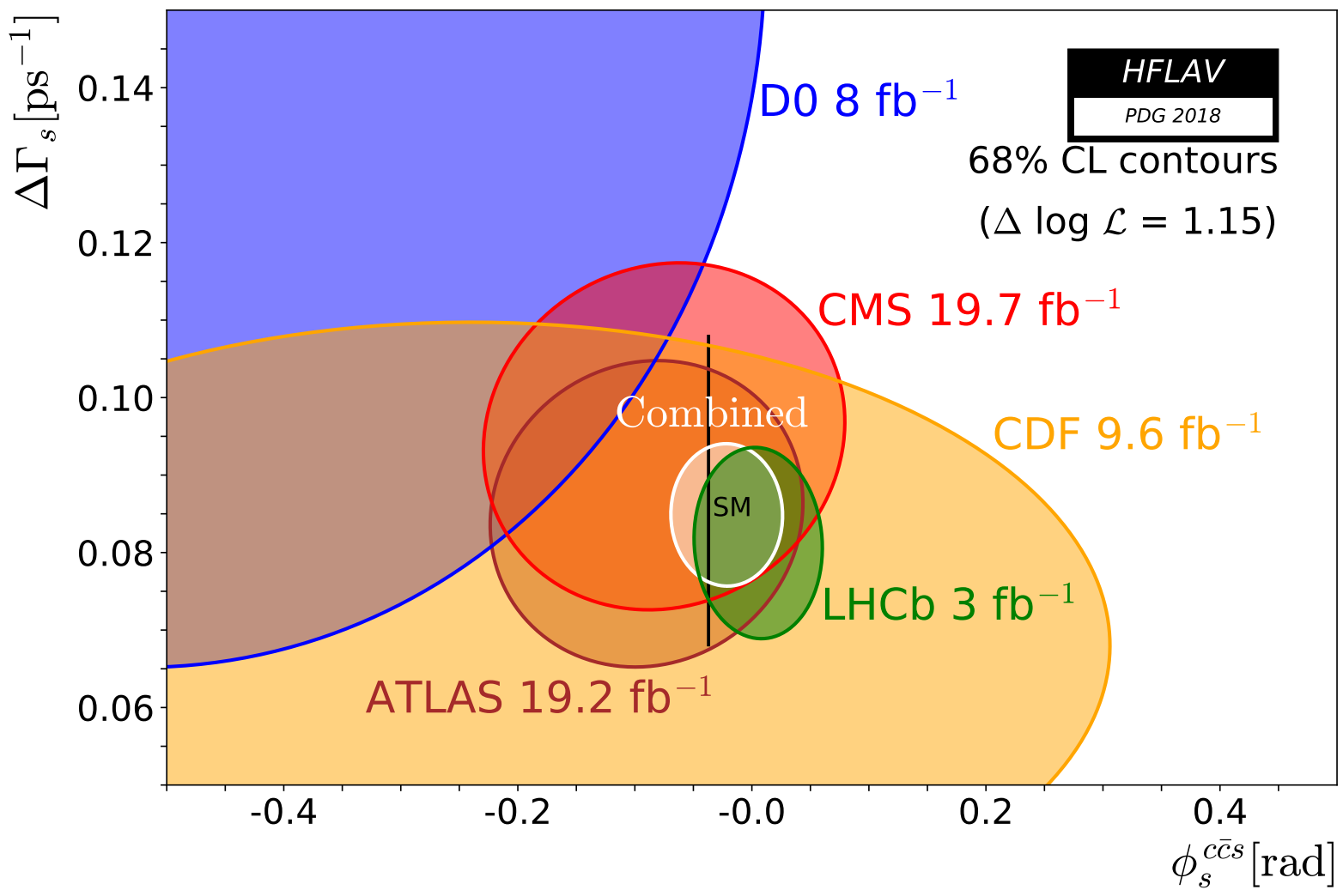

Figure 1: Current experimental summary of the $\phi_s$ measurements. Figure taken from [2, 3].

ATLAS measured in Run 1 [4] the phase $\phi_s$ and the corresponding width difference $\Delta\Gamma_s$ as:

$$\begin{aligned} \phi_s &= -0.090 \pm 0.078 \text{ (stat)} \pm 0.041 \text{ (syst)} \ \text{ rad} \\ \Delta\Gamma_s &= \ \ 0.085 \pm 0.011 \text{ (stat)} \pm 0.007 \text{ (syst)} \ \text{ ps}^{-1}. \end{aligned}$$

This document presents an estimate of the precision that ATLAS could achieve on $\phi_s$ using the $B_s^0 \to J/\psi\phi$ decay, during HL-LHC. ATLAS is a general purpose detector with central geometry, exploiting the full luminosity provided by the LHC to record, among others, a large amount of $J/\psi$ decays, leading to similarly large exclusive $B \to J/\psi X$ decay samples. In addition to the much higher number of events which will be available in comparison to the current set-up, ATLAS' impact on the $\phi_s$ measurement will be significantly affected by the upgrade foreseen for its inner detector.

The document is structuted as follows: the assumptions made in the extrapolations are explained in Section 2, followed by the description of the simulation, reconstruction and the selection of the $B_s^0 \to J/\psi\phi$ decay candidates in Section 3 and 4, and the resulting event yields in Section 5. The performance study of the upgraded detector for the $B_s^0 \to J/\psi\phi$ events in the high pile-up environment of HL-LHC is presented in Section 6. All characteristics are brought together in the Section 7 to provide an estimate of the precision for $\phi_s$ achievable with HL-LHC. The method used to estimate the $\phi_s$ precision is described in Section 7. Notes about the evolution of the systematics are summarized in Section 8. Finally, conclusions are given in Section 9.

## 2 Assumptions of the extrapolations

The experimental result on $B_s^0 \to J/\psi\phi$ is based on a simultaneous fit to the signal candidates time-dependent angular distribution. Without performing a complete analysis, the relative improvement in the $\phi_s$ precision can be projected from several key factors: the extrapolated number of signal and background events, the B-flavour tagging performance and the $B_s^0$ proper decay time resolution, all combined as described in Section 7.

The number of signal events and the signal fractions are estimated using real Run 2 data collected during 2015 and 2016, as described in Section 5. Three trigger scenarios are considered, all based on di-muon trigger targeting the $J/\psi \to \mu\mu$ decay. In the most optimistic case, the individual muon trigger transverse momentum ($p_{\mathrm{T}}$)[1] thresholds are kept at at 6 GeV, corresponding to lowest unprescaled Run 2 $B_s^0 \to J/\psi\phi$ trigger. An intermediate scenario requires in addition one of the muons to have $p_{\mathrm{T}}$ above 10 GeV. In the conservative case, only events with both muons of $p_{\mathrm{T}}$ above 10 GeV are collected. The three scenarios are hereafter denoted as μ6μ6, μ10μ6 and μ10μ10. The choice of the scenarios is based on considerations discussed in the ATLAS Phase II TDAQ TDR [5] in Section 2.10. The study here ignores possible inefficiency for close-by muons described in Section 3.2.2 of the TDAQ TDR, as this is likely to be addressed by future developments of the ATLAS upgrade triggers. In this paper, the trigger thresholds are emulated with corresponding offline-reconstruction cuts on the $p_{\mathrm{T}}$ of the muons, since more accurate simulations of the trigger selections are not yet available. Trigger muon momentum smearing relative to offline reconstruction could introduce additional effects.

The knowledge of the $B_s^0$ meson flavour at the time of production (or decay) dramatically increases sensitivity of the $B_s^0 \to J/\psi\phi$ analysis. In the current ATLAS analyses, opposite-side taggers have been developed, relying on reconstruction and flavour identification of the other $b$ hadron in the same event. Each tagging is characterized by its efficiency and purity (see Section 7). The tagging performance at HL-LHC is assumed to be similar or better than in Run 1 data, since in the future Runs new tagging algorithms (e.g. same-side flavour taggers) are expected to be developed.

[1] ATLAS uses a right-handed coordinate system with its origin at the nominal interaction point (IP) in the centre of the detector and the $z$-axis along the beam pipe. The $x$-axis points from the IP to the centre of the LHC ring, and the $y$-axis points upward. Cylindrical coordinates $(r, \Phi)$ are used in the transverse plane, $\Phi$ being the azimuthal angle around the $z$-axis. The pseudorapidity is defined in terms of the polar angle $\theta$ as $\eta = -\ln\tan(\theta/2)$.

Finally, the precision on $B_s^0$ candidates proper decay time is an important ingredient to the analysis. For the HL-LHC extrapolations this is extracted from the simulation of signal events. The resulting improvement will be demonstrated in Section 6.

## 3 Simulation samples

The projected $B_s^0 \to J/\psi\phi$ performance during the HL-LHC phase is based on signal samples simulated with the upgraded ATLAS detector layout. During the HL-LHC installation (Phase-II upgrade) the ATLAS detector systems will undergo several improvements and replacements. B-physics measurements are most effected by the replacement of the tracking system by a full-silicon detectors based tracker, ITk [6]. The pixel detector layer closest to the beam pipe is expected to consist of sensing elements of size 50 µm×50 µm and will have 39 mm radius.

The simulated samples include pile-up of 200 events, expected to be reached at the maximum HL-LHC instantaneous luminosity of $7 \times 10^{34}$ cm$^2$s$^{-1}$. The extrapolations in this paper assume a total integrated luminosity of 3000 fb$^{-1}$.

ATLAS Run 1 and Run 2 MC simulations of $B_s^0 \to J/\psi\phi$ decays are used when comparing performance with the HL-LHC set-up. In Run 2, the ATLAS tracking system was upgraded by installation of an additional layer of Pixel detectors closest to the beam pipe, and significantly improving the secondary vertex resolution [7, 8]. With ITk, the resolution is expected to further improve, as described in Section 6.

## 4 Event selection

In order to be included in the analysis, reconstructed $B_s^0$ MC events are required to contain at least one pair of oppositely charged muon candidates, reconstructed using information from the Muon Spectrometer (MS) and the Inner Detector (ID) [9]. In this analysis the muon parameters and kinematics are taken from the ID measurement alone, as the MS does not improve the precision in the momentum regime of this analysis. The pairs of tracks associated to muons are fitted to a common vertex and accepted in the analysis if the vertex fit results in $\chi^2$/d.o.f. < 10. The invariant mass of the muon pair is calculated from the track parameters after the vertex fit. The $J/\psi$ candidates are required to have a reconstructed mass within a window, that was selected to retain 99.8% of the $J/\psi$ candidates identified in the fits.

The $\phi \to K^+K^-$ candidates are reconstructed from all pairs of oppositely charged tracks with $p_{\mathrm{T}} > 1$ GeV and $|\eta| < 2.5$ that are not identified as muons and the invariant mass of the track pairs (using a Kaon mass hypothesis) falls within the interval 1.0085 GeV $< m(K^+K^-) <$ 1.0305 GeV. $B_s^0 \to J/\psi(\mu^+\mu^-)\phi(K^+K^-)$ candidates result from fitting the tracks for each combination of $J/\psi \to \mu^+\mu^-$ and $\phi \to K^+K^-$ to a common vertex. The fit is further constrained by fixing the invariant mass calculated from the two muon tracks to the world average $J/\psi$ mass [2]. These quadruplets of tracks are accepted for further analysis if the vertex fit has a $\chi^2$/d.o.f. < 3 and the invariant mass of the candidate $K^+K^-$ pair falls within the interval 1.0085 GeV $< m(K^+K^-) <$ 1.0305 GeV. $B_s^0$ candidates used in the fit are collected within a mass range of 5.15 GeV $< m(B_s^0) <$ 5.65 GeV.

The decay time of the $B_s^0$ meson is calculated as:

$$t = \frac{L_{xy}\, m_B}{c\, p_{\mathrm{T}}(B)} \tag{1}$$

where $p_{\mathrm{T}}(B)$ is the reconstructed transverse momentum of the $B_s^0$ meson candidate and $m_B$ denotes the mass of the $B_s^0$ meson, taken from Ref. [2]. The transverse decay length, $L_{xy}$, is the displacement in the transverse plane of the $B_s^0$ meson decay vertex with respect to the primary vertex, projected onto the direction of the $B_s^0$ transverse momentum. The events are simulated with the number of pile-up proton–proton interactions ranging between 190-210, corresponding to the HL-LHC conditions. The average number of reconstructed primary vertices is $N_{\mathrm{PV}} \sim 97$. The primary vertex originating the $B_s^0$ candidate needs to be identified. The variable used is the impact parameter $a_0$, which is calculated as the distance between the line extrapolated from the reconstructed $B_s^0$ meson vertex in the direction of the $B_s^0$ momentum, and each reconstructed primary vertex candidate. The chosen primary vertex is the one with the smallest $a_0$. The primary vertex position is then recalculated after removing any tracks used in the $B_s^0$ meson candidate to avoid biasing $L_{xy}$. In order to ascertain if the primary vertex selection procedure degrades the resolution the proper decay time of $B_s^0$ meson, the difference between true and the reconstructed proper decay times $\Delta_t^{\mathrm{MC-reco}} = t_{\mathrm{MC}} - t_{\mathrm{reconstructed}}$ has been studied as a function of the number of reconstructed primary vertices. The RMS of the $\Delta_t^{\mathrm{MC-reco}}$ values for each $N_{\mathrm{PV}}$ value are shown in Figure 2 (left), with the corresponding mean distribution reported on the Figure 2 (right). From these distributions it has been concluded that increasing levels of pile-up do not significantly bias or degrade resolution of the measurement of the proper time.

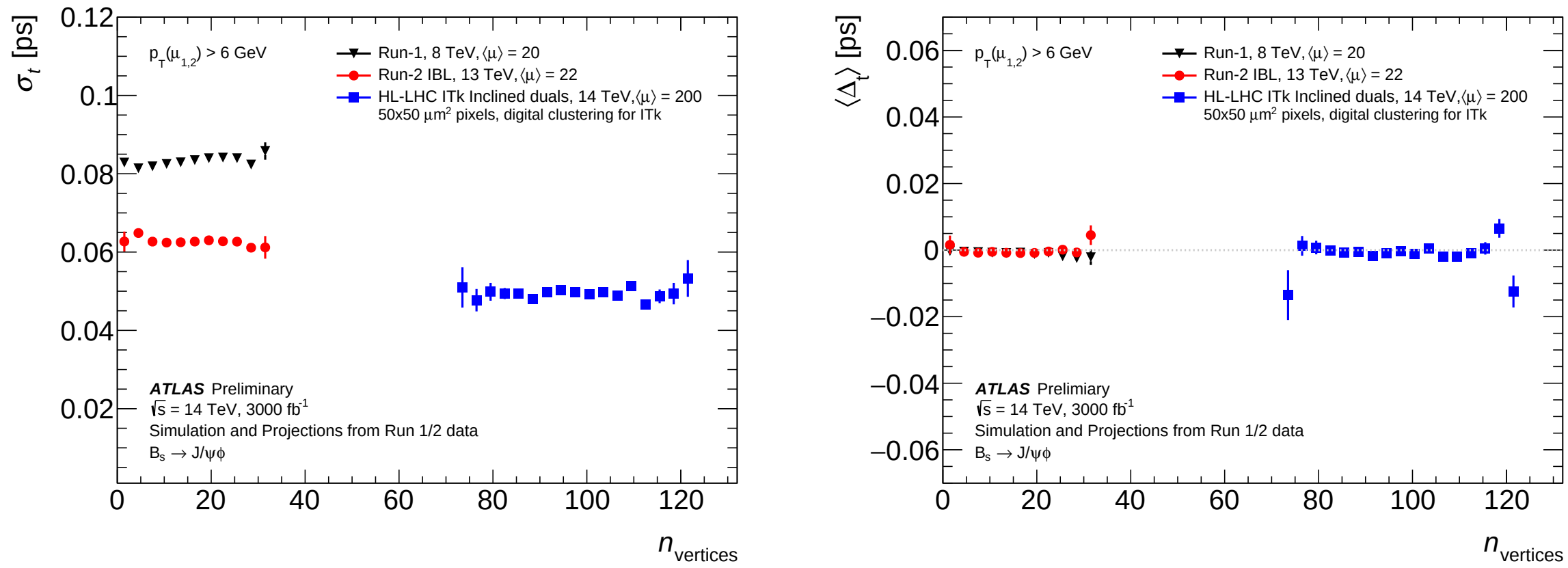

Figure 2: Dependence of the MC-true based proper decay time resolution (left) and bias of the proper decay time reconstruction (right) of the $B_s^0 \to J/\psi\phi$ on the number of reconstructed primary vertices. Run 1 (ID), Run 2 (IBL) and upgrade HL-LHC MC simulations are included for a comparison. All these samples use 6 GeV muon $p_{\mathrm{T}}$ cuts.

## 5 Event yields

The number of signal ($N_{\mathrm{sig}}$) and background events are extrapolated from real ATLAS data. In order to use as close as possible data taking conditions, the extrapolations are performed from preliminary Run 2 data: 36.2 fb$^{-1}$ collected during 2015 and 2016 $pp$ collisions at $\sqrt{s} = 13$ TeV, from Ref. [8]. The relative trigger efficiency in the trigger scenarios is determined from HL-LHC simulations. The level of background events is estimated from the Run 2 data signal fraction in events passing the three trigger conditions. The increase of $b\bar{b}$ production cross-section from 13 TeV to 14 TeV centre of mass energy is neglected. Similarly, to be conservative, the offline reconstruction efficiency of the $B_s^0 \to J/\psi(\mu^+\mu^-)\phi(K^+K^-)$ decay is assumed to be the same as in Run 2.

The resulting number of expected signal events $N_{\text{sig}}$ and the signal fractions $f_{\text{sig}}$ are shown in the Table 1 together with the results obtained in Section 7.

# 6 ATLAS Upgrade performance

The precision of the $CPV$ phase $\phi_s$ strongly depends on the $B_s^0$ meson proper decay time uncertainty. The upgraded ATLAS tracking system is expected to improve tracking and vertexing precision, as documented in the Pixel TDR [6]. These improvements are propagated to the $B_s^0$ meson reconstruction, using the dedicated HL-LHC $B_s^0 \to J/\psi\phi$ signal MC samples. The proper decay time resolution[2] as a function of the $B_s^0$ meson transverse momentum is extracted from HL-LHC simulations and compared in Figure 3 (left) with the Run 1 and Run 2 simulated detector performances. The ITk is expected to improve the proper decay time resolution by 21% and 39%, compared to Run 2 and Run 1 ATLAS tracking systems (with and without IBL). Stability of the resolution in the HL-LHC pile-up conditions ($\langle\mu\rangle = 200$) is demonstrated in the Figure 2 (left), showing the resolution as a function of the number of reconstructed primary vertices. The results shown in Figure 3 update previous studies from Ref. [6] with updated simulations of the detector geometry and more realistic material descriptions.

With the ITk is also expected to improve the $B_s^0$ invariant mass resolution by 30%. The effect of this improvement on the $\phi_s$ precision is not trivial and is neglected in this paper.

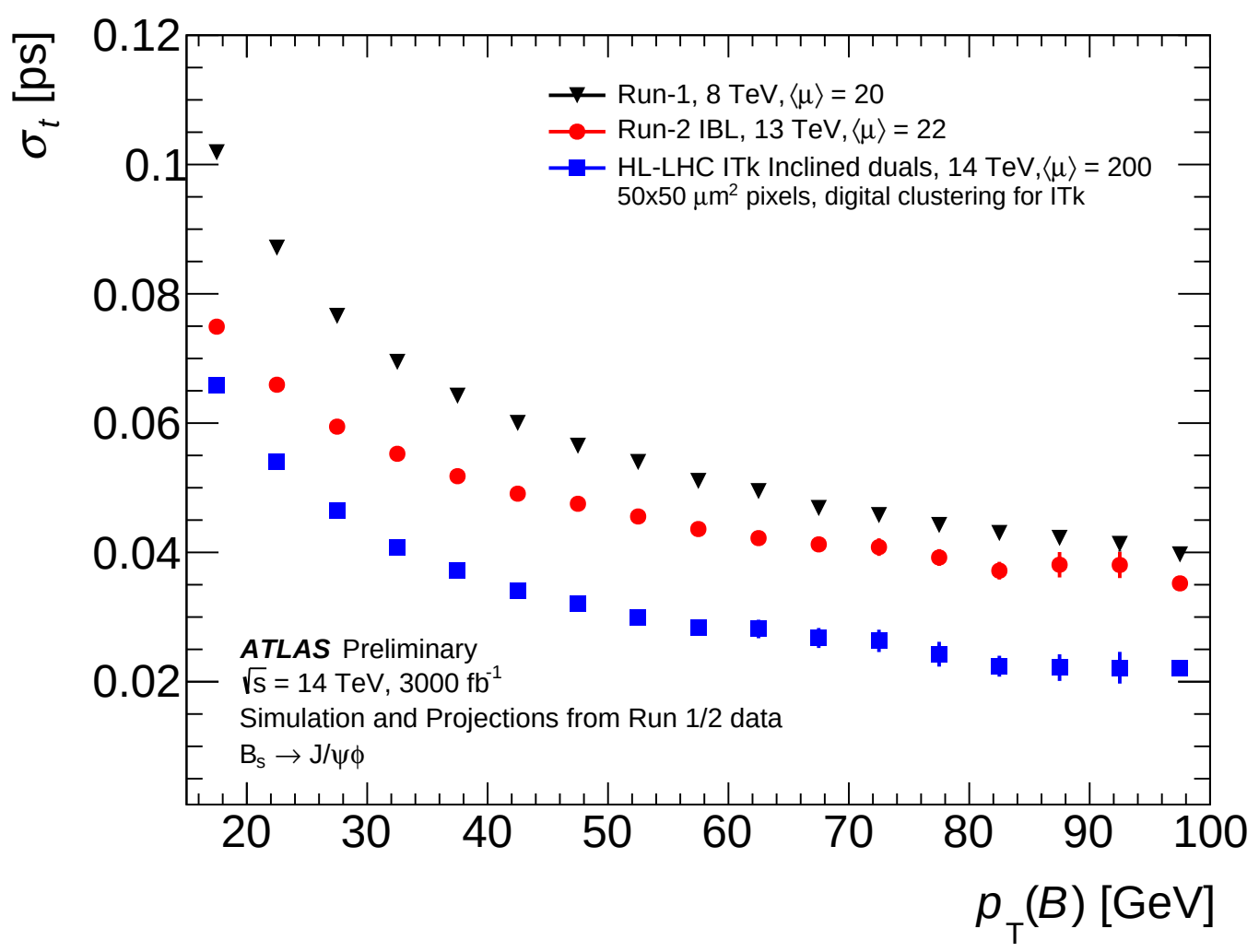

Figure 3: Dependence of the proper decay time resolution of the $B_s^0$ meson of the signal $B_s^0 \to J/\psi\phi$ decay on $B_s^0$ $p_{\text{T}}$. Per-candidate resolutions corrected for scale-factors are shown, comparing the performance in Run 1 (ID), Run 2 (IBL) and upgrade HL-LHC MC simulations. All samples use 6 GeV muon $p_{\text{T}}$ cuts.

[2] The ATLAS $B_s^0 \to J/\psi\phi$ analyses use per-candidate reconstructed proper decay time resolutions, where uncertainties on the track parameters are propagated to the proper decay time calculation. The presented proper decay time resolution is thus defined as the RMS of the multi-Gaussian distribution constructed from Gaussians with $\sigma$ equal to the per-candidate errors. Since the errors on the track parameters do not fully describe the real track resolution, the per-candidate errors are further corrected by a global scale-factor $S_t$. $S_t$ is extracted from the comparison of the per-candidate resolution and the MC-truth based resolution.

# 7 Extraction of $\phi_s$ and $\Delta\Gamma_s$ precision method and results

The main parameters driving the $\phi_s$ precision can be extracted from the events simulated and reconstructed in the HL-LHC ATLAS set-up as described in Ref. [6]. Following the estimation procedure detailed in [10], the expected precision on $\phi_s$ can be estimated from the $B_s^0 \to J/\psi\phi$ probability density function in the space of decay angles and time [4]. The estimate presented here integrates over the angular variables and assumes the world average value $\phi_s = -0.021 \pm 0.031$ rad [3], approximating $\cos(\phi_s) = 1$ and $\sin(\phi_s) = \phi_s$. Assuming also $\Gamma_H \approx \Gamma_L \approx \Gamma_s$, the following relationships hold:

$$\delta_{\phi_s}^{\text{stat}} \sim \left[\sqrt{TP} \cdot D_{\text{bck}} \cdot D_{\text{time}} \cdot \sqrt{N_{\text{sig}}}\right]^{-1} \tag{2}$$

$$TP = \epsilon \cdot (1 - 2W)^2 \tag{3}$$

$$D_{\text{bck}} = \frac{N_{\text{sig}}}{N_{\text{sig}} + N_{\text{bck}}} = f_{\text{sig}} \tag{4}$$

$$D_{\text{time}} = \exp\left[-\frac{1}{2}(\sigma_t \Delta m_s)^2\right] \tag{5}$$

where $TP$ and $D_{\text{bck}}$ account for the flavour tagging power and the signal purity, respectively. $\epsilon$ denotes the tagging efficiency, $W$ the wrong tag fraction, $N_{\text{sig}}$ and $N_{\text{bck}}$ the numbers of signal and background events and $f_{\text{sig}}$ the signal fraction. The factor $D_{\text{time}}$ accounts for dilution effects due to proper time resolution as a convolution of the time dependent part of the decay amplitude: $\exp(-\Gamma_s\tau) \cdot \sin(\Delta m_s \tau)$ with a Gaussian function of width $\sigma_t$ (the proper decay time measurement resolution); $\Delta m_s = 17.757 \pm 0.021$ ps$^{-1}$ is the mass difference between the heavy and light $B_s^0$ mass states.

The predicted $\delta_{\phi_s}^{\text{stat}}$ values are calculated relative to the value measured using 2012 data, published in [4].

$$\delta_{\phi_s}^{\text{stat}} = \delta_{\phi_s}^{\text{stat}}(12) \cdot \frac{\sqrt{TP(12)} \cdot f_{\text{sig}}(12) \cdot \exp\left[-\frac{1}{2}(\sigma_t(12)\Delta m_s)^2\right] \cdot \sqrt{N_{\text{sig}}(12)}}{\sqrt{TP} \cdot f_{\text{sig}} \cdot \exp\left[-\frac{1}{2}(\sigma_t \Delta m_s)^2\right] \cdot \sqrt{N_{\text{sig}}}} \tag{6}$$

While $f_{\text{sig}}$ has been verified to not depend appreciably on pile-up, it is in principle dependent on the centre of mass energy (since this affects the particles multiplicity per hard scatter and hence e.g. the primary vertex detection efficiency and accuracy). However this extrapolation is small when going from 13 TeV to 14 TeV and it is neglected. The prediction method has been validated using 2011 data, for which the predicted $\delta_{\phi_s}^{\text{stat}}$ (see Table 1), is consistent with the measurement, published in [11]. HL-LHC extrapolations are reported in Table 1 for the three different trigger muon thresholds: 6 GeV-6 GeV, 6 GeV-10 GeV and 10 GeV-10 GeV. Table 1 includes also the predicted precision on the measurement of $\Delta\Gamma_s$, which scales proportionally to $(\sqrt{N_{\text{sig}}})^{-1}$.

Since the performance of the flavour tagging methods is highly dependent on the environment, which can not yet be precisely extracted from current upgrade simulation and reconstruction tools (and must be typically calibrated on the data itself), and also the potential for development of further tagging algorithms, such as same-side tagging (not included in Run-1 measurements), the precision on $\phi_s$, $\delta_{\phi_s}^{\text{stat}}$, is presented in the Figure 4 for a broad range of $TP$ values for each of the trigger threshold scenarios considered. The lower $TP$ bound is given conservatively by muon tagger only performance in the HL-LHC simulation. The upper bound considers the potential increased $TP$ in comparison to Run 1 value due to inclusion of same-side tagging methods, which have not been included in Run 1. For the results in Table 1, and the contours of Figure 4, the nominal value of the $TP$, taken from Run 1, is assumed.

| Period | $L_{\text{int}}$ [fb$^{-1}$] | $N_{\text{sig}}$ | $f_{\text{sig}}$ | Tag Power [%] | $\sigma(\tau)$ [ps] | $\delta^{\text{stat}}_{\phi_s}$ [rad] measured (extrapolated) | $\delta^{\text{stat}}_{\Delta\Gamma_s}$ [ps$^{-1}$] measured (extrapolated) |
|---|---|---|---|---|---|---|---|
| 2012 | 14.3 | 73693 | 0.20 | 1.49 | 0.091 | 0.082 | 0.013 |
| 2011 | 4.9 | 22690 | 0.17 | 1.45 | 0.100 | 0.25 (0.22) | 0.021 (0.023) |
| | | | | | | $\delta^{\text{stat}}_{\phi_s}$ [rad] extrapolated | |
| HL-LHC | 3000 | | | | | | |
| Trigger μ6μ6 | | $9.72 \cdot 10^6$ | 0.17 | 1.49 | 0.048 | 0.004 | 0.0011 |
| Trigger μ10μ6 | | $5.93 \cdot 10^6$ | 0.17 | 1.49 | 0.044 | 0.005 | 0.0014 |
| Trigger μ10μ10 | | $1.75 \cdot 10^6$ | 0.15 | 1.49 | 0.038 | 0.009 | 0.003 |

Table 1: Table summarising $B_s^0 \to J/\psi\phi$ performance for existing data and predictions for HL-LHC. The precision on $\phi_s$ is statistical only.

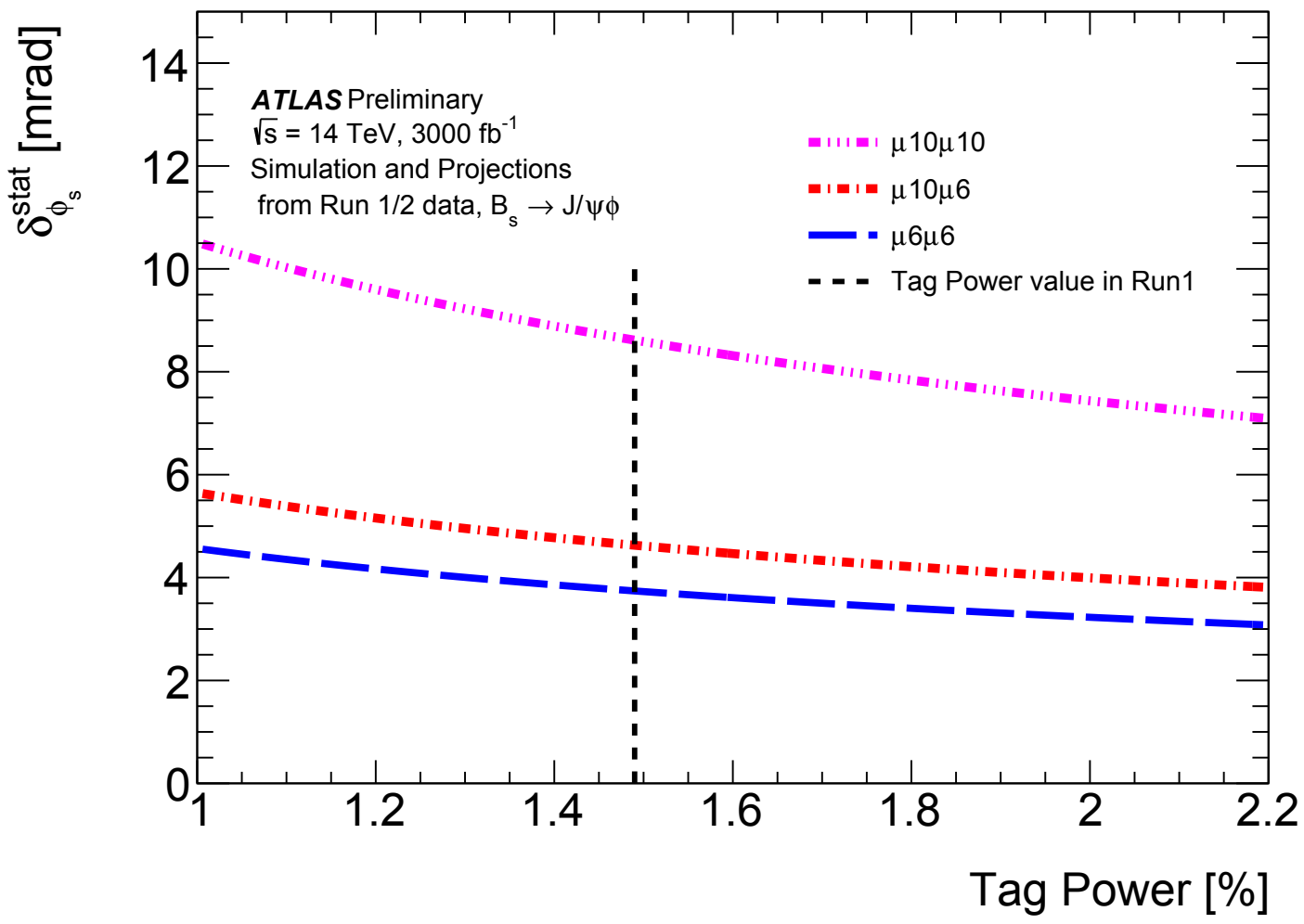

Figure 4: Dependence of the $\phi_s$ precision, $\delta^{\text{stat}}_{\phi_s}$, on Tag Power ($TP$), for a broad range of $TP$ values for each of the upgrade trigger threshold scenarios.

# 8 Systematic uncertainties

All the significant sources of systematic uncertainties considered in the Run 1 result can be potentially improved with the higher statistics of data collected at the HL-LHC:

- Fit model systematic uncertainties rely on the constraints of the background PDF coming from the signal sidebands in data and hence will be better constrained as signal statistics increases.
- The uncertainty due to flavour tagging is similarly data-driven, since the flavour tagger is calibrated on $B^{\pm}$ candidates in data.
- Detector acceptance uncertainty is determined by MC statistics as well as data-MC comparison, which will improve with larger data and MC samples.

- Detector alignment systematic uncertainties are determined from data-driven alignment techniques, thus possible improving with larger calibration samples and improved techniques.
- Systematic uncertainties due to peaking backgrounds (e.g. $B_d^0$ and $\Lambda_b$ decays) are driven by these modes' branching ratio uncertainties and are thus expected to improve with HL-LHC data collected by ATLAS and other experiments.

A realistic estimate of the overall systematics uncertainty can be thus made by scaling most of the systematics from the measurement at $\sqrt{s}$ = 8 TeV with the inverse square root of the ratio of the integrated luminosities. The MC-statistics dependent systematic uncertainties are assumed to be negligible. In a conservative approach, the detector alignment systematic uncertainties are kept at the value of the Run 1 analysis, yielding $\delta_{\phi_s}^{\text{syst}} \approx 0.006$ rad and $\delta_{\Delta\Gamma_s}^{\text{syst}} \approx 0.0005$ ps$^{-1}$. However, a preliminary method [12] of correcting for radial bias in the alignment has been developed for Run 2 and has the potential to reduce the detector alignment systematic by factor of $\sim 4\times$. Considering this improvement the total systematic errors yields $\delta_{\phi_s}^{\text{syst}} \approx 0.003$ rad and $\delta_{\Delta\Gamma_s}^{\text{syst}} \approx 0.0005$ ps$^{-1}$. The $B_s^0 \to J/\psi\phi$ analysis is thus expected to be limited mainly by the statistical precision.

Figure 5 shows the extrapolated ATLAS precision (combining statistical and systematic uncertainties) overlaid with the present precision on $\phi_s$ and $\Delta\Gamma_s$. Since the correlation between $\phi_s$ and $\Delta\Gamma_s$ in ATLAS Run 1 result has found to be smaller than 10%, the extrapolated contours of ATLAS HL-LHC are made with zero correlation.

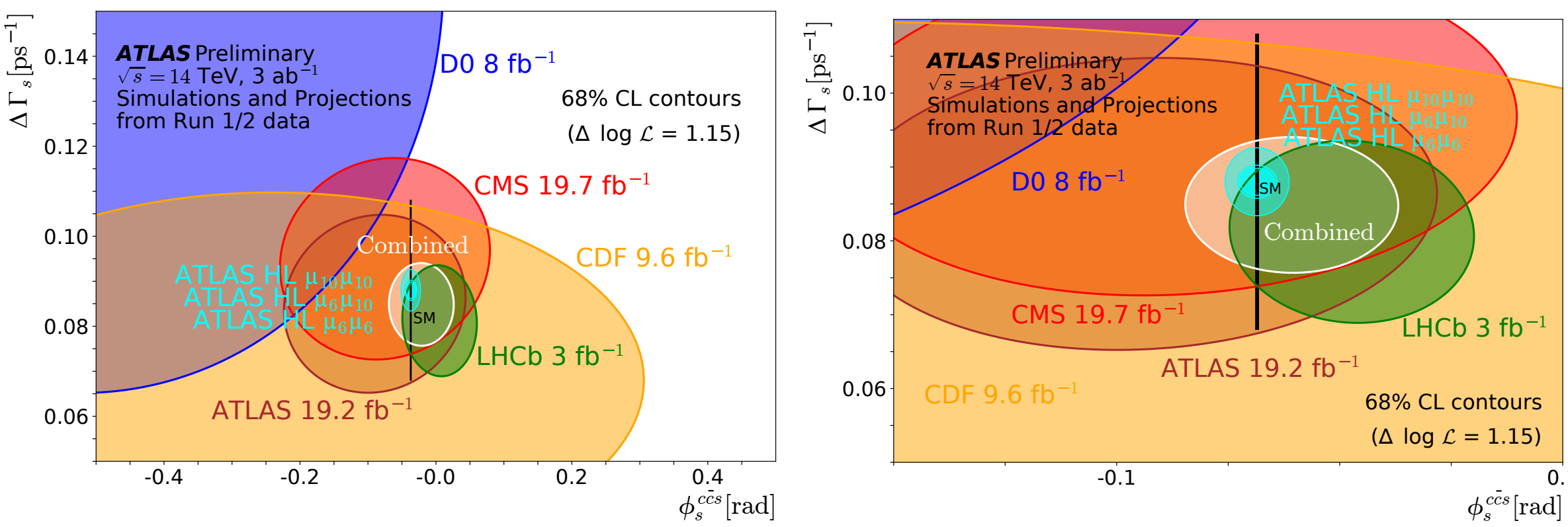

Figure 5: Current experimental summary of the $\phi_s$ measurements with superimposed ATLAS HL-LHC extrapolations, including both the projected statistical and systematic uncertainties. Modified figure, original taken from [2, 3].

## 9 Conclusions

The precision of the measurement of the $CP$-violating phase $\phi_s$ in the $B_s^0 \to J/\psi\phi$ decay at the upgraded ATLAS detector at High-Luminosity LHC is presented. The projections account for the most relevant improvements to the detector performance, particularly in relation to the $B_s^0$ meson proper decay time resolution. The size of the signal sample will strongly depend on the trigger thresholds. Three trigger scenarios are considered, providing optimistic (Run-2 like), intermediate and conservative estimates. The High-Luminosity LHC statistical precision on $\phi_s$ is expected to improve relative to the ATLAS Run 1 result by a factor ranging between 9× to 20× depending on the trigger scenario. Similarly, the improvement in the statistical precision on $\Delta\Gamma_s$ is ranging between 4× to 10×. The measurement remains dominated by the statistical precision. In the most optimistic trigger scenario, the total uncertainty on $\phi_s$ will be 8× larger than the current theoretical uncertainty and 7× smaller than the predicted $\phi_s$ value from the Standard Model.

# CMS Physics Analysis Summary

Contact: cms-phys-conveners-ftr@cern.ch 

# CP-violation studies at the HL-LHC with CMS using $B_s^0$ decays to $J/\psi\phi(1020)$

The CMS Collaboration

## Abstract

We have estimated the expected sensitivity on the CP-violating phase $\phi_s$ measured in the decay channel $B_s^0 \rightarrow J/\psi\phi(1020)$ in pp collisions with the CMS detector at the end of the HL-LHC data-taking with 3 ab$^{-1}$ of collected data. The sensitivity on $\phi_s$ mainly depends on the collected statistics, on the flavour-tagging power, and on the proper-decay-time resolution. The study is performed using fully simulated signal events and toy Monte Carlo experiments, for a few assumed tagging scenarios. The sensitivity on $\phi_s$ is expected to be in the 5-6 mrad range, which improves the current world average uncertainty by a factor of five.

# 1 Introduction and motivations

The $B_s^0 \to J/\psi\phi(1020)$ decay is considered the golden channel for the study of CP–violation in the $B_s^0$ sector. A measurable phase $\phi_s$ arises from the interference between the $B_s^0$-$\overline{B}_s^0$ oscillation and the decay of the neutral meson via the $b \to c\bar{c}s$ transition, which allows the final state to be the same for mixed and unmixed mesons. Neglecting sub–leading penguin contributions, the Standard Model (SM) CP–violating phase $\phi_s$ is predicted to be equal $-2\beta_s$, where $\beta_s = \arg[-(V_{ts}V_{tb}^*)/(V_{cs}V_{cb}^*)]$ and $V_{ij}$ are the CKM matrix elements.

SM global fits to experimental heavy flavour data allow to infer the value of $-2\beta_s$, which is currently determined to be $-36.86^{+0.96}_{-0.68}$ mrad [1]. Any significant deviation measured from such a precisely predicted value might be interpreted as evidence of physics beyond the SM, which would affect the CP–violating phase through the contribution of exotic particles in loop diagrams describing $B_s^0$ mixing. The theoretical prediction for the decay width difference $\Delta\Gamma_s$ between the light and heavy mass eigenstates is $\Delta\Gamma_s = (0.085 \pm 0.020)\ \mathrm{ps}^{-1}$ [2]. The current determination of $\phi_s$ from combined experimental $b \to \bar{c}cs$ measurements by CDF, D0, ATLAS, CMS, LHCb, is $-21 \pm 31$ mrad [3].

In CMS, $\phi_s$ has been measured [4] from a fit of the angular distribution of the $B_s^0 \to J/\psi\phi(1020)$ decay products as a function of the $B_s^0$ decay time, with the $J/\psi$ decaying to $\mu^+\mu^-$ and the $\phi(1020)$ meson decaying to $K^+K^-$. Since the $B_s^0$ is a pseudo–scalar decaying to two vector mesons, the final state is an admixture of CP–even and CP–odd states with orbital angular momentum of the two–mesons system $L = 0, 1$ or 2. The two kaons from the $\phi(1020)$ decay are produced in a $P$ wave due to angular momentum conservation, but an S–wave component can be present in the final state due to a non–resonant contribution. This leads to an angular distribution containing many terms, produced by the various angular momentum components in the final state, whose contributions have to be determined. The definition of the decay angles and the functional form of the expected contributions to the measured angular distribution are described in reference [4].

This note presents an estimate of the precision attainable by CMS in the measurement of the CP–violation angle $\phi_s$ by the end of the High–Luminosity LHC phase (HL–LHC, starting in 2026) considering an integrated luminosity of 3 $\mathrm{ab}^{-1}$ of pp collisions at 14 TeV. An estimate of the sensitivity $\sigma_{\phi_s}$= 3 mrad at the end of the HL–LHC phase has been published by the LHCb collaboration [5]. CMS estimate is carried out by using fully simulated signal events and toy Monte Carlo (MC) pseudo–experiments, for three different tagging scenarios described briefly hereafter. In the following, the HL–LHC era will also be referred to as “Phase 2”, while “Phase 1” refers to the 2017–2023 running period.

The CMS detector [6] will be substantially upgraded in order to fully exploit the physics potential offered by the increase in luminosity, and to cope with the demanding operational conditions at the Phase 2 [7–11]. The upgrade of the first level hardware trigger (L1) will allow for an increase of L1 rate and latency to about 750 kHz and 12.5 $\mu$s, respectively, and the high-level software trigger (HLT) is expected to reduce the rate by about a factor of 100 to 7.5 kHz. The entire pixel and strip tracker detectors will be replaced to increase the granularity, reduce the material budget in the tracking volume, improve the radiation hardness, and extend the geometrical coverage and provide efficient tracking up to pseudorapidities of about $|\eta| = 4$. The muon system will be enhanced by upgrading the electronics of the existing cathode strip chambers (CSC), resistive plate chambers (RPC) and drift tubes (DT). New muon detectors based on improved RPC and gas electron multiplier (GEM) technologies will be installed to add redundancy, increase the geometrical coverage up to about $|\eta| = 2.8$, and improve the trigger and reconstruction performance in the forward region. Finally, the addition of a new timing detector for minimum ionizing particles (MTD) in both barrel and endcap regions is envisaged to provide the capability for 4-dimensional reconstruction of interaction vertices that will significantly offset the CMS performance degradation due to high PU rates.

A detailed overview of the CMS detector upgrade program is presented in Ref. [7–11], while the expected performance of the reconstruction algorithms and pile-up mitigation with the CMS detector is summarised in Ref. [12].

## 2 Analysis strategy

To study the expected detector performance in Phase 2, a dedicated Monte Carlo sample generated with a centre–of–mass energy of 14 TeV and ideal Phase–2 detector conditions was used.

As far as this analysis is concerned, the most important upgrade of the CMS detector is undoubtedly the Silicon tracker (with its L1 trigger capabilities). The better hit resolution of the Phase 2 tracker and the reduction of the material budget ($\sim$ 50% of the current tracker) result in significant improvements of the momentum and transverse impact parameter resolution with respect to the Phase 1 detector. This in turn results in substantial improvements of the mass and lifetime resolutions, as well as more precise kinematic measurements, such as angular distributions [13]. In addition, the larger pseudorapidity coverage (up to $|\eta| = 4$) will increase the acceptance for track reconstruction.

The new L1 trigger capability to reconstruct charged tracks above 2 GeV in transverse momentum ($p_{\mathrm{T}}$) with almost offline–like resolutions, will be able to provide a clean $\mathrm{J}/\psi$ sample. Preliminary studies of the transverse impact parameter reconstruction at L1 indicate a resolution between 100 and 300 $\mu$m, which can be used to reduce the prompt $\mathrm{J}/\psi\phi(1020)$ component at L1 if needed. The Phase 2 L1 (hardware) and HLT (software) trigger performances are expected to be comparable to those during Run–2, and sustainable in terms of rates. The offline selections could therefore be identical to those used in the 2012 data analysis [4] and we assume no difference in the signal over background ratio with respect to 2012 data. The latter assumption is also motivated by the future presence of the timing layer, which will mitigate the background pollution due to tracks coming from pile–up vertices. Figure 1 shows the expected performance in $\mathrm{B}^0_\mathrm{s}$ invariant mass resolution and proper decay length uncertainty.

The figure of merit for the sensitivity for the Phase 2 analysis for tagged events can be estimated using the following relation [14]:

$$S \propto \sqrt{\frac{\epsilon D^2 N_S}{2}} \sqrt{\frac{N_S}{N_S + N_{BG}}} e^{-\frac{\sigma_t^2 \Delta m_s^2}{2}} \tag{1}$$

where $\epsilon$ is the flavour–tagging efficiency (for details on flavour tagging algorithms see for instance [15]), $D = 1 - 2\omega$ is the dilution factor, $\omega$ the wrong tag fraction, $N_S$ and $N_{BG}$ are the signal and background yields respectively, while $\sigma_t$ is the proper time resolution. Improvements are expected to come from the much larger signal sample expected in Phase 2 and sizable enhancement of the lifetime resolution. Improvements of the tagging algorithms are also foreseen for the Phase 2 upgrade, however no detailed, quantitative, study is available.

According to equation (1) the main ingredients in the sensitivity estimation are the number of signal events, the tagging performance (efficiency $\epsilon$ and mistag $w$), and the proper decay length uncertainty. In this study the expected sensitivity for Phase 2 is estimated using the toy MC

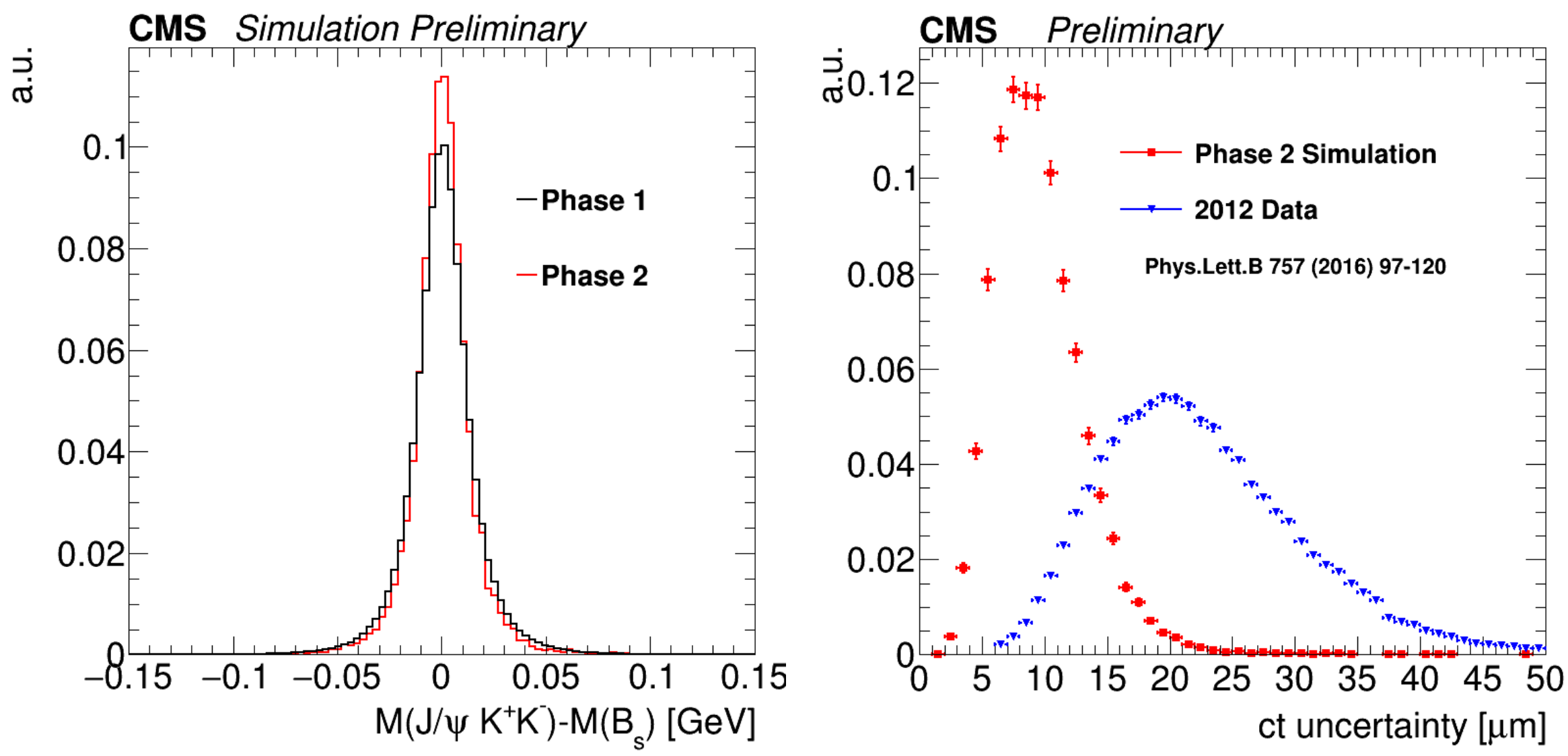

Figure 1: Left: invariant mass resolution in the Phase 2 sample compared with Phase 1 case. Right: $c\tau$ uncertainty distribution in 2012 data (blue) and Phase 2 MC (red) samples. The better performance of Phase 2 w.r.t. 2012 data is due to the Phase 2 tracker.

pseudo–experiment technique. A set of toy MC samples is generated using the signal model of $B_s^0 \rightarrow J/\psi\phi(1020)$ decay described in the 2012 data analysis document [4]. Each sample consists of about 9 million signal events, corresponding to the expected yield after 3 $ab^{-1}$ of integrated luminosity, which is a conservative assumption based on 2012 and 2018 data rates. The proper decay length uncertainty was estimated using a MC signal sample with a GEANT4 simulation of an ideal Phase 2 detector response, figure 1 (right). The flavour tagging dilution was included in the toy MC production by fixing it to its "effective" value (the constant value of the dilution which reproduces the same effect on the $\phi_s$ accuracy of the per–event dilution). The generated toy MC samples were fitted using the same signal model used for their production. The fit extracts $\phi_s$ and $\Delta\Gamma_s$ for each toy experiment. The resulting $\phi_s$ uncertainty distribution was fitted using a Landau function and the most probable value was used to determine the sensitivity on $\phi_s$ for the set of toy MC samples.

## 3 Systematics

The uncertainty on 2012 data CMS analysis [4] was driven by the statistical uncertainty. The main sources of systematic uncertainty on the CP–violating phase $\phi_s$ were the angular efficiency, the fitting model, and the discrepancy between the $p_T$-distribution of the kaons in the MC and the data. The angular efficiency systematic uncertainty was mainly due to the limited statistics of the signal MC sample used to estimate the efficiency function. This can be reduced to acceptable values by increasing the MC statistics, and through data–driven techniques to estimate/calibrate the angular efficiency. The fitting model uncertainties can be reduced by optimizing the choice of fitted parameters, and by improving the likelihood fit function.

The flavour tagging systematic uncertainty is directly correlated with the $\phi_s$ measurement. The systematic uncertainty due to flavour tagging was 3 mrad in the 2012 CMS analysis. That uncertainty depends mainly on the statistics of the calibration channel used to tune the flavour tagging tool. We do not know at present what the calibration channel statistics will be (in the 2012 analysis the $B^\pm \rightarrow J/\psi K^\pm$ channel was used) but we can reasonably assume the total statistics collected by the end of Phase 2 to be 10 times that of 2012, which will reduce the flavour tagging systematics to below 1 mrad. All these systematic uncertainties can be reasonably kept under control, and the total uncertainty on $\phi_s$ could still be statistically limited at the end of Phase 2.

# 4 Results

Three different scenarios have been tested: in scenario *a* we tested the performance of a flavour tagging based on muons and jet–charge; in scenario *b* we used muon and electron flavour tagging – as in the 2012 analysis – while scenario *c* assumed a well performing flavour tagging based on leptons, jet–charge, and same side jet–charge/kaon tagging. See Table 1 for the details of the flavour tagging performance in each scenario.

Figure 2 (left) shows the statistical uncertainty for the value of $\phi_s$ obtained in the different tagging scenarios. From the results of the studied scenarios we can estimate the $\phi_s$ uncertainty to be in the range 5–6 mrad.

The $\phi_s$ measurement is usually shown in a $\phi_s$–$\Delta\Gamma_s$ plane. Figure 2 (right) shows the expected sensitivity in the $\phi_s$–$\Delta\Gamma_s$ plane at the end of HL-LHC program obtained from the fit of a toy MC pseudo-experiment generated in the tagging scenario *c*. The contour combines the expected statistical and systematic uncertainties. The uncertainty on $\phi_s$ is expected to be dominated by statistics, while the systematics on $\Delta\Gamma_s$ is assumed to equal in size the statistical uncertainty.

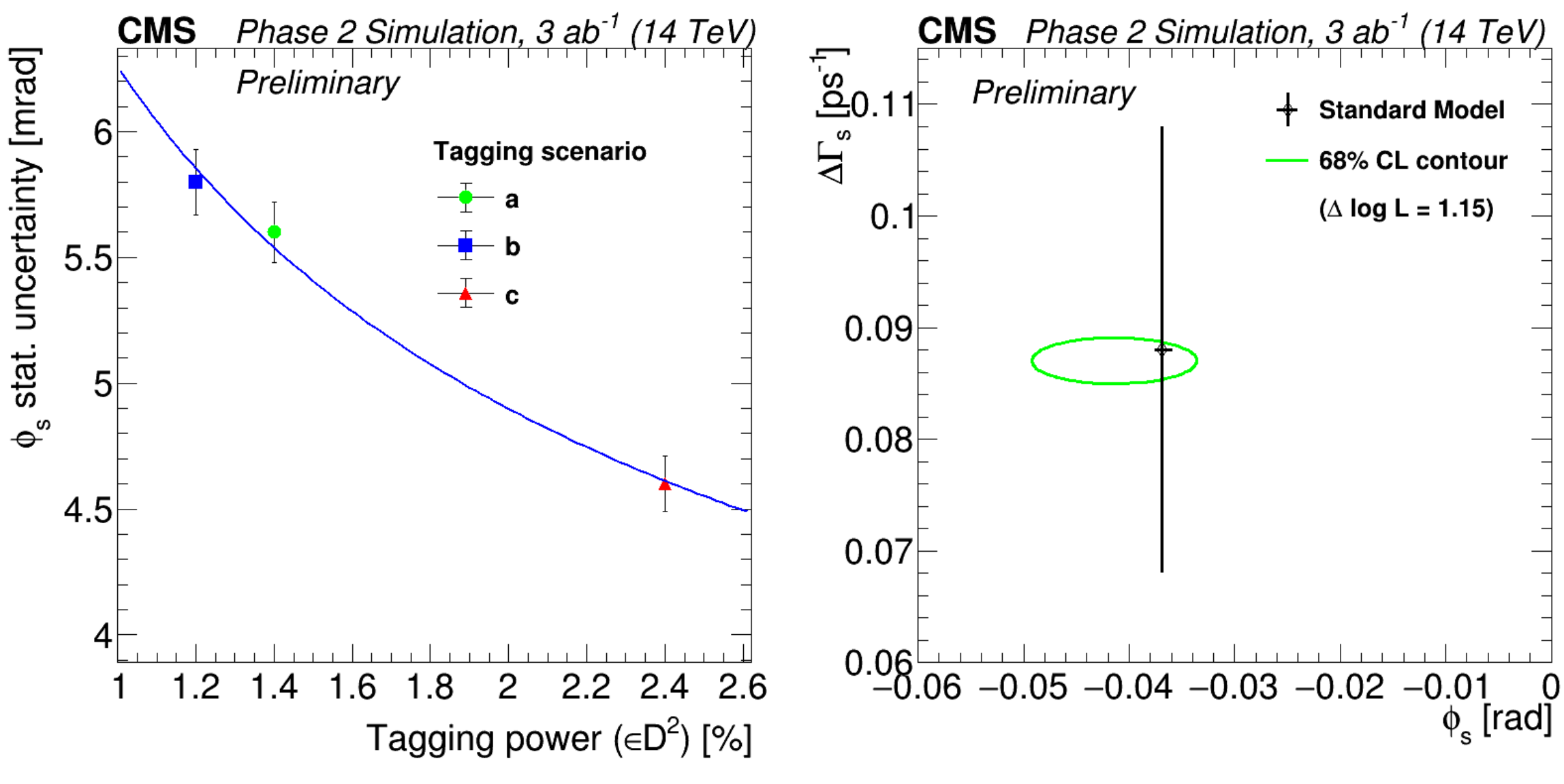

Figure 2: Left: variation of the $\phi_s$ statistical uncertainty as function of the tagging power ($\epsilon D^2$) see equation 1, measured in different flavour tagging scenarios. A function proportional to $1/\sqrt{\epsilon D^2}$ is shown to describe the behaviour of the $\phi_s$ uncertainty in the selected range. Right: 68% confidence level (CL) contour from the fit of a toy MC pseudo-experiment generated in the tagging scenario *c*. The contour combines statistical and systematic uncertainties. The black cross represents the SM expectations [2][1].

# 5 Conclusions

The CMS sensitivity for the measurement of the CP–violating phase $\phi_s$ in the HL–LHC era has been estimated using simulated data and MC toy pseudo–experiments corresponding to the 3 $ab^{-1}$ of integrated luminosity. The offline selection of signal events and the analysis strategy are similar to what was used in the past except for the tagging performance, for which three different scenarios have been considered. Assuming the new tagging power ($\epsilon D^2$) to be in the range 1.2–2.4% , and a total of 9 million $B_s^0$ candidates, we expect the $\phi_s$ statistical uncertainty to be 5–6 mrad at the end of Phase 2 data taking, which improves the current world average uncertainty by a factor of five.

**6**

| `scenario` | $\epsilon$ [%] | $\omega$ [%] | $\epsilon D^2$ [%] | $\sigma_{\phi_s}$ [mrad] |
|---|---|---|---|---|
| $a$ | 32 | 39.4 | 1.4 | 5.6 |
| $b$ | 8 | 30.2 | 1.2 | 5.8 |
| $c$ | 33 | 36.4 | 2.4 | 4.6 |

Table 1: Statistical uncertainty of $\phi_s$ obtained from toy MC pseudo-experiments for different scenarios of flavour tagging.

**ATLAS PUB Note**

ATL-PHYS-PUB-2019-003

18th January 2019

# $B_d^0 \to K^{*0}\mu\mu$ angular analysis prospects with the upgraded ATLAS detector at the HL-LHC

The ATLAS Collaboration

This note estimates the ATLAS detector performance in measuring the angular parameters describing the $B_d^0 \to K^{*0}\mu\mu$ decay angular distribution, during the whole HL-LHC campaign. The projections are based on the Run 1 analysis, while accounting for the most relevant ATLAS detector improvements.

# 1 Introduction

Flavour-changing neutral currents (FCNC) have played a significant role in the construction of the Standard Model of particle physics (SM). These processes are forbidden at tree level and can proceed only via loops, hence are characterized by small amplitudes. An important set of FCNC processes involves the decays of $b$-quark to $s\mu^+\mu^-$ final states mediated by electroweak box and penguin diagrams. Non-SM particles - even heavier than what can be directly probed with existing colliders - may contribute to FCNC decay amplitudes, affecting the measurement of observables related to the decay under study.

The $B_d^0 \to K^{*0}(892)\mu^+\mu^-$ decay is a semileptonic decay mediated by FCNC. [1] The observables sensitive to contributions from physics beyond the SM include the differential branching fraction, charge and isospin asymmetries, the angular distribution of decay products and - for some contributions that are lepton-flavour dependent - the ratio of decay rates into dimuon and dielectron final states.

The kinematics of the four particles in the final state of $B_d^0 \to K^*(K\pi)\mu^+\mu^-$ is described by the invariant mass of the dimuon system ($q^2$) and three helicity angles. The full angular differential decay rate can then be expressed as a function of $q^2$, helicity angles, seven angular coefficients $S_i$ and the fraction of longitudinally polarised $K^*$ mesons, $F_L$. Folding transformations based on symmetry of trigonometric functions can be used to simplify the distributions reducing the dependence to only on three parameters: $F_L$, $S_3$ and one of $S_4$, $S_5$, $S_7$ and $S_8$. A set of optimised parameters $P_i^{(\prime)}$ was also proposed [1, 2] to reduce the theoretical uncertainties that come from hadronic form factors. The $P_i^{(\prime)}$ parameters can be derived from the measured values of $S_i$ and $F_L$. The sign of $S_5$ and $S_8$ parameters depends on the flavour of the $B_d^0$ meson. Incorrect flavour tag assignment (mistag) thus leads to dilution effects.

The parameters of the angular distribution for $B_d^0 \to K^*\mu\mu$ with a subsequent $K^* \to K^+\pi^-$ decay have been measured recently by the Babar, Belle, CDF, CMS, LHCb and ATLAS collaborations [3–12]. The LHCb collaboration reported a potential hint of deviation from SM predictions [6] using their Run 1 dataset. The results from the Belle collaboration [10], CMS [8] and ATLAS analysis of 2012 data [12] are consistent with both the LHCb results and with the SM calculations.

This document presents an estimate of the precision that ATLAS could achieve in measurement of the same angular parameters using the dataset expected to be collected at the HL-LHC. This estimate is based on the extrapolation of the ATLAS result found in Ref. [12] and takes into account the extrapolated numbers of signal and background events as well as the expected improvements in mass resolution.

# 2 Assumptions of the projections

The core of the $B_d^0 \to K^*\mu\mu$ analysis is an extended unbinned maximum likelihood fit (UML) to the decay angles and the $B_d^0$ invariant mass (binned in the $q^2$ observable).

Monte Carlo simulations (toy-MC) are employed to estimate the ATLAS precision at HL-LHC in the $B_d^0 \to K^*\mu\mu$ analysis. Toy-MC parameters include the extrapolated number of signal ($N_{\text{sig}}$) and background ($N_{\text{bck}}$) events at the HL-LHC, and the effect of the performance of the ATLAS Upgrade tracker (ITk) [13]. The latter namely including the improvement in the 4-prong invariant mass resolutions, as studied with the $B_s^0 \to J/\psi\phi$ [14] decay channel.

[1] Hereafter, the $K^{*0}(892)$ is referred to as $K^*$ and charge conjugation is implied throughout, unless stated otherwise.

$N_{\text{sig}}$ and $N_{\text{bck}}$ are extrapolated from the simulations and the real data of the Run 1 $B_d^0 \to K^*\mu\mu$ analysis [12]. Three dimuon trigger scenarios are considered, with varying muons $p_T$ thresholds: a high-yield one requiring 6 GeV $p_T$ cuts on both muons, the intermediate one requiring in addition at least one of the muons to have $p_T$ above 10 GeV, and a low-statistics scenario requiring both muons to have $p_T$ above 10 GeV. Hereafter the scenarios are referred as μ6μ6, μ10μ6 and μ10μ10. The offline selection and reconstruction efficiencies are assumed to be similar to the ones observed in the Run 1 analysis.

The analysis of the simulated sample follows exactly the same fit model as in the Run 1 analysis as well as the same $q^2$ range and binning. A dilution correction due to $B_d^0$ flavor mistagging is applied assuming an effect identical to what measured in Run 1. While the trigger thresholds would have an effect on the detector angular acceptance shapes, their effect on the resulting precision is limited and thus same acceptance shape as in Run 1 is used.

## 3 Event yields

The extrapolation of the event yields is based on the measured signal and background yields of the Run 1 analysis (using 20.3 fb$^{-1}$ of $pp$ collisions at $\sqrt{s}$ = 8 TeV). It accounts for the total integrated luminosity expected at HL-LHC (3000 fb$^{-1}$), the increase of the production cross-section of $b$-hadrons between 8 and 14 TeV (1.7× according to Ref. [15]) and the expected trigger efficiencies.

The signal trigger efficiencies for the three trigger scenarios are extracted from the Run 1 simulation of the $B_d^0 \to K^*\mu\mu$ decay, emulating the trigger thresholds with offline muon $p_T$ cuts individually for each $q^2$ bin. The background rejection in the various trigger scenarios is measured on events from the Run 1 data sidebands (4.9 GeV < $m(B_d^0)$ < 5.058 GeV and 5.498 GeV < $m(B_d^0)$ < 5.7 GeV).

The extrapolated signal event yields are shown in the third column of Table 1. The signal to background ratios for the $q^2$ bins and trigger configurations range between 0.4 and 1.2 (not significantly different from the Run 1 measurement). Due to the low event yields in the Run 1 analysis, all the projected $N_{\text{sig}}$ and $N_{\text{bck}}$ suffer from statistical uncertainties up to ~ 25%.

## 4 Toy-MC simulations and results

The precision on the angular parameters is extracted using toy-MC simulations. The toy-MC and consequent UML fits in $q^2$ bins follow exactly the same configuration as in Run 1: three $q^2$ bins are analyzed: [0.04, 2.0] GeV$^2$, [2.0, 4.0] GeV$^2$ [4.0, 6.0] GeV$^2$. The same four folding transformations as in Run 1 are applied to the generated toy-MC data and correspondingly simplified angular distributions are fitted [12]. A two step fit is run: $B_d^0$ invariant mass only first to fix the mass background shape and signal fraction, followed by the simultaneous mass-angular fit. In the toy-MC generation phase, the signal angular distribution is set to follow the theory prediction (DHMV [16]). The detector acceptance and the background shapes are set same as in the Run 1 analysis. The signal mass distribution is narrowed w.r.t. the Run 1 analysis, corresponding to the ITk improvement in the mass resolution: the HL-LHC simulations of the $B_s^0 \to J/\psi\phi$ decay [14] channels show that the ITk improves resolution in the reconstructed 4-prong invariant mass by 30%.

Table 1 summarizes the projected statistical uncertainties on the angular parameters, as extracted from the fits to the generated toy-MC data. Ten toy-MC fits are run and the average statistical uncertainty out of these ten fits is reported in the Table 1. The toy-MC does not include effect of mistagging. In the Run 1 analysis, a post-fit correction has been applied to the UML fit results, affecting both the central values and the uncertainties of the $P_5'$ and $P_8'$ parameters. Following the same procedure, a dilution correction corresponding to a mistag fraction of 10% is applied to the $P_5'$ and $P_8'$ uncertainties.

The procedure of folding leads to replacement of a single UML fit to the full angular distribution, by four fits to four simplified angular distributions. Out of each of these four fits, $F_L$, $P_1$ and $P_i^{(\prime)}$ are extracted, with $P_i^{(\prime)}$ corresponding to $P_4'$, $P_5'$, $P_6'$ or $P_8'$. Therefore the values and uncertainties of the $F_L$ and $P_1$ parameters are ambiguous. The ambiguity is resolved by using the highest average statistical uncertainty out of these four fits.

| LHC phase | $q^2$ [GeV$^2$] | $N_{\rm sig} \pm \delta^{\rm stat}_{N_{\rm sig}}$ | $\delta^{\rm stat}_{F_L}$ | $\delta^{\rm stat}_{P_1}$ | $\delta^{\rm stat}_{P_4'}$ | $\delta^{\rm stat}_{P_5'}$ | $\delta^{\rm stat}_{P_6'}$ | $\delta^{\rm stat}_{P_8'}$ |
|---|---|---|---|---|---|---|---|---|
| Run 1 | [0.04, 2.0] | 128 ± 22 | 0.08 | 0.30 | 0.40 | 0.26 | 0.21 | 0.48 |
| | [2.0, 4.0] | 106 ± 23 | 0.11 | 0.51 | 0.31 | 0.31 | 0.28 | 0.41 |
| | [4.0, 6.0] | 114 ± 24 | 0.13 | 0.43 | 0.33 | 0.35 | 0.27 | 0.42 |
| HL-LHC μ6μ6 | [0.04, 2.0] | 15800 ± 190 | 0.007 | 0.025 | 0.030 | 0.024 | 0.018 | 0.038 |
| | [2.0, 4.0] | 15200 ± 180 | 0.007 | 0.055 | 0.030 | 0.028 | 0.020 | 0.037 |
| | [4.0, 6.0] | 14000 ± 200 | 0.009 | 0.063 | 0.031 | 0.034 | 0.027 | 0.039 |
| HL-LHC μ10μ6 | [0.04, 2.0] | 10000 ± 160 | 0.009 | 0.036 | 0.040 | 0.029 | 0.022 | 0.049 |
| | [2.0, 4.0] | 9700 ± 150 | 0.010 | 0.071 | 0.039 | 0.034 | 0.027 | 0.045 |
| | [4.0, 6.0] | 8900 ± 170 | 0.012 | 0.084 | 0.039 | 0.042 | 0.033 | 0.050 |
| HL-LHC μ10μ10 | [0.04, 2.0] | 3200 ± 90 | 0.017 | 0.065 | 0.072 | 0.052 | 0.040 | 0.090 |
| | [2.0, 4.0] | 3100 ± 90 | 0.017 | 0.13 | 0.069 | 0.063 | 0.048 | 0.080 |
| | [4.0, 6.0] | 2800 ± 100 | 0.022 | 0.16 | 0.074 | 0.075 | 0.060 | 0.088 |

Table 1: Statistical uncertainties of the $F_L$ and $P_i^{(\prime)}$ parameters from the 2012 data measurement and projected to the HL-LHC phase for the three trigger scenarios. The $N_{\rm sig}$ uncertainties $\delta^{\rm stat}_{N_{\rm sig}}$ for the HL-LHC projections are extracted from the fits of the toy-MC data.

## 5 Systematic uncertainties

All the systematic uncertainties considered in the Run 1 analysis have the potential to be improved with higher statistics of the data collected at HL-LHC:

- Fit-model systematic uncertainties rely on the precision of the model, which is determined with simulations and sidebands data. Thus will be better constrained with the larger HL-LHC dataset, approximately scaling by $1/\sqrt{L_{\rm int}}$.
- Similarly, systematics due to individual background modes will be better measured and simulated at the HL-LHC. These systematics are assumed to reduce as $1/\sqrt{L_{\rm int}}$. This assumption is quite conservative, since the Run 1 analysis accounted for these background components by studying the variation in the fit result when including or excluding the corresponding fit models or events in the dataset: potential room for improvement could come from a more careful evaluation of these systematic effects.

- The precision of the detector acceptance and the mistagging are MC driven and thus will be reduced with larger data and simulation samples. Further improvements would originate from the inclusion of the shapes of the 10% of mistagged events into the UML fit. Since these systematics are driven by the MC statistics, they are neglected in the HL-LHC extrapolation.
- Systematic due to neglecting $S$-wave will be significantly reduced with the inclusion of this component in the UML fit. Accounting for the fact that the $S$-wave contribution has already been measured with $\sim 20\%$ precision by LHCb [17], a reduction by a factor of $5\times$ of this uncertainty is used for the HL-LHC extrapolation.
- Detector alignment and B-field systematics uncertainties are obtained from data-driven techniques and in Run 1 were based on the measured radial distortion. These would improve with larger calibration samples and innovative techniques: already for Run 2 a method of correcting this distortion has been developed [18]. The precision of the method indicates the possibility to reduce the systematics in the $B_d^0 \to K^*\mu\mu$ measurement by a factor of $\sim 4\times$.

| LHC phase | $q^2$ [GeV$^2$] | $\delta^{syst}_{F_L}$ | $\delta^{syst}_{P_1}$ | $\delta^{syst}_{P'_4}$ | $\delta^{syst}_{P'_5}$ | $\delta^{syst}_{P'_6}$ | $\delta^{syst}_{P'_8}$ |
|---|---|---|---|---|---|---|---|
| Run 1 | [0.04, 2.0] | 0.07 | 0.08 | 0.20 | 0.16 | 0.04 | 0.18 |
| | [2.0, 4.0] | 0.05 | 0.34 | 0.21 | 0.13 | 0.19 | 0.39 |
| | [4.0, 6.0] | 0.12 | 0.26 | 0.18 | 0.18 | 0.13 | 0.09 |
| HL-LHC | [0.04, 2.0] | 0.007 | 0.010 | 0.021 | 0.028 | 0.007 | 0.025 |
| | [2.0, 4.0] | 0.004 | 0.075 | 0.027 | 0.025 | 0.035 | 0.060 |
| | [4.0, 6.0] | 0.013 | 0.054 | 0.007 | 0.032 | 0.019 | 0.015 |

Table 2: Systematic uncertainties of the $F_L$ and $P_i^{(\prime)}$ parameters from the 2012 data measurement and projected to the HL-LHC phase.

| LHC phase | $q^2$ [GeV$^2$] | $\delta^{tot}_{F_L}$ | $\delta^{tot}_{P_1}$ | $\delta^{tot}_{P'_4}$ | $\delta^{tot}_{P'_5}$ | $\delta^{tot}_{P'_6}$ | $\delta^{tot}_{P'_8}$ |
|---|---|---|---|---|---|---|---|
| Run 1 | [0.04, 2.0] | 0.11 | 0.31 | 0.45 | 0.31 | 0.21 | 0.51 |
| | [2.0, 4.0] | 0.12 | 0.61 | 0.37 | 0.34 | 0.34 | 0.57 |
| | [4.0, 6.0] | 0.18 | 0.50 | 0.38 | 0.39 | 0.30 | 0.43 |
| HL-LHC μ6μ6 | [0.04, 2.0] | 0.010 | 0.027 | 0.037 | 0.037 | 0.019 | 0.046 |
| | [2.0, 4.0] | 0.008 | 0.093 | 0.040 | 0.038 | 0.040 | 0.070 |
| | [4.0, 6.0] | 0.016 | 0.083 | 0.032 | 0.047 | 0.033 | 0.041 |
| HL-LHC μ10μ6 | [0.04, 2.0] | 0.011 | 0.037 | 0.046 | 0.040 | 0.023 | 0.055 |
| | [2.0, 4.0] | 0.011 | 0.103 | 0.047 | 0.042 | 0.044 | 0.075 |
| | [4.0, 6.0] | 0.018 | 0.100 | 0.040 | 0.053 | 0.038 | 0.052 |
| HL-LHC μ10μ10 | [0.04, 2.0] | 0.018 | 0.065 | 0.076 | 0.059 | 0.041 | 0.093 |
| | [2.0, 4.0] | 0.017 | 0.15 | 0.074 | 0.068 | 0.059 | 0.100 |
| | [4.0, 6.0] | 0.026 | 0.17 | 0.074 | 0.082 | 0.063 | 0.090 |

Table 3: Total uncertainties of the $F_L$ and $P_i^{(\prime)}$ parameters from the 2012 data measurement and projected to the HL-LHC phase for the three trigger scenarios.

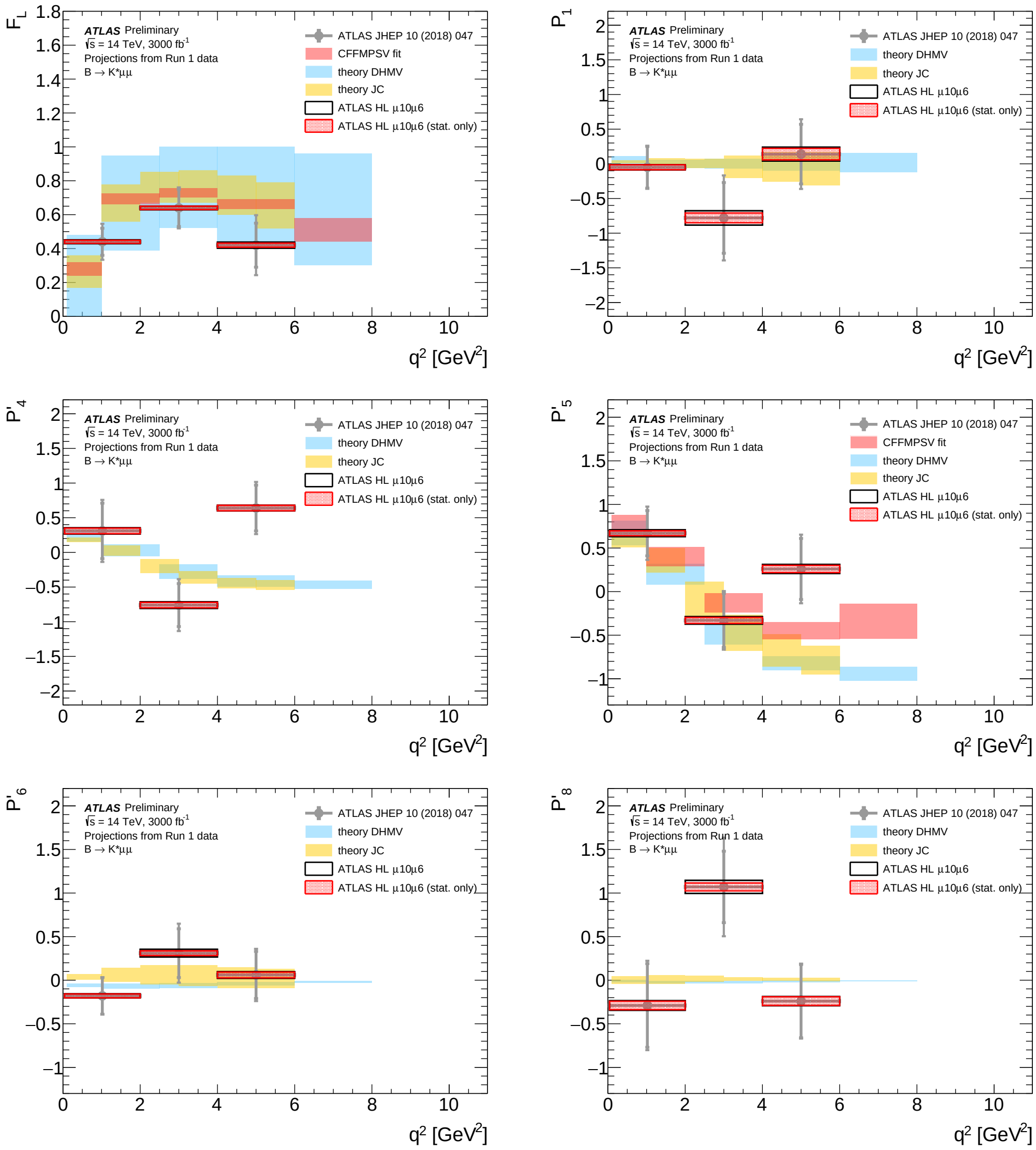

Figure 1: Projected ATLAS HL-LHC measurement precision in the $F_L$, $P_1$, $P_4'$, $P_5'$, $P_6'$ and $P_8'$ parameters for the intermediate μ10μ6 trigger scenario compared to the ATLAS Run 1 measurement. Alongside theory predictions (CFFMPSV [19], DHMV [16], JC [20]) are also shown. Both the projected statistical and the total (statistical and systematic) uncertainties are shown. While the HL-LHC toy-MC were generated with the DHMV central values of the $F_L$ and $P_i^{(\prime)}$ parameters, in these plots the central values are moved to the ATLAS Run 1 measurement for better visualization of the improvement in the precision.

This approach yields the HL-LHC systematic uncertainties shown in Table 2. Compared to the statistical uncertainties in Table 1, this shows that the analysis is not dominated by the systematic uncertainties, with few exceptions where at most the systematics is $\sim 1.6\times$ larger than the statistical precision. The HL-LHC result is expected to benefit from less conservative estimates of the dominant systematic uncertainties, and will thus likely be still limited by statistical uncertainties in all cases.

The projected statistical precision and the systematical uncertainties in measuring the angular parameters in the $B_d^0 \to K^*\mu\mu$ analysis at HL-LHC are presented in Figure 1 and summarized in Table 3.

## 6 Conclusions

The precision of the measurement of the angular parameters in the $B_d^0 \to K^{*0}(892)\mu^+\mu^-$ decay at the upgraded ATLAS detector at High-Luminosity LHC is presented. The projections extrapolate signal and background yields observed in Run 1 and employ toy-MC simulations and consequent fit to the decay angular distributions split in three $q^2$ bins in the range [0.04, 6.0] GeV$^2$. The toy-MC generation accounts for the improved performance of the ATLAS Upgrade tracking system and for the estimate of the expected number of signal and background events at the HL-LHC. Three trigger scenarios, affecting the signal yields, are considered, providing high-yield, intermediate and low-statistics estimates. Using the same $q^2$ binning as the Run 1 analysis, the precision in measuring a representative $P_5'$ parameter is expected to improve by factors of $\sim 9\times$, $\sim 8\times$, $\sim 5\times$ (correspondingly in the three trigger scenarios) relative to the Run 1 measurement.

# CMS Physics Analysis Summary

Contact: cms-phys-conveners-ftr@cern.ch

# Study of the expected sensitivity to the $P_5'$ parameter in the $B^0 \to K^{*0}\mu^+\mu^-$ decay at the HL-LHC

The CMS Collaboration

## Abstract

The expected sensitivity to the $P_5'$ parameter in $B^0 \to K^{*0}\mu^+\mu^-$ decays from an integrated luminosity of 300 and 3000 fb$^{-1}$ of pp collisions at a center-of-mass energy of 14 TeV at the HL-LHC is presented. Angular observables in the $B^0 \to K^{*0}\mu^+\mu^-$ decay, such as the $P_5'$ parameter, are of particular interest as their theoretical predictions are less affected by hadronic uncertainties. With an integrated luminosity of 3000 fb$^{-1}$, the uncertainties on the shape of the $P_5'$ parameter will improve by up to a factor of 15, depending on the dimuon mass squared region, compared to the published results from 20 fb$^{-1}$ at 8 TeV.

# 1 Introduction

The high-luminosity upgrade of the CERN LHC accelerator (HL-LHC) and the detectors will allow for the collection of an unprecedented amount of data. The expected integrated luminosity of 3000 $\mathrm{fb}^{-1}$ during 10 years of operation [1] will provide the ability to perform precision studies of rare decays of b hadrons. In particular, the $B^0 \to K^{*0}(K^+\pi^-)\mu^+\mu^-$ channel, whose branching ratio is at the level of $10^{-7}$, can be used to precisely measure important angular parameters, including the so-called $P_5'$ variable [2, 3].

The differential decay rate for the $B^0 \to K^{*0}\mu^+\mu^-$ channel can be written in terms of the dimuon invariant mass squared $q^2$ and three angular variables as a combination of spherical harmonics, weighted by $q^2$-dependent angular parameters. These angular parameters in turn depend upon complex decay amplitudes, which are described by Wilson coefficients in the Effective Field Theory (EFT) Hamiltonian. The LHCb Collaboration reported a discrepancy of about 3 standard deviations with respect to the standard model (SM) predictions for the parameter $P_5'$ [4], the Belle Collaboration reported a discrepancy almost as large [5], and CMS recently published a value consistent with the SM [6]. More precise measurements are needed to understand the tension between the measurements and the SM predictions. This measurement gains particular interest when considered in the more general framework of the "flavor anomalies", that suggest a possibility of Lepton Flavour Universality Violation [7].

The HL-LHC conditions present particular challenges for the collection, reconstruction, and analysis of b hadron decays. With an average of 200 proton-proton (pp) collisions per bunch crossing (pileup), the reconstruction of the relatively low-momentum charged tracks from b hadron decays and the assignment of the tracks to the correct vertex becomes quite challenging. In addition, being able to trigger on relatively low momentum muons, with the associated high data rate can be problematic.

On the other hand, the CMS detector will undergo many upgrades to handle the HL-LHC conditions. The relevant upgrades for this analysis are a new silicon tracker with finer granularity, extended coverage, and better radiation tolerance, improvements to the muon system, the ability to reconstruct and use tracks in the first stage of the trigger, and a data-acquisition system to allow for many more events to be stored. These improvements are designed to ensure that the CMS performance meets or exceeds the original performance even in the harsh environment of the HL-LHC. A detailed overview of the CMS detector upgrade program is presented in Ref. [8–10], while the expected performance of the reconstruction algorithms and pile-up mitigation with the CMS detector is summarized in Ref. [11].

In this paper the sensitivity for the measurement of the $P_5'$ parameter at HL-LHC is estimated. Starting from the existing CMS measurement [6] obtained from 8 TeV pp collision data, we use the expected improvements of the statistical and systematic uncertainties assuming a center-of-mass energy of 14 TeV and integrated luminosity of 3000 $\mathrm{fb}^{-1}$ to obtain the expected precision on $P_5'$ at the end of the HL-LHC period.

# 2 Summary of previous analysis

The CMS analysis of Run I data [6] is based on an integrated luminosity of 20 $\mathrm{fb}^{-1}$, collected at $\sqrt{s} = 8$ TeV in 2012. The analysis measures the $P_5'$ variable of the $B^0 \to K^{*0}\mu^+\mu^-$ decay as a function of $q^2$ in the range from 1 to 19 $\mathrm{GeV}^2$. CMS had previously exploited the same data set to measure two other angular parameters in the $B^0 \to K^{*0}\mu^+\mu^-$ decay as a function of $q^2$, the forward-backward asymmetry of the muons, $A_{FB}$, and the $K^{*0}$ longitudinal polarization frac-tion, $F_L$, as well as the differential branching fraction, $d\mathcal{B}/dq^2$ [12]. The decay is fully described as a function of the three angles $\theta_\ell$, $\theta_K$ and $\phi$, where $\theta_\ell$ is the angle between the momentum of the positive (negative) muon and the direction opposite to the $B^0$ ($\overline{B}^0$) in the dimuon rest frame; $\theta_K$ is the angle between the kaon momentum and the direction opposite to the $B^0$ ($\overline{B}^0$) in the $K^{*0}$ ($\overline{K}^{*0}$) rest frame; $\phi$ is the angle between the dimuon and the $K^+\pi^-$ decay planes in the $B^0$ rest frame. The expression describing the angular distribution can be found in Ref. [6]. The possible contribution from spinless (S-wave) $K^+\pi^-$ combinations is taken into account in the decay description with three terms: $F_S$, which is related to the S-wave fraction, and $A_S$ and $A_S^5$, which are the interference amplitudes between the S-wave and P-wave decays. The observables of interest are extracted for each $q^2$ bin from an unbinned extended maximum likelihood fit to $m(\mu^+\mu^-K^+\pi^-)$ and the three angular variables.

In this analysis the CP state assignment is of great importance because the angular observables behave oppositely for each one of the two CP eigenstates. In absence of a particle ID detector, and given that the ionization energy loss method for hadron identification is not applicable to the kinematic range of the particles involved in this process [13], the four-track candidate is identified as a $B^0$ or $\overline{B}^0$ based on the $K^+\pi^-$ or $K^-\pi^+$ invariant mass being closest to the nominal $K^{*0}$ mass. The fraction of candidates assigned to the incorrect state is estimated from the simulation to vary between 12 and 14% among the different $q^2$ bins.

The probability density function takes into account correctly and wrongly tagged signal events, background events, and the efficiency in the three angular variables. The efficiency, which is the product of the acceptance of the detector and the trigger, reconstruction, and selection efficiencies, is obtained from a Monte Carlo (MC) simulation, which reproduces the data taking conditions. It is determined, for each $q^2$ bin, as a function of the three angles $\cos\theta_\ell$, $\cos\theta_K$ and $\phi$.

The resonant $B^0 \to J/\psi K^{*0}$ and $B^0 \to \psi' K^{*0}$ decays are used as control channels (corresponding to the $q^2$ bins 8.68 – 10.09 and 12.90 – 14.18 $GeV^2$). Here, $\psi'$ denotes the $\psi(2S)$ meson.

The online event selection uses a hardware low-$p_T$ dimuon trigger and a High Level Trigger (HLT) selection based on the dimuon invariant mass and the compatibility of the two muons with a common vertex displaced from the pp collision region. The offline reconstruction requires that two oppositely charged muons and two oppositely charged hadrons are fit to a common vertex, and satisfy the set of kinematic and topological requirements described in Ref. [6]. In case multiple $B^0$ candidates per event are found, only the one with the largest $\chi^2$ fit probability is retained.

Contamination from the resonant $B^0 \to J/\psi K^{*0}$ and $B^0 \to \psi' K^{*0}$ decays is reduced using a combined selection on $m(\mu^+\mu^-)$ and $m(\mu^+\mu^-K^+\pi^-)$, i.e., rejecting events for which the condition $|(m(\mu^+\mu^-K^+\pi^-) - m(B^0)_{PDG}) - (m(\mu^+\mu^-) - m(J/\psi \text{ or } \psi')_{PDG})| \leq R_{rej}$ is satisfied. The $R_{rej}$ value and the use of $m(J/\psi)_{PDG}$ versus $m(\psi')_{PDG}$ [14] depend on the $m(\mu^+\mu^-)$ analyzed region [6].

## 3 Extrapolation to the HL-LHC

In order to extrapolate from the Run I results, some assumptions are made. We have not considered the effects of improvements in the analysis strategy (for instance the use of different selection criteria or fits). We have assumed that the trigger thresholds and efficiencies will remain the same. In fact, this is likely to be a conservative assumption as the availability of tracking information at the first level of the trigger may result in a higher efficiency than in Run I. The extrapolation method assumes that the signal-to-background is the same. Except as noted below regarding the mass resolution, this is expected to be the case as the primary source of background is from other b decays, whose cross section scales the same as the signal. Samples of simulated signal events were used to evaluate the effect of three important aspects of the analysis: mass resolution, CP mistagging rate, and the effect of pileup in order to justify the extrapolation method.

## 3.1 Mass resolution

For analyses with significant background, the mass resolution is an important aspect in obtaining a high signal to background. The left plot of Fig. 1 shows the $K^+\pi^-\mu^+\mu^-$ invariant mass distribution in a specific $q^2$ bin for the Run I and Phase-2 simulations, and the width of the $B^0$ signal for each $q^2$ bin is shown on the right. The width is measured by performing a fit to the $K^+\pi^-\mu^+\mu^-$ mass distribution in each $q^2$ bin, parametrizing the $B^0$ signal with the sum of two Gaussian distributions and taking the average of the two Gaussian widths (weighted by their relative contribution) as the $B^0$ width. The improvement in mass resolution with the Phase-2 conditions should improve the signal-to-background ratio from the Run I result. This improvement is not included in the extrapolation.

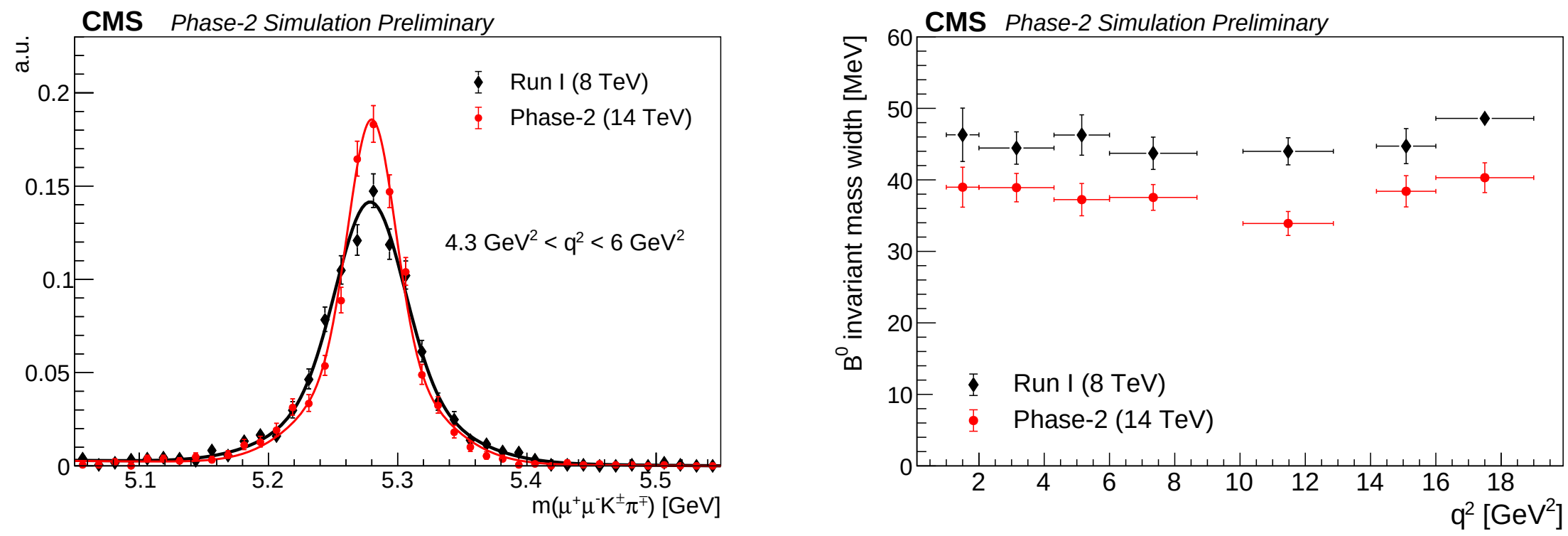

Figure 1: Left: the $K^+\pi^-\mu^+\mu^-$ invariant mass distribution for bin 2 from Run I (black diamonds) and Phase-2 (red circles) simulation. A fit with the sum of two Gaussian functions is superimposed to each distribution. Right: the $B^0$ signal width for each $q^2$ bin in the Run I and Phase-2 simulations.

## 3.2 Mistag rate

The assignment of the CP state is based on the distance of the invariant mass of the two hadrons from the $K^{*0}$ PDG mass [14]. Both mass hypotheses are computed, i.e. $K^+\pi^-$ and $K^-\pi^+$, but only the one closest to the $K^{*0}$ world average mass is retained, which also directly determines the CP state of the mother meson. The CP mistag fraction, defined as the ratio between the number of wrongly tagged events and the total number of signal events, is determined from simulation by counting the number of correctly and wrongly tagged events, where only truth-matched events passing all of the selection criteria are considered. The mistag fraction obtained from the Phase-2 MC simulation is found to be the same as in Run I.

## 3.3 Pileup effects

The analysis performance was proven not to be significantly affected by pileup during the studies performed for the previous publications [6, 12]. In particular, the event selection re-quirements do not depend on the primary vertex choice or utilize isolation information. Furthermore, requiring that each track in the decay have a distance of closest approach to the beamspot greater than 2 standard deviations helps to reduce the contamination from tracks originating from pileup vertices.

In order to have a more quantitative estimation of possible pileup effects, we compared the distributions of the more relevant variables between the samples with and without pileup: no significant degradation of the discriminating power was observed.

### 3.4 Expected yield and statistical uncertainty

For each $q^2$ bin, the expected $B^0 \to K^{*0}\mu^+\mu^-$ signal yields are obtained from a sample of simulated signal events generated with the Phase-2 conditions, including an average of 200 pileup. The yields in each $q^2$ bin are obtained from an extended unbinned maximum likelihood fit to the $K^+\pi^-\mu^+\mu^-$ invariant mass, parameterizing the signal with the sum of two Gaussian distributions and the (negligible) background with an exponential distribution. All parameters are freely varying in the fit. The yields are weighted by the trigger efficiencies measured in the Run I sample and scaled to luminosities of 300 and 3000 $fb^{-1}$. The total expected number of $B^0 \to K^{*0}\mu^+\mu^-$ signal events, excluding the $q^2$ regions associated with the resonant decays, is around 700K for an integrated luminosity of 3000 $fb^{-1}$. The estimated statistical uncertainty on the $P_5'$ parameter is obtained by scaling the statistical uncertainty measured in Run I by the square root of the ratio between the yields observed in the Run I data and the Phase-2 simulation:

$$\sigma_{P_5'}^{Phase2} = \sqrt{\frac{N^{RunI}}{N^{Phase2}}}\sigma_{P_5'}^{RunI} \tag{1}$$

## 4 Systematic uncertainties

The systematic uncertainties are also extrapolated from the Run I analysis. Improved understanding of theory and the experimental apparatus is expected to reduce many uncertainties by a factor of 2 in the Phase-2 scenario. These uncertainties are those related to contamination from resonant decays, signal mass shape, CP mistagging rate, efficiency, angular resolution, and other simulation modeling. The uncertainty on the description of the background mass distribution, the one associated with the propagation of the uncertainty on $F_L$, $F_S$ and $A_S$, and the fit bias introduced by the fitting procedure depend on the available amount of data. These uncertainties are therefore scaled the same as the statistical uncertainties. The uncertainty related to the limited number of simulated events is neglected, under the assumption that sufficiently large simulation samples will be available by the time the HL-LHC becomes operational.

## 5 Results

The Run I results and the projected statistical uncertainties and total uncertainties (statistical and systematic uncertainties added in quadrature) in each $q^2$ bin are shown in Figs. 2 and 3 for an integrated luminosity of 300 and 3000 $fb^{-1}$, respectively. The statistical and systematic uncertainties from Run I and for an integrated luminosity of 3000 $fb^{-1}$ in Phase-2 are also given in Table 1.

The increased amount of collected data foreseen for Phase-2 offers us the opportunity to perform the angular analysis in narrower $q^2$ bins, in order to measure the $P_5'$ shape as a function of $q^2$ with finer granularity. The $q^2$ region below the J/$\psi$ mass (squared), which is more sensitive to possible new physics effects, is considered. Each Run I $q^2$ bin is split into smaller and equal-size bins trying to achieve a statistical uncertainty of the order of the total systematic uncertainty in the same bin with the additional constraint of having a bin width at least 5 times larger than the dimuon mass resolution $\sigma_r$. If both conditions cannot be satisfied, then only the looser requirement on the $5\sigma_r$ bin width is imposed. The dimuon mass resolution is obtained from the MC simulation as a function of $q^2$. With respect to the Phase-2 systematic uncertainties with wider bins, the systematic uncertainties that were scaled the same as the statistical uncertainties are adjusted to account for less data in each bin while the other systematic uncertainties are unchanged. The resulting binning is given in Table 2, along with the projected statistical and systematic uncertainties. The lower two pads of Fig. 3 show the projected statistical and total uncertainties.

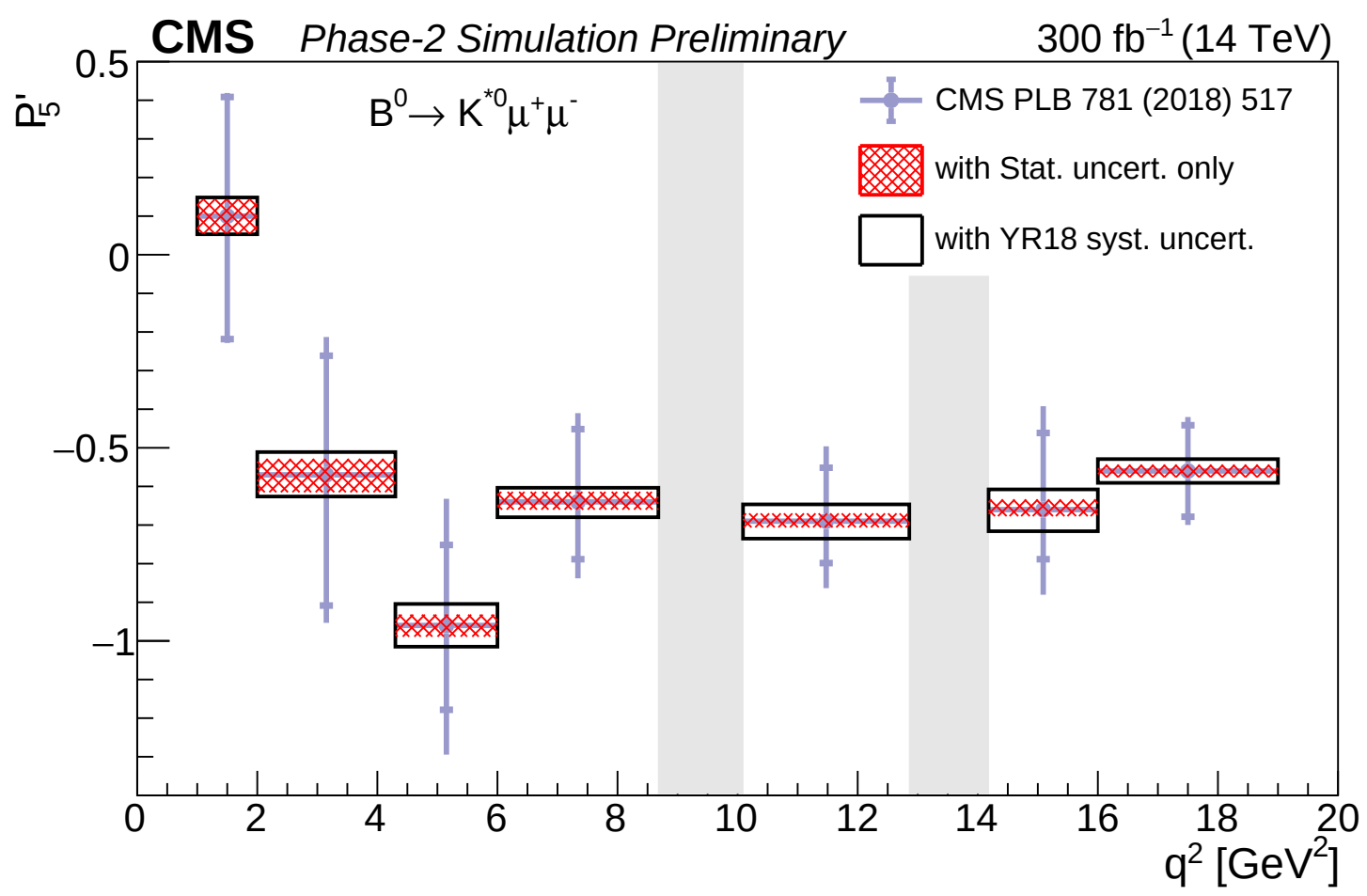

Figure 2: Projected statistical (hatched regions) and total (open boxes) uncertainties on the $P_5'$ parameter versus $q^2$ in the Phase-2 scenario with an integrated luminosity of 300 fb$^{-1}$. The CMS Run I measurement of $P_5'$ is shown by circles with inner vertical bars representing the statistical uncertainties and outer vertical bars representing the total uncertainties. The vertical shaded regions correspond to the J/$\psi$ and $\psi'$ resonances.

## 6 Conclusions

The large amount of data expected from the HL-LHC will allow CMS to investigate rare B physics decay channels and, in particular, precisely measure the $P_5'$ parameter shape in the $B^0 \to K^{*0}\mu^+\mu^-$ mode through an angular analysis. With the large data set of 3000 fb$^{-1}$, corresponding to around 700K fully reconstructed $B^0 \to K^{*0}\mu^+\mu^-$ events, the $P_5'$ uncertainties in the $q^2$ bins are estimated to improve by up to a factor of 15 compared to the CMS measurement from 20 fb$^{-1}$ of 8 TeV data. We also studied the possibility to perform the analysis of the angular observables in narrower $q^2$ bins, as a better determination of the $P_5'$ parameter shape will allow significant tests for both beyond Standard Model physics and between different Standard Model calculations. The future sensitivity of the $P_5'$ angular variable has been presented, however it is worth mentioning that, with the foreseen HL-LHC high statistics, CMS will have the capability to perform a full angular analysis of the $B^0 \to K^{*0}\mu^+\mu^-$ decay mode.

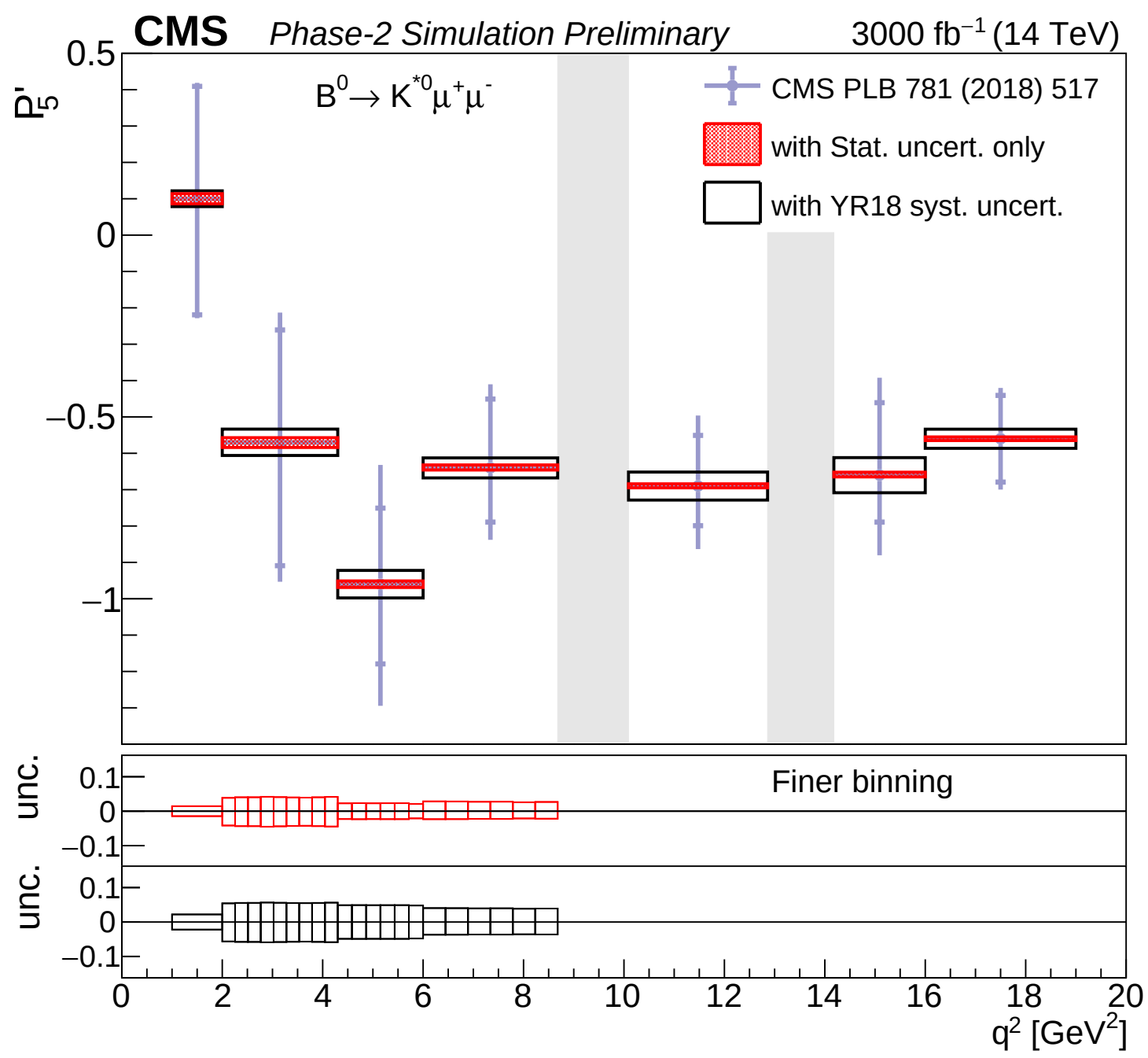

Figure 3: Projected statistical (hatched regions) and total (open boxes) uncertainties on the $P_5'$ parameter versus $q^2$ in the Phase-2 scenario with an integrated luminosity of 3000 fb$^{-1}$. The CMS Run I measurement of $P_5'$ is shown by circles with inner vertical bars representing the statistical uncertainties and outer vertical bars representing the total uncertainties. The vertical shaded regions correspond to the $J/\psi$ and $\psi'$ resonances. The two lower pads represent the statistical (upper pad) and total (lower pad) uncertainties with the finer $q^2$ binning.

Table 1: Statistical and systematic uncertainties in each $q^2$ bin from the Run I measurement [6] and the HL-LHC extrapolation to 3000 fb$^{-1}$.

| $q^2$ bin (GeV$^2$) | Run I | Phase-2 |
|---|---|---|
| $1.00 < q^2 < 2.00$ | $\sigma_{\text{stat}} = ^{+0.32}_{-0.31}$ <br> $\sigma_{\text{syst}} = \pm 0.07$ | $\sigma_{\text{stat}} = \pm 0.014$ <br> $\sigma_{\text{syst}} = \pm 0.017$ |
| $2.00 < q^2 < 4.30$ | $\sigma_{\text{stat}} = ^{+0.34}_{-0.31}$ <br> $\sigma_{\text{syst}} = \pm 0.18$ | $\sigma_{\text{stat}} = ^{+0.014}_{-0.013}$ <br> $\sigma_{\text{syst}} = \pm 0.034$ |
| $4.30 < q^2 < 6.00$ | $\sigma_{\text{stat}} = ^{+0.22}_{-0.21}$ <br> $\sigma_{\text{syst}} = \pm 0.25$ | $\sigma_{\text{stat}} = \pm 0.009$ <br> $\sigma_{\text{syst}} = \pm 0.037$ |
| $6.00 < q^2 < 8.68$ | $\sigma_{\text{stat}} = ^{+0.15}_{-0.19}$ <br> $\sigma_{\text{syst}} = \pm 0.13$ | $\sigma_{\text{stat}} = ^{+0.006}_{-0.008}$ <br> $\sigma_{\text{syst}} = \pm 0.026$ |
| $10.09 < q^2 < 12.86$ | $\sigma_{\text{stat}} = ^{+0.11}_{-0.14}$ <br> $\sigma_{\text{syst}} = \pm 0.13$ | $\sigma_{\text{stat}} = ^{+0.005}_{-0.006}$ <br> $\sigma_{\text{syst}} = \pm 0.038$ |
| $14.18 < q^2 < 16.00$ | $\sigma_{\text{stat}} = ^{+0.13}_{-0.20}$ <br> $\sigma_{\text{syst}} = \pm 0.18$ | $\sigma_{\text{stat}} = ^{+0.005}_{-0.008}$ <br> $\sigma_{\text{syst}} = \pm 0.048$ |
| $16.00 < q^2 < 19.00$ | $\sigma_{\text{stat}} = \pm 0.12$ <br> $\sigma_{\text{syst}} = \pm 0.07$ | $\sigma_{\text{stat}} = \pm 0.005$ <br> $\sigma_{\text{syst}} = \pm 0.026$ |

Table 2: Projected statistical and systematic uncertainties from 3000 fb$^{-1}$ HL-LHC with finer $q^2$ binning in the low $q^2$ region.

| Run I $q^2$ bin (GeV$^2$) | Finer $q^2$ bin (GeV$^2$) | Stat. uncertainty | Syst. uncertainty |
|---|---|---|---|
| $1.00 < q^2 < 2.00$ | $1.00 < q^2 < 2.00$ | $\sigma_{stat} = \pm 0.014$ | $\sigma_{syst} = \pm 0.017$ |
| $2.00 < q^2 < 4.30$ | $2.00 < q^2 < 2.26$ | $\sigma_{stat} = \pm 0.042$ | $\sigma_{syst} = \pm 0.038$ |
| | $2.26 < q^2 < 2.51$ | $\sigma_{stat} = \pm 0.044$ | |
| | $2.51 < q^2 < 2.77$ | $\sigma_{stat} = \pm 0.044$ | |
| | $2.77 < q^2 < 3.02$ | $\sigma_{stat} = \pm 0.045$ | |
| | $3.02 < q^2 < 3.28$ | $\sigma_{stat} = \pm 0.044$ | |
| | $3.28 < q^2 < 3.53$ | $\sigma_{stat} = \pm 0.043$ | |
| | $3.53 < q^2 < 3.79$ | $\sigma_{stat} = \pm 0.043$ | |
| | $3.79 < q^2 < 4.04$ | $\sigma_{stat} = \pm 0.043$ | |
| | $4.04 < q^2 < 4.30$ | $\sigma_{stat} = \pm 0.045$ | |
| $4.30 < q^2 < 6.00$ | $4.30 < q^2 < 4.58$ | $\sigma_{stat} = \pm 0.023$ | $\sigma_{syst} = \pm 0.043$ |
| | $4.58 < q^2 < 4.87$ | $\sigma_{stat} = \pm 0.023$ | |
| | $4.87 < q^2 < 5.15$ | $\sigma_{stat} = \pm 0.023$ | |
| | $5.15 < q^2 < 5.43$ | $\sigma_{stat} = \pm 0.023$ | |
| | $5.43 < q^2 < 5.72$ | $\sigma_{stat} = \pm 0.023$ | |
| | $5.72 < q^2 < 6.00$ | $\sigma_{stat} = \pm 0.021$ | |
| $6.00 < q^2 < 8.68$ | $6.00 < q^2 < 6.45$ | $\sigma_{stat} = \pm 0.028$ | $\sigma_{syst} = \pm 0.029$ |
| | $6.45 < q^2 < 6.89$ | $\sigma_{stat} = \pm 0.028$ | |
| | $6.89 < q^2 < 7.34$ | $\sigma_{stat} = \pm 0.027$ | |
| | $7.34 < q^2 < 7.79$ | $\sigma_{stat} = \pm 0.028$ | |
| | $7.79 < q^2 < 8.23$ | $\sigma_{stat} = \pm 0.026$ | |
| | $8.23 < q^2 < 8.68$ | $\sigma_{stat} = \pm 0.027$ | |

**ATLAS PUB Note**

# Prospects for lepton flavour violation measurements in $\tau \to 3\mu$ decays with the ATLAS detector at the HL-LHC

The ATLAS Collaboration

This note estimates the ATLAS detector sensitivity to measure the branching fraction of the lepton flavour violating decay $\tau \to 3\mu$ with 3000 fb$^{-1}$ of 14 TeV proton-proton collision data expected during the HL-LHC campaign. Two sources of $\tau$ leptons are considered: W-bosons and heavy flavour hadrons. The estimation is a projection of the measurement of the search for $\tau \to 3\mu$ decays in $W \to \tau\nu$ events performed by the ATLAS experiment with data collected in 2012. Several scenarios and expected upper limits at different acceptance and background rejection levels are discussed.

# 1 Introduction

This note documents a simulation-based study of the confidence regions for the search for lepton flavour violating decays (LFV) of $\tau$ leptons in the $\tau \to 3\mu$ channel expected with the HL-LHC data-taking campaign corresponding to an integrated luminosity of 3000 fb$^{-1}$ for the ATLAS detector [1]. Two different production channels are considered, the W-channel in which the $\tau$ lepton originates from W-boson decays and the heavy flavour (HF) channel in which the $\tau$ lepton is produced in decays of c- and b-hadrons, dominated by the $D_s \to \tau\nu$ decay. Background events arise dominantly from lepton fakes from hadrons ($c\bar{c}/b\bar{b} \to X\mu\mu$), with additional contributions due to pile-up.

Flavour violation has been observed in both the quark and the neutral lepton sectors, but not for charged leptons. In the SM charged lepton flavour violating branching fractions are heavily suppressed, with predicted values of the order of $10^{-55}$ [2], leaving no possibility to observe such SM processes at any collider based experiment. In new physics scenarios, such as non universal $Z'$ [3], SO(10) supersymmetric [4] or Type-II Seesaw [5] models with an extended Higgs sector, predicted branching fractions can be of the order of up to $10^{-10} - 10^{-8}$, opening the possibility of observing such effects. Observing $\tau \to 3\mu$ decays would provide an immediate proof of physics beyond the SM. The presented studies extrapolate the latest result from the ATLAS collaboration of the $\tau \to 3\mu$ search in the $W \to \tau\nu$ channel, which is based on the data collected during Run 1 [6].

# 2 Analysis procedure

The increased number of pile-up events in the extrapolation of the Run 1 result is assumed to have no strong impact on the performance of the analysis and thus has been neglected. The analysis workflow is taken to be unchanged w.r.t the Run 1 analysis: a loose cut-based pre-selection to select events with three muons originating from a common vertex and kinematics compatible with the ones expected of a $\tau$ lepton decay is applied. This loosely selected dataset is used to train a machine learning algorithm[1] using recorded data events from a mass sideband region as background model and Monte-Carlo (MC) simulation for the signal. Given the low number of expected background events (<1) in the 8 TeV analysis a complicated fitting procedure involving fits to the BDT and invariant three-muon mass shape has been developed. It is assumed that this will be repeated in this work's projections, although a higher background level might be expected in the HL-LHC scenario. The underlying assumption is that any change to this procedure will not influence the performance of the analysis. Lower systematic uncertainties on the background expectation are taken into account as detailed below. Additionally, the Run 1 analysis allowed to improve trigger selection and offline muon reconstruction, that was included in the Run 2 and HL-LHC MC simulation. Signal yields will be extrapolated from Run 2 MC simulation. The expected upper limit on the branching fraction will be extracted exploiting the relationship:

$$BR(\tau \to 3\mu) = \frac{N_S^{UL}}{\mathcal{A} \times \epsilon \cdot N_{X \to \tau\nu}} \quad (1)$$

with $N_S^{UL}$ being the upper limit on the expected number of signal events, $\mathcal{A} \times \epsilon$ the signal acceptance times efficiency of the selection, and $N_{X \to \tau\nu}$ the number of expected tau leptons produced via the decay of a given intial state X, being either a W-boson or charm/beauty hadron. The HL-LHC ATLAS tracker upgrades [7, 8] entail improvements in vertex and mass determination. This reflects in an improved mass resolution in both the W-channel and HF-channel. Figure 1 exemplifies this, comparing the reconstructed tau mass obtained from Run 2 simulations to the one obtained from simulating the HL-LHC detector and collision conditions. The reconstructed tau mass is fitted with a double-Gaussian with both means and width floating. The total width, $\sigma$, is obtained from the weighted average of the width of each single Gaussian. Signal mass windows different from the ones used in Run 1 are included as an improvement taking advantage of the reduced background contributions.

[1] In Run 1 a Boosted Decision Tree (BDT) was applied. The exact choice of the algorithm is subject to changes in the future, e.g. to Neural Networks (NN). In this note it is assumed that the performance of the classifier is at least as good as the BDT in Run 1.

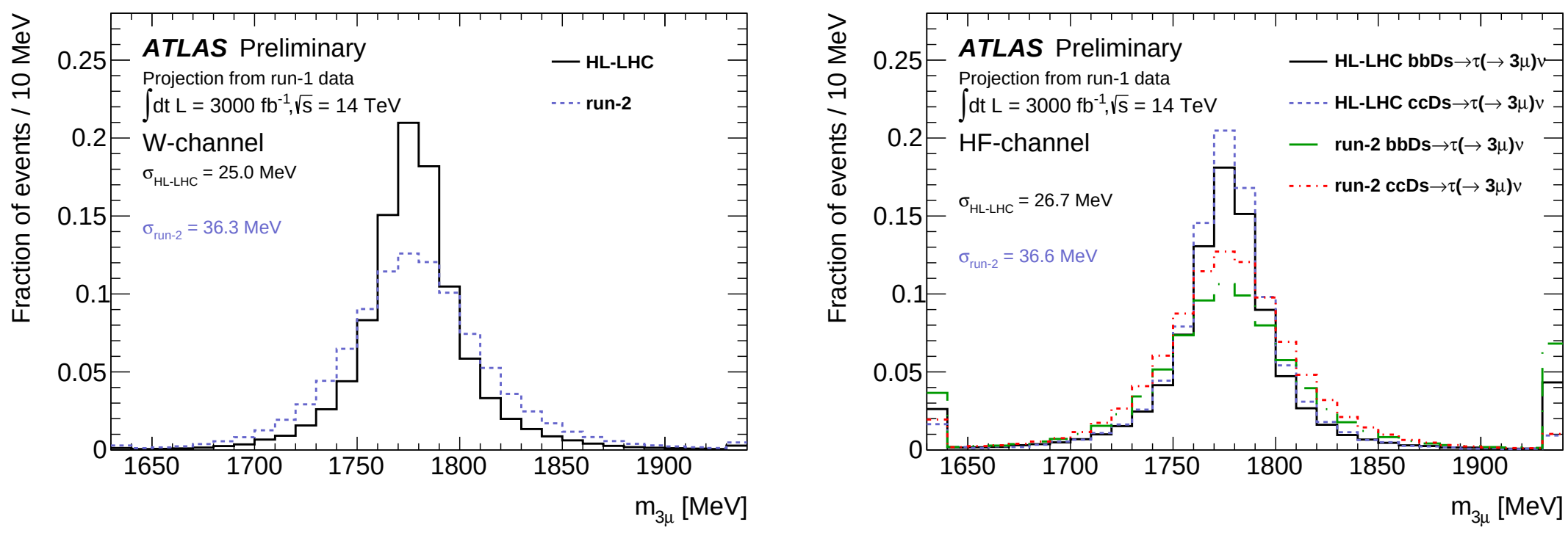

Figure 1: Comparison of tau mass resolutions in the W- (left) and HF-channel (right) in run-2 and under HL-LHC detector conditions. The quoted widths, $\sigma$, are obtained from a double Gaussian fit.

Given that the Run 1 analysis was performed in the W-channel only different extrapolation approaches are chosen for the two production modes and are summarised in the following two sections.

## 2.1 W-channel

The projection in the W-channel is based on the ATLAS Run 1 result [6]. The inclusive W production cross section at $\sqrt{s} = 14$ TeV has been calculated at NNLO using FEWZ [9, 10] and the MSTW2008NNLO pdf set [11] to $\sigma(pp \to W^{\pm}(\to l\nu)) = 21.66$ nb [12], thus in 3ab$^{-1}$ of collision data the tau lepton yield from W-boson decays increased by about a factor of 260 w.r.t the Run 1 statistics ($N^{\mathrm{HL-LHC}}_{W\to\tau\nu} = 6.50 \times 10^{10}$). Since the Run 1 result, several improvements in triggering on low momentum and close-by muons as well as their reconstruction have been developed and deployed. To estimate the impact of these improvements the combined trigger and reconstruction efficiencies are evaluated in Run 2 MC simulation relative to Run 1. The acceptance improves by a factor of 2.2 evaluated on Run 2 and this is confirmed using HL-LHC MC simulations. Scenarios corresponding to different assumed levels of analysis improvements, relative to the Run 1 benchmark are defined, taking into account different optimisations.

1. **Non-improved scenario:** Here no analysis or detector improvements are considered and the sensitivity is extrapolated scaling for the integrated luminosity and higher production cross section at $\sqrt{s} = 14$ TeV. The background yield of the Run 1 analysis ($\mathrm{N}^{\mathrm{Run\ 1}}_{\mathrm{bkg}} = 0.193$) is scaled by a factor of 260, while $\mathcal{A} \times \epsilon$ is considered to be the same as in Run 1. This is by far the most conservative approach of the three approaches.
2. **Intermediate scenario:** In this scenario the improvements in triggering and reconstruction of low $p_{\mathrm{T}}$ muons estimated from Run 2 MC are included in the projection, while no effects on further impacts of the machine learning (ML) selection or better resolution are considered. The net effect of these improvements is an increase of a factor 2.2 in the signal yield relative to Run 1. Additionally, the extrapolation factor of 260 accounting for the increased cross-section and integrated luminosity is applied.

3. **Improved scenario:** In this scenario the signal search window is tightened, taking into account expected improvements in mass resolution. The improvement is estimated fitting the HL-LHC MC signal three-muon invariant mass resolution with a double Gaussian model. The combined fit has a width 20% smaller than the Run 1 counterpart. The improved mass resolution and thus smaller signal region reflects in a 25% improvement in the S/B ratio which is applied to the projection. The improvements of the previous scenarios are applicable as well.

The expected sensitivity for each scenario is calculated based on a profile likelihood fit to the expected event yields in the signal region using HistFitter [13]. The likelihood function $L(\mu_s, \hat{\theta})$ consists of Poisson probabilities for the event yield in the signal region and a Gaussian distribution for the systematic uncertainties which are included in the fit as a nuisance parameter, $\hat{\theta}$. Signal and background yields are linearly scaled by the luminosity, cross section and improvements expected in each scenario. The parameter of interest in such a fit is the signal strength $\mu_s$, corresponding to a scale factor on the signal yield relative to the expected yield. The systematic uncertainty on the background yield in the Run 1 analysis evaluated to 67.9 %. The dominant sources arise from the variation of the three-$\mu$ mass side-band region used for the extrapolation towards the signal region and the uncertainty on the BDT fit which is mainly driven by the low number of background events. As the general strategy will remain the same, this is expected to be the dominant contributor also in future analyses. In particular the analysis is expected to be optimised such that the expected background yield is still small. However, given that no proper background model can be provided by means of simulation, it is impossible to retrieve a reasonable estimate of the machine learning algorithm and thus the $\mathcal{A} \times \epsilon$ in a small background scenario. Thus, in this projection the expected number of background events will be considered high as does the signal acceptance and efficiency.

In this scenario assuming a large systematic uncertainty is unreasonable and would introduce an additional penalty factor on top of the large background yield. As pointed out above in Run 1 the systematic uncertainty was dominated by statistics available for the background estimation. Scaling this down with the increased statistics corresponding to the HL-LHC luminosity but preserving constant terms for reconstruction efficiency systematics results in an assumed uncertainty of the 15%. The impact of varying the background systematic uncertainty by 5% has been evaluated to a change of the upper limit of 10%. Based on the expected upper limit on the signal yields, the limit on $\mathrm{BR}(\tau \rightarrow 3\mu)$ can be calculated according to Eq. 1. Fig 2 shows the $\mathrm{CL_s}$ curves for each discussed scenario. The expected 90% $\mathrm{CL_s}$ upper limits on the $\tau \rightarrow 3\mu$ branching fraction are obtained from the intersection with the red line. The corresponding projections are summarised together with the considered inputs in Tab. 1.

## 2.2 HF-channel

Although with a less clean signature, the majority of $\tau$ leptons produced at the LHC originate from heavy flavour meson decays, dominated by $D_s$ decays. The HF-channel has not been exploited by ATLAS so far, but the W-channel result can be used as the basis for a rough sensitivity extrapolation to the HF case. $D_s$ mesons are produced either promptly or in the decay of b-hadrons. The inclusive cross section for the production of HF decays to tau leptons has been calculated at Fixed Order Next to Leading Log (FONLL) [14–16] using the CTEQ 6.6 pdf set [17] and is found to be $\sigma(\mathrm{HF} \rightarrow \tau\nu) = \left(745^{+172}_{-130}\right)$ nb in the fiducial phase space defined by $p_{\mathrm{T}} > 10$ GeV and $|\eta| < 2.5$. Systematic uncertainties include variations of the scales and masses and are combined quadratically. Thus, in 3 $\mathrm{ab}^{-1}$ of pp collision data $N_{HF\to\tau\nu} = \left(2.23^{+0.52}_{-0.39}\right) \times 10^{12}$ are expected, which increases the dataset of recorded $\tau$ leptons by a factor of $\sim$ 40 compared to the W-channel. Tau leptons from HF decays tend to have lower $p_{\mathrm{T}}$ relative to the W-produced ones, resulting in a smaller acceptance. The trigger and reconstruction efficiencies are estimated from MC simulations of $c\bar{c}/b\bar{b} \to D_s(\tau \to 3\mu) + X$ events. The acceptance is extracted from a generator level study applying kinematic selections looser than the ones imposed by the following trigger scenarios:

1. low-$p_{\mathrm{T}}$ selection: all three muons have $p_{\mathrm{T}} > 3.5$ GeV;
2. high-$p_{\mathrm{T}}$ selection: three muons must pass $p_{\mathrm{T}} > (10.5, 5.5, 2)$ GeV, respectively.

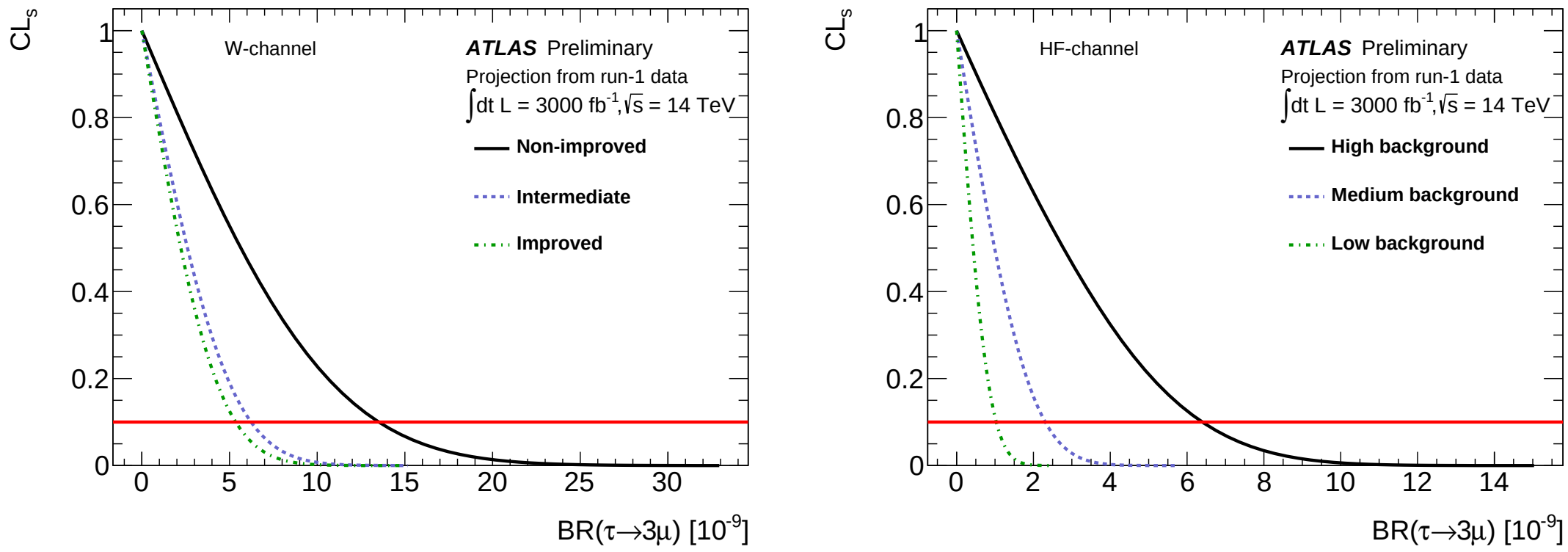

Figure 2: $\mathrm{CL_s}$ versus the $\tau \to 3\mu$ branching fraction, $\mathrm{BR}(\tau \to 3\mu)$ ,for each of the discussed scenarios in the W-channel (left) and HF-channel (right). The horizontal red line denotes the 90% CL. The limit is obtained from the intersection of the $\mathrm{CL_s}$ and this line.

| Scenario | $\mathcal{A} \times \epsilon$ [%] | $N_{\mathrm{bkg}}^{\mathrm{exp}}$ | 90% CL UL on $\mathrm{BR}(\tau \to 3\mu)$ $[10^{-9}]$ |
|---|---|---|---|
| Run 1 result | 2.31 | 0.19 | 276 |
| Non-improved | 2.31 | 50.71 | 13.52 |
| Intermediate | 5.01 | 50.71 | 6.23 |
| Improved | 5.01 | 40.06 | 5.36 |

Table 1: Summary of the inputs to the limit calculation, i.e. $\mathcal{A} \times \epsilon$ and number of expected background events, $N_{\mathrm{bkg}}^{\mathrm{exp}}$, for each scenario as well as the expected 90% $\mathrm{CL_s}$ upper limit on the $\tau \to 3\mu$ branching fraction for an assumed luminosity of 3 $\mathrm{ab}^{-1}$ of pp collisions at $\sqrt{s} = 14$ TeV in the W-channel.

The efficiency of events passing either of these selections is evaluated to 14%. Since no trigger decision is simulated in the HL-LHC samples, the trigger efficiency is obtained from Run 2 simulation. The resulting $\mathcal{A} \times \epsilon$ excluding the ML efficiency is 3.1%. Since no ML classifier can be trained due to the lack of a reliable background simulation, the BDT efficiency obtained in the Run 1 analysis of 28% is applied. The upper limit is estimated assuming three different background scenarios:

1. **High background scenario:** This scenario is the most conservative approach taking the background level one order of magnitude larger than in the Run 1 W-channel analysis. The background estimated in the Run 1 analysis is scaled by the increase in luminosity and data and an additional penalty factor of 10 is applied on top.

2. **Medium background scenario:** In this scenario it is considered that the background level in the HF-channel is a factor of 3 larger than in the W-channel. The scaling according to the increase in luminosity is applied.

3. **Low background scenario:** This is the most aggressive of the scenarios considered in the HF-channel. The background level is assumed to be the same as in the W-channel, still taking into account the increase in luminosity. This scenario provides a reference for the effect of potential analysis improvements.

The background variability considered above is meant to include also a potentially less effective background rejection relative to the W-channel, where the missing energy variable plays an important role. The expected upper limits are estimated with the same procedure used in the W-channel case. Figure 2 and Table 2 summarise the $\mathrm{CL_s}$ curves and the 90% $\mathrm{CL_s}$ limits for the three scenarios considered.

| Scenario | $\mathcal{A} \times \epsilon$ [%] | $N_{\mathrm{bkg}}^{\mathrm{exp}}$ | 90% CL UL on BR($\tau \to 3\mu$) [$10^{-9}$] |
|---|---|---|---|
| High background | 0.88 | 507.05 | 6.40 |
| Medium background | 0.88 | 152.12 | 2.31 |
| Low background | 0.88 | 50.71 | 1.03 |

Table 2: Summary of the inputs to the limit calculation for each scenario as well as the expected 90% $\mathrm{CL_s}$ upper limit on the LFV branching fraction for an assumed luminosity of 3 $\mathrm{ab^{-1}}$ of pp collisions at $\sqrt{s}$ = 14 TeV in the HF-channel.

## 3 Conclusion

A study of the ATLAS experiment reach in the search for lepton flavour violation in the charged sector by searching for $\tau \to 3\mu$ decays at the HL-LHC is presented. The study is based on the results of the $W \to \tau(\to 3\mu)\nu$ search performed on the data collected during Run 1 of LHC, and takes into account several aspects of the extrapolation such as trigger selections and efficiencies, detector performance effects, luminosity and collision energy conditions, which have been validated using Run 2 simulations. Systematic uncertainties are extrapolated from the Run 1 analysis. Extrapolations to the statistics ATLAS expects to collect at the HL-LHC are performed and upper limits on the $\tau \to 3\mu$ branching fraction are obtained for different levels of background and $\mathcal{A} \times \epsilon$. Exclusion limits at 90% CL on the $\tau \to 3\mu$ branching fraction below $10^{-8}$ in the W-channel and a few $10^{-9}$ in the HF-channel are foreseen. Both channels are complementary due to the different tau production channel, contributing backgrounds and systematic uncertainties and thus a combination will yield a stronger limit.

# High Density QCD Physics

# CMS Physics Analysis Summary

Contact: cms-future-conveners@cern.ch 

# Projected Heavy Ion Physics Performance at the High Luminosity LHC Era with the CMS Detector

The CMS Collaboration

## Abstract

In this note, the projected performance of heavy ion physics in the HL-LHC era with the CMS detector is presented. The results are based on data from the pp, pPb and PbPb data taken between 2010 and 2016. The extrapolated performance with PbPb data corresponding to a total integrated luminosity of 10 $\mathrm{nb}^{-1}$ at $\sqrt{s_{\mathrm{NN}}} = 5.02$ TeV shows a dramatic improvement in the accuracy of a number of selected measurements using jets, quarkonia, and identified heavy flavor hadrons.

# 1 Introduction

In this document a brief exploratory study is presented that investigates the physics prospects of heavy ion data analysis in the High Luminosity(HL) LHC era with the CMS experiment [1]. For the HL-LHC running period the heavy ion experiments have requested to deliver an integrated luminosity of 10 $\mathrm{nb}^{-1}$ of PbPb collisions. Based on this data sample size the physics reach of a few example physics channels are studied in the context of the general topics of particle spectra measurements, elliptic flow, high-$p_{\mathrm{T}}$ and dilepton physics. These studies aim to illustrate how the large available statistics will facilitate not only a higher $p_{\mathrm{T}}$ reach for certain observables, but most importantly more differential measurements of phenomena already observed in the Run-1 and Run-2 data set. Studies of hard probes differential in parton flavor, azimuthal orientation with respect to the collision reaction plane and collision centrality are crucially needed to study the details of parton energy loss mechanism in a strongly interacting medium which will allow the precise characterization of its properties.

# 2 Impact of Detector Upgrades

Up to long shutdown three (LS3), CMS will upgrade the pixel system, the trigger and the data acquisition systems, among other upgrade projects. These improvements will allow the CMS heavy-ion program to fully exploit the high luminosity heavy-ion running for jet quenching analyses and will augment the heavy-ion reconstruction performance to about the level of the current pp reconstruction algorithms.

The biggest improvement of the heavy ion charged particle reconstruction will be provided by the first PbPb data with a four layer pixel system. Compared to the previous three layer pixel system this will significantly enhance the seed quality of the pattern recognition algorithm, which should result in an improvement of the reconstruction efficiency and reduction of the rate of falsely reconstructed trajectories. The significantly lowered material budget will also improve the secondary vertex resolution which will make CMS ideally suited for the studies of heavy flavor mesons and heavy flavor tagged jets.

# 3 Physics Performance

## 3.1 Nuclear Modification Factors of Heavy Flavor Mesons

One of the proposed observables that reveal the flavor dependence of in-medium parton energy loss is the reduction of heavy flavor meson yield. This can be studied by measurements of nuclear modification factors ($R_{\mathrm{AA}}$), defined as the ratio of the yield in nucleus-nucleus collisions to that observed in pp collisions, scaled by the number of binary nucleon-nucleon collisions. At low $p_{\mathrm{T}}$, the production rate of heavy flavor mesons in PbPb is sensitive to the elastic energy loss of the heavy quark, the gluon shadowing effect in the nuclear parton distribution function, and the rate of recombination of the heavy quark and light flavors at the hadronization stage. At high $p_{\mathrm{T}}$, the size of the suppression is sensitive to the heavy quark radiative energy loss in the pQCD-based models. Precision measurement of the $R_{\mathrm{AA}}$ from intermediate $p_{\mathrm{T}}$ ($\sim$ 10 GeV) to very high $p_{\mathrm{T}}$ (200–400 GeV) in bins of event centrality could provide insights about the path length and momentum dependence of the heavy quark energy loss and potentially distinguish between models based on AdS/CFT and pQCD.

Figure 1 shows the expected performance with the data recorded in 2015 and the projected performance in 2018 and beyond. The central values of the $R_{\mathrm{AA}}$ for charged particles and nonprompt $J/\psi$ are taken from [2] and [3], respectively. In the $p_T > 2$ GeV interval, the $D^0$ $R_{AA}$ values are taken from [4] and the spectra is smoothed using a polynomial fit on the existing measurements. In the $p_T < 2$ GeV region, the central values of $D^0$ $R_{AA}$ are taken from [5]. Finally, the $B^+$ $R_{AA}$ values are from predictions in [6]. The systematical uncertainties are scaled down to account for the increased integrated luminosity in the HL-LHC era, with a minimum of 4% per charged track.

With the high statistics jet and heavy flavor triggered sample, the precision of the $R_{AA}$ measurements at high $p_T$ could be greatly improved. At the same time, with the L1 trigger rate upgrade, the much larger minimum-bias sample enables CMS to perform these studies down to the very low $p_T$ region. The expected precision of charged particle, $D^0$ and $B^+$ $R_{AA}$ measurements from low $p_T$ to high $p_T$ could provide a strong constraint on theoretical models, and the significant difference in the suppression magnitude between those mesons could be observed for the first time. In addition, the comparison between $B_s$, $B^0$ and $B^+ R_{AA}$ will become possible and the first measurements of flavor identified $B$ mesons has been performed in proton-lead collisions [7].

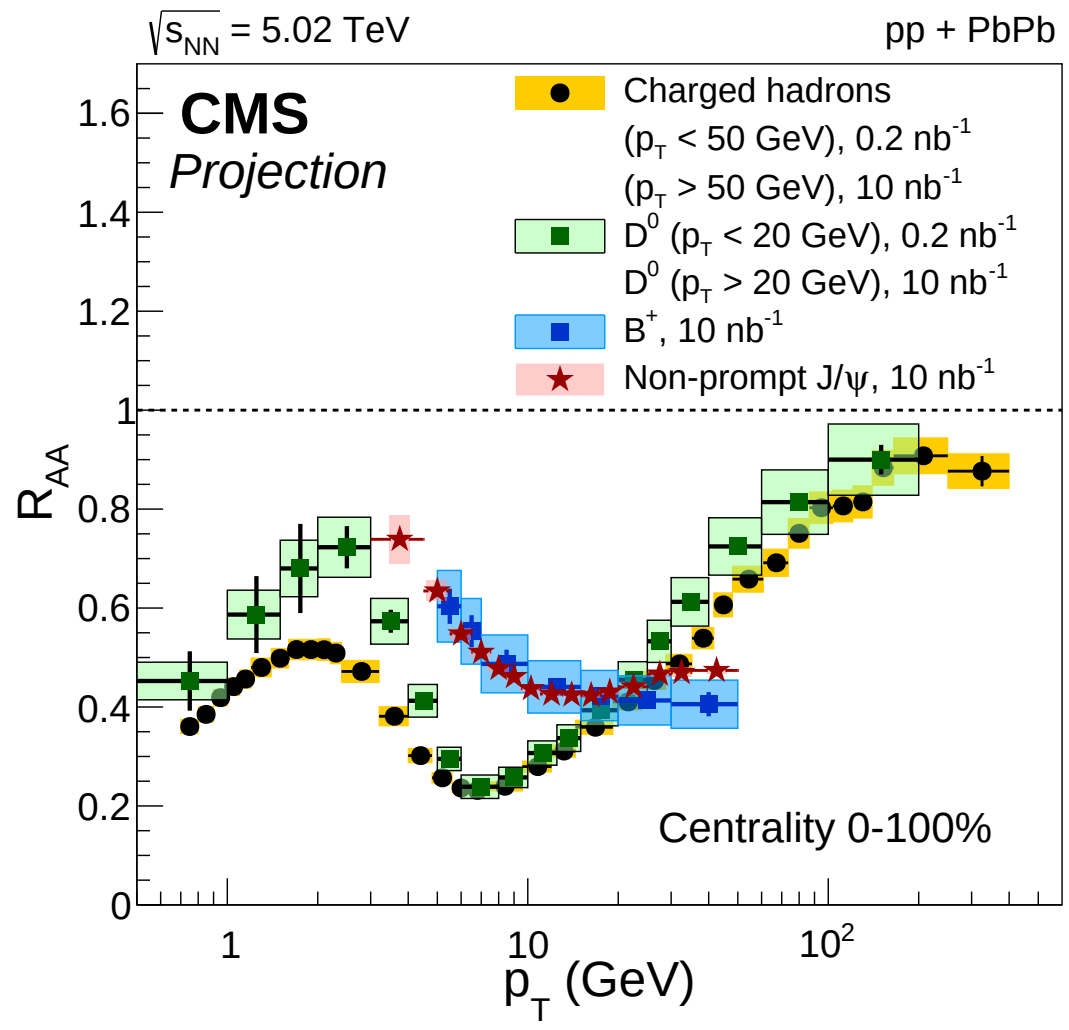

Figure 1: Nuclear modification factors of charged particles, $D^0$, $B^+$ and nonprompt $J/\psi$ with the PbPb statistics expected with 10 nb$^{-1}$.

## 3.2 Azimuthal anisotropy of Heavy Flavor Mesons

Azimuthal anisotropy of hadrons provides information about their production with respect to the reaction plane ($\Psi_n$). At high $p_T$, the larger in-medium path-length of the mother partons emitted in the direction of the reaction plane leads to a stronger suppression of the yield due to jet quenching. Therefore, measurements of the $v_n$ coefficients from Fourier expansion of the particle distributions $dN/d\psi$ are sensitive to path length dependence of the parton energy loss. At low $p_T$, a large $v_2$ (elliptic flow) signal is considered as evidence for collective hydrodynamical expansion of the medium. Measurements of heavy flavor meson $v_n$ could provide important information about the thermalization of the heavy quarks in the medium. Precision measurements of $v_n$ as a function of heavy flavor meson $p_T$ could teach us how the azimuthal anisotropy of the light flavor partons contribute to the observed anisotropy through the recombination of heavy quarks and light quarks. The predicted elliptic flow ($v_2$) signal covers a large range of values due to the difference in the treatments of in-medium parton transport and parton energy loss.

Figure 2 shows the expected performance of elliptic flow ($v_2$) measurements with the data recorded in 2015 and the projected performance in 2018 and beyond. The central values are taken from the previous CMS publications with Run 2 data [8, 9] and smoothed with a polynomial fit. The systematics are scaled down to account for the increased luminosity in the HL-LHC era. The new precise data will be able to constrain various components of the theoretical models to determine the heavy quark diffusion coefficient in QGP, and to reveal the possible flavor dependence of the parton energy loss.

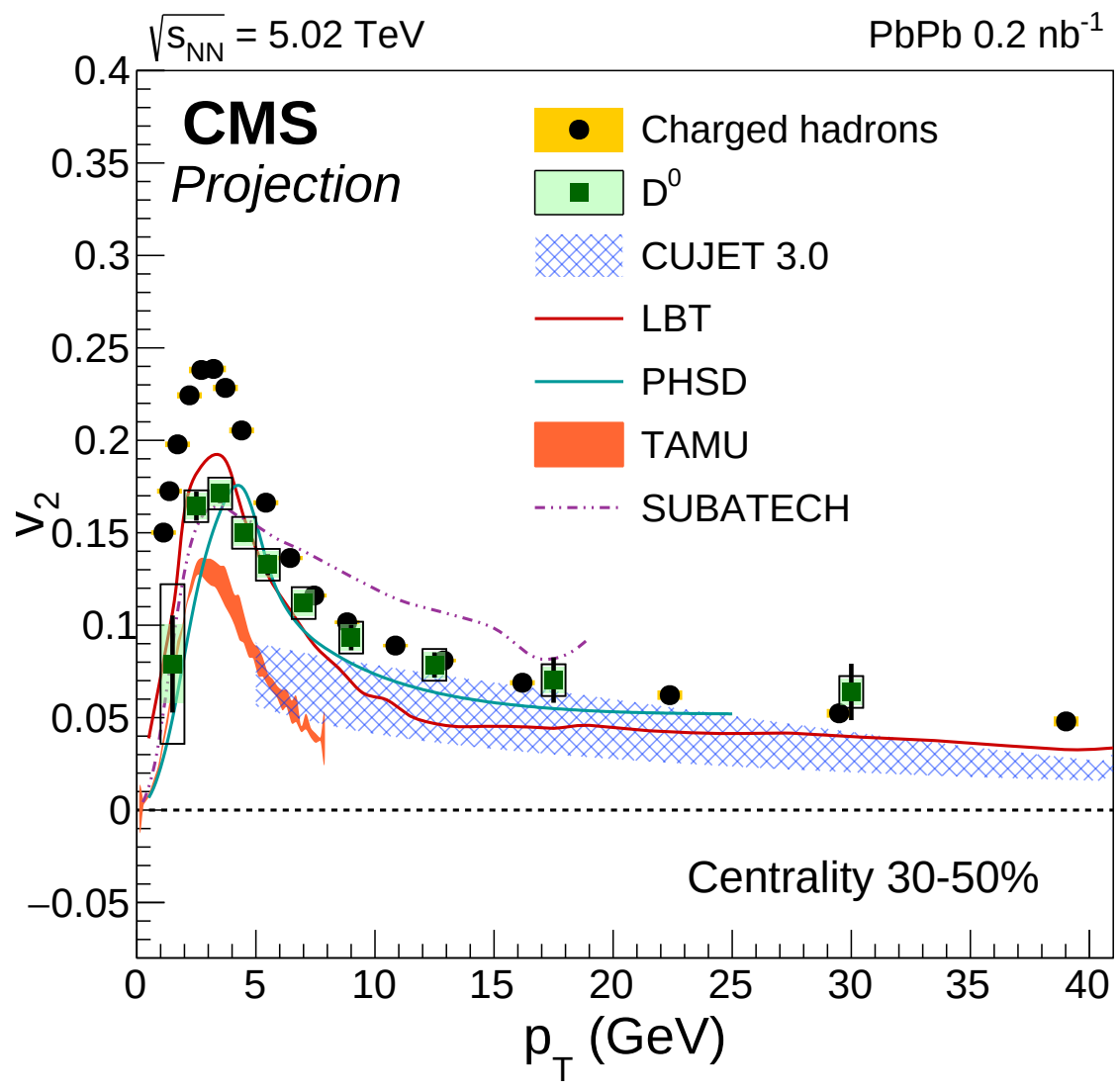

Figure 2: $v_2$ of charged particles, $D^0$ with the PbPb statistics expected with 10 nb$^{-1}$ compared to theoretical predictons of $D^0$ $v_2$.

## 3.3 Boson-Jet Transverse Momentum Balance

Important channels to study parton energy loss, which should be performed with high accuracy at the HL-LHC are the $\gamma$+jet and Z+jet transverse momentum balance. Comparing Z+jet and $\gamma$+jet observables to dijets [10, 11] will allow the studies of quark and gluon initiated jets. Figure 4 and 3 shows the expected performance at the HL-LHC. The central values are based on the smoothed data from the previous CMS publication [12, 13]. The systematical uncertainties are reduced by a factor of two with respect to the results with 2015 data due to the possible improvements on the jet energy scale and jet energy resolution uncertainties. The collected number of $\gamma$+jet events will also be sufficient to study the jet quenching as a function of the reaction plane for the first time.

## 3.4 Inclusive Jet substructure variables

Interaction with the hot QCD medium is expected to result in an increase in the gluon radiation probability of the propagating partons. It could also lead to modifications of the distribution of momentum of the parton shower as well as opening angle between the original and the radiated partons. Using the jet grooming algorithm "soft drop" [14], one could remove large angle soft radiation inside a jet [15–18], leaving the hard structure of the jet as two subjets. The momentum sharing and the opening angle can be measured through jet substructure variables such as jet splitting function (Figure 5) and groomed jet mass (Figure 6). The central values of the extrapolated splitting function and jet mass are from previous CMS publications [19, 20]. The systematical uncertainties are reduced by a factor of two with respect to the results with 2015 data due to the possible improvements on the jet energy scale and jet energy resolution uncertainties. With the HL-LHC data, those jet substructure observables could be measured with unprecedented accuracy and provide important constraints on the magnitude of the correlated medium response and the parton energy loss mechanism.

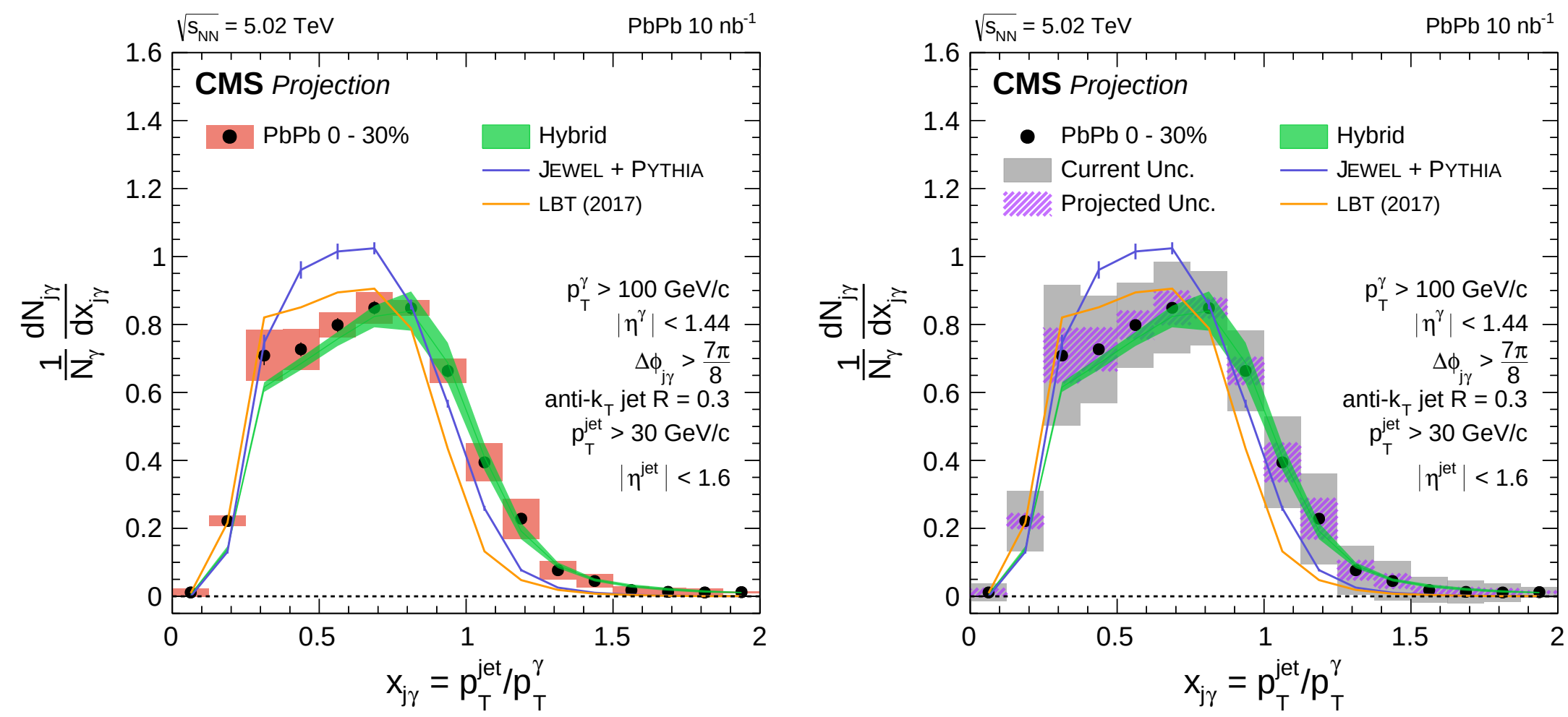

Figure 3: (Left Panel) $X_{j\gamma}$ distribution for isolated-photon+jets of $p_\gamma > 100$ GeV/c and $|\eta_\gamma| < 1.44$, $p_{jet} > 30$ GeV/c and $|\eta_{jet}| < 1.6$ in the HL-LHC data (Right Panel) Comparison between the current performance with 0.4 nb$^{-1}$ of PbPb data collected in 2015 and with HL-LHC data.

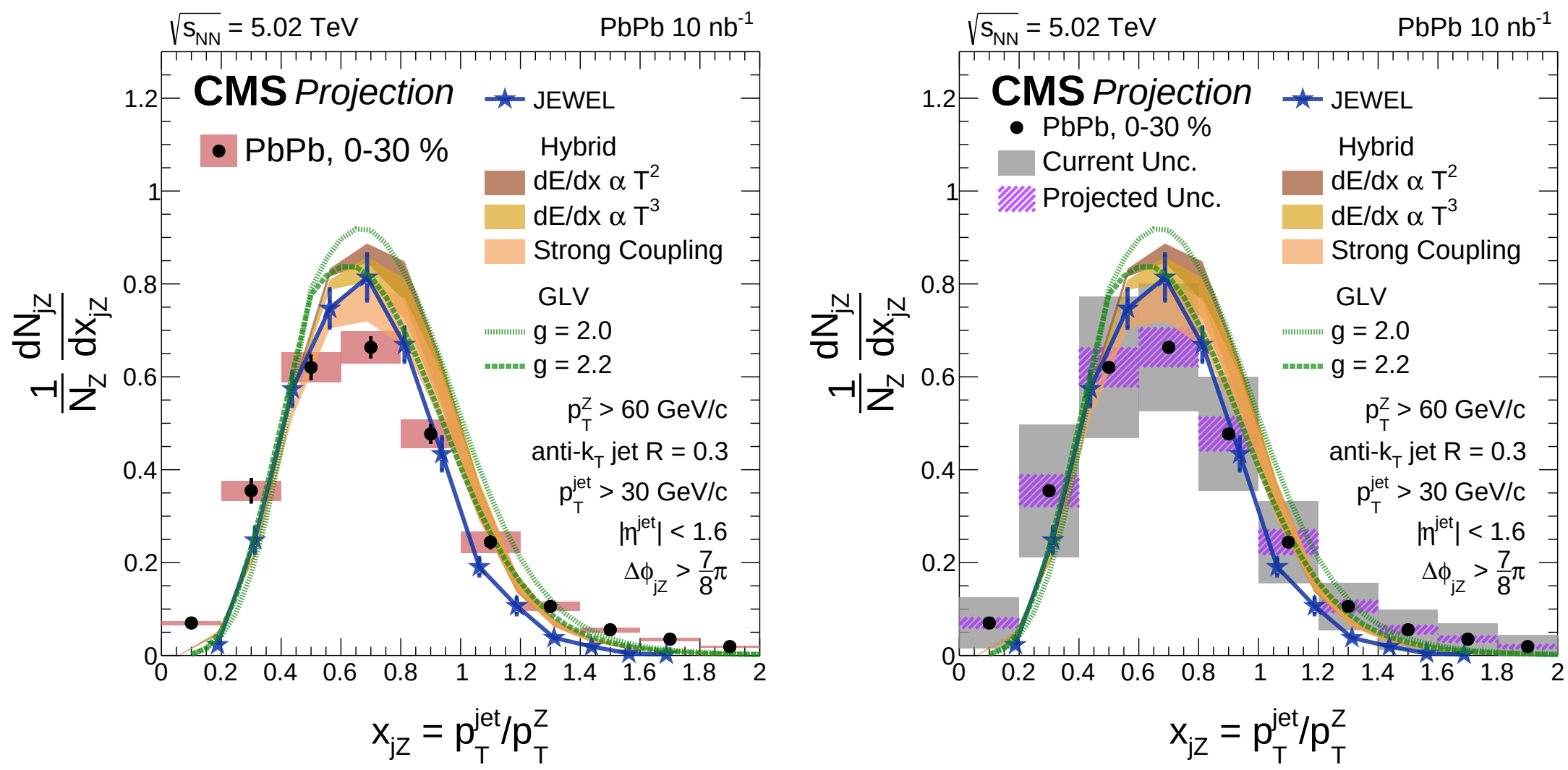

Figure 4: (Left Panel) $X_{jZ}$ distribution for isolated-photon+jets of $p_Z > 100$ GeV/c, $p_{jet} > 30$ GeV/c and $|\eta_{jet}| < 1.6$ in the HL-LHC data (Right Panel) Comparison between the current performance with 0.4 nb$^{-1}$ of PbPb data collected in 2015 and with HL-LHC data.

### 3.5 Jet substructure using photon-tagged jets

Significant amount of progress has been made in the LHC collaborations for the studies of jets and dijet pairs such as jet fragmentation functions [15], missing transverse momentum [10, 16] in dijet systems and jet-track correlation [17, 18]. However, understanding how properties of the measured jets relate to their parent partons is one of the key challenges of these measure-ments when selecting events based on jet kinematics since the amount of energy lost to the medium before forming the final-state particles comprising the jet cannot be unambiguously determined. One way to overcome this difficulty is to study the isolated photon-tagged jets, where the photon energy can be used to tagged the away-side and serve as an reference when comparing PbPb data to pp references. The expected performance of such kind of measurement with the HL-LHC data is shown in Figure 7. The central values of the extrapolated spectra are obtained by smoothing the results from [21]. The systematical uncertainties are reduced by a factor of two with respect to the results with 2015 data due to the possible improvements on the jet energy scale and jet energy resolution uncertainties. The photon-tagged fragmentation function could be measured with high precision and provide valuable insights about the modification of the jet substructure of quark initiated jets in the strongly interacting medium.

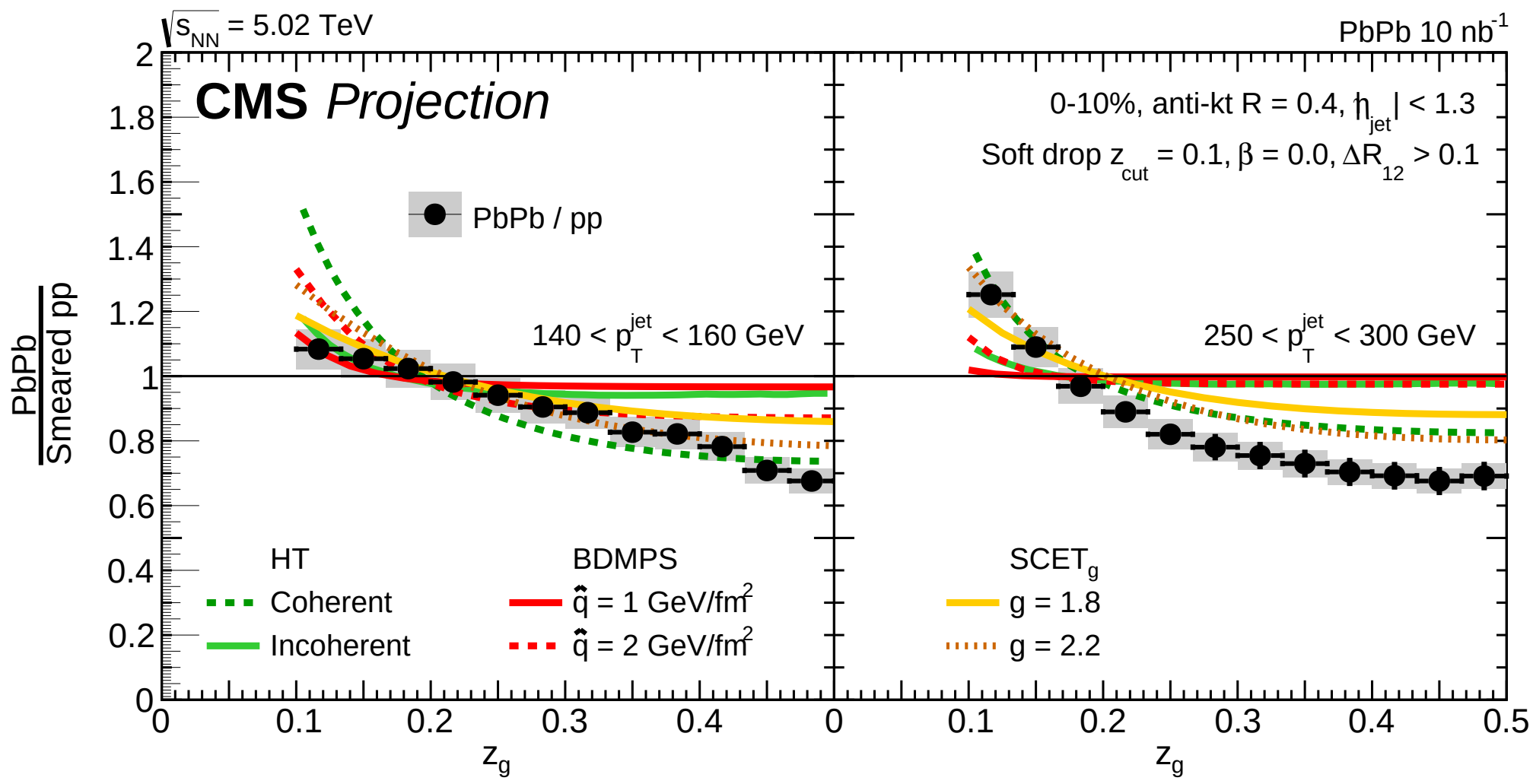

Figure 5: Performance of jet splitting function measurement with HL-LHC data in PbPb collisions

### 3.6 Quarkonia Production

CMS has an enormous potential in measuring the suppression patterns of the five quarkonium states as a function of transverse momentum, rapidity and centrality. These measurements provide crucial information to aid our understanding of the phase transition and using the melting temperature of each state to determine the medium temperature. Figure 8 shows the performance of charmonia and bottomia with HL-LHC PbPb data. The charmonia mass spectra are from previous publication [22]. The central values of the $\Upsilon$(nS) nuclear modification factors are obtained from the theoretical calculation in [23] and the systematical uncertainties are reduced by a factor of 3 compared to the CMS publication [24] due to the improvement coming from a larger dimuon control sample in HL-LHC era. Potentially a significant $\Upsilon(3S)$ signal could be observed for the first time in PbPb collisions.

### 3.7 W Boson Production in Proton-Lead collisions

Electroweak boson production in proton-nucleus collisions offers a unique opportunity to probe nuclear parton distribution functions (nPDFs). For $W$ boson production at the LHC, the dominant processes are $u\bar{d} \to W^+$ and $d\bar{u} \to W^-$, reflecting interactions that take place between valence quarks and sea antiquarks. The asymmetry of the pPb collision system allows for the measurement of forward-backward pseudorapidity asymmetries $A_{\mathrm{FB}}$, which are sensitive to the nuclear modifications of the PDFs. Figure 9 shows the expected performance in HL-LHC era which provides strong constraints for the nPDFs, where the systematical uncertainties are scaled down by a factor of 3 compared to the current measurement [25].

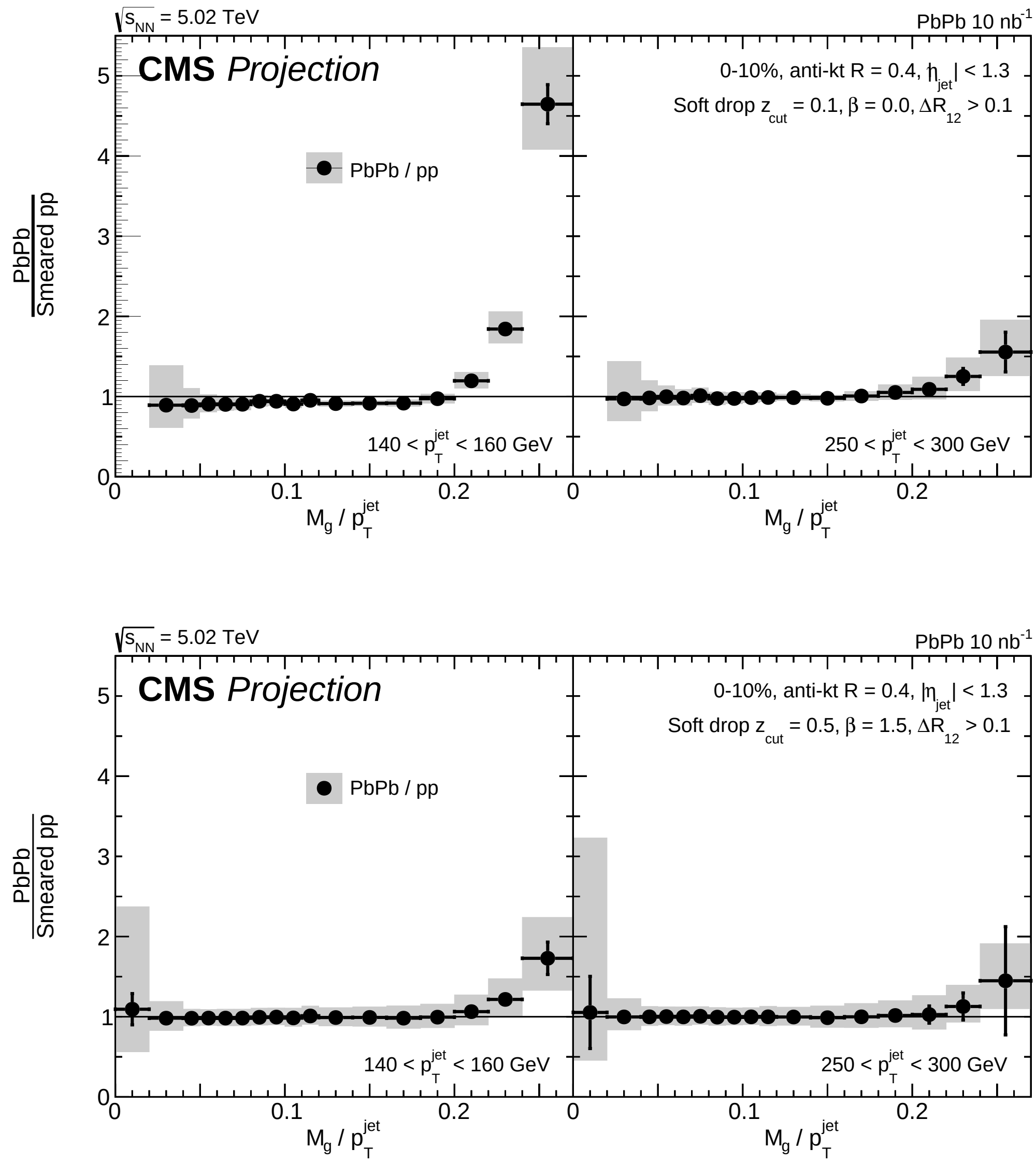

Figure 6: Jet Mass distribution with grooming setting $(z_{cut}, \beta) = (0.1, 0.0)$ (Upper plots) and $(z_{cut}, \beta) = (0.5, 1.5)$ (Lower plots)

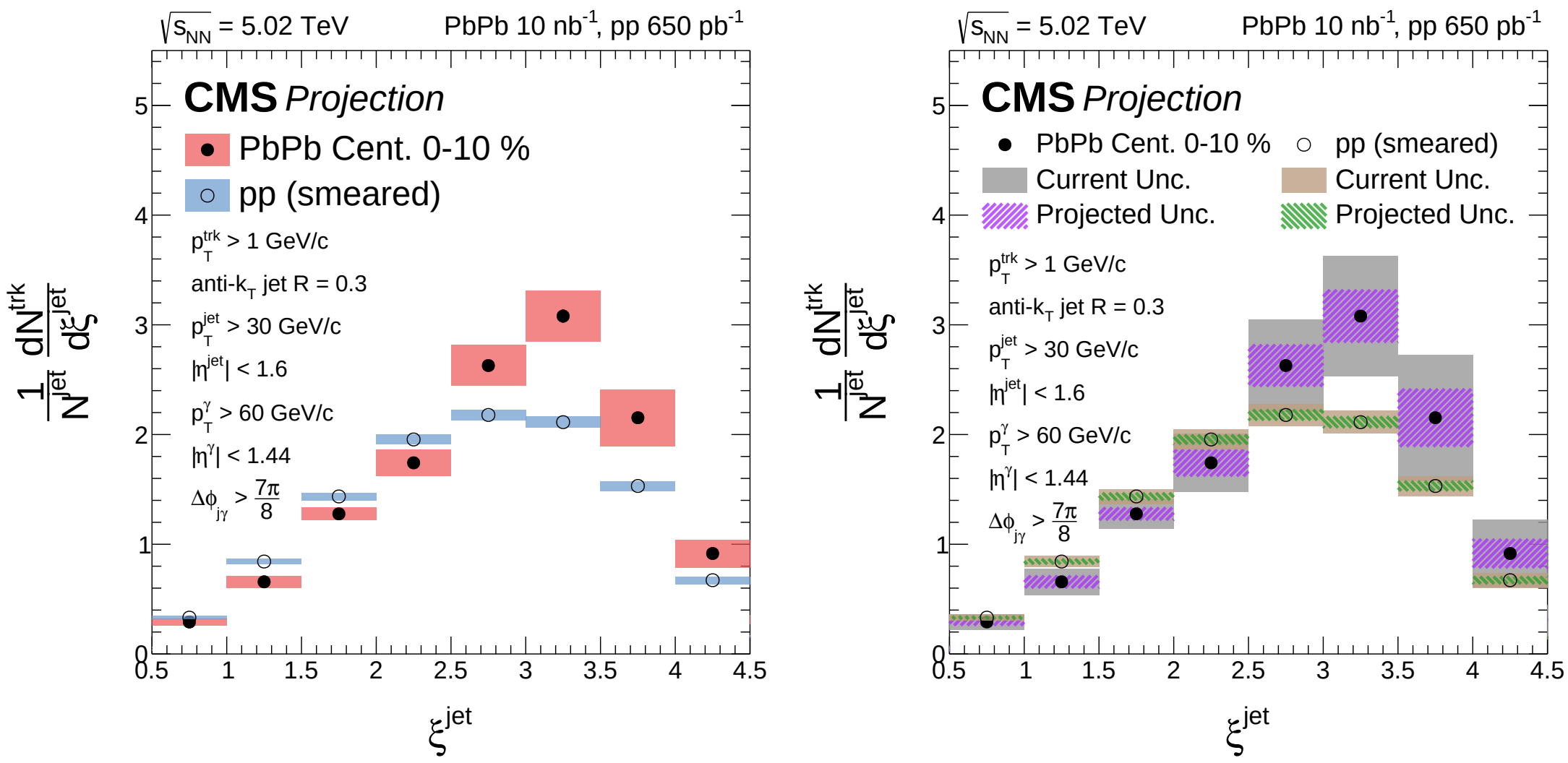

Figure 7: (Photon-tagged fragmentation function in the HL-LHC data (Left Panel) Comparison between the current performance with 0.4 nb$^{-1}$ of PbPb data collected in 2015 and with HL-LHC data. (Right Panel)

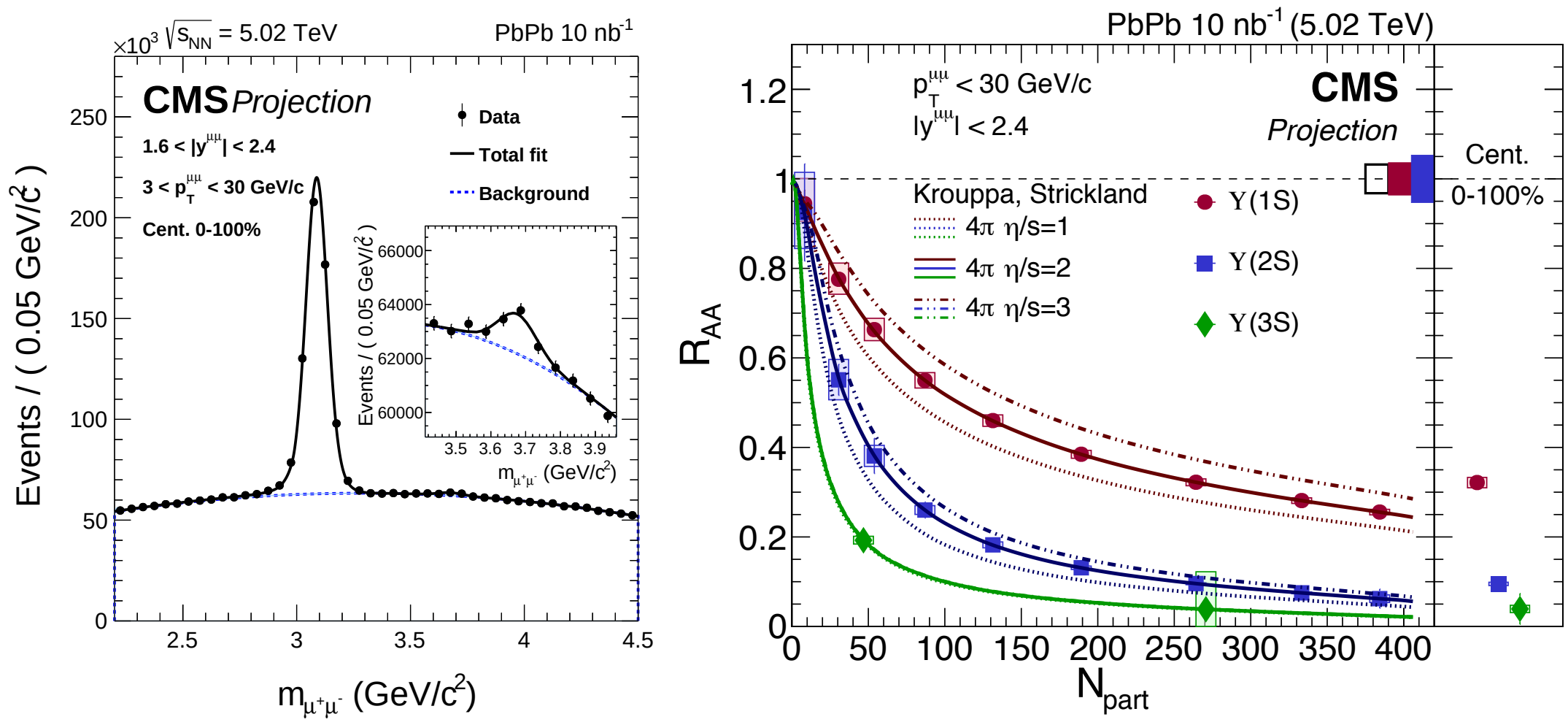

Figure 8: (Left panel) Dimuon mass distribution in the charmonia mass region (Right panel) Performance of upsilon nuclear modification factors

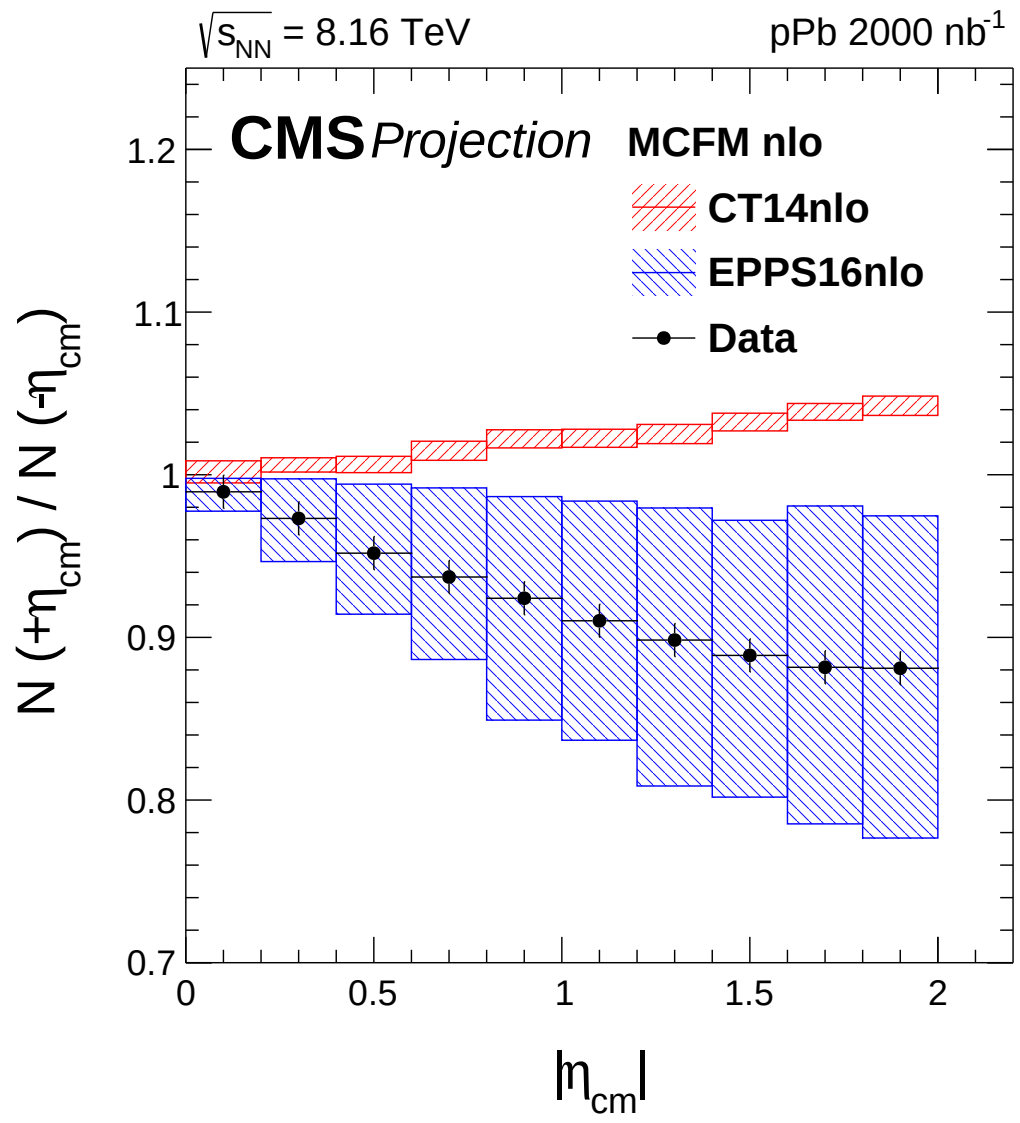

Figure 9: Forward-backward asymmetry of the W boson production in proton-lead collisions at the HL-LHC

# Projections for ATLAS Measurements of Bulk Properties of Pb+Pb, $p$+Pb, and $pp$ Collisions in LHC Runs 3 and 4

The ATLAS Collaboration

This note describes expected ATLAS measurements of bulk and flow properties of dense nuclear matter measured in heavy-ion collisions (AA and pA) in LHC Runs 3 and 4. In particular projections for heavy flavor flow coefficients measured using muons, and flow decorrelation measured with charged particles are presented for Pb+Pb collisions. For $p$+Pb collisions projections Hanbury-Brown Twiss radii and forward-backward multiplicity correlations are shown. Multi-particle cumulants measured in $pp$ collisions are also presented. The projections are made for a baseline of 10 nb$^{-1}$ of Pb+Pb collisions, as well as considering both 500 nb$^{-1}$ and 1 pb$^{-1}$ of $p$+Pb collisions, and 200 pb$^{-1}$ for $pp$ collisions. The performance of the ATLAS Inner Tracker upgrade in heavy-ion collisions is also discussed.

# 1 Introduction

Among the key features of the quark gluon plasma (QGP) created in heavy-ion collisions are the collective properties of produced particles, in particular azimuthal anisotropy of particle emission. Sophisticated analyses of the azimuthal angle distributions of charged particles in Pb+Pb collisions are effective tools to study the QGP [1]. They have been found to be largely consistent with a hydrodynamic description of the produced matter, and are used as input to constrain and tune models of the collision system [2]. In addition, in studies of *p*+Pb collisions, which were initially assumed not to display any collective flow, similar properties to Pb+Pb collisions are observed [3–5]. Future measurements in *p*+Pb collisions are expected to shed light on the relationship between the collective properties observed in the *p*+Pb system and the QGP.

This note presents several quantities of interest in Pb+Pb and *p*+Pb collisions and demonstrates expected improvements to the measurements which can be made with the luminosity and detector upgrades in the upcoming LHC Runs 3 and 4. In particular, the increased acceptance of the Inner Tracker (ITk) will be beneficial for measurements of both soft and hard scale in heavy ion collisions, such as measurement of multi-particle correlations or production of forward jets. The projections are made for a baseline expected luminosity of 10 $nb^{-1}$ (5 $nb^{-1}$ each in Runs 3 and 4) of Pb+Pb data and two luminosity scenarios for *p*+Pb: 500 $nb^{-1}$ and 1000 $nb^{-1}$. Many of the considered analyses use triggers that do not sample the entire delivered luminosity (*e.g.* a minimum bias trigger), it is assumed that the fraction of the delivered luminosity sampled by the relevant triggers will remain the same as in existing analyses.

# 2 ATLAS detector and Inner Tracker upgrade

The ATLAS detector is described in detail in Ref. [6]. Significant upgrades to the ATLAS detector are expected to be installed for Run 4, in particular the ITk [7] will be of great importance to the discussed measurements. The primary benefit of the ITk for the measurements discussed in this note is the increased tracking acceptance of charged hadrons. The new tracking detector will have a pseudorapidity[1], $\eta$, coverage of 8 units, $|\eta| < 4$ (for comparison the present ATLAS Inner Detector covers $|\eta| < 2.5$). The design and sensor layout of the ITk detector is described in Refs. [8] and [9]. The considered ITk layout is an all-silicon detector consisting of 5 pixel and 4 double-sided strip barrel layers. The end-cap system on each side consists of 6 strip discs arranged to provide optimal coverage and pixel disk arrangement in the forward direction covering the pseudorapidity range up to $|\eta| = 4.0$. The ITk is immersed in solenoidal magnetic field of 2 T to measure charged particle transverse momenta.

Performance expectations for the ITk in *pp* collisions are detailed in [7] while in this section results for 5.02 TeV Pb+Pb collisions are presented. In order to study the performance of the ITk detector, a minimum bias (0–100% centrality) sample of $10^4$ Pb+Pb Monte Carlo simulated events were generated using HIJING version 1.38b [10]. The generated sample is passed through a full simulation of the detector using Geant4 [11, 12], and the simulated events are reconstructed using the same software as in Ref. [7]. In this study the “Inclined duals” geometry layout is used, as described in Ref. [8]. The performance of the detector in simulated Pb+Pb collisions is found to be comparable to the performance in *pp* collisions.

[1] ATLAS uses a right-handed coordinate system with its origin at the nominal interaction point (IP) in the centre of the detector and the $z$-axis along the beam pipe. The $x$-axis points from the IP to the centre of the LHC ring, and the $y$-axis points upward. Cylindrical coordinates $(r, \phi)$ are used in the transverse plane, $\phi$ being the azimuthal angle around the $z$-axis. The pseudorapidity is defined in terms of the polar angle $\theta$ as $\eta = -\ln\tan(\theta/2)$.

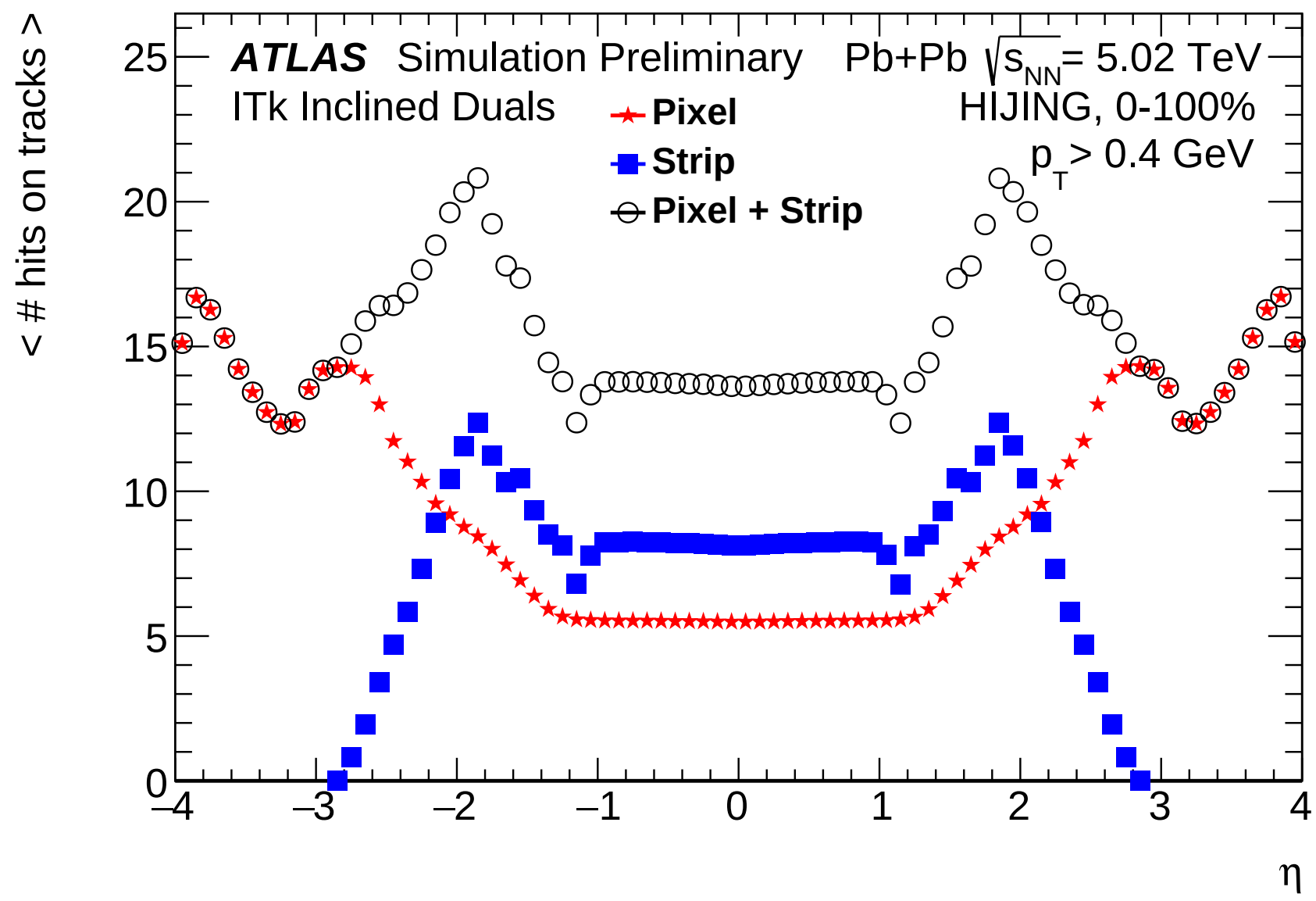

Figure 1: The mean number of hits per track as a function of pseudorapidity for the different detector sub-systems in minimum bias (0–100% centrality) Pb+Pb collisions with the ITk upgrade.

The mean number of silicon clusters (hits) per reconstructed track with the transverse momentum $p_T > 0.4$ GeV is shown in Figure 1 as a function of $\eta$ for the pixel tracker, the strip tracker, and both together. The ITk provides at least 12 silicon hits for a track with $|\eta| < 4$. In the barrel region 8 strip clusters are expected on average, and 5 pixel clusters. In the end-cap regions, the number of clusters changes over the pseudorapidity range $1 < |\eta| < 4$. A larger number of hits per track, up to $\approx 21$, is observed in the transition region $|\eta| \approx 2$.

The tracking efficiency is defined as a fraction of the number of reconstructed tracks matched to truth particles [7]. Figure 2 shows the efficiency to reconstruct tracks of charged particles, with $p_T > 0.4$ GeV, as a function of $\eta$ and $p_T$ in minimum bias Pb+Pb collisions. A reconstruction efficiency of 70-90% averaged over $p_T$ is attained. The efficiency increases with transverse momentum from 65% at lowest-$p_T$ to about 85% at $p_T$ of 1 GeV and 90% at $p_T$ of 10 GeV. The inefficiency is largely due to hadronic interactions in the detector material. Each reconstructed track is characterized by impact parameters, $d_0$ and $z_0$, where $d_0$ is the distance of the point of closest approach to the beam axis and $z_0$ is its $z$ coordinate. The resolution of track impact parameters $\sigma(d_0)$ and $\sigma(z_0)$ as a function of pseudorapidity in minimum bias Pb+Pb collisions is shown in Figure 3. The impact parameter resolution of low-momentum particles, $p_T \sim 0.4$ GeV, is dominated by the effects of multiple scattering. At $|\eta| = 0$, $\sigma(d_0) \approx 0.32$ mm and $\sigma(z_0) \approx 0.34$ mm and they increase with $|\eta|$ due to the larger impact of multiple scattering for particles impinging the detector at steeper angles.

## 3 Projections

In the following sub-sections projections are made for several expected observables in Pb+Pb and $p$+Pb collisions. The projections presented in this note focus on improvements in statistical uncertainties and detector acceptance. For the systematic uncertainties, some improvement is expected for the Runs 3 and 4 data based on advancement in measurement technique and improved understanding of the detector. However, at present, it is not possible to make a quantitative projection of these expectations for most of the analyses discussed in this note.

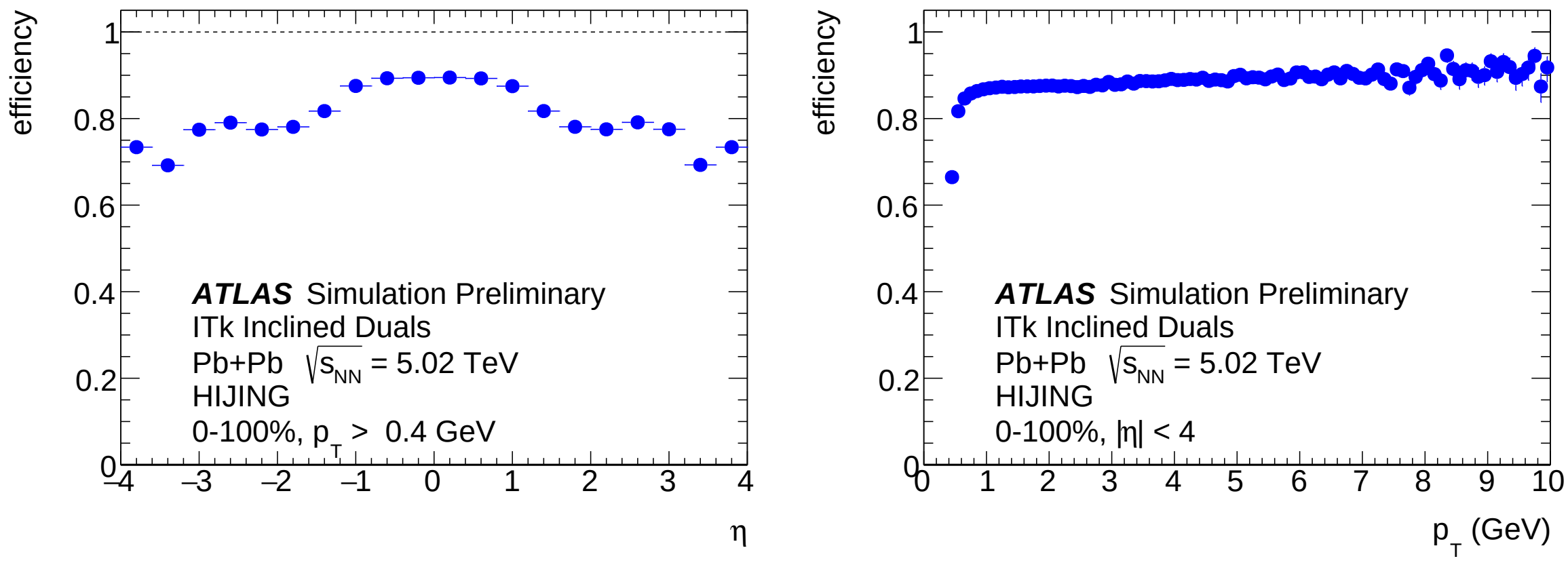

Figure 2: Track reconstruction efficiency as a function of pseudorapidity and transverse momentum in minimum bias (0–100% centrality) Pb+Pb collisions with the ITk upgrade.

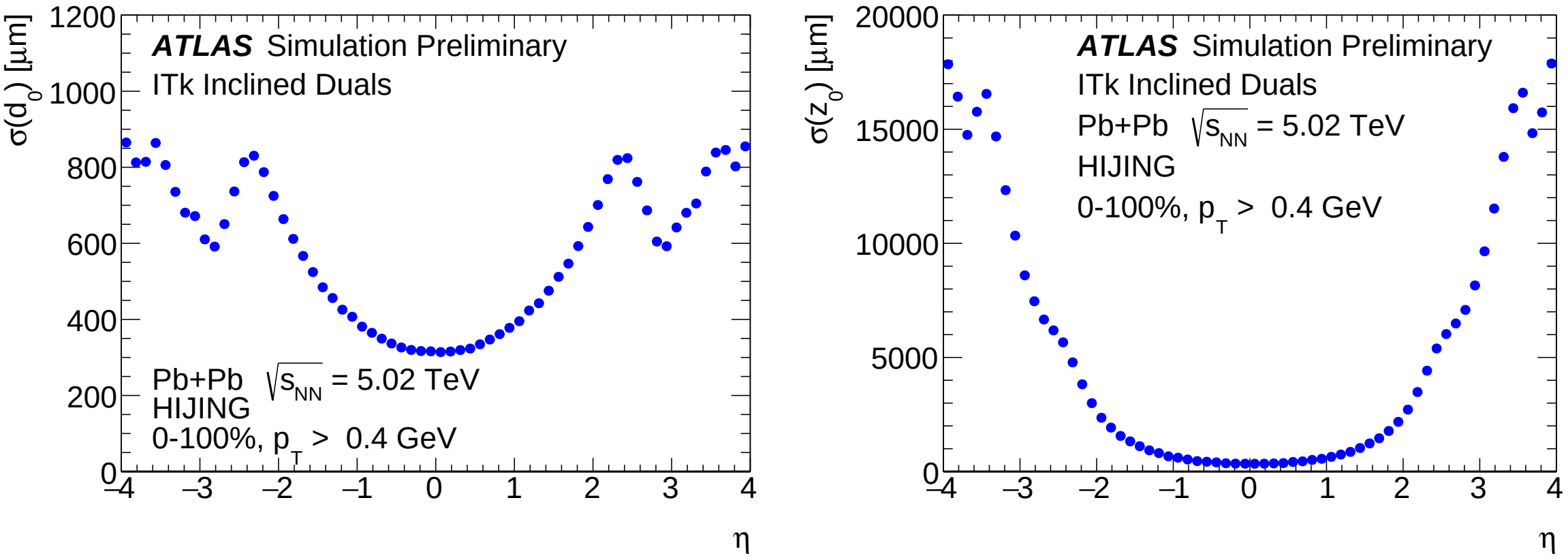

Figure 3: Resolution of track parameters $d_0$ (left) and $z_0$ (right) as a function of pseudorapidity for minimum track $p_T$ threshold 0.4 GeV in minimum bias (0–100% centrality) Pb+Pb collisions with the ITk upgrade.

## 3.1 Pb+Pb Measurements

### 3.1.1 Heavy-flavor flow

Understanding the interactions of heavy quarks in the quark gluon plasma can help to understand transport properties of the plasma [13, 14]. ATLAS has performed measurements of the elliptic anisotropy of muons from heavy-flavor decays, $v_2^{\mathrm{HF}\to\mu}$ [15], but these are of limited experimental precision. Figure 4 shows projections for the elliptic anisotropy of muons from heavy-flavor decays. The projections are made at $\sqrt{s}$ = 2.76 TeV assuming an integrated luminosity of 10 nb$^{-1}$ and that the muon trigger with a 4 GeV threshold will sample the entire luminosity. The existing measurements [15] using Run 1 data corresponding to a luminosity of 0.14 nb$^{-1}$ are also shown. The central values of the projections are obtained by fitting the present measurements with an exponential function. The statistical uncertainties in the projections are made by scaling down the present uncertainties to correspond to the expected 10 nb$^{-1}$. The projections are also compared to calculations from the DABMod model [16]. Figures 5 and 6 show similar distributions for $v_3^{HF\to\mu}$ and $v_4^{HF\to\mu}$, respectively.

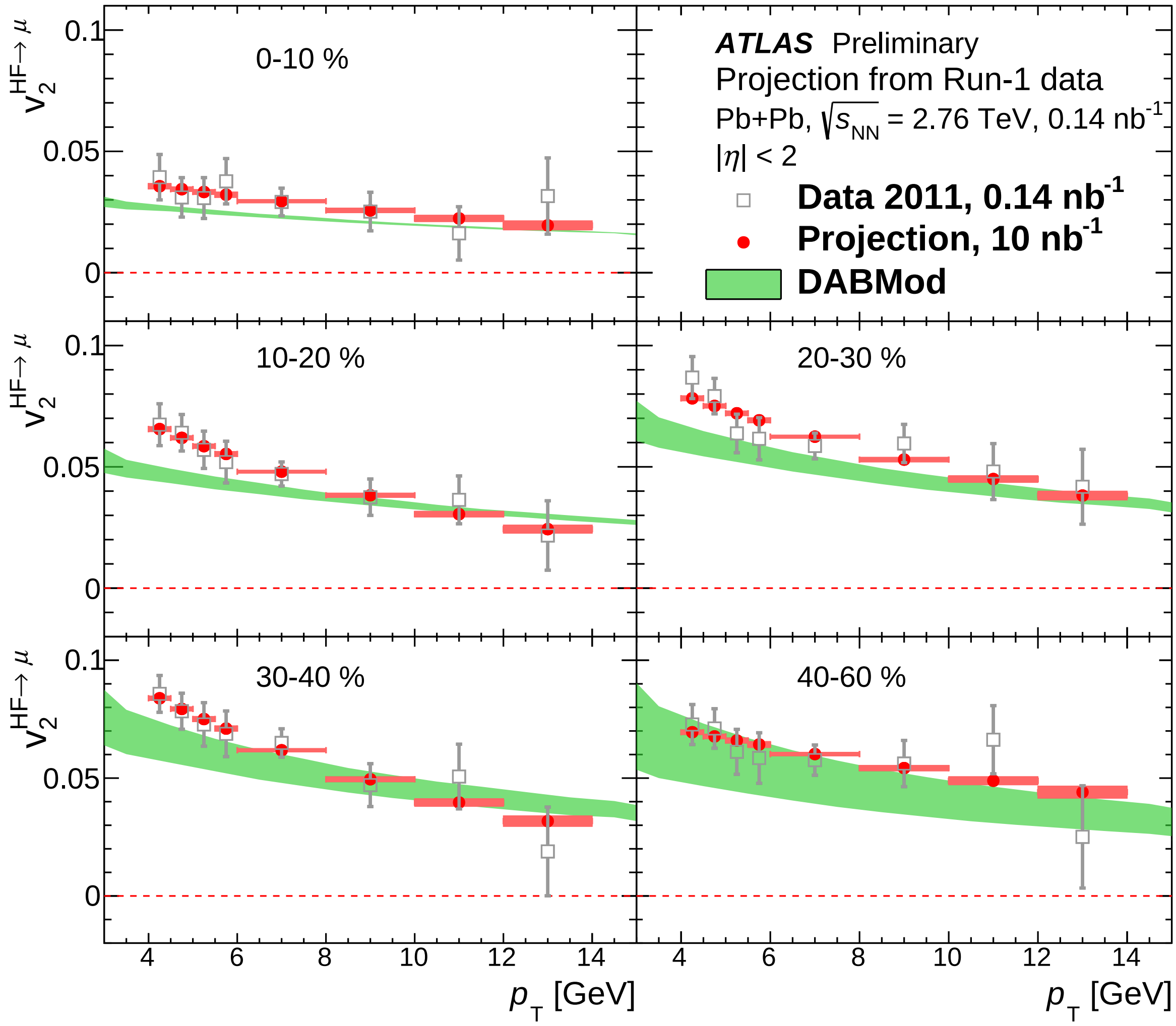

Figure 4: Projection of $v_2$ as a function of $p_T$ of muons from the decay of heavy-flavor quarks. Each panel corresponds to a different centrality interval. The present measurements are also shown for comparison. The error bars (shaded boxes) correspond to statistical uncertainties only. The projections are also compared to calculations from the DABMod model.

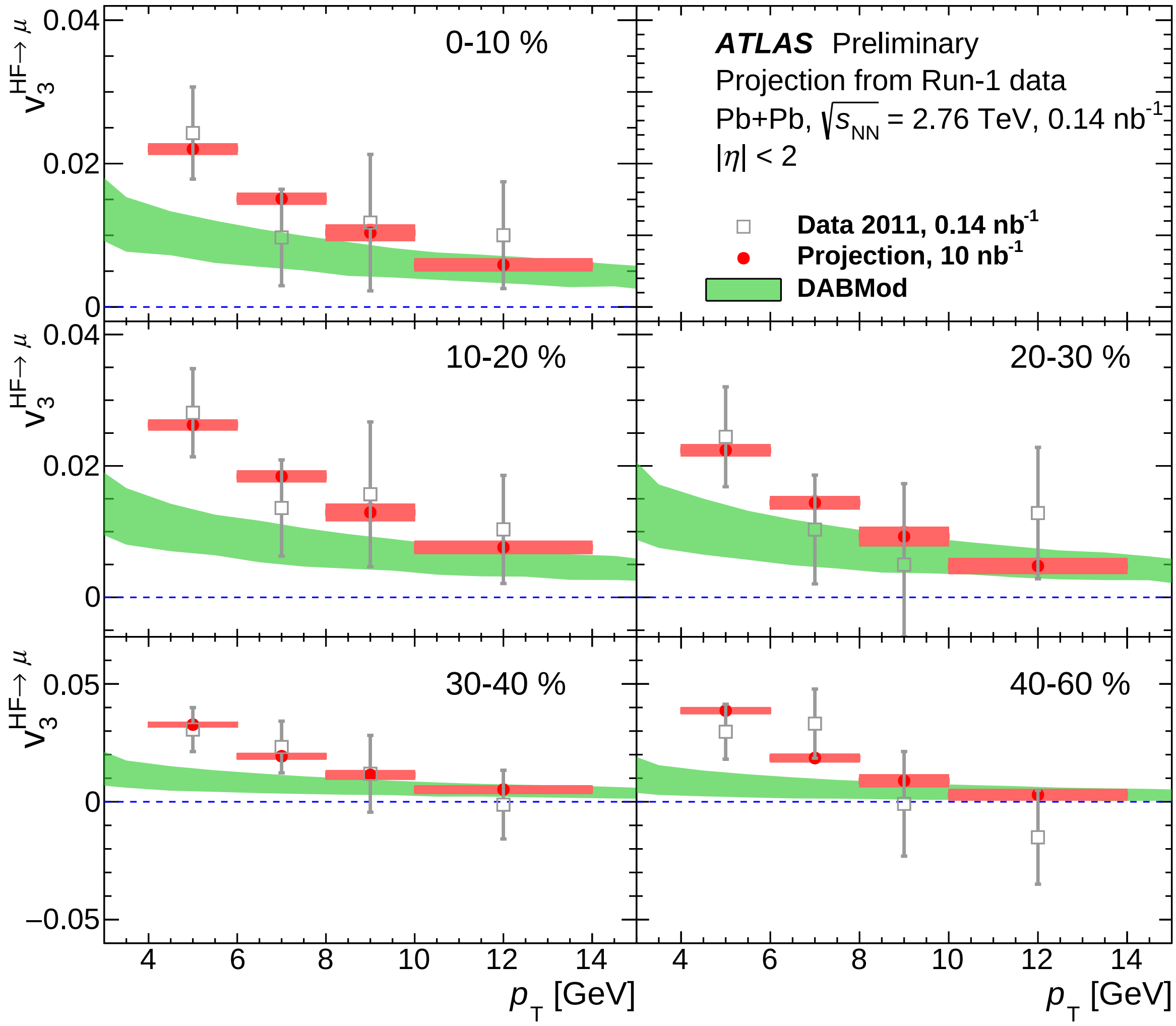

Figure 5: Projection of $v_3$ as a function of $p_T$ of muons from the decay of heavy-flavor quarks. Each panel corresponds to a different centrality interval. The present measurements are also shown for comparison. The error bars (shaded boxes) correspond to statistical uncertainties only. The projections are also compared to calculations from the DABMod model.

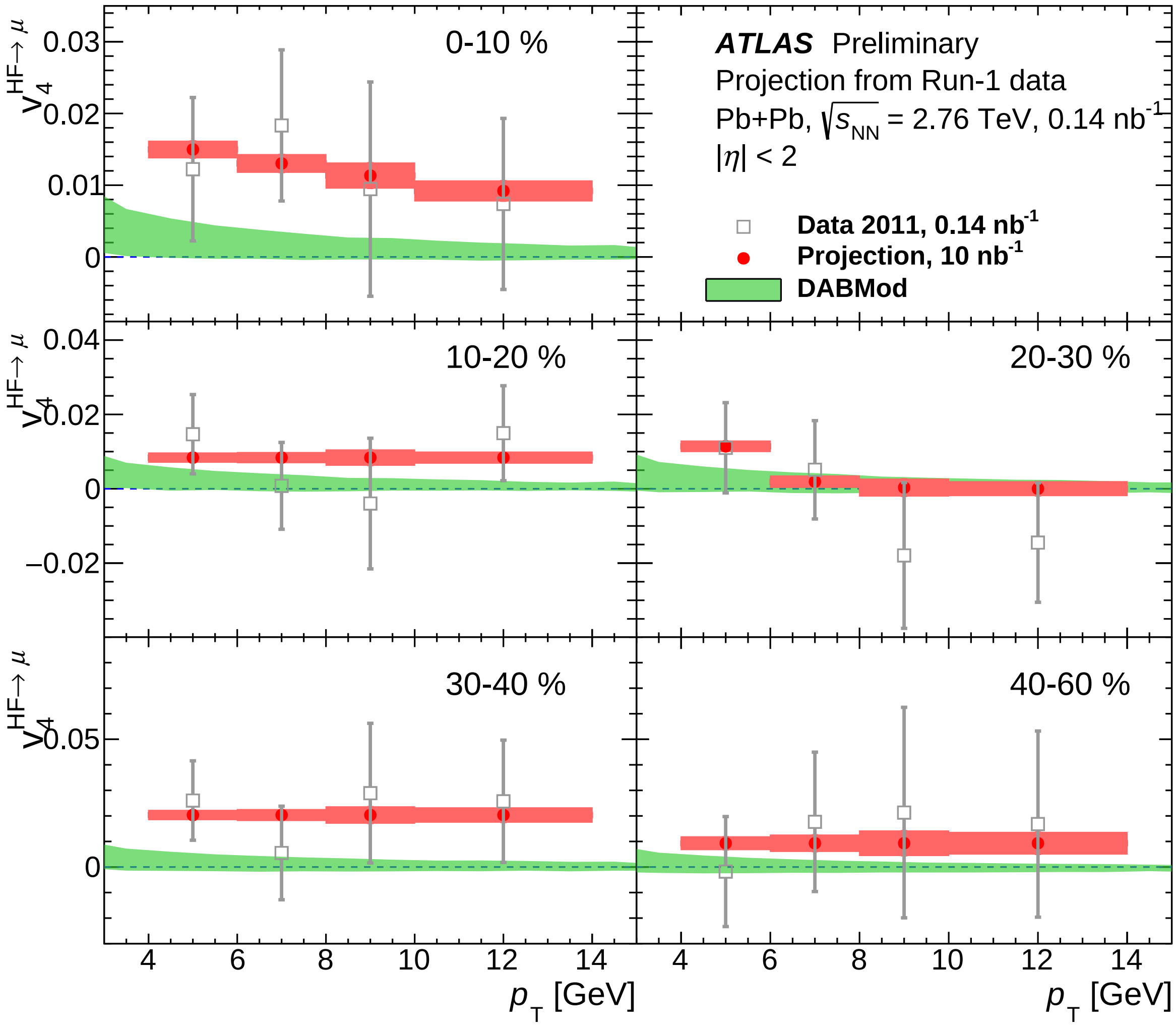

Figure 6: Projection of $v_4$ as a function of $p_T$ of muons from the decay of heavy-flavor quarks. Each panel corresponds to a different centrality interval. The present measurements are also shown for comparison. The error bars (shaded boxes) correspond to statistical uncertainties only. The projections are also compared to calculations from the DABMod model.

### 3.1.2 Flow decorrelation

ATLAS has observed that the factorization of two-particle azimuthal correlations into single-particle flow harmonics is broken [17]. Repeating this measurement in Run 4 will lead to significant improvement due to a larger dataset and especially due to increased tracking acceptance in pseudorapidity with the presence of the ITk detector.

The amount of factorization breaking, or decorrelation, is quantified by constructing the correlator $r_{n|n;1}$ for the first moment of the flow vectors defined as:

$$r_{n|n;1}(\eta) = \frac{\langle \mathbf{q}_n(-\eta)\mathbf{q}_n^*(\eta_{\text{ref}})\rangle}{\langle \mathbf{q}_n(\eta)\mathbf{q}_n^*(\eta_{\text{ref}})\rangle} \tag{1}$$

where $\mathbf{q}_n$ is the normalized flow vector, and $\eta_{\text{ref}}$ is the reference pseudorapidity [17]. The correlator, $r_{n|n;1}$, measures the relative difference between flow $v_n e^{in\Phi_n}$ at different $\eta$ and $-\eta$. When flow is boost-invariant, $r_{n|n;1}$ will always equal unity. However any difference in the $\eta$ dependence of the flow magnitude $v_n$ and the event plane angle $\Phi_n$ would result in $r_{n|n;1}$ smaller than 1. Here, the focus is on elliptic flow which corresponds to the $n = 2$ case. In Run 2, $r_{2|2;1}$ was measured as a function of $\eta$ for different event centrality classes and it was found that $r_{2|2;1}$ differ significantly between central and peripheral collisions [17]. In particular, it was found that:

- $r_{2|2;1}$ is much smaller in central collisions than in more peripheral collisions, which indicates stronger flow decorrelation.
- $r_{2|2;1}$ decreases faster in the lower $p_T$ region in central collisions than in mid-central collisions.
- $r_{2|2;1}$ decreases along a parabolic curve in central collisions, while it decreases linearly in mid-central collisions.

The results show good precision for measurements of $r_{2|2;1}$, where analysis is done in 5% centrality bins for minimum bias events, but is not viable for smaller centrality bins. By extrapolating from the luminosity used in the existing measurement, projections are made for Run 4 and shown in Figure 7 as dashed lines with color bands representing statistical uncertainties. The projections are made with a luminosity of 5 nb$^{-1}$ (estimated for Run 4 without Run 3). The reduced statistical uncertainty is smaller than the marker size. The ITk in Run 4 extends the $\eta$ range to ± 4 units, but the projected measurement is made to ± 3.5 units, in order to leave a gap between the ITk and the region of the forward calorimeter in which the reference measurement is made ($|\eta|$ = 4.4 – 4.9).

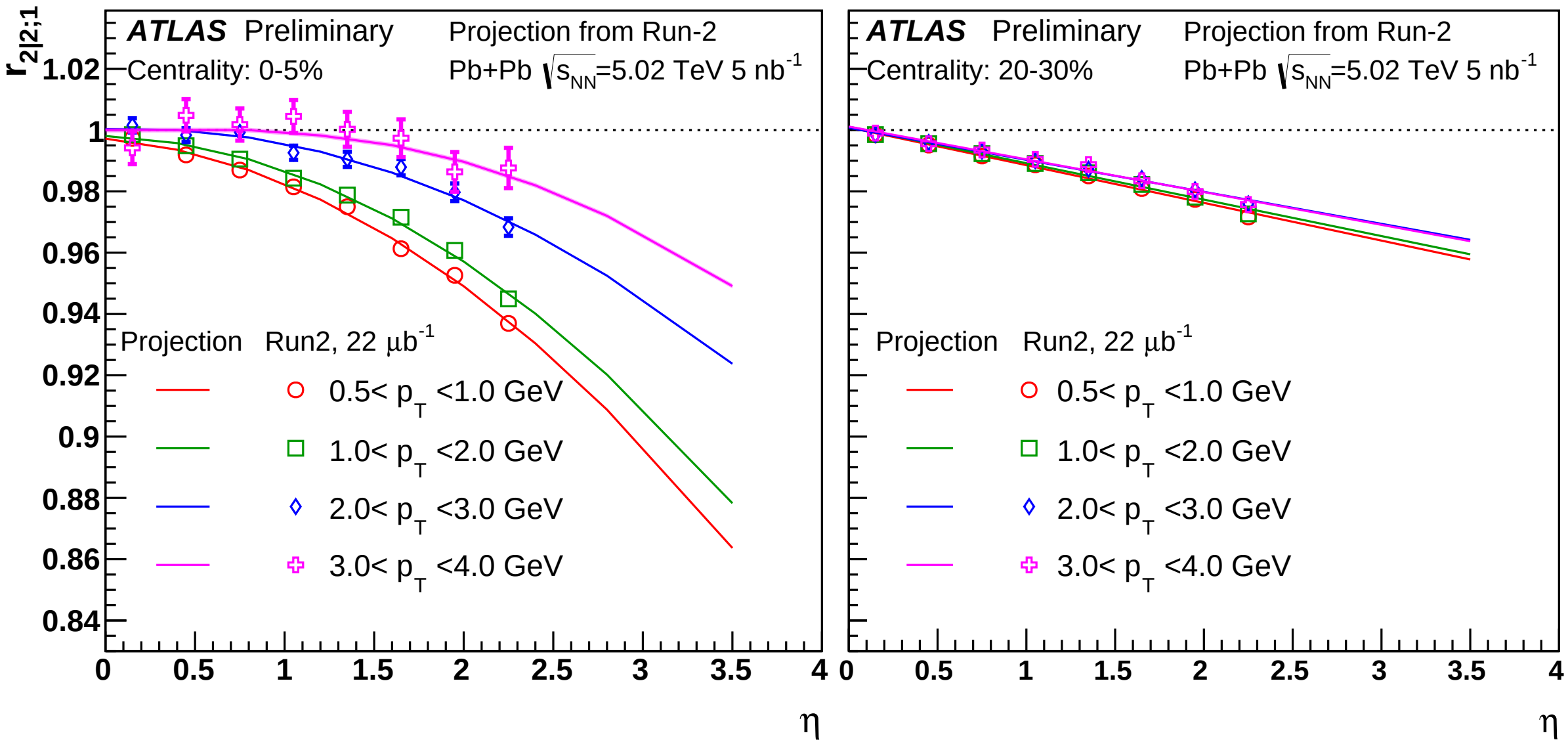

Figure 7: Projection for the flow correlator as expected to be measured in Run 4. The left plot is for 0 - 5% centrality and the right plot for 20-30% centrality. The width of the projection bands indicates the expected statistical uncertainty.

## 3.2 *pp* and *p*+Pb Measurements

### 3.2.1 Azimuthal femtoscopy

The freeze-out dimensions of nuclear collisions can be measured with femtoscopy [18]. Knowledge of the wavefunction between pairs of outgoing particles can be leveraged to provide an image of the source, and the Bose-Einstein correlation between identical bosons provides particularly good resolution image of the source. Charged pions are most commonly used, as the final-state interaction between same-charge pion pairs is well-described by a Bose-Einstein correlation with a correction taking into account Coulomb force interactions. The effect of this interaction is an enhancement in the two-particle momentum-space correlation functions that have characteristic widths in momentum space inversely proportional to the length scales of the source function. The extracted length scales of the source are referred to as Hanbury Brown and Twiss (HBT) radii.

With azimuthally-sensitive femtoscopy, the spatial ellipticity of the source at freeze-out can be measured. This measurement can be used to test the prediction of hydrodynamics that an initial source ellipticity leads to increased transverse momentum along the minor axis of the transverse source ellipse. Results at the LHC for Pb+Pb collisions are observed to be consistent with the hydrodynamic expansion of a short-lived source [19]. A preliminary result by ATLAS finds this behavior in central *p*+Pb collisions as well [20]. The transverse HBT radius $R_{\text{side}}$, defined by the transverse direction perpendicular to outgoing particle pairs, provides the cleanest signal for detecting an elliptic modulation of the source. Such a signal would be an independent observation of hydrodynamic-like properties in *p*+Pb collisions. The normalized second-order Fourier component of $R_{\text{side}}$ as a function of flow vector magnitude, $|\vec{q}_2|$, from the ATLAS preliminary result is shown in Figure 8. This figure also displays projected statistical uncertainties 1 pb$^{-1}$ of *p*+Pb collisions. These projected statistical uncertainties are obtained by scaling those from the preliminary result. This improvement in data quantity would allow the measurement to indicate the dependence of the modulation strength on $|\vec{q}_2|$, which is difficult to discern with the current data.

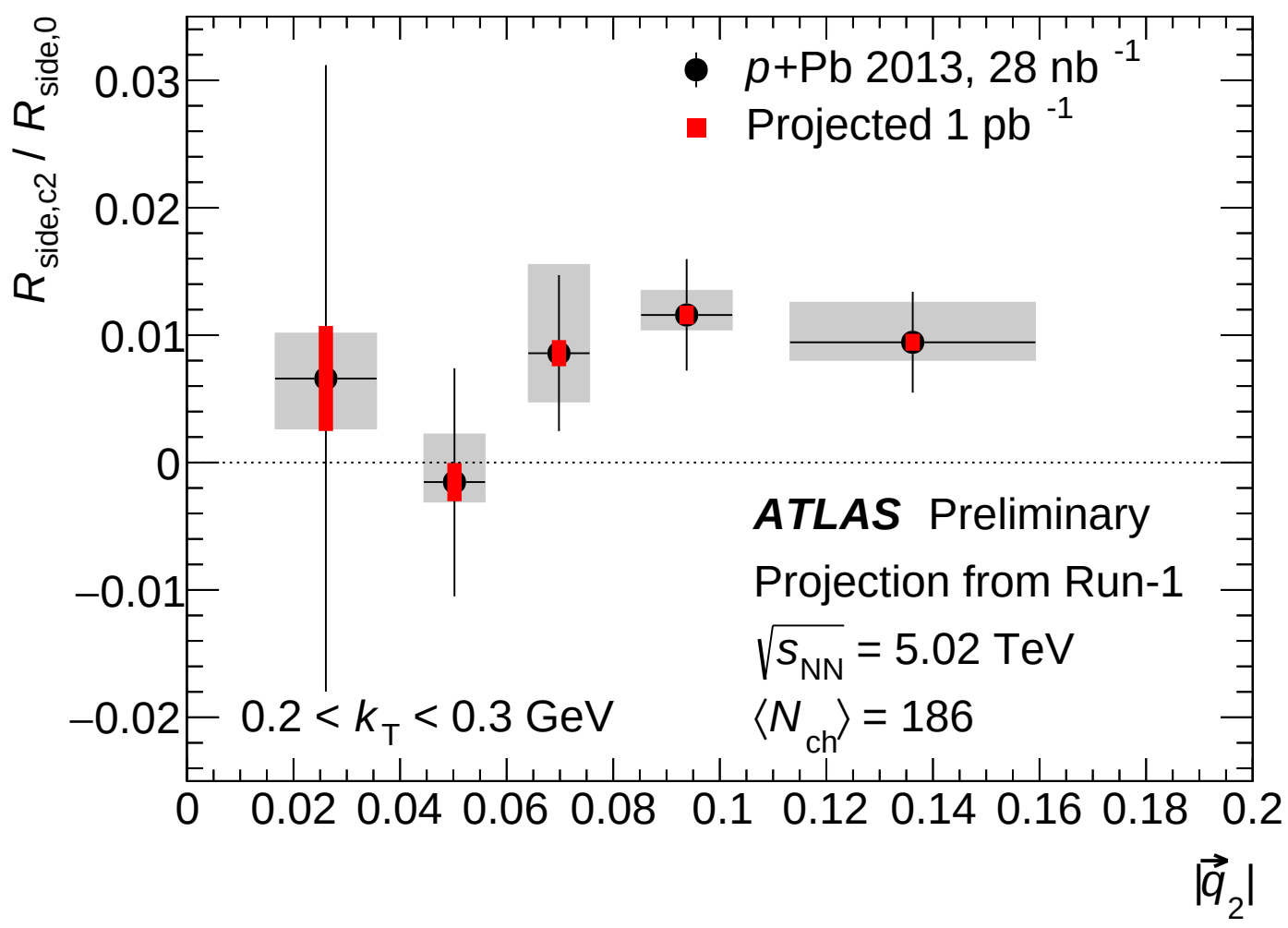

Figure 8: The second-order Fourier coefficient of the sideways HBT radius $R_{\text{side}}$ as a function of flow vector magnitude $|\vec{q}_2|$, normalized by the zeroth-order Fourier coefficient. The preliminary results from Ref. [20] are shown in the black points with statistical (bars) and systematic (shaded boxes) uncertainties, and projected statistical uncertainties from 1 pb$^{-1}$of future $p$+Pb running are indicated by the red bars.

### 3.2.2 Forward-backward multiplicity correlation

Forward-backward (FB) multiplicity correlations in the longitudinal direction are sensitive to early-time density fluctuations in $\eta$. These density fluctuations generate long-range correlations (LRC) at the early stages of the collision, well before the onset of any collective behavior, and appear as correlations of the multiplicity densities of produced particles separated in $\eta$. The event-by-event multiplicity density $\rho(\eta)$ in pseudorapidity interval $[-Y, Y]$ is quantified in terms of Legendre polynomials $P_n$:

$$\rho(\eta) \propto 1 + \sum_n a_n T_n(\eta),\ T_n(\eta) \equiv \sqrt{\frac{2n+1}{3}} Y P_n(\frac{\eta}{Y}) \tag{2}$$

where the coefficient $a_n$ is measured through two-particle correlation function $C(\eta_1, \eta_2)$:

$$\begin{aligned} C(\eta_1, \eta_2) &= \frac{\langle N(\eta_1) N(\eta_2) \rangle}{\langle N(\eta_1) \rangle \langle N(\eta_2) \rangle} \\ &= 1 + \sum_{n,m=1}^{\infty} \langle a_n a_m \rangle T_n(\eta_1) T_m(\eta_2) \end{aligned} \tag{3}$$

where $N(\eta)$ is the multiplicity density distribution in a single event and $\langle N(\eta) \rangle$ is the average distribution for a given event-multiplicity class [21].

Significant values are observed for $a_1$ in all multiplicity ranges of $p$+Pb collisions and higher-order coefficients are consistent with zero [21], which implies the FB multiplicity correlation is dominated by the linear component of the Legendre polynomials (Eq. 2). However, several theoretical studies suggest a non-linear component should exist in the region $|\eta| > 2.5$ [22, 23]. The increased tracking acceptance and increase in luminosity in Run 4 will provide a great opportunity to measure possible deviation beyond the linear component of the Legendre polynomials.

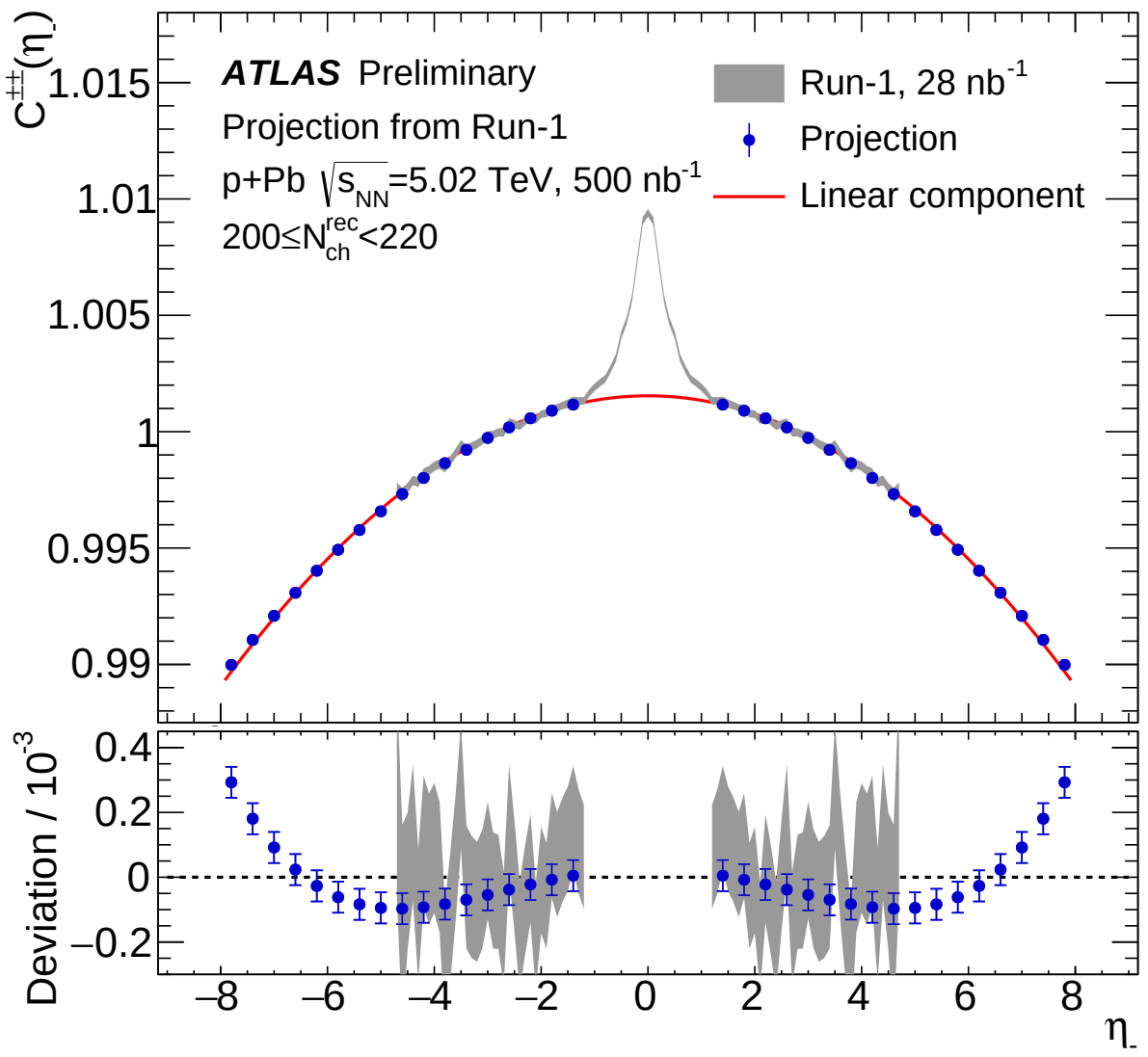

Figure 9: Projection of $C(\eta_1, \eta_2)$ distributions for same-charge pairs (±±) into one-dimensional $\eta_-(\equiv \eta_1 - \eta_2)$ distributions over a narrow slice $|\eta_1 + \eta_2| < 0.4$. In the top panel, the grey band represents the measurement using Run 1 $p$+Pb data and the red curve represents the linear component. The projected Run 4 results are indicated by the blue circles. The relative difference between $C^{\pm\pm}(\eta_-)$ and linear component is shown in the bottom panel.

The top panel of Figure 9 shows the projection of $C(\eta_1, \eta_2)$ distributions for same-charge pairs into one-dimensional $\eta_-(\equiv \eta_1 - \eta_2)$ distributions over a narrow slice $|\eta_1 + \eta_2| < 0.4$. The distributions are denoted by $C^{\pm\pm}(\eta_-)$. The red quadratic curve represents the linear component and the relative difference between $C^{\pm\pm}(\eta_-)$ and linear component is shown in the bottom panel. Short-range correlation contributes to the peak in the range $|\eta_-| < 1.0$, while in the long-range region, $|\eta_-| > 1.0$, $C^{\pm\pm}(\eta_-)$ is consistent with the linear component. To estimate the Run 4 projection, the magnitude of the first non-linear component, $a_2$, is assumed to be 15% of $a_1$. The statistical precision should be sufficient to quantitatively distinguish the possible non-linear component from the linear component. This projection suggests that Run 4 should bring a better understanding of the early-time density fluctuations in pseudorapidity.

### 3.2.3 Multi-particle azimuthal correlation

The measurement of multi-particle azimuthal correlations has led to initial observations of collective-like effects in small systems [5, 24, 25]. In $pp$ collisions at the LHC collective-like effects are of interest in two distinct regions: high-multiplicity collisions to compare to $p$+Pb and Pb+Pb collisions, and low-multiplicity collisions to search for the onset of these effects. Several performance estimates are presented here as examples for the rich physics which can be addressed with multi-particle azimuthal correlations in $pp$ and $p$+Pb collisions.

State of the art measured 4-particle cumulants of $v_3$ ($c_3\{4\}$) in $pp$ and $p$+Pb collisions [25] are presented in Figure 10 overlaid with the projections for Runs 3 and 4. The increase in luminosity in Runs 3 and 4, assumed here to be 200 pb$^{-1}$ for $pp$ collisions and 1000 nb$^{-1}$ for $p$+Pb (compared with 0.9 pb$^{-1}$ and 28

11

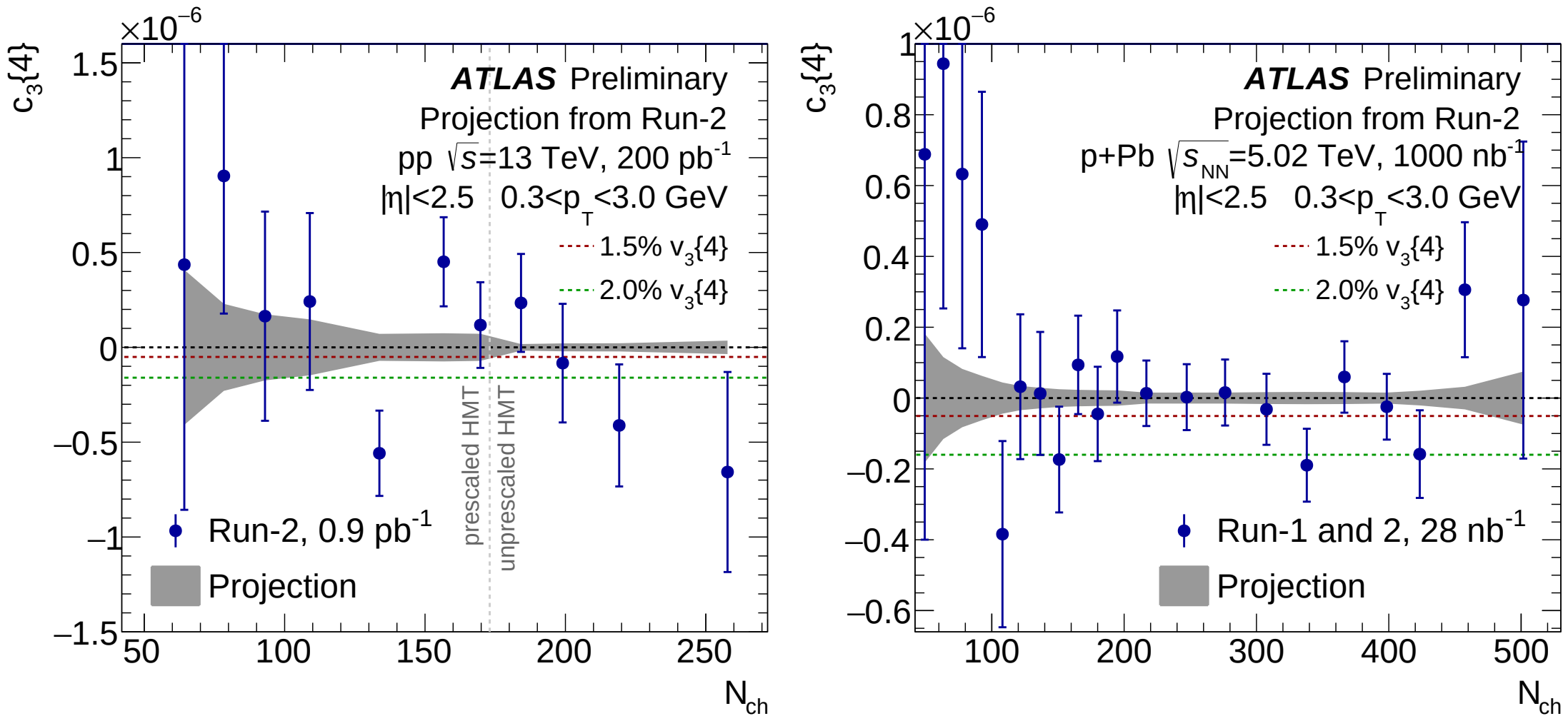

Figure 10: Projected 4-particle cumulants $c_3\{4\}$ with 3-subevent method for *pp* (left) and *p*+Pb (right) as a function of $N_{\text{ch}}$. Only statistical uncertainties are shown in the figure and the gray band represents the projected statistical uncertainty, with $c_3\{4\}$ assumed to be zero. The red and green dash lines represent 1.5% and 2.0% $v_3\{4\}$ signal separately. The vertical line in the left panel indicates the transition between prescaled high-multiplicity track triggered events and unprescaled ones.

nb$^{-1}$ in published *pp* and *p*+Pb measurements), provides a great opportunity to measure $c_3\{4\}$ with high precision. In order to remove the non-flow contributions, the 3-subevent method is applied [26]. In *pp* collisions, with the data collected in Run 2, the statistical uncertainties are large and the $c_3\{4\}$ values are consistent with zero in most of the $N_{\text{ch}}$ range [25]. On the other hand, in large systems, significantly non-zero $c_3\{4\}$ was measured, which reflects the nucleonic fluctuations in the initial state. If is of great interest to determine whether similar behavior is observed in small systems. With luminosity increased in *pp*, the statistics will be sufficient to measure a signal down to $v_3\{4\} = 1.5\%$ for $N_{\text{ch}} > 170$, while 2% signals are accessible with large significance over a wide multiplicity range ($N_{\text{ch}}$>100). Similarly, in *p*+Pb collisions, the existing result shows that $c_3\{4\}$ is consistent with zero [25], but increased statistics will help to detect a potential non-zero $c_3\{4\}$ smaller than 1.5% for $100 < N_{\text{ch}} < 500$. Similar projections for the case where the $\eta$ acceptance is extended to 4.0 by using the ITk is shown in Figure 11 and the statistical significance further increases; it is possible to measure $v_3\{4\}$ down to 1.5% even in the intermediate $N_{\text{ch}}$ range.

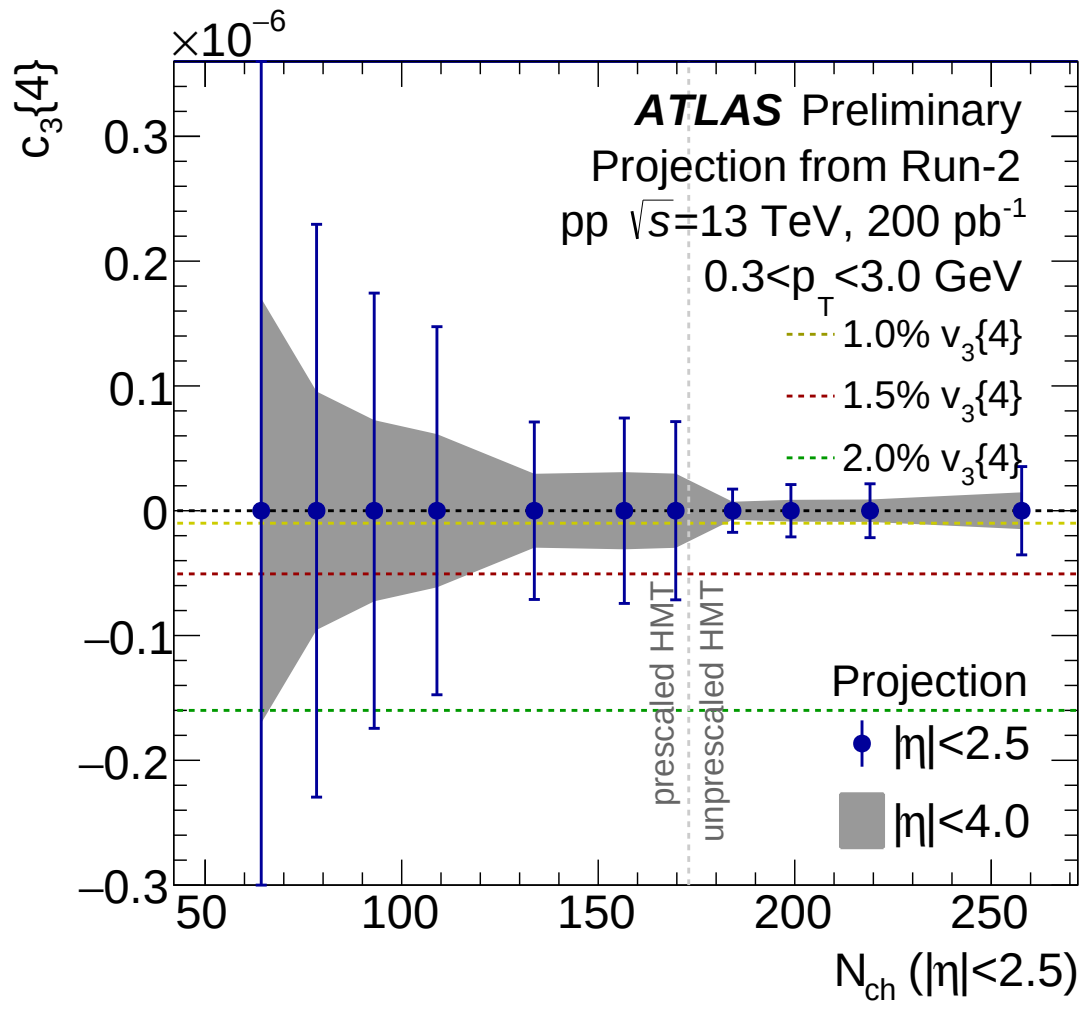

Figure 11: Projected 4-particle cumulants $c_3\{4\}$ with 3-subevent method for $pp$ as a function of $N_{\mathrm{ch}}$. The projections are estimated using particles from $|\eta| < 2.5$ and $|\eta| < 4.0$. In order for the Figure to be comparable with Figure 10 the x-axis, $N_{\mathrm{ch}}$, is defined with particles within $|\eta| < 2.5$. Only statistical uncertainties are shown in the figure, with $c_3\{4\}$ assumed to be zero. The yellow, red and green dash lines represent 1.0%, 1.5% and 2.0% $v_3\{4\}$ signals separately. The vertical line in the left panel indicates the transition between prescaled high-multiplicity track triggered events and unprescaled ones.

# Summary

This note presents projections for several ATLAS measurements of the bulk properties of the medium created in heavy-ion collisions at the LHC in the upcoming Runs 3 and 4. A key part of future measurements is the installation of the ITk for Run 4, and its performance in the heavy-ion environment is discussed. The tracking performance results demonstrate the excellent potential of the ITk detector for the future LHC heavy ion physics program. Projections are made for flow decorrelation and heavy flavor elliptic flow measurements in Pb+Pb collisions, as well as forward-backward multiplicity correlations and Hanbury-Brown Twiss radii in $p$+Pb collisions. These projections are representative of many measurements which can be performed with the ATLAS detector with the high luminosity available in Runs 3 and 4 and are sensitive to the properties of the QGP.

# CMS Physics Analysis Summary

Contact: cms-phys-conveners-ftr@cern.ch

# Constraining nuclear parton distributions with heavy ion collisions at the HL-LHC with the CMS experiment

The CMS Collaboration

## Abstract

Recent measurements in heavy ion collisions by the CERN LHC Collaborations have been used to assess nuclear effects and provide valuable data for nuclear parton distribution analyses. In this note, performance studies for measurements with the CMS detector at the High-Luminosity LHC (HL-LHC) are presented. These include the coherent $\Upsilon(1S)$ photoproduction in ultraperipheral lead-lead collisions, corresponding to a total integrated luminosity of 10 $\mathrm{nb}^{-1}$ at a nucleon-nucleon (NN) center-of-mass energy ($\sqrt{s_{\mathrm{NN}}}$) of 5.5 TeV. This note also presents the performance studies at the HL-LHC for analyses of inclusive Z boson, dijet, and top quark pair production in proton-lead collisions at $\sqrt{s_{\mathrm{NN}}} = 8.16$ TeV for an integrated luminosity of 2 $\mathrm{pb}^{-1}$.

# 1 Introduction

This note contains a series of performance studies that illustrate the physics potential using heavy ion data that will be recorded with the CMS experiment in the High-Luminosity LHC (HL-LHC) era in the near future [1]. For the HL-LHC phase, which is planned to operate from 2026, the LHC experiments have requested an integrated luminosity of about 10–13 $\text{nb}^{-1}$ and 2 $\text{pb}^{-1}$ using lead-lead (PbPb) and proton-lead (pPb) data at nucleon-nucleon (NN) center-of-mass energies ($\sqrt{s_{_\text{NN}}}$) of 5.5 and 8.8 TeV, respectively.

Based on these scenarios, performance studies for future measurements of coherent $\Upsilon(1S)$ photoproduction in PbPb collisions at $\sqrt{s_{_\text{NN}}} = 5.02$ TeV are presented. Performance results are also presented for analyses in pPb collisions at $\sqrt{s_{_\text{NN}}} = 8.16$ TeV, namely, studies of the inclusive Z boson production, differential cross sections of top quark pair ($\text{t}\bar{\text{t}}$) production, and the pseudorapidity distributions of dijets. Altogether, these studies show the potential of having large sample sizes to substantially reduce the statistical uncertainty in the measurements that will be carried out as discussed in this note, while opening up new opportunities to study nuclear parton distribution functions (nPDFs) and quantum chromodynamics (QCD) phenomena and its associated nuclear effects (nuclear gluon shadowing, among others).

# 2 Impact of detector upgrades

The CMS detector [1] will be substantially upgraded in order to fully exploit the physics potential offered by the increase in luminosity at the HL-LHC [2, 3], and to cope with the demanding operational conditions at the HL-LHC [4–8]. The upgrade of the first level hardware trigger (L1) will allow for an increase of L1 rate and latency to about 750 kHz and 12.5 $\mu$s, respectively, and the high-level software trigger (HLT) is expected to reduce the rate by about a factor of 100 to 7.5 kHz. The entire pixel and strip tracker detectors will be replaced to increase the granularity, reduce the material budget in the tracking volume, improve the radiation hardness, and extend the geometrical coverage and provide efficient tracking up to pseudorapidities of about $|\eta| = 4$. The muon system will be enhanced by upgrading the electronics of the existing cathode strip chambers (CSC), resistive plate chambers (RPC) and drift tubes (DT). New muon detectors based on improved RPC and gas electron multiplier (GEM) technologies will be installed to add redundancy, increase the geometrical coverage up to about $|\eta| = 2.8$, and improve the trigger and reconstruction performance in the forward region. The barrel electromagnetic calorimeter (ECAL) will feature the upgraded front-end electronics that will be able to exploit the information from single crystals at the L1 trigger level, to accommodate trigger latency and bandwidth requirements, and to provide 160 MHz sampling allowing high precision timing capability for photons. The hadronic calorimeter (HCAL), consisting in the barrel region of brass absorber plates and plastic scintillator layers, will be read out by silicon photomultipliers (SiPMs). The endcap electromagnetic and hadron calorimeters will be replaced with a new combined sampling calorimeter (HGCal) that will provide highly-segmented spatial information in both transverse and longitudinal directions, as well as high-precision timing information. Finally, the addition of a new timing detector for minimum ionizing particles (MTD) in both barrel and endcap region is envisaged to provide capability for 4-dimensional reconstruction of interaction vertices that will allow to significantly offset the CMS performance degradation due to high PU rates.

A detailed overview of the CMS detector upgrade program is presented in Ref. [4–8], while the expected performance of the reconstruction algorithms and pile-up mitigation with the CMS detector is summarized in Ref. [9].

In the following sections, physics performance studies for both PbPb and pPb collisions, based on the existing data and Monte Carlo (MC) simulations, are presented.

# 3 Coherent quarkonia photoproduction in ultraperipheral PbPb collisions at $\sqrt{s_{\mathrm{NN}}} = 5.5\,\mathrm{TeV}$

The data from ultra-peripheral collisions at the LHC have the potential to provide new constraints to the gluon PDFs in protons and nuclei. Photon-induced interactions can be studied in ultra-peripheral heavy ion collisions [10]. Both the ALICE and CMS Collaborations have recently carried out measurements on coherent photoproduction of $\rho^0$ mesons [11], J/$\psi$ [12–14] and $\psi(2S)$ [15] for the $\gamma + \mathrm{Pb} \rightarrow \mathrm{VM} + \mathrm{Pb}$ process, with "VM" denoting a vector meson. CMS has also results on the exclusive photoproduction of $\rho^0$ [16] and $\Upsilon(1S)$ [17] for the $\gamma + \mathrm{p} \rightarrow \mathrm{VM} + \mathrm{p}$ process. LHCb has also recent results in the exclusive photoproduction of J/$\psi$, $\psi(2S)$ and $\Upsilon$ in pp collisions [18, 19].

It was first suggested by [20, 21] that the photoproduction cross section of vector mesons is proportional to the squared gluon density at the scale $Q = m_{\mathrm{V}}/2$ at leading order QCD. By comparing results of the photoproduction cross section in both $\gamma + \mathrm{Pb}$ and $\gamma + \mathrm{p}$ interactions it is possible to extract information about the nuclear gluon density at various Bjorken-$x$ values for a given VM. Ref. [22] illustrates how this can be done by calculating the nuclear suppression factor ($R_{\mathrm{Pb}}(x)$) which is defined as the root squared of the ratio between the photoproduction cross section measured in $\gamma + \mathrm{Pb}$ ($\sigma_{\gamma\mathrm{Pb}}$) to the corresponding one in the Impulse Approximation (IA):

$$R_{\mathrm{Pb}}(x) = \sqrt{\left(\frac{\sigma_{\gamma\mathrm{Pb}}(x)}{\sigma_{\mathrm{IA}}(x)}\right)}, \qquad \text{where} \qquad x = \frac{m_{\mathrm{V}}}{\sqrt{s_{\mathrm{NN}}}}\exp(-y). \tag{1}$$

The Impulse Approximation is computed using data from the photoproduction of the vector meson in $\gamma + \mathrm{p}$ scaled by the integral over the squared Pb form factor as described in [22]. The impulse approximation calculation neglects all nuclear effects such as the expected modification of the gluon density in the lead nuclei compared to that of the proton. A recent CMS study of coherent J/$\psi$ photoproduction has followed this procedure to estimate the nuclear gluon shadowing as reported in [14].

The high luminosities envisaged for the HL-LHC will significantly extend the Bjorken-$x$ values that can be explored using coherent vector meson photoproduction. Although the gluon shadowing is smaller for $\Upsilon(1S)$ than that for J/$\psi$, having measurements from $\Upsilon(1S)$ photoproduction will serve as important tests to theoretical models that can describe the J/$\psi$ data from existing results by the ALICE and CMS collaborations.

We have used the calculations provided by V. Guzey *et al.* as described in [23] which takes into account nuclear gluon shadowing corrections [24]. We assume that the CMS experiment will have an improved level of detector and triggering performance during the HL-LHC operation by increasing the combined acceptance and efficiency from about 60% to 80% [17].

Physics performance projections for the nuclear suppression factor for $\Upsilon(1S)$ photoproduction are shown in Fig. 1. The error bars represent the statistical uncertainties, and the boxes the systematic ones.

For rapidity values different than zero there is a two-fold ambiguity in the photon direction of the $\gamma + \mathrm{Pb}$ system (either Pb can serve as photon emitter or photon target). To overcome this uncertainty we follow the prescription discussed in [23] that suggests studying the dependence of the vector meson photoproduction cross section on the associated production of forward or backward neutrons (break-up modes) to disentangle the two photon directions. In particular, this would require measuring coherent vector meson photoproduction in the configuration where the $\Upsilon(1S)$ is accompanied by at least one neutron in either the forward or backward direction from the interaction point using zero degree calorimeters (ZDC), and in the configuration where both ZDCs record no neutrons.

For a given rapidity interval different than zero there will be two solutions, one corresponding to low Bjorken-$x$ and another one for high Bjorken-$x$ values. Since such a procedure has not been reported in any measurements so far, although there are qualitatively evidence from existing CMS measurements [14] that this method is sound, we are not providing projections for the most forward/backward rapidity intervals that can be studied with the CMS detector where both the theoretical and systematic uncertainties are the largest. At the same time, this is the rapidity interval where measurements corresponding to Bjorken-$x$ below $10^{-4}$ for $\Upsilon(1S)$ photoproduction can be explored.

For the measurements with a rapidity value different than zero, the statistical uncertainty is larger for the lowest $x$ solution as shown in Fig. 1. The statistical uncertainty at mid-rapidity is negligible, corresponding to the middle point of the nuclear suppression factor as a function of Bjorken-$x$ as shown in Fig. 1.

This analysis assumes that there will be no significant improvement in the theoretical descriptions of relevant physics effects (photon flux uncertainty). Projected uncertainties on luminosity (4%), reference cross section in photon-proton interactions (5%) and photon flux (5%) result in $\sim 8\%$ systematic uncertainty on the ratio $\sigma_{\gamma Pb}(x)/\sigma_{IA}(x)$ and $\sim 4\%$ uncertainty on the nuclear suppression factor $R_{Pb}(x)$ are reported. Systematic uncertainties in the identification and isolation efficiencies for muons are expected to be reduced to around 0.5%.

The uncertainty in the integrated luminosity of the data sample could easily be reduced from 5% down to 4% in PbPb by a better understanding of the calibration and fit models employed in its determination, and making use of the finer granularity and improved electronics of the upgraded detectors. A luminosity uncertainty in the 3.2–3.5% range has recently been reported for pPb collisions in Run 2 [25].

## 4 Projections for inclusive Z boson production in pPb collisions at $\sqrt{s_{NN}} = 8.16$ TeV

The differential cross section of Z boson production as a function of its rapidity in the center-of-mass frame has also been studied. The integrated luminosity of 2 $pb^{-1}$ at $\sqrt{s_{NN}} = 8.16$ TeV has been considered. The MCFM program is used to generate the Z boson signal [27]. We have used the calculations from the CT14 proton PDF [28] and the EPPS16 nPDF for the lead ions [29]; the latter are used as central values for our projections.

The extrapolation assumes that the CMS experiment will have a similar level of detector and triggering performance during the HL-LHC operation as it provided during the LHC Run 2 period, which is quite a cautious assumption for pPb running. The acceptance and efficiency corrections are estimated using Run 2 simulation, while systematic uncertainties are reduced by a factor 3 with respect to the previous Z boson measurements in pPb collisions $\sqrt{s_{NN}} = 5.02$ TeV [30]. Figure 2 shows the results for the projected Z boson differential cross sections.

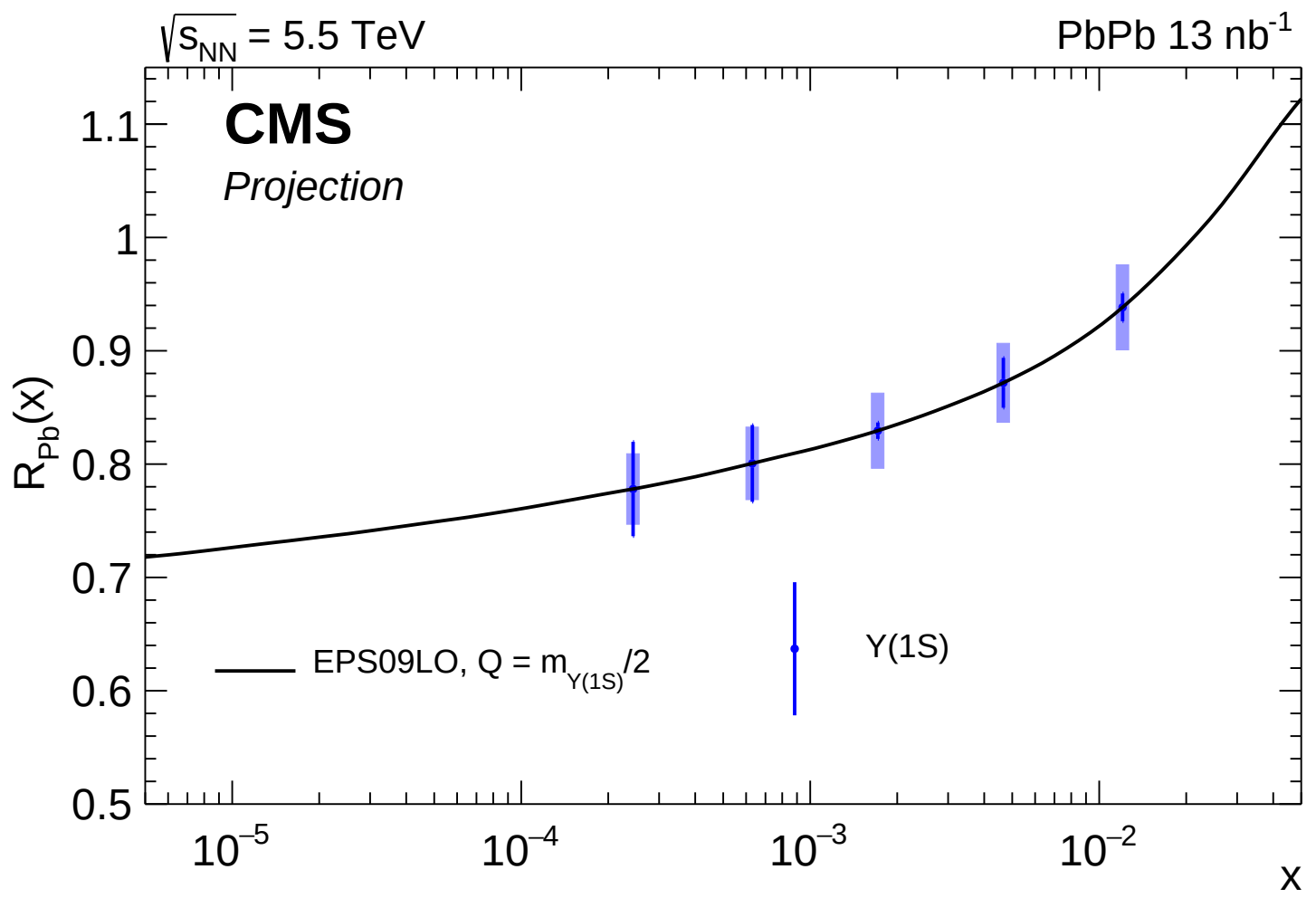

Figure 1: Projections for gluon shadowing factor measured with $\Upsilon(1S)$ photoproduction in ultraperipheral PbPb collisions at $\sqrt{s_{NN}} = 5.5\,\text{TeV}$. The error bars represent the statistical uncertainties, and the boxes the systematic ones. The projected data is compared to the central value of the EPS09 global fit [26]. The most dominant uncertainties are those of EPS09 (not shown).

## 5 Projections for $t\bar{t}$ differential cross sections in pPb collisions at $\sqrt{s_{NN}} = 8.16\,\text{TeV}$

In proton-nucleus collisions, the top quark is a novel and theoretically precise probe of the nuclear gluon density at high virtualities $Q^2 \approx m_t^2$ (where $m_t$ is the top quark mass [31]) in the less explored high Bjorken-$x$ region ($x \gtrsim 2m_t/\sqrt{s_{NN}} \approx 0.05$ at leading order in QCD). The first observation of the inclusive $t\bar{t}$ production ($\sigma_{t\bar{t}}$) in pPb collisions at $\sqrt{s_{NN}} = 8.16\,\text{TeV}$ [32] has been performed using $174{\pm}6\,\text{nb}^{-1}$ [25]. The measured cross section of $\sigma(t\bar{t}) = 45 \pm 8\,\text{nb}$ is consistent with predictions from perturbative QCD as well as with the expectations from scaled pp data. However, the total uncertainty of about 17% is not sufficient for imposing any constraints on current nPDF parameterizations: the PDF uncertainty in the theoretical prediction of $\sigma(\text{pPb} \to t\bar{t} + X) = 59.0 \pm 5.3\,(\text{PDF})\,^{+1.6}_{-2.1}\,(\text{scale})\,\text{nb}$ [32] is approximately 8%, corresponding to a 90% confidence level (CL). The prospects of measuring $\sigma_{t\bar{t}}$ differentially have recently been studied in Refs. [33, 34]. A simple feasibility study of the $\sigma_{t\bar{t}}$ measurement is therefore carried out as a function of the reconstructed lepton $p_T$ and rapidity, based on existing simulated events [32] of the Run 2 CMS detector.

Events are selected fulfilling the same requirements ("visible phase space") established in Ref. [32]. The baseline selection includes exactly one charged electron (e) or muon ($\mu$) with $p_T > 30\,\text{GeV}$ and $|\eta| < 2.1$, and at least four jets with $p_T > 25\,\text{GeV}$ and $|\eta| < 2.5$ are required, the latter reconstructed based on the anti-$k_T$ clustering algorithm [35] using a distance parameter of 0.4. The QCD multijet background is retained from the nonisolated control region obtained with the pPb data sample of $174\,\text{nb}^{-1}$ [32], whereas the signal ($m_t = 172.5$ GeV) is simulated at next-to-leading order with POWHEG (v2) [36–38] using PYTHIA (v8.205) [39] to simulate parton showering and hadronization. The existing MC samples [32] make use of PYTHIA (v6.424) [40] for simulating W+jets and Drell–Yan (DY) production of charged-lepton pairs with invariant mass larger than 30 GeV. The expectation of signal and background processes is scaled to $2\,\text{pb}^{-1}$.

The $\sigma_{t\bar{t}}$ measurement is performed fitting the mass, $m_{jj'}$, of the non b-tagged jets that are closest in the $\eta$–$\phi$ plane according to the $\Delta R = \sqrt{(\Delta\eta)^2 + (\Delta\phi)^2}$ criterion, where $\Delta\eta$ and $\Delta\phi$ are their separations in pseudorapidity and azimuthal angle. This dijet system is expected to be primarily found in $\mathrm{W} \to \mathrm{q\bar{q}}'$ decays, and hence expected to be of resonant and combinatorial nature for most of the $\mathrm{t\bar{t}}$ signal and background events, respectively. The fit is combined from different event categories depending on the flavor of the charged lepton and the b-tagging multiplicity. Figures 3 shows the dijet invariant mass and a proxy of the top quark mass, $m_{\mathrm{top}}$, defined as the invariant mass of the $\mathrm{t} \to \mathrm{jj'b}$ candidate formed by pairing the W candidate with a b-tagged jet [32].

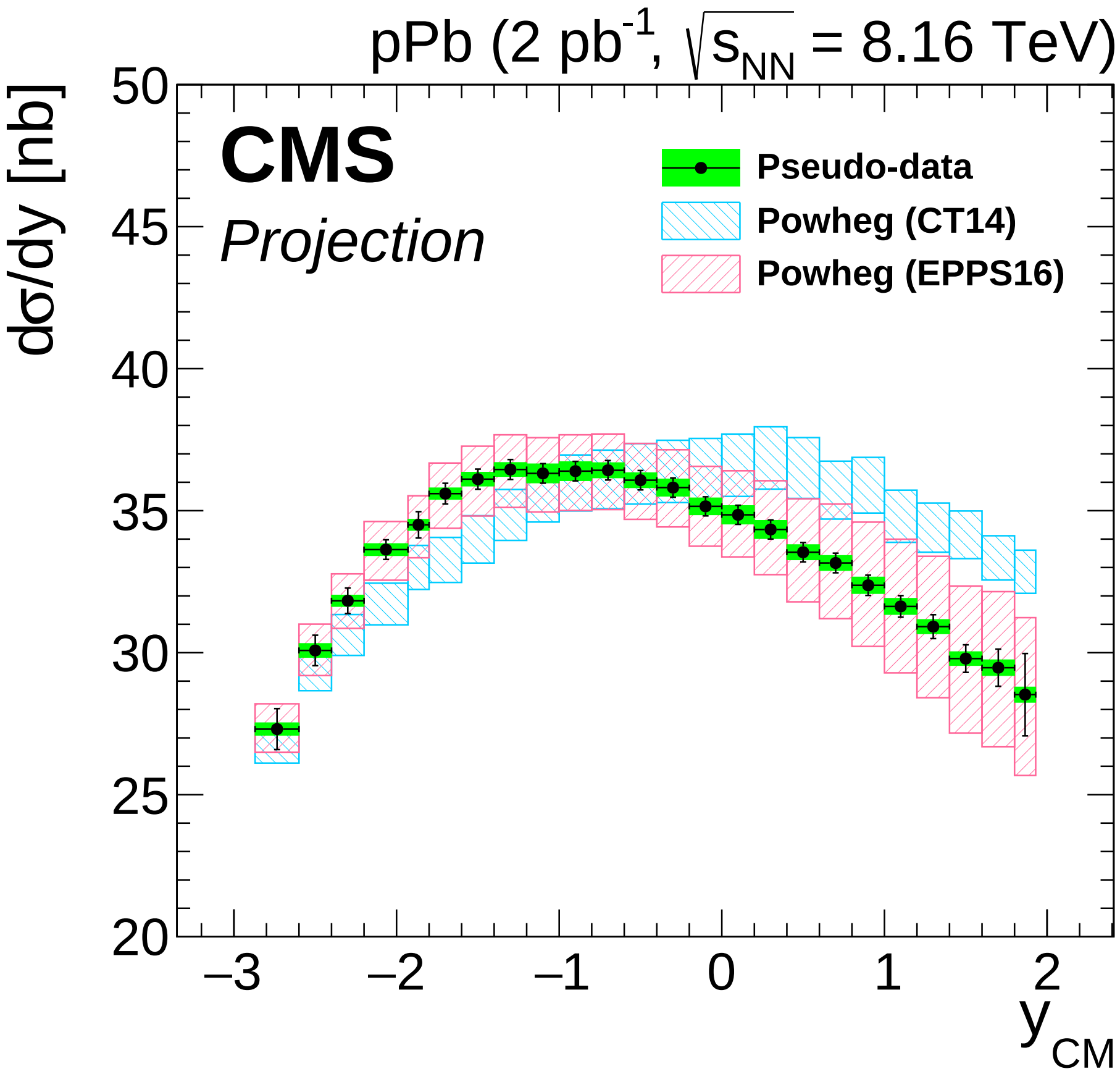

Figure 2: Projections for Z boson differential cross section in pPb collisions at $\sqrt{s_{NN}} = 8.16$ TeV as a function of the Z boson rapidity in the center-of-mass (CM) frame. The expectations from CT14 PDF and EPPS16 nPDF are also shown.

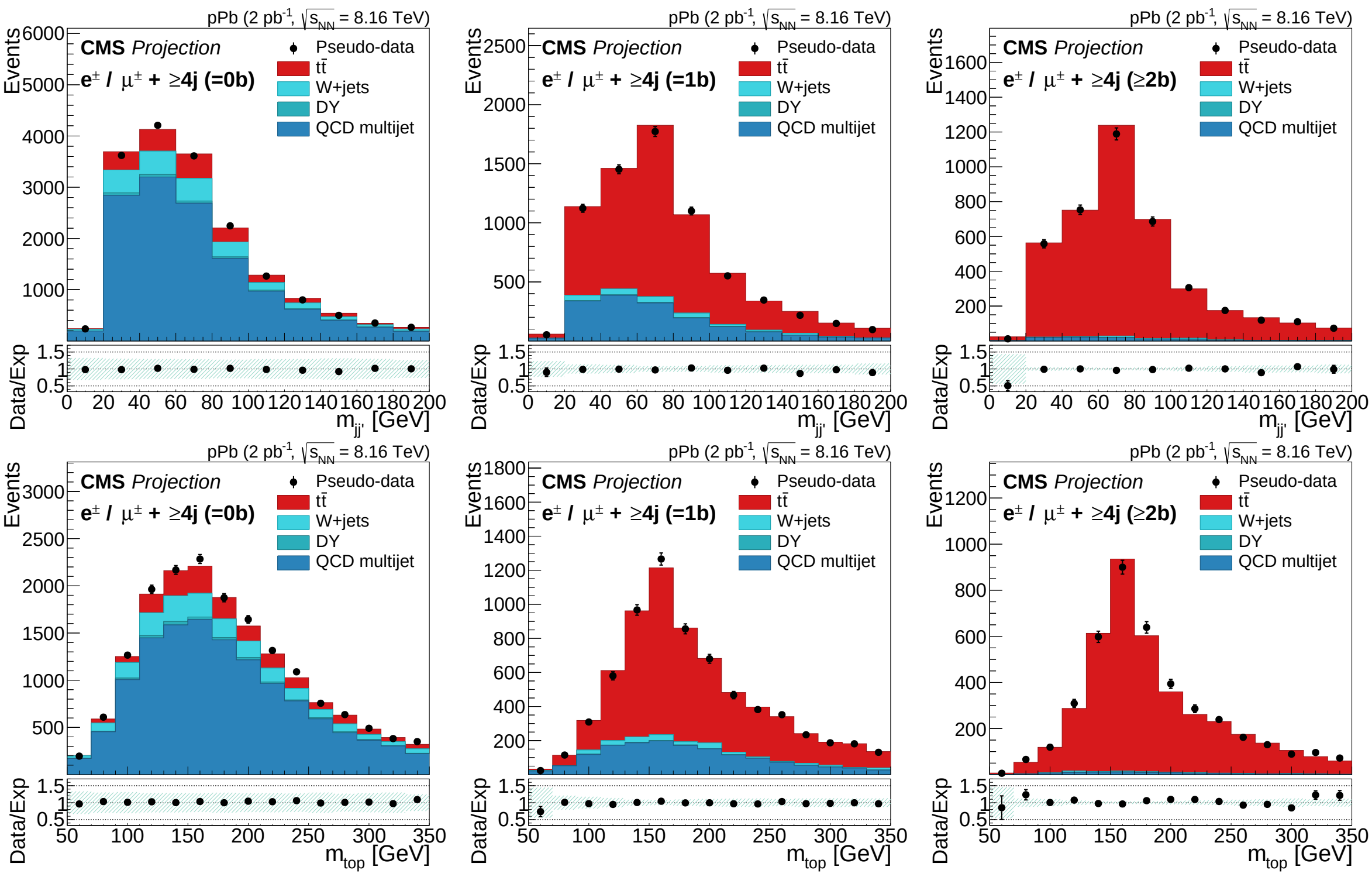

Figure 3: Distributions of the $m_{\mathrm{jj'}}$ (top) and $m_{\mathrm{top}}$ (bottom). From left to right the events are classified in the 0, 1, and 2 b-tagged jet categories. The sum of the predictions for the $\mathrm{t\bar{t}}$ signal and background is compared to pseudo-data (sampled randomly from the total of the predictions in each category). The bottom plots show the ratio between the pseudo-data and the sum of the predictions. The shaded band represents the relative uncertainty due to the limited event count in the simulated samples and the estimate of the normalization of the QCD multijet background.

To separate the signal from background contributions, and hence optimizing for the conjectured pPb data sample, we make use of the $_s\mathcal{P}lot$ technique [41]. The $m_{\mathrm{jj'}}$ ("discriminating") variable, used to extract the signal and background yields, does not correlate with the lepton kinematic ("control") variables, rendering it particularly suited to the $_s\mathcal{P}lot$ technique. Figure 4 displays the differential $\sigma_{\mathrm{t\bar{t}}}$ in the visible phase space as a function of the charged lepton $p_{\mathrm{T}}$ and rapidity at reconstruction ("detector") level. The relative statistical uncertainty in both variables is found to be at the level of 4–5% in each bin, and it is expected to be the dominant uncertainty. Despite the fact that most sources of systematic uncertainty are expected to cancel out in the normalized measurement of the differential $\sigma_{\mathrm{t\bar{t}}}$, a kinematic-independent systematic uncertainty of 5% is taken into account. The latter is considered as a conservative estimate, given the extrapolation—as described above—assumes similar future performance to Run 2, and it is partly motivated from Ref. [42]. No unfolding of the detector- to particle-level [43] dis-tributions is performed, although the physics reach should not be compromised significantly because of the high purity/stability of the response matrices.

The bottom panel of Figure 4 displays the ratio between the pseudo-data used in the study and the POWHEG+PYTHIA prediction employing the EPPS16 nPDFs. The comparison between the projected and the overall nPDF uncertainty is also shown. To that end, the nPDF uncertainty is scaled from 90 to 68% CL, and is computed using the prescription described in Ref. [44]: In the Hessian representation, a central PDF is given along with error sets, each of which corresponds to an eigenvector of the covariance matrix in parameter space.

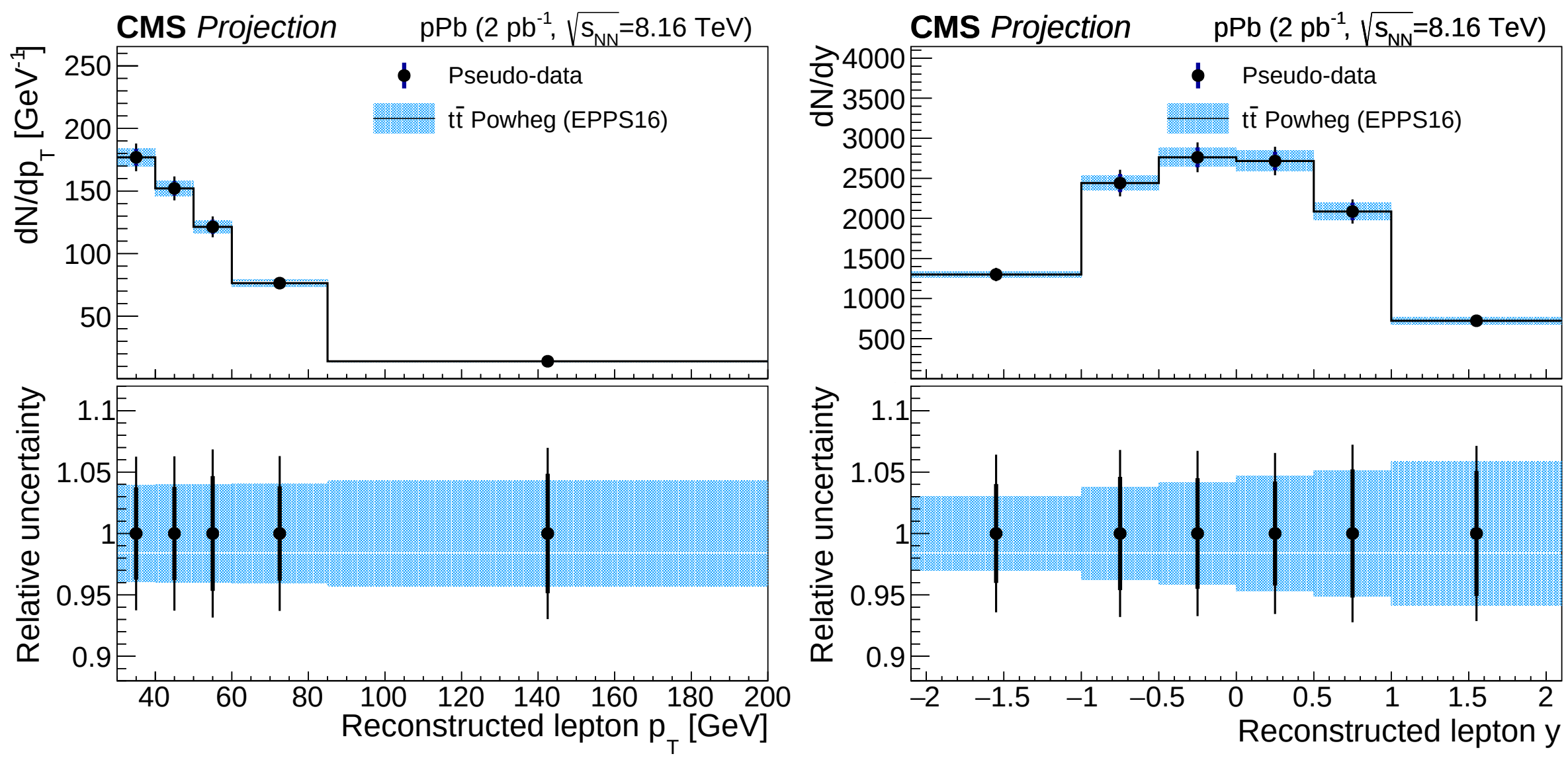

Figure 4: The top panels represent the differential $t\bar{t}$ production cross section in the visible phase space as a function of the charged lepton $p_T$ (left) and rapidity (right) at reconstruction level. The statistical uncertainty in the pseudo-data, represented by the inner error bars, is estimated through the application of the $_s\mathcal{P}lot$ technique [41]. The outer error bars represent the total uncertainty, assuming a conservative 5% systematic uncertainty envelope. The uncertainty in the POWHEG+PYTHIA [36–39] prediction is shown as a band corresponding to the 68% CL variation envelope of the EPPS16 [29] nPDF eigenvalues. The bottom panels represent the relative uncertainties in the pseudo-data and theory predictions.

# 6 Projections for dijet pseudorapidity distributions in pPb collisions

Theoretical calculations and recent experimental data have shown that dijet pseudorapidity distributions are sensitive to nuclear modifications of the gluon nPDFs [45, 46]. The expected luminosities for pPb collisions at $\sqrt{s_{NN}} = 8.8$ TeV during the HL-LHC phase will allow for an extension of the current measurements from pPb collisions at $\sqrt{s_{NN}} = 5.02$ TeV [46] by one bin to lower $\eta_{\text{dijet}}$ values.

The projected results for the ratio of the dijet pseudorapidity distributions between pPb and pp data, and the associated statistical and systematic uncertainties, are shown in Fig. 5.

The central values used in the projections are based on the existing data and smoothed by a third-order polynomial fit. The statistical uncertainty is scaled to a total integrated luminosity of 2 pb$^{-1}$. while the systematic uncertainties are reduced based on an assumption of a 50% reduction in the uncertainty in jet energy scale, consistent with other projections of jet measurements in heavy ion collisions. This result is a consequence of the large increase in available data, since the absolute jet energy scale is derived using a data-driven method based on photon- and Z-jet events. This technique is currently limited by the small total number of such events at large $\eta$. Hence, an improvement in the derivation of the jet energy scale corrections and its associated systematic uncertainty in that region can be expected with the higher luminosities.

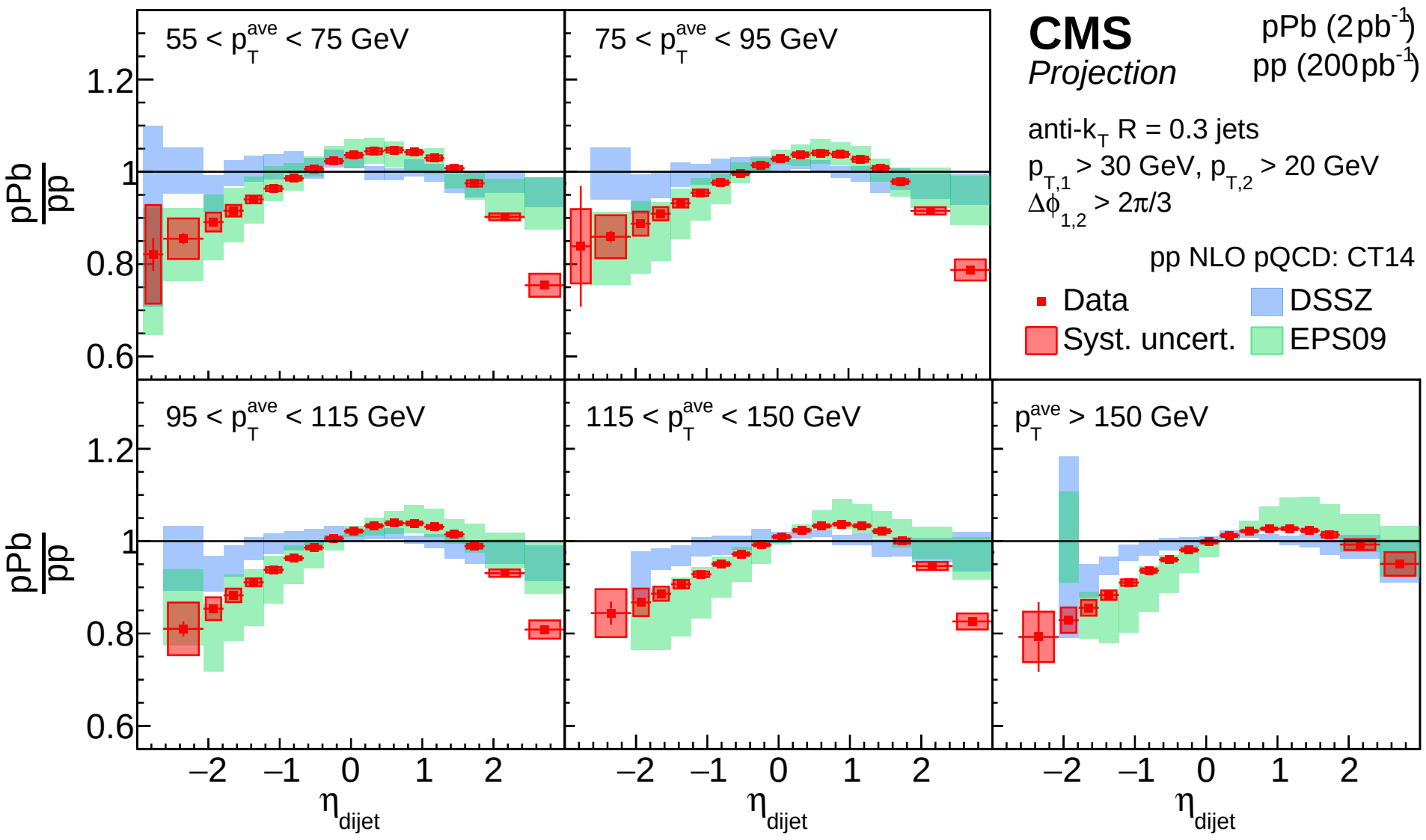

Figure 5: Projections for dijet pseudorapidity distributions for pPb collisions with a total integrated luminosity of 2 $\mathrm{pb}^{-1}$.

# 7 Summary

We have presented a series of performance studies for future measurements in both PbPb and pPb collisions for the High-Luminosity LHC project, putting special emphasis on a selected number of physics analyses that can serve to get insights into nuclear effects and nuclear parton distribution functions with the projected larger sample sizes that are envisaged.

**10**

**ATLAS PUB Note**

ATL-PHYS-PUB-2018-039

29th November 2018

# Expected ATLAS Measurement Capabilities of Observables Sensitive to Nuclear Parton Distributions

The ATLAS Collaboration

The study of proton-nucleus collisions at the LHC provides an opportunity to study QCD in regimes of high parton density and high centre-of-mass energy but with the expectations that final-state effects associated with a quark-gluon plasma will be small if present at all. Therefore, proton-nucleus collisions serve a twofold purpose within the heavy-ion program at the LHC: to understand the nature of initial-state nuclear effects at partonic scales, and as a control system for nucleus-nucleus collisions. This note discusses projections for representative ATLAS measurements exploring nuclear modifications to parton distributions in LHC Runs 3 and 4.

# 1 Introduction

The study of proton-nucleus collisions at the LHC provides an opportunity to study QCD in regimes of high parton density and high centre-of-mass energy with the expectations that final-state effects associated with a quark-gluon plasma will be small, if present at all. Furthermore, in nucleus-nucleus (A+A) collisions nuclear effects are present in both colliding objects, whereas in proton-nucleus ($p$+A) collisions the partons in the proton can be considered a relatively well understood probe of the nucleus. Therefore, $p$+A collisions serve a twofold purpose within the heavy-ion program at the LHC: to understand the nature of initial-state nuclear effects, and more generally Cold Nuclear Matter (CNM) effects, and as a control system for A+A collisions.

Existing measurements of electroweak bosons in proton-lead collision at the LHC have consistently shown that pQCD calculations that use free nucleon parton distribution functions (PDFs) as inputs do a poorer job of describing data than calculations that use modified nuclear parton distribution functions (nPDFs) [1, 2]. Similar results have been found with di-jet measurements [3]. However, the data, at the current level of precision, only weakly constrain the nPDFs.

In this note projections are made for measurements of $W$ and $Z$ bosons as well as photon-jet events in expected future proton-lead ($p$+Pb) collisions at the LHC with $\sqrt{s_{\mathrm{NN}}}$ =8.8 TeV. Three luminosity scenarios are considered: 0.5 pb$^{-1}$, 1 pb$^{-1}$, and 2 pb$^{-1}$. The first two scenarios correspond to lower and upper estimates of a single $p$+Pb running period at the LHC, and 2 pb$^{-1}$ corresponds to an optimistic scenario of two separate periods (likely one during Run 3 and one during Run 4). In the main body of the note projections are shown for 2 pb$^{-1}$, and they are repeated for the other luminosity scenarios in Appendix A. Appendix B shows $W$ boson projections for $\sqrt{s_{\mathrm{NN}}}$ =8.16 TeV $p$+Pb collsions.

# 2 ATLAS detector

The ATLAS detector is described in detail in [4]. Significant upgrades to the ATLAS detector are expected to be installed for Run 4, in particular the Inner Tracker (ITk) [5]. The performance of the ITk in Pb+Pb collisions is discussed in Ref. [6]. For this note, performance improvements due to the ITk are not taken into account.

# 3 Projections

## 3.1 $W$ Boson Measurements in $p$+Pb

The measurement of $W$ bosons in particular at forward rapidity in proton-nucleus collisions is an important input for nPDF global fits. To explore the resolution and prospects for increased statistical precision of the future data, different sets of predictions were made. For unmodified PDF predictions, events were generated with the Powheg generator [7, 8] and showered with Pythia8 [9] using the CT14 PDF set [10]. By convention, the direction of the proton defines the positive rapidity direction. Both $np$ and $pp$ collisions were generated and combined in accordance with the proton and neutron content of the Pb ion. The derived nucleon-nucleon cross sections were then scaled by the A=208 of the target nucleus. The number of generated events was chosen to match the expected number of events in the expected luminosity scenarios. For nPDF predictions, the EPPS16 set [11] was used in the parton-level Monte Carlo generator MCFM [12].

Figure 1 presents projected measurements of differential cross-sections for $W^+$ and $W^-$ boson production in $p$+Pb collisions at $\sqrt{s_{\mathrm{NN}}} = 8.8$ TeV with integrated luminosities of 2 $\mathrm{pb}^{-1}$. The cross-sections are calculated in the leptonic $W$ boson decay channels within a fiducial phase-space region: $p_{\mathrm{T}}^{\ell} > 25$ GeV, $p_{\mathrm{T}}^{\nu} > 25$ GeV, $m_{\mathrm{T}} > 40$ GeV. As in existing measurements, the decay neutrino kinematics are inferred from the missing transverse energy, $E_{\mathrm{T}}^{\mathrm{miss}}$, of the hadronic system recoiling from the $W$ boson.

The projected statistical uncertainties on the cross-sections include the effect of an applied scale factor of 70% to account for lepton trigger, reconstruction and identification efficiencies, while systematic uncertainties are estimated based on previous ATLAS measurements of $W$ boson production in $pp$, $p$+Pb and Pb+Pb collisions [13–15]. For the projected luminosity, systematic sources of uncertainty dominate over statistical uncertainty. Considered sources of systematic uncertainty are:

- $E_{\mathrm{T}}^{\mathrm{miss}}$ resolution – contributes 3% to the total systematic uncertainty
- Lepton reconstruction and identification – 0.4%
- Lepton trigger – 0.6%
- Estimation of QCD multi-jet background – 1.5%
- Normalization of electroweak backgrounds – 0.1%

In addition to these sources of uncertainty the luminosity uncertainty is expected to be similar to previous $p$+Pb running in which it was less than 3% [1] (not shown in the figures).

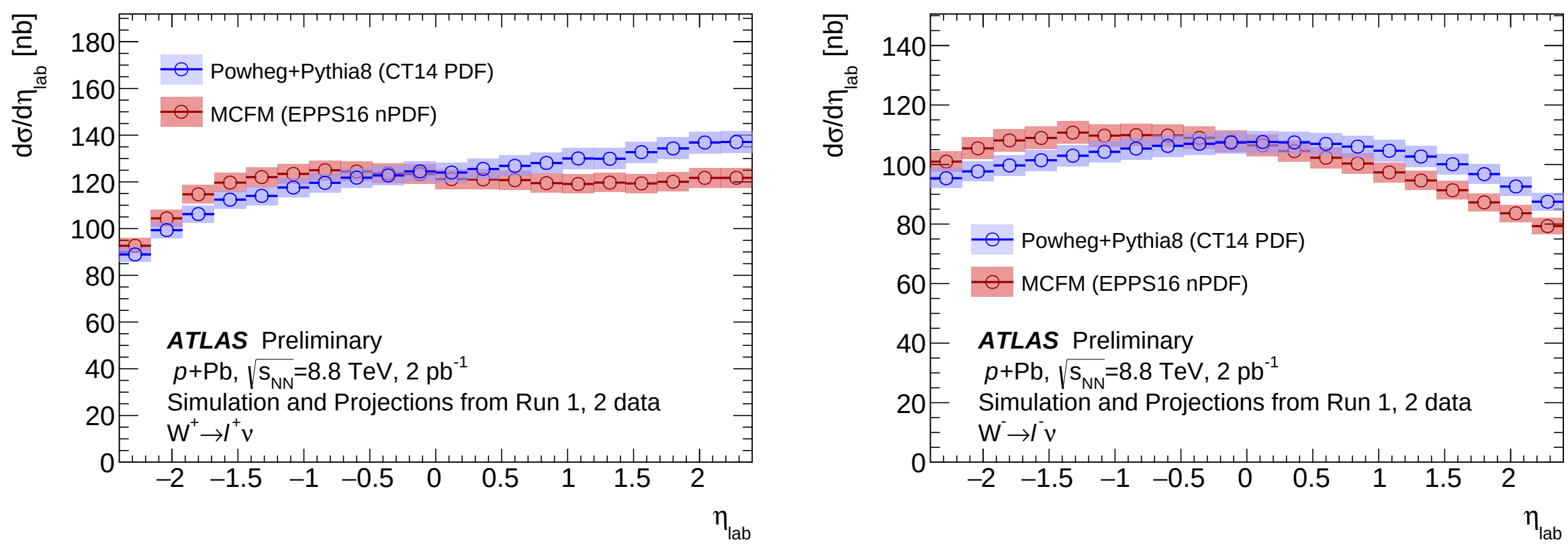

Figure 1: Fiducial cross-sections for $W^+$ (left) and $W^-$ (right) boson production in $p$+Pb collisions at $\sqrt{s_{\mathrm{NN}}} = 8.8$ TeV differential in the charged lepton pseudorapidity measured in the laboratory frame $\eta_{\mathrm{lab}}$. The cross-sections are projected with nuclear effects described by the EPPS16 nPDF set and without any nuclear effects using the CT14 PDF set with an integrated luminosity of 2 $\mathrm{pb}^{-1}$. The boxes represent the projected total uncertainties (quadratic sum of statistical and systematic uncertainties), while vertical bars represent statistical uncertainties (smaller than the marker size).

Figure 2 presents projected measurements of the forward-backward ratios of differential cross-sections for $W^+$ and $W^-$ boson production. These ratios are particularly sensitive to possible nuclear modifications of PDFs, since some of the systematic uncertainties are largely reduced due to their being correlated between forward and backward bins. In the presented projections, the systematic uncertainties coming from the $E_{\mathrm{T}}^{\mathrm{miss}}$ resolution and the estimation of QCD multi-jet background are assumed to cancel out in the ratios.

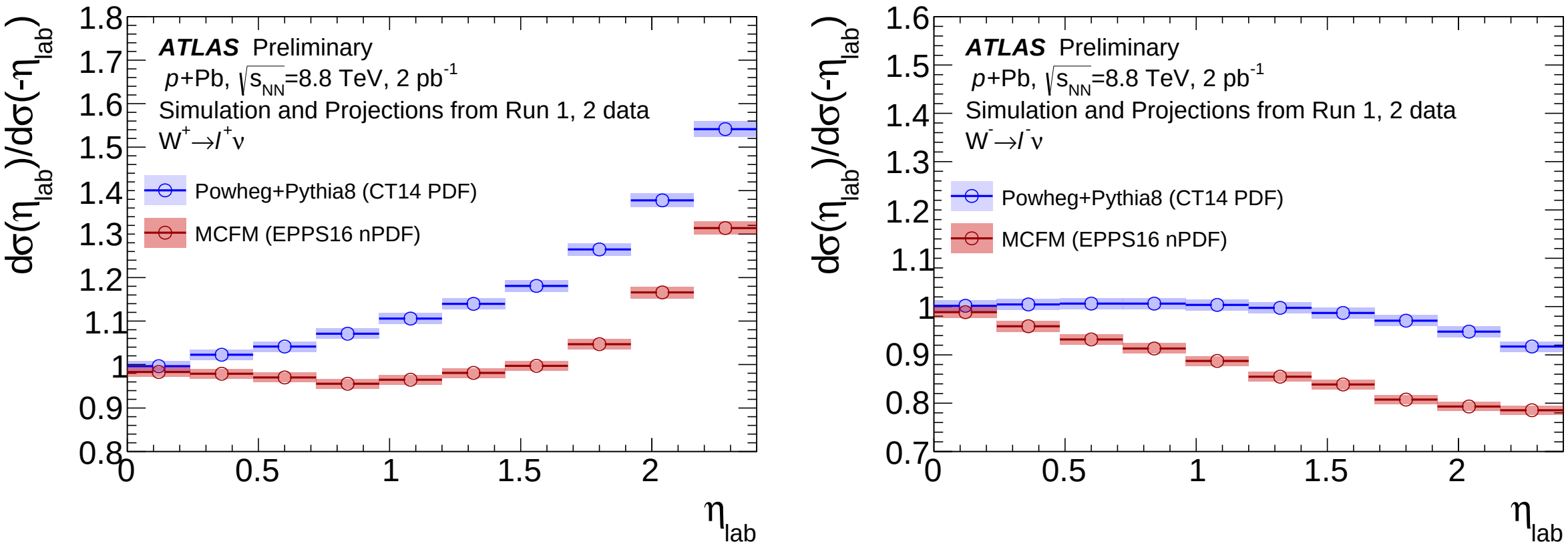

Figure 2: Forward-backward ratios of the differential fiducial cross-sections for $W^{+}$ (left) and $W^{-}$ (right) boson production in $p$+Pb collisions at $\sqrt{s_{\mathrm{NN}}}$ = 8.8 TeV. The ratios are projected with nuclear effects described by the EPPS16 nPDF set and without any nuclear effects using the CT14 PDF set assuming integrated luminosity of 2 pb$^{-1}$. The boxes represent the projected total uncertainties (quadratic sum of statistical and systematic uncertainties), while vertical bars represent statistical uncertainties.

Figure 3 presents projected uncertainties on measurements of the $W$ boson yield ratio of central to peripheral collisions $R_{\mathrm{CP}}$. In the $R_{\mathrm{CP}}$ all non-centrality dependent (nPDF or other) modifications will cancel, and so the observable is well suited to studying centrality dependent effects. The peripheral reference consists of events from the 80–90% centrality class, while the central selections are the 0–5%, 20–30% and 60–70% centrality classes. Geometric parameters for these centrality classes, such as the mean nuclear overlap function, $\langle T_{\mathrm{AB}}\rangle$, are calculated from the Glauber model [16] for $p$+Pb collisions at $\sqrt{s_{\mathrm{NN}}}$ =8.16 TeV. The statistical uncertainties on $R_{\mathrm{CP}}$ are estimated assuming that binary scaling of $W$ boson production holds and are dominated by the statistical precision of the peripheral reference. Systematic uncertainties contributed by all sources are assumed to vanish in the ratios, except for the contribution from the QCD multi-jet background estimation.

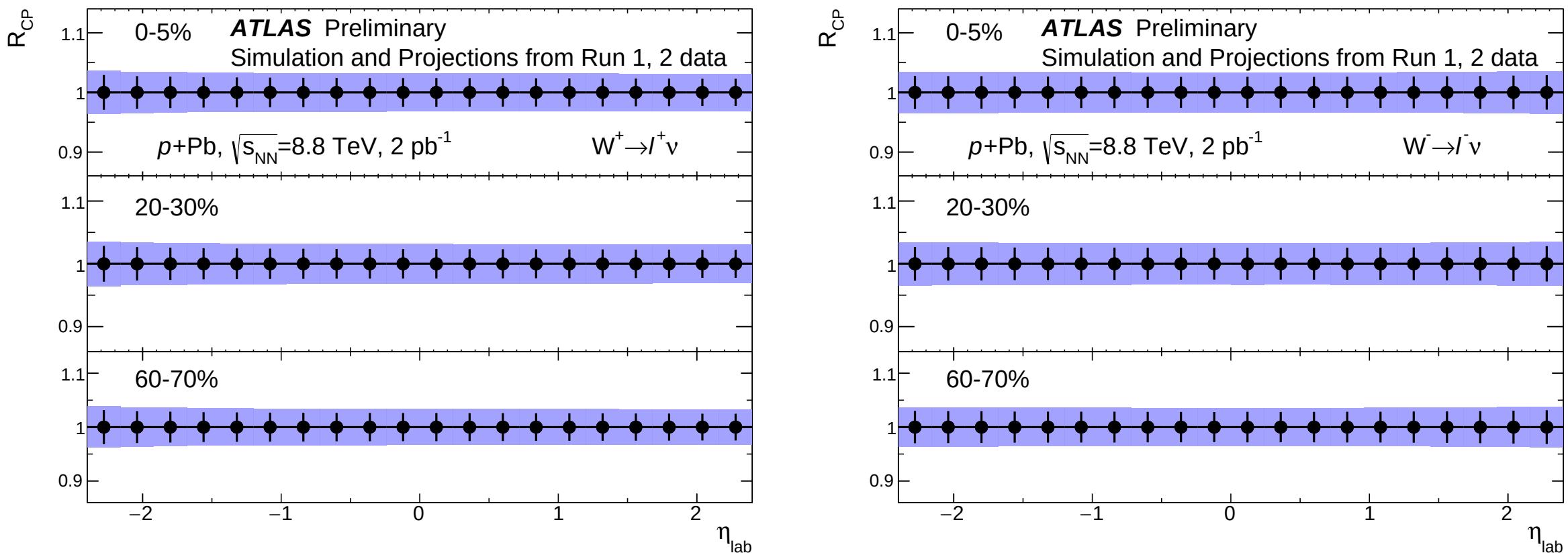

Figure 3: Uncertainties on measurments of the $W^+$ (left) and $W^-$ (right) boson yield ratio of central to peripheral collisions $R_{\rm CP}$. The uncertainties are projected for $p$+Pb collisions at $\sqrt{s_{\rm NN}}$ = 8.8 TeV with an integrated luminosity of 2 pb$^{-1}$. The boxes represent the projected total uncertainties (quadratic sum of statistical and systematic uncertainties), while vertical bars represent statistical uncertainties.

### 3.2 $Z$ Boson Measurements in $p$+Pb

Similar to $W$ bosons, the measurement of $Z$ boson production rates in $p$+Pb collisions is a precise probe of the lead nucleus. In particular, the rapidity distribution of $Z$ bosons is sensitive to nPDF effects since it is directly related to the Bjorken $x$ fraction of the nucleon momentum carried by the quark in the Drell-Yan process. Furthermore, the dependence of the $Z$ boson yields on $\langle T_{AB} \rangle$, offers the possibility to test various models of how the nuclear collision geometry affects particle production in $p$+Pb collisions.

Events were generated using Powheg+Pythia8 and MCFM were prepared similarly to the $W$ boson projections discussed in Section 3.1 (the CT10 PDF set [17] is used rather than CT14 with Powheg+Pythia8).

Figure 4 shows the rapidity differential cross for the free nucleon PDF (black markers) and nPDF set (open red markers). In the mid-rapidity region, $-3 < y^* < 2$ (where $y^*$ is the rapidity in the centre-of-mass frame) the statistical uncertainty is subdominant and the uncertainty is expected to be driven by the muon reconstruction uncertainty systematics on the per-mille level. A conservative 1% uncertainty is therefore plotted in that region. In the region $y^* < -3$ and $y^* > 2$ the measurement will rely solely on the reconstructed forward electron candidates. The acceptance for reconstruction of these electron candidates is expected to extend up to $|\eta| < 4.9$ of the forward calorimeter (FCal). A systematic uncertainty of 5% in this region was extracted as the lower limit of the systematic uncertainty measured in Run 1 [1], this does not take into consideration the expected improvements due to the ITk detector upgrade. The statistical uncertainty was estimated by scaling the measured uncertainty of the Run 1 data by an appropriate factor accounting for the luminosity and centre-of-mass energy increase.

Figure 5 illustrates the momentum reach of the projected measurement obtained with the free nucleon PDF set. Systematic uncertainty on the $p_T^Z$ differential cross-section is derived from the relative systematic uncertainty measured in the Run 1 data. For the high-$p_T^Z$ points where the Run 1 measurement does not exist, the relative systematic uncertainty of the last measured point is taken as a characteristic uncertainty of the projected spectra.

In order to study the anticipated data precision in exploring different implementations of the nucleon-nucleus overlap geometry, projected results are presented as the centrality-dependent $Z$ boson yields assuming binary collision scaling. Figure 6 shows the projected yield per inelastic nucleon-nucleon collision and divided by $\langle T_{AB} \rangle$. The projections are compared with the measurements in $\sqrt{s_{NN}}$ =5.02 TeV proton-lead collisions in Run 1 [1]. Uncertainties on $\langle T_{AB} \rangle$ are derived from the Run 1 measurement, assuming no change in the model precision. Systematic uncertainties on the projected yields are estimated to be 1%. Statistical uncertainties are less than the marker sizes in each of the bins. In the previous study the uncertainties from the measurement of the $Z$ boson yields were comparable to the uncertainties on $\langle T_{AB} \rangle$ whereas in the projected data the uncertainties of the boson analysis will be significantly smaller than those uncertainties. This may allow the $Z$ boson yields to be used as an independent estimate of the $p$+Pb parton-parton luminosity [18]. Further, the added precision will enable finer binning in centrality.

6

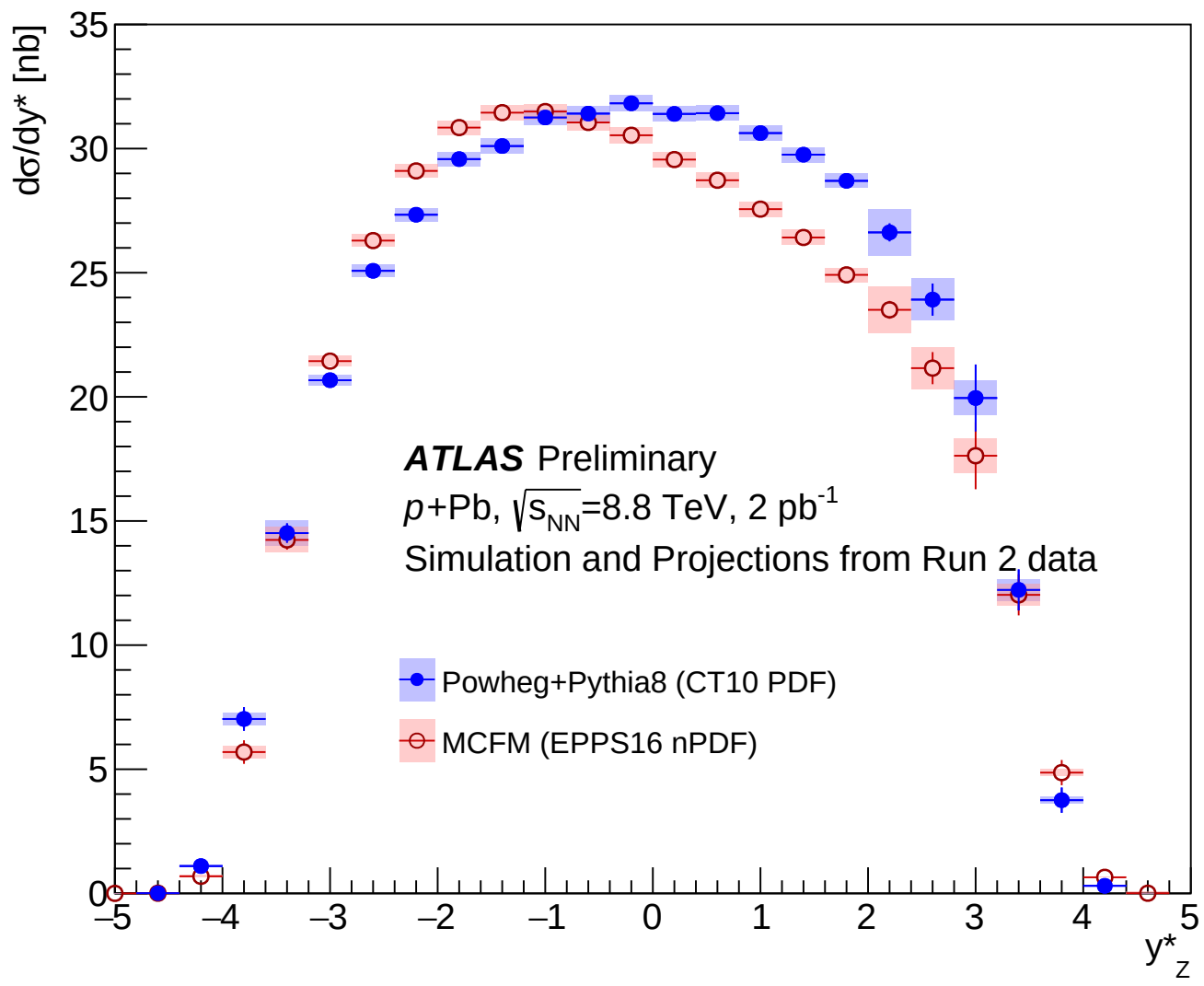

Figure 4: Projected $Z$ boson rapidity differential cross sections for a 2 $\text{pb}^{-1}$luminosity scenario. The free nucleon PDF projection is obtained with the CT10 PDF set and the nPDF with the EPPS16 set. Statistical uncertainties are smaller than the marker size in the mid-rapidity region. Projected systematic uncertainties are described in the text.

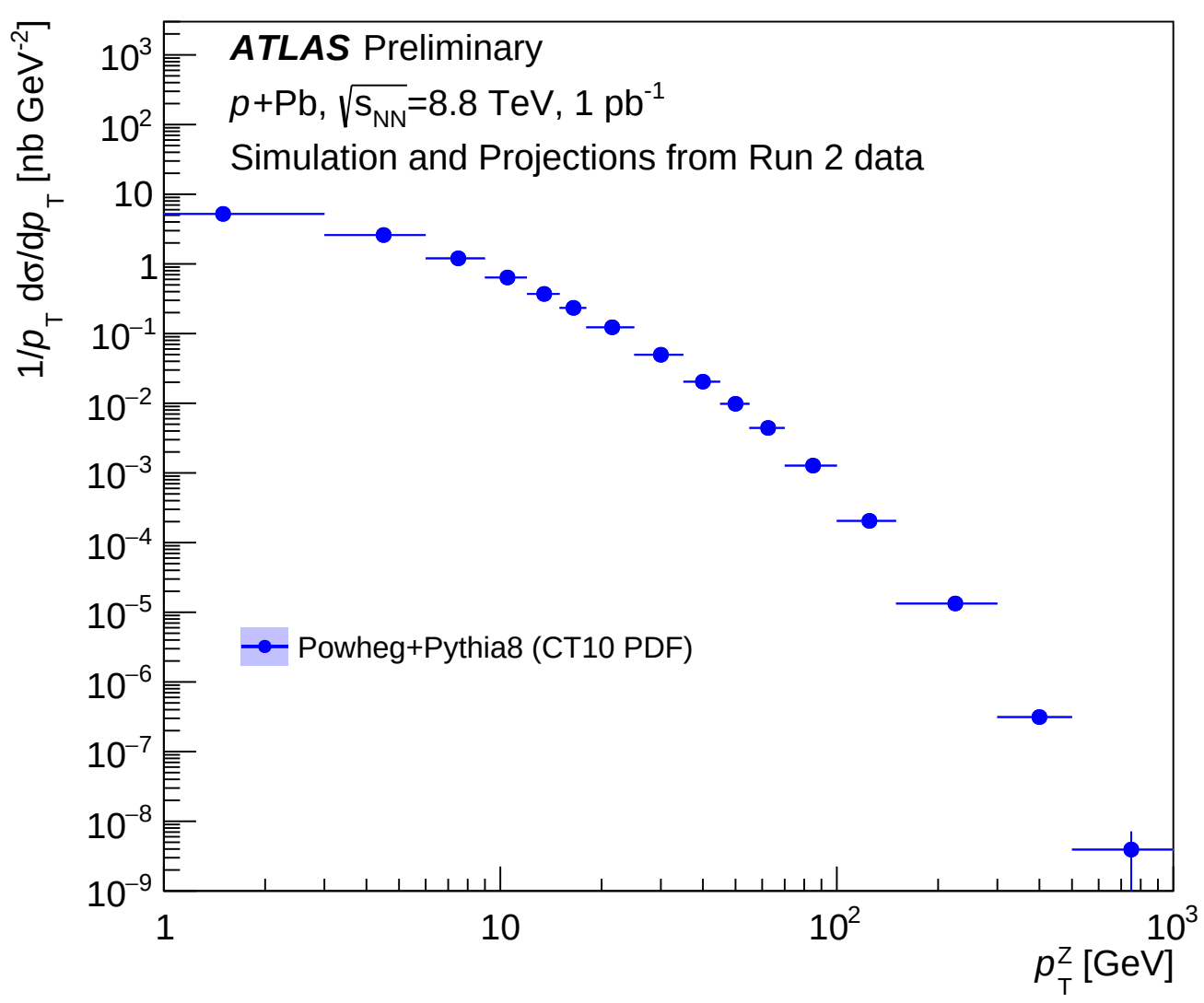

Figure 5: Projected $Z$ boson momentum differential cross sections for for a 2 $\text{pb}^{-1}$luminosity scenario. The free nucleon PDF projection is obtained with the CT10 PDF set. Statistical uncertainties are everywhere in the spectrum smaller than the marker size, except for the last bin. Projected systematic uncertainties are described in the text.

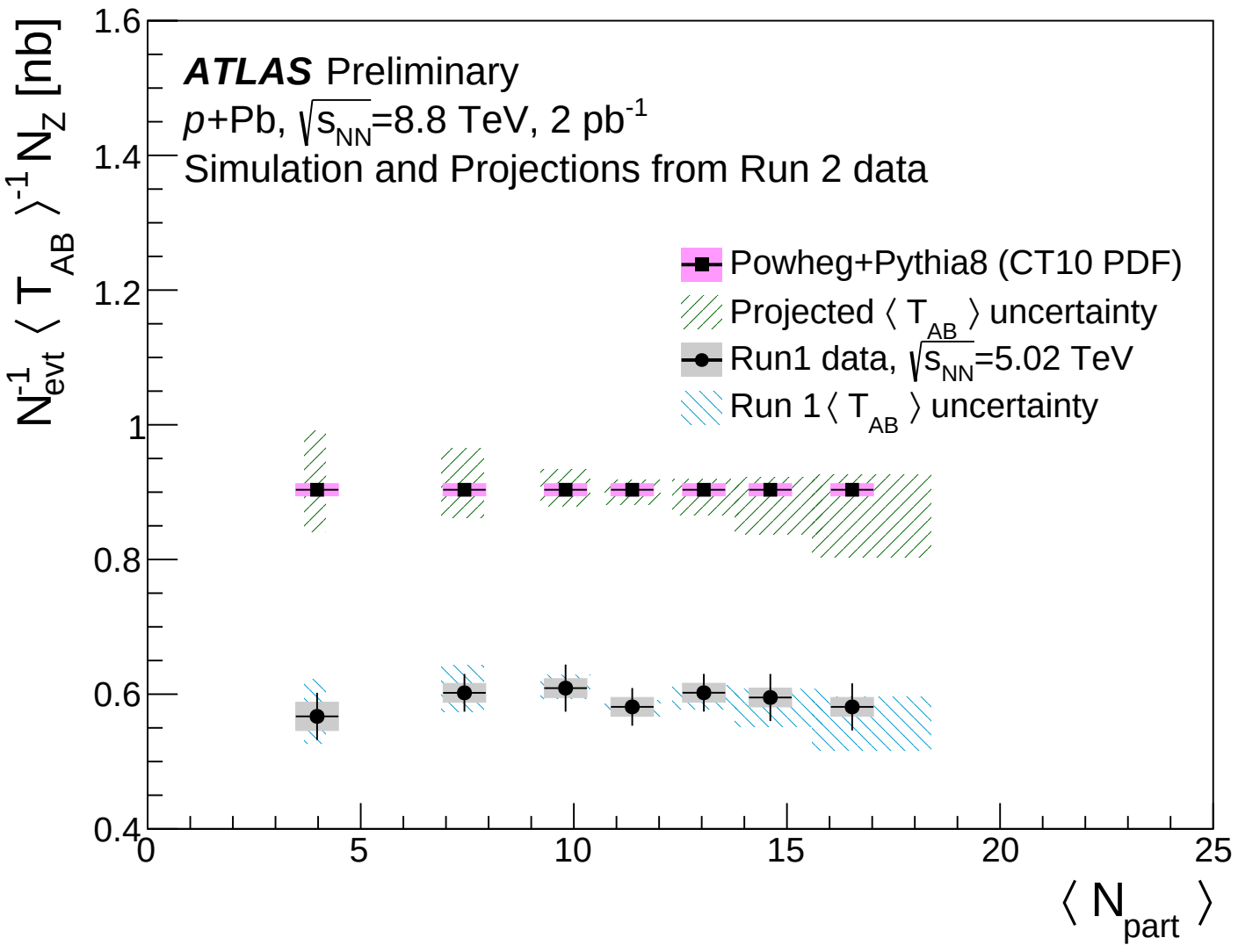

Figure 6: Projected $Z$ boson yields scaled by $\langle T_{\mathrm{AB}} \rangle$ as a function of the number of nucleon participants, $\langle N_{\mathrm{part}} \rangle$, for a 2 $\mathrm{pb}^{-1}$luminosity scenario. Statistical uncertainties are everywhere smaller than the marker size, and projected systematic uncertainties (pink shaded boxes) are described in the text. The projections are compared with a previous Run 1 analysis [1]. Glauber model uncertainties are shown with hashed boxes.

### 3.3 Photon + Jet Measurements in $p$+Pb

In addition to the measurements with heavy electroweak bosons described above, photon+jet events are an effective channel through which to probe nuclear effects in a precise manner. Using the kinematics of the high-$p_{\mathrm{T}}$ photon together with those of the highest-$p_{\mathrm{T}}$ jet in the opposite azimuthal hemisphere (the "leading jet") provides event-by-event sensitivity to the nuclear-$x$ and $Q^2$ of the underlying parton-parton scattering process.

Statistical projections are constructed using Pythia8 photon+jet events, using the NNPDF 2.3 LO parton distribution function set [19], and including both direct and fragmentation photon processes as is commonly done in ATLAS measurements of photon production [20–22]. Jets are defined by applying the anti-$k_t$ [23] algorithm with $R = 0.4$ to stable particles. Projections are shown as a function of photon $p_{\mathrm{T}}$ and pseudorapidity in the centre-of-mass frame ($\eta^*$) selections corresponding to the ATLAS electromagnetic calorimeter acceptance, with positive and negative sides shown separately, as these probe different regions of nuclear $x$. These projections include an estimated combined efficiency of 90% for photon reconstruction, identification and selection [22].

Figure 7 shows the expected double-differential photon yield accessible in 2 pb$^{-1}$ of 8.8 TeV $p$+Pb collisions. The closed circles denote the total double-differential photon yield. The open circles show the yield of photon+jet pairs in which the leading jet is required to have $p_{\mathrm{T}} > 20$ GeV, to be within the ATLAS acceptance of $|\eta| < 4.9$, and to be balanced by a photon with an azimuthal angle difference ($\Delta\phi$) greater than $3\pi/4$. Above $p_{\mathrm{T}}^{\gamma} > 50$ GeV, essentially all photon+jet pairs meet these criteria and can thus be used to tag the parton-level kinematics.

To approximate possible effects of nPDF modifications on the cross section, the Pythia events are reweighted on an event-by-event basis by the value given by the EPS09 nuclear PDF set [24] for the flavor, $x$, and $Q^2$, of the parton in the "nuclear" beam (in the particular generated event). Figure 8 demonstrates one way to evaluate how the photon+jet data is sensitive to nuclear PDF modification. Although the dependence of the cross-section on the PDFs is more complicated than the method used here, since the expected modifications are small and smoothly varying with the chosen kinematic variables, this approximation is sufficient to gauge the possible size of nPDF effects. The figure shows the projected statistical uncertainties on a measurement of the nuclear modification factor $R_{p\mathrm{Pb}}$, given as a function of the nuclear-$x$ values, $x_A$, taken directly from Pythia. The modification of the cross-section in each $\eta$ and $p_{\mathrm{T}}$ bin shown in Fig. 8 is plotted as a function of the mean $x_A$ in Pythia for events in each kinematic selection. The kinematic selections for the photon+jet pair are the same as those described above for Fig. 7. The central values are set to be equal to those predicted by the EPS09 nuclear PDF set, as a way of demonstrating the ability of the data to be sensitive to the magnitude of the nuclear modification effects.

In this study, only the expected statistical uncertainties have been evaluated and the magnitude of possible systematic uncertainties have not been explicitly estimated. Previous measurements of jet and photon production in $p$+Pb collisions in Runs 1 and 2 [22, 25] have achieved small (<5%) uncertainties on jet and photon production cross-sections in a broad kinematic ranges. Furthermore, many sources may be correlated with the comparison $pp$ data and thus cancel in ratios such as the $R_{p\mathrm{Pb}}$. Reference data from $pp$ collisions at $\sqrt{s} = 8.8$ TeV would be ideal for the $R_{p\mathrm{Pb}}$ measurement, uncertainties on the order of a few percent are expected for a reference based on an extrapolation of $\sqrt{s} = 8$ TeV $pp$ collisions. Further improvements to the systematic uncertainties are also expected or the Runs 3 and 4 measurements, where the larger event statistics will allow for more precise data-driven studies.

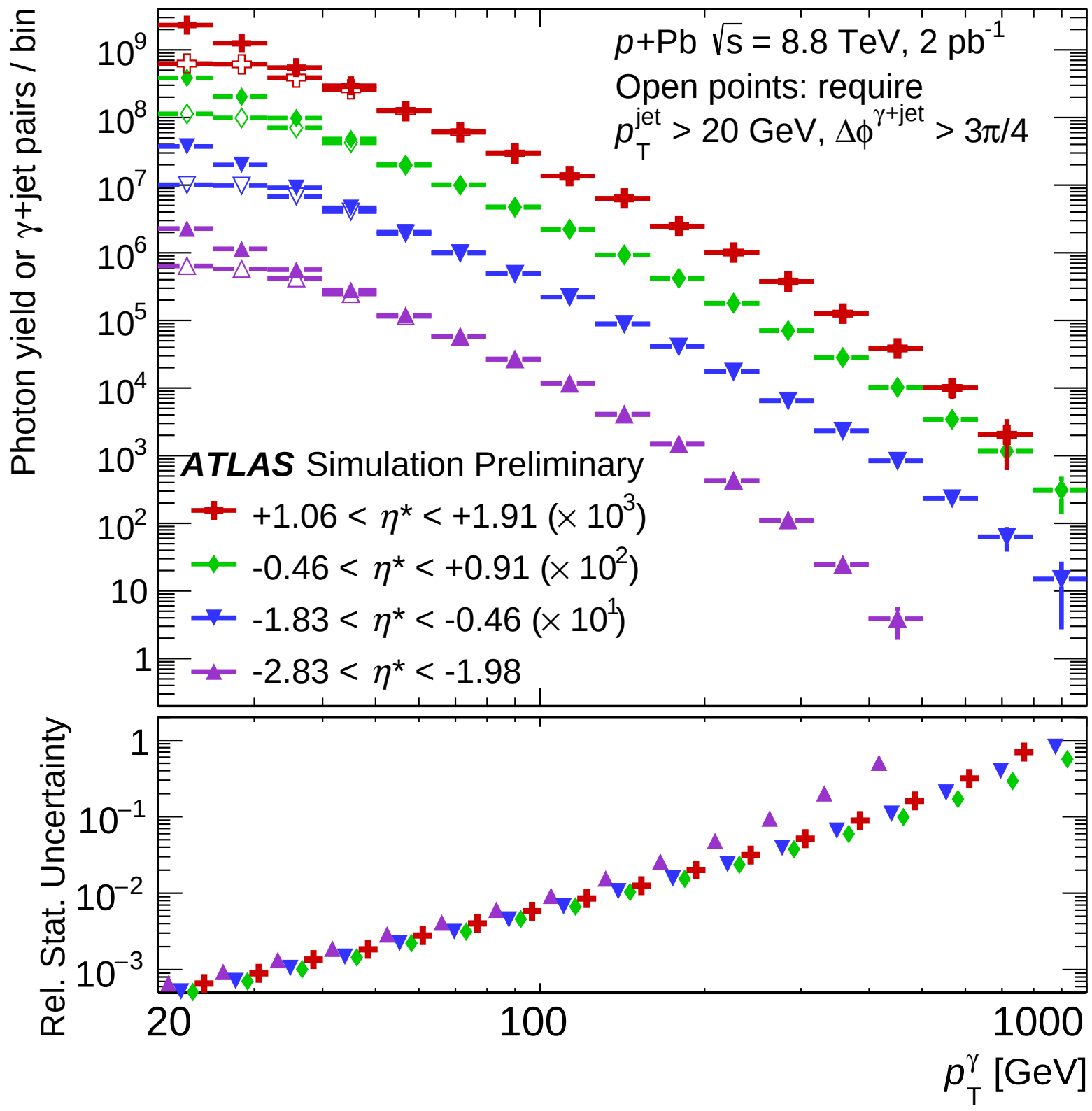

Figure 7: Projected yields of photons and photon+jet events in 2 $\text{pb}^{-1}$ of 8.8 TeV $p$+Pb data, shown as a function of photon $p_T$ with different selections on the photon pseudorapidity in the centre-of-mass frame shown in different colors. Closed markers denote the yield of events with a $p_T$ photon, while open markers denote the subset of those which also include a balancing jet above threshold in the ATLAS acceptance. The bottom panel below shows the projected statistical uncertainties.

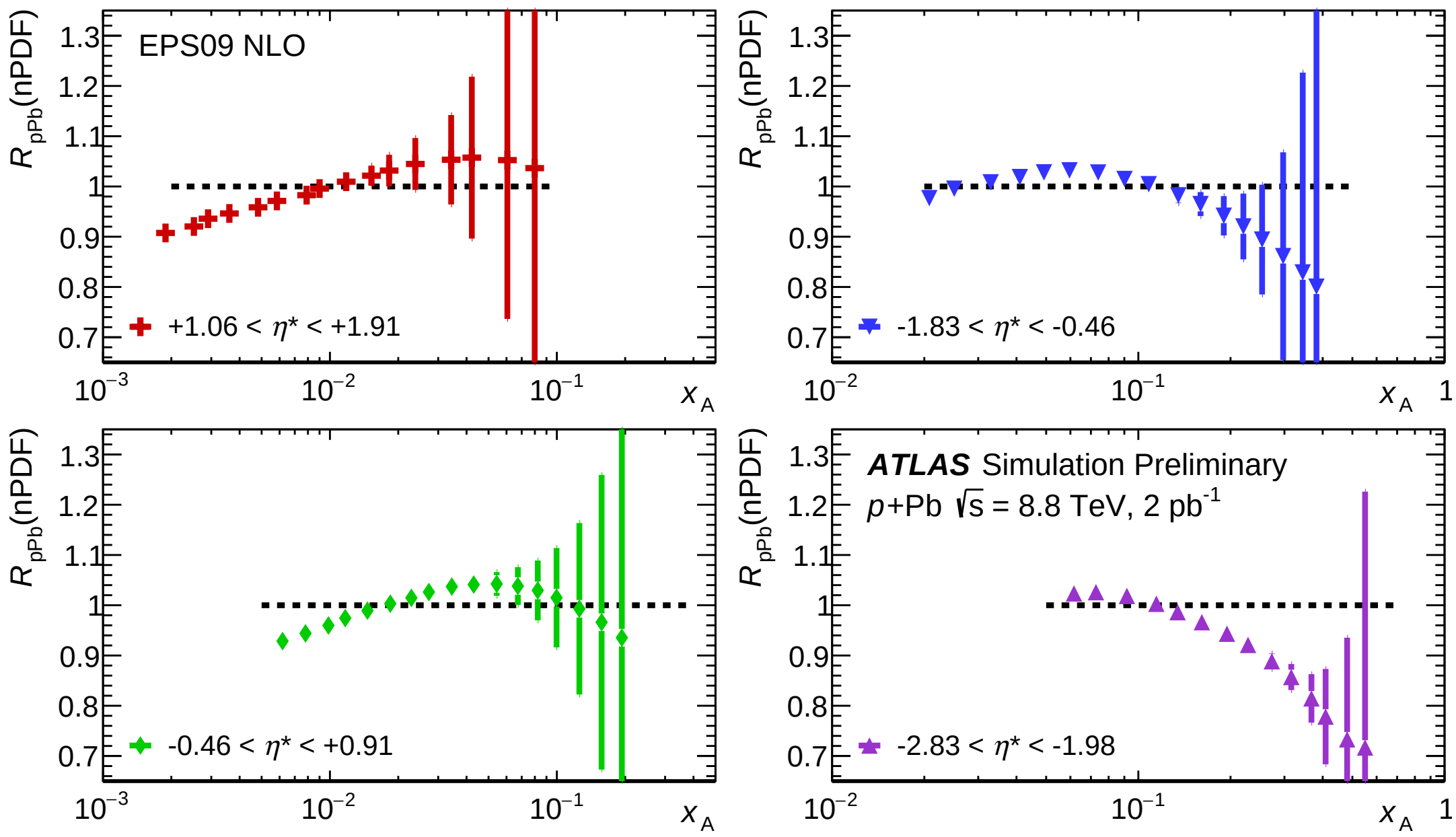

Figure 8: Projected statistical uncertainties for an $R_{p\mathrm{Pb}}$ measurement, plotted as a function of the average $x_A$, taken from Pythia, for each photon $p_{\mathrm{T}}$ bin in Fig. 7. The panels show projections for photon+jet events in different selections on the photon pseudorapidity measured in the centre-of-mass frame. The central values are set to be equal to the predictions of the EPS09 nuclear PDF set. Statistical uncertainties on the $pp$ reference are neglected.

### 3.4 $Z$ boson + Jet Measurements in $p$+Pb

In addition to photon+jet events in $p$+Pb collisions discussed in section 3.3, $Z$ boson+jet events may also be studied. Although the $Z$ boson production cross section is lower compared to photons, $Z$ boson+jet events have some advantages both theoretically, *e.g.* the large $Z$ boson mass sharply reduces fragmentation and decay contributions, and experimentally, *e.g.* smaller backgrounds. Therefore $Z$ boson+jet events are used to complementarily study similar questions as those studied with photon+jet events. The anticipated yield of the produced $Z$ bosons in the LHC Runs 3 and 4 allows for a precise measurement of the momentum balance between the produced boson and the opposite-side leading jet in the event. This is a valuable measurement in $p$+Pb collisions to study possible nPDF and CNM effects in general. Furthermore, there are some indications that nuclear effects not usually categorized as CNM can produce rapidity dependent modification of jet production [25], and so the momentum balance is presented for different selections on the jets' rapidity.

Leptonic $Z$ boson decays generated using Powheg+Pythia8 with the CT10 PDF set, prepared similarly to the event generation discussed in Sections 3.1 and 3.2, are used for these projections. The jets are clustered at truth level with the anti-$k_t$ algorithm with a radius parameter $R$=0.4. The $Z$ boson and lepton kinematics are defined at the Born level, excluding final-state radiation. The number of generated events corresponds to the expected number of $Z$ boson candidates with transverse momentum above 35 GeV for luminosity of 2 pb$^{-1}$ at $\sqrt{s_{\mathrm{NN}}}$ = 8.8 TeV. Events for the analysis are selected by applying the fiducial lepton cuts of $p_{\mathrm{T}}$ >20 GeV and $|\eta|$<2.5 in the laboratory frame. $Z$ boson candidates are further required to have $p_{\mathrm{T}}$ >60 GeV. The opposite side leading jet is required to be correlated in azimuth with $|\Delta\phi|$>7/8$\pi$ and with the transverse momentum above 30 GeV.

Figure 9 shows the distribution of the $Z$-jet momentum balance, $x_{\mathrm{J}Z} = p_{\mathrm{T}}^{\mathrm{jet}}/p_{\mathrm{T}}^{Z}$ for three different slices in the jet pseudorapidity: forward ($2 < \eta < 4.5$), mid-rapidity ($|\eta| < 1$) and backward jets ($-4.5 < \eta < -2$). The distributions are normalised per $Z$ boson candidate. The forward and backward jets show a slightly different $x_{\mathrm{J}Z}$ distribution compared to the mid-rapidity jets. It is expected that there may be modification of these distributions relative to the calculations without nPDF modification shown in Figure 9 related to the observations of rapidity dependence of jet $R_{p\mathrm{Pb}}$ in Ref. [25]. The systematic uncertainties are derived from the leading jet energy scale (JES) uncertainties typical for the Run 2 LHC measurement [26], and amount to 1% in the midrapidity region and 3% in the forward/backward region. Figure 10 shows the average of the distribution, $\langle x_{\mathrm{J}Z}\rangle$, as a function of the recoiling jet pseudorapidity in the centre-of-mass frame.

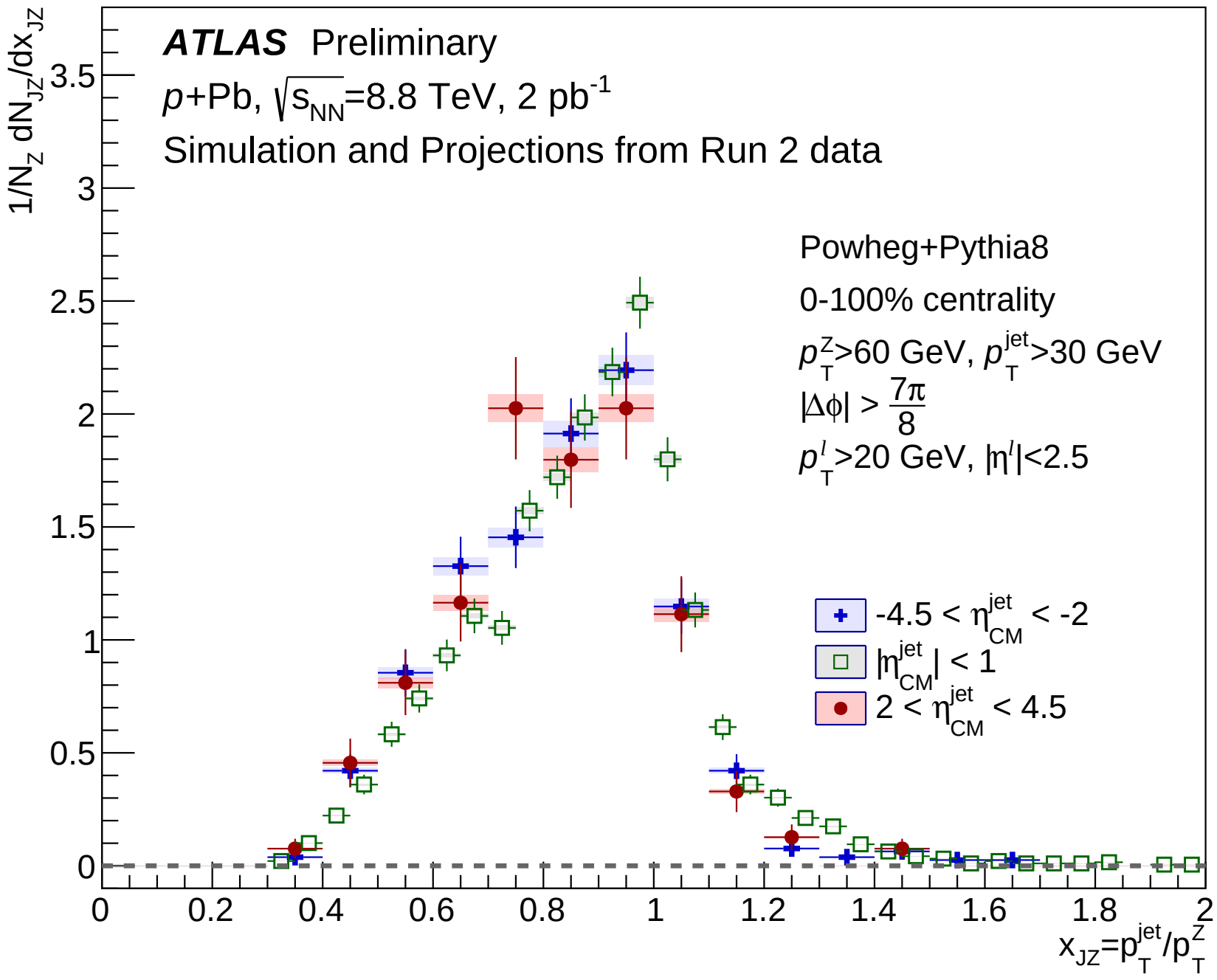

Figure 9: Distribution of the momentum balance between the $Z$ boson and the opposite side leading jet, $x_{JZ} = p_T^{\text{jet}}/p_T^Z$, normalised per $Z$ boson candidate in 2 pb$^{-1}$of 8.8 TeV $p$+Pb data. Different selections on the jet pseudorapidity in the centre-of-mass frame are shown in different colors. The $Z$ boson and leading jet are required to be correlated in azimuth with $|\Delta\phi|>7/8\pi$ and are selected with $p_T$ >60 GeV and $p_T$ >30 GeV for the $Z$ boson and the jet, respectively. Corresponding statistical (vertical bars) and systematic uncertainties (boxes) are shown.

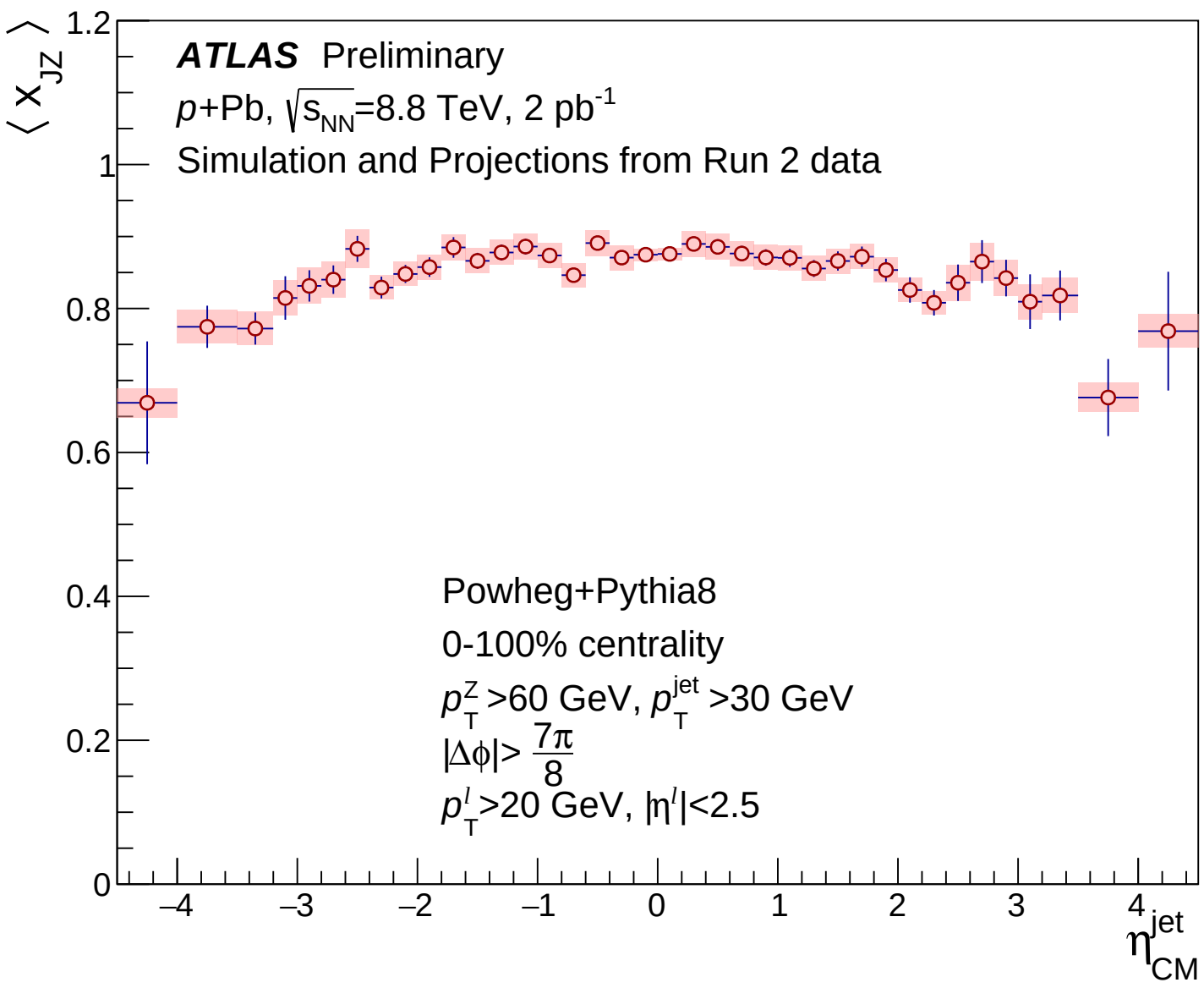

Figure 10: The average momentum balance between a $Z$ boson and the corresponding leading jet, $\langle x_{\mathrm{JZ}}\rangle = \langle p_{\mathrm{T}}^{\mathrm{jet}}/p_{\mathrm{T}}^{Z}\rangle$, as function of jet pseudorapidity in the centre-of-mass frame, projected for 2 pb$^{-1}$of 8.8 TeV $p$+Pb data. The $Z$ boson is selected in the lepton fiducial space defined with $p_{\mathrm{T}}$ >20 GeV and $|\eta|$ <2.5 in the laboratory frame. The $Z$ boson and leading jet are required to be correlated in azimuth with $|\Delta\phi|$>7/8$\pi$ and are selected with $p_{\mathrm{T}}$ >60 GeV and $p_{\mathrm{T}}$ >30 GeV for the $Z$ boson and the jet, respectively. Corresponding statistical (vertical bars) and systematic uncertainties (boxes) are shown.

## 4 Summary

This note presents projections for electroweak boson and photon+jet measurements in $p$+Pb collisions in LHC Runs 3 and 4 with the ATLAS detector. Cross-sections considered in rapidity, centrality, and $p_{\mathrm{T}}$ of $W$ and $Z$ bosons are presented. Photon+jet nuclear modification factors as a function of $x_A$, and $Z$ boson+jet momentum balance distributions are also presented. These projections are representative of future measurements sensitive to nuclear parton distributions.

# Appendix

## A Alternate Luminosity Scenarios

Projections are shown for $W$ and $Z$ boson, and photon + jet yields for 0.5 pb$^{-1}$and 1 pb$^{-1}$of $p$+Pb collisions with $\sqrt{s_{\mathrm{NN}}}$ = 8.8 TeV.

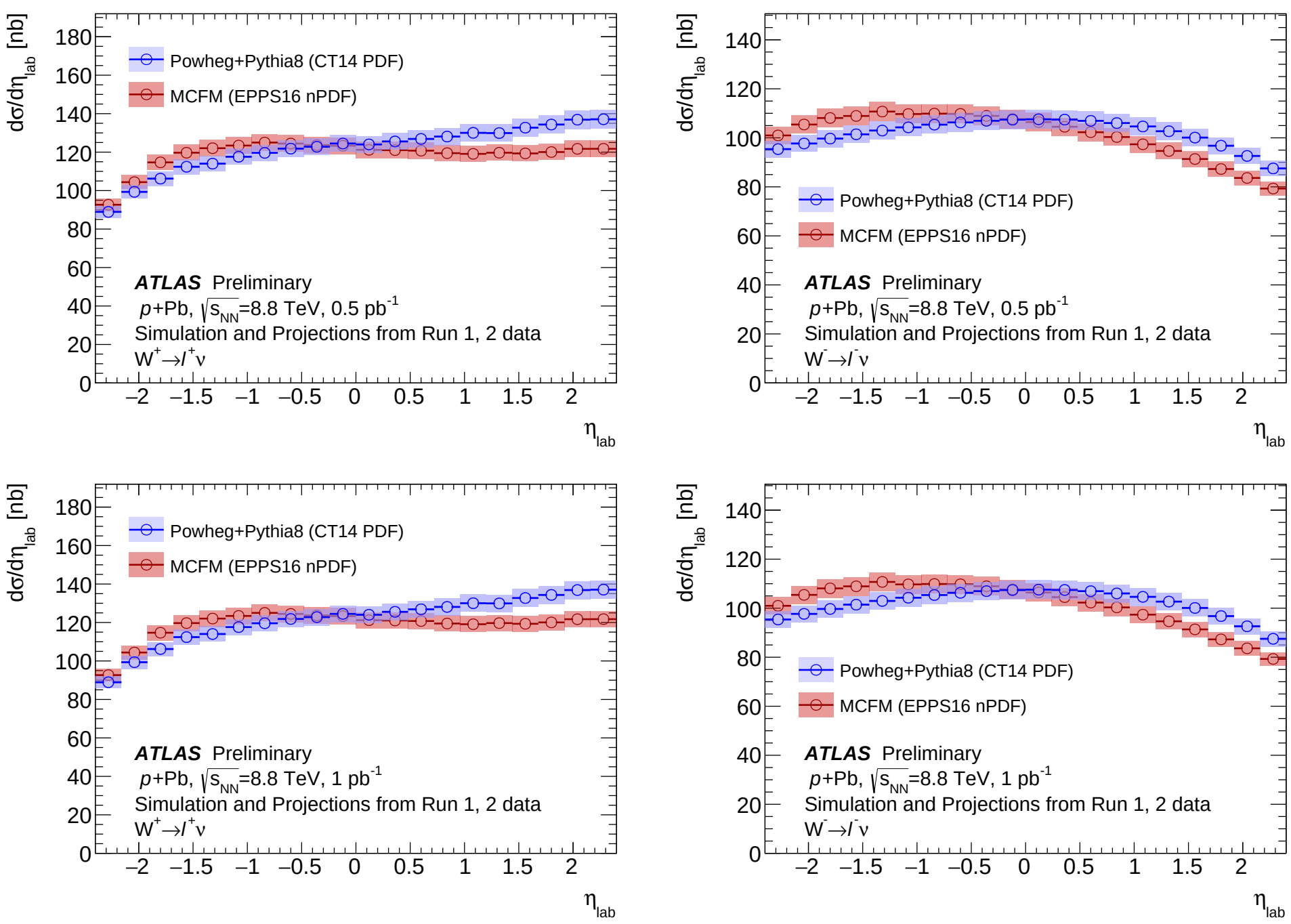

Figure 11: Fiducial cross-sections for $W^+$ (left) and $W^-$ (right) boson production in $p$+Pb collisions at $\sqrt{s_{\mathrm{NN}}}$ = 8.8 TeV differential in the charged lepton pseudorapidity measured in the laboratory frame $\eta_{\mathrm{lab}}$. The cross-sections are projected with nuclear effects described by the EPPS16 nPDF set and without any nuclear effects using the CT14 PDF set for an integrated luminosity of 0.5 pb$^{-1}$(top) and 1 pb$^{-1}$(bottom). The boxes represent the projected total uncertainties (quadratic sum of statistical and systematic uncertainties), while vertical bars represent statistical uncertainties (smaller than the marker size).

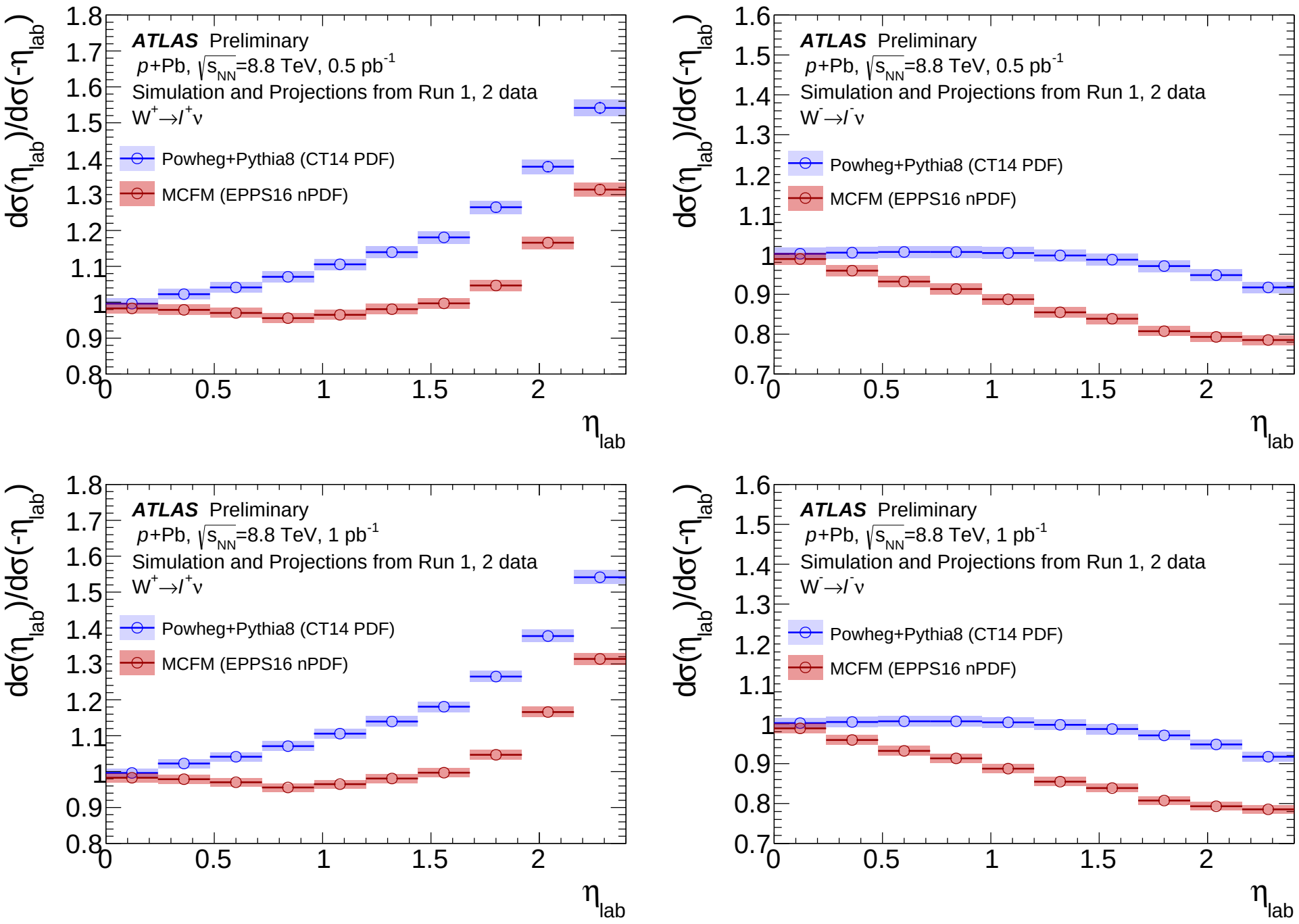

Figure 12: Forward-backward ratios of the differential fiducial cross-sections for $W^+$ (left) and $W^-$ (right) boson production in $p$+Pb collisions at $\sqrt{s_{\mathrm{NN}}}$ = 8.8 TeV. The ratios are projected with nuclear effects described by the EPPS16 nPDF set and without any nuclear effects using the CT14 PDF set for an integrated luminosity of 0.5 pb$^{-1}$(top) and 1 pb$^{-1}$(bottom). The boxes represent the projected total uncertainties (quadratic sum of statistical and systematic uncertainties), while vertical bars represent statistical uncertainties.

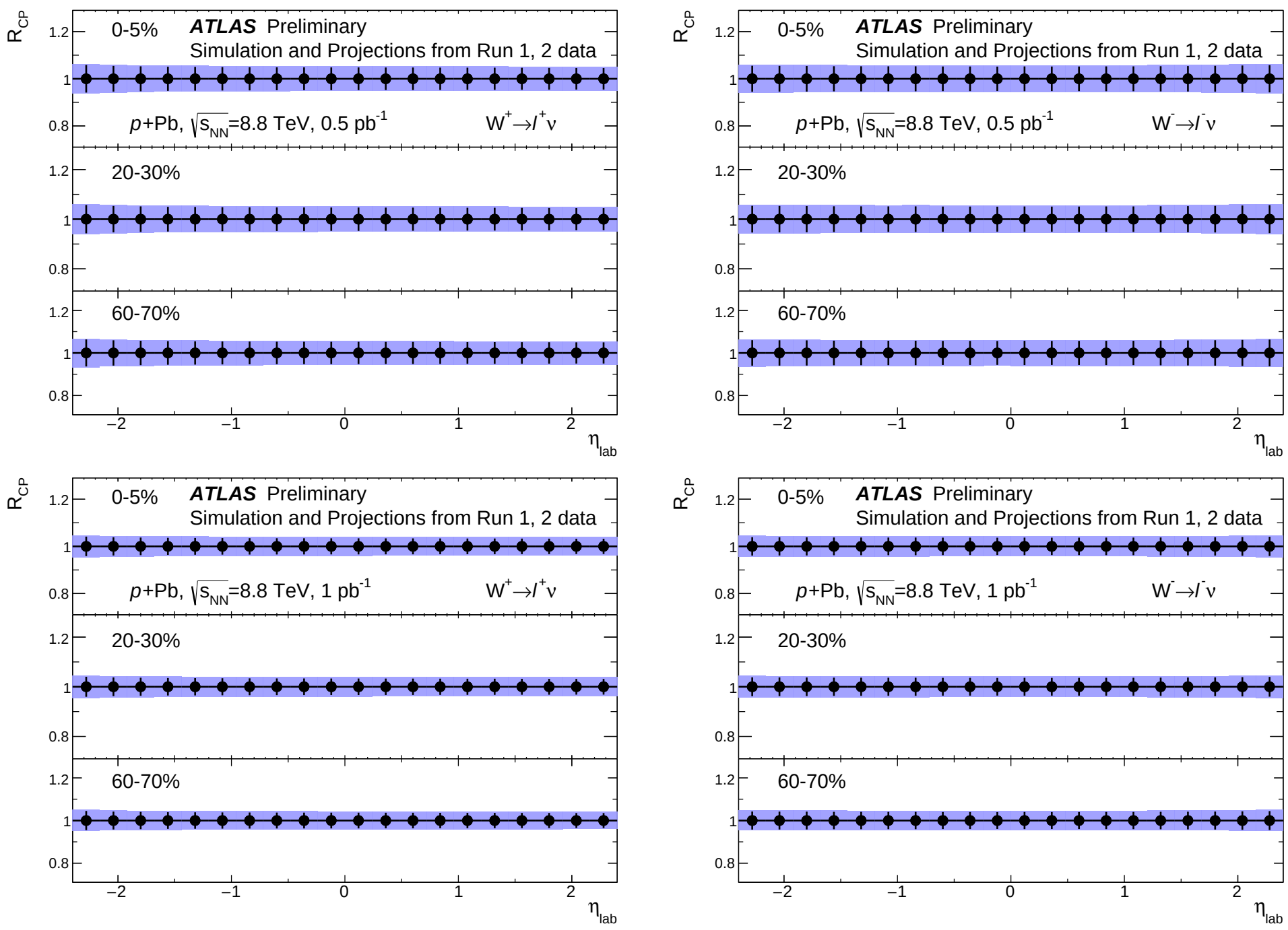

Figure 13: Uncertainties on measurments of the $W^+$ (left) and $W^-$ (right) boson yield ratio of central to peripheral collisions $R_{\mathrm{CP}}$. The uncertainties are projected for $p$+Pb collisions at $\sqrt{s_{\mathrm{NN}}} = 8.8$ TeV with an integrated luminosity of 0.5 pb$^{-1}$(top) and 1 pb$^{-1}$(bottom). The boxes represent the projected total uncertainties (quadratic sum of statistical and systematic uncertainties), while vertical bars represent statistical uncertainties.

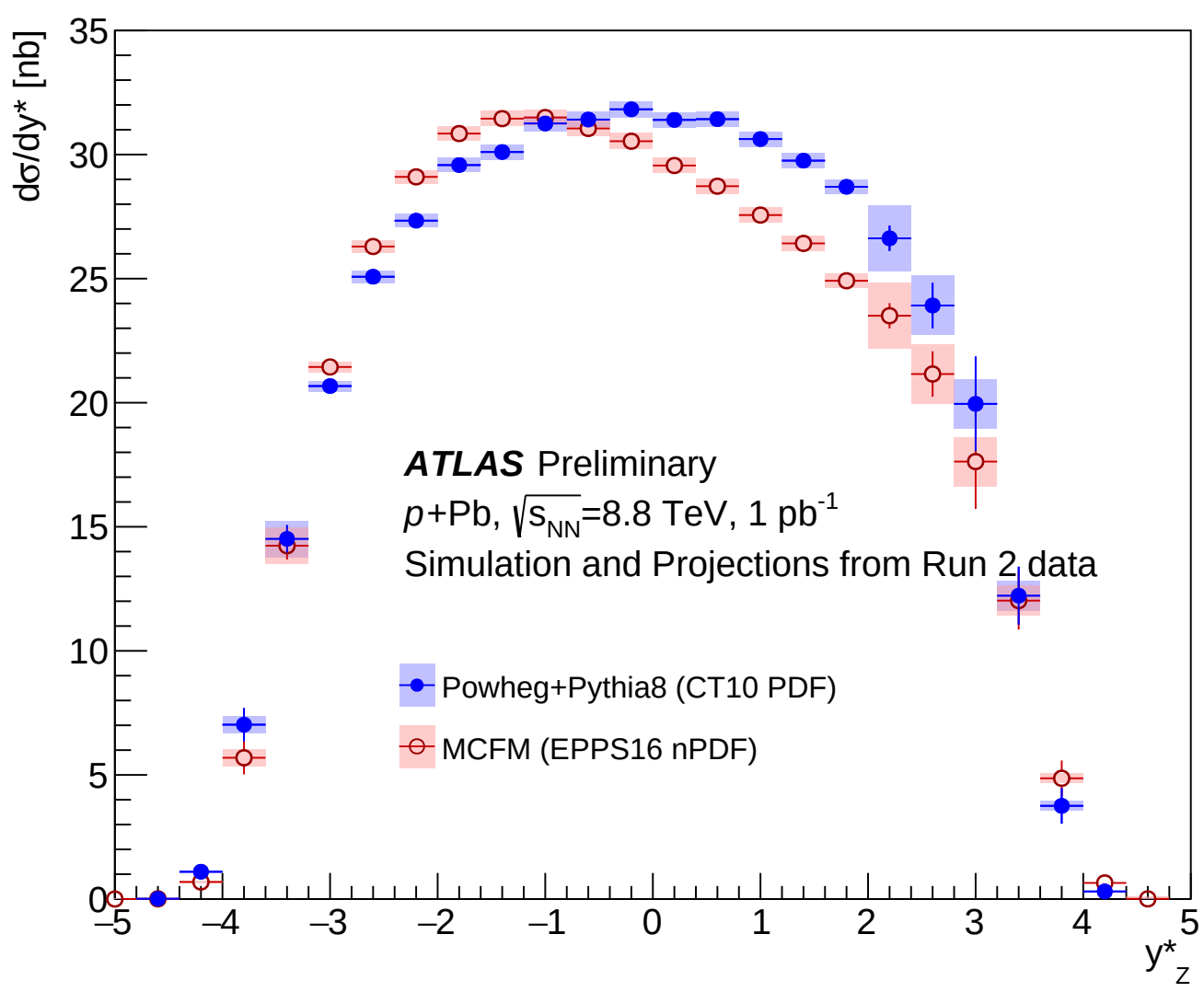

Figure 14: Projected $Z$ boson rapidity differential cross sections for a 1 pb$^{-1}$luminosity scenario. The free nucleon PDF projection is obtained with the CT10 PDF set (black markers) and the nPDF with the EPPS16 set (open red markers). Statistical uncertainties are less than the marker size in the mid-rapidity region. Projected systematic uncertainties (green and red boxes) are described in the text.

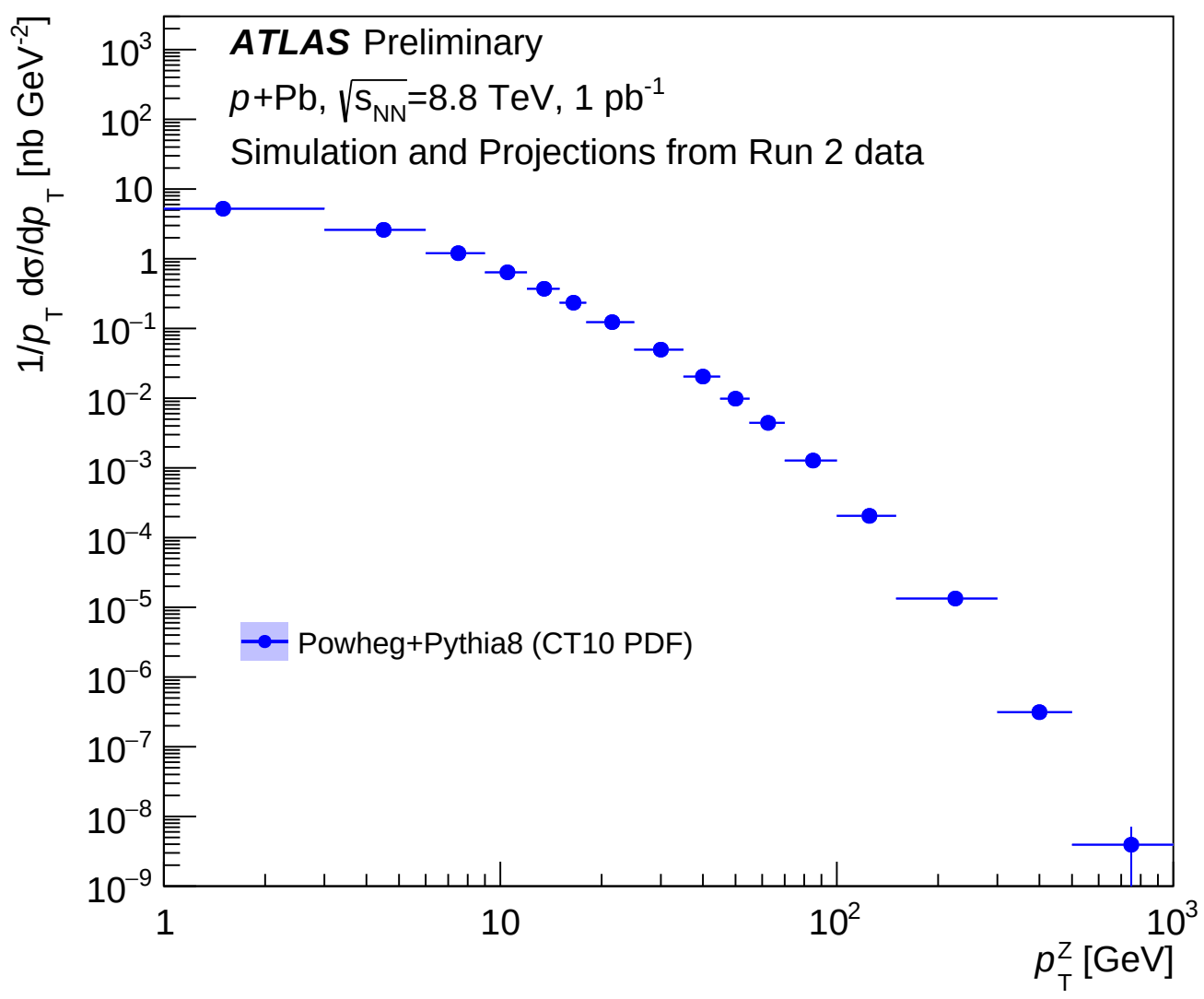

Figure 15: Projected $Z$ boson momentum differential cross sections for a 1 $pb^{-1}$luminosity scenario. The free nucleon PDF projection is obtained with the CT10 PDF set (black markers). Statistical uncertainties are everywhere in the spectrum less than the marker size, except for the last bin. Projected systematic uncertainties (green) are described in the text. The projections are compared to the Run 1 published results for which the statistical and systematic uncertainties are combined.

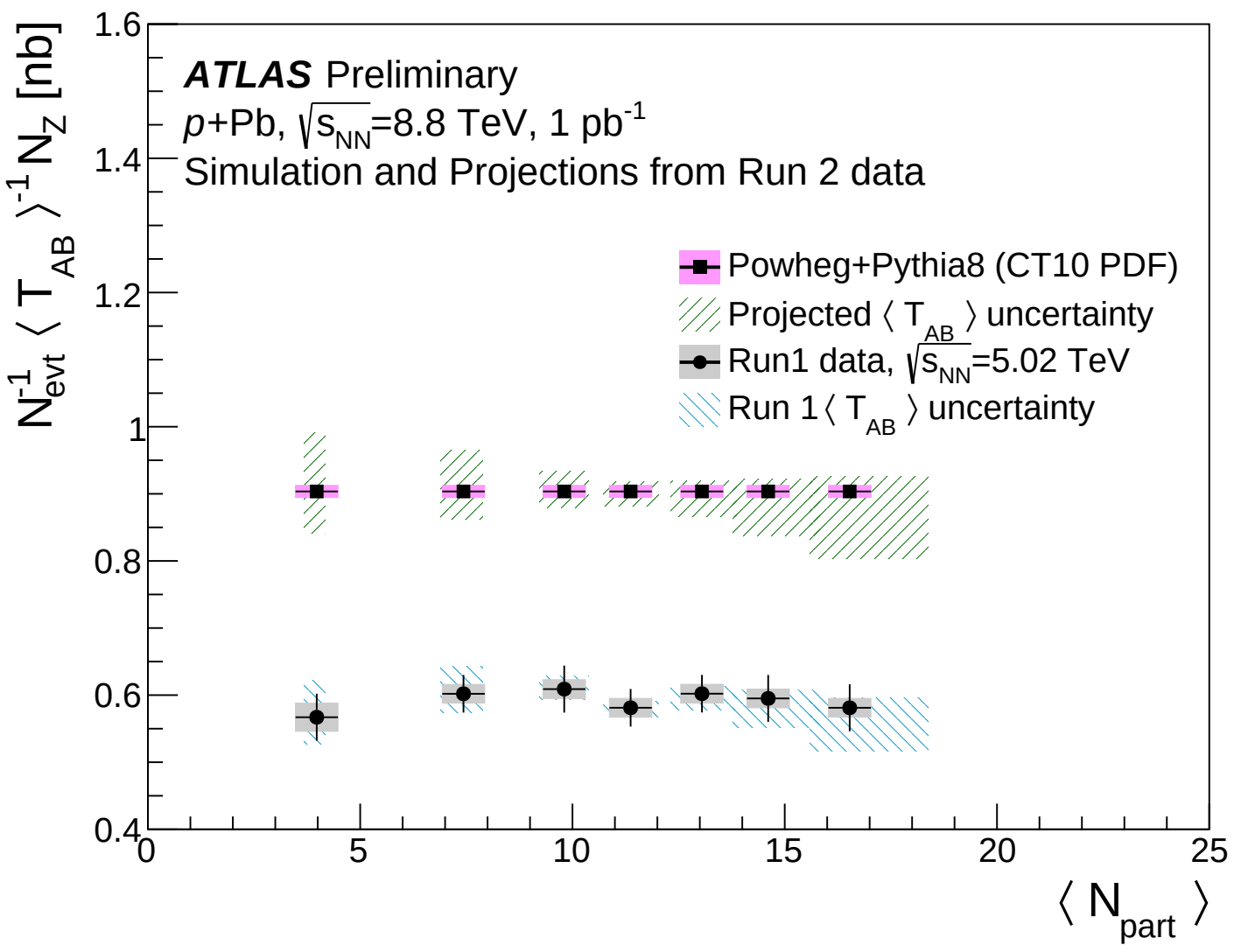

Figure 16: Projected $Z$ boson yields as a function of the number of nucleon participants for a 1 $pb^{-1}$luminosity scenario. Statistical uncertainties are everywhere smaller than the marker size, and projected systematic uncertainties (green) are described in the text. The projections are compared with a previous Run 1 analysis [1]. Glauber model uncertainties are shown with shaded boxes.

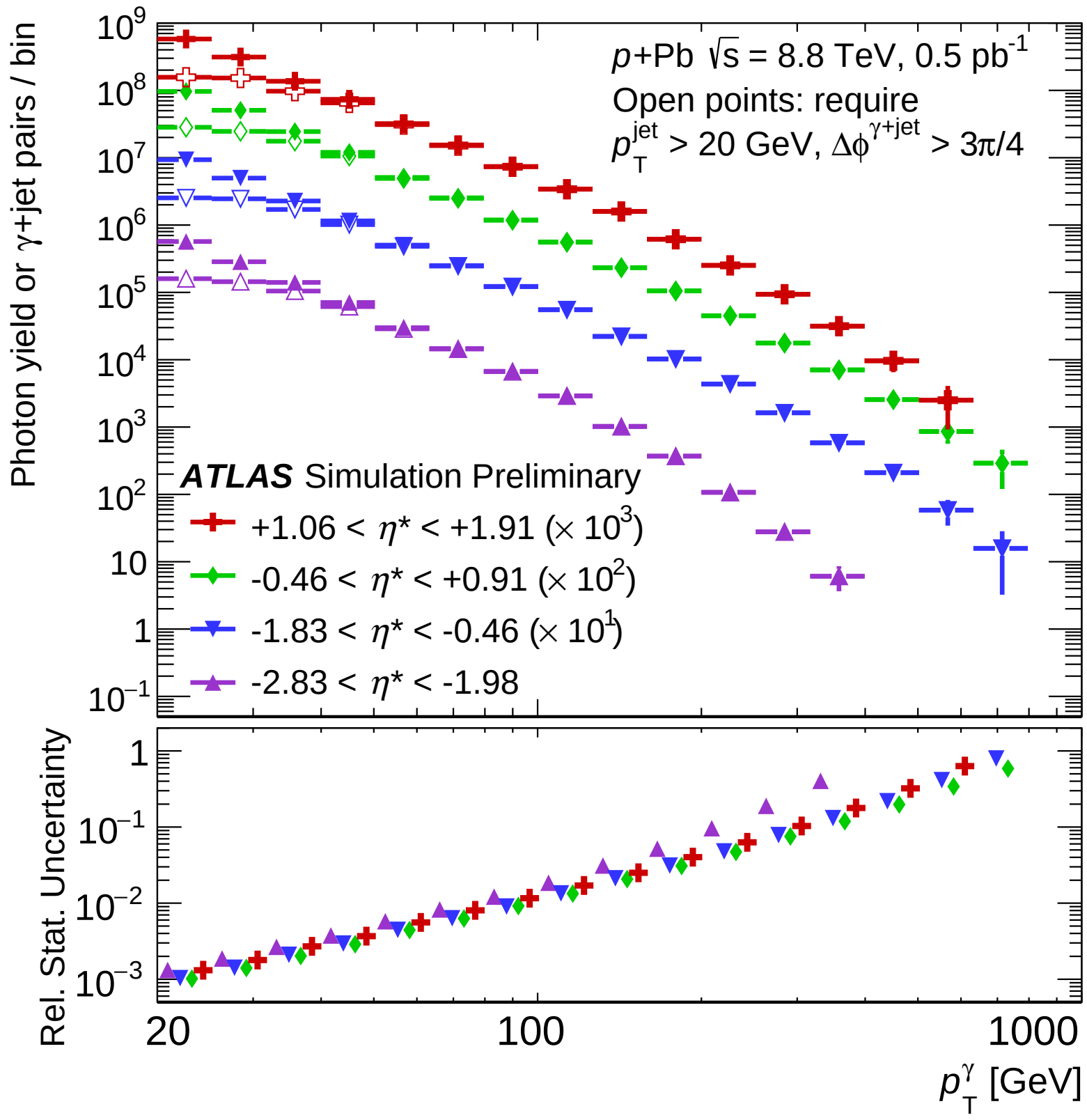

Figure 17: Projected yield of photon +jet events in 0.5 pb$^{-1}$ of 8.8 TeV $p$+Pb data, shown as a function of photon $p_{\mathrm{T}}$ with different selections on the photon pseudorapidity in the laboratory frame shown in different colors. Closed markers denote the yield of events with a $p_{\mathrm{T}}$ photon, while open markers denote the subset of those which also include a balancing jet above threshold in the ATLAS acceptance. The bottom panel below shows the projected statistical uncertainties.

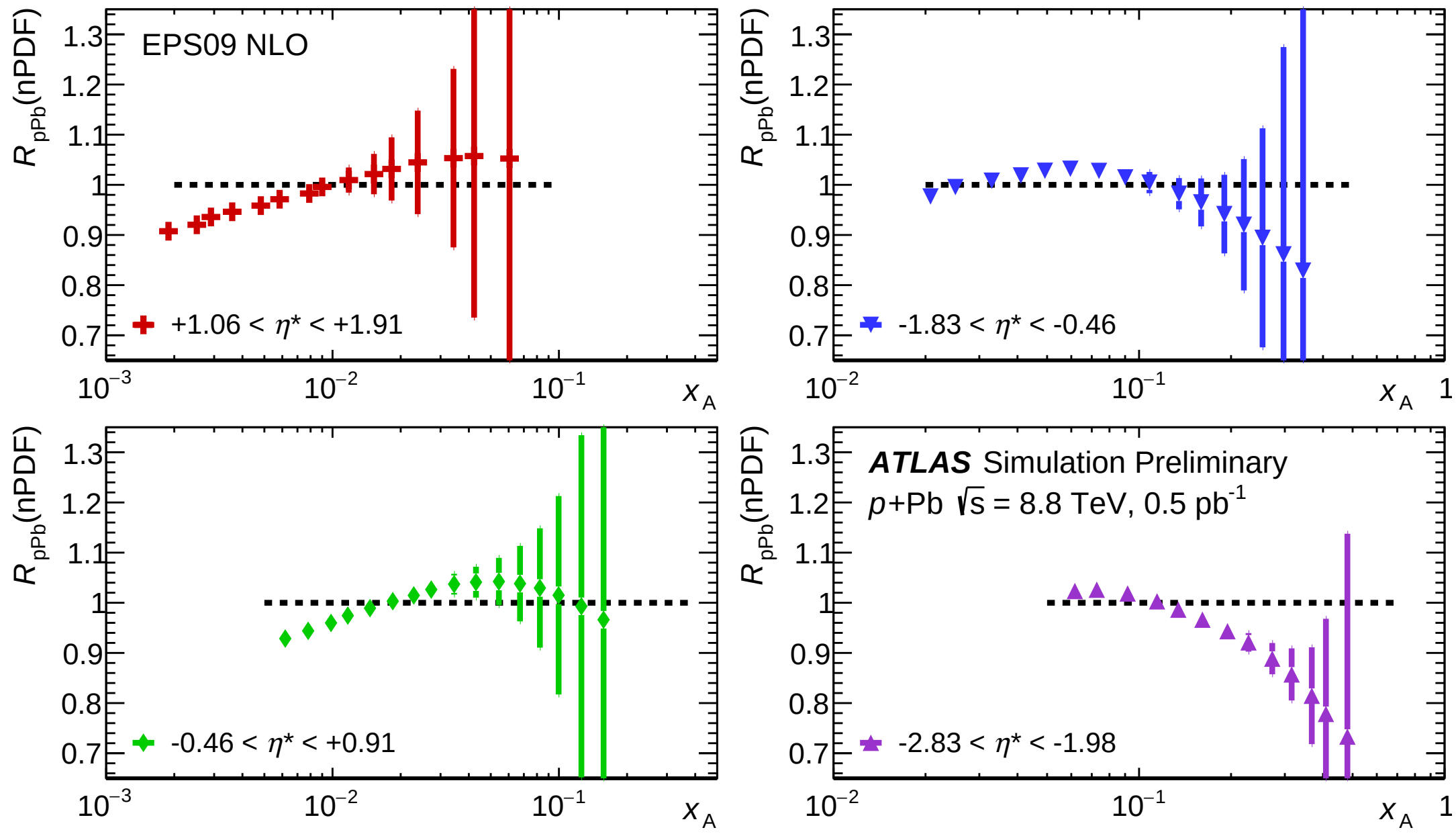

Figure 18: Projected statistical uncertainties for an $R_{p\mathrm{Pb}}$ measurement, plotted as a function of the average nuclear-$x$, taken from Pythia, for each photon $p_{\mathrm{T}}$ bin in Fig. 17. The panels show projections for photon+jet events in different selections on the photon pseudorapidity measured in the laboratory frame. The central values are set to be equal to the predictions of the EPS09 nuclear PDF set. The projections in this Figure use 0.5 pb$^{-1}$ of 8.8 TeV $p$+Pb data and assume negligible statistical uncertainties on the $pp$ reference.

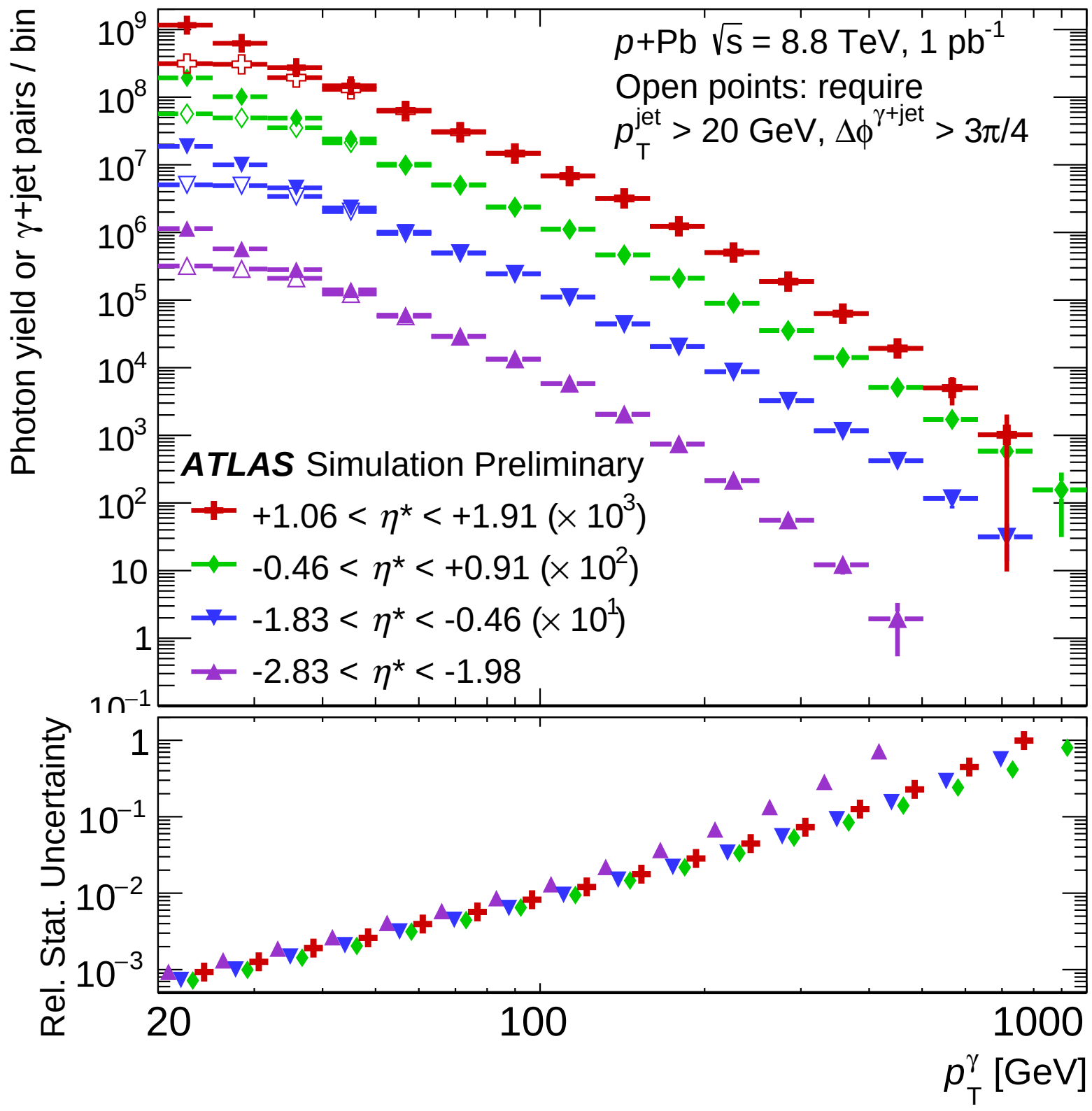

Figure 19: Projected yield of photon +jet events in 1 $\text{pb}^{-1}$ of 8.8 TeV $p$+Pb data, shown as a function of photon $p_{\mathrm{T}}$ with different selections on the photon pseudorapidity in the center-of-mass frame shown in different colors. Closed markers denote the yield of events with a $p_{\mathrm{T}}$ photon, while open markers denote the subset of those which also include a balancing jet above threshold in the ATLAS acceptance. The bottom panel below shows the projected statistical uncertainties.

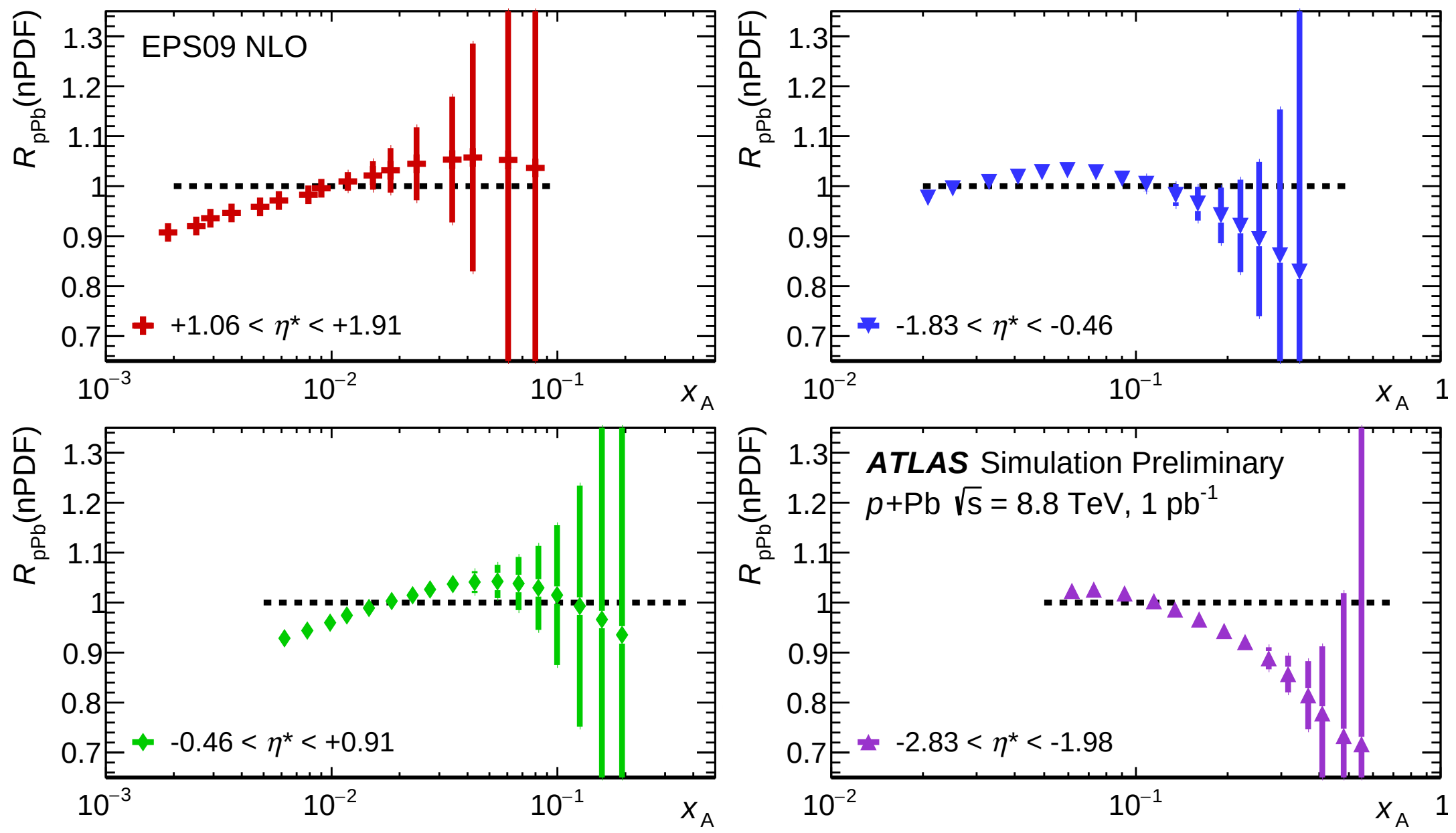

Figure 20: Projected statistical uncertainties for an $R_{p\mathrm{Pb}}$ measurement, plotted as a function of the average nuclear-$x$, taken from Pythia, for each photon $p_{\mathrm{T}}$ bin in Fig. 19.The panels show projections for photon+jet events in different selections on the photon pseudorapidity measured in the centre-of-mass frame. The central values are set to be equal to the predictions of the EPS09 nuclear PDF set. The projections in this Figure use 1 $\mathrm{pb}^{-1}$ of 8.8 TeV $p$+Pb data and assume negligible statistical uncertainties on the $pp$ reference.

# B $W$ Boson Projections with $\sqrt{s_{\mathrm{NN}}}$ = 8.16 TeV

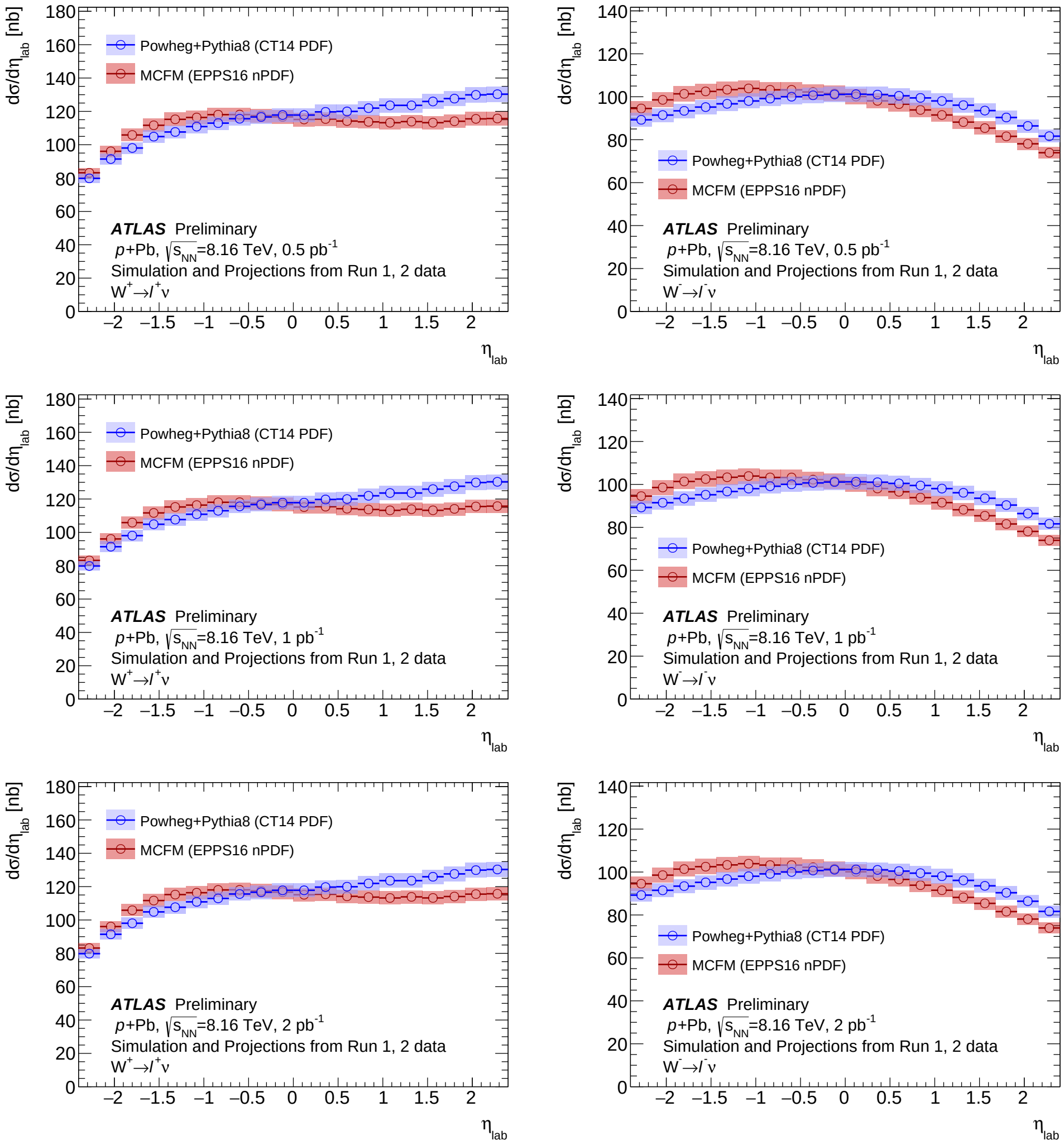

Figure 21: Fiducial cross-sections for $W^+$ (left) and $W^-$ (right) boson production in $p$+Pb collisions at $\sqrt{s_{\mathrm{NN}}}$ = 8.16 TeV differential in the charged lepton pseudorapidity measured in the laboratory frame $\eta_{\mathrm{lab}}$. The cross-sections are projected with nuclear effects described by the EPPS16 nPDF set and without any nuclear effects using the CT14 PDF set for an integrated luminosity of 0.5 pb$^{-1}$(top), 1 pb$^{-1}$(middle), and 2 pb$^{-1}$(bottom). The boxes represent the projected total uncertainties (quadratic sum of statistical and systematic uncertainties), while vertical bars represent statistical uncertainties (smaller than the marker size).

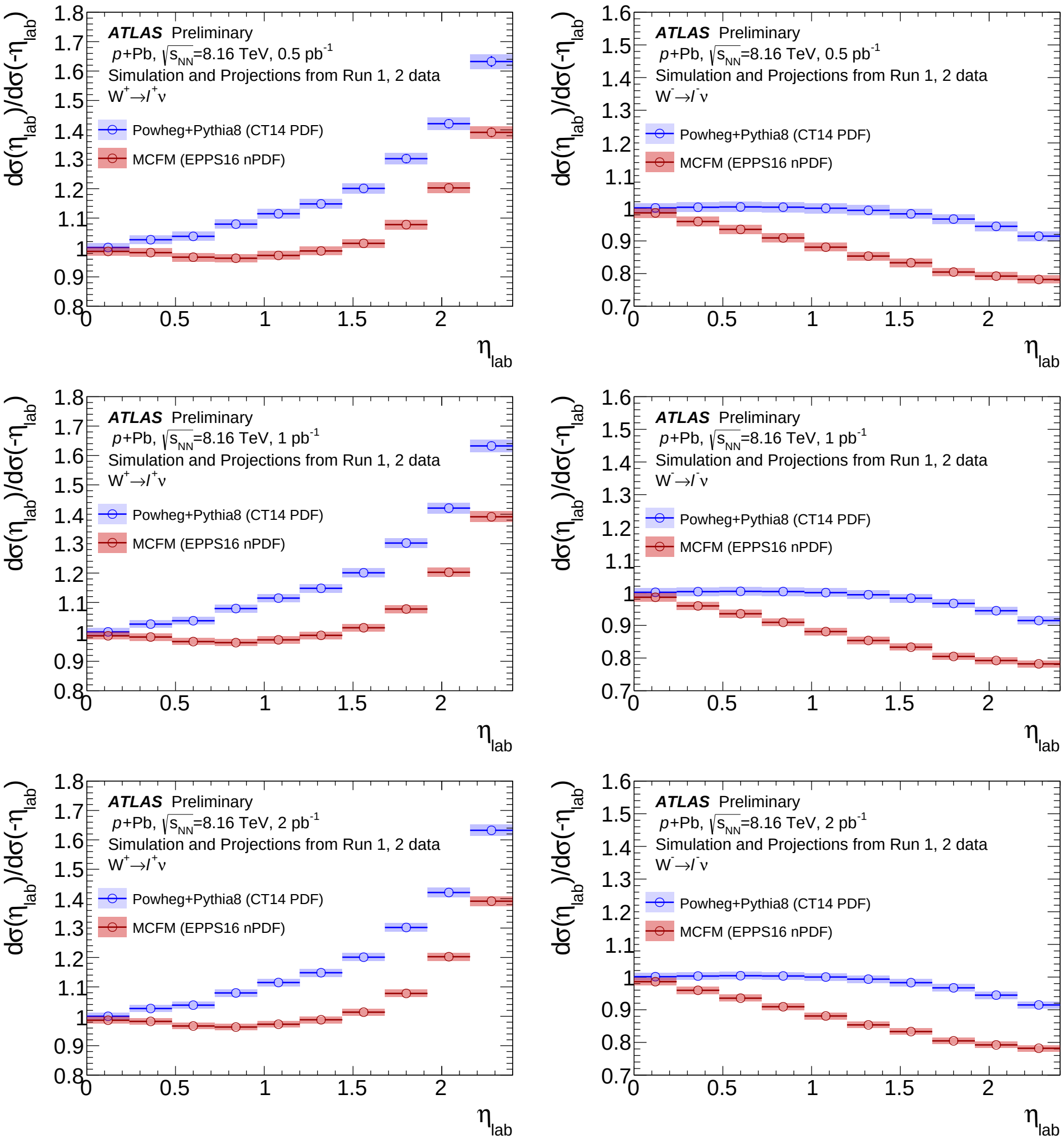

Figure 22: Forward-backward ratios of the differential fiducial cross-sections for $W^+$ (left) and $W^-$ (right) boson production in $p$+Pb collisions at $\sqrt{s_{\mathrm{NN}}} = 8.16$ TeV. The ratios are projected with nuclear effects described by the EPPS16 nPDF set and without any nuclear effects using the CT14 PDF set for an integrated luminosity of 0.5 pb$^{-1}$(top), 1 pb$^{-1}$(middle), and 2 pb$^{-1}$(bottom). The boxes represent the projected total uncertainties (quadratic sum of statistical and systematic uncertainties), while vertical bars represent statistical uncertainties.

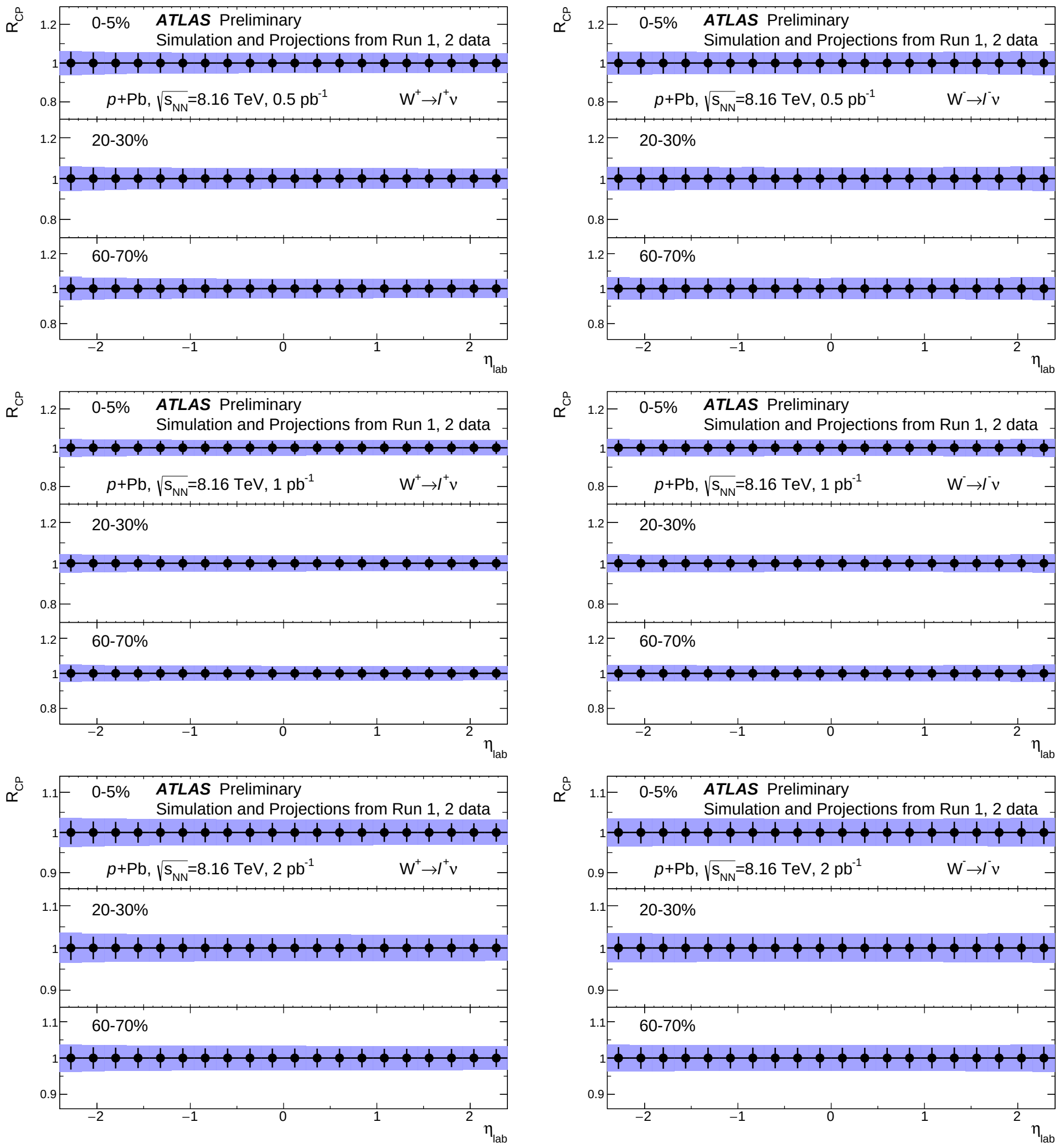

Figure 23: Uncertainties on measurments of the $W^+$ (left) and $W^-$ (right) boson yield ratio of central to peripheral collisions $R_{\mathrm{CP}}$. The uncertainties are projected for $p$+Pb collisions at $\sqrt{s_{\mathrm{NN}}} = 8.16$ TeV with an integrated luminosity of 0.5 pb$^{-1}$(top), 1 pb$^{-1}$(middle), and 2 pb$^{-1}$(bottom). The boxes represent the projected total uncertainties (quadratic sum of statistical and systematic uncertainties), while vertical bars represent statistical uncertainties.

# CMS Physics Analysis Summary

Contact: cms-phys-conveners-ftr@cern.ch 

# Performance of jet quenching measurements in pp and PbPb collisions with CMS at the HL-LHC

The CMS Collaboration

## Abstract

The projected performance of different jet measurements to be carried out with the large data samples expected in PbPb and pp collisions during the high-luminosity phase of the LHC (HL-LHC) is presented. The analyses are based on Monte Carlo simulations and extrapolations from data collected in pp and PbPb collisions between 2010 and 2016. The extrapolated performance for PbPb collisions, corresponding to an integrated luminosity of 10 $\mathrm{nb}^{-1}$ at $\sqrt{s_{\mathrm{NN}}} = 5.02$ TeV and an upgraded CMS detector, shows significant reductions in the uncertainties of the measurements of photon-tagged jet shapes and $\mathrm{D}^0$-jet correlations. The large proton-proton data sample at $\sqrt{s} = 14$ TeV, to be collected under low-pileup conditions in Run 3 and beyond, will also provide new opportunities for the study of jet quenching phenomena in small systems, as shown through $\gamma$, Z-jet transverse momentum imbalance distributions in high-multiplicity pp events.

# 1 Introduction

Studies of jet production and modifications in high-energy heavy ion compared to proton-proton (pp) collisions provide precise information on the properties of the Quark Gluon Plasma (QGP) [1]. This document reports several performance studies that illustrate the improved measurements of jet-quenching observables to be carried out with the upgraded CMS detector in heavy ion and pp collisions during the High Luminosity Large Hadron Collider (HL-LHC) phase [2]. For the HL-LHC running period, the LHC experiments expect an integrated luminosity of 10–13 $\mathrm{nb}^{-1}$ of PbPb data at 5.5 TeV. Based on this data sample size, Monte Carlo simulations, and extrapolations from measurements carried out in the last years, performance studies of different jet quenching observables are presented. These studies show that the large data samples, and the improved jet reconstruction thanks to detector upgrades, will result in significantly reduced statistical and systematic uncertainties for key observables such as photon-tagged jet shapes and $\mathrm{D}^0$-meson–jet correlations in PbPb collisions. In addition, 2000 $\mathrm{nb}^{-1}$ of pPb data at 8.8 TeV and low-pileup runs of pp collisions at 5.5 TeV are requested in order to study the evolution of jet-quenching phenomena with system size. Sensitivity studies using boson-jet transverse momentum imbalance distributions in high-multiplicity pp events, where jet quenching has not (yet) been observed, are presented. These studies show that such larger data samples will provide new opportunities for searching the onset of jet quenching in small final-state quark-gluon systems.

# 2 CMS detector upgrades

The CMS detector [2] will be substantially upgraded in order to fully exploit the physics potential offered by the increase in luminosity, and to cope with the demanding operational conditions, at the HL-LHC [3–7]. The upgrade of the first level hardware trigger (L1) will enable increases of the L1 rate and latency to about 750 kHz and 12.5 $\mu$s, respectively, whereas the high-level software trigger (HLT) is expected to reduce the final logging rate by about a factor of 100 to 7.5 kHz. The entire pixel and strip tracker detectors will be replaced to increase their granularity, reduce the material budget in the tracking volume, improve the radiation hardness, and extend the geometrical coverage thereby providing efficient tracking up to pseudorapidities of about $|\eta| = 4$. The muon system will be enhanced by upgrading the electronics of the existing cathode strip chambers (CSC), resistive plate chambers (RPC) and drift tubes (DT). New muon detectors based on improved RPC and gas electron multiplier (GEM) technologies will be installed to add redundancy, increase the geometrical coverage up to about $|\eta| = 2.8$, and improve the trigger and reconstruction performance in the forward region. The barrel electromagnetic calorimeter (ECAL) will feature upgraded front-end electronics that will be able to exploit the information from single crystals at the L1 trigger level, to accommodate trigger latency and bandwidth requirements, and to provide 160 MHz sampling allowing high precision timing capability for photons. The hadronic calorimeter (HCAL), consisting in the barrel region of brass absorber plates and plastic scintillator layers, will be read out by silicon photomultipliers (SiPMs). The endcap electromagnetic and hadron calorimeters will be replaced with a new combined sampling calorimeter (HGCal) that will provide highly-segmented spatial information in both transverse and longitudinal directions, as well as high-precision timing information. Finally, the addition of a new timing detector for minimum ionizing particles (MTD) in both barrel and endcap regions is envisaged to provide the capability for 4-dimensional reconstruction of interaction vertices that will significantly offset the CMS performance degradation due to high PU rates.

A detailed overview of the CMS detector upgrade program is presented in Refs. [3–7], while the expected performance of the reconstruction algorithms and pileup mitigation in pp collisions is summarized in Ref. [8].

# 3 Impact of CMS detector upgrades on heavy ion studies

The upgraded inner tracker, available after the LHC Long Shutdown 3, will provide a large acceptance for charged particles, up to $|\eta| < 4$ [3]. The improved L1 trigger and data acquisition rate (up to 60 GB/s) will enable to define more sophisticated triggers and to record a larger number of minimum-bias triggered events. In addition, the proposed MTD [9], located between the tracker and the electromagnetic calorimeters at about at $\sim$1.16 meters from the beam pipe, will provide a time resolution down to around 30 ps. Such timing capabilities, combined with information from other detectors, can be employed for proton, pion, and kaon separation in the $p_{\mathrm{T}} \approx 0.7$–2 GeV/$c$ range at midrapidity ($|\eta| < 1.5$). The expected detector upgrades will augment the heavy ion reconstruction performance to about the level of the current pp reconstruction algorithms. The biggest improvement will be provided by the four-layer pixel system. Compared to the previous (Run 1 and 2) three-layer pixel system, the new pixel tracking will significantly enhance the seed quality of the pattern recognition algorithm, resulting in an improvement of the track reconstruction efficiency and a reduction of the rate of falsely reconstructed trajectories. The significantly lowered material budget will also improve the secondary vertex resolution, thereby making CMS ideally suited for studies of heavy-flavor mesons, heavy-flavor-tagged jets, and their correlations. Such detector upgrades and the increased trigger performance will significantly improve the measurements of heavy-flavored mesons for all colliding systems, flow fluctuations in small systems, and jet spectra and jet substructure observables in heavy ion collisions.

For the studies reported here, the impact of the upgraded CMS detector capabilities is not explicitly estimated except through reasonable assumptions on the reduction of the associated systematic uncertainties thanks to the expected improved jet reconstruction performance. The enlarged tracking acceptance is not included and, therefore, the extrapolations presented can be considered as a conservative estimate of the final improvements to be expected.

# 4 Physics Performance

In this section, the physics performance based on the existing data and on Monte Carlo (MC) simulations from Run 2, is presented. The HL-LHC nucleon-nucleon center-of-mass (c.m.) energy will be $\sqrt{s_{\mathrm{NN}}} = 5.5\,\mathrm{TeV}$ but, however, to facilitate the comparison with the collected data at $\sqrt{s_{\mathrm{NN}}} = 5.02\,\mathrm{TeV}$, the latter c.m. energy is used for our HL-LHC PbPb estimates.

## 4.1 Jet substructure using photon-tagged jets

Significant progress has been made in jet quenching studies at the LHC via many single jets and dijet observables, including jet shapes [10], jet fragmentation functions [11, 12], missing transverse momentum in dijet systems [13, 14], and jet-track correlations [15, 16] among others. However, understanding how properties of the measured jets relate to those of their parent partons is still a key challenge for these measurements: When selecting events based on jet kinematics, the amount of energy lost by the parton into the medium, before fragmenting into final-state particles, cannot be unambiguously determined. One way to overcome this difficulty is to study jets tagged against recoiling isolated photons, where the photon energy can be used as reference for that of the parton before it suffered any energy loss. In addition, since isolated photons at the LHC mostly recoil against a quark, photon-tagging leads to enriched samples to study *quark* (rather than gluon) energy loss. Recently, the measurements of photon-tagged jet shapes [17] and jet fragmentation functions [18] have become possible for the first time. However, the sensitivity of the Run 2 data to the underlying jet-quenching phenomenon is limited by the statistical accuracy of the measurements.

The current (left) and projected (right) measurements of the modification of the radial distribution of particles in jets measured in PbPb and pp collisions is shown in Figure 1 for central PbPb collisions (0–10% event centrality), where the effects due to the presence of a strongly interacting medium are the largest. The central values of the extrapolated ratio are obtained by smoothing out the results from [17] with a third-order polynomial. The systematic uncertainties shown are those from the current measurements divided by a factor of two to take into account expected improvements on the jet energy scale and jet energy resolution uncertainties thanks to the detector upgrades. The results show that the photon-tagged shape will be measured with high precision, and thereby provide valuable insights on the medium-modified transverse structure of quark-initiated jets.

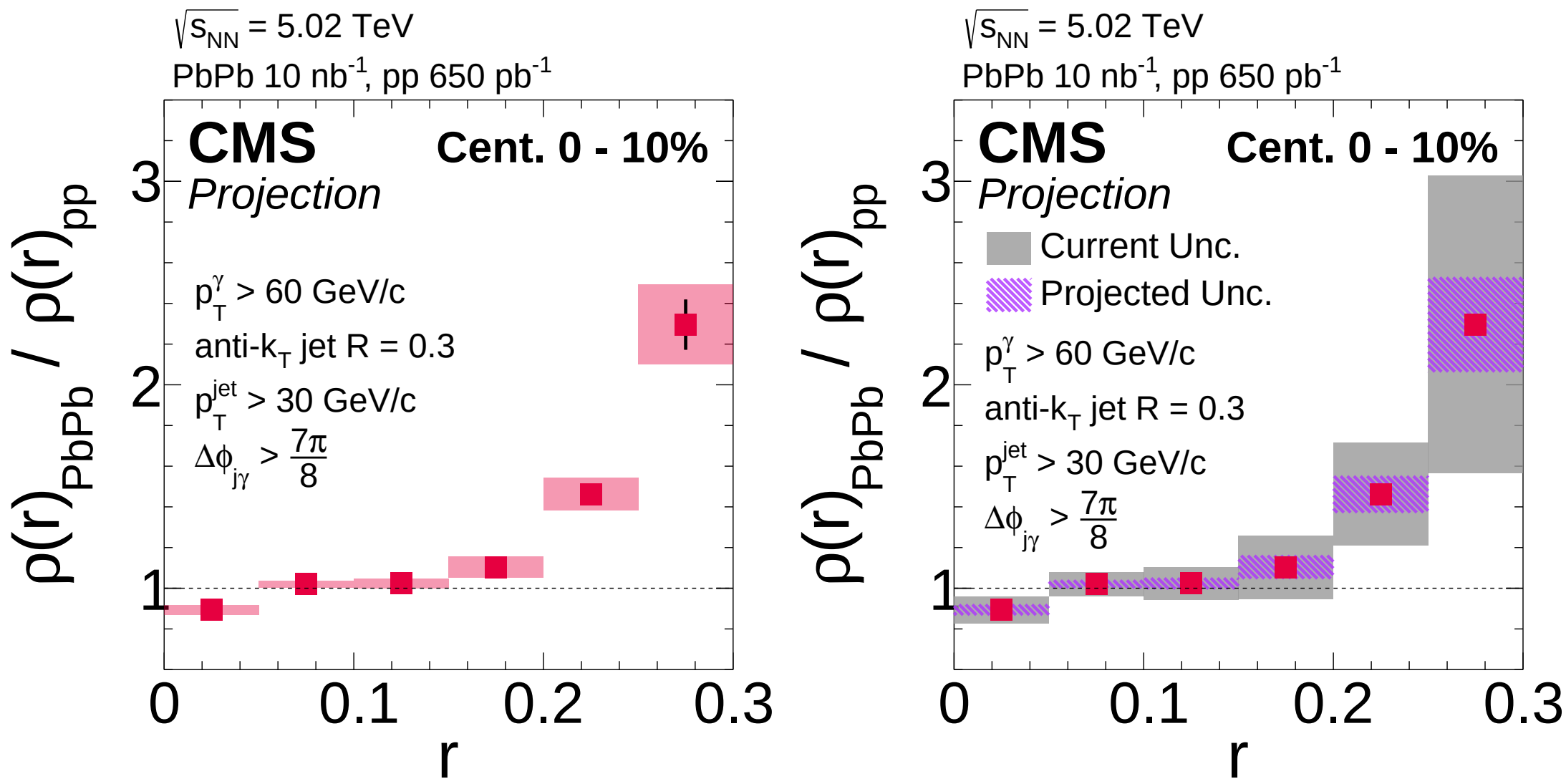

Figure 1: Ratio of the density of particles produced at a radius $r$ in photon-tagged jets in PbPb (0–10% most central) and pp collisions. The left panel shows the expected distribution at the HL-LHC with the error bars (bands) indicating statistical (systematic) uncertainties. The right panel shows a comparison of the current total uncertainty in the 2015 data (grey bands) [17] to the projected HL-LHC total uncertainty (violet bands).

## 4.2 Radial distribution of $D^0$ mesons in jets

The interest of studying charm meson production in jets is manifold. First, it can help to understand the origin of the enhanced production of low-$p_T$ hadrons at large angles with respect to the jet axis in PbPb compared to pp collisions (Fig. 1). Interpretations of this observation include medium-induced gluon radiation, modification of jet splitting functions, and medium response to the hard scattered partons. Since the heavy quark masses are larger than the typical scale given by the temperature of the QGP created at LHC energies, one can gain insights into the underlying parton dynamics by studying heavy flavor mesons associated with jets. Second, precise measurements of heavy flavor meson production in jets provide new information on charm and bottom quark radiation and fragmentation in both pp and PbPb collisions. Last but not least, angular correlation measurements between heavy flavor meson and jets can be used to constrain parton energy loss mechanisms and to measure heavy quark diffusion inside the medium [19–21] that are complementary to the measurements of inclusive heavy flavor meson spectra [22–26], heavy flavor meson azimuthal anisotropy [26–30], and heavy flavor tagged jets [31, 32].

CMS performed the first measurement of the radial distributions of $D^0$ mesons in jets using pp and PbPb data collected in 2015. Jets with $p_T > 60\,\text{GeV}/c$ were reconstructed with the anti-$k_T$ algorithm [33–35] with a distance parameter of 0.3. The radial distributions of $D^0$ mesons with transverse momenta $p_T > 4$ GeV/$c$ were measured with respect to the jet axis. The measured PbPb compared to the pp data, indicate that $D^0$ mesons are produced at relatively larger distances in the former system. Figure 2 shows the expected $D^0$ radial distributions at the HL-LHC for low-$p_T^D$ (top) and high-$p_T^D$ (bottom) $D^0$ meson $p_T$ intervals. The central positions of the projected data points are taken to match the current measurement, with the correspondingly reduced uncertainties shown in the left panel compared to those of the 2015 data. The uncertainties associated with yield extraction and jet energy scale, which were currently limited by the available statistics in the 2015 data analysis [36], are also scaled down to account for the larger dataset at the HL-LHC. Figure 3 shows the corresponding PbPb over pp ratios. The much larger HL-LHC PbPb and pp data samples will dramatically reduce the total uncertainties of the measurements.

### 4.3 Boson-jet transverse momentum balance in high-multiplicity pp collisions

The azimuth and transverse momentum correlations between bosons and jets, measured in $\gamma, Z + \text{jets}$ events, are valuable observables to study parton energy loss in the QGP. Imbalanced correlations are a signature of jet quenching. At the HL-LHC, these measurements can be performed for the first time in high-multiplicity pp collisions with large data samples under low-pileup conditions. Combined with dijet asymmetry studies [37] in pPb collisions and similar boson+jet [38, 39] measurements in PbPb collisions, they can be used to ascertain the system-size dependence of jet quenching phenomena.

The projected numbers of $\gamma, Z + \text{jets}$ events expected in pp collisions at $\sqrt{s_{NN}} = 14\,\text{TeV}$, in low-pileup samples at the LHC Run 3 and beyond, are obtained from theoretical 14-to-5.02 TeV cross section ratios that are then used to scale up the measured yields in the 5.02 TeV pp data. The ratio of Z+jet events produced at 5.02 and 14 TeV is found to be 4.5 from a next-to-leading-order FEWZ v3.1 [40] calculation for pp collisions. Similarly, JETPHOX [41] is used to extract the scaling factor for $\gamma$+jet events. In order to estimate the number of events in different pp multiplicity classes, additional scale factors based on multiplicity fluctuation studies from ALICE, CMS, and ATLAS are used. These scale factors have been agreed among the four major experiments (CMS, ATLAS, ALICE and LHCb) in order to use the same baseline for benchmarking the expected performance.

Figures 4 and 5 show, respectively, the expected distributions of the transverse momentum imbalance of $\gamma$+jet ($x_{j\gamma} = p_T^{\text{jet}}/p_T^{\gamma}$) and Z+jet ($x_{jZ} = p_T^{\text{jet}}/p_T^{Z}$) events in different event multiplicity bins ($\langle N_{ch} \rangle$) in pp collisions. The $\langle N_{ch} \rangle$ bins, defined as the mean number of reconstructed charged particles with $|\eta| < 2.4$ and $p_T > 1\,\text{GeV}/c$, passing offline selection criteria described in [42], are chosen so as to have an overlap with typical multiplicities found in pPb and peripheral PbPb collisions. The central values of the $x_{j\gamma}$ and $x_{jZ}$ distributions are derived from smoothed 5 TeV MC simulations of previous CMS minimum-bias pp results [38, 43]. For the extrapolation to the various pp multiplicity bins, predictions based on PYTHIA 8.212 [44] with tune CUETP8M1 [45] were used as the central value of the distributions. The systematic uncertainties are conservatively assumed to be the same as those from the 2015 pp result, although improvements on the jet energy scale and jet energy resolution determinations in the HL-LHC period will further reduce them. The 200 pb$^{-1}$ pp data will facilitate the first measurement of high-multiplicity Z+jet momentum imbalance, and the high-precision measurement of $x_{j\gamma}$ distributions. The collected number of $\gamma$+jet events will also be large enough to study the $x_{j\gamma}$ distribution as a function of the reaction plane angle for the first time. For a precise measurement of the $x_{jZ}$ distribution, a much larger sample of low-pileup pp data (2000 pb$^{-1}$) will be needed as shown in Figure 5.

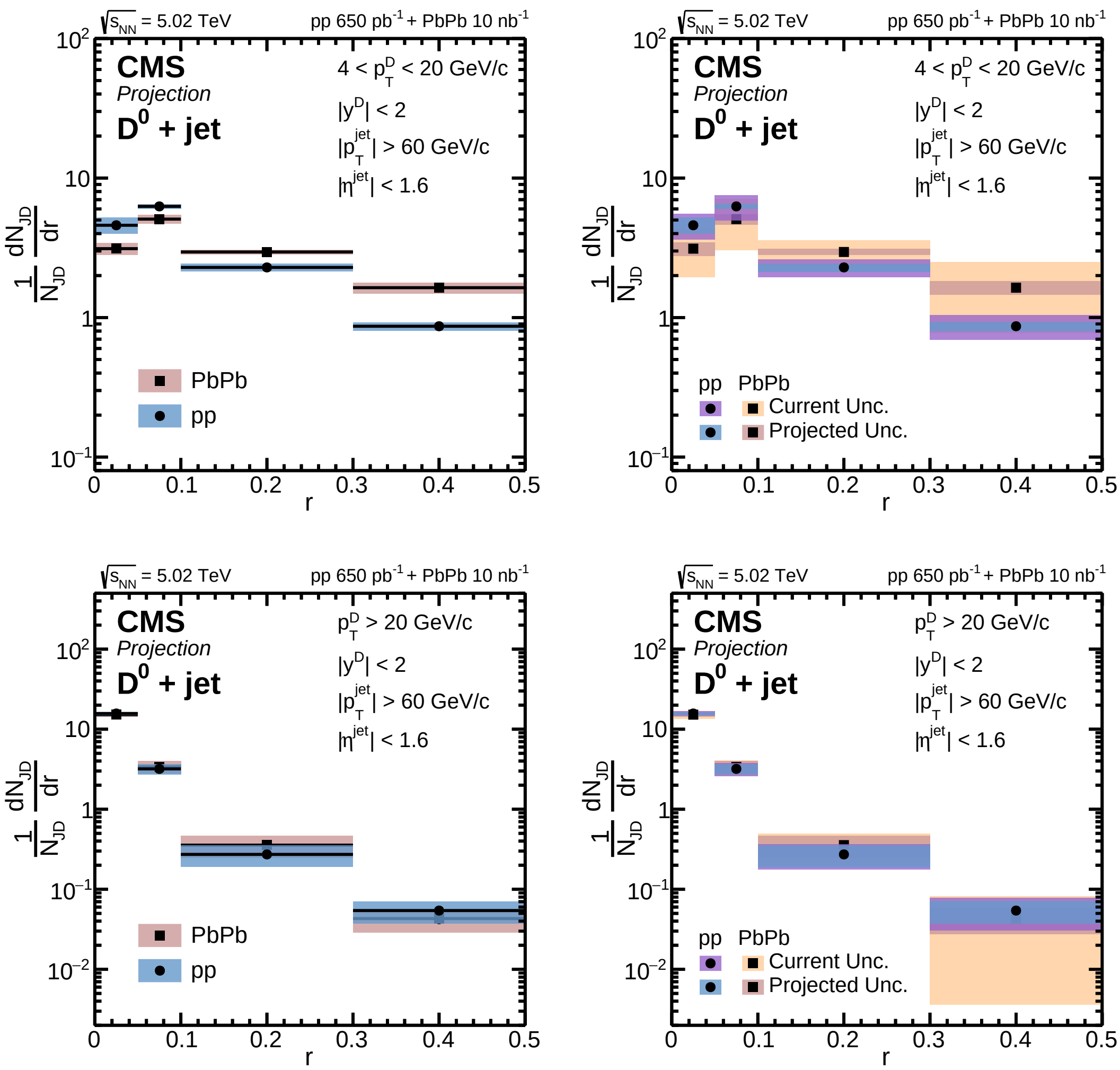

Figure 2: Left: Projected distributions of $\mathrm{D}^0$ mesons with $p_{\mathrm{T}}^{\mathrm{D}} = 4$–$20\,\mathrm{GeV}/c$ (top) and $p_{\mathrm{T}}^{\mathrm{D}} > 20\,\mathrm{GeV}/c$ (bottom), as a function of their distance $r$ from the jet axis (for jets of $p_{\mathrm{T}}^{\mathrm{jet}} > 60\ \mathrm{GeV}/c$ and $|\eta^{\mathrm{jet}}| < 1.6$) in pp and PbPb collisions at $\sqrt{s_{NN}} = 5.02$ TeV at the HL-LHC. The error bands (error bars, barely seen) indicate systematic (statistical) uncertainties. The horizontal bars indicate the size of the bin width. Right: Same distributions as on the left, comparing the current total uncertainty (wide bands) [36] and the projected total uncertainty (thin bands).

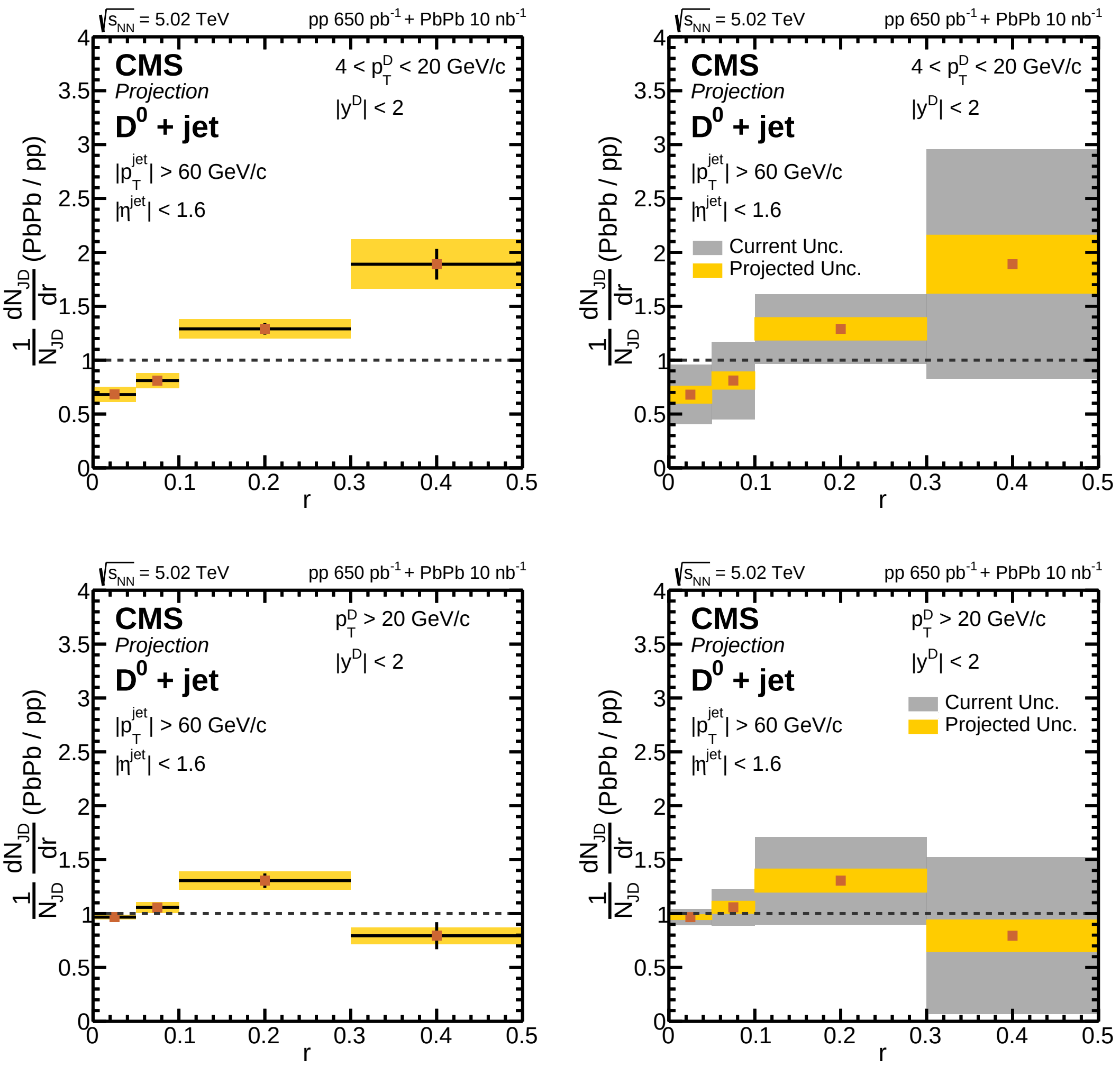

Figure 3: Left: Projected ratios at the HL-LHC of the radial distributions of $D^0$ mesons produced in PbPb over pp collisions at $\sqrt{s_{NN}} = 5.02\,\mathrm{TeV}$, for $p_{\mathrm{T}}^{\mathrm{D}} = 4\text{–}20\ \mathrm{GeV}/c$ (top) and $p_{\mathrm{T}}^{\mathrm{D}} > 20\,\mathrm{GeV}/c$ (bottom) The error bars (bands) indicate statistical (systematic) uncertainties. The horizontal bars indicate the size of the bin width. Rigth: Same distributions as on the left, comparing the current total uncertainty (grey bands) [36] and the projected total uncertainty at the HL-LHC (yellow bands).

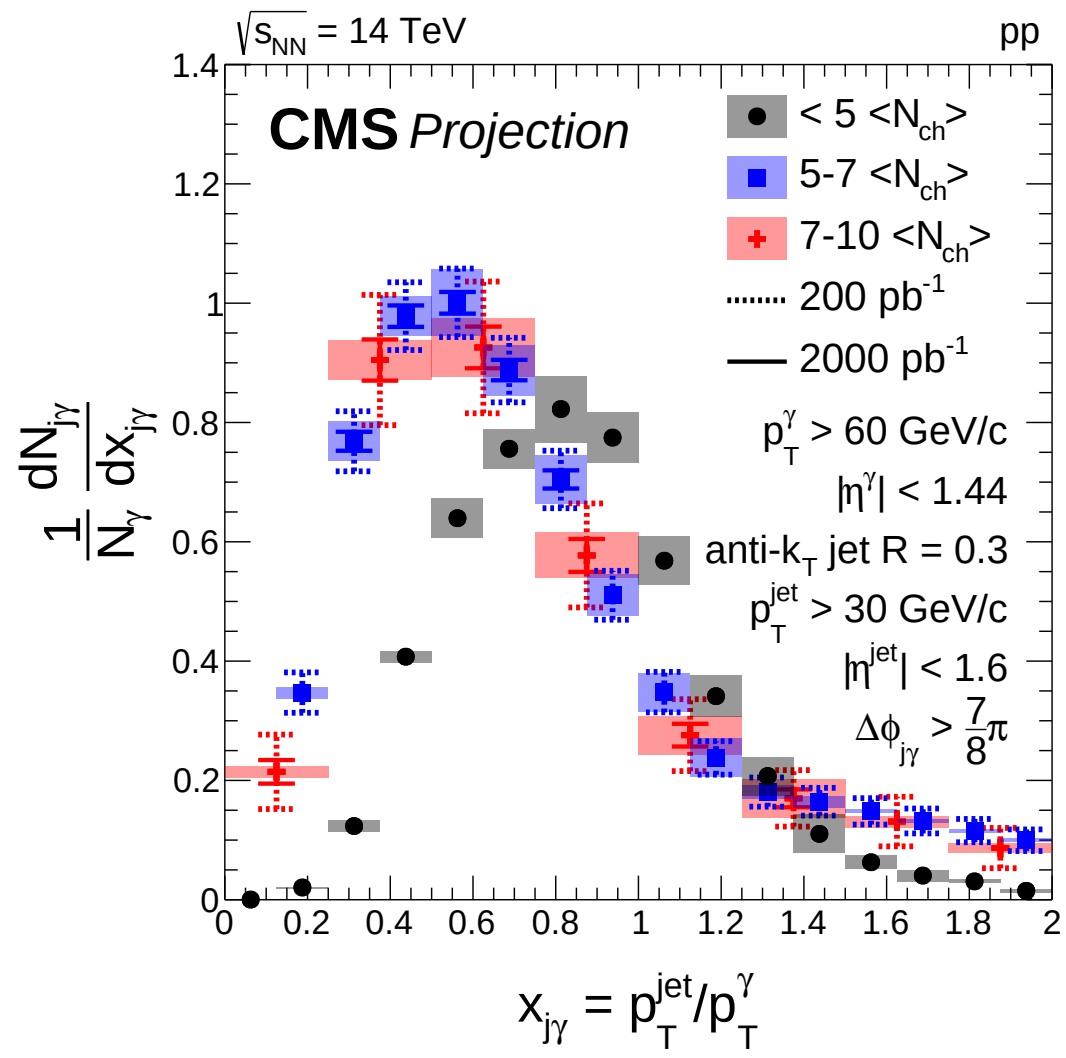

Figure 4: Distributions of transverse momentum imbalance, $x_{j\gamma}$, in three different centrality classes of pp isolated-photon+jet events ($p_T^\gamma > 60$ GeV/$c$ and $|\eta_\gamma| < 1.44$, $p_T^{jet} > 30$ GeV/$c$ and $|\eta_{jet}| < 1.6$) with 200 (dashed line) and 2000 (solid line) pb$^{-1}$ integrated luminosities expected to be taken under low-pileup conditions at HL-LHC. The central values of the distributions are derived from existing measurements [43] combined with PYTHIA 8 MC simulations. The error bars (bands) indicate statistical (systematic) uncertainties.

## 5 Summary

The projected performance of different jet measurements expected to be carried out with large data samples of PbPb and low-pileup pp collisions at HL-LHC are presented. The HL-LHC performance, extrapolated from existing PbPb data and Monte Carlo simulations, corresponding to a total integrated luminosity of 10 nb$^{-1}$ at $\sqrt{s_{NN}} = 5.02$ TeV, shows significantly improved precision in the measurements of photon-tagged jet shapes and D$^0$-jet correlations. These measurements will provide new constraints on models of jet quenching in the QGP, heavy quark production and diffusion inside the medium, and improved understanding of the medium response and medium excitation to hard probes. The large pp data samples at $\sqrt{s} = 14$ TeV with low pileup, expected to be collected in Run 3 and beyond, will also provide precise measurements of boson-jet transverse momentum imbalance distributions in (high-multiplicity) small systems, and thereby study the system-size dependence of jet quenching phenomena.

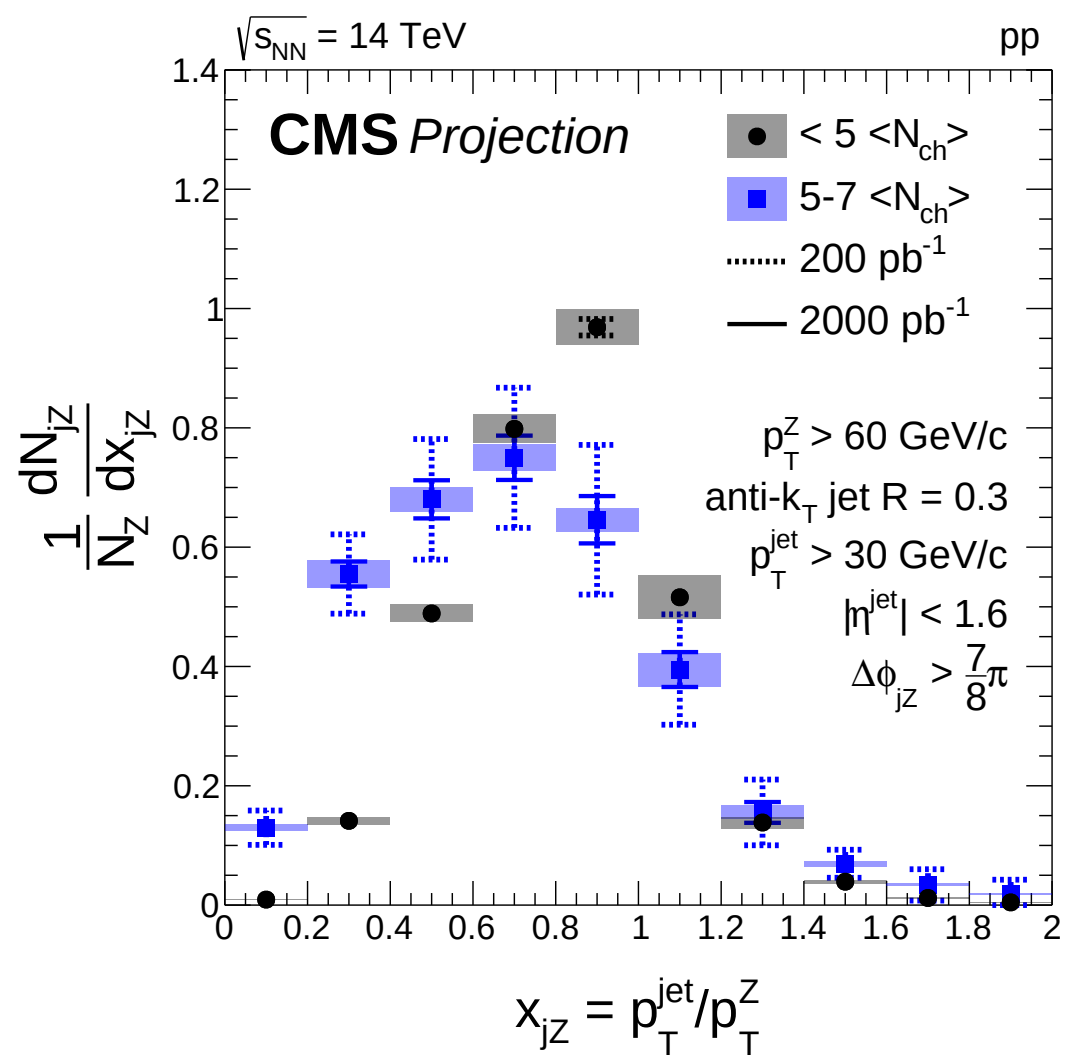

Figure 5: Distributions of transverse momentum imbalance, $x_{jZ}$, in two different centrality classes of pp Z+jet events ($p_{\mathrm{T}}^{\mathrm{Z}} > 60\,\mathrm{GeV}/c$, $p_{\mathrm{T}}^{\mathrm{jet}} > 30\,\mathrm{GeV}/c$ and $|\eta_{\mathrm{jet}}| < 1.6$) with 200 (dashed line) and 2000 (solid line) $\mathrm{pb}^{-1}$ integrated luminosities expected to be taken under low-pileup conditions at HL-LHC. The central values of the distributions are derived from existing measurements [38] combined with PYTHIA 8 MC. The error bars (bands) indicate statistical (systematic) uncertainties.

## 10

# Projections for ATLAS Measurements of Jet Modifications in Pb+Pb Collisions in LHC Runs 3 and 4

The ATLAS Collaboration

The primary goal of the heavy ion program at ATLAS is to study the properties of the deconfined, strongly interacting matter, the quark-gluon plasma (QGP). QCD jets are one of the primary tools used to study this matter. Jet properties are observed to be modified in Pb+Pb collisions compared to *pp* collisions and the modifications are sensitive to the interactions between the QGP and the jet. Jet measurements in heavy ion collisions at ATLAS aim to provide information about the nature of these interactions. This note describes expected ATLAS measurements of jet modification in heavy-ion collisions in the upcoming LHC Runs 3 and 4. In particular, projections are made for inclusive jet yield measurements, jet fragmentation function measurements, and photon-tagged jet measurements.

# 1 Introduction

The primary goal of the heavy ion program at the LHC is to study the properties of deconfined strongly-interacting matter, often referred to as a quark-gluon plasma (QGP), created in ultra-relativistic nuclear collisions. Jets are among the most powerful probes of this matter because they are sensitive to the short distance interactions between hard scattered-partons and the constituents of the QGP [1].

Results from jet measurements in Pb+Pb collisions at the LHC have shown that the transverse momentum ($p_{\mathrm{T}}$) balance of back-to-back produced jets and photon-jet pairs is modified [2, 3]. The number of measured jets is reduced by approximately a factor of two as compared to the jet yields measured in *pp* collisions scaled by the geometric nuclear overlap [4, 5]. Furthermore, the fragmentation of both inclusive jets and jets opposite a photon [6] are modified relative to *pp* collisions.

Current measurements are based on 0.49 $\mathrm{nb}^{-1}$of Pb+Pb collisions and 25 $\mathrm{pb}^{-1}$ of *pp* collisions collected in 2015. Jets are rarely produced and additional luminosity is needed to extend the kinematic range and improve the precision of the measurements. In this note, the expected performance for some jet related observables in the LHC Run 3 and 4 are presented. The assumed luminosity for Runs 3 and 4 is 10 $\mathrm{nb}^{-1}$, approximately a factor of 20 more than is available in current measurements. The projections presented here represent a selected list of measurements of interest using the Run 3 and 4 data. The selected suite of measurements shows clear examples of physics that could only be addressable with substantially increased luminosity.

# 2 ATLAS Detector

The ATLAS detector is described in detail in [7]. Significant upgrades to the ATLAS detector are expected to be installed for Run 4, in particular the Inner Tracker (ITk) [8]. The performance of the ITk in Pb+Pb collisions is discussed in [9]. In this note the projections using the ITk are not discussed.

# 3 Projections

The projections presented in this note account only for statistical uncertainties. For systematic uncertainties, some improvement is expected for the Run 3 and 4 data based on advancement in the measurement technique and improved understanding of the detector. These are discussed in section 4, however, at present, it is not possible to make a quantitative projection taking into account these expectations.

### 3.1 Inclusive Jet Measurements

The quantification of the amount of energy lost by the parton shower due to interactions with the deconfined QCD medium can be provided in terms of the nuclear modification factor defined as:

$$R_{\mathrm{AA}} = \frac{\left.\frac{1}{N_{\mathrm{evt}}^{\mathrm{tot}}} \frac{\mathrm{d}^2 N_{\mathrm{jet}}}{\mathrm{d}p_{\mathrm{T}}\mathrm{d}y}\right|_{\mathrm{cent}}}{\left.T_{\mathrm{AA}} \frac{\mathrm{d}^2 \sigma_{\mathrm{jet}}}{\mathrm{d}p_{\mathrm{T}}\mathrm{d}y}\right|_{pp}}, \tag{1}$$

where $N_{\mathrm{jet}}$ and $\sigma_{\mathrm{jet}}$ are the jet yield in Pb+Pb collisions and the jet cross section in $pp$ collisions, respectively, both measured as a function of transverse momentum, $p_{\mathrm{T}}$, and rapidity, $y$. $N_{\mathrm{evt}}^{\mathrm{tot}}$ and $T_{\mathrm{AA}}$ are the total number of Pb+Pb collisions within a chosen centrality interval and the nuclear thickness function, respectively. Measurement of the jet $R_{\mathrm{AA}}$ is a baseline measurement for comparisons with theoretical predictions of jet quenching. A value of $R_{\mathrm{AA}} \approx 0.5 - 0.6$ in central Pb+Pb collisions was reported in measurements at $\sqrt{s_{NN}} = 2.76$ TeV and $\sqrt{s_{NN}} = 5.02$ TeV [4, 10, 11]. This implies a suppression of jet yields by roughly a factor of two in central collisions to the expectation from scaled $pp$ collisions at the same center-of-mass energy. Two intriguing features were revealed by these studies: $R_{\mathrm{AA}}$ increases only very slowly with increasing jet $p_{\mathrm{T}}$ and seems to saturate at the value of 0.6 in the region of $p_{\mathrm{T}}$ between 400 GeV and 1 TeV; $R_{\mathrm{AA}}$ exhibits no dependence on the rapidity of the jet (except for the most forward region where the $R_{\mathrm{AA}}$ shows signs of decreasing with rapidity). Both of these features seem counter-intuitive since one would expect that jet quenching effects should play smaller role at high $p_{\mathrm{T}}$ compared to low $p_{\mathrm{T}}$ and since there is *a priori* no reason to expect rapidity independence given the differences in the spectral shape and differences in the flavor composition of jets at different rapidities. Providing a precise quantification of the $R_{\mathrm{AA}}$ at a TeV scale, along with detailing the evolution of $R_{\mathrm{AA}}$ with rapidity, are two primary goals for LHC Runs 3 and 4.

The data points from the measured jet $R_{\mathrm{AA}}$ at 5.02 TeV [11] with smoothed statistical variations are plotted in Figure 1 together with statistical uncertainties projected for 10 nb$^{-1}$. The left panel of Figure 1 shows the estimated gain in the statistical precision for jet $R_{\mathrm{AA}}$ measured in the rapidity interval of $|y| < 2.1$ reaching a factor of approximately 5. The estimate was also made using two different scenarios for the total integrated luminosity of $pp$ data, 600 pb$^{-1}$and 1.2 fb$^{-1}$. A negligible difference in the statistical precision was found between those two scenarios which demonstrates no need for $pp$ integrated luminosity beyond 600 pb$^{-1}$. Expected precision on the $R_{\mathrm{AA}}$ measurement is compared with several recent model predictions: Linear Boltzmann Transport model (LBT) [12], three calculations using the Soft Collinear Effective Theory (SCET) [13–16], and Effective Quenching model (EQ) [17]. The phenomenological effective quenching model [17] includes energy loss effects through downward shifts in the jet $p_{\mathrm{T}}$ spectrum that depends on the flavor of the parton that initiating a jet. The figure demonstrates that the higher precision data will provide means to constrain or falsify theoretical model predictions. The right panel of Figure 1 shows the estimated gain in the statistical precision in the forward rapidity region. The statistical precision should be sufficient to quantitatively assess the rapidity dependence of the $R_{\mathrm{AA}}$. Both of these predictions indicate that Runs 3 and 4 should bring a definitive understanding of the intriguing features of jet $R_{\mathrm{AA}}$ as seen in the current data.

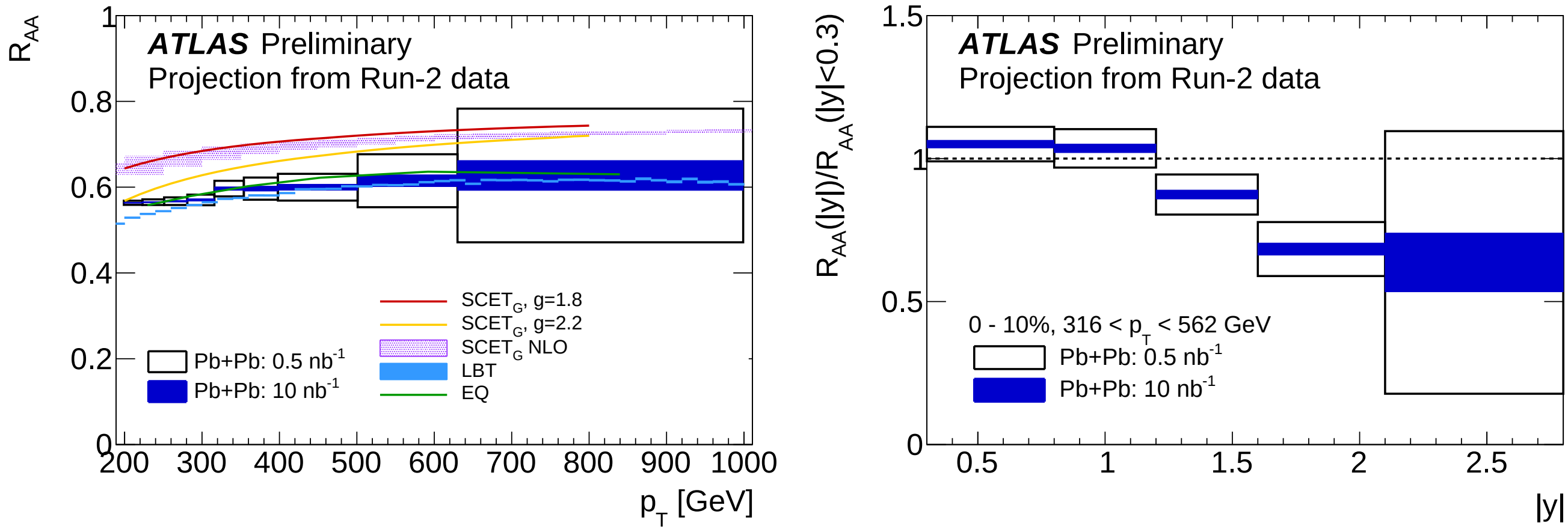

Figure 1: Left: Transverse momentum dependence of jet $R_{\mathrm{AA}}$ for $|y| < 2.1$. Right: Rapidity dependence of jet $R_{\mathrm{AA}}$ for $316 < p_{\mathrm{T}} < 562$ GeV. Boxes represent magnitude of statistical uncertainty. Open black boxes represent the current precision on the jet $R_{\mathrm{AA}}$ measurements while blue closed boxes represent a projection of statistical uncertainties towards 10 nb$^{-1}$of Pb+Pb data. Expected precision on the $R_{\mathrm{AA}}$ measurement is compared with several recent model predictions: Linear Boltzmann Transport model (LBT), three calculations using the Soft Collinear Effective Theory (SCET), and Effective Quenching model (EQ).

## 3.2 Fragmentation Function Measurements

In order to further explore the jet energy loss mechanism, the internal structure of jets is measured in heavy-ion collisions and the results are compared to similar measurements in $pp$ collisions. The measurement of the jet structure in the sample of inclusive jets provides higher precision and the ability to perform the measurement more differentially compared to the photon–tagged studies, which suffer from a limited statistical precision. The jet internal structure can be characterized by the charged-particle longitudinal fragmentation functions, where charged particles are associated with jets, and the longitudinal momentum fraction relative to the jet momentum is evaluated as:

$$z \equiv p_{\mathrm{T}} \cos \Delta R \; / \; p_{\mathrm{T}}^{\mathrm{jet}}. \tag{2}$$

The fragmentation function is then defined as:

$$D(z) \equiv \frac{1}{N_{\mathrm{jet}}} \frac{\mathrm{d}n_{\mathrm{ch}}}{\mathrm{d}z}, \tag{3}$$

where $N_{\mathrm{jet}}$ is the number of jets and $n_{\mathrm{ch}}$ is the number of charged particles associated with jets via an angular matching $\Delta R < 0.4$, where $\Delta R = \sqrt{\Delta\eta^2 + \Delta\phi^2}$ with $\Delta\eta$ and $\Delta\phi$ defined as the distances between the jet axis and the charged-particle direction in pseudorapidity and azimuth, respectively.

Charged-particle longitudinal fragmentation functions were observed to be modified in Pb+Pb collisions compared to the $pp$ reference both at 2.76 TeV [18–20] and 5.02 TeV [21]. The magnitude of the modification is quantified by the ratios of the fragmentation functions in the two colliding systems:

$$R_{D(z)} \equiv \frac{D(z)_{\mathrm{PbPb}}}{D(z)_{pp}}. \tag{4}$$

Measurements of jet fragmentation functions reported an excess in yield of hard (large $z$) and soft (small $z$) fragments and suppression in the region between these two excesses. The excess of soft fragments can be explained by the jet quenching process where the lost energy is transferred to soft particles within and around the jet [22, 23]. Gluon-initiated jets are expected to undergo larger energy loss compared to quark-initiated jets and the enhancement of the hard fragments yield may be due to the different relative shapes of the quark and gluon fragmentation functions [17].

The reduction of statistical uncertainties on the ratios of $R_{D(z)}$ distributions is demonstrated in Fig. 2 where the statistical uncertainties on the measurement using 0.49 nb$^{-1}$of Pb+Pb data are compared to those expected with 10 nb$^{-1}$ of Pb+Pb data. The higher luminosity data will significantly improve the precision of the measurement especially of yields of hard fragments that will strongly constrain theoretical models.

As the fraction of quark- and gluon-initiated jets changes with jet rapidity, measurements of the rapidity dependence of jet observables in Pb+Pb collisions are of great interest. Current measurements of the rapidity dependence of jet fragmentation functions in Pb+Pb are statistics-limited [20] and no significant rapidity dependence is observed. The rapidity dependence can be quantified as the ratio of $R_{D(z)}$ in the rapidity intervals 0.3–0.8 and 1.2–2.1 to the $|y| < 0.3$ reference. This observable benefits from a large cancellation of systematic uncertainties that are to a large extent correlated between rapidity intervals. Figure 3 shows these ratios for jets with $p_{\mathrm{T}}$ in intervals of 126–158 GeV and 200–251 GeV for the most central 0–10% collisions. The ratios are based on the measured distributions in Pb+Pb collisions at 5.02 TeV [21] with the smoothed statistical variations, which are then projected for the integrated luminosity of 10 nb$^{-1}$.

The projections are compared to theoretical calculations from Refs. [24] and [17]. The hybrid model [24] for jet quenching uses perturbative techniques for the high $Q^2$ processes in the jet evolution and strong coupling for the low momentum scales associated with the hot QCD matter. The study of the rapidity dependence shows a small enhancement of yields of fragments with low and intermediate $z$ and reduction of yields of high $z$ fragments for more forward jets.

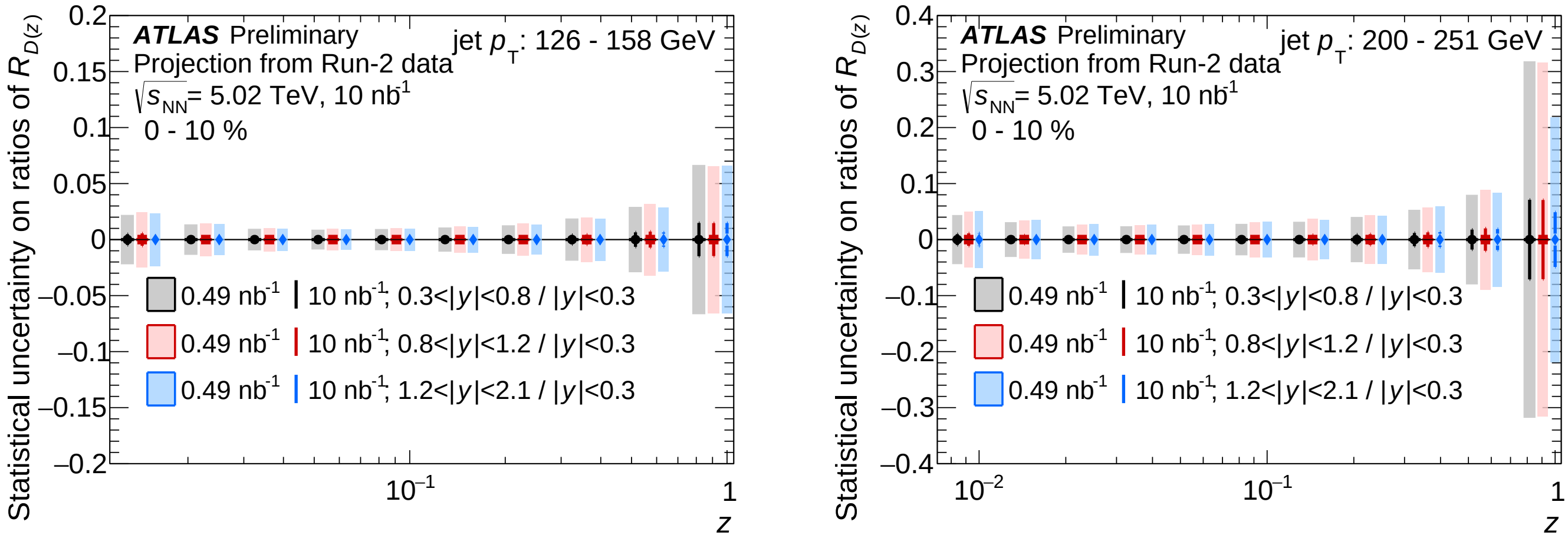

Figure 2: Comparison of the statistical uncertainties on the ratio of the rapidity selected $R_{D(z)}$ distributions to the $R_{D(z)}$ distributions measured in $|y| < 0.3$ in 0–10% centrality Pb+Pb collisions, for the jet $p_{\mathrm{T}}$ intervals 126–158 GeV (left) and 200–251 GeV (right). Shaded boxes are for 0.49 nb$^{-1}$while vertical error bars are for 10 nb$^{-1}$(many are smaller than the plotted marker size).

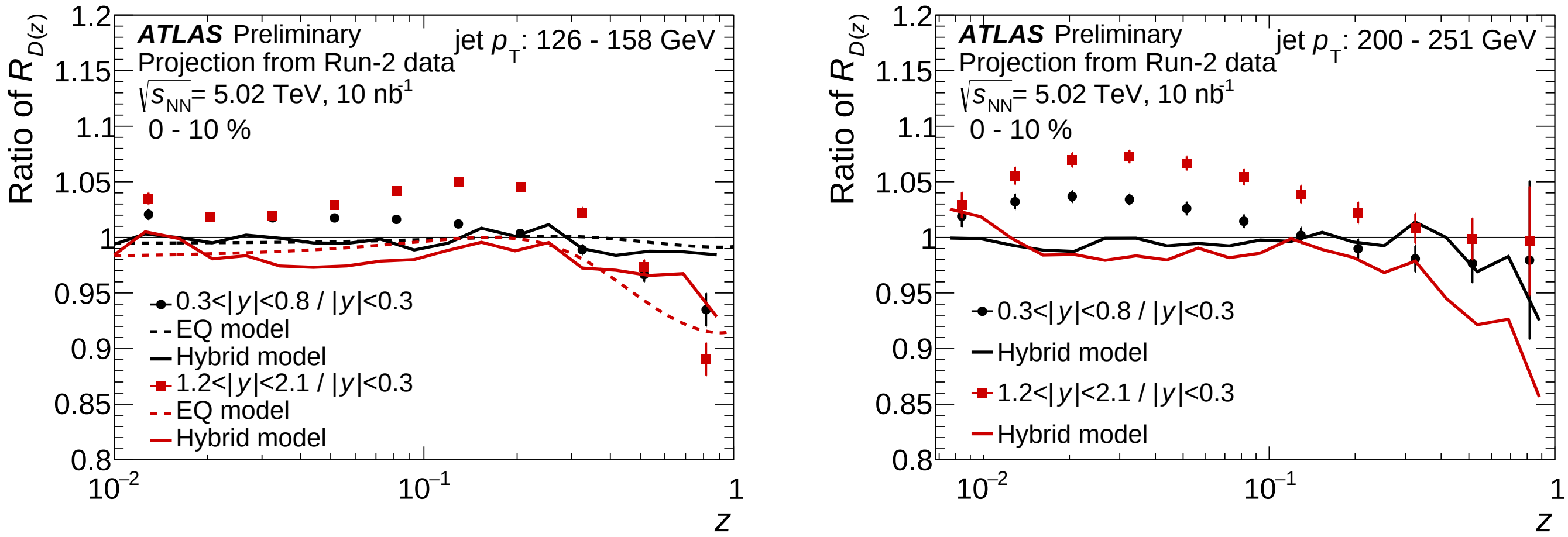

Figure 3: Comparison of the ratio of the rapidity selected $R_{D(z)}$ distributions to the $R_{D(z)}$ distributions measured in $|y| < 0.3$ and the same quantity evaluated in the hybrid model [24] with a reasonable choice of the resolution parameter ($R_{\mathrm{res}} = 3$) and in the EQ model [17] in 0–10% central collisions for 126–158 GeV (left) and 200–251 GeV (right) jet $p_{\mathrm{T}}$ interval. The comparison to the EQ model is shown only for 126–158 GeV jet $p_{\mathrm{T}}$ interval. The vertical bars on the data points indicate statistical uncertainties. In most cases, the statistical uncertainties are smaller than the marker size.

### 3.3 Photon-tagged Jet Measurements

Figure 4 shows the projected statistical uncertainties for measurements of photon-tagged jet fragmentation functions with 10 nb$^{-1}$of Pb+Pb data. The advantage of studying photon-tagged jet fragmentation functions, in contrast to the study of inclusive jets, includes the selection of jets not biased by how much energy the jet has lost, and that the selected sample is dominated by quark-initiated jets. Measurements of this quantity using the 2015 Pb+Pb data are statistics-limited [6]. In that measurement, photon-tagged jet fragmentation functions are found to have a stronger centrality dependence than the inclusive fragmentation functions. However, the limited statistics of the 2015 data necessitated that the analysis be done in very broad centrality selections (0–30% and 30–80%). With the Run 3 and 4 data, one of the highest priority goals would be to repeat this measurement in much finer centrality selections. The projected statistical precision of this measurement for 10 nb$^{-1}$is shown in Figure 4.

## 4 Systematic Uncertainties

To achieve an improvement in the precision of jet measurements, not only statistical uncertainties but also systematic uncertainties will need to be reduced. A full quantitative projection of this improvement is not available at this time, however we discuss expected improvements qualitatively using the inclusive jet fragmentation at high-z and jet structure in the gamma-jet system as examples. The dominant systematic uncertainties for these measurements stem from knowledge of the jet energy scale (JES) and the track reconstruction. For jet spectra measurements, most limiting uncertainties are due to the determination of the nuclear overlap function and the luminosity.

The JES uncertainty is expected to be reduced by general improvements on the proton-proton baseline systematic uncertainties which can be achieved by using large samples of gamma-jet or Z-jet events. Also important for the reduction is the use of MC generators which realistically simulate jet quenching phenomenon. The systematic uncertainty on track reconstruction is expected to be reduced by the use of the upgraded tracker (ITk). To improve the systematic uncertainty on the determination of the nuclear overlap function, data-driven techniques for the centrality determination need to be explored. These include use of the information from inclusive prompt photon, Z and W boson measurements as well as possible use of information from forward detectors. Recent progress in understanding the fluctuating nature of the nucleon-nucleon cross-section as well as factorization of soft and hard processes should also bring improvement to the modeling of centrality [25, 26].

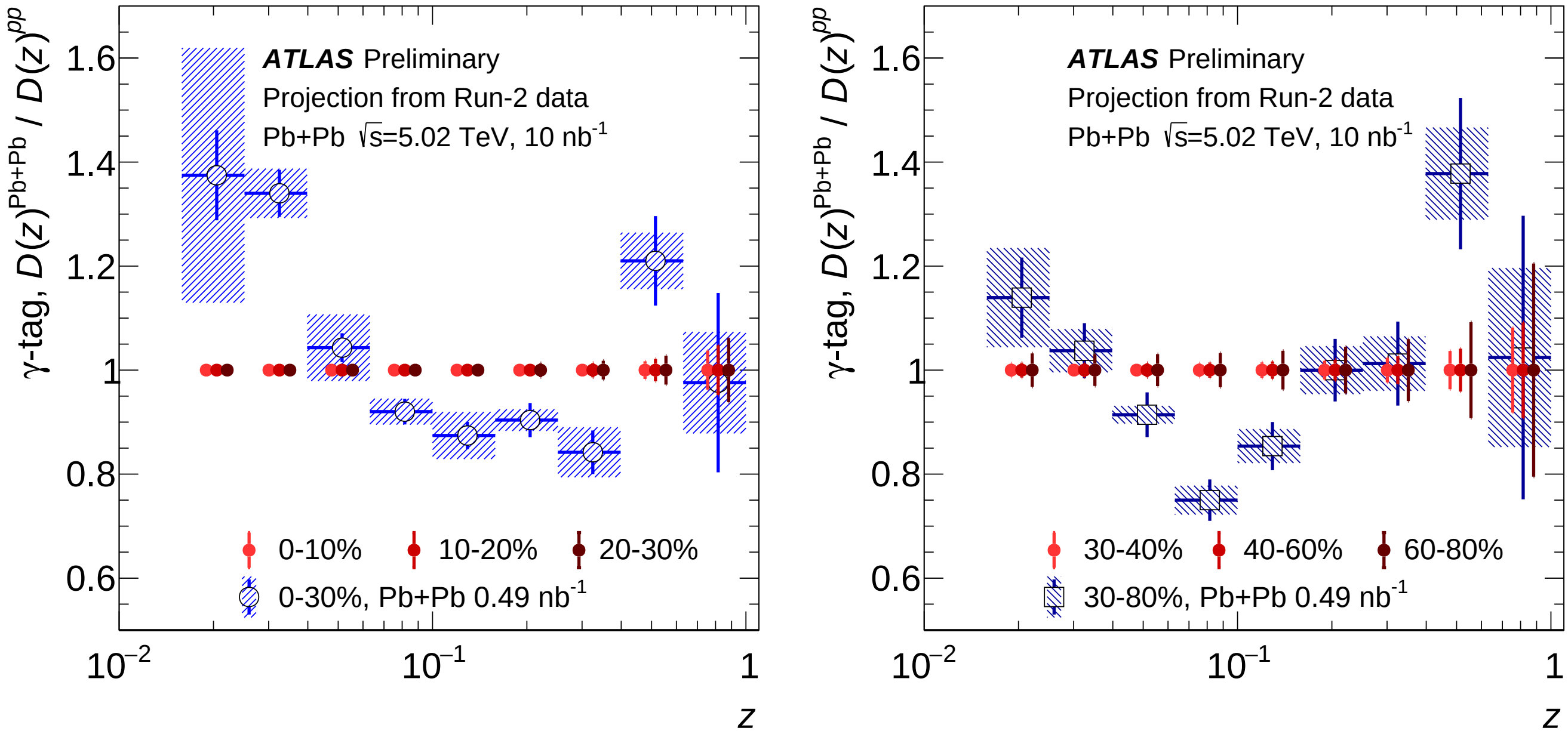

Figure 4: Statistical projection for measurements of photon-tagged jet fragmentation function modification in Pb+Pb collisions compared to *pp* collisions. Preliminary data is shown from 0.49 nb$^{-1}$of Pb+Pb collisions with statistical (bars) and systematic uncertainty (shaded boxes) and compared to projected relative statistical uncertainty for 10 nb$^{-1}$of Pb+Pb data in central (*left*) and peripheral (*right*) collisions. Many of the projected points have uncertainty bars smaller than the marker size.

## Summary

This note presents projections for jet measurements using 10 nb$^{-1}$of Pb+Pb data in Runs 3 and 4 with the ATLAS detector. Jet $R_{AA}$, inclusive jet fragmentation functions and the fragmentation functions of jets opposite a photon are presented. These representative measurements which can be performed with ATLAS are sensitive to the properties of the QGP and can be made with the high luminosity available in the future.

# CMS Physics Analysis Summary

Contact: cms-phys-conveners-ftr@cern.ch

# Open heavy flavor and quarkonia in heavy ion collisions at HL-LHC

The CMS Collaboration

## Abstract

The projected performance for heavy flavour hadrons and quarkonium measurements in the high luminosity phase of the LHC in pp and PbPb collisions at $\sqrt{s_{\mathrm{NN}}} = 5.02$ TeV is presented, focusing on J/$\psi$ and $\Upsilon$ states (including the elliptic flow $v_2$), as well as $\mathrm{B_s}$ mesons and $\Lambda_c^+$ baryons. Projections for the nuclear modification factor of $\mathrm{B^+}$, $\mathrm{B^0}$ and $\mathrm{B_s}$ mesons in pPb data at $\sqrt{s_{\mathrm{NN}}} = 5.02$ TeV are also reported.

# 1 Introduction

The first years of running of the LHC have brought major advances in the understanding of the dynamics of heavy quark production in heavy ion collisions and their evolution to the observed heavy flavour hadrons. In this document exploratory studies are presented that investigate the physics prospects of heavy ion data analysis in the High-Luminosity LHC (HL-LHC) era with the CMS experiment. For the HL-LHC running period, the requested integrated luminosities are $13\,\mathrm{nb}^{-1}$ of PbPb collisions, and $2\,\mathrm{pb}^{-1}$ of pPb collisions. Studies in this note complement those found in a previous note [1].

The CMS detector [2] will be substantially upgraded in order to fully exploit the physics potential offered by the increase in luminosity at the HL-LHC [3], and to cope with the demanding operational conditions at the HL-LHC [4–8]. The upgrade of the first level hardware trigger (L1) will allow for an increase of L1 rate and latency to about 750 kHz and 12.5 $\mu$s, respectively, and the high-level software trigger (HLT) is expected to reduce the rate by about a factor of 100 to 7.5 kHz. The entire pixel and strip tracker detectors will be replaced to increase the granularity, reduce the material budget in the tracking volume, improve the radiation hardness, and extend the geometrical coverage and provide efficient tracking up to pseudorapidities of about $|\eta| = 4$. The muon system will be enhanced by upgrading the electronics of the existing cathode strip chambers (CSC), resistive plate chambers (RPC) and drift tubes (DT). New muon detectors based on improved RPC and gas electron multiplier (GEM) technologies will be installed to add redundancy, increase the geometrical coverage up to about $|\eta| = 2.8$, and improve the trigger and reconstruction performance in the forward region. The barrel electromagnetic calorimeter (ECAL) will feature the upgraded front-end electronics that will be able to exploit the information from single crystals at the L1 trigger level, to accommodate trigger latency and bandwidth requirements, and to provide 160 MHz sampling allowing high precision timing capability for photons. The hadronic calorimeter (HCAL), consisting in the barrel region of brass absorber plates and plastic scintillator layers, will be read out by silicon photomultipliers (SiPMs). The endcap electromagnetic and hadron calorimeters will be replaced with a new combined sampling calorimeter (HGCal) that will provide highly-segmented spatial information in both transverse and longitudinal directions, as well as high-precision timing information. Finally, the addition of a new timing detector for minimum ionising particles (MTD) in both barrel and endcap region is envisaged to provide capability for 4-dimensional reconstruction of interaction vertices that will substantially mitigate the CMS performance degradation due to high PU rates. A detailed overview of the CMS detector upgrade program is presented in Ref. [4–8], while the expected performance of the reconstruction algorithms and pile-up mitigation with the CMS detector is summarised in Ref. [9].

The potential of the upgraded CMS experiment to perform heavy flavour and quarkonium measurements in PbPb at HL-LHC is estimated by extrapolating the performance of the existing CMS measurements [10–13], performed using 2015 5.02 TeV pp and PbPb data (with integrated luminosities of respectively $27.4\,\mathrm{pb}^{-1}$ and about $0.35\,\mathrm{nb}^{-1}$), to a larger data set of $10\,\mathrm{nb}^{-1}$ assuming the same center-of-mass energy of 5.02 TeV, per nucleon pair. The extrapolation of B meson measurements performed in 2013 pPb collisions at 5.02 TeV [14] from about $34\,\mathrm{nb}^{-1}$ to $2\,\mathrm{pb}^{-1}$ is also reported. This extrapolation assumes that the CMS experiment will have a similar level of detector and triggering performance during the HL-LHC operation as it provided during the LHC Run 2 period [4–8]: similar efficiency and resolution are assumed for all objects, such that extrapolations are based on Run 2 results, without additional corrections. The results of extrapolations, referred to as projections, are presented either without systematic uncertainties, when a projection is difficult to make, or in a scenario assuming that there will be further advances on the experimental methods at the HL-LHC. In the latter scenario, the exact assumptions are described in the corresponding section below, where systematic uncertainties are scaled down until they reach a defined lower limit based on estimates of the achievable accuracy with the upgraded detector [9], namely 0.5% per muon (1% per dimuon) for muon identification. No improvement over the current charged particle tracking uncertainty (4% per track) is however assumed. The intrinsic statistical uncertainty in the measurement is reduced by a factor $1/\sqrt{\mathrm{L}}$, where L is the projection integrated luminosity divided by that of the reference Run 2 analysis. Projections on nuclear modification factors $R_{\mathrm{AA}}$, defined as the ratio of the yield in nucleus-nucleus collisions to that observed in pp collisions, scaled by the number of binary nucleon-nucleon collisions, are also estimated. In these cases, this statistical uncertainty has contributions from both the pp and PbPb datasets, and both are assumed to scale in the same way (in other words, the size of the pp reference dataset at $\sqrt{s} = 5.02$ TeV is assumed to be of similar or larger size as the PbPb dataset, as is currently requested).

Though the projections in this document are assuming similar detector performance and pseudorapidity coverage as for Run 2, the impact of the expected improvement in the muon momentum resolution has been studied. Full simulation of the Phase-II detector shows that the dimuon mass resolution for the $\mathrm{B_s}$ meson will improve by about 30% [4], which has been propagated to the $\Upsilon$ system, using pseudo-experiments thrown by propagating this improvement to the measured dimuon mass spectrum in PbPb collisions during Run 2 [10]. The expected improvement in the relative statistical uncertainty, due to a better signal over background ratio, ranges from about 10% for the $\Upsilon$(3S) meson to about 25% for the $\Upsilon$(1S) meson.

## 2 The $p_{\mathrm{T}}$ reach for prompt J/$\psi$ and $\Upsilon$(1S)

The ATLAS [15] and CMS [11] collaborations have recently reported hints for an increase of the nuclear modification factor of prompt J/$\psi$ mesons at high $p_{\mathrm{T}}$, up to about $30 - 50\,\mathrm{GeV}/c$. This trend has been compared to the similar one observed for $\mathrm{D^0}$ mesons [16] and charged particles [17], consistent with a picture in which J/$\psi$ mesons are likely to be produced by parton fragmentation for $p_{\mathrm{T}} \gg m_{\mathrm{J}/\psi}c$, hence to be sensitive to the parton energy loss in the quark-gluon plasma [18, 19]. The interplay of such mechanism with Debye screening [20, 21] is still not clear, also in view of the lack of significant $p_{\mathrm{T}}$ dependence so far for $\Upsilon$(1S) mesons [10]. Whether it is due to the slower motion of bottomonia, to a smaller parton shower component than that of J/$\psi$ at fixed $p_{\mathrm{T}}$, or simply to present lack of precision, will be known by collecting more data to measure them to higher $p_{\mathrm{T}}$.

To estimate the maximum $p_{\mathrm{T}}$ reach of prompt J/$\psi$ and $\Upsilon$(1S) mesons as a function of luminosity, we compute the bin boundaries of the last $p_{\mathrm{T}}$ bin, assuming the same bin width as the one used for the PbPb analyses at 5.02 TeV with an integrated luminosity of $368\,\mu\mathrm{b}^{-1}$ [10, 11]. We scan the $[p_{\mathrm{T}}^{low}, p_{\mathrm{T}}^{up}]$ boundaries (keeping the width constant) until we find those for which $N_{\mathrm{new}} \approx N_0$, where $N_0$ is the uncorrected number of events in the highest $p_{\mathrm{T}}$ bin of the current measurement ($N_0 \approx 150$ for prompt J/$\psi$ and $N_0 \approx 840$ for $\Upsilon$(1S)), and $N_{\mathrm{new}}$ is the expected uncorrected number of measured quarkonia. This number $N_{\mathrm{new}}$ is deduced from a phenomenological fit to the $p_{\mathrm{T}}$-dependent uncorrected number of events from current measurements [10, 11]. The $[p_{\mathrm{T}}^{low}, p_{\mathrm{T}}^{up}]$ interval is determined for several luminosities, ranging up to the expected HL-LHC luminosity and beyond. The results of the $p_{\mathrm{T}}$ bin boundaries where we would find the same number of mesons as the one obtained in the analyses with a luminosity of $368\,\mu\mathrm{b}^{-1}$ are shown in Fig. 1 as a function of luminosity. The maximum $p_{\mathrm{T}}$ we expect to reach with an integrated luminosity of $10\,\mathrm{nb}^{-1}$ is about $80\,\mathrm{GeV}/c$ for prompt J/$\psi$ and $50\,\mathrm{GeV}/c$ for $\Upsilon$(1S). For comparison, these are similar to the current $p_{\mathrm{T}}$ reach for $\mathrm{D^0}$and prompt J/$\psi$ mesons, respectively. Measure-ments at such high $p_T$ will provide further insights into parton energy loss.

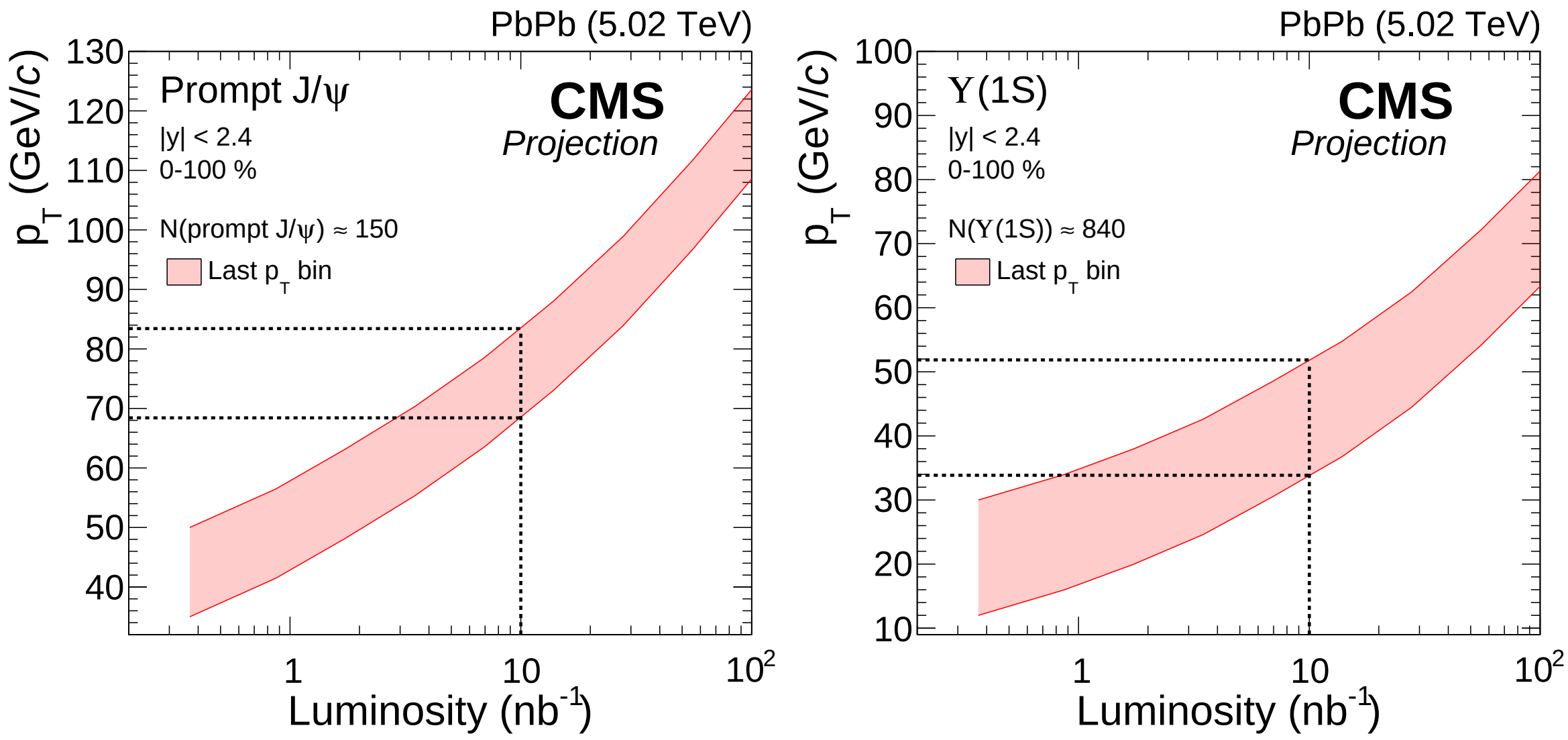

Figure 1: Prompt J/$\psi$ and Υ(1S) [$p_T^{low}$,$p_T^{up}$] boundaries as a function of luminosity. The boundaries are chosen in such a way the number of mesons in the bin for the corresponding luminosity equals the number of mesons found in the last $p_T$ bin of the analysis with a luminosity of 368 $\mu$b$^{-1}$.

# 3 Υ meson production

## 3.1 Nuclear modification factor $R_{AA}$

Further information regarding the mechanisms at play in bottomonium suppression in heavy ion collisions can be gained by measurement of the kinematic dependence of $R_{AA}$, with $p_T$ and $|y|$, in addition to the inclusive modification. For instance, there is ongoing debate on the importance of the recombination of uncorrelated quarks for bottomonia [22–26], which is expected to strongly depend on the meson $p_T$; this process is believed to be dominant at low $p_T$ for charmonia. In addition, cold nuclear matter effects (such as shadowing) are expected to depend on rapidity, as well as the temperature of the medium [25, 26]. The expected precision on the centrality-dependent $R_{AA}$ for the three Υ states with 10 nb$^{-1}$ of PbPb data has been previously reported [1], showing that very high precision will allow to constrain model parameters such as the shear viscosity or the initial temperature; the Υ(3S) meson may also be measured for the first time in PbPb collisions, depending on its exact suppression.

We report the expected precision on $R_{AA}$ of Υ(1S) and Υ(2S) as a function of $p_T$ and $|y|$, shown in Fig. 2. The statistical uncertainties are scaled from the current measurements [10] assuming a projected luminosity of 10 nb$^{-1}$. The total systematic uncertainty is chosen to be reduced by a factor of three, motivated by the fact that dominant systematic uncertainties on current measurements (invariant mass model, data-driven efficiency corrections) are largely correlated to the size of the data sample, though not directly. In addition, the uncertainty on backgroud modeling is assumed to become negligible for $p_T < 4\,\mathrm{GeV}/c$, where it is dominating the systematic uncertainty in the current measurement, due to imperfect knowledge limited by the data sample size. The projected uncertainties will allow to detect fine structures in the $p_T$ dependence, such as a possible bump at low $p_T$ because of flow or regeneration, or a possible increase at high $p_T$ as hinted for instance for J/$\psi$ mesons. The rapidity dependence will also be known up to a large precision, though models predict only a very small dependence of $R_{AA}$ with $|y|$ within the CMS acceptance. Both kinematic dependencies scan through different boosts for the bottomonia and different production angles with respect to the beam axis, providing information on the interaction of bottomonia with the medium, as well as its space and time evolution. For both the $p_T$ and $y$ dependence, projected experimental uncertainties are smaller than current ones on model calculations, showing that this data will constrain the model parameters.

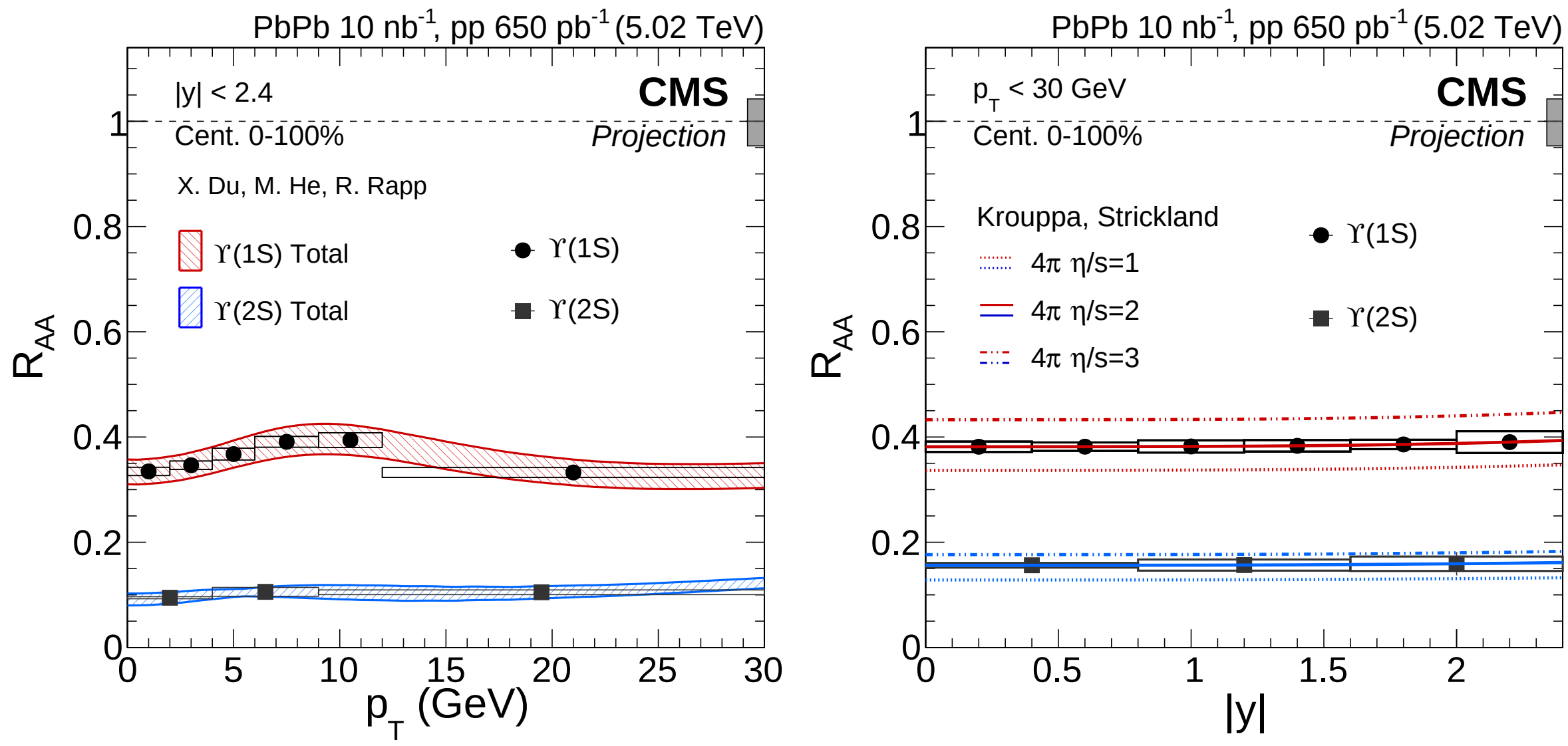

Figure 2: Projections of $R_{AA}$ of Υ(1S) and Υ(2S) as a function of $p_T$ (left) and $y$ (right), assuming 10 nb$^{-1}$, and reduction of the total systematic uncertainty by 1/3. Central values are taken from Ref. [25] for the $p_T$ dependence and Ref. [26] for the $y$ dependence.

## 3.2 Second Fourier coefficient $v_2$

The elliptic flow of Υ mesons brings additional information, complementary to $R_{AA}$, especially about the importance of the regeneration process, as well as about the strength of the coupling to the medium. However this measurement is difficult because of the small production cross section and heavy suppression of the Υ states.

One can project the expected precision of $v_2$ for Υ, assuming an expected central $v_2$ (from a theoretical model), the expected yield, and the signal over background ratio, by comparing the estimated numbers of events measured inside or outside the reaction plane. The validity of the procedure has been checked by reproducing the measured statistical uncertainties on the J/$\psi$ $v_2$ at $\sqrt{s_{NN}} = 2.76$ TeV, starting from the measured J/$\psi$ yields [27].

We can then proceed to making projections for Υ mesons for the HL-LHC. We make the following assumptions:

- the number of expected Υ(nS) mesons is obtained from scaling current measurements [10] up to a luminosity of 10 nb$^{-1}$;
- the central value of the $v_2$ is taken from a theoretical prediction [25];
- the signal over background ratio is assumed equal to that observed in current measurements;
- the centrality range assumed is 5-60%;

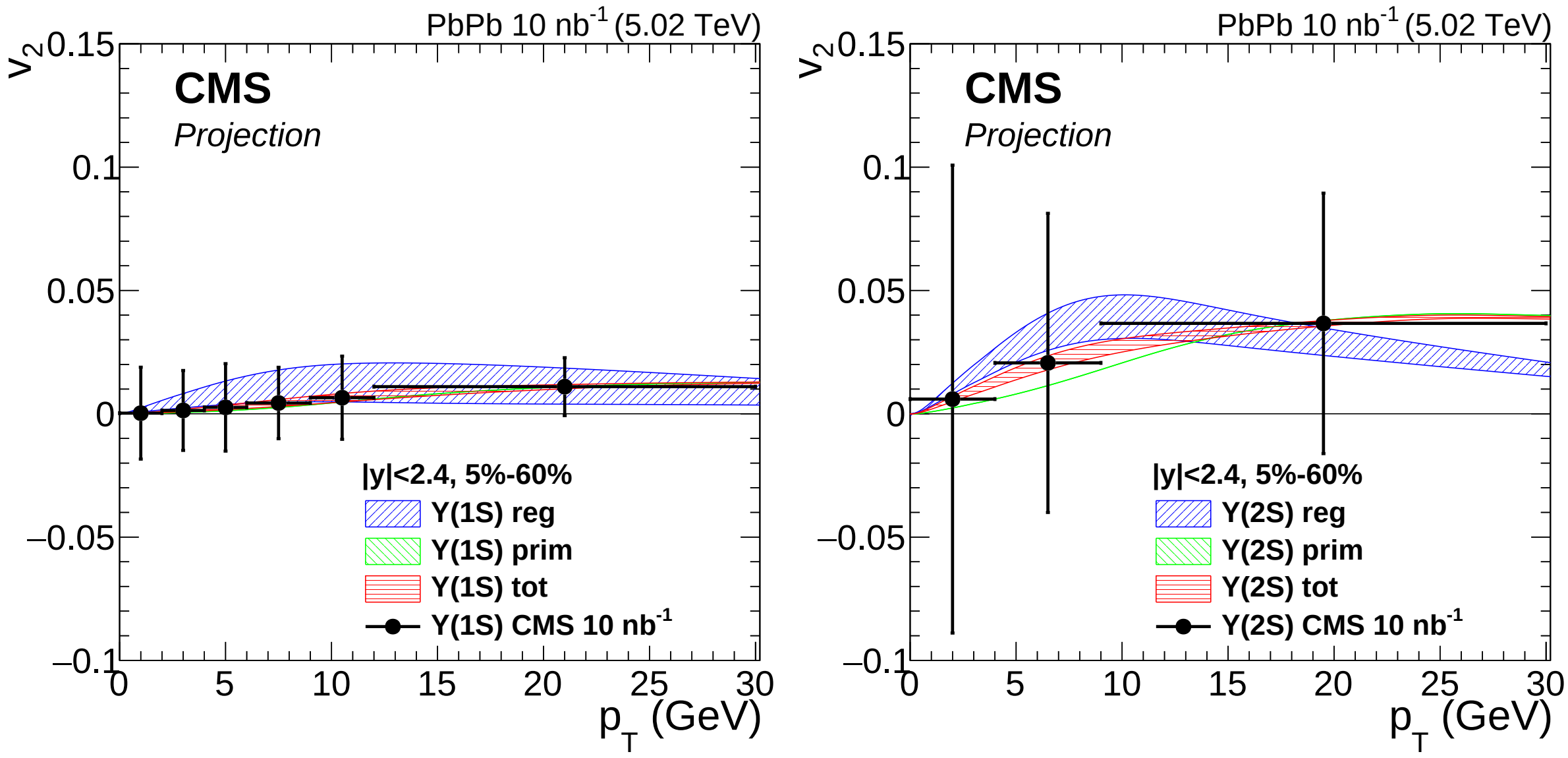

Figure 3: Projections for ϒ(1S) (left) and ϒ(2S) (right) $v_2$, assuming $10\,\mathrm{nb}^{-1}$. The projected data points are overlaid with the total theoretical prediction [25], where the primordial (green) and regenerated (blue) components are also shown separately.

- no systematic uncertainty is considered.

The projections can be found in Fig. 3. Systematic uncertainties expected on the $v_2$ measurements have not been estimated, and only projected statistical uncertainties are reported. It can be seen that the expected statistical precision is too low to make a conclusive statement on the ϒ $v_2$, assuming predictions from Ref. [25] are correct. A combination with other LHC experiments will be useful. It is also not unlikely that a larger-than-expected $v_2$ will be measured, given that this is the case for the J/$\psi$ meson in PbPb collisions [27–29], as well as in the smaller pPb system [30].

## 4 $\mathrm{B_s}$ meson production

Over the past few years, although many theoretical efforts have been taken to understand the transport properties of heavy flavors in the QGP, the hadronisation mechanisms are not understood well. Because of the interplay between the predicted enhancement of strange quark production [31] and the quenching mechanism of beauty quarks, the measurement of strange beauty particles is fundamentally important for studying the mechanisms of beauty hadronisation in heavy ion collisions. The measurement of $R_{\mathrm{AA}}$ of $\mathrm{B_s}$ mesons is a unique tool in that respect.

In Fig. 4, the left panel shows the performance of $R_{\mathrm{AA}}$ of $\mathrm{B^+}$, nonprompt J/$\psi$ and $\mathrm{B_s}$ in 2015 PbPb collisions in centrality range $0-100\%$, and the central values for $\mathrm{B_s}$ are taken from the TAMU model [32, 33]. Projections for $\mathrm{B^+}$ and nonprompt J/$\psi$ are taken from Ref. [1]. The uncertainties are extracted from the measurement of $\mathrm{B_s}$ in PbPb collisions at $\sqrt{s_{\mathrm{NN}}} = 5.02\,\mathrm{TeV}$ performed by CMS in 2015 [12]. To split into finer $p_{\mathrm{T}}$ bins, the statistical uncertainties for $15 < p_{\mathrm{T}} < 20\,\mathrm{GeV}/c$ and $20 < p_{\mathrm{T}} < 50\,\mathrm{GeV}/c$ are obtained by scaling production yields calculated by FONLL calculations [35–37]. In the right panel of Fig. 4, the projection of $R_{\mathrm{AA}}$ of $\mathrm{B^+}$, nonprompt J/$\psi$ and $\mathrm{B_s}$ in $10\,\mathrm{nb}^{-1}$ is presented. The statistical uncertainties are obtained by the scale of luminosity increasing, and the systematic uncertainties also follows the scale of luminosity increasing, with assuming a minimum systematical uncertainty from hadronic track efficiency (2.5% per track) and muon efficiency (0.5% per muon).

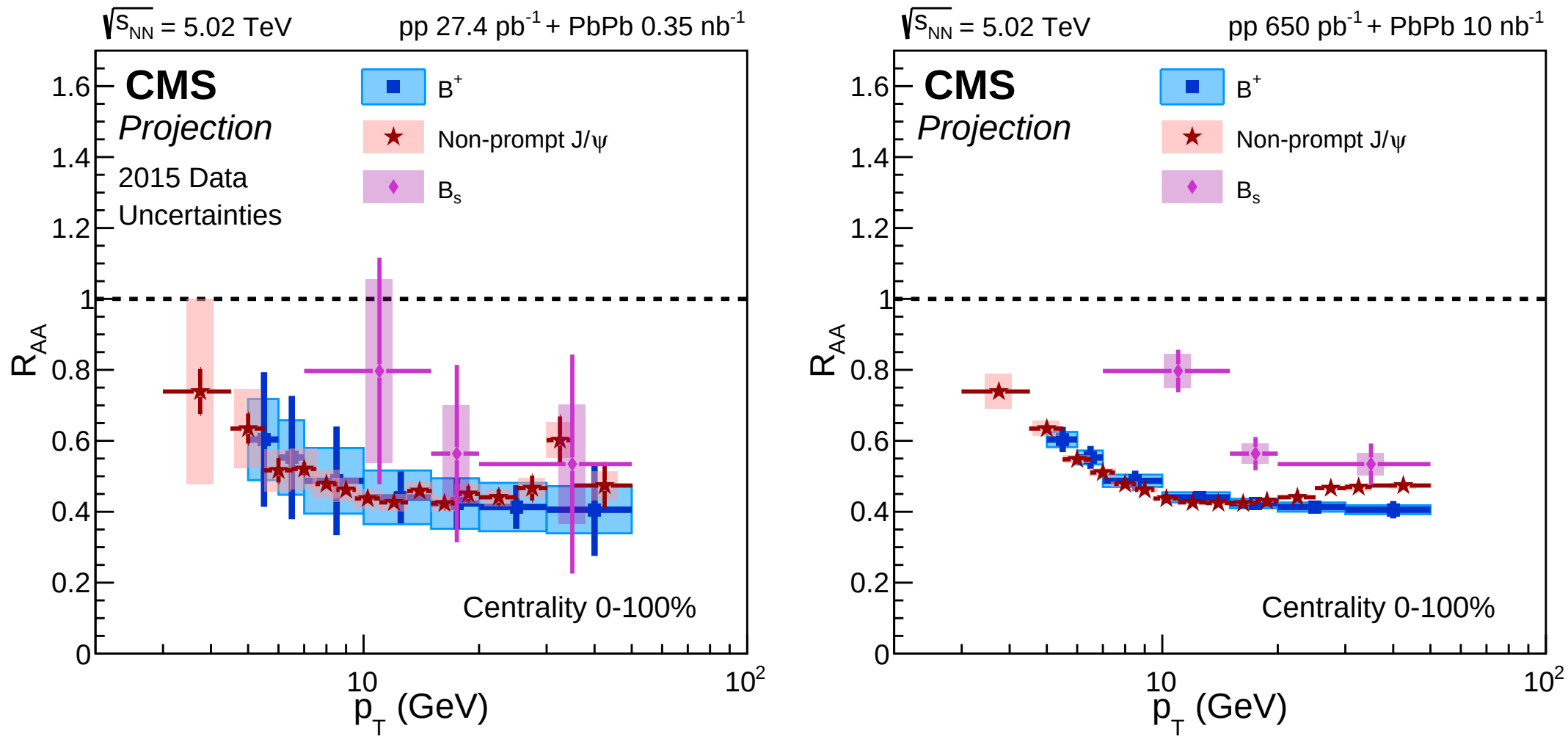

Figure 4: Current uncertainties on the $R_{AA}$ of $B_s$ in 2015 PbPb collisions [12] (left) and projection using $10\,nb^{-1}$ of PbPb data at $\sqrt{s_{NN}} = 5.02$ TeV (right). The central values are taken from the TAMU model [32, 33]. The $B^+$and nonprompt $J/\psi$ uncertainties from current measurements [11, 34] and their projection in $10\,nb^{-1}$ of PbPb data [1] are also shown.

The PbPb luminosity of $10\,nb^{-1}$ allows for separating strange and nonstrange B mesons and study the interplay between the predicted enhancement of strange quark production and the quenching mechanism of beauty quarks, as well as the mechanisms of beauty quark hadronisation in heavy ion collisions. In addition, the CMS Collaboration will also measure $D_s^+$ meson production in pp and PbPb collisions, over a wide $p_T$ range, expected approximately from 6 GeV/$c$ to 40 GeV/$c$. This additional probe will provide further information on a possible enhancement due to recombination with strange quarks and on the hadronisation of charm quarks.

## 5 $\Lambda_c^+$ baryon production

Comparison of $\Lambda_c^+$ production in pp and PbPb collisions can provide essential inputs to understanding two important physics processes, namely, the heavy quark transport in the QGP and the hadronic phase of the medium and heavy quark fragmentation via coalescence. Figure 5 shows the projected performance of $\Lambda_c^+$ signal extraction at the HL-LHC, in pp and PbPb data recorded with a minimum bias trigger (meaning with only loose requirements selecting hadronic events). These expected invariant mass distributions are obtained based on fits to 2015 pp and PbPb data [13], and generating toy experiments by scaling the number of events to $200\,nb^{-1}$ for pp and $0.2\,nb^{-1}$ for PbPb (the expected integrated luminosity for minimum bias events, smaller than the total PbPb luminosity for triggered events). The width of the signal component is also reduced, reflecting the expected improvement in track momentum resolution with the CMS Phase-II detector [4, 5]. The two-particle mass resolution will be reduced by about 2/3 of the current one, so the 3-prong mass is assumed to be multiplied by $(2/3)^{3/2}$. This expected raw yield in each $p_T$ interval is obtained using an unbinned maximum likelihood fit.

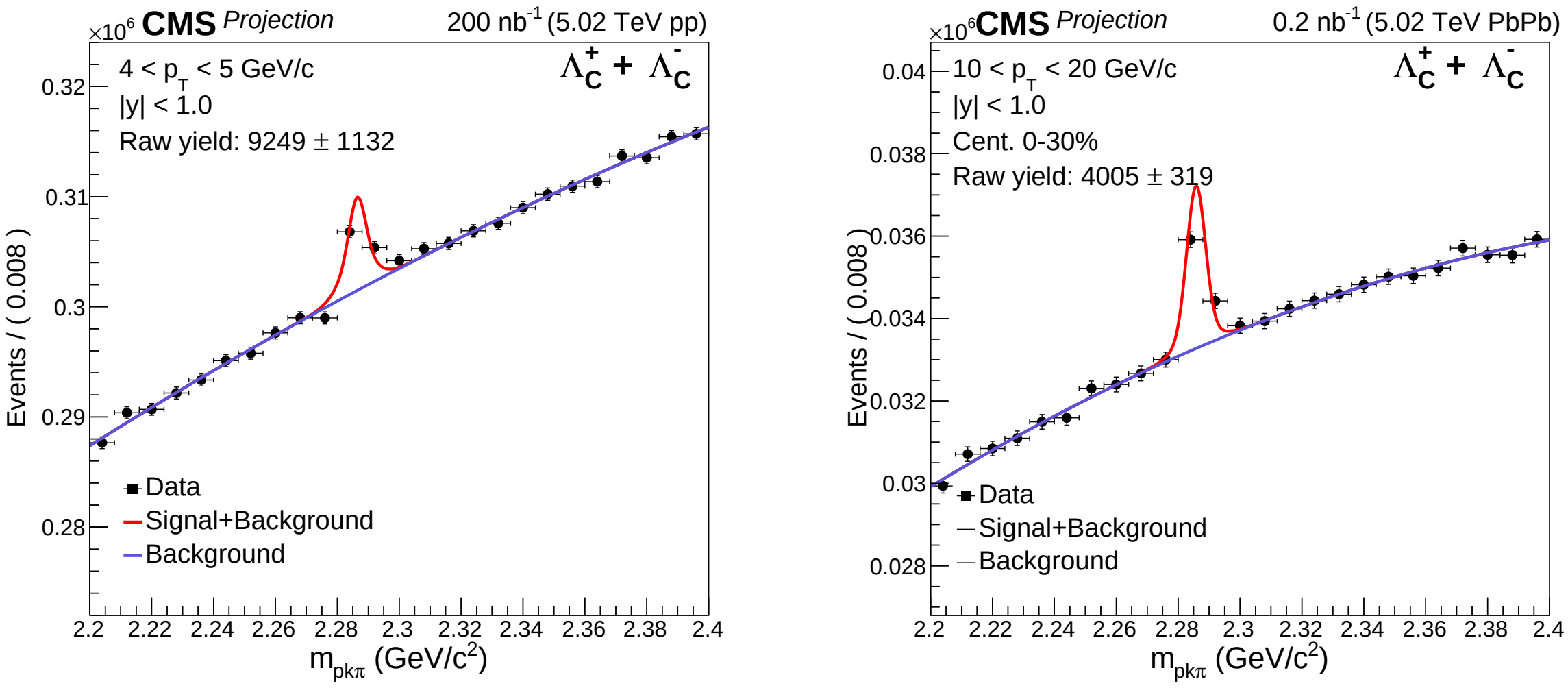

Figure 5: Expected $pK^+\pi^-$ invariant mass spectrum in pp ($4 < p_T < 5\,\mathrm{GeV}/c$, left) and PbPb ($10 < p_T < 20\,\mathrm{GeV}/c$, centrality range 0–30%, right) collisions. The red line represents the signal on top of the background and the blue line represents the background. The signal fit function is double Gaussian and the background fit function is the 2nd-order Chebychev polynomial function.

# 6 $B^+$, $B^0$ and $B_s$ mesons in pPb

Reduction of the measured yield of high-$p_T$ hadrons is observed in heavy-ion collisions, which is considered as consequence of parton energy loss in quark gluon plasma. However, other phenomena can affect the yield of heavy-flavor particles, independently of the presence of a deconfined partonic medium. For instance, modifications of the parton distribution functions (PDFs) in the nucleus with respect to nucleon PDFs [38–40] could change the production rate. In pPb collisions, measurements of heavy-flavoured meson production can provide important baselines for the understanding of heavy-quark energy loss in PbPb collisions. These studies can also provide useful constraints to the nuclear PDFs.

Fig. 6 shows the projection of $R_{pA}$ of $B^+$(top left), $B^0$(top right) and $B_s$ (bottom) in $2\,\mathrm{pb}^{-1}$ of pPb data at $\sqrt{s_{NN}} = 5.02\,\mathrm{TeV}$. The center values and uncertainties are based on measurements in pPb collisions at the same energy performed by CMS in 2013 [14]. The statistical uncertainties are obtained by the scale of luminosity increasing, and the systematic uncertainties also follows the scale of luminosity increasing, with assuming a minimum systematical uncertainty from hadronic track efficiency (2.5% per track) and muon efficiency (0.5% per muon). To split into finer $p_T$ bins, the statistical uncertainties are obtained by the scale of production yield calculated by FONLL calculations [35–37], while central values are kept to the Run 1 measurement value [14]. Predictions from POWLANG model [41] of beauty hadron $R_{pA}$ under different configurations are also presented on top of the projection. The purple line shows the result with no medium assumed. Green and orange lines present simulations with weak-coupling and lattice-QCD transport coefficients respectively. Results with different choices of the smearing of the initial condition are also shown with different styles of lines. As seen in the figure, to distinguish medium effect in pPb collisions, experimental measurement should be extended to $p_T < 10\,\mathrm{GeV}/c$. This is possibly realised with nonprompt D meson measurements in the future.

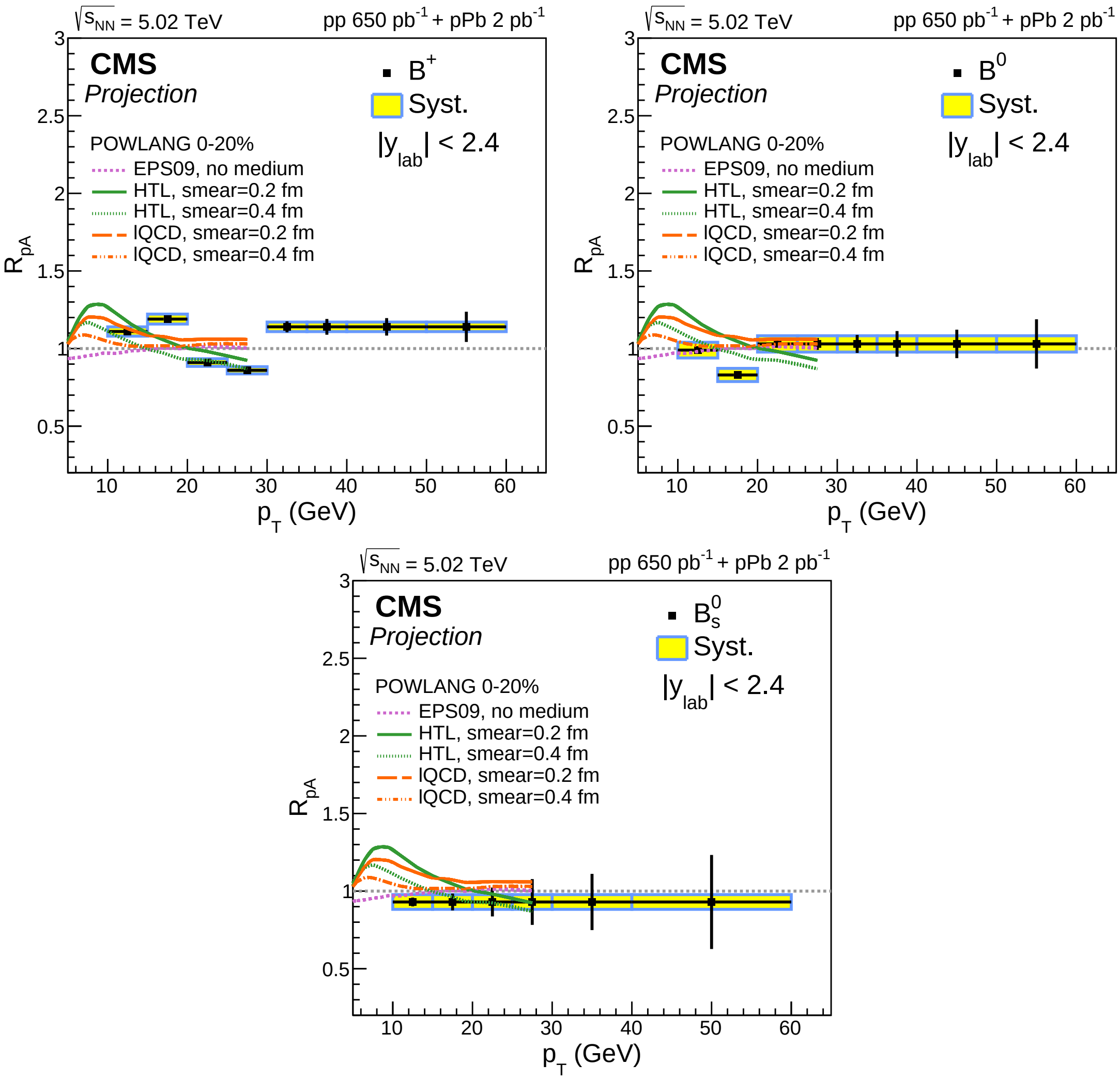

Figure 6: Projection of nuclear modification factors of $B^{+}$(top left), $B^{0}$(top right) and $B_s$ (bottom) in pPb collisions with 2 $pb^{-1}$ of pPb data. Predictions from POWLANG [41] model under different transport coefficients and the smearing of the initial condition.

# 7 Summary

Very precise and differential measurements of quarkonia and heavy flavour mesons will be made possible at the HL-LHC, benefiting from the very large data sample (10 $nb^{-1}$), combined with the excellent performance of the CMS detector in terms of pseudo-rapidity coverage, vertex reconstruction, muon tracking (identification and momentum resolution), and charged particle tracking. Quarkonia will be measured up to very high $p_T$, allowing for direct comparison to charged particles, and $D^0$ and B mesons, providing crucial information on parton energy loss. The precise measurement of $\Upsilon$(nS) $R_{AA}$ as a function of $p_T$ and $|y|$ will allow to better understand the ingredients to bottomonium suppression in PbPb collisions, in complement to the first $\Upsilon$(nS) $v_2$ measurements in PbPb. Despite their limited precision, $v_2$ measurements will provide crucial inputs to models and be sensitive to a possible large signal, not unexpected given existing measurements of J/$\psi$ $v_2$ in pPb and PbPb. $B_s$ meson production in pp and PbPb collisions will also be measured with sufficient precision to be compared to $B^{+}$meson suppression and investigate strangeness enhancement due to recombination with strange quarks. $\Lambda_c^{+}$

baryon production will also be measured in pp and PbPb collisions, providing an additional handle for the study of charm quark dynamics in the medium, as well as the charm quark hadronisation to $\Lambda_c^+$ baryons. Finally, precise measurements of $B^+$, $B^0$ and $B_s$ mesons in pPb collisions will provide a baseline for the study of in-medium b quark energy loss in PbPb collisions.

**10**

# CMS Physics Analysis Summary

Contact: cms-phys-conveners-ftr@cern.ch

# Predictions on the precision achievable for small system flow observables in the context of the HL-LHC

The CMS Collaboration

## Abstract

In this note, we discuss how the future HL-LHC program will enable highly precise measurements of flow observables in small systems. Projections of the statistical uncertainties achievable for symmetric cumulant analyses at $\sqrt{s} = 13$ TeV for pp collisions and at $\sqrt{s_{NN}} = 5.02$ TeV for pPb collisions are presented. The improvement in the symmetric cumulant precision by increasing the pp beam energy to 14 TeV, while extending the CMS tracker pseudorapidity coverage to $|\eta| < 4$, is also shown. In addition, we show how the HL-LHC will allow for elliptic flow measurements of $D^0$ and $J/\psi$ mesons in 8.16 TeV pPb collisions that are a factor of two more precise than currently possible.

# 1 Introduction

In heavy ion collisions, a quark-gluon plasma (QGP) state is created in the overlap region of the colliding ions. The hydrodynamic expansion of the QGP is reflected in the observed correlations of particles in the final state. These correlations are usually quantified by the Fourier harmonic coefficients ($v_n$) of the final-state particle azimuthal distributions. A key feature of such correlations in ultra-relativistic nucleus-nucleus collisions is a pronounced structure on the near side (relative azimuthal angle $|\Delta\phi| \approx 0$) that extends over a large range in relative pseudorapidity ($|\Delta\eta|$ up to 4 units or more). This feature is known as the “ridge” and is thought to be a consequence of the QGP medium reflecting higher-order terms in the overlap geometry of the collision (see, e.g., Ref. [1]). However, a ridge-like behavior has also been found at the LHC in high multiplicity events for small colliding systems, such as the pp and pPb systems. The collective nature of this correlation across all colliding systems may challenge the current accepted paradigm that describes both small and large colliding systems within a similar QCD framework. Despite a significant effort by the CMS, ATLAS, and ALICE experiments at the LHC to explore the ridge, its origin is still unknown. With the HL-LHC program, it will be possible to reach an unprecedented multiplicity regime and an experimental precision that will help to establish the origin of the ridge correlations in small systems. In this note, projections for the key observables of symmetric cumulant correlations and $v_2$ coefficients of heavy-flavor particles are presented for pp collisions at 13 and 14 TeV. In addition, projections for a future pPb run are provided. All of the projections are based on current CMS results presented in Refs. [2–5] and assume similar tracking performance and trigger efficiencies as found for these earlier analyses. It is also assumed that the data taking conditions will be similar, other than for the extended rapidity coverage of the CMS tracker that will be available with the HL-LHC.

# 2 Upgraded CMS Detector

The CMS detector [6] will be substantially upgraded in order to fully exploit the physics potential offered by the increase in luminosity, and to cope with the demanding operational conditions at the HL-LHC [7–11]. The upgrade of the first level hardware trigger (L1) will allow for an increase of L1 rate and latency to about 750 kHz and 12.5 $\mu$s, respectively, and the high-level software trigger (HLT) is expected to reduce the rate by about a factor of 100, to 7.5 kHz. The entire pixel and strip tracker detectors will be replaced to increase the granularity, reduce the material budget in the tracking volume, improve the radiation hardness, and extend the geometrical coverage and provide efficient tracking up to pseudorapidities of about $|\eta| = 4$. The muon system will be enhanced by upgrading the electronics of the existing cathode strip chambers (CSC), resistive plate chambers (RPC) and drift tubes (DT). New muon detectors based on improved RPC and gas electron multiplier (GEM) technologies will be installed to add redundancy, increase the geometrical coverage up to about $|\eta| = 2.8$, and improve the trigger and reconstruction performance in the forward region. The barrel electromagnetic calorimeter (ECAL) will feature the upgraded front-end electronics that will be able to exploit the information from single crystals at the L1 trigger level, to accommodate trigger latency and bandwidth requirements, and to provide 160 MHz sampling allowing high precision timing capability for photons. The hadronic calorimeter (HCAL), consisting in the barrel region of brass absorber plates and plastic scintillator layers, will be read out by silicon photomultipliers (SiPMs). The endcap electromagnetic and hadron calorimeters will be replaced with a new combined sampling calorimeter (HGCal) that will provide highly-segmented spatial information in both transverse and longitudinal directions, as well as high-precision timing information. Finally, the addition of a new timing detector for minimum ionizing particles (MTD) in both barrel and endcap regions is envisaged to provide the capability for 4-dimensional reconstruction of interaction vertices that will significantly offset the CMS performance degradation due to high pileup rates.

A detailed overview of the CMS detector upgrade program is presented in Ref. [7–11], while the expected performance of the reconstruction algorithms and pileup mitigation with the CMS detector is summarised in Ref. [12].

## 3 Projection for Symmetric Cumulants

The symmetric cumulants, denoted SC($m, n$), are based on 4-particle correlations and measure correlations between the Fourier coefficients $m$ and $n$. In this section, the improvement in statistical precision for these measurements that is achieved by increasing the center-of-mass energy and by extending the CMS tracker rapidity coverage is considered for pp collisions. The impact of increasing the integrated luminosity on the statistical precision of pPb symmetric cumulant results is also studied.

### 3.1 HL-LHC projections for 13 TeV pp collisions and for 5.02 TeV pPb collisions

In Fig. 1, the projection for the symmetric cumulant measurement SC(2,3) are shown for pp collisions 13 TeV and pPb collisions at 5.02 TeV. The data points are the CMS results using Run 1 and Run 2 data and are found using the 4-particle cumulant method. Only statistical errors are displayed. The results are expressed as a function of the total multiplicity, as corrected for efficiency and the experimental $p_T$ range. The vertical dashed line shows the multiplicity range above which data were collected using a high-multiplicity trigger. The rising trend of SC(2,3) observed in data when moving toward lower multiplicities is known to come from nonflow effects. At the highest multiplicities, SC(2,3) is weakly dependent on nonflow and a high precision measurement of a negative signal will further constrain the current interpretation of the ridge phenomenon in small colliding systems.

In addition, it has been shown that nonflow effects can be reduced, at the expense of statistical precision, by analyzing the data in multiple subevent regions in the intermediate and low multiplicity ranges [4]. The experimental precision that can be achieved with the HL-LHC for the 2-, 3-, and 4-subevent methods are also shown in Fig. 1. The projections are for pp collisions at 13 TeV (left) and for pPb collisions at 5.02 TeV (right), assuming integrated luminosities of 200 $\mathrm{pb}^{-1}$ and 1000 $\mathrm{nb}^{-1}$ for the two systems, respectively. For these projections, constant mean SC(2,3) values are assumed as a function of total multiplicity, based on existing high-multiplicity measurements for the pp [13] and pPb [4] systems. These projections indicate that the subevent analyses should have absolute uncertainties on the order of $10^{-7}$. As the SC(2,3) value is particularly sensitive to the initial state and its fluctuations, a precision measurement of this quantity will test our understanding this early stage of the collision.

### 3.2 HL-LHC projections for 14 TeV pp collisions

Figure 2 shows the same projections as in Fig. 1, but estimated for pp collisions at 14 TeV. The increase of the number of events for each multiplicity bin is estimated using the multiplicity distribution extrapolated using available data at various center-of-mass energies. Data from the 13 TeV pp analysis are displayed for comparison. With increasing center-of-mass energy, the multiplicity spectra get harder at high-multiplicity. This leads to a dramatic reduction in the experimental uncertainties, by at least an order of magnitude.

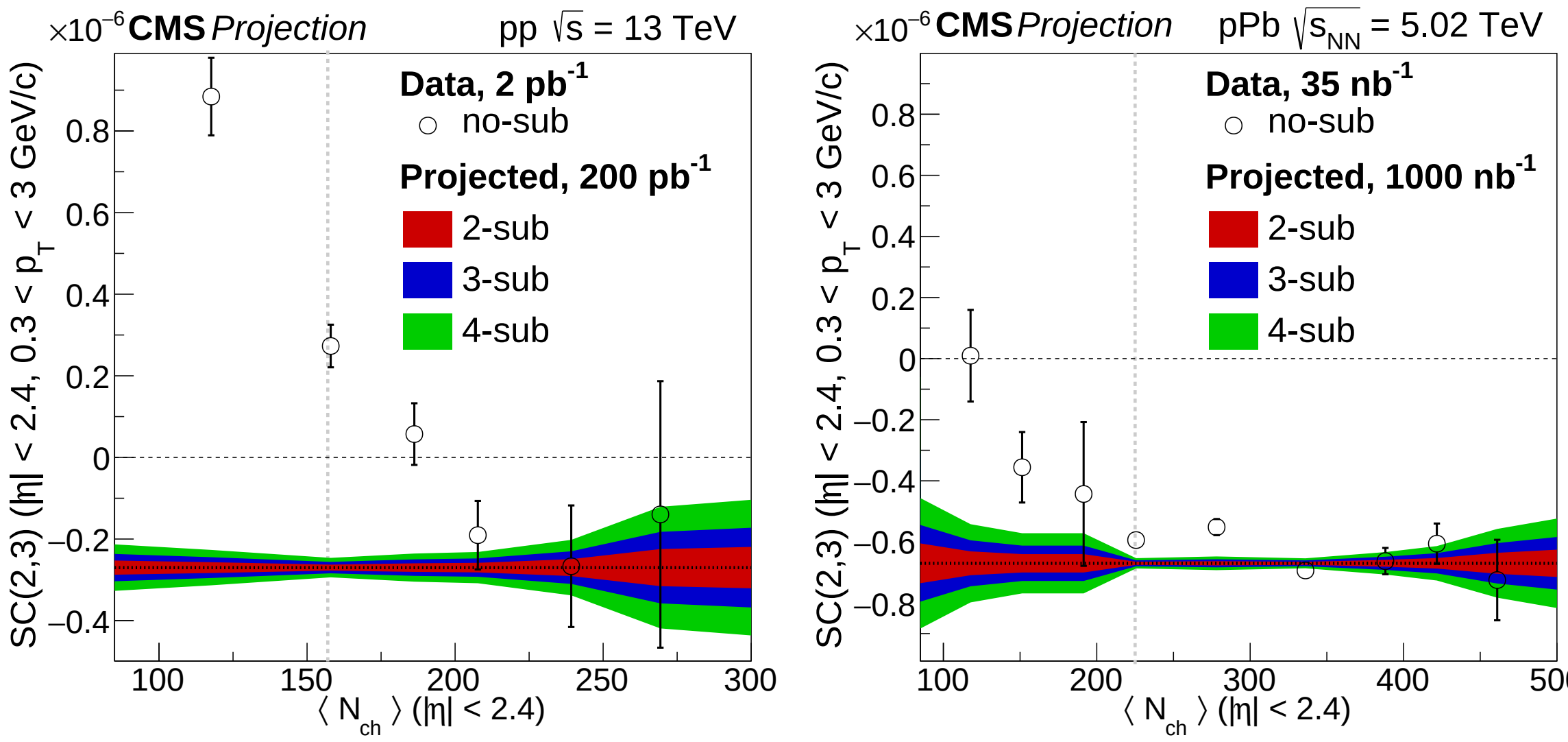

Figure 1: SC(2,3) as a function of total multiplicity in pp collisions at 13 TeV (left) and pPb collisions at 5.02 TeV (right). Only statistical uncertainties are displayed. The open circles show the current CMS results standard 4-particle cumulant method [2]. The vertical dashed line shows the multiplicity range above which data were collected using a high-multiplicity trigger. The color-shaded areas show the HL-LHC projections for 2-, 3- and 4-subevent symmetric cumulant analyses, as indicated.

### 3.3 CMS extended tracker coverage projection

For the HL-LHC runs, the CMS tracker acceptance will be extended to 4 units in pseudorapidity. The projected experimental uncertainties for pp collisions associated with this extended $\eta$ coverage are provided in Fig. 3 based on the 3-subevent symmetric cumulant analysis. With the extended pseudorapidity range, it is no longer possible to assume a flat multiplicity distribution in $\eta$. Consequently, the scaling factor is calculated using Monte Carlo simulations. Pythia and EPOS are used for this simulation and found to give consistent results. The experimental uncertainties for pp collisions at both 13 and 14 TeV are found to be significantly reduced using the increased pseudorapidity coverage.

## 4 HL-LHC projections for $D^0$ and J/Ψ elliptic flow

In this section, the statistical precision of elliptic flow measurement for heavy-flavor mesons with increasing integrated luminosity is studied. Figure 4 (left) shows the projections for the second Fourier harmonic coefficients as a function of $p_T$ for $D^0$ and J/Ψ mesons in pPb collisions at 8 TeV with integral luminosities of 500 nb$^{-1}$ and 2000 nb$^{-1}$. A factor of two improvement in the experimental uncertainties is observed compared to the existing experimental results. Figure 4 (right) shows the same projections as a function of transverse kinetic energy ($KE_T$) scaled by the number of constituent quarks ($n_{\mathrm{q}}$). The production mechanisms of open/hidden charm are poorly known. In addition, whether the charm quarks thermalize in the hot and dense environment created in high multiplicity pPb collisions is still an open question. Assuming a quark-gluon plasma production and partial thermalization of the charm quarks, the following ordering is expected at low $KE_T$: $v_2(\text{charged hadrons}) > v_2(D^0) > v_2 J/\Psi$. With the precision made possible by the HL-LHC, it will be possible to put stringent constraints on heavy particle production and thermalization within a high multiplicity envi-ronment.

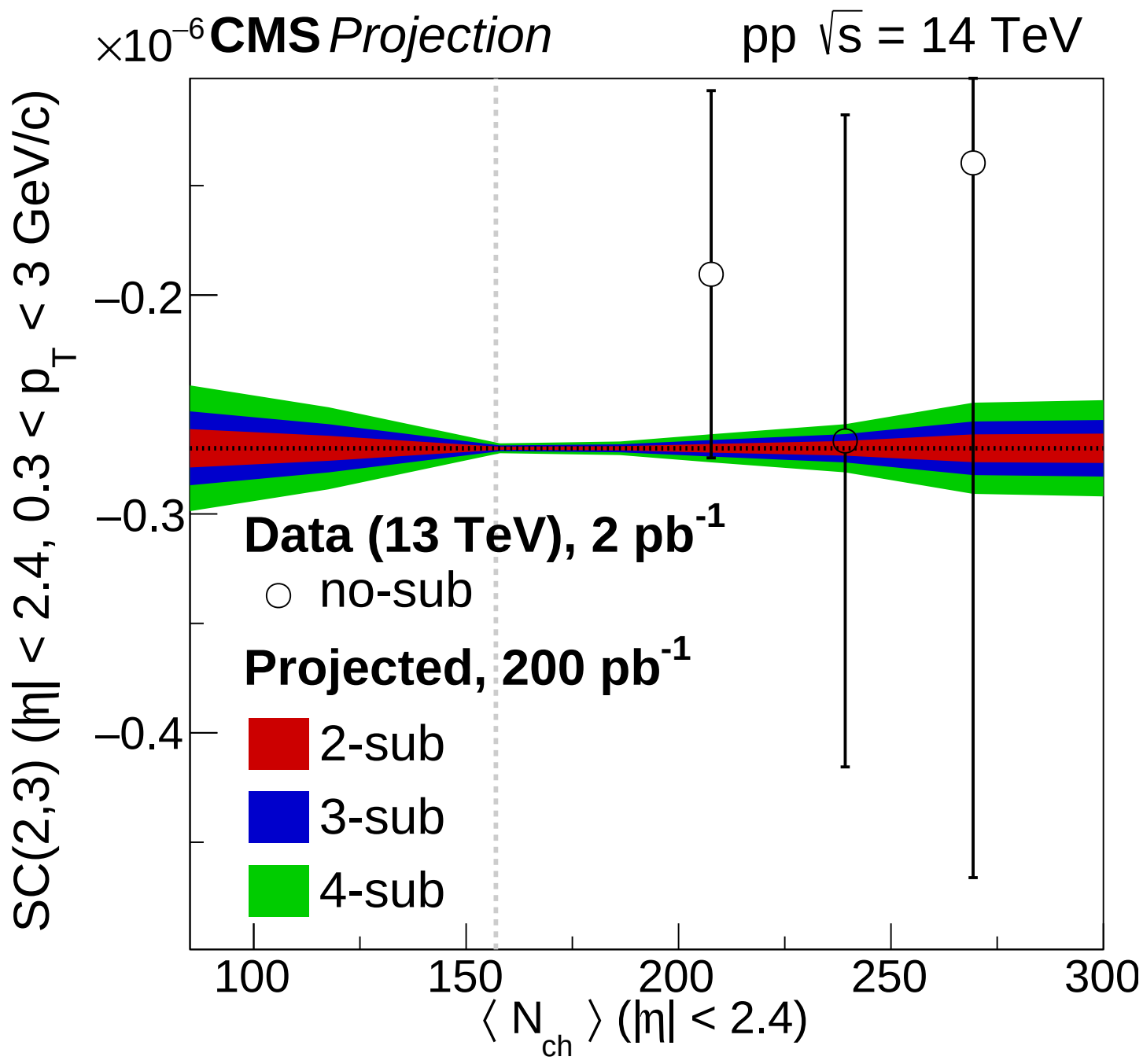

Figure 2: SC(2,3) as a function of total multiplicity in pp collisions at 14 TeV. Only statistical uncertainties are displayed. The vertical dashed line shows the multiplicity range above which data were collected using a high-multiplicity trigger. The open circles are the current CMS results at 13 TeV [2]. The color-shaded areas show the HL-LHC projections for 2-, 3- and 4-subevent symmetric cumulant analyses, as indicated.

# 5 Summary

In this note, we have presented projections for symmetric cumulant and heavy particle elliptic flow analyses in the context of the HL-LHC. The increase of luminosity significantly reduces the experimental uncertainties compared to existing results. Such measurements will provide a better understanding of the “ridge” structure observed in small colliding system. In terms of its theoretical understanding, this is among the most controversial behaviors found in relativistic heavy-ion collisions.

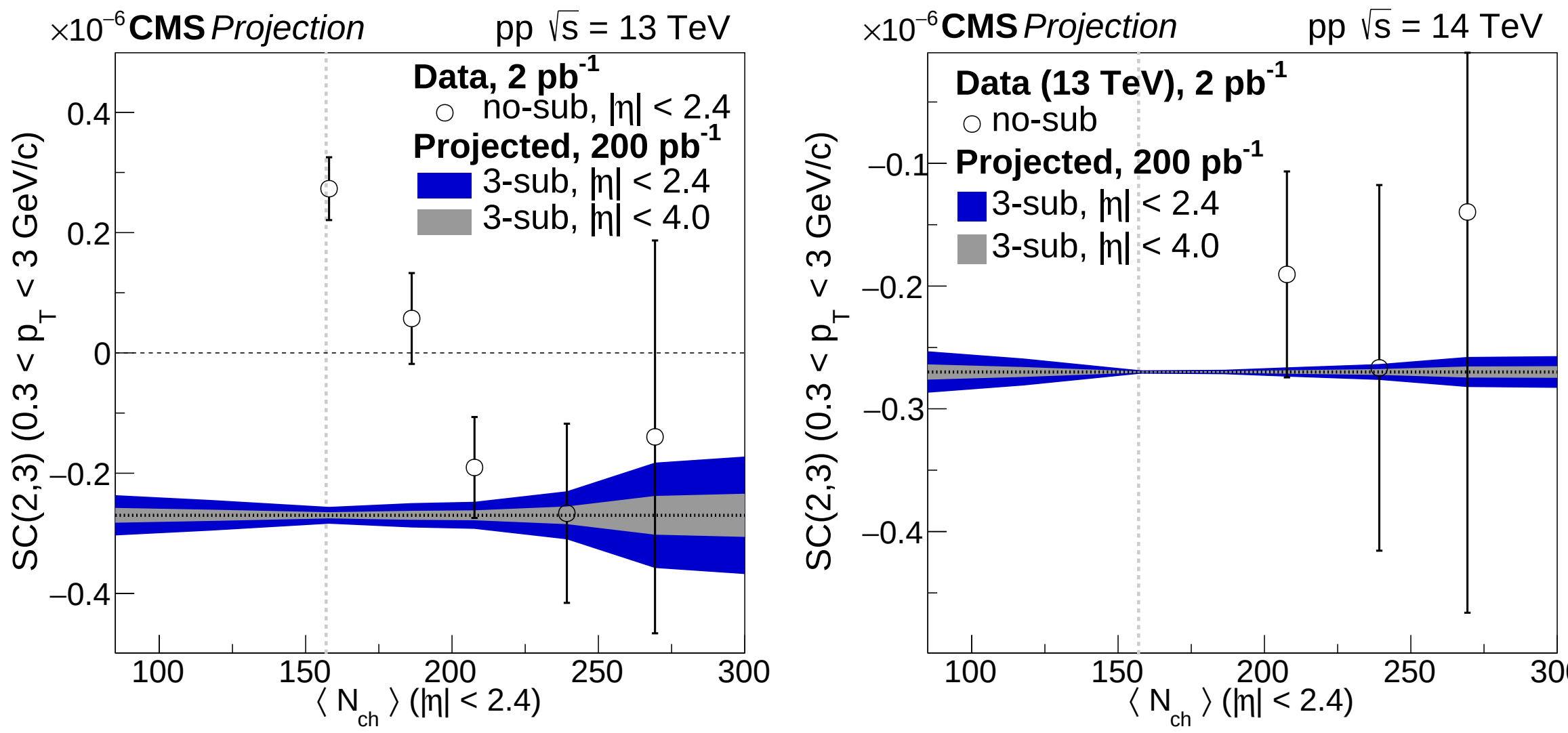

Figure 3: SC(2,3) as a function of total multiplicity in pp collisions at 13 and 14 TeV using the 3-subevent method. Only statistical uncertainties are displayed. The open circles show the current CMS results using the standard 4-particle cumulant method [2]. The vertical dashed line shows the multiplicity range above which data were collected using a high-multiplicity trigger. The blue shaded area is the projection for the current CMS tracker acceptance and the gray shaded area is the projection for CMS extended tracker acceptance.

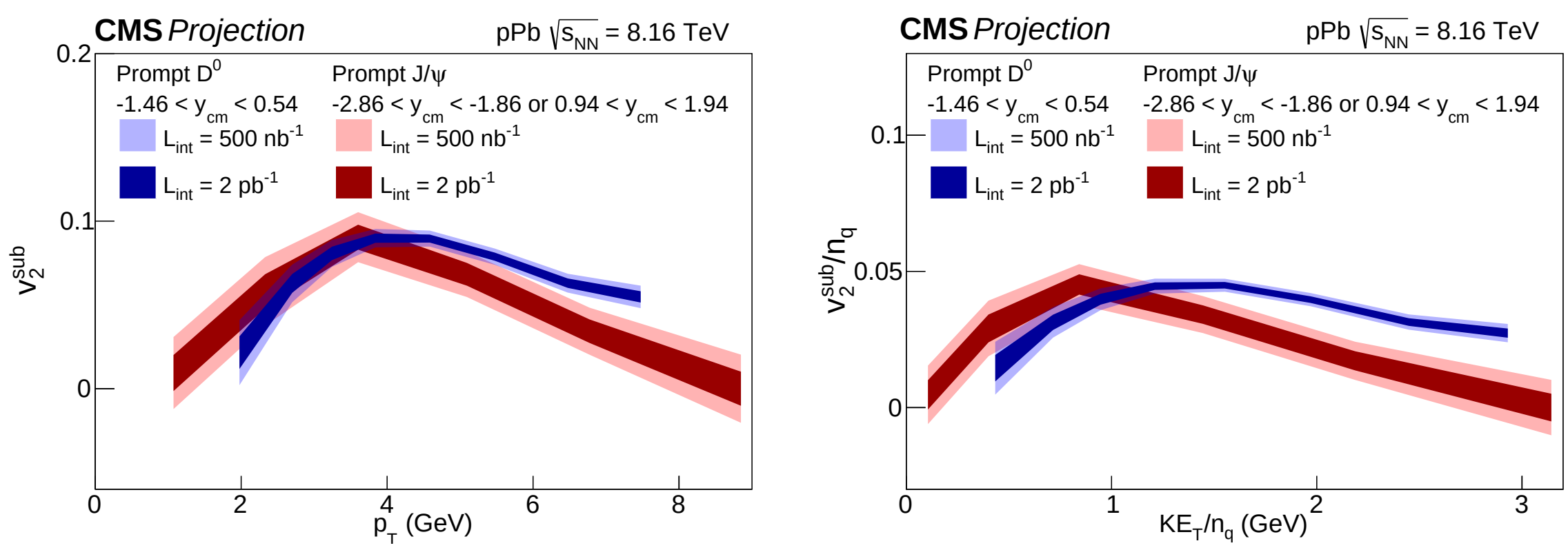

Figure 4: (Left) HL-LHC projections for $D^0$ and $J/\Psi$ elliptic flow as a function of $p_T$. (Right) Elliptic flow projections scaled by the number of constituent quarks ($n_q$) as a function of the similarly scaled transverse kinetic energy ($KE_T/n_q$). Only statistical uncertainties are displayed.

**6**

# Prospects for Measurements of Photon-Induced Processes in Ultra-Peripheral Collisions of Heavy Ions with the ATLAS Detector in the LHC Runs 3 and 4

The ATLAS Collaboration

This note describes results which could be expected for measurements of photon-photon interactions in lead-lead collisions by the ATLAS experiment in the LHC Runs 3 and 4. The potential of light-by-light scattering, exclusive di-muon production and axion-like particle searches are discussed with 10 $\mathrm{nb}^{-1}$ of data.

# 1 Introduction

One of the main goals of ultra-relativistic heavy-ion collisions is to study the properties of the deconfined matter, called Quark-Gluon Plasma (QGP) produced in the hadronic collisions of two nuclei. However, since nuclei have electric charge $Ze$ ($e$ is the electron charge and $Z$ is the atomic number) and are accelerated to nearly the speed of light, they generate extremely strong electromagnetic (EM) fields. These EM fields can also interact either with another nucleus or with its EM fields. Therefore, besides nuclear hadronic interactions, EM interations also occur in ultra-relativistic heavy-ion collisions. These EM interactions can be studied in the ultra-peripheral collisions (UPC) which occur when the distance between two nuclei in the transverse plane is larger than two times the nuclear radius, and hadronic interactions are thus suppressed [1].

The ATLAS Collaboration measured di-muons produced from two-photon interactions [2, 3] in 5.02 TeV lead-lead (Pb+Pb) collisions and established first evidence of light-by-light (LbyL) scattering [4] using a data set of 0.48 $\text{nb}^{-1}$ data collected in 2015. The latter allowed to put the most stringent limits to date on axion-like particle (ALP) production [5, 6] in $\gamma\gamma$ interactions in the invariant mass range of 10-100 GeV.

This note presents studies of LbyL scattering, exclusive production of di-muon pairs and the potential of ALP searches in UPC interactions of the Pb+Pb system in the upgraded ATLAS detector with an expected integrated luminosity of 10 $\text{nb}^{-1}$ delivered in LHC Runs 3 and 4. The projections presented in this note are derived for Pb+Pb collisions with $\sqrt{s_{\text{NN}}}$ =5.02 TeV and 10 $\text{nb}^{-1}$. The results presented here are not strongly sensitive to the slightly higher energy value, if these collisions are delivered at $\sqrt{s_{\text{NN}}} = 5.52$ TeV.

# 2 ATLAS Detector

The ATLAS detector is described in detail in Ref. [7]. Significant upgrades to the detector are expected to be performed for Run 4. In particular a new all-silicon tracking detector, the Inner Tracker (ITk) [8], will be installed for the high-luminosity LHC (HL-LHC) data taking. It is designed to cope with the very high pile-up rates, radiation doses, occupancies, and data transmission rates at the HL-LHC. The amount of inactive material will be significantly reduced in comparison to the current Inner Detector. This will lead to a smaller probability of photon conversions to electron-positron pairs. In the ITk the probability for a photon in the central barrel region $|\eta| < 1.0$ (barrel–endcap transition region $1.52 < |\eta| < 2.37$) to convert within a radius of 1.2 m from the interaction point is expected to be approximately 22% (38%) [8]. In comparison, in the current Inner Detector as many as 60% of the photons convert into an electron-positron pair before reaching the face of the calorimeter [7].

The ITk detector will cover eight units in pseudorapidity, $|\eta| < 4$, thus significantly extending the geometric tracking acceptance of the ATLAS detector beyond that available in the LHC Runs 1-3 ($|\eta| < 2.5$). This increased acceptance of the ITk will be beneficial for the measurement of charged particles produced in heavy-ion collisions. The design and sensor layout of the ITk detector is described in Refs. [9] and [10]. Tracking performance expectations for the ITk in 14 TeV $pp$ collisions are detailed in Ref. [8], and those for 5.02 TeV Pb+Pb collisions are discussed in Ref. [11].

## 3 Simulated Samples

Several simulated samples are used for the studies in this note. The LbyL signal, $\gamma\gamma \to \gamma\gamma$, and backgrounds originating from central exclusive production (CEP) of photon pairs, $gg \to \gamma\gamma$, are generated using SuperChic 2.06 [12, 13], while STARLight 2.0 [14] is used for exclusive production of di-lepton pairs, $\gamma\gamma \to \ell\ell$ with $\ell = e, \mu$, and the ALP signal, $\gamma\gamma \to \mathrm{a} \to \gamma\gamma$. In both generators, the maximum energy for coherent photons emitted from a relativistic nucleus is approximately $\gamma\hbar c/R$, where $\gamma$ is the Lorentz factor of the nucleus and $R$ is the nuclear radius. This is about 75 GeV in the 5.02 TeV Pb+Pb system ($\gamma = 2705$). For the ALP studies, several axion mass slices are generated in the region 5 GeV $< m_{\mathrm{a}} <$ 150 GeV.

The generated samples are passed through a full simulation of the ATLAS detector using GEANT4 [15, 16], and the simulated events are reconstructed using the same algorithms as used for $pp$ performance studies with a minimum track $p_{\mathrm{T}}$ requirement of 300 MeV as described in Ref. [8]. In this study the "Inclined duals" geometry layout of the ITk is used, as described in Ref. [9]. Trigger simulation is not available in the samples utilised for this note.

In this note two types of results are discussed. Projections based on generator-level information are denoted hereafter as truth-level information, and studies based on the full ATLAS simulation are referred to as reconstructed-level results.

## 4 Exclusive Di-Muon Production

The exclusive production of di-muon pairs, $\gamma\gamma \to \mu^+\mu^-$, was measured by ATLAS for invariant masses of the di-muon system between 10-100 GeV [2]. Given the substantially increased statistics, the measurement will be precision-like in the LHC Runs 3 and 4, thus it is supposed to provide a calibration of the initial photon flux and can be used to constrain predictions for the other processes covered in this note. The cross section at high pair mass is also sensitive to the nuclear geometry assumed in the calculations.

Both trigger and reconstruction efficiencies can be determined using data-driven techniques on the same dataset, which will lead to reduced systematic uncertainties given the expected statistics. In this note, the $\gamma\gamma \to \mu^+\mu^-$ process is calculated using truth-level quantities, as kinematic bin migration effects have been found to be small.

Figure 1 presents a differential cross section as a function of the invariant mass of the di-muon system in the range of 10-200 GeV with expected statistical uncertainties represented by two bands corresponding to integrated luminosities of 0.5 nb$^{-1}$ and 10 nb$^{-1}$. Two scenarios are considered for the nuclear geometry: a realistic skin depth of the nucleus or a hard sphere [17]. For the 10 nb$^{-1}$ scenario, a significant reduction of the statistical uncertainty is expected for the highest $m_{\mu\mu}$ bin which spans 100-200 GeV. This will help in reducing uncertainties from the modeling of the nuclear charge distributions. The expected upgrades of the Zero Degree Calorimeters (ZDC) in the LHC Run 3 will also be important for isolating the contributions from dissociation.

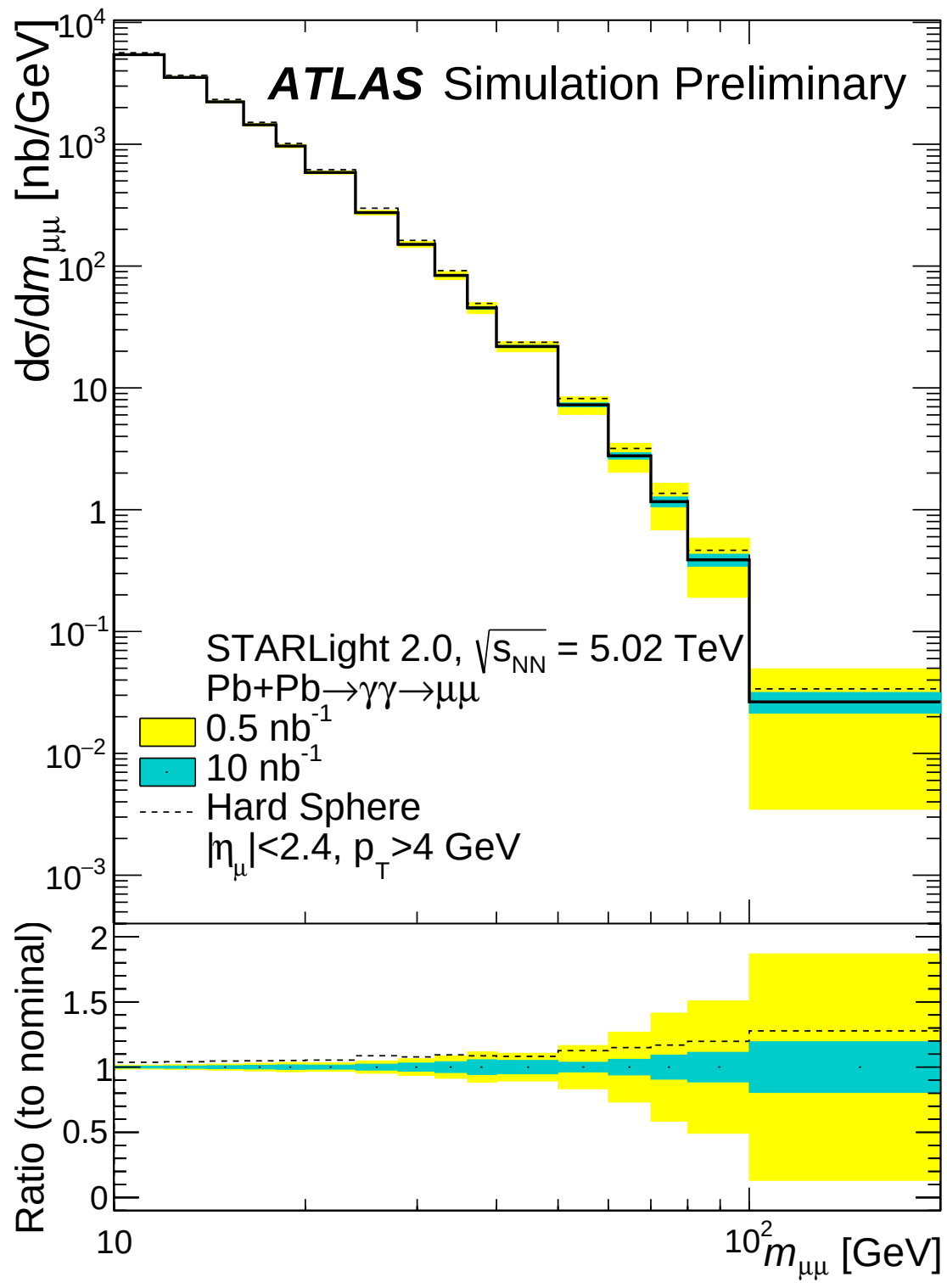

Figure 1: (Upper) Differential cross section for exclusive production of the di-muon pairs as a function of the di-muon mass for $10 < m_{\mu\mu} < 200$ GeV extracted from STARLight. Two scenarios are considered for the nuclear geometry: a realistic skin depth of the nucleus (solid line) or a hard sphere (dashed line). (Bottom) Ratio to nominal as a function of the di-muon mass, where "nominal" stands for the realistic skin depth of the nucleus. Shaded bands represent expected statistical uncertainties for integrated luminosity of 0.5 nb$^{-1}$ (yellow), and 10 nb$^{-1}$ (cyan).

## 5 Light-by-Light Scattering

The LbyL process is first studied using the truth-level quantities associated with signal photons provided in the simulated signal sample. Figure 2 presents a differential cross section as a function of the di-photon rapidity for LbyL scattering for photons with $p_T^\gamma > 2.5$ GeV or $p_T^\gamma > 2.0$ GeV, and $|\eta^\gamma| < 4$. LbyL scattering occurs in the central region: 91% of the integrated cross section is within $|\eta^\gamma| < 2.37$ . A strong dependence on the $p_T^\gamma$ is however observed. The cross section increases by a factor of two when the single photon $p_T^\gamma$ threshold is lowered by half a GeV from 2.5 to 2.0 GeV. The corresponding integrated cross sections in the fiducial region are 112 nb for $p_T^\gamma > 2.5$ GeV and 221 nb for $p_T^\gamma > 2.0$ GeV.

At the reconstructed-level, LbyL event candidates are selected using the requirements which were optimised for the LbyL analysis performed in 5.02 TeV Pb+Pb collisions by ATLAS [4], which are as follows:

- Two photon candidates passing loose identification criteria with $p_T^\gamma > 2.5$ GeV and $|\eta^\gamma| < 2.37$,
- Di-photon invariant mass ($m_{\gamma\gamma}$) greater than 5 GeV,

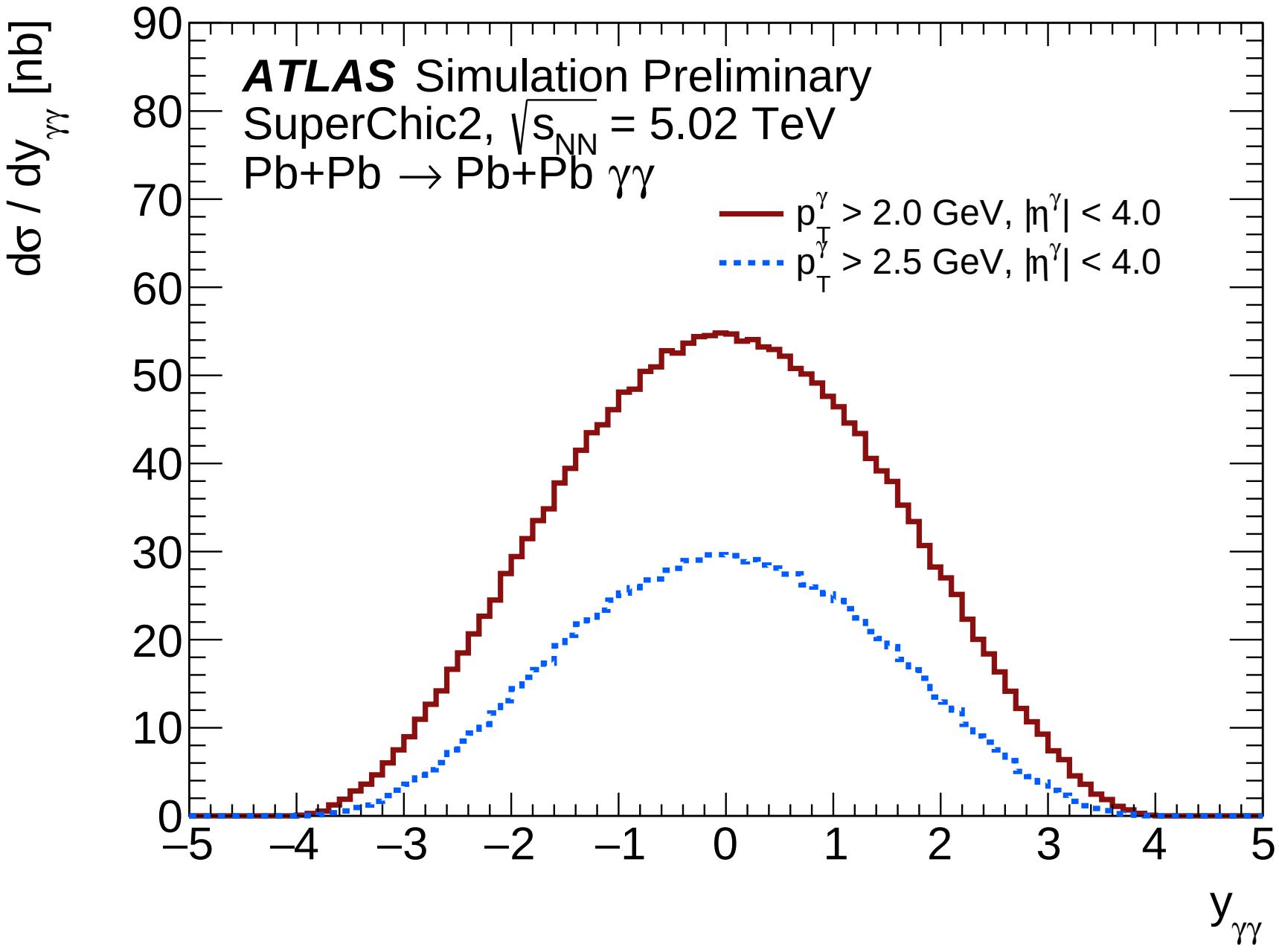

Figure 2: Predicted differential cross section as a function of the di-photon rapidity for LbyL scattering for photons with $p_{\mathrm{T}}^{\gamma} > 2.5$ GeV (dashed) or $p_{\mathrm{T}}^{\gamma} > 2.0$ GeV (solid), and $|\eta^{\gamma}| < 4$ extracted from SuperChic2.

- A veto on the presence of any charged-particle track with $p_{\mathrm{T}} > 300$ MeV, $|\eta| < 4.0$ and having at least one hit in the pixel detector,
- Transverse momentum of the di-photon system ($p_{\mathrm{T}}^{\gamma\gamma}$) below 2 GeV,
- Acoplanarity ($= |1 - \frac{\Delta\phi}{\pi}|$) smaller than 0.01, where $\Delta\phi$ is a difference in the azimuthal angle between two photons.

The track-veto requirement exploits the full acceptance of the ITk detector. Due to lack of the trigger information in the simulated samples, no trigger selection is imposed at the reconstructed-level. However, with the upgraded Run-3 trigger system [18], capabilities should be in place to allow triggering on low-$p_{\mathrm{T}}$ di-photon events with high efficiency.

Figure 3 shows acoplanarity and invariant mass distributions for the di-photon system from LbyL signal and two background components originating from exclusive production of di-electron pairs and di-photons produced in CEP reaction. The analysis is sensitive to a very particular subset of di-electron events where both electrons are mis-identified as photons in the ATLAS detector. In particular the acoplanarity distribution was proven to be powerful in discriminating between signal and background processes in the previous LbyL studies [4]. In the left panel of Figure 3 the acoplanarity distribution is shown.

About 640 LbyL events pass the selection requirements for acoplanarity below 0.01 in 5.02 TeV Pb+Pb collisions with an integrated luminosity of 10 nb$^{-1}$, in comparison to 13 events observed in the 2015 data analysis. The signal events are peaked at acoplanarities close to zero, while the background processes are distributed either uniformly (di-photons from CEP) or even grow with acoplanarity ($e^+e^-$ pairs from exclusive di-electron production). The rise of the number of di-electron pairs with acoplanarity, which was not seen in the 2015 LbyL analysis, is driven by the lower $p_{\mathrm{T}}^{\gamma}$ threshold of 2.5 GeV and the higher track $p_T$ requirement of 300 MeV on the track veto in comparison to the 3 GeV and 100 MeV requirements, respectively, imposed in the 2015 LbyL analysis.

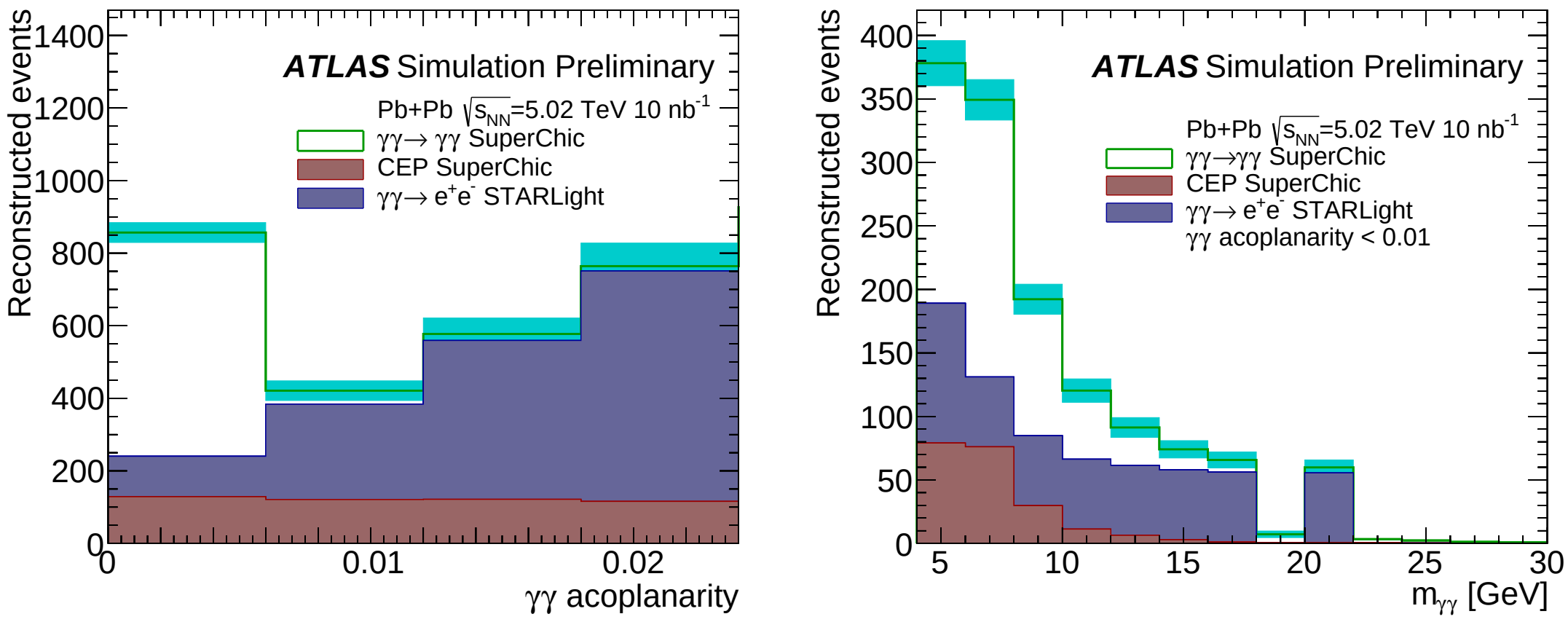

Figure 3: Reco-level acoplanarity (left) and invariant mass (right) distributions of the di-photon system for photons from the LbyL signal and background processes in 5.02 TeV Pb+Pb collisions with an integrated luminosity of 10 nb$^{-1}$. The shaded band in cyan represents expected statistical uncertainties.

## 6 Searching for Axion-like Particles

Axions and ALP are fundamental components of extensions of the Standard Model, occurring in most solutions of the strong CP problem [19, 20]. In addition, ALP with masses below the MeV scale could have a wide range of implications for cosmology and astrophysics. In particular, they are good candidates for cold dark matter [21], which could affect the thermal evolution of the universe, the Cosmic Microwave Background [22] or lead to astrophysical anomalies, such as the observed transparency of the universe to very high energy $\gamma$-rays [23].

Recently an increasing interest has also been paid to ALP masses above 1 GeV [24]. In this mass range, ALP are largely irrelevant for cosmology but they can have a number of implications for general physics. Indeed, the Higgs discovery has set spin zero particles in the spotlight of searches for new physics, with scalar and pseudo-scalar particles (elementary or not) as heralds of new phenomena. An interesting feature is that ALP (generically labeled as $a$ in the following) in this mass range would induce an anomalous contribution to the LbyL, via the reaction: $\gamma\gamma \to a \to \gamma\gamma$, under the condition that the magnitudes of the EM fields associated with the incident photon are large enough, typically $\left|\vec{E}\right| > 10^{18}$ V/m. This has triggered the study presented in Ref. [5], and then in Ref. [6] using the recent observation of LbyL scattering published by the ATLAS experiment in Pb+Pb collisions [4], where the electric field produced by the ultra-relativistic Pb is of the order of $10^{25}$ V/m (thus satisfying the above condition). Then, the photon-ALP interaction can be described by a Lagrangian density of the form [24]:

$$\mathcal{L}_{a\gamma\gamma} = \frac{1}{4\Lambda} a F^{\mu\nu} \tilde{F}_{\mu\nu} = \frac{1}{\Lambda} a \mathbf{E} \cdot \mathbf{B},$$

where $a$ is the massive scalar ALP field (of mass $m_a$) and $1/\Lambda$ is the coupling of the interaction (the dimension of $\Lambda$ is energy). This means in particular that the equation of motion of the field $a$ reads:

6

$(\partial_\mu \partial^\mu + m^2)a = -\frac{1}{\Lambda}\,\mathbf{E}\cdot\mathbf{B}$. This is the physical link between $a$ and the EM fields. Let us note that the EM fields themselves satisfy the Maxwell equations modified by the presence of the ALP field. In the high energy limit, for the narrow-width approximation, one can suppose that $\sigma_{\gamma\gamma\to a\to\gamma\gamma}$ is non-zero only when the invariant mass of the two photons is equal to $m_a \pm \Gamma$, where $\Gamma$ is the decay width of the ALP. The cross section can be shown to have the form [24]:

$$\sigma_{\gamma\gamma\to a\to\gamma\gamma} \propto \frac{1}{\Lambda^2}\mathcal{B}_{a\to\gamma\gamma},$$

where $\mathcal{B}_{a\to\gamma\gamma}$ is the branching ratio of the ALP into photons. Thus, one can write: $\mathcal{B}_{a\to\gamma\gamma} = \Gamma(a \to \gamma\gamma)/\Gamma$, where $\Gamma(a \to \gamma\gamma)$ is the minimal decay width of the ALP into photons given by $\Gamma = m_a^3/(64\pi\Lambda^2)$. In the extraction of the limit for the coupling in $1/\Lambda$ as a function of $m_a$, the convention used in the previous collider searches is followed. It assumes that $\mathcal{B}_{a\to\gamma\gamma} = 1$ [5, 24]. If this assumption is removed, the corresponding exclusion regions would correspond to smaller ranges of $1/\Lambda$. One would get a lower signal rate as the total decay width would increase for decreasing $\mathcal{B}_{a\to\gamma\gamma}$ (as $\Gamma(a \to \gamma\gamma)$ is fixed). Therefore, the region of the ALP (bump) search shrinks with $\mathcal{B}_{a\to\gamma\gamma}$ and the exclusion region is less optimal.

The potential of ALP searches in UPC Pb+Pb collisions is studied using reconstructed-level quantities with the selection requirements described in Section 5. In addition to these selection criteria, and in order to increase the sensitivity to the ALP signal, a requirement of $p_{\mathrm{T}}^{\gamma}/m_{\gamma\gamma} > 0.35$ is applied. The overall selection efficiency (times acceptance) relative to generated event increases from about 40% to 65% for ALP masses ranging from 7 GeV to 80 GeV. Also, the mass resolution varies from 0.5 GeV at low masses (below 15 GeV) up to 1 GeV for larger masses. The invariant mass distribution is used as the discriminating variable, with bin widths comparable to the expected resolution of a narrow resonant signal. A binned likelihood function is constructed in each bin of the $m_{\gamma\gamma}$ distribution from the Poisson probability of the sum of the contributions of the background and of a hypothetical signal of strength relative to the benchmark model. This likelihood function is used to set limits on the presence of a signal. A systematic uncertainty of 25% is considered for the shape of the LbyL background distribution as well as a systematic uncertainty of 20% on the normalisation of the total cross section for this background. Also, the uncertainty on the integrated luminosity is taken to be 6%. Uncertainties related to the knowledge of initial photon fluxes, potentially affecting the ALP signal acceptance and efficiency, are not included in this analysis. These normalisation uncertainties could be further constrained depending on results of the measurements presented in Sections 4 and 5. The systematic uncertainties enter as nuisance parameters with Gaussian or log-normal prior distributions, in convolution with the nominal background distribution.

In Figure 4 the expected mass distributions for three ALP signal mass values, and the main background from LbyL are shown. In this study, other sources of backgrounds are neglected, since they have been found to be small in the LbyL measurement [4].

Upper limits are set on the product of the production cross section of new resonances and their decay branching ratio into $\gamma\gamma$. Exclusion intervals are derived using the CLs method [25] in the asymptotic approximation. The limit set on the signal strength $\mu$ is then translated into a limit on the signal cross section times branching ratio and the coupling, as presented in Figure 5. The branching ratio is taken to be 1 (see above). These limits are found to be compatible with the expected limits estimated in Ref. [5].

In Figure 6 the exclusion limit extracted above along with the existing exclusion limits from the compilation presented in Ref. [26] are presented. Sensitivity of this analysis covers the range in ALP masses between 7 GeV and 140 GeV, where the previous analysis [5] is also shown (labeled as ATLAS 2016 in the figure).

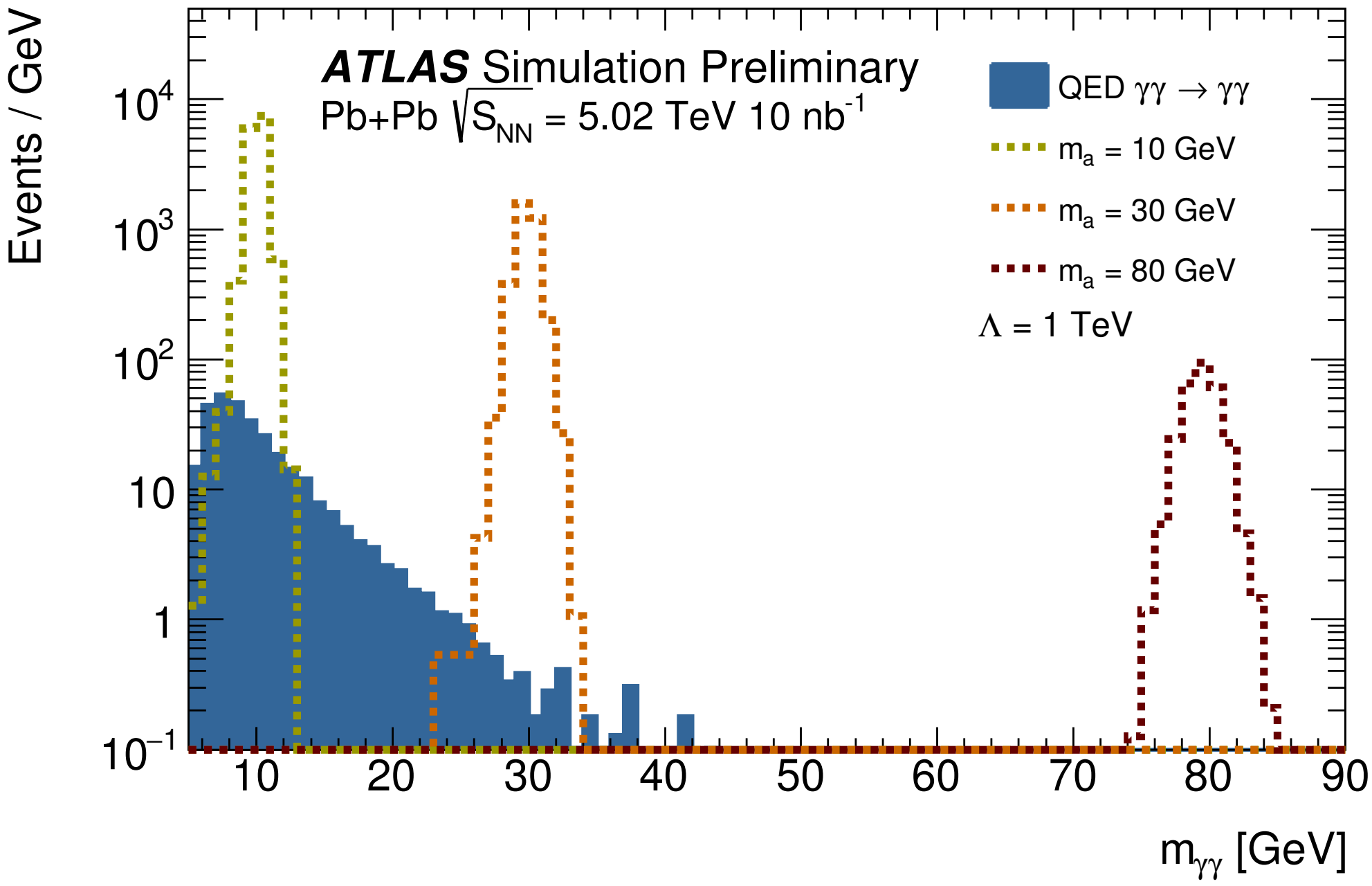

Figure 4: Mass distribution for the ALP signal shown for three values of the ALP mass: $m_a = 10, 30$ and 80 GeV (in red). Also shown (in blue) the LbyL background (see text). All ALP mass points are generated with $\Lambda = 1$ TeV.

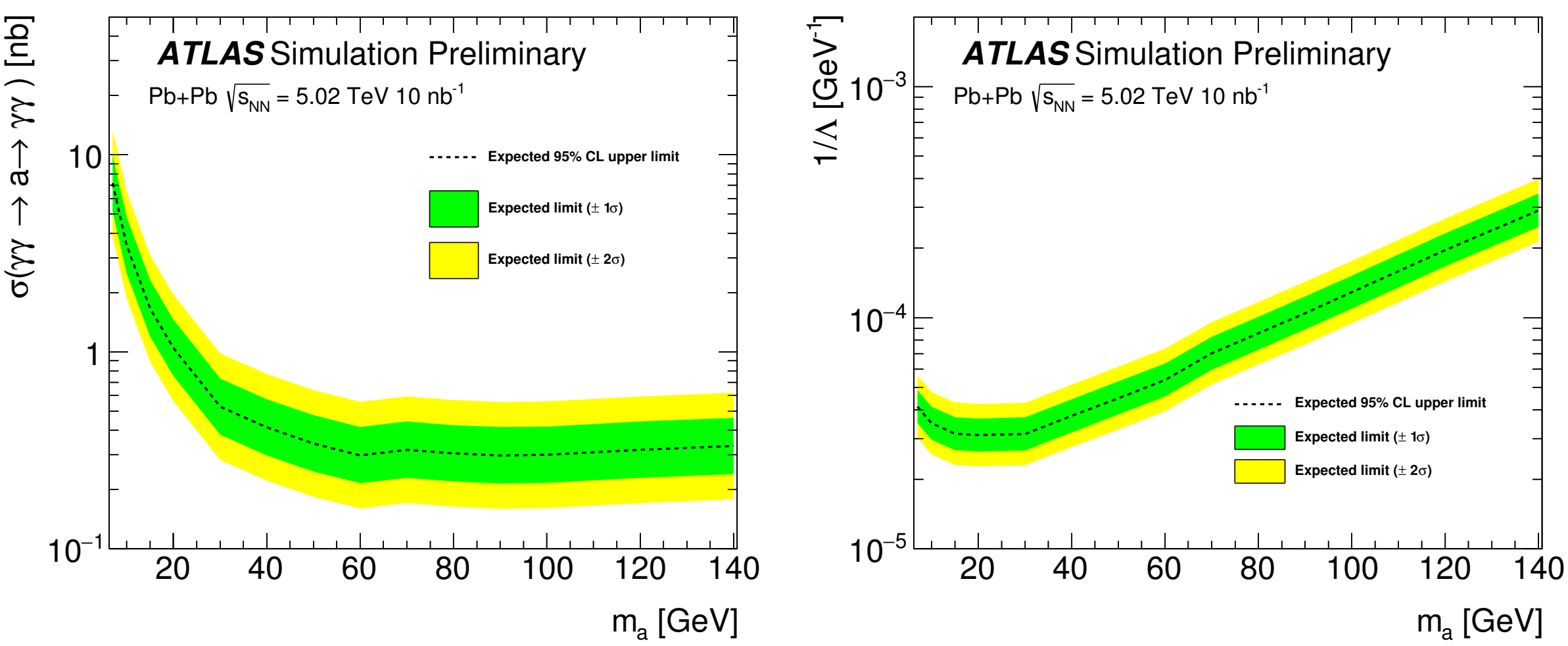

Figure 5: Expected 95% CLs upper limits on $\sigma_{a\to\gamma\gamma}$ (left) and the coupling (right).

8

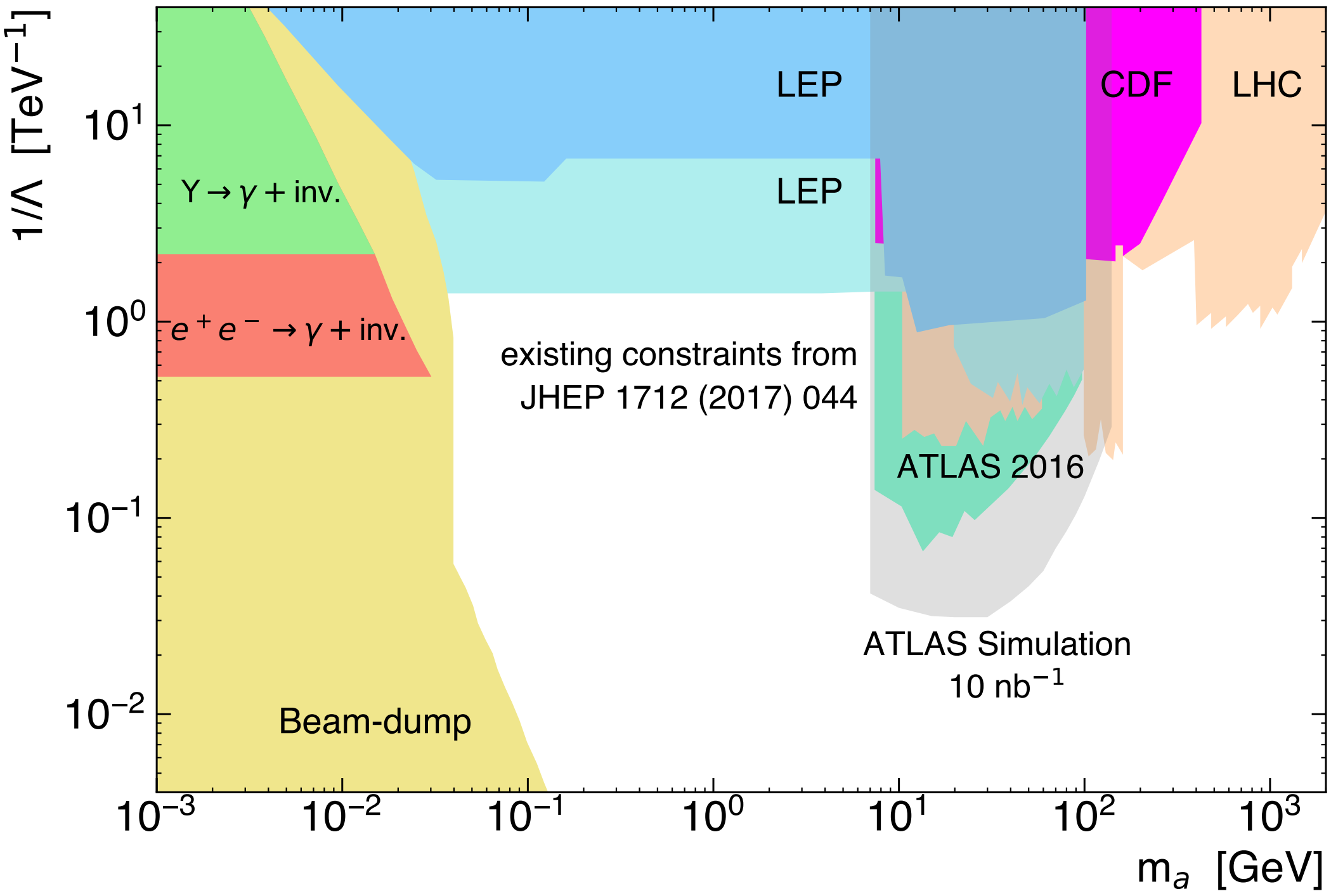

Figure 6: Compilation of exclusion limits obtained by different experiments (see text). ATLAS 2016 represents the exclusion limit derived from the recent LbyL cross section measured in Pb+Pb collisions by ATLAS. In light grey, ATLAS 10 $\text{nb}^{-1}$ is shown corresponding to the analysis described in this document. A more complete version of the existing constraints on ALPs masses versus coupling, including the constraints in the sub meV range from astrophysical observations and from dedicated experiments such as CAST can be found in Ref. [24].

# 7 Conclusion

Several processes involving $\gamma\gamma$ interactions in ultra-peripheral collisions of lead nuclei at 5.02 TeV are studied using the simulation of future upgrade of the ATLAS detector. The expected integrated luminosity of 10 $\text{nb}^{-1}$ leads to improvements in precision for measurements which suffer from lack of statistics in 2015 lead-lead collisions (e.g. light-by-light scattering). Also a potential of observing axion-like particles is discussed with axion masses covering the region of 7-140 GeV. Expected limits on the axion production cross section and coupling are provided.

# ATLAS and CMS Author Lists

## The ATLAS Collaboration

G. Aad[100], B. Abbott[126], D.C. Abbott[101], O. Abdinov[13,*], A. Abed Abud[69a,69b], D.K. Abhayasinghe[92], S.H. Abidi[165], O.S. AbouZeid[39], N.L. Abraham[154], H. Abramowicz[159], H. Abreu[158], Y. Abulaiti[6], B.S. Acharya[65a,65b,o], B.Achkar Achkar[52], S. Adachi[161], L. Adam[98], L. Adamczyk[82a], L. Adamek[165], J. Adelman[120], M. Adersberger[113], A. Adiguzel[12c,ag], S. Adorni[53], T. Adye[142], A.A. Affolder[144], Y. Afik[158], C. Agapopoulou[130], M.N. Agaras[37], A. Aggarwal[118], C. Agheorghiesei[27c], J.A. Aguilar-Saavedra[138f,138a,af], F. Ahmadov[78], X. Ai[15a], G. Aielli[72a,72b], S. Akatsuka[84], T.P.A. Åkesson[95], E. Akilli[53], A.V. Akimov[109], K. Al Khoury[130], G.L. Alberghi[23b,23a], J. Albert[174], M.J. Alconada Verzini[87], S. Alderweireldt[118], M. Aleksa[35], I.N. Aleksandrov[78], C. Alexa[27b], D. Alexandre[19], T. Alexopoulos[10], A. Alfonsi[119], M. Alhroob[126], B. Ali[140], G. Alimonti[67a], J. Alison[36], S.P. Alkire[146], C. Allaire[130], B.M.M. Allbrooke[154], B.W. Allen[129], P.P. Allport[21], A. Aloisio[68a,68b], A. Alonso[39], F. Alonso[87], C. Alpigiani[146], A.A. Alshehri[56], M. Alvarez Estevez[97], B. Alvarez Gonzalez[35], D. Álvarez Piqueras[172], M.G. Alviggi[68a,68b], Y. Amaral Coutinho[79b], A. Ambler[102], L. Ambroz[133], C. Amelung[26], D. Amidei[104], S.P. Amor Dos Santos[138a], S. Amoroso[45], C.S. Amrouche[53], F. An[77], C. Anastopoulos[147], N. Andari[143], T. Andeen[11], C.F. Anders[60b], J.K. Anders[20], A. Andreazza[67a,67b], V. Andrei[60a], C.R. Anelli[174], S. Angelidakis[37], I. Angelozzi[119], A. Angerami[38], A.V. Anisenkov[121b,121a], A. Annovi[70a], C. Antel[60a], M.T. Anthony[147], M. Antonelli[50], D.J.A. Antrim[169], F. Anulli[71a], M. Aoki[80], J.A. Aparisi Pozo[172], L. Aperio Bella[35], G. Arabidze[105], J.P. Araque[138a], V. Araujo Ferraz[79b], R. Araujo Pereira[79b], C. Arcangeletti[50], A.T.H. Arce[48], F.A. Arduh[87], J-F. Arguin[108], S. Argyropoulos[76], J.-H. Arling[45], A.J. Armbruster[35], L.J. Armitage[91], A. Armstrong[169], O. Arnaez[165], H. Arnold[119], A. Artamonov[110,*], G. Artoni[133], S. Artz[98], S. Asai[161], N. Asbah[58], E.M. Asimakopoulou[170], L. Asquith[154], K. Assamagan[29], R. Astalos[28a], R.J. Atkin[32a], M. Atkinson[171], N.B. Atlay[149], H. Atmani[130], K. Augsten[140], G. Avolio[35], R. Avramidou[59a], M.K. Ayoub[15a], A.M. Azoulay[166b], G. Azuelos[108,au], M.J. Baca[21], H. Bachacou[143], K. Bachas[66a,66b], M. Backes[133], F. Backman[44a,44b], P. Bagnaia[71a,71b], M. Bahmani[83], H. Bahrasemani[150], A.J. Bailey[172], V.R. Bailey[171], J.T. Baines[142], M. Bajic[39], C. Bakalis[10], O.K. Baker[181], P.J. Bakker[119], D. Bakshi Gupta[8], S. Balaji[155], E.M. Baldin[121b,121a], P. Balek[178], F. Balli[143], W.K. Balunas[133], J. Balz[98], E. Banas[83], A. Bandyopadhyay[24], S. Banerjee[179,k], A.A.E. Bannoura[180], L. Barak[159], W.M. Barbe[37], E.L. Barberio[103], D. Barberis[54b,54a], M. Barbero[100], T. Barillari[114], M-S. Barisits[35], J. Barkeloo[129], T. Barklow[151], R. Barnea[158], S.L. Barnes[59c], B.M. Barnett[142], R.M. Barnett[18], Z. Barnovska-Blenessy[59a], A. Baroncelli[59a], G. Barone[29], A.J. Barr[133], L. Barranco Navarro[172], F. Barreiro[97], J. Barreiro Guimarães da Costa[15a], R. Bartoldus[151], G. Bartolini[100], A.E. Barton[88], P. Bartos[28a], A. Basalaev[45], A. Bassalat[130], R.L. Bates[56], S.J. Batista[165], S. Batlamous[34e], J.R. Batley[31], B. Batool[149], M. Battaglia[144], M. Bauce[71a,71b], F. Bauer[143], K.T. Bauer[169], H.S. Bawa[151], J.B. Beacham[124], T. Beau[134], P.H. Beauchemin[168], F. Becherer[51], P. Bechtle[24], H.C. Beck[52], H.P. Beck[20,r], K. Becker[51], M. Becker[98], C. Becot[45], A. Beddall[12d], A.J. Beddall[12a], V.A. Bednyakov[78], M. Bedognetti[119], C.P. Bee[153], T.A. Beermann[75], M. Begalli[79b], M. Begel[29], A. Behera[153], J.K. Behr[45], F. Beisiegel[24], A.S. Bell[93], G. Bella[159], L. Bellagamba[23b], A. Bellerive[33], P. Bellos[9], K. Beloborodov[121b,121a], K. Belotskiy[111], N.L. Belyaev[111], D. Benchekroun[34a], N. Benekos[10], Y. Benhammou[159], D.P. Benjamin[6], M. Benoit[53],

J.R. Bensinger[26], S. Bentvelsen[119], L. Beresford[133], M. Beretta[50], D. Berge[45],
E. Bergeaas Kuutmann[170], N. Berger[5], B. Bergmann[140], L.J. Bergsten[26], J. Beringer[18],
S. Berlendis[7], N.R. Bernard[101], G. Bernardi[134], C. Bernius[151], F.U. Bernlochner[24], T. Berry[92],
P. Berta[98], C. Bertella[15a], I.A. Bertram[88], G.J. Besjes[39], O. Bessidskaia Bylund[180], N. Besson[143],
A. Bethani[99], S. Bethke[114], A. Betti[24], A.J. Bevan[91], J. Beyer[114], R. Bi[137], R.M. Bianchi[137],
O. Biebel[113], D. Biedermann[19], R. Bielski[35], K. Bierwagen[98], N.V. Biesuz[70a,70b], M. Biglietti[73a],
T.R.V. Billoud[108], M. Bindi[52], A. Bingul[12d], C. Bini[71a,71b], S. Biondi[23b,23a], M. Birman[178],
T. Bisanz[52], J.P. Biswal[159], A. Bitadze[99], C. Bittrich[47], K. Bjørke[132], K.M. Black[25], T. Blazek[28a],
I. Bloch[45], C. Blocker[26], A. Blue[56], U. Blumenschein[91], G.J. Bobbink[119],
V.S. Bobrovnikov[121b,121a], S.S. Bocchetta[95], A. Bocci[48], D. Boerner[45], D. Bogavac[14],
A.G. Bogdanchikov[121b,121a], C. Bohm[44a], V. Boisvert[92], P. Bokan[52,170], T. Bold[82a],
A.S. Boldyrev[112], A.E. Bolz[60b], M. Bomben[134], M. Bona[91], J.S. Bonilla[129], M. Boonekamp[143],
H.M. Borecka-Bielska[89], A. Borisov[122], G. Borissov[88], J. Bortfeldt[35], D. Bortoletto[133],
V. Bortolotto[72a,72b], D. Boscherini[23b], M. Bosman[14], J.D. Bossio Sola[102], K. Bouaouda[34a],
J. Boudreau[137], E.V. Bouhova-Thacker[88], D. Boumediene[37], C. Bourdarios[130], S.K. Boutle[56],
A. Boveia[124], J. Boyd[35], D. Boye[32b,ao], I.R. Boyko[78], A.J. Bozson[92], J. Bracinik[21], N. Brahimi[100],
G. Brandt[180], O. Brandt[60a], F. Braren[45], U. Bratzler[162], B. Brau[101], J.E. Brau[129],
W.D. Breaden Madden[56], K. Brendlinger[45], L. Brenner[45], R. Brenner[170], S. Bressler[178],
B. Brickwedde[98], D.L. Briglin[21], D. Britton[56], D. Britzger[114], I. Brock[24], R. Brock[105],
G. Brooijmans[38], W.K. Brooks[145b], E. Brost[120], J.H Broughton[21], P.A. Bruckman de Renstrom[83],
D. Bruncko[28b], A. Bruni[23b], G. Bruni[23b], L.S. Bruni[119], S. Bruno[72a,72b], B.H. Brunt[31],
M. Bruschi[23b], N. Bruscino[137], P. Bryant[36], L. Bryngemark[95], T. Buanes[17], Q. Buat[35],
P. Buchholz[149], A.G. Buckley[56], I.A. Budagov[78], M.K. Bugge[132], F. Bührer[51], O. Bulekov[111],
T.J. Burch[120], S. Burdin[89], C.D. Burgard[119], A.M. Burger[127], B. Burghgrave[8], K. Burka[83],
J.T.P. Burr[45], V. Büscher[98], E. Buschmann[52], P. Bussey[56], J.M. Butler[25], C.M. Buttar[56],
J.M. Butterworth[93], P. Butti[35], W. Buttinger[35], A. Buzatu[156], A.R. Buzykaev[121b,121a],
G. Cabras[23b,23a], S. Cabrera Urbán[172], D. Caforio[55], H. Cai[171], V.M.M. Cairo[151], O. Cakir[4a],
N. Calace[35], P. Calafiura[18], A. Calandri[100], G. Calderini[134], P. Calfayan[64], G. Callea[56],
L.P. Caloba[79b], S. Calvente Lopez[97], D. Calvet[37], S. Calvet[37], T.P. Calvet[153], M. Calvetti[70a,70b],
R. Camacho Toro[134], S. Camarda[35], D. Camarero Munoz[97], P. Camarri[72a,72b], D. Cameron[132],
R. Caminal Armadans[101], C. Camincher[35], S. Campana[35], M. Campanelli[93], A. Camplani[39],
A. Campoverde[149], V. Canale[68a,68b], A. Canesse[102], M. Cano Bret[59c], J. Cantero[127], T. Cao[159],
Y. Cao[171], M.D.M. Capeans Garrido[35], M. Capua[40b,40a], R. Cardarelli[72a], F.C. Cardillo[147],
I. Carli[141], T. Carli[35], G. Carlino[68a], B.T. Carlson[137], L. Carminati[67a,67b], R.M.D. Carney[44a,44b],
S. Caron[118], E. Carquin[145b], S. Carrá[67a,67b], J.W.S. Carter[165], M.P. Casado[14,g], A.F. Casha[165],
D.W. Casper[169], R. Castelijn[119], F.L. Castillo[172], V. Castillo Gimenez[172], N.F. Castro[138a,138e],
A. Catinaccio[35], J.R. Catmore[132], A. Cattai[35], J. Caudron[24], V. Cavaliere[29], E. Cavallaro[14],
D. Cavalli[67a], M. Cavalli-Sforza[14], V. Cavasinni[70a,70b], E. Celebi[12b], L. Cerda Alberich[172],
A.S. Cerqueira[79a], A. Cerri[154], L. Cerrito[72a,72b], F. Cerutti[18], A. Cervelli[23b,23a], S.A. Cetin[12b],
D. Chakraborty[120], S.K. Chan[58], W.S. Chan[119], W.Y. Chan[89], J.D. Chapman[31],
B. Chargeishvili[157b], D.G. Charlton[21], T.P. Charman[91], C.C. Chau[33], S. Che[124],
A. Chegwidden[105], S. Chekanov[6], S.V. Chekulaev[166a], G.A. Chelkov[78,at], M.A. Chelstowska[35],
B. Chen[77], C. Chen[59a], C.H. Chen[77], H. Chen[29], J. Chen[59a], J. Chen[38], S. Chen[135], S.J. Chen[15c],

2

X. Chen[15b,as], Y. Chen[81], Y-H. Chen[45], H.C. Cheng[62a], H.J. Cheng[15d], A. Cheplakov[78], E. Cheremushkina[122], R. Cherkaoui El Moursli[34e], E. Cheu[7], K. Cheung[63], T.J.A. Chevalérias[143], L. Chevalier[143], V. Chiarella[50], G. Chiarelli[70a], G. Chiodini[66a], A.S. Chisholm[35,21], A. Chitan[27b], I. Chiu[161], Y.H. Chiu[174], M.V. Chizhov[78], K. Choi[64], A.R. Chomont[130], S. Chouridou[160], Y.S. Chow[119], M.C. Chu[62a], J. Chudoba[139], A.J. Chuinard[102], J.J. Chwastowski[83], L. Chytka[128], K.M. Ciesla[83], D. Cinca[46], V. Cindro[90], I.A. Cioară[27b], A. Ciocio[18], F. Cirotto[68a,68b], Z.H. Citron[178], M. Citterio[67a], D.A. Ciubotaru[27b], B.M. Ciungu[165], A. Clark[53], M.R. Clark[38], P.J. Clark[49], C. Clement[44a,44b], Y. Coadou[100], M. Cobal[65a,65c], A. Coccaro[54b], J. Cochran[77], H. Cohen[159], A.E.C. Coimbra[35], L. Colasurdo[118], B. Cole[38], A.P. Colijn[119], J. Collot[57], P. Conde Muiño[138a,h], E. Coniavitis[51], S.H. Connell[32b], I.A. Connelly[56], S. Constantinescu[27b], F. Conventi[68a,av], A.M. Cooper-Sarkar[133], F. Cormier[173], K.J.R. Cormier[165], L.D. Corpe[93], M. Corradi[71a,71b], E.E. Corrigan[95], F. Corriveau[102,ab], A. Cortes-Gonzalez[35], M.J. Costa[172], F. Costanza[5], D. Costanzo[147], G. Cowan[92], J.W. Cowley[31], J. Crane[99], K. Cranmer[123], S.J. Crawley[56], R.A. Creager[135], S. Crépé-Renaudin[57], F. Crescioli[134], M. Cristinziani[24], V. Croft[119], G. Crosetti[40b,40a], A. Cueto[5], T. Cuhadar Donszelmann[147], A.R. Cukierman[151], S. Czekierda[83], P. Czodrowski[35], M.J. Da Cunha Sargedas De Sousa[59b], J.V. Da Fonseca Pinto[79b], C. Da Via[99], W. Dabrowski[82a], T. Dado[28a], S. Dahbi[34e], T. Dai[104], C. Dallapiccola[101], M. Dam[39], G. D'amen[23b,23a], V. D'Amico[73a,73b], J. Damp[98], J.R. Dandoy[135], M.F. Daneri[30], N.P. Dang[179,k], N.D Dann[99], M. Danninger[173], V. Dao[35], G. Darbo[54b], O. Dartsi[5], A. Dattagupta[129], T. Daubney[45], S. D'Auria[67a,67b], W. Davey[24], C. David[45], T. Davidek[141], D.R. Davis[48], E. Dawe[103], I. Dawson[147], K. De[8], R. De Asmundis[68a], M. De Beurs[119], S. De Castro[23b,23a], S. De Cecco[71a,71b], N. De Groot[118], P. de Jong[119], H. De la Torre[105], A. De Maria[15c], D. De Pedis[71a], A. De Salvo[71a], U. De Sanctis[72a,72b], M. De Santis[72a,72b], A. De Santo[154], K. De Vasconcelos Corga[100], J.B. De Vivie De Regie[130], C. Debenedetti[144], D.V. Dedovich[78], M. Del Gaudio[40b,40a], J. Del Peso[97], Y. Delabat Diaz[45], D. Delgove[130], F. Deliot[143], C.M. Delitzsch[7], M. Della Pietra[68a,68b], D. Della Volpe[53], A. Dell'Acqua[35], L. Dell'Asta[72a,72b], M. Delmastro[5], C. Delporte[130], P.A. Delsart[57], D.A. DeMarco[165], S. Demers[181], M. Demichev[78], G. Demontigny[108], S.P. Denisov[122], D. Denysiuk[119], L. D'Eramo[134], D. Derendarz[83], J.E. Derkaoui[34d], F. Derue[134], P. Dervan[89], K. Desch[24], C. Deterre[45], K. Dette[165], C. Deutsch[24], M.R. Devesa[30], P.O. Deviveiros[35], A. Dewhurst[142], S. Dhaliwal[26], F.A. Di Bello[53], A. Di Ciaccio[72a,72b], L. Di Ciaccio[5], W.K. Di Clemente[135], C. Di Donato[68a,68b], A. Di Girolamo[35], G. Di Gregorio[70a,70b], B. Di Micco[73a,73b], R. Di Nardo[101], K.F. Di Petrillo[58], R. Di Sipio[165], D. Di Valentino[33], C. Diaconu[100], F.A. Dias[39], T. Dias Do Vale[138a], M.A. Diaz[145a], J. Dickinson[18], E.B. Diehl[104], J. Dietrich[19], S. Díez Cornell[45], A. Dimitrievska[18], W. Ding[15b], J. Dingfelder[24], F. Dittus[35], F. Djama[100], T. Djobava[157b], J.I. Djuvsland[17], M.A.B. Do Vale[79c], M. Dobre[27b], D. Dodsworth[26], C. Doglioni[95], J. Dolejsi[141], Z. Dolezal[141], M. Donadelli[79d], J. Donini[37], A. D'onofrio[91], M. D'Onofrio[89], J. Dopke[142], A. Doria[68a], M.T. Dova[87], A.T. Doyle[56], E. Drechsler[150], E. Dreyer[150], T. Dreyer[52], Y. Duan[59b], F. Dubinin[109], M. Dubovsky[28a], A. Dubreuil[53], E. Duchovni[178], G. Duckeck[113], A. Ducourthial[134], O.A. Ducu[108], D. Duda[114], A. Dudarev[35], A.C. Dudder[98], E.M. Duffield[18], L. Duflot[130], M. Dührssen[35], C. Dülsen[180], M. Dumancic[178], A.E. Dumitriu[27b], A.K. Duncan[56], M. Dunford[60a], A. Duperrin[100], H. Duran Yildiz[4a], M. Düren[55], A. Durglishvili[157b], D. Duschinger[47], B. Dutta[45], D. Duvnjak[1], G. Dyckes[135], M. Dyndal[35], S. Dysch[99], B.S. Dziedzic[83], K.M. Ecker[114], R.C. Edgar[104],

T. Eifert[35], G. Eigen[17], K. Einsweiler[18], T. Ekelof[170], M. El Kacimi[34c], R. El Kosseifi[100], V. Ellajosyula[170], M. Ellert[170], F. Ellinghaus[180], A.A. Elliot[91], N. Ellis[35], J. Elmsheuser[29], M. Elsing[35], D. Emeliyanov[142], A. Emerman[38], Y. Enari[161], J.S. Ennis[176], M.B. Epland[48], J. Erdmann[46], A. Ereditato[20], M. Errenst[35], M. Escalier[130], C. Escobar[172], O. Estrada Pastor[172], E. Etzion[159], H. Evans[64], A. Ezhilov[136], F. Fabbri[56], L. Fabbri[23b,23a], V. Fabiani[118], G. Facini[93], R.M. Faisca Rodrigues Pereira[138a], R.M. Fakhrutdinov[122], S. Falciano[71a], P.J. Falke[5], S. Falke[5], J. Faltova[141], Y. Fang[15a], Y. Fang[15a], G. Fanourakis[43], M. Fanti[67a,67b], A. Farbin[8], A. Farilla[73a], E.M. Farina[69a,69b], T. Farooque[105], S. Farrell[18], S.M. Farrington[176], P. Farthouat[35], F. Fassi[34e], P. Fassnacht[35], D. Fassouliotis[9], M. Faucci Giannelli[49], W.J. Fawcett[31], L. Fayard[130], O.L. Fedin[136,p], W. Fedorko[173], M. Feickert[41], S. Feigl[132], L. Feligioni[100], A. Fell[147], C. Feng[59b], E.J. Feng[35], M. Feng[48], M.J. Fenton[56], A.B. Fenyuk[122], J. Ferrando[45], A. Ferrante[171], A. Ferrari[170], P. Ferrari[119], R. Ferrari[69a], D.E. Ferreira de Lima[60b], A. Ferrer[172], D. Ferrere[53], C. Ferretti[104], F. Fiedler[98], A. Filipčič[90], F. Filthaut[118], K.D. Finelli[25], M.C.N. Fiolhais[138a,a], L. Fiorini[172], F.F. Fischer[113], W.C. Fisher[105], I. Fleck[149], P. Fleischmann[104], R.R.M. Fletcher[135], T. Flick[180], B.M. Flierl[113], L.M. Flores[135], L.R. Flores Castillo[62a], F.M. Follega[74a,74b], N. Fomin[17], G.T. Forcolin[74a,74b], A. Formica[143], F.A. Förster[14], A.C. Forti[99], A.G. Foster[21], M.G. Foti[133], D. Fournier[130], H. Fox[88], P. Francavilla[70a,70b], M. Franchini[23b,23a], S. Franchino[60a], D. Francis[35], L. Franconi[20], M. Franklin[58], A.N. Fray[91], B. Freund[108], W.S. Freund[79b], E.M. Freundlich[46], D.C. Frizzell[126], D. Froidevaux[35], J.A. Frost[133], C. Fukunaga[162], E. Fullana Torregrosa[172], E. Fumagalli[54b,54a], T. Fusayasu[115], J. Fuster[172], A. Gabrielli[23b,23a], A. Gabrielli[18], G.P. Gach[82a], S. Gadatsch[53], P. Gadow[114], G. Gagliardi[54b,54a], L.G. Gagnon[108], C. Galea[27b], B. Galhardo[138a], G.E. Gallardo[133], E.J. Gallas[133], B.J. Gallop[142], P. Gallus[140], G. Galster[39], R. Gamboa Goni[91], K.K. Gan[124], S. Ganguly[178], J. Gao[59a], Y. Gao[89], Y.S. Gao[151,m], C. García[172], J.E. García Navarro[172], J.A. García Pascual[15a], C. Garcia-Argos[51], M. Garcia-Sciveres[18], R.W. Gardner[36], N. Garelli[151], S. Gargiulo[51], V. Garonne[132], A. Gaudiello[54b,54a], G. Gaudio[69a], I.L. Gavrilenko[109], A. Gavrilyuk[110], C. Gay[173], G. Gaycken[24], E.N. Gazis[10], A.A. Geanta[27b], C.N.P. Gee[142], J. Geisen[52], M. Geisen[98], M.P. Geisler[60a], C. Gemme[54b], M.H. Genest[57], C. Geng[104], S. Gentile[71a,71b], S. George[92], T. Geralis[43], D. Gerbaudo[14], L.O. Gerlach[52], P. Gessinger-Befurt[98], G. Gessner[46], S. Ghasemi[149], M. Ghasemi Bostanabad[174], M. Ghneimat[24], A. Ghosh[76], B. Giacobbe[23b], S. Giagu[71a,71b], N. Giangiacomi[23b,23a], P. Giannetti[70a], A. Giannini[68a,68b], S.M. Gibson[92], M. Gignac[144], D. Gillberg[33], G. Gilles[180], D.M. Gingrich[3,au], M.P. Giordani[65a,65c], F.M. Giorgi[23b], P.F. Giraud[143], G. Giugliarelli[65a,65c], D. Giugni[67a], F. Giuli[72a,72b], S. Gkaitatzis[160], I. Gkialas[9,j], E.L. Gkougkousis[14], P. Gkountoumis[10], L.K. Gladilin[112], C. Glasman[97], J. Glatzer[14], P.C.F. Glaysher[45], A. Glazov[45], M. Goblirsch-Kolb[26], S. Goldfarb[103], T. Golling[53], D. Golubkov[122], A. Gomes[138a,138b], R. Goncalves Gama[52], R. Gonçalo[138a,138b], G. Gonella[51], L. Gonella[21], A. Gongadze[78], F. Gonnella[21], J.L. Gonski[58], S. González de la Hoz[172], S. Gonzalez-Sevilla[53], G.R. Gonzalvo Rodriguez[172], L. Goossens[35], P.A. Gorbounov[110], H.A. Gordon[29], B. Gorini[35], E. Gorini[66a,66b], A. Gorišek[90], A.T. Goshaw[48], C. Gössling[46], M.I. Gostkin[78], C.A. Gottardo[24], M. Gouighri[34a], D. Goujdami[34c], A.G. Goussiou[146], N. Govender[32b,c], C. Goy[5], E. Gozani[158], I. Grabowska-Bold[82a], E.C. Graham[89], J. Gramling[169], E. Gramstad[132], S. Grancagnolo[19], M. Grandi[154], V. Gratchev[136], P.M. Gravila[27f], F.G. Gravili[66a,66b], C. Gray[56], H.M. Gray[18], C. Grefe[24], K. Gregersen[95], I.M. Gregor[45], P. Grenier[151], K. Grevtsov[45], N.A. Grieser[126], J. Griffiths[8], A.A. Grillo[144],

4

K. Grimm[151,b], S. Grinstein[14,w], J.-F. Grivaz[130], S. Groh[98], E. Gross[178], J. Grosse-Knetter[52], Z.J. Grout[93], C. Grud[104], A. Grummer[117], L. Guan[104], W. Guan[179], J. Guenther[35], A. Guerguichon[130], F. Guescini[166a], D. Guest[169], R. Gugel[51], B. Gui[124], T. Guillemin[5], S. Guindon[35], U. Gul[56], J. Guo[59c], W. Guo[104], Y. Guo[59a,s], Z. Guo[100], R. Gupta[45], S. Gurbuz[12c], G. Gustavino[126], P. Gutierrez[126], C. Gutschow[93], C. Guyot[143], M.P. Guzik[82a], C. Gwenlan[133], C.B. Gwilliam[89], A. Haas[123], C. Haber[18], H.K. Hadavand[8], N. Haddad[34e], A. Hadef[59a], S. Hageböck[35], M. Hagihara[167], M. Haleem[175], J. Haley[127], G. Halladjian[105], G.D. Hallewell[100], K. Hamacher[180], P. Hamal[128], K. Hamano[174], H. Hamdaoui[34e], G.N. Hamity[147], K. Han[59a,ai], L. Han[59a], S. Han[15d], K. Hanagaki[80,u], M. Hance[144], D.M. Handl[113], B. Haney[135], R. Hankache[134], P. Hanke[60a], E. Hansen[95], J.B. Hansen[39], J.D. Hansen[39], M.C. Hansen[24], P.H. Hansen[39], E.C. Hanson[99], K. Hara[167], A.S. Hard[179], T. Harenberg[180], S. Harkusha[106], P.F. Harrison[176], N.M. Hartmann[113], Y. Hasegawa[148], A. Hasib[49], S. Hassani[143], S. Haug[20], R. Hauser[105], L.B. Havener[38], M. Havranek[140], C.M. Hawkes[21], R.J. Hawkings[35], D. Hayden[105], C. Hayes[153], R.L. Hayes[173], C.P. Hays[133], J.M. Hays[91], H.S. Hayward[89], S.J. Haywood[142], F. He[59a], M.P. Heath[49], V. Hedberg[95], L. Heelan[8], S. Heer[24], K.K. Heidegger[51], J. Heilman[33], S. Heim[45], T. Heim[18], B. Heinemann[45,ap], J.J. Heinrich[129], L. Heinrich[123], C. Heinz[55], J. Hejbal[139], L. Helary[60b], A. Held[173], S. Hellesund[132], C.M. Helling[144], S. Hellman[44a,44b], C. Helsens[35], R.C.W. Henderson[88], Y. Heng[179], S. Henkelmann[173], A.M. Henriques Correia[35], G.H. Herbert[19], H. Herde[26], V. Herget[175], Y. Hernández Jiménez[32c], H. Herr[98], M.G. Herrmann[113], T. Herrmann[47], G. Herten[51], R. Hertenberger[113], L. Hervas[35], T.C. Herwig[135], G.G. Hesketh[93], N.P. Hessey[166a], A. Higashida[161], S. Higashino[80], E. Higón-Rodriguez[172], K. Hildebrand[36], E. Hill[174], J.C. Hill[31], K.K. Hill[29], K.H. Hiller[45], S.J. Hillier[21], M. Hils[47], I. Hinchliffe[18], F. Hinterkeuser[24], M. Hirose[131], S. Hirose[51], D. Hirschbuehl[180], B. Hiti[90], O. Hladik[139], D.R. Hlaluku[32c], X. Hoad[49], J. Hobbs[153], N. Hod[178], M.C. Hodgkinson[147], A. Hoecker[35], F. Hoenig[113], D. Hohn[51], D. Hohov[130], T.R. Holmes[36], M. Holzbock[113], L.B.A.H Hommels[31], S. Honda[167], T. Honda[80], T.M. Hong[137], A. Hönle[114], B.H. Hooberman[171], W.H. Hopkins[6], Y. Horii[116], P. Horn[47], A.J. Horton[150], L.A. Horyn[36], J-Y. Hostachy[57], A. Hostiuc[146], S. Hou[156], A. Hoummada[34a], J. Howarth[99], J. Hoya[87], M. Hrabovsky[128], J. Hrdinka[75], I. Hristova[19], J. Hrivnac[130], A. Hrynevich[107], T. Hryn'ova[5], P.J. Hsu[63], S.-C. Hsu[146], Q. Hu[29], S. Hu[59c], Y. Huang[15a], Z. Hubacek[140], F. Hubaut[100], M. Huebner[24], F. Huegging[24], T.B. Huffman[133], M. Huhtinen[35], R.F.H. Hunter[33], P. Huo[153], A.M. Hupe[33], N. Huseynov[78,ad], J. Huston[105], J. Huth[58], R. Hyneman[104], S. Hyrych[28a], G. Iacobucci[53], G. Iakovidis[29], I. Ibragimov[149], L. Iconomidou-Fayard[130], Z. Idrissi[34e], P. Iengo[35], R. Ignazzi[39], O. Igonkina[119,y], R. Iguchi[161], T. Iizawa[53], Y. Ikegami[80], M. Ikeno[80], D. Iliadis[160], N. Ilic[118], F. Iltzsche[47], G. Introzzi[69a,69b], M. Iodice[73a], K. Iordanidou[38], V. Ippolito[71a,71b], M.F. Isacson[170], N. Ishijima[131], M. Ishino[161], M. Ishitsuka[163], W. Islam[127], C. Issever[133], S. Istin[158], F. Ito[167], J.M. Iturbe Ponce[62a], R. Iuppa[74a,74b], A. Ivina[178], H. Iwasaki[80], J.M. Izen[42], V. Izzo[68a], P. Jacka[139], P. Jackson[1], R.M. Jacobs[24], V. Jain[2], G. Jäkel[180], K.B. Jakobi[98], K. Jakobs[51], S. Jakobsen[75], T. Jakoubek[139], J. Jamieson[56], R. Jansky[53], J. Janssen[24], M. Janus[52], P.A. Janus[82a], G. Jarlskog[95], N. Javadov[78,ad], T. Javůrek[35], M. Javurkova[51], F. Jeanneau[143], L. Jeanty[129], J. Jejelava[157a,ae], A. Jelinskas[176], P. Jenni[51,d], J. Jeong[45], N. Jeong[45], S. Jézéquel[5], H. Ji[179], J. Jia[153], H. Jiang[77], Y. Jiang[59a], Z. Jiang[151,q], S. Jiggins[51], F.A. Jimenez Morales[37], J. Jimenez Pena[172], S. Jin[15c], A. Jinaru[27b], O. Jinnouchi[163], H. Jivan[32c], P. Johansson[147], K.A. Johns[7], C.A. Johnson[64],

R.W.L. Jones[88], S.D. Jones[154], S. Jones[7], T.J. Jones[89], J. Jongmanns[60a], P.M. Jorge[138a], J. Jovicevic[166a], X. Ju[18], J.J. Junggeburth[114], A. Juste Rozas[14,w], A. Kaczmarska[83], M. Kado[130], H. Kagan[124], M. Kagan[151], T. Kaji[177], E. Kajomovitz[158], C.W. Kalderon[95], A. Kaluza[98], A. Kamenshchikov[122], L. Kanjir[90], Y. Kano[161], V.A. Kantserov[111], J. Kanzaki[80], L.S. Kaplan[179], D. Kar[32c], M.J. Kareem[166b], E. Karentzos[10], S.N. Karpov[78], Z.M. Karpova[78], V. Kartvelishvili[88], A.N. Karyukhin[122], L. Kashif[179], R.D. Kass[124], A. Kastanas[44a,44b], Y. Kataoka[161], C. Kato[59d,59c], J. Katzy[45], K. Kawade[81], K. Kawagoe[86], T. Kawaguchi[116], T. Kawamoto[161], G. Kawamura[52], E.F. Kay[174], V.F. Kazanin[121b,121a], R. Keeler[174], R. Kehoe[41], J.S. Keller[33], E. Kellermann[95], J.J. Kempster[21], J. Kendrick[21], O. Kepka[139], S. Kersten[180], B.P. Kerševan[90], S. Ketabchi Haghighat[165], M. Khader[171], F. Khalil-Zada[13], M.K. Khandoga[143], A. Khanov[127], A.G. Kharlamov[121b,121a], T. Kharlamova[121b,121a], E.E. Khoda[173], A. Khodinov[164], T.J. Khoo[53], E. Khramov[78], J. Khubua[157b], S. Kido[81], M. Kiehn[53], C.R. Kilby[92], Y.K. Kim[36], N. Kimura[65a,65c], O.M. Kind[19], B.T. King[89], D. Kirchmeier[47], J. Kirk[142], A.E. Kiryunin[114], T. Kishimoto[161], D.P. Kisliuk[165], V. Kitali[45], O. Kivernyk[5], E. Kladiva[28b,*], T. Klapdor-Kleingrothaus[51], M.H. Klein[104], M. Klein[89], U. Klein[89], K. Kleinknecht[98], P. Klimek[120], A. Klimentov[29], T. Klingl[24], T. Klioutchnikova[35], F.F. Klitzner[113], P. Kluit[119], S. Kluth[114], E. Kneringer[75], E.B.F.G. Knoops[100], A. Knue[51], D. Kobayashi[86], T. Kobayashi[161], M. Kobel[47], M. Kocian[151], P. Kodys[141], P.T. Koenig[24], T. Koffas[33], N.M. Köhler[114], T. Koi[151], M. Kolb[60b], I. Koletsou[5], T. Komarek[128], T. Kondo[80], N. Kondrashova[59c], K. Köneke[51], A.C. König[118], T. Kono[80], R. Konoplich[123,al], V. Konstantinides[93], N. Konstantinidis[93], B. Konya[95], R. Kopeliansky[64], S. Koperny[82a], K. Korcyl[83], K. Kordas[160], G. Koren[159], A. Korn[93], I. Korolkov[14], E.V. Korolkova[147], N. Korotkova[112], O. Kortner[114], S. Kortner[114], T. Kosek[141], V.V. Kostyukhin[24], A. Kotwal[48], A. Koulouris[10], A. Kourkoumeli-Charalampidi[69a,69b], C. Kourkoumelis[9], E. Kourlitis[147], V. Kouskoura[29], A.B. Kowalewska[83], R. Kowalewski[174], C. Kozakai[161], W. Kozanecki[143], A.S. Kozhin[122], V.A. Kramarenko[112], G. Kramberger[90], D. Krasnopevtsev[59a], M.W. Krasny[134], A. Krasznahorkay[35], D. Krauss[114], J.A. Kremer[82a], J. Kretzschmar[89], P. Krieger[165], F. Krieter[113], A. Krishnan[60b], K. Krizka[18], K. Kroeninger[46], H. Kroha[114], J. Kroll[139], J. Kroll[135], J. Krstic[16], U. Kruchonak[78], H. Krüger[24], N. Krumnack[77], M.C. Kruse[48], T. Kubota[103], S. Kuday[4b], J.T. Kuechler[45], S. Kuehn[35], A. Kugel[60a], T. Kuhl[45], V. Kukhtin[78], R. Kukla[100], Y. Kulchitsky[106,ah], S. Kuleshov[145b], Y.P. Kulinich[171], M. Kuna[57], T. Kunigo[84], A. Kupco[139], T. Kupfer[46], O. Kuprash[51], H. Kurashige[81], L.L. Kurchaninov[166a], Y.A. Kurochkin[106], A. Kurova[111], M.G. Kurth[15d], E.S. Kuwertz[35], M. Kuze[163], A.K. Kvam[146], J. Kvita[128], T. Kwan[102], A. La Rosa[114], L. La Rotonda[40b,40a], F. La Ruffa[40b,40a], C. Lacasta[172], F. Lacava[71a,71b], D.P.J. Lack[99], H. Lacker[19], D. Lacour[134], E. Ladygin[78], R. Lafaye[5], B. Laforge[134], T. Lagouri[32c], S. Lai[52], S. Lammers[64], W. Lampl[7], C. Lampoudis[160], E. Lançon[29], U. Landgraf[51], M.P.J. Landon[91], M.C. Lanfermann[53], V.S. Lang[45], J.C. Lange[52], R.J. Langenberg[35], A.J. Lankford[169], F. Lanni[29], K. Lantzsch[24], A. Lanza[69a], A. Lapertosa[54b,54a], S. Laplace[134], J.F. Laporte[143], T. Lari[67a], F. Lasagni Manghi[23b,23a], M. Lassnig[35], T.S. Lau[62a], A. Laudrain[130], A. Laurier[33], M. Lavorgna[68a,68b], M. Lazzaroni[67a,67b], B. Le[103], O. Le Dortz[134], E. Le Guirriec[100], M. LeBlanc[7], T. LeCompte[6], F. Ledroit-Guillon[57], C.A. Lee[29], G.R. Lee[17], L. Lee[58], S.C. Lee[156], S.J. Lee[33], B. Lefebvre[166a], M. Lefebvre[174], F. Legger[113], C. Leggett[18], K. Lehmann[150], N. Lehmann[180], G. Lehmann Miotto[35], W.A. Leight[45], A. Leisos[160,v], M.A.L. Leite[79d], R. Leitner[141], D. Lellouch[178], K.J.C. Leney[41], T. Lenz[24], B. Lenzi[35], R. Leone[7],

S. Leone[70a], C. Leonidopoulos[49], A. Leopold[134], G. Lerner[154], C. Leroy[108], R. Les[165],
C.G. Lester[31], M. Levchenko[136], J. Levêque[5], D. Levin[104], L.J. Levinson[178], D.J. Lewis[21], B. Li[15b],
B. Li[104], C-Q. Li[59a], F. Li[59c], H. Li[59a], H. Li[59b], J. Li[59c], K. Li[151], L. Li[59c], M. Li[15a], Q. Li[15d],
Q.Y. Li[59a], S. Li[59d,59c], X. Li[45], Y. Li[45], Z. Li[59b], Z. Liang[15a], B. Liberti[72a], A. Liblong[165],
K. Lie[62c], S. Liem[119], C.Y. Lin[31], K. Lin[105], T.H. Lin[98], R.A. Linck[64], J.H. Lindon[21],
A.L. Lionti[53], E. Lipeles[135], A. Lipniacka[17], M. Lisovyi[60b], T.M. Liss[171,ar], A. Lister[173],
A.M. Litke[144], J.D. Little[8], B. Liu[77,aa], B.L Liu[6], H.B. Liu[29], H. Liu[104], J.B. Liu[59a], J.K.K. Liu[133],
K. Liu[134], M. Liu[59a], P. Liu[18], Y. Liu[15a], Y.L. Liu[104], Y.W. Liu[59a], M. Livan[69a,69b], A. Lleres[57],
J. Llorente Merino[15a], S.L. Lloyd[91], C.Y. Lo[62b], F. Lo Sterzo[41], E.M. Lobodzinska[45], P. Loch[7],
S. Loffredo[72a,72b], T. Lohse[19], K. Lohwasser[147], M. Lokajicek[139], J.D. Long[171], R.E. Long[88],
L. Longo[35], K.A. Looper[124], J.A. Lopez[145b], I. Lopez Paz[99], A. Lopez Solis[147], J. Lorenz[113],
N. Lorenzo Martinez[5], M. Losada[22], P.J. Lösel[113], A. Lösle[51], X. Lou[45], X. Lou[15a], A. Lounis[130],
J. Love[6], P.A. Love[88], J.J. Lozano Bahilo[172], M. Lu[59a], Y.J. Lu[63], H.J. Lubatti[146], C. Luci[71a,71b],
A. Lucotte[57], C. Luedtke[51], F. Luehring[64], I. Luise[134], L. Luminari[71a], B. Lund-Jensen[152],
M.S. Lutz[101], D. Lynn[29], R. Lysak[139], E. Lytken[95], F. Lyu[15a], V. Lyubushkin[78],
T. Lyubushkina[78], H. Ma[29], L.L. Ma[59b], Y. Ma[59b], G. Maccarrone[50], A. Macchiolo[114],
C.M. Macdonald[147], J. Machado Miguens[135], D. Madaffari[172], R. Madar[37], W.F. Mader[47],
N. Madysa[47], J. Maeda[81], K. Maekawa[161], S. Maeland[17], T. Maeno[29], M. Maerker[47],
A.S. Maevskiy[112], V. Magerl[51], N. Magini[77], D.J. Mahon[38], C. Maidantchik[79b], T. Maier[113],
A. Maio[138a,138b,138d], O. Majersky[28a], S. Majewski[129], Y. Makida[80], N. Makovec[130],
B. Malaescu[134], Pa. Malecki[83], V.P. Maleev[136], F. Malek[57], U. Mallik[76], D. Malon[6], C. Malone[31],
S. Maltezos[10], S. Malyukov[35], J. Mamuzic[172], G. Mancini[50], I. Mandić[90],
L. Manhaes de Andrade Filho[79a], I.M. Maniatis[160], J. Manjarres Ramos[47], K.H. Mankinen[95],
A. Mann[113], A. Manousos[75], B. Mansoulie[143], I. Manthos[160], S. Manzoni[119], A. Marantis[160],
G. Marceca[30], L. Marchese[133], G. Marchiori[134], M. Marcisovsky[139], C. Marcon[95],
C.A. Marin Tobon[35], M. Marjanovic[37], F. Marroquim[79b], Z. Marshall[18], M.U.F Martensson[170],
S. Marti-Garcia[172], C.B. Martin[124], T.A. Martin[176], V.J. Martin[49], B. Martin dit Latour[17],
L. Martinelli[73a,73b], M. Martinez[14,w], V.I. Martinez Outschoorn[101], S. Martin-Haugh[142],
V.S. Martoiu[27b], A.C. Martyniuk[93], A. Marzin[35], L. Masetti[98], T. Mashimo[161], R. Mashinistov[109],
J. Masik[99], A.L. Maslennikov[121b,121a], L.H. Mason[103], L. Massa[72a,72b], P. Massarotti[68a,68b],
P. Mastrandrea[70a,70b], A. Mastroberardino[40b,40a], T. Masubuchi[161], A. Matic[113], P. Mättig[24],
J. Maurer[27b], B. Maček[90], S.J. Maxfield[89], D.A. Maximov[121b,121a], R. Mazini[156], I. Maznas[160],
S.M. Mazza[144], S.P. Mc Kee[104], A. McCarn, Deiana[41], T.G. McCarthy[114], L.I. McClymont[93],
W.P. McCormack[18], E.F. McDonald[103], J.A. Mcfayden[35], G. Mchedlidze[52], M.A. McKay[41],
K.D. McLean[174], S.J. McMahon[142], P.C. McNamara[103], C.J. McNicol[176], R.A. McPherson[174,ab],
J.E. Mdhluli[32c], Z.A. Meadows[101], S. Meehan[146], T.M. Megy[51], S. Mehlhase[113], A. Mehta[89],
T. Meideck[57], B. Meirose[42], D. Melini[172], B.R. Mellado Garcia[32c], J.D. Mellenthin[52], M. Melo[28a],
F. Meloni[45], A. Melzer[24], S.B. Menary[99], E.D. Mendes Gouveia[138a,138e], L. Meng[35], X.T. Meng[104],
S. Menke[114], E. Meoni[40b,40a], S. Mergelmeyer[19], S.A.M. Merkt[137], C. Merlassino[20], P. Mermod[53],
L. Merola[68a,68b], C. Meroni[67a], O. Meshkov[112,109], J.K.R. Meshreki[149], A. Messina[71a,71b],
J. Metcalfe[6], A.S. Mete[169], C. Meyer[64], J. Meyer[158], J-P. Meyer[143], H. Meyer Zu Theenhausen[60a],
F. Miano[154], R.P. Middleton[142], L. Mijović[49], G. Mikenberg[178], M. Mikestikova[139], M. Mikuž[90],
H. Mildner[147], M. Milesi[103], A. Milic[165], D.A. Millar[91], D.W. Miller[36], A. Milov[178],

D.A. Milstead[44a,44b], R.A. Mina[151,q], A.A. Minaenko[122], M. Miñano Moya[172], I.A. Minashvili[157b], A.I. Mincer[123], B. Mindur[82a], M. Mineev[78], Y. Minegishi[161], Y. Ming[179], L.M. Mir[14], A. Mirto[66a,66b], K.P. Mistry[135], T. Mitani[177], J. Mitrevski[113], V.A. Mitsou[172], M. Mittal[59c], A. Miucci[20], P.S. Miyagawa[147], A. Mizukami[80], J.U. Mjörnmark[95], T. Mkrtchyan[182], M. Mlynarikova[141], T. Moa[44a,44b], K. Mochizuki[108], P. Mogg[51], S. Mohapatra[38], R. Moles-Valls[24], M.C. Mondragon[105], K. Mönig[45], J. Monk[39], E. Monnier[100], A. Montalbano[150], J. Montejo Berlingen[35], M. Montella[93], F. Monticelli[87], S. Monzani[67a], N. Morange[130], D. Moreno[22], M. Moreno Llácer[35], P. Morettini[54b], M. Morgenstern[119], S. Morgenstern[47], D. Mori[150], M. Morii[58], M. Morinaga[177], V. Morisbak[132], A.K. Morley[35], G. Mornacchi[35], A.P. Morris[93], L. Morvaj[153], P. Moschovakos[10], B. Moser[119], M. Mosidze[157b], T. Moskalets[143], H.J. Moss[147], J. Moss[151,n], K. Motohashi[163], E. Mountricha[35], E.J.W. Moyse[101], S. Muanza[100], J. Mueller[137], R.S.P. Mueller[113], D. Muenstermann[88], G.A. Mullier[95], J.L. Munoz Martinez[14], F.J. Munoz Sanchez[99], P. Murin[28b], W.J. Murray[176,142], A. Murrone[67a,67b], M. Muškinja[18], C. Mwewa[32a], A.G. Myagkov[122,am], J. Myers[129], M. Myska[140], B.P. Nachman[18], O. Nackenhorst[46], A.Nag Nag[47], K. Nagai[133], K. Nagano[80], Y. Nagasaka[61], M. Nagel[51], E. Nagy[100], A.M. Nairz[35], Y. Nakahama[116], K. Nakamura[80], T. Nakamura[161], I. Nakano[125], H. Nanjo[131], F. Napolitano[60a], R.F. Naranjo Garcia[45], R. Narayan[11], D.I. Narrias Villar[60a], I. Naryshkin[136], T. Naumann[45], G. Navarro[22], H.A. Neal[104,*], P.Y. Nechaeva[109], F. Nechansky[45], T.J. Neep[21], A. Negri[69a,69b], M. Negrini[23b], C. Nellist[52], M.E. Nelson[133], S. Nemecek[139], P. Nemethy[123], M. Nessi[35,f], M.S. Neubauer[171], M. Neumann[180], P.R. Newman[21], T.Y. Ng[62c], Y.S. Ng[19], Y.W.Y. Ng[169], H.D.N. Nguyen[100], T. Nguyen Manh[108], E. Nibigira[37], R.B. Nickerson[133], R. Nicolaidou[143], D.S. Nielsen[39], J. Nielsen[144], N. Nikiforou[11], V. Nikolaenko[122,am], I. Nikolic-Audit[134], K. Nikolopoulos[21], P. Nilsson[29], H.R. Nindhito[53], Y. Ninomiya[80], A. Nisati[71a], N. Nishu[59c], R. Nisius[114], I. Nitsche[46], T. Nitta[177], T. Nobe[161], Y. Noguchi[84], M. Nomachi[131], I. Nomidis[134], M.A. Nomura[29], M. Nordberg[35], N. Norjoharuddeen[133], T. Novak[90], O. Novgorodova[47], R. Novotny[140], L. Nozka[128], K. Ntekas[169], E. Nurse[93], F.G. Oakham[33,au], H. Oberlack[114], J. Ocariz[134], A. Ochi[81], I. Ochoa[38], J.P. Ochoa-Ricoux[145a], K. O'Connor[26], S. Oda[86], S. Odaka[80], S. Oerdek[52], A. Ogrodnik[82a], A. Oh[99], S.H. Oh[48], C.C. Ohm[152], H. Oide[54b,54a], M.L. Ojeda[165], H. Okawa[167], Y. Okazaki[84], Y. Okumura[161], T. Okuyama[80], A. Olariu[27b], L.F. Oleiro Seabra[138a], S.A. Olivares Pino[145a], D. Oliveira Damazio[29], J.L. Oliver[1], M.J.R. Olsson[169], A. Olszewski[83], J. Olszowska[83], D.C. O'Neil[150], A. Onofre[138a,138e], K. Onogi[116], P.U.E. Onyisi[11], H. Oppen[132], M.J. Oreglia[36], G.E. Orellana[87], Y. Oren[159], D. Orestano[73a,73b], N. Orlando[14], R.S. Orr[165], V. O'Shea[56], R. Ospanov[59a], G. Otero y Garzon[30], H. Otono[86], M. Ouchrif[34d], F. Ould-Saada[132], A. Ouraou[143], Q. Ouyang[15a], M. Owen[56], R.E. Owen[21], V.E. Ozcan[12c], N. Ozturk[8], J. Pacalt[128], H.A. Pacey[31], K. Pachal[48], A. Pacheco Pages[14], C. Padilla Aranda[14], S. Pagan Griso[18], M. Paganini[181], G. Palacino[64], S. Palazzo[49], S. Palestini[35], M. Palka[82b], D. Pallin[37], I. Panagoulias[10], C.E. Pandini[35], J.G. Panduro Vazquez[92], P. Pani[45], G. Panizzo[65a,65c], L. Paolozzi[53], C. Papadatos[108], K. Papageorgiou[9,j], A. Paramonov[6], D. Paredes Hernandez[62b], S.R. Paredes Saenz[133], B. Parida[164], T.H. Park[165], A.J. Parker[88], M.A. Parker[31], F. Parodi[54b,54a], E.W.P. Parrish[120], J.A. Parsons[38], U. Parzefall[51], L. Pascual Dominguez[134], V.R. Pascuzzi[165], J.M.P. Pasner[144], E. Pasqualucci[71a], S. Passaggio[54b], F. Pastore[92], P. Pasuwan[44a,44b], S. Pataraia[98], J.R. Pater[99], A. Pathak[179,k], T. Pauly[35], B. Pearson[114], M. Pedersen[132], L. Pedraza Diaz[118], R. Pedro[138a], T. Peiffer[52], S.V. Peleganchuk[121b,121a], O. Penc[139], H. Peng[59a],

8

B.S. Peralva[79a], M.M. Perego[130], A.P. Pereira Peixoto[138a], D.V. Perepelitsa[29], F. Peri[19], L. Perini[67a,67b], H. Pernegger[35], S. Perrella[68a,68b], V.D. Peshekhonov[78,*], K. Peters[45], R.F.Y. Peters[99], B.A. Petersen[35], T.C. Petersen[39], E. Petit[57], A. Petridis[1], C. Petridou[160], P. Petroff[130], M. Petrov[133], F. Petrucci[73a,73b], M. Pettee[181], N.E. Pettersson[101], K. Petukhova[141], A. Peyaud[143], R. Pezoa[145b], L. Pezzotti[69a,69b], T. Pham[103], F.H. Phillips[105], P.W. Phillips[142], M.W. Phipps[171], G. Piacquadio[153], E. Pianori[18], A. Picazio[101], R.H. Pickles[99], R. Piegaia[30], D. Pietreanu[27b], J.E. Pilcher[36], A.D. Pilkington[99], M. Pinamonti[72a,72b], J.L. Pinfold[3], M. Pitt[178], L. Pizzimento[72a,72b], M.-A. Pleier[29], V. Pleskot[141], E. Plotnikova[78], D. Pluth[77], P. Podberezko[121b,121a], R. Poettgen[95], R. Poggi[53], L. Poggioli[130], I. Pogrebnyak[105], D. Pohl[24], I. Pokharel[52], G. Polesello[69a], A. Poley[18], A. Policicchio[71a,71b], R. Polifka[141], A. Polini[23b], C.S. Pollard[45], V. Polychronakos[29], D. Ponomarenko[111], L. Pontecorvo[35], S. Popa[27a], G.A. Popeneciu[27d], D.M. Portillo Quintero[57], S. Pospisil[140], K. Potamianos[45], I.N. Potrap[78], C.J. Potter[31], H. Potti[11], T. Poulsen[95], J. Poveda[35], T.D. Powell[147], G. Pownall[45], M.E. Pozo Astigarraga[35], P. Pralavorio[100], S. Prell[77], D. Price[99], M. Primavera[66a], S. Prince[102], M.L. Proffitt[146], N. Proklova[111], K. Prokofiev[62c], F. Prokoshin[145b], S. Protopopescu[29], J. Proudfoot[6], M. Przybycien[82a], A. Puri[171], P. Puzo[130], J. Qian[104], Y. Qin[99], A. Quadt[52], M. Queitsch-Maitland[45], A. Qureshi[1], P. Rados[103], F. Ragusa[67a,67b], G. Rahal[96], J.A. Raine[53], S. Rajagopalan[29], A. Ramirez Morales[91], K. Ran[15a], T. Rashid[130], S. Raspopov[5], M.G. Ratti[67a,67b], D.M. Rauch[45], F. Rauscher[113], S. Rave[98], B. Ravina[147], I. Ravinovich[178], J.H. Rawling[99], M. Raymond[35], A.L. Read[132], N.P. Readioff[57], M. Reale[66a,66b], D.M. Rebuzzi[69a,69b], A. Redelbach[175], G. Redlinger[29], K. Reeves[42], L. Rehnisch[19], J. Reichert[135], D. Reikher[159], A. Reiss[98], A. Rej[149], C. Rembser[35], M. Renda[27b], M. Rescigno[71a], S. Resconi[67a], E.D. Resseguie[135], S. Rettie[173], E. Reynolds[21], O.L. Rezanova[121b,121a], P. Reznicek[141], E. Ricci[74a,74b], R. Richter[114], S. Richter[45], E. Richter-Was[82b], O. Ricken[24], M. Ridel[134], P. Rieck[114], C.J. Riegel[180], O. Rifki[45], M. Rijssenbeek[153], A. Rimoldi[69a,69b], M. Rimoldi[20], L. Rinaldi[23b], G. Ripellino[152], B. Ristić[88], E. Ritsch[35], I. Riu[14], J.C. Rivera Vergara[145a], F. Rizatdinova[127], E. Rizvi[91], C. Rizzi[35], R.T. Roberts[99], S.H. Robertson[102,ab], M. Robin[45], D. Robinson[31], J.E.M. Robinson[45], A. Robson[56], E. Rocco[98], C. Roda[70a,70b], S. Rodriguez Bosca[172], A. Rodriguez Perez[14], D. Rodriguez Rodriguez[172], A.M. Rodríguez Vera[166b], S. Roe[35], O. Røhne[132], R. Röhrig[114], C.P.A. Roland[64], J. Roloff[58], A. Romaniouk[111], M. Romano[23b,23a], N. Rompotis[89], M. Ronzani[123], L. Roos[134], S. Rosati[71a], K. Rosbach[51], N-A. Rosien[52], G. Rosin[101], B.J. Rosser[135], E. Rossi[45], E. Rossi[73a,73b], E. Rossi[68a,68b], L.P. Rossi[54b], L. Rossini[67a,67b], R. Rosten[14], M. Rotaru[27b], J. Rothberg[146], D. Rousseau[130], G. Rovelli[69a,69b], D. Roy[32c], A. Rozanov[100], Y. Rozen[158], X. Ruan[32c], F. Rubbo[151], F. Rühr[51], A. Ruiz-Martinez[172], A. Rummler[35], Z. Rurikova[51], N.A. Rusakovich[78], H.L. Russell[102], L. Rustige[37,46], J.P. Rutherfoord[7], E.M. Rüttinger[45,l], Y.F. Ryabov[136], M. Rybar[38], G. Rybkin[130], A. Ryzhov[122], G.F. Rzehorz[52], P. Sabatini[52], G. Sabato[119], S. Sacerdoti[130], H.F-W. Sadrozinski[144], R. Sadykov[78], F. Safai Tehrani[71a], B. Safarzadeh Samani[154], P. Saha[120], S. Saha[102], M. Sahinsoy[60a], A. Sahu[180], M. Saimpert[45], M. Saito[161], T. Saito[161], H. Sakamoto[161], A. Sakharov[123,al], D. Salamani[53], G. Salamanna[73a,73b], J.E. Salazar Loyola[145b], P.H. Sales De Bruin[170], D. Salihagic[114,*], A. Salnikov[151], J. Salt[172], D. Salvatore[40b,40a], F. Salvatore[154], A. Salvucci[62a,62b,62c], A. Salzburger[35], J. Samarati[35], D. Sammel[51], D. Sampsonidis[160], D. Sampsonidou[160], J. Sánchez[172], A. Sanchez Pineda[65a,65c],

H. Sandaker[132], C.O. Sander[45], I.G. Sanderswood[88], M. Sandhoff[180], C. Sandoval[22],
D.P.C. Sankey[142], M. Sannino[54b,54a], Y. Sano[116], A. Sansoni[50], C. Santoni[37], H. Santos[138a,138b],
S.N. Santpur[18], A. Santra[172], A. Sapronov[78], J.G. Saraiva[138a,138d], O. Sasaki[80], K. Sato[167],
E. Sauvan[5], P. Savard[165,au], N. Savic[114], R. Sawada[161], C. Sawyer[142], L. Sawyer[94,aj], C. Sbarra[23b],
A. Sbrizzi[23a], T. Scanlon[93], J. Schaarschmidt[146], P. Schacht[114], B.M. Schachtner[113], D. Schaefer[36],
L. Schaefer[135], J. Schaeffer[98], S. Schaepe[35], U. Schäfer[98], A.C. Schaffer[130], D. Schaile[113],
R.D. Schamberger[153], N. Scharmberg[99], V.A. Schegelsky[136], D. Scheirich[141], F. Schenck[19],
M. Schernau[169], C. Schiavi[54b,54a], S. Schier[144], L.K. Schildgen[24], Z.M. Schillaci[26], E.J. Schioppa[35],
M. Schioppa[40b,40a], K.E. Schleicher[51], S. Schlenker[35], K.R. Schmidt-Sommerfeld[114],
K. Schmieden[35], C. Schmitt[98], S. Schmitt[45], S. Schmitz[98], J.C. Schmoeckel[45], U. Schnoor[51],
L. Schoeffel[143], A. Schoening[60b], E. Schopf[133], M. Schott[98], J.F.P. Schouwenberg[118],
J. Schovancova[35], S. Schramm[53], F. Schroeder[180], A. Schulte[98], H-C. Schultz-Coulon[60a],
M. Schumacher[51], B.A. Schumm[144], Ph. Schune[143], A. Schwartzman[151], T.A. Schwarz[104],
Ph. Schwemling[143], R. Schwienhorst[105], A. Sciandra[144], G. Sciolla[26], M. Scornajenghi[40b,40a],
F. Scuri[70a], F. Scutti[103], L.M. Scyboz[114], C.D. Sebastiani[71a,71b], P. Seema[19], S.C. Seidel[117],
A. Seiden[144], T. Seiss[36], J.M. Seixas[79b], G. Sekhniaidze[68a], K. Sekhon[104], S.J. Sekula[41],
N. Semprini-Cesari[23b,23a], S. Sen[48], S. Senkin[37], C. Serfon[75], L. Serin[130], L. Serkin[65a,65b],
M. Sessa[59a], H. Severini[126], F. Sforza[168], A. Sfyrla[53], E. Shabalina[52], J.D. Shahinian[144],
N.W. Shaikh[44a,44b], D. Shaked Renous[178], L.Y. Shan[15a], R. Shang[171], J.T. Shank[25], M. Shapiro[18],
A.S. Sharma[1], A. Sharma[133], P.B. Shatalov[110], K. Shaw[154], S.M. Shaw[99], A. Shcherbakova[136],
Y. Shen[126], N. Sherafati[33], A.D. Sherman[25], P. Sherwood[93], L. Shi[156,aq], S. Shimizu[80],
C.O. Shimmin[181], Y. Shimogama[177], M. Shimojima[115], I.P.J. Shipsey[133], S. Shirabe[86],
M. Shiyakova[78], J. Shlomi[178], A. Shmeleva[109], M.J. Shochet[36], S. Shojaii[103], D.R. Shope[126],
S. Shrestha[124], E. Shulga[111], P. Sicho[139], A.M. Sickles[171], P.E. Sidebo[152], E. Sideras Haddad[32c],
O. Sidiropoulou[35], A. Sidoti[23b,23a], F. Siegert[47], Dj. Sijacki[16], M. Silva Jr.[179],
M.V. Silva Oliveira[79a], S.B. Silverstein[44a], S. Simion[130], E. Simioni[98], R. Simoniello[98],
P. Sinervo[165], N.B. Sinev[129], M. Sioli[23b,23a], I. Siral[104], S.Yu. Sivoklokov[112], J. Sjölin[44a,44b],
E. Skorda[95], P. Skubic[126], M. Slawinska[83], K. Sliwa[168], R. Slovak[141], V. Smakhtin[178],
B.H. Smart[142], J. Smiesko[28a], N. Smirnov[111], S.Yu. Smirnov[111], Y. Smirnov[111], L.N. Smirnova[112],
O. Smirnova[95], J.W. Smith[52], M. Smizanska[88], K. Smolek[140], A. Smykiewicz[83], A.A. Snesarev[109],
H.L. Snoek[119], I.M. Snyder[129], S. Snyder[29], R. Sobie[174,ab], A.M. Soffa[169], A. Soffer[159],
A. Søgaard[49], F. Sohns[52], C.A. Solans Sanchez[35], E.Yu. Soldatov[111], U. Soldevila[172],
A.A. Solodkov[122], A. Soloshenko[78], O.V. Solovyanov[122], V. Solovyev[136], P. Sommer[147], H. Son[168],
W. Song[142], W.Y. Song[166b], A. Sopczak[140], F. Sopkova[28b], C.L. Sotiropoulou[70a,70b],
S. Sottocornola[69a,69b], R. Soualah[65a,65c,i], A.M. Soukharev[121b,121a], D. South[45], S. Spagnolo[66a,66b],
M. Spalla[114], M. Spangenberg[176], F. Spanò[92], D. Sperlich[51], T.M. Spieker[60a], R. Spighi[23b],
G. Spigo[35], L.A. Spiller[103], M. Spina[154], D.P. Spiteri[56], M. Spousta[141], A. Stabile[67a,67b],
B.L. Stamas[120], R. Stamen[60a], M. Stamenkovic[119], E. Stanecka[83], R.W. Stanek[6], B. Stanislaus[133],
M.M. Stanitzki[45], M. Stankaityte[133], B. Stapf[119], E.A. Starchenko[122], G.H. Stark[144], J. Stark[57],
S.H Stark[39], P. Staroba[139], P. Starovoitov[60a], S. Stärz[102], R. Staszewski[83], G. Stavropoulos[43],
M. Stegler[45], P. Steinberg[29], A.L. Steinhebel[129], B. Stelzer[150], H.J. Stelzer[137],
O. Stelzer-Chilton[166a], H. Stenzel[55], T.J. Stevenson[154], G.A. Stewart[35], M.C. Stockton[35],
G. Stoicea[27b], M. Stolarski[138a], P. Stolte[52], S. Stonjek[114], A. Straessner[47], J. Strandberg[152],

S. Strandberg[44a,44b], M. Strauss[126], P. Strizenec[28b], R. Ströhmer[175], D.M. Strom[129], R. Stroynowski[41], A. Strubig[49], S.A. Stucci[29], B. Stugu[17], J. Stupak[126], N.A. Styles[45], D. Su[151], S. Suchek[60a], Y. Sugaya[131], V.V. Sulin[109], M.J. Sullivan[89], D.M.S. Sultan[53], S. Sultansoy[4c], T. Sumida[84], S. Sun[104], X. Sun[3], K. Suruliz[154], C.J.E. Suster[155], M.R. Sutton[154], S. Suzuki[80], M. Svatos[139], M. Swiatlowski[36], S.P. Swift[2], T. Swirski[175], A. Sydorenko[98], I. Sykora[28a], M. Sykora[141], T. Sykora[141], D. Ta[98], K. Tackmann[45,x], J. Taenzer[159], A. Taffard[169], R. Tafirout[166a], E. Tahirovic[91], H. Takai[29], R. Takashima[85], K. Takeda[81], T. Takeshita[148], E.P. Takeva[49], Y. Takubo[80], M. Talby[100], A.A. Talyshev[121b,121a], N.M. Tamir[159], J. Tanaka[161], M. Tanaka[163], R. Tanaka[130], B.B. Tannenwald[124], S. Tapia Araya[171], S. Tapprogge[98], A. Tarek Abouelfadl Mohamed[134], S. Tarem[158], G. Tarna[27b,e], G.F. Tartarelli[67a], P. Tas[141], M. Tasevsky[139], T. Tashiro[84], E. Tassi[40b,40a], A. Tavares Delgado[138a,138b], Y. Tayalati[34e], A.J. Taylor[49], G.N. Taylor[103], W. Taylor[166b], A.S. Tee[88], R. Teixeira De Lima[151], P. Teixeira-Dias[92], H. Ten Kate[35], J.J. Teoh[119], S. Terada[80], K. Terashi[161], J. Terron[97], S. Terzo[14], M. Testa[50], R.J. Teuscher[165,ab], S.J. Thais[181], T. Theveneaux-Pelzer[45], F. Thiele[39], D.W. Thomas[92], J.O. Thomas[41], J.P. Thomas[21], A.S. Thompson[56], P.D. Thompson[21], L.A. Thomsen[181], E. Thomson[135], Y. Tian[38], R.E. Ticse Torres[52], V.O. Tikhomirov[109,an], Yu.A. Tikhonov[121b,121a], S. Timoshenko[111], P. Tipton[181], S. Tisserant[100], K. Todome[23b,23a], S. Todorova-Nova[5], S. Todt[47], J. Tojo[86], S. Tokár[28a], K. Tokushuku[80], E. Tolley[124], K.G. Tomiwa[32c], M. Tomoto[116], L. Tompkins[151,q], K. Toms[117], B. Tong[58], P. Tornambe[101], E. Torrence[129], H. Torres[47], E. Torró Pastor[146], C. Tosciri[133], J. Toth[100,z], D.R. Tovey[147], C.J. Treado[123], T. Trefzger[175], F. Tresoldi[154], A. Tricoli[29], I.M. Trigger[166a], S. Trincaz-Duvoid[134], W. Trischuk[165], B. Trocmé[57], A. Trofymov[130], C. Troncon[67a], M. Trovatelli[174], F. Trovato[154], L. Truong[32b], M. Trzebinski[83], A. Trzupek[83], F. Tsai[45], J.C-L. Tseng[133], P.V. Tsiareshka[106,ah], A. Tsirigotis[160], N. Tsirintanis[9], V. Tsiskaridze[153], E.G. Tskhadadze[157a], M. Tsopoulou[160], I.I. Tsukerman[110], V. Tsulaia[18], S. Tsuno[80], D. Tsybychev[153,164], Y. Tu[62b], A. Tudorache[27b], V. Tudorache[27b], T.T. Tulbure[27a], A.N. Tuna[58], S. Turchikhin[78], D. Turgeman[178], I. Turk Cakir[4b,t], R.J. Turner[21], R.T. Turra[67a], P.M. Tuts[38], S Tzamarias[160], E. Tzovara[98], G. Ucchielli[46], I. Ueda[80], M. Ughetto[44a,44b], F. Ukegawa[167], G. Unal[35], A. Undrus[29], G. Unel[169], F.C. Ungaro[103], Y. Unno[80], K. Uno[161], J. Urban[28b], P. Urquijo[103], G. Usai[8], J. Usui[80], L. Vacavant[100], V. Vacek[140], B. Vachon[102], K.O.H. Vadla[132], A. Vaidya[93], C. Valderanis[113], E. Valdes Santurio[44a,44b], M. Valente[53], S. Valentinetti[23b,23a], A. Valero[172], L. Valéry[45], R.A. Vallance[21], A. Vallier[35], J.A. Valls Ferrer[172], T.R. Van Daalen[14], P. Van Gemmeren[6], I. Van Vulpen[119], M. Vanadia[72a,72b], W. Vandelli[35], A. Vaniachine[164], D. Vannicola[71a,71b], R. Vari[71a], E.W. Varnes[7], C. Varni[54b,54a], T. Varol[41], D. Varouchas[130], K.E. Varvell[155], M.E. Vasile[27b], G.A. Vasquez[174], J.G. Vasquez[181], F. Vazeille[37], D. Vazquez Furelos[14], T. Vazquez Schroeder[35], J. Veatch[52], V. Vecchio[73a,73b], M.J. Veen[119], L.M. Veloce[165], F. Veloso[138a,138c], S. Veneziano[71a], A. Ventura[66a,66b], N. Venturi[35], A. Verbytskyi[114], V. Vercesi[69a], M. Verducci[73a,73b], C.M. Vergel Infante[77], C. Vergis[24], W. Verkerke[119], A.T. Vermeulen[119], J.C. Vermeulen[119], M.C. Vetterli[150,au], N. Viaux Maira[145b], M. Vicente Barreto Pinto[53], I. Vichou[171,*], T. Vickey[147], O.E. Vickey Boeriu[147], G.H.A. Viehhauser[133], L. Vigani[133], M. Villa[23b,23a], M. Villaplana Perez[67a,67b], E. Vilucchi[50], M.G. Vincter[33], V.B. Vinogradov[78], A. Vishwakarma[45], C. Vittori[23b,23a], I. Vivarelli[154], M. Vogel[180], P. Vokac[140], S.E. von Buddenbrock[32c], E. Von Toerne[24], V. Vorobel[141], K. Vorobev[111], M. Vos[172],

J.H. Vossebeld[89], N. Vranjes[16], M. Vranjes Milosavljevic[16], V. Vrba[140], M. Vreeswijk[119], T. Šfiligoj[90], R. Vuillermet[35], I. Vukotic[36], T. Ženiš[28a], L. Živković[16], P. Wagner[24], W. Wagner[180], J. Wagner-Kuhr[113], H. Wahlberg[87], K. Wakamiya[81], V.M. Walbrecht[114], J. Walder[88], R. Walker[113], S.D. Walker[92], W. Walkowiak[149], V. Wallangen[44a,44b], A.M. Wang[58], C. Wang[59b], F. Wang[179], H. Wang[18], H. Wang[3], J. Wang[155], J. Wang[60b], P. Wang[41], Q. Wang[126], R.-J. Wang[98], R. Wang[59a], R. Wang[6], S.M. Wang[156], W.T. Wang[59a], W. Wang[15c,ac], W.X. Wang[59a,ac], Y. Wang[59a,ak], Z. Wang[59c], C. Wanotayaroj[45], A. Warburton[102], C.P. Ward[31], D.R. Wardrope[93], A. Washbrook[49], A.T. Watson[21], M.F. Watson[21], G. Watts[146], B.M. Waugh[93], A.F. Webb[11], S. Webb[98], C. Weber[181], M.S. Weber[20], S.A. Weber[33], S.M. Weber[60a], A.R. Weidberg[133], J. Weingarten[46], M. Weirich[98], C. Weiser[51], P.S. Wells[35], T. Wenaus[29], T. Wengler[35], S. Wenig[35], N. Wermes[24], M.D. Werner[77], P. Werner[35], M. Wessels[60a], T.D. Weston[20], K. Whalen[129], N.L. Whallon[146], A.M. Wharton[88], A.S. White[104], A. White[8], M.J. White[1], D. Whiteson[169], B.W. Whitmore[88], F.J. Wickens[142], W. Wiedenmann[179], M. Wielers[142], N. Wieseotte[98], C. Wiglesworth[39], L.A.M. Wiik-Fuchs[51], F. Wilk[99], H.G. Wilkens[35], L.J. Wilkins[92], H.H. Williams[135], S. Williams[31], C. Willis[105], S. Willocq[101], J.A. Wilson[21], I. Wingerter-Seez[5], E. Winkels[154], F. Winklmeier[129], O.J. Winston[154], B.T. Winter[51], M. Wittgen[151], M. Wobisch[94], A. Wolf[98], T.M.H. Wolf[119], R. Wolff[100], R.W. Wölker[133], J. Wollrath[51], M.W. Wolter[83], H. Wolters[138a,138c], V.W.S. Wong[173], N.L. Woods[144], S.D. Worm[21], B.K. Wosiek[83], K.W. Woźniak[83], K. Wraight[56], S.L. Wu[179], X. Wu[53], Y. Wu[59a], T.R. Wyatt[99], B.M. Wynne[49], S. Xella[39], Z. Xi[104], L. Xia[176], D. Xu[15a], H. Xu[59a,e], L. Xu[29], T. Xu[143], W. Xu[104], Z. Xu[59b], Z. Xu[151], B. Yabsley[155], S. Yacoob[32a], K. Yajima[131], D.P. Yallup[93], D. Yamaguchi[163], Y. Yamaguchi[163], A. Yamamoto[80], T. Yamanaka[161], F. Yamane[81], M. Yamatani[161], T. Yamazaki[161], Y. Yamazaki[81], Z. Yan[25], H.J. Yang[59c,59d], H.T. Yang[18], S. Yang[76], X. Yang[59b,57], Y. Yang[161], Z. Yang[17], W-M. Yao[18], Y.C. Yap[45], Y. Yasu[80], E. Yatsenko[59c,59d], J. Ye[41], S. Ye[29], I. Yeletskikh[78], M.R. Yexley[88], E. Yigitbasi[25], E. Yildirim[98], K. Yorita[177], K. Yoshihara[135], C.J.S. Young[35], C. Young[151], J. Yu[77], R. Yuan[59b], X. Yue[60a], S.P.Y. Yuen[24], B. Zabinski[83], G. Zacharis[10], E. Zaffaroni[53], J. Zahreddine[134], R. Zaidan[14], A.M. Zaitsev[122,am], T. Zakareishvili[157b], N. Zakharchuk[33], S. Zambito[58], D. Zanzi[35], D.R. Zaripovas[56], S.V. Zeißner[46], C. Zeitnitz[180], G. Zemaityte[133], J.C. Zeng[171], O. Zenin[122], D. Zerwas[130], M. Zgubič[133], D.F. Zhang[15b], F. Zhang[179], G. Zhang[59a], G. Zhang[15b], H. Zhang[15c], J. Zhang[6], L. Zhang[15c], L. Zhang[59a], M. Zhang[171], R. Zhang[59a], R. Zhang[24], X. Zhang[59b], Y. Zhang[15d], Z. Zhang[62a], Z. Zhang[130], P. Zhao[48], Y. Zhao[59b], Z. Zhao[59a], A. Zhemchugov[78], Z. Zheng[104], D. Zhong[171], B. Zhou[104], C. Zhou[179], M.S. Zhou[15d], M. Zhou[153], N. Zhou[59c], Y. Zhou[7], C.G. Zhu[59b], H.L. Zhu[59a], H. Zhu[15a], J. Zhu[104], Y. Zhu[59a], X. Zhuang[15a], K. Zhukov[109], V. Zhulanov[121b,121a], D. Zieminska[64], N.I. Zimine[78], S. Zimmermann[51], Z. Zinonos[114], M. Ziolkowski[149], G. Zobernig[179], A. Zoccoli[23b,23a], K. Zoch[52], T.G. Zorbas[147], R. Zou[36], L. Zwalinski[35].

[1]Department of Physics, University of Adelaide, Adelaide; Australia.
[2]Physics Department, SUNY Albany, Albany NY; United States of America.
[3]Department of Physics, University of Alberta, Edmonton AB; Canada.
[4](a)Department of Physics, Ankara University, Ankara;(b)Istanbul Aydin University, Istanbul;(c)Division of Physics, TOBB University of Economics and Technology, Ankara; Turkey.
[5]LAPP, Université Grenoble Alpes, Université Savoie Mont Blanc, CNRS/IN2P3, Annecy; France.

[6]High Energy Physics Division, Argonne National Laboratory, Argonne IL; United States of America.
[7]Department of Physics, University of Arizona, Tucson AZ; United States of America.
[8]Department of Physics, University of Texas at Arlington, Arlington TX; United States of America.
[9]Physics Department, National and Kapodistrian University of Athens, Athens; Greece.
[10]Physics Department, National Technical University of Athens, Zografou; Greece.
[11]Department of Physics, University of Texas at Austin, Austin TX; United States of America.
[12(a)]Bahcesehir University, Faculty of Engineering and Natural Sciences, Istanbul;[(b)]Istanbul Bilgi University, Faculty of Engineering and Natural Sciences, Istanbul;[(c)]Department of Physics, Bogazici University, Istanbul;[(d)]Department of Physics Engineering, Gaziantep University, Gaziantep; Turkey.
[13]Institute of Physics, Azerbaijan Academy of Sciences, Baku; Azerbaijan.
[14]Institut de Física d'Altes Energies (IFAE), Barcelona Institute of Science and Technology, Barcelona; Spain.
[15(a)]Institute of High Energy Physics, Chinese Academy of Sciences, Beijing;[(b)]Physics Department, Tsinghua University, Beijing;[(c)]Department of Physics, Nanjing University, Nanjing;[(d)]University of Chinese Academy of Science (UCAS), Beijing; China.
[16]Institute of Physics, University of Belgrade, Belgrade; Serbia.
[17]Department for Physics and Technology, University of Bergen, Bergen; Norway.
[18]Physics Division, Lawrence Berkeley National Laboratory and University of California, Berkeley CA; United States of America.
[19]Institut für Physik, Humboldt Universität zu Berlin, Berlin; Germany.
[20]Albert Einstein Center for Fundamental Physics and Laboratory for High Energy Physics, University of Bern, Bern; Switzerland.
[21]School of Physics and Astronomy, University of Birmingham, Birmingham; United Kingdom.
[22]Facultad de Ciencias y Centro de Investigaciónes, Universidad Antonio Nariño, Bogota; Colombia.
[23(a)]INFN Bologna and Universita' di Bologna, Dipartimento di Fisica;[(b)]INFN Sezione di Bologna; Italy.
[24]Physikalisches Institut, Universität Bonn, Bonn; Germany.
[25]Department of Physics, Boston University, Boston MA; United States of America.
[26]Department of Physics, Brandeis University, Waltham MA; United States of America.
[27(a)]Transilvania University of Brasov, Brasov;[(b)]Horia Hulubei National Institute of Physics and Nuclear Engineering, Bucharest;[(c)]Department of Physics, Alexandru Ioan Cuza University of Iasi, Iasi;[(d)]National Institute for Research and Development of Isotopic and Molecular Technologies, Physics Department, Cluj-Napoca;[(e)]University Politehnica Bucharest, Bucharest;[(f)]West University in Timisoara, Timisoara; Romania.
[28(a)]Faculty of Mathematics, Physics and Informatics, Comenius University, Bratislava;[(b)]Department of Subnuclear Physics, Institute of Experimental Physics of the Slovak Academy of Sciences, Kosice; Slovak Republic.
[29]Physics Department, Brookhaven National Laboratory, Upton NY; United States of America.
[30]Departamento de Física, Universidad de Buenos Aires, Buenos Aires; Argentina.

[31]Cavendish Laboratory, University of Cambridge, Cambridge; United Kingdom.
[32(a)]Department of Physics, University of Cape Town, Cape Town;[(b)]Department of Mechanical Engineering Science, University of Johannesburg, Johannesburg;[(c)]School of Physics, University of the Witwatersrand, Johannesburg; South Africa.
[33]Department of Physics, Carleton University, Ottawa ON; Canada.
[34(a)]Faculté des Sciences Ain Chock, Réseau Universitaire de Physique des Hautes Energies - Université Hassan II, Casablanca;[(b)]Centre National de l'Energie des Sciences Techniques Nucleaires (CNESTEN), Rabat;[(c)]Faculté des Sciences Semlalia, Université Cadi Ayyad, LPHEA-Marrakech;[(d)]Faculté des Sciences, Université Mohamed Premier and LPTPM, Oujda;[(e)]Faculté des sciences, Université Mohammed V, Rabat; Morocco.
[35]CERN, Geneva; Switzerland.
[36]Enrico Fermi Institute, University of Chicago, Chicago IL; United States of America.
[37]LPC, Université Clermont Auvergne, CNRS/IN2P3, Clermont-Ferrand; France.
[38]Nevis Laboratory, Columbia University, Irvington NY; United States of America.
[39]Niels Bohr Institute, University of Copenhagen, Copenhagen; Denmark.
[40(a)]Dipartimento di Fisica, Università della Calabria, Rende;[(b)]INFN Gruppo Collegato di Cosenza, Laboratori Nazionali di Frascati; Italy.
[41]Physics Department, Southern Methodist University, Dallas TX; United States of America.
[42]Physics Department, University of Texas at Dallas, Richardson TX; United States of America.
[43]National Centre for Scientific Research "Demokritos", Agia Paraskevi; Greece.
[44(a)]Department of Physics, Stockholm University;[(b)]Oskar Klein Centre, Stockholm; Sweden.
[45]Deutsches Elektronen-Synchrotron DESY, Hamburg and Zeuthen; Germany.
[46]Lehrstuhl für Experimentelle Physik IV, Technische Universität Dortmund, Dortmund; Germany.
[47]Institut für Kern- und Teilchenphysik, Technische Universität Dresden, Dresden; Germany.
[48]Department of Physics, Duke University, Durham NC; United States of America.
[49]SUPA - School of Physics and Astronomy, University of Edinburgh, Edinburgh; United Kingdom.
[50]INFN e Laboratori Nazionali di Frascati, Frascati; Italy.
[51]Physikalisches Institut, Albert-Ludwigs-Universität Freiburg, Freiburg; Germany.
[52]II. Physikalisches Institut, Georg-August-Universität Göttingen, Göttingen; Germany.
[53]Département de Physique Nucléaire et Corpusculaire, Université de Genève, Genève; Switzerland.
[54(a)]Dipartimento di Fisica, Università di Genova, Genova;[(b)]INFN Sezione di Genova; Italy.
[55]II. Physikalisches Institut, Justus-Liebig-Universität Giessen, Giessen; Germany.
[56]SUPA - School of Physics and Astronomy, University of Glasgow, Glasgow; United Kingdom.
[57]LPSC, Université Grenoble Alpes, CNRS/IN2P3, Grenoble INP, Grenoble; France.
[58]Laboratory for Particle Physics and Cosmology, Harvard University, Cambridge MA; United States of America.
[59(a)]Department of Modern Physics and State Key Laboratory of Particle Detection and Electronics, University of Science and Technology of China, Hefei;[(b)]Institute of Frontier and Interdisciplinary Science and Key Laboratory of Particle Physics and Particle Irradiation (MOE), Shandong University, Qingdao;[(c)]School of Physics and Astronomy, Shanghai Jiao Tong

University, KLPPAC-MoE, SKLPPC, Shanghai;[(d)]Tsung-Dao Lee Institute, Shanghai; China.
[60(a)]Kirchhoff-Institut für Physik, Ruprecht-Karls-Universität Heidelberg, Heidelberg;[(b)]Physikalisches Institut, Ruprecht-Karls-Universität Heidelberg, Heidelberg; Germany.
[61]Faculty of Applied Information Science, Hiroshima Institute of Technology, Hiroshima; Japan.
[62(a)]Department of Physics, Chinese University of Hong Kong, Shatin, N.T., Hong Kong;[(b)]Department of Physics, University of Hong Kong, Hong Kong;[(c)]Department of Physics and Institute for Advanced Study, Hong Kong University of Science and Technology, Clear Water Bay, Kowloon, Hong Kong; China.
[63]Department of Physics, National Tsing Hua University, Hsinchu; Taiwan.
[64]Department of Physics, Indiana University, Bloomington IN; United States of America.
[65(a)]INFN Gruppo Collegato di Udine, Sezione di Trieste, Udine;[(b)]ICTP, Trieste;[(c)]Dipartimento Politecnico di Ingegneria e Architettura, Università di Udine, Udine; Italy.
[66(a)]INFN Sezione di Lecce;[(b)]Dipartimento di Matematica e Fisica, Università del Salento, Lecce; Italy.
[67(a)]INFN Sezione di Milano;[(b)]Dipartimento di Fisica, Università di Milano, Milano; Italy.
[68(a)]INFN Sezione di Napoli;[(b)]Dipartimento di Fisica, Università di Napoli, Napoli; Italy.
[69(a)]INFN Sezione di Pavia;[(b)]Dipartimento di Fisica, Università di Pavia, Pavia; Italy.
[70(a)]INFN Sezione di Pisa;[(b)]Dipartimento di Fisica E. Fermi, Università di Pisa, Pisa; Italy.
[71(a)]INFN Sezione di Roma;[(b)]Dipartimento di Fisica, Sapienza Università di Roma, Roma; Italy.
[72(a)]INFN Sezione di Roma Tor Vergata;[(b)]Dipartimento di Fisica, Università di Roma Tor Vergata, Roma; Italy.
[73(a)]INFN Sezione di Roma Tre;[(b)]Dipartimento di Matematica e Fisica, Università Roma Tre, Roma; Italy.
[74(a)]INFN-TIFPA;[(b)]Università degli Studi di Trento, Trento; Italy.
[75]Institut für Astro- und Teilchenphysik, Leopold-Franzens-Universität, Innsbruck; Austria.
[76]University of Iowa, Iowa City IA; United States of America.
[77]Department of Physics and Astronomy, Iowa State University, Ames IA; United States of America.
[78]Joint Institute for Nuclear Research, Dubna; Russia.
[79(a)]Departamento de Engenharia Elétrica, Universidade Federal de Juiz de Fora (UFJF), Juiz de Fora;[(b)]Universidade Federal do Rio De Janeiro COPPE/EE/IF, Rio de Janeiro;[(c)]Universidade Federal de São João del Rei (UFSJ), São João del Rei;[(d)]Instituto de Física, Universidade de São Paulo, São Paulo; Brazil.
[80]KEK, High Energy Accelerator Research Organization, Tsukuba; Japan.
[81]Graduate School of Science, Kobe University, Kobe; Japan.
[82(a)]AGH University of Science and Technology, Faculty of Physics and Applied Computer Science, Krakow;[(b)]Marian Smoluchowski Institute of Physics, Jagiellonian University, Krakow; Poland.
[83]Institute of Nuclear Physics Polish Academy of Sciences, Krakow; Poland.
[84]Faculty of Science, Kyoto University, Kyoto; Japan.
[85]Kyoto University of Education, Kyoto; Japan.
[86]Research Center for Advanced Particle Physics and Department of Physics, Kyushu University, Fukuoka ; Japan.
[87]Instituto de Física La Plata, Universidad Nacional de La Plata and CONICET, La Plata;

Argentina.
[88]Physics Department, Lancaster University, Lancaster; United Kingdom.
[89]Oliver Lodge Laboratory, University of Liverpool, Liverpool; United Kingdom.
[90]Department of Experimental Particle Physics, Jožef Stefan Institute and Department of Physics, University of Ljubljana, Ljubljana; Slovenia.
[91]School of Physics and Astronomy, Queen Mary University of London, London; United Kingdom.
[92]Department of Physics, Royal Holloway University of London, Egham; United Kingdom.
[93]Department of Physics and Astronomy, University College London, London; United Kingdom.
[94]Louisiana Tech University, Ruston LA; United States of America.
[95]Fysiska institutionen, Lunds universitet, Lund; Sweden.
[96]Centre de Calcul de l'Institut National de Physique Nucléaire et de Physique des Particules (IN2P3), Villeurbanne; France.
[97]Departamento de Física Teorica C-15 and CIAFF, Universidad Autónoma de Madrid, Madrid; Spain.
[98]Institut für Physik, Universität Mainz, Mainz; Germany.
[99]School of Physics and Astronomy, University of Manchester, Manchester; United Kingdom.
[100]CPPM, Aix-Marseille Université, CNRS/IN2P3, Marseille; France.
[101]Department of Physics, University of Massachusetts, Amherst MA; United States of America.
[102]Department of Physics, McGill University, Montreal QC; Canada.
[103]School of Physics, University of Melbourne, Victoria; Australia.
[104]Department of Physics, University of Michigan, Ann Arbor MI; United States of America.
[105]Department of Physics and Astronomy, Michigan State University, East Lansing MI; United States of America.
[106]B.I. Stepanov Institute of Physics, National Academy of Sciences of Belarus, Minsk; Belarus.
[107]Research Institute for Nuclear Problems of Byelorussian State University, Minsk; Belarus.
[108]Group of Particle Physics, University of Montreal, Montreal QC; Canada.
[109]P.N. Lebedev Physical Institute of the Russian Academy of Sciences, Moscow; Russia.
[110]Institute for Theoretical and Experimental Physics of the National Research Centre Kurchatov Institute, Moscow; Russia.
[111]National Research Nuclear University MEPhI, Moscow; Russia.
[112]D.V. Skobeltsyn Institute of Nuclear Physics, M.V. Lomonosov Moscow State University, Moscow; Russia.
[113]Fakultät für Physik, Ludwig-Maximilians-Universität München, München; Germany.
[114]Max-Planck-Institut für Physik (Werner-Heisenberg-Institut), München; Germany.
[115]Nagasaki Institute of Applied Science, Nagasaki; Japan.
[116]Graduate School of Science and Kobayashi-Maskawa Institute, Nagoya University, Nagoya; Japan.
[117]Department of Physics and Astronomy, University of New Mexico, Albuquerque NM; United States of America.
[118]Institute for Mathematics, Astrophysics and Particle Physics, Radboud University Nijmegen/Nikhef, Nijmegen; Netherlands.
[119]Nikhef National Institute for Subatomic Physics and University of Amsterdam, Amsterdam; Netherlands.

[120]Department of Physics, Northern Illinois University, DeKalb IL; United States of America.
[121(a)]Budker Institute of Nuclear Physics and NSU, SB RAS, Novosibirsk;[(b)]Novosibirsk State University Novosibirsk; Russia.
[122]Institute for High Energy Physics of the National Research Centre Kurchatov Institute, Protvino; Russia.
[123]Department of Physics, New York University, New York NY; United States of America.
[124]Ohio State University, Columbus OH; United States of America.
[125]Faculty of Science, Okayama University, Okayama; Japan.
[126]Homer L. Dodge Department of Physics and Astronomy, University of Oklahoma, Norman OK; United States of America.
[127]Department of Physics, Oklahoma State University, Stillwater OK; United States of America.
[128]Palacký University, RCPTM, Joint Laboratory of Optics, Olomouc; Czech Republic.
[129]Center for High Energy Physics, University of Oregon, Eugene OR; United States of America.
[130]LAL, Université Paris-Sud, CNRS/IN2P3, Université Paris-Saclay, Orsay; France.
[131]Graduate School of Science, Osaka University, Osaka; Japan.
[132]Department of Physics, University of Oslo, Oslo; Norway.
[133]Department of Physics, Oxford University, Oxford; United Kingdom.
[134]LPNHE, Sorbonne Université, Paris Diderot Sorbonne Paris Cité, CNRS/IN2P3, Paris; France.
[135]Department of Physics, University of Pennsylvania, Philadelphia PA; United States of America.
[136]Konstantinov Nuclear Physics Institute of National Research Centre "Kurchatov Institute", PNPI, St. Petersburg; Russia.
[137]Department of Physics and Astronomy, University of Pittsburgh, Pittsburgh PA; United States of America.
[138(a)]Laboratório de Instrumentação e Física Experimental de Partículas - LIP;[(b)]Departamento de Física, Faculdade de Ciências, Universidade de Lisboa, Lisboa;[(c)]Departamento de Física, Universidade de Coimbra, Coimbra;[(d)]Centro de Física Nuclear da Universidade de Lisboa, Lisboa;[(e)]Departamento de Física, Universidade do Minho, Braga;[(f)]Universidad de Granada, Granada (Spain);[(g)]Dep Física and CEFITEC of Faculdade de Ciências e Tecnologia, Universidade Nova de Lisboa, Caparica; Portugal.
[139]Institute of Physics of the Czech Academy of Sciences, Prague; Czech Republic.
[140]Czech Technical University in Prague, Prague; Czech Republic.
[141]Charles University, Faculty of Mathematics and Physics, Prague; Czech Republic.
[142]Particle Physics Department, Rutherford Appleton Laboratory, Didcot; United Kingdom.
[143]IRFU, CEA, Université Paris-Saclay, Gif-sur-Yvette; France.
[144]Santa Cruz Institute for Particle Physics, University of California Santa Cruz, Santa Cruz CA; United States of America.
[145(a)]Departamento de Física, Pontificia Universidad Católica de Chile, Santiago;[(b)]Departamento de Física, Universidad Técnica Federico Santa María, Valparaíso; Chile.
[146]Department of Physics, University of Washington, Seattle WA; United States of America.
[147]Department of Physics and Astronomy, University of Sheffield, Sheffield; United Kingdom.
[148]Department of Physics, Shinshu University, Nagano; Japan.
[149]Department Physik, Universität Siegen, Siegen; Germany.
[150]Department of Physics, Simon Fraser University, Burnaby BC; Canada.

[151]SLAC National Accelerator Laboratory, Stanford CA; United States of America.
[152]Physics Department, Royal Institute of Technology, Stockholm; Sweden.
[153]Departments of Physics and Astronomy, Stony Brook University, Stony Brook NY; United States of America.
[154]Department of Physics and Astronomy, University of Sussex, Brighton; United Kingdom.
[155]School of Physics, University of Sydney, Sydney; Australia.
[156]Institute of Physics, Academia Sinica, Taipei; Taiwan.
[157(a)]E. Andronikashvili Institute of Physics, Iv. Javakhishvili Tbilisi State University, Tbilisi;[(b)]High Energy Physics Institute, Tbilisi State University, Tbilisi; Georgia.
[158]Department of Physics, Technion, Israel Institute of Technology, Haifa; Israel.
[159]Raymond and Beverly Sackler School of Physics and Astronomy, Tel Aviv University, Tel Aviv; Israel.
[160]Department of Physics, Aristotle University of Thessaloniki, Thessaloniki; Greece.
[161]International Center for Elementary Particle Physics and Department of Physics, University of Tokyo, Tokyo; Japan.
[162]Graduate School of Science and Technology, Tokyo Metropolitan University, Tokyo; Japan.
[163]Department of Physics, Tokyo Institute of Technology, Tokyo; Japan.
[164]Tomsk State University, Tomsk; Russia.
[165]Department of Physics, University of Toronto, Toronto ON; Canada.
[166(a)]TRIUMF, Vancouver BC;[(b)]Department of Physics and Astronomy, York University, Toronto ON; Canada.
[167]Division of Physics and Tomonaga Center for the History of the Universe, Faculty of Pure and Applied Sciences, University of Tsukuba, Tsukuba; Japan.
[168]Department of Physics and Astronomy, Tufts University, Medford MA; United States of America.
[169]Department of Physics and Astronomy, University of California Irvine, Irvine CA; United States of America.
[170]Department of Physics and Astronomy, University of Uppsala, Uppsala; Sweden.
[171]Department of Physics, University of Illinois, Urbana IL; United States of America.
[172]Instituto de Física Corpuscular (IFIC), Centro Mixto Universidad de Valencia - CSIC, Valencia; Spain.
[173]Department of Physics, University of British Columbia, Vancouver BC; Canada.
[174]Department of Physics and Astronomy, University of Victoria, Victoria BC; Canada.
[175]Fakultät für Physik und Astronomie, Julius-Maximilians-Universität Würzburg, Würzburg; Germany.
[176]Department of Physics, University of Warwick, Coventry; United Kingdom.
[177]Waseda University, Tokyo; Japan.
[178]Department of Particle Physics, Weizmann Institute of Science, Rehovot; Israel.
[179]Department of Physics, University of Wisconsin, Madison WI; United States of America.
[180]Fakultät für Mathematik und Naturwissenschaften, Fachgruppe Physik, Bergische Universität Wuppertal, Wuppertal; Germany.
[181]Department of Physics, Yale University, New Haven CT; United States of America.
[182]Yerevan Physics Institute, Yerevan; Armenia.

[a] Also at Borough of Manhattan Community College, City University of New York, NY; United States of America.
[b] Also at California State University, East Bay; United States of America.
[c] Also at Centre for High Performance Computing, CSIR Campus, Rosebank, Cape Town; South Africa.
[d] Also at CERN, Geneva; Switzerland.
[e] Also at CPPM, Aix-Marseille Université, CNRS/IN2P3, Marseille; France.
[f] Also at Département de Physique Nucléaire et Corpusculaire, Université de Genève, Genève; Switzerland.
[g] Also at Departament de Fisica de la Universitat Autonoma de Barcelona, Barcelona; Spain.
[h] Also at Departamento de FÃŋsica, Instituto Superior TÃľcnico, Universidade de Lisboa, Lisboa; Portugal.
[i] Also at Department of Applied Physics and Astronomy, University of Sharjah, Sharjah; United Arab Emirates.
[j] Also at Department of Financial and Management Engineering, University of the Aegean, Chios; Greece.
[k] Also at Department of Physics and Astronomy, University of Louisville, Louisville, KY; United States of America.
[l] Also at Department of Physics and Astronomy, University of Sheffield, Sheffield; United Kingdom.
[m] Also at Department of Physics, California State University, Fresno CA; United States of America.
[n] Also at Department of Physics, California State University, Sacramento CA; United States of America.
[o] Also at Department of Physics, King's College London, London; United Kingdom.
[p] Also at Department of Physics, St. Petersburg State Polytechnical University, St. Petersburg; Russia.
[q] Also at Department of Physics, Stanford University; United States of America.
[r] Also at Department of Physics, University of Fribourg, Fribourg; Switzerland.
[s] Also at Department of Physics, University of Michigan, Ann Arbor MI; United States of America.
[t] Also at Giresun University, Faculty of Engineering, Giresun; Turkey.
[u] Also at Graduate School of Science, Osaka University, Osaka; Japan.
[v] Also at Hellenic Open University, Patras; Greece.
[w] Also at Institucio Catalana de Recerca i Estudis Avancats, ICREA, Barcelona; Spain.
[x] Also at Institut für Experimentalphysik, Universität Hamburg, Hamburg; Germany.
[y] Also at Institute for Mathematics, Astrophysics and Particle Physics, Radboud University Nijmegen/Nikhef, Nijmegen; Netherlands.
[z] Also at Institute for Particle and Nuclear Physics, Wigner Research Centre for Physics, Budapest; Hungary.
[aa] Also at Institute of High Energy Physics, Chinese Academy of Sciences, Beijing; China.
[ab] Also at Institute of Particle Physics (IPP); Canada.
[ac] Also at Institute of Physics, Academia Sinica, Taipei; Taiwan.

[ad] Also at Institute of Physics, Azerbaijan Academy of Sciences, Baku; Azerbaijan.
[ae] Also at Institute of Theoretical Physics, Ilia State University, Tbilisi; Georgia.
[af] Also at Instituto de FÃŋsica TeÃşrica de la Universidad AutÃşnoma de Madrid; Spain.
[ag] Also at Istanbul University, Dept. of Physics, Istanbul; Turkey.
[ah] Also at Joint Institute for Nuclear Research, Dubna; Russia.
[ai] Also at LAL, Université Paris-Sud, CNRS/IN2P3, Université Paris-Saclay, Orsay; France.
[aj] Also at Louisiana Tech University, Ruston LA; United States of America.
[ak] Also at LPNHE, Sorbonne Université, Paris Diderot Sorbonne Paris Cité, CNRS/IN2P3, Paris; France.
[al] Also at Manhattan College, New York NY; United States of America.
[am] Also at Moscow Institute of Physics and Technology State University, Dolgoprudny; Russia.
[an] Also at National Research Nuclear University MEPhI, Moscow; Russia.
[ao] Also at Physics Dept, University of South Africa, Pretoria; South Africa.
[ap] Also at Physikalisches Institut, Albert-Ludwigs-Universität Freiburg, Freiburg; Germany.
[aq] Also at School of Physics, Sun Yat-sen University, Guangzhou; China.
[ar] Also at The City College of New York, New York NY; United States of America.
[as] Also at The Collaborative Innovation Center of Quantum Matter (CICQM), Beijing; China.
[at] Also at Tomsk State University, Tomsk, and Moscow Institute of Physics and Technology State University, Dolgoprudny; Russia.
[au] Also at TRIUMF, Vancouver BC; Canada.
[av] Also at Universita di Napoli Parthenope, Napoli; Italy.
* Deceased

# The CMS Collaboration

**Yerevan Physics Institute, Yerevan, Armenia**
A.M. Sirunyan†, A. Tumasyan
**Institut für Hochenergiephysik, Wien, Austria**
W. Adam, F. Ambrogi, T. Bergauer, J. Brandstetter, M. Dragicevic, J. Erö, A. Escalante Del Valle, M. Flechl, R. Frühwirth[1], M. Jeitler[1], N. Krammer, I. Krätschmer, S. Kulkarni, L. Lechner, D. Liko, T. Madlener, I. Mikulec, N. Rad, J. Schieck[1], R. Schöfbeck, M. Spanring, D. Spitzbart, W. Waltenberger, J. Wittmann, C.-E. Wulz[1], M. Zarucki
**Institute for Nuclear Problems, Minsk, Belarus**
V. Drugakov, V. Mossolov, J. Suarez Gonzalez
**Universiteit Antwerpen, Antwerpen, Belgium**
M.R. Darwish, E.A. De Wolf, D. Di Croce, X. Janssen, J. Lauwers, A. Lelek, M. Pieters, H. Rejeb Sfar, H. Van Haevermaet, P. Van Mechelen, S. Van Putte, N. Van Remortel
**Vrije Universiteit Brussel, Brussel, Belgium**
F. Blekman, E.S. Bols, S.S. Chhibra, J. D'Hondt, J. De Clercq, D. Lontkovskyi, S. Lowette, I. Marchesini, S. Moortgat, L. Moreels, Q. Python, K. Skovpen, S. Tavernier, W. Van Doninck, P. Van Mulders, I. Van Parijs
**Université Libre de Bruxelles, Bruxelles, Belgium**
D. Beghin, B. Bilin, H. Brun, B. Clerbaux, G. De Lentdecker, H. Delannoy, B. Dorney, L. Favart, A. Grebenyuk, A.K. Kalsi, J. Luetic, A. Popov, N. Postiau, E. Starling, L. Thomas, C. Vander Velde, P. Vanlaer, D. Vannerom, Q. Wang
**Ghent University, Ghent, Belgium**
T. Cornelis, D. Dobur, I. Khvastunov[2], C. Roskas, D. Trocino, M. Tytgat, W. Verbeke, B. Vermassen, M. Vit, N. Zaganidis
**Université Catholique de Louvain, Louvain-la-Neuve, Belgium**
O. Bondu, G. Bruno, C. Caputo, P. David, C. Delaere, M. Delcourt, A. Giammanco, G. Krintiras, V. Lemaitre, A. Magitteri, J. Prisciandaro, A. Saggio, M. Vidal Marono, P. Vischia, J. Zobec
**Centro Brasileiro de Pesquisas Fisicas, Rio de Janeiro, Brazil**
F.L. Alves, G.A. Alves, G. Correia Silva, C. Hensel, A. Moraes, P. Rebello Teles
**Universidade do Estado do Rio de Janeiro, Rio de Janeiro, Brazil**
E. Belchior Batista Das Chagas, W. Carvalho, J. Chinellato[3], E. Coelho, E.M. Da Costa, G.G. Da Silveira[4], D. De Jesus Damiao, C. De Oliveira Martins, S. Fonseca De Souza, L.M. Huertas Guativa, H. Malbouisson, J. Martins[5], D. Matos Figueiredo, M. Medina Jaime[6], M. Melo De Almeida, C. Mora Herrera, L. Mundim, H. Nogima, W.L. Prado Da Silva, L.J. Sanchez Rosas, A. Santoro, A. Sznajder, M. Thiel, E.J. Tonelli Manganote[3], F. Torres Da Silva De Araujo, A. Vilela Pereira
**Universidade Estadual Paulista [a], Universidade Federal do ABC [b], São Paulo, Brazil**
S. Ahuja[a], C.A. Bernardes[a], L. Calligaris[a], T.R. Fernandez Perez Tomei[a], E.M. Gregores[b], D.S. Lemos, P.G. Mercadante[b], S.F. Novaes[a], S.S. Padula[a]
**Institute for Nuclear Research and Nuclear Energy, Bulgarian Academy of Sciences, Sofia, Bulgaria**
A. Aleksandrov, G. Antchev, R. Hadjiiska, P. Iaydjiev, A. Marinov, M. Misheva, M. Rodozov, M. Shopova, G. Sultanov
**University of Sofia, Sofia, Bulgaria**
A. Dimitrov, L. Litov, B. Pavlov, P. Petkov
**Beihang University, Beijing, China**
W. Fang[7], X. Gao[7], L. Yuan
**Institute of High Energy Physics, Beijing, China**
M. Ahmad, G.M. Chen, H.S. Chen, M. Chen, C.H. Jiang, D. Leggat, H. Liao, Z. Liu, S.M. Shaheen[8], A. Spiezia, J. Tao, E. Yazgan, H. Zhang, S. Zhang[8], J. Zhao
**State Key Laboratory of Nuclear Physics and Technology, Peking University, Beijing, China**
Y. Ban, G. Chen, J. Li, L. Li, Q. Li, Y. Mao, S.J. Qian, D. Wang
**Tsinghua University, Beijing, China**
Z. Hu, Y. Wang
**Universidad de Los Andes, Bogota, Colombia**
C. Avila, A. Cabrera, L.F. Chaparro Sierra, C. Florez, C.F. González Hernández, M.A. Segura Delgado

**Universidad de Antioquia, Medellin, Colombia**
J. Mejia Guisao, J.D. Ruiz Alvarez, C.A. Salazar González, N. Vanegas Arbelaez
**University of Split, Faculty of Electrical Engineering, Mechanical Engineering and Naval Architecture, Split, Croatia**
D. Giljanović, N. Godinovic, D. Lelas, I. Puljak, T. Sculac
**University of Split, Faculty of Science, Split, Croatia**
Z. Antunovic, M. Kovac
**Institute Rudjer Boskovic, Zagreb, Croatia**
V. Brigljevic, S. Ceci, D. Ferencek, K. Kadija, B. Mesic, M. Roguljic, A. Starodumov[9], T. Susa
**University of Cyprus, Nicosia, Cyprus**
M.W. Ather, A. Attikis, E. Erodotou, A. Ioannou, M. Kolosova, S. Konstantinou, G. Mavromanolakis, J. Mousa, C. Nicolaou, F. Ptochos, P.A. Razis, H. Rykaczewski, D. Tsiakkouri
**Charles University, Prague, Czech Republic**
M. Finger[10], M. Finger Jr.[10], A. Kveton, J. Tomsa
**Escuela Politecnica Nacional, Quito, Ecuador**
E. Ayala
**Universidad San Francisco de Quito, Quito, Ecuador**
E. Carrera Jarrin
**Academy of Scientific Research and Technology of the Arab Republic of Egypt, Egyptian Network of High Energy Physics, Cairo, Egypt**
Y. Assran[11,12], S. Elgammal[12]
**National Institute of Chemical Physics and Biophysics, Tallinn, Estonia**
S. Bhowmik, A. Carvalho Antunes De Oliveira, R.K. Dewanjee, K. Ehataht, M. Kadastik, M. Raidal, C. Veelken
**Department of Physics, University of Helsinki, Helsinki, Finland**
P. Eerola, L. Forthomme, H. Kirschenmann, K. Osterberg, J. Pekkanen, M. Voutilainen
**Helsinki Institute of Physics, Helsinki, Finland**
F. Garcia, J. Havukainen, J.K. Heikkilä, T. Järvinen, V. Karimäki, R. Kinnunen, T. Lampén, K. Lassila-Perini, S. Laurila, S. Lehti, T. Lindén, P. Luukka, T. Mäenpää, H. Siikonen, E. Tuominen, J. Tuominiemi
**Lappeenranta University of Technology, Lappeenranta, Finland**
T. Tuuva
**IRFU, CEA, Université Paris-Saclay, Gif-sur-Yvette, France**
M. Besancon, F. Couderc, M. Dejardin, D. Denegri, B. Fabbro, J.L. Faure, F. Ferri, S. Ganjour, A. Givernaud, P. Gras, G. Hamel de Monchenault, P. Jarry, C. Leloup, E. Locci, J. Malcles, J. Rander, A. Rosowsky, M.Ö. Sahin, A. Savoy-Navarro[13], M. Titov
**Laboratoire Leprince-Ringuet, Ecole polytechnique, CNRS/IN2P3, Université Paris-Saclay, Palaiseau, France**
C. Amendola, F. Beaudette, P. Busson, C. Charlot, B. Diab, R. Granier de Cassagnac, I. Kucher, A. Lobanov, C. Martin Perez, M. Nguyen, C. Ochando, P. Paganini, J. Rembser, R. Salerno, J.B. Sauvan, Y. Sirois, A. Zabi, A. Zghiche
**Université de Strasbourg, CNRS, IPHC UMR 7178, Strasbourg, France**
J.-L. Agram[14], J. Andrea, D. Bloch, G. Bourgatte, J.-M. Brom, E.C. Chabert, C. Collard, E. Conte[14], J.-C. Fontaine[14], D. Gelé, U. Goerlach, M. Jansová, A.-C. Le Bihan, N. Tonon, P. Van Hove
**Centre de Calcul de l'Institut National de Physique Nucleaire et de Physique des Particules, CNRS/IN2P3, Villeurbanne, France**
S. Gadrat
**Université de Lyon, Université Claude Bernard Lyon 1, CNRS-IN2P3, Institut de Physique Nucléaire de Lyon, Villeurbanne, France**
S. Beauceron, C. Bernet, G. Boudoul, C. Camen, N. Chanon, R. Chierici, D. Contardo, P. Depasse, H. El Mamouni, J. Fay, S. Gascon, M. Gouzevitch, B. Ille, F. Lagarde, I.B. Laktineh, H. Lattaud, M. Lethuillier, L. Mirabito, S. Perries, V. Sordini, G. Touquet, M. Vander Donckt, S. Viret
**Georgian Technical University, Tbilisi, Georgia**
A. Khvedelidze[10]
**Tbilisi State University, Tbilisi, Georgia**
Z. Tsamalaidze[10]
**RWTH Aachen University, I. Physikalisches Institut, Aachen, Germany**
C. Autermann, L. Feld, M.K. Kiesel, K. Klein, M. Lipinski, D. Meuser, A. Pauls, M. Preuten, M.P. Rauch, C. Schomakers, M. Teroerde, B. Wittmer

**RWTH Aachen University, III. Physikalisches Institut A, Aachen, Germany**
A. Albert, M. Erdmann, S. Erdweg, T. Esch, B. Fischer, R. Fischer, S. Ghosh, T. Hebbeker, K. Hoepfner, H. Keller, L. Mastrolorenzo, M. Merschmeyer, A. Meyer, P. Millet, G. Mocellin, S. Mondal, S. Mukherjee, D. Noll, A. Novak, T. Pook, A. Pozdnyakov, T. Quast, M. Radziej, Y. Rath, H. Reithler, M. Rieger, A. Schmidt, S.C. Schuler, A. Sharma, S. Thüer, S. Wiedenbeck
**RWTH Aachen University, III. Physikalisches Institut B, Aachen, Germany**
G. Flügge, W. Haj Ahmad[15], O. Hlushchenko, T. Kress, T. Müller, A. Nehrkorn, A. Nowack, C. Pistone, O. Pooth, D. Roy, H. Sert, A. Stahl[16]
**Deutsches Elektronen-Synchrotron, Hamburg, Germany**
M. Aldaya Martin, C. Asawatangtrakuldee, P. Asmuss, I. Babounikau, H. Bakhshiansohi, K. Beernaert, O. Behnke, U. Behrens, A. Bermúdez Martínez, D. Bertsche, A.A. Bin Anuar, K. Borras[17], V. Botta, A. Campbell, A. Cardini, P. Connor, S. Consuegra Rodríguez, C. Contreras-Campana, V. Danilov, A. De Wit, M.M. Defranchis, C. Diez Pardos, D. Domínguez Damiani, G. Eckerlin, D. Eckstein, T. Eichhorn, A. Elwood, E. Eren, E. Gallo[18], A. Geiser, J.M. Grados Luyando, A. Grohsjean, M. Guthoff, M. Haranko, A. Harb, A. Jafari, N.Z. Jomhari, H. Jung, A. Kasem[17], M. Kasemann, J. Keaveney, C. Kleinwort, J. Knolle, D. Krücker, W. Lange, T. Lenz, J. Leonard, J. Lidrych, K. Lipka, W. Lohmann[19], R. Mankel, I.-A. Melzer-Pellmann, A.B. Meyer, M. Meyer, M. Missiroli, G. Mittag, J. Mnich, A. Mussgiller, V. Myronenko, D. Pérez Adán, S.K. Pflitsch, D. Pitzl, A. Raspereza, A. Saibel, M. Savitskyi, V. Scheurer, P. Schütze, C. Schwanenberger, R. Shevchenko, A. Singh, H. Tholen, O. Turkot, A. Vagnerini, M. Van De Klundert, G.P. Van Onsem, R. Walsh, Y. Wen, K. Wichmann, C. Wissing, O. Zenaiev, R. Zlebcik
**University of Hamburg, Hamburg, Germany**
R. Aggleton, S. Bein, L. Benato, A. Benecke, V. Blobel, T. Dreyer, A. Ebrahimi, A. Fröhlich, C. Garbers, E. Garutti, D. Gonzalez, P. Gunnellini, J. Haller, A. Hinzmann, A. Karavdina, G. Kasieczka, R. Klanner, R. Kogler, N. Kovalchuk, S. Kurz, V. Kutzner, J. Lange, T. Lange, A. Malara, D. Marconi, J. Multhaup, M. Niedziela, C.E.N. Niemeyer, D. Nowatschin, A. Perieanu, A. Reimers, O. Rieger, C. Scharf, P. Schleper, S. Schumann, J. Schwandt, J. Sonneveld, H. Stadie, G. Steinbrück, F.M. Stober, M. Stöver, B. Vormwald, I. Zoi
**Karlsruher Institut fuer Technologie, Karlsruhe, Germany**
M. Akbiyik, C. Barth, M. Baselga, S. Baur, T. Berger, E. Butz, T. Chwalek, W. De Boer, A. Dierlamm, K. El Morabit, N. Faltermann, M. Giffels, P. Goldenzweig, A. Gottmann, M.A. Harrendorf, F. Hartmann[16], U. Husemann, P. Keicher, S. Kudella, S. Mitra, M.U. Mozer, Th. Müller, M. Musich, A. Nürnberg, G. Quast, K. Rabbertz, M. Schröder, I. Shvetsov, H.J. Simonis, R. Ulrich, M. Weber, C. Wöhrmann, R. Wolf
**Institute of Nuclear and Particle Physics (INPP), NCSR Demokritos, Aghia Paraskevi, Greece**
G. Anagnostou, P. Asenov, G. Daskalakis, T. Geralis, A. Kyriakis, D. Loukas, G. Paspalaki
**National and Kapodistrian University of Athens, Athens, Greece**
M. Diamantopoulou, G. Karathanasis, P. Kontaxakis, A. Panagiotou, I. Papavergou, N. Saoulidou, A. Stakia, K. Theofilatos, K. Vellidis
**National Technical University of Athens, Athens, Greece**
G. Bakas, K. Kousouris, I. Papakrivopoulos, G. Tsipolitis
**University of Ioánnina, Ioánnina, Greece**
I. Evangelou, C. Foudas, P. Gianneios, P. Katsoulis, P. Kokkas, S. Mallios, K. Manitara, N. Manthos, I. Papadopoulos, J. Strologas, F.A. Triantis, D. Tsitsonis
**MTA-ELTE Lendület CMS Particle and Nuclear Physics Group, Eötvös Loránd University, Budapest, Hungary**
M. Bartók[20], M. Csanad, P. Major, K. Mandal, A. Mehta, M.I. Nagy, G. Pasztor, O. Surányi, G.I. Veres
**Wigner Research Centre for Physics, Budapest, Hungary**
G. Bencze, C. Hajdu, D. Horvath[21], F. Sikler, T.Á. Vámi, V. Veszpremi, G. Vesztergombi[†]
**Institute of Nuclear Research ATOMKI, Debrecen, Hungary**
N. Beni, S. Czellar, J. Karancsi[20], A. Makovec, J. Molnar, Z. Szillasi
**Institute of Physics, University of Debrecen, Debrecen, Hungary**
P. Raics, D. Teyssier, B. Ujvari, G. Zilizi
**Eszterhazy Karoly University, Karoly Robert Campus, Gyongyos, Hungary**
T.F. Csorgo, W.J. Metzger, F. Nemes, T. Novak
**Indian Institute of Science (IISc), Bangalore, India**
S. Choudhury, J.R. Komaragiri, L. Panwar, P.C. Tiwari

**National Institute of Science Education and Research, HBNI, Bhubaneswar, India**
S. Bahinipati[23], C. Kar, P. Mal, V.K. Muraleedharan Nair Bindhu, A. Nayak[24], S. Roy Chowdhury, D.K. Sahoo[23], S.K. Swain

**Panjab University, Chandigarh, India**
S. Bansal, S.B. Beri, V. Bhatnagar, S. Chauhan, R. Chawla, N. Dhingra, R. Gupta, A. Kaur, M. Kaur, S. Kaur, P. Kumari, M. Lohan, M. Meena, K. Sandeep, S. Sharma, J.B. Singh, A.K. Virdi

**University of Delhi, Delhi, India**
A. Bhardwaj, B.C. Choudhary, R.B. Garg, M. Gola, S. Keshri, Ashok Kumar, S. Malhotra, M. Naimuddin, P. Priyanka, K. Ranjan, Aashaq Shah, R. Sharma

**Saha Institute of Nuclear Physics, HBNI, Kolkata, India**
R. Bhardwaj[25], M. Bharti[25], R. Bhattacharya, S. Bhattacharya, U. Bhawandeep[25], D. Bhowmik, S. Dey, S. Dutta, S. Ghosh, M. Maity[26], K. Mondal, S. Nandan, A. Purohit, P.K. Rout, A. Roy, G. Saha, S. Sarkar, T. Sarkar[26], M. Sharan, B. Singh[25], S. Thakur[25]

**Indian Institute of Technology Madras, Madras, India**
P.K. Behera, P. Kalbhor, A. Muhammad, P.R. Pujahari, A. Sharma, A.K. Sikdar

**Bhabha Atomic Research Centre, Mumbai, India**
R. Chudasama, D. Dutta, V. Jha, V. Kumar, D.K. Mishra, P.K. Netrakanti, L.M. Pant, P. Shukla

**Tata Institute of Fundamental Research-A, Mumbai, India**
T. Aziz, M.A. Bhat, S. Dugad, G.B. Mohanty, N. Sur, RavindraKumar Verma

**Tata Institute of Fundamental Research-B, Mumbai, India**
S. Banerjee, S. Bhattacharya, S. Chatterjee, P. Das, M. Guchait, S. Karmakar, S. Kumar, G. Majumder, K. Mazumdar, N. Sahoo, S. Sawant

**Indian Institute of Science Education and Research (IISER), Pune, India**
S. Chauhan, S. Dube, V. Hegde, A. Kapoor, K. Kothekar, S. Pandey, A. Rane, A. Rastogi, S. Sharma

**Institute for Research in Fundamental Sciences (IPM), Tehran, Iran**
S. Chenarani[27], E. Eskandari Tadavani, S.M. Etesami[27], M. Khakzad, M. Mohammadi Najafabadi, M. Naseri, F. Rezaei Hosseinabadi

**University College Dublin, Dublin, Ireland**
M. Felcini, M. Grunewald

**INFN Sezione di Bari [a], Università di Bari [b], Politecnico di Bari [c], Bari, Italy**
M. Abbrescia[a,b], C. Calabria[a,b], A. Colaleo[a], D. Creanza[a,c], L. Cristella[a,b], N. De Filippis[a,c], M. De Palma[a,b], A. Di Florio[a,b], L. Fiore[a], A. Gelmi[a,b], G. Iaselli[a,c], M. Ince[a,b], S. Lezki[a,b], G. Maggi[a,c], M. Maggi[a], G. Miniello[a,b], S. My[a,b], S. Nuzzo[a,b], A. Pompili[a,b], G. Pugliese[a,c], R. Radogna[a], A. Ranieri[a], G. Selvaggi[a,b], L. Silvestris[a], R. Venditti[a], P. Verwilligen[a]

**INFN Sezione di Bologna [a], Università di Bologna [b], Bologna, Italy**
G. Abbiendi[a], C. Battilana[a,b], D. Bonacorsi[a,b], L. Borgonovi[a,b], S. Braibant-Giacomelli[a,b], R. Campanini[a,b], P. Capiluppi[a,b], A. Castro[a,b], F.R. Cavallo[a], C. Ciocca[a], G. Codispoti[a,b], M. Cuffiani[a,b], G.M. Dallavalle[a], F. Fabbri[a], A. Fanfani[a,b], E. Fontanesi, P. Giacomelli[a], C. Grandi[a], L. Guiducci[a,b], F. Iemmi[a,b], S. Lo Meo[a,28], S. Marcellini[a], G. Masetti[a], F.L. Navarria[a,b], A. Perrotta[a], F. Primavera[a,b], A.M. Rossi[a,b], T. Rovelli[a,b], G.P. Siroli[a,b], N. Tosi[a]

**INFN Sezione di Catania [a], Università di Catania [b], Catania, Italy**
S. Albergo[a,b,29], S. Costa[a,b], A. Di Mattia[a], R. Potenza[a,b], A. Tricomi[a,b,29], C. Tuve[a,b]

**INFN Sezione di Firenze [a], Università di Firenze [b], Firenze, Italy**
G. Barbagli[a], R. Ceccarelli, K. Chatterjee[a,b], V. Ciulli[a,b], C. Civinini[a], R. D'Alessandro[a,b], E. Focardi[a,b], G. Latino, P. Lenzi[a,b], M. Meschini[a], S. Paoletti[a], G. Sguazzoni[a], D. Strom[a], L. Viliani[a]

**INFN Laboratori Nazionali di Frascati, Frascati, Italy**
L. Benussi, S. Bianco, D. Piccolo

**INFN Sezione di Genova [a], Università di Genova [b], Genova, Italy**
M. Bozzo[a,b], F. Ferro[a], R. Mulargia[a,b], E. Robutti[a], S. Tosi[a,b]

**INFN Sezione di Milano-Bicocca [a], Università di Milano-Bicocca [b], Milano, Italy**
A. Benaglia[a], A. Beschi[a,b], F. Brivio[a,b], V. Ciriolo[a,b,16], S. Di Guida[a,b,16], M.E. Dinardo[a,b], P. Dini[a], S. Fiorendi[a,b], S. Gennai[a], A. Ghezzi[a,b], P. Govoni[a,b], L. Guzzi[a,b], M. Malberti[a], S. Malvezzi[a], D. Menasce[a], F. Monti[a,b], L. Moroni[a], G. Ortona[a,b], M. Paganoni[a,b], D. Pedrini[a], S. Ragazzi[a,b], T. Tabarelli de Fatis[a,b], D. Zuolo[a,b]

**INFN Sezione di Napoli [a], Università di Napoli 'Federico II' [b], Napoli, Italy, Università della Basilicata [c], Potenza, Italy, Università G. Marconi [d], Roma, Italy**
S. Buontempo[a], N. Cavallo[a,c], A. De Iorio[a,b], A. Di Crescenzo[a,b], F. Fabozzi[a,c], F. Fienga[a], G. Galati[a], A.O.M. Iorio[a,b], L. Lista[a,b], S. Meola[a,d,16], P. Paolucci[a,16], B. Rossi[a], C. Sciacca[a,b], E. Voevodina[a,b]

**INFN Sezione di Padova [a], Università di Padova [b], Padova, Italy, Università di Trento[c], Trento, Italy**
P. Azzi[a], N. Bacchetta[a], D. Bisello[a,b], A. Boletti[a,b], A. Bragagnolo, R. Carlin[a,b], P. Checchia[a], P. De Castro Manzano[a], T. Dorigo[a], U. Dosselli[a], F. Gasparini[a,b], U. Gasparini[a,b], A. Gozzelino[a], S.Y. Hoh, P. Lujan, M. Margoni[a,b], A.T. Meneguzzo[a,b], J. Pazzini[a,b], M. Presilla[b], P. Ronchese[a,b], R. Rossin[a,b], F. Simonetto[a,b], A. Tiko, M. Tosi[a,b], M. Zanetti[a,b], P. Zotto[a,b], G. Zumerle[a,b]

**INFN Sezione di Pavia [a], Università di Pavia [b], Pavia, Italy**
A. Braghieri[a], P. Montagna[a,b], S.P. Ratti[a,b], V. Re[a], M. Ressegotti[a,b], C. Riccardi[a,b], P. Salvini[a], I. Vai[a,b], P. Vitulo[a,b]

**INFN Sezione di Perugia [a], Università di Perugia [b], Perugia, Italy**
M. Biasini[a,b], G.M. Bilei[a], C. Cecchi[a,b], D. Ciangottini[a,b], L. Fanò[a,b], P. Lariccia[a,b], R. Leonardi[a,b], E. Manoni[a], G. Mantovani[a,b], V. Mariani[a,b], M. Menichelli[a], A. Rossi[a,b], A. Santocchia[a,b], D. Spiga[a]

**INFN Sezione di Pisa [a], Università di Pisa [b], Scuola Normale Superiore di Pisa [c], Pisa, Italy**
K. Androsov[a], P. Azzurri[a], G. Bagliesi[a], V. Bertacchi[a,c], L. Bianchini[a], T. Boccali[a], R. Castaldi[a], M.A. Ciocci[a,b], R. Dell'Orso[a], G. Fedi[a], F. Fiori[a,c], L. Giannini[a,c], A. Giassi[a], M.T. Grippo[a], F. Ligabue[a,c], E. Manca[a,c], G. Mandorli[a,c], A. Messineo[a,b], F. Palla[a], A. Rizzi[a,b], G. Rolandi[30], A. Scribano[a], P. Spagnolo[a], R. Tenchini[a], G. Tonelli[a,b], N. Turini, A. Venturi[a], P.G. Verdini[a]

**INFN Sezione di Roma [a], Sapienza Università di Roma [b], Rome, Italy**
F. Cavallari[a], M. Cipriani[a,b], D. Del Re[a,b], E. Di Marco[a,b], M. Diemoz[a], E. Longo[a,b], B. Marzocchi[a,b], P. Meridiani[a], G. Organtini[a,b], F. Pandolfi[a], R. Paramatti[a,b], C. Quaranta[a,b], S. Rahatlou[a,b], C. Rovelli[a], F. Santanastasio[a,b]

**INFN Sezione di Torino [a], Università di Torino [b], Torino, Italy, Università del Piemonte Orientale[c], Novara, Italy**
N. Amapane[a,b], R. Arcidiacono[a,c], S. Argiro[a,b], M. Arneodo[a,c], N. Bartosik[a], R. Bellan[a,b], C. Biino[a], A. Cappati[a,b], N. Cartiglia[a], S. Cometti[a], M. Costa[a,b], R. Covarelli[a,b], N. Demaria[a], B. Kiani[a,b], C. Mariotti[a], S. Maselli[a], E. Migliore[a,b], V. Monaco[a,b], E. Monteil[a,b], M. Monteno[a], M.M. Obertino[a,b], L. Pacher[a,b], N. Pastrone[a], M. Pelliccioni[a], G.L. Pinna Angioni[a,b], A. Romero[a,b], M. Ruspa[a,c], R. Sacchi[a,b], R. Salvatico[a,b], K. Shchelina[a,b], V. Sola[a], A. Solano[a,b], D. Soldi[a,b], A. Staiano[a]

**INFN Sezione di Trieste [a], Università di Trieste [b], Trieste, Italy**
S. Belforte[a], V. Candelise[a,b], M. Casarsa[a], F. Cossutti[a], A. Da Rold[a,b], G. Della Ricca[a,b], F. Vazzoler[a,b], A. Zanetti[a]

**Kyungpook National University, Daegu, Korea**
B. Kim, D.H. Kim, G.N. Kim, J. Lee, S.W. Lee, C.S. Moon, Y.D. Oh, S.I. Pak, S. Sekmen, D.C. Son, Y.C. Yang, R. Ye

**Chonnam National University, Institute for Universe and Elementary Particles, Kwangju, Korea**
H. Kim, D.H. Moon, G. Oh

**Hanyang University, Seoul, Korea**
B. Francois, T.J. Kim, J. Park

**Korea University, Seoul, Korea**
S. Cho, S. Choi, Y. Go, D. Gyun, S. Ha, B. Hong, K. Lee, K.S. Lee, J. Lim, J. Park, S.K. Park, Y. Roh

**Kyung Hee University, Department of Physics**
J. Goh

**Sejong University, Seoul, Korea**
H.S. Kim

**Seoul National University, Seoul, Korea**
J. Almond, J.H. Bhyun, J. Choi, S. Jeon, J. Kim, J.S. Kim, H. Lee, K. Lee, S. Lee, K. Nam, S.B. Oh, B.C. Radburn-Smith, S.h. Seo, U.K. Yang, H.D. Yoo, I. Yoon, G.B. Yu

**University of Seoul, Seoul, Korea**
D. Jeon, H. Kim, J.H. Kim, J.S.H. Lee, I.C. Park, I. Watson

**Sungkyunkwan University, Suwon, Korea**
Y. Choi, C. Hwang, Y. Jeong, J. Lee, Y. Lee, I. Yu

**Riga Technical University, Riga, Latvia**
V. Veckalns[31]

**Vilnius University, Vilnius, Lithuania**
V. Dudenas, A. Juodagalvis, J. Vaitkus

**National Centre for Particle Physics, Universiti Malaya, Kuala Lumpur, Malaysia**
Z.A. Ibrahim, F. Mohamad Idris[32], W.A.T. Wan Abdullah, M.N. Yusli, Z. Zolkapli
**Universidad de Sonora (UNISON), Hermosillo, Mexico**
J.F. Benitez, A. Castaneda Hernandez, J.A. Murillo Quijada, L. Valencia Palomo
**Centro de Investigacion y de Estudios Avanzados del IPN, Mexico City, Mexico**
H. Castilla-Valdez, E. De La Cruz-Burelo, I. Heredia-De La Cruz[33], R. Lopez-Fernandez, A. Sanchez-Hernandez
**Universidad Iberoamericana, Mexico City, Mexico**
S. Carrillo Moreno, C. Oropeza Barrera, M. Ramirez-Garcia, F. Vazquez Valencia
**Benemerita Universidad Autonoma de Puebla, Puebla, Mexico**
J. Eysermans, I. Pedraza, H.A. Salazar Ibarguen, C. Uribe Estrada
**Universidad Autónoma de San Luis Potosí, San Luis Potosí, Mexico**
A. Morelos Pineda
**University of Montenegro, Podgorica, Montenegro**
N. Raicevic
**University of Auckland, Auckland, New Zealand**
D. Krofcheck
**University of Canterbury, Christchurch, New Zealand**
S. Bheesette, P.H. Butler
**National Centre for Physics, Quaid-I-Azam University, Islamabad, Pakistan**
A. Ahmad, M. Ahmad, Q. Hassan, H.R. Hoorani, W.A. Khan, M.A. Shah, M. Shoaib, M. Waqas
**AGH University of Science and Technology Faculty of Computer Science, Electronics and Telecommunications, Krakow, Poland**
V. Avati, L. Grzanka, M. Malawski
**National Centre for Nuclear Research, Swierk, Poland**
H. Bialkowska, M. Bluj, B. Boimska, M. Górski, M. Kazana, M. Szleper, P. Zalewski
**Institute of Experimental Physics, Faculty of Physics, University of Warsaw, Warsaw, Poland**
K. Bunkowski, A. Byszuk[34], K. Doroba, A. Kalinowski, M. Konecki, J. Krolikowski, M. Misiura, M. Olszewski, A. Pyskir, M. Walczak
**Laboratório de Instrumentação e Física Experimental de Partículas, Lisboa, Portugal**
M. Araujo, P. Bargassa, D. Bastos, M. Bengala, A. Di Francesco, P. Faccioli, B. Galinhas, M. Gallinaro, J. Hollar, R. José Santo, N. Leonardo, J. Seixas, G. Strong, O. Toldaiev, J. Varela
**Joint Institute for Nuclear Research, Dubna, Russia**
S. Afanasiev, P. Bunin, M. Gavrilenko, I. Golutvin, I. Gorbunov, A. Kamenev, V. Karjavine, A. Lanev, A. Malakhov, V. Matveev[35,36], P. Moisenz, V. Palichik, V. Perelygin, M. Savina, S. Shmatov, S. Shulha, N. Skatchkov, V. Smirnov, N. Voytishin, A. Zarubin
**Petersburg Nuclear Physics Institute, Gatchina (St. Petersburg), Russia**
L. Chtchipounov, V. Golovtsov, Y. Ivanov, V. Kim[37], E. Kuznetsova[38], P. Levchenko, V. Murzin, V. Oreshkin, I. Smirnov, D. Sosnov, V. Sulimov, L. Uvarov, A. Vorobyev
**Institute for Nuclear Research, Moscow, Russia**
Yu. Andreev, A. Dermenev, S. Gninenko, N. Golubev, A. Karneyeu, M. Kirsanov, N. Krasnikov, A. Pashenkov, D. Tlisov, A. Toropin
**Institute for Theoretical and Experimental Physics, Moscow, Russia**
V. Epshteyn, V. Gavrilov, N. Lychkovskaya, A. Nikitenko[39], V. Popov, I. Pozdnyakov, G. Safronov, A. Spiridonov, A. Stepennov, M. Toms, E. Vlasov, A. Zhokin
**Moscow Institute of Physics and Technology, Moscow, Russia**
T. Aushev
**National Research Nuclear University 'Moscow Engineering Physics Institute' (MEPhI), Moscow, Russia**
M. Chadeeva[40], M. Danilov[40], P. Parygin, D. Philippov, E. Popova
**P.N. Lebedev Physical Institute, Moscow, Russia**
V. Andreev, M. Azarkin, I. Dremin[36], M. Kirakosyan, A. Terkulov
**Skobeltsyn Institute of Nuclear Physics, Lomonosov Moscow State University, Moscow, Russia**
A. Baskakov, A. Belyaev, E. Boos, V. Bunichev, M. Dubinin[41], L. Dudko, V. Klyukhin, O. Kodolova, I. Lokhtin, S. Obraztsov, M. Perfilov, S. Petrushanko, V. Savrin
**Novosibirsk State University (NSU), Novosibirsk, Russia**
A. Barnyakov[42], V. Blinov[42], T. Dimova[42], L. Kardapoltsev[42], Y. Skovpen[42]

**Institute for High Energy Physics of National Research Centre 'Kurchatov Institute', Protvino, Russia**
I. Azhgirey, I. Bayshev, S. Bitioukov, V. Kachanov, D. Konstantinov, P. Mandrik, V. Petrov, R. Ryutin, S. Slabospitskii, A. Sobol, S. Troshin, N. Tyurin, A. Uzunian, A. Volkov
**National Research Tomsk Polytechnic University, Tomsk, Russia**
A. Babaev, A. Iuzhakov, V. Okhotnikov
**Tomsk State University**
V. Borchsh, V. Ivanchenko, E. Tcherniaev
**University of Belgrade: Faculty of Physics and VINCA Institute of Nuclear Sciences**
P. Adzic[43], P. Cirkovic, D. Devetak, M. Dordevic, P. Milenovic, J. Milosevic, M. Stojanovic
**Centro de Investigaciones Energéticas Medioambientales y Tecnológicas (CIEMAT), Madrid, Spain**
M. Aguilar-Benitez, J. Alcaraz Maestre, A. Álvarez Fernández, I. Bachiller, M. Barrio Luna, J.A. Brochero Cifuentes, C.A. Carrillo Montoya, M. Cepeda, M. Cerrada, N. Colino, B. De La Cruz, A. Delgado Peris, C. Fernandez Bedoya, J.P. Fernández Ramos, J. Flix, M.C. Fouz, O. Gonzalez Lopez, S. Goy Lopez, J.M. Hernandez, M.I. Josa, D. Moran, Á. Navarro Tobar, A. Pérez-Calero Yzquierdo, J. Puerta Pelayo, I. Redondo, L. Romero, S. Sánchez Navas, M.S. Soares, A. Triossi, C. Willmott
**Universidad Autónoma de Madrid, Madrid, Spain**
C. Albajar, J.F. de Trocóniz
**Universidad de Oviedo, Oviedo, Spain**
B. Alvarez Gonzalez, J. Cuevas, C. Erice, J. Fernandez Menendez, S. Folgueras, I. Gonzalez Caballero, J.R. González Fernández, E. Palencia Cortezon, V. Rodríguez Bouza, S. Sanchez Cruz
**Instituto de Física de Cantabria (IFCA), CSIC-Universidad de Cantabria, Santander, Spain**
I.J. Cabrillo, A. Calderon, B. Chazin Quero, J. Duarte Campderros, M. Fernandez, P.J. Fernández Manteca, A. García Alonso, G. Gomez, C. Martinez Rivero, P. Martinez Ruiz del Arbol, F. Matorras, J. Piedra Gomez, C. Prieels, T. Rodrigo, A. Ruiz-Jimeno, L. Russo[44], L. Scodellaro, N. Trevisani, I. Vila, J.M. Vizan Garcia
**University of Colombo, Colombo, Sri Lanka**
K. Malagalage
**University of Ruhuna, Department of Physics, Matara, Sri Lanka**
W.G.D. Dharmaratna, N. Wickramage
**CERN, European Organization for Nuclear Research, Geneva, Switzerland**
D. Abbaneo, B. Akgun, E. Auffray, G. Auzinger, J. Baechler, P. Baillon, A.H. Ball, D. Barney, J. Bendavid, M. Bianco, A. Bocci, E. Bossini, C. Botta, E. Brondolin, T. Camporesi, A. Caratelli, G. Cerminara, E. Chapon, G. Cucciati, D. d'Enterria, A. Dabrowski, N. Daci, V. Daponte, A. David, A. De Roeck, N. Deelen, M. Deile, M. Dobson, M. Dünser, N. Dupont, A. Elliott-Peisert, F. Fallavollita[45], D. Fasanella, G. Franzoni, J. Fulcher, W. Funk, S. Giani, D. Gigi, A. Gilbert, K. Gill, F. Glege, M. Gruchala, M. Guilbaud, D. Gulhan, J. Hegeman, C. Heidegger, Y. Iiyama, V. Innocente, P. Janot, O. Karacheban[19], J. Kaspar, J. Kieseler, M. Krammer[1], C. Lange, P. Lecoq, C. Lourenço, L. Malgeri, M. Mannelli, A. Massironi, F. Meijers, J.A. Merlin, S. Mersi, E. Meschi, F. Moortgat, M. Mulders, J. Ngadiuba, S. Nourbakhsh, S. Orfanelli, L. Orsini, F. Pantaleo[16], L. Pape, E. Perez, M. Peruzzi, A. Petrilli, G. Petrucciani, A. Pfeiffer, M. Pierini, F.M. Pitters, M. Quinto, D. Rabady, A. Racz, M. Rovere, H. Sakulin, C. Schäfer, C. Schwick, M. Selvaggi, A. Sharma, P. Silva, W. Snoeys, P. Sphicas[46], J. Steggemann, V.R. Tavolaro, D. Treille, A. Tsirou, A. Vartak, M. Verzetti, W.D. Zeuner
**Paul Scherrer Institut, Villigen, Switzerland**
L. Caminada[47], K. Deiters, W. Erdmann, R. Horisberger, Q. Ingram, H.C. Kaestli, D. Kotlinski, U. Langenegger, T. Rohe, S.A. Wiederkehr
**ETH Zurich - Institute for Particle Physics and Astrophysics (IPA), Zurich, Switzerland**
M. Backhaus, P. Berger, N. Chernyavskaya, G. Dissertori, M. Dittmar, M. Donegà, C. Dorfer, T.A. Gómez Espinosa, C. Grab, D. Hits, T. Klijnsma, W. Lustermann, R.A. Manzoni, M. Marionneau, M.T. Meinhard, F. Micheli, P. Musella, F. Nessi-Tedaldi, F. Pauss, G. Perrin, L. Perrozzi, S. Pigazzini, M. Reichmann, C. Reissel, T. Reitenspiess, D. Ruini, D.A. Sanz Becerra, M. Schönenberger, L. Shchutska, M.L. Vesterbacka Olsson, R. Wallny, D.H. Zhu
**Universität Zürich, Zurich, Switzerland**
T.K. Aarrestad, C. Amsler[48], D. Brzhechko, M.F. Canelli, A. De Cosa, R. Del Burgo, S. Donato, B. Kilminster, S. Leontsinis, V.M. Mikuni, I. Neutelings, G. Rauco, P. Robmann, D. Salerno, K. Schweiger, C. Seitz, Y. Takahashi, S. Wertz, A. Zucchetta

**National Central University, Chung-Li, Taiwan**
T.H. Doan, C.M. Kuo, W. Lin, S.S. Yu

**National Taiwan University (NTU), Taipei, Taiwan**
P. Chang, Y. Chao, K.F. Chen, P.H. Chen, W.-S. Hou, Y.y. Li, R.-S. Lu, E. Paganis, A. Psallidas, A. Steen

**Chulalongkorn University, Faculty of Science, Department of Physics, Bangkok, Thailand**
B. Asavapibhop, N. Srimanobhas, N. Suwonjandee

**Çukurova University, Physics Department, Science and Art Faculty, Adana, Turkey**
A. Bat, F. Boran, S. Cerci[49], S. Damarseckin[50], Z.S. Demiroglu, F. Dolek, C. Dozen, I. Dumanoglu, G. Gokbulut, E. Gurpinar Guler[51], Y. Guler, I. Hos[52], C. Isik, E.E. Kangal[53], O. Kara, A. Kayis Topaksu, U. Kiminsu, M. Oglakci, G. Onengut, K. Ozdemir[54], S. Ozturk[55], A.E. Simsek, D. Sunar Cerci[49], U.G. Tok, S. Turkcapar, I.S. Zorbakir, C. Zorbilmez

**Middle East Technical University, Physics Department, Ankara, Turkey**
B. Isildak[56], G. Karapinar[57], M. Yalvac

**Bogazici University, Istanbul, Turkey**
I.O. Atakisi, E. Gülmez, O. Kaya[58], B. Kaynak, Ö. Özçelik, S. Ozkorucuklu[59], S. Tekten, E.A. Yetkin[60]

**Istanbul Technical University, Istanbul, Turkey**
A. Cakir, Y. Komurcu, S. Sen[61]

**Institute for Scintillation Materials of National Academy of Science of Ukraine, Kharkov, Ukraine**
B. Grynyov

**National Scientific Center, Kharkov Institute of Physics and Technology, Kharkov, Ukraine**
L. Levchuk

**University of Bristol, Bristol, United Kingdom**
F. Ball, E. Bhal, S. Bologna, J.J. Brooke, D. Burns, E. Clement, D. Cussans, O. Davignon, H. Flacher, J. Goldstein, G.P. Heath, H.F. Heath, L. Kreczko, S. Paramesvaran, B. Penning, T. Sakuma, S. Seif El Nasr-Storey, D. Smith, V.J. Smith, J. Taylor, A. Titterton

**Rutherford Appleton Laboratory, Didcot, United Kingdom**
K.W. Bell, A. Belyaev[62], C. Brew, R.M. Brown, D. Cieri, D.J.A. Cockerill, J.A. Coughlan, K. Harder, S. Harper, J. Linacre, K. Manolopoulos, D.M. Newbold, E. Olaiya, D. Petyt, T. Reis, T. Schuh, C.H. Shepherd-Themistocleous, A. Thea, I.R. Tomalin, T. Williams, W.J. Womersley

**Imperial College, London, United Kingdom**
R. Bainbridge, P. Bloch, J. Borg, S. Breeze, O. Buchmuller, A. Bundock, G.S. Chahal[63], D. Colling, P. Dauncey, G. Davies, M. Della Negra, R. Di Maria, P. Everaerts, G. Hall, G. Iles, T. James, M. Komm, C. Laner, J. Langford, L. Lyons, A.-M. Magnan, S. Malik, A. Martelli, V. Milosevic, J. Nash[64], V. Palladino, M. Pesaresi, D.M. Raymond, A. Richards, A. Rose, E. Scott, C. Seez, A. Shtipliyski, M. Stoye, T. Strebler, S. Summers, A. Tapper, K. Uchida, T. Virdee[16], N. Wardle, D. Winterbottom, J. Wright, A.G. Zecchinelli, S.C. Zenz

**Brunel University, Uxbridge, United Kingdom**
J.E. Cole, P.R. Hobson, A. Khan, P. Kyberd, C.K. Mackay, A. Morton, I.D. Reid, L. Teodorescu, S. Zahid

**Baylor University, Waco, USA**
K. Call, J. Dittmann, K. Hatakeyama, C. Madrid, B. McMaster, N. Pastika, C. Smith

**Catholic University of America, Washington, DC, USA**
R. Bartek, A. Dominguez, R. Uniyal

**The University of Alabama, Tuscaloosa, USA**
A. Buccilli, S.I. Cooper, C. Henderson, P. Rumerio, C. West

**Boston University, Boston, USA**
D. Arcaro, T. Bose, Z. Demiragli, D. Gastler, S. Girgis, D. Pinna, C. Richardson, J. Rohlf, D. Sperka, I. Suarez, L. Sulak, D. Zou

**Brown University, Providence, USA**
G. Benelli, B. Burkle, X. Coubez, D. Cutts, M. Hadley, J. Hakala, U. Heintz, J.M. Hogan[65], K.H.M. Kwok, E. Laird, G. Landsberg, J. Lee, Z. Mao, M. Narain, S. Sagir[66], R. Syarif, E. Usai, D. Yu, W. Zhang

**University of California, Davis, Davis, USA**
R. Band, C. Brainerd, R. Breedon, M. Calderon De La Barca Sanchez, M. Chertok, J. Conway, R. Conway, P.T. Cox, R. Erbacher, C. Flores, G. Funk, F. Jensen, W. Ko, O. Kukral, R. Lander, M. Mulhearn, D. Pellett, J. Pilot, M. Shi, D. Stolp, D. Taylor, K. Tos, M. Tripathi, Z. Wang, F. Zhang

**University of California, Los Angeles, USA**
M. Bachtis, C. Bravo, R. Cousins, A. Dasgupta, A. Florent, J. Hauser, M. Ignatenko, N. Mccoll, S. Regnard, D. Saltzberg, C. Schnaible, V. Valuev
**University of California, Riverside, Riverside, USA**
K. Burt, R. Clare, J.W. Gary, S.M.A. Ghiasi Shirazi, G. Hanson, G. Karapostoli, E. Kennedy, O.R. Long, M. Olmedo Negrete, M.I. Paneva, W. Si, L. Wang, H. Wei, S. Wimpenny, B.R. Yates, Y. Zhang
**University of California, San Diego, La Jolla, USA**
J.G. Branson, P. Chang, S. Cittolin, M. Derdzinski, R. Gerosa, D. Gilbert, B. Hashemi, D. Klein, V. Krutelyov, J. Letts, M. Masciovecchio, S. May, S. Padhi, M. Pieri, V. Sharma, M. Tadel, F. Würthwein, A. Yagil, G. Zevi Della Porta
**University of California, Santa Barbara - Department of Physics, Santa Barbara, USA**
N. Amin, R. Bhandari, C. Campagnari, M. Citron, V. Dutta, M. Franco Sevilla, L. Gouskos, J. Incandela, B. Marsh, H. Mei, A. Ovcharova, H. Qu, J. Richman, U. Sarica, D. Stuart, S. Wang, J. Yoo
**California Institute of Technology, Pasadena, USA**
D. Anderson, A. Bornheim, J.M. Lawhorn, N. Lu, H.B. Newman, T.Q. Nguyen, J. Pata, M. Spiropulu, J.R. Vlimant, S. Xie, Z. Zhang, R.Y. Zhu
**Carnegie Mellon University, Pittsburgh, USA**
M.B. Andrews, T. Ferguson, T. Mudholkar, M. Paulini, M. Sun, I. Vorobiev, M. Weinberg
**University of Colorado Boulder, Boulder, USA**
J.P. Cumalat, W.T. Ford, A. Johnson, E. MacDonald, T. Mulholland, R. Patel, A. Perloff, K. Stenson, K.A. Ulmer, S.R. Wagner
**Cornell University, Ithaca, USA**
J. Alexander, J. Chaves, Y. Cheng, J. Chu, A. Datta, A. Frankenthal, K. Mcdermott, N. Mirman, J.R. Patterson, D. Quach, A. Rinkevicius[67], A. Ryd, S.M. Tan, Z. Tao, J. Thom, P. Wittich, M. Zientek
**Fermi National Accelerator Laboratory, Batavia, USA**
S. Abdullin, M. Albrow, M. Alyari, G. Apollinari, A. Apresyan, A. Apyan, S. Banerjee, L.A.T. Bauerdick, A. Beretvas, J. Berryhill, P.C. Bhat, K. Burkett, J.N. Butler, A. Canepa, G.B. Cerati, H.W.K. Cheung, F. Chlebana, M. Cremonesi, J. Duarte, V.D. Elvira, J. Freeman, Z. Gecse, E. Gottschalk, L. Gray, D. Green, S. Grünendahl, O. Gutsche, AllisonReinsvold Hall, J. Hanlon, R.M. Harris, S. Hasegawa, R. Heller, J. Hirschauer, B. Jayatilaka, S. Jindariani, M. Johnson, U. Joshi, B. Klima, M.J. Kortelainen, B. Kreis, S. Lammel, J. Lewis, D. Lincoln, R. Lipton, M. Liu, T. Liu, J. Lykken, K. Maeshima, J.M. Marraffino, D. Mason, P. McBride, P. Merkel, S. Mrenna, S. Nahn, V. O'Dell, V. Papadimitriou, K. Pedro, C. Pena, G. Rakness, F. Ravera, L. Ristori, B. Schneider, E. Sexton-Kennedy, N. Smith, A. Soha, W.J. Spalding, L. Spiegel, S. Stoynev, J. Strait, N. Strobbe, L. Taylor, S. Tkaczyk, N.V. Tran, L. Uplegger, E.W. Vaandering, C. Vernieri, M. Verzocchi, R. Vidal, M. Wang, H.A. Weber
**University of Florida, Gainesville, USA**
D. Acosta, P. Avery, P. Bortignon, D. Bourilkov, A. Brinkerhoff, L. Cadamuro, A. Carnes, V. Cherepanov, D. Curry, F. Errico, R.D. Field, S.V. Gleyzer, B.M. Joshi, M. Kim, J. Konigsberg, A. Korytov, K.H. Lo, P. Ma, K. Matchev, N. Menendez, G. Mitselmakher, D. Rosenzweig, K. Shi, J. Wang, S. Wang, X. Zuo
**Florida International University, Miami, USA**
Y.R. Joshi
**Florida State University, Tallahassee, USA**
T. Adams, A. Askew, S. Hagopian, V. Hagopian, K.F. Johnson, R. Khurana, T. Kolberg, G. Martinez, T. Perry, H. Prosper, C. Schiber, R. Yohay, J. Zhang
**Florida Institute of Technology, Melbourne, USA**
M.M. Baarmand, V. Bhopatkar, M. Hohlmann, D. Noonan, M. Rahmani, M. Saunders, F. Yumiceva
**University of Illinois at Chicago (UIC), Chicago, USA**
M.R. Adams, L. Apanasevich, D. Berry, R.R. Betts, R. Cavanaugh, X. Chen, S. Dittmer, O. Evdokimov, C.E. Gerber, D.A. Hangal, D.J. Hofman, K. Jung, C. Mills, T. Roy, M.B. Tonjes, N. Varelas, H. Wang, X. Wang, Z. Wu
**The University of Iowa, Iowa City, USA**
M. Alhusseini, B. Bilki[51], W. Clarida, K. Dilsiz[68], S. Durgut, R.P. Gandrajula, M. Haytmyradov, V. Khristenko, O.K. Köseyan, J.-P. Merlo, A. Mestvirishvili[69], A. Moeller, J. Nachtman, H. Ogul[70], Y. Onel, F. Ozok[71], A. Penzo, C. Snyder, E. Tiras, J. Wetzel
**Johns Hopkins University, Baltimore, USA**
B. Blumenfeld, A. Cocoros, N. Eminizer, D. Fehling, L. Feng, A.V. Gritsan, W.T. Hung, P. Maksimovic, J. Roskes, M. Swartz, M. Xiao

**The University of Kansas, Lawrence, USA**
C. Baldenegro Barrera, P. Baringer, A. Bean, S. Boren, J. Bowen, A. Bylinkin, T. Isidori, S. Khalil, J. King, A. Kropivnitskaya, C. Lindsey, D. Majumder, W. Mcbrayer, N. Minafra, M. Murray, C. Rogan, C. Royon, S. Sanders, E. Schmitz, J.D. Tapia Takaki, Q. Wang, J. Williams
**Kansas State University, Manhattan, USA**
S. Duric, A. Ivanov, K. Kaadze, D. Kim, Y. Maravin, D.R. Mendis, T. Mitchell, A. Modak, A. Mohammadi
**Lawrence Livermore National Laboratory, Livermore, USA**
F. Rebassoo, D. Wright
**University of Maryland, College Park, USA**
A. Baden, O. Baron, A. Belloni, S.C. Eno, Y. Feng, N.J. Hadley, S. Jabeen, G.Y. Jeng, R.G. Kellogg, J. Kunkle, A.C. Mignerey, S. Nabili, F. Ricci-Tam, M. Seidel, Y.H. Shin, A. Skuja, S.C. Tonwar, K. Wong
**Massachusetts Institute of Technology, Cambridge, USA**
D. Abercrombie, B. Allen, A. Baty, R. Bi, S. Brandt, W. Busza, I.A. Cali, M. D'Alfonso, G. Gomez Ceballos, M. Goncharov, P. Harris, D. Hsu, M. Hu, M. Klute, D. Kovalskyi, Y.-J. Lee, P.D. Luckey, B. Maier, A.C. Marini, C. Mcginn, C. Mironov, S. Narayanan, X. Niu, C. Paus, D. Rankin, C. Roland, G. Roland, Z. Shi, G.S.F. Stephans, K. Sumorok, K. Tatar, D. Velicanu, J. Wang, T.W. Wang, B. Wyslouch
**University of Minnesota, Minneapolis, USA**
A.C. Benvenuti[†], R.M. Chatterjee, A. Evans, S. Guts, P. Hansen, J. Hiltbrand, S. Kalafut, Y. Kubota, Z. Lesko, J. Mans, R. Rusack, M.A. Wadud
**University of Mississippi, Oxford, USA**
J.G. Acosta, S. Oliveros
**University of Nebraska-Lincoln, Lincoln, USA**
K. Bloom, D.R. Claes, C. Fangmeier, L. Finco, F. Golf, R. Gonzalez Suarez, R. Kamalieddin, I. Kravchenko, J.E. Siado, G.R. Snow, B. Stieger
**State University of New York at Buffalo, Buffalo, USA**
C. Harrington, I. Iashvili, A. Kharchilava, C. Mclean, D. Nguyen, A. Parker, S. Rappoccio, B. Roozbahani
**Northeastern University, Boston, USA**
G. Alverson, E. Barberis, C. Freer, Y. Haddad, A. Hortiangtham, G. Madigan, D.M. Morse, T. Orimoto, L. Skinnari, A. Tishelman-Charny, T. Wamorkar, B. Wang, A. Wisecarver, D. Wood
**Northwestern University, Evanston, USA**
S. Bhattacharya, J. Bueghly, T. Gunter, K.A. Hahn, N. Odell, M.H. Schmitt, K. Sung, M. Trovato, M. Velasco
**University of Notre Dame, Notre Dame, USA**
R. Bucci, N. Dev, R. Goldouzian, M. Hildreth, K. Hurtado Anampa, C. Jessop, D.J. Karmgard, K. Lannon, W. Li, N. Loukas, N. Marinelli, I. Mcalister, F. Meng, C. Mueller, Y. Musienko[35], M. Planer, R. Ruchti, P. Siddireddy, G. Smith, S. Taroni, M. Wayne, A. Wightman, M. Wolf, A. Woodard
**The Ohio State University, Columbus, USA**
J. Alimena, B. Bylsma, L.S. Durkin, S. Flowers, B. Francis, C. Hill, W. Ji, A. Lefeld, T.Y. Ling, B.L. Winer
**Princeton University, Princeton, USA**
S. Cooperstein, G. Dezoort, P. Elmer, J. Hardenbrook, N. Haubrich, S. Higginbotham, A. Kalogeropoulos, S. Kwan, D. Lange, M.T. Lucchini, J. Luo, D. Marlow, K. Mei, I. Ojalvo, J. Olsen, C. Palmer, P. Piroué, J. Salfeld-Nebgen, D. Stickland, C. Tully, Z. Wang
**University of Puerto Rico, Mayaguez, USA**
S. Malik, S. Norberg
**Purdue University, West Lafayette, USA**
A. Barker, V.E. Barnes, S. Das, L. Gutay, M. Jones, A.W. Jung, A. Khatiwada, B. Mahakud, D.H. Miller, G. Negro, N. Neumeister, C.C. Peng, S. Piperov, H. Qiu, J.F. Schulte, J. Sun, F. Wang, R. Xiao, W. Xie
**Purdue University Northwest, Hammond, USA**
T. Cheng, J. Dolen, N. Parashar
**Rice University, Houston, USA**
K.M. Ecklund, S. Freed, F.J.M. Geurts, M. Kilpatrick, Arun Kumar, W. Li, B.P. Padley, R. Redjimi, J. Roberts, J. Rorie, W. Shi, A.G. Stahl Leiton, Z. Tu, A. Zhang

**University of Rochester, Rochester, USA**
A. Bodek, P. de Barbaro, R. Demina, Y.t. Duh, J.L. Dulemba, C. Fallon, T. Ferbel, M. Galanti, A. Garcia-Bellido, J. Han, O. Hindrichs, A. Khukhunaishvili, E. Ranken, P. Tan, R. Taus

**Rutgers, The State University of New Jersey, Piscataway, USA**
B. Chiarito, J.P. Chou, A. Gandrakota, Y. Gershtein, E. Halkiadakis, A. Hart, M. Heindl, E. Hughes, S. Kaplan, S. Kyriacou, I. Laflotte, A. Lath, R. Montalvo, K. Nash, M. Osherson, H. Saka, S. Salur, S. Schnetzer, D. Sheffield, S. Somalwar, R. Stone, S. Thomas, P. Thomassen

**University of Tennessee, Knoxville, USA**
H. Acharya, A.G. Delannoy, J. Heideman, G. Riley, S. Spanier

**Texas A&M University, College Station, USA**
O. Bouhali[72], A. Celik, M. Dalchenko, M. De Mattia, A. Delgado, S. Dildick, R. Eusebi, J. Gilmore, T. Huang, T. Kamon[73], S. Luo, D. Marley, R. Mueller, D. Overton, L. Perniè, D. Rathjens, A. Safonov

**Texas Tech University, Lubbock, USA**
N. Akchurin, J. Damgov, F. De Guio, S. Kunori, K. Lamichhane, S.W. Lee, T. Mengke, S. Muthumuni, T. Peltola, S. Undleeb, I. Volobouev, Z. Wang, A. Whitbeck

**Vanderbilt University, Nashville, USA**
S. Greene, A. Gurrola, R. Janjam, W. Johns, C. Maguire, H. Ni, F. Romeo, P. Sheldon, S. Tuo, J. Velkovska, M. Verweij

**University of Virginia, Charlottesville, USA**
M.W. Arenton, P. Barria, B. Cox, G. Cummings, R. Hirosky, M. Joyce, A. Ledovskoy, C. Neu, B. Tannenwald, Y. Wang, E. Wolfe, F. Xia

**Wayne State University, Detroit, USA**
R. Harr, P.E. Karchin, N. Poudyal, J. Sturdy, P. Thapa, S. Zaleski

**University of Wisconsin - Madison, Madison, WI, USA**
J. Buchanan, C. Caillol, D. Carlsmith, S. Dasu, I. De Bruyn, L. Dodd, C. Galloni, B. Gomber[74], M. Herndon, A. Hervé, U. Hussain, P. Klabbers, A. Lanaro, A. Loeliger, K. Long, R. Loveless, J. Madhusudanan Sreekala, T. Ruggles, A. Savin, V. Sharma, W.H. Smith, D. Teague, S. Trembath-reichert, N. Woods

†: Deceased
1: Also at Vienna University of Technology, Vienna, Austria
2: Also at IRFU, CEA, Université Paris-Saclay, Gif-sur-Yvette, France
3: Also at Universidade Estadual de Campinas, Campinas, Brazil
4: Also at Federal University of Rio Grande do Sul, Porto Alegre, Brazil
5: Also at UFMS/CPNA, Federal University of Mato Grosso do Sul/Campus of Nova Andradina, Nova Andradina, Brazil
6: Also at Universidade Federal de Pelotas, Pelotas, Brazil
7: Also at Université Libre de Bruxelles, Bruxelles, Belgium
8: Also at University of Chinese Academy of Sciences, Beijing, China
9: Also at Institute for Theoretical and Experimental Physics, Moscow, Russia
10: Also at Joint Institute for Nuclear Research, Dubna, Russia
11: Also at Suez University, Suez, Egypt
12: Now at British University in Egypt, Cairo, Egypt
13: Also at Purdue University, West Lafayette, USA
14: Also at Université de Haute Alsace, Mulhouse, France
15: Also at Erzincan Binali Yildirim University, Erzincan, Turkey
16: Also at CERN, European Organization for Nuclear Research, Geneva, Switzerland
17: Also at RWTH Aachen University, III. Physikalisches Institut A, Aachen, Germany
18: Also at University of Hamburg, Hamburg, Germany
19: Also at Brandenburg University of Technology, Cottbus, Germany
20: Also at Institute of Physics, University of Debrecen, Debrecen, Hungary
21: Also at Institute of Nuclear Research ATOMKI, Debrecen, Hungary
22: Also at MTA-ELTE Lendület CMS Particle and Nuclear Physics Group, Eötvös Loránd University, Budapest, Hungary
23: Also at Indian Institute of Technology Bhubaneswar, Bhubaneswar, India
24: Also at Institute of Physics, Bhubaneswar, India
25: Also at Shoolini University, Solan, India
26: Also at University of Visva-Bharati, Santiniketan, India
27: Also at Isfahan University of Technology, Isfahan, Iran

28: Also at Italian National Agency for New Technologies, Energy and Sustainable Economic Development, Bologna, Italy
29: Also at Centro Siciliano di Fisica Nucleare e di Struttura Della Materia, Catania, Italy
30: Also at Scuola Normale e Sezione dell'INFN, Pisa, Italy
31: Also at Riga Technical University, Riga, Latvia
32: Also at Malaysian Nuclear Agency, MOSTI, Kajang, Malaysia
33: Also at Consejo Nacional de Ciencia y Tecnología, Mexico City, Mexico
34: Also at Warsaw University of Technology, Institute of Electronic Systems, Warsaw, Poland
35: Also at Institute for Nuclear Research, Moscow, Russia
36: Now at National Research Nuclear University 'Moscow Engineering Physics Institute' (MEPhI), Moscow, Russia
37: Also at St. Petersburg State Polytechnical University, St. Petersburg, Russia
38: Also at University of Florida, Gainesville, USA
39: Also at Imperial College, London, United Kingdom
40: Also at P.N. Lebedev Physical Institute, Moscow, Russia
41: Also at California Institute of Technology, Pasadena, USA
42: Also at Budker Institute of Nuclear Physics, Novosibirsk, Russia
43: Also at Faculty of Physics, University of Belgrade, Belgrade, Serbia
44: Also at Università degli Studi di Siena, Siena, Italy
45: Also at INFN Sezione di Pavia [a], Università di Pavia [b], Pavia, Italy
46: Also at National and Kapodistrian University of Athens, Athens, Greece
47: Also at Universität Zürich, Zurich, Switzerland
48: Also at Stefan Meyer Institute for Subatomic Physics (SMI), Vienna, Austria
49: Also at Adiyaman University, Adiyaman, Turkey
50: Also at Sirnak University, SIRNAK, Turkey
51: Also at Beykent University, Istanbul, Turkey
52: Also at Istanbul Aydin University, Istanbul, Turkey
53: Also at Mersin University, Mersin, Turkey
54: Also at Piri Reis University, Istanbul, Turkey
55: Also at Gaziosmanpasa University, Tokat, Turkey
56: Also at Ozyegin University, Istanbul, Turkey
57: Also at Izmir Institute of Technology, Izmir, Turkey
58: Also at Kafkas University, Kars, Turkey
59: Also at Istanbul University, Faculty of Science, Istanbul, Turkey
60: Also at Istanbul Bilgi University, Istanbul, Turkey
61: Also at Hacettepe University, Ankara, Turkey
62: Also at School of Physics and Astronomy, University of Southampton, Southampton, United Kingdom
63: Also at Institute for Particle Physics Phenomenology Durham University, Durham, United Kingdom
64: Also at Monash University, Faculty of Science, Clayton, Australia
65: Also at Bethel University, St. Paul, USA
66: Also at Karamanoğlu Mehmetbey University, Karaman, Turkey
67: Also at Vilnius University, Vilnius, Lithuania
68: Also at Bingol University, Bingol, Turkey
69: Also at Georgian Technical University, Tbilisi, Georgia
70: Also at Sinop University, Sinop, Turkey
71: Also at Mimar Sinan University, Istanbul, Istanbul, Turkey
72: Also at Texas A&M University at Qatar, Doha, Qatar
73: Also at Kyungpook National University, Daegu, Korea
74: Also at University of Hyderabad, Hyderabad, India